# The AFLOW Library of Crystallographic Prototypes

Michael J. Mehl[a], David Hicks[b,c], Cormac Toher[b,c], Ohad Levy[b,c,d],
Robert M. Hanson[e], Gus Hart[f], Stefano Curtarolo[b,g]

[a]*Center for Materials Physics and Technology, Code 6390, U.S. Naval Research Laboratory, Washington DC 20375*
[b] *Center for Materials Genomics, Duke University, Durham, NC 27708, USA*
[c]*Department of Mechanical Engineering and Materials Science, Duke University, Durham NC 27708*
[d]*Department of Physics, NRCN, P.O. Box 9001, Beer-Sheva 84190, Israel*
[e]*Department of Chemistry, St. Olaf College, Northfield, Minnesota 55057*
[f]*Department of Physics and Astronomy, Brigham Young University, Provo UT 84602*
[g]*Materials Science, Electrical Engineering, Physics and Chemistry, Duke University, Durham, North Carolina 27708*

**Abstract**

An easily available resource of common crystal structures is essential for researchers, teachers, and students. For many years this was provided by the U.S. Naval Research Laboratory's *Crystal Lattice Structures* web page, which contained nearly 300 crystal structures, including a majority of those which were given *Strukturbericht* designations. This article presents the updated version of the database, now including 288 standardized structures in 92 space groups. Similar to what was available on the web page before, we present a complete description of each structure, including the formulas for the primitive vectors, all of the basis vectors, and the AFLOW commands to generate the standardized cells. We also present a brief discussion of crystal systems, space groups, primitive and conventional lattices, Wyckoff positions, Pearson symbols and *Strukturbericht* designations.

*Email addresses:* `mmehl@.usna.edu` (Michael J. Mehl), `stefano@duke.edu` (Stefano Curtarolo)

## Table of Contents

**Prototypes**

*Hydrophilite, $\eta$-$Fe_2C$, and marcasite have the same AFLOW prototype label. They are generated by the same symmetry operations with different sets of parameters.

†$Co_2Si$, $HgCl_2$, and cotunnite have the same AFLOW prototype label. They are generated by the same symmetry operations with different sets of parameters.

‡GeS, MnP, FeB, and SnS have the same AFLOW prototype label. They are generated by the same symmetry operations with different sets of parameters.

§$\alpha$-Ga, black phosphorus, and molecular iodine have the same AFLOW prototype label. They are generated by the same symmetry operations with different sets of parameters.

[¶]In and $\alpha$-Pa have the same AFLOW prototype label. They are generated by the same symmetry operations with different sets of parameters.

[‖]$\alpha$-As, rhombohedral graphite, and $\beta$-O have the same AFLOW prototype label. They are generated by the same symmetry operations with different sets of parameters.

[**]$\beta$-Po and $\alpha$-Hg have the same AFLOW prototype label. They are generated by the same symmetry operations with different sets of parameters.

[††]Paraelectric $LiNbO_3$ and calcite have the same AFLOW prototype label. They are generated by the same symmetry operations with different sets of parameters.

[‡‡]$\alpha$-CO and FeSi have the same AFLOW prototype label. They are generated by the same symmetry operations with different sets of parameters.

## 1. Introduction

In 1913, W. H. and W. L. Bragg [1] determined the crystal structure of diamond by X-ray diffraction. This was followed by many other structural determinations, some using rather unique techniques [2]. Soon large amounts of structural data were being generated, beyond the ability of the average scientist to collect, much less critically evaluate and organize.

The first systematic attempt to organize this data was the *Strukturbericht* series [3], first edited by P. P. Ewald and continued by others into the middle of the Second World War. The *Strukturbericht* volumes gave each crystal structure a letter designation followed by a number. *A* designated single element structures (A1 ≡ Cu, A2 ≡ W, A3 ≡ Mg, *etc.*), *B* binary compounds AB (B1 ≡ NaCl, B2 ≡ CsCl, ...), *C* for binary $AB_2$ compounds (C1 ≡ $CaF_2$ fluorite, C2 ≡ $FeS_2$ pyrite, ...), *L* for alloy related structures, and so on. The *Strukturbericht* designation did not apply to a single compound, *e.g.* all single element face-centered crystals were designated A1, all salts similar to sodium chloride were listed as B1, and so on. Although these designations are still in use today, they quickly became unwieldy, requiring both numerical ($L1_2$ ≡ $Cu_3Au$) [4] and alphabetic ($D5_a$ ≡ $Si_2U_3$) [512] subscripts. Following the war, *Strukturbericht's* designations were dropped by its successor, the International Union of Crystallography's *Structure Reports* [6].

*Strukturbericht* was not the only compilation of crystal structures. In 1924, R. W. G. Wyckoff published the first edition of *The Structure of Crystals*, which eventually became a six volume set [7] describing hundreds of different crystals, organizing structures by prototype, *e.g.* ZnS for compounds with the zincblende/sphalerite structure [8].

Post-war, in 1958 W. B. Pearson published the first edition [9] of what is now known as *Pearson's Handbook*, a collection of crystallographic data for metals and intermetallic alloys. This classified the structures by prototype compound, space group, Pearson symbol, and, when available, *Strukturbericht* designation. This monumental work was updated by Villars and Calvert, whose second edition [10] contains more than 50,000 entries. In the intervening years, Pearson published *The Crystal Chemistry and Physics of Metals and Alloys* [11], which described a variety of crystal structures, categorizing them by physical and geometrical considerations.

After the turn of the century, the wide availability of computer storage and high speed internet connections made large electronic databases possible. The Inorganic Crystal Structure Database (ICSD), while not strictly organized by prototypes, is a useful and well-known online materials database that contains structural data for over 185,000 materials [12]. 2003 saw the first publication of *The American Mineralogist Crystal Structure Database* (AMCD) [13], which lists crystallographic data for well over two thousand minerals, often with multiple entries, drawn from *The American Mineralogist*, *Acta Crystallographica*, and other sources, including the compilations listed above. At the beginning of this decade, Pierre Villars and others made an enormous collection of data known as the Pauling File [14] available through the Springer Materials website. The crystallographic part of this database is essentially an extension of *Pearson's Handbook*, going well beyond metals and intermetallic alloys. More recently in 2007, a website *The Structure of Materials* [15] provided data for approximately 100 different structure types.

On a much more modest level, in 1995 one of us, with help from summer student R. Benjamin Young, first made a web page called *Crystal Lattice Structures* available on the World Wide Web. While it contained information that could also be found in the above sources, it provided information that was useful to researchers unfamiliar with crystallographic conventions. For example, the AMCD lists the structure of fluorite ($CaF_2$) as

```
5.4631 5.4631 5.4631 90 90 90 Fm-3m
atom x y z
Ca 0 0 0
F .25 .25 .25
```

Those familiar with crystallographic conventions would immediately recognize that the primitive unit cell of this system was face-centered cubic (from the space group label, $Fm\overline{3}m$), with cubic lattice constant 5.4631Å. A calcium atom is at the origin, and a fluorine atom is at the position $(1/4a, 1/4a, 1/4a)$. The researcher would then go to the *International Tables for Crystallography* [16] or the online Bilbao Crystallographic Server [17, 18, 19] to determine the complete set of atomic positions.

On the other hand, the *Crystal Lattice Structures* page included all of the above information, but also explicitly showed the primitive vectors of the face-centered cubic unit cell,

$$\begin{aligned} \mathbf{a}_1 &= \frac{a}{2}\,\hat{\mathbf{y}} + \frac{a}{2}\,\hat{\mathbf{z}} \\ \mathbf{a}_2 &= \frac{a}{2}\,\hat{\mathbf{x}} + \frac{a}{2}\,\hat{\mathbf{z}} \\ \mathbf{a}_3 &= \frac{a}{2}\,\hat{\mathbf{x}} + \frac{a}{2}\,\hat{\mathbf{y}}, \end{aligned}$$

(where $\hat{\mathbf{x}}, \hat{\mathbf{y}}$ and $\hat{\mathbf{z}}$ are the Cartesian unit vectors), as well as the atomic positions of all of the atoms, in terms of both the primitive lattice and Cartesian vectors:

$$\begin{aligned} \mathbf{B}_1 &= 0\,\mathbf{a}_1 + 0\,\mathbf{a}_2 + 0\,\mathbf{a}_3 \\ &= 0\,\hat{\mathbf{x}} + 0\,\hat{\mathbf{y}} + 0\,\hat{\mathbf{z}} \qquad (4a) \qquad \text{Ca} \\ \mathbf{B}_2 &= \frac{1}{4}\,\mathbf{a}_1 + \frac{1}{4}\,\mathbf{a}_2 + \frac{1}{4}\,\mathbf{a}_3 \\ &= \frac{a}{4}\,\hat{\mathbf{x}} + \frac{a}{4}\,\hat{\mathbf{y}} + \frac{a}{4}\,\hat{\mathbf{z}} \qquad (8c) \qquad \text{F} \\ \mathbf{B}_3 &= \frac{3}{4}\,\mathbf{a}_1 + \frac{3}{4}\,\mathbf{a}_2 + \frac{3}{4}\,\mathbf{a}_3 \\ &= \frac{3a}{4}\,\hat{\mathbf{x}} + \frac{3a}{4}\,\hat{\mathbf{y}} + \frac{3a}{4}\,\hat{\mathbf{z}}. \qquad (8c) \qquad \text{F} \end{aligned}$$

This shows all of the atoms in the primitive unit cell of the system. The web page also offered views of the system from several angles, and in later iterations allowed the user to rotate the unit cell to see the crystal from any angle. Finally, all of the structures in the database were classified by type (face- or body-centered cubic, hexagonal close-packed, $sp^3$ bonding, *etc.*), Pearson Symbol, and space group. These features made the *Crystal Lattice Structures* page very popular with students and researchers.

While popular, the web site was never properly supported. It grew in a haphazard and piecemeal fashion, so that the format of one page might be different from another page. While many pages listed the original reference for a structure, many others did not. For these and other reasons the web site was removed for redesign in 2010.

Recent advances in computational materials science present novel opportunities for structure discovery and optimization, including uncovering of unsuspected compounds and metastable structures, electronic structure, surface, and nano-particle properties. These opportunities largely depend on the ability to apply modern high-throughput computational methods to analyze the properties of large data sets of structures and requires systematic generation and classification of the relevant computational data by high-throughput methods [20] and data repositories, such as AFLOW [21, 22, 23, 24], Materials Project [25], OQMD [26], NoMaD [27], and AiiDA [28]. It has become imperative to make the data from these structure databases more accessible to the growing community of computational materials scientists. Such exposure should provide an easy route to use the crystallographic data included in these compilations in advanced software frameworks, such as AFLOW [21, 22, 23, 24], for high-throughput calculation of crystal structure properties of alloys, intermetallics and inorganic compounds. These frameworks decorate structural prototypes, often sourced from databases such as those described in this work, with different species to perform automated high-throughput materials discovery and characterization. This synergy would provide materials scientists with a powerful tool for efficient quantum computational materials discovery and characterization [29, 30].

This article, then, describes a new version of the database, designed with this synergy in mind, and renamed as *The AFLOW Library of Crystallographic Prototypes*. The web version of this database will be located at `http://aflow.org/CrystalDatabase`.

The format of this article is as follows: Section 2 discusses the basics of three dimensional periodic systems. Section 3 discusses the seven crystal systems and fourteen Bravais lattices that can exist in three dimensions as well as the definition of space groups, and gives our standard representation of the primitive vectors of each lattice. Section 11 shows how to take Wyckoff positions from the International Space Group tables and transform them into lattice coordinates for a given crystal system. Section 12 explains the format of the pages of the database, the online Crystallographic Information File (CIF) [31] for the structure, and the online POSCAR[1] file that summarizes the structural information.

## 2. Periodic Three-Dimensional Systems

In this section we give a brief review of the mathematics of three-dimensional periodic systems, describing the notation used in the database. Expanded descriptions of this topic, with an emphasis on condensed matter physics, can be found in Lax [33], Ashcroft and Mermin [34], Barrett and Massalski [35], and the various editions of Kittel [36].

In condensed matter physics, a crystal structure is a periodic system characterized by three primitive lattice vectors, $(\mathbf{a}_1, \mathbf{a}_2, \mathbf{a}_3)$. These vectors must not be co-planar, so that $\mathbf{a}_1 \cdot (\mathbf{a}_2 \times \mathbf{a}_3) \neq 0$. If the atomic nuclei are located at basis vectors $\mathbf{B}_i$, then their periodic replicas are located at

$$\mathbf{B}_i + N_1\,\mathbf{a}_1 + N_2\,\mathbf{a}_2 + N_3\,\mathbf{a}_3 \tag{1}$$

for all positive and negative integers $(N_1, N_2, N_3)$. Furthermore, everything in the crystal is periodic, including continuous functions such as the electronic density $\rho(\mathbf{r})$:

$$\rho\,(\mathbf{r}) = \rho\,(\mathbf{r} + N_1\,\mathbf{a}_1 + N_2\,\mathbf{a}_2 + N_3\,\mathbf{a}_3)\,. \tag{2}$$

A unit cell is a volume which, when translated through by all vectors of the form $N_1\,\mathbf{a}_1 + N_2\,\mathbf{a}_2 + N_3\mathbf{a}_3$ (the set points known as the Bravais lattice), fills all space. One possible unit cell is the set of all points [34]

$$\mathbf{r} = u_1\,\mathbf{a}_1 + u_2\,\mathbf{a}_2 + u_3\,\mathbf{a}_3 \tag{3}$$

such that $0 \leq u_i < 1$. This is obviously not unique, *e.g.* we could pick the interval $-1/2 < u_i \leq 1/2$. Indeed, the choice of primitive vectors is not unique. For any choice of $\mathbf{a}_I$, we can find alternative vectors $\mathbf{a}'_i$ which are related to the original primitive vectors by

$$\begin{pmatrix} \mathbf{a}'_1 \\ \mathbf{a}'_2 \\ \mathbf{a}'_3 \end{pmatrix} = \begin{pmatrix} n_{11} & n_{12} & n_{13} \\ n_{21} & n_{22} & n_{23} \\ n_{31} & n_{32} & n_{33} \end{pmatrix} \cdot \begin{pmatrix} \mathbf{a}_1 \\ \mathbf{a}_2 \\ \mathbf{a}_3 \end{pmatrix}, \tag{4}$$

where the $n_{ij}$ are integers. This will produce an identical Bravais lattice provided the determinant of the $n$ matrix is plus or minus unity:

$$\begin{vmatrix} n_{11} & n_{12} & n_{13} \\ n_{21} & n_{22} & n_{23} \\ n_{31} & n_{32} & n_{33} \end{vmatrix} = \pm 1. \tag{5}$$

No matter what choice we make, the volume of the unit cell is given by

$$V = \mathbf{a}_1 \cdot (\mathbf{a}_2 \times \mathbf{a}_3)\,. \tag{6}$$

[1] A POSCAR file is used to describe the primitive vectors and atomic positions in the *Vienna Ab-Initio Simulation Package* (VASP) [32].

We will always assume what is known as a right-handed coordinate system, so that $V > 0$.

Since the periodic replicas of a unit cell fill all space, any point $\mathbf{r}$ in space may be defined by its Cartesian coordinates

$$\mathbf{r} = x_1\,\hat{\mathbf{x}} + x_2\,\hat{\mathbf{y}} + x_3\,\hat{\mathbf{z}}, \tag{7}$$

where $(\hat{\mathbf{x}}, \hat{\mathbf{y}}, \hat{\mathbf{z}})$ are orthogonal vectors with unit length, or by its lattice coordinates,

$$\mathbf{r} = u_1\,\mathbf{a}_1 + u_2\,\mathbf{a}_2 + u_3\,\mathbf{a}_3. \tag{8}$$

Lattice coordinates are often called fractional coordinates. The transformation between the Cartesian coordinates ($x_1$, $y_1$, $z_1$) and lattice coordinates ($u_1$, $u_2$, $u_3$) is most easily accomplished by defining a periodic reciprocal lattice, defined by vectors ($\mathbf{b}_1$, $\mathbf{b}_2$, $\mathbf{b}_3$), which are chosen so that

$$\mathbf{a}_i \cdot \mathbf{b}_j = 2\pi\,\delta_{ij}, \tag{9}$$

where $\delta_{ij}$ is the Kronecker[2] $\delta$. The reciprocal lattice vectors can be determined from the primitive lattice vectors via the formula

$$\mathbf{b}_i = \left(\frac{2\pi}{V}\right)\mathbf{a}_j \times \mathbf{a}_k, \tag{10}$$

where $(ijk) = (123), (231)$, or $(312)$. Given this definition, we see that we can obtain the lattice coordinates for any vector $\mathbf{r}$ via

$$u_i = \frac{\mathbf{b}_i \cdot \mathbf{r}}{2\pi}. \tag{11}$$

This gives us a general procedure for translating between Cartesian and lattice coordinates: given the lattice coordinates of a point, we can find the Cartesian coordinates via (7). Given the Cartesian coordinates, we can find the lattice coordinates via (11).

It should also be noted that any function exhibiting the symmetry of the crystal can be written as a periodic function of the lattice coordinates with a period of 1 in each direction. In other words, in lattice coordinates equation (2) becomes

$$\rho\,(u_1, u_2, u_3) = \rho\,(u_1 + N_1, u_2 + N_2, u_3 + N_3)\,, \tag{12}$$

where $(u_1, u_2, u_3)$ are given by (11). This means that we know everything about the crystal if we can describe it for values of $u_i$ in the range $[\alpha_i, \alpha_i + 1]$, where $\alpha_i$ can be chosen arbitrarily. Usual values are 0 or 1/2, so that the range is $[0, 1]$ or $[-1/2, 1/2]$.

Lastly, a rotation of a crystal about any point does not change its physical properties, only its orientation in space. We could, for example replace the unit vectors $(\hat{\mathbf{x}}, \hat{\mathbf{y}}, \hat{\mathbf{z}})$ in (7) by any other three orthogonal unit vectors and still describe the same crystal. For that reason it is useful to define the lattice by the lengths of its primitive vectors and the angles between them. This requires six values conventionally chosen as $(a, b, c)$ to describe the lengths and $(\alpha, \beta, \gamma)$ to describe the angles, where

$$\begin{aligned} a &= |\mathbf{a}_1| \\ b &= |\mathbf{a}_2| \\ c &= |\mathbf{a}_3|, \end{aligned} \tag{13}$$

and the cosines of the angles are given by

$$\begin{aligned} \cos\alpha &= \frac{\mathbf{a}_2 \cdot \mathbf{a}_3}{b\,c} \\ \cos\beta &= \frac{\mathbf{a}_3 \cdot \mathbf{a}_1}{c\,a} \\ \cos\gamma &= \frac{\mathbf{a}_1 \cdot \mathbf{a}_2}{a\,b}. \end{aligned} \tag{14}$$

The primitive cell of a crystal is uniquely specified by these values (up to an arbitrary rotation), and crystallographic articles report the structures in this form.

## 3. Crystal Systems, Lattices, Space Groups and Standard Lattice Vectors

Having defined what we mean by a lattice, we now discuss the possible lattices that can exist in a three dimensional space, and some of their properties. Here we define our terms following Lax [37], paraphrasing his discussion.

1. A *Crystal* is a periodic array of physical objects. In this article we discuss crystals made of periodic arrays of atoms and their associated electrons.
2. A *Crystal Structure* is the complete description of the crystal including its periodic structure and the contents of the unit cell. In our case, we obtain a complete description of the crystal by specifying the primitive vectors of the periodic lattice $\mathbf{a}_i, (i = 1, 2, 3)$ and the positions $\mathbf{B}_j, (j = 1, 2, 3, \ldots, N)$ of the $N$ atoms in a unit cell. The ground state electronic charge density, if desired, can then be computed from these atomic positions.
3. A *Space Group* is the set of all operations (translations, rotations, and reflections) that restore a crystal to itself. In three dimensional space there are 230 space groups.
4. A *Crystal Class* is the point group of the crystal. This includes all possible rotations and reflections (but not translations) that leave the shape of the crystal unchanged. This does not mean that the crystal is transformed into itself, only the point group. In three dimensions there are 32 crystal classes.

[2] The physical and mathematical motivation for the definition of the reciprocal lattice, including the $2\pi$ factor, is beyond the scope of this article. More information can be found in any of the texts given at the beginning of this section. Here we simply use the reciprocal lattice to translate between Cartesian and lattice coordinates.

5. A *Bravais Lattice* is a collection of points

$$\{t_1\,\mathbf{a}_1 + t_2\,\mathbf{a}_2 + t_3\,\mathbf{a}_3\}\,, \tag{15}$$

where the $t_i$ are integers and the $\mathbf{a}_i$ are not co-planar, *i.e.* the volume (6) is non-zero. Equation (4) allows some freedom in the choice of $\mathbf{a}_i$, but all choices lead to the same points for a given Bravais lattice. In three dimensions a given crystal class has at least one and a maximum of four Bravais lattices.

6. The *holohedry* of a Bravais lattice is the point group that describes its rotational symmetry.

7. A *Crystal System* is the set of all Bravais lattices that have the same holohedry. In three dimensions there are seven crystal systems, many of which contain multiple Bravais lattices.

In 1891 E. S. Federov [38] and A. Schönflies [39] determined the 230 space groups allowed in three dimensions. Wyckoff [40] tabulated all of these groups, and determined the special atomic coordinates (the Wyckoff positions) allowed for each space group. Here we briefly describe the properties of each crystal system and its associated Bravais lattices, and list the space groups associated with each lattice. In general we will start with the lowest symmetry and go to increasingly higher symmetries. Each space group will be labeled by the International symbol associated with its standard orientation as defined in the International Tables [16]. Alternative orientations of the space groups will lead to different labels. Cockcroft [41] has a complete list of these online.

In the following, we will frequently refer to conventional lattices and primitive, or Bravais, lattices. The basic definition of a Bravais lattice is that it describes the periodicity of a particular system. A conventional lattice, on the other hand, describes the holohedry of all of the Bravais lattices in a given crystal system. Each crystal system has a Bravais lattice that is identical with the conventional lattice.

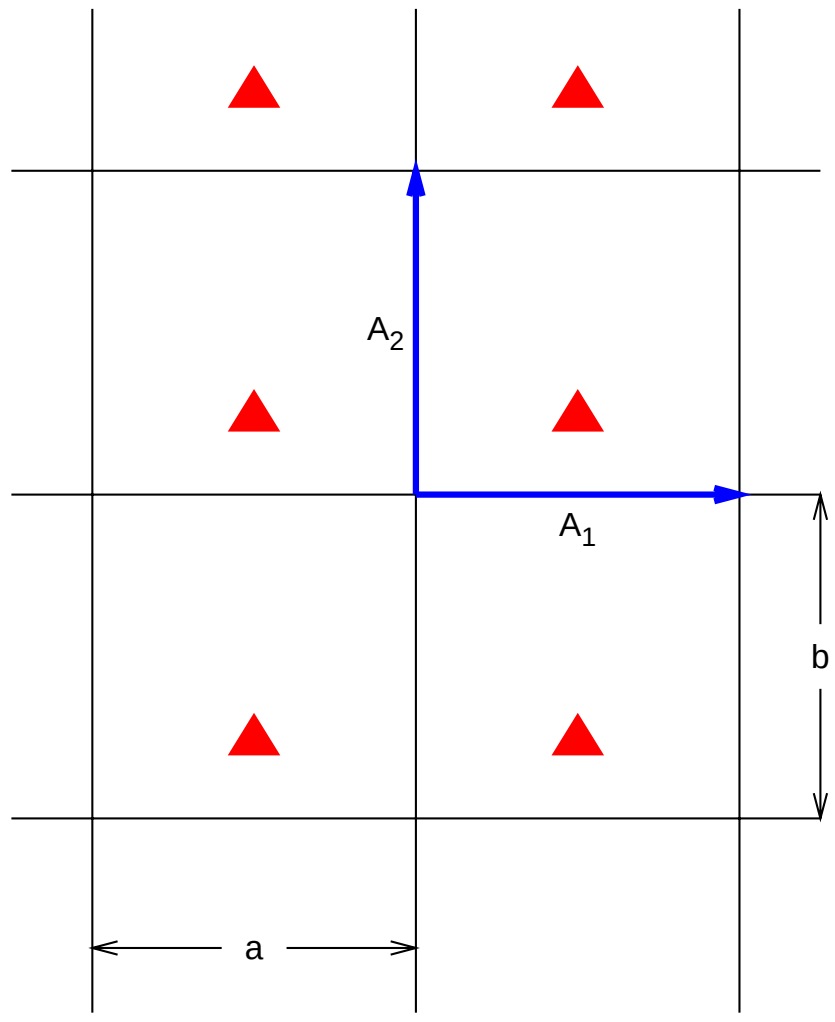

Figure 1: A two-dimensional rectangular system, with primitive vectors $\mathbf{A}_1$ and $\mathbf{A}_2$ given by (16). The solid lines denote the edges of the unit cell.

This is most easily seen in two dimensions. Figure 1 shows a rectangular periodic structure. The periodicity can be described by the primitive vectors

$$\begin{aligned} \mathbf{A}_1 &= a\,\hat{\mathbf{x}} \\ \mathbf{A}_2 &= b\,\hat{\mathbf{y}}. \end{aligned} \tag{16}$$

The solid lines mark the edges of the unit cell for this system.

Next consider the structure shown in Figure 2. It can obviously be described as a periodic structure with primitive vectors (16) and a unit cell bounded by the solid lines. However, it can also be described by the primitive vectors

$$\begin{aligned} \mathbf{a}_1 &= \frac{a}{2}\,\hat{\mathbf{x}} - \frac{b}{2}\,\hat{\mathbf{y}} \\ \mathbf{a}_2 &= \frac{a}{2}\,\hat{\mathbf{x}} + \frac{b}{2}\,\hat{\mathbf{y}}. \end{aligned} \tag{17}$$

These primitive vectors are shown in Figure 2, and the unit cells associated with these vectors are bounded by the dashed lines.

Both of these structures have the same holohedry, belonging to the two-dimensional rectangular crystal system. They have different Bravais lattices. We can call these lattices *simple* rectangular, shown in Figure 1, and *centered* rectangular, shown in Figure 2.[3]

[3] This is called a centered lattice because the primitive vectors (17) point to the center of the rectangular unit cell.

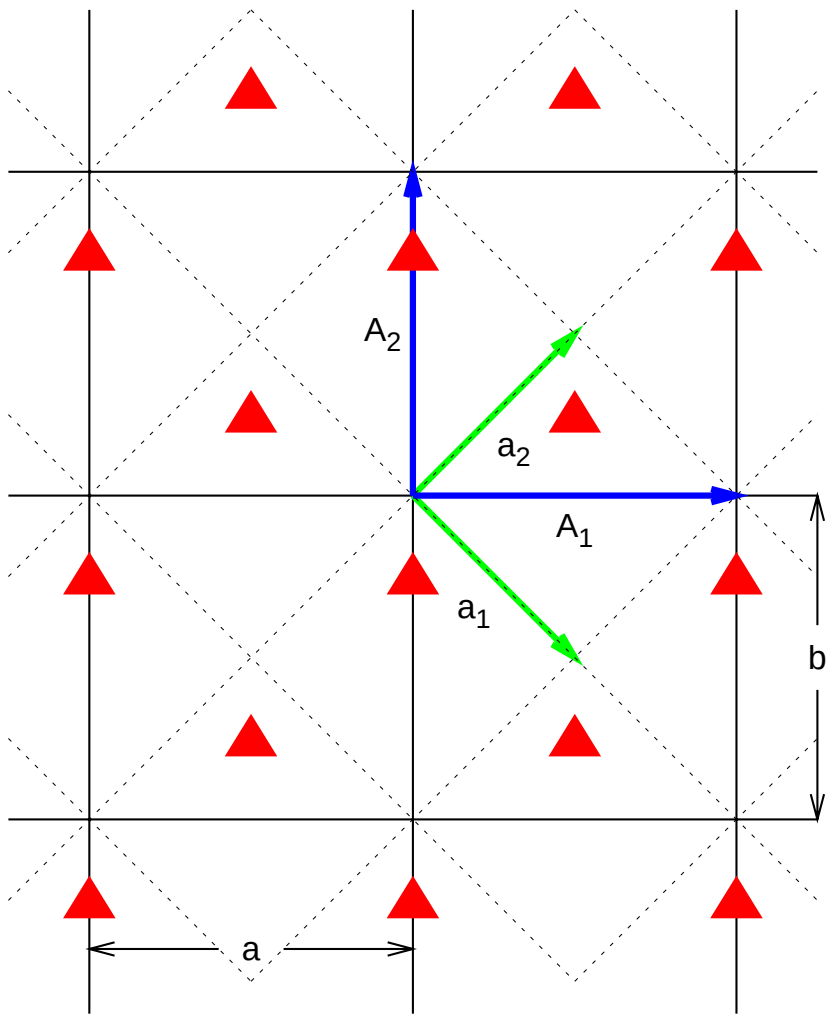

Figure 2: A two-dimensional centered rectangular system, with primitive vectors $\mathbf{a}_1$ and $\mathbf{a}_2$ given by 17. The dashed lines denote the edges of the primitive unit cell, while the solid lines denote the edges of the conventional unit cell.

Both the holohedry and translational symmetry of these structures can be described by the vectors (16), although the centered cell has another translation. Equations (16) then define the *conventional* unit cell for the two-dimensional rectangular crystal system.

The area of the conventional Bravais lattice, *ab*, is twice that of the centered Bravais lattice, as can be seen from Figure 2, which also shows that the conventional lattice has twice as many triangles (atoms) per unit cell as the Bravais lattice. In three dimensions, as we will see, the conventional lattice can hold one, two, three or four times as many atoms as the underlying Bravais lattice.

Going back to three dimensions, standard crystallographic practice is to report the lattice parameters ($a$, $b$, $c$, $\alpha, \beta, \gamma$) of (13-14) using the conventional lattice, rather than the Bravais lattice. While this may seem arbitrary, the primitive vectors

$$\begin{aligned} \mathbf{a}_1 &= a\,\hat{\mathbf{x}} \\ \mathbf{a}_2 &= \frac{a}{2}\,\hat{\mathbf{x}} + \frac{b}{2}\,\hat{\mathbf{y}} \end{aligned}$$

describe Figure 2 just as well as (17), but have different lengths and angles, and there are a multitude of other possible sets. There is, however, only one logical way to describe the conventional cell, (16). This happens in three dimensions as well. As a general rule,[4] ($a$, $b$, $c$, $\alpha$, $\beta$, $\gamma$), and even the number of atoms in a unit cell, are given for the conventional lattice. The size of the primitive cell has to be inferred from knowledge of the space group.

We now consider the seven crystal systems, including the Bravais lattice, and the space groups associated with each Bravais lattice.

[4]The one exception to this rule is the rhombohedral lattice, which we shall discuss below.

As noted above, there are an infinite number of choices for a set of primitive vectors describing a unit cell. In general we follow the choices made by Setyawan and Curtarolo [42]. Differences occur in the monoclinic, base-centered orthorhombic, and rhombohedral lattices, and are discussed in the footnotes.

## 4. The Triclinic Crystal System

The triclinic is the most general crystal system. All of the other crystal systems can be considered special cases of the triclinic. The primitive vectors are also completely general: their lengths ($a$, $b$, $c$) and angles ($\alpha, \beta, \gamma$) may have arbitrary values. The triclinic system has one Bravais lattice, which is also the conventional lattice for this system.

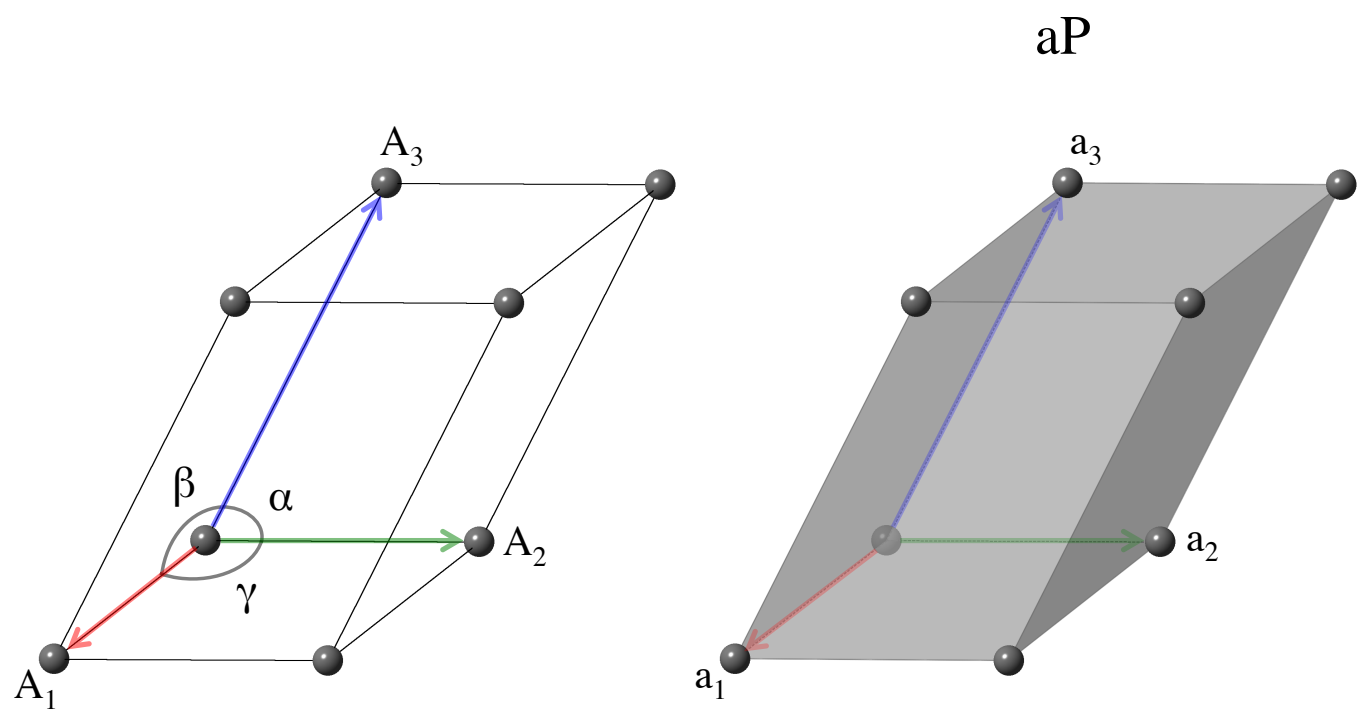

Figure 3: The conventional and simple unit cells for the triclinic crystal system.

### 4.1. Lattice 1: Triclinic

There are many choices for the primitive vectors in the triclinic system. We make the choice

$$\begin{aligned} \mathbf{a}_1 &= a\,\hat{\mathbf{x}} \\ \mathbf{a}_2 &= b\,\cos\gamma\,\hat{\mathbf{x}} + b\,\sin\gamma\,\hat{\mathbf{y}} \\ \mathbf{a}_3 &= c_x\,\hat{\mathbf{x}} + c_y\,\hat{\mathbf{y}} + c_z\,\hat{\mathbf{z}}, \end{aligned} \tag{18}$$

where

$$\begin{aligned} c_x &= c\,\cos\beta \\ c_y &= \frac{c\,(\cos\alpha - \cos\beta\,\cos\gamma)}{\sin\gamma} \end{aligned}$$

and

$$c_z = \sqrt{c^2 - c_x^2 - c_y^2}.$$

The volume of the triclinic unit cell is

$$V = a\,b\,c_z\,\sin\gamma. \tag{19}$$

The space groups associated with the triclinic lattice are given in Table 1.

Table 1: The space groups associated with the triclinic Bravais lattice (18) are

| | |
|---|---|
| 1. $P1$ | 2. $P\bar{1}$ |

## 5. The Monoclinic Crystal System

In the monoclinic crystal system, the conventional unit cell is defined by primitive vectors of arbitrary length, where one of the vectors is perpendicular to the other two. Modern convention chooses this vector to be the one with length $b$ (or "unique axis $b$" in the literature), so that $\alpha = \gamma = \pi/2$ and $\beta \neq \pi/2$.[5]

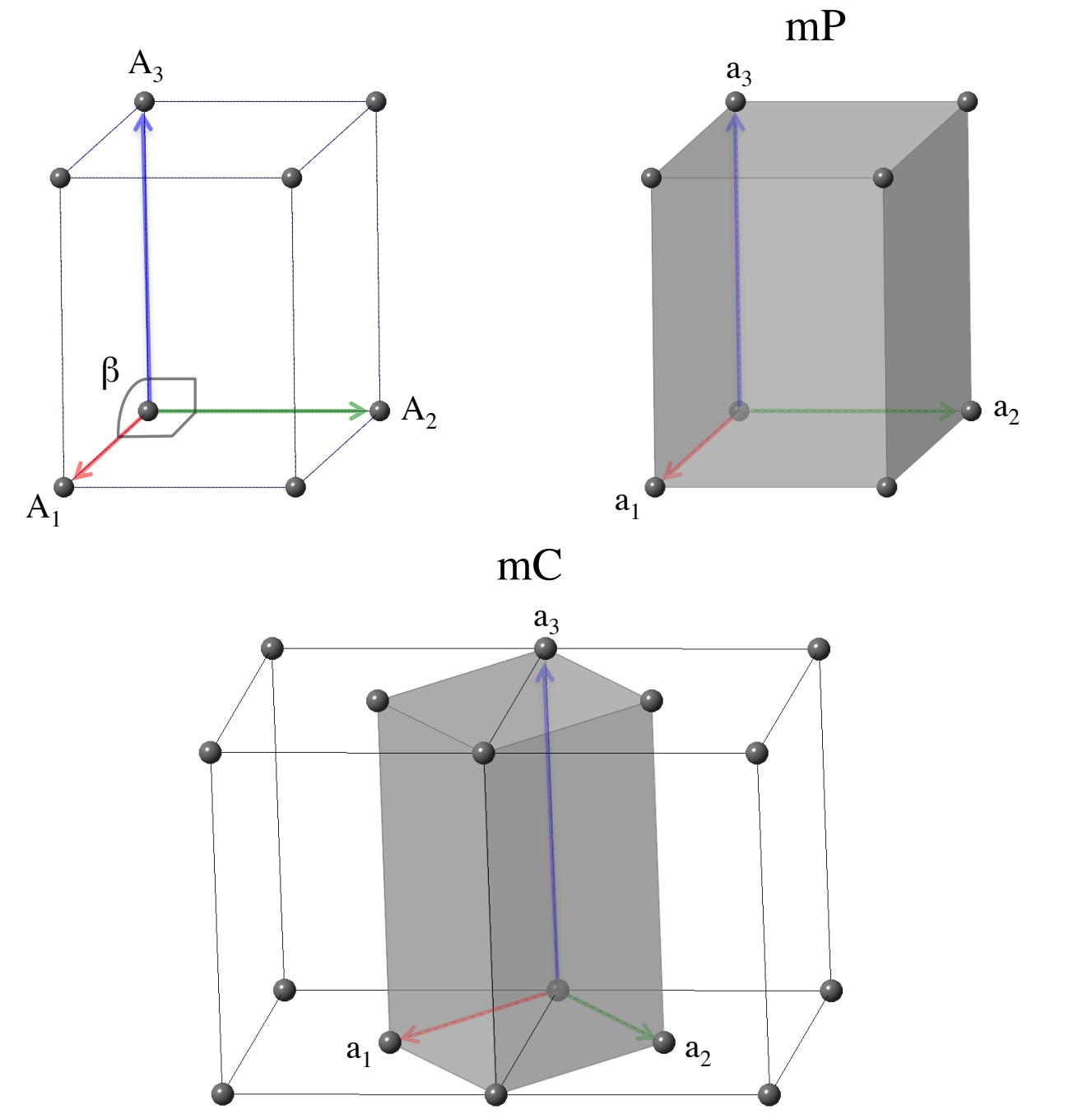

Figure 4: The conventional, simple, and base-centered unit cells for the monoclinic crystal system.

The conventional unit cell can be described by the vectors

$$
\begin{aligned}
\mathbf{A}_1 &= a\,\hat{\mathbf{x}} \\
\mathbf{A}_2 &= b\,\hat{\mathbf{y}} \\
\mathbf{A}_3 &= c\,\cos\beta\,\hat{\mathbf{x}} + c\,\sin\beta\,\hat{\mathbf{z}},
\end{aligned} \tag{20}
$$

and the volume of a conventional unit cell is

$$V = a\,b\,c\,\sin\beta. \tag{21}$$

### 5.1. Lattice 2: Simple Monoclinic

The simple monoclinic cell is identical to the conventional cell. Its primitive vectors are identical to (20)

$$
\begin{aligned}
\mathbf{a}_1 &= a\,\hat{\mathbf{x}} \\
\mathbf{a}_2 &= b\,\hat{\mathbf{y}} \\
\mathbf{a}_3 &= c\,\cos\beta\,\hat{\mathbf{x}} + c\,\sin\beta\,\hat{\mathbf{z}},
\end{aligned} \tag{22}
$$

and the cell volume is just

$$V = a\,b\,c\,\sin\beta. \tag{23}$$

[5]Note that this orientation differs from that of Setyawan and Curtarolo [42], who used an unique axis "$a$" setting. Their angle $\alpha$ would be $\gamma$ in our notation.

The space groups associated with the simple monoclinic lattice are given in Table 2.

Table 2: The space groups associated with the simple monoclinic Bravais lattice (22) are

| | | |
|---|---|---|
| 3. $P2$ | 4. $P2_1$ | 6. $Pm$ |
| 7. $Pc$ | 10. $P2/m$ | 11. $P2_1/m$ |
| 13. $P2/c$ | 14. $P2_1/c$ | |

### 5.2. Lattice 3: Base-Centered Monoclinic

The base-centered monoclinic lattice is in the same crystal system as the monoclinic lattice, but its periodicity allows an additional translation in the plane defined by $\mathbf{a}_1$ and $\mathbf{a}_2$, much as in (17). The primitive vectors for the base-centered monoclinic lattice can be written

$$
\begin{aligned}
\mathbf{a}_1 &= \frac{a}{2}\,\hat{\mathbf{x}} - \frac{b}{2}\,\hat{\mathbf{y}} \\
\mathbf{a}_2 &= \frac{a}{2}\,\hat{\mathbf{x}} + \frac{b}{2}\,\hat{\mathbf{y}} \\
\mathbf{a}_3 &= c\,\cos\beta\,\hat{\mathbf{x}} + c\,\sin\beta\,\hat{\mathbf{z}}.
\end{aligned} \tag{24}
$$

The volume of the base-centered monoclinic unit cell is

$$V = \left(\frac{1}{2}\right) a\,b\,c\,\sin\beta, \tag{25}$$

half that of the conventional unit cell.

The space groups associated with the base-centered monoclinic lattice are given in Table 3. The labels for these space groups all begin with $C$, indicating the base-centered translation associated with these groups. This differs from the labels for space groups in Table 1 and Table 2, which begin with $P$, indicating that the primitive lattice is the conventional lattice.

The *International Tables* offer two representations of the base-centered monoclinic space groups, one for "unique axis $b$" and one for "unique axis $c$," where $\alpha \neq \pi/2$ and $\beta = \pi/2$. Space group 5 is then listed as "B2" or "C2" depending on this choice. Most authors ignore this distinction, as will we.

Table 3: The space groups associated with the base-centered monoclinic Bravais lattice (24) are

| | | |
|---|---|---|
| 5. $C2$ | 8. $Cm$ | 9. $Cc$ |
| 12. $C2/m$ | 15. $C2/c$ | |

## 6. The Orthorhombic Crystal System

In the orthorhombic system, the conventional unit cell is a parallelepiped, defined by three mutually orthogonal vectors of unequal length:

$$
\begin{aligned}
\mathbf{A}_1 &= a\,\hat{\mathbf{x}} \\
\mathbf{A}_2 &= b\,\hat{\mathbf{y}} \\
\mathbf{A}_3 &= c\,\hat{\mathbf{z}},
\end{aligned} \tag{26}
$$

so that $a \neq b \neq c$, but $\alpha = \beta = \gamma = \pi/2$. It is a limiting case of the conventional monoclinic crystal with $\beta \rightarrow \pi/2$. The volume of the conventional unit cell is

$$V = a\,b\,c. \tag{27}$$

There are four Bravais lattices in the orthorhombic system.

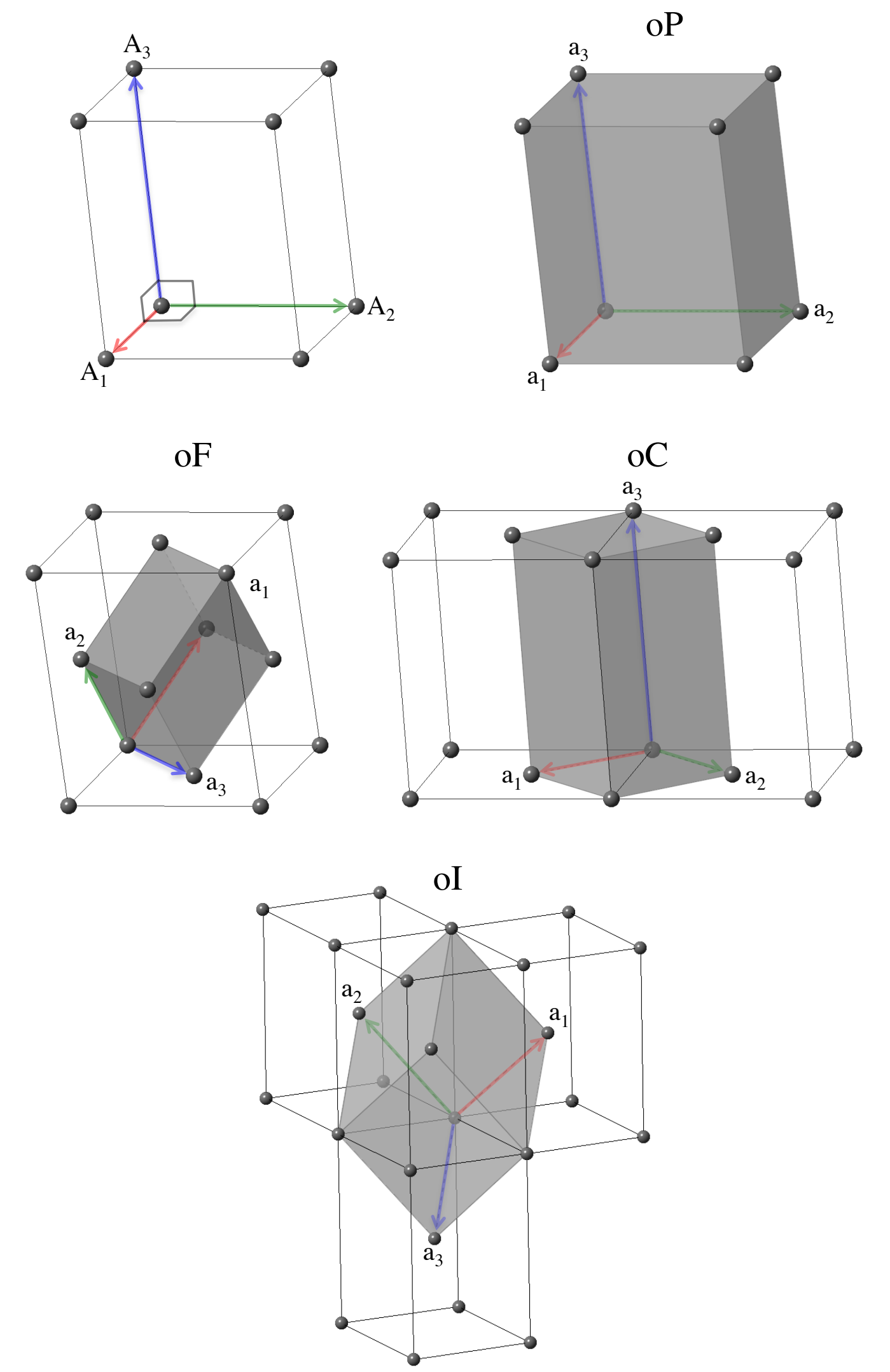

Figure 5: The conventional, simple, face-centered, base-centered, and body-centered unit cells for the orthorhombic crystal system.

### 6.1. *Lattice 4: Simple Orthorhombic*

The simple orthorhombic Bravais lattice is identical to the conventional cell

$$\begin{aligned} \mathbf{a}_1 &= a\,\hat{\mathbf{x}} \\ \mathbf{a}_2 &= b\,\hat{\mathbf{y}} \\ \mathbf{a}_3 &= c\,\hat{\mathbf{z}}, \end{aligned} \tag{28}$$

with volume

$$V = a\,b\,c\,. \tag{29}$$

The space groups associated with the simple orthorhombic lattice are given in Table 4.

Table 4: The space groups associated with the simple orthorhombic lattice (28) are

| | | |
|---|---|---|
| 16. $P222$ | 17. $P222_1$ | 18. $P2_12_12$ |
| 19. $P2_12_12_1$ | 25. $Pmm2$ | 26. $Pmc2_1$ |
| 27. $Pcc2$ | 28. $Pma2$ | 29. $Pca2_1$ |
| 30. $Pnc2$ | 31. $Pmn2_1$ | 32. $Pba2$ |
| 33. $Pna2_1$ | 34. $Pnn2$ | 47. $Pmmm$ |
| 48. $Pnnn$ | 49. $Pccm$ | 50. $Pban$ |
| 51. $Pmma$ | 52. $Pnna$ | 53. $Pmna$ |
| 54. $Pcca$ | 55. $Pbam$ | 56. $Pccn$ |
| 57. $Pbcm$ | 58. $Pnnm$ | 59. $Pmmn$ |
| 60. $Pbcn$ | 61. $Pbca$ | 62. $Pnma$ |

### 6.2. *Lattice 5: Base-Centered Orthorhombic*

Like the base-centered monoclinic lattice, the base-centered orthorhombic system allows a translation in one of the base planes. Unfortunately, the standard plane chosen depends on the space group, as shown in Table 5. Space groups beginning with C put the translation in the $a - b$ plane, that is, the plane defined by $\mathbf{A}_1$ and $\mathbf{A}_2$ (26). In this case the primitive vectors can be taken to be

$$\begin{aligned} \mathbf{a}_1 &= \frac{a}{2}\,\hat{\mathbf{x}} - \frac{b}{2}\,\hat{\mathbf{y}} \\ \mathbf{a}_2 &= \frac{a}{2}\,\hat{\mathbf{x}} + \frac{b}{2}\,\hat{\mathbf{y}} \\ \mathbf{a}_3 &= c\,\hat{\mathbf{z}}\,. \end{aligned} \tag{30}$$

Space groups beginning with A put the translation in the $b - c$ plane, defined by $\mathbf{A}_2$ and $\mathbf{A}_3$. We use the primitive vectors[6]

$$\begin{aligned} \mathbf{a}_1 &= a\,\hat{\mathbf{x}} \\ \mathbf{a}_2 &= \frac{b}{2}\,\hat{\mathbf{y}} - \frac{c}{2}\,\hat{\mathbf{z}} \\ \mathbf{a}_3 &= \frac{b}{2}\,\hat{\mathbf{y}} + \frac{c}{2}\,\hat{\mathbf{z}}. \end{aligned} \tag{31}$$

In both cases the volume of the primitive unit cell is

$$V = \frac{a\,b\,c}{2}. \tag{32}$$

There are two primitive base-centered orthorhombic unit cells in the conventional orthorhombic unit cell.

[6]Orientation (31) is not used by Setyawan and Curtarolo [42], who only considered centering in the "C" plane defined by $\mathbf{a}_2$ and $\mathbf{a}_3$. A simple rotation brings the vectors into agreement.

Table 5: The space groups associated with the base-centered orthorhombic lattice. Space groups beginning with C place the base-translation in the $a-b$ plane and use primitive vectors (30), while space groups beginning with A put the translation in the $b-c$ plane and use the primitive vectors (31).

| | | |
|---|---|---|
| 20. $C222_1$ | 21. $C222$ | 35. $Cmm2$ |
| 36. $Cmc2_1$ | 37. $Ccc2$ | 38. $Amm2$ |
| 39. $Abm2$ | 40. $Ama2$ | 41. $Aba2$ |
| 63. $Cmcm$ | 64. $Cmca$ | 65. $Cmmm$ |
| 66. $Cccm$ | 67. $Cmma$ | 68. $Ccca$ |

### 6.3. Lattice 6: Body-Centered Orthorhombic

The body-centered orthorhombic lattice has the same point group and translational symmetry as the simple orthorhombic system, with the addition of a translation to the center of the parallelepiped defined by the vectors (26). Our standard form of the primitive vectors is

$$\begin{aligned} \mathbf{a}_1 &= -\frac{a}{2}\,\hat{\mathbf{x}} + \frac{b}{2}\,\hat{\mathbf{y}} + \frac{c}{2}\,\hat{\mathbf{z}} \\ \mathbf{a}_2 &= \frac{a}{2}\,\hat{\mathbf{x}} - \frac{b}{2}\,\hat{\mathbf{y}} + \frac{c}{2}\,\hat{\mathbf{z}} \\ \mathbf{a}_3 &= \frac{a}{2}\,\hat{\mathbf{x}} + \frac{b}{2}\,\hat{\mathbf{y}} - \frac{c}{2}\,\hat{\mathbf{z}}. \end{aligned} \tag{33}$$

The volume of the primitive body-centered orthorhombic unit cell is

$$V = \frac{a\,b\,c}{2}. \tag{34}$$

There are two primitive body-centered orthorhombic unit cells in the conventional orthorhombic unit cell. The space groups associated with this lattice, all of which begin with $I$ in standard notation, are given in Table 6.

Table 6: The space groups associated with the body-centered orthorhombic lattice (33).

| | | |
|---|---|---|
| 23. $I222$ | 24. $I2_12_12_1$ | 44. $Imm2$ |
| 45. $Iba2$ | 46. $Ima2$ | 71. $Immm$ |
| 72. $Ibam$ | 73. $Ibca$ | 74. $Imma$ |

### 6.4. Lattice 7: Face-Centered Orthorhombic

While the base-centered monoclinic lattice allows translations to one base plane, the face-centered orthorhombic lattice allows translations to any of the base planes. Our standard choice for the primitive vectors of this system are given by

$$\mathbf{a}_1 = \frac{b}{2}\,\hat{\mathbf{y}} + \frac{c}{2}\,\hat{\mathbf{z}} \tag{35}$$

$$\mathbf{a}_2 = \frac{a}{2}\,\hat{\mathbf{x}} + \frac{c}{2}\,\hat{\mathbf{z}} \tag{36}$$

$$\mathbf{a}_3 = \frac{a}{2}\,\hat{\mathbf{x}} + \frac{b}{2}\,\hat{\mathbf{y}}. \tag{37}$$

The volume of the primitive face-centered orthorhombic unit cell is

$$V = \frac{a\,b\,c}{4}, \tag{38}$$

so that there are four primitive body-centered orthorhombic unit cells in the conventional orthorhombic unit cell. The space groups associated with this lattice, all of which begin with $F$ in standard notation, are given in Table 7.

Table 7: The space groups associated with the face-centered orthorhombic lattice (37).

| | | |
|---|---|---|
| 22. $F222$ | 42. $Fmm2$ | 43. $Fdd2$ |
| 69. $Fmmm$ | 70. $Fddd$ | |

## 7. The Tetragonal Crystal System

In the tetragonal system, like the orthorhombic, the conventional unit cell is a parallelepiped, but two sides are equal, so that $a = b$ and $c \neq a$, while $\alpha = \beta = \gamma = \pi/2$, and this is a special case of the orthorhombic system. The primitive vectors of the conventional unit cell are

$$\begin{aligned} \mathbf{A}_1 &= a\,\hat{\mathbf{x}} \\ \mathbf{A}_2 &= a\,\hat{\mathbf{y}} \\ \mathbf{A}_3 &= c\,\hat{\mathbf{z}}. \end{aligned} \tag{39}$$

The volume of the conventional unit cell is

$$V = a^2\,c. \tag{40}$$

Given the similarity between the tetragonal and orthorhombic crystal system, we might expect that the tetragonal system would have four Bravais lattices as well, but the additional symmetry generated because $b = a$ reduces this to two. When $b \rightarrow a$, the base-centered orthorhombic Bravais lattice (30) becomes a simple tetragonal lattice, while the face-centered orthorhombic lattice (37) can be shown to be identical to a body-centered tetragonal cell [43].

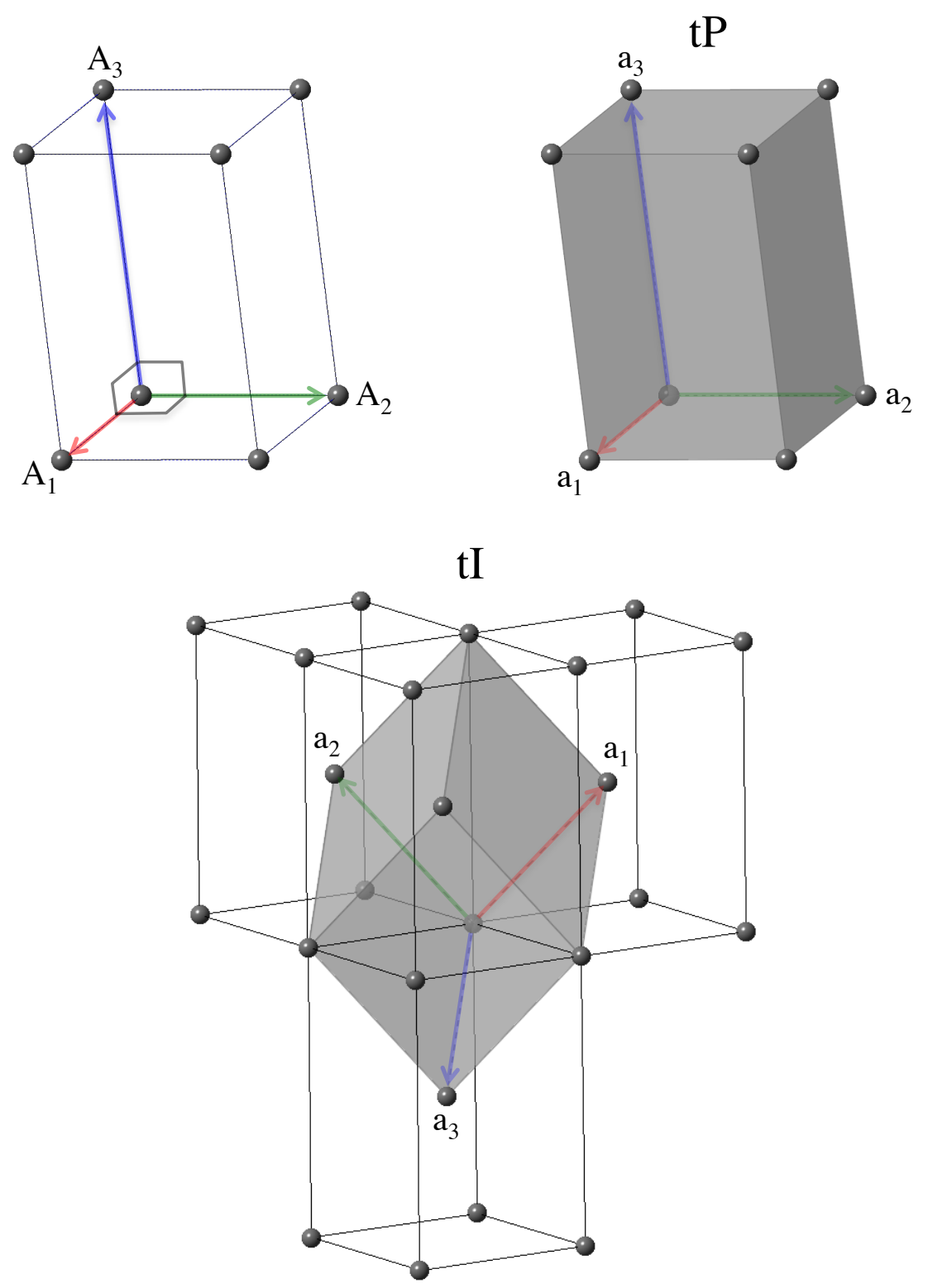

Figure 6: The conventional, simple, and body-centered unit cells for the tetragonal crystal system.

### 7.1. Lattice 8: Simple Tetragonal

The simple tetragonal Bravais lattice is identical to the conventional cell

$$\begin{aligned} \mathbf{a}_1 &= a\,\hat{\mathbf{x}} \\ \mathbf{a}_2 &= a\,\hat{\mathbf{y}} \\ \mathbf{a}_3 &= c\,\hat{\mathbf{z}}, \end{aligned} \tag{41}$$

with volume

$$V = a^2\,c. \tag{42}$$

The space groups associated with the simple tetragonal lattice are given in Table 8.

Table 8: The space groups associated with the simple tetragonal lattice (41).

| | | |
|---|---|---|
| 75. $P4$ | 76. $P4_1$ | 77. $P4_2$ |
| 78. $P4_3$ | 81. $P\bar{4}$ | 83. $P4/m$ |
| 84. $P4_2/m$ | 85. $P4/n$ | 86. $P4_2/n$ |
| 89. $P422$ | 90. $P42_12$ | 91. $P4_122$ |
| 92. $P4_12_12$ | 93. $P4_222$ | 94. $P4_22_12$ |
| 95. $P4_322$ | 96. $P4_32_12$ | 99. $P4mm$ |
| 100. $P4bm$ | 101. $P4_2cm$ | 102. $P4_2nm$ |
| 103. $P4cc$ | 104. $P4nc$ | 105. $P4_2mc$ |
| 106. $P4_2bc$ | 111. $P\bar{4}2m$ | 112. $P\bar{4}2c$ |
| 113. $P\bar{4}2_1m$ | 114. $P\bar{4}2_1c$ | 115. $P\bar{4}m2$ |
| 116. $P\bar{4}c2$ | 117. $P\bar{4}b2$ | 118. $P\bar{4}n2$ |
| 123. $P4/mmm$ | 124. $P4/mcc$ | 125. $P4/nbm$ |
| 126. $P4/nnc$ | 127. $P4/mbm$ | 128. $P4/mnc$ |
| 129. $P4/nmm$ | 130. $P4/ncc$ | 131. $P4_2/mmc$ |
| 132. $P4_2/mcm$ | 133. $P4_2/nbc$ | 134. $P4_2/nnm$ |
| 135. $P4_2/mbc$ | 136. $P4_2/mnm$ | 137. $P4_2/nmc$ |
| 138. $P4_2/ncm$ | | |

### 7.2. Lattice 9: Body-Centered Tetragonal

The body-centered tetragonal system has the same point group and translational symmetry as the simple tetragonal system, with the addition of a translation to the center of the parallelepiped defined by the vectors (39). Our standard form of the primitive vectors is

$$\begin{aligned} \mathbf{a}_1 &= -\frac{a}{2}\,\hat{\mathbf{x}} + \frac{a}{2}\,\hat{\mathbf{y}} + \frac{c}{2}\,\hat{\mathbf{z}} \\ \mathbf{a}_2 &= \frac{a}{2}\,\hat{\mathbf{x}} - \frac{a}{2}\,\hat{\mathbf{y}} + \frac{c}{2}\,\hat{\mathbf{z}} \\ \mathbf{a}_3 &= \frac{a}{2}\,\hat{\mathbf{x}} + \frac{a}{2}\,\hat{\mathbf{y}} - \frac{c}{2}\,\hat{\mathbf{z}}. \end{aligned} \tag{43}$$

The volume of the primitive body-centered tetragonal unit cell is

$$V = \frac{a^2\,c}{2}. \tag{44}$$

There are two primitive body-centered tetragonal unit cells in the conventional tetragonal unit cell. The space groups associated with this lattice, all of which begin with $I$ in standard notation, are given in Table 9.

Table 9: The space groups associated with the body-centered tetragonal lattice (43).

| | | |
|---|---|---|
| 79. $I4$ | 80. $I4_1$ | 82. $I\bar{4}$ |
| 87. $I4/m$ | 88. $I4_1/a$ | 97. $I422$ |
| 98. $I4_122$ | 107. $I4mm$ | 108. $I4cm$ |
| 109. $I4_1md$ | 110. $I4_1cd$ | 119. $I\bar{4}m2$ |
| 120. $I\bar{4}c2$ | 121. $I\bar{4}2m$ | 122. $I\bar{4}2d$ |
| 139. $I4/mmm$ | 140. $I4/mcm$ | 141. $I4_1/amd$ |
| 142. $I4_1/acd$ | | |

## 8. The Trigonal Crystal System

The trigonal crystal system is defined by a three-fold rotation axis, and can be generated from the cubic crystal system (Section 10) by stretching the cube along its diagonal. The symmetry requires the primitive vectors to have the form $a = b$, $\alpha = \beta = \pi/2$, $\gamma = 120°$.[7] The trigonal system is a limiting case of the simple monoclinic Bravais lattice (22), with $\beta = 120°$. It can also be obtained from the base-centered orthorhombic Bravais lattice (30) with $b = \sqrt{3}a$. The conventional unit cell is described by the vectors

$$\begin{aligned} \mathbf{A}_1 &= \frac{a}{2}\,\hat{\mathbf{x}} - \frac{\sqrt{3}}{2}\,a\,\hat{\mathbf{y}} \\ \mathbf{A}_2 &= \frac{a}{2}\,\hat{\mathbf{x}} + \frac{\sqrt{3}}{2}\,a\,\hat{\mathbf{y}} \\ \mathbf{A}_3 &= c\,\hat{\mathbf{z}}. \end{aligned} \tag{45}$$

There are two Bravais lattices in the trigonal system.

[7] We could take $\gamma = 60°$, but in that case the three-fold rotation axis is not obvious from the primitive vectors.

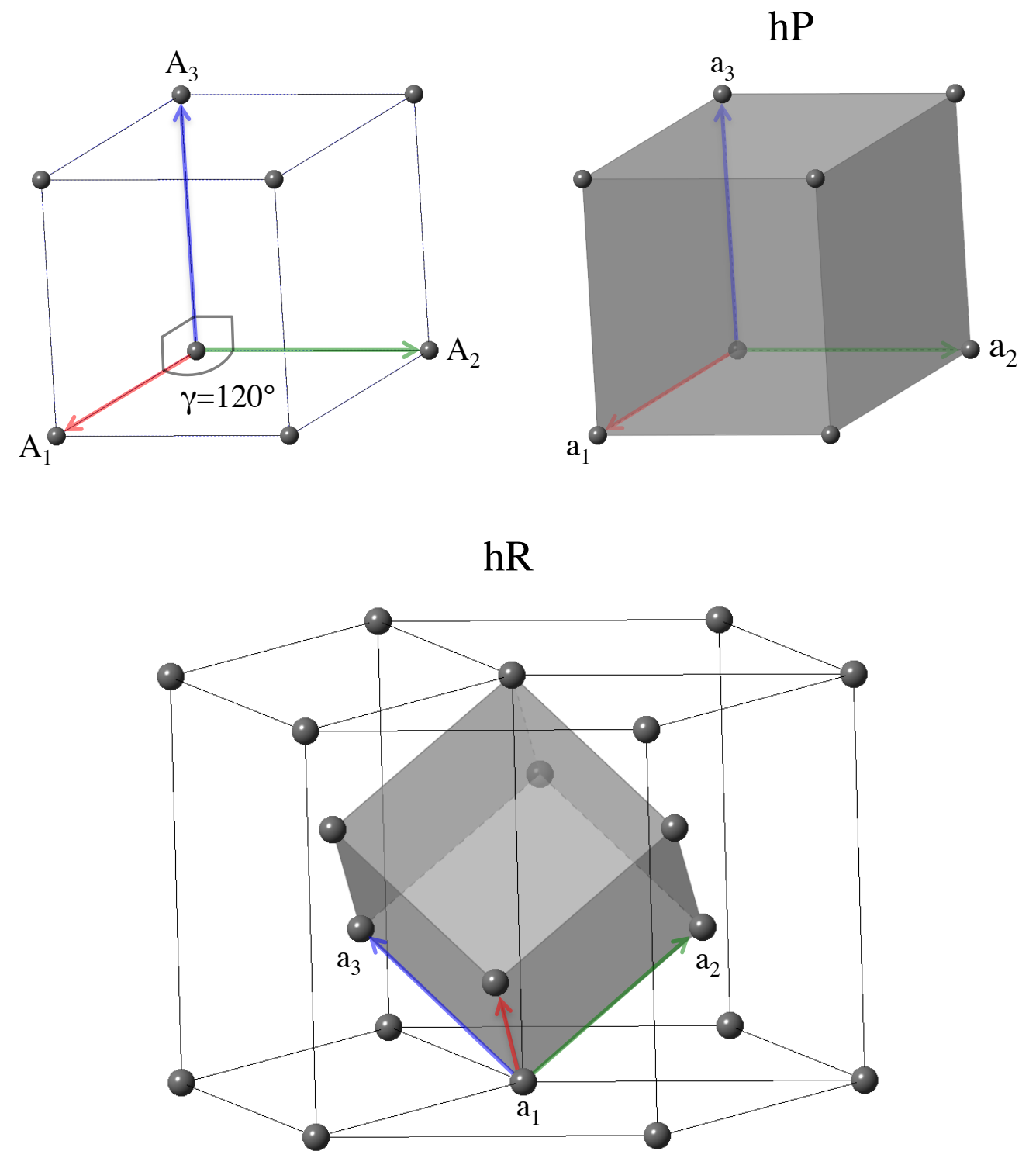

Figure 7: The conventional, simple (hexagonal), and rhombohedral unit cells for the trigonal crystal system.

### 8.1. Lattice 10: Hexagonal

Somewhat confusingly, what might be called the simple trigonal Bravais lattice is known as the hexagonal lattice. It shares the same primitive vectors, but not point operations, as the hexagonal crystal system (9). The primitive vectors are identical to those of the conventional cell,

$$\begin{aligned} \mathbf{a}_1 &= \frac{a}{2}\,\hat{\mathbf{x}} - \frac{\sqrt{3}}{2}\,a\,\hat{\mathbf{y}} \\ \mathbf{a}_2 &= \frac{a}{2}\,\hat{\mathbf{x}} + \frac{\sqrt{3}}{2}\,a\,\hat{\mathbf{y}} \\ \mathbf{a}_3 &= c\,\hat{\mathbf{z}}. \end{aligned} \tag{46}$$

The volume of the primitive cell is

$$V = \left(\frac{\sqrt{3}}{2}\right) a^2\,c. \tag{47}$$

The space groups associated with the (trigonal) hexagonal lattice are given in Table 10.

Table 10: The space groups associated with the (trigonal) hexagonal lattice (46).

| | | |
|---|---|---|
| 143. $P3$ | 144. $P3_1$ | 145. $P3_2$ |
| 147. $P\bar{3}$ | 149. $P312$ | 150. $P321$ |
| 151. $P3_112$ | 152. $P3_121$ | 153. $P3_212$ |
| 154. $P3_221$ | 156. $P3m1$ | 157. $P31m$ |
| 158. $P3c1$ | 159. $P31c$ | 162. $P\bar{3}1m$ |
| 163. $P\bar{3}1c$ | 164. $P\bar{3}m1$ | 165. $P\bar{3}c1$ |

### 8.2. Lattice 11: Rhombohedral

The rhombohedral Bravais lattice has the periodicity of the conventional trigonal cell (45), with the addition of two translation vectors, $2/3\mathbf{A}_1 + 1/3\mathbf{A}_2 + 1/3\mathbf{A}_3$ and $1/3\mathbf{A}_1 + 2/3\mathbf{A}_2 + 2/3\mathbf{A}_3$.

The primitive vectors can be taken in the form

$$\begin{aligned} \mathbf{a}_1 &= \frac{a}{2}\,\hat{\mathbf{x}} - \frac{a}{\left(2\sqrt{3}\right)}\,\hat{\mathbf{y}} + \frac{c}{3}\,\hat{\mathbf{z}} \\ \mathbf{a}_2 &= \frac{a}{\sqrt{3}}\,\hat{\mathbf{y}} + \frac{c}{3}\,\hat{\mathbf{z}} \\ \mathbf{a}_3 &= -\frac{a}{2}\,\hat{\mathbf{x}} - \frac{a}{\left(2\sqrt{3}\right)}\,\hat{\mathbf{y}} + \frac{c}{3}\,\hat{\mathbf{z}}, \end{aligned} \tag{48}$$

and the volume of the primitive cell is one-third that of the conventional cell,

$$V = \left(\frac{2}{\sqrt{3}}\right) a^2\, c. \tag{49}$$

The vectors (48) are all of identical length,

$$|\mathbf{a}_1| = |\mathbf{a}_2| = |\mathbf{a}_3| = \sqrt{\frac{a^2}{3} + \frac{c^2}{9}} \equiv a', \tag{50}$$

or, equivalently, $a = b = c \equiv a'$, where we designate the common length as $a'$ to distinguish it from the length of the first two vectors in the conventional lattice. The vectors also make equal angles with each other

$$\alpha = \beta = \gamma = \cos^{-1}\left(\frac{2c^2 - 3a^2}{2\left(c^2 + 3a^2\right)}\right). \tag{51}$$

Equations (50) and (51) provide another definition of the rhombohedral lattice. We can show this by writing the primitive vectors in a form that depends only on the common length and separation angle,[8]

$$\begin{aligned} \mathbf{a}_1 &= a' \begin{pmatrix} \sin\frac{\alpha}{2}\,\hat{\mathbf{x}} \\ -\left(\frac{1}{\sqrt{3}}\right)\sin\frac{\alpha}{2}\,\hat{\mathbf{y}} \\ +\sqrt{\frac{1}{3}\left(4\cos^2\frac{\alpha}{2} - 1\right)}\,\hat{\mathbf{z}} \end{pmatrix} \\ \mathbf{a}_2 &= a' \begin{pmatrix} \left(2/\sqrt{3}\right)\sin\frac{\alpha}{2}\,\hat{\mathbf{y}} \\ +\sqrt{\frac{1}{3}\left(4\cos^2\frac{\alpha}{2} - 1\right)}\,\hat{\mathbf{z}} \end{pmatrix} \\ \mathbf{a}_3 &= a' \begin{pmatrix} -\sin\frac{\alpha}{2}\,\hat{\mathbf{x}} \\ -\left(\frac{1}{\sqrt{3}}\right)\sin\frac{\alpha}{2}\,\hat{\mathbf{y}} \\ +\sqrt{\frac{1}{3}\left(4\cos^2\frac{\alpha}{2} - 1\right)}\,\hat{\mathbf{z}} \end{pmatrix}. \end{aligned} \tag{52}$$

[8] An alternative orientation is given by Setyawan and Curtarolo [42], who only give the primitive vectors in this $(a', \alpha)$ setting. The primitive vectors used for their rhombohedral cell (section A.11) differ from (52) only by the orientation of the vectors relative to the Cartesian axes. Their choice is simpler for computational purposes, but does not show the relationship between (48) and (52).

We can define the rhombohedral lattice in two ways: as a trigonal lattice with additional translational vectors, or as a "simple" lattice with equal primitive vectors making equal angles with one another. The *International Tables* addresses this ambiguity by listing atomic positions for the rhombohedral lattice in a "hexagonal setting," where all coordinates are referenced to the conventional cell (45), and in a "rhombohedral setting," where the coordinates are referenced to (52). To further confuse matters, the unit cell's dimensions might be reported in terms of $(a, c)$ from (45), or in terms of $(a', \alpha)$ from (52). An article might say that there were N atoms in the rhombohedral cell, or 3N atoms in the conventional cell. One has to pay attention to the context.

In the database, we will report the lattice parameters of the system by giving $a$ and $c$, since that is the usual crystallographic practice. However, we will record atomic positions using the primitive vectors (48), since computer calculations work best with the smallest number of atoms needed to describe the system.

The space groups associated with the rhombohedral lattice are given in Table 11.

Table 11: The space groups associated with the rhombohedral lattice (48).

| | | |
|---|---|---|
| 146. $R3$ | 148. $R\bar{3}$ | 155. $R32$ |
| 160. $R3m$ | 161. $R3c$ | 166. $R\bar{3}m$ |
| 167. $R\bar{3}c$ | | |

## 9. The Hexagonal Crystal System

The hexagonal crystal system has a six-fold rotation axis. There is only one Bravais lattice in this system, the hexagonal Bravais lattice given by (45) and (46), so the conventional and primitive lattices are equivalent.

The space groups associated with the hexagonal crystal system and lattice are given in Table 12.

Table 12: The space groups associated with the hexagonal crystal system and lattice.

| | | |
|---|---|---|
| 168. $P6$ | 169. $P6_1$ | 170. $P6_5$ |
| 171. $P6_2$ | 172. $P6_4$ | 173. $P6_3$ |
| 174. $P\bar{6}$ | 175. $P6/m$ | 176. $P6_3/m$ |
| 177. $P622$ | 178. $P6_122$ | 179. $P6_522$ |
| 180. $P6_222$ | 181. $P6_422$ | 182. $P6_322$ |
| 183. $P6mm$ | 184. $P6cc$ | 185. $P6_3cm$ |
| 186. $P6_3mc$ | 187. $P\bar{6}m2$ | 188. $P\bar{6}c2$ |
| 189. $P\bar{6}2m$ | 190. $P\bar{6}2c$ | 191. $P6/mmm$ |
| 192. $P6/mcc$ | 193. $P6_3/mcm$ | 194. $P6_3/mmc$ |

## 10. The Cubic Crystal System

The cubic crystal system is defined as having the symmetry of a cube: the conventional unit cell can be rotated by 90° about any axis, or by 180° around an axis running through the center of two opposing cube edges, or by 120° around a body diagonal, and retain the same shape. The conventional cell then takes the form

$$\begin{aligned} \mathbf{A}_1 &= a\,\hat{\mathbf{x}} \\ \mathbf{A}_2 &= a\,\hat{\mathbf{y}} \\ \mathbf{A}_3 &= a\,\hat{\mathbf{z}}, \end{aligned} \tag{53}$$

with unit cell volume

$$V = a^3. \tag{54}$$

This is the limiting case of both the orthorhombic (26) and tetragonal (39) systems when all primitive vectors are equal in length. There are three Bravais lattices in the cubic system.

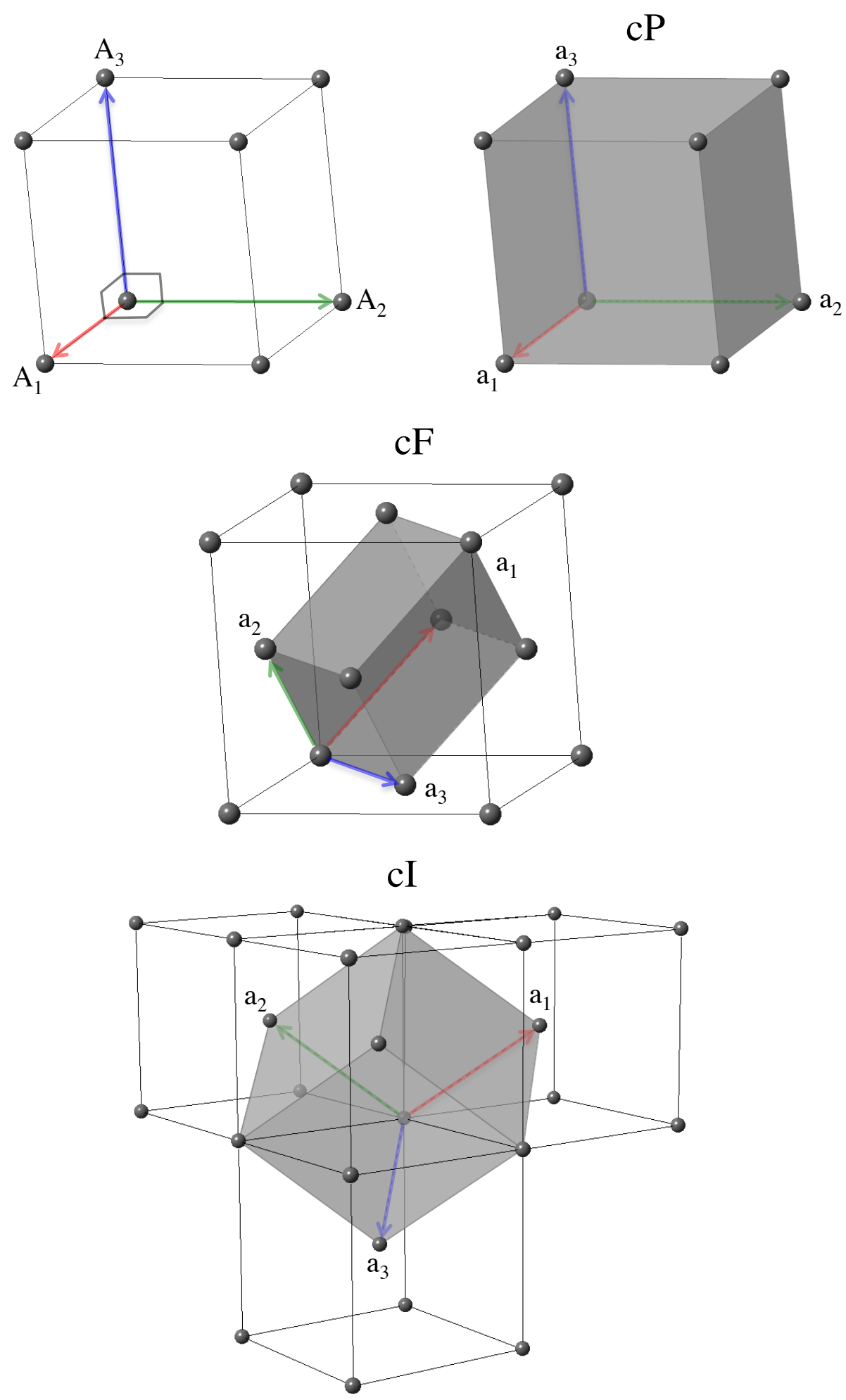

Figure 8: The conventional, simple, face-centered, and and body-centered unit cells for the cubic crystal system.

### 10.1. *Lattice 12: Simple Cubic*

The simple cubic system is identical to the conventional cubic unit cell

$$\begin{aligned} \mathbf{a}_1 &= a\,\hat{\mathbf{x}} \\ \mathbf{a}_2 &= a\,\hat{\mathbf{y}} \\ \mathbf{a}_3 &= a\,\hat{\mathbf{z}}, \end{aligned} \tag{55}$$

with volume

$$V = a^3. \tag{56}$$

This can also be considered as a rhombohedral lattice (48) with $\alpha = \pi/2$. The space groups associated with this lattice are given in Table 13.

Table 13: The space groups associated with the simple cubic lattice (55).

| | | |
|---|---|---|
| 195. $P23$ | 198. $P2_13$ | 200. $Pm\bar{3}$ |
| 201. $Pn\bar{3}$ | 205. $Pa\bar{3}$ | 207. $P432$ |
| 208. $P4_232$ | 212. $P4_332$ | 213. $P4_132$ |
| 215. $P\bar{4}3m$ | 218. $P\bar{4}3n$ | 221. $Pm\bar{3}m$ |
| 222. $Pn\bar{3}n$ | 223. $Pm\bar{3}n$ | 224. $Pn\bar{3}m$ |

### 10.2. *Lattice 13: Face-Centered Cubic*

The face-centered cubic lattice can be derived from its predecessors in the orthorhombic and tetragonal systems, having the same periodicity as its simple cubic parent with the addition of a translation from one corner of the cube to the center of any face. Our standard face-centered cubic primitive vectors have the form

$$\mathbf{a}_1 = \frac{a}{2}\,\hat{\mathbf{y}} + \frac{a}{2}\,\hat{\mathbf{z}} \tag{57}$$

$$\mathbf{a}_2 = \frac{a}{2}\,\hat{\mathbf{x}} + \frac{a}{2}\,\hat{\mathbf{z}} \tag{58}$$

$$\mathbf{a}_3 = \frac{a}{2}\,\hat{\mathbf{x}} + \frac{a}{2}\,\hat{\mathbf{y}}, \tag{59}$$

and the primitive cell volume is

$$V = \frac{a^3}{4}. \tag{60}$$

There are four face-centered cubic primitive cells in the conventional cubic cell. The face-centered cubic lattice can be considered as a rhombohedral lattice where $\alpha = 60°$. The space groups associated with this lattice are given in Table 14.

Table 14: The space groups associated with the face-centered cubic lattice (57).

| | | |
|---|---|---|
| 196. $F23$ | 202. $Fm\bar{3}$ | 203. $Fd\bar{3}$ |
| 209. $F432$ | 210. $F4_132$ | 216. $F\bar{4}3m$ |
| 219. $F\bar{4}3c$ | 225. $Fm\bar{3}m$ | 226. $Fm\bar{3}c$ |
| 227. $Fd\bar{3}m$ | 228. $Fd\bar{3}c$ | |

*10.3. Lattice 14: Body-Centered Cubic*

Like its predecessors in the orthorhombic and tetragonal systems, the body-centered cubic crystal has the same periodicity as its parent with the addition of a translation from one corner of the cube to its center. Our standard body-centered cubic primitive vectors have the form

$$\mathbf{a}_1 = -\frac{a}{2}\,\hat{\mathbf{x}} + \frac{a}{2}\,\hat{\mathbf{y}} + \frac{a}{2}\,\hat{\mathbf{z}} \quad (61)$$
$$\mathbf{a}_2 = \frac{a}{2}\,\hat{\mathbf{x}} - \frac{a}{2}\,\hat{\mathbf{y}} + \frac{a}{2}\,\hat{\mathbf{z}} \quad (62)$$
$$\mathbf{a}_3 = \frac{a}{2}\,\hat{\mathbf{x}} + \frac{a}{2}\,\hat{\mathbf{y}} - \frac{a}{2}\,\hat{\mathbf{z}}, \quad (63)$$

and the primitive cell volume is

$$V = \frac{a^3}{2}. \quad (64)$$

There are two body-centered cubic primitive cells in the conventional cubic cell. The body-centered cubic lattice can be considered as a rhombohedral lattice where $\alpha = \cos^{-1}(-1/3) \approx 109.47°$. The space groups associated with this lattice are given in Table 15.

Table 15: The space groups associated with the body-centered cubic lattice (61).

| | | |
|---|---|---|
| 197. $I23$ | 199. $I2_13$ | 204. $Im\overline{3}$ |
| 206. $Ia\overline{3}$ | 211. $I432$ | 214. $I4_132$ |
| 217. $I\overline{4}3m$ | 220. $I\overline{4}3d$ | 229. $Im\overline{3}m$ |
| 230. $Ia\overline{3}d$ | | |

## 11. Locating the atoms in the unit cell

Section 3 describes the Bravais lattices that occur in three dimensional space. Just describing the lattice, however, does not describe the complete crystal system. We must also find the positions of the atoms in the primitive (or conventional) unit cell. These positions are restricted by the crystal system, Bravais lattice, and space group that the system is in.

We will illustrate this using our two-dimensional centered rectangular lattice (17). There are seventeen plane groups in two dimensions [44]. Two are centered rectangular plane groups, $c1m1$ (#5) and $c2mm$ (#9). If we look at the International Tables [16] or Bilbao server [17], we will find a table that looks much like Table 16.

Table 16: The Wyckoff positions for the plane group $c1m1$ (#5). This is a somewhat simplified version of the table, as we neglect the site symmetries of each point. See Refs. [16] and [17] for complete information.

| Wyckoff Position | Coordinates +(1/2, 1/2) |
|---|---|
| (4b) | (x,y) (-x, y) |
| (2a) | (0,y) |

This table gives a set of Wyckoff positions, so called because Wyckoff denoted all possible positions for the 230 three dimensional space groups [40]. The first set of points (4b) refer to the general points for the $c1m1$ system, and refer to atomic positions based on the conventional rectangular unit cell (16).[9] This says that there are atoms located at two basis vectors

$$\mathbf{B}_1 = x\,\mathbf{A}_1 + y\,\mathbf{A}_2 = x\,a\,\hat{\mathbf{x}} + y\,b\,\hat{\mathbf{y}}$$
$$\mathbf{B}_2 = -x\,\mathbf{A}_1 + y\,\mathbf{A}_2 = -x\,a\,\hat{\mathbf{x}} + y\,b\,\hat{\mathbf{y}}. \quad (65)$$

Note, however, the "Coordinates +(1/2, 1/2)" entry in Table 16. This means that each position $(x, y)$ has a duplicate position at $(1/2 + x,\ 1/2 + y)$, giving rise to two more atomic positions

$$\begin{aligned}
\mathbf{B}_1' &= \left(\frac{1}{2} + x\right)\mathbf{A}_1 + \left(\frac{1}{2} + y\right)\mathbf{A}_2 \\
&= \left(\frac{1}{2} + x\right)a\,\hat{\mathbf{x}} + \left(\frac{1}{2} + y\right)b\,\hat{\mathbf{y}} \\
\mathbf{B}_2' &= \left(\frac{1}{2} - x\right)\mathbf{A}_1 + \left(\frac{1}{2} + y\right)\mathbf{A}_2 \\
&= \left(\frac{1}{2} - x\right)a\,\hat{\mathbf{x}} + \left(\frac{1}{2} + y\right)b\,\hat{\mathbf{y}}. \quad (66)
\end{aligned}$$

This extra shift of (1/2, 1/2) occurs because plane group $c1m1$ is defined on a centered rectangular lattice, and this shift gives the extra atomic positions in the conventional unit cell. We can see this by setting $x = y = 0$ in (66), as then both vectors correspond to primitive vector $\mathbf{a}_2$ in the Bravais lattice (17). Indeed, we can express the atomic positions in terms of $\mathbf{a}_1$ and $\mathbf{a}_2$:

$$\begin{aligned}
\mathbf{B}_1 &= (x_1 - y_1)\,\mathbf{a}_1 + (x_1 + y_1)\,\mathbf{a}_2 \\
&= x_1\,a\,\hat{\mathbf{x}} + y_1\,b\,\hat{\mathbf{y}} \\
\mathbf{B}_2 &= -(x_1 + y_1)\,\mathbf{a}_1 + (y_1 - x_1)\,\mathbf{a}_2 \\
&= \left(\frac{1}{2} - x_1\right)a\,\hat{\mathbf{x}} + \left(\frac{1}{2} + y_1\right)b\,\hat{\mathbf{y}}. \quad (67)
\end{aligned}$$

What about the other line in Table 16? If we set $x = 0$, the (4b) Wyckoff positions both become $(0, y)$, which means that instead of two atoms in the primitive cell, only one is allowed. In the conventional cell there are two such atoms, so this is known as the (2a) Wyckoff position. In terms of the primitive Bravais lattice this atom is located at

$$\mathbf{B}_3 = -y_2\,\mathbf{a}_1 + y_2\,\mathbf{a}_2 = y_2\,b\,\hat{\mathbf{y}}. \quad (68)$$

[9] Wyckoff positions are labeled by letter in descending order, from the most general symmetry to the most restrictive. Space group $Pm\overline{3}n$ (#223), for example, has twelve Wyckoff positions. The most general, which has forty-eight operations, is labeled (48l). Space group $Pmmm$ (#47) has twenty seven Wyckoff positions, so its most general one is denoted (8A).

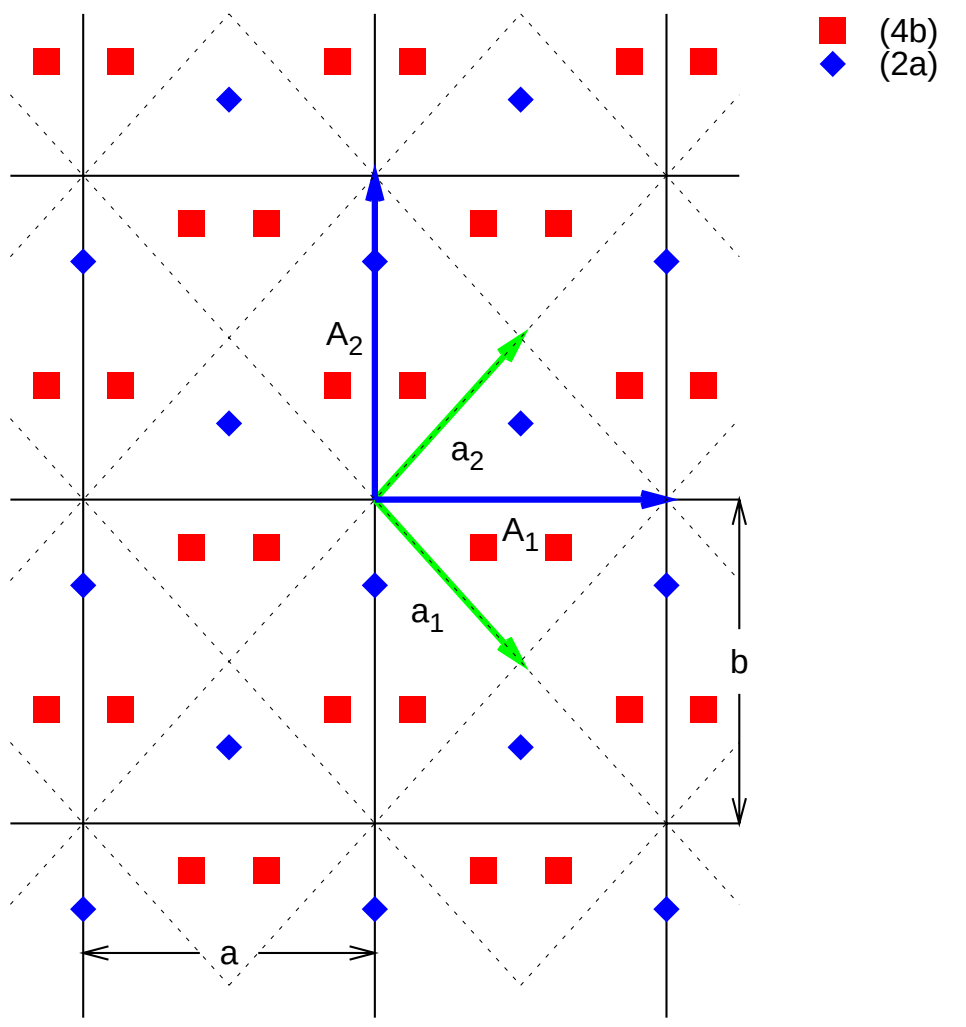

Figure 9: A two-dimensional centered rectangular crystal system in plane group $c1m1$.

One possible arrangement of this crystal system is shown in Figure 9. The squares represent (4b) atoms, while the diamonds are (2a) atoms. There are two (2a) atoms and four (4b) atoms in each conventional unit cell, bounded by the solid lines, and one (2a) atom and two (4b) atoms in each primitive cell, bounded by the dashed lines.

Table 16 could easily have been written with (4b) positions $(x, y)$ $(x, -y)$ and (4a) positions $(x, 0)$, *i.e.* we could have rotated the lattice by 90°. The choice made here is a matter of convention. This convention persists in the three dimensional case, and explains why some base-centered orthorhombic space groups (Table 5) use the "A" orientation rather than the "C" orientation. For example, space group Amm2 (#38) has the (2a) Wyckoff position $(00z)$. If we used the equivalent space group C2mm, the resulting Wyckoff position would be $(x00)$. Keeping with convention demands that we rotate the system.

In the listings below, we will follow convention by giving the lattice coordinates $(a, b, c, \alpha, \beta, \gamma)$ in terms of the conventional lattice for the crystal system, and the atomic positions in terms of the coordinates of the corresponding Wyckoff position. However, we will describe the unit cell using the Bravais vectors of the primitive lattices in Section 3, and describe the lattice coordinates of an atom relative to the these vectors. The lattice coordinates of atom $\mathbf{B}_1$, above, will be given as $(x - y), (x + y)$. Finally, we will only list atomic positions in the primitive cell, not the conventional unit cell. This means, *e.g.*, that we will only give the positions of four atoms for a Wyckoff position of (8c) for a body-centered orthorhombic cell.

## 12. Description of a Database Entry

The current database consists of 288 entries. This section describes the format of a database entry, as well as an outline of the CIF and POSCAR files.

### *12.1. The Database*

For each structure in the database, we have

**Title:** The title gives a brief description of the entry. If available this will include the mineral name, chemical formula, and the *Strukturbericht* Designation. In the case where an element or compound exists in multiple phases, the title may also list the space group, if that is necessary to distinguish one phase from another.

**Diagrams:** Four different views are shown for the structure: along each axis, and a tilted view. The boundaries of the conventional unit cell are also shown. For cells with trigonal (including rhombohedral) and hexagonal symmetry we also show two views of the structure containing three conventional cells. This allows us to see the full trigonal/hexagonal symmetry of the system.

**Prototype:** Each structure has a prototype compound which represents all compounds with that structure. For example, the prototype of the rock salt structure is NaCl.

**AFLOW prototype label:** The standardized name of a prototype, including stoichiometry, Pearson symbol, space group, and Wyckoff positions of the atoms.

***Strukturbericht* designation:** This gives the *Strukturbericht* symbol for the structure, if one is defined.

**Pearson symbol:** These were defined by W. B. Pearson [9] to give a shorthand description of a crystal structure. A Pearson symbol has three parts

1. *The crystal system* designates each crystal system by a letter:
   1.1. *a*: The triclinic system
   1.2. *m*: The monoclinic system
   1.3. *o*: The orthorhombic system
   1.4. *t*: The tetragonal system
   1.5. *h*: The trigonal and hexagonal system, combined
   1.6. *c*: The cubic system
2. *The lattice type* designates the type of lattice in the crystal system:
   2.1. *P*: A primitive (simple) lattice. Used when the primitive and conventional lattices are identical
   2.2. *C*: A base-centered lattice. The label could be A, B, or C, depending upon the base chosen, but C is usually chosen for all orientations. To avoid confusion, the symbol "S" is often used in place of C.
   2.3. *I*: A body-centered lattice
   2.4. *F*: A face-centered lattice

2.5. *R*: A rhombohedral lattice

3. *The number of atoms* in the cell is given in the final field. For most systems this is *in the conventional cell*. As an example, consider copper, which is the prototype for the face-centered cubic (*Strukturbericht* A1) system. It has a face-centered cubic lattice, with one atom in the primitive cell. This means that there are four atoms in the conventional cubic cell, so the Pearson symbol of copper is $cF4$.

   An exception arises for the rhombohedral structure. If a rhombohedral system has five atoms in the primitive cell, then the Pearson Symbol might be hR5 or hR15, the number of atoms in the conventional hexagonal unit cell. This depends upon the author. We will use the number of atoms in the primitive cell. The AFLOW framework does the same [21, 22, 23, 24].

   A further complication arises because the number of atoms in the Pearson symbol might be replaced by the number of formula units. Wurtzite (*Strukturbericht* B4) is a hexagonal crystal, with two atoms of sulfur and two atoms of zinc. Its Pearson symbol is hR4. However, there are two Zn-S units in the cell, so it is often referred to as 2H. While this is not strictly a Pearson symbol, it is Pearson-like, and one must be aware of the difference.

**Space group number:** The crystal's space group in the International Tables.

**Space group symbol:** We use the International Tables' space group symbol, in the standard orientation.

**AFLOW prototype command:** The command used to generate this structure in the AFLOW [21, 22, 23, 24] software system. The choice of the prototype is performed with the keyword `--proto=label` where the label is the standardized name. The prototype's degrees of freedom, internal and external, are specified with the keyword `--params=parameter_1,parameter_2,...` with the parameters separated by commas. Note that while most prototype labels are unique, there are a few cases where the same geometrical structure could be generated through more than one prototype by modifying the parameters. By default, the `--proto` keyword generates structures where the species are fictitiously taken as *A*, *B*, *C*, etc. The user can override their order using the following syntax: `--proto=label:A:B:C...` (after the colon ":" any permutation of *A*, *B*, *C*... separated by colons ":"). The user can also associate the same species with two or more different types using the following syntax: `--proto=label:A:A:B...` or use real element names using the following syntax: `--proto=label:Ag:Cd:Zr...`. After the generation of the prototype, the software will alphabetically reorder the atoms. Note that the total number of specified species (if provided) and parameters need to correspond to the exact requirements of the prototype otherwise AFLOW will prematurely terminate with an error message.[10]

**Rhombohedral systems** : All crystal structures are generated in AFLOW using the specifications of this document. Rhombohedral systems can be represented in either the rhombohedral or hexagonal settings. While the default is rhombohedral, a user can obtain the hexagonal setting by using the optional keyword `--hex`.

**Reference:** The journal article for the structure. If the structure was found in one of the standard locations (Pearson's Handbook, Pauling File, AMCD, *etc.*) then the reference to that entry is also given, under the label **Found in**.

**Other compounds (elements) with this structure:** One prototype phase may be observed with many different chemical compositions. This item provides a partial list for some compounds. *Pearson's Handbook* [10] provides a detailed list for intermetallic compounds.

**Primitive vectors:** Primitive vectors are given in the standard orientations listed in Section 2. A small figure next to the listing of the vectors shows the primitive vectors as well as outlining the primitive and conventional cells of the structure.

**Basis vectors:** The position of each atom in the primitive cell is defined by its

1. *Lattice Vector*: The label $\mathbf{B}_N$, where $N$ is the number of the atom in the primitive cell.
2. *Lattice Coordinates:* Lattice coordinates are determined from the Cartesian coordinates of the system using (11).
3. *Cartesian Coordinates:* These are determined from the Wyckoff positions and the lattice parameters of the conventional primitive cell. If the Wyckoff position of a particular site is ($x_n$, $y_n$, $z_n$), the Cartesian coordinates will be determined from the position

$$x_n\,\mathbf{A}_1 + y_n\,\mathbf{A}_2 + z_n\,\mathbf{A}_3,$$

[10] Generation of the prototypes presented in this article is supported by the AFLOW software v3.2 and above.

where the $\mathbf{A}_i$, describe the conventional unit cell. An exception is again made for rhombohedral structures, where we use the rhombohedral setting of the space group and vectors (48) to expand into Cartesian coordinates.

Each Wyckoff position in the crystal is given a subscript, *e.g.* the third Wyckoff position in a given structure will have coordinates $(x_3, y_3, z_3)$. A coordinate fixed by symmetry is replaced by its value, so $x_2$ is replaced by zero in (68).

4. *The Wyckoff Position:* From the International Tables. There may be multiple instances of a given Wyckoff position in a structure. Although the number in the Wyckoff position indicates the number of atoms in the conventional cell, we will only give the basis vectors for atoms in the primitive cell.

5. *The Atomic Label:* The chemical symbol for the atom at the current site. If there are multiple Wyckoff positions with the same species of atom, a Roman numeral appears after the atom type. For example, if a structure has oxygen atoms on (2e) and (4f) Wyckoff positions, the oxygen atoms on the (2e) site will be labeled "O I," while the atoms on the (4f) site will be labeled "O II."

### *12.2. Visualization*

Each structure is accompanied by a graphic showing the conventional unit cell of the structure looking down each of the cell's primitive vectors, and in an oblique representation which provides an overall view of the structure. These figures were drawn from the accompanying CIF file using `Jmol` [45].

### *12.3. The Crystallographic Information File (CIF)*

The Crystallographic Information File [31] (CIF) is the standard method for crystallographic information interchange. There is a CIF file for the prototype material of each structure, generated by `FINDSYM` [46], and modified by us to include the references in CIF format. The CIF file can be used for visualization, and programs such as CIF2cell [47] can use it to generate input for electronic structure programs. Note, the coordinates in the CIF may not exactly match the coordinates in the reference. Some coordinates may have been shifted for visualization purposes. We use the following features of the CIF format:

**_chemical_name_mineral:** The common name of the structure, if one exists

**_chemical_formula_sum:** The chemical formula of the structure, in computer readable form, *e.g.* $KClO_3$ would be written K Cl O3.

**_publ_author_name:** The authors of the reference

**_journal_...:** Entries describing the journal reference, including name, and publication year, first and last page.

**_publ_Section_title:** The title of the article.

**# Found in:** If we found the reference via another source, that source is given here. This is not part of the CIF standard, hence the hashtag.

**_aflow_proto:** The AFLOW command to generate the structure.

**_aflow_params:** The AFLOW parameters indicating the degrees of freedom required to generate the structure.

**_aflow_params_values:** The values of the AFLOW parameters used to generate the particular structure prototype.

**_aflow_Strukturbericht:** The Strukturbericht designation of the structure.

**_aflow_Pearson:** The Pearson symbol of the structure.

**_symmetry_space_group_name_Hall:** The name of the space group in Hall notation.

**_symmetry_space_group_name_H-M:** The name of the space group in Hermann-Mauguin notation.

**_symmetry_Int_Tables_number:** The number of the space group in the International Tables.

**_cell_length_(abc):** The lengths (13) of the primitive vectors of the unit cell, in Ångstroms.

**_cell_angle_(alpha beta gamma):** The angles (14) between the primitive vectors, in degrees.

**_space_group_symop_operation_xyz:** The combination of rotations, reflections, and translations allowed for the space group. For example, an entry for the two-dimensional $c1m1$ space group in Table 16 would have the form

```
1 x,y
2 -x,y
3 x+1/2,y+1/2
4 -x+1/2,y+1/2
```

so that the complete crystal structure can be determined from this table and the atomic Wyckoff positions at the end of the CIF file.

**_atom_site_...:** This describes the atomic positions of the atoms at each Wyckoff site in the crystal. There is one entry for each occupied Wyckoff position, the remaining atoms are derived from the space group operations above. Each line contains:

1. A label, usually the atom type with a number.
2. The atomic symbol for the atom occupying this position in the crystal.
3. The multiplicity of the Wyckoff site, for example, if the site was "2e" the number here would be "2."
4. The Wyckoff label, in our example "e."
5. The $x$, $y$, and $z$ coordinates of the atom, usually written to five decimal places.
6. The occupancy, to allow for partially filled sites.

A proper CIF file completely describes a crystal. The standard even allows for a complete article to be submitted as a CIF file.

### *12.4. The POSCAR File*

By default, the prototype is generated in the POSCAR format, which is the standard description of a crystallographic system used in VASP [32]. While the information it provides is identical to that in the CIF file, we list it here because it explicitly shows the construction of the primitive unit cell and the positions of all the atoms in the unit cell. The POSCAR files provided in the database are an annotated version of the VASP format for the prototype material, and can easily be edited for use with other materials with the same space group and occupied Wyckoff positions. A typical example is for pyrite, $FeS_2$:

```
AB2_cP12_205_a_c & \\
a,x2 --params=5.417,0.3851 & Pa-3 T_h^6 \\
#205 (ac) & cP12 & C2 & FeS2 & Pyrite & \\
Bayliss, Am. Min. 62, 1168-72 (1977)
1.0000000000000000
5.41700 0.00000 0.00000
0.00000 5.41700 0.00000
0.00000 0.00000 5.41700
Fe S
4 8
Direct
0.00000 0.00000 0.00000 Fe (4a)
0.00000 0.50000 0.50000 Fe (4a)
0.50000 0.00000 0.50000 Fe (4a)
0.50000 0.50000 0.00000 Fe (4a)
0.11490 0.61490 0.88510 S (8c)
0.11490 0.88510 0.38510 S (8c)
0.38510 0.11490 0.88510 S (8c)
0.38510 0.38510 0.38510 S (8c)
0.61490 0.61490 0.61490 S (8c)
0.61490 0.88510 0.11490 S (8c)
0.88510 0.11490 0.61490 S (8c)
0.88510 0.38510 0.11490 S (8c)
```

1. The first line, separated into four lines here, has eight fields, separated by ampersands:
   1.1. The first field contains the AFLOW prototype label for this structure.
   1.2. The second field contains the AFLOW parameters that have degrees of freedom and the values needed to create this structure.
   1.3. The third field contains the space group name, both in International and Schönflies formats, the space group number, and the Wyckoff positions occupied in this structure.
   1.4. The fourth field is the Pearson symbol.
   1.5. The fifth field is the *Strukturbericht* designation, if any.
   1.6. The sixth field is the prototype's chemical formula.
   1.7. The seventh field is the mineral name, if known; or, in the case of substances with multiple phases, the phase name, *e.g.* "alpha" or "beta".
   1.8. The final field is an abbreviated reference for the structure.
2. The second line is a scale factor. All Cartesian components in the lines below are multiplied by this factor.
3. The third, fourth, and fifth lines contain the Cartesian components of the primitive vectors, one per line. If the scale factor in line two is unity, these distances are in Ångstroms.
4. The sixth line (only in VASP version 5 and above) gives the names of the elements in the unit cell. We list the elements in alphabetical order by chemical symbol. The VASP POTCAR file must also list the elements in this order.
5. The seventh line contains the number of atoms of each element type, in the same order as the previous line.
6. `Direct` in line eight indicates that the atomic positions are in lattice coordinates. We will always use this form.
7. The remaining lines list the positions of the atoms in the primitive cell, one line per atom. The three numbers are the lattice coordinates. In our annotated file this is followed by the atomic label, and then the Wyckoff position occupied. The atoms are ordered by species, in agreement with lines six and seven.

While the POSCAR file is unique to VASP other electronic structure codes use similar formats and the information in the POSCAR can be easily converted into the required form. If this is not possible, programs such as CIF2cell [47] can use the CIF file to generate the electronic structure input in a variety of forms.

### *12.5.* QUANTUM ESPRESSO*, ABINIT, and FHI-AIMS Formats*

The user can generate prototypes in the QUANTUM ESPRESSO[48], ABINIT [49], and FHI-AIMS [50] formats

with the keywords `--qe`, `--abinit`, or `--aims`, respectively. Example: `aflow --proto=label --params=... --qe`. Other formats will be added in the future.

## 13. Conclusion

This article presents *The AFLOW Library of Crystallographic Prototypes*, an updated version of the original *Crystal Lattice Structures* web page. We present a complete description of 288 crystal structures, including the space group, Pearson and *Strukturbericht* symbols, primitive vectors, basis vectors, and references to the literature.

## 14. Acknowledgments

The authors would like to thank D. A. Papaconstantopoulos, who first proposed the *Crystal Lattice Structures* database, and R. Benjamin Young, who help set up the original web site in the summer of 1995. Special thanks are due to the H. Stokes, for providing updates to the `FINDSYM` code, and to the librarians at the U.S. Naval Research Laboratory Ruth H. Hooker Research Library, who tracked down many dozens of research articles which are not yet available online. M. J. Mehl is supported under contract from Duke University. We also acknowledge support by the by DOD-ONR (N00014-13-1-0635, N00014-15-1-2863, N00014-16-1-2781). C. Toher and S. Curtarolo acknowledge partial support from the DOE (DE-AC02-05CH11231), specifically the Basic Energy Sciences program under Grant #ED-CBEE. D. Hicks acknowledges support from the Department of Defense through the National Defense Science and Engineering Graduate (NDSEG) Fellowship Program. The AFLOW consortium would like to acknowledge the Duke University Center for Materials Genomics and the CRAY corporation for computational support.

# $FeS_2$ (P1) Structure: AB2_aP12_1_4a_8a

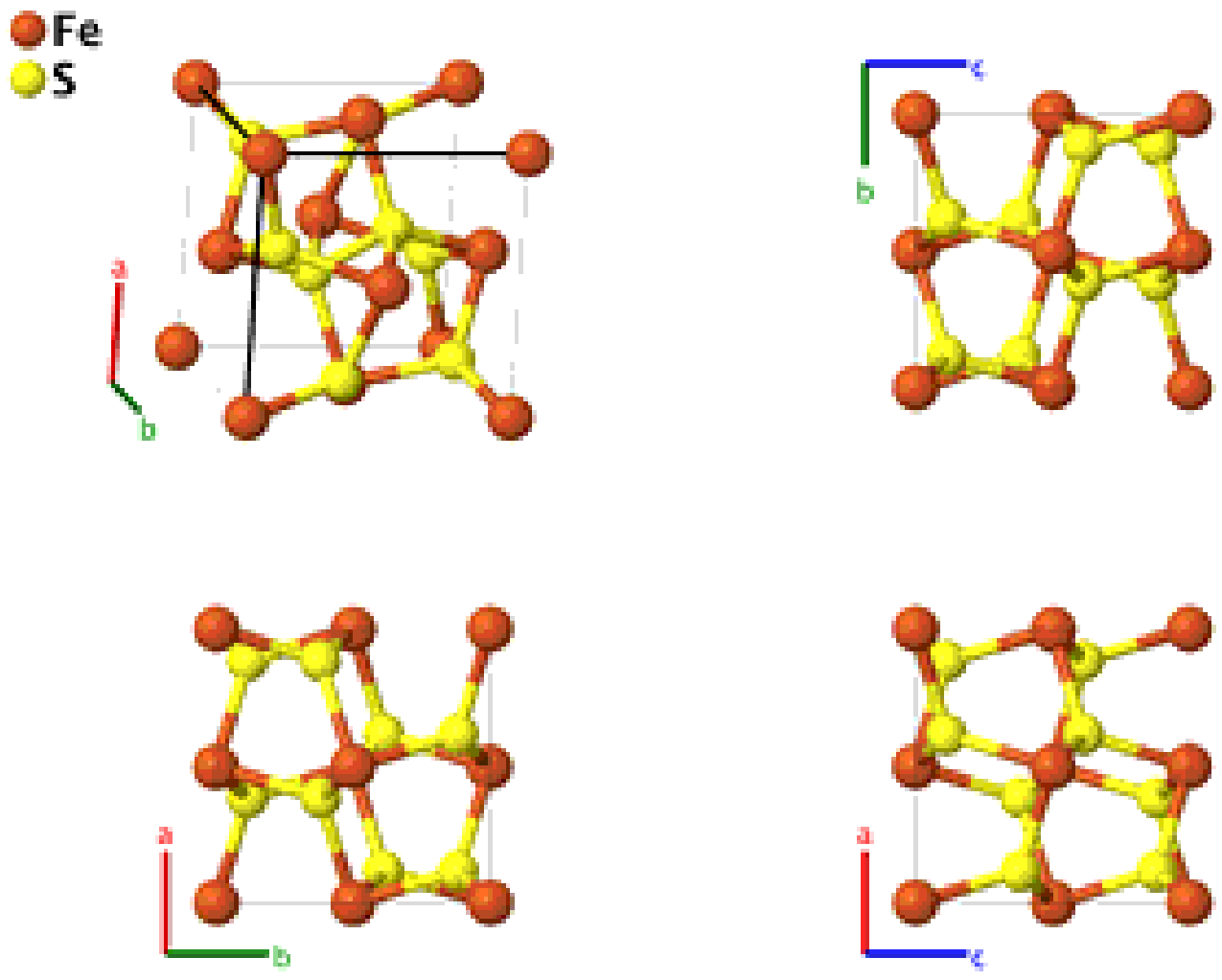

| | | |
|---|---|---|
| **Prototype** | : | $FeS_2$ |
| **AFLOW prototype label** | : | AB2_aP12_1_4a_8a |
| ***Strukturbericht* designation** | : | None |
| **Pearson symbol** | : | aP12 |
| **Space group number** | : | 1 |
| **Space group symbol** | : | P1 |
| **AFLOW prototype command** | : | `aflow --proto=AB2_aP12_1_4a_8a` `--params=`$a, b/a, c/a, \alpha, \beta, \gamma, x_1, y_1, z_1, x_2, y_2, z_2, x_3, y_3, z_3, x_4, y_4, z_4, x_5, y_5, z_5, x_6, y_6, z_6, x_7, y_7, z_7, x_8, y_8, z_8, x_9, y_9, z_9, x_{10}, y_{10}, z_{10}, x_{11}, y_{11}, z_{11}, x_{12}, y_{12}, z_{12}$ |

- This structure is just a slightly distorted version of pyrite (C2), with no rotational symmetry whatsoever.

**Triclinic primitive vectors:**

$$\mathbf{a}_1 = a\,\hat{\mathbf{x}}$$
$$\mathbf{a}_2 = b\cos\gamma\,\hat{\mathbf{x}} + b\sin\gamma\,\hat{\mathbf{y}}$$
$$\mathbf{a}_3 = c_x\hat{\mathbf{x}} + c_y\,\hat{\mathbf{y}} + c_z\,\hat{\mathbf{z}}$$

$$c_x = c\,\cos\beta$$
$$c_y = c\,(\cos\alpha - \cos\beta\cos\gamma)/\sin\gamma$$
$$c_z = \sqrt{c^2 - c_x^2 - c_y^2}$$

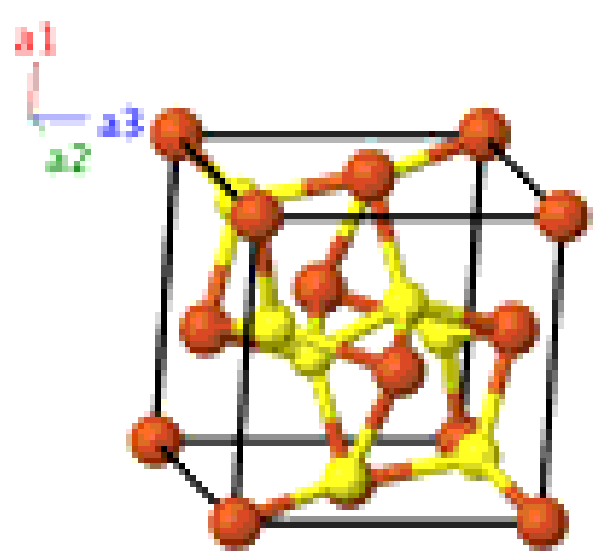

**Basis vectors:**

| | | Lattice Coordinates | | Cartesian Coordinates | Wyckoff Position | Atom Type |
|---|---|---|---|---|---|---|
| $\mathbf{B}_1$ | = | $x_1\,\mathbf{a}_1 + y_1\,\mathbf{a}_2 + z_1\,\mathbf{a}_3$ | = | $(x_1\,a + y_1\,b\,\cos\gamma + z_1\,c_x)\,\hat{\mathbf{x}} + \left(y_1\,b\,\sin\gamma + z_1\,c_y\right)\,\hat{\mathbf{y}} + z_1\,c_z\,\hat{\mathbf{z}}$ | $(1a)$ | Fe I |
| $\mathbf{B}_2$ | = | $x_2\,\mathbf{a}_1 + y_2\,\mathbf{a}_2 + z_2\,\mathbf{a}_3$ | = | $(x_2\,a + y_2\,b\,\cos\gamma + z_2\,c_x)\,\hat{\mathbf{x}} + \left(y_2\,b\,\sin\gamma + z_2\,c_y\right)\,\hat{\mathbf{y}} + z_2\,c_z\,\hat{\mathbf{z}}$ | $(1a)$ | Fe II |
| $\mathbf{B}_{3}$ | = | $x_{3}\,\mathbf{a}_{1}+y_{3}\,\mathbf{a}_{2}+z_{3}\,\mathbf{a}_{3}$ | = | $\left(x_{3}\,a+y_{3}\,b\,\cos\gamma\,+z_{3}\,c_{x}\right)\,\mathbf{\hat{x}}+\left(y_{3}\,b\,\sin\gamma+z_{3}\,c_{y}\right)\,\mathbf{\hat{y}}+z_{3}\,c_{z}\,\mathbf{\hat{z}}$ | (1$a$) | Fe III |
| $\mathbf{B}_{4}$ | = | $x_{4}\,\mathbf{a}_{1}+y_{4}\,\mathbf{a}_{2}+z_{4}\,\mathbf{a}_{3}$ | = | $\left(x_{4}\,a+y_{4}\,b\,\cos\gamma\,+z_{4}\,c_{x}\right)\,\mathbf{\hat{x}}+\left(y_{4}\,b\,\sin\gamma+z_{4}\,c_{y}\right)\,\mathbf{\hat{y}}+z_{4}\,c_{z}\,\mathbf{\hat{z}}$ | (1$a$) | Fe IV |
| $\mathbf{B}_{5}$ | = | $x_{5}\,\mathbf{a}_{1}+y_{5}\,\mathbf{a}_{2}+z_{5}\,\mathbf{a}_{3}$ | = | $\left(x_{5}\,a+y_{5}\,b\,\cos\gamma\,+z_{5}\,c_{x}\right)\,\mathbf{\hat{x}}+\left(y_{5}\,b\,\sin\gamma+z_{5}\,c_{y}\right)\,\mathbf{\hat{y}}+z_{5}\,c_{z}\,\mathbf{\hat{z}}$ | (1$a$) | S I |
| $\mathbf{B}_{6}$ | = | $x_{6}\,\mathbf{a}_{1}+y_{6}\,\mathbf{a}_{2}+z_{6}\,\mathbf{a}_{3}$ | = | $\left(x_{6}\,a+y_{6}\,b\,\cos\gamma\,+z_{6}\,c_{x}\right)\,\mathbf{\hat{x}}+\left(y_{6}\,b\,\sin\gamma+z_{6}\,c_{y}\right)\,\mathbf{\hat{y}}+z_{6}\,c_{z}\,\mathbf{\hat{z}}$ | (1$a$) | S II |
| $\mathbf{B}_{7}$ | = | $x_{7}\,\mathbf{a}_{1}+y_{7}\,\mathbf{a}_{2}+z_{7}\,\mathbf{a}_{3}$ | = | $\left(x_{7}\,a+y_{7}\,b\,\cos\gamma\,+z_{7}\,c_{x}\right)\,\mathbf{\hat{x}}+\left(y_{7}\,b\,\sin\gamma+z_{7}\,c_{y}\right)\,\mathbf{\hat{y}}+z_{7}\,c_{z}\,\mathbf{\hat{z}}$ | (1$a$) | S III |
| $\mathbf{B}_{8}$ | = | $x_{8}\,\mathbf{a}_{1}+y_{8}\,\mathbf{a}_{2}+z_{8}\,\mathbf{a}_{3}$ | = | $\left(x_{8}\,a+y_{8}\,b\,\cos\gamma\,+z_{8}\,c_{x}\right)\,\mathbf{\hat{x}}+\left(y_{8}\,b\,\sin\gamma+z_{8}\,c_{y}\right)\,\mathbf{\hat{y}}+z_{8}\,c_{z}\,\mathbf{\hat{z}}$ | (1$a$) | S IV |
| $\mathbf{B}_{9}$ | = | $x_{9}\,\mathbf{a}_{1}+y_{9}\,\mathbf{a}_{2}+z_{9}\,\mathbf{a}_{3}$ | = | $\left(x_{9}\,a+y_{9}\,b\,\cos\gamma\,+z_{9}\,c_{x}\right)\,\mathbf{\hat{x}}+\left(y_{9}\,b\,\sin\gamma+z_{9}\,c_{y}\right)\,\mathbf{\hat{y}}+z_{9}\,c_{z}\,\mathbf{\hat{z}}$ | (1$a$) | S V |
| $\mathbf{B}_{10}$ | = | $x_{10}\,\mathbf{a}_{1}+y_{10}\,\mathbf{a}_{2}+z_{10}\,\mathbf{a}_{3}$ | = | $\left(x_{10}\,a+y_{10}\,b\,\cos\gamma\,+z_{10}\,c_{x}\right)\,\mathbf{\hat{x}}+\left(y_{10}\,b\,\sin\gamma+z_{10}\,c_{y}\right)\,\mathbf{\hat{y}}+z_{10}\,c_{z}\,\mathbf{\hat{z}}$ | (1$a$) | S VI |
| $\mathbf{B}_{11}$ | = | $x_{11}\,\mathbf{a}_{1}+y_{11}\,\mathbf{a}_{2}+z_{11}\,\mathbf{a}_{3}$ | = | $\left(x_{11}\,a+y_{11}\,b\,\cos\gamma\,+z_{11}\,c_{x}\right)\,\mathbf{\hat{x}}+\left(y_{11}\,b\,\sin\gamma+z_{11}\,c_{y}\right)\,\mathbf{\hat{y}}+z_{11}\,c_{z}\,\mathbf{\hat{z}}$ | (1$a$) | S VII |
| $\mathbf{B}_{12}$ | = | $x_{12}\,\mathbf{a}_{1}+y_{12}\,\mathbf{a}_{2}+z_{12}\,\mathbf{a}_{3}$ | = | $\left(x_{12}\,a+y_{12}\,b\,\cos\gamma\,+z_{12}\,c_{x}\right)\,\mathbf{\hat{x}}+\left(y_{12}\,b\,\sin\gamma+z_{12}\,c_{y}\right)\,\mathbf{\hat{y}}+z_{12}\,c_{z}\,\mathbf{\hat{z}}$ | (1$a$) | S VIII |

**Geometry files:**

- CIF: pp. 641
- POSCAR: pp. 641

# $AsKSe_2$ (P1) Structure: ABC2_aP16_1_4a_4a_8a

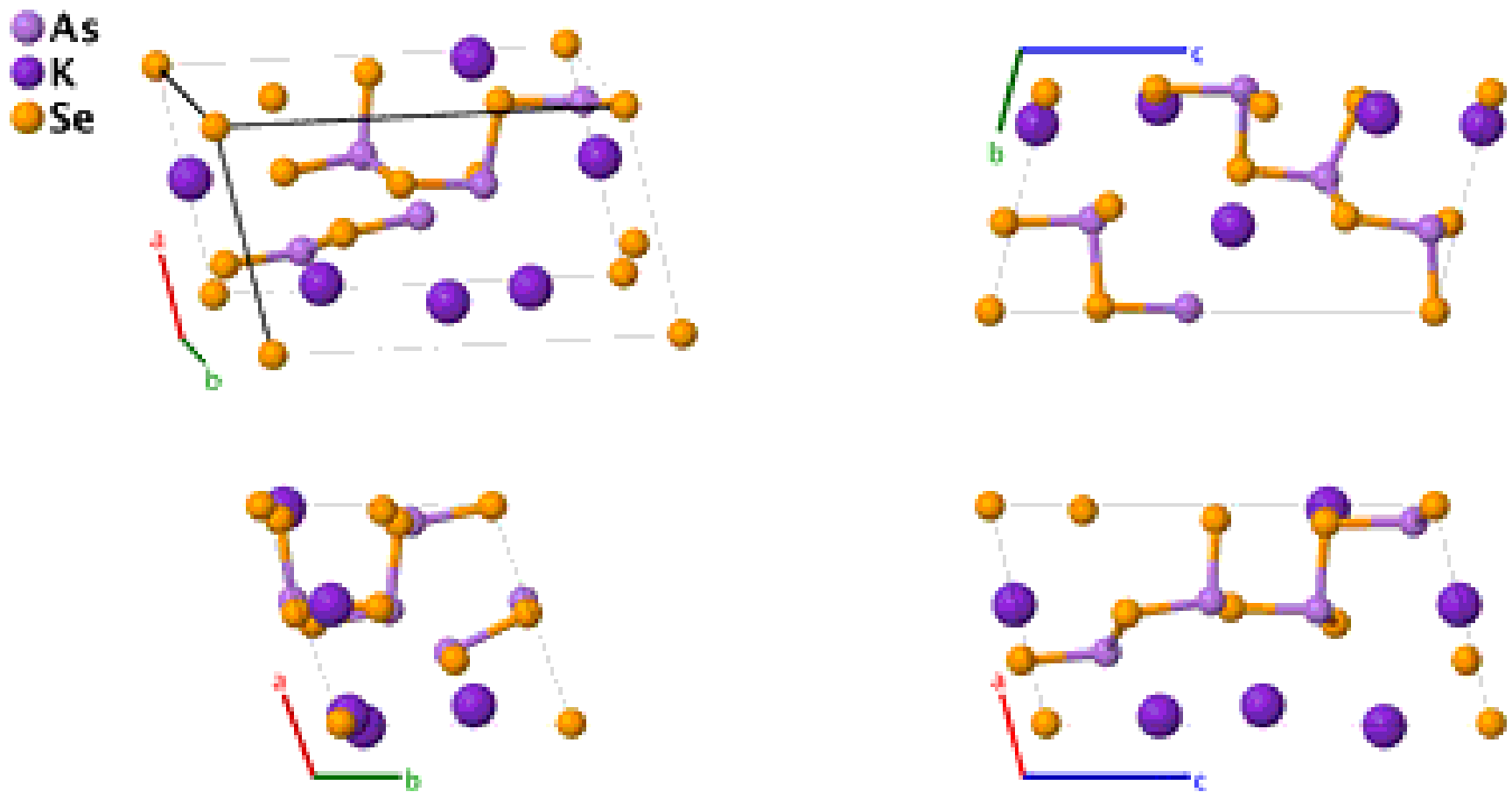

**Prototype** : $AsKSe_2$

**AFLOW prototype label** : ABC2_aP16_1_4a_4a_8a

***Strukturbericht* designation** : None

**Pearson symbol** : aP16

**Space group number** : 1

**Space group symbol** : P1

**AFLOW prototype command** : `aflow --proto=ABC2_aP16_1_4a_4a_8a`
`--params=`$a, b/a, c/a, \alpha, \beta, \gamma, x_1, y_1, z_1, x_2, y_2, z_2, x_3, y_3, z_3, x_4, y_4, z_4, x_5, y_5, z_5, x_6, y_6, z_6, x_7, y_7, z_7, x_8, y_8, z_8, x_9, y_9, z_9, x_{10}, y_{10}, z_{10}, x_{11}, y_{11}, z_{11}, x_{12}, y_{12}, z_{12}, x_{13}, y_{13}, z_{13}, x_{14}, y_{14}, z_{14}, x_{15}, y_{15}, z_{15}, x_{16}, y_{16}, z_{16}$

---

**Triclinic primitive vectors:**

$$\mathbf{a}_1 = a\,\hat{\mathbf{x}}$$
$$\mathbf{a}_2 = b\cos\gamma\,\hat{\mathbf{x}} + b\sin\gamma\,\hat{\mathbf{y}}$$
$$\mathbf{a}_3 = c_x\hat{\mathbf{x}} + c_y\,\hat{\mathbf{y}} + c_z\,\hat{\mathbf{z}}$$

$$c_x = c\,\cos\beta$$
$$c_y = c\,(\cos\alpha - \cos\beta\cos\gamma)/\sin\gamma$$
$$c_z = \sqrt{c^2 - c_x^2 - c_y^2}$$

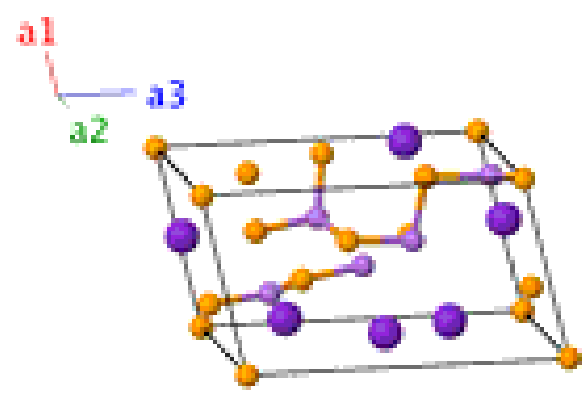

---

**Basis vectors:**

| | | Lattice Coordinates | | Cartesian Coordinates | Wyckoff Position | Atom Type |
|---|---|---|---|---|---|---|
| $\mathbf{B}_1$ | = | $x_1\,\mathbf{a}_1 + y_1\,\mathbf{a}_2 + z_1\,\mathbf{a}_3$ | = | $(x_1\,a + y_1\,b\,\cos\gamma + z_1\,c_x)\,\hat{\mathbf{x}} + \left(y_1\,b\,\sin\gamma + z_1\,c_y\right)\,\hat{\mathbf{y}} + z_1\,c_z\,\hat{\mathbf{z}}$ | $(1a)$ | As I |
| $\mathbf{B}_2$ | = | $x_2\,\mathbf{a}_1 + y_2\,\mathbf{a}_2 + z_2\,\mathbf{a}_3$ | = | $(x_2\,a + y_2\,b\,\cos\gamma + z_2\,c_x)\,\hat{\mathbf{x}} + \left(y_2\,b\,\sin\gamma + z_2\,c_y\right)\,\hat{\mathbf{y}} + z_2\,c_z\,\hat{\mathbf{z}}$ | $(1a)$ | As II |
| $\mathbf{B}_3$ | = | $x_3\,\mathbf{a}_1 + y_3\,\mathbf{a}_2 + z_3\,\mathbf{a}_3$ | = | $(x_3\,a + y_3\,b\,\cos\gamma + z_3\,c_x)\,\hat{\mathbf{x}} + \left(y_3\,b\,\sin\gamma + z_3\,c_y\right)\,\hat{\mathbf{y}} + z_3\,c_z\,\hat{\mathbf{z}}$ | $(1a)$ | As III |
| $\mathbf{B}_4$ | = | $x_4\,\mathbf{a}_1 + y_4\,\mathbf{a}_2 + z_4\,\mathbf{a}_3$ | = | $(x_4\,a + y_4\,b\,\cos\gamma + z_4\,c_x)\,\hat{\mathbf{x}} + \left(y_4\,b\,\sin\gamma + z_4\,c_y\right)\,\hat{\mathbf{y}} + z_4\,c_z\,\hat{\mathbf{z}}$ | $(1a)$ | As IV |
| $\mathbf{B_5}$ | = | $x_5\,\mathbf{a_1} + y_5\,\mathbf{a_2} + z_5\,\mathbf{a_3}$ | = | $(x_5\,a + y_5\,b\,\cos\gamma\, + z_5\,c_x)\,\mathbf{\hat{x}} + \left(y_5\,b\,\sin\gamma + z_5\,c_y\right)\,\mathbf{\hat{y}} + z_5\,c_z\,\mathbf{\hat{z}}$ | (1$a$) | K I |
| $\mathbf{B_6}$ | = | $x_6\,\mathbf{a_1} + y_6\,\mathbf{a_2} + z_6\,\mathbf{a_3}$ | = | $(x_6\,a + y_6\,b\,\cos\gamma\, + z_6\,c_x)\,\mathbf{\hat{x}} + \left(y_6\,b\,\sin\gamma + z_6\,c_y\right)\,\mathbf{\hat{y}} + z_6\,c_z\,\mathbf{\hat{z}}$ | (1$a$) | K II |
| $\mathbf{B_7}$ | = | $x_7\,\mathbf{a_1} + y_7\,\mathbf{a_2} + z_7\,\mathbf{a_3}$ | = | $(x_7\,a + y_7\,b\,\cos\gamma\, + z_7\,c_x)\,\mathbf{\hat{x}} + \left(y_7\,b\,\sin\gamma + z_7\,c_y\right)\,\mathbf{\hat{y}} + z_7\,c_z\,\mathbf{\hat{z}}$ | (1$a$) | K III |
| $\mathbf{B_8}$ | = | $x_8\,\mathbf{a_1} + y_8\,\mathbf{a_2} + z_8\,\mathbf{a_3}$ | = | $(x_8\,a + y_8\,b\,\cos\gamma\, + z_8\,c_x)\,\mathbf{\hat{x}} + \left(y_8\,b\,\sin\gamma + z_8\,c_y\right)\,\mathbf{\hat{y}} + z_8\,c_z\,\mathbf{\hat{z}}$ | (1$a$) | K IV |
| $\mathbf{B_9}$ | = | $x_9\,\mathbf{a_1} + y_9\,\mathbf{a_2} + z_9\,\mathbf{a_3}$ | = | $(x_9\,a + y_9\,b\,\cos\gamma\, + z_9\,c_x)\,\mathbf{\hat{x}} + \left(y_9\,b\,\sin\gamma + z_9\,c_y\right)\,\mathbf{\hat{y}} + z_9\,c_z\,\mathbf{\hat{z}}$ | (1$a$) | Se I |
| $\mathbf{B_{10}}$ | = | $x_{10}\,\mathbf{a_1} + y_{10}\,\mathbf{a_2} + z_{10}\,\mathbf{a_3}$ | = | $(x_{10}\,a + y_{10}\,b\,\cos\gamma\, + z_{10}\,c_x)\,\mathbf{\hat{x}} + \left(y_{10}\,b\,\sin\gamma + z_{10}\,c_y\right)\,\mathbf{\hat{y}} + z_{10}\,c_z\,\mathbf{\hat{z}}$ | (1$a$) | Se II |
| $\mathbf{B_{11}}$ | = | $x_{11}\,\mathbf{a_1} + y_{11}\,\mathbf{a_2} + z_{11}\,\mathbf{a_3}$ | = | $(x_{11}\,a + y_{11}\,b\,\cos\gamma\, + z_{11}\,c_x)\,\mathbf{\hat{x}} + \left(y_{11}\,b\,\sin\gamma + z_{11}\,c_y\right)\,\mathbf{\hat{y}} + z_{11}\,c_z\,\mathbf{\hat{z}}$ | (1$a$) | Se III |
| $\mathbf{B_{12}}$ | = | $x_{12}\,\mathbf{a_1} + y_{12}\,\mathbf{a_2} + z_{12}\,\mathbf{a_3}$ | = | $(x_{12}\,a + y_{12}\,b\,\cos\gamma\, + z_{12}\,c_x)\,\mathbf{\hat{x}} + \left(y_{12}\,b\,\sin\gamma + z_{12}\,c_y\right)\,\mathbf{\hat{y}} + z_{12}\,c_z\,\mathbf{\hat{z}}$ | (1$a$) | Se IV |
| $\mathbf{B_{13}}$ | = | $x_{13}\,\mathbf{a_1} + y_{13}\,\mathbf{a_2} + z_{13}\,\mathbf{a_3}$ | = | $(x_{13}\,a + y_{13}\,b\,\cos\gamma\, + z_{13}\,c_x)\,\mathbf{\hat{x}} + \left(y_{13}\,b\,\sin\gamma + z_{13}\,c_y\right)\,\mathbf{\hat{y}} + z_{13}\,c_z\,\mathbf{\hat{z}}$ | (1$a$) | Se V |
| $\mathbf{B_{14}}$ | = | $x_{14}\,\mathbf{a_1} + y_{14}\,\mathbf{a_2} + z_{14}\,\mathbf{a_3}$ | = | $(x_{14}\,a + y_{14}\,b\,\cos\gamma\, + z_{14}\,c_x)\,\mathbf{\hat{x}} + \left(y_{14}\,b\,\sin\gamma + z_{14}\,c_y\right)\,\mathbf{\hat{y}} + z_{14}\,c_z\,\mathbf{\hat{z}}$ | (1$a$) | Se VI |
| $\mathbf{B_{15}}$ | = | $x_{15}\,\mathbf{a_1} + y_{15}\,\mathbf{a_2} + z_{15}\,\mathbf{a_3}$ | = | $(x_{15}\,a + y_{15}\,b\,\cos\gamma\, + z_{15}\,c_x)\,\mathbf{\hat{x}} + \left(y_{15}\,b\,\sin\gamma + z_{15}\,c_y\right)\,\mathbf{\hat{y}} + z_{15}\,c_z\,\mathbf{\hat{z}}$ | (1$a$) | Se VII |
| $\mathbf{B_{16}}$ | = | $x_{16}\,\mathbf{a_1} + y_{16}\,\mathbf{a_2} + z_{16}\,\mathbf{a_3}$ | = | $(x_{16}\,a + y_{16}\,b\,\cos\gamma\, + z_{16}\,c_x)\,\mathbf{\hat{x}} + \left(y_{16}\,b\,\sin\gamma + z_{16}\,c_y\right)\,\mathbf{\hat{y}} + z_{16}\,c_z\,\mathbf{\hat{z}}$ | (1$a$) | Se VIII |

---

**Found in:**

---

**Geometry files:**

- CIF: pp. 641

- POSCAR: pp. 641

# $P_2I_4$ Structure: A2B_aP6_2_2i_i

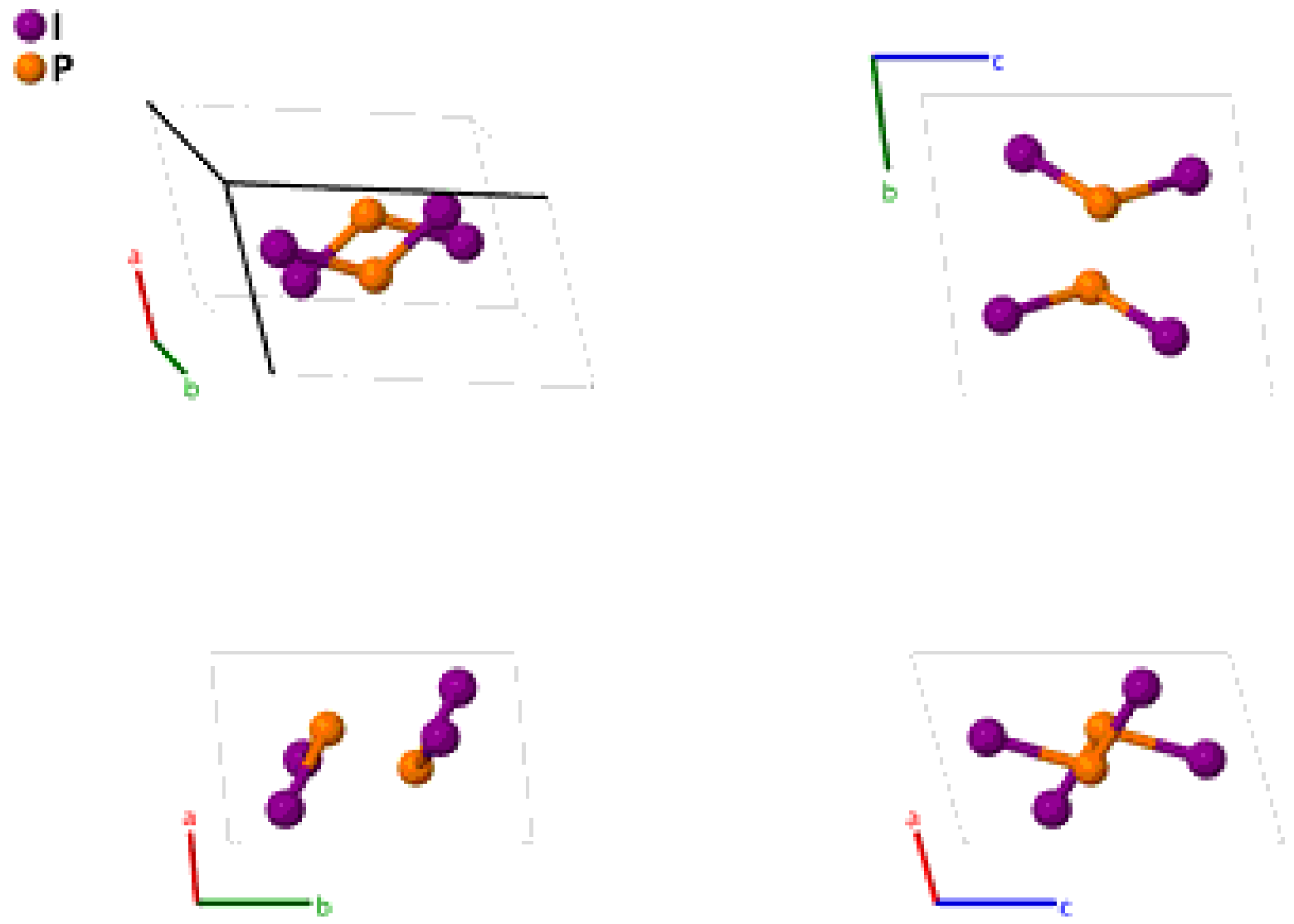

**Prototype** : $P_2I_4$

**AFLOW prototype label** : A2B_aP6_2_2i_i

***Strukturbericht* designation** : None

**Pearson symbol** : aP6

**Space group number** : 2

**Space group symbol** : $P\bar{1}$

**AFLOW prototype command** : `aflow --proto=A2B_aP6_2_2i_i`
`--params=`$a, b/a, c/a, \alpha, \beta, \gamma, x_1, y_1, z_1, x_2, y_2, z_2, x_3, y_3, z_3$

---

**Triclinic primitive vectors:**

$$\mathbf{a}_1 = a\hat{\mathbf{x}}$$
$$\mathbf{a}_2 = b\cos\gamma\,\hat{\mathbf{x}} + b\sin\gamma\,\hat{\mathbf{y}}$$
$$\mathbf{a}_3 = c_x\hat{\mathbf{x}} + c_y\,\hat{\mathbf{y}} + c_z\,\hat{\mathbf{z}}$$

$$c_x = c\,\cos\beta$$
$$c_y = c\,(\cos\alpha - \cos\beta\cos\gamma)/\sin\gamma$$
$$c_z = \sqrt{c^2 - c_x^2 - c_y^2}$$

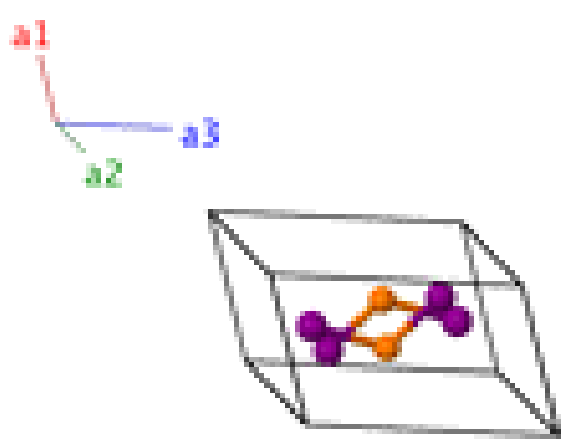

---

**Basis vectors:**

| | | Lattice Coordinates | | Cartesian Coordinates | Wyckoff Position | Atom Type |
|---|---|---|---|---|---|---|
| $\mathbf{B_1}$ | = | $x_1\,\mathbf{a_1} + y_1\,\mathbf{a_2} + z_1\,\mathbf{a_3}$ | = | $(x_1\,a + y_1\,b\,\cos\gamma\, + z_1\,c_x)\,\hat{\mathbf{x}} + \left(y_1\,b\,\sin\gamma + z_1\,c_y\right)\,\hat{\mathbf{y}} + z_1\,c_z\,\hat{\mathbf{z}}$ | (2$i$) | I I |
| $\mathbf{B_2}$ | = | $-x_1\,\mathbf{a_1} - y_1\,\mathbf{a_2} - z_1\,\mathbf{a_3}$ | = | $-(x_1\,a + y_1\,b\,\cos\gamma\, + z_1\,c_x)\,\hat{\mathbf{x}} - \left(y_1\,b\,\sin\gamma + z_1\,c_y\right)\,\hat{\mathbf{y}} - z_1\,c_z\,\hat{\mathbf{z}}$ | (2$i$) | I I |
| $\mathbf{B_3}$ | = | $x_2\,\mathbf{a_1} + y_2\,\mathbf{a_2} + z_2\,\mathbf{a_3}$ | = | $(x_2\,a + y_2\,b\,\cos\gamma\, + z_2\,c_x)\,\hat{\mathbf{x}} + \left(y_2\,b\,\sin\gamma + z_2\,c_y\right)\,\hat{\mathbf{y}} + z_2\,c_z\,\hat{\mathbf{z}}$ | (2$i$) | I II |
| $\mathbf{B_4}$ | = | $-x_2\,\mathbf{a_1} - y_2\,\mathbf{a_2} - z_2\,\mathbf{a_3}$ | = | $-\left(x_2\,a + y_2\,b\,\cos\gamma + z_2\,c_x\right)\,\mathbf{\hat{x}} - \left(y_2\,b\,\sin\gamma + z_2\,c_y\right)\,\mathbf{\hat{y}} - z_2\,c_z\,\mathbf{\hat{z}}$ | (2$i$) | I II |
| $\mathbf{B_5}$ | = | $x_3\,\mathbf{a_1} + y_3\,\mathbf{a_2} + z_3\,\mathbf{a_3}$ | = | $\left(x_3\,a + y_3\,b\,\cos\gamma + z_3\,c_x\right)\,\mathbf{\hat{x}} + \left(y_3\,b\,\sin\gamma + z_3\,c_y\right)\,\mathbf{\hat{y}} + z_3\,c_z\,\mathbf{\hat{z}}$ | (2$i$) | P |
| $\mathbf{B_6}$ | = | $-x_3\,\mathbf{a_1} - y_3\,\mathbf{a_2} - z_3\,\mathbf{a_3}$ | = | $-\left(x_3\,a + y_3\,b\,\cos\gamma + z_3\,c_x\right)\,\mathbf{\hat{x}} - \left(y_3\,b\,\sin\gamma + z_3\,c_y\right)\,\mathbf{\hat{y}} - z_3\,c_z\,\mathbf{\hat{z}}$ | (2$i$) | P |

**Geometry files:**

- CIF: pp. 642
- POSCAR: pp. 642

# Cf Structure: A_aP4_2_aci

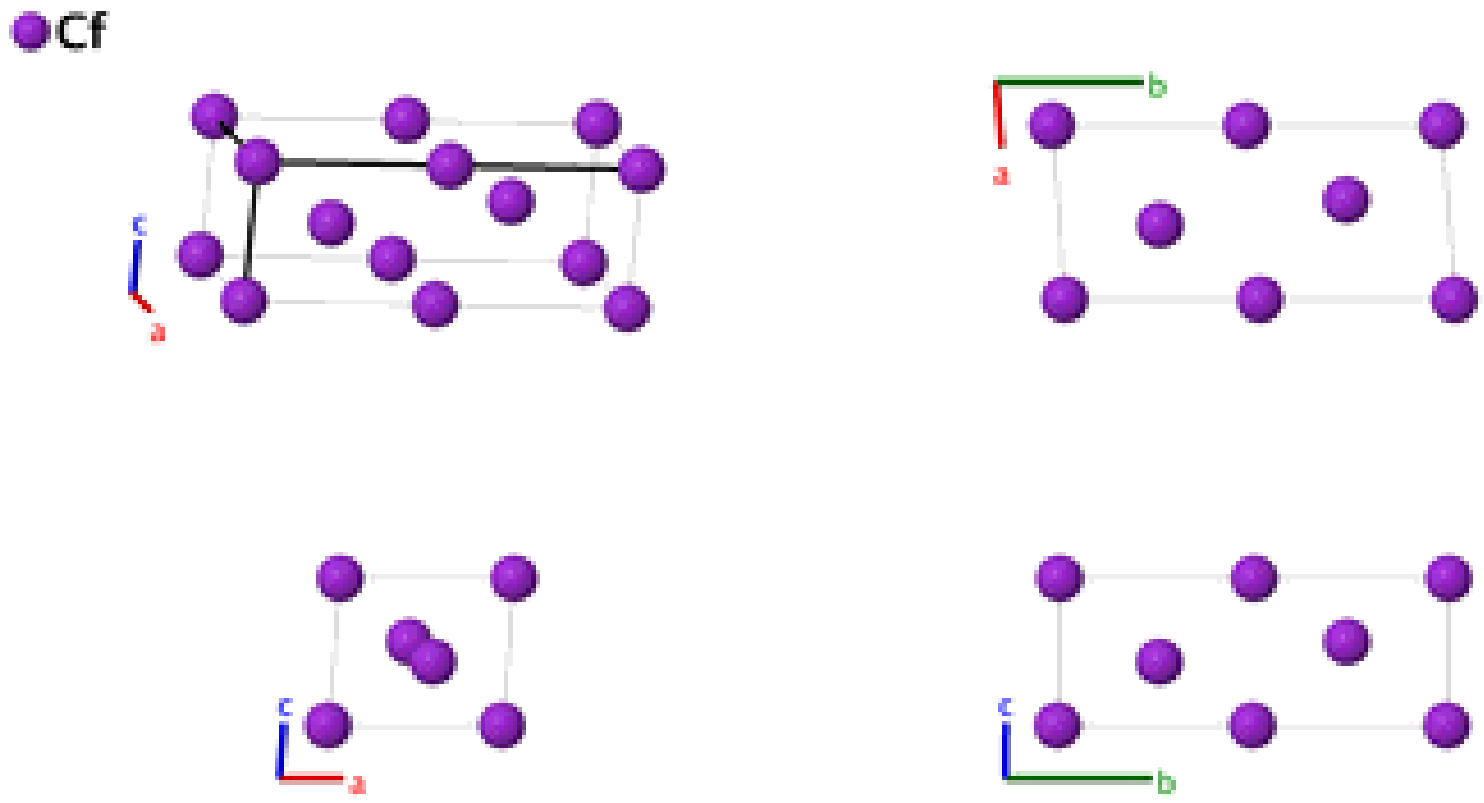

| | | |
|---|---|---|
| **Prototype** | : | Cf |
| **AFLOW prototype label** | : | A_aP4_2_aci |
| ***Strukturbericht* designation** | : | None |
| **Pearson symbol** | : | aP4 |
| **Space group number** | : | 2 |
| **Space group symbol** | : | $P\bar{1}$ |
| **AFLOW prototype command** | : | `aflow --proto=A_aP4_2_aci`<br>`--params=`$a, b/a, c/a, \alpha, \beta, \gamma, x_3, y_3, z_3$ |

- This is a high-pressure phase, observed between 30 and 40 GPa.

**Triclinic primitive vectors:**

$$\mathbf{a}_1 = a\,\hat{\mathbf{x}}$$
$$\mathbf{a}_2 = b\cos\gamma\,\hat{\mathbf{x}} + b\sin\gamma\,\hat{\mathbf{y}}$$
$$\mathbf{a}_3 = c_x\hat{\mathbf{x}} + c_y\,\hat{\mathbf{y}} + c_z\,\hat{\mathbf{z}}$$

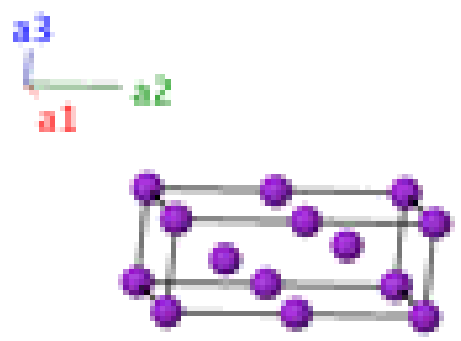

$$c_x = c\,\cos\beta$$
$$c_y = c\,(\cos\alpha - \cos\beta\cos\gamma)/\sin\gamma$$
$$c_z = \sqrt{c^2 - c_x^2 - c_y^2}$$

**Basis vectors:**

| | | Lattice Coordinates | | Cartesian Coordinates | Wyckoff Position | Atom Type |
|---|---|---|---|---|---|---|
| $\mathbf{B}_1$ | = | $0\,\mathbf{a}_1 + 0\,\mathbf{a}_2 + 0\,\mathbf{a}_3$ | = | $0\hat{\mathbf{x}} + 0\hat{\mathbf{y}} + 0\hat{\mathbf{z}}$ | (1$a$) | Cf I |
| $\mathbf{B}_2$ | = | $\frac{1}{2}\mathbf{a}_2$ | = | $\frac{1}{2}\,b\,\cos\gamma\,\hat{\mathbf{x}} + \frac{1}{2}\,b\,\sin\gamma\,\hat{\mathbf{y}}$ | (1$c$) | Cf II |
| $\mathbf{B}_3$ | = | $x_3\,\mathbf{a}_1 + y_3\,\mathbf{a}_2 + z_3\,\mathbf{a}_3$ | = | $(x_3\,a + y_3\,b\,\cos\gamma\, + z_3\,c_x)\,\hat{\mathbf{x}} + \left(y_3\,b\,\sin\gamma + z_3\,c_y\right)\,\hat{\mathbf{y}} + z_3\,c_z\,\hat{\mathbf{z}}$ | (2$i$) | Cf III |
| $\mathbf{B}_4$ | = | $-x_3\,\mathbf{a}_1 - y_3\,\mathbf{a}_2 - z_3\,\mathbf{a}_3$ | = | $-(x_3\,a + y_3\,b\,\cos\gamma\, + z_3\,c_x)\,\hat{\mathbf{x}} - \left(y_3\,b\,\sin\gamma + z_3\,c_y\right)\,\hat{\mathbf{y}} - z_3\,c_z\,\hat{\mathbf{z}}$ | (2$i$) | Cf III |

**References:**

- R. B. Roof, *Concerning the Structure of a High Pressure Phase in Californium Metal*, J. Less-Common Met. **120**, 345–349 (1986), doi:10.1016/0022-5088(86)90660-0.

**Found in:**

- P. Villars and L. Calvert, *Pearson's Handbook of Crystallographic Data for Intermetallic Phases* (ASM International, Materials Park, OH, 1991), 2nd edn, pp. 2332.

---

**Geometry files:**

- CIF: pp. 642
- POSCAR: pp. 642

# $SiO_2$ (P2) Structure: A2B_mP12_3_bc3e_2e

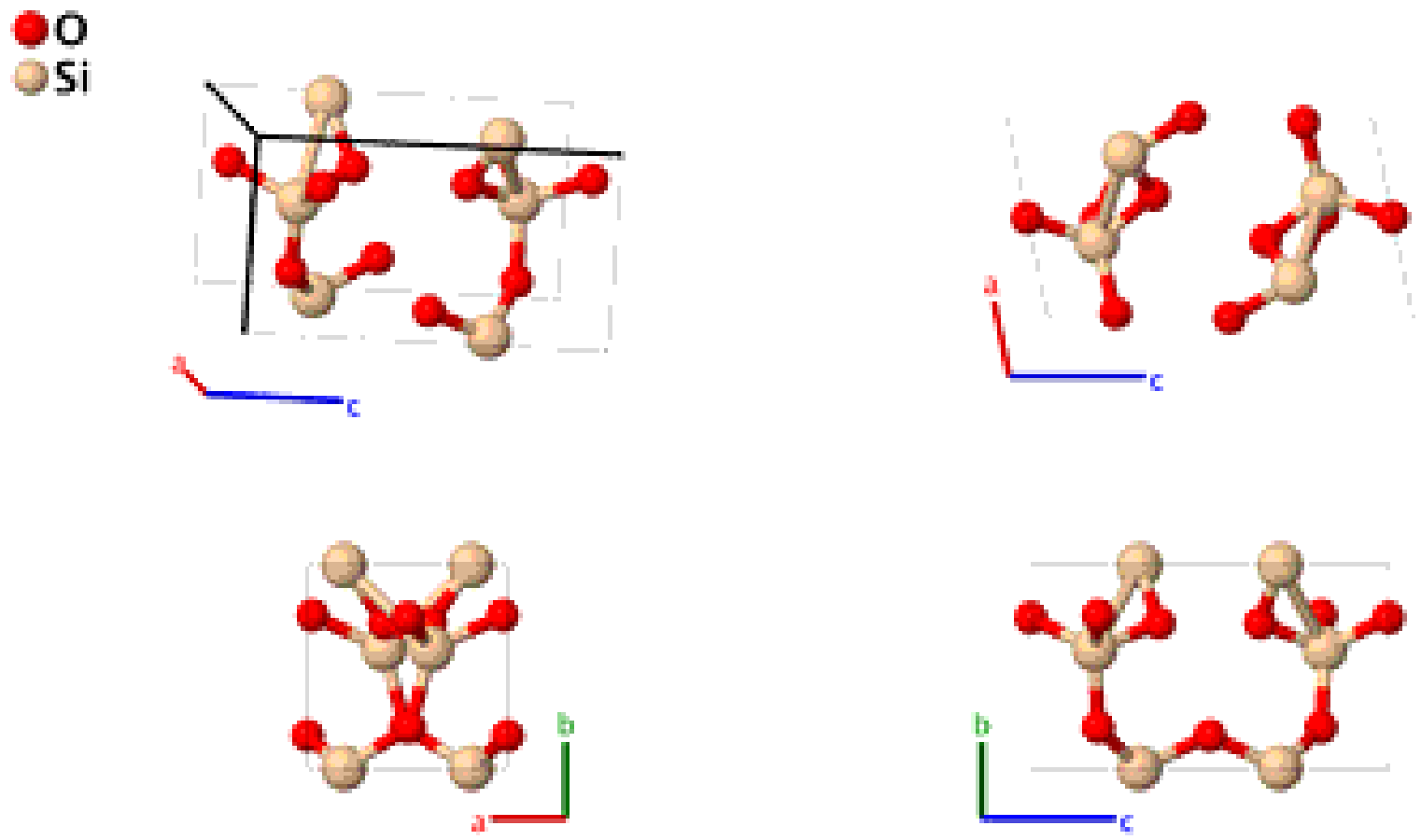

| | | |
|---|---|---|
| **Prototype** | : | $SiO_2$ |
| **AFLOW prototype label** | : | A2B_mP12_3_bc3e_2e |
| ***Strukturbericht* designation** | : | None |
| **Pearson symbol** | : | mP12 |
| **Space group number** | : | 3 |
| **Space group symbol** | : | P2 |
| **AFLOW prototype command** | : | `aflow --proto=A2B_mP12_3_bc3e_2e`<br>`--params=`$a, b/a, c/a, \beta, y_1, y_2, x_3, y_3, z_3, x_4, y_4, z_4, x_5, y_5, z_5, x_6, y_6, z_6, x_7, y_7, z_7$ |

- This structure is the result of simulations of $SiO_2$ structures from a potential fitted to the $H_6Si_2O_7$ molecule. As such, we do not believe it has been seen in nature. It does, however, describe a structure in space group P2 (#3).

**Simple Monoclinic primitive vectors:**

$$\mathbf{a}_1 = a\,\hat{\mathbf{x}}$$
$$\mathbf{a}_2 = b\,\hat{\mathbf{y}}$$
$$\mathbf{a}_3 = c\cos\beta\,\hat{\mathbf{x}} + c\sin\beta\,\hat{\mathbf{z}}$$

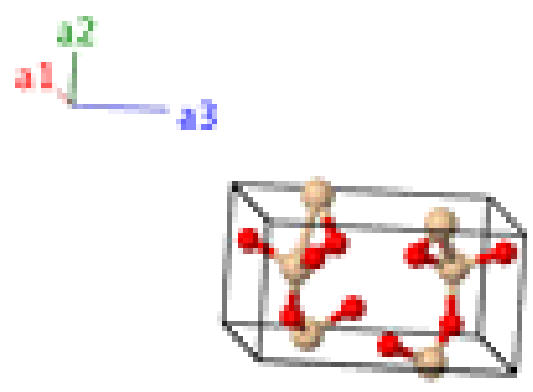

**Basis vectors:**

| | | Lattice Coordinates | | Cartesian Coordinates | Wyckoff Position | Atom Type |
|---|---|---|---|---|---|---|
| $\mathbf{B_1}$ | = | $y_1\,\mathbf{a_2} + \frac{1}{2}\,\mathbf{a_3}$ | = | $\frac{1}{2}\,c\,\cos\beta\,\hat{\mathbf{x}} + y_1 b\,\hat{\mathbf{y}} + \frac{1}{2}\,c\,\sin\beta\,\hat{\mathbf{z}}$ | (1$b$) | O I |
| $\mathbf{B_2}$ | = | $\frac{1}{2}\,\mathbf{a_1} + y_2\,\mathbf{a_2}$ | = | $\frac{1}{2}\,a\,\hat{\mathbf{x}} + y_2 b\,\hat{\mathbf{y}}$ | (1$c$) | O II |
| $\mathbf{B_3}$ | = | $x_3\,\mathbf{a_1} + y_3\,\mathbf{a_2} + z_3\,\mathbf{a_3}$ | = | $(x_3\,a + z_3\,c\,\cos\beta)\,\hat{\mathbf{x}} + y_3\,b\,\hat{\mathbf{y}} + z_3\,c\,\sin\beta\,\hat{\mathbf{z}}$ | (2$e$) | O III |
| $\mathbf{B_4}$ | = | $-x_3\,\mathbf{a_1} + y_3\,\mathbf{a_2} - z_3\,\mathbf{a_3}$ | = | $-(x_3\,a + z_3\,c\,\cos\beta)\,\hat{\mathbf{x}} + y_3\,b\,\hat{\mathbf{y}} - z_3\,c\,\sin\beta\,\hat{\mathbf{z}}$ | (2$e$) | O III |
| $\mathbf{B_5}$ | = | $x_4\,\mathbf{a_1} + y_4\,\mathbf{a_2} + z_4\,\mathbf{a_3}$ | = | $(x_4\,a + z_4\,c\,\cos\beta)\,\hat{\mathbf{x}} + y_4\,b\,\hat{\mathbf{y}} + z_4\,c\,\sin\beta\,\hat{\mathbf{z}}$ | (2$e$) | O IV |
| $\mathbf{B_6}$ | = | $-x_4\,\mathbf{a_1} + y_4\,\mathbf{a_2} - z_4\,\mathbf{a_3}$ | = | $-(x_4\,a + z_4\,c\,\cos\beta)\,\hat{\mathbf{x}} + y_4\,b\,\hat{\mathbf{y}} - z_4\,c\,\sin\beta\,\hat{\mathbf{z}}$ | (2$e$) | O IV |
| $\mathbf{B_7}$ | = | $x_5\,\mathbf{a_1} + y_5\,\mathbf{a_2} + z_5\,\mathbf{a_3}$ | = | $(x_5\,a + z_5\,c\,\cos\beta)\,\hat{\mathbf{x}} + y_5\,b\,\hat{\mathbf{y}} + z_5\,c\,\sin\beta\,\hat{\mathbf{z}}$ | $(2e)$ | O V |
| $\mathbf{B_8}$ | = | $-x_5\,\mathbf{a_1} + y_5\,\mathbf{a_2} - z_5\,\mathbf{a_3}$ | = | $-(x_5\,a + z_5\,c\,\cos\beta)\,\hat{\mathbf{x}} + y_5\,b\,\hat{\mathbf{y}} - z_5\,c\,\sin\beta\,\hat{\mathbf{z}}$ | $(2e)$ | O V |
| $\mathbf{B_9}$ | = | $x_6\,\mathbf{a_1} + y_6\,\mathbf{a_2} + z_6\,\mathbf{a_3}$ | = | $(x_6\,a + z_6\,c\,\cos\beta)\,\hat{\mathbf{x}} + y_6\,b\,\hat{\mathbf{y}} + z_6\,c\,\sin\beta\,\hat{\mathbf{z}}$ | $(2e)$ | Si I |
| $\mathbf{B_{10}}$ | = | $-x_6\,\mathbf{a_1} + y_6\,\mathbf{a_2} - z_6\,\mathbf{a_3}$ | = | $-(x_6\,a + z_6\,c\,\cos\beta)\,\hat{\mathbf{x}} + y_6\,b\,\hat{\mathbf{y}} - z_6\,c\,\sin\beta\,\hat{\mathbf{z}}$ | $(2e)$ | Si I |
| $\mathbf{B_{11}}$ | = | $x_7\,\mathbf{a_1} + y_7\,\mathbf{a_2} + z_7\,\mathbf{a_3}$ | = | $(x_7\,a + z_7\,c\,\cos\beta)\,\hat{\mathbf{x}} + y_7\,b\,\hat{\mathbf{y}} + z_7\,c\,\sin\beta\,\hat{\mathbf{z}}$ | $(2e)$ | Si II |
| $\mathbf{B_{12}}$ | = | $-x_7\,\mathbf{a_1} + y_7\,\mathbf{a_2} - z_7\,\mathbf{a_3}$ | = | $-(x_7\,a + z_7\,c\,\cos\beta)\,\hat{\mathbf{x}} + y_7\,b\,\hat{\mathbf{y}} - z_7\,c\,\sin\beta\,\hat{\mathbf{z}}$ | $(2e)$ | Si II |

---

---

**Geometry files:**

- CIF: pp. 642
- POSCAR: pp. 643

# High-Pressure Te Structure: A_mP4_4_2a

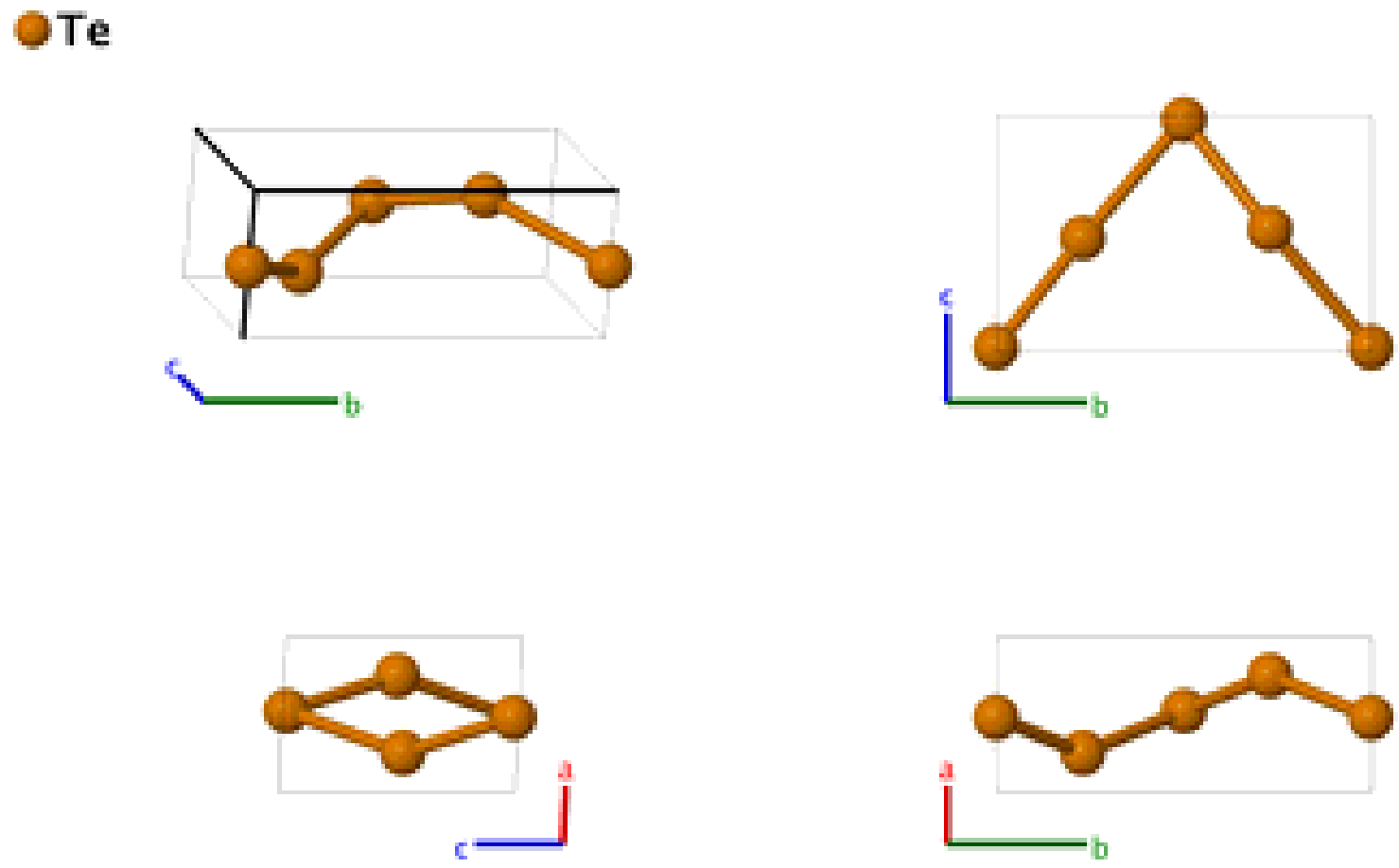

**Prototype** : Te

**AFLOW prototype label** : A_mP4_4_2a

***Strukturbericht* designation** : None

**Pearson symbol** : mP4

**Space group number** : 4

**Space group symbol** : $P2_1$

**AFLOW prototype command** : `aflow --proto=A_mP4_4_2a`
`--params=`$a, b/a, c/a, \beta, x_1, y_1, z_1, x_2, y_2, z_2$

---

- This is a high-pressure phase of Te, stable in the 4-7 GPa range. The ground state of Te appears to be $\gamma$-Se (A8).

---

**Simple Monoclinic primitive vectors:**

$$\mathbf{a}_1 = a\,\hat{\mathbf{x}}$$
$$\mathbf{a}_2 = b\,\hat{\mathbf{y}}$$
$$\mathbf{a}_3 = c\cos\beta\,\hat{\mathbf{x}} + c\sin\beta\,\hat{\mathbf{z}}$$

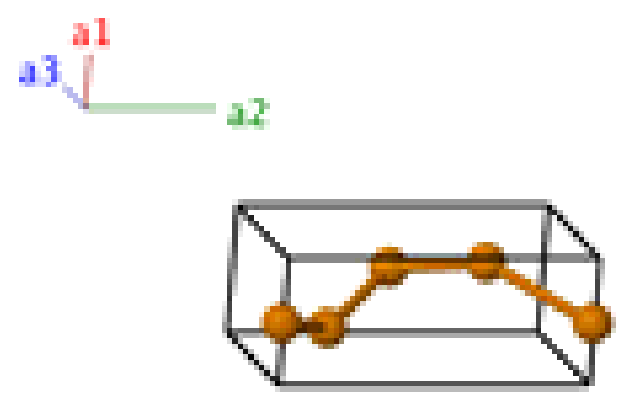

---

**Basis vectors:**

| | | Lattice Coordinates | | Cartesian Coordinates | Wyckoff Position | Atom Type |
|---|---|---|---|---|---|---|
| $\mathbf{B}_1$ | = | $x_1\,\mathbf{a}_1 + y_1\,\mathbf{a}_2 + z_1\,\mathbf{a}_3$ | = | $(x_1\,a + z_1\,c\,\cos\beta)\,\hat{\mathbf{x}} + y_1\,b\,\hat{\mathbf{y}} + z_1\,c\,\sin\beta\,\hat{\mathbf{z}}$ | (2a) | Te I |
| $\mathbf{B}_2$ | = | $-x_1\,\mathbf{a}_1 + \left(\frac{1}{2}+y_1\right)\mathbf{a}_2 - z_1\,\mathbf{a}_3$ | = | $-(x_1\,a + z_1\,c\,\cos\beta)\,\hat{\mathbf{x}} + \left(\frac{1}{2}+y_1\right)b\,\hat{\mathbf{y}} - z_1\,c\,\sin\beta\,\hat{\mathbf{z}}$ | (2a) | Te I |
| $\mathbf{B}_3$ | = | $x_2\,\mathbf{a}_1 + y_2\,\mathbf{a}_2 + z_2\,\mathbf{a}_3$ | = | $(x_2\,a + z_2\,c\,\cos\beta)\,\hat{\mathbf{x}} + y_2\,b\,\hat{\mathbf{y}} + z_2\,c\,\sin\beta\,\hat{\mathbf{z}}$ | (2a) | Te II |
| $\mathbf{B}_4$ | = | $-x_2\,\mathbf{a}_1 + \left(\frac{1}{2}+y_2\right)\mathbf{a}_2 - z_2\,\mathbf{a}_3$ | = | $-(x_2\,a + z_2\,c\,\cos\beta)\,\hat{\mathbf{x}} + \left(\frac{1}{2}+y_2\right)b\,\hat{\mathbf{y}} - z_2\,c\,\sin\beta\,\hat{\mathbf{z}}$ | (2a) | Te II |

**Geometry files:**

- CIF: pp. 643
- POSCAR: pp. 643

# Po (A19) Structure: A_mC12_5_3c

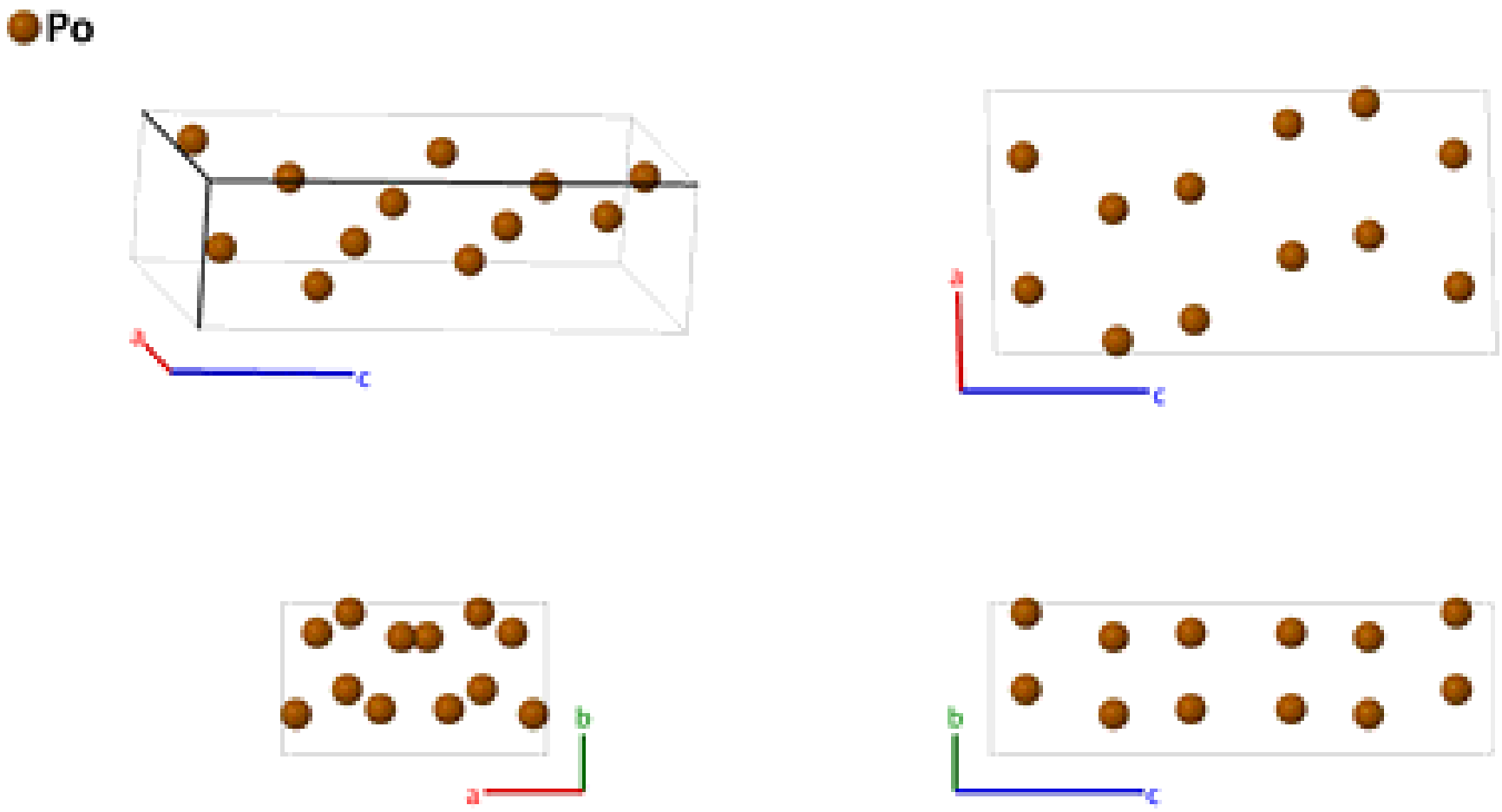

**Prototype** : Po

**AFLOW prototype label** : A_mC12_5_3c

***Strukturbericht* designation** : A19

**Pearson symbol** : mC12

**Space group number** : 5

**Space group symbol** : C2

**AFLOW prototype command** : `aflow --proto=A_mC12_5_3c`
`--params=`$a, b/a, c/a, \beta, x_1, y_1, z_1, x_2, y_2, z_2, x_3, y_3, z_3$

---

- This was the original determination of the structure of Po, and given the Strukturbericht designation A19. (Gottfried, 1938, pp. 4-5). Eventually it was determined that the sample used here was a mixture of $\alpha$-Po ($A_h$) and $\beta$-Po ($A_i$) (Donohue, 1982, pp. 390). We retain the A19 page for historical interest.

---

**Base-centered Monoclinic primitive vectors:**

$$\mathbf{a}_1 = \frac{1}{2}\,a\,\hat{\mathbf{x}} - \frac{1}{2}\,b\,\hat{\mathbf{y}}$$

$$\mathbf{a}_2 = \frac{1}{2}\,a\,\hat{\mathbf{x}} + \frac{1}{2}\,b\,\hat{\mathbf{y}}$$

$$\mathbf{a}_3 = c\cos\beta\,\hat{\mathbf{x}} + c\sin\beta\,\hat{\mathbf{z}}$$

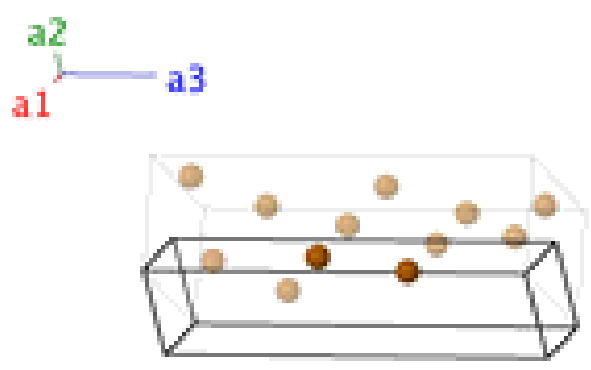

---

**Basis vectors:**

| | | Lattice Coordinates | | Cartesian Coordinates | Wyckoff Position | Atom Type |
|---|---|---|---|---|---|---|
| $\mathbf{B}_1$ | = | $(x_1 - y_1)\,\mathbf{a}_1 + (x_1 + y_1)\,\mathbf{a}_2 + z_1\,\mathbf{a}_3$ | = | $(x_1\,a + z_1\,c\,\cos\beta)\,\hat{\mathbf{x}} + y_1\,b\,\hat{\mathbf{y}} + z_1\,c\,\sin\beta\,\hat{\mathbf{z}}$ | (4$c$) | Po I |
| $\mathbf{B}_2$ | = | $-(x_1 + y_1)\,\mathbf{a}_1 + (y_1 - x_1)\,\mathbf{a}_2 - z_1\,\mathbf{a}_3$ | = | $-(x_1\,a + z_1\,c\,\cos\beta)\,\hat{\mathbf{x}} + y_1\,b\,\hat{\mathbf{y}} - z_1\,c\,\sin\beta\,\hat{\mathbf{z}}$ | (4$c$) | Po I |
| $\mathbf{B}_3$ | = | $(x_2 - y_2)\,\mathbf{a}_1 + (x_2 + y_2)\,\mathbf{a}_2 + z_2\,\mathbf{a}_3$ | = | $(x_2\,a + z_2\,c\,\cos\beta)\,\hat{\mathbf{x}} + y_2\,b\,\hat{\mathbf{y}} + z_2\,c\,\sin\beta\,\hat{\mathbf{z}}$ | (4$c$) | Po II |
| $\mathbf{B_4}$ | = | $-(x_2+y_2)\,\mathbf{a_1}+(y_2-x_2)\,\mathbf{a_2}-z_2\,\mathbf{a_3}$ | = | $-(x_2\,a+z_2\,c\,\cos\beta)\,\hat{\mathbf{x}}+y_2\,b\,\hat{\mathbf{y}}-z_2\,c\,\sin\beta\,\hat{\mathbf{z}}$ | (4$c$) | Po II |
| $\mathbf{B_5}$ | = | $(x_3-y_3)\,\mathbf{a_1}+(x_3+y_3)\,\mathbf{a_2}+z_3\,\mathbf{a_3}$ | = | $(x_3\,a+z_3\,c\,\cos\beta)\,\hat{\mathbf{x}}+y_3\,b\,\hat{\mathbf{y}}+z_3\,c\,\sin\beta\,\hat{\mathbf{z}}$ | (4$c$) | Po III |
| $\mathbf{B_6}$ | = | $-(x_3+y_3)\,\mathbf{a_1}+(y_3-x_3)\,\mathbf{a_2}-z_3\,\mathbf{a_3}$ | = | $-(x_3\,a+z_3\,c\,\cos\beta)\,\hat{\mathbf{x}}+y_3\,b\,\hat{\mathbf{y}}-z_3\,c\,\sin\beta\,\hat{\mathbf{z}}$ | (4$c$) | Po III |

---

**Found in:**

---

**Geometry files:**

- CIF: pp. 643
- POSCAR: pp. 644

# Monoclinic PZT [$Pb(Zr_xTi_{1-x})O_3$] Structure: A3BC_mC10_8_ab_a_a

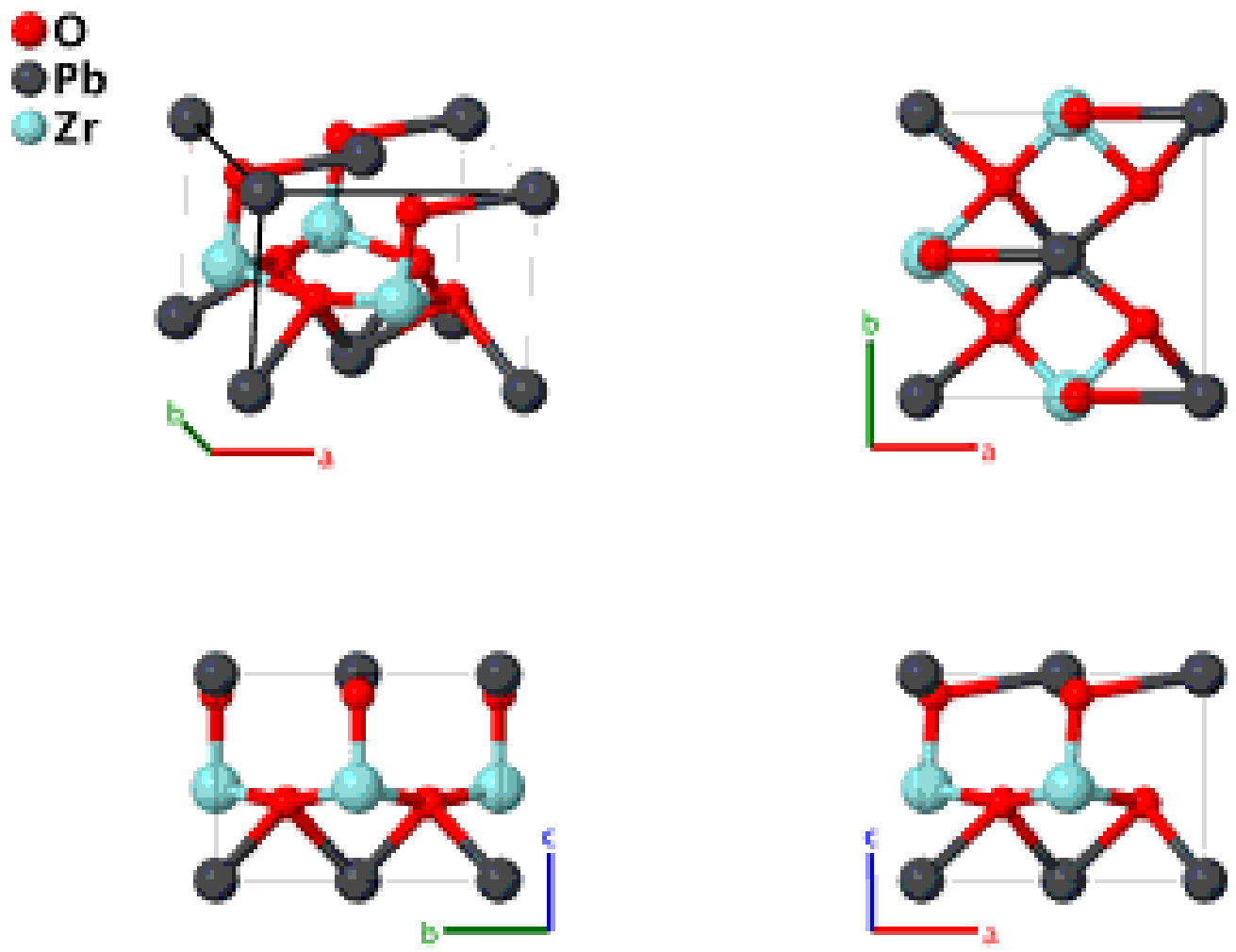

| | | |
|---|---|---|
| **Prototype** | : | $Pb(Zr_{0.52}Ti_{0.48})O_3$ |
| **AFLOW prototype label** | : | A3BC_mC10_8_ab_a_a |
| ***Strukturbericht* designation** | : | None |
| **Pearson symbol** | : | mC10 |
| **Space group number** | : | 8 |
| **Space group symbol** | : | Cm |
| **AFLOW prototype command** | : | `aflow --proto=A3BC_mC10_8_ab_a_a`<br>`--params=`$a, b/a, c/a, \beta, x_1, z_1, x_2, z_2, x_3, z_3, x_4, y_4, z_4$ |

- This is a monoclinic ferroelectric distortion of the perovskite structure. In $Pb(Zr_xTi_{1-x})O_3$ (aka PZT) it is found only when $x = 0.52$. Although the second (2a) site is nearly equally occupied by Zr and Ti atoms, the pictures use Zr atoms. Compare this to the tetragonal PZT structure.

**Base-centered Monoclinic primitive vectors:**

$$\mathbf{a}_1 = \frac{1}{2}\,a\,\hat{\mathbf{x}} - \frac{1}{2}\,b\,\hat{\mathbf{y}}$$
$$\mathbf{a}_2 = \frac{1}{2}\,a\,\hat{\mathbf{x}} + \frac{1}{2}\,b\,\hat{\mathbf{y}}$$
$$\mathbf{a}_3 = c\cos\beta\,\hat{\mathbf{x}} + c\sin\beta\,\hat{\mathbf{z}}$$

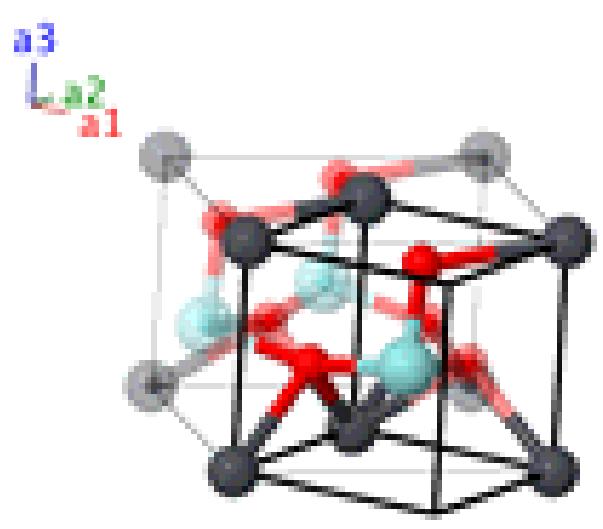

**Basis vectors:**

| | | Lattice Coordinates | | Cartesian Coordinates | Wyckoff Position | Atom Type |
|---|---|---|---|---|---|---|
| $\mathbf{B_1}$ | = | $x_1\,\mathbf{a_1} + x_1\,\mathbf{a_2} + z_1\,\mathbf{a_3}$ | = | $(x_1\,a + z_1\,c\,\cos\beta)\,\hat{\mathbf{x}} + z_1\,c\,\sin\beta\,\hat{\mathbf{z}}$ | (2a) | O I |
| $\mathbf{B_2}$ | = | $x_2\,\mathbf{a_1} + x_2\,\mathbf{a_2} + z_2\,\mathbf{a_3}$ | = | $(x_2\,a + z_2\,c\,\cos\beta)\,\hat{\mathbf{x}} + z_2\,c\,\sin\beta\,\hat{\mathbf{z}}$ | (2a) | Pb |
| $\mathbf{B_3}$ | = | $x_3\,\mathbf{a_1} + x_3\,\mathbf{a_2} + z_3\,\mathbf{a_3}$ | = | $(x_3\,a + z_3\,c\,\cos\beta)\,\hat{\mathbf{x}} + z_3\,c\,\sin\beta\,\hat{\mathbf{z}}$ | (2$a$) | Zr |
| $\mathbf{B_4}$ | = | $(x_4 - y_4)\,\mathbf{a_1} + (x_4 + y_4)\,\mathbf{a_2} + z_4\,\mathbf{a_3}$ | = | $(x_4\,a + z_4\,c\,\cos\beta)\,\hat{\mathbf{x}} + y_4\,b\,\hat{\mathbf{y}} + z_4\,c\,\sin\beta\,\hat{\mathbf{z}}$ | (4$b$) | O II |
| $\mathbf{B_5}$ | = | $(x_4 + y_4)\,\mathbf{a_1} + (x_4 - y_4)\,\mathbf{a_2} + z_4\,\mathbf{a_3}$ | = | $(x_4\,a + z_4\,c\,\cos\beta)\,\hat{\mathbf{x}} - y_4\,b\,\hat{\mathbf{y}} + z_4\,c\,\sin\beta\,\hat{\mathbf{z}}$ | (4$b$) | O II |

---

---

**Geometry files:**

- CIF: pp. 644
- POSCAR: pp. 644

# Monoclinic (Cc) Low Tridymite ($SiO_2$) Structure: A2B_mC144_9_24a_12a

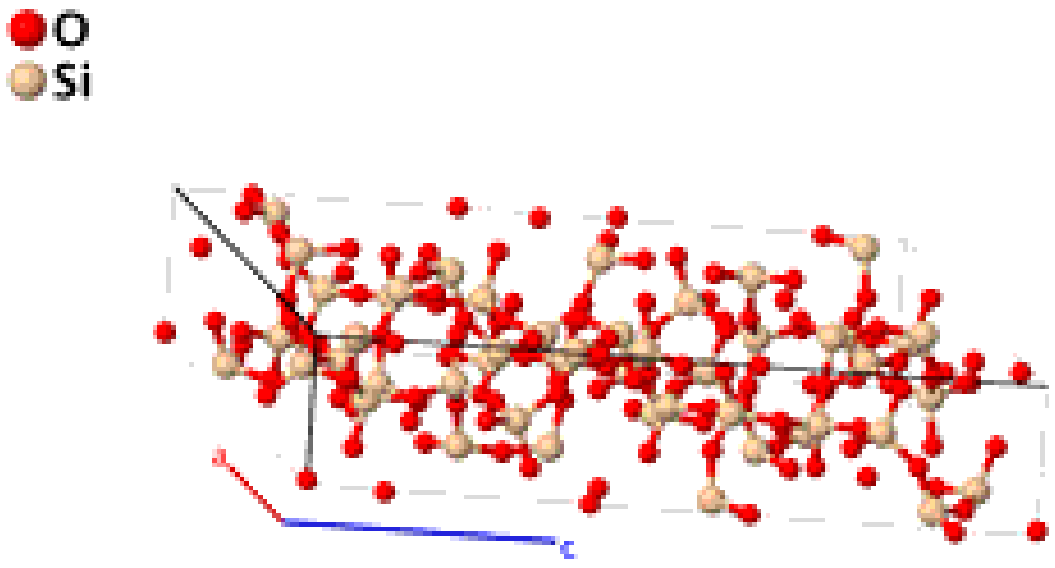

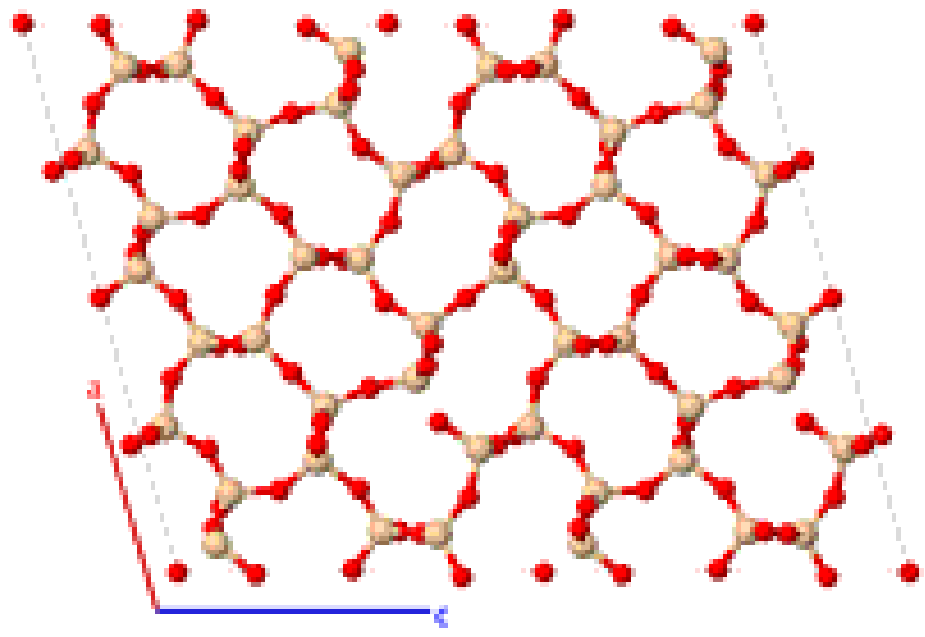

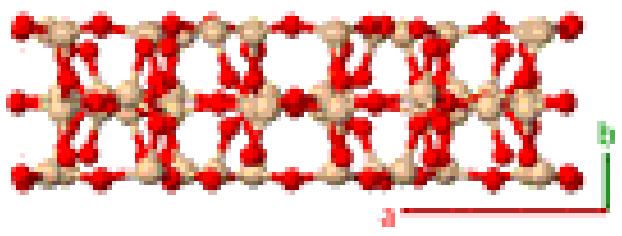

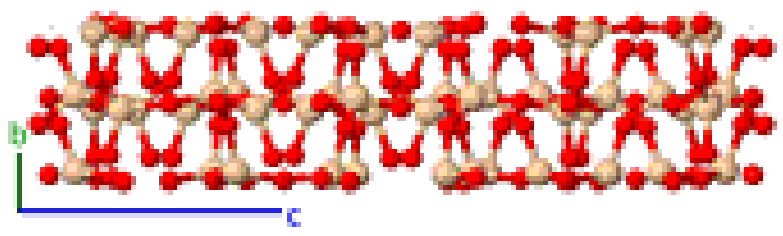

| | | |
|---|---|---|
| **Prototype** | : | $SiO_2$ |
| **AFLOW prototype label** | : | A2B_mC144_9_24a_12a |
| ***Strukturbericht* designation** | : | None |
| **Pearson symbol** | : | mC144 |
| **Space group number** | : | 9 |
| **Space group symbol** | : | Cc |
| **AFLOW prototype command** | : | `aflow --proto=A2B_mC144_9_24a_12a` `--params=`$a, b/a, c/a, \beta, x_1, y_1, z_1, x_2, y_2, z_2, x_3, y_3, z_3, x_4, y_4, z_4, x_5, y_5, z_5, x_6, y_6, z_6, x_7, y_7, z_7, x_8, y_8, z_8, x_9, y_9, z_9, x_{10}, y_{10}, z_{10}, x_{11}, y_{11}, z_{11}, x_{12}, y_{12}, z_{12}, x_{13}, y_{13}, z_{13}, x_{14}, y_{14}, z_{14}, x_{15}, y_{15}, z_{15}, x_{16}, y_{16}, z_{16}, x_{17}, y_{17}, z_{17}, x_{18}, y_{18}, z_{18}, x_{19}, y_{19}, z_{19}, x_{20}, y_{20}, z_{20}, x_{21}, y_{21}, z_{21}, x_{22}, y_{22}, z_{22}, x_{23}, y_{23}, z_{23}, x_{24}, y_{24}, z_{24}, x_{25}, y_{25}, z_{25}, x_{26}, y_{26}, z_{26}, x_{27}, y_{27}, z_{27}, x_{28}, y_{28}, z_{28}, x_{29}, y_{29}, z_{29}, x_{30}, y_{30}, z_{30}, x_{31}, y_{31}, z_{31}, x_{32}, y_{32}, z_{32}, x_{33}, y_{33}, z_{33}, x_{34}, y_{34}, z_{34}, x_{35}, y_{35}, z_{35}, x_{36}, y_{36}, z_{36}$ |

**Base-centered Monoclinic primitive vectors:**

$$\mathbf{a}_1 = \frac{1}{2}\,a\,\hat{\mathbf{x}} - \frac{1}{2}\,b\,\hat{\mathbf{y}}$$
$$\mathbf{a}_2 = \frac{1}{2}\,a\,\hat{\mathbf{x}} + \frac{1}{2}\,b\,\hat{\mathbf{y}}$$
$$\mathbf{a}_3 = c\cos\beta\,\hat{\mathbf{x}} + c\sin\beta\,\hat{\mathbf{z}}$$

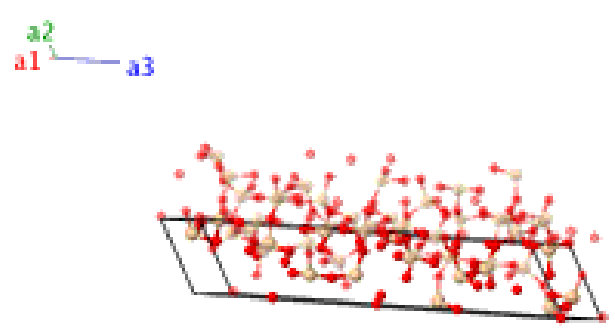

**Basis vectors:**

| | | Lattice Coordinates | | Cartesian Coordinates | Wyckoff Position | Atom Type |
|---|---|---|---|---|---|---|
| $\mathbf{B_1}$ | = | $(x_1 - y_1)\,\mathbf{a_1} + (x_1 + y_1)\,\mathbf{a_2} + z_1\,\mathbf{a_3}$ | = | $(x_1\,a + z_1\,c\,\cos\beta)\,\mathbf{\hat{x}} + y_1\,b\,\mathbf{\hat{y}} + z_1\,c\,\sin\beta\,\mathbf{\hat{z}}$ | (4*a*) | O I |
| $\mathbf{B_2}$ | = | $(x_1 + y_1)\,\mathbf{a_1} + (x_1 - y_1)\,\mathbf{a_2} + \left(\frac{1}{2} + z_1\right)\mathbf{a_3}$ | = | $\left(x_1\,a + \left(\frac{1}{2} + z_1\right)c\,\cos\beta\right)\mathbf{\hat{x}} - y_1\,b\,\mathbf{\hat{y}} + \left(\frac{1}{2} + z_1\right)c\,\sin\beta\,\mathbf{\hat{z}}$ | (4*a*) | O I |
| $\mathbf{B_3}$ | = | $(x_2 - y_2)\,\mathbf{a_1} + (x_2 + y_2)\,\mathbf{a_2} + z_2\,\mathbf{a_3}$ | = | $(x_2\,a + z_2\,c\,\cos\beta)\,\mathbf{\hat{x}} + y_2\,b\,\mathbf{\hat{y}} + z_2\,c\,\sin\beta\,\mathbf{\hat{z}}$ | (4*a*) | O II |
| $\mathbf{B_4}$ | = | $(x_2 + y_2)\,\mathbf{a_1} + (x_2 - y_2)\,\mathbf{a_2} + \left(\frac{1}{2} + z_2\right)\mathbf{a_3}$ | = | $\left(x_2\,a + \left(\frac{1}{2} + z_2\right)c\,\cos\beta\right)\mathbf{\hat{x}} - y_2\,b\,\mathbf{\hat{y}} + \left(\frac{1}{2} + z_2\right)c\,\sin\beta\,\mathbf{\hat{z}}$ | (4*a*) | O II |
| $\mathbf{B_5}$ | = | $(x_3 - y_3)\,\mathbf{a_1} + (x_3 + y_3)\,\mathbf{a_2} + z_3\,\mathbf{a_3}$ | = | $(x_3\,a + z_3\,c\,\cos\beta)\,\mathbf{\hat{x}} + y_3\,b\,\mathbf{\hat{y}} + z_3\,c\,\sin\beta\,\mathbf{\hat{z}}$ | (4*a*) | O III |
| $\mathbf{B_6}$ | = | $(x_3 + y_3)\,\mathbf{a_1} + (x_3 - y_3)\,\mathbf{a_2} + \left(\frac{1}{2} + z_3\right)\mathbf{a_3}$ | = | $\left(x_3\,a + \left(\frac{1}{2} + z_3\right)c\,\cos\beta\right)\mathbf{\hat{x}} - y_3\,b\,\mathbf{\hat{y}} + \left(\frac{1}{2} + z_3\right)c\,\sin\beta\,\mathbf{\hat{z}}$ | (4*a*) | O III |
| $\mathbf{B_7}$ | = | $(x_4 - y_4)\,\mathbf{a_1} + (x_4 + y_4)\,\mathbf{a_2} + z_4\,\mathbf{a_3}$ | = | $(x_4\,a + z_4\,c\,\cos\beta)\,\mathbf{\hat{x}} + y_4\,b\,\mathbf{\hat{y}} + z_4\,c\,\sin\beta\,\mathbf{\hat{z}}$ | (4*a*) | O IV |
| $\mathbf{B_8}$ | = | $(x_4 + y_4)\,\mathbf{a_1} + (x_4 - y_4)\,\mathbf{a_2} + \left(\frac{1}{2} + z_4\right)\mathbf{a_3}$ | = | $\left(x_4\,a + \left(\frac{1}{2} + z_4\right)c\,\cos\beta\right)\mathbf{\hat{x}} - y_4\,b\,\mathbf{\hat{y}} + \left(\frac{1}{2} + z_4\right)c\,\sin\beta\,\mathbf{\hat{z}}$ | (4*a*) | O IV |
| $\mathbf{B_9}$ | = | $(x_5 - y_5)\,\mathbf{a_1} + (x_5 + y_5)\,\mathbf{a_2} + z_5\,\mathbf{a_3}$ | = | $(x_5\,a + z_5\,c\,\cos\beta)\,\mathbf{\hat{x}} + y_5\,b\,\mathbf{\hat{y}} + z_5\,c\,\sin\beta\,\mathbf{\hat{z}}$ | (4*a*) | O V |
| $\mathbf{B_{10}}$ | = | $(x_5 + y_5)\,\mathbf{a_1} + (x_5 - y_5)\,\mathbf{a_2} + \left(\frac{1}{2} + z_5\right)\mathbf{a_3}$ | = | $\left(x_5\,a + \left(\frac{1}{2} + z_5\right)c\,\cos\beta\right)\mathbf{\hat{x}} - y_5\,b\,\mathbf{\hat{y}} + \left(\frac{1}{2} + z_5\right)c\,\sin\beta\,\mathbf{\hat{z}}$ | (4*a*) | O V |
| $\mathbf{B_{11}}$ | = | $(x_6 - y_6)\,\mathbf{a_1} + (x_6 + y_6)\,\mathbf{a_2} + z_6\,\mathbf{a_3}$ | = | $(x_6\,a + z_6\,c\,\cos\beta)\,\mathbf{\hat{x}} + y_6\,b\,\mathbf{\hat{y}} + z_6\,c\,\sin\beta\,\mathbf{\hat{z}}$ | (4*a*) | O VI |
| $\mathbf{B_{12}}$ | = | $(x_6 + y_6)\,\mathbf{a_1} + (x_6 - y_6)\,\mathbf{a_2} + \left(\frac{1}{2} + z_6\right)\mathbf{a_3}$ | = | $\left(x_6\,a + \left(\frac{1}{2} + z_6\right)c\,\cos\beta\right)\mathbf{\hat{x}} - y_6\,b\,\mathbf{\hat{y}} + \left(\frac{1}{2} + z_6\right)c\,\sin\beta\,\mathbf{\hat{z}}$ | (4*a*) | O VI |
| $\mathbf{B_{13}}$ | = | $(x_7 - y_7)\,\mathbf{a_1} + (x_7 + y_7)\,\mathbf{a_2} + z_7\,\mathbf{a_3}$ | = | $(x_7\,a + z_7\,c\,\cos\beta)\,\mathbf{\hat{x}} + y_7\,b\,\mathbf{\hat{y}} + z_7\,c\,\sin\beta\,\mathbf{\hat{z}}$ | (4*a*) | O VII |
| $\mathbf{B_{14}}$ | = | $(x_7 + y_7)\,\mathbf{a_1} + (x_7 - y_7)\,\mathbf{a_2} + \left(\frac{1}{2} + z_7\right)\mathbf{a_3}$ | = | $\left(x_7\,a + \left(\frac{1}{2} + z_7\right)c\,\cos\beta\right)\mathbf{\hat{x}} - y_7\,b\,\mathbf{\hat{y}} + \left(\frac{1}{2} + z_7\right)c\,\sin\beta\,\mathbf{\hat{z}}$ | (4*a*) | O VII |
| $\mathbf{B_{15}}$ | = | $(x_8 - y_8)\,\mathbf{a_1} + (x_8 + y_8)\,\mathbf{a_2} + z_8\,\mathbf{a_3}$ | = | $(x_8\,a + z_8\,c\,\cos\beta)\,\mathbf{\hat{x}} + y_8\,b\,\mathbf{\hat{y}} + z_8\,c\,\sin\beta\,\mathbf{\hat{z}}$ | (4*a*) | O VIII |
| $\mathbf{B_{16}}$ | = | $(x_8 + y_8)\,\mathbf{a_1} + (x_8 - y_8)\,\mathbf{a_2} + \left(\frac{1}{2} + z_8\right)\mathbf{a_3}$ | = | $\left(x_8\,a + \left(\frac{1}{2} + z_8\right)c\,\cos\beta\right)\mathbf{\hat{x}} - y_8\,b\,\mathbf{\hat{y}} + \left(\frac{1}{2} + z_8\right)c\,\sin\beta\,\mathbf{\hat{z}}$ | (4*a*) | O VIII |
| $\mathbf{B_{17}}$ | = | $(x_9 - y_9)\,\mathbf{a_1} + (x_9 + y_9)\,\mathbf{a_2} + z_9\,\mathbf{a_3}$ | = | $(x_9\,a + z_9\,c\,\cos\beta)\,\mathbf{\hat{x}} + y_9\,b\,\mathbf{\hat{y}} + z_9\,c\,\sin\beta\,\mathbf{\hat{z}}$ | (4*a*) | O IX |
| $\mathbf{B_{18}}$ | = | $(x_9 + y_9)\,\mathbf{a_1} + (x_9 - y_9)\,\mathbf{a_2} + \left(\frac{1}{2} + z_9\right)\mathbf{a_3}$ | = | $\left(x_9\,a + \left(\frac{1}{2} + z_9\right)c\,\cos\beta\right)\mathbf{\hat{x}} - y_9\,b\,\mathbf{\hat{y}} + \left(\frac{1}{2} + z_9\right)c\,\sin\beta\,\mathbf{\hat{z}}$ | (4*a*) | O IX |
| $\mathbf{B_{19}}$ | = | $(x_{10} - y_{10})\,\mathbf{a_1} + (x_{10} + y_{10})\,\mathbf{a_2} + z_{10}\,\mathbf{a_3}$ | = | $(x_{10}\,a + z_{10}\,c\,\cos\beta)\,\mathbf{\hat{x}} + y_{10}\,b\,\mathbf{\hat{y}} + z_{10}\,c\,\sin\beta\,\mathbf{\hat{z}}$ | (4*a*) | O X |
| $\mathbf{B_{20}}$ | = | $(x_{10} + y_{10})\,\mathbf{a_1} + (x_{10} - y_{10})\,\mathbf{a_2} + \left(\frac{1}{2} + z_{10}\right)\mathbf{a_3}$ | = | $\left(x_{10}\,a + \left(\frac{1}{2} + z_{10}\right)c\,\cos\beta\right)\mathbf{\hat{x}} - y_{10}\,b\,\mathbf{\hat{y}} + \left(\frac{1}{2} + z_{10}\right)c\,\sin\beta\,\mathbf{\hat{z}}$ | (4*a*) | O X |
| $\mathbf{B_{21}}$ | = | $(x_{11} - y_{11})\,\mathbf{a_1} + (x_{11} + y_{11})\,\mathbf{a_2} + z_{11}\,\mathbf{a_3}$ | = | $(x_{11}\,a + z_{11}\,c\,\cos\beta)\,\mathbf{\hat{x}} + y_{11}\,b\,\mathbf{\hat{y}} + z_{11}\,c\,\sin\beta\,\mathbf{\hat{z}}$ | (4*a*) | O XI |
| $\mathbf{B}_{22}$ | = | $(x_{11}+y_{11})\,\mathbf{a}_1 + (x_{11}-y_{11})\,\mathbf{a}_2 + \left(\frac{1}{2}+z_{11}\right)\,\mathbf{a}_3$ | = | $\left(x_{11}\,a + \left(\frac{1}{2}+z_{11}\right)\,c\,\cos\beta\right)\,\hat{\mathbf{x}} - y_{11}\,b\,\hat{\mathbf{y}} + \left(\frac{1}{2}+z_{11}\right)\,c\,\sin\beta\,\hat{\mathbf{z}}$ | (4$a$) | O XI |
| $\mathbf{B}_{23}$ | = | $(x_{12}-y_{12})\,\mathbf{a}_1 + (x_{12}+y_{12})\,\mathbf{a}_2 + z_{12}\,\mathbf{a}_3$ | = | $(x_{12}\,a + z_{12}\,c\,\cos\beta)\,\hat{\mathbf{x}} + y_{12}\,b\,\hat{\mathbf{y}} + z_{12}\,c\,\sin\beta\,\hat{\mathbf{z}}$ | (4$a$) | O XII |
| $\mathbf{B}_{24}$ | = | $(x_{12}+y_{12})\,\mathbf{a}_1 + (x_{12}-y_{12})\,\mathbf{a}_2 + \left(\frac{1}{2}+z_{12}\right)\,\mathbf{a}_3$ | = | $\left(x_{12}\,a + \left(\frac{1}{2}+z_{12}\right)\,c\,\cos\beta\right)\,\hat{\mathbf{x}} - y_{12}\,b\,\hat{\mathbf{y}} + \left(\frac{1}{2}+z_{12}\right)\,c\,\sin\beta\,\hat{\mathbf{z}}$ | (4$a$) | O XII |
| $\mathbf{B}_{25}$ | = | $(x_{13}-y_{13})\,\mathbf{a}_1 + (x_{13}+y_{13})\,\mathbf{a}_2 + z_{13}\,\mathbf{a}_3$ | = | $(x_{13}\,a + z_{13}\,c\,\cos\beta)\,\hat{\mathbf{x}} + y_{13}\,b\,\hat{\mathbf{y}} + z_{13}\,c\,\sin\beta\,\hat{\mathbf{z}}$ | (4$a$) | O XIII |
| $\mathbf{B}_{26}$ | = | $(x_{13}+y_{13})\,\mathbf{a}_1 + (x_{13}-y_{13})\,\mathbf{a}_2 + \left(\frac{1}{2}+z_{13}\right)\,\mathbf{a}_3$ | = | $\left(x_{13}\,a + \left(\frac{1}{2}+z_{13}\right)\,c\,\cos\beta\right)\,\hat{\mathbf{x}} - y_{13}\,b\,\hat{\mathbf{y}} + \left(\frac{1}{2}+z_{13}\right)\,c\,\sin\beta\,\hat{\mathbf{z}}$ | (4$a$) | O XIII |
| $\mathbf{B}_{27}$ | = | $(x_{14}-y_{14})\,\mathbf{a}_1 + (x_{14}+y_{14})\,\mathbf{a}_2 + z_{14}\,\mathbf{a}_3$ | = | $(x_{14}\,a + z_{14}\,c\,\cos\beta)\,\hat{\mathbf{x}} + y_{14}\,b\,\hat{\mathbf{y}} + z_{14}\,c\,\sin\beta\,\hat{\mathbf{z}}$ | (4$a$) | O XIV |
| $\mathbf{B}_{28}$ | = | $(x_{14}+y_{14})\,\mathbf{a}_1 + (x_{14}-y_{14})\,\mathbf{a}_2 + \left(\frac{1}{2}+z_{14}\right)\,\mathbf{a}_3$ | = | $\left(x_{14}\,a + \left(\frac{1}{2}+z_{14}\right)\,c\,\cos\beta\right)\,\hat{\mathbf{x}} - y_{14}\,b\,\hat{\mathbf{y}} + \left(\frac{1}{2}+z_{14}\right)\,c\,\sin\beta\,\hat{\mathbf{z}}$ | (4$a$) | O XIV |
| $\mathbf{B}_{29}$ | = | $(x_{15}-y_{15})\,\mathbf{a}_1 + (x_{15}+y_{15})\,\mathbf{a}_2 + z_{15}\,\mathbf{a}_3$ | = | $(x_{15}\,a + z_{15}\,c\,\cos\beta)\,\hat{\mathbf{x}} + y_{15}\,b\,\hat{\mathbf{y}} + z_{15}\,c\,\sin\beta\,\hat{\mathbf{z}}$ | (4$a$) | O XV |
| $\mathbf{B}_{30}$ | = | $(x_{15}+y_{15})\,\mathbf{a}_1 + (x_{15}-y_{15})\,\mathbf{a}_2 + \left(\frac{1}{2}+z_{15}\right)\,\mathbf{a}_3$ | = | $\left(x_{15}\,a + \left(\frac{1}{2}+z_{15}\right)\,c\,\cos\beta\right)\,\hat{\mathbf{x}} - y_{15}\,b\,\hat{\mathbf{y}} + \left(\frac{1}{2}+z_{15}\right)\,c\,\sin\beta\,\hat{\mathbf{z}}$ | (4$a$) | O XV |
| $\mathbf{B}_{31}$ | = | $(x_{16}-y_{16})\,\mathbf{a}_1 + (x_{16}+y_{16})\,\mathbf{a}_2 + z_{16}\,\mathbf{a}_3$ | = | $(x_{16}\,a + z_{16}\,c\,\cos\beta)\,\hat{\mathbf{x}} + y_{16}\,b\,\hat{\mathbf{y}} + z_{16}\,c\,\sin\beta\,\hat{\mathbf{z}}$ | (4$a$) | O XVI |
| $\mathbf{B}_{32}$ | = | $(x_{16}+y_{16})\,\mathbf{a}_1 + (x_{16}-y_{16})\,\mathbf{a}_2 + \left(\frac{1}{2}+z_{16}\right)\,\mathbf{a}_3$ | = | $\left(x_{16}\,a + \left(\frac{1}{2}+z_{16}\right)\,c\,\cos\beta\right)\,\hat{\mathbf{x}} - y_{16}\,b\,\hat{\mathbf{y}} + \left(\frac{1}{2}+z_{16}\right)\,c\,\sin\beta\,\hat{\mathbf{z}}$ | (4$a$) | O XVI |
| $\mathbf{B}_{33}$ | = | $(x_{17}-y_{17})\,\mathbf{a}_1 + (x_{17}+y_{17})\,\mathbf{a}_2 + z_{17}\,\mathbf{a}_3$ | = | $(x_{17}\,a + z_{17}\,c\,\cos\beta)\,\hat{\mathbf{x}} + y_{17}\,b\,\hat{\mathbf{y}} + z_{17}\,c\,\sin\beta\,\hat{\mathbf{z}}$ | (4$a$) | O XVII |
| $\mathbf{B}_{34}$ | = | $(x_{17}+y_{17})\,\mathbf{a}_1 + (x_{17}-y_{17})\,\mathbf{a}_2 + \left(\frac{1}{2}+z_{17}\right)\,\mathbf{a}_3$ | = | $\left(x_{17}\,a + \left(\frac{1}{2}+z_{17}\right)\,c\,\cos\beta\right)\,\hat{\mathbf{x}} - y_{17}\,b\,\hat{\mathbf{y}} + \left(\frac{1}{2}+z_{17}\right)\,c\,\sin\beta\,\hat{\mathbf{z}}$ | (4$a$) | O XVII |
| $\mathbf{B}_{35}$ | = | $(x_{18}-y_{18})\,\mathbf{a}_1 + (x_{18}+y_{18})\,\mathbf{a}_2 + z_{18}\,\mathbf{a}_3$ | = | $(x_{18}\,a + z_{18}\,c\,\cos\beta)\,\hat{\mathbf{x}} + y_{18}\,b\,\hat{\mathbf{y}} + z_{18}\,c\,\sin\beta\,\hat{\mathbf{z}}$ | (4$a$) | O XVIII |
| $\mathbf{B}_{36}$ | = | $(x_{18}+y_{18})\,\mathbf{a}_1 + (x_{18}-y_{18})\,\mathbf{a}_2 + \left(\frac{1}{2}+z_{18}\right)\,\mathbf{a}_3$ | = | $\left(x_{18}\,a + \left(\frac{1}{2}+z_{18}\right)\,c\,\cos\beta\right)\,\hat{\mathbf{x}} - y_{18}\,b\,\hat{\mathbf{y}} + \left(\frac{1}{2}+z_{18}\right)\,c\,\sin\beta\,\hat{\mathbf{z}}$ | (4$a$) | O XVIII |
| $\mathbf{B}_{37}$ | = | $(x_{19}-y_{19})\,\mathbf{a}_1 + (x_{19}+y_{19})\,\mathbf{a}_2 + z_{19}\,\mathbf{a}_3$ | = | $(x_{19}\,a + z_{19}\,c\,\cos\beta)\,\hat{\mathbf{x}} + y_{19}\,b\,\hat{\mathbf{y}} + z_{19}\,c\,\sin\beta\,\hat{\mathbf{z}}$ | (4$a$) | O XIX |
| $\mathbf{B}_{38}$ | = | $(x_{19}+y_{19})\,\mathbf{a}_1 + (x_{19}-y_{19})\,\mathbf{a}_2 + \left(\frac{1}{2}+z_{19}\right)\,\mathbf{a}_3$ | = | $\left(x_{19}\,a + \left(\frac{1}{2}+z_{19}\right)\,c\,\cos\beta\right)\,\hat{\mathbf{x}} - y_{19}\,b\,\hat{\mathbf{y}} + \left(\frac{1}{2}+z_{19}\right)\,c\,\sin\beta\,\hat{\mathbf{z}}$ | (4$a$) | O XIX |
| $\mathbf{B}_{39}$ | = | $(x_{20}-y_{20})\,\mathbf{a}_1 + (x_{20}+y_{20})\,\mathbf{a}_2 + z_{20}\,\mathbf{a}_3$ | = | $(x_{20}\,a + z_{20}\,c\,\cos\beta)\,\hat{\mathbf{x}} + y_{20}\,b\,\hat{\mathbf{y}} + z_{20}\,c\,\sin\beta\,\hat{\mathbf{z}}$ | (4$a$) | O XX |
| $\mathbf{B}_{40}$ | = | $(x_{20}+y_{20})\,\mathbf{a}_1 + (x_{20}-y_{20})\,\mathbf{a}_2 + \left(\frac{1}{2}+z_{20}\right)\,\mathbf{a}_3$ | = | $\left(x_{20}\,a + \left(\frac{1}{2}+z_{20}\right)\,c\,\cos\beta\right)\,\hat{\mathbf{x}} - y_{20}\,b\,\hat{\mathbf{y}} + \left(\frac{1}{2}+z_{20}\right)\,c\,\sin\beta\,\hat{\mathbf{z}}$ | (4$a$) | O XX |
| $\mathbf{B}_{41}$ | = | $(x_{21}-y_{21})\,\mathbf{a}_1 + (x_{21}+y_{21})\,\mathbf{a}_2 + z_{21}\,\mathbf{a}_3$ | = | $(x_{21}\,a + z_{21}\,c\,\cos\beta)\,\hat{\mathbf{x}} + y_{21}\,b\,\hat{\mathbf{y}} + z_{21}\,c\,\sin\beta\,\hat{\mathbf{z}}$ | (4$a$) | O XXI |
| $\mathbf{B}_{42}$ | = | $(x_{21}+y_{21})\,\mathbf{a}_1 + (x_{21}-y_{21})\,\mathbf{a}_2 + \left(\frac{1}{2}+z_{21}\right)\,\mathbf{a}_3$ | = | $\left(x_{21}\,a + \left(\frac{1}{2}+z_{21}\right)\,c\,\cos\beta\right)\,\hat{\mathbf{x}} - y_{21}\,b\,\hat{\mathbf{y}} + \left(\frac{1}{2}+z_{21}\right)\,c\,\sin\beta\,\hat{\mathbf{z}}$ | (4$a$) | O XXI |
| $\mathbf{B}_{43}$ | = | $(x_{22}-y_{22})\,\mathbf{a}_1 + (x_{22}+y_{22})\,\mathbf{a}_2 + z_{22}\,\mathbf{a}_3$ | = | $(x_{22}\,a + z_{22}\,c\,\cos\beta)\,\hat{\mathbf{x}} + y_{22}\,b\,\hat{\mathbf{y}} + z_{22}\,c\,\sin\beta\,\hat{\mathbf{z}}$ | (4$a$) | O XXII |
| $\mathbf{B}_{44}$ | = | $(x_{22}+y_{22})\,\mathbf{a}_1+(x_{22}-y_{22})\,\mathbf{a}_2+\left(\frac{1}{2}+z_{22}\right)\mathbf{a}_3$ | = | $\left(x_{22}\,a+\left(\frac{1}{2}+z_{22}\right)c\,\cos\beta\right)\hat{\mathbf{x}}-y_{22}\,b\,\hat{\mathbf{y}}+\left(\frac{1}{2}+z_{22}\right)c\,\sin\beta\,\hat{\mathbf{z}}$ | (4*a*) | O XXII |
| $\mathbf{B}_{45}$ | = | $(x_{23}-y_{23})\,\mathbf{a}_1+(x_{23}+y_{23})\,\mathbf{a}_2+z_{23}\,\mathbf{a}_3$ | = | $(x_{23}\,a+z_{23}\,c\,\cos\beta)\,\hat{\mathbf{x}}+y_{23}\,b\,\hat{\mathbf{y}}+z_{23}\,c\,\sin\beta\,\hat{\mathbf{z}}$ | (4*a*) | O XXIII |
| $\mathbf{B}_{46}$ | = | $(x_{23}+y_{23})\,\mathbf{a}_1+(x_{23}-y_{23})\,\mathbf{a}_2+\left(\frac{1}{2}+z_{23}\right)\mathbf{a}_3$ | = | $\left(x_{23}\,a+\left(\frac{1}{2}+z_{23}\right)c\,\cos\beta\right)\hat{\mathbf{x}}-y_{23}\,b\,\hat{\mathbf{y}}+\left(\frac{1}{2}+z_{23}\right)c\,\sin\beta\,\hat{\mathbf{z}}$ | (4*a*) | O XXIII |
| $\mathbf{B}_{47}$ | = | $(x_{24}-y_{24})\,\mathbf{a}_1+(x_{24}+y_{24})\,\mathbf{a}_2+z_{24}\,\mathbf{a}_3$ | = | $(x_{24}\,a+z_{24}\,c\,\cos\beta)\,\hat{\mathbf{x}}+y_{24}\,b\,\hat{\mathbf{y}}+z_{24}\,c\,\sin\beta\,\hat{\mathbf{z}}$ | (4*a*) | O XXIV |
| $\mathbf{B}_{48}$ | = | $(x_{24}+y_{24})\,\mathbf{a}_1+(x_{24}-y_{24})\,\mathbf{a}_2+\left(\frac{1}{2}+z_{24}\right)\mathbf{a}_3$ | = | $\left(x_{24}\,a+\left(\frac{1}{2}+z_{24}\right)c\,\cos\beta\right)\hat{\mathbf{x}}-y_{24}\,b\,\hat{\mathbf{y}}+\left(\frac{1}{2}+z_{24}\right)c\,\sin\beta\,\hat{\mathbf{z}}$ | (4*a*) | O XXIV |
| $\mathbf{B}_{49}$ | = | $(x_{25}-y_{25})\,\mathbf{a}_1+(x_{25}+y_{25})\,\mathbf{a}_2+z_{25}\,\mathbf{a}_3$ | = | $(x_{25}\,a+z_{25}\,c\,\cos\beta)\,\hat{\mathbf{x}}+y_{25}\,b\,\hat{\mathbf{y}}+z_{25}\,c\,\sin\beta\,\hat{\mathbf{z}}$ | (4*a*) | Si I |
| $\mathbf{B}_{50}$ | = | $(x_{25}+y_{25})\,\mathbf{a}_1+(x_{25}-y_{25})\,\mathbf{a}_2+\left(\frac{1}{2}+z_{25}\right)\mathbf{a}_3$ | = | $\left(x_{25}\,a+\left(\frac{1}{2}+z_{25}\right)c\,\cos\beta\right)\hat{\mathbf{x}}-y_{25}\,b\,\hat{\mathbf{y}}+\left(\frac{1}{2}+z_{25}\right)c\,\sin\beta\,\hat{\mathbf{z}}$ | (4*a*) | Si I |
| $\mathbf{B}_{51}$ | = | $(x_{26}-y_{26})\,\mathbf{a}_1+(x_{26}+y_{26})\,\mathbf{a}_2+z_{26}\,\mathbf{a}_3$ | = | $(x_{26}\,a+z_{26}\,c\,\cos\beta)\,\hat{\mathbf{x}}+y_{26}\,b\,\hat{\mathbf{y}}+z_{26}\,c\,\sin\beta\,\hat{\mathbf{z}}$ | (4*a*) | Si II |
| $\mathbf{B}_{52}$ | = | $(x_{26}+y_{26})\,\mathbf{a}_1+(x_{26}-y_{26})\,\mathbf{a}_2+\left(\frac{1}{2}+z_{26}\right)\mathbf{a}_3$ | = | $\left(x_{26}\,a+\left(\frac{1}{2}+z_{26}\right)c\,\cos\beta\right)\hat{\mathbf{x}}-y_{26}\,b\,\hat{\mathbf{y}}+\left(\frac{1}{2}+z_{26}\right)c\,\sin\beta\,\hat{\mathbf{z}}$ | (4*a*) | Si II |
| $\mathbf{B}_{53}$ | = | $(x_{27}-y_{27})\,\mathbf{a}_1+(x_{27}+y_{27})\,\mathbf{a}_2+z_{27}\,\mathbf{a}_3$ | = | $(x_{27}\,a+z_{27}\,c\,\cos\beta)\,\hat{\mathbf{x}}+y_{27}\,b\,\hat{\mathbf{y}}+z_{27}\,c\,\sin\beta\,\hat{\mathbf{z}}$ | (4*a*) | Si III |
| $\mathbf{B}_{54}$ | = | $(x_{27}+y_{27})\,\mathbf{a}_1+(x_{27}-y_{27})\,\mathbf{a}_2+\left(\frac{1}{2}+z_{27}\right)\mathbf{a}_3$ | = | $\left(x_{27}\,a+\left(\frac{1}{2}+z_{27}\right)c\,\cos\beta\right)\hat{\mathbf{x}}-y_{27}\,b\,\hat{\mathbf{y}}+\left(\frac{1}{2}+z_{27}\right)c\,\sin\beta\,\hat{\mathbf{z}}$ | (4*a*) | Si III |
| $\mathbf{B}_{55}$ | = | $(x_{28}-y_{28})\,\mathbf{a}_1+(x_{28}+y_{28})\,\mathbf{a}_2+z_{28}\,\mathbf{a}_3$ | = | $(x_{28}\,a+z_{28}\,c\,\cos\beta)\,\hat{\mathbf{x}}+y_{28}\,b\,\hat{\mathbf{y}}+z_{28}\,c\,\sin\beta\,\hat{\mathbf{z}}$ | (4*a*) | Si IV |
| $\mathbf{B}_{56}$ | = | $(x_{28}+y_{28})\,\mathbf{a}_1+(x_{28}-y_{28})\,\mathbf{a}_2+\left(\frac{1}{2}+z_{28}\right)\mathbf{a}_3$ | = | $\left(x_{28}\,a+\left(\frac{1}{2}+z_{28}\right)c\,\cos\beta\right)\hat{\mathbf{x}}-y_{28}\,b\,\hat{\mathbf{y}}+\left(\frac{1}{2}+z_{28}\right)c\,\sin\beta\,\hat{\mathbf{z}}$ | (4*a*) | Si IV |
| $\mathbf{B}_{57}$ | = | $(x_{29}-y_{29})\,\mathbf{a}_1+(x_{29}+y_{29})\,\mathbf{a}_2+z_{29}\,\mathbf{a}_3$ | = | $(x_{29}\,a+z_{29}\,c\,\cos\beta)\,\hat{\mathbf{x}}+y_{29}\,b\,\hat{\mathbf{y}}+z_{29}\,c\,\sin\beta\,\hat{\mathbf{z}}$ | (4*a*) | Si V |
| $\mathbf{B}_{58}$ | = | $(x_{29}+y_{29})\,\mathbf{a}_1+(x_{29}-y_{29})\,\mathbf{a}_2+\left(\frac{1}{2}+z_{29}\right)\mathbf{a}_3$ | = | $\left(x_{29}\,a+\left(\frac{1}{2}+z_{29}\right)c\,\cos\beta\right)\hat{\mathbf{x}}-y_{29}\,b\,\hat{\mathbf{y}}+\left(\frac{1}{2}+z_{29}\right)c\,\sin\beta\,\hat{\mathbf{z}}$ | (4*a*) | Si V |
| $\mathbf{B}_{59}$ | = | $(x_{30}-y_{30})\,\mathbf{a}_1+(x_{30}+y_{30})\,\mathbf{a}_2+z_{30}\,\mathbf{a}_3$ | = | $(x_{30}\,a+z_{30}\,c\,\cos\beta)\,\hat{\mathbf{x}}+y_{30}\,b\,\hat{\mathbf{y}}+z_{30}\,c\,\sin\beta\,\hat{\mathbf{z}}$ | (4*a*) | Si VI |
| $\mathbf{B}_{60}$ | = | $(x_{30}+y_{30})\,\mathbf{a}_1+(x_{30}-y_{30})\,\mathbf{a}_2+\left(\frac{1}{2}+z_{30}\right)\mathbf{a}_3$ | = | $\left(x_{30}\,a+\left(\frac{1}{2}+z_{30}\right)c\,\cos\beta\right)\hat{\mathbf{x}}-y_{30}\,b\,\hat{\mathbf{y}}+\left(\frac{1}{2}+z_{30}\right)c\,\sin\beta\,\hat{\mathbf{z}}$ | (4*a*) | Si VI |
| $\mathbf{B}_{61}$ | = | $(x_{31}-y_{31})\,\mathbf{a}_1+(x_{31}+y_{31})\,\mathbf{a}_2+z_{31}\,\mathbf{a}_3$ | = | $(x_{31}\,a+z_{31}\,c\,\cos\beta)\,\hat{\mathbf{x}}+y_{31}\,b\,\hat{\mathbf{y}}+z_{31}\,c\,\sin\beta\,\hat{\mathbf{z}}$ | (4*a*) | Si VII |
| $\mathbf{B}_{62}$ | = | $(x_{31}+y_{31})\,\mathbf{a}_1+(x_{31}-y_{31})\,\mathbf{a}_2+\left(\frac{1}{2}+z_{31}\right)\mathbf{a}_3$ | = | $\left(x_{31}\,a+\left(\frac{1}{2}+z_{31}\right)c\,\cos\beta\right)\hat{\mathbf{x}}-y_{31}\,b\,\hat{\mathbf{y}}+\left(\frac{1}{2}+z_{31}\right)c\,\sin\beta\,\hat{\mathbf{z}}$ | (4*a*) | Si VII |
| $\mathbf{B}_{63}$ | = | $(x_{32}-y_{32})\,\mathbf{a}_1+(x_{32}+y_{32})\,\mathbf{a}_2+z_{32}\,\mathbf{a}_3$ | = | $(x_{32}\,a+z_{32}\,c\,\cos\beta)\,\hat{\mathbf{x}}+y_{32}\,b\,\hat{\mathbf{y}}+z_{32}\,c\,\sin\beta\,\hat{\mathbf{z}}$ | (4*a*) | Si VIII |
| $\mathbf{B}_{64}$ | = | $(x_{32}+y_{32})\,\mathbf{a}_1+(x_{32}-y_{32})\,\mathbf{a}_2+\left(\frac{1}{2}+z_{32}\right)\mathbf{a}_3$ | = | $\left(x_{32}\,a+\left(\frac{1}{2}+z_{32}\right)c\,\cos\beta\right)\hat{\mathbf{x}}-y_{32}\,b\,\hat{\mathbf{y}}+\left(\frac{1}{2}+z_{32}\right)c\,\sin\beta\,\hat{\mathbf{z}}$ | (4*a*) | Si VIII |
| $\mathbf{B}_{65}$ | = | $(x_{33}-y_{33})\,\mathbf{a}_1+(x_{33}+y_{33})\,\mathbf{a}_2+z_{33}\,\mathbf{a}_3$ | = | $(x_{33}\,a+z_{33}\,c\,\cos\beta)\,\hat{\mathbf{x}}+y_{33}\,b\,\hat{\mathbf{y}}+z_{33}\,c\,\sin\beta\,\hat{\mathbf{z}}$ | (4*a*) | Si IX |
| $\mathbf{B_{66}}$ | = | $(x_{33}+y_{33})\,\mathbf{a_1}+(x_{33}-y_{33})\,\mathbf{a_2}+\left(\frac{1}{2}+z_{33}\right)\mathbf{a_3}$ | = | $\left(x_{33}\,a+\left(\frac{1}{2}+z_{33}\right)\,c\,\cos\beta\right)\,\hat{\mathbf{x}}-y_{33}\,b\,\hat{\mathbf{y}}+\left(\frac{1}{2}+z_{33}\right)\,c\,\sin\beta\,\hat{\mathbf{z}}$ | (4$a$) | Si IX |
| $\mathbf{B_{67}}$ | = | $(x_{34}-y_{34})\,\mathbf{a_1}+(x_{34}+y_{34})\,\mathbf{a_2}+z_{34}\,\mathbf{a_3}$ | = | $(x_{34}\,a+z_{34}\,c\,\cos\beta)\,\hat{\mathbf{x}}+y_{34}\,b\,\hat{\mathbf{y}}+z_{34}\,c\,\sin\beta\,\hat{\mathbf{z}}$ | (4$a$) | Si X |
| $\mathbf{B_{68}}$ | = | $(x_{34}+y_{34})\,\mathbf{a_1}+(x_{34}-y_{34})\,\mathbf{a_2}+\left(\frac{1}{2}+z_{34}\right)\mathbf{a_3}$ | = | $\left(x_{34}\,a+\left(\frac{1}{2}+z_{34}\right)\,c\,\cos\beta\right)\,\hat{\mathbf{x}}-y_{34}\,b\,\hat{\mathbf{y}}+\left(\frac{1}{2}+z_{34}\right)\,c\,\sin\beta\,\hat{\mathbf{z}}$ | (4$a$) | Si X |
| $\mathbf{B_{69}}$ | = | $(x_{35}-y_{35})\,\mathbf{a_1}+(x_{35}+y_{35})\,\mathbf{a_2}+z_{35}\,\mathbf{a_3}$ | = | $(x_{35}\,a+z_{35}\,c\,\cos\beta)\,\hat{\mathbf{x}}+y_{35}\,b\,\hat{\mathbf{y}}+z_{35}\,c\,\sin\beta\,\hat{\mathbf{z}}$ | (4$a$) | Si XI |
| $\mathbf{B_{70}}$ | = | $(x_{35}+y_{35})\,\mathbf{a_1}+(x_{35}-y_{35})\,\mathbf{a_2}+\left(\frac{1}{2}+z_{35}\right)\mathbf{a_3}$ | = | $\left(x_{35}\,a+\left(\frac{1}{2}+z_{35}\right)\,c\,\cos\beta\right)\,\hat{\mathbf{x}}-y_{35}\,b\,\hat{\mathbf{y}}+\left(\frac{1}{2}+z_{35}\right)\,c\,\sin\beta\,\hat{\mathbf{z}}$ | (4$a$) | Si XI |
| $\mathbf{B_{71}}$ | = | $(x_{36}-y_{36})\,\mathbf{a_1}+(x_{36}+y_{36})\,\mathbf{a_2}+z_{36}\,\mathbf{a_3}$ | = | $(x_{36}\,a+z_{36}\,c\,\cos\beta)\,\hat{\mathbf{x}}+y_{36}\,b\,\hat{\mathbf{y}}+z_{36}\,c\,\sin\beta\,\hat{\mathbf{z}}$ | (4$a$) | Si XII |
| $\mathbf{B_{72}}$ | = | $(x_{36}+y_{36})\,\mathbf{a_1}+(x_{36}-y_{36})\,\mathbf{a_2}+\left(\frac{1}{2}+z_{36}\right)\mathbf{a_3}$ | = | $\left(x_{36}\,a+\left(\frac{1}{2}+z_{36}\right)\,c\,\cos\beta\right)\,\hat{\mathbf{x}}-y_{36}\,b\,\hat{\mathbf{y}}+\left(\frac{1}{2}+z_{36}\right)\,c\,\sin\beta\,\hat{\mathbf{z}}$ | (4$a$) | Si XII |

**Geometry files:**

- CIF: pp. 644
- POSCAR: pp. 645

# NiTi Structure: AB_mP4_11_e_e

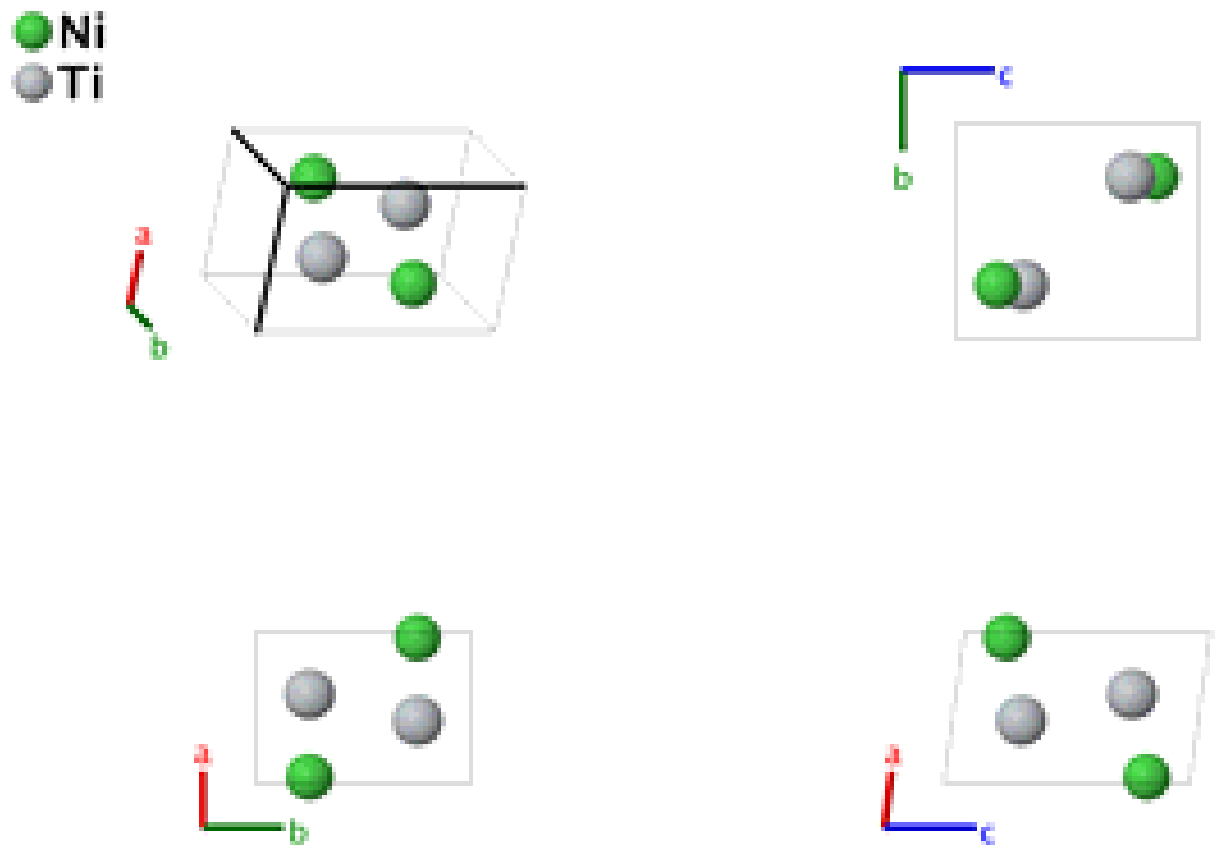

| | | |
|---|---|---|
| **Prototype** | : | NiTi |
| **AFLOW prototype label** | : | AB_mP4_11_e_e |
| ***Strukturbericht* designation** | : | None |
| **Pearson symbol** | : | mP4 |
| **Space group number** | : | 11 |
| **Space group symbol** | : | $P2_1/m$ |
| **AFLOW prototype command** | : | `aflow --proto=AB_mP4_11_e_e` <br> `--params=`$a, b/a, c/a, \beta, x_1, z_1, x_2, z_2$ |

**Simple Monoclinic primitive vectors:**

$$\mathbf{a}_1 = a\,\hat{\mathbf{x}}$$

$$\mathbf{a}_2 = b\,\hat{\mathbf{y}}$$

$$\mathbf{a}_3 = c\cos\beta\,\hat{\mathbf{x}} + c\sin\beta\,\hat{\mathbf{z}}$$

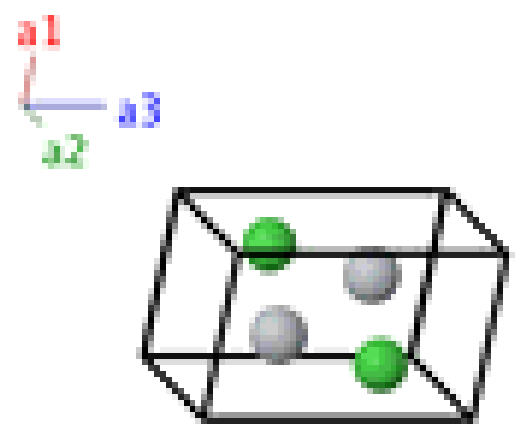

**Basis vectors:**

| | | Lattice Coordinates | | Cartesian Coordinates | Wyckoff Position | Atom Type |
|---|---|---|---|---|---|---|
| $\mathbf{B}_1$ | = | $x_1\,\mathbf{a}_1 + \frac{1}{4}\,\mathbf{a}_2 + z_1\,\mathbf{a}_3$ | = | $(x_1\,a + z_1\,c\,\cos\beta)\,\hat{\mathbf{x}} + \frac{1}{4}\,b\,\hat{\mathbf{y}} + z_1\,c\,\sin\beta\,\hat{\mathbf{z}}$ | (2e) | Ni |
| $\mathbf{B}_2$ | = | $-x_1\,\mathbf{a}_1 + \frac{3}{4}\,\mathbf{a}_2 - z_1\,\mathbf{a}_3$ | = | $-(x_1\,a + z_1\,c\,\cos\beta)\,\hat{\mathbf{x}} + \frac{3}{4}\,b\,\hat{\mathbf{y}} - z_1\,c\,\sin\beta\,\hat{\mathbf{z}}$ | (2e) | Ni |
| $\mathbf{B}_3$ | = | $x_2\,\mathbf{a}_1 + \frac{1}{4}\,\mathbf{a}_2 + z_2\,\mathbf{a}_3$ | = | $(x_2\,a + z_2\,c\,\cos\beta)\,\hat{\mathbf{x}} + \frac{1}{4}\,b\,\hat{\mathbf{y}} + z_2\,c\,\sin\beta\,\hat{\mathbf{z}}$ | (2e) | Ti |
| $\mathbf{B}_4$ | = | $-x_2\,\mathbf{a}_1 + \frac{3}{4}\,\mathbf{a}_2 - z_2\,\mathbf{a}_3$ | = | $-(x_2\,a + z_2\,c\,\cos\beta)\,\hat{\mathbf{x}} + \frac{3}{4}\,b\,\hat{\mathbf{y}} - z_2\,c\,\sin\beta\,\hat{\mathbf{z}}$ | (2e) | Ti |

**Found in:**

---

**Geometry files:**

- CIF: pp. 645
- POSCAR: pp. 646

# $KClO_3$ ($G0_6$) Structure: ABC3_mP10_11_e_e_ef

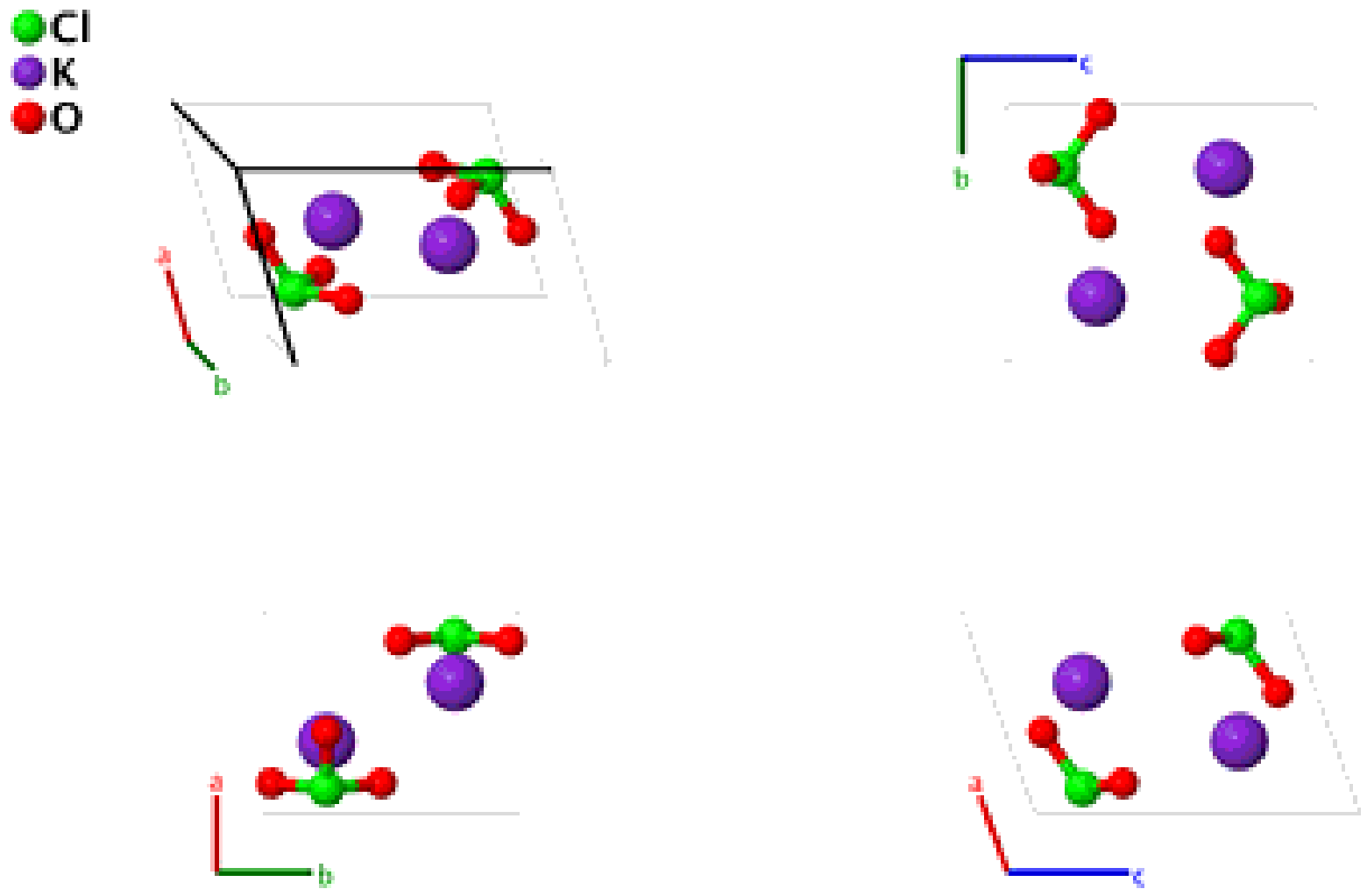

| | | |
|---|---|---|
| **Prototype** | : | $KClO_3$ |
| **AFLOW prototype label** | : | ABC3_mP10_11_e_e_ef |
| ***Strukturbericht* designation** | : | $G0_6$ |
| **Pearson symbol** | : | mP10 |
| **Space group number** | : | 11 |
| **Space group symbol** | : | $P2_1/m$ |
| **AFLOW prototype command** | : | `aflow --proto=ABC3_mP10_11_e_e_ef`<br>`--params=`$a, b/a, c/a, \beta, x_1, z_1, x_2, z_2, x_3, z_3, x_4, y_4, z_4$ |

**Simple Monoclinic primitive vectors:**

$$\mathbf{a}_1 = a\,\hat{\mathbf{x}}$$
$$\mathbf{a}_2 = b\,\hat{\mathbf{y}}$$
$$\mathbf{a}_3 = c\cos\beta\,\hat{\mathbf{x}} + c\sin\beta\,\hat{\mathbf{z}}$$

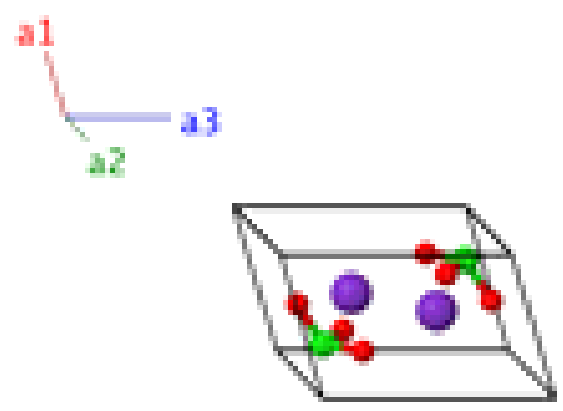

**Basis vectors:**

| | | Lattice Coordinates | | Cartesian Coordinates | Wyckoff Position | Atom Type |
|---|---|---|---|---|---|---|
| $\mathbf{B}_1$ | = | $x_1\,\mathbf{a}_1 + \frac{1}{4}\,\mathbf{a}_2 + z_1\,\mathbf{a}_3$ | = | $(x_1\,a + z_1\,c\,\cos\beta)\,\hat{\mathbf{x}} + \frac{1}{4}\,b\,\hat{\mathbf{y}} + z_1\,c\,\sin\beta\,\hat{\mathbf{z}}$ | (2e) | Cl |
| $\mathbf{B}_2$ | = | $-x_1\,\mathbf{a}_1 + \frac{3}{4}\,\mathbf{a}_2 - z_1\,\mathbf{a}_3$ | = | $-(x_1\,a + z_1\,c\,\cos\beta)\,\hat{\mathbf{x}} + \frac{3}{4}\,b\,\hat{\mathbf{y}} - z_1\,c\,\sin\beta\,\hat{\mathbf{z}}$ | (2e) | Cl |
| $\mathbf{B}_3$ | = | $x_2\,\mathbf{a}_1 + \frac{1}{4}\,\mathbf{a}_2 + z_2\,\mathbf{a}_3$ | = | $(x_2\,a + z_2\,c\,\cos\beta)\,\hat{\mathbf{x}} + \frac{1}{4}\,b\,\hat{\mathbf{y}} + z_2\,c\,\sin\beta\,\hat{\mathbf{z}}$ | (2e) | K |
| $\mathbf{B}_4$ | = | $-x_2\,\mathbf{a}_1 + \frac{3}{4}\,\mathbf{a}_2 - z_2\,\mathbf{a}_3$ | = | $-(x_2\,a + z_2\,c\,\cos\beta)\,\hat{\mathbf{x}} + \frac{3}{4}\,b\,\hat{\mathbf{y}} - z_2\,c\,\sin\beta\,\hat{\mathbf{z}}$ | (2e) | K |
| $\mathbf{B}_5$ | = | $x_3\,\mathbf{a}_1 + \frac{1}{4}\,\mathbf{a}_2 + z_3\,\mathbf{a}_3$ | = | $(x_3\,a + z_3\,c\,\cos\beta)\,\hat{\mathbf{x}} + \frac{1}{4}\,b\,\hat{\mathbf{y}} + z_3\,c\,\sin\beta\,\hat{\mathbf{z}}$ | (2e) | O I |
| $\mathbf{B}_6$ | = | $-x_3\,\mathbf{a}_1 + \frac{3}{4}\,\mathbf{a}_2 - z_3\,\mathbf{a}_3$ | = | $-(x_3\,a + z_3\,c\,\cos\beta)\,\hat{\mathbf{x}} + \frac{3}{4}\,b\,\hat{\mathbf{y}} - z_3\,c\,\sin\beta\,\hat{\mathbf{z}}$ | (2e) | O I |
| $\mathbf{B}_7$ | = | $x_4\,\mathbf{a}_1 + y_4\,\mathbf{a}_2 + z_4\,\mathbf{a}_3$ | = | $(x_4\,a + z_4\,c\,\cos\beta)\,\hat{\mathbf{x}} + y_4\,b\,\hat{\mathbf{y}} + z_4\,c\,\sin\beta\,\hat{\mathbf{z}}$ | (4f) | O II |
| $\mathbf{B_8}$ | = | $-x_4\,\mathbf{a_1} + \left(\frac{1}{2} + y_4\right)\,\mathbf{a_2} - z_4\,\mathbf{a_3}$ | = | $-\left(x_4\,a + z_4\,c\,\cos\beta\right)\,\mathbf{\hat{x}} + \left(\frac{1}{2} + y_4\right)\,b\,\mathbf{\hat{y}} - z_4\,c\,\sin\beta\,\mathbf{\hat{z}}$ | $(4f)$ | O II |
| $\mathbf{B_9}$ | = | $-x_4\,\mathbf{a_1} - y_4\,\mathbf{a_2} - z_4\,\mathbf{a_3}$ | = | $-\left(x_4\,a + z_4\,c\,\cos\beta\right)\,\mathbf{\hat{x}} - y_4\,b\,\mathbf{\hat{y}} - z_4\,c\,\sin\beta\,\mathbf{\hat{z}}$ | $(4f)$ | O II |
| $\mathbf{B_{10}}$ | = | $x_4\,\mathbf{a_1} + \left(\frac{1}{2} - y_4\right)\,\mathbf{a_2} + z_4\,\mathbf{a_3}$ | = | $\left(x_4\,a + z_4\,c\,\cos\beta\right)\,\mathbf{\hat{x}} + \left(\frac{1}{2} - y_4\right)\,b\,\mathbf{\hat{y}} + z_4\,c\,\sin\beta\,\mathbf{\hat{z}}$ | $(4f)$ | O II |

---

---

**Geometry files:**

- CIF: pp. 646
- POSCAR: pp. 646

# $\alpha$-Pu Structure: A_mP16_11_8e

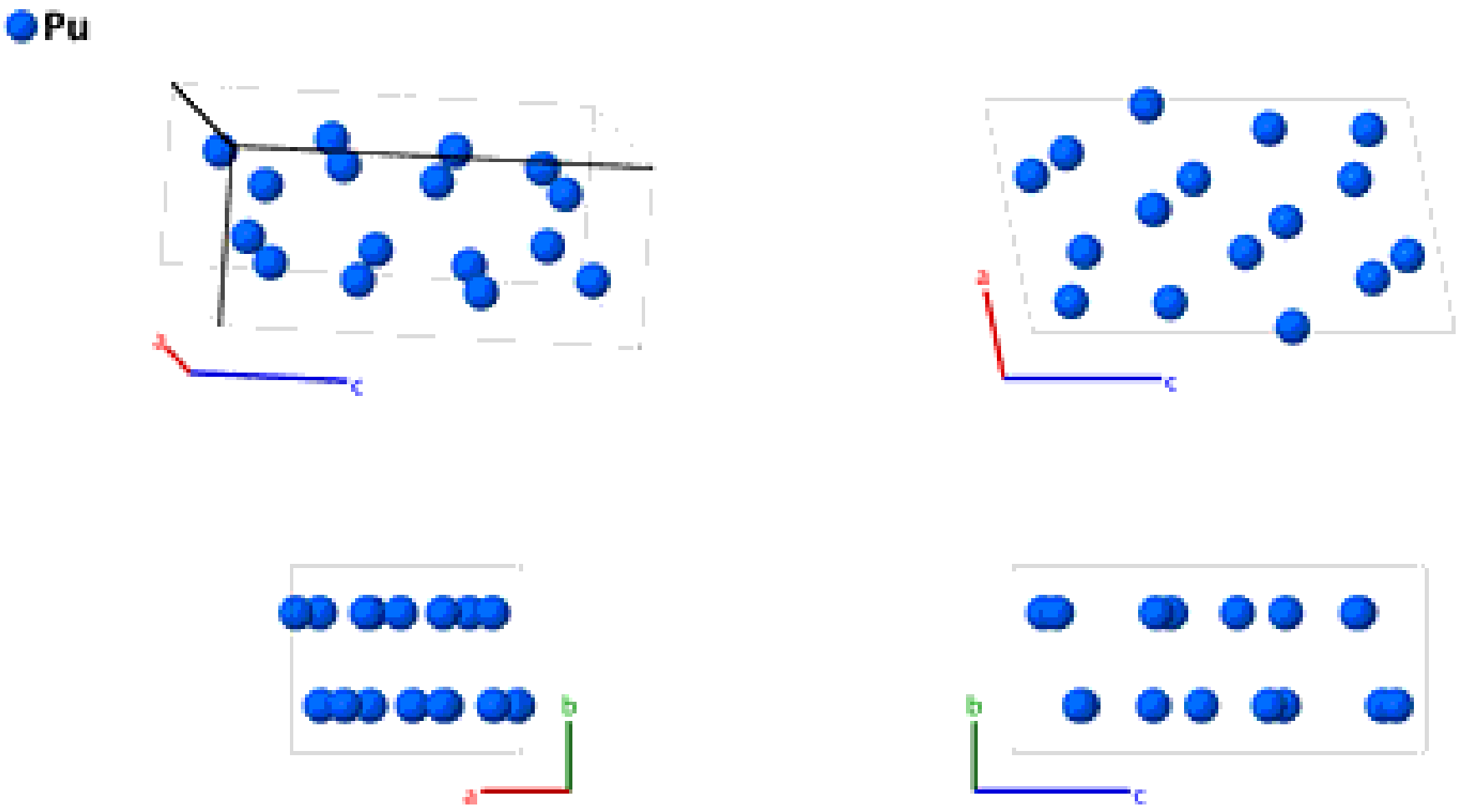

| | | |
|---|---|---|
| **Prototype** | : | $\alpha$-Pu |
| **AFLOW prototype label** | : | A_mP16_11_8e |
| ***Strukturbericht* designation** | : | None |
| **Pearson symbol** | : | mP16 |
| **Space group number** | : | 11 |
| **Space group symbol** | : | $P2_1/m$ |
| **AFLOW prototype command** | : | `aflow --proto=A_mP16_11_8e`<br>`--params=`$a, b/a, c/a, \beta, x_1, z_1, x_2, z_2, x_3, z_3, x_4, z_4, x_5, z_5, x_6, z_6, x_7, z_7, x_8, z_8$ |

**Simple Monoclinic primitive vectors:**

$$\mathbf{a}_1 = a\,\hat{\mathbf{x}}$$
$$\mathbf{a}_2 = b\,\hat{\mathbf{y}}$$
$$\mathbf{a}_3 = c\cos\beta\,\hat{\mathbf{x}} + c\sin\beta\,\hat{\mathbf{z}}$$

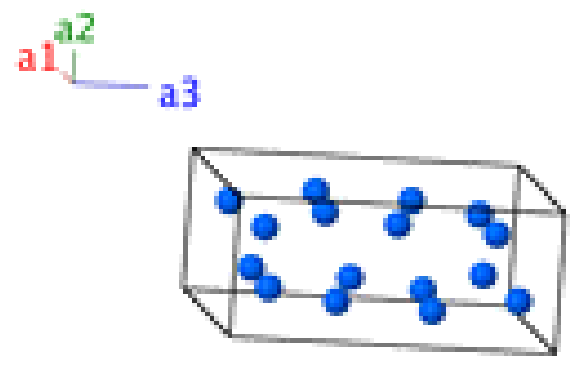

**Basis vectors:**

| | | Lattice Coordinates | | Cartesian Coordinates | Wyckoff Position | Atom Type |
|---|---|---|---|---|---|---|
| $\mathbf{B}_1$ | = | $x_1\,\mathbf{a}_1 + \frac{1}{4}\,\mathbf{a}_2 + z_1\,\mathbf{a}_3$ | = | $(x_1\,a + z_1\,c\,\cos\beta)\,\hat{\mathbf{x}} + \frac{1}{4}\,b\,\hat{\mathbf{y}} + z_1\,c\,\sin\beta\,\hat{\mathbf{z}}$ | (2e) | Pu I |
| $\mathbf{B}_2$ | = | $-x_1\,\mathbf{a}_1 + \frac{3}{4}\,\mathbf{a}_2 - z_1\,\mathbf{a}_3$ | = | $-(x_1\,a + z_1\,c\,\cos\beta)\,\hat{\mathbf{x}} + \frac{3}{4}\,b\,\hat{\mathbf{y}} - z_1\,c\,\sin\beta\,\hat{\mathbf{z}}$ | (2e) | Pu I |
| $\mathbf{B}_3$ | = | $x_2\,\mathbf{a}_1 + \frac{1}{4}\,\mathbf{a}_2 + z_2\,\mathbf{a}_3$ | = | $(x_2\,a + z_2\,c\,\cos\beta)\,\hat{\mathbf{x}} + \frac{1}{4}\,b\,\hat{\mathbf{y}} + z_2\,c\,\sin\beta\,\hat{\mathbf{z}}$ | (2e) | Pu II |
| $\mathbf{B}_4$ | = | $-x_2\,\mathbf{a}_1 + \frac{3}{4}\,\mathbf{a}_2 - z_2\,\mathbf{a}_3$ | = | $-(x_2\,a + z_2\,c\,\cos\beta)\,\hat{\mathbf{x}} + \frac{3}{4}\,b\,\hat{\mathbf{y}} - z_2\,c\,\sin\beta\,\hat{\mathbf{z}}$ | (2e) | Pu II |
| $\mathbf{B}_5$ | = | $x_3\,\mathbf{a}_1 + \frac{1}{4}\,\mathbf{a}_2 + z_3\,\mathbf{a}_3$ | = | $(x_3\,a + z_3\,c\,\cos\beta)\,\hat{\mathbf{x}} + \frac{1}{4}\,b\,\hat{\mathbf{y}} + z_3\,c\,\sin\beta\,\hat{\mathbf{z}}$ | (2e) | Pu III |
| $\mathbf{B}_6$ | = | $-x_3\,\mathbf{a}_1 + \frac{3}{4}\,\mathbf{a}_2 - z_3\,\mathbf{a}_3$ | = | $-(x_3\,a + z_3\,c\,\cos\beta)\,\hat{\mathbf{x}} + \frac{3}{4}\,b\,\hat{\mathbf{y}} - z_3\,c\,\sin\beta\,\hat{\mathbf{z}}$ | (2e) | Pu III |
| $\mathbf{B}_7$ | = | $x_4\,\mathbf{a}_1 + \frac{1}{4}\,\mathbf{a}_2 + z_4\,\mathbf{a}_3$ | = | $(x_4\,a + z_4\,c\,\cos\beta)\,\hat{\mathbf{x}} + \frac{1}{4}\,b\,\hat{\mathbf{y}} + z_4\,c\,\sin\beta\,\hat{\mathbf{z}}$ | (2e) | Pu IV |
| $\mathbf{B}_8$ | = | $-x_4\,\mathbf{a}_1 + \frac{3}{4}\,\mathbf{a}_2 - z_4\,\mathbf{a}_3$ | = | $-(x_4\,a + z_4\,c\,\cos\beta)\,\hat{\mathbf{x}} + \frac{3}{4}\,b\,\hat{\mathbf{y}} - z_4\,c\,\sin\beta\,\hat{\mathbf{z}}$ | (2e) | Pu IV |
| $\mathbf{B_9}$ | = | $x_5\,\mathbf{a_1} + \frac{1}{4}\,\mathbf{a_2} + z_5\,\mathbf{a_3}$ | = | $(x_5\,a + z_5\,c\,\cos\beta)\,\hat{\mathbf{x}} + \frac{1}{4}\,b\,\hat{\mathbf{y}} + z_5\,c\,\sin\beta\,\hat{\mathbf{z}}$ | (2$e$) | Pu V |
| $\mathbf{B_{10}}$ | = | $-x_5\,\mathbf{a_1} + \frac{3}{4}\,\mathbf{a_2} - z_5\,\mathbf{a_3}$ | = | $-(x_5\,a + z_5\,c\,\cos\beta)\,\hat{\mathbf{x}} + \frac{3}{4}\,b\,\hat{\mathbf{y}} - z_5\,c\,\sin\beta\,\hat{\mathbf{z}}$ | (2$e$) | Pu V |
| $\mathbf{B_{11}}$ | = | $x_6\,\mathbf{a_1} + \frac{1}{4}\,\mathbf{a_2} + z_6\,\mathbf{a_3}$ | = | $(x_6\,a + z_6\,c\,\cos\beta)\,\hat{\mathbf{x}} + \frac{1}{4}\,b\,\hat{\mathbf{y}} + z_6\,c\,\sin\beta\,\hat{\mathbf{z}}$ | (2$e$) | Pu VI |
| $\mathbf{B_{12}}$ | = | $-x_6\,\mathbf{a_1} + \frac{3}{4}\,\mathbf{a_2} - z_6\,\mathbf{a_3}$ | = | $-(x_6\,a + z_6\,c\,\cos\beta)\,\hat{\mathbf{x}} + \frac{3}{4}\,b\,\hat{\mathbf{y}} - z_6\,c\,\sin\beta\,\hat{\mathbf{z}}$ | (2$e$) | Pu VI |
| $\mathbf{B_{13}}$ | = | $x_7\,\mathbf{a_1} + \frac{1}{4}\,\mathbf{a_2} + z_7\,\mathbf{a_3}$ | = | $(x_7\,a + z_7\,c\,\cos\beta)\,\hat{\mathbf{x}} + \frac{1}{4}\,b\,\hat{\mathbf{y}} + z_7\,c\,\sin\beta\,\hat{\mathbf{z}}$ | (2$e$) | Pu VII |
| $\mathbf{B_{14}}$ | = | $-x_7\,\mathbf{a_1} + \frac{3}{4}\,\mathbf{a_2} - z_7\,\mathbf{a_3}$ | = | $-(x_7\,a + z_7\,c\,\cos\beta)\,\hat{\mathbf{x}} + \frac{3}{4}\,b\,\hat{\mathbf{y}} - z_7\,c\,\sin\beta\,\hat{\mathbf{z}}$ | (2$e$) | Pu VII |
| $\mathbf{B_{15}}$ | = | $x_8\,\mathbf{a_1} + \frac{1}{4}\,\mathbf{a_2} + z_8\,\mathbf{a_3}$ | = | $(x_8\,a + z_8\,c\,\cos\beta)\,\hat{\mathbf{x}} + \frac{1}{4}\,b\,\hat{\mathbf{y}} + z_8\,c\,\sin\beta\,\hat{\mathbf{z}}$ | (2$e$) | Pu VIII |
| $\mathbf{B_{16}}$ | = | $-x_8\,\mathbf{a_1} + \frac{3}{4}\,\mathbf{a_2} - z_8\,\mathbf{a_3}$ | = | $-(x_8\,a + z_8\,c\,\cos\beta)\,\hat{\mathbf{x}} + \frac{3}{4}\,b\,\hat{\mathbf{y}} - z_8\,c\,\sin\beta\,\hat{\mathbf{z}}$ | (2$e$) | Pu VIII |

---

**Found in:**

---

**Geometry files:**

- CIF: pp. 646
- POSCAR: pp. 646

# Calaverite ($AuTe_2$, C34) Structure: AB2_mC6_12_a_i

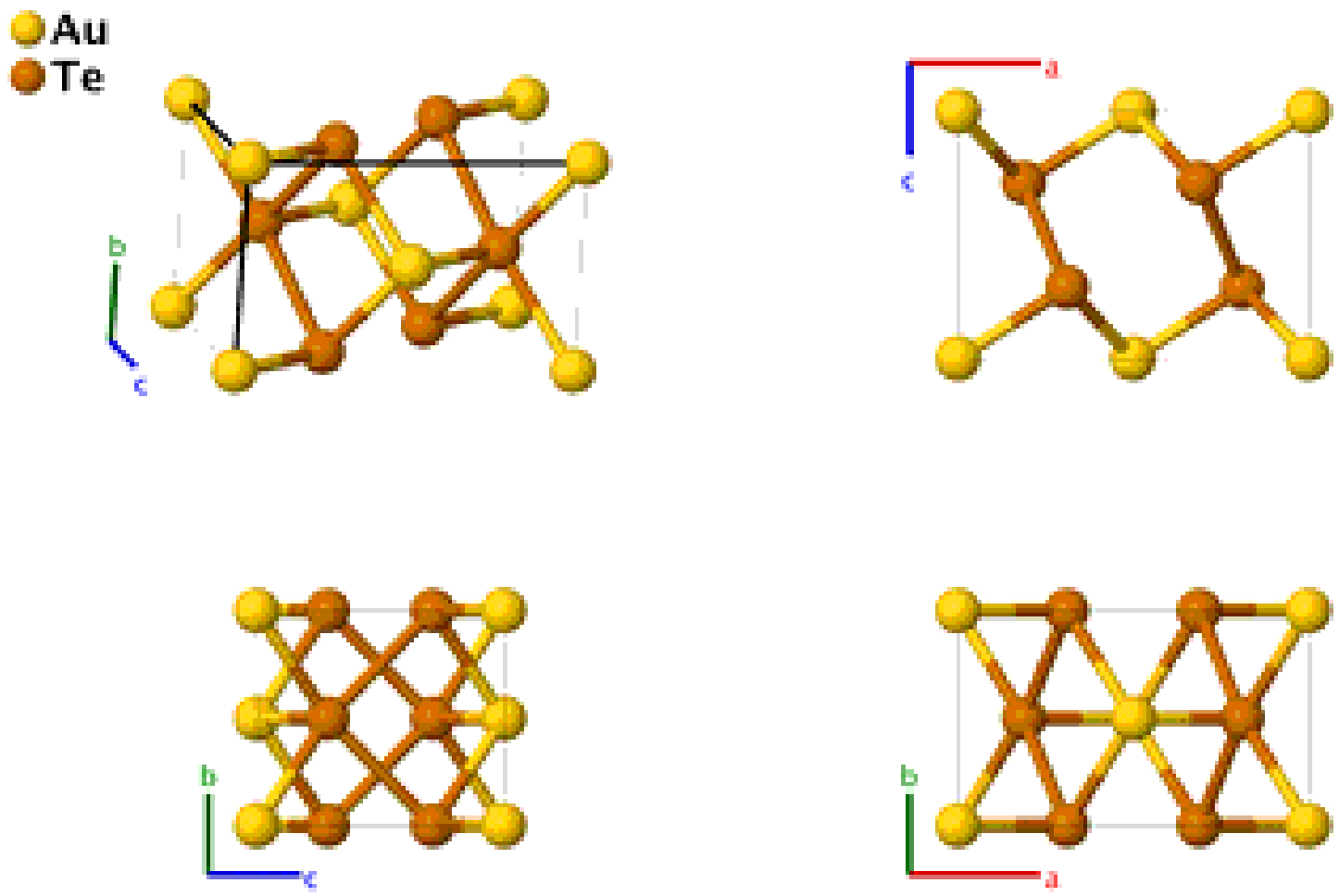

| | | |
|---|---|---|
| **Prototype** | : | $AuTe_2$ |
| **AFLOW prototype label** | : | AB2_mC6_12_a_i |
| ***Strukturbericht* designation** | : | C34 |
| **Pearson symbol** | : | mC6 |
| **Space group number** | : | 12 |
| **Space group symbol** | : | C2/m |
| **AFLOW prototype command** | : | `aflow --proto=AB2_mC6_12_a_i`<br>`--params=`$a, b/a, c/a, \beta, x_2, z_2$ |

**Other compounds with this structure:**

- $Au_{10}Se_3Te_{17}$

**Base-centered Monoclinic primitive vectors:**

$$
\begin{aligned}
\mathbf{a}_1 &= \tfrac{1}{2}\,a\,\hat{\mathbf{x}} - \tfrac{1}{2}\,b\,\hat{\mathbf{y}} \\
\mathbf{a}_2 &= \tfrac{1}{2}\,a\,\hat{\mathbf{x}} + \tfrac{1}{2}\,b\,\hat{\mathbf{y}} \\
\mathbf{a}_3 &= c\cos\beta\,\hat{\mathbf{x}} + c\sin\beta\,\hat{\mathbf{z}}
\end{aligned}
$$

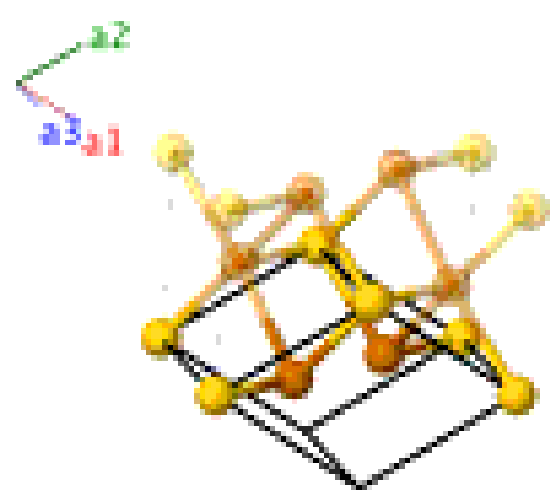

**Basis vectors:**

| | | Lattice Coordinates | | Cartesian Coordinates | Wyckoff Position | Atom Type |
|---|---|---|---|---|---|---|
| $\mathbf{B_1}$ | = | $0\,\mathbf{a_1} + 0\,\mathbf{a_2} + 0\,\mathbf{a_3}$ | = | $0\hat{\mathbf{x}} + 0\hat{\mathbf{y}} + 0\hat{\mathbf{z}}$ | (2$a$) | Au |
| $\mathbf{B_2}$ | = | $x_2\,\mathbf{a_1} + x_2\,\mathbf{a_2} + z_2\,\mathbf{a_3}$ | = | $(x_2\,a + z_2\,c\,\cos\beta)\,\hat{\mathbf{x}} + z_2\,c\,\sin\beta\,\hat{\mathbf{z}}$ | (4$i$) | Te |
| $\mathbf{B_3}$ | = | $-x_2\,\mathbf{a_1} - x_2\,\mathbf{a_2} - z_2\,\mathbf{a_3}$ | = | $-(x_2\,a + z_2\,c\,\cos\beta)\,\hat{\mathbf{x}} - z_2\,c\,\sin\beta\,\hat{\mathbf{z}}$ | (4$i$) | Te |

---

**Geometry files:**

- CIF: pp. 647
- POSCAR: pp. 647

# $\beta$-Pu Structure: A_mC34_12_ah3i2j

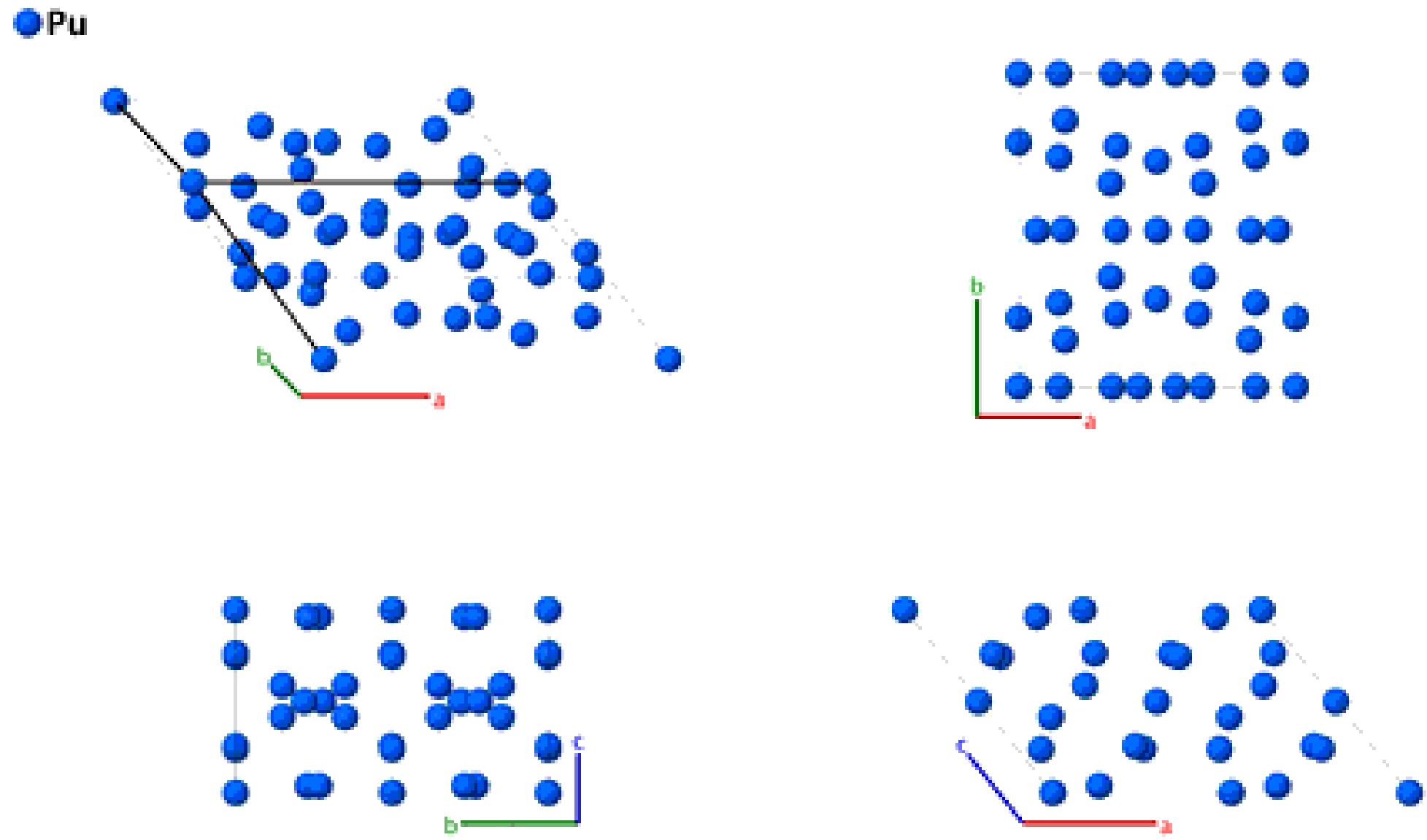

| | | |
|---|---|---|
| **Prototype** | : | $\beta$-Pu |
| **AFLOW prototype label** | : | A_mC34_12_ah3i2j |
| ***Strukturbericht* designation** | : | None |
| **Pearson symbol** | : | mC34 |
| **Space group number** | : | 12 |
| **Space group symbol** | : | C2/m |
| **AFLOW prototype command** | : | `aflow --proto=A_mC34_12_ah3i2j` `--params=`$a, b/a, c/a, \beta, y_2, x_3, z_3, x_4, z_4, x_5, z_5, x_6, y_6, z_6, x_7, y_7, z_7$ |

**Base-centered Monoclinic primitive vectors:**

$$\mathbf{a}_1 = \frac{1}{2}\,a\,\hat{\mathbf{x}} - \frac{1}{2}\,b\,\hat{\mathbf{y}}$$
$$\mathbf{a}_2 = \frac{1}{2}\,a\,\hat{\mathbf{x}} + \frac{1}{2}\,b\,\hat{\mathbf{y}}$$
$$\mathbf{a}_3 = c\cos\beta\,\hat{\mathbf{x}} + c\sin\beta\,\hat{\mathbf{z}}$$

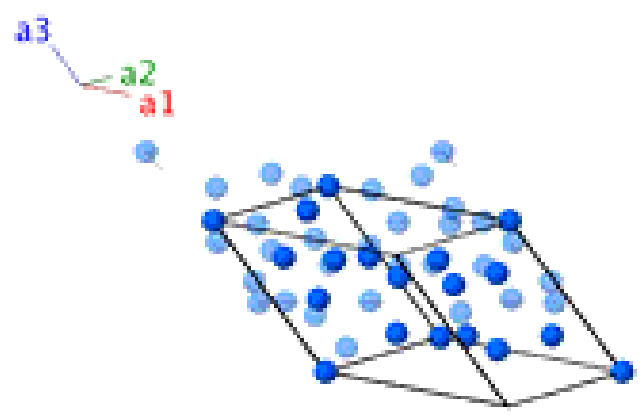

**Basis vectors:**

| | | Lattice Coordinates | | Cartesian Coordinates | Wyckoff Position | Atom Type |
|---|---|---|---|---|---|---|
| $\mathbf{B}_1$ | = | $0\,\mathbf{a}_1 + 0\,\mathbf{a}_2 + 0\,\mathbf{a}_3$ | = | $0\hat{\mathbf{x}} + 0\hat{\mathbf{y}} + 0\hat{\mathbf{z}}$ | (2$a$) | Pu I |
| $\mathbf{B}_2$ | = | $-y_2\,\mathbf{a}_1 + y_2\,\mathbf{a}_2 + \frac{1}{2}\,\mathbf{a}_3$ | = | $\frac{1}{2}\,c\,\cos\beta\,\hat{\mathbf{x}} + y_2\,b\,\hat{\mathbf{y}} + \frac{1}{2}\,c\,\sin\beta\,\hat{\mathbf{z}}$ | (4$h$) | Pu II |
| $\mathbf{B}_3$ | = | $y_2\,\mathbf{a}_1 - y_2\,\mathbf{a}_2 + \frac{1}{2}\,\mathbf{a}_3$ | = | $\frac{1}{2}\,c\,\cos\beta\,\hat{\mathbf{x}} - y_2\,b\,\hat{\mathbf{y}} + \frac{1}{2}\,c\,\sin\beta\,\hat{\mathbf{z}}$ | (4$h$) | Pu II |
| $\mathbf{B}_4$ | = | $x_3\,\mathbf{a}_1 + x_3\,\mathbf{a}_2 + z_3\,\mathbf{a}_3$ | = | $(x_3\,a + z_3\,c\,\cos\beta)\,\hat{\mathbf{x}} + z_3\,c\,\sin\beta\,\hat{\mathbf{z}}$ | (4$i$) | Pu III |
| $\mathbf{B}_5$ | = | $-x_3\,\mathbf{a}_1 - x_3\,\mathbf{a}_2 - z_3\,\mathbf{a}_3$ | = | $-(x_3\,a + z_3\,c\,\cos\beta)\,\hat{\mathbf{x}} - z_3\,c\,\sin\beta\,\hat{\mathbf{z}}$ | (4$i$) | Pu III |
| $\mathbf{B}_6$ | = | $x_4\,\mathbf{a}_1 + x_4\,\mathbf{a}_2 + z_4\,\mathbf{a}_3$ | = | $(x_4\,a + z_4\,c\,\cos\beta)\,\hat{\mathbf{x}} + z_4\,c\,\sin\beta\,\hat{\mathbf{z}}$ | (4$i$) | Pu IV |
| $\mathbf{B}_7$ | = | $-x_4\,\mathbf{a}_1 - x_4\,\mathbf{a}_2 - z_4\,\mathbf{a}_3$ | = | $-(x_4\,a + z_4\,c\,\cos\beta)\,\hat{\mathbf{x}} - z_4\,c\,\sin\beta\,\hat{\mathbf{z}}$ | (4$i$) | Pu IV |
| $\mathbf{B}_8$ | = | $x_5\,\mathbf{a}_1 + x_5\,\mathbf{a}_2 + z_5\,\mathbf{a}_3$ | = | $(x_5\,a + z_5\,c\,\cos\beta)\,\hat{\mathbf{x}} + z_5\,c\,\sin\beta\,\hat{\mathbf{z}}$ | (4$i$) | Pu V |
| $\mathbf{B}_9$ | = | $-x_5\,\mathbf{a}_1 - x_5\,\mathbf{a}_2 - z_5\,\mathbf{a}_3$ | = | $-(x_5\,a + z_5\,c\,\cos\beta)\,\hat{\mathbf{x}} - z_5\,c\,\sin\beta\,\hat{\mathbf{z}}$ | (4$i$) | Pu V |
| $\mathbf{B}_{10}$ | = | $(x_6 - y_6)\,\mathbf{a}_1 + (x_6 + y_6)\,\mathbf{a}_2 + z_6\,\mathbf{a}_3$ | = | $(x_6\,a + z_6\,c\,\cos\beta)\,\hat{\mathbf{x}} + y_6\,b\,\hat{\mathbf{y}} + z_6\,c\,\sin\beta\,\hat{\mathbf{z}}$ | (8$j$) | Pu VI |
| $\mathbf{B}_{11}$ | = | $(x_6 + y_6)\,\mathbf{a}_1 + (x_6 - y_6)\,\mathbf{a}_2 + z_6\,\mathbf{a}_3$ | = | $(x_6\,a + z_6\,c\,\cos\beta)\,\hat{\mathbf{x}} - y_6\,b\,\hat{\mathbf{y}} + z_6\,c\,\sin\beta\,\hat{\mathbf{z}}$ | (8$j$) | Pu VI |
| $\mathbf{B}_{12}$ | = | $(y_6 - x_6)\,\mathbf{a}_1 - (x_6 + y_6)\,\mathbf{a}_2 - z_6\,\mathbf{a}_3$ | = | $-(x_6\,a + z_6\,c\,\cos\beta)\,\hat{\mathbf{x}} - y_6\,b\,\hat{\mathbf{y}} - z_6\,c\,\sin\beta\,\hat{\mathbf{z}}$ | (8$j$) | Pu VI |
| $\mathbf{B}_{13}$ | = | $-(x_6 + y_6)\,\mathbf{a}_1 + (y_6 - x_6)\,\mathbf{a}_2 - z_6\,\mathbf{a}_3$ | = | $-(x_6\,a + z_6\,c\,\cos\beta)\,\hat{\mathbf{x}} + y_6\,b\,\hat{\mathbf{y}} - z_6\,c\,\sin\beta\,\hat{\mathbf{z}}$ | (8$j$) | Pu VI |
| $\mathbf{B}_{14}$ | = | $(x_7 - y_7)\,\mathbf{a}_1 + (x_7 + y_7)\,\mathbf{a}_2 + z_7\,\mathbf{a}_3$ | = | $(x_7\,a + z_7\,c\,\cos\beta)\,\hat{\mathbf{x}} + y_7\,b\,\hat{\mathbf{y}} + z_7\,c\,\sin\beta\,\hat{\mathbf{z}}$ | (8$j$) | Pu VII |
| $\mathbf{B}_{15}$ | = | $(x_7 + y_7)\,\mathbf{a}_1 + (x_7 - y_7)\,\mathbf{a}_2 + z_7\,\mathbf{a}_3$ | = | $(x_7\,a + z_7\,c\,\cos\beta)\,\hat{\mathbf{x}} - y_7\,b\,\hat{\mathbf{y}} + z_7\,c\,\sin\beta\,\hat{\mathbf{z}}$ | (8$j$) | Pu VII |
| $\mathbf{B}_{16}$ | = | $(y_7 - x_7)\,\mathbf{a}_1 - (x_7 + y_7)\,\mathbf{a}_2 - z_7\,\mathbf{a}_3$ | = | $-(x_7\,a + z_7\,c\,\cos\beta)\,\hat{\mathbf{x}} - y_7\,b\,\hat{\mathbf{y}} - z_7\,c\,\sin\beta\,\hat{\mathbf{z}}$ | (8$j$) | Pu VII |
| $\mathbf{B}_{17}$ | = | $-(x_7 + y_7)\,\mathbf{a}_1 + (y_7 - x_7)\,\mathbf{a}_2 - z_7\,\mathbf{a}_3$ | = | $-(x_7\,a + z_7\,c\,\cos\beta)\,\hat{\mathbf{x}} + y_7\,b\,\hat{\mathbf{y}} - z_7\,c\,\sin\beta\,\hat{\mathbf{z}}$ | (8$j$) | Pu VII |

---

**Found in:**

---

**Geometry files:**

- CIF: pp. 647
- POSCAR: pp. 647

# $AlCl_3$ ($D0_{15}$) Structure: AB3_mC16_12_g_ij

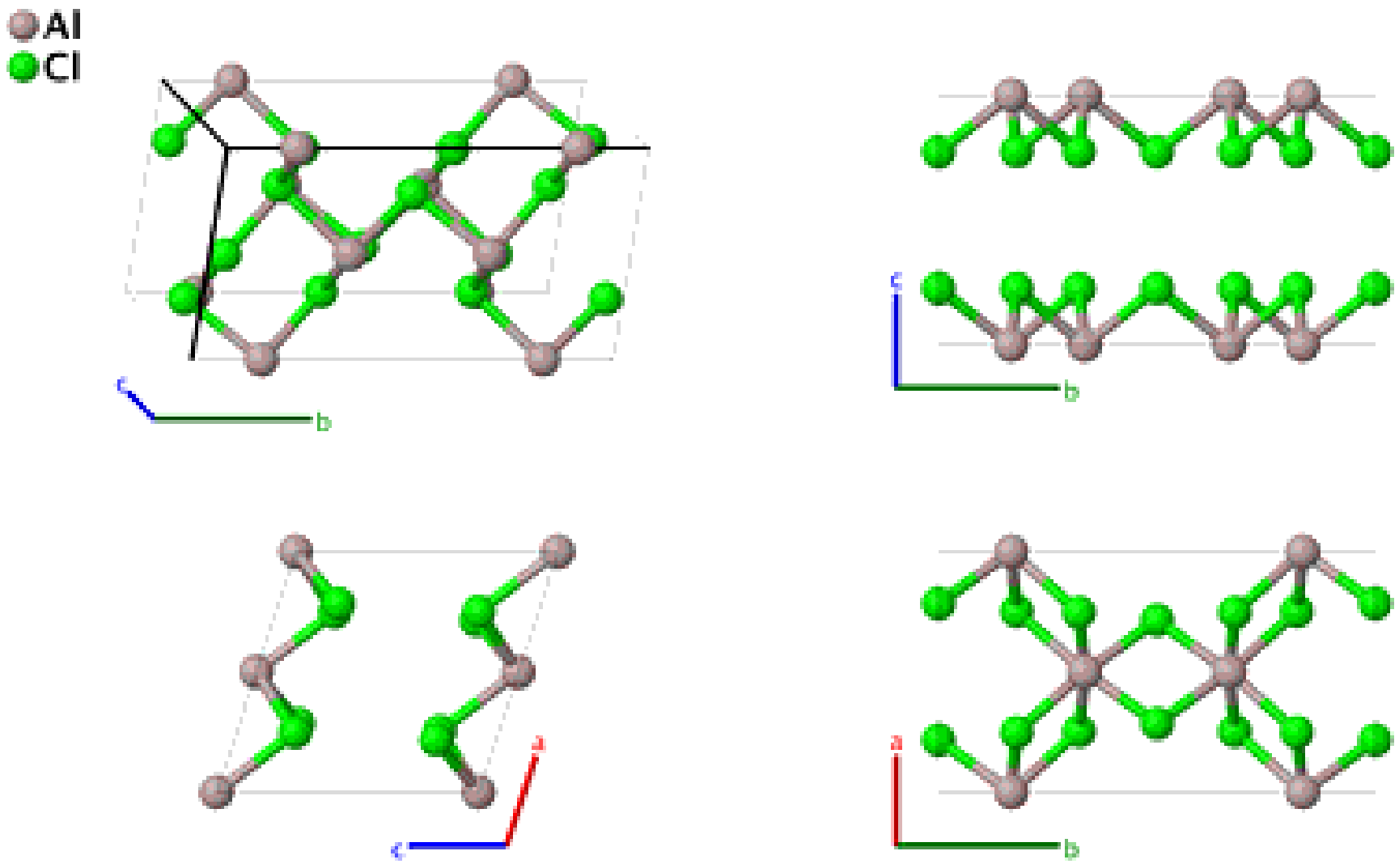

| | | |
|---|---|---|
| **Prototype** | : | $AlCl_3$ |
| **AFLOW prototype label** | : | AB3_mC16_12_g_ij |
| ***Strukturbericht* designation** | : | $D0_{15}$ |
| **Pearson symbol** | : | mC16 |
| **Space group number** | : | 12 |
| **Space group symbol** | : | C2/m |
| **AFLOW prototype command** | : | `aflow --proto=AB3_mC16_12_g_ij`<br>`--params=`$a, b/a, c/a, \beta, y_1, x_2, z_2, x_3, y_3, z_3$ |

**Other compounds with this structure:**

- $DyCl_3$, $ErCl_3$, $HoCl_3$, $InCl_3$, $LuCl_3$, $TlCl_3$, $TmCl_3$, $YbCl_3$

- This structure has a somewhat complicated history. Strukturbericht Volume II lists the space group as either $P3_112$ or $P3_212$. This structure was later refined by Ketelaar in space group C2/m. See (Villars, 2008) for more information.

**Base-centered Monoclinic primitive vectors:**

$$\mathbf{a}_1 = \frac{1}{2}\,a\,\hat{\mathbf{x}} - \frac{1}{2}\,b\,\hat{\mathbf{y}}$$
$$\mathbf{a}_2 = \frac{1}{2}\,a\,\hat{\mathbf{x}} + \frac{1}{2}\,b\,\hat{\mathbf{y}}$$
$$\mathbf{a}_3 = c\cos\beta\,\hat{\mathbf{x}} + c\sin\beta\,\hat{\mathbf{z}}$$

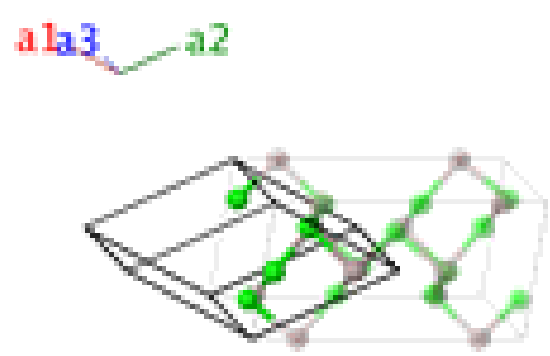

**Basis vectors:**

| | | Lattice Coordinates | | Cartesian Coordinates | Wyckoff Position | Atom Type |
|---|---|---|---|---|---|---|
| $\mathbf{B_1}$ | = | $-y_1\,\mathbf{a_1} + y_1\,\mathbf{a_2}$ | = | $y_1\,b\,\hat{\mathbf{y}}$ | (4$g$) | Al |
| $\mathbf{B_2}$ | = | $y_1\,\mathbf{a_1} - y_1\,\mathbf{a_2}$ | = | $-y_1\,b\,\hat{\mathbf{y}}$ | (4$g$) | Al |
| $\mathbf{B_3}$ | = | $x_2\,\mathbf{a_1} + x_2\,\mathbf{a_2} + z_2\,\mathbf{a_3}$ | = | $(x_2\,a + z_2\,c\,\cos\beta)\,\mathbf{\hat{x}} + z_2\,c\,\sin\beta\,\mathbf{\hat{z}}$ | (4$i$) | Cl I |
| $\mathbf{B_4}$ | = | $-x_2\,\mathbf{a_1} - x_2\,\mathbf{a_2} - z_2\,\mathbf{a_3}$ | = | $-(x_2\,a + z_2\,c\,\cos\beta)\,\mathbf{\hat{x}} - z_2\,c\,\sin\beta\,\mathbf{\hat{z}}$ | (4$i$) | Cl I |
| $\mathbf{B_5}$ | = | $(x_3 - y_3)\,\mathbf{a_1} + (x_3 + y_3)\,\mathbf{a_2} + z_3\,\mathbf{a_3}$ | = | $(x_3\,a + z_3\,c\,\cos\beta)\,\mathbf{\hat{x}} + y_3\,b\,\mathbf{\hat{y}} + z_3\,c\,\sin\beta\,\mathbf{\hat{z}}$ | (8$j$) | Cl II |
| $\mathbf{B_6}$ | = | $-(x_3 + y_3)\,\mathbf{a_1} + (y_3 - x_3)\,\mathbf{a_2} - z_3\,\mathbf{a_3}$ | = | $-(x_3\,a + z_3\,c\,\cos\beta)\,\mathbf{\hat{x}} + y_3\,b\,\mathbf{\hat{y}} - z_3\,c\,\sin\beta\,\mathbf{\hat{z}}$ | (8$j$) | Cl II |
| $\mathbf{B_7}$ | = | $(y_3 - x_3)\,\mathbf{a_1} - (x_3 + y_3)\,\mathbf{a_2} - z_3\,\mathbf{a_3}$ | = | $-(x_3\,a + z_3\,c\,\cos\beta)\,\mathbf{\hat{x}} - y_3\,b\,\mathbf{\hat{y}} - z_3\,c\,\sin\beta\,\mathbf{\hat{z}}$ | (8$j$) | Cl II |
| $\mathbf{B_8}$ | = | $(x_3 + y_3)\,\mathbf{a_1} + (x_3 - y_3)\,\mathbf{a_2} + z_3\,\mathbf{a_3}$ | = | $(x_3\,a + z_3\,c\,\cos\beta)\,\mathbf{\hat{x}} - y_3\,b\,\mathbf{\hat{y}} + z_3\,c\,\sin\beta\,\mathbf{\hat{z}}$ | (8$j$) | Cl II |

---

**Found in:**

---

**Geometry files:**

- CIF: pp. 648
- POSCAR: pp. 648

# $Au_5Mn_2$ Structure: A5B2_mC14_12_a2i_i

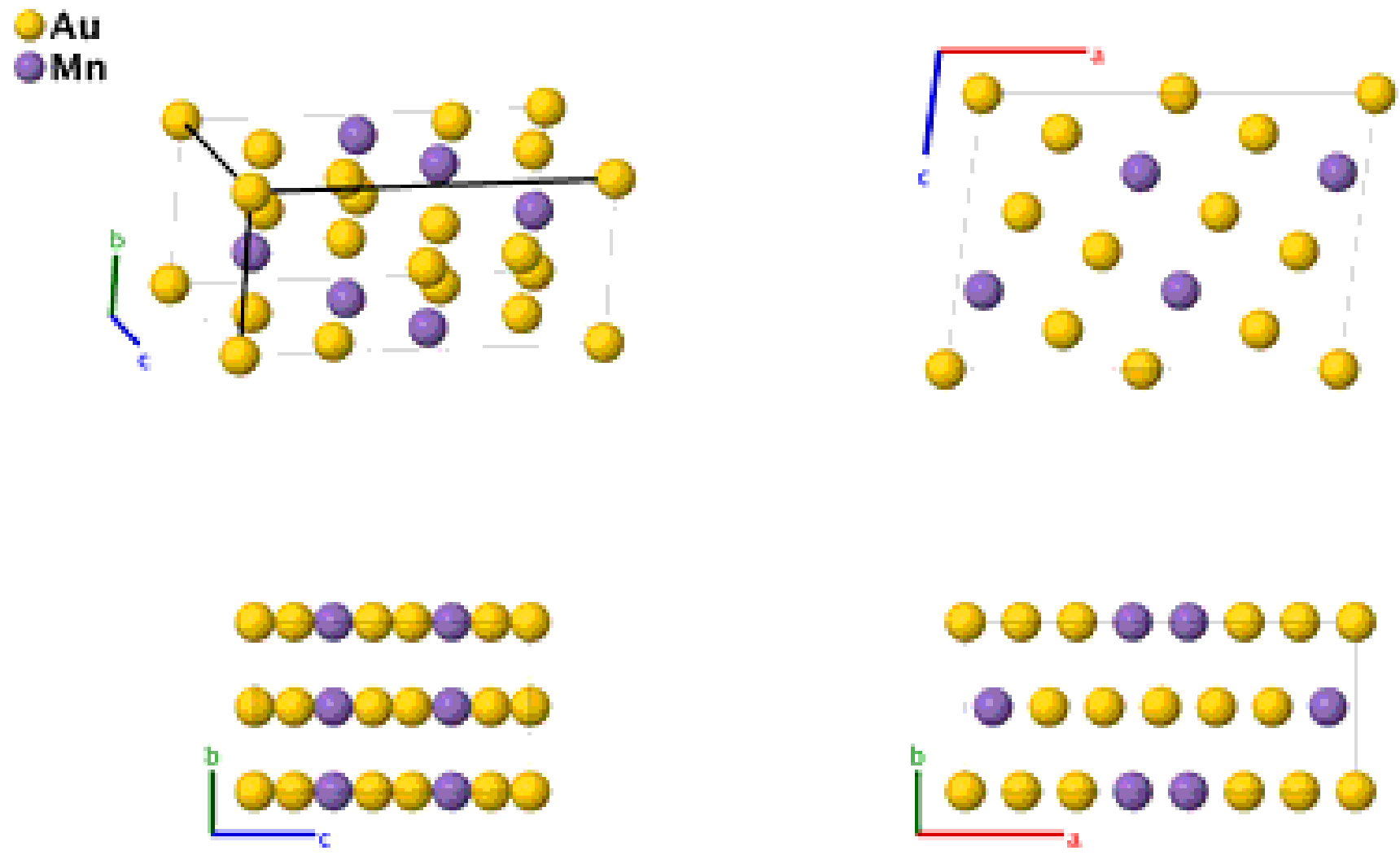

| | | |
|---|---|---|
| **Prototype** | : | $Au_5Mn_2$ |
| **AFLOW prototype label** | : | A5B2_mC14_12_a2i_i |
| ***Strukturbericht* designation** | : | None |
| **Pearson symbol** | : | mC14 |
| **Space group number** | : | 12 |
| **Space group symbol** | : | C2/m |
| **AFLOW prototype command** | : | `aflow --proto=A5B2_mC14_12_a2i_i` <br> `--params=`$a, b/a, c/a, \beta, x_2, z_2, x_3, z_3, x_4, z_4$ |

- As noted by (Pearson, 1972), this structure is very nearly cubic close-packed. As such, it is frequently used for cluster expansion models.

**Base-centered Monoclinic primitive vectors:**

$$\mathbf{a}_1 = \frac{1}{2}\,a\,\mathbf{\hat{x}} - \frac{1}{2}\,b\,\mathbf{\hat{y}}$$
$$\mathbf{a}_2 = \frac{1}{2}\,a\,\mathbf{\hat{x}} + \frac{1}{2}\,b\,\mathbf{\hat{y}}$$
$$\mathbf{a}_3 = c\cos\beta\,\mathbf{\hat{x}} + c\sin\beta\,\mathbf{\hat{z}}$$

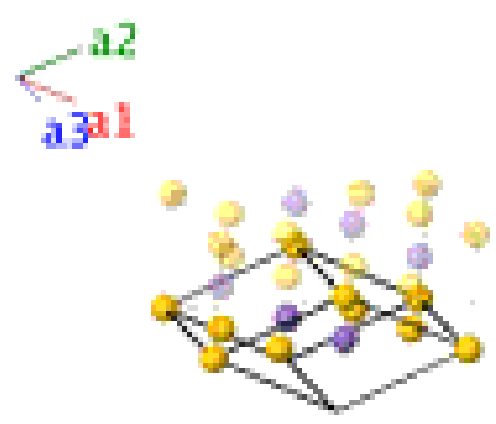

**Basis vectors:**

| | | Lattice Coordinates | | Cartesian Coordinates | Wyckoff Position | Atom Type |
|---|---|---|---|---|---|---|
| $\mathbf{B}_1$ | = | $0\,\mathbf{a}_1 + 0\,\mathbf{a}_2 + 0\,\mathbf{a}_3$ | = | $0\mathbf{\hat{x}} + 0\mathbf{\hat{y}} + 0\mathbf{\hat{z}}$ | $(2a)$ | Au I |
| $\mathbf{B}_2$ | = | $x_2\,\mathbf{a}_1 + x_2\,\mathbf{a}_2 + z_2\,\mathbf{a}_3$ | = | $(x_2\,a + z_2\,c\cos\beta)\,\mathbf{\hat{x}} + z_2\,c\sin\beta\,\mathbf{\hat{z}}$ | $(4i)$ | Au II |
| $\mathbf{B}_3$ | = | $-x_2\,\mathbf{a}_1 - x_2\,\mathbf{a}_2 - z_2\,\mathbf{a}_3$ | = | $-(x_2\,a + z_2\,c\cos\beta)\,\mathbf{\hat{x}} - z_2\,c\sin\beta\,\mathbf{\hat{z}}$ | $(4i)$ | Au II |
| $\mathbf{B}_4$ | = | $x_3\,\mathbf{a}_1 + x_3\,\mathbf{a}_2 + z_3\,\mathbf{a}_3$ | = | $(x_3\,a + z_3\,c\cos\beta)\,\mathbf{\hat{x}} + z_3\,c\sin\beta\,\mathbf{\hat{z}}$ | $(4i)$ | Au III |
| $\mathbf{B_5}$ | = | $-x_3\,\mathbf{a_1} - x_3\,\mathbf{a_2} - z_3\,\mathbf{a_3}$ | = | $-(x_3\,a + z_3\,c\,\cos\beta)\,\hat{\mathbf{x}} - z_3\,c\,\sin\beta\,\hat{\mathbf{z}}$ | (4$i$) | Au III |
| $\mathbf{B_6}$ | = | $x_4\,\mathbf{a_1} + x_4\,\mathbf{a_2} + z_4\,\mathbf{a_3}$ | = | $(x_4\,a + z_4\,c\,\cos\beta)\,\hat{\mathbf{x}} + z_4\,c\,\sin\beta\,\hat{\mathbf{z}}$ | (4$i$) | Mn |
| $\mathbf{B_7}$ | = | $-x_4\,\mathbf{a_1} - x_4\,\mathbf{a_2} - z_4\,\mathbf{a_3}$ | = | $-(x_4\,a + z_4\,c\,\cos\beta)\,\hat{\mathbf{x}} - z_4\,c\,\sin\beta\,\hat{\mathbf{z}}$ | (4$i$) | Mn |

---

---

**Geometry files:**

- CIF: pp. 648
- POSCAR: pp. 648

# $\alpha$-O Structure: A_mC4_12_i

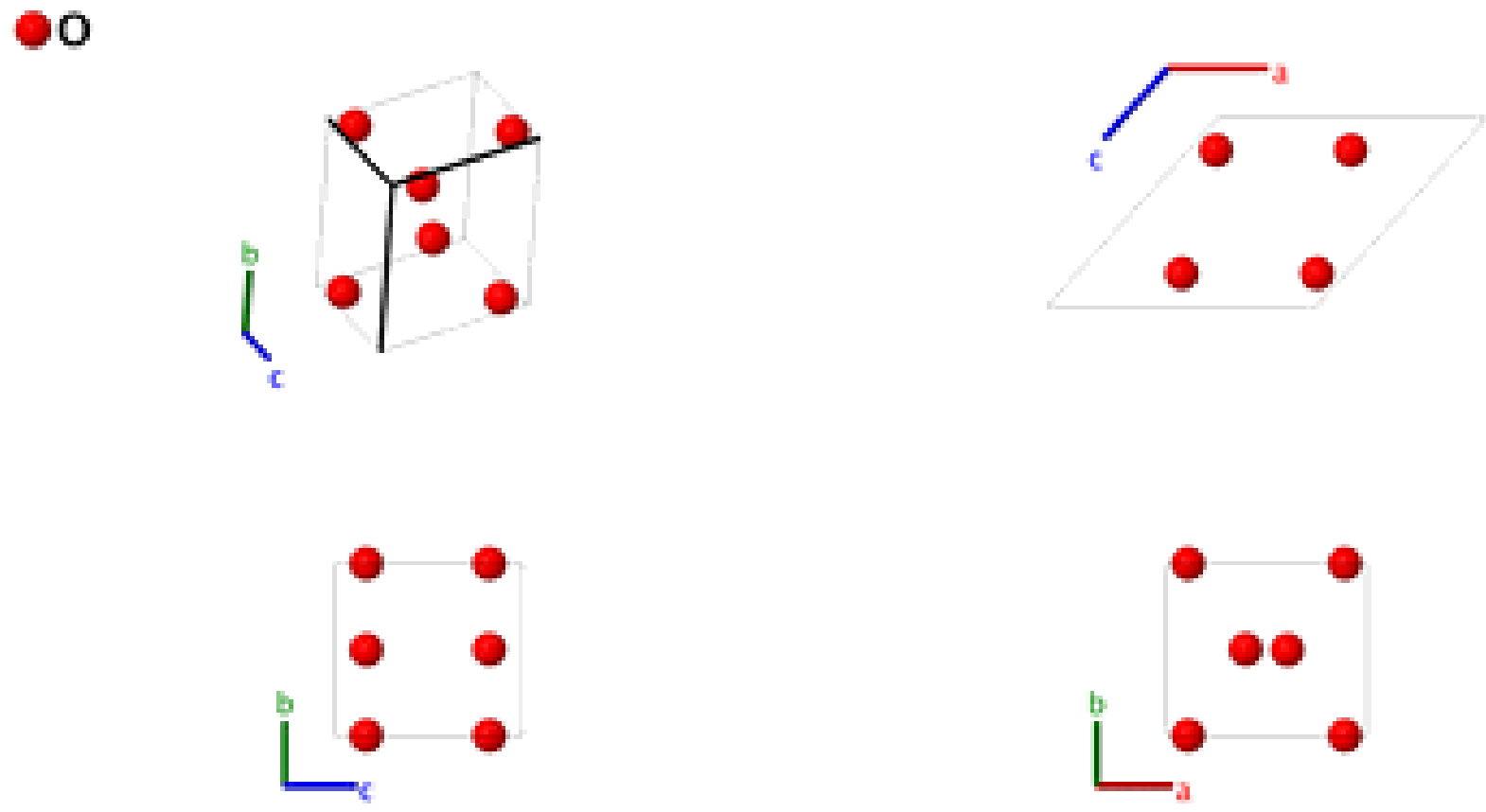

| | | |
|---|---|---|
| **Prototype** | : | $\alpha$-O |
| **AFLOW prototype label** | : | A_mC4_12_i |
| ***Strukturbericht* designation** | : | None |
| **Pearson symbol** | : | mC4 |
| **Space group number** | : | 12 |
| **Space group symbol** | : | C2/m |
| **AFLOW prototype command** | : | `aflow --proto=A_mC4_12_i`<br>`--params=`$a, b/a, c/a, \beta, x_1, z_1$ |

---

**Base-centered Monoclinic primitive vectors:**

$$\mathbf{a}_1 = \frac{1}{2}\,a\,\hat{\mathbf{x}} - \frac{1}{2}\,b\,\hat{\mathbf{y}}$$
$$\mathbf{a}_2 = \frac{1}{2}\,a\,\hat{\mathbf{x}} + \frac{1}{2}\,b\,\hat{\mathbf{y}}$$
$$\mathbf{a}_3 = c\cos\beta\,\hat{\mathbf{x}} + c\sin\beta\,\hat{\mathbf{z}}$$

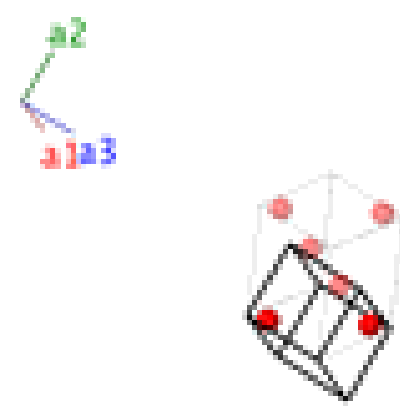

---

**Basis vectors:**

| | | Lattice Coordinates | | Cartesian Coordinates | Wyckoff Position | Atom Type |
|---|---|---|---|---|---|---|
| $\mathbf{B_1}$ | = | $x_1\,\mathbf{a_1} + x_1\,\mathbf{a_2} + z_1\,\mathbf{a_3}$ | = | $(x_1\,a + z_1\,c\,\cos\beta)\,\hat{\mathbf{x}} + z_1\,c\,\sin\beta\,\hat{\mathbf{z}}$ | (4$i$) | O |
| $\mathbf{B_2}$ | = | $-x_1\,\mathbf{a_1} - x_1\,\mathbf{a_2} - z_1\,\mathbf{a_3}$ | = | $-(x_1\,a + z_1\,c\,\cos\beta)\,\hat{\mathbf{x}} - z_1\,c\,\sin\beta\,\hat{\mathbf{z}}$ | (4$i$) | O |

---

---

**Geometry files:**

- CIF: pp. 648
- POSCAR: pp. 649

# Sylvanite ($AgAuTe_4$, $E1_b$) Structure: ABC4_mP12_13_e_a_2g

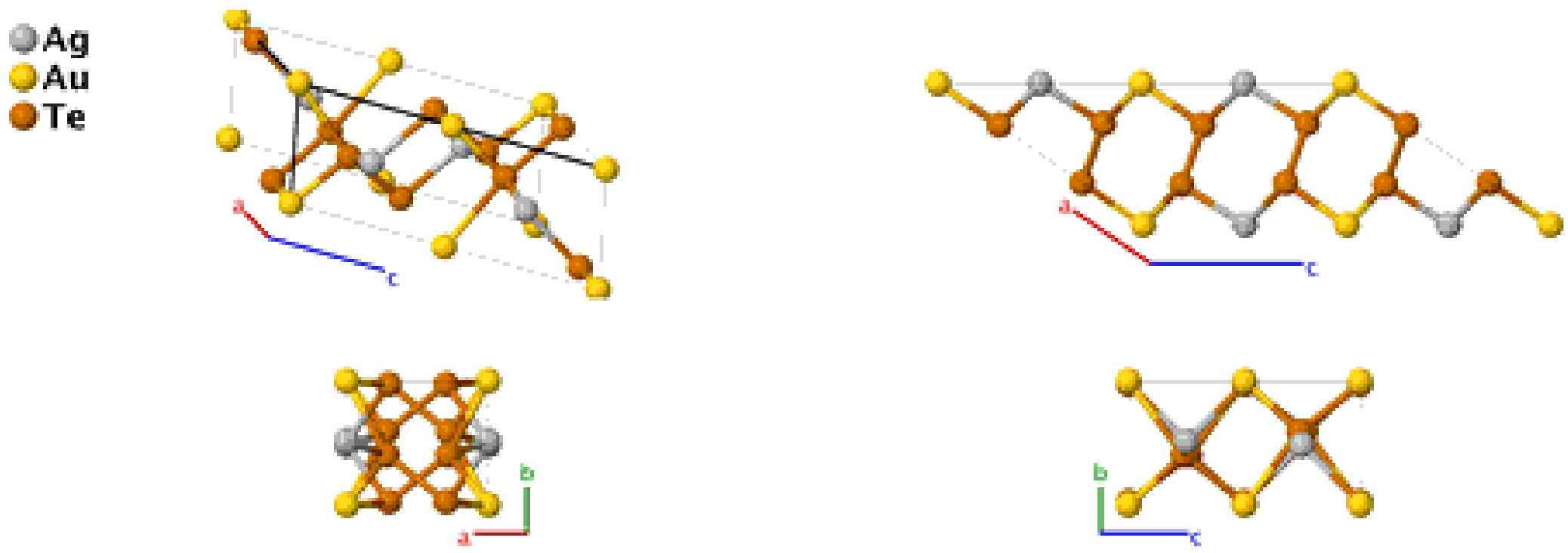

| | | |
|---|---|---|
| **Prototype** | : | $AgAuTe_4$ |
| **AFLOW prototype label** | : | ABC4_mP12_13_e_a_2g |
| ***Strukturbericht* designation** | : | $E1_b$ |
| **Pearson symbol** | : | mP12 |
| **Space group number** | : | 13 |
| **Space group symbol** | : | P2/c |
| **AFLOW prototype command** | : | `aflow --proto=ABC4_mP12_13_e_a_2g --params=`$a, b/a, c/a, \beta, y_2, x_3, y_3, z_3, x_4, y_4, z_4$ |

**Simple Monoclinic primitive vectors:**

$$\mathbf{a}_1 = a\,\hat{\mathbf{x}}$$
$$\mathbf{a}_2 = b\,\hat{\mathbf{y}}$$
$$\mathbf{a}_3 = c\cos\beta\,\hat{\mathbf{x}} + c\sin\beta\,\hat{\mathbf{z}}$$

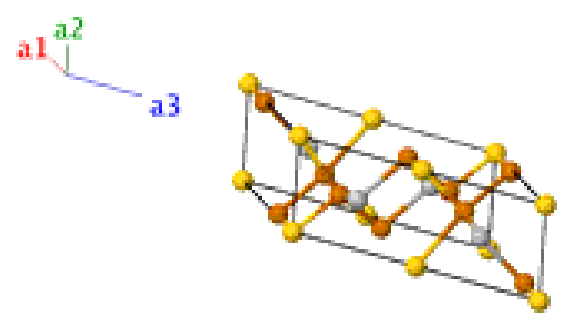

**Basis vectors:**

| | | Lattice Coordinates | | Cartesian Coordinates | Wyckoff Position | Atom Type |
|---|---|---|---|---|---|---|
| $\mathbf{B}_1$ | = | $0\,\mathbf{a}_1 + 0\,\mathbf{a}_2 + 0\,\mathbf{a}_3$ | = | $0\hat{\mathbf{x}} + 0\hat{\mathbf{y}} + 0\hat{\mathbf{z}}$ | $(2a)$ | Au |
| $\mathbf{B}_2$ | = | $\frac{1}{2}\,\mathbf{a}_3$ | = | $\frac{1}{2}\,c\cos\beta\,\hat{\mathbf{x}} + \frac{1}{2}\,c\sin\beta\,\hat{\mathbf{z}}$ | $(2a)$ | Au |
| $\mathbf{B}_3$ | = | $y_2\,\mathbf{a}_2 + \frac{1}{4}\,\mathbf{a}_3$ | = | $\frac{1}{4}\,c\cos\beta\,\hat{\mathbf{x}} + y_2\,b\,\hat{\mathbf{y}} + \frac{1}{4}\,c\sin\beta\,\hat{\mathbf{z}}$ | $(2e)$ | Ag |
| $\mathbf{B}_4$ | = | $-y_2\,\mathbf{a}_2 + \frac{3}{4}\,\mathbf{a}_3$ | = | $\frac{3}{4}\,c\cos\beta\,\hat{\mathbf{x}} - y_2\,b\,\hat{\mathbf{y}} + \frac{3}{4}\,c\sin\beta\,\hat{\mathbf{z}}$ | $(2e)$ | Ag |
| $\mathbf{B}_5$ | = | $x_3\,\mathbf{a}_1 + y_3\,\mathbf{a}_2 + z_3\,\mathbf{a}_3$ | = | $(x_3\,a + z_3\,c\,\cos\beta)\,\hat{\mathbf{x}} + y_3\,b\,\hat{\mathbf{y}} + z_3\,c\,\sin\beta\,\hat{\mathbf{z}}$ | $(4g)$ | Te I |
| $\mathbf{B}_6$ | = | $-x_3\,\mathbf{a}_1 + y_3\,\mathbf{a}_2 + \left(\frac{1}{2} - z_3\right)\mathbf{a}_3$ | = | $\left(-x_3\,a + \left(\frac{1}{2} - z_3\right)\,c\,\cos\beta\right)\,\hat{\mathbf{x}} + y_3\,b\,\hat{\mathbf{y}} + \left(\frac{1}{2} - z_3\right)\,c\,\sin\beta\,\hat{\mathbf{z}}$ | $(4g)$ | Te I |
| $\mathbf{B}_7$ | = | $-x_3\,\mathbf{a}_1 - y_3\,\mathbf{a}_2 - z_3\,\mathbf{a}_3$ | = | $-(x_3\,a + z_3\,c\,\cos\beta)\,\hat{\mathbf{x}} - y_3\,b\,\hat{\mathbf{y}} - z_3\,c\,\sin\beta\,\hat{\mathbf{z}}$ | $(4g)$ | Te I |
| $\mathbf{B}_8$ | = | $x_3\,\mathbf{a}_1 - y_3\,\mathbf{a}_2 + \left(\frac{1}{2} + z_3\right)\mathbf{a}_3$ | = | $\left(x_3\,a + \left(\frac{1}{2} + z_3\right)\,c\,\cos\beta\right)\,\hat{\mathbf{x}} - y_3\,b\,\hat{\mathbf{y}} + \left(\frac{1}{2} + z_3\right)\,c\,\sin\beta\,\hat{\mathbf{z}}$ | $(4g)$ | Te I |
| $\mathbf{B}_9$ | = | $x_4\,\mathbf{a}_1 + y_4\,\mathbf{a}_2 + z_4\,\mathbf{a}_3$ | = | $(x_4\,a + z_4\,c\,\cos\beta)\,\hat{\mathbf{x}} + y_4\,b\,\hat{\mathbf{y}} + z_4\,c\,\sin\beta\,\hat{\mathbf{z}}$ | $(4g)$ | Te II |
| $\mathbf{B}_{10}$ | = | $-x_4\,\mathbf{a}_1 + y_4\,\mathbf{a}_2 + \left(\frac{1}{2} - z_4\right)\mathbf{a}_3$ | = | $\left(-x_4\,a + \left(\frac{1}{2} - z_4\right)\,c\,\cos\beta\right)\,\hat{\mathbf{x}} + y_4\,b\,\hat{\mathbf{y}} + \left(\frac{1}{2} - z_4\right)\,c\,\sin\beta\,\hat{\mathbf{z}}$ | $(4g)$ | Te II |
| $\mathbf{B_{11}}$ | = | $-x_4\,\mathbf{a_1} - y_4\,\mathbf{a_2} - z_4\,\mathbf{a_3}$ | = | $-\left(x_4\,a + z_4\,c\,\cos\beta\right)\,\hat{\mathbf{x}} - y_4\,b\,\hat{\mathbf{y}} - z_4\,c\,\sin\beta\,\hat{\mathbf{z}}$ | (4g) | Te II |
| $\mathbf{B_{12}}$ | = | $x_4\,\mathbf{a_1} - y_4\,\mathbf{a_2} + \left(\frac{1}{2} + z_4\right)\,\mathbf{a_3}$ | = | $\left(x_4\,a + \left(\frac{1}{2} + z_4\right)\,c\,\cos\beta\right)\,\hat{\mathbf{x}} - y_4\,b\,\hat{\mathbf{y}} + \left(\frac{1}{2} + z_4\right)\,c\,\sin\beta\,\hat{\mathbf{z}}$ | (4g) | Te II |

---

**Found in:**

---

**Geometry files:**

- CIF: pp. 649
- POSCAR: pp. 649

# Monoclinic (Hittorf's) Phosphorus Structure: A_mP84_13_21g

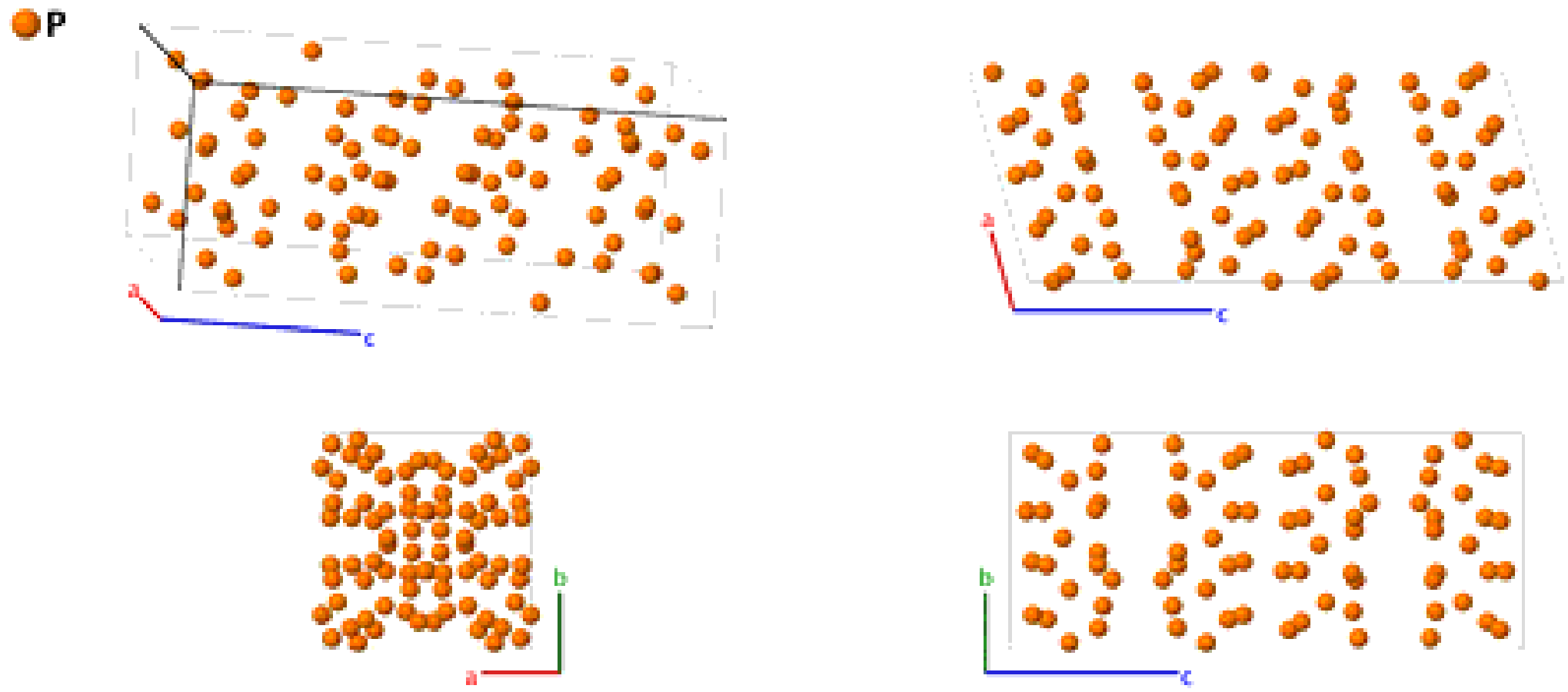

| | | |
|---|---|---|
| **Prototype** | : | P |
| **AFLOW prototype label** | : | A_mP84_13_21g |
| ***Strukturbericht* designation** | : | None |
| **Pearson symbol** | : | mP84 |
| **Space group number** | : | 13 |
| **Space group symbol** | : | P2/c |
| **AFLOW prototype command** | : | `aflow --proto=A_mP84_13_21g` `--params=`$a, b/a, c/a, \beta, x_1, y_1, z_1, x_2, y_2, z_2, x_3, y_3, z_3, x_4, y_4, z_4, x_5, y_5, z_5, x_6, y_6, z_6, x_7, y_7, z_7, x_8, y_8, z_8, x_9, y_9, z_9, x_{10}, y_{10}, z_{10}, x_{11}, y_{11}, z_{11}, x_{12}, y_{12}, z_{12}, x_{13}, y_{13}, z_{13}, x_{14}, y_{14}, z_{14}, x_{15}, y_{15}, z_{15}, x_{16}, y_{16}, z_{16}, x_{17}, y_{17}, z_{17}, x_{18}, y_{18}, z_{18}, x_{19}, y_{19}, z_{19}, x_{20}, y_{20}, z_{20}, x_{21}, y_{21}, z_{21}$ |

**Simple Monoclinic primitive vectors:**

$$\mathbf{a}_1 = a\,\hat{\mathbf{x}}$$
$$\mathbf{a}_2 = b\,\hat{\mathbf{y}}$$
$$\mathbf{a}_3 = c\cos\beta\,\hat{\mathbf{x}} + c\sin\beta\,\hat{\mathbf{z}}$$

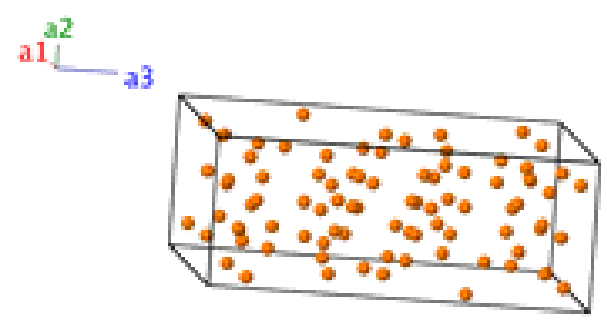

**Basis vectors:**

| | | Lattice Coordinates | | Cartesian Coordinates | Wyckoff Position | Atom Type |
|---|---|---|---|---|---|---|
| $\mathbf{B}_1$ | = | $x_1\,\mathbf{a}_1 + y_1\,\mathbf{a}_2 + z_1\,\mathbf{a}_3$ | = | $(x_1\,a + z_1\,c\,\cos\beta)\,\hat{\mathbf{x}} + y_1\,b\,\hat{\mathbf{y}} + z_1\,c\,\sin\beta\,\hat{\mathbf{z}}$ | (4$g$) | P I |
| $\mathbf{B}_2$ | = | $-x_1\,\mathbf{a}_1 + y_1\,\mathbf{a}_2 + \left(\frac{1}{2} - z_1\right)\,\mathbf{a}_3$ | = | $\left(-x_1\,a + \left(\frac{1}{2} - z_1\right)\,c\,\cos\beta\right)\,\hat{\mathbf{x}} + y_1\,b\,\hat{\mathbf{y}} + \left(\frac{1}{2} - z_1\right)\,c\,\sin\beta\,\hat{\mathbf{z}}$ | (4$g$) | P I |
| $\mathbf{B}_3$ | = | $-x_1\,\mathbf{a}_1 - y_1\,\mathbf{a}_2 - z_1\,\mathbf{a}_3$ | = | $-(x_1\,a + z_1\,c\,\cos\beta)\,\hat{\mathbf{x}} - y_1\,b\,\hat{\mathbf{y}} - z_1\,c\,\sin\beta\,\hat{\mathbf{z}}$ | (4$g$) | P I |
| $\mathbf{B}_4$ | = | $x_1\,\mathbf{a}_1 - y_1\,\mathbf{a}_2 + \left(\frac{1}{2} + z_1\right)\,\mathbf{a}_3$ | = | $\left(x_1\,a + \left(\frac{1}{2} + z_1\right)\,c\,\cos\beta\right)\,\hat{\mathbf{x}} - y_1\,b\,\hat{\mathbf{y}} + \left(\frac{1}{2} + z_1\right)\,c\,\sin\beta\,\hat{\mathbf{z}}$ | (4$g$) | P I |
| $\mathbf{B}_5$ | = | $x_2\,\mathbf{a}_1 + y_2\,\mathbf{a}_2 + z_2\,\mathbf{a}_3$ | = | $(x_2\,a + z_2\,c\,\cos\beta)\,\hat{\mathbf{x}} + y_2\,b\,\hat{\mathbf{y}} + z_2\,c\,\sin\beta\,\hat{\mathbf{z}}$ | (4$g$) | P II |
| $\mathbf{B_6}$ | = | $-x_2\,\mathbf{a_1} + y_2\,\mathbf{a_2} + \left(\frac{1}{2} - z_2\right)\,\mathbf{a_3}$ | = | $\left(-x_2\,a + \left(\frac{1}{2} - z_2\right)\,c\,\cos\beta\right)\,\mathbf{\hat{x}} + y_2\,b\,\mathbf{\hat{y}} + \left(\frac{1}{2} - z_2\right)\,c\,\sin\beta\,\mathbf{\hat{z}}$ | (4$g$) | P II |
| $\mathbf{B_7}$ | = | $-x_2\,\mathbf{a_1} - y_2\,\mathbf{a_2} - z_2\,\mathbf{a_3}$ | = | $-\left(x_2\,a + z_2\,c\,\cos\beta\right)\,\mathbf{\hat{x}} - y_2\,b\,\mathbf{\hat{y}} - z_2\,c\,\sin\beta\,\mathbf{\hat{z}}$ | (4$g$) | P II |
| $\mathbf{B_8}$ | = | $x_2\,\mathbf{a_1} - y_2\,\mathbf{a_2} + \left(\frac{1}{2} + z_2\right)\,\mathbf{a_3}$ | = | $\left(x_2\,a + \left(\frac{1}{2} + z_2\right)\,c\,\cos\beta\right)\,\mathbf{\hat{x}} - y_2\,b\,\mathbf{\hat{y}} + \left(\frac{1}{2} + z_2\right)\,c\,\sin\beta\,\mathbf{\hat{z}}$ | (4$g$) | P II |
| $\mathbf{B_9}$ | = | $x_3\,\mathbf{a_1} + y_3\,\mathbf{a_2} + z_3\,\mathbf{a_3}$ | = | $\left(x_3\,a + z_3\,c\,\cos\beta\right)\,\mathbf{\hat{x}} + y_3\,b\,\mathbf{\hat{y}} + z_3\,c\,\sin\beta\,\mathbf{\hat{z}}$ | (4$g$) | P III |
| $\mathbf{B_{10}}$ | = | $-x_3\,\mathbf{a_1} + y_3\,\mathbf{a_2} + \left(\frac{1}{2} - z_3\right)\,\mathbf{a_3}$ | = | $\left(-x_3\,a + \left(\frac{1}{2} - z_3\right)\,c\,\cos\beta\right)\,\mathbf{\hat{x}} + y_3\,b\,\mathbf{\hat{y}} + \left(\frac{1}{2} - z_3\right)\,c\,\sin\beta\,\mathbf{\hat{z}}$ | (4$g$) | P III |
| $\mathbf{B_{11}}$ | = | $-x_3\,\mathbf{a_1} - y_3\,\mathbf{a_2} - z_3\,\mathbf{a_3}$ | = | $-\left(x_3\,a + z_3\,c\,\cos\beta\right)\,\mathbf{\hat{x}} - y_3\,b\,\mathbf{\hat{y}} - z_3\,c\,\sin\beta\,\mathbf{\hat{z}}$ | (4$g$) | P III |
| $\mathbf{B_{12}}$ | = | $x_3\,\mathbf{a_1} - y_3\,\mathbf{a_2} + \left(\frac{1}{2} + z_3\right)\,\mathbf{a_3}$ | = | $\left(x_3\,a + \left(\frac{1}{2} + z_3\right)\,c\,\cos\beta\right)\,\mathbf{\hat{x}} - y_3\,b\,\mathbf{\hat{y}} + \left(\frac{1}{2} + z_3\right)\,c\,\sin\beta\,\mathbf{\hat{z}}$ | (4$g$) | P III |
| $\mathbf{B_{13}}$ | = | $x_4\,\mathbf{a_1} + y_4\,\mathbf{a_2} + z_4\,\mathbf{a_3}$ | = | $\left(x_4\,a + z_4\,c\,\cos\beta\right)\,\mathbf{\hat{x}} + y_4\,b\,\mathbf{\hat{y}} + z_4\,c\,\sin\beta\,\mathbf{\hat{z}}$ | (4$g$) | P IV |
| $\mathbf{B_{14}}$ | = | $-x_4\,\mathbf{a_1} + y_4\,\mathbf{a_2} + \left(\frac{1}{2} - z_4\right)\,\mathbf{a_3}$ | = | $\left(-x_4\,a + \left(\frac{1}{2} - z_4\right)\,c\,\cos\beta\right)\,\mathbf{\hat{x}} + y_4\,b\,\mathbf{\hat{y}} + \left(\frac{1}{2} - z_4\right)\,c\,\sin\beta\,\mathbf{\hat{z}}$ | (4$g$) | P IV |
| $\mathbf{B_{15}}$ | = | $-x_4\,\mathbf{a_1} - y_4\,\mathbf{a_2} - z_4\,\mathbf{a_3}$ | = | $-\left(x_4\,a + z_4\,c\,\cos\beta\right)\,\mathbf{\hat{x}} - y_4\,b\,\mathbf{\hat{y}} - z_4\,c\,\sin\beta\,\mathbf{\hat{z}}$ | (4$g$) | P IV |
| $\mathbf{B_{16}}$ | = | $x_4\,\mathbf{a_1} - y_4\,\mathbf{a_2} + \left(\frac{1}{2} + z_4\right)\,\mathbf{a_3}$ | = | $\left(x_4\,a + \left(\frac{1}{2} + z_4\right)\,c\,\cos\beta\right)\,\mathbf{\hat{x}} - y_4\,b\,\mathbf{\hat{y}} + \left(\frac{1}{2} + z_4\right)\,c\,\sin\beta\,\mathbf{\hat{z}}$ | (4$g$) | P IV |
| $\mathbf{B_{17}}$ | = | $x_5\,\mathbf{a_1} + y_5\,\mathbf{a_2} + z_5\,\mathbf{a_3}$ | = | $\left(x_5\,a + z_5\,c\,\cos\beta\right)\,\mathbf{\hat{x}} + y_5\,b\,\mathbf{\hat{y}} + z_5\,c\,\sin\beta\,\mathbf{\hat{z}}$ | (4$g$) | P V |
| $\mathbf{B_{18}}$ | = | $-x_5\,\mathbf{a_1} + y_5\,\mathbf{a_2} + \left(\frac{1}{2} - z_5\right)\,\mathbf{a_3}$ | = | $\left(-x_5\,a + \left(\frac{1}{2} - z_5\right)\,c\,\cos\beta\right)\,\mathbf{\hat{x}} + y_5\,b\,\mathbf{\hat{y}} + \left(\frac{1}{2} - z_5\right)\,c\,\sin\beta\,\mathbf{\hat{z}}$ | (4$g$) | P V |
| $\mathbf{B_{19}}$ | = | $-x_5\,\mathbf{a_1} - y_5\,\mathbf{a_2} - z_5\,\mathbf{a_3}$ | = | $-\left(x_5\,a + z_5\,c\,\cos\beta\right)\,\mathbf{\hat{x}} - y_5\,b\,\mathbf{\hat{y}} - z_5\,c\,\sin\beta\,\mathbf{\hat{z}}$ | (4$g$) | P V |
| $\mathbf{B_{20}}$ | = | $x_5\,\mathbf{a_1} - y_5\,\mathbf{a_2} + \left(\frac{1}{2} + z_5\right)\,\mathbf{a_3}$ | = | $\left(x_5\,a + \left(\frac{1}{2} + z_5\right)\,c\,\cos\beta\right)\,\mathbf{\hat{x}} - y_5\,b\,\mathbf{\hat{y}} + \left(\frac{1}{2} + z_5\right)\,c\,\sin\beta\,\mathbf{\hat{z}}$ | (4$g$) | P V |
| $\mathbf{B_{21}}$ | = | $x_6\,\mathbf{a_1} + y_6\,\mathbf{a_2} + z_6\,\mathbf{a_3}$ | = | $\left(x_6\,a + z_6\,c\,\cos\beta\right)\,\mathbf{\hat{x}} + y_6\,b\,\mathbf{\hat{y}} + z_6\,c\,\sin\beta\,\mathbf{\hat{z}}$ | (4$g$) | P VI |
| $\mathbf{B_{22}}$ | = | $-x_6\,\mathbf{a_1} + y_6\,\mathbf{a_2} + \left(\frac{1}{2} - z_6\right)\,\mathbf{a_3}$ | = | $\left(-x_6\,a + \left(\frac{1}{2} - z_6\right)\,c\,\cos\beta\right)\,\mathbf{\hat{x}} + y_6\,b\,\mathbf{\hat{y}} + \left(\frac{1}{2} - z_6\right)\,c\,\sin\beta\,\mathbf{\hat{z}}$ | (4$g$) | P VI |
| $\mathbf{B_{23}}$ | = | $-x_6\,\mathbf{a_1} - y_6\,\mathbf{a_2} - z_6\,\mathbf{a_3}$ | = | $-\left(x_6\,a + z_6\,c\,\cos\beta\right)\,\mathbf{\hat{x}} - y_6\,b\,\mathbf{\hat{y}} - z_6\,c\,\sin\beta\,\mathbf{\hat{z}}$ | (4$g$) | P VI |
| $\mathbf{B_{24}}$ | = | $x_6\,\mathbf{a_1} - y_6\,\mathbf{a_2} + \left(\frac{1}{2} + z_6\right)\,\mathbf{a_3}$ | = | $\left(x_6\,a + \left(\frac{1}{2} + z_6\right)\,c\,\cos\beta\right)\,\mathbf{\hat{x}} - y_6\,b\,\mathbf{\hat{y}} + \left(\frac{1}{2} + z_6\right)\,c\,\sin\beta\,\mathbf{\hat{z}}$ | (4$g$) | P VI |
| $\mathbf{B_{25}}$ | = | $x_7\,\mathbf{a_1} + y_7\,\mathbf{a_2} + z_7\,\mathbf{a_3}$ | = | $\left(x_7\,a + z_7\,c\,\cos\beta\right)\,\mathbf{\hat{x}} + y_7\,b\,\mathbf{\hat{y}} + z_7\,c\,\sin\beta\,\mathbf{\hat{z}}$ | (4$g$) | P VII |
| $\mathbf{B_{26}}$ | = | $-x_7\,\mathbf{a_1} + y_7\,\mathbf{a_2} + \left(\frac{1}{2} - z_7\right)\,\mathbf{a_3}$ | = | $\left(-x_7\,a + \left(\frac{1}{2} - z_7\right)\,c\,\cos\beta\right)\,\mathbf{\hat{x}} + y_7\,b\,\mathbf{\hat{y}} + \left(\frac{1}{2} - z_7\right)\,c\,\sin\beta\,\mathbf{\hat{z}}$ | (4$g$) | P VII |
| $\mathbf{B_{27}}$ | = | $-x_7\,\mathbf{a_1} - y_7\,\mathbf{a_2} - z_7\,\mathbf{a_3}$ | = | $-\left(x_7\,a + z_7\,c\,\cos\beta\right)\,\mathbf{\hat{x}} - y_7\,b\,\mathbf{\hat{y}} - z_7\,c\,\sin\beta\,\mathbf{\hat{z}}$ | (4$g$) | P VII |
| $\mathbf{B}_{28}$ | = | $x_{7}\,\mathbf{a}_{1} - y_{7}\,\mathbf{a}_{2} + \left(\frac{1}{2} + z_{7}\right)\mathbf{a}_{3}$ | = | $\left(x_{7}\,a + \left(\frac{1}{2} + z_{7}\right) c\,\cos\beta\right)\hat{\mathbf{x}} - y_{7}\,b\,\hat{\mathbf{y}} + \left(\frac{1}{2} + z_{7}\right) c\,\sin\beta\,\hat{\mathbf{z}}$ | (4g) | P VII |
| $\mathbf{B}_{29}$ | = | $x_{8}\,\mathbf{a}_{1} + y_{8}\,\mathbf{a}_{2} + z_{8}\,\mathbf{a}_{3}$ | = | $(x_{8}\,a + z_{8}\,c\,\cos\beta)\,\hat{\mathbf{x}} + y_{8}\,b\,\hat{\mathbf{y}} + z_{8}\,c\,\sin\beta\,\hat{\mathbf{z}}$ | (4g) | P VIII |
| $\mathbf{B}_{30}$ | = | $-x_{8}\,\mathbf{a}_{1} + y_{8}\,\mathbf{a}_{2} + \left(\frac{1}{2} - z_{8}\right)\mathbf{a}_{3}$ | = | $\left(-x_{8}\,a + \left(\frac{1}{2} - z_{8}\right) c\,\cos\beta\right)\hat{\mathbf{x}} + y_{8}\,b\,\hat{\mathbf{y}} + \left(\frac{1}{2} - z_{8}\right) c\,\sin\beta\,\hat{\mathbf{z}}$ | (4g) | P VIII |
| $\mathbf{B}_{31}$ | = | $-x_{8}\,\mathbf{a}_{1} - y_{8}\,\mathbf{a}_{2} - z_{8}\,\mathbf{a}_{3}$ | = | $-(x_{8}\,a + z_{8}\,c\,\cos\beta)\,\hat{\mathbf{x}} - y_{8}\,b\,\hat{\mathbf{y}} - z_{8}\,c\,\sin\beta\,\hat{\mathbf{z}}$ | (4g) | P VIII |
| $\mathbf{B}_{32}$ | = | $x_{8}\,\mathbf{a}_{1} - y_{8}\,\mathbf{a}_{2} + \left(\frac{1}{2} + z_{8}\right)\mathbf{a}_{3}$ | = | $\left(x_{8}\,a + \left(\frac{1}{2} + z_{8}\right) c\,\cos\beta\right)\hat{\mathbf{x}} - y_{8}\,b\,\hat{\mathbf{y}} + \left(\frac{1}{2} + z_{8}\right) c\,\sin\beta\,\hat{\mathbf{z}}$ | (4g) | P VIII |
| $\mathbf{B}_{33}$ | = | $x_{9}\,\mathbf{a}_{1} + y_{9}\,\mathbf{a}_{2} + z_{9}\,\mathbf{a}_{3}$ | = | $(x_{9}\,a + z_{9}\,c\,\cos\beta)\,\hat{\mathbf{x}} + y_{9}\,b\,\hat{\mathbf{y}} + z_{9}\,c\,\sin\beta\,\hat{\mathbf{z}}$ | (4g) | P IX |
| $\mathbf{B}_{34}$ | = | $-x_{9}\,\mathbf{a}_{1} + y_{9}\,\mathbf{a}_{2} + \left(\frac{1}{2} - z_{9}\right)\mathbf{a}_{3}$ | = | $\left(-x_{9}\,a + \left(\frac{1}{2} - z_{9}\right) c\,\cos\beta\right)\hat{\mathbf{x}} + y_{9}\,b\,\hat{\mathbf{y}} + \left(\frac{1}{2} - z_{9}\right) c\,\sin\beta\,\hat{\mathbf{z}}$ | (4g) | P IX |
| $\mathbf{B}_{35}$ | = | $-x_{9}\,\mathbf{a}_{1} - y_{9}\,\mathbf{a}_{2} - z_{9}\,\mathbf{a}_{3}$ | = | $-(x_{9}\,a + z_{9}\,c\,\cos\beta)\,\hat{\mathbf{x}} - y_{9}\,b\,\hat{\mathbf{y}} - z_{9}\,c\,\sin\beta\,\hat{\mathbf{z}}$ | (4g) | P IX |
| $\mathbf{B}_{36}$ | = | $x_{9}\,\mathbf{a}_{1} - y_{9}\,\mathbf{a}_{2} + \left(\frac{1}{2} + z_{9}\right)\mathbf{a}_{3}$ | = | $\left(x_{9}\,a + \left(\frac{1}{2} + z_{9}\right) c\,\cos\beta\right)\hat{\mathbf{x}} - y_{9}\,b\,\hat{\mathbf{y}} + \left(\frac{1}{2} + z_{9}\right) c\,\sin\beta\,\hat{\mathbf{z}}$ | (4g) | P IX |
| $\mathbf{B}_{37}$ | = | $x_{10}\,\mathbf{a}_{1} + y_{10}\,\mathbf{a}_{2} + z_{10}\,\mathbf{a}_{3}$ | = | $(x_{10}\,a + z_{10}\,c\,\cos\beta)\,\hat{\mathbf{x}} + y_{10}\,b\,\hat{\mathbf{y}} + z_{10}\,c\,\sin\beta\,\hat{\mathbf{z}}$ | (4g) | P X |
| $\mathbf{B}_{38}$ | = | $-x_{10}\,\mathbf{a}_{1} + y_{10}\,\mathbf{a}_{2} + \left(\frac{1}{2} - z_{10}\right)\mathbf{a}_{3}$ | = | $\left(-x_{10}\,a + \left(\frac{1}{2} - z_{10}\right) c\,\cos\beta\right)\hat{\mathbf{x}} + y_{10}\,b\,\hat{\mathbf{y}} + \left(\frac{1}{2} - z_{10}\right) c\,\sin\beta\,\hat{\mathbf{z}}$ | (4g) | P X |
| $\mathbf{B}_{39}$ | = | $-x_{10}\,\mathbf{a}_{1} - y_{10}\,\mathbf{a}_{2} - z_{10}\,\mathbf{a}_{3}$ | = | $-(x_{10}\,a + z_{10}\,c\,\cos\beta)\,\hat{\mathbf{x}} - y_{10}\,b\,\hat{\mathbf{y}} - z_{10}\,c\,\sin\beta\,\hat{\mathbf{z}}$ | (4g) | P X |
| $\mathbf{B}_{40}$ | = | $x_{10}\,\mathbf{a}_{1} - y_{10}\,\mathbf{a}_{2} + \left(\frac{1}{2} + z_{10}\right)\mathbf{a}_{3}$ | = | $\left(x_{10}\,a + \left(\frac{1}{2} + z_{10}\right) c\,\cos\beta\right)\hat{\mathbf{x}} - y_{10}\,b\,\hat{\mathbf{y}} + \left(\frac{1}{2} + z_{10}\right) c\,\sin\beta\,\hat{\mathbf{z}}$ | (4g) | P X |
| $\mathbf{B}_{41}$ | = | $x_{11}\,\mathbf{a}_{1} + y_{11}\,\mathbf{a}_{2} + z_{11}\,\mathbf{a}_{3}$ | = | $(x_{11}\,a + z_{11}\,c\,\cos\beta)\,\hat{\mathbf{x}} + y_{11}\,b\,\hat{\mathbf{y}} + z_{11}\,c\,\sin\beta\,\hat{\mathbf{z}}$ | (4g) | P XI |
| $\mathbf{B}_{42}$ | = | $-x_{11}\,\mathbf{a}_{1} + y_{11}\,\mathbf{a}_{2} + \left(\frac{1}{2} - z_{11}\right)\mathbf{a}_{3}$ | = | $\left(-x_{11}\,a + \left(\frac{1}{2} - z_{11}\right) c\,\cos\beta\right)\hat{\mathbf{x}} + y_{11}\,b\,\hat{\mathbf{y}} + \left(\frac{1}{2} - z_{11}\right) c\,\sin\beta\,\hat{\mathbf{z}}$ | (4g) | P XI |
| $\mathbf{B}_{43}$ | = | $-x_{11}\,\mathbf{a}_{1} - y_{11}\,\mathbf{a}_{2} - z_{11}\,\mathbf{a}_{3}$ | = | $-(x_{11}\,a + z_{11}\,c\,\cos\beta)\,\hat{\mathbf{x}} - y_{11}\,b\,\hat{\mathbf{y}} - z_{11}\,c\,\sin\beta\,\hat{\mathbf{z}}$ | (4g) | P XI |
| $\mathbf{B}_{44}$ | = | $x_{11}\,\mathbf{a}_{1} - y_{11}\,\mathbf{a}_{2} + \left(\frac{1}{2} + z_{11}\right)\mathbf{a}_{3}$ | = | $\left(x_{11}\,a + \left(\frac{1}{2} + z_{11}\right) c\,\cos\beta\right)\hat{\mathbf{x}} - y_{11}\,b\,\hat{\mathbf{y}} + \left(\frac{1}{2} + z_{11}\right) c\,\sin\beta\,\hat{\mathbf{z}}$ | (4g) | P XI |
| $\mathbf{B}_{45}$ | = | $x_{12}\,\mathbf{a}_{1} + y_{12}\,\mathbf{a}_{2} + z_{12}\,\mathbf{a}_{3}$ | = | $(x_{12}\,a + z_{12}\,c\,\cos\beta)\,\hat{\mathbf{x}} + y_{12}\,b\,\hat{\mathbf{y}} + z_{12}\,c\,\sin\beta\,\hat{\mathbf{z}}$ | (4g) | P XII |
| $\mathbf{B}_{46}$ | = | $-x_{12}\,\mathbf{a}_{1} + y_{12}\,\mathbf{a}_{2} + \left(\frac{1}{2} - z_{12}\right)\mathbf{a}_{3}$ | = | $\left(-x_{12}\,a + \left(\frac{1}{2} - z_{12}\right) c\,\cos\beta\right)\hat{\mathbf{x}} + y_{12}\,b\,\hat{\mathbf{y}} + \left(\frac{1}{2} - z_{12}\right) c\,\sin\beta\,\hat{\mathbf{z}}$ | (4g) | P XII |
| $\mathbf{B}_{47}$ | = | $-x_{12}\,\mathbf{a}_{1} - y_{12}\,\mathbf{a}_{2} - z_{12}\,\mathbf{a}_{3}$ | = | $-(x_{12}\,a + z_{12}\,c\,\cos\beta)\,\hat{\mathbf{x}} - y_{12}\,b\,\hat{\mathbf{y}} - z_{12}\,c\,\sin\beta\,\hat{\mathbf{z}}$ | (4g) | P XII |
| $\mathbf{B}_{48}$ | = | $x_{12}\,\mathbf{a}_{1} - y_{12}\,\mathbf{a}_{2} + \left(\frac{1}{2} + z_{12}\right)\mathbf{a}_{3}$ | = | $\left(x_{12}\,a + \left(\frac{1}{2} + z_{12}\right) c\,\cos\beta\right)\hat{\mathbf{x}} - y_{12}\,b\,\hat{\mathbf{y}} + \left(\frac{1}{2} + z_{12}\right) c\,\sin\beta\,\hat{\mathbf{z}}$ | (4g) | P XII |
| $\mathbf{B}_{49}$ | = | $x_{13}\,\mathbf{a}_{1} + y_{13}\,\mathbf{a}_{2} + z_{13}\,\mathbf{a}_{3}$ | = | $(x_{13}\,a + z_{13}\,c\,\cos\beta)\,\hat{\mathbf{x}} + y_{13}\,b\,\hat{\mathbf{y}} + z_{13}\,c\,\sin\beta\,\hat{\mathbf{z}}$ | (4g) | P XIII |
| $\mathbf{B}_{50}$ | = | $-x_{13}\,\mathbf{a}_{1}+y_{13}\,\mathbf{a}_{2}+\left(\frac{1}{2}-z_{13}\right)\,\mathbf{a}_{3}$ | = | $\left(-x_{13}\,a+\left(\frac{1}{2}-z_{13}\right)\,c\,\cos\beta\right)\,\mathbf{\hat{x}}+y_{13}\,b\,\mathbf{\hat{y}}+\left(\frac{1}{2}-z_{13}\right)\,c\,\sin\beta\,\mathbf{\hat{z}}$ | (4$g$) | P XIII |
| $\mathbf{B}_{51}$ | = | $-x_{13}\,\mathbf{a}_{1}-y_{13}\,\mathbf{a}_{2}-z_{13}\,\mathbf{a}_{3}$ | = | $-\left(x_{13}\,a+z_{13}\,c\,\cos\beta\right)\,\mathbf{\hat{x}}-y_{13}\,b\,\mathbf{\hat{y}}-z_{13}\,c\,\sin\beta\,\mathbf{\hat{z}}$ | (4$g$) | P XIII |
| $\mathbf{B}_{52}$ | = | $x_{13}\,\mathbf{a}_{1}-y_{13}\,\mathbf{a}_{2}+\left(\frac{1}{2}+z_{13}\right)\,\mathbf{a}_{3}$ | = | $\left(x_{13}\,a+\left(\frac{1}{2}+z_{13}\right)\,c\,\cos\beta\right)\,\mathbf{\hat{x}}-y_{13}\,b\,\mathbf{\hat{y}}+\left(\frac{1}{2}+z_{13}\right)\,c\,\sin\beta\,\mathbf{\hat{z}}$ | (4$g$) | P XIII |
| $\mathbf{B}_{53}$ | = | $x_{14}\,\mathbf{a}_{1}+y_{14}\,\mathbf{a}_{2}+z_{14}\,\mathbf{a}_{3}$ | = | $\left(x_{14}\,a+z_{14}\,c\,\cos\beta\right)\,\mathbf{\hat{x}}+y_{14}\,b\,\mathbf{\hat{y}}+z_{14}\,c\,\sin\beta\,\mathbf{\hat{z}}$ | (4$g$) | P XIV |
| $\mathbf{B}_{54}$ | = | $-x_{14}\,\mathbf{a}_{1}+y_{14}\,\mathbf{a}_{2}+\left(\frac{1}{2}-z_{14}\right)\,\mathbf{a}_{3}$ | = | $\left(-x_{14}\,a+\left(\frac{1}{2}-z_{14}\right)\,c\,\cos\beta\right)\,\mathbf{\hat{x}}+y_{14}\,b\,\mathbf{\hat{y}}+\left(\frac{1}{2}-z_{14}\right)\,c\,\sin\beta\,\mathbf{\hat{z}}$ | (4$g$) | P XIV |
| $\mathbf{B}_{55}$ | = | $-x_{14}\,\mathbf{a}_{1}-y_{14}\,\mathbf{a}_{2}-z_{14}\,\mathbf{a}_{3}$ | = | $-\left(x_{14}\,a+z_{14}\,c\,\cos\beta\right)\,\mathbf{\hat{x}}-y_{14}\,b\,\mathbf{\hat{y}}-z_{14}\,c\,\sin\beta\,\mathbf{\hat{z}}$ | (4$g$) | P XIV |
| $\mathbf{B}_{56}$ | = | $x_{14}\,\mathbf{a}_{1}-y_{14}\,\mathbf{a}_{2}+\left(\frac{1}{2}+z_{14}\right)\,\mathbf{a}_{3}$ | = | $\left(x_{14}\,a+\left(\frac{1}{2}+z_{14}\right)\,c\,\cos\beta\right)\,\mathbf{\hat{x}}-y_{14}\,b\,\mathbf{\hat{y}}+\left(\frac{1}{2}+z_{14}\right)\,c\,\sin\beta\,\mathbf{\hat{z}}$ | (4$g$) | P XIV |
| $\mathbf{B}_{57}$ | = | $x_{15}\,\mathbf{a}_{1}+y_{15}\,\mathbf{a}_{2}+z_{15}\,\mathbf{a}_{3}$ | = | $\left(x_{15}\,a+z_{15}\,c\,\cos\beta\right)\,\mathbf{\hat{x}}+y_{15}\,b\,\mathbf{\hat{y}}+z_{15}\,c\,\sin\beta\,\mathbf{\hat{z}}$ | (4$g$) | P XV |
| $\mathbf{B}_{58}$ | = | $-x_{15}\,\mathbf{a}_{1}+y_{15}\,\mathbf{a}_{2}+\left(\frac{1}{2}-z_{15}\right)\,\mathbf{a}_{3}$ | = | $\left(-x_{15}\,a+\left(\frac{1}{2}-z_{15}\right)\,c\,\cos\beta\right)\,\mathbf{\hat{x}}+y_{15}\,b\,\mathbf{\hat{y}}+\left(\frac{1}{2}-z_{15}\right)\,c\,\sin\beta\,\mathbf{\hat{z}}$ | (4$g$) | P XV |
| $\mathbf{B}_{59}$ | = | $-x_{15}\,\mathbf{a}_{1}-y_{15}\,\mathbf{a}_{2}-z_{15}\,\mathbf{a}_{3}$ | = | $-\left(x_{15}\,a+z_{15}\,c\,\cos\beta\right)\,\mathbf{\hat{x}}-y_{15}\,b\,\mathbf{\hat{y}}-z_{15}\,c\,\sin\beta\,\mathbf{\hat{z}}$ | (4$g$) | P XV |
| $\mathbf{B}_{60}$ | = | $x_{15}\,\mathbf{a}_{1}-y_{15}\,\mathbf{a}_{2}+\left(\frac{1}{2}+z_{15}\right)\,\mathbf{a}_{3}$ | = | $\left(x_{15}\,a+\left(\frac{1}{2}+z_{15}\right)\,c\,\cos\beta\right)\,\mathbf{\hat{x}}-y_{15}\,b\,\mathbf{\hat{y}}+\left(\frac{1}{2}+z_{15}\right)\,c\,\sin\beta\,\mathbf{\hat{z}}$ | (4$g$) | P XV |
| $\mathbf{B}_{61}$ | = | $x_{16}\,\mathbf{a}_{1}+y_{16}\,\mathbf{a}_{2}+z_{16}\,\mathbf{a}_{3}$ | = | $\left(x_{16}\,a+z_{16}\,c\,\cos\beta\right)\,\mathbf{\hat{x}}+y_{16}\,b\,\mathbf{\hat{y}}+z_{16}\,c\,\sin\beta\,\mathbf{\hat{z}}$ | (4$g$) | P XVI |
| $\mathbf{B}_{62}$ | = | $-x_{16}\,\mathbf{a}_{1}+y_{16}\,\mathbf{a}_{2}+\left(\frac{1}{2}-z_{16}\right)\,\mathbf{a}_{3}$ | = | $\left(-x_{16}\,a+\left(\frac{1}{2}-z_{16}\right)\,c\,\cos\beta\right)\,\mathbf{\hat{x}}+y_{16}\,b\,\mathbf{\hat{y}}+\left(\frac{1}{2}-z_{16}\right)\,c\,\sin\beta\,\mathbf{\hat{z}}$ | (4$g$) | P XVI |
| $\mathbf{B}_{63}$ | = | $-x_{16}\,\mathbf{a}_{1}-y_{16}\,\mathbf{a}_{2}-z_{16}\,\mathbf{a}_{3}$ | = | $-\left(x_{16}\,a+z_{16}\,c\,\cos\beta\right)\,\mathbf{\hat{x}}-y_{16}\,b\,\mathbf{\hat{y}}-z_{16}\,c\,\sin\beta\,\mathbf{\hat{z}}$ | (4$g$) | P XVI |
| $\mathbf{B}_{64}$ | = | $x_{16}\,\mathbf{a}_{1}-y_{16}\,\mathbf{a}_{2}+\left(\frac{1}{2}+z_{16}\right)\,\mathbf{a}_{3}$ | = | $\left(x_{16}\,a+\left(\frac{1}{2}+z_{16}\right)\,c\,\cos\beta\right)\,\mathbf{\hat{x}}-y_{16}\,b\,\mathbf{\hat{y}}+\left(\frac{1}{2}+z_{16}\right)\,c\,\sin\beta\,\mathbf{\hat{z}}$ | (4$g$) | P XVI |
| $\mathbf{B}_{65}$ | = | $x_{17}\,\mathbf{a}_{1}+y_{17}\,\mathbf{a}_{2}+z_{17}\,\mathbf{a}_{3}$ | = | $\left(x_{17}\,a+z_{17}\,c\,\cos\beta\right)\,\mathbf{\hat{x}}+y_{17}\,b\,\mathbf{\hat{y}}+z_{17}\,c\,\sin\beta\,\mathbf{\hat{z}}$ | (4$g$) | P XVII |
| $\mathbf{B}_{66}$ | = | $-x_{17}\,\mathbf{a}_{1}+y_{17}\,\mathbf{a}_{2}+\left(\frac{1}{2}-z_{17}\right)\,\mathbf{a}_{3}$ | = | $\left(-x_{17}\,a+\left(\frac{1}{2}-z_{17}\right)\,c\,\cos\beta\right)\,\mathbf{\hat{x}}+y_{17}\,b\,\mathbf{\hat{y}}+\left(\frac{1}{2}-z_{17}\right)\,c\,\sin\beta\,\mathbf{\hat{z}}$ | (4$g$) | P XVII |
| $\mathbf{B}_{67}$ | = | $-x_{17}\,\mathbf{a}_{1}-y_{17}\,\mathbf{a}_{2}-z_{17}\,\mathbf{a}_{3}$ | = | $-\left(x_{17}\,a+z_{17}\,c\,\cos\beta\right)\,\mathbf{\hat{x}}-y_{17}\,b\,\mathbf{\hat{y}}-z_{17}\,c\,\sin\beta\,\mathbf{\hat{z}}$ | (4$g$) | P XVII |
| $\mathbf{B}_{68}$ | = | $x_{17}\,\mathbf{a}_{1}-y_{17}\,\mathbf{a}_{2}+\left(\frac{1}{2}+z_{17}\right)\,\mathbf{a}_{3}$ | = | $\left(x_{17}\,a+\left(\frac{1}{2}+z_{17}\right)\,c\,\cos\beta\right)\,\mathbf{\hat{x}}-y_{17}\,b\,\mathbf{\hat{y}}+\left(\frac{1}{2}+z_{17}\right)\,c\,\sin\beta\,\mathbf{\hat{z}}$ | (4$g$) | P XVII |
| $\mathbf{B}_{69}$ | = | $x_{18}\,\mathbf{a}_{1}+y_{18}\,\mathbf{a}_{2}+z_{18}\,\mathbf{a}_{3}$ | = | $\left(x_{18}\,a+z_{18}\,c\,\cos\beta\right)\,\mathbf{\hat{x}}+y_{18}\,b\,\mathbf{\hat{y}}+z_{18}\,c\,\sin\beta\,\mathbf{\hat{z}}$ | (4$g$) | P XVIII |
| $\mathbf{B}_{70}$ | = | $-x_{18}\,\mathbf{a}_{1}+y_{18}\,\mathbf{a}_{2}+\left(\frac{1}{2}-z_{18}\right)\,\mathbf{a}_{3}$ | = | $\left(-x_{18}\,a+\left(\frac{1}{2}-z_{18}\right)\,c\,\cos\beta\right)\,\mathbf{\hat{x}}+y_{18}\,b\,\mathbf{\hat{y}}+\left(\frac{1}{2}-z_{18}\right)\,c\,\sin\beta\,\mathbf{\hat{z}}$ | (4$g$) | P XVIII |
| $\mathbf{B}_{71}$ | = | $-x_{18}\,\mathbf{a}_{1}-y_{18}\,\mathbf{a}_{2}-z_{18}\,\mathbf{a}_{3}$ | = | $-\left(x_{18}\,a+z_{18}\,c\,\cos\beta\right)\,\mathbf{\hat{x}}-y_{18}\,b\,\mathbf{\hat{y}}-z_{18}\,c\,\sin\beta\,\mathbf{\hat{z}}$ | (4$g$) | P XVIII |
| $\mathbf{B}_{72}$ | = | $x_{18}\,\mathbf{a}_1 - y_{18}\,\mathbf{a}_2 + \left(\frac{1}{2} + z_{18}\right)\,\mathbf{a}_3$ | = | $\left(x_{18}\,a + \left(\frac{1}{2} + z_{18}\right)\,c\,\cos\beta\right)\,\hat{\mathbf{x}} - y_{18}\,b\,\hat{\mathbf{y}} + \left(\frac{1}{2} + z_{18}\right)\,c\,\sin\beta\,\hat{\mathbf{z}}$ | (4$g$) | P XVIII |
| $\mathbf{B}_{73}$ | = | $x_{19}\,\mathbf{a}_1 + y_{19}\,\mathbf{a}_2 + z_{19}\,\mathbf{a}_3$ | = | $(x_{19}\,a + z_{19}\,c\,\cos\beta)\,\hat{\mathbf{x}} + y_{19}\,b\,\hat{\mathbf{y}} + z_{19}\,c\,\sin\beta\,\hat{\mathbf{z}}$ | (4$g$) | P XIX |
| $\mathbf{B}_{74}$ | = | $-x_{19}\,\mathbf{a}_1 + y_{19}\,\mathbf{a}_2 + \left(\frac{1}{2} - z_{19}\right)\,\mathbf{a}_3$ | = | $\left(-x_{19}\,a + \left(\frac{1}{2} - z_{19}\right)\,c\,\cos\beta\right)\,\hat{\mathbf{x}} + y_{19}\,b\,\hat{\mathbf{y}} + \left(\frac{1}{2} - z_{19}\right)\,c\,\sin\beta\,\hat{\mathbf{z}}$ | (4$g$) | P XIX |
| $\mathbf{B}_{75}$ | = | $-x_{19}\,\mathbf{a}_1 - y_{19}\,\mathbf{a}_2 - z_{19}\,\mathbf{a}_3$ | = | $-(x_{19}\,a + z_{19}\,c\,\cos\beta)\,\hat{\mathbf{x}} - y_{19}\,b\,\hat{\mathbf{y}} - z_{19}\,c\,\sin\beta\,\hat{\mathbf{z}}$ | (4$g$) | P XIX |
| $\mathbf{B}_{76}$ | = | $x_{19}\,\mathbf{a}_1 - y_{19}\,\mathbf{a}_2 + \left(\frac{1}{2} + z_{19}\right)\,\mathbf{a}_3$ | = | $\left(x_{19}\,a + \left(\frac{1}{2} + z_{19}\right)\,c\,\cos\beta\right)\,\hat{\mathbf{x}} - y_{19}\,b\,\hat{\mathbf{y}} + \left(\frac{1}{2} + z_{19}\right)\,c\,\sin\beta\,\hat{\mathbf{z}}$ | (4$g$) | P XIX |
| $\mathbf{B}_{77}$ | = | $x_{20}\,\mathbf{a}_1 + y_{20}\,\mathbf{a}_2 + z_{20}\,\mathbf{a}_3$ | = | $(x_{20}\,a + z_{20}\,c\,\cos\beta)\,\hat{\mathbf{x}} + y_{20}\,b\,\hat{\mathbf{y}} + z_{20}\,c\,\sin\beta\,\hat{\mathbf{z}}$ | (4$g$) | P XX |
| $\mathbf{B}_{78}$ | = | $-x_{20}\,\mathbf{a}_1 + y_{20}\,\mathbf{a}_2 + \left(\frac{1}{2} - z_{20}\right)\,\mathbf{a}_3$ | = | $\left(-x_{20}\,a + \left(\frac{1}{2} - z_{20}\right)\,c\,\cos\beta\right)\,\hat{\mathbf{x}} + y_{20}\,b\,\hat{\mathbf{y}} + \left(\frac{1}{2} - z_{20}\right)\,c\,\sin\beta\,\hat{\mathbf{z}}$ | (4$g$) | P XX |
| $\mathbf{B}_{79}$ | = | $-x_{20}\,\mathbf{a}_1 - y_{20}\,\mathbf{a}_2 - z_{20}\,\mathbf{a}_3$ | = | $-(x_{20}\,a + z_{20}\,c\,\cos\beta)\,\hat{\mathbf{x}} - y_{20}\,b\,\hat{\mathbf{y}} - z_{20}\,c\,\sin\beta\,\hat{\mathbf{z}}$ | (4$g$) | P XX |
| $\mathbf{B}_{80}$ | = | $x_{20}\,\mathbf{a}_1 - y_{20}\,\mathbf{a}_2 + \left(\frac{1}{2} + z_{20}\right)\,\mathbf{a}_3$ | = | $\left(x_{20}\,a + \left(\frac{1}{2} + z_{20}\right)\,c\,\cos\beta\right)\,\hat{\mathbf{x}} - y_{20}\,b\,\hat{\mathbf{y}} + \left(\frac{1}{2} + z_{20}\right)\,c\,\sin\beta\,\hat{\mathbf{z}}$ | (4$g$) | P XX |
| $\mathbf{B}_{81}$ | = | $x_{21}\,\mathbf{a}_1 + y_{21}\,\mathbf{a}_2 + z_{21}\,\mathbf{a}_3$ | = | $(x_{21}\,a + z_{21}\,c\,\cos\beta)\,\hat{\mathbf{x}} + y_{21}\,b\,\hat{\mathbf{y}} + z_{21}\,c\,\sin\beta\,\hat{\mathbf{z}}$ | (4$g$) | P XXI |
| $\mathbf{B}_{82}$ | = | $-x_{21}\,\mathbf{a}_1 + y_{21}\,\mathbf{a}_2 + \left(\frac{1}{2} - z_{21}\right)\,\mathbf{a}_3$ | = | $\left(-x_{21}\,a + \left(\frac{1}{2} - z_{21}\right)\,c\,\cos\beta\right)\,\hat{\mathbf{x}} + y_{21}\,b\,\hat{\mathbf{y}} + \left(\frac{1}{2} - z_{21}\right)\,c\,\sin\beta\,\hat{\mathbf{z}}$ | (4$g$) | P XXI |
| $\mathbf{B}_{83}$ | = | $-x_{21}\,\mathbf{a}_1 - y_{21}\,\mathbf{a}_2 - z_{21}\,\mathbf{a}_3$ | = | $-(x_{21}\,a + z_{21}\,c\,\cos\beta)\,\hat{\mathbf{x}} - y_{21}\,b\,\hat{\mathbf{y}} - z_{21}\,c\,\sin\beta\,\hat{\mathbf{z}}$ | (4$g$) | P XXI |
| $\mathbf{B}_{84}$ | = | $x_{21}\,\mathbf{a}_1 - y_{21}\,\mathbf{a}_2 + \left(\frac{1}{2} + z_{21}\right)\,\mathbf{a}_3$ | = | $\left(x_{21}\,a + \left(\frac{1}{2} + z_{21}\right)\,c\,\cos\beta\right)\,\hat{\mathbf{x}} - y_{21}\,b\,\hat{\mathbf{y}} + \left(\frac{1}{2} + z_{21}\right)\,c\,\sin\beta\,\hat{\mathbf{z}}$ | (4$g$) | P XXI |

**Geometry files:**

- CIF: pp. 649
- POSCAR: pp. 650

# Baddeleyite ($ZrO_2$, C43) Structure: A2B_mP12_14_2e_e

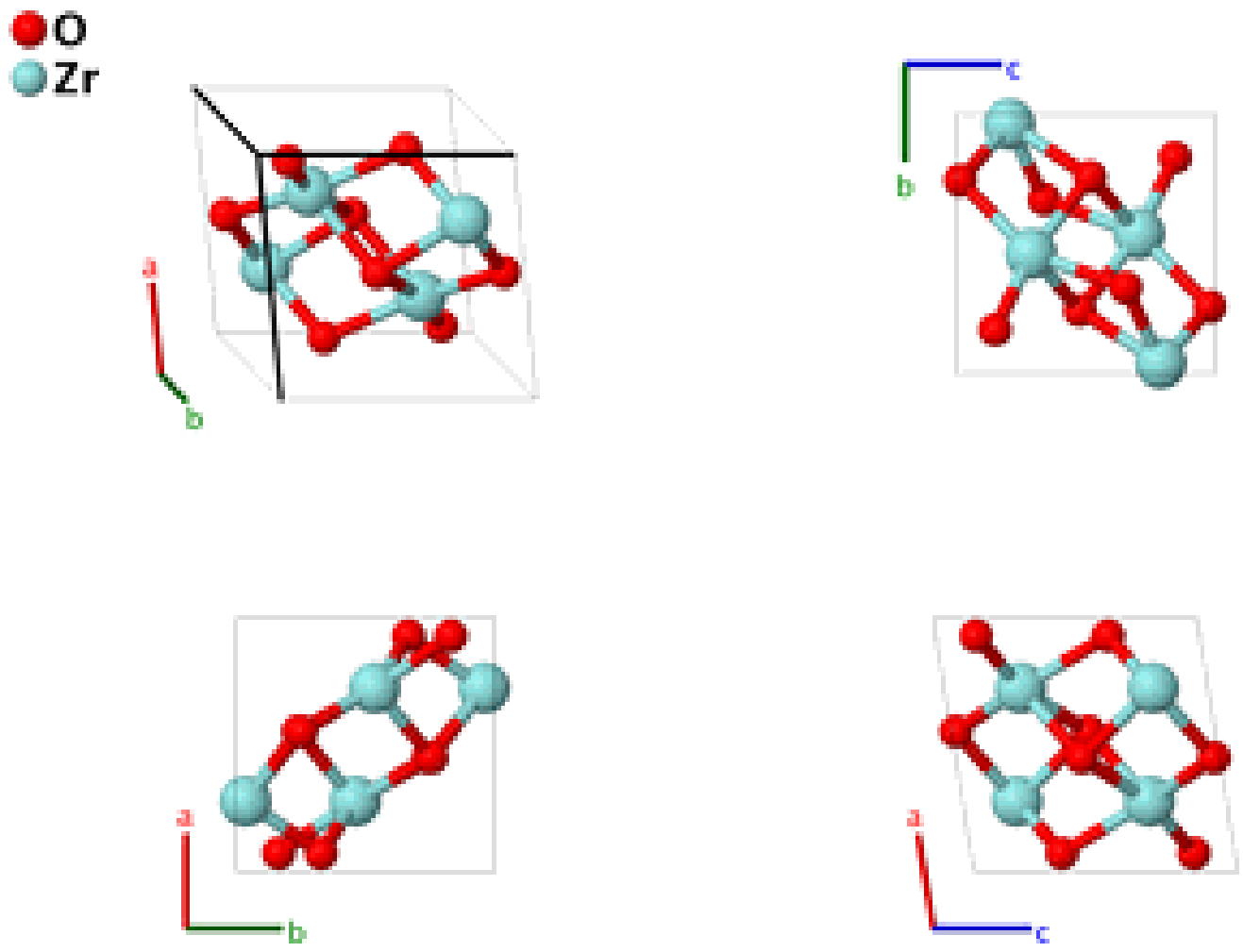

**Prototype** : $ZrO_2$

**AFLOW prototype label** : A2B_mP12_14_2e_e

***Strukturbericht* designation** : C43

**Pearson symbol** : mP12

**Space group number** : 14

**Space group symbol** : $P2_1/c$

**AFLOW prototype command** : `aflow --proto=A2B_mP12_14_2e_e`
`--params=`$a, b/a, c/a, \beta, x_1, y_1, z_1, x_2, y_2, z_2, x_3, y_3, z_3$

---

**Other compounds with this structure:**

- $HfO_2$, $CoSb_2$, $Ag_2Te$

---

**Simple Monoclinic primitive vectors:**

$$\mathbf{a}_1 = a\,\hat{\mathbf{x}}$$
$$\mathbf{a}_2 = b\,\hat{\mathbf{y}}$$
$$\mathbf{a}_3 = c\cos\beta\,\hat{\mathbf{x}} + c\sin\beta\,\hat{\mathbf{z}}$$

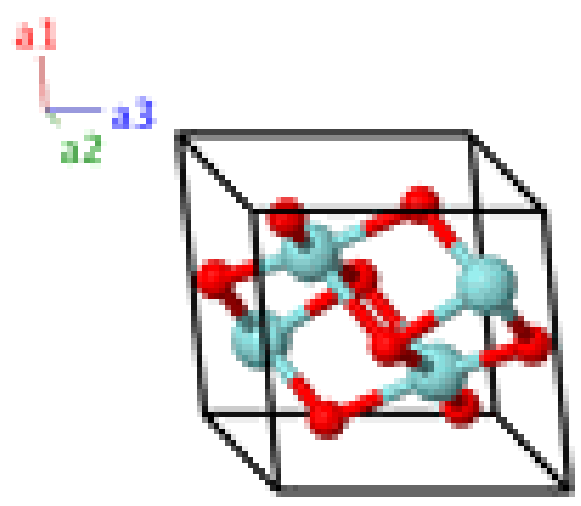

---

**Basis vectors:**

| | | Lattice Coordinates | | Cartesian Coordinates | Wyckoff Position | Atom Type |
|---|---|---|---|---|---|---|
| $\mathbf{B}_1$ | = | $x_1\,\mathbf{a}_1 + y_1\,\mathbf{a}_2 + z_1\,\mathbf{a}_3$ | = | $(x_1\,a + z_1\,c\,\cos\beta)\,\hat{\mathbf{x}} + y_1\,b\,\hat{\mathbf{y}} + z_1\,c\,\sin\beta\,\hat{\mathbf{z}}$ | (4e) | O I |
| $\mathbf{B}_2$ | = | $-x_1\,\mathbf{a}_1 + \left(\frac{1}{2} + y_1\right)\mathbf{a}_2 + \left(\frac{1}{2} - z_1\right)\mathbf{a}_3$ | = | $\left(\left(\frac{1}{2} - z_1\right) c\,\cos\beta - x_1\,a\right)\hat{\mathbf{x}} + \left(\frac{1}{2} + y_1\right) b\,\hat{\mathbf{y}} + \left(\frac{1}{2} - z_1\right) c\,\sin\beta\,\hat{\mathbf{z}}$ | (4e) | O I |
| $\mathbf{B}_3$ | = | $-x_1\,\mathbf{a}_1 - y_1\,\mathbf{a}_2 - z_1\,\mathbf{a}_3$ | = | $-(x_1\,a + z_1\,c\,\cos\beta)\,\hat{\mathbf{x}} - y_1\,b\,\hat{\mathbf{y}} - z_1\,c\,\sin\beta\,\hat{\mathbf{z}}$ | (4e) | O I |
| $\mathbf{B_{4}}$ | = | $x_{1}\,\mathbf{a_{1}}+\left(\frac{1}{2}-y_{1}\right)\,\mathbf{a_{2}}+\left(\frac{1}{2}+z_{1}\right)\,\mathbf{a_{3}}$ | = | $\left(\left(\frac{1}{2}+z_{1}\right)\,c\,\cos\beta+x_{1}\,a\right)\,\mathbf{\hat{x}}+\left(\frac{1}{2}-y_{1}\right)\,b\,\mathbf{\hat{y}}+\left(\frac{1}{2}+z_{1}\right)\,c\,\sin\beta\,\mathbf{\hat{z}}$ | (4$e$) | O I |
| $\mathbf{B_{5}}$ | = | $x_{2}\,\mathbf{a_{1}}+y_{2}\,\mathbf{a_{2}}+z_{2}\,\mathbf{a_{3}}$ | = | $(x_{2}\,a+z_{2}\,c\,\cos\beta)\,\mathbf{\hat{x}}+y_{2}\,b\,\mathbf{\hat{y}}+z_{2}\,c\,\sin\beta\,\mathbf{\hat{z}}$ | (4$e$) | O II |
| $\mathbf{B_{6}}$ | = | $-x_{2}\,\mathbf{a_{1}}+\left(\frac{1}{2}+y_{2}\right)\,\mathbf{a_{2}}+\left(\frac{1}{2}-z_{2}\right)\,\mathbf{a_{3}}$ | = | $\left(\left(\frac{1}{2}-z_{2}\right)\,c\,\cos\beta-x_{2}\,a\right)\,\mathbf{\hat{x}}+\left(\frac{1}{2}+y_{2}\right)\,b\,\mathbf{\hat{y}}+\left(\frac{1}{2}-z_{2}\right)\,c\,\sin\beta\,\mathbf{\hat{z}}$ | (4$e$) | O II |
| $\mathbf{B_{7}}$ | = | $-x_{2}\,\mathbf{a_{1}}-y_{2}\,\mathbf{a_{2}}-z_{2}\,\mathbf{a_{3}}$ | = | $-(x_{2}\,a+z_{2}\,c\,\cos\beta)\,\mathbf{\hat{x}}-y_{2}\,b\,\mathbf{\hat{y}}-z_{2}\,c\,\sin\beta\,\mathbf{\hat{z}}$ | (4$e$) | O II |
| $\mathbf{B_{8}}$ | = | $x_{2}\,\mathbf{a_{1}}+\left(\frac{1}{2}-y_{2}\right)\,\mathbf{a_{2}}+\left(\frac{1}{2}+z_{2}\right)\,\mathbf{a_{3}}$ | = | $\left(\left(\frac{1}{2}+z_{2}\right)\,c\,\cos\beta+x_{2}\,a\right)\,\mathbf{\hat{x}}+\left(\frac{1}{2}-y_{2}\right)\,b\,\mathbf{\hat{y}}+\left(\frac{1}{2}+z_{2}\right)\,c\,\sin\beta\,\mathbf{\hat{z}}$ | (4$e$) | O II |
| $\mathbf{B_{9}}$ | = | $x_{3}\,\mathbf{a_{1}}+y_{3}\,\mathbf{a_{2}}+z_{3}\,\mathbf{a_{3}}$ | = | $(x_{3}\,a+z_{3}\,c\,\cos\beta)\,\mathbf{\hat{x}}+y_{3}\,b\,\mathbf{\hat{y}}+z_{3}\,c\,\sin\beta\,\mathbf{\hat{z}}$ | (4$e$) | Zr |
| $\mathbf{B_{10}}$ | = | $-x_{3}\,\mathbf{a_{1}}+\left(\frac{1}{2}+y_{3}\right)\,\mathbf{a_{2}}+\left(\frac{1}{2}-z_{3}\right)\,\mathbf{a_{3}}$ | = | $\left(\left(\frac{1}{2}-z_{3}\right)\,c\,\cos\beta-x_{3}\,a\right)\,\mathbf{\hat{x}}+\left(\frac{1}{2}+y_{3}\right)\,b\,\mathbf{\hat{y}}+\left(\frac{1}{2}-z_{3}\right)\,c\,\sin\beta\,\mathbf{\hat{z}}$ | (4$e$) | Zr |
| $\mathbf{B_{11}}$ | = | $-x_{3}\,\mathbf{a_{1}}-y_{3}\,\mathbf{a_{2}}-z_{3}\,\mathbf{a_{3}}$ | = | $-(x_{3}\,a+z_{3}\,c\,\cos\beta)\,\mathbf{\hat{x}}-y_{3}\,b\,\mathbf{\hat{y}}-z_{3}\,c\,\sin\beta\,\mathbf{\hat{z}}$ | (4$e$) | Zr |
| $\mathbf{B_{12}}$ | = | $x_{3}\,\mathbf{a_{1}}+\left(\frac{1}{2}-y_{3}\right)\,\mathbf{a_{2}}+\left(\frac{1}{2}+z_{3}\right)\,\mathbf{a_{3}}$ | = | $\left(\left(\frac{1}{2}+z_{3}\right)\,c\,\cos\beta+x_{3}\,a\right)\,\mathbf{\hat{x}}+\left(\frac{1}{2}-y_{3}\right)\,b\,\mathbf{\hat{y}}+\left(\frac{1}{2}+z_{3}\right)\,c\,\sin\beta\,\mathbf{\hat{z}}$ | (4$e$) | Zr |

**Geometry files:**

- CIF: pp. 650
- POSCAR: pp. 651

# $\beta$-Se ($A_l$) Structure: A_mP32_14_8e

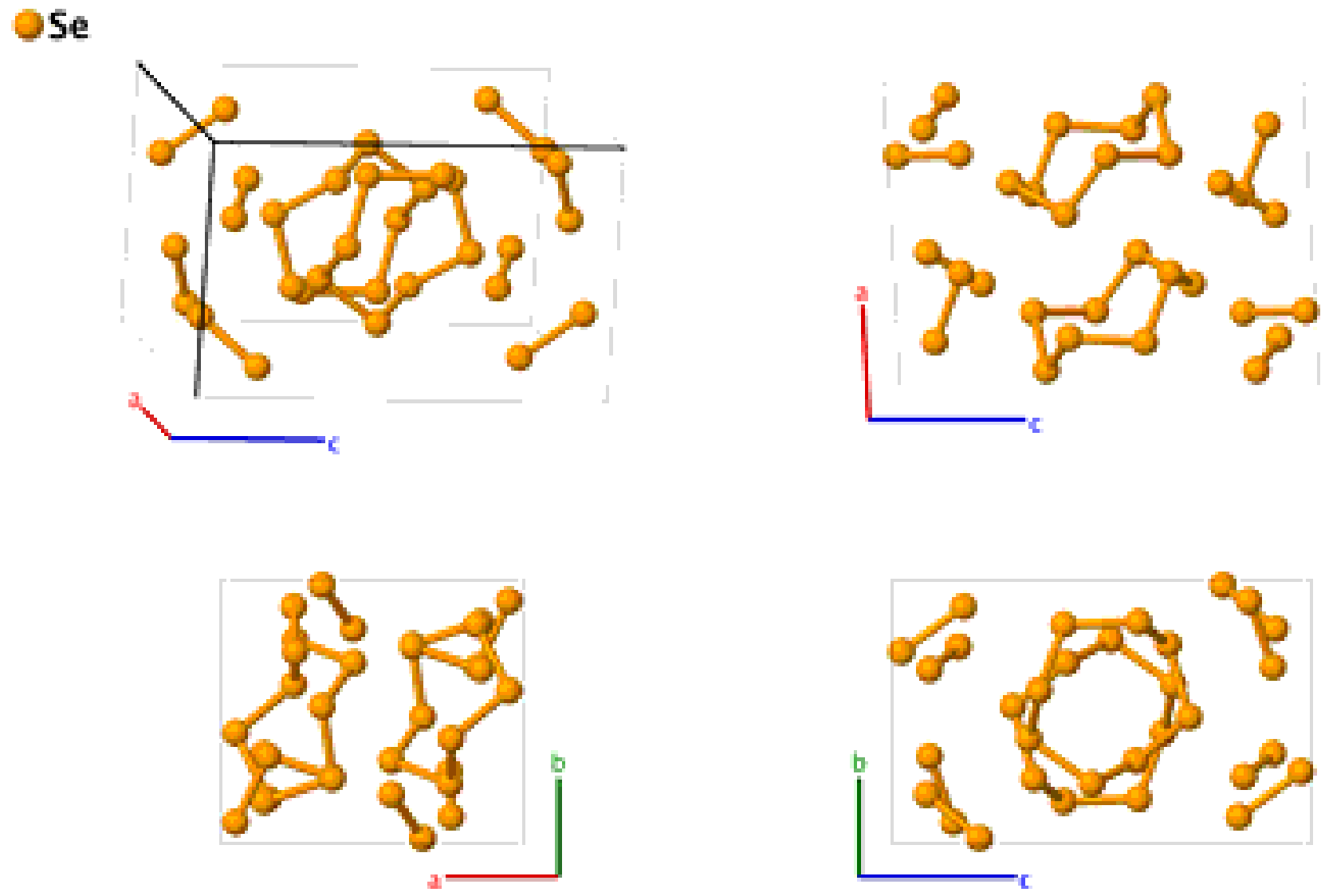

**Prototype** : $\beta$-Se

**AFLOW prototype label** : A_mP32_14_8e

***Strukturbericht* designation** : $A_l$

**Pearson symbol** : mP32

**Space group number** : 14

**Space group symbol** : $P2_1/c$

**AFLOW prototype command** : `aflow --proto=A_mP32_14_8e`
`--params=`$a, b/a, c/a, \beta, x_1, y_1, z_1, x_2, y_2, z_2, x_3, y_3, z_3, x_4, y_4, z_4, x_5, y_5, z_5, x_6, y_6, z_6, x_7, y_7, z_7, x_8, y_8, z_8$

---

- Donohue (1982) refers to this as the “monoclinic $\beta$-Se structure”.

---

**Simple Monoclinic primitive vectors:**

$$
\begin{aligned}
\mathbf{a}_1 &= a\,\hat{\mathbf{x}} \\
\mathbf{a}_2 &= b\,\hat{\mathbf{y}} \\
\mathbf{a}_3 &= c\cos\beta\,\hat{\mathbf{x}} + c\sin\beta\,\hat{\mathbf{z}}
\end{aligned}
$$

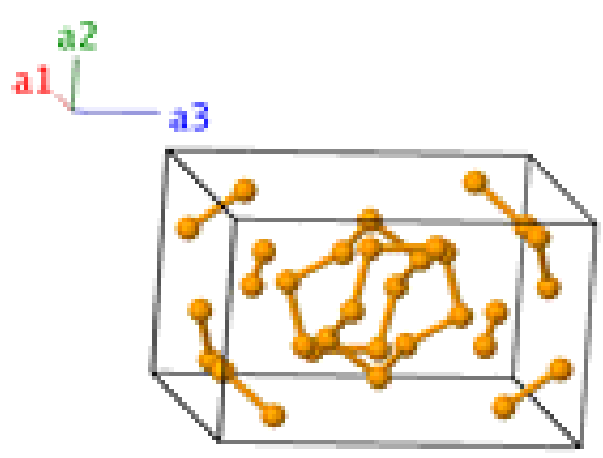

---

**Basis vectors:**

| | | Lattice Coordinates | | Cartesian Coordinates | Wyckoff Position | Atom Type |
|---|---|---|---|---|---|---|
| $\mathbf{B}_1$ | = | $x_1\,\mathbf{a}_1 + y_1\,\mathbf{a}_2 + z_1\,\mathbf{a}_3$ | = | $(x_1\,a + z_1\,c\,\cos\beta)\,\hat{\mathbf{x}} + y_1\,b\,\hat{\mathbf{y}} + z_1\,c\,\sin\beta\,\hat{\mathbf{z}}$ | (4$e$) | Se I |
| $\mathbf{B}_2$ | = | $-x_1\,\mathbf{a}_1 + \left(\frac{1}{2} + y_1\right)\mathbf{a}_2 + \left(\frac{1}{2} - z_1\right)\mathbf{a}_3$ | = | $\left(\left(\frac{1}{2} - z_1\right)c\,\cos\beta - x_1\,a\right)\hat{\mathbf{x}} + \left(\frac{1}{2} + y_1\right)b\,\hat{\mathbf{y}} + \left(\frac{1}{2} - z_1\right)c\,\sin\beta\,\hat{\mathbf{z}}$ | (4$e$) | Se I |
| $\mathbf{B_3}$ | = | $-x_1\,\mathbf{a_1} - y_1\,\mathbf{a_2} - z_1\,\mathbf{a_3}$ | = | $-\left(x_1\,a + z_1\,c\,\cos\beta\right)\,\mathbf{\hat{x}} - y_1\,b\,\mathbf{\hat{y}} - z_1\,c\,\sin\beta\,\mathbf{\hat{z}}$ | (4*e*) | Se I |
| $\mathbf{B_4}$ | = | $x_1\,\mathbf{a_1} + \left(\frac{1}{2} - y_1\right)\,\mathbf{a_2} + \left(\frac{1}{2} + z_1\right)\,\mathbf{a_3}$ | = | $\left(\left(\frac{1}{2} + z_1\right)\,c\,\cos\beta + x_1\,a\right)\,\mathbf{\hat{x}} + \left(\frac{1}{2} - y_1\right)\,b\,\mathbf{\hat{y}} + \left(\frac{1}{2} + z_1\right)\,c\,\sin\beta\,\mathbf{\hat{z}}$ | (4*e*) | Se I |
| $\mathbf{B_5}$ | = | $x_2\,\mathbf{a_1} + y_2\,\mathbf{a_2} + z_2\,\mathbf{a_3}$ | = | $\left(x_2\,a + z_2\,c\,\cos\beta\right)\,\mathbf{\hat{x}} + y_2\,b\,\mathbf{\hat{y}} + z_2\,c\,\sin\beta\,\mathbf{\hat{z}}$ | (4*e*) | Se II |
| $\mathbf{B_6}$ | = | $-x_2\,\mathbf{a_1} + \left(\frac{1}{2} + y_2\right)\,\mathbf{a_2} + \left(\frac{1}{2} - z_2\right)\,\mathbf{a_3}$ | = | $\left(\left(\frac{1}{2} - z_2\right)\,c\,\cos\beta - x_2\,a\right)\,\mathbf{\hat{x}} + \left(\frac{1}{2} + y_2\right)\,b\,\mathbf{\hat{y}} + \left(\frac{1}{2} - z_2\right)\,c\,\sin\beta\,\mathbf{\hat{z}}$ | (4*e*) | Se II |
| $\mathbf{B_7}$ | = | $-x_2\,\mathbf{a_1} - y_2\,\mathbf{a_2} - z_2\,\mathbf{a_3}$ | = | $-\left(x_2\,a + z_2\,c\,\cos\beta\right)\,\mathbf{\hat{x}} - y_2\,b\,\mathbf{\hat{y}} - z_2\,c\,\sin\beta\,\mathbf{\hat{z}}$ | (4*e*) | Se II |
| $\mathbf{B_8}$ | = | $x_2\,\mathbf{a_1} + \left(\frac{1}{2} - y_2\right)\,\mathbf{a_2} + \left(\frac{1}{2} + z_2\right)\,\mathbf{a_3}$ | = | $\left(\left(\frac{1}{2} + z_2\right)\,c\,\cos\beta + x_2\,a\right)\,\mathbf{\hat{x}} + \left(\frac{1}{2} - y_2\right)\,b\,\mathbf{\hat{y}} + \left(\frac{1}{2} + z_2\right)\,c\,\sin\beta\,\mathbf{\hat{z}}$ | (4*e*) | Se II |
| $\mathbf{B_9}$ | = | $x_3\,\mathbf{a_1} + y_3\,\mathbf{a_2} + z_3\,\mathbf{a_3}$ | = | $\left(x_3\,a + z_3\,c\,\cos\beta\right)\,\mathbf{\hat{x}} + y_3\,b\,\mathbf{\hat{y}} + z_3\,c\,\sin\beta\,\mathbf{\hat{z}}$ | (4*e*) | Se III |
| $\mathbf{B_{10}}$ | = | $-x_3\,\mathbf{a_1} + \left(\frac{1}{2} + y_3\right)\,\mathbf{a_2} + \left(\frac{1}{2} - z_3\right)\,\mathbf{a_3}$ | = | $\left(\left(\frac{1}{2} - z_3\right)\,c\,\cos\beta - x_3\,a\right)\,\mathbf{\hat{x}} + \left(\frac{1}{2} + y_3\right)\,b\,\mathbf{\hat{y}} + \left(\frac{1}{2} - z_3\right)\,c\,\sin\beta\,\mathbf{\hat{z}}$ | (4*e*) | Se III |
| $\mathbf{B_{11}}$ | = | $-x_3\,\mathbf{a_1} - y_3\,\mathbf{a_2} - z_3\,\mathbf{a_3}$ | = | $-\left(x_3\,a + z_3\,c\,\cos\beta\right)\,\mathbf{\hat{x}} - y_3\,b\,\mathbf{\hat{y}} - z_3\,c\,\sin\beta\,\mathbf{\hat{z}}$ | (4*e*) | Se III |
| $\mathbf{B_{12}}$ | = | $x_3\,\mathbf{a_1} + \left(\frac{1}{2} - y_3\right)\,\mathbf{a_2} + \left(\frac{1}{2} + z_3\right)\,\mathbf{a_3}$ | = | $\left(\left(\frac{1}{2} + z_3\right)\,c\,\cos\beta + x_3\,a\right)\,\mathbf{\hat{x}} + \left(\frac{1}{2} - y_3\right)\,b\,\mathbf{\hat{y}} + \left(\frac{1}{2} + z_3\right)\,c\,\sin\beta\,\mathbf{\hat{z}}$ | (4*e*) | Se III |
| $\mathbf{B_{13}}$ | = | $x_4\,\mathbf{a_1} + y_4\,\mathbf{a_2} + z_4\,\mathbf{a_3}$ | = | $\left(x_4\,a + z_4\,c\,\cos\beta\right)\,\mathbf{\hat{x}} + y_4\,b\,\mathbf{\hat{y}} + z_4\,c\,\sin\beta\,\mathbf{\hat{z}}$ | (4*e*) | Se IV |
| $\mathbf{B_{14}}$ | = | $-x_4\,\mathbf{a_1} + \left(\frac{1}{2} + y_4\right)\,\mathbf{a_2} + \left(\frac{1}{2} - z_4\right)\,\mathbf{a_3}$ | = | $\left(\left(\frac{1}{2} - z_4\right)\,c\,\cos\beta - x_4\,a\right)\,\mathbf{\hat{x}} + \left(\frac{1}{2} + y_4\right)\,b\,\mathbf{\hat{y}} + \left(\frac{1}{2} - z_4\right)\,c\,\sin\beta\,\mathbf{\hat{z}}$ | (4*e*) | Se IV |
| $\mathbf{B_{15}}$ | = | $-x_4\,\mathbf{a_1} - y_4\,\mathbf{a_2} - z_4\,\mathbf{a_3}$ | = | $-\left(x_4\,a + z_4\,c\,\cos\beta\right)\,\mathbf{\hat{x}} - y_4\,b\,\mathbf{\hat{y}} - z_4\,c\,\sin\beta\,\mathbf{\hat{z}}$ | (4*e*) | Se IV |
| $\mathbf{B_{16}}$ | = | $x_4\,\mathbf{a_1} + \left(\frac{1}{2} - y_4\right)\,\mathbf{a_2} + \left(\frac{1}{2} + z_4\right)\,\mathbf{a_3}$ | = | $\left(\left(\frac{1}{2} + z_4\right)\,c\,\cos\beta + x_4\,a\right)\,\mathbf{\hat{x}} + \left(\frac{1}{2} - y_4\right)\,b\,\mathbf{\hat{y}} + \left(\frac{1}{2} + z_4\right)\,c\,\sin\beta\,\mathbf{\hat{z}}$ | (4*e*) | Se IV |
| $\mathbf{B_{17}}$ | = | $x_5\,\mathbf{a_1} + y_5\,\mathbf{a_2} + z_5\,\mathbf{a_3}$ | = | $\left(x_5\,a + z_5\,c\,\cos\beta\right)\,\mathbf{\hat{x}} + y_5\,b\,\mathbf{\hat{y}} + z_5\,c\,\sin\beta\,\mathbf{\hat{z}}$ | (4*e*) | Se V |
| $\mathbf{B_{18}}$ | = | $-x_5\,\mathbf{a_1} + \left(\frac{1}{2} + y_5\right)\,\mathbf{a_2} + \left(\frac{1}{2} - z_5\right)\,\mathbf{a_3}$ | = | $\left(\left(\frac{1}{2} - z_5\right)\,c\,\cos\beta - x_5\,a\right)\,\mathbf{\hat{x}} + \left(\frac{1}{2} + y_5\right)\,b\,\mathbf{\hat{y}} + \left(\frac{1}{2} - z_5\right)\,c\,\sin\beta\,\mathbf{\hat{z}}$ | (4*e*) | Se V |
| $\mathbf{B_{19}}$ | = | $-x_5\,\mathbf{a_1} - y_5\,\mathbf{a_2} - z_5\,\mathbf{a_3}$ | = | $-\left(x_5\,a + z_5\,c\,\cos\beta\right)\,\mathbf{\hat{x}} - y_5\,b\,\mathbf{\hat{y}} - z_5\,c\,\sin\beta\,\mathbf{\hat{z}}$ | (4*e*) | Se V |
| $\mathbf{B_{20}}$ | = | $x_5\,\mathbf{a_1} + \left(\frac{1}{2} - y_5\right)\,\mathbf{a_2} + \left(\frac{1}{2} + z_5\right)\,\mathbf{a_3}$ | = | $\left(\left(\frac{1}{2} + z_5\right)\,c\,\cos\beta + x_5\,a\right)\,\mathbf{\hat{x}} + \left(\frac{1}{2} - y_5\right)\,b\,\mathbf{\hat{y}} + \left(\frac{1}{2} + z_5\right)\,c\,\sin\beta\,\mathbf{\hat{z}}$ | (4*e*) | Se V |
| $\mathbf{B_{21}}$ | = | $x_6\,\mathbf{a_1} + y_6\,\mathbf{a_2} + z_6\,\mathbf{a_3}$ | = | $\left(x_6\,a + z_6\,c\,\cos\beta\right)\,\mathbf{\hat{x}} + y_6\,b\,\mathbf{\hat{y}} + z_6\,c\,\sin\beta\,\mathbf{\hat{z}}$ | (4*e*) | Se VI |
| $\mathbf{B_{22}}$ | = | $-x_6\,\mathbf{a_1} + \left(\frac{1}{2} + y_6\right)\,\mathbf{a_2} + \left(\frac{1}{2} - z_6\right)\,\mathbf{a_3}$ | = | $\left(\left(\frac{1}{2} - z_6\right)\,c\,\cos\beta - x_6\,a\right)\,\mathbf{\hat{x}} + \left(\frac{1}{2} + y_6\right)\,b\,\mathbf{\hat{y}} + \left(\frac{1}{2} - z_6\right)\,c\,\sin\beta\,\mathbf{\hat{z}}$ | (4*e*) | Se VI |
| $\mathbf{B_{23}}$ | = | $-x_6\,\mathbf{a_1} - y_6\,\mathbf{a_2} - z_6\,\mathbf{a_3}$ | = | $-\left(x_6\,a + z_6\,c\,\cos\beta\right)\,\mathbf{\hat{x}} - y_6\,b\,\mathbf{\hat{y}} - z_6\,c\,\sin\beta\,\mathbf{\hat{z}}$ | (4*e*) | Se VI |
| $\mathbf{B_{24}}$ | = | $x_6\,\mathbf{a_1} + \left(\frac{1}{2} - y_6\right)\,\mathbf{a_2} + \left(\frac{1}{2} + z_6\right)\,\mathbf{a_3}$ | = | $\left(\left(\frac{1}{2} + z_6\right)\,c\,\cos\beta + x_6\,a\right)\,\mathbf{\hat{x}} + \left(\frac{1}{2} - y_6\right)\,b\,\mathbf{\hat{y}} + \left(\frac{1}{2} + z_6\right)\,c\,\sin\beta\,\mathbf{\hat{z}}$ | (4*e*) | Se VI |
| $\mathbf{B}_{25}$ | = | $x_7\,\mathbf{a}_1 + y_7\,\mathbf{a}_2 + z_7\,\mathbf{a}_3$ | = | $(x_7\,a + z_7\,c\,\cos\beta)\,\hat{\mathbf{x}} + y_7\,b\,\hat{\mathbf{y}} + z_7\,c\,\sin\beta\,\hat{\mathbf{z}}$ | (4$e$) | Se VII |
| $\mathbf{B}_{26}$ | = | $-x_7\,\mathbf{a}_1 + \left(\frac{1}{2} + y_7\right)\,\mathbf{a}_2 + \left(\frac{1}{2} - z_7\right)\,\mathbf{a}_3$ | = | $\left(\left(\frac{1}{2} - z_7\right)\,c\,\cos\beta - x_7\,a\right)\,\hat{\mathbf{x}} + \left(\frac{1}{2} + y_7\right)\,b\,\hat{\mathbf{y}} + \left(\frac{1}{2} - z_7\right)\,c\,\sin\beta\,\hat{\mathbf{z}}$ | (4$e$) | Se VII |
| $\mathbf{B}_{27}$ | = | $-x_7\,\mathbf{a}_1 - y_7\,\mathbf{a}_2 - z_7\,\mathbf{a}_3$ | = | $-(x_7\,a + z_7\,c\,\cos\beta)\,\hat{\mathbf{x}} - y_7\,b\,\hat{\mathbf{y}} - z_7\,c\,\sin\beta\,\hat{\mathbf{z}}$ | (4$e$) | Se VII |
| $\mathbf{B}_{28}$ | = | $x_7\,\mathbf{a}_1 + \left(\frac{1}{2} - y_7\right)\,\mathbf{a}_2 + \left(\frac{1}{2} + z_7\right)\,\mathbf{a}_3$ | = | $\left(\left(\frac{1}{2} + z_7\right)\,c\,\cos\beta + x_7\,a\right)\,\hat{\mathbf{x}} + \left(\frac{1}{2} - y_7\right)\,b\,\hat{\mathbf{y}} + \left(\frac{1}{2} + z_7\right)\,c\,\sin\beta\,\hat{\mathbf{z}}$ | (4$e$) | Se VII |
| $\mathbf{B}_{29}$ | = | $x_8\,\mathbf{a}_1 + y_8\,\mathbf{a}_2 + z_8\,\mathbf{a}_3$ | = | $(x_8\,a + z_8\,c\,\cos\beta)\,\hat{\mathbf{x}} + y_8\,b\,\hat{\mathbf{y}} + z_8\,c\,\sin\beta\,\hat{\mathbf{z}}$ | (4$e$) | Se VIII |
| $\mathbf{B}_{30}$ | = | $-x_8\,\mathbf{a}_1 + \left(\frac{1}{2} + y_8\right)\,\mathbf{a}_2 + \left(\frac{1}{2} - z_8\right)\,\mathbf{a}_3$ | = | $\left(\left(\frac{1}{2} - z_8\right)\,c\,\cos\beta - x_8\,a\right)\,\hat{\mathbf{x}} + \left(\frac{1}{2} + y_8\right)\,b\,\hat{\mathbf{y}} + \left(\frac{1}{2} - z_8\right)\,c\,\sin\beta\,\hat{\mathbf{z}}$ | (4$e$) | Se VIII |
| $\mathbf{B}_{31}$ | = | $-x_8\,\mathbf{a}_1 - y_8\,\mathbf{a}_2 - z_8\,\mathbf{a}_3$ | = | $-(x_8\,a + z_8\,c\,\cos\beta)\,\hat{\mathbf{x}} - y_8\,b\,\hat{\mathbf{y}} - z_8\,c\,\sin\beta\,\hat{\mathbf{z}}$ | (4$e$) | Se VIII |
| $\mathbf{B}_{32}$ | = | $x_8\,\mathbf{a}_1 + \left(\frac{1}{2} - y_8\right)\,\mathbf{a}_2 + \left(\frac{1}{2} + z_8\right)\,\mathbf{a}_3$ | = | $\left(\left(\frac{1}{2} + z_8\right)\,c\,\cos\beta + x_8\,a\right)\,\hat{\mathbf{x}} + \left(\frac{1}{2} - y_8\right)\,b\,\hat{\mathbf{y}} + \left(\frac{1}{2} + z_8\right)\,c\,\sin\beta\,\hat{\mathbf{z}}$ | (4$e$) | Se VIII |

---

**Found in:**

---

**Geometry files:**

- CIF: pp. 651
- POSCAR: pp. 651

# Se ($A_k$) Structure: A_mP64_14_16e

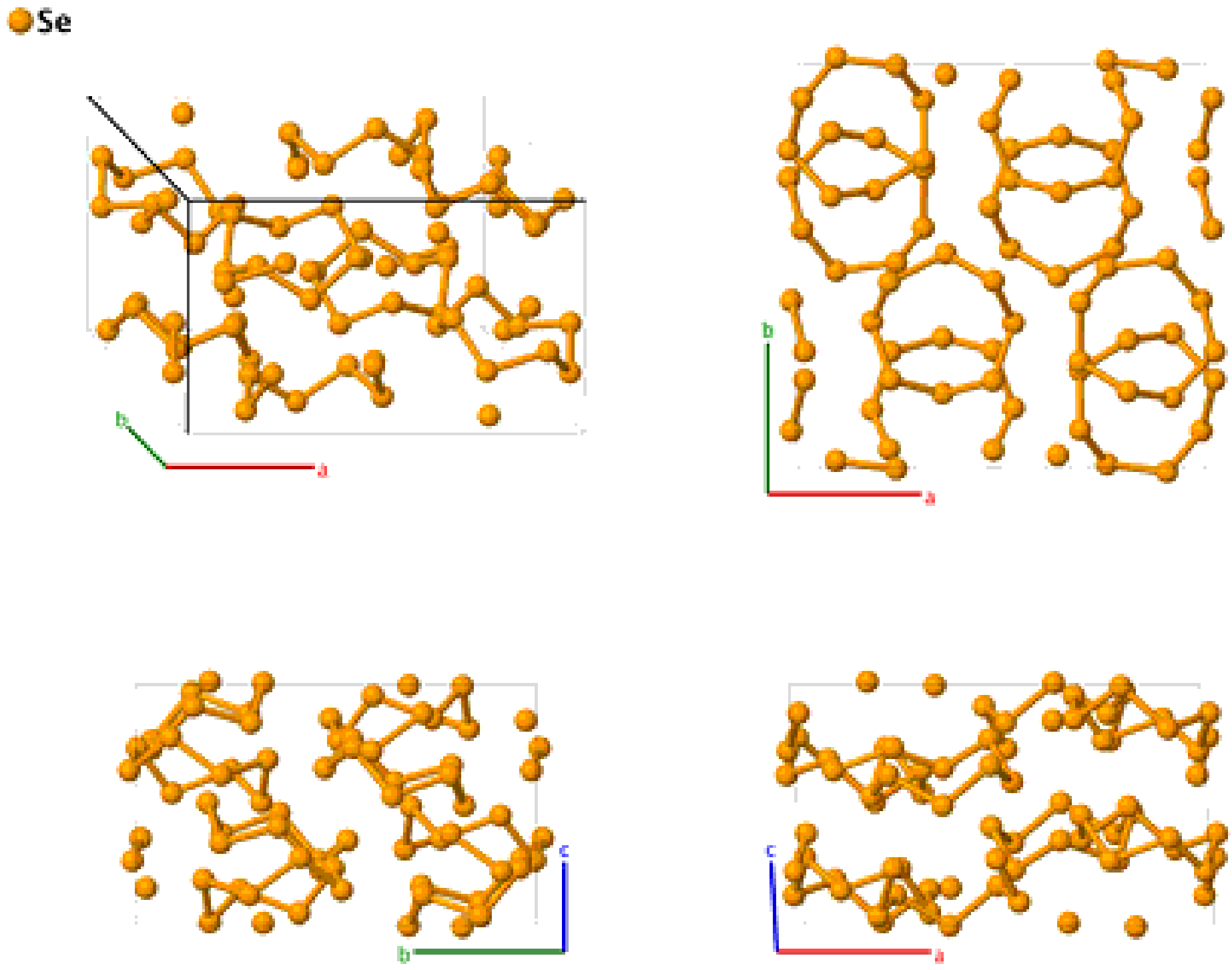

| | | |
|---|---|---|
| **Prototype** | : | Se |
| **AFLOW prototype label** | : | A_mP64_14_16e |
| ***Strukturbericht* designation** | : | $A_k$ |
| **Pearson symbol** | : | mP64 |
| **Space group number** | : | 14 |
| **Space group symbol** | : | $P2_1/c$ |
| **AFLOW prototype command** | : | `aflow --proto=A_mP64_14_16e`<br>`--params=`$a, b/a, c/a, \beta, x_1, y_1, z_1, x_2, y_2, z_2, x_3, y_3, z_3, x_4, y_4, z_4, x_5, y_5, z_5, x_6, y_6, z_6, x_7, y_7, z_7, x_8, y_8, z_8, x_9, y_9, z_9, x_{10}, y_{10}, z_{10}, x_{11}, y_{11}, z_{11}, x_{12}, y_{12}, z_{12}, x_{13}, y_{13}, z_{13}, x_{14}, y_{14}, z_{14}, x_{15}, y_{15}, z_{15}, x_{16}, y_{16}, z_{16}$ |

- We follow Villars (1991) and give this structure the $A_k$ designation. As noted in Villars (1991), the atomic coordinates are not provided in the referenced paper, but were given to the editors by the authors. We use those coordinates. Downs (2003) has the notation “gamma-monoclinic selenium is allotrope of cyclo-octaselenium”. Despite that, note that this is not what we refer to as $\gamma$-Se.

**Simple Monoclinic primitive vectors:**

$$
\begin{aligned}
\mathbf{a}_1 &= a\,\hat{\mathbf{x}} \\
\mathbf{a}_2 &= b\,\hat{\mathbf{y}} \\
\mathbf{a}_3 &= c\cos\beta\,\hat{\mathbf{x}} + c\sin\beta\,\hat{\mathbf{z}}
\end{aligned}
$$

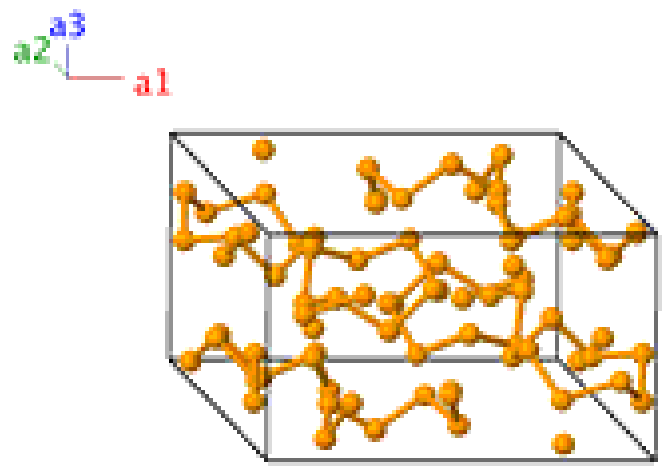

**Basis vectors:**

| | | Lattice Coordinates | | Cartesian Coordinates | Wyckoff Position | Atom Type |
|---|---|---|---|---|---|---|
| $\mathbf{B}_1$ | = | $x_1\,\mathbf{a}_1 + y_1\,\mathbf{a}_2 + z_1\,\mathbf{a}_3$ | = | $(x_1\,a + z_1\,c\,\cos\beta)\,\hat{\mathbf{x}} + y_1\,b\,\hat{\mathbf{y}} + z_1\,c\,\sin\beta\,\hat{\mathbf{z}}$ | (4$e$) | Se I |
| $\mathbf{B}_2$ | = | $-x_1\,\mathbf{a}_1 + \left(\frac{1}{2} + y_1\right)\mathbf{a}_2 + \left(\frac{1}{2} - z_1\right)\mathbf{a}_3$ | = | $\left(\left(\frac{1}{2} - z_1\right)c\,\cos\beta - x_1\,a\right)\hat{\mathbf{x}} + \left(\frac{1}{2} + y_1\right)b\,\hat{\mathbf{y}} + \left(\frac{1}{2} - z_1\right)c\,\sin\beta\,\hat{\mathbf{z}}$ | (4$e$) | Se I |
| $\mathbf{B}_3$ | = | $-x_1\,\mathbf{a}_1 - y_1\,\mathbf{a}_2 - z_1\,\mathbf{a}_3$ | = | $-(x_1\,a + z_1\,c\,\cos\beta)\,\hat{\mathbf{x}} - y_1\,b\,\hat{\mathbf{y}} - z_1\,c\,\sin\beta\,\hat{\mathbf{z}}$ | (4$e$) | Se I |
| $\mathbf{B}_4$ | = | $x_1\,\mathbf{a}_1 + \left(\frac{1}{2} - y_1\right)\mathbf{a}_2 + \left(\frac{1}{2} + z_1\right)\mathbf{a}_3$ | = | $\left(\left(\frac{1}{2} + z_1\right)c\,\cos\beta + x_1\,a\right)\hat{\mathbf{x}} + \left(\frac{1}{2} - y_1\right)b\,\hat{\mathbf{y}} + \left(\frac{1}{2} + z_1\right)c\,\sin\beta\,\hat{\mathbf{z}}$ | (4$e$) | Se I |
| $\mathbf{B}_5$ | = | $x_2\,\mathbf{a}_1 + y_2\,\mathbf{a}_2 + z_2\,\mathbf{a}_3$ | = | $(x_2\,a + z_2\,c\,\cos\beta)\,\hat{\mathbf{x}} + y_2\,b\,\hat{\mathbf{y}} + z_2\,c\,\sin\beta\,\hat{\mathbf{z}}$ | (4$e$) | Se II |
| $\mathbf{B}_6$ | = | $-x_2\,\mathbf{a}_1 + \left(\frac{1}{2} + y_2\right)\mathbf{a}_2 + \left(\frac{1}{2} - z_2\right)\mathbf{a}_3$ | = | $\left(\left(\frac{1}{2} - z_2\right)c\,\cos\beta - x_2\,a\right)\hat{\mathbf{x}} + \left(\frac{1}{2} + y_2\right)b\,\hat{\mathbf{y}} + \left(\frac{1}{2} - z_2\right)c\,\sin\beta\,\hat{\mathbf{z}}$ | (4$e$) | Se II |
| $\mathbf{B}_7$ | = | $-x_2\,\mathbf{a}_1 - y_2\,\mathbf{a}_2 - z_2\,\mathbf{a}_3$ | = | $-(x_2\,a + z_2\,c\,\cos\beta)\,\hat{\mathbf{x}} - y_2\,b\,\hat{\mathbf{y}} - z_2\,c\,\sin\beta\,\hat{\mathbf{z}}$ | (4$e$) | Se II |
| $\mathbf{B}_8$ | = | $x_2\,\mathbf{a}_1 + \left(\frac{1}{2} - y_2\right)\mathbf{a}_2 + \left(\frac{1}{2} + z_2\right)\mathbf{a}_3$ | = | $\left(\left(\frac{1}{2} + z_2\right)c\,\cos\beta + x_2\,a\right)\hat{\mathbf{x}} + \left(\frac{1}{2} - y_2\right)b\,\hat{\mathbf{y}} + \left(\frac{1}{2} + z_2\right)c\,\sin\beta\,\hat{\mathbf{z}}$ | (4$e$) | Se II |
| $\mathbf{B}_9$ | = | $x_3\,\mathbf{a}_1 + y_3\,\mathbf{a}_2 + z_3\,\mathbf{a}_3$ | = | $(x_3\,a + z_3\,c\,\cos\beta)\,\hat{\mathbf{x}} + y_3\,b\,\hat{\mathbf{y}} + z_3\,c\,\sin\beta\,\hat{\mathbf{z}}$ | (4$e$) | Se III |
| $\mathbf{B}_{10}$ | = | $-x_3\,\mathbf{a}_1 + \left(\frac{1}{2} + y_3\right)\mathbf{a}_2 + \left(\frac{1}{2} - z_3\right)\mathbf{a}_3$ | = | $\left(\left(\frac{1}{2} - z_3\right)c\,\cos\beta - x_3\,a\right)\hat{\mathbf{x}} + \left(\frac{1}{2} + y_3\right)b\,\hat{\mathbf{y}} + \left(\frac{1}{2} - z_3\right)c\,\sin\beta\,\hat{\mathbf{z}}$ | (4$e$) | Se III |
| $\mathbf{B}_{11}$ | = | $-x_3\,\mathbf{a}_1 - y_3\,\mathbf{a}_2 - z_3\,\mathbf{a}_3$ | = | $-(x_3\,a + z_3\,c\,\cos\beta)\,\hat{\mathbf{x}} - y_3\,b\,\hat{\mathbf{y}} - z_3\,c\,\sin\beta\,\hat{\mathbf{z}}$ | (4$e$) | Se III |
| $\mathbf{B}_{12}$ | = | $x_3\,\mathbf{a}_1 + \left(\frac{1}{2} - y_3\right)\mathbf{a}_2 + \left(\frac{1}{2} + z_3\right)\mathbf{a}_3$ | = | $\left(\left(\frac{1}{2} + z_3\right)c\,\cos\beta + x_3\,a\right)\hat{\mathbf{x}} + \left(\frac{1}{2} - y_3\right)b\,\hat{\mathbf{y}} + \left(\frac{1}{2} + z_3\right)c\,\sin\beta\,\hat{\mathbf{z}}$ | (4$e$) | Se III |
| $\mathbf{B}_{13}$ | = | $x_4\,\mathbf{a}_1 + y_4\,\mathbf{a}_2 + z_4\,\mathbf{a}_3$ | = | $(x_4\,a + z_4\,c\,\cos\beta)\,\hat{\mathbf{x}} + y_4\,b\,\hat{\mathbf{y}} + z_4\,c\,\sin\beta\,\hat{\mathbf{z}}$ | (4$e$) | Se IV |
| $\mathbf{B}_{14}$ | = | $-x_4\,\mathbf{a}_1 + \left(\frac{1}{2} + y_4\right)\mathbf{a}_2 + \left(\frac{1}{2} - z_4\right)\mathbf{a}_3$ | = | $\left(\left(\frac{1}{2} - z_4\right)c\,\cos\beta - x_4\,a\right)\hat{\mathbf{x}} + \left(\frac{1}{2} + y_4\right)b\,\hat{\mathbf{y}} + \left(\frac{1}{2} - z_4\right)c\,\sin\beta\,\hat{\mathbf{z}}$ | (4$e$) | Se IV |
| $\mathbf{B}_{15}$ | = | $-x_4\,\mathbf{a}_1 - y_4\,\mathbf{a}_2 - z_4\,\mathbf{a}_3$ | = | $-(x_4\,a + z_4\,c\,\cos\beta)\,\hat{\mathbf{x}} - y_4\,b\,\hat{\mathbf{y}} - z_4\,c\,\sin\beta\,\hat{\mathbf{z}}$ | (4$e$) | Se IV |
| $\mathbf{B}_{16}$ | = | $x_4\,\mathbf{a}_1 + \left(\frac{1}{2} - y_4\right)\mathbf{a}_2 + \left(\frac{1}{2} + z_4\right)\mathbf{a}_3$ | = | $\left(\left(\frac{1}{2} + z_4\right)c\,\cos\beta + x_4\,a\right)\hat{\mathbf{x}} + \left(\frac{1}{2} - y_4\right)b\,\hat{\mathbf{y}} + \left(\frac{1}{2} + z_4\right)c\,\sin\beta\,\hat{\mathbf{z}}$ | (4$e$) | Se IV |
| $\mathbf{B_{17}}$ | = | $x_{5} \, \mathbf{a}_{1} + y_{5} \, \mathbf{a}_{2} + z_{5} \, \mathbf{a}_{3}$ | = | $(x_{5} \, a + z_{5} \, c \, \cos\beta) \, \mathbf{\hat{x}} + y_{5} \, b \, \mathbf{\hat{y}} + z_{5} \, c \, \sin\beta \, \mathbf{\hat{z}}$ | (4*e*) | Se V |
| $\mathbf{B_{18}}$ | = | $-x_{5} \, \mathbf{a}_{1} + \left(\frac{1}{2} + y_{5}\right) \, \mathbf{a}_{2} + \left(\frac{1}{2} - z_{5}\right) \, \mathbf{a}_{3}$ | = | $\left(\left(\frac{1}{2} - z_{5}\right) \, c \, \cos\beta - x_{5} \, a\right) \, \mathbf{\hat{x}} + \left(\frac{1}{2} + y_{5}\right) \, b \, \mathbf{\hat{y}} + \left(\frac{1}{2} - z_{5}\right) \, c \, \sin\beta \, \mathbf{\hat{z}}$ | (4*e*) | Se V |
| $\mathbf{B_{19}}$ | = | $-x_{5} \, \mathbf{a}_{1} - y_{5} \, \mathbf{a}_{2} - z_{5} \, \mathbf{a}_{3}$ | = | $-(x_{5} \, a + z_{5} \, c \, \cos\beta) \, \mathbf{\hat{x}} - y_{5} \, b \, \mathbf{\hat{y}} - z_{5} \, c \, \sin\beta \, \mathbf{\hat{z}}$ | (4*e*) | Se V |
| $\mathbf{B_{20}}$ | = | $x_{5} \, \mathbf{a}_{1} + \left(\frac{1}{2} - y_{5}\right) \, \mathbf{a}_{2} + \left(\frac{1}{2} + z_{5}\right) \, \mathbf{a}_{3}$ | = | $\left(\left(\frac{1}{2} + z_{5}\right) \, c \, \cos\beta + x_{5} \, a\right) \, \mathbf{\hat{x}} + \left(\frac{1}{2} - y_{5}\right) \, b \, \mathbf{\hat{y}} + \left(\frac{1}{2} + z_{5}\right) \, c \, \sin\beta \, \mathbf{\hat{z}}$ | (4*e*) | Se V |
| $\mathbf{B_{21}}$ | = | $x_{6} \, \mathbf{a}_{1} + y_{6} \, \mathbf{a}_{2} + z_{6} \, \mathbf{a}_{3}$ | = | $(x_{6} \, a + z_{6} \, c \, \cos\beta) \, \mathbf{\hat{x}} + y_{6} \, b \, \mathbf{\hat{y}} + z_{6} \, c \, \sin\beta \, \mathbf{\hat{z}}$ | (4*e*) | Se VI |
| $\mathbf{B_{22}}$ | = | $-x_{6} \, \mathbf{a}_{1} + \left(\frac{1}{2} + y_{6}\right) \, \mathbf{a}_{2} + \left(\frac{1}{2} - z_{6}\right) \, \mathbf{a}_{3}$ | = | $\left(\left(\frac{1}{2} - z_{6}\right) \, c \, \cos\beta - x_{6} \, a\right) \, \mathbf{\hat{x}} + \left(\frac{1}{2} + y_{6}\right) \, b \, \mathbf{\hat{y}} + \left(\frac{1}{2} - z_{6}\right) \, c \, \sin\beta \, \mathbf{\hat{z}}$ | (4*e*) | Se VI |
| $\mathbf{B_{23}}$ | = | $-x_{6} \, \mathbf{a}_{1} - y_{6} \, \mathbf{a}_{2} - z_{6} \, \mathbf{a}_{3}$ | = | $-(x_{6} \, a + z_{6} \, c \, \cos\beta) \, \mathbf{\hat{x}} - y_{6} \, b \, \mathbf{\hat{y}} - z_{6} \, c \, \sin\beta \, \mathbf{\hat{z}}$ | (4*e*) | Se VI |
| $\mathbf{B_{24}}$ | = | $x_{6} \, \mathbf{a}_{1} + \left(\frac{1}{2} - y_{6}\right) \, \mathbf{a}_{2} + \left(\frac{1}{2} + z_{6}\right) \, \mathbf{a}_{3}$ | = | $\left(\left(\frac{1}{2} + z_{6}\right) \, c \, \cos\beta + x_{6} \, a\right) \, \mathbf{\hat{x}} + \left(\frac{1}{2} - y_{6}\right) \, b \, \mathbf{\hat{y}} + \left(\frac{1}{2} + z_{6}\right) \, c \, \sin\beta \, \mathbf{\hat{z}}$ | (4*e*) | Se VI |
| $\mathbf{B_{25}}$ | = | $x_{7} \, \mathbf{a}_{1} + y_{7} \, \mathbf{a}_{2} + z_{7} \, \mathbf{a}_{3}$ | = | $(x_{7} \, a + z_{7} \, c \, \cos\beta) \, \mathbf{\hat{x}} + y_{7} \, b \, \mathbf{\hat{y}} + z_{7} \, c \, \sin\beta \, \mathbf{\hat{z}}$ | (4*e*) | Se VII |
| $\mathbf{B_{26}}$ | = | $-x_{7} \, \mathbf{a}_{1} + \left(\frac{1}{2} + y_{7}\right) \, \mathbf{a}_{2} + \left(\frac{1}{2} - z_{7}\right) \, \mathbf{a}_{3}$ | = | $\left(\left(\frac{1}{2} - z_{7}\right) \, c \, \cos\beta - x_{7} \, a\right) \, \mathbf{\hat{x}} + \left(\frac{1}{2} + y_{7}\right) \, b \, \mathbf{\hat{y}} + \left(\frac{1}{2} - z_{7}\right) \, c \, \sin\beta \, \mathbf{\hat{z}}$ | (4*e*) | Se VII |
| $\mathbf{B_{27}}$ | = | $-x_{7} \, \mathbf{a}_{1} - y_{7} \, \mathbf{a}_{2} - z_{7} \, \mathbf{a}_{3}$ | = | $-(x_{7} \, a + z_{7} \, c \, \cos\beta) \, \mathbf{\hat{x}} - y_{7} \, b \, \mathbf{\hat{y}} - z_{7} \, c \, \sin\beta \, \mathbf{\hat{z}}$ | (4*e*) | Se VII |
| $\mathbf{B_{28}}$ | = | $x_{7} \, \mathbf{a}_{1} + \left(\frac{1}{2} - y_{7}\right) \, \mathbf{a}_{2} + \left(\frac{1}{2} + z_{7}\right) \, \mathbf{a}_{3}$ | = | $\left(\left(\frac{1}{2} + z_{7}\right) \, c \, \cos\beta + x_{7} \, a\right) \, \mathbf{\hat{x}} + \left(\frac{1}{2} - y_{7}\right) \, b \, \mathbf{\hat{y}} + \left(\frac{1}{2} + z_{7}\right) \, c \, \sin\beta \, \mathbf{\hat{z}}$ | (4*e*) | Se VII |
| $\mathbf{B_{29}}$ | = | $x_{8} \, \mathbf{a}_{1} + y_{8} \, \mathbf{a}_{2} + z_{8} \, \mathbf{a}_{3}$ | = | $(x_{8} \, a + z_{8} \, c \, \cos\beta) \, \mathbf{\hat{x}} + y_{8} \, b \, \mathbf{\hat{y}} + z_{8} \, c \, \sin\beta \, \mathbf{\hat{z}}$ | (4*e*) | Se VIII |
| $\mathbf{B_{30}}$ | = | $-x_{8} \, \mathbf{a}_{1} + \left(\frac{1}{2} + y_{8}\right) \, \mathbf{a}_{2} + \left(\frac{1}{2} - z_{8}\right) \, \mathbf{a}_{3}$ | = | $\left(\left(\frac{1}{2} - z_{8}\right) \, c \, \cos\beta - x_{8} \, a\right) \, \mathbf{\hat{x}} + \left(\frac{1}{2} + y_{8}\right) \, b \, \mathbf{\hat{y}} + \left(\frac{1}{2} - z_{8}\right) \, c \, \sin\beta \, \mathbf{\hat{z}}$ | (4*e*) | Se VIII |
| $\mathbf{B_{31}}$ | = | $-x_{8} \, \mathbf{a}_{1} - y_{8} \, \mathbf{a}_{2} - z_{8} \, \mathbf{a}_{3}$ | = | $-(x_{8} \, a + z_{8} \, c \, \cos\beta) \, \mathbf{\hat{x}} - y_{8} \, b \, \mathbf{\hat{y}} - z_{8} \, c \, \sin\beta \, \mathbf{\hat{z}}$ | (4*e*) | Se VIII |
| $\mathbf{B_{32}}$ | = | $x_{8} \, \mathbf{a}_{1} + \left(\frac{1}{2} - y_{8}\right) \, \mathbf{a}_{2} + \left(\frac{1}{2} + z_{8}\right) \, \mathbf{a}_{3}$ | = | $\left(\left(\frac{1}{2} + z_{8}\right) \, c \, \cos\beta + x_{8} \, a\right) \, \mathbf{\hat{x}} + \left(\frac{1}{2} - y_{8}\right) \, b \, \mathbf{\hat{y}} + \left(\frac{1}{2} + z_{8}\right) \, c \, \sin\beta \, \mathbf{\hat{z}}$ | (4*e*) | Se VIII |
| $\mathbf{B_{33}}$ | = | $x_{9} \, \mathbf{a}_{1} + y_{9} \, \mathbf{a}_{2} + z_{9} \, \mathbf{a}_{3}$ | = | $(x_{9} \, a + z_{9} \, c \, \cos\beta) \, \mathbf{\hat{x}} + y_{9} \, b \, \mathbf{\hat{y}} + z_{9} \, c \, \sin\beta \, \mathbf{\hat{z}}$ | (4*e*) | Se IX |
| $\mathbf{B_{34}}$ | = | $-x_{9} \, \mathbf{a}_{1} + \left(\frac{1}{2} + y_{9}\right) \, \mathbf{a}_{2} + \left(\frac{1}{2} - z_{9}\right) \, \mathbf{a}_{3}$ | = | $\left(\left(\frac{1}{2} - z_{9}\right) \, c \, \cos\beta - x_{9} \, a\right) \, \mathbf{\hat{x}} + \left(\frac{1}{2} + y_{9}\right) \, b \, \mathbf{\hat{y}} + \left(\frac{1}{2} - z_{9}\right) \, c \, \sin\beta \, \mathbf{\hat{z}}$ | (4*e*) | Se IX |
| $\mathbf{B_{35}}$ | = | $-x_{9} \, \mathbf{a}_{1} - y_{9} \, \mathbf{a}_{2} - z_{9} \, \mathbf{a}_{3}$ | = | $-(x_{9} \, a + z_{9} \, c \, \cos\beta) \, \mathbf{\hat{x}} - y_{9} \, b \, \mathbf{\hat{y}} - z_{9} \, c \, \sin\beta \, \mathbf{\hat{z}}$ | (4*e*) | Se IX |
| $\mathbf{B_{36}}$ | = | $x_{9} \, \mathbf{a}_{1} + \left(\frac{1}{2} - y_{9}\right) \, \mathbf{a}_{2} + \left(\frac{1}{2} + z_{9}\right) \, \mathbf{a}_{3}$ | = | $\left(\left(\frac{1}{2} + z_{9}\right) \, c \, \cos\beta + x_{9} \, a\right) \, \mathbf{\hat{x}} + \left(\frac{1}{2} - y_{9}\right) \, b \, \mathbf{\hat{y}} + \left(\frac{1}{2} + z_{9}\right) \, c \, \sin\beta \, \mathbf{\hat{z}}$ | (4*e*) | Se IX |
| $\mathbf{B_{37}}$ | = | $x_{10} \, \mathbf{a}_{1} + y_{10} \, \mathbf{a}_{2} + z_{10} \, \mathbf{a}_{3}$ | = | $(x_{10} \, a + z_{10} \, c \, \cos\beta) \, \mathbf{\hat{x}} + y_{10} \, b \, \mathbf{\hat{y}} + z_{10} \, c \, \sin\beta \, \mathbf{\hat{z}}$ | (4*e*) | Se X |
| $\mathbf{B_{38}}$ | = | $-x_{10} \, \mathbf{a}_{1} + \left(\frac{1}{2} + y_{10}\right) \, \mathbf{a}_{2} + \left(\frac{1}{2} - z_{10}\right) \, \mathbf{a}_{3}$ | = | $\left(\left(\frac{1}{2} - z_{10}\right) \, c \, \cos\beta - x_{10} \, a\right) \, \mathbf{\hat{x}} + \left(\frac{1}{2} + y_{10}\right) \, b \, \mathbf{\hat{y}} + \left(\frac{1}{2} - z_{10}\right) \, c \, \sin\beta \, \mathbf{\hat{z}}$ | (4*e*) | Se X |
| $\mathbf{B_{39}}$ | = | $-x_{10}\,\mathbf{a_1} - y_{10}\,\mathbf{a_2} - z_{10}\,\mathbf{a_3}$ | = | $-\left(x_{10}\,a + z_{10}\,c\,\cos\beta\right)\,\mathbf{\hat{x}} - y_{10}\,b\,\mathbf{\hat{y}} - z_{10}\,c\,\sin\beta\,\mathbf{\hat{z}}$ | (4$e$) | Se X |
| $\mathbf{B_{40}}$ | = | $x_{10}\,\mathbf{a_1} + \left(\frac{1}{2} - y_{10}\right)\,\mathbf{a_2} + \left(\frac{1}{2} + z_{10}\right)\,\mathbf{a_3}$ | = | $\left(\left(\frac{1}{2} + z_{10}\right)\,c\,\cos\beta + x_{10}\,a\right)\,\mathbf{\hat{x}} + \left(\frac{1}{2} - y_{10}\right)\,b\,\mathbf{\hat{y}} + \left(\frac{1}{2} + z_{10}\right)\,c\,\sin\beta\,\mathbf{\hat{z}}$ | (4$e$) | Se X |
| $\mathbf{B_{41}}$ | = | $x_{11}\,\mathbf{a_1} + y_{11}\,\mathbf{a_2} + z_{11}\,\mathbf{a_3}$ | = | $\left(x_{11}\,a + z_{11}\,c\,\cos\beta\right)\,\mathbf{\hat{x}} + y_{11}\,b\,\mathbf{\hat{y}} + z_{11}\,c\,\sin\beta\,\mathbf{\hat{z}}$ | (4$e$) | Se XI |
| $\mathbf{B_{42}}$ | = | $-x_{11}\,\mathbf{a_1} + \left(\frac{1}{2} + y_{11}\right)\,\mathbf{a_2} + \left(\frac{1}{2} - z_{11}\right)\,\mathbf{a_3}$ | = | $\left(\left(\frac{1}{2} - z_{11}\right)\,c\,\cos\beta - x_{11}\,a\right)\,\mathbf{\hat{x}} + \left(\frac{1}{2} + y_{11}\right)\,b\,\mathbf{\hat{y}} + \left(\frac{1}{2} - z_{11}\right)\,c\,\sin\beta\,\mathbf{\hat{z}}$ | (4$e$) | Se XI |
| $\mathbf{B_{43}}$ | = | $-x_{11}\,\mathbf{a_1} - y_{11}\,\mathbf{a_2} - z_{11}\,\mathbf{a_3}$ | = | $-\left(x_{11}\,a + z_{11}\,c\,\cos\beta\right)\,\mathbf{\hat{x}} - y_{11}\,b\,\mathbf{\hat{y}} - z_{11}\,c\,\sin\beta\,\mathbf{\hat{z}}$ | (4$e$) | Se XI |
| $\mathbf{B_{44}}$ | = | $x_{11}\,\mathbf{a_1} + \left(\frac{1}{2} - y_{11}\right)\,\mathbf{a_2} + \left(\frac{1}{2} + z_{11}\right)\,\mathbf{a_3}$ | = | $\left(\left(\frac{1}{2} + z_{11}\right)\,c\,\cos\beta + x_{11}\,a\right)\,\mathbf{\hat{x}} + \left(\frac{1}{2} - y_{11}\right)\,b\,\mathbf{\hat{y}} + \left(\frac{1}{2} + z_{11}\right)\,c\,\sin\beta\,\mathbf{\hat{z}}$ | (4$e$) | Se XI |
| $\mathbf{B_{45}}$ | = | $x_{12}\,\mathbf{a_1} + y_{12}\,\mathbf{a_2} + z_{12}\,\mathbf{a_3}$ | = | $\left(x_{12}\,a + z_{12}\,c\,\cos\beta\right)\,\mathbf{\hat{x}} + y_{12}\,b\,\mathbf{\hat{y}} + z_{12}\,c\,\sin\beta\,\mathbf{\hat{z}}$ | (4$e$) | Se XII |
| $\mathbf{B_{46}}$ | = | $-x_{12}\,\mathbf{a_1} + \left(\frac{1}{2} + y_{12}\right)\,\mathbf{a_2} + \left(\frac{1}{2} - z_{12}\right)\,\mathbf{a_3}$ | = | $\left(\left(\frac{1}{2} - z_{12}\right)\,c\,\cos\beta - x_{12}\,a\right)\,\mathbf{\hat{x}} + \left(\frac{1}{2} + y_{12}\right)\,b\,\mathbf{\hat{y}} + \left(\frac{1}{2} - z_{12}\right)\,c\,\sin\beta\,\mathbf{\hat{z}}$ | (4$e$) | Se XII |
| $\mathbf{B_{47}}$ | = | $-x_{12}\,\mathbf{a_1} - y_{12}\,\mathbf{a_2} - z_{12}\,\mathbf{a_3}$ | = | $-\left(x_{12}\,a + z_{12}\,c\,\cos\beta\right)\,\mathbf{\hat{x}} - y_{12}\,b\,\mathbf{\hat{y}} - z_{12}\,c\,\sin\beta\,\mathbf{\hat{z}}$ | (4$e$) | Se XII |
| $\mathbf{B_{48}}$ | = | $x_{12}\,\mathbf{a_1} + \left(\frac{1}{2} - y_{12}\right)\,\mathbf{a_2} + \left(\frac{1}{2} + z_{12}\right)\,\mathbf{a_3}$ | = | $\left(\left(\frac{1}{2} + z_{12}\right)\,c\,\cos\beta + x_{12}\,a\right)\,\mathbf{\hat{x}} + \left(\frac{1}{2} - y_{12}\right)\,b\,\mathbf{\hat{y}} + \left(\frac{1}{2} + z_{12}\right)\,c\,\sin\beta\,\mathbf{\hat{z}}$ | (4$e$) | Se XII |
| $\mathbf{B_{49}}$ | = | $x_{13}\,\mathbf{a_1} + y_{13}\,\mathbf{a_2} + z_{13}\,\mathbf{a_3}$ | = | $\left(x_{13}\,a + z_{13}\,c\,\cos\beta\right)\,\mathbf{\hat{x}} + y_{13}\,b\,\mathbf{\hat{y}} + z_{13}\,c\,\sin\beta\,\mathbf{\hat{z}}$ | (4$e$) | Se XIII |
| $\mathbf{B_{50}}$ | = | $-x_{13}\,\mathbf{a_1} + \left(\frac{1}{2} + y_{13}\right)\,\mathbf{a_2} + \left(\frac{1}{2} - z_{13}\right)\,\mathbf{a_3}$ | = | $\left(\left(\frac{1}{2} - z_{13}\right)\,c\,\cos\beta - x_{13}\,a\right)\,\mathbf{\hat{x}} + \left(\frac{1}{2} + y_{13}\right)\,b\,\mathbf{\hat{y}} + \left(\frac{1}{2} - z_{13}\right)\,c\,\sin\beta\,\mathbf{\hat{z}}$ | (4$e$) | Se XIII |
| $\mathbf{B_{51}}$ | = | $-x_{13}\,\mathbf{a_1} - y_{13}\,\mathbf{a_2} - z_{13}\,\mathbf{a_3}$ | = | $-\left(x_{13}\,a + z_{13}\,c\,\cos\beta\right)\,\mathbf{\hat{x}} - y_{13}\,b\,\mathbf{\hat{y}} - z_{13}\,c\,\sin\beta\,\mathbf{\hat{z}}$ | (4$e$) | Se XIII |
| $\mathbf{B_{52}}$ | = | $x_{13}\,\mathbf{a_1} + \left(\frac{1}{2} - y_{13}\right)\,\mathbf{a_2} + \left(\frac{1}{2} + z_{13}\right)\,\mathbf{a_3}$ | = | $\left(\left(\frac{1}{2} + z_{13}\right)\,c\,\cos\beta + x_{13}\,a\right)\,\mathbf{\hat{x}} + \left(\frac{1}{2} - y_{13}\right)\,b\,\mathbf{\hat{y}} + \left(\frac{1}{2} + z_{13}\right)\,c\,\sin\beta\,\mathbf{\hat{z}}$ | (4$e$) | Se XIII |
| $\mathbf{B_{53}}$ | = | $x_{14}\,\mathbf{a_1} + y_{14}\,\mathbf{a_2} + z_{14}\,\mathbf{a_3}$ | = | $\left(x_{14}\,a + z_{14}\,c\,\cos\beta\right)\,\mathbf{\hat{x}} + y_{14}\,b\,\mathbf{\hat{y}} + z_{14}\,c\,\sin\beta\,\mathbf{\hat{z}}$ | (4$e$) | Se XIV |
| $\mathbf{B_{54}}$ | = | $-x_{14}\,\mathbf{a_1} + \left(\frac{1}{2} + y_{14}\right)\,\mathbf{a_2} + \left(\frac{1}{2} - z_{14}\right)\,\mathbf{a_3}$ | = | $\left(\left(\frac{1}{2} - z_{14}\right)\,c\,\cos\beta - x_{14}\,a\right)\,\mathbf{\hat{x}} + \left(\frac{1}{2} + y_{14}\right)\,b\,\mathbf{\hat{y}} + \left(\frac{1}{2} - z_{14}\right)\,c\,\sin\beta\,\mathbf{\hat{z}}$ | (4$e$) | Se XIV |
| $\mathbf{B_{55}}$ | = | $-x_{14}\,\mathbf{a_1} - y_{14}\,\mathbf{a_2} - z_{14}\,\mathbf{a_3}$ | = | $-\left(x_{14}\,a + z_{14}\,c\,\cos\beta\right)\,\mathbf{\hat{x}} - y_{14}\,b\,\mathbf{\hat{y}} - z_{14}\,c\,\sin\beta\,\mathbf{\hat{z}}$ | (4$e$) | Se XIV |
| $\mathbf{B_{56}}$ | = | $x_{14}\,\mathbf{a_1} + \left(\frac{1}{2} - y_{14}\right)\,\mathbf{a_2} + \left(\frac{1}{2} + z_{14}\right)\,\mathbf{a_3}$ | = | $\left(\left(\frac{1}{2} + z_{14}\right)\,c\,\cos\beta + x_{14}\,a\right)\,\mathbf{\hat{x}} + \left(\frac{1}{2} - y_{14}\right)\,b\,\mathbf{\hat{y}} + \left(\frac{1}{2} + z_{14}\right)\,c\,\sin\beta\,\mathbf{\hat{z}}$ | (4$e$) | Se XIV |
| $\mathbf{B_{57}}$ | = | $x_{15}\,\mathbf{a_1} + y_{15}\,\mathbf{a_2} + z_{15}\,\mathbf{a_3}$ | = | $\left(x_{15}\,a + z_{15}\,c\,\cos\beta\right)\,\mathbf{\hat{x}} + y_{15}\,b\,\mathbf{\hat{y}} + z_{15}\,c\,\sin\beta\,\mathbf{\hat{z}}$ | (4$e$) | Se XV |
| $\mathbf{B_{58}}$ | = | $-x_{15}\,\mathbf{a_1} + \left(\frac{1}{2} + y_{15}\right)\,\mathbf{a_2} + \left(\frac{1}{2} - z_{15}\right)\,\mathbf{a_3}$ | = | $\left(\left(\frac{1}{2} - z_{15}\right)\,c\,\cos\beta - x_{15}\,a\right)\,\mathbf{\hat{x}} + \left(\frac{1}{2} + y_{15}\right)\,b\,\mathbf{\hat{y}} + \left(\frac{1}{2} - z_{15}\right)\,c\,\sin\beta\,\mathbf{\hat{z}}$ | (4$e$) | Se XV |
| $\mathbf{B_{59}}$ | = | $-x_{15}\,\mathbf{a_1} - y_{15}\,\mathbf{a_2} - z_{15}\,\mathbf{a_3}$ | = | $-\left(x_{15}\,a + z_{15}\,c\,\cos\beta\right)\,\mathbf{\hat{x}} - y_{15}\,b\,\mathbf{\hat{y}} - z_{15}\,c\,\sin\beta\,\mathbf{\hat{z}}$ | (4$e$) | Se XV |
| $\mathbf{B_{60}}$ | = | $x_{15}\,\mathbf{a_1} + \left(\frac{1}{2} - y_{15}\right)\,\mathbf{a_2} + \left(\frac{1}{2} + z_{15}\right)\,\mathbf{a_3}$ | = | $\left(\left(\frac{1}{2} + z_{15}\right)\,c\,\cos\beta + x_{15}\,a\right)\,\mathbf{\hat{x}} + \left(\frac{1}{2} - y_{15}\right)\,b\,\mathbf{\hat{y}} + \left(\frac{1}{2} + z_{15}\right)\,c\,\sin\beta\,\mathbf{\hat{z}}$ | (4$e$) | Se XV |
| $\mathbf{B_{61}}$ | = | $x_{16}\,\mathbf{a_1} + y_{16}\,\mathbf{a_2} + z_{16}\,\mathbf{a_3}$ | = | $(x_{16}\,a + z_{16}\,c\,\cos\beta)\,\hat{\mathbf{x}} + y_{16}\,b\,\hat{\mathbf{y}} + z_{16}\,c\,\sin\beta\,\hat{\mathbf{z}}$ | (4$e$) | Se XVI |
| $\mathbf{B_{62}}$ | = | $-x_{16}\,\mathbf{a_1} + \left(\frac{1}{2} + y_{16}\right)\,\mathbf{a_2} + \left(\frac{1}{2} - z_{16}\right)\,\mathbf{a_3}$ | = | $\left(\left(\frac{1}{2} - z_{16}\right)\,c\,\cos\beta - x_{16}\,a\right)\,\hat{\mathbf{x}} + \left(\frac{1}{2} + y_{16}\right)\,b\,\hat{\mathbf{y}} + \left(\frac{1}{2} - z_{16}\right)\,c\,\sin\beta\,\hat{\mathbf{z}}$ | (4$e$) | Se XVI |
| $\mathbf{B_{63}}$ | = | $-x_{16}\,\mathbf{a_1} - y_{16}\,\mathbf{a_2} - z_{16}\,\mathbf{a_3}$ | = | $-(x_{16}\,a + z_{16}\,c\,\cos\beta)\,\hat{\mathbf{x}} - y_{16}\,b\,\hat{\mathbf{y}} - z_{16}\,c\,\sin\beta\,\hat{\mathbf{z}}$ | (4$e$) | Se XVI |
| $\mathbf{B_{64}}$ | = | $x_{16}\,\mathbf{a_1} + \left(\frac{1}{2} - y_{16}\right)\,\mathbf{a_2} + \left(\frac{1}{2} + z_{16}\right)\,\mathbf{a_3}$ | = | $\left(\left(\frac{1}{2} + z_{16}\right)\,c\,\cos\beta + x_{16}\,a\right)\,\hat{\mathbf{x}} + \left(\frac{1}{2} - y_{16}\right)\,b\,\hat{\mathbf{y}} + \left(\frac{1}{2} + z_{16}\right)\,c\,\sin\beta\,\hat{\mathbf{z}}$ | (4$e$) | Se XVI |

---

**Found in:**

---

**Geometry files:**

- CIF: pp. 651
- POSCAR: pp. 652

# $B_2Pd_5$ Structure: A2B5_mC28_15_f_e2f

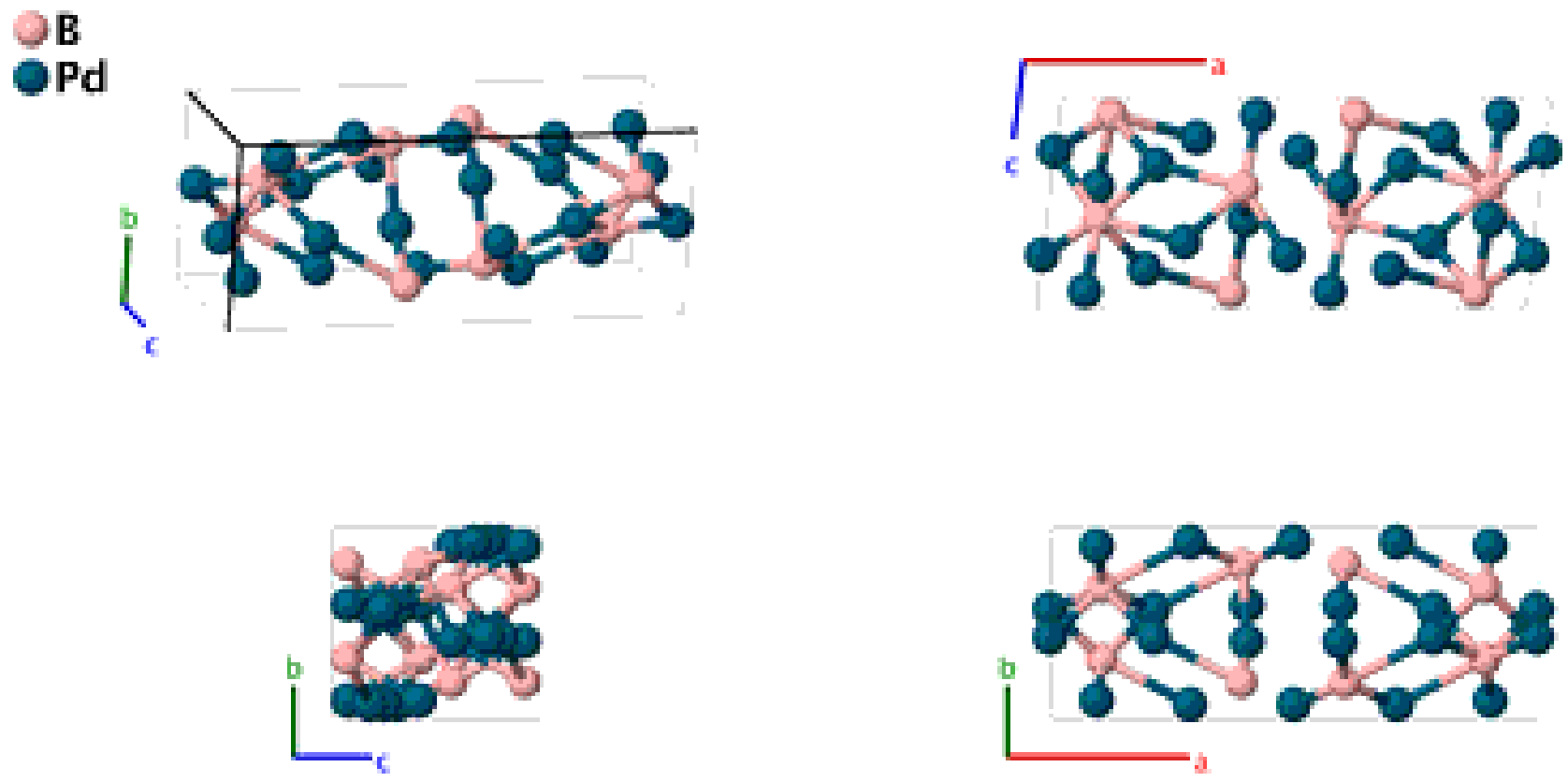

| | | |
|---|---|---|
| **Prototype** | : | $B_2Pd_5$ |
| **AFLOW prototype label** | : | A2B5_mC28_15_f_e2f |
| ***Strukturbericht* designation** | : | None |
| **Pearson symbol** | : | mC28 |
| **Space group number** | : | 15 |
| **Space group symbol** | : | C2/c |
| **AFLOW prototype command** | : | `aflow --proto=A2B5_mC28_15_f_e2f`<br>`--params=`$a, b/a, c/a, \beta, y_1, x_2, y_2, z_2, x_3, y_3, z_3, x_4, y_4, z_4$ |

**Base-centered Monoclinic primitive vectors:**

$$\mathbf{a}_1 = \frac{1}{2}\,a\,\hat{\mathbf{x}} - \frac{1}{2}\,b\,\hat{\mathbf{y}}$$
$$\mathbf{a}_2 = \frac{1}{2}\,a\,\hat{\mathbf{x}} + \frac{1}{2}\,b\,\hat{\mathbf{y}}$$
$$\mathbf{a}_3 = c\cos\beta\,\hat{\mathbf{x}} + c\sin\beta\,\hat{\mathbf{z}}$$

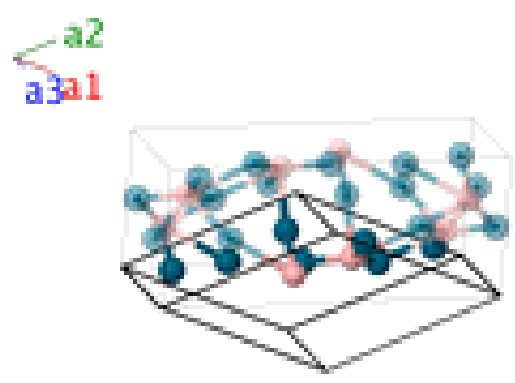

**Basis vectors:**

| | | Lattice Coordinates | | Cartesian Coordinates | Wyckoff Position | Atom Type |
|---|---|---|---|---|---|---|
| $\mathbf{B_1}$ | = | $-y_1\,\mathbf{a_1} + y_1\,\mathbf{a_2} + \frac{1}{4}\,\mathbf{a_3}$ | = | $\frac{1}{4}\,c\,\cos\beta\,\hat{\mathbf{x}} + y_1\,b\,\hat{\mathbf{y}} + \frac{1}{4}\,c\,\sin\beta\,\hat{\mathbf{z}}$ | (4$e$) | Pd I |
| $\mathbf{B_2}$ | = | $y_1\,\mathbf{a_1} - y_1\,\mathbf{a_2} + \frac{3}{4}\,\mathbf{a_3}$ | = | $\frac{3}{4}\,c\,\cos\beta\,\hat{\mathbf{x}} - y_1\,b\,\hat{\mathbf{y}} + \frac{3}{4}\,c\,\sin\beta\,\hat{\mathbf{z}}$ | (4$e$) | Pd I |
| $\mathbf{B_3}$ | = | $(x_2 - y_2)\,\mathbf{a_1} + (x_2 + y_2)\,\mathbf{a_2} + z_2\,\mathbf{a_3}$ | = | $(x_2\,a + z_2\,c\,\cos\beta)\,\hat{\mathbf{x}} + y_2\,b\,\hat{\mathbf{y}} + z_2\,c\,\sin\beta\,\hat{\mathbf{z}}$ | (8$f$) | B |
| $\mathbf{B_4}$ | = | $-(x_2 + y_2)\,\mathbf{a_1} + (y_2 - x_2)\,\mathbf{a_2} + \left(\frac{1}{2} - z_2\right)\,\mathbf{a_3}$ | = | $\left(-x_2\,a + \left(\frac{1}{2} - z_2\right)\,c\,\cos\beta\right)\,\hat{\mathbf{x}} + y_2\,b\,\hat{\mathbf{y}} + \left(\frac{1}{2} - z_2\right)\,c\,\sin\beta\,\hat{\mathbf{z}}$ | (8$f$) | B |
| $\mathbf{B_5}$ | = | $(y_2 - x_2)\,\mathbf{a_1} - (x_2 + y_2)\,\mathbf{a_2} - z_2\,\mathbf{a_3}$ | = | $-(x_2\,a + z_2\,c\,\cos\beta)\,\hat{\mathbf{x}} - y_2\,b\,\hat{\mathbf{y}} - z_2\,c\,\sin\beta\,\hat{\mathbf{z}}$ | (8$f$) | B |
| $\mathbf{B_6}$ | = | $(x_2 + y_2)\,\mathbf{a_1} + (x_2 - y_2)\,\mathbf{a_2} + \left(\frac{1}{2} + z_2\right)\,\mathbf{a_3}$ | = | $\left(x_2\,a + \left(\frac{1}{2} + z_2\right)\,c\,\cos\beta\right)\,\hat{\mathbf{x}} - y_2\,b\,\hat{\mathbf{y}} + \left(\frac{1}{2} + z_2\right)\,c\,\sin\beta\,\hat{\mathbf{z}}$ | (8$f$) | B |
| $\mathbf{B_7}$ | = | $(x_3 - y_3)\,\mathbf{a_1} + (x_3 + y_3)\,\mathbf{a_2} + z_3\,\mathbf{a_3}$ | = | $(x_3\,a + z_3\,c\,\cos\beta)\,\hat{\mathbf{x}} + y_3\,b\,\hat{\mathbf{y}} + z_3\,c\,\sin\beta\,\hat{\mathbf{z}}$ | (8$f$) | Pd II |
| $\mathbf{B_8}$ | = | $-\left(x_3+y_3\right)\,\mathbf{a_1}+\left(y_3-x_3\right)\,\mathbf{a_2}+\left(\frac{1}{2}-z_3\right)\,\mathbf{a_3}$ | = | $\left(-x_3\,a+\left(\frac{1}{2}-z_3\right)\,c\,\cos\beta\right)\,\mathbf{\hat{x}}+y_3\,b\,\mathbf{\hat{y}}+\left(\frac{1}{2}-z_3\right)\,c\,\sin\beta\,\mathbf{\hat{z}}$ | $(8f)$ | Pd II |
| $\mathbf{B_9}$ | = | $\left(y_3-x_3\right)\,\mathbf{a_1}-\left(x_3+y_3\right)\,\mathbf{a_2}-z_3\,\mathbf{a_3}$ | = | $-\left(x_3\,a+z_3\,c\,\cos\beta\right)\,\mathbf{\hat{x}}-y_3\,b\,\mathbf{\hat{y}}-z_3\,c\,\sin\beta\,\mathbf{\hat{z}}$ | $(8f)$ | Pd II |
| $\mathbf{B_{10}}$ | = | $\left(x_3+y_3\right)\,\mathbf{a_1}+\left(x_3-y_3\right)\,\mathbf{a_2}+\left(\frac{1}{2}+z_3\right)\,\mathbf{a_3}$ | = | $\left(x_3\,a+\left(\frac{1}{2}+z_3\right)\,c\,\cos\beta\right)\,\mathbf{\hat{x}}-y_3\,b\,\mathbf{\hat{y}}+\left(\frac{1}{2}+z_3\right)\,c\,\sin\beta\,\mathbf{\hat{z}}$ | $(8f)$ | Pd II |
| $\mathbf{B_{11}}$ | = | $\left(x_4-y_4\right)\,\mathbf{a_1}+\left(x_4+y_4\right)\,\mathbf{a_2}+z_4\,\mathbf{a_3}$ | = | $\left(x_4\,a+z_4\,c\,\cos\beta\right)\,\mathbf{\hat{x}}+y_4\,b\,\mathbf{\hat{y}}+z_4\,c\,\sin\beta\,\mathbf{\hat{z}}$ | $(8f)$ | Pd III |
| $\mathbf{B_{12}}$ | = | $-\left(x_4+y_4\right)\,\mathbf{a_1}+\left(y_4-x_4\right)\,\mathbf{a_2}+\left(\frac{1}{2}-z_4\right)\,\mathbf{a_3}$ | = | $\left(-x_4\,a+\left(\frac{1}{2}-z_4\right)\,c\,\cos\beta\right)\,\mathbf{\hat{x}}+y_4\,b\,\mathbf{\hat{y}}+\left(\frac{1}{2}-z_4\right)\,c\,\sin\beta\,\mathbf{\hat{z}}$ | $(8f)$ | Pd III |
| $\mathbf{B_{13}}$ | = | $\left(y_4-x_4\right)\,\mathbf{a_1}-\left(x_4+y_4\right)\,\mathbf{a_2}-z_4\,\mathbf{a_3}$ | = | $-\left(x_4\,a+z_4\,c\,\cos\beta\right)\,\mathbf{\hat{x}}-y_4\,b\,\mathbf{\hat{y}}-z_4\,c\,\sin\beta\,\mathbf{\hat{z}}$ | $(8f)$ | Pd III |
| $\mathbf{B_{14}}$ | = | $\left(x_4+y_4\right)\,\mathbf{a_1}+\left(x_4-y_4\right)\,\mathbf{a_2}+\left(\frac{1}{2}+z_4\right)\,\mathbf{a_3}$ | = | $\left(x_4\,a+\left(\frac{1}{2}+z_4\right)\,c\,\cos\beta\right)\,\mathbf{\hat{x}}-y_4\,b\,\mathbf{\hat{y}}+\left(\frac{1}{2}+z_4\right)\,c\,\sin\beta\,\mathbf{\hat{z}}$ | $(8f)$ | Pd III |

**Geometry files:**

- CIF: pp. 652
- POSCAR: pp. 653

# Tenorite (CuO, B26) Structure: AB_mC8_15_c_e

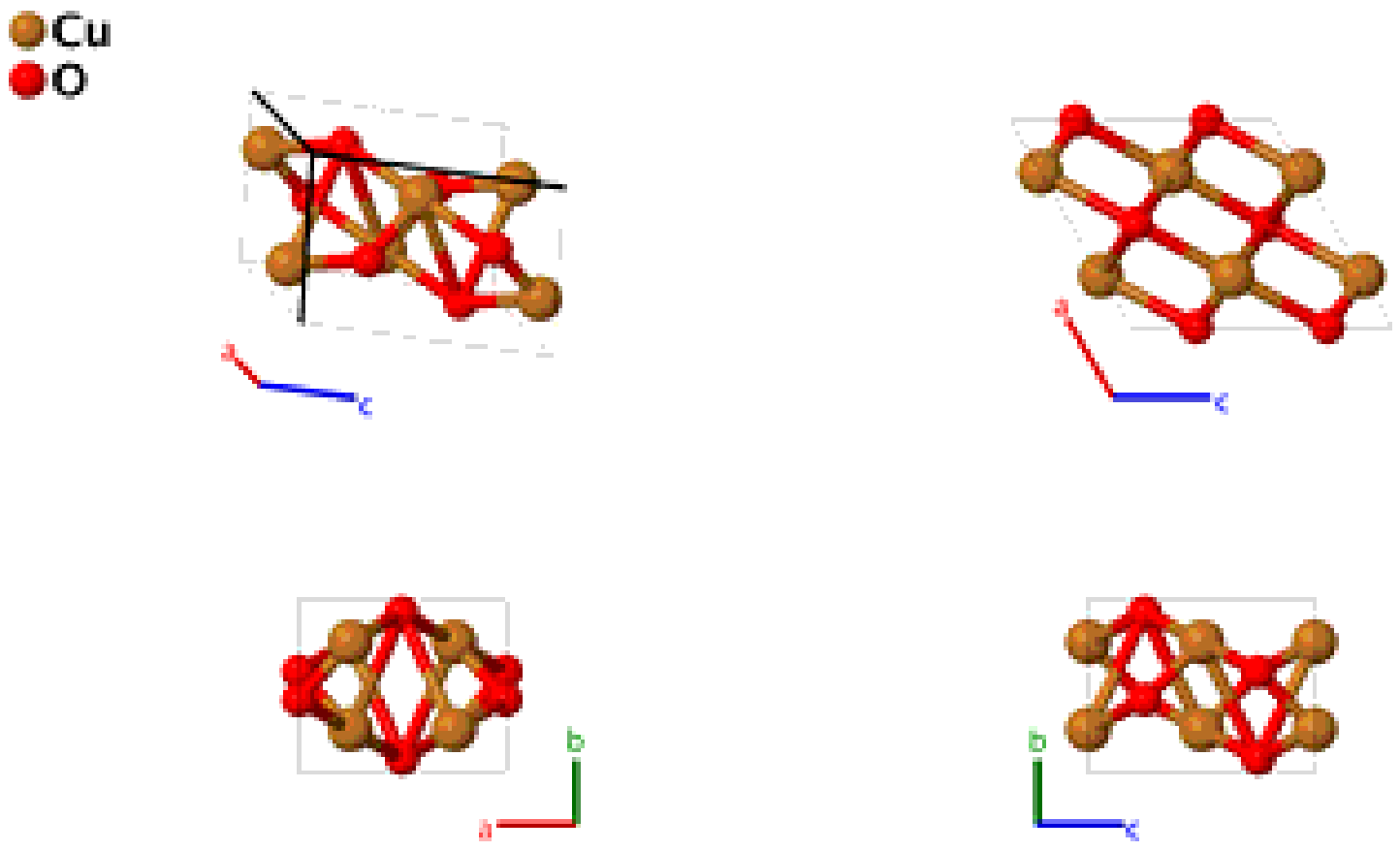

**Prototype** : CuO

**AFLOW prototype label** : AB_mC8_15_c_e

***Strukturbericht* designation** : B26

**Pearson symbol** : mC8

**Space group number** : 15

**Space group symbol** : C2/c

**AFLOW prototype command** : `aflow --proto=AB_mC8_15_c_e`
`--params=`$a, b/a, c/a, \beta, y_2$

---

**Base-centered Monoclinic primitive vectors:**

$$\mathbf{a}_1 = \frac{1}{2}\,a\,\hat{\mathbf{x}} - \frac{1}{2}\,b\,\hat{\mathbf{y}}$$
$$\mathbf{a}_2 = \frac{1}{2}\,a\,\hat{\mathbf{x}} + \frac{1}{2}\,b\,\hat{\mathbf{y}}$$
$$\mathbf{a}_3 = c\cos\beta\,\hat{\mathbf{x}} + c\sin\beta\,\hat{\mathbf{z}}$$

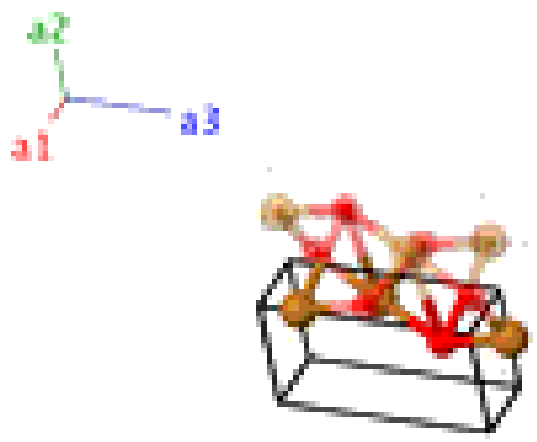

---

**Basis vectors:**

| | | Lattice Coordinates | | Cartesian Coordinates | Wyckoff Position | Atom Type |
|---|---|---|---|---|---|---|
| $\mathbf{B}_1$ | = | $\frac{1}{2}\,\mathbf{a}_2$ | = | $\frac{1}{4}\,a\,\hat{\mathbf{x}} + \frac{1}{4}\,b\,\hat{\mathbf{y}}$ | (4$c$) | Cu |
| $\mathbf{B}_2$ | = | $\frac{1}{2}\,\mathbf{a}_1 + \frac{1}{2}\,\mathbf{a}_3$ | = | $\left(\frac{1}{4}\,a + \frac{1}{2}\,c\,\cos\beta\right)\hat{\mathbf{x}} + \frac{3}{4}\,b\,\hat{\mathbf{y}} + \frac{1}{2}\,c\,\sin\beta\,\hat{\mathbf{z}}$ | (4$c$) | Cu |
| $\mathbf{B}_3$ | = | $-y_2\,\mathbf{a}_1 + y_2\,\mathbf{a}_2 + \frac{1}{4}\,\mathbf{a}_3$ | = | $\frac{1}{4}\,c\,\cos\beta\,\hat{\mathbf{x}} + y_2\,b\,\hat{\mathbf{y}} + \frac{1}{4}\,c\,\sin\beta\,\hat{\mathbf{z}}$ | (4$e$) | O |
| $\mathbf{B}_4$ | = | $y_2\,\mathbf{a}_1 - y_2\,\mathbf{a}_2 + \frac{3}{4}\,\mathbf{a}_3$ | = | $\frac{3}{4}\,c\,\cos\beta\,\hat{\mathbf{x}} - y_2\,b\,\hat{\mathbf{y}} + \frac{3}{4}\,c\,\sin\beta\,\hat{\mathbf{z}}$ | (4$e$) | O |

---

---

**Geometry files:**

- CIF: pp. 653
- POSCAR: pp. 653

# Coesite ($SiO_2$) Structure: A2B_mC48_15_ae3f_2f

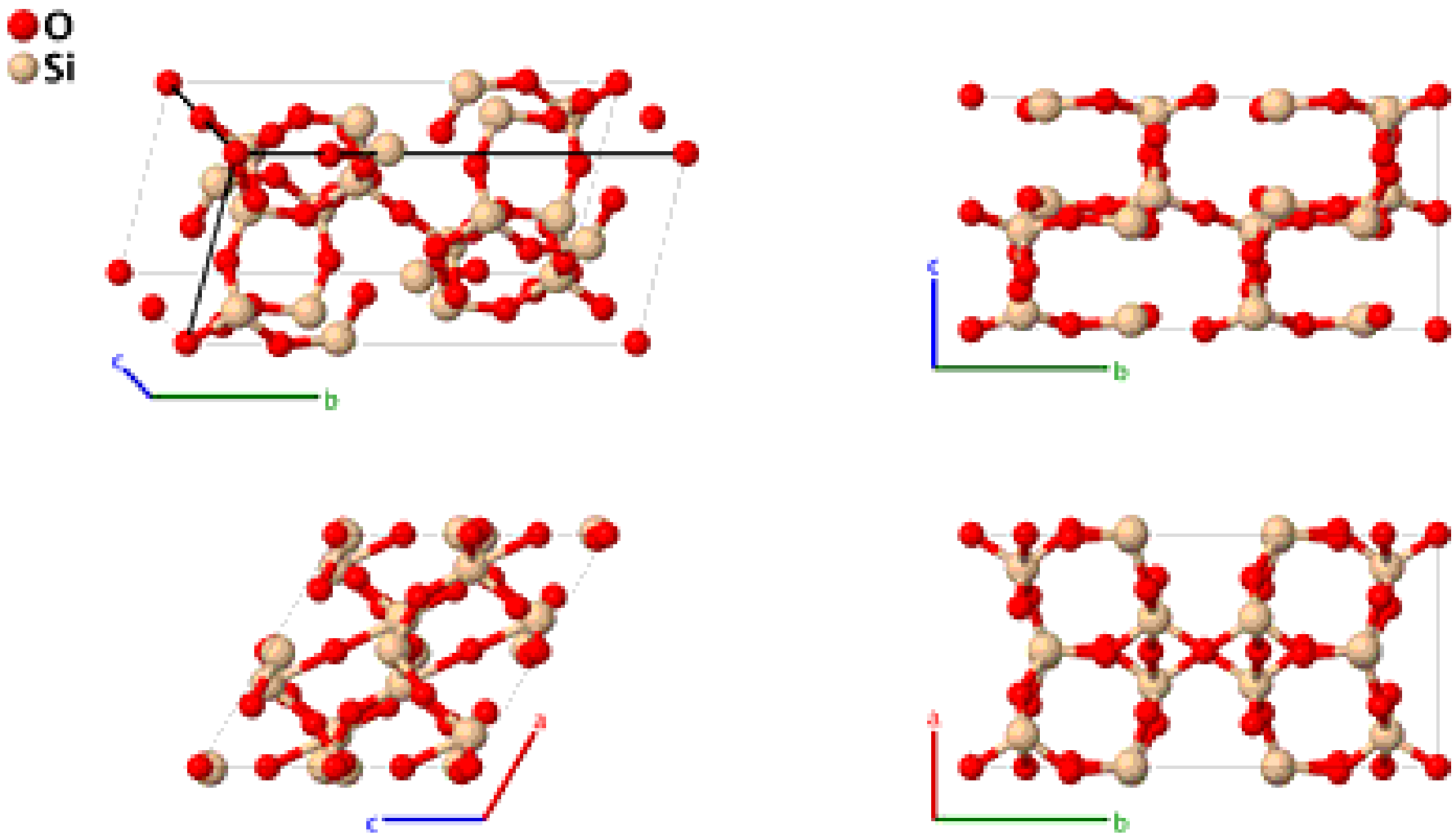

| | | |
|---|---|---|
| **Prototype** | : | $SiO_2$ |
| **AFLOW prototype label** | : | A2B_mC48_15_ae3f_2f |
| ***Strukturbericht* designation** | : | None |
| **Pearson symbol** | : | mC48 |
| **Space group number** | : | 15 |
| **Space group symbol** | : | C2/c |
| **AFLOW prototype command** | : | `aflow --proto=A2B_mC48_15_ae3f_2f` <br> `--params=`$a, b/a, c/a, \beta, y_2, x_3, y_3, z_3, x_4, y_4, z_4, x_5, y_5, z_5, x_6, y_6, z_6, x_7, y_7, z_7$ |

---

**Base-centered Monoclinic primitive vectors:**

$$\mathbf{a}_1 = \frac{1}{2}\,a\,\mathbf{\hat{x}} - \frac{1}{2}\,b\,\mathbf{\hat{y}}$$
$$\mathbf{a}_2 = \frac{1}{2}\,a\,\mathbf{\hat{x}} + \frac{1}{2}\,b\,\mathbf{\hat{y}}$$
$$\mathbf{a}_3 = c\cos\beta\,\mathbf{\hat{x}} + c\sin\beta\,\mathbf{\hat{z}}$$

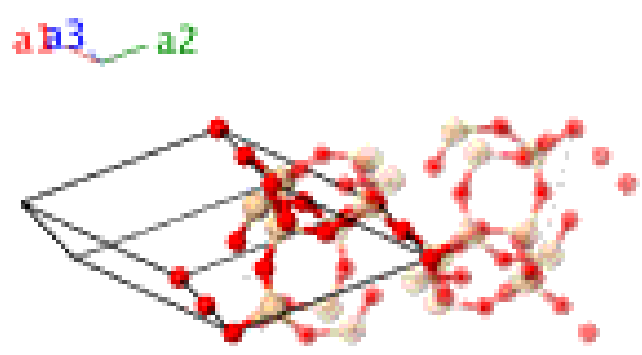

---

**Basis vectors:**

| | | Lattice Coordinates | | Cartesian Coordinates | Wyckoff Position | Atom Type |
|---|---|---|---|---|---|---|
| $\mathbf{B}_1$ | = | $0\,\mathbf{a}_1 + 0\,\mathbf{a}_2 + 0\,\mathbf{a}_3$ | = | $0\mathbf{\hat{x}} + 0\mathbf{\hat{y}} + 0\mathbf{\hat{z}}$ | (4$a$) | O I |
| $\mathbf{B}_2$ | = | $\frac{1}{2}\,\mathbf{a}_3$ | = | $\frac{1}{2}\,c\,\cos\beta\,\mathbf{\hat{x}} + \frac{1}{2}\,c\,\sin\beta\mathbf{\hat{z}}$ | (4$a$) | O I |
| $\mathbf{B}_3$ | = | $-y_2\,\mathbf{a}_1 + y_2\,\mathbf{a}_2 + \frac{1}{4}\,\mathbf{a}_3$ | = | $\frac{1}{4}\,c\,\cos\beta\,\mathbf{\hat{x}} + y_2\,b\,\mathbf{\hat{y}} + \frac{1}{4}\,c\,\sin\beta\,\mathbf{\hat{z}}$ | (4$e$) | O II |
| $\mathbf{B}_4$ | = | $y_2\,\mathbf{a}_1 - y_2\,\mathbf{a}_2 + \frac{3}{4}\,\mathbf{a}_3$ | = | $\frac{3}{4}\,c\,\cos\beta\,\mathbf{\hat{x}} - y_2\,b\,\mathbf{\hat{y}} + \frac{3}{4}\,c\,\sin\beta\,\mathbf{\hat{z}}$ | (4$e$) | O II |
| $\mathbf{B}_5$ | = | $(x_3 - y_3)\,\mathbf{a}_1 + (x_3 + y_3)\,\mathbf{a}_2 + z_3\,\mathbf{a}_3$ | = | $(x_3\,a + z_3\,c\,\cos\beta)\,\mathbf{\hat{x}} + y_3\,b\,\mathbf{\hat{y}} + z_3\,c\,\sin\beta\,\mathbf{\hat{z}}$ | (8$f$) | O III |
| $\mathbf{B}_6$ | = | $-(x_3 + y_3)\,\mathbf{a}_1 + (y_3 - x_3)\,\mathbf{a}_2 + \left(\frac{1}{2} - z_3\right)\,\mathbf{a}_3$ | = | $\left(-x_3\,a + \left(\frac{1}{2} - z_3\right)\,c\,\cos\beta\right)\,\mathbf{\hat{x}} + y_3\,b\,\mathbf{\hat{y}} + \left(\frac{1}{2} - z_3\right)\,c\,\sin\beta\,\mathbf{\hat{z}}$ | (8$f$) | O III |
| $\mathbf{B_{7}}$ | = | $\left(y_{3}-x_{3}\right)\,\mathbf{a_{1}}-\left(x_{3}+y_{3}\right)\,\mathbf{a_{2}}-z_{3}\,\mathbf{a_{3}}$ | = | $-\left(x_{3}\,a+z_{3}\,c\,\cos\beta\right)\,\mathbf{\hat{x}}-y_{3}\,b\,\mathbf{\hat{y}}-z_{3}\,c\,\sin\beta\,\mathbf{\hat{z}}$ | $(8f)$ | O III |
| $\mathbf{B_{8}}$ | = | $\left(x_{3}+y_{3}\right)\,\mathbf{a_{1}}+\left(x_{3}-y_{3}\right)\,\mathbf{a_{2}}+\left(\frac{1}{2}+z_{3}\right)\,\mathbf{a_{3}}$ | = | $\left(x_{3}\,a+\left(\frac{1}{2}+z_{3}\right)\,c\,\cos\beta\right)\,\mathbf{\hat{x}}-y_{3}\,b\,\mathbf{\hat{y}}+\left(\frac{1}{2}+z_{3}\right)\,c\,\sin\beta\,\mathbf{\hat{z}}$ | $(8f)$ | O III |
| $\mathbf{B_{9}}$ | = | $\left(x_{4}-y_{4}\right)\,\mathbf{a_{1}}+\left(x_{4}+y_{4}\right)\,\mathbf{a_{2}}+z_{4}\,\mathbf{a_{3}}$ | = | $\left(x_{4}\,a+z_{4}\,c\,\cos\beta\right)\,\mathbf{\hat{x}}+y_{4}\,b\,\mathbf{\hat{y}}+z_{4}\,c\,\sin\beta\,\mathbf{\hat{z}}$ | $(8f)$ | O IV |
| $\mathbf{B_{10}}$ | = | $-\left(x_{4}+y_{4}\right)\,\mathbf{a_{1}}+\left(y_{4}-x_{4}\right)\,\mathbf{a_{2}}+\left(\frac{1}{2}-z_{4}\right)\,\mathbf{a_{3}}$ | = | $\left(-x_{4}\,a+\left(\frac{1}{2}-z_{4}\right)\,c\,\cos\beta\right)\,\mathbf{\hat{x}}+y_{4}\,b\,\mathbf{\hat{y}}+\left(\frac{1}{2}-z_{4}\right)\,c\,\sin\beta\,\mathbf{\hat{z}}$ | $(8f)$ | O IV |
| $\mathbf{B_{11}}$ | = | $\left(y_{4}-x_{4}\right)\,\mathbf{a_{1}}-\left(x_{4}+y_{4}\right)\,\mathbf{a_{2}}-z_{4}\,\mathbf{a_{3}}$ | = | $-\left(x_{4}\,a+z_{4}\,c\,\cos\beta\right)\,\mathbf{\hat{x}}-y_{4}\,b\,\mathbf{\hat{y}}-z_{4}\,c\,\sin\beta\,\mathbf{\hat{z}}$ | $(8f)$ | O IV |
| $\mathbf{B_{12}}$ | = | $\left(x_{4}+y_{4}\right)\,\mathbf{a_{1}}+\left(x_{4}-y_{4}\right)\,\mathbf{a_{2}}+\left(\frac{1}{2}+z_{4}\right)\,\mathbf{a_{3}}$ | = | $\left(x_{4}\,a+\left(\frac{1}{2}+z_{4}\right)\,c\,\cos\beta\right)\,\mathbf{\hat{x}}-y_{4}\,b\,\mathbf{\hat{y}}+\left(\frac{1}{2}+z_{4}\right)\,c\,\sin\beta\,\mathbf{\hat{z}}$ | $(8f)$ | O IV |
| $\mathbf{B_{13}}$ | = | $\left(x_{5}-y_{5}\right)\,\mathbf{a_{1}}+\left(x_{5}+y_{5}\right)\,\mathbf{a_{2}}+z_{5}\,\mathbf{a_{3}}$ | = | $\left(x_{5}\,a+z_{5}\,c\,\cos\beta\right)\,\mathbf{\hat{x}}+y_{5}\,b\,\mathbf{\hat{y}}+z_{5}\,c\,\sin\beta\,\mathbf{\hat{z}}$ | $(8f)$ | O V |
| $\mathbf{B_{14}}$ | = | $-\left(x_{5}+y_{5}\right)\,\mathbf{a_{1}}+\left(y_{5}-x_{5}\right)\,\mathbf{a_{2}}+\left(\frac{1}{2}-z_{5}\right)\,\mathbf{a_{3}}$ | = | $\left(-x_{5}\,a+\left(\frac{1}{2}-z_{5}\right)\,c\,\cos\beta\right)\,\mathbf{\hat{x}}+y_{5}\,b\,\mathbf{\hat{y}}+\left(\frac{1}{2}-z_{5}\right)\,c\,\sin\beta\,\mathbf{\hat{z}}$ | $(8f)$ | O V |
| $\mathbf{B_{15}}$ | = | $\left(y_{5}-x_{5}\right)\,\mathbf{a_{1}}-\left(x_{5}+y_{5}\right)\,\mathbf{a_{2}}-z_{5}\,\mathbf{a_{3}}$ | = | $-\left(x_{5}\,a+z_{5}\,c\,\cos\beta\right)\,\mathbf{\hat{x}}-y_{5}\,b\,\mathbf{\hat{y}}-z_{5}\,c\,\sin\beta\,\mathbf{\hat{z}}$ | $(8f)$ | O V |
| $\mathbf{B_{16}}$ | = | $\left(x_{5}+y_{5}\right)\,\mathbf{a_{1}}+\left(x_{5}-y_{5}\right)\,\mathbf{a_{2}}+\left(\frac{1}{2}+z_{5}\right)\,\mathbf{a_{3}}$ | = | $\left(x_{5}\,a+\left(\frac{1}{2}+z_{5}\right)\,c\,\cos\beta\right)\,\mathbf{\hat{x}}-y_{5}\,b\,\mathbf{\hat{y}}+\left(\frac{1}{2}+z_{5}\right)\,c\,\sin\beta\,\mathbf{\hat{z}}$ | $(8f)$ | O V |
| $\mathbf{B_{17}}$ | = | $\left(x_{6}-y_{6}\right)\,\mathbf{a_{1}}+\left(x_{6}+y_{6}\right)\,\mathbf{a_{2}}+z_{6}\,\mathbf{a_{3}}$ | = | $\left(x_{6}\,a+z_{6}\,c\,\cos\beta\right)\,\mathbf{\hat{x}}+y_{6}\,b\,\mathbf{\hat{y}}+z_{6}\,c\,\sin\beta\,\mathbf{\hat{z}}$ | $(8f)$ | Si I |
| $\mathbf{B_{18}}$ | = | $-\left(x_{6}+y_{6}\right)\,\mathbf{a_{1}}+\left(y_{6}-x_{6}\right)\,\mathbf{a_{2}}+\left(\frac{1}{2}-z_{6}\right)\,\mathbf{a_{3}}$ | = | $\left(-x_{6}\,a+\left(\frac{1}{2}-z_{6}\right)\,c\,\cos\beta\right)\,\mathbf{\hat{x}}+y_{6}\,b\,\mathbf{\hat{y}}+\left(\frac{1}{2}-z_{6}\right)\,c\,\sin\beta\,\mathbf{\hat{z}}$ | $(8f)$ | Si I |
| $\mathbf{B_{19}}$ | = | $\left(y_{6}-x_{6}\right)\,\mathbf{a_{1}}-\left(x_{6}+y_{6}\right)\,\mathbf{a_{2}}-z_{6}\,\mathbf{a_{3}}$ | = | $-\left(x_{6}\,a+z_{6}\,c\,\cos\beta\right)\,\mathbf{\hat{x}}-y_{6}\,b\,\mathbf{\hat{y}}-z_{6}\,c\,\sin\beta\,\mathbf{\hat{z}}$ | $(8f)$ | Si I |
| $\mathbf{B_{20}}$ | = | $\left(x_{6}+y_{6}\right)\,\mathbf{a_{1}}+\left(x_{6}-y_{6}\right)\,\mathbf{a_{2}}+\left(\frac{1}{2}+z_{6}\right)\,\mathbf{a_{3}}$ | = | $\left(x_{6}\,a+\left(\frac{1}{2}+z_{6}\right)\,c\,\cos\beta\right)\,\mathbf{\hat{x}}-y_{6}\,b\,\mathbf{\hat{y}}+\left(\frac{1}{2}+z_{6}\right)\,c\,\sin\beta\,\mathbf{\hat{z}}$ | $(8f)$ | Si I |
| $\mathbf{B_{21}}$ | = | $\left(x_{7}-y_{7}\right)\,\mathbf{a_{1}}+\left(x_{7}+y_{7}\right)\,\mathbf{a_{2}}+z_{7}\,\mathbf{a_{3}}$ | = | $\left(x_{7}\,a+z_{7}\,c\,\cos\beta\right)\,\mathbf{\hat{x}}+y_{7}\,b\,\mathbf{\hat{y}}+z_{7}\,c\,\sin\beta\,\mathbf{\hat{z}}$ | $(8f)$ | Si II |
| $\mathbf{B_{22}}$ | = | $-\left(x_{7}+y_{7}\right)\,\mathbf{a_{1}}+\left(y_{7}-x_{7}\right)\,\mathbf{a_{2}}+\left(\frac{1}{2}-z_{7}\right)\,\mathbf{a_{3}}$ | = | $\left(-x_{7}\,a+\left(\frac{1}{2}-z_{7}\right)\,c\,\cos\beta\right)\,\mathbf{\hat{x}}+y_{7}\,b\,\mathbf{\hat{y}}+\left(\frac{1}{2}-z_{7}\right)\,c\,\sin\beta\,\mathbf{\hat{z}}$ | $(8f)$ | Si II |
| $\mathbf{B_{23}}$ | = | $\left(y_{7}-x_{7}\right)\,\mathbf{a_{1}}-\left(x_{7}+y_{7}\right)\,\mathbf{a_{2}}-z_{7}\,\mathbf{a_{3}}$ | = | $-\left(x_{7}\,a+z_{7}\,c\,\cos\beta\right)\,\mathbf{\hat{x}}-y_{7}\,b\,\mathbf{\hat{y}}-z_{7}\,c\,\sin\beta\,\mathbf{\hat{z}}$ | $(8f)$ | Si II |
| $\mathbf{B_{24}}$ | = | $\left(x_{7}+y_{7}\right)\,\mathbf{a_{1}}+\left(x_{7}-y_{7}\right)\,\mathbf{a_{2}}+\left(\frac{1}{2}+z_{7}\right)\,\mathbf{a_{3}}$ | = | $\left(x_{7}\,a+\left(\frac{1}{2}+z_{7}\right)\,c\,\cos\beta\right)\,\mathbf{\hat{x}}-y_{7}\,b\,\mathbf{\hat{y}}+\left(\frac{1}{2}+z_{7}\right)\,c\,\sin\beta\,\mathbf{\hat{z}}$ | $(8f)$ | Si II |

**Geometry files:**

- CIF: pp. 653
- POSCAR: pp. 653

# Esseneite Structure: ABC6D2_mC40_15_e_e_3f_f

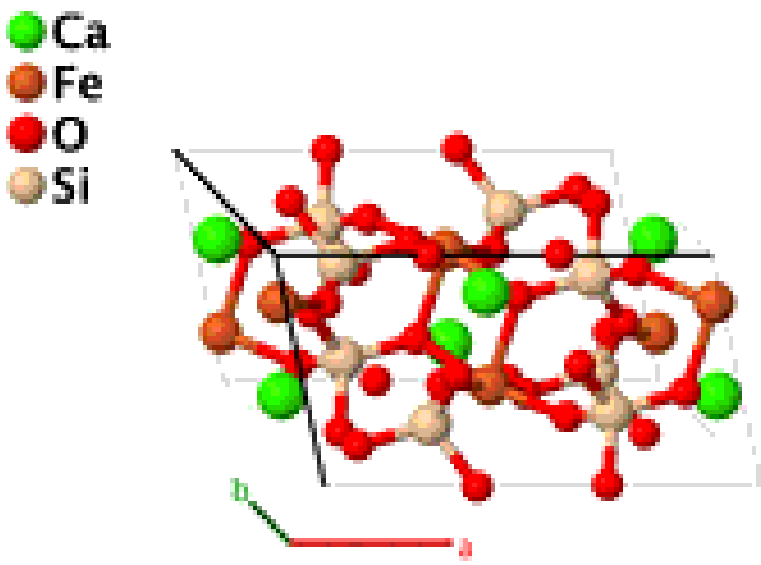

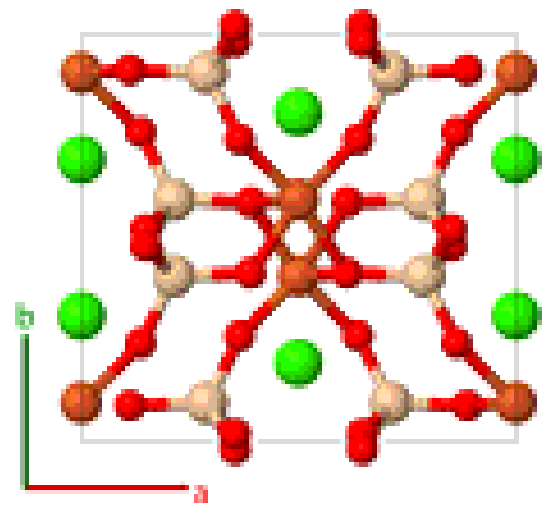

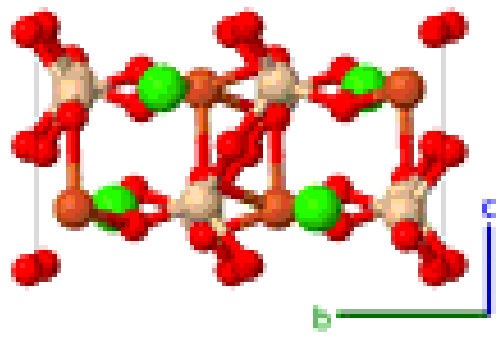

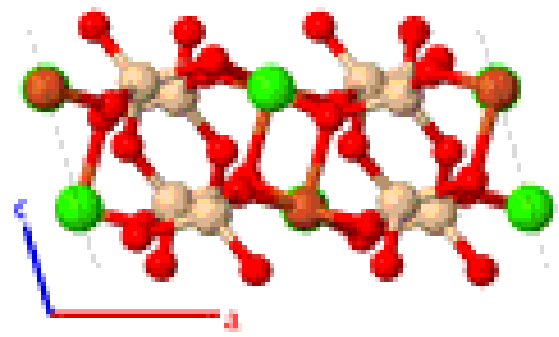

| | | |
|---|---|---|
| **Prototype** | : | $CaFeO_6Si_2$ |
| **AFLOW prototype label** | : | ABC6D2_mC40_15_e_e_3f_f |
| ***Strukturbericht* designation** | : | None |
| **Pearson symbol** | : | mC40 |
| **Space group number** | : | 15 |
| **Space group symbol** | : | C2/c |
| **AFLOW prototype command** | : | `aflow --proto=ABC6D2_mC40_15_e_e_3f_f` `--params=`$a, b/a, c/a, \beta, y_1, y_2, x_3, y_3, z_3, x_4, y_4, z_4, x_5, y_5, z_5, x_6, y_6, z_6$ |

- Named for University of Michigan geologist Eric Essene (1939-2010). (Cosca, 1987) gives the composition as $(Ca_{0.97}Fe_{0.03})(Fe_{0.58}Al_{0.42})O_6(Si_{0.54}Al_{0.46})_2$. We will use the majority atom at each site to draw the structure.

**Base-centered Monoclinic primitive vectors:**

$$\mathbf{a}_1 = \frac{1}{2}\,a\,\mathbf{\hat{x}} - \frac{1}{2}\,b\,\mathbf{\hat{y}}$$
$$\mathbf{a}_2 = \frac{1}{2}\,a\,\mathbf{\hat{x}} + \frac{1}{2}\,b\,\mathbf{\hat{y}}$$
$$\mathbf{a}_3 = c\cos\beta\,\mathbf{\hat{x}} + c\sin\beta\,\mathbf{\hat{z}}$$

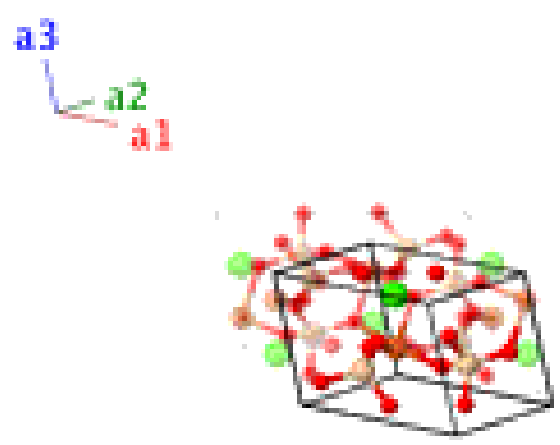

**Basis vectors:**

| | | Lattice Coordinates | | Cartesian Coordinates | Wyckoff Position | Atom Type |
|---|---|---|---|---|---|---|
| $\mathbf{B_1}$ | = | $-y_1\,\mathbf{a_1} + y_1\,\mathbf{a_2} + \frac{1}{4}\,\mathbf{a_3}$ | = | $\frac{1}{4}\,c\,\cos\beta\,\mathbf{\hat{x}} + y_1\,b\,\mathbf{\hat{y}} + \frac{1}{4}\,c\,\sin\beta\,\mathbf{\hat{z}}$ | (4$e$) | Ca |
| $\mathbf{B_2}$ | = | $y_1\,\mathbf{a_1} - y_1\,\mathbf{a_2} + \frac{3}{4}\,\mathbf{a_3}$ | = | $\frac{3}{4}\,c\,\cos\beta\,\mathbf{\hat{x}} - y_1\,b\,\mathbf{\hat{y}} + \frac{3}{4}\,c\,\sin\beta\,\mathbf{\hat{z}}$ | (4$e$) | Ca |
| $\mathbf{B_3}$ | = | $-y_2\,\mathbf{a_1} + y_2\,\mathbf{a_2} + \frac{1}{4}\,\mathbf{a_3}$ | = | $\frac{1}{4}\,c\,\cos\beta\,\mathbf{\hat{x}} + y_2\,b\,\mathbf{\hat{y}} + \frac{1}{4}\,c\,\sin\beta\,\mathbf{\hat{z}}$ | (4e) | Fe |
| $\mathbf{B_4}$ | = | $y_2\,\mathbf{a_1} - y_2\,\mathbf{a_2} + \frac{3}{4}\,\mathbf{a_3}$ | = | $\frac{3}{4}\,c\,\cos\beta\,\mathbf{\hat{x}} - y_2\,b\,\mathbf{\hat{y}} + \frac{3}{4}\,c\,\sin\beta\,\mathbf{\hat{z}}$ | (4e) | Fe |
| $\mathbf{B_5}$ | = | $(x_3 - y_3)\,\mathbf{a_1} + (x_3 + y_3)\,\mathbf{a_2} + z_3\,\mathbf{a_3}$ | = | $(x_3\,a + z_3\,c\,\cos\beta)\,\mathbf{\hat{x}} + y_3\,b\,\mathbf{\hat{y}} + z_3\,c\,\sin\beta\,\mathbf{\hat{z}}$ | (8f) | O I |
| $\mathbf{B_6}$ | = | $-(x_3 + y_3)\,\mathbf{a_1} + (y_3 - x_3)\,\mathbf{a_2} + \left(\frac{1}{2} - z_3\right)\,\mathbf{a_3}$ | = | $\left(-x_3\,a + \left(\frac{1}{2} - z_3\right)\,c\,\cos\beta\right)\,\mathbf{\hat{x}} + y_3\,b\,\mathbf{\hat{y}} + \left(\frac{1}{2} - z_3\right)\,c\,\sin\beta\,\mathbf{\hat{z}}$ | (8f) | O I |
| $\mathbf{B_7}$ | = | $(y_3 - x_3)\,\mathbf{a_1} - (x_3 + y_3)\,\mathbf{a_2} - z_3\,\mathbf{a_3}$ | = | $-(x_3\,a + z_3\,c\,\cos\beta)\,\mathbf{\hat{x}} - y_3\,b\,\mathbf{\hat{y}} - z_3\,c\,\sin\beta\,\mathbf{\hat{z}}$ | (8f) | O I |
| $\mathbf{B_8}$ | = | $(x_3 + y_3)\,\mathbf{a_1} + (x_3 - y_3)\,\mathbf{a_2} + \left(\frac{1}{2} + z_3\right)\,\mathbf{a_3}$ | = | $\left(x_3\,a + \left(\frac{1}{2} + z_3\right)\,c\,\cos\beta\right)\,\mathbf{\hat{x}} - y_3\,b\,\mathbf{\hat{y}} + \left(\frac{1}{2} + z_3\right)\,c\,\sin\beta\,\mathbf{\hat{z}}$ | (8f) | O I |
| $\mathbf{B_9}$ | = | $(x_4 - y_4)\,\mathbf{a_1} + (x_4 + y_4)\,\mathbf{a_2} + z_4\,\mathbf{a_3}$ | = | $(x_4\,a + z_4\,c\,\cos\beta)\,\mathbf{\hat{x}} + y_4\,b\,\mathbf{\hat{y}} + z_4\,c\,\sin\beta\,\mathbf{\hat{z}}$ | (8f) | O II |
| $\mathbf{B_{10}}$ | = | $-(x_4 + y_4)\,\mathbf{a_1} + (y_4 - x_4)\,\mathbf{a_2} + \left(\frac{1}{2} - z_4\right)\,\mathbf{a_3}$ | = | $\left(-x_4\,a + \left(\frac{1}{2} - z_4\right)\,c\,\cos\beta\right)\,\mathbf{\hat{x}} + y_4\,b\,\mathbf{\hat{y}} + \left(\frac{1}{2} - z_4\right)\,c\,\sin\beta\,\mathbf{\hat{z}}$ | (8f) | O II |
| $\mathbf{B_{11}}$ | = | $(y_4 - x_4)\,\mathbf{a_1} - (x_4 + y_4)\,\mathbf{a_2} - z_4\,\mathbf{a_3}$ | = | $-(x_4\,a + z_4\,c\,\cos\beta)\,\mathbf{\hat{x}} - y_4\,b\,\mathbf{\hat{y}} - z_4\,c\,\sin\beta\,\mathbf{\hat{z}}$ | (8f) | O II |
| $\mathbf{B_{12}}$ | = | $(x_4 + y_4)\,\mathbf{a_1} + (x_4 - y_4)\,\mathbf{a_2} + \left(\frac{1}{2} + z_4\right)\,\mathbf{a_3}$ | = | $\left(x_4\,a + \left(\frac{1}{2} + z_4\right)\,c\,\cos\beta\right)\,\mathbf{\hat{x}} - y_4\,b\,\mathbf{\hat{y}} + \left(\frac{1}{2} + z_4\right)\,c\,\sin\beta\,\mathbf{\hat{z}}$ | (8f) | O II |
| $\mathbf{B_{13}}$ | = | $(x_5 - y_5)\,\mathbf{a_1} + (x_5 + y_5)\,\mathbf{a_2} + z_5\,\mathbf{a_3}$ | = | $(x_5\,a + z_5\,c\,\cos\beta)\,\mathbf{\hat{x}} + y_5\,b\,\mathbf{\hat{y}} + z_5\,c\,\sin\beta\,\mathbf{\hat{z}}$ | (8f) | O III |
| $\mathbf{B_{14}}$ | = | $-(x_5 + y_5)\,\mathbf{a_1} + (y_5 - x_5)\,\mathbf{a_2} + \left(\frac{1}{2} - z_5\right)\,\mathbf{a_3}$ | = | $\left(-x_5\,a + \left(\frac{1}{2} - z_5\right)\,c\,\cos\beta\right)\,\mathbf{\hat{x}} + y_5\,b\,\mathbf{\hat{y}} + \left(\frac{1}{2} - z_5\right)\,c\,\sin\beta\,\mathbf{\hat{z}}$ | (8f) | O III |
| $\mathbf{B_{15}}$ | = | $(y_5 - x_5)\,\mathbf{a_1} - (x_5 + y_5)\,\mathbf{a_2} - z_5\,\mathbf{a_3}$ | = | $-(x_5\,a + z_5\,c\,\cos\beta)\,\mathbf{\hat{x}} - y_5\,b\,\mathbf{\hat{y}} - z_5\,c\,\sin\beta\,\mathbf{\hat{z}}$ | (8f) | O III |
| $\mathbf{B_{16}}$ | = | $(x_5 + y_5)\,\mathbf{a_1} + (x_5 - y_5)\,\mathbf{a_2} + \left(\frac{1}{2} + z_5\right)\,\mathbf{a_3}$ | = | $\left(x_5\,a + \left(\frac{1}{2} + z_5\right)\,c\,\cos\beta\right)\,\mathbf{\hat{x}} - y_5\,b\,\mathbf{\hat{y}} + \left(\frac{1}{2} + z_5\right)\,c\,\sin\beta\,\mathbf{\hat{z}}$ | (8f) | O III |
| $\mathbf{B_{17}}$ | = | $(x_6 - y_6)\,\mathbf{a_1} + (x_6 + y_6)\,\mathbf{a_2} + z_6\,\mathbf{a_3}$ | = | $(x_6\,a + z_6\,c\,\cos\beta)\,\mathbf{\hat{x}} + y_6\,b\,\mathbf{\hat{y}} + z_6\,c\,\sin\beta\,\mathbf{\hat{z}}$ | (8f) | Si |
| $\mathbf{B_{18}}$ | = | $-(x_6 + y_6)\,\mathbf{a_1} + (y_6 - x_6)\,\mathbf{a_2} + \left(\frac{1}{2} - z_6\right)\,\mathbf{a_3}$ | = | $\left(-x_6\,a + \left(\frac{1}{2} - z_6\right)\,c\,\cos\beta\right)\,\mathbf{\hat{x}} + y_6\,b\,\mathbf{\hat{y}} + \left(\frac{1}{2} - z_6\right)\,c\,\sin\beta\,\mathbf{\hat{z}}$ | (8f) | Si |
| $\mathbf{B_{19}}$ | = | $(y_6 - x_6)\,\mathbf{a_1} - (x_6 + y_6)\,\mathbf{a_2} - z_6\,\mathbf{a_3}$ | = | $-(x_6\,a + z_6\,c\,\cos\beta)\,\mathbf{\hat{x}} - y_6\,b\,\mathbf{\hat{y}} - z_6\,c\,\sin\beta\,\mathbf{\hat{z}}$ | (8f) | Si |
| $\mathbf{B_{20}}$ | = | $(x_6 + y_6)\,\mathbf{a_1} + (x_6 - y_6)\,\mathbf{a_2} + \left(\frac{1}{2} + z_6\right)\,\mathbf{a_3}$ | = | $\left(x_6\,a + \left(\frac{1}{2} + z_6\right)\,c\,\cos\beta\right)\,\mathbf{\hat{x}} - y_6\,b\,\mathbf{\hat{y}} + \left(\frac{1}{2} + z_6\right)\,c\,\sin\beta\,\mathbf{\hat{z}}$ | (8f) | Si |

**Geometry files:**

- CIF: pp. 654
- POSCAR: pp. 654

# $AlPS_4$ Structure: ABC4_oP12_16_ag_cd_2u

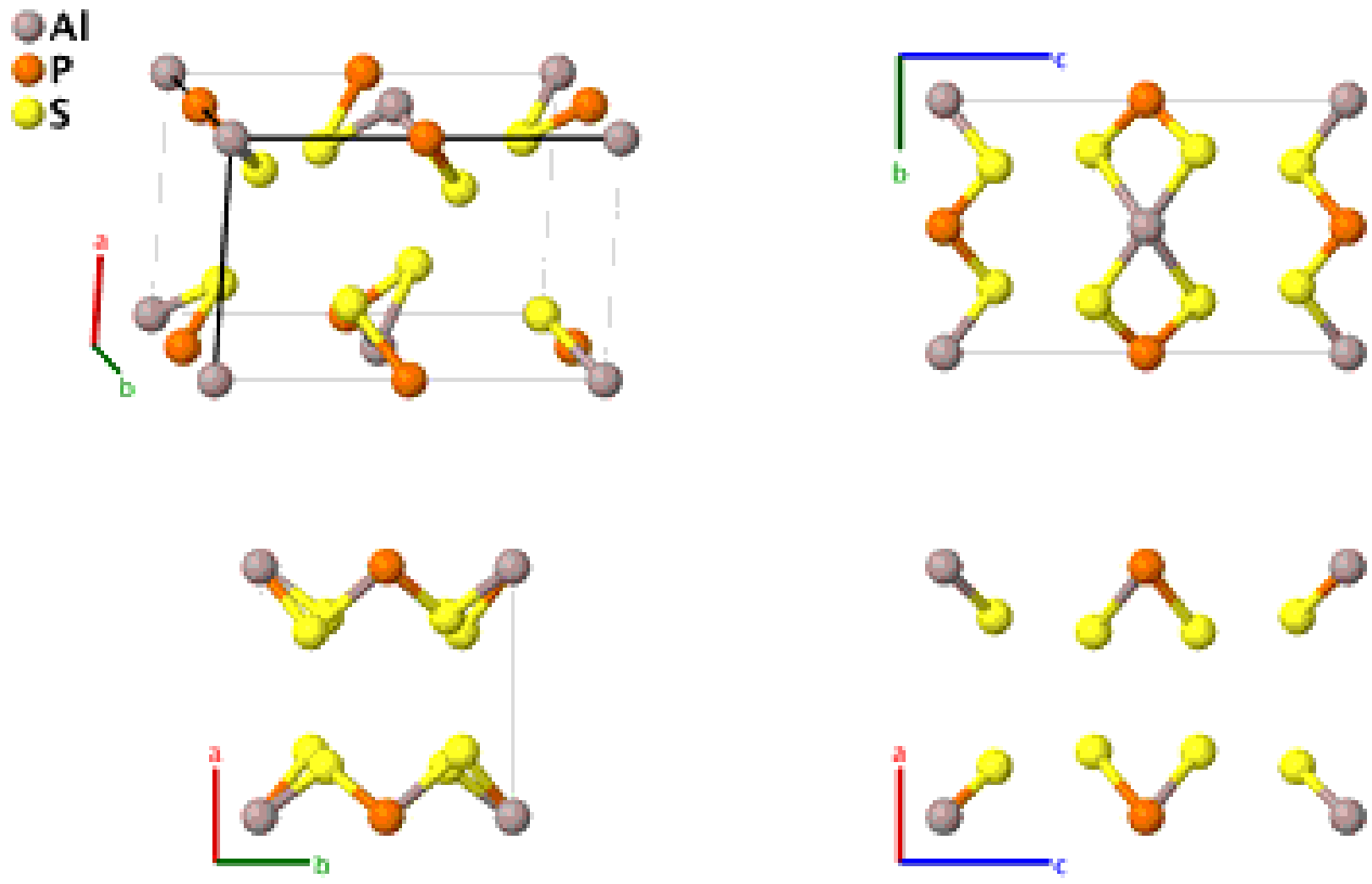

| | | |
|---|---|---|
| **Prototype** | : | $AlPS_4$ |
| **AFLOW prototype label** | : | ABC4_oP12_16_ag_cd_2u |
| ***Strukturbericht* designation** | : | None |
| **Pearson symbol** | : | oP12 |
| **Space group number** | : | 16 |
| **Space group symbol** | : | P222 |
| **AFLOW prototype command** | : | `aflow --proto=ABC4_oP12_16_ag_cd_2u` <br> `--params=`$a, b/a, c/a, x_5, y_5, z_5, x_6, y_6, z_6$ |

**Simple Orthorhombic primitive vectors:**

$$\mathbf{a}_1 = a\,\hat{\mathbf{x}}$$
$$\mathbf{a}_2 = b\,\hat{\mathbf{y}}$$
$$\mathbf{a}_3 = c\,\hat{\mathbf{z}}$$

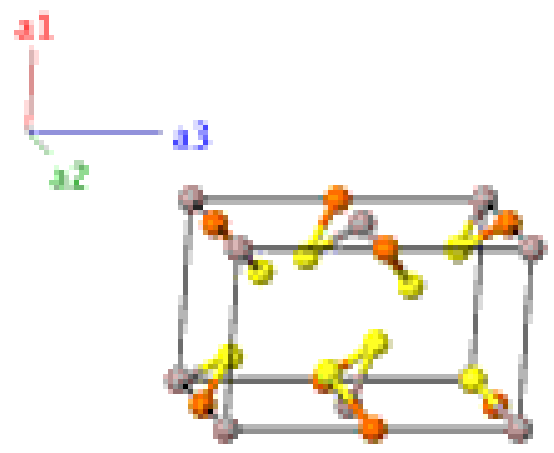

**Basis vectors:**

| | | Lattice Coordinates | | Cartesian Coordinates | Wyckoff Position | Atom Type |
|---|---|---|---|---|---|---|
| $\mathbf{B}_1$ | = | $0\,\mathbf{a}_1 + 0\,\mathbf{a}_2 + 0\,\mathbf{a}_3$ | = | $0\hat{\mathbf{x}} + 0\hat{\mathbf{y}} + 0\hat{\mathbf{z}}$ | $(1a)$ | Al I |
| $\mathbf{B}_2$ | = | $\frac{1}{2}\mathbf{a}_2$ | = | $\frac{1}{2}\,b\,\hat{\mathbf{y}}$ | $(1c)$ | P I |
| $\mathbf{B}_3$ | = | $\frac{1}{2}\mathbf{a}_3$ | = | $\frac{1}{2}\,c\,\hat{\mathbf{z}}$ | $(1d)$ | P II |
| $\mathbf{B}_4$ | = | $\frac{1}{2}\mathbf{a}_2 + \frac{1}{2}\mathbf{a}_3$ | = | $\frac{1}{2}\,b\,\hat{\mathbf{y}} + \frac{1}{2}\,c\,\hat{\mathbf{z}}$ | $(1g)$ | Al II |
| $\mathbf{B}_5$ | = | $x_5\mathbf{a}_1 + y_5\mathbf{a}_2 + z_5\mathbf{a}_3$ | = | $x_5\,a\,\hat{\mathbf{x}} + y_5\,b\,\hat{\mathbf{y}} + z_5\,c\,\hat{\mathbf{z}}$ | $(4u)$ | S I |
| $\mathbf{B}_6$ | = | $-x_5\mathbf{a}_1 - y_5\mathbf{a}_2 + z_5\mathbf{a}_3$ | = | $-x_5\,a\,\hat{\mathbf{x}} - y_5\,b\,\hat{\mathbf{y}} + z_5\,c\,\hat{\mathbf{z}}$ | $(4u)$ | S I |
| $\mathbf{B}_7$ | = | $-x_5\mathbf{a}_1 + y_5\mathbf{a}_2 - z_5\mathbf{a}_3$ | = | $-x_5\,a\,\hat{\mathbf{x}} + y_5\,b\,\hat{\mathbf{y}} - z_5\,c\,\hat{\mathbf{z}}$ | $(4u)$ | S I |
| $\mathbf{B}_8$ | = | $x_5\mathbf{a}_1 - y_5\mathbf{a}_2 - z_5\mathbf{a}_3$ | = | $x_5\,a\,\hat{\mathbf{x}} - y_5\,b\,\hat{\mathbf{y}} - z_5\,c\,\hat{\mathbf{z}}$ | $(4u)$ | S I |
| $\mathbf{B}_9$ | = | $x_6\mathbf{a}_1 + y_6\mathbf{a}_2 + z_6\mathbf{a}_3$ | = | $x_6\,a\,\hat{\mathbf{x}} + y_6\,b\,\hat{\mathbf{y}} + z_6\,c\,\hat{\mathbf{z}}$ | $(4u)$ | S II |
| $\mathbf{B}_{10}$ | = | $-x_6\mathbf{a}_1 - y_6\mathbf{a}_2 + z_6\mathbf{a}_3$ | = | $-x_6\,a\,\hat{\mathbf{x}} - y_6\,b\,\hat{\mathbf{y}} + z_6\,c\,\hat{\mathbf{z}}$ | $(4u)$ | S II |
| $\mathbf{B}_{11}$ | = | $-x_6\mathbf{a}_1 + y_6\mathbf{a}_2 - z_6\mathbf{a}_3$ | = | $-x_6\,a\,\hat{\mathbf{x}} + y_6\,b\,\hat{\mathbf{y}} - z_6\,c\,\hat{\mathbf{z}}$ | $(4u)$ | S II |
| $\mathbf{B}_{12}$ | = | $x_6\mathbf{a}_1 - y_6\mathbf{a}_2 - z_6\mathbf{a}_3$ | = | $x_6\,a\,\hat{\mathbf{x}} - y_6\,b\,\hat{\mathbf{y}} - z_6\,c\,\hat{\mathbf{z}}$ | $(4u)$ | S II |

**Geometry files:**

- CIF: pp. 654
- POSCAR: pp. 655

# $BaS_3$ Structure: AB3_oP16_18_ab_3c

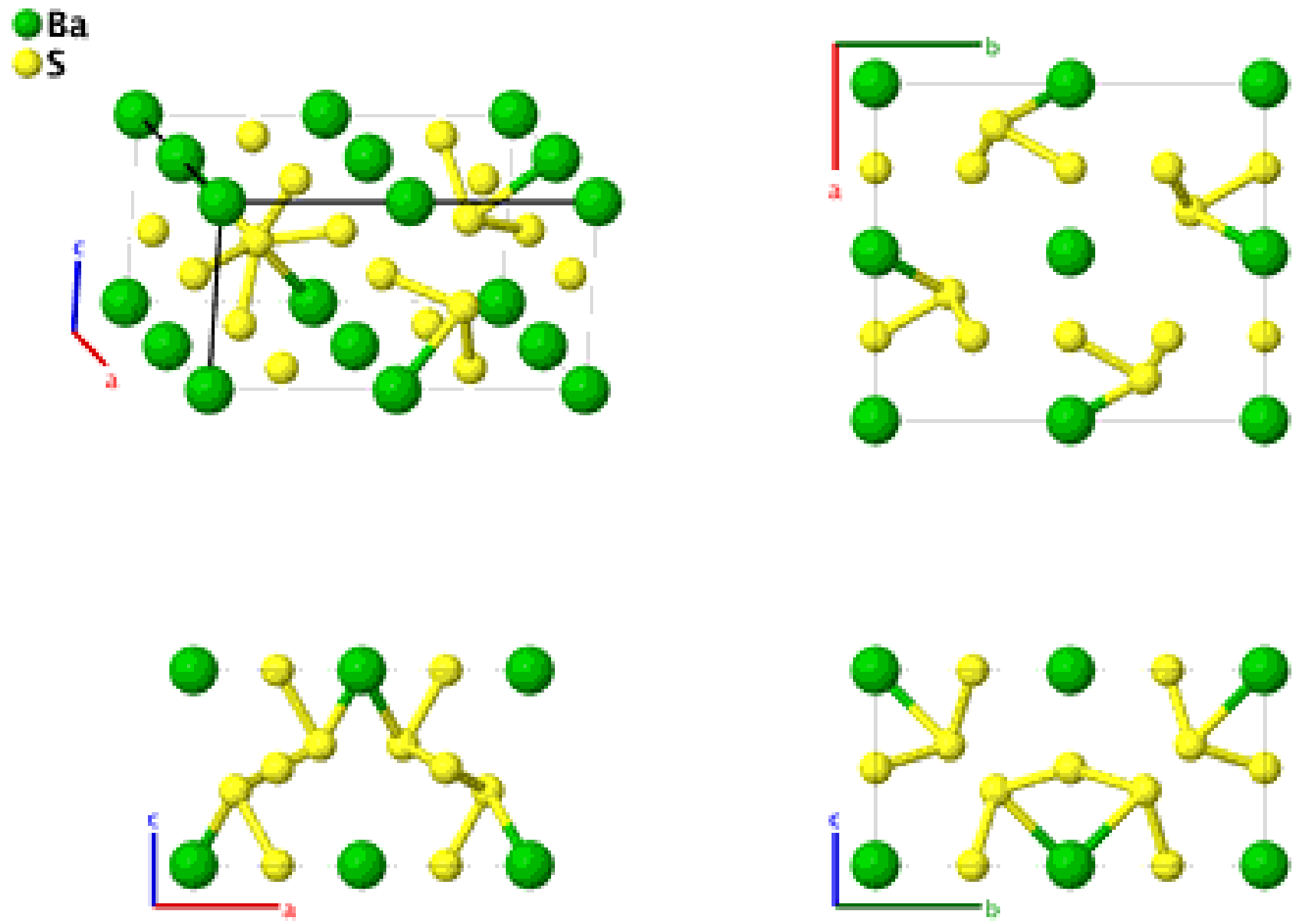

| | | |
|---|---|---|
| **Prototype** | : | $BaS_3$ |
| **AFLOW prototype label** | : | AB3_oP16_18_ab_3c |
| ***Strukturbericht* designation** | : | None |
| **Pearson symbol** | : | oP16 |
| **Space group number** | : | 18 |
| **Space group symbol** | : | $P2_12_12$ |
| **AFLOW prototype command** | : | `aflow --proto=AB3_oP16_18_ab_3c`<br>`--params=`$a, b/a, c/a, z_1, z_2, x_3, y_3, z_3, x_4, y_4, z_4, x_5, y_5, z_5$ |

- Not to be confused with the other $BaS_3$ ($D0_{17}$) structure, which has space group $P\bar{4}2_1m$ (#113).

**Simple Orthorhombic primitive vectors:**

$$\mathbf{a}_1 = a\,\hat{\mathbf{x}}$$
$$\mathbf{a}_2 = b\,\hat{\mathbf{y}}$$
$$\mathbf{a}_3 = c\,\hat{\mathbf{z}}$$

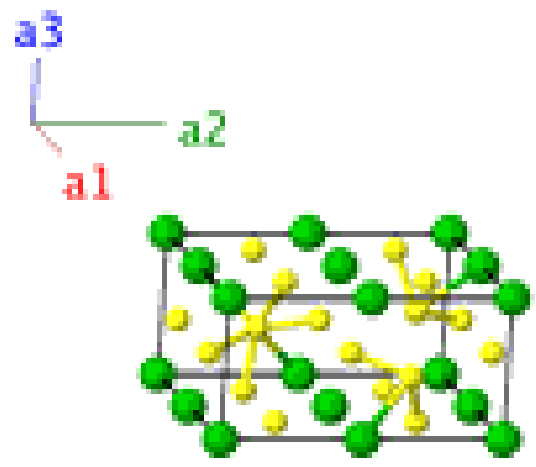

**Basis vectors:**

| | | Lattice Coordinates | | Cartesian Coordinates | Wyckoff Position | Atom Type |
|---|---|---|---|---|---|---|
| $\mathbf{B}_1$ | = | $z_1\,\mathbf{a}_3$ | = | $z_1\,c\,\hat{\mathbf{z}}$ | (2$a$) | Ba I |
| $\mathbf{B}_2$ | = | $\frac{1}{2}\,\mathbf{a}_1 + \frac{1}{2}\,\mathbf{a}_2 - z_1\,\mathbf{a}_3$ | = | $\frac{1}{2}\,a\,\hat{\mathbf{x}} + \frac{1}{2}\,b\,\hat{\mathbf{y}} - z_1\,c\,\hat{\mathbf{z}}$ | (2$a$) | Ba I |
| $\mathbf{B}_3$ | = | $\frac{1}{2}\,\mathbf{a}_2 + z_2\,\mathbf{a}_3$ | = | $\frac{1}{2}\,b\,\hat{\mathbf{y}} + z_2\,c\,\hat{\mathbf{z}}$ | (2$b$) | Ba II |
| $\mathbf{B}_4$ | = | $\frac{1}{2}\,\mathbf{a}_1 - z_2\,\mathbf{a}_3$ | = | $\frac{1}{2}\,a\,\hat{\mathbf{x}} - z_2\,c\,\hat{\mathbf{z}}$ | (2$b$) | Ba II |
| $\mathbf{B}_5$ | = | $x_3\,\mathbf{a}_1 + y_3\,\mathbf{a}_2 + z_3\,\mathbf{a}_3$ | = | $x_3\,a\,\hat{\mathbf{x}} + y_3\,b\,\hat{\mathbf{y}} + z_3\,c\,\hat{\mathbf{z}}$ | (4$c$) | S I |
| $\mathbf{B}_6$ | = | $-x_3\,\mathbf{a}_1 - y_3\,\mathbf{a}_2 + z_3\,\mathbf{a}_3$ | = | $-x_3\,a\,\hat{\mathbf{x}} - y_3\,b\,\hat{\mathbf{y}} + z_3\,c\,\hat{\mathbf{z}}$ | (4$c$) | S I |
| $\mathbf{B}_7$ | = | $\left(\frac{1}{2} - x_3\right)\,\mathbf{a}_1 + \left(\frac{1}{2} + y_3\right)\,\mathbf{a}_2 - z_3\,\mathbf{a}_3$ | = | $\left(\frac{1}{2} - x_3\right)\,a\,\hat{\mathbf{x}} + \left(\frac{1}{2} + y_3\right)\,b\,\hat{\mathbf{y}} - z_3\,c\,\hat{\mathbf{z}}$ | (4$c$) | S I |
| $\mathbf{B}_8$ | = | $\left(\frac{1}{2} + x_3\right)\,\mathbf{a}_1 + \left(\frac{1}{2} - y_3\right)\,\mathbf{a}_2 - z_3\,\mathbf{a}_3$ | = | $\left(\frac{1}{2} + x_3\right)\,a\,\hat{\mathbf{x}} + \left(\frac{1}{2} - y_3\right)\,b\,\hat{\mathbf{y}} - z_3\,c\,\hat{\mathbf{z}}$ | (4$c$) | S I |
| $\mathbf{B}_9$ | = | $x_4\,\mathbf{a}_1 + y_4\,\mathbf{a}_2 + z_4\,\mathbf{a}_3$ | = | $x_4\,a\,\hat{\mathbf{x}} + y_4\,b\,\hat{\mathbf{y}} + z_4\,c\,\hat{\mathbf{z}}$ | (4$c$) | S II |
| $\mathbf{B}_{10}$ | = | $-x_4\,\mathbf{a}_1 - y_4\,\mathbf{a}_2 + z_4\,\mathbf{a}_3$ | = | $-x_4\,a\,\hat{\mathbf{x}} - y_4\,b\,\hat{\mathbf{y}} + z_4\,c\,\hat{\mathbf{z}}$ | (4$c$) | S II |
| $\mathbf{B}_{11}$ | = | $\left(\frac{1}{2} - x_4\right)\,\mathbf{a}_1 + \left(\frac{1}{2} + y_4\right)\,\mathbf{a}_2 - z_4\,\mathbf{a}_3$ | = | $\left(\frac{1}{2} - x_4\right)\,a\,\hat{\mathbf{x}} + \left(\frac{1}{2} + y_4\right)\,b\,\hat{\mathbf{y}} - z_4\,c\,\hat{\mathbf{z}}$ | (4$c$) | S II |
| $\mathbf{B}_{12}$ | = | $\left(\frac{1}{2} + x_4\right)\,\mathbf{a}_1 + \left(\frac{1}{2} - y_4\right)\,\mathbf{a}_2 - z_4\,\mathbf{a}_3$ | = | $\left(\frac{1}{2} + x_4\right)\,a\,\hat{\mathbf{x}} + \left(\frac{1}{2} - y_4\right)\,b\,\hat{\mathbf{y}} - z_4\,c\,\hat{\mathbf{z}}$ | (4$c$) | S II |
| $\mathbf{B}_{13}$ | = | $x_5\,\mathbf{a}_1 + y_5\,\mathbf{a}_2 + z_5\,\mathbf{a}_3$ | = | $x_5\,a\,\hat{\mathbf{x}} + y_5\,b\,\hat{\mathbf{y}} + z_5\,c\,\hat{\mathbf{z}}$ | (4$c$) | S III |
| $\mathbf{B}_{14}$ | = | $-x_5\,\mathbf{a}_1 - y_5\,\mathbf{a}_2 + z_5\,\mathbf{a}_3$ | = | $-x_5\,a\,\hat{\mathbf{x}} - y_5\,b\,\hat{\mathbf{y}} + z_5\,c\,\hat{\mathbf{z}}$ | (4$c$) | S III |
| $\mathbf{B}_{15}$ | = | $\left(\frac{1}{2} - x_5\right)\,\mathbf{a}_1 + \left(\frac{1}{2} + y_5\right)\,\mathbf{a}_2 - z_5\,\mathbf{a}_3$ | = | $\left(\frac{1}{2} - x_5\right)\,a\,\hat{\mathbf{x}} + \left(\frac{1}{2} + y_5\right)\,b\,\hat{\mathbf{y}} - z_5\,c\,\hat{\mathbf{z}}$ | (4$c$) | S III |
| $\mathbf{B}_{16}$ | = | $\left(\frac{1}{2} + x_5\right)\,\mathbf{a}_1 + \left(\frac{1}{2} - y_5\right)\,\mathbf{a}_2 - z_5\,\mathbf{a}_3$ | = | $\left(\frac{1}{2} + x_5\right)\,a\,\hat{\mathbf{x}} + \left(\frac{1}{2} - y_5\right)\,b\,\hat{\mathbf{y}} - z_5\,c\,\hat{\mathbf{z}}$ | (4$c$) | S III |

**Geometry files:**

- CIF: pp. 655
- POSCAR: pp. 655

# Naumannite ($Ag_2Se$) Structure: A2B_oP12_19_2a_a

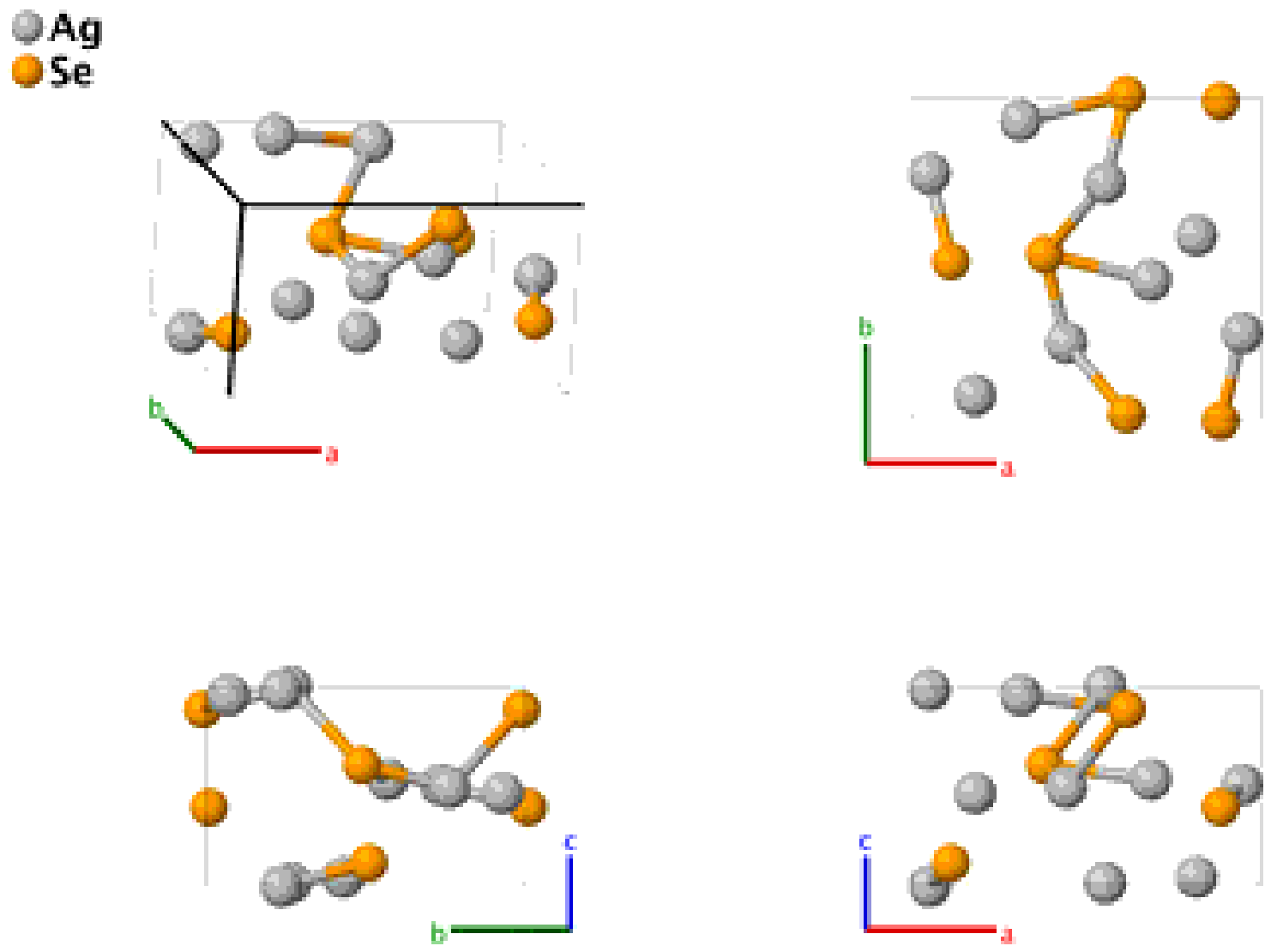

| | | |
|---|---|---|
| **Prototype** | : | $Ag_2Se$ |
| **AFLOW prototype label** | : | A2B_oP12_19_2a_a |
| ***Strukturbericht* designation** | : | None |
| **Pearson symbol** | : | oP12 |
| **Space group number** | : | 19 |
| **Space group symbol** | : | $P2_12_12_1$ |
| **AFLOW prototype command** | : | `aflow --proto=A2B_oP12_19_2a_a`<br>`--params=`$a, b/a, c/a, x_1, y_1, z_1, x_2, y_2, z_2, x_3, y_3, z_3$ |

**Simple Orthorhombic primitive vectors:**

$$\mathbf{a}_1 = a\,\hat{\mathbf{x}}$$
$$\mathbf{a}_2 = b\,\hat{\mathbf{y}}$$
$$\mathbf{a}_3 = c\,\hat{\mathbf{z}}$$

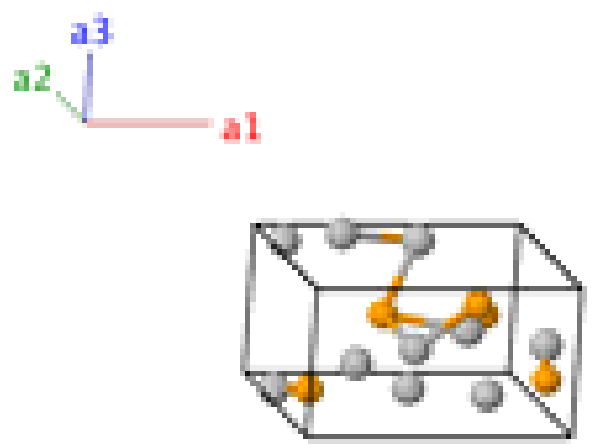

**Basis vectors:**

| | | Lattice Coordinates | | Cartesian Coordinates | Wyckoff Position | Atom Type |
|---|---|---|---|---|---|---|
| $\mathbf{B}_1$ | = | $x_1\,\mathbf{a}_1 + y_1\,\mathbf{a}_2 + z_1\,\mathbf{a}_3$ | = | $x_1\,a\,\hat{\mathbf{x}} + y_1\,b\,\hat{\mathbf{y}} + z_1\,c\,\hat{\mathbf{z}}$ | (4a) | Ag I |
| $\mathbf{B}_2$ | = | $\left(\frac{1}{2} - x_1\right)\mathbf{a}_1 - y_1\,\mathbf{a}_2 + \left(\frac{1}{2} + z_1\right)\mathbf{a}_3$ | = | $\left(\frac{1}{2} - x_1\right)a\,\hat{\mathbf{x}} - y_1\,b\,\hat{\mathbf{y}} + \left(\frac{1}{2} + z_1\right)c\,\hat{\mathbf{z}}$ | (4a) | Ag I |
| $\mathbf{B}_3$ | = | $-x_1\,\mathbf{a}_1 + \left(\frac{1}{2} + y_1\right)\mathbf{a}_2 + \left(\frac{1}{2} - z_1\right)\mathbf{a}_3$ | = | $-x_1\,a\,\hat{\mathbf{x}} + \left(\frac{1}{2} + y_1\right)b\,\hat{\mathbf{y}} + \left(\frac{1}{2} - z_1\right)c\,\hat{\mathbf{z}}$ | (4a) | Ag I |
| $\mathbf{B}_4$ | = | $\left(\frac{1}{2} + x_1\right)\mathbf{a}_1 + \left(\frac{1}{2} - y_1\right)\mathbf{a}_2 - z_1\,\mathbf{a}_3$ | = | $\left(\frac{1}{2} + x_1\right)a\,\hat{\mathbf{x}} + \left(\frac{1}{2} - y_1\right)b\,\hat{\mathbf{y}} - z_1\,c\,\hat{\mathbf{z}}$ | (4a) | Ag I |
| $\mathbf{B}_5$ | = | $x_2\,\mathbf{a}_1 + y_2\,\mathbf{a}_2 + z_2\,\mathbf{a}_3$ | = | $x_2\,a\,\hat{\mathbf{x}} + y_2\,b\,\hat{\mathbf{y}} + z_2\,c\,\hat{\mathbf{z}}$ | (4a) | Ag II |
| $\mathbf{B_6}$ | = | $\left(\frac{1}{2} - x_2\right)\mathbf{a_1} - y_2\,\mathbf{a_2} + \left(\frac{1}{2} + z_2\right)\mathbf{a_3}$ | = | $\left(\frac{1}{2} - x_2\right) a\,\hat{\mathbf{x}} - y_2\,b\,\hat{\mathbf{y}} + \left(\frac{1}{2} + z_2\right) c\,\hat{\mathbf{z}}$ | (4$a$) | Ag II |
| $\mathbf{B_7}$ | = | $-x_2\,\mathbf{a_1} + \left(\frac{1}{2} + y_2\right)\mathbf{a_2} + \left(\frac{1}{2} - z_2\right)\mathbf{a_3}$ | = | $-x_2\,a\,\hat{\mathbf{x}} + \left(\frac{1}{2} + y_2\right) b\,\hat{\mathbf{y}} + \left(\frac{1}{2} - z_2\right) c\,\hat{\mathbf{z}}$ | (4$a$) | Ag II |
| $\mathbf{B_8}$ | = | $\left(\frac{1}{2} + x_2\right)\mathbf{a_1} + \left(\frac{1}{2} - y_2\right)\mathbf{a_2} - z_2\,\mathbf{a_3}$ | = | $\left(\frac{1}{2} + x_2\right) a\,\hat{\mathbf{x}} + \left(\frac{1}{2} - y_2\right) b\,\hat{\mathbf{y}} - z_2\,c\,\hat{\mathbf{z}}$ | (4$a$) | Ag II |
| $\mathbf{B_9}$ | = | $x_3\,\mathbf{a_1} + y_3\,\mathbf{a_2} + z_3\,\mathbf{a_3}$ | = | $x_3\,a\,\hat{\mathbf{x}} + y_3\,b\,\hat{\mathbf{y}} + z_3\,c\,\hat{\mathbf{z}}$ | (4$a$) | Se |
| $\mathbf{B_{10}}$ | = | $\left(\frac{1}{2} - x_3\right)\mathbf{a_1} - y_3\,\mathbf{a_2} + \left(\frac{1}{2} + z_3\right)\mathbf{a_3}$ | = | $\left(\frac{1}{2} - x_3\right) a\,\hat{\mathbf{x}} - y_3\,b\,\hat{\mathbf{y}} + \left(\frac{1}{2} + z_3\right) c\,\hat{\mathbf{z}}$ | (4$a$) | Se |
| $\mathbf{B_{11}}$ | = | $-x_3\,\mathbf{a_1} + \left(\frac{1}{2} + y_3\right)\mathbf{a_2} + \left(\frac{1}{2} - z_3\right)\mathbf{a_3}$ | = | $-x_3\,a\,\hat{\mathbf{x}} + \left(\frac{1}{2} + y_3\right) b\,\hat{\mathbf{y}} + \left(\frac{1}{2} - z_3\right) c\,\hat{\mathbf{z}}$ | (4$a$) | Se |
| $\mathbf{B_{12}}$ | = | $\left(\frac{1}{2} + x_3\right)\mathbf{a_1} + \left(\frac{1}{2} - y_3\right)\mathbf{a_2} - z_3\,\mathbf{a_3}$ | = | $\left(\frac{1}{2} + x_3\right) a\,\hat{\mathbf{x}} + \left(\frac{1}{2} - y_3\right) b\,\hat{\mathbf{y}} - z_3\,c\,\hat{\mathbf{z}}$ | (4$a$) | Se |

---

**Found in:**

---

**Geometry files:**

- CIF: pp. 655
- POSCAR: pp. 655

# Orthorhombic Tridymite ($SiO_2$) Structure: A2B_oC24_20_abc_c

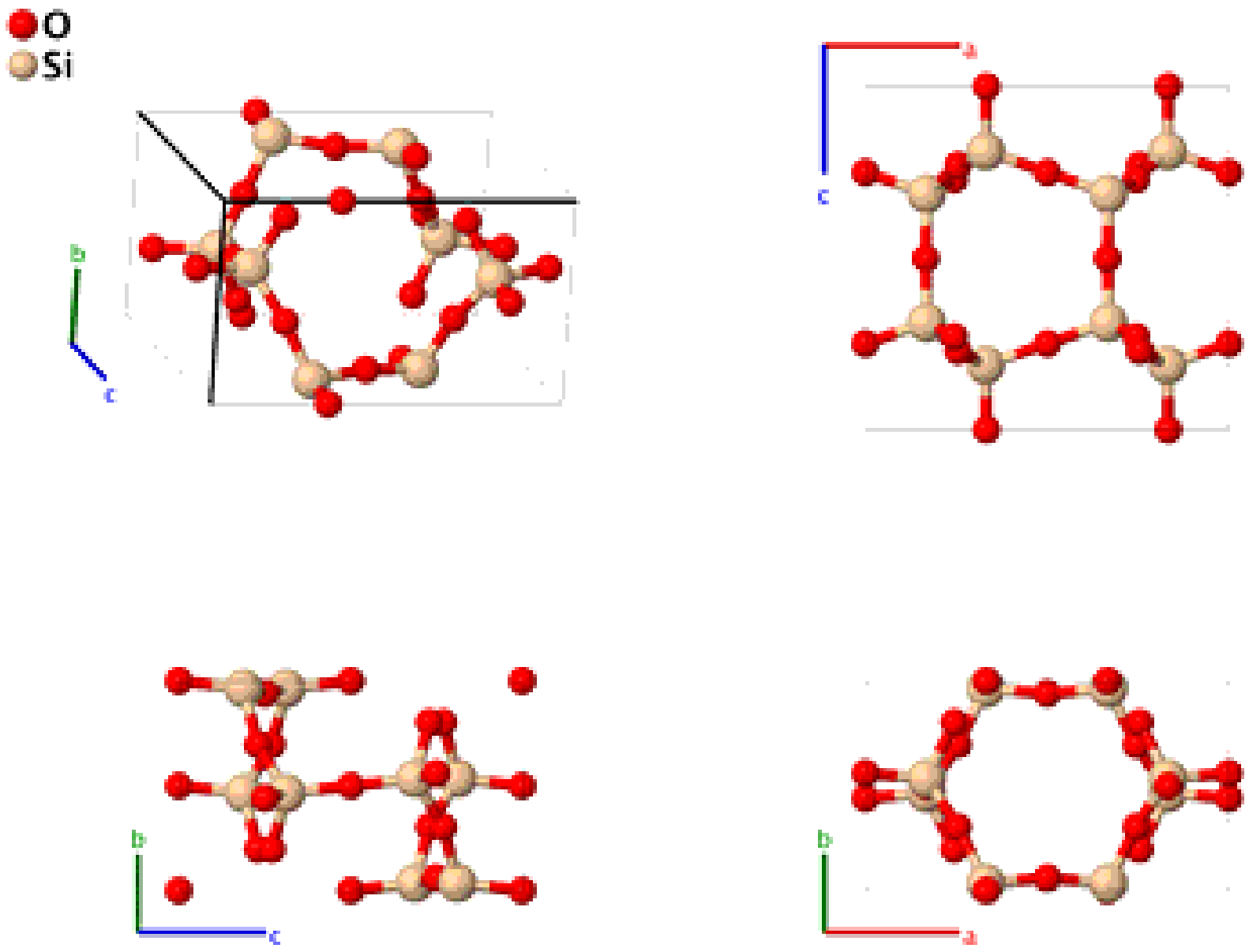

| | | |
|---|---|---|
| **Prototype** | : | $SiO_2$ |
| **AFLOW prototype label** | : | A2B_oC24_20_abc_c |
| ***Strukturbericht* designation** | : | None |
| **Pearson symbol** | : | oC24 |
| **Space group number** | : | 20 |
| **Space group symbol** | : | $C222_1$ |
| **AFLOW prototype command** | : | `aflow --proto=A2B_oC24_20_abc_c`<br>`--params=`$a, b/a, c/a, x_1, y_2, x_3, y_3, z_3, x_4, y_4, z_4$ |

**Base-centered Orthorhombic primitive vectors:**

$$\mathbf{a}_1 = \frac{1}{2}\,a\,\hat{\mathbf{x}} - \frac{1}{2}\,b\,\hat{\mathbf{y}}$$
$$\mathbf{a}_2 = \frac{1}{2}\,a\,\hat{\mathbf{x}} + \frac{1}{2}\,b\,\hat{\mathbf{y}}$$
$$\mathbf{a}_3 = c\,\hat{\mathbf{z}}$$

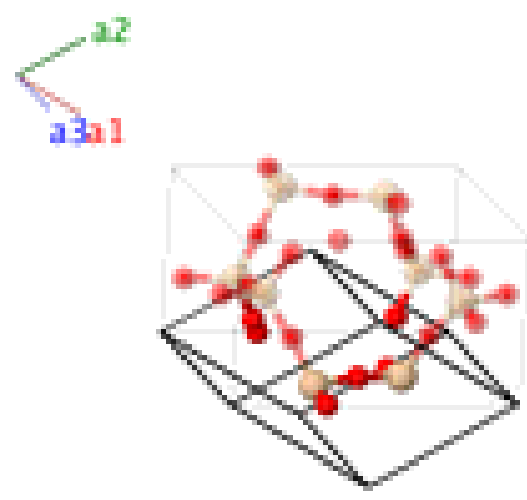

**Basis vectors:**

| | | Lattice Coordinates | | Cartesian Coordinates | Wyckoff Position | Atom Type |
|---|---|---|---|---|---|---|
| $\mathbf{B_1}$ | = | $x_1\,\mathbf{a_1} + x_1\,\mathbf{a_2}$ | = | $x_1\,a\,\hat{\mathbf{x}}$ | (4$a$) | O I |
| $\mathbf{B_2}$ | = | $-x_1\,\mathbf{a_1} - x_1\,\mathbf{a_2} + \frac{1}{2}\,\mathbf{a_3}$ | = | $-x_1\,a\,\hat{\mathbf{x}} + \frac{1}{2}\,c\hat{\mathbf{z}}$ | (4$a$) | O I |
| $\mathbf{B_3}$ | = | $-y_2\,\mathbf{a_1} + y_2\,\mathbf{a_2} + \frac{1}{4}\,\mathbf{a_3}$ | = | $y_2\,b\,\hat{\mathbf{y}} + \frac{1}{4}\,c\hat{\mathbf{z}}$ | (4$b$) | O II |
| $\mathbf{B_4}$ | = | $y_2\,\mathbf{a_1} - y_2\,\mathbf{a_2} + \frac{3}{4}\,\mathbf{a_3}$ | = | $-y_2\,b\,\hat{\mathbf{y}} + \frac{3}{4}\,c\hat{\mathbf{z}}$ | (4$b$) | O II |
| $\mathbf{B_5}$ | = | $(x_3 - y_3)\,\mathbf{a_1} + (x_3 + y_3)\,\mathbf{a_2} + z_3\,\mathbf{a_3}$ | = | $x_3\,a\hat{\mathbf{x}} + y_3\,b\,\hat{\mathbf{y}} + z_3\,c\hat{\mathbf{z}}$ | (8$c$) | O III |
| $\mathbf{B_6}$ | = | $(y_3 - x_3)\,\mathbf{a_1} - (x_3 + y_3)\,\mathbf{a_2} + \left(\frac{1}{2} + z_3\right)\,\mathbf{a_3}$ | = | $-x_3\,a\hat{\mathbf{x}} - y_3\,b\,\hat{\mathbf{y}} + \left(\frac{1}{2} + z_3\right)\,c\hat{\mathbf{z}}$ | (8$c$) | O III |
| $\mathbf{B_7}$ | = | $-(x_3 + y_3)\,\mathbf{a_1} + (y_3 - x_3)\,\mathbf{a_2} + \left(\frac{1}{2} - z_3\right)\,\mathbf{a_3}$ | = | $-x_3\,a\hat{\mathbf{x}} + y_3\,b\,\hat{\mathbf{y}} + \left(\frac{1}{2} - z_3\right)\,c\hat{\mathbf{z}}$ | (8$c$) | O III |
| $\mathbf{B_8}$ | = | $(x_3 + y_3)\,\mathbf{a_1} + (x_3 - y_3)\,\mathbf{a_2} - z_3\,\mathbf{a_3}$ | = | $x_3\,a\hat{\mathbf{x}} - y_3\,b\,\hat{\mathbf{y}} - z_3\,c\hat{\mathbf{z}}$ | (8$c$) | O III |
| $\mathbf{B_9}$ | = | $(x_4 - y_4)\,\mathbf{a_1} + (x_4 + y_4)\,\mathbf{a_2} + z_4\,\mathbf{a_3}$ | = | $x_4\,a\hat{\mathbf{x}} + y_4\,b\,\hat{\mathbf{y}} + z_4\,c\hat{\mathbf{z}}$ | (8$c$) | Si |
| $\mathbf{B_{10}}$ | = | $(y_4 - x_4)\,\mathbf{a_1} - (x_4 + y_4)\,\mathbf{a_2} + \left(\frac{1}{2} + z_4\right)\,\mathbf{a_3}$ | = | $-x_4\,a\hat{\mathbf{x}} - y_4\,b\,\hat{\mathbf{y}} + \left(\frac{1}{2} + z_4\right)\,c\hat{\mathbf{z}}$ | (8$c$) | Si |
| $\mathbf{B_{11}}$ | = | $-(x_4 + y_4)\,\mathbf{a_1} + (y_4 - x_4)\,\mathbf{a_2} + \left(\frac{1}{2} - z_4\right)\,\mathbf{a_3}$ | = | $-x_4\,a\hat{\mathbf{x}} + y_4\,b\,\hat{\mathbf{y}} + \left(\frac{1}{2} - z_4\right)\,c\hat{\mathbf{z}}$ | (8$c$) | Si |
| $\mathbf{B_{12}}$ | = | $(x_4 + y_4)\,\mathbf{a_1} + (x_4 - y_4)\,\mathbf{a_2} - z_4\,\mathbf{a_3}$ | = | $x_4\,a\hat{\mathbf{x}} - y_4\,b\,\hat{\mathbf{y}} - z_4\,c\hat{\mathbf{z}}$ | (8$c$) | Si |

**Geometry files:**

- CIF: pp. 656
- POSCAR: pp. 656

# High-Pressure CdTe Structure: AB_oP2_25_b_a

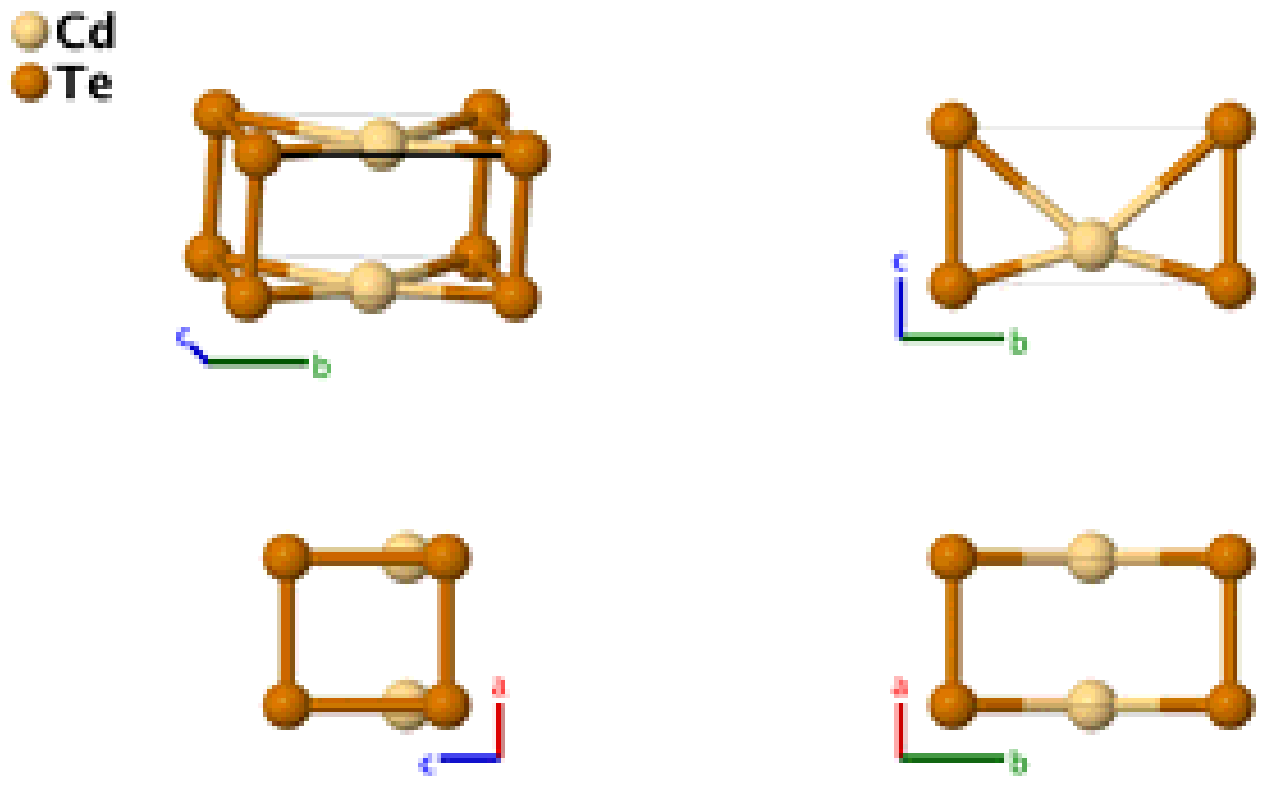

| | | |
|---|---|---|
| **Prototype** | : | CdTe |
| **AFLOW prototype label** | : | AB_oP2_25_b_a |
| ***Strukturbericht* designation** | : | None |
| **Pearson symbol** | : | oP2 |
| **Space group number** | : | 25 |
| **Space group symbol** | : | Pmm2 |
| **AFLOW prototype command** | : | `aflow --proto=AB_oP2_25_b_a` `--params=`$a, b/a, c/a, z_1, z_2$ |

- This is a high-pressure phase of CdTe. We use the data given for a pressure of 19.3 GPa.

**Simple Orthorhombic primitive vectors:**

$$\mathbf{a}_1 = a\,\hat{\mathbf{x}}$$
$$\mathbf{a}_2 = b\,\hat{\mathbf{y}}$$
$$\mathbf{a}_3 = c\,\hat{\mathbf{z}}$$

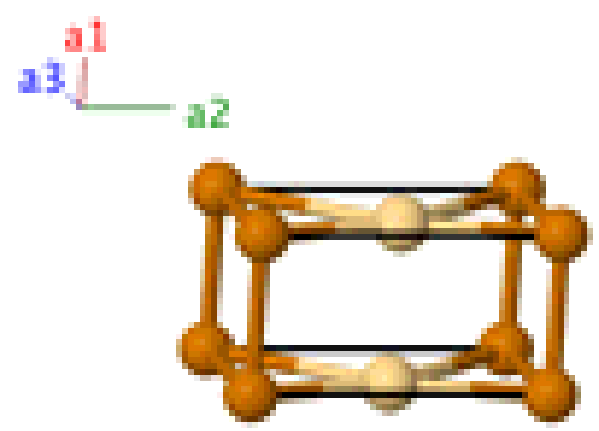

**Basis vectors:**

| | | Lattice Coordinates | | Cartesian Coordinates | Wyckoff Position | Atom Type |
|---|---|---|---|---|---|---|
| $\mathbf{B}_1$ | = | $z_1\,\mathbf{a}_3$ | = | $z_1\,c\,\hat{\mathbf{z}}$ | (1$a$) | Te |
| $\mathbf{B}_2$ | = | $\frac{1}{2}\,\mathbf{a}_2 + z_2\,\mathbf{a}_3$ | = | $\frac{1}{2}\,b\,\hat{\mathbf{y}} + z_2\,c\,\hat{\mathbf{z}}$ | (1$b$) | Cd |

**Geometry files:**

- CIF: pp. 656
- POSCAR: pp. 656

# Krennerite ($AuTe_2$, C46) Structure: AB2_oP24_28_acd_2c3d

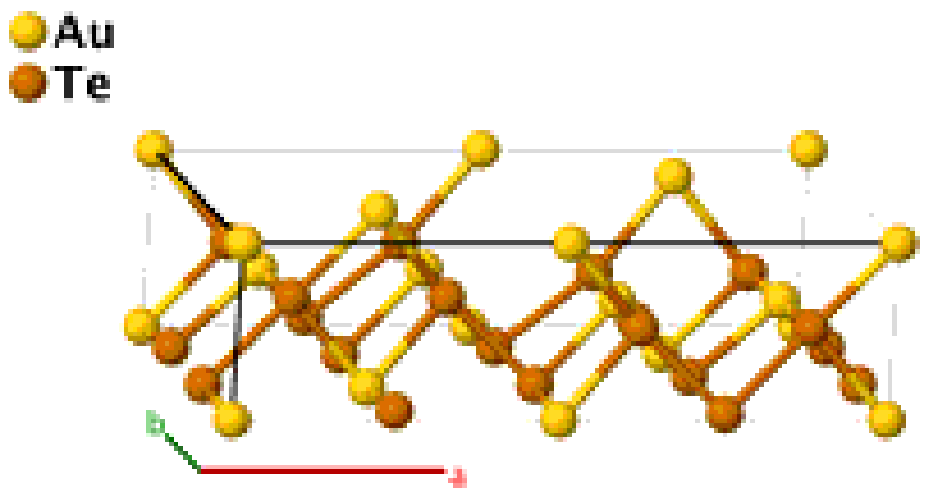

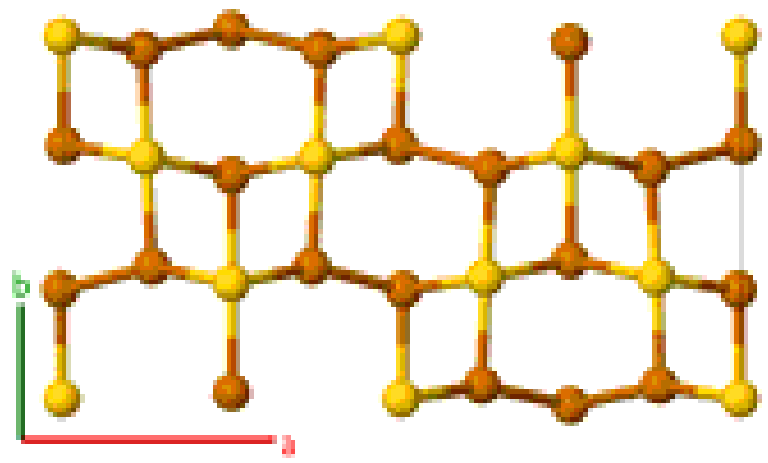

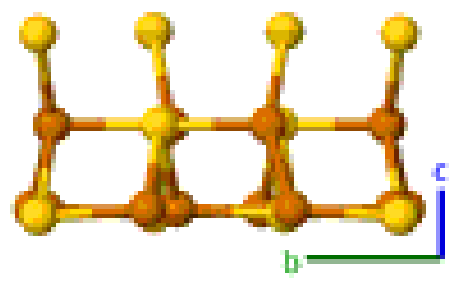

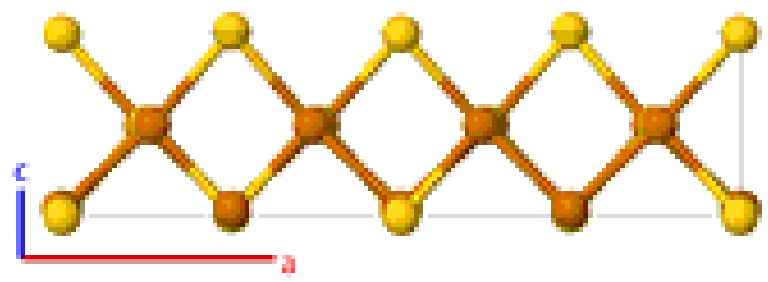

| | | |
|---|---|---|
| **Prototype** | : | $AuTe_2$ |
| **AFLOW prototype label** | : | AB2_oP24_28_acd_2c3d |
| ***Strukturbericht* designation** | : | C46 |
| **Pearson symbol** | : | oP24 |
| **Space group number** | : | 28 |
| **Space group symbol** | : | Pma2 |
| **AFLOW prototype command** | : | `aflow --proto=AB2_oP24_28_acd_2c3d` <br> `--params=`$a, b/a, c/a, z_1, y_2, z_2, y_3, z_3, y_4, z_4, x_5, y_5, z_5, x_6, y_6, z_6, x_7, y_7, z_7, x_8, y_8, z_8$ |

- The sample studied had composition ($Au_{0.88}$,$Ag_{0.12}$)$Te_2$. For simplicity we make all of the Au/Ag sites Au. (Pearson, 1972) states that this is a distortion of the trigonal $\omega$ phase. Note that $AuTe_2$ also exists in the C34 structure.

**Simple Orthorhombic primitive vectors:**

$$\mathbf{a}_1 = a\,\hat{\mathbf{x}}$$
$$\mathbf{a}_2 = b\,\hat{\mathbf{y}}$$
$$\mathbf{a}_3 = c\,\hat{\mathbf{z}}$$

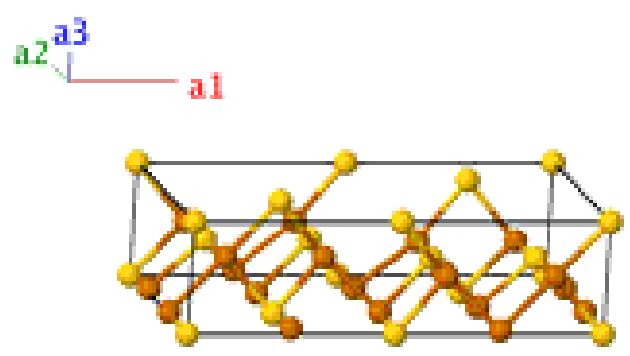

**Basis vectors:**

| | | Lattice Coordinates | | Cartesian Coordinates | Wyckoff Position | Atom Type |
|---|---|---|---|---|---|---|
| $\mathbf{B_1}$ | = | $z_1\,\mathbf{a_3}$ | = | $z_1\,c\,\hat{\mathbf{z}}$ | (2$a$) | Au I |
| $\mathbf{B_2}$ | = | $\frac{1}{2}\,\mathbf{a_1} + z_1\,\mathbf{a_3}$ | = | $\frac{1}{2}\,a\,\hat{\mathbf{x}} + z_1\,c\,\hat{\mathbf{z}}$ | (2$a$) | Au I |
| $\mathbf{B_3}$ | = | $\frac{1}{4}\,\mathbf{a_1} + y_2\,\mathbf{a_2} + z_2\,\mathbf{a_3}$ | = | $\frac{1}{4}\,a\,\hat{\mathbf{x}} + y_2\,b\,\hat{\mathbf{y}} + z_2\,c\,\hat{\mathbf{z}}$ | (2$c$) | Au II |
| $\mathbf{B_4}$ | = | $\frac{3}{4}\,\mathbf{a_1} - y_2\,\mathbf{a_2} + z_2\,\mathbf{a_3}$ | = | $\frac{3}{4}\,a\,\hat{\mathbf{x}} - y_2\,b\,\hat{\mathbf{y}} + z_2\,c\,\hat{\mathbf{z}}$ | (2$c$) | Au II |
| $\mathbf{B_5}$ | = | $\frac{1}{4}\,\mathbf{a_1} + y_3\,\mathbf{a_2} + z_3\,\mathbf{a_3}$ | = | $\frac{1}{4}\,a\,\hat{\mathbf{x}} + y_3\,b\,\hat{\mathbf{y}} + z_3\,c\,\hat{\mathbf{z}}$ | (2$c$) | Te I |
| $\mathbf{B}_6$ | = | $\frac{3}{4}\,\mathbf{a}_1 - y_3\,\mathbf{a}_2 + z_3\,\mathbf{a}_3$ | = | $\frac{3}{4}\,a\,\hat{\mathbf{x}} - y_3\,b\,\hat{\mathbf{y}} + z_3\,c\,\hat{\mathbf{z}}$ | $(2c)$ | Te I |
| $\mathbf{B}_7$ | = | $\frac{1}{4}\,\mathbf{a}_1 + y_4\,\mathbf{a}_2 + z_4\,\mathbf{a}_3$ | = | $\frac{1}{4}\,a\,\hat{\mathbf{x}} + y_4\,b\,\hat{\mathbf{y}} + z_4\,c\,\hat{\mathbf{z}}$ | $(2c)$ | Te II |
| $\mathbf{B}_8$ | = | $\frac{3}{4}\,\mathbf{a}_1 - y_4\,\mathbf{a}_2 + z_4\,\mathbf{a}_3$ | = | $\frac{3}{4}\,a\,\hat{\mathbf{x}} - y_4\,b\,\hat{\mathbf{y}} + z_4\,c\,\hat{\mathbf{z}}$ | $(2c)$ | Te II |
| $\mathbf{B}_9$ | = | $x_5\,\mathbf{a}_1 + y_5\,\mathbf{a}_2 + z_5\,\mathbf{a}_3$ | = | $x_5\,a\,\hat{\mathbf{x}} + y_5\,b\,\hat{\mathbf{y}} + z_5\,c\,\hat{\mathbf{z}}$ | $(4d)$ | Au III |
| $\mathbf{B}_{10}$ | = | $-x_5\,\mathbf{a}_1 - y_5\,\mathbf{a}_2 + z_5\,\mathbf{a}_3$ | = | $-x_5\,a\,\hat{\mathbf{x}} - y_5\,b\,\hat{\mathbf{y}} + z_5\,c\,\hat{\mathbf{z}}$ | $(4d)$ | Au III |
| $\mathbf{B}_{11}$ | = | $\left(\frac{1}{2} + x_5\right)\,\mathbf{a}_1 - y_5\,\mathbf{a}_2 + z_5\,\mathbf{a}_3$ | = | $\left(\frac{1}{2} + x_5\right)\,a\,\hat{\mathbf{x}} - y_5\,b\,\hat{\mathbf{y}} + z_5\,c\,\hat{\mathbf{z}}$ | $(4d)$ | Au III |
| $\mathbf{B}_{12}$ | = | $\left(\frac{1}{2} - x_5\right)\,\mathbf{a}_1 + y_5\,\mathbf{a}_2 + z_5\,\mathbf{a}_3$ | = | $\left(\frac{1}{2} - x_5\right)\,a\,\hat{\mathbf{x}} + y_5\,b\,\hat{\mathbf{y}} + z_5\,c\,\hat{\mathbf{z}}$ | $(4d)$ | Au III |
| $\mathbf{B}_{13}$ | = | $x_6\,\mathbf{a}_1 + y_6\,\mathbf{a}_2 + z_6\,\mathbf{a}_3$ | = | $x_6\,a\,\hat{\mathbf{x}} + y_6\,b\,\hat{\mathbf{y}} + z_6\,c\,\hat{\mathbf{z}}$ | $(4d)$ | Te III |
| $\mathbf{B}_{14}$ | = | $-x_6\,\mathbf{a}_1 - y_6\,\mathbf{a}_2 + z_6\,\mathbf{a}_3$ | = | $-x_6\,a\,\hat{\mathbf{x}} - y_6\,b\,\hat{\mathbf{y}} + z_6\,c\,\hat{\mathbf{z}}$ | $(4d)$ | Te III |
| $\mathbf{B}_{15}$ | = | $\left(\frac{1}{2} + x_6\right)\,\mathbf{a}_1 - y_6\,\mathbf{a}_2 + z_6\,\mathbf{a}_3$ | = | $\left(\frac{1}{2} + x_6\right)\,a\,\hat{\mathbf{x}} - y_6\,b\,\hat{\mathbf{y}} + z_6\,c\,\hat{\mathbf{z}}$ | $(4d)$ | Te III |
| $\mathbf{B}_{16}$ | = | $\left(\frac{1}{2} - x_6\right)\,\mathbf{a}_1 + y_6\,\mathbf{a}_2 + z_6\,\mathbf{a}_3$ | = | $\left(\frac{1}{2} - x_6\right)\,a\,\hat{\mathbf{x}} + y_6\,b\,\hat{\mathbf{y}} + z_6\,c\,\hat{\mathbf{z}}$ | $(4d)$ | Te III |
| $\mathbf{B}_{17}$ | = | $x_7\,\mathbf{a}_1 + y_7\,\mathbf{a}_2 + z_7\,\mathbf{a}_3$ | = | $x_7\,a\,\hat{\mathbf{x}} + y_7\,b\,\hat{\mathbf{y}} + z_7\,c\,\hat{\mathbf{z}}$ | $(4d)$ | Te IV |
| $\mathbf{B}_{18}$ | = | $-x_7\,\mathbf{a}_1 - y_7\,\mathbf{a}_2 + z_7\,\mathbf{a}_3$ | = | $-x_7\,a\,\hat{\mathbf{x}} - y_7\,b\,\hat{\mathbf{y}} + z_7\,c\,\hat{\mathbf{z}}$ | $(4d)$ | Te IV |
| $\mathbf{B}_{19}$ | = | $\left(\frac{1}{2} + x_7\right)\,\mathbf{a}_1 - y_7\,\mathbf{a}_2 + z_7\,\mathbf{a}_3$ | = | $\left(\frac{1}{2} + x_7\right)\,a\,\hat{\mathbf{x}} - y_7\,b\,\hat{\mathbf{y}} + z_7\,c\,\hat{\mathbf{z}}$ | $(4d)$ | Te IV |
| $\mathbf{B}_{20}$ | = | $\left(\frac{1}{2} - x_7\right)\,\mathbf{a}_1 + y_7\,\mathbf{a}_2 + z_7\,\mathbf{a}_3$ | = | $\left(\frac{1}{2} - x_7\right)\,a\,\hat{\mathbf{x}} + y_7\,b\,\hat{\mathbf{y}} + z_7\,c\,\hat{\mathbf{z}}$ | $(4d)$ | Te IV |
| $\mathbf{B}_{21}$ | = | $x_8\,\mathbf{a}_1 + y_8\,\mathbf{a}_2 + z_8\,\mathbf{a}_3$ | = | $x_8\,a\,\hat{\mathbf{x}} + y_8\,b\,\hat{\mathbf{y}} + z_8\,c\,\hat{\mathbf{z}}$ | $(4d)$ | Te V |
| $\mathbf{B}_{22}$ | = | $-x_8\,\mathbf{a}_1 - y_8\,\mathbf{a}_2 + z_8\,\mathbf{a}_3$ | = | $-x_8\,a\,\hat{\mathbf{x}} - y_8\,b\,\hat{\mathbf{y}} + z_8\,c\,\hat{\mathbf{z}}$ | $(4d)$ | Te V |
| $\mathbf{B}_{23}$ | = | $\left(\frac{1}{2} + x_8\right)\,\mathbf{a}_1 - y_8\,\mathbf{a}_2 + z_8\,\mathbf{a}_3$ | = | $\left(\frac{1}{2} + x_8\right)\,a\,\hat{\mathbf{x}} - y_8\,b\,\hat{\mathbf{y}} + z_8\,c\,\hat{\mathbf{z}}$ | $(4d)$ | Te V |
| $\mathbf{B}_{24}$ | = | $\left(\frac{1}{2} - x_8\right)\,\mathbf{a}_1 + y_8\,\mathbf{a}_2 + z_8\,\mathbf{a}_3$ | = | $\left(\frac{1}{2} - x_8\right)\,a\,\hat{\mathbf{x}} + y_8\,b\,\hat{\mathbf{y}} + z_8\,c\,\hat{\mathbf{z}}$ | $(4d)$ | Te V |

**Geometry files:**

- CIF: pp. 656
- POSCAR: pp. 657

# Enargite ($AsCu_3S_4$, $H2_5$) Structure: AB3C4_oP16_31_a_ab_2ab

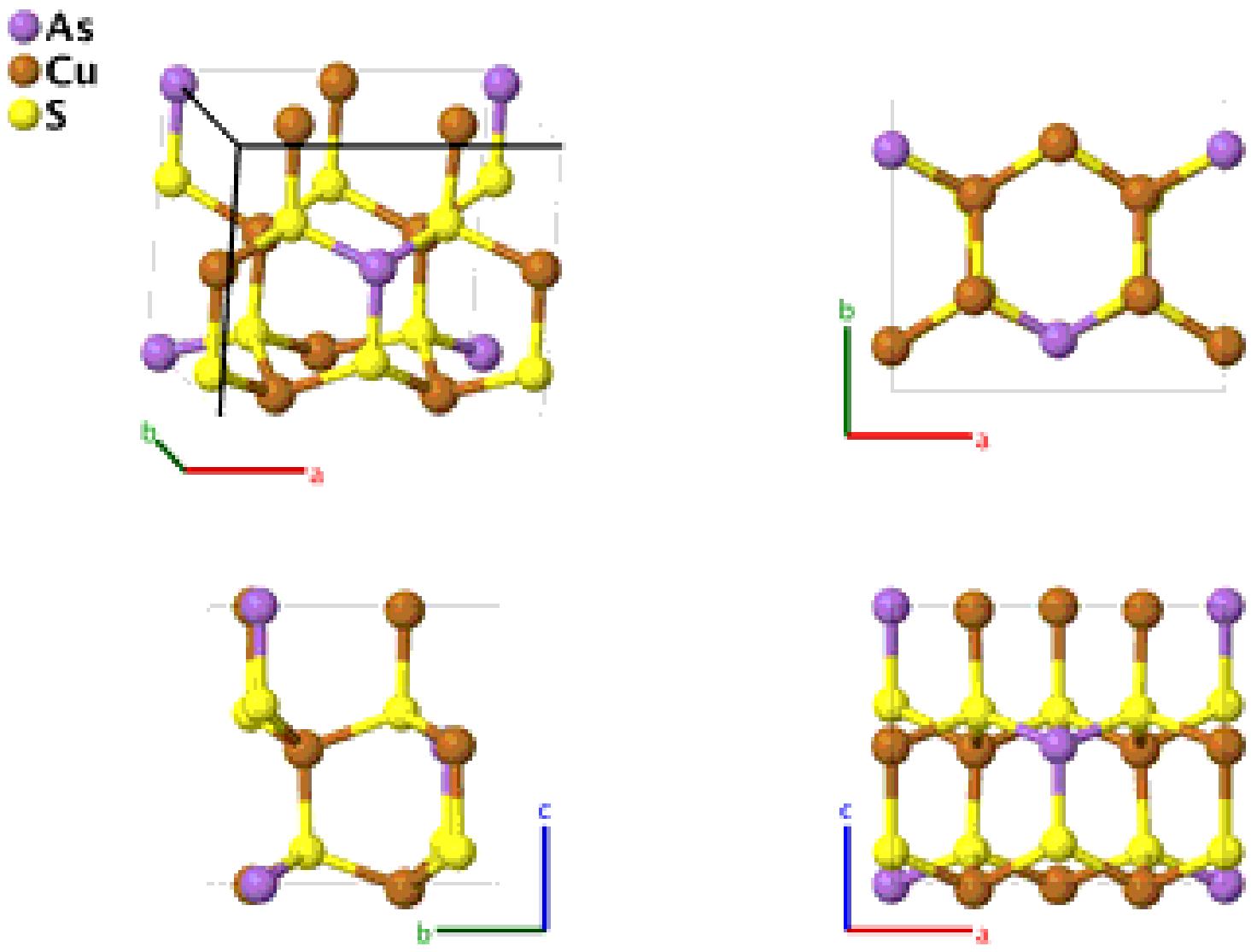

**Prototype** : $AsCu_3S_4$

**AFLOW prototype label** : AB3C4_oP16_31_a_ab_2ab

***Strukturbericht* designation** : $H2_5$

**Pearson symbol** : oP16

**Space group number** : 31

**Space group symbol** : $Pmn2_1$

**AFLOW prototype command** : `aflow --proto=AB3C4_oP16_31_a_ab_2ab`
`--params=`$a, b/a, c/a, y_1, z_1, y_2, z_2, y_3, z_3, y_4, z_4, x_5, y_5, z_5, x_6, y_6, z_6$

---

- This structure should not be confused with the lazarevićite form of $AsCu_3S_4$, which is related to an $sp^3$ cubic structure.

---

**Simple Orthorhombic primitive vectors:**

$$\mathbf{a}_1 = a\,\hat{\mathbf{x}}$$
$$\mathbf{a}_2 = b\,\hat{\mathbf{y}}$$
$$\mathbf{a}_3 = c\,\hat{\mathbf{z}}$$

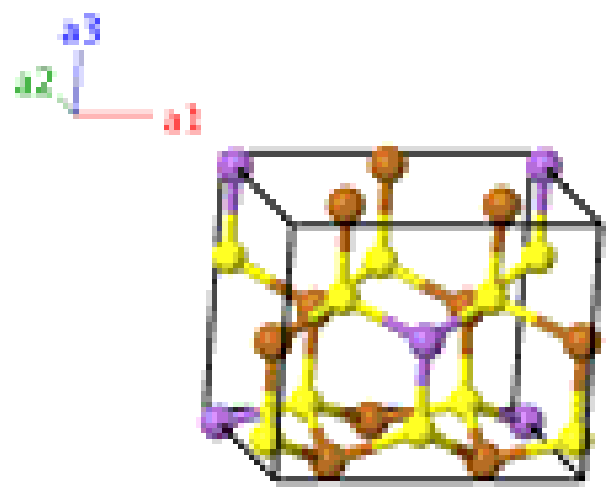

---

**Basis vectors:**

| | | Lattice Coordinates | | Cartesian Coordinates | Wyckoff Position | Atom Type |
|---|---|---|---|---|---|---|
| $\mathbf{B}_1$ | = | $y_1\,\mathbf{a}_2 + z_1\,\mathbf{a}_3$ | = | $y_1\,b\,\hat{\mathbf{y}} + z_1\,c\,\hat{\mathbf{z}}$ | (2$a$) | As |
| $\mathbf{B}_2$ | = | $\frac{1}{2}\,\mathbf{a}_1 - y_1\,\mathbf{a}_2 + \left(\frac{1}{2} + z_1\right)\,\mathbf{a}_3$ | = | $\frac{1}{2}\,a\,\hat{\mathbf{x}} - y_1\,b\,\hat{\mathbf{y}} + \left(\frac{1}{2} + z_1\right)\,c\,\hat{\mathbf{z}}$ | (2$a$) | As |
| $\mathbf{B}_3$ | = | $y_2\,\mathbf{a}_2 + z_2\,\mathbf{a}_3$ | = | $y_2\,b\,\hat{\mathbf{y}} + z_2\,c\,\hat{\mathbf{z}}$ | (2$a$) | Cu I |
| $\mathbf{B}_4$ | = | $\frac{1}{2}\,\mathbf{a}_1 - y_2\,\mathbf{a}_2 + \left(\frac{1}{2} + z_2\right)\,\mathbf{a}_3$ | = | $\frac{1}{2}\,a\,\hat{\mathbf{x}} - y_2\,b\,\hat{\mathbf{y}} + \left(\frac{1}{2} + z_2\right)\,c\,\hat{\mathbf{z}}$ | (2$a$) | Cu I |
| $\mathbf{B}_5$ | = | $y_3\,\mathbf{a}_2 + z_3\,\mathbf{a}_3$ | = | $y_3\,b\,\mathbf{\hat{y}} + z_3\,c\,\mathbf{\hat{z}}$ | $(2a)$ | S I |
| $\mathbf{B}_6$ | = | $\frac{1}{2}\,\mathbf{a}_1 - y_3\,\mathbf{a}_2 + \left(\frac{1}{2} + z_3\right)\,\mathbf{a}_3$ | = | $\frac{1}{2}\,a\,\mathbf{\hat{x}} - y_3\,b\,\mathbf{\hat{y}} + \left(\frac{1}{2} + z_3\right)\,c\,\mathbf{\hat{z}}$ | $(2a)$ | S I |
| $\mathbf{B}_7$ | = | $y_4\,\mathbf{a}_2 + z_4\,\mathbf{a}_3$ | = | $y_4\,b\,\mathbf{\hat{y}} + z_4\,c\,\mathbf{\hat{z}}$ | $(2a)$ | S II |
| $\mathbf{B}_8$ | = | $\frac{1}{2}\,\mathbf{a}_1 - y_4\,\mathbf{a}_2 + \left(\frac{1}{2} + z_4\right)\,\mathbf{a}_3$ | = | $\frac{1}{2}\,a\,\mathbf{\hat{x}} - y_4\,b\,\mathbf{\hat{y}} + \left(\frac{1}{2} + z_4\right)\,c\,\mathbf{\hat{z}}$ | $(2a)$ | S II |
| $\mathbf{B}_9$ | = | $x_5\,\mathbf{a}_1 + y_5\,\mathbf{a}_2 + z_5\,\mathbf{a}_3$ | = | $x_5\,a\,\mathbf{\hat{x}} + y_5\,b\,\mathbf{\hat{y}} + z_5\,c\,\mathbf{\hat{z}}$ | $(4b)$ | Cu II |
| $\mathbf{B}_{10}$ | = | $\left(\frac{1}{2} - x_5\right)\,\mathbf{a}_1 - y_5\,\mathbf{a}_2 + \left(\frac{1}{2} + z_5\right)\,\mathbf{a}_3$ | = | $\left(\frac{1}{2} - x_5\right)\,a\,\mathbf{\hat{x}} - y_5\,b\,\mathbf{\hat{y}} + \left(\frac{1}{2} + z_5\right)\,c\,\mathbf{\hat{z}}$ | $(4b)$ | Cu II |
| $\mathbf{B}_{11}$ | = | $\left(\frac{1}{2} + x_5\right)\,\mathbf{a}_1 - y_5\,\mathbf{a}_2 + \left(\frac{1}{2} + z_5\right)\,\mathbf{a}_3$ | = | $\left(\frac{1}{2} + x_5\right)\,a\,\mathbf{\hat{x}} - y_5\,b\,\mathbf{\hat{y}} + \left(\frac{1}{2} + z_5\right)\,c\,\mathbf{\hat{z}}$ | $(4b)$ | Cu II |
| $\mathbf{B}_{12}$ | = | $-x_5\,\mathbf{a}_1 + y_5\,\mathbf{a}_2 + z_5\,\mathbf{a}_3$ | = | $-x_5\,a\,\mathbf{\hat{x}} + y_5\,b\,\mathbf{\hat{y}} + z_5\,c\,\mathbf{\hat{z}}$ | $(4b)$ | Cu II |
| $\mathbf{B}_{13}$ | = | $x_6\,\mathbf{a}_1 + y_6\,\mathbf{a}_2 + z_6\,\mathbf{a}_3$ | = | $x_6\,a\,\mathbf{\hat{x}} + y_6\,b\,\mathbf{\hat{y}} + z_6\,c\,\mathbf{\hat{z}}$ | $(4b)$ | S III |
| $\mathbf{B}_{14}$ | = | $\left(\frac{1}{2} - x_6\right)\,\mathbf{a}_1 - y_6\,\mathbf{a}_2 + \left(\frac{1}{2} + z_6\right)\,\mathbf{a}_3$ | = | $\left(\frac{1}{2} - x_6\right)\,a\,\mathbf{\hat{x}} - y_6\,b\,\mathbf{\hat{y}} + \left(\frac{1}{2} + z_6\right)\,c\,\mathbf{\hat{z}}$ | $(4b)$ | S III |
| $\mathbf{B}_{15}$ | = | $\left(\frac{1}{2} + x_6\right)\,\mathbf{a}_1 - y_6\,\mathbf{a}_2 + \left(\frac{1}{2} + z_6\right)\,\mathbf{a}_3$ | = | $\left(\frac{1}{2} + x_6\right)\,a\,\mathbf{\hat{x}} - y_6\,b\,\mathbf{\hat{y}} + \left(\frac{1}{2} + z_6\right)\,c\,\mathbf{\hat{z}}$ | $(4b)$ | S III |
| $\mathbf{B}_{16}$ | = | $-x_6\,\mathbf{a}_1 + y_6\,\mathbf{a}_2 + z_6\,\mathbf{a}_3$ | = | $-x_6\,a\,\mathbf{\hat{x}} + y_6\,b\,\mathbf{\hat{y}} + z_6\,c\,\mathbf{\hat{z}}$ | $(4b)$ | S III |

---

---

**Geometry files:**

- CIF: pp. 657
- POSCAR: pp. 657

# Modderite (CoAs) Structure: AB_oP8_33_a_a

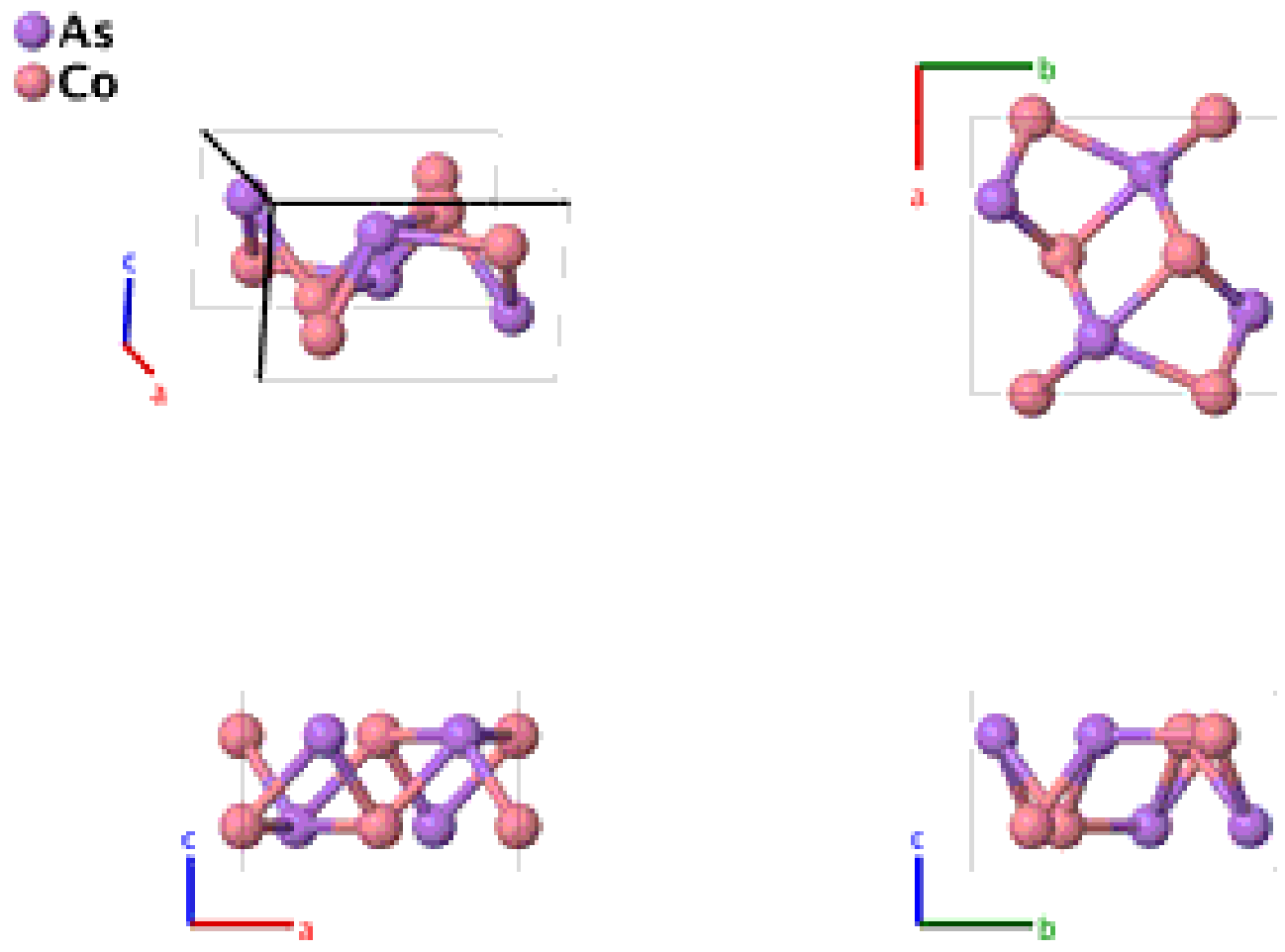

| | | |
|---|---|---|
| **Prototype** | : | CoAs |
| **AFLOW prototype label** | : | AB_oP8_33_a_a |
| ***Strukturbericht* designation** | : | None |
| **Pearson symbol** | : | oP8 |
| **Space group number** | : | 33 |
| **Space group symbol** | : | $Pna2_1$ |
| **AFLOW prototype command** | : | `aflow --proto=AB_oP8_33_a_a`<br>`--params=`$a, b/a, c/a, x_1, y_1, z_1, x_2, y_2, z_2$ |

---

**Other compounds with this structure:**

- FeAs

---

- (Lyman, 1984) arbitrarily set $z_2 = 1/4$, which is allowed for this space group. When $z_1 = z_2 = 1/4$, the space group becomes Pnma and the structure is equivalent to MnP (B31). (Lyman, 1984) lists both space groups for both CoAs and FeAs, and prefers the MnP structure for these compounds.

---

**Simple Orthorhombic primitive vectors:**

$$\mathbf{a}_1 = a\,\hat{\mathbf{x}}$$
$$\mathbf{a}_2 = b\,\hat{\mathbf{y}}$$
$$\mathbf{a}_3 = c\,\hat{\mathbf{z}}$$

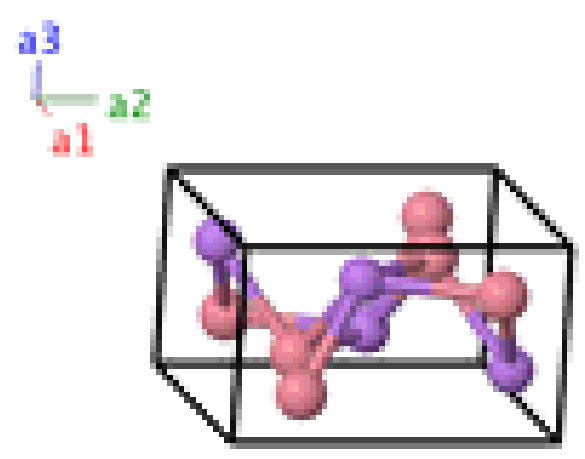

---

**Basis vectors:**

| | | Lattice Coordinates | | Cartesian Coordinates | Wyckoff Position | Atom Type |
|---|---|---|---|---|---|---|
| $\mathbf{B}_1$ | = | $x_1\,\mathbf{a}_1 + y_1\,\mathbf{a}_2 + z_1\,\mathbf{a}_3$ | = | $x_1\,a\,\hat{\mathbf{x}} + y_1\,b\,\hat{\mathbf{y}} + z_1\,c\,\hat{\mathbf{z}}$ | (4a) | As |

$$
\begin{array}{llllll}
\mathbf{B_2} & = & -x_1\,\mathbf{a_1} - y_1\,\mathbf{a_2} + \left(\frac{1}{2} + z_1\right)\,\mathbf{a_3} & = & -x_1\,a\,\mathbf{\hat{x}} - y_1\,b\,\mathbf{\hat{y}} + \left(\frac{1}{2} + z_1\right)\,c\,\mathbf{\hat{z}} & (4a) & \text{As} \\
\mathbf{B_3} & = & \left(\frac{1}{2} + x_1\right)\,\mathbf{a_1} + \left(\frac{1}{2} - y_1\right)\,\mathbf{a_2} + z_1\,\mathbf{a_3} & = & \left(\frac{1}{2} + x_1\right)\,a\,\mathbf{\hat{x}} + \left(\frac{1}{2} - y_1\right)\,b\,\mathbf{\hat{y}} + z_1\,c\,\mathbf{\hat{z}} & (4a) & \text{As} \\
\mathbf{B_4} & = & \left(\frac{1}{2} - x_1\right)\,\mathbf{a_1} + \left(\frac{1}{2} + y_1\right)\,\mathbf{a_2} + \left(\frac{1}{2} + z_1\right)\,\mathbf{a_3} & = & \left(\frac{1}{2} - x_1\right)\,a\,\mathbf{\hat{x}} + \left(\frac{1}{2} + y_1\right)\,b\,\mathbf{\hat{y}} + \left(\frac{1}{2} + z_1\right)\,c\,\mathbf{\hat{z}} & (4a) & \text{As} \\
\mathbf{B_5} & = & x_2\,\mathbf{a_1} + y_2\,\mathbf{a_2} + z_2\,\mathbf{a_3} & = & x_2\,a\,\mathbf{\hat{x}} + y_2\,b\,\mathbf{\hat{y}} + z_2\,c\,\mathbf{\hat{z}} & (4a) & \text{Co} \\
\mathbf{B_6} & = & -x_2\,\mathbf{a_1} - y_2\,\mathbf{a_2} + \left(\frac{1}{2} + z_2\right)\,\mathbf{a_3} & = & -x_2\,a\,\mathbf{\hat{x}} - y_2\,b\,\mathbf{\hat{y}} + \left(\frac{1}{2} + z_2\right)\,c\,\mathbf{\hat{z}} & (4a) & \text{Co} \\
\mathbf{B_7} & = & \left(\frac{1}{2} + x_2\right)\,\mathbf{a_1} + \left(\frac{1}{2} - y_2\right)\,\mathbf{a_2} + z_2\,\mathbf{a_3} & = & \left(\frac{1}{2} + x_2\right)\,a\,\mathbf{\hat{x}} + \left(\frac{1}{2} - y_2\right)\,b\,\mathbf{\hat{y}} + z_2\,c\,\mathbf{\hat{z}} & (4a) & \text{Co} \\
\mathbf{B_8} & = & \left(\frac{1}{2} - x_2\right)\,\mathbf{a_1} + \left(\frac{1}{2} + y_2\right)\,\mathbf{a_2} + \left(\frac{1}{2} + z_2\right)\,\mathbf{a_3} & = & \left(\frac{1}{2} - x_2\right)\,a\,\mathbf{\hat{x}} + \left(\frac{1}{2} + y_2\right)\,b\,\mathbf{\hat{y}} + \left(\frac{1}{2} + z_2\right)\,c\,\mathbf{\hat{z}} & (4a) & \text{Co}
\end{array}
$$

---

---

**Geometry files:**

- CIF: pp. 658
- POSCAR: pp. 658

# $AsK_3S_4$ Structure: AB3C4_oP32_33_a_3a_4a

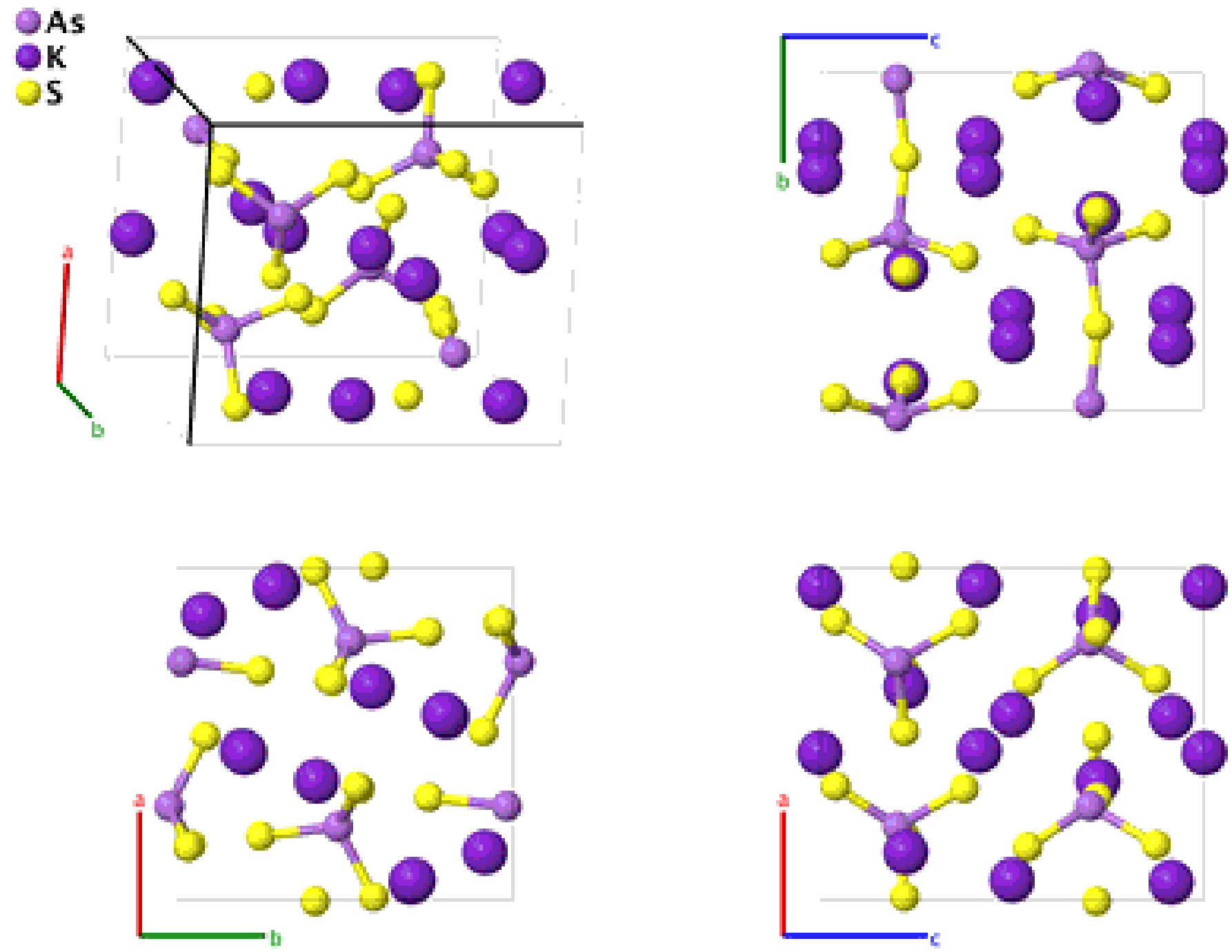

| | | |
|---|---|---|
| **Prototype** | : | $AsK_3S_4$ |
| **AFLOW prototype label** | : | AB3C4_oP32_33_a_3a_4a |
| ***Strukturbericht* designation** | : | None |
| **Pearson symbol** | : | oP32 |
| **Space group number** | : | 33 |
| **Space group symbol** | : | $Pna2_1$ |
| **AFLOW prototype command** | : | `aflow --proto=AB3C4_oP32_33_a_3a_4a` `--params=`$a, b/a, c/a, x_1, y_1, z_1, x_2, y_2, z_2, x_3, y_3, z_3, x_4, y_4, z_4, x_5, y_5, z_5, x_6, y_6, z_6, x_7, y_7, z_7, x_8, y_8, z_8$ |

- Note that the authors arbitrarily set $z_2 = 1/4$, as is allowed by this space group.

**Simple Orthorhombic primitive vectors:**

$$\mathbf{a}_1 = a\,\hat{\mathbf{x}}$$
$$\mathbf{a}_2 = b\,\hat{\mathbf{y}}$$
$$\mathbf{a}_3 = c\,\hat{\mathbf{z}}$$

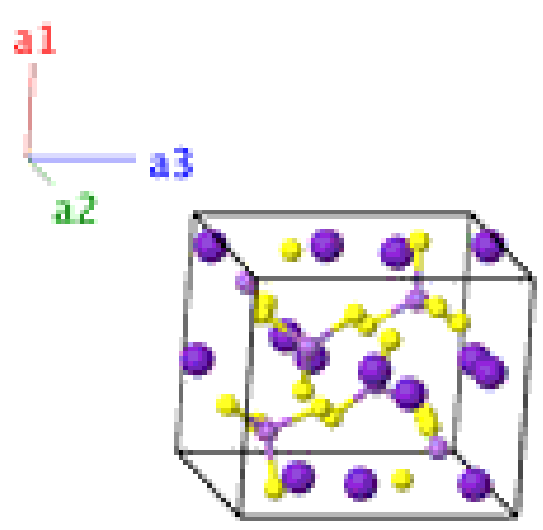

**Basis vectors:**

| | | Lattice Coordinates | | Cartesian Coordinates | Wyckoff Position | Atom Type |
|---|---|---|---|---|---|---|
| $\mathbf{B}_1$ | = | $x_1\,\mathbf{a}_1 + y_1\,\mathbf{a}_2 + z_1\,\mathbf{a}_3$ | = | $x_1\,a\,\hat{\mathbf{x}} + y_1\,b\,\hat{\mathbf{y}} + z_1\,c\,\hat{\mathbf{z}}$ | (4*a*) | As |
| $\mathbf{B_{2}}$ | = | $-x_{1}\,\mathbf{a_{1}} - y_{1}\,\mathbf{a_{2}} + \left(\frac{1}{2} + z_{1}\right)\,\mathbf{a_{3}}$ | = | $-x_{1}\,a\,\mathbf{\hat{x}} - y_{1}\,b\,\mathbf{\hat{y}} + \left(\frac{1}{2} + z_{1}\right)\,c\,\mathbf{\hat{z}}$ | (4$a$) | As |
| $\mathbf{B_{3}}$ | = | $\left(\frac{1}{2} + x_{1}\right)\,\mathbf{a_{1}} + \left(\frac{1}{2} - y_{1}\right)\,\mathbf{a_{2}} + z_{1}\,\mathbf{a_{3}}$ | = | $\left(\frac{1}{2} + x_{1}\right)\,a\,\mathbf{\hat{x}} + \left(\frac{1}{2} - y_{1}\right)\,b\,\mathbf{\hat{y}} + z_{1}\,c\,\mathbf{\hat{z}}$ | (4$a$) | As |
| $\mathbf{B_{4}}$ | = | $\left(\frac{1}{2} - x_{1}\right)\,\mathbf{a_{1}} + \left(\frac{1}{2} + y_{1}\right)\,\mathbf{a_{2}} + \left(\frac{1}{2} + z_{1}\right)\,\mathbf{a_{3}}$ | = | $\left(\frac{1}{2} - x_{1}\right)\,a\,\mathbf{\hat{x}} + \left(\frac{1}{2} + y_{1}\right)\,b\,\mathbf{\hat{y}} + \left(\frac{1}{2} + z_{1}\right)\,c\,\mathbf{\hat{z}}$ | (4$a$) | As |
| $\mathbf{B_{5}}$ | = | $x_{2}\,\mathbf{a_{1}} + y_{2}\,\mathbf{a_{2}} + z_{2}\,\mathbf{a_{3}}$ | = | $x_{2}\,a\,\mathbf{\hat{x}} + y_{2}\,b\,\mathbf{\hat{y}} + z_{2}\,c\,\mathbf{\hat{z}}$ | (4$a$) | K I |
| $\mathbf{B_{6}}$ | = | $-x_{2}\,\mathbf{a_{1}} - y_{2}\,\mathbf{a_{2}} + \left(\frac{1}{2} + z_{2}\right)\,\mathbf{a_{3}}$ | = | $-x_{2}\,a\,\mathbf{\hat{x}} - y_{2}\,b\,\mathbf{\hat{y}} + \left(\frac{1}{2} + z_{2}\right)\,c\,\mathbf{\hat{z}}$ | (4$a$) | K I |
| $\mathbf{B_{7}}$ | = | $\left(\frac{1}{2} + x_{2}\right)\,\mathbf{a_{1}} + \left(\frac{1}{2} - y_{2}\right)\,\mathbf{a_{2}} + z_{2}\,\mathbf{a_{3}}$ | = | $\left(\frac{1}{2} + x_{2}\right)\,a\,\mathbf{\hat{x}} + \left(\frac{1}{2} - y_{2}\right)\,b\,\mathbf{\hat{y}} + z_{2}\,c\,\mathbf{\hat{z}}$ | (4$a$) | K I |
| $\mathbf{B_{8}}$ | = | $\left(\frac{1}{2} - x_{2}\right)\,\mathbf{a_{1}} + \left(\frac{1}{2} + y_{2}\right)\,\mathbf{a_{2}} + \left(\frac{1}{2} + z_{2}\right)\,\mathbf{a_{3}}$ | = | $\left(\frac{1}{2} - x_{2}\right)\,a\,\mathbf{\hat{x}} + \left(\frac{1}{2} + y_{2}\right)\,b\,\mathbf{\hat{y}} + \left(\frac{1}{2} + z_{2}\right)\,c\,\mathbf{\hat{z}}$ | (4$a$) | K I |
| $\mathbf{B_{9}}$ | = | $x_{3}\,\mathbf{a_{1}} + y_{3}\,\mathbf{a_{2}} + z_{3}\,\mathbf{a_{3}}$ | = | $x_{3}\,a\,\mathbf{\hat{x}} + y_{3}\,b\,\mathbf{\hat{y}} + z_{3}\,c\,\mathbf{\hat{z}}$ | (4$a$) | K II |
| $\mathbf{B_{10}}$ | = | $-x_{3}\,\mathbf{a_{1}} - y_{3}\,\mathbf{a_{2}} + \left(\frac{1}{2} + z_{3}\right)\,\mathbf{a_{3}}$ | = | $-x_{3}\,a\,\mathbf{\hat{x}} - y_{3}\,b\,\mathbf{\hat{y}} + \left(\frac{1}{2} + z_{3}\right)\,c\,\mathbf{\hat{z}}$ | (4$a$) | K II |
| $\mathbf{B_{11}}$ | = | $\left(\frac{1}{2} + x_{3}\right)\,\mathbf{a_{1}} + \left(\frac{1}{2} - y_{3}\right)\,\mathbf{a_{2}} + z_{3}\,\mathbf{a_{3}}$ | = | $\left(\frac{1}{2} + x_{3}\right)\,a\,\mathbf{\hat{x}} + \left(\frac{1}{2} - y_{3}\right)\,b\,\mathbf{\hat{y}} + z_{3}\,c\,\mathbf{\hat{z}}$ | (4$a$) | K II |
| $\mathbf{B_{12}}$ | = | $\left(\frac{1}{2} - x_{3}\right)\,\mathbf{a_{1}} + \left(\frac{1}{2} + y_{3}\right)\,\mathbf{a_{2}} + \left(\frac{1}{2} + z_{3}\right)\,\mathbf{a_{3}}$ | = | $\left(\frac{1}{2} - x_{3}\right)\,a\,\mathbf{\hat{x}} + \left(\frac{1}{2} + y_{3}\right)\,b\,\mathbf{\hat{y}} + \left(\frac{1}{2} + z_{3}\right)\,c\,\mathbf{\hat{z}}$ | (4$a$) | K II |
| $\mathbf{B_{13}}$ | = | $x_{4}\,\mathbf{a_{1}} + y_{4}\,\mathbf{a_{2}} + z_{4}\,\mathbf{a_{3}}$ | = | $x_{4}\,a\,\mathbf{\hat{x}} + y_{4}\,b\,\mathbf{\hat{y}} + z_{4}\,c\,\mathbf{\hat{z}}$ | (4$a$) | K III |
| $\mathbf{B_{14}}$ | = | $-x_{4}\,\mathbf{a_{1}} - y_{4}\,\mathbf{a_{2}} + \left(\frac{1}{2} + z_{4}\right)\,\mathbf{a_{3}}$ | = | $-x_{4}\,a\,\mathbf{\hat{x}} - y_{4}\,b\,\mathbf{\hat{y}} + \left(\frac{1}{2} + z_{4}\right)\,c\,\mathbf{\hat{z}}$ | (4$a$) | K III |
| $\mathbf{B_{15}}$ | = | $\left(\frac{1}{2} + x_{4}\right)\,\mathbf{a_{1}} + \left(\frac{1}{2} - y_{4}\right)\,\mathbf{a_{2}} + z_{4}\,\mathbf{a_{3}}$ | = | $\left(\frac{1}{2} + x_{4}\right)\,a\,\mathbf{\hat{x}} + \left(\frac{1}{2} - y_{4}\right)\,b\,\mathbf{\hat{y}} + z_{4}\,c\,\mathbf{\hat{z}}$ | (4$a$) | K III |
| $\mathbf{B_{16}}$ | = | $\left(\frac{1}{2} - x_{4}\right)\,\mathbf{a_{1}} + \left(\frac{1}{2} + y_{4}\right)\,\mathbf{a_{2}} + \left(\frac{1}{2} + z_{4}\right)\,\mathbf{a_{3}}$ | = | $\left(\frac{1}{2} - x_{4}\right)\,a\,\mathbf{\hat{x}} + \left(\frac{1}{2} + y_{4}\right)\,b\,\mathbf{\hat{y}} + \left(\frac{1}{2} + z_{4}\right)\,c\,\mathbf{\hat{z}}$ | (4$a$) | K III |
| $\mathbf{B_{17}}$ | = | $x_{5}\,\mathbf{a_{1}} + y_{5}\,\mathbf{a_{2}} + z_{5}\,\mathbf{a_{3}}$ | = | $x_{5}\,a\,\mathbf{\hat{x}} + y_{5}\,b\,\mathbf{\hat{y}} + z_{5}\,c\,\mathbf{\hat{z}}$ | (4$a$) | S I |
| $\mathbf{B_{18}}$ | = | $-x_{5}\,\mathbf{a_{1}} - y_{5}\,\mathbf{a_{2}} + \left(\frac{1}{2} + z_{5}\right)\,\mathbf{a_{3}}$ | = | $-x_{5}\,a\,\mathbf{\hat{x}} - y_{5}\,b\,\mathbf{\hat{y}} + \left(\frac{1}{2} + z_{5}\right)\,c\,\mathbf{\hat{z}}$ | (4$a$) | S I |
| $\mathbf{B_{19}}$ | = | $\left(\frac{1}{2} + x_{5}\right)\,\mathbf{a_{1}} + \left(\frac{1}{2} - y_{5}\right)\,\mathbf{a_{2}} + z_{5}\,\mathbf{a_{3}}$ | = | $\left(\frac{1}{2} + x_{5}\right)\,a\,\mathbf{\hat{x}} + \left(\frac{1}{2} - y_{5}\right)\,b\,\mathbf{\hat{y}} + z_{5}\,c\,\mathbf{\hat{z}}$ | (4$a$) | S I |
| $\mathbf{B_{20}}$ | = | $\left(\frac{1}{2} - x_{5}\right)\,\mathbf{a_{1}} + \left(\frac{1}{2} + y_{5}\right)\,\mathbf{a_{2}} + \left(\frac{1}{2} + z_{5}\right)\,\mathbf{a_{3}}$ | = | $\left(\frac{1}{2} - x_{5}\right)\,a\,\mathbf{\hat{x}} + \left(\frac{1}{2} + y_{5}\right)\,b\,\mathbf{\hat{y}} + \left(\frac{1}{2} + z_{5}\right)\,c\,\mathbf{\hat{z}}$ | (4$a$) | S I |
| $\mathbf{B_{21}}$ | = | $x_{6}\,\mathbf{a_{1}} + y_{6}\,\mathbf{a_{2}} + z_{6}\,\mathbf{a_{3}}$ | = | $x_{6}\,a\,\mathbf{\hat{x}} + y_{6}\,b\,\mathbf{\hat{y}} + z_{6}\,c\,\mathbf{\hat{z}}$ | (4$a$) | S II |
| $\mathbf{B_{22}}$ | = | $-x_{6}\,\mathbf{a_{1}} - y_{6}\,\mathbf{a_{2}} + \left(\frac{1}{2} + z_{6}\right)\,\mathbf{a_{3}}$ | = | $-x_{6}\,a\,\mathbf{\hat{x}} - y_{6}\,b\,\mathbf{\hat{y}} + \left(\frac{1}{2} + z_{6}\right)\,c\,\mathbf{\hat{z}}$ | (4$a$) | S II |
| $\mathbf{B_{23}}$ | = | $\left(\frac{1}{2} + x_{6}\right)\,\mathbf{a_{1}} + \left(\frac{1}{2} - y_{6}\right)\,\mathbf{a_{2}} + z_{6}\,\mathbf{a_{3}}$ | = | $\left(\frac{1}{2} + x_{6}\right)\,a\,\mathbf{\hat{x}} + \left(\frac{1}{2} - y_{6}\right)\,b\,\mathbf{\hat{y}} + z_{6}\,c\,\mathbf{\hat{z}}$ | (4$a$) | S II |
| $\mathbf{B_{24}}$ | = | $\left(\frac{1}{2} - x_{6}\right)\,\mathbf{a_{1}} + \left(\frac{1}{2} + y_{6}\right)\,\mathbf{a_{2}} + \left(\frac{1}{2} + z_{6}\right)\,\mathbf{a_{3}}$ | = | $\left(\frac{1}{2} - x_{6}\right)\,a\,\mathbf{\hat{x}} + \left(\frac{1}{2} + y_{6}\right)\,b\,\mathbf{\hat{y}} + \left(\frac{1}{2} + z_{6}\right)\,c\,\mathbf{\hat{z}}$ | (4$a$) | S II |
| $\mathbf{B_{25}}$ | = | $x_{7}\,\mathbf{a_{1}} + y_{7}\,\mathbf{a_{2}} + z_{7}\,\mathbf{a_{3}}$ | = | $x_{7}\,a\,\mathbf{\hat{x}} + y_{7}\,b\,\mathbf{\hat{y}} + z_{7}\,c\,\mathbf{\hat{z}}$ | (4$a$) | S III |
| $\mathbf{B_{26}}$ | = | $-x_{7}\,\mathbf{a_{1}} - y_{7}\,\mathbf{a_{2}} + \left(\frac{1}{2} + z_{7}\right)\,\mathbf{a_{3}}$ | = | $-x_{7}\,a\,\mathbf{\hat{x}} - y_{7}\,b\,\mathbf{\hat{y}} + \left(\frac{1}{2} + z_{7}\right)\,c\,\mathbf{\hat{z}}$ | (4$a$) | S III |
| $\mathbf{B_{27}}$ | = | $\left(\frac{1}{2} + x_{7}\right)\,\mathbf{a_{1}} + \left(\frac{1}{2} - y_{7}\right)\,\mathbf{a_{2}} + z_{7}\,\mathbf{a_{3}}$ | = | $\left(\frac{1}{2} + x_{7}\right)\,a\,\mathbf{\hat{x}} + \left(\frac{1}{2} - y_{7}\right)\,b\,\mathbf{\hat{y}} + z_{7}\,c\,\mathbf{\hat{z}}$ | (4$a$) | S III |
| $\mathbf{B_{28}}$ | = | $\left(\frac{1}{2} - x_{7}\right)\,\mathbf{a_{1}} + \left(\frac{1}{2} + y_{7}\right)\,\mathbf{a_{2}} + \left(\frac{1}{2} + z_{7}\right)\,\mathbf{a_{3}}$ | = | $\left(\frac{1}{2} - x_{7}\right)\,a\,\mathbf{\hat{x}} + \left(\frac{1}{2} + y_{7}\right)\,b\,\mathbf{\hat{y}} + \left(\frac{1}{2} + z_{7}\right)\,c\,\mathbf{\hat{z}}$ | (4$a$) | S III |
| $\mathbf{B_{29}}$ | = | $x_{8}\,\mathbf{a_{1}} + y_{8}\,\mathbf{a_{2}} + z_{8}\,\mathbf{a_{3}}$ | = | $x_{8}\,a\,\mathbf{\hat{x}} + y_{8}\,b\,\mathbf{\hat{y}} + z_{8}\,c\,\mathbf{\hat{z}}$ | (4$a$) | S IV |
| $\mathbf{B_{30}}$ | = | $-x_{8}\,\mathbf{a_{1}} - y_{8}\,\mathbf{a_{2}} + \left(\frac{1}{2} + z_{8}\right)\,\mathbf{a_{3}}$ | = | $-x_{8}\,a\,\mathbf{\hat{x}} - y_{8}\,b\,\mathbf{\hat{y}} + \left(\frac{1}{2} + z_{8}\right)\,c\,\mathbf{\hat{z}}$ | (4$a$) | S IV |
| $\mathbf{B_{31}}$ | = | $\left(\frac{1}{2} + x_{8}\right)\,\mathbf{a_{1}} + \left(\frac{1}{2} - y_{8}\right)\,\mathbf{a_{2}} + z_{8}\,\mathbf{a_{3}}$ | = | $\left(\frac{1}{2} + x_{8}\right)\,a\,\mathbf{\hat{x}} + \left(\frac{1}{2} - y_{8}\right)\,b\,\mathbf{\hat{y}} + z_{8}\,c\,\mathbf{\hat{z}}$ | (4$a$) | S IV |
| $\mathbf{B_{32}}$ | = | $\left(\frac{1}{2} - x_8\right)\mathbf{a_1} + \left(\frac{1}{2} + y_8\right)\mathbf{a_2} + \left(\frac{1}{2} + z_8\right)\mathbf{a_3}$ | = | $\left(\frac{1}{2} - x_8\right)a\,\mathbf{\hat{x}} + \left(\frac{1}{2} + y_8\right)b\,\mathbf{\hat{y}} + \left(\frac{1}{2} + z_8\right)c\,\mathbf{\hat{z}}$ | (4*a*) | S IV |

---

**Found in:**

---

**Geometry files:**

- CIF: pp. 658
- POSCAR: pp. 658

# $HgBr_2$ (C24) Structure: A2B_oC12_36_2a_a

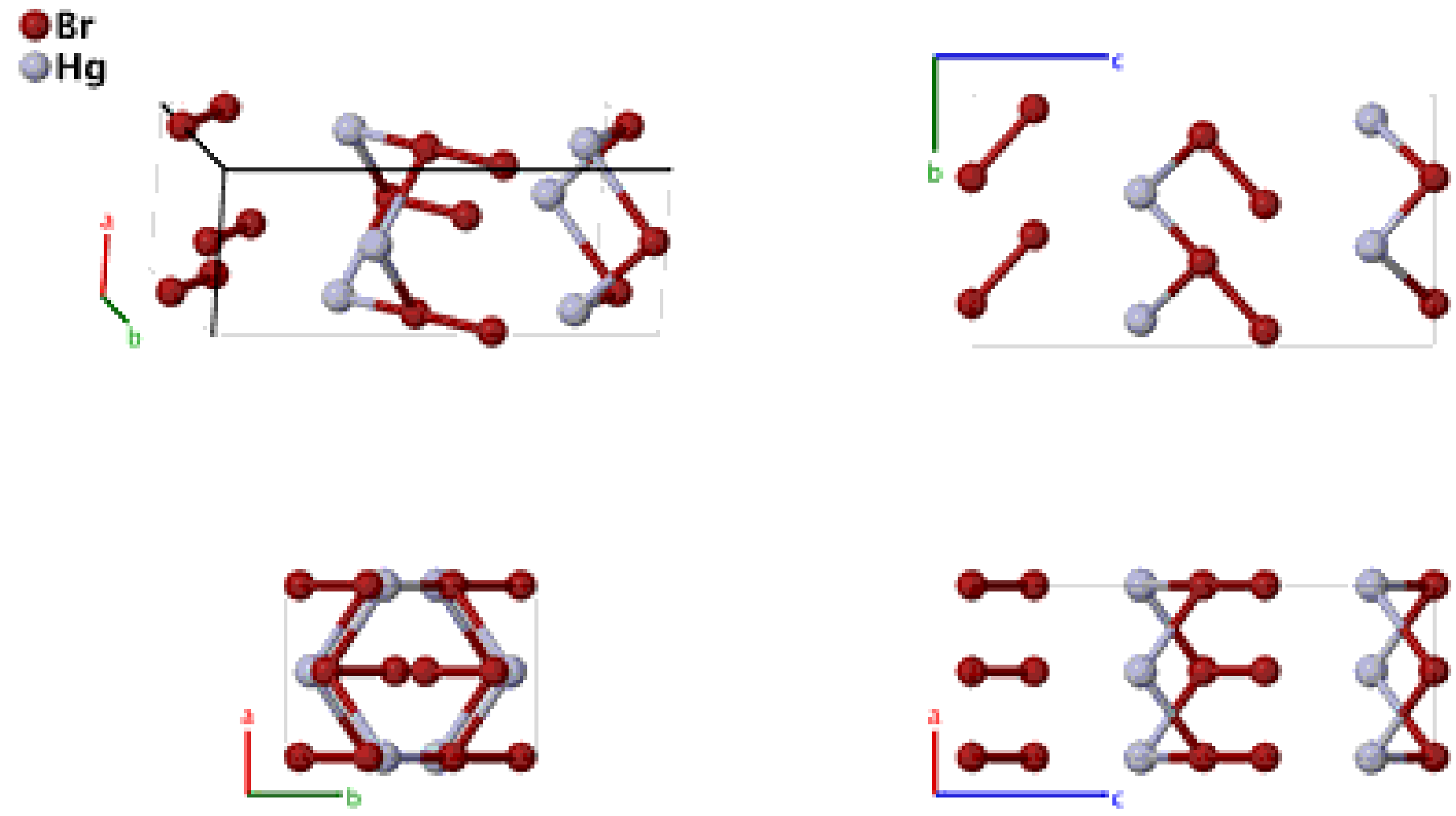

| | | |
|---|---|---|
| **Prototype** | : | $HgBr_2$ |
| **AFLOW prototype label** | : | A2B_oC12_36_2a_a |
| ***Strukturbericht* designation** | : | C24 |
| **Pearson symbol** | : | oC12 |
| **Space group number** | : | 36 |
| **Space group symbol** | : | $Cmc2_1$ |
| **AFLOW prototype command** | : | `aflow --proto=A2B_oC12_36_2a_a`<br>`--params=`$a, b/a, c/a, y_1, z_1, y_2, z_2, y_3, z_3$ |

**Base-centered Orthorhombic primitive vectors:**

$$
\begin{aligned}
\mathbf{a}_1 &= \tfrac{1}{2}\,a\,\mathbf{\hat{x}} - \tfrac{1}{2}\,b\,\mathbf{\hat{y}} \\
\mathbf{a}_2 &= \tfrac{1}{2}\,a\,\mathbf{\hat{x}} + \tfrac{1}{2}\,b\,\mathbf{\hat{y}} \\
\mathbf{a}_3 &= c\,\mathbf{\hat{z}}
\end{aligned}
$$

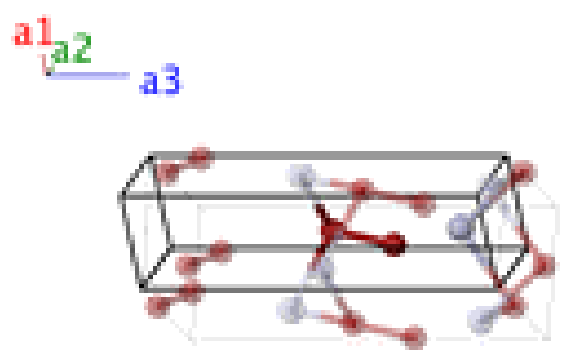

**Basis vectors:**

| | | Lattice Coordinates | | Cartesian Coordinates | Wyckoff Position | Atom Type |
|---|---|---|---|---|---|---|
| $\mathbf{B}_1$ | = | $-y_1\,\mathbf{a}_1 + y_1\,\mathbf{a}_2 + z_1\,\mathbf{a}_3$ | = | $y_1\,b\,\mathbf{\hat{y}} + z_1\,c\mathbf{\hat{z}}$ | (4$a$) | Br I |
| $\mathbf{B}_2$ | = | $y_1\,\mathbf{a}_1 - y_1\,\mathbf{a}_2 + \left(\frac{1}{2} + z_1\right)\,\mathbf{a}_3$ | = | $-y_1\,b\,\mathbf{\hat{y}} + \left(\frac{1}{2} + z_1\right)\,c\mathbf{\hat{z}}$ | (4$a$) | Br I |
| $\mathbf{B}_3$ | = | $-y_2\,\mathbf{a}_1 + y_2\,\mathbf{a}_2 + z_2\,\mathbf{a}_3$ | = | $y_2\,b\,\mathbf{\hat{y}} + z_2\,c\mathbf{\hat{z}}$ | (4$a$) | Br II |
| $\mathbf{B}_4$ | = | $y_2\,\mathbf{a}_1 - y_2\,\mathbf{a}_2 + \left(\frac{1}{2} + z_2\right)\,\mathbf{a}_3$ | = | $-y_2\,b\,\mathbf{\hat{y}} + \left(\frac{1}{2} + z_2\right)\,c\mathbf{\hat{z}}$ | (4$a$) | Br II |
| $\mathbf{B}_5$ | = | $-y_3\,\mathbf{a}_1 + y_3\,\mathbf{a}_2 + z_3\,\mathbf{a}_3$ | = | $y_3\,b\,\mathbf{\hat{y}} + z_3\,c\mathbf{\hat{z}}$ | (4$a$) | Hg |
| $\mathbf{B}_6$ | = | $y_3\,\mathbf{a}_1 - y_3\,\mathbf{a}_2 + \left(\frac{1}{2} + z_3\right)\,\mathbf{a}_3$ | = | $-y_3\,b\,\mathbf{\hat{y}} + \left(\frac{1}{2} + z_3\right)\,c\mathbf{\hat{z}}$ | (4$a$) | Hg |

**Found in:**

---

**Geometry files:**

- CIF: pp. 659
- POSCAR: pp. 659

# $C_2CeNi$ Structure: A2BC_oC8_38_e_a_b

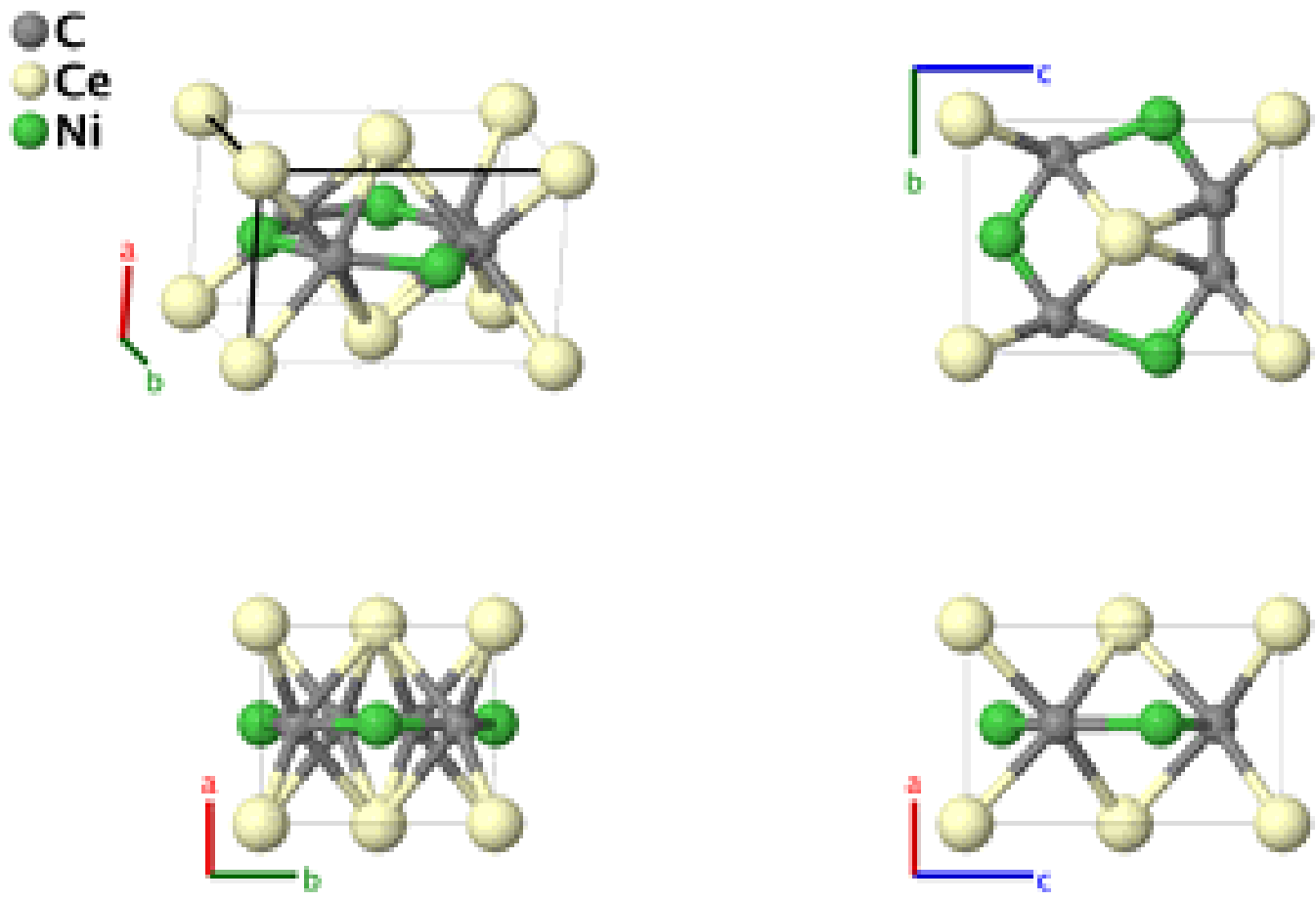

| | | |
|---|---|---|
| **Prototype** | : | $C_2CeNi$ |
| **AFLOW prototype label** | : | A2BC_oC8_38_e_a_b |
| ***Strukturbericht* designation** | : | None |
| **Pearson symbol** | : | oC8 |
| **Space group number** | : | 38 |
| **Space group symbol** | : | Amm2 |
| **AFLOW prototype command** | : | `aflow --proto=A2BC_oC8_38_e_a_b`<br>`--params=`$a, b/a, c/a, z_1, z_2, y_3, z_3$ |

**Other compounds with this structure:**

- $C_2CoDy$, $C_2ErFe$, $C_2FeSm$, $C_2NiPa$, $C_2NiYb$, $C_2PrRh$, many other $C_2XY$

**Base-centered Orthorhombic primitive vectors:**

$$\mathbf{a}_1 = a\,\hat{\mathbf{x}}$$
$$\mathbf{a}_2 = \frac{1}{2}\,b\,\hat{\mathbf{y}} - \frac{1}{2}\,c\,\hat{\mathbf{z}}$$
$$\mathbf{a}_3 = \frac{1}{2}\,b\,\hat{\mathbf{y}} + \frac{1}{2}\,c\,\hat{\mathbf{z}}$$

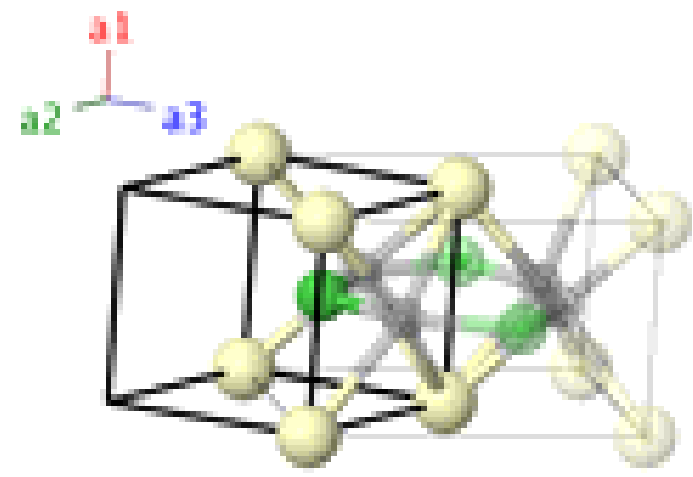

**Basis vectors:**

| | | Lattice Coordinates | | Cartesian Coordinates | Wyckoff Position | Atom Type |
|---|---|---|---|---|---|---|
| $\mathbf{B_1}$ | = | $-z_1\,\mathbf{a_2} + z_1\,\mathbf{a_3}$ | = | $z_1\,c\,\hat{\mathbf{z}}$ | (2$a$) | Ce |
| $\mathbf{B_2}$ | = | $\frac{1}{2}\,\mathbf{a_1} - z_2\,\mathbf{a_2} + z_2\,\mathbf{a_3}$ | = | $\frac{1}{2}\,a\,\hat{\mathbf{x}} + z_2\,c\,\hat{\mathbf{z}}$ | (2$b$) | Ni |
| $\mathbf{B_3}$ | = | $\frac{1}{2}\,\mathbf{a_1} + (y_3 - z_3)\,\mathbf{a_2} + (y_3 + z_3)\,\mathbf{a_3}$ | = | $\frac{1}{2}\,a\,\hat{\mathbf{x}} + y_3\,b\,\hat{\mathbf{y}} + z_3\,c\,\hat{\mathbf{z}}$ | (4$e$) | C |
| $\mathbf{B_4}$ | = | $\frac{1}{2}\,\mathbf{a_1} - (y_3 + z_3)\,\mathbf{a_2} + (z_3 - y_3)\,\mathbf{a_3}$ | = | $\frac{1}{2}\,a\,\hat{\mathbf{x}} - y_3\,b\,\hat{\mathbf{y}} + z_3\,c\,\hat{\mathbf{z}}$ | (4$e$) | C |

**Found in:**

---

**Geometry files:**

- CIF: pp. 659
- POSCAR: pp. 659

# $Au_2V$ Structure: A2B_oC12_38_de_ab

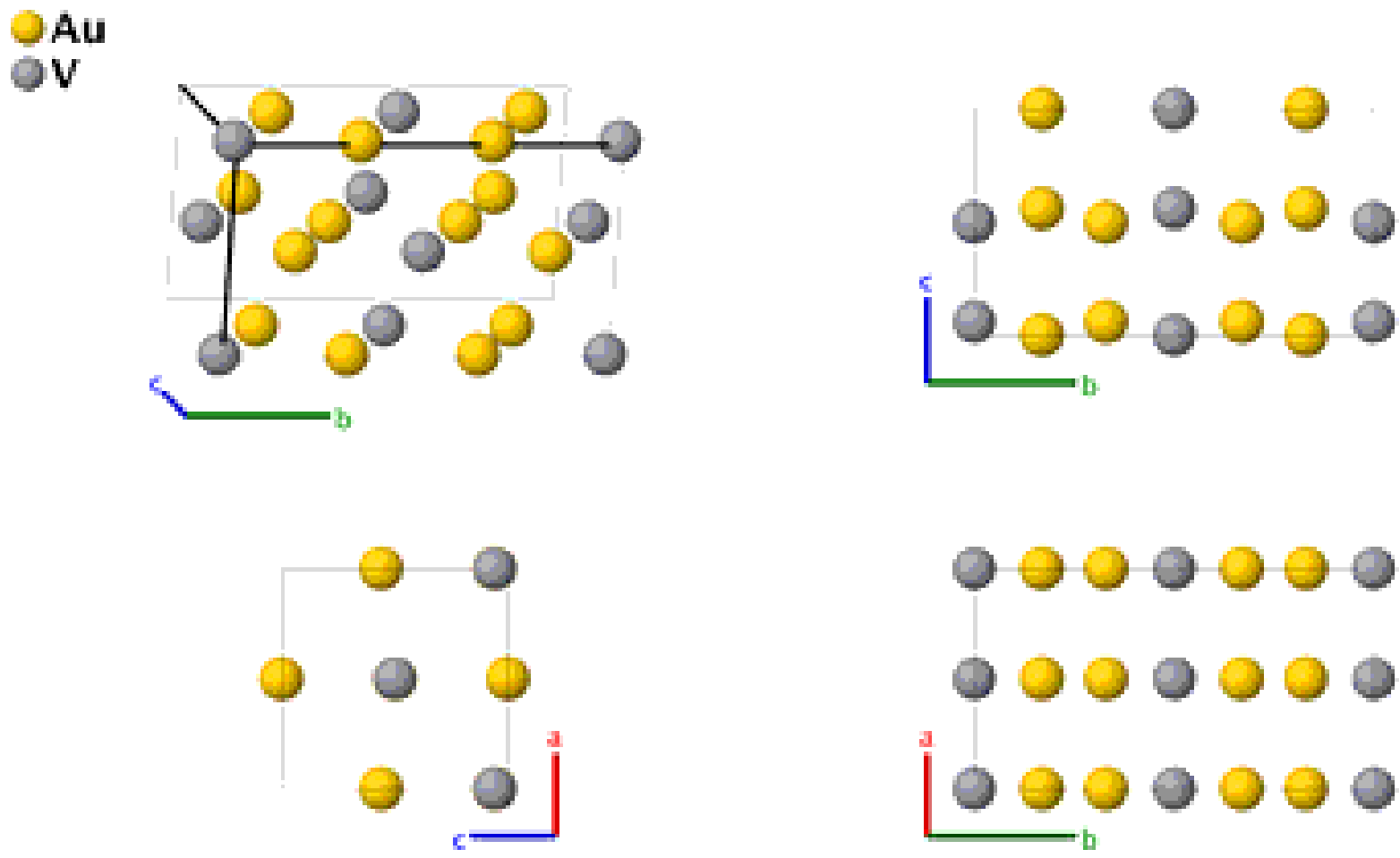

| | | |
|---|---|---|
| **Prototype** | : | $Au_2V$ |
| **AFLOW prototype label** | : | A2B_oC12_38_de_ab |
| ***Strukturbericht* designation** | : | None |
| **Pearson symbol** | : | oC12 |
| **Space group number** | : | 38 |
| **Space group symbol** | : | Amm2 |
| **AFLOW prototype command** | : | `aflow --proto=A2B_oC12_38_de_ab` <br> `--params=`$a, b/a, c/a, z_1, z_2, y_3, z_3, y_4, z_4$ |

**Other compounds with this structure:**

- $Cu_2Ti$, $Pt_2Ta$

- Note that the published atomic positions put the system in the Cmcm space group, despite the author's statement that the system is in the Amm2 space group. We forced this system into the Amm2 space group by slightly shifting the $y_4$ coordinate. If $y_3 = y_4$ then the space group becomes Cmcm.

**Base-centered Orthorhombic primitive vectors:**

$$\mathbf{a}_1 = a\,\hat{\mathbf{x}}$$
$$\mathbf{a}_2 = \frac{1}{2}\,b\,\hat{\mathbf{y}} - \frac{1}{2}\,c\,\hat{\mathbf{z}}$$
$$\mathbf{a}_3 = \frac{1}{2}\,b\,\hat{\mathbf{y}} + \frac{1}{2}\,c\,\hat{\mathbf{z}}$$

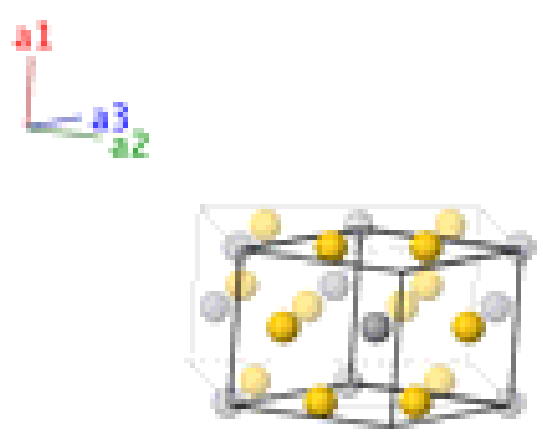

**Basis vectors:**

| | | Lattice Coordinates | | Cartesian Coordinates | Wyckoff Position | Atom Type |
|---|---|---|---|---|---|---|
| $\mathbf{B_1}$ | = | $-z_1\,\mathbf{a_2} + z_1\,\mathbf{a_3}$ | = | $z_1\,c\,\hat{\mathbf{z}}$ | (2$a$) | V I |
| $\mathbf{B_2}$ | = | $\frac{1}{2}\,\mathbf{a_1} - z_2\,\mathbf{a_2} + z_2\,\mathbf{a_3}$ | = | $\frac{1}{2}\,a\,\hat{\mathbf{x}} + z_2\,c\,\hat{\mathbf{z}}$ | (2$b$) | V II |
| $\mathbf{B_3}$ | = | $(y_3 - z_3)\,\mathbf{a_2} + (y_3 + z_3)\,\mathbf{a_3}$ | = | $y_3\,b\,\hat{\mathbf{y}} + z_3\,c\,\hat{\mathbf{z}}$ | $(4d)$ | Au I |
| $\mathbf{B_4}$ | = | $-(y_3 + z_3)\,\mathbf{a_2} + (z_3 - y_3)\,\mathbf{a_3}$ | = | $-y_3\,b\,\hat{\mathbf{y}} + z_3\,c\,\hat{\mathbf{z}}$ | $(4d)$ | Au I |
| $\mathbf{B_5}$ | = | $\frac{1}{2}\,\mathbf{a_1} + (y_4 - z_4)\,\mathbf{a_2} + (y_4 + z_4)\,\mathbf{a_3}$ | = | $\frac{1}{2}\,a\,\hat{\mathbf{x}} + y_4\,b\,\hat{\mathbf{y}} + z_4\,c\,\hat{\mathbf{z}}$ | $(4e)$ | Au II |
| $\mathbf{B_6}$ | = | $\frac{1}{2}\,\mathbf{a_1} - (y_4 + z_4)\,\mathbf{a_2} + (z_4 - y_4)\,\mathbf{a_3}$ | = | $\frac{1}{2}\,a\,\hat{\mathbf{x}} - y_4\,b\,\hat{\mathbf{y}} + z_4\,c\,\hat{\mathbf{z}}$ | $(4e)$ | Au II |

---

**Found in:**

---

**Geometry files:**

- CIF: pp. 659
- POSCAR: pp. 660

# $PtSn_4$ Structure: AB4_oC20_41_a_2b

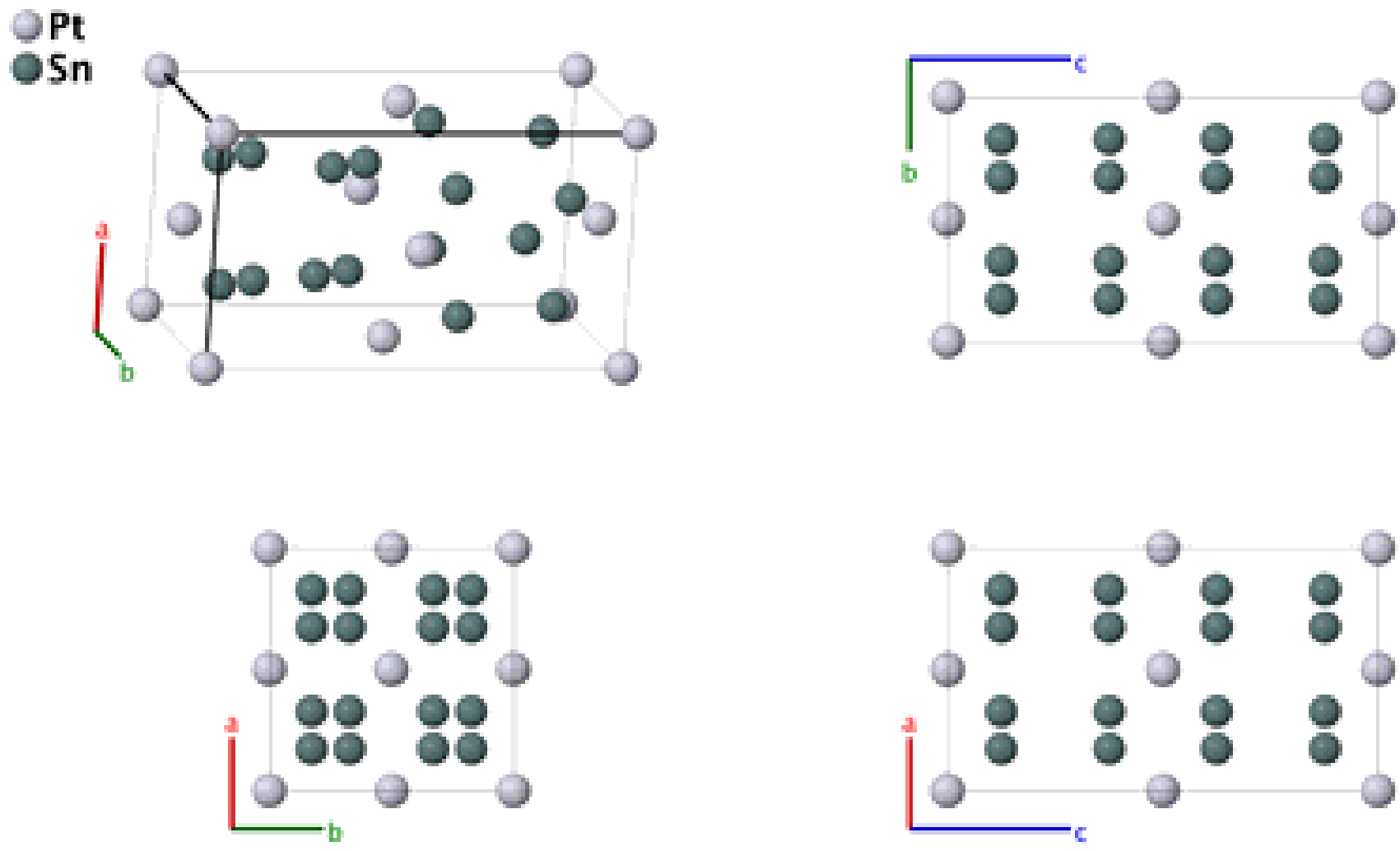

| | | |
|---|---|---|
| **Prototype** | : | $PtSn_4$ |
| **AFLOW prototype label** | : | AB4_oC20_41_a_2b |
| ***Strukturbericht* designation** | : | $D1_c$ |
| **Pearson symbol** | : | oC20 |
| **Space group number** | : | 41 |
| **Space group symbol** | : | Aba2 |
| **AFLOW prototype command** | : | `aflow --proto=AB4_oC20_41_a_2b` <br> `--params=`$a, b/a, c/a, z_1, x_2, y_2, z_2, x_3, y_3, z_3$ |

**Other compounds with this structure:**

- $AuSn_4$, $IrSn_4$, $PdSn_4$

- The published atomic positions have $x_2 = y_3$, $x_3 = y_2$ and $z_2 = -z_3$. This puts the system into space group Ccca. To get space group Aba2 we shifted the $z_3$ position slightly.

**Base-centered Orthorhombic primitive vectors:**

$$\mathbf{a}_1 = a\,\hat{\mathbf{x}}$$
$$\mathbf{a}_2 = \frac{1}{2}\,b\,\hat{\mathbf{y}} - \frac{1}{2}\,c\,\hat{\mathbf{z}}$$
$$\mathbf{a}_3 = \frac{1}{2}\,b\,\hat{\mathbf{y}} + \frac{1}{2}\,c\,\hat{\mathbf{z}}$$

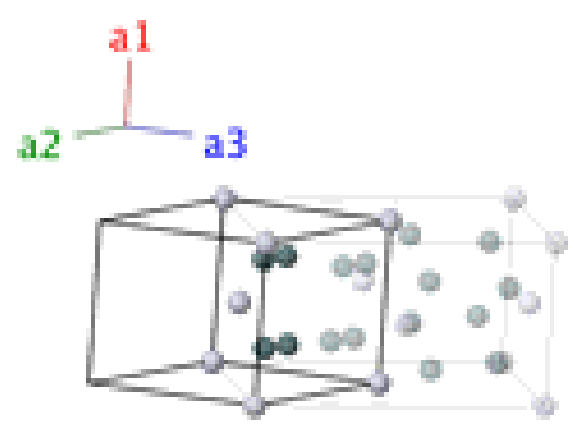

**Basis vectors:**

| | | Lattice Coordinates | | Cartesian Coordinates | Wyckoff Position | Atom Type |
|---|---|---|---|---|---|---|
| $\mathbf{B_1}$ | = | $-z_1\,\mathbf{a_2} + z_1\,\mathbf{a_3}$ | = | $z_1\,c\,\hat{\mathbf{z}}$ | (4$a$) | Pt |
| $\mathbf{B_2}$ | = | $\frac{1}{2}\,\mathbf{a_1} + \left(\frac{1}{2} - z_1\right)\,\mathbf{a_2} + \left(\frac{1}{2} + z_1\right)\,\mathbf{a_3}$ | = | $\frac{1}{2}\,a\,\hat{\mathbf{x}} + \frac{1}{2}\,b\,\hat{\mathbf{y}} + z_1\,c\,\hat{\mathbf{z}}$ | (4$a$) | Pt |
| $\mathbf{B_3}$ | = | $x_2\,\mathbf{a_1} + (y_2 - z_2)\,\mathbf{a_2} + (y_2 + z_2)\,\mathbf{a_3}$ | = | $x_2\,a\,\hat{\mathbf{x}} + y_2\,b\,\hat{\mathbf{y}} + z_2\,c\,\hat{\mathbf{z}}$ | (8$b$) | Sn I |
| $\mathbf{B_4}$ | = | $-x_2\,\mathbf{a_1} - (y_2 + z_2)\,\mathbf{a_2} + (z_2 - y_2)\,\mathbf{a_3}$ | = | $-x_2\,a\,\hat{\mathbf{x}} - y_2\,b\,\hat{\mathbf{y}} + z_2\,c\,\hat{\mathbf{z}}$ | (8$b$) | Sn I |
| $\mathbf{B_5}$ | = | $\left(\frac{1}{2} + x_2\right)\mathbf{a_1} + \left(\frac{1}{2} - y_2 - z_2\right)\mathbf{a_2} + \left(\frac{1}{2} - y_2 + z_2\right)\mathbf{a_3}$ | = | $\left(\frac{1}{2} + x_2\right)a\,\hat{\mathbf{x}} + \left(\frac{1}{2} - y_2\right)b\,\hat{\mathbf{y}} + z_2\,c\,\hat{\mathbf{z}}$ | (8$b$) | Sn I |
| $\mathbf{B_6}$ | = | $\left(\frac{1}{2} - x_2\right)\mathbf{a_1} + \left(\frac{1}{2} + y_2 - z_2\right)\mathbf{a_2} + \left(\frac{1}{2} + y_2 + z_2\right)\mathbf{a_3}$ | = | $\left(\frac{1}{2} - x_2\right)a\,\hat{\mathbf{x}} + \left(\frac{1}{2} + y_2\right)b\,\hat{\mathbf{y}} + z_2\,c\,\hat{\mathbf{z}}$ | (8$b$) | Sn I |
| $\mathbf{B_7}$ | = | $x_3\,\mathbf{a_1} + (y_3 - z_3)\,\mathbf{a_2} + (y_3 + z_3)\,\mathbf{a_3}$ | = | $x_3\,a\,\hat{\mathbf{x}} + y_3\,b\,\hat{\mathbf{y}} + z_3\,c\,\hat{\mathbf{z}}$ | (8$b$) | Sn II |
| $\mathbf{B_8}$ | = | $-x_3\,\mathbf{a_1} - (y_3 + z_3)\,\mathbf{a_2} + (z_3 - y_3)\,\mathbf{a_3}$ | = | $-x_3\,a\,\hat{\mathbf{x}} - y_3\,b\,\hat{\mathbf{y}} + z_3\,c\,\hat{\mathbf{z}}$ | (8$b$) | Sn II |
| $\mathbf{B_9}$ | = | $\left(\frac{1}{2} + x_3\right)\mathbf{a_1} + \left(\frac{1}{2} - y_3 - z_3\right)\mathbf{a_2} + \left(\frac{1}{2} - y_3 + z_3\right)\mathbf{a_3}$ | = | $\left(\frac{1}{2} + x_3\right)a\,\hat{\mathbf{x}} + \left(\frac{1}{2} - y_3\right)b\,\hat{\mathbf{y}} + z_3\,c\,\hat{\mathbf{z}}$ | (8$b$) | Sn II |
| $\mathbf{B_{10}}$ | = | $\left(\frac{1}{2} - x_3\right)\mathbf{a_1} + \left(\frac{1}{2} + y_3 - z_3\right)\mathbf{a_2} + \left(\frac{1}{2} + y_3 + z_3\right)\mathbf{a_3}$ | = | $\left(\frac{1}{2} - x_3\right)a\,\hat{\mathbf{x}} + \left(\frac{1}{2} + y_3\right)b\,\hat{\mathbf{y}} + z_3\,c\,\hat{\mathbf{z}}$ | (8$b$) | Sn II |

---

**Found in:**

---

**Geometry files:**

- CIF: pp. 660
- POSCAR: pp. 660

# $PdSn_2$ ($C_e$) Structure: AB2_oC24_41_2a_2b

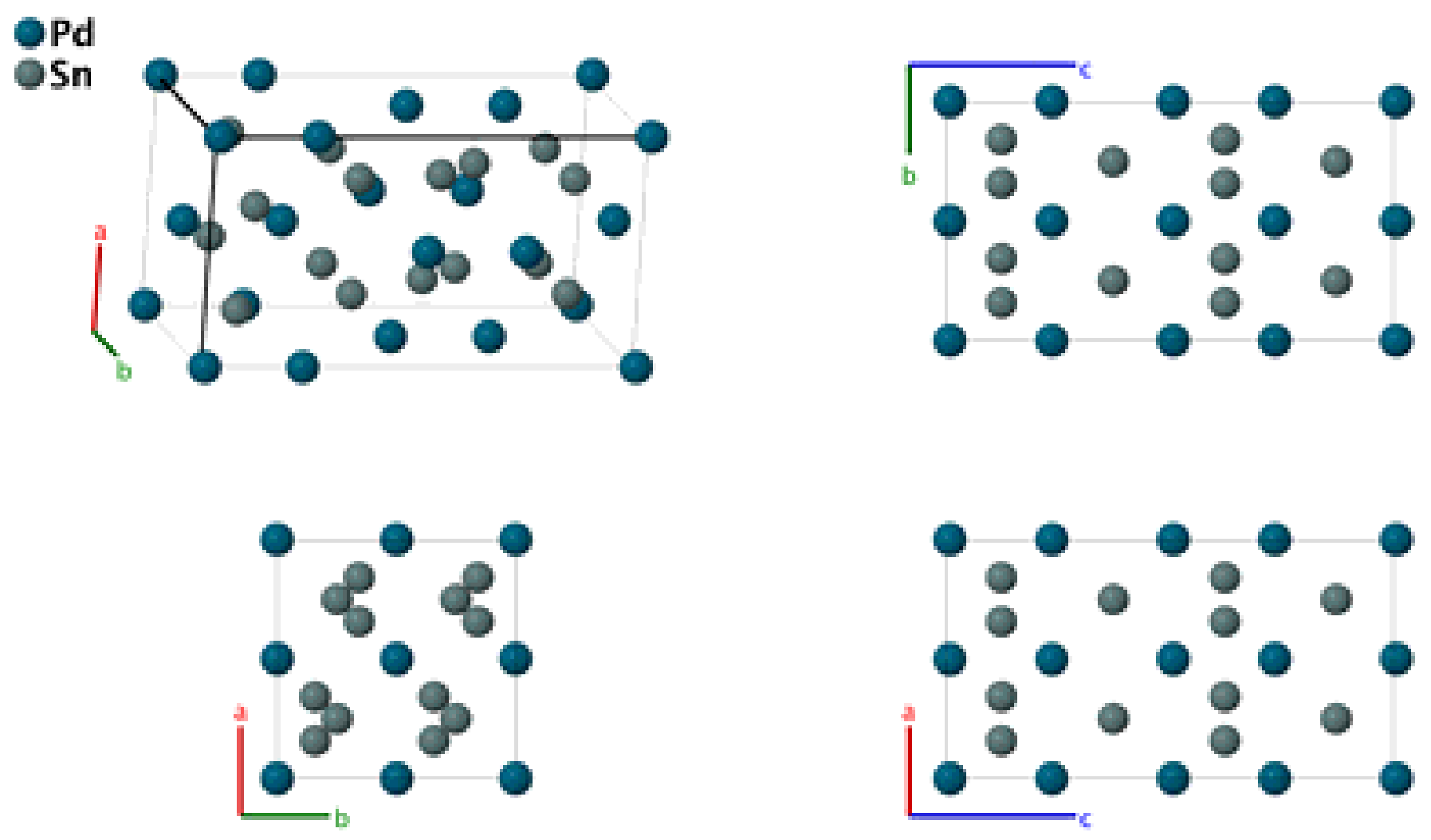

| | | |
|---|---|---|
| **Prototype** | : | $PdSn_2$ |
| **AFLOW prototype label** | : | AB2_oC24_41_2a_2b |
| ***Strukturbericht* designation** | : | $C_e$ |
| **Pearson symbol** | : | oC24 |
| **Space group number** | : | 41 |
| **Space group symbol** | : | Aba2 |
| **AFLOW prototype command** | : | `aflow --proto=AB2_oC24_41_2a_2b` <br> `--params=`$a, b/a, c/a, z_1, z_2, x_3, y_3, z_3, x_4, y_4, z_4$ |

**Other compounds with this structure:**

- $CoGe_2$, $GaGe_3Ni_2$, $RhSn_2$

**Base-centered Orthorhombic primitive vectors:**

$$\mathbf{a}_1 = a\,\hat{\mathbf{x}}$$
$$\mathbf{a}_2 = \frac{1}{2}\,b\,\hat{\mathbf{y}} - \frac{1}{2}\,c\,\hat{\mathbf{z}}$$
$$\mathbf{a}_3 = \frac{1}{2}\,b\,\hat{\mathbf{y}} + \frac{1}{2}\,c\,\hat{\mathbf{z}}$$

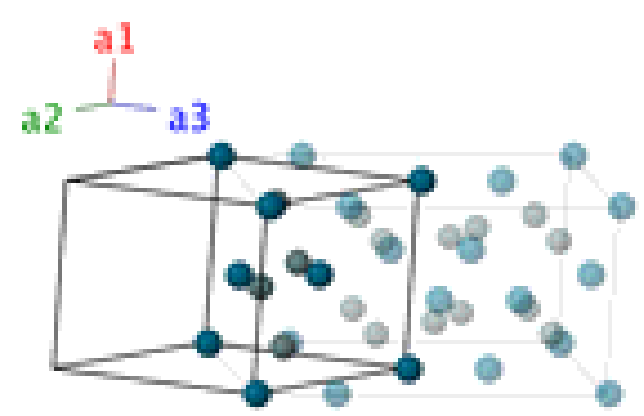

**Basis vectors:**

| | | Lattice Coordinates | | Cartesian Coordinates | Wyckoff Position | Atom Type |
|---|---|---|---|---|---|---|
| $\mathbf{B}_1$ | = | $-z_1\,\mathbf{a}_2 + z_1\,\mathbf{a}_3$ | = | $z_1\,c\,\hat{\mathbf{z}}$ | (4a) | Pd I |
| $\mathbf{B}_2$ | = | $\frac{1}{2}\,\mathbf{a}_1 + \left(\frac{1}{2} - z_1\right)\,\mathbf{a}_2 + \left(\frac{1}{2} + z_1\right)\,\mathbf{a}_3$ | = | $\frac{1}{2}\,a\,\hat{\mathbf{x}} + \frac{1}{2}\,b\,\hat{\mathbf{y}} + z_1\,c\,\hat{\mathbf{z}}$ | (4a) | Pd I |
| $\mathbf{B}_3$ | = | $-z_2\,\mathbf{a}_2 + z_2\,\mathbf{a}_3$ | = | $z_2\,c\,\hat{\mathbf{z}}$ | (4a) | Pd II |
| $\mathbf{B}_4$ | = | $\frac{1}{2}\,\mathbf{a}_1 + \left(\frac{1}{2} - z_2\right)\,\mathbf{a}_2 + \left(\frac{1}{2} + z_2\right)\,\mathbf{a}_3$ | = | $\frac{1}{2}\,a\,\hat{\mathbf{x}} + \frac{1}{2}\,b\,\hat{\mathbf{y}} + z_2\,c\,\hat{\mathbf{z}}$ | (4a) | Pd II |
| $\mathbf{B}_5$ | = | $x_3\,\mathbf{a}_1 + (y_3 - z_3)\,\mathbf{a}_2 + (y_3 + z_3)\,\mathbf{a}_3$ | = | $x_3\,a\,\hat{\mathbf{x}} + y_3\,b\,\hat{\mathbf{y}} + z_3\,c\,\hat{\mathbf{z}}$ | (8b) | Sn I |
| $\mathbf{B_6}$ | = | $-x_3\,\mathbf{a_1} - (y_3+z_3)\,\mathbf{a_2} + (z_3-y_3)\,\mathbf{a_3}$ | = | $-x_3\,a\,\mathbf{\hat{x}} - y_3\,b\,\mathbf{\hat{y}} + z_3\,c\,\mathbf{\hat{z}}$ | (8$b$) | Sn I |
| $\mathbf{B_7}$ | = | $\left(\frac{1}{2}+x_3\right)\mathbf{a_1} + \left(\frac{1}{2}-y_3-z_3\right)\mathbf{a_2} + \left(\frac{1}{2}-y_3+z_3\right)\mathbf{a_3}$ | = | $\left(\frac{1}{2}+x_3\right)a\,\mathbf{\hat{x}} + \left(\frac{1}{2}-y_3\right)b\,\mathbf{\hat{y}} + z_3\,c\,\mathbf{\hat{z}}$ | (8$b$) | Sn I |
| $\mathbf{B_8}$ | = | $\left(\frac{1}{2}-x_3\right)\mathbf{a_1} + \left(\frac{1}{2}+y_3-z_3\right)\mathbf{a_2} + \left(\frac{1}{2}+y_3+z_3\right)\mathbf{a_3}$ | = | $\left(\frac{1}{2}-x_3\right)a\,\mathbf{\hat{x}} + \left(\frac{1}{2}+y_3\right)b\,\mathbf{\hat{y}} + z_3\,c\,\mathbf{\hat{z}}$ | (8$b$) | Sn I |
| $\mathbf{B_9}$ | = | $x_4\,\mathbf{a_1} + (y_4-z_4)\,\mathbf{a_2} + (y_4+z_4)\,\mathbf{a_3}$ | = | $x_4\,a\,\mathbf{\hat{x}} + y_4\,b\,\mathbf{\hat{y}} + z_4\,c\,\mathbf{\hat{z}}$ | (8$b$) | Sn II |
| $\mathbf{B_{10}}$ | = | $-x_4\,\mathbf{a_1} - (y_4+z_4)\,\mathbf{a_2} + (z_4-y_4)\,\mathbf{a_3}$ | = | $-x_4\,a\,\mathbf{\hat{x}} - y_4\,b\,\mathbf{\hat{y}} + z_4\,c\,\mathbf{\hat{z}}$ | (8$b$) | Sn II |
| $\mathbf{B_{11}}$ | = | $\left(\frac{1}{2}+x_4\right)\mathbf{a_1} + \left(\frac{1}{2}-y_4-z_4\right)\mathbf{a_2} + \left(\frac{1}{2}-y_4+z_4\right)\mathbf{a_3}$ | = | $\left(\frac{1}{2}+x_4\right)a\,\mathbf{\hat{x}} + \left(\frac{1}{2}-y_4\right)b\,\mathbf{\hat{y}} + z_4\,c\,\mathbf{\hat{z}}$ | (8$b$) | Sn II |
| $\mathbf{B_{12}}$ | = | $\left(\frac{1}{2}-x_4\right)\mathbf{a_1} + \left(\frac{1}{2}+y_4-z_4\right)\mathbf{a_2} + \left(\frac{1}{2}+y_4+z_4\right)\mathbf{a_3}$ | = | $\left(\frac{1}{2}-x_4\right)a\,\mathbf{\hat{x}} + \left(\frac{1}{2}+y_4\right)b\,\mathbf{\hat{y}} + z_4\,c\,\mathbf{\hat{z}}$ | (8$b$) | Sn II |

**Geometry files:**

- CIF: pp. 660
- POSCAR: pp. 661

# $GeS_2$ (C44) Structure: AB2_oF72_43_ab_3b

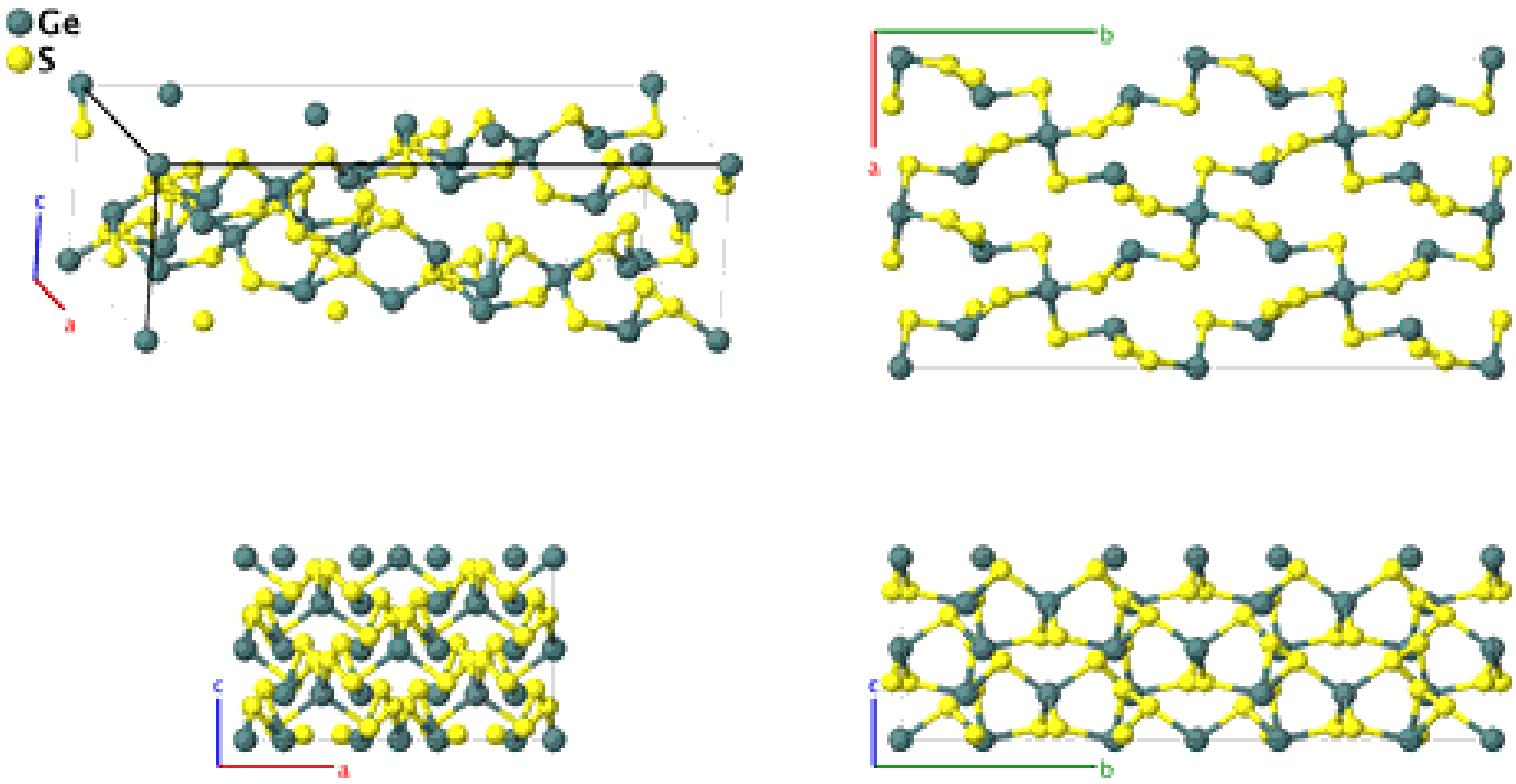

| | | |
|---|---|---|
| **Prototype** | : | $GeS_2$ |
| **AFLOW prototype label** | : | AB2_oF72_43_ab_3b |
| ***Strukturbericht* designation** | : | C44 |
| **Pearson symbol** | : | oF72 |
| **Space group number** | : | 43 |
| **Space group symbol** | : | Fdd2 |
| **AFLOW prototype command** | : | `aflow --proto=AB2_oF72_43_ab_3b` `--params=`$a, b/a, c/a, z_1, x_2, y_2, z_2, x_3, y_3, z_3, x_4, y_4, z_4, x_5, y_5, z_5$ |

**Face-centered Orthorhombic primitive vectors:**

$$\mathbf{a}_1 = \frac{1}{2}\,b\,\hat{\mathbf{y}} + \frac{1}{2}\,c\,\hat{\mathbf{z}}$$
$$\mathbf{a}_2 = \frac{1}{2}\,a\,\hat{\mathbf{x}} + \frac{1}{2}\,c\,\hat{\mathbf{z}}$$
$$\mathbf{a}_3 = \frac{1}{2}\,a\,\hat{\mathbf{x}} + \frac{1}{2}\,b\,\hat{\mathbf{y}}$$

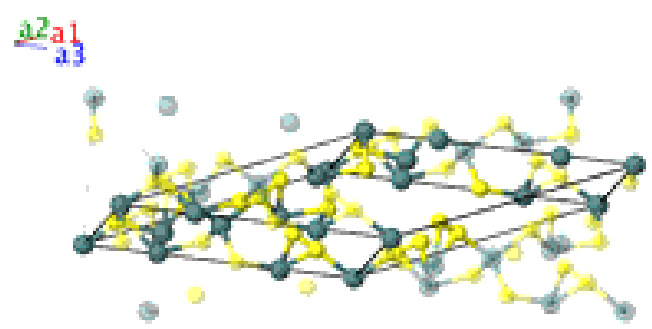

**Basis vectors:**

| | | Lattice Coordinates | | Cartesian Coordinates | Wyckoff Position | Atom Type |
|---|---|---|---|---|---|---|
| $\mathbf{B}_1$ | = | $z_1\,\mathbf{a}_1 + z_1\,\mathbf{a}_2 - z_1\,\mathbf{a}_3$ | = | $z_1\,c\,\hat{\mathbf{z}}$ | (8$a$) | Ge I |
| $\mathbf{B}_2$ | = | $\left(\frac{1}{4}+z_1\right)\mathbf{a}_1 + \left(\frac{1}{4}+z_1\right)\mathbf{a}_2 + \left(\frac{1}{4}-z_1\right)\mathbf{a}_3$ | = | $\frac{1}{4}\,a\,\hat{\mathbf{x}} + \frac{1}{4}\,b\,\hat{\mathbf{y}} + \left(\frac{1}{4}+z_1\right)c\,\hat{\mathbf{z}}$ | (8$a$) | Ge I |
| $\mathbf{B}_3$ | = | $(-x_2+y_2+z_2)\,\mathbf{a}_1 + (x_2-y_2+z_2)\,\mathbf{a}_2 + (x_2+y_2-z_2)\,\mathbf{a}_3$ | = | $x_2\,a\,\hat{\mathbf{x}} + y_2\,b\,\hat{\mathbf{y}} + z_2\,c\,\hat{\mathbf{z}}$ | (16$b$) | Ge II |
| $\mathbf{B}_4$ | = | $(x_2-y_2+z_2)\,\mathbf{a}_1 + (-x_2+y_2+z_2)\,\mathbf{a}_2 - (x_2+y_2+z_2)\,\mathbf{a}_3$ | = | $-x_2\,a\,\hat{\mathbf{x}} - y_2\,b\,\hat{\mathbf{y}} + z_2\,c\,\hat{\mathbf{z}}$ | (16$b$) | Ge II |
| $\mathbf{B_5}$ | = | $\left(\frac{1}{4} - x_2 - y_2 + z_2\right) \mathbf{a_1} + \left(\frac{1}{4} + x_2 + y_2 + z_2\right) \mathbf{a_2} + \left(\frac{1}{4} + x_2 - y_2 - z_2\right) \mathbf{a_3}$ | = | $\left(\frac{1}{4} + x_2\right) a \mathbf{\hat{x}} + \left(\frac{1}{4} - y_2\right) b \mathbf{\hat{y}} + \left(\frac{1}{4} + z_2\right) c \mathbf{\hat{z}}$ | (16$b$) | Ge II |
| $\mathbf{B_6}$ | = | $\left(\frac{1}{4} + x_2 + y_2 + z_2\right) \mathbf{a_1} + \left(\frac{1}{4} - x_2 - y_2 + z_2\right) \mathbf{a_2} + \left(\frac{1}{4} - x_2 + y_2 - z_2\right) \mathbf{a_3}$ | = | $\left(\frac{1}{4} - x_2\right) a \mathbf{\hat{x}} + \left(\frac{1}{4} + y_2\right) b \mathbf{\hat{y}} + \left(\frac{1}{4} + z_2\right) c \mathbf{\hat{z}}$ | (16$b$) | Ge II |
| $\mathbf{B_7}$ | = | $(-x_3 + y_3 + z_3) \mathbf{a_1} + (x_3 - y_3 + z_3) \mathbf{a_2} + (x_3 + y_3 - z_3) \mathbf{a_3}$ | = | $x_3 a \mathbf{\hat{x}} + y_3 b \mathbf{\hat{y}} + z_3 c \mathbf{\hat{z}}$ | (16$b$) | S I |
| $\mathbf{B_8}$ | = | $(x_3 - y_3 + z_3) \mathbf{a_1} + (-x_3 + y_3 + z_3) \mathbf{a_2} - (x_3 + y_3 + z_3) \mathbf{a_3}$ | = | $-x_3 a \mathbf{\hat{x}} - y_3 b \mathbf{\hat{y}} + z_3 c \mathbf{\hat{z}}$ | (16$b$) | S I |
| $\mathbf{B_9}$ | = | $\left(\frac{1}{4} - x_3 - y_3 + z_3\right) \mathbf{a_1} + \left(\frac{1}{4} + x_3 + y_3 + z_3\right) \mathbf{a_2} + \left(\frac{1}{4} + x_3 - y_3 - z_3\right) \mathbf{a_3}$ | = | $\left(\frac{1}{4} + x_3\right) a \mathbf{\hat{x}} + \left(\frac{1}{4} - y_3\right) b \mathbf{\hat{y}} + \left(\frac{1}{4} + z_3\right) c \mathbf{\hat{z}}$ | (16$b$) | S I |
| $\mathbf{B_{10}}$ | = | $\left(\frac{1}{4} + x_3 + y_3 + z_3\right) \mathbf{a_1} + \left(\frac{1}{4} - x_3 - y_3 + z_3\right) \mathbf{a_2} + \left(\frac{1}{4} - x_3 + y_3 - z_3\right) \mathbf{a_3}$ | = | $\left(\frac{1}{4} - x_3\right) a \mathbf{\hat{x}} + \left(\frac{1}{4} + y_3\right) b \mathbf{\hat{y}} + \left(\frac{1}{4} + z_3\right) c \mathbf{\hat{z}}$ | (16$b$) | S I |
| $\mathbf{B_{11}}$ | = | $(-x_4 + y_4 + z_4) \mathbf{a_1} + (x_4 - y_4 + z_4) \mathbf{a_2} + (x_4 + y_4 - z_4) \mathbf{a_3}$ | = | $x_4 a \mathbf{\hat{x}} + y_4 b \mathbf{\hat{y}} + z_4 c \mathbf{\hat{z}}$ | (16$b$) | S II |
| $\mathbf{B_{12}}$ | = | $(x_4 - y_4 + z_4) \mathbf{a_1} + (-x_4 + y_4 + z_4) \mathbf{a_2} - (x_4 + y_4 + z_4) \mathbf{a_3}$ | = | $-x_4 a \mathbf{\hat{x}} - y_4 b \mathbf{\hat{y}} + z_4 c \mathbf{\hat{z}}$ | (16$b$) | S II |
| $\mathbf{B_{13}}$ | = | $\left(\frac{1}{4} - x_4 - y_4 + z_4\right) \mathbf{a_1} + \left(\frac{1}{4} + x_4 + y_4 + z_4\right) \mathbf{a_2} + \left(\frac{1}{4} + x_4 - y_4 - z_4\right) \mathbf{a_3}$ | = | $\left(\frac{1}{4} + x_4\right) a \mathbf{\hat{x}} + \left(\frac{1}{4} - y_4\right) b \mathbf{\hat{y}} + \left(\frac{1}{4} + z_4\right) c \mathbf{\hat{z}}$ | (16$b$) | S II |
| $\mathbf{B_{14}}$ | = | $\left(\frac{1}{4} + x_4 + y_4 + z_4\right) \mathbf{a_1} + \left(\frac{1}{4} - x_4 - y_4 + z_4\right) \mathbf{a_2} + \left(\frac{1}{4} - x_4 + y_4 - z_4\right) \mathbf{a_3}$ | = | $\left(\frac{1}{4} - x_4\right) a \mathbf{\hat{x}} + \left(\frac{1}{4} + y_4\right) b \mathbf{\hat{y}} + \left(\frac{1}{4} + z_4\right) c \mathbf{\hat{z}}$ | (16$b$) | S II |
| $\mathbf{B_{15}}$ | = | $(-x_5 + y_5 + z_5) \mathbf{a_1} + (x_5 - y_5 + z_5) \mathbf{a_2} + (x_5 + y_5 - z_5) \mathbf{a_3}$ | = | $x_5 a \mathbf{\hat{x}} + y_5 b \mathbf{\hat{y}} + z_5 c \mathbf{\hat{z}}$ | (16$b$) | S III |
| $\mathbf{B_{16}}$ | = | $(x_5 - y_5 + z_5) \mathbf{a_1} + (-x_5 + y_5 + z_5) \mathbf{a_2} - (x_5 + y_5 + z_5) \mathbf{a_3}$ | = | $-x_5 a \mathbf{\hat{x}} - y_5 b \mathbf{\hat{y}} + z_5 c \mathbf{\hat{z}}$ | (16$b$) | S III |
| $\mathbf{B_{17}}$ | = | $\left(\frac{1}{4} - x_5 - y_5 + z_5\right) \mathbf{a_1} + \left(\frac{1}{4} + x_5 + y_5 + z_5\right) \mathbf{a_2} + \left(\frac{1}{4} + x_5 - y_5 - z_5\right) \mathbf{a_3}$ | = | $\left(\frac{1}{4} + x_5\right) a \mathbf{\hat{x}} + \left(\frac{1}{4} - y_5\right) b \mathbf{\hat{y}} + \left(\frac{1}{4} + z_5\right) c \mathbf{\hat{z}}$ | (16$b$) | S III |
| $\mathbf{B_{18}}$ | = | $\left(\frac{1}{4} + x_5 + y_5 + z_5\right) \mathbf{a_1} + \left(\frac{1}{4} - x_5 - y_5 + z_5\right) \mathbf{a_2} + \left(\frac{1}{4} - x_5 + y_5 - z_5\right) \mathbf{a_3}$ | = | $\left(\frac{1}{4} - x_5\right) a \mathbf{\hat{x}} + \left(\frac{1}{4} + y_5\right) b \mathbf{\hat{y}} + \left(\frac{1}{4} + z_5\right) c \mathbf{\hat{z}}$ | (16$b$) | S III |

doi:10.1063/1.1749915.

**Found in:**

---

**Geometry files:**

- CIF: pp. 661
- POSCAR: pp. 661

# High-pressure GaAs Structure: AB_oI4_44_a_b

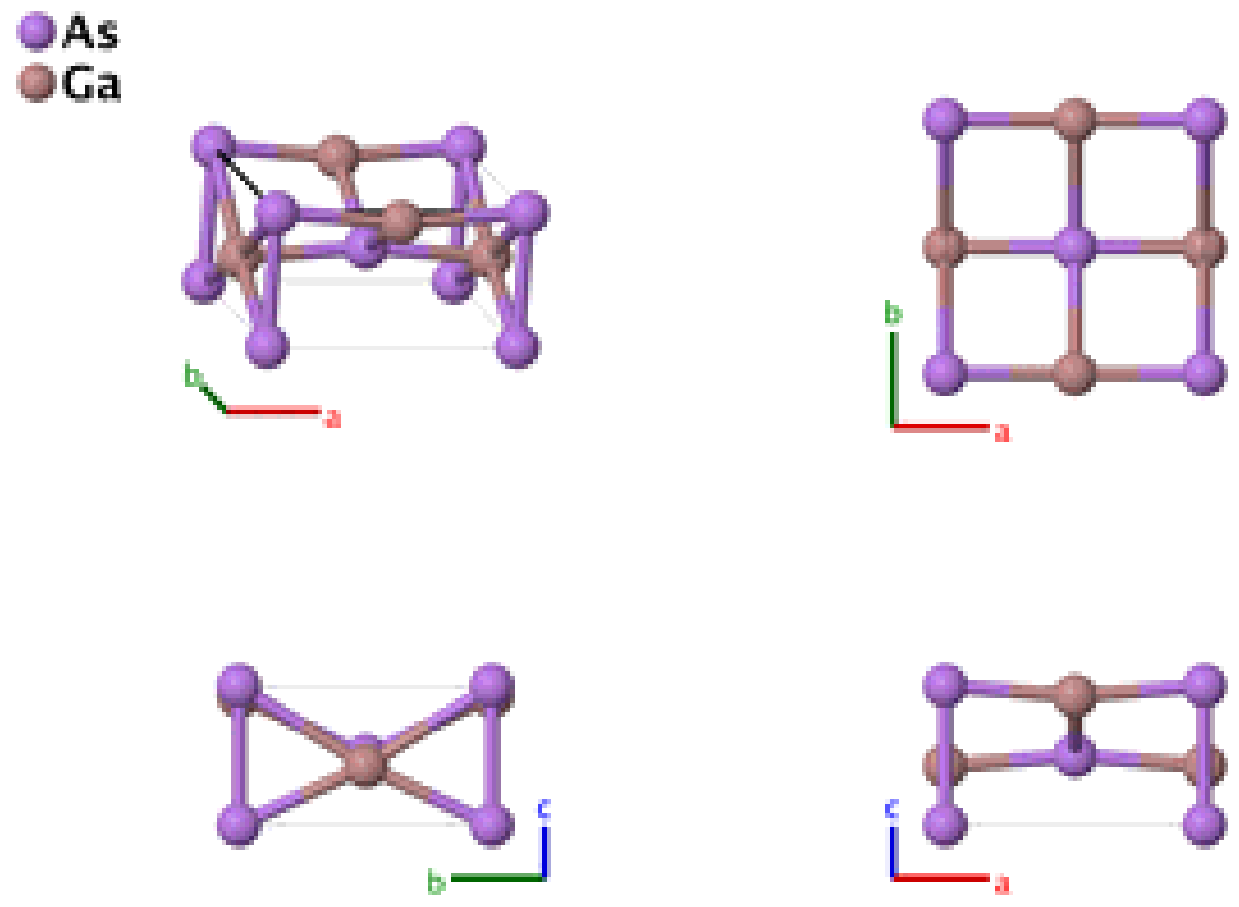

| | | |
|---|---|---|
| **Prototype** | : | GaAs |
| **AFLOW prototype label** | : | AB_oI4_44_a_b |
| ***Strukturbericht* designation** | : | None |
| **Pearson symbol** | : | oI4 |
| **Space group number** | : | 44 |
| **Space group symbol** | : | Imm2 |
| **AFLOW prototype command** | : | `aflow --proto=AB_oI4_44_a_b` `--params=`$a, b/a, c/a, z_1, z_2$ |

- This is a high-pressure phase of GaAs, stable above 24 GPa. The experimental data used here was taken at a pressure of 28.1 GPa. Without loss of generality we can take $z_1 = 0$. When $a = b$ and $z_2 = z_1 + 1/4$ this structure becomes the $\beta$-Sn (A5) structure.

**Body-centered Orthorhombic primitive vectors:**

$$\mathbf{a}_1 = -\frac{1}{2}\,a\,\hat{\mathbf{x}} + \frac{1}{2}\,b\,\hat{\mathbf{y}} + \frac{1}{2}\,c\,\hat{\mathbf{z}}$$
$$\mathbf{a}_2 = \frac{1}{2}\,a\,\hat{\mathbf{x}} - \frac{1}{2}\,b\,\hat{\mathbf{y}} + \frac{1}{2}\,c\,\hat{\mathbf{z}}$$
$$\mathbf{a}_3 = \frac{1}{2}\,a\,\hat{\mathbf{x}} + \frac{1}{2}\,b\,\hat{\mathbf{y}} - \frac{1}{2}\,c\,\hat{\mathbf{z}}$$

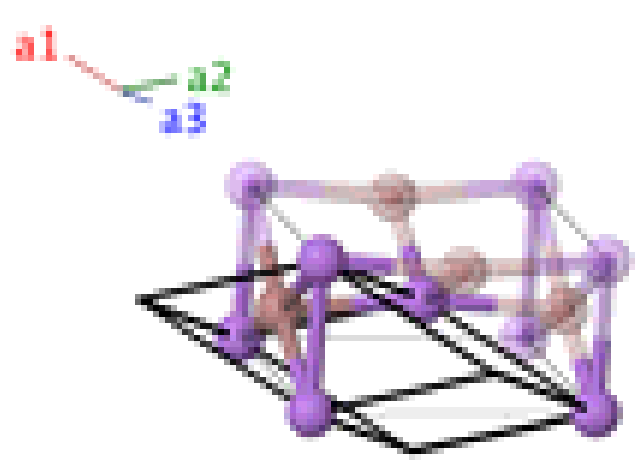

**Basis vectors:**

| | | Lattice Coordinates | | Cartesian Coordinates | Wyckoff Position | Atom Type |
|---|---|---|---|---|---|---|
| $\mathbf{B}_1$ | = | $z_1\,\mathbf{a}_1 + z_1\,\mathbf{a}_2$ | = | $z_1\,c\,\hat{\mathbf{z}}$ | (2$a$) | As |
| $\mathbf{B}_2$ | = | $\left(\frac{1}{2} + z_2\right)\mathbf{a}_1 + z_2\,\mathbf{a}_2 + \frac{1}{2}\,\mathbf{a}_3$ | = | $\frac{1}{2}\,b\,\hat{\mathbf{y}} + z_2\,c\,\hat{\mathbf{z}}$ | (2$b$) | Ga |

**Found in:**

---

**Geometry files:**

- CIF: pp. 661
- POSCAR: pp. 662

# 1212C [$YBa_2Cu_3O_{7-x}$] High-$T_c$ Structure: A2B3C7D_oP13_47_t_aq_eqrs_h

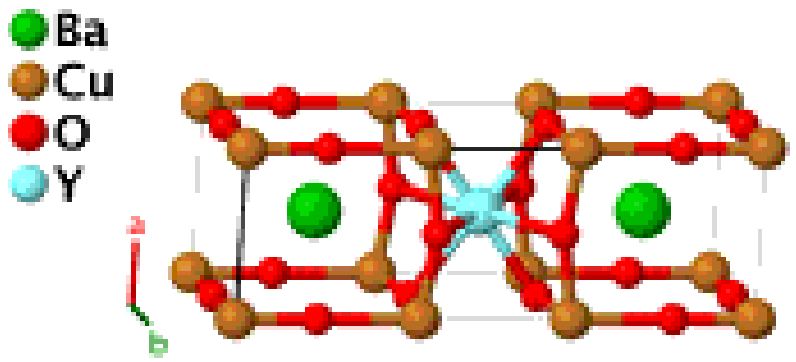

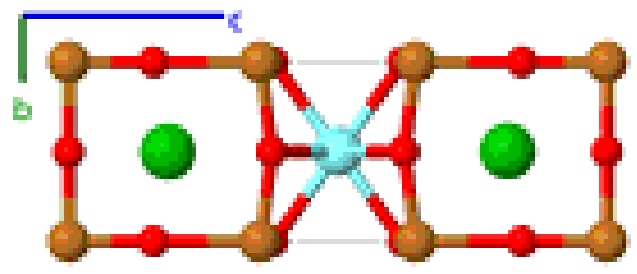
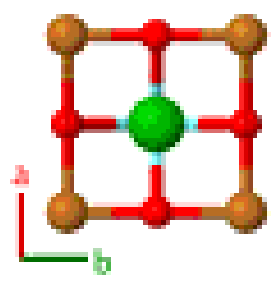
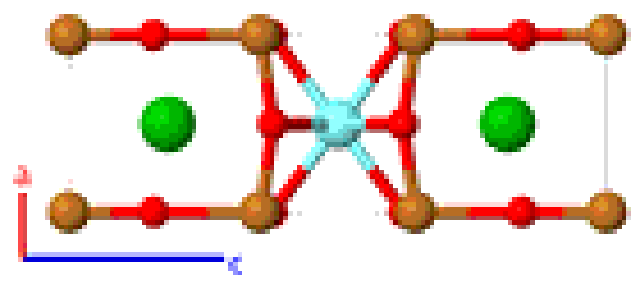

| | | |
|---|---|---|
| **Prototype** | : | $YBa_2Cu_3O_{7-x}$ |
| **AFLOW prototype label** | : | A2B3C7D_oP13_47_t_aq_eqrs_h |
| ***Strukturbericht* designation** | : | None |
| **Pearson symbol** | : | oP13 |
| **Space group number** | : | 47 |
| **Space group symbol** | : | Pmmm |
| **AFLOW prototype command** | : | `aflow --proto=A2B3C7D_oP13_47_t_aq_eqrs_h` <br> `--params=`$a, b/a, c/a, z_4, z_5, z_6, z_7, z_8$ |

**Other compounds with this structure:**

- $GaSr_2(Y,Ca)Cu_2O_7$

- The designation 1212C is from (Shaked, 1994). We will assume that the oxygen concentration is exactly 7. In experiment the O (2s) site is 92% occupied.

**Simple Orthorhombic primitive vectors:**

$$\mathbf{a}_1 = a\,\hat{\mathbf{x}}$$
$$\mathbf{a}_2 = b\,\hat{\mathbf{y}}$$
$$\mathbf{a}_3 = c\,\hat{\mathbf{z}}$$

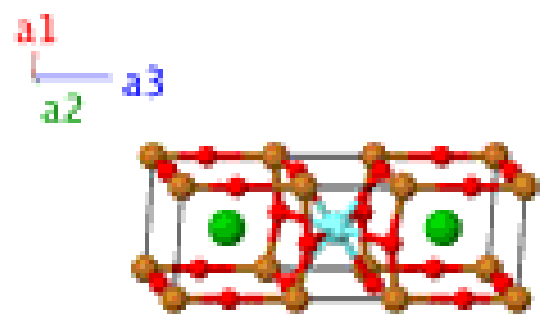

**Basis vectors:**

| | | Lattice Coordinates | | Cartesian Coordinates | Wyckoff Position | Atom Type |
|---|---|---|---|---|---|---|
| $\mathbf{B}_1$ | = | $0\,\mathbf{a}_1 + 0\,\mathbf{a}_2 + 0\,\mathbf{a}_3$ | = | $0\hat{\mathbf{x}} + 0\hat{\mathbf{y}} + 0\hat{\mathbf{z}}$ | (1$a$) | Cu I |
| $\mathbf{B}_2$ | = | $\frac{1}{2}\,\mathbf{a}_2$ | = | $\frac{1}{2}\,b\,\hat{\mathbf{y}}$ | (1$e$) | O I |
| $\mathbf{B}_3$ | = | $\frac{1}{2}\,\mathbf{a}_1 + \frac{1}{2}\,\mathbf{a}_2 + \frac{1}{2}\,\mathbf{a}_3$ | = | $\frac{1}{2}\,a\,\hat{\mathbf{x}} + \frac{1}{2}\,b\,\hat{\mathbf{y}} + \frac{1}{2}\,c\,\hat{\mathbf{z}}$ | (1$h$) | Y |
| $\mathbf{B}_4$ | = | $z_4\,\mathbf{a}_3$ | = | $z_4\,c\,\hat{\mathbf{z}}$ | (2$q$) | Cu II |
| $\mathbf{B_5}$ | = | $-z_4\,\mathbf{a_3}$ | = | $-z_4\,c\,\hat{\mathbf{z}}$ | (2$q$) | Cu II |
| $\mathbf{B_6}$ | = | $z_5\,\mathbf{a_3}$ | = | $z_5\,c\,\hat{\mathbf{z}}$ | (2$q$) | O II |
| $\mathbf{B_7}$ | = | $-z_5\,\mathbf{a_3}$ | = | $-z_5\,c\,\hat{\mathbf{z}}$ | (2$q$) | O II |
| $\mathbf{B_8}$ | = | $\frac{1}{2}\,\mathbf{a_2} + z_6\,\mathbf{a_3}$ | = | $\frac{1}{2}\,b\,\hat{\mathbf{y}} + z_6\,c\,\hat{\mathbf{z}}$ | (2$r$) | O III |
| $\mathbf{B_9}$ | = | $\frac{1}{2}\,\mathbf{a_2} - z_6\,\mathbf{a_3}$ | = | $\frac{1}{2}\,b\,\hat{\mathbf{y}} - z_6\,c\,\hat{\mathbf{z}}$ | (2$r$) | O III |
| $\mathbf{B_{10}}$ | = | $\frac{1}{2}\,\mathbf{a_1} + z_7\,\mathbf{a_3}$ | = | $\frac{1}{2}\,a\,\hat{\mathbf{x}} + z_7\,c\,\hat{\mathbf{z}}$ | (2$s$) | O IV |
| $\mathbf{B_{11}}$ | = | $\frac{1}{2}\,\mathbf{a_1} - z_7\,\mathbf{a_3}$ | = | $\frac{1}{2}\,a\,\hat{\mathbf{x}} - z_7\,c\,\hat{\mathbf{z}}$ | (2$s$) | O IV |
| $\mathbf{B_{12}}$ | = | $\frac{1}{2}\,\mathbf{a_1} + \frac{1}{2}\,\mathbf{a_2} + z_8\,\mathbf{a_3}$ | = | $\frac{1}{2}\,a\,\hat{\mathbf{x}} + \frac{1}{2}\,b\,\hat{\mathbf{y}} + z_8\,c\,\hat{\mathbf{z}}$ | (2$t$) | Ba |
| $\mathbf{B_{13}}$ | = | $\frac{1}{2}\,\mathbf{a_1} + \frac{1}{2}\,\mathbf{a_2} - z_8\,\mathbf{a_3}$ | = | $\frac{1}{2}\,a\,\hat{\mathbf{x}} + \frac{1}{2}\,b\,\hat{\mathbf{y}} - z_8\,c\,\hat{\mathbf{z}}$ | (2$t$) | Ba |

---

---

**Geometry files:**

- CIF: pp. 662
- POSCAR: pp. 662

# $\beta'$-AuCd (B19) Structure: AB_oP4_51_e_f

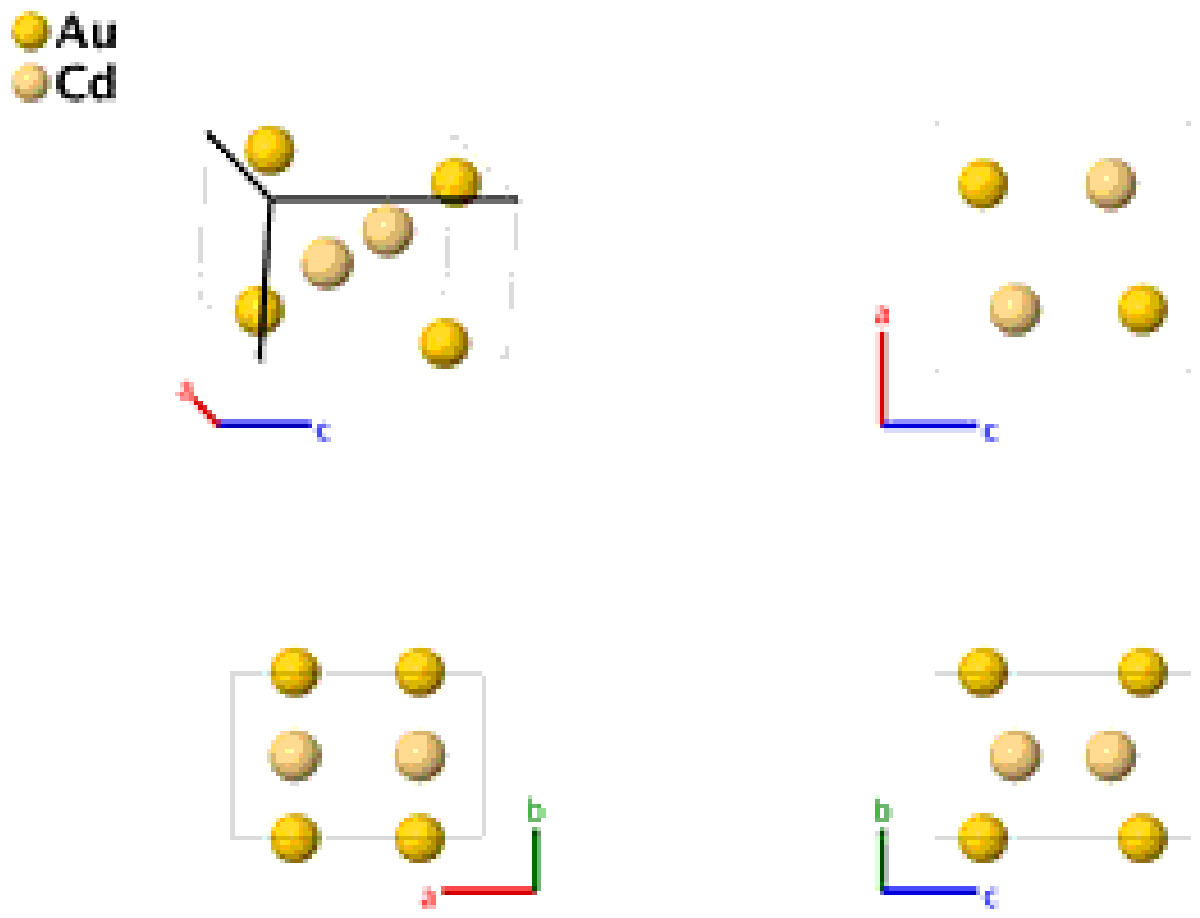

**Prototype** : $\beta'$-AuCd

**AFLOW prototype label** : AB_oP4_51_e_f

***Strukturbericht* designation** : B19

**Pearson symbol** : oP4

**Space group number** : 51

**Space group symbol** : Pmma

**AFLOW prototype command** : `aflow --proto=AB_oP4_51_e_f`
`--params=`$a, b/a, c/a, z_1, z_2$

---

- When $a = b = c$, $z_1 = 1/4$, and $z_2 = 3/4$ the atoms are on the sites of a face-centered cubic lattice. When $a = c$, $z_1 = 1/4$, and $z_2 = 3/4$ the system reduces to the L1$_0$ (AuCu) structure. When $a/b = (8/3)^{2/3}$, $c/b = 3^{1/2}$, $z_1 = 1/3$, and $z_2 = 5/6$, the atoms are on the sites of the hcp structure. Finally, when $z_2 = 1/2 + z_1$ the atoms are at the positions of the $\alpha$-U structure.

---

**Simple Orthorhombic primitive vectors:**

$$\mathbf{a}_1 = a\,\mathbf{\hat{x}}$$
$$\mathbf{a}_2 = b\,\mathbf{\hat{y}}$$
$$\mathbf{a}_3 = c\,\mathbf{\hat{z}}$$

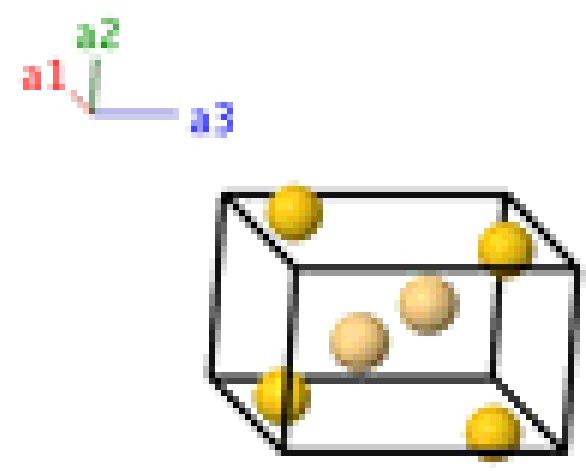

---

**Basis vectors:**

| | | Lattice Coordinates | | Cartesian Coordinates | Wyckoff Position | Atom Type |
|---|---|---|---|---|---|---|
| $\mathbf{B}_1$ | = | $\frac{1}{4}\,\mathbf{a}_1 + z_1\,\mathbf{a}_3$ | = | $\frac{1}{4}\,a\,\mathbf{\hat{x}} + z_1\,c\,\mathbf{\hat{z}}$ | (2$e$) | Au |
| $\mathbf{B}_2$ | = | $\frac{3}{4}\,\mathbf{a}_1 - z_1\,\mathbf{a}_3$ | = | $\frac{3}{4}\,a\,\mathbf{\hat{x}} - z_1\,c\,\mathbf{\hat{z}}$ | (2$e$) | Au |
| $\mathbf{B}_3$ | = | $\frac{1}{4}\,\mathbf{a}_1 + \frac{1}{2}\,\mathbf{a}_2 + z_2\,\mathbf{a}_3$ | = | $\frac{1}{4}\,a\,\mathbf{\hat{x}} + \frac{1}{2}\,b\,\mathbf{\hat{y}} + z_2\,c\,\mathbf{\hat{z}}$ | (2$f$) | Cd |
| $\mathbf{B}_4$ | = | $\frac{3}{4}\,\mathbf{a}_1 + \frac{1}{2}\,\mathbf{a}_2 - z_2\,\mathbf{a}_3$ | = | $\frac{3}{4}\,a\,\mathbf{\hat{x}} + \frac{1}{2}\,b\,\mathbf{\hat{y}} - z_2\,c\,\mathbf{\hat{z}}$ | (2$f$) | Cd |

**Found in:**

---

**Geometry files:**

- CIF: pp. 662
- POSCAR: pp. 663

# $Sb_2O_3$ ($D5_{11}$) Structure: A3B2_oP20_56_ce_e

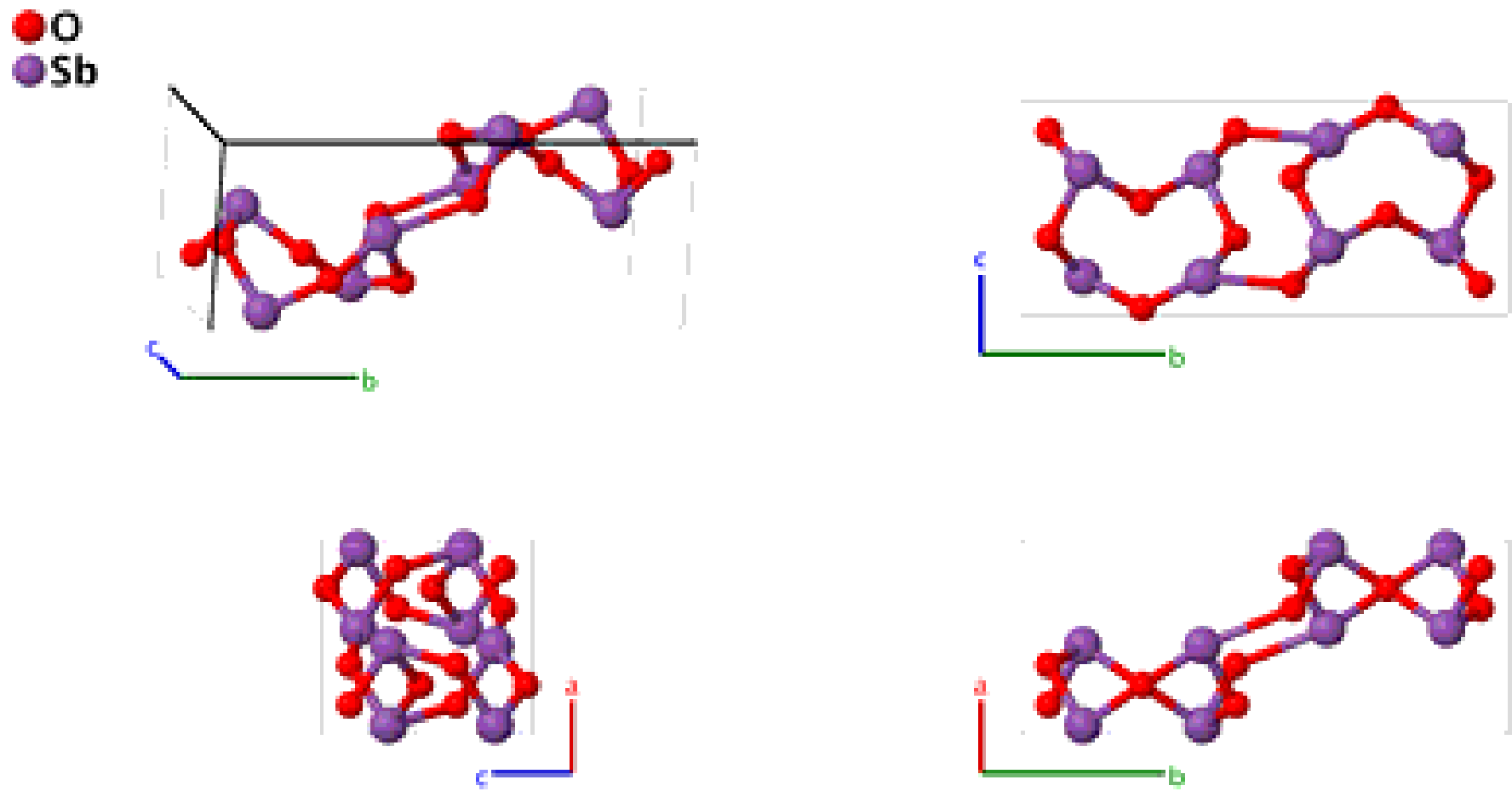

| | | |
|---|---|---|
| **Prototype** | : | $Sb_2O_3$ |
| **AFLOW prototype label** | : | A3B2_oP20_56_ce_e |
| ***Strukturbericht* designation** | : | $D5_{11}$ |
| **Pearson symbol** | : | oP20 |
| **Space group number** | : | 56 |
| **Space group symbol** | : | Pccn |
| **AFLOW prototype command** | : | `aflow --proto=A3B2_oP20_56_ce_e`<br>`--params=`$a, b/a, c/a, z_1, x_2, y_2, z_2, x_3, y_3, z_3$ |

**Simple Orthorhombic primitive vectors:**

$$\mathbf{a}_1 = a\,\hat{\mathbf{x}}$$
$$\mathbf{a}_2 = b\,\hat{\mathbf{y}}$$
$$\mathbf{a}_3 = c\,\hat{\mathbf{z}}$$

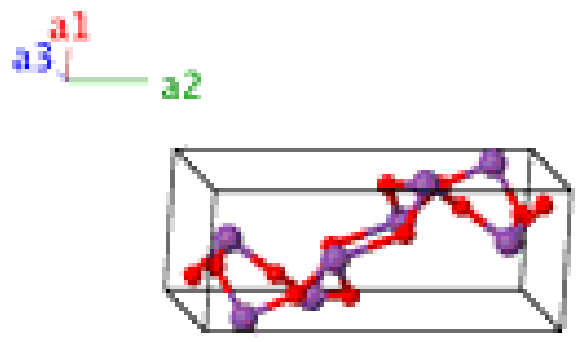

**Basis vectors:**

| | | Lattice Coordinates | | Cartesian Coordinates | Wyckoff Position | Atom Type |
|---|---|---|---|---|---|---|
| $\mathbf{B}_1$ | = | $\frac{1}{4}\,\mathbf{a}_1 + \frac{1}{4}\,\mathbf{a}_2 + z_1\,\mathbf{a}_3$ | = | $\frac{1}{4}\,a\,\hat{\mathbf{x}} + \frac{1}{4}\,b\,\hat{\mathbf{y}} + z_1\,c\,\hat{\mathbf{z}}$ | (4c) | O I |
| $\mathbf{B}_2$ | = | $\frac{3}{4}\,\mathbf{a}_1 + \frac{3}{4}\,\mathbf{a}_2 + \left(\frac{1}{2} - z_1\right)\,\mathbf{a}_3$ | = | $\frac{3}{4}\,a\,\hat{\mathbf{x}} + \frac{3}{4}\,b\,\hat{\mathbf{y}} + \left(\frac{1}{2} - z_1\right)\,c\,\hat{\mathbf{z}}$ | (4c) | O I |
| $\mathbf{B}_3$ | = | $\frac{3}{4}\,\mathbf{a}_1 + \frac{3}{4}\,\mathbf{a}_2 - z_1\,\mathbf{a}_3$ | = | $\frac{3}{4}\,a\,\hat{\mathbf{x}} + \frac{3}{4}\,b\,\hat{\mathbf{y}} - z_1\,c\,\hat{\mathbf{z}}$ | (4c) | O I |
| $\mathbf{B}_4$ | = | $\frac{1}{4}\,\mathbf{a}_1 + \frac{1}{4}\,\mathbf{a}_2 + \left(\frac{1}{2} + z_1\right)\,\mathbf{a}_3$ | = | $\frac{1}{4}\,a\,\hat{\mathbf{x}} + \frac{1}{4}\,b\,\hat{\mathbf{y}} + \left(\frac{1}{2} + z_1\right)\,c\,\hat{\mathbf{z}}$ | (4c) | O I |
| $\mathbf{B}_5$ | = | $x_2\,\mathbf{a}_1 + y_2\,\mathbf{a}_2 + z_2\,\mathbf{a}_3$ | = | $x_2\,a\,\hat{\mathbf{x}} + y_2\,b\,\hat{\mathbf{y}} + z_2\,c\,\hat{\mathbf{z}}$ | (8e) | O II |
| $\mathbf{B}_6$ | = | $\left(\frac{1}{2} - x_2\right)\,\mathbf{a}_1 + \left(\frac{1}{2} - y_2\right)\,\mathbf{a}_2 + z_2\,\mathbf{a}_3$ | = | $\left(\frac{1}{2} - x_2\right)\,a\,\hat{\mathbf{x}} + \left(\frac{1}{2} - y_2\right)\,b\,\hat{\mathbf{y}} + z_2\,c\,\hat{\mathbf{z}}$ | (8e) | O II |
| $\mathbf{B}_7$ | = | $-x_2\,\mathbf{a}_1 + \left(\frac{1}{2} + y_2\right)\,\mathbf{a}_2 + \left(\frac{1}{2} - z_2\right)\,\mathbf{a}_3$ | = | $-x_2\,a\,\hat{\mathbf{x}} + \left(\frac{1}{2} + y_2\right)\,b\,\hat{\mathbf{y}} + \left(\frac{1}{2} - z_2\right)\,c\,\hat{\mathbf{z}}$ | (8e) | O II |
| $\mathbf{B}_8$ | = | $\left(\frac{1}{2} + x_2\right)\,\mathbf{a}_1 - y_2\,\mathbf{a}_2 + \left(\frac{1}{2} - z_2\right)\,\mathbf{a}_3$ | = | $\left(\frac{1}{2} + x_2\right)\,a\,\hat{\mathbf{x}} - y_2\,b\,\hat{\mathbf{y}} + \left(\frac{1}{2} - z_2\right)\,c\,\hat{\mathbf{z}}$ | (8e) | O II |
| $\mathbf{B}_9$ | = | $-x_2\,\mathbf{a}_1 - y_2\,\mathbf{a}_2 - z_2\,\mathbf{a}_3$ | = | $-x_2\,a\,\hat{\mathbf{x}} - y_2\,b\,\hat{\mathbf{y}} - z_2\,c\,\hat{\mathbf{z}}$ | (8e) | O II |
| $\mathbf{B_{10}}$ | = | $\left(\frac{1}{2}+x_2\right)\mathbf{a_1}+\left(\frac{1}{2}+y_2\right)\mathbf{a_2}-z_2\,\mathbf{a_3}$ | = | $\left(\frac{1}{2}+x_2\right)a\,\mathbf{\hat{x}}+\left(\frac{1}{2}+y_2\right)b\,\mathbf{\hat{y}}-z_2\,c\,\mathbf{\hat{z}}$ | (8$e$) | O II |
| $\mathbf{B_{11}}$ | = | $x_2\,\mathbf{a_1}+\left(\frac{1}{2}-y_2\right)\mathbf{a_2}+\left(\frac{1}{2}+z_2\right)\mathbf{a_3}$ | = | $x_2\,a\,\mathbf{\hat{x}}+\left(\frac{1}{2}-y_2\right)b\,\mathbf{\hat{y}}+\left(\frac{1}{2}+z_2\right)c\,\mathbf{\hat{z}}$ | (8$e$) | O II |
| $\mathbf{B_{12}}$ | = | $\left(\frac{1}{2}-x_2\right)\mathbf{a_1}+y_2\,\mathbf{a_2}+\left(\frac{1}{2}+z_2\right)\mathbf{a_3}$ | = | $\left(\frac{1}{2}-x_2\right)a\,\mathbf{\hat{x}}+y_2\,b\,\mathbf{\hat{y}}+\left(\frac{1}{2}+z_2\right)c\,\mathbf{\hat{z}}$ | (8$e$) | O II |
| $\mathbf{B_{13}}$ | = | $x_3\,\mathbf{a_1}+y_3\,\mathbf{a_2}+z_3\,\mathbf{a_3}$ | = | $x_3\,a\,\mathbf{\hat{x}}+y_3\,b\,\mathbf{\hat{y}}+z_3\,c\,\mathbf{\hat{z}}$ | (8$e$) | Sb |
| $\mathbf{B_{14}}$ | = | $\left(\frac{1}{2}-x_3\right)\mathbf{a_1}+\left(\frac{1}{2}-y_3\right)\mathbf{a_2}+z_3\,\mathbf{a_3}$ | = | $\left(\frac{1}{2}-x_3\right)a\,\mathbf{\hat{x}}+\left(\frac{1}{2}-y_3\right)b\,\mathbf{\hat{y}}+z_3\,c\,\mathbf{\hat{z}}$ | (8$e$) | Sb |
| $\mathbf{B_{15}}$ | = | $-x_3\,\mathbf{a_1}+\left(\frac{1}{2}+y_3\right)\mathbf{a_2}+\left(\frac{1}{2}-z_3\right)\mathbf{a_3}$ | = | $-x_3\,a\,\mathbf{\hat{x}}+\left(\frac{1}{2}+y_3\right)b\,\mathbf{\hat{y}}+\left(\frac{1}{2}-z_3\right)c\,\mathbf{\hat{z}}$ | (8$e$) | Sb |
| $\mathbf{B_{16}}$ | = | $\left(\frac{1}{2}+x_3\right)\mathbf{a_1}-y_3\,\mathbf{a_2}+\left(\frac{1}{2}-z_3\right)\mathbf{a_3}$ | = | $\left(\frac{1}{2}+x_3\right)a\,\mathbf{\hat{x}}-y_3\,b\,\mathbf{\hat{y}}+\left(\frac{1}{2}-z_3\right)c\,\mathbf{\hat{z}}$ | (8$e$) | Sb |
| $\mathbf{B_{17}}$ | = | $-x_3\,\mathbf{a_1}-y_3\,\mathbf{a_2}-z_3\,\mathbf{a_3}$ | = | $-x_3\,a\,\mathbf{\hat{x}}-y_3\,b\,\mathbf{\hat{y}}-z_3\,c\,\mathbf{\hat{z}}$ | (8$e$) | Sb |
| $\mathbf{B_{18}}$ | = | $\left(\frac{1}{2}+x_3\right)\mathbf{a_1}+\left(\frac{1}{2}+y_3\right)\mathbf{a_2}-z_3\,\mathbf{a_3}$ | = | $\left(\frac{1}{2}+x_3\right)a\,\mathbf{\hat{x}}+\left(\frac{1}{2}+y_3\right)b\,\mathbf{\hat{y}}-z_3\,c\,\mathbf{\hat{z}}$ | (8$e$) | Sb |
| $\mathbf{B_{19}}$ | = | $x_3\,\mathbf{a_1}+\left(\frac{1}{2}-y_3\right)\mathbf{a_2}+\left(\frac{1}{2}+z_3\right)\mathbf{a_3}$ | = | $x_3\,a\,\mathbf{\hat{x}}+\left(\frac{1}{2}-y_3\right)b\,\mathbf{\hat{y}}+\left(\frac{1}{2}+z_3\right)c\,\mathbf{\hat{z}}$ | (8$e$) | Sb |
| $\mathbf{B_{20}}$ | = | $\left(\frac{1}{2}-x_3\right)\mathbf{a_1}+y_3\,\mathbf{a_2}+\left(\frac{1}{2}+z_3\right)\mathbf{a_3}$ | = | $\left(\frac{1}{2}-x_3\right)a\,\mathbf{\hat{x}}+y_3\,b\,\mathbf{\hat{y}}+\left(\frac{1}{2}+z_3\right)c\,\mathbf{\hat{z}}$ | (8$e$) | Sb |

**Found in:**

**Geometry files:**

- CIF: pp. 663
- POSCAR: pp. 663

# KCNS ($F5_9$) Structure: ABCD_oP16_57_d_c_d_d

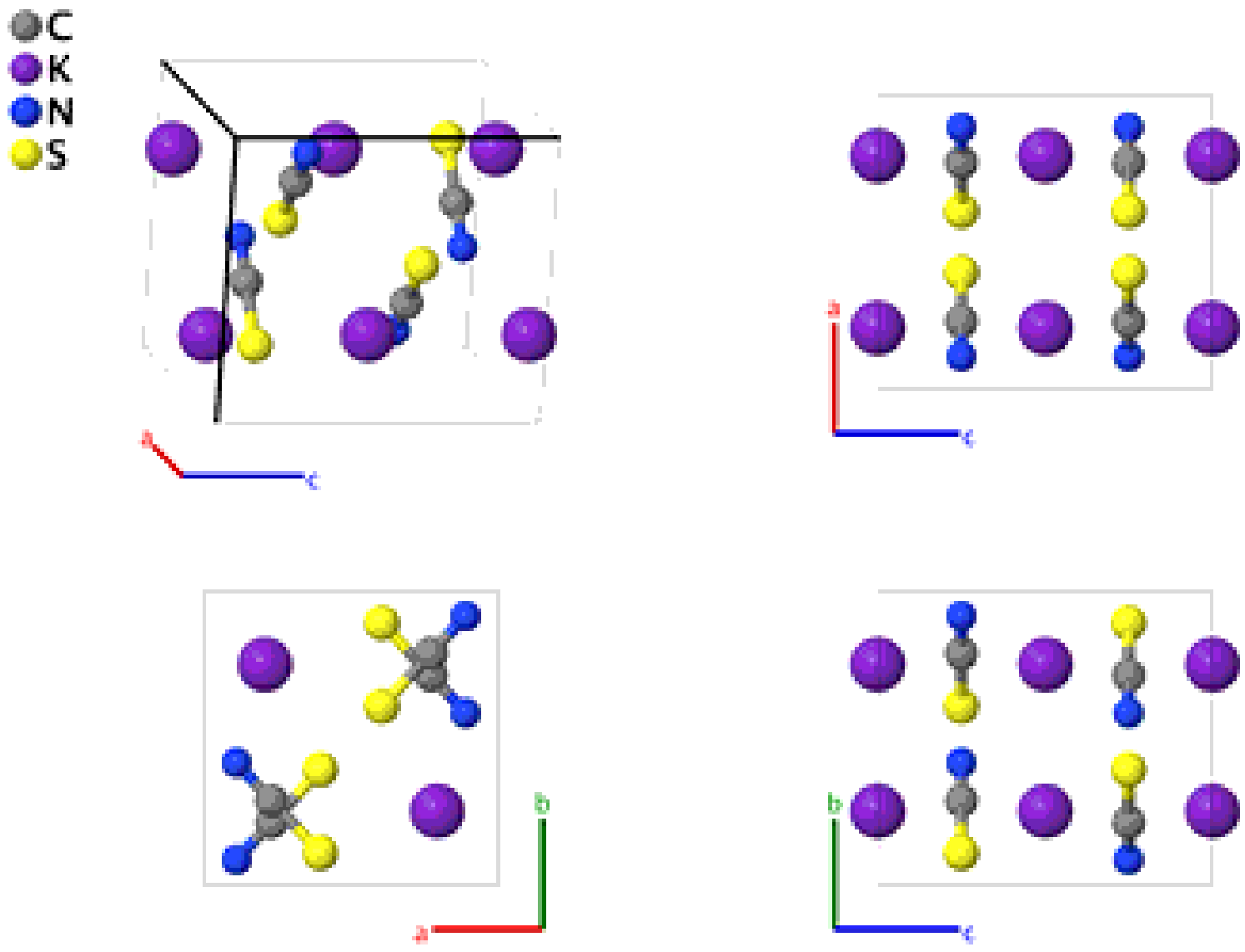

| | | |
|---|---|---|
| **Prototype** | : | KCNS |
| **AFLOW prototype label** | : | ABCD_oP16_57_d_c_d_d |
| ***Strukturbericht* designation** | : | $F5_9$ |
| **Pearson symbol** | : | oP16 |
| **Space group number** | : | 57 |
| **Space group symbol** | : | Pbcm |
| **AFLOW prototype command** | : | `aflow --proto=ABCD_oP16_57_d_c_d_d`<br>`--params=`$a, b/a, c/a, x_1, x_2, y_2, x_3, y_3, x_4, y_4$ |

**Simple Orthorhombic primitive vectors:**

$$\mathbf{a}_1 = a\,\hat{\mathbf{x}}$$
$$\mathbf{a}_2 = b\,\hat{\mathbf{y}}$$
$$\mathbf{a}_3 = c\,\hat{\mathbf{z}}$$

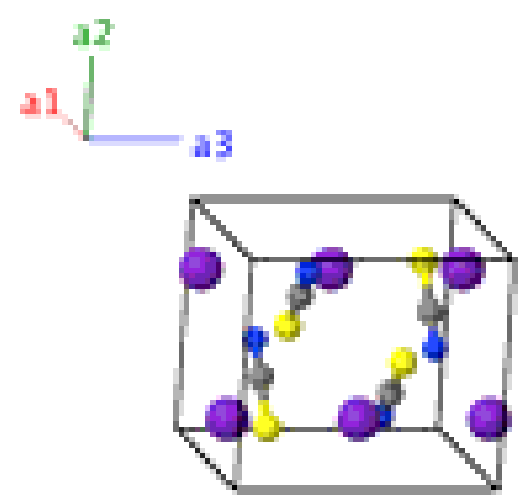

**Basis vectors:**

| | | Lattice Coordinates | | Cartesian Coordinates | Wyckoff Position | Atom Type |
|---|---|---|---|---|---|---|
| $\mathbf{B}_1$ | = | $x_1\,\mathbf{a}_1 + \frac{1}{4}\,\mathbf{a}_2$ | = | $x_1\,a\,\hat{\mathbf{x}} + \frac{1}{4}\,b\,\hat{\mathbf{y}}$ | (4c) | K |
| $\mathbf{B}_2$ | = | $-x_1\,\mathbf{a}_1 + \frac{3}{4}\,\mathbf{a}_2 + \frac{1}{2}\,\mathbf{a}_3$ | = | $-x_1\,a\,\hat{\mathbf{x}} + \frac{3}{4}\,b\,\hat{\mathbf{y}} + \frac{1}{2}\,c\,\hat{\mathbf{z}}$ | (4c) | K |
| $\mathbf{B}_3$ | = | $-x_1\,\mathbf{a}_1 + \frac{3}{4}\,\mathbf{a}_2$ | = | $-x_1\,a\,\hat{\mathbf{x}} + \frac{3}{4}\,b\,\hat{\mathbf{y}}$ | (4c) | K |
| $\mathbf{B}_4$ | = | $x_1\,\mathbf{a}_1 + \frac{1}{4}\,\mathbf{a}_2 + \frac{1}{2}\,\mathbf{a}_3$ | = | $x_1\,a\,\hat{\mathbf{x}} + \frac{1}{4}\,b\,\hat{\mathbf{y}} + \frac{1}{2}\,c\,\hat{\mathbf{z}}$ | (4c) | K |
| $\mathbf{B}_5$ | = | $x_2\,\mathbf{a}_1 + y_2\,\mathbf{a}_2 + \frac{1}{4}\,\mathbf{a}_3$ | = | $x_2\,a\,\hat{\mathbf{x}} + y_2\,b\,\hat{\mathbf{y}} + \frac{1}{4}\,c\,\hat{\mathbf{z}}$ | (4d) | C |
| $\mathbf{B}_6$ | = | $-x_2\,\mathbf{a}_1 - y_2\,\mathbf{a}_2 + \frac{3}{4}\,\mathbf{a}_3$ | = | $-x_2\,a\,\hat{\mathbf{x}} - y_2\,b\,\hat{\mathbf{y}} + \frac{3}{4}\,c\,\hat{\mathbf{z}}$ | (4d) | C |
| $\mathbf{B_7}$ | = | $-x_2\,\mathbf{a_1} + \left(\frac{1}{2} + y_2\right)\,\mathbf{a_2} + \frac{1}{4}\,\mathbf{a_3}$ | = | $-x_2\,a\,\hat{\mathbf{x}} + \left(\frac{1}{2} + y_2\right)\,b\,\hat{\mathbf{y}} + \frac{1}{4}\,c\,\hat{\mathbf{z}}$ | (4$d$) | C |
| $\mathbf{B_8}$ | = | $x_2\,\mathbf{a_1} + \left(\frac{1}{2} - y_2\right)\,\mathbf{a_2} + \frac{3}{4}\,\mathbf{a_3}$ | = | $x_2\,a\,\hat{\mathbf{x}} + \left(\frac{1}{2} - y_2\right)\,b\,\hat{\mathbf{y}} + \frac{3}{4}\,c\,\hat{\mathbf{z}}$ | (4$d$) | C |
| $\mathbf{B_9}$ | = | $x_3\,\mathbf{a_1} + y_3\,\mathbf{a_2} + \frac{1}{4}\,\mathbf{a_3}$ | = | $x_3\,a\,\hat{\mathbf{x}} + y_3\,b\,\hat{\mathbf{y}} + \frac{1}{4}\,c\,\hat{\mathbf{z}}$ | (4$d$) | N |
| $\mathbf{B_{10}}$ | = | $-x_3\,\mathbf{a_1} - y_3\,\mathbf{a_2} + \frac{3}{4}\,\mathbf{a_3}$ | = | $-x_3\,a\,\hat{\mathbf{x}} - y_3\,b\,\hat{\mathbf{y}} + \frac{3}{4}\,c\,\hat{\mathbf{z}}$ | (4$d$) | N |
| $\mathbf{B_{11}}$ | = | $-x_3\,\mathbf{a_1} + \left(\frac{1}{2} + y_3\right)\,\mathbf{a_2} + \frac{1}{4}\,\mathbf{a_3}$ | = | $-x_3\,a\,\hat{\mathbf{x}} + \left(\frac{1}{2} + y_3\right)\,b\,\hat{\mathbf{y}} + \frac{1}{4}\,c\,\hat{\mathbf{z}}$ | (4$d$) | N |
| $\mathbf{B_{12}}$ | = | $x_3\,\mathbf{a_1} + \left(\frac{1}{2} - y_3\right)\,\mathbf{a_2} + \frac{3}{4}\,\mathbf{a_3}$ | = | $x_3\,a\,\hat{\mathbf{x}} + \left(\frac{1}{2} - y_3\right)\,b\,\hat{\mathbf{y}} + \frac{3}{4}\,c\,\hat{\mathbf{z}}$ | (4$d$) | N |
| $\mathbf{B_{13}}$ | = | $x_4\,\mathbf{a_1} + y_4\,\mathbf{a_2} + \frac{1}{4}\,\mathbf{a_3}$ | = | $x_4\,a\,\hat{\mathbf{x}} + y_4\,b\,\hat{\mathbf{y}} + \frac{1}{4}\,c\,\hat{\mathbf{z}}$ | (4$d$) | S |
| $\mathbf{B_{14}}$ | = | $-x_4\,\mathbf{a_1} - y_4\,\mathbf{a_2} + \frac{3}{4}\,\mathbf{a_3}$ | = | $-x_4\,a\,\hat{\mathbf{x}} - y_4\,b\,\hat{\mathbf{y}} + \frac{3}{4}\,c\,\hat{\mathbf{z}}$ | (4$d$) | S |
| $\mathbf{B_{15}}$ | = | $-x_4\,\mathbf{a_1} + \left(\frac{1}{2} + y_4\right)\,\mathbf{a_2} + \frac{1}{4}\,\mathbf{a_3}$ | = | $-x_4\,a\,\hat{\mathbf{x}} + \left(\frac{1}{2} + y_4\right)\,b\,\hat{\mathbf{y}} + \frac{1}{4}\,c\,\hat{\mathbf{z}}$ | (4$d$) | S |
| $\mathbf{B_{16}}$ | = | $x_4\,\mathbf{a_1} + \left(\frac{1}{2} - y_4\right)\,\mathbf{a_2} + \frac{3}{4}\,\mathbf{a_3}$ | = | $x_4\,a\,\hat{\mathbf{x}} + \left(\frac{1}{2} - y_4\right)\,b\,\hat{\mathbf{y}} + \frac{3}{4}\,c\,\hat{\mathbf{z}}$ | (4$d$) | S |

**Found in:**

**Geometry files:**

- CIF: pp. 663
- POSCAR: pp. 664

# TlF-II Structure: AB_oP8_57_d_d

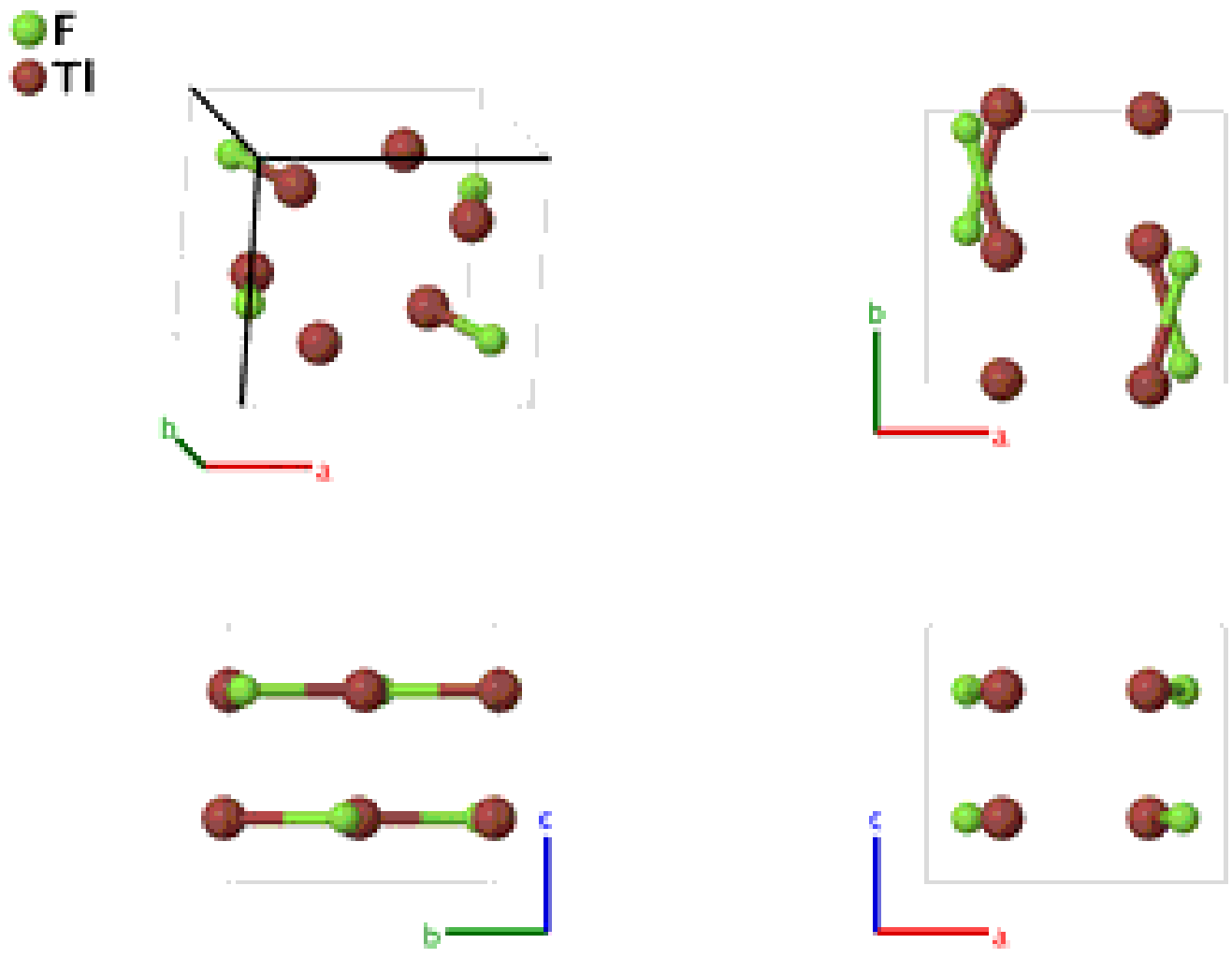

| | | |
|---|---|---|
| **Prototype** | : | TlF |
| **AFLOW prototype label** | : | AB_oP8_57_d_d |
| ***Strukturbericht* designation** | : | None |
| **Pearson symbol** | : | oP8 |
| **Space group number** | : | 57 |
| **Space group symbol** | : | Pbcm |
| **AFLOW prototype command** | : | `aflow --proto=AB_oP8_57_d_d`<br>`--params=`$a, b/a, c/a, x_1, y_1, x_2, y_2$ |

- This is the true low-temperature ground state of TlF. Like the B24 structure, it is a distortion of the rock salt (B1) structure.

**Simple Orthorhombic primitive vectors:**

$$\mathbf{a}_1 = a\,\hat{\mathbf{x}}$$
$$\mathbf{a}_2 = b\,\hat{\mathbf{y}}$$
$$\mathbf{a}_3 = c\,\hat{\mathbf{z}}$$

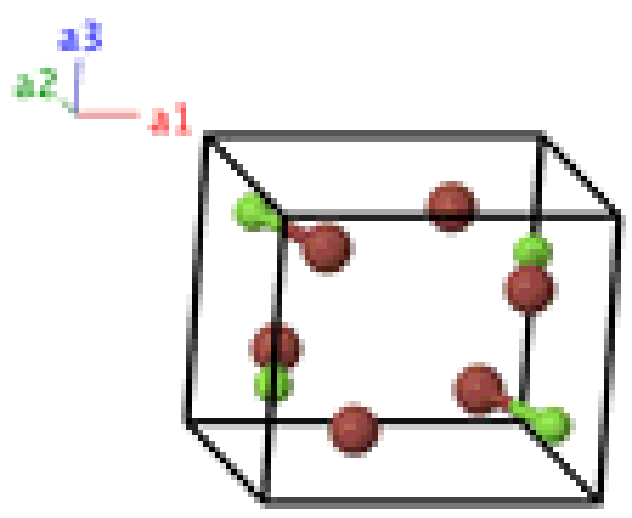

**Basis vectors:**

| | | Lattice Coordinates | | Cartesian Coordinates | Wyckoff Position | Atom Type |
|---|---|---|---|---|---|---|
| $\mathbf{B_1}$ | = | $x_1\,\mathbf{a_1} + y_1\,\mathbf{a_2} + \frac{1}{4}\,\mathbf{a_3}$ | = | $x_1\,a\,\hat{\mathbf{x}} + y_1\,b\,\hat{\mathbf{y}} + \frac{1}{4}\,c\,\hat{\mathbf{z}}$ | (4$d$) | F |
| $\mathbf{B_2}$ | = | $-x_1\,\mathbf{a_1} - y_1\,\mathbf{a_2} + \frac{3}{4}\,\mathbf{a_3}$ | = | $-x_1\,a\,\hat{\mathbf{x}} - y_1\,b\,\hat{\mathbf{y}} + \frac{3}{4}\,c\,\hat{\mathbf{z}}$ | (4$d$) | F |
| $\mathbf{B_3}$ | = | $-x_1\,\mathbf{a_1} + \left(\frac{1}{2} + y_1\right)\,\mathbf{a_2} + \frac{1}{4}\,\mathbf{a_3}$ | = | $-x_1\,a\,\hat{\mathbf{x}} + \left(\frac{1}{2} + y_1\right)\,b\,\hat{\mathbf{y}} + \frac{1}{4}\,c\,\hat{\mathbf{z}}$ | (4$d$) | F |
| $\mathbf{B_4}$ | = | $x_1\,\mathbf{a_1} + \left(\frac{1}{2} - y_1\right)\,\mathbf{a_2} + \frac{3}{4}\,\mathbf{a_3}$ | = | $x_1\,a\,\hat{\mathbf{x}} + \left(\frac{1}{2} - y_1\right)\,b\,\hat{\mathbf{y}} + \frac{3}{4}\,c\,\hat{\mathbf{z}}$ | (4$d$) | F |

$$\mathbf{B_5} = x_2\,\mathbf{a_1} + y_2\,\mathbf{a_2} + \frac{1}{4}\,\mathbf{a_3} = x_2\,a\,\mathbf{\hat{x}} + y_2\,b\,\mathbf{\hat{y}} + \frac{1}{4}\,c\,\mathbf{\hat{z}} \qquad (4d) \qquad \text{Tl}$$

$$\mathbf{B_6} = -x_2\,\mathbf{a_1} - y_2\,\mathbf{a_2} + \frac{3}{4}\,\mathbf{a_3} = -x_2\,a\,\mathbf{\hat{x}} - y_2\,b\,\mathbf{\hat{y}} + \frac{3}{4}\,c\,\mathbf{\hat{z}} \qquad (4d) \qquad \text{Tl}$$

$$\mathbf{B_7} = -x_2\,\mathbf{a_1} + \left(\frac{1}{2} + y_2\right)\,\mathbf{a_2} + \frac{1}{4}\,\mathbf{a_3} = -x_2\,a\,\mathbf{\hat{x}} + \left(\frac{1}{2} + y_2\right)\,b\,\mathbf{\hat{y}} + \frac{1}{4}\,c\,\mathbf{\hat{z}} \qquad (4d) \qquad \text{Tl}$$

$$\mathbf{B_8} = x_2\,\mathbf{a_1} + \left(\frac{1}{2} - y_2\right)\,\mathbf{a_2} + \frac{3}{4}\,\mathbf{a_3} = x_2\,a\,\mathbf{\hat{x}} + \left(\frac{1}{2} - y_2\right)\,b\,\mathbf{\hat{y}} + \frac{3}{4}\,c\,\mathbf{\hat{z}} \qquad (4d) \qquad \text{Tl}$$

---

---

**Geometry files:**

- CIF: pp. 664
- POSCAR: pp. 664

# Hydrophilite ($CaCl_2$, C35) Structure: AB2_oP6_58_a_g

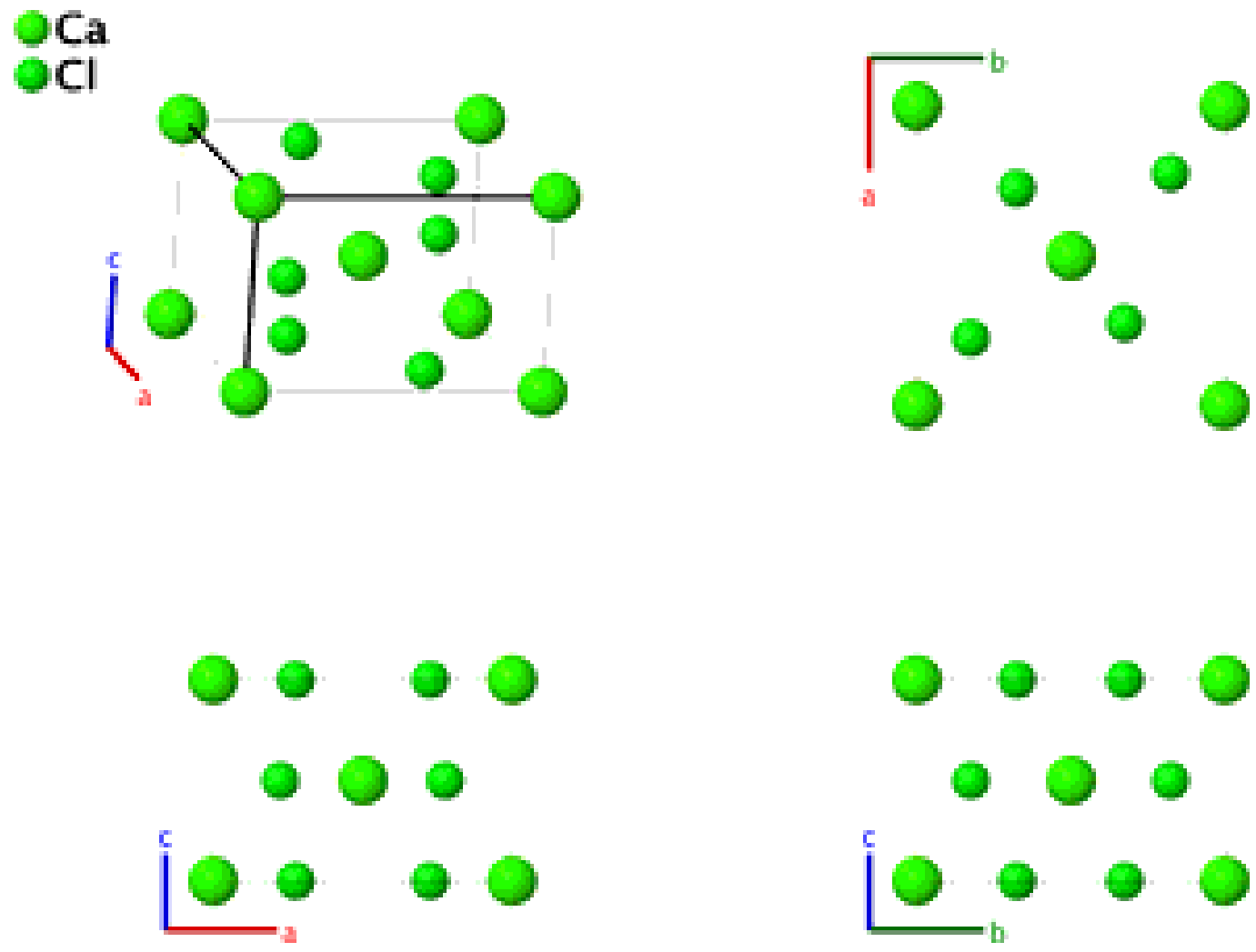

| | | |
|---|---|---|
| **Prototype** | : | $CaCl_2$ |
| **AFLOW prototype label** | : | AB2_oP6_58_a_g |
| ***Strukturbericht* designation** | : | C35 |
| **Pearson symbol** | : | oP6 |
| **Space group number** | : | 58 |
| **Space group symbol** | : | Pnnm |
| **AFLOW prototype command** | : | `aflow --proto=AB2_oP6_58_a_g`<br>`--params=`$a, b/a, c/a, x_2, y_2$ |

- Note that hydrophilite (pp. 133), $\eta$-$Fe_2C$ (pp. 135), and marcasite (pp. 137) have the same AFLOW prototype label. They are generated by the same symmetry operations with different sets of parameters (`--params`) specified in their corresponding CIF files.

**Simple Orthorhombic primitive vectors:**

$$\mathbf{a}_1 = a\,\hat{\mathbf{x}}$$
$$\mathbf{a}_2 = b\,\hat{\mathbf{y}}$$
$$\mathbf{a}_3 = c\,\hat{\mathbf{z}}$$

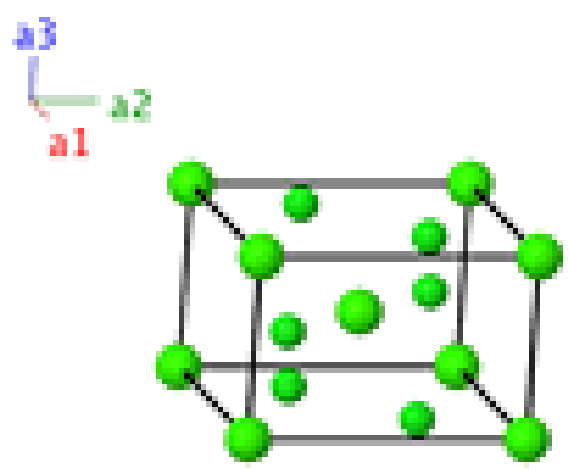

**Basis vectors:**

| | | Lattice Coordinates | | Cartesian Coordinates | Wyckoff Position | Atom Type |
|---|---|---|---|---|---|---|
| $\mathbf{B}_1$ | = | $0\,\mathbf{a}_1 + 0\,\mathbf{a}_2 + 0\,\mathbf{a}_3$ | = | $0\hat{\mathbf{x}} + 0\hat{\mathbf{y}} + 0\hat{\mathbf{z}}$ | (2$a$) | Ca |
| $\mathbf{B}_2$ | = | $\frac{1}{2}\,\mathbf{a}_1 + \frac{1}{2}\,\mathbf{a}_2 + \frac{1}{2}\,\mathbf{a}_3$ | = | $\frac{1}{2}\,a\,\hat{\mathbf{x}} + \frac{1}{2}\,b\,\hat{\mathbf{y}} + \frac{1}{2}\,c\,\hat{\mathbf{z}}$ | (2$a$) | Ca |
| $\mathbf{B}_3$ | = | $x_2\,\mathbf{a}_1 + y_2\,\mathbf{a}_2$ | = | $x_2\,a\,\hat{\mathbf{x}} + y_2\,b\,\hat{\mathbf{y}}$ | (4$g$) | Cl |
| $\mathbf{B_4}$ | = | $-x_2\,\mathbf{a_1} - y_2\,\mathbf{a_2}$ | = | $-x_2\,a\,\mathbf{\hat{x}} - y_2\,b\,\mathbf{\hat{y}}$ | (4$g$) | Cl |
| $\mathbf{B_5}$ | = | $\left(\frac{1}{2} - x_2\right)\mathbf{a_1} + \left(\frac{1}{2} + y_2\right)\mathbf{a_2} + \frac{1}{2}\,\mathbf{a_3}$ | = | $\left(\frac{1}{2} - x_2\right)a\,\mathbf{\hat{x}} + \left(\frac{1}{2} + y_2\right)b\,\mathbf{\hat{y}} + \frac{1}{2}\,c\,\mathbf{\hat{z}}$ | (4$g$) | Cl |
| $\mathbf{B_6}$ | = | $\left(\frac{1}{2} + x_2\right)\mathbf{a_1} + \left(\frac{1}{2} - y_2\right)\mathbf{a_2} + \frac{1}{2}\,\mathbf{a_3}$ | = | $\left(\frac{1}{2} + x_2\right)a\,\mathbf{\hat{x}} + \left(\frac{1}{2} - y_2\right)b\,\mathbf{\hat{y}} + \frac{1}{2}\,c\,\mathbf{\hat{z}}$ | (4$g$) | Cl |

---

**Found in:**

---

**Geometry files:**

- CIF: pp. 664
- POSCAR: pp. 665

# $\eta$-$Fe_2C$ Structure: AB2_oP6_58_a_g

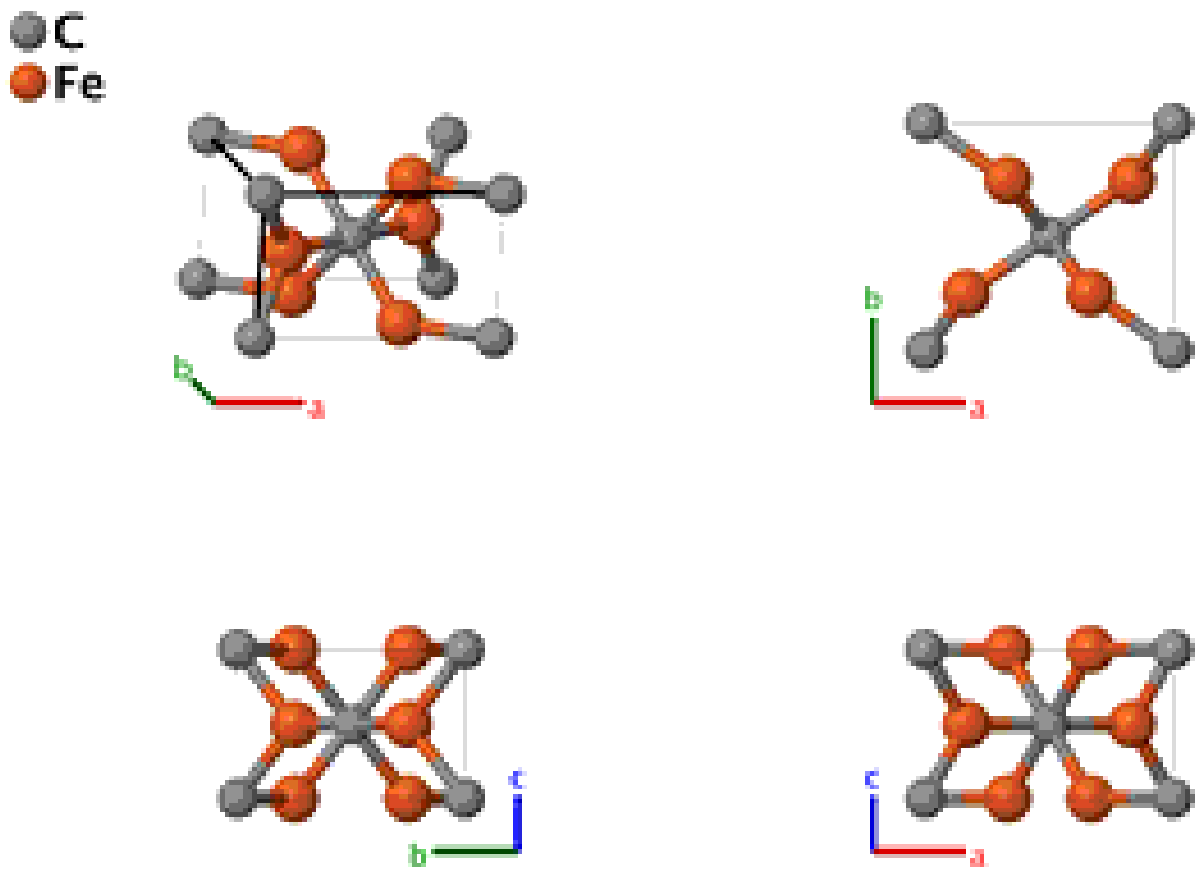

| | | |
|---|---|---|
| **Prototype** | : | $\eta$-$Fe_2C$ |
| **AFLOW prototype label** | : | AB2_oP6_58_a_g |
| ***Strukturbericht* designation** | : | None |
| **Pearson symbol** | : | oP6 |
| **Space group number** | : | 58 |
| **Space group symbol** | : | Pnnm |
| **AFLOW prototype command** | : | `aflow --proto=AB2_oP6_58_a_g` <br> `--params=`$a, b/a, c/a, x_2, y_2$ |

- Classified as bcc-related by Hellner and Schwarz (Westbrook, 1995), Vol. I, Chap. 13. Note that hydrophilite (pp. 133), $\eta$-$Fe_2C$ (pp. 135), and marcasite (pp. 137) have the same AFLOW prototype label. They are generated by the same symmetry operations with different sets of parameters (`--params`) specified in their corresponding CIF files.

**Simple Orthorhombic primitive vectors:**

$$\mathbf{a}_1 = a\,\hat{\mathbf{x}}$$
$$\mathbf{a}_2 = b\,\hat{\mathbf{y}}$$
$$\mathbf{a}_3 = c\,\hat{\mathbf{z}}$$

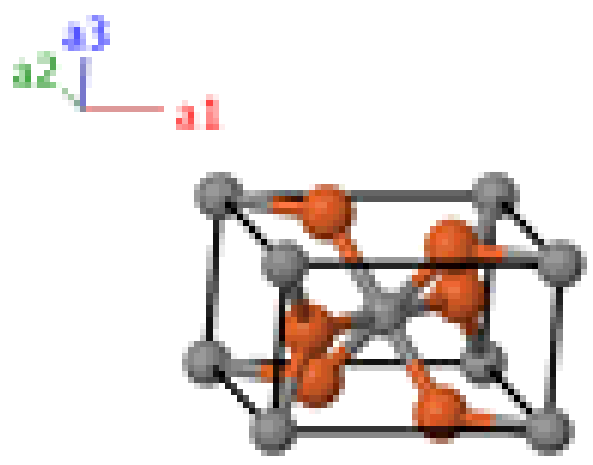

**Basis vectors:**

| | | Lattice Coordinates | | Cartesian Coordinates | Wyckoff Position | Atom Type |
|---|---|---|---|---|---|---|
| $\mathbf{B}_1$ | = | $0\,\mathbf{a}_1 + 0\,\mathbf{a}_2 + 0\,\mathbf{a}_3$ | = | $0\hat{\mathbf{x}} + 0\hat{\mathbf{y}} + 0\hat{\mathbf{z}}$ | (2$a$) | C |
| $\mathbf{B}_2$ | = | $\frac{1}{2}\,\mathbf{a}_1 + \frac{1}{2}\,\mathbf{a}_2 + \frac{1}{2}\,\mathbf{a}_3$ | = | $\frac{1}{2}\,a\,\hat{\mathbf{x}} + \frac{1}{2}\,b\,\hat{\mathbf{y}} + \frac{1}{2}\,c\,\hat{\mathbf{z}}$ | (2$a$) | C |
| $\mathbf{B}_3$ | = | $x_2\,\mathbf{a}_1 + y_2\,\mathbf{a}_2$ | = | $x_2\,a\,\hat{\mathbf{x}} + y_2\,b\,\hat{\mathbf{y}}$ | (4$g$) | Fe |
| $\mathbf{B}_4$ | = | $-x_2\,\mathbf{a}_1 - y_2\,\mathbf{a}_2$ | = | $-x_2\,a\,\hat{\mathbf{x}} - y_2\,b\,\hat{\mathbf{y}}$ | (4$g$) | Fe |
| $\mathbf{B}_5$ | = | $\left(\frac{1}{2} - x_2\right)\,\mathbf{a}_1 + \left(\frac{1}{2} + y_2\right)\,\mathbf{a}_2 + \frac{1}{2}\,\mathbf{a}_3$ | = | $\left(\frac{1}{2} - x_2\right)\,a\,\hat{\mathbf{x}} + \left(\frac{1}{2} + y_2\right)\,b\,\hat{\mathbf{y}} + \frac{1}{2}\,c\,\hat{\mathbf{z}}$ | (4$g$) | Fe |
| $\mathbf{B}_6$ | = | $\left(\frac{1}{2} + x_2\right)\,\mathbf{a}_1 + \left(\frac{1}{2} - y_2\right)\,\mathbf{a}_2 + \frac{1}{2}\,\mathbf{a}_3$ | = | $\left(\frac{1}{2} + x_2\right)\,a\,\hat{\mathbf{x}} + \left(\frac{1}{2} - y_2\right)\,b\,\hat{\mathbf{y}} + \frac{1}{2}\,c\,\hat{\mathbf{z}}$ | (4$g$) | Fe |

**Geometry files:**

- CIF: pp. 665
- POSCAR: pp. 665

# Marcasite ($FeS_2$, C18) Structure: AB2_oP6_58_a_g

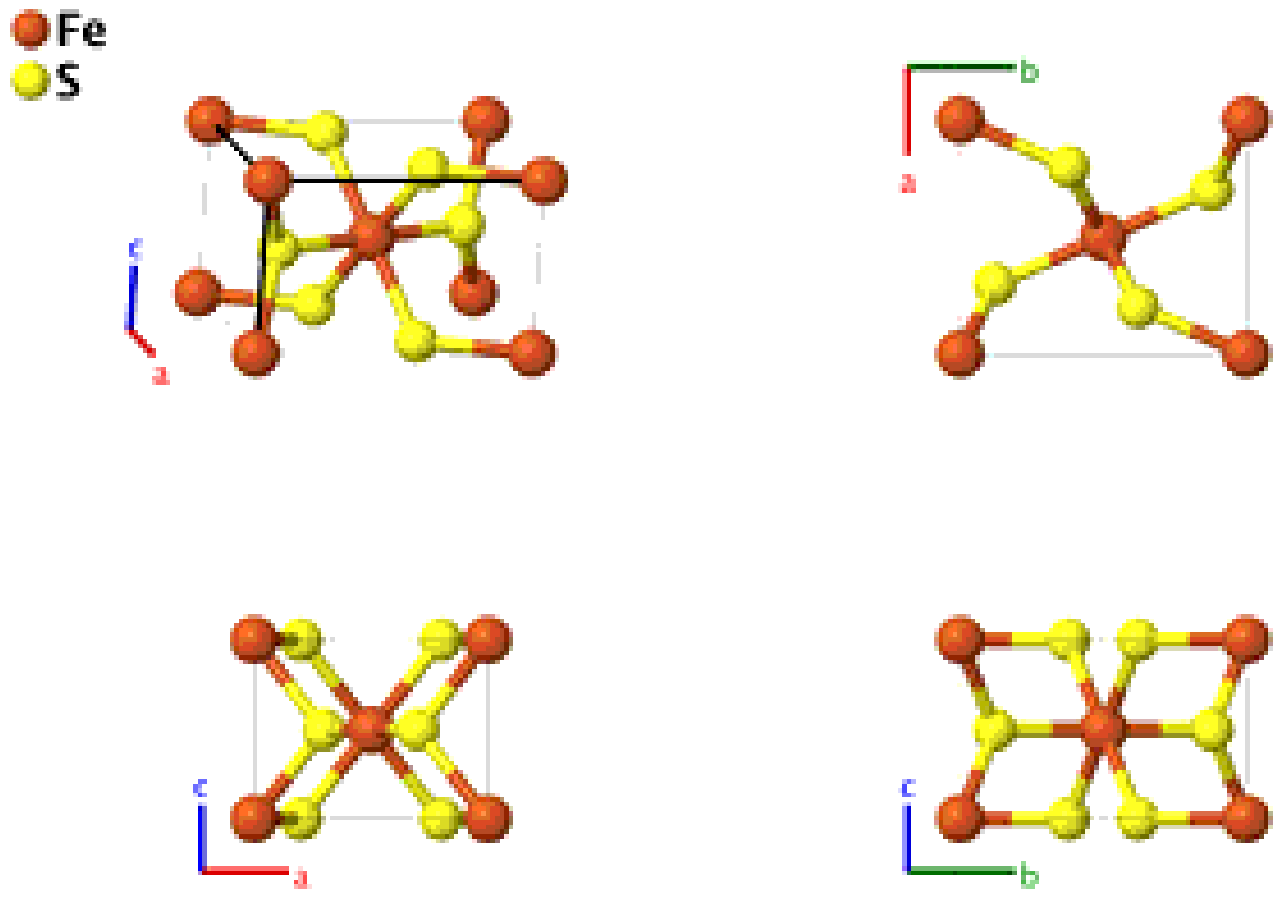

| | | |
|---|---|---|
| **Prototype** | : | $FeS_2$ |
| **AFLOW prototype label** | : | AB2_oP6_58_a_g |
| ***Strukturbericht* designation** | : | C18 |
| **Pearson symbol** | : | oP6 |
| **Space group number** | : | 58 |
| **Space group symbol** | : | Pnnm |
| **AFLOW prototype command** | : | `aflow --proto=AB2_oP6_58_a_g`<br>`--params=`$a, b/a, c/a, x_2, y_2$ |

**Other compounds with this structure:**

- $As_2Co$, $CrSb_2$, $NiSb_2$, $CuS_2$, $FeP_2$, $RuTe_2$, many more

- Note that hydrophilite (pp. 133), $\eta$-$Fe_2C$ (pp. 135), and marcasite (pp. 137) have the same AFLOW prototype label. They are generated by the same symmetry operations with different sets of parameters (`--params`) specified in their corresponding CIF files.

**Simple Orthorhombic primitive vectors:**

$$\mathbf{a}_1 = a\,\hat{\mathbf{x}}$$
$$\mathbf{a}_2 = b\,\hat{\mathbf{y}}$$
$$\mathbf{a}_3 = c\,\hat{\mathbf{z}}$$

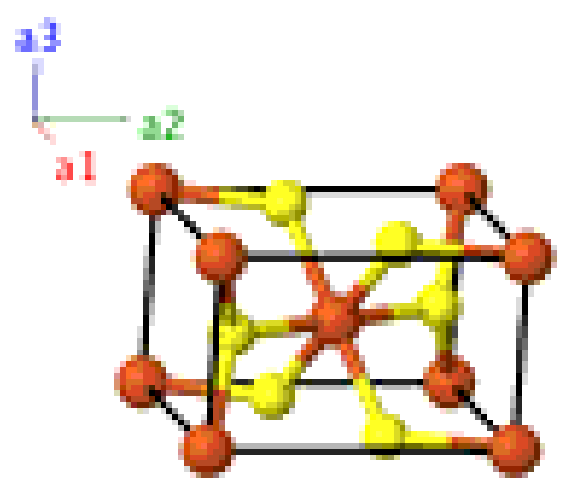

**Basis vectors:**

| | | Lattice Coordinates | | Cartesian Coordinates | Wyckoff Position | Atom Type |
|---|---|---|---|---|---|---|
| $\mathbf{B}_1$ | = | $0\,\mathbf{a}_1 + 0\,\mathbf{a}_2 + 0\,\mathbf{a}_3$ | = | $0\hat{\mathbf{x}} + 0\hat{\mathbf{y}} + 0\hat{\mathbf{z}}$ | $(2a)$ | Fe |
| $\mathbf{B}_2$ | = | $\frac{1}{2}\,\mathbf{a}_1 + \frac{1}{2}\,\mathbf{a}_2 + \frac{1}{2}\,\mathbf{a}_3$ | = | $\frac{1}{2}\,a\,\hat{\mathbf{x}} + \frac{1}{2}\,b\,\hat{\mathbf{y}} + \frac{1}{2}\,c\,\hat{\mathbf{z}}$ | $(2a)$ | Fe |
| $\mathbf{B_3}$ | = | $x_2\,\mathbf{a_1} + y_2\,\mathbf{a_2}$ | = | $x_2\,a\,\mathbf{\hat{x}} + y_2\,b\,\mathbf{\hat{y}}$ | (4$g$) | S |
| $\mathbf{B_4}$ | = | $-x_2\,\mathbf{a_1} - y_2\,\mathbf{a_2}$ | = | $-x_2\,a\,\mathbf{\hat{x}} - y_2\,b\,\mathbf{\hat{y}}$ | (4$g$) | S |
| $\mathbf{B_5}$ | = | $\left(\frac{1}{2} - x_2\right)\,\mathbf{a_1} + \left(\frac{1}{2} + y_2\right)\,\mathbf{a_2} + \frac{1}{2}\,\mathbf{a_3}$ | = | $\left(\frac{1}{2} - x_2\right)\,a\,\mathbf{\hat{x}} + \left(\frac{1}{2} + y_2\right)\,b\,\mathbf{\hat{y}} + \frac{1}{2}\,c\,\mathbf{\hat{z}}$ | (4$g$) | S |
| $\mathbf{B_6}$ | = | $\left(\frac{1}{2} + x_2\right)\,\mathbf{a_1} + \left(\frac{1}{2} - y_2\right)\,\mathbf{a_2} + \frac{1}{2}\,\mathbf{a_3}$ | = | $\left(\frac{1}{2} + x_2\right)\,a\,\mathbf{\hat{x}} + \left(\frac{1}{2} - y_2\right)\,b\,\mathbf{\hat{y}} + \frac{1}{2}\,c\,\mathbf{\hat{z}}$ | (4$g$) | S |

**Found in:**

**Geometry files:**

- CIF: pp. 665
- POSCAR: pp. 665

# Vulcanite (CuTe) Structure: AB_oP4_59_a_b

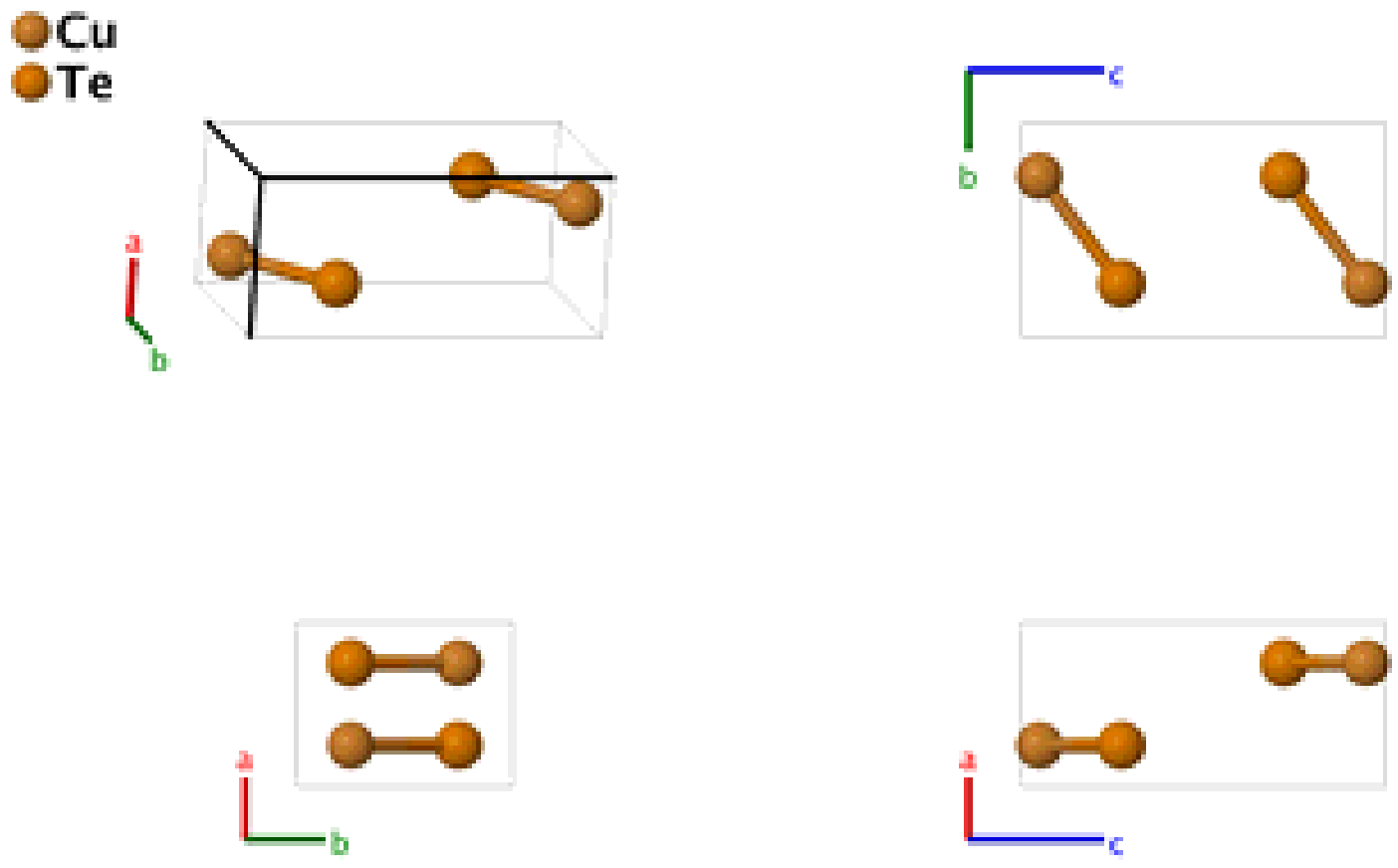

**Prototype** : CuTe

**AFLOW prototype label** : AB_oP4_59_a_b

***Strukturbericht* designation** : None

**Pearson symbol** : oP4

**Space group number** : 59

**Space group symbol** : Pmmn

**AFLOW prototype command** : `aflow --proto=AB_oP4_59_a_b`
`--params=`$a, b/a, c/a, z_1, z_2$

---

**Simple Orthorhombic primitive vectors:**

$$\mathbf{a}_1 = a\,\hat{\mathbf{x}}$$
$$\mathbf{a}_2 = b\,\hat{\mathbf{y}}$$
$$\mathbf{a}_3 = c\,\hat{\mathbf{z}}$$

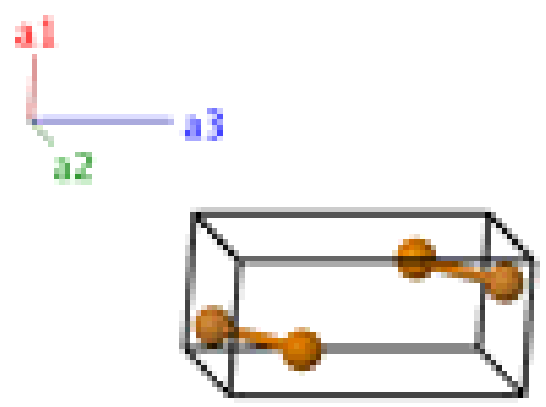

---

**Basis vectors:**

| | | Lattice Coordinates | | Cartesian Coordinates | Wyckoff Position | Atom Type |
|---|---|---|---|---|---|---|
| $\mathbf{B}_1$ | = | $\frac{1}{4}\,\mathbf{a}_1 + \frac{1}{4}\,\mathbf{a}_2 + z_1\,\mathbf{a}_3$ | = | $\frac{1}{4}\,a\,\hat{\mathbf{x}} + \frac{1}{4}\,b\,\hat{\mathbf{y}} + z_1\,c\,\hat{\mathbf{z}}$ | (2$a$) | Cu |
| $\mathbf{B}_2$ | = | $\frac{3}{4}\,\mathbf{a}_1 + \frac{3}{4}\,\mathbf{a}_2 - z_1\,\mathbf{a}_3$ | = | $\frac{3}{4}\,a\,\hat{\mathbf{x}} + \frac{3}{4}\,b\,\hat{\mathbf{y}} - z_1\,c\,\hat{\mathbf{z}}$ | (2$a$) | Cu |
| $\mathbf{B}_3$ | = | $\frac{1}{4}\,\mathbf{a}_1 + \frac{3}{4}\,\mathbf{a}_2 + z_2\,\mathbf{a}_3$ | = | $\frac{1}{4}\,a\,\hat{\mathbf{x}} + \frac{3}{4}\,b\,\hat{\mathbf{y}} + z_2\,c\,\hat{\mathbf{z}}$ | (2$b$) | Te |
| $\mathbf{B}_4$ | = | $\frac{3}{4}\,\mathbf{a}_1 + \frac{1}{4}\,\mathbf{a}_2 - z_2\,\mathbf{a}_3$ | = | $\frac{3}{4}\,a\,\hat{\mathbf{x}} + \frac{1}{4}\,b\,\hat{\mathbf{y}} - z_2\,c\,\hat{\mathbf{z}}$ | (2$b$) | Te |

---

**Found in:**

---

**Geometry files:**

- CIF: pp. 665
- POSCAR: pp. 666

# CNCl Structure: ABC_oP6_59_a_a_a

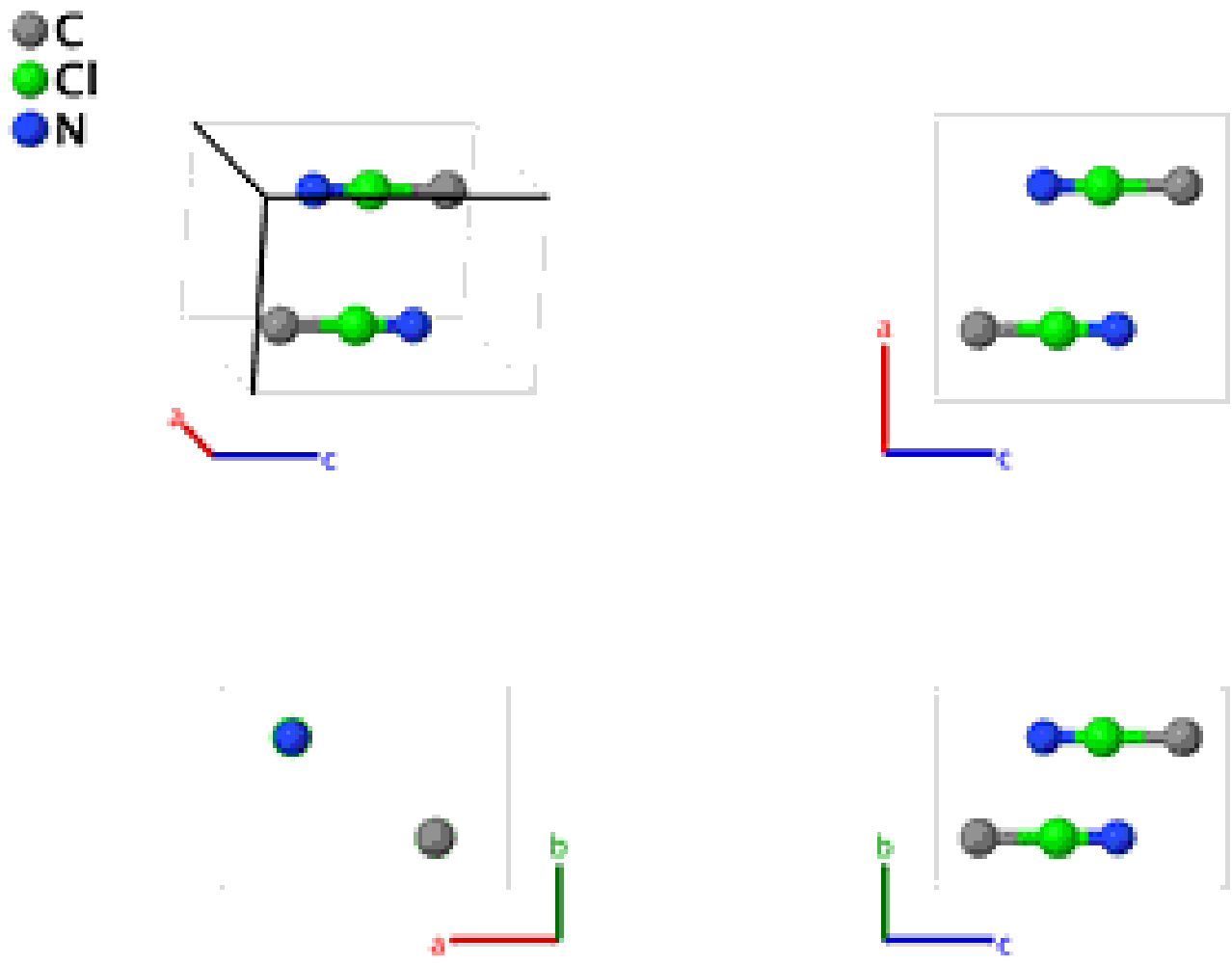

| | | |
|---|---|---|
| **Prototype** | : | CNCl |
| **AFLOW prototype label** | : | ABC_oP6_59_a_a_a |
| ***Strukturbericht* designation** | : | None |
| **Pearson symbol** | : | oP6 |
| **Space group number** | : | 59 |
| **Space group symbol** | : | Pmmn |
| **AFLOW prototype command** | : | `aflow --proto=ABC_oP6_59_a_a_a`<br>`--params=`$a, b/a, c/a, z_1, z_2, z_3$ |

**Simple Orthorhombic primitive vectors:**

$$\mathbf{a}_1 = a\,\hat{\mathbf{x}}$$
$$\mathbf{a}_2 = b\,\hat{\mathbf{y}}$$
$$\mathbf{a}_3 = c\,\hat{\mathbf{z}}$$

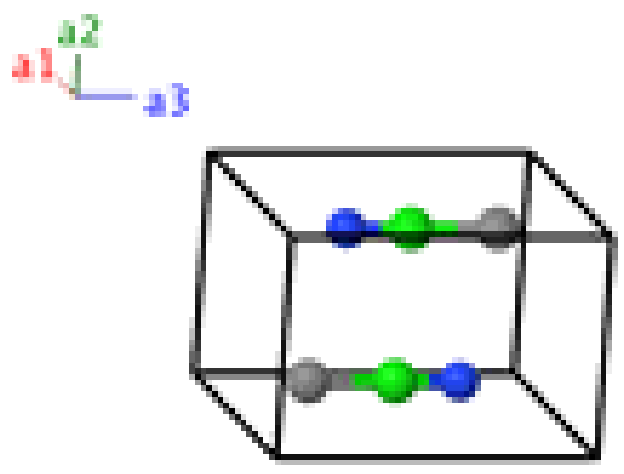

**Basis vectors:**

| | | Lattice Coordinates | | Cartesian Coordinates | Wyckoff Position | Atom Type |
|---|---|---|---|---|---|---|
| $\mathbf{B}_1$ | = | $\frac{1}{4}\,\mathbf{a}_1 + \frac{1}{4}\,\mathbf{a}_2 + z_1\,\mathbf{a}_3$ | = | $\frac{1}{4}\,a\,\hat{\mathbf{x}} + \frac{1}{4}\,b\,\hat{\mathbf{y}} + z_1\,c\,\hat{\mathbf{z}}$ | (2$a$) | C |
| $\mathbf{B}_2$ | = | $\frac{3}{4}\,\mathbf{a}_1 + \frac{3}{4}\,\mathbf{a}_2 - z_1\,\mathbf{a}_3$ | = | $\frac{3}{4}\,a\,\hat{\mathbf{x}} + \frac{3}{4}\,b\,\hat{\mathbf{y}} - z_1\,c\,\hat{\mathbf{z}}$ | (2$a$) | C |
| $\mathbf{B}_3$ | = | $\frac{1}{4}\,\mathbf{a}_1 + \frac{1}{4}\,\mathbf{a}_2 + z_2\,\mathbf{a}_3$ | = | $\frac{1}{4}\,a\,\hat{\mathbf{x}} + \frac{1}{4}\,b\,\hat{\mathbf{y}} + z_2\,c\,\hat{\mathbf{z}}$ | (2$a$) | Cl |
| $\mathbf{B}_4$ | = | $\frac{3}{4}\,\mathbf{a}_1 + \frac{3}{4}\,\mathbf{a}_2 - z_2\,\mathbf{a}_3$ | = | $\frac{3}{4}\,a\,\hat{\mathbf{x}} + \frac{3}{4}\,b\,\hat{\mathbf{y}} - z_2\,c\,\hat{\mathbf{z}}$ | (2$a$) | Cl |
| $\mathbf{B}_5$ | = | $\frac{1}{4}\,\mathbf{a}_1 + \frac{1}{4}\,\mathbf{a}_2 + z_3\,\mathbf{a}_3$ | = | $\frac{1}{4}\,a\,\hat{\mathbf{x}} + \frac{1}{4}\,b\,\hat{\mathbf{y}} + z_3\,c\,\hat{\mathbf{z}}$ | (2$a$) | N |
| $\mathbf{B}_6$ | = | $\frac{3}{4}\,\mathbf{a}_1 + \frac{3}{4}\,\mathbf{a}_2 - z_3\,\mathbf{a}_3$ | = | $\frac{3}{4}\,a\,\hat{\mathbf{x}} + \frac{3}{4}\,b\,\hat{\mathbf{y}} - z_3\,c\,\hat{\mathbf{z}}$ | (2$a$) | N |

**Found in:**

---

**Geometry files:**

- CIF: pp. 666
- POSCAR: pp. 666

# $\beta$-TiCu$_3$ (D0$_a$) Structure: A3B_oP8_59_bf_a

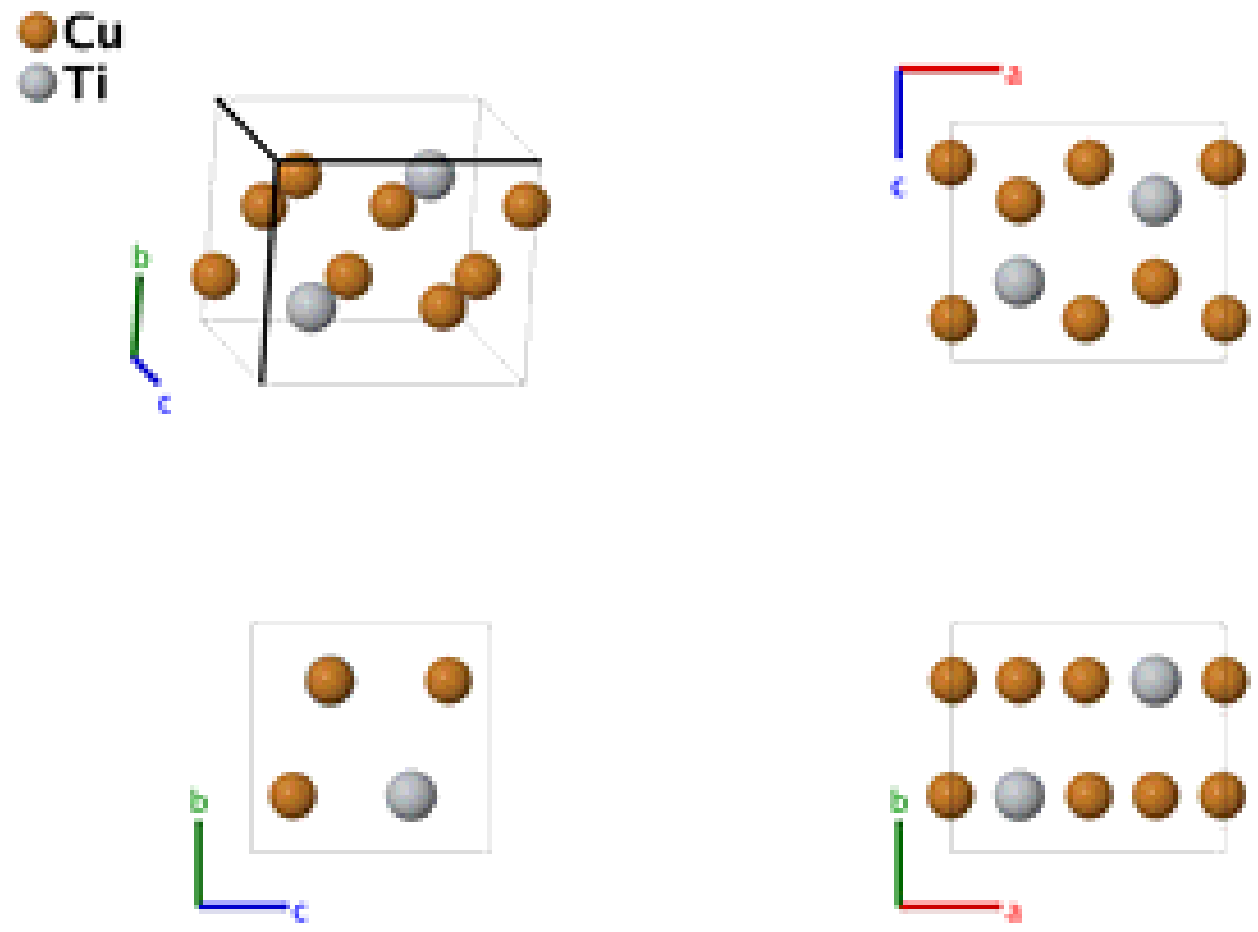

| | | |
|---|---|---|
| **Prototype** | : | $\beta$-TiCu$_3$ |
| **AFLOW prototype label** | : | A3B_oP8_59_bf_a |
| ***Strukturbericht* designation** | : | D0$_a$ |
| **Pearson symbol** | : | oP8 |
| **Space group number** | : | 59 |
| **Space group symbol** | : | Pmmn |
| **AFLOW prototype command** | : | `aflow --proto=A3B_oP8_59_bf_a`<br>`--params=`$a, b/a, c/a, z_1, z_2, x_3, z_3$ |

- We have been so far unable to obtain the original reference (Karlsson, 1951), and Pearson does not give the exact atomic coordinates. Wyckoff positions have been deduced from the structure of Cu$_3$Sb, which Villars (1991) lists as having the TiCu$_3$ structure. Atomic positions are set to give the approximate nearest-neighbor distances listed in Pearson. (Giessen, 1971) says that (Karlsson, 1951) structure of $\beta$-TiCu$_3$ is mistaken. They do find a metastable $\beta$-TiCu$_3$ phase which has the same space group and Wyckoff positions, but substantially different lattice constants than the original determination for $\beta$-TiCu$_3$.

**Simple Orthorhombic primitive vectors:**

$$\mathbf{a}_1 = a\,\hat{\mathbf{x}}$$
$$\mathbf{a}_2 = b\,\hat{\mathbf{y}}$$
$$\mathbf{a}_3 = c\,\hat{\mathbf{z}}$$

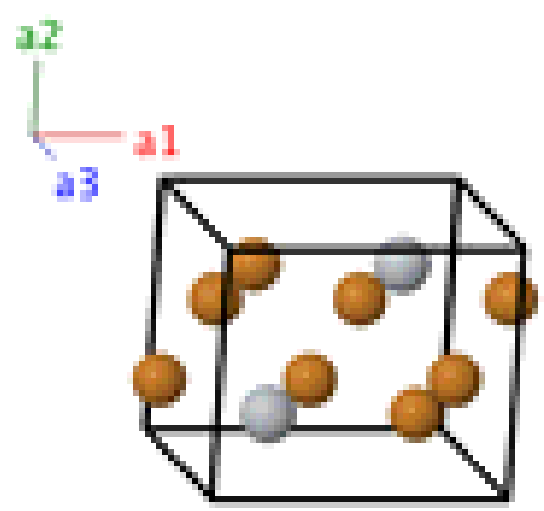

**Basis vectors:**

| | | Lattice Coordinates | | Cartesian Coordinates | Wyckoff Position | Atom Type |
|---|---|---|---|---|---|---|
| $\mathbf{B}_1$ | = | $\frac{1}{4}\,\mathbf{a}_1 + \frac{1}{4}\,\mathbf{a}_2 + z_1\,\mathbf{a}_3$ | = | $\frac{1}{4}\,a\,\hat{\mathbf{x}} + \frac{1}{4}\,b\,\hat{\mathbf{y}} + z_1\,c\,\hat{\mathbf{z}}$ | (2$a$) | Ti |
| $\mathbf{B}_2$ | = | $\frac{3}{4}\,\mathbf{a}_1 + \frac{3}{4}\,\mathbf{a}_2 - z_1\,\mathbf{a}_3$ | = | $\frac{3}{4}\,a\,\hat{\mathbf{x}} + \frac{3}{4}\,b\,\hat{\mathbf{y}} - z_1\,c\,\hat{\mathbf{z}}$ | (2$a$) | Ti |
| $\mathbf{B_3}$ | = | $\frac{1}{4}\,\mathbf{a_1} + \frac{3}{4}\,\mathbf{a_2} + z_2\,\mathbf{a_3}$ | = | $\frac{1}{4}\,a\,\mathbf{\hat{x}} + \frac{3}{4}\,b\,\mathbf{\hat{y}} + z_2\,c\,\mathbf{\hat{z}}$ | $(2b)$ | Cu I |
| $\mathbf{B_4}$ | = | $\frac{3}{4}\,\mathbf{a_1} + \frac{1}{4}\,\mathbf{a_2} - z_2\,\mathbf{a_3}$ | = | $\frac{3}{4}\,a\,\mathbf{\hat{x}} + \frac{1}{4}\,b\,\mathbf{\hat{y}} - z_2\,c\,\mathbf{\hat{z}}$ | $(2b)$ | Cu I |
| $\mathbf{B_5}$ | = | $x_3\,\mathbf{a_1} + \frac{1}{4}\,\mathbf{a_2} + z_3\,\mathbf{a_3}$ | = | $x_3\,a\,\mathbf{\hat{x}} + \frac{1}{4}\,b\,\mathbf{\hat{y}} + z_3\,c\,\mathbf{\hat{z}}$ | $(4f)$ | Cu II |
| $\mathbf{B_6}$ | = | $\left(\frac{1}{2} - x_3\right)\,\mathbf{a_1} + \frac{1}{4}\,\mathbf{a_2} + z_3\,\mathbf{a_3}$ | = | $\left(\frac{1}{2} - x_3\right)\,a\,\mathbf{\hat{x}} + \frac{1}{4}\,b\,\mathbf{\hat{y}} + z_3\,c\,\mathbf{\hat{z}}$ | $(4f)$ | Cu II |
| $\mathbf{B_7}$ | = | $-x_3\,\mathbf{a_1} + \frac{3}{4}\,\mathbf{a_2} - z_3\,\mathbf{a_3}$ | = | $-x_3\,a\,\mathbf{\hat{x}} + \frac{3}{4}\,b\,\mathbf{\hat{y}} - z_3\,c\,\mathbf{\hat{z}}$ | $(4f)$ | Cu II |
| $\mathbf{B_8}$ | = | $\left(\frac{1}{2} + x_3\right)\,\mathbf{a_1} + \frac{3}{4}\,\mathbf{a_2} - z_3\,\mathbf{a_3}$ | = | $\left(\frac{1}{2} + x_3\right)\,a\,\mathbf{\hat{x}} + \frac{3}{4}\,b\,\mathbf{\hat{y}} - z_3\,c\,\mathbf{\hat{z}}$ | $(4f)$ | Cu II |

---

**Found in:**

---

**Geometry files:**

- CIF: pp. 666
- POSCAR: pp. 667

# CdSb ($B_e$) Structure: AB_oP16_61_c_c

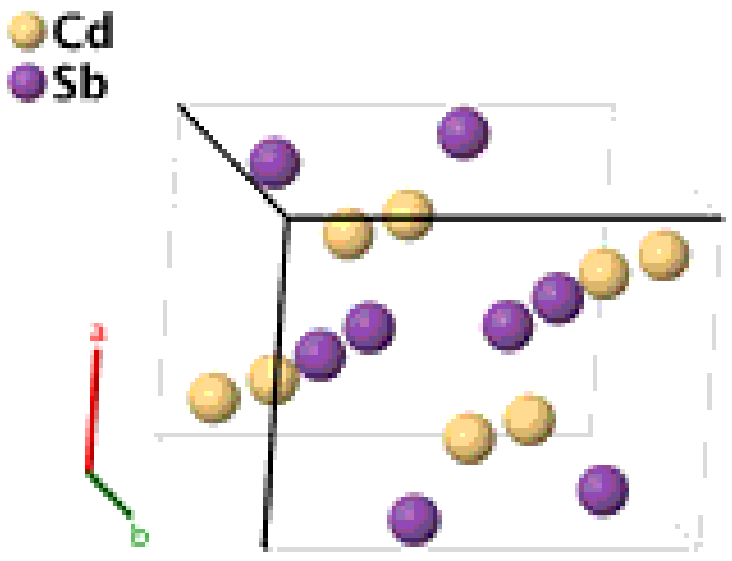

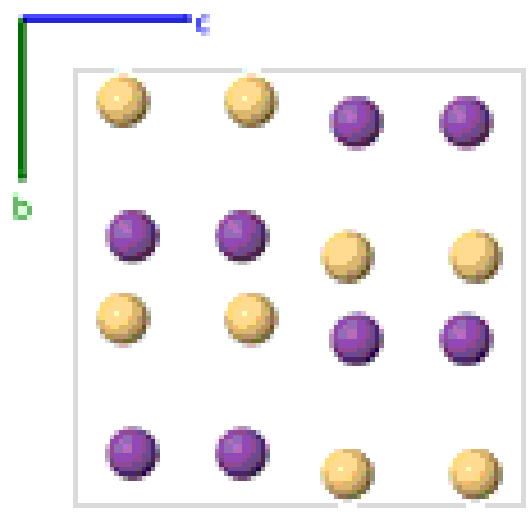

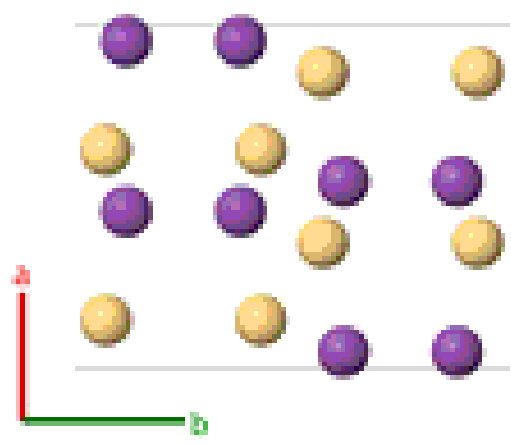

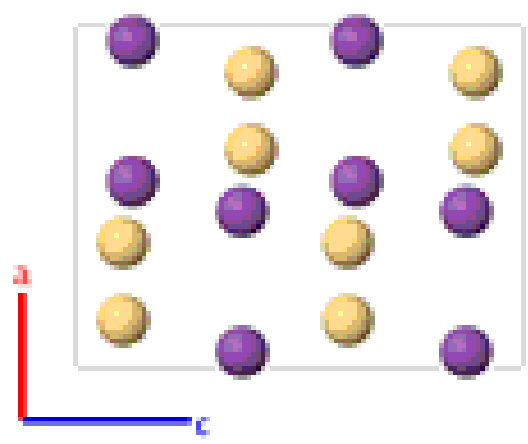

**Prototype** : CdSb

**AFLOW prototype label** : AB_oP16_61_c_c

***Strukturbericht* designation** : $B_e$

**Pearson symbol** : oP16

**Space group number** : 61

**Space group symbol** : Pbca

**AFLOW prototype command** : `aflow --proto=AB_oP16_61_c_c`
`--params=`$a, b/a, c/a, x_1, y_1, z_1, x_2, y_2, z_2$

---

**Other compounds with this structure:**

- AsCd, AsZn, SbZn

---

**Simple Orthorhombic primitive vectors:**

$$\mathbf{a}_1 = a\,\hat{\mathbf{x}}$$
$$\mathbf{a}_2 = b\,\hat{\mathbf{y}}$$
$$\mathbf{a}_3 = c\,\hat{\mathbf{z}}$$

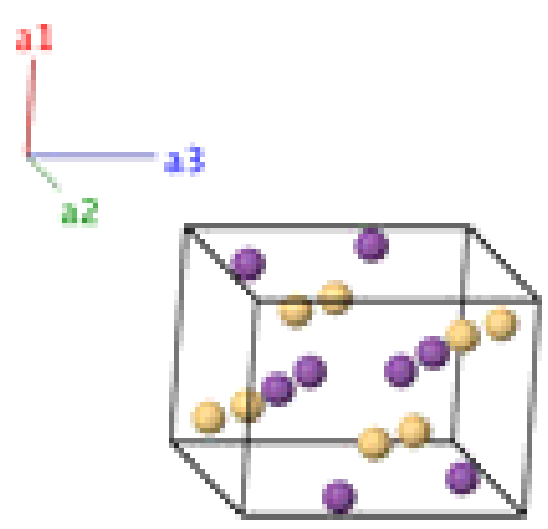

---

**Basis vectors:**

| | | Lattice Coordinates | | Cartesian Coordinates | Wyckoff Position | Atom Type |
|---|---|---|---|---|---|---|
| $\mathbf{B}_1$ | = | $x_1\,\mathbf{a}_1 + y_1\,\mathbf{a}_2 + z_1\,\mathbf{a}_3$ | = | $x_1\,a\,\hat{\mathbf{x}} + y_1\,b\,\hat{\mathbf{y}} + z_1\,c\,\hat{\mathbf{z}}$ | (8$c$) | Cd |
| $\mathbf{B_2}$ | = | $\left(\frac{1}{2} - x_1\right)\,\mathbf{a_1} - y_1\,\mathbf{a_2} + \left(\frac{1}{2} + z_1\right)\,\mathbf{a_3}$ | = | $\left(\frac{1}{2} - x_1\right)\,a\,\hat{\mathbf{x}} - y_1\,b\,\hat{\mathbf{y}} + \left(\frac{1}{2} + z_1\right)\,c\,\hat{\mathbf{z}}$ | (8$c$) | Cd |
| $\mathbf{B_3}$ | = | $-x_1\,\mathbf{a_1} + \left(\frac{1}{2} + y_1\right)\,\mathbf{a_2} + \left(\frac{1}{2} - z_1\right)\,\mathbf{a_3}$ | = | $-x_1\,a\,\hat{\mathbf{x}} + \left(\frac{1}{2} + y_1\right)\,b\,\hat{\mathbf{y}} + \left(\frac{1}{2} - z_1\right)\,c\,\hat{\mathbf{z}}$ | (8$c$) | Cd |
| $\mathbf{B_4}$ | = | $\left(\frac{1}{2} + x_1\right)\,\mathbf{a_1} + \left(\frac{1}{2} - y_1\right)\,\mathbf{a_2} - z_1\,\mathbf{a_3}$ | = | $\left(\frac{1}{2} + x_1\right)\,a\,\hat{\mathbf{x}} + \left(\frac{1}{2} - y_1\right)\,b\,\hat{\mathbf{y}} - z_1\,c\,\hat{\mathbf{z}}$ | (8$c$) | Cd |
| $\mathbf{B_5}$ | = | $-x_1\,\mathbf{a_1} - y_1\,\mathbf{a_2} - z_1\,\mathbf{a_3}$ | = | $-x_1\,a\,\hat{\mathbf{x}} - y_1\,b\,\hat{\mathbf{y}} - z_1\,c\,\hat{\mathbf{z}}$ | (8$c$) | Cd |
| $\mathbf{B_6}$ | = | $\left(\frac{1}{2} + x_1\right)\,\mathbf{a_1} + y_1\,\mathbf{a_2} + \left(\frac{1}{2} - z_1\right)\,\mathbf{a_3}$ | = | $\left(\frac{1}{2} + x_1\right)\,a\,\hat{\mathbf{x}} + y_1\,b\,\hat{\mathbf{y}} + \left(\frac{1}{2} - z_1\right)\,c\,\hat{\mathbf{z}}$ | (8$c$) | Cd |
| $\mathbf{B_7}$ | = | $x_1\,\mathbf{a_1} + \left(\frac{1}{2} - y_1\right)\,\mathbf{a_2} + \left(\frac{1}{2} + z_1\right)\,\mathbf{a_3}$ | = | $x_1\,a\,\hat{\mathbf{x}} + \left(\frac{1}{2} - y_1\right)\,b\,\hat{\mathbf{y}} + \left(\frac{1}{2} + z_1\right)\,c\,\hat{\mathbf{z}}$ | (8$c$) | Cd |
| $\mathbf{B_8}$ | = | $\left(\frac{1}{2} - x_1\right)\,\mathbf{a_1} + \left(\frac{1}{2} + y_1\right)\,\mathbf{a_2} + z_1\,\mathbf{a_3}$ | = | $\left(\frac{1}{2} - x_1\right)\,a\,\hat{\mathbf{x}} + \left(\frac{1}{2} + y_1\right)\,b\,\hat{\mathbf{y}} + z_1\,c\,\hat{\mathbf{z}}$ | (8$c$) | Cd |
| $\mathbf{B_9}$ | = | $x_2\,\mathbf{a_1} + y_2\,\mathbf{a_2} + z_2\,\mathbf{a_3}$ | = | $x_2\,a\,\hat{\mathbf{x}} + y_2\,b\,\hat{\mathbf{y}} + z_2\,c\,\hat{\mathbf{z}}$ | (8$c$) | Sb |
| $\mathbf{B_{10}}$ | = | $\left(\frac{1}{2} - x_2\right)\,\mathbf{a_1} - y_2\,\mathbf{a_2} + \left(\frac{1}{2} + z_2\right)\,\mathbf{a_3}$ | = | $\left(\frac{1}{2} - x_2\right)\,a\,\hat{\mathbf{x}} - y_2\,b\,\hat{\mathbf{y}} + \left(\frac{1}{2} + z_2\right)\,c\,\hat{\mathbf{z}}$ | (8$c$) | Sb |
| $\mathbf{B_{11}}$ | = | $-x_2\,\mathbf{a_1} + \left(\frac{1}{2} + y_2\right)\,\mathbf{a_2} + \left(\frac{1}{2} - z_2\right)\,\mathbf{a_3}$ | = | $-x_2\,a\,\hat{\mathbf{x}} + \left(\frac{1}{2} + y_2\right)\,b\,\hat{\mathbf{y}} + \left(\frac{1}{2} - z_2\right)\,c\,\hat{\mathbf{z}}$ | (8$c$) | Sb |
| $\mathbf{B_{12}}$ | = | $\left(\frac{1}{2} + x_2\right)\,\mathbf{a_1} + \left(\frac{1}{2} - y_2\right)\,\mathbf{a_2} - z_2\,\mathbf{a_3}$ | = | $\left(\frac{1}{2} + x_2\right)\,a\,\hat{\mathbf{x}} + \left(\frac{1}{2} - y_2\right)\,b\,\hat{\mathbf{y}} - z_2\,c\,\hat{\mathbf{z}}$ | (8$c$) | Sb |
| $\mathbf{B_{13}}$ | = | $-x_2\,\mathbf{a_1} - y_2\,\mathbf{a_2} - z_2\,\mathbf{a_3}$ | = | $-x_2\,a\,\hat{\mathbf{x}} - y_2\,b\,\hat{\mathbf{y}} - z_2\,c\,\hat{\mathbf{z}}$ | (8$c$) | Sb |
| $\mathbf{B_{14}}$ | = | $\left(\frac{1}{2} + x_2\right)\,\mathbf{a_1} + y_2\,\mathbf{a_2} + \left(\frac{1}{2} - z_2\right)\,\mathbf{a_3}$ | = | $\left(\frac{1}{2} + x_2\right)\,a\,\hat{\mathbf{x}} + y_2\,b\,\hat{\mathbf{y}} + \left(\frac{1}{2} - z_2\right)\,c\,\hat{\mathbf{z}}$ | (8$c$) | Sb |
| $\mathbf{B_{15}}$ | = | $x_2\,\mathbf{a_1} + \left(\frac{1}{2} - y_2\right)\,\mathbf{a_2} + \left(\frac{1}{2} + z_2\right)\,\mathbf{a_3}$ | = | $x_2\,a\,\hat{\mathbf{x}} + \left(\frac{1}{2} - y_2\right)\,b\,\hat{\mathbf{y}} + \left(\frac{1}{2} + z_2\right)\,c\,\hat{\mathbf{z}}$ | (8$c$) | Sb |
| $\mathbf{B_{16}}$ | = | $\left(\frac{1}{2} - x_2\right)\,\mathbf{a_1} + \left(\frac{1}{2} + y_2\right)\,\mathbf{a_2} + z_2\,\mathbf{a_3}$ | = | $\left(\frac{1}{2} - x_2\right)\,a\,\hat{\mathbf{x}} + \left(\frac{1}{2} + y_2\right)\,b\,\hat{\mathbf{y}} + z_2\,c\,\hat{\mathbf{z}}$ | (8$c$) | Sb |

**Geometry files:**

- CIF: pp. 667
- POSCAR: pp. 667

# Brookite ($TiO_2$, C21) Structure: A2B_oP24_61_2c_c

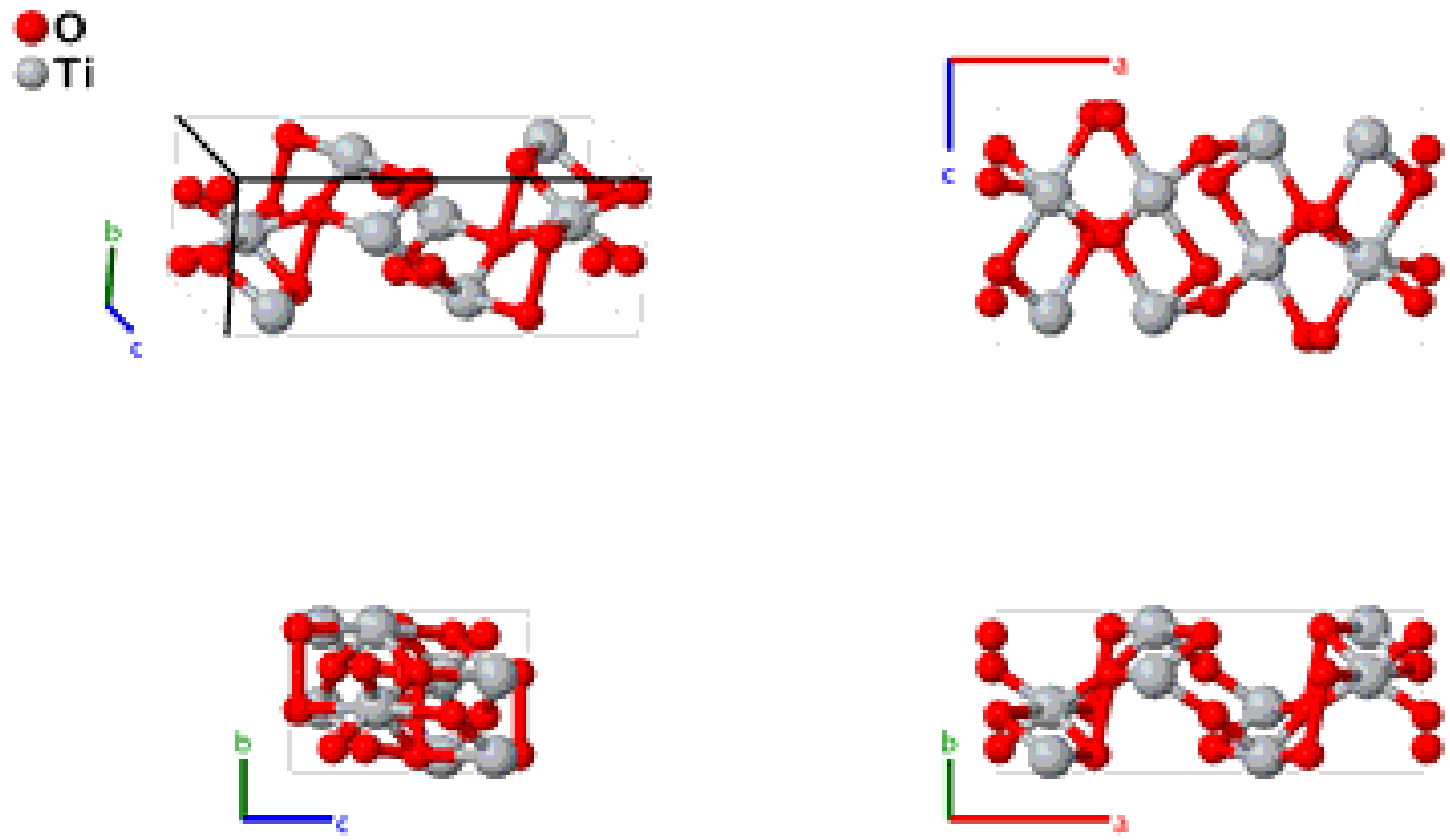

| | | |
|---|---|---|
| **Prototype** | : | $TiO_2$ |
| **AFLOW prototype label** | : | A2B_oP24_61_2c_c |
| ***Strukturbericht* designation** | : | C21 |
| **Pearson symbol** | : | oP24 |
| **Space group number** | : | 61 |
| **Space group symbol** | : | Pbca |
| **AFLOW prototype command** | : | `aflow --proto=A2B_oP24_61_2c_c` `--params=`$a, b/a, c/a, x_1, y_1, z_1, x_2, y_2, z_2, x_3, y_3, z_3$ |

**Other compounds with this structure:**

- $TeO_2$ tellurite

**Simple Orthorhombic primitive vectors:**

$$
\begin{aligned}
\mathbf{a}_1 &= a\,\mathbf{\hat{x}} \\
\mathbf{a}_2 &= b\,\mathbf{\hat{y}} \\
\mathbf{a}_3 &= c\,\mathbf{\hat{z}}
\end{aligned}
$$

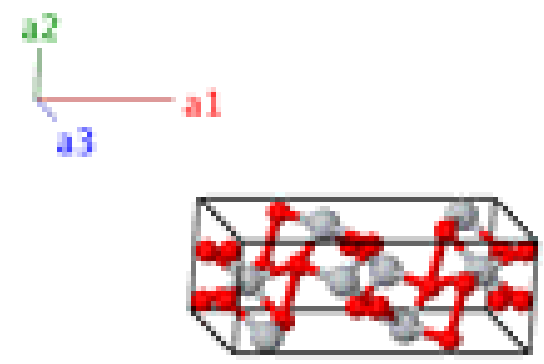

**Basis vectors:**

| | | Lattice Coordinates | | Cartesian Coordinates | Wyckoff Position | Atom Type |
|---|---|---|---|---|---|---|
| $\mathbf{B}_1$ | = | $x_1\,\mathbf{a}_1 + y_1\,\mathbf{a}_2 + z_1\,\mathbf{a}_3$ | = | $x_1\,a\,\mathbf{\hat{x}} + y_1\,b\,\mathbf{\hat{y}} + z_1\,c\,\mathbf{\hat{z}}$ | (8c) | O I |
| $\mathbf{B}_2$ | = | $\left(\frac{1}{2} - x_1\right)\mathbf{a}_1 - y_1\,\mathbf{a}_2 + \left(\frac{1}{2} + z_1\right)\mathbf{a}_3$ | = | $\left(\frac{1}{2} - x_1\right)a\,\mathbf{\hat{x}} - y_1\,b\,\mathbf{\hat{y}} + \left(\frac{1}{2} + z_1\right)c\,\mathbf{\hat{z}}$ | (8c) | O I |
| $\mathbf{B}_3$ | = | $-x_1\,\mathbf{a}_1 + \left(\frac{1}{2} + y_1\right)\mathbf{a}_2 + \left(\frac{1}{2} - z_1\right)\mathbf{a}_3$ | = | $-x_1\,a\,\mathbf{\hat{x}} + \left(\frac{1}{2} + y_1\right)b\,\mathbf{\hat{y}} + \left(\frac{1}{2} - z_1\right)c\,\mathbf{\hat{z}}$ | (8c) | O I |
| $\mathbf{B}_4$ | = | $\left(\frac{1}{2} + x_1\right)\mathbf{a}_1 + \left(\frac{1}{2} - y_1\right)\mathbf{a}_2 - z_1\,\mathbf{a}_3$ | = | $\left(\frac{1}{2} + x_1\right)a\,\mathbf{\hat{x}} + \left(\frac{1}{2} - y_1\right)b\,\mathbf{\hat{y}} - z_1\,c\,\mathbf{\hat{z}}$ | (8c) | O I |
| $\mathbf{B}_5$ | = | $-x_1\,\mathbf{a}_1 - y_1\,\mathbf{a}_2 - z_1\,\mathbf{a}_3$ | = | $-x_1\,a\,\mathbf{\hat{x}} - y_1\,b\,\mathbf{\hat{y}} - z_1\,c\,\mathbf{\hat{z}}$ | (8c) | O I |
| $\mathbf{B_{6}}$ | = | $\left(\frac{1}{2}+x_1\right)\mathbf{a_1}+y_1\,\mathbf{a_2}+\left(\frac{1}{2}-z_1\right)\mathbf{a_3}$ | = | $\left(\frac{1}{2}+x_1\right)a\,\hat{\mathbf{x}}+y_1\,b\,\hat{\mathbf{y}}+\left(\frac{1}{2}-z_1\right)c\,\hat{\mathbf{z}}$ | (8$c$) | O I |
| $\mathbf{B_{7}}$ | = | $x_1\,\mathbf{a_1}+\left(\frac{1}{2}-y_1\right)\mathbf{a_2}+\left(\frac{1}{2}+z_1\right)\mathbf{a_3}$ | = | $x_1\,a\,\hat{\mathbf{x}}+\left(\frac{1}{2}-y_1\right)b\,\hat{\mathbf{y}}+\left(\frac{1}{2}+z_1\right)c\,\hat{\mathbf{z}}$ | (8$c$) | O I |
| $\mathbf{B_{8}}$ | = | $\left(\frac{1}{2}-x_1\right)\mathbf{a_1}+\left(\frac{1}{2}+y_1\right)\mathbf{a_2}+z_1\,\mathbf{a_3}$ | = | $\left(\frac{1}{2}-x_1\right)a\,\hat{\mathbf{x}}+\left(\frac{1}{2}+y_1\right)b\,\hat{\mathbf{y}}+z_1\,c\,\hat{\mathbf{z}}$ | (8$c$) | O I |
| $\mathbf{B_{9}}$ | = | $x_2\,\mathbf{a_1}+y_2\,\mathbf{a_2}+z_2\,\mathbf{a_3}$ | = | $x_2\,a\,\hat{\mathbf{x}}+y_2\,b\,\hat{\mathbf{y}}+z_2\,c\,\hat{\mathbf{z}}$ | (8$c$) | O II |
| $\mathbf{B_{10}}$ | = | $\left(\frac{1}{2}-x_2\right)\mathbf{a_1}-y_2\,\mathbf{a_2}+\left(\frac{1}{2}+z_2\right)\mathbf{a_3}$ | = | $\left(\frac{1}{2}-x_2\right)a\,\hat{\mathbf{x}}-y_2\,b\,\hat{\mathbf{y}}+\left(\frac{1}{2}+z_2\right)c\,\hat{\mathbf{z}}$ | (8$c$) | O II |
| $\mathbf{B_{11}}$ | = | $-x_2\,\mathbf{a_1}+\left(\frac{1}{2}+y_2\right)\mathbf{a_2}+\left(\frac{1}{2}-z_2\right)\mathbf{a_3}$ | = | $-x_2\,a\,\hat{\mathbf{x}}+\left(\frac{1}{2}+y_2\right)b\,\hat{\mathbf{y}}+\left(\frac{1}{2}-z_2\right)c\,\hat{\mathbf{z}}$ | (8$c$) | O II |
| $\mathbf{B_{12}}$ | = | $\left(\frac{1}{2}+x_2\right)\mathbf{a_1}+\left(\frac{1}{2}-y_2\right)\mathbf{a_2}-z_2\,\mathbf{a_3}$ | = | $\left(\frac{1}{2}+x_2\right)a\,\hat{\mathbf{x}}+\left(\frac{1}{2}-y_2\right)b\,\hat{\mathbf{y}}-z_2\,c\,\hat{\mathbf{z}}$ | (8$c$) | O II |
| $\mathbf{B_{13}}$ | = | $-x_2\,\mathbf{a_1}-y_2\,\mathbf{a_2}-z_2\,\mathbf{a_3}$ | = | $-x_2\,a\,\hat{\mathbf{x}}-y_2\,b\,\hat{\mathbf{y}}-z_2\,c\,\hat{\mathbf{z}}$ | (8$c$) | O II |
| $\mathbf{B_{14}}$ | = | $\left(\frac{1}{2}+x_2\right)\mathbf{a_1}+y_2\,\mathbf{a_2}+\left(\frac{1}{2}-z_2\right)\mathbf{a_3}$ | = | $\left(\frac{1}{2}+x_2\right)a\,\hat{\mathbf{x}}+y_2\,b\,\hat{\mathbf{y}}+\left(\frac{1}{2}-z_2\right)c\,\hat{\mathbf{z}}$ | (8$c$) | O II |
| $\mathbf{B_{15}}$ | = | $x_2\,\mathbf{a_1}+\left(\frac{1}{2}-y_2\right)\mathbf{a_2}+\left(\frac{1}{2}+z_2\right)\mathbf{a_3}$ | = | $x_2\,a\,\hat{\mathbf{x}}+\left(\frac{1}{2}-y_2\right)b\,\hat{\mathbf{y}}+\left(\frac{1}{2}+z_2\right)c\,\hat{\mathbf{z}}$ | (8$c$) | O II |
| $\mathbf{B_{16}}$ | = | $\left(\frac{1}{2}-x_2\right)\mathbf{a_1}+\left(\frac{1}{2}+y_2\right)\mathbf{a_2}+z_2\,\mathbf{a_3}$ | = | $\left(\frac{1}{2}-x_2\right)a\,\hat{\mathbf{x}}+\left(\frac{1}{2}+y_2\right)b\,\hat{\mathbf{y}}+z_2\,c\,\hat{\mathbf{z}}$ | (8$c$) | O II |
| $\mathbf{B_{17}}$ | = | $x_3\,\mathbf{a_1}+y_3\,\mathbf{a_2}+z_3\,\mathbf{a_3}$ | = | $x_3\,a\,\hat{\mathbf{x}}+y_3\,b\,\hat{\mathbf{y}}+z_3\,c\,\hat{\mathbf{z}}$ | (8$c$) | Ti |
| $\mathbf{B_{18}}$ | = | $\left(\frac{1}{2}-x_3\right)\mathbf{a_1}-y_3\,\mathbf{a_2}+\left(\frac{1}{2}+z_3\right)\mathbf{a_3}$ | = | $\left(\frac{1}{2}-x_3\right)a\,\hat{\mathbf{x}}-y_3\,b\,\hat{\mathbf{y}}+\left(\frac{1}{2}+z_3\right)c\,\hat{\mathbf{z}}$ | (8$c$) | Ti |
| $\mathbf{B_{19}}$ | = | $-x_3\,\mathbf{a_1}+\left(\frac{1}{2}+y_3\right)\mathbf{a_2}+\left(\frac{1}{2}-z_3\right)\mathbf{a_3}$ | = | $-x_3\,a\,\hat{\mathbf{x}}+\left(\frac{1}{2}+y_3\right)b\,\hat{\mathbf{y}}+\left(\frac{1}{2}-z_3\right)c\,\hat{\mathbf{z}}$ | (8$c$) | Ti |
| $\mathbf{B_{20}}$ | = | $\left(\frac{1}{2}+x_3\right)\mathbf{a_1}+\left(\frac{1}{2}-y_3\right)\mathbf{a_2}-z_3\,\mathbf{a_3}$ | = | $\left(\frac{1}{2}+x_3\right)a\,\hat{\mathbf{x}}+\left(\frac{1}{2}-y_3\right)b\,\hat{\mathbf{y}}-z_3\,c\,\hat{\mathbf{z}}$ | (8$c$) | Ti |
| $\mathbf{B_{21}}$ | = | $-x_3\,\mathbf{a_1}-y_3\,\mathbf{a_2}-z_3\,\mathbf{a_3}$ | = | $-x_3\,a\,\hat{\mathbf{x}}-y_3\,b\,\hat{\mathbf{y}}-z_3\,c\,\hat{\mathbf{z}}$ | (8$c$) | Ti |
| $\mathbf{B_{22}}$ | = | $\left(\frac{1}{2}+x_3\right)\mathbf{a_1}+y_3\,\mathbf{a_2}+\left(\frac{1}{2}-z_3\right)\mathbf{a_3}$ | = | $\left(\frac{1}{2}+x_3\right)a\,\hat{\mathbf{x}}+y_3\,b\,\hat{\mathbf{y}}+\left(\frac{1}{2}-z_3\right)c\,\hat{\mathbf{z}}$ | (8$c$) | Ti |
| $\mathbf{B_{23}}$ | = | $x_3\,\mathbf{a_1}+\left(\frac{1}{2}-y_3\right)\mathbf{a_2}+\left(\frac{1}{2}+z_3\right)\mathbf{a_3}$ | = | $x_3\,a\,\hat{\mathbf{x}}+\left(\frac{1}{2}-y_3\right)b\,\hat{\mathbf{y}}+\left(\frac{1}{2}+z_3\right)c\,\hat{\mathbf{z}}$ | (8$c$) | Ti |
| $\mathbf{B_{24}}$ | = | $\left(\frac{1}{2}-x_3\right)\mathbf{a_1}+\left(\frac{1}{2}+y_3\right)\mathbf{a_2}+z_3\,\mathbf{a_3}$ | = | $\left(\frac{1}{2}-x_3\right)a\,\hat{\mathbf{x}}+\left(\frac{1}{2}+y_3\right)b\,\hat{\mathbf{y}}+z_3\,c\,\hat{\mathbf{z}}$ | (8$c$) | Ti |

---

---

**Geometry files:**

- CIF: pp. 667
- POSCAR: pp. 668

# Stibnite ($Sb_2S_3$, $D5_8$) Structure: A3B2_oP20_62_3c_2c

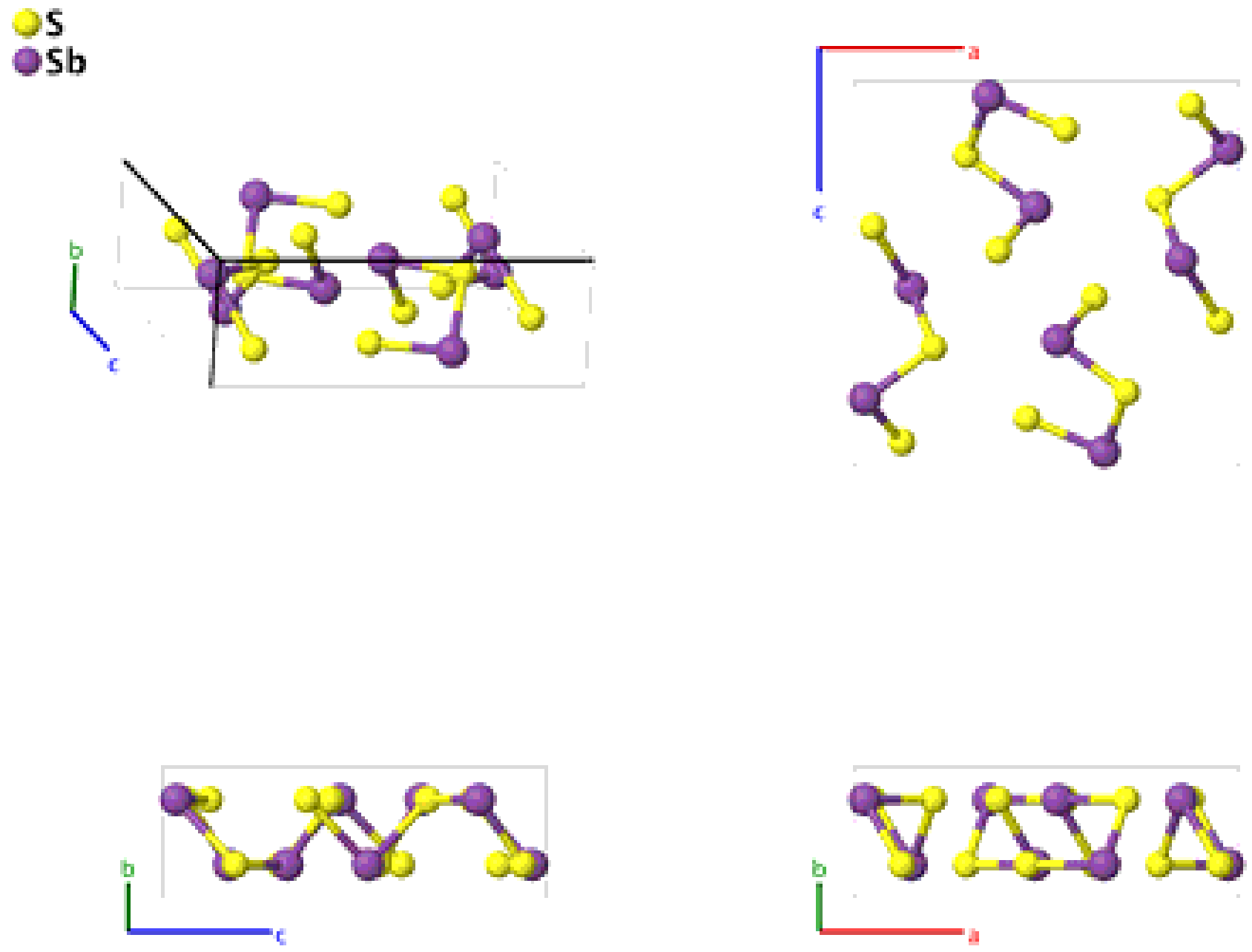

| | | |
|---|---|---|
| **Prototype** | : | $Sb_2S_3$ |
| **AFLOW prototype label** | : | A3B2_oP20_62_3c_2c |
| ***Strukturbericht* designation** | : | $D5_8$ |
| **Pearson symbol** | : | oP20 |
| **Space group number** | : | 62 |
| **Space group symbol** | : | Pnma |
| **AFLOW prototype command** | : | `aflow --proto=A3B2_oP20_62_3c_2c` `--params=`$a, b/a, c/a, x_1, z_1, x_2, z_2, x_3, z_3, x_4, z_4, x_5, z_5$ |

**Simple Orthorhombic primitive vectors:**

$$\mathbf{a}_1 = a\,\hat{\mathbf{x}}$$
$$\mathbf{a}_2 = b\,\hat{\mathbf{y}}$$
$$\mathbf{a}_3 = c\,\hat{\mathbf{z}}$$

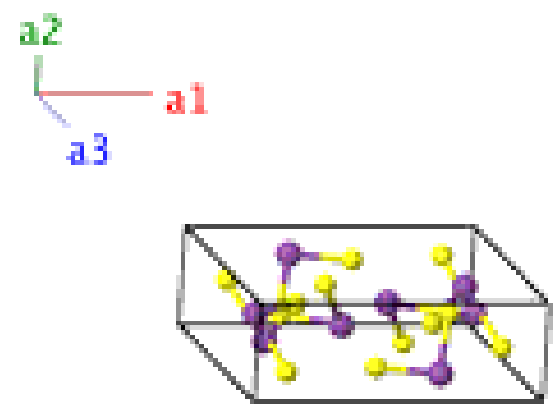

**Basis vectors:**

| | | Lattice Coordinates | | Cartesian Coordinates | Wyckoff Position | Atom Type |
|---|---|---|---|---|---|---|
| $\mathbf{B}_1$ | = | $x_1\,\mathbf{a}_1 + \frac{1}{4}\,\mathbf{a}_2 + z_1\,\mathbf{a}_3$ | = | $x_1\,a\,\hat{\mathbf{x}} + \frac{1}{4}\,b\,\hat{\mathbf{y}} + z_1\,c\,\hat{\mathbf{z}}$ | (4c) | S I |
| $\mathbf{B}_2$ | = | $\left(\frac{1}{2} - x_1\right)\mathbf{a}_1 + \frac{3}{4}\,\mathbf{a}_2 + \left(\frac{1}{2} + z_1\right)\mathbf{a}_3$ | = | $\left(\frac{1}{2} - x_1\right)a\,\hat{\mathbf{x}} + \frac{3}{4}\,b\,\hat{\mathbf{y}} + \left(\frac{1}{2} + z_1\right)c\,\hat{\mathbf{z}}$ | (4c) | S I |
| $\mathbf{B_3}$ | = | $-x_1\,\mathbf{a_1} + \frac{3}{4}\,\mathbf{a_2} - z_1\,\mathbf{a_3}$ | = | $-x_1\,a\,\mathbf{\hat{x}} + \frac{3}{4}\,b\,\mathbf{\hat{y}} - z_1\,c\,\mathbf{\hat{z}}$ | (4$c$) | S I |
| $\mathbf{B_4}$ | = | $\left(\frac{1}{2} + x_1\right)\mathbf{a_1} + \frac{1}{4}\,\mathbf{a_2} + \left(\frac{1}{2} - z_1\right)\mathbf{a_3}$ | = | $\left(\frac{1}{2} + x_1\right)a\,\mathbf{\hat{x}} + \frac{1}{4}\,b\,\mathbf{\hat{y}} + \left(\frac{1}{2} - z_1\right)c\,\mathbf{\hat{z}}$ | (4$c$) | S I |
| $\mathbf{B_5}$ | = | $x_2\,\mathbf{a_1} + \frac{1}{4}\,\mathbf{a_2} + z_2\,\mathbf{a_3}$ | = | $x_2\,a\,\mathbf{\hat{x}} + \frac{1}{4}\,b\,\mathbf{\hat{y}} + z_2\,c\,\mathbf{\hat{z}}$ | (4$c$) | S II |
| $\mathbf{B_6}$ | = | $\left(\frac{1}{2} - x_2\right)\mathbf{a_1} + \frac{3}{4}\,\mathbf{a_2} + \left(\frac{1}{2} + z_2\right)\mathbf{a_3}$ | = | $\left(\frac{1}{2} - x_2\right)a\,\mathbf{\hat{x}} + \frac{3}{4}\,b\,\mathbf{\hat{y}} + \left(\frac{1}{2} + z_2\right)c\,\mathbf{\hat{z}}$ | (4$c$) | S II |
| $\mathbf{B_7}$ | = | $-x_2\,\mathbf{a_1} + \frac{3}{4}\,\mathbf{a_2} - z_2\,\mathbf{a_3}$ | = | $-x_2\,a\,\mathbf{\hat{x}} + \frac{3}{4}\,b\,\mathbf{\hat{y}} - z_2\,c\,\mathbf{\hat{z}}$ | (4$c$) | S II |
| $\mathbf{B_8}$ | = | $\left(\frac{1}{2} + x_2\right)\mathbf{a_1} + \frac{1}{4}\,\mathbf{a_2} + \left(\frac{1}{2} - z_2\right)\mathbf{a_3}$ | = | $\left(\frac{1}{2} + x_2\right)a\,\mathbf{\hat{x}} + \frac{1}{4}\,b\,\mathbf{\hat{y}} + \left(\frac{1}{2} - z_2\right)c\,\mathbf{\hat{z}}$ | (4$c$) | S II |
| $\mathbf{B_9}$ | = | $x_3\,\mathbf{a_1} + \frac{1}{4}\,\mathbf{a_2} + z_3\,\mathbf{a_3}$ | = | $x_3\,a\,\mathbf{\hat{x}} + \frac{1}{4}\,b\,\mathbf{\hat{y}} + z_3\,c\,\mathbf{\hat{z}}$ | (4$c$) | S III |
| $\mathbf{B_{10}}$ | = | $\left(\frac{1}{2} - x_3\right)\mathbf{a_1} + \frac{3}{4}\,\mathbf{a_2} + \left(\frac{1}{2} + z_3\right)\mathbf{a_3}$ | = | $\left(\frac{1}{2} - x_3\right)a\,\mathbf{\hat{x}} + \frac{3}{4}\,b\,\mathbf{\hat{y}} + \left(\frac{1}{2} + z_3\right)c\,\mathbf{\hat{z}}$ | (4$c$) | S III |
| $\mathbf{B_{11}}$ | = | $-x_3\,\mathbf{a_1} + \frac{3}{4}\,\mathbf{a_2} - z_3\,\mathbf{a_3}$ | = | $-x_3\,a\,\mathbf{\hat{x}} + \frac{3}{4}\,b\,\mathbf{\hat{y}} - z_3\,c\,\mathbf{\hat{z}}$ | (4$c$) | S III |
| $\mathbf{B_{12}}$ | = | $\left(\frac{1}{2} + x_3\right)\mathbf{a_1} + \frac{1}{4}\,\mathbf{a_2} + \left(\frac{1}{2} - z_3\right)\mathbf{a_3}$ | = | $\left(\frac{1}{2} + x_3\right)a\,\mathbf{\hat{x}} + \frac{1}{4}\,b\,\mathbf{\hat{y}} + \left(\frac{1}{2} - z_3\right)c\,\mathbf{\hat{z}}$ | (4$c$) | S III |
| $\mathbf{B_{13}}$ | = | $x_4\,\mathbf{a_1} + \frac{1}{4}\,\mathbf{a_2} + z_4\,\mathbf{a_3}$ | = | $x_4\,a\,\mathbf{\hat{x}} + \frac{1}{4}\,b\,\mathbf{\hat{y}} + z_4\,c\,\mathbf{\hat{z}}$ | (4$c$) | Sb I |
| $\mathbf{B_{14}}$ | = | $\left(\frac{1}{2} - x_4\right)\mathbf{a_1} + \frac{3}{4}\,\mathbf{a_2} + \left(\frac{1}{2} + z_4\right)\mathbf{a_3}$ | = | $\left(\frac{1}{2} - x_4\right)a\,\mathbf{\hat{x}} + \frac{3}{4}\,b\,\mathbf{\hat{y}} + \left(\frac{1}{2} + z_4\right)c\,\mathbf{\hat{z}}$ | (4$c$) | Sb I |
| $\mathbf{B_{15}}$ | = | $-x_4\,\mathbf{a_1} + \frac{3}{4}\,\mathbf{a_2} - z_4\,\mathbf{a_3}$ | = | $-x_4\,a\,\mathbf{\hat{x}} + \frac{3}{4}\,b\,\mathbf{\hat{y}} - z_4\,c\,\mathbf{\hat{z}}$ | (4$c$) | Sb I |
| $\mathbf{B_{16}}$ | = | $\left(\frac{1}{2} + x_4\right)\mathbf{a_1} + \frac{1}{4}\,\mathbf{a_2} + \left(\frac{1}{2} - z_4\right)\mathbf{a_3}$ | = | $\left(\frac{1}{2} + x_4\right)a\,\mathbf{\hat{x}} + \frac{1}{4}\,b\,\mathbf{\hat{y}} + \left(\frac{1}{2} - z_4\right)c\,\mathbf{\hat{z}}$ | (4$c$) | Sb I |
| $\mathbf{B_{17}}$ | = | $x_5\,\mathbf{a_1} + \frac{1}{4}\,\mathbf{a_2} + z_5\,\mathbf{a_3}$ | = | $x_5\,a\,\mathbf{\hat{x}} + \frac{1}{4}\,b\,\mathbf{\hat{y}} + z_5\,c\,\mathbf{\hat{z}}$ | (4$c$) | Sb II |
| $\mathbf{B_{18}}$ | = | $\left(\frac{1}{2} - x_5\right)\mathbf{a_1} + \frac{3}{4}\,\mathbf{a_2} + \left(\frac{1}{2} + z_5\right)\mathbf{a_3}$ | = | $\left(\frac{1}{2} - x_5\right)a\,\mathbf{\hat{x}} + \frac{3}{4}\,b\,\mathbf{\hat{y}} + \left(\frac{1}{2} + z_5\right)c\,\mathbf{\hat{z}}$ | (4$c$) | Sb II |
| $\mathbf{B_{19}}$ | = | $-x_5\,\mathbf{a_1} + \frac{3}{4}\,\mathbf{a_2} - z_5\,\mathbf{a_3}$ | = | $-x_5\,a\,\mathbf{\hat{x}} + \frac{3}{4}\,b\,\mathbf{\hat{y}} - z_5\,c\,\mathbf{\hat{z}}$ | (4$c$) | Sb II |
| $\mathbf{B_{20}}$ | = | $\left(\frac{1}{2} + x_5\right)\mathbf{a_1} + \frac{1}{4}\,\mathbf{a_2} + \left(\frac{1}{2} - z_5\right)\mathbf{a_3}$ | = | $\left(\frac{1}{2} + x_5\right)a\,\mathbf{\hat{x}} + \frac{1}{4}\,b\,\mathbf{\hat{y}} + \left(\frac{1}{2} - z_5\right)c\,\mathbf{\hat{z}}$ | (4$c$) | Sb II |

---

**Found in:**

---

**Geometry files:**

- CIF: pp. 668
- POSCAR: pp. 668

# $CaTiO_3$ Pnma Perovskite Structure: AB3C_oP20_62_c_cd_a

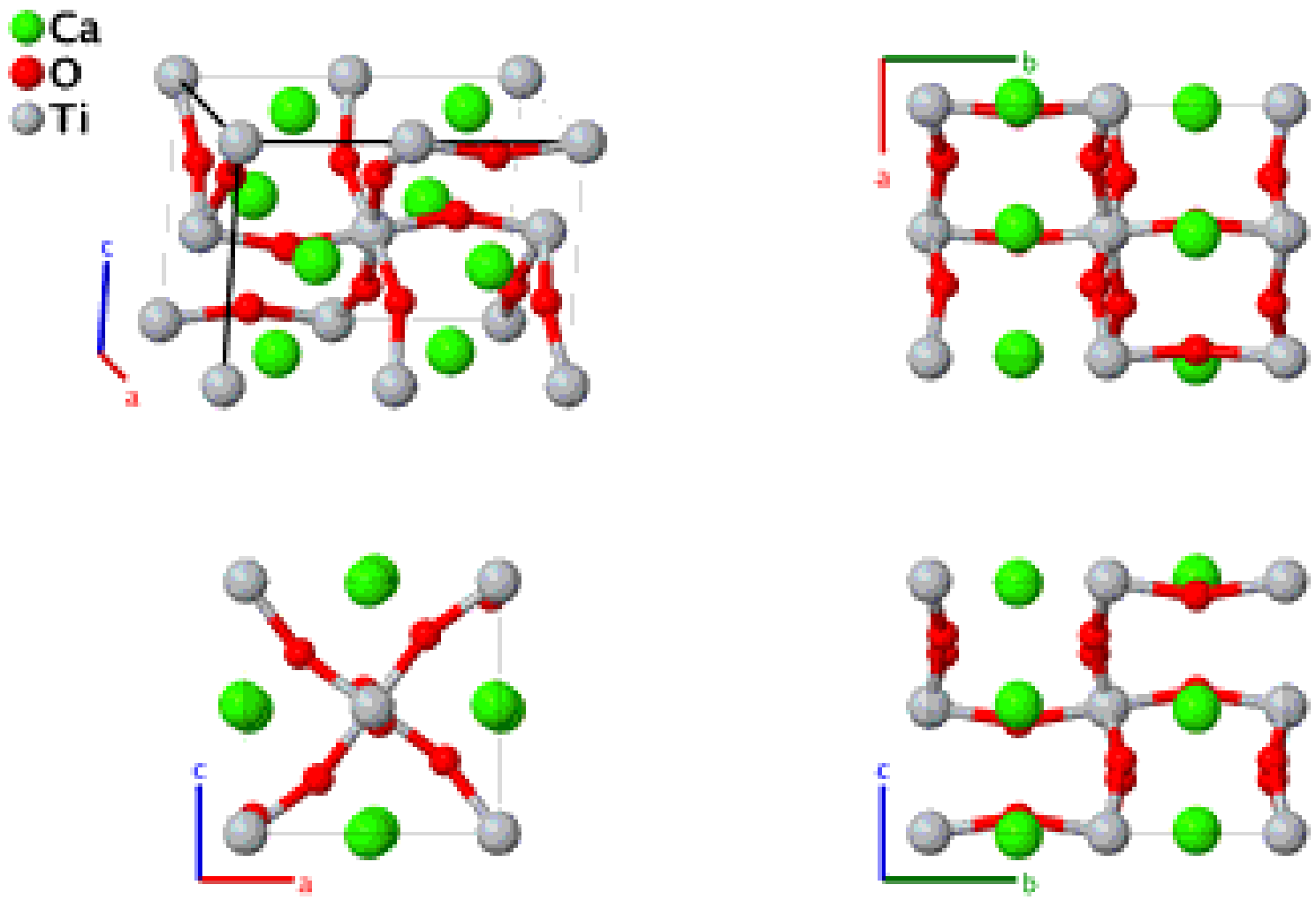

| | | |
|---|---|---|
| **Prototype** | : | $CaTiO_3$ |
| **AFLOW prototype label** | : | AB3C_oP20_62_c_cd_a |
| ***Strukturbericht* designation** | : | None |
| **Pearson symbol** | : | oP20 |
| **Space group number** | : | 62 |
| **Space group symbol** | : | Pnma |
| **AFLOW prototype command** | : | `aflow --proto=AB3C_oP20_62_c_cd_a`<br>`--params=`$a, b/a, c/a, x_2, z_2, x_3, z_3, x_4, y_4, z_4$ |

---

**Other compounds with this structure:**

- $LaMnO_3$, $YAlO_3$, $RFeO_3$ (R = La, Pr, Nd, Sm, Eu, Y)

---

- This is the true ground state of the prototype perovskite, $CaTiO_3$. The standard cubic structure, $E2_1$, is a high-temperature phase.

---

**Simple Orthorhombic primitive vectors:**

$$\mathbf{a}_1 = a\,\hat{\mathbf{x}}$$
$$\mathbf{a}_2 = b\,\hat{\mathbf{y}}$$
$$\mathbf{a}_3 = c\,\hat{\mathbf{z}}$$

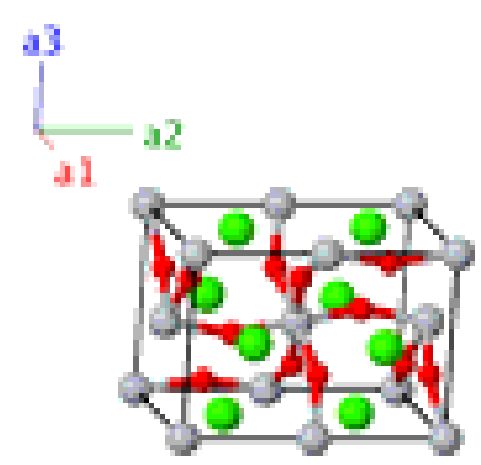

---

**Basis vectors:**

| | | Lattice Coordinates | | Cartesian Coordinates | Wyckoff Position | Atom Type |
|---|---|---|---|---|---|---|
| $\mathbf{B}_1$ | = | $0\,\mathbf{a}_1 + 0\,\mathbf{a}_2 + 0\,\mathbf{a}_3$ | = | $0\hat{\mathbf{x}} + 0\hat{\mathbf{y}} + 0\hat{\mathbf{z}}$ | (4$a$) | Ti |
| $\mathbf{B}_2$ | = | $\frac{1}{2}\,\mathbf{a}_1 + \frac{1}{2}\,\mathbf{a}_3$ | = | $\frac{1}{2}\,a\,\hat{\mathbf{x}} + \frac{1}{2}\,c\,\hat{\mathbf{z}}$ | (4$a$) | Ti |
| $\mathbf{B}_{3}$ | = | $\frac{1}{2}\,\mathbf{a}_{2}$ | = | $\frac{1}{2}\,b\,\hat{\mathbf{y}}$ | (4*a*) | Ti |
| $\mathbf{B}_{4}$ | = | $\frac{1}{2}\,\mathbf{a}_{1}+\frac{1}{2}\,\mathbf{a}_{2}+\frac{1}{2}\,\mathbf{a}_{3}$ | = | $\frac{1}{2}\,a\,\hat{\mathbf{x}}+\frac{1}{2}\,b\,\hat{\mathbf{y}}+\frac{1}{2}\,c\,\hat{\mathbf{z}}$ | (4*a*) | Ti |
| $\mathbf{B}_{5}$ | = | $x_{2}\,\mathbf{a}_{1}+\frac{1}{4}\,\mathbf{a}_{2}+z_{2}\,\mathbf{a}_{3}$ | = | $x_{2}\,a\,\hat{\mathbf{x}}+\frac{1}{4}\,b\,\hat{\mathbf{y}}+z_{2}\,c\,\hat{\mathbf{z}}$ | (4*c*) | Ca |
| $\mathbf{B}_{6}$ | = | $\left(\frac{1}{2}-x_{2}\right)\,\mathbf{a}_{1}+\frac{3}{4}\,\mathbf{a}_{2}+\left(\frac{1}{2}+z_{2}\right)\,\mathbf{a}_{3}$ | = | $\left(\frac{1}{2}-x_{2}\right)\,a\,\hat{\mathbf{x}}+\frac{3}{4}\,b\,\hat{\mathbf{y}}+\left(\frac{1}{2}+z_{2}\right)\,c\,\hat{\mathbf{z}}$ | (4*c*) | Ca |
| $\mathbf{B}_{7}$ | = | $-x_{2}\,\mathbf{a}_{1}+\frac{3}{4}\,\mathbf{a}_{2}-z_{2}\,\mathbf{a}_{3}$ | = | $-x_{2}\,a\,\hat{\mathbf{x}}+\frac{3}{4}\,b\,\hat{\mathbf{y}}-z_{2}\,c\,\hat{\mathbf{z}}$ | (4*c*) | Ca |
| $\mathbf{B}_{8}$ | = | $\left(\frac{1}{2}+x_{2}\right)\,\mathbf{a}_{1}+\frac{1}{4}\,\mathbf{a}_{2}+\left(\frac{1}{2}-z_{2}\right)\,\mathbf{a}_{3}$ | = | $\left(\frac{1}{2}+x_{2}\right)\,a\,\hat{\mathbf{x}}+\frac{1}{4}\,b\,\hat{\mathbf{y}}+\left(\frac{1}{2}-z_{2}\right)\,c\,\hat{\mathbf{z}}$ | (4*c*) | Ca |
| $\mathbf{B}_{9}$ | = | $x_{3}\,\mathbf{a}_{1}+\frac{1}{4}\,\mathbf{a}_{2}+z_{3}\,\mathbf{a}_{3}$ | = | $x_{3}\,a\,\hat{\mathbf{x}}+\frac{1}{4}\,b\,\hat{\mathbf{y}}+z_{3}\,c\,\hat{\mathbf{z}}$ | (4*c*) | O I |
| $\mathbf{B}_{10}$ | = | $\left(\frac{1}{2}-x_{3}\right)\,\mathbf{a}_{1}+\frac{3}{4}\,\mathbf{a}_{2}+\left(\frac{1}{2}+z_{3}\right)\,\mathbf{a}_{3}$ | = | $\left(\frac{1}{2}-x_{3}\right)\,a\,\hat{\mathbf{x}}+\frac{3}{4}\,b\,\hat{\mathbf{y}}+\left(\frac{1}{2}+z_{3}\right)\,c\,\hat{\mathbf{z}}$ | (4*c*) | O I |
| $\mathbf{B}_{11}$ | = | $-x_{3}\,\mathbf{a}_{1}+\frac{3}{4}\,\mathbf{a}_{2}-z_{3}\,\mathbf{a}_{3}$ | = | $-x_{3}\,a\,\hat{\mathbf{x}}+\frac{3}{4}\,b\,\hat{\mathbf{y}}-z_{3}\,c\,\hat{\mathbf{z}}$ | (4*c*) | O I |
| $\mathbf{B}_{12}$ | = | $\left(\frac{1}{2}+x_{3}\right)\,\mathbf{a}_{1}+\frac{1}{4}\,\mathbf{a}_{2}+\left(\frac{1}{2}-z_{3}\right)\,\mathbf{a}_{3}$ | = | $\left(\frac{1}{2}+x_{3}\right)\,a\,\hat{\mathbf{x}}+\frac{1}{4}\,b\,\hat{\mathbf{y}}+\left(\frac{1}{2}-z_{3}\right)\,c\,\hat{\mathbf{z}}$ | (4*c*) | O I |
| $\mathbf{B}_{13}$ | = | $x_{4}\,\mathbf{a}_{1}+y_{4}\,\mathbf{a}_{2}+z_{4}\,\mathbf{a}_{3}$ | = | $x_{4}\,a\,\hat{\mathbf{x}}+y_{4}\,b\,\hat{\mathbf{y}}+z_{4}\,c\,\hat{\mathbf{z}}$ | (8*d*) | O II |
| $\mathbf{B}_{14}$ | = | $\left(\frac{1}{2}-x_{4}\right)\,\mathbf{a}_{1}-y_{4}\,\mathbf{a}_{2}+\left(\frac{1}{2}+z_{4}\right)\,\mathbf{a}_{3}$ | = | $\left(\frac{1}{2}-x_{4}\right)\,a\,\hat{\mathbf{x}}-y_{4}\,b\,\hat{\mathbf{y}}+\left(\frac{1}{2}+z_{4}\right)\,c\,\hat{\mathbf{z}}$ | (8*d*) | O II |
| $\mathbf{B}_{15}$ | = | $-x_{4}\,\mathbf{a}_{1}+\left(\frac{1}{2}+y_{4}\right)\,\mathbf{a}_{2}-z_{4}\,\mathbf{a}_{3}$ | = | $-x_{4}\,a\,\hat{\mathbf{x}}+\left(\frac{1}{2}+y_{4}\right)\,b\,\hat{\mathbf{y}}-z_{4}\,c\,\hat{\mathbf{z}}$ | (8*d*) | O II |
| $\mathbf{B}_{16}$ | = | $\left(\frac{1}{2}+x_{4}\right)\,\mathbf{a}_{1}+\left(\frac{1}{2}-y_{4}\right)\,\mathbf{a}_{2}+\left(\frac{1}{2}-z_{4}\right)\,\mathbf{a}_{3}$ | = | $\left(\frac{1}{2}+x_{4}\right)\,a\,\hat{\mathbf{x}}+\left(\frac{1}{2}-y_{4}\right)\,b\,\hat{\mathbf{y}}+\left(\frac{1}{2}-z_{4}\right)\,c\,\hat{\mathbf{z}}$ | (8*d*) | O II |
| $\mathbf{B}_{17}$ | = | $-x_{4}\,\mathbf{a}_{1}-y_{4}\,\mathbf{a}_{2}-z_{4}\,\mathbf{a}_{3}$ | = | $-x_{4}\,a\,\hat{\mathbf{x}}-y_{4}\,b\,\hat{\mathbf{y}}-z_{4}\,c\,\hat{\mathbf{z}}$ | (8*d*) | O II |
| $\mathbf{B}_{18}$ | = | $\left(\frac{1}{2}+x_{4}\right)\,\mathbf{a}_{1}+y_{4}\,\mathbf{a}_{2}+\left(\frac{1}{2}-z_{4}\right)\,\mathbf{a}_{3}$ | = | $\left(\frac{1}{2}+x_{4}\right)\,a\,\hat{\mathbf{x}}+y_{4}\,b\,\hat{\mathbf{y}}+\left(\frac{1}{2}-z_{4}\right)\,c\,\hat{\mathbf{z}}$ | (8*d*) | O II |
| $\mathbf{B}_{19}$ | = | $x_{4}\,\mathbf{a}_{1}+\left(\frac{1}{2}-y_{4}\right)\,\mathbf{a}_{2}+z_{4}\,\mathbf{a}_{3}$ | = | $x_{4}\,a\,\hat{\mathbf{x}}+\left(\frac{1}{2}-y_{4}\right)\,b\,\hat{\mathbf{y}}+z_{4}\,c\,\hat{\mathbf{z}}$ | (8*d*) | O II |
| $\mathbf{B}_{20}$ | = | $\left(\frac{1}{2}-x_{4}\right)\,\mathbf{a}_{1}+\left(\frac{1}{2}+y_{4}\right)\,\mathbf{a}_{2}+\left(\frac{1}{2}+z_{4}\right)\,\mathbf{a}_{3}$ | = | $\left(\frac{1}{2}-x_{4}\right)\,a\,\hat{\mathbf{x}}+\left(\frac{1}{2}+y_{4}\right)\,b\,\hat{\mathbf{y}}+\left(\frac{1}{2}+z_{4}\right)\,c\,\hat{\mathbf{z}}$ | (8*d*) | O II |

**Geometry files:**

- CIF: pp. 668
- POSCAR: pp. 669

# $MgB_4$ Structure: A4B_oP20_62_2cd_c

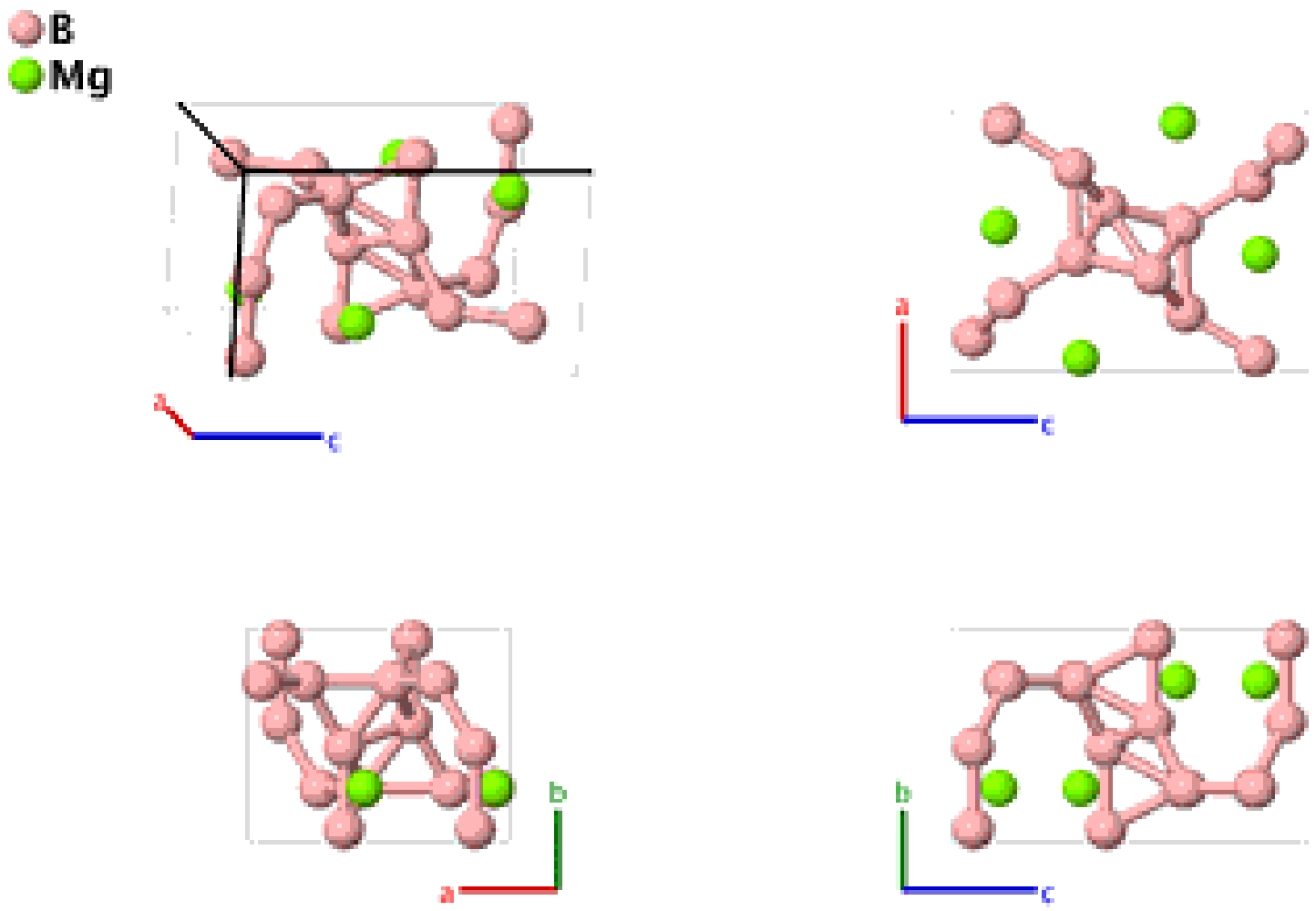

**Prototype** : $MgB_4$

**AFLOW prototype label** : A4B_oP20_62_2cd_c

***Strukturbericht* designation** : None

**Pearson symbol** : oP20

**Space group number** : 62

**Space group symbol** : Pnma

**AFLOW prototype command** : `aflow --proto=A4B_oP20_62_2cd_c --params=`$a, b/a, c/a, x_1, z_1, x_2, z_2, x_3, z_3, x_4, y_4, z_4$

---

**Simple Orthorhombic primitive vectors:**

$$\mathbf{a}_1 = a\,\hat{\mathbf{x}}$$
$$\mathbf{a}_2 = b\,\hat{\mathbf{y}}$$
$$\mathbf{a}_3 = c\,\hat{\mathbf{z}}$$

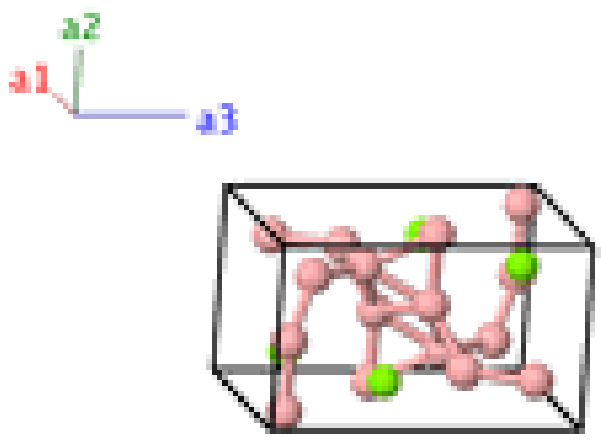

---

**Basis vectors:**

| | | Lattice Coordinates | | Cartesian Coordinates | Wyckoff Position | Atom Type |
|---|---|---|---|---|---|---|
| $\mathbf{B}_1$ | = | $x_1\,\mathbf{a}_1 + \frac{1}{4}\,\mathbf{a}_2 + z_1\,\mathbf{a}_3$ | = | $x_1\,a\,\hat{\mathbf{x}} + \frac{1}{4}\,b\,\hat{\mathbf{y}} + z_1\,c\,\hat{\mathbf{z}}$ | (4$c$) | B I |
| $\mathbf{B}_2$ | = | $\left(\frac{1}{2} - x_1\right)\,\mathbf{a}_1 + \frac{3}{4}\,\mathbf{a}_2 + \left(\frac{1}{2} + z_1\right)\,\mathbf{a}_3$ | = | $\left(\frac{1}{2} - x_1\right)\,a\,\hat{\mathbf{x}} + \frac{3}{4}\,b\,\hat{\mathbf{y}} + \left(\frac{1}{2} + z_1\right)\,c\,\hat{\mathbf{z}}$ | (4$c$) | B I |
| $\mathbf{B}_3$ | = | $-x_1\,\mathbf{a}_1 + \frac{3}{4}\,\mathbf{a}_2 - z_1\,\mathbf{a}_3$ | = | $-x_1\,a\,\hat{\mathbf{x}} + \frac{3}{4}\,b\,\hat{\mathbf{y}} - z_1\,c\,\hat{\mathbf{z}}$ | (4$c$) | B I |
| $\mathbf{B}_4$ | = | $\left(\frac{1}{2} + x_1\right)\,\mathbf{a}_1 + \frac{1}{4}\,\mathbf{a}_2 + \left(\frac{1}{2} - z_1\right)\,\mathbf{a}_3$ | = | $\left(\frac{1}{2} + x_1\right)\,a\,\hat{\mathbf{x}} + \frac{1}{4}\,b\,\hat{\mathbf{y}} + \left(\frac{1}{2} - z_1\right)\,c\,\hat{\mathbf{z}}$ | (4$c$) | B I |
| $\mathbf{B}_5$ | = | $x_2\,\mathbf{a}_1 + \frac{1}{4}\,\mathbf{a}_2 + z_2\,\mathbf{a}_3$ | = | $x_2\,a\,\hat{\mathbf{x}} + \frac{1}{4}\,b\,\hat{\mathbf{y}} + z_2\,c\,\hat{\mathbf{z}}$ | (4$c$) | B II |
| $\mathbf{B}_6$ | = | $\left(\frac{1}{2} - x_2\right)\,\mathbf{a}_1 + \frac{3}{4}\,\mathbf{a}_2 + \left(\frac{1}{2} + z_2\right)\,\mathbf{a}_3$ | = | $\left(\frac{1}{2} - x_2\right)\,a\,\hat{\mathbf{x}} + \frac{3}{4}\,b\,\hat{\mathbf{y}} + \left(\frac{1}{2} + z_2\right)\,c\,\hat{\mathbf{z}}$ | (4$c$) | B II |
| $\mathbf{B}_7$ | = | $-x_2\,\mathbf{a}_1 + \frac{3}{4}\,\mathbf{a}_2 - z_2\,\mathbf{a}_3$ | = | $-x_2\,a\,\hat{\mathbf{x}} + \frac{3}{4}\,b\,\hat{\mathbf{y}} - z_2\,c\,\hat{\mathbf{z}}$ | (4$c$) | B II |
| $\mathbf{B_{8}}$ | = | $\left(\frac{1}{2}+x_{2}\right)\mathbf{a_{1}}+\frac{1}{4}\mathbf{a_{2}}+\left(\frac{1}{2}-z_{2}\right)\mathbf{a_{3}}$ | = | $\left(\frac{1}{2}+x_{2}\right)a\,\hat{\mathbf{x}}+\frac{1}{4}b\,\hat{\mathbf{y}}+\left(\frac{1}{2}-z_{2}\right)c\,\hat{\mathbf{z}}$ | (4$c$) | B II |
| $\mathbf{B_{9}}$ | = | $x_{3}\mathbf{a_{1}}+\frac{1}{4}\mathbf{a_{2}}+z_{3}\mathbf{a_{3}}$ | = | $x_{3}a\,\hat{\mathbf{x}}+\frac{1}{4}b\,\hat{\mathbf{y}}+z_{3}c\,\hat{\mathbf{z}}$ | (4$c$) | Mg |
| $\mathbf{B_{10}}$ | = | $\left(\frac{1}{2}-x_{3}\right)\mathbf{a_{1}}+\frac{3}{4}\mathbf{a_{2}}+\left(\frac{1}{2}+z_{3}\right)\mathbf{a_{3}}$ | = | $\left(\frac{1}{2}-x_{3}\right)a\,\hat{\mathbf{x}}+\frac{3}{4}b\,\hat{\mathbf{y}}+\left(\frac{1}{2}+z_{3}\right)c\,\hat{\mathbf{z}}$ | (4$c$) | Mg |
| $\mathbf{B_{11}}$ | = | $-x_{3}\mathbf{a_{1}}+\frac{3}{4}\mathbf{a_{2}}-z_{3}\mathbf{a_{3}}$ | = | $-x_{3}a\,\hat{\mathbf{x}}+\frac{3}{4}b\,\hat{\mathbf{y}}-z_{3}c\,\hat{\mathbf{z}}$ | (4$c$) | Mg |
| $\mathbf{B_{12}}$ | = | $\left(\frac{1}{2}+x_{3}\right)\mathbf{a_{1}}+\frac{1}{4}\mathbf{a_{2}}+\left(\frac{1}{2}-z_{3}\right)\mathbf{a_{3}}$ | = | $\left(\frac{1}{2}+x_{3}\right)a\,\hat{\mathbf{x}}+\frac{1}{4}b\,\hat{\mathbf{y}}+\left(\frac{1}{2}-z_{3}\right)c\,\hat{\mathbf{z}}$ | (4$c$) | Mg |
| $\mathbf{B_{13}}$ | = | $x_{4}\mathbf{a_{1}}+y_{4}\mathbf{a_{2}}+z_{4}\mathbf{a_{3}}$ | = | $x_{4}a\,\hat{\mathbf{x}}+y_{4}b\,\hat{\mathbf{y}}+z_{4}c\,\hat{\mathbf{z}}$ | (8$d$) | B III |
| $\mathbf{B_{14}}$ | = | $\left(\frac{1}{2}-x_{4}\right)\mathbf{a_{1}}-y_{4}\mathbf{a_{2}}+\left(\frac{1}{2}+z_{4}\right)\mathbf{a_{3}}$ | = | $\left(\frac{1}{2}-x_{4}\right)a\,\hat{\mathbf{x}}-y_{4}b\,\hat{\mathbf{y}}+\left(\frac{1}{2}+z_{4}\right)c\,\hat{\mathbf{z}}$ | (8$d$) | B III |
| $\mathbf{B_{15}}$ | = | $-x_{4}\mathbf{a_{1}}+\left(\frac{1}{2}+y_{4}\right)\mathbf{a_{2}}-z_{4}\mathbf{a_{3}}$ | = | $-x_{4}a\,\hat{\mathbf{x}}+\left(\frac{1}{2}+y_{4}\right)b\,\hat{\mathbf{y}}-z_{4}c\,\hat{\mathbf{z}}$ | (8$d$) | B III |
| $\mathbf{B_{16}}$ | = | $\left(\frac{1}{2}+x_{4}\right)\mathbf{a_{1}}+\left(\frac{1}{2}-y_{4}\right)\mathbf{a_{2}}+\left(\frac{1}{2}-z_{4}\right)\mathbf{a_{3}}$ | = | $\left(\frac{1}{2}+x_{4}\right)a\,\hat{\mathbf{x}}+\left(\frac{1}{2}-y_{4}\right)b\,\hat{\mathbf{y}}+\left(\frac{1}{2}-z_{4}\right)c\,\hat{\mathbf{z}}$ | (8$d$) | B III |
| $\mathbf{B_{17}}$ | = | $-x_{4}\mathbf{a_{1}}-y_{4}\mathbf{a_{2}}-z_{4}\mathbf{a_{3}}$ | = | $-x_{4}a\,\hat{\mathbf{x}}-y_{4}b\,\hat{\mathbf{y}}-z_{4}c\,\hat{\mathbf{z}}$ | (8$d$) | B III |
| $\mathbf{B_{18}}$ | = | $\left(\frac{1}{2}+x_{4}\right)\mathbf{a_{1}}+y_{4}\mathbf{a_{2}}+\left(\frac{1}{2}-z_{4}\right)\mathbf{a_{3}}$ | = | $\left(\frac{1}{2}+x_{4}\right)a\,\hat{\mathbf{x}}+y_{4}b\,\hat{\mathbf{y}}+\left(\frac{1}{2}-z_{4}\right)c\,\hat{\mathbf{z}}$ | (8$d$) | B III |
| $\mathbf{B_{19}}$ | = | $x_{4}\mathbf{a_{1}}+\left(\frac{1}{2}-y_{4}\right)\mathbf{a_{2}}+z_{4}\mathbf{a_{3}}$ | = | $x_{4}a\,\hat{\mathbf{x}}+\left(\frac{1}{2}-y_{4}\right)b\,\hat{\mathbf{y}}+z_{4}c\,\hat{\mathbf{z}}$ | (8$d$) | B III |
| $\mathbf{B_{20}}$ | = | $\left(\frac{1}{2}-x_{4}\right)\mathbf{a_{1}}+\left(\frac{1}{2}+y_{4}\right)\mathbf{a_{2}}+\left(\frac{1}{2}+z_{4}\right)\mathbf{a_{3}}$ | = | $\left(\frac{1}{2}-x_{4}\right)a\,\hat{\mathbf{x}}+\left(\frac{1}{2}+y_{4}\right)b\,\hat{\mathbf{y}}+\left(\frac{1}{2}+z_{4}\right)c\,\hat{\mathbf{z}}$ | (8$d$) | B III |

**Geometry files:**

- CIF: pp. 669
- POSCAR: pp. 669

# Chalcostibite ($CuSbS_2$, $F5_6$): AB2C_oP16_62_c_2c_c

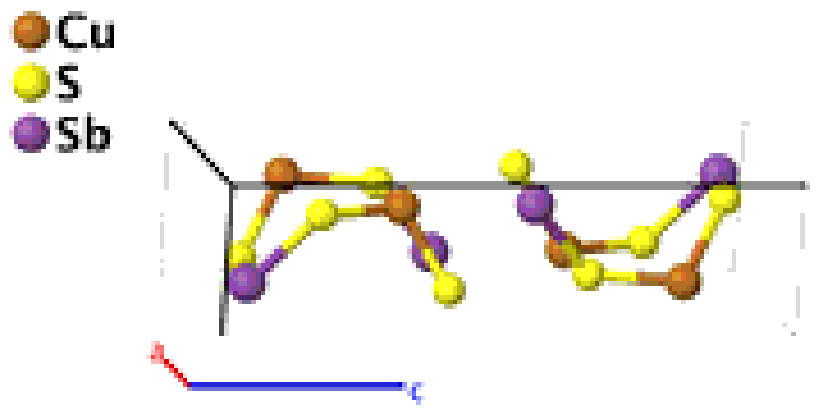

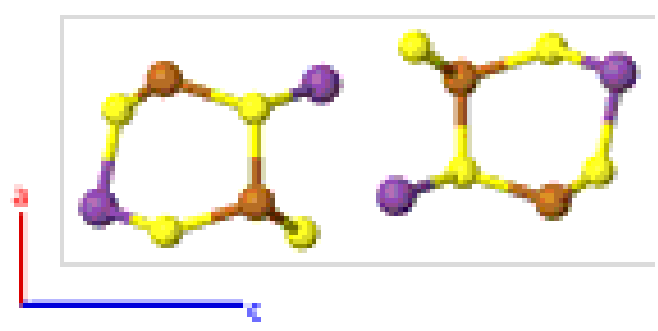

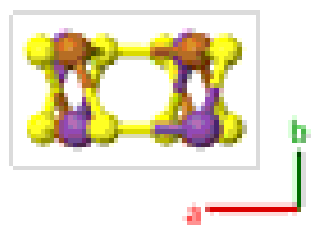

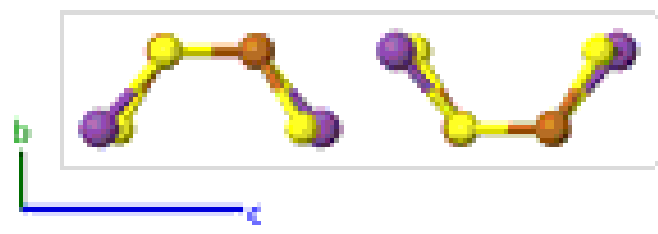

| | | |
|---|---|---|
| **Prototype** | : | $CuSbS_2$ |
| **AFLOW prototype label** | : | AB2C_oP16_62_c_2c_c |
| ***Strukturbericht* designation** | : | $F5_6$ |
| **Pearson symbol** | : | oP16 |
| **Space group number** | : | 62 |
| **Space group symbol** | : | Pnma |
| **AFLOW prototype command** | : | `aflow --proto=AB2C_oP16_62_c_2c_c`<br>`--params=`$a, b/a, c/a, x_1, z_1, x_2, z_2, x_3, z_3, x_4, z_4$ |

**Other compounds with this structure:**

- $CuBiS_2$

**Simple Orthorhombic primitive vectors:**

$$\mathbf{a}_1 = a\,\hat{\mathbf{x}}$$
$$\mathbf{a}_2 = b\,\hat{\mathbf{y}}$$
$$\mathbf{a}_3 = c\,\hat{\mathbf{z}}$$

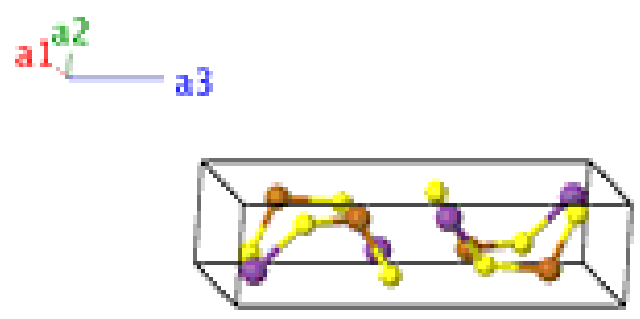

**Basis vectors:**

| | | Lattice Coordinates | | Cartesian Coordinates | Wyckoff Position | Atom Type |
|---|---|---|---|---|---|---|
| $\mathbf{B}_1$ | = | $x_1\,\mathbf{a}_1 + \frac{1}{4}\,\mathbf{a}_2 + z_1\,\mathbf{a}_3$ | = | $x_1\,a\,\hat{\mathbf{x}} + \frac{1}{4}\,b\,\hat{\mathbf{y}} + z_1\,c\,\hat{\mathbf{z}}$ | (4c) | Cu |
| $\mathbf{B}_2$ | = | $\left(\frac{1}{2} - x_1\right)\mathbf{a}_1 + \frac{3}{4}\,\mathbf{a}_2 + \left(\frac{1}{2} + z_1\right)\mathbf{a}_3$ | = | $\left(\frac{1}{2} - x_1\right)a\,\hat{\mathbf{x}} + \frac{3}{4}\,b\,\hat{\mathbf{y}} + \left(\frac{1}{2} + z_1\right)c\,\hat{\mathbf{z}}$ | (4c) | Cu |
| $\mathbf{B}_3$ | = | $-x_1\,\mathbf{a}_1 + \frac{3}{4}\,\mathbf{a}_2 - z_1\,\mathbf{a}_3$ | = | $-x_1\,a\,\hat{\mathbf{x}} + \frac{3}{4}\,b\,\hat{\mathbf{y}} - z_1\,c\,\hat{\mathbf{z}}$ | (4c) | Cu |
| $\mathbf{B}_4$ | = | $\left(\frac{1}{2} + x_1\right)\mathbf{a}_1 + \frac{1}{4}\,\mathbf{a}_2 + \left(\frac{1}{2} - z_1\right)\mathbf{a}_3$ | = | $\left(\frac{1}{2} + x_1\right)a\,\hat{\mathbf{x}} + \frac{1}{4}\,b\,\hat{\mathbf{y}} + \left(\frac{1}{2} - z_1\right)c\,\hat{\mathbf{z}}$ | (4c) | Cu |
| $\mathbf{B}_5$ | = | $x_2\,\mathbf{a}_1 + \frac{1}{4}\,\mathbf{a}_2 + z_2\,\mathbf{a}_3$ | = | $x_2\,a\,\hat{\mathbf{x}} + \frac{1}{4}\,b\,\hat{\mathbf{y}} + z_2\,c\,\hat{\mathbf{z}}$ | (4c) | S I |
| $\mathbf{B}_6$ | = | $\left(\frac{1}{2} - x_2\right)\mathbf{a}_1 + \frac{3}{4}\,\mathbf{a}_2 + \left(\frac{1}{2} + z_2\right)\mathbf{a}_3$ | = | $\left(\frac{1}{2} - x_2\right)a\,\hat{\mathbf{x}} + \frac{3}{4}\,b\,\hat{\mathbf{y}} + \left(\frac{1}{2} + z_2\right)c\,\hat{\mathbf{z}}$ | (4c) | S I |
| $\mathbf{B}_7$ | = | $-x_2\,\mathbf{a}_1 + \frac{3}{4}\,\mathbf{a}_2 - z_2\,\mathbf{a}_3$ | = | $-x_2\,a\,\hat{\mathbf{x}} + \frac{3}{4}\,b\,\hat{\mathbf{y}} - z_2\,c\,\hat{\mathbf{z}}$ | (4c) | S I |
| $\mathbf{B_8}$ | = | $\left(\frac{1}{2}+x_2\right)\mathbf{a_1}+\frac{1}{4}\mathbf{a_2}+\left(\frac{1}{2}-z_2\right)\mathbf{a_3}$ | = | $\left(\frac{1}{2}+x_2\right)a\,\mathbf{\hat{x}}+\frac{1}{4}b\,\mathbf{\hat{y}}+\left(\frac{1}{2}-z_2\right)c\,\mathbf{\hat{z}}$ | (4c) | S I |
| $\mathbf{B_9}$ | = | $x_3\,\mathbf{a_1}+\frac{1}{4}\mathbf{a_2}+z_3\,\mathbf{a_3}$ | = | $x_3\,a\,\mathbf{\hat{x}}+\frac{1}{4}b\,\mathbf{\hat{y}}+z_3\,c\,\mathbf{\hat{z}}$ | (4c) | S II |
| $\mathbf{B_{10}}$ | = | $\left(\frac{1}{2}-x_3\right)\mathbf{a_1}+\frac{3}{4}\mathbf{a_2}+\left(\frac{1}{2}+z_3\right)\mathbf{a_3}$ | = | $\left(\frac{1}{2}-x_3\right)a\,\mathbf{\hat{x}}+\frac{3}{4}b\,\mathbf{\hat{y}}+\left(\frac{1}{2}+z_3\right)c\,\mathbf{\hat{z}}$ | (4c) | S II |
| $\mathbf{B_{11}}$ | = | $-x_3\,\mathbf{a_1}+\frac{3}{4}\mathbf{a_2}-z_3\,\mathbf{a_3}$ | = | $-x_3\,a\,\mathbf{\hat{x}}+\frac{3}{4}b\,\mathbf{\hat{y}}-z_3\,c\,\mathbf{\hat{z}}$ | (4c) | S II |
| $\mathbf{B_{12}}$ | = | $\left(\frac{1}{2}+x_3\right)\mathbf{a_1}+\frac{1}{4}\mathbf{a_2}+\left(\frac{1}{2}-z_3\right)\mathbf{a_3}$ | = | $\left(\frac{1}{2}+x_3\right)a\,\mathbf{\hat{x}}+\frac{1}{4}b\,\mathbf{\hat{y}}+\left(\frac{1}{2}-z_3\right)c\,\mathbf{\hat{z}}$ | (4c) | S II |
| $\mathbf{B_{13}}$ | = | $x_4\,\mathbf{a_1}+\frac{1}{4}\mathbf{a_2}+z_4\,\mathbf{a_3}$ | = | $x_4\,a\,\mathbf{\hat{x}}+\frac{1}{4}b\,\mathbf{\hat{y}}+z_4\,c\,\mathbf{\hat{z}}$ | (4c) | Sb |
| $\mathbf{B_{14}}$ | = | $\left(\frac{1}{2}-x_4\right)\mathbf{a_1}+\frac{3}{4}\mathbf{a_2}+\left(\frac{1}{2}+z_4\right)\mathbf{a_3}$ | = | $\left(\frac{1}{2}-x_4\right)a\,\mathbf{\hat{x}}+\frac{3}{4}b\,\mathbf{\hat{y}}+\left(\frac{1}{2}+z_4\right)c\,\mathbf{\hat{z}}$ | (4c) | Sb |
| $\mathbf{B_{15}}$ | = | $-x_4\,\mathbf{a_1}+\frac{3}{4}\mathbf{a_2}-z_4\,\mathbf{a_3}$ | = | $-x_4\,a\,\mathbf{\hat{x}}+\frac{3}{4}b\,\mathbf{\hat{y}}-z_4\,c\,\mathbf{\hat{z}}$ | (4c) | Sb |
| $\mathbf{B_{16}}$ | = | $\left(\frac{1}{2}+x_4\right)\mathbf{a_1}+\frac{1}{4}\mathbf{a_2}+\left(\frac{1}{2}-z_4\right)\mathbf{a_3}$ | = | $\left(\frac{1}{2}+x_4\right)a\,\mathbf{\hat{x}}+\frac{1}{4}b\,\mathbf{\hat{y}}+\left(\frac{1}{2}-z_4\right)c\,\mathbf{\hat{z}}$ | (4c) | Sb |

---

---

**Geometry files:**

- CIF: pp. 669
- POSCAR: pp. 670

# $Co_2Si$ (C37) Structure: A2B_oP12_62_2c_c

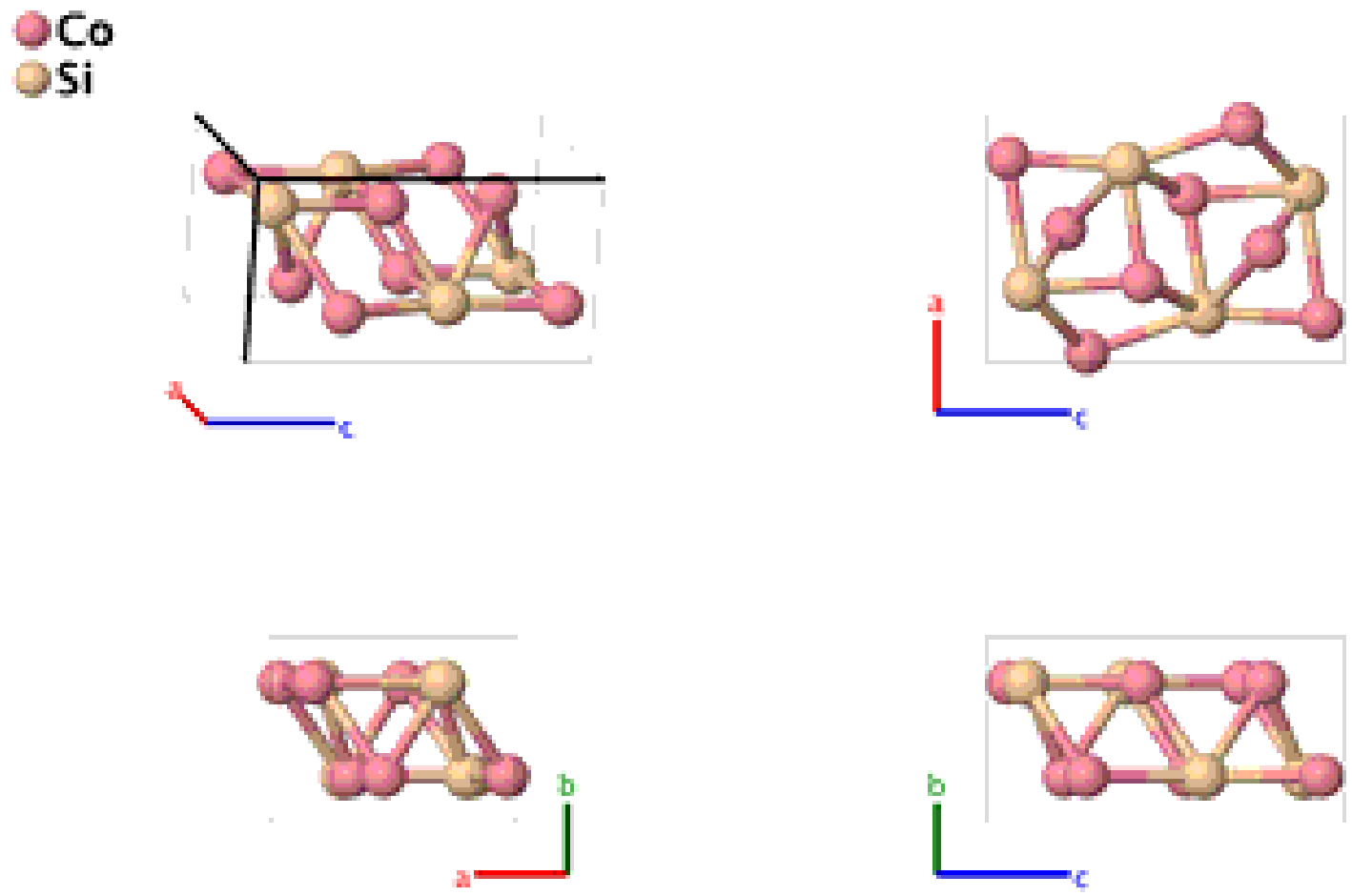

| | | |
|---|---|---|
| **Prototype** | : | $Co_2Si$ |
| **AFLOW prototype label** | : | A2B_oP12_62_2c_c |
| ***Strukturbericht* designation** | : | C37 |
| **Pearson symbol** | : | oP12 |
| **Space group number** | : | 62 |
| **Space group symbol** | : | Pnma |
| **AFLOW prototype command** | : | `aflow --proto=A2B_oP12_62_2c_c` <br> `--params=`$a, b/a, c/a, x_1, z_1, x_2, z_2, x_3, z_3$ |

- Note that $Co_2Si$ (pp. 157), $HgCl_2$ (pp. 159), and cotunnite (pp. 161) have the same AFLOW prototype label. They are generated by the same symmetry operations with different sets of parameters (`--params`) specified in their corresponding CIF files.

**Simple Orthorhombic primitive vectors:**

$$\mathbf{a}_1 = a\,\hat{\mathbf{x}}$$
$$\mathbf{a}_2 = b\,\hat{\mathbf{y}}$$
$$\mathbf{a}_3 = c\,\hat{\mathbf{z}}$$

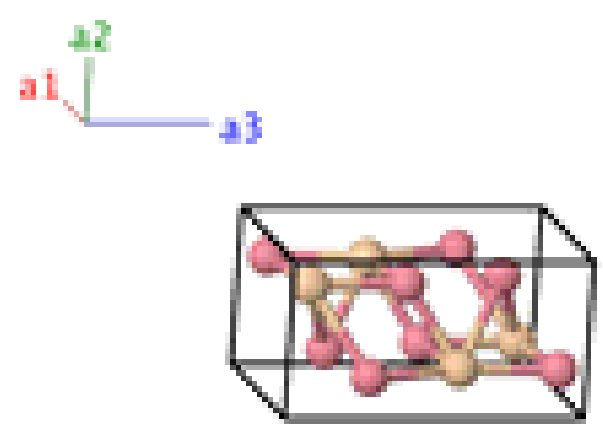

**Basis vectors:**

| | | Lattice Coordinates | | Cartesian Coordinates | Wyckoff Position | Atom Type |
|---|---|---|---|---|---|---|
| $\mathbf{B}_1$ | = | $x_1\,\mathbf{a}_1 + \frac{1}{4}\,\mathbf{a}_2 + z_1\,\mathbf{a}_3$ | = | $x_1\,a\,\hat{\mathbf{x}} + \frac{1}{4}\,b\,\hat{\mathbf{y}} + z_1\,c\,\hat{\mathbf{z}}$ | (4c) | Co I |
| $\mathbf{B}_2$ | = | $\left(\frac{1}{2} - x_1\right)\mathbf{a}_1 + \frac{3}{4}\,\mathbf{a}_2 + \left(\frac{1}{2} + z_1\right)\mathbf{a}_3$ | = | $\left(\frac{1}{2} - x_1\right)a\,\hat{\mathbf{x}} + \frac{3}{4}\,b\,\hat{\mathbf{y}} + \left(\frac{1}{2} + z_1\right)c\,\hat{\mathbf{z}}$ | (4c) | Co I |
| $\mathbf{B}_3$ | = | $-x_1\,\mathbf{a}_1 + \frac{3}{4}\,\mathbf{a}_2 - z_1\,\mathbf{a}_3$ | = | $-x_1\,a\,\hat{\mathbf{x}} + \frac{3}{4}\,b\,\hat{\mathbf{y}} - z_1\,c\,\hat{\mathbf{z}}$ | (4c) | Co I |
| $\mathbf{B}_4$ | = | $\left(\frac{1}{2} + x_1\right)\mathbf{a}_1 + \frac{1}{4}\,\mathbf{a}_2 + \left(\frac{1}{2} - z_1\right)\mathbf{a}_3$ | = | $\left(\frac{1}{2} + x_1\right)a\,\hat{\mathbf{x}} + \frac{1}{4}\,b\,\hat{\mathbf{y}} + \left(\frac{1}{2} - z_1\right)c\,\hat{\mathbf{z}}$ | (4c) | Co I |
| $\mathbf{B_5}$ | = | $x_2\,\mathbf{a_1} + \frac{1}{4}\,\mathbf{a_2} + z_2\,\mathbf{a_3}$ | = | $x_2\,a\,\mathbf{\hat{x}} + \frac{1}{4}\,b\,\mathbf{\hat{y}} + z_2\,c\,\mathbf{\hat{z}}$ | (4c) | Co II |
| $\mathbf{B_6}$ | = | $\left(\frac{1}{2} - x_2\right)\,\mathbf{a_1} + \frac{3}{4}\,\mathbf{a_2} + \left(\frac{1}{2} + z_2\right)\,\mathbf{a_3}$ | = | $\left(\frac{1}{2} - x_2\right)\,a\,\mathbf{\hat{x}} + \frac{3}{4}\,b\,\mathbf{\hat{y}} + \left(\frac{1}{2} + z_2\right)\,c\,\mathbf{\hat{z}}$ | (4c) | Co II |
| $\mathbf{B_7}$ | = | $-x_2\,\mathbf{a_1} + \frac{3}{4}\,\mathbf{a_2} - z_2\,\mathbf{a_3}$ | = | $-x_2\,a\,\mathbf{\hat{x}} + \frac{3}{4}\,b\,\mathbf{\hat{y}} - z_2\,c\,\mathbf{\hat{z}}$ | (4c) | Co II |
| $\mathbf{B_8}$ | = | $\left(\frac{1}{2} + x_2\right)\,\mathbf{a_1} + \frac{1}{4}\,\mathbf{a_2} + \left(\frac{1}{2} - z_2\right)\,\mathbf{a_3}$ | = | $\left(\frac{1}{2} + x_2\right)\,a\,\mathbf{\hat{x}} + \frac{1}{4}\,b\,\mathbf{\hat{y}} + \left(\frac{1}{2} - z_2\right)\,c\,\mathbf{\hat{z}}$ | (4c) | Co II |
| $\mathbf{B_9}$ | = | $x_3\,\mathbf{a_1} + \frac{1}{4}\,\mathbf{a_2} + z_3\,\mathbf{a_3}$ | = | $x_3\,a\,\mathbf{\hat{x}} + \frac{1}{4}\,b\,\mathbf{\hat{y}} + z_3\,c\,\mathbf{\hat{z}}$ | (4c) | Si |
| $\mathbf{B_{10}}$ | = | $\left(\frac{1}{2} - x_3\right)\,\mathbf{a_1} + \frac{3}{4}\,\mathbf{a_2} + \left(\frac{1}{2} + z_3\right)\,\mathbf{a_3}$ | = | $\left(\frac{1}{2} - x_3\right)\,a\,\mathbf{\hat{x}} + \frac{3}{4}\,b\,\mathbf{\hat{y}} + \left(\frac{1}{2} + z_3\right)\,c\,\mathbf{\hat{z}}$ | (4c) | Si |
| $\mathbf{B_{11}}$ | = | $-x_3\,\mathbf{a_1} + \frac{3}{4}\,\mathbf{a_2} - z_3\,\mathbf{a_3}$ | = | $-x_3\,a\,\mathbf{\hat{x}} + \frac{3}{4}\,b\,\mathbf{\hat{y}} - z_3\,c\,\mathbf{\hat{z}}$ | (4c) | Si |
| $\mathbf{B_{12}}$ | = | $\left(\frac{1}{2} + x_3\right)\,\mathbf{a_1} + \frac{1}{4}\,\mathbf{a_2} + \left(\frac{1}{2} - z_3\right)\,\mathbf{a_3}$ | = | $\left(\frac{1}{2} + x_3\right)\,a\,\mathbf{\hat{x}} + \frac{1}{4}\,b\,\mathbf{\hat{y}} + \left(\frac{1}{2} - z_3\right)\,c\,\mathbf{\hat{z}}$ | (4c) | Si |

**Geometry files:**

- CIF: pp. 670
- POSCAR: pp. 670

# $HgCl_2$ (C25) Structure: A2B_oP12_62_2c_c

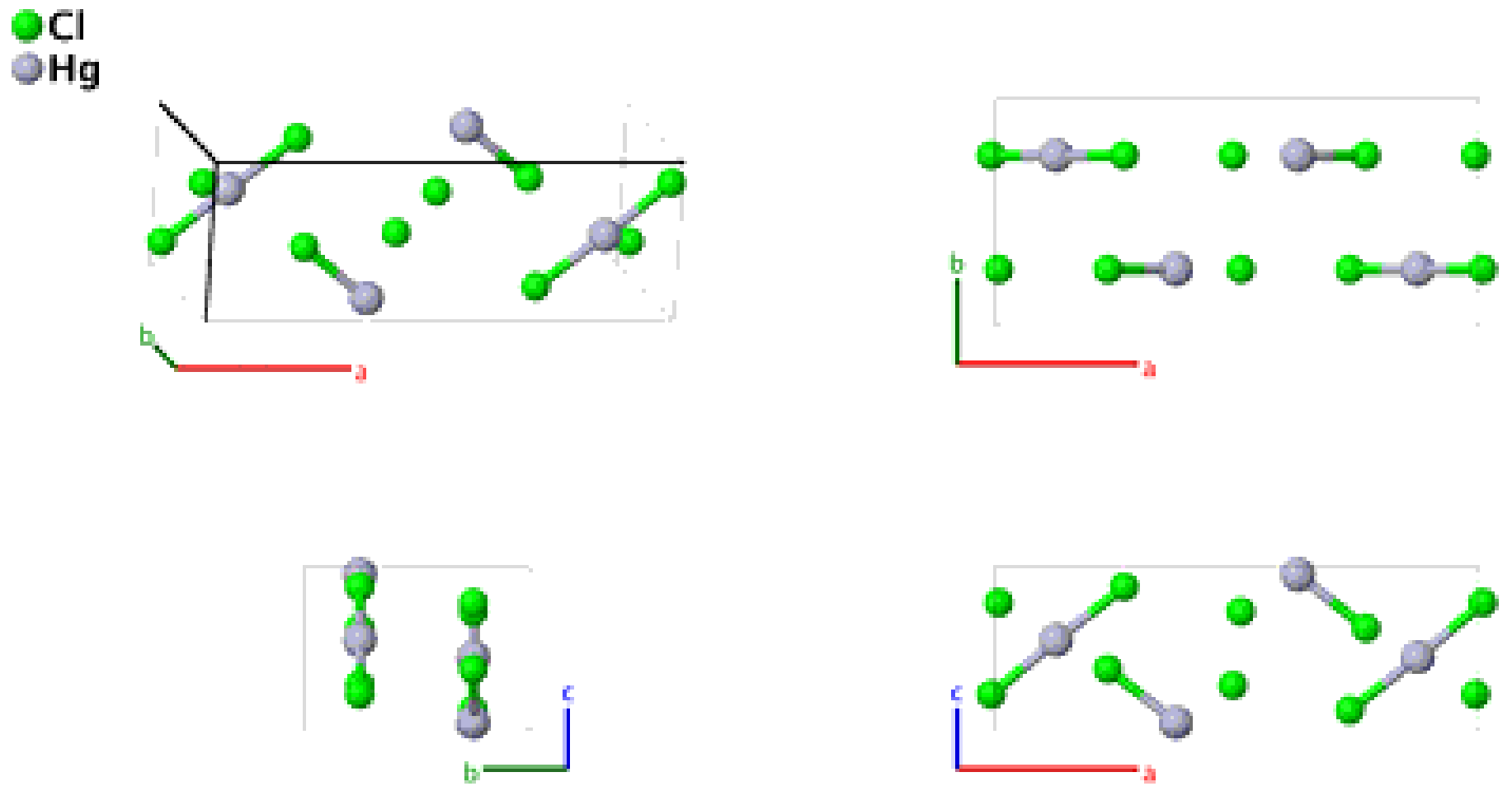

| | | |
|---|---|---|
| **Prototype** | : | $HgCl_2$ |
| **AFLOW prototype label** | : | A2B_oP12_62_2c_c |
| ***Strukturbericht* designation** | : | C25 |
| **Pearson symbol** | : | oP12 |
| **Space group number** | : | 62 |
| **Space group symbol** | : | Pnma |
| **AFLOW prototype command** | : | `aflow --proto=A2B_oP12_62_2c_c` <br> `--params=`$a, b/a, c/a, x_1, z_1, x_2, z_2, x_3, z_3$ |

**Other compounds with this structure:**

- $FeO_2$ (Goethite)

- Note that $Co_2Si$ (pp. 157), $HgCl_2$ (pp. 159), and cotunnite (pp. 161) have the same AFLOW prototype label. They are generated by the same symmetry operations with different sets of parameters (`--params`) specified in their corresponding CIF files.

**Simple Orthorhombic primitive vectors:**

$$\mathbf{a}_1 = a\,\hat{\mathbf{x}}$$
$$\mathbf{a}_2 = b\,\hat{\mathbf{y}}$$
$$\mathbf{a}_3 = c\,\hat{\mathbf{z}}$$

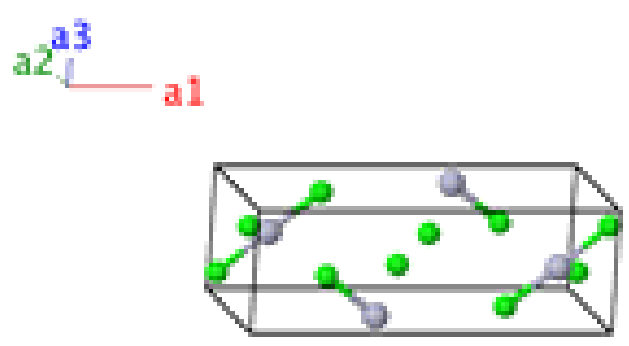

**Basis vectors:**

| | | Lattice Coordinates | | Cartesian Coordinates | Wyckoff Position | Atom Type |
|---|---|---|---|---|---|---|
| $\mathbf{B}_1$ | = | $x_1\,\mathbf{a}_1 + \frac{1}{4}\,\mathbf{a}_2 + z_1\,\mathbf{a}_3$ | = | $x_1\,a\,\hat{\mathbf{x}} + \frac{1}{4}\,b\,\hat{\mathbf{y}} + z_1\,c\,\hat{\mathbf{z}}$ | (4c) | Cl I |
| $\mathbf{B}_2$ | = | $\left(\frac{1}{2} - x_1\right)\mathbf{a}_1 + \frac{3}{4}\,\mathbf{a}_2 + \left(\frac{1}{2} + z_1\right)\mathbf{a}_3$ | = | $\left(\frac{1}{2} - x_1\right)a\,\hat{\mathbf{x}} + \frac{3}{4}\,b\,\hat{\mathbf{y}} + \left(\frac{1}{2} + z_1\right)c\,\hat{\mathbf{z}}$ | (4c) | Cl I |
| $\mathbf{B_3}$ | = | $-x_1\,\mathbf{a_1} + \frac{3}{4}\,\mathbf{a_2} - z_1\,\mathbf{a_3}$ | = | $-x_1\,a\,\mathbf{\hat{x}} + \frac{3}{4}\,b\,\mathbf{\hat{y}} - z_1\,c\,\mathbf{\hat{z}}$ | (4$c$) | Cl I |
| $\mathbf{B_4}$ | = | $\left(\frac{1}{2} + x_1\right)\mathbf{a_1} + \frac{1}{4}\,\mathbf{a_2} + \left(\frac{1}{2} - z_1\right)\mathbf{a_3}$ | = | $\left(\frac{1}{2} + x_1\right)a\,\mathbf{\hat{x}} + \frac{1}{4}\,b\,\mathbf{\hat{y}} + \left(\frac{1}{2} - z_1\right)c\,\mathbf{\hat{z}}$ | (4$c$) | Cl I |
| $\mathbf{B_5}$ | = | $x_2\,\mathbf{a_1} + \frac{1}{4}\,\mathbf{a_2} + z_2\,\mathbf{a_3}$ | = | $x_2\,a\,\mathbf{\hat{x}} + \frac{1}{4}\,b\,\mathbf{\hat{y}} + z_2\,c\,\mathbf{\hat{z}}$ | (4$c$) | Cl II |
| $\mathbf{B_6}$ | = | $\left(\frac{1}{2} - x_2\right)\mathbf{a_1} + \frac{3}{4}\,\mathbf{a_2} + \left(\frac{1}{2} + z_2\right)\mathbf{a_3}$ | = | $\left(\frac{1}{2} - x_2\right)a\,\mathbf{\hat{x}} + \frac{3}{4}\,b\,\mathbf{\hat{y}} + \left(\frac{1}{2} + z_2\right)c\,\mathbf{\hat{z}}$ | (4$c$) | Cl II |
| $\mathbf{B_7}$ | = | $-x_2\,\mathbf{a_1} + \frac{3}{4}\,\mathbf{a_2} - z_2\,\mathbf{a_3}$ | = | $-x_2\,a\,\mathbf{\hat{x}} + \frac{3}{4}\,b\,\mathbf{\hat{y}} - z_2\,c\,\mathbf{\hat{z}}$ | (4$c$) | Cl II |
| $\mathbf{B_8}$ | = | $\left(\frac{1}{2} + x_2\right)\mathbf{a_1} + \frac{1}{4}\,\mathbf{a_2} + \left(\frac{1}{2} - z_2\right)\mathbf{a_3}$ | = | $\left(\frac{1}{2} + x_2\right)a\,\mathbf{\hat{x}} + \frac{1}{4}\,b\,\mathbf{\hat{y}} + \left(\frac{1}{2} - z_2\right)c\,\mathbf{\hat{z}}$ | (4$c$) | Cl II |
| $\mathbf{B_9}$ | = | $x_3\,\mathbf{a_1} + \frac{1}{4}\,\mathbf{a_2} + z_3\,\mathbf{a_3}$ | = | $x_3\,a\,\mathbf{\hat{x}} + \frac{1}{4}\,b\,\mathbf{\hat{y}} + z_3\,c\,\mathbf{\hat{z}}$ | (4$c$) | Hg |
| $\mathbf{B_{10}}$ | = | $\left(\frac{1}{2} - x_3\right)\mathbf{a_1} + \frac{3}{4}\,\mathbf{a_2} + \left(\frac{1}{2} + z_3\right)\mathbf{a_3}$ | = | $\left(\frac{1}{2} - x_3\right)a\,\mathbf{\hat{x}} + \frac{3}{4}\,b\,\mathbf{\hat{y}} + \left(\frac{1}{2} + z_3\right)c\,\mathbf{\hat{z}}$ | (4$c$) | Hg |
| $\mathbf{B_{11}}$ | = | $-x_3\,\mathbf{a_1} + \frac{3}{4}\,\mathbf{a_2} - z_3\,\mathbf{a_3}$ | = | $-x_3\,a\,\mathbf{\hat{x}} + \frac{3}{4}\,b\,\mathbf{\hat{y}} - z_3\,c\,\mathbf{\hat{z}}$ | (4$c$) | Hg |
| $\mathbf{B_{12}}$ | = | $\left(\frac{1}{2} + x_3\right)\mathbf{a_1} + \frac{1}{4}\,\mathbf{a_2} + \left(\frac{1}{2} - z_3\right)\mathbf{a_3}$ | = | $\left(\frac{1}{2} + x_3\right)a\,\mathbf{\hat{x}} + \frac{1}{4}\,b\,\mathbf{\hat{y}} + \left(\frac{1}{2} - z_3\right)c\,\mathbf{\hat{z}}$ | (4$c$) | Hg |

**Geometry files:**

- CIF: pp. 670
- POSCAR: pp. 671

# Cotunnite ($PbCl_2$, C23) Structure: A2B_oP12_62_2c_c

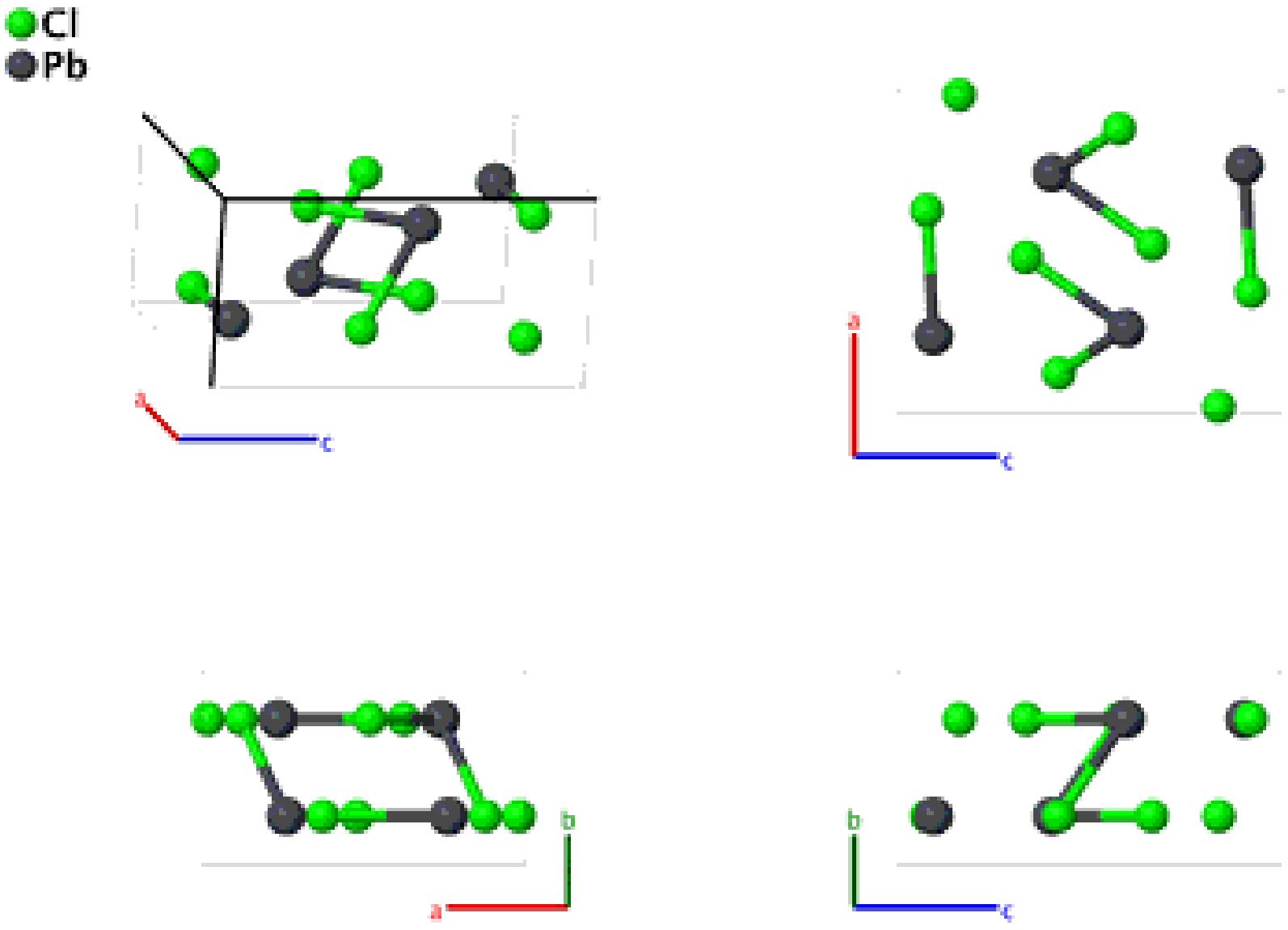

| | | |
|---|---|---|
| **Prototype** | : | $PbCl_2$ |
| **AFLOW prototype label** | : | A2B_oP12_62_2c_c |
| ***Strukturbericht* designation** | : | C23 |
| **Pearson symbol** | : | oP12 |
| **Space group number** | : | 62 |
| **Space group symbol** | : | Pnma |
| **AFLOW prototype command** | : | `aflow --proto=A2B_oP12_62_2c_c`<br>`--params=`$a, b/a, c/a, x_1, z_1, x_2, z_2, x_3, z_3$ |

**Other compounds with this structure:**

- $PbO_2$, $ZrO_2$, $TiO_2$

- Note that $Co_2Si$ (pp. 157), $HgCl_2$ (pp. 159), and cotunnite (pp. 161) have the same AFLOW prototype label. They are generated by the same symmetry operations with different sets of parameters (`--params`) specified in their corresponding CIF files.

**Simple Orthorhombic primitive vectors:**

$$
\begin{aligned}
\mathbf{a}_1 &= a\,\hat{\mathbf{x}} \\
\mathbf{a}_2 &= b\,\hat{\mathbf{y}} \\
\mathbf{a}_3 &= c\,\hat{\mathbf{z}}
\end{aligned}
$$

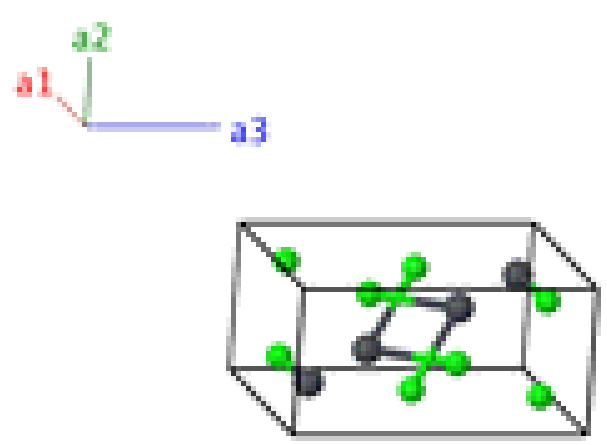

**Basis vectors:**

| | | Lattice Coordinates | | Cartesian Coordinates | Wyckoff Position | Atom Type |
|---|---|---|---|---|---|---|
| $\mathbf{B_1}$ | = | $x_1\,\mathbf{a_1} + \frac{1}{4}\,\mathbf{a_2} + z_1\,\mathbf{a_3}$ | = | $x_1\,a\,\hat{\mathbf{x}} + \frac{1}{4}\,b\,\hat{\mathbf{y}} + z_1\,c\,\hat{\mathbf{z}}$ | (4$c$) | Cl I |
| $\mathbf{B_2}$ | = | $\left(\frac{1}{2} - x_1\right)\mathbf{a_1} + \frac{3}{4}\,\mathbf{a_2} + \left(\frac{1}{2} + z_1\right)\mathbf{a_3}$ | = | $\left(\frac{1}{2} - x_1\right)a\,\hat{\mathbf{x}} + \frac{3}{4}\,b\,\hat{\mathbf{y}} + \left(\frac{1}{2} + z_1\right)c\,\hat{\mathbf{z}}$ | (4$c$) | Cl I |
| $\mathbf{B_3}$ | = | $-x_1\,\mathbf{a_1} + \frac{3}{4}\,\mathbf{a_2} - z_1\,\mathbf{a_3}$ | = | $-x_1\,a\,\hat{\mathbf{x}} + \frac{3}{4}\,b\,\hat{\mathbf{y}} - z_1\,c\,\hat{\mathbf{z}}$ | (4$c$) | Cl I |
| $\mathbf{B_4}$ | = | $\left(\frac{1}{2} + x_1\right)\mathbf{a_1} + \frac{1}{4}\,\mathbf{a_2} + \left(\frac{1}{2} - z_1\right)\mathbf{a_3}$ | = | $\left(\frac{1}{2} + x_1\right)a\,\hat{\mathbf{x}} + \frac{1}{4}\,b\,\hat{\mathbf{y}} + \left(\frac{1}{2} - z_1\right)c\,\hat{\mathbf{z}}$ | (4$c$) | Cl I |
| $\mathbf{B_5}$ | = | $x_2\,\mathbf{a_1} + \frac{1}{4}\,\mathbf{a_2} + z_2\,\mathbf{a_3}$ | = | $x_2\,a\,\hat{\mathbf{x}} + \frac{1}{4}\,b\,\hat{\mathbf{y}} + z_2\,c\,\hat{\mathbf{z}}$ | (4$c$) | Cl II |
| $\mathbf{B_6}$ | = | $\left(\frac{1}{2} - x_2\right)\mathbf{a_1} + \frac{3}{4}\,\mathbf{a_2} + \left(\frac{1}{2} + z_2\right)\mathbf{a_3}$ | = | $\left(\frac{1}{2} - x_2\right)a\,\hat{\mathbf{x}} + \frac{3}{4}\,b\,\hat{\mathbf{y}} + \left(\frac{1}{2} + z_2\right)c\,\hat{\mathbf{z}}$ | (4$c$) | Cl II |
| $\mathbf{B_7}$ | = | $-x_2\,\mathbf{a_1} + \frac{3}{4}\,\mathbf{a_2} - z_2\,\mathbf{a_3}$ | = | $-x_2\,a\,\hat{\mathbf{x}} + \frac{3}{4}\,b\,\hat{\mathbf{y}} - z_2\,c\,\hat{\mathbf{z}}$ | (4$c$) | Cl II |
| $\mathbf{B_8}$ | = | $\left(\frac{1}{2} + x_2\right)\mathbf{a_1} + \frac{1}{4}\,\mathbf{a_2} + \left(\frac{1}{2} - z_2\right)\mathbf{a_3}$ | = | $\left(\frac{1}{2} + x_2\right)a\,\hat{\mathbf{x}} + \frac{1}{4}\,b\,\hat{\mathbf{y}} + \left(\frac{1}{2} - z_2\right)c\,\hat{\mathbf{z}}$ | (4$c$) | Cl II |
| $\mathbf{B_9}$ | = | $x_3\,\mathbf{a_1} + \frac{1}{4}\,\mathbf{a_2} + z_3\,\mathbf{a_3}$ | = | $x_3\,a\,\hat{\mathbf{x}} + \frac{1}{4}\,b\,\hat{\mathbf{y}} + z_3\,c\,\hat{\mathbf{z}}$ | (4$c$) | Pb |
| $\mathbf{B_{10}}$ | = | $\left(\frac{1}{2} - x_3\right)\mathbf{a_1} + \frac{3}{4}\,\mathbf{a_2} + \left(\frac{1}{2} + z_3\right)\mathbf{a_3}$ | = | $\left(\frac{1}{2} - x_3\right)a\,\hat{\mathbf{x}} + \frac{3}{4}\,b\,\hat{\mathbf{y}} + \left(\frac{1}{2} + z_3\right)c\,\hat{\mathbf{z}}$ | (4$c$) | Pb |
| $\mathbf{B_{11}}$ | = | $-x_3\,\mathbf{a_1} + \frac{3}{4}\,\mathbf{a_2} - z_3\,\mathbf{a_3}$ | = | $-x_3\,a\,\hat{\mathbf{x}} + \frac{3}{4}\,b\,\hat{\mathbf{y}} - z_3\,c\,\hat{\mathbf{z}}$ | (4$c$) | Pb |
| $\mathbf{B_{12}}$ | = | $\left(\frac{1}{2} + x_3\right)\mathbf{a_1} + \frac{1}{4}\,\mathbf{a_2} + \left(\frac{1}{2} - z_3\right)\mathbf{a_3}$ | = | $\left(\frac{1}{2} + x_3\right)a\,\hat{\mathbf{x}} + \frac{1}{4}\,b\,\hat{\mathbf{y}} + \left(\frac{1}{2} - z_3\right)c\,\hat{\mathbf{z}}$ | (4$c$) | Pb |

---

---

**Geometry files:**

- CIF: pp. 671
- POSCAR: pp. 671

# GeS (B16) Structure: AB_oP8_62_c_c

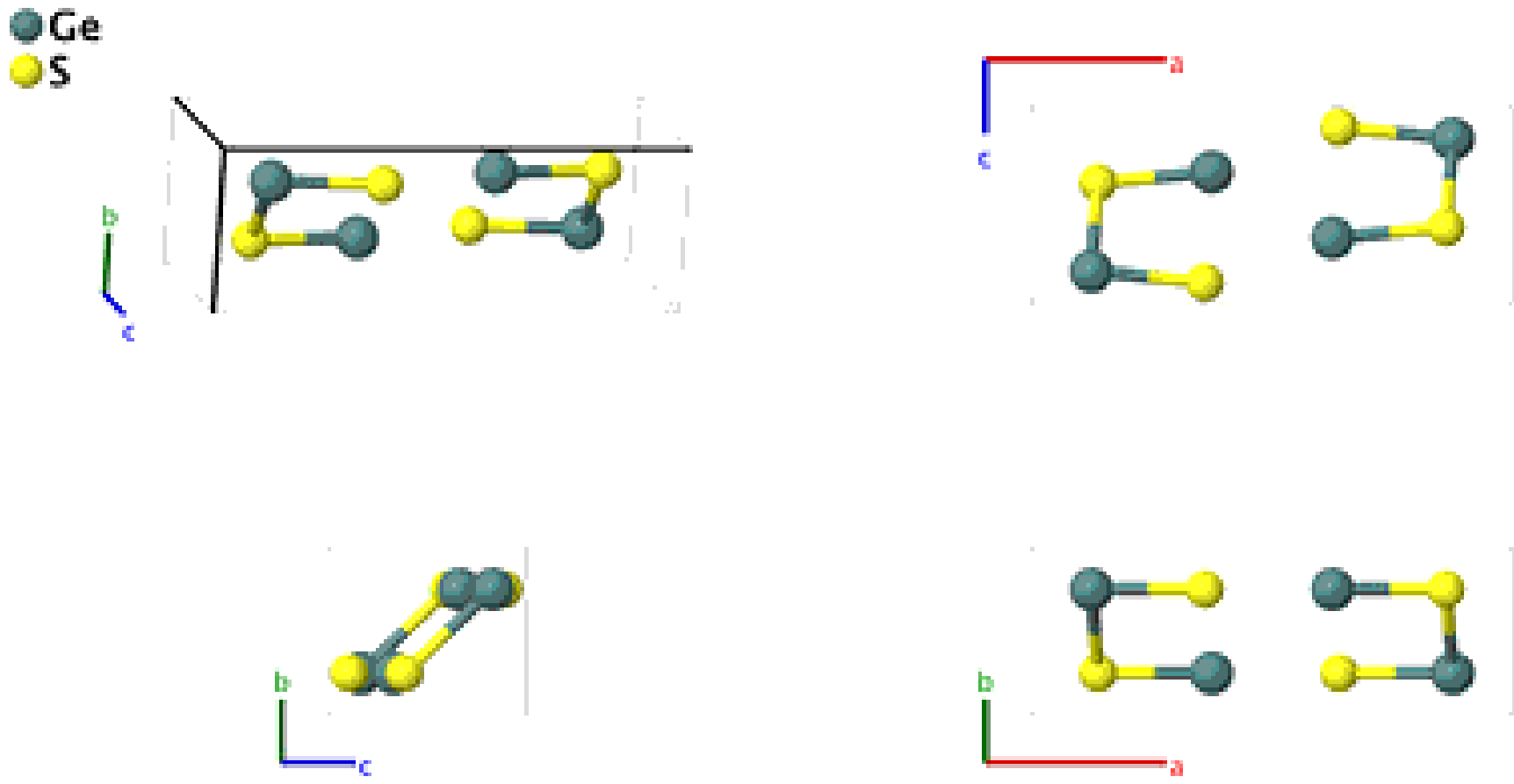

| | | |
|---|---|---|
| **Prototype** | : | GeS |
| **AFLOW prototype label** | : | AB_oP8_62_c_c |
| ***Strukturbericht* designation** | : | B16 |
| **Pearson symbol** | : | oP8 |
| **Space group number** | : | 62 |
| **Space group symbol** | : | Pnma |
| **AFLOW prototype command** | : | `aflow --proto=AB_oP8_62_c_c` `--params=`$a, b/a, c/a, x_1, z_1, x_2, z_2$ |

**Other compounds with this structure:**

- GeSe, GeS, GeTe, PbS, PbSe, PbTe, SnS, SnSe, SnTe

- Also see the closely related B29 (SnS) structure. (Parthé, 1993) prefers the B16 designation for both structures. Note that GeS (pp. 163), MnP (pp. 165), FeB (pp. 174), and SnS (pp. 176) have the same AFLOW prototype label. They are generated by the same symmetry operations with different sets of parameters (`--params`) specified in their corresponding CIF files.

**Simple Orthorhombic primitive vectors:**

$$\mathbf{a}_1 = a\,\hat{\mathbf{x}}$$
$$\mathbf{a}_2 = b\,\hat{\mathbf{y}}$$
$$\mathbf{a}_3 = c\,\hat{\mathbf{z}}$$

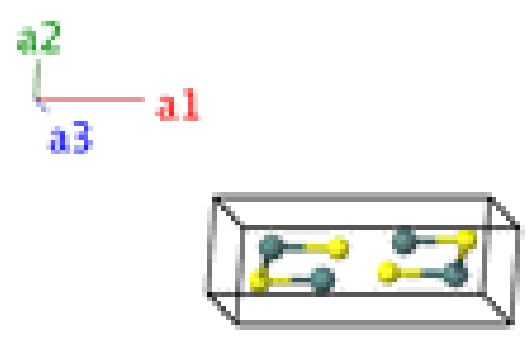

**Basis vectors:**

| | | Lattice Coordinates | | Cartesian Coordinates | Wyckoff Position | Atom Type |
|---|---|---|---|---|---|---|
| $\mathbf{B}_1$ | = | $x_1\,\mathbf{a}_1 + \frac{1}{4}\,\mathbf{a}_2 + z_1\,\mathbf{a}_3$ | = | $x_1\,a\,\hat{\mathbf{x}} + \frac{1}{4}\,b\,\hat{\mathbf{y}} + z_1\,c\,\hat{\mathbf{z}}$ | (4$c$) | Ge |
| $\mathbf{B}_2$ | = | $\left(\frac{1}{2} - x_1\right)\,\mathbf{a}_1 + \frac{3}{4}\,\mathbf{a}_2 + \left(\frac{1}{2} + z_1\right)\,\mathbf{a}_3$ | = | $\left(\frac{1}{2} - x_1\right)\,a\,\hat{\mathbf{x}} + \frac{3}{4}\,b\,\hat{\mathbf{y}} + \left(\frac{1}{2} + z_1\right)\,c\,\hat{\mathbf{z}}$ | (4$c$) | Ge |
| $\mathbf{B}_3$ | = | $-x_1\,\mathbf{a}_1 + \frac{3}{4}\,\mathbf{a}_2 - z_1\,\mathbf{a}_3$ | = | $-x_1\,a\,\hat{\mathbf{x}} + \frac{3}{4}\,b\,\hat{\mathbf{y}} - z_1\,c\,\hat{\mathbf{z}}$ | (4$c$) | Ge |
| $\mathbf{B_4}$ | = | $\left(\frac{1}{2}+x_1\right)\mathbf{a_1}+\frac{1}{4}\mathbf{a_2}+\left(\frac{1}{2}-z_1\right)\mathbf{a_3}$ | = | $\left(\frac{1}{2}+x_1\right)a\,\hat{\mathbf{x}}+\frac{1}{4}b\,\hat{\mathbf{y}}+\left(\frac{1}{2}-z_1\right)c\,\hat{\mathbf{z}}$ | (4$c$) | Ge |
| $\mathbf{B_5}$ | = | $x_2\,\mathbf{a_1}+\frac{1}{4}\mathbf{a_2}+z_2\,\mathbf{a_3}$ | = | $x_2\,a\,\hat{\mathbf{x}}+\frac{1}{4}b\,\hat{\mathbf{y}}+z_2\,c\,\hat{\mathbf{z}}$ | (4$c$) | S |
| $\mathbf{B_6}$ | = | $\left(\frac{1}{2}-x_2\right)\mathbf{a_1}+\frac{3}{4}\mathbf{a_2}+\left(\frac{1}{2}+z_2\right)\mathbf{a_3}$ | = | $\left(\frac{1}{2}-x_2\right)a\,\hat{\mathbf{x}}+\frac{3}{4}b\,\hat{\mathbf{y}}+\left(\frac{1}{2}+z_2\right)c\,\hat{\mathbf{z}}$ | (4$c$) | S |
| $\mathbf{B_7}$ | = | $-x_2\,\mathbf{a_1}+\frac{3}{4}\mathbf{a_2}-z_2\,\mathbf{a_3}$ | = | $-x_2\,a\,\hat{\mathbf{x}}+\frac{3}{4}b\,\hat{\mathbf{y}}-z_2\,c\,\hat{\mathbf{z}}$ | (4$c$) | S |
| $\mathbf{B_8}$ | = | $\left(\frac{1}{2}+x_2\right)\mathbf{a_1}+\frac{1}{4}\mathbf{a_2}+\left(\frac{1}{2}-z_2\right)\mathbf{a_3}$ | = | $\left(\frac{1}{2}+x_2\right)a\,\hat{\mathbf{x}}+\frac{1}{4}b\,\hat{\mathbf{y}}+\left(\frac{1}{2}-z_2\right)c\,\hat{\mathbf{z}}$ | (4$c$) | S |

---

---

**Geometry files:**
- CIF: pp. 671
- POSCAR: pp. 672

# MnP (B31) Structure: AB_oP8_62_c_c

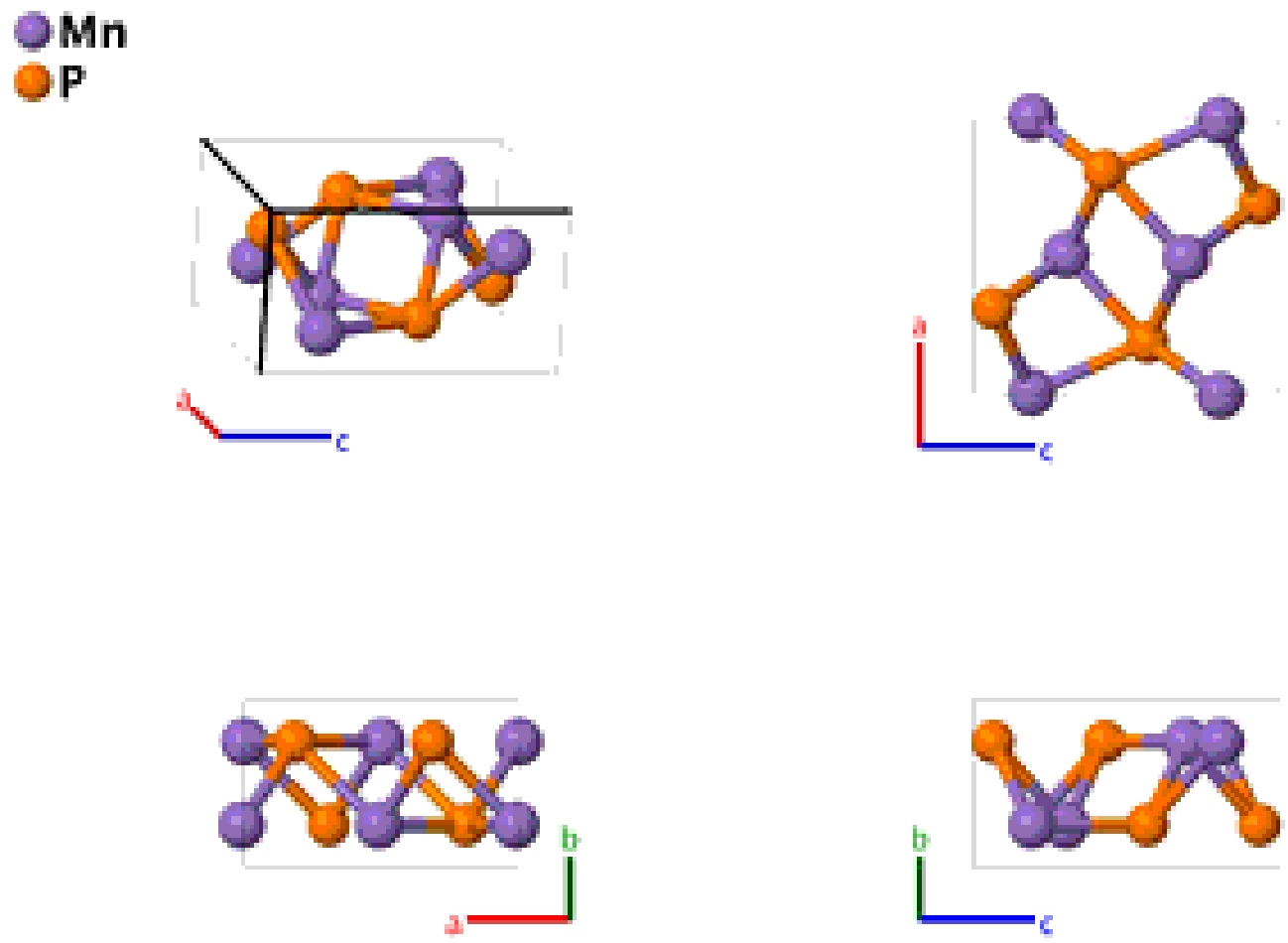

| | | |
|---|---|---|
| **Prototype** | : | MnP |
| **AFLOW prototype label** | : | AB_oP8_62_c_c |
| ***Strukturbericht* designation** | : | B31 |
| **Pearson symbol** | : | oP8 |
| **Space group number** | : | 62 |
| **Space group symbol** | : | Pnma |
| **AFLOW prototype command** | : | `aflow --proto=AB_oP8_62_c_c`<br>`--params=`$a, b/a, c/a, x_1, z_1, x_2, z_2$ |

**Other compounds with this structure:**

- AsCo, AsCr, AsFe, AsV, AsMo, CoP, CrP, FeP, FeS, GeNi, GeIr, GeRh, IrSi, RhSi, SeTi, many more

- (Hermann, 1937), pp. 7, assigns the prototype FeAs and the Strukturbericht designation B14 to this structure. This was superseded by the similar MnP structure in (Gottfried, 1937) pp. 17-18, where it is designated B31. Note that (Parthé, 1993) prefers the B14 designation. Note that GeS (pp. 163), MnP (pp. 165), FeB (pp. 174), and SnS (pp. 176) have the same AFLOW prototype label. They are generated by the same symmetry operations with different sets of parameters (`--params`) specified in their corresponding CIF files.

**Simple Orthorhombic primitive vectors:**

$$\mathbf{a}_1 = a\,\hat{\mathbf{x}}$$
$$\mathbf{a}_2 = b\,\hat{\mathbf{y}}$$
$$\mathbf{a}_3 = c\,\hat{\mathbf{z}}$$

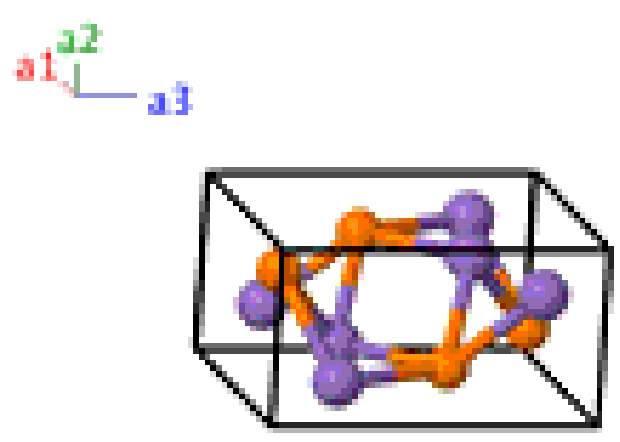

**Basis vectors:**

| | Lattice Coordinates | Cartesian Coordinates | Wyckoff Position | Atom Type |
|---|---|---|---|---|

| | | | | | | |
|---|---|---|---|---|---|---|
| $\mathbf{B_1}$ | = | $x_1\,\mathbf{a_1} + \frac{1}{4}\,\mathbf{a_2} + z_1\,\mathbf{a_3}$ | = | $x_1\,a\,\mathbf{\hat{x}} + \frac{1}{4}\,b\,\mathbf{\hat{y}} + z_1\,c\,\mathbf{\hat{z}}$ | $(4c)$ | Mn |
| $\mathbf{B_2}$ | = | $\left(\frac{1}{2} - x_1\right)\mathbf{a_1} + \frac{3}{4}\,\mathbf{a_2} + \left(\frac{1}{2} + z_1\right)\mathbf{a_3}$ | = | $\left(\frac{1}{2} - x_1\right)a\,\mathbf{\hat{x}} + \frac{3}{4}\,b\,\mathbf{\hat{y}} + \left(\frac{1}{2} + z_1\right)c\,\mathbf{\hat{z}}$ | $(4c)$ | Mn |
| $\mathbf{B_3}$ | = | $-x_1\,\mathbf{a_1} + \frac{3}{4}\,\mathbf{a_2} - z_1\,\mathbf{a_3}$ | = | $-x_1\,a\,\mathbf{\hat{x}} + \frac{3}{4}\,b\,\mathbf{\hat{y}} - z_1\,c\,\mathbf{\hat{z}}$ | $(4c)$ | Mn |
| $\mathbf{B_4}$ | = | $\left(\frac{1}{2} + x_1\right)\mathbf{a_1} + \frac{1}{4}\,\mathbf{a_2} + \left(\frac{1}{2} - z_1\right)\mathbf{a_3}$ | = | $\left(\frac{1}{2} + x_1\right)a\,\mathbf{\hat{x}} + \frac{1}{4}\,b\,\mathbf{\hat{y}} + \left(\frac{1}{2} - z_1\right)c\,\mathbf{\hat{z}}$ | $(4c)$ | Mn |
| $\mathbf{B_5}$ | = | $x_2\,\mathbf{a_1} + \frac{1}{4}\,\mathbf{a_2} + z_2\,\mathbf{a_3}$ | = | $x_2\,a\,\mathbf{\hat{x}} + \frac{1}{4}\,b\,\mathbf{\hat{y}} + z_2\,c\,\mathbf{\hat{z}}$ | $(4c)$ | P |
| $\mathbf{B_6}$ | = | $\left(\frac{1}{2} - x_2\right)\mathbf{a_1} + \frac{3}{4}\,\mathbf{a_2} + \left(\frac{1}{2} + z_2\right)\mathbf{a_3}$ | = | $\left(\frac{1}{2} - x_2\right)a\,\mathbf{\hat{x}} + \frac{3}{4}\,b\,\mathbf{\hat{y}} + \left(\frac{1}{2} + z_2\right)c\,\mathbf{\hat{z}}$ | $(4c)$ | P |
| $\mathbf{B_7}$ | = | $-x_2\,\mathbf{a_1} + \frac{3}{4}\,\mathbf{a_2} - z_2\,\mathbf{a_3}$ | = | $-x_2\,a\,\mathbf{\hat{x}} + \frac{3}{4}\,b\,\mathbf{\hat{y}} - z_2\,c\,\mathbf{\hat{z}}$ | $(4c)$ | P |
| $\mathbf{B_8}$ | = | $\left(\frac{1}{2} + x_2\right)\mathbf{a_1} + \frac{1}{4}\,\mathbf{a_2} + \left(\frac{1}{2} - z_2\right)\mathbf{a_3}$ | = | $\left(\frac{1}{2} + x_2\right)a\,\mathbf{\hat{x}} + \frac{1}{4}\,b\,\mathbf{\hat{y}} + \left(\frac{1}{2} - z_2\right)c\,\mathbf{\hat{z}}$ | $(4c)$ | P |

---

---

**Geometry files:**

- CIF: pp. 672
- POSCAR: pp. 672

# Cementite ($Fe_3C$, $D0_{11}$) Structure: AB3_oP16_62_c_cd

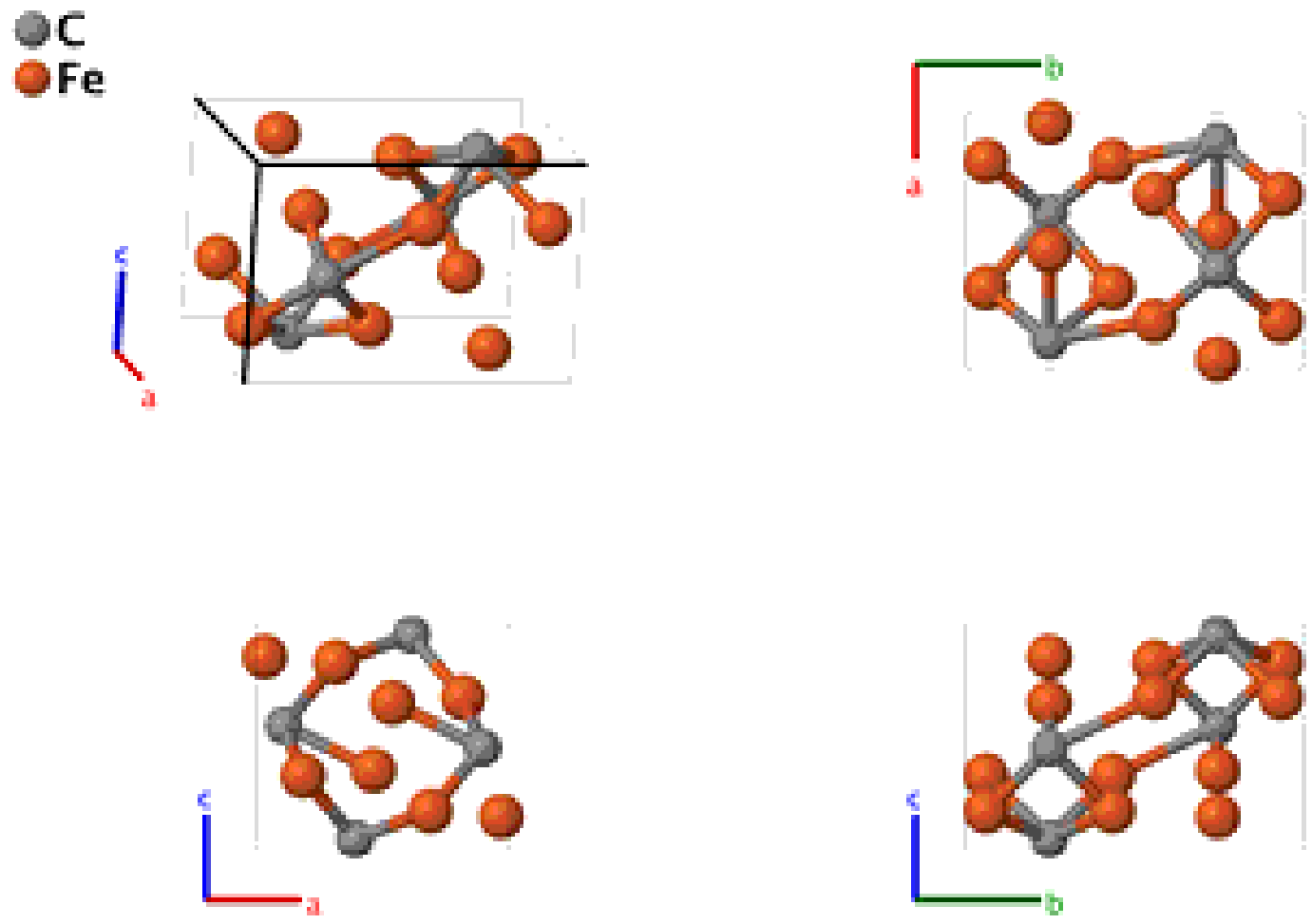

| | | |
|---|---|---|
| **Prototype** | : | $Fe_3C$ |
| **AFLOW prototype label** | : | AB3_oP16_62_c_cd |
| ***Strukturbericht* designation** | : | $D0_{11}$ |
| **Pearson symbol** | : | oP16 |
| **Space group number** | : | 62 |
| **Space group symbol** | : | Pnma |
| **AFLOW prototype command** | : | `aflow --proto=AB3_oP16_62_c_cd` <br> `--params=`$a, b/a, c/a, x_1, z_1, x_2, z_2, x_3, y_3, z_3$ |

**Simple Orthorhombic primitive vectors:**

$$\mathbf{a}_1 = a\,\hat{\mathbf{x}}$$
$$\mathbf{a}_2 = b\,\hat{\mathbf{y}}$$
$$\mathbf{a}_3 = c\,\hat{\mathbf{z}}$$

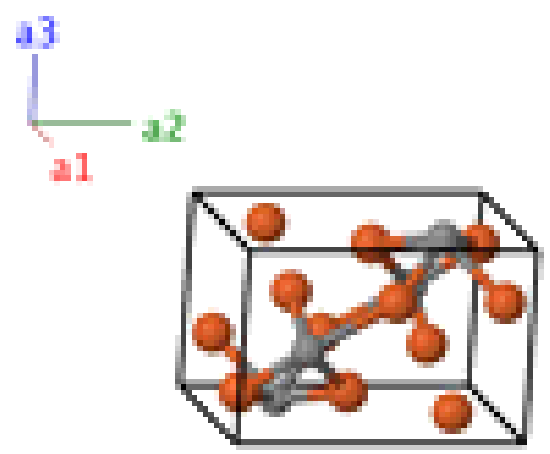

**Basis vectors:**

| | | Lattice Coordinates | | Cartesian Coordinates | Wyckoff Position | Atom Type |
|---|---|---|---|---|---|---|
| $\mathbf{B}_1$ | = | $x_1\,\mathbf{a}_1 + \frac{1}{4}\,\mathbf{a}_2 + z_1\,\mathbf{a}_3$ | = | $x_1\,a\,\hat{\mathbf{x}} + \frac{1}{4}\,b\,\hat{\mathbf{y}} + z_1\,c\,\hat{\mathbf{z}}$ | (4$c$) | C |
| $\mathbf{B}_2$ | = | $\left(\frac{1}{2} - x_1\right)\mathbf{a}_1 + \frac{3}{4}\,\mathbf{a}_2 + \left(\frac{1}{2} + z_1\right)\mathbf{a}_3$ | = | $\left(\frac{1}{2} - x_1\right)a\,\hat{\mathbf{x}} + \frac{3}{4}\,b\,\hat{\mathbf{y}} + \left(\frac{1}{2} + z_1\right)c\,\hat{\mathbf{z}}$ | (4$c$) | C |
| $\mathbf{B}_3$ | = | $-x_1\,\mathbf{a}_1 + \frac{3}{4}\,\mathbf{a}_2 - z_1\,\mathbf{a}_3$ | = | $-x_1\,a\,\hat{\mathbf{x}} + \frac{3}{4}\,b\,\hat{\mathbf{y}} - z_1\,c\,\hat{\mathbf{z}}$ | (4$c$) | C |
| $\mathbf{B}_4$ | = | $\left(\frac{1}{2} + x_1\right)\mathbf{a}_1 + \frac{1}{4}\,\mathbf{a}_2 + \left(\frac{1}{2} - z_1\right)\mathbf{a}_3$ | = | $\left(\frac{1}{2} + x_1\right)a\,\hat{\mathbf{x}} + \frac{1}{4}\,b\,\hat{\mathbf{y}} + \left(\frac{1}{2} - z_1\right)c\,\hat{\mathbf{z}}$ | (4$c$) | C |
| $\mathbf{B}_5$ | = | $x_2\,\mathbf{a}_1 + \frac{1}{4}\,\mathbf{a}_2 + z_2\,\mathbf{a}_3$ | = | $x_2\,a\,\hat{\mathbf{x}} + \frac{1}{4}\,b\,\hat{\mathbf{y}} + z_2\,c\,\hat{\mathbf{z}}$ | (4$c$) | Fe I |
| $\mathbf{B}_6$ | = | $\left(\frac{1}{2} - x_2\right)\mathbf{a}_1 + \frac{3}{4}\,\mathbf{a}_2 + \left(\frac{1}{2} + z_2\right)\mathbf{a}_3$ | = | $\left(\frac{1}{2} - x_2\right)a\,\hat{\mathbf{x}} + \frac{3}{4}\,b\,\hat{\mathbf{y}} + \left(\frac{1}{2} + z_2\right)c\,\hat{\mathbf{z}}$ | (4$c$) | Fe I |
| $\mathbf{B}_7$ | = | $-x_2\,\mathbf{a}_1 + \frac{3}{4}\,\mathbf{a}_2 - z_2\,\mathbf{a}_3$ | = | $-x_2\,a\,\hat{\mathbf{x}} + \frac{3}{4}\,b\,\hat{\mathbf{y}} - z_2\,c\,\hat{\mathbf{z}}$ | (4$c$) | Fe I |
| $\mathbf{B_{8}}$ | = | $\left(\frac{1}{2}+x_{2}\right)\,\mathbf{a_{1}}+\frac{1}{4}\,\mathbf{a_{2}}+\left(\frac{1}{2}-z_{2}\right)\,\mathbf{a_{3}}$ | = | $\left(\frac{1}{2}+x_{2}\right)\,a\,\mathbf{\hat{x}}+\frac{1}{4}\,b\,\mathbf{\hat{y}}+\left(\frac{1}{2}-z_{2}\right)\,c\,\mathbf{\hat{z}}$ | (4$c$) | Fe I |
| $\mathbf{B_{9}}$ | = | $x_{3}\,\mathbf{a_{1}}+y_{3}\,\mathbf{a_{2}}+z_{3}\,\mathbf{a_{3}}$ | = | $x_{3}\,a\,\mathbf{\hat{x}}+y_{3}\,b\,\mathbf{\hat{y}}+z_{3}\,c\,\mathbf{\hat{z}}$ | (8$d$) | Fe II |
| $\mathbf{B_{10}}$ | = | $\left(\frac{1}{2}-x_{3}\right)\,\mathbf{a_{1}}-y_{3}\,\mathbf{a_{2}}+\left(\frac{1}{2}+z_{3}\right)\,\mathbf{a_{3}}$ | = | $\left(\frac{1}{2}-x_{3}\right)\,a\,\mathbf{\hat{x}}-y_{3}\,b\,\mathbf{\hat{y}}+\left(\frac{1}{2}+z_{3}\right)\,c\,\mathbf{\hat{z}}$ | (8$d$) | Fe II |
| $\mathbf{B_{11}}$ | = | $-x_{3}\,\mathbf{a_{1}}+\left(\frac{1}{2}+y_{3}\right)\,\mathbf{a_{2}}-z_{3}\,\mathbf{a_{3}}$ | = | $-x_{3}\,a\,\mathbf{\hat{x}}+\left(\frac{1}{2}+y_{3}\right)\,b\,\mathbf{\hat{y}}-z_{3}\,c\,\mathbf{\hat{z}}$ | (8$d$) | Fe II |
| $\mathbf{B_{12}}$ | = | $\left(\frac{1}{2}+x_{3}\right)\,\mathbf{a_{1}}+\left(\frac{1}{2}-y_{3}\right)\,\mathbf{a_{2}}+\left(\frac{1}{2}-z_{3}\right)\,\mathbf{a_{3}}$ | = | $\left(\frac{1}{2}+x_{3}\right)\,a\,\mathbf{\hat{x}}+\left(\frac{1}{2}-y_{3}\right)\,b\,\mathbf{\hat{y}}+\left(\frac{1}{2}-z_{3}\right)\,c\,\mathbf{\hat{z}}$ | (8$d$) | Fe II |
| $\mathbf{B_{13}}$ | = | $-x_{3}\,\mathbf{a_{1}}-y_{3}\,\mathbf{a_{2}}-z_{3}\,\mathbf{a_{3}}$ | = | $-x_{3}\,a\,\mathbf{\hat{x}}-y_{3}\,b\,\mathbf{\hat{y}}-z_{3}\,c\,\mathbf{\hat{z}}$ | (8$d$) | Fe II |
| $\mathbf{B_{14}}$ | = | $\left(\frac{1}{2}+x_{3}\right)\,\mathbf{a_{1}}+y_{3}\,\mathbf{a_{2}}+\left(\frac{1}{2}-z_{3}\right)\,\mathbf{a_{3}}$ | = | $\left(\frac{1}{2}+x_{3}\right)\,a\,\mathbf{\hat{x}}+y_{3}\,b\,\mathbf{\hat{y}}+\left(\frac{1}{2}-z_{3}\right)\,c\,\mathbf{\hat{z}}$ | (8$d$) | Fe II |
| $\mathbf{B_{15}}$ | = | $x_{3}\,\mathbf{a_{1}}+\left(\frac{1}{2}-y_{3}\right)\,\mathbf{a_{2}}+z_{3}\,\mathbf{a_{3}}$ | = | $x_{3}\,a\,\mathbf{\hat{x}}+\left(\frac{1}{2}-y_{3}\right)\,b\,\mathbf{\hat{y}}+z_{3}\,c\,\mathbf{\hat{z}}$ | (8$d$) | Fe II |
| $\mathbf{B_{16}}$ | = | $\left(\frac{1}{2}-x_{3}\right)\,\mathbf{a_{1}}+\left(\frac{1}{2}+y_{3}\right)\,\mathbf{a_{2}}+\left(\frac{1}{2}+z_{3}\right)\,\mathbf{a_{3}}$ | = | $\left(\frac{1}{2}-x_{3}\right)\,a\,\mathbf{\hat{x}}+\left(\frac{1}{2}+y_{3}\right)\,b\,\mathbf{\hat{y}}+\left(\frac{1}{2}+z_{3}\right)\,c\,\mathbf{\hat{z}}$ | (8$d$) | Fe II |

**Geometry files:**

- CIF: pp. 672
- POSCAR: pp. 672

# $C_3Cr_7$ ($D10_1$) Structure: A3B7_oP40_62_cd_3c2d

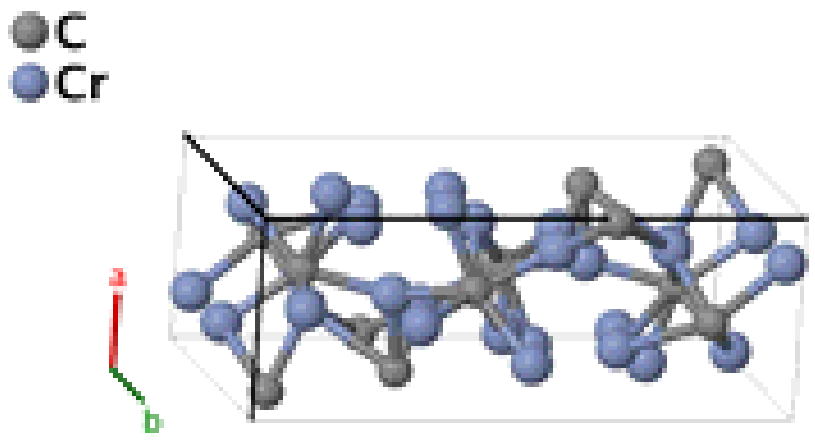

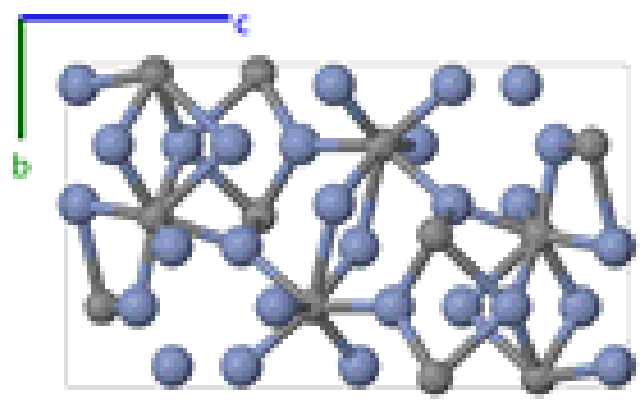

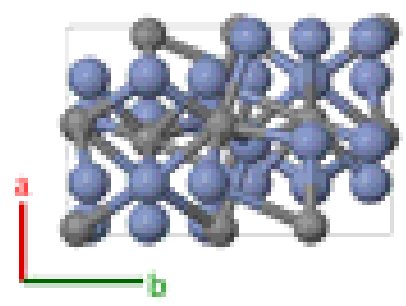

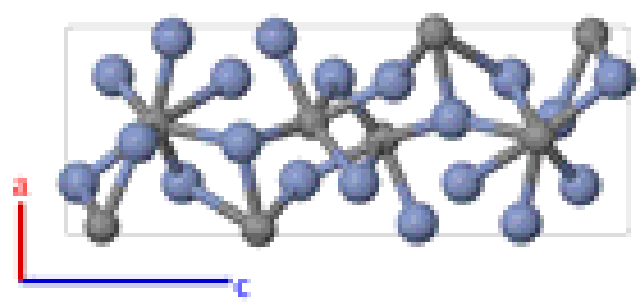

| | | |
|---|---|---|
| **Prototype** | : | $C_3Cr_7$ |
| **AFLOW prototype label** | : | A3B7_oP40_62_cd_3c2d |
| ***Strukturbericht* designation** | : | $D10_1$ |
| **Pearson symbol** | : | oP40 |
| **Space group number** | : | 62 |
| **Space group symbol** | : | Pnma |
| **AFLOW prototype command** | : | `aflow --proto=A3B7_oP40_62_cd_3c2d`<br>`--params=`$a, b/a, c/a, x_1, z_1, x_2, z_2, x_3, z_3, x_4, z_4, x_5, y_5, z_5, x_6, y_6, z_6, x_7, y_7, z_7$ |

**Simple Orthorhombic primitive vectors:**

$$\mathbf{a}_1 = a\,\mathbf{\hat{x}}$$
$$\mathbf{a}_2 = b\,\mathbf{\hat{y}}$$
$$\mathbf{a}_3 = c\,\mathbf{\hat{z}}$$

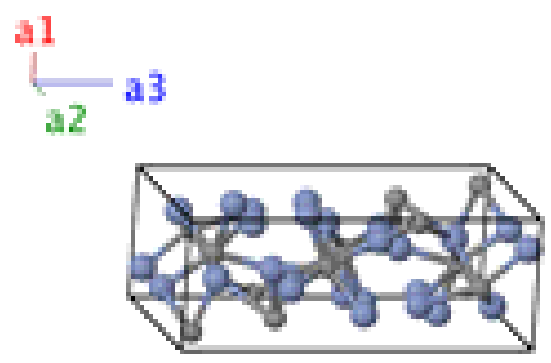

**Basis vectors:**

| | | Lattice Coordinates | | Cartesian Coordinates | Wyckoff Position | Atom Type |
|---|---|---|---|---|---|---|
| $\mathbf{B}_1$ | = | $x_1\,\mathbf{a}_1 + \frac{1}{4}\,\mathbf{a}_2 + z_1\,\mathbf{a}_3$ | = | $x_1\,a\,\mathbf{\hat{x}} + \frac{1}{4}\,b\,\mathbf{\hat{y}} + z_1\,c\,\mathbf{\hat{z}}$ | (4c) | C I |
| $\mathbf{B}_2$ | = | $\left(\frac{1}{2} - x_1\right)\mathbf{a}_1 + \frac{3}{4}\,\mathbf{a}_2 + \left(\frac{1}{2} + z_1\right)\mathbf{a}_3$ | = | $\left(\frac{1}{2} - x_1\right)a\,\mathbf{\hat{x}} + \frac{3}{4}\,b\,\mathbf{\hat{y}} + \left(\frac{1}{2} + z_1\right)c\,\mathbf{\hat{z}}$ | (4c) | C I |
| $\mathbf{B}_3$ | = | $-x_1\,\mathbf{a}_1 + \frac{3}{4}\,\mathbf{a}_2 - z_1\,\mathbf{a}_3$ | = | $-x_1\,a\,\mathbf{\hat{x}} + \frac{3}{4}\,b\,\mathbf{\hat{y}} - z_1\,c\,\mathbf{\hat{z}}$ | (4c) | C I |
| $\mathbf{B}_4$ | = | $\left(\frac{1}{2} + x_1\right)\mathbf{a}_1 + \frac{1}{4}\,\mathbf{a}_2 + \left(\frac{1}{2} - z_1\right)\mathbf{a}_3$ | = | $\left(\frac{1}{2} + x_1\right)a\,\mathbf{\hat{x}} + \frac{1}{4}\,b\,\mathbf{\hat{y}} + \left(\frac{1}{2} - z_1\right)c\,\mathbf{\hat{z}}$ | (4c) | C I |
| $\mathbf{B}_5$ | = | $x_2\,\mathbf{a}_1 + \frac{1}{4}\,\mathbf{a}_2 + z_2\,\mathbf{a}_3$ | = | $x_2\,a\,\mathbf{\hat{x}} + \frac{1}{4}\,b\,\mathbf{\hat{y}} + z_2\,c\,\mathbf{\hat{z}}$ | (4c) | Cr I |
| $\mathbf{B}_6$ | = | $\left(\frac{1}{2} - x_2\right)\mathbf{a}_1 + \frac{3}{4}\,\mathbf{a}_2 + \left(\frac{1}{2} + z_2\right)\mathbf{a}_3$ | = | $\left(\frac{1}{2} - x_2\right)a\,\mathbf{\hat{x}} + \frac{3}{4}\,b\,\mathbf{\hat{y}} + \left(\frac{1}{2} + z_2\right)c\,\mathbf{\hat{z}}$ | (4c) | Cr I |
| $\mathbf{B}_7$ | = | $-x_2\,\mathbf{a}_1 + \frac{3}{4}\,\mathbf{a}_2 - z_2\,\mathbf{a}_3$ | = | $-x_2\,a\,\mathbf{\hat{x}} + \frac{3}{4}\,b\,\mathbf{\hat{y}} - z_2\,c\,\mathbf{\hat{z}}$ | (4c) | Cr I |
| $\mathbf{B_{8}}$ | = | $\left(\frac{1}{2}+x_{2}\right)\mathbf{a_{1}}+\frac{1}{4}\mathbf{a_{2}}+\left(\frac{1}{2}-z_{2}\right)\mathbf{a_{3}}$ | = | $\left(\frac{1}{2}+x_{2}\right)a\,\mathbf{\hat{x}}+\frac{1}{4}b\,\mathbf{\hat{y}}+\left(\frac{1}{2}-z_{2}\right)c\,\mathbf{\hat{z}}$ | (4$c$) | Cr I |
| $\mathbf{B_{9}}$ | = | $x_{3}\,\mathbf{a_{1}}+\frac{1}{4}\mathbf{a_{2}}+z_{3}\,\mathbf{a_{3}}$ | = | $x_{3}\,a\,\mathbf{\hat{x}}+\frac{1}{4}b\,\mathbf{\hat{y}}+z_{3}\,c\,\mathbf{\hat{z}}$ | (4$c$) | Cr II |
| $\mathbf{B_{10}}$ | = | $\left(\frac{1}{2}-x_{3}\right)\mathbf{a_{1}}+\frac{3}{4}\mathbf{a_{2}}+\left(\frac{1}{2}+z_{3}\right)\mathbf{a_{3}}$ | = | $\left(\frac{1}{2}-x_{3}\right)a\,\mathbf{\hat{x}}+\frac{3}{4}b\,\mathbf{\hat{y}}+\left(\frac{1}{2}+z_{3}\right)c\,\mathbf{\hat{z}}$ | (4$c$) | Cr II |
| $\mathbf{B_{11}}$ | = | $-x_{3}\,\mathbf{a_{1}}+\frac{3}{4}\mathbf{a_{2}}-z_{3}\,\mathbf{a_{3}}$ | = | $-x_{3}\,a\,\mathbf{\hat{x}}+\frac{3}{4}b\,\mathbf{\hat{y}}-z_{3}\,c\,\mathbf{\hat{z}}$ | (4$c$) | Cr II |
| $\mathbf{B_{12}}$ | = | $\left(\frac{1}{2}+x_{3}\right)\mathbf{a_{1}}+\frac{1}{4}\mathbf{a_{2}}+\left(\frac{1}{2}-z_{3}\right)\mathbf{a_{3}}$ | = | $\left(\frac{1}{2}+x_{3}\right)a\,\mathbf{\hat{x}}+\frac{1}{4}b\,\mathbf{\hat{y}}+\left(\frac{1}{2}-z_{3}\right)c\,\mathbf{\hat{z}}$ | (4$c$) | Cr II |
| $\mathbf{B_{13}}$ | = | $x_{4}\,\mathbf{a_{1}}+\frac{1}{4}\mathbf{a_{2}}+z_{4}\,\mathbf{a_{3}}$ | = | $x_{4}\,a\,\mathbf{\hat{x}}+\frac{1}{4}b\,\mathbf{\hat{y}}+z_{4}\,c\,\mathbf{\hat{z}}$ | (4$c$) | Cr III |
| $\mathbf{B_{14}}$ | = | $\left(\frac{1}{2}-x_{4}\right)\mathbf{a_{1}}+\frac{3}{4}\mathbf{a_{2}}+\left(\frac{1}{2}+z_{4}\right)\mathbf{a_{3}}$ | = | $\left(\frac{1}{2}-x_{4}\right)a\,\mathbf{\hat{x}}+\frac{3}{4}b\,\mathbf{\hat{y}}+\left(\frac{1}{2}+z_{4}\right)c\,\mathbf{\hat{z}}$ | (4$c$) | Cr III |
| $\mathbf{B_{15}}$ | = | $-x_{4}\,\mathbf{a_{1}}+\frac{3}{4}\mathbf{a_{2}}-z_{4}\,\mathbf{a_{3}}$ | = | $-x_{4}\,a\,\mathbf{\hat{x}}+\frac{3}{4}b\,\mathbf{\hat{y}}-z_{4}\,c\,\mathbf{\hat{z}}$ | (4$c$) | Cr III |
| $\mathbf{B_{16}}$ | = | $\left(\frac{1}{2}+x_{4}\right)\mathbf{a_{1}}+\frac{1}{4}\mathbf{a_{2}}+\left(\frac{1}{2}-z_{4}\right)\mathbf{a_{3}}$ | = | $\left(\frac{1}{2}+x_{4}\right)a\,\mathbf{\hat{x}}+\frac{1}{4}b\,\mathbf{\hat{y}}+\left(\frac{1}{2}-z_{4}\right)c\,\mathbf{\hat{z}}$ | (4$c$) | Cr III |
| $\mathbf{B_{17}}$ | = | $x_{5}\,\mathbf{a_{1}}+y_{5}\,\mathbf{a_{2}}+z_{5}\,\mathbf{a_{3}}$ | = | $x_{5}\,a\,\mathbf{\hat{x}}+y_{5}\,b\,\mathbf{\hat{y}}+z_{5}\,c\,\mathbf{\hat{z}}$ | (8$d$) | C II |
| $\mathbf{B_{18}}$ | = | $\left(\frac{1}{2}-x_{5}\right)\mathbf{a_{1}}-y_{5}\,\mathbf{a_{2}}+\left(\frac{1}{2}+z_{5}\right)\mathbf{a_{3}}$ | = | $\left(\frac{1}{2}-x_{5}\right)a\,\mathbf{\hat{x}}-y_{5}\,b\,\mathbf{\hat{y}}+\left(\frac{1}{2}+z_{5}\right)c\,\mathbf{\hat{z}}$ | (8$d$) | C II |
| $\mathbf{B_{19}}$ | = | $-x_{5}\,\mathbf{a_{1}}+\left(\frac{1}{2}+y_{5}\right)\mathbf{a_{2}}-z_{5}\,\mathbf{a_{3}}$ | = | $-x_{5}\,a\,\mathbf{\hat{x}}+\left(\frac{1}{2}+y_{5}\right)b\,\mathbf{\hat{y}}-z_{5}\,c\,\mathbf{\hat{z}}$ | (8$d$) | C II |
| $\mathbf{B_{20}}$ | = | $\left(\frac{1}{2}+x_{5}\right)\mathbf{a_{1}}+\left(\frac{1}{2}-y_{5}\right)\mathbf{a_{2}}+\left(\frac{1}{2}-z_{5}\right)\mathbf{a_{3}}$ | = | $\left(\frac{1}{2}+x_{5}\right)a\,\mathbf{\hat{x}}+\left(\frac{1}{2}-y_{5}\right)b\,\mathbf{\hat{y}}+\left(\frac{1}{2}-z_{5}\right)c\,\mathbf{\hat{z}}$ | (8$d$) | C II |
| $\mathbf{B_{21}}$ | = | $-x_{5}\,\mathbf{a_{1}}-y_{5}\,\mathbf{a_{2}}-z_{5}\,\mathbf{a_{3}}$ | = | $-x_{5}\,a\,\mathbf{\hat{x}}-y_{5}\,b\,\mathbf{\hat{y}}-z_{5}\,c\,\mathbf{\hat{z}}$ | (8$d$) | C II |
| $\mathbf{B_{22}}$ | = | $\left(\frac{1}{2}+x_{5}\right)\mathbf{a_{1}}+y_{5}\,\mathbf{a_{2}}+\left(\frac{1}{2}-z_{5}\right)\mathbf{a_{3}}$ | = | $\left(\frac{1}{2}+x_{5}\right)a\,\mathbf{\hat{x}}+y_{5}\,b\,\mathbf{\hat{y}}+\left(\frac{1}{2}-z_{5}\right)c\,\mathbf{\hat{z}}$ | (8$d$) | C II |
| $\mathbf{B_{23}}$ | = | $x_{5}\,\mathbf{a_{1}}+\left(\frac{1}{2}-y_{5}\right)\mathbf{a_{2}}+z_{5}\,\mathbf{a_{3}}$ | = | $x_{5}\,a\,\mathbf{\hat{x}}+\left(\frac{1}{2}-y_{5}\right)b\,\mathbf{\hat{y}}+z_{5}\,c\,\mathbf{\hat{z}}$ | (8$d$) | C II |
| $\mathbf{B_{24}}$ | = | $\left(\frac{1}{2}-x_{5}\right)\mathbf{a_{1}}+\left(\frac{1}{2}+y_{5}\right)\mathbf{a_{2}}+\left(\frac{1}{2}+z_{5}\right)\mathbf{a_{3}}$ | = | $\left(\frac{1}{2}-x_{5}\right)a\,\mathbf{\hat{x}}+\left(\frac{1}{2}+y_{5}\right)b\,\mathbf{\hat{y}}+\left(\frac{1}{2}+z_{5}\right)c\,\mathbf{\hat{z}}$ | (8$d$) | C II |
| $\mathbf{B_{25}}$ | = | $x_{6}\,\mathbf{a_{1}}+y_{6}\,\mathbf{a_{2}}+z_{6}\,\mathbf{a_{3}}$ | = | $x_{6}\,a\,\mathbf{\hat{x}}+y_{6}\,b\,\mathbf{\hat{y}}+z_{6}\,c\,\mathbf{\hat{z}}$ | (8$d$) | Cr IV |
| $\mathbf{B_{26}}$ | = | $\left(\frac{1}{2}-x_{6}\right)\mathbf{a_{1}}-y_{6}\,\mathbf{a_{2}}+\left(\frac{1}{2}+z_{6}\right)\mathbf{a_{3}}$ | = | $\left(\frac{1}{2}-x_{6}\right)a\,\mathbf{\hat{x}}-y_{6}\,b\,\mathbf{\hat{y}}+\left(\frac{1}{2}+z_{6}\right)c\,\mathbf{\hat{z}}$ | (8$d$) | Cr IV |
| $\mathbf{B_{27}}$ | = | $-x_{6}\,\mathbf{a_{1}}+\left(\frac{1}{2}+y_{6}\right)\mathbf{a_{2}}-z_{6}\,\mathbf{a_{3}}$ | = | $-x_{6}\,a\,\mathbf{\hat{x}}+\left(\frac{1}{2}+y_{6}\right)b\,\mathbf{\hat{y}}-z_{6}\,c\,\mathbf{\hat{z}}$ | (8$d$) | Cr IV |
| $\mathbf{B_{28}}$ | = | $\left(\frac{1}{2}+x_{6}\right)\mathbf{a_{1}}+\left(\frac{1}{2}-y_{6}\right)\mathbf{a_{2}}+\left(\frac{1}{2}-z_{6}\right)\mathbf{a_{3}}$ | = | $\left(\frac{1}{2}+x_{6}\right)a\,\mathbf{\hat{x}}+\left(\frac{1}{2}-y_{6}\right)b\,\mathbf{\hat{y}}+\left(\frac{1}{2}-z_{6}\right)c\,\mathbf{\hat{z}}$ | (8$d$) | Cr IV |
| $\mathbf{B_{29}}$ | = | $-x_{6}\,\mathbf{a_{1}}-y_{6}\,\mathbf{a_{2}}-z_{6}\,\mathbf{a_{3}}$ | = | $-x_{6}\,a\,\mathbf{\hat{x}}-y_{6}\,b\,\mathbf{\hat{y}}-z_{6}\,c\,\mathbf{\hat{z}}$ | (8$d$) | Cr IV |
| $\mathbf{B_{30}}$ | = | $\left(\frac{1}{2}+x_{6}\right)\mathbf{a_{1}}+y_{6}\,\mathbf{a_{2}}+\left(\frac{1}{2}-z_{6}\right)\mathbf{a_{3}}$ | = | $\left(\frac{1}{2}+x_{6}\right)a\,\mathbf{\hat{x}}+y_{6}\,b\,\mathbf{\hat{y}}+\left(\frac{1}{2}-z_{6}\right)c\,\mathbf{\hat{z}}$ | (8$d$) | Cr IV |
| $\mathbf{B_{31}}$ | = | $x_{6}\,\mathbf{a_{1}}+\left(\frac{1}{2}-y_{6}\right)\mathbf{a_{2}}+z_{6}\,\mathbf{a_{3}}$ | = | $x_{6}\,a\,\mathbf{\hat{x}}+\left(\frac{1}{2}-y_{6}\right)b\,\mathbf{\hat{y}}+z_{6}\,c\,\mathbf{\hat{z}}$ | (8$d$) | Cr IV |
| $\mathbf{B_{32}}$ | = | $\left(\frac{1}{2}-x_{6}\right)\mathbf{a_{1}}+\left(\frac{1}{2}+y_{6}\right)\mathbf{a_{2}}+\left(\frac{1}{2}+z_{6}\right)\mathbf{a_{3}}$ | = | $\left(\frac{1}{2}-x_{6}\right)a\,\mathbf{\hat{x}}+\left(\frac{1}{2}+y_{6}\right)b\,\mathbf{\hat{y}}+\left(\frac{1}{2}+z_{6}\right)c\,\mathbf{\hat{z}}$ | (8$d$) | Cr IV |
| $\mathbf{B_{33}}$ | = | $x_{7}\,\mathbf{a_{1}}+y_{7}\,\mathbf{a_{2}}+z_{7}\,\mathbf{a_{3}}$ | = | $x_{7}\,a\,\mathbf{\hat{x}}+y_{7}\,b\,\mathbf{\hat{y}}+z_{7}\,c\,\mathbf{\hat{z}}$ | (8$d$) | Cr V |
| $\mathbf{B_{34}}$ | = | $\left(\frac{1}{2}-x_{7}\right)\mathbf{a_{1}}-y_{7}\,\mathbf{a_{2}}+\left(\frac{1}{2}+z_{7}\right)\mathbf{a_{3}}$ | = | $\left(\frac{1}{2}-x_{7}\right)a\,\mathbf{\hat{x}}-y_{7}\,b\,\mathbf{\hat{y}}+\left(\frac{1}{2}+z_{7}\right)c\,\mathbf{\hat{z}}$ | (8$d$) | Cr V |
| $\mathbf{B_{35}}$ | = | $-x_{7}\,\mathbf{a_{1}}+\left(\frac{1}{2}+y_{7}\right)\mathbf{a_{2}}-z_{7}\,\mathbf{a_{3}}$ | = | $-x_{7}\,a\,\mathbf{\hat{x}}+\left(\frac{1}{2}+y_{7}\right)b\,\mathbf{\hat{y}}-z_{7}\,c\,\mathbf{\hat{z}}$ | (8$d$) | Cr V |
| $\mathbf{B_{36}}$ | = | $\left(\frac{1}{2}+x_{7}\right)\mathbf{a_{1}}+\left(\frac{1}{2}-y_{7}\right)\mathbf{a_{2}}+\left(\frac{1}{2}-z_{7}\right)\mathbf{a_{3}}$ | = | $\left(\frac{1}{2}+x_{7}\right)a\,\mathbf{\hat{x}}+\left(\frac{1}{2}-y_{7}\right)b\,\mathbf{\hat{y}}+\left(\frac{1}{2}-z_{7}\right)c\,\mathbf{\hat{z}}$ | (8$d$) | Cr V |
| $\mathbf{B_{37}}$ | = | $-x_{7}\,\mathbf{a_{1}}-y_{7}\,\mathbf{a_{2}}-z_{7}\,\mathbf{a_{3}}$ | = | $-x_{7}\,a\,\mathbf{\hat{x}}-y_{7}\,b\,\mathbf{\hat{y}}-z_{7}\,c\,\mathbf{\hat{z}}$ | (8$d$) | Cr V |
| $\mathbf{B_{38}}$ | = | $\left(\frac{1}{2}+x_{7}\right)\mathbf{a_{1}}+y_{7}\,\mathbf{a_{2}}+\left(\frac{1}{2}-z_{7}\right)\mathbf{a_{3}}$ | = | $\left(\frac{1}{2}+x_{7}\right)a\,\mathbf{\hat{x}}+y_{7}\,b\,\mathbf{\hat{y}}+\left(\frac{1}{2}-z_{7}\right)c\,\mathbf{\hat{z}}$ | (8$d$) | Cr V |
| $\mathbf{B_{39}}$ | = | $x_{7}\,\mathbf{a_{1}}+\left(\frac{1}{2}-y_{7}\right)\mathbf{a_{2}}+z_{7}\,\mathbf{a_{3}}$ | = | $x_{7}\,a\,\mathbf{\hat{x}}+\left(\frac{1}{2}-y_{7}\right)b\,\mathbf{\hat{y}}+z_{7}\,c\,\mathbf{\hat{z}}$ | (8$d$) | Cr V |
| $\mathbf{B_{40}}$ | = | $\left(\frac{1}{2} - x_7\right)\mathbf{a_1} + \left(\frac{1}{2} + y_7\right)\mathbf{a_2} + \left(\frac{1}{2} + z_7\right)\mathbf{a_3}$ | = | $\left(\frac{1}{2} - x_7\right) a\,\mathbf{\hat{x}} + \left(\frac{1}{2} + y_7\right) b\,\mathbf{\hat{y}} + \left(\frac{1}{2} + z_7\right) c\,\mathbf{\hat{z}}$ | (8*d*) | Cr V |

---

**Found in:**

---

**Geometry files:**

- CIF: pp. 673
- POSCAR: pp. 673

# $\alpha$-Np ($A_c$) Structure: A_oP8_62_2c

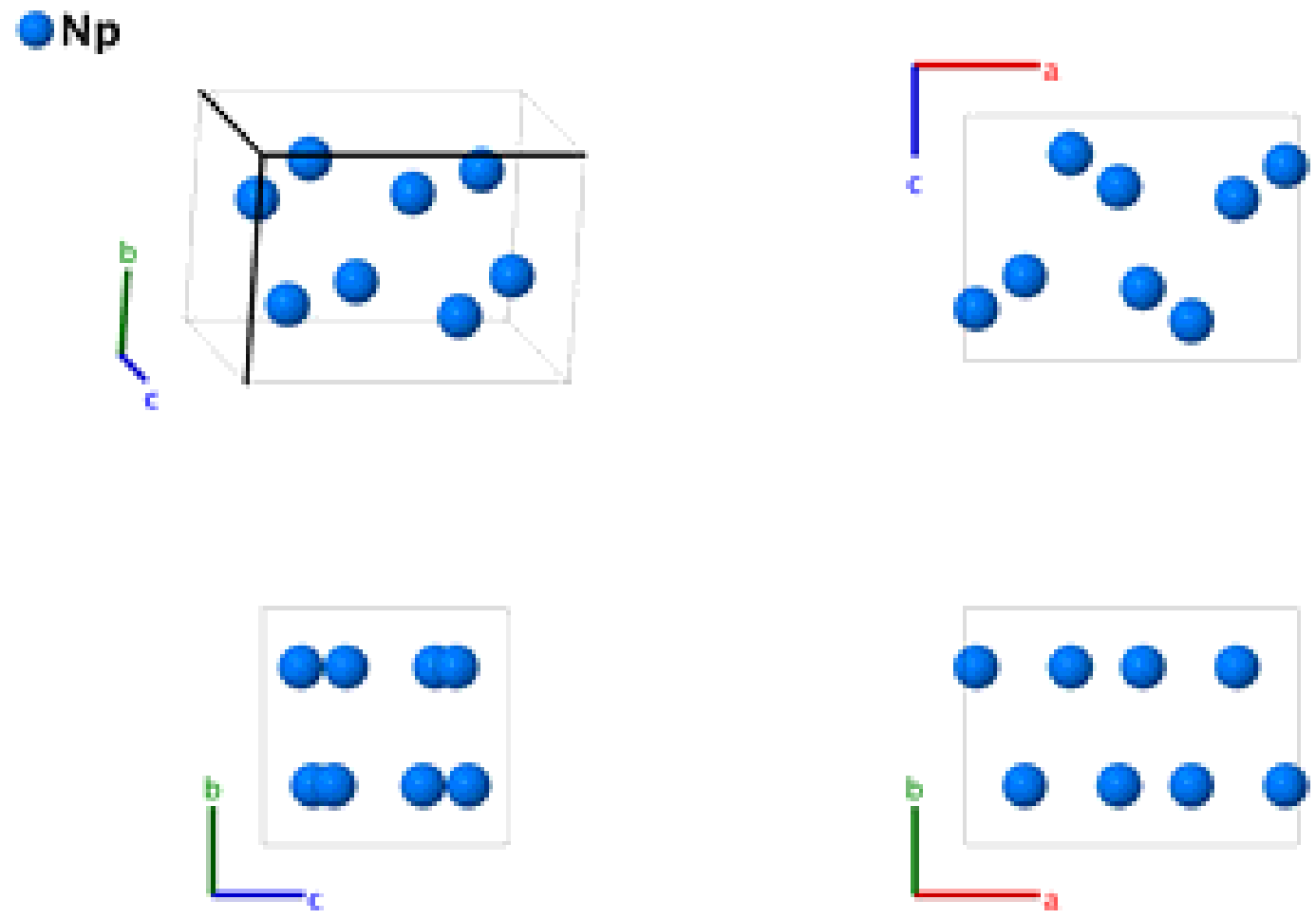

| | | |
|---|---|---|
| **Prototype** | : | $\alpha$-Np |
| **AFLOW prototype label** | : | A_oP8_62_2c |
| ***Strukturbericht* designation** | : | $A_c$ |
| **Pearson symbol** | : | oP8 |
| **Space group number** | : | 62 |
| **Space group symbol** | : | Pnma |
| **AFLOW prototype command** | : | `aflow --proto=A_oP8_62_2c` <br> `--params=`$a, b/a, c/a, x_1, z_1, x_2, z_2$ |

**Simple Orthorhombic primitive vectors:**

$$\mathbf{a}_1 = a\,\hat{\mathbf{x}}$$
$$\mathbf{a}_2 = b\,\hat{\mathbf{y}}$$
$$\mathbf{a}_3 = c\,\hat{\mathbf{z}}$$

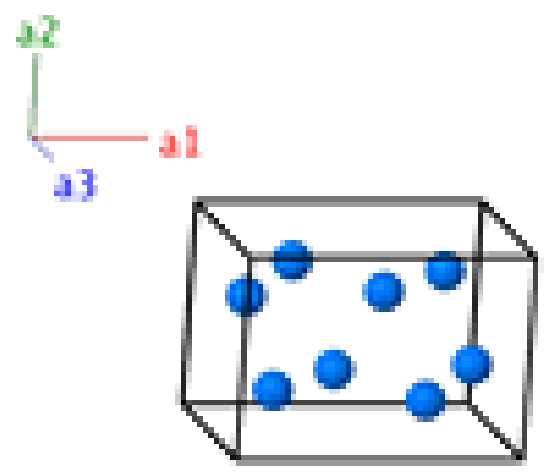

**Basis vectors:**

| | | Lattice Coordinates | | Cartesian Coordinates | Wyckoff Position | Atom Type |
|---|---|---|---|---|---|---|
| $\mathbf{B}_1$ | = | $x_1\,\mathbf{a}_1 + \frac{1}{4}\,\mathbf{a}_2 + z_1\,\mathbf{a}_3$ | = | $x_1\,a\,\hat{\mathbf{x}} + \frac{1}{4}\,b\,\hat{\mathbf{y}} + z_1\,c\,\hat{\mathbf{z}}$ | (4$c$) | Np I |
| $\mathbf{B}_2$ | = | $\left(\frac{1}{2} - x_1\right)\mathbf{a}_1 + \frac{3}{4}\,\mathbf{a}_2 + \left(\frac{1}{2} + z_1\right)\mathbf{a}_3$ | = | $\left(\frac{1}{2} - x_1\right)a\,\hat{\mathbf{x}} + \frac{3}{4}\,b\,\hat{\mathbf{y}} + \left(\frac{1}{2} + z_1\right)c\,\hat{\mathbf{z}}$ | (4$c$) | Np I |
| $\mathbf{B}_3$ | = | $-x_1\,\mathbf{a}_1 + \frac{3}{4}\,\mathbf{a}_2 - z_1\,\mathbf{a}_3$ | = | $-x_1\,a\,\hat{\mathbf{x}} + \frac{3}{4}\,b\,\hat{\mathbf{y}} - z_1\,c\,\hat{\mathbf{z}}$ | (4$c$) | Np I |
| $\mathbf{B}_4$ | = | $\left(\frac{1}{2} + x_1\right)\mathbf{a}_1 + \frac{1}{4}\,\mathbf{a}_2 + \left(\frac{1}{2} - z_1\right)\mathbf{a}_3$ | = | $\left(\frac{1}{2} + x_1\right)a\,\hat{\mathbf{x}} + \frac{1}{4}\,b\,\hat{\mathbf{y}} + \left(\frac{1}{2} - z_1\right)c\,\hat{\mathbf{z}}$ | (4$c$) | Np I |
| $\mathbf{B}_5$ | = | $x_2\,\mathbf{a}_1 + \frac{1}{4}\,\mathbf{a}_2 + z_2\,\mathbf{a}_3$ | = | $x_2\,a\,\hat{\mathbf{x}} + \frac{1}{4}\,b\,\hat{\mathbf{y}} + z_2\,c\,\hat{\mathbf{z}}$ | (4$c$) | Np II |
| $\mathbf{B}_6$ | = | $\left(\frac{1}{2} - x_2\right)\mathbf{a}_1 + \frac{3}{4}\,\mathbf{a}_2 + \left(\frac{1}{2} + z_2\right)\mathbf{a}_3$ | = | $\left(\frac{1}{2} - x_2\right)a\,\hat{\mathbf{x}} + \frac{3}{4}\,b\,\hat{\mathbf{y}} + \left(\frac{1}{2} + z_2\right)c\,\hat{\mathbf{z}}$ | (4$c$) | Np II |
| $\mathbf{B}_7$ | = | $-x_2\,\mathbf{a}_1 + \frac{3}{4}\,\mathbf{a}_2 - z_2\,\mathbf{a}_3$ | = | $-x_2\,a\,\hat{\mathbf{x}} + \frac{3}{4}\,b\,\hat{\mathbf{y}} - z_2\,c\,\hat{\mathbf{z}}$ | (4$c$) | Np II |
| $\mathbf{B}_8$ | = | $\left(\frac{1}{2} + x_2\right)\mathbf{a}_1 + \frac{1}{4}\,\mathbf{a}_2 + \left(\frac{1}{2} - z_2\right)\mathbf{a}_3$ | = | $\left(\frac{1}{2} + x_2\right)a\,\hat{\mathbf{x}} + \frac{1}{4}\,b\,\hat{\mathbf{y}} + \left(\frac{1}{2} - z_2\right)c\,\hat{\mathbf{z}}$ | (4$c$) | Np II |

**Found in:**

**Geometry files:**

- CIF: pp. 673
- POSCAR: pp. 674

# FeB (B27) Structure: AB_oP8_62_c_c

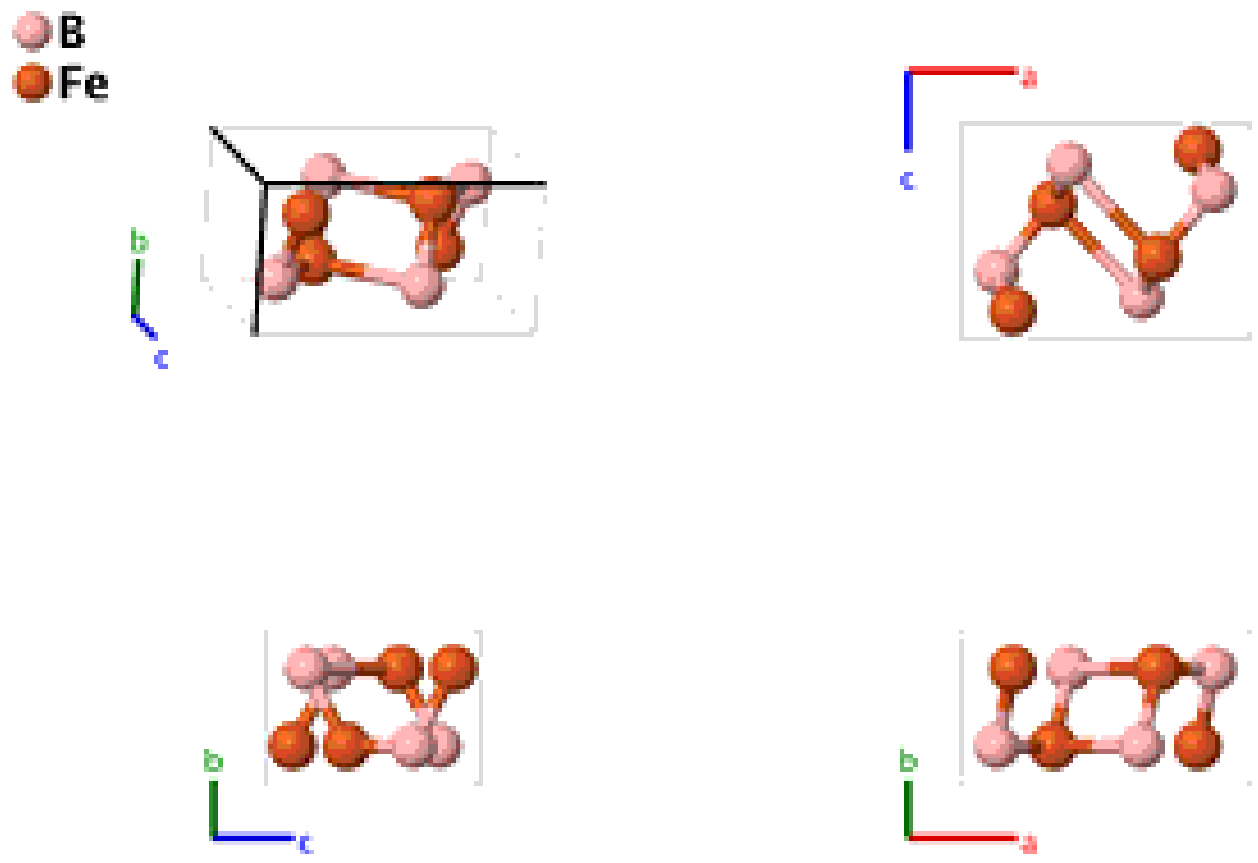

| | | |
|---|---|---|
| **Prototype** | : | FeB |
| **AFLOW prototype label** | : | AB_oP8_62_c_c |
| ***Strukturbericht* designation** | : | B27 |
| **Pearson symbol** | : | oP8 |
| **Space group number** | : | 62 |
| **Space group symbol** | : | Pnma |
| **AFLOW prototype command** | : | `aflow --proto=AB_oP8_62_c_c` `--params=`$a, b/a, c/a, x_1, z_1, x_2, z_2$ |

**Other compounds with this structure:**

- GeSe, GeS, GeTe, PbS, PbSe, PbTe, SnS, SnSe, SnTe

---

- (Hermann, 1937), pp. 7, assigns the Strukturbericht designation B15 to this structure. This was superseded by the current B27 structure in in (Gottfried, 1937) pp. 12. Here we follow (Parthé, 1993), who prefers the B27 designation. Note that GeS (pp. 163), MnP (pp. 165), FeB (pp. 174), and SnS (pp. 176) have the same AFLOW prototype label. They are generated by the same symmetry operations with different sets of parameters (`--params`) specified in their corresponding CIF files.

---

**Simple Orthorhombic primitive vectors:**

$$\mathbf{a}_1 = a\,\hat{\mathbf{x}}$$
$$\mathbf{a}_2 = b\,\hat{\mathbf{y}}$$
$$\mathbf{a}_3 = c\,\hat{\mathbf{z}}$$

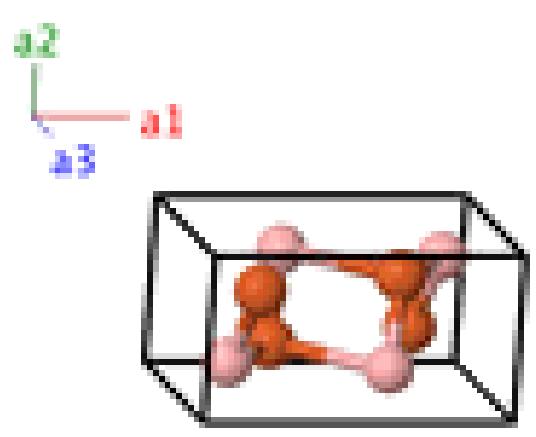

---

**Basis vectors:**

| | | Lattice Coordinates | | Cartesian Coordinates | Wyckoff Position | Atom Type |
|---|---|---|---|---|---|---|
| $\mathbf{B}_1$ | = | $x_1\,\mathbf{a}_1 + \frac{1}{4}\,\mathbf{a}_2 + z_1\,\mathbf{a}_3$ | = | $x_1\,a\,\hat{\mathbf{x}} + \frac{1}{4}\,b\,\hat{\mathbf{y}} + z_1\,c\,\hat{\mathbf{z}}$ | (4$c$) | B |
| $\mathbf{B_2}$ | = | $\left(\frac{1}{2} - x_1\right)\mathbf{a_1} + \frac{3}{4}\mathbf{a_2} + \left(\frac{1}{2} + z_1\right)\mathbf{a_3}$ | = | $\left(\frac{1}{2} - x_1\right) a\,\mathbf{\hat{x}} + \frac{3}{4}\,b\,\mathbf{\hat{y}} + \left(\frac{1}{2} + z_1\right) c\,\mathbf{\hat{z}}$ | (4c) | B |
| $\mathbf{B_3}$ | = | $-x_1\,\mathbf{a_1} + \frac{3}{4}\mathbf{a_2} - z_1\,\mathbf{a_3}$ | = | $-x_1\,a\,\mathbf{\hat{x}} + \frac{3}{4}\,b\,\mathbf{\hat{y}} - z_1\,c\,\mathbf{\hat{z}}$ | (4c) | B |
| $\mathbf{B_4}$ | = | $\left(\frac{1}{2} + x_1\right)\mathbf{a_1} + \frac{1}{4}\mathbf{a_2} + \left(\frac{1}{2} - z_1\right)\mathbf{a_3}$ | = | $\left(\frac{1}{2} + x_1\right) a\,\mathbf{\hat{x}} + \frac{1}{4}\,b\,\mathbf{\hat{y}} + \left(\frac{1}{2} - z_1\right) c\,\mathbf{\hat{z}}$ | (4c) | B |
| $\mathbf{B_5}$ | = | $x_2\,\mathbf{a_1} + \frac{1}{4}\mathbf{a_2} + z_2\,\mathbf{a_3}$ | = | $x_2\,a\,\mathbf{\hat{x}} + \frac{1}{4}\,b\,\mathbf{\hat{y}} + z_2\,c\,\mathbf{\hat{z}}$ | (4c) | Fe |
| $\mathbf{B_6}$ | = | $\left(\frac{1}{2} - x_2\right)\mathbf{a_1} + \frac{3}{4}\mathbf{a_2} + \left(\frac{1}{2} + z_2\right)\mathbf{a_3}$ | = | $\left(\frac{1}{2} - x_2\right) a\,\mathbf{\hat{x}} + \frac{3}{4}\,b\,\mathbf{\hat{y}} + \left(\frac{1}{2} + z_2\right) c\,\mathbf{\hat{z}}$ | (4c) | Fe |
| $\mathbf{B_7}$ | = | $-x_2\,\mathbf{a_1} + \frac{3}{4}\mathbf{a_2} - z_2\,\mathbf{a_3}$ | = | $-x_2\,a\,\mathbf{\hat{x}} + \frac{3}{4}\,b\,\mathbf{\hat{y}} - z_2\,c\,\mathbf{\hat{z}}$ | (4c) | Fe |
| $\mathbf{B_8}$ | = | $\left(\frac{1}{2} + x_2\right)\mathbf{a_1} + \frac{1}{4}\mathbf{a_2} + \left(\frac{1}{2} - z_2\right)\mathbf{a_3}$ | = | $\left(\frac{1}{2} + x_2\right) a\,\mathbf{\hat{x}} + \frac{1}{4}\,b\,\mathbf{\hat{y}} + \left(\frac{1}{2} - z_2\right) c\,\mathbf{\hat{z}}$ | (4c) | Fe |

---

**Found in:**

---

**Geometry files:**

- CIF: pp. 674
- POSCAR: pp. 674

# SnS (B29) Structure: AB_oP8_62_c_c

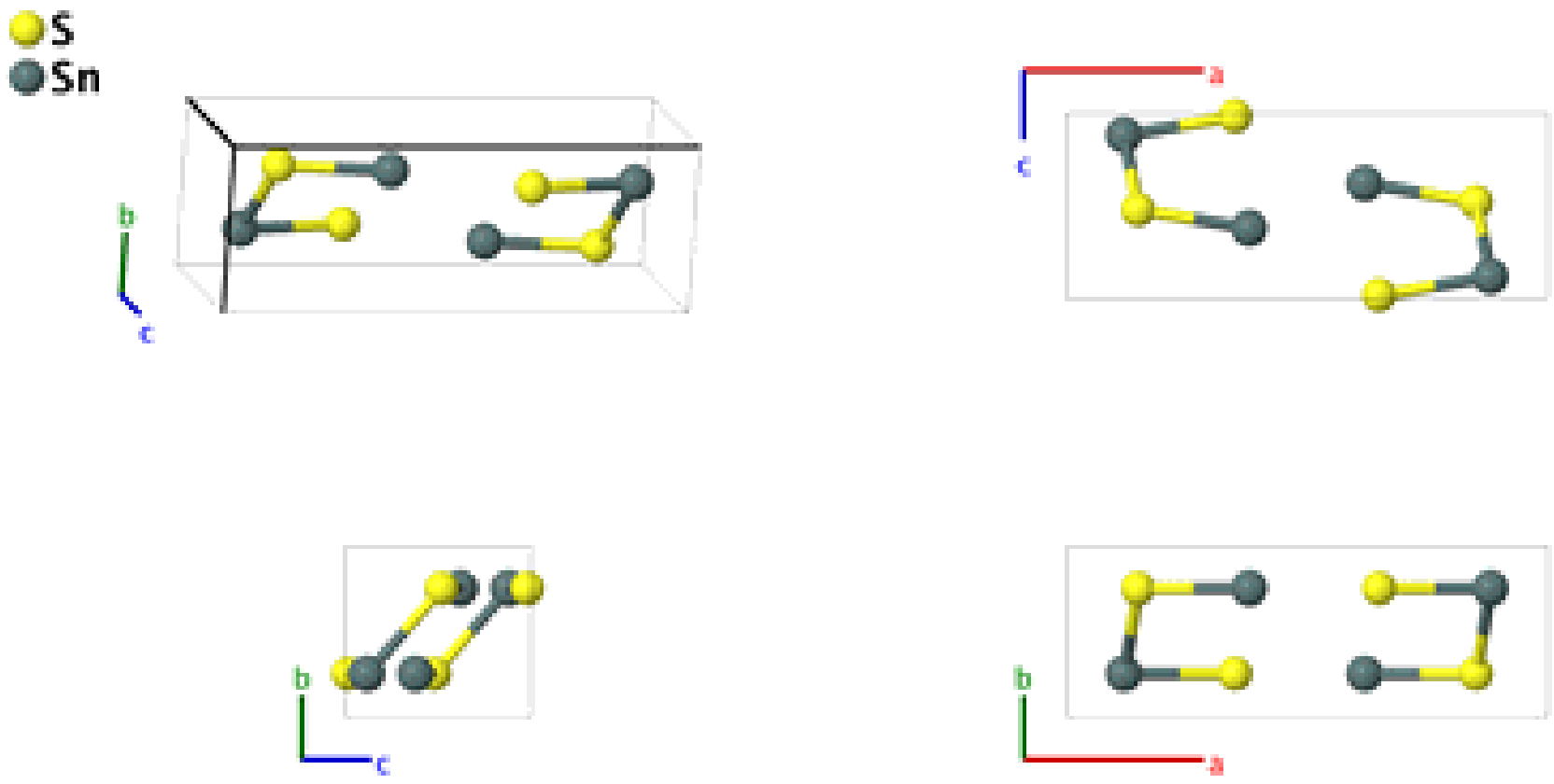

| | | |
|---|---|---|
| **Prototype** | : | SnS |
| **AFLOW prototype label** | : | AB_oP8_62_c_c |
| ***Strukturbericht* designation** | : | B29 |
| **Pearson symbol** | : | oP8 |
| **Space group number** | : | 62 |
| **Space group symbol** | : | Pnma |
| **AFLOW prototype command** | : | `aflow --proto=AB_oP8_62_c_c` <br> `--params=`$a, b/a, c/a, x_1, z_1, x_2, z_2$ |

- This structure is closely related to the B16 (GeS) structure. (Parthé, 1993) prefers the B16 designation for this structure. Note that GeS (pp. 163), MnP (pp. 165), FeB (pp. 174), and SnS (pp. 176) have the same AFLOW prototype label. They are generated by the same symmetry operations with different sets of parameters (`--params`) specified in their corresponding CIF files.

**Simple Orthorhombic primitive vectors:**

$$\mathbf{a}_1 = a\,\hat{\mathbf{x}}$$
$$\mathbf{a}_2 = b\,\hat{\mathbf{y}}$$
$$\mathbf{a}_3 = c\,\hat{\mathbf{z}}$$

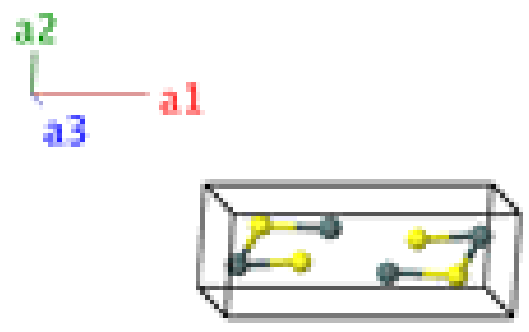

**Basis vectors:**

| | | Lattice Coordinates | | Cartesian Coordinates | Wyckoff Position | Atom Type |
|---|---|---|---|---|---|---|
| $\mathbf{B}_1$ | = | $x_1\,\mathbf{a}_1 + \frac{1}{4}\,\mathbf{a}_2 + z_1\,\mathbf{a}_3$ | = | $x_1\,a\,\hat{\mathbf{x}} + \frac{1}{4}\,b\,\hat{\mathbf{y}} + z_1\,c\,\hat{\mathbf{z}}$ | (4c) | S |
| $\mathbf{B}_2$ | = | $\left(\frac{1}{2} - x_1\right)\mathbf{a}_1 + \frac{3}{4}\,\mathbf{a}_2 + \left(\frac{1}{2} + z_1\right)\mathbf{a}_3$ | = | $\left(\frac{1}{2} - x_1\right)a\,\hat{\mathbf{x}} + \frac{3}{4}\,b\,\hat{\mathbf{y}} + \left(\frac{1}{2} + z_1\right)c\,\hat{\mathbf{z}}$ | (4c) | S |
| $\mathbf{B}_3$ | = | $-x_1\,\mathbf{a}_1 + \frac{3}{4}\,\mathbf{a}_2 - z_1\,\mathbf{a}_3$ | = | $-x_1\,a\,\hat{\mathbf{x}} + \frac{3}{4}\,b\,\hat{\mathbf{y}} - z_1\,c\,\hat{\mathbf{z}}$ | (4c) | S |
| $\mathbf{B}_4$ | = | $\left(\frac{1}{2} + x_1\right)\mathbf{a}_1 + \frac{1}{4}\,\mathbf{a}_2 + \left(\frac{1}{2} - z_1\right)\mathbf{a}_3$ | = | $\left(\frac{1}{2} + x_1\right)a\,\hat{\mathbf{x}} + \frac{1}{4}\,b\,\hat{\mathbf{y}} + \left(\frac{1}{2} - z_1\right)c\,\hat{\mathbf{z}}$ | (4c) | S |
| $\mathbf{B}_5$ | = | $x_2\,\mathbf{a}_1 + \frac{1}{4}\,\mathbf{a}_2 + z_2\,\mathbf{a}_3$ | = | $x_2\,a\,\hat{\mathbf{x}} + \frac{1}{4}\,b\,\hat{\mathbf{y}} + z_2\,c\,\hat{\mathbf{z}}$ | (4c) | Sn |
| $\mathbf{B}_6$ | = | $\left(\frac{1}{2} - x_2\right)\mathbf{a}_1 + \frac{3}{4}\,\mathbf{a}_2 + \left(\frac{1}{2} + z_2\right)\mathbf{a}_3$ | = | $\left(\frac{1}{2} - x_2\right)a\,\hat{\mathbf{x}} + \frac{3}{4}\,b\,\hat{\mathbf{y}} + \left(\frac{1}{2} + z_2\right)c\,\hat{\mathbf{z}}$ | (4c) | Sn |
| $\mathbf{B_7}$ | = | $-x_2\,\mathbf{a_1} + \frac{3}{4}\,\mathbf{a_2} - z_2\,\mathbf{a_3}$ | = | $-x_2\,a\,\hat{\mathbf{x}} + \frac{3}{4}\,b\,\hat{\mathbf{y}} - z_2\,c\,\hat{\mathbf{z}}$ | (4$c$) | Sn |
| $\mathbf{B_8}$ | = | $\left(\frac{1}{2} + x_2\right)\mathbf{a_1} + \frac{1}{4}\,\mathbf{a_2} + \left(\frac{1}{2} - z_2\right)\mathbf{a_3}$ | = | $\left(\frac{1}{2} + x_2\right)a\,\hat{\mathbf{x}} + \frac{1}{4}\,b\,\hat{\mathbf{y}} + \left(\frac{1}{2} - z_2\right)c\,\hat{\mathbf{z}}$ | (4$c$) | Sn |

---

---

**Geometry files:**

- CIF: pp. 674
- POSCAR: pp. 675

# $SrCuO_2$ Structure: AB2C_oC16_63_c_2c_c

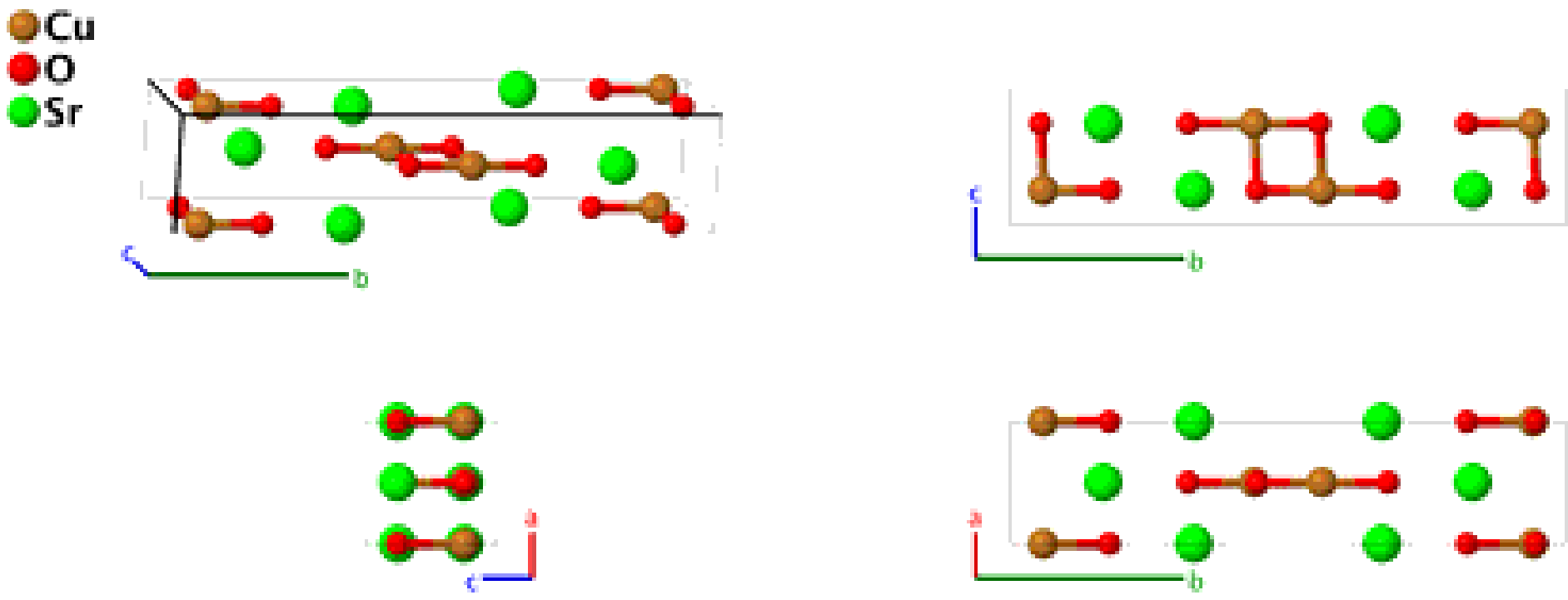

| | | |
|---|---|---|
| **Prototype** | : | $SrCuO_2$ |
| **AFLOW prototype label** | : | AB2C_oC16_63_c_2c_c |
| ***Strukturbericht* designation** | : | None |
| **Pearson symbol** | : | oC16 |
| **Space group number** | : | 63 |
| **Space group symbol** | : | Cmcm |
| **AFLOW prototype command** | : | `aflow --proto=AB2C_oC16_63_c_2c_c` <br> `--params=`$a, b/a, c/a, y_1, y_2, y_3, y_4$ |

**Base-centered Orthorhombic primitive vectors:**

$$\mathbf{a}_1 = \frac{1}{2}\,a\,\hat{\mathbf{x}} - \frac{1}{2}\,b\,\hat{\mathbf{y}}$$
$$\mathbf{a}_2 = \frac{1}{2}\,a\,\hat{\mathbf{x}} + \frac{1}{2}\,b\,\hat{\mathbf{y}}$$
$$\mathbf{a}_3 = c\,\hat{\mathbf{z}}$$

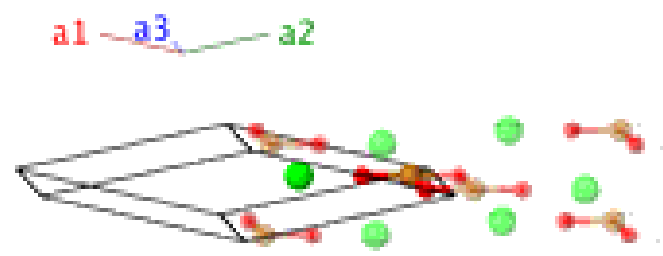

**Basis vectors:**

| | | Lattice Coordinates | | Cartesian Coordinates | Wyckoff Position | Atom Type |
|---|---|---|---|---|---|---|
| $\mathbf{B}_1$ | = | $-y_1\,\mathbf{a}_1 + y_1\,\mathbf{a}_2 + \frac{1}{4}\,\mathbf{a}_3$ | = | $y_1\,b\,\hat{\mathbf{y}} + \frac{1}{4}\,c\hat{\mathbf{z}}$ | (4$c$) | Cu |
| $\mathbf{B}_2$ | = | $y_1\,\mathbf{a}_1 - y_1\,\mathbf{a}_2 + \frac{3}{4}\,\mathbf{a}_3$ | = | $-y_1\,b\,\hat{\mathbf{y}} + \frac{3}{4}\,c\hat{\mathbf{z}}$ | (4$c$) | Cu |
| $\mathbf{B}_3$ | = | $-y_2\,\mathbf{a}_1 + y_2\,\mathbf{a}_2 + \frac{1}{4}\,\mathbf{a}_3$ | = | $y_2\,b\,\hat{\mathbf{y}} + \frac{1}{4}\,c\hat{\mathbf{z}}$ | (4$c$) | O I |
| $\mathbf{B}_4$ | = | $y_2\,\mathbf{a}_1 - y_2\,\mathbf{a}_2 + \frac{3}{4}\,\mathbf{a}_3$ | = | $-y_2\,b\,\hat{\mathbf{y}} + \frac{3}{4}\,c\hat{\mathbf{z}}$ | (4$c$) | O I |
| $\mathbf{B}_5$ | = | $-y_3\,\mathbf{a}_1 + y_3\,\mathbf{a}_2 + \frac{1}{4}\,\mathbf{a}_3$ | = | $y_3\,b\,\hat{\mathbf{y}} + \frac{1}{4}\,c\hat{\mathbf{z}}$ | (4$c$) | O II |
| $\mathbf{B}_6$ | = | $y_3\,\mathbf{a}_1 - y_3\,\mathbf{a}_2 + \frac{3}{4}\,\mathbf{a}_3$ | = | $-y_3\,b\,\hat{\mathbf{y}} + \frac{3}{4}\,c\hat{\mathbf{z}}$ | (4$c$) | O II |
| $\mathbf{B}_7$ | = | $-y_4\,\mathbf{a}_1 + y_4\,\mathbf{a}_2 + \frac{1}{4}\,\mathbf{a}_3$ | = | $y_4\,b\,\hat{\mathbf{y}} + \frac{1}{4}\,c\hat{\mathbf{z}}$ | (4$c$) | Sr |
| $\mathbf{B}_8$ | = | $y_4\,\mathbf{a}_1 - y_4\,\mathbf{a}_2 + \frac{3}{4}\,\mathbf{a}_3$ | = | $-y_4\,b\,\hat{\mathbf{y}} + \frac{3}{4}\,c\hat{\mathbf{z}}$ | (4$c$) | Sr |

**Geometry files:**

- CIF: pp. 675
- POSCAR: pp. 675

# $ZrSi_2$ (C49) Structure: A2B_oC12_63_2c_c

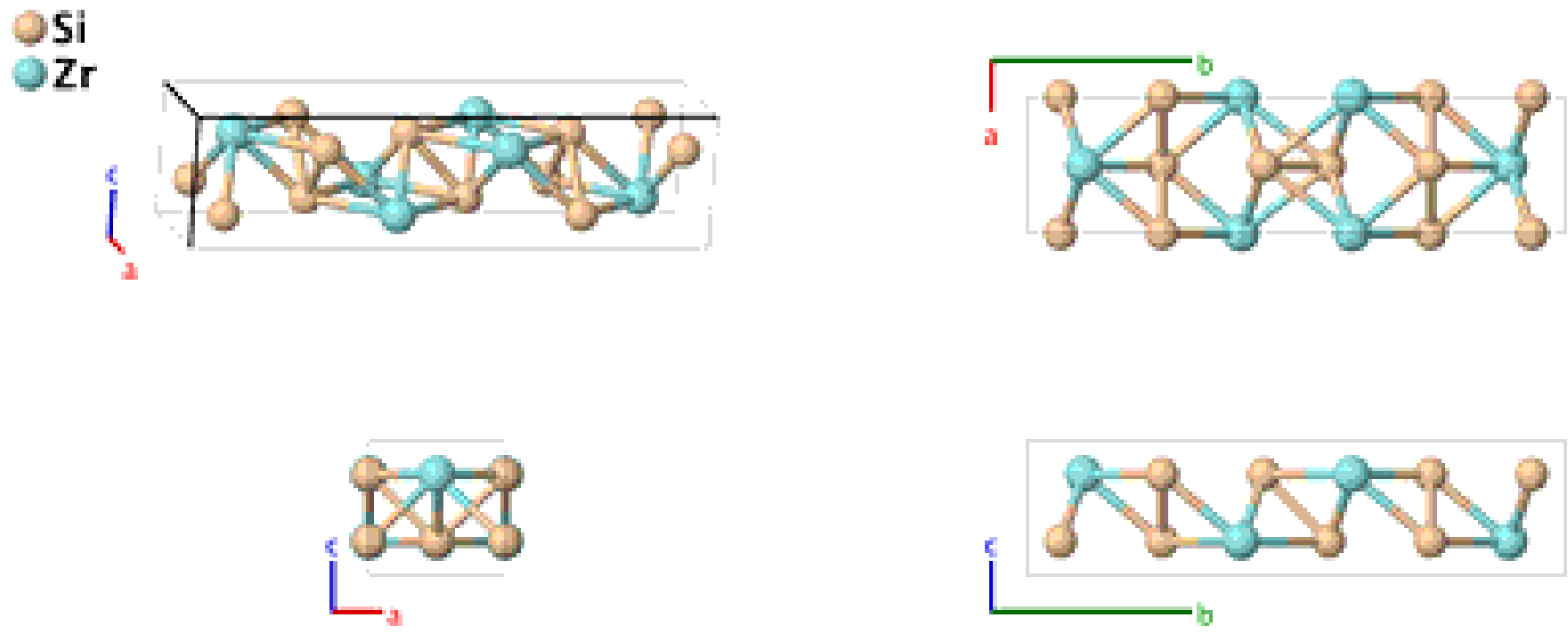

| | | |
|---|---|---|
| **Prototype** | : | $ZrSi_2$ |
| **AFLOW prototype label** | : | A2B_oC12_63_2c_c |
| ***Strukturbericht* designation** | : | C49 |
| **Pearson symbol** | : | oC12 |
| **Space group number** | : | 63 |
| **Space group symbol** | : | Cmcm |
| **AFLOW prototype command** | : | `aflow --proto=A2B_oC12_63_2c_c`<br>`--params=`$a, b/a, c/a, y_1, y_2, y_3$ |

**Base-centered Orthorhombic primitive vectors:**

$$\begin{aligned} \mathbf{a}_1 &= \tfrac{1}{2}\,a\,\hat{\mathbf{x}} - \tfrac{1}{2}\,b\,\hat{\mathbf{y}} \\ \mathbf{a}_2 &= \tfrac{1}{2}\,a\,\hat{\mathbf{x}} + \tfrac{1}{2}\,b\,\hat{\mathbf{y}} \\ \mathbf{a}_3 &= c\,\hat{\mathbf{z}} \end{aligned}$$

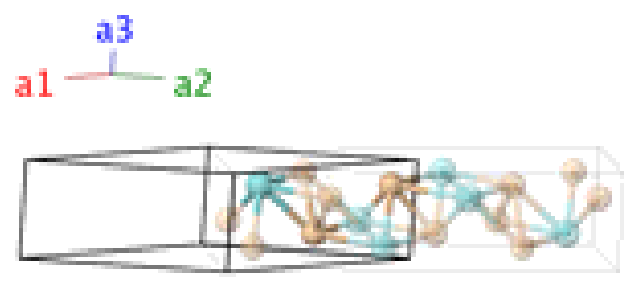

**Basis vectors:**

| | | Lattice Coordinates | | Cartesian Coordinates | Wyckoff Position | Atom Type |
|---|---|---|---|---|---|---|
| $\mathbf{B}_1$ | = | $-y_1\,\mathbf{a}_1 + y_1\,\mathbf{a}_2 + \frac{1}{4}\,\mathbf{a}_3$ | = | $y_1\,b\,\hat{\mathbf{y}} + \frac{1}{4}\,c\hat{\mathbf{z}}$ | (4$c$) | Si I |
| $\mathbf{B}_2$ | = | $y_1\,\mathbf{a}_1 - y_1\,\mathbf{a}_2 + \frac{3}{4}\,\mathbf{a}_3$ | = | $-y_1\,b\,\hat{\mathbf{y}} + \frac{3}{4}\,c\hat{\mathbf{z}}$ | (4$c$) | Si I |
| $\mathbf{B}_3$ | = | $-y_2\,\mathbf{a}_1 + y_2\,\mathbf{a}_2 + \frac{1}{4}\,\mathbf{a}_3$ | = | $y_2\,b\,\hat{\mathbf{y}} + \frac{1}{4}\,c\hat{\mathbf{z}}$ | (4$c$) | Si II |
| $\mathbf{B}_4$ | = | $y_2\,\mathbf{a}_1 - y_2\,\mathbf{a}_2 + \frac{3}{4}\,\mathbf{a}_3$ | = | $-y_2\,b\,\hat{\mathbf{y}} + \frac{3}{4}\,c\hat{\mathbf{z}}$ | (4$c$) | Si II |
| $\mathbf{B}_5$ | = | $-y_3\,\mathbf{a}_1 + y_3\,\mathbf{a}_2 + \frac{1}{4}\,\mathbf{a}_3$ | = | $y_3\,b\,\hat{\mathbf{y}} + \frac{1}{4}\,c\hat{\mathbf{z}}$ | (4$c$) | Zr |
| $\mathbf{B}_6$ | = | $y_3\,\mathbf{a}_1 - y_3\,\mathbf{a}_2 + \frac{3}{4}\,\mathbf{a}_3$ | = | $-y_3\,b\,\hat{\mathbf{y}} + \frac{3}{4}\,c\hat{\mathbf{z}}$ | (4$c$) | Zr |

**Geometry files:**

- CIF: pp. 675
- POSCAR: pp. 676

# CrB (B33) Structure: AB_oC8_63_c_c

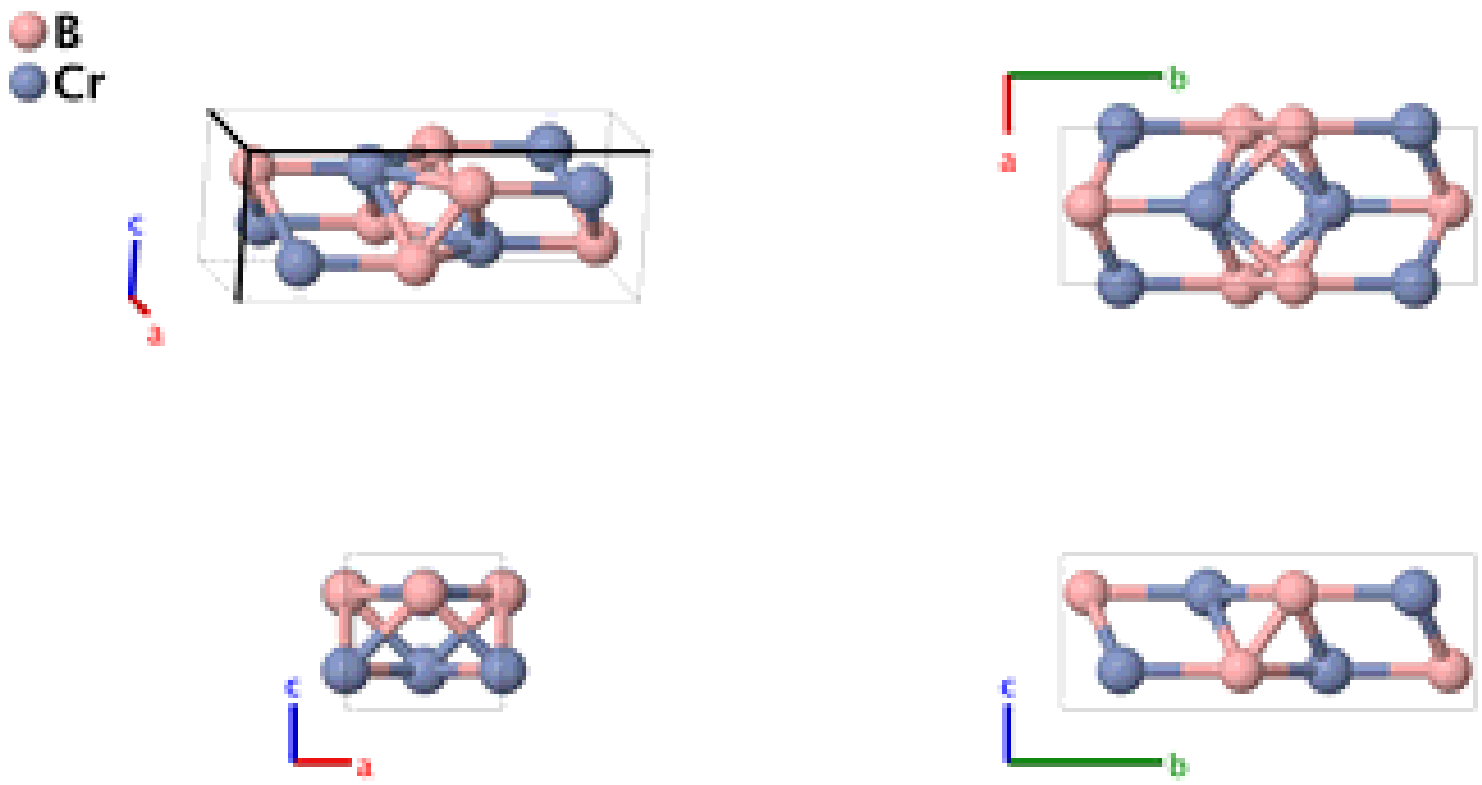

| | | |
|---|---|---|
| **Prototype** | : | CrB |
| **AFLOW prototype label** | : | AB_oC8_63_c_c |
| ***Strukturbericht* designation** | : | B33 |
| **Pearson symbol** | : | oC8 |
| **Space group number** | : | 63 |
| **Space group symbol** | : | Cmcm |
| **AFLOW prototype command** | : | `aflow --proto=AB_oC8_63_c_c` <br> `--params=`$a, b/a, c/a, y_1, y_2$ |

- Note that removing either the Cr or B atoms transforms this into the $\alpha$-U (A20) structure.

**Base-centered Orthorhombic primitive vectors:**

$$\mathbf{a}_1 = \frac{1}{2}\,a\,\hat{\mathbf{x}} - \frac{1}{2}\,b\,\hat{\mathbf{y}}$$
$$\mathbf{a}_2 = \frac{1}{2}\,a\,\hat{\mathbf{x}} + \frac{1}{2}\,b\,\hat{\mathbf{y}}$$
$$\mathbf{a}_3 = c\,\hat{\mathbf{z}}$$

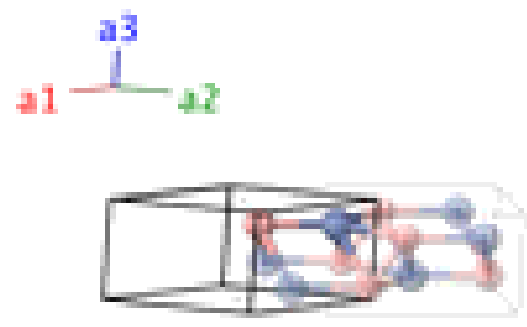

**Basis vectors:**

| | | Lattice Coordinates | | Cartesian Coordinates | Wyckoff Position | Atom Type |
|---|---|---|---|---|---|---|
| $\mathbf{B_1}$ | = | $-y_1\,\mathbf{a_1} + y_1\,\mathbf{a_2} + \frac{1}{4}\,\mathbf{a_3}$ | = | $y_1\,b\,\hat{\mathbf{y}} + \frac{1}{4}\,c\hat{\mathbf{z}}$ | (4$c$) | B |
| $\mathbf{B_2}$ | = | $y_1\,\mathbf{a_1} - y_1\,\mathbf{a_2} + \frac{3}{4}\,\mathbf{a_3}$ | = | $-y_1\,b\,\hat{\mathbf{y}} + \frac{3}{4}\,c\hat{\mathbf{z}}$ | (4$c$) | B |
| $\mathbf{B_3}$ | = | $-y_2\,\mathbf{a_1} + y_2\,\mathbf{a_2} + \frac{1}{4}\,\mathbf{a_3}$ | = | $y_2\,b\,\hat{\mathbf{y}} + \frac{1}{4}\,c\hat{\mathbf{z}}$ | (4$c$) | Cr |
| $\mathbf{B_4}$ | = | $y_2\,\mathbf{a_1} - y_2\,\mathbf{a_2} + \frac{3}{4}\,\mathbf{a_3}$ | = | $-y_2\,b\,\hat{\mathbf{y}} + \frac{3}{4}\,c\hat{\mathbf{z}}$ | (4$c$) | Cr |

**Found in:**

---

**Geometry files:**

- CIF: pp. 676
- POSCAR: pp. 676

# $\alpha$-U (A20) Structure: A_oC4_63_c

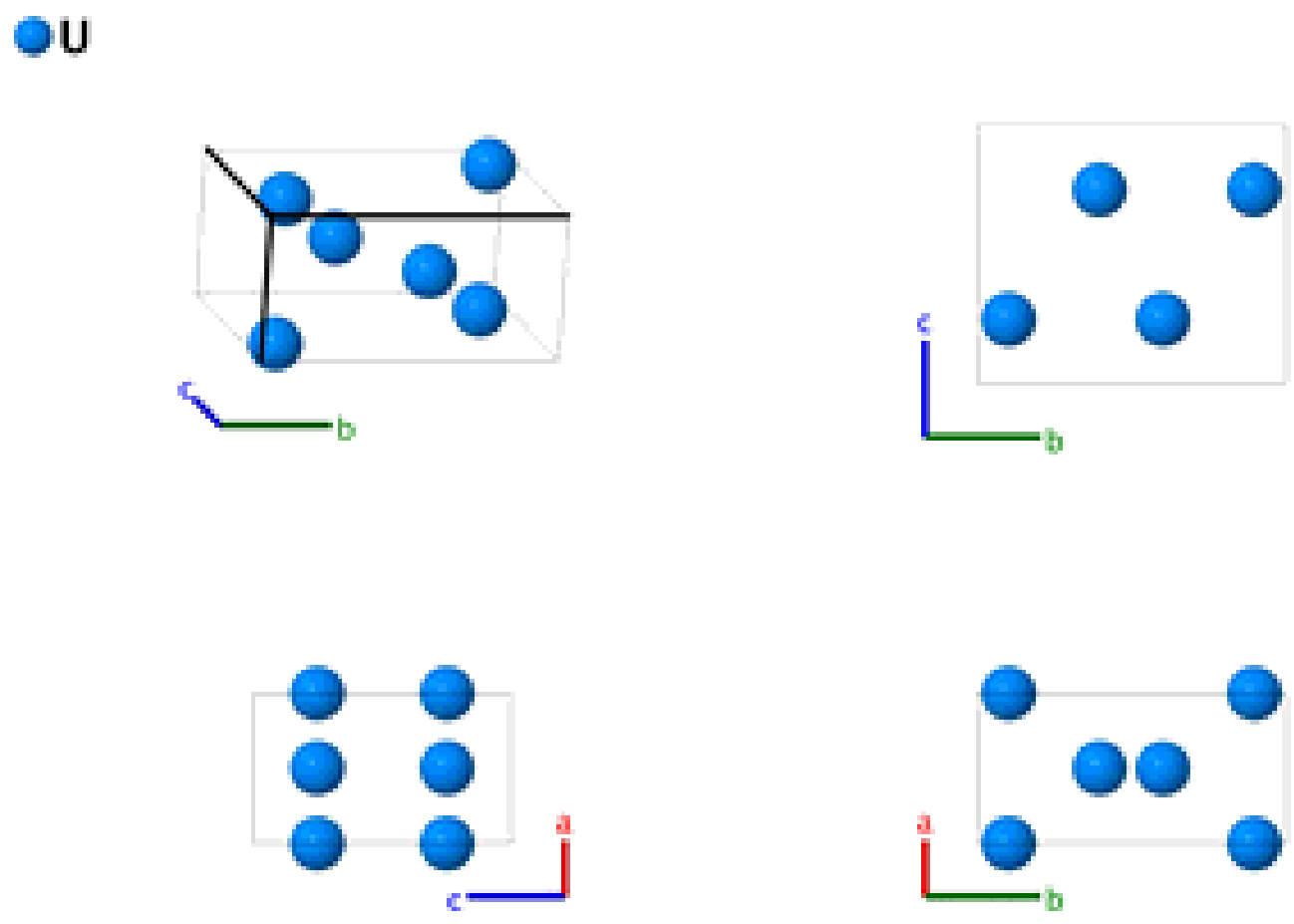

| | | |
|---|---|---|
| **Prototype** | : | $\alpha$-U |
| **AFLOW prototype label** | : | A_oC4_63_c |
| ***Strukturbericht* designation** | : | A20 |
| **Pearson symbol** | : | oC4 |
| **Space group number** | : | 63 |
| **Space group symbol** | : | Cmcm |
| **AFLOW prototype command** | : | `aflow --proto=A_oC4_63_c --params=`$a, b/a, c/a, y_1$ |

**Other elements and compounds with this structure:**

- Tb, Dy, Ge (metastable), AgCd (random alloy), $\gamma$-Ti

- Using data for the $\alpha$-U structure at 4.2K. (Vohra, 2001) showed that at pressures above 116 GPa titanium transforms from the hexagonal omega (C32) phase to this phase. This structure was studied by (Wentzcovitch, 1987) as a possible pathway for the pressure-induced transformation of magnesium from the hcp (A3) to the bcc (A2) phase. Much like the trigonal omega phase (C6), we can generate several high-symmetry structures from this phase by the appropriate choice of parameters.

| Lattice parameter | hcp | bcc | fcc | simple cubic |
|---|---|---|---|---|
| a | $a_{hcp}$ | $a_{bcc}$ | $a_{fcc}$ | $a_{sc}$ |
| b | $\sqrt{3}a_{hcp}$ | $\sqrt{2}a_{bcc}$ | $a_{fcc}$ | $a_{sc}$ |
| c | $c_{hcp}$ | $\sqrt{2}a_{bcc}$ | $a_{fcc}$ | $2a_{sc}$ |
| y | $\frac{1}{6}$ | $\frac{1}{4}$ | $\frac{1}{4}$ | 0 |

| Lattice parameter | hcp | bcc | fcc | simple cubic |
|---|---|---|---|---|
| *Strukturbericht* | A3 | A2 | A1 | $A_h$ |
| Pearson symbol | hP2 | cI2 | cF4 | cP1 |
| Space group | $P6_3/mmc$ | $Im\bar{3}m$ | $Fm\bar{3}m$ | $Pm\bar{3}m$ |

**Base-centered Orthorhombic primitive vectors:**

$$
\begin{aligned}
\mathbf{a}_1 &= \tfrac{1}{2}\,a\,\hat{\mathbf{x}} - \tfrac{1}{2}\,b\,\hat{\mathbf{y}} \\
\mathbf{a}_2 &= \tfrac{1}{2}\,a\,\hat{\mathbf{x}} + \tfrac{1}{2}\,b\,\hat{\mathbf{y}} \\
\mathbf{a}_3 &= c\,\hat{\mathbf{z}}
\end{aligned}
$$

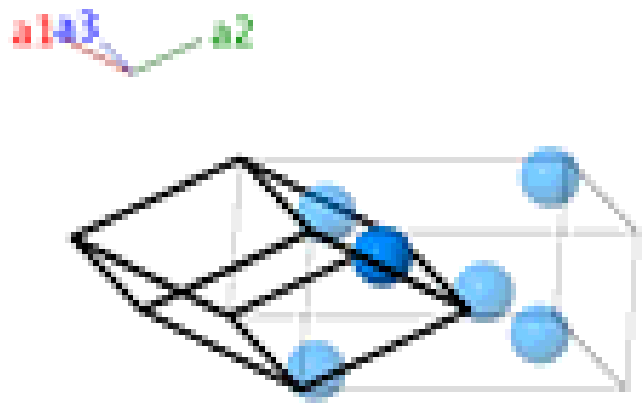

**Basis vectors:**

| | | Lattice Coordinates | | Cartesian Coordinates | Wyckoff Position | Atom Type |
|---|---|---|---|---|---|---|
| $\mathbf{B_1}$ | = | $-y_1\,\mathbf{a_1} + y_1\,\mathbf{a_2} + \frac{1}{4}\,\mathbf{a_3}$ | = | $y_1\,b\,\hat{\mathbf{y}} + \frac{1}{4}\,c\hat{\mathbf{z}}$ | (4$c$) | U |
| $\mathbf{B_2}$ | = | $y_1\,\mathbf{a_1} - y_1\,\mathbf{a_2} + \frac{3}{4}\,\mathbf{a_3}$ | = | $-y_1\,b\,\hat{\mathbf{y}} + \frac{3}{4}\,c\hat{\mathbf{z}}$ | (4$c$) | U |

**Geometry files:**

- CIF: pp. 676
- POSCAR: pp. 676

# $\alpha$-Ga (A11) Structure: A_oC8_64_f

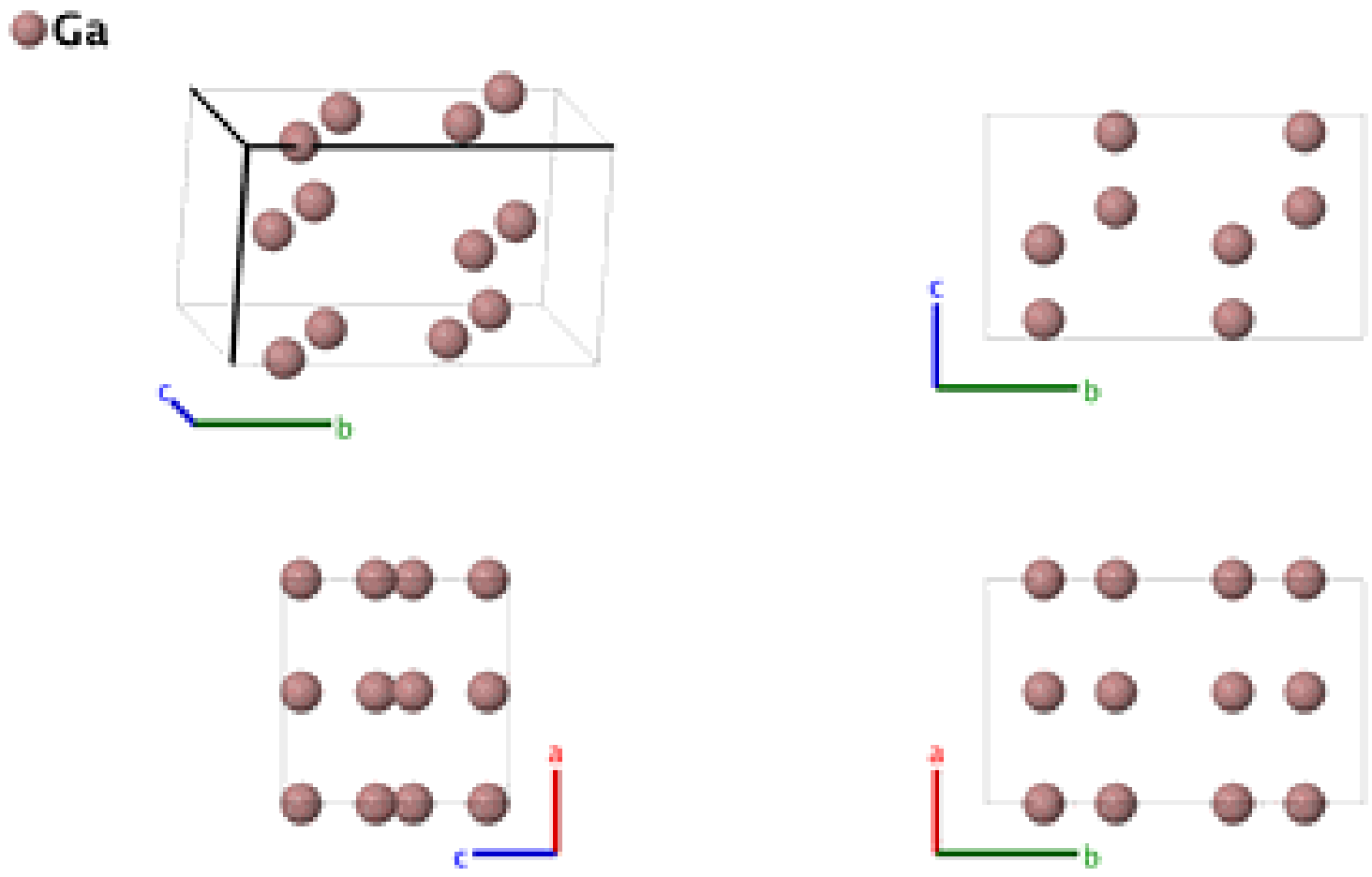

| | | |
|---|---|---|
| **Prototype** | : | $\alpha$-Ga |
| **AFLOW prototype label** | : | A_oC8_64_f |
| ***Strukturbericht* designation** | : | A11 |
| **Pearson symbol** | : | oC8 |
| **Space group number** | : | 64 |
| **Space group symbol** | : | Cmca |
| **AFLOW prototype command** | : | `aflow --proto=A_oC8_64_f --params=`$a, b/a, c/a, y_1, z_1$ |

- Note that $\alpha$-Ga (pp. 186), black phosphorus (pp. 191), and molecular iodine (pp. 193) have the same AFLOW prototype label. They are generated by the same symmetry operations with different sets of parameters (`--params`) specified in their corresponding CIF files.

**Base-centered Orthorhombic primitive vectors:**

$$\mathbf{a}_1 = \tfrac{1}{2}\,a\,\hat{\mathbf{x}} - \tfrac{1}{2}\,b\,\hat{\mathbf{y}}$$
$$\mathbf{a}_2 = \tfrac{1}{2}\,a\,\hat{\mathbf{x}} + \tfrac{1}{2}\,b\,\hat{\mathbf{y}}$$
$$\mathbf{a}_3 = c\,\hat{\mathbf{z}}$$

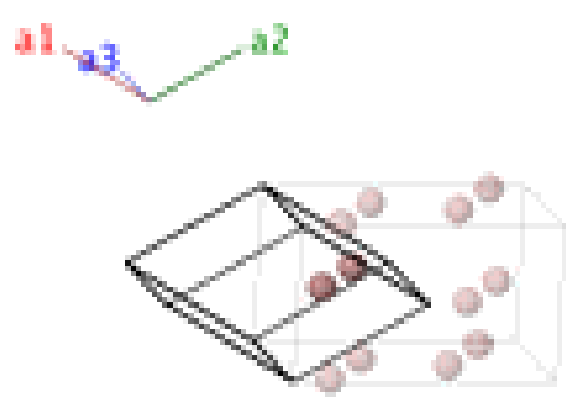

**Basis vectors:**

| | | Lattice Coordinates | | Cartesian Coordinates | Wyckoff Position | Atom Type |
|---|---|---|---|---|---|---|
| $\mathbf{B_1}$ | = | $-y_1\,\mathbf{a_1} + y_1\,\mathbf{a_2} + z_1\,\mathbf{a_3}$ | = | $y_1\,b\,\hat{\mathbf{y}} + z_1\,c\hat{\mathbf{z}}$ | $(8f)$ | Ga |
| $\mathbf{B_2}$ | = | $\left(\frac{1}{2} + y_1\right)\mathbf{a_1} + \left(\frac{1}{2} - y_1\right)\mathbf{a_2} + \left(\frac{1}{2} + z_1\right)\mathbf{a_3}$ | = | $\frac{1}{2}\,a\,\hat{\mathbf{x}} - y_1\,b\,\hat{\mathbf{y}} + \left(\frac{1}{2} + z_1\right)c\hat{\mathbf{z}}$ | $(8f)$ | Ga |
| $\mathbf{B_3}$ | = | $\left(\frac{1}{2} - y_1\right)\mathbf{a_1} + \left(\frac{1}{2} + y_1\right)\mathbf{a_2} + \left(\frac{1}{2} - z_1\right)\mathbf{a_3}$ | = | $\frac{1}{2}\,a\,\hat{\mathbf{x}} + y_1\,b\,\hat{\mathbf{y}} + \left(\frac{1}{2} - z_1\right)c\hat{\mathbf{z}}$ | $(8f)$ | Ga |
| $\mathbf{B_4}$ | = | $y_1\,\mathbf{a_1} - y_1\,\mathbf{a_2} - z_1\,\mathbf{a_3}$ | = | $-y_1\,b\,\hat{\mathbf{y}} - z_1\,c\hat{\mathbf{z}}$ | $(8f)$ | Ga |

**Geometry files:**
- CIF: pp. 677
- POSCAR: pp. 677

# $MgB_2C_2$ Crystal Structure: A2B2C_oC80_64_efg_efg_df

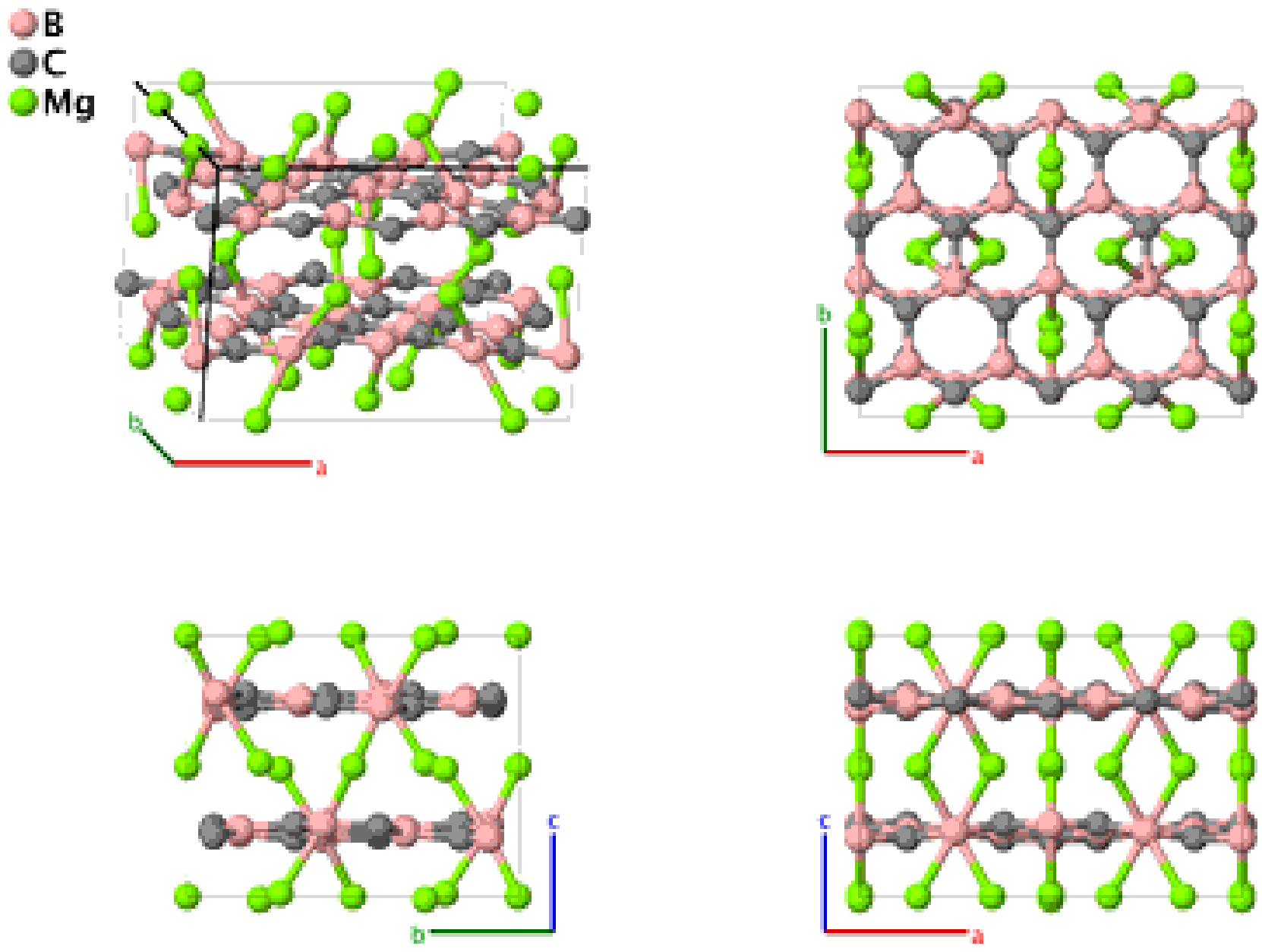

| | | |
|---|---|---|
| **Prototype** | : | $MgB_2C_2$ |
| **AFLOW prototype label** | : | A2B2C_oC80_64_efg_efg_df |
| ***Strukturbericht* designation** | : | None |
| **Pearson symbol** | : | oC80 |
| **Space group number** | : | 64 |
| **Space group symbol** | : | Cmca |
| **AFLOW prototype command** | : | `aflow --proto=A2B2C_oC80_64_efg_efg_df` `--params=`$a, b/a, c/a, x_1, y_2, y_3, y_4, z_4, y_5, z_5, y_6, z_6, x_7, y_7, z_7, x_8, y_8, z_8$ |

**Base-centered Orthorhombic primitive vectors:**

$$\mathbf{a}_1 = \frac{1}{2}\,a\,\hat{\mathbf{x}} - \frac{1}{2}\,b\,\hat{\mathbf{y}}$$
$$\mathbf{a}_2 = \frac{1}{2}\,a\,\hat{\mathbf{x}} + \frac{1}{2}\,b\,\hat{\mathbf{y}}$$
$$\mathbf{a}_3 = c\,\hat{\mathbf{z}}$$

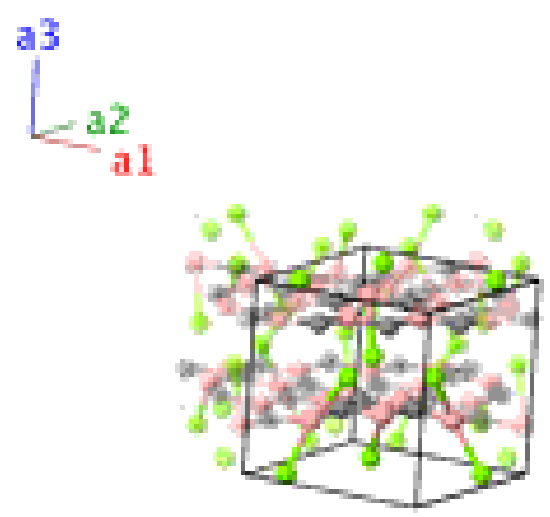

**Basis vectors:**

| | | Lattice Coordinates | | Cartesian Coordinates | Wyckoff Position | Atom Type |
|---|---|---|---|---|---|---|
| $\mathbf{B_1}$ | = | $x_1\,\mathbf{a_1} + x_1\,\mathbf{a_2}$ | = | $x_1\,a\,\hat{\mathbf{x}}$ | (8$d$) | Mg I |
| $\mathbf{B_2}$ | = | $\left(\frac{1}{2} - x_1\right)\mathbf{a_1} + \left(\frac{1}{2} - x_1\right)\mathbf{a_2} + \frac{1}{2}\,\mathbf{a_3}$ | = | $\left(\frac{1}{2} - x_1\right)a\,\hat{\mathbf{x}} + \frac{1}{2}\,c\,\hat{\mathbf{z}}$ | (8$d$) | Mg I |
| $\mathbf{B_3}$ | = | $-x_1\,\mathbf{a_1} - x_1\,\mathbf{a_2}$ | = | $-x_1\,a\,\hat{\mathbf{x}}$ | (8$d$) | Mg I |
| $\mathbf{B_4}$ | = | $\left(\frac{1}{2} + x_1\right)\mathbf{a_1} + \left(\frac{1}{2} + x_1\right)\mathbf{a_2} + \frac{1}{2}\,\mathbf{a_3}$ | = | $\left(\frac{1}{2} + x_1\right)a\,\hat{\mathbf{x}} + \frac{1}{2}\,c\,\hat{\mathbf{z}}$ | (8$d$) | Mg I |
| $\mathbf{B_5}$ | = | $\left(\frac{1}{4}-y_2\right)\mathbf{a_1}+\left(\frac{1}{4}+y_2\right)\mathbf{a_2}+\frac{1}{4}\mathbf{a_3}$ | = | $\frac{1}{4}a\,\hat{\mathbf{x}}+y_2\,b\,\hat{\mathbf{y}}+\frac{1}{4}c\,\hat{\mathbf{z}}$ | (8*e*) | B I |
| $\mathbf{B_6}$ | = | $\left(\frac{1}{4}+y_2\right)\mathbf{a_1}+\left(\frac{1}{4}-y_2\right)\mathbf{a_2}+\frac{3}{4}\mathbf{a_3}$ | = | $\frac{1}{4}a\,\hat{\mathbf{x}}-y_2\,b\,\hat{\mathbf{y}}+\frac{3}{4}c\,\hat{\mathbf{z}}$ | (8*e*) | B I |
| $\mathbf{B_7}$ | = | $\left(\frac{3}{4}+y_2\right)\mathbf{a_1}+\left(\frac{3}{4}-y_2\right)\mathbf{a_2}+\frac{3}{4}\mathbf{a_3}$ | = | $\frac{3}{4}a\,\hat{\mathbf{x}}-y_2\,b\,\hat{\mathbf{y}}+\frac{3}{4}c\,\hat{\mathbf{z}}$ | (8*e*) | B I |
| $\mathbf{B_8}$ | = | $\left(\frac{3}{4}-y_2\right)\mathbf{a_1}+\left(\frac{3}{4}+y_2\right)\mathbf{a_2}+\frac{1}{4}\mathbf{a_3}$ | = | $\frac{3}{4}a\,\hat{\mathbf{x}}+y_2\,b\,\hat{\mathbf{y}}+\frac{1}{4}c\,\hat{\mathbf{z}}$ | (8*e*) | B I |
| $\mathbf{B_9}$ | = | $\left(\frac{1}{4}-y_3\right)\mathbf{a_1}+\left(\frac{1}{4}+y_3\right)\mathbf{a_2}+\frac{1}{4}\mathbf{a_3}$ | = | $\frac{1}{4}a\,\hat{\mathbf{x}}+y_3\,b\,\hat{\mathbf{y}}+\frac{1}{4}c\,\hat{\mathbf{z}}$ | (8*e*) | C I |
| $\mathbf{B_{10}}$ | = | $\left(\frac{1}{4}+y_3\right)\mathbf{a_1}+\left(\frac{1}{4}-y_3\right)\mathbf{a_2}+\frac{3}{4}\mathbf{a_3}$ | = | $\frac{1}{4}a\,\hat{\mathbf{x}}-y_3\,b\,\hat{\mathbf{y}}+\frac{3}{4}c\,\hat{\mathbf{z}}$ | (8*e*) | C I |
| $\mathbf{B_{11}}$ | = | $\left(\frac{3}{4}+y_3\right)\mathbf{a_1}+\left(\frac{3}{4}-y_3\right)\mathbf{a_2}+\frac{3}{4}\mathbf{a_3}$ | = | $\frac{3}{4}a\,\hat{\mathbf{x}}-y_3\,b\,\hat{\mathbf{y}}+\frac{3}{4}c\,\hat{\mathbf{z}}$ | (8*e*) | C I |
| $\mathbf{B_{12}}$ | = | $\left(\frac{3}{4}-y_3\right)\mathbf{a_1}+\left(\frac{3}{4}+y_3\right)\mathbf{a_2}+\frac{1}{4}\mathbf{a_3}$ | = | $\frac{3}{4}a\,\hat{\mathbf{x}}+y_3\,b\,\hat{\mathbf{y}}+\frac{1}{4}c\,\hat{\mathbf{z}}$ | (8*e*) | C I |
| $\mathbf{B_{13}}$ | = | $-y_4\,\mathbf{a_1}+y_4\,\mathbf{a_2}+z_4\,\mathbf{a_3}$ | = | $y_4\,b\,\hat{\mathbf{y}}+z_4\,c\,\hat{\mathbf{z}}$ | (8*f*) | B II |
| $\mathbf{B_{14}}$ | = | $\left(\frac{1}{2}+y_4\right)\mathbf{a_1}+\left(\frac{1}{2}-y_4\right)\mathbf{a_2}+\left(\frac{1}{2}+z_4\right)\mathbf{a_3}$ | = | $\frac{1}{2}a\,\hat{\mathbf{x}}-y_4\,b\,\hat{\mathbf{y}}+\left(\frac{1}{2}+z_4\right)c\,\hat{\mathbf{z}}$ | (8*f*) | B II |
| $\mathbf{B_{15}}$ | = | $\left(\frac{1}{2}-y_4\right)\mathbf{a_1}+\left(\frac{1}{2}+y_4\right)\mathbf{a_2}+\left(\frac{1}{2}-z_4\right)\mathbf{a_3}$ | = | $\frac{1}{2}a\,\hat{\mathbf{x}}+y_4\,b\,\hat{\mathbf{y}}+\left(\frac{1}{2}-z_4\right)c\,\hat{\mathbf{z}}$ | (8*f*) | B II |
| $\mathbf{B_{16}}$ | = | $y_4\,\mathbf{a_1}-y_4\,\mathbf{a_2}-z_4\,\mathbf{a_3}$ | = | $-y_4\,b\,\hat{\mathbf{y}}-z_4\,c\,\hat{\mathbf{z}}$ | (8*f*) | B II |
| $\mathbf{B_{17}}$ | = | $-y_5\,\mathbf{a_1}+y_5\,\mathbf{a_2}+z_5\,\mathbf{a_3}$ | = | $y_5\,b\,\hat{\mathbf{y}}+z_5\,c\,\hat{\mathbf{z}}$ | (8*f*) | C II |
| $\mathbf{B_{18}}$ | = | $\left(\frac{1}{2}+y_5\right)\mathbf{a_1}+\left(\frac{1}{2}-y_5\right)\mathbf{a_2}+\left(\frac{1}{2}+z_5\right)\mathbf{a_3}$ | = | $\frac{1}{2}a\,\hat{\mathbf{x}}-y_5\,b\,\hat{\mathbf{y}}+\left(\frac{1}{2}+z_5\right)c\,\hat{\mathbf{z}}$ | (8*f*) | C II |
| $\mathbf{B_{19}}$ | = | $\left(\frac{1}{2}-y_5\right)\mathbf{a_1}+\left(\frac{1}{2}+y_5\right)\mathbf{a_2}+\left(\frac{1}{2}-z_5\right)\mathbf{a_3}$ | = | $\frac{1}{2}a\,\hat{\mathbf{x}}+y_5\,b\,\hat{\mathbf{y}}+\left(\frac{1}{2}-z_5\right)c\,\hat{\mathbf{z}}$ | (8*f*) | C II |
| $\mathbf{B_{20}}$ | = | $y_5\,\mathbf{a_1}-y_5\,\mathbf{a_2}-z_5\,\mathbf{a_3}$ | = | $-y_5\,b\,\hat{\mathbf{y}}-z_5\,c\,\hat{\mathbf{z}}$ | (8*f*) | C II |
| $\mathbf{B_{21}}$ | = | $-y_6\,\mathbf{a_1}+y_6\,\mathbf{a_2}+z_6\,\mathbf{a_3}$ | = | $y_6\,b\,\hat{\mathbf{y}}+z_6\,c\,\hat{\mathbf{z}}$ | (8*f*) | Mg II |
| $\mathbf{B_{22}}$ | = | $\left(\frac{1}{2}+y_6\right)\mathbf{a_1}+\left(\frac{1}{2}-y_6\right)\mathbf{a_2}+\left(\frac{1}{2}+z_6\right)\mathbf{a_3}$ | = | $\frac{1}{2}a\,\hat{\mathbf{x}}-y_6\,b\,\hat{\mathbf{y}}+\left(\frac{1}{2}+z_6\right)c\,\hat{\mathbf{z}}$ | (8*f*) | Mg II |
| $\mathbf{B_{23}}$ | = | $\left(\frac{1}{2}-y_6\right)\mathbf{a_1}+\left(\frac{1}{2}+y_6\right)\mathbf{a_2}+\left(\frac{1}{2}-z_6\right)\mathbf{a_3}$ | = | $\frac{1}{2}a\,\hat{\mathbf{x}}+y_6\,b\,\hat{\mathbf{y}}+\left(\frac{1}{2}-z_6\right)c\,\hat{\mathbf{z}}$ | (8*f*) | Mg II |
| $\mathbf{B_{24}}$ | = | $y_6\,\mathbf{a_1}-y_6\,\mathbf{a_2}-z_6\,\mathbf{a_3}$ | = | $-y_6\,b\,\hat{\mathbf{y}}-z_6\,c\,\hat{\mathbf{z}}$ | (8*f*) | Mg II |
| $\mathbf{B_{25}}$ | = | $(x_7-y_7)\,\mathbf{a_1}+(x_7+y_7)\,\mathbf{a_2}+z_7\,\mathbf{a_3}$ | = | $x_7\,a\,\hat{\mathbf{x}}+y_7\,b\,\hat{\mathbf{y}}+z_7\,c\,\hat{\mathbf{z}}$ | (16*g*) | B III |
| $\mathbf{B_{26}}$ | = | $\left(y_7-x_7+\frac{1}{2}\right)\mathbf{a_1}+\left(\frac{1}{2}-x_7-y_7\right)\mathbf{a_2}+\left(\frac{1}{2}+z_7\right)\mathbf{a_3}$ | = | $\left(\frac{1}{2}-x_7\right)a\,\hat{\mathbf{x}}-y_7\,b\,\hat{\mathbf{y}}+\left(\frac{1}{2}+z_7\right)c\,\hat{\mathbf{z}}$ | (16*g*) | B III |
| $\mathbf{B_{27}}$ | = | $\left(\frac{1}{2}-x_7-y_7\right)\mathbf{a_1}+\left(\frac{1}{2}-x_7+y_7\right)\mathbf{a_2}+\left(\frac{1}{2}-z_7\right)\mathbf{a_3}$ | = | $\left(\frac{1}{2}-x_7\right)a\,\hat{\mathbf{x}}+y_7\,b\,\hat{\mathbf{y}}+\left(\frac{1}{2}-z_7\right)c\,\hat{\mathbf{z}}$ | (16*g*) | B III |
| $\mathbf{B_{28}}$ | = | $(x_7+y_7)\,\mathbf{a_1}+(x_7-y_7)\,\mathbf{a_2}-z_7\,\mathbf{a_3}$ | = | $x_7\,a\,\hat{\mathbf{x}}-y_7\,b\,\hat{\mathbf{y}}-z_7\,c\,\hat{\mathbf{z}}$ | (16*g*) | B III |
| $\mathbf{B_{29}}$ | = | $(y_7-x_7)\,\mathbf{a_1}-(x_7+y_7)\,\mathbf{a_2}-z_7\,\mathbf{a_3}$ | = | $-x_7\,a\,\hat{\mathbf{x}}-y_7\,b\,\hat{\mathbf{y}}-z_7\,c\,\hat{\mathbf{z}}$ | (16*g*) | B III |
| $\mathbf{B_{30}}$ | = | $\left(\frac{1}{2}+x_7-y_7\right)\mathbf{a_1}+\left(\frac{1}{2}+x_7+y_7\right)\mathbf{a_2}+\left(\frac{1}{2}-z_7\right)\mathbf{a_3}$ | = | $\left(\frac{1}{2}+x_7\right)a\,\hat{\mathbf{x}}+y_7\,b\,\hat{\mathbf{y}}+\left(\frac{1}{2}-z_7\right)c\,\hat{\mathbf{z}}$ | (16*g*) | B III |
| $\mathbf{B_{31}}$ | = | $\left(\frac{1}{2}+x_7+y_7\right)\mathbf{a_1}+\left(\frac{1}{2}+x_7-y_7\right)\mathbf{a_2}+\left(\frac{1}{2}+z_7\right)\mathbf{a_3}$ | = | $\left(\frac{1}{2}+x_7\right)a\,\hat{\mathbf{x}}-y_7\,b\,\hat{\mathbf{y}}+\left(\frac{1}{2}+z_7\right)c\,\hat{\mathbf{z}}$ | (16*g*) | B III |
| $\mathbf{B_{32}}$ | = | $-(x_7+y_7)\,\mathbf{a_1}+(y_7-x_7)\,\mathbf{a_2}+z_7\,\mathbf{a_3}$ | = | $-x_7\,a\,\hat{\mathbf{x}}+y_7\,b\,\hat{\mathbf{y}}+z_7\,c\,\hat{\mathbf{z}}$ | (16*g*) | B III |
| $\mathbf{B_{33}}$ | = | $(x_8 - y_8)\,\mathbf{a_1} + (x_8 + y_8)\,\mathbf{a_2} + z_8\,\mathbf{a_3}$ | = | $x_8\,a\,\hat{\mathbf{x}} + y_8\,b\,\hat{\mathbf{y}} + z_8\,c\hat{\mathbf{z}}$ | (16$g$) | C III |
| $\mathbf{B_{34}}$ | = | $\left(\frac{1}{2} + y_8 - x_8\right)\mathbf{a_1} + \left(\frac{1}{2} - x_8 - y_8\right)\mathbf{a_2} + \left(\frac{1}{2} + z_8\right)\mathbf{a_3}$ | = | $\left(\frac{1}{2} - x_8\right)a\,\hat{\mathbf{x}} - y_8\,b\,\hat{\mathbf{y}} + \left(\frac{1}{2} + z_8\right)c\hat{\mathbf{z}}$ | (16$g$) | C III |
| $\mathbf{B_{35}}$ | = | $\left(\frac{1}{2} - x_8 - y_8\right)\mathbf{a_1} + \left(\frac{1}{2} - x_8 + y_8\right)\mathbf{a_2} + \left(\frac{1}{2} - z_8\right)\mathbf{a_3}$ | = | $\left(\frac{1}{2} - x_8\right)a\,\hat{\mathbf{x}} + y_8\,b\,\hat{\mathbf{y}} + \left(\frac{1}{2} - z_8\right)c\hat{\mathbf{z}}$ | (16$g$) | C III |
| $\mathbf{B_{36}}$ | = | $(x_8 + y_8)\,\mathbf{a_1} + (x_8 - y_8)\,\mathbf{a_2} - z_8\,\mathbf{a_3}$ | = | $x_8\,a\,\hat{\mathbf{x}} - y_8\,b\,\hat{\mathbf{y}} - z_8\,c\hat{\mathbf{z}}$ | (16$g$) | C III |
| $\mathbf{B_{37}}$ | = | $(y_8 - x_8)\,\mathbf{a_1} - (x_8 + y_8)\,\mathbf{a_2} - z_8\,\mathbf{a_3}$ | = | $-x_8\,a\,\hat{\mathbf{x}} - y_8\,b\,\hat{\mathbf{y}} - z_8\,c\hat{\mathbf{z}}$ | (16$g$) | C III |
| $\mathbf{B_{38}}$ | = | $\left(\frac{1}{2} + x_8 - y_8\right)\mathbf{a_1} + \left(\frac{1}{2} + x_8 + y_8\right)\mathbf{a_2} + \left(\frac{1}{2} - z_8\right)\mathbf{a_3}$ | = | $\left(\frac{1}{2} + x_8\right)a\,\hat{\mathbf{x}} + y_8\,b\,\hat{\mathbf{y}} + \left(\frac{1}{2} - z_8\right)c\hat{\mathbf{z}}$ | (16$g$) | C III |
| $\mathbf{B_{39}}$ | = | $\left(\frac{1}{2} + x_8 + y_8\right)\mathbf{a_1} + \left(\frac{1}{2} + x_8 - y_8\right)\mathbf{a_2} + \left(\frac{1}{2} + z_8\right)\mathbf{a_3}$ | = | $\left(\frac{1}{2} + x_8\right)a\,\hat{\mathbf{x}} - y_8\,b\,\hat{\mathbf{y}} + \left(\frac{1}{2} + z_8\right)c\hat{\mathbf{z}}$ | (16$g$) | C III |
| $\mathbf{B_{40}}$ | = | $-(x_8 + y_8)\,\mathbf{a_1} + (y_8 - x_8)\,\mathbf{a_2} + z_8\,\mathbf{a_3}$ | = | $-x_8\,a\,\hat{\mathbf{x}} + y_8\,b\,\hat{\mathbf{y}} + z_8\,c\hat{\mathbf{z}}$ | (16$g$) | C III |

---

---

**Geometry files:**

- CIF: pp. 677
- POSCAR: pp. 677

# Black Phosphorus (A17) Crystal Structure: A_oC8_64_f

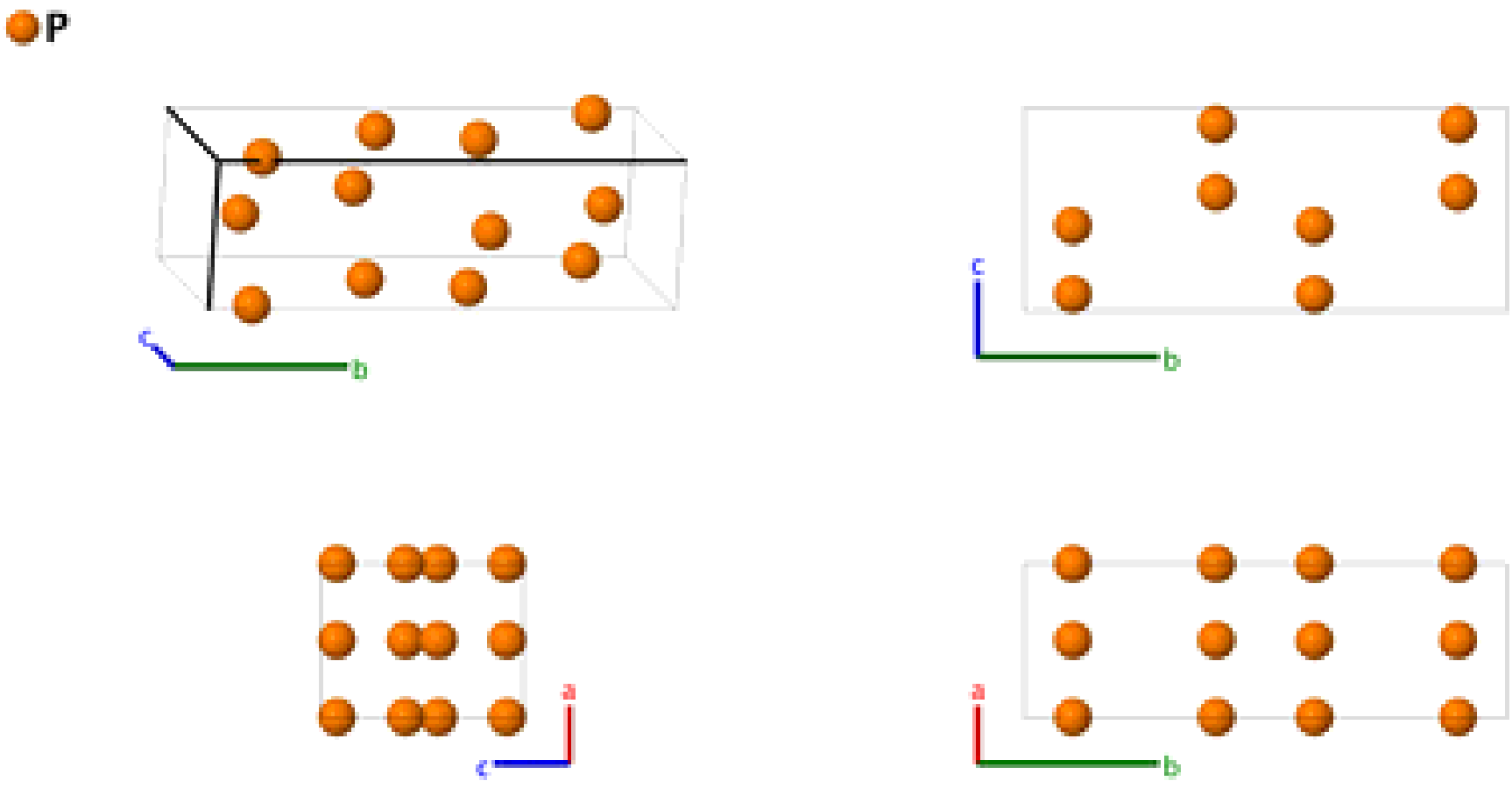

| | | |
|---|---|---|
| **Prototype** | : | P |
| **AFLOW prototype label** | : | A_oC8_64_f |
| ***Strukturbericht* designation** | : | A17 |
| **Pearson symbol** | : | oC8 |
| **Space group number** | : | 64 |
| **Space group symbol** | : | Cmca |
| **AFLOW prototype command** | : | `aflow --proto=A_oC8_64_f`<br>`--params=`$a, b/a, c/a, y_1, z_1$ |

- Note that $\alpha$-Ga (pp. 186), black phosphorus (pp. 191), and molecular iodine (pp. 193) have the same AFLOW prototype label. They are generated by the same symmetry operations with different sets of parameters (`--params`) specified in their corresponding CIF files.

**Base-centered Orthorhombic primitive vectors:**

$$\mathbf{a}_1 = \frac{1}{2}\,a\,\hat{\mathbf{x}} - \frac{1}{2}\,b\,\hat{\mathbf{y}}$$
$$\mathbf{a}_2 = \frac{1}{2}\,a\,\hat{\mathbf{x}} + \frac{1}{2}\,b\,\hat{\mathbf{y}}$$
$$\mathbf{a}_3 = c\,\hat{\mathbf{z}}$$

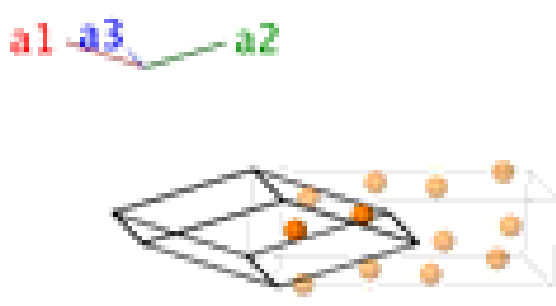

**Basis vectors:**

| | | Lattice Coordinates | | Cartesian Coordinates | Wyckoff Position | Atom Type |
|---|---|---|---|---|---|---|
| $\mathbf{B}_1$ | = | $-y_1\,\mathbf{a}_1 + y_1\,\mathbf{a}_2 + z_1\,\mathbf{a}_3$ | = | $y_1\,b\,\hat{\mathbf{y}} + z_1\,c\hat{\mathbf{z}}$ | (8$f$) | P |
| $\mathbf{B}_2$ | = | $\left(\frac{1}{2}+y_1\right)\mathbf{a}_1 + \left(\frac{1}{2}-y_1\right)\mathbf{a}_2 + \left(\frac{1}{2}+z_1\right)\mathbf{a}_3$ | = | $\frac{1}{2}\,a\,\hat{\mathbf{x}} - y_1\,b\,\hat{\mathbf{y}} + \left(\frac{1}{2}+z_1\right)c\hat{\mathbf{z}}$ | (8$f$) | P |
| $\mathbf{B}_3$ | = | $\left(\frac{1}{2}-y_1\right)\mathbf{a}_1 + \left(\frac{1}{2}+y_1\right)\mathbf{a}_2 + \left(\frac{1}{2}-z_1\right)\mathbf{a}_3$ | = | $\frac{1}{2}\,a\,\hat{\mathbf{x}} + y_1\,b\,\hat{\mathbf{y}} + \left(\frac{1}{2}-z_1\right)c\hat{\mathbf{z}}$ | (8$f$) | P |
| $\mathbf{B}_4$ | = | $y_1\,\mathbf{a}_1 - y_1\,\mathbf{a}_2 - z_1\,\mathbf{a}_3$ | = | $-y_1\,b\,\hat{\mathbf{y}} - z_1\,c\hat{\mathbf{z}}$ | (8$f$) | P |

---

**Geometry files:**

- CIF: pp. 678
- POSCAR: pp. 678

# Molecular Iodine (I) Crystal Structure (A14): A_oC8_64_f

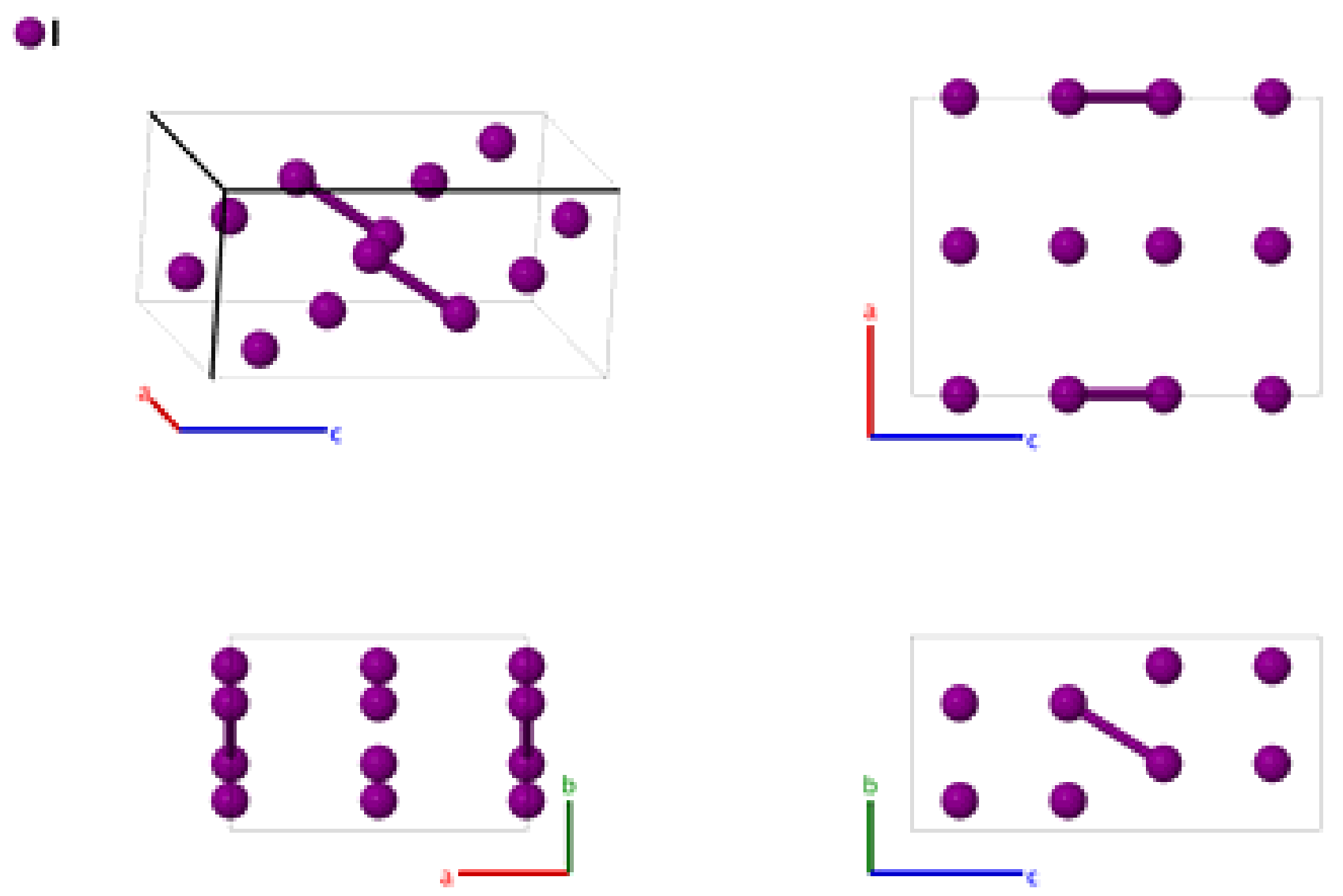

| | | |
|---|---|---|
| **Prototype** | : | I |
| **AFLOW prototype label** | : | A_oC8_64_f |
| ***Strukturbericht* designation** | : | A14 |
| **Pearson symbol** | : | oC8 |
| **Space group number** | : | 64 |
| **Space group symbol** | : | Cmca |
| **AFLOW prototype command** | : | `aflow --proto=A_oC8_64_f`<br>`--params=`$a, b/a, c/a, y_1, z_1$ |

- Note that $\alpha$-Ga (pp. 186), black phosphorus (pp. 191), and molecular iodine (pp. 193) have the same AFLOW prototype label. They are generated by the same symmetry operations with different sets of parameters (`--params`) specified in their corresponding CIF files.

**Base-centered Orthorhombic primitive vectors:**

$$\mathbf{a}_1 = \frac{1}{2}\,a\,\hat{\mathbf{x}} - \frac{1}{2}\,b\,\hat{\mathbf{y}}$$
$$\mathbf{a}_2 = \frac{1}{2}\,a\,\hat{\mathbf{x}} + \frac{1}{2}\,b\,\hat{\mathbf{y}}$$
$$\mathbf{a}_3 = c\,\hat{\mathbf{z}}$$

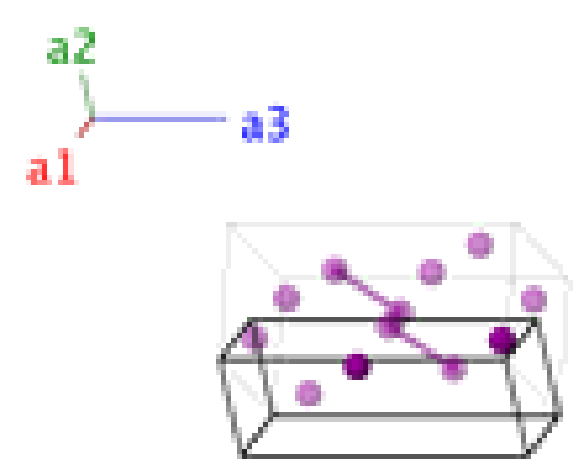

**Basis vectors:**

| | | Lattice Coordinates | | Cartesian Coordinates | Wyckoff Position | Atom Type |
|---|---|---|---|---|---|---|
| $\mathbf{B}_1$ | = | $-y_1\,\mathbf{a}_1 + y_1\,\mathbf{a}_2 + z_1\,\mathbf{a}_3$ | = | $y_1\,b\,\hat{\mathbf{y}} + z_1\,c\hat{\mathbf{z}}$ | $(8f)$ | I |
| $\mathbf{B}_2$ | = | $\left(\frac{1}{2} + y_1\right)\mathbf{a}_1 + \left(\frac{1}{2} - y_1\right)\mathbf{a}_2 + \left(\frac{1}{2} + z_1\right)\mathbf{a}_3$ | = | $\frac{1}{2}\,a\,\hat{\mathbf{x}} - y_1\,b\,\hat{\mathbf{y}} + \left(\frac{1}{2} + z_1\right)c\hat{\mathbf{z}}$ | $(8f)$ | I |
| $\mathbf{B_3}$ | = | $\left(\frac{1}{2} - y_1\right) \mathbf{a_1} + \left(\frac{1}{2} + y_1\right) \mathbf{a_2} + \left(\frac{1}{2} - z_1\right) \mathbf{a_3}$ | = | $\frac{1}{2} a \, \mathbf{\hat{x}} + y_1 \, b \, \mathbf{\hat{y}} + \left(\frac{1}{2} - z_1\right) c \mathbf{\hat{z}}$ | (8$f$) | I |
| $\mathbf{B_4}$ | = | $y_1 \, \mathbf{a_1} - y_1 \, \mathbf{a_2} - z_1 \, \mathbf{a_3}$ | = | $-y_1 \, b \, \mathbf{\hat{y}} - z_1 \, c \mathbf{\hat{z}}$ | (8$f$) | I |

---

**Found in:**

---

**Geometry files:**

- CIF: pp. 678
- POSCAR: pp. 678

# $\alpha$-IrV Crystal Structure: AB_oC8_65_j_g

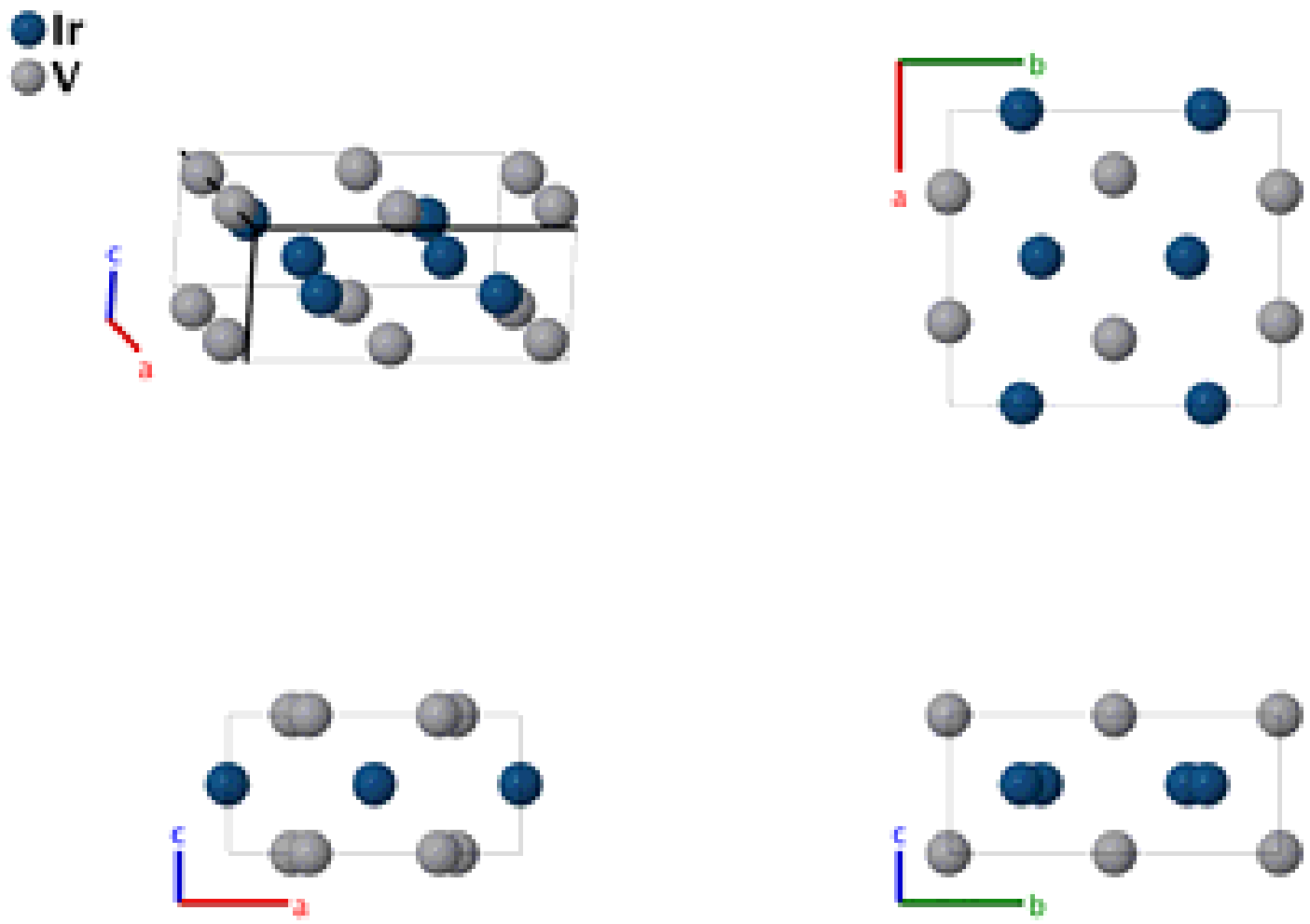

| | | |
|---|---|---|
| **Prototype** | : | $\alpha$-IrV |
| **AFLOW prototype label** | : | AB_oC8_65_j_g |
| ***Strukturbericht* designation** | : | None |
| **Pearson symbol** | : | oC8 |
| **Space group number** | : | 65 |
| **Space group symbol** | : | Cmmm |
| **AFLOW prototype command** | : | `aflow --proto=AB_oC8_65_j_g` <br> `--params=`$a, b/a, c/a, x_1, y_2$ |

**Base-centered Orthorhombic primitive vectors:**

$$\mathbf{a}_1 = \frac{1}{2} a \, \mathbf{\hat{x}} - \frac{1}{2} b \, \mathbf{\hat{y}}$$
$$\mathbf{a}_2 = \frac{1}{2} a \, \mathbf{\hat{x}} + \frac{1}{2} b \, \mathbf{\hat{y}}$$
$$\mathbf{a}_3 = c \, \mathbf{\hat{z}}$$

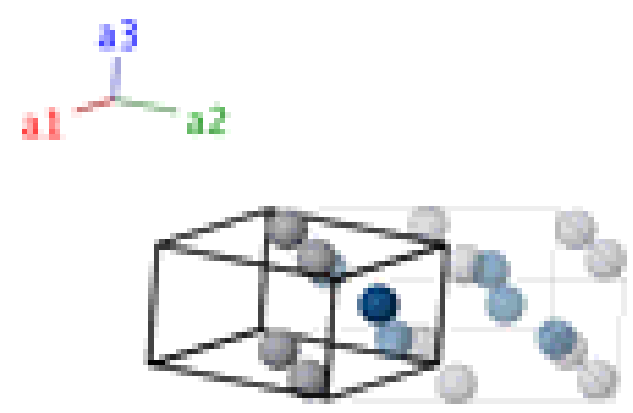

**Basis vectors:**

| | | Lattice Coordinates | | Cartesian Coordinates | Wyckoff Position | Atom Type |
|---|---|---|---|---|---|---|
| $\mathbf{B_1}$ | = | $x_1 \, \mathbf{a_1} + x_1 \, \mathbf{a_2}$ | = | $x_1 \, a \, \mathbf{\hat{x}}$ | (4$g$) | V |
| $\mathbf{B_2}$ | = | $-x_1 \, \mathbf{a_1} - x_1 \, \mathbf{a_2}$ | = | $-x_1 \, a \, \mathbf{\hat{x}}$ | (4$g$) | V |
| $\mathbf{B_3}$ | = | $-y_2 \, \mathbf{a_1} + y_2 \, \mathbf{a_2} + \frac{1}{2} \, \mathbf{a_3}$ | = | $y_2 \, b \, \mathbf{\hat{y}} + \frac{1}{2} \, c \, \mathbf{\hat{z}}$ | (4$j$) | Ir |
| $\mathbf{B_4}$ | = | $y_2 \, \mathbf{a_1} - y_2 \, \mathbf{a_2} + \frac{1}{2} \, \mathbf{a_3}$ | = | $-y_2 \, b \, \mathbf{\hat{y}} + \frac{1}{2} \, c \, \mathbf{\hat{z}}$ | (4$j$) | Ir |

**Found in:**

---

**Geometry files:**

- CIF: pp. 679
- POSCAR: pp. 679

# $Ga_3Pt_5$ Structure: A3B5_oC16_65_ah_bej

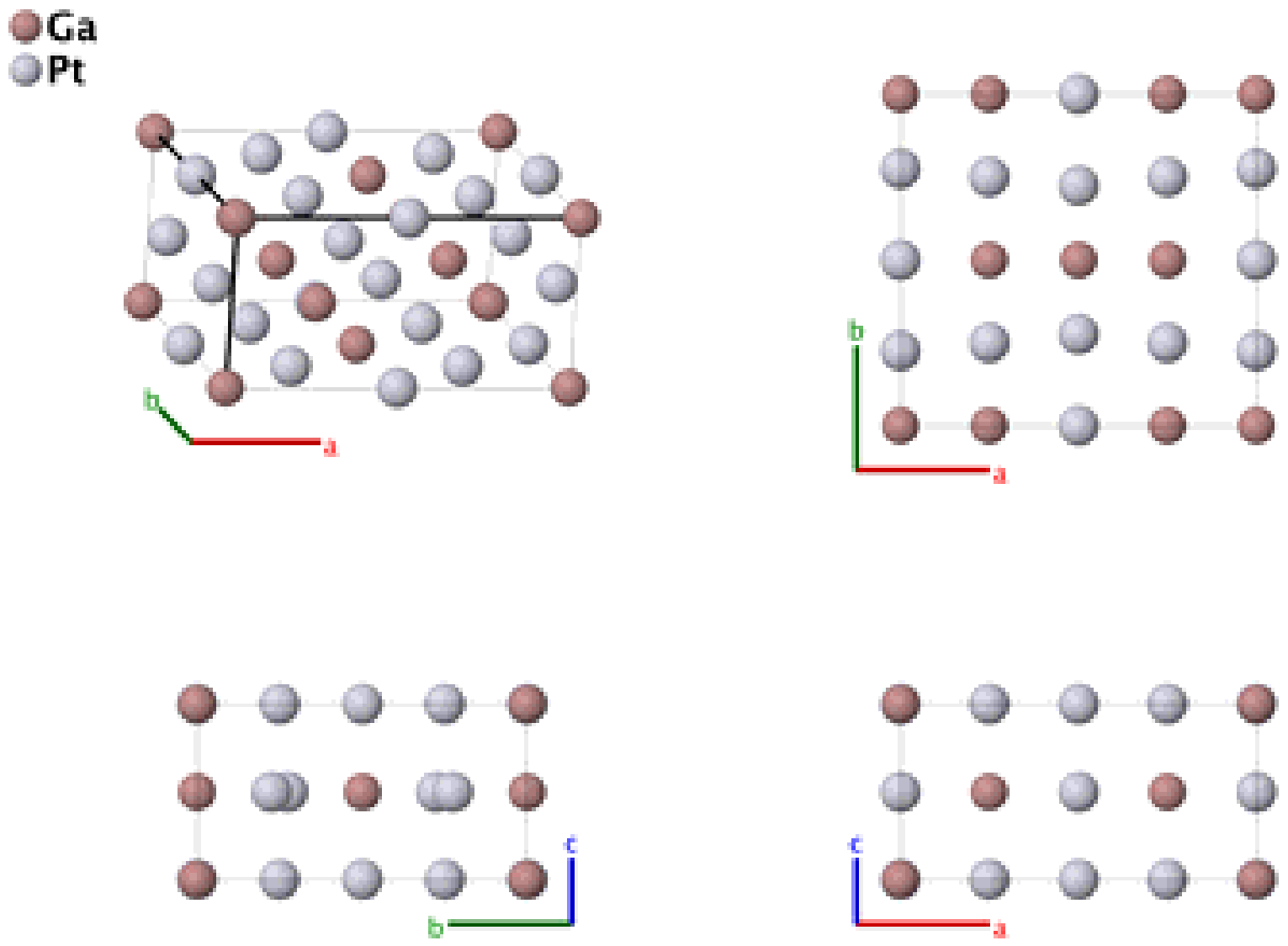

| | | |
|---|---|---|
| **Prototype** | : | $Ga_3Pt_5$ |
| **AFLOW prototype label** | : | A3B5_oC16_65_ah_bej |
| ***Strukturbericht* designation** | : | None |
| **Pearson symbol** | : | oC16 |
| **Space group number** | : | 65 |
| **Space group symbol** | : | Cmmm |
| **AFLOW prototype command** | : | `aflow --proto=A3B5_oC16_65_ah_bej` <br> `--params=`$a, b/a, c/a, x_4, y_5$ |

---

**Base-centered Orthorhombic primitive vectors:**

$$\mathbf{a}_1 = \frac{1}{2}\,a\,\mathbf{\hat{x}} - \frac{1}{2}\,b\,\mathbf{\hat{y}}$$
$$\mathbf{a}_2 = \frac{1}{2}\,a\,\mathbf{\hat{x}} + \frac{1}{2}\,b\,\mathbf{\hat{y}}$$
$$\mathbf{a}_3 = c\,\mathbf{\hat{z}}$$

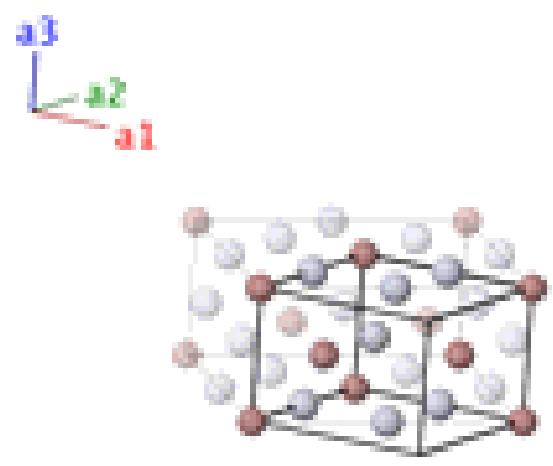

---

**Basis vectors:**

| | | Lattice Coordinates | | Cartesian Coordinates | Wyckoff Position | Atom Type |
|---|---|---|---|---|---|---|
| $\mathbf{B}_1$ | = | $0\,\mathbf{a}_1 + 0\,\mathbf{a}_2 + 0\,\mathbf{a}_3$ | = | $0\mathbf{\hat{x}} + 0\mathbf{\hat{y}} + 0\mathbf{\hat{z}}$ | (2$a$) | Ga I |
| $\mathbf{B}_2$ | = | $\frac{1}{2}\,\mathbf{a}_1 + \frac{1}{2}\,\mathbf{a}_2$ | = | $\frac{1}{2}\,a\,\mathbf{\hat{x}}$ | (2$b$) | Pt I |
| $\mathbf{B}_3$ | = | $\frac{1}{2}\,\mathbf{a}_2$ | = | $\frac{1}{4}\,a\,\mathbf{\hat{x}} + \frac{1}{4}\,b\,\mathbf{\hat{y}}$ | (4$e$) | Pt II |
| $\mathbf{B}_4$ | = | $\frac{1}{2}\,\mathbf{a}_1$ | = | $\frac{1}{4}\,a\,\mathbf{\hat{x}} + \frac{3}{4}\,b\,\mathbf{\hat{y}}$ | (4$e$) | Pt II |
| $\mathbf{B}_5$ | = | $x_4\,\mathbf{a}_1 + x_4\,\mathbf{a}_2 + \frac{1}{2}\,\mathbf{a}_3$ | = | $x_4\,a\,\mathbf{\hat{x}} + \frac{1}{2}\,c\mathbf{\hat{z}}$ | (4$h$) | Ga II |
| $\mathbf{B_6}$ | = | $-x_4\,\mathbf{a_1} - x_4\,\mathbf{a_2} + \frac{1}{2}\,\mathbf{a_3}$ | = | $-x_4\,a\,\hat{\mathbf{x}} + \frac{1}{2}\,c\hat{\mathbf{z}}$ | $(4h)$ | Ga II |
| $\mathbf{B_7}$ | = | $-y_5\,\mathbf{a_1} + y_5\,\mathbf{a_2} + \frac{1}{2}\,\mathbf{a_3}$ | = | $y_5\,b\,\hat{\mathbf{y}} + \frac{1}{2}\,c\hat{\mathbf{z}}$ | $(4j)$ | Pt III |
| $\mathbf{B_8}$ | = | $y_5\,\mathbf{a_1} - y_5\,\mathbf{a_2} + \frac{1}{2}\,\mathbf{a_3}$ | = | $-y_5\,b\,\hat{\mathbf{y}} + \frac{1}{2}\,c\hat{\mathbf{z}}$ | $(4j)$ | Pt III |

---

**Found in:**

---

**Geometry files:**

- CIF: pp. 679
- POSCAR: pp. 679

# Predicted $CdPt_3$ ("$L1_3$") Structure: AB3_oC8_65_a_bf

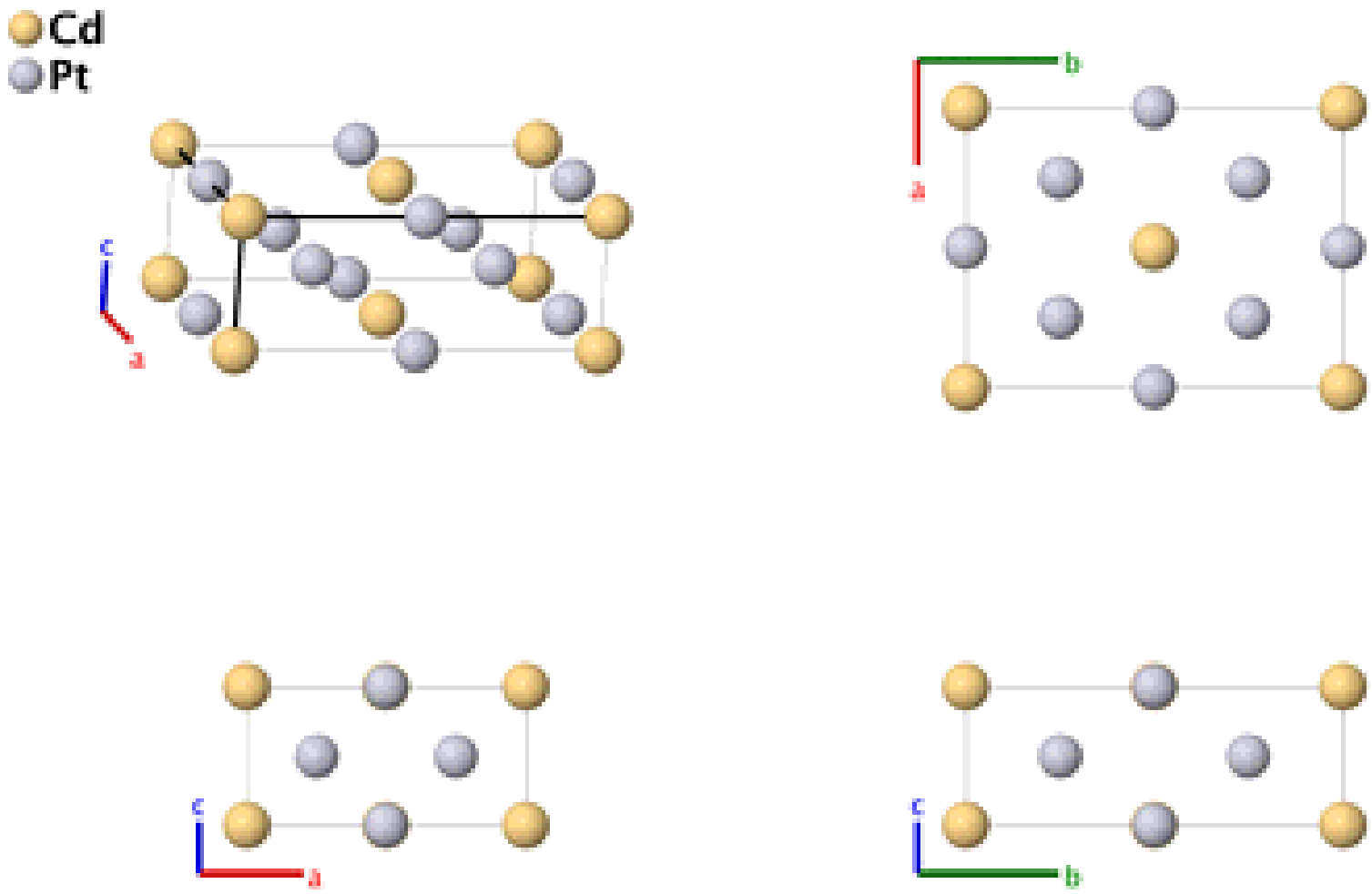

| | | |
|---|---|---|
| **Prototype** | : | $CdPt_3$ |
| **AFLOW prototype label** | : | AB3_oC8_65_a_bf |
| ***Strukturbericht* designation** | : | $L1_3$ |
| **Pearson symbol** | : | oC8 |
| **Space group number** | : | 65 |
| **Space group symbol** | : | Cmmm |
| **AFLOW prototype command** | : | `aflow --proto=AB3_oC8_65_a_bf --params=`$a, b/a, c/a$ |

- This structure has not been experimentally confirmed, but it has frequently been predicted as a low energy structure. (Hart, 2009) has a review of these calculations. The $L1_3$ designation is not official, but we use it here for consistency with previous literature. Data for this structure comes from the supplemental material of (Hart, 2013).

**Base-centered Orthorhombic primitive vectors:**

$$\begin{aligned}
\mathbf{a}_1 &= \tfrac{1}{2}\,a\,\hat{\mathbf{x}} - \tfrac{1}{2}\,b\,\hat{\mathbf{y}} \\
\mathbf{a}_2 &= \tfrac{1}{2}\,a\,\hat{\mathbf{x}} + \tfrac{1}{2}\,b\,\hat{\mathbf{y}} \\
\mathbf{a}_3 &= c\,\hat{\mathbf{z}}
\end{aligned}$$

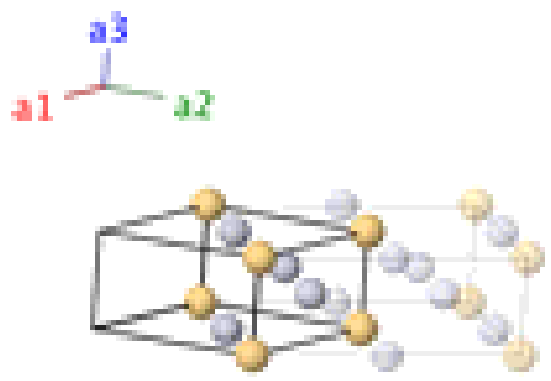

**Basis vectors:**

| | | Lattice Coordinates | | Cartesian Coordinates | Wyckoff Position | Atom Type |
|---|---|---|---|---|---|---|
| $\mathbf{B}_1$ | = | $0\,\mathbf{a}_1 + 0\,\mathbf{a}_2 + 0\,\mathbf{a}_3$ | = | $0\hat{\mathbf{x}} + 0\hat{\mathbf{y}} + 0\hat{\mathbf{z}}$ | $(2a)$ | Cd |
| $\mathbf{B}_2$ | = | $\tfrac{1}{2}\,\mathbf{a}_1 + \tfrac{1}{2}\,\mathbf{a}_2$ | = | $\tfrac{1}{2}\,a\,\hat{\mathbf{x}}$ | $(2b)$ | Pt I |
| $\mathbf{B}_3$ | = | $\tfrac{1}{2}\,\mathbf{a}_2 + \tfrac{1}{2}\,\mathbf{a}_3$ | = | $\tfrac{1}{4}\,a\,\hat{\mathbf{x}} + \tfrac{1}{4}\,b\,\hat{\mathbf{y}} + \tfrac{1}{2}\,c\,\hat{\mathbf{z}}$ | $(4f)$ | Pt II |
| $\mathbf{B}_4$ | = | $\tfrac{1}{2}\,\mathbf{a}_1 + \tfrac{1}{2}\,\mathbf{a}_3$ | = | $\tfrac{1}{4}\,a\,\hat{\mathbf{x}} + \tfrac{3}{4}\,b\,\hat{\mathbf{y}} + \tfrac{1}{2}\,c\,\hat{\mathbf{z}}$ | $(4f)$ | Pt II |

**Geometry files:**

- CIF: pp. 680
- POSCAR: pp. 680

# TlF (B24) Structure: AB_oF8_69_a_b

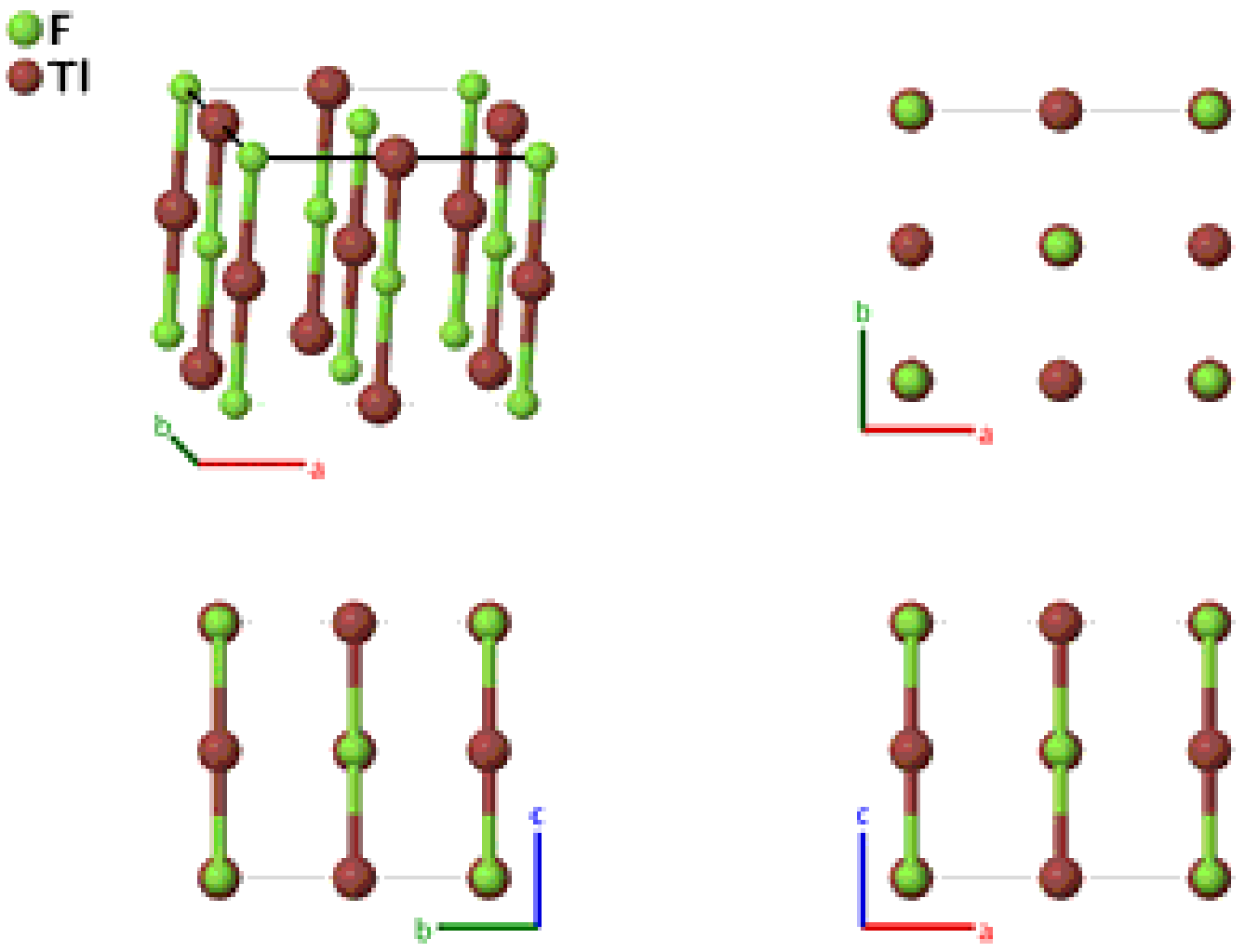

**Prototype** : TlF

**AFLOW prototype label** : AB_oF8_69_a_b

***Strukturbericht* designation** : B24

**Pearson symbol** : oF8

**Space group number** : 69

**Space group symbol** : Fmmm

**AFLOW prototype command** : `aflow --proto=AB_oF8_69_a_b --params=`$a, b/a, c/a$

---

- Although this is the B24 structure defined in Strukturbericht, it is not the currently accepted structure for TlF. See (Berastegui, 2000) and the TlF-II page. This is a slight distortion of the rock salt (B1) structure.

---

**Face-centered Orthorhombic primitive vectors:**

$$\mathbf{a}_1 = \tfrac{1}{2}\,b\,\hat{\mathbf{y}} + \tfrac{1}{2}\,c\,\hat{\mathbf{z}}$$
$$\mathbf{a}_2 = \tfrac{1}{2}\,a\,\hat{\mathbf{x}} + \tfrac{1}{2}\,c\,\hat{\mathbf{z}}$$
$$\mathbf{a}_3 = \tfrac{1}{2}\,a\,\hat{\mathbf{x}} + \tfrac{1}{2}\,b\,\hat{\mathbf{y}}$$

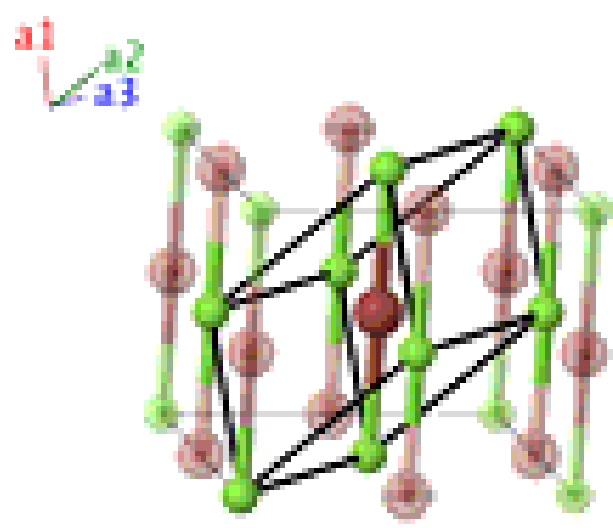

---

**Basis vectors:**

| | | Lattice Coordinates | | Cartesian Coordinates | Wyckoff Position | Atom Type |
|---|---|---|---|---|---|---|
| $\mathbf{B}_1$ | = | $0\,\mathbf{a}_1 + 0\,\mathbf{a}_2 + 0\,\mathbf{a}_3$ | = | $0\hat{\mathbf{x}} + 0\hat{\mathbf{y}} + 0\hat{\mathbf{z}}$ | (4$a$) | F |
| $\mathbf{B}_2$ | = | $\frac{1}{2}\,\mathbf{a}_1 + \frac{1}{2}\,\mathbf{a}_2 + \frac{1}{2}\,\mathbf{a}_3$ | = | $\frac{1}{2}\,a\,\hat{\mathbf{x}} + \frac{1}{2}\,b\,\hat{\mathbf{y}} + \frac{1}{2}\,c\,\hat{\mathbf{z}}$ | (4$b$) | Tl |

---

**References:**

- J. A. A. Ketelaar, *Die Kristallstruktur des Thallofluorids*, Zeitschrift für Kristallographie - Crystalline Materials **92**, 30–38 (1935), doi:10.1524/zkri.1935.92.1.30.

**Found in:**

- P. Berastegui and S. Hull, *The Crystal Structures of Thallium(I) Fluoride*, J. Solid State Chem. **150**, 266–275 (2000), doi:10.1006/jssc.1999.8587.

---

**Geometry files:**

- CIF: pp. 680
- POSCAR: pp. 680

# $\gamma$-Pu Structure: A_oF8_70_a

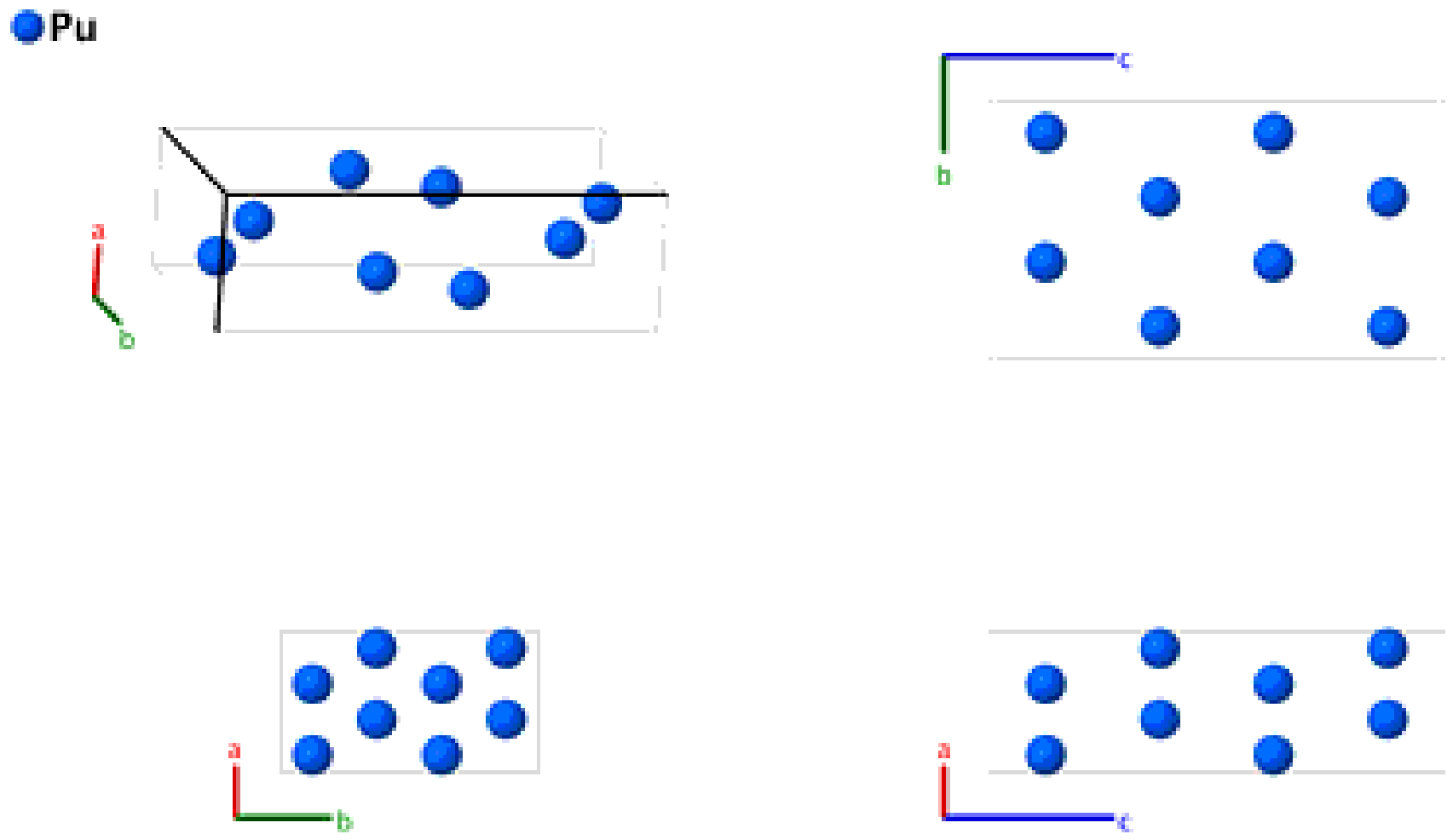

| | | |
|---|---|---|
| **Prototype** | : | $\gamma$-Pu |
| **AFLOW prototype label** | : | A_oF8_70_a |
| ***Strukturbericht* designation** | : | None |
| **Pearson symbol** | : | oF8 |
| **Space group number** | : | 70 |
| **Space group symbol** | : | Fddd |
| **AFLOW prototype command** | : | `aflow --proto=A_oF8_70_a` `--params=`$a, b/a, c/a$ |

- It is obvious from the coordinates that this is an extremely distorted diamond (A4) structure, but, as noted by (Donohue, 1982), it can also be considered as a distorted hcp (A3) structure.

**Face-centered Orthorhombic primitive vectors:**

$$\mathbf{a}_1 = \frac{1}{2}\,b\,\hat{\mathbf{y}} + \frac{1}{2}\,c\,\hat{\mathbf{z}}$$
$$\mathbf{a}_2 = \frac{1}{2}\,a\,\hat{\mathbf{x}} + \frac{1}{2}\,c\,\hat{\mathbf{z}}$$
$$\mathbf{a}_3 = \frac{1}{2}\,a\,\hat{\mathbf{x}} + \frac{1}{2}\,b\,\hat{\mathbf{y}}$$

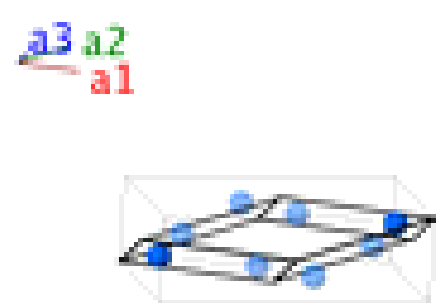

**Basis vectors:**

| | | Lattice Coordinates | | Cartesian Coordinates | Wyckoff Position | Atom Type |
|---|---|---|---|---|---|---|
| $\mathbf{B}_1$ | = | $\frac{1}{8}\,\mathbf{a}_1 + \frac{1}{8}\,\mathbf{a}_2 + \frac{1}{8}\,\mathbf{a}_3$ | = | $\frac{1}{8}\,a\,\hat{\mathbf{x}} + \frac{1}{8}\,b\,\hat{\mathbf{y}} + \frac{1}{8}\,c\,\hat{\mathbf{z}}$ | (8$a$) | Pu |
| $\mathbf{B}_2$ | = | $\frac{7}{8}\,\mathbf{a}_1 + \frac{7}{8}\,\mathbf{a}_2 + \frac{7}{8}\,\mathbf{a}_3$ | = | $\frac{7}{8}\,a\,\hat{\mathbf{x}} + \frac{7}{8}\,b\,\hat{\mathbf{y}} + \frac{7}{8}\,c\,\hat{\mathbf{z}}$ | (8$a$) | Pu |

---

**Geometry files:**
- CIF: pp. 680
- POSCAR: pp. 681

# $TiSi_2$ (C54) Structure: A2B_oF24_70_e_a

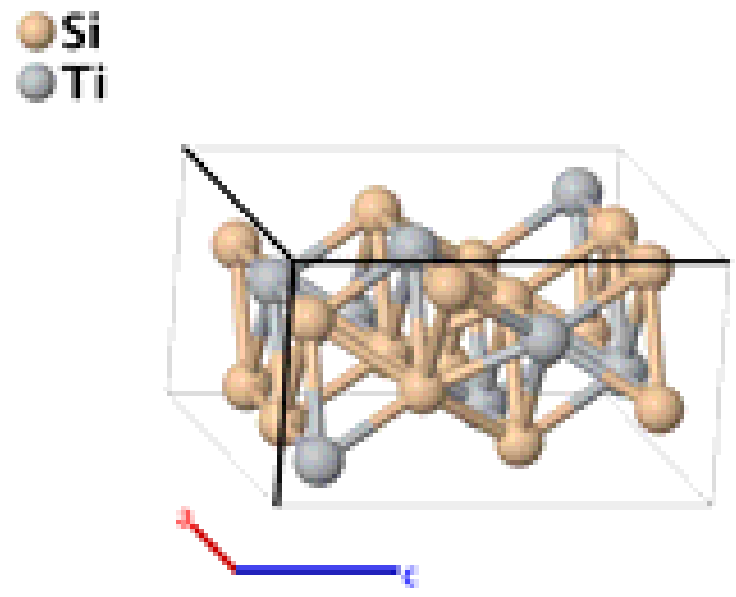

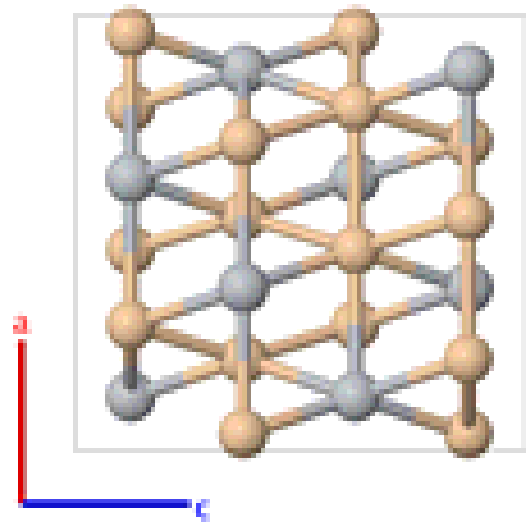

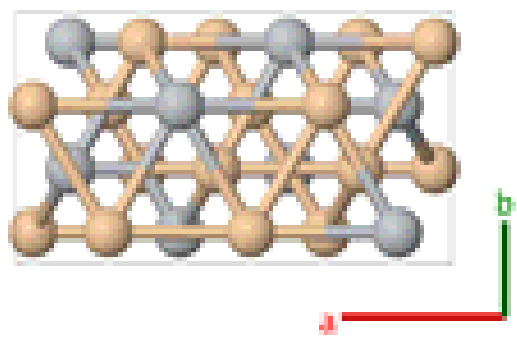

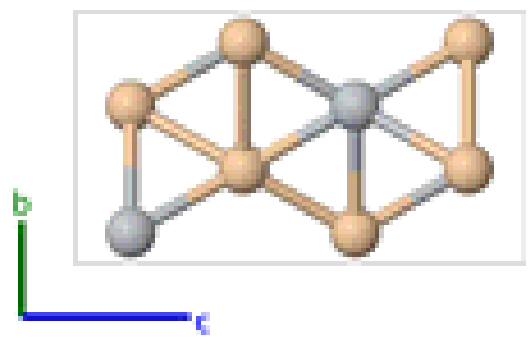

| | | |
|---|---|---|
| **Prototype** | : | $TiSi_2$ |
| **AFLOW prototype label** | : | A2B_oF24_70_e_a |
| ***Strukturbericht* designation** | : | C54 |
| **Pearson symbol** | : | oF24 |
| **Space group number** | : | 70 |
| **Space group symbol** | : | Fddd |
| **AFLOW prototype command** | : | `aflow --proto=A2B_oF24_70_e_a` `--params=`$a, b/a, c/a, x_2$ |

**Face-centered Orthorhombic primitive vectors:**

$$\mathbf{a}_1 = \frac{1}{2}\,b\,\hat{\mathbf{y}} + \frac{1}{2}\,c\,\hat{\mathbf{z}}$$
$$\mathbf{a}_2 = \frac{1}{2}\,a\,\hat{\mathbf{x}} + \frac{1}{2}\,c\,\hat{\mathbf{z}}$$
$$\mathbf{a}_3 = \frac{1}{2}\,a\,\hat{\mathbf{x}} + \frac{1}{2}\,b\,\hat{\mathbf{y}}$$

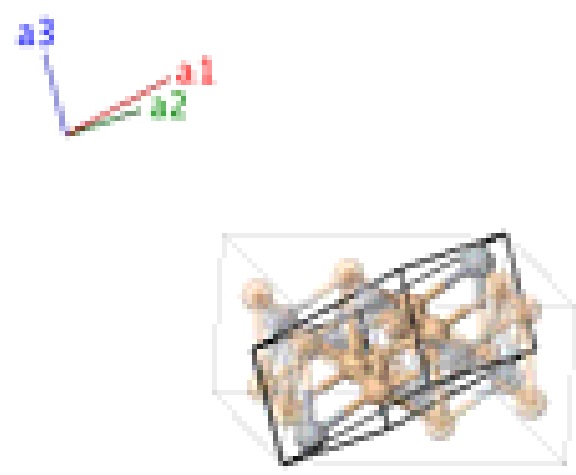

**Basis vectors:**

| | | Lattice Coordinates | | Cartesian Coordinates | Wyckoff Position | Atom Type |
|---|---|---|---|---|---|---|
| $\mathbf{B}_1$ | = | $\frac{1}{8}\,\mathbf{a}_1 + \frac{1}{8}\,\mathbf{a}_2 + \frac{1}{8}\,\mathbf{a}_3$ | = | $\frac{1}{8}\,a\,\hat{\mathbf{x}} + \frac{1}{8}\,b\,\hat{\mathbf{y}} + \frac{1}{8}\,c\,\hat{\mathbf{z}}$ | (8$a$) | Ti |
| $\mathbf{B}_2$ | = | $\frac{7}{8}\,\mathbf{a}_1 + \frac{7}{8}\,\mathbf{a}_2 + \frac{7}{8}\,\mathbf{a}_3$ | = | $\frac{7}{8}\,a\,\hat{\mathbf{x}} + \frac{7}{8}\,b\,\hat{\mathbf{y}} + \frac{7}{8}\,c\,\hat{\mathbf{z}}$ | (8$a$) | Ti |
| $\mathbf{B}_3$ | = | $\left(\frac{1}{4} - x_2\right)\,\mathbf{a}_1 + x_2\,\mathbf{a}_2 + x_2\,\mathbf{a}_3$ | = | $x_2\,a\,\hat{\mathbf{x}} + \frac{1}{8}\,b\,\hat{\mathbf{y}} + \frac{1}{8}\,c\,\hat{\mathbf{z}}$ | (16$e$) | Si |
| $\mathbf{B}_4$ | = | $x_2\,\mathbf{a}_1 + \left(\frac{1}{4} - x_2\right)\,\mathbf{a}_2 + \left(\frac{1}{4} - x_2\right)\,\mathbf{a}_3$ | = | $\left(\frac{1}{4} - x_2\right)\,a\,\hat{\mathbf{x}} + \frac{1}{8}\,b\,\hat{\mathbf{y}} + \frac{1}{8}\,c\,\hat{\mathbf{z}}$ | (16$e$) | Si |
| $\mathbf{B_5}$ | = | $\left(\frac{3}{4}+x_2\right)\mathbf{a_1} - x_2\,\mathbf{a_2} + -x_2\,\mathbf{a_3}$ | = | $-x_2\,a\,\hat{\mathbf{x}} + \frac{3}{8}\,b\,\hat{\mathbf{y}} + \frac{3}{8}\,c\,\hat{\mathbf{z}}$ | (16$e$) | Si |
| $\mathbf{B_6}$ | = | $-x_2\,\mathbf{a_1} + \left(\frac{3}{4}+x_2\right)\mathbf{a_2} + \left(\frac{3}{4}+x_2\right)\mathbf{a_3}$ | = | $\left(\frac{3}{4}+x_2\right)a\,\hat{\mathbf{x}} + \frac{3}{8}\,b\,\hat{\mathbf{y}} + \frac{3}{8}\,c\,\hat{\mathbf{z}}$ | (16$e$) | Si |

---

---

**Geometry files:**

- CIF: pp. 681
- POSCAR: pp. 681

# $\alpha$-S (A16) Structure: A_oF128_70_4h

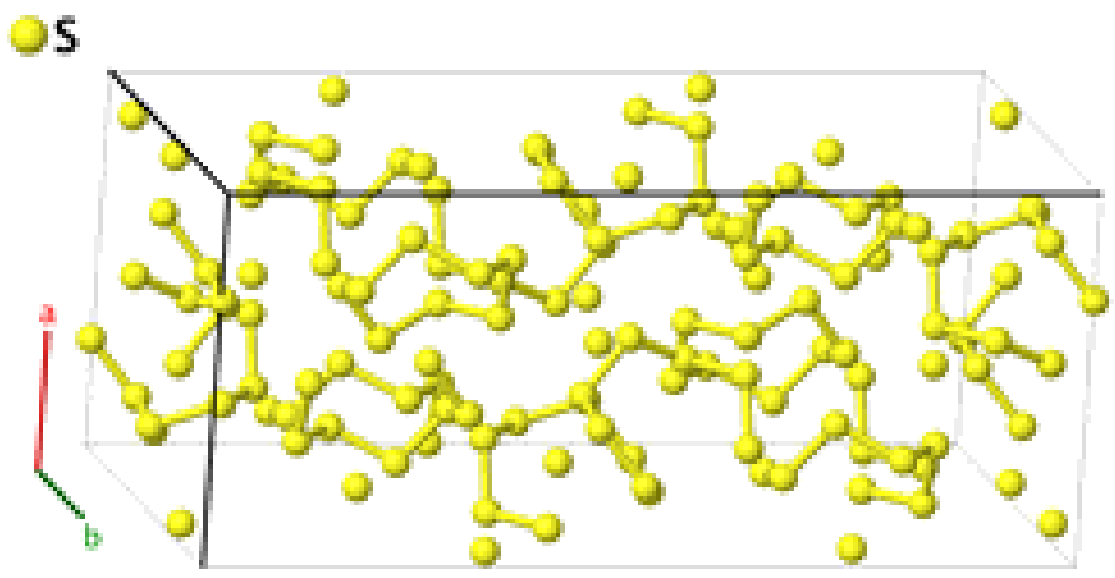

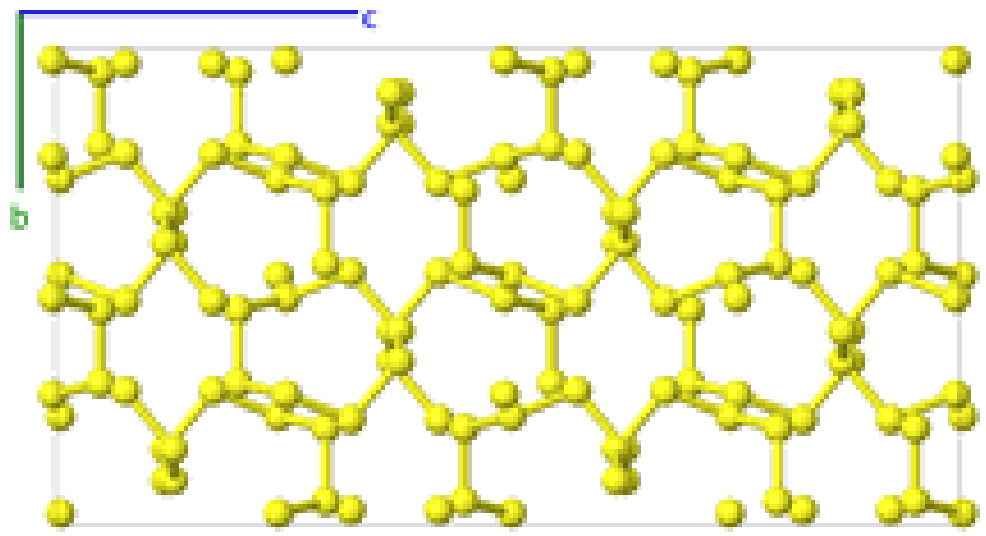

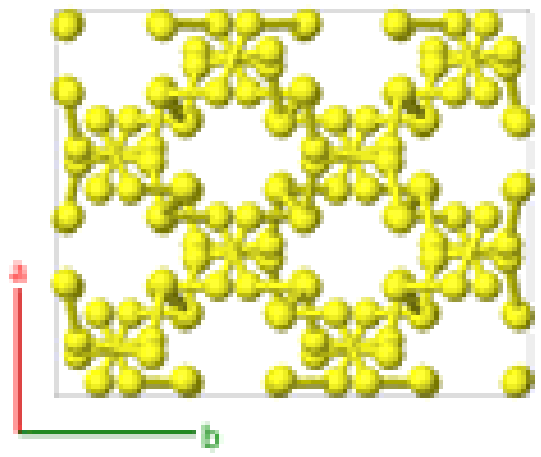

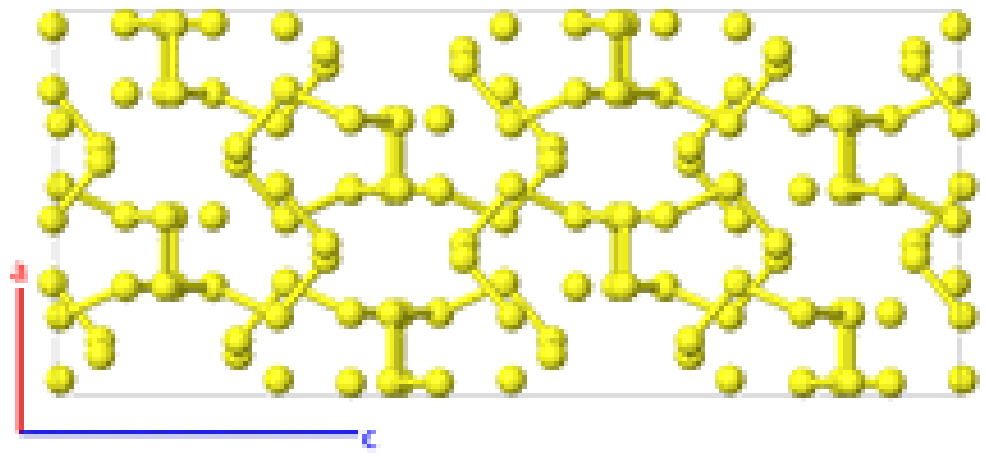

| | | |
|---|---|---|
| **Prototype** | : | $\alpha$-S |
| **AFLOW prototype label** | : | A_oF128_70_4h |
| ***Strukturbericht* designation** | : | A16 |
| **Pearson symbol** | : | oF128 |
| **Space group number** | : | 70 |
| **Space group symbol** | : | Fddd |
| **AFLOW prototype command** | : | `aflow --proto=A_oF128_70_4h` <br> `--params=`$a, b/a, c/a, x_1, y_1, z_1, x_2, y_2, z_2, x_3, y_3, z_3, x_4, y_4, z_4$ |

**Face-centered Orthorhombic primitive vectors:**

$$\mathbf{a}_1 = \frac{1}{2}\, b\, \hat{\mathbf{y}} + \frac{1}{2}\, c\, \hat{\mathbf{z}}$$
$$\mathbf{a}_2 = \frac{1}{2}\, a\, \hat{\mathbf{x}} + \frac{1}{2}\, c\, \hat{\mathbf{z}}$$
$$\mathbf{a}_3 = \frac{1}{2}\, a\, \hat{\mathbf{x}} + \frac{1}{2}\, b\, \hat{\mathbf{y}}$$

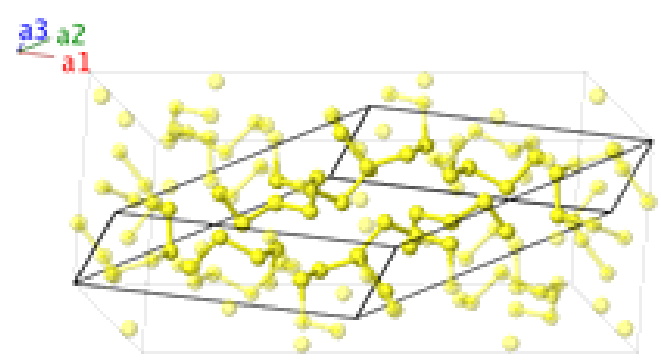

**Basis vectors:**

| | | Lattice Coordinates | | Cartesian Coordinates | Wyckoff Position | Atom Type |
|---|---|---|---|---|---|---|
| $\mathbf{B}_1$ | = | $(y_1 + z_1 - x_1)\, \mathbf{a}_1 + (z_1 + x_1 - y_1)\, \mathbf{a}_2 + (x_1 + y_1 - z_1)\, \mathbf{a}_3$ | = | $x_1\, a\, \hat{\mathbf{x}} + y_1\, b\, \hat{\mathbf{y}} + z_1\, c\, \hat{\mathbf{z}}$ | (32$h$) | S I |
| $\mathbf{B}_2$ | = | $(x_1 - y_1 + z_1)\, \mathbf{a}_1 + (y_1 + z_1 - x_1)\, \mathbf{a}_2 + \left(\frac{1}{2} - x_1 - y_1 - z_1\right) \mathbf{a}_3$ | = | $\left(\frac{1}{4} - x_1\right) a\, \hat{\mathbf{x}} + \left(\frac{1}{4} - y_1\right) b\, \hat{\mathbf{y}} + z_1\, c\, \hat{\mathbf{z}}$ | (32$h$) | S I |
| $\mathbf{B}_3$ | = | $(x_1 + y_1 - z_1)\, \mathbf{a}_1 + \left(\frac{1}{2} - x_1 - y_1 - z_1\right) \mathbf{a}_2 + (y_1 + z_1 - x_1)\, \mathbf{a}_3$ | = | $\left(\frac{1}{4} - x_1\right) a\, \hat{\mathbf{x}} + y_1\, b\, \hat{\mathbf{y}} + \left(\frac{1}{4} - z_1\right) c\, \hat{\mathbf{z}}$ | (32$h$) | S I |
| $\mathbf{B_4}$ | = | $\left(\frac{1}{2} - x_1 - y_1 - z_1\right) \mathbf{a_1} + (x_1 + y_1 - z_1) \mathbf{a_2} + (x_1 - y_1 + z_1) \mathbf{a_3}$ | = | $x_1 a \hat{\mathbf{x}} + \left(\frac{1}{4} - y_1\right) b \hat{\mathbf{y}} + \left(\frac{1}{4} - z_1\right) c \hat{\mathbf{z}}$ | (32$h$) | S I |
| $\mathbf{B_5}$ | = | $(x_1 - y_1 - z_1) \mathbf{a_1} + (y_1 - z_1 - x_1) \mathbf{a_2} + (z_1 - x_1 - y_1) \mathbf{a_3}$ | = | $-x_1 a \hat{\mathbf{x}} - y_1 b \hat{\mathbf{y}} - z_1 c \hat{\mathbf{z}}$ | (32$h$) | S I |
| $\mathbf{B_6}$ | = | $(y_1 - z_1 - x_1) \mathbf{a_1} + (x_1 - y_1 - z_1) \mathbf{a_2} + \left(\frac{1}{2} + x_1 + y_1 + z_1\right) \mathbf{a_3}$ | = | $\left(\frac{1}{4} + x_1\right) a \hat{\mathbf{x}} + \left(\frac{1}{4} + y_1\right) b \hat{\mathbf{y}} - z_1 c \hat{\mathbf{z}}$ | (32$h$) | S I |
| $\mathbf{B_7}$ | = | $(z_1 - x_1 - y_1) \mathbf{a_1} + \left(\frac{1}{2} + x_1 + y_1 + z_1\right) \mathbf{a_2} + (x_1 - y_1 - z_1) \mathbf{a_3}$ | = | $\left(\frac{1}{4} + x_1\right) a \hat{\mathbf{x}} - y_1 b \hat{\mathbf{y}} + \left(\frac{1}{4} + z_1\right) c \hat{\mathbf{z}}$ | (32$h$) | S I |
| $\mathbf{B_8}$ | = | $\left(\frac{1}{2} + x_1 + y_1 + z_1\right) \mathbf{a_1} + (z_1 - x_1 - y_1) \mathbf{a_2} + (y_1 - z_1 - x_1) \mathbf{a_3}$ | = | $-x_1 a \hat{\mathbf{x}} + \left(\frac{1}{4} + y_1\right) b \hat{\mathbf{y}} + \left(\frac{1}{4} + z_1\right) c \hat{\mathbf{z}}$ | (32$h$) | S I |
| $\mathbf{B_9}$ | = | $(y_2 + z_2 - x_2) \mathbf{a_1} + (z_2 + x_2 - y_2) \mathbf{a_2} + (x_2 + y_2 - z_2) \mathbf{a_3}$ | = | $x_2 a \hat{\mathbf{x}} + y_2 b \hat{\mathbf{y}} + z_2 c \hat{\mathbf{z}}$ | (32$h$) | S II |
| $\mathbf{B_{10}}$ | = | $(x_2 - y_2 + z_2) \mathbf{a_1} + (y_2 + z_2 - x_2) \mathbf{a_2} + \left(\frac{1}{2} - x_2 - y_2 - z_2\right) \mathbf{a_3}$ | = | $\left(\frac{1}{4} - x_2\right) a \hat{\mathbf{x}} + \left(\frac{1}{4} - y_2\right) b \hat{\mathbf{y}} + z_2 c \hat{\mathbf{z}}$ | (32$h$) | S II |
| $\mathbf{B_{11}}$ | = | $(x_2 + y_2 - z_2) \mathbf{a_1} + \left(\frac{1}{2} - x_2 - y_2 - z_2\right) \mathbf{a_2} + (y_2 + z_2 - x_2) \mathbf{a_3}$ | = | $\left(\frac{1}{4} - x_2\right) a \hat{\mathbf{x}} + y_2 b \hat{\mathbf{y}} + \left(\frac{1}{4} - z_2\right) c \hat{\mathbf{z}}$ | (32$h$) | S II |
| $\mathbf{B_{12}}$ | = | $\left(\frac{1}{2} - x_2 - y_2 - z_2\right) \mathbf{a_1} + (x_2 + y_2 - z_2) \mathbf{a_2} + (x_2 - y_2 + z_2) \mathbf{a_3}$ | = | $x_2 a \hat{\mathbf{x}} + \left(\frac{1}{4} - y_2\right) b \hat{\mathbf{y}} + \left(\frac{1}{4} - z_2\right) c \hat{\mathbf{z}}$ | (32$h$) | S II |
| $\mathbf{B_{13}}$ | = | $(x_2 - y_2 - z_2) \mathbf{a_1} + (y_2 - z_2 - x_2) \mathbf{a_2} + (z_2 - x_2 - y_2) \mathbf{a_3}$ | = | $-x_2 a \hat{\mathbf{x}} - y_2 b \hat{\mathbf{y}} - z_2 c \hat{\mathbf{z}}$ | (32$h$) | S II |
| $\mathbf{B_{14}}$ | = | $(y_2 - z_2 - x_2) \mathbf{a_1} + (x_2 - y_2 - z_2) \mathbf{a_2} + \left(\frac{1}{2} + x_2 + y_2 + z_2\right) \mathbf{a_3}$ | = | $\left(\frac{1}{4} + x_2\right) a \hat{\mathbf{x}} + \left(\frac{1}{4} + y_2\right) b \hat{\mathbf{y}} - z_2 c \hat{\mathbf{z}}$ | (32$h$) | S II |
| $\mathbf{B_{15}}$ | = | $(z_2 - x_2 - y_2) \mathbf{a_1} + \left(\frac{1}{2} + x_2 + y_2 + z_2\right) \mathbf{a_2} + (x_2 - y_2 - z_2) \mathbf{a_3}$ | = | $\left(\frac{1}{4} + x_2\right) a \hat{\mathbf{x}} - y_2 b \hat{\mathbf{y}} + \left(\frac{1}{4} + z_2\right) c \hat{\mathbf{z}}$ | (32$h$) | S II |
| $\mathbf{B_{16}}$ | = | $\left(\frac{1}{2} + x_2 + y_2 + z_2\right) \mathbf{a_1} + (z_2 - x_2 - y_2) \mathbf{a_2} + (y_2 - z_2 - x_2) \mathbf{a_3}$ | = | $-x_2 a \hat{\mathbf{x}} + \left(\frac{1}{4} + y_2\right) b \hat{\mathbf{y}} + \left(\frac{1}{4} + z_2\right) c \hat{\mathbf{z}}$ | (32$h$) | S II |
| $\mathbf{B_{17}}$ | = | $(y_3 + z_3 - x_3) \mathbf{a_1} + (z_3 + x_3 - y_3) \mathbf{a_2} + (x_3 + y_3 - z_3) \mathbf{a_3}$ | = | $x_3 a \hat{\mathbf{x}} + y_3 b \hat{\mathbf{y}} + z_3 c \hat{\mathbf{z}}$ | (32$h$) | S III |
| $\mathbf{B_{18}}$ | = | $(x_3 - y_3 + z_3) \mathbf{a_1} + (y_3 + z_3 - x_3) \mathbf{a_2} + \left(\frac{1}{2} - x_3 - y_3 - z_3\right) \mathbf{a_3}$ | = | $\left(\frac{1}{4} - x_3\right) a \hat{\mathbf{x}} + \left(\frac{1}{4} - y_3\right) b \hat{\mathbf{y}} + z_3 c \hat{\mathbf{z}}$ | (32$h$) | S III |
| $\mathbf{B_{19}}$ | = | $(x_3 + y_3 - z_3) \mathbf{a_1} + \left(\frac{1}{2} - x_3 - y_3 - z_3\right) \mathbf{a_2} + (y_3 + z_3 - x_3) \mathbf{a_3}$ | = | $\left(\frac{1}{4} - x_3\right) a \hat{\mathbf{x}} + y_3 b \hat{\mathbf{y}} + \left(\frac{1}{4} - z_3\right) c \hat{\mathbf{z}}$ | (32$h$) | S III |
| $\mathbf{B}_{20}$ | = | $\left(\frac{1}{2} - x_3 - y_3 - z_3\right)\,\mathbf{a}_1 + (x_3 + y_3 - z_3)\,\mathbf{a}_2 + (x_3 - y_3 + z_3)\,\mathbf{a}_3$ | = | $x_3\,a\,\mathbf{\hat{x}} + \left(\frac{1}{4} - y_3\right)\,b\,\mathbf{\hat{y}} + \left(\frac{1}{4} - z_3\right)\,c\,\mathbf{\hat{z}}$ | (32$h$) | S III |
| $\mathbf{B}_{21}$ | = | $(x_3 - y_3 - z_3)\,\mathbf{a}_1 + (y_3 - z_3 - x_3)\,\mathbf{a}_2 + (z_3 - x_3 - y_3)\,\mathbf{a}_3$ | = | $-x_3\,a\,\mathbf{\hat{x}} - y_3\,b\,\mathbf{\hat{y}} - z_3\,c\,\mathbf{\hat{z}}$ | (32$h$) | S III |
| $\mathbf{B}_{22}$ | = | $(y_3 - z_3 - x_3)\,\mathbf{a}_1 + (x_3 - y_3 - z_3)\,\mathbf{a}_2 + \left(\frac{1}{2} + x_3 + y_3 + z_3\right)\,\mathbf{a}_3$ | = | $\left(\frac{1}{4} + x_3\right)\,a\,\mathbf{\hat{x}} + \left(\frac{1}{4} + y_3\right)\,b\,\mathbf{\hat{y}} - z_3\,c\,\mathbf{\hat{z}}$ | (32$h$) | S III |
| $\mathbf{B}_{23}$ | = | $(z_3 - x_3 - y_3)\,\mathbf{a}_1 + \left(\frac{1}{2} + x_3 + y_3 + z_3\right)\,\mathbf{a}_2 + (x_3 - y_3 - z_3)\,\mathbf{a}_3$ | = | $\left(\frac{1}{4} + x_3\right)\,a\,\mathbf{\hat{x}} - y_3\,b\,\mathbf{\hat{y}} + \left(\frac{1}{4} + z_3\right)\,c\,\mathbf{\hat{z}}$ | (32$h$) | S III |
| $\mathbf{B}_{24}$ | = | $\left(\frac{1}{2} + x_3 + y_3 + z_3\right)\,\mathbf{a}_1 + (z_3 - x_3 - y_3)\,\mathbf{a}_2 + (y_3 - z_3 - x_3)\,\mathbf{a}_3$ | = | $-x_3\,a\,\mathbf{\hat{x}} + \left(\frac{1}{4} + y_3\right)\,b\,\mathbf{\hat{y}} + \left(\frac{1}{4} + z_3\right)\,c\,\mathbf{\hat{z}}$ | (32$h$) | S III |
| $\mathbf{B}_{25}$ | = | $(y_4 + z_4 - x_4)\,\mathbf{a}_1 + (z_4 + x_4 - y_4)\,\mathbf{a}_2 + (x_4 + y_4 - z_4)\,\mathbf{a}_3$ | = | $x_4\,a\,\mathbf{\hat{x}} + y_4\,b\,\mathbf{\hat{y}} + z_4\,c\,\mathbf{\hat{z}}$ | (32$h$) | S IV |
| $\mathbf{B}_{26}$ | = | $(x_4 - y_4 + z_4)\,\mathbf{a}_1 + (y_4 + z_4 - x_4)\,\mathbf{a}_2 + \left(\frac{1}{2} - x_4 - y_4 - z_4\right)\,\mathbf{a}_3$ | = | $\left(\frac{1}{4} - x_4\right)\,a\,\mathbf{\hat{x}} + \left(\frac{1}{4} - y_4\right)\,b\,\mathbf{\hat{y}} + z_4\,c\,\mathbf{\hat{z}}$ | (32$h$) | S IV |
| $\mathbf{B}_{27}$ | = | $(x_4 + y_4 - z_4)\,\mathbf{a}_1 + \left(\frac{1}{2} - x_4 - y_4 - z_4\right)\,\mathbf{a}_2 + (y_4 + z_4 - x_4)\,\mathbf{a}_3$ | = | $\left(\frac{1}{4} - x_4\right)\,a\,\mathbf{\hat{x}} + y_4\,b\,\mathbf{\hat{y}} + \left(\frac{1}{4} - z_4\right)\,c\,\mathbf{\hat{z}}$ | (32$h$) | S IV |
| $\mathbf{B}_{28}$ | = | $\left(\frac{1}{2} - x_4 - y_4 - z_4\right)\,\mathbf{a}_1 + (x_4 + y_4 - z_4)\,\mathbf{a}_2 + (x_4 - y_4 + z_4)\,\mathbf{a}_3$ | = | $x_4\,a\,\mathbf{\hat{x}} + \left(\frac{1}{4} - y_4\right)\,b\,\mathbf{\hat{y}} + \left(\frac{1}{4} - z_4\right)\,c\,\mathbf{\hat{z}}$ | (32$h$) | S IV |
| $\mathbf{B}_{29}$ | = | $(x_4 - y_4 - z_4)\,\mathbf{a}_1 + (y_4 - z_4 - x_4)\,\mathbf{a}_2 + (z_4 - x_4 - y_4)\,\mathbf{a}_3$ | = | $-x_4\,a\,\mathbf{\hat{x}} - y_4\,b\,\mathbf{\hat{y}} - z_4\,c\,\mathbf{\hat{z}}$ | (32$h$) | S IV |
| $\mathbf{B}_{30}$ | = | $(y_4 - z_4 - x_4)\,\mathbf{a}_1 + (x_4 - y_4 - z_4)\,\mathbf{a}_2 + \left(\frac{1}{2} + x_4 + y_4 + z_4\right)\,\mathbf{a}_3$ | = | $\left(\frac{1}{4} + x_4\right)\,a\,\mathbf{\hat{x}} + \left(\frac{1}{4} + y_4\right)\,b\,\mathbf{\hat{y}} - z_4\,c\,\mathbf{\hat{z}}$ | (32$h$) | S IV |
| $\mathbf{B}_{31}$ | = | $(z_4 - x_4 - y_4)\,\mathbf{a}_1 + \left(\frac{1}{2} + x_4 + y_4 + z_4\right)\,\mathbf{a}_2 + (x_4 - y_4 - z_4)\,\mathbf{a}_3$ | = | $\left(\frac{1}{4} + x_4\right)\,a\,\mathbf{\hat{x}} - y_4\,b\,\mathbf{\hat{y}} + \left(\frac{1}{4} + z_4\right)\,c\,\mathbf{\hat{z}}$ | (32$h$) | S IV |
| $\mathbf{B}_{32}$ | = | $\left(\frac{1}{2} + x_4 + y_4 + z_4\right)\,\mathbf{a}_1 + (z_4 - x_4 - y_4)\,\mathbf{a}_2 + (y_4 - z_4 - x_4)\,\mathbf{a}_3$ | = | $-x_4\,a\,\mathbf{\hat{x}} + \left(\frac{1}{4} + y_4\right)\,b\,\mathbf{\hat{y}} + \left(\frac{1}{4} + z_4\right)\,c\,\mathbf{\hat{z}}$ | (32$h$) | S IV |

**Geometry files:**

- CIF: pp. 682
- POSCAR: pp. 682

# $ReSi_2$ Structure: AB2_oI6_71_a_i

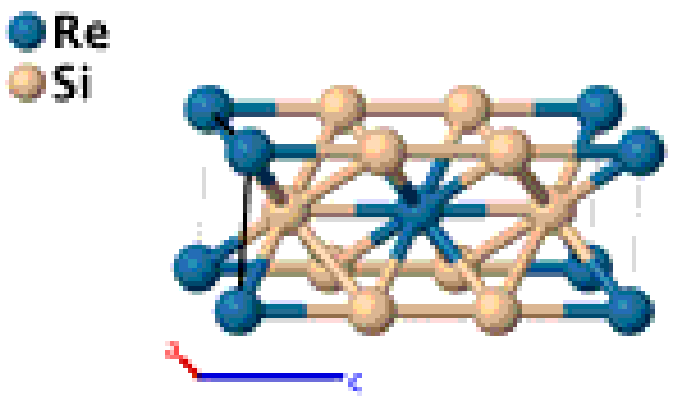

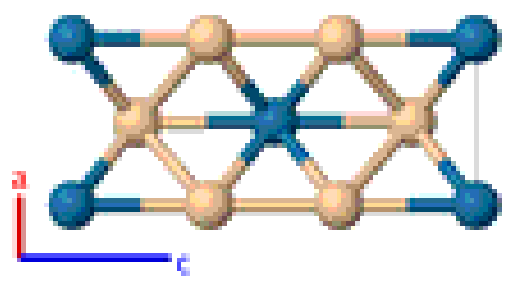

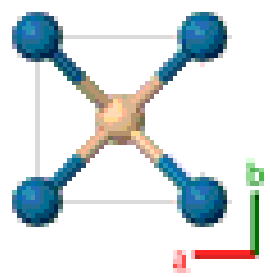

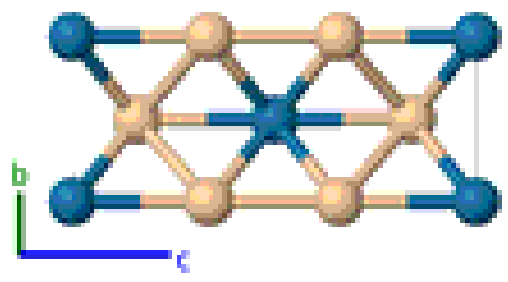

| | | |
|---|---|---|
| **Prototype** | : | $ReSi_2$ |
| **AFLOW prototype label** | : | AB2_oI6_71_a_i |
| ***Strukturbericht* designation** | : | None |
| **Pearson symbol** | : | oI6 |
| **Space group number** | : | 71 |
| **Space group symbol** | : | Immm |
| **AFLOW prototype command** | : | `aflow --proto=AB2_oI6_71_a_i --params=`$a, b/a, c/a, z_2$ |

**Body-centered Orthorhombic primitive vectors:**

$$\mathbf{a}_1 = -\frac{1}{2}\,a\,\hat{\mathbf{x}} + \frac{1}{2}\,b\,\hat{\mathbf{y}} + \frac{1}{2}\,c\,\hat{\mathbf{z}}$$
$$\mathbf{a}_2 = \frac{1}{2}\,a\,\hat{\mathbf{x}} - \frac{1}{2}\,b\,\hat{\mathbf{y}} + \frac{1}{2}\,c\,\hat{\mathbf{z}}$$
$$\mathbf{a}_3 = \frac{1}{2}\,a\,\hat{\mathbf{x}} + \frac{1}{2}\,b\,\hat{\mathbf{y}} - \frac{1}{2}\,c\,\hat{\mathbf{z}}$$

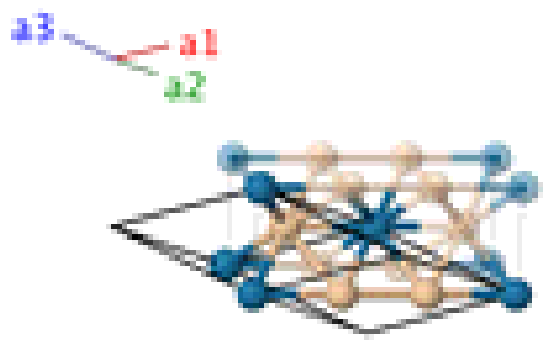

**Basis vectors:**

| | | Lattice Coordinates | | Cartesian Coordinates | Wyckoff Position | Atom Type |
|---|---|---|---|---|---|---|
| $\mathbf{B_1}$ | = | $0\,\mathbf{a}_1 + 0\,\mathbf{a}_2 + 0\,\mathbf{a}_3$ | = | $0\hat{\mathbf{x}} + 0\hat{\mathbf{y}} + 0\hat{\mathbf{z}}$ | (2$a$) | Re |
| $\mathbf{B_2}$ | = | $z_2\,\mathbf{a}_1 + z_2\,\mathbf{a}_2$ | = | $z_2\,c\,\hat{\mathbf{z}}$ | (4$i$) | Si |
| $\mathbf{B_3}$ | = | $-z_2\,\mathbf{a}_1 - z_2\,\mathbf{a}_2$ | = | $-z_2\,c\,\hat{\mathbf{z}}$ | (4$i$) | Si |

**Geometry files:**

- CIF: pp. 682
- POSCAR: pp. 683

# $MoPt_2$ Structure: AB2_oI6_71_a_g

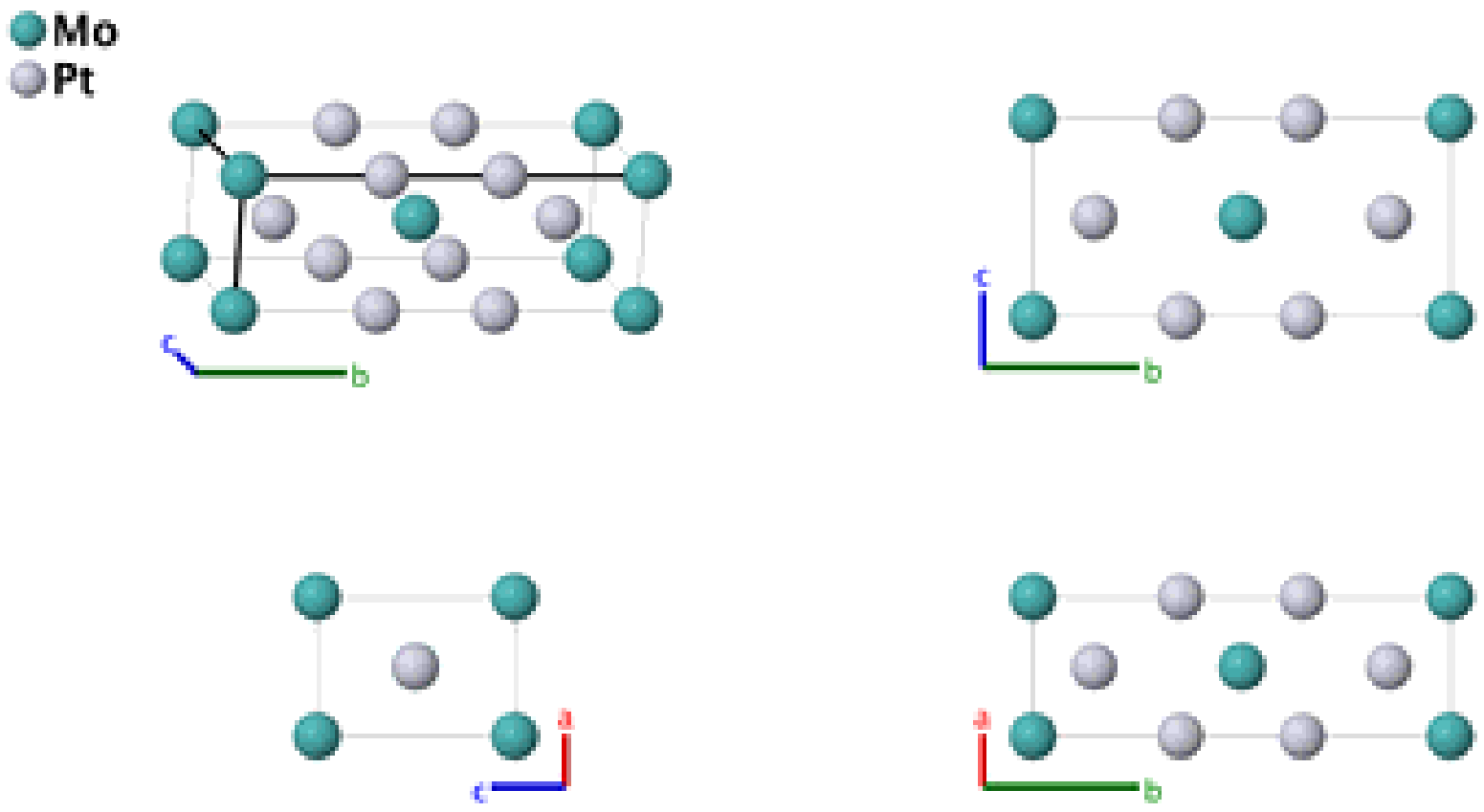

| | | |
|---|---|---|
| **Prototype** | : | $MoPt_2$ |
| **AFLOW prototype label** | : | AB2_oI6_71_a_g |
| ***Strukturbericht* designation** | : | None |
| **Pearson symbol** | : | oI6 |
| **Space group number** | : | 71 |
| **Space group symbol** | : | Immm |
| **AFLOW prototype command** | : | `aflow --proto=AB2_oI6_71_a_g`<br>`--params=`$a, b/a, c/a, y_2$ |

**Body-centered Orthorhombic primitive vectors:**

$$\mathbf{a}_1 = -\frac{1}{2}\,a\,\hat{\mathbf{x}} + \frac{1}{2}\,b\,\hat{\mathbf{y}} + \frac{1}{2}\,c\,\hat{\mathbf{z}}$$

$$\mathbf{a}_2 = \frac{1}{2}\,a\,\hat{\mathbf{x}} - \frac{1}{2}\,b\,\hat{\mathbf{y}} + \frac{1}{2}\,c\,\hat{\mathbf{z}}$$

$$\mathbf{a}_3 = \frac{1}{2}\,a\,\hat{\mathbf{x}} + \frac{1}{2}\,b\,\hat{\mathbf{y}} - \frac{1}{2}\,c\,\hat{\mathbf{z}}$$

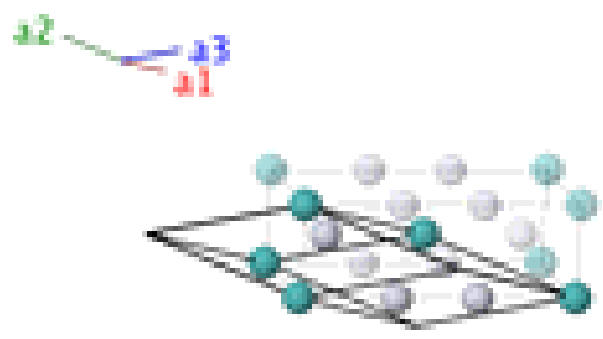

**Basis vectors:**

| | | Lattice Coordinates | | Cartesian Coordinates | Wyckoff Position | Atom Type |
|---|---|---|---|---|---|---|
| $\mathbf{B}_1$ | = | $0\,\mathbf{a}_1 + 0\,\mathbf{a}_2 + 0\,\mathbf{a}_3$ | = | $0\hat{\mathbf{x}} + 0\hat{\mathbf{y}} + 0\hat{\mathbf{z}}$ | (2$a$) | Mo |
| $\mathbf{B}_2$ | = | $y_2\,\mathbf{a}_1 + y_2\,\mathbf{a}_3$ | = | $y_2\,b\,\hat{\mathbf{y}}$ | (4$g$) | Pt |
| $\mathbf{B}_3$ | = | $-y_2\,\mathbf{a}_1 - y_2\,\mathbf{a}_3$ | = | $-y_2\,b\,\hat{\mathbf{y}}$ | (4$g$) | Pt |

**Found in:**

**Geometry files:**

- CIF: pp. 683
- POSCAR: pp. 683

# $SiS_2$ Structure: A2B_oI12_72_j_a

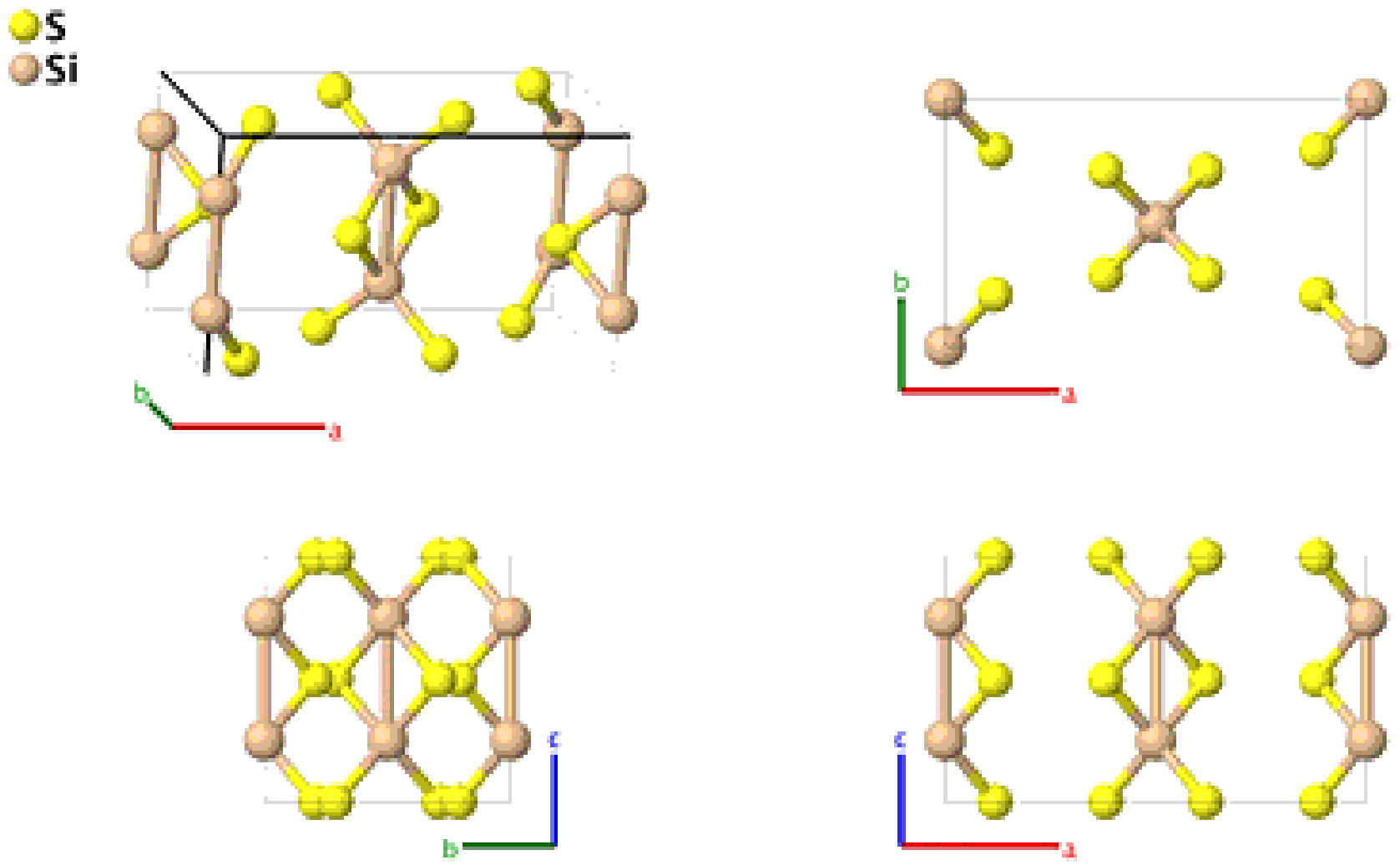

**Prototype** : $SiS_2$

**AFLOW prototype label** : A2B_oI12_72_j_a

***Strukturbericht* designation** : C42

**Pearson symbol** : oI12

**Space group number** : 72

**Space group symbol** : Ibam

**AFLOW prototype command** : `aflow --proto=A2B_oI12_72_j_a --params=`$a, b/a, c/a, x_2, y_2$

---

**Other compounds with this structure:**

- $SeS_2$

---

**Body-centered Orthorhombic primitive vectors:**

$$
\begin{aligned}
\mathbf{a}_1 &= -\frac{1}{2}\,a\,\mathbf{\hat{x}} + \frac{1}{2}\,b\,\mathbf{\hat{y}} + \frac{1}{2}\,c\,\mathbf{\hat{z}} \\
\mathbf{a}_2 &= \frac{1}{2}\,a\,\mathbf{\hat{x}} - \frac{1}{2}\,b\,\mathbf{\hat{y}} + \frac{1}{2}\,c\,\mathbf{\hat{z}} \\
\mathbf{a}_3 &= \frac{1}{2}\,a\,\mathbf{\hat{x}} + \frac{1}{2}\,b\,\mathbf{\hat{y}} - \frac{1}{2}\,c\,\mathbf{\hat{z}}
\end{aligned}
$$

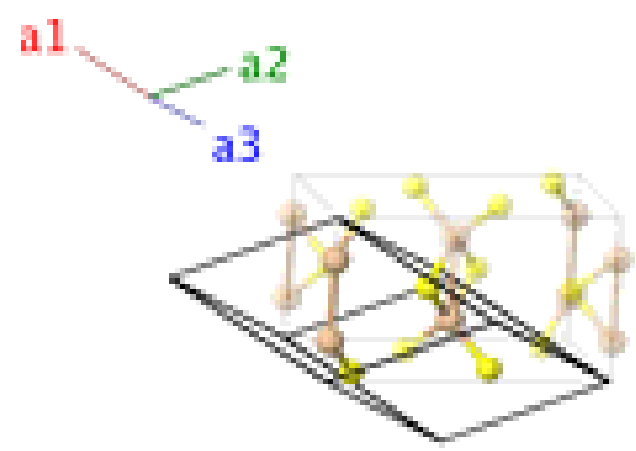

---

**Basis vectors:**

| | | Lattice Coordinates | | Cartesian Coordinates | Wyckoff Position | Atom Type |
|---|---|---|---|---|---|---|
| $\mathbf{B}_1$ | = | $\frac{1}{4}\,\mathbf{a}_1 + \frac{1}{4}\,\mathbf{a}_2$ | = | $\frac{1}{4}\,c\,\mathbf{\hat{z}}$ | (4$a$) | Si |
| $\mathbf{B}_2$ | = | $\frac{3}{4}\,\mathbf{a}_1 + \frac{3}{4}\,\mathbf{a}_2$ | = | $\frac{3}{4}\,c\,\mathbf{\hat{z}}$ | (4$a$) | Si |
| $\mathbf{B}_3$ | = | $y_2\,\mathbf{a}_1 + x_2\,\mathbf{a}_2 + (x_2 + y_2)\,\mathbf{a}_3$ | = | $x_2\,a\,\mathbf{\hat{x}} + y_2\,b\,\mathbf{\hat{y}}$ | (8$j$) | S |
| $\mathbf{B}_4$ | = | $-y_2\,\mathbf{a}_1 - x_2\,\mathbf{a}_2 - (x_2 + y_2)\,\mathbf{a}_3$ | = | $-x_2\,a\,\mathbf{\hat{x}} - y_2\,b\,\mathbf{\hat{y}}$ | (8$j$) | S |
| $\mathbf{B}_5$ | = | $\left(\frac{1}{2} + y_2\right)\,\mathbf{a}_1 + \left(\frac{1}{2} - x_2\right)\,\mathbf{a}_2 + (-x_2 + y_2)\,\mathbf{a}_3$ | = | $-x_2\,a\,\mathbf{\hat{x}}y_2\,b\,\mathbf{\hat{y}} + \frac{1}{2}\,c\,\mathbf{\hat{z}}$ | (8$j$) | S |
| $\mathbf{B}_6$ | = | $\left(\frac{1}{2} - y_2\right)\,\mathbf{a}_1 + \left(\frac{1}{2} + x_2\right)\,\mathbf{a}_2 + (x_2 - y_2)\,\mathbf{a}_3$ | = | $x_2\,a\,\mathbf{\hat{x}} - y_2\,b\,\mathbf{\hat{y}} + \frac{1}{2}\,c\,\mathbf{\hat{z}}$ | (8$j$) | S |

**Geometry files:**

- CIF: pp. 683
- POSCAR: pp. 684

# $BPO_4$ ($H0_7$) Structure: AB4C_tI12_82_c_g_a

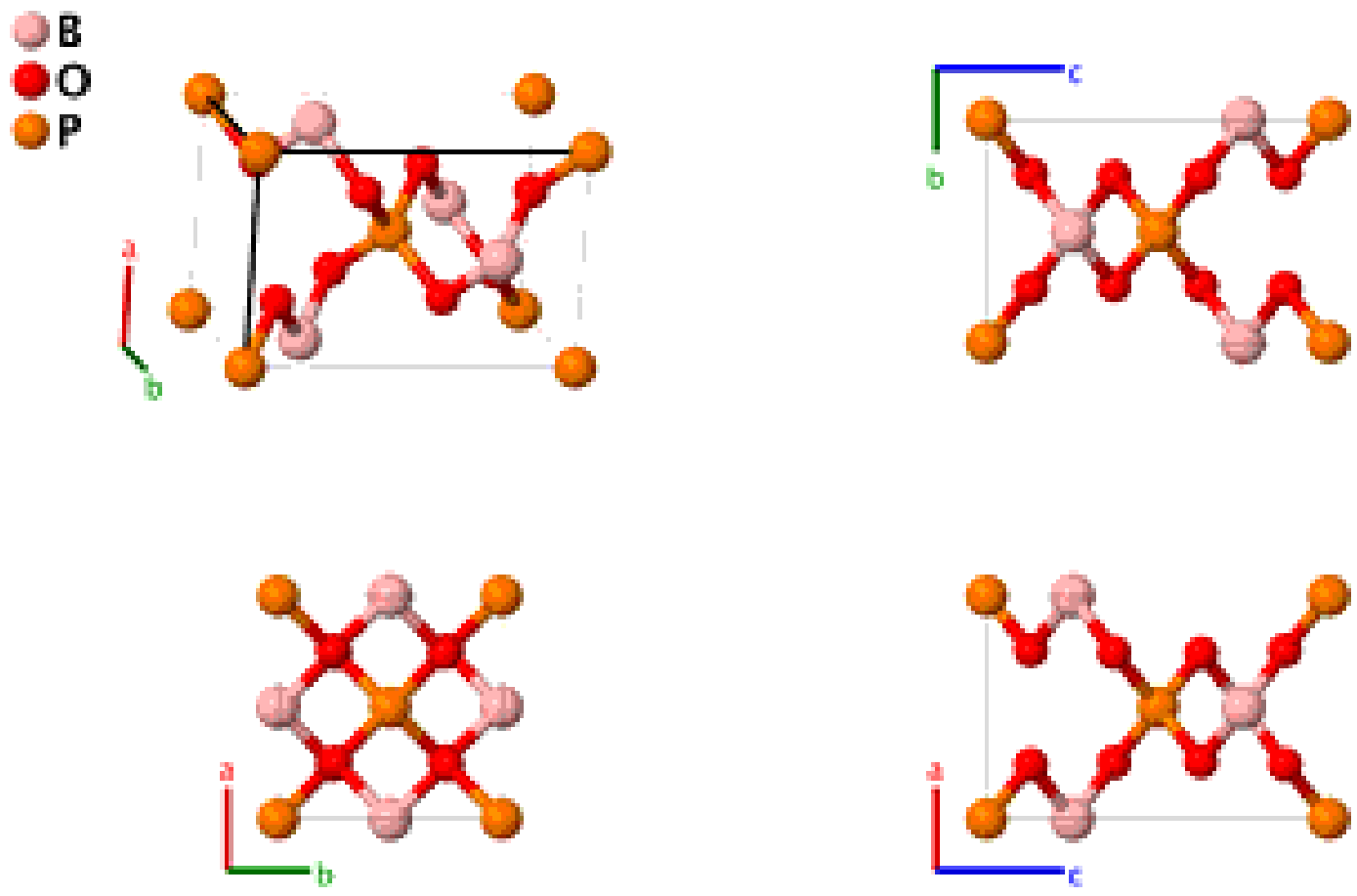

| | | |
|---|---|---|
| **Prototype** | : | $BPO_4$ |
| **AFLOW prototype label** | : | AB4C_tI12_82_c_g_a |
| ***Strukturbericht* designation** | : | $H0_7$ |
| **Pearson symbol** | : | tI12 |
| **Space group number** | : | 82 |
| **Space group symbol** | : | $I\bar{4}$ |
| **AFLOW prototype command** | : | `aflow --proto=AB4C_tI12_82_c_g_a`<br>`--params=`$a, c/a, x_3, y_3, z_3$ |

---

**Body-centered Tetragonal primitive vectors:**

$$\mathbf{a}_1 = -\frac{1}{2}\,a\,\hat{\mathbf{x}} + \frac{1}{2}\,a\,\hat{\mathbf{y}} + \frac{1}{2}\,c\,\hat{\mathbf{z}}$$
$$\mathbf{a}_2 = \frac{1}{2}\,a\,\hat{\mathbf{x}} - \frac{1}{2}\,a\,\hat{\mathbf{y}} + \frac{1}{2}\,c\,\hat{\mathbf{z}}$$
$$\mathbf{a}_3 = \frac{1}{2}\,a\,\hat{\mathbf{x}} + \frac{1}{2}\,a\,\hat{\mathbf{y}} - \frac{1}{2}\,c\,\hat{\mathbf{z}}$$

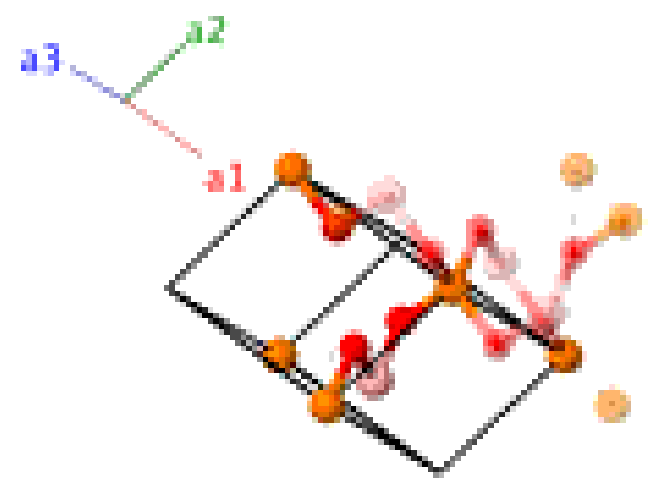

---

**Basis vectors:**

| | | Lattice Coordinates | | Cartesian Coordinates | Wyckoff Position | Atom Type |
|---|---|---|---|---|---|---|
| $\mathbf{B}_1$ | = | $0\,\mathbf{a}_1 + 0\,\mathbf{a}_2 + 0\,\mathbf{a}_3$ | = | $0\hat{\mathbf{x}} + 0\hat{\mathbf{y}} + 0\hat{\mathbf{z}}$ | (2$a$) | P |
| $\mathbf{B}_2$ | = | $\frac{3}{4}\,\mathbf{a}_1 + \frac{1}{4}\,\mathbf{a}_2 + \frac{1}{2}\,\mathbf{a}_3$ | = | $\frac{1}{2}\,a\,\hat{\mathbf{y}} + \frac{1}{4}\,c\,\hat{\mathbf{z}}$ | (2$c$) | B |
| $\mathbf{B}_3$ | = | $(y_3 + z_3)\,\mathbf{a}_1 + (z_3 + x_3)\,\mathbf{a}_2 + (x_3 + y_3)\,\mathbf{a}_3$ | = | $x_3\,a\,\hat{\mathbf{x}} + y_3\,a\,\hat{\mathbf{y}} + z_3\,c\,\hat{\mathbf{z}}$ | (8$g$) | O |
| $\mathbf{B}_4$ | = | $(z_3 - y_3)\,\mathbf{a}_1 + (z_3 - x_3)\,\mathbf{a}_2 - (x_3 + y_3)\,\mathbf{a}_3$ | = | $-x_3\,a\,\hat{\mathbf{x}} - y_3\,a\,\hat{\mathbf{y}} + z_3\,c\,\hat{\mathbf{z}}$ | (8$g$) | O |
| $\mathbf{B}_5$ | = | $-(z_3 + x_3)\,\mathbf{a}_1 + (y_3 - z_3)\,\mathbf{a}_2 + (y_3 - x_3)\,\mathbf{a}_3$ | = | $y_3\,a\,\hat{\mathbf{x}} - x_3\,a\,\hat{\mathbf{y}} - z_3\,c\,\hat{\mathbf{z}}$ | (8$g$) | O |
| $\mathbf{B}_6$ | = | $(x_3 - z_3)\,\mathbf{a}_1 - (y_3 + z_3)\,\mathbf{a}_2 + (x_3 - y_3)\,\mathbf{a}_3$ | = | $-y_3\,a\,\hat{\mathbf{x}} + x_3\,a\,\hat{\mathbf{y}} - z_3\,c\,\hat{\mathbf{z}}$ | (8$g$) | O |

---

---

**Geometry files:**

- CIF: pp. 684
- POSCAR: pp. 684

# $CdAl_2S_4$ (E3) Structure: A2BC4_tI14_82_bc_a_g

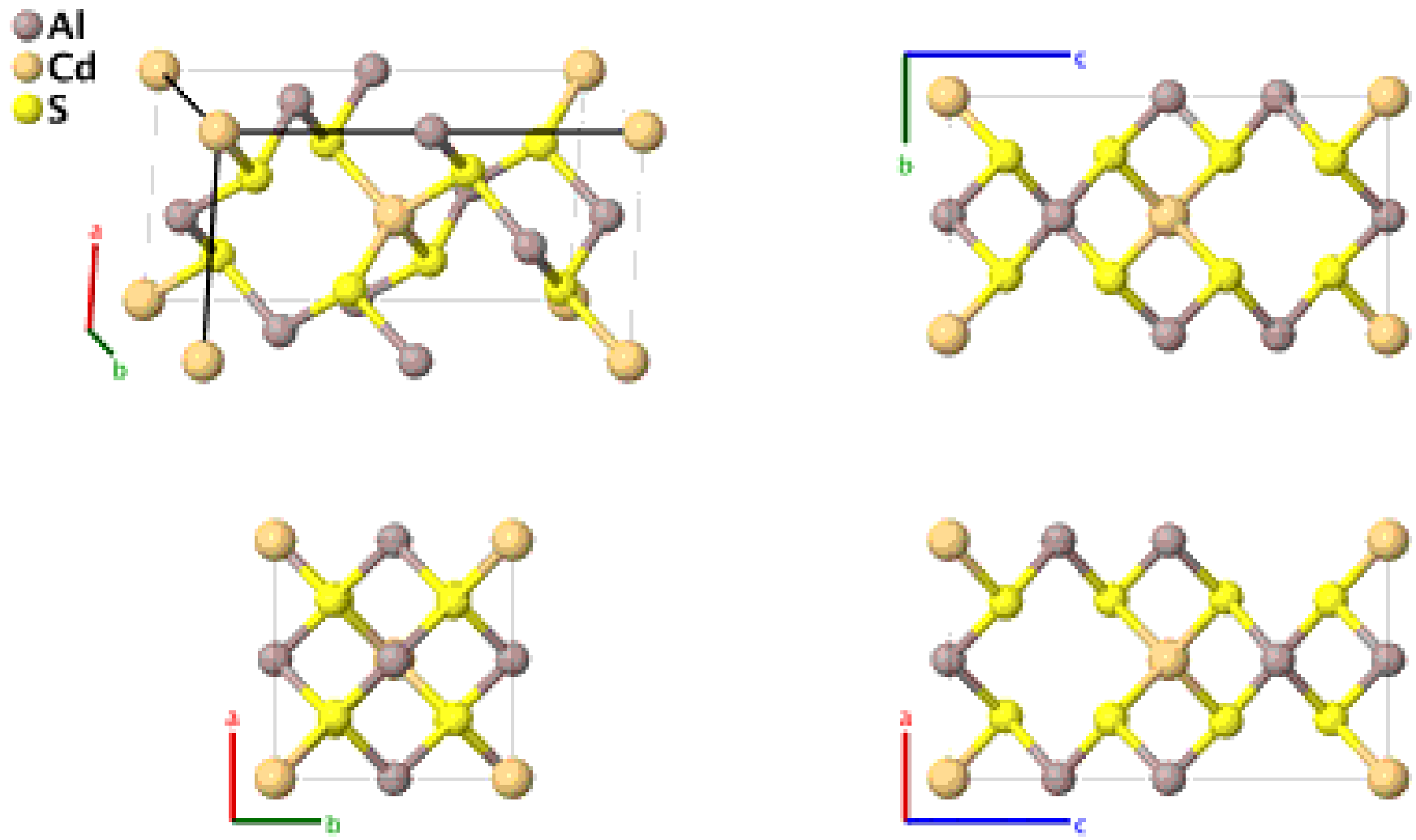

| | | |
|---|---|---|
| **Prototype** | : | $CdAl_2S_4$ |
| **AFLOW prototype label** | : | A2BC4_tI14_82_bc_a_g |
| ***Strukturbericht* designation** | : | E3 |
| **Pearson symbol** | : | tI14 |
| **Space group number** | : | 82 |
| **Space group symbol** | : | $I\bar{4}$ |
| **AFLOW prototype command** | : | `aflow --proto=A2BC4_tI14_82_bc_a_g` <br> `--params=`$a, c/a, x_4, y_4, z_4$ |

**Other compounds with this structure:**

- $CoGa_2S_4$, $FeGa_2S_4$, $HgGa_2Te_4$, $ZnGa_2S_4$, $HgAl_2S_4$, numerous others.

- When $c = 2a$ and $x = y = 1/4$, and $z = 1/8$ the atoms are on the sites of the diamond (A4) structure, but of course there are defects. Removing the Al-I (2b) atom transforms this to the $BPO_4$ ($H0_7$) structure.

**Body-centered Tetragonal primitive vectors:**

$$\mathbf{a}_1 = -\frac{1}{2}\,a\,\hat{\mathbf{x}} + \frac{1}{2}\,a\,\hat{\mathbf{y}} + \frac{1}{2}\,c\,\hat{\mathbf{z}}$$
$$\mathbf{a}_2 = \frac{1}{2}\,a\,\hat{\mathbf{x}} - \frac{1}{2}\,a\,\hat{\mathbf{y}} + \frac{1}{2}\,c\,\hat{\mathbf{z}}$$
$$\mathbf{a}_3 = \frac{1}{2}\,a\,\hat{\mathbf{x}} + \frac{1}{2}\,a\,\hat{\mathbf{y}} - \frac{1}{2}\,c\,\hat{\mathbf{z}}$$

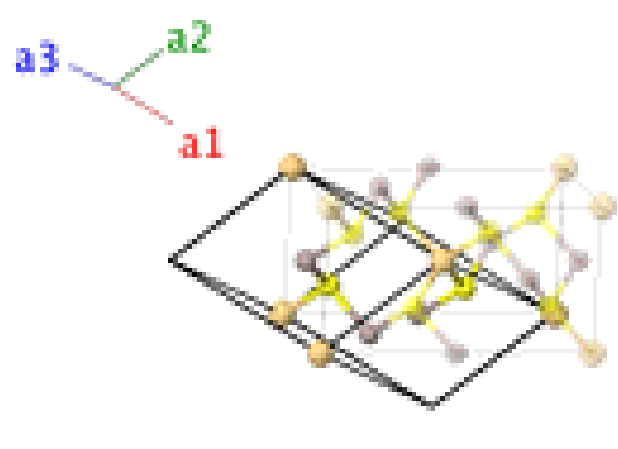

**Basis vectors:**

| | | Lattice Coordinates | | Cartesian Coordinates | Wyckoff Position | Atom Type |
|---|---|---|---|---|---|---|
| $\mathbf{B}_1$ | = | $0\,\mathbf{a}_1 + 0\,\mathbf{a}_2 + 0\,\mathbf{a}_3$ | = | $0\hat{\mathbf{x}} + 0\hat{\mathbf{y}} + 0\hat{\mathbf{z}}$ | (2$a$) | Cd |
| $\mathbf{B}_2$ | = | $\frac{1}{2}\,\mathbf{a}_1 + \frac{1}{2}\,\mathbf{a}_2$ | = | $\frac{1}{2}\,c\,\hat{\mathbf{z}}$ | (2$b$) | Al I |
| $\mathbf{B_3}$ | = | $\frac{3}{4}\,\mathbf{a_1} + \frac{1}{4}\,\mathbf{a_2} + \frac{1}{2}\,\mathbf{a_3}$ | = | $\frac{1}{2}\,a\,\mathbf{\hat{y}} + \frac{1}{4}\,c\,\mathbf{\hat{z}}$ | (2c) | Al II |
| $\mathbf{B_4}$ | = | $(y_4 + z_4)\,\mathbf{a_1} + (z_4 + x_4)\,\mathbf{a_2} + (x_4 + y_4)\,\mathbf{a_3}$ | = | $x_4\,a\,\mathbf{\hat{x}} + y_4\,a\,\mathbf{\hat{y}} + z_4\,c\,\mathbf{\hat{z}}$ | (8g) | S |
| $\mathbf{B_5}$ | = | $(z_4 - y_4)\,\mathbf{a_1} + (z_4 - x_4)\,\mathbf{a_2} - (x_4 + y_4)\,\mathbf{a_3}$ | = | $-x_4\,a\,\mathbf{\hat{x}} - y_4\,a\,\mathbf{\hat{y}} + z_4\,c\,\mathbf{\hat{z}}$ | (8g) | S |
| $\mathbf{B_6}$ | = | $-(z_4 + x_4)\,\mathbf{a_1} + (y_4 - z_4)\,\mathbf{a_2} + (y_4 - x_4)\,\mathbf{a_3}$ | = | $y_4\,a\,\mathbf{\hat{x}} - x_4\,a\,\mathbf{\hat{y}} - z_4\,c\,\mathbf{\hat{z}}$ | (8g) | S |
| $\mathbf{B_7}$ | = | $(x_4 - z_4)\,\mathbf{a_1} - (y_4 + z_4)\,\mathbf{a_2} + (x_4 - y_4)\,\mathbf{a_3}$ | = | $-y_4\,a\,\mathbf{\hat{x}} + x_4\,a\,\mathbf{\hat{y}} - z_4\,c\,\mathbf{\hat{z}}$ | (8g) | S |

**Geometry files:**

- CIF: pp. 684
- POSCAR: pp. 684

# PdS (B34) Structure: AB_tP16_84_cej_k

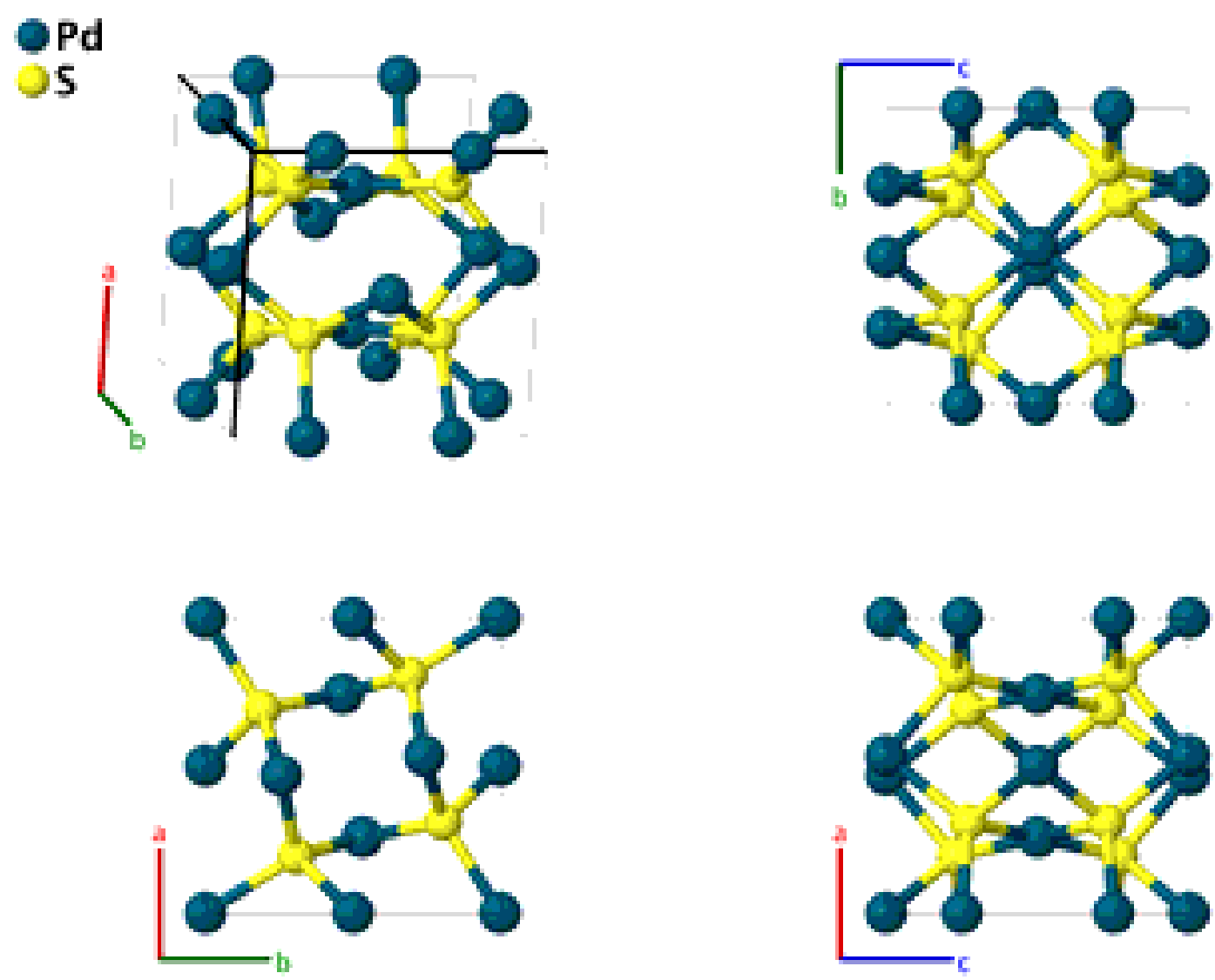

| | | |
|---|---|---|
| **Prototype** | : | PdS |
| **AFLOW prototype label** | : | AB_tP16_84_cej_k |
| ***Strukturbericht* designation** | : | B34 |
| **Pearson symbol** | : | tP16 |
| **Space group number** | : | 84 |
| **Space group symbol** | : | $P4_2/m$ |
| **AFLOW prototype command** | : | `aflow --proto=AB_tP16_84_cej_k`<br>`--params=`$a, c/a, x_3, y_3, x_4, y_4, z_4$ |

**Simple Tetragonal primitive vectors:**

$$\mathbf{a}_1 = a\,\hat{\mathbf{x}}$$
$$\mathbf{a}_2 = a\,\hat{\mathbf{y}}$$
$$\mathbf{a}_3 = c\,\hat{\mathbf{z}}$$

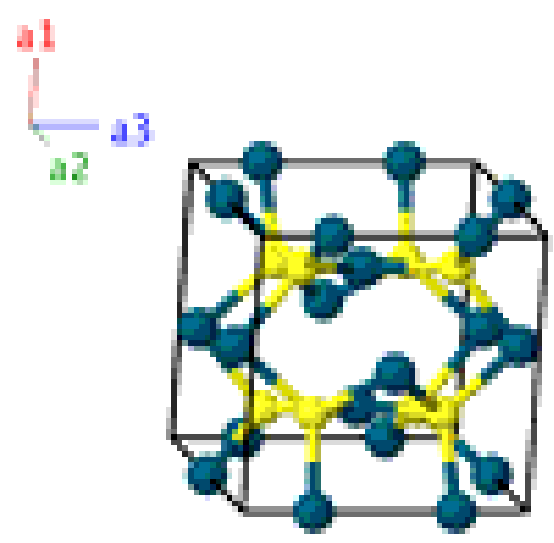

**Basis vectors:**

| | | Lattice Coordinates | | Cartesian Coordinates | Wyckoff Position | Atom Type |
|---|---|---|---|---|---|---|
| $\mathbf{B}_1$ | = | $\frac{1}{2}\,\mathbf{a}_2$ | = | $\frac{1}{2}\,a\,\hat{\mathbf{y}}$ | (2c) | Pd I |
| $\mathbf{B}_2$ | = | $\frac{1}{2}\,\mathbf{a}_1 + \frac{1}{2}\,\mathbf{a}_3$ | = | $\frac{1}{2}\,a\,\hat{\mathbf{x}} + \frac{1}{2}\,c\,\hat{\mathbf{z}}$ | (2c) | Pd I |
| $\mathbf{B}_3$ | = | $\frac{1}{4}\,\mathbf{a}_3$ | = | $\frac{1}{4}\,c\,\hat{\mathbf{z}}$ | (2e) | Pd II |
| $\mathbf{B}_4$ | = | $\frac{3}{4}\,\mathbf{a}_3$ | = | $\frac{3}{4}\,c\,\hat{\mathbf{z}}$ | (2e) | Pd II |
| $\mathbf{B}_5$ | = | $x_3\,\mathbf{a}_1 + y_3\,\mathbf{a}_2$ | = | $x_3\,a\,\hat{\mathbf{x}} + y_3\,a\,\hat{\mathbf{y}}$ | (4j) | Pd III |
| $\mathbf{B}_6$ | = | $-x_3\,\mathbf{a}_1 - y_3\,\mathbf{a}_2$ | = | $-x_3\,a\,\hat{\mathbf{x}} - y_3\,a\,\hat{\mathbf{y}}$ | (4j) | Pd III |
| $\mathbf{B}_{7}$ | = | $-y_3\,\mathbf{a}_1 + x_3\,\mathbf{a}_2 + \frac{1}{2}\,\mathbf{a}_3$ | = | $-y_3\,a\,\hat{\mathbf{x}} + x_3\,a\,\hat{\mathbf{y}} + \frac{1}{2}\,c\,\hat{\mathbf{z}}$ | $(4j)$ | Pd III |
| $\mathbf{B}_{8}$ | = | $y_3\,\mathbf{a}_1 - x_3\,\mathbf{a}_2 + \frac{1}{2}\,\mathbf{a}_3$ | = | $y_3\,a\,\hat{\mathbf{x}} - x_3\,a\,\hat{\mathbf{y}} + \frac{1}{2}\,c\,\hat{\mathbf{z}}$ | $(4j)$ | Pd III |
| $\mathbf{B}_{9}$ | = | $x_4\,\mathbf{a}_1 + y_4\,\mathbf{a}_2 + z_4\,\mathbf{a}_3$ | = | $x_4\,a\,\hat{\mathbf{x}} + y_4\,a\,\hat{\mathbf{y}} + z_4\,c\,\hat{\mathbf{z}}$ | $(8k)$ | S |
| $\mathbf{B}_{10}$ | = | $-x_4\,\mathbf{a}_1 - y_4\,\mathbf{a}_2 + z_4\,\mathbf{a}_3$ | = | $-x_4\,a\,\hat{\mathbf{x}} - y_4\,a\,\hat{\mathbf{y}} + z_4\,c\,\hat{\mathbf{z}}$ | $(8k)$ | S |
| $\mathbf{B}_{11}$ | = | $-y_4\,\mathbf{a}_1 + x_4\,\mathbf{a}_2 + \left(\frac{1}{2} + z_4\right)\,\mathbf{a}_3$ | = | $-y_4\,a\,\hat{\mathbf{x}} + x_4\,a\,\hat{\mathbf{y}} + \left(\frac{1}{2} + z_4\right)\,c\,\hat{\mathbf{z}}$ | $(8k)$ | S |
| $\mathbf{B}_{12}$ | = | $y_4\,\mathbf{a}_1 - x_4\,\mathbf{a}_2 + \left(\frac{1}{2} + z_4\right)\,\mathbf{a}_3$ | = | $y_4\,a\,\hat{\mathbf{x}} - x_4\,a\,\hat{\mathbf{y}} + \left(\frac{1}{2} + z_4\right)\,c\,\hat{\mathbf{z}}$ | $(8k)$ | S |
| $\mathbf{B}_{13}$ | = | $-x_4\,\mathbf{a}_1 - y_4\,\mathbf{a}_2 - z_4\,\mathbf{a}_3$ | = | $-x_4\,a\,\hat{\mathbf{x}} - y_4\,a\,\hat{\mathbf{y}} - z_4\,c\,\hat{\mathbf{z}}$ | $(8k)$ | S |
| $\mathbf{B}_{14}$ | = | $x_4\,\mathbf{a}_1 + y_4\,\mathbf{a}_2 - z_4\,\mathbf{a}_3$ | = | $x_4\,a\,\hat{\mathbf{x}} + y_4\,a\,\hat{\mathbf{y}} - z_4\,c\,\hat{\mathbf{z}}$ | $(8k)$ | S |
| $\mathbf{B}_{15}$ | = | $y_4\,\mathbf{a}_1 - x_4\,\mathbf{a}_2 + \left(\frac{1}{2} - z_4\right)\,\mathbf{a}_3$ | = | $y_4\,a\,\hat{\mathbf{x}} - x_4\,a\,\hat{\mathbf{y}} + \left(\frac{1}{2} - z_4\right)\,c\,\hat{\mathbf{z}}$ | $(8k)$ | S |
| $\mathbf{B}_{16}$ | = | $-y_4\,\mathbf{a}_1 + x_4\,\mathbf{a}_2 + \left(\frac{1}{2} - z_4\right)\,\mathbf{a}_3$ | = | $-y_4\,a\,\hat{\mathbf{x}} + x_4\,a\,\hat{\mathbf{y}} + \left(\frac{1}{2} - z_4\right)\,c\,\hat{\mathbf{z}}$ | $(8k)$ | S |

---

---

**Geometry files:**

- CIF: pp. 685
- POSCAR: pp. 685

# $Ti_5Te_4$ Structure: A4B5_tI18_87_h_ah

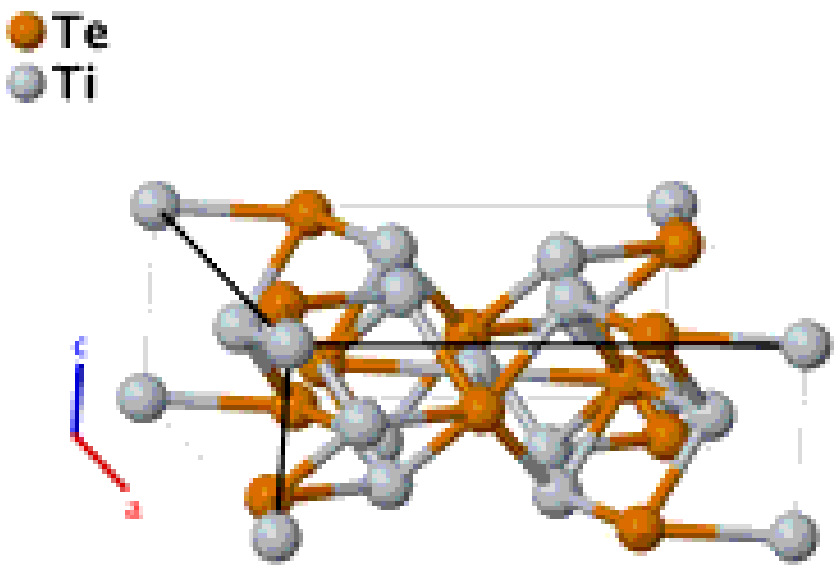

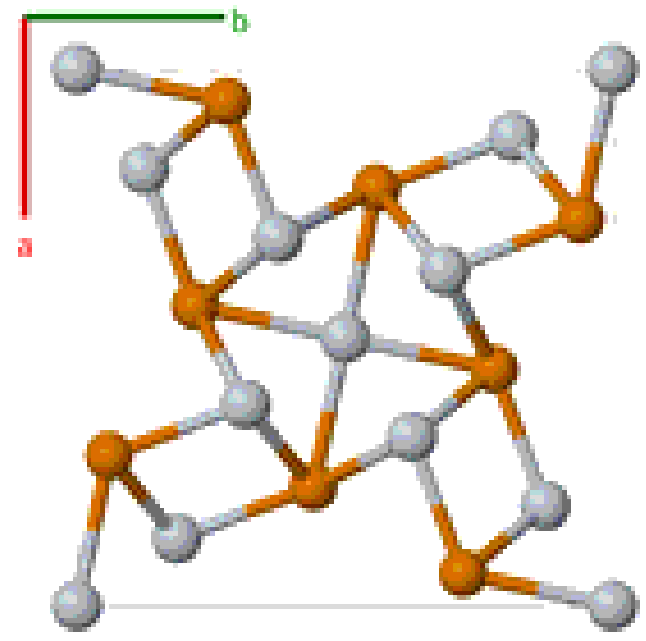

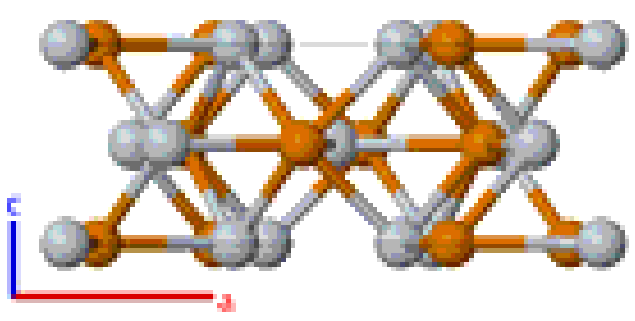

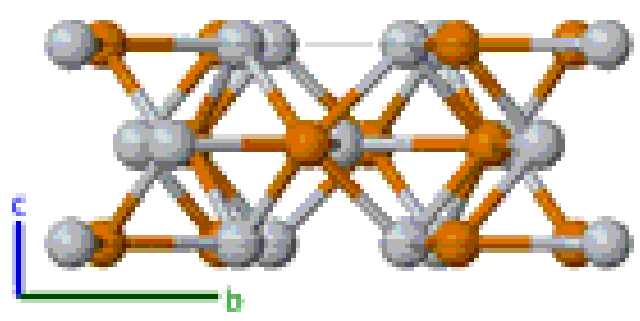

| | | |
|---|---|---|
| **Prototype** | : | $Ti_5Te_4$ |
| **AFLOW prototype label** | : | A4B5_tI18_87_h_ah |
| ***Strukturbericht* designation** | : | None |
| **Pearson symbol** | : | tI18 |
| **Space group number** | : | 87 |
| **Space group symbol** | : | I4/m |
| **AFLOW prototype command** | : | `aflow --proto=A4B5_tI18_87_h_ah` <br> `--params=`$a, c/a, x_2, y_2, x_3, y_3$ |

**Body-centered Tetragonal primitive vectors:**

$$\mathbf{a}_1 = -\frac{1}{2}\,a\,\mathbf{\hat{x}} + \frac{1}{2}\,a\,\mathbf{\hat{y}} + \frac{1}{2}\,c\,\mathbf{\hat{z}}$$
$$\mathbf{a}_2 = \frac{1}{2}\,a\,\mathbf{\hat{x}} - \frac{1}{2}\,a\,\mathbf{\hat{y}} + \frac{1}{2}\,c\,\mathbf{\hat{z}}$$
$$\mathbf{a}_3 = \frac{1}{2}\,a\,\mathbf{\hat{x}} + \frac{1}{2}\,a\,\mathbf{\hat{y}} - \frac{1}{2}\,c\,\mathbf{\hat{z}}$$

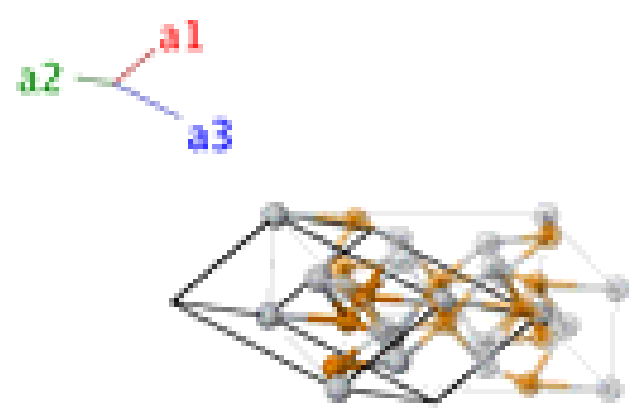

**Basis vectors:**

| | | Lattice Coordinates | | Cartesian Coordinates | Wyckoff Position | Atom Type |
|---|---|---|---|---|---|---|
| $\mathbf{B}_1$ | = | $0\,\mathbf{a}_1 + 0\,\mathbf{a}_2 + 0\,\mathbf{a}_3$ | = | $0\mathbf{\hat{x}} + 0\mathbf{\hat{y}} + 0\mathbf{\hat{z}}$ | (2$a$) | Ti I |
| $\mathbf{B}_2$ | = | $y_2\,\mathbf{a}_1 + x_2\,\mathbf{a}_2 + (x_2 + y_2)\,\mathbf{a}_3$ | = | $x_2\,a\,\mathbf{\hat{x}} + y_2\,a\,\mathbf{\hat{y}}$ | (8$h$) | Te |
| $\mathbf{B}_3$ | = | $-y_2\,\mathbf{a}_1 - x_2\,\mathbf{a}_2 - (x_2 + y_2)\,\mathbf{a}_3$ | = | $-x_2\,a\,\hat{\mathbf{x}} - y_2\,a\,\hat{\mathbf{y}}$ | (8$h$) | Te |
| $\mathbf{B}_4$ | = | $x_2\,\mathbf{a}_1 - y_2\,\mathbf{a}_2 + (x_2 - y_2)\,\mathbf{a}_3$ | = | $-y_2\,a\,\hat{\mathbf{x}} + x_2\,a\,\hat{\mathbf{y}}$ | (8$h$) | Te |
| $\mathbf{B}_5$ | = | $-x_2\,\mathbf{a}_1 + y_2\,\mathbf{a}_2 + (y_2 - x_2)\,\mathbf{a}_3$ | = | $y_2\,a\,\hat{\mathbf{x}} - x_2\,a\,\hat{\mathbf{y}}$ | (8$h$) | Te |
| $\mathbf{B}_6$ | = | $y_3\,\mathbf{a}_1 + x_3\,\mathbf{a}_2 + (x_3 + y_3)\,\mathbf{a}_3$ | = | $x_3\,a\,\hat{\mathbf{x}} + y_3\,a\,\hat{\mathbf{y}}$ | (8$h$) | Ti II |
| $\mathbf{B}_7$ | = | $-y_3\,\mathbf{a}_1 - x_3\,\mathbf{a}_2 - (x_3 + y_3)\,\mathbf{a}_3$ | = | $-x_3\,a\,\hat{\mathbf{x}} - y_3\,a\,\hat{\mathbf{y}}$ | (8$h$) | Ti II |
| $\mathbf{B}_8$ | = | $x_3\,\mathbf{a}_1 - y_3\,\mathbf{a}_2 + (x_3 - y_3)\,\mathbf{a}_3$ | = | $-y_3\,a\,\hat{\mathbf{x}} + x_3\,a\,\hat{\mathbf{y}}$ | (8$h$) | Ti II |
| $\mathbf{B}_9$ | = | $-x_3\,\mathbf{a}_1 + y_3\,\mathbf{a}_2 + (y_3 - x_3)\,\mathbf{a}_3$ | = | $y_3\,a\,\hat{\mathbf{x}} - x_3\,a\,\hat{\mathbf{y}}$ | (8$h$) | Ti II |

---

**Found in:**

---

**Geometry files:**

- CIF: pp. 685
- POSCAR: pp. 685

# $Ni_4Mo$ ($D1_a$) Structure: AB4_tI10_87_a_h

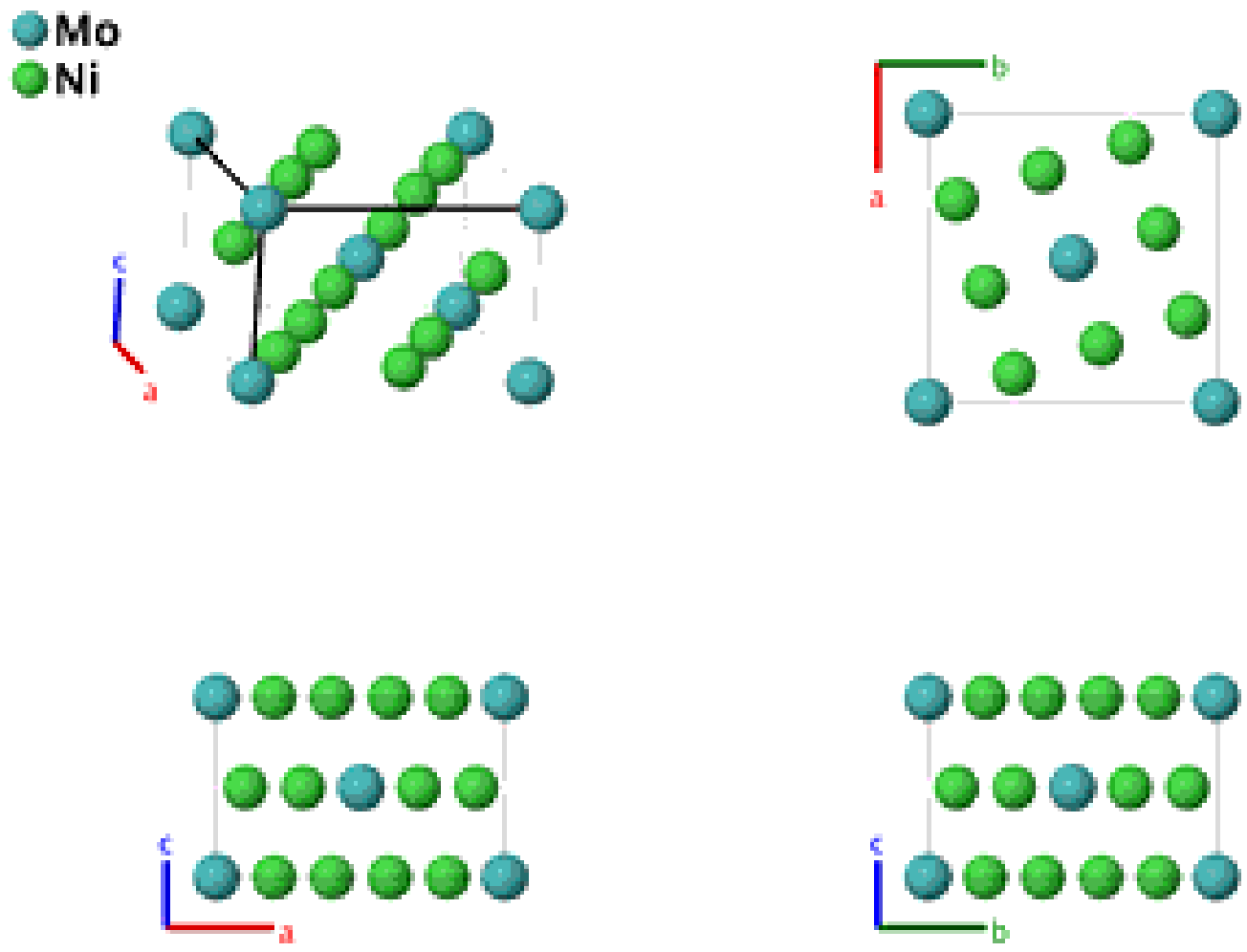

| | | |
|---|---|---|
| **Prototype** | : | $Ni_4Mo$ |
| **AFLOW prototype label** | : | AB4_tI10_87_a_h |
| ***Strukturbericht* designation** | : | $D1_a$ |
| **Pearson symbol** | : | tI10 |
| **Space group number** | : | 87 |
| **Space group symbol** | : | I4/m |
| **AFLOW prototype command** | : | `aflow --proto=AB4_tI10_87_a_h` `--params=`$a, c/a, x_2, y_2$ |

**Other compounds with this structure:**

- $Ag_4Lu$, $Ag_4Sc$, $Au_4Cr$, $Au_4Er$, $Au_4Ho$, $Au_4Lu$, $Au_4Mn$, $Au_4V$, $Au_4Yb$, $Ni_4W$

**Body-centered Tetragonal primitive vectors:**

$$\begin{aligned}
\mathbf{a}_1 &= -\frac{1}{2}\,a\,\hat{\mathbf{x}} + \frac{1}{2}\,a\,\hat{\mathbf{y}} + \frac{1}{2}\,c\,\hat{\mathbf{z}} \\
\mathbf{a}_2 &= \frac{1}{2}\,a\,\hat{\mathbf{x}} - \frac{1}{2}\,a\,\hat{\mathbf{y}} + \frac{1}{2}\,c\,\hat{\mathbf{z}} \\
\mathbf{a}_3 &= \frac{1}{2}\,a\,\hat{\mathbf{x}} + \frac{1}{2}\,a\,\hat{\mathbf{y}} - \frac{1}{2}\,c\,\hat{\mathbf{z}}
\end{aligned}$$

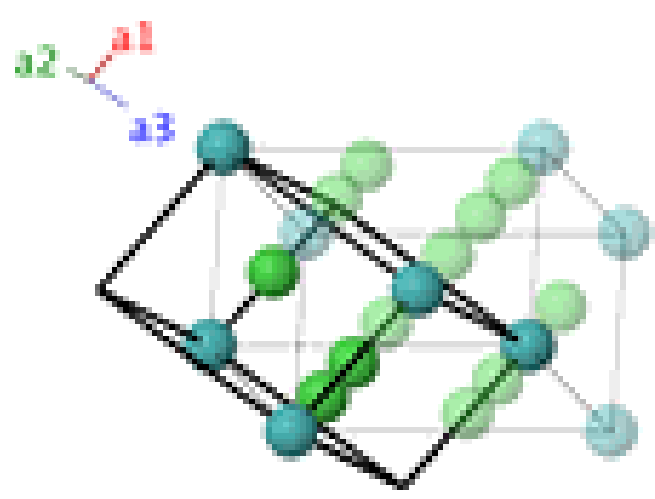

**Basis vectors:**

| | | Lattice Coordinates | | Cartesian Coordinates | Wyckoff Position | Atom Type |
|---|---|---|---|---|---|---|
| $\mathbf{B}_1$ | = | $0\,\mathbf{a}_1 + 0\,\mathbf{a}_2 + 0\,\mathbf{a}_3$ | = | $0\hat{\mathbf{x}} + 0\hat{\mathbf{y}} + 0\hat{\mathbf{z}}$ | (2$a$) | Mo |
| $\mathbf{B}_2$ | = | $y_2\,\mathbf{a}_1 + x_2\,\mathbf{a}_2 + (x_2 + y_2)\,\mathbf{a}_3$ | = | $x_2\,a\,\hat{\mathbf{x}} + y_2\,a\,\hat{\mathbf{y}}$ | (8$h$) | Ni |
| $\mathbf{B}_3$ | = | $-y_2\,\mathbf{a}_1 - x_2\,\mathbf{a}_2 - (x_2 + y_2)\,\mathbf{a}_3$ | = | $-x_2\,a\,\hat{\mathbf{x}} - y_2\,a\,\hat{\mathbf{y}}$ | (8$h$) | Ni |
| $\mathbf{B}_4$ | = | $x_2\,\mathbf{a}_1 - y_2\,\mathbf{a}_2 + (x_2 - y_2)\,\mathbf{a}_3$ | = | $-y_2\,a\,\hat{\mathbf{x}} + x_2\,a\,\hat{\mathbf{y}}$ | (8$h$) | Ni |
| $\mathbf{B}_5$ | = | $-x_2\,\mathbf{a}_1 + y_2\,\mathbf{a}_2 + (y_2 - x_2)\,\mathbf{a}_3$ | = | $y_2\,a\,\hat{\mathbf{x}} - x_2\,a\,\hat{\mathbf{y}}$ | (8$h$) | Ni |

**Geometry files:**

- CIF: pp. 686
- POSCAR: pp. 686

# $\alpha$-Cristobalite ($SiO_2$, low) Structure: A2B_tP12_92_b_a

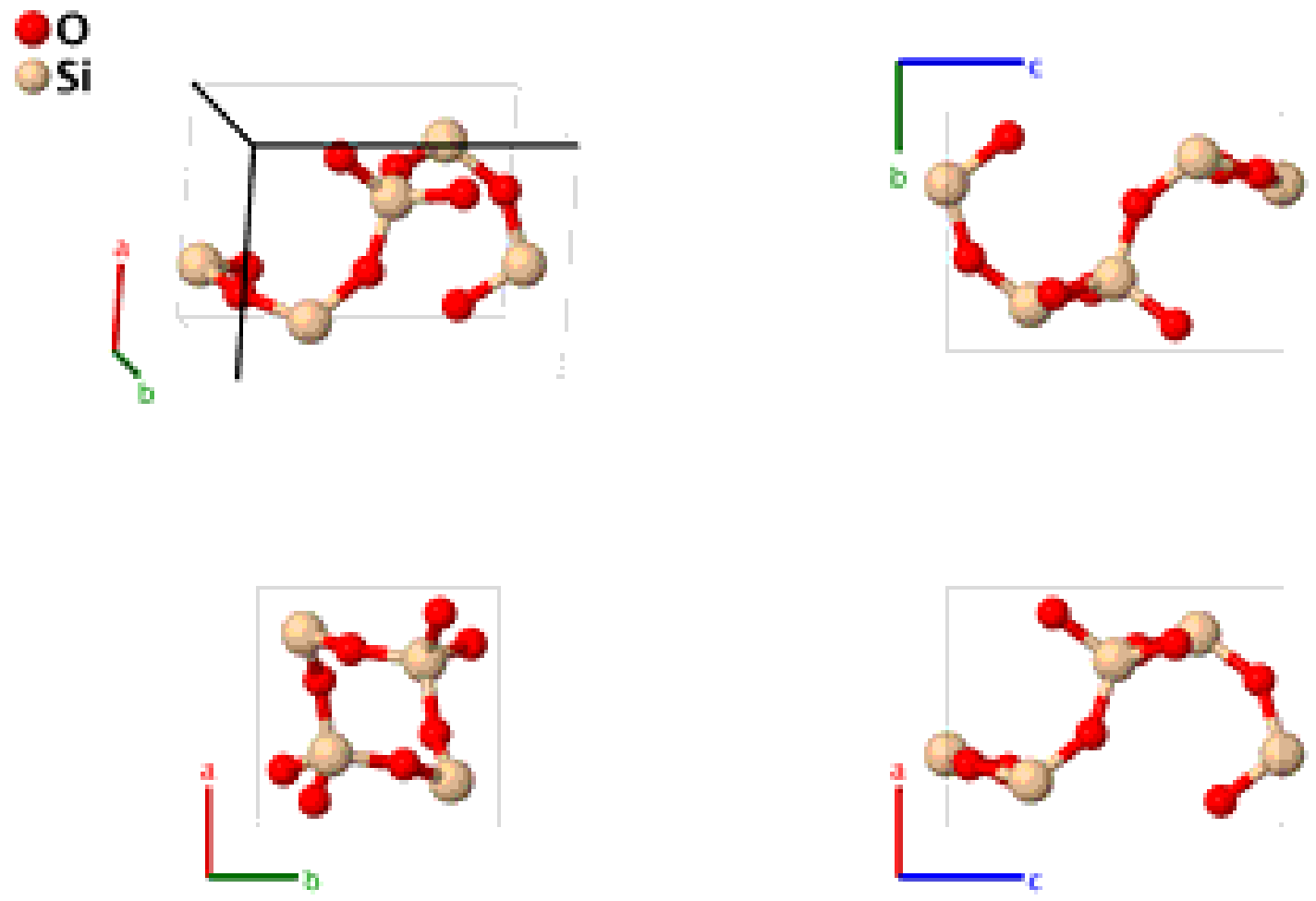

| | | |
|---|---|---|
| **Prototype** | : | $SiO_2$ |
| **AFLOW prototype label** | : | A2B_tP12_92_b_a |
| ***Strukturbericht* designation** | : | None |
| **Pearson symbol** | : | tP12 |
| **Space group number** | : | 92 |
| **Space group symbol** | : | $P4_12_12$ |
| **AFLOW prototype command** | : | `aflow --proto=A2B_tP12_92_b_a --params=`$a, c/a, x_1, x_2, y_2, z_2$ |

**Other compounds with this structure:**

- $TeO_2$ paratellurite, $BeF_2$

**Simple Tetragonal primitive vectors:**

$$\mathbf{a}_1 = a\,\hat{\mathbf{x}}$$
$$\mathbf{a}_2 = a\,\hat{\mathbf{y}}$$
$$\mathbf{a}_3 = c\,\hat{\mathbf{z}}$$

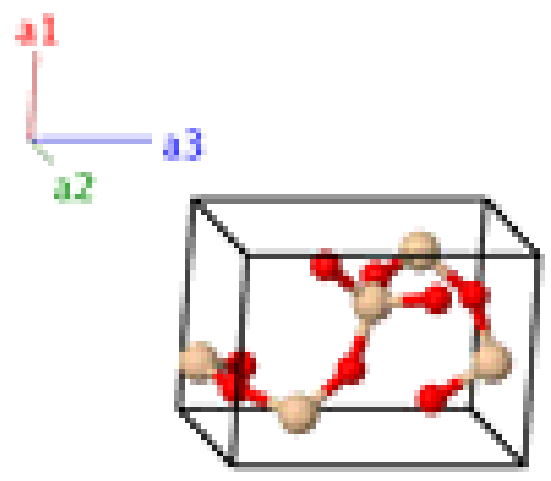

**Basis vectors:**

| | | Lattice Coordinates | | Cartesian Coordinates | Wyckoff Position | Atom Type |
|---|---|---|---|---|---|---|
| $\mathbf{B}_1$ | = | $x_1\,\mathbf{a}_1 + x_1\,\mathbf{a}_2$ | = | $x_1\,a\,\hat{\mathbf{x}} + x_1\,a\,\hat{\mathbf{y}}$ | (4$a$) | Si |
| $\mathbf{B}_2$ | = | $-x_1\,\mathbf{a}_1 - x_1\,\mathbf{a}_2 + \frac{1}{2}\,\mathbf{a}_3$ | = | $-x_1\,a\,\hat{\mathbf{x}} - x_1\,a\,\hat{\mathbf{y}} + \frac{1}{2}\,c\,\hat{\mathbf{z}}$ | (4$a$) | Si |
| $\mathbf{B}_3$ | = | $\left(\frac{1}{2} - x_1\right)\mathbf{a}_1 + \left(\frac{1}{2} + x_1\right)\mathbf{a}_2 + \frac{1}{4}\,\mathbf{a}_3$ | = | $\left(\frac{1}{2} - x_1\right)a\,\hat{\mathbf{x}} + \left(\frac{1}{2} + x_1\right)a\,\hat{\mathbf{y}} + \frac{1}{4}\,c\,\hat{\mathbf{z}}$ | (4$a$) | Si |
| $\mathbf{B}_4$ | = | $\left(\frac{1}{2} + x_1\right)\mathbf{a}_1 + \left(\frac{1}{2} - x_1\right)\mathbf{a}_2 + \frac{3}{4}\,\mathbf{a}_3$ | = | $\left(\frac{1}{2} + x_1\right)a\,\hat{\mathbf{x}} + \left(\frac{1}{2} - x_1\right)a\,\hat{\mathbf{y}} + \frac{3}{4}\,c\,\hat{\mathbf{z}}$ | (4$a$) | Si |
| $\mathbf{B}_5$ | = | $x_2\,\mathbf{a}_1 + y_2\,\mathbf{a}_2 + z_2\,\mathbf{a}_3$ | = | $x_2\,a\,\hat{\mathbf{x}} + y_2\,a\,\hat{\mathbf{y}} + z_2\,c\,\hat{\mathbf{z}}$ | (8$b$) | O |
| $\mathbf{B_6}$ | = | $-x_2\,\mathbf{a_1} - y_2\,\mathbf{a_2} + \left(\frac{1}{2} + z_2\right)\,\mathbf{a_3}$ | = | $-x_2\,a\,\mathbf{\hat{x}} - y_2\,a\,\mathbf{\hat{y}} + \left(\frac{1}{2} + z_2\right)\,c\,\mathbf{\hat{z}}$ | (8*b*) | O |
| $\mathbf{B_7}$ | = | $\left(\frac{1}{2} - y_2\right)\,\mathbf{a_1} + \left(\frac{1}{2} + x_2\right)\,\mathbf{a_2} + \left(\frac{1}{4} + z_2\right)\,\mathbf{a_3}$ | = | $\left(\frac{1}{2} - y_2\right)\,a\,\mathbf{\hat{x}} + \left(\frac{1}{2} + x_2\right)\,a\,\mathbf{\hat{y}} + \left(\frac{1}{4} + z_2\right)\,c\,\mathbf{\hat{z}}$ | (8*b*) | O |
| $\mathbf{B_8}$ | = | $\left(\frac{1}{2} + y_2\right)\,\mathbf{a_1} + \left(\frac{1}{2} - x_2\right)\,\mathbf{a_2} + \left(\frac{3}{4} + z_2\right)\,\mathbf{a_3}$ | = | $\left(\frac{1}{2} + y_2\right)\,a\,\mathbf{\hat{x}} + \left(\frac{1}{2} - x_2\right)\,a\,\mathbf{\hat{y}} + \left(\frac{3}{4} + z_2\right)\,c\,\mathbf{\hat{z}}$ | (8*b*) | O |
| $\mathbf{B_9}$ | = | $\left(\frac{1}{2} - x_2\right)\,\mathbf{a_1} + \left(\frac{1}{2} + y_2\right)\,\mathbf{a_2} + \left(\frac{1}{4} - z_2\right)\,\mathbf{a_3}$ | = | $\left(\frac{1}{2} - x_2\right)\,a\,\mathbf{\hat{x}} + \left(\frac{1}{2} + y_2\right)\,a\,\mathbf{\hat{y}} + \left(\frac{1}{4} - z_2\right)\,c\,\mathbf{\hat{z}}$ | (8*b*) | O |
| $\mathbf{B_{10}}$ | = | $\left(\frac{1}{2} + x_2\right)\,\mathbf{a_1} + \left(\frac{1}{2} - y_2\right)\,\mathbf{a_2} + \left(\frac{3}{4} - z_2\right)\,\mathbf{a_3}$ | = | $\left(\frac{1}{2} + x_2\right)\,a\,\mathbf{\hat{x}} + \left(\frac{1}{2} - y_2\right)\,a\,\mathbf{\hat{y}} + \left(\frac{3}{4} - z_2\right)\,c\,\mathbf{\hat{z}}$ | (8*b*) | O |
| $\mathbf{B_{11}}$ | = | $y_2\,\mathbf{a_1} + x_2\,\mathbf{a_2} - z_2\,\mathbf{a_3}$ | = | $y_2\,a\,\mathbf{\hat{x}} + x_2\,a\,\mathbf{\hat{y}} - z_2\,c\,\mathbf{\hat{z}}$ | (8*b*) | O |
| $\mathbf{B_{12}}$ | = | $-y_2\,\mathbf{a_1} - x_2\,\mathbf{a_2} + \left(\frac{1}{2} - z_2\right)\,\mathbf{a_3}$ | = | $-y_2\,a\,\mathbf{\hat{x}} - x_2\,a\,\mathbf{\hat{y}} + \left(\frac{1}{2} - z_2\right)\,c\,\mathbf{\hat{z}}$ | (8*b*) | O |

**Found in:**

**Geometry files:**

- CIF: pp. 686
- POSCAR: pp. 686

# Keatite ($SiO_2$) Structure: A2B_tP36_96_3b_ab

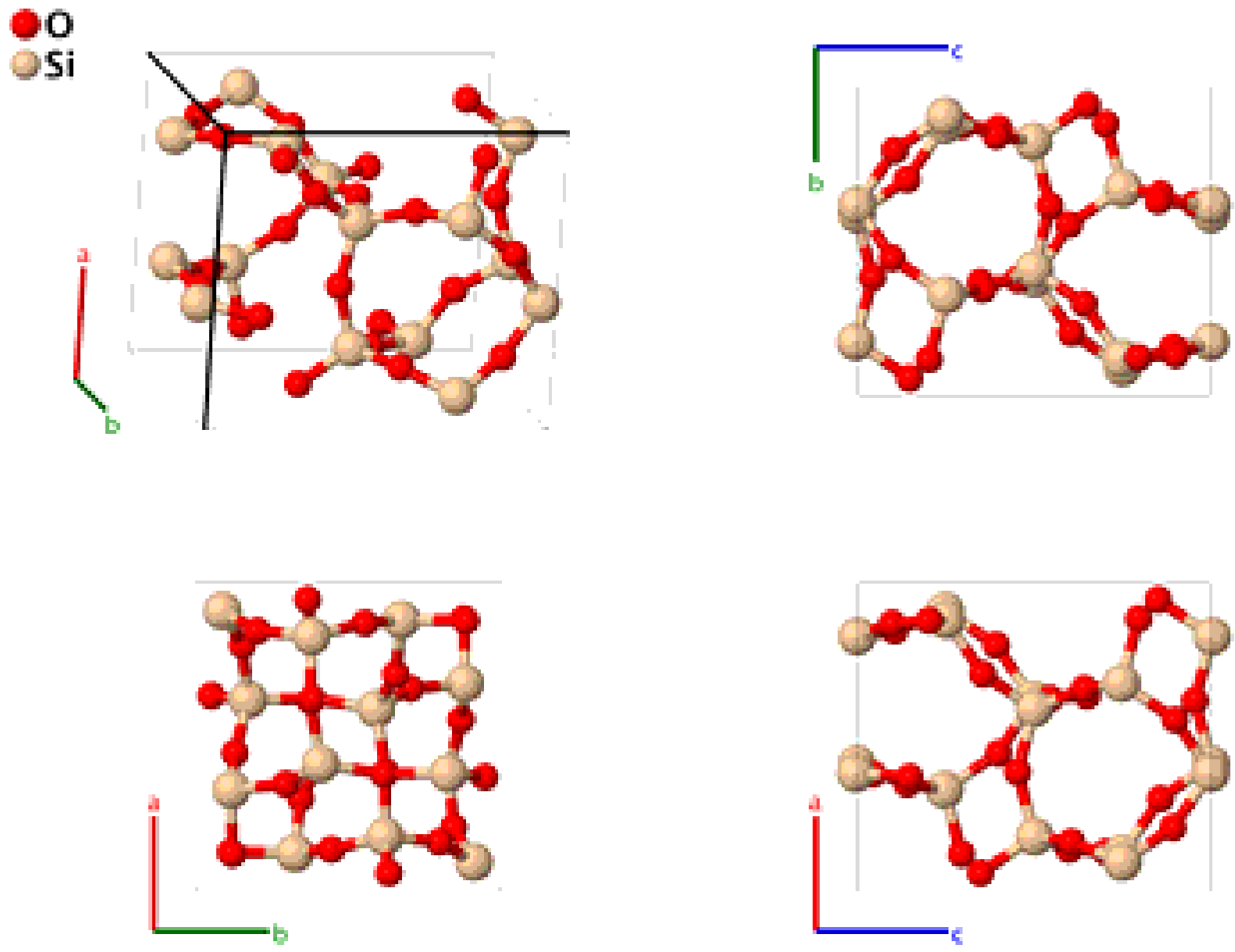

| | | |
|---|---|---|
| **Prototype** | : | $SiO_2$ |
| **AFLOW prototype label** | : | A2B_tP36_96_3b_ab |
| ***Strukturbericht* designation** | : | None |
| **Pearson symbol** | : | tP36 |
| **Space group number** | : | 96 |
| **Space group symbol** | : | $P4_32_12$ |
| **AFLOW prototype command** | : | `aflow --proto=A2B_tP36_96_3b_ab`<br>`--params=`$a, c/a, x_1, x_2, y_2, z_2, x_3, y_3, z_3, x_4, y_4, z_4, x_5, y_5, z_5$ |

- All references, including (Wyckoff, 1963), (Shropshire, 1959) and (Demuth, 1999) note that keatite can exist in both space group $P4_12_12$-$D_4^4$ (#92) and its enantiomorph $P4_32_12$-$D_4^8$ (#96). Wyckoff uses the coordinates proposed by Shropshire and assumes the space group is $P4_12_12$. He then notes that one of the Si-O bonds in this structure is very long (3.69 Å), and is "so improbable that there is something wrong either with the parameters as stated or the structure itself". If we use space group $P4_32_12$ while retaining Shropshire's coordinates we obtain a much more convincing structure, one that looks much like the structure in Shropshire's Fig. 3. For this reason we place this structure in $P4_32_12$.

**Simple Tetragonal primitive vectors:**

$$\mathbf{a}_1 = a\,\hat{\mathbf{x}}$$
$$\mathbf{a}_2 = a\,\hat{\mathbf{y}}$$
$$\mathbf{a}_3 = c\,\hat{\mathbf{z}}$$

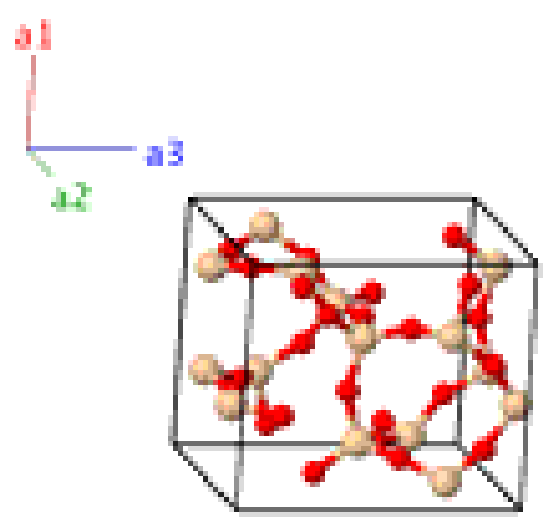

**Basis vectors:**

Lattice Coordinates | Cartesian Coordinates | Wyckoff Position | Atom Type

| | | | | | | |
|---|---|---|---|---|---|---|
| $\mathbf{B}_{1}$ | = | $x_{1}\,\mathbf{a}_{1} + x_{1}\,\mathbf{a}_{2}$ | = | $x_{1}\,a\,\hat{\mathbf{x}} + x_{1}\,a\,\hat{\mathbf{y}}$ | (4*a*) | Si I |
| $\mathbf{B}_{2}$ | = | $-x_{1}\,\mathbf{a}_{1} - x_{1}\,\mathbf{a}_{2} + \frac{1}{2}\,\mathbf{a}_{3}$ | = | $-x_{1}\,a\,\hat{\mathbf{x}} - x_{1}\,a\,\hat{\mathbf{y}} + \frac{1}{2}\,c\,\hat{\mathbf{z}}$ | (4*a*) | Si I |
| $\mathbf{B}_{3}$ | = | $\left(\frac{1}{2} - x_{1}\right)\mathbf{a}_{1} + \left(\frac{1}{2} + x_{1}\right)\mathbf{a}_{2} + \frac{3}{4}\,\mathbf{a}_{3}$ | = | $\left(\frac{1}{2} - x_{1}\right)a\,\hat{\mathbf{x}} + \left(\frac{1}{2} + x_{1}\right)a\,\hat{\mathbf{y}} + \frac{3}{4}\,c\,\hat{\mathbf{z}}$ | (4*a*) | Si I |
| $\mathbf{B}_{4}$ | = | $\left(\frac{1}{2} + x_{1}\right)\mathbf{a}_{1} + \left(\frac{1}{2} - x_{1}\right)\mathbf{a}_{2} + \frac{1}{4}\,\mathbf{a}_{3}$ | = | $\left(\frac{1}{2} + x_{1}\right)a\,\hat{\mathbf{x}} + \left(\frac{1}{2} - x_{1}\right)a\,\hat{\mathbf{y}} + \frac{1}{4}\,c\,\hat{\mathbf{z}}$ | (4*a*) | Si I |
| $\mathbf{B}_{5}$ | = | $x_{2}\,\mathbf{a}_{1} + y_{2}\,\mathbf{a}_{2} + z_{2}\,\mathbf{a}_{3}$ | = | $x_{2}\,a\,\hat{\mathbf{x}} + y_{2}\,a\,\hat{\mathbf{y}} + z_{2}\,c\,\hat{\mathbf{z}}$ | (8*b*) | O I |
| $\mathbf{B}_{6}$ | = | $-x_{2}\,\mathbf{a}_{1} - y_{2}\,\mathbf{a}_{2} + \left(\frac{1}{2} + z_{2}\right)\mathbf{a}_{3}$ | = | $-x_{2}\,a\,\hat{\mathbf{x}} - y_{2}\,a\,\hat{\mathbf{y}} + \left(\frac{1}{2} + z_{2}\right)c\,\hat{\mathbf{z}}$ | (8*b*) | O I |
| $\mathbf{B}_{7}$ | = | $\left(\frac{1}{2} - y_{2}\right)\mathbf{a}_{1} + \left(\frac{1}{2} + x_{2}\right)\mathbf{a}_{2} + \left(\frac{3}{4} + z_{2}\right)\mathbf{a}_{3}$ | = | $\left(\frac{1}{2} - y_{2}\right)a\,\hat{\mathbf{x}} + \left(\frac{1}{2} + x_{2}\right)a\,\hat{\mathbf{y}} + \left(\frac{3}{4} + z_{2}\right)c\,\hat{\mathbf{z}}$ | (8*b*) | O I |
| $\mathbf{B}_{8}$ | = | $\left(\frac{1}{2} + y_{2}\right)\mathbf{a}_{1} + \left(\frac{1}{2} - x_{2}\right)\mathbf{a}_{2} + \left(\frac{1}{4} + z_{2}\right)\mathbf{a}_{3}$ | = | $\left(\frac{1}{2} + y_{2}\right)a\,\hat{\mathbf{x}} + \left(\frac{1}{2} - x_{2}\right)a\,\hat{\mathbf{y}} + \left(\frac{1}{4} + z_{2}\right)c\,\hat{\mathbf{z}}$ | (8*b*) | O I |
| $\mathbf{B}_{9}$ | = | $\left(\frac{1}{2} - x_{2}\right)\mathbf{a}_{1} + \left(\frac{1}{2} + y_{2}\right)\mathbf{a}_{2} + \left(\frac{3}{4} - z_{2}\right)\mathbf{a}_{3}$ | = | $\left(\frac{1}{2} - x_{2}\right)a\,\hat{\mathbf{x}} + \left(\frac{1}{2} + y_{2}\right)a\,\hat{\mathbf{y}} + \left(\frac{3}{4} - z_{2}\right)c\,\hat{\mathbf{z}}$ | (8*b*) | O I |
| $\mathbf{B}_{10}$ | = | $\left(\frac{1}{2} + x_{2}\right)\mathbf{a}_{1} + \left(\frac{1}{2} - y_{2}\right)\mathbf{a}_{2} + \left(\frac{1}{4} - z_{2}\right)\mathbf{a}_{3}$ | = | $\left(\frac{1}{2} + x_{2}\right)a\,\hat{\mathbf{x}} + \left(\frac{1}{2} - y_{2}\right)a\,\hat{\mathbf{y}} + \left(\frac{1}{4} - z_{2}\right)c\,\hat{\mathbf{z}}$ | (8*b*) | O I |
| $\mathbf{B}_{11}$ | = | $y_{2}\,\mathbf{a}_{1} + x_{2}\,\mathbf{a}_{2} - z_{2}\,\mathbf{a}_{3}$ | = | $y_{2}\,a\,\hat{\mathbf{x}} + x_{2}\,a\,\hat{\mathbf{y}} - z_{2}\,c\,\hat{\mathbf{z}}$ | (8*b*) | O I |
| $\mathbf{B}_{12}$ | = | $-y_{2}\,\mathbf{a}_{1} - x_{2}\,\mathbf{a}_{2} + \left(\frac{1}{2} - z_{2}\right)\mathbf{a}_{3}$ | = | $-y_{2}\,a\,\hat{\mathbf{x}} - x_{2}\,a\,\hat{\mathbf{y}} + \left(\frac{1}{2} - z_{2}\right)c\,\hat{\mathbf{z}}$ | (8*b*) | O I |
| $\mathbf{B}_{13}$ | = | $x_{3}\,\mathbf{a}_{1} + y_{3}\,\mathbf{a}_{2} + z_{3}\,\mathbf{a}_{3}$ | = | $x_{3}\,a\,\hat{\mathbf{x}} + y_{3}\,a\,\hat{\mathbf{y}} + z_{3}\,c\,\hat{\mathbf{z}}$ | (8*b*) | O II |
| $\mathbf{B}_{14}$ | = | $-x_{3}\,\mathbf{a}_{1} - y_{3}\,\mathbf{a}_{2} + \left(\frac{1}{2} + z_{3}\right)\mathbf{a}_{3}$ | = | $-x_{3}\,a\,\hat{\mathbf{x}} - y_{3}\,a\,\hat{\mathbf{y}} + \left(\frac{1}{2} + z_{3}\right)c\,\hat{\mathbf{z}}$ | (8*b*) | O II |
| $\mathbf{B}_{15}$ | = | $\left(\frac{1}{2} - y_{3}\right)\mathbf{a}_{1} + \left(\frac{1}{2} + x_{3}\right)\mathbf{a}_{2} + \left(\frac{3}{4} + z_{3}\right)\mathbf{a}_{3}$ | = | $\left(\frac{1}{2} - y_{3}\right)a\,\hat{\mathbf{x}} + \left(\frac{1}{2} + x_{3}\right)a\,\hat{\mathbf{y}} + \left(\frac{3}{4} + z_{3}\right)c\,\hat{\mathbf{z}}$ | (8*b*) | O II |
| $\mathbf{B}_{16}$ | = | $\left(\frac{1}{2} + y_{3}\right)\mathbf{a}_{1} + \left(\frac{1}{2} - x_{3}\right)\mathbf{a}_{2} + \left(\frac{1}{4} + z_{3}\right)\mathbf{a}_{3}$ | = | $\left(\frac{1}{2} + y_{3}\right)a\,\hat{\mathbf{x}} + \left(\frac{1}{2} - x_{3}\right)a\,\hat{\mathbf{y}} + \left(\frac{1}{4} + z_{3}\right)c\,\hat{\mathbf{z}}$ | (8*b*) | O II |
| $\mathbf{B}_{17}$ | = | $\left(\frac{1}{2} - x_{3}\right)\mathbf{a}_{1} + \left(\frac{1}{2} + y_{3}\right)\mathbf{a}_{2} + \left(\frac{3}{4} - z_{3}\right)\mathbf{a}_{3}$ | = | $\left(\frac{1}{2} - x_{3}\right)a\,\hat{\mathbf{x}} + \left(\frac{1}{2} + y_{3}\right)a\,\hat{\mathbf{y}} + \left(\frac{3}{4} - z_{3}\right)c\,\hat{\mathbf{z}}$ | (8*b*) | O II |
| $\mathbf{B}_{18}$ | = | $\left(\frac{1}{2} + x_{3}\right)\mathbf{a}_{1} + \left(\frac{1}{2} - y_{3}\right)\mathbf{a}_{2} + \left(\frac{1}{4} - z_{3}\right)\mathbf{a}_{3}$ | = | $\left(\frac{1}{2} + x_{3}\right)a\,\hat{\mathbf{x}} + \left(\frac{1}{2} - y_{3}\right)a\,\hat{\mathbf{y}} + \left(\frac{1}{4} - z_{3}\right)c\,\hat{\mathbf{z}}$ | (8*b*) | O II |
| $\mathbf{B}_{19}$ | = | $y_{3}\,\mathbf{a}_{1} + x_{3}\,\mathbf{a}_{2} - z_{3}\,\mathbf{a}_{3}$ | = | $y_{3}\,a\,\hat{\mathbf{x}} + x_{3}\,a\,\hat{\mathbf{y}} - z_{3}\,c\,\hat{\mathbf{z}}$ | (8*b*) | O II |
| $\mathbf{B}_{20}$ | = | $-y_{3}\,\mathbf{a}_{1} - x_{3}\,\mathbf{a}_{2} + \left(\frac{1}{2} - z_{3}\right)\mathbf{a}_{3}$ | = | $-y_{3}\,a\,\hat{\mathbf{x}} - x_{3}\,a\,\hat{\mathbf{y}} + \left(\frac{1}{2} - z_{3}\right)c\,\hat{\mathbf{z}}$ | (8*b*) | O II |
| $\mathbf{B}_{21}$ | = | $x_{4}\,\mathbf{a}_{1} + y_{4}\,\mathbf{a}_{2} + z_{4}\,\mathbf{a}_{3}$ | = | $x_{4}\,a\,\hat{\mathbf{x}} + y_{4}\,a\,\hat{\mathbf{y}} + z_{4}\,c\,\hat{\mathbf{z}}$ | (8*b*) | O III |
| $\mathbf{B}_{22}$ | = | $-x_{4}\,\mathbf{a}_{1} - y_{4}\,\mathbf{a}_{2} + \left(\frac{1}{2} + z_{4}\right)\mathbf{a}_{3}$ | = | $-x_{4}\,a\,\hat{\mathbf{x}} - y_{4}\,a\,\hat{\mathbf{y}} + \left(\frac{1}{2} + z_{4}\right)c\,\hat{\mathbf{z}}$ | (8*b*) | O III |
| $\mathbf{B}_{23}$ | = | $\left(\frac{1}{2} - y_{4}\right)\mathbf{a}_{1} + \left(\frac{1}{2} + x_{4}\right)\mathbf{a}_{2} + \left(\frac{3}{4} + z_{4}\right)\mathbf{a}_{3}$ | = | $\left(\frac{1}{2} - y_{4}\right)a\,\hat{\mathbf{x}} + \left(\frac{1}{2} + x_{4}\right)a\,\hat{\mathbf{y}} + \left(\frac{3}{4} + z_{4}\right)c\,\hat{\mathbf{z}}$ | (8*b*) | O III |
| $\mathbf{B}_{24}$ | = | $\left(\frac{1}{2} + y_{4}\right)\mathbf{a}_{1} + \left(\frac{1}{2} - x_{4}\right)\mathbf{a}_{2} + \left(\frac{1}{4} + z_{4}\right)\mathbf{a}_{3}$ | = | $\left(\frac{1}{2} + y_{4}\right)a\,\hat{\mathbf{x}} + \left(\frac{1}{2} - x_{4}\right)a\,\hat{\mathbf{y}} + \left(\frac{1}{4} + z_{4}\right)c\,\hat{\mathbf{z}}$ | (8*b*) | O III |
| $\mathbf{B}_{25}$ | = | $\left(\frac{1}{2} - x_{4}\right)\mathbf{a}_{1} + \left(\frac{1}{2} + y_{4}\right)\mathbf{a}_{2} + \left(\frac{3}{4} - z_{4}\right)\mathbf{a}_{3}$ | = | $\left(\frac{1}{2} - x_{4}\right)a\,\hat{\mathbf{x}} + \left(\frac{1}{2} + y_{4}\right)a\,\hat{\mathbf{y}} + \left(\frac{3}{4} - z_{4}\right)c\,\hat{\mathbf{z}}$ | (8*b*) | O III |
| $\mathbf{B}_{26}$ | = | $\left(\frac{1}{2} + x_{4}\right)\mathbf{a}_{1} + \left(\frac{1}{2} - y_{4}\right)\mathbf{a}_{2} + \left(\frac{1}{4} - z_{4}\right)\mathbf{a}_{3}$ | = | $\left(\frac{1}{2} + x_{4}\right)a\,\hat{\mathbf{x}} + \left(\frac{1}{2} - y_{4}\right)a\,\hat{\mathbf{y}} + \left(\frac{1}{4} - z_{4}\right)c\,\hat{\mathbf{z}}$ | (8*b*) | O III |
| $\mathbf{B}_{27}$ | = | $y_{4}\,\mathbf{a}_{1} + x_{4}\,\mathbf{a}_{2} - z_{4}\,\mathbf{a}_{3}$ | = | $y_{4}\,a\,\hat{\mathbf{x}} + x_{4}\,a\,\hat{\mathbf{y}} - z_{4}\,c\,\hat{\mathbf{z}}$ | (8*b*) | O III |
| $\mathbf{B_{28}}$ | = | $-y_4\,\mathbf{a_1} - x_4\,\mathbf{a_2} + \left(\frac{1}{2} - z_4\right)\,\mathbf{a_3}$ | = | $-y_4\,a\,\mathbf{\hat{x}} - x_4\,a\,\mathbf{\hat{y}} + \left(\frac{1}{2} - z_4\right)\,c\,\mathbf{\hat{z}}$ | (8$b$) | O III |
| $\mathbf{B_{29}}$ | = | $x_5\,\mathbf{a_1} + y_5\,\mathbf{a_2} + z_5\,\mathbf{a_3}$ | = | $x_5\,a\,\mathbf{\hat{x}} + y_5\,a\,\mathbf{\hat{y}} + z_5\,c\,\mathbf{\hat{z}}$ | (8$b$) | Si II |
| $\mathbf{B_{30}}$ | = | $-x_5\,\mathbf{a_1} - y_5\,\mathbf{a_2} + \left(\frac{1}{2} + z_5\right)\,\mathbf{a_3}$ | = | $-x_5\,a\,\mathbf{\hat{x}} - y_5\,a\,\mathbf{\hat{y}} + \left(\frac{1}{2} + z_5\right)\,c\,\mathbf{\hat{z}}$ | (8$b$) | Si II |
| $\mathbf{B_{31}}$ | = | $\left(\frac{1}{2} - y_5\right)\,\mathbf{a_1} + \left(\frac{1}{2} + x_5\right)\,\mathbf{a_2} + \left(\frac{3}{4} + z_5\right)\,\mathbf{a_3}$ | = | $\left(\frac{1}{2} - y_5\right)\,a\,\mathbf{\hat{x}} + \left(\frac{1}{2} + x_5\right)\,a\,\mathbf{\hat{y}} + \left(\frac{3}{4} + z_5\right)\,c\,\mathbf{\hat{z}}$ | (8$b$) | Si II |
| $\mathbf{B_{32}}$ | = | $\left(\frac{1}{2} + y_5\right)\,\mathbf{a_1} + \left(\frac{1}{2} - x_5\right)\,\mathbf{a_2} + \left(\frac{1}{4} + z_5\right)\,\mathbf{a_3}$ | = | $\left(\frac{1}{2} + y_5\right)\,a\,\mathbf{\hat{x}} + \left(\frac{1}{2} - x_5\right)\,a\,\mathbf{\hat{y}} + \left(\frac{1}{4} + z_5\right)\,c\,\mathbf{\hat{z}}$ | (8$b$) | Si II |
| $\mathbf{B_{33}}$ | = | $\left(\frac{1}{2} - x_5\right)\,\mathbf{a_1} + \left(\frac{1}{2} + y_5\right)\,\mathbf{a_2} + \left(\frac{3}{4} - z_5\right)\,\mathbf{a_3}$ | = | $\left(\frac{1}{2} - x_5\right)\,a\,\mathbf{\hat{x}} + \left(\frac{1}{2} + y_5\right)\,a\,\mathbf{\hat{y}} + \left(\frac{3}{4} - z_5\right)\,c\,\mathbf{\hat{z}}$ | (8$b$) | Si II |
| $\mathbf{B_{34}}$ | = | $\left(\frac{1}{2} + x_5\right)\,\mathbf{a_1} + \left(\frac{1}{2} - y_5\right)\,\mathbf{a_2} + \left(\frac{1}{4} - z_5\right)\,\mathbf{a_3}$ | = | $\left(\frac{1}{2} + x_5\right)\,a\,\mathbf{\hat{x}} + \left(\frac{1}{2} - y_5\right)\,a\,\mathbf{\hat{y}} + \left(\frac{1}{4} - z_5\right)\,c\,\mathbf{\hat{z}}$ | (8$b$) | Si II |
| $\mathbf{B_{35}}$ | = | $y_5\,\mathbf{a_1} + x_5\,\mathbf{a_2} - z_5\,\mathbf{a_3}$ | = | $y_5\,a\,\mathbf{\hat{x}} + x_5\,a\,\mathbf{\hat{y}} - z_5\,c\,\mathbf{\hat{z}}$ | (8$b$) | Si II |
| $\mathbf{B_{36}}$ | = | $-y_5\,\mathbf{a_1} - x_5\,\mathbf{a_2} + \left(\frac{1}{2} - z_5\right)\,\mathbf{a_3}$ | = | $-y_5\,a\,\mathbf{\hat{x}} - x_5\,a\,\mathbf{\hat{y}} + \left(\frac{1}{2} - z_5\right)\,c\,\mathbf{\hat{z}}$ | (8$b$) | Si II |

---

**Found in:**

---

**Geometry files:**

- CIF: pp. 686
- POSCAR: pp. 687

# “ST12” Structure of Si: A_tP12_96_ab

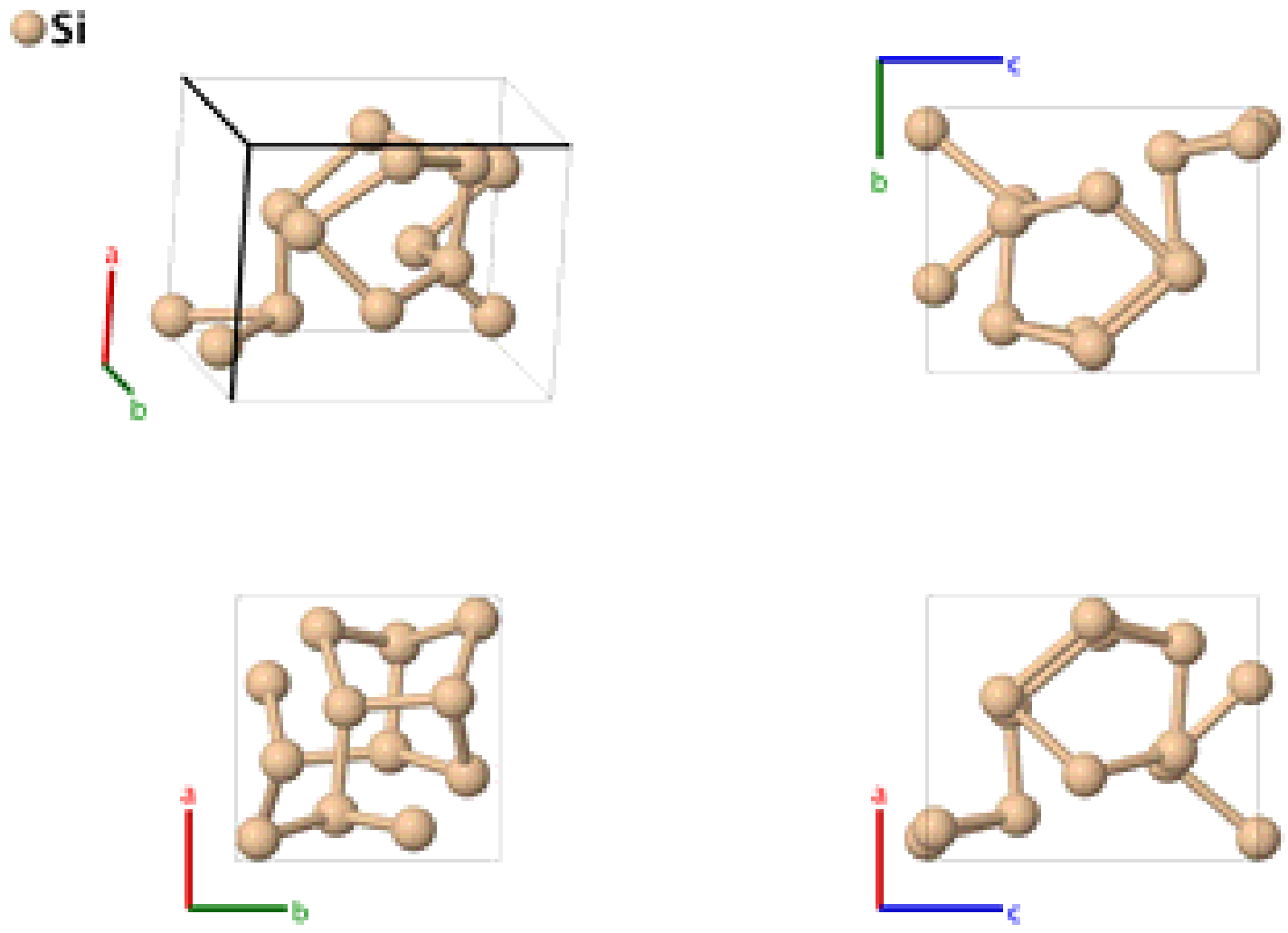

| | | |
|---|---|---|
| **Prototype** | : | Si |
| **AFLOW prototype label** | : | A_tP12_96_ab |
| ***Strukturbericht* designation** | : | None |
| **Pearson symbol** | : | tP12 |
| **Space group number** | : | 96 |
| **Space group symbol** | : | $P4_32_12$ |
| **AFLOW prototype command** | : | `aflow --proto=A_tP12_96_ab`<br>`--params=`$a, c/a, x_1, x_2, y_2, z_2$ |

- This is a tetragonally bonded structure which packs more efficiently than diamond. It is seen experimentally in some silicon and germanium samples and is a staple for testing silicon potentials and first-principles calculations. The structure shown here is taken from the calculations in (Crain, 1994).

**Simple Tetragonal primitive vectors:**

$$\mathbf{a}_1 = a\,\hat{\mathbf{x}}$$
$$\mathbf{a}_2 = a\,\hat{\mathbf{y}}$$
$$\mathbf{a}_3 = c\,\hat{\mathbf{z}}$$

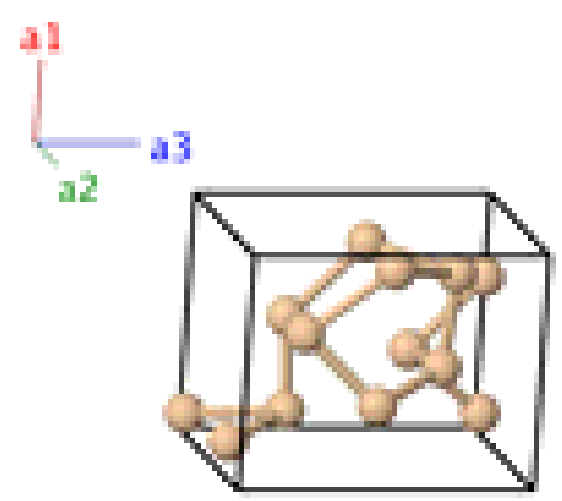

**Basis vectors:**

| | | Lattice Coordinates | | Cartesian Coordinates | Wyckoff Position | Atom Type |
|---|---|---|---|---|---|---|
| $\mathbf{B}_1$ | = | $x_1\,\mathbf{a}_1 + x_1\,\mathbf{a}_2$ | = | $x_1\,a\,\hat{\mathbf{x}} + x_1\,a\,\hat{\mathbf{y}}$ | (4$a$) | Si I |
| $\mathbf{B}_2$ | = | $-x_1\,\mathbf{a}_1 - x_1\,\mathbf{a}_2 + \frac{1}{2}\,\mathbf{a}_3$ | = | $-x_1\,a\,\hat{\mathbf{x}} - x_1\,a\,\hat{\mathbf{y}} + \frac{1}{2}\,c\,\hat{\mathbf{z}}$ | (4$a$) | Si I |
| $\mathbf{B}_3$ | = | $\left(\frac{1}{2} - x_1\right)\mathbf{a}_1 + \left(\frac{1}{2} + x_1\right)\mathbf{a}_2 + \frac{3}{4}\,\mathbf{a}_3$ | = | $\left(\frac{1}{2} - x_1\right)a\,\hat{\mathbf{x}} + \left(\frac{1}{2} + x_1\right)a\,\hat{\mathbf{y}} + \frac{3}{4}\,c\,\hat{\mathbf{z}}$ | (4$a$) | Si I |
| $\mathbf{B}_4$ | = | $\left(\frac{1}{2} + x_1\right)\mathbf{a}_1 + \left(\frac{1}{2} - x_1\right)\mathbf{a}_2 + \frac{1}{4}\,\mathbf{a}_3$ | = | $\left(\frac{1}{2} + x_1\right)a\,\hat{\mathbf{x}} + \left(\frac{1}{2} - x_1\right)a\,\hat{\mathbf{y}} + \frac{1}{4}\,c\,\hat{\mathbf{z}}$ | (4$a$) | Si I |
| $\mathbf{B}_5$ | = | $x_2\,\mathbf{a}_1 + y_2\,\mathbf{a}_2 + z_2\,\mathbf{a}_3$ | = | $x_2\,a\,\hat{\mathbf{x}} + y_2\,a\,\hat{\mathbf{y}} + z_2\,c\,\hat{\mathbf{z}}$ | (8$b$) | Si II |
| $\mathbf{B}_6$ | = | $-x_2\,\mathbf{a}_1 - y_2\,\mathbf{a}_2 + \left(\frac{1}{2} + z_2\right)\,\mathbf{a}_3$ | = | $-x_2\,a\,\hat{\mathbf{x}} - y_2\,a\,\hat{\mathbf{y}} + \left(\frac{1}{2} + z_2\right)\,c\,\hat{\mathbf{z}}$ | (8$b$) | Si II |
| $\mathbf{B}_7$ | = | $\left(\frac{1}{2} - y_2\right)\,\mathbf{a}_1 + \left(\frac{1}{2} + x_2\right)\,\mathbf{a}_2 + \left(\frac{3}{4} + z_2\right)\,\mathbf{a}_3$ | = | $\left(\frac{1}{2} - y_2\right)\,a\,\hat{\mathbf{x}} + \left(\frac{1}{2} + x_2\right)\,a\,\hat{\mathbf{y}} + \left(\frac{3}{4} + z_2\right)\,c\,\hat{\mathbf{z}}$ | (8$b$) | Si II |
| $\mathbf{B}_8$ | = | $\left(\frac{1}{2} + y_2\right)\,\mathbf{a}_1 + \left(\frac{1}{2} - x_2\right)\,\mathbf{a}_2 + \left(\frac{1}{4} + z_2\right)\,\mathbf{a}_3$ | = | $\left(\frac{1}{2} + y_2\right)\,a\,\hat{\mathbf{x}} + \left(\frac{1}{2} - x_2\right)\,a\,\hat{\mathbf{y}} + \left(\frac{1}{4} + z_2\right)\,c\,\hat{\mathbf{z}}$ | (8$b$) | Si II |
| $\mathbf{B}_9$ | = | $\left(\frac{1}{2} - x_2\right)\,\mathbf{a}_1 + \left(\frac{1}{2} + y_2\right)\,\mathbf{a}_2 + \left(\frac{3}{4} - z_2\right)\,\mathbf{a}_3$ | = | $\left(\frac{1}{2} - x_2\right)\,a\,\hat{\mathbf{x}} + \left(\frac{1}{2} + y_2\right)\,a\,\hat{\mathbf{y}} + \left(\frac{3}{4} - z_2\right)\,c\,\hat{\mathbf{z}}$ | (8$b$) | Si II |
| $\mathbf{B}_{10}$ | = | $\left(\frac{1}{2} + x_2\right)\,\mathbf{a}_1 + \left(\frac{1}{2} - y_2\right)\,\mathbf{a}_2 + \left(\frac{1}{4} - z_2\right)\,\mathbf{a}_3$ | = | $\left(\frac{1}{2} + x_2\right)\,a\,\hat{\mathbf{x}} + \left(\frac{1}{2} - y_2\right)\,a\,\hat{\mathbf{y}} + \left(\frac{1}{4} - z_2\right)\,c\,\hat{\mathbf{z}}$ | (8$b$) | Si II |
| $\mathbf{B}_{11}$ | = | $y_2\,\mathbf{a}_1 + x_2\,\mathbf{a}_2 - z_2\,\mathbf{a}_3$ | = | $y_2\,a\,\hat{\mathbf{x}} + x_2\,a\,\hat{\mathbf{y}} - z_2\,c\,\hat{\mathbf{z}}$ | (8$b$) | Si II |
| $\mathbf{B}_{12}$ | = | $-y_2\,\mathbf{a}_1 - x_2\,\mathbf{a}_2 + \left(\frac{1}{2} - z_2\right)\,\mathbf{a}_3$ | = | $-y_2\,a\,\hat{\mathbf{x}} - x_2\,a\,\hat{\mathbf{y}} + \left(\frac{1}{2} - z_2\right)\,c\,\hat{\mathbf{z}}$ | (8$b$) | Si II |

**Geometry files:**

- CIF: pp. 687
- POSCAR: pp. 687

# Tetragonal PZT [$Pb(Zr_xTi_{1-x})O_3$] Structure: A3BC_tP5_99_bc_a_b

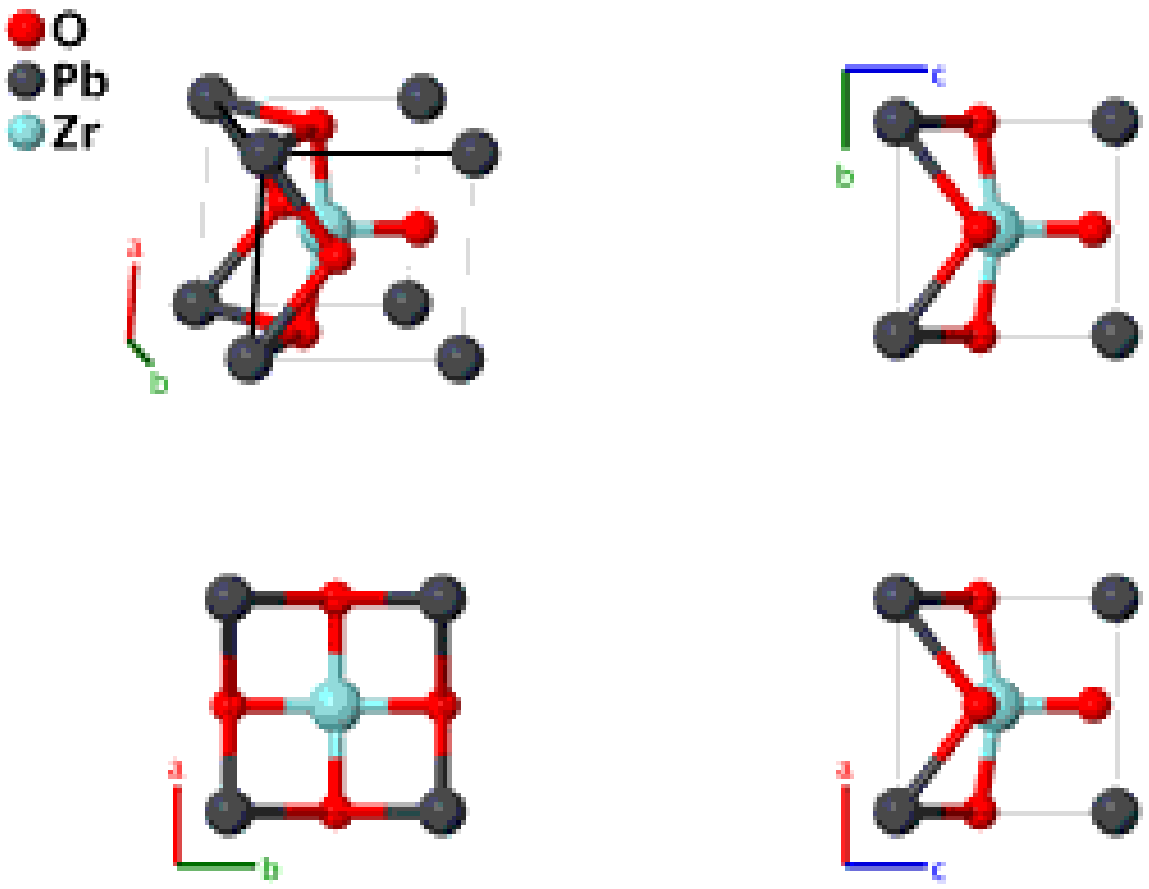

| | | |
|---|---|---|
| **Prototype** | : | $Pb(Zr_{0.52}Ti_{0.48})O_3$ |
| **AFLOW prototype label** | : | A3BC_tP5_99_bc_a_b |
| ***Strukturbericht* designation** | : | None |
| **Pearson symbol** | : | tP5 |
| **Space group number** | : | 99 |
| **Space group symbol** | : | P4mm |
| **AFLOW prototype command** | : | `aflow --proto=A3BC_tP5_99_bc_a_b` `--params=`$a, c/a, z_1, z_2, z_3, z_4$ |

- This is a tetragonal ferroelectric distortion of the perovskite structure. In $PbZr_xTi_{1-x}O_3$ (aka PZT) it is found for $x < 0.52$. Although the first (2b) site is nearly equally occupied by Zr and Ti atoms, the pictures use Zr atoms. Compare this to the monoclinic PZT structure. To recover the cubic perovskite structure, take $c = a$, $z_1 = 0$, $z_2 = 1/2$, $z_3 = 0$, $z_4 = 1/2$.

**Simple Tetragonal primitive vectors:**

$$\mathbf{a}_1 = a\,\hat{\mathbf{x}}$$
$$\mathbf{a}_2 = a\,\hat{\mathbf{y}}$$
$$\mathbf{a}_3 = c\,\hat{\mathbf{z}}$$

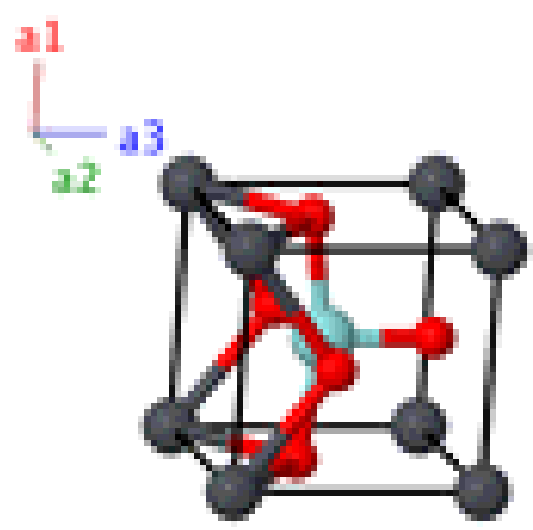

**Basis vectors:**

| | | Lattice Coordinates | | Cartesian Coordinates | Wyckoff Position | Atom Type |
|---|---|---|---|---|---|---|
| $\mathbf{B}_1$ | = | $z_1\,\mathbf{a}_3$ | = | $z_1\,c\,\hat{\mathbf{z}}$ | (1$a$) | Pb |
| $\mathbf{B}_2$ | = | $\frac{1}{2}\,\mathbf{a}_1 + \frac{1}{2}\,\mathbf{a}_2 + z_2\,\mathbf{a}_3$ | = | $\frac{1}{2}\,a\,\hat{\mathbf{x}} + \frac{1}{2}\,a\,\hat{\mathbf{y}} + z_2\,c\,\hat{\mathbf{z}}$ | (1$b$) | O I |
| $\mathbf{B}_3$ | = | $\frac{1}{2}\,\mathbf{a}_1 + \frac{1}{2}\,\mathbf{a}_2 + z_3\,\mathbf{a}_3$ | = | $\frac{1}{2}\,a\,\hat{\mathbf{x}} + \frac{1}{2}\,a\,\hat{\mathbf{y}} + z_3\,c\,\hat{\mathbf{z}}$ | (1$b$) | Zr |
| $\mathbf{B_4}$ | = | $\frac{1}{2}\,\mathbf{a_1} + z_4\,\mathbf{a_3}$ | = | $\frac{1}{2}\,a\,\mathbf{\hat{x}} + z_4\,c\,\mathbf{\hat{z}}$ | (2$c$) | O II |
| $\mathbf{B_5}$ | = | $\frac{1}{2}\,\mathbf{a_2} + z_4\,\mathbf{a_3}$ | = | $\frac{1}{2}\,a\,\mathbf{\hat{y}} + z_4\,c\,\mathbf{\hat{z}}$ | (2$c$) | O II |

---

---

**Geometry files:**

- CIF: pp. 688
- POSCAR: pp. 688

# $BaS_3$ ($D0_{17}$) Structure: AB3_tP8_113_a_ce

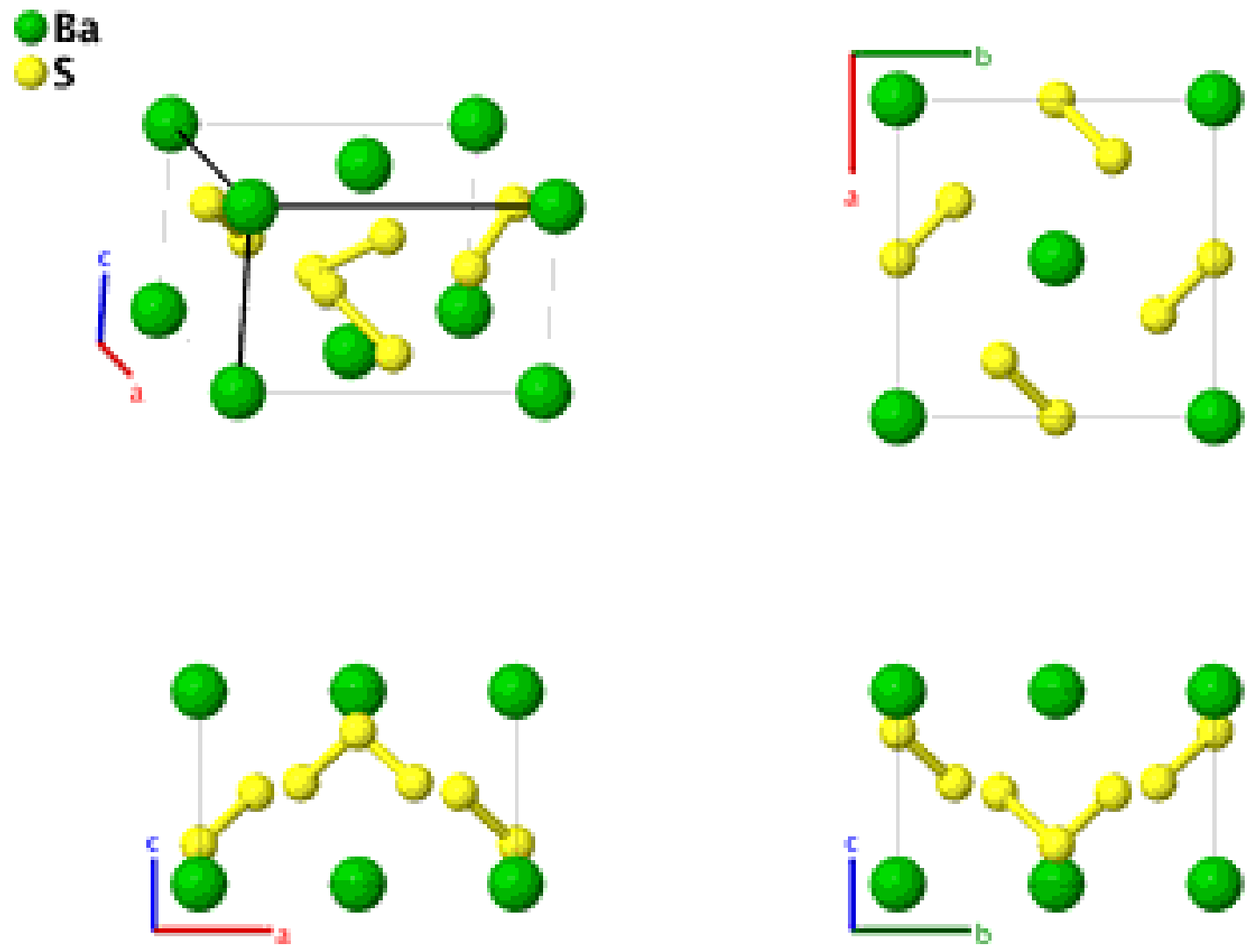

| | | |
|---|---|---|
| **Prototype** | : | $BaS_3$ |
| **AFLOW prototype label** | : | AB3_tP8_113_a_ce |
| ***Strukturbericht* designation** | : | $D0_{17}$ |
| **Pearson symbol** | : | tP8 |
| **Space group number** | : | 113 |
| **Space group symbol** | : | $P\bar{4}2_1m$ |
| **AFLOW prototype command** | : | `aflow --proto=AB3_tP8_113_a_ce`<br>`--params=`$a, c/a, z_2, x_3, z_3$ |

**Other compounds with this structure:**

- $AgDyTe_2$, $AgHgTe_2$, $AgErTe_2$, $AgTe_2Tm$, $AgGdTe_2$, $AgTe_2Y$, $BaSe_3$, $BaTe_3$

- Not to be confused with the other $BaS_3$ structure, which has space group $P2_12_12$ (#18).

**Simple Tetragonal primitive vectors:**

$$\mathbf{a}_1 = a\,\hat{\mathbf{x}}$$
$$\mathbf{a}_2 = a\,\hat{\mathbf{y}}$$
$$\mathbf{a}_3 = c\,\hat{\mathbf{z}}$$

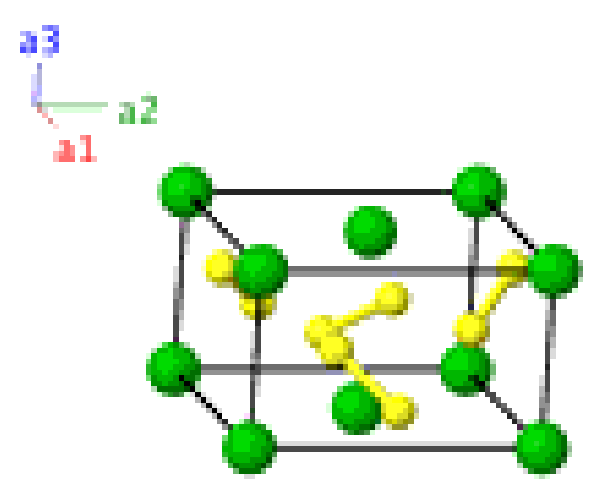

**Basis vectors:**

| | | Lattice Coordinates | | Cartesian Coordinates | Wyckoff Position | Atom Type |
|---|---|---|---|---|---|---|
| $\mathbf{B}_1$ | = | $0\,\mathbf{a}_1 + 0\,\mathbf{a}_2 + 0\,\mathbf{a}_3$ | = | $0\hat{\mathbf{x}} + 0\hat{\mathbf{y}} + 0\hat{\mathbf{z}}$ | $(2a)$ | Ba |
| $\mathbf{B}_2$ | = | $\frac{1}{2}\,\mathbf{a}_1 + \frac{1}{2}\,\mathbf{a}_2$ | = | $\frac{1}{2}\,a\,\hat{\mathbf{x}} + \frac{1}{2}\,a\,\hat{\mathbf{y}}$ | (2$a$) | Ba |
| $\mathbf{B}_3$ | = | $\frac{1}{2}\,\mathbf{a}_2 + z_2\,\mathbf{a}_3$ | = | $\frac{1}{2}\,a\,\hat{\mathbf{y}} + z_2\,c\,\hat{\mathbf{z}}$ | (2$c$) | S I |
| $\mathbf{B}_4$ | = | $\frac{1}{2}\,\mathbf{a}_1 - z_2\,\mathbf{a}_3$ | = | $\frac{1}{2}\,a\,\hat{\mathbf{x}} - z_2\,c\,\hat{\mathbf{z}}$ | (2$c$) | S I |
| $\mathbf{B}_5$ | = | $x_3\,\mathbf{a}_1 + \left(\frac{1}{2} + x_3\right)\,\mathbf{a}_2 + z_3\,\mathbf{a}_3$ | = | $x_3\,a\,\hat{\mathbf{x}} + \left(\frac{1}{2} + x_3\right)\,a\,\hat{\mathbf{y}} + z_3\,c\,\hat{\mathbf{z}}$ | (4$e$) | S II |
| $\mathbf{B}_6$ | = | $-x_3\,\mathbf{a}_1 + \left(\frac{1}{2} - x_3\right)\,\mathbf{a}_2 + z_3\,\mathbf{a}_3$ | = | $-x_3\,a\,\hat{\mathbf{x}} + \left(\frac{1}{2} - x_3\right)\,a\,\hat{\mathbf{y}} + z_3\,c\,\hat{\mathbf{z}}$ | (4$e$) | S II |
| $\mathbf{B}_7$ | = | $\left(\frac{1}{2} + x_3\right)\,\mathbf{a}_1 - x_3\,\mathbf{a}_2 - z_3\,\mathbf{a}_3$ | = | $\left(\frac{1}{2} + x_3\right)\,a\,\hat{\mathbf{x}} - x_3\,a\,\hat{\mathbf{y}} - z_3\,c\,\hat{\mathbf{z}}$ | (4$e$) | S II |
| $\mathbf{B}_8$ | = | $\left(\frac{1}{2} - x_3\right)\,\mathbf{a}_1 + x_3\,\mathbf{a}_2 - z_3\,\mathbf{a}_3$ | = | $\left(\frac{1}{2} - x_3\right)\,a\,\hat{\mathbf{x}} + x_3\,a\,\hat{\mathbf{y}} - z_3\,c\,\hat{\mathbf{z}}$ | (4$e$) | S II |

---

**Found in:**

---

**Geometry files:**

- CIF: pp. 688
- POSCAR: pp. 688

# Stannite ($Cu_2FeS_4Sn$, $H2_6$) Structure: A2BC4D_tI16_121_d_a_i_b

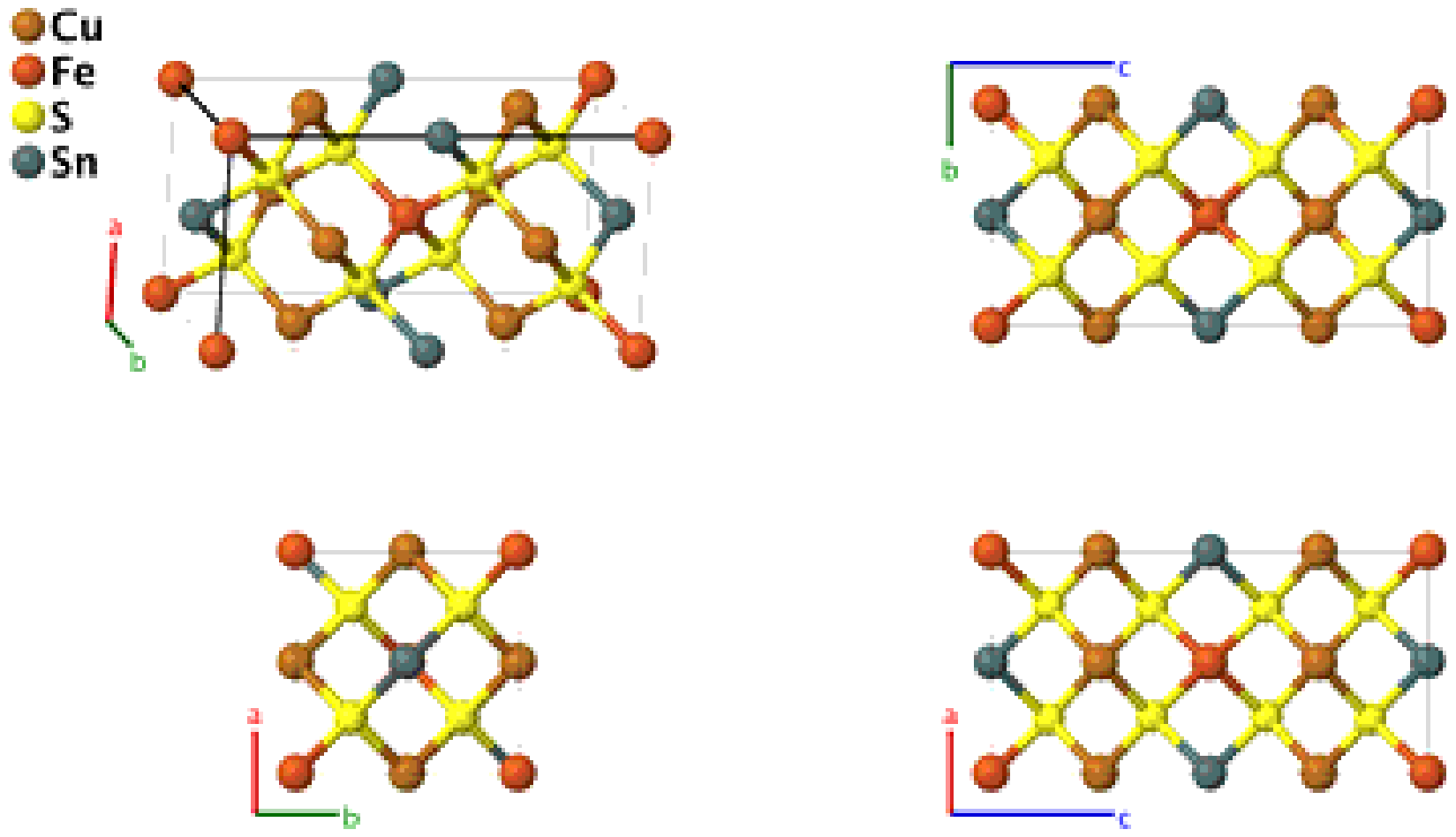

**Prototype** : $Cu_2FeS_4Sn$

**AFLOW prototype label** : A2BC4D_tI16_121_d_a_i_b

***Strukturbericht* designation** : $H2_6$

**Pearson symbol** : tI16

**Space group number** : 121

**Space group symbol** : $I\bar{4}2m$

**AFLOW prototype command** : `aflow --proto=A2BC4D_tI16_121_d_a_i_b --params=`$a, c/a, x_4, z_4$

---

**Other compounds with this structure:**

- $Cu_2CdSe_4Sn$, $CoCu_2S_4Sn$, $Cu_2GeHgS_4$, $Cu_2HgS_4Sn$, $Ag_2FeS_4Sn$

---

- If $c = 2a$, $x = 1/4$, and $z = 3/8$, the atoms are on the sites of the diamond (A4) structure. If, in addition, the Cu, Fe, and Sn atoms are replaced by a single atom type, the crystal reduces to the zincblende (B3) structure.

---

**Body-centered Tetragonal primitive vectors:**

$$\mathbf{a}_1 = -\frac{1}{2}\,a\,\hat{\mathbf{x}} + \frac{1}{2}\,a\,\hat{\mathbf{y}} + \frac{1}{2}\,c\,\hat{\mathbf{z}}$$
$$\mathbf{a}_2 = \frac{1}{2}\,a\,\hat{\mathbf{x}} - \frac{1}{2}\,a\,\hat{\mathbf{y}} + \frac{1}{2}\,c\,\hat{\mathbf{z}}$$
$$\mathbf{a}_3 = \frac{1}{2}\,a\,\hat{\mathbf{x}} + \frac{1}{2}\,a\,\hat{\mathbf{y}} - \frac{1}{2}\,c\,\hat{\mathbf{z}}$$

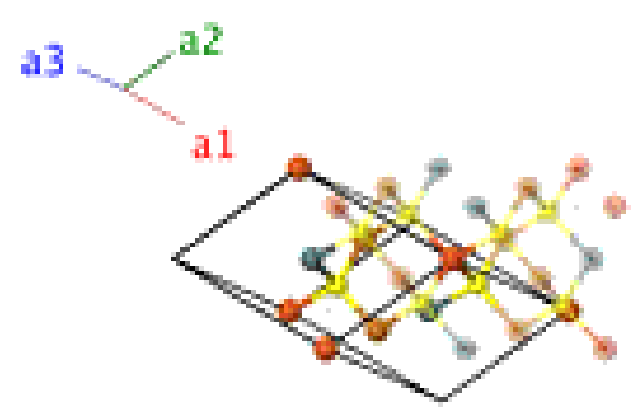

---

**Basis vectors:**

| | | Lattice Coordinates | | Cartesian Coordinates | Wyckoff Position | Atom Type |
|---|---|---|---|---|---|---|
| $\mathbf{B}_1$ | = | $0\,\mathbf{a}_1 + 0\,\mathbf{a}_2 + 0\,\mathbf{a}_3$ | = | $0\hat{\mathbf{x}} + 0\hat{\mathbf{y}} + 0\hat{\mathbf{z}}$ | $(2a)$ | Fe |
| $\mathbf{B}_2$ | = | $\frac{1}{2}\,\mathbf{a}_1 + \frac{1}{2}\,\mathbf{a}_2$ | = | $+\frac{1}{2}\,c\,\hat{\mathbf{z}}$ | $(2b)$ | Sn |
| $\mathbf{B_3}$ | = | $\frac{3}{4}\,\mathbf{a_1} + \frac{1}{4}\,\mathbf{a_2} + \frac{1}{2}\,\mathbf{a_3}$ | = | $\frac{1}{2}\,a\,\mathbf{\hat{y}} + \frac{1}{4}\,c\,\mathbf{\hat{z}}$ | (4$d$) | Cu |
| $\mathbf{B_4}$ | = | $\frac{1}{4}\,\mathbf{a_1} + \frac{3}{4}\,\mathbf{a_2} + \frac{1}{2}\,\mathbf{a_3}$ | = | $\frac{1}{2}\,a\,\mathbf{\hat{x}} + \frac{1}{4}\,c\,\mathbf{\hat{z}}$ | (4$d$) | Cu |
| $\mathbf{B_5}$ | = | $(x_4 + z_4)\,\mathbf{a_1} + (x_4 + z_4)\,\mathbf{a_2} + 2x_4\,\mathbf{a_3}$ | = | $x_4\,a\,\mathbf{\hat{x}} + x_4\,a\,\mathbf{\hat{y}} + z_4\,c\,\mathbf{\hat{z}}$ | (8$i$) | S |
| $\mathbf{B_6}$ | = | $(z_4 - x_4)\,\mathbf{a_1} + (z_4 - x_4)\,\mathbf{a_2} - 2x_4\,\mathbf{a_3}$ | = | $-x_4\,a\,\mathbf{\hat{x}} - x_4\,a\,\mathbf{\hat{y}} + z_4\,c\,\mathbf{\hat{z}}$ | (8$i$) | S |
| $\mathbf{B_7}$ | = | $-(x_4 + z_4)\,\mathbf{a_1} + (x_4 - z_4)\,\mathbf{a_2}$ | = | $x_4\,a\,\mathbf{\hat{x}} - x_4\,a\,\mathbf{\hat{y}} - z_4\,c\,\mathbf{\hat{z}}$ | (8$i$) | S |
| $\mathbf{B_8}$ | = | $(x_4 - z_4)\,\mathbf{a_1} - (x_4 + z_4)\,\mathbf{a_2}$ | = | $-x_4\,a\,\mathbf{\hat{x}} + x_4\,a\,\mathbf{\hat{y}} - z_4\,c\,\mathbf{\hat{z}}$ | (8$i$) | S |

---

---

**Geometry files:**

- CIF: pp. 688
- POSCAR: pp. 689

# Chalcopyrite ($CuFeS_2$, $E1_1$) Structure: ABC2_tI16_122_a_b_d

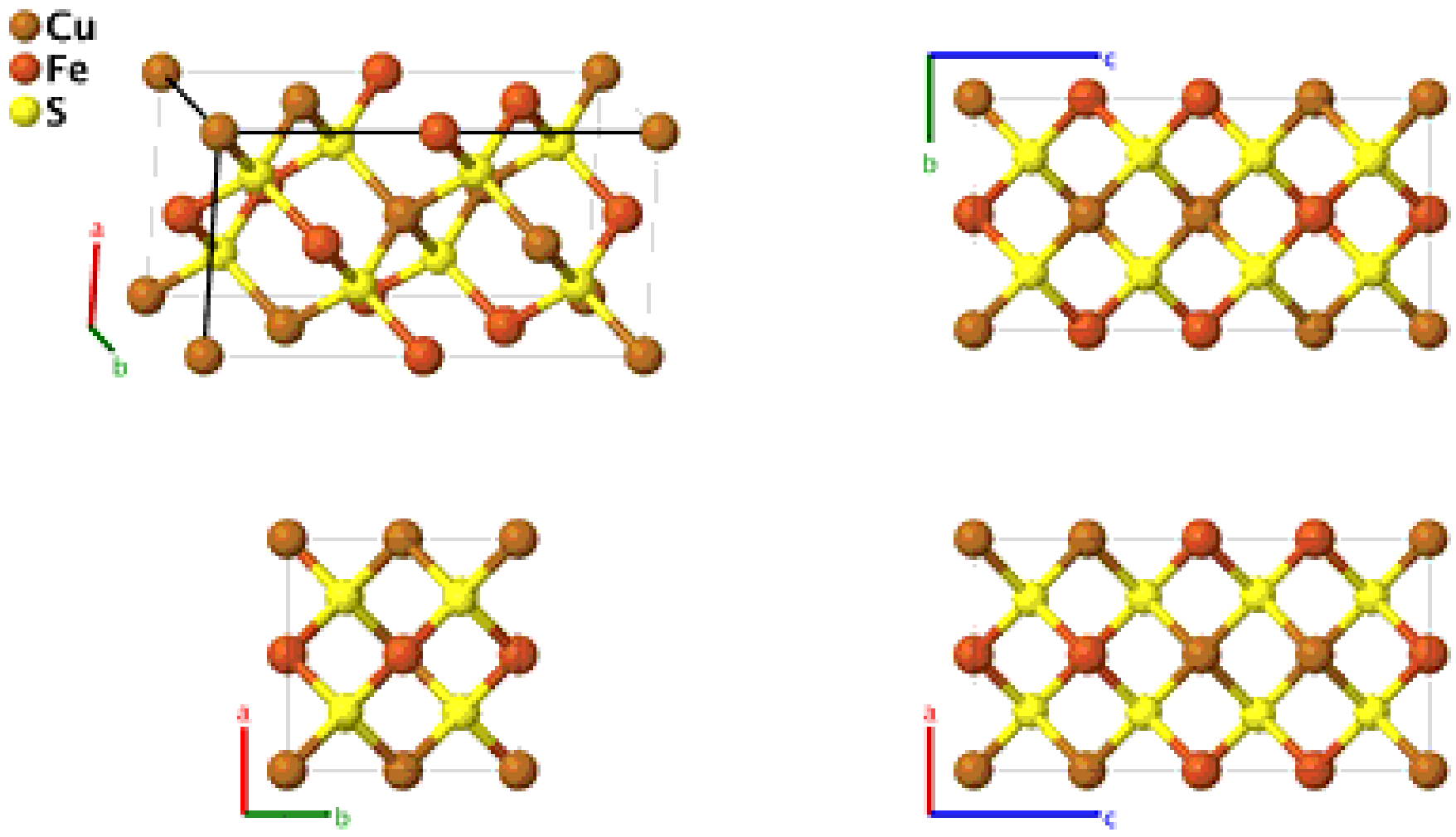

| | | |
|---|---|---|
| **Prototype** | : | $CuFeS_2$ |
| **AFLOW prototype label** | : | ABC2_tI16_122_a_b_d |
| ***Strukturbericht* designation** | : | $E1_1$ |
| **Pearson symbol** | : | tI16 |
| **Space group number** | : | 122 |
| **Space group symbol** | : | $I\bar{4}2d$ |
| **AFLOW prototype command** | : | `aflow --proto=ABC2_tI16_122_a_b_d --params=`$a, c/a, x_3$ |

**Other compounds with this structure:**

- $CuInS_2$, $CuInSe_2$

- When $c = 2a$ and $x_3 = 1/8$ the atoms are on the sites of the diamond (A4) structure. In this case, if we replace the Fe atoms by Cu, we get the zincblende (B3) structure.

**Body-centered Tetragonal primitive vectors:**

$$
\begin{aligned}
\mathbf{a}_1 &= -\tfrac{1}{2}\,a\,\hat{\mathbf{x}} + \tfrac{1}{2}\,a\,\hat{\mathbf{y}} + \tfrac{1}{2}\,c\,\hat{\mathbf{z}} \\
\mathbf{a}_2 &= \tfrac{1}{2}\,a\,\hat{\mathbf{x}} - \tfrac{1}{2}\,a\,\hat{\mathbf{y}} + \tfrac{1}{2}\,c\,\hat{\mathbf{z}} \\
\mathbf{a}_3 &= \tfrac{1}{2}\,a\,\hat{\mathbf{x}} + \tfrac{1}{2}\,a\,\hat{\mathbf{y}} - \tfrac{1}{2}\,c\,\hat{\mathbf{z}}
\end{aligned}
$$

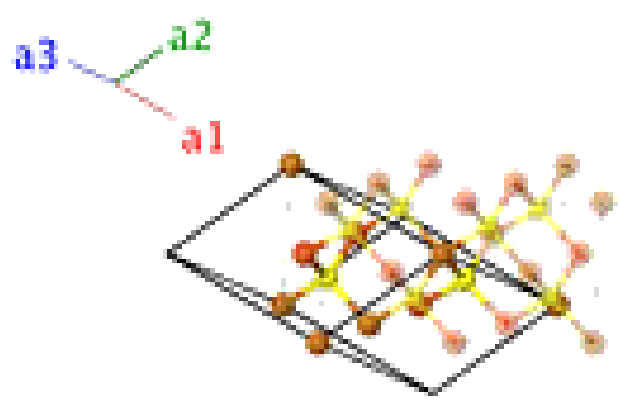

**Basis vectors:**

| | | Lattice Coordinates | | Cartesian Coordinates | Wyckoff Position | Atom Type |
|---|---|---|---|---|---|---|
| $\mathbf{B_1}$ | = | $0\,\mathbf{a_1} + 0\,\mathbf{a_2} + 0\,\mathbf{a_3}$ | = | $0\hat{\mathbf{x}} + 0\hat{\mathbf{y}} + 0\hat{\mathbf{z}}$ | $(4a)$ | Cu |
| $\mathbf{B_2}$ | = | $\frac{3}{4}\,\mathbf{a_1} + \frac{1}{4}\,\mathbf{a_2} + \frac{1}{2}\,\mathbf{a_3}$ | = | $\frac{1}{2}\,a\,\hat{\mathbf{y}} + \frac{1}{4}\,c\,\hat{\mathbf{z}}$ | $(4a)$ | Cu |
| $\mathbf{B_3}$ | = | $\frac{1}{2}\,\mathbf{a_1} + \frac{1}{2}\,\mathbf{a_2}$ | = | $\frac{1}{2}\,c\,\hat{\mathbf{z}}$ | $(4b)$ | Fe |
| $\mathbf{B_4}$ | = | $\frac{1}{4}\,\mathbf{a_1} + \frac{3}{4}\,\mathbf{a_2} + \frac{1}{2}\,\mathbf{a_3}$ | = | $\frac{1}{2}\,a\,\hat{\mathbf{x}} + \frac{1}{4}\,c\,\hat{\mathbf{z}}$ | (4$b$) | Fe |
| $\mathbf{B_5}$ | = | $\frac{3}{8}\,\mathbf{a_1} + \left(\frac{1}{8} + x_3\right)\,\mathbf{a_2} + \left(\frac{1}{4} + x_3\right)\,\mathbf{a_3}$ | = | $x_3\,a\,\hat{\mathbf{x}} + \frac{1}{4}\,a\,\hat{\mathbf{y}} + \frac{1}{8}\,c\,\hat{\mathbf{z}}$ | (8$d$) | S |
| $\mathbf{B_6}$ | = | $\frac{7}{8}\,\mathbf{a_1} + \left(\frac{1}{8} - x_3\right)\,\mathbf{a_2} + \left(\frac{3}{4} - x_3\right)\,\mathbf{a_3}$ | = | $-x_3\,a\,\hat{\mathbf{x}} + \frac{3}{4}\,a\,\hat{\mathbf{y}} + \frac{1}{8}\,c\,\hat{\mathbf{z}}$ | (8$d$) | S |
| $\mathbf{B_7}$ | = | $\left(\frac{7}{8} - x_3\right)\,\mathbf{a_1} + \frac{1}{8}\,\mathbf{a_2} + \left(\frac{1}{4} - x_3\right)\,\mathbf{a_3}$ | = | $\frac{3}{4}\,a\,\hat{\mathbf{x}} + \left(\frac{1}{2} - x_3\right)\,a\,\hat{\mathbf{y}} + \frac{3}{8}\,c\,\hat{\mathbf{z}}$ | (8$d$) | S |
| $\mathbf{B_8}$ | = | $\left(\frac{7}{8} + x_3\right)\,\mathbf{a_1} + \frac{5}{8}\,\mathbf{a_2} + \left(\frac{3}{4} + x_3\right)\,\mathbf{a_3}$ | = | $\frac{1}{4}\,a\,\hat{\mathbf{x}} + \left(\frac{1}{2} + x_3\right)\,a\,\hat{\mathbf{y}} + \frac{3}{8}\,c\,\hat{\mathbf{z}}$ | (8$d$) | S |

---

**Found in:**

---

**Geometry files:**

- CIF: pp. 689
- POSCAR: pp. 689

# $HoCoGa_5$ Structure: AB5C_tP7_123_b_ci_a

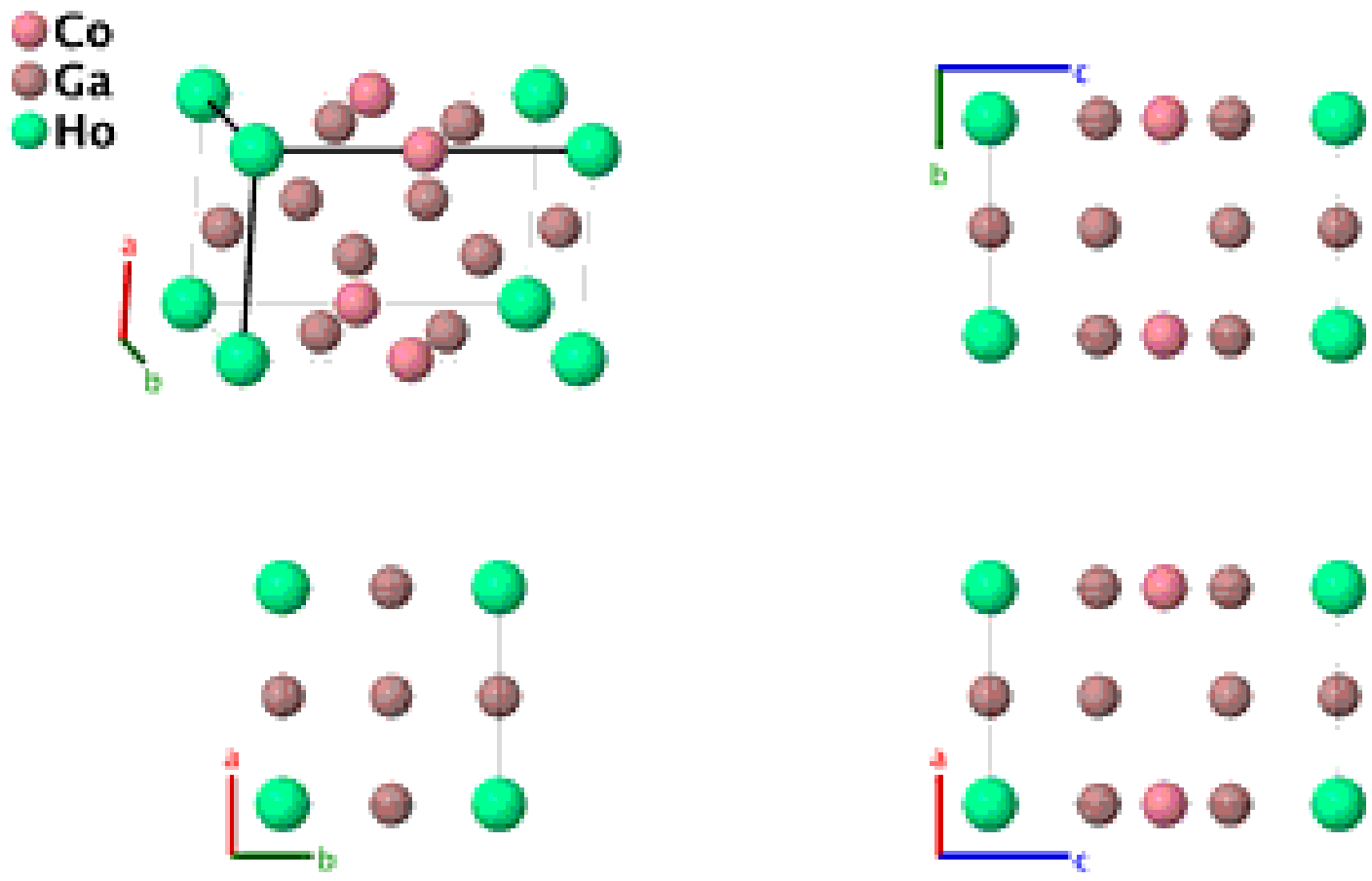

| | | |
|---|---|---|
| **Prototype** | : | $HoCoGa_5$ |
| **AFLOW prototype label** | : | AB5C_tP7_123_b_ci_a |
| ***Strukturbericht* designation** | : | None |
| **Pearson symbol** | : | tP7 |
| **Space group number** | : | 123 |
| **Space group symbol** | : | P4/mmm |
| **AFLOW prototype command** | : | `aflow --proto=AB5C_tP7_123_b_ci_a --params=`$a, c/a, z_4$ |

**Simple Tetragonal primitive vectors:**

$$\mathbf{a}_1 = a\,\hat{\mathbf{x}}$$
$$\mathbf{a}_2 = a\,\hat{\mathbf{y}}$$
$$\mathbf{a}_3 = c\,\hat{\mathbf{z}}$$

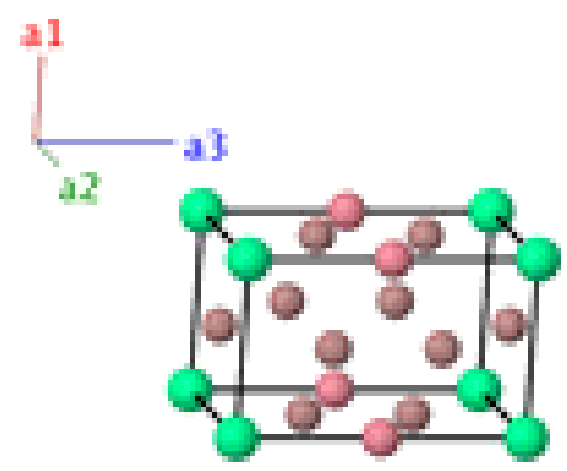

**Basis vectors:**

| | | Lattice Coordinates | | Cartesian Coordinates | Wyckoff Position | Atom Type |
|---|---|---|---|---|---|---|
| $\mathbf{B}_1$ | = | $0\,\mathbf{a}_1 + 0\,\mathbf{a}_2 + 0\,\mathbf{a}_3$ | = | $0\hat{\mathbf{x}} + 0\hat{\mathbf{y}} + 0\hat{\mathbf{z}}$ | (1*a*) | Ho |
| $\mathbf{B}_2$ | = | $\frac{1}{2}\,\mathbf{a}_3$ | = | $\frac{1}{2}\,c\,\hat{\mathbf{z}}$ | (1*b*) | Co |
| $\mathbf{B}_3$ | = | $\frac{1}{2}\,\mathbf{a}_1 + \frac{1}{2}\,\mathbf{a}_2$ | = | $\frac{1}{2}\,a\,\hat{\mathbf{x}} + \frac{1}{2}\,a\,\hat{\mathbf{y}}$ | (1*c*) | Ga I |
| $\mathbf{B}_4$ | = | $\frac{1}{2}\,\mathbf{a}_2 + z_4\,\mathbf{a}_3$ | = | $\frac{1}{2}\,a\,\hat{\mathbf{y}} + z_4\,c\,\hat{\mathbf{z}}$ | (4*i*) | Ga II |
| $\mathbf{B}_5$ | = | $\frac{1}{2}\,\mathbf{a}_1 + z_4\,\mathbf{a}_3$ | = | $\frac{1}{2}\,a\,\hat{\mathbf{x}} + z_4\,c\,\hat{\mathbf{z}}$ | (4*i*) | Ga II |
| $\mathbf{B}_6$ | = | $\frac{1}{2}\,\mathbf{a}_2 - z_4\,\mathbf{a}_3$ | = | $\frac{1}{2}\,a\,\hat{\mathbf{y}} - z_4\,c\,\hat{\mathbf{z}}$ | (4*i*) | Ga II |
| $\mathbf{B}_7$ | = | $\frac{1}{2}\,\mathbf{a}_1 - z_4\,\mathbf{a}_3$ | = | $\frac{1}{2}\,a\,\hat{\mathbf{x}} - z_4\,c\,\hat{\mathbf{z}}$ | (4*i*) | Ga II |

**Found in:**

---

**Geometry files:**

- CIF: pp. 689
- POSCAR: pp. 690

# $CuTi_3$ ($L6_0$) Structure: AB3_tP4_123_a_ce

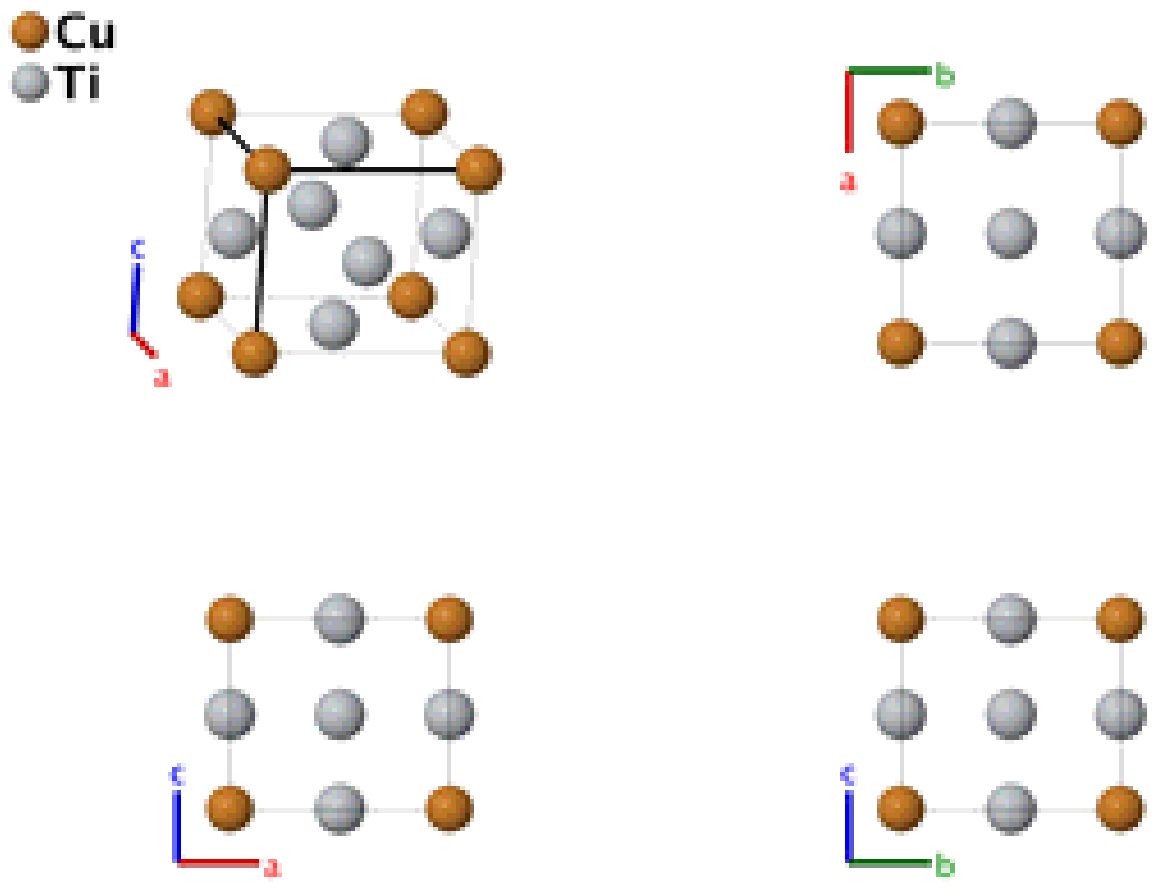

| | | |
|---|---|---|
| **Prototype** | : | $CuTi_3$ |
| **AFLOW prototype label** | : | AB3_tP4_123_a_ce |
| ***Strukturbericht* designation** | : | $L6_0$ |
| **Pearson symbol** | : | tP4 |
| **Space group number** | : | 123 |
| **Space group symbol** | : | P4/mmm |
| **AFLOW prototype command** | : | `aflow --proto=AB3_tP4_123_a_ce --params=`$a, c/a$ |

- This is a tetragonal distortion of the $L1_2$ ($Cu_3Au$) structure. When $c = a$ the atoms are at the positions of a face-centered cubic lattice. If we replace the Ti I atom by Cu, then the system reduces to the $L1_0$ (CuAu) structure. Interestingly, (Massalski, 1986) lists no stable or metastable structures with composition $CuTi_3$.

**Simple Tetragonal primitive vectors:**

$$\mathbf{a}_1 = a\,\hat{\mathbf{x}}$$
$$\mathbf{a}_2 = a\,\hat{\mathbf{y}}$$
$$\mathbf{a}_3 = c\,\hat{\mathbf{z}}$$

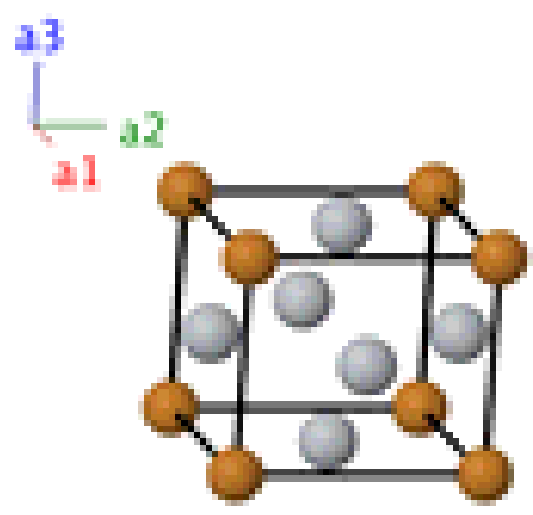

**Basis vectors:**

| | | Lattice Coordinates | | Cartesian Coordinates | Wyckoff Position | Atom Type |
|---|---|---|---|---|---|---|
| $\mathbf{B}_1$ | = | $0\,\mathbf{a}_1 + 0\,\mathbf{a}_2 + 0\,\mathbf{a}_3$ | = | $0\hat{\mathbf{x}} + 0\hat{\mathbf{y}} + 0\hat{\mathbf{z}}$ | $(1a)$ | Cu |
| $\mathbf{B}_2$ | = | $\frac{1}{2}\,\mathbf{a}_1 + \frac{1}{2}\,\mathbf{a}_2$ | = | $\frac{1}{2}\,a\,\hat{\mathbf{x}} + \frac{1}{2}\,a\,\hat{\mathbf{y}}$ | $(1c)$ | Ti I |
| $\mathbf{B}_3$ | = | $\frac{1}{2}\,\mathbf{a}_2 + \frac{1}{2}\,\mathbf{a}_3$ | = | $\frac{1}{2}\,a\,\hat{\mathbf{y}} + \frac{1}{2}\,c\,\hat{\mathbf{z}}$ | $(2e)$ | Ti II |
| $\mathbf{B}_4$ | = | $\frac{1}{2}\,\mathbf{a}_1 + \frac{1}{2}\,\mathbf{a}_3$ | = | $\frac{1}{2}\,a\,\hat{\mathbf{x}} + \frac{1}{2}\,c\,\hat{\mathbf{z}}$ | $(2e)$ | Ti II |

**Geometry files:**

- CIF: pp. 690
- POSCAR: pp. 690

# CuAu ($L1_0$) Structure: AB_tP2_123_a_d

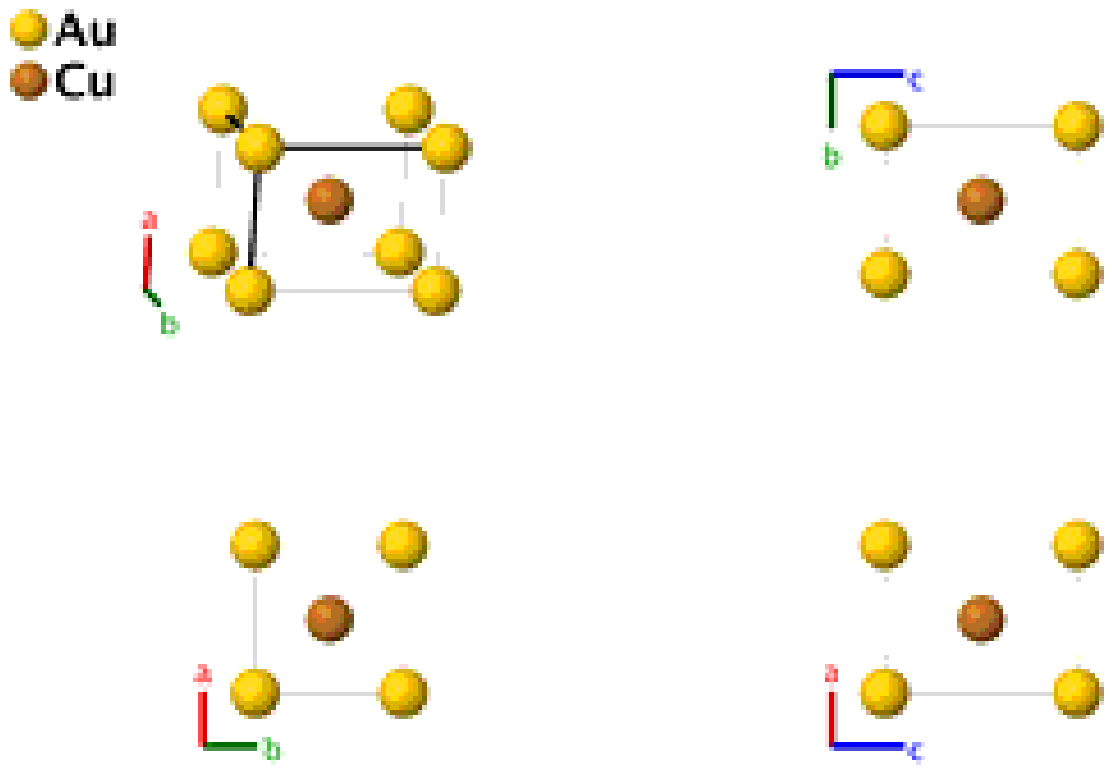

| | | |
|---|---|---|
| **Prototype** | : | CuAu |
| **AFLOW prototype label** | : | AB_tP2_123_a_d |
| ***Strukturbericht* designation** | : | $L1_0$ |
| **Pearson symbol** | : | tP2 |
| **Space group number** | : | 123 |
| **Space group symbol** | : | P4/mmm |
| **AFLOW prototype command** | : | `aflow --proto=AB_tP2_123_a_d --params=`$a, c/a$ |

- When $c = \sqrt{2}a$ the atoms are at the positions of a face-centered cubic lattice. When $c = a$ the atoms are at the positions of a body-centered cubic lattice.

**Simple Tetragonal primitive vectors:**

$$\mathbf{a}_1 = a\,\hat{\mathbf{x}}$$
$$\mathbf{a}_2 = a\,\hat{\mathbf{y}}$$
$$\mathbf{a}_3 = c\,\hat{\mathbf{z}}$$

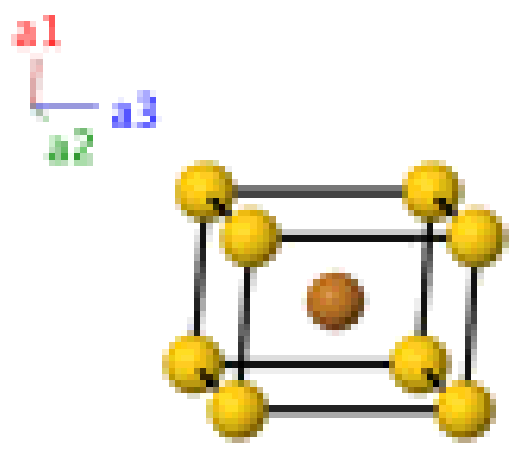

**Basis vectors:**

| | | Lattice Coordinates | | Cartesian Coordinates | Wyckoff Position | Atom Type |
|---|---|---|---|---|---|---|
| $\mathbf{B}_1$ | = | $0\,\mathbf{a}_1 + 0\,\mathbf{a}_2 + 0\,\mathbf{a}_3$ | = | $0\hat{\mathbf{x}} + 0\hat{\mathbf{y}} + 0\hat{\mathbf{z}}$ | $(1a)$ | Au |
| $\mathbf{B}_2$ | = | $\frac{1}{2}\,\mathbf{a}_1 + \frac{1}{2}\,\mathbf{a}_2 + \frac{1}{2}\,\mathbf{a}_3$ | = | $\frac{1}{2}\,a\,\hat{\mathbf{x}} + \frac{1}{2}\,a\,\hat{\mathbf{y}} + \frac{1}{2}\,c\,\hat{\mathbf{z}}$ | $(1d)$ | Cu |

**Found in:**

---

**Geometry files:**

- CIF: pp. 690
- POSCAR: pp. 691

# $CaCuO_2$ Structure: ABC2_tP4_123_d_a_f

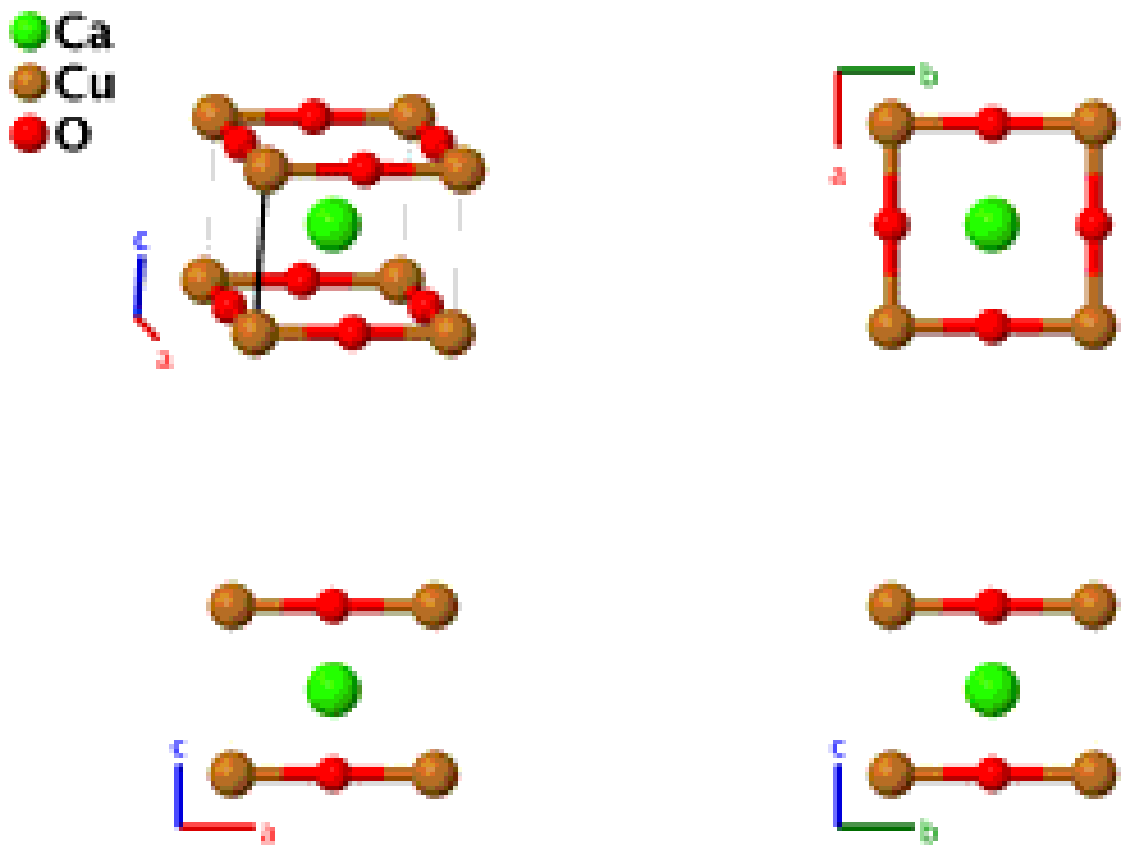

| | | |
|---|---|---|
| **Prototype** | : | $CaCuO_2$ |
| **AFLOW prototype label** | : | ABC2_tP4_123_d_a_f |
| ***Strukturbericht* designation** | : | None |
| **Pearson symbol** | : | tP4 |
| **Space group number** | : | 123 |
| **Space group symbol** | : | P4/mmm |
| **AFLOW prototype command** | : | `aflow --proto=ABC2_tP4_123_d_a_f`<br>`--params=`$a, c/a$ |

- As noted in (Siegrist, 1988) this is the parent structure of the high-temperature cuprate superconductors.

**Simple Tetragonal primitive vectors:**

$$\mathbf{a}_1 = a\,\hat{\mathbf{x}}$$
$$\mathbf{a}_2 = a\,\hat{\mathbf{y}}$$
$$\mathbf{a}_3 = c\,\hat{\mathbf{z}}$$

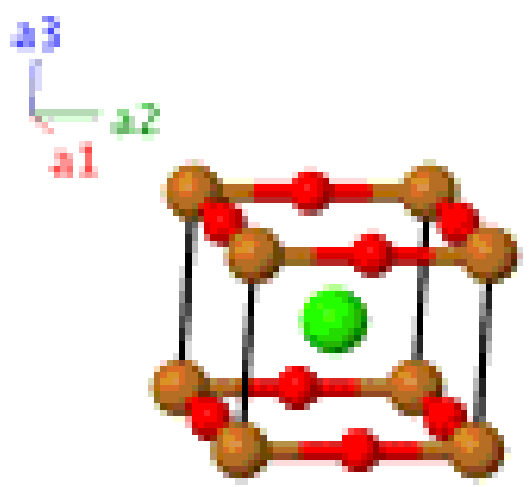

**Basis vectors:**

| | | Lattice Coordinates | | Cartesian Coordinates | Wyckoff Position | Atom Type |
|---|---|---|---|---|---|---|
| $\mathbf{B}_1$ | = | $0\,\mathbf{a}_1 + 0\,\mathbf{a}_2 + 0\,\mathbf{a}_3$ | = | $0\hat{\mathbf{x}} + 0\hat{\mathbf{y}} + 0\hat{\mathbf{z}}$ | (1$a$) | Cu |
| $\mathbf{B}_2$ | = | $\frac{1}{2}\,\mathbf{a}_1 + \frac{1}{2}\,\mathbf{a}_2 + \frac{1}{2}\,\mathbf{a}_3$ | = | $\frac{1}{2}\,a\,\hat{\mathbf{x}} + \frac{1}{2}\,a\,\hat{\mathbf{y}} + \frac{1}{2}\,c\,\hat{\mathbf{z}}$ | (1$d$) | Ca |
| $\mathbf{B}_3$ | = | $\frac{1}{2}\,\mathbf{a}_2$ | = | $\frac{1}{2}\,a\,\hat{\mathbf{y}}$ | (2$f$) | O |
| $\mathbf{B}_4$ | = | $\frac{1}{2}\,\mathbf{a}_1$ | = | $\frac{1}{2}\,a\,\hat{\mathbf{x}}$ | (2$f$) | O |

---

**Geometry files:**

- CIF: pp. 691
- POSCAR: pp. 691

# $Si_2U_3$ ($D5_a$) Structure: A2B3_tP10_127_g_ah

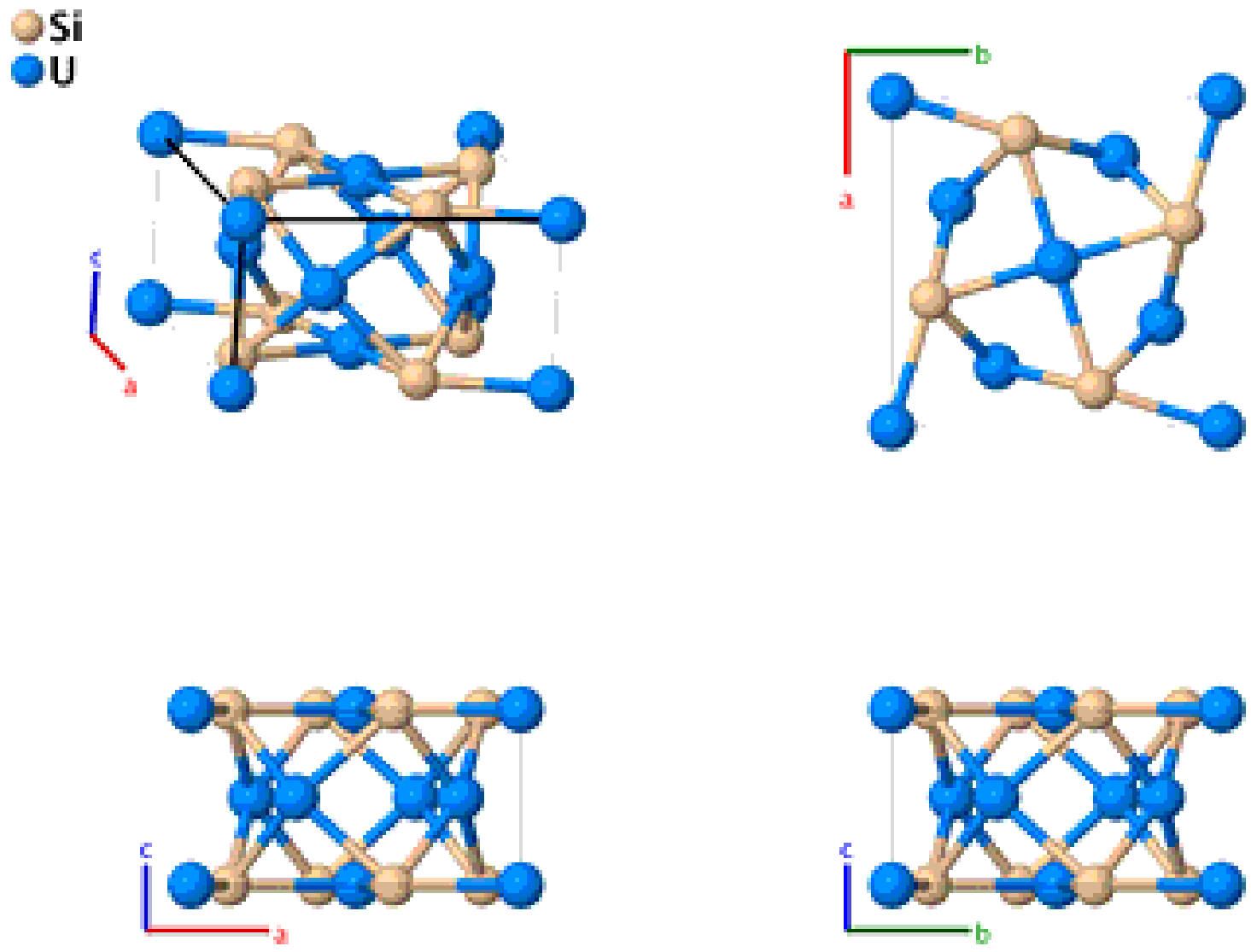

| | | |
|---|---|---|
| **Prototype** | : | $Si_2U_3$ |
| **AFLOW prototype label** | : | A2B3_tP10_127_g_ah |
| ***Strukturbericht* designation** | : | $D5_a$ |
| **Pearson symbol** | : | tP10 |
| **Space group number** | : | 127 |
| **Space group symbol** | : | P4/mbm |
| **AFLOW prototype command** | : | `aflow --proto=A2B3_tP10_127_g_ah --params=`$a, c/a, x_2, x_3$ |

- If we consider the $Si_2$ dimers as a pseudo-atom, then this is a tetragonal distortion of the $Cu_3Au$ ($L1_2$) structure.

**Simple Tetragonal primitive vectors:**

$$\mathbf{a}_1 = a\,\hat{\mathbf{x}}$$
$$\mathbf{a}_2 = a\,\hat{\mathbf{y}}$$
$$\mathbf{a}_3 = c\,\hat{\mathbf{z}}$$

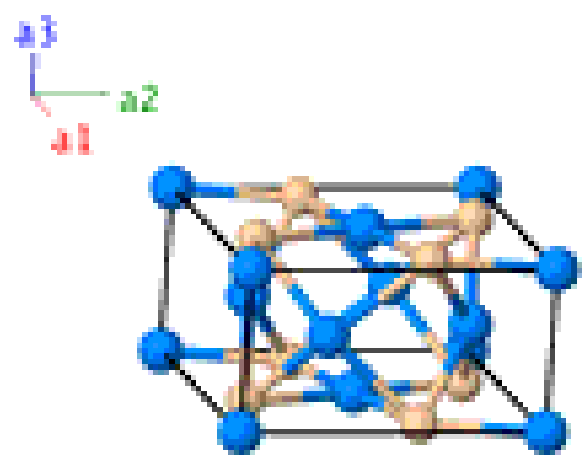

**Basis vectors:**

| | | Lattice Coordinates | | Cartesian Coordinates | Wyckoff Position | Atom Type |
|---|---|---|---|---|---|---|
| $\mathbf{B}_1$ | = | $0\,\mathbf{a}_1 + 0\,\mathbf{a}_2 + 0\,\mathbf{a}_3$ | = | $0\hat{\mathbf{x}} + 0\hat{\mathbf{y}} + 0\hat{\mathbf{z}}$ | $(2a)$ | U I |
| $\mathbf{B}_2$ | = | $\frac{1}{2}\,\mathbf{a}_1 + \frac{1}{2}\,\mathbf{a}_2$ | = | $\frac{1}{2}\,a\,\hat{\mathbf{x}} + \frac{1}{2}\,a\,\hat{\mathbf{y}}$ | $(2a)$ | U I |
| $\mathbf{B}_3$ | = | $x_2\,\mathbf{a}_1 + \left(\frac{1}{2} + x_2\right)\,\mathbf{a}_2$ | = | $x_2\,a\,\hat{\mathbf{x}} + \left(\frac{1}{2} + x_2\right)\,a\,\hat{\mathbf{y}}$ | $(4g)$ | Si |
| $\mathbf{B_4}$ | = | $-x_2\,\mathbf{a_1} + \left(\frac{1}{2} - x_2\right)\,\mathbf{a_2}$ | = | $-x_2\,a\,\mathbf{\hat{x}} + \left(\frac{1}{2} - x_2\right)\,a\,\mathbf{\hat{y}}$ | $(4g)$ | Si |
| $\mathbf{B_5}$ | = | $\left(\frac{1}{2} - x_2\right)\,\mathbf{a_1} + x_2\,\mathbf{a_2}$ | = | $\left(\frac{1}{2} - x_2\right)\,a\,\mathbf{\hat{x}} + x_2\,a\,\mathbf{\hat{y}}$ | $(4g)$ | Si |
| $\mathbf{B_6}$ | = | $\left(\frac{1}{2} + x_2\right)\,\mathbf{a_1} - x_2\,\mathbf{a_2}$ | = | $\left(\frac{1}{2} + x_2\right)\,a\,\mathbf{\hat{x}} - x_2\,a\,\mathbf{\hat{y}}$ | $(4g)$ | Si |
| $\mathbf{B_7}$ | = | $x_3\,\mathbf{a_1} + \left(\frac{1}{2} + x_3\right)\,\mathbf{a_2} + \frac{1}{2}\,\mathbf{a_3}$ | = | $x_3\,a\,\mathbf{\hat{x}} + \left(\frac{1}{2} + x_3\right)\,a\,\mathbf{\hat{y}} + \frac{1}{2}\,c\,\mathbf{\hat{z}}$ | $(4h)$ | U II |
| $\mathbf{B_8}$ | = | $-x_3\,\mathbf{a_1} + \left(\frac{1}{2} - x_3\right)\,\mathbf{a_2} + \frac{1}{2}\,\mathbf{a_3}$ | = | $-x_3\,a\,\mathbf{\hat{x}} + \left(\frac{1}{2} - x_3\right)\,a\,\mathbf{\hat{y}} + \frac{1}{2}\,c\,\mathbf{\hat{z}}$ | $(4h)$ | U II |
| $\mathbf{B_9}$ | = | $\left(\frac{1}{2} - x_3\right)\,\mathbf{a_1} + x_3\,\mathbf{a_2} + \frac{1}{2}\,\mathbf{a_3}$ | = | $\left(\frac{1}{2} - x_3\right)\,a\,\mathbf{\hat{x}} + x_3\,a\,\mathbf{\hat{y}} + \frac{1}{2}\,c\,\mathbf{\hat{z}}$ | $(4h)$ | U II |
| $\mathbf{B_{10}}$ | = | $\left(\frac{1}{2} + x_3\right)\,\mathbf{a_1} - x_3\,\mathbf{a_2} + \frac{1}{2}\,\mathbf{a_3}$ | = | $\left(\frac{1}{2} + x_3\right)\,a\,\mathbf{\hat{x}} - x_3\,a\,\mathbf{\hat{y}} + \frac{1}{2}\,c\,\mathbf{\hat{z}}$ | $(4h)$ | U II |

**Geometry files:**

- CIF: pp. 691
- POSCAR: pp. 692

# AsCuSiZr Structure: ABCD_tP8_129_c_b_a_c

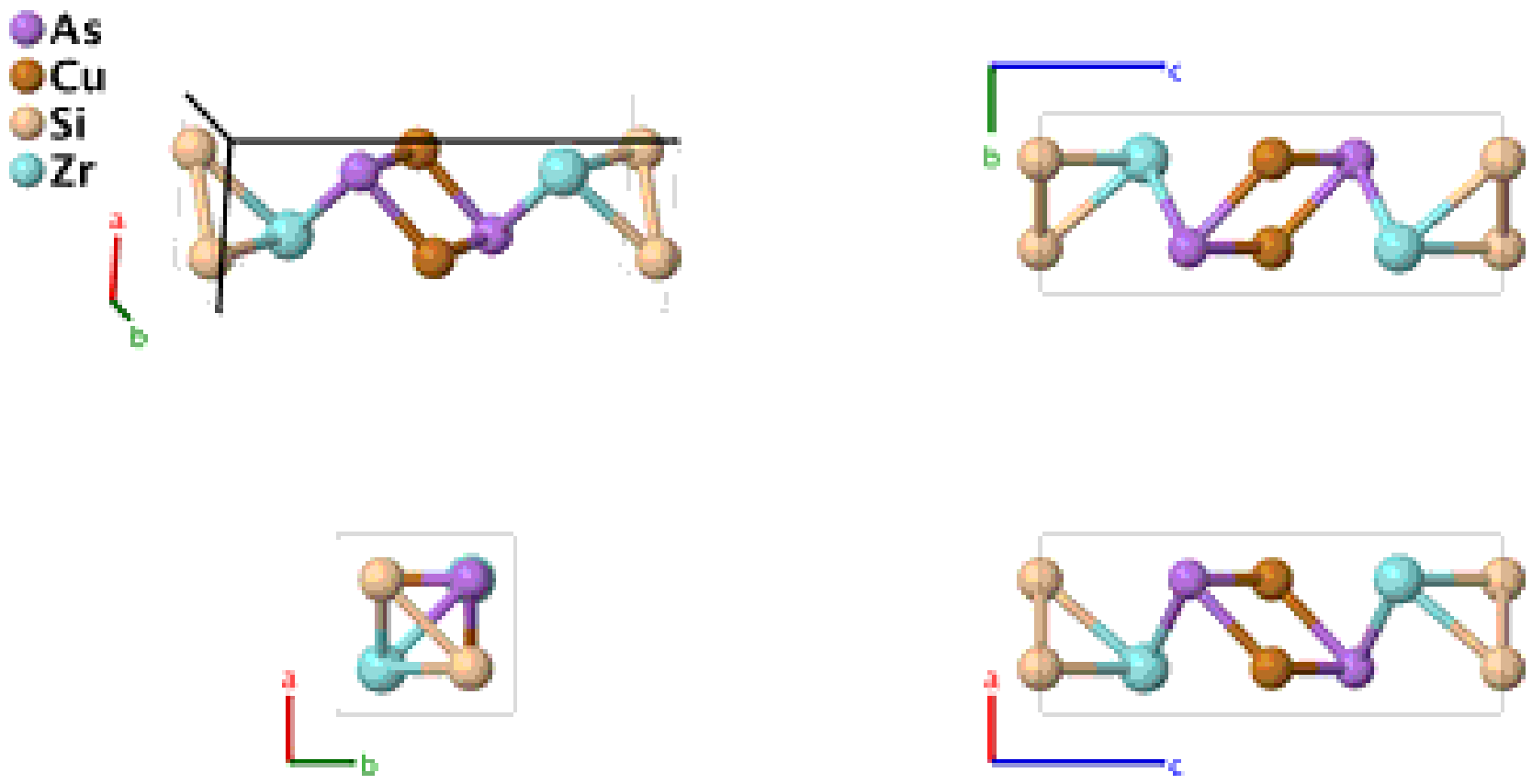

**Prototype** : AsCuSiZr

**AFLOW prototype label** : ABCD_tP8_129_c_b_a_c

***Strukturbericht* designation** : None

**Pearson symbol** : tP8

**Space group number** : 129

**Space group symbol** : P4/nmm

**AFLOW prototype command** : `aflow --proto=ABCD_tP8_129_c_b_a_c`
`--params=`$a, c/a, z_3, z_4$

---

**Other compounds with this structure:**

- LaOFeAs, AsCuHfSi, $As_2CuU$, $Bi_2CuMn$, $CoLiSb_2$, $CuGe_2Hf$, $LaMnSb_2$, $NiPrSb_2$

---

- This is the parent structure for the iron-pnictide superconductors.

---

**Simple Tetragonal primitive vectors:**

$$\mathbf{a}_1 = a\,\hat{\mathbf{x}}$$
$$\mathbf{a}_2 = a\,\hat{\mathbf{y}}$$
$$\mathbf{a}_3 = c\,\hat{\mathbf{z}}$$

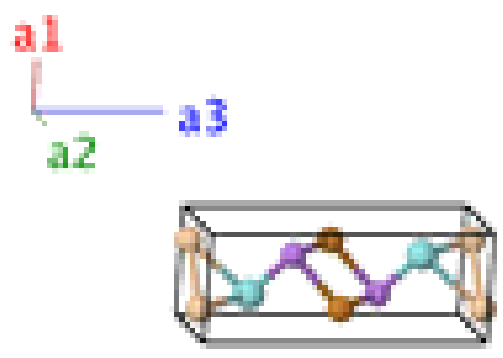

---

**Basis vectors:**

| | | Lattice Coordinates | | Cartesian Coordinates | Wyckoff Position | Atom Type |
|---|---|---|---|---|---|---|
| $\mathbf{B}_1$ | = | $\frac{3}{4}\,\mathbf{a}_1 + \frac{1}{4}\,\mathbf{a}_2$ | = | $\frac{3}{4}\,a\,\hat{\mathbf{x}} + \frac{1}{4}\,a\,\hat{\mathbf{y}}$ | (2$a$) | Si |
| $\mathbf{B}_2$ | = | $\frac{1}{4}\,\mathbf{a}_1 + \frac{3}{4}\,\mathbf{a}_2$ | = | $\frac{1}{4}\,a\,\hat{\mathbf{x}} + \frac{3}{4}\,a\,\hat{\mathbf{y}}$ | (2$a$) | Si |
| $\mathbf{B}_3$ | = | $\frac{3}{4}\,\mathbf{a}_1 + \frac{1}{4}\,\mathbf{a}_2 + \frac{1}{2}\,\mathbf{a}_3$ | = | $\frac{3}{4}\,a\,\hat{\mathbf{x}} + \frac{1}{4}\,a\,\hat{\mathbf{y}} + \frac{1}{2}\,c\,\hat{\mathbf{z}}$ | (2$b$) | Cu |
| $\mathbf{B}_4$ | = | $\frac{1}{4}\,\mathbf{a}_1 + \frac{3}{4}\,\mathbf{a}_2 + \frac{1}{2}\,\mathbf{a}_3$ | = | $\frac{1}{4}\,a\,\hat{\mathbf{x}} + \frac{3}{4}\,a\,\hat{\mathbf{y}} + \frac{1}{2}\,c\,\hat{\mathbf{z}}$ | (2$b$) | Cu |
| $\mathbf{B}_5$ | = | $\frac{1}{4}\,\mathbf{a}_1 + \frac{1}{4}\,\mathbf{a}_2 + z_3\,\mathbf{a}_3$ | = | $\frac{1}{4}\,a\,\hat{\mathbf{x}} + \frac{1}{4}\,a\,\hat{\mathbf{y}} + z_3\,c\,\hat{\mathbf{z}}$ | (2$c$) | As |
| $\mathbf{B_6}$ | = | $\frac{3}{4}\,\mathbf{a_1} + \frac{3}{4}\,\mathbf{a_2} - z_3\,\mathbf{a_3}$ | = | $\frac{3}{4}\,a\,\hat{\mathbf{x}} + \frac{3}{4}\,a\,\hat{\mathbf{y}} - z_3\,c\,\hat{\mathbf{z}}$ | (2$c$) | As |
| $\mathbf{B_7}$ | = | $\frac{1}{4}\,\mathbf{a_1} + \frac{1}{4}\,\mathbf{a_2} + z_4\,\mathbf{a_3}$ | = | $\frac{1}{4}\,a\,\hat{\mathbf{x}} + \frac{1}{4}\,a\,\hat{\mathbf{y}} + z_4\,c\,\hat{\mathbf{z}}$ | (2$c$) | Zr |
| $\mathbf{B_8}$ | = | $\frac{3}{4}\,\mathbf{a_1} + \frac{3}{4}\,\mathbf{a_2} - z_4\,\mathbf{a_3}$ | = | $\frac{3}{4}\,a\,\hat{\mathbf{x}} + \frac{3}{4}\,a\,\hat{\mathbf{y}} - z_4\,c\,\hat{\mathbf{z}}$ | (2$c$) | Zr |

**Geometry files:**

- CIF: pp. 692
- POSCAR: pp. 692

# $\beta$-Np ($A_d$) Structure: A_tP4_129_ac

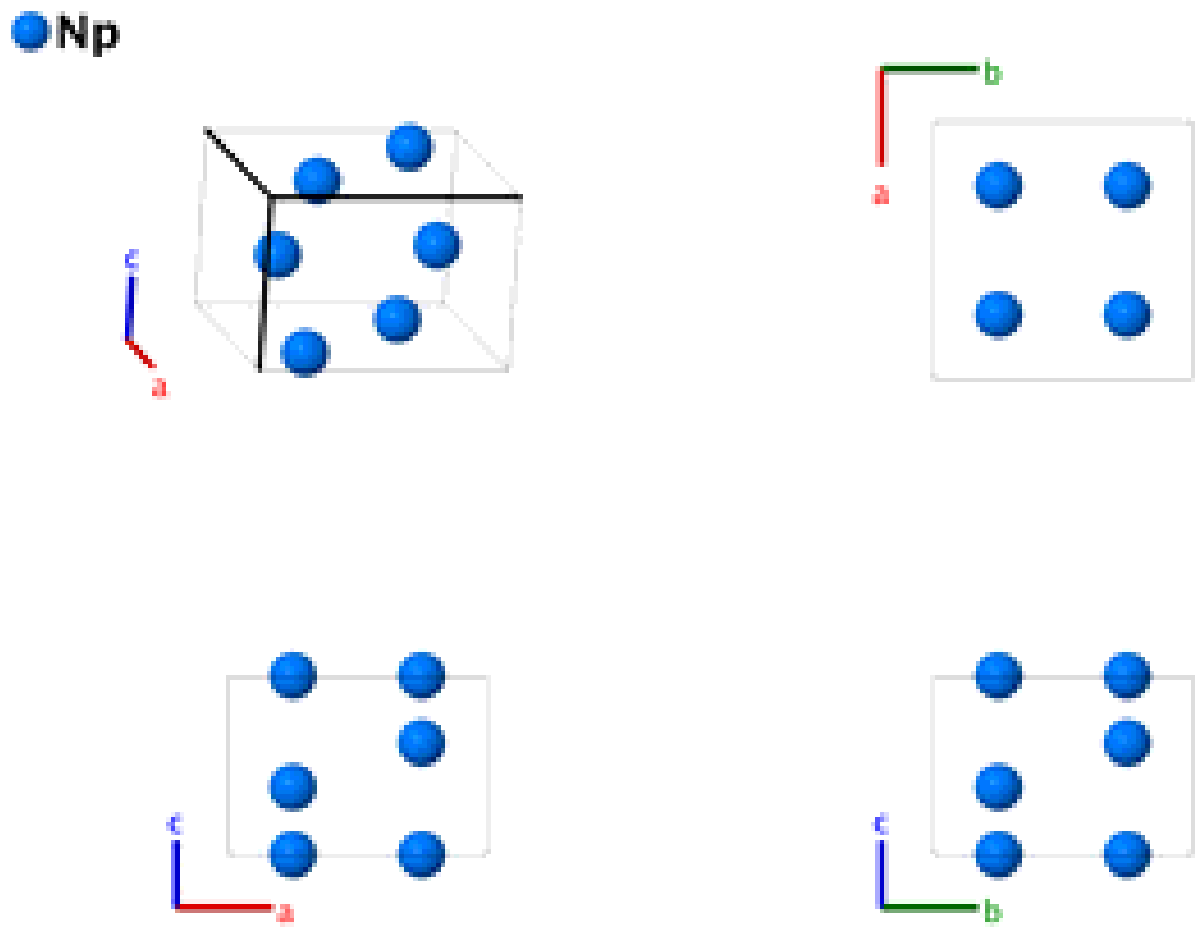

| | | |
|---|---|---|
| **Prototype** | : | $\beta$-Np |
| **AFLOW prototype label** | : | A_tP4_129_ac |
| ***Strukturbericht* designation** | : | $A_d$ |
| **Pearson symbol** | : | tP4 |
| **Space group number** | : | 129 |
| **Space group symbol** | : | P4/nmm |
| **AFLOW prototype command** | : | `aflow --proto=A_tP4_129_ac`<br>`--params=`$a, c/a, z_2$ |

- When $z = 1/2$ the atoms in this structure are in the $L1_0$ (CuAu) or the A6 (indium) structure. This structure is identical to the B10 (PbO) structure. Pearson's Handbook, along with the original papers, give the space group as $P42_1$. However, as noted by Structure Reports 15, 121 (1951), the correct space group is P4/nmm. $P42_1$ is a subgroup of P4/nmm.

**Simple Tetragonal primitive vectors:**

$$\mathbf{a}_1 = a\,\hat{\mathbf{x}}$$
$$\mathbf{a}_2 = a\,\hat{\mathbf{y}}$$
$$\mathbf{a}_3 = c\,\hat{\mathbf{z}}$$

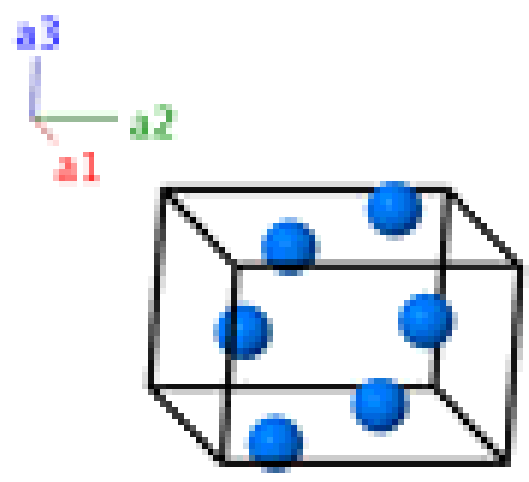

**Basis vectors:**

| | | Lattice Coordinates | | Cartesian Coordinates | Wyckoff Position | Atom Type |
|---|---|---|---|---|---|---|
| $\mathbf{B}_1$ | = | $\frac{3}{4}\,\mathbf{a}_1 + \frac{1}{4}\,\mathbf{a}_2$ | = | $\frac{3}{4}\,a\,\hat{\mathbf{x}} + \frac{1}{4}\,a\,\hat{\mathbf{y}}$ | (2$a$) | Np I |
| $\mathbf{B}_2$ | = | $\frac{1}{4}\,\mathbf{a}_1 + \frac{3}{4}\,\mathbf{a}_2$ | = | $\frac{1}{4}\,a\,\hat{\mathbf{x}} + \frac{3}{4}\,a\,\hat{\mathbf{y}}$ | (2$a$) | Np I |
| $\mathbf{B}_3$ | = | $\frac{1}{4}\,\mathbf{a}_1 + \frac{1}{4}\,\mathbf{a}_2 + z_2\,\mathbf{a}_3$ | = | $\frac{1}{4}\,a\,\hat{\mathbf{x}} + \frac{1}{4}\,a\,\hat{\mathbf{y}} + z_2\,c\,\hat{\mathbf{z}}$ | (2$c$) | Np II |
| $\mathbf{B}_4$ | = | $\frac{3}{4}\,\mathbf{a}_1 + \frac{3}{4}\,\mathbf{a}_2 - z_2\,\mathbf{a}_3$ | = | $\frac{3}{4}\,a\,\hat{\mathbf{x}} + \frac{3}{4}\,a\,\hat{\mathbf{y}} - z_2\,c\,\hat{\mathbf{z}}$ | (2$c$) | Np II |

**Geometry files:**

- CIF: pp. 692
- POSCAR: pp. 693

# Matlockite ($E0_1$, PbFCl) Structure: ABC_tP6_129_c_a_c

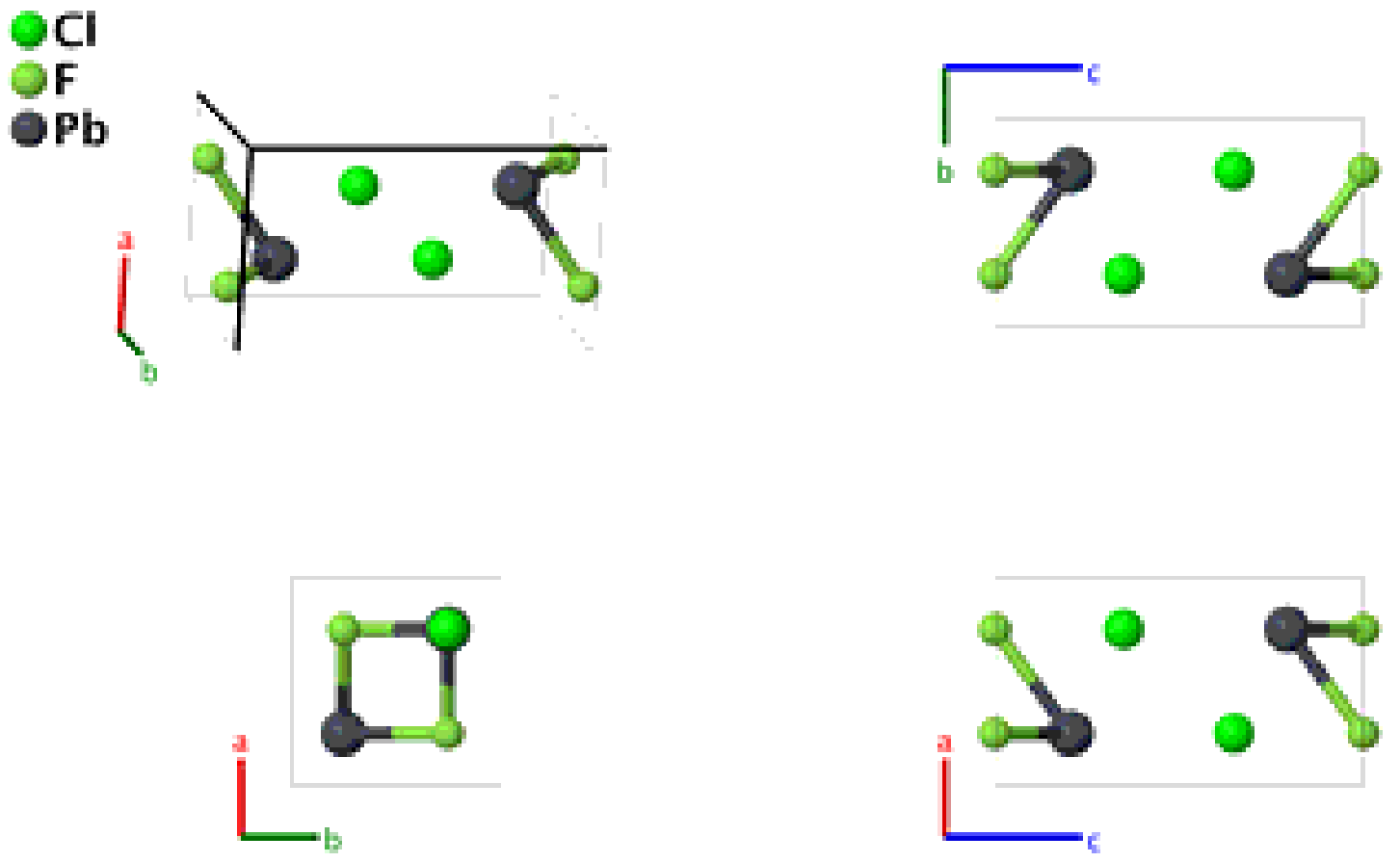

| | | |
|---|---|---|
| **Prototype** | : | PbFCl |
| **AFLOW prototype label** | : | ABC_tP6_129_c_a_c |
| ***Strukturbericht* designation** | : | $E0_1$ |
| **Pearson symbol** | : | tP6 |
| **Space group number** | : | 129 |
| **Space group symbol** | : | P4/nmm |
| **AFLOW prototype command** | : | `aflow --proto=ABC_tP6_129_c_a_c`<br>`--params=`$a, c/a, z_2, z_3$ |

**Other compounds with this structure:**

- AcOBr, AmOCl, BaHCl, BiOBr, BiOI, CaHBr, CeOCl, DyOCl, LaOI, NdOCl, NpOS, PbFBr, PrOCl, SmOI, SrHI, ThOTe, UOS, $UTe_2$, YbOI, others.

**Simple Tetragonal primitive vectors:**

$$\mathbf{a}_1 = a\,\hat{\mathbf{x}}$$
$$\mathbf{a}_2 = a\,\hat{\mathbf{y}}$$
$$\mathbf{a}_3 = c\,\hat{\mathbf{z}}$$

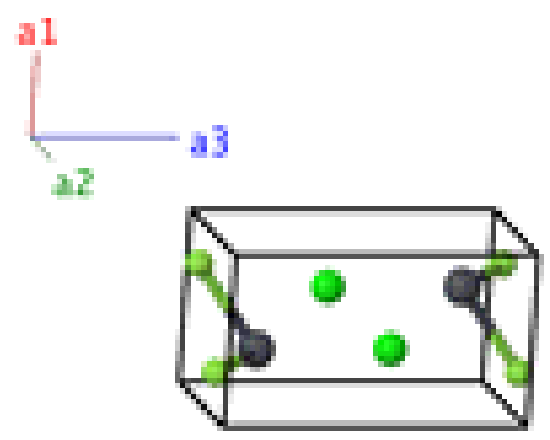

**Basis vectors:**

| | | Lattice Coordinates | | Cartesian Coordinates | Wyckoff Position | Atom Type |
|---|---|---|---|---|---|---|
| $\mathbf{B}_1$ | = | $\frac{3}{4}\,\mathbf{a}_1 + \frac{1}{4}\,\mathbf{a}_2$ | = | $\frac{3}{4}\,a\,\hat{\mathbf{x}} + \frac{1}{4}\,a\,\hat{\mathbf{y}}$ | (2$a$) | F |
| $\mathbf{B}_2$ | = | $\frac{1}{4}\,\mathbf{a}_1 + \frac{3}{4}\,\mathbf{a}_2$ | = | $\frac{1}{4}\,a\,\hat{\mathbf{x}} + \frac{3}{4}\,a\,\hat{\mathbf{y}}$ | (2$a$) | F |
| $\mathbf{B}_3$ | = | $\frac{1}{4}\,\mathbf{a}_1 + \frac{1}{4}\,\mathbf{a}_2 + z_2\,\mathbf{a}_3$ | = | $\frac{1}{4}\,a\,\hat{\mathbf{x}} + \frac{1}{4}\,a\,\hat{\mathbf{y}} + z_2\,c\,\hat{\mathbf{z}}$ | (2$c$) | Cl |
| $\mathbf{B}_4$ | = | $\frac{3}{4}\,\mathbf{a}_1 + \frac{3}{4}\,\mathbf{a}_2 - z_2\,\mathbf{a}_3$ | = | $\frac{3}{4}\,a\,\hat{\mathbf{x}} + \frac{3}{4}\,a\,\hat{\mathbf{y}} - z_2\,c\,\hat{\mathbf{z}}$ | (2$c$) | Cl |
| $\mathbf{B}_5$ | = | $\frac{1}{4}\,\mathbf{a}_1 + \frac{1}{4}\,\mathbf{a}_2 + z_3\,\mathbf{a}_3$ | = | $\frac{1}{4}\,a\,\hat{\mathbf{x}} + \frac{1}{4}\,a\,\hat{\mathbf{y}} + z_3\,c\,\hat{\mathbf{z}}$ | (2$c$) | Pb |
| $\mathbf{B}_6$ | = | $\frac{3}{4}\,\mathbf{a}_1 + \frac{3}{4}\,\mathbf{a}_2 - z_3\,\mathbf{a}_3$ | = | $\frac{3}{4}\,a\,\hat{\mathbf{x}} + \frac{3}{4}\,a\,\hat{\mathbf{y}} - z_3\,c\,\hat{\mathbf{z}}$ | (2$c$) | Pb |

**Found in:**

**Geometry files:**

- CIF: pp. 693
- POSCAR: pp. 693

# $Cu_2Sb$ (C38) Structure: A2B_tP6_129_ac_c

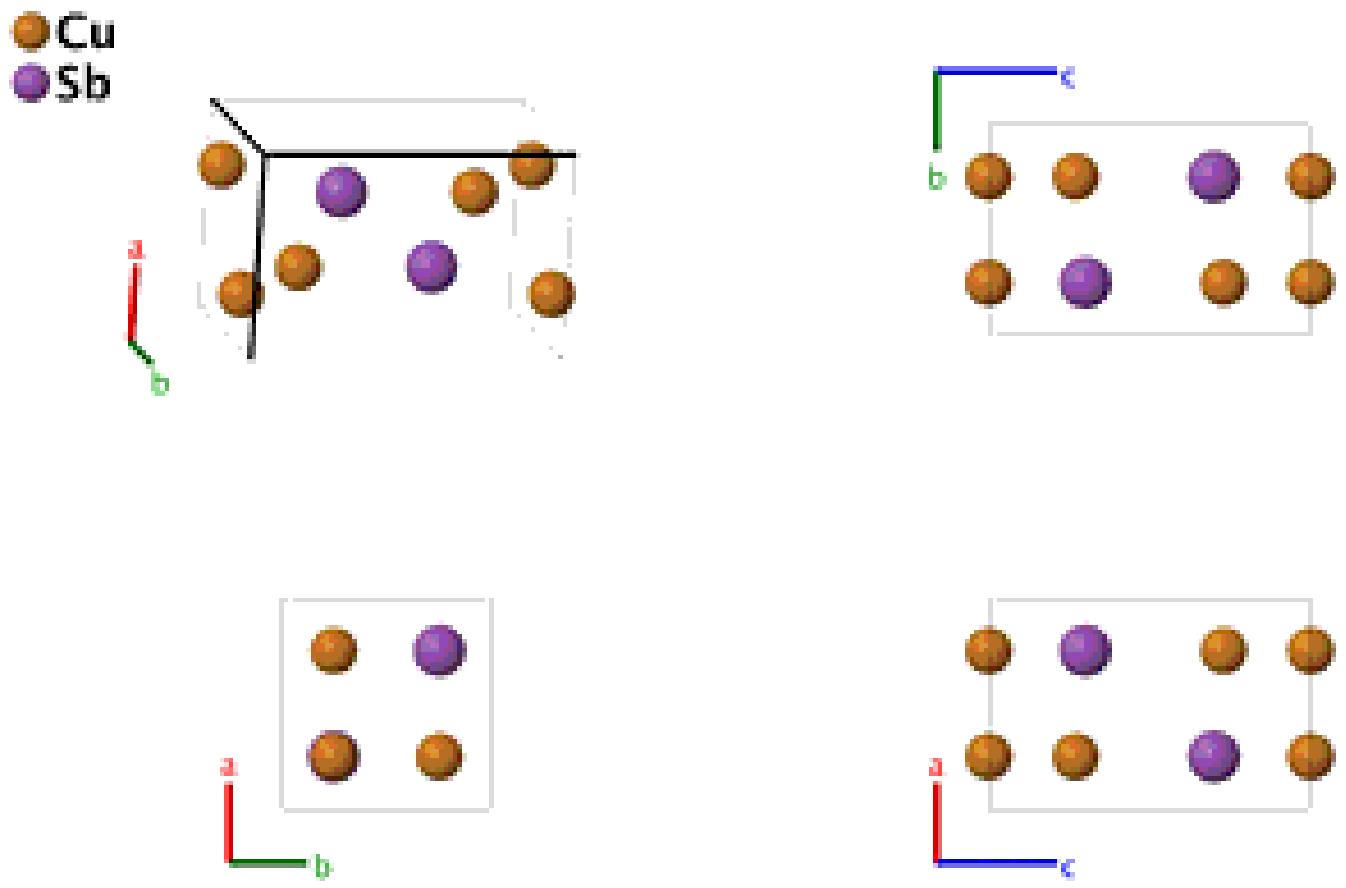

| | | |
|---|---|---|
| **Prototype** | : | $Cu_2Sb$ |
| **AFLOW prototype label** | : | A2B_tP6_129_ac_c |
| ***Strukturbericht* designation** | : | C38 |
| **Pearson symbol** | : | tP6 |
| **Space group number** | : | 129 |
| **Space group symbol** | : | P4/nmm |
| **AFLOW prototype command** | : | `aflow --proto=A2B_tP6_129_ac_c`<br>`--params=`$a, c/a, z_2, z_3$ |

**Other compounds with this structure:**

- $AsCu_2$, $As_2U$, $CeSe_2$, $HoSe_2$, $GdO_2$, $HFSb_2$, $Te_2U$, $S_2Yb$, KMgP, AsKMn, AlGeMn, GeNbSb, SnTeU, numerous others

**Simple Tetragonal primitive vectors:**

$$\mathbf{a}_1 = a\,\hat{\mathbf{x}}$$
$$\mathbf{a}_2 = a\,\hat{\mathbf{y}}$$
$$\mathbf{a}_3 = c\,\hat{\mathbf{z}}$$

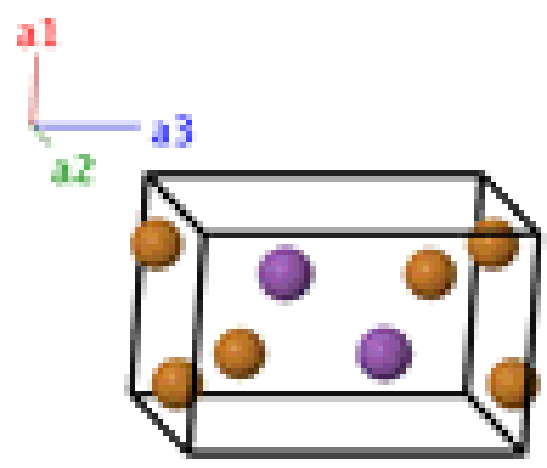

**Basis vectors:**

| | | Lattice Coordinates | | Cartesian Coordinates | Wyckoff Position | Atom Type |
|---|---|---|---|---|---|---|
| $\mathbf{B}_1$ | = | $\frac{3}{4}\,\mathbf{a}_1 + \frac{1}{4}\,\mathbf{a}_2$ | = | $\frac{3}{4}\,a\,\hat{\mathbf{x}} + \frac{1}{4}\,a\,\hat{\mathbf{y}}$ | (2$a$) | Cu I |
| $\mathbf{B}_2$ | = | $\frac{1}{4}\,\mathbf{a}_1 + \frac{3}{4}\,\mathbf{a}_2$ | = | $\frac{1}{4}\,a\,\hat{\mathbf{x}} + \frac{3}{4}\,a\,\hat{\mathbf{y}}$ | (2$a$) | Cu I |
| $\mathbf{B}_3$ | = | $\frac{1}{4}\,\mathbf{a}_1 + \frac{1}{4}\,\mathbf{a}_2 + z_2\,\mathbf{a}_3$ | = | $\frac{1}{4}\,a\,\hat{\mathbf{x}} + \frac{1}{4}\,a\,\hat{\mathbf{y}} + z_2\,c\,\hat{\mathbf{z}}$ | (2$c$) | Cu II |
| $\mathbf{B}_4$ | = | $\frac{3}{4}\,\mathbf{a}_1 + \frac{3}{4}\,\mathbf{a}_2 - z_2\,\mathbf{a}_3$ | = | $\frac{3}{4}\,a\,\hat{\mathbf{x}} + \frac{3}{4}\,a\,\hat{\mathbf{y}} - z_2\,c\,\hat{\mathbf{z}}$ | (2$c$) | Cu II |
| $\mathbf{B}_5$ | = | $\frac{1}{4}\,\mathbf{a}_1 + \frac{1}{4}\,\mathbf{a}_2 + z_3\,\mathbf{a}_3$ | = | $\frac{1}{4}\,a\,\hat{\mathbf{x}} + \frac{1}{4}\,a\,\hat{\mathbf{y}} + z_3\,c\,\hat{\mathbf{z}}$ | (2$c$) | Sb |
| $\mathbf{B}_6$ | = | $\frac{3}{4}\,\mathbf{a}_1 + \frac{3}{4}\,\mathbf{a}_2 - z_3\,\mathbf{a}_3$ | = | $\frac{3}{4}\,a\,\hat{\mathbf{x}} + \frac{3}{4}\,a\,\hat{\mathbf{y}} - z_3\,c\,\hat{\mathbf{z}}$ | (2$c$) | Sb |

**Geometry files:**

- CIF: pp. 693
- POSCAR: pp. 694

# PbO (B10) Structure: AB_tP4_129_a_c

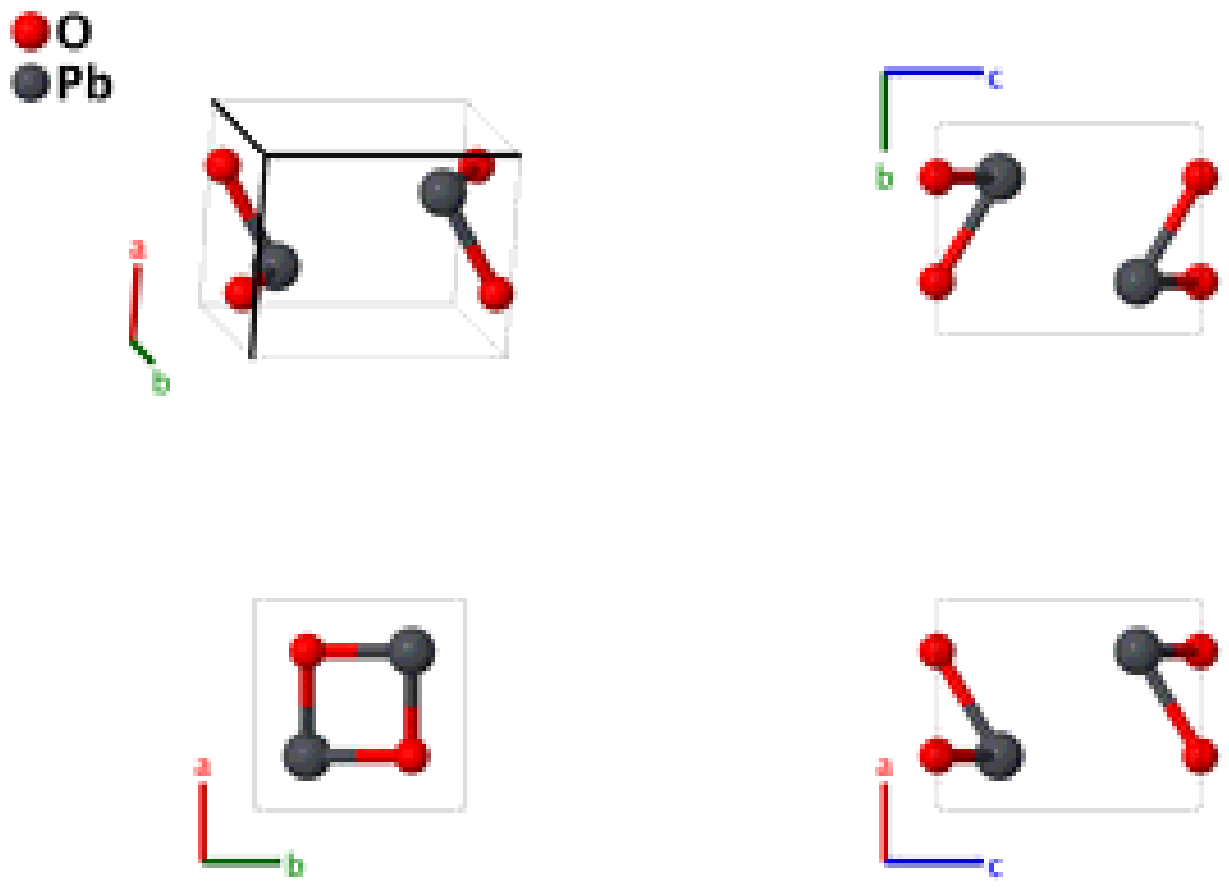

| | | |
|---|---|---|
| **Prototype** | : | PbO |
| **AFLOW prototype label** | : | AB_tP4_129_a_c |
| ***Strukturbericht* designation** | : | B10 |
| **Pearson symbol** | : | tP4 |
| **Space group number** | : | 129 |
| **Space group symbol** | : | P4/nmm |
| **AFLOW prototype command** | : | `aflow --proto=AB_tP4_129_a_c`<br>`--params=`$a, c/a, z_2$ |

- When $z = 1/2$ the atoms in this structure are in the $L1_0$ (CuAu) or the A6 (indium) structure. This structure is identical to the $A_d$ ($\beta$-Np) structure.

**Simple Tetragonal primitive vectors:**

$$\mathbf{a}_1 = a\,\hat{\mathbf{x}}$$
$$\mathbf{a}_2 = a\,\hat{\mathbf{y}}$$
$$\mathbf{a}_3 = c\,\hat{\mathbf{z}}$$

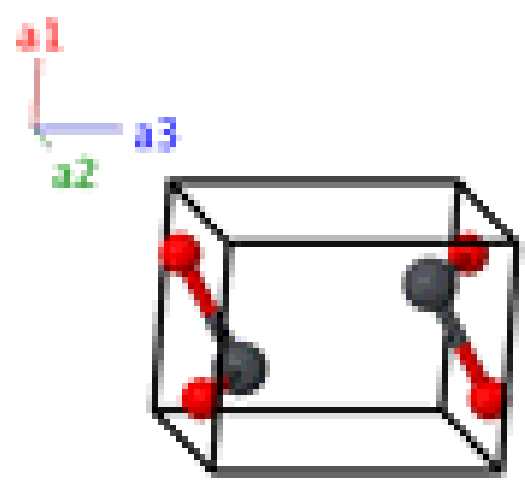

**Basis vectors:**

| | | Lattice Coordinates | | Cartesian Coordinates | Wyckoff Position | Atom Type |
|---|---|---|---|---|---|---|
| $\mathbf{B}_1$ | = | $\frac{3}{4}\,\mathbf{a}_1 + \frac{1}{4}\,\mathbf{a}_2$ | = | $\frac{3}{4}\,a\,\hat{\mathbf{x}} + \frac{1}{4}\,a\,\hat{\mathbf{y}}$ | (2$a$) | O |
| $\mathbf{B}_2$ | = | $\frac{1}{4}\,\mathbf{a}_1 + \frac{3}{4}\,\mathbf{a}_2$ | = | $\frac{1}{4}\,a\,\hat{\mathbf{x}} + \frac{3}{4}\,a\,\hat{\mathbf{y}}$ | (2$a$) | O |
| $\mathbf{B}_3$ | = | $\frac{1}{4}\,\mathbf{a}_1 + \frac{1}{4}\,\mathbf{a}_2 + z_2\,\mathbf{a}_3$ | = | $\frac{1}{4}\,a\,\hat{\mathbf{x}} + \frac{1}{4}\,a\,\hat{\mathbf{y}} + z_2\,c\,\hat{\mathbf{z}}$ | (2$c$) | Pb |
| $\mathbf{B}_4$ | = | $\frac{3}{4}\,\mathbf{a}_1 + \frac{3}{4}\,\mathbf{a}_2 - z_2\,\mathbf{a}_3$ | = | $\frac{3}{4}\,a\,\hat{\mathbf{x}} + \frac{3}{4}\,a\,\hat{\mathbf{y}} - z_2\,c\,\hat{\mathbf{z}}$ | (2$c$) | Pb |

---

**Geometry files:**

- CIF: pp. 694
- POSCAR: pp. 694

# $\gamma$-CuTi (B11) Structure: AB_tP4_129_c_c

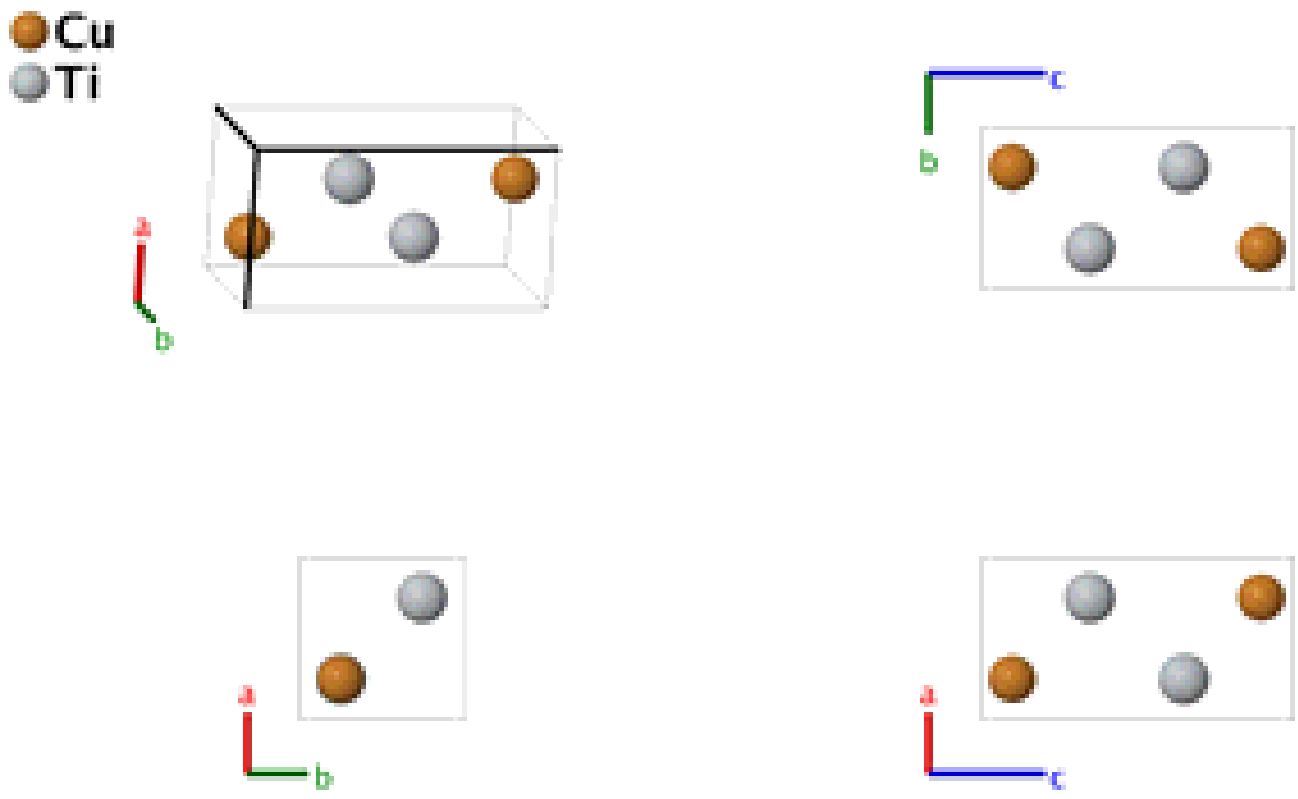

| | | |
|---|---|---|
| **Prototype** | : | $\gamma$-CuTi |
| **AFLOW prototype label** | : | AB_tP4_129_c_c |
| ***Strukturbericht* designation** | : | B11 |
| **Pearson symbol** | : | tP4 |
| **Space group number** | : | 129 |
| **Space group symbol** | : | P4/nmm |
| **AFLOW prototype command** | : | `aflow --proto=AB_tP4_129_c_c` `--params=`$a, c/a, z_1, z_2$ |

**Other compounds with this structure:**

- AuCu, AlRe, TlF-I (high-temperature)

- When $c = 2a$, $z_1 = 1/8$, and $z_2 = 5/8$, the atoms are on the sites of a body-centered cubic lattice. If, on the other hand, $c = 2\sqrt{2}a$, with the same $x_i$, the atoms are on the site of a face-centered cubic lattice. This is the phase that Lu et al. refer to as "Z2".

**Simple Tetragonal primitive vectors:**

$$\mathbf{a}_1 = a\,\hat{\mathbf{x}}$$
$$\mathbf{a}_2 = a\,\hat{\mathbf{y}}$$
$$\mathbf{a}_3 = c\,\hat{\mathbf{z}}$$

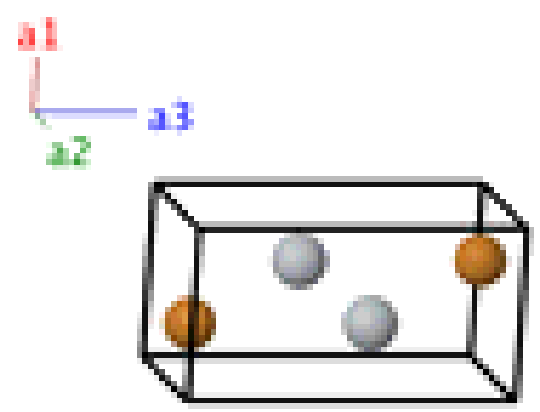

**Basis vectors:**

| | | Lattice Coordinates | | Cartesian Coordinates | Wyckoff Position | Atom Type |
|---|---|---|---|---|---|---|
| $\mathbf{B}_1$ | = | $\frac{1}{4}\mathbf{a}_1 + \frac{1}{4}\mathbf{a}_2 + z_1\,\mathbf{a}_3$ | = | $\frac{1}{4}a\,\hat{\mathbf{x}} + \frac{1}{4}a\,\hat{\mathbf{y}} + z_1\,c\,\hat{\mathbf{z}}$ | (2$c$) | Cu |
| $\mathbf{B}_2$ | = | $\frac{3}{4}\mathbf{a}_1 + \frac{3}{4}\mathbf{a}_2 - z_1\,\mathbf{a}_3$ | = | $\frac{3}{4}a\,\hat{\mathbf{x}} + \frac{3}{4}a\,\hat{\mathbf{y}} - z_1\,c\,\hat{\mathbf{z}}$ | (2$c$) | Cu |
| $\mathbf{B}_3$ | = | $\frac{1}{4}\mathbf{a}_1 + \frac{1}{4}\mathbf{a}_2 + z_2\,\mathbf{a}_3$ | = | $\frac{1}{4}a\,\hat{\mathbf{x}} + \frac{1}{4}a\,\hat{\mathbf{y}} + z_2\,c\,\hat{\mathbf{z}}$ | (2$c$) | Ti |
| $\mathbf{B}_4$ | = | $\frac{3}{4}\mathbf{a}_1 + \frac{3}{4}\mathbf{a}_2 - z_2\,\mathbf{a}_3$ | = | $\frac{3}{4}a\,\hat{\mathbf{x}} + \frac{3}{4}a\,\hat{\mathbf{y}} - z_2\,c\,\hat{\mathbf{z}}$ | (2$c$) | Ti |

**Geometry files:**

- CIF: pp. 694
- POSCAR: pp. 694

# PtS (B17) Structure: AB_tP4_131_c_e

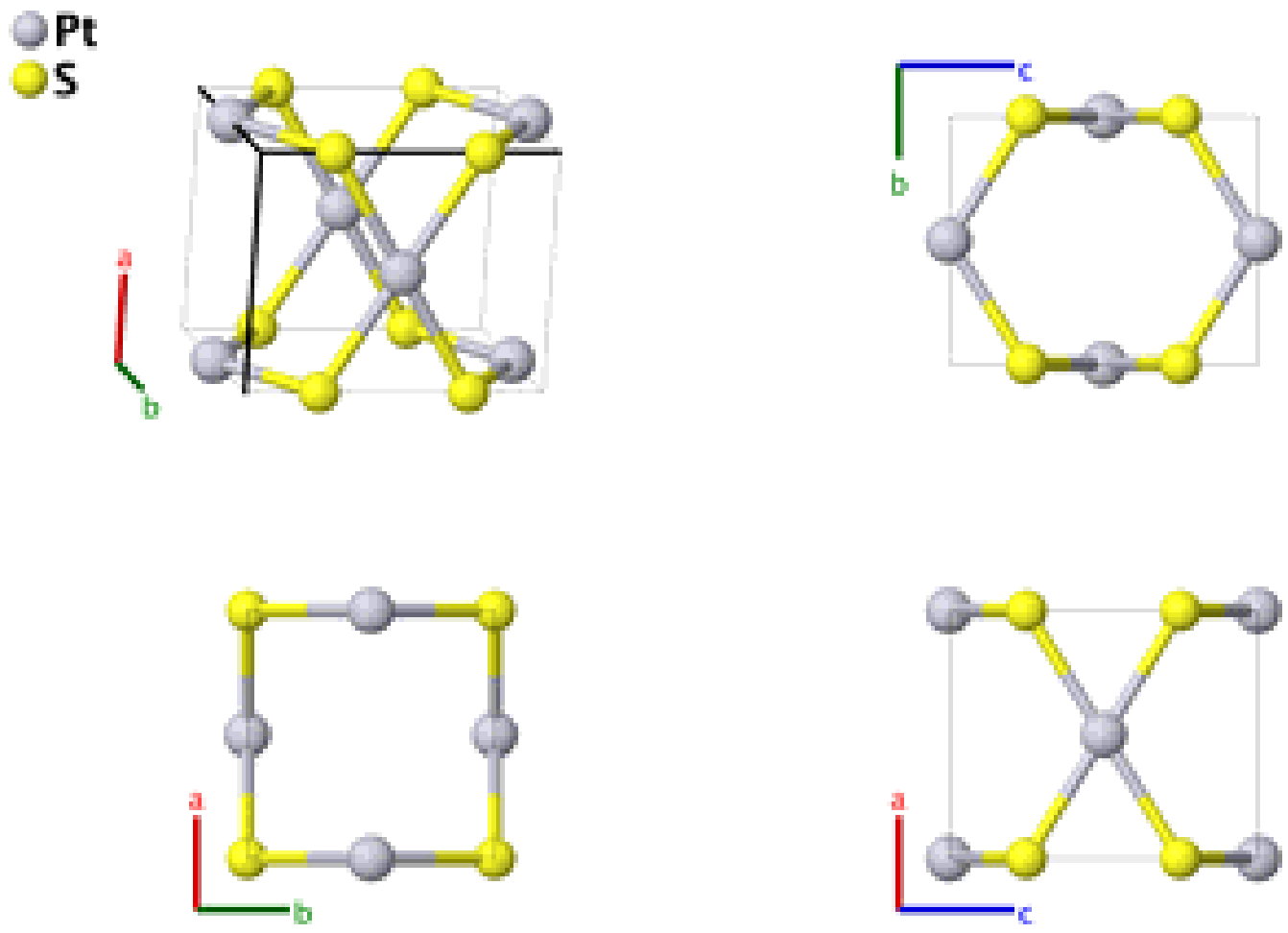

| | | |
|---|---|---|
| **Prototype** | : | PtS |
| **AFLOW prototype label** | : | AB_tP4_131_c_e |
| ***Strukturbericht* designation** | : | B17 |
| **Pearson symbol** | : | tP4 |
| **Space group number** | : | 131 |
| **Space group symbol** | : | $P4_2/mmc$ |
| **AFLOW prototype command** | : | `aflow --proto=AB_tP4_131_c_e --params=`$a, c/a$ |

**Simple Tetragonal primitive vectors:**

$$
\begin{aligned}
\mathbf{a}_1 &= a\,\hat{\mathbf{x}} \\
\mathbf{a}_2 &= a\,\hat{\mathbf{y}} \\
\mathbf{a}_3 &= c\,\hat{\mathbf{z}}
\end{aligned}
$$

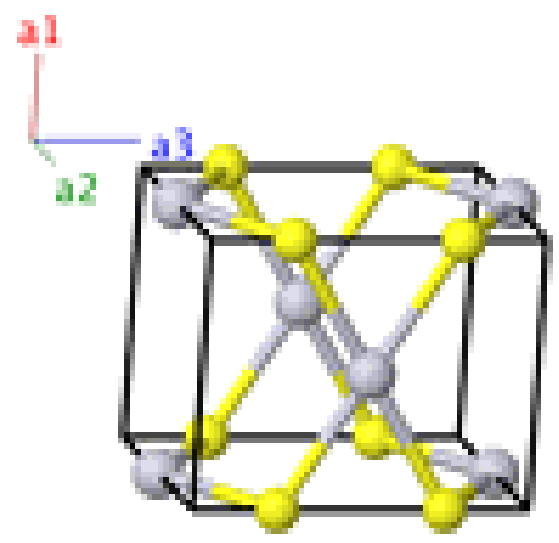

**Basis vectors:**

| | | Lattice Coordinates | | Cartesian Coordinates | Wyckoff Position | Atom Type |
|---|---|---|---|---|---|---|
| $\mathbf{B}_1$ | = | $\frac{1}{2}\,\mathbf{a}_2$ | = | $\frac{1}{2}\,a\,\hat{\mathbf{y}}$ | (2$c$) | Pt |
| $\mathbf{B}_2$ | = | $\frac{1}{2}\,\mathbf{a}_1 + \frac{1}{2}\,\mathbf{a}_3$ | = | $\frac{1}{2}\,a\,\hat{\mathbf{x}} + \frac{1}{2}\,c\,\hat{\mathbf{z}}$ | (2$c$) | Pt |
| $\mathbf{B}_3$ | = | $\frac{1}{4}\,\mathbf{a}_3$ | = | $\frac{1}{4}\,c\,\hat{\mathbf{z}}$ | (2$e$) | S |
| $\mathbf{B}_4$ | = | $\frac{3}{4}\,\mathbf{a}_3$ | = | $\frac{3}{4}\,c\,\hat{\mathbf{z}}$ | (2$e$) | S |

**Geometry files:**

- CIF: pp. 695
- POSCAR: pp. 695

# T-50 B ($A_g$) Structure: A_tP50_134_b2m2n

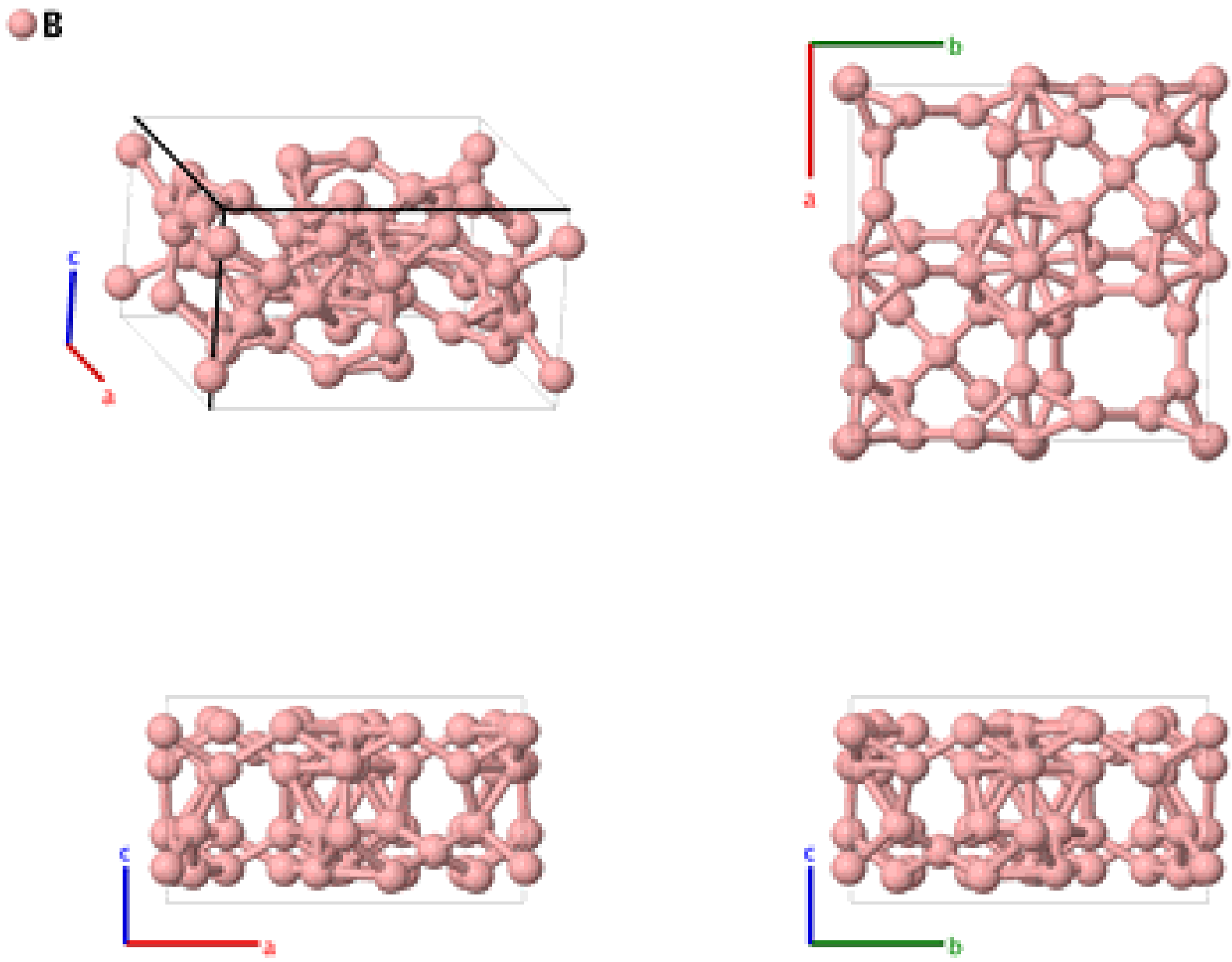

| | | |
|---|---|---|
| **Prototype** | : | B |
| **AFLOW prototype label** | : | A_tP50_134_b2m2n |
| ***Strukturbericht* designation** | : | $A_g$ |
| **Pearson symbol** | : | tP50 |
| **Space group number** | : | 134 |
| **Space group symbol** | : | $P4_2/nnm$ |
| **AFLOW prototype command** | : | `aflow --proto=A_tP50_134_b2m2n`<br>`--params=`$a, c/a, x_2, z_2, x_3, z_3, x_4, y_4, z_4, x_5, y_5, z_5$ |

- This is apparently the most common form of boron. At least, it's listed first in (Donohue, 1982). Note that the basic building block is a slightly distorted icosahedron. This icosahedron also appears in $\alpha$-B (R12) and $\beta$-B (R105).

**Simple Tetragonal primitive vectors:**

$$\mathbf{a}_1 = a\,\hat{\mathbf{x}}$$
$$\mathbf{a}_2 = a\,\hat{\mathbf{y}}$$
$$\mathbf{a}_3 = c\,\hat{\mathbf{z}}$$

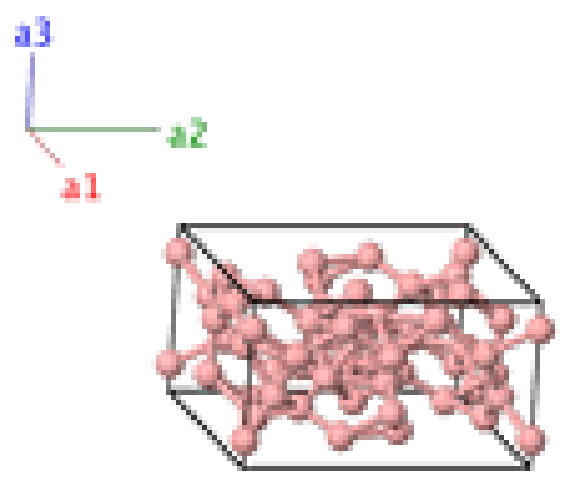

**Basis vectors:**

| | | Lattice Coordinates | | Cartesian Coordinates | Wyckoff Position | Atom Type |
|---|---|---|---|---|---|---|
| $\mathbf{B}_1$ | = | $\frac{3}{4}\,\mathbf{a}_1 + \frac{1}{4}\,\mathbf{a}_2 + \frac{1}{4}\,\mathbf{a}_3$ | = | $\frac{3}{4}\,a\,\hat{\mathbf{x}} + \frac{1}{4}\,a\,\hat{\mathbf{y}} + \frac{1}{4}\,c\,\hat{\mathbf{z}}$ | (2b) | B I |
| $\mathbf{B}_{2}$ | = | $\frac{1}{4}\,\mathbf{a}_{1}+\frac{3}{4}\,\mathbf{a}_{2}+\frac{3}{4}\,\mathbf{a}_{3}$ | = | $\frac{1}{4}\,a\,\mathbf{\hat{x}}+\frac{3}{4}\,a\,\mathbf{\hat{y}}+\frac{3}{4}\,c\,\mathbf{\hat{z}}$ | (2*b*) | B I |
| $\mathbf{B}_{3}$ | = | $x_{2}\,\mathbf{a}_{1}-x_{2}\,\mathbf{a}_{2}+z_{2}\,\mathbf{a}_{3}$ | = | $x_{2}\,a\,\mathbf{\hat{x}}-x_{2}\,a\,\mathbf{\hat{y}}+z_{2}\,c\,\mathbf{\hat{z}}$ | (8*m*) | B II |
| $\mathbf{B}_{4}$ | = | $\left(\frac{1}{2}-x_{2}\right)\mathbf{a}_{1}+\left(\frac{1}{2}+x_{2}\right)\mathbf{a}_{2}+z_{2}\,\mathbf{a}_{3}$ | = | $\left(\frac{1}{2}-x_{2}\right)a\,\mathbf{\hat{x}}+\left(\frac{1}{2}+x_{2}\right)a\,\mathbf{\hat{y}}+z_{2}\,c\,\mathbf{\hat{z}}$ | (8*m*) | B II |
| $\mathbf{B}_{5}$ | = | $\left(\frac{1}{2}+x_{2}\right)\mathbf{a}_{1}+x_{2}\,\mathbf{a}_{2}+\left(\frac{1}{2}+z_{2}\right)\mathbf{a}_{3}$ | = | $\left(\frac{1}{2}+x_{2}\right)a\,\mathbf{\hat{x}}+x_{2}\,a\,\mathbf{\hat{y}}+\left(\frac{1}{2}+z_{2}\right)c\,\mathbf{\hat{z}}$ | (8*m*) | B II |
| $\mathbf{B}_{6}$ | = | $-x_{2}\,\mathbf{a}_{1}+\left(\frac{1}{2}-x_{2}\right)\mathbf{a}_{2}+\left(\frac{1}{2}+z_{2}\right)\mathbf{a}_{3}$ | = | $-x_{2}\,a\,\mathbf{\hat{x}}+\left(\frac{1}{2}-x_{2}\right)a\,\mathbf{\hat{y}}+\left(\frac{1}{2}+z_{2}\right)c\,\mathbf{\hat{z}}$ | (8*m*) | B II |
| $\mathbf{B}_{7}$ | = | $\left(\frac{1}{2}-x_{2}\right)\mathbf{a}_{1}-x_{2}\,\mathbf{a}_{2}+\left(\frac{1}{2}-z_{2}\right)\mathbf{a}_{3}$ | = | $\left(\frac{1}{2}-x_{2}\right)a\,\mathbf{\hat{x}}-x_{2}\,a\,\mathbf{\hat{y}}+\left(\frac{1}{2}-z_{2}\right)c\,\mathbf{\hat{z}}$ | (8*m*) | B II |
| $\mathbf{B}_{8}$ | = | $x_{2}\,\mathbf{a}_{1}+\left(\frac{1}{2}+x_{2}\right)\mathbf{a}_{2}+\left(\frac{1}{2}-z_{2}\right)\mathbf{a}_{3}$ | = | $x_{2}\,a\,\mathbf{\hat{x}}+\left(\frac{1}{2}+x_{2}\right)a\,\mathbf{\hat{y}}+\left(\frac{1}{2}-z_{2}\right)c\,\mathbf{\hat{z}}$ | (8*m*) | B II |
| $\mathbf{B}_{9}$ | = | $-x_{2}\,\mathbf{a}_{1}+x_{2}\,\mathbf{a}_{2}-z_{2}\,\mathbf{a}_{3}$ | = | $-x_{2}\,a\,\mathbf{\hat{x}}+x_{2}\,a\,\mathbf{\hat{y}}-z_{2}\,c\,\mathbf{\hat{z}}$ | (8*m*) | B II |
| $\mathbf{B}_{10}$ | = | $\left(\frac{1}{2}+x_{2}\right)\mathbf{a}_{1}+\left(\frac{1}{2}-x_{2}\right)\mathbf{a}_{2}-z_{2}\,\mathbf{a}_{3}$ | = | $\left(\frac{1}{2}+x_{2}\right)a\,\mathbf{\hat{x}}+\left(\frac{1}{2}-x_{2}\right)a\,\mathbf{\hat{y}}-z_{2}\,c\,\mathbf{\hat{z}}$ | (8*m*) | B II |
| $\mathbf{B}_{11}$ | = | $x_{3}\,\mathbf{a}_{1}-x_{3}\,\mathbf{a}_{2}+z_{3}\,\mathbf{a}_{3}$ | = | $x_{3}\,a\,\mathbf{\hat{x}}-x_{3}\,a\,\mathbf{\hat{y}}+z_{3}\,c\,\mathbf{\hat{z}}$ | (8*m*) | B III |
| $\mathbf{B}_{12}$ | = | $\left(\frac{1}{2}-x_{3}\right)\mathbf{a}_{1}+\left(\frac{1}{2}+x_{3}\right)\mathbf{a}_{2}+z_{3}\,\mathbf{a}_{3}$ | = | $\left(\frac{1}{2}-x_{3}\right)a\,\mathbf{\hat{x}}+\left(\frac{1}{2}+x_{3}\right)a\,\mathbf{\hat{y}}+z_{3}\,c\,\mathbf{\hat{z}}$ | (8*m*) | B III |
| $\mathbf{B}_{13}$ | = | $\left(\frac{1}{2}+x_{3}\right)\mathbf{a}_{1}+x_{3}\,\mathbf{a}_{2}+\left(\frac{1}{2}+z_{3}\right)\mathbf{a}_{3}$ | = | $\left(\frac{1}{2}+x_{3}\right)a\,\mathbf{\hat{x}}+x_{3}\,a\,\mathbf{\hat{y}}+\left(\frac{1}{2}+z_{3}\right)c\,\mathbf{\hat{z}}$ | (8*m*) | B III |
| $\mathbf{B}_{14}$ | = | $-x_{3}\,\mathbf{a}_{1}+\left(\frac{1}{2}-x_{3}\right)\mathbf{a}_{2}+\left(\frac{1}{2}+z_{3}\right)\mathbf{a}_{3}$ | = | $-x_{3}\,a\,\mathbf{\hat{x}}+\left(\frac{1}{2}-x_{3}\right)a\,\mathbf{\hat{y}}+\left(\frac{1}{2}+z_{3}\right)c\,\mathbf{\hat{z}}$ | (8*m*) | B III |
| $\mathbf{B}_{15}$ | = | $\left(\frac{1}{2}-x_{3}\right)\mathbf{a}_{1}-x_{3}\,\mathbf{a}_{2}+\left(\frac{1}{2}-z_{3}\right)\mathbf{a}_{3}$ | = | $\left(\frac{1}{2}-x_{3}\right)a\,\mathbf{\hat{x}}-x_{3}\,a\,\mathbf{\hat{y}}+\left(\frac{1}{2}-z_{3}\right)c\,\mathbf{\hat{z}}$ | (8*m*) | B III |
| $\mathbf{B}_{16}$ | = | $x_{3}\,\mathbf{a}_{1}+\left(\frac{1}{2}+x_{3}\right)\mathbf{a}_{2}+\left(\frac{1}{2}-z_{3}\right)\mathbf{a}_{3}$ | = | $x_{3}\,a\,\mathbf{\hat{x}}+\left(\frac{1}{2}+x_{3}\right)a\,\mathbf{\hat{y}}+\left(\frac{1}{2}-z_{3}\right)c\,\mathbf{\hat{z}}$ | (8*m*) | B III |
| $\mathbf{B}_{17}$ | = | $-x_{3}\,\mathbf{a}_{1}+x_{3}\,\mathbf{a}_{2}-z_{3}\,\mathbf{a}_{3}$ | = | $-x_{3}\,a\,\mathbf{\hat{x}}+x_{3}\,a\,\mathbf{\hat{y}}-z_{3}\,c\,\mathbf{\hat{z}}$ | (8*m*) | B III |
| $\mathbf{B}_{18}$ | = | $\left(\frac{1}{2}+x_{3}\right)\mathbf{a}_{1}+\left(\frac{1}{2}-x_{3}\right)\mathbf{a}_{2}-z_{3}\,\mathbf{a}_{3}$ | = | $\left(\frac{1}{2}+x_{3}\right)a\,\mathbf{\hat{x}}+\left(\frac{1}{2}-x_{3}\right)a\,\mathbf{\hat{y}}-z_{3}\,c\,\mathbf{\hat{z}}$ | (8*m*) | B III |
| $\mathbf{B}_{19}$ | = | $x_{4}\,\mathbf{a}_{1}+y_{4}\,\mathbf{a}_{2}+z_{4}\,\mathbf{a}_{3}$ | = | $x_{4}\,a\,\mathbf{\hat{x}}+y_{4}\,a\,\mathbf{\hat{y}}+z_{4}\,c\,\mathbf{\hat{z}}$ | (16*n*) | B IV |
| $\mathbf{B}_{20}$ | = | $\left(\frac{1}{2}-x_{4}\right)\mathbf{a}_{1}+\left(\frac{1}{2}-y_{4}\right)\mathbf{a}_{2}+z_{4}\,\mathbf{a}_{3}$ | = | $\left(\frac{1}{2}-x_{4}\right)a\,\mathbf{\hat{x}}+\left(\frac{1}{2}-y_{4}\right)a\,\mathbf{\hat{y}}+z_{4}\,c\,\mathbf{\hat{z}}$ | (16*n*) | B IV |
| $\mathbf{B}_{21}$ | = | $\left(\frac{1}{2}-y_{4}\right)\mathbf{a}_{1}+x_{4}\,\mathbf{a}_{2}+\left(\frac{1}{2}+z_{4}\right)\mathbf{a}_{3}$ | = | $\left(\frac{1}{2}-y_{4}\right)a\,\mathbf{\hat{x}}+x_{4}\,a\,\mathbf{\hat{y}}+\left(\frac{1}{2}+z_{4}\right)c\,\mathbf{\hat{z}}$ | (16*n*) | B IV |
| $\mathbf{B}_{22}$ | = | $y_{4}\,\mathbf{a}_{1}+\left(\frac{1}{2}-x_{4}\right)\mathbf{a}_{2}+\left(\frac{1}{2}+z_{4}\right)\mathbf{a}_{3}$ | = | $y_{4}\,a\,\mathbf{\hat{x}}+\left(\frac{1}{2}-x_{4}\right)a\,\mathbf{\hat{y}}+\left(\frac{1}{2}+z_{4}\right)c\,\mathbf{\hat{z}}$ | (16*n*) | B IV |
| $\mathbf{B}_{23}$ | = | $\left(\frac{1}{2}-x_{4}\right)\mathbf{a}_{1}+y_{4}\,\mathbf{a}_{2}+\left(\frac{1}{2}-z_{4}\right)\mathbf{a}_{3}$ | = | $\left(\frac{1}{2}-x_{4}\right)a\,\mathbf{\hat{x}}+y_{4}\,a\,\mathbf{\hat{y}}+\left(\frac{1}{2}-z_{4}\right)c\,\mathbf{\hat{z}}$ | (16*n*) | B IV |
| $\mathbf{B}_{24}$ | = | $x_{4}\,\mathbf{a}_{1}+\left(\frac{1}{2}-y_{4}\right)\mathbf{a}_{2}+\left(\frac{1}{2}-z_{4}\right)\mathbf{a}_{3}$ | = | $x_{4}\,a\,\mathbf{\hat{x}}+\left(\frac{1}{2}-y_{4}\right)a\,\mathbf{\hat{y}}+\left(\frac{1}{2}-z_{4}\right)c\,\mathbf{\hat{z}}$ | (16*n*) | B IV |
| $\mathbf{B}_{25}$ | = | $y_{4}\,\mathbf{a}_{1}+x_{4}\,\mathbf{a}_{2}-z_{4}\,\mathbf{a}_{3}$ | = | $y_{4}\,a\,\mathbf{\hat{x}}+x_{4}\,a\,\mathbf{\hat{y}}-z_{4}\,c\,\mathbf{\hat{z}}$ | (16*n*) | B IV |
| $\mathbf{B}_{26}$ | = | $\left(\frac{1}{2}-y_{4}\right)\mathbf{a}_{1}+\left(\frac{1}{2}-x_{4}\right)\mathbf{a}_{2}-z_{4}\,\mathbf{a}_{3}$ | = | $\left(\frac{1}{2}-y_{4}\right)a\,\mathbf{\hat{x}}+\left(\frac{1}{2}-x_{4}\right)a\,\mathbf{\hat{y}}-z_{4}\,c\,\mathbf{\hat{z}}$ | (16*n*) | B IV |
| $\mathbf{B}_{27}$ | = | $-x_{4}\,\mathbf{a}_{1}-y_{4}\,\mathbf{a}_{2}-z_{4}\,\mathbf{a}_{3}$ | = | $-x_{4}\,a\,\mathbf{\hat{x}}-y_{4}\,a\,\mathbf{\hat{y}}-z_{4}\,c\,\mathbf{\hat{z}}$ | (16*n*) | B IV |
| $\mathbf{B}_{28}$ | = | $\left(\frac{1}{2}+x_{4}\right)\mathbf{a}_{1}+\left(\frac{1}{2}+y_{4}\right)\mathbf{a}_{2}-z_{4}\,\mathbf{a}_{3}$ | = | $\left(\frac{1}{2}+x_{4}\right)a\,\mathbf{\hat{x}}+\left(\frac{1}{2}+y_{4}\right)a\,\mathbf{\hat{y}}-z_{4}\,c\,\mathbf{\hat{z}}$ | (16*n*) | B IV |
| $\mathbf{B}_{29}$ | = | $\left(\frac{1}{2}+y_{4}\right)\mathbf{a}_{1}-x_{4}\,\mathbf{a}_{2}+\left(\frac{1}{2}-z_{4}\right)\mathbf{a}_{3}$ | = | $\left(\frac{1}{2}+y_{4}\right)a\,\mathbf{\hat{x}}-x_{4}\,a\,\mathbf{\hat{y}}+\left(\frac{1}{2}-z_{4}\right)c\,\mathbf{\hat{z}}$ | (16*n*) | B IV |
| $\mathbf{B}_{30}$ | = | $-y_{4}\,\mathbf{a}_{1}+\left(\frac{1}{2}+x_{4}\right)\mathbf{a}_{2}+\left(\frac{1}{2}-z_{4}\right)\mathbf{a}_{3}$ | = | $-y_{4}\,a\,\mathbf{\hat{x}}+\left(\frac{1}{2}+x_{4}\right)a\,\mathbf{\hat{y}}+\left(\frac{1}{2}-z_{4}\right)c\,\mathbf{\hat{z}}$ | (16*n*) | B IV |
| $\mathbf{B}_{31}$ | = | $\left(\frac{1}{2}+x_{4}\right)\mathbf{a}_{1}-y_{4}\,\mathbf{a}_{2}+\left(\frac{1}{2}+z_{4}\right)\mathbf{a}_{3}$ | = | $\left(\frac{1}{2}+x_{4}\right)a\,\mathbf{\hat{x}}-y_{4}\,a\,\mathbf{\hat{y}}+\left(\frac{1}{2}+z_{4}\right)c\,\mathbf{\hat{z}}$ | (16*n*) | B IV |
| $\mathbf{B}_{32}$ | = | $-x_{4}\,\mathbf{a}_{1}+\left(\frac{1}{2}+y_{4}\right)\mathbf{a}_{2}+\left(\frac{1}{2}+z_{4}\right)\mathbf{a}_{3}$ | = | $-x_{4}\,a\,\mathbf{\hat{x}}+\left(\frac{1}{2}+y_{4}\right)a\,\mathbf{\hat{y}}+\left(\frac{1}{2}+z_{4}\right)c\,\mathbf{\hat{z}}$ | (16*n*) | B IV |
| $\mathbf{B}_{33}$ | = | $-y_{4}\,\mathbf{a}_{1}-x_{4}\,\mathbf{a}_{2}+z_{4}\,\mathbf{a}_{3}$ | = | $-y_{4}\,a\,\mathbf{\hat{x}}-x_{4}\,a\,\mathbf{\hat{y}}+z_{4}\,c\,\mathbf{\hat{z}}$ | (16*n*) | B IV |
| $\mathbf{B}_{34}$ | = | $\left(\frac{1}{2}+y_{4}\right)\mathbf{a}_{1}+\left(\frac{1}{2}+x_{4}\right)\mathbf{a}_{2}+z_{4}\,\mathbf{a}_{3}$ | = | $\left(\frac{1}{2}+y_{4}\right)a\,\mathbf{\hat{x}}+\left(\frac{1}{2}+x_{4}\right)a\,\mathbf{\hat{y}}+z_{4}\,c\,\mathbf{\hat{z}}$ | (16*n*) | B IV |
| $\mathbf{B}_{35}$ | = | $x_{5}\,\mathbf{a}_{1}+y_{5}\,\mathbf{a}_{2}+z_{5}\,\mathbf{a}_{3}$ | = | $x_{5}\,a\,\mathbf{\hat{x}}+y_{5}\,a\,\mathbf{\hat{y}}+z_{5}\,c\,\mathbf{\hat{z}}$ | (16*n*) | B V |
| $\mathbf{B}_{36}$ | = | $\left(\frac{1}{2}-x_{5}\right)\mathbf{a}_{1}+\left(\frac{1}{2}-y_{5}\right)\mathbf{a}_{2}+z_{5}\,\mathbf{a}_{3}$ | = | $\left(\frac{1}{2}-x_{5}\right)a\,\mathbf{\hat{x}}+\left(\frac{1}{2}-y_{5}\right)a\,\mathbf{\hat{y}}+z_{5}\,c\,\mathbf{\hat{z}}$ | (16*n*) | B V |
| $\mathbf{B}_{37}$ | = | $\left(\frac{1}{2}-y_{5}\right)\mathbf{a}_{1}+x_{5}\,\mathbf{a}_{2}+\left(\frac{1}{2}+z_{5}\right)\mathbf{a}_{3}$ | = | $\left(\frac{1}{2}-y_{5}\right)a\,\mathbf{\hat{x}}+x_{5}\,a\,\mathbf{\hat{y}}+\left(\frac{1}{2}+z_{5}\right)c\,\mathbf{\hat{z}}$ | (16*n*) | B V |
| $\mathbf{B_{38}}$ | = | $y_5\,\mathbf{a_1} + \left(\frac{1}{2} - x_5\right)\,\mathbf{a_2} + \left(\frac{1}{2} + z_5\right)\,\mathbf{a_3}$ | = | $y_5\,a\,\mathbf{\hat{x}} + \left(\frac{1}{2} - x_5\right)\,a\,\mathbf{\hat{y}} + \left(\frac{1}{2} + z_5\right)\,c\,\mathbf{\hat{z}}$ | $(16n)$ | B V |
| $\mathbf{B_{39}}$ | = | $\left(\frac{1}{2} - x_5\right)\,\mathbf{a_1} + y_5\,\mathbf{a_2} + \left(\frac{1}{2} - z_5\right)\,\mathbf{a_3}$ | = | $\left(\frac{1}{2} - x_5\right)\,a\,\mathbf{\hat{x}} + y_5\,a\,\mathbf{\hat{y}} + \left(\frac{1}{2} - z_5\right)\,c\,\mathbf{\hat{z}}$ | $(16n)$ | B V |
| $\mathbf{B_{40}}$ | = | $x_5\,\mathbf{a_1} + \left(\frac{1}{2} - y_5\right)\,\mathbf{a_2} + \left(\frac{1}{2} - z_5\right)\,\mathbf{a_3}$ | = | $x_5\,a\,\mathbf{\hat{x}} + \left(\frac{1}{2} - y_5\right)\,a\,\mathbf{\hat{y}} + \left(\frac{1}{2} - z_5\right)\,c\,\mathbf{\hat{z}}$ | $(16n)$ | B V |
| $\mathbf{B_{41}}$ | = | $y_5\,\mathbf{a_1} + x_5\,\mathbf{a_2} - z_5\,\mathbf{a_3}$ | = | $y_5\,a\,\mathbf{\hat{x}} + x_5\,a\,\mathbf{\hat{y}} - z_5\,c\,\mathbf{\hat{z}}$ | $(16n)$ | B V |
| $\mathbf{B_{42}}$ | = | $\left(\frac{1}{2} - y_5\right)\,\mathbf{a_1} + \left(\frac{1}{2} - x_5\right)\,\mathbf{a_2} - z_5\,\mathbf{a_3}$ | = | $\left(\frac{1}{2} - y_5\right)\,a\,\mathbf{\hat{x}} + \left(\frac{1}{2} - x_5\right)\,a\,\mathbf{\hat{y}} - z_5\,c\,\mathbf{\hat{z}}$ | $(16n)$ | B V |
| $\mathbf{B_{43}}$ | = | $-x_5\,\mathbf{a_1} - y_5\,\mathbf{a_2} - z_5\,\mathbf{a_3}$ | = | $-x_5\,a\,\mathbf{\hat{x}} - y_5\,a\,\mathbf{\hat{y}} - z_5\,c\,\mathbf{\hat{z}}$ | $(16n)$ | B V |
| $\mathbf{B_{44}}$ | = | $\left(\frac{1}{2} + x_5\right)\,\mathbf{a_1} + \left(\frac{1}{2} + y_5\right)\,\mathbf{a_2} - z_5\,\mathbf{a_3}$ | = | $\left(\frac{1}{2} + x_5\right)\,a\,\mathbf{\hat{x}} + \left(\frac{1}{2} + y_5\right)\,a\,\mathbf{\hat{y}} - z_5\,c\,\mathbf{\hat{z}}$ | $(16n)$ | B V |
| $\mathbf{B_{45}}$ | = | $\left(\frac{1}{2} + y_5\right)\,\mathbf{a_1} - x_5\,\mathbf{a_2} + \left(\frac{1}{2} - z_5\right)\,\mathbf{a_3}$ | = | $\left(\frac{1}{2} + y_5\right)\,a\,\mathbf{\hat{x}} - x_5\,a\,\mathbf{\hat{y}} + \left(\frac{1}{2} - z_5\right)\,c\,\mathbf{\hat{z}}$ | $(16n)$ | B V |
| $\mathbf{B_{46}}$ | = | $-y_5\,\mathbf{a_1} + \left(\frac{1}{2} + x_5\right)\,\mathbf{a_2} + \left(\frac{1}{2} - z_5\right)\,\mathbf{a_3}$ | = | $-y_5\,a\,\mathbf{\hat{x}}+\left(\frac{1}{2} + x_5\right)\,a\,\mathbf{\hat{y}}+\left(\frac{1}{2} - z_5\right)\,c\,\mathbf{\hat{z}}$ | $(16n)$ | B V |
| $\mathbf{B_{47}}$ | = | $\left(\frac{1}{2} + x_5\right)\,\mathbf{a_1} - y_5\,\mathbf{a_2} + \left(\frac{1}{2} + z_5\right)\,\mathbf{a_3}$ | = | $\left(\frac{1}{2} + x_5\right)\,a\,\mathbf{\hat{x}} - y_5\,a\,\mathbf{\hat{y}} + \left(\frac{1}{2} + z_5\right)\,c\,\mathbf{\hat{z}}$ | $(16n)$ | B V |
| $\mathbf{B_{48}}$ | = | $-x_5\,\mathbf{a_1} + \left(\frac{1}{2} + y_5\right)\,\mathbf{a_2} + \left(\frac{1}{2} + z_5\right)\,\mathbf{a_3}$ | = | $-x_5\,a\,\mathbf{\hat{x}}+\left(\frac{1}{2} + y_5\right)\,a\,\mathbf{\hat{y}}+\left(\frac{1}{2} + z_5\right)\,c\,\mathbf{\hat{z}}$ | $(16n)$ | B V |
| $\mathbf{B_{49}}$ | = | $-y_5\,\mathbf{a_1} - x_5\,\mathbf{a_2} + z_5\,\mathbf{a_3}$ | = | $-y_5\,a\,\mathbf{\hat{x}} - x_5\,a\,\mathbf{\hat{y}} + z_5\,c\,\mathbf{\hat{z}}$ | $(16n)$ | B V |
| $\mathbf{B_{50}}$ | = | $\left(\frac{1}{2} + y_5\right)\,\mathbf{a_1} + \left(\frac{1}{2} + x_5\right)\,\mathbf{a_2} + z_5\,\mathbf{a_3}$ | = | $\left(\frac{1}{2} + y_5\right)\,a\,\mathbf{\hat{x}} + \left(\frac{1}{2} + x_5\right)\,a\,\mathbf{\hat{y}} + z_5\,c\,\mathbf{\hat{z}}$ | $(16n)$ | B V |

---

**Found in:**

---

**Geometry files:**

- CIF: pp. 695
- POSCAR: pp. 695

# $\beta$-U ($A_b$) Structure: A_tP30_136_bf2ij

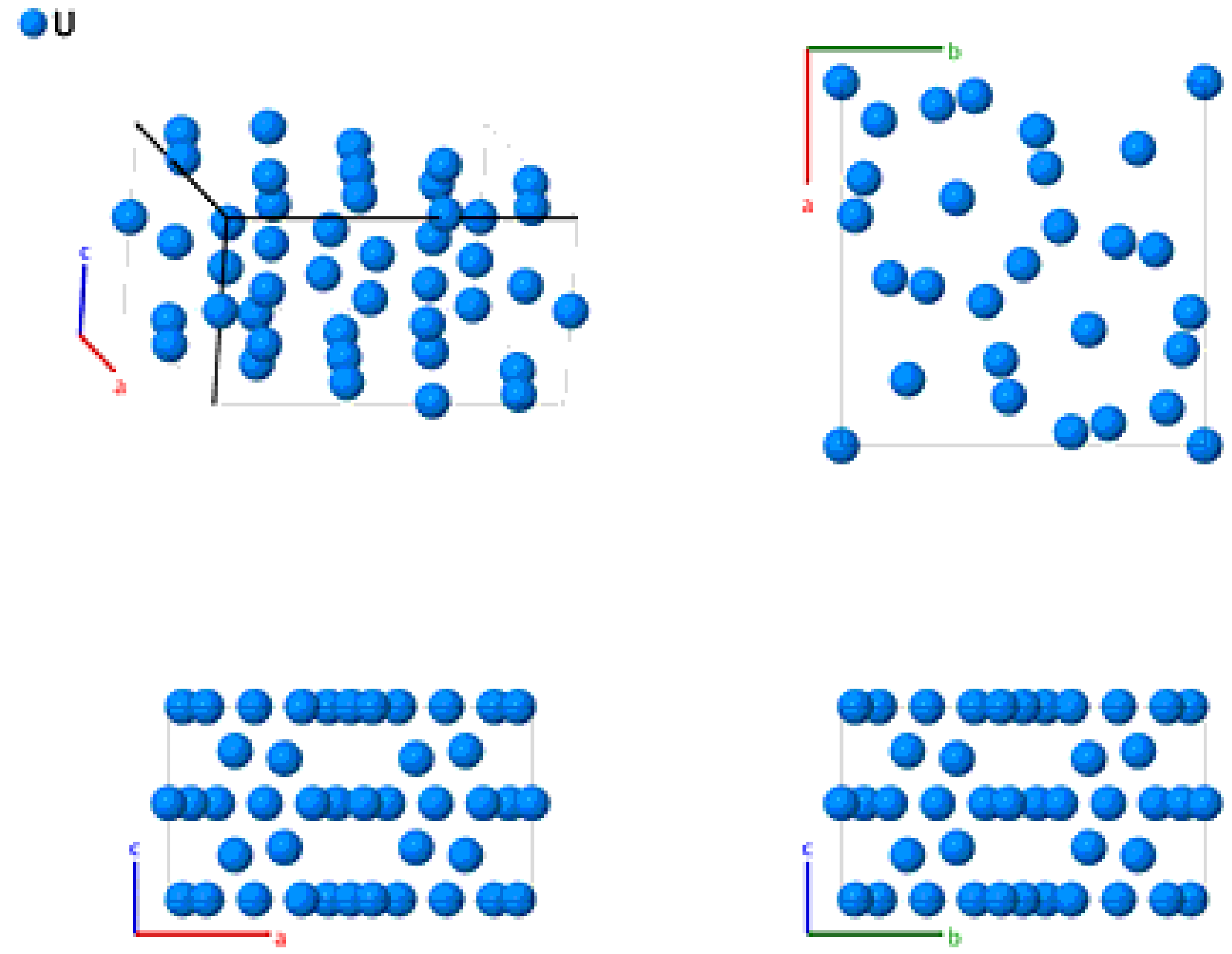

**Prototype** : $\beta$-U

**AFLOW prototype label** : A_tP30_136_bf2ij

***Strukturbericht* designation** : $A_b$

**Pearson symbol** : tP30

**Space group number** : 136

**Space group symbol** : $P4_2/mnm$

**AFLOW prototype command** : `aflow --proto=A_tP30_136_bf2ij`
`--params=`$a, c/a, x_2, x_3, y_3, x_4, y_4, x_5, z_5$

---

- According to (Donohue, 1982), there are three possible space groups which fit the diffraction data for $\beta$-U. This is the highest symmetry space group of the three. Except for a shift of the origin, this structure is crystallographically equivalent to $\sigma$-CrFe ($D8_b$).

---

**Simple Tetragonal primitive vectors:**

$$\mathbf{a}_1 = a\,\hat{\mathbf{x}}$$
$$\mathbf{a}_2 = a\,\hat{\mathbf{y}}$$
$$\mathbf{a}_3 = c\,\hat{\mathbf{z}}$$

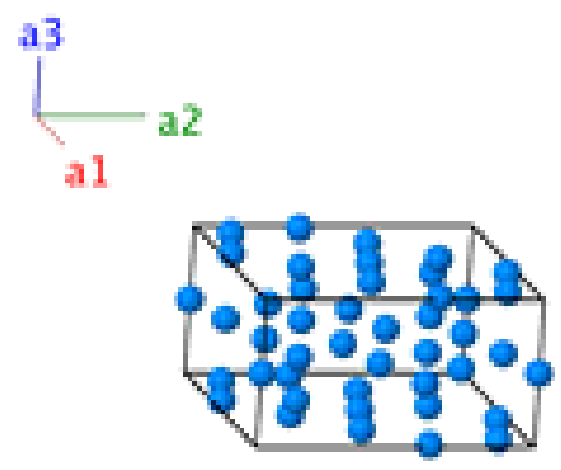

---

**Basis vectors:**

| | | Lattice Coordinates | | Cartesian Coordinates | Wyckoff Position | Atom Type |
|---|---|---|---|---|---|---|
| $\mathbf{B_1}$ | = | $\frac{1}{2}\,\mathbf{a_3}$ | = | $\frac{1}{2}c\,\hat{\mathbf{z}}$ | (2$b$) | U I |
| $\mathbf{B_2}$ | = | $\frac{1}{2}\,\mathbf{a_1} + \frac{1}{2}\,\mathbf{a_2}$ | = | $\frac{1}{2}\,a\,\hat{\mathbf{x}} + \frac{1}{2}\,a\,\hat{\mathbf{y}}$ | (2$b$) | U I |
| $\mathbf{B_3}$ | = | $x_2\,\mathbf{a_1} + x_2\,\mathbf{a_2}$ | = | $x_2\,a\,\hat{\mathbf{x}} + x_2\,a\,\hat{\mathbf{y}}$ | (4$f$) | U II |
| $\mathbf{B_4}$ | = | $-x_2\,\mathbf{a_1} - x_2\,\mathbf{a_2}$ | = | $-x_2\,a\,\hat{\mathbf{x}} - x_2\,a\,\hat{\mathbf{y}}$ | (4$f$) | U II |
| $\mathbf{B_5}$ | = | $\left(\frac{1}{2} - x_2\right)\,\mathbf{a_1} + \left(\frac{1}{2} + x_2\right)\,\mathbf{a_2} + \frac{1}{2}\,\mathbf{a_3}$ | = | $\left(\frac{1}{2} - x_2\right)\,a\,\hat{\mathbf{x}} + \left(\frac{1}{2} + x_2\right)\,a\,\hat{\mathbf{y}} + \frac{1}{2}\,c\,\hat{\mathbf{z}}$ | (4$f$) | U II |
| $\mathbf{B_6}$ | = | $\left(\frac{1}{2} + x_2\right)\,\mathbf{a_1} + \left(\frac{1}{2} - x_2\right)\,\mathbf{a_2} + \frac{1}{2}\,\mathbf{a_3}$ | = | $\left(\frac{1}{2} + x_2\right)\,a\,\hat{\mathbf{x}} + \left(\frac{1}{2} - x_2\right)\,a\,\hat{\mathbf{y}} + \frac{1}{2}\,c\,\hat{\mathbf{z}}$ | (4$f$) | U II |
| $\mathbf{B_7}$ | = | $x_3\,\mathbf{a_1} + y_3\,\mathbf{a_2}$ | = | $x_3\,a\,\hat{\mathbf{x}} + y_3\,a\,\hat{\mathbf{y}}$ | (8$i$) | U III |
| $\mathbf{B_8}$ | = | $-x_3\,\mathbf{a_1} - y_3\,\mathbf{a_2}$ | = | $-x_3\,a\,\hat{\mathbf{x}} - y_3\,a\,\hat{\mathbf{y}}$ | (8$i$) | U III |
| $\mathbf{B_9}$ | = | $\left(\frac{1}{2} - y_3\right)\,\mathbf{a_1} + \left(\frac{1}{2} + x_3\right)\,\mathbf{a_2} + \frac{1}{2}\,\mathbf{a_3}$ | = | $\left(\frac{1}{2} - y_3\right)\,a\,\hat{\mathbf{x}} + \left(\frac{1}{2} + x_3\right)\,a\,\hat{\mathbf{y}} + \frac{1}{2}\,c\,\hat{\mathbf{z}}$ | (8$i$) | U III |
| $\mathbf{B_{10}}$ | = | $\left(\frac{1}{2} + y_3\right)\,\mathbf{a_1} + \left(\frac{1}{2} - x_3\right)\,\mathbf{a_2} + \frac{1}{2}\,\mathbf{a_3}$ | = | $\left(\frac{1}{2} + y_3\right)\,a\,\hat{\mathbf{x}} + \left(\frac{1}{2} - x_3\right)\,a\,\hat{\mathbf{y}} + \frac{1}{2}\,c\,\hat{\mathbf{z}}$ | (8$i$) | U III |
| $\mathbf{B_{11}}$ | = | $\left(\frac{1}{2} - x_3\right)\,\mathbf{a_1} + \left(\frac{1}{2} + y_3\right)\,\mathbf{a_2} + \frac{1}{2}\,\mathbf{a_3}$ | = | $\left(\frac{1}{2} - x_3\right)\,a\,\hat{\mathbf{x}} + \left(\frac{1}{2} + y_3\right)\,a\,\hat{\mathbf{y}} + \frac{1}{2}\,c\,\hat{\mathbf{z}}$ | (8$i$) | U III |
| $\mathbf{B_{12}}$ | = | $\left(\frac{1}{2} + x_3\right)\,\mathbf{a_1} + \left(\frac{1}{2} - y_3\right)\,\mathbf{a_2} + \frac{1}{2}\,\mathbf{a_3}$ | = | $\left(\frac{1}{2} + x_3\right)\,a\,\hat{\mathbf{x}} + \left(\frac{1}{2} - y_3\right)\,a\,\hat{\mathbf{y}} + \frac{1}{2}\,c\,\hat{\mathbf{z}}$ | (8$i$) | U III |
| $\mathbf{B_{13}}$ | = | $y_3\,\mathbf{a_1} + x_3\,\mathbf{a_2}$ | = | $y_3\,a\,\hat{\mathbf{x}} + x_3\,a\,\hat{\mathbf{y}}$ | (8$i$) | U III |
| $\mathbf{B_{14}}$ | = | $-y_3\,\mathbf{a_1} - x_3\,\mathbf{a_2}$ | = | $-y_3\,a\,\hat{\mathbf{x}} - x_3\,a\,\hat{\mathbf{y}}$ | (8$i$) | U III |
| $\mathbf{B_{15}}$ | = | $x_4\,\mathbf{a_1} + y_4\,\mathbf{a_2}$ | = | $x_4\,a\,\hat{\mathbf{x}} + y_4\,a\,\hat{\mathbf{y}}$ | (8$i$) | U IV |
| $\mathbf{B_{16}}$ | = | $-x_4\,\mathbf{a_1} - y_4\,\mathbf{a_2}$ | = | $-x_4\,a\,\hat{\mathbf{x}} - y_4\,a\,\hat{\mathbf{y}}$ | (8$i$) | U IV |
| $\mathbf{B_{17}}$ | = | $\left(\frac{1}{2} - y_4\right)\,\mathbf{a_1} + \left(\frac{1}{2} + x_4\right)\,\mathbf{a_2} + \frac{1}{2}\,\mathbf{a_3}$ | = | $\left(\frac{1}{2} - y_4\right)\,a\,\hat{\mathbf{x}} + \left(\frac{1}{2} + x_4\right)\,a\,\hat{\mathbf{y}} + \frac{1}{2}\,c\,\hat{\mathbf{z}}$ | (8$i$) | U IV |
| $\mathbf{B_{18}}$ | = | $\left(\frac{1}{2} + y_4\right)\,\mathbf{a_1} + \left(\frac{1}{2} - x_4\right)\,\mathbf{a_2} + \frac{1}{2}\,\mathbf{a_3}$ | = | $\left(\frac{1}{2} + y_4\right)\,a\,\hat{\mathbf{x}} + \left(\frac{1}{2} - x_4\right)\,a\,\hat{\mathbf{y}} + \frac{1}{2}\,c\,\hat{\mathbf{z}}$ | (8$i$) | U IV |
| $\mathbf{B_{19}}$ | = | $\left(\frac{1}{2} - x_4\right)\,\mathbf{a_1} + \left(\frac{1}{2} + y_4\right)\,\mathbf{a_2} + \frac{1}{2}\,\mathbf{a_3}$ | = | $\left(\frac{1}{2} - x_4\right)\,a\,\hat{\mathbf{x}} + \left(\frac{1}{2} + y_4\right)\,a\,\hat{\mathbf{y}} + \frac{1}{2}\,c\,\hat{\mathbf{z}}$ | (8$i$) | U IV |
| $\mathbf{B_{20}}$ | = | $\left(\frac{1}{2} + x_4\right)\,\mathbf{a_1} + \left(\frac{1}{2} - y_4\right)\,\mathbf{a_2} + \frac{1}{2}\,\mathbf{a_3}$ | = | $\left(\frac{1}{2} + x_4\right)\,a\,\hat{\mathbf{x}} + \left(\frac{1}{2} - y_4\right)\,a\,\hat{\mathbf{y}} + \frac{1}{2}\,c\,\hat{\mathbf{z}}$ | (8$i$) | U IV |
| $\mathbf{B_{21}}$ | = | $y_4\,\mathbf{a_1} + x_4\,\mathbf{a_2}$ | = | $y_4\,a\,\hat{\mathbf{x}} + x_4\,a\,\hat{\mathbf{y}}$ | (8$i$) | U IV |
| $\mathbf{B_{22}}$ | = | $-y_4\,\mathbf{a_1} - x_4\,\mathbf{a_2}$ | = | $-y_4\,a\,\hat{\mathbf{x}} - x_4\,a\,\hat{\mathbf{y}}$ | (8$i$) | U IV |
| $\mathbf{B_{23}}$ | = | $x_5\,\mathbf{a_1} + x_5\,\mathbf{a_2} + z_5\,\mathbf{a_3}$ | = | $x_5\,a\,\hat{\mathbf{x}} + x_5\,a\,\hat{\mathbf{y}} + z_5\,c\,\hat{\mathbf{z}}$ | (8$j$) | U V |
| $\mathbf{B_{24}}$ | = | $-x_5\,\mathbf{a_1} - x_5\,\mathbf{a_2} + z_5\,\mathbf{a_3}$ | = | $-x_5\,a\,\hat{\mathbf{x}} - x_5\,a\,\hat{\mathbf{y}} + z_5\,c\,\hat{\mathbf{z}}$ | (8$j$) | U V |
| $\mathbf{B_{25}}$ | = | $\left(\frac{1}{2} - x_5\right)\,\mathbf{a_1} + \left(\frac{1}{2} + x_5\right)\,\mathbf{a_2} + \left(\frac{1}{2} + z_5\right)\,\mathbf{a_3}$ | = | $\left(\frac{1}{2} - x_5\right)\,a\,\hat{\mathbf{x}} + \left(\frac{1}{2} + x_5\right)\,a\,\hat{\mathbf{y}} + \left(\frac{1}{2} + z_5\right)\,c\,\hat{\mathbf{z}}$ | (8$j$) | U V |
| $\mathbf{B_{26}}$ | = | $\left(\frac{1}{2} + x_5\right)\,\mathbf{a_1} + \left(\frac{1}{2} - x_5\right)\,\mathbf{a_2} + \left(\frac{1}{2} + z_5\right)\,\mathbf{a_3}$ | = | $\left(\frac{1}{2} + x_5\right)\,a\,\hat{\mathbf{x}} + \left(\frac{1}{2} - x_5\right)\,a\,\hat{\mathbf{y}} + \left(\frac{1}{2} + z_5\right)\,c\,\hat{\mathbf{z}}$ | (8$j$) | U V |
| $\mathbf{B_{27}}$ | = | $\left(\frac{1}{2} - x_5\right)\,\mathbf{a_1} + \left(\frac{1}{2} + x_5\right)\,\mathbf{a_2} + \left(\frac{1}{2} - z_5\right)\,\mathbf{a_3}$ | = | $\left(\frac{1}{2} - x_5\right)\,a\,\hat{\mathbf{x}} + \left(\frac{1}{2} + x_5\right)\,a\,\hat{\mathbf{y}} + \left(\frac{1}{2} - z_5\right)\,c\,\hat{\mathbf{z}}$ | (8$j$) | U V |
| $\mathbf{B_{28}}$ | = | $\left(\frac{1}{2} + x_5\right)\,\mathbf{a_1} + \left(\frac{1}{2} - x_5\right)\,\mathbf{a_2} + \left(\frac{1}{2} - z_5\right)\,\mathbf{a_3}$ | = | $\left(\frac{1}{2} + x_5\right)\,a\,\hat{\mathbf{x}} + \left(\frac{1}{2} - x_5\right)\,a\,\hat{\mathbf{y}} + \left(\frac{1}{2} - z_5\right)\,c\,\hat{\mathbf{z}}$ | (8$j$) | U V |
| $\mathbf{B_{29}}$ | = | $x_5\,\mathbf{a_1} + x_5\,\mathbf{a_2} - z_5\,\mathbf{a_3}$ | = | $x_5\,a\,\hat{\mathbf{x}} + x_5\,a\,\hat{\mathbf{y}} - z_5\,c\,\hat{\mathbf{z}}$ | (8$j$) | U V |
| $\mathbf{B_{30}}$ | = | $-x_5\,\mathbf{a_1} - x_5\,\mathbf{a_2} - z_5\,\mathbf{a_3}$ | = | $-x_5\,a\,\hat{\mathbf{x}} - x_5\,a\,\hat{\mathbf{y}} - z_5\,c\,\hat{\mathbf{z}}$ | (8$j$) | U V |

(1953), doi:10.1107/S0365110X53002167.

**Found in:**

- J. Donohue, *The Structure of the Elements* (Robert E. Krieger Publishing Company, Malabar, Florida, 1982), pp. 134-147.

---

**Geometry files:**

- CIF: pp. 696
- POSCAR: pp. 696

# $\beta$-BeO Structure: AB_tP8_136_g_f

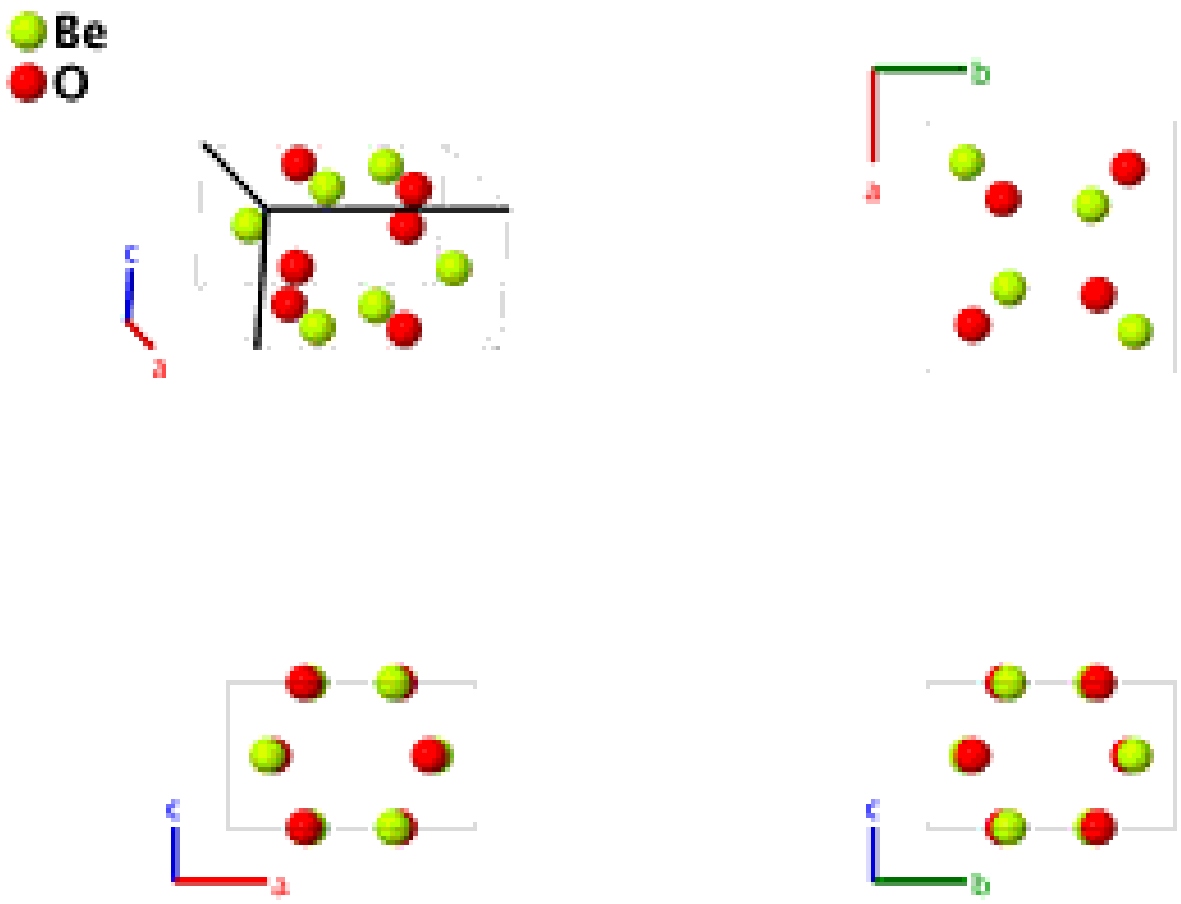

| | | |
|---|---|---|
| **Prototype** | : | $\beta$-BeO |
| **AFLOW prototype label** | : | AB_tP8_136_g_f |
| ***Strukturbericht* designation** | : | None |
| **Pearson symbol** | : | tP8 |
| **Space group number** | : | 136 |
| **Space group symbol** | : | $P4_2/mnm$ |
| **AFLOW prototype command** | : | `aflow --proto=AB_tP8_136_g_f`<br>`--params=`$a, c/a, x_1, x_2$ |

**Other compounds with this structure:**

- ZnO

**Simple Tetragonal primitive vectors:**

$$\mathbf{a}_1 = a\,\hat{\mathbf{x}}$$
$$\mathbf{a}_2 = a\,\hat{\mathbf{y}}$$
$$\mathbf{a}_3 = c\,\hat{\mathbf{z}}$$

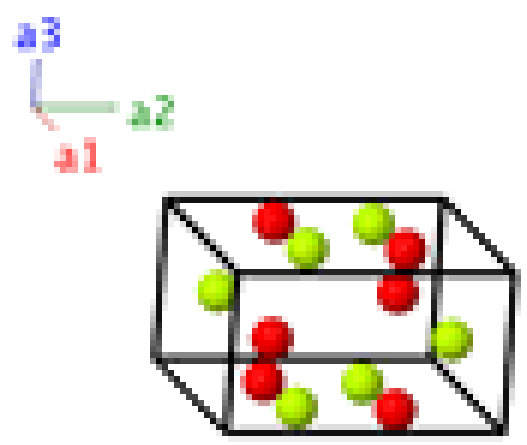

**Basis vectors:**

| | | Lattice Coordinates | | Cartesian Coordinates | Wyckoff Position | Atom Type |
|---|---|---|---|---|---|---|
| $\mathbf{B_1}$ | = | $x_1\,\mathbf{a_1} + x_1\,\mathbf{a_2}$ | = | $x_1\,a\,\hat{\mathbf{x}} + x_1\,a\,\hat{\mathbf{y}}$ | $(4f)$ | O |
| $\mathbf{B_2}$ | = | $-x_1\,\mathbf{a_1} - x_1\,\mathbf{a_2}$ | = | $-x_1\,a\,\hat{\mathbf{x}} - x_1\,a\,\hat{\mathbf{y}}$ | $(4f)$ | O |
| $\mathbf{B_3}$ | = | $\left(\frac{1}{2} - x_1\right)\mathbf{a_1} + \left(\frac{1}{2} + x_1\right)\mathbf{a_2} + \frac{1}{2}\,\mathbf{a_3}$ | = | $\left(\frac{1}{2} - x_1\right)a\,\hat{\mathbf{x}} + \left(\frac{1}{2} + x_1\right)a\,\hat{\mathbf{y}} + \frac{1}{2}\,c\,\hat{\mathbf{z}}$ | $(4f)$ | O |
| $\mathbf{B_4}$ | = | $\left(\frac{1}{2} + x_1\right)\mathbf{a_1} + \left(\frac{1}{2} - x_1\right)\mathbf{a_2} + \frac{1}{2}\,\mathbf{a_3}$ | = | $\left(\frac{1}{2} + x_1\right)a\,\hat{\mathbf{x}} + \left(\frac{1}{2} - x_1\right)a\,\hat{\mathbf{y}} + \frac{1}{2}\,c\,\hat{\mathbf{z}}$ | $(4f)$ | O |
| $\mathbf{B_5}$ | = | $x_2\,\mathbf{a_1} - x_2\,\mathbf{a_2}$ | = | $x_2\,a\,\hat{\mathbf{x}} - x_2\,a\,\hat{\mathbf{y}}$ | $(4g)$ | Be |
| $\mathbf{B_6}$ | = | $-x_2\,\mathbf{a_1} + x_2\,\mathbf{a_2}$ | = | $-x_2\,a\,\mathbf{\hat{x}} + x_2\,a\,\mathbf{\hat{y}}$ | (4$g$) | Be |
| $\mathbf{B_7}$ | = | $\left(\frac{1}{2} + x_2\right)\mathbf{a_1} + \left(\frac{1}{2} + x_2\right)\mathbf{a_2} + \frac{1}{2}\,\mathbf{a_3}$ | = | $\left(\frac{1}{2} + x_2\right)a\,\mathbf{\hat{x}} + \left(\frac{1}{2} + x_2\right)a\,\mathbf{\hat{y}} + \frac{1}{2}\,c\,\mathbf{\hat{z}}$ | (4$g$) | Be |
| $\mathbf{B_8}$ | = | $\left(\frac{1}{2} - x_2\right)\mathbf{a_1} + \left(\frac{1}{2} - x_2\right)\mathbf{a_2} + \frac{1}{2}\,\mathbf{a_3}$ | = | $\left(\frac{1}{2} - x_2\right)a\,\mathbf{\hat{x}} + \left(\frac{1}{2} - x_2\right)a\,\mathbf{\hat{y}} + \frac{1}{2}\,c\,\mathbf{\hat{z}}$ | (4$g$) | Be |

---

---

**Geometry files:**

- CIF: pp. 696
- POSCAR: pp. 697

# Rutile ($TiO_2$, C4) Structure: A2B_tP6_136_f_a

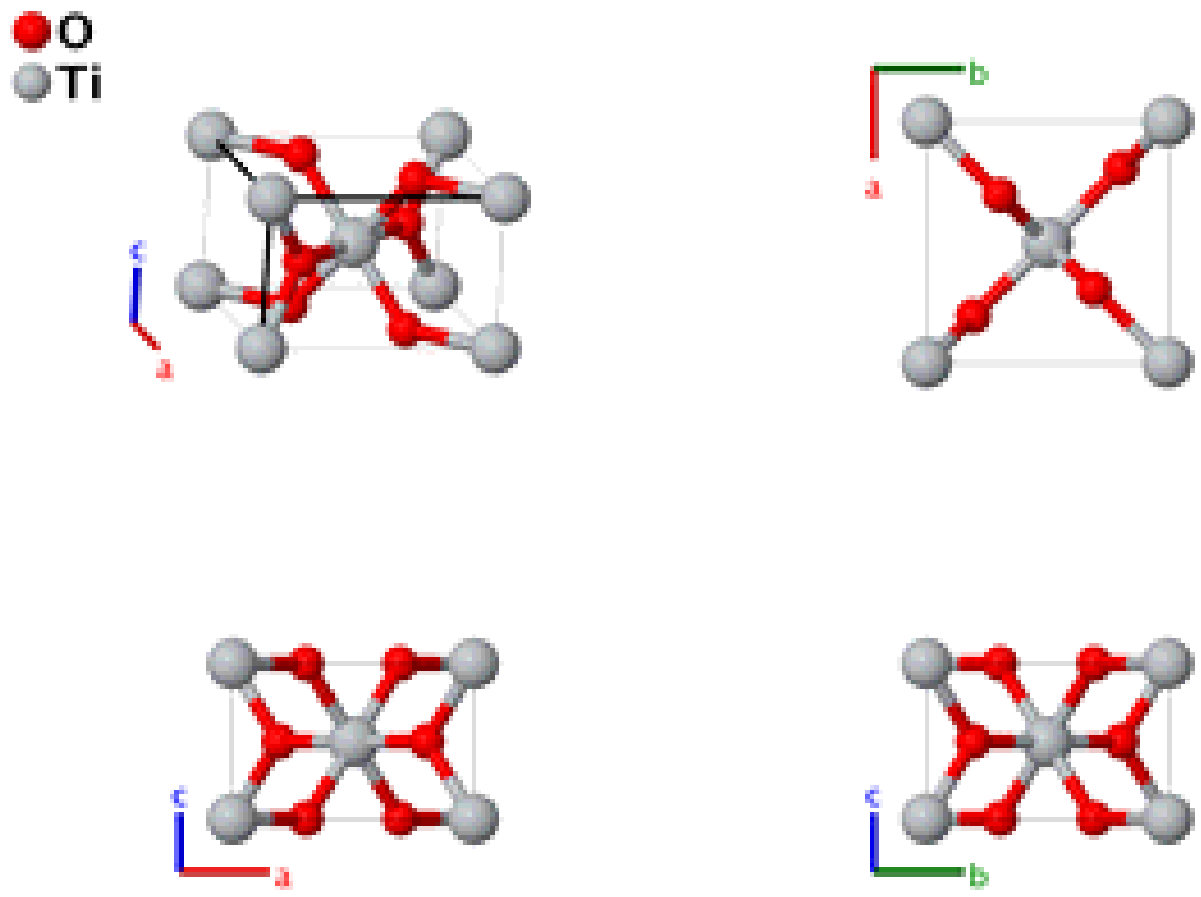

| | | |
|---|---|---|
| **Prototype** | : | $TiO_2$ |
| **AFLOW prototype label** | : | A2B_tP6_136_f_a |
| ***Strukturbericht* designation** | : | C4 |
| **Pearson symbol** | : | tP6 |
| **Space group number** | : | 136 |
| **Space group symbol** | : | $P4_2/mnm$ |
| **AFLOW prototype command** | : | `aflow --proto=A2B_tP6_136_f_a`<br>`--params=`$a, c/a, x_2$ |

**Other compounds with this structure:**

- $CoF_2$, $MgF_2$, $MnF_2$, $NiF_2$, $ZnF_2$, $GeO_2$, $IrO_2$, $MoO_2$, $PbO_2$, $SiO_2$ (stishovite), $SnO_2$ (cassiterite), $TaO_2$, $WO_2$

**Simple Tetragonal primitive vectors:**

$$\mathbf{a}_1 = a\,\mathbf{\hat{x}}$$
$$\mathbf{a}_2 = a\,\mathbf{\hat{y}}$$
$$\mathbf{a}_3 = c\,\mathbf{\hat{z}}$$

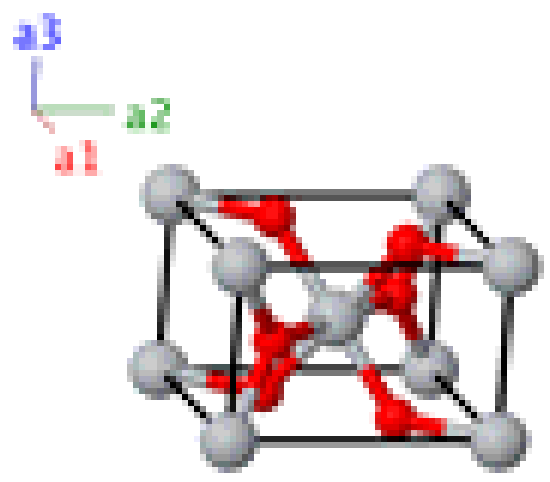

**Basis vectors:**

| | | Lattice Coordinates | | Cartesian Coordinates | Wyckoff Position | Atom Type |
|---|---|---|---|---|---|---|
| $\mathbf{B}_1$ | = | $0\,\mathbf{a}_1 + 0\,\mathbf{a}_2 + 0\,\mathbf{a}_3$ | = | $0\mathbf{\hat{x}} + 0\mathbf{\hat{y}} + 0\mathbf{\hat{z}}$ | (2$a$) | Ti |
| $\mathbf{B}_2$ | = | $\frac{1}{2}\,\mathbf{a}_1 + \frac{1}{2}\,\mathbf{a}_2 + \frac{1}{2}\,\mathbf{a}_3$ | = | $\frac{1}{2}\,a\,\mathbf{\hat{x}} + \frac{1}{2}\,a\,\mathbf{\hat{y}} + \frac{1}{2}\,c\,\mathbf{\hat{z}}$ | (2$a$) | Ti |
| $\mathbf{B}_3$ | = | $x_2\,\mathbf{a}_1 + x_2\,\mathbf{a}_2$ | = | $x_2\,a\,\mathbf{\hat{x}} + x_2\,a\,\mathbf{\hat{y}}$ | (4$f$) | O |
| $\mathbf{B}_4$ | = | $-x_2\,\mathbf{a}_1 - x_2\,\mathbf{a}_2$ | = | $-x_2\,a\,\mathbf{\hat{x}} - x_2\,a\,\mathbf{\hat{y}}$ | (4$f$) | O |
| $\mathbf{B}_5$ | = | $\left(\frac{1}{2} - x_2\right)\,\mathbf{a}_1 + \left(\frac{1}{2} + x_2\right)\,\mathbf{a}_2 + \frac{1}{2}\,\mathbf{a}_3$ | = | $\left(\frac{1}{2} - x_2\right)\,a\,\mathbf{\hat{x}} + \left(\frac{1}{2} + x_2\right)\,a\,\mathbf{\hat{y}} + \frac{1}{2}\,c\,\mathbf{\hat{z}}$ | (4$f$) | O |
| $\mathbf{B}_6$ | = | $\left(\frac{1}{2} + x_2\right)\,\mathbf{a}_1 + \left(\frac{1}{2} - x_2\right)\,\mathbf{a}_2 + \frac{1}{2}\,\mathbf{a}_3$ | = | $\left(\frac{1}{2} + x_2\right)\,a\,\mathbf{\hat{x}} + \left(\frac{1}{2} - x_2\right)\,a\,\mathbf{\hat{y}} + \frac{1}{2}\,c\,\mathbf{\hat{z}}$ | (4$f$) | O |

**Found in:**

**Geometry files:**

- CIF: pp. 697
- POSCAR: pp. 697

# $\sigma$-CrFe ($D8_b$) Structure: sigma_tP30_136_bf2ij

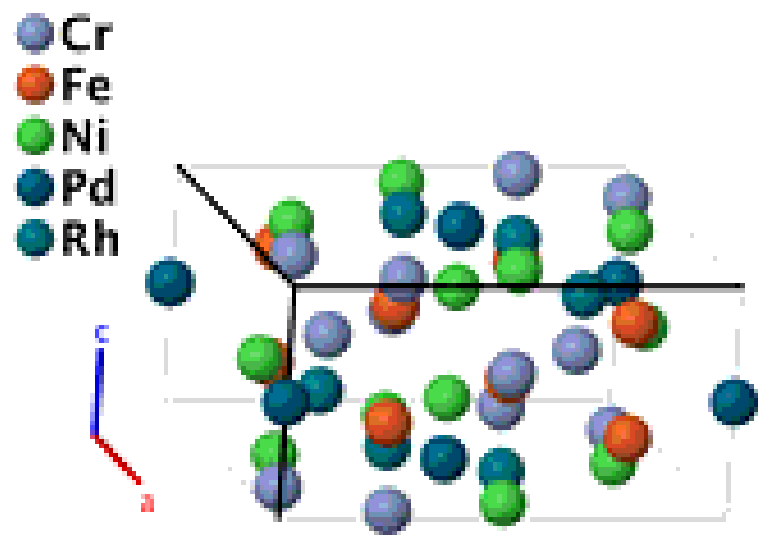

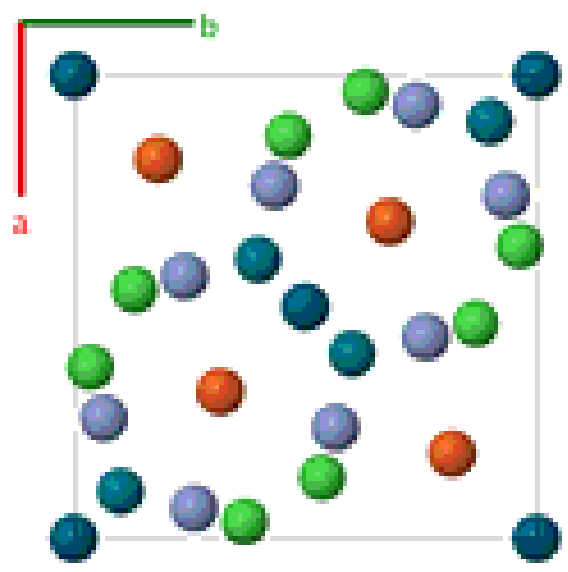

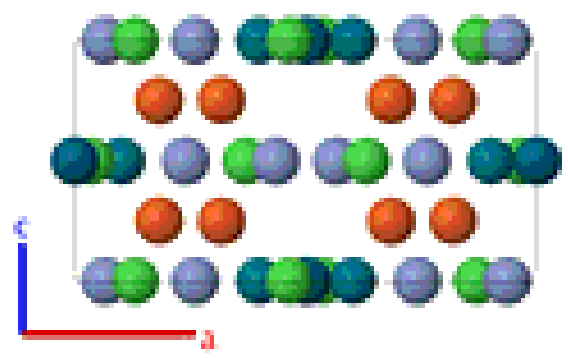

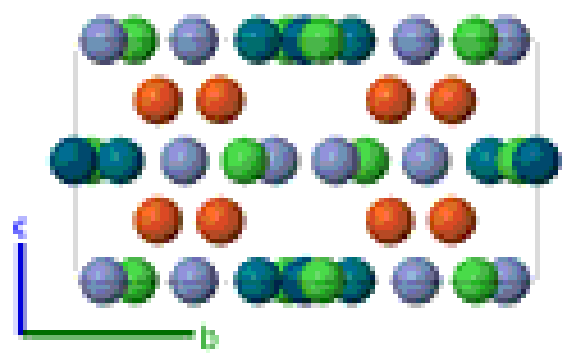

| | | |
|---|---|---|
| **Prototype** | : | $\sigma$-CrFe |
| **AFLOW prototype label** | : | sigma_tP30_136_bf2ij |
| ***Strukturbericht* designation** | : | $D8_b$ |
| **Pearson symbol** | : | tP30 |
| **Space group number** | : | 136 |
| **Space group symbol** | : | $P4_2/mnm$ |
| **AFLOW prototype command** | : | `aflow --proto=sigma_tP30_136_bf2ij` `--params=`$a, c/a, x_2, x_3, y_3, x_4, y_4, x_5, z_5$ |

**Other compounds with this structure:**

- $Al_3CoNb_6$, $AlCrNb_3$, $Co_2Mo_3$, $Ta_3V_7$, $PdTa_3$, $IrMo_2$, $IrW_3$, many others.

- The atoms in this lattice are completely disordered, that is, the Cr and Fe atoms are distributed randomly on the sites in the unit cell. This seems to be the case for all of the compounds listed below. We have chosen several of the atoms near Fe and Cr in the periodic table to color the above pictures. Except for a shift of the origin, this structure is crystallographically equivalent to $\beta$-U ($A_b$).

**Simple Tetragonal primitive vectors:**

$$\mathbf{a}_1 = a\,\hat{\mathbf{x}}$$
$$\mathbf{a}_2 = a\,\hat{\mathbf{y}}$$
$$\mathbf{a}_3 = c\,\hat{\mathbf{z}}$$

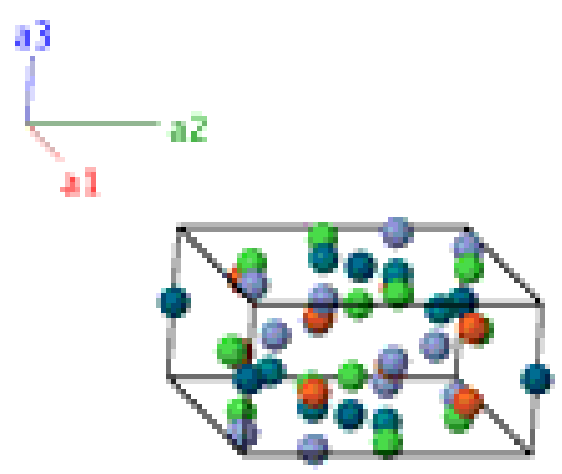

**Basis vectors:**

| | | Lattice Coordinates | | Cartesian Coordinates | Wyckoff Position | Atom Type |
|---|---|---|---|---|---|---|
| $\mathbf{B_1}$ | = | $\frac{1}{2}\,\mathbf{a_3}$ | = | $\frac{1}{2}c\,\hat{\mathbf{z}}$ | $(2b)$ | M I |
| $\mathbf{B_2}$ | = | $\frac{1}{2}\,\mathbf{a_1} + \frac{1}{2}\,\mathbf{a_2}$ | = | $\frac{1}{2}\,a\,\hat{\mathbf{x}} + \frac{1}{2}\,a\,\hat{\mathbf{y}}$ | $(2b)$ | M I |
| $\mathbf{B_3}$ | = | $x_2\,\mathbf{a_1} + x_2\,\mathbf{a_2}$ | = | $x_2\,a\,\hat{\mathbf{x}} + x_2\,a\,\hat{\mathbf{y}}$ | $(4f)$ | M II |
| $\mathbf{B_4}$ | = | $-x_2\,\mathbf{a_1} - x_2\,\mathbf{a_2}$ | = | $-x_2\,a\,\hat{\mathbf{x}} - x_2\,a\,\hat{\mathbf{y}}$ | $(4f)$ | M II |
| $\mathbf{B_5}$ | = | $\left(\frac{1}{2} - x_2\right)\,\mathbf{a_1} + \left(\frac{1}{2} + x_2\right)\,\mathbf{a_2} + \frac{1}{2}\,\mathbf{a_3}$ | = | $\left(\frac{1}{2} - x_2\right)\,a\,\hat{\mathbf{x}} + \left(\frac{1}{2} + x_2\right)\,a\,\hat{\mathbf{y}} + \frac{1}{2}\,c\,\hat{\mathbf{z}}$ | $(4f)$ | M II |
| $\mathbf{B_6}$ | = | $\left(\frac{1}{2} + x_2\right)\,\mathbf{a_1} + \left(\frac{1}{2} - x_2\right)\,\mathbf{a_2} + \frac{1}{2}\,\mathbf{a_3}$ | = | $\left(\frac{1}{2} + x_2\right)\,a\,\hat{\mathbf{x}} + \left(\frac{1}{2} - x_2\right)\,a\,\hat{\mathbf{y}} + \frac{1}{2}\,c\,\hat{\mathbf{z}}$ | $(4f)$ | M II |
| $\mathbf{B_7}$ | = | $x_3\,\mathbf{a_1} + y_3\,\mathbf{a_2}$ | = | $x_3\,a\,\hat{\mathbf{x}} + y_3\,a\,\hat{\mathbf{y}}$ | $(8i)$ | M III |
| $\mathbf{B_8}$ | = | $-x_3\,\mathbf{a_1} - y_3\,\mathbf{a_2}$ | = | $-x_3\,a\,\hat{\mathbf{x}} - y_3\,a\,\hat{\mathbf{y}}$ | $(8i)$ | M III |
| $\mathbf{B_9}$ | = | $\left(\frac{1}{2} - y_3\right)\,\mathbf{a_1} + \left(\frac{1}{2} + x_3\right)\,\mathbf{a_2} + \frac{1}{2}\,\mathbf{a_3}$ | = | $\left(\frac{1}{2} - y_3\right)\,a\,\hat{\mathbf{x}} + \left(\frac{1}{2} + x_3\right)\,a\,\hat{\mathbf{y}} + \frac{1}{2}\,c\,\hat{\mathbf{z}}$ | $(8i)$ | M III |
| $\mathbf{B_{10}}$ | = | $\left(\frac{1}{2} + y_3\right)\,\mathbf{a_1} + \left(\frac{1}{2} - x_3\right)\,\mathbf{a_2} + \frac{1}{2}\,\mathbf{a_3}$ | = | $\left(\frac{1}{2} + y_3\right)\,a\,\hat{\mathbf{x}} + \left(\frac{1}{2} - x_3\right)\,a\,\hat{\mathbf{y}} + \frac{1}{2}\,c\,\hat{\mathbf{z}}$ | $(8i)$ | M III |
| $\mathbf{B_{11}}$ | = | $\left(\frac{1}{2} - x_3\right)\,\mathbf{a_1} + \left(\frac{1}{2} + y_3\right)\,\mathbf{a_2} + \frac{1}{2}\,\mathbf{a_3}$ | = | $\left(\frac{1}{2} - x_3\right)\,a\,\hat{\mathbf{x}} + \left(\frac{1}{2} + y_3\right)\,a\,\hat{\mathbf{y}} + \frac{1}{2}\,c\,\hat{\mathbf{z}}$ | $(8i)$ | M III |
| $\mathbf{B_{12}}$ | = | $\left(\frac{1}{2} + x_3\right)\,\mathbf{a_1} + \left(\frac{1}{2} - y_3\right)\,\mathbf{a_2} + \frac{1}{2}\,\mathbf{a_3}$ | = | $\left(\frac{1}{2} + x_3\right)\,a\,\hat{\mathbf{x}} + \left(\frac{1}{2} - y_3\right)\,a\,\hat{\mathbf{y}} + \frac{1}{2}\,c\,\hat{\mathbf{z}}$ | $(8i)$ | M III |
| $\mathbf{B_{13}}$ | = | $y_3\,\mathbf{a_1} + x_3\,\mathbf{a_2}$ | = | $y_3\,a\,\hat{\mathbf{x}} + x_3\,a\,\hat{\mathbf{y}}$ | $(8i)$ | M III |
| $\mathbf{B_{14}}$ | = | $-y_3\,\mathbf{a_1} - x_3\,\mathbf{a_2}$ | = | $-y_3\,a\,\hat{\mathbf{x}} - x_3\,a\,\hat{\mathbf{y}}$ | $(8i)$ | M III |
| $\mathbf{B_{15}}$ | = | $x_4\,\mathbf{a_1} + y_4\,\mathbf{a_2}$ | = | $x_4\,a\,\hat{\mathbf{x}} + y_4\,a\,\hat{\mathbf{y}}$ | $(8i)$ | M IV |
| $\mathbf{B_{16}}$ | = | $-x_4\,\mathbf{a_1} - y_4\,\mathbf{a_2}$ | = | $-x_4\,a\,\hat{\mathbf{x}} - y_4\,a\,\hat{\mathbf{y}}$ | $(8i)$ | M IV |
| $\mathbf{B_{17}}$ | = | $\left(\frac{1}{2} - y_4\right)\,\mathbf{a_1} + \left(\frac{1}{2} + x_4\right)\,\mathbf{a_2} + \frac{1}{2}\,\mathbf{a_3}$ | = | $\left(\frac{1}{2} - y_4\right)\,a\,\hat{\mathbf{x}} + \left(\frac{1}{2} + x_4\right)\,a\,\hat{\mathbf{y}} + \frac{1}{2}\,c\,\hat{\mathbf{z}}$ | $(8i)$ | M IV |
| $\mathbf{B_{18}}$ | = | $\left(\frac{1}{2} + y_4\right)\,\mathbf{a_1} + \left(\frac{1}{2} - x_4\right)\,\mathbf{a_2} + \frac{1}{2}\,\mathbf{a_3}$ | = | $\left(\frac{1}{2} + y_4\right)\,a\,\hat{\mathbf{x}} + \left(\frac{1}{2} - x_4\right)\,a\,\hat{\mathbf{y}} + \frac{1}{2}\,c\,\hat{\mathbf{z}}$ | $(8i)$ | M IV |
| $\mathbf{B_{19}}$ | = | $\left(\frac{1}{2} - x_4\right)\,\mathbf{a_1} + \left(\frac{1}{2} + y_4\right)\,\mathbf{a_2} + \frac{1}{2}\,\mathbf{a_3}$ | = | $\left(\frac{1}{2} - x_4\right)\,a\,\hat{\mathbf{x}} + \left(\frac{1}{2} + y_4\right)\,a\,\hat{\mathbf{y}} + \frac{1}{2}\,c\,\hat{\mathbf{z}}$ | $(8i)$ | M IV |
| $\mathbf{B_{20}}$ | = | $\left(\frac{1}{2} + x_4\right)\,\mathbf{a_1} + \left(\frac{1}{2} - y_4\right)\,\mathbf{a_2} + \frac{1}{2}\,\mathbf{a_3}$ | = | $\left(\frac{1}{2} + x_4\right)\,a\,\hat{\mathbf{x}} + \left(\frac{1}{2} - y_4\right)\,a\,\hat{\mathbf{y}} + \frac{1}{2}\,c\,\hat{\mathbf{z}}$ | $(8i)$ | M IV |
| $\mathbf{B_{21}}$ | = | $y_4\,\mathbf{a_1} + x_4\,\mathbf{a_2}$ | = | $y_4\,a\,\hat{\mathbf{x}} + x_4\,a\,\hat{\mathbf{y}}$ | $(8i)$ | M IV |
| $\mathbf{B_{22}}$ | = | $-y_4\,\mathbf{a_1} - x_4\,\mathbf{a_2}$ | = | $-y_4\,a\,\hat{\mathbf{x}} - x_4\,a\,\hat{\mathbf{y}}$ | $(8i)$ | M IV |
| $\mathbf{B_{23}}$ | = | $x_5\,\mathbf{a_1} + x_5\,\mathbf{a_2} + z_5\,\mathbf{a_3}$ | = | $x_5\,a\,\hat{\mathbf{x}} + x_5\,a\,\hat{\mathbf{y}} + z_5\,c\,\hat{\mathbf{z}}$ | $(8j)$ | M V |
| $\mathbf{B_{24}}$ | = | $-x_5\,\mathbf{a_1} - x_5\,\mathbf{a_2} + z_5\,\mathbf{a_3}$ | = | $-x_5\,a\,\hat{\mathbf{x}} - x_5\,a\,\hat{\mathbf{y}} + z_5\,c\,\hat{\mathbf{z}}$ | $(8j)$ | M V |
| $\mathbf{B_{25}}$ | = | $\left(\frac{1}{2} - x_5\right)\,\mathbf{a_1} + \left(\frac{1}{2} + x_5\right)\,\mathbf{a_2} + \left(\frac{1}{2} + z_5\right)\,\mathbf{a_3}$ | = | $\left(\frac{1}{2} - x_5\right)\,a\,\hat{\mathbf{x}} + \left(\frac{1}{2} + x_5\right)\,a\,\hat{\mathbf{y}} + \left(\frac{1}{2} + z_5\right)\,c\,\hat{\mathbf{z}}$ | $(8j)$ | M V |
| $\mathbf{B_{26}}$ | = | $\left(\frac{1}{2} + x_5\right)\,\mathbf{a_1} + \left(\frac{1}{2} - x_5\right)\,\mathbf{a_2} + \left(\frac{1}{2} + z_5\right)\,\mathbf{a_3}$ | = | $\left(\frac{1}{2} + x_5\right)\,a\,\hat{\mathbf{x}} + \left(\frac{1}{2} - x_5\right)\,a\,\hat{\mathbf{y}} + \left(\frac{1}{2} + z_5\right)\,c\,\hat{\mathbf{z}}$ | $(8j)$ | M V |
| $\mathbf{B_{27}}$ | = | $\left(\frac{1}{2} - x_5\right)\,\mathbf{a_1} + \left(\frac{1}{2} + x_5\right)\,\mathbf{a_2} + \left(\frac{1}{2} - z_5\right)\,\mathbf{a_3}$ | = | $\left(\frac{1}{2} - x_5\right)\,a\,\hat{\mathbf{x}} + \left(\frac{1}{2} + x_5\right)\,a\,\hat{\mathbf{y}} + \left(\frac{1}{2} - z_5\right)\,c\,\hat{\mathbf{z}}$ | $(8j)$ | M V |
| $\mathbf{B_{28}}$ | = | $\left(\frac{1}{2} + x_5\right)\,\mathbf{a_1} + \left(\frac{1}{2} - x_5\right)\,\mathbf{a_2} + \left(\frac{1}{2} - z_5\right)\,\mathbf{a_3}$ | = | $\left(\frac{1}{2} + x_5\right)\,a\,\hat{\mathbf{x}} + \left(\frac{1}{2} - x_5\right)\,a\,\hat{\mathbf{y}} + \left(\frac{1}{2} - z_5\right)\,c\,\hat{\mathbf{z}}$ | $(8j)$ | M V |
| $\mathbf{B_{29}}$ | = | $x_5\,\mathbf{a_1} + x_5\,\mathbf{a_2} - z_5\,\mathbf{a_3}$ | = | $x_5\,a\,\hat{\mathbf{x}} + x_5\,a\,\hat{\mathbf{y}} - z_5\,c\,\hat{\mathbf{z}}$ | $(8j)$ | M V |
| $\mathbf{B_{30}}$ | = | $-x_5\,\mathbf{a_1} - x_5\,\mathbf{a_2} - z_5\,\mathbf{a_3}$ | = | $-x_5\,a\,\hat{\mathbf{x}} - x_5\,a\,\hat{\mathbf{y}} - z_5\,c\,\hat{\mathbf{z}}$ | $(8j)$ | M V |

**Found in:**

---

**Geometry files:**

- CIF: pp. 697
- POSCAR: pp. 698

# $\gamma$-N Structure: A_tP4_136_f

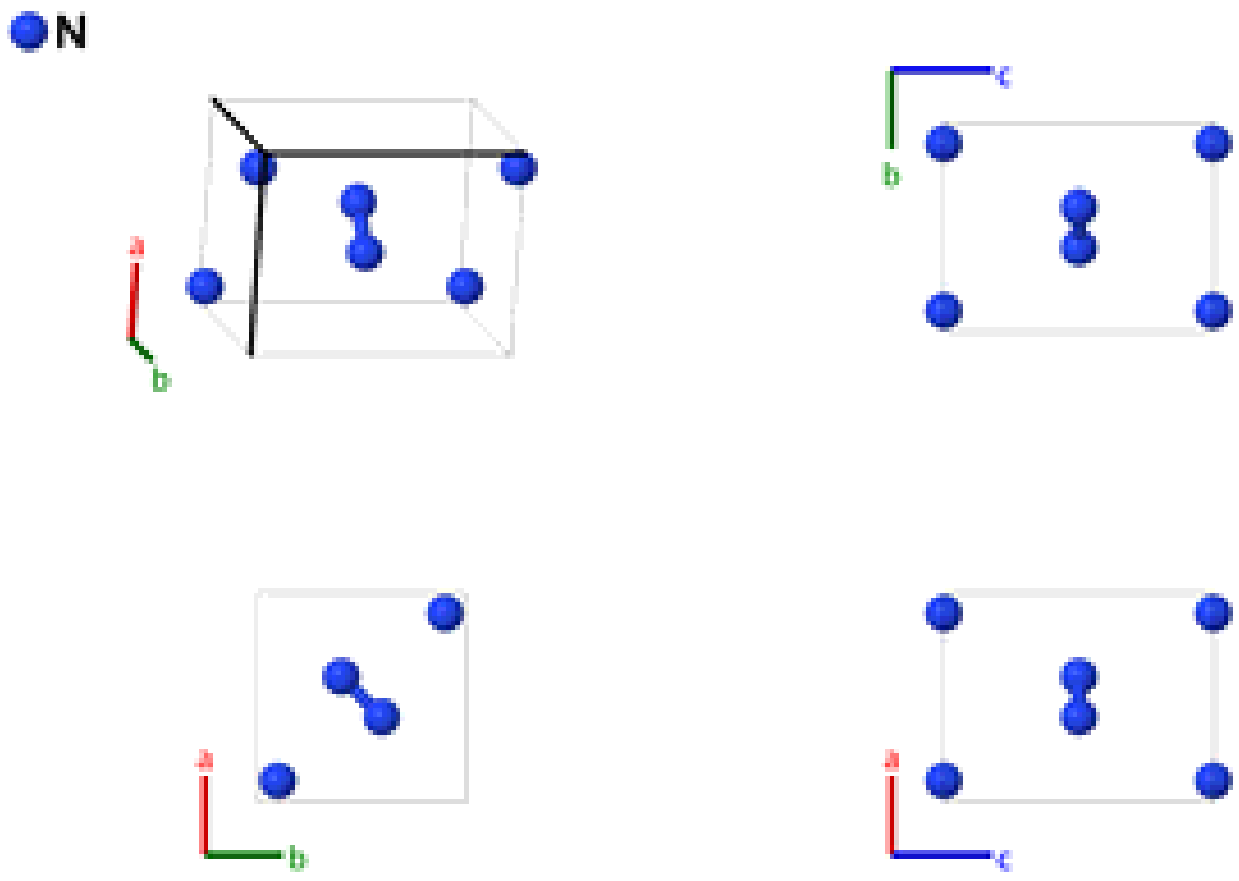

| | | |
|---|---|---|
| **Prototype** | : | $\gamma$-N |
| **AFLOW prototype label** | : | A_tP4_136_f |
| ***Strukturbericht* designation** | : | None |
| **Pearson symbol** | : | tP4 |
| **Space group number** | : | 136 |
| **Space group symbol** | : | $P4_2/mnm$ |
| **AFLOW prototype command** | : | `aflow --proto=A_tP4_136_f` <br> `--params=`$a, c/a, x_1$ |

**Simple Tetragonal primitive vectors:**

$$
\begin{aligned}
\mathbf{a}_1 &= a\,\hat{\mathbf{x}} \\
\mathbf{a}_2 &= a\,\hat{\mathbf{y}} \\
\mathbf{a}_3 &= c\,\hat{\mathbf{z}}
\end{aligned}
$$

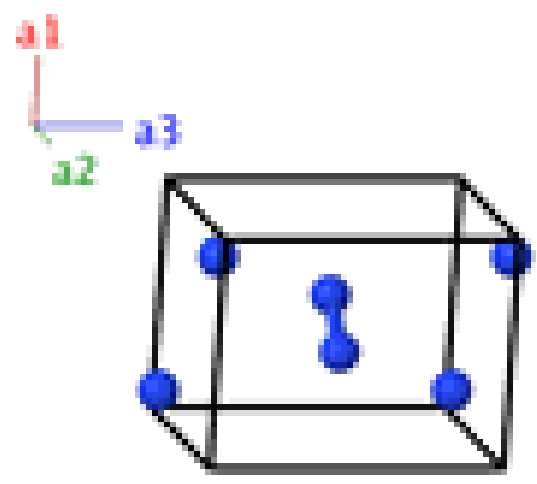

**Basis vectors:**

| | | Lattice Coordinates | | Cartesian Coordinates | Wyckoff Position | Atom Type |
|---|---|---|---|---|---|---|
| $\mathbf{B}_1$ | = | $x_1\,\mathbf{a}_1 + x_1\,\mathbf{a}_2$ | = | $x_1\,a\,\hat{\mathbf{x}} + x_1\,a\,\hat{\mathbf{y}}$ | (4$f$) | N |
| $\mathbf{B}_2$ | = | $-x_1\,\mathbf{a}_1 - x_1\,\mathbf{a}_2$ | = | $-x_1\,a\,\hat{\mathbf{x}} - x_1\,a\,\hat{\mathbf{y}}$ | (4$f$) | N |
| $\mathbf{B}_3$ | = | $\left(\frac{1}{2} - x_1\right)\mathbf{a}_1 + \left(\frac{1}{2} + x_1\right)\mathbf{a}_2 + \frac{1}{2}\,\mathbf{a}_3$ | = | $\left(\frac{1}{2} - x_1\right)a\,\hat{\mathbf{x}} + \left(\frac{1}{2} + x_1\right)a\,\hat{\mathbf{y}} + \frac{1}{2}\,c\,\hat{\mathbf{z}}$ | (4$f$) | N |
| $\mathbf{B}_4$ | = | $\left(\frac{1}{2} + x_1\right)\mathbf{a}_1 + \left(\frac{1}{2} - x_1\right)\mathbf{a}_2 + \frac{1}{2}\,\mathbf{a}_3$ | = | $\left(\frac{1}{2} + x_1\right)a\,\hat{\mathbf{x}} + \left(\frac{1}{2} - x_1\right)a\,\hat{\mathbf{y}} + \frac{1}{2}\,c\,\hat{\mathbf{z}}$ | (4$f$) | N |

**Found in:**

---

**Geometry files:**

- CIF: pp. 698
- POSCAR: pp. 698

# Cl (A18) Structure: A_tP16_138_j

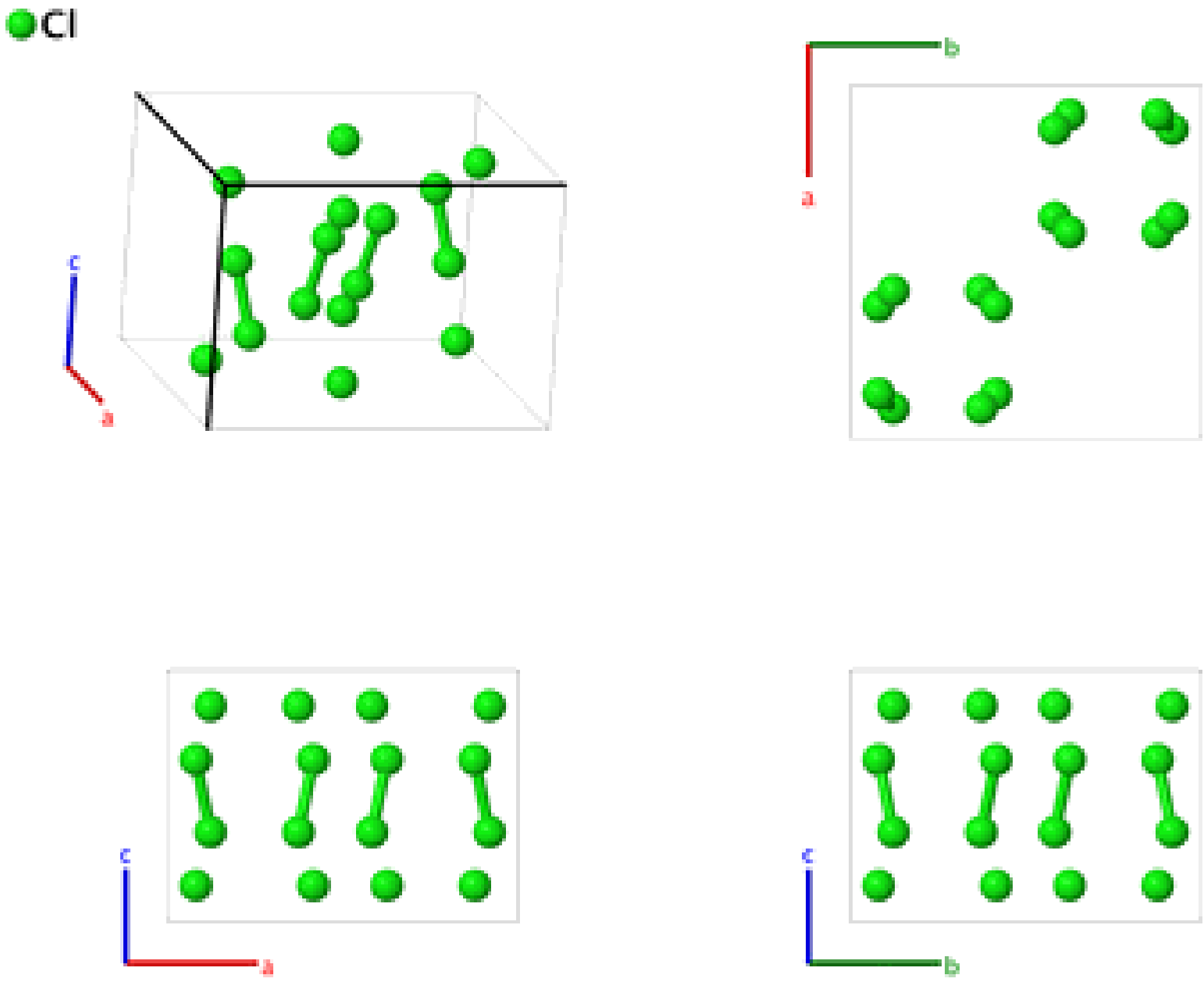

| | | |
|---|---|---|
| **Prototype** | : | Cl |
| **AFLOW prototype label** | : | A_tP16_138_j |
| ***Strukturbericht* designation** | : | A18 |
| **Pearson symbol** | : | tP16 |
| **Space group number** | : | 138 |
| **Space group symbol** | : | $P4_2/ncm$ |
| **AFLOW prototype command** | : | `aflow --proto=A_tP16_138_j`<br>`--params=`$a, c/a, x_1, y_1, z_1$ |

- As given, this structure has a Cl-Cl bond distance of 1.82Å, far too small for chlorine. The structure was eventually reanalyzed, and found to be similar to molecular iodine (A14). See (Donohue, 1982, pp. 396) for details. We retain this structure for its historical interest. Note that all atoms are on the general sites of space group $P4_2/ncm$.

**Simple Tetragonal primitive vectors:**

$$
\begin{aligned}
\mathbf{a}_1 &= a\,\hat{\mathbf{x}} \\
\mathbf{a}_2 &= a\,\hat{\mathbf{y}} \\
\mathbf{a}_3 &= c\,\hat{\mathbf{z}}
\end{aligned}
$$

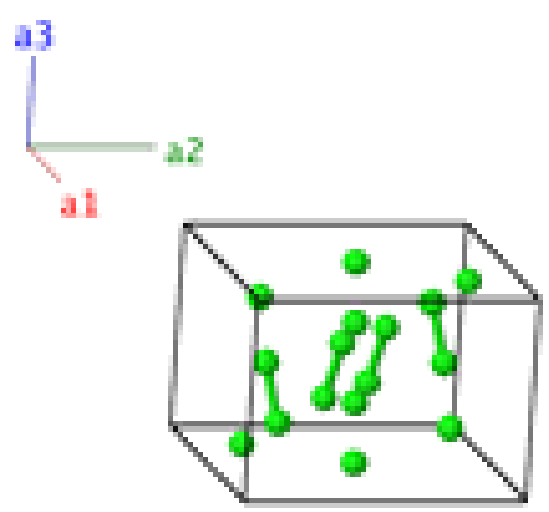

**Basis vectors:**

| Lattice Coordinates | Cartesian Coordinates | Wyckoff Position | Atom Type |
|---|---|---|---|

| | | | | | | |
|---|---|---|---|---|---|---|
| $\mathbf{B_1}$ | = | $x_1\,\mathbf{a_1} + y_1\,\mathbf{a_2} + z_1\,\mathbf{a_3}$ | = | $x_1\,a\,\mathbf{\hat{x}} + y_1\,a\,\mathbf{\hat{y}} + z_1\,c\,\mathbf{\hat{z}}$ | (16$j$) | Cl |
| $\mathbf{B_2}$ | = | $\left(\frac{1}{2} - x_1\right)\mathbf{a_1} + \left(\frac{1}{2} - y_1\right)\mathbf{a_2} + z_1\,\mathbf{a_3}$ | = | $\left(\frac{1}{2} - x_1\right)a\,\mathbf{\hat{x}} + \left(\frac{1}{2} - y_1\right)a\,\mathbf{\hat{y}} + z_1\,c\,\mathbf{\hat{z}}$ | (16$j$) | Cl |
| $\mathbf{B_3}$ | = | $\left(\frac{1}{2} - y_1\right)\mathbf{a_1} + x_1\,\mathbf{a_2} + \left(\frac{1}{2} + z_1\right)\mathbf{a_3}$ | = | $\left(\frac{1}{2} - y_1\right)a\,\mathbf{\hat{x}} + x_1\,a\,\mathbf{\hat{y}} + \left(\frac{1}{2} + z_1\right)c\,\mathbf{\hat{z}}$ | (16$j$) | Cl |
| $\mathbf{B_4}$ | = | $y_1\,\mathbf{a_1} + \left(\frac{1}{2} - x_1\right)\mathbf{a_2} + \left(\frac{1}{2} + z_1\right)\mathbf{a_3}$ | = | $y_1\,a\,\mathbf{\hat{x}} + \left(\frac{1}{2} - x_1\right)a\,\mathbf{\hat{y}} + \left(\frac{1}{2} + z_1\right)c\,\mathbf{\hat{z}}$ | (16$j$) | Cl |
| $\mathbf{B_5}$ | = | $-x_1\,\mathbf{a_1} + \left(\frac{1}{2} + y_1\right)\mathbf{a_2} + \left(\frac{1}{2} - z_1\right)\mathbf{a_3}$ | = | $-x_1\,a\,\mathbf{\hat{x}} + \left(\frac{1}{2} + y_1\right)a\,\mathbf{\hat{y}} + \left(\frac{1}{2} - z_1\right)c\,\mathbf{\hat{z}}$ | (16$j$) | Cl |
| $\mathbf{B_6}$ | = | $\left(\frac{1}{2} + x_1\right)\mathbf{a_1} - y_1\,\mathbf{a_2} + \left(\frac{1}{2} - z_1\right)\mathbf{a_3}$ | = | $\left(\frac{1}{2} + x_1\right)a\,\mathbf{\hat{x}} - y_1\,a\,\mathbf{\hat{y}} + \left(\frac{1}{2} - z_1\right)c\,\mathbf{\hat{z}}$ | (16$j$) | Cl |
| $\mathbf{B_7}$ | = | $\left(\frac{1}{2} + y_1\right)\mathbf{a_1} + \left(\frac{1}{2} + x_1\right)\mathbf{a_2} - z_1\,\mathbf{a_3}$ | = | $\left(\frac{1}{2} + y_1\right)a\,\mathbf{\hat{x}} + \left(\frac{1}{2} + x_1\right)a\,\mathbf{\hat{y}} - z_1\,c\,\mathbf{\hat{z}}$ | (16$j$) | Cl |
| $\mathbf{B_8}$ | = | $-y_1\,\mathbf{a_1} - x_1\,\mathbf{a_2} - z_1\,\mathbf{a_3}$ | = | $-y_1\,a\,\mathbf{\hat{x}} - x_1\,a\,\mathbf{\hat{y}} - z_1\,c\,\mathbf{\hat{z}}$ | (16$j$) | Cl |
| $\mathbf{B_9}$ | = | $-x_1\,\mathbf{a_1} - y_1\,\mathbf{a_2} - z_1\,\mathbf{a_3}$ | = | $-x_1\,a\,\mathbf{\hat{x}} - y_1\,a\,\mathbf{\hat{y}} - z_1\,c\,\mathbf{\hat{z}}$ | (16$j$) | Cl |
| $\mathbf{B_{10}}$ | = | $\left(\frac{1}{2} + x_1\right)\mathbf{a_1} + \left(\frac{1}{2} + y_1\right)\mathbf{a_2} - z_1\,\mathbf{a_3}$ | = | $\left(\frac{1}{2} + x_1\right)a\,\mathbf{\hat{x}} + \left(\frac{1}{2} + y_1\right)a\,\mathbf{\hat{y}} - z_1\,c\,\mathbf{\hat{z}}$ | (16$j$) | Cl |
| $\mathbf{B_{11}}$ | = | $\left(\frac{1}{2} + y_1\right)\mathbf{a_1} - x_1\,\mathbf{a_2} + \left(\frac{1}{2} - z_1\right)\mathbf{a_3}$ | = | $\left(\frac{1}{2} + y_1\right)a\,\mathbf{\hat{x}} - x_1\,a\,\mathbf{\hat{y}} + \left(\frac{1}{2} - z_1\right)c\,\mathbf{\hat{z}}$ | (16$j$) | Cl |
| $\mathbf{B_{12}}$ | = | $-y_1\,\mathbf{a_1} + \left(\frac{1}{2} + x_1\right)\mathbf{a_2} + \left(\frac{1}{2} - z_1\right)\mathbf{a_3}$ | = | $-y_1\,a\,\mathbf{\hat{x}} + \left(\frac{1}{2} + x_1\right)a\,\mathbf{\hat{y}} + \left(\frac{1}{2} - z_1\right)c\,\mathbf{\hat{z}}$ | (16$j$) | Cl |
| $\mathbf{B_{13}}$ | = | $x_1\,\mathbf{a_1} + \left(\frac{1}{2} - y_1\right)\mathbf{a_2} + \left(\frac{1}{2} + z_1\right)\mathbf{a_3}$ | = | $x_1\,a\,\mathbf{\hat{x}} + \left(\frac{1}{2} - y_1\right)a\,\mathbf{\hat{y}} + \left(\frac{1}{2} + z_1\right)c\,\mathbf{\hat{z}}$ | (16$j$) | Cl |
| $\mathbf{B_{14}}$ | = | $\left(\frac{1}{2} - x_1\right)\mathbf{a_1} + y_1\,\mathbf{a_2} + \left(\frac{1}{2} + z_1\right)\mathbf{a_3}$ | = | $\left(\frac{1}{2} - x_1\right)a\,\mathbf{\hat{x}} + y_1\,a\,\mathbf{\hat{y}} + \left(\frac{1}{2} + z_1\right)c\,\mathbf{\hat{z}}$ | (16$j$) | Cl |
| $\mathbf{B_{15}}$ | = | $\left(\frac{1}{2} - y_1\right)\mathbf{a_1} + \left(\frac{1}{2} - x_1\right)\mathbf{a_2} + z_1\,\mathbf{a_3}$ | = | $\left(\frac{1}{2} - y_1\right)a\,\mathbf{\hat{x}} + \left(\frac{1}{2} - x_1\right)a\,\mathbf{\hat{y}} + z_1\,c\,\mathbf{\hat{z}}$ | (16$j$) | Cl |
| $\mathbf{B_{16}}$ | = | $y_1\,\mathbf{a_1} + x_1\,\mathbf{a_2} + z_1\,\mathbf{a_3}$ | = | $y_1\,a\,\mathbf{\hat{x}} + x_1\,a\,\mathbf{\hat{y}} + z_1\,c\,\mathbf{\hat{z}}$ | (16$j$) | Cl |

**Geometry files:**

- CIF: pp. 698
- POSCAR: pp. 699

# $Al_3Zr$ ($D0_{23}$) Structure: A3B_tI16_139_cde_e

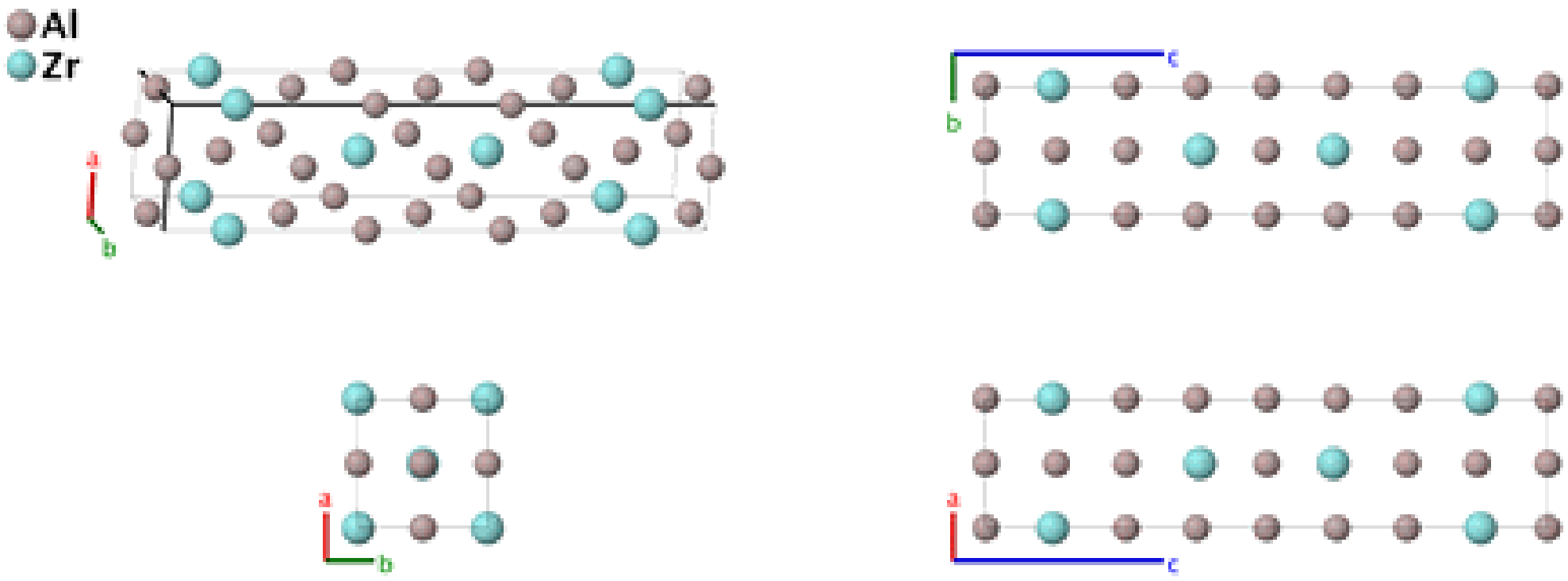

**Prototype** : $Al_3Zr$

**AFLOW prototype label** : A3B_tI16_139_cde_e

***Strukturbericht* designation** : $D0_{23}$

**Pearson symbol** : tI16

**Space group number** : 139

**Space group symbol** : I4/mmm

**AFLOW prototype command** : `aflow --proto=A3B_tI16_139_cde_e --params=`$a, c/a, z_3, z_4$

---

- When $c = 4a$, $z_3 = 3/8$, and $z_4 = 1/8$ the atoms are on the sites of a face-centered cubic lattice. This phase can also be described as a set of alternating $L1_2$ and $D0_{22}$ lattices.

---

**Body-centered Tetragonal primitive vectors:**

$$\mathbf{a}_1 = -\frac{1}{2}\,a\,\hat{\mathbf{x}} + \frac{1}{2}\,a\,\hat{\mathbf{y}} + \frac{1}{2}\,c\,\hat{\mathbf{z}}$$
$$\mathbf{a}_2 = \frac{1}{2}\,a\,\hat{\mathbf{x}} - \frac{1}{2}\,a\,\hat{\mathbf{y}} + \frac{1}{2}\,c\,\hat{\mathbf{z}}$$
$$\mathbf{a}_3 = \frac{1}{2}\,a\,\hat{\mathbf{x}} + \frac{1}{2}\,a\,\hat{\mathbf{y}} - \frac{1}{2}\,c\,\hat{\mathbf{z}}$$

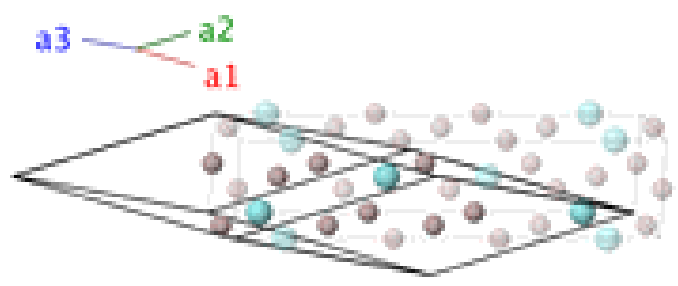

---

**Basis vectors:**

| | | Lattice Coordinates | | Cartesian Coordinates | Wyckoff Position | Atom Type |
|---|---|---|---|---|---|---|
| $\mathbf{B}_1$ | = | $\frac{1}{2}\,\mathbf{a}_1 + \frac{1}{2}\,\mathbf{a}_3$ | = | $\frac{1}{2}\,a\,\hat{\mathbf{y}}$ | (4$c$) | Al I |
| $\mathbf{B}_2$ | = | $\frac{1}{2}\,\mathbf{a}_2 + \frac{1}{2}\,\mathbf{a}_3$ | = | $\frac{1}{2}\,a\,\hat{\mathbf{x}}$ | (4$c$) | Al I |
| $\mathbf{B}_3$ | = | $\frac{3}{4}\,\mathbf{a}_1 + \frac{1}{4}\,\mathbf{a}_2 + \frac{1}{2}\,\mathbf{a}_3$ | = | $\frac{1}{2}\,a\,\hat{\mathbf{y}} + \frac{1}{4}\,c\,\hat{\mathbf{z}}$ | (4$d$) | Al II |
| $\mathbf{B}_4$ | = | $\frac{1}{4}\,\mathbf{a}_1 + \frac{3}{4}\,\mathbf{a}_2 + \frac{1}{2}\,\mathbf{a}_3$ | = | $\frac{1}{2}\,a\,\hat{\mathbf{x}} + \frac{1}{4}\,c\,\hat{\mathbf{z}}$ | (4$d$) | Al II |
| $\mathbf{B}_5$ | = | $z_3\,\mathbf{a}_1 + z_3\,\mathbf{a}_2$ | = | $+z_3\,c\,\hat{\mathbf{z}}$ | (4$e$) | Al III |
| $\mathbf{B}_6$ | = | $-z_3\,\mathbf{a}_1 - z_3\,\mathbf{a}_2$ | = | $-z_3\,c\,\hat{\mathbf{z}}$ | (4$e$) | Al III |
| $\mathbf{B}_7$ | = | $z_4\,\mathbf{a}_1 + z_4\,\mathbf{a}_2$ | = | $+z_4\,c\,\hat{\mathbf{z}}$ | (4$e$) | Zr |
| $\mathbf{B}_8$ | = | $-z_4\,\mathbf{a}_1 - z_4\,\mathbf{a}_2$ | = | $-z_4\,c\,\hat{\mathbf{z}}$ | (4$e$) | Zr |

**Found in:**

**Geometry files:**

- CIF: pp. 699
- POSCAR: pp. 699

# Hypothetical BCT5 Si Structure: A_tI4_139_e

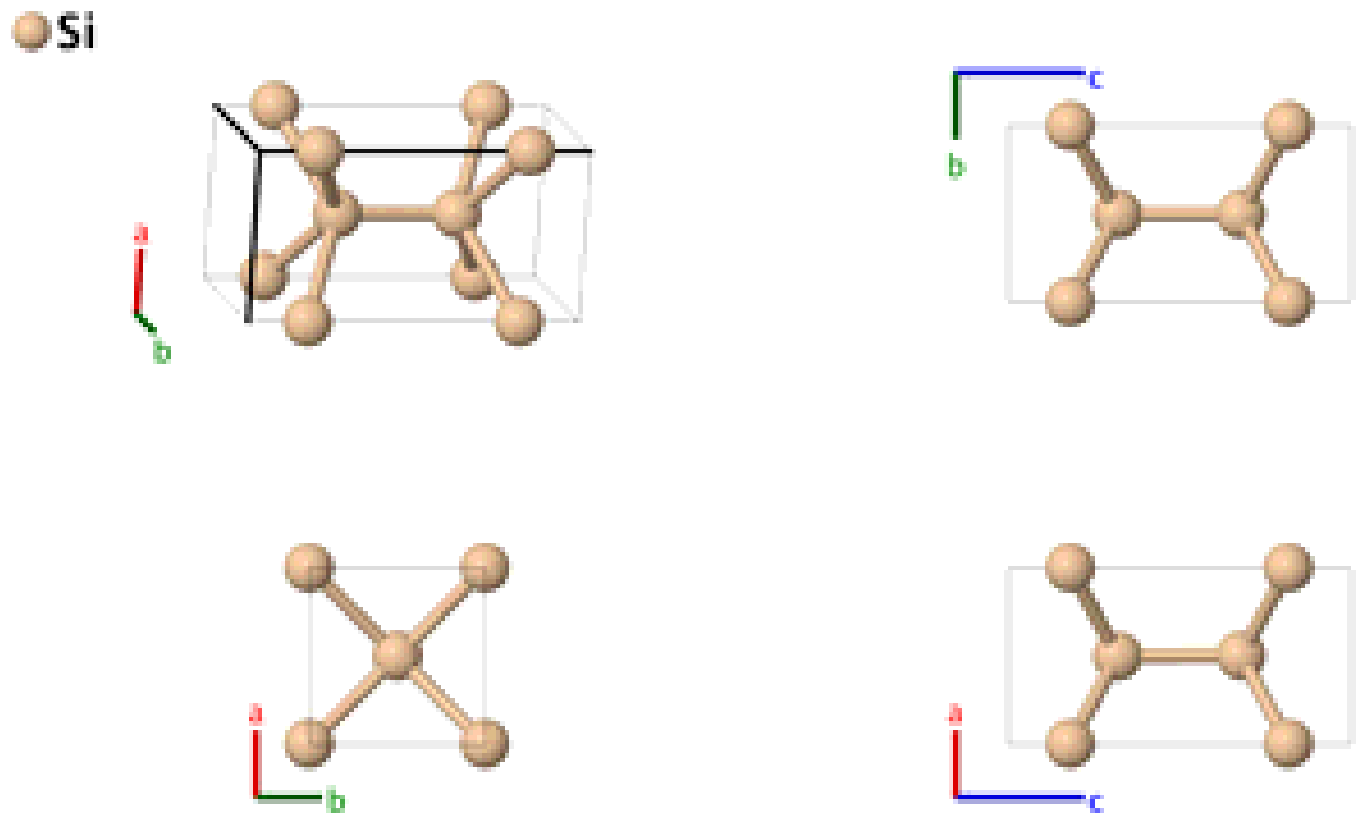

| | | |
|---|---|---|
| **Prototype** | : | Si |
| **AFLOW prototype label** | : | A_tI4_139_e |
| ***Strukturbericht* designation** | : | None |
| **Pearson symbol** | : | tI4 |
| **Space group number** | : | 139 |
| **Space group symbol** | : | I4/mmm |
| **AFLOW prototype command** | : | `aflow --proto=A_tI4_139_e` `--params=`$a, c/a, z_1$ |

- The bct5 structure is a tetragonal analog of the diamond (A4) structure, with 5-fold coordination. It was proposed in (Boyer, 1991) as a low energy, metastable phase of silicon, based on first-principles calculations and model potentials. To the best of our knowledge, this has not been observed experimentally. A search of Pearson's Handbook does not show any compound with this structure.

**Body-centered Tetragonal primitive vectors:**

$$\mathbf{a}_1 = -\frac{1}{2}\,a\,\hat{\mathbf{x}} + \frac{1}{2}\,a\,\hat{\mathbf{y}} + \frac{1}{2}\,c\,\hat{\mathbf{z}}$$
$$\mathbf{a}_2 = \frac{1}{2}\,a\,\hat{\mathbf{x}} - \frac{1}{2}\,a\,\hat{\mathbf{y}} + \frac{1}{2}\,c\,\hat{\mathbf{z}}$$
$$\mathbf{a}_3 = \frac{1}{2}\,a\,\hat{\mathbf{x}} + \frac{1}{2}\,a\,\hat{\mathbf{y}} - \frac{1}{2}\,c\,\hat{\mathbf{z}}$$

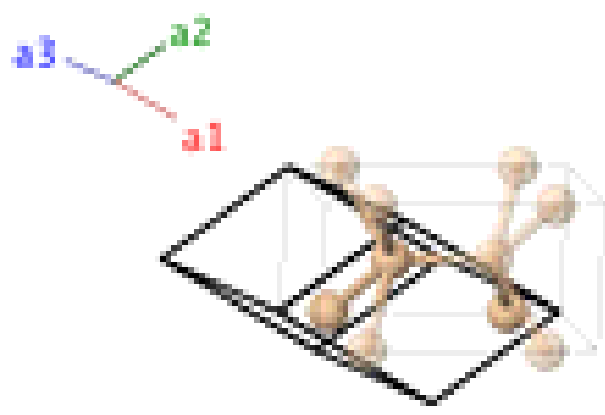

**Basis vectors:**

| | | Lattice Coordinates | | Cartesian Coordinates | Wyckoff Position | Atom Type |
|---|---|---|---|---|---|---|
| $\mathbf{B}_1$ | = | $z_1\,\mathbf{a}_1 + z_1\,\mathbf{a}_2$ | = | $+z_1\,c\,\hat{\mathbf{z}}$ | (4$e$) | Si |
| $\mathbf{B}_2$ | = | $-z_1\,\mathbf{a}_1 - z_1\,\mathbf{a}_2$ | = | $-z_1\,c\,\hat{\mathbf{z}}$ | (4$e$) | Si |

**Geometry files:**
- CIF: pp. 699
- POSCAR: pp. 700

# 0201 [$(La,Ba)_2CuO_4$] High-$T_c$ Structure: AB2C4_tI14_139_a_e_ce

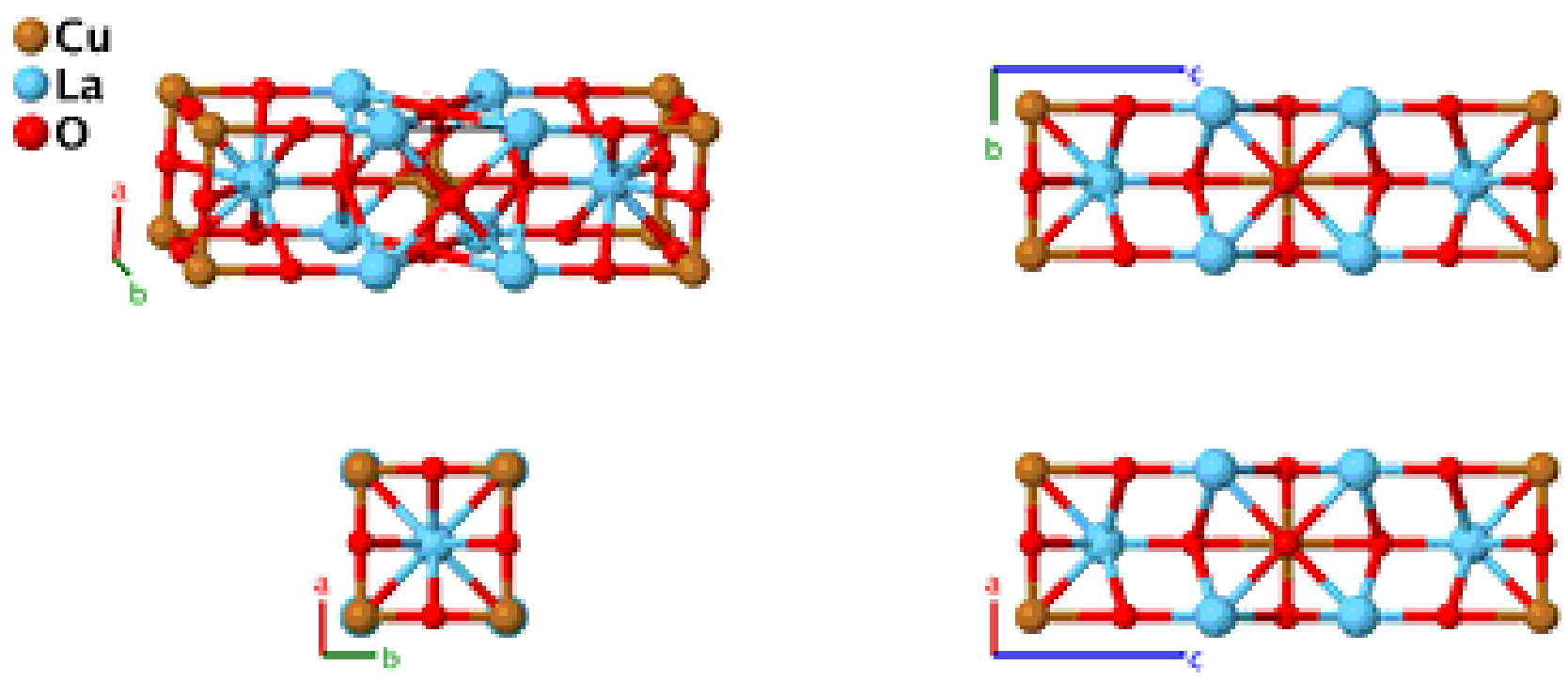

| | | |
|---|---|---|
| **Prototype** | : | $(La,Ba)_2CuO_4$ |
| **AFLOW prototype label** | : | AB2C4_tI14_139_a_e_ce |
| ***Strukturbericht* designation** | : | None |
| **Pearson symbol** | : | tI14 |
| **Space group number** | : | 139 |
| **Space group symbol** | : | I4/mmm |
| **AFLOW prototype command** | : | `aflow --proto=AB2C4_tI14_139_a_e_ce --params=`$a, c/a, z_3, z_4$ |

- The original "high"-temperature (30K) superconductor found by Bednorz and Mueller. Barium and lanthanum atoms are distributed randomly on the lanthanum sublattice. The ground state structure of the parent compound, $La_2CuO_4$, is an orthorhombic distortion of this unit cell.

**Body-centered Tetragonal primitive vectors:**

$$\mathbf{a}_1 = -\frac{1}{2}\,a\,\hat{\mathbf{x}} + \frac{1}{2}\,a\,\hat{\mathbf{y}} + \frac{1}{2}\,c\,\hat{\mathbf{z}}$$
$$\mathbf{a}_2 = \frac{1}{2}\,a\,\hat{\mathbf{x}} - \frac{1}{2}\,a\,\hat{\mathbf{y}} + \frac{1}{2}\,c\,\hat{\mathbf{z}}$$
$$\mathbf{a}_3 = \frac{1}{2}\,a\,\hat{\mathbf{x}} + \frac{1}{2}\,a\,\hat{\mathbf{y}} - \frac{1}{2}\,c\,\hat{\mathbf{z}}$$

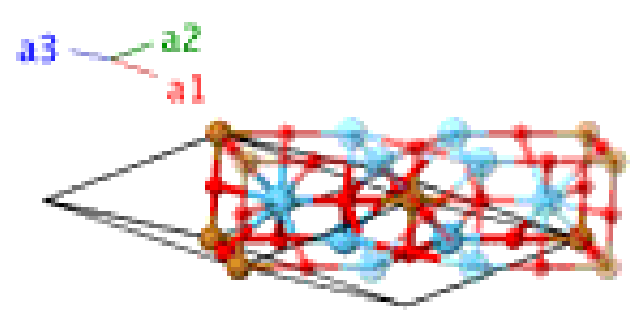

**Basis vectors:**

| | | Lattice Coordinates | | Cartesian Coordinates | Wyckoff Position | Atom Type |
|---|---|---|---|---|---|---|
| $\mathbf{B}_1$ | = | $0\,\mathbf{a}_1 + 0\,\mathbf{a}_2 + 0\,\mathbf{a}_3$ | = | $0\hat{\mathbf{x}} + 0\hat{\mathbf{y}} + 0\hat{\mathbf{z}}$ | (2$a$) | Cu |
| $\mathbf{B}_2$ | = | $\frac{1}{2}\,\mathbf{a}_1 + \frac{1}{2}\,\mathbf{a}_3$ | = | $\frac{1}{2}\,a\,\hat{\mathbf{y}}$ | (4$c$) | O I |
| $\mathbf{B}_3$ | = | $\frac{1}{2}\,\mathbf{a}_2 + \frac{1}{2}\,\mathbf{a}_3$ | = | $\frac{1}{2}\,a\,\hat{\mathbf{x}}$ | (4$c$) | O I |
| $\mathbf{B}_4$ | = | $z_3\,\mathbf{a}_1 + z_3\,\mathbf{a}_2$ | = | $z_3\,c\,\hat{\mathbf{z}}$ | (4$e$) | La/Ba |
| $\mathbf{B}_5$ | = | $-z_3\,\mathbf{a}_1 - z_3\,\mathbf{a}_2$ | = | $-z_3\,c\,\hat{\mathbf{z}}$ | (4$e$) | La/Ba |
| $\mathbf{B}_6$ | = | $z_4\,\mathbf{a}_1 + z_4\,\mathbf{a}_2$ | = | $z_4\,c\,\hat{\mathbf{z}}$ | (4$e$) | O II |
| $\mathbf{B}_7$ | = | $-z_4\,\mathbf{a}_1 - z_4\,\mathbf{a}_2$ | = | $-z_4\,c\,\hat{\mathbf{z}}$ | (4$e$) | O II |

**Found in:**

**Geometry files:**

- CIF: pp. 700
- POSCAR: pp. 700

# $Mn_{12}Th$ ($D2_b$) Structure: A12B_tI26_139_fij_a

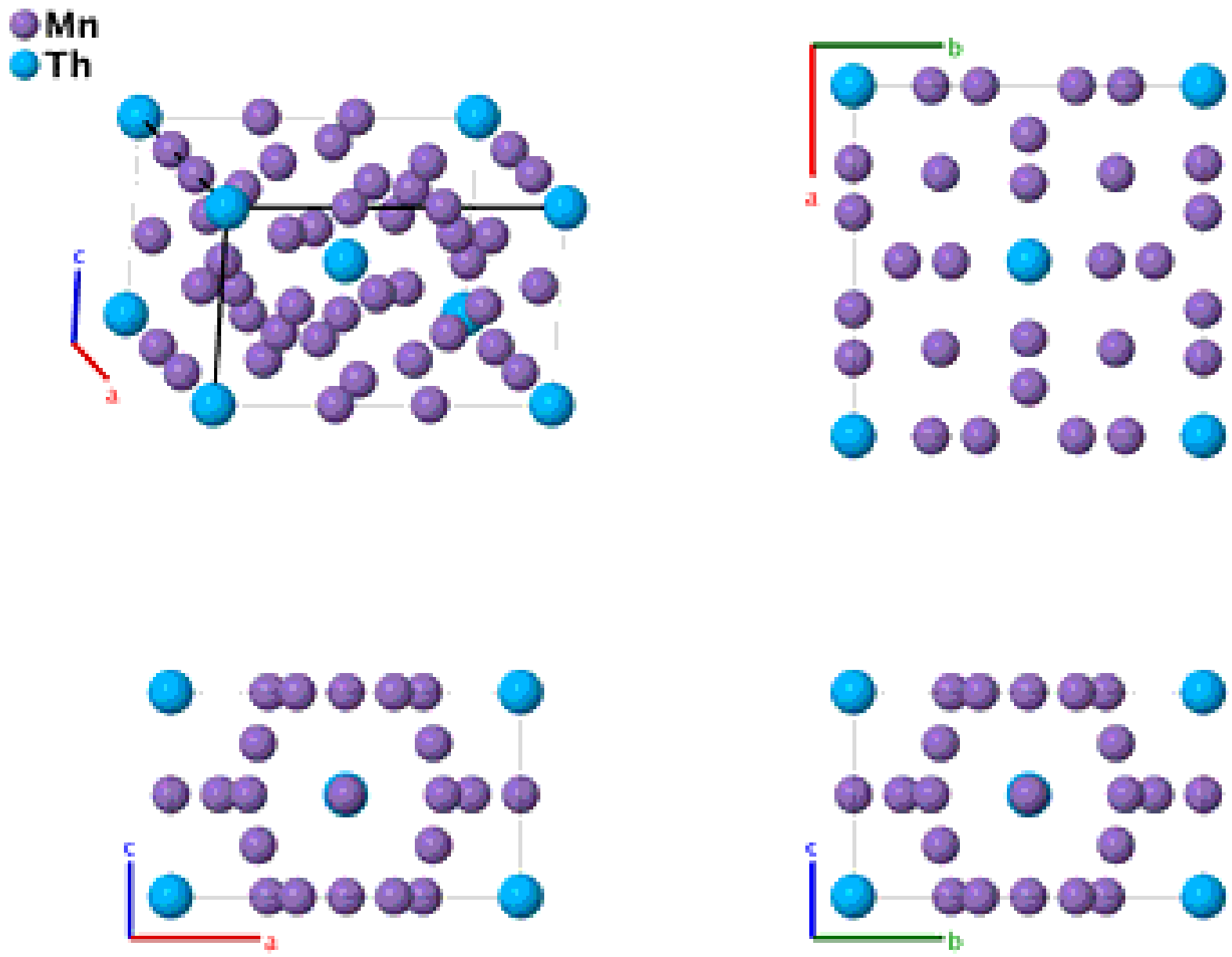

| | | |
|---|---|---|
| **Prototype** | : | $Mn_{12}Th$ |
| **AFLOW prototype label** | : | A12B_tI26_139_fij_a |
| ***Strukturbericht* designation** | : | $D2_b$ |
| **Pearson symbol** | : | tI26 |
| **Space group number** | : | 139 |
| **Space group symbol** | : | I4/mmm |
| **AFLOW prototype command** | : | `aflow --proto=A12B_tI26_139_fij_a --params=`$a, c/a, x_3, x_4$ |

**Other compounds with this structure:**

- $AgBe_{12}$, $Al_8Cr_4Er$, $Fe_4Mn_8$, $Fe_7Mn_5$, others.

**Body-centered Tetragonal primitive vectors:**

$$
\begin{aligned}
\mathbf{a}_1 &= -\tfrac{1}{2}\,a\,\hat{\mathbf{x}} + \tfrac{1}{2}\,a\,\hat{\mathbf{y}} + \tfrac{1}{2}\,c\,\hat{\mathbf{z}} \\
\mathbf{a}_2 &= \tfrac{1}{2}\,a\,\hat{\mathbf{x}} - \tfrac{1}{2}\,a\,\hat{\mathbf{y}} + \tfrac{1}{2}\,c\,\hat{\mathbf{z}} \\
\mathbf{a}_3 &= \tfrac{1}{2}\,a\,\hat{\mathbf{x}} + \tfrac{1}{2}\,a\,\hat{\mathbf{y}} - \tfrac{1}{2}\,c\,\hat{\mathbf{z}}
\end{aligned}
$$

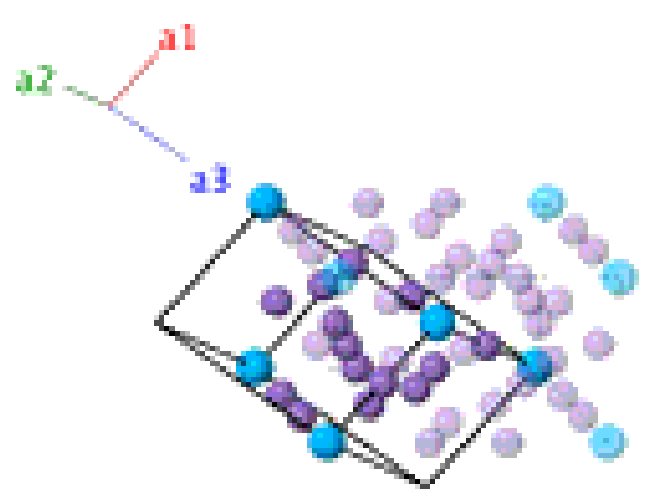

**Basis vectors:**

| | | Lattice Coordinates | | Cartesian Coordinates | Wyckoff Position | Atom Type |
|---|---|---|---|---|---|---|
| $\mathbf{B}_1$ | = | $0\,\mathbf{a}_1 + 0\,\mathbf{a}_2 + 0\,\mathbf{a}_3$ | = | $0\hat{\mathbf{x}} + 0\hat{\mathbf{y}} + 0\hat{\mathbf{z}}$ | (2a) | Th |
| $\mathbf{B}_2$ | = | $\frac{1}{2}\,\mathbf{a}_1 + \frac{1}{2}\,\mathbf{a}_2 + \frac{1}{2}\,\mathbf{a}_3$ | = | $\frac{1}{4}\,a\,\hat{\mathbf{x}} + \frac{1}{4}\,a\,\hat{\mathbf{y}} + \frac{1}{4}\,c\,\hat{\mathbf{z}}$ | $(8f)$ | Mn I |
| $\mathbf{B}_3$ | = | $\frac{1}{2}\,\mathbf{a}_3$ | = | $\frac{1}{4}\,a\,\hat{\mathbf{x}} + \frac{1}{4}\,a\,\hat{\mathbf{y}} + \frac{3}{4}\,c\,\hat{\mathbf{z}}$ | $(8f)$ | Mn I |
| $\mathbf{B}_4$ | = | $\frac{1}{2}\,\mathbf{a}_1$ | = | $\frac{3}{4}\,a\,\hat{\mathbf{x}} + \frac{1}{4}\,a\,\hat{\mathbf{y}} + \frac{1}{4}\,c\,\hat{\mathbf{z}}$ | $(8f)$ | Mn I |
| $\mathbf{B}_5$ | = | $\frac{1}{2}\,\mathbf{a}_2$ | = | $\frac{1}{4}\,a\,\hat{\mathbf{x}} + \frac{3}{4}\,a\,\hat{\mathbf{y}} + \frac{1}{4}\,c\,\hat{\mathbf{z}}$ | $(8f)$ | Mn I |
| $\mathbf{B}_6$ | = | $x_3\,\mathbf{a}_2 + x_3\,\mathbf{a}_3$ | = | $x_3\,a\,\hat{\mathbf{x}}$ | $(8i)$ | Mn II |
| $\mathbf{B}_7$ | = | $-x_3\,\mathbf{a}_2 - x_3\,\mathbf{a}_3$ | = | $-x_3\,a\,\hat{\mathbf{x}}$ | $(8i)$ | Mn II |
| $\mathbf{B}_8$ | = | $x_3\,\mathbf{a}_1 + x_3\,\mathbf{a}_3$ | = | $x_3\,a\,\hat{\mathbf{y}}$ | $(8i)$ | Mn II |
| $\mathbf{B}_9$ | = | $-x_3\,\mathbf{a}_1 - x_3\,\mathbf{a}_3$ | = | $-x_3\,a\,\hat{\mathbf{y}}$ | $(8i)$ | Mn II |
| $\mathbf{B}_{10}$ | = | $\frac{1}{2}\,\mathbf{a}_1 + x_4\,\mathbf{a}_2 + \left(\frac{1}{2} + x_4\right)\,\mathbf{a}_3$ | = | $x_4\,a\,\hat{\mathbf{x}} + \frac{1}{2}\,a\,\hat{\mathbf{y}}$ | $(8j)$ | Mn III |
| $\mathbf{B}_{11}$ | = | $\frac{1}{2}\,\mathbf{a}_1 - x_4\,\mathbf{a}_2 + \left(\frac{1}{2} - x_4\right)\,\mathbf{a}_3$ | = | $-x_4\,a\,\hat{\mathbf{x}} + \frac{1}{2}\,a\,\hat{\mathbf{y}}$ | $(8j)$ | Mn III |
| $\mathbf{B}_{12}$ | = | $x_4\,\mathbf{a}_1 + \frac{1}{2}\,\mathbf{a}_2 + \left(\frac{1}{2} + x_4\right)\,\mathbf{a}_3$ | = | $\frac{1}{2}\,a\,\hat{\mathbf{x}} + x_4\,a\,\hat{\mathbf{y}}$ | $(8j)$ | Mn III |
| $\mathbf{B}_{13}$ | = | $-x_4\,\mathbf{a}_1 + \frac{1}{2}\,\mathbf{a}_2 + \left(\frac{1}{2} - x_4\right)\,\mathbf{a}_3$ | = | $\frac{1}{2}\,a\,\hat{\mathbf{x}} - x_4\,a\,\hat{\mathbf{y}}$ | $(8j)$ | Mn III |

---

---

**Geometry files:**

- CIF: pp. 700
- POSCAR: pp. 701

# In (A6) Structure: A_tI2_139_a

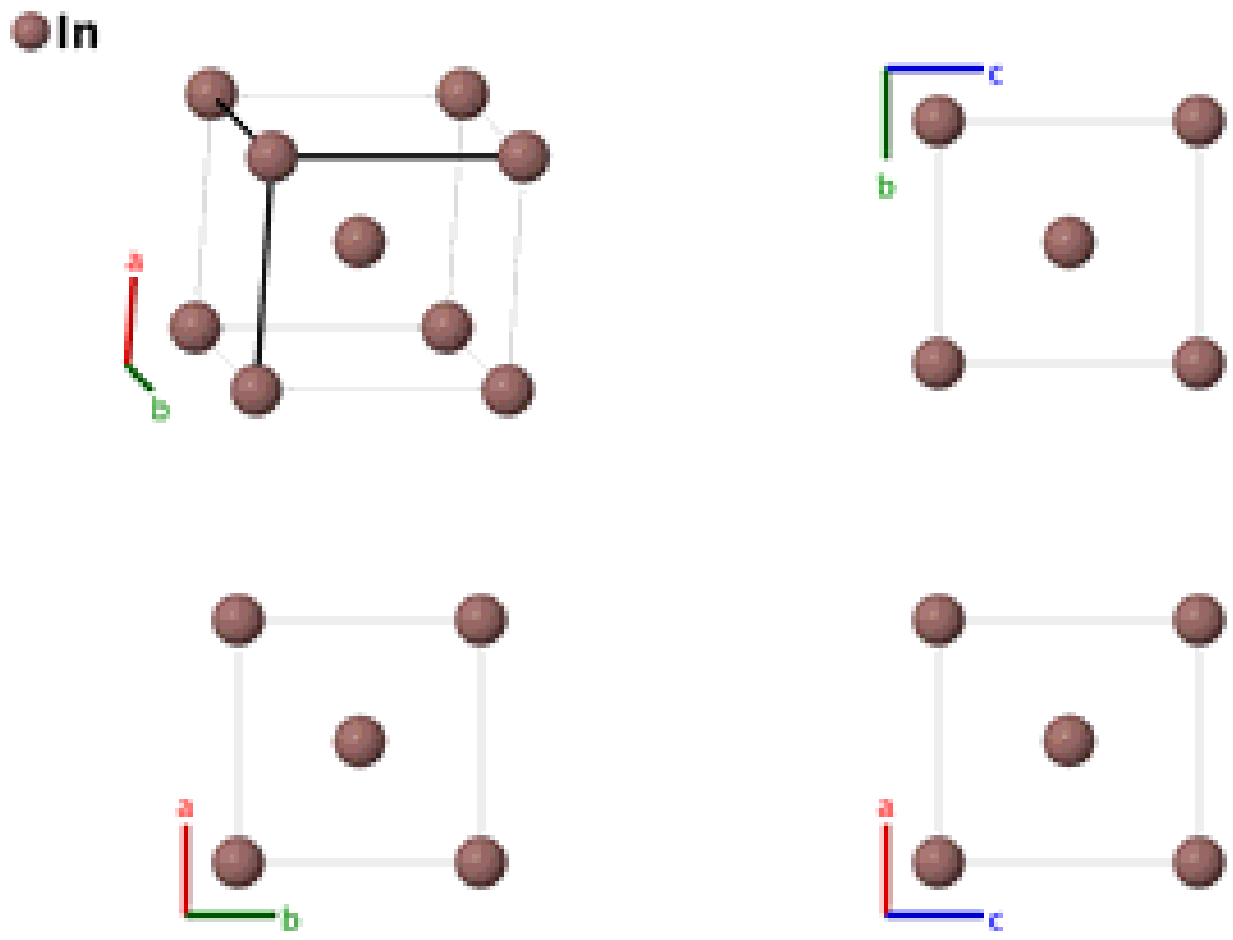

| | | |
|---|---|---|
| **Prototype** | : | In |
| **AFLOW prototype label** | : | A_tI2_139_a |
| ***Strukturbericht* designation** | : | A6 |
| **Pearson symbol** | : | tI2 |
| **Space group number** | : | 139 |
| **Space group symbol** | : | I4/mmm |
| **AFLOW prototype command** | : | `aflow --proto=A_tI2_139_a` <br> `--params=`$a, c/a$ |

- This is an example of a "face"-centered tetragonal (fct) lattice, a distortion of the fcc lattice. Note that this structure is actually a body-centered tetragonal lattice, since in the tetragonal system there is no distinction between face- and body-centered structures. In the A6 structure, $c/a$ is near the fcc ratio of $\sqrt{2}$, while in the $A_a$ structure, $c/a$ is near the bcc ratio of 1. Note that In (pp. 289) and $\alpha$-Pa (pp. 305) have the same AFLOW prototype label. They are generated by the same symmetry operations with different sets of parameters (`--params`) specified in their corresponding CIF files.

**Body-centered Tetragonal primitive vectors:**

$$\mathbf{a}_1 = -\frac{1}{2}\,a\,\hat{\mathbf{x}} + \frac{1}{2}\,a\,\hat{\mathbf{y}} + \frac{1}{2}\,c\,\hat{\mathbf{z}}$$
$$\mathbf{a}_2 = \frac{1}{2}\,a\,\hat{\mathbf{x}} - \frac{1}{2}\,a\,\hat{\mathbf{y}} + \frac{1}{2}\,c\,\hat{\mathbf{z}}$$
$$\mathbf{a}_3 = \frac{1}{2}\,a\,\hat{\mathbf{x}} + \frac{1}{2}\,a\,\hat{\mathbf{y}} - \frac{1}{2}\,c\,\hat{\mathbf{z}}$$

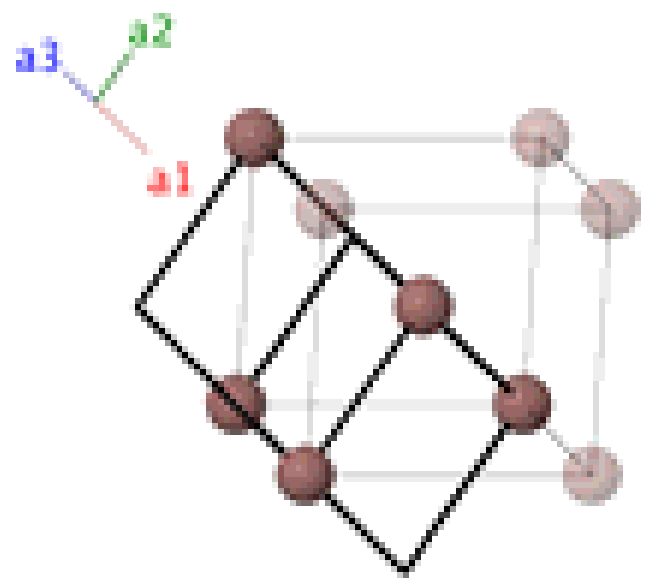

**Basis vectors:**

| | | Lattice Coordinates | | Cartesian Coordinates | Wyckoff Position | Atom Type |
|---|---|---|---|---|---|---|
| $\mathbf{B}_1$ | = | $0\,\mathbf{a}_1 + 0\,\mathbf{a}_2 + 0\,\mathbf{a}_3$ | = | $0\hat{\mathbf{x}} + 0\hat{\mathbf{y}} + 0\hat{\mathbf{z}}$ | $(2a)$ | In |

**Found in:**

---

**Geometry files:**

- CIF: pp. 701
- POSCAR: pp. 702

# Hypothetical Tetrahedrally Bonded Carbon with 4-Member Rings: A_tI8_139_h

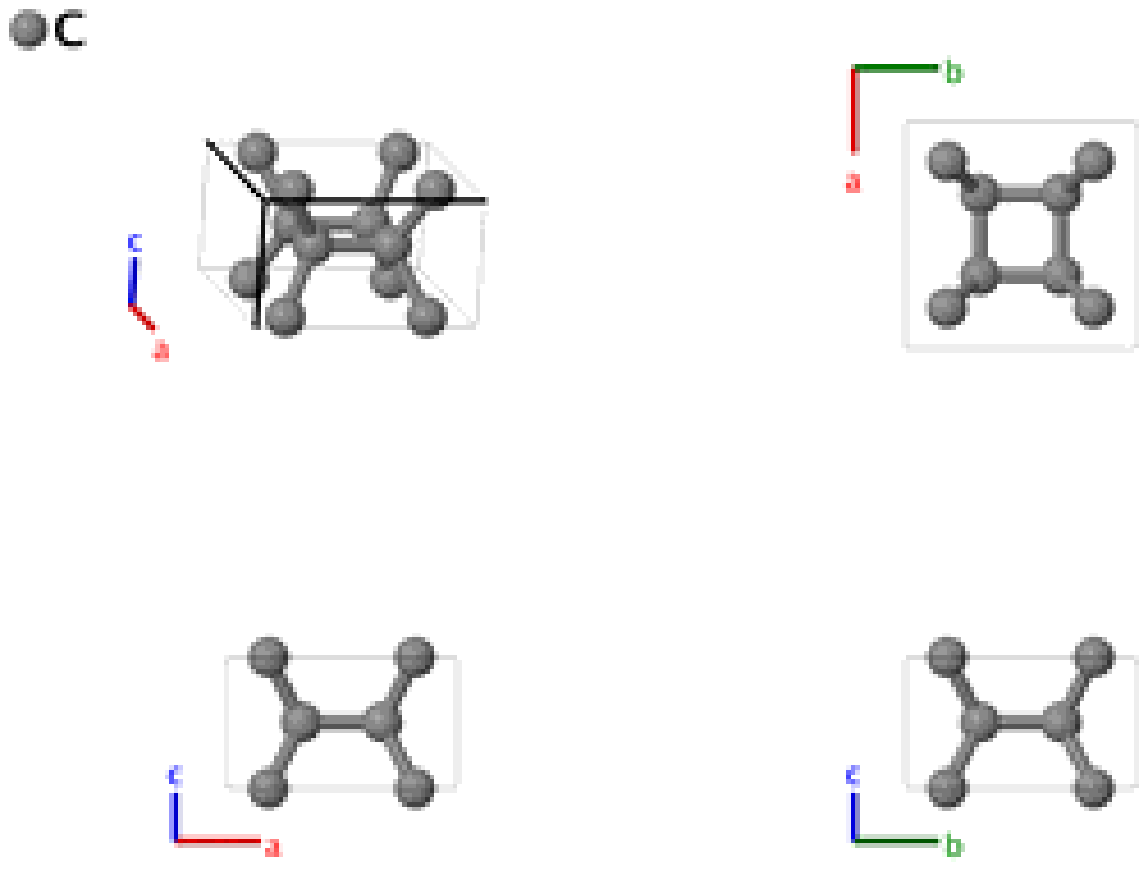

**Prototype** : C

**AFLOW prototype label** : A_tI8_139_h

***Strukturbericht* designation** : None

**Pearson symbol** : tI8

**Space group number** : 139

**Space group symbol** : I4/mmm

**AFLOW prototype command** : `aflow --proto=A_tI8_139_h`
`--params=`$a, c/a, x_1$

---

- This structure was proposed in (Schultz, 1999) to show that it was energetically possible to form four-member rings in amorphous $sp^3$ carbon structures.

---

**Body-centered Tetragonal primitive vectors:**

$$
\begin{aligned}
\mathbf{a}_1 &= -\tfrac{1}{2}\,a\,\hat{\mathbf{x}} + \tfrac{1}{2}\,a\,\hat{\mathbf{y}} + \tfrac{1}{2}\,c\,\hat{\mathbf{z}} \\
\mathbf{a}_2 &= \tfrac{1}{2}\,a\,\hat{\mathbf{x}} - \tfrac{1}{2}\,a\,\hat{\mathbf{y}} + \tfrac{1}{2}\,c\,\hat{\mathbf{z}} \\
\mathbf{a}_3 &= \tfrac{1}{2}\,a\,\hat{\mathbf{x}} + \tfrac{1}{2}\,a\,\hat{\mathbf{y}} - \tfrac{1}{2}\,c\,\hat{\mathbf{z}}
\end{aligned}
$$

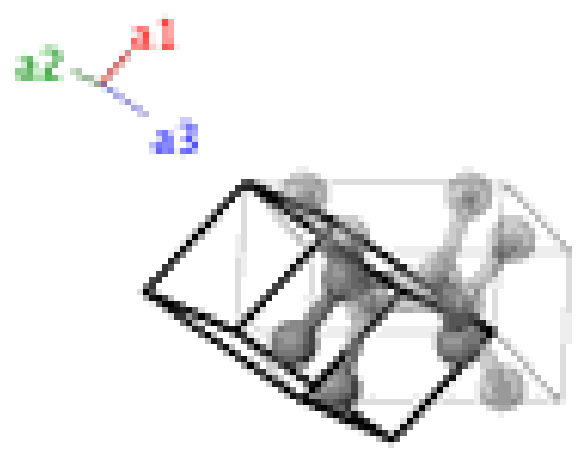

---

**Basis vectors:**

| | | Lattice Coordinates | | Cartesian Coordinates | Wyckoff Position | Atom Type |
|---|---|---|---|---|---|---|
| $\mathbf{B}_1$ | = | $x_1\,\mathbf{a}_1 + x_1\,\mathbf{a}_2 + 2x_1\,\mathbf{a}_3$ | = | $x_1\,a\,\hat{\mathbf{x}} + x_1\,a\,\hat{\mathbf{y}}$ | (8$h$) | C |
| $\mathbf{B}_2$ | = | $-x_1\,\mathbf{a}_1 - x_1\,\mathbf{a}_2 - 2x_1\,\mathbf{a}_3$ | = | $-x_1\,a\,\hat{\mathbf{x}} - x_1\,a\,\hat{\mathbf{y}}$ | (8$h$) | C |
| $\mathbf{B}_3$ | = | $x_1\,\mathbf{a}_1 - x_1\,\mathbf{a}_2$ | = | $-x_1\,a\,\hat{\mathbf{x}} + x_1\,a\,\hat{\mathbf{y}}$ | (8$h$) | C |
| $\mathbf{B}_4$ | = | $-x_1\,\mathbf{a}_1 + x_1\,\mathbf{a}_2$ | = | $x_1\,a\,\hat{\mathbf{x}} - x_1\,a\,\hat{\mathbf{y}}$ | (8$h$) | C |

**Geometry files:**

- CIF: pp. 702
- POSCAR: pp. 702

# $Al_3Ti$ ($D0_{22}$) Structure: A3B_tI8_139_bd_a

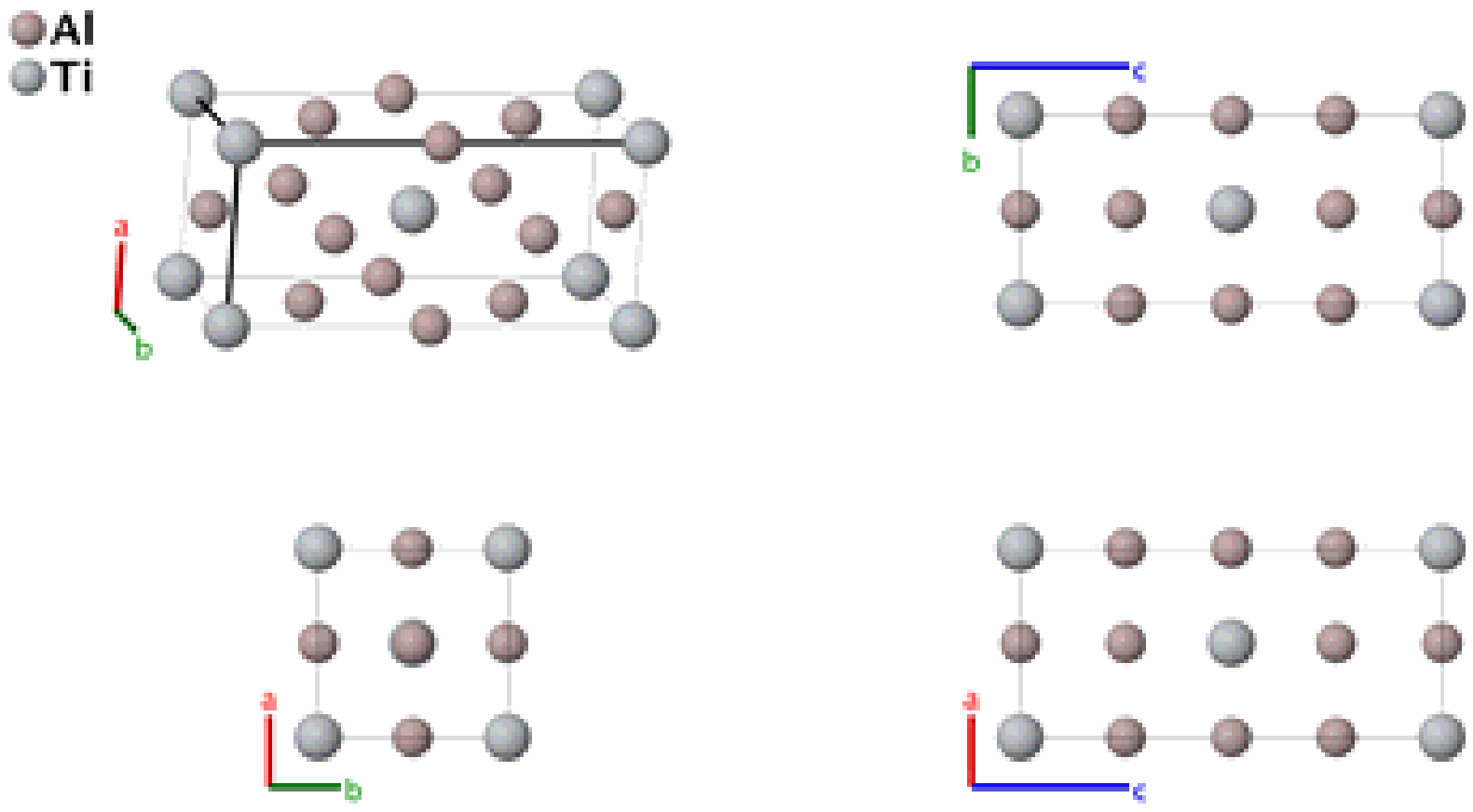

| | | |
|---|---|---|
| **Prototype** | : | $Al_3Ti$ |
| **AFLOW prototype label** | : | A3B_tI8_139_bd_a |
| ***Strukturbericht* designation** | : | $D0_{22}$ |
| **Pearson symbol** | : | tI8 |
| **Space group number** | : | 139 |
| **Space group symbol** | : | I4/mmm |
| **AFLOW prototype command** | : | `aflow --proto=A3B_tI8_139_bd_a`<br>`--params=`$a, c/a$ |

- When $c = 2a$ the atoms are on the sites of a face-centered cubic lattice. When $c/a = 1/\sqrt{2}$, this becomes the cubic $D0_3$ structure.

**Body-centered Tetragonal primitive vectors:**

$$\mathbf{a}_1 = -\frac{1}{2}\,a\,\hat{\mathbf{x}} + \frac{1}{2}\,a\,\hat{\mathbf{y}} + \frac{1}{2}\,c\,\hat{\mathbf{z}}$$
$$\mathbf{a}_2 = \frac{1}{2}\,a\,\hat{\mathbf{x}} - \frac{1}{2}\,a\,\hat{\mathbf{y}} + \frac{1}{2}\,c\,\hat{\mathbf{z}}$$
$$\mathbf{a}_3 = \frac{1}{2}\,a\,\hat{\mathbf{x}} + \frac{1}{2}\,a\,\hat{\mathbf{y}} - \frac{1}{2}\,c\,\hat{\mathbf{z}}$$

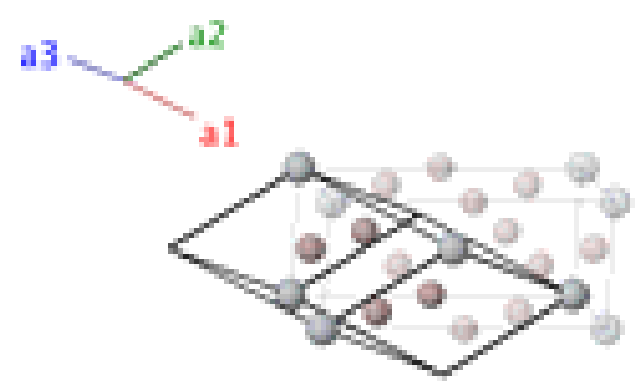

**Basis vectors:**

| | | Lattice Coordinates | | Cartesian Coordinates | Wyckoff Position | Atom Type |
|---|---|---|---|---|---|---|
| $\mathbf{B}_1$ | = | $0\,\mathbf{a}_1 + 0\,\mathbf{a}_2 + 0\,\mathbf{a}_3$ | = | $0\hat{\mathbf{x}} + 0\hat{\mathbf{y}} + 0\hat{\mathbf{z}}$ | $(2a)$ | Ti |
| $\mathbf{B}_2$ | = | $\frac{1}{2}\,\mathbf{a}_1 + \frac{1}{2}\,\mathbf{a}_2$ | = | $\frac{1}{2}\,c\,\hat{\mathbf{z}}$ | $(2b)$ | Al I |
| $\mathbf{B}_3$ | = | $\frac{3}{4}\,\mathbf{a}_1 + \frac{1}{4}\,\mathbf{a}_2 + \frac{1}{2}\,\mathbf{a}_3$ | = | $\frac{1}{2}\,a\,\hat{\mathbf{y}} + \frac{1}{4}\,c\,\hat{\mathbf{z}}$ | $(4d)$ | Al II |
| $\mathbf{B}_4$ | = | $\frac{1}{4}\,\mathbf{a}_1 + \frac{3}{4}\,\mathbf{a}_2 + \frac{1}{2}\,\mathbf{a}_3$ | = | $\frac{1}{2}\,a\,\hat{\mathbf{x}} + \frac{1}{4}\,c\,\hat{\mathbf{z}}$ | $(4d)$ | Al II |

**Found in:**

---

**Geometry files:**

- CIF: pp. 702
- POSCAR: pp. 703

# $MoSi_2$ ($C11_b$) Structure: AB2_tI6_139_a_e

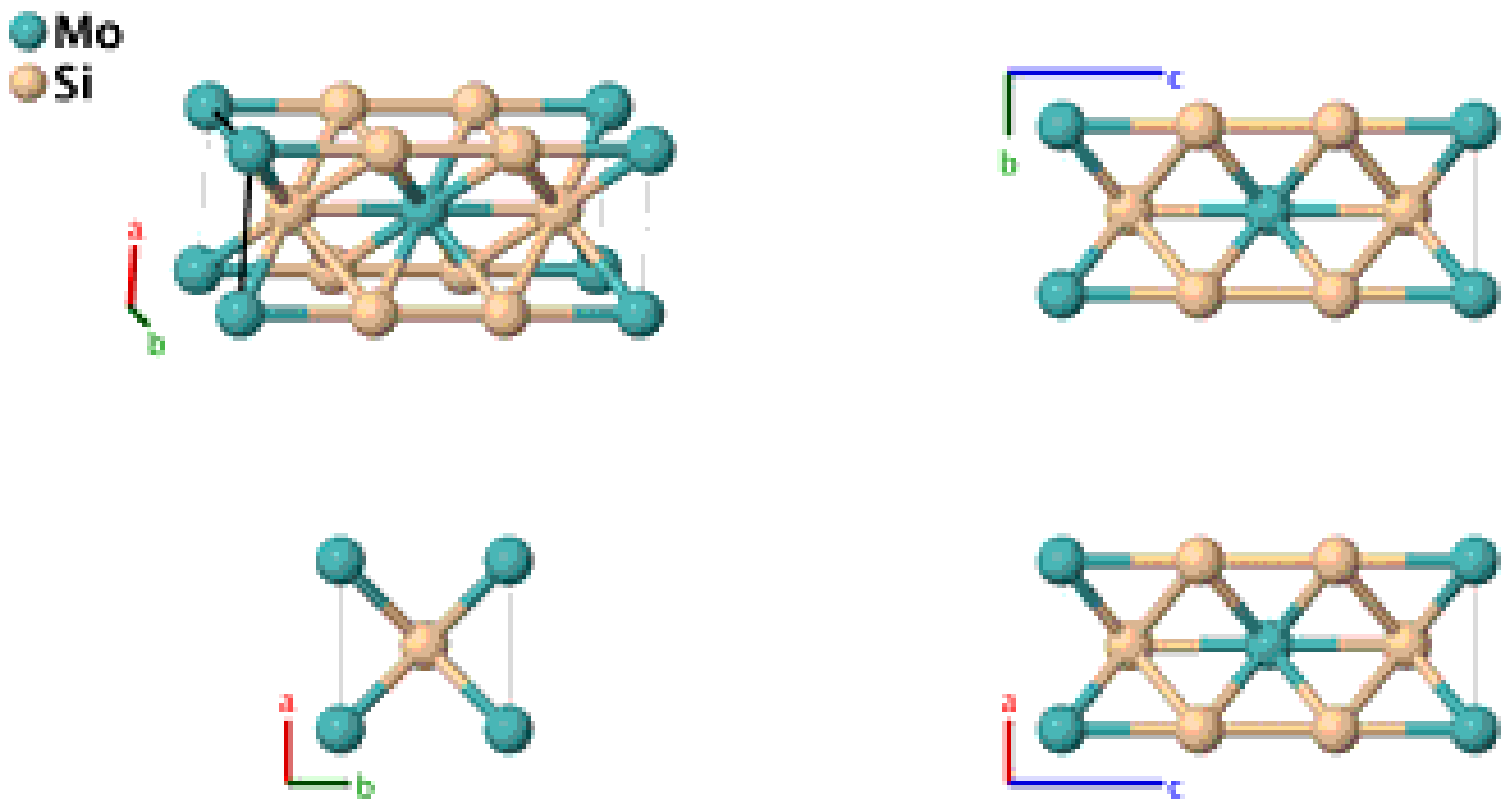

| | | |
|---|---|---|
| **Prototype** | : | $MoSi_2$ |
| **AFLOW prototype label** | : | AB2_tI6_139_a_e |
| ***Strukturbericht* designation** | : | $C11_b$ |
| **Pearson symbol** | : | tI6 |
| **Space group number** | : | 139 |
| **Space group symbol** | : | I4/mmm |
| **AFLOW prototype command** | : | `aflow --proto=AB2_tI6_139_a_e --params=`$a, c/a, z_2$ |

**Other compounds with this structure:**

- $CaC_2$, $CdTi_2$

- When $c = 3a$ and $z_2 = 1/3$ the atoms are on the sites of a body-centered cubic lattice. For $MoSi_2$ itself, (Harada, 1998) gives $c/a = 2.45$ and $z_2 = 0.3353$. Other compounds in this structure have very different values of $c/a$ and even $z_2$.

**Body-centered Tetragonal primitive vectors:**

$$\mathbf{a}_1 = -\frac{1}{2}\,a\,\hat{\mathbf{x}} + \frac{1}{2}\,a\,\hat{\mathbf{y}} + \frac{1}{2}\,c\,\hat{\mathbf{z}}$$
$$\mathbf{a}_2 = \frac{1}{2}\,a\,\hat{\mathbf{x}} - \frac{1}{2}\,a\,\hat{\mathbf{y}} + \frac{1}{2}\,c\,\hat{\mathbf{z}}$$
$$\mathbf{a}_3 = \frac{1}{2}\,a\,\hat{\mathbf{x}} + \frac{1}{2}\,a\,\hat{\mathbf{y}} - \frac{1}{2}\,c\,\hat{\mathbf{z}}$$

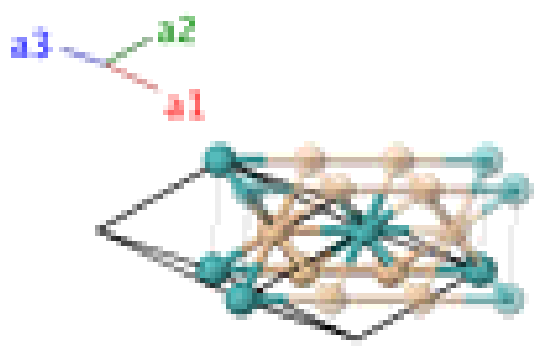

**Basis vectors:**

| | | Lattice Coordinates | | Cartesian Coordinates | Wyckoff Position | Atom Type |
|---|---|---|---|---|---|---|
| $\mathbf{B}_1$ | = | $0\,\mathbf{a}_1 + 0\,\mathbf{a}_2 + 0\,\mathbf{a}_3$ | = | $0\hat{\mathbf{x}} + 0\hat{\mathbf{y}} + 0\hat{\mathbf{z}}$ | (2a) | Mo |
| $\mathbf{B}_2$ | = | $z_2\,\mathbf{a}_1 + z_2\,\mathbf{a}_2$ | = | $z_2\,c\,\hat{\mathbf{z}}$ | (4e) | Si |
| $\mathbf{B}_3$ | = | $-z_2\,\mathbf{a}_1 - z_2\,\mathbf{a}_2$ | = | $-z_2\,c\,\hat{\mathbf{z}}$ | (4e) | Si |

---

**Geometry files:**

- CIF: pp. 703
- POSCAR: pp. 703

# $V_4Zn_5$ Structure: A4B5_tI18_139_i_ah

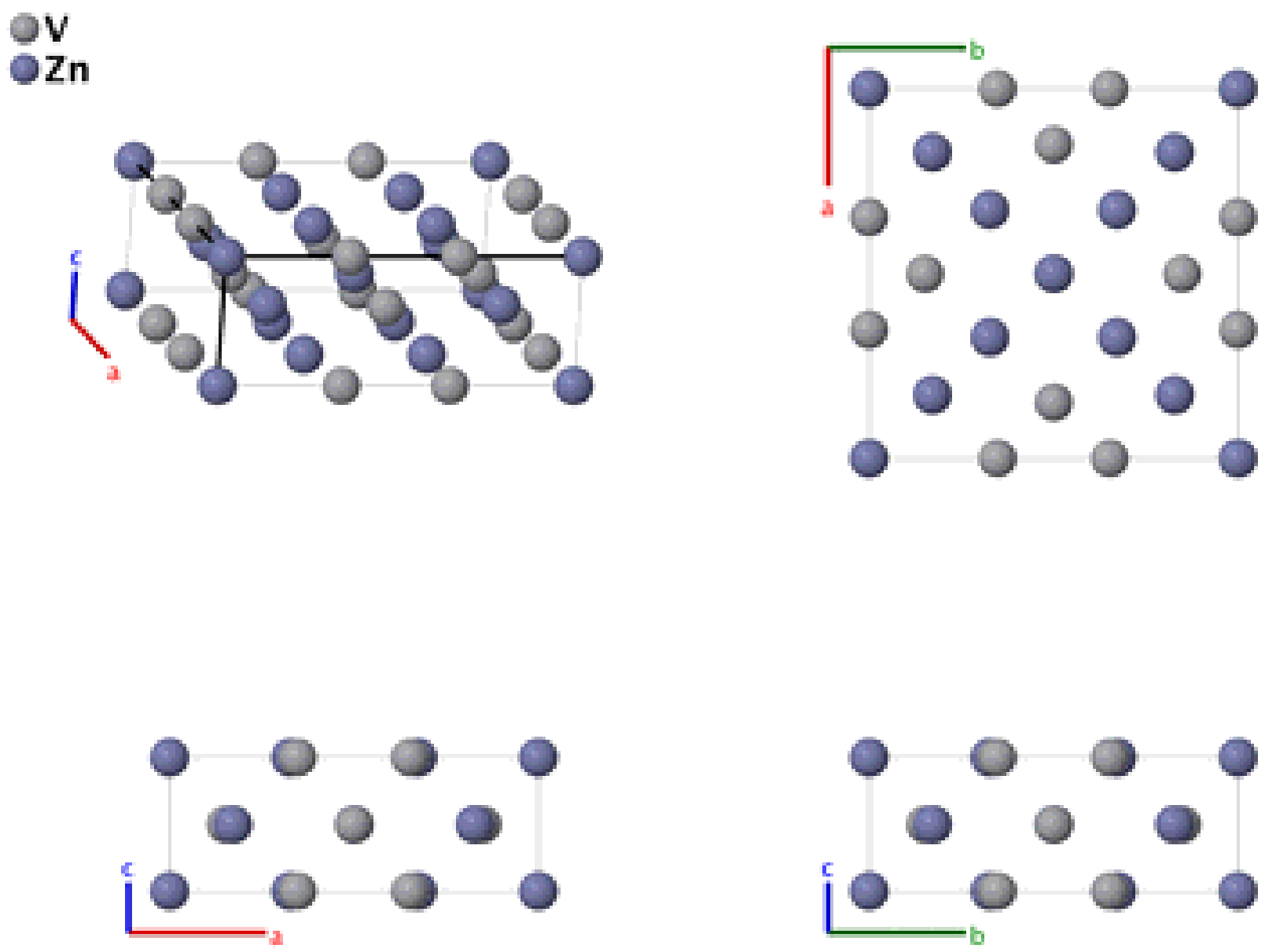

| | | |
|---|---|---|
| **Prototype** | : | $V_4Zn_5$ |
| **AFLOW prototype label** | : | A4B5_tI18_139_i_ah |
| ***Strukturbericht* designation** | : | None |
| **Pearson symbol** | : | tI18 |
| **Space group number** | : | 139 |
| **Space group symbol** | : | I4/mmm |
| **AFLOW prototype command** | : | `aflow --proto=A4B5_tI18_139_i_ah`<br>`--params=`$a, c/a, x_2, x_3$ |

**Other compounds with this structure:**

- $Pt_8Ti$

- This is very similar to the $Pt_8Ti$ structure.

**Body-centered Tetragonal primitive vectors:**

$$\mathbf{a}_1 = -\frac{1}{2}\,a\,\hat{\mathbf{x}} + \frac{1}{2}\,a\,\hat{\mathbf{y}} + \frac{1}{2}\,c\,\hat{\mathbf{z}}$$
$$\mathbf{a}_2 = \frac{1}{2}\,a\,\hat{\mathbf{x}} - \frac{1}{2}\,a\,\hat{\mathbf{y}} + \frac{1}{2}\,c\,\hat{\mathbf{z}}$$
$$\mathbf{a}_3 = \frac{1}{2}\,a\,\hat{\mathbf{x}} + \frac{1}{2}\,a\,\hat{\mathbf{y}} - \frac{1}{2}\,c\,\hat{\mathbf{z}}$$

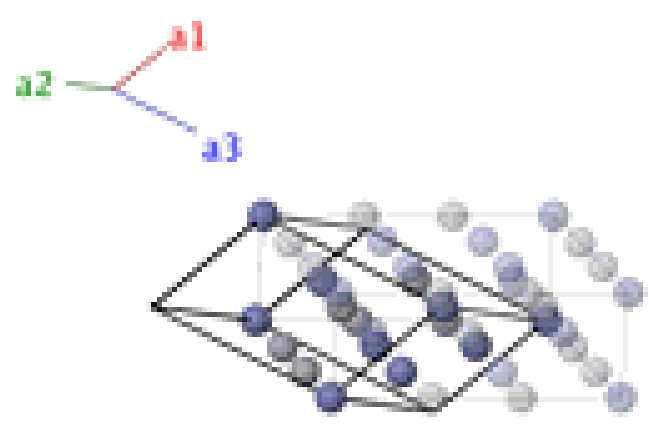

**Basis vectors:**

| | | Lattice Coordinates | | Cartesian Coordinates | Wyckoff Position | Atom Type |
|---|---|---|---|---|---|---|
| $\mathbf{B_1}$ | = | $0\,\mathbf{a_1} + 0\,\mathbf{a_2} + 0\,\mathbf{a_3}$ | = | $0\hat{\mathbf{x}} + 0\hat{\mathbf{y}} + 0\hat{\mathbf{z}}$ | $(2a)$ | Zn I |
| $\mathbf{B_2}$ | = | $x_2\,\mathbf{a_1} + x_2\,\mathbf{a_2} + 2x_2\,\mathbf{a_3}$ | = | $x_2\,a\,\hat{\mathbf{x}} + x_2\,a\,\hat{\mathbf{y}}$ | $(8h)$ | Zn II |
| $\mathbf{B_3}$ | = | $-x_2\,\mathbf{a_1} - x_2\,\mathbf{a_2} - 2x_2\,\mathbf{a_3}$ | = | $-x_2\,a\,\hat{\mathbf{x}} - x_2\,a\,\hat{\mathbf{y}}$ | $(8h)$ | Zn II |
| $\mathbf{B_4}$ | = | $x_2\,\mathbf{a_1} - x_2\,\mathbf{a_2}$ | = | $-x_2\,a\,\hat{\mathbf{x}} + x_2\,a\,\hat{\mathbf{y}}$ | $(8h)$ | Zn II |
| $\mathbf{B_5}$ | = | $-x_2\,\mathbf{a_1} + x_2\,\mathbf{a_2}$ | = | $x_2\,a\,\hat{\mathbf{x}} - x_2\,a\,\hat{\mathbf{y}}$ | $(8h)$ | Zn II |
| $\mathbf{B_6}$ | = | $x_3\,\mathbf{a_2} + x_3\,\mathbf{a_3}$ | = | $x_3\,a\,\hat{\mathbf{x}}$ | $(8i)$ | V |
| $\mathbf{B_7}$ | = | $-x_3\,\mathbf{a_2} - x_3\,\mathbf{a_3}$ | = | $-x_3\,a\,\hat{\mathbf{x}}$ | $(8i)$ | V |
| $\mathbf{B_8}$ | = | $x_3\,\mathbf{a_1} + x_3\,\mathbf{a_3}$ | = | $x_3\,a\,\hat{\mathbf{y}}$ | $(8i)$ | V |
| $\mathbf{B_9}$ | = | $-x_3\,\mathbf{a_1} - x_3\,\mathbf{a_3}$ | = | $-x_3\,a\,\hat{\mathbf{y}}$ | $(8i)$ | V |

**Geometry files:**

- CIF: pp. 703
- POSCAR: pp. 704

# $Al_4Ba$ ($D1_3$) Structure: A4B_tI10_139_de_a

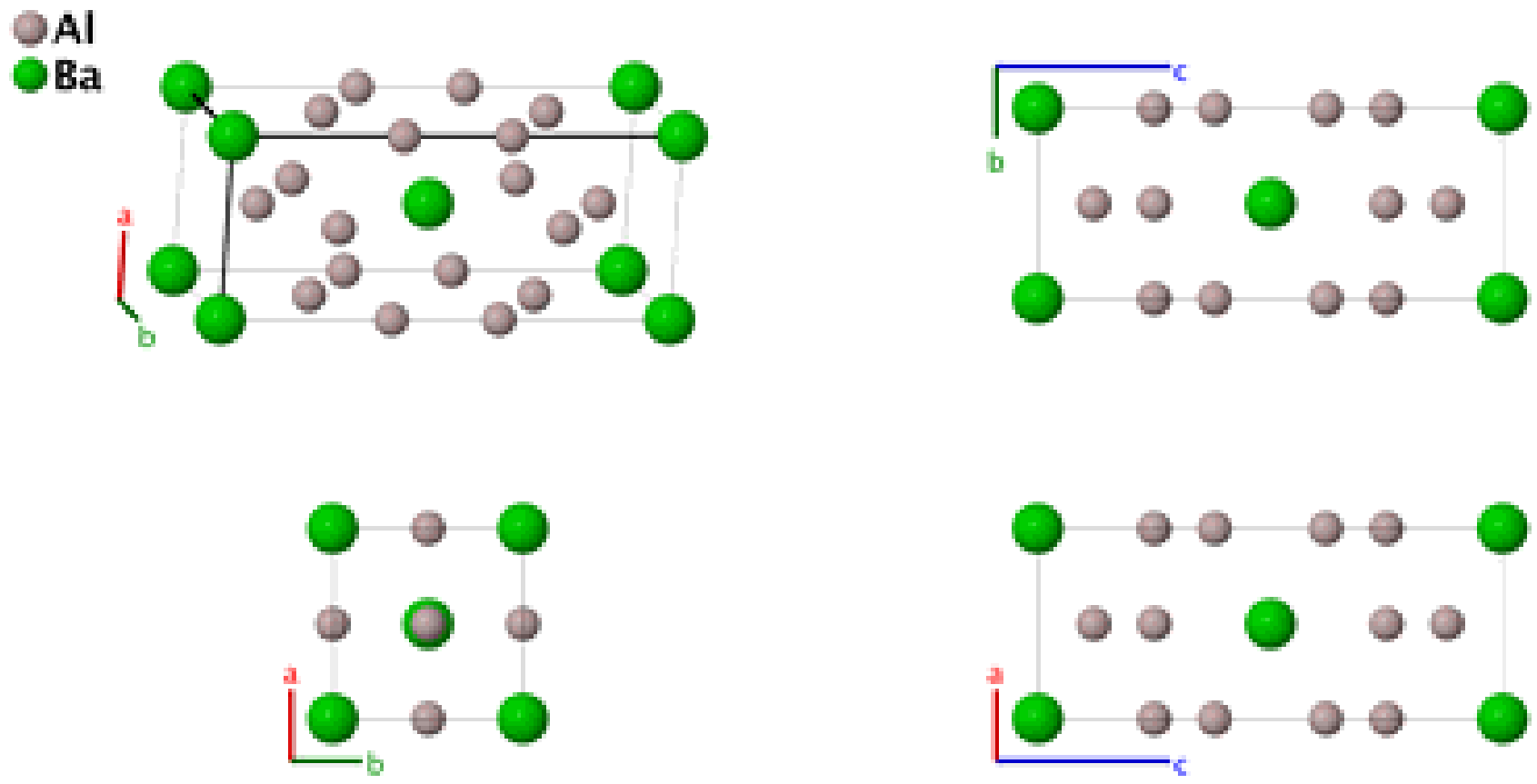

| | | |
|---|---|---|
| **Prototype** | : | $Al_4Ba$ |
| **AFLOW prototype label** | : | A4B_tI10_139_de_a |
| ***Strukturbericht* designation** | : | $D1_3$ |
| **Pearson symbol** | : | tI10 |
| **Space group number** | : | 139 |
| **Space group symbol** | : | I4/mmm |
| **AFLOW prototype command** | : | `aflow --proto=A4B_tI10_139_de_a`<br>`--params=`$a, c/a, z_3$ |

**Other compounds with this structure:**

- $ThCu_2Si_2$, $ThCr_2Si_2$, $BaFe_2As_2$, $KFe_2As_2$, hundreds more

---

- Removing the Al-I atoms transforms this to the $MoSi_2$ ($C11_b$) structure.

**Body-centered Tetragonal primitive vectors:**

$$
\begin{aligned}
\mathbf{a}_1 &= -\tfrac{1}{2}\,a\,\hat{\mathbf{x}} + \tfrac{1}{2}\,a\,\hat{\mathbf{y}} + \tfrac{1}{2}\,c\,\hat{\mathbf{z}} \\
\mathbf{a}_2 &= \tfrac{1}{2}\,a\,\hat{\mathbf{x}} - \tfrac{1}{2}\,a\,\hat{\mathbf{y}} + \tfrac{1}{2}\,c\,\hat{\mathbf{z}} \\
\mathbf{a}_3 &= \tfrac{1}{2}\,a\,\hat{\mathbf{x}} + \tfrac{1}{2}\,a\,\hat{\mathbf{y}} - \tfrac{1}{2}\,c\,\hat{\mathbf{z}}
\end{aligned}
$$

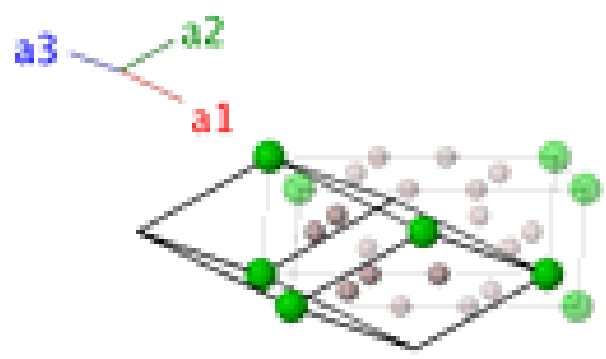

**Basis vectors:**

| | | Lattice Coordinates | | Cartesian Coordinates | Wyckoff Position | Atom Type |
|---|---|---|---|---|---|---|
| $\mathbf{B}_1$ | = | $0\,\mathbf{a}_1 + 0\,\mathbf{a}_2 + 0\,\mathbf{a}_3$ | = | $0\hat{\mathbf{x}} + 0\hat{\mathbf{y}} + 0\hat{\mathbf{z}}$ | (2$a$) | Ba |
| $\mathbf{B}_2$ | = | $\frac{3}{4}\,\mathbf{a}_1 + \frac{1}{4}\,\mathbf{a}_2 + \frac{1}{2}\,\mathbf{a}_3$ | = | $\frac{1}{2}\,a\,\hat{\mathbf{y}} + \frac{1}{4}\,c\,\hat{\mathbf{z}}$ | (4$d$) | Al I |
| $\mathbf{B}_3$ | = | $\frac{1}{4}\,\mathbf{a}_1 + \frac{3}{4}\,\mathbf{a}_2 + \frac{1}{2}\,\mathbf{a}_3$ | = | $\frac{1}{2}\,a\,\hat{\mathbf{x}} + \frac{1}{4}\,c\,\hat{\mathbf{z}}$ | (4$d$) | Al I |
| $\mathbf{B}_4$ | = | $z_3\,\mathbf{a}_1 + z_3\,\mathbf{a}_2$ | = | $z_3\,c\,\hat{\mathbf{z}}$ | (4$e$) | Al II |
| $\mathbf{B}_5$ | = | $-z_3\,\mathbf{a}_1 - z_3\,\mathbf{a}_2$ | = | $-z_3\,c\,\hat{\mathbf{z}}$ | (4$e$) | Al II |

**Geometry files:**

- CIF: pp. 704
- POSCAR: pp. 704

# $Pt_8Ti$ Structure: A8B_tI18_139_hi_a

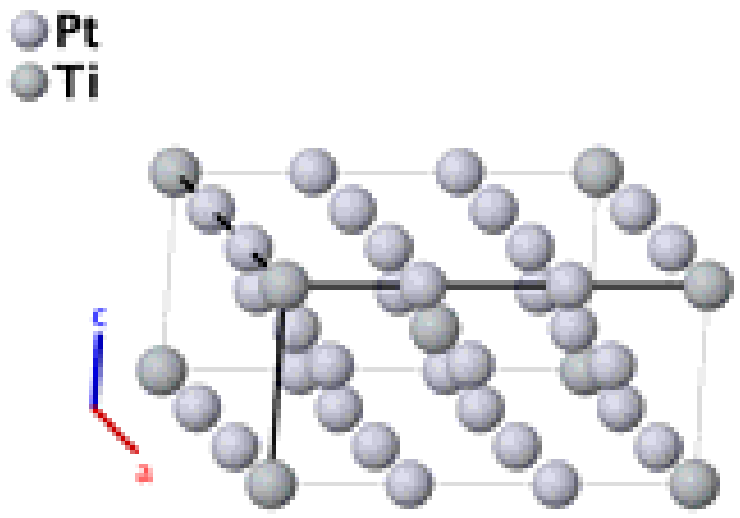

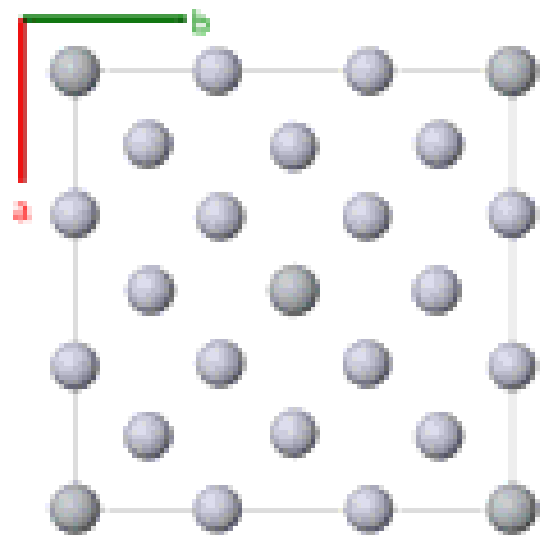

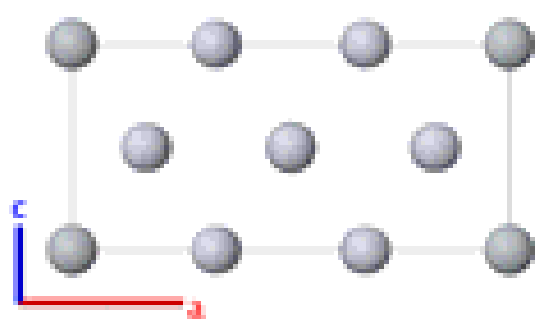

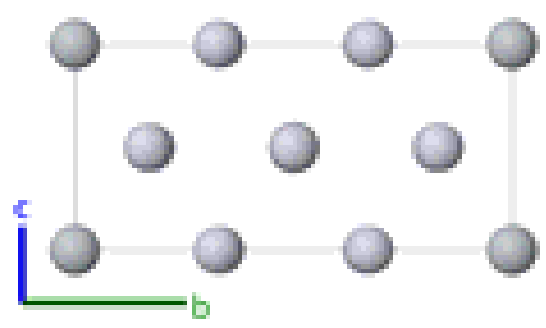

| | | |
|---|---|---|
| **Prototype** | : | $Pt_8Ti$ |
| **AFLOW prototype label** | : | A8B_tI18_139_hi_a |
| ***Strukturbericht* designation** | : | None |
| **Pearson symbol** | : | tI18 |
| **Space group number** | : | 139 |
| **Space group symbol** | : | I4/mmm |
| **AFLOW prototype command** | : | `aflow --proto=A8B_tI18_139_hi_a` `--params=`$a, c/a, x_2, x_3$ |

**Other compounds with this structure:**

- $NbNi_8$

- $a = 3/\left(\sqrt{2}\right) a_{fcc}$, $c = a_{fcc}$, $x_2 = 1/3$, and $x_3 = 1/3$, the atoms are on the sites of the fcc lattice. The pictures here are drawn with these parameters, with $a_{fcc}$ appropriate for nickel. Compare this to the very similar $V_4Zn_5$ structure.

**Body-centered Tetragonal primitive vectors:**

$$
\begin{aligned}
\mathbf{a}_1 &= -\frac{1}{2}\,a\,\hat{\mathbf{x}} + \frac{1}{2}\,a\,\hat{\mathbf{y}} + \frac{1}{2}\,c\,\hat{\mathbf{z}} \\
\mathbf{a}_2 &= \frac{1}{2}\,a\,\hat{\mathbf{x}} - \frac{1}{2}\,a\,\hat{\mathbf{y}} + \frac{1}{2}\,c\,\hat{\mathbf{z}} \\
\mathbf{a}_3 &= \frac{1}{2}\,a\,\hat{\mathbf{x}} + \frac{1}{2}\,a\,\hat{\mathbf{y}} - \frac{1}{2}\,c\,\hat{\mathbf{z}}
\end{aligned}
$$

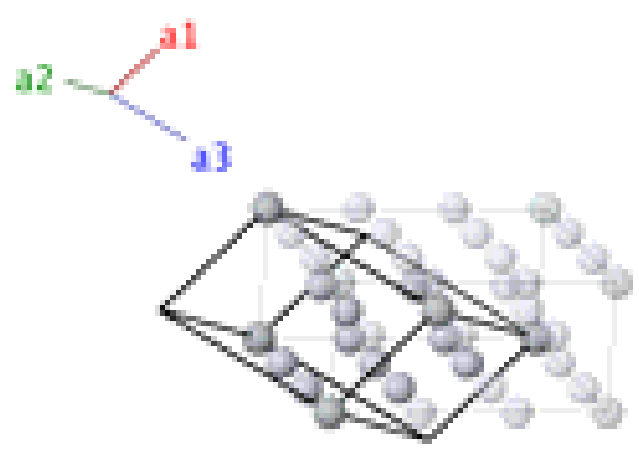

**Basis vectors:**

| | | Lattice Coordinates | | Cartesian Coordinates | Wyckoff Position | Atom Type |
|---|---|---|---|---|---|---|
| $\mathbf{B_1}$ | = | $0\,\mathbf{a_1} + 0\,\mathbf{a_2} + 0\,\mathbf{a_3}$ | = | $0\hat{\mathbf{x}} + 0\hat{\mathbf{y}} + 0\hat{\mathbf{z}}$ | $(2a)$ | Ti |
| $\mathbf{B_2}$ | = | $x_2\,\mathbf{a_1} + x_2\,\mathbf{a_2} + 2x_2\,\mathbf{a_3}$ | = | $x_2\,a\,\hat{\mathbf{x}} + x_2\,a\,\hat{\mathbf{y}}$ | $(8h)$ | Pt I |
| $\mathbf{B_3}$ | = | $-x_2\,\mathbf{a_1} - x_2\,\mathbf{a_2} - 2x_2\,\mathbf{a_3}$ | = | $-x_2\,a\,\hat{\mathbf{x}} - x_2\,a\,\hat{\mathbf{y}}$ | $(8h)$ | Pt I |
| $\mathbf{B_4}$ | = | $x_2\,\mathbf{a_1} - x_2\,\mathbf{a_2}$ | = | $-x_2\,a\,\hat{\mathbf{x}} + x_2\,a\,\hat{\mathbf{y}}$ | $(8h)$ | Pt I |
| $\mathbf{B_5}$ | = | $-x_2\,\mathbf{a_1} + x_2\,\mathbf{a_2}$ | = | $x_2\,a\,\hat{\mathbf{x}} - x_2\,a\,\hat{\mathbf{y}}$ | $(8h)$ | Pt I |
| $\mathbf{B_6}$ | = | $x_3\,\mathbf{a_2} + x_3\,\mathbf{a_3}$ | = | $x_3\,a\,\hat{\mathbf{x}}$ | $(8i)$ | Pt II |
| $\mathbf{B_7}$ | = | $-x_3\,\mathbf{a_2} - x_3\,\mathbf{a_3}$ | = | $-x_3\,a\,\hat{\mathbf{x}}$ | $(8i)$ | Pt II |
| $\mathbf{B_8}$ | = | $x_3\,\mathbf{a_1} + x_3\,\mathbf{a_3}$ | = | $x_3\,a\,\hat{\mathbf{y}}$ | $(8i)$ | Pt II |
| $\mathbf{B_9}$ | = | $-x_3\,\mathbf{a_1} - x_3\,\mathbf{a_3}$ | = | $-x_3\,a\,\hat{\mathbf{y}}$ | $(8i)$ | Pt II |

---

**Found in:**

---

**Geometry files:**

- CIF: pp. 704
- POSCAR: pp. 705

# $ThH_2$ (L'2) Structure: A2B_tI6_139_d_a

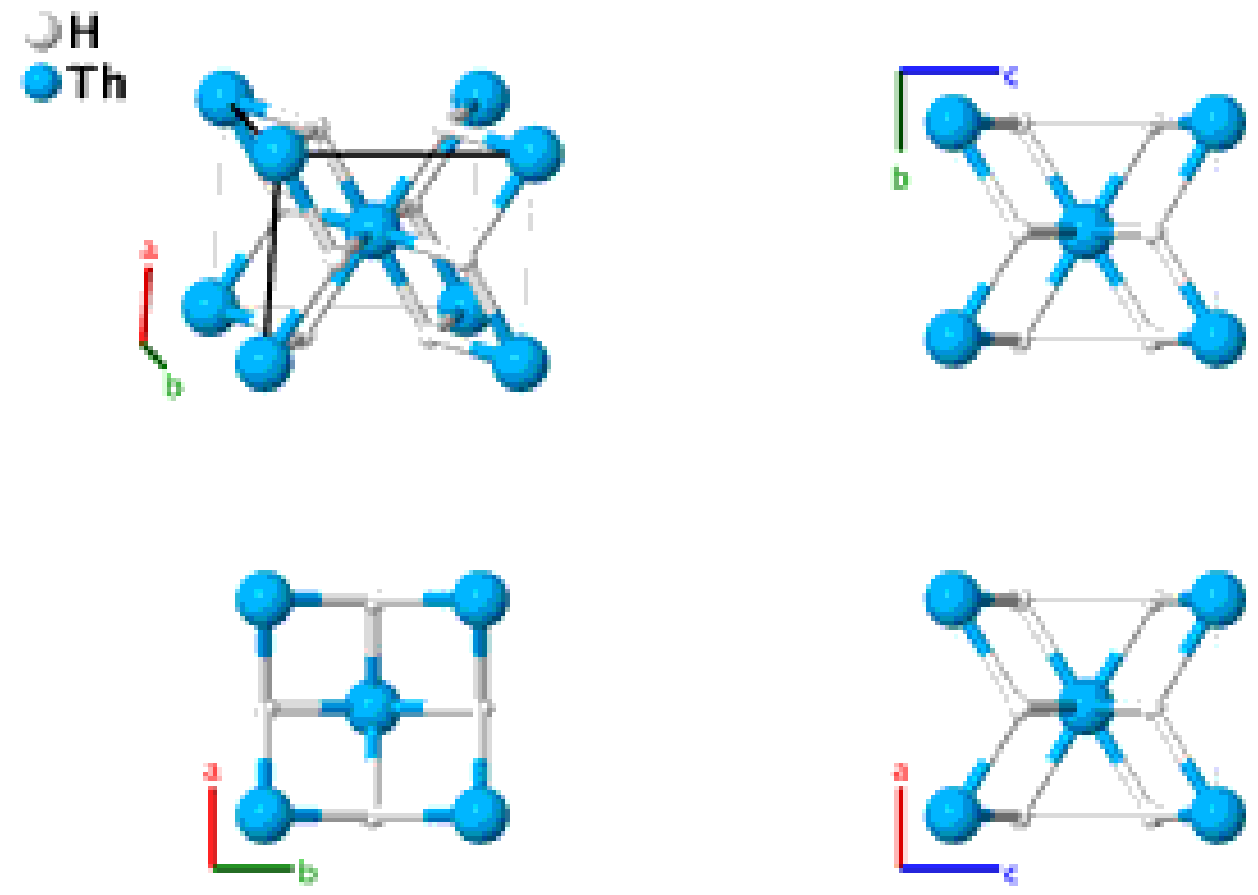

**Prototype** : $ThH_2$

**AFLOW prototype label** : A2B_tI6_139_d_a

***Strukturbericht* designation** : L'2

**Pearson symbol** : tI6

**Space group number** : 139

**Space group symbol** : I4/mmm

**AFLOW prototype command** : `aflow --proto=A2B_tI6_139_d_a --params=`$a, c/a$

---

**Other compounds with this structure:**

- $SiPt_2$, $TiH_2$, $ZrH_2$

---

**Body-centered Tetragonal primitive vectors:**

$$\mathbf{a}_1 = -\frac{1}{2}\,a\,\hat{\mathbf{x}} + \frac{1}{2}\,a\,\hat{\mathbf{y}} + \frac{1}{2}\,c\,\hat{\mathbf{z}}$$
$$\mathbf{a}_2 = \frac{1}{2}\,a\,\hat{\mathbf{x}} - \frac{1}{2}\,a\,\hat{\mathbf{y}} + \frac{1}{2}\,c\,\hat{\mathbf{z}}$$
$$\mathbf{a}_3 = \frac{1}{2}\,a\,\hat{\mathbf{x}} + \frac{1}{2}\,a\,\hat{\mathbf{y}} - \frac{1}{2}\,c\,\hat{\mathbf{z}}$$

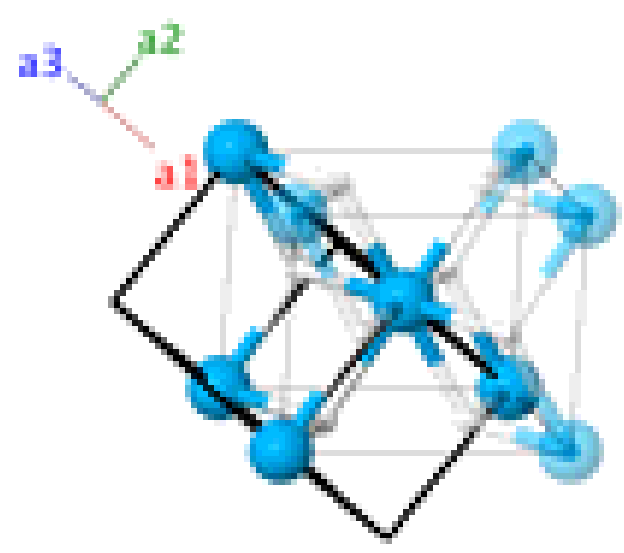

---

**Basis vectors:**

| | | Lattice Coordinates | | Cartesian Coordinates | Wyckoff Position | Atom Type |
|---|---|---|---|---|---|---|
| $\mathbf{B}_1$ | = | $0\,\mathbf{a}_1 + 0\,\mathbf{a}_2 + 0\,\mathbf{a}_3$ | = | $0\hat{\mathbf{x}} + 0\hat{\mathbf{y}} + 0\hat{\mathbf{z}}$ | (2$a$) | Th |
| $\mathbf{B}_2$ | = | $\frac{3}{4}\,\mathbf{a}_1 + \frac{1}{4}\,\mathbf{a}_2 + \frac{1}{2}\,\mathbf{a}_3$ | = | $\frac{1}{2}\,a\,\hat{\mathbf{y}} + \frac{1}{4}\,c\,\hat{\mathbf{z}}$ | (4$d$) | H |
| $\mathbf{B}_3$ | = | $\frac{1}{4}\,\mathbf{a}_1 + \frac{3}{4}\,\mathbf{a}_2 + \frac{1}{2}\,\mathbf{a}_3$ | = | $\frac{1}{2}\,a\,\hat{\mathbf{x}} + \frac{1}{4}\,c\,\hat{\mathbf{z}}$ | (4$d$) | H |

---

---

**Geometry files:**

- CIF: pp. 705
- POSCAR: pp. 705

# $\alpha$-Pa ($A_a$) Structure: A_tI2_139_a

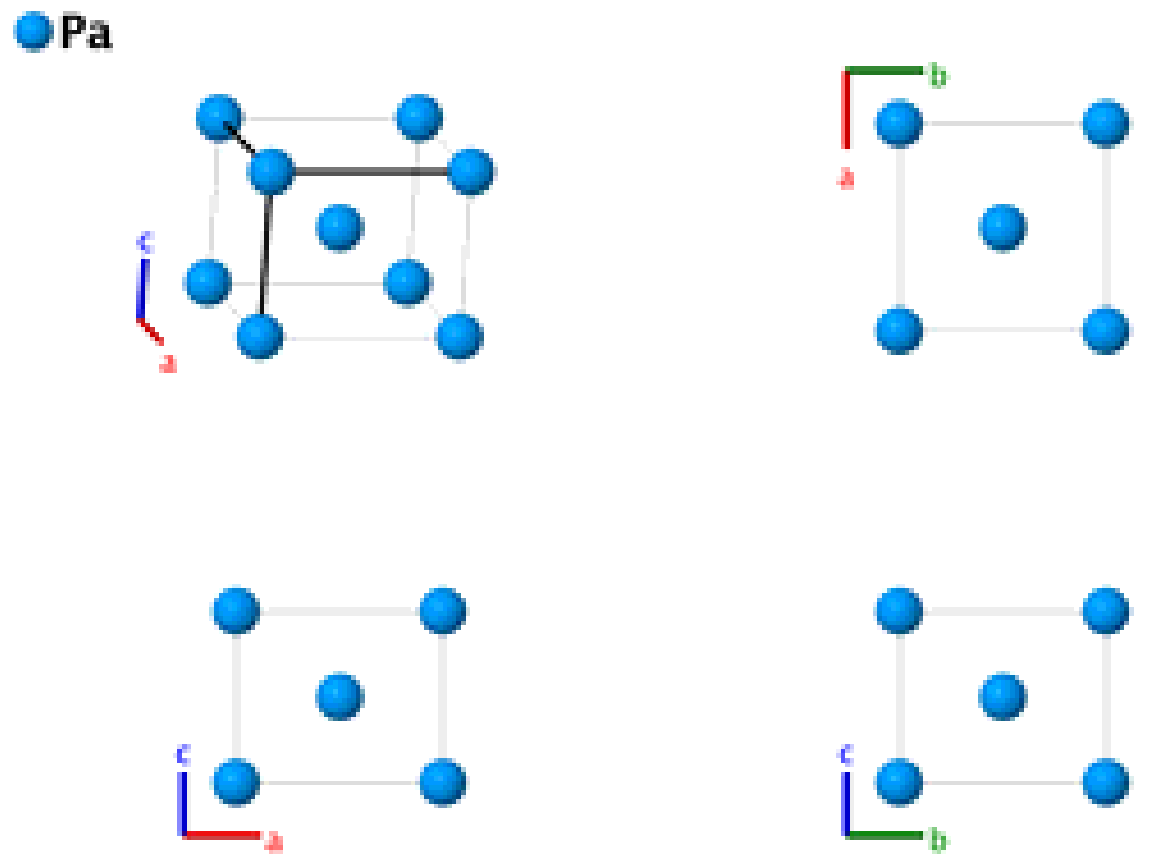

| | | |
|---|---|---|
| **Prototype** | : | $\alpha$-Pa |
| **AFLOW prototype label** | : | A_tI2_139_a |
| ***Strukturbericht* designation** | : | $A_a$ |
| **Pearson symbol** | : | tI2 |
| **Space group number** | : | 139 |
| **Space group symbol** | : | I4/mmm |
| **AFLOW prototype command** | : | `aflow --proto=A_tI2_139_a --params=`$a, c/a$ |

- This is an example of a body-centered tetragonal (bct) lattice, a distortion of the bcc lattice. Note that In (pp. 289) and $\alpha$-Pa (pp. 305) have the same AFLOW prototype label. They are generated by the same symmetry operations with different sets of parameters (--params) specified in their corresponding CIF files. In the A6 structure, c/a is near the fcc ratio of $\sqrt{2}$, while in the $A_a$ structure, c/a is near the bcc ratio of 1. When $c/a = \sqrt{2/3} \approx 0.816$ the coordination number of this system is 10. In Pa the c/a ratio is 0.827.

**Body-centered Tetragonal primitive vectors:**

$$\begin{aligned}
\mathbf{a}_1 &= -\tfrac{1}{2}\,a\,\hat{\mathbf{x}} + \tfrac{1}{2}\,a\,\hat{\mathbf{y}} + \tfrac{1}{2}\,c\,\hat{\mathbf{z}} \\
\mathbf{a}_2 &= \tfrac{1}{2}\,a\,\hat{\mathbf{x}} - \tfrac{1}{2}\,a\,\hat{\mathbf{y}} + \tfrac{1}{2}\,c\,\hat{\mathbf{z}} \\
\mathbf{a}_3 &= \tfrac{1}{2}\,a\,\hat{\mathbf{x}} + \tfrac{1}{2}\,a\,\hat{\mathbf{y}} - \tfrac{1}{2}\,c\,\hat{\mathbf{z}}
\end{aligned}$$

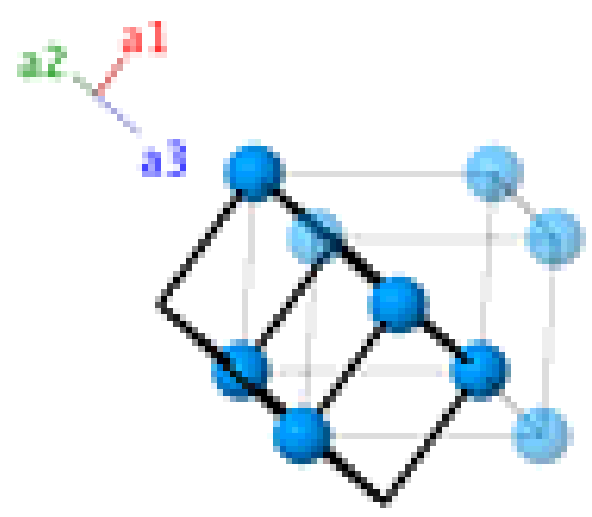

**Basis vectors:**

| | | Lattice Coordinates | | Cartesian Coordinates | Wyckoff Position | Atom Type |
|---|---|---|---|---|---|---|
| $\mathbf{B}_1$ | = | $0\,\mathbf{a}_1 + 0\,\mathbf{a}_2 + 0\,\mathbf{a}_3$ | = | $0\hat{\mathbf{x}} + 0\hat{\mathbf{y}} + 0\hat{\mathbf{z}}$ | $(2a)$ | Pa |

**Found in:**

---

**Geometry files:**
- CIF: pp. 705
- POSCAR: pp. 706

# Khatyrkite ($Al_2Cu$, C16) Structure: A2B_tI12_140_h_a

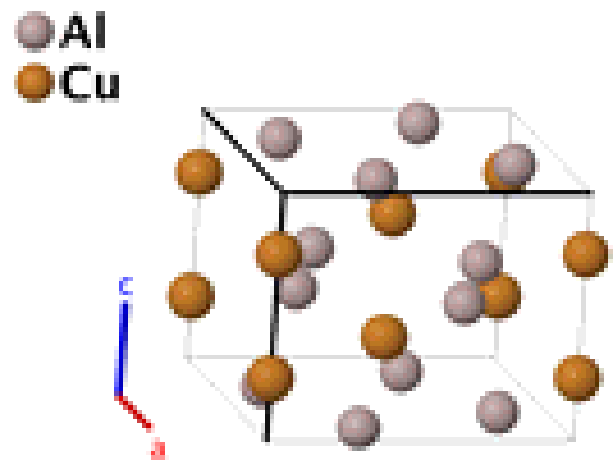

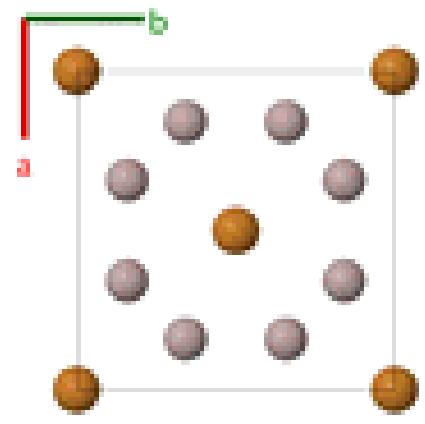

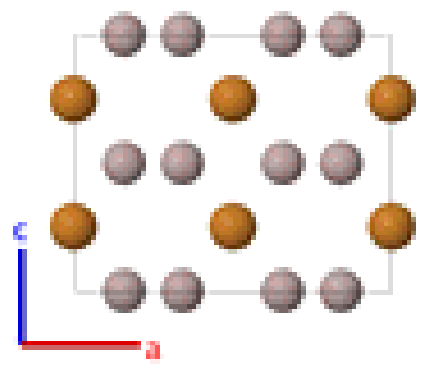

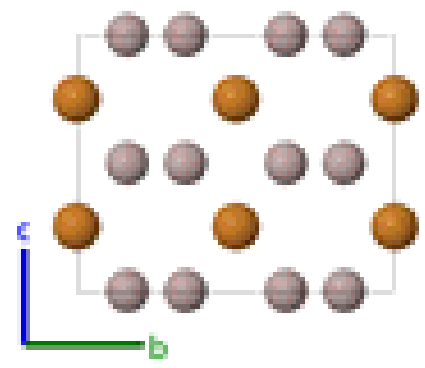

| | | |
|---|---|---|
| **Prototype** | : | $Al_2Cu$ |
| **AFLOW prototype label** | : | A2B_tI12_140_h_a |
| ***Strukturbericht* designation** | : | C16 |
| **Pearson symbol** | : | tI12 |
| **Space group number** | : | 140 |
| **Space group symbol** | : | I4/mcm |
| **AFLOW prototype command** | : | `aflow --proto=A2B_tI12_140_h_a` <br> `--params=`$a, c/a, x_2$ |

**Body-centered Tetragonal primitive vectors:**

$$\mathbf{a}_1 = -\frac{1}{2}\,a\,\hat{\mathbf{x}} + \frac{1}{2}\,a\,\hat{\mathbf{y}} + \frac{1}{2}\,c\,\hat{\mathbf{z}}$$
$$\mathbf{a}_2 = \frac{1}{2}\,a\,\hat{\mathbf{x}} - \frac{1}{2}\,a\,\hat{\mathbf{y}} + \frac{1}{2}\,c\,\hat{\mathbf{z}}$$
$$\mathbf{a}_3 = \frac{1}{2}\,a\,\hat{\mathbf{x}} + \frac{1}{2}\,a\,\hat{\mathbf{y}} - \frac{1}{2}\,c\,\hat{\mathbf{z}}$$

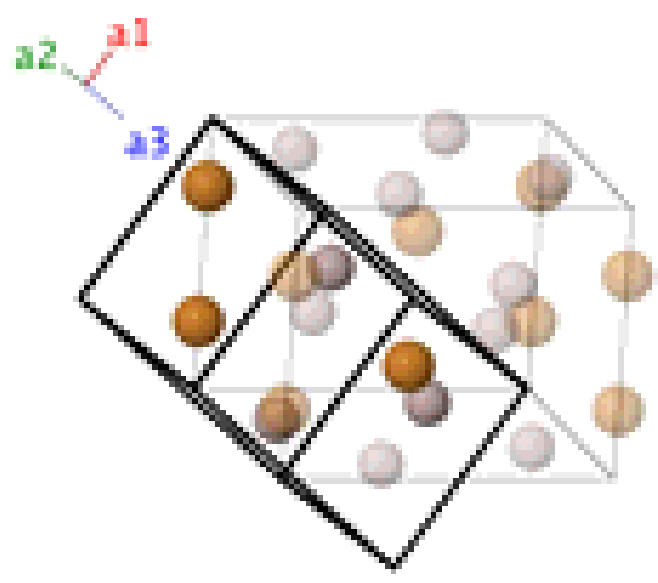

**Basis vectors:**

| | | Lattice Coordinates | | Cartesian Coordinates | Wyckoff Position | Atom Type |
|---|---|---|---|---|---|---|
| $\mathbf{B}_1$ | = | $\frac{1}{4}\,\mathbf{a}_1 + \frac{1}{4}\,\mathbf{a}_2$ | = | $\frac{1}{4}\,c\,\hat{\mathbf{z}}$ | (4$a$) | Cu |
| $\mathbf{B}_2$ | = | $\frac{3}{4}\,\mathbf{a}_1 + \frac{3}{4}\,\mathbf{a}_2$ | = | $\frac{3}{4}\,c\,\hat{\mathbf{z}}$ | (4$a$) | Cu |
| $\mathbf{B}_3$ | = | $\left(\frac{1}{2} + x_2\right)\,\mathbf{a}_1 + x_2\,\mathbf{a}_2 + \left(\frac{1}{2} + 2x_2\right)\,\mathbf{a}_3$ | = | $x_2\,a\,\hat{\mathbf{x}} + \left(\frac{1}{2} + x_2\right)\,a\,\hat{\mathbf{y}}$ | (8$h$) | Al |
| $\mathbf{B}_4$ | = | $\left(\frac{1}{2} - x_2\right)\,\mathbf{a}_1 - x_2\,\mathbf{a}_2 + \left(\frac{1}{2} - 2x_2\right)\,\mathbf{a}_3$ | = | $-x_2\,a\,\hat{\mathbf{x}} + \left(\frac{1}{2} - x_2\right)\,a\,\hat{\mathbf{y}}$ | (8$h$) | Al |
| $\mathbf{B}_5$ | = | $x_2\,\mathbf{a}_1 + \left(\frac{1}{2} - x_2\right)\,\mathbf{a}_2 + \frac{1}{2}\,\mathbf{a}_3$ | = | $\left(\frac{1}{2} - x_2\right)\,a\,\hat{\mathbf{x}} + x_2\,a\,\hat{\mathbf{y}}$ | (8$h$) | Al |
| $\mathbf{B}_6$ | = | $-x_2\,\mathbf{a}_1 + \left(\frac{1}{2} + x_2\right)\,\mathbf{a}_2 + \frac{1}{2}\,\mathbf{a}_3$ | = | $\left(\frac{1}{2} + x_2\right)\,a\,\hat{\mathbf{x}} - x_2\,a\,\hat{\mathbf{y}}$ | (8$h$) | Al |

**Geometry files:**

- CIF: pp. 706
- POSCAR: pp. 706

# $SiU_3$ ($D0_c$) Structure: AB3_tI16_140_b_ah

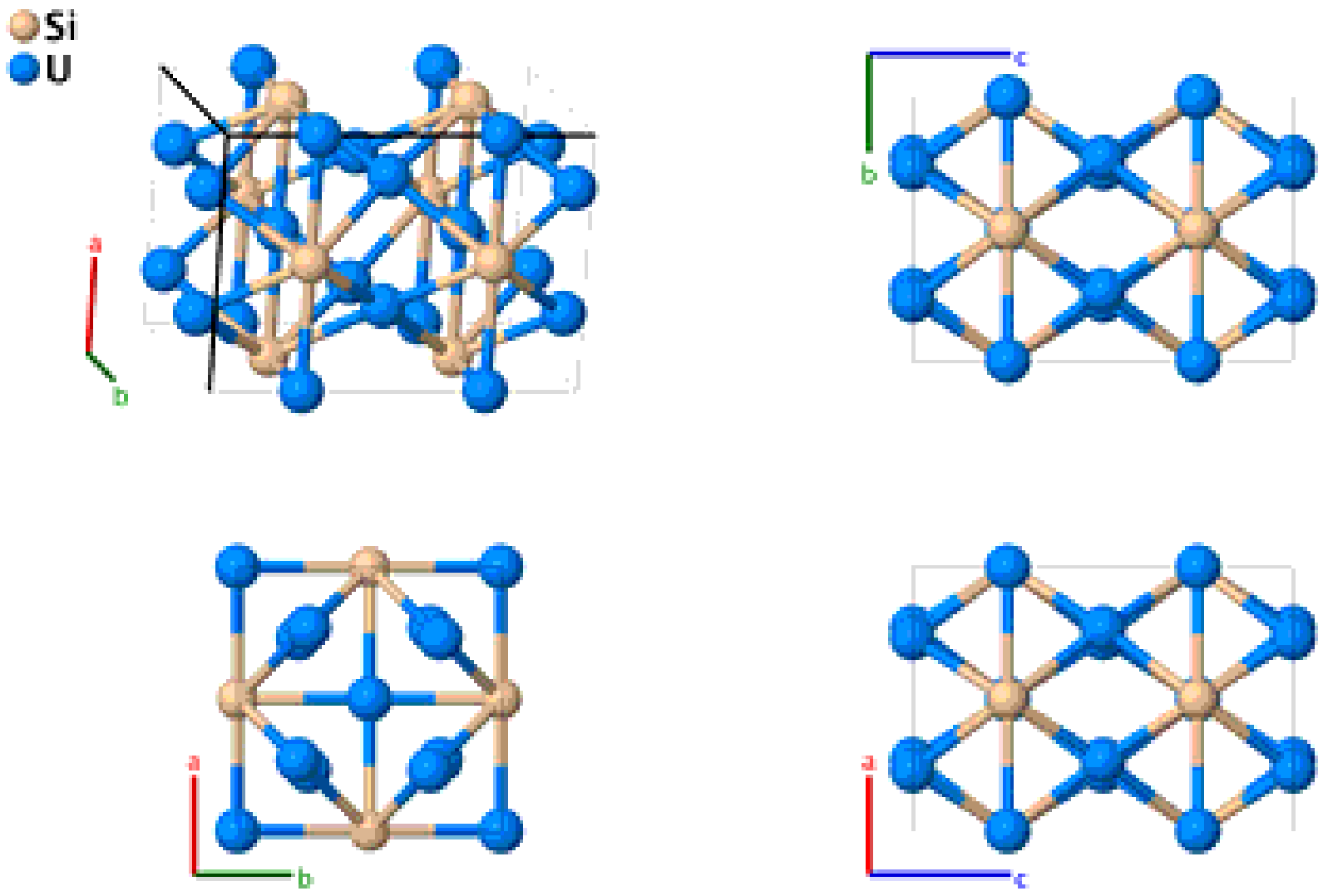

| | | |
|---|---|---|
| **Prototype** | : | $SiU_3$ |
| **AFLOW prototype label** | : | AB3_tI16_140_b_ah |
| ***Strukturbericht* designation** | : | $D0_c$ |
| **Pearson symbol** | : | tI16 |
| **Space group number** | : | 140 |
| **Space group symbol** | : | I4/mcm |
| **AFLOW prototype command** | : | `aflow --proto=AB3_tI16_140_b_ah --params=`$a, c/a, x_3$ |

**Other compounds with this structure:**

- $AlPt_3$, $GaPt_3$, $Pt_3Si$, $GePt_3$, $Ir_3Si$

- When $c = \sqrt{2}a$ and $x_3 = 1/4$ the atoms are at the positions of the $Cu_3Au$ ($L1_2$) structure. Many references define both a $D0_c$ and a D0'$_3$ ($Ir_3Si$) structure. The primary difference seems to be positioning the Si atoms on the (2a) or (2b) sites. Since this is merely an origin shift we will ignore the D0'$_3$ structure.

**Body-centered Tetragonal primitive vectors:**

$$\mathbf{a}_1 = -\frac{1}{2}\,a\,\hat{\mathbf{x}} + \frac{1}{2}\,a\,\hat{\mathbf{y}} + \frac{1}{2}\,c\,\hat{\mathbf{z}}$$
$$\mathbf{a}_2 = \frac{1}{2}\,a\,\hat{\mathbf{x}} - \frac{1}{2}\,a\,\hat{\mathbf{y}} + \frac{1}{2}\,c\,\hat{\mathbf{z}}$$
$$\mathbf{a}_3 = \frac{1}{2}\,a\,\hat{\mathbf{x}} + \frac{1}{2}\,a\,\hat{\mathbf{y}} - \frac{1}{2}\,c\,\hat{\mathbf{z}}$$

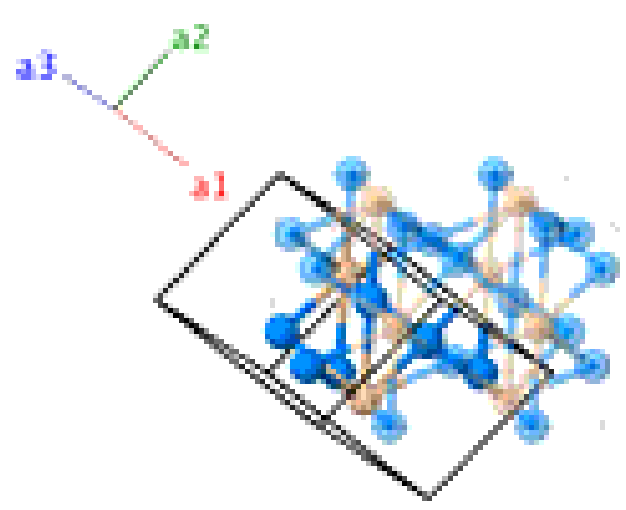

**Basis vectors:**

| | | Lattice Coordinates | | Cartesian Coordinates | Wyckoff Position | Atom Type |
|---|---|---|---|---|---|---|
| $\mathbf{B}_1$ | = | $\frac{1}{4}\,\mathbf{a}_1 + \frac{1}{4}\,\mathbf{a}_2$ | = | $\frac{1}{4}\,c\,\hat{\mathbf{z}}$ | (4$a$) | U I |
| $\mathbf{B_2}$ | = | $\frac{3}{4}\,\mathbf{a_1} + \frac{3}{4}\,\mathbf{a_2}$ | = | $\frac{3}{4}\,c\,\mathbf{\hat{z}}$ | (4$a$) | U I |
| $\mathbf{B_3}$ | = | $\frac{3}{4}\,\mathbf{a_1} + \frac{1}{4}\,\mathbf{a_2} + \frac{1}{2}\mathbf{a_3}$ | = | $\frac{1}{2}\,a\,\mathbf{\hat{y}} + \frac{1}{4}\,c\,\mathbf{\hat{z}}$ | (4$b$) | Si |
| $\mathbf{B_4}$ | = | $\frac{1}{4}\,\mathbf{a_1} + \frac{3}{4}\,\mathbf{a_2} + \frac{1}{2}\,\mathbf{a_3}$ | = | $\frac{1}{2}\,a\,\mathbf{\hat{x}} + \frac{1}{4}\,c\,\mathbf{\hat{z}}$ | (4$b$) | Si |
| $\mathbf{B_5}$ | = | $\left(\frac{1}{2} + x_3\right)\,\mathbf{a_1} + x_3\,\mathbf{a_2} + \left(\frac{1}{2} + 2x_3\right)\,\mathbf{a_3}$ | = | $x_3\,a\,\mathbf{\hat{x}} + \left(\frac{1}{2} + x_3\right)\,a\,\mathbf{\hat{y}}$ | (8$h$) | U II |
| $\mathbf{B_6}$ | = | $\left(\frac{1}{2} - x_3\right)\,\mathbf{a_1} - x_3\,\mathbf{a_2} + \left(\frac{1}{2} - 2x_3\right)\,\mathbf{a_3}$ | = | $-x_3\,a\,\mathbf{\hat{x}} + \left(\frac{1}{2} - x_3\right)\,a\,\mathbf{\hat{y}}$ | (8$h$) | U II |
| $\mathbf{B_7}$ | = | $x_3\,\mathbf{a_1} + \left(\frac{1}{2} - x_3\right)\,\mathbf{a_2} + \frac{1}{2}\,\mathbf{a_3}$ | = | $\left(\frac{1}{2} - x_3\right)\,a\,\mathbf{\hat{x}} + x_3\,a\,\mathbf{\hat{y}}$ | (8$h$) | U II |
| $\mathbf{B_8}$ | = | $-x_3\,\mathbf{a_1} + \left(\frac{1}{2} + x_3\right)\,\mathbf{a_2} + \frac{1}{2}\,\mathbf{a_3}$ | = | $\left(\frac{1}{2} + x_3\right)\,a\,\mathbf{\hat{x}} - x_3\,a\,\mathbf{\hat{y}}$ | (8$h$) | U II |

**Geometry files:**

- CIF: pp. 706
- POSCAR: pp. 707

# SeTl (B37) Structure: AB_tI16_140_ab_h

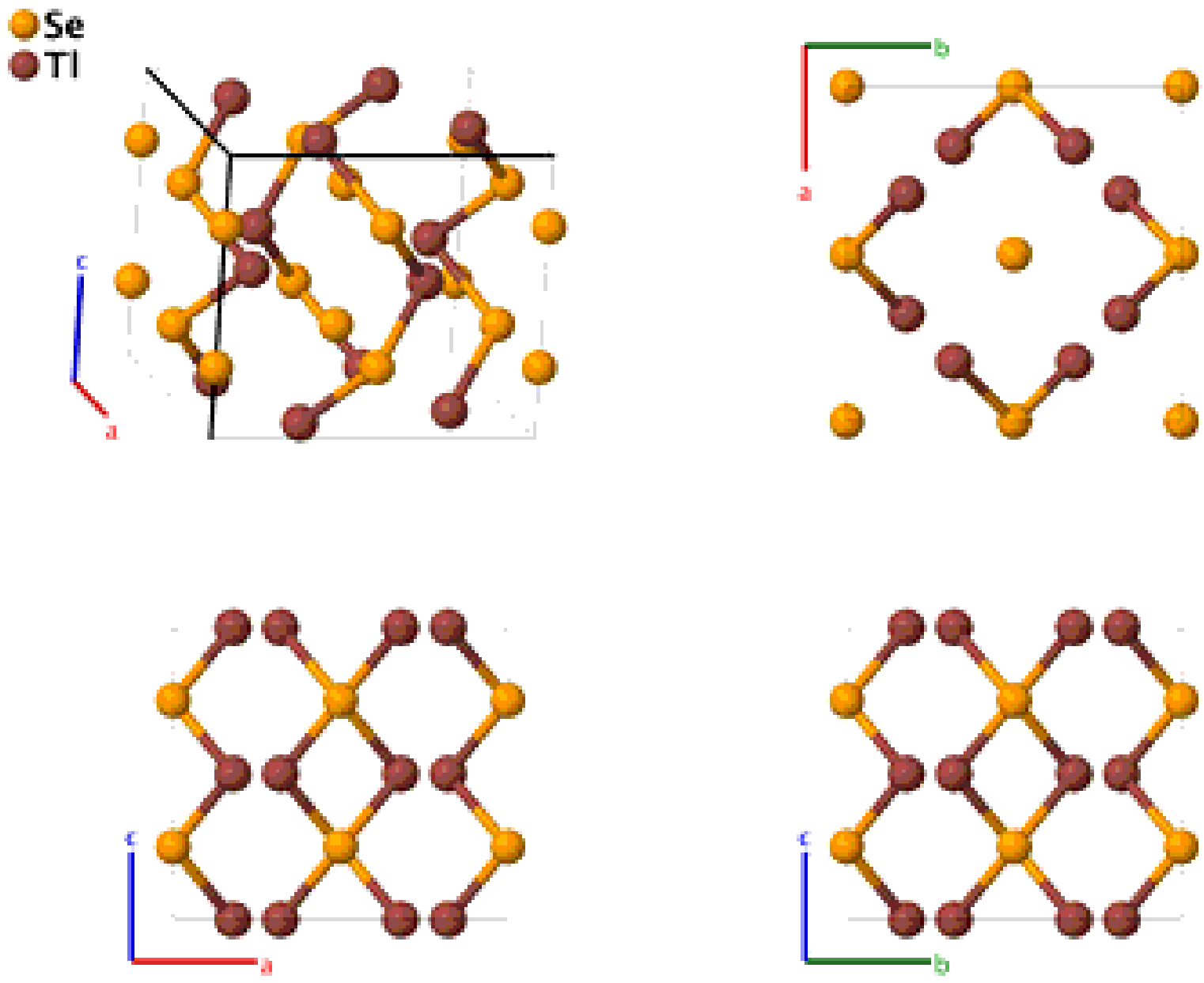

| | | |
|---|---|---|
| **Prototype** | : | SeTl |
| **AFLOW prototype label** | : | AB_tI16_140_ab_h |
| ***Strukturbericht* designation** | : | B37 |
| **Pearson symbol** | : | tI16 |
| **Space group number** | : | 140 |
| **Space group symbol** | : | I4/mcm |
| **AFLOW prototype command** | : | `aflow --proto=AB_tI16_140_ab_h`<br>`--params=`$a, c/a, x_3$ |

**Other compounds with this structure:**

- $AlKTe_2$, $AlNaSe_2$, $AlNaTe_2$, $GaNaTe_2$, $InKTe_2$, $InS_2Te$, $InNaTe_2$, $GaTe_2Tl$

- When $c = a$ and $x = 1/4$ the atoms are at the positions of the body-centered cubic (A2) lattice.

**Body-centered Tetragonal primitive vectors:**

$$\mathbf{a}_1 = -\frac{1}{2}\,a\,\hat{\mathbf{x}} + \frac{1}{2}\,a\,\hat{\mathbf{y}} + \frac{1}{2}\,c\,\hat{\mathbf{z}}$$
$$\mathbf{a}_2 = \frac{1}{2}\,a\,\hat{\mathbf{x}} - \frac{1}{2}\,a\,\hat{\mathbf{y}} + \frac{1}{2}\,c\,\hat{\mathbf{z}}$$
$$\mathbf{a}_3 = \frac{1}{2}\,a\,\hat{\mathbf{x}} + \frac{1}{2}\,a\,\hat{\mathbf{y}} - \frac{1}{2}\,c\,\hat{\mathbf{z}}$$

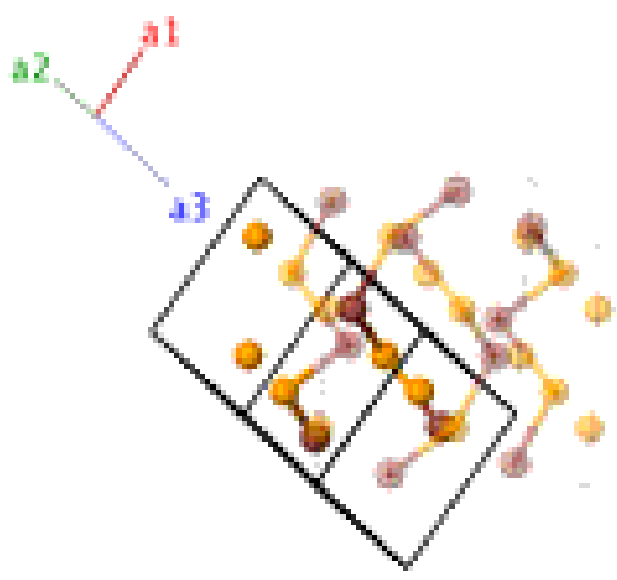

**Basis vectors:**

| | Lattice Coordinates | Cartesian Coordinates | Wyckoff Position | Atom Type |
|---|---|---|---|---|

| | | | | | | |
|---|---|---|---|---|---|---|
| $\mathbf{B_1}$ | = | $\frac{1}{4}\,\mathbf{a_1} + \frac{1}{4}\,\mathbf{a_2}$ | = | $\frac{1}{4}\,c\,\hat{\mathbf{z}}$ | (4$a$) | Se I |
| $\mathbf{B_2}$ | = | $\frac{3}{4}\,\mathbf{a_1} + \frac{3}{4}\,\mathbf{a_2}$ | = | $\frac{3}{4}\,c\,\hat{\mathbf{z}}$ | (4$a$) | Se I |
| $\mathbf{B_3}$ | = | $\frac{3}{4}\,\mathbf{a_1} + \frac{1}{4}\,\mathbf{a_2} + \frac{1}{2}\mathbf{a_3}$ | = | $\frac{1}{2}\,a\,\hat{\mathbf{y}} + \frac{1}{4}\,c\,\hat{\mathbf{z}}$ | (4$b$) | Se II |
| $\mathbf{B_4}$ | = | $\frac{1}{4}\,\mathbf{a_1} + \frac{3}{4}\,\mathbf{a_2} + \frac{1}{2}\,\mathbf{a_3}$ | = | $\frac{1}{2}\,a\,\hat{\mathbf{x}} + \frac{1}{4}\,c\,\hat{\mathbf{z}}$ | (4$b$) | Se II |
| $\mathbf{B_5}$ | = | $\left(\frac{1}{2} + x_3\right)\,\mathbf{a_1} + x_3\,\mathbf{a_2} + \left(\frac{1}{2} + 2x_3\right)\,\mathbf{a_3}$ | = | $x_3\,a\,\hat{\mathbf{x}} + \left(\frac{1}{2} + x_3\right)\,a\,\hat{\mathbf{y}}$ | (8$h$) | Tl |
| $\mathbf{B_6}$ | = | $\left(\frac{1}{2} - x_3\right)\,\mathbf{a_1} - x_3\,\mathbf{a_2} + \left(\frac{1}{2} - 2x_3\right)\,\mathbf{a_3}$ | = | $-x_3\,a\,\hat{\mathbf{x}} + \left(\frac{1}{2} - x_3\right)\,a\,\hat{\mathbf{y}}$ | (8$h$) | Tl |
| $\mathbf{B_7}$ | = | $x_3\,\mathbf{a_1} + \left(\frac{1}{2} - x_3\right)\,\mathbf{a_2} + \frac{1}{2}\,\mathbf{a_3}$ | = | $\left(\frac{1}{2} - x_3\right)\,a\,\hat{\mathbf{x}} + x_3\,a\,\hat{\mathbf{y}}$ | (8$h$) | Tl |
| $\mathbf{B_8}$ | = | $-x_3\,\mathbf{a_1} + \left(\frac{1}{2} + x_3\right)\,\mathbf{a_2} + \frac{1}{2}\,\mathbf{a_3}$ | = | $\left(\frac{1}{2} + x_3\right)\,a\,\hat{\mathbf{x}} - x_3\,a\,\hat{\mathbf{y}}$ | (8$h$) | Tl |

**Found in:**

**Geometry files:**

- CIF: pp. 707
- POSCAR: pp. 707

# Zircon ($ZrSiO_4$) Structure: A4BC_tI24_141_h_b_a

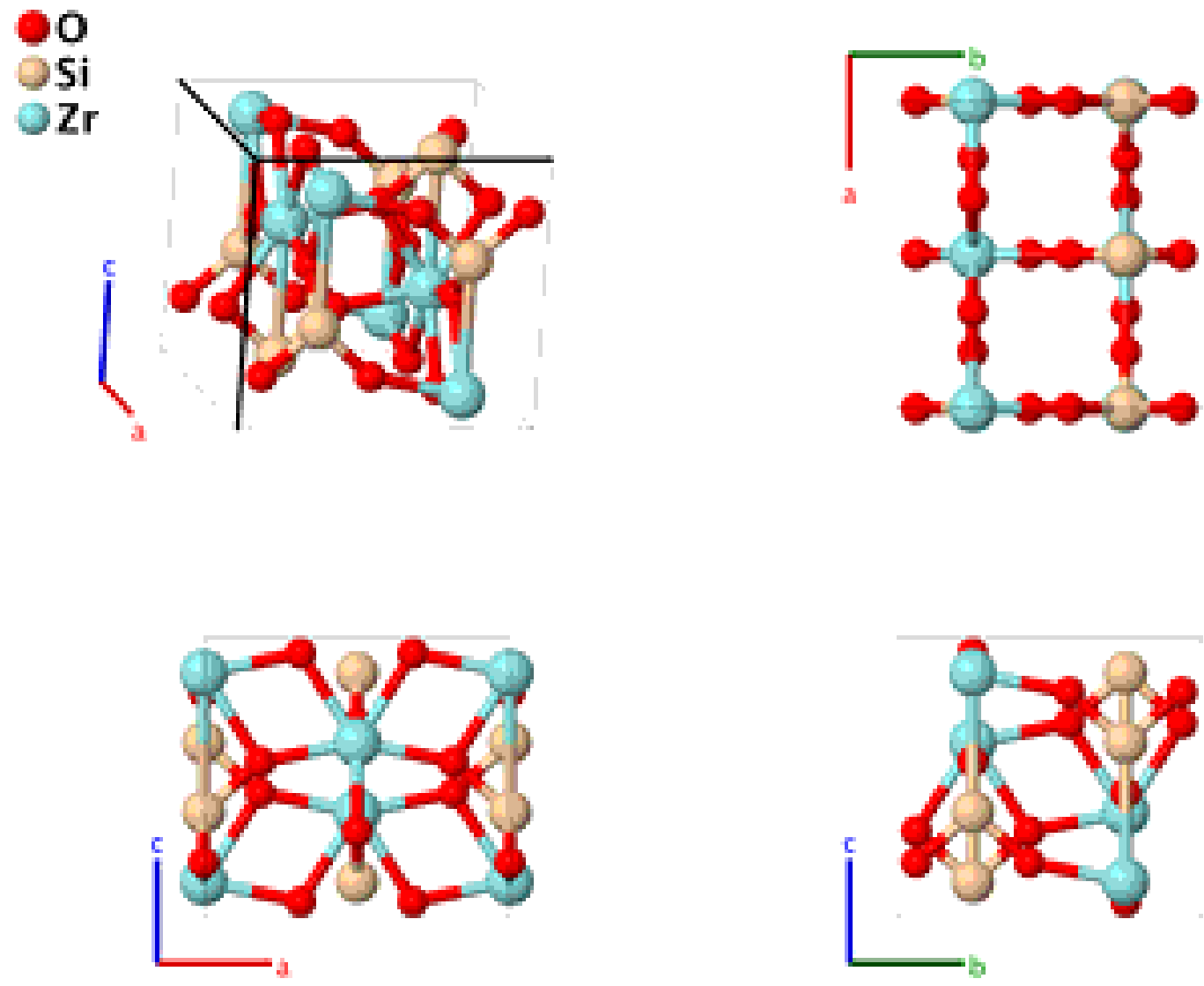

| | | |
|---|---|---|
| **Prototype** | : | $ZrSiO_4$ |
| **AFLOW prototype label** | : | A4BC_tI24_141_h_b_a |
| ***Strukturbericht* designation** | : | None |
| **Pearson symbol** | : | tI24 |
| **Space group number** | : | 141 |
| **Space group symbol** | : | $I4_1/amd$ |
| **AFLOW prototype command** | : | `aflow --proto=A4BC_tI24_141_h_b_a --params=`$a, c/a, y_3, z_3$ |

- We have filed this under quartz and related structures since the Si atoms have the same type of Si-O bonding.

**Body-centered Tetragonal primitive vectors:**

$$\mathbf{a}_1 = -\frac{1}{2}\,a\,\hat{\mathbf{x}} + \frac{1}{2}\,a\,\hat{\mathbf{y}} + \frac{1}{2}\,c\,\hat{\mathbf{z}}$$
$$\mathbf{a}_2 = \frac{1}{2}\,a\,\hat{\mathbf{x}} - \frac{1}{2}\,a\,\hat{\mathbf{y}} + \frac{1}{2}\,c\,\hat{\mathbf{z}}$$
$$\mathbf{a}_3 = \frac{1}{2}\,a\,\hat{\mathbf{x}} + \frac{1}{2}\,a\,\hat{\mathbf{y}} - \frac{1}{2}\,c\,\hat{\mathbf{z}}$$

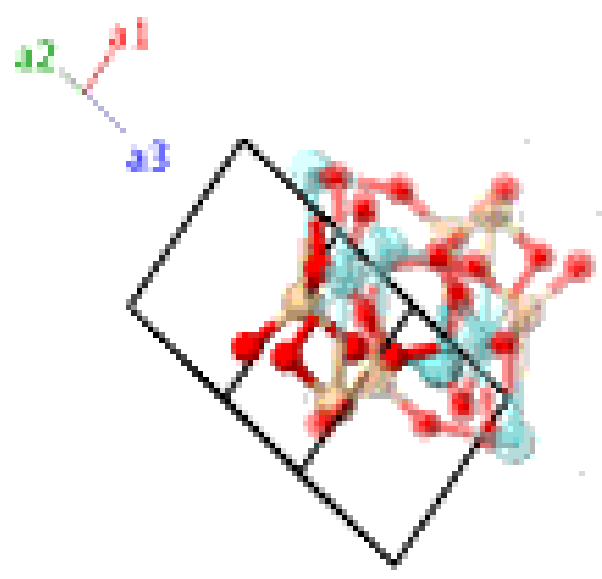

**Basis vectors:**

| | | Lattice Coordinates | | Cartesian Coordinates | Wyckoff Position | Atom Type |
|---|---|---|---|---|---|---|
| $\mathbf{B}_1$ | = | $\frac{7}{8}\,\mathbf{a}_1 + \frac{1}{8}\,\mathbf{a}_2 + \frac{3}{4}\,\mathbf{a}_3$ | = | $\frac{3}{4}\,a\,\hat{\mathbf{y}} + \frac{1}{8}\,c\,\hat{\mathbf{z}}$ | (4$a$) | Zr |
| $\mathbf{B}_2$ | = | $\frac{1}{8}\,\mathbf{a}_1 + \frac{7}{8}\,\mathbf{a}_2 + \frac{1}{4}\,\mathbf{a}_3$ | = | $\frac{1}{2}\,a\,\hat{\mathbf{x}} + \frac{3}{4}\,a\,\hat{\mathbf{y}} + \frac{3}{8}\,c\,\hat{\mathbf{z}}$ | (4$a$) | Zr |
| $\mathbf{B}_3$ | = | $\frac{5}{8}\,\mathbf{a}_1 + \frac{3}{8}\,\mathbf{a}_2 + \frac{1}{4}\,\mathbf{a}_3$ | = | $\frac{1}{4}\,a\,\hat{\mathbf{y}} + \frac{3}{8}\,c\,\hat{\mathbf{z}}$ | (4$b$) | Si |
| $\mathbf{B}_4$ | = | $\frac{3}{8}\,\mathbf{a}_1 + \frac{5}{8}\,\mathbf{a}_2 + \frac{3}{4}\,\mathbf{a}_3$ | = | $\frac{1}{2}\,a\,\hat{\mathbf{x}} + \frac{1}{4}\,a\,\hat{\mathbf{y}} + \frac{1}{8}\,c\,\hat{\mathbf{z}}$ | (4$b$) | Si |
| $\mathbf{B_5}$ | = | $(y_3 + z_3)\,\mathbf{a_1} + z_3\,\mathbf{a_2} + y_3\,\mathbf{a_3}$ | = | $y_3\,a\,\hat{\mathbf{y}} + z_3\,c\,\hat{\mathbf{z}}$ | (16$h$) | O |
| $\mathbf{B_6}$ | = | $\left(\frac{1}{2} - y_3 + z_3\right)\mathbf{a_1} + z_3\,\mathbf{a_2} + \left(\frac{1}{2} - y_3\right)\mathbf{a_3}$ | = | $\left(\frac{1}{2} - y_3\right)a\,\hat{\mathbf{y}} + z_3\,c\,\hat{\mathbf{z}}$ | (16$h$) | O |
| $\mathbf{B_7}$ | = | $z_3\,\mathbf{a_1} + \left(\frac{1}{2} - y_3 + z_3\right)\mathbf{a_2} - y_3\,\mathbf{a_3}$ | = | $\left(\frac{1}{4} - y_3\right)a\,\hat{\mathbf{x}} + \frac{3}{4}\,a\,\hat{\mathbf{y}} + \left(\frac{1}{4} + z_3\right)c\,\hat{\mathbf{z}}$ | (16$h$) | O |
| $\mathbf{B_8}$ | = | $z_3\,\mathbf{a_1} + (y_3 + z_3)\,\mathbf{a_2} + \left(\frac{1}{2} + y_3\right)\mathbf{a_3}$ | = | $\left(\frac{1}{4} + y_3\right)a\,\hat{\mathbf{x}} + \frac{1}{4}\,a\,\hat{\mathbf{y}} + \left(\frac{3}{4} + z_3\right)c\,\hat{\mathbf{z}}$ | (16$h$) | O |
| $\mathbf{B_9}$ | = | $\left(\frac{1}{2} + y_3 - z_3\right)\mathbf{a_1} - z_3\,\mathbf{a_2} + \left(\frac{1}{2} + y_3\right)\mathbf{a_3}$ | = | $\left(\frac{1}{2} + y_3\right)a\,\hat{\mathbf{y}} - z_3\,c\,\hat{\mathbf{z}}$ | (16$h$) | O |
| $\mathbf{B_{10}}$ | = | $-(y_3 + z_3)\,\mathbf{a_1} - z_3\,\mathbf{a_2} - y_3\,\mathbf{a_3}$ | = | $-y_3\,a\,\hat{\mathbf{y}} - z_3\,c\,\hat{\mathbf{z}}$ | (16$h$) | O |
| $\mathbf{B_{11}}$ | = | $-z_3\,\mathbf{a_1} + \left(\frac{1}{2} + y_3 - z_3\right)\mathbf{a_2} + y_3\,\mathbf{a_3}$ | = | $\left(\frac{1}{4} + y_3\right)a\,\hat{\mathbf{x}} + \frac{3}{4}\,a\,\hat{\mathbf{y}} + \left(\frac{1}{4} - z_3\right)c\,\hat{\mathbf{z}}$ | (16$h$) | O |
| $\mathbf{B_{12}}$ | = | $-z_3\,\mathbf{a_1} - (y_3 + z_3)\,\mathbf{a_2} + \left(\frac{1}{2} - y_3\right)\mathbf{a_3}$ | = | $\left(\frac{1}{4} - y_3\right)a\,\hat{\mathbf{x}} + \frac{1}{4}\,a\,\hat{\mathbf{y}} + \left(\frac{3}{4} - z_3\right)c\,\hat{\mathbf{z}}$ | (16$h$) | O |

**Found in:**

**Geometry files:**

- CIF: pp. 708
- POSCAR: pp. 708

# $\beta$-Sn (A5) Structure: A_tI4_141_a

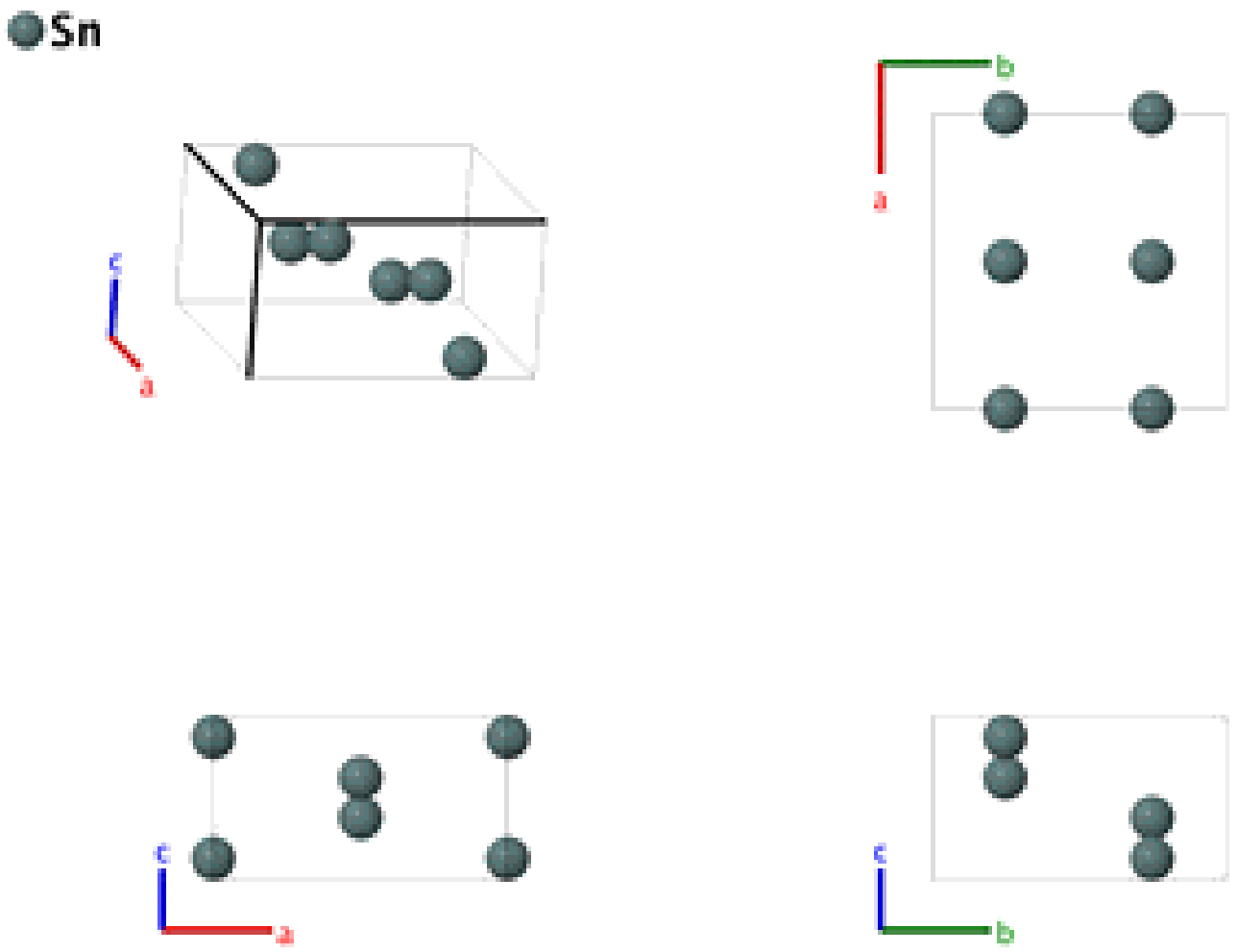

| | | |
|---|---|---|
| **Prototype** | : | $\beta$-Sn |
| **AFLOW prototype label** | : | A_tI4_141_a |
| ***Strukturbericht* designation** | : | A5 |
| **Pearson symbol** | : | tI4 |
| **Space group number** | : | 141 |
| **Space group symbol** | : | $I4_1/amd$ |
| **AFLOW prototype command** | : | `aflow --proto=A_tI4_141_a --params=`$a, c/a$ |

- When $c/a = \sqrt{2}$ this structure is equivalent to diamond (A4).

**Body-centered Tetragonal primitive vectors:**

$$\mathbf{a}_1 = -\frac{1}{2}\,a\,\hat{\mathbf{x}} + \frac{1}{2}\,a\,\hat{\mathbf{y}} + \frac{1}{2}\,c\,\hat{\mathbf{z}}$$
$$\mathbf{a}_2 = \frac{1}{2}\,a\,\hat{\mathbf{x}} - \frac{1}{2}\,a\,\hat{\mathbf{y}} + \frac{1}{2}\,c\,\hat{\mathbf{z}}$$
$$\mathbf{a}_3 = \frac{1}{2}\,a\,\hat{\mathbf{x}} + \frac{1}{2}\,a\,\hat{\mathbf{y}} - \frac{1}{2}\,c\,\hat{\mathbf{z}}$$

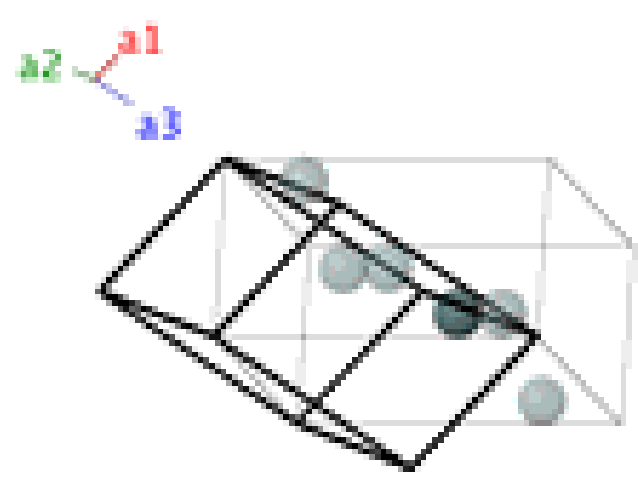

**Basis vectors:**

| | | Lattice Coordinates | | Cartesian Coordinates | Wyckoff Position | Atom Type |
|---|---|---|---|---|---|---|
| $\mathbf{B}_1$ | = | $\frac{7}{8}\,\mathbf{a}_1 + \frac{1}{8}\,\mathbf{a}_2 + \frac{3}{4}\,\mathbf{a}_3$ | = | $\frac{3}{4}\,a\,\hat{\mathbf{y}} + \frac{1}{8}\,c\,\hat{\mathbf{z}}$ | $(4a)$ | Sn |
| $\mathbf{B}_2$ | = | $\frac{1}{8}\,\mathbf{a}_1 + \frac{7}{8}\,\mathbf{a}_2 + \frac{1}{4}\,\mathbf{a}_3$ | = | $\frac{1}{2}\,a\,\hat{\mathbf{x}} + \frac{3}{4}\,a\,\hat{\mathbf{y}} + \frac{3}{8}\,c\,\hat{\mathbf{z}}$ | $(4a)$ | Sn |

**Found in:**

---

**Geometry files:**

- CIF: pp. 708
- POSCAR: pp. 709

# Hausmannite ($Mn_3O_4$) Structure: A3B4_tI28_141_ad_h

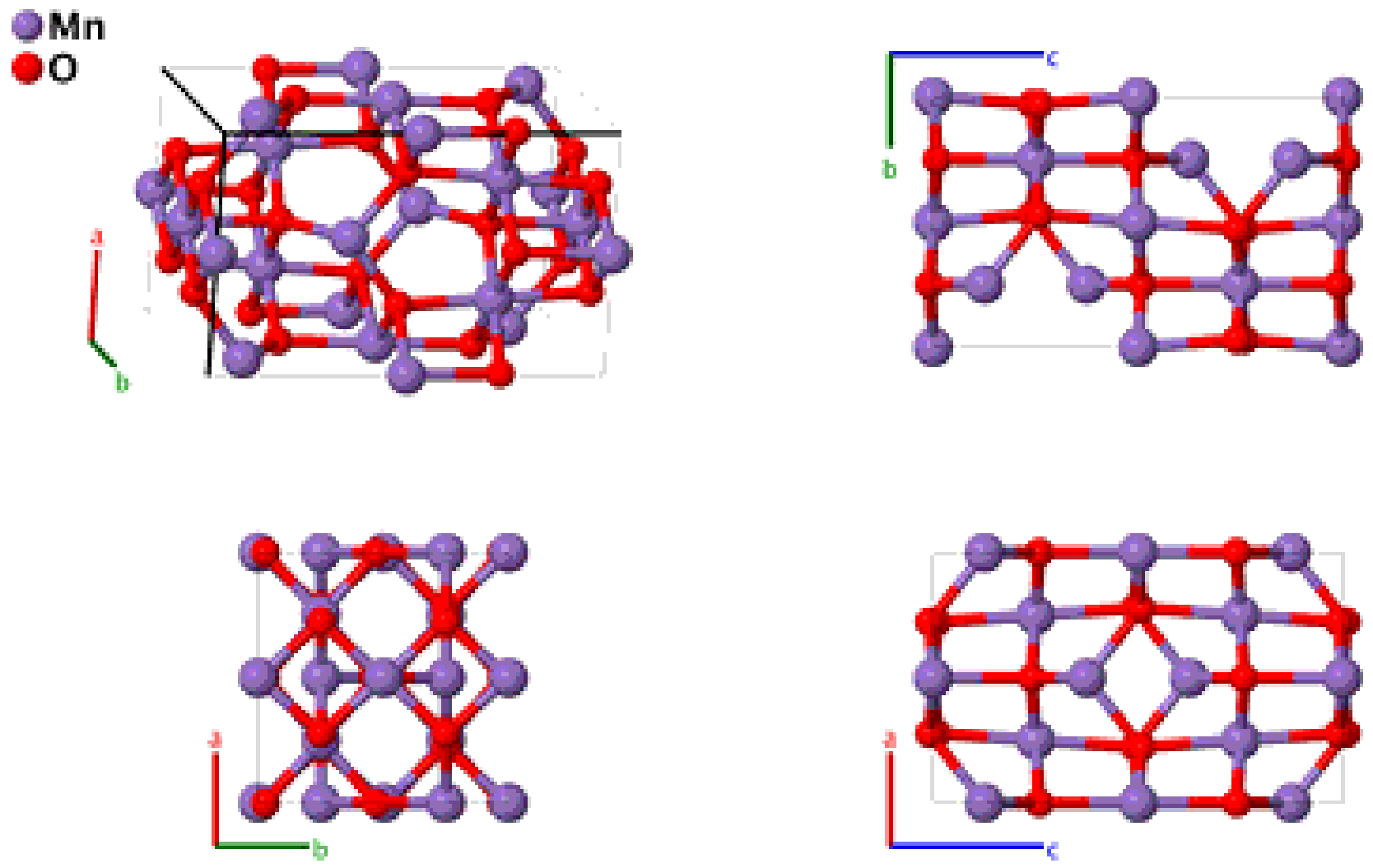

| | | |
|---|---|---|
| **Prototype** | : | $Mn_3O_4$ |
| **AFLOW prototype label** | : | A3B4_tI28_141_ad_h |
| ***Strukturbericht* designation** | : | None |
| **Pearson symbol** | : | tI28 |
| **Space group number** | : | 141 |
| **Space group symbol** | : | $I4_1/amd$ |
| **AFLOW prototype command** | : | `aflow --proto=A3B4_tI28_141_ad_h`<br>`--params=`$a, c/a, y_3, z_3$ |

---

**Body-centered Tetragonal primitive vectors:**

$$\mathbf{a}_1 = -\frac{1}{2}\,a\,\hat{\mathbf{x}} + \frac{1}{2}\,a\,\hat{\mathbf{y}} + \frac{1}{2}\,c\,\hat{\mathbf{z}}$$
$$\mathbf{a}_2 = \frac{1}{2}\,a\,\hat{\mathbf{x}} - \frac{1}{2}\,a\,\hat{\mathbf{y}} + \frac{1}{2}\,c\,\hat{\mathbf{z}}$$
$$\mathbf{a}_3 = \frac{1}{2}\,a\,\hat{\mathbf{x}} + \frac{1}{2}\,a\,\hat{\mathbf{y}} - \frac{1}{2}\,c\,\hat{\mathbf{z}}$$

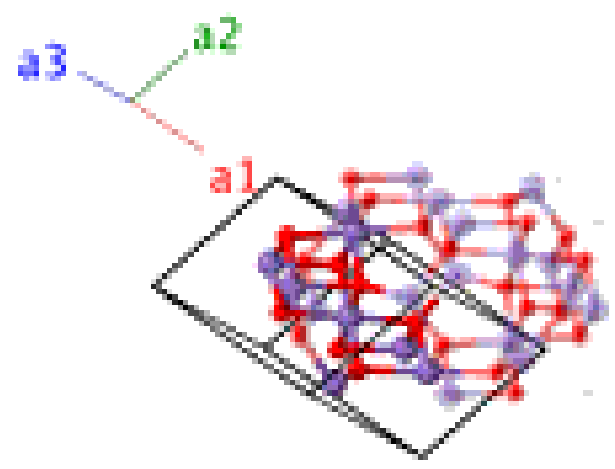

---

**Basis vectors:**

| | | Lattice Coordinates | | Cartesian Coordinates | Wyckoff Position | Atom Type |
|---|---|---|---|---|---|---|
| $\mathbf{B}_1$ | = | $\frac{7}{8}\,\mathbf{a}_1 + \frac{1}{8}\,\mathbf{a}_2 + \frac{3}{4}\,\mathbf{a}_3$ | = | $\frac{3}{4}\,a\,\hat{\mathbf{y}} + \frac{1}{8}\,c\,\hat{\mathbf{z}}$ | (4$a$) | Mn I |
| $\mathbf{B}_2$ | = | $\frac{1}{8}\,\mathbf{a}_1 + \frac{7}{8}\,\mathbf{a}_2 + \frac{1}{4}\,\mathbf{a}_3$ | = | $\frac{1}{2}\,a\,\hat{\mathbf{x}} + \frac{3}{4}\,a\,\hat{\mathbf{y}} + \frac{3}{8}\,c\,\hat{\mathbf{z}}$ | (4$a$) | Mn I |
| $\mathbf{B}_3$ | = | $\frac{1}{2}\,\mathbf{a}_1 + \frac{1}{2}\,\mathbf{a}_2$ | = | $\frac{1}{2}\,c\,\hat{\mathbf{z}}$ | (8$d$) | Mn II |
| $\mathbf{B}_4$ | = | $\frac{1}{2}\,\mathbf{a}_2 + \frac{1}{2}\,\mathbf{a}_3$ | = | $\frac{1}{2}\,a\,\hat{\mathbf{x}}$ | (8$d$) | Mn II |
| $\mathbf{B}_5$ | = | $\frac{1}{2}\,\mathbf{a}_1$ | = | $\frac{3}{4}\,a\,\hat{\mathbf{x}} + \frac{1}{4}\,a\,\hat{\mathbf{y}} + \frac{1}{4}\,c\,\hat{\mathbf{z}}$ | (8$d$) | Mn II |
| $\mathbf{B}_6$ | = | $\frac{1}{2}\,\mathbf{a}_1 + \frac{1}{2}\,\mathbf{a}_2 + \frac{1}{2}\,\mathbf{a}_3$ | = | $\frac{1}{4}\,a\,\hat{\mathbf{x}} + \frac{1}{4}\,a\,\hat{\mathbf{y}} + \frac{1}{4}\,c\,\hat{\mathbf{z}}$ | (8$d$) | Mn II |
| $\mathbf{B}_7$ | = | $(y_3 + z_3)\,\mathbf{a}_1 + z_3\,\mathbf{a}_2 + y_3\,\mathbf{a}_3$ | = | $y_3\,a\,\hat{\mathbf{y}} + z_3\,c\,\hat{\mathbf{z}}$ | (16$h$) | O |
| $\mathbf{B_8}$ | = | $\left(\frac{1}{2} - y_3 + z_3\right) \mathbf{a_1} + z_3 \mathbf{a_2} + \left(\frac{1}{2} - y_3\right) \mathbf{a_3}$ | = | $\left(\frac{1}{2} - y_3\right) a \hat{\mathbf{y}} + z_3 c \hat{\mathbf{z}}$ | (16$h$) | O |
| $\mathbf{B_9}$ | = | $z_3 \mathbf{a_1} + \left(\frac{1}{2} - y_3 + z_3\right) \mathbf{a_2} - y_3 \mathbf{a_3}$ | = | $\left(\frac{1}{4} - y_3\right) a \hat{\mathbf{x}} + \frac{3}{4} a \hat{\mathbf{y}} + \left(\frac{1}{4} + z_3\right) c \hat{\mathbf{z}}$ | (16$h$) | O |
| $\mathbf{B_{10}}$ | = | $z_3 \mathbf{a_1} + (y_3 + z_3) \mathbf{a_2} + \left(\frac{1}{2} + y_3\right) \mathbf{a_3}$ | = | $\left(\frac{1}{4} + y_3\right) a \hat{\mathbf{x}} + \frac{1}{4} a \hat{\mathbf{y}} + \left(\frac{3}{4} + z_3\right) c \hat{\mathbf{z}}$ | (16$h$) | O |
| $\mathbf{B_{11}}$ | = | $\left(\frac{1}{2} + y_3 - z_3\right) \mathbf{a_1} - z_3 \mathbf{a_2} + \left(\frac{1}{2} + y_3\right) \mathbf{a_3}$ | = | $\left(\frac{1}{2} + y_3\right) a \hat{\mathbf{y}} - z_3 c \hat{\mathbf{z}}$ | (16$h$) | O |
| $\mathbf{B_{12}}$ | = | $-(y_3 + z_3) \mathbf{a_1} - z_3 \mathbf{a_2} - y_3 \mathbf{a_3}$ | = | $-y_3 a \hat{\mathbf{y}} - z_3 c \hat{\mathbf{z}}$ | (16$h$) | O |
| $\mathbf{B_{13}}$ | = | $-z_3 \mathbf{a_1} + \left(\frac{1}{2} + y_3 - z_3\right) \mathbf{a_2} + y_3 \mathbf{a_3}$ | = | $\left(\frac{1}{4} + y_3\right) a \hat{\mathbf{x}} + \frac{3}{4} a \hat{\mathbf{y}} + \left(\frac{1}{4} - z_3\right) c \hat{\mathbf{z}}$ | (16$h$) | O |
| $\mathbf{B_{14}}$ | = | $-z_3 \mathbf{a_1} - (y_3 + z_3) \mathbf{a_2} + \left(\frac{1}{2} - y_3\right) \mathbf{a_3}$ | = | $\left(\frac{1}{4} - y_3\right) a \hat{\mathbf{x}} + \frac{1}{4} a \hat{\mathbf{y}} + \left(\frac{3}{4} - z_3\right) c \hat{\mathbf{z}}$ | (16$h$) | O |

---

**Found in:**

---

**Geometry files:**

- CIF: pp. 709
- POSCAR: pp. 709

# Anatase ($TiO_2$, C5) Structure: A2B_tI12_141_e_a

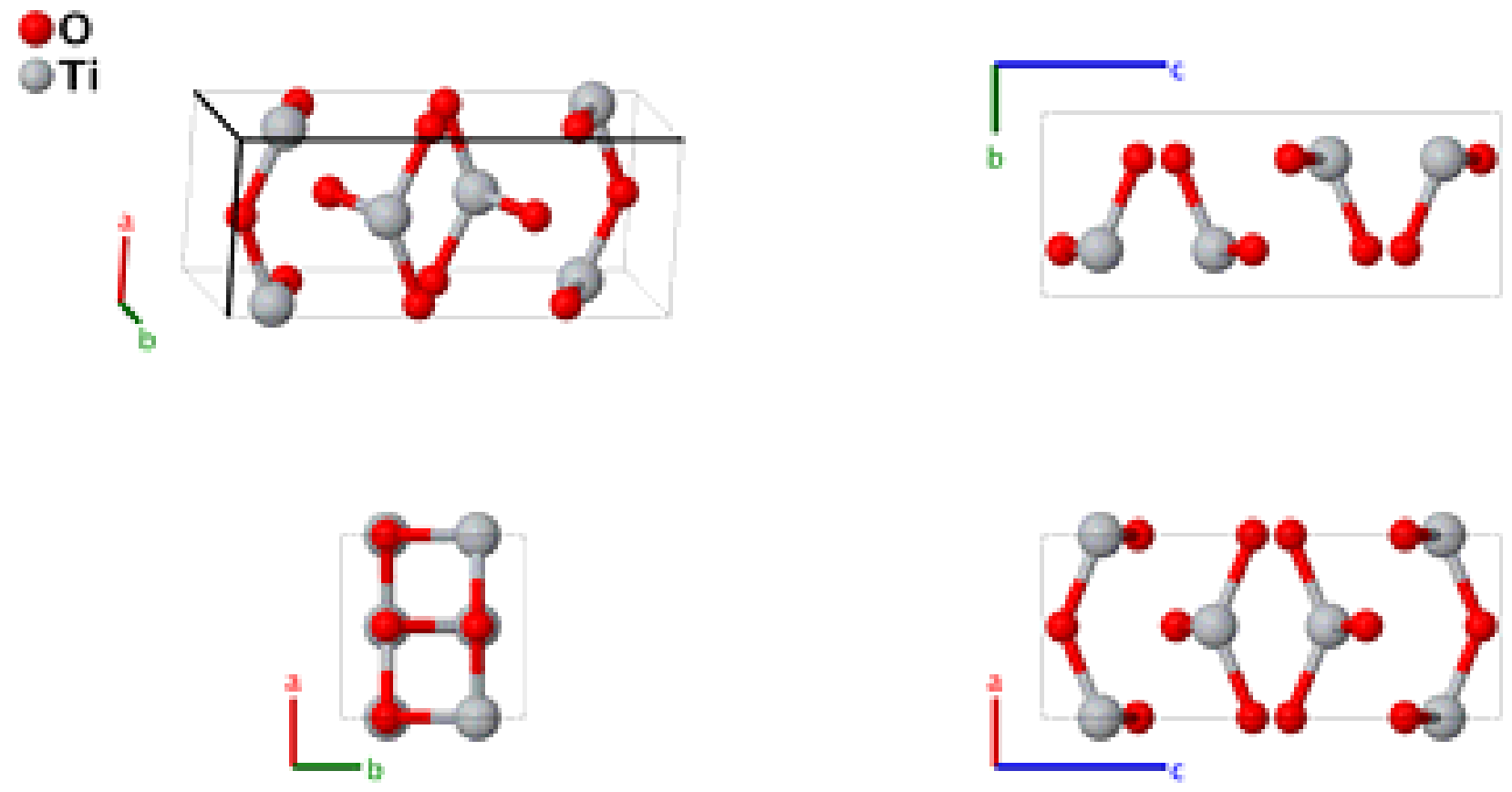

| | | |
|---|---|---|
| **Prototype** | : | $TiO_2$ |
| **AFLOW prototype label** | : | A2B_tI12_141_e_a |
| ***Strukturbericht* designation** | : | C5 |
| **Pearson symbol** | : | tI12 |
| **Space group number** | : | 141 |
| **Space group symbol** | : | $I4_1/amd$ |
| **AFLOW prototype command** | : | `aflow --proto=A2B_tI12_141_e_a`<br>`--params=`$a, c/a, z_2$ |

**Body-centered Tetragonal primitive vectors:**

$$\mathbf{a}_1 = -\frac{1}{2}\,a\,\hat{\mathbf{x}} + \frac{1}{2}\,a\,\hat{\mathbf{y}} + \frac{1}{2}\,c\,\hat{\mathbf{z}}$$

$$\mathbf{a}_2 = \frac{1}{2}\,a\,\hat{\mathbf{x}} - \frac{1}{2}\,a\,\hat{\mathbf{y}} + \frac{1}{2}\,c\,\hat{\mathbf{z}}$$

$$\mathbf{a}_3 = \frac{1}{2}\,a\,\hat{\mathbf{x}} + \frac{1}{2}\,a\,\hat{\mathbf{y}} - \frac{1}{2}\,c\,\hat{\mathbf{z}}$$

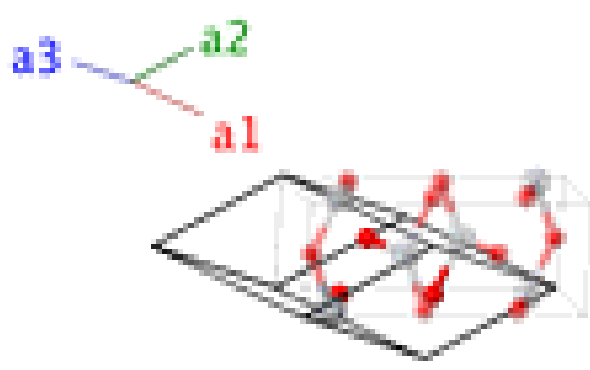

**Basis vectors:**

| | | Lattice Coordinates | | Cartesian Coordinates | Wyckoff Position | Atom Type |
|---|---|---|---|---|---|---|
| $\mathbf{B}_1$ | = | $\frac{7}{8}\,\mathbf{a}_1 + \frac{1}{8}\,\mathbf{a}_2 + \frac{3}{4}\,\mathbf{a}_3$ | = | $\frac{3}{4}\,a\,\hat{\mathbf{y}} + \frac{1}{8}\,c\,\hat{\mathbf{z}}$ | (4$a$) | Ti |
| $\mathbf{B}_2$ | = | $\frac{1}{8}\,\mathbf{a}_1 + \frac{7}{8}\,\mathbf{a}_2 + \frac{1}{4}\,\mathbf{a}_3$ | = | $\frac{1}{2}\,a\,\hat{\mathbf{x}} + \frac{3}{4}\,a\,\hat{\mathbf{y}} + \frac{3}{8}\,c\,\hat{\mathbf{z}}$ | (4$a$) | Ti |
| $\mathbf{B}_3$ | = | $\left(\frac{1}{4} + z_2\right)\,\mathbf{a}_1 + z_2\,\mathbf{a}_2 + \frac{1}{4}\,\mathbf{a}_3$ | = | $\frac{1}{4}\,a\,\hat{\mathbf{y}} + z_2\,c\,\hat{\mathbf{z}}$ | (8$e$) | O |
| $\mathbf{B}_4$ | = | $z_2\,\mathbf{a}_1 + \left(\frac{1}{4} + z_2\right)\,\mathbf{a}_2 + \frac{3}{4}\,\mathbf{a}_3$ | = | $\frac{1}{2}\,a\,\hat{\mathbf{x}} + \frac{1}{4}\,a\,\hat{\mathbf{y}} + \left(\frac{3}{4} + z_2\right)\,c\,\hat{\mathbf{z}}$ | (8$e$) | O |
| $\mathbf{B}_5$ | = | $\left(\frac{3}{4} - z_2\right)\,\mathbf{a}_1 - z_2\,\mathbf{a}_2 + \frac{3}{4}\,\mathbf{a}_3$ | = | $\frac{3}{4}\,a\,\hat{\mathbf{y}} - z_2\,c\,\hat{\mathbf{z}}$ | (8$e$) | O |
| $\mathbf{B}_6$ | = | $-z_2\,\mathbf{a}_1 + \left(\frac{3}{4} - z_2\right)\,\mathbf{a}_2 + \frac{1}{4}\,\mathbf{a}_3$ | = | $\frac{1}{2}\,a\,\hat{\mathbf{x}} + \frac{3}{4}\,a\,\hat{\mathbf{y}} + \left(\frac{1}{4} - z_2\right)\,c\,\hat{\mathbf{z}}$ | (8$e$) | O |

**Geometry files:**
- CIF: pp. 709
- POSCAR: pp. 710

# MoB ($B_g$) Structure: AB_tI16_141_e_e

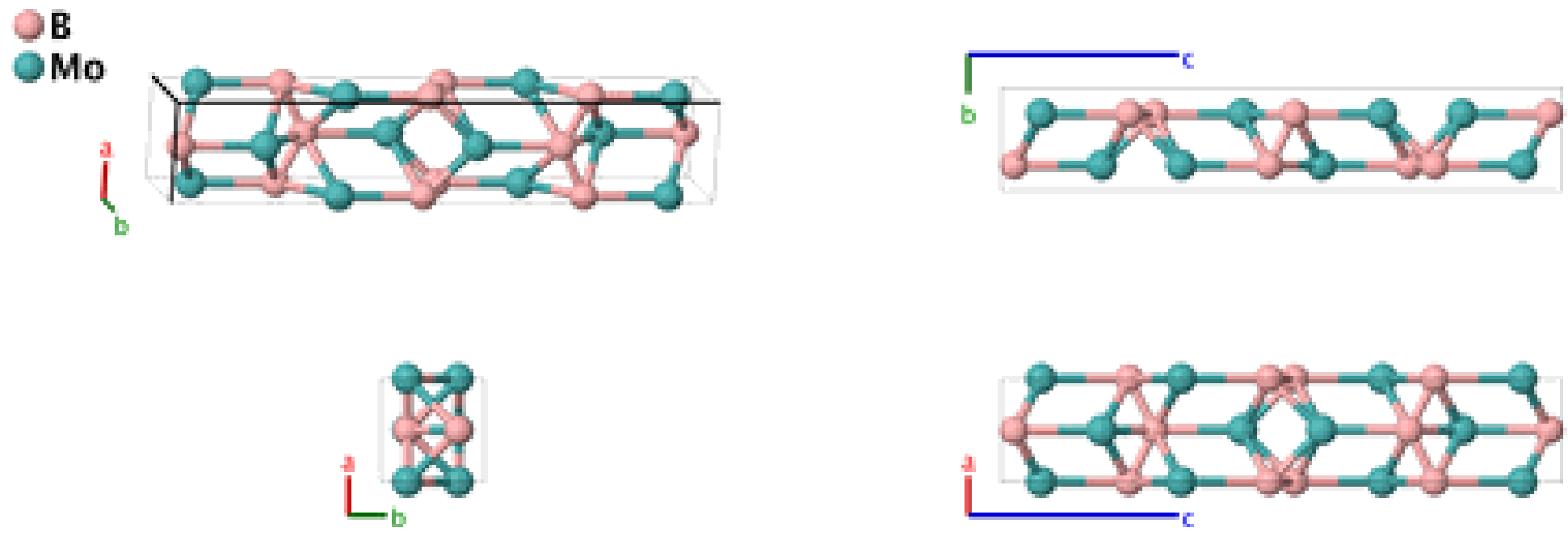

| | | |
|---|---|---|
| **Prototype** | : | MoB |
| **AFLOW prototype label** | : | AB_tI16_141_e_e |
| ***Strukturbericht* designation** | : | $B_g$ |
| **Pearson symbol** | : | tI16 |
| **Space group number** | : | 141 |
| **Space group symbol** | : | $I4_1/amd$ |
| **AFLOW prototype command** | : | `aflow --proto=AB_tI16_141_e_e --params=`$a, c/a, z_1, z_2$ |

**Other compounds with this structure:**

- BCr, GaZr, $B_5Re_3V_2$, $Co_3Er_5Ni_2$, $Ga_3Hf_2Sc$

**Body-centered Tetragonal primitive vectors:**

$$\mathbf{a}_1 = -\frac{1}{2}\,a\,\hat{\mathbf{x}} + \frac{1}{2}\,a\,\hat{\mathbf{y}} + \frac{1}{2}\,c\,\hat{\mathbf{z}}$$
$$\mathbf{a}_2 = \frac{1}{2}\,a\,\hat{\mathbf{x}} - \frac{1}{2}\,a\,\hat{\mathbf{y}} + \frac{1}{2}\,c\,\hat{\mathbf{z}}$$
$$\mathbf{a}_3 = \frac{1}{2}\,a\,\hat{\mathbf{x}} + \frac{1}{2}\,a\,\hat{\mathbf{y}} - \frac{1}{2}\,c\,\hat{\mathbf{z}}$$

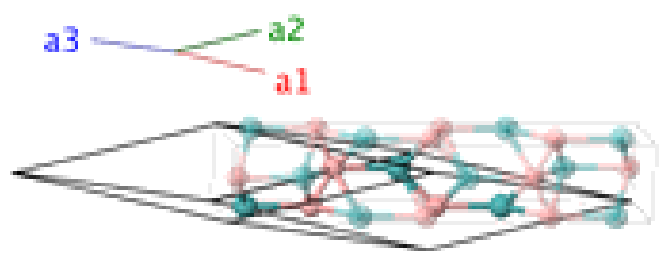

**Basis vectors:**

| | | Lattice Coordinates | | Cartesian Coordinates | Wyckoff Position | Atom Type |
|---|---|---|---|---|---|---|
| $\mathbf{B}_1$ | = | $\left(\frac{1}{4}+z_1\right)\mathbf{a}_1 + z_1\,\mathbf{a}_2 + \frac{1}{4}\,\mathbf{a}_3$ | = | $\frac{1}{4}\,a\,\hat{\mathbf{y}} + z_1\,c\,\hat{\mathbf{z}}$ | (8$e$) | B |
| $\mathbf{B}_2$ | = | $z_1\,\mathbf{a}_1 + \left(\frac{1}{4}+z_1\right)\mathbf{a}_2 + \frac{3}{4}\,\mathbf{a}_3$ | = | $\frac{1}{2}\,a\,\hat{\mathbf{x}} + \frac{1}{4}\,a\,\hat{\mathbf{y}} + \left(\frac{3}{4}+z_1\right)c\,\hat{\mathbf{z}}$ | (8$e$) | B |
| $\mathbf{B}_3$ | = | $\left(\frac{3}{4}-z_1\right)\mathbf{a}_1 - z_1\,\mathbf{a}_2 + \frac{3}{4}\,\mathbf{a}_3$ | = | $\frac{3}{4}\,a\,\hat{\mathbf{y}} - z_1\,c\,\hat{\mathbf{z}}$ | (8$e$) | B |
| $\mathbf{B}_4$ | = | $-z_1\,\mathbf{a}_1 + \left(\frac{3}{4}-z_1\right)\mathbf{a}_2 + \frac{1}{4}\,\mathbf{a}_3$ | = | $\frac{1}{2}\,a\,\hat{\mathbf{x}} + \frac{3}{4}\,a\,\hat{\mathbf{y}} + \left(\frac{1}{4}-z_1\right)c\,\hat{\mathbf{z}}$ | (8$e$) | B |
| $\mathbf{B}_5$ | = | $\left(\frac{1}{4}+z_2\right)\mathbf{a}_1 + z_2\,\mathbf{a}_2 + \frac{1}{4}\,\mathbf{a}_3$ | = | $\frac{1}{4}\,a\,\hat{\mathbf{y}} + z_2\,c\,\hat{\mathbf{z}}$ | (8$e$) | Mo |
| $\mathbf{B}_6$ | = | $z_2\,\mathbf{a}_1 + \left(\frac{1}{4}+z_2\right)\mathbf{a}_2 + \frac{3}{4}\,\mathbf{a}_3$ | = | $\frac{1}{2}\,a\,\hat{\mathbf{x}} + \frac{1}{4}\,a\,\hat{\mathbf{y}} + \left(\frac{3}{4}+z_2\right)c\,\hat{\mathbf{z}}$ | (8$e$) | Mo |
| $\mathbf{B}_7$ | = | $\left(\frac{3}{4}-z_2\right)\mathbf{a}_1 - z_2\,\mathbf{a}_2 + \frac{3}{4}\,\mathbf{a}_3$ | = | $\frac{3}{4}\,a\,\hat{\mathbf{y}} - z_2\,c\,\hat{\mathbf{z}}$ | (8$e$) | Mo |
| $\mathbf{B}_8$ | = | $-z_2\,\mathbf{a}_1 + \left(\frac{3}{4}-z_2\right)\mathbf{a}_2 + \frac{1}{4}\,\mathbf{a}_3$ | = | $\frac{1}{2}\,a\,\hat{\mathbf{x}} + \frac{3}{4}\,a\,\hat{\mathbf{y}} + \left(\frac{1}{4}-z_2\right)c\,\hat{\mathbf{z}}$ | (8$e$) | Mo |

**Geometry files:**

- CIF: pp. 710
- POSCAR: pp. 710

# $Ga_2Hf$ Structure: A2B_tI24_141_2e_e

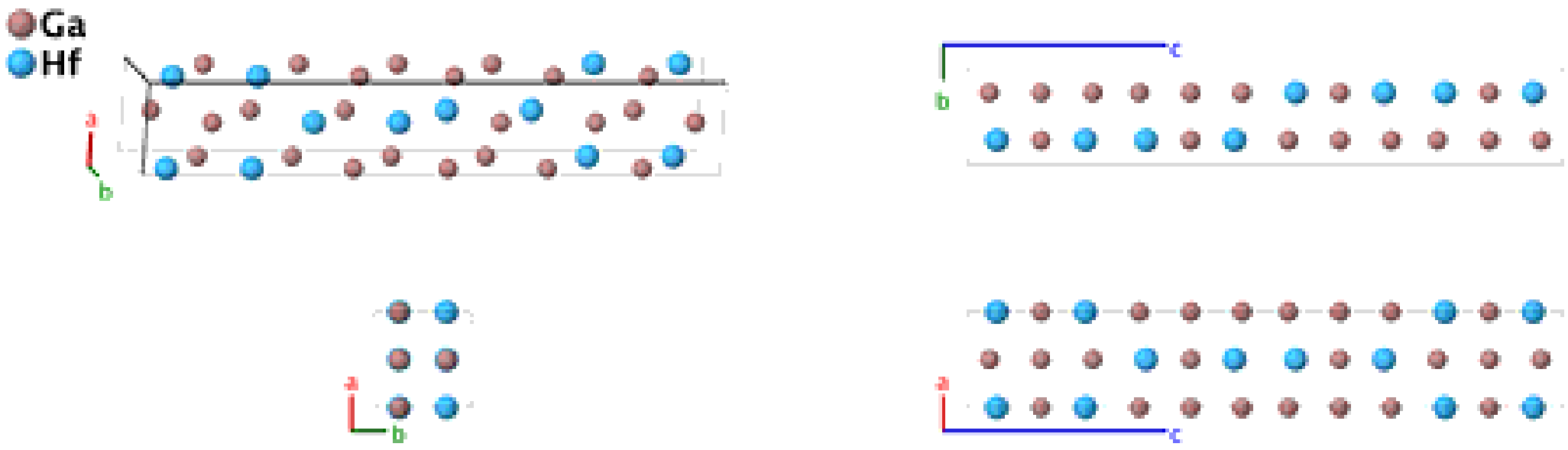

| | | |
|---|---|---|
| **Prototype** | : | $Ga_2Hf$ |
| **AFLOW prototype label** | : | A2B_tI24_141_2e_e |
| ***Strukturbericht* designation** | : | None |
| **Pearson symbol** | : | tI24 |
| **Space group number** | : | 141 |
| **Space group symbol** | : | $I4_1/amd$ |
| **AFLOW prototype command** | : | `aflow --proto=A2B_tI24_141_2e_e`<br>`--params=`$a, c/a, z_1, z_2, z_3$ |

**Other compounds with this structure:**

- $Al_2Mg$, $Al_2Ti$, $Ga_2Ti$, $In_2Zr$, $Pb_2Pr$, $Pb_2Pu$, $PuSn_2$

- When $z_1 = 1/4$, $z_2 = 5/12$, and $z_3 = 1/12$, the atoms are on the sites of the indium (A6) lattice. If, in this case, $c = 6a$, the atoms are on the sites of a face-centered cubic lattice, and if $c = a/\sqrt{2}$, the atoms are on the site of a body-centered cubic lattice. This lattice is placed with the face-centered cubic lattices because most known structures have $c$ near $6a$.

**Body-centered Tetragonal primitive vectors:**

$$\mathbf{a}_1 = -\frac{1}{2}\,a\,\hat{\mathbf{x}} + \frac{1}{2}\,a\,\hat{\mathbf{y}} + \frac{1}{2}\,c\,\hat{\mathbf{z}}$$
$$\mathbf{a}_2 = \frac{1}{2}\,a\,\hat{\mathbf{x}} - \frac{1}{2}\,a\,\hat{\mathbf{y}} + \frac{1}{2}\,c\,\hat{\mathbf{z}}$$
$$\mathbf{a}_3 = \frac{1}{2}\,a\,\hat{\mathbf{x}} + \frac{1}{2}\,a\,\hat{\mathbf{y}} - \frac{1}{2}\,c\,\hat{\mathbf{z}}$$

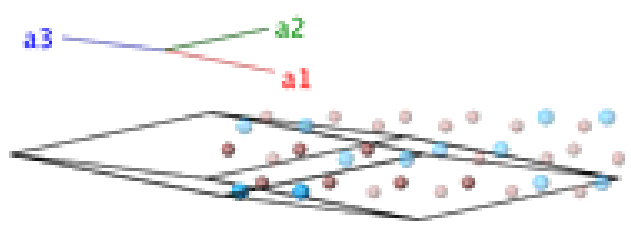

**Basis vectors:**

| | | Lattice Coordinates | | Cartesian Coordinates | Wyckoff Position | Atom Type |
|---|---|---|---|---|---|---|
| $\mathbf{B}_1$ | = | $\left(\frac{1}{4} + z_1\right)\mathbf{a}_1 + z_1\,\mathbf{a}_2 + \frac{1}{4}\,\mathbf{a}_3$ | = | $\frac{1}{4}\,a\,\hat{\mathbf{y}} + z_1\,c\,\hat{\mathbf{z}}$ | (8e) | Ga I |
| $\mathbf{B}_2$ | = | $z_1\,\mathbf{a}_1 + \left(\frac{1}{4} + z_1\right)\mathbf{a}_2 + \frac{3}{4}\,\mathbf{a}_3$ | = | $\frac{1}{2}\,a\,\hat{\mathbf{x}} + \frac{1}{4}\,a\,\hat{\mathbf{y}} + \left(\frac{3}{4} + z_1\right)c\,\hat{\mathbf{z}}$ | (8e) | Ga I |
| $\mathbf{B}_3$ | = | $\left(\frac{3}{4} - z_1\right)\mathbf{a}_1 - z_1\,\mathbf{a}_2 + \frac{3}{4}\,\mathbf{a}_3$ | = | $\frac{3}{4}\,a\,\hat{\mathbf{y}} - z_1\,c\,\hat{\mathbf{z}}$ | (8e) | Ga I |
| $\mathbf{B}_4$ | = | $-z_1\,\mathbf{a}_1 + \left(\frac{3}{4} - z_1\right)\mathbf{a}_2 + \frac{1}{4}\,\mathbf{a}_3$ | = | $\frac{1}{2}\,a\,\hat{\mathbf{x}} + \frac{3}{4}\,a\,\hat{\mathbf{y}} + \left(\frac{1}{4} - z_1\right)c\,\hat{\mathbf{z}}$ | (8e) | Ga I |
| $\mathbf{B}_5$ | = | $\left(\frac{1}{4} + z_2\right)\mathbf{a}_1 + z_2\,\mathbf{a}_2 + \frac{1}{4}\,\mathbf{a}_3$ | = | $\frac{1}{4}\,a\,\hat{\mathbf{y}} + z_2\,c\,\hat{\mathbf{z}}$ | (8e) | Ga II |
| $\mathbf{B}_6$ | = | $z_2\,\mathbf{a}_1 + \left(\frac{1}{4} + z_2\right)\mathbf{a}_2 + \frac{3}{4}\,\mathbf{a}_3$ | = | $\frac{1}{2}\,a\,\hat{\mathbf{x}} + \frac{1}{4}\,a\,\hat{\mathbf{y}} + \left(\frac{3}{4} + z_2\right)c\,\hat{\mathbf{z}}$ | (8e) | Ga II |
| $\mathbf{B}_7$ | = | $\left(\frac{3}{4} - z_2\right)\mathbf{a}_1 - z_2\,\mathbf{a}_2 + \frac{3}{4}\,\mathbf{a}_3$ | = | $\frac{3}{4}\,a\,\hat{\mathbf{y}} - z_2\,c\,\hat{\mathbf{z}}$ | (8e) | Ga II |
| $\mathbf{B}_8$ | = | $-z_2\,\mathbf{a}_1 + \left(\frac{3}{4} - z_2\right)\mathbf{a}_2 + \frac{1}{4}\,\mathbf{a}_3$ | = | $\frac{1}{2}\,a\,\hat{\mathbf{x}} + \frac{3}{4}\,a\,\hat{\mathbf{y}} + \left(\frac{1}{4} - z_2\right)c\,\hat{\mathbf{z}}$ | (8e) | Ga II |
| $\mathbf{B_9}$ | = | $\left(\frac{1}{4} + z_3\right) \mathbf{a_1} + z_3 \mathbf{a_2} + \frac{1}{4} \mathbf{a_3}$ | = | $\frac{1}{4} a \mathbf{\hat{y}} + z_3 c \mathbf{\hat{z}}$ | (8$e$) | Hf |
| $\mathbf{B_{10}}$ | = | $z_3 \mathbf{a_1} + \left(\frac{1}{4} + z_3\right) \mathbf{a_2} + \frac{3}{4} \mathbf{a_3}$ | = | $\frac{1}{2} a \mathbf{\hat{x}} + \frac{1}{4} a \mathbf{\hat{y}} + \left(\frac{3}{4} + z_3\right) c \mathbf{\hat{z}}$ | (8$e$) | Hf |
| $\mathbf{B_{11}}$ | = | $\left(\frac{3}{4} - z_3\right) \mathbf{a_1} - z_3 \mathbf{a_2} + \frac{3}{4} \mathbf{a_3}$ | = | $\frac{3}{4} a \mathbf{\hat{y}} - z_3 c \mathbf{\hat{z}}$ | (8$e$) | Hf |
| $\mathbf{B_{12}}$ | = | $-z_3 \mathbf{a_1} + \left(\frac{3}{4} - z_3\right) \mathbf{a_2} + \frac{1}{4} \mathbf{a_3}$ | = | $\frac{1}{2} a \mathbf{\hat{x}} + \frac{3}{4} a \mathbf{\hat{y}} + \left(\frac{1}{4} - z_3\right) c \mathbf{\hat{z}}$ | (8$e$) | Hf |

---

**Found in:**

---

**Geometry files:**

- CIF: pp. 710
- POSCAR: pp. 711

# NbP ("40") Structure: AB_tI8_141_a_b

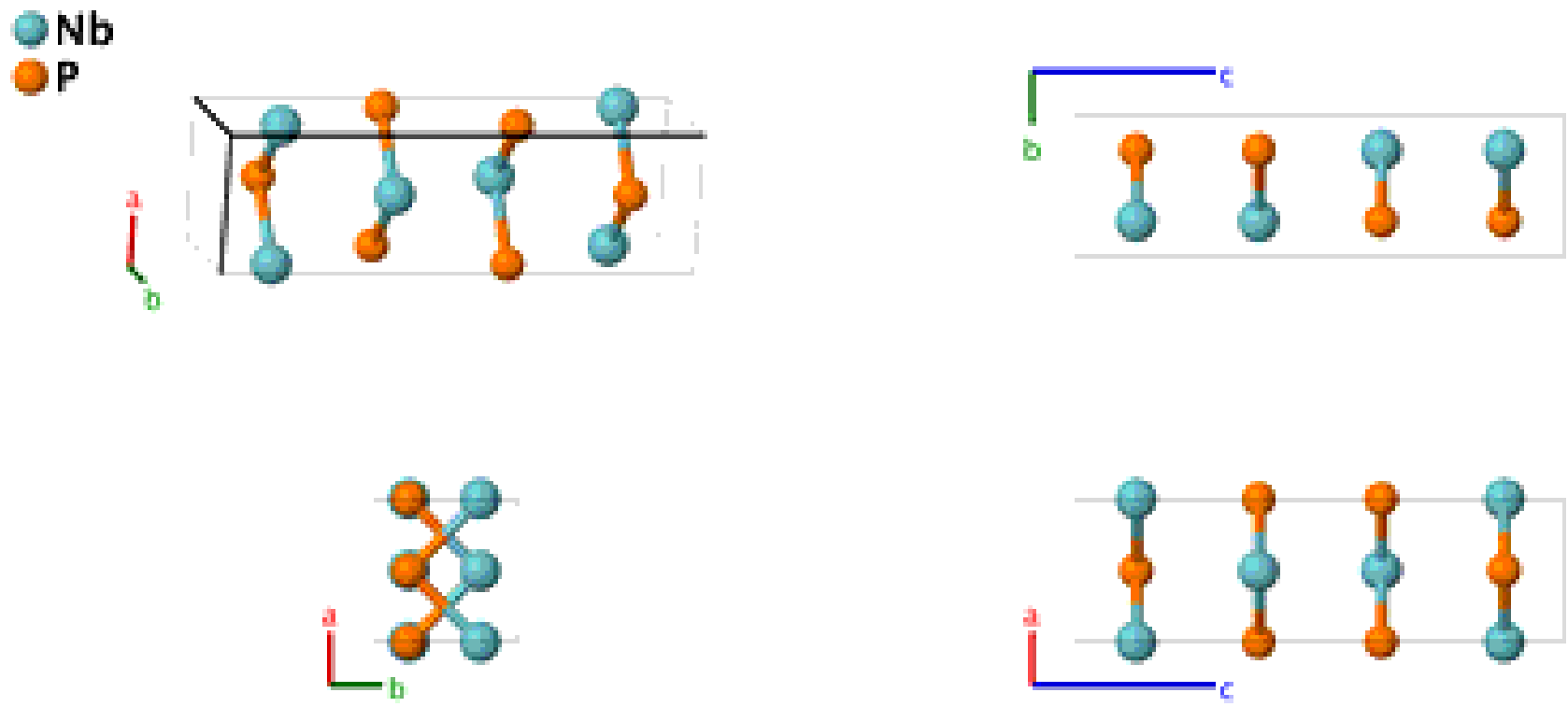

| | | |
|---|---|---|
| **Prototype** | : | NbP |
| **AFLOW prototype label** | : | AB_tI8_141_a_b |
| ***Strukturbericht* designation** | : | "40" |
| **Pearson symbol** | : | tI8 |
| **Space group number** | : | 141 |
| **Space group symbol** | : | $I4_1/amd$ |
| **AFLOW prototype command** | : | `aflow --proto=AB_tI8_141_a_b`<br>`--params=`$a, c/a$ |

- When $c/a = 2$, the atoms in this lattice are on the sites of the face-centered cubic lattice. Thus (Lu, 1991) use this lattice for their structural stability studies, and arbitrarily assign this lattice a Strukturbericht designation of "40". Note that (Schönberg, 1954) gives the space group as $I4_122$, but as (Villars, 1991) notes, the coordinates given increase the symmetry to $I4_1/amd$

**Body-centered Tetragonal primitive vectors:**

$$\mathbf{a}_1 = -\frac{1}{2}\,a\,\mathbf{\hat{x}} + \frac{1}{2}\,a\,\mathbf{\hat{y}} + \frac{1}{2}\,c\,\mathbf{\hat{z}}$$
$$\mathbf{a}_2 = \frac{1}{2}\,a\,\mathbf{\hat{x}} - \frac{1}{2}\,a\,\mathbf{\hat{y}} + \frac{1}{2}\,c\,\mathbf{\hat{z}}$$
$$\mathbf{a}_3 = \frac{1}{2}\,a\,\mathbf{\hat{x}} + \frac{1}{2}\,a\,\mathbf{\hat{y}} - \frac{1}{2}\,c\,\mathbf{\hat{z}}$$

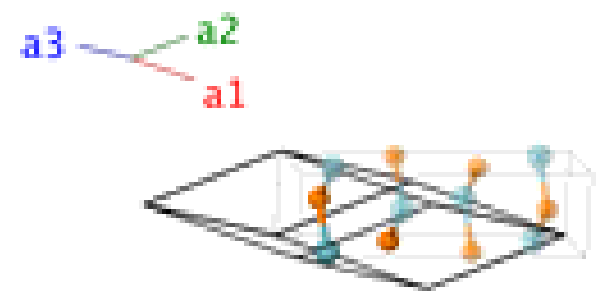

**Basis vectors:**

| | | Lattice Coordinates | | Cartesian Coordinates | Wyckoff Position | Atom Type |
|---|---|---|---|---|---|---|
| $\mathbf{B}_1$ | = | $\frac{7}{8}\,\mathbf{a}_1 + \frac{1}{8}\,\mathbf{a}_2 + \frac{3}{4}\,\mathbf{a}_3$ | = | $\frac{3}{4}\,a\,\mathbf{\hat{y}} + \frac{1}{8}\,c\,\mathbf{\hat{z}}$ | (4*a*) | Nb |
| $\mathbf{B}_2$ | = | $\frac{1}{8}\,\mathbf{a}_1 + \frac{7}{8}\,\mathbf{a}_2 + \frac{1}{4}\,\mathbf{a}_3$ | = | $\frac{1}{2}\,a\,\mathbf{\hat{x}} + \frac{3}{4}\,a\,\mathbf{\hat{y}} + \frac{3}{8}\,c\,\mathbf{\hat{z}}$ | (4*a*) | Nb |
| $\mathbf{B}_3$ | = | $\frac{5}{8}\,\mathbf{a}_1 + \frac{3}{8}\,\mathbf{a}_2 + \frac{1}{4}\,\mathbf{a}_3$ | = | $\frac{1}{4}\,a\,\mathbf{\hat{y}} + \frac{3}{8}\,c\,\mathbf{\hat{z}}$ | (4*b*) | P |
| $\mathbf{B}_4$ | = | $\frac{3}{8}\,\mathbf{a}_1 + \frac{5}{8}\,\mathbf{a}_2 + \frac{3}{4}\,\mathbf{a}_3$ | = | $\frac{1}{2}\,a\,\mathbf{\hat{x}} + \frac{1}{4}\,a\,\mathbf{\hat{y}} + \frac{1}{8}\,c\,\mathbf{\hat{z}}$ | (4*b*) | P |

**Found in:**

---

**Geometry files:**

- CIF: pp. 711
- POSCAR: pp. 711

# $\beta$-$In_2S_3$ Crystal Structure: A2B3_tI80_141_ceh_3h

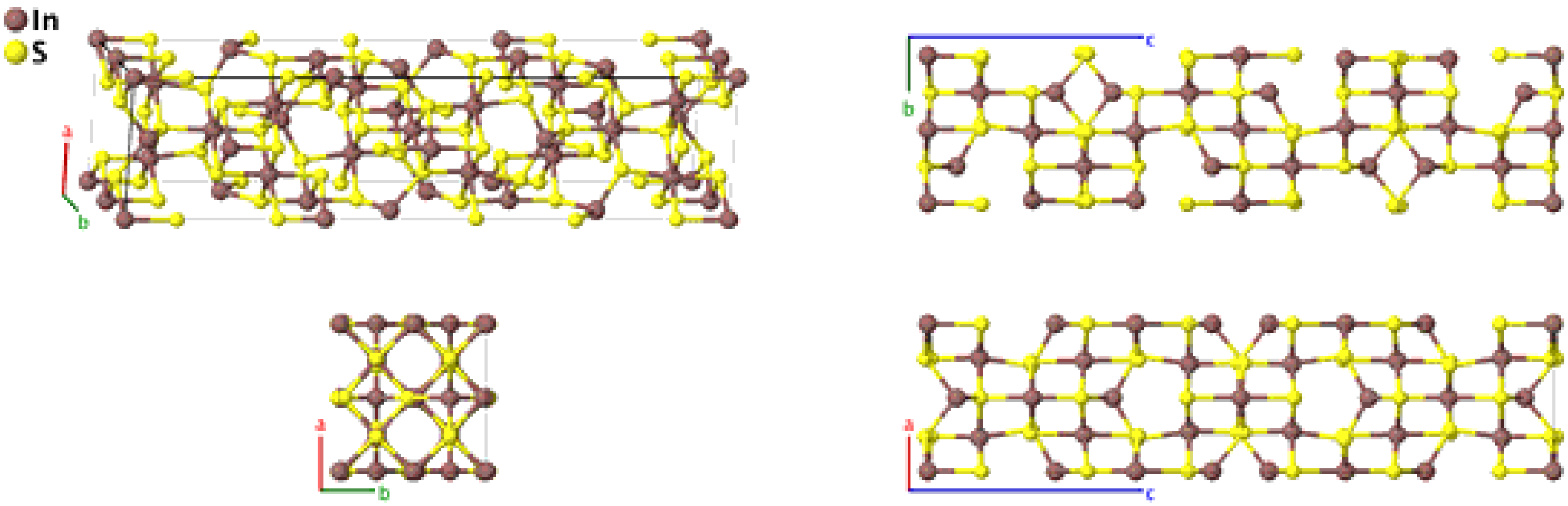

| | | |
|---|---|---|
| **Prototype** | : | $\beta$-$In_2S_3$ |
| **AFLOW prototype label** | : | A2B3_tI80_141_ceh_3h |
| ***Strukturbericht* designation** | : | None |
| **Pearson symbol** | : | tI80 |
| **Space group number** | : | 141 |
| **Space group symbol** | : | $I4_1/amd$ |
| **AFLOW prototype command** | : | `aflow --proto=A2B3_tI80_141_ceh_3h --params=`$a, c/a, z_2, y_3, z_3, y_4, z_4, y_5, z_5, y_6, z_6$ |

- This is a spinel structure with ordered defects.

**Body-centered Tetragonal primitive vectors:**

$$\mathbf{a}_1 = -\frac{1}{2}\,a\,\hat{\mathbf{x}} + \frac{1}{2}\,a\,\hat{\mathbf{y}} + \frac{1}{2}\,c\,\hat{\mathbf{z}}$$

$$\mathbf{a}_2 = \frac{1}{2}\,a\,\hat{\mathbf{x}} - \frac{1}{2}\,a\,\hat{\mathbf{y}} + \frac{1}{2}\,c\,\hat{\mathbf{z}}$$

$$\mathbf{a}_3 = \frac{1}{2}\,a\,\hat{\mathbf{x}} + \frac{1}{2}\,a\,\hat{\mathbf{y}} - \frac{1}{2}\,c\,\hat{\mathbf{z}}$$

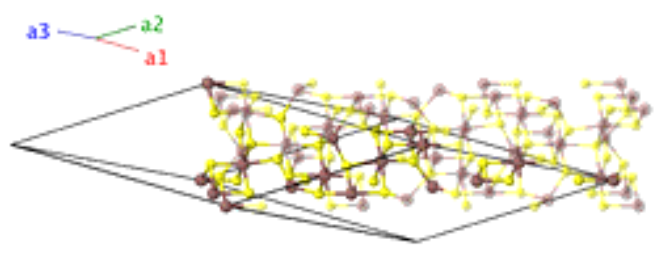

**Basis vectors:**

| | | Lattice Coordinates | | Cartesian Coordinates | Wyckoff Position | Atom Type |
|---|---|---|---|---|---|---|
| $\mathbf{B}_1$ | = | $0\,\mathbf{a}_1 + 0\,\mathbf{a}_2 + 0\,\mathbf{a}_3$ | = | $0\hat{\mathbf{x}} + 0\hat{\mathbf{y}} + 0\hat{\mathbf{z}}$ | (8$c$) | In I |
| $\mathbf{B}_2$ | = | $\frac{1}{2}\,\mathbf{a}_1 + \frac{1}{2}\,\mathbf{a}_3$ | = | $\frac{1}{2}\,a\,\hat{\mathbf{y}}$ | (8$c$) | In I |
| $\mathbf{B}_3$ | = | $\frac{1}{2}\,\mathbf{a}_2$ | = | $\frac{1}{4}\,a\,\hat{\mathbf{x}} + \frac{3}{4}\,a\,\hat{\mathbf{y}} + \frac{1}{4}\,c\,\hat{\mathbf{z}}$ | (8$c$) | In I |
| $\mathbf{B}_4$ | = | $\frac{1}{2}\,\mathbf{a}_3$ | = | $\frac{1}{4}\,a\,\hat{\mathbf{x}} + \frac{1}{4}\,a\,\hat{\mathbf{y}} + \frac{3}{4}\,c\,\hat{\mathbf{z}}$ | (8$c$) | In I |
| $\mathbf{B}_5$ | = | $\left(\frac{1}{4} + z_2\right)\,\mathbf{a}_1 + z_2\,\mathbf{a}_2 + \frac{1}{4}\,\mathbf{a}_3$ | = | $\frac{1}{4}\,a\,\hat{\mathbf{y}} + z_2\,c\,\hat{\mathbf{z}}$ | (8$e$) | In II |
| $\mathbf{B}_6$ | = | $z_2\,\mathbf{a}_1 + \left(\frac{1}{4} + z_2\right)\,\mathbf{a}_2 + \frac{3}{4}\,\mathbf{a}_3$ | = | $\frac{1}{2}\,a\,\hat{\mathbf{x}} + \frac{1}{4}\,a\,\hat{\mathbf{y}} + \left(\frac{3}{4} + z_2\right)\,c\,\hat{\mathbf{z}}$ | (8$e$) | In II |
| $\mathbf{B}_7$ | = | $\left(\frac{3}{4} - z_2\right)\,\mathbf{a}_1 - z_2\,\mathbf{a}_2 + \frac{3}{4}\,\mathbf{a}_3$ | = | $\frac{3}{4}\,a\,\hat{\mathbf{y}} - z_2\,c\,\hat{\mathbf{z}}$ | (8$e$) | In II |
| $\mathbf{B}_8$ | = | $-z_2\,\mathbf{a}_1 + \left(\frac{3}{4} - z_2\right)\,\mathbf{a}_2 + \frac{1}{4}\,\mathbf{a}_3$ | = | $\frac{1}{2}\,a\,\hat{\mathbf{x}} + \frac{3}{4}\,a\,\hat{\mathbf{y}} + \left(\frac{1}{4} - z_2\right)\,c\,\hat{\mathbf{z}}$ | (8$e$) | In II |
| $\mathbf{B}_9$ | = | $(y_3 + z_3)\,\mathbf{a}_1 + z_3\,\mathbf{a}_2 + y_3\,\mathbf{a}_3$ | = | $y_3\,a\,\hat{\mathbf{y}} + z_3\,c\,\hat{\mathbf{z}}$ | (16$h$) | In III |
| $\mathbf{B}_{10}$ | = | $\left(\frac{1}{2} - y_3 + z_3\right)\,\mathbf{a}_1 + z_3\,\mathbf{a}_2 + \left(\frac{1}{2} - y_3\right)\,\mathbf{a}_3$ | = | $\left(\frac{1}{2} - y_3\right)\,a\,\hat{\mathbf{y}} + z_3\,c\,\hat{\mathbf{z}}$ | (16$h$) | In III |
| $\mathbf{B_{11}}$ | = | $z_3\,\mathbf{a_1} + \left(\frac{1}{2} - y_3 + z_3\right)\,\mathbf{a_2} - y_3\,\mathbf{a_3}$ | = | $\left(\frac{1}{4} - y_3\right)\,a\,\mathbf{\hat{x}} + \frac{3}{4}\,a\,\mathbf{\hat{y}} + \left(\frac{1}{4} + z_3\right)\,c\,\mathbf{\hat{z}}$ | (16$h$) | In III |
| $\mathbf{B_{12}}$ | = | $z_3\,\mathbf{a_1} + (y_3 + z_3)\,\mathbf{a_2} + \left(\frac{1}{2} + y_3\right)\,\mathbf{a_3}$ | = | $\left(\frac{1}{4} + y_3\right)\,a\,\mathbf{\hat{x}} + \frac{1}{4}\,a\,\mathbf{\hat{y}} + \left(\frac{3}{4} + z_3\right)\,c\,\mathbf{\hat{z}}$ | (16$h$) | In III |
| $\mathbf{B_{13}}$ | = | $\left(\frac{1}{2} + y_3 - z_3\right)\,\mathbf{a_1} - z_3\,\mathbf{a_2} + \left(\frac{1}{2} + y_3\right)\,\mathbf{a_3}$ | = | $\left(\frac{1}{2} + y_3\right)\,a\,\mathbf{\hat{y}} - z_3\,c\,\mathbf{\hat{z}}$ | (16$h$) | In III |
| $\mathbf{B_{14}}$ | = | $-(y_3 + z_3)\,\mathbf{a_1} - z_3\,\mathbf{a_2} - y_3\,\mathbf{a_3}$ | = | $-y_3\,a\,\mathbf{\hat{y}} - z_3\,c\,\mathbf{\hat{z}}$ | (16$h$) | In III |
| $\mathbf{B_{15}}$ | = | $-z_3\,\mathbf{a_1} + \left(\frac{1}{2} + y_3 - z_3\right)\,\mathbf{a_2} + y_3\,\mathbf{a_3}$ | = | $\left(\frac{1}{4} + y_3\right)\,a\,\mathbf{\hat{x}} + \frac{3}{4}\,a\,\mathbf{\hat{y}} + \left(\frac{1}{4} - z_3\right)\,c\,\mathbf{\hat{z}}$ | (16$h$) | In III |
| $\mathbf{B_{16}}$ | = | $-z_3\,\mathbf{a_1} - (y_3 + z_3)\,\mathbf{a_2} + \left(\frac{1}{2} - y_3\right)\,\mathbf{a_3}$ | = | $\left(\frac{1}{4} - y_3\right)\,a\,\mathbf{\hat{x}} + \frac{1}{4}\,a\,\mathbf{\hat{y}} + \left(\frac{3}{4} - z_3\right)\,c\,\mathbf{\hat{z}}$ | (16$h$) | In III |
| $\mathbf{B_{17}}$ | = | $(y_4 + z_4)\,\mathbf{a_1} + z_4\,\mathbf{a_2} + y_4\,\mathbf{a_3}$ | = | $y_4\,a\,\mathbf{\hat{y}} + z_4\,c\,\mathbf{\hat{z}}$ | (16$h$) | S I |
| $\mathbf{B_{18}}$ | = | $\left(\frac{1}{2} - y_4 + z_4\right)\,\mathbf{a_1} + z_4\,\mathbf{a_2} + \left(\frac{1}{2} - y_4\right)\,\mathbf{a_3}$ | = | $\left(\frac{1}{2} - y_4\right)\,a\,\mathbf{\hat{y}} + z_4\,c\,\mathbf{\hat{z}}$ | (16$h$) | S I |
| $\mathbf{B_{19}}$ | = | $z_4\,\mathbf{a_1} + \left(\frac{1}{2} - y_4 + z_4\right)\,\mathbf{a_2} - y_4\,\mathbf{a_3}$ | = | $\left(\frac{1}{4} - y_4\right)\,a\,\mathbf{\hat{x}} + \frac{3}{4}\,a\,\mathbf{\hat{y}} + \left(\frac{1}{4} + z_4\right)\,c\,\mathbf{\hat{z}}$ | (16$h$) | S I |
| $\mathbf{B_{20}}$ | = | $z_4\,\mathbf{a_1} + (y_4 + z_4)\,\mathbf{a_2} + \left(\frac{1}{2} + y_4\right)\,\mathbf{a_3}$ | = | $\left(\frac{1}{4} + y_4\right)\,a\,\mathbf{\hat{x}} + \frac{1}{4}\,a\,\mathbf{\hat{y}} + \left(\frac{3}{4} + z_4\right)\,c\,\mathbf{\hat{z}}$ | (16$h$) | S I |
| $\mathbf{B_{21}}$ | = | $\left(\frac{1}{2} + y_4 - z_4\right)\,\mathbf{a_1} - z_4\,\mathbf{a_2} + \left(\frac{1}{2} + y_4\right)\,\mathbf{a_3}$ | = | $\left(\frac{1}{2} + y_4\right)\,a\,\mathbf{\hat{y}} - z_4\,c\,\mathbf{\hat{z}}$ | (16$h$) | S I |
| $\mathbf{B_{22}}$ | = | $-(y_4 + z_4)\,\mathbf{a_1} - z_4\,\mathbf{a_2} - y_4\,\mathbf{a_3}$ | = | $-y_4\,a\,\mathbf{\hat{y}} - z_4\,c\,\mathbf{\hat{z}}$ | (16$h$) | S I |
| $\mathbf{B_{23}}$ | = | $-z_4\,\mathbf{a_1} + \left(\frac{1}{2} + y_4 - z_4\right)\,\mathbf{a_2} + y_4\,\mathbf{a_3}$ | = | $\left(\frac{1}{4} + y_4\right)\,a\,\mathbf{\hat{x}} + \frac{3}{4}\,a\,\mathbf{\hat{y}} + \left(\frac{1}{4} - z_4\right)\,c\,\mathbf{\hat{z}}$ | (16$h$) | S I |
| $\mathbf{B_{24}}$ | = | $-z_4\,\mathbf{a_1} - (y_4 + z_4)\,\mathbf{a_2} + \left(\frac{1}{2} - y_4\right)\,\mathbf{a_3}$ | = | $\left(\frac{1}{4} - y_4\right)\,a\,\mathbf{\hat{x}} + \frac{1}{4}\,a\,\mathbf{\hat{y}} + \left(\frac{3}{4} - z_4\right)\,c\,\mathbf{\hat{z}}$ | (16$h$) | S I |
| $\mathbf{B_{25}}$ | = | $(y_5 + z_5)\,\mathbf{a_1} + z_5\,\mathbf{a_2} + y_5\,\mathbf{a_3}$ | = | $y_5\,a\,\mathbf{\hat{y}} + z_5\,c\,\mathbf{\hat{z}}$ | (16$h$) | S II |
| $\mathbf{B_{26}}$ | = | $\left(\frac{1}{2} - y_5 + z_5\right)\,\mathbf{a_1} + z_5\,\mathbf{a_2} + \left(\frac{1}{2} - y_5\right)\,\mathbf{a_3}$ | = | $\left(\frac{1}{2} - y_5\right)\,a\,\mathbf{\hat{y}} + z_5\,c\,\mathbf{\hat{z}}$ | (16$h$) | S II |
| $\mathbf{B_{27}}$ | = | $z_5\,\mathbf{a_1} + \left(\frac{1}{2} - y_5 + z_5\right)\,\mathbf{a_2} - y_5\,\mathbf{a_3}$ | = | $\left(\frac{1}{4} - y_5\right)\,a\,\mathbf{\hat{x}} + \frac{3}{4}\,a\,\mathbf{\hat{y}} + \left(\frac{1}{4} + z_5\right)\,c\,\mathbf{\hat{z}}$ | (16$h$) | S II |
| $\mathbf{B_{28}}$ | = | $z_5\,\mathbf{a_1} + (y_5 + z_5)\,\mathbf{a_2} + \left(\frac{1}{2} + y_5\right)\,\mathbf{a_3}$ | = | $\left(\frac{1}{4} + y_5\right)\,a\,\mathbf{\hat{x}} + \frac{1}{4}\,a\,\mathbf{\hat{y}} + \left(\frac{3}{4} + z_5\right)\,c\,\mathbf{\hat{z}}$ | (16$h$) | S II |
| $\mathbf{B_{29}}$ | = | $\left(\frac{1}{2} + y_5 - z_5\right)\,\mathbf{a_1} - z_5\,\mathbf{a_2} + \left(\frac{1}{2} + y_5\right)\,\mathbf{a_3}$ | = | $\left(\frac{1}{2} + y_5\right)\,a\,\mathbf{\hat{y}} - z_5\,c\,\mathbf{\hat{z}}$ | (16$h$) | S II |
| $\mathbf{B_{30}}$ | = | $-(y_5 + z_5)\,\mathbf{a_1} - z_5\,\mathbf{a_2} - y_5\,\mathbf{a_3}$ | = | $-y_5\,a\,\mathbf{\hat{y}} - z_5\,c\,\mathbf{\hat{z}}$ | (16$h$) | S II |
| $\mathbf{B_{31}}$ | = | $-z_5\,\mathbf{a_1} + \left(\frac{1}{2} + y_5 - z_5\right)\,\mathbf{a_2} + y_5\,\mathbf{a_3}$ | = | $\left(\frac{1}{4} + y_5\right)\,a\,\mathbf{\hat{x}} + \frac{3}{4}\,a\,\mathbf{\hat{y}} + \left(\frac{1}{4} - z_5\right)\,c\,\mathbf{\hat{z}}$ | (16$h$) | S II |
| $\mathbf{B_{32}}$ | = | $-z_5\,\mathbf{a_1} - (y_5 + z_5)\,\mathbf{a_2} + \left(\frac{1}{2} - y_5\right)\,\mathbf{a_3}$ | = | $\left(\frac{1}{4} - y_5\right)\,a\,\mathbf{\hat{x}} + \frac{1}{4}\,a\,\mathbf{\hat{y}} + \left(\frac{3}{4} - z_5\right)\,c\,\mathbf{\hat{z}}$ | (16$h$) | S II |
| $\mathbf{B_{33}}$ | = | $(y_6 + z_6)\,\mathbf{a_1} + z_6\,\mathbf{a_2} + y_6\,\mathbf{a_3}$ | = | $y_6\,a\,\mathbf{\hat{y}} + z_6\,c\,\mathbf{\hat{z}}$ | (16$h$) | S III |
| $\mathbf{B_{34}}$ | = | $\left(\frac{1}{2} - y_6 + z_6\right)\,\mathbf{a_1} + z_6\,\mathbf{a_2} + \left(\frac{1}{2} - y_6\right)\,\mathbf{a_3}$ | = | $\left(\frac{1}{2} - y_6\right)\,a\,\mathbf{\hat{y}} + z_6\,c\,\mathbf{\hat{z}}$ | (16$h$) | S III |
| $\mathbf{B_{35}}$ | = | $z_6\,\mathbf{a_1} + \left(\frac{1}{2} - y_6 + z_6\right)\,\mathbf{a_2} - y_6\,\mathbf{a_3}$ | = | $\left(\frac{1}{4} - y_6\right)\,a\,\mathbf{\hat{x}} + \frac{3}{4}\,a\,\mathbf{\hat{y}} + \left(\frac{1}{4} + z_6\right)\,c\,\mathbf{\hat{z}}$ | (16$h$) | S III |
| $\mathbf{B_{36}}$ | = | $z_6\,\mathbf{a_1} + (y_6 + z_6)\,\mathbf{a_2} + \left(\frac{1}{2} + y_6\right)\,\mathbf{a_3}$ | = | $\left(\frac{1}{4} + y_6\right)\,a\,\mathbf{\hat{x}} + \frac{1}{4}\,a\,\mathbf{\hat{y}} + \left(\frac{3}{4} + z_6\right)\,c\,\mathbf{\hat{z}}$ | (16$h$) | S III |
| $\mathbf{B_{37}}$ | = | $\left(\frac{1}{2} + y_6 - z_6\right)\,\mathbf{a_1} - z_6\,\mathbf{a_2} + \left(\frac{1}{2} + y_6\right)\,\mathbf{a_3}$ | = | $\left(\frac{1}{2} + y_6\right)\,a\,\mathbf{\hat{y}} - z_6\,c\,\mathbf{\hat{z}}$ | (16$h$) | S III |
| $\mathbf{B_{38}}$ | = | $-(y_6 + z_6)\,\mathbf{a_1} - z_6\,\mathbf{a_2} - y_6\,\mathbf{a_3}$ | = | $-y_6\,a\,\mathbf{\hat{y}} - z_6\,c\,\mathbf{\hat{z}}$ | (16$h$) | S III |
| $\mathbf{B_{39}}$ | = | $-z_6\,\mathbf{a_1} + \left(\frac{1}{2} + y_6 - z_6\right)\,\mathbf{a_2} + y_6\,\mathbf{a_3}$ | = | $\left(\frac{1}{4} + y_6\right)\,a\,\mathbf{\hat{x}} + \frac{3}{4}\,a\,\mathbf{\hat{y}} + \left(\frac{1}{4} - z_6\right)\,c\,\mathbf{\hat{z}}$ | (16$h$) | S III |
| $\mathbf{B_{40}}$ | = | $-z_6\,\mathbf{a_1} - (y_6 + z_6)\,\mathbf{a_2} + \left(\frac{1}{2} - y_6\right)\,\mathbf{a_3}$ | = | $\left(\frac{1}{4} - y_6\right)\,a\,\mathbf{\hat{x}} + \frac{1}{4}\,a\,\mathbf{\hat{y}} + \left(\frac{3}{4} - z_6\right)\,c\,\mathbf{\hat{z}}$ | (16$h$) | S III |

**Geometry files:**

- CIF: pp. 711
- POSCAR: pp. 712

# $PPrS_4$ Structure: ABC4_tI96_142_e_ab_2g

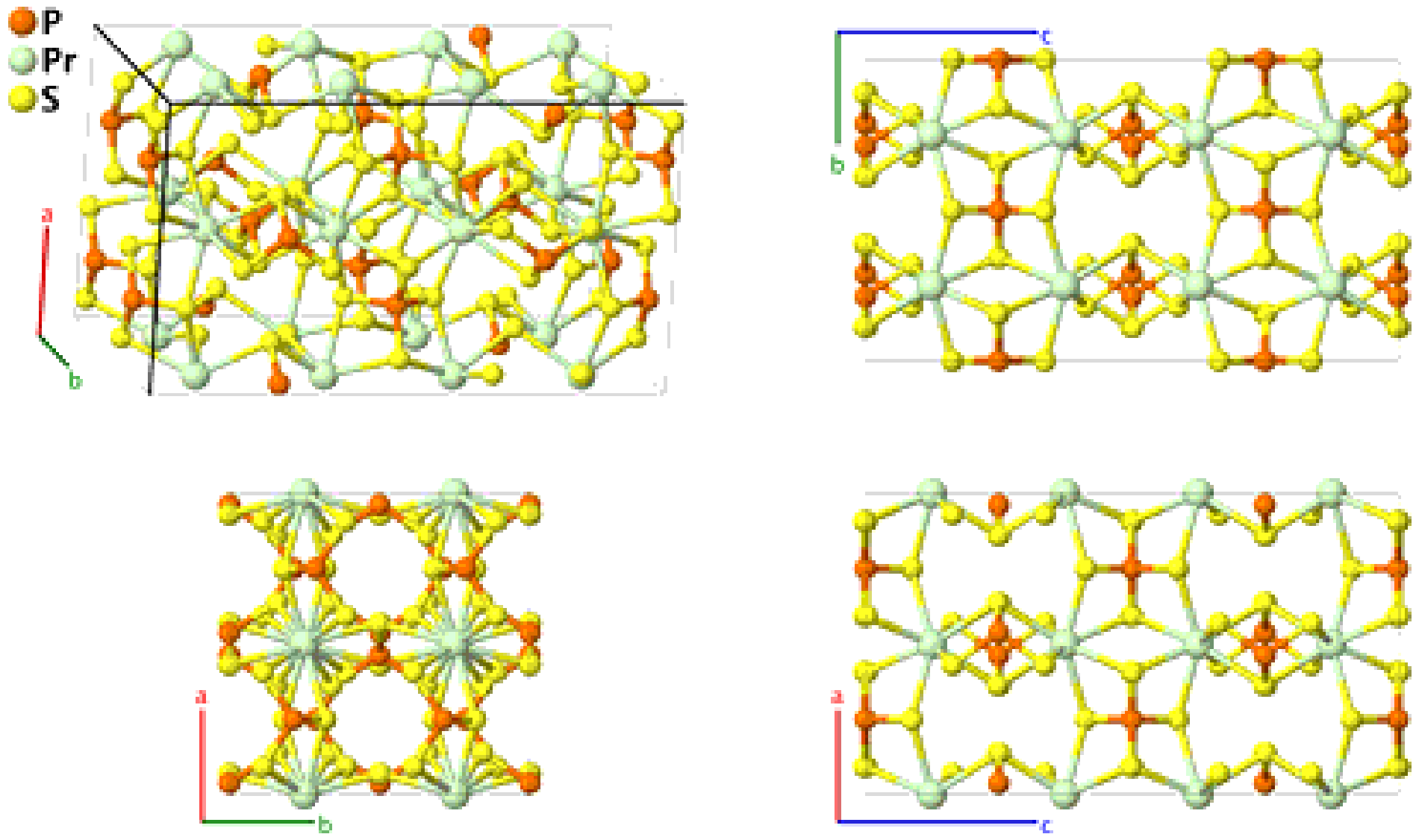

| | | |
|---|---|---|
| **Prototype** | : | $PPrS_4$ |
| **AFLOW prototype label** | : | ABC4_tI96_142_e_ab_2g |
| ***Strukturbericht* designation** | : | None |
| **Pearson symbol** | : | tI96 |
| **Space group number** | : | 142 |
| **Space group symbol** | : | $I4_1/acd$ |
| **AFLOW prototype command** | : | `aflow --proto=ABC4_tI96_142_e_ab_2g`<br>`--params=`$a, c/a, x_3, x_4, y_4, z_4, x_5, y_5, z_5$ |

---

**Body-centered Tetragonal primitive vectors:**

$$\mathbf{a}_1 = -\frac{1}{2}\,a\,\hat{\mathbf{x}} + \frac{1}{2}\,a\,\hat{\mathbf{y}} + \frac{1}{2}\,c\,\hat{\mathbf{z}}$$
$$\mathbf{a}_2 = \frac{1}{2}\,a\,\hat{\mathbf{x}} - \frac{1}{2}\,a\,\hat{\mathbf{y}} + \frac{1}{2}\,c\,\hat{\mathbf{z}}$$
$$\mathbf{a}_3 = \frac{1}{2}\,a\,\hat{\mathbf{x}} + \frac{1}{2}\,a\,\hat{\mathbf{y}} - \frac{1}{2}\,c\,\hat{\mathbf{z}}$$

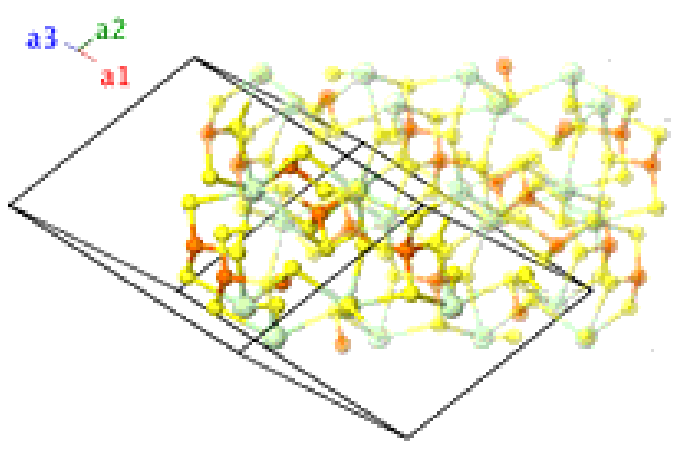

---

**Basis vectors:**

| | | Lattice Coordinates | | Cartesian Coordinates | Wyckoff Position | Atom Type |
|---|---|---|---|---|---|---|
| $\mathbf{B}_1$ | = | $\frac{5}{8}\,\mathbf{a}_1 + \frac{3}{8}\,\mathbf{a}_2 + \frac{1}{4}\,\mathbf{a}_3$ | = | $\frac{1}{4}\,a\,\hat{\mathbf{y}} + \frac{3}{8}\,c\,\hat{\mathbf{z}}$ | (8$a$) | Pr I |
| $\mathbf{B}_2$ | = | $\frac{3}{8}\,\mathbf{a}_1 + \frac{5}{8}\,\mathbf{a}_2 + \frac{3}{4}\,\mathbf{a}_3$ | = | $\frac{1}{2}\,a\,\hat{\mathbf{x}} + \frac{1}{4}\,a\,\hat{\mathbf{y}} + \frac{1}{8}\,c\,\hat{\mathbf{z}}$ | (8$a$) | Pr I |
| $\mathbf{B}_3$ | = | $\frac{7}{8}\,\mathbf{a}_1 + \frac{1}{8}\,\mathbf{a}_2 + \frac{3}{4}\,\mathbf{a}_3$ | = | $\frac{3}{4}\,a\,\hat{\mathbf{y}} + \frac{1}{8}\,c\,\hat{\mathbf{z}}$ | (8$a$) | Pr I |
| $\mathbf{B}_4$ | = | $\frac{1}{8}\,\mathbf{a}_1 + \frac{7}{8}\,\mathbf{a}_2 + \frac{1}{4}\,\mathbf{a}_3$ | = | $\frac{1}{2}\,a\,\hat{\mathbf{x}} + \frac{3}{4}\,a\,\hat{\mathbf{y}} + \frac{3}{8}\,c\,\hat{\mathbf{z}}$ | (8$a$) | Pr I |
| $\mathbf{B}_5$ | = | $\frac{3}{8}\,\mathbf{a}_1 + \frac{1}{8}\,\mathbf{a}_2 + \frac{1}{4}\,\mathbf{a}_3$ | = | $\frac{1}{4}\,a\,\hat{\mathbf{y}} + \frac{1}{8}\,c\,\hat{\mathbf{z}}$ | (8$b$) | Pr II |
| $\mathbf{B}_6$ | = | $\frac{1}{8}\,\mathbf{a}_1 + \frac{3}{8}\,\mathbf{a}_2 + \frac{3}{4}\,\mathbf{a}_3$ | = | $\frac{1}{2}\,a\,\hat{\mathbf{x}} + \frac{1}{4}\,a\,\hat{\mathbf{y}} + \frac{7}{8}\,c\,\hat{\mathbf{z}}$ | (8$b$) | Pr II |
| $\mathbf{B}_7$ | = | $\frac{5}{8}\,\mathbf{a}_1 + \frac{7}{8}\,\mathbf{a}_2 + \frac{3}{4}\,\mathbf{a}_3$ | = | $\frac{1}{2}\,a\,\hat{\mathbf{x}} + \frac{1}{4}\,a\,\hat{\mathbf{y}} + \frac{3}{8}\,c\,\hat{\mathbf{z}}$ | (8$b$) | Pr II |
| $\mathbf{B}_8$ | = | $\frac{7}{8}\,\mathbf{a}_1 + \frac{5}{8}\,\mathbf{a}_2 + \frac{1}{4}\,\mathbf{a}_3$ | = | $\frac{1}{4}\,a\,\hat{\mathbf{y}} + \frac{5}{8}\,c\,\hat{\mathbf{z}}$ | (8$b$) | Pr II |
| $\mathbf{B_{9}}$ | = | $\frac{1}{4}\,\mathbf{a_1} + \left(\frac{1}{4} + x_3\right)\,\mathbf{a_2} + x_3\,\mathbf{a_3}$ | = | $x_3\,a\,\mathbf{\hat{x}} + \frac{1}{4}\,c\,\mathbf{\hat{z}}$ | (16$e$) | P |
| $\mathbf{B_{10}}$ | = | $\frac{3}{4}\,\mathbf{a_1} + \left(\frac{1}{4} - x_3\right)\,\mathbf{a_2} + \left(\frac{1}{2} - x_3\right)\,\mathbf{a_3}$ | = | $-x_3\,a\,\mathbf{\hat{x}} + \frac{1}{2}\,a\,\mathbf{\hat{y}} + \frac{1}{4}\,c\,\mathbf{\hat{z}}$ | (16$e$) | P |
| $\mathbf{B_{11}}$ | = | $\left(\frac{1}{4} + x_3\right)\,\mathbf{a_1} + \frac{3}{4}\,\mathbf{a_2} + x_3\,\mathbf{a_3}$ | = | $\frac{1}{4}\,a\,\mathbf{\hat{x}} + \left(\frac{3}{4} + x_3\right)\,a\,\mathbf{\hat{y}} + \frac{1}{2}\,c\,\mathbf{\hat{z}}$ | (16$e$) | P |
| $\mathbf{B_{12}}$ | = | $\left(\frac{1}{4} - x_3\right)\,\mathbf{a_1} + \frac{1}{4}\,\mathbf{a_2} + \left(\frac{1}{2} - x_3\right)\,\mathbf{a_3}$ | = | $\frac{1}{4}\,a\,\mathbf{\hat{x}} + \left(\frac{1}{4} - x_3\right)\,a\,\mathbf{\hat{y}}$ | (16$e$) | P |
| $\mathbf{B_{13}}$ | = | $\frac{3}{4}\,\mathbf{a_1} + \left(\frac{3}{4} - x_3\right)\,\mathbf{a_2} - x_3\,\mathbf{a_3}$ | = | $-x_3\,a\,\mathbf{\hat{x}} + \frac{3}{4}\,c\,\mathbf{\hat{z}}$ | (16$e$) | P |
| $\mathbf{B_{14}}$ | = | $\frac{1}{4}\,\mathbf{a_1} + \left(\frac{3}{4} + x_3\right)\,\mathbf{a_2} + \left(\frac{1}{2} + x_3\right)\,\mathbf{a_3}$ | = | $\left(\frac{1}{2} + x_3\right)\,a\,\mathbf{\hat{x}} + \frac{1}{4}\,c\,\mathbf{\hat{z}}$ | (16$e$) | P |
| $\mathbf{B_{15}}$ | = | $\left(\frac{3}{4} - x_3\right)\,\mathbf{a_1} + \frac{1}{4}\,\mathbf{a_2} - x_3\,\mathbf{a_3}$ | = | $\frac{3}{4}\,a\,\mathbf{\hat{x}} + \left(\frac{1}{4} - x_3\right)\,a\,\mathbf{\hat{y}} + \frac{1}{2}\,c\,\mathbf{\hat{z}}$ | (16$e$) | P |
| $\mathbf{B_{16}}$ | = | $\left(\frac{3}{4} + x_3\right)\,\mathbf{a_1} + \frac{3}{4}\,\mathbf{a_2} + \left(\frac{1}{2} + x_3\right)\,\mathbf{a_3}$ | = | $\frac{1}{4}\,a\,\mathbf{\hat{x}} + \left(\frac{1}{4} + x_3\right)\,a\,\mathbf{\hat{y}} + \frac{1}{2}\,c\,\mathbf{\hat{z}}$ | (16$e$) | P |
| $\mathbf{B_{17}}$ | = | $(y_4 + z_4)\,\mathbf{a_1} + (x_4 + z_4)\,\mathbf{a_2} + (x_4 + y_4)\,\mathbf{a_3}$ | = | $x_4\,a\,\mathbf{\hat{x}} + y_4\,a\,\mathbf{\hat{y}} + z_4\,c\,\mathbf{\hat{z}}$ | (32$g$) | S I |
| $\mathbf{B_{18}}$ | = | $\left(\frac{1}{2} - y_4 + z_4\right)\,\mathbf{a_1} + (z_4 - x_4)\,\mathbf{a_2} + \left(\frac{1}{2} - x_4 - y_4\right)\,\mathbf{a_3}$ | = | $-x_4\,a\,\mathbf{\hat{x}} + \left(\frac{1}{2} - y_4\right)\,a\,\mathbf{\hat{y}} + z_4\,c\,\mathbf{\hat{z}}$ | (32$g$) | S I |
| $\mathbf{B_{19}}$ | = | $(x_4 + z_4)\,\mathbf{a_1} + \left(\frac{1}{2} - y_4 + z_4\right)\,\mathbf{a_2} + (x_4 - y_4)\,\mathbf{a_3}$ | = | $\left(\frac{1}{4} - y_4\right)\,a\,\mathbf{\hat{x}} + \left(\frac{3}{4} + x_4\right)\,a\,\mathbf{\hat{y}} + \left(\frac{1}{4} + z_4\right)\,c\,\mathbf{\hat{z}}$ | (32$g$) | S I |
| $\mathbf{B_{20}}$ | = | $(z_4 - x_4)\,\mathbf{a_1} + (y_4 + z_4)\,\mathbf{a_2} + \left(\frac{1}{2} - x_4 + y_4\right)\,\mathbf{a_3}$ | = | $\left(\frac{1}{4} + y_4\right)\,a\,\mathbf{\hat{x}} + \left(\frac{1}{4} - x_4\right)\,a\,\mathbf{\hat{y}} + \left(\frac{3}{4} + z_4\right)\,c\,\mathbf{\hat{z}}$ | (32$g$) | S I |
| $\mathbf{B_{21}}$ | = | $(y_4 - z_4)\,\mathbf{a_1} + \left(\frac{1}{2} - x_4 - z_4\right)\,\mathbf{a_2} + \left(\frac{1}{2} - x_4 + y_4\right)\,\mathbf{a_3}$ | = | $\left(\frac{1}{2} - x_4\right)\,a\,\mathbf{\hat{x}} + y_4\,a\,\mathbf{\hat{y}} - z_4\,c\,\mathbf{\hat{z}}$ | (32$g$) | S I |
| $\mathbf{B_{22}}$ | = | $\left(\frac{1}{2} - y_4 - z_4\right)\,\mathbf{a_1} + \left(\frac{1}{2} + x_4 - z_4\right)\,\mathbf{a_2} + (x_4 - y_4)\,\mathbf{a_3}$ | = | $x_4\,a\,\mathbf{\hat{x}} - y_4\,a\,\mathbf{\hat{y}} + \left(\frac{1}{2} - z_4\right)\,c\,\mathbf{\hat{z}}$ | (32$g$) | S I |
| $\mathbf{B_{23}}$ | = | $\left(\frac{1}{2} + x_4 - z_4\right)\,\mathbf{a_1} + (y_4 - z_4)\,\mathbf{a_2} + (x_4 + y_4)\,\mathbf{a_3}$ | = | $\left(\frac{3}{4} + y_4\right)\,a\,\mathbf{\hat{x}} + \left(\frac{1}{4} + x_4\right)\,a\,\mathbf{\hat{y}} + \left(\frac{1}{4} - z_4\right)\,c\,\mathbf{\hat{z}}$ | (32$g$) | S I |
| $\mathbf{B_{24}}$ | = | $\left(\frac{1}{2} - x_4 - z_4\right)\,\mathbf{a_1} + \left(\frac{1}{2} - y_4 - z_4\right)\,\mathbf{a_2} + \left(\frac{1}{2} - x_4 - y_4\right)\,\mathbf{a_3}$ | = | $\left(\frac{1}{4} - y_4\right)\,a\,\mathbf{\hat{x}} + \left(\frac{1}{4} - x_4\right)\,a\,\mathbf{\hat{y}} + \left(\frac{1}{4} - z_4\right)\,c\,\mathbf{\hat{z}}$ | (32$g$) | S I |
| $\mathbf{B_{25}}$ | = | $-(y_4 + z_4)\,\mathbf{a_1} - (x_4 + z_4)\,\mathbf{a_2} - (x_4 + y_4)\,\mathbf{a_3}$ | = | $-x_4\,a\,\mathbf{\hat{x}} - y_4\,a\,\mathbf{\hat{y}} - z_4\,c\,\mathbf{\hat{z}}$ | (32$g$) | S I |
| $\mathbf{B_{26}}$ | = | $\left(\frac{1}{2} + y_4 - z_4\right)\,\mathbf{a_1} + (x_4 - z_4)\,\mathbf{a_2} + \left(\frac{1}{2} + x_4 + y_4\right)\,\mathbf{a_3}$ | = | $x_4\,a\,\mathbf{\hat{x}} + \left(\frac{1}{2} + y_4\right)\,a\,\mathbf{\hat{y}} - z_4\,c\,\mathbf{\hat{z}}$ | (32$g$) | S I |
| $\mathbf{B_{27}}$ | = | $-(x_4 + z_4)\,\mathbf{a_1} + \left(\frac{1}{2} + y_4 - z_4\right)\,\mathbf{a_2} + (y_4 - x_4)\,\mathbf{a_3}$ | = | $\left(\frac{1}{4} + y_4\right)\,a\,\mathbf{\hat{x}} + \left(\frac{3}{4} - x_4\right)\,a\,\mathbf{\hat{y}} + \left(\frac{1}{4} - z_4\right)\,c\,\mathbf{\hat{z}}$ | (32$g$) | S I |
| $\mathbf{B_{28}}$ | = | $(x_4 - z_4)\,\mathbf{a_1} - (y_4 + z_4)\,\mathbf{a_2} + \left(\frac{1}{2} + x_4 - y_4\right)\,\mathbf{a_3}$ | = | $\left(\frac{1}{4} - y_4\right)\,a\,\mathbf{\hat{x}} + \left(\frac{1}{4} + x_4\right)\,a\,\mathbf{\hat{y}} + \left(\frac{3}{4} - z_4\right)\,c\,\mathbf{\hat{z}}$ | (32$g$) | S I |
| $\mathbf{B_{29}}$ | = | $(z_4 - y_4)\,\mathbf{a_1} + \left(\frac{1}{2} + x_4 + z_4\right)\,\mathbf{a_2} + \left(\frac{1}{2} + x_4 - y_4\right)\,\mathbf{a_3}$ | = | $\left(\frac{1}{2} + x_4\right)\,a\,\mathbf{\hat{x}} - y_4\,a\,\mathbf{\hat{y}} + z_4\,c\,\mathbf{\hat{z}}$ | (32$g$) | S I |
| $\mathbf{B_{30}}$ | = | $\left(\frac{1}{2} + y_4 + z_4\right)\,\mathbf{a_1} + \left(\frac{1}{2} - x_4 + z_4\right)\,\mathbf{a_2} + (y_4 - x_4)\,\mathbf{a_3}$ | = | $-x_4\,a\,\mathbf{\hat{x}} + y_4\,a\,\mathbf{\hat{y}} + \left(\frac{1}{2} + z_4\right)\,c\,\mathbf{\hat{z}}$ | (32$g$) | S I |
| $\mathbf{B_{31}}$ | = | $\left(\frac{1}{2} - x_4 + z_4\right)\,\mathbf{a_1} + (z_4 - y_4)\,\mathbf{a_2} - (x_4 + y_4)\,\mathbf{a_3}$ | = | $\left(\frac{3}{4} - y_4\right)\,a\,\mathbf{\hat{x}} + \left(\frac{1}{4} - x_4\right)\,a\,\mathbf{\hat{y}} + \left(\frac{1}{4} + z_4\right)\,c\,\mathbf{\hat{z}}$ | (32$g$) | S I |
| $\mathbf{B_{32}}$ | = | $\left(\frac{1}{2} + x_4 + z_4\right)\,\mathbf{a_1} + \left(\frac{1}{2} + y_4 + z_4\right)\,\mathbf{a_2} + \left(\frac{1}{2} + x_4 + y_4\right)\,\mathbf{a_3}$ | = | $\left(\frac{1}{4} + y_4\right)\,a\,\mathbf{\hat{x}} + \left(\frac{1}{4} + x_4\right)\,a\,\mathbf{\hat{y}} + \left(\frac{1}{4} + z_4\right)\,c\,\mathbf{\hat{z}}$ | (32$g$) | S I |
| $\mathbf{B_{33}}$ | = | $\left(y_5 + z_5\right)\,\mathbf{a_1} + \left(x_5 + z_5\right)\,\mathbf{a_2} + \left(x_5 + y_5\right)\,\mathbf{a_3}$ | = | $x_5\,a\,\mathbf{\hat{x}} + y_5\,a\,\mathbf{\hat{y}} + z_5\,c\,\mathbf{\hat{z}}$ | (32$g$) | S II |
| $\mathbf{B_{34}}$ | = | $\left(\frac{1}{2} - y_5 + z_5\right)\,\mathbf{a_1} + \left(z_5 - x_5\right)\,\mathbf{a_2} + \left(\frac{1}{2} - x_5 - y_5\right)\,\mathbf{a_3}$ | = | $-x_5\,a\,\mathbf{\hat{x}} + \left(\frac{1}{2} - y_5\right)\,a\,\mathbf{\hat{y}} + z_5\,c\,\mathbf{\hat{z}}$ | (32$g$) | S II |
| $\mathbf{B_{35}}$ | = | $\left(x_5 + z_5\right)\,\mathbf{a_1} + \left(\frac{1}{2} - y_5 + z_5\right)\,\mathbf{a_2} + \left(x_5 - y_5\right)\,\mathbf{a_3}$ | = | $\left(\frac{1}{4} - y_5\right)\,a\,\mathbf{\hat{x}} + \left(\frac{3}{4} + x_5\right)\,a\,\mathbf{\hat{y}} + \left(\frac{1}{4} + z_5\right)\,c\,\mathbf{\hat{z}}$ | (32$g$) | S II |
| $\mathbf{B_{36}}$ | = | $\left(z_5 - x_5\right)\,\mathbf{a_1} + \left(y_5 + z_5\right)\,\mathbf{a_2} + \left(\frac{1}{2} - x_5 + y_5\right)\,\mathbf{a_3}$ | = | $\left(\frac{1}{4} + y_5\right)\,a\,\mathbf{\hat{x}} + \left(\frac{1}{4} - x_5\right)\,a\,\mathbf{\hat{y}} + \left(\frac{3}{4} + z_5\right)\,c\,\mathbf{\hat{z}}$ | (32$g$) | S II |
| $\mathbf{B_{37}}$ | = | $\left(y_5 - z_5\right)\,\mathbf{a_1} + \left(\frac{1}{2} - x_5 - z_5\right)\,\mathbf{a_2} + \left(\frac{1}{2} - x_5 + y_5\right)\,\mathbf{a_3}$ | = | $\left(\frac{1}{2} - x_5\right)\,a\,\mathbf{\hat{x}} + y_5\,a\,\mathbf{\hat{y}} - z_5\,c\,\mathbf{\hat{z}}$ | (32$g$) | S II |
| $\mathbf{B_{38}}$ | = | $\left(\frac{1}{2} - y_5 - z_5\right)\,\mathbf{a_1} + \left(\frac{1}{2} + x_5 - z_5\right)\,\mathbf{a_2} + \left(x_5 - y_5\right)\,\mathbf{a_3}$ | = | $x_5\,a\,\mathbf{\hat{x}} - y_5\,a\,\mathbf{\hat{y}} + \left(\frac{1}{2} - z_5\right)\,c\,\mathbf{\hat{z}}$ | (32$g$) | S II |
| $\mathbf{B_{39}}$ | = | $\left(\frac{1}{2} + x_5 - z_5\right)\,\mathbf{a_1} + \left(y_5 - z_5\right)\,\mathbf{a_2} + \left(x_5 + y_5\right)\,\mathbf{a_3}$ | = | $\left(\frac{3}{4} + y_5\right)\,a\,\mathbf{\hat{x}} + \left(\frac{1}{4} + x_5\right)\,a\,\mathbf{\hat{y}} + \left(\frac{1}{4} - z_5\right)\,c\,\mathbf{\hat{z}}$ | (32$g$) | S II |
| $\mathbf{B_{40}}$ | = | $\left(\frac{1}{2} - x_5 - z_5\right)\,\mathbf{a_1} + \left(\frac{1}{2} - y_5 - z_5\right)\,\mathbf{a_2} + \left(\frac{1}{2} - x_5 - y_5\right)\,\mathbf{a_3}$ | = | $\left(\frac{1}{4} - y_5\right)\,a\,\mathbf{\hat{x}} + \left(\frac{1}{4} - x_5\right)\,a\,\mathbf{\hat{y}} + \left(\frac{1}{4} - z_5\right)\,c\,\mathbf{\hat{z}}$ | (32$g$) | S II |
| $\mathbf{B_{41}}$ | = | $-\left(y_5 + z_5\right)\,\mathbf{a_1} - \left(x_5 + z_5\right)\,\mathbf{a_2} - \left(x_5 + y_5\right)\,\mathbf{a_3}$ | = | $-x_5\,a\,\mathbf{\hat{x}} - y_5\,a\,\mathbf{\hat{y}} - z_5\,c\,\mathbf{\hat{z}}$ | (32$g$) | S II |
| $\mathbf{B_{42}}$ | = | $\left(\frac{1}{2} + y_5 - z_5\right)\,\mathbf{a_1} + \left(x_5 - z_5\right)\,\mathbf{a_2} + \left(\frac{1}{2} + x_5 + y_5\right)\,\mathbf{a_3}$ | = | $x_5\,a\,\mathbf{\hat{x}} + \left(\frac{1}{2} + y_5\right)\,a\,\mathbf{\hat{y}} - z_5\,c\,\mathbf{\hat{z}}$ | (32$g$) | S II |
| $\mathbf{B_{43}}$ | = | $-\left(x_5 + z_5\right)\,\mathbf{a_1} + \left(\frac{1}{2} + y_5 - z_5\right)\,\mathbf{a_2} + \left(y_5 - x_5\right)\,\mathbf{a_3}$ | = | $\left(\frac{1}{4} + y_5\right)\,a\,\mathbf{\hat{x}} + \left(\frac{3}{4} - x_5\right)\,a\,\mathbf{\hat{y}} + \left(\frac{1}{4} - z_5\right)\,c\,\mathbf{\hat{z}}$ | (32$g$) | S II |
| $\mathbf{B_{44}}$ | = | $\left(x_5 - z_5\right)\,\mathbf{a_1} - \left(y_5 + z_5\right)\,\mathbf{a_2} + \left(\frac{1}{2} + x_5 - y_5\right)\,\mathbf{a_3}$ | = | $\left(\frac{1}{4} - y_5\right)\,a\,\mathbf{\hat{x}} + \left(\frac{1}{4} + x_5\right)\,a\,\mathbf{\hat{y}} + \left(\frac{3}{4} - z_5\right)\,c\,\mathbf{\hat{z}}$ | (32$g$) | S II |
| $\mathbf{B_{45}}$ | = | $\left(z_5 - y_5\right)\,\mathbf{a_1} + \left(\frac{1}{2} + x_5 + z_5\right)\,\mathbf{a_2} + \left(\frac{1}{2} + x_5 - y_5\right)\,\mathbf{a_3}$ | = | $\left(\frac{1}{2} + x_5\right)\,a\,\mathbf{\hat{x}} - y_5\,a\,\mathbf{\hat{y}} + z_5\,c\,\mathbf{\hat{z}}$ | (32$g$) | S II |
| $\mathbf{B_{46}}$ | = | $\left(\frac{1}{2} + y_5 + z_5\right)\,\mathbf{a_1} + \left(\frac{1}{2} - x_5 + z_5\right)\,\mathbf{a_2} + \left(y_5 - x_5\right)\,\mathbf{a_3}$ | = | $-x_5\,a\,\mathbf{\hat{x}} + y_5\,a\,\mathbf{\hat{y}} + \left(\frac{1}{2} + z_5\right)\,c\,\mathbf{\hat{z}}$ | (32$g$) | S II |
| $\mathbf{B_{47}}$ | = | $\left(\frac{1}{2} - x_5 + z_5\right)\,\mathbf{a_1} + \left(z_5 - y_5\right)\,\mathbf{a_2} - \left(x_5 + y_5\right)\,\mathbf{a_3}$ | = | $\left(\frac{3}{4} - y_5\right)\,a\,\mathbf{\hat{x}} + \left(\frac{1}{4} - x_5\right)\,a\,\mathbf{\hat{y}} + \left(\frac{1}{4} + z_5\right)\,c\,\mathbf{\hat{z}}$ | (32$g$) | S II |
| $\mathbf{B_{48}}$ | = | $\left(\frac{1}{2} + x_5 + z_5\right)\,\mathbf{a_1} + \left(\frac{1}{2} + y_5 + z_5\right)\,\mathbf{a_2} + \left(\frac{1}{2} + x_5 + y_5\right)\,\mathbf{a_3}$ | = | $\left(\frac{1}{4} + y_5\right)\,a\,\mathbf{\hat{x}} + \left(\frac{1}{4} + x_5\right)\,a\,\mathbf{\hat{y}} + \left(\frac{1}{4} + z_5\right)\,c\,\mathbf{\hat{z}}$ | (32$g$) | S II |

**Found in:**

**Geometry files:**

- CIF: pp. 712
- POSCAR: pp. 713

# $\zeta$-AgZn ($B_b$) Structure: A2B_hP9_147_g_ad

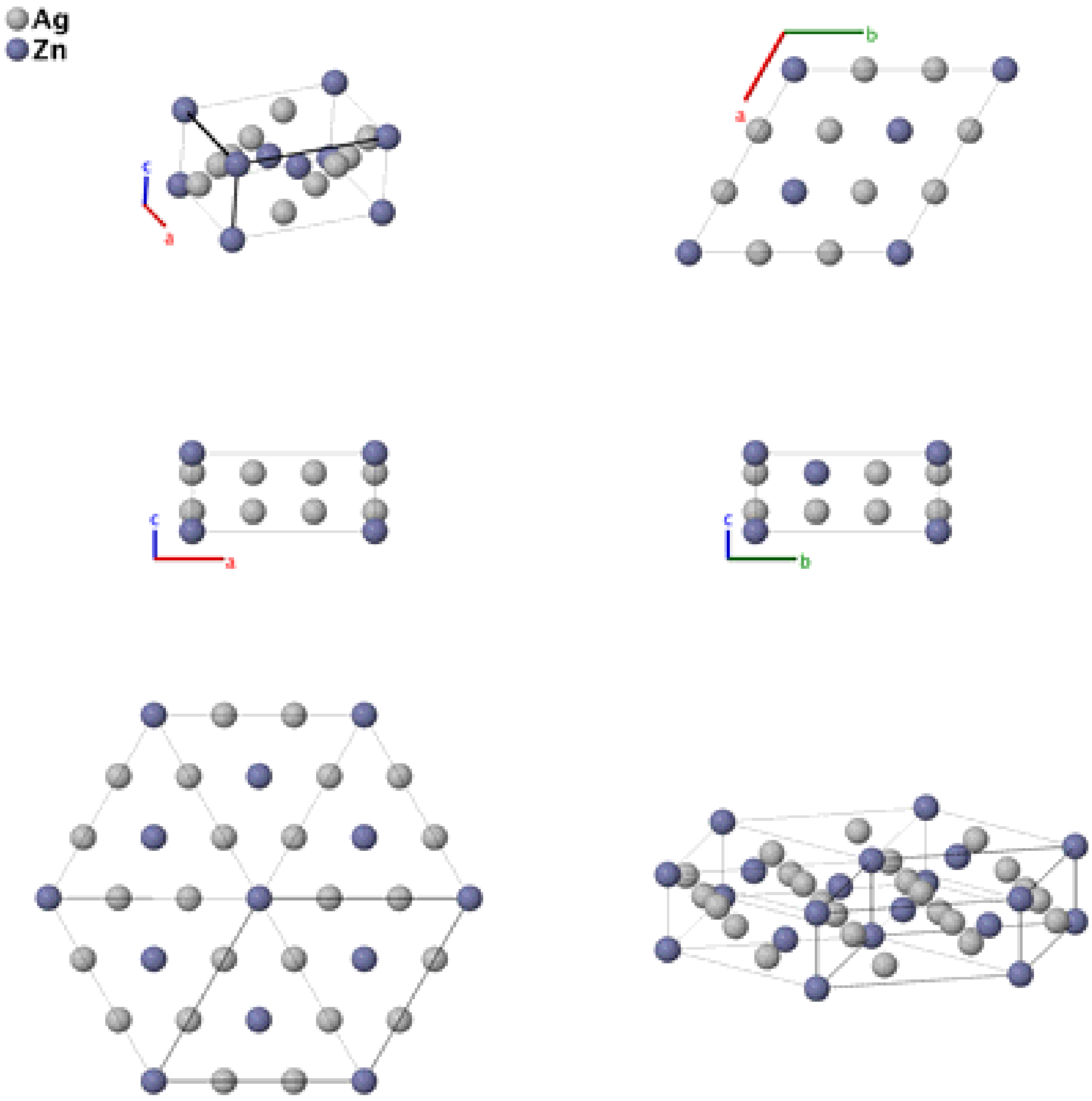

| | | |
|---|---|---|
| **Prototype** | : | $\zeta$-AgZn |
| **AFLOW prototype label** | : | A2B_hP9_147_g_ad |
| ***Strukturbericht* designation** | : | $B_b$ |
| **Pearson symbol** | : | hP9 |
| **Space group number** | : | 147 |
| **Space group symbol** | : | $P\bar{3}$ |
| **AFLOW prototype command** | : | `aflow --proto=A2B_hP9_147_g_ad`<br>`--params=`$a, c/a, z_2, x_3, y_3, z_3$ |

---

**Other compounds with this structure:**

- $Ag_{10}CdZn_9$, $Ag_{50}MgZn_{49}$

---

- When $z_2 = 0$, $x_3 = 1/3$, $y_3 = 0$, and $z_3 = 1/2$, this structure becomes the hexagonal omega (C32) structure. This is an alloy phase. The (1a) and (2d) sites are pure Zn, but the (6g) site is a mixture of Ag and Zn, so we designate it as “M”. If the system is stoichiometric then M = $(Ag_{4.5},Zn_{1.5})$.

**Trigonal Hexagonal primitive vectors:**

$$\mathbf{a}_1 = \frac{1}{2}\, a\,\hat{\mathbf{x}} - \frac{\sqrt{3}}{2}\, a\,\hat{\mathbf{y}}$$
$$\mathbf{a}_2 = \frac{1}{2}\, a\,\hat{\mathbf{x}} + \frac{\sqrt{3}}{2}\, a\,\hat{\mathbf{y}}$$
$$\mathbf{a}_3 = c\,\hat{\mathbf{z}}$$

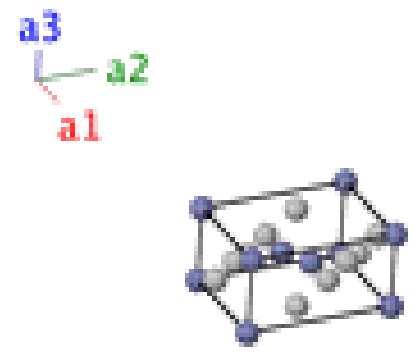

**Basis vectors:**

| | | Lattice Coordinates | | Cartesian Coordinates | Wyckoff Position | Atom Type |
|---|---|---|---|---|---|---|
| $\mathbf{B}_1$ | = | $0\,\mathbf{a}_1 + 0\,\mathbf{a}_2 + 0\,\mathbf{a}_3$ | = | $0\hat{\mathbf{x}} + 0\hat{\mathbf{y}} + 0\hat{\mathbf{z}}$ | (1$a$) | Zn |
| $\mathbf{B}_2$ | = | $\frac{1}{3}\,\mathbf{a}_1 + \frac{2}{3}\,\mathbf{a}_2 + z_2\,\mathbf{a}_3$ | = | $\frac{1}{2}\,a\,\hat{\mathbf{x}} + \frac{1}{2\sqrt{3}}\,a\,\hat{\mathbf{y}} + z_2\,c\,\hat{\mathbf{z}}$ | (2$d$) | Zn |
| $\mathbf{B}_3$ | = | $\frac{2}{3}\,\mathbf{a}_1 + \frac{1}{3}\,\mathbf{a}_2 - z_2\,\mathbf{a}_3$ | = | $\frac{1}{2}\,a\,\hat{\mathbf{x}} - \frac{1}{2\sqrt{3}}\,a\,\hat{\mathbf{y}} - z_2\,c\,\hat{\mathbf{z}}$ | (2$d$) | Zn |
| $\mathbf{B}_4$ | = | $x_3\,\mathbf{a}_1 + y_3\,\mathbf{a}_2 + z_3\,\mathbf{a}_3$ | = | $\frac{1}{2}\,(x_3 + y_3)\,a\,\hat{\mathbf{x}} + \frac{\sqrt{3}}{2}\,(y_3 - x_3)\,a\,\hat{\mathbf{y}} + z_3\,c\,\hat{\mathbf{z}}$ | (6$g$) | M |
| $\mathbf{B}_5$ | = | $-y_3\,\mathbf{a}_1 + (x_3 - y_3)\,\mathbf{a}_2 + z_3\,\mathbf{a}_3$ | = | $\frac{1}{2}\,(x_3 - 2y_3)\,a\,\hat{\mathbf{x}} + \frac{\sqrt{3}}{2}\,x_3\,a\,\hat{\mathbf{y}} + z_3\,c\,\hat{\mathbf{z}}$ | (6$g$) | M |
| $\mathbf{B}_6$ | = | $(y_3 - x_3)\,\mathbf{a}_1 - x_3\,\mathbf{a}_2 + z_3\,\mathbf{a}_3$ | = | $\frac{1}{2}\,(y_3 - 2x_3)\,a\,\hat{\mathbf{x}} - \frac{\sqrt{3}}{2}\,y_3\,a\,\hat{\mathbf{y}} + z_3\,c\,\hat{\mathbf{z}}$ | (6$g$) | M |
| $\mathbf{B}_7$ | = | $-x_3\,\mathbf{a}_1 - y_3\,\mathbf{a}_2 - z_3\,\mathbf{a}_3$ | = | $-\frac{1}{2}\,(x_3 + y_3)\,a\,\hat{\mathbf{x}} + \frac{\sqrt{3}}{2}\,(x_3 - y_3)\,a\,\hat{\mathbf{y}} - z_3\,c\,\hat{\mathbf{z}}$ | (6$g$) | M |
| $\mathbf{B}_8$ | = | $y_3\,\mathbf{a}_1 + (y_3 - x_3)\,\mathbf{a}_2 - z_3\,\mathbf{a}_3$ | = | $\frac{1}{2}\,(2y_3 - x_3)\,a\,\hat{\mathbf{x}} - \frac{\sqrt{3}}{2}\,x_3\,a\,\hat{\mathbf{y}} - z_3\,c\,\hat{\mathbf{z}}$ | (6$g$) | M |
| $\mathbf{B}_9$ | = | $(x_3 - y_3)\,\mathbf{a}_1 + x_3\,\mathbf{a}_2 - z_3\,\mathbf{a}_3$ | = | $\frac{1}{2}\,(2x_3 - y_3)\,a\,\hat{\mathbf{x}} + \frac{\sqrt{3}}{2}\,y_3\,a\,\hat{\mathbf{y}} - z_3\,c\,\hat{\mathbf{z}}$ | (6$g$) | M |

**Geometry files:**

- CIF: pp. 713
- POSCAR: pp. 713

# Solid Cubane ($C_8H_8$) Structure: AB_hR16_148_cf_cf

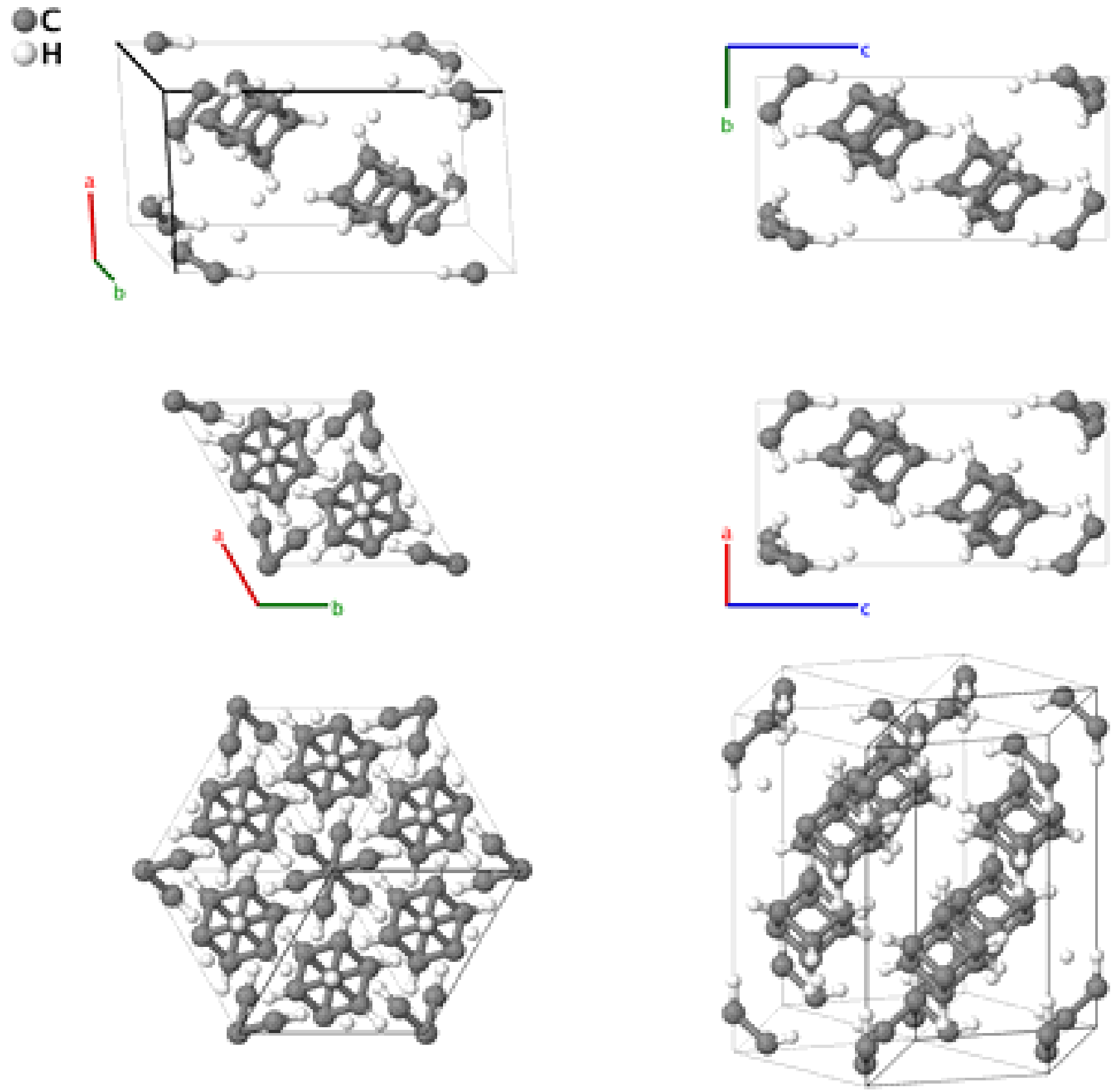

| | | |
|---|---|---|
| **Prototype** | : | $C_8H_8$ |
| **AFLOW prototype label** | : | AB_hR16_148_cf_cf |
| ***Strukturbericht* designation** | : | None |
| **Pearson symbol** | : | hR16 |
| **Space group number** | : | 148 |
| **Space group symbol** | : | $R\bar{3}$ |
| **AFLOW prototype command** | : | `aflow --proto=AB_hR16_148_cf_cf [--hex]`<br>`--params=`$a, c/a, x_1, x_2, x_3, y_3, z_3, x_4, y_4, z_4$ |

- Hexagonal settings of this structure can be obtained with the option `--hex`.

**Rhombohedral primitive vectors:**

$$
\begin{aligned}
\mathbf{a}_1 &= \tfrac{1}{2}\,a\,\hat{\mathbf{x}} - \tfrac{1}{2\sqrt{3}}\,a\,\hat{\mathbf{y}} + \tfrac{1}{3}\,c\,\hat{\mathbf{z}} \\
\mathbf{a}_2 &= \tfrac{1}{\sqrt{3}}\,a\,\hat{\mathbf{y}} + \tfrac{1}{3}\,c\,\hat{\mathbf{z}} \\
\mathbf{a}_3 &= -\tfrac{1}{2}\,a\,\hat{\mathbf{x}} - \tfrac{1}{2\sqrt{3}}\,a\,\hat{\mathbf{y}} + \tfrac{1}{3}\,c\,\hat{\mathbf{z}}
\end{aligned}
$$

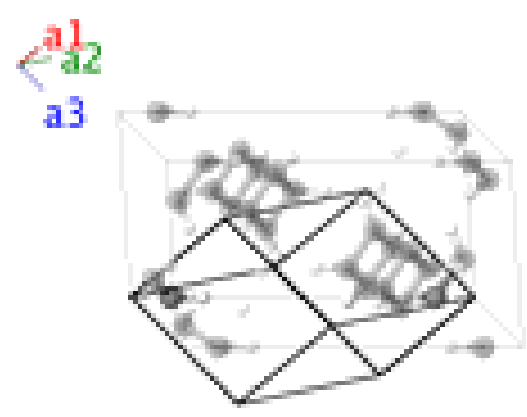

**Basis vectors:**

Lattice Coordinates | Cartesian Coordinates | Wyckoff Position | Atom Type

| | | | | | | |
|---|---|---|---|---|---|---|
| $\mathbf{B}_1$ | = | $x_1\,\mathbf{a}_1 + x_1\,\mathbf{a}_2 + x_1\,\mathbf{a}_3$ | = | $x_1\,c\,\hat{\mathbf{z}}$ | (2c) | C I |
| $\mathbf{B}_2$ | = | $-x_1\,\mathbf{a}_1 - x_1\,\mathbf{a}_2 - x_1\,\mathbf{a}_3$ | = | $-x_1\,c\,\hat{\mathbf{z}}$ | (2c) | C I |
| $\mathbf{B}_3$ | = | $x_2\,\mathbf{a}_1 + x_2\,\mathbf{a}_2 + x_2\,\mathbf{a}_3$ | = | $x_2\,c\,\hat{\mathbf{z}}$ | (2c) | H I |
| $\mathbf{B}_4$ | = | $-x_2\,\mathbf{a}_1 - x_2\,\mathbf{a}_2 - x_2\,\mathbf{a}_3$ | = | $-x_2\,c\,\hat{\mathbf{z}}$ | (2c) | H I |
| $\mathbf{B}_5$ | = | $x_3\,\mathbf{a}_1 + y_3\,\mathbf{a}_2 + z_3\,\mathbf{a}_3$ | = | $\frac{1}{2}\,(x_3 - z_3)\,a\,\hat{\mathbf{x}} + \frac{1}{2\sqrt{3}}\,(2y_3 - z_3 - x_3)\,a\,\hat{\mathbf{y}} + \frac{1}{3}\,(x_3 + y_3 + z_3)\,c\,\hat{\mathbf{z}}$ | (6f) | C II |
| $\mathbf{B}_6$ | = | $z_3\,\mathbf{a}_1 + x_3\,\mathbf{a}_2 + y_3\,\mathbf{a}_3$ | = | $\frac{1}{2}\,(z_3 - y_3)\,a\,\hat{\mathbf{x}} + \frac{1}{2\sqrt{3}}\,(2x_3 - y_3 - z_3)\,a\,\hat{\mathbf{y}} + \frac{1}{3}\,(x_3 + y_3 + z_3)\,c\,\hat{\mathbf{z}}$ | (6f) | C II |
| $\mathbf{B}_7$ | = | $y_3\,\mathbf{a}_1 + z_3\,\mathbf{a}_2 + x_3\,\mathbf{a}_3$ | = | $\frac{1}{2}\,(y_3 - x_3)\,a\,\hat{\mathbf{x}} + \frac{1}{2\sqrt{3}}\,(2z_3 - x_3 - y_3)\,a\,\hat{\mathbf{y}} + \frac{1}{3}\,(x_3 + y_3 + z_3)\,c\,\hat{\mathbf{z}}$ | (6f) | C II |
| $\mathbf{B}_8$ | = | $-x_3\,\mathbf{a}_1 - y_3\,\mathbf{a}_2 - z_3\,\mathbf{a}_3$ | = | $\frac{1}{2}\,(z_3 - x_3)\,a\,\hat{\mathbf{x}} + \frac{1}{2\sqrt{3}}\,(z_3 + x_3 - 2y_3)\,a\,\hat{\mathbf{y}} - \frac{1}{3}\,(x_3 + y_3 + z_3)\,c\,\hat{\mathbf{z}}$ | (6f) | C II |
| $\mathbf{B}_9$ | = | $-z_3\,\mathbf{a}_1 - x_3\,\mathbf{a}_2 - y_3\,\mathbf{a}_3$ | = | $\frac{1}{2}\,(y_3 - z_3)\,a\,\hat{\mathbf{x}} + \frac{1}{2\sqrt{3}}\,(y_3 + z_3 - 2x_3)\,a\,\hat{\mathbf{y}} - \frac{1}{3}\,(x_3 + y_3 + z_3)\,c\,\hat{\mathbf{z}}$ | (6f) | C II |
| $\mathbf{B}_{10}$ | = | $-y_3\,\mathbf{a}_1 - z_3\,\mathbf{a}_2 - x_3\,\mathbf{a}_3$ | = | $\frac{1}{2}\,(x_3 - y_3)\,a\,\hat{\mathbf{x}} + \frac{1}{2\sqrt{3}}\,(x_3 + y_3 - 2z_3)\,a\,\hat{\mathbf{y}} - \frac{1}{3}\,(x_3 + y_3 + z_3)\,c\,\hat{\mathbf{z}}$ | (6f) | C II |
| $\mathbf{B}_{11}$ | = | $x_4\,\mathbf{a}_1 + y_4\,\mathbf{a}_2 + z_4\,\mathbf{a}_3$ | = | $\frac{1}{2}\,(x_4 - z_4)\,a\,\hat{\mathbf{x}} + \frac{1}{2\sqrt{3}}\,(2y_4 - z_4 - x_4)\,a\,\hat{\mathbf{y}} + \frac{1}{3}\,(x_4 + y_4 + z_4)\,c\,\hat{\mathbf{z}}$ | (6f) | H II |
| $\mathbf{B}_{12}$ | = | $z_4\,\mathbf{a}_1 + x_4\,\mathbf{a}_2 + y_4\,\mathbf{a}_3$ | = | $\frac{1}{2}\,(z_4 - y_4)\,a\,\hat{\mathbf{x}} + \frac{1}{2\sqrt{3}}\,(2x_4 - y_4 - z_4)\,a\,\hat{\mathbf{y}} + \frac{1}{3}\,(x_4 + y_4 + z_4)\,c\,\hat{\mathbf{z}}$ | (6f) | H II |
| $\mathbf{B}_{13}$ | = | $y_4\,\mathbf{a}_1 + z_4\,\mathbf{a}_2 + x_4\,\mathbf{a}_3$ | = | $\frac{1}{2}\,(y_4 - x_4)\,a\,\hat{\mathbf{x}} + \frac{1}{2\sqrt{3}}\,(2z_4 - x_4 - y_4)\,a\,\hat{\mathbf{y}} + \frac{1}{3}\,(x_4 + y_4 + z_4)\,c\,\hat{\mathbf{z}}$ | (6f) | H II |
| $\mathbf{B}_{14}$ | = | $-x_4\,\mathbf{a}_1 - y_4\,\mathbf{a}_2 - z_4\,\mathbf{a}_3$ | = | $\frac{1}{2}\,(z_4 - x_4)\,a\,\hat{\mathbf{x}} + \frac{1}{2\sqrt{3}}\,(z_4 + x_4 - 2y_4)\,a\,\hat{\mathbf{y}} - \frac{1}{3}\,(x_4 + y_4 + z_4)\,c\,\hat{\mathbf{z}}$ | (6f) | H II |
| $\mathbf{B}_{15}$ | = | $-z_4\,\mathbf{a}_1 - x_4\,\mathbf{a}_2 - y_4\,\mathbf{a}_3$ | = | $\frac{1}{2}\,(y_4 - z_4)\,a\,\hat{\mathbf{x}} + \frac{1}{2\sqrt{3}}\,(y_4 + z_4 - 2x_4)\,a\,\hat{\mathbf{y}} - \frac{1}{3}\,(x_4 + y_4 + z_4)\,c\,\hat{\mathbf{z}}$ | (6f) | H II |
| $\mathbf{B}_{16}$ | = | $-y_4\,\mathbf{a}_1 - z_4\,\mathbf{a}_2 - x_4\,\mathbf{a}_3$ | = | $\frac{1}{2}\,(x_4 - y_4)\,a\,\hat{\mathbf{x}} + \frac{1}{2\sqrt{3}}\,(x_4 + y_4 - 2z_4)\,a\,\hat{\mathbf{y}} - \frac{1}{3}\,(x_4 + y_4 + z_4)\,c\,\hat{\mathbf{z}}$ | (6f) | H II |

**Geometry files:**

- CIF: pp. 713
- POSCAR: pp. 714

# $BiI_3$ ($D0_5$) Structure: AB3_hR8_148_c_f

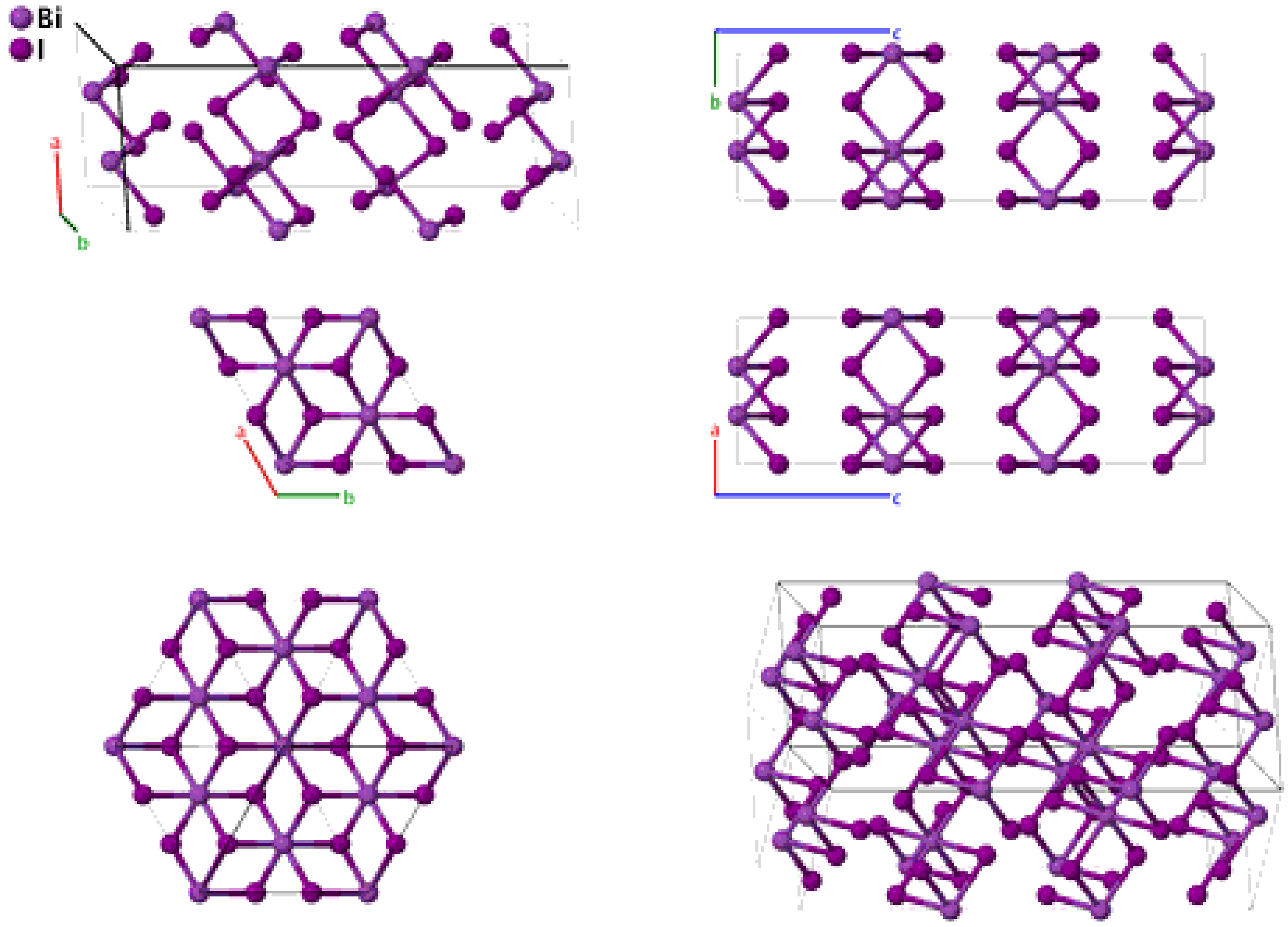

| | | |
|---|---|---|
| **Prototype** | : | $BiI_3$ |
| **AFLOW prototype label** | : | AB3_hR8_148_c_f |
| ***Strukturbericht* designation** | : | $D0_5$ |
| **Pearson symbol** | : | hR8 |
| **Space group number** | : | 148 |
| **Space group symbol** | : | $R\bar{3}$ |
| **AFLOW prototype command** | : | `aflow --proto=AB3_hR8_148_c_f [--hex]` `--params=`$a, c/a, x_1, x_2, y_2, z_2$ |

**Other compounds with this structure:**

- $SbI_3$, $AsI_3$, $FeCl_3$, $CrBr_3$

---

- Hexagonal settings of this structure can be obtained with the option `--hex`.

**Rhombohedral primitive vectors:**

$$
\begin{aligned}
\mathbf{a}_1 &= \tfrac{1}{2}\,a\,\hat{\mathbf{x}} - \tfrac{1}{2\sqrt{3}}\,a\,\hat{\mathbf{y}} + \tfrac{1}{3}\,c\,\hat{\mathbf{z}} \\
\mathbf{a}_2 &= \tfrac{1}{\sqrt{3}}\,a\,\hat{\mathbf{y}} + \tfrac{1}{3}\,c\,\hat{\mathbf{z}} \\
\mathbf{a}_3 &= -\tfrac{1}{2}\,a\,\hat{\mathbf{x}} - \tfrac{1}{2\sqrt{3}}\,a\,\hat{\mathbf{y}} + \tfrac{1}{3}\,c\,\hat{\mathbf{z}}
\end{aligned}
$$

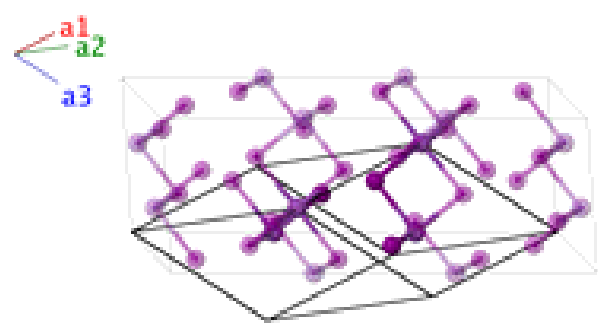

**Basis vectors:**

| Lattice Coordinates | Cartesian Coordinates | Wyckoff Position | Atom Type |
|---|---|---|---|

| | | | | | | |
|---|---|---|---|---|---|---|
| $\mathbf{B_1}$ | = | $x_1\,\mathbf{a_1} + x_1\,\mathbf{a_2} + x_1\,\mathbf{a_3}$ | = | $x_1\,c\,\hat{\mathbf{z}}$ | $(2c)$ | Bi |
| $\mathbf{B_2}$ | = | $-x_1\,\mathbf{a_1} - x_1\,\mathbf{a_2} - x_1\,\mathbf{a_3}$ | = | $-x_1\,c\,\hat{\mathbf{z}}$ | $(2c)$ | Bi |
| $\mathbf{B_3}$ | = | $x_2\,\mathbf{a_1} + y_2\,\mathbf{a_2} + z_2\,\mathbf{a_3}$ | = | $\frac{1}{2}\,(x_2 - z_2)\,a\,\hat{\mathbf{x}} + \frac{1}{2\sqrt{3}}\,(2y_2 - z_2 - x_2)\,a\,\hat{\mathbf{y}} + \frac{1}{3}\,(x_2 + y_2 + z_2)\,c\,\hat{\mathbf{z}}$ | $(6f)$ | I |
| $\mathbf{B_4}$ | = | $z_2\,\mathbf{a_1} + x_2\,\mathbf{a_2} + y_2\,\mathbf{a_3}$ | = | $\frac{1}{2}\,(z_2 - y_2)\,a\,\hat{\mathbf{x}} + \frac{1}{2\sqrt{3}}\,(2x_2 - y_2 - z_2)\,a\,\hat{\mathbf{y}} + \frac{1}{3}\,(x_2 + y_2 + z_2)\,c\,\hat{\mathbf{z}}$ | $(6f)$ | I |
| $\mathbf{B_5}$ | = | $y_2\,\mathbf{a_1} + z_2\,\mathbf{a_2} + x_2\,\mathbf{a_3}$ | = | $\frac{1}{2}\,(y_2 - x_2)\,a\,\hat{\mathbf{x}} + \frac{1}{2\sqrt{3}}\,(2z_2 - x_2 - y_2)\,a\,\hat{\mathbf{y}} + \frac{1}{3}\,(x_2 + y_2 + z_2)\,c\,\hat{\mathbf{z}}$ | $(6f)$ | I |
| $\mathbf{B_6}$ | = | $-x_2\,\mathbf{a_1} - y_2\,\mathbf{a_2} - z_2\,\mathbf{a_3}$ | = | $\frac{1}{2}\,(z_2 - x_2)\,a\,\hat{\mathbf{x}} + \frac{1}{2\sqrt{3}}\,(z_2 + x_2 - 2y_2)\,a\,\hat{\mathbf{y}} - \frac{1}{3}\,(x_2 + y_2 + z_2)\,c\,\hat{\mathbf{z}}$ | $(6f)$ | I |
| $\mathbf{B_7}$ | = | $-z_2\,\mathbf{a_1} - x_2\,\mathbf{a_2} - y_2\,\mathbf{a_3}$ | = | $\frac{1}{2}\,(y_2 - z_2)\,a\,\hat{\mathbf{x}} + \frac{1}{2\sqrt{3}}\,(y_2 + z_2 - 2x_2)\,a\,\hat{\mathbf{y}} - \frac{1}{3}\,(x_2 + y_2 + z_2)\,c\,\hat{\mathbf{z}}$ | $(6f)$ | I |
| $\mathbf{B_8}$ | = | $-y_2\,\mathbf{a_1} - z_2\,\mathbf{a_2} - x_2\,\mathbf{a_3}$ | = | $\frac{1}{2}\,(x_2 - y_2)\,a\,\hat{\mathbf{x}} + \frac{1}{2\sqrt{3}}\,(x_2 + y_2 - 2z_2)\,a\,\hat{\mathbf{y}} - \frac{1}{3}\,(x_2 + y_2 + z_2)\,c\,\hat{\mathbf{z}}$ | $(6f)$ | I |

**Geometry files:**

- CIF: pp. 714
- POSCAR: pp. 714

# PdAl Structure: AB_hR26_148_b2f_a2f

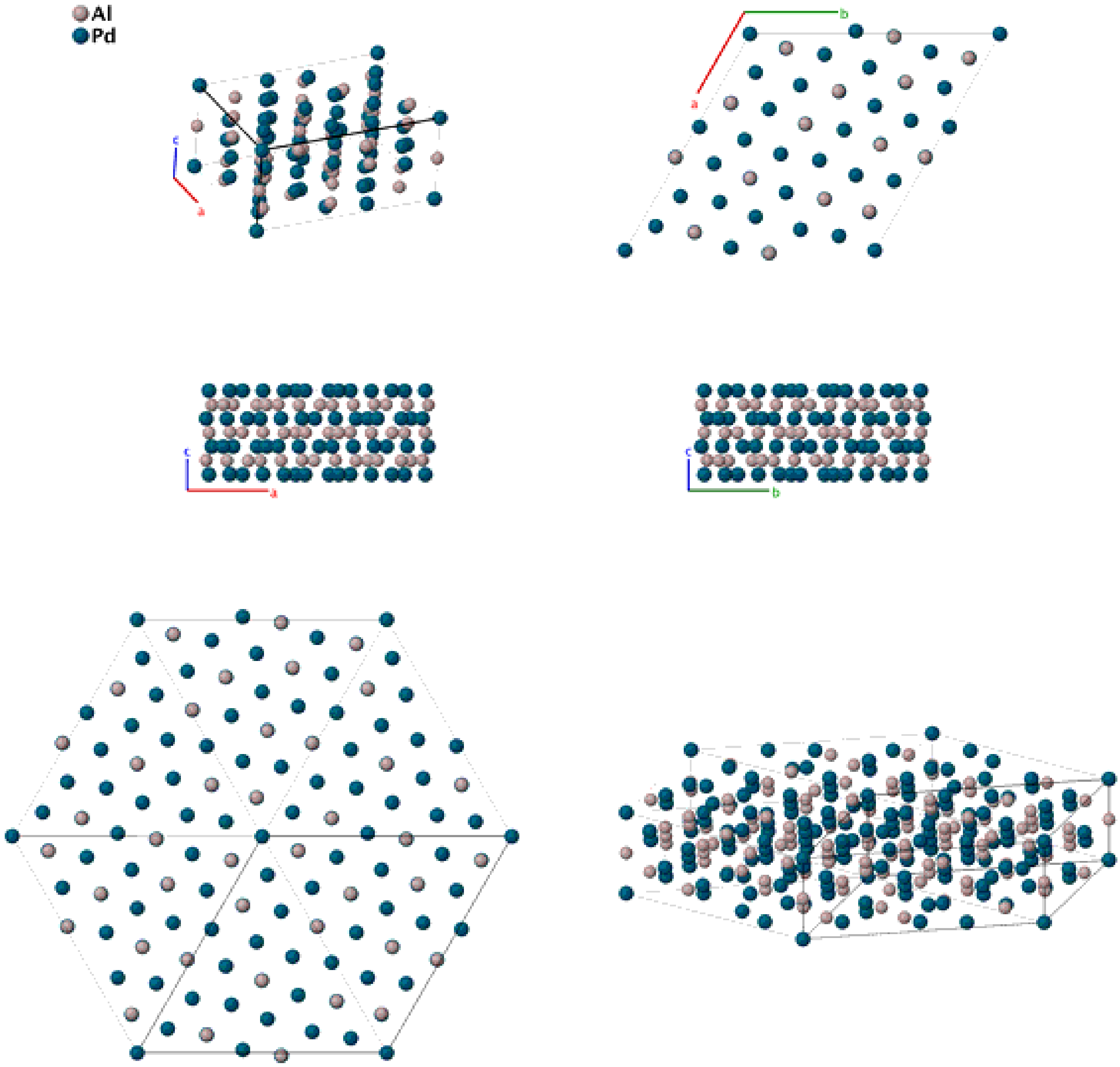

| | | |
|---|---|---|
| **Prototype** | : | PdAl |
| **AFLOW prototype label** | : | AB_hR26_148_b2f_a2f |
| ***Strukturbericht* designation** | : | None |
| **Pearson symbol** | : | hR26 |
| **Space group number** | : | 148 |
| **Space group symbol** | : | $R\bar{3}$ |
| **AFLOW prototype command** | : | `aflow --proto=AB_hR26_148_b2f_a2f [--hex]`<br>`--params=`$a, c/a, x_3, y_3, z_3, x_4, y_4, z_4, x_5, y_5, z_5, x_6, y_6, z_6$ |

- Hexagonal settings of this structure can be obtained with the option `--hex`.

## Rhombohedral primitive vectors:

$$
\begin{aligned}
\mathbf{a}_1 &= \tfrac{1}{2}\,a\,\hat{\mathbf{x}} - \tfrac{1}{2\sqrt{3}}\,a\,\hat{\mathbf{y}} + \tfrac{1}{3}\,c\,\hat{\mathbf{z}} \\
\mathbf{a}_2 &= \tfrac{1}{\sqrt{3}}\,a\,\hat{\mathbf{y}} + \tfrac{1}{3}\,c\,\hat{\mathbf{z}} \\
\mathbf{a}_3 &= -\tfrac{1}{2}\,a\,\hat{\mathbf{x}} - \tfrac{1}{2\sqrt{3}}\,a\,\hat{\mathbf{y}} + \tfrac{1}{3}\,c\,\hat{\mathbf{z}}
\end{aligned}
$$

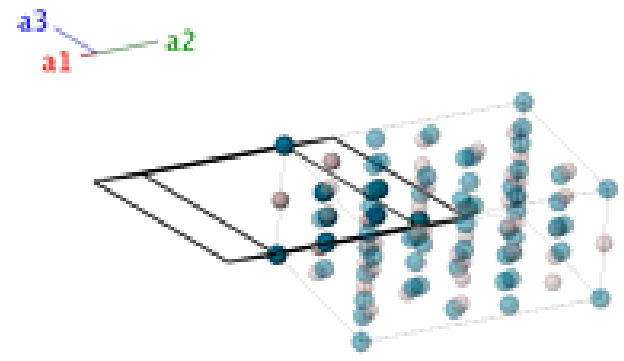

## Basis vectors:

| | | Lattice Coordinates | | Cartesian Coordinates | Wyckoff Position | Atom Type |
|---|---|---|---|---|---|---|
| $\mathbf{B}_1$ | = | $0\,\mathbf{a}_1 + 0\,\mathbf{a}_2 + 0\,\mathbf{a}_3$ | = | $0\hat{\mathbf{x}} + 0\hat{\mathbf{y}} + 0\hat{\mathbf{z}}$ | (1*a*) | Pd I |
| $\mathbf{B}_2$ | = | $\frac{1}{2}\,\mathbf{a}_1 + \frac{1}{2}\,\mathbf{a}_2 + \frac{1}{2}\,\mathbf{a}_3$ | = | $\frac{1}{2}\,c\,\hat{\mathbf{z}}$ | (1*b*) | Al I |
| $\mathbf{B}_3$ | = | $x_3\,\mathbf{a}_1 + y_3\,\mathbf{a}_2 + z_3\,\mathbf{a}_3$ | = | $\frac{1}{2}\,(x_3 - z_3)\,a\,\hat{\mathbf{x}} + \frac{1}{2\sqrt{3}}\,(2y_3 - z_3 - x_3)\,a\,\hat{\mathbf{y}} + \frac{1}{3}\,(x_3 + y_3 + z_3)\,c\,\hat{\mathbf{z}}$ | (6*f*) | Al II |
| $\mathbf{B}_4$ | = | $z_3\,\mathbf{a}_1 + x_3\,\mathbf{a}_2 + y_3\,\mathbf{a}_3$ | = | $\frac{1}{2}\,(z_3 - y_3)\,a\,\hat{\mathbf{x}} + \frac{1}{2\sqrt{3}}\,(2x_3 - y_3 - z_3)\,a\,\hat{\mathbf{y}} + \frac{1}{3}\,(x_3 + y_3 + z_3)\,c\,\hat{\mathbf{z}}$ | (6*f*) | Al II |
| $\mathbf{B}_5$ | = | $y_3\,\mathbf{a}_1 + z_3\,\mathbf{a}_2 + x_3\,\mathbf{a}_3$ | = | $\frac{1}{2}\,(y_3 - x_3)\,a\,\hat{\mathbf{x}} + \frac{1}{2\sqrt{3}}\,(2z_3 - x_3 - y_3)\,a\,\hat{\mathbf{y}} + \frac{1}{3}\,(x_3 + y_3 + z_3)\,c\,\hat{\mathbf{z}}$ | (6*f*) | Al II |
| $\mathbf{B}_6$ | = | $-x_3\,\mathbf{a}_1 - y_3\,\mathbf{a}_2 - z_3\,\mathbf{a}_3$ | = | $\frac{1}{2}\,(z_3 - x_3)\,a\,\hat{\mathbf{x}} + \frac{1}{2\sqrt{3}}\,(z_3 + x_3 - 2y_3)\,a\,\hat{\mathbf{y}} - \frac{1}{3}\,(x_3 + y_3 + z_3)\,c\,\hat{\mathbf{z}}$ | (6*f*) | Al II |
| $\mathbf{B}_7$ | = | $-z_3\,\mathbf{a}_1 - x_3\,\mathbf{a}_2 - y_3\,\mathbf{a}_3$ | = | $\frac{1}{2}\,(y_3 - z_3)\,a\,\hat{\mathbf{x}} + \frac{1}{2\sqrt{3}}\,(y_3 + z_3 - 2x_3)\,a\,\hat{\mathbf{y}} - \frac{1}{3}\,(x_3 + y_3 + z_3)\,c\,\hat{\mathbf{z}}$ | (6*f*) | Al II |
| $\mathbf{B}_8$ | = | $-y_3\,\mathbf{a}_1 - z_3\,\mathbf{a}_2 - x_3\,\mathbf{a}_3$ | = | $\frac{1}{2}\,(x_3 - y_3)\,a\,\hat{\mathbf{x}} + \frac{1}{2\sqrt{3}}\,(x_3 + y_3 - 2z_3)\,a\,\hat{\mathbf{y}} - \frac{1}{3}\,(x_3 + y_3 + z_3)\,c\,\hat{\mathbf{z}}$ | (6*f*) | Al II |
| $\mathbf{B}_9$ | = | $x_4\,\mathbf{a}_1 + y_4\,\mathbf{a}_2 + z_4\,\mathbf{a}_3$ | = | $\frac{1}{2}\,(x_4 - z_4)\,a\,\hat{\mathbf{x}} + \frac{1}{2\sqrt{3}}\,(2y_4 - z_4 - x_4)\,a\,\hat{\mathbf{y}} + \frac{1}{3}\,(x_4 + y_4 + z_4)\,c\,\hat{\mathbf{z}}$ | (6*f*) | Al III |
| $\mathbf{B}_{10}$ | = | $z_4\,\mathbf{a}_1 + x_4\,\mathbf{a}_2 + y_4\,\mathbf{a}_3$ | = | $\frac{1}{2}\,(z_4 - y_4)\,a\,\hat{\mathbf{x}} + \frac{1}{2\sqrt{3}}\,(2x_4 - y_4 - z_4)\,a\,\hat{\mathbf{y}} + \frac{1}{3}\,(x_4 + y_4 + z_4)\,c\,\hat{\mathbf{z}}$ | (6*f*) | Al III |
| $\mathbf{B}_{11}$ | = | $y_4\,\mathbf{a}_1 + z_4\,\mathbf{a}_2 + x_4\,\mathbf{a}_3$ | = | $\frac{1}{2}\,(y_4 - x_4)\,a\,\hat{\mathbf{x}} + \frac{1}{2\sqrt{3}}\,(2z_4 - x_4 - y_4)\,a\,\hat{\mathbf{y}} + \frac{1}{3}\,(x_4 + y_4 + z_4)\,c\,\hat{\mathbf{z}}$ | (6*f*) | Al III |
| $\mathbf{B}_{12}$ | = | $-x_4\,\mathbf{a}_1 - y_4\,\mathbf{a}_2 - z_4\,\mathbf{a}_3$ | = | $\frac{1}{2}\,(z_4 - x_4)\,a\,\hat{\mathbf{x}} + \frac{1}{2\sqrt{3}}\,(z_4 + x_4 - 2y_4)\,a\,\hat{\mathbf{y}} - \frac{1}{3}\,(x_4 + y_4 + z_4)\,c\,\hat{\mathbf{z}}$ | (6*f*) | Al III |
| $\mathbf{B}_{13}$ | = | $-z_4\,\mathbf{a}_1 - x_4\,\mathbf{a}_2 - y_4\,\mathbf{a}_3$ | = | $\frac{1}{2}\,(y_4 - z_4)\,a\,\hat{\mathbf{x}} + \frac{1}{2\sqrt{3}}\,(y_4 + z_4 - 2x_4)\,a\,\hat{\mathbf{y}} - \frac{1}{3}\,(x_4 + y_4 + z_4)\,c\,\hat{\mathbf{z}}$ | (6*f*) | Al III |
| $\mathbf{B}_{14}$ | = | $-y_4\,\mathbf{a}_1 - z_4\,\mathbf{a}_2 - x_4\,\mathbf{a}_3$ | = | $\frac{1}{2}\,(x_4 - y_4)\,a\,\hat{\mathbf{x}} + \frac{1}{2\sqrt{3}}\,(x_4 + y_4 - 2z_4)\,a\,\hat{\mathbf{y}} - \frac{1}{3}\,(x_4 + y_4 + z_4)\,c\,\hat{\mathbf{z}}$ | (6*f*) | Al III |
| $\mathbf{B}_{15}$ | = | $x_5\,\mathbf{a}_1 + y_5\,\mathbf{a}_2 + z_5\,\mathbf{a}_3$ | = | $\frac{1}{2}\,(x_5 - z_5)\,a\,\hat{\mathbf{x}} + \frac{1}{2\sqrt{3}}\,(2y_5 - z_5 - x_5)\,a\,\hat{\mathbf{y}} + \frac{1}{3}\,(x_5 + y_5 + z_5)\,c\,\hat{\mathbf{z}}$ | (6*f*) | Pd II |
| $\mathbf{B}_{16}$ | = | $z_5\,\mathbf{a}_1 + x_5\,\mathbf{a}_2 + y_5\,\mathbf{a}_3$ | = | $\frac{1}{2}\,(z_5 - y_5)\,a\,\hat{\mathbf{x}} + \frac{1}{2\sqrt{3}}\,(2x_5 - y_5 - z_5)\,a\,\hat{\mathbf{y}} + \frac{1}{3}\,(x_5 + y_5 + z_5)\,c\,\hat{\mathbf{z}}$ | (6*f*) | Pd II |
| $\mathbf{B}_{17}$ | = | $y_5\,\mathbf{a}_1 + z_5\,\mathbf{a}_2 + x_5\,\mathbf{a}_3$ | = | $\frac{1}{2}\,(y_5 - x_5)\,a\,\hat{\mathbf{x}} + \frac{1}{2\sqrt{3}}\,(2z_5 - x_5 - y_5)\,a\,\hat{\mathbf{y}} + \frac{1}{3}\,(x_5 + y_5 + z_5)\,c\,\hat{\mathbf{z}}$ | (6*f*) | Pd II |
| $\mathbf{B}_{18}$ | = | $-x_5\,\mathbf{a}_1 - y_5\,\mathbf{a}_2 - z_5\,\mathbf{a}_3$ | = | $\frac{1}{2}\left(z_5 - x_5\right)a\,\hat{\mathbf{x}} + \frac{1}{2\sqrt{3}}\left(z_5 + x_5 - 2y_5\right)a\,\hat{\mathbf{y}} - \frac{1}{3}\left(x_5 + y_5 + z_5\right)c\,\hat{\mathbf{z}}$ | (6*f*) | Pd II |
| $\mathbf{B}_{19}$ | = | $-z_5\,\mathbf{a}_1 - x_5\,\mathbf{a}_2 - y_5\,\mathbf{a}_3$ | = | $\frac{1}{2}\left(y_5 - z_5\right)a\,\hat{\mathbf{x}} + \frac{1}{2\sqrt{3}}\left(y_5 + z_5 - 2x_5\right)a\,\hat{\mathbf{y}} - \frac{1}{3}\left(x_5 + y_5 + z_5\right)c\,\hat{\mathbf{z}}$ | (6*f*) | Pd II |
| $\mathbf{B}_{20}$ | = | $-y_5\,\mathbf{a}_1 - z_5\,\mathbf{a}_2 - x_5\,\mathbf{a}_3$ | = | $\frac{1}{2}\left(x_5 - y_5\right)a\,\hat{\mathbf{x}} + \frac{1}{2\sqrt{3}}\left(x_5 + y_5 - 2z_5\right)a\,\hat{\mathbf{y}} - \frac{1}{3}\left(x_5 + y_5 + z_5\right)c\,\hat{\mathbf{z}}$ | (6*f*) | Pd II |
| $\mathbf{B}_{21}$ | = | $x_6\,\mathbf{a}_1 + y_6\,\mathbf{a}_2 + z_6\,\mathbf{a}_3$ | = | $\frac{1}{2}\left(x_6 - z_6\right)a\,\hat{\mathbf{x}} + \frac{1}{2\sqrt{3}}\left(2y_6 - z_6 - x_6\right)a\,\hat{\mathbf{y}} + \frac{1}{3}\left(x_6 + y_6 + z_6\right)c\,\hat{\mathbf{z}}$ | (6*f*) | Pd III |
| $\mathbf{B}_{22}$ | = | $z_6\,\mathbf{a}_1 + x_6\,\mathbf{a}_2 + y_6\,\mathbf{a}_3$ | = | $\frac{1}{2}\left(z_6 - y_6\right)a\,\hat{\mathbf{x}} + \frac{1}{2\sqrt{3}}\left(2x_6 - y_6 - z_6\right)a\,\hat{\mathbf{y}} + \frac{1}{3}\left(x_6 + y_6 + z_6\right)c\,\hat{\mathbf{z}}$ | (6*f*) | Pd III |
| $\mathbf{B}_{23}$ | = | $y_6\,\mathbf{a}_1 + z_6\,\mathbf{a}_2 + x_6\,\mathbf{a}_3$ | = | $\frac{1}{2}\left(y_6 - x_6\right)a\,\hat{\mathbf{x}} + \frac{1}{2\sqrt{3}}\left(2z_6 - x_6 - y_6\right)a\,\hat{\mathbf{y}} + \frac{1}{3}\left(x_6 + y_6 + z_6\right)c\,\hat{\mathbf{z}}$ | (6*f*) | Pd III |
| $\mathbf{B}_{24}$ | = | $-x_6\,\mathbf{a}_1 - y_6\,\mathbf{a}_2 - z_6\,\mathbf{a}_3$ | = | $\frac{1}{2}\left(z_6 - x_6\right)a\,\hat{\mathbf{x}} + \frac{1}{2\sqrt{3}}\left(z_6 + x_6 - 2y_6\right)a\,\hat{\mathbf{y}} - \frac{1}{3}\left(x_6 + y_6 + z_6\right)c\,\hat{\mathbf{z}}$ | (6*f*) | Pd III |
| $\mathbf{B}_{25}$ | = | $-z_6\,\mathbf{a}_1 - x_6\,\mathbf{a}_2 - y_6\,\mathbf{a}_3$ | = | $\frac{1}{2}\left(y_6 - z_6\right)a\,\hat{\mathbf{x}} + \frac{1}{2\sqrt{3}}\left(y_6 + z_6 - 2x_6\right)a\,\hat{\mathbf{y}} - \frac{1}{3}\left(x_6 + y_6 + z_6\right)c\,\hat{\mathbf{z}}$ | (6*f*) | Pd III |
| $\mathbf{B}_{26}$ | = | $-y_6\,\mathbf{a}_1 - z_6\,\mathbf{a}_2 - x_6\,\mathbf{a}_3$ | = | $\frac{1}{2}\left(x_6 - y_6\right)a\,\hat{\mathbf{x}} + \frac{1}{2\sqrt{3}}\left(x_6 + y_6 - 2z_6\right)a\,\hat{\mathbf{y}} - \frac{1}{3}\left(x_6 + y_6 + z_6\right)c\,\hat{\mathbf{z}}$ | (6*f*) | Pd III |

---

**Found in:**

---

**Geometry files:**

- CIF: pp. 714
- POSCAR: pp. 715

# Ilmenite ($FeTiO_3$) Structure: AB3C_hR10_148_c_f_c

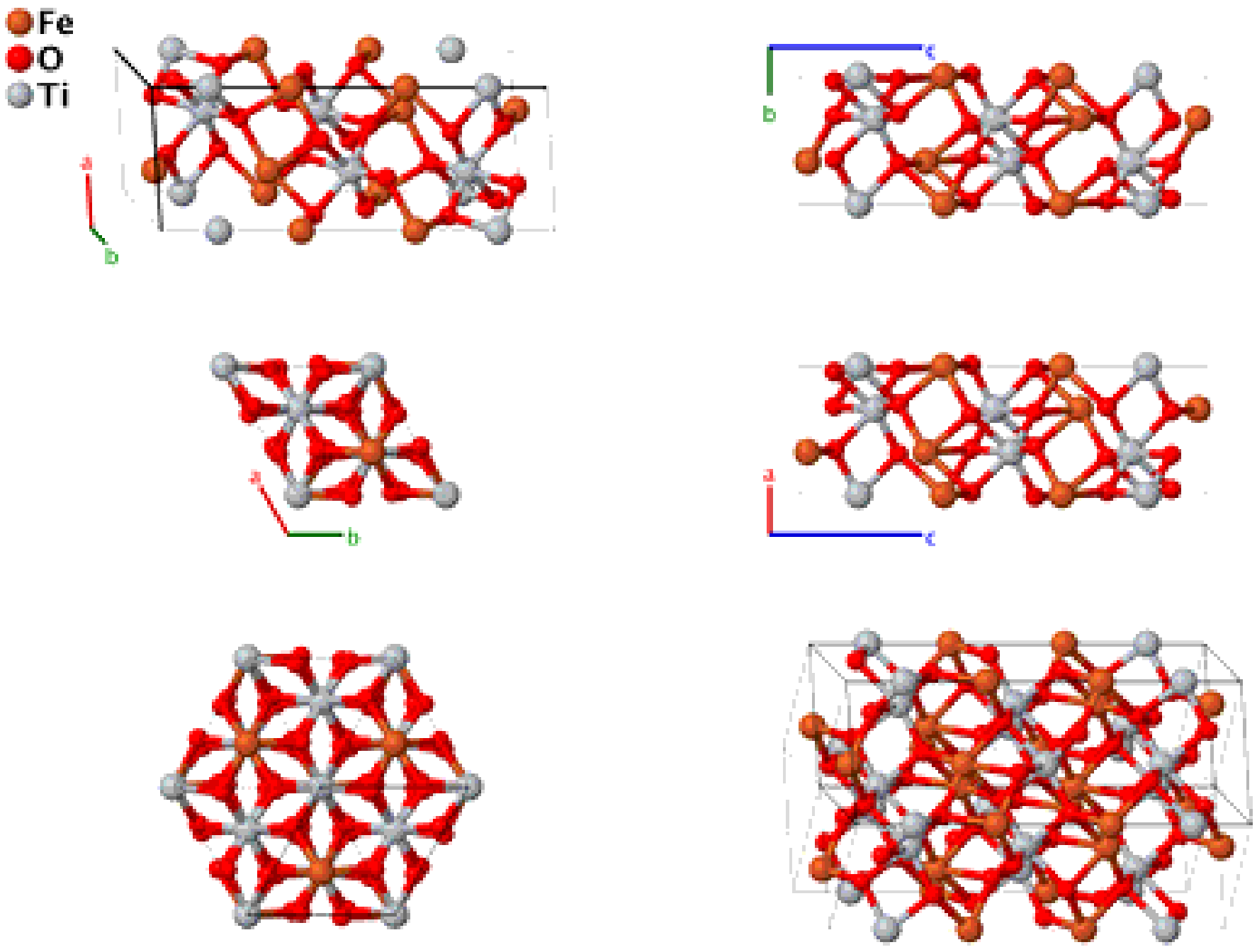

| | | |
|---|---|---|
| **Prototype** | : | $FeTiO_3$ |
| **AFLOW prototype label** | : | AB3C_hR10_148_c_f_c |
| ***Strukturbericht* designation** | : | None |
| **Pearson symbol** | : | hR10 |
| **Space group number** | : | 148 |
| **Space group symbol** | : | R$\bar{3}$ |
| **AFLOW prototype command** | : | `aflow --proto=AB3C_hR10_148_c_f_c [--hex]` <br> `--params=`$a, c/a, x_1, x_2, x_3, y_3, z_3$ |

- The oxygen atoms form a nearly close-packed system, hence the classifications of this structure. Hexagonal settings of this structure can be obtained with the option `--hex`.

**Rhombohedral primitive vectors:**

$$\mathbf{a}_1 = \frac{1}{2}\,a\,\hat{\mathbf{x}} - \frac{1}{2\sqrt{3}}\,a\,\hat{\mathbf{y}} + \frac{1}{3}\,c\,\hat{\mathbf{z}}$$

$$\mathbf{a}_2 = \frac{1}{\sqrt{3}}\,a\,\hat{\mathbf{y}} + \frac{1}{3}\,c\,\hat{\mathbf{z}}$$

$$\mathbf{a}_3 = -\frac{1}{2}\,a\,\hat{\mathbf{x}} - \frac{1}{2\sqrt{3}}\,a\,\hat{\mathbf{y}} + \frac{1}{3}\,c\,\hat{\mathbf{z}}$$

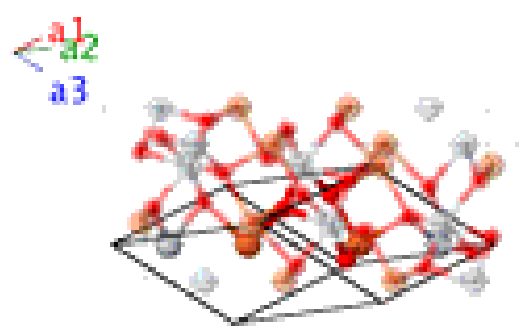

**Basis vectors:**

| | | Lattice Coordinates | | Cartesian Coordinates | Wyckoff Position | Atom Type |
|---|---|---|---|---|---|---|
| $\mathbf{B}_1$ | = | $x_1\,\mathbf{a}_1 + x_1\,\mathbf{a}_2 + x_1\,\mathbf{a}_3$ | = | $x_1\,c\,\hat{\mathbf{z}}$ | (2$c$) | Fe |
| $\mathbf{B}_2$ | = | $-x_1\,\mathbf{a}_1 - x_1\,\mathbf{a}_2 - x_1\,\mathbf{a}_3$ | = | $-x_1\,c\,\hat{\mathbf{z}}$ | (2$c$) | Fe |
| $\mathbf{B}_3$ | = | $x_2\,\mathbf{a}_1 + x_2\,\mathbf{a}_2 + x_2\,\mathbf{a}_3$ | = | $x_2\,c\,\hat{\mathbf{z}}$ | (2$c$) | Ti |

$$\mathbf{B}_4 = -x_2\,\mathbf{a}_1 - x_2\,\mathbf{a}_2 - x_2\,\mathbf{a}_3 = -x_2\,c\,\hat{\mathbf{z}} \qquad (2c) \qquad \text{Ti}$$

$$\mathbf{B}_5 = x_3\,\mathbf{a}_1 + y_3\,\mathbf{a}_2 + z_3\,\mathbf{a}_3 = \frac{1}{2}\,(x_3 - z_3)\,a\,\hat{\mathbf{x}} + \frac{1}{2\sqrt{3}}\,(2y_3 - z_3 - x_3)\,a\,\hat{\mathbf{y}} + \frac{1}{3}\,(x_3 + y_3 + z_3)\,c\,\hat{\mathbf{z}} \qquad (6f) \qquad \text{O}$$

$$\mathbf{B}_6 = z_3\,\mathbf{a}_1 + x_3\,\mathbf{a}_2 + y_3\,\mathbf{a}_3 = \frac{1}{2}\,(z_3 - y_3)\,a\,\hat{\mathbf{x}} + \frac{1}{2\sqrt{3}}\,(2x_3 - y_3 - z_3)\,a\,\hat{\mathbf{y}} + \frac{1}{3}\,(x_3 + y_3 + z_3)\,c\,\hat{\mathbf{z}} \qquad (6f) \qquad \text{O}$$

$$\mathbf{B}_7 = y_3\,\mathbf{a}_1 + z_3\,\mathbf{a}_2 + x_3\,\mathbf{a}_3 = \frac{1}{2}\,(y_3 - x_3)\,a\,\hat{\mathbf{x}} + \frac{1}{2\sqrt{3}}\,(2z_3 - x_3 - y_3)\,a\,\hat{\mathbf{y}} + \frac{1}{3}\,(x_3 + y_3 + z_3)\,c\,\hat{\mathbf{z}} \qquad (6f) \qquad \text{O}$$

$$\mathbf{B}_8 = -x_3\,\mathbf{a}_1 - y_3\,\mathbf{a}_2 - z_3\,\mathbf{a}_3 = \frac{1}{2}\,(z_3 - x_3)\,a\,\hat{\mathbf{x}} + \frac{1}{2\sqrt{3}}\,(z_3 + x_3 - 2y_3)\,a\,\hat{\mathbf{y}} - \frac{1}{3}\,(x_3 + y_3 + z_3)\,c\,\hat{\mathbf{z}} \qquad (6f) \qquad \text{O}$$

$$\mathbf{B}_9 = -z_3\,\mathbf{a}_1 - x_3\,\mathbf{a}_2 - y_3\,\mathbf{a}_3 = \frac{1}{2}\,(y_3 - z_3)\,a\,\hat{\mathbf{x}} + \frac{1}{2\sqrt{3}}\,(y_3 + z_3 - 2x_3)\,a\,\hat{\mathbf{y}} - \frac{1}{3}\,(x_3 + y_3 + z_3)\,c\,\hat{\mathbf{z}} \qquad (6f) \qquad \text{O}$$

$$\mathbf{B}_{10} = -y_3\,\mathbf{a}_1 - z_3\,\mathbf{a}_2 - x_3\,\mathbf{a}_3 = \frac{1}{2}\,(x_3 - y_3)\,a\,\hat{\mathbf{x}} + \frac{1}{2\sqrt{3}}\,(x_3 + y_3 - 2z_3)\,a\,\hat{\mathbf{y}} - \frac{1}{3}\,(x_3 + y_3 + z_3)\,c\,\hat{\mathbf{z}} \qquad (6f) \qquad \text{O}$$

---

---

**Geometry files:**

- CIF: pp. 715
- POSCAR: pp. 715

# Original $Fe_2P$ (C22) Structure: A2B_hP9_150_ef_bd

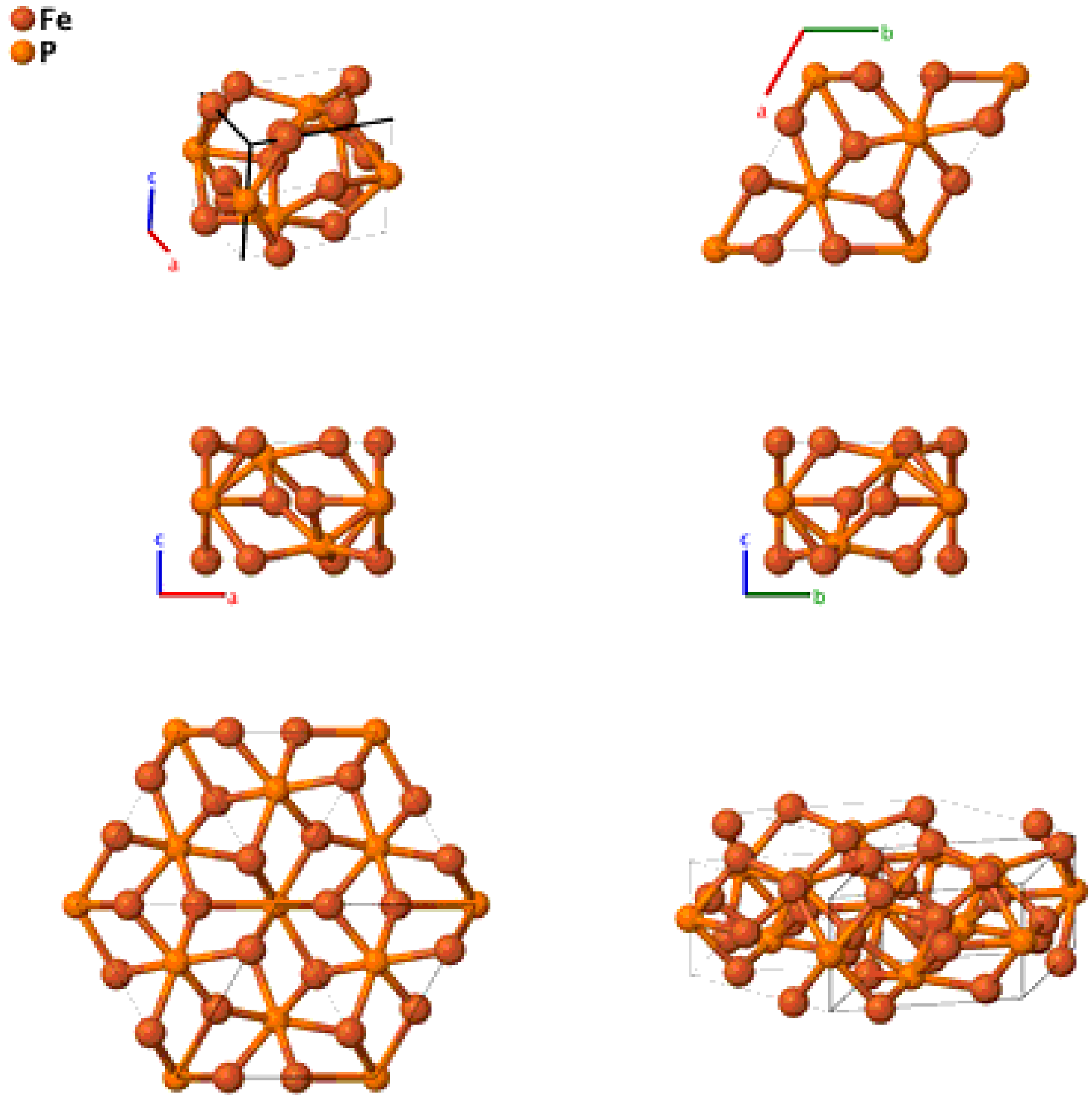

| | | |
|---|---|---|
| **Prototype** | : | $Fe_2P$ |
| **AFLOW prototype label** | : | A2B_hP9_150_ef_bd |
| ***Strukturbericht* designation** | : | C22 |
| **Pearson symbol** | : | hP9 |
| **Space group number** | : | 150 |
| **Space group symbol** | : | P321 |
| **AFLOW prototype command** | : | `aflow --proto=A2B_hP9_150_ef_bd`<br>`--params=`$a, c/a, z_2, x_3, x_4$ |

- This is the structure given in Strukturbericht Vol. II. As noted by Wyckoff, the structure, which was "generally accepted for years, has recently been shown to be incorrect." (Vol I., pp. 360) This corrected structure, as given in Pearson's Handbook, is given in the revised $Fe_2P$ page. When $z_2$ is set to zero this structure reverts to the revised $Fe_2P$ structure.

**Trigonal Hexagonal primitive vectors:**

$$\mathbf{a}_1 = \frac{1}{2}\,a\,\hat{\mathbf{x}} - \frac{\sqrt{3}}{2}\,a\,\hat{\mathbf{y}}$$
$$\mathbf{a}_2 = \frac{1}{2}\,a\,\hat{\mathbf{x}} + \frac{\sqrt{3}}{2}\,a\,\hat{\mathbf{y}}$$
$$\mathbf{a}_3 = c\,\hat{\mathbf{z}}$$

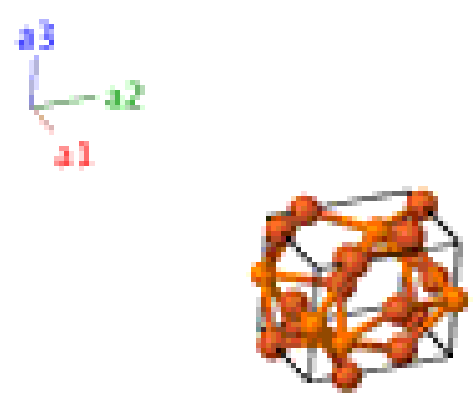

**Basis vectors:**

| | | Lattice Coordinates | | Cartesian Coordinates | Wyckoff Position | Atom Type |
|---|---|---|---|---|---|---|
| $\mathbf{B_1}$ | = | $\frac{1}{2}\,\mathbf{a_3}$ | = | $\frac{1}{2}\,c\,\mathbf{\hat{z}}$ | $(1b)$ | P I |
| $\mathbf{B_2}$ | = | $\frac{1}{3}\,\mathbf{a_1} + \frac{2}{3}\,\mathbf{a_2} + z_2\,\mathbf{a_3}$ | = | $\frac{1}{2}\,a\,\mathbf{\hat{x}} + \frac{1}{2\sqrt{3}}\,a\,\mathbf{\hat{y}} + z_2\,c\,\mathbf{\hat{z}}$ | $(2d)$ | P II |
| $\mathbf{B_3}$ | = | $\frac{2}{3}\,\mathbf{a_1} + \frac{1}{3}\,\mathbf{a_2} - z_2\,\mathbf{a_3}$ | = | $\frac{1}{2}\,a\,\mathbf{\hat{x}} - \frac{1}{2\sqrt{3}}\,a\,\mathbf{\hat{y}} - z_2\,c\,\mathbf{\hat{z}}$ | $(2d)$ | P II |
| $\mathbf{B_4}$ | = | $x_3\,\mathbf{a_1}$ | = | $\frac{1}{2}\,x_3\,a\,\mathbf{\hat{x}} - \frac{\sqrt{3}}{2}\,x_3\,a\,\mathbf{\hat{y}}$ | $(3e)$ | Fe I |
| $\mathbf{B_5}$ | = | $x_3\,\mathbf{a_2}$ | = | $\frac{1}{2}\,x_3\,a\,\mathbf{\hat{x}} + \frac{\sqrt{3}}{2}\,x_3\,a\,\mathbf{\hat{y}}$ | $(3e)$ | Fe I |
| $\mathbf{B_6}$ | = | $-x_3\,\mathbf{a_1} - x_3\,\mathbf{a_2}$ | = | $-x_3\,a\,\mathbf{\hat{x}}$ | $(3e)$ | Fe I |
| $\mathbf{B_7}$ | = | $x_4\,\mathbf{a_1} + \frac{1}{2}\,\mathbf{a_3}$ | = | $\frac{1}{2}\,x_4\,a\,\mathbf{\hat{x}} - \frac{\sqrt{3}}{2}\,x_4\,a\,\mathbf{\hat{y}} + \frac{1}{2}\,c\,\mathbf{\hat{z}}$ | $(3f)$ | Fe II |
| $\mathbf{B_8}$ | = | $x_4\,\mathbf{a_2} + \frac{1}{2}\,\mathbf{a_3}$ | = | $\frac{1}{2}\,x_4\,a\,\mathbf{\hat{x}} + \frac{\sqrt{3}}{2}\,x_4\,a\,\mathbf{\hat{y}} + \frac{1}{2}\,c\,\mathbf{\hat{z}}$ | $(3f)$ | Fe II |
| $\mathbf{B_9}$ | = | $-x_4\,\mathbf{a_1} - x_4\,\mathbf{a_2} + \frac{1}{2}\,\mathbf{a_3}$ | = | $-x_4\,a\,\mathbf{\hat{x}} + \frac{1}{2}\,c\,\mathbf{\hat{z}}$ | $(3f)$ | Fe II |

---

---

**Geometry files:**

- CIF: pp. 715
- POSCAR: pp. 716

# $CrCl_3$ ($D0_4$) Crystal Structure: A3B_hP24_151_3c_2a

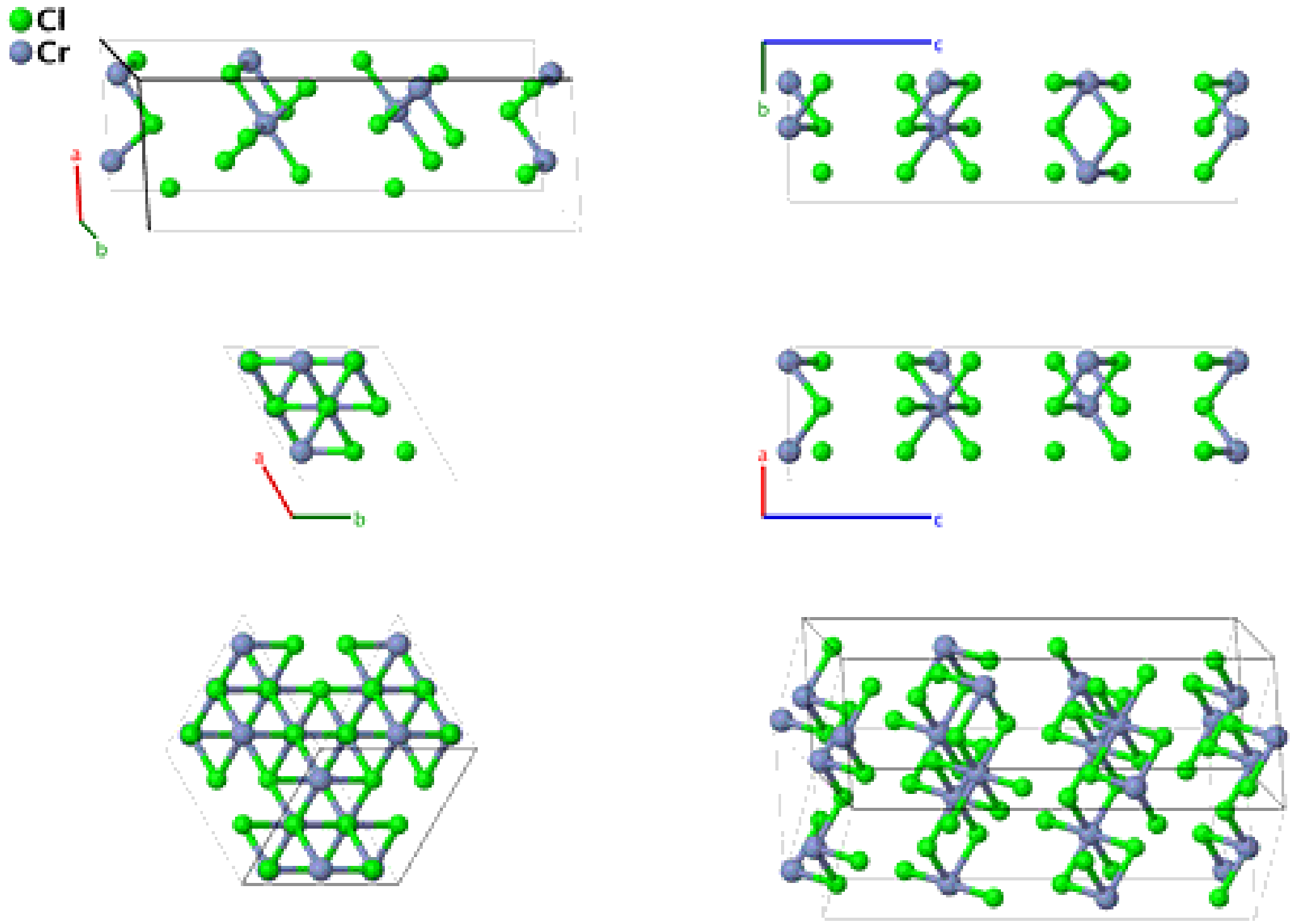

**Prototype** : $CrCl_3$

**AFLOW prototype label** : A3B_hP24_151_3c_2a

***Strukturbericht* designation** : $D0_4$

**Pearson symbol** : hP24

**Space group number** : 151

**Space group symbol** : $P3_112$

**AFLOW prototype command** : `aflow --proto=A3B_hP24_151_3c_2a`
`--params=`$a, c/a, x_1, x_2, x_3, y_3, z_3, x_4, y_4, z_4, x_5, y_5, z_5$

---

**Trigonal Hexagonal primitive vectors:**

$$\mathbf{a}_1 = \frac{1}{2}\, a\, \mathbf{\hat{x}} - \frac{\sqrt{3}}{2}\, a\, \mathbf{\hat{y}}$$
$$\mathbf{a}_2 = \frac{1}{2}\, a\, \mathbf{\hat{x}} + \frac{\sqrt{3}}{2}\, a\, \mathbf{\hat{y}}$$
$$\mathbf{a}_3 = c\, \mathbf{\hat{z}}$$

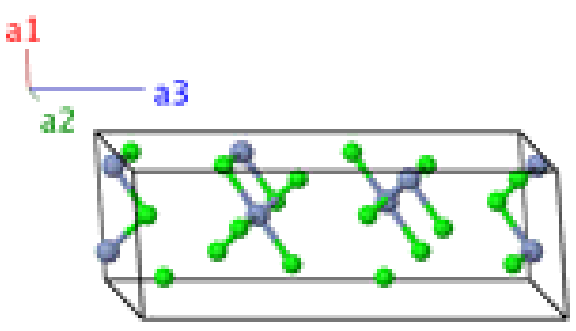

---

**Basis vectors:**

| | | Lattice Coordinates | | Cartesian Coordinates | Wyckoff Position | Atom Type |
|---|---|---|---|---|---|---|
| $\mathbf{B}_1$ | = | $x_1\,\mathbf{a}_1 - x_1\,\mathbf{a}_2 + \frac{1}{3}\,\mathbf{a}_3$ | = | $-\sqrt{3}\,x_1\,a\,\mathbf{\hat{y}} + \frac{1}{3}\,c\,\mathbf{\hat{z}}$ | (3a) | Cr I |
| $\mathbf{B}_2$ | = | $x_1\,\mathbf{a}_1 + 2x_1\,\mathbf{a}_2 + \frac{2}{3}\,\mathbf{a}_3$ | = | $\frac{3}{2}\,x_1\,a\,\mathbf{\hat{x}} + \frac{\sqrt{3}}{2}\,x_1\,a\,\mathbf{\hat{y}} + \frac{2}{3}\,c\,\mathbf{\hat{z}}$ | (3a) | Cr I |
| $\mathbf{B}_3$ | = | $-2x_1\,\mathbf{a}_1 - x_1\,\mathbf{a}_2$ | = | $-\frac{3}{2}\,x_1\,a\,\mathbf{\hat{x}} + \frac{\sqrt{3}}{2}\,x_1\,a\,\mathbf{\hat{y}}$ | (3a) | Cr I |
| $\mathbf{B}_4$ | = | $x_2\,\mathbf{a}_1 - x_2\,\mathbf{a}_2 + \frac{1}{3}\,\mathbf{a}_3$ | = | $-\sqrt{3}\,x_2\,a\,\mathbf{\hat{y}} + \frac{1}{3}\,c\,\mathbf{\hat{z}}$ | (3a) | Cr II |
| $\mathbf{B}_5$ | = | $x_2\,\mathbf{a}_1 + 2x_2\,\mathbf{a}_2 + \frac{2}{3}\,\mathbf{a}_3$ | = | $\frac{3}{2}\,x_2\,a\,\mathbf{\hat{x}} + \frac{\sqrt{3}}{2}\,x_2\,a\,\mathbf{\hat{y}} + \frac{2}{3}\,c\,\mathbf{\hat{z}}$ | (3a) | Cr II |
| $\mathbf{B_6}$ | = | $-2x_2\,\mathbf{a_1} - x_2\,\mathbf{a_2}$ | = | $-\frac{3}{2}\,x_2\,a\,\mathbf{\hat{x}} + \frac{\sqrt{3}}{2}\,x_2\,a\,\mathbf{\hat{y}}$ | (3a) | Cr II |
| $\mathbf{B_7}$ | = | $x_3\,\mathbf{a_1} + y_3\,\mathbf{a_2} + z_3\,\mathbf{a_3}$ | = | $\frac{1}{2}\,(x_3 + y_3)\,a\,\mathbf{\hat{x}} + \frac{\sqrt{3}}{2}\,(y_3 - x_3)\,a\,\mathbf{\hat{y}} + z_3\,c\,\mathbf{\hat{z}}$ | (6c) | Cl I |
| $\mathbf{B_8}$ | = | $-y_3\,\mathbf{a_1} + (x_3 - y_3)\,\mathbf{a_2} + \left(\frac{1}{3} + z_3\right)\,\mathbf{a_3}$ | = | $\frac{1}{2}\,(x_3 - 2y_3)\,a\,\mathbf{\hat{x}} + \frac{\sqrt{3}}{2}\,x_3\,a\,\mathbf{\hat{y}} + \left(\frac{1}{3} + z_3\right)\,c\,\mathbf{\hat{z}}$ | (6c) | Cl I |
| $\mathbf{B_9}$ | = | $(y_3 - x_3)\,\mathbf{a_1} - x_3\,\mathbf{a_2} + \left(\frac{2}{3} + z_3\right)\,\mathbf{a_3}$ | = | $\frac{1}{2}\,(y_3 - 2x_3)\,a\,\mathbf{\hat{x}} - \frac{\sqrt{3}}{2}\,y_3\,a\,\mathbf{\hat{y}} + \left(\frac{2}{3} + z_3\right)\,c\,\mathbf{\hat{z}}$ | (6c) | Cl I |
| $\mathbf{B_{10}}$ | = | $-y_3\,\mathbf{a_1} - x_3\,\mathbf{a_2} + \left(\frac{2}{3} - z_3\right)\,\mathbf{a_3}$ | = | $-\frac{1}{2}\,(x_3 + y_3)\,a\,\mathbf{\hat{x}} + \frac{\sqrt{3}}{2}\,(y_3 - x_3)\,a\,\mathbf{\hat{y}} + \left(\frac{2}{3} - z_3\right)\,c\,\mathbf{\hat{z}}$ | (6c) | Cl I |
| $\mathbf{B_{11}}$ | = | $(y_3 - x_3)\,\mathbf{a_1} + y_3\,\mathbf{a_2} + \left(\frac{1}{3} - z_3\right)\,\mathbf{a_3}$ | = | $\frac{1}{2}\,(2y_3 - x_3)\,a\,\mathbf{\hat{x}} + \frac{\sqrt{3}}{2}\,x_3\,a\,\mathbf{\hat{y}} + \left(\frac{1}{3} - z_3\right)\,c\,\mathbf{\hat{z}}$ | (6c) | Cl I |
| $\mathbf{B_{12}}$ | = | $x_3\,\mathbf{a_1} + (x_3 - y_3)\,\mathbf{a_2} - z_3\,\mathbf{a_3}$ | = | $\frac{1}{2}\,(2x_3 - y_3)\,a\,\mathbf{\hat{x}} - \frac{\sqrt{3}}{2}\,y_3\,a\,\mathbf{\hat{y}} - z_3\,c\,\mathbf{\hat{z}}$ | (6c) | Cl I |
| $\mathbf{B_{13}}$ | = | $x_4\,\mathbf{a_1} + y_4\,\mathbf{a_2} + z_4\,\mathbf{a_3}$ | = | $\frac{1}{2}\,(x_4 + y_4)\,a\,\mathbf{\hat{x}} + \frac{\sqrt{3}}{2}\,(y_4 - x_4)\,a\,\mathbf{\hat{y}} + z_4\,c\,\mathbf{\hat{z}}$ | (6c) | Cl II |
| $\mathbf{B_{14}}$ | = | $-y_4\,\mathbf{a_1} + (x_4 - y_4)\,\mathbf{a_2} + \left(\frac{1}{3} + z_4\right)\,\mathbf{a_3}$ | = | $\frac{1}{2}\,(x_4 - 2y_4)\,a\,\mathbf{\hat{x}} + \frac{\sqrt{3}}{2}\,x_4\,a\,\mathbf{\hat{y}} + \left(\frac{1}{3} + z_4\right)\,c\,\mathbf{\hat{z}}$ | (6c) | Cl II |
| $\mathbf{B_{15}}$ | = | $(y_4 - x_4)\,\mathbf{a_1} - x_4\,\mathbf{a_2} + \left(\frac{2}{3} + z_4\right)\,\mathbf{a_3}$ | = | $\frac{1}{2}\,(y_4 - 2x_4)\,a\,\mathbf{\hat{x}} - \frac{\sqrt{3}}{2}\,y_4\,a\,\mathbf{\hat{y}} + \left(\frac{2}{3} + z_4\right)\,c\,\mathbf{\hat{z}}$ | (6c) | Cl II |
| $\mathbf{B_{16}}$ | = | $-y_4\,\mathbf{a_1} - x_4\,\mathbf{a_2} + \left(\frac{2}{3} - z_4\right)\,\mathbf{a_3}$ | = | $-\frac{1}{2}\,(x_4 + y_4)\,a\,\mathbf{\hat{x}} + \frac{\sqrt{3}}{2}\,(y_4 - x_4)\,a\,\mathbf{\hat{y}} + \left(\frac{2}{3} - z_4\right)\,c\,\mathbf{\hat{z}}$ | (6c) | Cl II |
| $\mathbf{B_{17}}$ | = | $(y_4 - x_4)\,\mathbf{a_1} + y_4\,\mathbf{a_2} + \left(\frac{1}{3} - z_4\right)\,\mathbf{a_3}$ | = | $\frac{1}{2}\,(2y_4 - x_4)\,a\,\mathbf{\hat{x}} + \frac{\sqrt{3}}{2}\,x_4\,a\,\mathbf{\hat{y}} + \left(\frac{1}{3} - z_4\right)\,c\,\mathbf{\hat{z}}$ | (6c) | Cl II |
| $\mathbf{B_{18}}$ | = | $x_4\,\mathbf{a_1} + (x_4 - y_4)\,\mathbf{a_2} - z_4\,\mathbf{a_3}$ | = | $\frac{1}{2}\,(2x_4 - y_4)\,a\,\mathbf{\hat{x}} - \frac{\sqrt{3}}{2}\,y_4\,a\,\mathbf{\hat{y}} - z_4\,c\,\mathbf{\hat{z}}$ | (6c) | Cl II |
| $\mathbf{B_{19}}$ | = | $x_5\,\mathbf{a_1} + y_5\,\mathbf{a_2} + z_5\,\mathbf{a_3}$ | = | $\frac{1}{2}\,(x_5 + y_5)\,a\,\mathbf{\hat{x}} + \frac{\sqrt{3}}{2}\,(y_5 - x_5)\,a\,\mathbf{\hat{y}} + z_5\,c\,\mathbf{\hat{z}}$ | (6c) | Cl III |
| $\mathbf{B_{20}}$ | = | $-y_5\,\mathbf{a_1} + (x_5 - y_5)\,\mathbf{a_2} + \left(\frac{1}{3} + z_5\right)\,\mathbf{a_3}$ | = | $\frac{1}{2}\,(x_5 - 2y_5)\,a\,\mathbf{\hat{x}} + \frac{\sqrt{3}}{2}\,x_5\,a\,\mathbf{\hat{y}} + \left(\frac{1}{3} + z_5\right)\,c\,\mathbf{\hat{z}}$ | (6c) | Cl III |
| $\mathbf{B_{21}}$ | = | $(y_5 - x_5)\,\mathbf{a_1} - x_5\,\mathbf{a_2} + \left(\frac{2}{3} + z_5\right)\,\mathbf{a_3}$ | = | $\frac{1}{2}\,(y_5 - 2x_5)\,a\,\mathbf{\hat{x}} - \frac{\sqrt{3}}{2}\,y_5\,a\,\mathbf{\hat{y}} + \left(\frac{2}{3} + z_5\right)\,c\,\mathbf{\hat{z}}$ | (6c) | Cl III |
| $\mathbf{B_{22}}$ | = | $-y_5\,\mathbf{a_1} - x_5\,\mathbf{a_2} + \left(\frac{2}{3} - z_5\right)\,\mathbf{a_3}$ | = | $-\frac{1}{2}\,(x_5 + y_5)\,a\,\mathbf{\hat{x}} + \frac{\sqrt{3}}{2}\,(y_5 - x_5)\,a\,\mathbf{\hat{y}} + \left(\frac{2}{3} - z_5\right)\,c\,\mathbf{\hat{z}}$ | (6c) | Cl III |
| $\mathbf{B_{23}}$ | = | $(y_5 - x_5)\,\mathbf{a_1} + y_5\,\mathbf{a_2} + \left(\frac{1}{3} - z_5\right)\,\mathbf{a_3}$ | = | $\frac{1}{2}\,(2y_5 - x_5)\,a\,\mathbf{\hat{x}} + \frac{\sqrt{3}}{2}\,x_5\,a\,\mathbf{\hat{y}} + \left(\frac{1}{3} - z_5\right)\,c\,\mathbf{\hat{z}}$ | (6c) | Cl III |
| $\mathbf{B_{24}}$ | = | $x_5\,\mathbf{a_1} + (x_5 - y_5)\,\mathbf{a_2} - z_5\,\mathbf{a_3}$ | = | $\frac{1}{2}\,(2x_5 - y_5)\,a\,\mathbf{\hat{x}} - \frac{\sqrt{3}}{2}\,y_5\,a\,\mathbf{\hat{y}} - z_5\,c\,\mathbf{\hat{z}}$ | (6c) | Cl III |

**Geometry files:**

- CIF: pp. 716
- POSCAR: pp. 716

# $\alpha$-Quartz (low Quartz) Structure: A2B_hP9_152_c_a

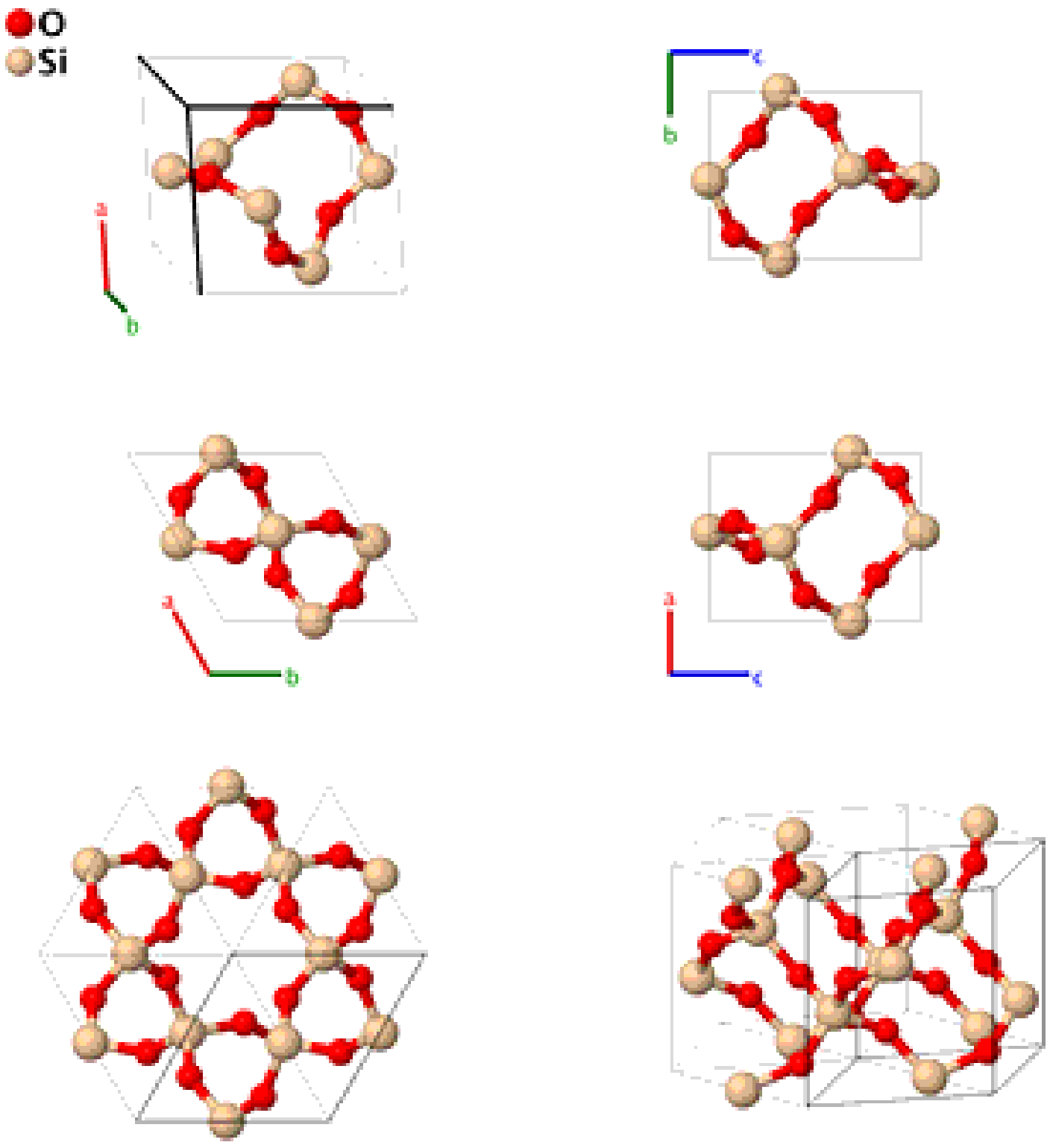

**Prototype** : $SiO_2$

**AFLOW prototype label** : A2B_hP9_152_c_a

***Strukturbericht* designation** : None

**Pearson symbol** : hP9

**Space group number** : 152

**Space group symbol** : $P3_121$

**AFLOW prototype command** : `aflow --proto=A2B_hP9_152_c_a`
`--params=`$a, c/a, x_1, x_2, y_2, z_2$

---

- When $x_1 = 1/2$, $y_2 = 2x_2$, and $z_2 = 1/2$, this tranforms into the high quartz (C8) structure. This structure is sometimes given using the enantiomorphic space groups $P3_221$ (#154).

---

**Trigonal Hexagonal primitive vectors:**

$$\mathbf{a}_1 = \frac{1}{2}\,a\,\hat{\mathbf{x}} - \frac{\sqrt{3}}{2}\,a\,\hat{\mathbf{y}}$$

$$\mathbf{a}_2 = \frac{1}{2}\,a\,\hat{\mathbf{x}} + \frac{\sqrt{3}}{2}\,a\,\hat{\mathbf{y}}$$

$$\mathbf{a}_3 = c\,\hat{\mathbf{z}}$$

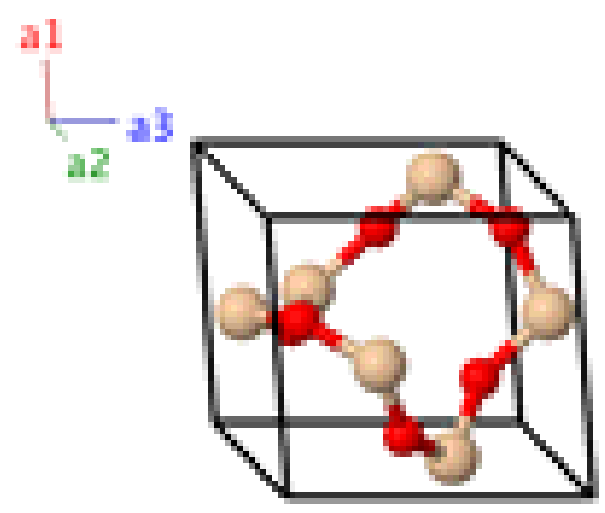

---

**Basis vectors:**

| | | Lattice Coordinates | | Cartesian Coordinates | Wyckoff Position | Atom Type |
|---|---|---|---|---|---|---|
| $\mathbf{B_1}$ | = | $x_1\,\mathbf{a_1} + \frac{1}{3}\,\mathbf{a_3}$ | = | $\frac{1}{2}\,x_1\,a\,\mathbf{\hat{x}} - \frac{\sqrt{3}}{2}\,x_1\,a\,\mathbf{\hat{y}} + \frac{1}{3}\,c\,\mathbf{\hat{z}}$ | $(3a)$ | Si |
| $\mathbf{B_2}$ | = | $x_1\,\mathbf{a_2} + \frac{2}{3}\,\mathbf{a_3}$ | = | $\frac{1}{2}\,x_1\,a\,\mathbf{\hat{x}} + \frac{\sqrt{3}}{2}\,x_1\,a\,\mathbf{\hat{y}} + \frac{2}{3}\,c\,\mathbf{\hat{z}}$ | $(3a)$ | Si |
| $\mathbf{B_3}$ | = | $-x_1\,\mathbf{a_1} - x_1\,\mathbf{a_2}$ | = | $-x_1\,a\,\mathbf{\hat{x}}$ | $(3a)$ | Si |
| $\mathbf{B_4}$ | = | $x_2\,\mathbf{a_1} + y_2\,\mathbf{a_2} + z_2\,\mathbf{a_3}$ | = | $\frac{1}{2}\,(x_2 + y_2)\,a\,\mathbf{\hat{x}} + \frac{\sqrt{3}}{2}\,(y_2 - x_2)\,a\,\mathbf{\hat{y}} + z_2\,c\,\mathbf{\hat{z}}$ | $(6c)$ | O |
| $\mathbf{B_5}$ | = | $-y_2\,\mathbf{a_1} + (x_2 - y_2)\,\mathbf{a_2} + \left(\frac{1}{3} + z_2\right)\,\mathbf{a_3}$ | = | $\frac{1}{2}\,(x_2 - 2y_2)\,a\,\mathbf{\hat{x}} + \frac{\sqrt{3}}{2}x_2\,a\,\mathbf{\hat{y}} + \left(\frac{1}{3} + z_2\right)\,c\,\mathbf{\hat{z}}$ | $(6c)$ | O |
| $\mathbf{B_6}$ | = | $(y_2 - x_2)\,\mathbf{a_1} - x_2\,\mathbf{a_2} + \left(\frac{2}{3} + z_2\right)\,\mathbf{a_3}$ | = | $\frac{1}{2}\,(y_2 - 2x_2)\,a\,\mathbf{\hat{x}} - \frac{\sqrt{3}}{2}y_2\,a\,\mathbf{\hat{y}} + \left(\frac{2}{3} + z_2\right)\,c\,\mathbf{\hat{z}}$ | $(6c)$ | O |
| $\mathbf{B_7}$ | = | $y_2\,\mathbf{a_1} + x_2\,\mathbf{a_2} - z_2\,\mathbf{a_3}$ | = | $\frac{1}{2}\,(x_2 + y_2)\,a\,\mathbf{\hat{x}} + \frac{\sqrt{3}}{2}\,(x_2 - y_2)\,a\,\mathbf{\hat{y}} - z_2\,c\,\mathbf{\hat{z}}$ | $(6c)$ | O |
| $\mathbf{B_8}$ | = | $(x_2 - y_2)\,\mathbf{a_1} - y_2\,\mathbf{a_2} + \left(\frac{2}{3} - z_2\right)\,\mathbf{a_3}$ | = | $\frac{1}{2}\,(x_2 - 2y_2)\,a\,\mathbf{\hat{x}} - \frac{\sqrt{3}}{2}x_2\,a\,\mathbf{\hat{y}} + \left(\frac{2}{3} - z_2\right)\,c\,\mathbf{\hat{z}}$ | $(6c)$ | O |
| $\mathbf{B_9}$ | = | $-x_2\,\mathbf{a_1} + (y_2 - x_2)\,\mathbf{a_2} + \left(\frac{1}{3} - z_2\right)\,\mathbf{a_3}$ | = | $\frac{1}{2}\,(y_2 - 2x_2)\,a\,\mathbf{\hat{x}} + \frac{\sqrt{3}}{2}y_2\,a\,\mathbf{\hat{y}} + \left(\frac{1}{3} - z_2\right)\,c\,\mathbf{\hat{z}}$ | $(6c)$ | O |

---

**Found in:**

---

**Geometry files:**

- CIF: pp. 716
- POSCAR: pp. 717

# $\gamma$-Se (A8) Structure: A_hP3_152_a

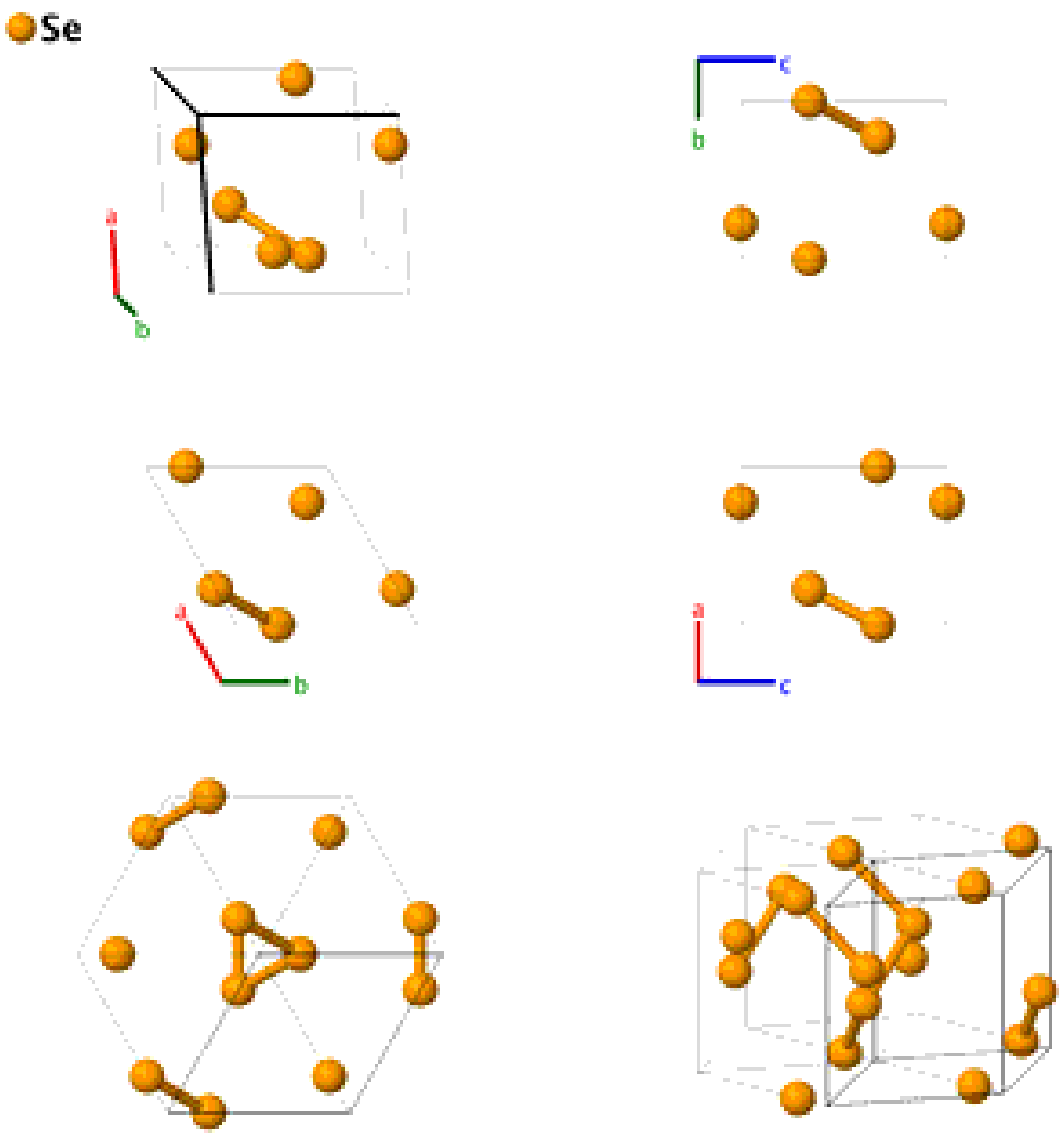

| | | |
|---|---|---|
| **Prototype** | : | $\gamma$-Se |
| **AFLOW prototype label** | : | A_hP3_152_a |
| ***Strukturbericht* designation** | : | A8 |
| **Pearson symbol** | : | hP3 |
| **Space group number** | : | 152 |
| **Space group symbol** | : | $P3_121$ |
| **AFLOW prototype command** | : | `aflow --proto=A_hP3_152_a`<br>`--params=`$a, c/a, x_1$ |

**Other compounds with this structure:**

- Te, SeTe, $Se_3Te$

- (Donohue, 1982) refers to this as the $\alpha$-Se structure, calling what we note as $\alpha$-Se and $\beta$-Se as "monoclinic $\alpha$" and "monoclinic $\beta$," respectively. When $x = 1/3$ this reduces to the $A_i$ ($\beta$-Po) or A10 ($\alpha$-Hg) structure. If, in addition, $c = \sqrt{6}a$, then the structure becomes fcc (A1). On the other hand, if $c = \sqrt{3/2}a$, then the structure becomes simple cubic ($A_h$).

**Trigonal Hexagonal primitive vectors:**

$$
\begin{aligned}
\mathbf{a}_1 &= \tfrac{1}{2}\,a\,\mathbf{\hat{x}} - \tfrac{\sqrt{3}}{2}\,a\,\mathbf{\hat{y}} \\
\mathbf{a}_2 &= \tfrac{1}{2}\,a\,\mathbf{\hat{x}} + \tfrac{\sqrt{3}}{2}\,a\,\mathbf{\hat{y}} \\
\mathbf{a}_3 &= c\,\mathbf{\hat{z}}
\end{aligned}
$$

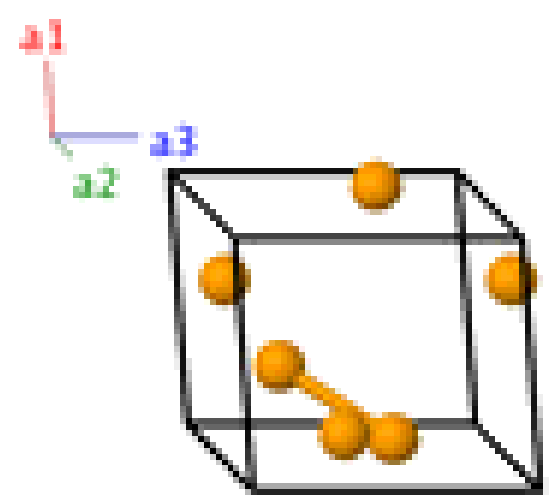

**Basis vectors:**

| | | Lattice Coordinates | | Cartesian Coordinates | Wyckoff Position | Atom Type |
|---|---|---|---|---|---|---|
| $\mathbf{B_1}$ | = | $x_1\,\mathbf{a_1} + \frac{1}{3}\,\mathbf{a_3}$ | = | $\frac{1}{2}\,x_1\,a\,\mathbf{\hat{x}} - \frac{\sqrt{3}}{2}\,x_1\,a\,\mathbf{\hat{y}} + \frac{1}{3}\,c\,\mathbf{\hat{z}}$ | $(3a)$ | Se |
| $\mathbf{B_2}$ | = | $x_1\,\mathbf{a_2} + \frac{2}{3}\,\mathbf{a_3}$ | = | $\frac{1}{2}\,x_1\,a\,\mathbf{\hat{x}} + \frac{\sqrt{3}}{2}\,x_1\,a\,\mathbf{\hat{y}} + \frac{2}{3}\,c\,\mathbf{\hat{z}}$ | $(3a)$ | Se |
| $\mathbf{B_3}$ | = | $-x_1\,\mathbf{a_1} - x_1\,\mathbf{a_2}$ | = | $-x_1\,a\,\mathbf{\hat{x}}$ | $(3a)$ | Se |

**Found in:**

**Geometry files:**

- CIF: pp. 717
- POSCAR: pp. 717

# Cinnabar (B9) Structure: AB_hP6_154_a_b

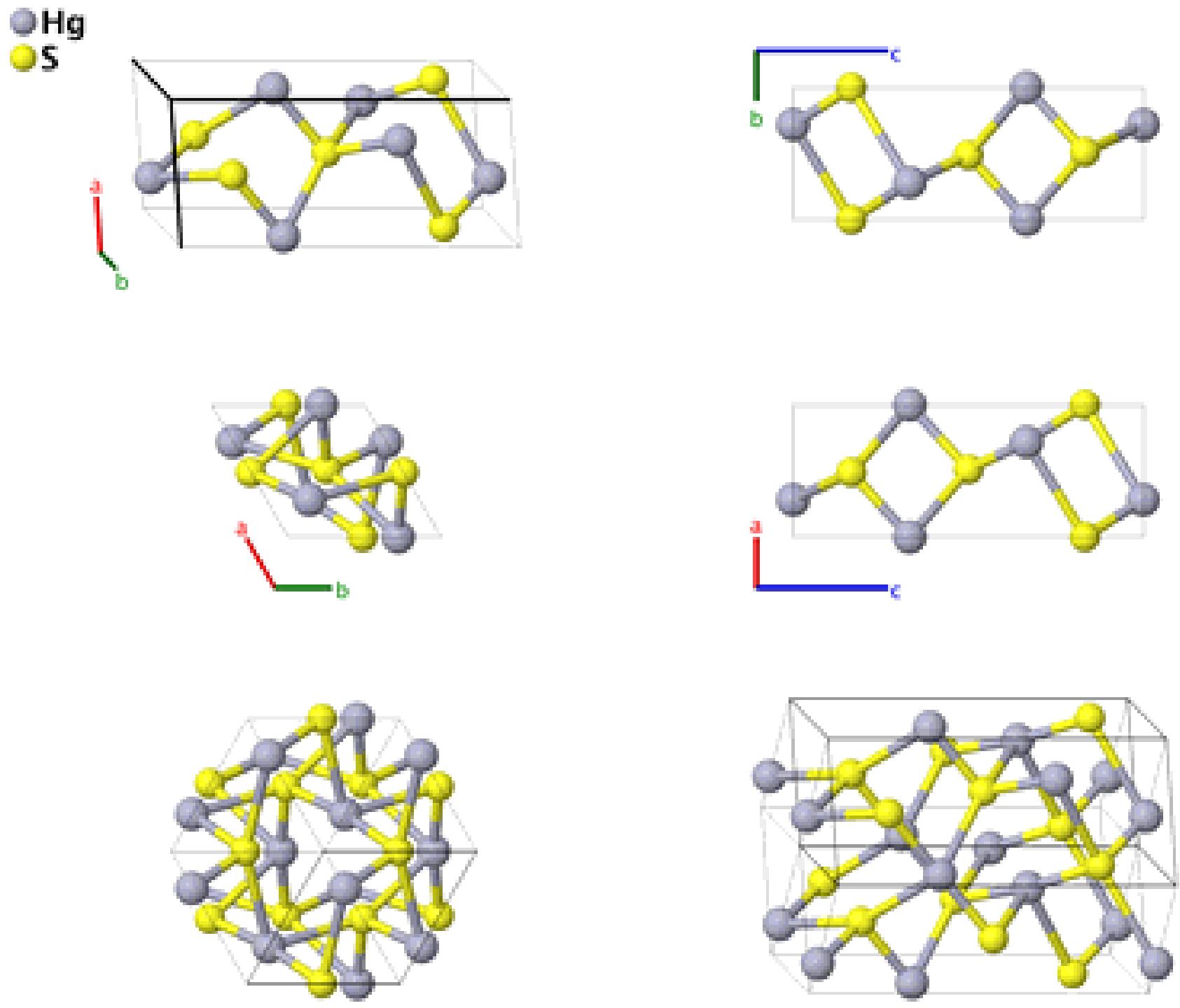

| | | |
|---|---|---|
| **Prototype** | : | HgS |
| **AFLOW prototype label** | : | AB_hP6_154_a_b |
| ***Strukturbericht* designation** | : | B9 |
| **Pearson symbol** | : | hP6 |
| **Space group number** | : | 154 |
| **Space group symbol** | : | $P3_221$ |
| **AFLOW prototype command** | : | `aflow --proto=AB_hP6_154_a_b`<br>`--params=`$a, c/a, x_1, x_2$ |

**Other compounds with this structure:**

- HgO

**Trigonal Hexagonal primitive vectors:**

$$\mathbf{a}_1 = \frac{1}{2}\,a\,\hat{\mathbf{x}} - \frac{\sqrt{3}}{2}\,a\,\hat{\mathbf{y}}$$
$$\mathbf{a}_2 = \frac{1}{2}\,a\,\hat{\mathbf{x}} + \frac{\sqrt{3}}{2}\,a\,\hat{\mathbf{y}}$$
$$\mathbf{a}_3 = c\,\hat{\mathbf{z}}$$

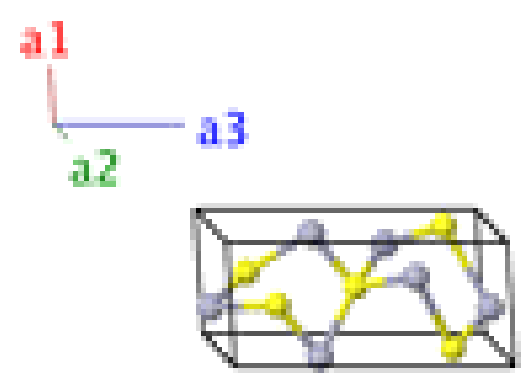

**Basis vectors:**

| | | Lattice Coordinates | | Cartesian Coordinates | Wyckoff Position | Atom Type |
|---|---|---|---|---|---|---|
| $\mathbf{B}_1$ | = | $x_1\,\mathbf{a}_1 + \frac{2}{3}\,\mathbf{a}_3$ | = | $\frac{1}{2}\,x_1\,a\,\hat{\mathbf{x}} - \frac{\sqrt{3}}{2}\,x_1\,a\,\hat{\mathbf{y}} + \frac{2}{3}\,c\,\hat{\mathbf{z}}$ | (3a) | Hg |
| $\mathbf{B_2}$ | = | $x_1\,\mathbf{a_2} + \frac{1}{3}\,\mathbf{a_3}$ | = | $\frac{1}{2}\,x_1\,a\,\mathbf{\hat{x}} + \frac{\sqrt{3}}{2}\,x_1\,a\,\mathbf{\hat{y}} + \frac{1}{3}\,c\,\mathbf{\hat{z}}$ | (3$a$) | Hg |
| $\mathbf{B_3}$ | = | $-x_1\,\mathbf{a_1} - x_1\,\mathbf{a_2}$ | = | $-x_1\,a\,\mathbf{\hat{x}}$ | (3$a$) | Hg |
| $\mathbf{B_4}$ | = | $x_2\,\mathbf{a_1} + \frac{1}{6}\,\mathbf{a_3}$ | = | $\frac{1}{2}\,x_2\,a\,\mathbf{\hat{x}} - \frac{\sqrt{3}}{2}\,x_2\,a\,\mathbf{\hat{y}} + \frac{1}{6}\,c\,\mathbf{\hat{z}}$ | (3$b$) | S |
| $\mathbf{B_5}$ | = | $x_2\,\mathbf{a_2} + \frac{5}{6}\,\mathbf{a_3}$ | = | $\frac{1}{2}\,x_2\,a\,\mathbf{\hat{x}} + \frac{\sqrt{3}}{2}\,x_2\,a\,\mathbf{\hat{y}} + \frac{5}{6}\,c\,\mathbf{\hat{z}}$ | (3$b$) | S |
| $\mathbf{B_6}$ | = | $-x_2\,\mathbf{a_1} - x_2\,\mathbf{a_2} + \frac{1}{2}\,\mathbf{a_3}$ | = | $-x_2\,a\,\mathbf{\hat{x}} + \frac{1}{2}\,c\,\mathbf{\hat{z}}$ | (3$b$) | S |

**Found in:**

**Geometry files:**

- CIF: pp. 717
- POSCAR: pp. 718

# $AlF_3$ ($D0_{14}$) Structure: AB3_hR8_155_c_de

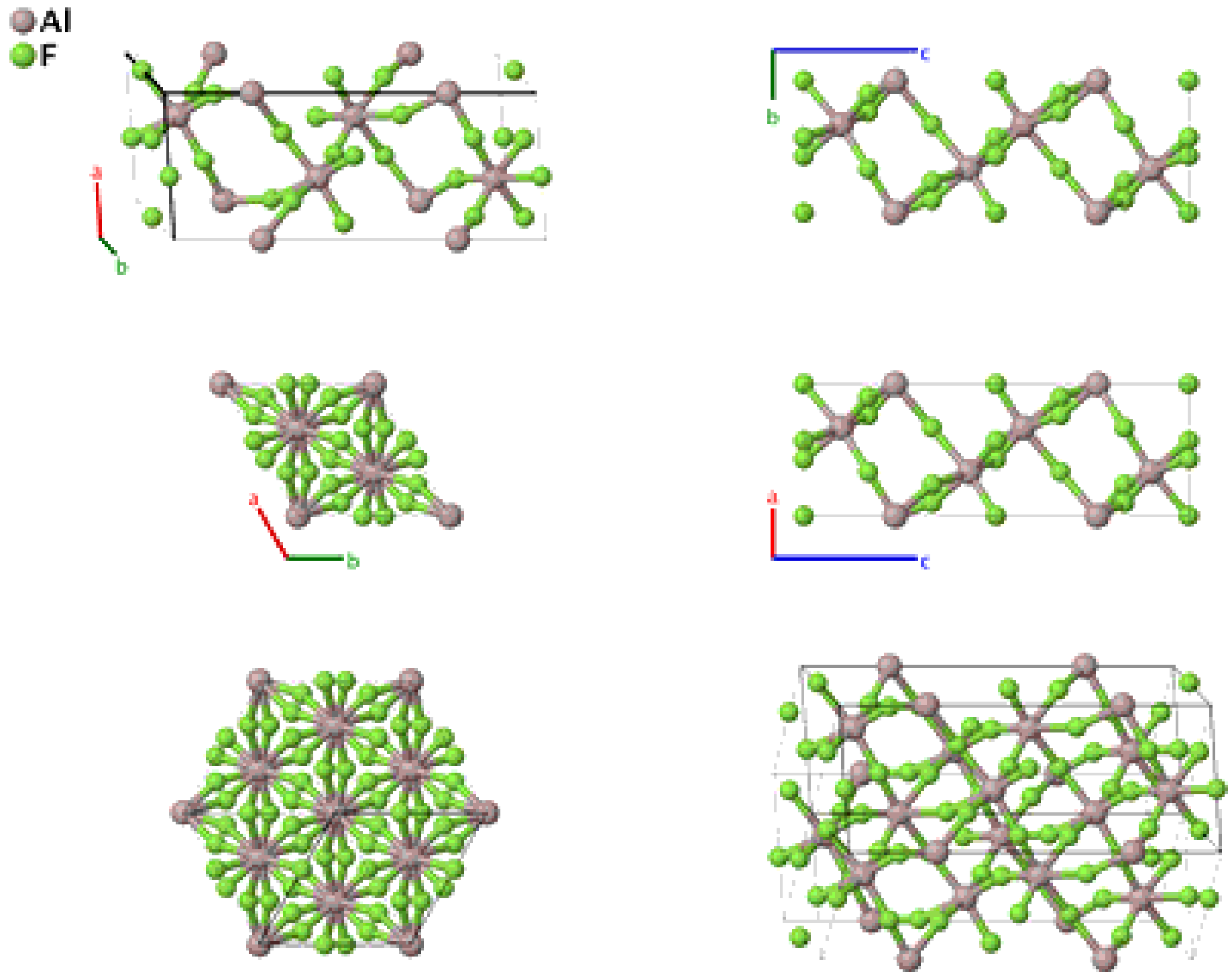

| | | |
|---|---|---|
| **Prototype** | : | $AlF_3$ |
| **AFLOW prototype label** | : | AB3_hR8_155_c_de |
| ***Strukturbericht* designation** | : | $D0_{14}$ |
| **Pearson symbol** | : | hR8 |
| **Space group number** | : | 155 |
| **Space group symbol** | : | R32 |
| **AFLOW prototype command** | : | `aflow --proto=AB3_hR8_155_c_de [--hex]` <br> `--params=`$a, c/a, x_1, y_2, y_3$ |

**Other compounds with this structure:**

- $FeF_3$

- Hexagonal settings of this structure can be obtained with the option `--hex`.

**Rhombohedral primitive vectors:**

$$\mathbf{a}_1 = \frac{1}{2}\,a\,\hat{\mathbf{x}} - \frac{1}{2\sqrt{3}}\,a\,\hat{\mathbf{y}} + \frac{1}{3}\,c\,\hat{\mathbf{z}}$$
$$\mathbf{a}_2 = \frac{1}{\sqrt{3}}\,a\,\hat{\mathbf{y}} + \frac{1}{3}\,c\,\hat{\mathbf{z}}$$
$$\mathbf{a}_3 = -\frac{1}{2}\,a\,\hat{\mathbf{x}} - \frac{1}{2\sqrt{3}}\,a\,\hat{\mathbf{y}} + \frac{1}{3}\,c\,\hat{\mathbf{z}}$$

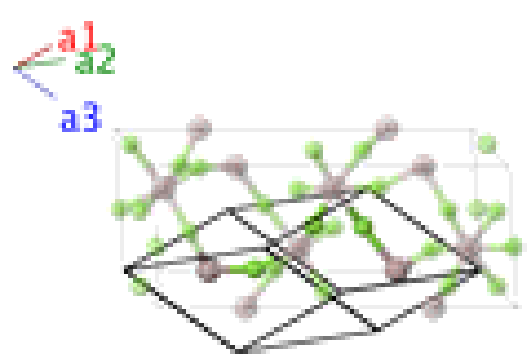

**Basis vectors:**

Lattice Coordinates | Cartesian Coordinates | Wyckoff Position | Atom Type

| | | | | | | |
|---|---|---|---|---|---|---|
| $\mathbf{B_1}$ | = | $x_1\,\mathbf{a_1} + x_1\,\mathbf{a_2} + x_1\,\mathbf{a_3}$ | = | $x_1\,c\,\hat{\mathbf{z}}$ | (2c) | Al |
| $\mathbf{B_2}$ | = | $-x_1\,\mathbf{a_1} - x_1\,\mathbf{a_2} - x_1\,\mathbf{a_3}$ | = | $-x_1\,c\,\hat{\mathbf{z}}$ | (2c) | Al |
| $\mathbf{B_3}$ | = | $y_2\,\mathbf{a_2} - y_2\,\mathbf{a_3}$ | = | $\frac{1}{2}\,y_2\,a\,\hat{\mathbf{x}} + \frac{\sqrt{3}}{2}\,y_2\,a\,\hat{\mathbf{y}}$ | (3d) | F I |
| $\mathbf{B_4}$ | = | $-y_2\,\mathbf{a_1} + y_2\,\mathbf{a_3}$ | = | $-y_2\,a\,\hat{\mathbf{x}}$ | (3d) | F I |
| $\mathbf{B_5}$ | = | $y_2\,\mathbf{a_1} - y_2\,\mathbf{a_2}$ | = | $\frac{1}{2}\,y_2\,a\,\hat{\mathbf{x}} - \frac{\sqrt{3}}{2}\,y_2\,a\,\hat{\mathbf{y}}$ | (3d) | F I |
| $\mathbf{B_6}$ | = | $\frac{1}{2}\,\mathbf{a_1} + y_3\,\mathbf{a_2} - y_3\,\mathbf{a_3}$ | = | $\frac{1}{4}\,(2\,y_3 + 1)\,a\,\hat{\mathbf{x}} + \frac{1}{4\sqrt{3}}\,(6\,y_3 - 1)\,a\,\hat{\mathbf{y}} + \frac{1}{6}\,c\,\hat{\mathbf{z}}$ | (3e) | F II |
| $\mathbf{B_7}$ | = | $-y_3\,\mathbf{a_1} + \frac{1}{2}\,\mathbf{a_2} + y_3\,\mathbf{a_3}$ | = | $-y_3\,a\,\hat{\mathbf{x}} + \frac{1}{2\sqrt{3}}\,a\,\hat{\mathbf{y}} + \frac{1}{6}\,c\,\hat{\mathbf{z}}$ | (3e) | F II |
| $\mathbf{B_8}$ | = | $y_3\,\mathbf{a_1} - y_3\,\mathbf{a_2} + \frac{1}{2}\,\mathbf{a_3}$ | = | $\frac{1}{4}\,(2\,y_3 - 1)\,a\,\hat{\mathbf{x}} - \frac{1}{4\sqrt{3}}\,(6\,y_3 + 1)\,a\,\hat{\mathbf{y}} + \frac{1}{6}\,c\,\hat{\mathbf{z}}$ | (3e) | F II |

**Found in:**

**Geometry files:**

- CIF: pp. 718
- POSCAR: pp. 718

# Hazelwoodite ($Ni_3S_2$, $D5_e$) Structure: A3B2_hR5_155_e_c

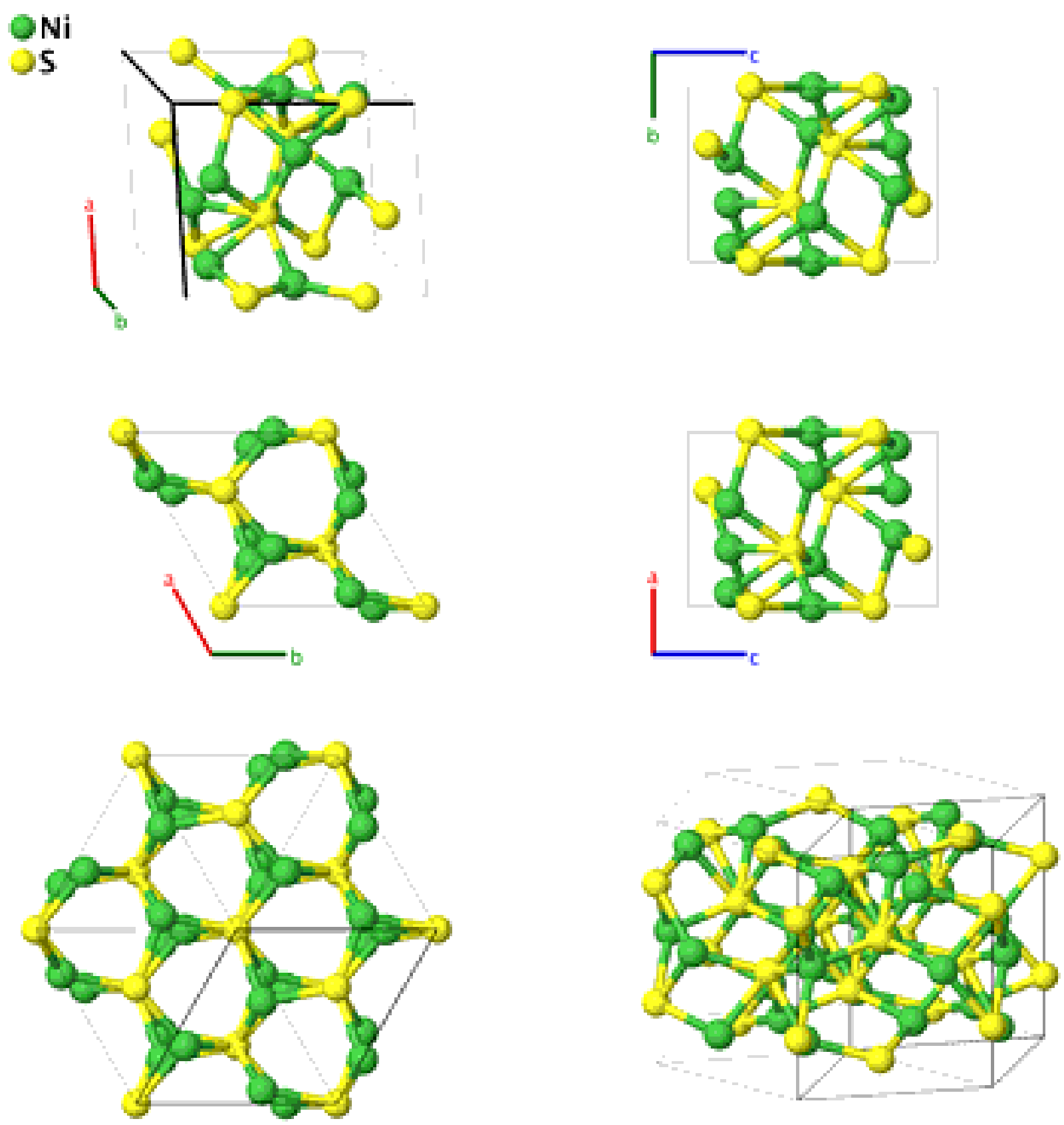

| | | |
|---|---|---|
| **Prototype** | : | $Ni_3S_2$ |
| **AFLOW prototype label** | : | A3B2_hR5_155_e_c |
| ***Strukturbericht* designation** | : | $D5_e$ |
| **Pearson symbol** | : | hR5 |
| **Space group number** | : | 155 |
| **Space group symbol** | : | R32 |
| **AFLOW prototype command** | : | `aflow --proto=A3B2_hR5_155_e_c [--hex]` <br> `--params=`$a, c/a, x_1, y_2$ |

**Other compounds with this structure:**

- $Ni_3Se_2$

- This can be considered as a prototype for a high concentration of ordered vacancies in the hcp structure. We get the ideal hcp atomic positions when $z_1 = 1/3$ and $y_2 = 1/6$, leaving a vacancy at the origin. Hexagonal settings of this structure can be obtained with the option `--hex`.

## Rhombohedral primitive vectors:

$$
\begin{aligned}
\mathbf{a}_1 &= \frac{1}{2}\,a\,\hat{\mathbf{x}} - \frac{1}{2\sqrt{3}}\,a\,\hat{\mathbf{y}} + \frac{1}{3}\,c\,\hat{\mathbf{z}} \\
\mathbf{a}_2 &= \frac{1}{\sqrt{3}}\,a\,\hat{\mathbf{y}} + \frac{1}{3}\,c\,\hat{\mathbf{z}} \\
\mathbf{a}_3 &= -\frac{1}{2}\,a\,\hat{\mathbf{x}} - \frac{1}{2\sqrt{3}}\,a\,\hat{\mathbf{y}} + \frac{1}{3}\,c\,\hat{\mathbf{z}}
\end{aligned}
$$

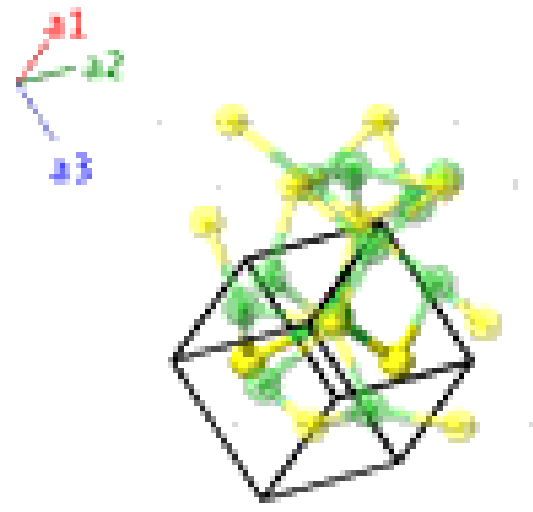

## Basis vectors:

| | | Lattice Coordinates | | Cartesian Coordinates | Wyckoff Position | Atom Type |
|---|---|---|---|---|---|---|
| $\mathbf{B}_1$ | = | $x_1\,\mathbf{a}_1 + x_1\,\mathbf{a}_2 + x_1\,\mathbf{a}_3$ | = | $x_1\,c\,\hat{\mathbf{z}}$ | (2$c$) | S |
| $\mathbf{B}_2$ | = | $-x_1\,\mathbf{a}_1 - x_1\,\mathbf{a}_2 - x_1\,\mathbf{a}_3$ | = | $-x_1\,c\,\hat{\mathbf{z}}$ | (2$c$) | S |
| $\mathbf{B}_3$ | = | $\frac{1}{2}\,\mathbf{a}_1 + y_2\,\mathbf{a}_2 - y_2\,\mathbf{a}_3$ | = | $\frac{1}{4}\,(1 + 2\,y_2)\,a\,\hat{\mathbf{x}} + \frac{1}{4\sqrt{3}}\,(6\,y_2 - 1)\,a\,\hat{\mathbf{y}} + \frac{1}{6}\,c\,\hat{\mathbf{z}}$ | (3$e$) | Ni |
| $\mathbf{B}_4$ | = | $-y_2\,\mathbf{a}_1 + \frac{1}{2}\,\mathbf{a}_2 + y_2\,\mathbf{a}_3$ | = | $-y_2\,a\,\hat{\mathbf{x}} + \frac{1}{2\sqrt{3}}\,a\,\hat{\mathbf{y}} + \frac{1}{6}\,c\,\hat{\mathbf{z}}$ | (3$e$) | Ni |
| $\mathbf{B}_5$ | = | $y_2\,\mathbf{a}_1 - y_2\,\mathbf{a}_2 + \frac{1}{2}\,\mathbf{a}_3$ | = | $\frac{1}{4}\,(2\,y_2 - 1)\,a\,\hat{\mathbf{x}} - \frac{1}{4\sqrt{3}}\,(1 + 6\,y_2)\,a\,\hat{\mathbf{y}} + \frac{1}{6}\,c\,\hat{\mathbf{z}}$ | (3$e$) | Ni |

## Geometry files:

- CIF: pp. 718
- POSCAR: pp. 719

# Millerite (NiS, B13) Structure: AB_hR6_160_b_b

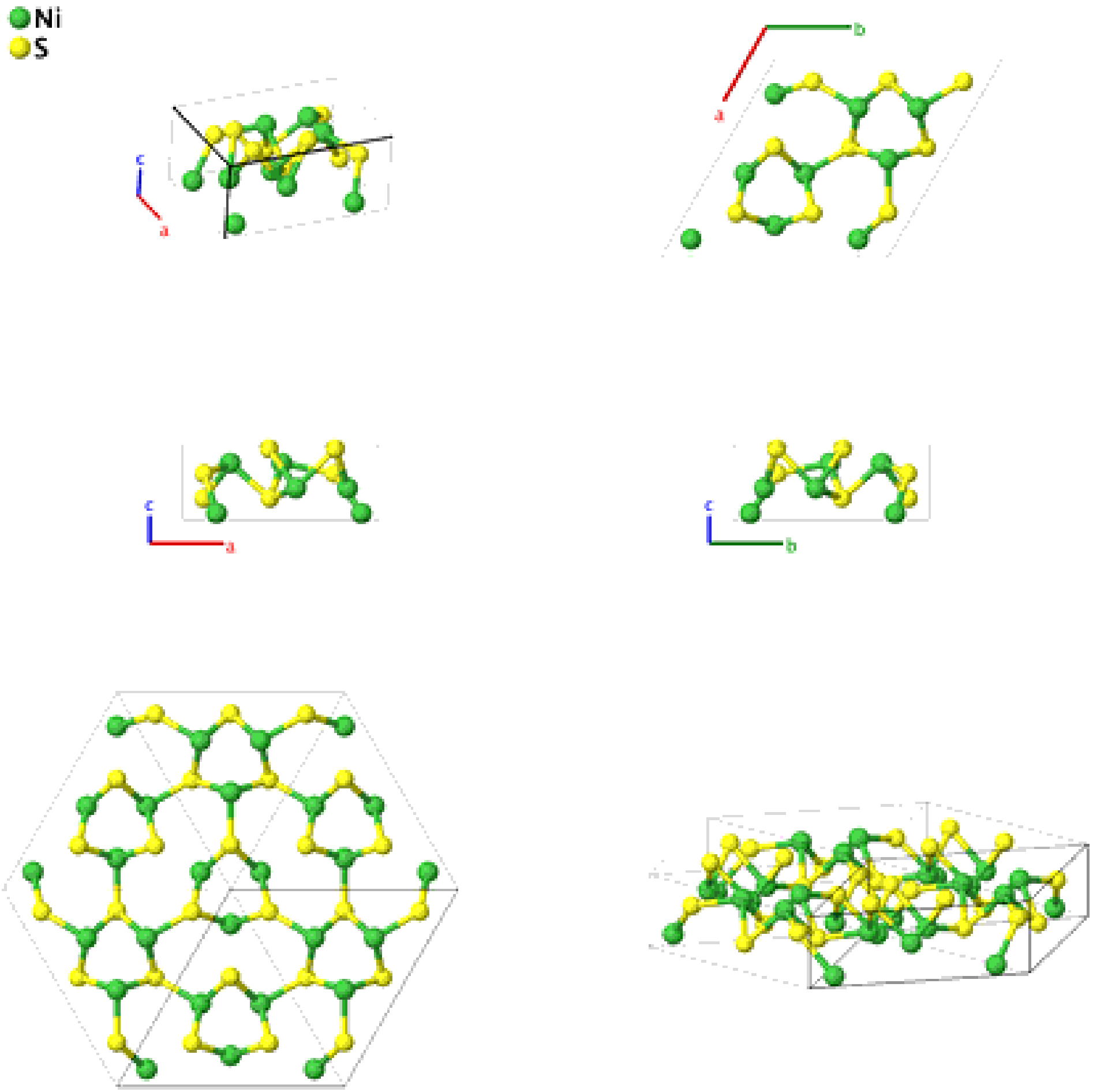

| | | |
|---|---|---|
| **Prototype** | : | NiS |
| **AFLOW prototype label** | : | AB_hR6_160_b_b |
| ***Strukturbericht* designation** | : | B13 |
| **Pearson symbol** | : | hR6 |
| **Space group number** | : | 160 |
| **Space group symbol** | : | R3m |
| **AFLOW prototype command** | : | `aflow --proto=AB_hR6_160_b_b [--hex]`<br>`--params=`$a, c/a, x_1, z_1, x_2, z_2$ |

**Other compounds with this structure:**

- $\beta$-FeS

- Hexagonal settings of this structure can be obtained with the option `--hex`.

**Rhombohedral primitive vectors:**

$$\mathbf{a}_1 = \frac{1}{2}\,a\,\hat{\mathbf{x}} - \frac{1}{2\sqrt{3}}\,a\,\hat{\mathbf{y}} + \frac{1}{3}\,c\,\hat{\mathbf{z}}$$
$$\mathbf{a}_2 = \frac{1}{\sqrt{3}}\,a\,\hat{\mathbf{y}} + \frac{1}{3}\,c\,\hat{\mathbf{z}}$$
$$\mathbf{a}_3 = -\frac{1}{2}\,a\,\hat{\mathbf{x}} - \frac{1}{2\sqrt{3}}\,a\,\hat{\mathbf{y}} + \frac{1}{3}\,c\,\hat{\mathbf{z}}$$

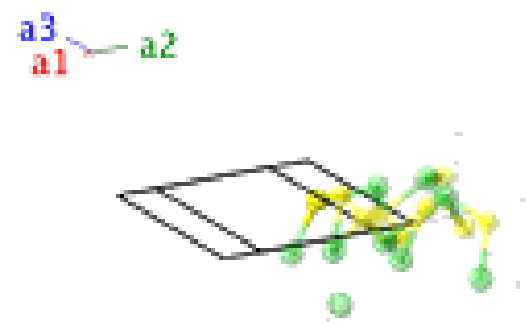

**Basis vectors:**

| | | Lattice Coordinates | | Cartesian Coordinates | Wyckoff Position | Atom Type |
|---|---|---|---|---|---|---|
| $\mathbf{B_1}$ | = | $x_1\,\mathbf{a_1} + x_1\,\mathbf{a_2} + z_1\,\mathbf{a_3}$ | = | $\frac{1}{2}\,(x_1 - z_1)\,a\,\hat{\mathbf{x}} + \frac{1}{2\sqrt{3}}\,(x_1 - z_1)\,a\,\hat{\mathbf{y}} + \frac{1}{3}\,(2x_1 + z_1)\,c\,\hat{\mathbf{z}}$ | $(3b)$ | Ni |
| $\mathbf{B_2}$ | = | $z_1\,\mathbf{a_1} + x_1\,\mathbf{a_2} + x_1\,\mathbf{a_3}$ | = | $\frac{1}{2}\,(z_1 - x_1)\,a\,\hat{\mathbf{x}} + \frac{1}{2\sqrt{3}}\,(x_1 - z_1)\,a\,\hat{\mathbf{y}} + \frac{1}{3}\,(2x_1 + z_1)\,c\,\hat{\mathbf{z}}$ | $(3b)$ | Ni |
| $\mathbf{B_3}$ | = | $x_1\,\mathbf{a_1} + z_1\,\mathbf{a_2} + x_1\,\mathbf{a_3}$ | = | $\frac{1}{\sqrt{3}}\,(z_1 - x_1)\,a\,\hat{\mathbf{y}} + \frac{1}{3}\,(2x_1 + z_1)\,c\,\hat{\mathbf{z}}$ | $(3b)$ | Ni |
| $\mathbf{B_4}$ | = | $x_2\,\mathbf{a_1} + x_2\,\mathbf{a_2} + z_2\,\mathbf{a_3}$ | = | $\frac{1}{2}\,(x_2 - z_2)\,a\,\hat{\mathbf{x}} + \frac{1}{2\sqrt{3}}\,(x_2 - z_2)\,a\,\hat{\mathbf{y}} + \frac{1}{3}\,(2x_2 + z_2)\,c\,\hat{\mathbf{z}}$ | $(3b)$ | S |
| $\mathbf{B_5}$ | = | $z_2\,\mathbf{a_1} + x_2\,\mathbf{a_2} + x_2\,\mathbf{a_3}$ | = | $\frac{1}{2}\,(z_2 - x_2)\,a\,\hat{\mathbf{x}} + \frac{1}{2\sqrt{3}}\,(x_2 - z_2)\,a\,\hat{\mathbf{y}} + \frac{1}{3}\,(2x_2 + z_2)\,c\,\hat{\mathbf{z}}$ | $(3b)$ | S |
| $\mathbf{B_6}$ | = | $x_2\,\mathbf{a_1} + z_2\,\mathbf{a_2} + x_2\,\mathbf{a_3}$ | = | $\frac{1}{\sqrt{3}}\,(z_2 - x_2)\,a\,\hat{\mathbf{y}} + \frac{1}{3}\,(2x_2 + z_2)\,c\,\hat{\mathbf{z}}$ | $(3b)$ | S |

**Found in:**

**Geometry files:**

- CIF: pp. 719
- POSCAR: pp. 719

# Moissanite 9R Crystal Structure: AB_hR6_160_3a_3a

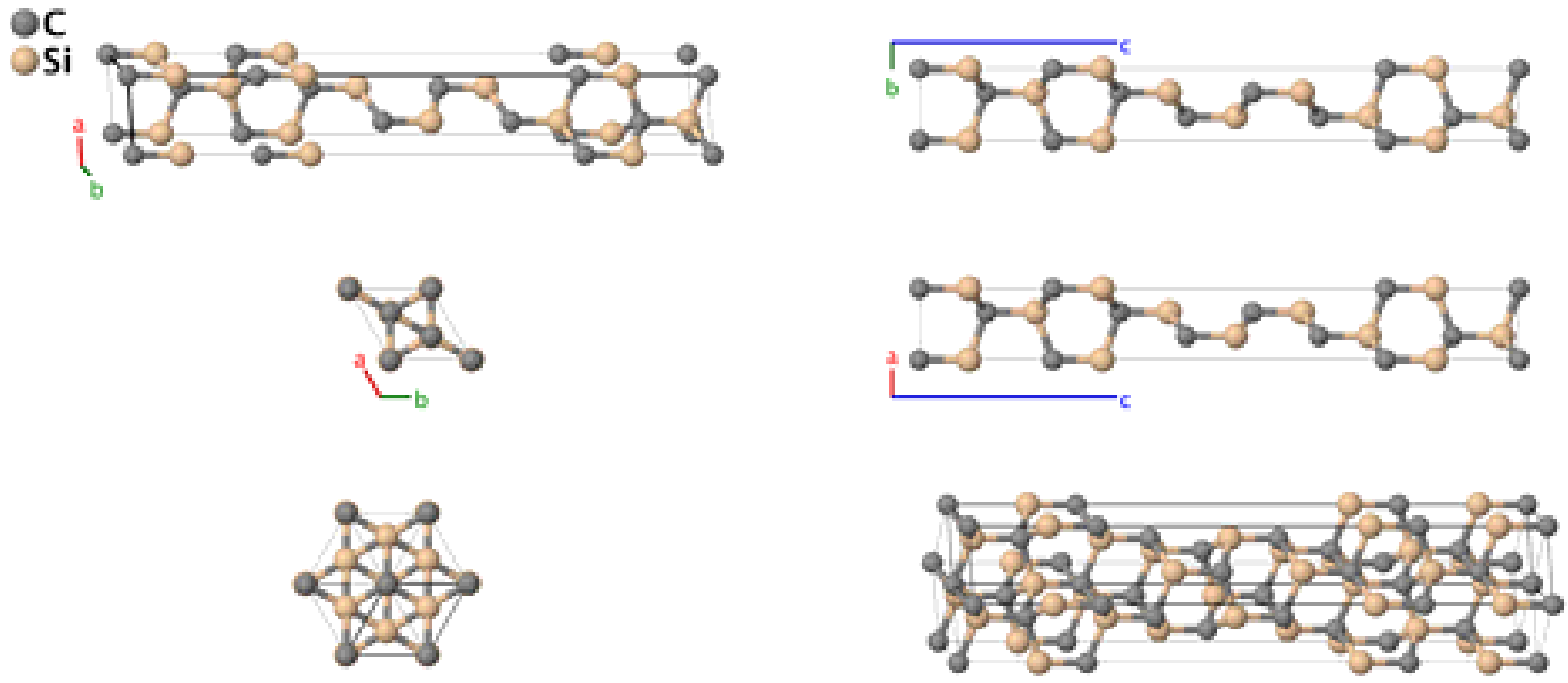

**Prototype** : CSi

**AFLOW prototype label** : AB_hR6_160_3a_3a

***Strukturbericht* designation** : None

**Pearson symbol** : hR6

**Space group number** : 160

**Space group symbol** : R3m

**AFLOW prototype command** : `aflow --proto=AB_hR6_160_3a_3a [--hex]`
`--params=`$a, c/a, x_1, x_2, x_3, x_4, x_5, x_6$

---

- We will loosely use the name moissanite to describe any tetrahedrally bonded silicon carbide compound that does not have another name. The labels 4H, 6H, 9R, etc., refer to the repeat stacking distance in the hexagonal unit cell, while H and R refer to the primitive hexagonal and rhombohedral lattices, respectively. The label C refers to a cubic unit cell, which is a special case of R. Note that 2, 3, 6, 9, etc., refers to the number of C-Si dimers that are stacked. Moissanite 9R is a hypothetical alternate stacking (ABCBCACAB) for tetrahedral structures. Compare this to wurtzite (ABABAB, 2H), zincblende (ABCABC, 3C), moissanite 4H (ABAC) and moissanite 6H (ABCACB). Hexagonal settings of this structure can be obtained with the option `--hex`.

---

**Rhombohedral primitive vectors:**

$$\mathbf{a}_1 = \frac{1}{2}\,a\,\mathbf{\hat{x}} - \frac{1}{2\sqrt{3}}\,a\,\mathbf{\hat{y}} + \frac{1}{3}\,c\,\mathbf{\hat{z}}$$
$$\mathbf{a}_2 = \frac{1}{\sqrt{3}}\,a\,\mathbf{\hat{y}} + \frac{1}{3}\,c\,\mathbf{\hat{z}}$$
$$\mathbf{a}_3 = -\frac{1}{2}\,a\,\mathbf{\hat{x}} - \frac{1}{2\sqrt{3}}\,a\,\mathbf{\hat{y}} + \frac{1}{3}\,c\,\mathbf{\hat{z}}$$

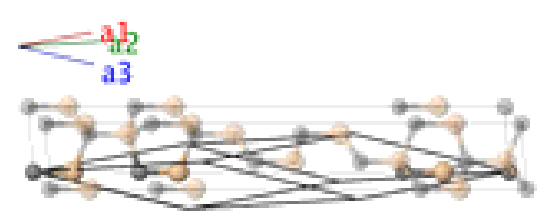

---

**Basis vectors:**

| | | Lattice Coordinates | | Cartesian Coordinates | Wyckoff Position | Atom Type |
|---|---|---|---|---|---|---|
| $\mathbf{B_1}$ | = | $x_1\,\mathbf{a_1} + x_1\,\mathbf{a_2} + x_1\,\mathbf{a_3}$ | = | $x_1\,c\,\mathbf{\hat{z}}$ | (1$a$) | C I |
| $\mathbf{B_2}$ | = | $x_2\,\mathbf{a_1} + x_2\,\mathbf{a_2} + x_2\,\mathbf{a_3}$ | = | $x_2\,c\,\mathbf{\hat{z}}$ | (1$a$) | C II |
| $\mathbf{B_3}$ | = | $x_3\,\mathbf{a_1} + x_3\,\mathbf{a_2} + x_3\,\mathbf{a_3}$ | = | $x_3\,c\,\mathbf{\hat{z}}$ | (1$a$) | C III |
| $\mathbf{B_4}$ | = | $x_4\,\mathbf{a_1} + x_4\,\mathbf{a_2} + x_4\,\mathbf{a_3}$ | = | $x_4\,c\,\mathbf{\hat{z}}$ | (1$a$) | Si I |
| $\mathbf{B_5}$ | = | $x_5\,\mathbf{a_1} + x_5\,\mathbf{a_2} + x_5\,\mathbf{a_3}$ | = | $x_5\,c\,\mathbf{\hat{z}}$ | (1$a$) | Si II |
| $\mathbf{B_6}$ | = | $x_6\,\mathbf{a_1} + x_6\,\mathbf{a_2} + x_6\,\mathbf{a_3}$ | = | $x_6\,c\,\mathbf{\hat{z}}$ | (1$a$) | Si III |

**Geometry files:**

- CIF: pp. 719
- POSCAR: pp. 719

# Ferroelectric $LiNbO_3$ Structure: ABC3_hR10_161_a_a_b

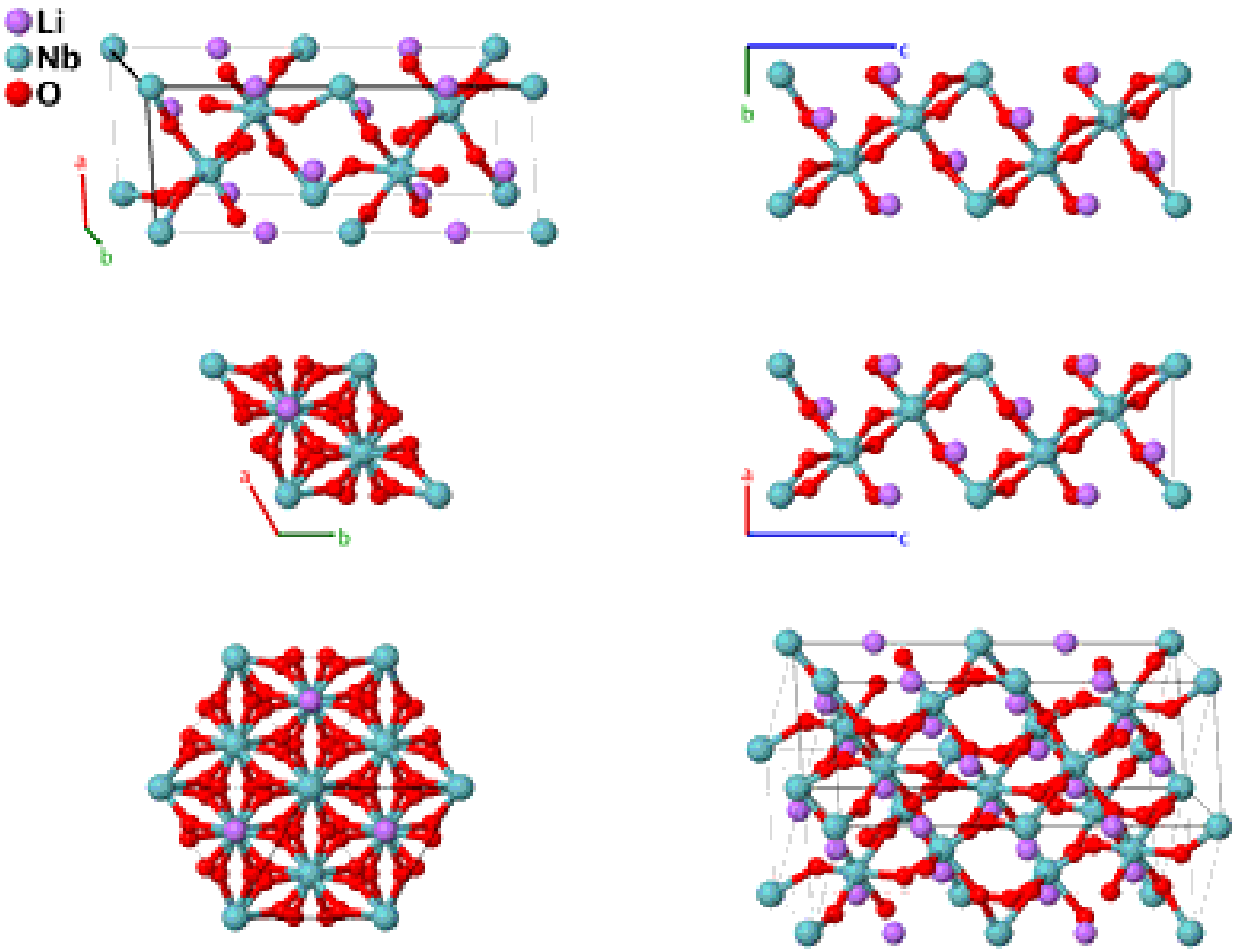

**Prototype** : $LiNbO_3$

**AFLOW prototype label** : ABC3_hR10_161_a_a_b

***Strukturbericht* designation** : None

**Pearson symbol** : hR10

**Space group number** : 161

**Space group symbol** : R3c

**AFLOW prototype command** : `aflow --proto=ABC3_hR10_161_a_a_b [--hex]`
`--params=`$a, c/a, x_1, x_2, x_3, y_3, z_3$

---

- This is the ferroelectric phase of $LiNbO_3$, which exists below 1430K. There is also a high-temperature paraelectric phase. This reduces to a double unit cell version of the cubic perovskite structure in the special case:
  - $c/a = \sqrt{6}$: This sets the angle between the rhombohedral primitive vectors to $60^o$. Experimentally the value is about $56^o$.
  - $z_1 = 1/4$
  - $z_2 = 0$
  - $x_3 = 1/2$
  - $y_3 = 0$
  - $z_3 = 0$

  Hexagonal settings of this structure can be obtained with the option `--hex`.

## Rhombohedral primitive vectors:

$$\mathbf{a}_1 = \frac{1}{2}\,a\,\hat{\mathbf{x}} - \frac{1}{2\sqrt{3}}\,a\,\hat{\mathbf{y}} + \frac{1}{3}\,c\,\hat{\mathbf{z}}$$
$$\mathbf{a}_2 = \frac{1}{\sqrt{3}}\,a\,\hat{\mathbf{y}} + \frac{1}{3}\,c\,\hat{\mathbf{z}}$$
$$\mathbf{a}_3 = -\frac{1}{2}\,a\,\hat{\mathbf{x}} - \frac{1}{2\sqrt{3}}\,a\,\hat{\mathbf{y}} + \frac{1}{3}\,c\,\hat{\mathbf{z}}$$

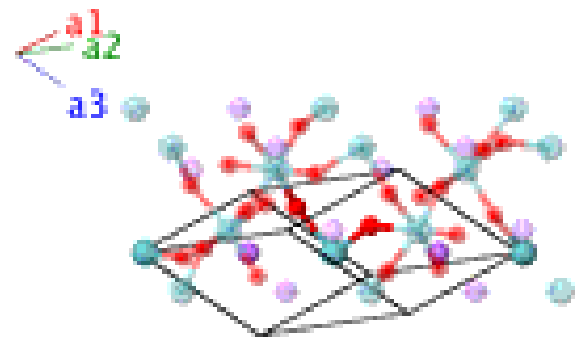

## Basis vectors:

| | | Lattice Coordinates | | Cartesian Coordinates | Wyckoff Position | Atom Type |
|---|---|---|---|---|---|---|
| $\mathbf{B_1}$ | = | $x_1\,\mathbf{a_1} + x_1\,\mathbf{a_2} + x_1\,\mathbf{a_3}$ | = | $x_1\,c\,\hat{\mathbf{z}}$ | (2$a$) | Li |
| $\mathbf{B_2}$ | = | $\left(\frac{1}{2} + x_1\right)\mathbf{a_1} + \left(\frac{1}{2} + x_1\right)\mathbf{a_2} + \left(\frac{1}{2} + x_1\right)\mathbf{a_3}$ | = | $\left(\frac{1}{2} + x_1\right)c\,\hat{\mathbf{z}}$ | (2$a$) | Li |
| $\mathbf{B_3}$ | = | $x_2\,\mathbf{a_1} + x_2\,\mathbf{a_2} + x_2\,\mathbf{a_3}$ | = | $x_2\,c\,\hat{\mathbf{z}}$ | (2$a$) | Nb |
| $\mathbf{B_4}$ | = | $\left(\frac{1}{2} + x_2\right)\mathbf{a_1} + \left(\frac{1}{2} + x_2\right)\mathbf{a_2} + \left(\frac{1}{2} + x_2\right)\mathbf{a_3}$ | = | $\left(\frac{1}{2} + x_2\right)c\,\hat{\mathbf{z}}$ | (2$a$) | Nb |
| $\mathbf{B_5}$ | = | $x_3\,\mathbf{a_1} + y_3\,\mathbf{a_2} + z_3\,\mathbf{a_3}$ | = | $\frac{1}{2}(x_3 - z_3)\,a\,\hat{\mathbf{x}} - \frac{1}{2\sqrt{3}}(x_3 - 2y_3 + z_3)\,a\,\hat{\mathbf{y}} + \frac{1}{3}(x_3 + y_3 + z_3)\,c\,\hat{\mathbf{z}}$ | (6$b$) | O |
| $\mathbf{B_6}$ | = | $z_3\,\mathbf{a_1} + x_3\,\mathbf{a_2} + y_3\,\mathbf{a_3}$ | = | $\frac{1}{2}(z_3 - y_3)\,a\,\hat{\mathbf{x}} - \frac{1}{2\sqrt{3}}(z_3 - 2x_3 + y_3)\,a\,\hat{\mathbf{y}} + \frac{1}{3}(x_3 + y_3 + z_3)\,c\,\hat{\mathbf{z}}$ | (6$b$) | O |
| $\mathbf{B_7}$ | = | $y_3\,\mathbf{a_1} + z_3\,\mathbf{a_2} + x_3\,\mathbf{a_3}$ | = | $\frac{1}{2}(y_3 - x_3)\,a\,\hat{\mathbf{x}} - \frac{1}{2\sqrt{3}}(y_3 - 2z_3 + x_3)\,a\,\hat{\mathbf{y}} + \frac{1}{3}(x_3 + y_3 + z_3)\,c\,\hat{\mathbf{z}}$ | (6$b$) | O |
| $\mathbf{B_8}$ | = | $\left(\frac{1}{2} + y_3\right)\mathbf{a_1} + \left(\frac{1}{2} + x_3\right)\mathbf{a_2} + \left(\frac{1}{2} + z_3\right)\mathbf{a_3}$ | = | $\frac{1}{2}(y_3 - z_3)\,a\,\hat{\mathbf{x}} - \frac{1}{2\sqrt{3}}(z3 - 2x_3 + y_3)\,a\,\hat{\mathbf{y}} + \frac{1}{6}(3 + 2x_3 + 2y_3 + 2z_3)\,c\,\hat{\mathbf{z}}$ | (6$b$) | O |
| $\mathbf{B_9}$ | = | $\left(\frac{1}{2} + x_3\right)\mathbf{a_1} + \left(\frac{1}{2} + z_3\right)\mathbf{a_2} + \left(\frac{1}{2} + y_3\right)\mathbf{a_3}$ | = | $\frac{1}{2}(x_3 - y_3)\,a\,\hat{\mathbf{x}} - \frac{1}{2\sqrt{3}}(y_3 - 2z_3 + x_3)\,a\,\hat{\mathbf{y}} + \frac{1}{6}(3 + 2x_3 + 2y_3 + 2z_3)\,c\,\hat{\mathbf{z}}$ | (6$b$) | O |
| $\mathbf{B_{10}}$ | = | $\left(\frac{1}{2} + z_3\right)\mathbf{a_1} + \left(\frac{1}{2} + y_3\right)\mathbf{a_2} + \left(\frac{1}{2} + x_3\right)\mathbf{a_3}$ | = | $\frac{1}{2}(z_3 - x_3)\,a\,\hat{\mathbf{x}} - \frac{1}{2\sqrt{3}}(x_3 - 2y_3 + z_3)\,a\,\hat{\mathbf{y}} + \frac{1}{6}(3 + 2x_3 + 2y_3 + 2z_3)\,c\,\hat{\mathbf{z}}$ | (6$b$) | O |

## Geometry files:

- CIF: pp. 720
- POSCAR: pp. 720

# $\beta$-$V_2N$ Structure: AB2_hP9_162_ad_k

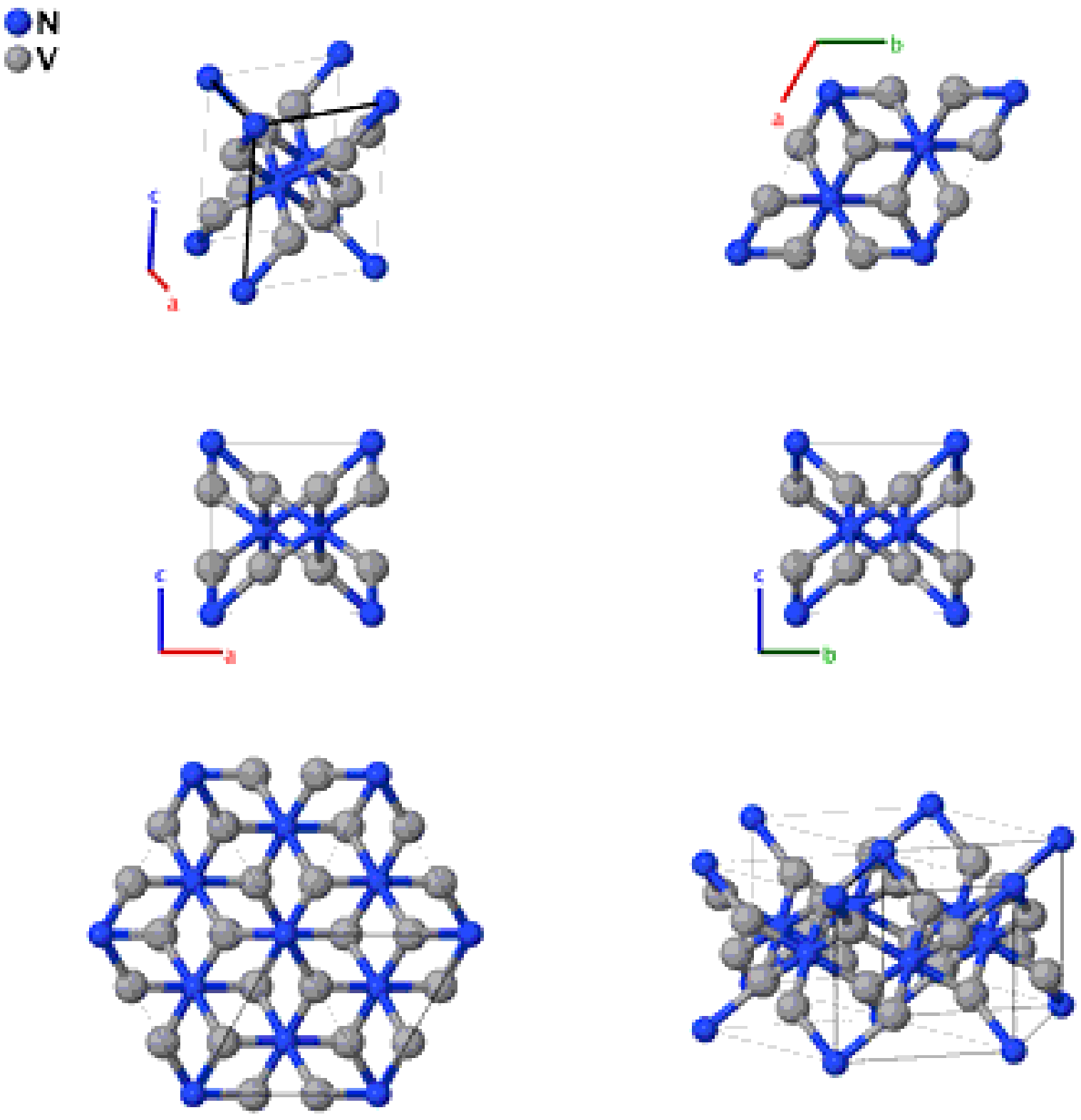

| | | |
|---|---|---|
| **Prototype** | : | $\beta$-$V_2N$ |
| **AFLOW prototype label** | : | AB2_hP9_162_ad_k |
| ***Strukturbericht* designation** | : | None |
| **Pearson symbol** | : | hP9 |
| **Space group number** | : | 162 |
| **Space group symbol** | : | P$\bar{3}$1m |
| **AFLOW prototype command** | : | `aflow --proto=AB2_hP9_162_ad_k`<br>`--params=`$a, c/a, x_3, z_3$ |

- Note that our reference (Christensen, 1979) states that $\epsilon$-$Fe_2N$ is the prototype for this structure. We will instead follow (Villars, 1991), which uses $\beta$-$V_2N$ as the prototype.

**Trigonal Hexagonal primitive vectors:**

$$
\begin{aligned}
\mathbf{a}_1 &= \frac{1}{2}\,a\,\hat{\mathbf{x}} - \frac{\sqrt{3}}{2}\,a\,\hat{\mathbf{y}} \\
\mathbf{a}_2 &= \frac{1}{2}\,a\,\hat{\mathbf{x}} + \frac{\sqrt{3}}{2}\,a\,\hat{\mathbf{y}} \\
\mathbf{a}_3 &= c\,\hat{\mathbf{z}}
\end{aligned}
$$

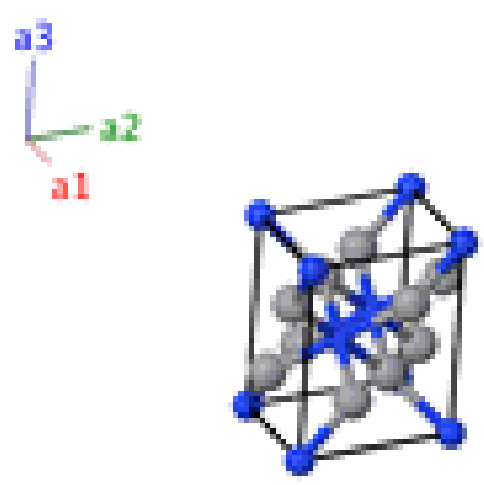

**Basis vectors:**

| | | Lattice Coordinates | | Cartesian Coordinates | Wyckoff Position | Atom Type |
|---|---|---|---|---|---|---|
| $\mathbf{B_1}$ | = | $0\,\mathbf{a_1} + 0\,\mathbf{a_2} + 0\,\mathbf{a_3}$ | = | $0\hat{\mathbf{x}} + 0\hat{\mathbf{y}} + 0\hat{\mathbf{z}}$ | (1$a$) | N I |
| $\mathbf{B_2}$ | = | $\frac{1}{3}\,\mathbf{a_1} + \frac{2}{3}\,\mathbf{a_2} + \frac{1}{2}\,\mathbf{a_3}$ | = | $\frac{1}{2}\,a\,\hat{\mathbf{x}} + \frac{1}{2\sqrt{3}}\,a\,\hat{\mathbf{y}} + \frac{1}{2}\,c\,\hat{\mathbf{z}}$ | (2$d$) | N II |
| $\mathbf{B_3}$ | = | $\frac{2}{3}\,\mathbf{a_1} + \frac{1}{3}\,\mathbf{a_2} + \frac{1}{2}\,\mathbf{a_3}$ | = | $\frac{1}{2}\,a\,\hat{\mathbf{x}} - \frac{1}{2\sqrt{3}}\,a\,\hat{\mathbf{y}} + \frac{1}{2}\,c\,\hat{\mathbf{z}}$ | (2$d$) | N II |
| $\mathbf{B_4}$ | = | $x_3\,\mathbf{a_1} + z_3\,\mathbf{a_3}$ | = | $\frac{1}{2}\,x_3\,a\,\hat{\mathbf{x}} - \frac{\sqrt{3}}{2}\,x_3\,a\,\hat{\mathbf{y}} + z_3\,c\,\hat{\mathbf{z}}$ | (6$k$) | V |
| $\mathbf{B_5}$ | = | $x_3\,\mathbf{a_2} + z_3\,\mathbf{a_3}$ | = | $\frac{1}{2}\,x_3\,a\,\hat{\mathbf{x}} + \frac{\sqrt{3}}{2}\,x_3\,a\,\hat{\mathbf{y}} + z_3\,c\,\hat{\mathbf{z}}$ | (6$k$) | V |
| $\mathbf{B_6}$ | = | $-x_3\,\mathbf{a_1} - x_3\,\mathbf{a_2} + z_3\,\mathbf{a_3}$ | = | $-x_3\,a\,\hat{\mathbf{x}} + z_3\,c\,\hat{\mathbf{z}}$ | (6$k$) | V |
| $\mathbf{B_7}$ | = | $-x_3\,\mathbf{a_2} - z_3\,\mathbf{a_3}$ | = | $-\frac{1}{2}\,x_3\,a\,\hat{\mathbf{x}} - \frac{\sqrt{3}}{2}\,x_3\,a\,\hat{\mathbf{y}} - z_3\,c\,\hat{\mathbf{z}}$ | (6$k$) | V |
| $\mathbf{B_8}$ | = | $-x_3\,\mathbf{a_1} - z_3\,\mathbf{a_3}$ | = | $-\frac{1}{2}\,x_3\,a\,\hat{\mathbf{x}} + \frac{\sqrt{3}}{2}\,x_3\,a\,\hat{\mathbf{y}} - z_3\,c\,\hat{\mathbf{z}}$ | (6$k$) | V |
| $\mathbf{B_9}$ | = | $x_3\,\mathbf{a_1} + x_3\,\mathbf{a_2} - z_3\,\mathbf{a_3}$ | = | $x_3\,a\,\hat{\mathbf{x}} - z_3\,c\,\hat{\mathbf{z}}$ | (6$k$) | V |

**Found in:**

**Geometry files:**

- CIF: pp. 720
- POSCAR: pp. 720

# $KAg(CN)_2$ ($F5_{10}$) Structure: AB2CD2_hP36_163_h_i_bf_i

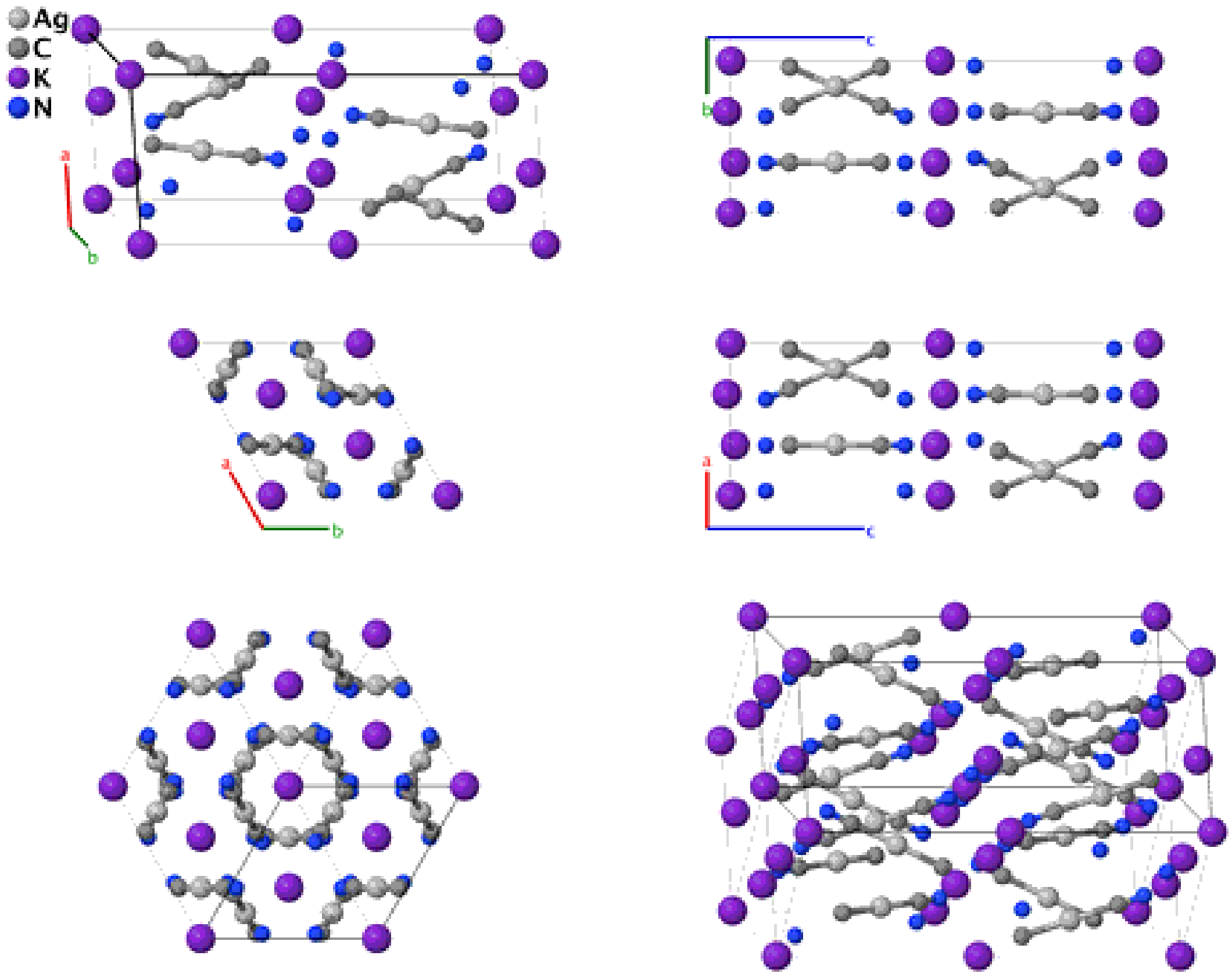

| | | |
|---|---|---|
| **Prototype** | : | $KAg(CN)_2$ |
| **AFLOW prototype label** | : | AB2CD2_hP36_163_h_i_bf_i |
| ***Strukturbericht* designation** | : | $F5_{10}$ |
| **Pearson symbol** | : | hP36 |
| **Space group number** | : | 163 |
| **Space group symbol** | : | $P\bar{3}1c$ |
| **AFLOW prototype command** | : | `aflow --proto=AB2CD2_hP36_163_h_i_bf_i`<br>`--params=`$a, c/a, z_2, x_3, x_4, y_4, z_4, x_5, y_5, z_5$ |

**Trigonal Hexagonal primitive vectors:**

$$\mathbf{a}_1 = \frac{1}{2}\,a\,\hat{\mathbf{x}} - \frac{\sqrt{3}}{2}\,a\,\hat{\mathbf{y}}$$
$$\mathbf{a}_2 = \frac{1}{2}\,a\,\hat{\mathbf{x}} + \frac{\sqrt{3}}{2}\,a\,\hat{\mathbf{y}}$$
$$\mathbf{a}_3 = c\,\hat{\mathbf{z}}$$

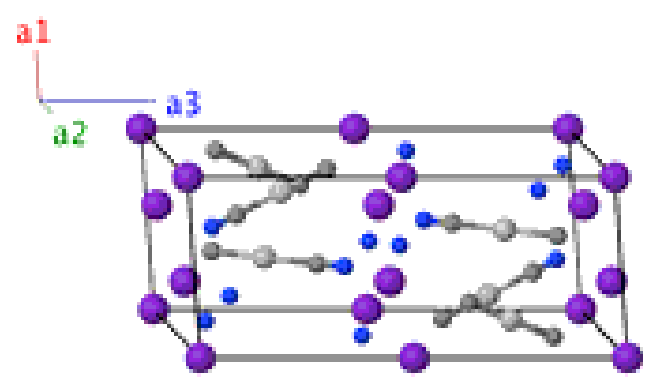

**Basis vectors:**

| | | Lattice Coordinates | | Cartesian Coordinates | Wyckoff Position | Atom Type |
|---|---|---|---|---|---|---|
| $\mathbf{B}_1$ | = | $0\,\mathbf{a}_1 + 0\,\mathbf{a}_2 + 0\,\mathbf{a}_3$ | = | $0\hat{\mathbf{x}} + 0\hat{\mathbf{y}} + 0\hat{\mathbf{z}}$ | (2$b$) | K I |
| $\mathbf{B}_2$ | = | $\frac{1}{2}\,\mathbf{a}_3$ | = | $\frac{1}{2}\,c\,\hat{\mathbf{z}}$ | (2$b$) | K I |
| $\mathbf{B_3}$ | = | $\frac{1}{3}\,\mathbf{a_1} + \frac{2}{3}\,\mathbf{a_2} + z_2\,\mathbf{a_3}$ | = | $\frac{1}{2}\,a\,\mathbf{\hat{x}} + \frac{1}{2\sqrt{3}}\,a\,\mathbf{\hat{y}} + z_2\,c\,\mathbf{\hat{z}}$ | (4$f$) | K II |
| $\mathbf{B_4}$ | = | $\frac{1}{3}\,\mathbf{a_1} + \frac{2}{3}\,\mathbf{a_2} + \left(\frac{1}{2} - z_2\right)\,\mathbf{a_3}$ | = | $\frac{1}{2}\,a\,\mathbf{\hat{x}} + \frac{1}{2\sqrt{3}}\,a\,\mathbf{\hat{y}} + \left(\frac{1}{2} - z_2\right)\,c\,\mathbf{\hat{z}}$ | (4$f$) | K II |
| $\mathbf{B_5}$ | = | $\frac{2}{3}\,\mathbf{a_1} + \frac{1}{3}\,\mathbf{a_2} - z_2\,\mathbf{a_3}$ | = | $\frac{1}{2}\,a\,\mathbf{\hat{x}} - \frac{1}{2\sqrt{3}}\,a\,\mathbf{\hat{y}} - z_2\,c\,\mathbf{\hat{z}}$ | (4$f$) | K II |
| $\mathbf{B_6}$ | = | $\frac{2}{3}\,\mathbf{a_1} + \frac{1}{3}\,\mathbf{a_2} + \left(\frac{1}{2} + z_2\right)\,\mathbf{a_3}$ | = | $\frac{1}{2}\,a\,\mathbf{\hat{x}} - \frac{1}{2\sqrt{3}}\,a\,\mathbf{\hat{y}} + \left(\frac{1}{2} + z_2\right)\,c\,\mathbf{\hat{z}}$ | (4$f$) | K II |
| $\mathbf{B_7}$ | = | $x_3\,\mathbf{a_1} - x_3\,\mathbf{a_2} + \frac{1}{4}\,\mathbf{a_3}$ | = | $-\sqrt{3}\,x_3\,a\,\mathbf{\hat{y}} + \frac{1}{4}\,c\,\mathbf{\hat{z}}$ | (6$h$) | Ag |
| $\mathbf{B_8}$ | = | $x_3\,\mathbf{a_1} + 2x_3\,\mathbf{a_2} + \frac{1}{4}\,\mathbf{a_3}$ | = | $\frac{3}{2}\,x_3\,a\mathbf{\hat{x}} + \frac{\sqrt{3}}{2}\,x_3\,a\,\mathbf{\hat{y}} + \frac{1}{4}\,c\,\mathbf{\hat{z}}$ | (6$h$) | Ag |
| $\mathbf{B_9}$ | = | $-2x_3\,\mathbf{a_1} - x_3\,\mathbf{a_2} + \frac{1}{4}\,\mathbf{a_3}$ | = | $-\frac{3}{2}\,x_3\,a\mathbf{\hat{x}} + \frac{\sqrt{3}}{2}\,x_3\,a\,\mathbf{\hat{y}} + \frac{1}{4}\,c\,\mathbf{\hat{z}}$ | (6$h$) | Ag |
| $\mathbf{B_{10}}$ | = | $-x_3\,\mathbf{a_1} + x_3\,\mathbf{a_2} + \frac{3}{4}\,\mathbf{a_3}$ | = | $\sqrt{3}\,x_3\,a\,\mathbf{\hat{y}} + \frac{3}{4}\,c\,\mathbf{\hat{z}}$ | (6$h$) | Ag |
| $\mathbf{B_{11}}$ | = | $-x_3\,\mathbf{a_1} - 2x_3\,\mathbf{a_2} + \frac{3}{4}\,\mathbf{a_3}$ | = | $-\frac{3}{2}\,x_3\,a\mathbf{\hat{x}} - \frac{\sqrt{3}}{2}\,x_3\,a\,\mathbf{\hat{y}} + \frac{3}{4}\,c\,\mathbf{\hat{z}}$ | (6$h$) | Ag |
| $\mathbf{B_{12}}$ | = | $2x_3\,\mathbf{a_1} + x_3\,\mathbf{a_2} + \frac{3}{4}\,\mathbf{a_3}$ | = | $\frac{3}{2}\,x_3\,a\mathbf{\hat{x}} - \frac{\sqrt{3}}{2}\,x_3\,a\,\mathbf{\hat{y}} + \frac{3}{4}\,c\,\mathbf{\hat{z}}$ | (6$h$) | Ag |
| $\mathbf{B_{13}}$ | = | $x_4\,\mathbf{a_1} + y_4\,\mathbf{a_2} + z_4\,\mathbf{a_3}$ | = | $\frac{1}{2}\,(x_4 + y_4)\,a\,\mathbf{\hat{x}} + \frac{\sqrt{3}}{2}\,(y_4 - x_4)\,a\,\mathbf{\hat{y}} + z_4\,c\,\mathbf{\hat{z}}$ | (12$i$) | C |
| $\mathbf{B_{14}}$ | = | $-y_4\,\mathbf{a_1} + (x_4 - y_4)\,\mathbf{a_2} + z_4\,\mathbf{a_3}$ | = | $\frac{1}{2}\,(x_4 - 2y_4)\,a\,\mathbf{\hat{x}} + \frac{\sqrt{3}}{2}\,x_4\,a\,\mathbf{\hat{y}} + z_4\,c\,\mathbf{\hat{z}}$ | (12$i$) | C |
| $\mathbf{B_{15}}$ | = | $(y_4 - x_4)\,\mathbf{a_1} - x_4\,\mathbf{a_2} + z_4\,\mathbf{a_3}$ | = | $\frac{1}{2}\,(y_4 - 2x_4)\,a\,\mathbf{\hat{x}} - \frac{\sqrt{3}}{2}\,y_4\,a\,\mathbf{\hat{y}} + z_4\,c\,\mathbf{\hat{z}}$ | (12$i$) | C |
| $\mathbf{B_{16}}$ | = | $-y_4\,\mathbf{a_1} - x_4\,\mathbf{a_2} + \left(\frac{1}{2} - z_4\right)\,\mathbf{a_3}$ | = | $-\frac{1}{2}\,(x_4 + y_4)\,a\,\mathbf{\hat{x}} + \frac{\sqrt{3}}{2}\,(y_4 - x_4)\,a\,\mathbf{\hat{y}} + \left(\frac{1}{2} - z_4\right)\,c\,\mathbf{\hat{z}}$ | (12$i$) | C |
| $\mathbf{B_{17}}$ | = | $(y_4 - x_4)\,\mathbf{a_1} + y_4\,\mathbf{a_2} + \left(\frac{1}{2} - z_4\right)\,\mathbf{a_3}$ | = | $\frac{1}{2}\,(2y_4 - x_4)\,a\,\mathbf{\hat{x}} + \frac{\sqrt{3}}{2}\,x_4\,a\,\mathbf{\hat{y}} + \left(\frac{1}{2} - z_4\right)\,c\,\mathbf{\hat{z}}$ | (12$i$) | C |
| $\mathbf{B_{18}}$ | = | $x_4\,\mathbf{a_1} + (x_4 - y_4)\,\mathbf{a_2} + \left(\frac{1}{2} - z_4\right)\,\mathbf{a_3}$ | = | $\frac{1}{2}\,(2x_4 - y_4)\,a\,\mathbf{\hat{x}} - \frac{\sqrt{3}}{2}\,y_4\,a\,\mathbf{\hat{y}} + \left(\frac{1}{2} - z_4\right)\,c\,\mathbf{\hat{z}}$ | (12$i$) | C |
| $\mathbf{B_{19}}$ | = | $-x_4\,\mathbf{a_1} - y_4\,\mathbf{a_2} - z_4\,\mathbf{a_3}$ | = | $-\frac{1}{2}\,(x_4 + y_4)\,a\,\mathbf{\hat{x}} + \frac{\sqrt{3}}{2}\,(x_4 - y_4)\,a\,\mathbf{\hat{y}} - z_4\,c\,\mathbf{\hat{z}}$ | (12$i$) | C |
| $\mathbf{B_{20}}$ | = | $y_4\,\mathbf{a_1} + (y_4 - x_4)\,\mathbf{a_2} - z_4\,\mathbf{a_3}$ | = | $\frac{1}{2}\,(2y_4 - x_4)\,a\,\mathbf{\hat{x}} - \frac{\sqrt{3}}{2}\,x_4\,a\,\mathbf{\hat{y}} - z_4\,c\,\mathbf{\hat{z}}$ | (12$i$) | C |
| $\mathbf{B_{21}}$ | = | $(x_4 - y_4)\,\mathbf{a_1} + x_4\,\mathbf{a_2} - z_4\,\mathbf{a_3}$ | = | $\frac{1}{2}\,(2x_4 - y_4)\,a\,\mathbf{\hat{x}} + \frac{\sqrt{3}}{2}\,y_4\,a\,\mathbf{\hat{y}} - z_4\,c\,\mathbf{\hat{z}}$ | (12$i$) | C |
| $\mathbf{B_{22}}$ | = | $y_4\,\mathbf{a_1} + x_4\,\mathbf{a_2} + \left(\frac{1}{2} + z_4\right)\,\mathbf{a_3}$ | = | $\frac{1}{2}\,(x_4 + y_4)\,a\,\mathbf{\hat{x}} + \frac{\sqrt{3}}{2}\,(x_4 - y_4)\,a\,\mathbf{\hat{y}} + \left(\frac{1}{2} + z_4\right)\,c\,\mathbf{\hat{z}}$ | (12$i$) | C |
| $\mathbf{B_{23}}$ | = | $(x_4 - y_4)\,\mathbf{a_1} - y_4\,\mathbf{a_2} + \left(\frac{1}{2} + z_4\right)\,\mathbf{a_3}$ | = | $\frac{1}{2}\,(x_4 - 2y_4)\,a\,\mathbf{\hat{x}} - \frac{\sqrt{3}}{2}\,x_4\,a\,\mathbf{\hat{y}} + \left(\frac{1}{2} + z_4\right)\,c\,\mathbf{\hat{z}}$ | (12$i$) | C |
| $\mathbf{B_{24}}$ | = | $-x_4\,\mathbf{a_1} + (y_4 - x_4)\,\mathbf{a_2} + \left(\frac{1}{2} + z_4\right)\,\mathbf{a_3}$ | = | $\frac{1}{2}\,(y_4 - 2x_4)\,a\,\mathbf{\hat{x}} + \frac{\sqrt{3}}{2}\,y_4\,a\,\mathbf{\hat{y}} + \left(\frac{1}{2} + z_4\right)\,c\,\mathbf{\hat{z}}$ | (12$i$) | C |
| $\mathbf{B_{25}}$ | = | $x_5\,\mathbf{a_1} + y_5\,\mathbf{a_2} + z_5\,\mathbf{a_3}$ | = | $\frac{1}{2}\,(x_5 + y_5)\,a\,\mathbf{\hat{x}} + \frac{\sqrt{3}}{2}\,(y_5 - x_5)\,a\,\mathbf{\hat{y}} + z_5\,c\,\mathbf{\hat{z}}$ | (12$i$) | N |
| $\mathbf{B_{26}}$ | = | $-y_5\,\mathbf{a_1} + (x_5 - y_5)\,\mathbf{a_2} + z_5\,\mathbf{a_3}$ | = | $\frac{1}{2}\,(x_5 - 2y_5)\,a\,\mathbf{\hat{x}} + \frac{\sqrt{3}}{2}\,x_5\,a\,\mathbf{\hat{y}} + z_5\,c\,\mathbf{\hat{z}}$ | (12$i$) | N |
| $\mathbf{B_{27}}$ | = | $(y_5 - x_5)\,\mathbf{a_1} - x_5\,\mathbf{a_2} + z_5\,\mathbf{a_3}$ | = | $\frac{1}{2}\,(y_5 - 2x_5)\,a\,\mathbf{\hat{x}} - \frac{\sqrt{3}}{2}\,y_5\,a\,\mathbf{\hat{y}} + z_5\,c\,\mathbf{\hat{z}}$ | (12$i$) | N |
| $\mathbf{B_{28}}$ | = | $-y_5\,\mathbf{a_1} - x_5\,\mathbf{a_2} + \left(\frac{1}{2} - z_5\right)\,\mathbf{a_3}$ | = | $-\frac{1}{2}\,(x_5 + y_5)\,a\,\mathbf{\hat{x}} + \frac{\sqrt{3}}{2}\,(y_5 - x_5)\,a\,\mathbf{\hat{y}} + \left(\frac{1}{2} - z_5\right)\,c\,\mathbf{\hat{z}}$ | (12$i$) | N |
| $\mathbf{B_{29}}$ | = | $(y_5 - x_5)\,\mathbf{a_1} + y_5\,\mathbf{a_2} + \left(\frac{1}{2} - z_5\right)\,\mathbf{a_3}$ | = | $\frac{1}{2}\,(2y_5 - x_5)\,a\,\mathbf{\hat{x}} + \frac{\sqrt{3}}{2}\,x_5\,a\,\mathbf{\hat{y}} + \left(\frac{1}{2} - z_5\right)\,c\,\mathbf{\hat{z}}$ | (12$i$) | N |
| $\mathbf{B}_{30}$ | = | $x_5\,\mathbf{a}_1 + (x_5 - y_5)\,\mathbf{a}_2 + \left(\frac{1}{2} - z_5\right)\,\mathbf{a}_3$ | = | $\frac{1}{2}\,(2x_5 - y_5)\,a\,\hat{\mathbf{x}} - \frac{\sqrt{3}}{2}\,y_5\,a\,\hat{\mathbf{y}} + \left(\frac{1}{2} - z_5\right)\,c\,\hat{\mathbf{z}}$ | (12$i$) | N |
| $\mathbf{B}_{31}$ | = | $-x_5\,\mathbf{a}_1 - y_5\,\mathbf{a}_2 - z_5\,\mathbf{a}_3$ | = | $-\frac{1}{2}\,(x_5 + y_5)\,a\,\hat{\mathbf{x}} + \frac{\sqrt{3}}{2}\,(x_5 - y_5)\,a\,\hat{\mathbf{y}} - z_5\,c\,\hat{\mathbf{z}}$ | (12$i$) | N |
| $\mathbf{B}_{32}$ | = | $y_5\,\mathbf{a}_1 + (y_5 - x_5)\,\mathbf{a}_2 - z_5\,\mathbf{a}_3$ | = | $\frac{1}{2}\,(2y_5 - x_5)\,a\,\hat{\mathbf{x}} - \frac{\sqrt{3}}{2}\,x_5\,a\,\hat{\mathbf{y}} - z_5\,c\,\hat{\mathbf{z}}$ | (12$i$) | N |
| $\mathbf{B}_{33}$ | = | $(x_5 - y_5)\,\mathbf{a}_1 + x_5\,\mathbf{a}_2 - z_5\,\mathbf{a}_3$ | = | $\frac{1}{2}\,(2x_5 - y_5)\,a\,\hat{\mathbf{x}} + \frac{\sqrt{3}}{2}\,y_5\,a\,\hat{\mathbf{y}} - z_5\,c\,\hat{\mathbf{z}}$ | (12$i$) | N |
| $\mathbf{B}_{34}$ | = | $y_5\,\mathbf{a}_1 + x_5\,\mathbf{a}_2 + \left(\frac{1}{2} + z_5\right)\,\mathbf{a}_3$ | = | $\frac{1}{2}\,(x_5 + y_5)\,a\,\hat{\mathbf{x}} + \frac{\sqrt{3}}{2}\,(x_5 - y_5)\,a\,\hat{\mathbf{y}} + \left(\frac{1}{2} + z_5\right)\,c\,\hat{\mathbf{z}}$ | (12$i$) | N |
| $\mathbf{B}_{35}$ | = | $(x_5 - y_5)\,\mathbf{a}_1 - y_5\,\mathbf{a}_2 + \left(\frac{1}{2} + z_5\right)\,\mathbf{a}_3$ | = | $\frac{1}{2}\,(x_5 - 2y_5)\,a\,\hat{\mathbf{x}} - \frac{\sqrt{3}}{2}\,x_5\,a\,\hat{\mathbf{y}} + \left(\frac{1}{2} + z_5\right)\,c\,\hat{\mathbf{z}}$ | (12$i$) | N |
| $\mathbf{B}_{36}$ | = | $-x_5\,\mathbf{a}_1 + (y_5 - x_5)\,\mathbf{a}_2 + \left(\frac{1}{2} + z_5\right)\,\mathbf{a}_3$ | = | $\frac{1}{2}\,(y_5 - 2x_5)\,a\,\hat{\mathbf{x}} + \frac{\sqrt{3}}{2}\,y_5\,a\,\hat{\mathbf{y}} + \left(\frac{1}{2} + z_5\right)\,c\,\hat{\mathbf{z}}$ | (12$i$) | N |

**Found in:**

**Geometry files:**

- CIF: pp. 721
- POSCAR: pp. 721

# $Al_3Ni_2$ ($D5_{13}$) Structure: A3B2_hP5_164_ad_d

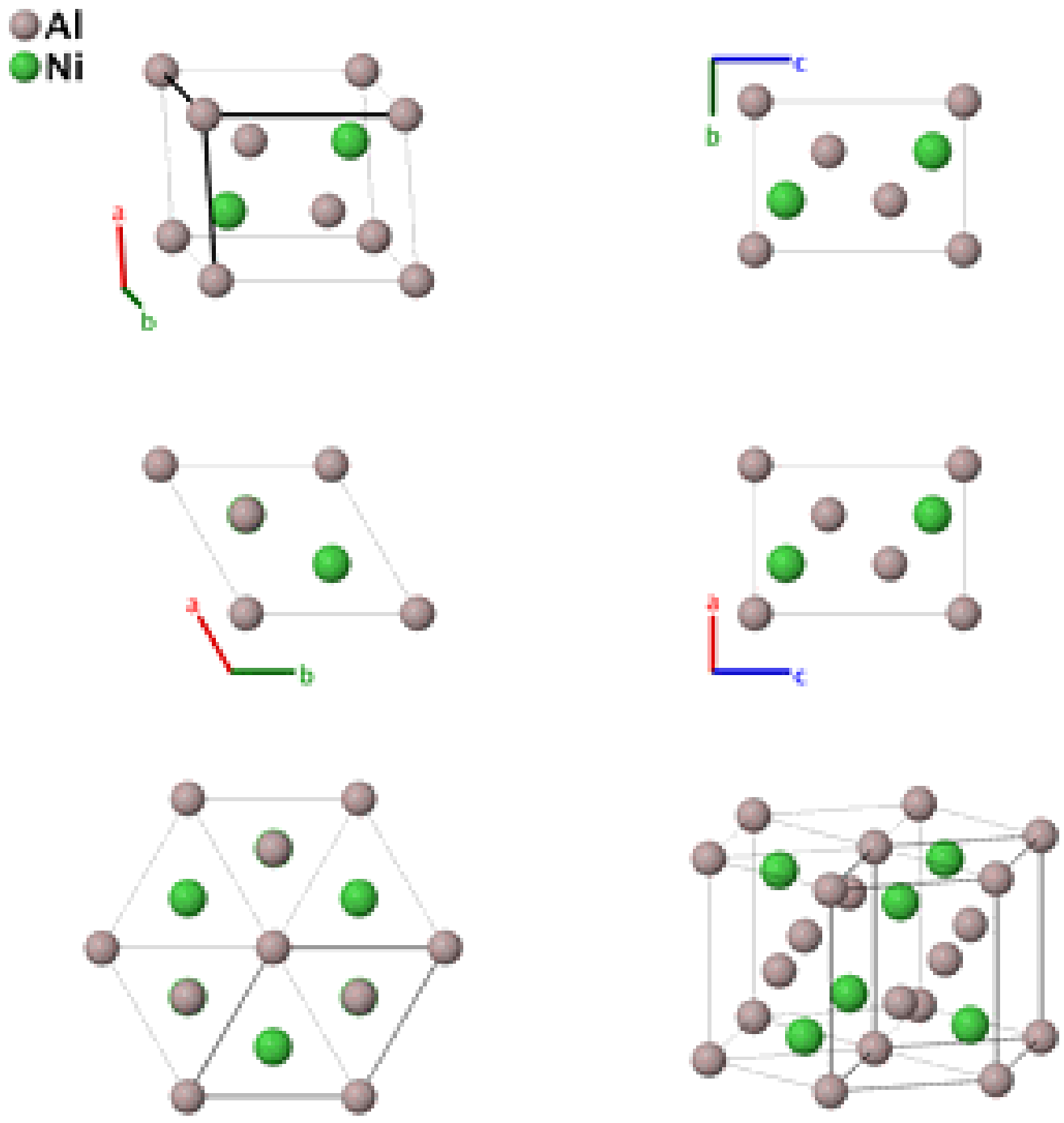

| | | |
|---|---|---|
| **Prototype** | : | $Al_3Ni_2$ |
| **AFLOW prototype label** | : | A3B2_hP5_164_ad_d |
| ***Strukturbericht* designation** | : | $D5_{13}$ |
| **Pearson symbol** | : | hP5 |
| **Space group number** | : | 164 |
| **Space group symbol** | : | $P\bar{3}m1$ |
| **AFLOW prototype command** | : | `aflow --proto=A3B2_hP5_164_ad_d` <br> `--params=`$a, c/a, z_2, z_3$ |

**Other compounds with this structure:**

- $Al_3Cu_2$, $Al_3Pd_2$, $Al_3Pt_2$, $Al_3In_2$, $Al_3Tc_2$, $In_3Al_2$, $In_3Pd_2$, $In_3Pt_2$, $Ga_3Pt_2$

- Either the 3 Al atoms or Al (1a) and the Ni atoms form a trigonal omega structure. Using the choices of internal parameters for $Al_3Ni_2$, this can be viewed as a five-layer close-packed unit cell with stacking ABCBCA.

**Trigonal Hexagonal primitive vectors:**

$$\mathbf{a}_1 = \frac{1}{2}\,a\,\hat{\mathbf{x}} - \frac{\sqrt{3}}{2}\,a\,\hat{\mathbf{y}}$$
$$\mathbf{a}_2 = \frac{1}{2}\,a\,\hat{\mathbf{x}} + \frac{\sqrt{3}}{2}\,a\,\hat{\mathbf{y}}$$
$$\mathbf{a}_3 = c\,\hat{\mathbf{z}}$$

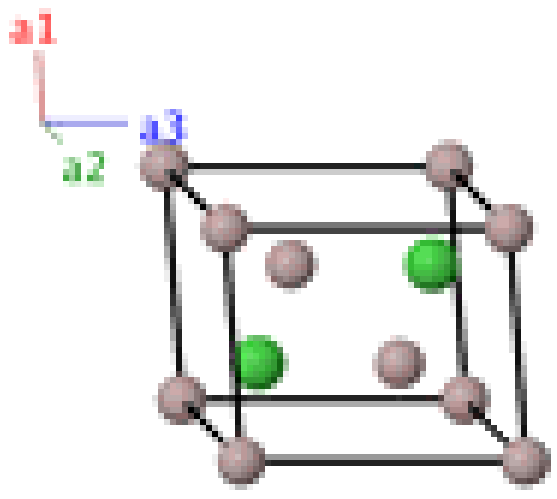

**Basis vectors:**

| | | Lattice Coordinates | | Cartesian Coordinates | Wyckoff Position | Atom Type |
|---|---|---|---|---|---|---|
| $\mathbf{B_1}$ | = | $0\,\mathbf{a_1} + 0\,\mathbf{a_2} + 0\,\mathbf{a_3}$ | = | $0\hat{\mathbf{x}} + 0\hat{\mathbf{y}} + 0\hat{\mathbf{z}}$ | $(1a)$ | Al I |
| $\mathbf{B_2}$ | = | $\frac{1}{3}\,\mathbf{a_1} + \frac{2}{3}\,\mathbf{a_2} + z_2\,\mathbf{a_3}$ | = | $\frac{1}{2}\,a\,\hat{\mathbf{x}} + \frac{1}{2\sqrt{3}}\,a\,\hat{\mathbf{y}} + z_2\,c\,\hat{\mathbf{z}}$ | $(2d)$ | Al II |
| $\mathbf{B_3}$ | = | $\frac{2}{3}\,\mathbf{a_1} + \frac{1}{3}\,\mathbf{a_2} - z_2\,\mathbf{a_3}$ | = | $\frac{1}{2}\,a\,\hat{\mathbf{x}} - \frac{1}{2\sqrt{3}}\,a\,\hat{\mathbf{y}} - z_2\,c\,\hat{\mathbf{z}}$ | $(2d)$ | Al II |
| $\mathbf{B_4}$ | = | $\frac{1}{3}\,\mathbf{a_1} + \frac{2}{3}\,\mathbf{a_2} + z_3\,\mathbf{a_3}$ | = | $\frac{1}{2}\,a\,\hat{\mathbf{x}} + \frac{1}{2\sqrt{3}}\,a\,\hat{\mathbf{y}} + z_3\,c\,\hat{\mathbf{z}}$ | $(2d)$ | Ni |
| $\mathbf{B_5}$ | = | $\frac{2}{3}\,\mathbf{a_1} + \frac{1}{3}\,\mathbf{a_2} - z_3\,\mathbf{a_3}$ | = | $\frac{1}{2}\,a\,\hat{\mathbf{x}} - \frac{1}{2\sqrt{3}}\,a\,\hat{\mathbf{y}} - z_3\,c\,\hat{\mathbf{z}}$ | $(2d)$ | Ni |

**Geometry files:**

- CIF: pp. 721
- POSCAR: pp. 721

# $\omega$ (C6) Phase: AB2_hP3_164_a_d

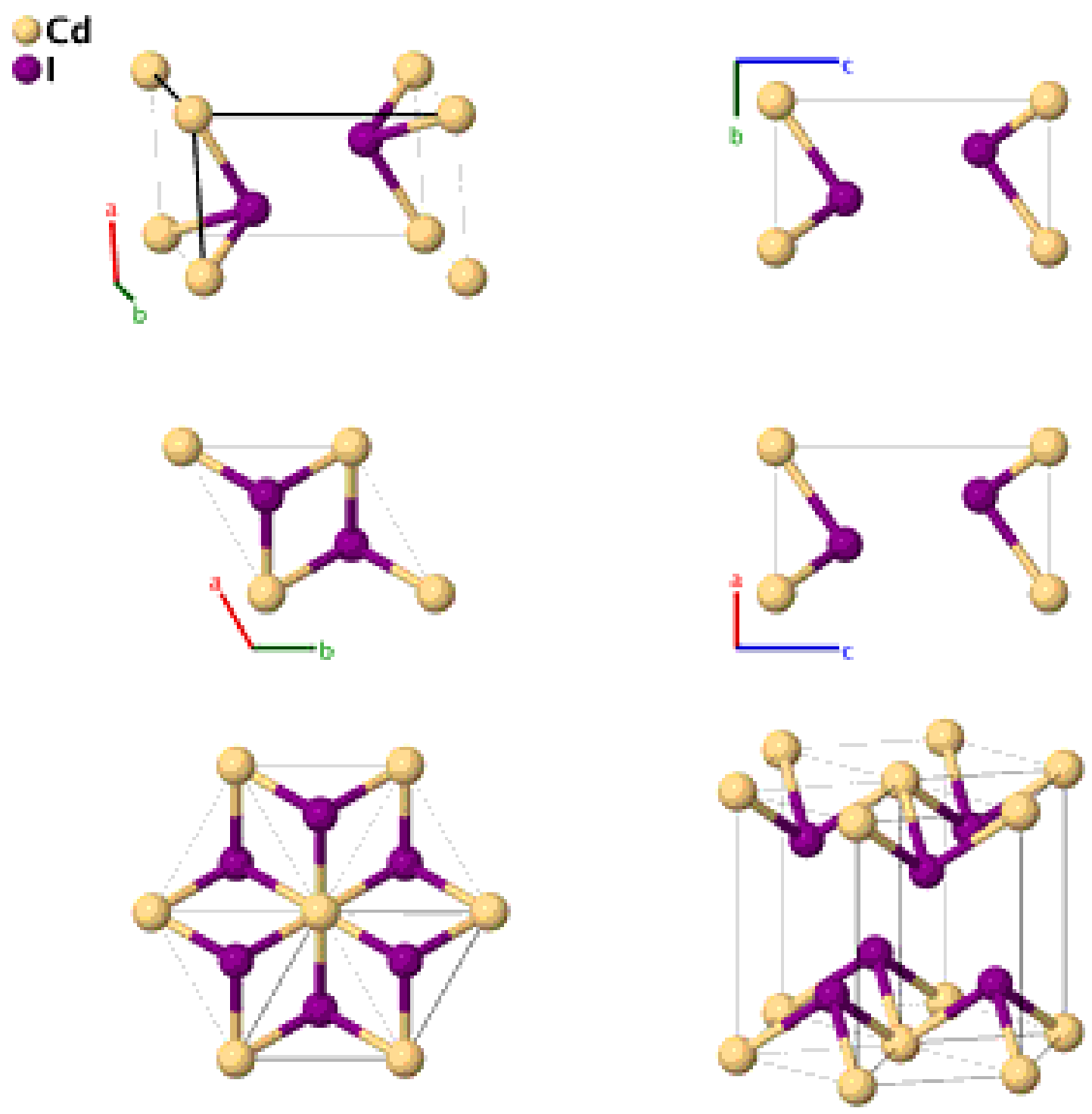

| | | |
|---|---|---|
| **Prototype** | : | $CdI_2$ |
| **AFLOW prototype label** | : | AB2_hP3_164_a_d |
| ***Strukturbericht* designation** | : | C6 |
| **Pearson symbol** | : | hP3 |
| **Space group number** | : | 164 |
| **Space group symbol** | : | $P\bar{3}m1$ |
| **AFLOW prototype command** | : | `aflow --proto=AB2_hP3_164_a_d`<br>`--params=`$a, c/a, z_2$ |

**Other compounds with this structure:**

- Ti, Zr, Hf, ZrNb, TiNb, TiV

- The $\omega$ phase can be either hexagonal or trigonal (shown here). The trigonal $\omega$ phase transforms into several high-symmetry structures under certain conditions:

| $c/a$ | $z$ | **Lattice** |
|---|---|---|
| Arbitrary | 0 | Ideal Omega (C32) |
| $\sqrt{\frac{3}{8}}$ | $\frac{1}{6}$ | Body-Centered Cubic (A2) |
| $\sqrt{\frac{3}{2}}$ | $\frac{1}{6}$ | Simple Cubic($A_h$) |
| $\sqrt{6}$ | $\frac{1}{6}$ | Face-Centered Cubic (A1) |
| Arbitrary | $\frac{1}{2}$ | Simple Hexagonal Structure ($A_f$) |

For more details about the omega phase and materials which form in the omega phase, see (Sikka, 1982) . As noted there, most omega phase intermetallic alloys are disordered. Although the "$\omega$" label comes from $\omega$-CrTi, (Ewald, 1931) lists the prototype for Strukturbericht designation C6 as CdI2.

## Trigonal Hexagonal primitive vectors:

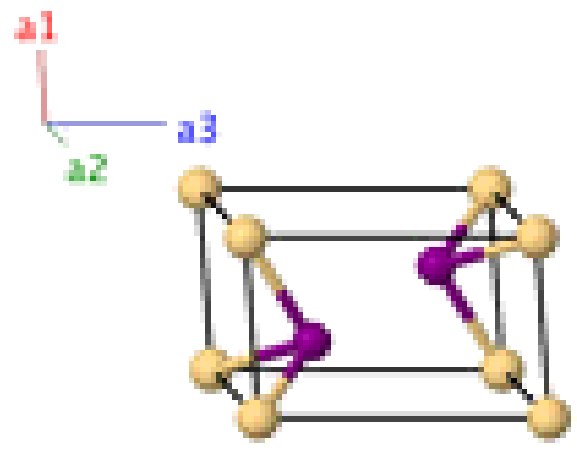

$$
\begin{aligned}
\mathbf{a}_1 &= \tfrac{1}{2}\,a\,\hat{\mathbf{x}} - \tfrac{\sqrt{3}}{2}\,a\,\hat{\mathbf{y}} \\
\mathbf{a}_2 &= \tfrac{1}{2}\,a\,\hat{\mathbf{x}} + \tfrac{\sqrt{3}}{2}\,a\,\hat{\mathbf{y}} \\
\mathbf{a}_3 &= c\,\hat{\mathbf{z}}
\end{aligned}
$$

## Basis vectors:

| | | Lattice Coordinates | | Cartesian Coordinates | Wyckoff Position | Atom Type |
|---|---|---|---|---|---|---|
| $\mathbf{B}_1$ | = | $0\,\mathbf{a}_1 + 0\,\mathbf{a}_2 + 0\,\mathbf{a}_3$ | = | $0\hat{\mathbf{x}} + 0\hat{\mathbf{y}} + 0\hat{\mathbf{z}}$ | (1$a$) | Cd |
| $\mathbf{B}_2$ | = | $\frac{1}{3}\,\mathbf{a}_1 + \frac{2}{3}\,\mathbf{a}_2 + z_2\,\mathbf{a}_3$ | = | $\frac{1}{2}\,a\,\hat{\mathbf{x}} + \frac{1}{2\sqrt{3}}\,a\,\hat{\mathbf{y}} + z_2\,c\,\hat{\mathbf{z}}$ | (2$d$) | I |
| $\mathbf{B}_3$ | = | $\frac{2}{3}\,\mathbf{a}_1 + \frac{1}{3}\,\mathbf{a}_2 - z_2\,\mathbf{a}_3$ | = | $\frac{1}{2}\,a\,\hat{\mathbf{x}} - \frac{1}{2\sqrt{3}}\,a\,\hat{\mathbf{y}} - z_2\,c\,\hat{\mathbf{z}}$ | (2$d$) | I |

## Found in:

## Geometry files:

- CIF: pp. 722
- POSCAR: pp. 722

# $H_3Ho$ Structure: A3B_hP24_165_adg_f

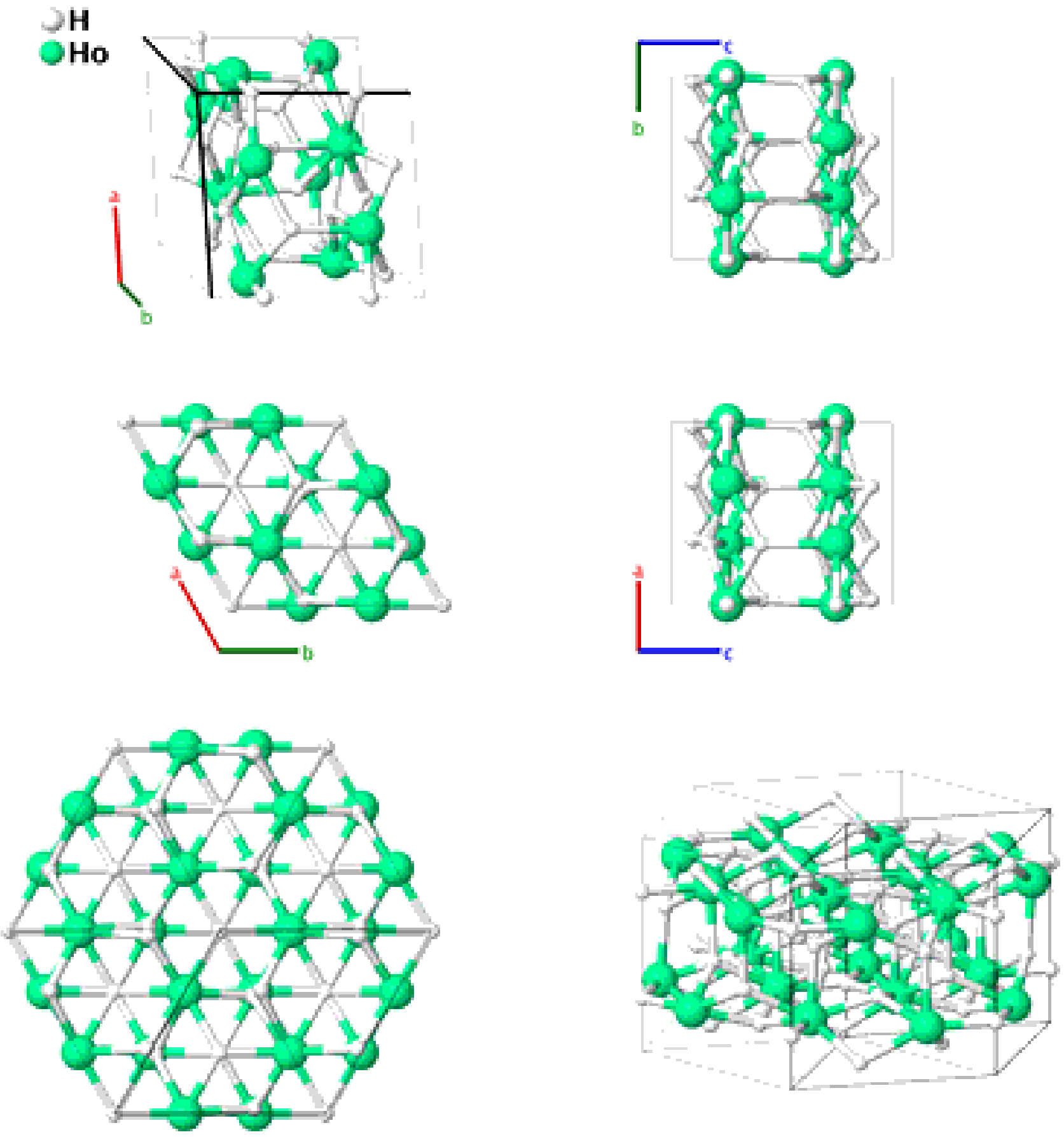

| | | |
|---|---|---|
| **Prototype** | : | $H_3Ho$ |
| **AFLOW prototype label** | : | A3B_hP24_165_adg_f |
| ***Strukturbericht* designation** | : | None |
| **Pearson symbol** | : | hP24 |
| **Space group number** | : | 165 |
| **Space group symbol** | : | $P\bar{3}c1$ |
| **AFLOW prototype command** | : | `aflow --proto=A3B_hP24_165_adg_f`<br>`--params=`$a, c/a, z_2, x_3, x_4, y_4, z_4$ |

**Other compounds with this structure:**

- $H_3Dy$, $H_3Er$, $H_3Gd$, $H_3Lu$, $H_3Sm$, $H_3Tb$, $H_3Tm$, $H_3Y$, $AuCu_3$, $AuMg_3$, $Cu_3P$

- As with all compounds involving hydrogen, structural determinations were made with deuterium.

**Trigonal Hexagonal primitive vectors:**

$$\mathbf{a}_1 = \frac{1}{2}\,a\,\hat{\mathbf{x}} - \frac{\sqrt{3}}{2}\,a\,\hat{\mathbf{y}}$$
$$\mathbf{a}_2 = \frac{1}{2}\,a\,\hat{\mathbf{x}} + \frac{\sqrt{3}}{2}\,a\,\hat{\mathbf{y}}$$
$$\mathbf{a}_3 = c\,\hat{\mathbf{z}}$$

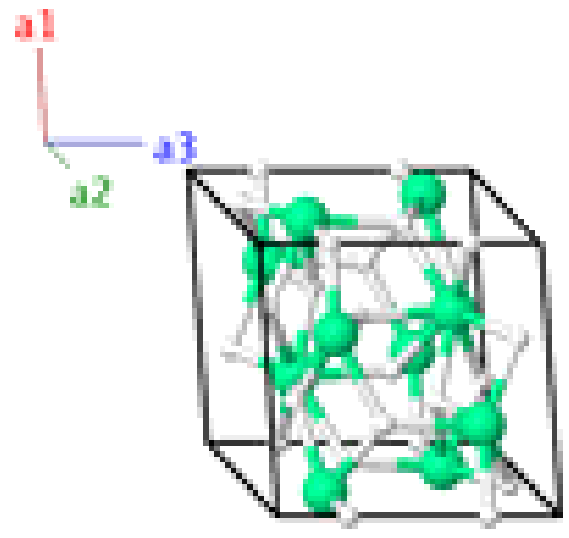

**Basis vectors:**

| | | Lattice Coordinates | | Cartesian Coordinates | Wyckoff Position | Atom Type |
|---|---|---|---|---|---|---|
| $\mathbf{B}_1$ | = | $\frac{1}{4}\,\mathbf{a}_3$ | = | $\frac{1}{4}\,c\,\hat{\mathbf{z}}$ | (2$a$) | H I |
| $\mathbf{B}_2$ | = | $\frac{3}{4}\,\mathbf{a}_3$ | = | $\frac{3}{4}\,c\,\hat{\mathbf{z}}$ | (2$a$) | H I |
| $\mathbf{B}_3$ | = | $\frac{1}{3}\,\mathbf{a}_1 + \frac{2}{3}\,\mathbf{a}_2 + z_2\,\mathbf{a}_3$ | = | $\frac{1}{2}\,a\,\hat{\mathbf{x}} + \frac{1}{2\sqrt{3}}\,a\,\hat{\mathbf{y}} + z_2\,c\,\hat{\mathbf{z}}$ | (4$d$) | H II |
| $\mathbf{B}_4$ | = | $\frac{2}{3}\,\mathbf{a}_1 + \frac{1}{3}\,\mathbf{a}_2 + \left(\frac{1}{2} - z_2\right)\,\mathbf{a}_3$ | = | $\frac{1}{2}\,a\,\hat{\mathbf{x}} - \frac{1}{2\sqrt{3}}\,a\,\hat{\mathbf{y}} + \left(\frac{1}{2} - z_2\right)\,c\,\hat{\mathbf{z}}$ | (4$d$) | H II |
| $\mathbf{B}_5$ | = | $\frac{2}{3}\,\mathbf{a}_1 + \frac{1}{3}\,\mathbf{a}_2 - z_2\,\mathbf{a}_3$ | = | $\frac{1}{2}\,a\,\hat{\mathbf{x}} - \frac{1}{2\sqrt{3}}\,a\,\hat{\mathbf{y}} - z_2\,c\,\hat{\mathbf{z}}$ | (4$d$) | H II |
| $\mathbf{B}_6$ | = | $\frac{1}{3}\,\mathbf{a}_1 + \frac{2}{3}\,\mathbf{a}_2 + \left(\frac{1}{2} + z_2\right)\,\mathbf{a}_3$ | = | $\frac{1}{2}\,a\,\hat{\mathbf{x}} + \frac{1}{2\sqrt{3}}\,a\,\hat{\mathbf{y}} + \left(\frac{1}{2} + z_2\right)\,c\,\hat{\mathbf{z}}$ | (4$d$) | H II |
| $\mathbf{B}_7$ | = | $x_3\,\mathbf{a}_1 + \frac{1}{4}\,\mathbf{a}_3$ | = | $\frac{1}{2}\,x_3\,a\,\hat{\mathbf{x}} - \frac{\sqrt{3}}{2}\,x_3\,a\,\hat{\mathbf{y}} + \frac{1}{4}\,c\,\hat{\mathbf{z}}$ | (6$f$) | Ho |
| $\mathbf{B}_8$ | = | $x_3\,\mathbf{a}_2 + \frac{1}{4}\,\mathbf{a}_3$ | = | $\frac{1}{2}\,x_3\,a\,\hat{\mathbf{x}} + \frac{\sqrt{3}}{2}\,x_3\,a\,\hat{\mathbf{y}} + \frac{1}{4}\,c\,\hat{\mathbf{z}}$ | (6$f$) | Ho |
| $\mathbf{B}_9$ | = | $-x_3\,\mathbf{a}_1 - x_3\,\mathbf{a}_2 + \frac{1}{4}\,\mathbf{a}_3$ | = | $-x_3\,a\,\hat{\mathbf{x}} + \frac{1}{4}\,c\,\hat{\mathbf{z}}$ | (6$f$) | Ho |
| $\mathbf{B}_{10}$ | = | $-x_3\,\mathbf{a}_1 + \frac{3}{4}\,\mathbf{a}_3$ | = | $-\frac{1}{2}\,x_3\,a\,\hat{\mathbf{x}} + \frac{\sqrt{3}}{2}\,x_3\,a\,\hat{\mathbf{y}} + \frac{3}{4}\,c\,\hat{\mathbf{z}}$ | (6$f$) | Ho |
| $\mathbf{B}_{11}$ | = | $-x_3\,\mathbf{a}_2 + \frac{3}{4}\,\mathbf{a}_3$ | = | $-\frac{1}{2}\,x_3\,a\,\hat{\mathbf{x}} - \frac{\sqrt{3}}{2}\,x_3\,a\,\hat{\mathbf{y}} + \frac{3}{4}\,c\,\hat{\mathbf{z}}$ | (6$f$) | Ho |
| $\mathbf{B}_{12}$ | = | $x_3\,\mathbf{a}_1 + x_3\,\mathbf{a}_2 + \frac{3}{4}\,\mathbf{a}_3$ | = | $x_3\,a\,\hat{\mathbf{x}} + \frac{3}{4}\,c\,\hat{\mathbf{z}}$ | (6$f$) | Ho |
| $\mathbf{B}_{13}$ | = | $x_4\,\mathbf{a}_1 + y_4\,\mathbf{a}_2 + z_4\,\mathbf{a}_3$ | = | $\frac{1}{2}\,(x_4 + y_4)\,a\,\hat{\mathbf{x}} + \frac{\sqrt{3}}{2}\,(y_4 - x_4)\,a\,\hat{\mathbf{y}} + z_4\,c\,\hat{\mathbf{z}}$ | (12$g$) | H III |
| $\mathbf{B}_{14}$ | = | $-y_4\,\mathbf{a}_1 + (x_4 - y_4)\,\mathbf{a}_2 + z_4\,\mathbf{a}_3$ | = | $\frac{1}{2}\,(x_4 - 2y_4)\,a\,\hat{\mathbf{x}} + \frac{\sqrt{3}}{2}x_4\,a\,\hat{\mathbf{y}} + z_4\,c\,\hat{\mathbf{z}}$ | (12$g$) | H III |
| $\mathbf{B}_{15}$ | = | $(y_4 - x_4)\,\mathbf{a}_1 - x_4\,\mathbf{a}_2 + z_4\,\mathbf{a}_3$ | = | $\frac{1}{2}\,(y_4 - 2x_4)\,a\,\hat{\mathbf{x}} - \frac{\sqrt{3}}{2}y_4\,a\,\hat{\mathbf{y}} + z_4\,c\,\hat{\mathbf{z}}$ | (12$g$) | H III |
| $\mathbf{B}_{16}$ | = | $y_4\,\mathbf{a}_1 + x_4\,\mathbf{a}_2 + \left(\frac{1}{2} - z_4\right)\,\mathbf{a}_3$ | = | $\frac{1}{2}\,(x_4 + y_4)\,a\,\hat{\mathbf{x}} + \frac{\sqrt{3}}{2}\,(x_4 - y_4)\,a\,\hat{\mathbf{y}} + \left(\frac{1}{2} - z_4\right)\,c\,\hat{\mathbf{z}}$ | (12$g$) | H III |
| $\mathbf{B}_{17}$ | = | $(x_4 - y_4)\,\mathbf{a}_1 - y_4\,\mathbf{a}_2 + \left(\frac{1}{2} - z_4\right)\,\mathbf{a}_3$ | = | $\frac{1}{2}\,(x_4 - 2y_4)\,a\,\hat{\mathbf{x}} - \frac{\sqrt{3}}{2}x_4\,a\,\hat{\mathbf{y}} + \left(\frac{1}{2} - z_4\right)\,c\,\hat{\mathbf{z}}$ | (12$g$) | H III |
| $\mathbf{B}_{18}$ | = | $-x_4\,\mathbf{a}_1 + (y_4 - x_4)\,\mathbf{a}_2 + \left(\frac{1}{2} - z_4\right)\,\mathbf{a}_3$ | = | $\frac{1}{2}\,(y_4 - 2x_4)\,a\,\hat{\mathbf{x}} + \frac{\sqrt{3}}{2}y_4\,a\,\hat{\mathbf{y}} + \left(\frac{1}{2} - z_4\right)\,c\,\hat{\mathbf{z}}$ | (12$g$) | H III |
| $\mathbf{B}_{19}$ | = | $-x_4\,\mathbf{a}_1 - y_4\,\mathbf{a}_2 - z_4\,\mathbf{a}_3$ | = | $-\frac{1}{2}\,(x_4 + y_4)\,a\,\hat{\mathbf{x}} + \frac{\sqrt{3}}{2}\,(x_4 - y_4)\,a\,\hat{\mathbf{y}} - z_4\,c\,\hat{\mathbf{z}}$ | (12$g$) | H III |
| $\mathbf{B}_{20}$ | = | $y_4\,\mathbf{a}_1 + (y_4 - x_4)\,\mathbf{a}_2 - z_4\,\mathbf{a}_3$ | = | $\frac{1}{2}\,(2y_4 - x_4)\,a\,\hat{\mathbf{x}} - \frac{\sqrt{3}}{2}x_4\,a\,\hat{\mathbf{y}} - z_4\,c\,\hat{\mathbf{z}}$ | (12$g$) | H III |
| $\mathbf{B}_{21}$ | = | $(x_4 - y_4)\,\mathbf{a}_1 + x_4\,\mathbf{a}_2 - z_4\,\mathbf{a}_3$ | = | $\frac{1}{2}\,(2x_4 - y_4)\,a\,\hat{\mathbf{x}} + \frac{\sqrt{3}}{2}y_4\,a\,\hat{\mathbf{y}} - z_4\,c\,\hat{\mathbf{z}}$ | (12$g$) | H III |
| $\mathbf{B}_{22}$ | = | $-y_4\,\mathbf{a}_1 - x_4\,\mathbf{a}_2 + \left(\frac{1}{2} + z_4\right)\,\mathbf{a}_3$ | = | $-\frac{1}{2}\,(x_4 + y_4)\,a\,\hat{\mathbf{x}} + \frac{\sqrt{3}}{2}\,(y_4 - x_4)\,a\,\hat{\mathbf{y}} + \left(\frac{1}{2} + z_4\right)\,c\,\hat{\mathbf{z}}$ | (12$g$) | H III |
| $\mathbf{B_{23}}$ | = | $(y_4 - x_4)\,\mathbf{a_1} + y_4\,\mathbf{a_2} + \left(\frac{1}{2} + z_4\right)\mathbf{a_3}$ | = | $\frac{1}{2}(2y_4 - x_4)\,a\,\mathbf{\hat{x}} + \frac{\sqrt{3}}{2}x_4\,a\,\mathbf{\hat{y}} + \left(\frac{1}{2} + z_4\right)c\,\mathbf{\hat{z}}$ | $(12g)$ | H III |
| $\mathbf{B_{24}}$ | = | $x_4\,\mathbf{a_1} + (x_4 - y_4)\,\mathbf{a_2} + \left(\frac{1}{2} + z_4\right)\mathbf{a_3}$ | = | $\frac{1}{2}(2x_4 - y_4)\,a\,\mathbf{\hat{x}} - \frac{\sqrt{3}}{2}y_4\,a\,\mathbf{\hat{y}} + \left(\frac{1}{2} + z_4\right)c\,\mathbf{\hat{z}}$ | $(12g)$ | H III |

**Geometry files:**

- CIF: pp. 722
- POSCAR: pp. 722

# CuPt ($L1_1$) Structure: AB_hR2_166_a_b

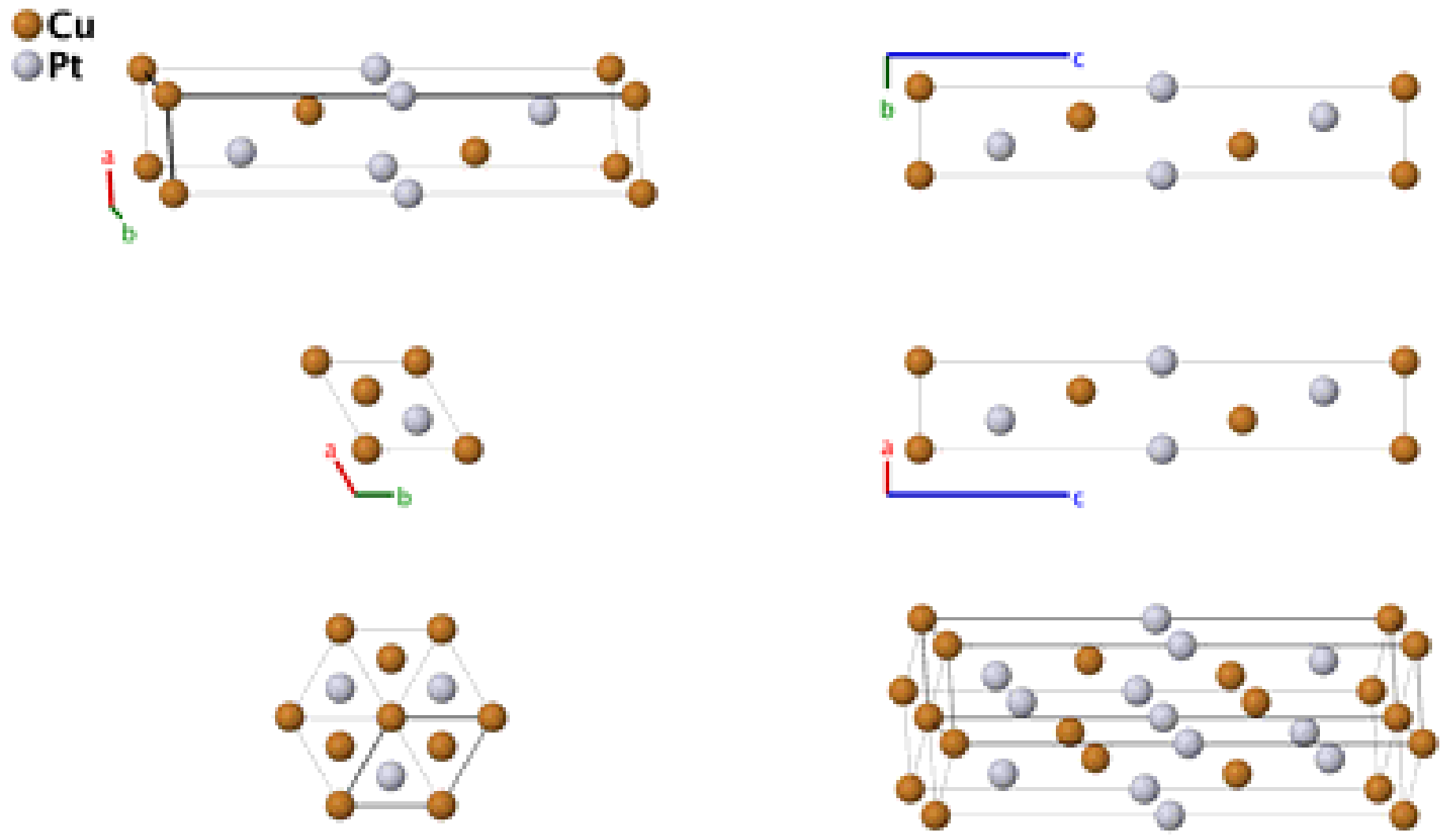

| | | |
|---|---|---|
| **Prototype** | : | CuPt |
| **AFLOW prototype label** | : | AB_hR2_166_a_b |
| ***Strukturbericht* designation** | : | $L1_1$ |
| **Pearson symbol** | : | hR2 |
| **Space group number** | : | 166 |
| **Space group symbol** | : | $R\bar{3}m$ |
| **AFLOW prototype command** | : | `aflow --proto=AB_hR2_166_a_b [--hex]` `--params=`$a, c/a$ |

- For notes on this structure, see the original reference, (Johansson, 1927), and the discussion in (Villars, 2007). We use the structure deduced by Villars et. al. As noted by (Barrett, 1980), even slight additions of Pt above stoichiometry will cause a change in the crystal structure. Hexagonal settings of this structure can be obtained with the option `--hex`.

**Rhombohedral primitive vectors:**

$$\mathbf{a}_1 = \frac{1}{2}\,a\,\hat{\mathbf{x}} - \frac{1}{2\sqrt{3}}\,a\,\hat{\mathbf{y}} + \frac{1}{3}\,c\,\hat{\mathbf{z}}$$
$$\mathbf{a}_2 = \frac{1}{\sqrt{3}}\,a\,\hat{\mathbf{y}} + \frac{1}{3}\,c\,\hat{\mathbf{z}}$$
$$\mathbf{a}_3 = -\frac{1}{2}\,a\,\hat{\mathbf{x}} - \frac{1}{2\sqrt{3}}\,a\,\hat{\mathbf{y}} + \frac{1}{3}\,c\,\hat{\mathbf{z}}$$

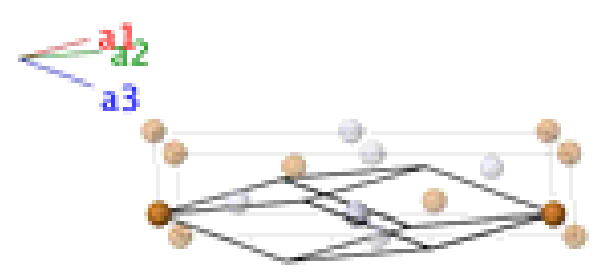

**Basis vectors:**

| | | Lattice Coordinates | | Cartesian Coordinates | Wyckoff Position | Atom Type |
|---|---|---|---|---|---|---|
| $\mathbf{B}_1$ | = | $0\,\mathbf{a}_1 + 0\,\mathbf{a}_2 + 0\,\mathbf{a}_3$ | = | $0\hat{\mathbf{x}} + 0\hat{\mathbf{y}} + 0\hat{\mathbf{z}}$ | (1$a$) | Cu |
| $\mathbf{B}_2$ | = | $\frac{1}{2}\,\mathbf{a}_1 + \frac{1}{2}\,\mathbf{a}_2 + \frac{1}{2}\,\mathbf{a}_3$ | = | $\frac{1}{2}\,c\,\hat{\mathbf{z}}$ | (1$b$) | Pt |

**Found in:**

---

**Geometry files:**

- CIF: pp. 723
- POSCAR: pp. 723

# $\alpha$-As (A7) Structure: A_hR2_166_c

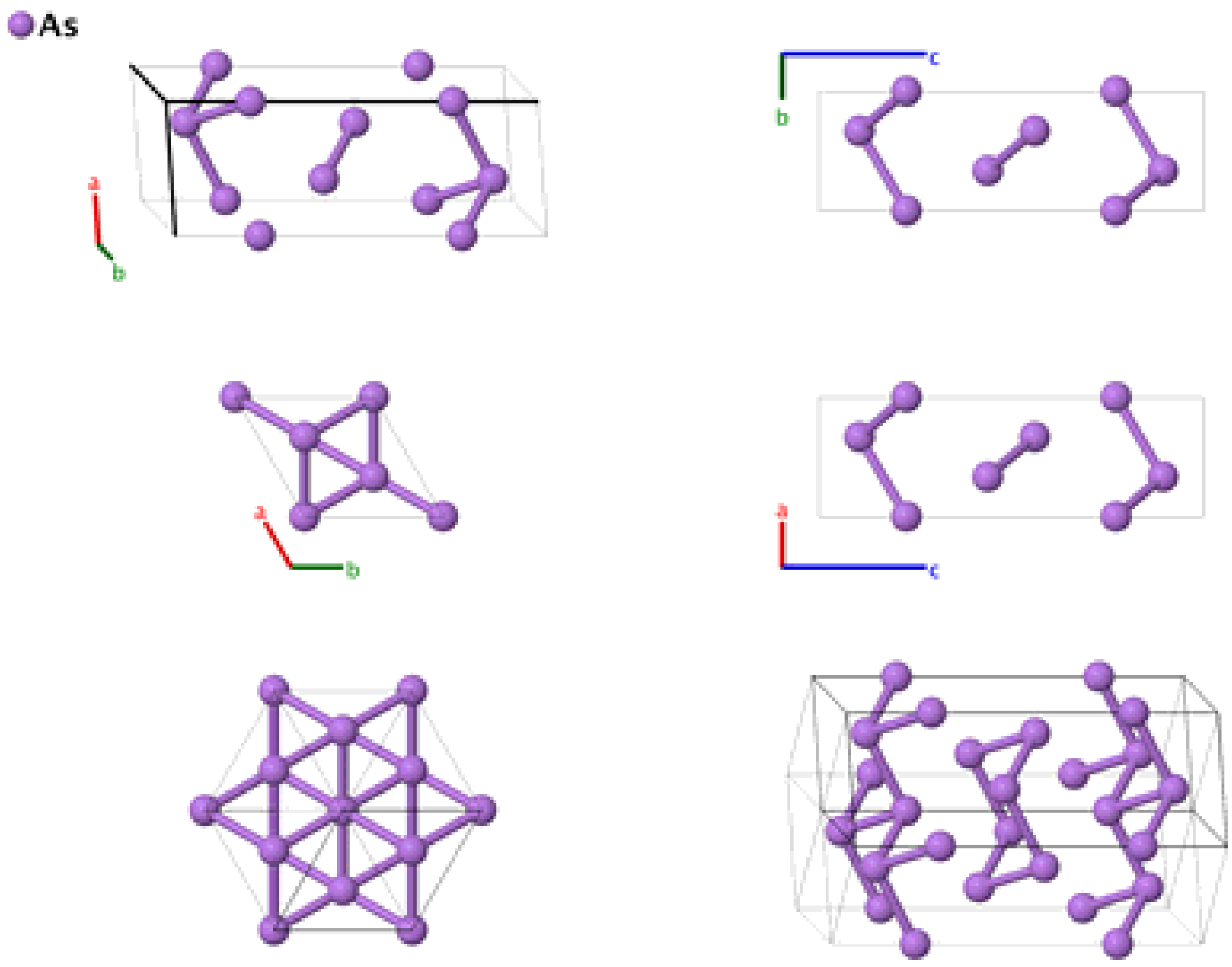

| | | |
|---|---|---|
| **Prototype** | : | $\alpha$-As |
| **AFLOW prototype label** | : | A_hR2_166_c |
| ***Strukturbericht* designation** | : | A7 |
| **Pearson symbol** | : | hR2 |
| **Space group number** | : | 166 |
| **Space group symbol** | : | R$\bar{3}$m |
| **AFLOW prototype command** | : | `aflow --proto=A_hR2_166_c [--hex]`<br>`--params=`$a, c/a, x_1$ |

**Other elements with this structure:**

- Sb, Bi

- When $c/a = \sqrt{6}$ and $z_1 = 1/8$ this becomes the diamond (A4) structure. Note that $\alpha$-As (pp. 378), rhombohedral graphite (pp. 392), and $\beta$-O (pp. 398) have the same AFLOW prototype label. They are generated by the same symmetry operations with different sets of parameters (`--params`) specified in their corresponding CIF files. Hexagonal settings of this structure can be obtained with the option `--hex`.

**Rhombohedral primitive vectors:**

$$\mathbf{a}_1 = \frac{1}{2}\,a\,\hat{\mathbf{x}} - \frac{1}{2\sqrt{3}}\,a\,\hat{\mathbf{y}} + \frac{1}{3}\,c\,\hat{\mathbf{z}}$$
$$\mathbf{a}_2 = \frac{1}{\sqrt{3}}\,a\,\hat{\mathbf{y}} + \frac{1}{3}\,c\,\hat{\mathbf{z}}$$
$$\mathbf{a}_3 = -\frac{1}{2}\,a\,\hat{\mathbf{x}} - \frac{1}{2\sqrt{3}}\,a\,\hat{\mathbf{y}} + \frac{1}{3}\,c\,\hat{\mathbf{z}}$$

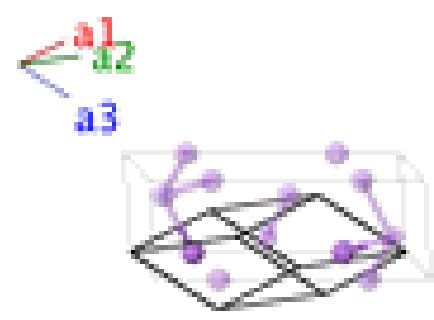

**Basis vectors:**

| | | Lattice Coordinates | | Cartesian Coordinates | Wyckoff Position | Atom Type |
|---|---|---|---|---|---|---|
| $\mathbf{B_1}$ | = | $x_1\,\mathbf{a_1} + x_1\,\mathbf{a_2} + x_1\,\mathbf{a_3}$ | = | $x_1\,c\,\hat{\mathbf{z}}$ | (2$c$) | As |
| $\mathbf{B_2}$ | = | $-x_1\,\mathbf{a_1} - x_1\,\mathbf{a_2} - x_1\,\mathbf{a_3}$ | = | $-x_1\,c\,\hat{\mathbf{z}}$ | (2$c$) | As |

**Geometry files:**

- CIF: pp. 723
- POSCAR: pp. 724

# $\beta$-Po ($A_i$) Structure: A_hR1_166_a

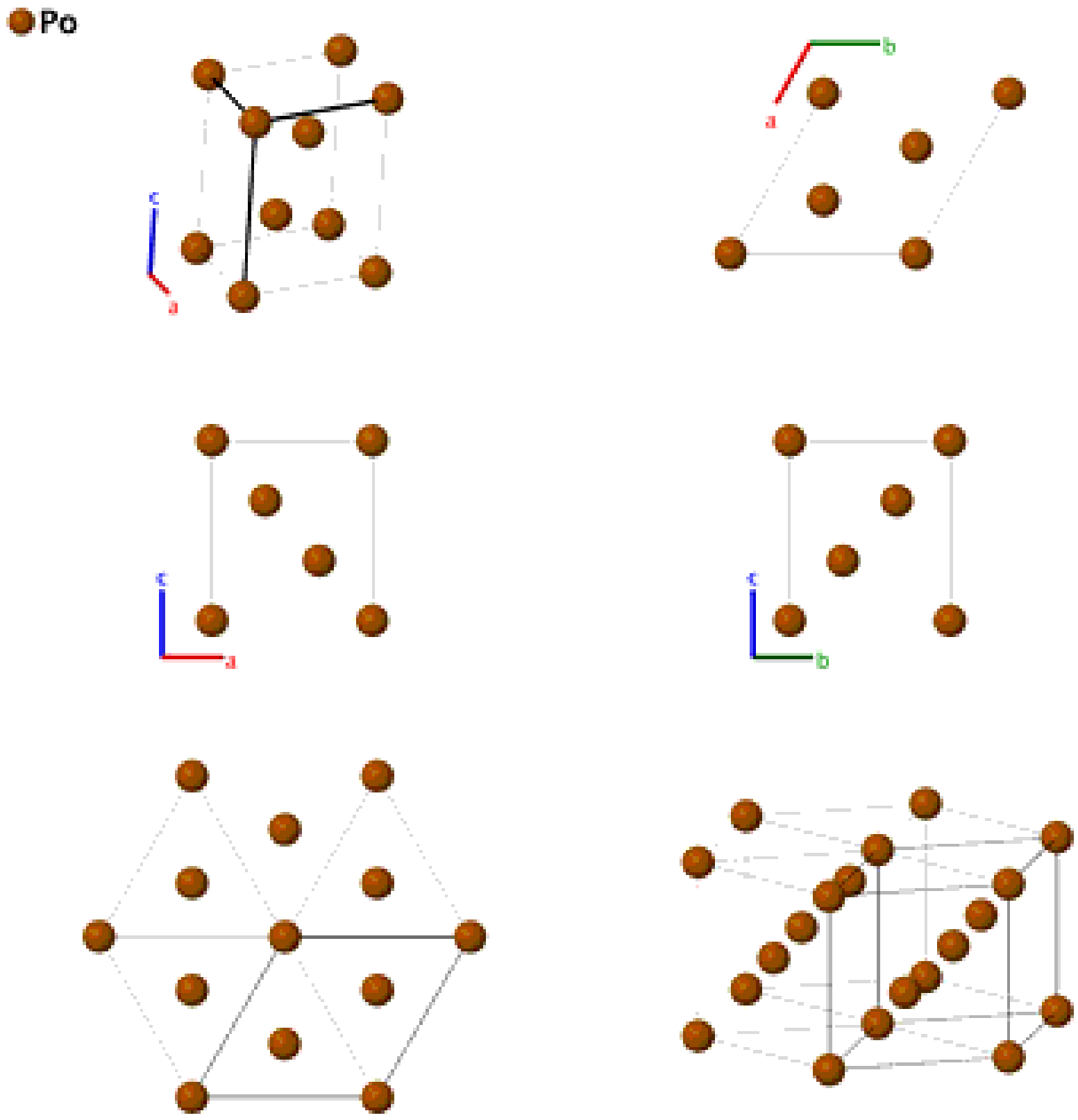

| | | |
|---|---|---|
| **Prototype** | : | $\beta$-Po |
| **AFLOW prototype label** | : | A_hR1_166_a |
| ***Strukturbericht* designation** | : | $A_i$ |
| **Pearson symbol** | : | hR1 |
| **Space group number** | : | 166 |
| **Space group symbol** | : | $R\bar{3}m$ |
| **AFLOW prototype command** | : | `aflow --proto=A_hR1_166_a [--hex]` `--params=`$a, c/a$ |

- This rhombohedral structure becomes cubic at various values of $c/a$ (or $\alpha$) to wit,

| $c/a$ | $\alpha$ | **Cubic Lattice** |
|---|---|---|
| $\sqrt{6}$ | $60^o$ | Face-Centered Cubic |
| $\sqrt{\frac{3}{2}}$ | $90^o$ | Simple Cubic |
| $\sqrt{\frac{3}{8}}$ | $109.47^o$ | Body-Centered Cubic |

Note that $\beta$-Po (pp. 380) and $\alpha$-Hg (pp. 388) have the same AFLOW prototype label. They are generated by the same symmetry operations with different sets of parameters (`--params`) specified in their corresponding CIF files. Experimentally, $\beta$-Po ($A_i$) has c/a near 1, or $\alpha > 90^o$, while $\alpha$-Hg (A10) has c/a near 2, or $\alpha < 90^o$. Originally, Po was assigned Strukturbericht designation: A19, which is now considered to be incorrect. (Donohue, 1982, pp. 390) Hexagonal settings of this structure can be obtained with the option `--hex`.

**Rhombohedral primitive vectors:**

$$\mathbf{a}_1 = \frac{1}{2}\,a\,\hat{\mathbf{x}} - \frac{1}{2\sqrt{3}}\,a\,\hat{\mathbf{y}} + \frac{1}{3}\,c\,\hat{\mathbf{z}}$$
$$\mathbf{a}_2 = \frac{1}{\sqrt{3}}\,a\,\hat{\mathbf{y}} + \frac{1}{3}\,c\,\hat{\mathbf{z}}$$
$$\mathbf{a}_3 = -\frac{1}{2}\,a\,\hat{\mathbf{x}} - \frac{1}{2\sqrt{3}}\,a\,\hat{\mathbf{y}} + \frac{1}{3}\,c\,\hat{\mathbf{z}}$$

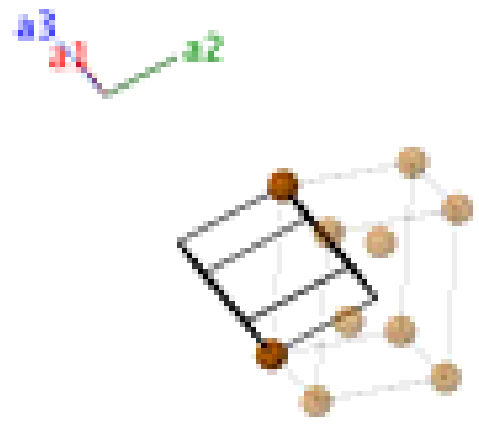

**Basis vectors:**

| | | Lattice Coordinates | | Cartesian Coordinates | Wyckoff Position | Atom Type |
|---|---|---|---|---|---|---|
| $\mathbf{B}_1$ | = | $0\,\mathbf{a}_1 + 0\,\mathbf{a}_2 + 0\,\mathbf{a}_3$ | = | $0\hat{\mathbf{x}} + 0\hat{\mathbf{y}} + 0\hat{\mathbf{z}}$ | $(1a)$ | Po |

**Found in:**

**Geometry files:**

- CIF: pp. 724
- POSCAR: pp. 724

# $Fe_7W_6$ ($D8_5$) $\mu$-phase: A7B6_hR13_166_ah_3c

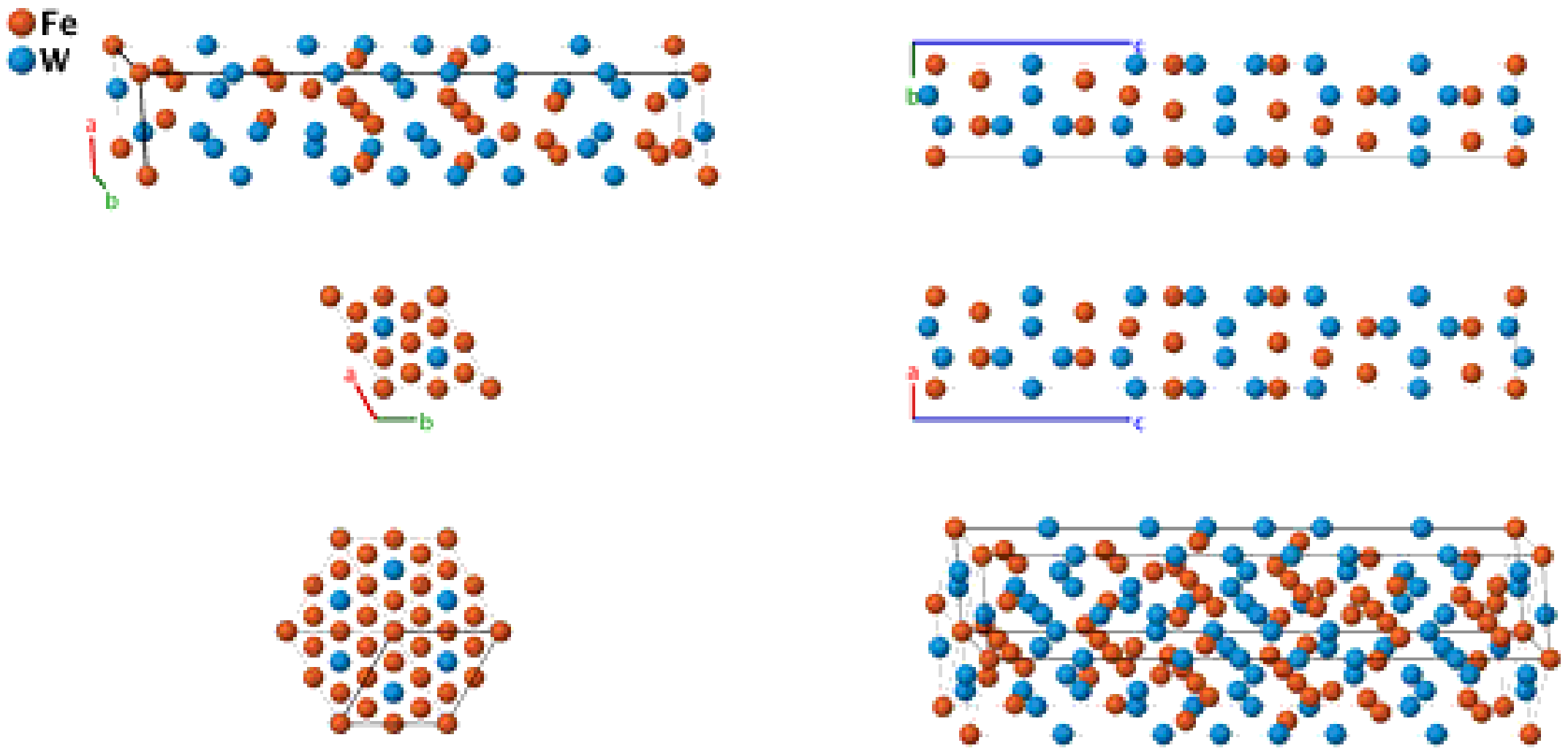

| | | |
|---|---|---|
| **Prototype** | : | $Fe_7W_6$ |
| **AFLOW prototype label** | : | A7B6_hR13_166_ah_3c |
| ***Strukturbericht* designation** | : | $D8_5$ |
| **Pearson symbol** | : | hR13 |
| **Space group number** | : | 166 |
| **Space group symbol** | : | $R\bar{3}m$ |
| **AFLOW prototype command** | : | `aflow --proto=A7B6_hR13_166_ah_3c [--hex]` <br> `--params=`$a, c/a, x_2, x_3, x_4, x_5, z_5$ |

**Other compounds with this structure:**

- $Co_7Mo_6$, $Co_6Mo_7$, $Co_7W_6$, $Co_6Re_6Si$, $Fe_6Ta_7$, $Fe_7Nb_6$, $Fe_7Mo_6$, $Fe_7Ta_6$, $Ta_6Zn_7$, $Mn_6Si_7$, etc.

- For more information on the $\mu$-phase, see (Pearson, 1972) pp. 664. There it is referred to as a tetrahedrally close-packed Frank-Kasper structure. We have been unable to obtain a copy of the original reference for this structure, (Arnfeldt, 1935), so we use the structure from (Villars, 1991) pp. 3415, which itself is taken from a secondary reference. Hexagonal settings of this structure can be obtained with the option `--hex`.

**Rhombohedral primitive vectors:**

$$\mathbf{a}_1 = \frac{1}{2}\,a\,\hat{\mathbf{x}} - \frac{1}{2\sqrt{3}}\,a\,\hat{\mathbf{y}} + \frac{1}{3}\,c\,\hat{\mathbf{z}}$$

$$\mathbf{a}_2 = \frac{1}{\sqrt{3}}\,a\,\hat{\mathbf{y}} + \frac{1}{3}\,c\,\hat{\mathbf{z}}$$

$$\mathbf{a}_3 = -\frac{1}{2}\,a\,\hat{\mathbf{x}} - \frac{1}{2\sqrt{3}}\,a\,\hat{\mathbf{y}} + \frac{1}{3}\,c\,\hat{\mathbf{z}}$$

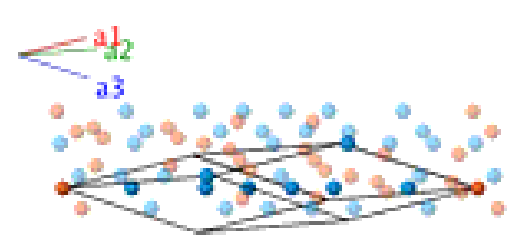

**Basis vectors:**

| | | Lattice Coordinates | | Cartesian Coordinates | Wyckoff Position | Atom Type |
|---|---|---|---|---|---|---|
| $\mathbf{B}_1$ | = | $0\,\mathbf{a}_1 + 0\,\mathbf{a}_2 + 0\,\mathbf{a}_3$ | = | $0\hat{\mathbf{x}} + 0\hat{\mathbf{y}} + 0\hat{\mathbf{z}}$ | (1$a$) | Fe I |
| $\mathbf{B}_2$ | = | $x_2\,\mathbf{a}_1 + x_2\,\mathbf{a}_2 + x_2\,\mathbf{a}_3$ | = | $x_2\,c\,\hat{\mathbf{z}}$ | (2$c$) | W I |
| $\mathbf{B}_3$ | = | $-x_2\,\mathbf{a}_1 - x_2\,\mathbf{a}_2 - x_2\,\mathbf{a}_3$ | = | $-x_2\,c\,\hat{\mathbf{z}}$ | (2$c$) | W I |
| $\mathbf{B}_4$ | = | $x_3\,\mathbf{a}_1 + x_3\,\mathbf{a}_2 + x_3\,\mathbf{a}_3$ | = | $x_3\,c\,\hat{\mathbf{z}}$ | (2$c$) | W II |
| $\mathbf{B}_5$ | = | $-x_3\,\mathbf{a}_1 - x_3\,\mathbf{a}_2 - x_3\,\mathbf{a}_3$ | = | $-x_3\,c\,\hat{\mathbf{z}}$ | (2$c$) | W II |
| $\mathbf{B}_6$ | = | $x_4\,\mathbf{a}_1 + x_4\,\mathbf{a}_2 + x_4\,\mathbf{a}_3$ | = | $x_4\,c\,\hat{\mathbf{z}}$ | (2$c$) | W III |
| $\mathbf{B}_7$ | = | $-x_4\,\mathbf{a}_1 - x_4\,\mathbf{a}_2 - x_4\,\mathbf{a}_3$ | = | $-x_4\,c\,\hat{\mathbf{z}}$ | (2$c$) | W III |
| $\mathbf{B}_8$ | = | $x_5\,\mathbf{a}_1 + x_5\,\mathbf{a}_2 + z_5\,\mathbf{a}_3$ | = | $\frac{1}{2}\,(x_5 - z_5)\,a\,\hat{\mathbf{x}} + \frac{1}{2\sqrt{3}}\,(x_5 - z_5)\,a\,\hat{\mathbf{y}} + \frac{1}{3}\,(2x_5 + z_5)\,c\,\hat{\mathbf{z}}$ | (6$h$) | Fe II |
| $\mathbf{B}_9$ | = | $z_5\,\mathbf{a}_1 + x_5\,\mathbf{a}_2 + x_5\,\mathbf{a}_3$ | = | $\frac{1}{2}\,(z_5 - x_5)\,a\,\hat{\mathbf{x}} + \frac{1}{2\sqrt{3}}\,(x_5 - z_5)\,a\,\hat{\mathbf{y}} + \frac{1}{3}\,(2x_5 + z_5)\,c\,\hat{\mathbf{z}}$ | (6$h$) | Fe II |
| $\mathbf{B}_{10}$ | = | $x_5\,\mathbf{a}_1 + z_5\,\mathbf{a}_2 + x_5\,\mathbf{a}_3$ | = | $\frac{1}{\sqrt{3}}\,(z_5 - x_5)\,a\,\hat{\mathbf{y}} + \frac{1}{3}\,(2x_5 + z_5)\,c\,\hat{\mathbf{z}}$ | (6$h$) | Fe II |
| $\mathbf{B}_{11}$ | = | $-x_5\,\mathbf{a}_1 - x_5\,\mathbf{a}_2 - z_5\,\mathbf{a}_3$ | = | $\frac{1}{2}\,(z_5 - x_5)\,a\,\hat{\mathbf{x}} + \frac{1}{2\sqrt{3}}\,(z_5 - x_5)\,a\,\hat{\mathbf{y}} - \frac{1}{3}\,(2x_5 + z_5)\,c\,\hat{\mathbf{z}}$ | (6$h$) | Fe II |
| $\mathbf{B}_{12}$ | = | $-z_5\,\mathbf{a}_1 - x_5\,\mathbf{a}_2 - x_5\,\mathbf{a}_3$ | = | $\frac{1}{2}\,(x_5 - z_5)\,a\,\hat{\mathbf{x}} + \frac{1}{2\sqrt{3}}\,(z_5 - x_5)\,a\,\hat{\mathbf{y}} - \frac{1}{3}\,(2x_5 + z_5)\,c\,\hat{\mathbf{z}}$ | (6$h$) | Fe II |
| $\mathbf{B}_{13}$ | = | $-x_5\,\mathbf{a}_1 - z_5\,\mathbf{a}_2 - x_5\,\mathbf{a}_3$ | = | $\frac{1}{\sqrt{3}}\,(x_5 - z_5)\,a\,\hat{\mathbf{y}} - \frac{1}{3}\,(2x_5 + z_5)\,c\,\hat{\mathbf{z}}$ | (6$h$) | Fe II |

**Geometry files:**

- CIF: pp. 724
- POSCAR: pp. 725

# $\alpha$-Sm (C19) Structure: A_hR3_166_ac

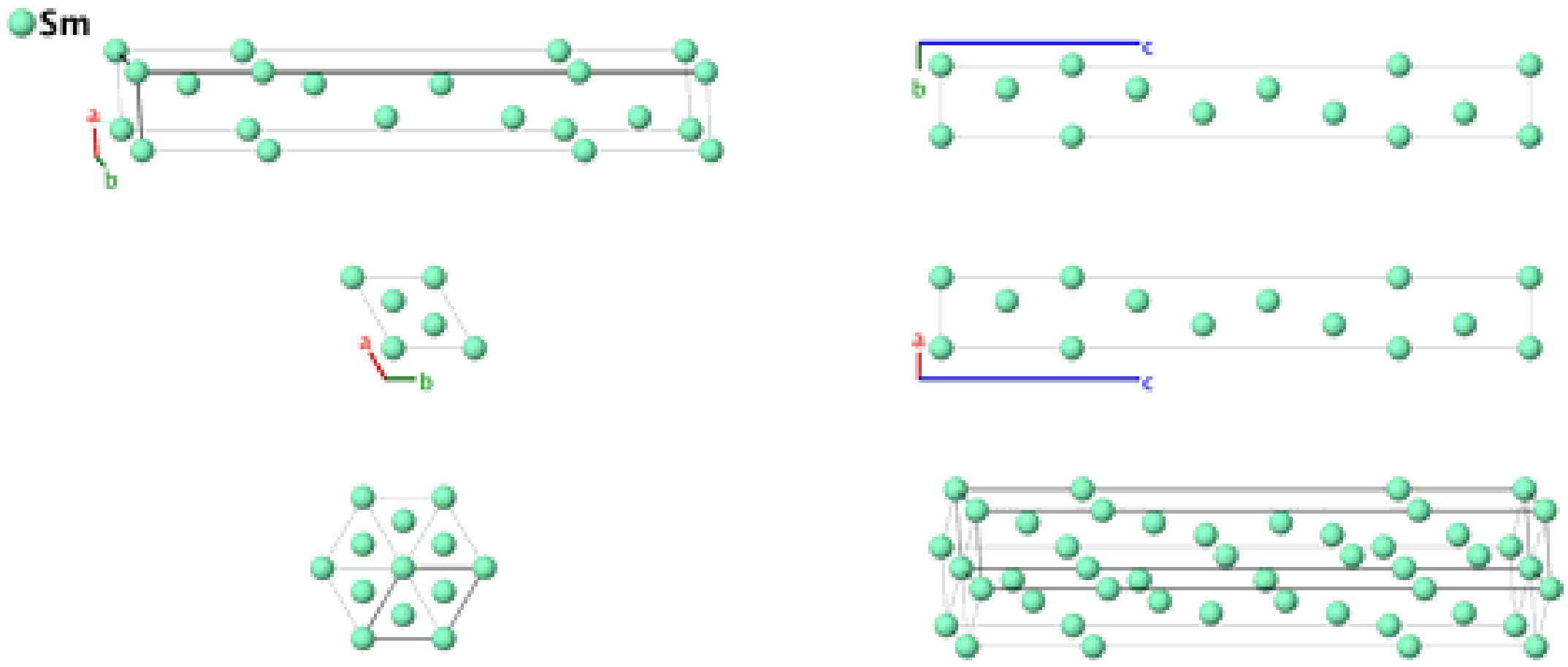

| | | |
|---|---|---|
| **Prototype** | : | $\alpha$-Sm |
| **AFLOW prototype label** | : | A_hR3_166_ac |
| ***Strukturbericht* designation** | : | C19 |
| **Pearson symbol** | : | hR3 |
| **Space group number** | : | 166 |
| **Space group symbol** | : | R$\bar{3}$m |
| **AFLOW prototype command** | : | `aflow --proto=A_hR3_166_ac [--hex]`<br>`--params=`$a, c/a, x_2$ |

**Other elements with this structure:**

- Li (Overhauser, 1984).

- Note that this is a close-packed system, with stacking ABCBCACAB, in contrast to the ABAB stacking of the hexagonal close-packed structure and the ABCABC stacking of the face-centered cubic structure. Hexagonal settings of this structure can be obtained with the option `--hex`.

**Rhombohedral primitive vectors:**

$$\mathbf{a}_1 = \frac{1}{2}\,a\,\mathbf{\hat{x}} - \frac{1}{2\sqrt{3}}\,a\,\mathbf{\hat{y}} + \frac{1}{3}\,c\,\mathbf{\hat{z}}$$
$$\mathbf{a}_2 = \frac{1}{\sqrt{3}}\,a\,\mathbf{\hat{y}} + \frac{1}{3}\,c\,\mathbf{\hat{z}}$$
$$\mathbf{a}_3 = -\frac{1}{2}\,a\,\mathbf{\hat{x}} - \frac{1}{2\sqrt{3}}\,a\,\mathbf{\hat{y}} + \frac{1}{3}\,c\,\mathbf{\hat{z}}$$

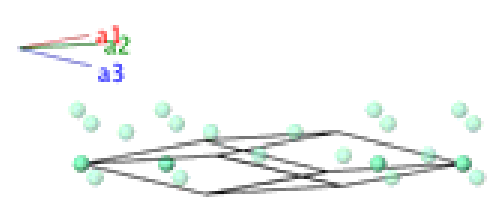

**Basis vectors:**

| | | Lattice Coordinates | | Cartesian Coordinates | Wyckoff Position | Atom Type |
|---|---|---|---|---|---|---|
| $\mathbf{B_1}$ | = | $0\,\mathbf{a_1} + 0\,\mathbf{a_2} + 0\,\mathbf{a_3}$ | = | $0\mathbf{\hat{x}} + 0\mathbf{\hat{y}} + 0\mathbf{\hat{z}}$ | $(1a)$ | Sm I |
| $\mathbf{B_2}$ | = | $x_2\,\mathbf{a_1} + x_2\,\mathbf{a_2} + x_2\,\mathbf{a_3}$ | = | $x_2\,c\,\mathbf{\hat{z}}$ | $(2c)$ | Sm II |
| $\mathbf{B_3}$ | = | $-x_2\,\mathbf{a_1} - x_2\,\mathbf{a_2} - x_2\,\mathbf{a_3}$ | = | $-x_2\,c\,\mathbf{\hat{z}}$ | $(2c)$ | Sm II |

---

**Geometry files:**

- CIF: pp. 725
- POSCAR: pp. 725

# $Bi_2Te_3$ Structure (C33): A2B3_hR5_166_c_ac

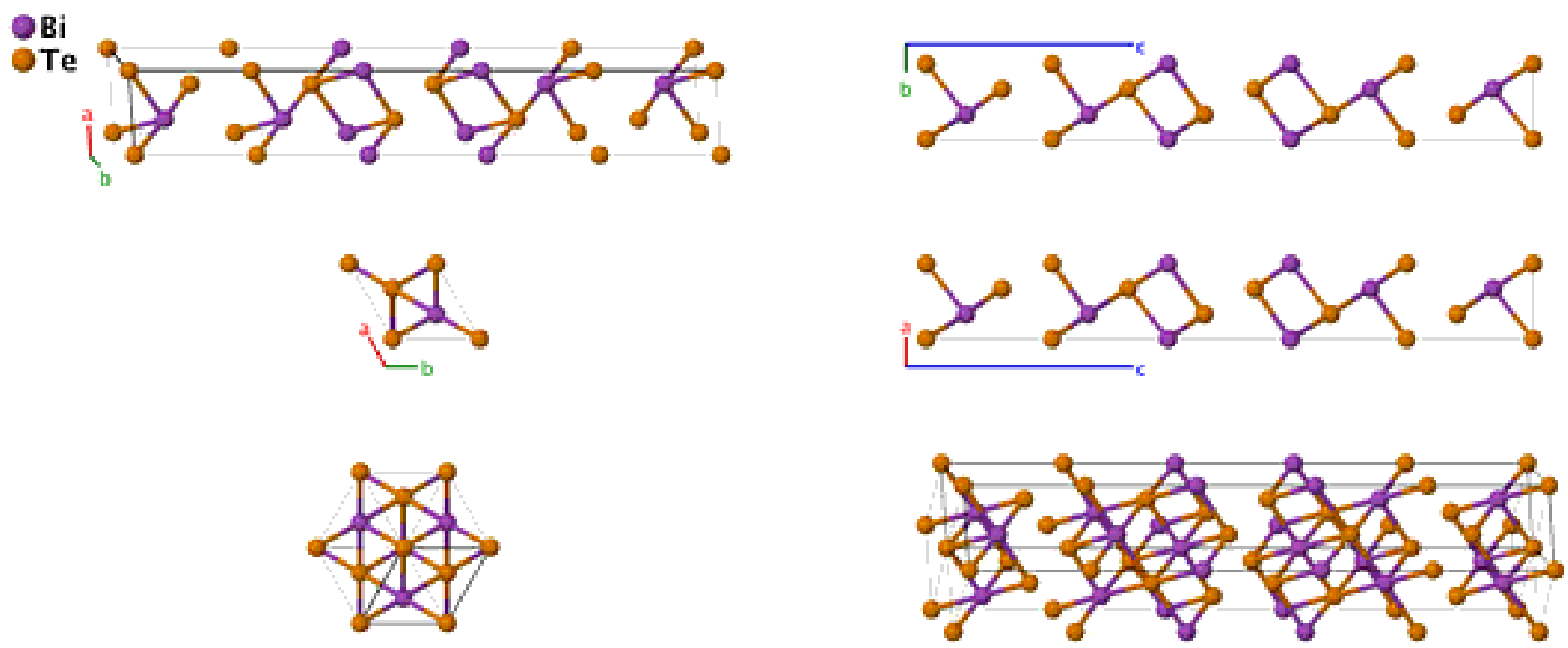

| | | |
|---|---|---|
| **Prototype** | : | $Bi_2Te_3$ |
| **AFLOW prototype label** | : | A2B3_hR5_166_c_ac |
| ***Strukturbericht* designation** | : | C33 |
| **Pearson symbol** | : | hR5 |
| **Space group number** | : | 166 |
| **Space group symbol** | : | $R\bar{3}m$ |
| **AFLOW prototype command** | : | `aflow --proto=A2B3_hR5_166_c_ac [--hex]` `--params=`$a, c/a, x_2, x_3$ |

**Other compounds with this structure:**

- $Be_2Te_2S$, $Sb_2Te_3$, $Bi_2Te_2Se$, $Bi_2Te_3$, $Bi_2Se_3$

- Hexagonal settings of this structure can be obtained with the option `--hex`.

**Rhombohedral primitive vectors:**

$$\mathbf{a}_1 = \frac{1}{2}\,a\,\hat{\mathbf{x}} - \frac{1}{2\sqrt{3}}\,a\,\hat{\mathbf{y}} + \frac{1}{3}\,c\,\hat{\mathbf{z}}$$
$$\mathbf{a}_2 = \frac{1}{\sqrt{3}}\,a\,\hat{\mathbf{y}} + \frac{1}{3}\,c\,\hat{\mathbf{z}}$$
$$\mathbf{a}_3 = -\frac{1}{2}\,a\,\hat{\mathbf{x}} - \frac{1}{2\sqrt{3}}\,a\,\hat{\mathbf{y}} + \frac{1}{3}\,c\,\hat{\mathbf{z}}$$

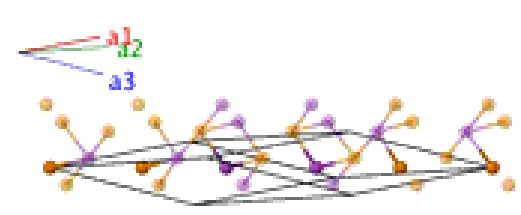

**Basis vectors:**

| | | Lattice Coordinates | | Cartesian Coordinates | Wyckoff Position | Atom Type |
|---|---|---|---|---|---|---|
| $\mathbf{B}_1$ | = | $0\,\mathbf{a}_1 + 0\,\mathbf{a}_2 + 0\,\mathbf{a}_3$ | = | $0\hat{\mathbf{x}} + 0\hat{\mathbf{y}} + 0\hat{\mathbf{z}}$ | (1*a*) | Te I |
| $\mathbf{B}_2$ | = | $x_2\,\mathbf{a}_1 + x_2\,\mathbf{a}_2 + x_2\,\mathbf{a}_3$ | = | $x_2\,c\,\hat{\mathbf{z}}$ | (2*c*) | Bi |
| $\mathbf{B}_3$ | = | $-x_2\,\mathbf{a}_1 - x_2\,\mathbf{a}_2 - x_2\,\mathbf{a}_3$ | = | $-x_2\,c\,\hat{\mathbf{z}}$ | (2*c*) | Bi |
| $\mathbf{B}_4$ | = | $x_3\,\mathbf{a}_1 + x_3\,\mathbf{a}_2 + x_3\,\mathbf{a}_3$ | = | $x_3\,c\,\hat{\mathbf{z}}$ | (2*c*) | Te II |
| $\mathbf{B}_5$ | = | $-x_3\,\mathbf{a}_1 - x_3\,\mathbf{a}_2 - x_3\,\mathbf{a}_3$ | = | $-x_3\,c\,\hat{\mathbf{z}}$ | (2*c*) | Te II |

**Geometry files:**

- CIF: pp. 725
- POSCAR: pp. 726

# $\alpha$-Hg (A10) Structure: A_hR1_166_a

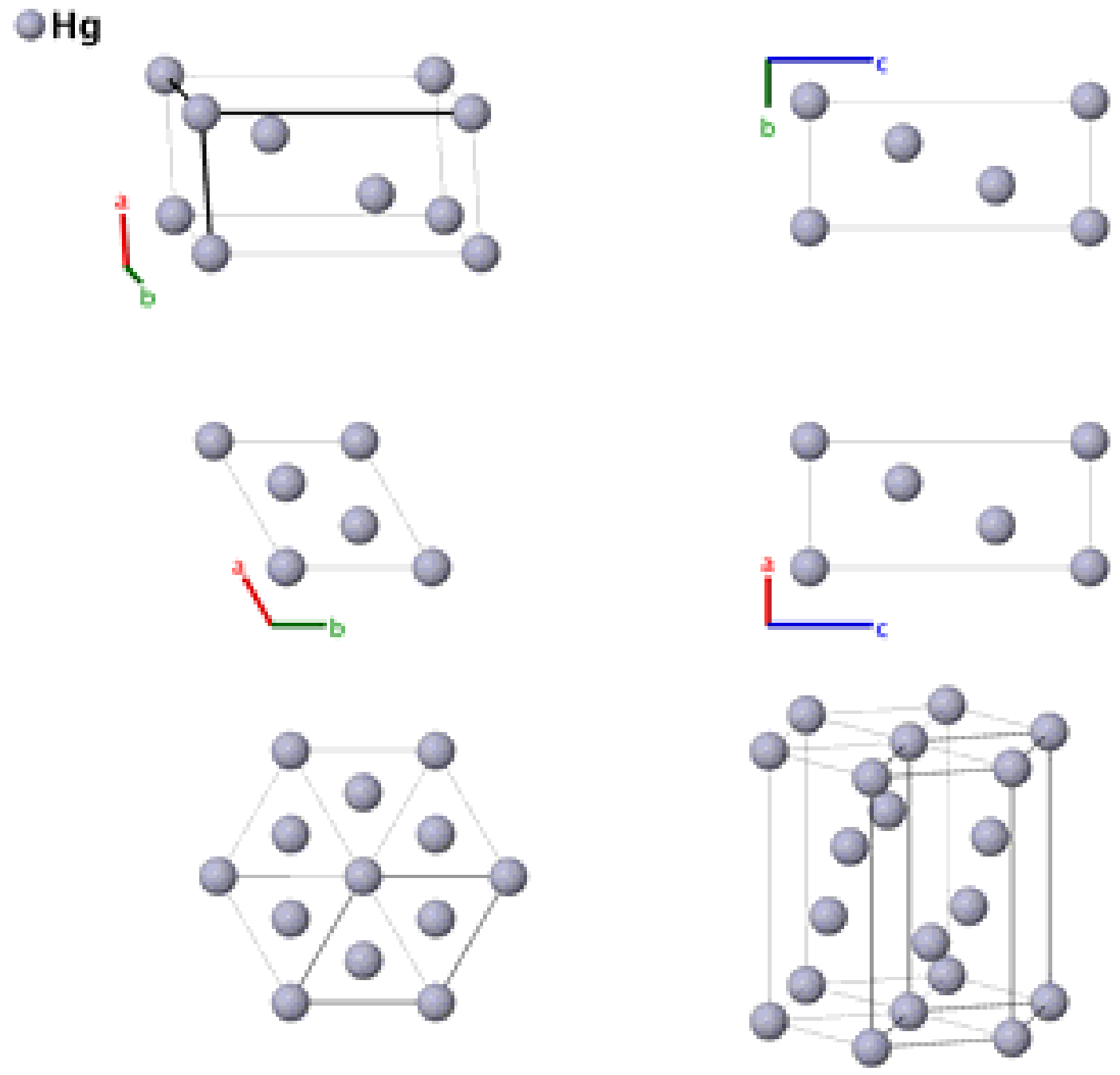

**Prototype** : $\alpha$-Hg

**AFLOW prototype label** : A_hR1_166_a

***Strukturbericht* designation** : A10

**Pearson symbol** : hR1

**Space group number** : 166

**Space group symbol** : $R\bar{3}m$

**AFLOW prototype command** : `aflow --proto=A_hR1_166_a [--hex]`
`--params=`$a, c/a$

- This rhombohedral structure becomes cubic at various values of $c/a$ (or $\alpha$) to wit,

| $c/a$ | $\alpha$ | **Cubic Lattice** |
|---|---|---|
| $\sqrt{6}$ | $60^o$ | Face-Centered Cubic |
| $\sqrt{\frac{3}{2}}$ | $90^o$ | Simple Cubic |
| $\sqrt{\frac{3}{8}}$ | $109.47^o$ | Body-Centered Cubic |

Note that $\beta$-Po (pp. 380) and $\alpha$-Hg (pp. 388) have the same AFLOW prototype label. They are generated by the same symmetry operations with different sets of parameters (`--params`) specified in their corresponding CIF files. Experimentally, $\beta$-Po ($A_i$) has c/a near 1, or $\alpha > 90^o$, while $\alpha$-Hg (A10) has c/a near 2, or $\alpha < 90^o$. Hexagonal settings of this structure can be obtained with the option `--hex`.

**Rhombohedral primitive vectors:**

$$\mathbf{a}_1 = \frac{1}{2}\,a\,\hat{\mathbf{x}} - \frac{1}{2\sqrt{3}}\,a\,\hat{\mathbf{y}} + \frac{1}{3}\,c\,\hat{\mathbf{z}}$$
$$\mathbf{a}_2 = \frac{1}{\sqrt{3}}\,a\,\hat{\mathbf{y}} + \frac{1}{3}\,c\,\hat{\mathbf{z}}$$
$$\mathbf{a}_3 = -\frac{1}{2}\,a\,\hat{\mathbf{x}} - \frac{1}{2\sqrt{3}}\,a\,\hat{\mathbf{y}} + \frac{1}{3}\,c\,\hat{\mathbf{z}}$$

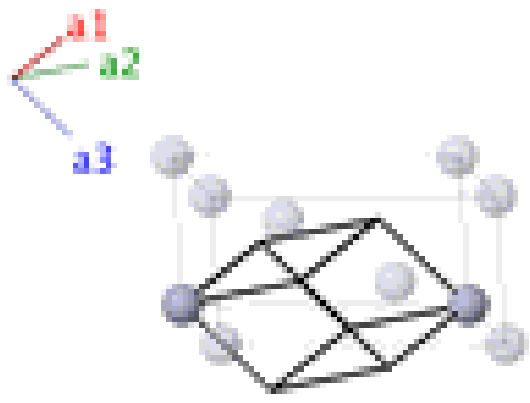

**Basis vectors:**

| | | Lattice Coordinates | | Cartesian Coordinates | Wyckoff Position | Atom Type |
|---|---|---|---|---|---|---|
| $\mathbf{B_1}$ | = | $0\,\mathbf{a_1} + 0\,\mathbf{a_2} + 0\,\mathbf{a_3}$ | = | $0\hat{\mathbf{x}} + 0\hat{\mathbf{y}} + 0\hat{\mathbf{z}}$ | $(1a)$ | Hg |

**Found in:**

**Geometry files:**

- CIF: pp. 726
- POSCAR: pp. 726

# $Mo_2B_5$ ($D8_i$) Structure: A5B2_hR7_166_a2c_c

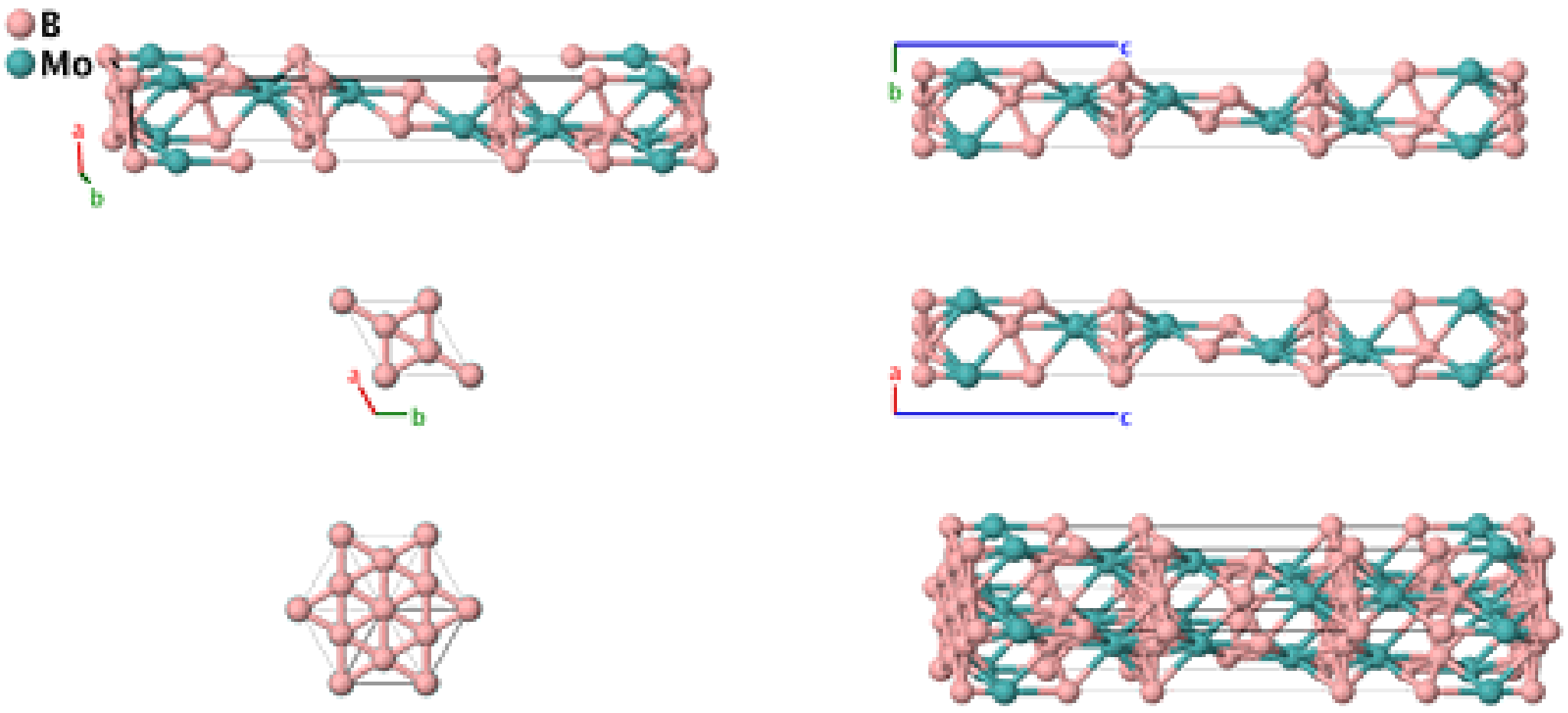

| | | |
|---|---|---|
| **Prototype** | : | $Mo_2B_5$ |
| **AFLOW prototype label** | : | A5B2_hR7_166_a2c_c |
| ***Strukturbericht* designation** | : | $D8_i$ |
| **Pearson symbol** | : | hR7 |
| **Space group number** | : | 166 |
| **Space group symbol** | : | $R\bar{3}m$ |
| **AFLOW prototype command** | : | `aflow --proto=A5B2_hR7_166_a2c_c [--hex]` `--params=`$a, c/a, x_2, x_3, x_4$ |

**Other compounds with this structure:**

- $V_2B_5$, $InL_5Tl$, $Li_5Sn_2$, $Li_5Tl_2$

- The boron atoms form buckled graphitic sheets, making this the rhombohedral form of $D8_h$. (Frotscher, 2007) suggest that the stable composition in this part of the molybdenum nitride system might be $Mo_2B_4$, but here we will describe the $D8_i$ structure, with the warning that this might not be the experimental structure. Hexagonal settings of this structure can be obtained with the option `--hex`.

**Rhombohedral primitive vectors:**

$$\mathbf{a}_1 = \frac{1}{2}\,a\,\hat{\mathbf{x}} - \frac{1}{2\sqrt{3}}\,a\,\hat{\mathbf{y}} + \frac{1}{3}\,c\,\hat{\mathbf{z}}$$
$$\mathbf{a}_2 = \frac{1}{\sqrt{3}}\,a\,\hat{\mathbf{y}} + \frac{1}{3}\,c\,\hat{\mathbf{z}}$$
$$\mathbf{a}_3 = -\frac{1}{2}\,a\,\hat{\mathbf{x}} - \frac{1}{2\sqrt{3}}\,a\,\hat{\mathbf{y}} + \frac{1}{3}\,c\,\hat{\mathbf{z}}$$

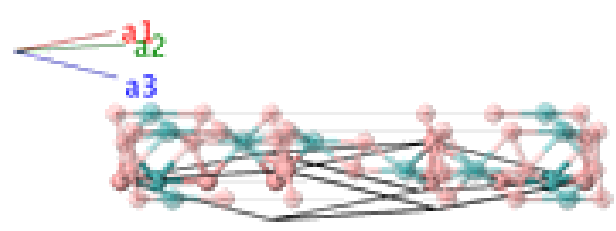

**Basis vectors:**

| | | Lattice Coordinates | | Cartesian Coordinates | Wyckoff Position | Atom Type |
|---|---|---|---|---|---|---|
| $\mathbf{B_1}$ | = | $0\,\mathbf{a_1} + 0\,\mathbf{a_2} + 0\,\mathbf{a_3}$ | = | $0\hat{\mathbf{x}} + 0\hat{\mathbf{y}} + 0\hat{\mathbf{z}}$ | $(1a)$ | B I |
| $\mathbf{B_2}$ | = | $x_2\,\mathbf{a_1} + x_2\,\mathbf{a_2} + x_2\,\mathbf{a_3}$ | = | $x_2\,c\,\hat{\mathbf{z}}$ | $(2c)$ | B II |
| $\mathbf{B_3}$ | = | $-x_2\,\mathbf{a_1} - x_2\,\mathbf{a_2} - x_2\,\mathbf{a_3}$ | = | $-x_2\,c\,\hat{\mathbf{z}}$ | $(2c)$ | B II |
| $\mathbf{B_4}$ | = | $x_3\,\mathbf{a_1} + x_3\,\mathbf{a_2} + x_3\,\mathbf{a_3}$ | = | $x_3\,c\,\hat{\mathbf{z}}$ | $(2c)$ | B III |
| $\mathbf{B_5}$ | = | $-x_3\,\mathbf{a_1} - x_3\,\mathbf{a_2} - x_3\,\mathbf{a_3}$ | = | $-x_3\,c\,\hat{\mathbf{z}}$ | (2$c$) | B III |
| $\mathbf{B_6}$ | = | $x_4\,\mathbf{a_1} + x_4\,\mathbf{a_2} + x_4\,\mathbf{a_3}$ | = | $x_4\,c\,\hat{\mathbf{z}}$ | (2$c$) | Mo |
| $\mathbf{B_7}$ | = | $-x_4\,\mathbf{a_1} - x_4\,\mathbf{a_2} - x_4\,\mathbf{a_3}$ | = | $-x_4\,c\,\hat{\mathbf{z}}$ | (2$c$) | Mo |

**Geometry files:**

- CIF: pp. 726
- POSCAR: pp. 727

# Rhombohedral Graphite Structure: A_hR2_166_c

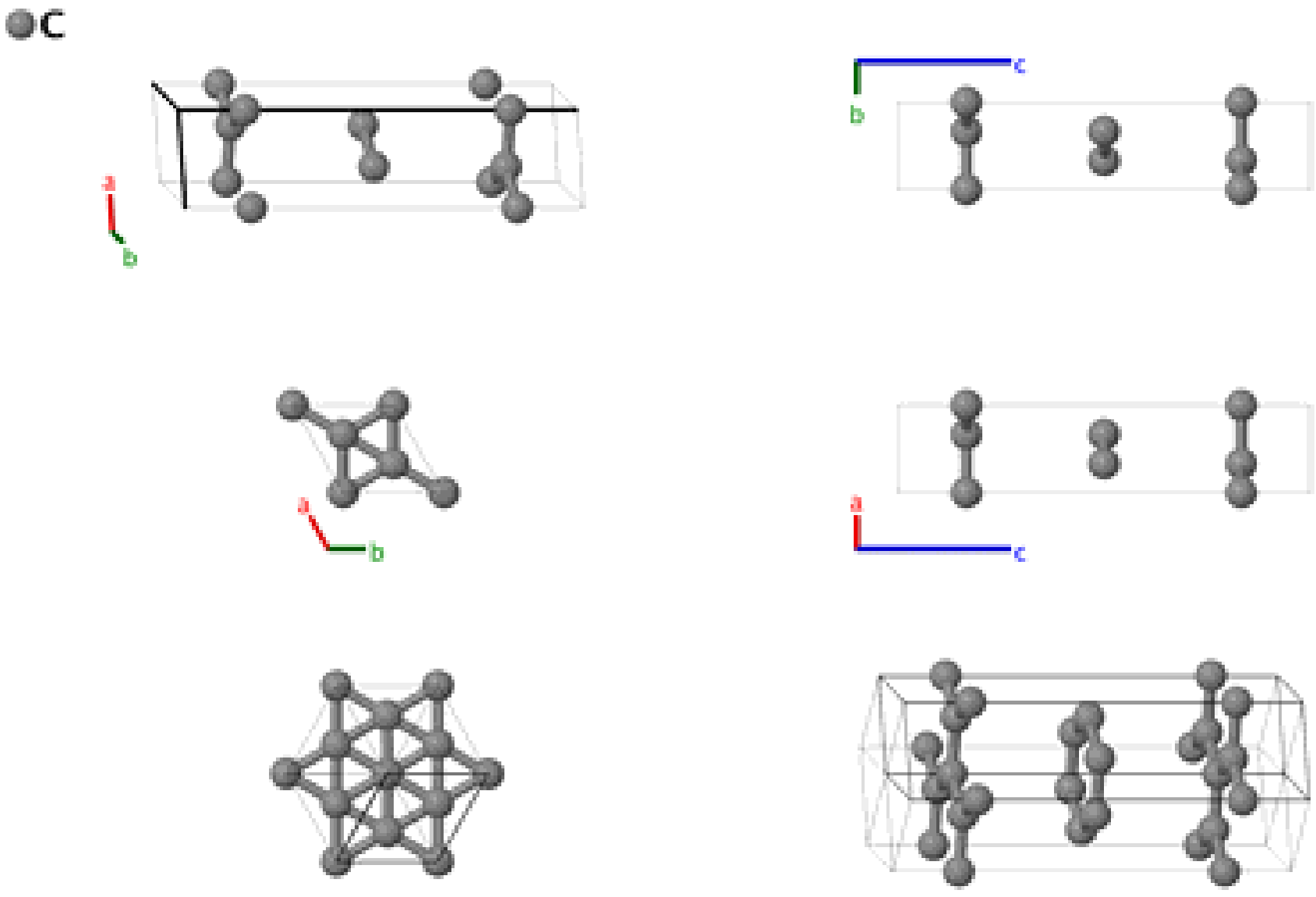

| | | |
|---|---|---|
| **Prototype** | : | C |
| **AFLOW prototype label** | : | A_hR2_166_c |
| ***Strukturbericht* designation** | : | None |
| **Pearson symbol** | : | hR2 |
| **Space group number** | : | 166 |
| **Space group symbol** | : | R$\bar{3}$m |
| **AFLOW prototype command** | : | `aflow --proto=A_hR2_166_c [--hex]`<br>`--params=`$a, c/a, x_1$ |

- Graphite also comes in a hexagonal form, which may be either flat (A9) or buckled. When $x_1 = 1/6$ the graphite sheets are flat. However this does not produce a change in symmetry, as it does in the hexagonal graphite structures. Note that $\alpha$-As (pp. 378), rhombohedral graphite (pp. 392), and $\beta$-O (pp. 398) have the same AFLOW prototype label. They are generated by the same symmetry operations with different sets of parameters (`--params`) specified in their corresponding CIF files. Hexagonal settings of this structure can be obtained with the option `--hex`.

**Rhombohedral primitive vectors:**

$$\mathbf{a}_1 = \frac{1}{2}\,a\,\hat{\mathbf{x}} - \frac{1}{2\sqrt{3}}\,a\,\hat{\mathbf{y}} + \frac{1}{3}\,c\,\hat{\mathbf{z}}$$
$$\mathbf{a}_2 = \frac{1}{\sqrt{3}}\,a\,\hat{\mathbf{y}} + \frac{1}{3}\,c\,\hat{\mathbf{z}}$$
$$\mathbf{a}_3 = -\frac{1}{2}\,a\,\hat{\mathbf{x}} - \frac{1}{2\sqrt{3}}\,a\,\hat{\mathbf{y}} + \frac{1}{3}\,c\,\hat{\mathbf{z}}$$

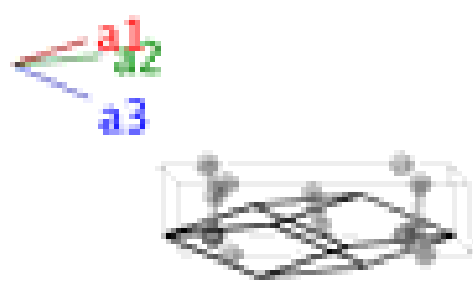

**Basis vectors:**

| | | Lattice Coordinates | | Cartesian Coordinates | Wyckoff Position | Atom Type |
|---|---|---|---|---|---|---|
| $\mathbf{B_1}$ | = | $x_1\,\mathbf{a_1} + x_1\,\mathbf{a_2} + x_1\,\mathbf{a_3}$ | = | $x_1\,c\,\hat{\mathbf{z}}$ | (2c) | C |
| $\mathbf{B_2}$ | = | $-x_1\,\mathbf{a_1} - x_1\,\mathbf{a_2} - x_1\,\mathbf{a_3}$ | = | $-x_1\,c\,\hat{\mathbf{z}}$ | (2c) | C |

**Geometry files:**

- CIF: pp. 727
- POSCAR: pp. 727

# $\alpha$-B (hR12) Structure: A_hR12_166_2h

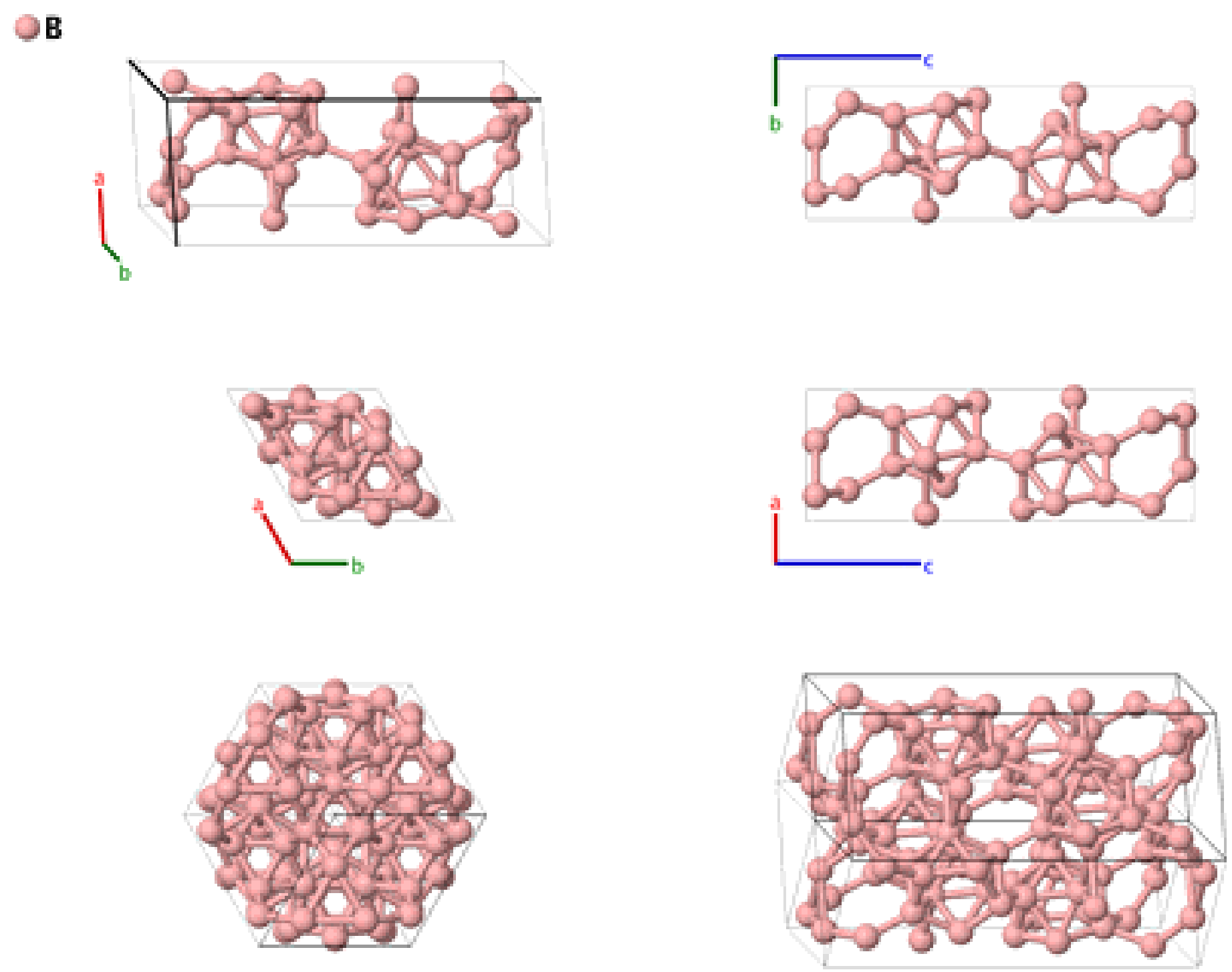

| | | |
|---|---|---|
| **Prototype** | : | $\alpha$-B |
| **AFLOW prototype label** | : | A_hR12_166_2h |
| ***Strukturbericht* designation** | : | None |
| **Pearson symbol** | : | hR12 |
| **Space group number** | : | 166 |
| **Space group symbol** | : | $R\bar{3}m$ |
| **AFLOW prototype command** | : | `aflow --proto=A_hR12_166_2h [--hex]` <br> `--params=`$a, c/a, x_1, z_1, x_2, z_2$ |

- This is a metastable phase of boron, and the simplest known phase (the ground state, $\beta$-B, has 105 or 320 atoms in the unit cell). Note the relationship between the icosahedra in this structure, in T-50 B, and in $\beta$-B. (Donohue, 1982) refers to this as rhombohedral-12 boron. Hexagonal settings of this structure can be obtained with the option `--hex`.

**Rhombohedral primitive vectors:**

$$\mathbf{a}_1 = \frac{1}{2}\,a\,\hat{\mathbf{x}} - \frac{1}{2\sqrt{3}}\,a\,\hat{\mathbf{y}} + \frac{1}{3}\,c\,\hat{\mathbf{z}}$$
$$\mathbf{a}_2 = \frac{1}{\sqrt{3}}\,a\,\hat{\mathbf{y}} + \frac{1}{3}\,c\,\hat{\mathbf{z}}$$
$$\mathbf{a}_3 = -\frac{1}{2}\,a\,\hat{\mathbf{x}} - \frac{1}{2\sqrt{3}}\,a\,\hat{\mathbf{y}} + \frac{1}{3}\,c\,\hat{\mathbf{z}}$$

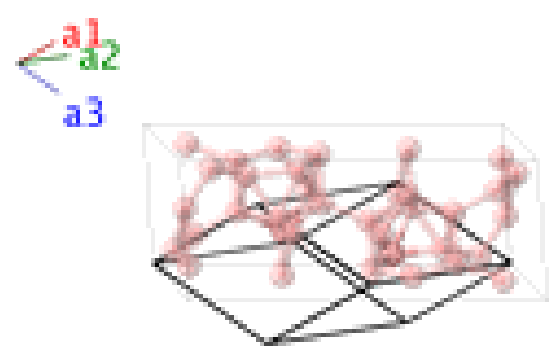

**Basis vectors:**

| | | Lattice Coordinates | | Cartesian Coordinates | Wyckoff Position | Atom Type |
|---|---|---|---|---|---|---|
| $\mathbf{B_1}$ | = | $x_1\,\mathbf{a}_1 + x_1\,\mathbf{a}_2 + z_1\,\mathbf{a}_3$ | = | $\frac{1}{2}\,(x_1 - z_1)\,a\,\hat{\mathbf{x}} + \frac{1}{2\sqrt{3}}\,(x_1 - z_1)\,a\,\hat{\mathbf{y}} + \frac{1}{3}\,(2x_1 + z_1)\,c\,\hat{\mathbf{z}}$ | (6$h$) | B I |
| $\mathbf{B_2}$ | = | $z_1\,\mathbf{a_1} + x_1\,\mathbf{a_2} + x_1\,\mathbf{a_3}$ | = | $\frac{1}{2}\,(z_1 - x_1)\,a\,\mathbf{\hat{x}} + \frac{1}{2\sqrt{3}}\,(x_1 - z_1)\,a\,\mathbf{\hat{y}} + \frac{1}{3}\,(2x_1 + z_1)\,c\,\mathbf{\hat{z}}$ | (6*h*) | B I |
| $\mathbf{B_3}$ | = | $x_1\,\mathbf{a_1} + z_1\,\mathbf{a_2} + x_1\,\mathbf{a_3}$ | = | $\frac{1}{\sqrt{3}}\,(z_1 - x_1)\,a\,\mathbf{\hat{y}} + \frac{1}{3}\,(2x_1 + z_1)\,c\,\mathbf{\hat{z}}$ | (6*h*) | B I |
| $\mathbf{B_4}$ | = | $-x_1\,\mathbf{a_1} - x_1\,\mathbf{a_2} - z_1\,\mathbf{a_3}$ | = | $\frac{1}{2}\,(z_1 - x_1)\,a\,\mathbf{\hat{x}} + \frac{1}{2\sqrt{3}}\,(z_1 - x_1)\,a\,\mathbf{\hat{y}} - \frac{1}{3}\,(2x_1 + z_1)\,c\,\mathbf{\hat{z}}$ | (6*h*) | B I |
| $\mathbf{B_5}$ | = | $-z_1\,\mathbf{a_1} - x_1\,\mathbf{a_2} - x_1\,\mathbf{a_3}$ | = | $\frac{1}{2}\,(x_1 - z_1)\,a\,\mathbf{\hat{x}} + \frac{1}{2\sqrt{3}}\,(z_1 - x_1)\,a\,\mathbf{\hat{y}} - \frac{1}{3}\,(2x_1 + z_1)\,c\,\mathbf{\hat{z}}$ | (6*h*) | B I |
| $\mathbf{B_6}$ | = | $-x_1\,\mathbf{a_1} - z_1\,\mathbf{a_2} - x_1\,\mathbf{a_3}$ | = | $\frac{1}{\sqrt{3}}\,(x_1 - z_1)\,a\,\mathbf{\hat{y}} - \frac{1}{3}\,(2x_1 + z_1)\,c\,\mathbf{\hat{z}}$ | (6*h*) | B I |
| $\mathbf{B_7}$ | = | $x_2\,\mathbf{a_1} + x_2\,\mathbf{a_2} + z_2\,\mathbf{a_3}$ | = | $\frac{1}{2}\,(x_2 - z_2)\,a\,\mathbf{\hat{x}} + \frac{1}{2\sqrt{3}}\,(x_2 - z_2)\,a\,\mathbf{\hat{y}} + \frac{1}{3}\,(2x_2 + z_2)\,c\,\mathbf{\hat{z}}$ | (6*h*) | B II |
| $\mathbf{B_8}$ | = | $z_2\,\mathbf{a_1} + x_2\,\mathbf{a_2} + x_2\,\mathbf{a_3}$ | = | $\frac{1}{2}\,(z_2 - x_2)\,a\,\mathbf{\hat{x}} + \frac{1}{2\sqrt{3}}\,(x_2 - z_2)\,a\,\mathbf{\hat{y}} + \frac{1}{3}\,(2x_2 + z_2)\,c\,\mathbf{\hat{z}}$ | (6*h*) | B II |
| $\mathbf{B_9}$ | = | $x_2\,\mathbf{a_1} + z_2\,\mathbf{a_2} + x_2\,\mathbf{a_3}$ | = | $\frac{1}{\sqrt{3}}\,(z_2 - x_2)\,a\,\mathbf{\hat{y}} + \frac{1}{3}\,(2x_2 + z_2)\,c\,\mathbf{\hat{z}}$ | (6*h*) | B II |
| $\mathbf{B_{10}}$ | = | $-x_2\,\mathbf{a_1} - x_2\,\mathbf{a_2} - z_2\,\mathbf{a_3}$ | = | $\frac{1}{2}\,(z_2 - x_2)\,a\,\mathbf{\hat{x}} + \frac{1}{2\sqrt{3}}\,(z_2 - x_2)\,a\,\mathbf{\hat{y}} - \frac{1}{3}\,(2x_2 + z_2)\,c\,\mathbf{\hat{z}}$ | (6*h*) | B II |
| $\mathbf{B_{11}}$ | = | $-z_2\,\mathbf{a_1} - x_2\,\mathbf{a_2} - x_2\,\mathbf{a_3}$ | = | $\frac{1}{2}\,(x_2 - z_2)\,a\,\mathbf{\hat{x}} + \frac{1}{2\sqrt{3}}\,(z_2 - x_2)\,a\,\mathbf{\hat{y}} - \frac{1}{3}\,(2x_2 + z_2)\,c\,\mathbf{\hat{z}}$ | (6*h*) | B II |
| $\mathbf{B_{12}}$ | = | $-x_2\,\mathbf{a_1} - z_2\,\mathbf{a_2} - x_2\,\mathbf{a_3}$ | = | $\frac{1}{\sqrt{3}}\,(x_2 - z_2)\,a\,\mathbf{\hat{y}} - \frac{1}{3}\,(2x_2 + z_2)\,c\,\mathbf{\hat{z}}$ | (6*h*) | B II |

---

**Found in:**

---

**Geometry files:**

- CIF: pp. 727
- POSCAR: pp. 728

# Caswellsilverite ($CrNaS_2$, $F5_1$) Crystal Structure: ABC2_hR4_166_a_b_c

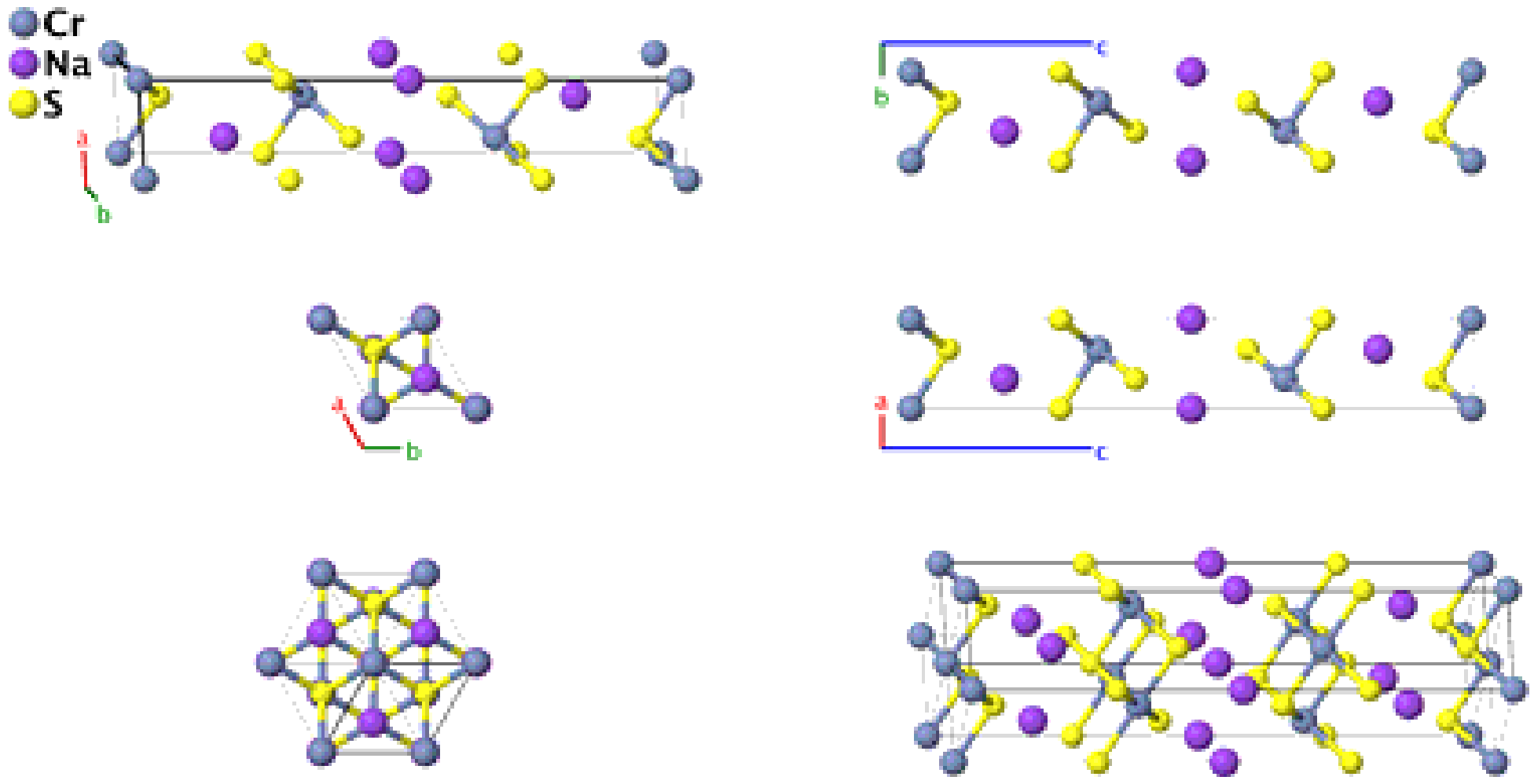

| | | |
|---|---|---|
| **Prototype** | : | $CrNaS_2$ |
| **AFLOW prototype label** | : | ABC2_hR4_166_a_b_c |
| ***Strukturbericht* designation** | : | $F5_1$ |
| **Pearson symbol** | : | hR4 |
| **Space group number** | : | 166 |
| **Space group symbol** | : | $R\bar{3}m$ |
| **AFLOW prototype command** | : | `aflow --proto=ABC2_hR4_166_a_b_c [--hex]`<br>`--params=`$a, c/a, x_3$ |

**Other compounds with this structure:**

- $AgAsSe_2$, $HoS_2Tl$, $AlCV_2$, $Te_2TlY$, many others

- This mineral did not obtain a name until it was discovered in nature (Okada, 1982). Hexagonal settings of this structure can be obtained with the option `--hex`.

**Rhombohedral primitive vectors:**

$$\mathbf{a}_1 = \frac{1}{2}\,a\,\mathbf{\hat{x}} - \frac{1}{2\sqrt{3}}\,a\,\mathbf{\hat{y}} + \frac{1}{3}\,c\,\mathbf{\hat{z}}$$
$$\mathbf{a}_2 = \frac{1}{\sqrt{3}}\,a\,\mathbf{\hat{y}} + \frac{1}{3}\,c\,\mathbf{\hat{z}}$$
$$\mathbf{a}_3 = -\frac{1}{2}\,a\,\mathbf{\hat{x}} - \frac{1}{2\sqrt{3}}\,a\,\mathbf{\hat{y}} + \frac{1}{3}\,c\,\mathbf{\hat{z}}$$

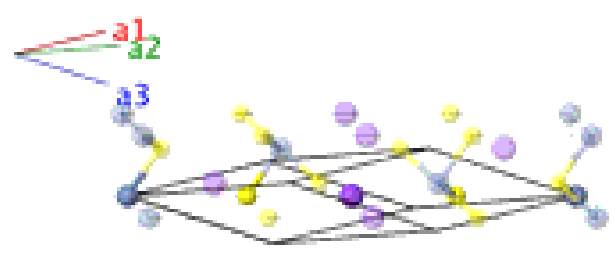

**Basis vectors:**

| | | Lattice Coordinates | | Cartesian Coordinates | Wyckoff Position | Atom Type |
|---|---|---|---|---|---|---|
| $\mathbf{B}_1$ | = | $0\,\mathbf{a}_1 + 0\,\mathbf{a}_2 + 0\,\mathbf{a}_3$ | = | $0\mathbf{\hat{x}} + 0\mathbf{\hat{y}} + 0\mathbf{\hat{z}}$ | $(1a)$ | Cr |
| $\mathbf{B}_2$ | = | $\frac{1}{2}\,\mathbf{a}_1 + \frac{1}{2}\,\mathbf{a}_2 + \frac{1}{2}\,\mathbf{a}_3$ | = | $\frac{1}{2}\,c\,\mathbf{\hat{z}}$ | $(1b)$ | Na |
| $\mathbf{B}_3$ | = | $x_3\,\mathbf{a}_1 + x_3\,\mathbf{a}_2 + x_3\,\mathbf{a}_3$ | = | $x_3\,c\,\mathbf{\hat{z}}$ | $(2c)$ | S |
| $\mathbf{B}_4$ | = | $-x_3\,\mathbf{a}_1 - x_3\,\mathbf{a}_2 - x_3\,\mathbf{a}_3$ | = | $-x_3\,c\,\mathbf{\hat{z}}$ | $(2c)$ | S |

**Geometry files:**

- CIF: pp. 728
- POSCAR: pp. 729

# $\beta$-O Structure: A_hR2_166_c

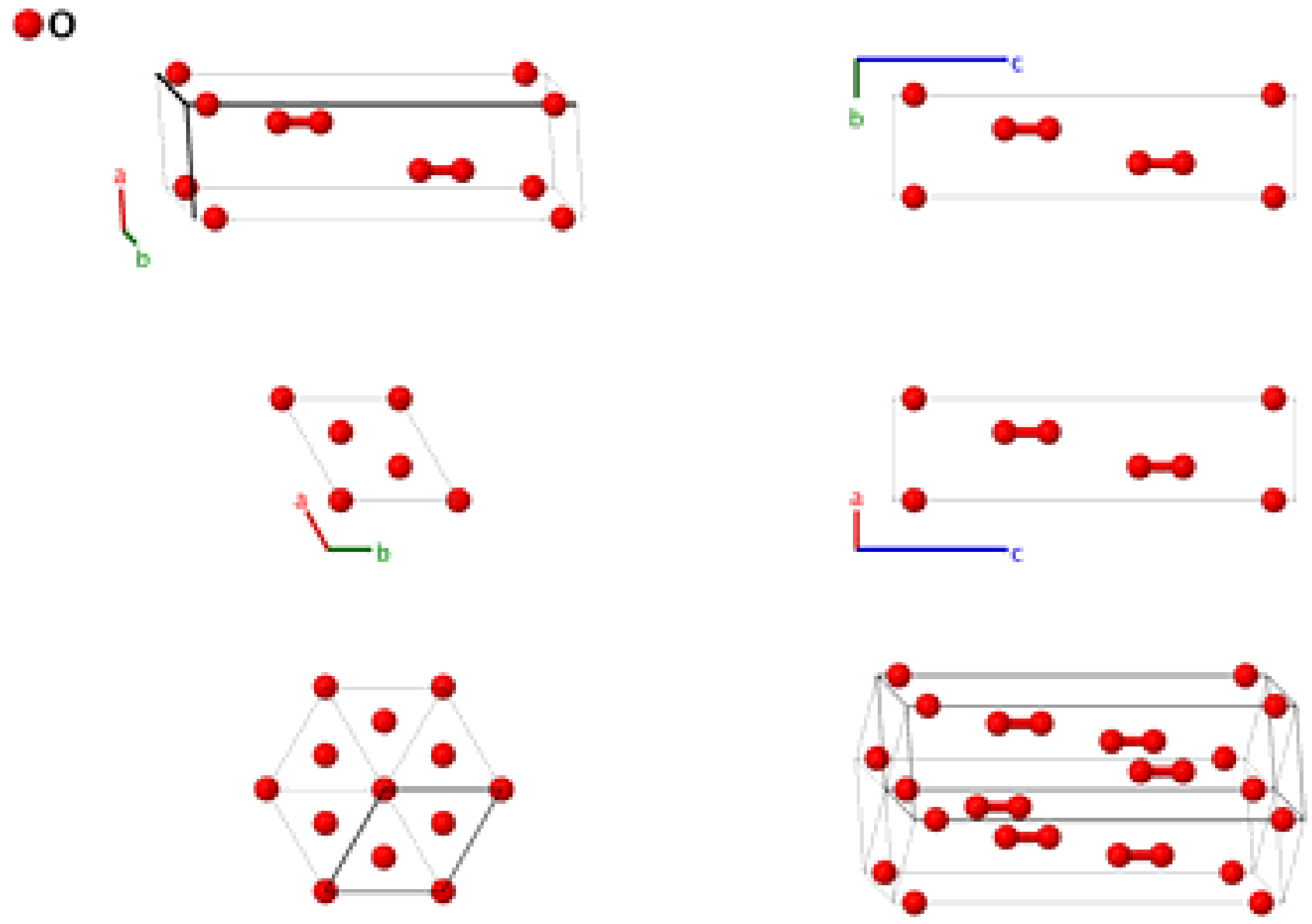

| | | |
|---|---|---|
| **Prototype** | : | $\beta$-O |
| **AFLOW prototype label** | : | A_hR2_166_c |
| ***Strukturbericht* designation** | : | None |
| **Pearson symbol** | : | hR2 |
| **Space group number** | : | 166 |
| **Space group symbol** | : | R$\bar{3}$m |
| **AFLOW prototype command** | : | `aflow --proto=A_hR2_166_c [--hex]`<br>`--params=`$a, c/a, x_1$ |

- Note that $\alpha$-As (pp. 378), rhombohedral graphite (pp. 392), and $\beta$-O (pp. 398) have the same AFLOW prototype label. They are generated by the same symmetry operations with different sets of parameters (`--params`) specified in their corresponding CIF files. Hexagonal settings of this structure can be obtained with the option `--hex`.

**Rhombohedral primitive vectors:**

$$
\begin{aligned}
\mathbf{a}_1 &= \tfrac{1}{2}\,a\,\hat{\mathbf{x}} - \tfrac{1}{2\sqrt{3}}\,a\,\hat{\mathbf{y}} + \tfrac{1}{3}\,c\,\hat{\mathbf{z}} \\
\mathbf{a}_2 &= \tfrac{1}{\sqrt{3}}\,a\,\hat{\mathbf{y}} + \tfrac{1}{3}\,c\,\hat{\mathbf{z}} \\
\mathbf{a}_3 &= -\tfrac{1}{2}\,a\,\hat{\mathbf{x}} - \tfrac{1}{2\sqrt{3}}\,a\,\hat{\mathbf{y}} + \tfrac{1}{3}\,c\,\hat{\mathbf{z}}
\end{aligned}
$$

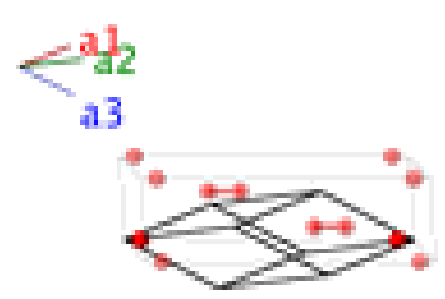

**Basis vectors:**

| | | Lattice Coordinates | | Cartesian Coordinates | Wyckoff Position | Atom Type |
|---|---|---|---|---|---|---|
| $\mathbf{B}_1$ | = | $x_1\,\mathbf{a}_1 + x_1\,\mathbf{a}_2 + x_1\,\mathbf{a}_3$ | = | $x_1\,c\,\hat{\mathbf{z}}$ | (2$c$) | O |
| $\mathbf{B}_2$ | = | $-x_1\,\mathbf{a}_1 - x_1\,\mathbf{a}_2 - x_1\,\mathbf{a}_3$ | = | $-x_1\,c\,\hat{\mathbf{z}}$ | (2$c$) | O |

---

**Geometry files:**

- CIF: pp. 729
- POSCAR: pp. 729

# $\beta$-B (R-105) Structure: A_hR105_166_bc9h4i

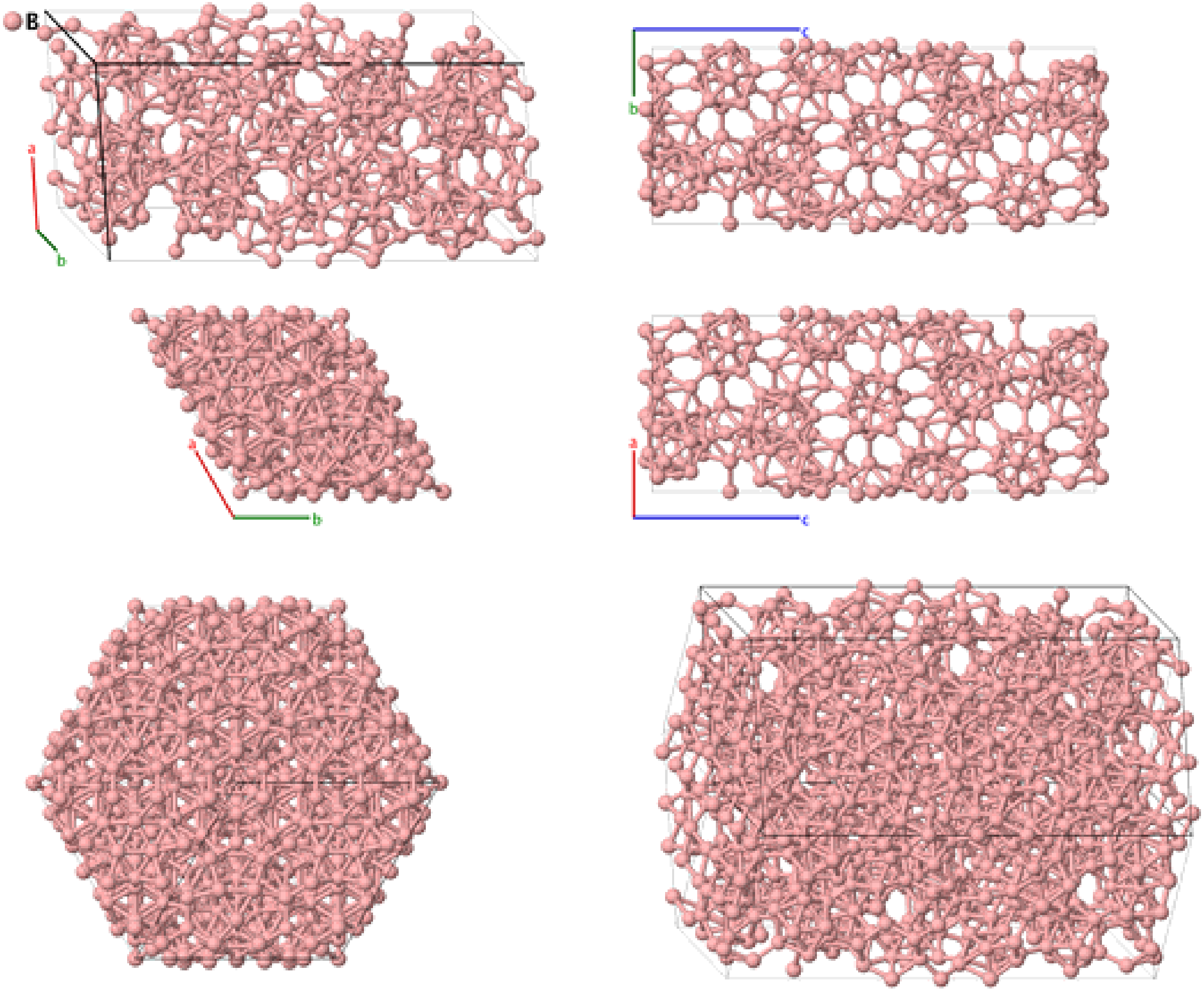

| | | |
|---|---|---|
| **Prototype** | : | $\beta$-B |
| **AFLOW prototype label** | : | A_hR105_166_bc9h4i |
| ***Strukturbericht* designation** | : | None |
| **Pearson symbol** | : | hR105 |
| **Space group number** | : | 166 |
| **Space group symbol** | : | R$\bar{3}$m |
| **AFLOW prototype command** | : | `aflow --proto=A_hR105_166_bc9h4i [--hex]` `--params=`$a, c/a, x_2, x_3, z_3, x_4, z_4, x_5, z_5, x_6, z_6, x_7, z_7, x_8, z_8, x_9, z_9, x_{10}, z_{10}, x_{11}, z_{11}, x_{12}, y_{12}, z_{12}, x_{13}, y_{13}, z_{13}, x_{14}, y_{14}, z_{14}, x_{15}, y_{15}, z_{15}$ |

- This is apparently the ground state of boron, with 105 atoms in the unit cell. Note the relationship between the icosahedra in this structure, $\alpha$-B and T-50 B. (Donohue, 1982) gives two possible sets of internal coordinates for the atoms on page 64. We use the second set (Geist, 1970), as it has no partially filled sites. Hexagonal settings of this structure can be obtained with the option `--hex`.

**Rhombohedral primitive vectors:**

$$\mathbf{a}_1 = \frac{1}{2}\,a\,\hat{\mathbf{x}} - \frac{1}{2\sqrt{3}}\,a\,\hat{\mathbf{y}} + \frac{1}{3}\,c\,\hat{\mathbf{z}}$$
$$\mathbf{a}_2 = \frac{1}{\sqrt{3}}\,a\,\hat{\mathbf{y}} + \frac{1}{3}\,c\,\hat{\mathbf{z}}$$
$$\mathbf{a}_3 = -\frac{1}{2}\,a\,\hat{\mathbf{x}} - \frac{1}{2\sqrt{3}}\,a\,\hat{\mathbf{y}} + \frac{1}{3}\,c\,\hat{\mathbf{z}}$$

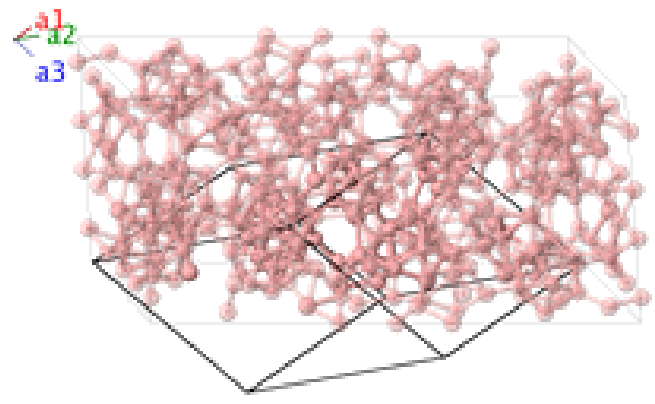

**Basis vectors:**

| | | Lattice Coordinates | | Cartesian Coordinates | Wyckoff Position | Atom Type |
|---|---|---|---|---|---|---|
| $\mathbf{B}_1$ | = | $\frac{1}{2}\,\mathbf{a}_1 + \frac{1}{2}\,\mathbf{a}_2 + \frac{1}{2}\,\mathbf{a}_3$ | = | $\frac{1}{2}\,c\,\hat{\mathbf{z}}$ | (1$b$) | B I |
| $\mathbf{B}_2$ | = | $x_2\,\mathbf{a}_1 + x_2\,\mathbf{a}_2 + x_2\,\mathbf{a}_3$ | = | $x_2\,c\,\hat{\mathbf{z}}$ | (2$c$) | B II |
| $\mathbf{B}_3$ | = | $-x_2\,\mathbf{a}_1 - x_2\,\mathbf{a}_2 - x_2\,\mathbf{a}_3$ | = | $-x_2\,c\,\hat{\mathbf{z}}$ | (2$c$) | B II |
| $\mathbf{B}_4$ | = | $x_3\,\mathbf{a}_1 + x_3\,\mathbf{a}_2 + z_3\,\mathbf{a}_3$ | = | $\frac{1}{2}\,(x_3 - z_3)\,a\,\hat{\mathbf{x}} + \frac{1}{2\sqrt{3}}\,(x_3 - z_3)\,a\,\hat{\mathbf{y}} + \frac{1}{3}\,(2x_3 + z_3)\,c\,\hat{\mathbf{z}}$ | (6$h$) | B III |
| $\mathbf{B}_5$ | = | $z_3\,\mathbf{a}_1 + x_3\,\mathbf{a}_2 + x_3\,\mathbf{a}_3$ | = | $\frac{1}{2}\,(z_3 - x_3)\,a\,\hat{\mathbf{x}} + \frac{1}{2\sqrt{3}}\,(x_3 - z_3)\,a\,\hat{\mathbf{y}} + \frac{1}{3}\,(2x_3 + z_3)\,c\,\hat{\mathbf{z}}$ | (6$h$) | B III |
| $\mathbf{B}_6$ | = | $x_3\,\mathbf{a}_1 + z_3\,\mathbf{a}_2 + x_3\,\mathbf{a}_3$ | = | $\frac{1}{\sqrt{3}}\,(z_3 - x_3)\,a\,\hat{\mathbf{y}} + \frac{1}{3}\,(2x_3 + z_3)\,c\,\hat{\mathbf{z}}$ | (6$h$) | B III |
| $\mathbf{B}_7$ | = | $-x_3\,\mathbf{a}_1 - x_3\,\mathbf{a}_2 - z_3\,\mathbf{a}_3$ | = | $\frac{1}{2}\,(z_3 - x_3)\,a\,\hat{\mathbf{x}} + \frac{1}{2\sqrt{3}}\,(z_3 - x_3)\,a\,\hat{\mathbf{y}} - \frac{1}{3}\,(2x_3 + z_3)\,c\,\hat{\mathbf{z}}$ | (6$h$) | B III |
| $\mathbf{B}_8$ | = | $-z_3\,\mathbf{a}_1 - x_3\,\mathbf{a}_2 - x_3\,\mathbf{a}_3$ | = | $\frac{1}{2}\,(x_3 - z_3)\,a\,\hat{\mathbf{x}} + \frac{1}{2\sqrt{3}}\,(z_3 - x_3)\,a\,\hat{\mathbf{y}} - \frac{1}{3}\,(2x_3 + z_3)\,c\,\hat{\mathbf{z}}$ | (6$h$) | B III |
| $\mathbf{B}_9$ | = | $-x_3\,\mathbf{a}_1 - z_3\,\mathbf{a}_2 - x_3\,\mathbf{a}_3$ | = | $\frac{1}{\sqrt{3}}\,(x_3 - z_3)\,a\,\hat{\mathbf{y}} - \frac{1}{3}\,(2x_3 + z_3)\,c\,\hat{\mathbf{z}}$ | (6$h$) | B III |
| $\mathbf{B}_{10}$ | = | $x_4\,\mathbf{a}_1 + x_4\,\mathbf{a}_2 + z_4\,\mathbf{a}_3$ | = | $\frac{1}{2}\,(x_4 - z_4)\,a\,\hat{\mathbf{x}} + \frac{1}{2\sqrt{3}}\,(x_4 - z_4)\,a\,\hat{\mathbf{y}} + \frac{1}{3}\,(2x_4 + z_4)\,c\,\hat{\mathbf{z}}$ | (6$h$) | B IV |
| $\mathbf{B}_{11}$ | = | $z_4\,\mathbf{a}_1 + x_4\,\mathbf{a}_2 + x_4\,\mathbf{a}_3$ | = | $\frac{1}{2}\,(z_4 - x_4)\,a\,\hat{\mathbf{x}} + \frac{1}{2\sqrt{3}}\,(x_4 - z_4)\,a\,\hat{\mathbf{y}} + \frac{1}{3}\,(2x_4 + z_4)\,c\,\hat{\mathbf{z}}$ | (6$h$) | B IV |
| $\mathbf{B}_{12}$ | = | $x_4\,\mathbf{a}_1 + z_4\,\mathbf{a}_2 + x_4\,\mathbf{a}_3$ | = | $\frac{1}{\sqrt{3}}\,(z_4 - x_4)\,a\,\hat{\mathbf{y}} + \frac{1}{3}\,(2x_4 + z_4)\,c\,\hat{\mathbf{z}}$ | (6$h$) | B IV |
| $\mathbf{B}_{13}$ | = | $-x_4\,\mathbf{a}_1 - x_4\,\mathbf{a}_2 - z_4\,\mathbf{a}_3$ | = | $\frac{1}{2}\,(z_4 - x_4)\,a\,\hat{\mathbf{x}} + \frac{1}{2\sqrt{3}}\,(z_4 - x_4)\,a\,\hat{\mathbf{y}} - \frac{1}{3}\,(2x_4 + z_4)\,c\,\hat{\mathbf{z}}$ | (6$h$) | B IV |
| $\mathbf{B}_{14}$ | = | $-z_4\,\mathbf{a}_1 - x_4\,\mathbf{a}_2 - x_4\,\mathbf{a}_3$ | = | $\frac{1}{2}\,(x_4 - z_4)\,a\,\hat{\mathbf{x}} + \frac{1}{2\sqrt{3}}\,(z_4 - x_4)\,a\,\hat{\mathbf{y}} - \frac{1}{3}\,(2x_4 + z_4)\,c\,\hat{\mathbf{z}}$ | (6$h$) | B IV |
| $\mathbf{B}_{15}$ | = | $-x_4\,\mathbf{a}_1 - z_4\,\mathbf{a}_2 - x_4\,\mathbf{a}_3$ | = | $\frac{1}{\sqrt{3}}\,(x_4 - z_4)\,a\,\hat{\mathbf{y}} - \frac{1}{3}\,(2x_4 + z_4)\,c\,\hat{\mathbf{z}}$ | (6$h$) | B IV |
| $\mathbf{B}_{16}$ | = | $x_5\,\mathbf{a}_1 + x_5\,\mathbf{a}_2 + z_5\,\mathbf{a}_3$ | = | $\frac{1}{2}\,(x_5 - z_5)\,a\,\hat{\mathbf{x}} + \frac{1}{2\sqrt{3}}\,(x_5 - z_5)\,a\,\hat{\mathbf{y}} + \frac{1}{3}\,(2x_5 + z_5)\,c\,\hat{\mathbf{z}}$ | (6$h$) | B V |
| $\mathbf{B}_{17}$ | = | $z_5\,\mathbf{a}_1 + x_5\,\mathbf{a}_2 + x_5\,\mathbf{a}_3$ | = | $\frac{1}{2}\,(z_5 - x_5)\,a\,\hat{\mathbf{x}} + \frac{1}{2\sqrt{3}}\,(x_5 - z_5)\,a\,\hat{\mathbf{y}} + \frac{1}{3}\,(2x_5 + z_5)\,c\,\hat{\mathbf{z}}$ | (6$h$) | B V |
| $\mathbf{B}_{18}$ | = | $x_5\,\mathbf{a}_1 + z_5\,\mathbf{a}_2 + x_5\,\mathbf{a}_3$ | = | $\frac{1}{\sqrt{3}}\,(z_5 - x_5)\,a\,\hat{\mathbf{y}} + \frac{1}{3}\,(2x_5 + z_5)\,c\,\hat{\mathbf{z}}$ | (6$h$) | B V |
| $\mathbf{B}_{19}$ | = | $-x_5\,\mathbf{a}_1 - x_5\,\mathbf{a}_2 - z_5\,\mathbf{a}_3$ | = | $\frac{1}{2}\,(z_5 - x_5)\,a\,\hat{\mathbf{x}} + \frac{1}{2\sqrt{3}}\,(z_5 - x_5)\,a\,\hat{\mathbf{y}} - \frac{1}{3}\,(2x_5 + z_5)\,c\,\hat{\mathbf{z}}$ | (6$h$) | B V |
| $\mathbf{B}_{20}$ | = | $-z_5\,\mathbf{a}_1 - x_5\,\mathbf{a}_2 - x_5\,\mathbf{a}_3$ | = | $\frac{1}{2}\,(x_5 - z_5)\,a\,\hat{\mathbf{x}} + \frac{1}{2\sqrt{3}}\,(z_5 - x_5)\,a\,\hat{\mathbf{y}} - \frac{1}{3}\,(2x_5 + z_5)\,c\,\hat{\mathbf{z}}$ | (6$h$) | B V |
| $\mathbf{B_{21}}$ | = | $-x_5\,\mathbf{a_1} - z_5\,\mathbf{a_2} - x_5\,\mathbf{a_3}$ | = | $\frac{1}{\sqrt{3}}\,(x_5 - z_5)\,a\,\hat{\mathbf{y}} - \frac{1}{3}\,(2x_5 + z_5)\,c\,\hat{\mathbf{z}}$ | (6*h*) | B V |
| $\mathbf{B_{22}}$ | = | $x_6\,\mathbf{a_1} + x_6\,\mathbf{a_2} + z_6\,\mathbf{a_3}$ | = | $\frac{1}{2}\,(x_6 - z_6)\,a\,\hat{\mathbf{x}} + \frac{1}{2\sqrt{3}}\,(x_6 - z_6)\,a\,\hat{\mathbf{y}} + \frac{1}{3}\,(2x_6 + z_6)\,c\,\hat{\mathbf{z}}$ | (6*h*) | B VI |
| $\mathbf{B_{23}}$ | = | $z_6\,\mathbf{a_1} + x_6\,\mathbf{a_2} + x_6\,\mathbf{a_3}$ | = | $\frac{1}{2}\,(z_6 - x_6)\,a\,\hat{\mathbf{x}} + \frac{1}{2\sqrt{3}}\,(x_6 - z_6)\,a\,\hat{\mathbf{y}} + \frac{1}{3}\,(2x_6 + z_6)\,c\,\hat{\mathbf{z}}$ | (6*h*) | B VI |
| $\mathbf{B_{24}}$ | = | $x_6\,\mathbf{a_1} + z_6\,\mathbf{a_2} + x_6\,\mathbf{a_3}$ | = | $\frac{1}{\sqrt{3}}\,(z_6 - x_6)\,a\,\hat{\mathbf{y}} + \frac{1}{3}\,(2x_6 + z_6)\,c\,\hat{\mathbf{z}}$ | (6*h*) | B VI |
| $\mathbf{B_{25}}$ | = | $-x_6\,\mathbf{a_1} - x_6\,\mathbf{a_2} - z_6\,\mathbf{a_3}$ | = | $\frac{1}{2}\,(z_6 - x_6)\,a\,\hat{\mathbf{x}} + \frac{1}{2\sqrt{3}}\,(z_6 - x_6)\,a\,\hat{\mathbf{y}} - \frac{1}{3}\,(2x_6 + z_6)\,c\,\hat{\mathbf{z}}$ | (6*h*) | B VI |
| $\mathbf{B_{26}}$ | = | $-z_6\,\mathbf{a_1} - x_6\,\mathbf{a_2} - x_6\,\mathbf{a_3}$ | = | $\frac{1}{2}\,(x_6 - z_6)\,a\,\hat{\mathbf{x}} + \frac{1}{2\sqrt{3}}\,(z_6 - x_6)\,a\,\hat{\mathbf{y}} - \frac{1}{3}\,(2x_6 + z_6)\,c\,\hat{\mathbf{z}}$ | (6*h*) | B VI |
| $\mathbf{B_{27}}$ | = | $-x_6\,\mathbf{a_1} - z_6\,\mathbf{a_2} - x_6\,\mathbf{a_3}$ | = | $\frac{1}{\sqrt{3}}\,(x_6 - z_6)\,a\,\hat{\mathbf{y}} - \frac{1}{3}\,(2x_6 + z_6)\,c\,\hat{\mathbf{z}}$ | (6*h*) | B VI |
| $\mathbf{B_{28}}$ | = | $x_7\,\mathbf{a_1} + x_7\,\mathbf{a_2} + z_7\,\mathbf{a_3}$ | = | $\frac{1}{2}\,(x_7 - z_7)\,a\,\hat{\mathbf{x}} + \frac{1}{2\sqrt{3}}\,(x_7 - z_7)\,a\,\hat{\mathbf{y}} + \frac{1}{3}\,(2x_7 + z_7)\,c\,\hat{\mathbf{z}}$ | (6*h*) | B VII |
| $\mathbf{B_{29}}$ | = | $z_7\,\mathbf{a_1} + x_7\,\mathbf{a_2} + x_7\,\mathbf{a_3}$ | = | $\frac{1}{2}\,(z_7 - x_7)\,a\,\hat{\mathbf{x}} + \frac{1}{2\sqrt{3}}\,(x_7 - z_7)\,a\,\hat{\mathbf{y}} + \frac{1}{3}\,(2x_7 + z_7)\,c\,\hat{\mathbf{z}}$ | (6*h*) | B VII |
| $\mathbf{B_{30}}$ | = | $x_7\,\mathbf{a_1} + z_7\,\mathbf{a_2} + x_7\,\mathbf{a_3}$ | = | $\frac{1}{\sqrt{3}}\,(z_7 - x_7)\,a\,\hat{\mathbf{y}} + \frac{1}{3}\,(2x_7 + z_7)\,c\,\hat{\mathbf{z}}$ | (6*h*) | B VII |
| $\mathbf{B_{31}}$ | = | $-x_7\,\mathbf{a_1} - x_7\,\mathbf{a_2} - z_7\,\mathbf{a_3}$ | = | $\frac{1}{2}\,(z_7 - x_7)\,a\,\hat{\mathbf{x}} + \frac{1}{2\sqrt{3}}\,(z_7 - x_7)\,a\,\hat{\mathbf{y}} - \frac{1}{3}\,(2x_7 + z_7)\,c\,\hat{\mathbf{z}}$ | (6*h*) | B VII |
| $\mathbf{B_{32}}$ | = | $-z_7\,\mathbf{a_1} - x_7\,\mathbf{a_2} - x_7\,\mathbf{a_3}$ | = | $\frac{1}{2}\,(x_7 - z_7)\,a\,\hat{\mathbf{x}} + \frac{1}{2\sqrt{3}}\,(z_7 - x_7)\,a\,\hat{\mathbf{y}} - \frac{1}{3}\,(2x_7 + z_7)\,c\,\hat{\mathbf{z}}$ | (6*h*) | B VII |
| $\mathbf{B_{33}}$ | = | $-x_7\,\mathbf{a_1} - z_7\,\mathbf{a_2} - x_7\,\mathbf{a_3}$ | = | $\frac{1}{\sqrt{3}}\,(x_7 - z_7)\,a\,\hat{\mathbf{y}} - \frac{1}{3}\,(2x_7 + z_7)\,c\,\hat{\mathbf{z}}$ | (6*h*) | B VII |
| $\mathbf{B_{34}}$ | = | $x_8\,\mathbf{a_1} + x_8\,\mathbf{a_2} + z_8\,\mathbf{a_3}$ | = | $\frac{1}{2}\,(x_8 - z_8)\,a\,\hat{\mathbf{x}} + \frac{1}{2\sqrt{3}}\,(x_8 - z_8)\,a\,\hat{\mathbf{y}} + \frac{1}{3}\,(2x_8 + z_8)\,c\,\hat{\mathbf{z}}$ | (6*h*) | B VIII |
| $\mathbf{B_{35}}$ | = | $z_8\,\mathbf{a_1} + x_8\,\mathbf{a_2} + x_8\,\mathbf{a_3}$ | = | $\frac{1}{2}\,(z_8 - x_8)\,a\,\hat{\mathbf{x}} + \frac{1}{2\sqrt{3}}\,(x_8 - z_8)\,a\,\hat{\mathbf{y}} + \frac{1}{3}\,(2x_8 + z_8)\,c\,\hat{\mathbf{z}}$ | (6*h*) | B VIII |
| $\mathbf{B_{36}}$ | = | $x_8\,\mathbf{a_1} + z_8\,\mathbf{a_2} + x_8\,\mathbf{a_3}$ | = | $\frac{1}{\sqrt{3}}\,(z_8 - x_8)\,a\,\hat{\mathbf{y}} + \frac{1}{3}\,(2x_8 + z_8)\,c\,\hat{\mathbf{z}}$ | (6*h*) | B VIII |
| $\mathbf{B_{37}}$ | = | $-x_8\,\mathbf{a_1} - x_8\,\mathbf{a_2} - z_8\,\mathbf{a_3}$ | = | $\frac{1}{2}\,(z_8 - x_8)\,a\,\hat{\mathbf{x}} + \frac{1}{2\sqrt{3}}\,(z_8 - x_8)\,a\,\hat{\mathbf{y}} - \frac{1}{3}\,(2x_8 + z_8)\,c\,\hat{\mathbf{z}}$ | (6*h*) | B VIII |
| $\mathbf{B_{38}}$ | = | $-z_8\,\mathbf{a_1} - x_8\,\mathbf{a_2} - x_8\,\mathbf{a_3}$ | = | $\frac{1}{2}\,(x_8 - z_8)\,a\,\hat{\mathbf{x}} + \frac{1}{2\sqrt{3}}\,(z_8 - x_8)\,a\,\hat{\mathbf{y}} - \frac{1}{3}\,(2x_8 + z_8)\,c\,\hat{\mathbf{z}}$ | (6*h*) | B VIII |
| $\mathbf{B_{39}}$ | = | $-x_8\,\mathbf{a_1} - z_8\,\mathbf{a_2} - x_8\,\mathbf{a_3}$ | = | $\frac{1}{\sqrt{3}}\,(x_8 - z_8)\,a\,\hat{\mathbf{y}} - \frac{1}{3}\,(2x_8 + z_8)\,c\,\hat{\mathbf{z}}$ | (6*h*) | B VIII |
| $\mathbf{B_{40}}$ | = | $x_9\,\mathbf{a_1} + x_9\,\mathbf{a_2} + z_9\,\mathbf{a_3}$ | = | $\frac{1}{2}\,(x_9 - z_9)\,a\,\hat{\mathbf{x}} + \frac{1}{2\sqrt{3}}\,(x_9 - z_9)\,a\,\hat{\mathbf{y}} + \frac{1}{3}\,(2x_9 + z_9)\,c\,\hat{\mathbf{z}}$ | (6*h*) | B IX |
| $\mathbf{B_{41}}$ | = | $z_9\,\mathbf{a_1} + x_9\,\mathbf{a_2} + x_9\,\mathbf{a_3}$ | = | $\frac{1}{2}\,(z_9 - x_9)\,a\,\hat{\mathbf{x}} + \frac{1}{2\sqrt{3}}\,(x_9 - z_9)\,a\,\hat{\mathbf{y}} + \frac{1}{3}\,(2x_9 + z_9)\,c\,\hat{\mathbf{z}}$ | (6*h*) | B IX |
| $\mathbf{B_{42}}$ | = | $x_9\,\mathbf{a_1} + z_9\,\mathbf{a_2} + x_9\,\mathbf{a_3}$ | = | $\frac{1}{\sqrt{3}}\,(z_9 - x_9)\,a\,\hat{\mathbf{y}} + \frac{1}{3}\,(2x_9 + z_9)\,c\,\hat{\mathbf{z}}$ | (6*h*) | B IX |
| $\mathbf{B_{43}}$ | = | $-x_9\,\mathbf{a_1} - x_9\,\mathbf{a_2} - z_9\,\mathbf{a_3}$ | = | $\frac{1}{2}\,(z_9 - x_9)\,a\,\hat{\mathbf{x}} + \frac{1}{2\sqrt{3}}\,(z_9 - x_9)\,a\,\hat{\mathbf{y}} - \frac{1}{3}\,(2x_9 + z_9)\,c\,\hat{\mathbf{z}}$ | (6*h*) | B IX |
| $\mathbf{B_{44}}$ | = | $-z_9\,\mathbf{a_1} - x_9\,\mathbf{a_2} - x_9\,\mathbf{a_3}$ | = | $\frac{1}{2}\,(x_9 - z_9)\,a\,\hat{\mathbf{x}} + \frac{1}{2\sqrt{3}}\,(z_9 - x_9)\,a\,\hat{\mathbf{y}} - \frac{1}{3}\,(2x_9 + z_9)\,c\,\hat{\mathbf{z}}$ | (6*h*) | B IX |
| $\mathbf{B}_{45}$ | = | $-x_9\,\mathbf{a}_1 - z_9\,\mathbf{a}_2 - x_9\,\mathbf{a}_3$ | = | $\frac{1}{\sqrt{3}}\,(x_9 - z_9)\,a\,\mathbf{\hat{y}} - \frac{1}{3}\,(2x_9 + z_9)\,c\,\mathbf{\hat{z}}$ | (6*h*) | B IX |
| $\mathbf{B}_{46}$ | = | $x_{10}\,\mathbf{a}_1 + x_{10}\,\mathbf{a}_2 + z_{10}\,\mathbf{a}_3$ | = | $\frac{1}{2}\,(x_{10} - z_{10})\,a\,\mathbf{\hat{x}} + \frac{1}{2\sqrt{3}}\,(x_{10} - z_{10})\,a\,\mathbf{\hat{y}} + \frac{1}{3}\,(2x_{10} + z_{10})\,c\,\mathbf{\hat{z}}$ | (6*h*) | B X |
| $\mathbf{B}_{47}$ | = | $z_{10}\,\mathbf{a}_1 + x_{10}\,\mathbf{a}_2 + x_{10}\,\mathbf{a}_3$ | = | $\frac{1}{2}\,(z_{10} - x_{10})\,a\,\mathbf{\hat{x}} + \frac{1}{2\sqrt{3}}\,(x_{10} - z_{10})\,a\,\mathbf{\hat{y}} + \frac{1}{3}\,(2x_{10} + z_{10})\,c\,\mathbf{\hat{z}}$ | (6*h*) | B X |
| $\mathbf{B}_{48}$ | = | $x_{10}\,\mathbf{a}_1 + z_{10}\,\mathbf{a}_2 + x_{10}\,\mathbf{a}_3$ | = | $\frac{1}{\sqrt{3}}\,(z_{10} - x_{10})\,a\,\mathbf{\hat{y}} + \frac{1}{3}\,(2x_{10} + z_{10})\,c\,\mathbf{\hat{z}}$ | (6*h*) | B X |
| $\mathbf{B}_{49}$ | = | $-x_{10}\,\mathbf{a}_1 - x_{10}\,\mathbf{a}_2 - z_{10}\,\mathbf{a}_3$ | = | $\frac{1}{2}\,(z_{10} - x_{10})\,a\,\mathbf{\hat{x}} + \frac{1}{2\sqrt{3}}\,(z_{10} - x_{10})\,a\,\mathbf{\hat{y}} - \frac{1}{3}\,(2x_{10} + z_{10})\,c\,\mathbf{\hat{z}}$ | (6*h*) | B X |
| $\mathbf{B}_{50}$ | = | $-z_{10}\,\mathbf{a}_1 - x_{10}\,\mathbf{a}_2 - x_{10}\,\mathbf{a}_3$ | = | $\frac{1}{2}\,(x_{10} - z_{10})\,a\,\mathbf{\hat{x}} + \frac{1}{2\sqrt{3}}\,(z_{10} - x_{10})\,a\,\mathbf{\hat{y}} - \frac{1}{3}\,(2x_{10} + z_{10})\,c\,\mathbf{\hat{z}}$ | (6*h*) | B X |
| $\mathbf{B}_{51}$ | = | $-x_{10}\,\mathbf{a}_1 - z_{10}\,\mathbf{a}_2 - x_{10}\,\mathbf{a}_3$ | = | $\frac{1}{\sqrt{3}}\,(x_{10} - z_{10})\,a\,\mathbf{\hat{y}} - \frac{1}{3}\,(2x_{10} + z_{10})\,c\,\mathbf{\hat{z}}$ | (6*h*) | B X |
| $\mathbf{B}_{52}$ | = | $x_{11}\,\mathbf{a}_1 + x_{11}\,\mathbf{a}_2 + z_{11}\,\mathbf{a}_3$ | = | $\frac{1}{2}\,(x_{11} - z_{11})\,a\,\mathbf{\hat{x}} + \frac{1}{2\sqrt{3}}\,(x_{11} - z_{11})\,a\,\mathbf{\hat{y}} + \frac{1}{3}\,(2x_{11} + z_{11})\,c\,\mathbf{\hat{z}}$ | (6*h*) | B XI |
| $\mathbf{B}_{53}$ | = | $z_{11}\,\mathbf{a}_1 + x_{11}\,\mathbf{a}_2 + x_{11}\,\mathbf{a}_3$ | = | $\frac{1}{2}\,(z_{11} - x_{11})\,a\,\mathbf{\hat{x}} + \frac{1}{2\sqrt{3}}\,(x_{11} - z_{11})\,a\,\mathbf{\hat{y}} + \frac{1}{3}\,(2x_{11} + z_{11})\,c\,\mathbf{\hat{z}}$ | (6*h*) | B XI |
| $\mathbf{B}_{54}$ | = | $x_{11}\,\mathbf{a}_1 + z_{11}\,\mathbf{a}_2 + x_{11}\,\mathbf{a}_3$ | = | $\frac{1}{\sqrt{3}}\,(z_{11} - x_{11})\,a\,\mathbf{\hat{y}} + \frac{1}{3}\,(2x_{11} + z_{11})\,c\,\mathbf{\hat{z}}$ | (6*h*) | B XI |
| $\mathbf{B}_{55}$ | = | $-x_{11}\,\mathbf{a}_1 - x_{11}\,\mathbf{a}_2 - z_{11}\,\mathbf{a}_3$ | = | $\frac{1}{2}\,(z_{11} - x_{11})\,a\,\mathbf{\hat{x}} + \frac{1}{2\sqrt{3}}\,(z_{11} - x_{11})\,a\,\mathbf{\hat{y}} - \frac{1}{3}\,(2x_{11} + z_{11})\,c\,\mathbf{\hat{z}}$ | (6*h*) | B XI |
| $\mathbf{B}_{56}$ | = | $-z_{11}\,\mathbf{a}_1 - x_{11}\,\mathbf{a}_2 - x_{11}\,\mathbf{a}_3$ | = | $\frac{1}{2}\,(x_{11} - z_{11})\,a\,\mathbf{\hat{x}} + \frac{1}{2\sqrt{3}}\,(z_{11} - x_{11})\,a\,\mathbf{\hat{y}} - \frac{1}{3}\,(2x_{11} + z_{11})\,c\,\mathbf{\hat{z}}$ | (6*h*) | B XI |
| $\mathbf{B}_{57}$ | = | $-x_{11}\,\mathbf{a}_1 - z_{11}\,\mathbf{a}_2 - x_{11}\,\mathbf{a}_3$ | = | $\frac{1}{\sqrt{3}}\,(x_{11} - z_{11})\,a\,\mathbf{\hat{y}} - \frac{1}{3}\,(2x_{11} + z_{11})\,c\,\mathbf{\hat{z}}$ | (6*h*) | B XI |
| $\mathbf{B}_{58}$ | = | $x_{12}\,\mathbf{a}_1 + y_{12}\,\mathbf{a}_2 + z_{12}\,\mathbf{a}_3$ | = | $\frac{1}{2}\,(x_{12} - z_{12})\,a\,\mathbf{\hat{x}} + \frac{1}{2\sqrt{3}}\,(2y_{12} - x_{12} - z_{12})\,a\,\mathbf{\hat{y}} + \frac{1}{3}\,(x_{12} + y_{12} + z_{12})\,c\,\mathbf{\hat{z}}$ | (12*i*) | B XII |
| $\mathbf{B}_{59}$ | = | $z_{12}\,\mathbf{a}_1 + x_{12}\,\mathbf{a}_2 + y_{12}\,\mathbf{a}_3$ | = | $\frac{1}{2}\,(z_{12} - y_{12})\,a\,\mathbf{\hat{x}} + \frac{1}{2\sqrt{3}}\,(2x_{12} - y_{12} - z_{12})\,a\,\mathbf{\hat{y}} + \frac{1}{3}\,(x_{12} + y_{12} + z_{12})\,c\,\mathbf{\hat{z}}$ | (12*i*) | B XII |
| $\mathbf{B}_{60}$ | = | $y_{12}\,\mathbf{a}_1 + z_{12}\,\mathbf{a}_2 + x_{12}\,\mathbf{a}_3$ | = | $\frac{1}{2}\,(y_{12} - x_{12})\,a\,\mathbf{\hat{x}} + \frac{1}{2\sqrt{3}}\,(2z_{12} - x_{12} - y_{12})\,a\,\mathbf{\hat{y}} + \frac{1}{3}\,(x_{12} + y_{12} + z_{12})\,c\,\mathbf{\hat{z}}$ | (12*i*) | B XII |
| $\mathbf{B}_{61}$ | = | $-y_{12}\,\mathbf{a}_1 - x_{12}\,\mathbf{a}_2 - z_{12}\,\mathbf{a}_3$ | = | $\frac{1}{2}\,(z_{12} - y_{12})\,a\,\mathbf{\hat{x}} + \frac{1}{2\sqrt{3}}\,(y_{12} + z_{12} - 2x_{12})\,a\,\mathbf{\hat{y}} - \frac{1}{3}\,(x_{12} + y_{12} + z_{12})\,c\,\mathbf{\hat{z}}$ | (12*i*) | B XII |
| $\mathbf{B}_{62}$ | = | $-x_{12}\,\mathbf{a}_1 - z_{12}\,\mathbf{a}_2 - y_{12}\,\mathbf{a}_3$ | = | $\frac{1}{2}\,(y_{12} - x_{12})\,a\,\mathbf{\hat{x}} + \frac{1}{2\sqrt{3}}\,(x_{12} + y_{12} - 2z_{12})\,a\,\mathbf{\hat{y}} - \frac{1}{3}\,(x_{12} + y_{12} + z_{12})\,c\,\mathbf{\hat{z}}$ | (12*i*) | B XII |
| $\mathbf{B}_{63}$ | = | $-z_{12}\,\mathbf{a}_1 - y_{12}\,\mathbf{a}_2 - x_{12}\,\mathbf{a}_3$ | = | $\frac{1}{2}\,(x_{12} - z_{12})\,a\,\mathbf{\hat{x}} + \frac{1}{2\sqrt{3}}\,(x_{12} + z_{12} - 2y_{12})\,a\,\mathbf{\hat{y}} - \frac{1}{3}\,(x_{12} + y_{12} + z_{12})\,c\,\mathbf{\hat{z}}$ | (12*i*) | B XII |
| $\mathbf{B}_{64}$ | = | $-x_{12}\,\mathbf{a}_1 - y_{12}\,\mathbf{a}_2 - z_{12}\,\mathbf{a}_3$ | = | $\frac{1}{2}\,(z_{12} - x_{12})\,a\,\mathbf{\hat{x}} + \frac{1}{2\sqrt{3}}\,(x_{12} + z_{12} - 2y_{12})\,a\,\mathbf{\hat{y}} - \frac{1}{3}\,(x_{12} + y_{12} + z_{12})\,c\,\mathbf{\hat{z}}$ | (12*i*) | B XII |
| $\mathbf{B_{65}}$ | = | $-z_{12}\,\mathbf{a_1} - x_{12}\,\mathbf{a_2} - y_{12}\,\mathbf{a_3}$ | = | $\frac{1}{2}\,(y_{12} - z_{12})\,a\,\hat{\mathbf{x}} + \frac{1}{2\sqrt{3}}\,(y_{12} + z_{12} - 2x_{12})\,a\,\hat{\mathbf{y}} - \frac{1}{3}\,(x_{12} + y_{12} + z_{12})\,c\,\hat{\mathbf{z}}$ | (12$i$) | B XII |
| $\mathbf{B_{66}}$ | = | $-y_{12}\,\mathbf{a_1} - z_{12}\,\mathbf{a_2} - x_{12}\,\mathbf{a_3}$ | = | $\frac{1}{2}\,(x_{12} - y_{12})\,a\,\hat{\mathbf{x}} + \frac{1}{2\sqrt{3}}\,(x_{12} + y_{12} - 2z_{12})\,a\,\hat{\mathbf{y}} - \frac{1}{3}\,(x_{12} + y_{12} + z_{12})\,c\,\hat{\mathbf{z}}$ | (12$i$) | B XII |
| $\mathbf{B_{67}}$ | = | $y_{12}\,\mathbf{a_1} + x_{12}\,\mathbf{a_2} + z_{12}\,\mathbf{a_3}$ | = | $\frac{1}{2}\,(y_{12} - z_{12})\,a\,\hat{\mathbf{x}} + \frac{1}{2\sqrt{3}}\,(2x_{12} - y_{12} - z_{12})\,a\,\hat{\mathbf{y}} + \frac{1}{3}\,(x_{12} + y_{12} + z_{12})\,c\,\hat{\mathbf{z}}$ | (12$i$) | B XII |
| $\mathbf{B_{68}}$ | = | $x_{12}\,\mathbf{a_1} + z_{12}\,\mathbf{a_2} + y_{12}\,\mathbf{a_3}$ | = | $\frac{1}{2}\,(x_{12} - y_{12})\,a\,\hat{\mathbf{x}} + \frac{1}{2\sqrt{3}}\,(2z_{12} - x_{12} - y_{12})\,a\,\hat{\mathbf{y}} + \frac{1}{3}\,(x_{12} + y_{12} + z_{12})\,c\,\hat{\mathbf{z}}$ | (12$i$) | B XII |
| $\mathbf{B_{69}}$ | = | $z_{12}\,\mathbf{a_1} + y_{12}\,\mathbf{a_2} + x_{12}\,\mathbf{a_3}$ | = | $\frac{1}{2}\,(z_{12} - x_{12})\,a\,\hat{\mathbf{x}} + \frac{1}{2\sqrt{3}}\,(2y_{12} - x_{12} - z_{12})\,a\,\hat{\mathbf{y}} + \frac{1}{3}\,(x_{12} + y_{12} + z_{12})\,c\,\hat{\mathbf{z}}$ | (12$i$) | B XII |
| $\mathbf{B_{70}}$ | = | $x_{13}\,\mathbf{a_1} + y_{13}\,\mathbf{a_2} + z_{13}\,\mathbf{a_3}$ | = | $\frac{1}{2}\,(x_{13} - z_{13})\,a\,\hat{\mathbf{x}} + \frac{1}{2\sqrt{3}}\,(2y_{13} - x_{13} - z_{13})\,a\,\hat{\mathbf{y}} + \frac{1}{3}\,(x_{13} + y_{13} + z_{13})\,c\,\hat{\mathbf{z}}$ | (12$i$) | B XIII |
| $\mathbf{B_{71}}$ | = | $z_{13}\,\mathbf{a_1} + x_{13}\,\mathbf{a_2} + y_{13}\,\mathbf{a_3}$ | = | $\frac{1}{2}\,(z_{13} - y_{13})\,a\,\hat{\mathbf{x}} + \frac{1}{2\sqrt{3}}\,(2x_{13} - y_{13} - z_{13})\,a\,\hat{\mathbf{y}} + \frac{1}{3}\,(x_{13} + y_{13} + z_{13})\,c\,\hat{\mathbf{z}}$ | (12$i$) | B XIII |
| $\mathbf{B_{72}}$ | = | $y_{13}\,\mathbf{a_1} + z_{13}\,\mathbf{a_2} + x_{13}\,\mathbf{a_3}$ | = | $\frac{1}{2}\,(y_{13} - x_{13})\,a\,\hat{\mathbf{x}} + \frac{1}{2\sqrt{3}}\,(2z_{13} - x_{13} - y_{13})\,a\,\hat{\mathbf{y}} + \frac{1}{3}\,(x_{13} + y_{13} + z_{13})\,c\,\hat{\mathbf{z}}$ | (12$i$) | B XIII |
| $\mathbf{B_{73}}$ | = | $-y_{13}\,\mathbf{a_1} - x_{13}\,\mathbf{a_2} - z_{13}\,\mathbf{a_3}$ | = | $\frac{1}{2}\,(z_{13} - y_{13})\,a\,\hat{\mathbf{x}} + \frac{1}{2\sqrt{3}}\,(y_{13} + z_{13} - 2x_{13})\,a\,\hat{\mathbf{y}} - \frac{1}{3}\,(x_{13} + y_{13} + z_{13})\,c\,\hat{\mathbf{z}}$ | (12$i$) | B XIII |
| $\mathbf{B_{74}}$ | = | $-x_{13}\,\mathbf{a_1} - z_{13}\,\mathbf{a_2} - y_{13}\,\mathbf{a_3}$ | = | $\frac{1}{2}\,(y_{13} - x_{13})\,a\,\hat{\mathbf{x}} + \frac{1}{2\sqrt{3}}\,(x_{13} + y_{13} - 2z_{13})\,a\,\hat{\mathbf{y}} - \frac{1}{3}\,(x_{13} + y_{13} + z_{13})\,c\,\hat{\mathbf{z}}$ | (12$i$) | B XIII |
| $\mathbf{B_{75}}$ | = | $-z_{13}\,\mathbf{a_1} - y_{13}\,\mathbf{a_2} - x_{13}\,\mathbf{a_3}$ | = | $\frac{1}{2}\,(x_{13} - z_{13})\,a\,\hat{\mathbf{x}} + \frac{1}{2\sqrt{3}}\,(x_{13} + z_{13} - 2y_{13})\,a\,\hat{\mathbf{y}} - \frac{1}{3}\,(x_{13} + y_{13} + z_{13})\,c\,\hat{\mathbf{z}}$ | (12$i$) | B XIII |
| $\mathbf{B_{76}}$ | = | $-x_{13}\,\mathbf{a_1} - y_{13}\,\mathbf{a_2} - z_{13}\,\mathbf{a_3}$ | = | $\frac{1}{2}\,(z_{13} - x_{13})\,a\,\hat{\mathbf{x}} + \frac{1}{2\sqrt{3}}\,(x_{13} + z_{13} - 2y_{13})\,a\,\hat{\mathbf{y}} - \frac{1}{3}\,(x_{13} + y_{13} + z_{13})\,c\,\hat{\mathbf{z}}$ | (12$i$) | B XIII |
| $\mathbf{B_{77}}$ | = | $-z_{13}\,\mathbf{a_1} - x_{13}\,\mathbf{a_2} - y_{13}\,\mathbf{a_3}$ | = | $\frac{1}{2}\,(y_{13} - z_{13})\,a\,\hat{\mathbf{x}} + \frac{1}{2\sqrt{3}}\,(y_{13} + z_{13} - 2x_{13})\,a\,\hat{\mathbf{y}} - \frac{1}{3}\,(x_{13} + y_{13} + z_{13})\,c\,\hat{\mathbf{z}}$ | (12$i$) | B XIII |
| $\mathbf{B_{78}}$ | = | $-y_{13}\,\mathbf{a_1} - z_{13}\,\mathbf{a_2} - x_{13}\,\mathbf{a_3}$ | = | $\frac{1}{2}\,(x_{13} - y_{13})\,a\,\hat{\mathbf{x}} + \frac{1}{2\sqrt{3}}\,(x_{13} + y_{13} - 2z_{13})\,a\,\hat{\mathbf{y}} - \frac{1}{3}\,(x_{13} + y_{13} + z_{13})\,c\,\hat{\mathbf{z}}$ | (12$i$) | B XIII |
| $\mathbf{B_{79}}$ | = | $y_{13}\,\mathbf{a_1} + x_{13}\,\mathbf{a_2} + z_{13}\,\mathbf{a_3}$ | = | $\frac{1}{2}\,(y_{13} - z_{13})\,a\,\hat{\mathbf{x}} + \frac{1}{2\sqrt{3}}\,(2x_{13} - y_{13} - z_{13})\,a\,\hat{\mathbf{y}} + \frac{1}{3}\,(x_{13} + y_{13} + z_{13})\,c\,\hat{\mathbf{z}}$ | (12$i$) | B XIII |

$$\mathbf{B}_{80} = x_{13}\,\mathbf{a}_1 + z_{13}\,\mathbf{a}_2 + y_{13}\,\mathbf{a}_3 = \frac{1}{2}\left(x_{13} - y_{13}\right) a\,\hat{\mathbf{x}} + \frac{1}{2\sqrt{3}}\left(2z_{13} - x_{13} - y_{13}\right) a\,\hat{\mathbf{y}} + \frac{1}{3}\left(x_{13} + y_{13} + z_{13}\right) c\,\hat{\mathbf{z}} \quad (12i) \quad \text{B XIII}$$

$$\mathbf{B}_{81} = z_{13}\,\mathbf{a}_1 + y_{13}\,\mathbf{a}_2 + x_{13}\,\mathbf{a}_3 = \frac{1}{2}\left(z_{13} - x_{13}\right) a\,\hat{\mathbf{x}} + \frac{1}{2\sqrt{3}}\left(2y_{13} - x_{13} - z_{13}\right) a\,\hat{\mathbf{y}} + \frac{1}{3}\left(x_{13} + y_{13} + z_{13}\right) c\,\hat{\mathbf{z}} \quad (12i) \quad \text{B XIII}$$

$$\mathbf{B}_{82} = x_{14}\,\mathbf{a}_1 + y_{14}\,\mathbf{a}_2 + z_{14}\,\mathbf{a}_3 = \frac{1}{2}\left(x_{14} - z_{14}\right) a\,\hat{\mathbf{x}} + \frac{1}{2\sqrt{3}}\left(2y_{14} - x_{14} - z_{14}\right) a\,\hat{\mathbf{y}} + \frac{1}{3}\left(x_{14} + y_{14} + z_{14}\right) c\,\hat{\mathbf{z}} \quad (12i) \quad \text{B XIV}$$

$$\mathbf{B}_{83} = z_{14}\,\mathbf{a}_1 + x_{14}\,\mathbf{a}_2 + y_{14}\,\mathbf{a}_3 = \frac{1}{2}\left(z_{14} - y_{14}\right) a\,\hat{\mathbf{x}} + \frac{1}{2\sqrt{3}}\left(2x_{14} - y_{14} - z_{14}\right) a\,\hat{\mathbf{y}} + \frac{1}{3}\left(x_{14} + y_{14} + z_{14}\right) c\,\hat{\mathbf{z}} \quad (12i) \quad \text{B XIV}$$

$$\mathbf{B}_{84} = y_{14}\,\mathbf{a}_1 + z_{14}\,\mathbf{a}_2 + x_{14}\,\mathbf{a}_3 = \frac{1}{2}\left(y_{14} - x_{14}\right) a\,\hat{\mathbf{x}} + \frac{1}{2\sqrt{3}}\left(2z_{14} - x_{14} - y_{14}\right) a\,\hat{\mathbf{y}} + \frac{1}{3}\left(x_{14} + y_{14} + z_{14}\right) c\,\hat{\mathbf{z}} \quad (12i) \quad \text{B XIV}$$

$$\mathbf{B}_{85} = -y_{14}\,\mathbf{a}_1 - x_{14}\,\mathbf{a}_2 - z_{14}\,\mathbf{a}_3 = \frac{1}{2}\left(z_{14} - y_{14}\right) a\,\hat{\mathbf{x}} + \frac{1}{2\sqrt{3}}\left(y_{14} + z_{14} - 2x_{14}\right) a\,\hat{\mathbf{y}} - \frac{1}{3}\left(x_{14} + y_{14} + z_{14}\right) c\,\hat{\mathbf{z}} \quad (12i) \quad \text{B XIV}$$

$$\mathbf{B}_{86} = -x_{14}\,\mathbf{a}_1 - z_{14}\,\mathbf{a}_2 - y_{14}\,\mathbf{a}_3 = \frac{1}{2}\left(y_{14} - x_{14}\right) a\,\hat{\mathbf{x}} + \frac{1}{2\sqrt{3}}\left(x_{14} + y_{14} - 2z_{14}\right) a\,\hat{\mathbf{y}} - \frac{1}{3}\left(x_{14} + y_{14} + z_{14}\right) c\,\hat{\mathbf{z}} \quad (12i) \quad \text{B XIV}$$

$$\mathbf{B}_{87} = -z_{14}\,\mathbf{a}_1 - y_{14}\,\mathbf{a}_2 - x_{14}\,\mathbf{a}_3 = \frac{1}{2}\left(x_{14} - z_{14}\right) a\,\hat{\mathbf{x}} + \frac{1}{2\sqrt{3}}\left(x_{14} + z_{14} - 2y_{14}\right) a\,\hat{\mathbf{y}} - \frac{1}{3}\left(x_{14} + y_{14} + z_{14}\right) c\,\hat{\mathbf{z}} \quad (12i) \quad \text{B XIV}$$

$$\mathbf{B}_{88} = -x_{14}\,\mathbf{a}_1 - y_{14}\,\mathbf{a}_2 - z_{14}\,\mathbf{a}_3 = \frac{1}{2}\left(z_{14} - x_{14}\right) a\,\hat{\mathbf{x}} + \frac{1}{2\sqrt{3}}\left(x_{14} + z_{14} - 2y_{14}\right) a\,\hat{\mathbf{y}} - \frac{1}{3}\left(x_{14} + y_{14} + z_{14}\right) c\,\hat{\mathbf{z}} \quad (12i) \quad \text{B XIV}$$

$$\mathbf{B}_{89} = -z_{14}\,\mathbf{a}_1 - x_{14}\,\mathbf{a}_2 - y_{14}\,\mathbf{a}_3 = \frac{1}{2}\left(y_{14} - z_{14}\right) a\,\hat{\mathbf{x}} + \frac{1}{2\sqrt{3}}\left(y_{14} + z_{14} - 2x_{14}\right) a\,\hat{\mathbf{y}} - \frac{1}{3}\left(x_{14} + y_{14} + z_{14}\right) c\,\hat{\mathbf{z}} \quad (12i) \quad \text{B XIV}$$

$$\mathbf{B}_{90} = -y_{14}\,\mathbf{a}_1 - z_{14}\,\mathbf{a}_2 - x_{14}\,\mathbf{a}_3 = \frac{1}{2}\left(x_{14} - y_{14}\right) a\,\hat{\mathbf{x}} + \frac{1}{2\sqrt{3}}\left(x_{14} + y_{14} - 2z_{14}\right) a\,\hat{\mathbf{y}} - \frac{1}{3}\left(x_{14} + y_{14} + z_{14}\right) c\,\hat{\mathbf{z}} \quad (12i) \quad \text{B XIV}$$

$$\mathbf{B}_{91} = y_{14}\,\mathbf{a}_1 + x_{14}\,\mathbf{a}_2 + z_{14}\,\mathbf{a}_3 = \frac{1}{2}\left(y_{14} - z_{14}\right) a\,\hat{\mathbf{x}} + \frac{1}{2\sqrt{3}}\left(2x_{14} - y_{14} - z_{14}\right) a\,\hat{\mathbf{y}} + \frac{1}{3}\left(x_{14} + y_{14} + z_{14}\right) c\,\hat{\mathbf{z}} \quad (12i) \quad \text{B XIV}$$

$$\mathbf{B}_{92} = x_{14}\,\mathbf{a}_1 + z_{14}\,\mathbf{a}_2 + y_{14}\,\mathbf{a}_3 = \frac{1}{2}\left(x_{14} - y_{14}\right) a\,\hat{\mathbf{x}} + \frac{1}{2\sqrt{3}}\left(2z_{14} - x_{14} - y_{14}\right) a\,\hat{\mathbf{y}} + \frac{1}{3}\left(x_{14} + y_{14} + z_{14}\right) c\,\hat{\mathbf{z}} \quad (12i) \quad \text{B XIV}$$

$$\mathbf{B}_{93} = z_{14}\,\mathbf{a}_1 + y_{14}\,\mathbf{a}_2 + x_{14}\,\mathbf{a}_3 = \frac{1}{2}\left(z_{14} - x_{14}\right) a\,\hat{\mathbf{x}} + \frac{1}{2\sqrt{3}}\left(2y_{14} - x_{14} - z_{14}\right) a\,\hat{\mathbf{y}} + \frac{1}{3}\left(x_{14} + y_{14} + z_{14}\right) c\,\hat{\mathbf{z}} \quad (12i) \quad \text{B XIV}$$

$$\mathbf{B}_{94} = x_{15}\,\mathbf{a}_1 + y_{15}\,\mathbf{a}_2 + z_{15}\,\mathbf{a}_3 = \frac{1}{2}\left(x_{15} - z_{15}\right) a\,\hat{\mathbf{x}} + \frac{1}{2\sqrt{3}}\left(2y_{15} - x_{15} - z_{15}\right) a\,\hat{\mathbf{y}} + \frac{1}{3}\left(x_{15} + y_{15} + z_{15}\right) c\,\hat{\mathbf{z}} \quad (12i) \quad \text{B XV}$$

| | | | | | | |
|---|---|---|---|---|---|---|
| $\mathbf{B}_{95}$ | = | $z_{15}\,\mathbf{a}_1 + x_{15}\,\mathbf{a}_2 + y_{15}\,\mathbf{a}_3$ | = | $\frac{1}{2}\left(z_{15} - y_{15}\right) a\,\hat{\mathbf{x}} + \frac{1}{2\sqrt{3}}\left(2x_{15} - y_{15} - z_{15}\right) a\,\hat{\mathbf{y}} + \frac{1}{3}\left(x_{15} + y_{15} + z_{15}\right) c\,\hat{\mathbf{z}}$ | (12$i$) | B XV |
| $\mathbf{B}_{96}$ | = | $y_{15}\,\mathbf{a}_1 + z_{15}\,\mathbf{a}_2 + x_{15}\,\mathbf{a}_3$ | = | $\frac{1}{2}\left(y_{15} - x_{15}\right) a\,\hat{\mathbf{x}} + \frac{1}{2\sqrt{3}}\left(2z_{15} - x_{15} - y_{15}\right) a\,\hat{\mathbf{y}} + \frac{1}{3}\left(x_{15} + y_{15} + z_{15}\right) c\,\hat{\mathbf{z}}$ | (12$i$) | B XV |
| $\mathbf{B}_{97}$ | = | $-y_{15}\,\mathbf{a}_1 - x_{15}\,\mathbf{a}_2 - z_{15}\,\mathbf{a}_3$ | = | $\frac{1}{2}\left(z_{15} - y_{15}\right) a\,\hat{\mathbf{x}} + \frac{1}{2\sqrt{3}}\left(y_{15} + z_{15} - 2x_{15}\right) a\,\hat{\mathbf{y}} - \frac{1}{3}\left(x_{15} + y_{15} + z_{15}\right) c\,\hat{\mathbf{z}}$ | (12$i$) | B XV |
| $\mathbf{B}_{98}$ | = | $-x_{15}\,\mathbf{a}_1 - z_{15}\,\mathbf{a}_2 - y_{15}\,\mathbf{a}_3$ | = | $\frac{1}{2}\left(y_{15} - x_{15}\right) a\,\hat{\mathbf{x}} + \frac{1}{2\sqrt{3}}\left(x_{15} + y_{15} - 2z_{15}\right) a\,\hat{\mathbf{y}} - \frac{1}{3}\left(x_{15} + y_{15} + z_{15}\right) c\,\hat{\mathbf{z}}$ | (12$i$) | B XV |
| $\mathbf{B}_{99}$ | = | $-z_{15}\,\mathbf{a}_1 - y_{15}\,\mathbf{a}_2 - x_{15}\,\mathbf{a}_3$ | = | $\frac{1}{2}\left(x_{15} - z_{15}\right) a\,\hat{\mathbf{x}} + \frac{1}{2\sqrt{3}}\left(x_{15} + z_{15} - 2y_{15}\right) a\,\hat{\mathbf{y}} - \frac{1}{3}\left(x_{15} + y_{15} + z_{15}\right) c\,\hat{\mathbf{z}}$ | (12$i$) | B XV |
| $\mathbf{B}_{100}$ | = | $-x_{15}\,\mathbf{a}_1 - y_{15}\,\mathbf{a}_2 - z_{15}\,\mathbf{a}_3$ | = | $\frac{1}{2}\left(z_{15} - x_{15}\right) a\,\hat{\mathbf{x}} + \frac{1}{2\sqrt{3}}\left(x_{15} + z_{15} - 2y_{15}\right) a\,\hat{\mathbf{y}} - \frac{1}{3}\left(x_{15} + y_{15} + z_{15}\right) c\,\hat{\mathbf{z}}$ | (12$i$) | B XV |
| $\mathbf{B}_{101}$ | = | $-z_{15}\,\mathbf{a}_1 - x_{15}\,\mathbf{a}_2 - y_{15}\,\mathbf{a}_3$ | = | $\frac{1}{2}\left(y_{15} - z_{15}\right) a\,\hat{\mathbf{x}} + \frac{1}{2\sqrt{3}}\left(y_{15} + z_{15} - 2x_{15}\right) a\,\hat{\mathbf{y}} - \frac{1}{3}\left(x_{15} + y_{15} + z_{15}\right) c\,\hat{\mathbf{z}}$ | (12$i$) | B XV |
| $\mathbf{B}_{102}$ | = | $-y_{15}\,\mathbf{a}_1 - z_{15}\,\mathbf{a}_2 - x_{15}\,\mathbf{a}_3$ | = | $\frac{1}{2}\left(x_{15} - y_{15}\right) a\,\hat{\mathbf{x}} + \frac{1}{2\sqrt{3}}\left(x_{15} + y_{15} - 2z_{15}\right) a\,\hat{\mathbf{y}} - \frac{1}{3}\left(x_{15} + y_{15} + z_{15}\right) c\,\hat{\mathbf{z}}$ | (12$i$) | B XV |
| $\mathbf{B}_{103}$ | = | $y_{15}\,\mathbf{a}_1 + x_{15}\,\mathbf{a}_2 + z_{15}\,\mathbf{a}_3$ | = | $\frac{1}{2}\left(y_{15} - z_{15}\right) a\,\hat{\mathbf{x}} + \frac{1}{2\sqrt{3}}\left(2x_{15} - y_{15} - z_{15}\right) a\,\hat{\mathbf{y}} + \frac{1}{3}\left(x_{15} + y_{15} + z_{15}\right) c\,\hat{\mathbf{z}}$ | (12$i$) | B XV |
| $\mathbf{B}_{104}$ | = | $x_{15}\,\mathbf{a}_1 + z_{15}\,\mathbf{a}_2 + y_{15}\,\mathbf{a}_3$ | = | $\frac{1}{2}\left(x_{15} - y_{15}\right) a\,\hat{\mathbf{x}} + \frac{1}{2\sqrt{3}}\left(2z_{15} - x_{15} - y_{15}\right) a\,\hat{\mathbf{y}} + \frac{1}{3}\left(x_{15} + y_{15} + z_{15}\right) c\,\hat{\mathbf{z}}$ | (12$i$) | B XV |
| $\mathbf{B}_{105}$ | = | $z_{15}\,\mathbf{a}_1 + y_{15}\,\mathbf{a}_2 + x_{15}\,\mathbf{a}_3$ | = | $\frac{1}{2}\left(z_{15} - x_{15}\right) a\,\hat{\mathbf{x}} + \frac{1}{2\sqrt{3}}\left(2y_{15} - x_{15} - z_{15}\right) a\,\hat{\mathbf{y}} + \frac{1}{3}\left(x_{15} + y_{15} + z_{15}\right) c\,\hat{\mathbf{z}}$ | (12$i$) | B XV |

---

**Found in:**

---

**Geometry files:**

- CIF: pp. 729
- POSCAR: pp. 730

# $CaC_6$ Structure: A6B_hR7_166_g_a

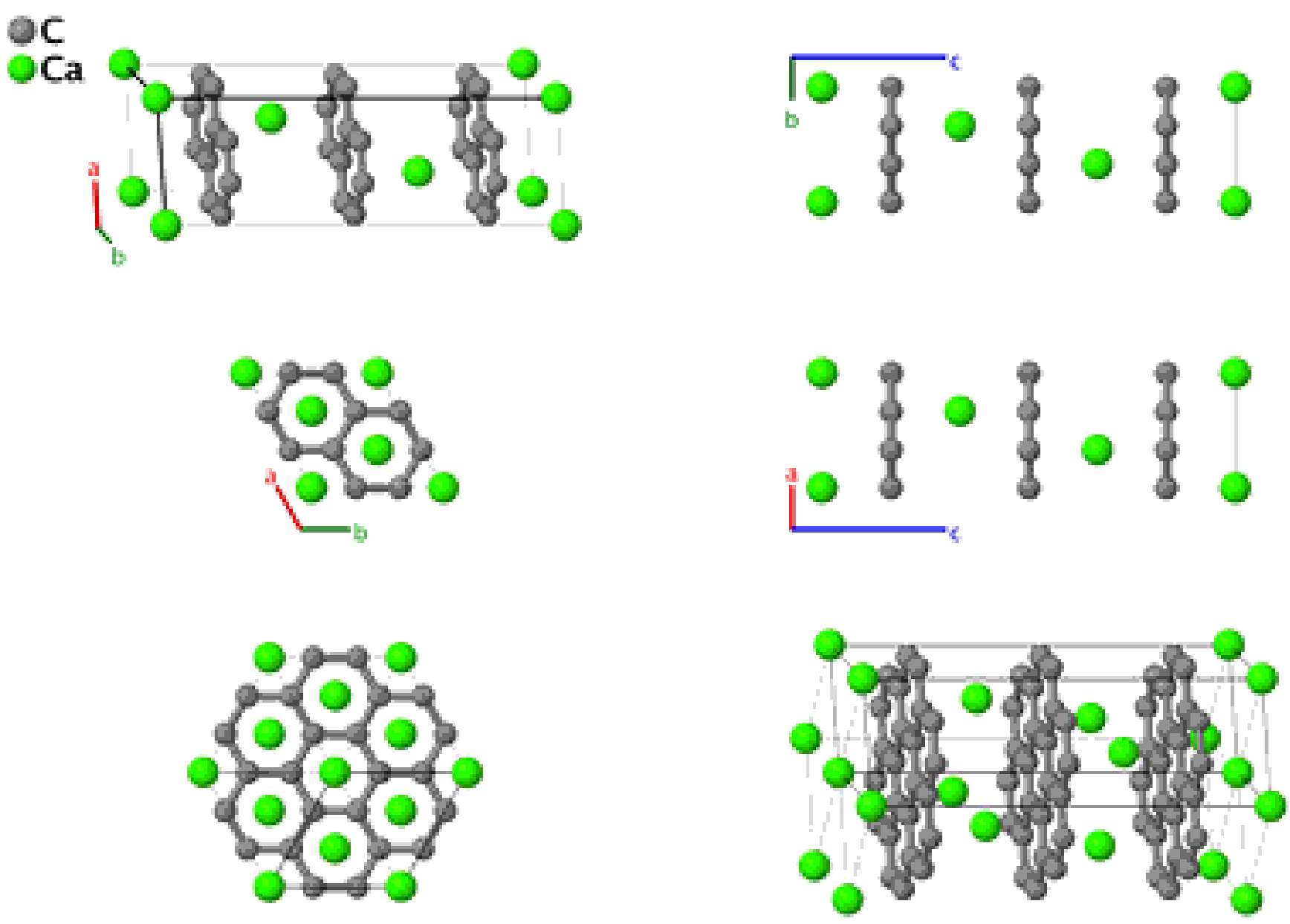

| | | |
|---|---|---|
| **Prototype** | : | $CaC_6$ |
| **AFLOW prototype label** | : | A6B_hR7_166_g_a |
| ***Strukturbericht* designation** | : | None |
| **Pearson symbol** | : | hR7 |
| **Space group number** | : | 166 |
| **Space group symbol** | : | $R\bar{3}m$ |
| **AFLOW prototype command** | : | `aflow --proto=A6B_hR7_166_g_a [--hex]` `--params=`$a, c/a, x_2$ |

- Superconducting structure, $T_c = 11.5$K. Hexagonal settings of this structure can be obtained with the option `--hex`.

**Rhombohedral primitive vectors:**

$$\mathbf{a}_1 = \frac{1}{2}\,a\,\mathbf{\hat{x}} - \frac{1}{2\sqrt{3}}\,a\,\mathbf{\hat{y}} + \frac{1}{3}\,c\,\mathbf{\hat{z}}$$
$$\mathbf{a}_2 = \frac{1}{\sqrt{3}}\,a\,\mathbf{\hat{y}} + \frac{1}{3}\,c\,\mathbf{\hat{z}}$$
$$\mathbf{a}_3 = -\frac{1}{2}\,a\,\mathbf{\hat{x}} - \frac{1}{2\sqrt{3}}\,a\,\mathbf{\hat{y}} + \frac{1}{3}\,c\,\mathbf{\hat{z}}$$

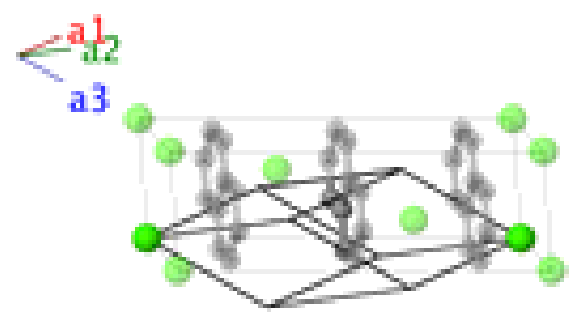

**Basis vectors:**

| | | Lattice Coordinates | | Cartesian Coordinates | Wyckoff Position | Atom Type |
|---|---|---|---|---|---|---|
| $\mathbf{B}_1$ | = | $0\,\mathbf{a}_1 + 0\,\mathbf{a}_2 + 0\,\mathbf{a}_3$ | = | $0\mathbf{\hat{x}} + 0\mathbf{\hat{y}} + 0\mathbf{\hat{z}}$ | $(1a)$ | Ca |
| $\mathbf{B}_2$ | = | $x_2\,\mathbf{a}_1 - x_2\,\mathbf{a}_2 + \frac{1}{2}\,\mathbf{a}_3$ | = | $\left(\frac{1}{2}x_2 + \frac{3}{4}\right)a\,\mathbf{\hat{x}} - \frac{1}{4\sqrt{3}}\left(1 + 6x_2\right)a\,\mathbf{\hat{y}} + \frac{1}{6}\,c\,\mathbf{\hat{z}}$ | $(6g)$ | C |
| $\mathbf{B}_3$ | = | $\frac{1}{2}\,\mathbf{a}_1 + x_2\,\mathbf{a}_2 - x_2\,\mathbf{a}_3$ | = | $\left(\frac{1}{2}x_2 + \frac{1}{4}\right)a\,\mathbf{\hat{x}} - \frac{1}{4\sqrt{3}}\left(1 - 6x_2\right)a\,\mathbf{\hat{y}} + \frac{1}{6}\,c\,\mathbf{\hat{z}}$ | $(6g)$ | C |
| $\mathbf{B}_4$ | = | $-x_2\,\mathbf{a}_1 + \frac{1}{2}\,\mathbf{a}_2 + x_2\,\mathbf{a}_3$ | = | $-x_2\,a\,\mathbf{\hat{x}} + \frac{1}{2\sqrt{3}}\,a\,\mathbf{\hat{y}} + \frac{1}{6}\,c\,\mathbf{\hat{z}}$ | $(6g)$ | C |
| $\mathbf{B}_5$ | = | $-x_2\,\mathbf{a}_1 + x_2\,\mathbf{a}_2 + \frac{1}{2}\,\mathbf{a}_3$ | = | $\left(\frac{3}{4} - \frac{1}{2}x_2\right)a\,\mathbf{\hat{x}} - \frac{1}{4\sqrt{3}}\left(1 - 6x_2\right)a\,\mathbf{\hat{y}} + \frac{1}{6}\,c\,\mathbf{\hat{z}}$ | $(6g)$ | C |
| $\mathbf{B_6}$ | = | $\frac{1}{2}\,\mathbf{a_1} - x_2\,\mathbf{a_2} + x_2\,\mathbf{a_3}$ | = | $\left(\frac{1}{4} - \frac{1}{2}x_2\right)\,a\,\hat{\mathbf{x}} - \frac{1}{4\sqrt{3}}\,(1 + 6x_2)\,a\,\hat{\mathbf{y}} + \frac{1}{6}\,c\,\hat{\mathbf{z}}$ | (6$g$) | C |
| $\mathbf{B_7}$ | = | $x_2\,\mathbf{a_1} + \frac{1}{2}\,\mathbf{a_2} - x_2\,\mathbf{a_3}$ | = | $x_2\,a\,\hat{\mathbf{x}} + \frac{1}{2\sqrt{3}}\,a\,\hat{\mathbf{y}} + \frac{1}{6}\,c\,\hat{\mathbf{z}}$ | (6$g$) | C |

---

---

**Geometry files:**

- CIF: pp. 730
- POSCAR: pp. 731

# Paraelectric $LiNbO_3$ Structure: ABC3_hR10_167_a_b_e

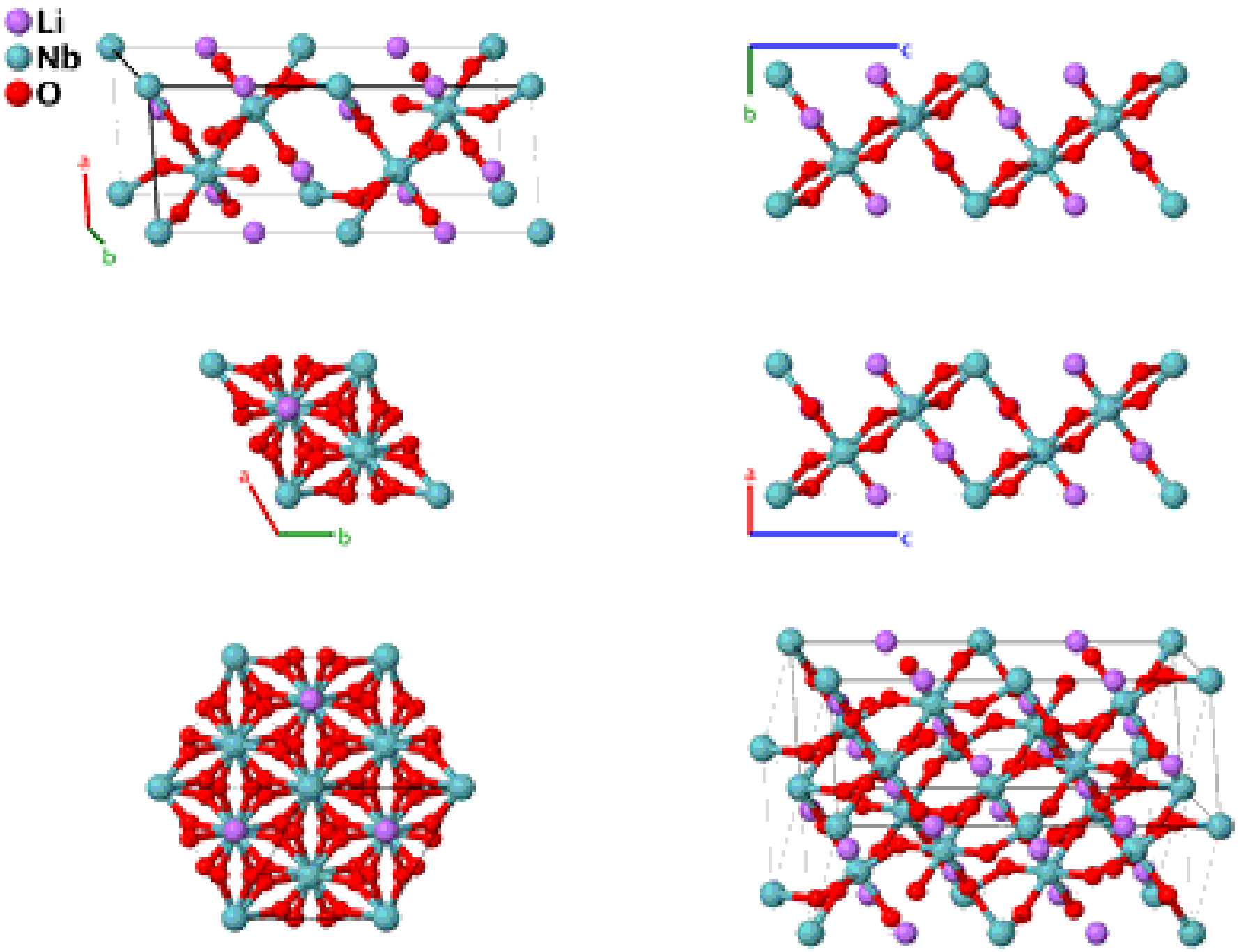

| | | |
|---|---|---|
| **Prototype** | : | $LiNbO_3$ |
| **AFLOW prototype label** | : | ABC3_hR10_167_a_b_e |
| ***Strukturbericht* designation** | : | None |
| **Pearson symbol** | : | hR10 |
| **Space group number** | : | 167 |
| **Space group symbol** | : | $R\bar{3}c$ |
| **AFLOW prototype command** | : | `aflow --proto=ABC3_hR10_167_a_b_e [--hex]`<br>`--params=`$a, c/a, x_3$ |

- This is the paraelectric phase, which exists above 1430K. There is also a ferroelectric phase. Note that paraelectric $LiNbO_3$ (pp. 409) and calcite (pp. 411) have the same AFLOW prototype label. They are generated by the same symmetry operations with different sets of parameters (`--params`) specified in their corresponding CIF files. Hexagonal settings of this structure can be obtained with the option `--hex`.

**Rhombohedral primitive vectors:**

$$\mathbf{a}_1 = \frac{1}{2}\,a\,\hat{\mathbf{x}} - \frac{1}{2\sqrt{3}}\,a\,\hat{\mathbf{y}} + \frac{1}{3}\,c\,\hat{\mathbf{z}}$$

$$\mathbf{a}_2 = \frac{1}{\sqrt{3}}\,a\,\hat{\mathbf{y}} + \frac{1}{3}\,c\,\hat{\mathbf{z}}$$

$$\mathbf{a}_3 = -\frac{1}{2}\,a\,\hat{\mathbf{x}} - \frac{1}{2\sqrt{3}}\,a\,\hat{\mathbf{y}} + \frac{1}{3}\,c\,\hat{\mathbf{z}}$$

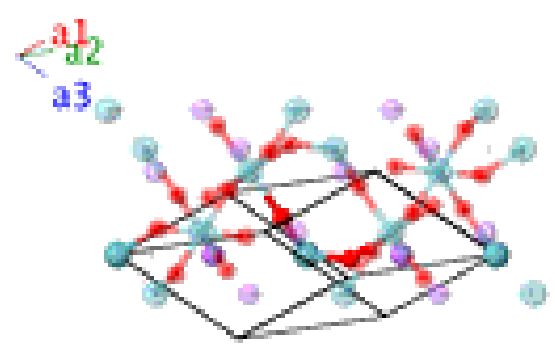

**Basis vectors:**

| | | Lattice Coordinates | | Cartesian Coordinates | Wyckoff Position | Atom Type |
|---|---|---|---|---|---|---|
| $\mathbf{B}_1$ | = | $\frac{1}{4}\,\mathbf{a}_1 + \frac{1}{4}\,\mathbf{a}_2 + \frac{1}{4}\,\mathbf{a}_3$ | = | $\frac{1}{4}c\,\hat{\mathbf{z}}$ | (2a) | Li |
| $\mathbf{B_2}$ | = | $\frac{3}{4}\,\mathbf{a_1} + \frac{3}{4}\,\mathbf{a_2} + \frac{3}{4}\,\mathbf{a_3}$ | = | $\frac{3}{4}c\,\mathbf{\hat{z}}$ | (2*a*) | Li |
| $\mathbf{B_3}$ | = | $0\,\mathbf{a_1} + 0\,\mathbf{a_2} + 0\,\mathbf{a_3}$ | = | $0\mathbf{\hat{x}} + 0\mathbf{\hat{y}} + 0\mathbf{\hat{z}}$ | (2*b*) | Nb |
| $\mathbf{B_4}$ | = | $\frac{1}{2}\,\mathbf{a_1} + \frac{1}{2}\,\mathbf{a_2} + \frac{1}{2}\,\mathbf{a_3}$ | = | $\frac{1}{2}c\,\mathbf{\hat{z}}$ | (2*b*) | Nb |
| $\mathbf{B_5}$ | = | $x_3\,\mathbf{a_1} + \left(\frac{1}{2} - x_3\right)\,\mathbf{a_2} + \frac{1}{4}\,\mathbf{a_3}$ | = | $-\frac{1}{8}\,(1 - 4x_3)\;a\,\mathbf{\hat{x}} + \frac{\sqrt{3}}{8}\,(1 - 4x_3)\;a\mathbf{\hat{y}} + \frac{1}{4}\,c\,\mathbf{\hat{z}}$ | (6*e*) | O |
| $\mathbf{B_6}$ | = | $\frac{1}{4}\,\mathbf{a_1} + x_3\,\mathbf{a_2} + \left(\frac{1}{2} - x_3\right)\,\mathbf{a_3}$ | = | $-\frac{1}{8}\,(1 - 4x_3)\;a\,\mathbf{\hat{x}} - \frac{\sqrt{3}}{8}\,(1 - 4x_3)\;a\mathbf{\hat{y}} + \frac{1}{4}\,c\,\mathbf{\hat{z}}$ | (6*e*) | O |
| $\mathbf{B_7}$ | = | $\left(\frac{1}{2} - x_3\right)\,\mathbf{a_1} + \frac{1}{4}\,\mathbf{a_2} + x_3\,\mathbf{a_3}$ | = | $\frac{1}{4}\,(1 - 4x_3)\;a\,\mathbf{\hat{x}} + \frac{1}{4}\,c\,\mathbf{\hat{z}}$ | (6*e*) | O |
| $\mathbf{B_8}$ | = | $-x_3\,\mathbf{a_1} + \left(\frac{1}{2} + x_3\right)\,\mathbf{a_2} + \frac{3}{4}\,\mathbf{a_3}$ | = | $-\frac{1}{8}\,(3 + 4x_3)\;a\,\mathbf{\hat{x}} + \frac{1}{8\sqrt{3}}\,(1 + 12x_3)\;a\mathbf{\hat{y}} + \frac{5}{12}\,c\,\mathbf{\hat{z}}$ | (6*e*) | O |
| $\mathbf{B_9}$ | = | $\frac{3}{4}\,\mathbf{a_1} - x_3\,\mathbf{a_2} + \left(\frac{1}{2} + x_3\right)\,\mathbf{a_3}$ | = | $\frac{1}{8}\,(1 - 4x_3)\;a\,\mathbf{\hat{x}} - \frac{1}{8\sqrt{3}}\,(5 + 12x_3)\;a\mathbf{\hat{y}} + \frac{5}{12}\,c\,\mathbf{\hat{z}}$ | (6*e*) | O |
| $\mathbf{B_{10}}$ | = | $\left(\frac{1}{2} + x_3\right)\,\mathbf{a_1} + \frac{3}{4}\,\mathbf{a_2} - x_3\,\mathbf{a_3}$ | = | $\frac{1}{4}\,(1 + 4x_3)\;a\,\mathbf{\hat{x}} + \frac{1}{2\sqrt{3}}\,a\mathbf{\hat{y}} + \frac{5}{12}\,c\,\mathbf{\hat{z}}$ | (6*e*) | O |

**Geometry files:**

- CIF: pp. 731
- POSCAR: pp. 731

# Calcite ($CaCO_3$, $G0_1$) Structure: ABC3_hR10_167_a_b_e

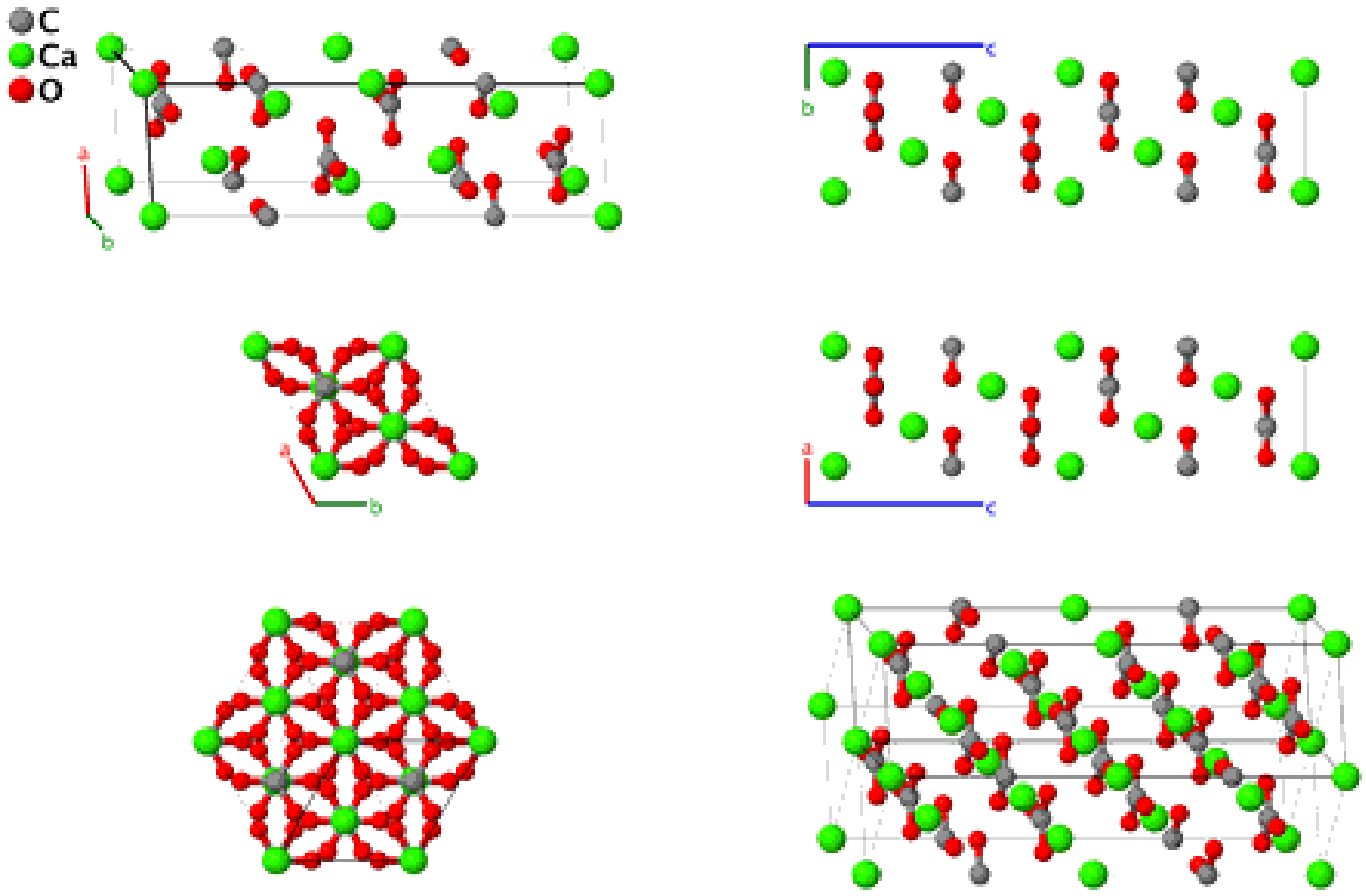

**Prototype** : $CaCO_3$

**AFLOW prototype label** : ABC3_hR10_167_a_b_e

***Strukturbericht* designation** : $G0_1$

**Pearson symbol** : hR10

**Space group number** : 167

**Space group symbol** : $R\bar{3}c$

**AFLOW prototype command** : `aflow --proto=ABC3_hR10_167_a_b_e [--hex]`
`--params=`$a, c/a, x_3$

---

- Strukturbericht Band I, (Ewald, 1931), pp. 292-295, gives this the designation G1, but the index in Band II, (Hermann, 1937) lists this as $G0_1$. Note that paraelectric $LiNbO_3$ (pp. 409) and calcite (pp. 411) have the same AFLOW prototype label. They are generated by the same symmetry operations with different sets of parameters (`--params`) specified in their corresponding CIF files. Hexagonal settings of this structure can be obtained with the option `--hex`.

---

**Rhombohedral primitive vectors:**

$$
\begin{aligned}
\mathbf{a}_1 &= \tfrac{1}{2}\,a\,\mathbf{\hat{x}} - \tfrac{1}{2\sqrt{3}}\,a\,\mathbf{\hat{y}} + \tfrac{1}{3}\,c\,\mathbf{\hat{z}} \\
\mathbf{a}_2 &= \tfrac{1}{\sqrt{3}}\,a\,\mathbf{\hat{y}} + \tfrac{1}{3}\,c\,\mathbf{\hat{z}} \\
\mathbf{a}_3 &= -\tfrac{1}{2}\,a\,\mathbf{\hat{x}} - \tfrac{1}{2\sqrt{3}}\,a\,\mathbf{\hat{y}} + \tfrac{1}{3}\,c\,\mathbf{\hat{z}}
\end{aligned}
$$

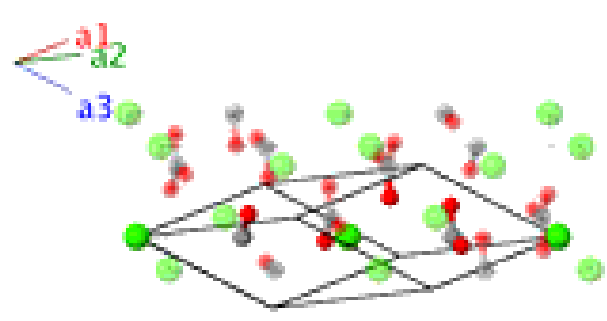

---

**Basis vectors:**

| | | Lattice Coordinates | | Cartesian Coordinates | Wyckoff Position | Atom Type |
|---|---|---|---|---|---|---|
| $\mathbf{B}_1$ | = | $\frac{1}{4}\,\mathbf{a}_1 + \frac{1}{4}\,\mathbf{a}_2 + \frac{1}{4}\,\mathbf{a}_3$ | = | $\frac{1}{4}c\,\mathbf{\hat{z}}$ | (2$a$) | C |
| $\mathbf{B}_2$ | = | $\frac{3}{4}\,\mathbf{a}_1 + \frac{3}{4}\,\mathbf{a}_2 + \frac{3}{4}\,\mathbf{a}_3$ | = | $\frac{3}{4}c\,\mathbf{\hat{z}}$ | (2$a$) | C |
| $\mathbf{B_3}$ | = | $0\,\mathbf{a_1} + 0\,\mathbf{a_2} + 0\,\mathbf{a_3}$ | = | $0\hat{\mathbf{x}} + 0\hat{\mathbf{y}} + 0\hat{\mathbf{z}}$ | (2$b$) | Ca |
| $\mathbf{B_4}$ | = | $\frac{1}{2}\,\mathbf{a_1} + \frac{1}{2}\,\mathbf{a_2} + \frac{1}{2}\,\mathbf{a_3}$ | = | $\frac{1}{2}c\,\hat{\mathbf{z}}$ | (2$b$) | Ca |
| $\mathbf{B_5}$ | = | $x_3\,\mathbf{a_1} + \left(\frac{1}{2} - x_3\right)\,\mathbf{a_2} + \frac{1}{4}\,\mathbf{a_3}$ | = | $-\frac{1}{8}\left(1 - 4x_3\right)\,a\,\hat{\mathbf{x}} + \frac{\sqrt{3}}{8}\left(1 - 4x_3\right)\,a\hat{\mathbf{y}} + \frac{1}{4}\,c\,\hat{\mathbf{z}}$ | (6$e$) | O |
| $\mathbf{B_6}$ | = | $\frac{1}{4}\,\mathbf{a_1} + x_3\,\mathbf{a_2} + \left(\frac{1}{2} - x_3\right)\,\mathbf{a_3}$ | = | $-\frac{1}{8}\left(1 - 4x_3\right)\,a\,\hat{\mathbf{x}} - \frac{\sqrt{3}}{8}\left(1 - 4x_3\right)\,a\hat{\mathbf{y}} + \frac{1}{4}\,c\,\hat{\mathbf{z}}$ | (6$e$) | O |
| $\mathbf{B_7}$ | = | $\left(\frac{1}{2} - x_3\right)\,\mathbf{a_1} + \frac{1}{4}\,\mathbf{a_2} + x_3\,\mathbf{a_3}$ | = | $\frac{1}{4}\left(1 - 4x_3\right)\,a\,\hat{\mathbf{x}} + \frac{1}{4}\,c\,\hat{\mathbf{z}}$ | (6$e$) | O |
| $\mathbf{B_8}$ | = | $-x_3\,\mathbf{a_1} + \left(\frac{1}{2} + x_3\right)\,\mathbf{a_2} + \frac{3}{4}\,\mathbf{a_3}$ | = | $-\frac{1}{8}\left(3 + 4x_3\right)\,a\,\hat{\mathbf{x}} + \frac{1}{8\sqrt{3}}\left(1 + 12x_3\right)\,a\hat{\mathbf{y}} + \frac{5}{12}\,c\,\hat{\mathbf{z}}$ | (6$e$) | O |
| $\mathbf{B_9}$ | = | $\frac{3}{4}\,\mathbf{a_1} - x_3\,\mathbf{a_2} + \left(\frac{1}{2} + x_3\right)\,\mathbf{a_3}$ | = | $\frac{1}{8}\left(1 - 4x_3\right)\,a\,\hat{\mathbf{x}} - \frac{1}{8\sqrt{3}}\left(5 + 12x_3\right)\,a\hat{\mathbf{y}} + \frac{5}{12}\,c\,\hat{\mathbf{z}}$ | (6$e$) | O |
| $\mathbf{B_{10}}$ | = | $\left(\frac{1}{2} + x_3\right)\,\mathbf{a_1} + \frac{3}{4}\,\mathbf{a_2} - x_3\,\mathbf{a_3}$ | = | $\frac{1}{4}\left(1 + 4x_3\right)\,a\,\hat{\mathbf{x}} + \frac{1}{2\sqrt{3}}\,a\hat{\mathbf{y}} + \frac{5}{12}\,c\,\hat{\mathbf{z}}$ | (6$e$) | O |

**Found in:**

**Geometry files:**

- CIF: pp. 731
- POSCAR: pp. 732

# Corundum ($Al_2O_3$, $D5_1$) Structure: A2B3_hR10_167_c_e

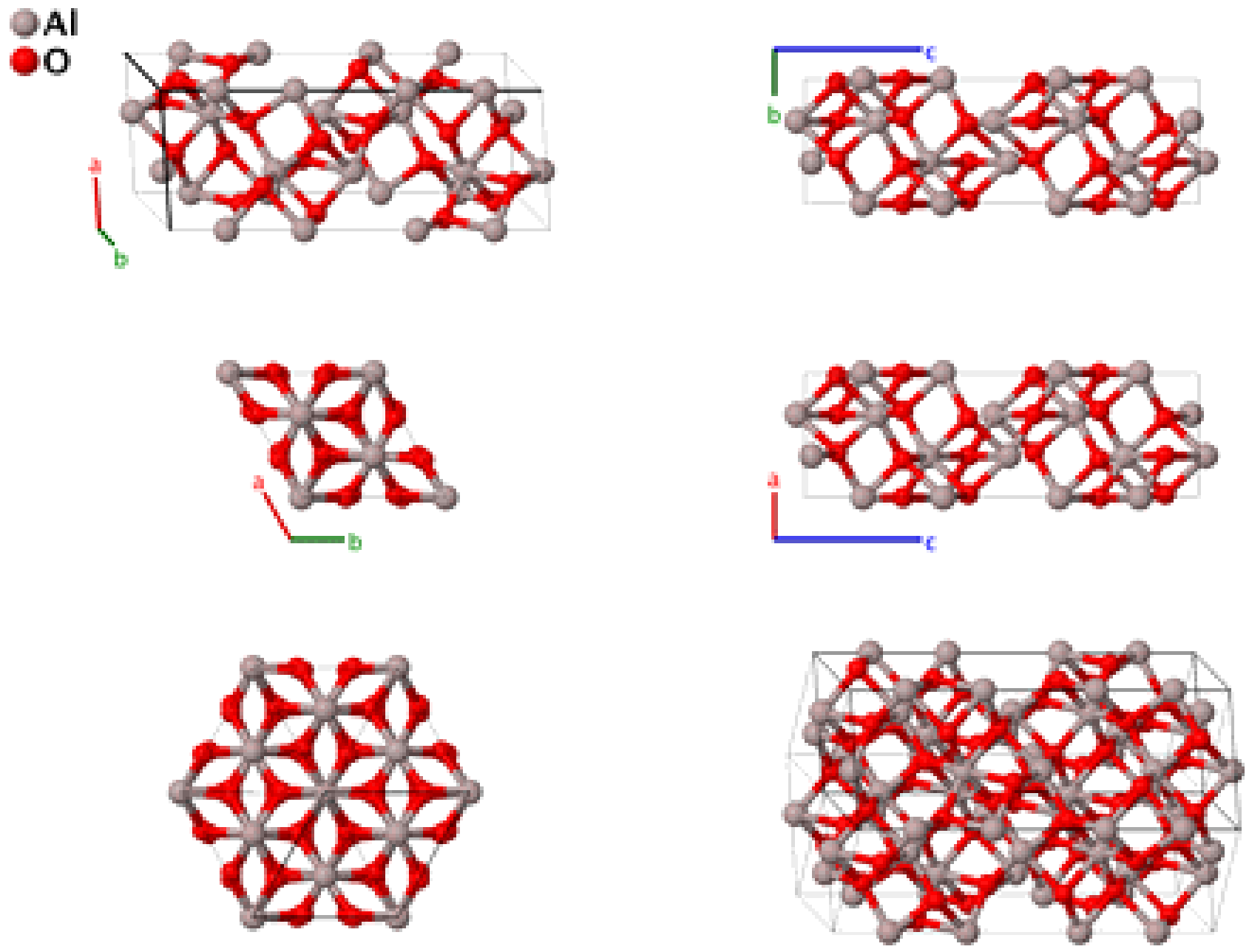

| | | |
|---|---|---|
| **Prototype** | : | $Al_2O_3$ |
| **AFLOW prototype label** | : | A2B3_hR10_167_c_e |
| ***Strukturbericht* designation** | : | $D5_1$ |
| **Pearson symbol** | : | hR10 |
| **Space group number** | : | 167 |
| **Space group symbol** | : | $R\bar{3}c$ |
| **AFLOW prototype command** | : | `aflow --proto=A2B3_hR10_167_c_e [--hex]`<br>`--params=`$a, c/a, x_1, x_2$ |

- Hexagonal settings of this structure can be obtained with the option `--hex`.

**Rhombohedral primitive vectors:**

$$\mathbf{a}_1 = \frac{1}{2}\,a\,\hat{\mathbf{x}} - \frac{1}{2\sqrt{3}}\,a\,\hat{\mathbf{y}} + \frac{1}{3}\,c\,\hat{\mathbf{z}}$$

$$\mathbf{a}_2 = \frac{1}{\sqrt{3}}\,a\,\hat{\mathbf{y}} + \frac{1}{3}\,c\,\hat{\mathbf{z}}$$

$$\mathbf{a}_3 = -\frac{1}{2}\,a\,\hat{\mathbf{x}} - \frac{1}{2\sqrt{3}}\,a\,\hat{\mathbf{y}} + \frac{1}{3}\,c\,\hat{\mathbf{z}}$$

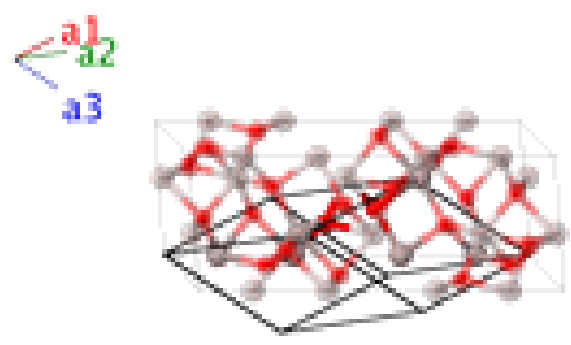

**Basis vectors:**

| | | Lattice Coordinates | | Cartesian Coordinates | Wyckoff Position | Atom Type |
|---|---|---|---|---|---|---|
| $\mathbf{B}_1$ | = | $x_1\,\mathbf{a}_1 + x_1\,\mathbf{a}_2 + x_1\,\mathbf{a}_3$ | = | $x_1 c\,\hat{\mathbf{z}}$ | (4$c$) | Al |
| $\mathbf{B}_2$ | = | $\left(\frac{1}{2} - x_1\right)\mathbf{a}_1 + \left(\frac{1}{2} - x_1\right)\mathbf{a}_2 + \left(\frac{1}{2} - x_1\right)\mathbf{a}_3$ | = | $\left(\frac{1}{2} - x_1\right) c\,\hat{\mathbf{z}}$ | (4$c$) | Al |
| $\mathbf{B}_3$ | = | $-x_1\,\mathbf{a}_1 - x_1\,\mathbf{a}_2 - x_1\,\mathbf{a}_3$ | = | $-x_1 c\,\hat{\mathbf{z}}$ | (4$c$) | Al |
| $\mathbf{B_4}$ | = | $\left(\frac{1}{2}+x_1\right)\mathbf{a_1}+\left(\frac{1}{2}+x_1\right)\mathbf{a_2}+\left(\frac{1}{2}+x_1\right)\mathbf{a_3}$ | = | $\left(\frac{1}{2}+x_1\right)c\,\mathbf{\hat{z}}$ | (4$c$) | Al |
| $\mathbf{B_5}$ | = | $x_2\,\mathbf{a_1}+\left(\frac{1}{2}-x_2\right)\mathbf{a_2}+\frac{1}{4}\,\mathbf{a_3}$ | = | $\frac{1}{8}\left(4\,x_2-1\right)a\,\mathbf{\hat{x}}+\frac{\sqrt{3}}{8}\left(1-4x_2\right)a\mathbf{\hat{y}}+\frac{1}{4}\,c\,\mathbf{\hat{z}}$ | (6$e$) | O |
| $\mathbf{B_6}$ | = | $\frac{1}{4}\,\mathbf{a_1}+x_2\,\mathbf{a_2}+\left(\frac{1}{2}-x_2\right)\mathbf{a_3}$ | = | $\frac{1}{8}\left(4x_2-1\right)a\,\mathbf{\hat{x}}-\frac{\sqrt{3}}{8}\left(1-4x_2\right)a\mathbf{\hat{y}}+\frac{1}{4}\,c\,\mathbf{\hat{z}}$ | (6$e$) | O |
| $\mathbf{B_7}$ | = | $\left(\frac{1}{2}-x_2\right)\mathbf{a_1}+\frac{1}{4}\,\mathbf{a_2}+x_2\,\mathbf{a_3}$ | = | $-\frac{1}{4}\left(4x_2-1\right)a\,\mathbf{\hat{x}}+\frac{1}{4}\,c\,\mathbf{\hat{z}}$ | (6$e$) | O |
| $\mathbf{B_8}$ | = | $-x_2\,\mathbf{a_1}+\left(\frac{1}{2}+x_2\right)\mathbf{a_2}+\frac{3}{4}\,\mathbf{a_3}$ | = | $-\frac{1}{8}\left(4x_2+3\right)a\,\mathbf{\hat{x}}+\frac{1}{8\sqrt{3}}\left(1+12x_2\right)a\mathbf{\hat{y}}+\frac{5}{12}\,c\,\mathbf{\hat{z}}$ | (6$e$) | O |
| $\mathbf{B_9}$ | = | $\frac{3}{4}\,\mathbf{a_1}-x_2\,\mathbf{a_2}+\left(\frac{1}{2}+x_2\right)\mathbf{a_3}$ | = | $-\frac{1}{8}\left(4x_2-1\right)a\,\mathbf{\hat{x}}-\frac{1}{8\sqrt{3}}\left(5+12x_2\right)a\mathbf{\hat{y}}+\frac{5}{12}\,c\,\mathbf{\hat{z}}$ | (6$e$) | O |
| $\mathbf{B_{10}}$ | = | $\left(\frac{1}{2}+x_2\right)\mathbf{a_1}+\frac{3}{4}\,\mathbf{a_2}-x_2\,\mathbf{a_3}$ | = | $\frac{1}{4}\left(4x_2+1\right)a\,\mathbf{\hat{x}}+\frac{1}{2\sqrt{3}}\,a\mathbf{\hat{y}}+\frac{5}{12}\,c\,\mathbf{\hat{z}}$ | (6$e$) | O |

---

---

**Geometry files:**

- CIF: pp. 732
- POSCAR: pp. 733

# $Mg_2Ni$ ($C_a$) Structure: A2B_hP18_180_fi_bd

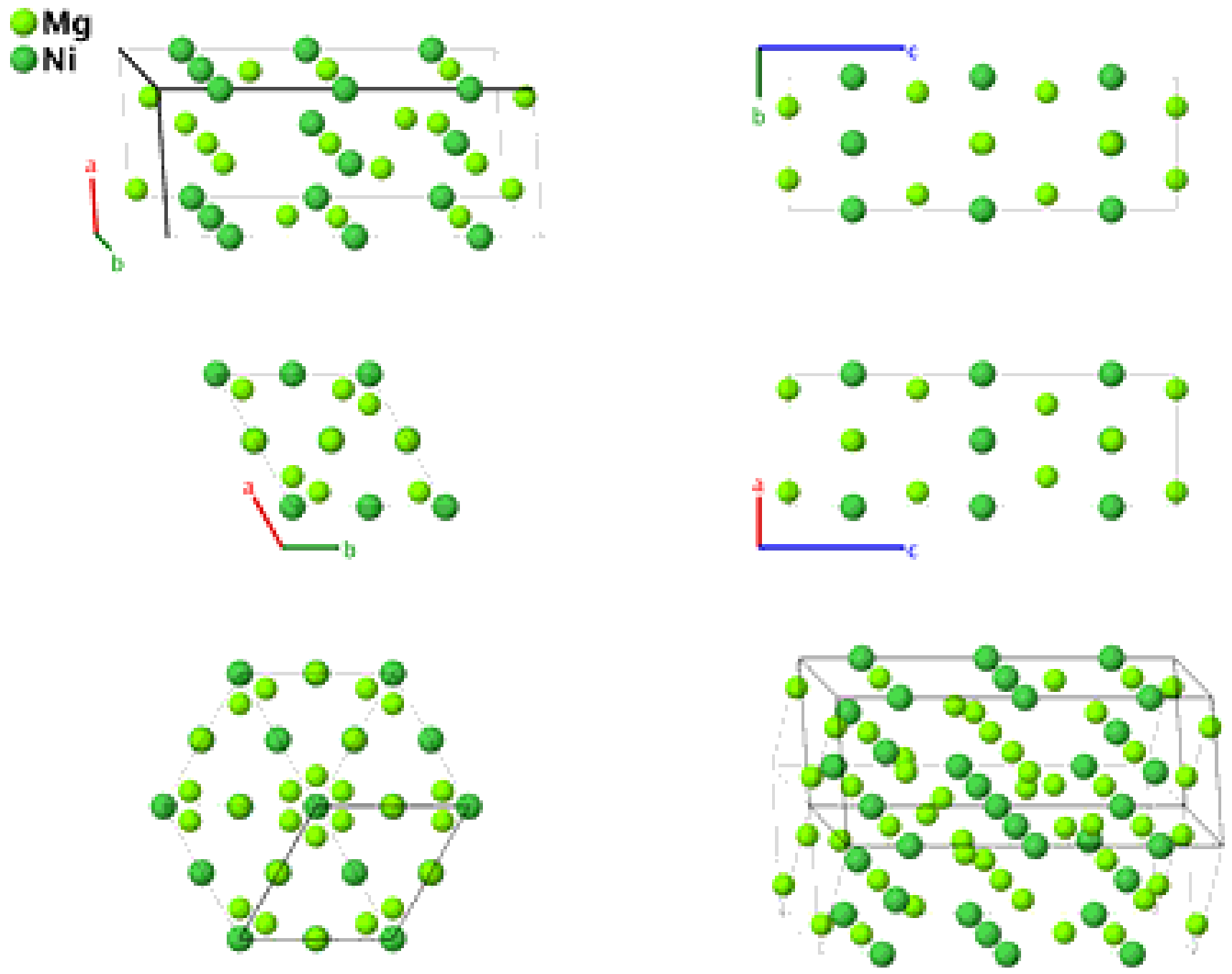

| | | |
|---|---|---|
| **Prototype** | : | $Mg_2Ni$ |
| **AFLOW prototype label** | : | A2B_hP18_180_fi_bd |
| ***Strukturbericht* designation** | : | $C_a$ |
| **Pearson symbol** | : | hP18 |
| **Space group number** | : | 180 |
| **Space group symbol** | : | $P6_222$ |
| **AFLOW prototype command** | : | `aflow --proto=A2B_hP18_180_fi_bd --params=`$a, c/a, z_3, x_4$ |

**Other compounds with this structure:**

- $CuMg_4Ni$

**Hexagonal primitive vectors:**

$$
\begin{aligned}
\mathbf{a}_1 &= \tfrac{1}{2}\,a\,\mathbf{\hat{x}} - \tfrac{\sqrt{3}}{2}\,a\,\mathbf{\hat{y}} \\
\mathbf{a}_2 &= \tfrac{1}{2}\,a\,\mathbf{\hat{x}} + \tfrac{\sqrt{3}}{2}\,a\,\mathbf{\hat{y}} \\
\mathbf{a}_3 &= c\,\mathbf{\hat{z}}
\end{aligned}
$$

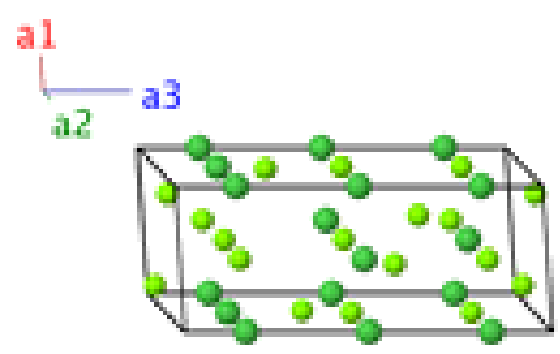

**Basis vectors:**

| | | Lattice Coordinates | | Cartesian Coordinates | Wyckoff Position | Atom Type |
|---|---|---|---|---|---|---|
| $\mathbf{B}_1$ | = | $\frac{1}{2}\,\mathbf{a}_3$ | = | $\frac{1}{2}\,c\,\mathbf{\hat{z}}$ | (3$b$) | Ni I |
| $\mathbf{B}_2$ | = | $\frac{1}{6}\,\mathbf{a}_3$ | = | $\frac{1}{6}\,c\,\mathbf{\hat{z}}$ | (3$b$) | Ni I |
| $\mathbf{B}_3$ | = | $\frac{5}{6}\,\mathbf{a}_3$ | = | $\frac{5}{6}\,c\,\mathbf{\hat{z}}$ | (3$b$) | Ni I |
| $\mathbf{B_4}$ | = | $\frac{1}{2}\,\mathbf{a_1} + \frac{1}{2}\,\mathbf{a_3}$ | = | $\frac{1}{4}\,a\,\hat{\mathbf{x}} - \frac{\sqrt{3}}{4}\,a\,\hat{\mathbf{y}} + \frac{1}{2}\,c\,\hat{\mathbf{z}}$ | (3*d*) | Ni II |
| $\mathbf{B_5}$ | = | $\frac{1}{2}\,\mathbf{a_2} + \frac{1}{6}\,\mathbf{a_3}$ | = | $\frac{1}{4}\,a\,\hat{\mathbf{x}} + \frac{\sqrt{3}}{4}\,a\,\hat{\mathbf{y}} + \frac{1}{6}\,c\,\hat{\mathbf{z}}$ | (3*d*) | Ni II |
| $\mathbf{B_6}$ | = | $\frac{1}{2}\,\mathbf{a_1} + \frac{1}{2}\,\mathbf{a_2} + \frac{5}{6}\,\mathbf{a_3}$ | = | $\frac{1}{2}\,a\,\hat{\mathbf{x}} + \frac{5}{6}\,c\,\hat{\mathbf{z}}$ | (3*d*) | Ni II |
| $\mathbf{B_7}$ | = | $\frac{1}{2}\,\mathbf{a_1} + z_3\,\mathbf{a_3}$ | = | $\frac{1}{4}\,a\,\hat{\mathbf{x}} - \frac{\sqrt{3}}{4}\,a\,\hat{\mathbf{y}} + z_3\,c\,\hat{\mathbf{z}}$ | (6*f*) | Mg I |
| $\mathbf{B_8}$ | = | $\frac{1}{2}\,\mathbf{a_2} + \left(\frac{2}{3} + z_3\right)\,\mathbf{a_3}$ | = | $\frac{1}{4}\,a\,\hat{\mathbf{x}} + \frac{\sqrt{3}}{4}\,a\,\hat{\mathbf{y}} + \left(\frac{2}{3} + z_3\right)\,c\,\hat{\mathbf{z}}$ | (6*f*) | Mg I |
| $\mathbf{B_9}$ | = | $\frac{1}{2}\,\mathbf{a_1} + \frac{1}{2}\,\mathbf{a_2} + \left(\frac{1}{3} + z_3\right)\,\mathbf{a_3}$ | = | $\frac{1}{2}\,a\,\hat{\mathbf{x}} + \left(\frac{1}{3} + z_3\right)\,c\,\hat{\mathbf{z}}$ | (6*f*) | Mg I |
| $\mathbf{B_{10}}$ | = | $\frac{1}{2}\,\mathbf{a_1} - z_3\,\mathbf{a_3}$ | = | $\frac{1}{4}\,a\,\hat{\mathbf{x}} - \frac{\sqrt{3}}{4}\,a\,\hat{\mathbf{y}} - z_3\,c\,\hat{\mathbf{z}}$ | (6*f*) | Mg I |
| $\mathbf{B_{11}}$ | = | $\frac{1}{2}\,\mathbf{a_2} + \left(\frac{2}{3} - z_3\right)\,\mathbf{a_3}$ | = | $\frac{1}{4}\,a\,\hat{\mathbf{x}} + \frac{\sqrt{3}}{4}\,a\,\hat{\mathbf{y}} + \left(\frac{2}{3} - z_3\right)\,c\,\hat{\mathbf{z}}$ | (6*f*) | Mg I |
| $\mathbf{B_{12}}$ | = | $\frac{1}{2}\,\mathbf{a_1} + \frac{1}{2}\,\mathbf{a_2} + \left(\frac{1}{3} - z_3\right)\,\mathbf{a_3}$ | = | $\frac{1}{2}\,a\,\hat{\mathbf{x}} + \left(\frac{1}{3} - z_3\right)\,c\,\hat{\mathbf{z}}$ | (6*f*) | Mg I |
| $\mathbf{B_{13}}$ | = | $x_4\,\mathbf{a_1} + 2x_4\,\mathbf{a_2}$ | = | $\frac{3}{2}\,x_4\,a\,\hat{\mathbf{x}} + \frac{\sqrt{3}}{2}\,x_4\,a\,\hat{\mathbf{y}}$ | (6*i*) | Mg II |
| $\mathbf{B_{14}}$ | = | $-2x_4\,\mathbf{a_1} - x_4\,\mathbf{a_2} + \frac{2}{3}\,\mathbf{a_3}$ | = | $-\frac{3}{2}\,x_4\,a\,\hat{\mathbf{x}} + \frac{\sqrt{3}}{2}\,x_4\,a\,\hat{\mathbf{y}} + \frac{2}{3}\,c\,\hat{\mathbf{z}}$ | (6*i*) | Mg II |
| $\mathbf{B_{15}}$ | = | $x_4\,\mathbf{a_1} - x_4\,\mathbf{a_2} + \frac{1}{3}\,\mathbf{a_3}$ | = | $-\sqrt{3}\,x_4\,a\,\hat{\mathbf{y}} + \frac{1}{3}\,c\,\hat{\mathbf{z}}$ | (6*i*) | Mg II |
| $\mathbf{B_{16}}$ | = | $-x_4\,\mathbf{a_1} - 2x_4\,\mathbf{a_2}$ | = | $-\frac{3}{2}\,x_4\,a\,\hat{\mathbf{x}} - \frac{\sqrt{3}}{2}\,x_4\,a\,\hat{\mathbf{y}}$ | (6*i*) | Mg II |
| $\mathbf{B_{17}}$ | = | $2x_4\,\mathbf{a_1} + x_4\,\mathbf{a_2} + \frac{2}{3}\,\mathbf{a_3}$ | = | $\frac{3}{2}\,x_4\,a\,\hat{\mathbf{x}} - \frac{\sqrt{3}}{2}\,x_4\,a\,\hat{\mathbf{y}} + \frac{2}{3}\,c\,\hat{\mathbf{z}}$ | (6*i*) | Mg II |
| $\mathbf{B_{18}}$ | = | $-x_4\,\mathbf{a_1} + x_4\,\mathbf{a_2} + \frac{1}{3}\,\mathbf{a_3}$ | = | $\sqrt{3}\,x_4\,a\,\hat{\mathbf{y}} + \frac{1}{3}\,c\,\hat{\mathbf{z}}$ | (6*i*) | Mg II |

**Found in:**

**Geometry files:**

- CIF: pp. 733
- POSCAR: pp. 733

# $CrSi_2$ (C40) Structure: AB2_hP9_180_d_j

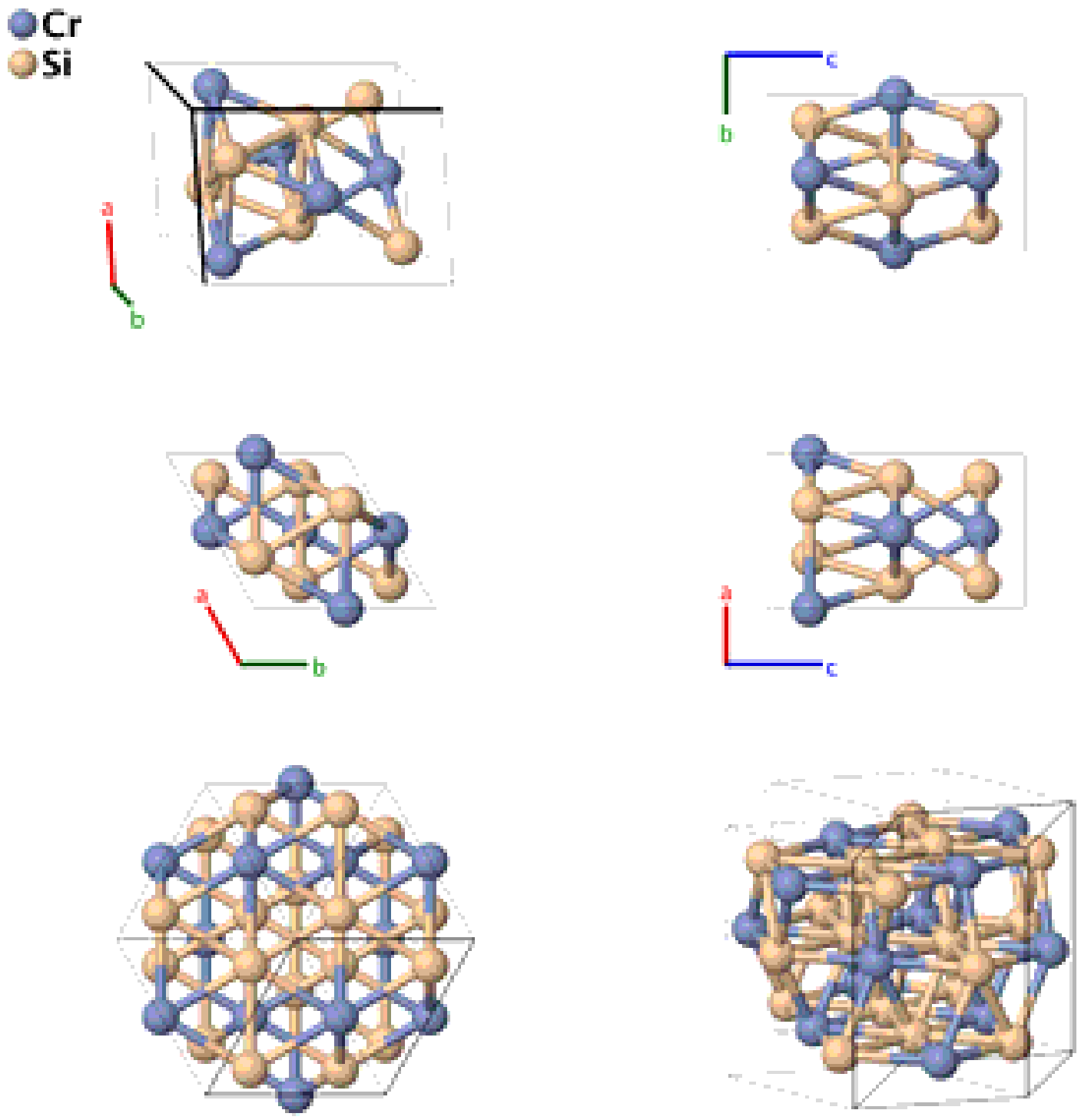

**Prototype** : $CrSi_2$

**AFLOW prototype label** : AB2_hP9_180_d_j

***Strukturbericht* designation** : C40

**Pearson symbol** : hP9

**Space group number** : 180

**Space group symbol** : $P6_222$

**AFLOW prototype command** : `aflow --proto=AB2_hP9_180_d_j --params=`$a, c/a, x_2$

---

**Other compounds with this structure:**

- $Ge_2Ta$, $Ge_2V$, $HfSn_2$, $Ge_2Nb$, $MoSi_2$, $Si_2Ta$, $Si_2V$, $Si_2W$

---

**Hexagonal primitive vectors:**

$$\mathbf{a}_1 = \frac{1}{2}\,a\,\hat{\mathbf{x}} - \frac{\sqrt{3}}{2}\,a\,\hat{\mathbf{y}}$$
$$\mathbf{a}_2 = \frac{1}{2}\,a\,\hat{\mathbf{x}} + \frac{\sqrt{3}}{2}\,a\,\hat{\mathbf{y}}$$
$$\mathbf{a}_3 = c\,\hat{\mathbf{z}}$$

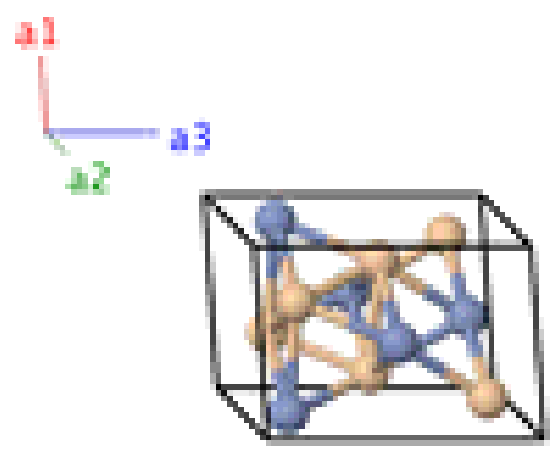

---

**Basis vectors:**

| | | Lattice Coordinates | | Cartesian Coordinates | Wyckoff Position | Atom Type |
|---|---|---|---|---|---|---|
| $\mathbf{B}_1$ | = | $\frac{1}{2}\,\mathbf{a}_1 + \frac{1}{2}\,\mathbf{a}_3$ | = | $\frac{1}{4}\,a\,\hat{\mathbf{x}} - \frac{\sqrt{3}}{4}\,a\,\hat{\mathbf{y}} + \frac{1}{2}\,c\,\hat{\mathbf{z}}$ | (3$d$) | Cr |
| $\mathbf{B_2}$ | = | $\frac{1}{2}\,\mathbf{a_2} + \frac{1}{6}\,\mathbf{a_3}$ | = | $\frac{1}{4}\,a\,\mathbf{\hat{x}} + \frac{\sqrt{3}}{4}\,a\,\mathbf{\hat{y}} + \frac{1}{6}\,c\,\mathbf{\hat{z}}$ | $(3d)$ | Cr |
| $\mathbf{B_3}$ | = | $\frac{1}{2}\,\mathbf{a_1} + \frac{1}{2}\,\mathbf{a_2} + \frac{5}{6}\,\mathbf{a_3}$ | = | $\frac{1}{2}\,a\,\mathbf{\hat{x}} + \frac{5}{6}\,c\,\mathbf{\hat{z}}$ | $(3d)$ | Cr |
| $\mathbf{B_4}$ | = | $x_2\,\mathbf{a_1} + 2x_2\,\mathbf{a_2} + \frac{1}{2}\,\mathbf{a_3}$ | = | $\frac{3}{2}\,x_2\,a\,\mathbf{\hat{x}} + \frac{\sqrt{3}}{2}\,x_2\,a\,\mathbf{\hat{y}} + \frac{1}{2}\,c\,\mathbf{\hat{z}}$ | $(6j)$ | Si |
| $\mathbf{B_5}$ | = | $-2x_2\,\mathbf{a_1} - x_2\,\mathbf{a_2} + \frac{1}{6}\,\mathbf{a_3}$ | = | $-\frac{3}{2}\,x_2\,a\,\mathbf{\hat{x}} + \frac{\sqrt{3}}{2}\,x_2\,a\,\mathbf{\hat{y}} + \frac{1}{6}\,c\,\mathbf{\hat{z}}$ | $(6j)$ | Si |
| $\mathbf{B_6}$ | = | $x_2\,\mathbf{a_1} - x_2\,\mathbf{a_2} + \frac{5}{6}\,\mathbf{a_3}$ | = | $-\sqrt{3}\,x_2\,a\,\mathbf{\hat{y}} + \frac{5}{6}\,c\,\mathbf{\hat{z}}$ | $(6j)$ | Si |
| $\mathbf{B_7}$ | = | $-x_2\,\mathbf{a_1} - 2x_2\,\mathbf{a_2} + \frac{1}{2}\,\mathbf{a_3}$ | = | $-\frac{3}{2}\,x_2\,a\,\mathbf{\hat{x}} - \frac{\sqrt{3}}{2}\,x_2\,a\,\mathbf{\hat{y}} + \frac{1}{2}\,c\,\mathbf{\hat{z}}$ | $(6j)$ | Si |
| $\mathbf{B_8}$ | = | $2x_2\,\mathbf{a_1} + x_2\,\mathbf{a_2} + \frac{1}{6}\,\mathbf{a_3}$ | = | $\frac{3}{2}\,x_2\,a\,\mathbf{\hat{x}} - \frac{\sqrt{3}}{2}\,x_2\,a\,\mathbf{\hat{y}} + \frac{1}{6}\,c\,\mathbf{\hat{z}}$ | $(6j)$ | Si |
| $\mathbf{B_9}$ | = | $-x_2\,\mathbf{a_1} + x_2\,\mathbf{a_2} + \frac{5}{6}\,\mathbf{a_3}$ | = | $\sqrt{3}\,x_2\,a\,\mathbf{\hat{y}} + \frac{5}{6}\,c\,\mathbf{\hat{z}}$ | $(6j)$ | Si |

**Geometry files:**

- CIF: pp. 733
- POSCAR: pp. 734

# $\beta$-Quartz ($SiO_2$, C8): A2B_hP9_180_j_c

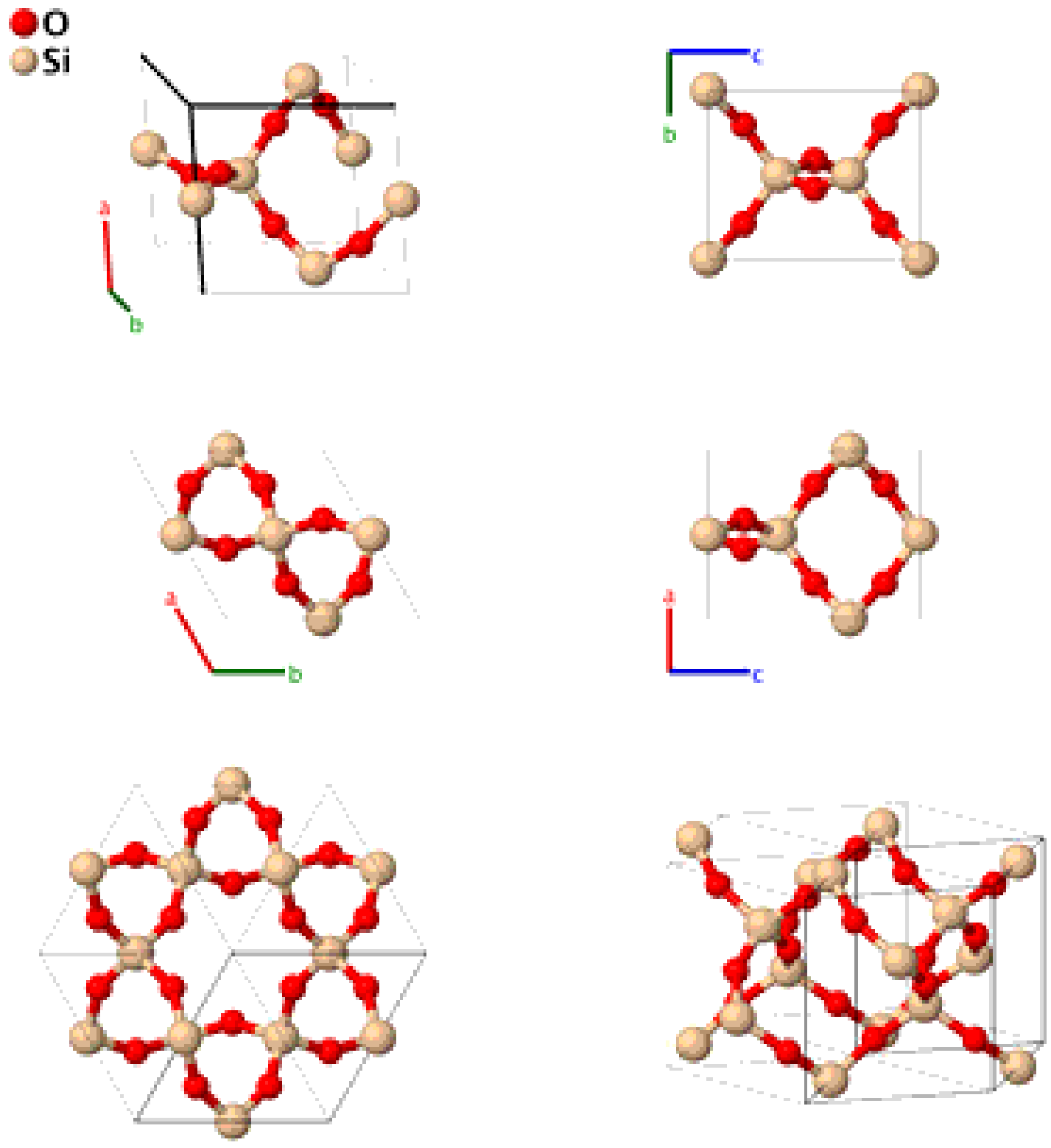

| | | |
|---|---|---|
| **Prototype** | : | $SiO_2$ |
| **AFLOW prototype label** | : | A2B_hP9_180_j_c |
| ***Strukturbericht* designation** | : | C8 |
| **Pearson symbol** | : | hP9 |
| **Space group number** | : | 180 |
| **Space group symbol** | : | $P6_222$ |
| **AFLOW prototype command** | : | `aflow --proto=A2B_hP9_180_j_c`<br>`--params=`$a, c/a, x_2$ |

- This is the high-temperature phase of $\alpha$-quartz.

**Hexagonal primitive vectors:**

$$
\begin{aligned}
\mathbf{a}_1 &= \tfrac{1}{2}\, a\,\hat{\mathbf{x}} - \tfrac{\sqrt{3}}{2}\, a\,\hat{\mathbf{y}} \\
\mathbf{a}_2 &= \tfrac{1}{2}\, a\,\hat{\mathbf{x}} + \tfrac{\sqrt{3}}{2}\, a\,\hat{\mathbf{y}} \\
\mathbf{a}_3 &= c\,\hat{\mathbf{z}}
\end{aligned}
$$

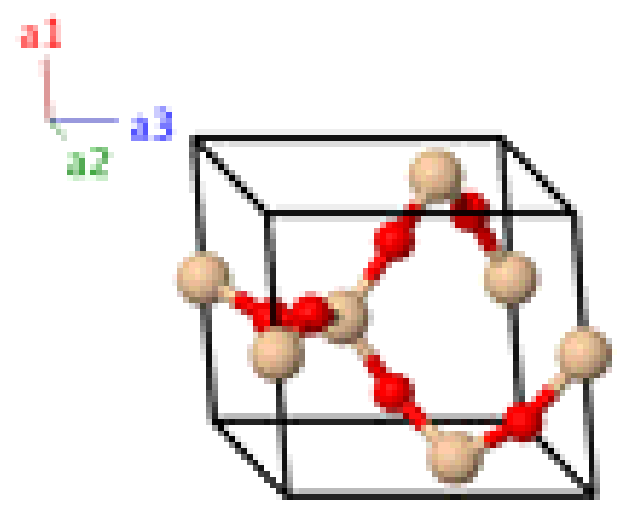

**Basis vectors:**

| Lattice Coordinates | Cartesian Coordinates | Wyckoff Position | Atom Type |
|---|---|---|---|

| | | | | | | |
|---|---|---|---|---|---|---|
| $\mathbf{B_1}$ | = | $\frac{1}{2}\,\mathbf{a_1}$ | = | $\frac{1}{4}\,a\,\mathbf{\hat{x}} - \frac{\sqrt{3}}{4}\,a\,\mathbf{\hat{y}}$ | (3$c$) | Si |
| $\mathbf{B_2}$ | = | $\frac{1}{2}\,\mathbf{a_2} + \frac{2}{3}\,\mathbf{a_3}$ | = | $\frac{1}{4}\,a\,\mathbf{\hat{x}} + \frac{\sqrt{3}}{4}\,a\,\mathbf{\hat{y}} + \frac{2}{3}\,c\,\mathbf{\hat{z}}$ | (3$c$) | Si |
| $\mathbf{B_3}$ | = | $\frac{1}{2}\,\mathbf{a_1} + \frac{1}{2}\,\mathbf{a_2} + \frac{1}{3}\,\mathbf{a_3}$ | = | $\frac{1}{2}\,a\,\mathbf{\hat{x}} + \frac{1}{3}\,c\,\mathbf{\hat{z}}$ | (3$c$) | Si |
| $\mathbf{B_4}$ | = | $x_2\,\mathbf{a_1} + 2x_2\,\mathbf{a_2} + \frac{1}{2}\,\mathbf{a_3}$ | = | $\frac{3}{2}\,x_2\,a\,\mathbf{\hat{x}} + \frac{\sqrt{3}}{2}\,x_2\,a\,\mathbf{\hat{y}} + \frac{1}{2}\,c\,\mathbf{\hat{z}}$ | (6$j$) | O |
| $\mathbf{B_5}$ | = | $-2x_2\,\mathbf{a_1} - x_2\,\mathbf{a_2} + \frac{1}{6}\,\mathbf{a_3}$ | = | $-\frac{3}{2}\,x_2\,a\,\mathbf{\hat{x}} + \frac{\sqrt{3}}{2}\,x_2\,a\,\mathbf{\hat{y}} + \frac{1}{6}\,c\,\mathbf{\hat{z}}$ | (6$j$) | O |
| $\mathbf{B_6}$ | = | $x_2\,\mathbf{a_1} - x_2\,\mathbf{a_2} + \frac{5}{6}\,\mathbf{a_3}$ | = | $-\sqrt{3}\,x_2\,a\,\mathbf{\hat{y}} + \frac{5}{6}\,c\,\mathbf{\hat{z}}$ | (6$j$) | O |
| $\mathbf{B_7}$ | = | $-x_2\,\mathbf{a_1} - 2x_2\,\mathbf{a_2} + \frac{1}{2}\,\mathbf{a_3}$ | = | $-\frac{3}{2}\,x_2\,a\,\mathbf{\hat{x}} - \frac{\sqrt{3}}{2}\,x_2\,a\,\mathbf{\hat{y}} + \frac{1}{2}\,c\,\mathbf{\hat{z}}$ | (6$j$) | O |
| $\mathbf{B_8}$ | = | $2x_2\,\mathbf{a_1} + x_2\,\mathbf{a_2} + \frac{1}{6}\,\mathbf{a_3}$ | = | $\frac{3}{2}\,x_2\,a\,\mathbf{\hat{x}} - \frac{\sqrt{3}}{2}\,x_2\,a\,\mathbf{\hat{y}} + \frac{1}{6}\,c\,\mathbf{\hat{z}}$ | (6$j$) | O |
| $\mathbf{B_9}$ | = | $-x_2\,\mathbf{a_1} + x_2\,\mathbf{a_2} + \frac{5}{6}\,\mathbf{a_3}$ | = | $\sqrt{3}\,x_2\,a\,\mathbf{\hat{y}} + \frac{5}{6}\,c\,\mathbf{\hat{z}}$ | (6$j$) | O |

**Geometry files:**

- CIF: pp. 734
- POSCAR: pp. 734

# Bainite ($Fe_3C$) Structure: AB3_hP8_182_c_g

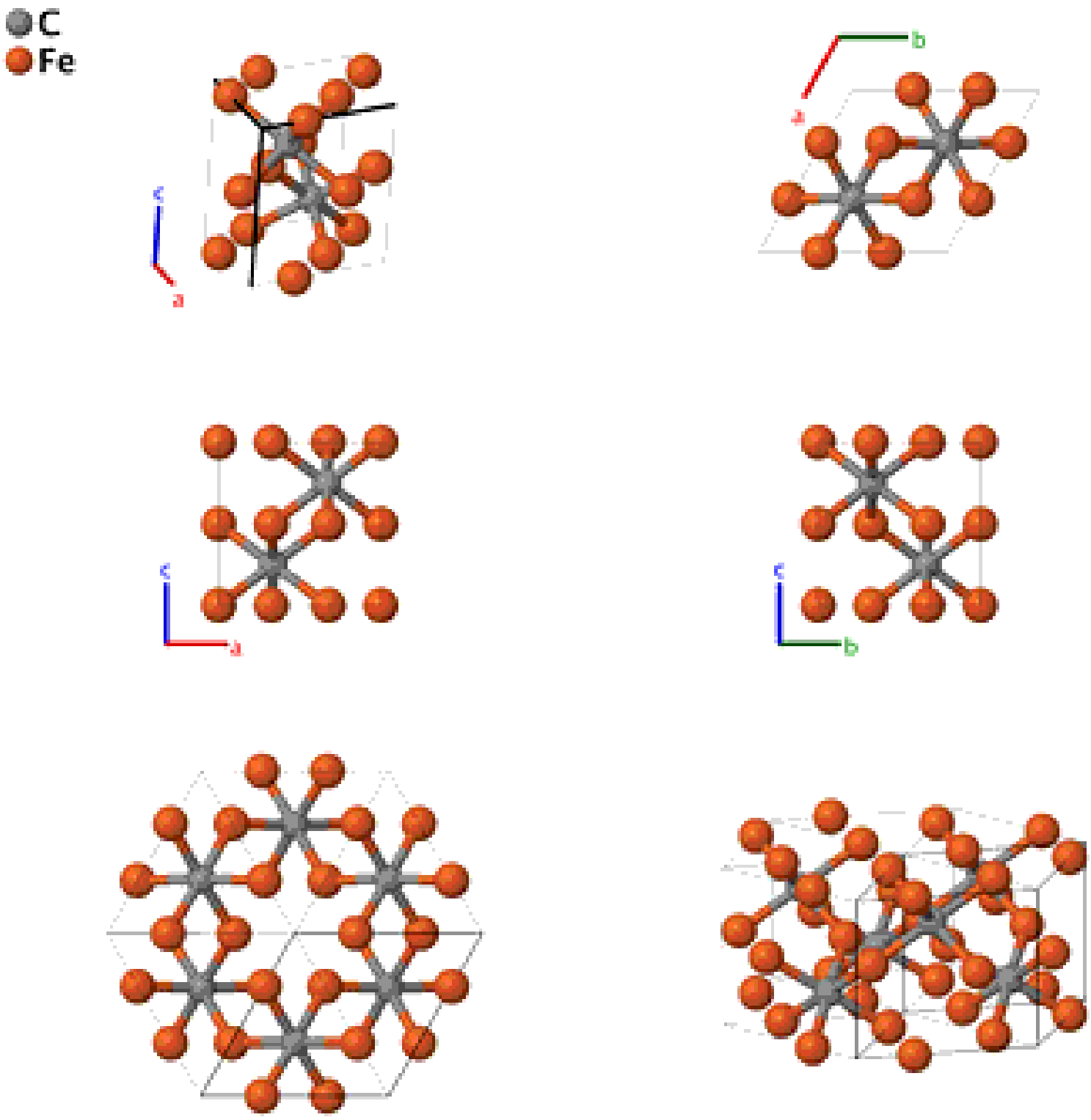

| | | |
|---|---|---|
| **Prototype** | : | $Fe_3C$ |
| **AFLOW prototype label** | : | AB3_hP8_182_c_g |
| ***Strukturbericht* designation** | : | None |
| **Pearson symbol** | : | hP8 |
| **Space group number** | : | 182 |
| **Space group symbol** | : | $P6_322$ |
| **AFLOW prototype command** | : | `aflow --proto=AB3_hP8_182_c_g`<br>`--params=`$a, c/a, x_2$ |

- Strictly speaking, bainite is a microstructure. However, (Villars, 1991) Vol. II, pp. 1894, refers to this crystal structure as upper bainite, and (Villars, 2014) refers to this as bainite.

**Hexagonal primitive vectors:**

$$\mathbf{a}_1 = \frac{1}{2}\,a\,\hat{\mathbf{x}} - \frac{\sqrt{3}}{2}\,a\,\hat{\mathbf{y}}$$
$$\mathbf{a}_2 = \frac{1}{2}\,a\,\hat{\mathbf{x}} + \frac{\sqrt{3}}{2}\,a\,\hat{\mathbf{y}}$$
$$\mathbf{a}_3 = c\,\hat{\mathbf{z}}$$

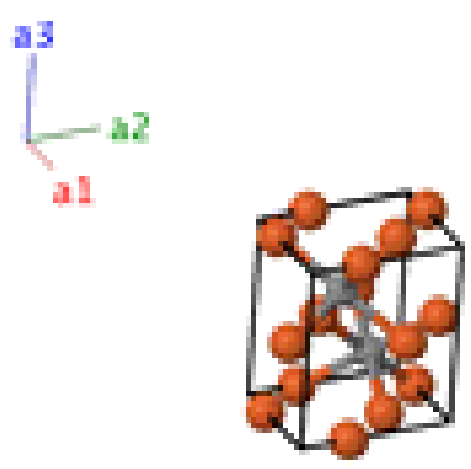

**Basis vectors:**

| Lattice Coordinates | Cartesian Coordinates | Wyckoff Position | Atom Type |
|---|---|---|---|

| | | | | | | |
|---|---|---|---|---|---|---|
| $\mathbf{B_1}$ | = | $\frac{1}{3}\,\mathbf{a_1} + \frac{2}{3}\,\mathbf{a_2} + \frac{1}{4}\,\mathbf{a_3}$ | = | $\frac{1}{2}\,a\,\hat{\mathbf{x}} + \frac{1}{2\sqrt{3}}\,a\,\hat{\mathbf{y}} + \frac{1}{4}\,c\,\hat{\mathbf{z}}$ | (2$c$) | C |
| $\mathbf{B_2}$ | = | $\frac{2}{3}\,\mathbf{a_1} + \frac{1}{3}\,\mathbf{a_2} + \frac{3}{4}\,\mathbf{a_3}$ | = | $\frac{1}{2}\,a\,\hat{\mathbf{x}} - \frac{1}{2\sqrt{3}}\,a\,\hat{\mathbf{y}} + \frac{3}{4}\,c\,\hat{\mathbf{z}}$ | (2$c$) | C |
| $\mathbf{B_3}$ | = | $x_2\,\mathbf{a_1}$ | = | $\frac{1}{2}\,x_2\,a\,\hat{\mathbf{x}} - \frac{\sqrt{3}}{2}\,x_2\,a\hat{\mathbf{y}}$ | (6$g$) | Fe |
| $\mathbf{B_4}$ | = | $x_2\,\mathbf{a_2}$ | = | $\frac{1}{2}\,x_2\,a\,\hat{\mathbf{x}} + \frac{\sqrt{3}}{2}\,x_2\,a\hat{\mathbf{y}}$ | (6$g$) | Fe |
| $\mathbf{B_5}$ | = | $-x_2\,\mathbf{a_1} - x_2\,\mathbf{a_2}$ | = | $-x_2\,a\,\hat{\mathbf{x}}$ | (6$g$) | Fe |
| $\mathbf{B_6}$ | = | $-x_2\,\mathbf{a_1} + \frac{1}{2}\,\mathbf{a_3}$ | = | $-\frac{1}{2}\,x_2\,a\,\hat{\mathbf{x}} + \frac{\sqrt{3}}{2}\,x_2\,a\hat{\mathbf{y}} + \frac{1}{2}\,c\,\hat{\mathbf{z}}$ | (6$g$) | Fe |
| $\mathbf{B_7}$ | = | $-x_2\,\mathbf{a_2} + \frac{1}{2}\,\mathbf{a_3}$ | = | $-\frac{1}{2}\,x_2\,a\,\hat{\mathbf{x}} - \frac{\sqrt{3}}{2}\,x_2\,a\hat{\mathbf{y}} + \frac{1}{2}\,c\,\hat{\mathbf{z}}$ | (6$g$) | Fe |
| $\mathbf{B_8}$ | = | $x_2\,\mathbf{a_1} + x_2\,\mathbf{a_2} + \frac{1}{2}\,\mathbf{a_3}$ | = | $+x_2\,a\,\hat{\mathbf{x}} + \frac{1}{2}\,c\,\hat{\mathbf{z}}$ | (6$g$) | Fe |

**Found in:**

**Geometry files:**

- CIF: pp. 734
- POSCAR: pp. 734

# Buckled Graphite Structure: A_hP4_186_ab

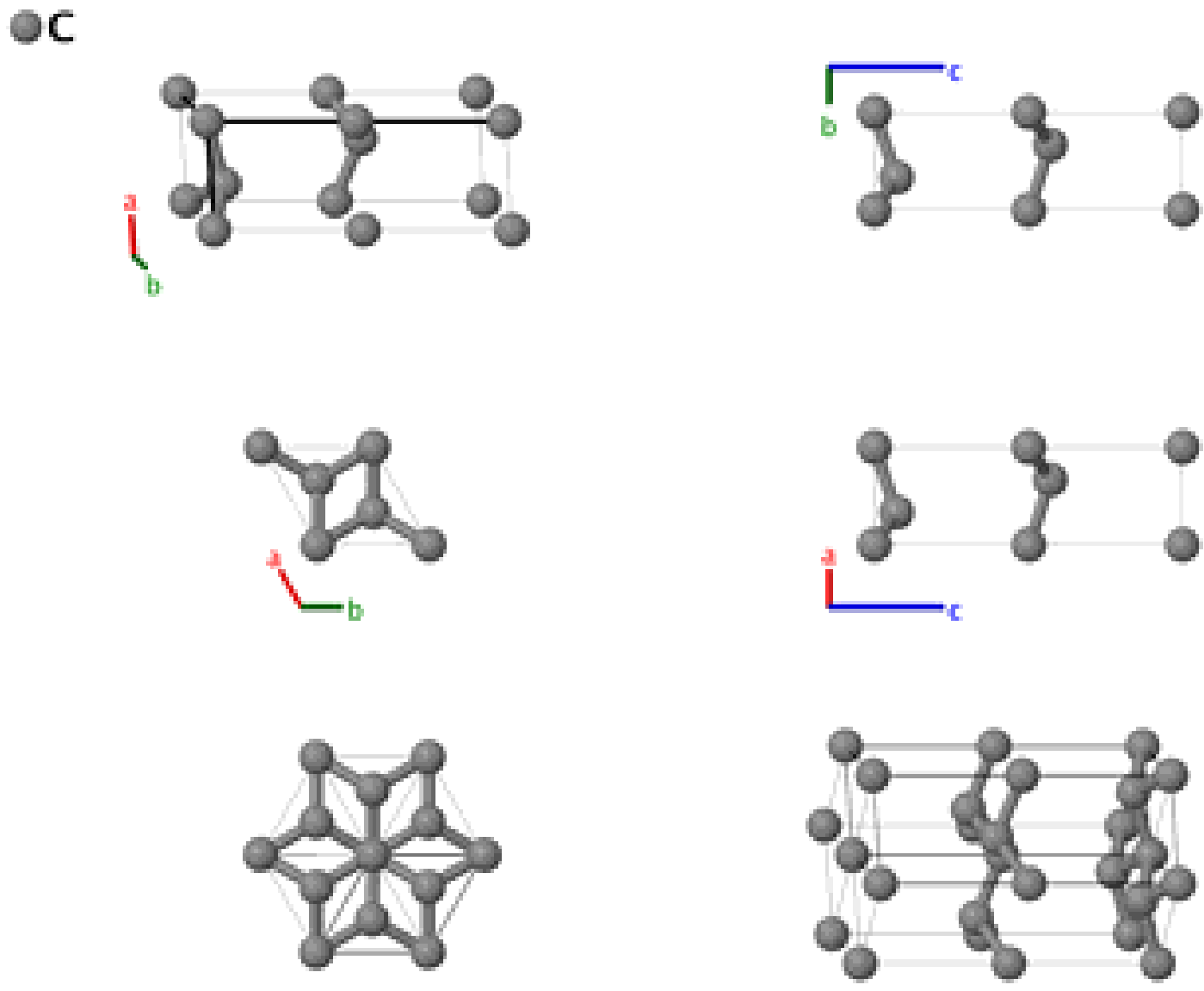

| | | |
|---|---|---|
| **Prototype** | : | C |
| **AFLOW prototype label** | : | A_hP4_186_ab |
| ***Strukturbericht* designation** | : | None |
| **Pearson symbol** | : | hP4 |
| **Space group number** | : | 186 |
| **Space group symbol** | : | $P6_3mc$ |
| **AFLOW prototype command** | : | `aflow --proto=A_hP4_186_ab` `--params=`$a, c/a, z_1, z_2$ |

- According to (Wyckoff, 1963), hexagonal graphite may be either flat, space group $P6_3/mmc$ (#194) or buckled, space group $P6_3mc$ (#186). "If it is buckled, the buckling parameter is small, less than 1/20 of the 'c' parameter of the hexagonal unit cell." We will assign the A9 *Strukturbericht* designation to the unbuckled structure. Experimentally, a rhombohedral ($R\bar{3}m$) graphite structure is also observed. In the pictures above we give $z_2$ the exaggerated value of 0.1 When $z_2 = 0$, this structure is equivalent to unbuckled (A9) hexagonal graphite.

**Hexagonal primitive vectors:**

$$\mathbf{a}_1 = \frac{1}{2}\,a\,\hat{\mathbf{x}} - \frac{\sqrt{3}}{2}\,a\,\hat{\mathbf{y}}$$
$$\mathbf{a}_2 = \frac{1}{2}\,a\,\hat{\mathbf{x}} + \frac{\sqrt{3}}{2}\,a\,\hat{\mathbf{y}}$$
$$\mathbf{a}_3 = c\,\hat{\mathbf{z}}$$

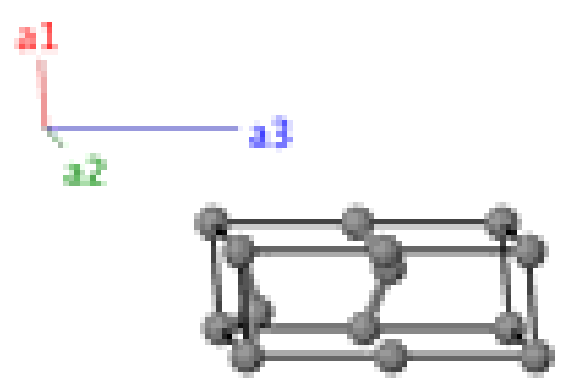

**Basis vectors:**

| | | Lattice Coordinates | | Cartesian Coordinates | Wyckoff Position | Atom Type |
|---|---|---|---|---|---|---|
| $\mathbf{B}_1$ | = | $z_1\,\mathbf{a}_3$ | = | $z_1\,c\,\hat{\mathbf{z}}$ | (2$a$) | C I |
| $\mathbf{B}_2$ | = | $\left(\frac{1}{2} + z_1\right)\,\mathbf{a}_3$ | = | $\left(\frac{1}{2} + z_1\right)\,c\,\hat{\mathbf{z}}$ | (2$a$) | C I |
| $\mathbf{B_3}$ | = | $\frac{1}{3}\,\mathbf{a_1} + \frac{2}{3}\,\mathbf{a_2} + z_2\,\mathbf{a_3}$ | = | $\frac{1}{2}\,a\,\mathbf{\hat{x}} + \frac{1}{2\sqrt{3}}\,a\,\mathbf{\hat{y}} + z_2\,c\,\mathbf{\hat{z}}$ | (2$b$) | C II |
| $\mathbf{B_4}$ | = | $\frac{2}{3}\,\mathbf{a_1} + \frac{1}{3}\,\mathbf{a_2} + \left(\frac{1}{2} + z_2\right)\,\mathbf{a_3}$ | = | $\frac{1}{2}\,a\,\mathbf{\hat{x}} - \frac{1}{2\sqrt{3}}\,a\,\mathbf{\hat{y}} + \left(\frac{1}{2} + z_2\right)\,c\,\mathbf{\hat{z}}$ | (2$b$) | C II |

**Found in:**

**Geometry files:**

- CIF: pp. 735
- POSCAR: pp. 735

# Moissanite-4H SiC (B5) Structure: AB_hP8_186_ab_ab

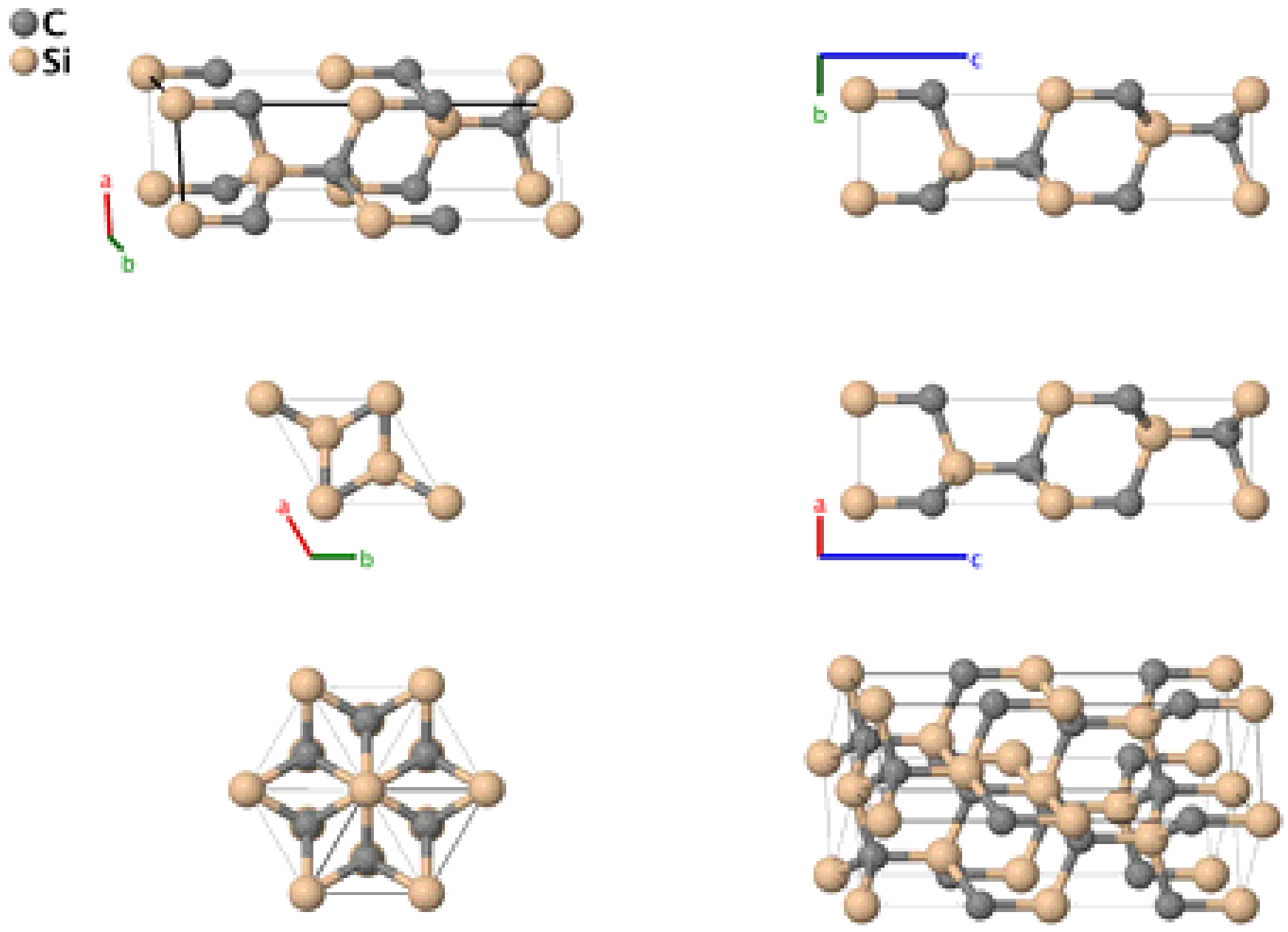

| | | |
|---|---|---|
| **Prototype** | : | SiC |
| **AFLOW prototype label** | : | AB_hP8_186_ab_ab |
| ***Strukturbericht* designation** | : | B5 |
| **Pearson symbol** | : | hP8 |
| **Space group number** | : | 186 |
| **Space group symbol** | : | $P6_3mc$ |
| **AFLOW prototype command** | : | `aflow --proto=AB_hP8_186_ab_ab`<br>`--params=`$a, c/a, z_1, z_2, z_3, z_4$ |

- This is one possible stacking (ABAC) for tetrahedral structures. Compare this to zincblende (ABCABC), wurtzite (ABABAB), 6H (ABCACB), and 9R (ABCBCACAB). The 4H refers to the fact that there are 4 CSi dimers in a hexagonal unit cell. Zincblende is denoted 3C, and wurtzite is 2H. This structure is related to the $\alpha$-La (A3') structure in the same way that zincblende (B3) is related to the fcc (A1) lattice. Without loss of generality, we can take any of the $z_i$ to be zero. In the pictures here we take $z_1 = 0$.

**Hexagonal primitive vectors:**

$$\begin{aligned}
\mathbf{a}_1 &= \tfrac{1}{2}\,a\,\hat{\mathbf{x}} - \tfrac{\sqrt{3}}{2}\,a\,\hat{\mathbf{y}} \\
\mathbf{a}_2 &= \tfrac{1}{2}\,a\,\hat{\mathbf{x}} + \tfrac{\sqrt{3}}{2}\,a\,\hat{\mathbf{y}} \\
\mathbf{a}_3 &= c\,\hat{\mathbf{z}}
\end{aligned}$$

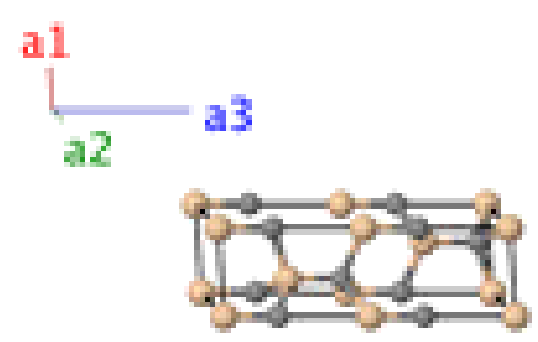

**Basis vectors:**

| | | Lattice Coordinates | | Cartesian Coordinates | Wyckoff Position | Atom Type |
|---|---|---|---|---|---|---|
| $\mathbf{B}_1$ | = | $z_1\,\mathbf{a}_3$ | = | $z_1\,c\,\hat{\mathbf{z}}$ | (2a) | C I |
| $\mathbf{B}_2$ | = | $\left(\frac{1}{2} + z_1\right)\mathbf{a}_3$ | = | $\left(\frac{1}{2} + z_1\right)c\,\hat{\mathbf{z}}$ | (2a) | C I |
| $\mathbf{B_3}$ | = | $z_2\,\mathbf{a_3}$ | = | $z_2\,c\,\hat{\mathbf{z}}$ | (2$a$) | Si I |
| $\mathbf{B_4}$ | = | $\left(\frac{1}{2}+z_2\right)\mathbf{a_3}$ | = | $\left(\frac{1}{2}+z_2\right)c\,\hat{\mathbf{z}}$ | (2$a$) | Si I |
| $\mathbf{B_5}$ | = | $\frac{1}{3}\,\mathbf{a_1}+\frac{2}{3}\,\mathbf{a_2}+z_3\,\mathbf{a_3}$ | = | $\frac{1}{2}\,a\,\hat{\mathbf{x}}+\frac{1}{2\sqrt{3}}\,a\,\hat{\mathbf{y}}+z_3\,c\,\hat{\mathbf{z}}$ | (2$b$) | C II |
| $\mathbf{B_6}$ | = | $\frac{2}{3}\,\mathbf{a_1}+\frac{1}{3}\,\mathbf{a_2}+\left(\frac{1}{2}+z_3\right)\mathbf{a_3}$ | = | $\frac{1}{2}\,a\,\hat{\mathbf{x}}-\frac{1}{2\sqrt{3}}\,a\,\hat{\mathbf{y}}+\left(\frac{1}{2}+z_3\right)c\,\hat{\mathbf{z}}$ | (2$b$) | C II |
| $\mathbf{B_7}$ | = | $\frac{1}{3}\,\mathbf{a_1}+\frac{2}{3}\,\mathbf{a_2}+z_4\,\mathbf{a_3}$ | = | $\frac{1}{2}\,a\,\hat{\mathbf{x}}+\frac{1}{2\sqrt{3}}\,a\,\hat{\mathbf{y}}+z_4\,c\,\hat{\mathbf{z}}$ | (2$b$) | Si II |
| $\mathbf{B_8}$ | = | $\frac{2}{3}\,\mathbf{a_1}+\frac{1}{3}\,\mathbf{a_2}+\left(\frac{1}{2}+z_4\right)\mathbf{a_3}$ | = | $\frac{1}{2}\,a\,\hat{\mathbf{x}}-\frac{1}{2\sqrt{3}}\,a\,\hat{\mathbf{y}}+\left(\frac{1}{2}+z_4\right)c\,\hat{\mathbf{z}}$ | (2$b$) | Si II |

**Geometry files:**

- CIF: pp. 735
- POSCAR: pp. 735

# Wurtzite (ZnS, B4) Structure: AB_hP4_186_b_b

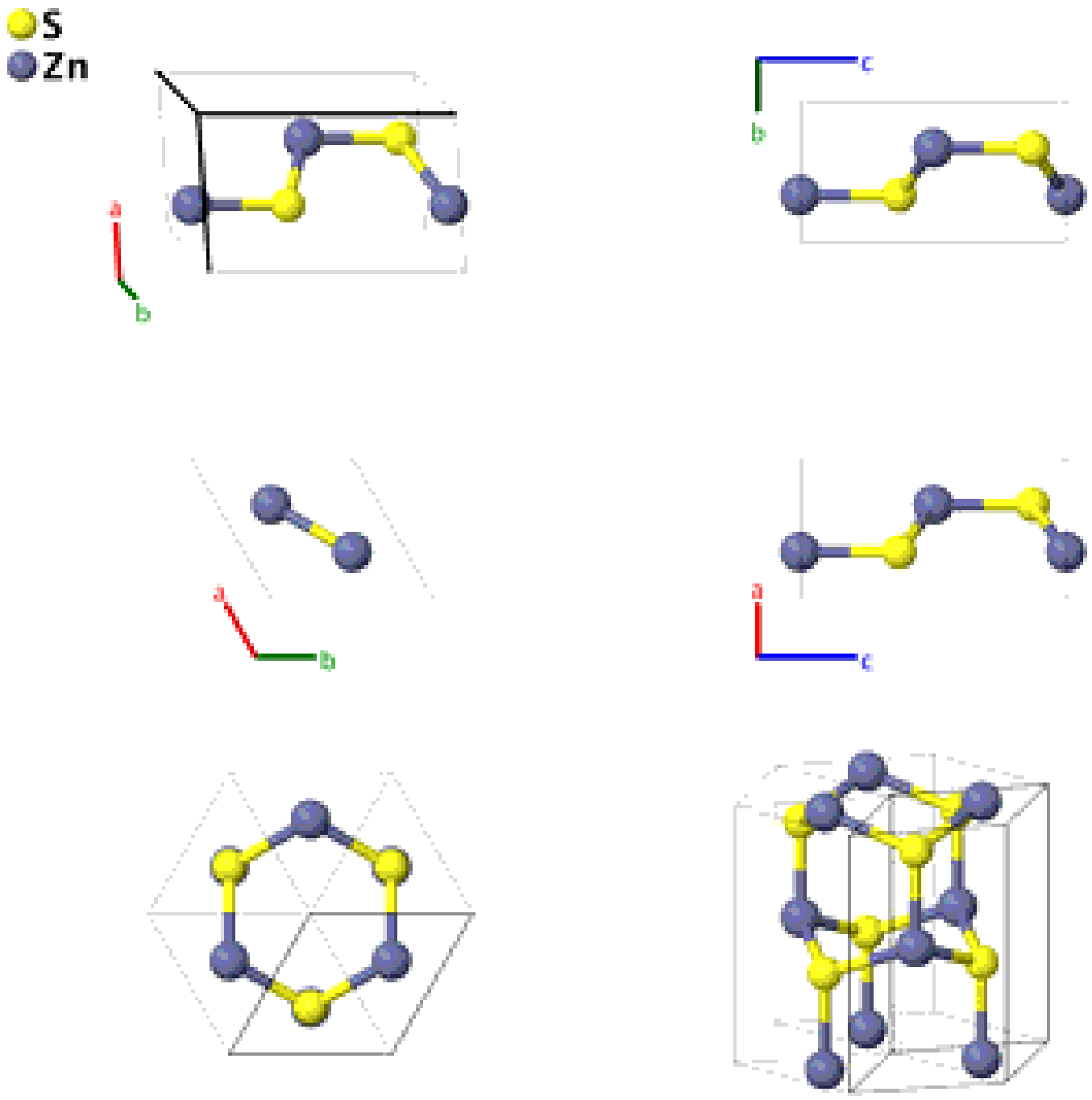

| | | |
|---|---|---|
| **Prototype** | : | ZnS |
| **AFLOW prototype label** | : | AB_hP4_186_b_b |
| ***Strukturbericht* designation** | : | B4 |
| **Pearson symbol** | : | hP4 |
| **Space group number** | : | 186 |
| **Space group symbol** | : | $P6_3mc$ |
| **AFLOW prototype command** | : | `aflow --proto=AB_hP4_186_b_b` `--params=`$a, c/a, z_1, z_2$ |

**Other compounds with this structure:**

- ZnO, SiC, AlN, CdSe, BN, C (hexagonal diamond)

- This is the hexagonal analog of the zincblende lattice, i.e. the stacking of the ZnS dimers is ABABAB... Replacing both the Zn and S atoms by C (or Si) gives the hexagonal diamond structure. The "ideal" structure, where the nearest-neighbor environment of each atom is the same as in zincblende, is achieved when we take $c/a = \sqrt{8/3}$ and $z_2 = 1/8$. In the extreme case $z_2 = 1/2$ this structure becomes the $B_k$ (BN) structure. Note that we have arbitrarily chosen the $z_1$ parameter for the zinc atoms to be zero.

**Hexagonal primitive vectors:**

$$
\begin{aligned}
\mathbf{a}_1 &= \tfrac{1}{2}\,a\,\hat{\mathbf{x}} - \tfrac{\sqrt{3}}{2}\,a\,\hat{\mathbf{y}} \\
\mathbf{a}_2 &= \tfrac{1}{2}\,a\,\hat{\mathbf{x}} + \tfrac{\sqrt{3}}{2}\,a\,\hat{\mathbf{y}} \\
\mathbf{a}_3 &= c\,\hat{\mathbf{z}}
\end{aligned}
$$

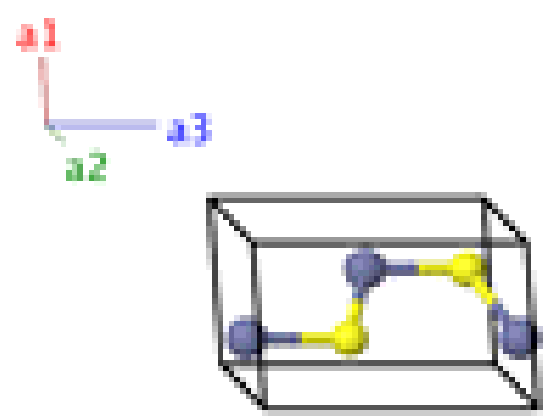

**Basis vectors:**

| | | Lattice Coordinates | | Cartesian Coordinates | Wyckoff Position | Atom Type |
|---|---|---|---|---|---|---|
| $\mathbf{B_1}$ | = | $\frac{1}{3}\,\mathbf{a_1} + \frac{2}{3}\,\mathbf{a_2} + z_1\,\mathbf{a_3}$ | = | $\frac{1}{2}\,a\,\hat{\mathbf{x}} + \frac{1}{2\sqrt{3}}\,a\,\hat{\mathbf{y}} + z_1\,c\,\hat{\mathbf{z}}$ | (2$b$) | S |
| $\mathbf{B_2}$ | = | $\frac{2}{3}\,\mathbf{a_1} + \frac{1}{3}\,\mathbf{a_2} + \left(\frac{1}{2} + z_1\right)\,\mathbf{a_3}$ | = | $\frac{1}{2}\,a\,\hat{\mathbf{x}} - \frac{1}{2\sqrt{3}}\,a\,\hat{\mathbf{y}} + \left(\frac{1}{2} + z_1\right)\,c\,\hat{\mathbf{z}}$ | (2$b$) | S |
| $\mathbf{B_3}$ | = | $\frac{1}{3}\,\mathbf{a_1} + \frac{2}{3}\,\mathbf{a_2} + z_2\,\mathbf{a_3}$ | = | $\frac{1}{2}\,a\,\hat{\mathbf{x}} + \frac{1}{2\sqrt{3}}\,a\,\hat{\mathbf{y}} + z_2\,c\,\hat{\mathbf{z}}$ | (2$b$) | Zn |
| $\mathbf{B_4}$ | = | $\frac{2}{3}\,\mathbf{a_1} + \frac{1}{3}\,\mathbf{a_2} + \left(\frac{1}{2} + z_2\right)\,\mathbf{a_3}$ | = | $\frac{1}{2}\,a\,\hat{\mathbf{x}} - \frac{1}{2\sqrt{3}}\,a\,\hat{\mathbf{y}} + \left(\frac{1}{2} + z_2\right)\,c\,\hat{\mathbf{z}}$ | (2$b$) | Zn |

**Found in:**

**Geometry files:**

- CIF: pp. 736
- POSCAR: pp. 736

# Moissanite-6H SiC (B6) Structure: AB_hP12_186_a2b_a2b

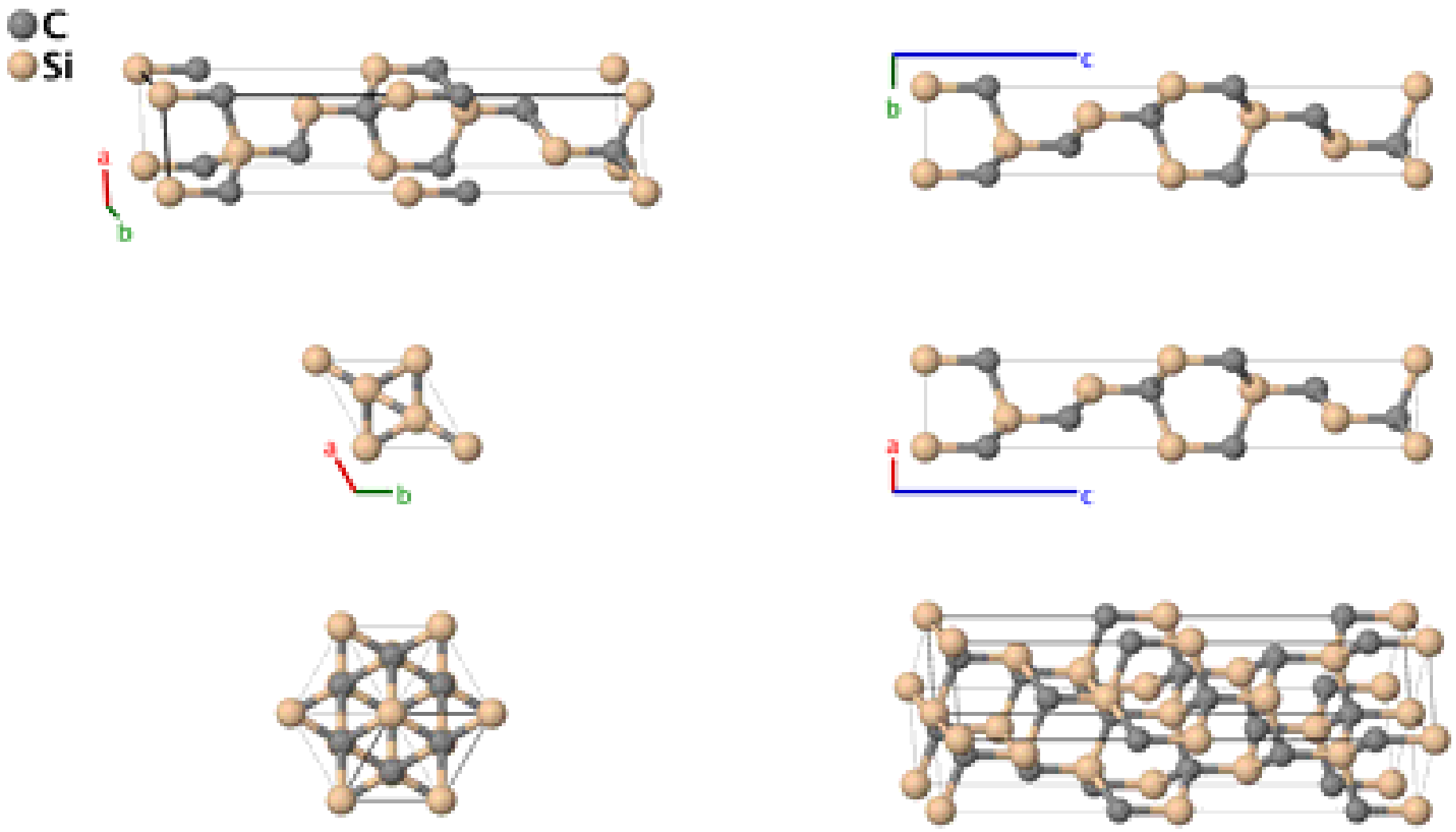

| | | |
|---|---|---|
| **Prototype** | : | SiC |
| **AFLOW prototype label** | : | AB_hP12_186_a2b_a2b |
| ***Strukturbericht* designation** | : | B6 |
| **Pearson symbol** | : | hP12 |
| **Space group number** | : | 186 |
| **Space group symbol** | : | $P6_3mc$ |
| **AFLOW prototype command** | : | `aflow --proto=AB_hP12_186_a2b_a2b` <br> `--params=`$a, c/a, z_1, z_2, z_3, z_4, z_5, z_6$ |

- This is an alternate stacking (ABCACB) for tetrahedral structures. Compare this to zincblende (ABCABC), moissanite-4H (ABAC), and wurtzite (ABABAB). The 6H refers to the fact that there are 6 CSi dimers in a hexagonal unit cell. Zincblende is denoted 3C, and wurtzite is 2H. Without loss of generality, we can take any of the $z_i$ to be zero. In the pictures here we take $z_1 = 0$.

**Hexagonal primitive vectors:**

$$\mathbf{a}_1 = \frac{1}{2}\,a\,\hat{\mathbf{x}} - \frac{\sqrt{3}}{2}\,a\,\hat{\mathbf{y}}$$
$$\mathbf{a}_2 = \frac{1}{2}\,a\,\hat{\mathbf{x}} + \frac{\sqrt{3}}{2}\,a\,\hat{\mathbf{y}}$$
$$\mathbf{a}_3 = c\,\hat{\mathbf{z}}$$

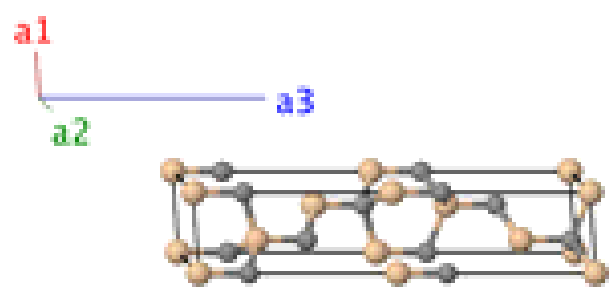

**Basis vectors:**

| | | Lattice Coordinates | | Cartesian Coordinates | Wyckoff Position | Atom Type |
|---|---|---|---|---|---|---|
| $\mathbf{B}_1$ | = | $z_1\,\mathbf{a}_3$ | = | $z_1\,c\,\hat{\mathbf{z}}$ | (2$a$) | C I |
| $\mathbf{B}_2$ | = | $\left(\frac{1}{2}+z_1\right)\mathbf{a}_3$ | = | $\left(\frac{1}{2}+z_1\right)c\,\hat{\mathbf{z}}$ | (2$a$) | C I |
| $\mathbf{B}_3$ | = | $z_2\,\mathbf{a}_3$ | = | $z_2\,c\,\hat{\mathbf{z}}$ | (2$a$) | Si I |
| $\mathbf{B}_4$ | = | $\left(\frac{1}{2}+z_2\right)\mathbf{a}_3$ | = | $\left(\frac{1}{2}+z_2\right)c\,\hat{\mathbf{z}}$ | (2$a$) | Si I |
| $\mathbf{B}_5$ | = | $\frac{1}{3}\,\mathbf{a}_1 + \frac{2}{3}\,\mathbf{a}_2 + z_3\,\mathbf{a}_3$ | = | $\frac{1}{2}\,a\,\hat{\mathbf{x}} + \frac{1}{2\sqrt{3}}\,a\,\hat{\mathbf{y}} + z_3\,c\,\hat{\mathbf{z}}$ | (2$b$) | C II |
| $\mathbf{B_6}$ | = | $\frac{2}{3}\,\mathbf{a_1} + \frac{1}{3}\,\mathbf{a_2} + \left(\frac{1}{2} + z_3\right)\,\mathbf{a_3}$ | = | $\frac{1}{2}\,a\,\hat{\mathbf{x}} - \frac{1}{2\sqrt{3}}\,a\,\hat{\mathbf{y}} + \left(\frac{1}{2} + z_3\right)\,c\,\hat{\mathbf{z}}$ | (2$b$) | C II |
| $\mathbf{B_7}$ | = | $\frac{1}{3}\,\mathbf{a_1} + \frac{2}{3}\,\mathbf{a_2} + z_4\,\mathbf{a_3}$ | = | $\frac{1}{2}\,a\,\hat{\mathbf{x}} + \frac{1}{2\sqrt{3}}\,a\,\hat{\mathbf{y}} + z_4\,c\,\hat{\mathbf{z}}$ | (2$b$) | C III |
| $\mathbf{B_8}$ | = | $\frac{2}{3}\,\mathbf{a_1} + \frac{1}{3}\,\mathbf{a_2} + \left(\frac{1}{2} + z_4\right)\,\mathbf{a_3}$ | = | $\frac{1}{2}\,a\,\hat{\mathbf{x}} - \frac{1}{2\sqrt{3}}\,a\,\hat{\mathbf{y}} + \left(\frac{1}{2} + z_4\right)\,c\,\hat{\mathbf{z}}$ | (2$b$) | C III |
| $\mathbf{B_9}$ | = | $\frac{1}{3}\,\mathbf{a_1} + \frac{2}{3}\,\mathbf{a_2} + z_5\,\mathbf{a_3}$ | = | $\frac{1}{2}\,a\,\hat{\mathbf{x}} + \frac{1}{2\sqrt{3}}\,a\,\hat{\mathbf{y}} + z_5\,c\,\hat{\mathbf{z}}$ | (2$b$) | Si II |
| $\mathbf{B_{10}}$ | = | $\frac{2}{3}\,\mathbf{a_1} + \frac{1}{3}\,\mathbf{a_2} + \left(\frac{1}{2} + z_5\right)\,\mathbf{a_3}$ | = | $\frac{1}{2}\,a\,\hat{\mathbf{x}} - \frac{1}{2\sqrt{3}}\,a\,\hat{\mathbf{y}} + \left(\frac{1}{2} + z_5\right)\,c\,\hat{\mathbf{z}}$ | (2$b$) | Si II |
| $\mathbf{B_{11}}$ | = | $\frac{1}{3}\,\mathbf{a_1} + \frac{2}{3}\,\mathbf{a_2} + z_6\,\mathbf{a_3}$ | = | $\frac{1}{2}\,a\,\hat{\mathbf{x}} + \frac{1}{2\sqrt{3}}\,a\,\hat{\mathbf{y}} + z_6\,c\,\hat{\mathbf{z}}$ | (2$b$) | Si III |
| $\mathbf{B_{12}}$ | = | $\frac{2}{3}\,\mathbf{a_1} + \frac{1}{3}\,\mathbf{a_2} + \left(\frac{1}{2} + z_6\right)\,\mathbf{a_3}$ | = | $\frac{1}{2}\,a\,\hat{\mathbf{x}} - \frac{1}{2\sqrt{3}}\,a\,\hat{\mathbf{y}} + \left(\frac{1}{2} + z_6\right)\,c\,\hat{\mathbf{z}}$ | (2$b$) | Si III |

**Geometry files:**

- CIF: pp. 736
- POSCAR: pp. 736

# $Al_5C_3N$ ($E9_4$) Structure: A5B3C_hP18_186_2a3b_2ab_b

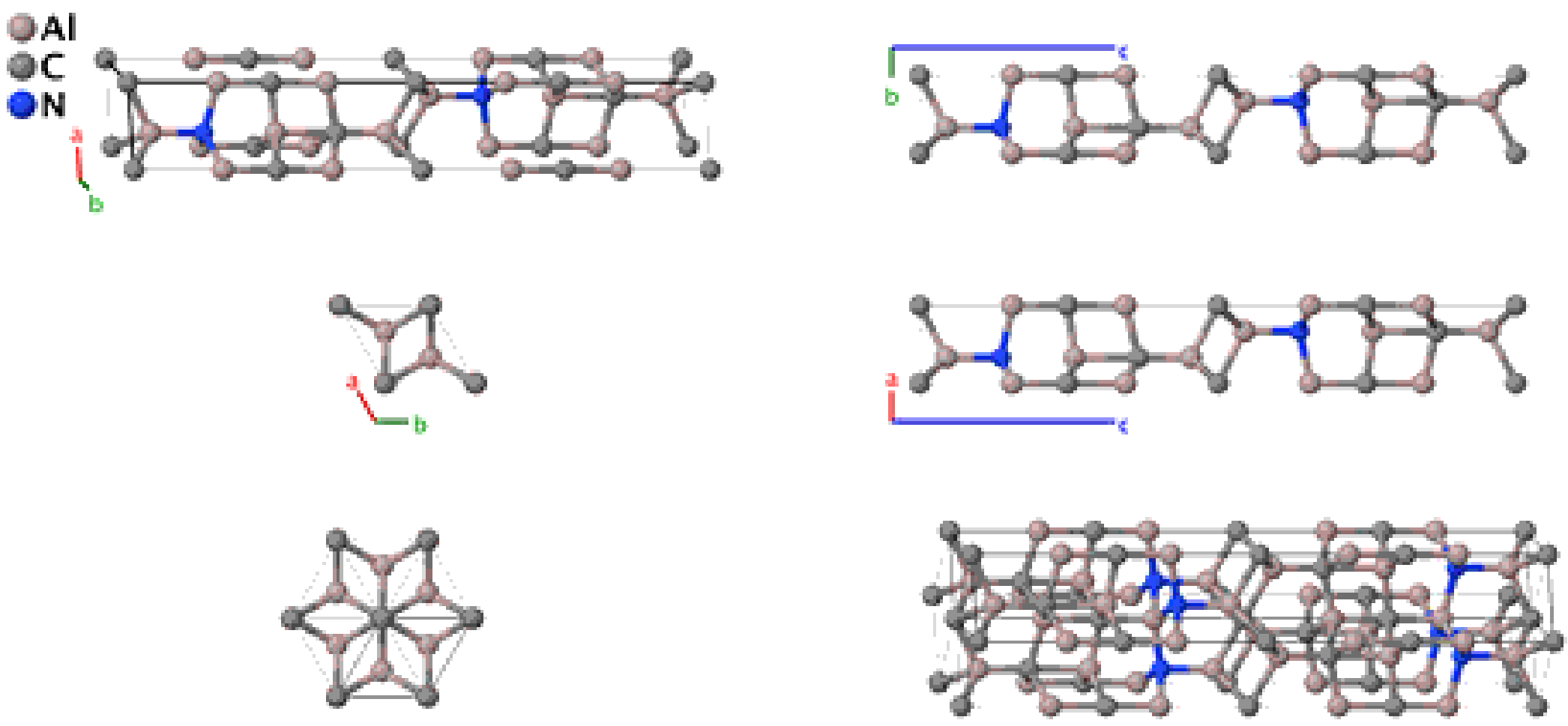

| | | |
|---|---|---|
| **Prototype** | : | $Al_5C_3N$ |
| **AFLOW prototype label** | : | A5B3C_hP18_186_2a3b_2ab_b |
| ***Strukturbericht* designation** | : | $E9_4$ |
| **Pearson symbol** | : | hP18 |
| **Space group number** | : | 186 |
| **Space group symbol** | : | $P6_3mc$ |
| **AFLOW prototype command** | : | `aflow --proto=A5B3C_hP18_186_2a3b_2ab_b` <br> `--params=`$a, c/a, z_1, z_2, z_3, z_4, z_5, z_6, z_7, z_8, z_9$ |

- Since space group #186 has no $z = 0$ mirror plane, we are free to uniformly shift the $z$ coordinates of the atoms. We have done this so that the first carbon atom is at the origin.

**Hexagonal primitive vectors:**

$$\mathbf{a}_1 = \frac{1}{2}\,a\,\hat{\mathbf{x}} - \frac{\sqrt{3}}{2}\,a\,\hat{\mathbf{y}}$$
$$\mathbf{a}_2 = \frac{1}{2}\,a\,\hat{\mathbf{x}} + \frac{\sqrt{3}}{2}\,a\,\hat{\mathbf{y}}$$
$$\mathbf{a}_3 = c\,\hat{\mathbf{z}}$$

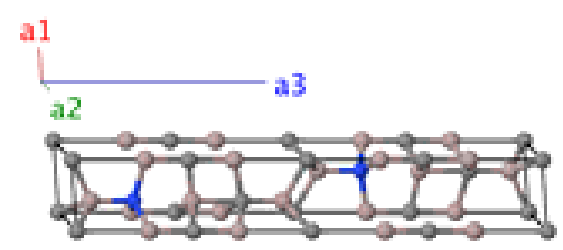

**Basis vectors:**

| | | Lattice Coordinates | | Cartesian Coordinates | Wyckoff Position | Atom Type |
|---|---|---|---|---|---|---|
| $\mathbf{B}_1$ | = | $z_1\,\mathbf{a}_3$ | = | $z_1\,c\,\hat{\mathbf{z}}$ | (2$a$) | Al I |
| $\mathbf{B}_2$ | = | $\left(\frac{1}{2}+z_1\right)\mathbf{a}_3$ | = | $\left(\frac{1}{2}+z_1\right)\,c\,\hat{\mathbf{z}}$ | (2$a$) | Al I |
| $\mathbf{B}_3$ | = | $z_2\,\mathbf{a}_3$ | = | $z_2\,c\,\hat{\mathbf{z}}$ | (2$a$) | Al II |
| $\mathbf{B}_4$ | = | $\left(\frac{1}{2}+z_2\right)\mathbf{a}_3$ | = | $\left(\frac{1}{2}+z_2\right)\,c\,\hat{\mathbf{z}}$ | (2$a$) | Al II |
| $\mathbf{B}_5$ | = | $z_3\,\mathbf{a}_3$ | = | $z_3\,c\,\hat{\mathbf{z}}$ | (2$a$) | C I |
| $\mathbf{B}_6$ | = | $\left(\frac{1}{2}+z_3\right)\mathbf{a}_3$ | = | $\left(\frac{1}{2}+z_3\right)\,c\,\hat{\mathbf{z}}$ | (2$a$) | C I |
| $\mathbf{B}_7$ | = | $z_4\,\mathbf{a}_3$ | = | $z_4\,c\,\hat{\mathbf{z}}$ | (2$a$) | C II |
| $\mathbf{B}_8$ | = | $\left(\frac{1}{2}+z_4\right)\mathbf{a}_3$ | = | $\left(\frac{1}{2}+z_4\right)\,c\,\hat{\mathbf{z}}$ | (2$a$) | C II |
| $\mathbf{B_9}$ | = | $\frac{1}{3}\mathbf{a_1} + \frac{2}{3}\mathbf{a_2} + z_5\mathbf{a_3}$ | = | $\frac{1}{2}\,a\,\hat{\mathbf{x}} + \frac{1}{2\sqrt{3}}\,a\,\hat{\mathbf{y}} + z_5\,c\,\hat{\mathbf{z}}$ | (2$b$) | Al III |
| $\mathbf{B_{10}}$ | = | $\frac{2}{3}\mathbf{a_1} + \frac{1}{3}\mathbf{a_2} + \left(\frac{1}{2} + z_5\right)\mathbf{a_3}$ | = | $\frac{1}{2}\,a\,\hat{\mathbf{x}} - \frac{1}{2\sqrt{3}}\,a\,\hat{\mathbf{y}} + \left(\frac{1}{2} + z_5\right)\,c\,\hat{\mathbf{z}}$ | (2$b$) | Al III |
| $\mathbf{B_{11}}$ | = | $\frac{1}{3}\mathbf{a_1} + \frac{2}{3}\mathbf{a_2} + z_6\mathbf{a_3}$ | = | $\frac{1}{2}\,a\,\hat{\mathbf{x}} + \frac{1}{2\sqrt{3}}\,a\,\hat{\mathbf{y}} + z_6\,c\,\hat{\mathbf{z}}$ | (2$b$) | Al IV |
| $\mathbf{B_{12}}$ | = | $\frac{2}{3}\mathbf{a_1} + \frac{1}{3}\mathbf{a_2} + \left(\frac{1}{2} + z_6\right)\mathbf{a_3}$ | = | $\frac{1}{2}\,a\,\hat{\mathbf{x}} - \frac{1}{2\sqrt{3}}\,a\,\hat{\mathbf{y}} + \left(\frac{1}{2} + z_6\right)\,c\,\hat{\mathbf{z}}$ | (2$b$) | Al IV |
| $\mathbf{B_{13}}$ | = | $\frac{1}{3}\mathbf{a_1} + \frac{2}{3}\mathbf{a_2} + z_7\mathbf{a_3}$ | = | $\frac{1}{2}\,a\,\hat{\mathbf{x}} + \frac{1}{2\sqrt{3}}\,a\,\hat{\mathbf{y}} + z_7\,c\,\hat{\mathbf{z}}$ | (2$b$) | Al V |
| $\mathbf{B_{14}}$ | = | $\frac{2}{3}\mathbf{a_1} + \frac{1}{3}\mathbf{a_2} + \left(\frac{1}{2} + z_7\right)\mathbf{a_3}$ | = | $\frac{1}{2}\,a\,\hat{\mathbf{x}} - \frac{1}{2\sqrt{3}}\,a\,\hat{\mathbf{y}} + \left(\frac{1}{2} + z_7\right)\,c\,\hat{\mathbf{z}}$ | (2$b$) | Al V |
| $\mathbf{B_{15}}$ | = | $\frac{1}{3}\mathbf{a_1} + \frac{2}{3}\mathbf{a_2} + z_8\mathbf{a_3}$ | = | $\frac{1}{2}\,a\,\hat{\mathbf{x}} + \frac{1}{2\sqrt{3}}\,a\,\hat{\mathbf{y}} + z_8\,c\,\hat{\mathbf{z}}$ | (2$b$) | C III |
| $\mathbf{B_{16}}$ | = | $\frac{2}{3}\mathbf{a_1} + \frac{1}{3}\mathbf{a_2} + \left(\frac{1}{2} + z_8\right)\mathbf{a_3}$ | = | $\frac{1}{2}\,a\,\hat{\mathbf{x}} - \frac{1}{2\sqrt{3}}\,a\,\hat{\mathbf{y}} + \left(\frac{1}{2} + z_8\right)\,c\,\hat{\mathbf{z}}$ | (2$b$) | C III |
| $\mathbf{B_{17}}$ | = | $\frac{1}{3}\mathbf{a_1} + \frac{2}{3}\mathbf{a_2} + z_9\mathbf{a_3}$ | = | $\frac{1}{2}\,a\,\hat{\mathbf{x}} + \frac{1}{2\sqrt{3}}\,a\,\hat{\mathbf{y}} + z_9\,c\,\hat{\mathbf{z}}$ | (2$b$) | N |
| $\mathbf{B_{18}}$ | = | $\frac{2}{3}\mathbf{a_1} + \frac{1}{3}\mathbf{a_2} + \left(\frac{1}{2} + z_9\right)\mathbf{a_3}$ | = | $\frac{1}{2}\,a\,\hat{\mathbf{x}} - \frac{1}{2\sqrt{3}}\,a\,\hat{\mathbf{y}} + \left(\frac{1}{2} + z_9\right)\,c\,\hat{\mathbf{z}}$ | (2$b$) | N |

---

---

**Geometry files:**

- CIF: pp. 736
- POSCAR: pp. 737

# Original BN (B12) Structure: AB_hP4_186_b_a

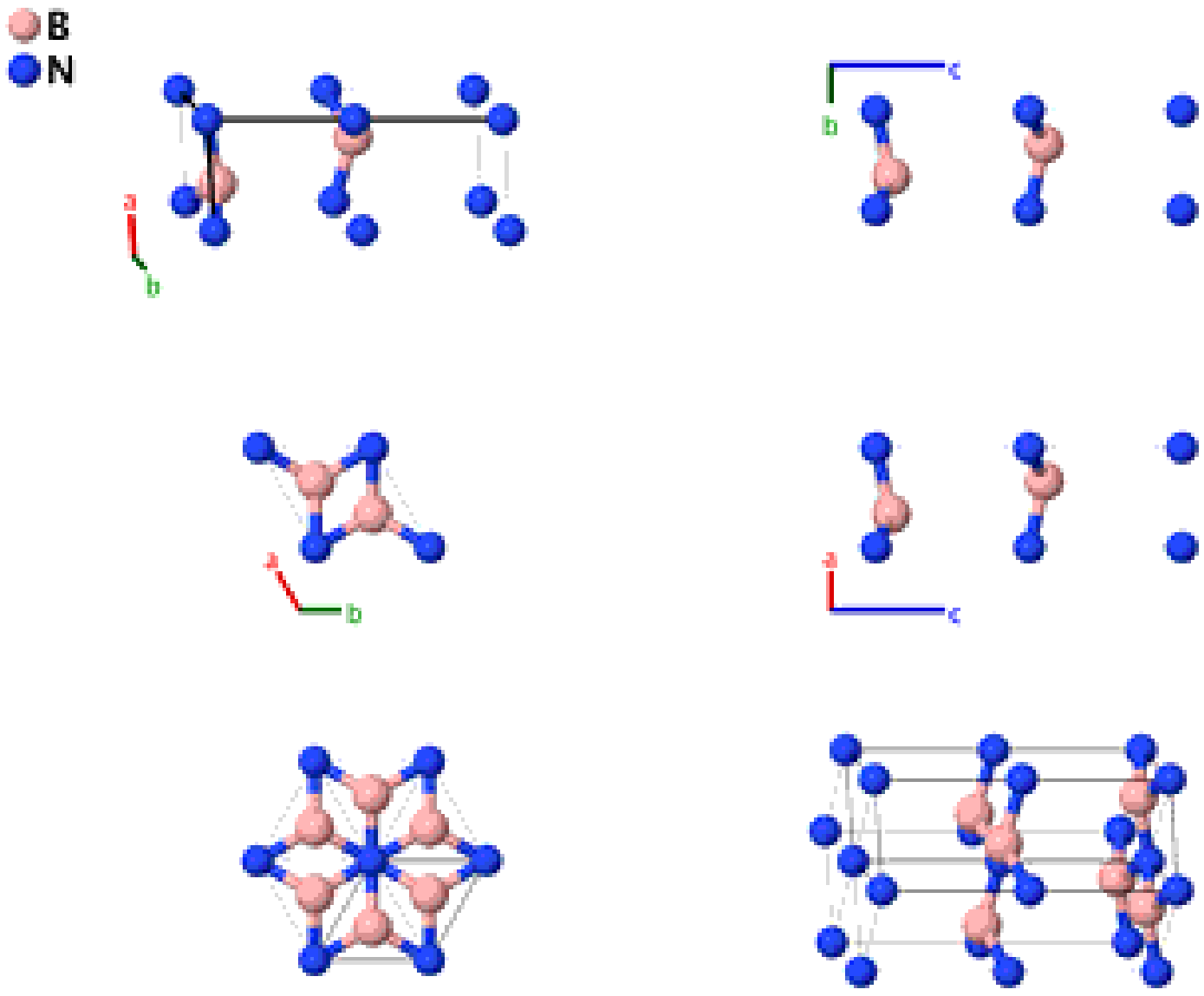

| | | |
|---|---|---|
| **Prototype** | : | BN |
| **AFLOW prototype label** | : | AB_hP4_186_b_a |
| ***Strukturbericht* designation** | : | B12 |
| **Pearson symbol** | : | hP4 |
| **Space group number** | : | 186 |
| **Space group symbol** | : | $P6_3mc$ |
| **AFLOW prototype command** | : | `aflow --proto=AB_hP4_186_b_a`<br>`--params=`$a, c/a, z_1, z_2$ |

- This is the BN structure found in (Ewald, 1931) pp. 95 and (Wilson, 1961) pp. 125-126. (Pease, 1950) later determined that the true boron nitride structure is what is now known as the $B_k$ structure. We leave this structure here for historical reasons. Note that it is crystallographically equivalent to the buckled graphite structure.

**Hexagonal primitive vectors:**

$$\mathbf{a}_1 = \frac{1}{2}\,a\,\hat{\mathbf{x}} - \frac{\sqrt{3}}{2}\,a\,\hat{\mathbf{y}}$$
$$\mathbf{a}_2 = \frac{1}{2}\,a\,\hat{\mathbf{x}} + \frac{\sqrt{3}}{2}\,a\,\hat{\mathbf{y}}$$
$$\mathbf{a}_3 = c\,\hat{\mathbf{z}}$$

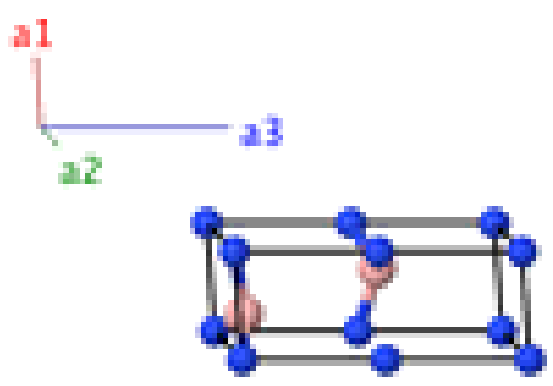

**Basis vectors:**

| | | Lattice Coordinates | | Cartesian Coordinates | Wyckoff Position | Atom Type |
|---|---|---|---|---|---|---|
| $\mathbf{B}_1$ | = | $z_1\,\mathbf{a}_3$ | = | $z_1\,c\,\hat{\mathbf{z}}$ | (2a) | N |
| $\mathbf{B}_2$ | = | $\left(\frac{1}{2} + z_1\right)\mathbf{a}_3$ | = | $\left(\frac{1}{2} + z_1\right)c\,\hat{\mathbf{z}}$ | (2a) | N |
| $\mathbf{B}_3$ | = | $\frac{1}{3}\,\mathbf{a}_1 + \frac{2}{3}\,\mathbf{a}_2 + z_2\,\mathbf{a}_3$ | = | $\frac{1}{2}\,a\,\hat{\mathbf{x}} + \frac{1}{2\sqrt{3}}\,a\,\hat{\mathbf{y}} + z_2\,c\,\hat{\mathbf{z}}$ | (2b) | B |
| $\mathbf{B}_4$ | = | $\frac{2}{3}\,\mathbf{a}_1 + \frac{1}{3}\,\mathbf{a}_2 + \left(\frac{1}{2} + z_2\right)\mathbf{a}_3$ | = | $\frac{1}{2}\,a\,\hat{\mathbf{x}} - \frac{1}{2\sqrt{3}}\,a\,\hat{\mathbf{y}} + \left(\frac{1}{2} + z_2\right)c\,\hat{\mathbf{z}}$ | (2b) | B |

**Geometry files:**

- CIF: pp. 737
- POSCAR: pp. 737

# BaPtSb Structure: ABC_hP3_187_a_d_f

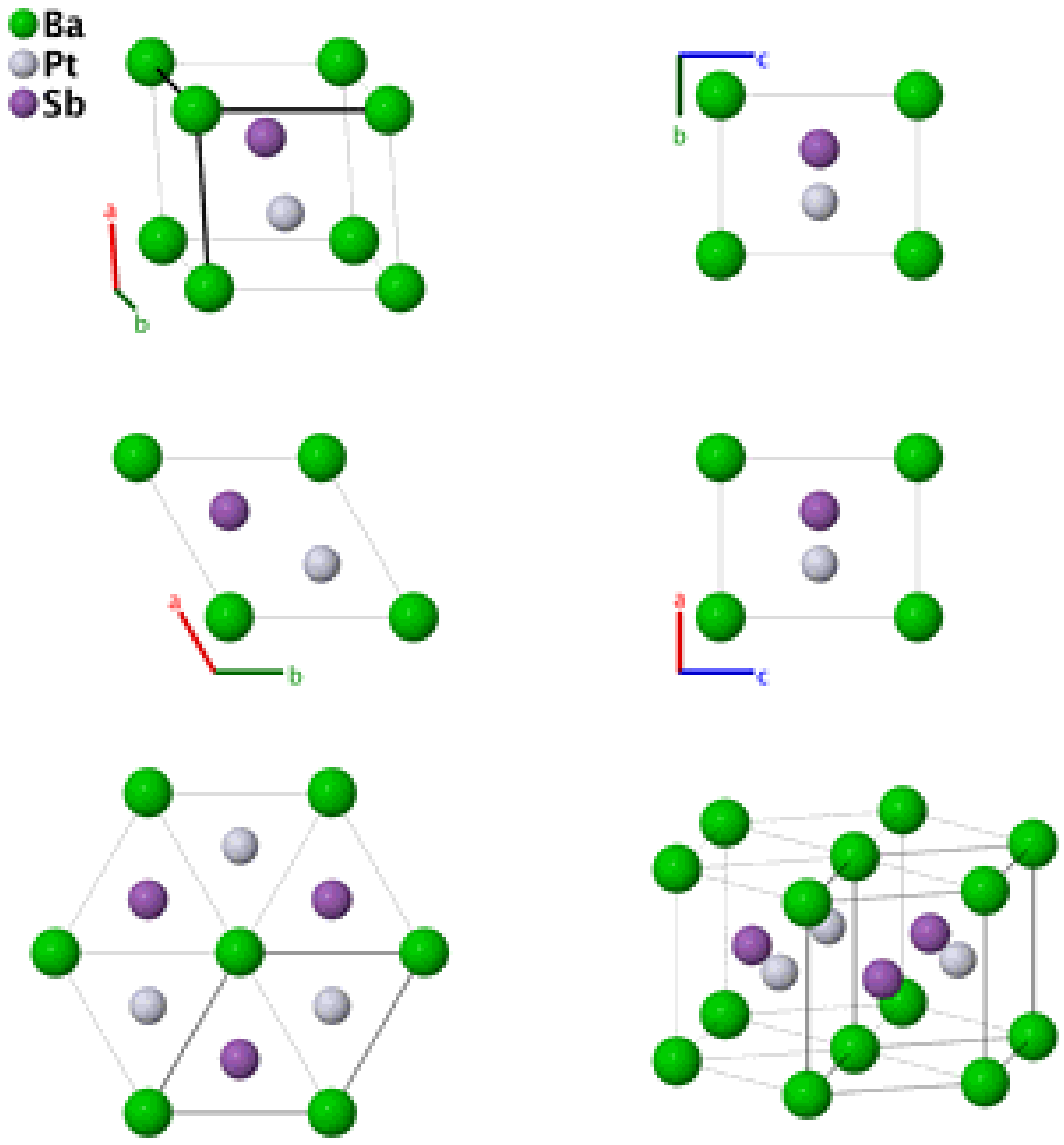

| | | |
|---|---|---|
| **Prototype** | : | BaPtSb |
| **AFLOW prototype label** | : | ABC_hP3_187_a_d_f |
| ***Strukturbericht* designation** | : | None |
| **Pearson symbol** | : | hP3 |
| **Space group number** | : | 187 |
| **Space group symbol** | : | P$\bar{6}$m2 |
| **AFLOW prototype command** | : | `aflow --proto=ABC_hP3_187_a_d_f --params=`$a, c/a$ |

**Other compounds with this structure:**

- AsKZn, PtSbSr, DyPPt, GdPPt, KSbZn, LuPPt, PPtSm, PPtTb, PPtTm, PPtY, PPtYb

**Hexagonal primitive vectors:**

$$
\begin{aligned}
\mathbf{a}_1 &= \tfrac{1}{2}\,a\,\hat{\mathbf{x}} - \tfrac{\sqrt{3}}{2}\,a\,\hat{\mathbf{y}} \\
\mathbf{a}_2 &= \tfrac{1}{2}\,a\,\hat{\mathbf{x}} + \tfrac{\sqrt{3}}{2}\,a\,\hat{\mathbf{y}} \\
\mathbf{a}_3 &= c\,\hat{\mathbf{z}}
\end{aligned}
$$

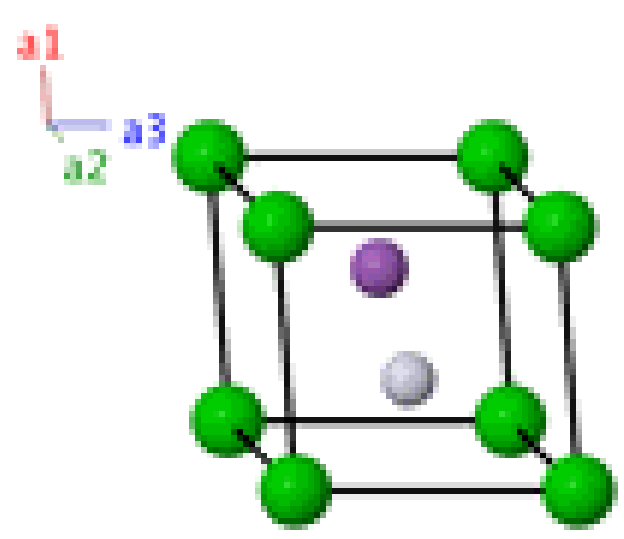

**Basis vectors:**

| Lattice Coordinates | Cartesian Coordinates | Wyckoff Position | Atom Type |
|---|---|---|---|

| | | | | | | |
|---|---|---|---|---|---|---|
| $\mathbf{B_1}$ | = | $0\,\mathbf{a_1} + 0\,\mathbf{a_2} + 0\,\mathbf{a_3}$ | = | $0\hat{\mathbf{x}} + 0\hat{\mathbf{y}} + 0\hat{\mathbf{z}}$ | $(1a)$ | Ba |
| $\mathbf{B_2}$ | = | $\frac{1}{3}\,\mathbf{a_1} + \frac{2}{3}\,\mathbf{a_2} + \frac{1}{2}\,\mathbf{a_3}$ | = | $\frac{1}{2}\,a\,\hat{\mathbf{x}} + \frac{1}{2\sqrt{3}}\,a\,\hat{\mathbf{y}} + \frac{1}{2}\,c\,\hat{\mathbf{z}}$ | $(1d)$ | Pt |
| $\mathbf{B_3}$ | = | $\frac{2}{3}\,\mathbf{a_1} + \frac{1}{3}\,\mathbf{a_2} + \frac{1}{2}\,\mathbf{a_3}$ | = | $\frac{1}{2}\,a\,\hat{\mathbf{x}} - \frac{1}{2\sqrt{3}}\,a\,\hat{\mathbf{y}} + \frac{1}{2}\,c\,\hat{\mathbf{z}}$ | $(1f)$ | Sb |

**Found in:**

**Geometry files:**

- CIF: pp. 737
- POSCAR: pp. 738

# Tungsten Carbide ($B_h$) Structure: AB_hP2_187_d_a

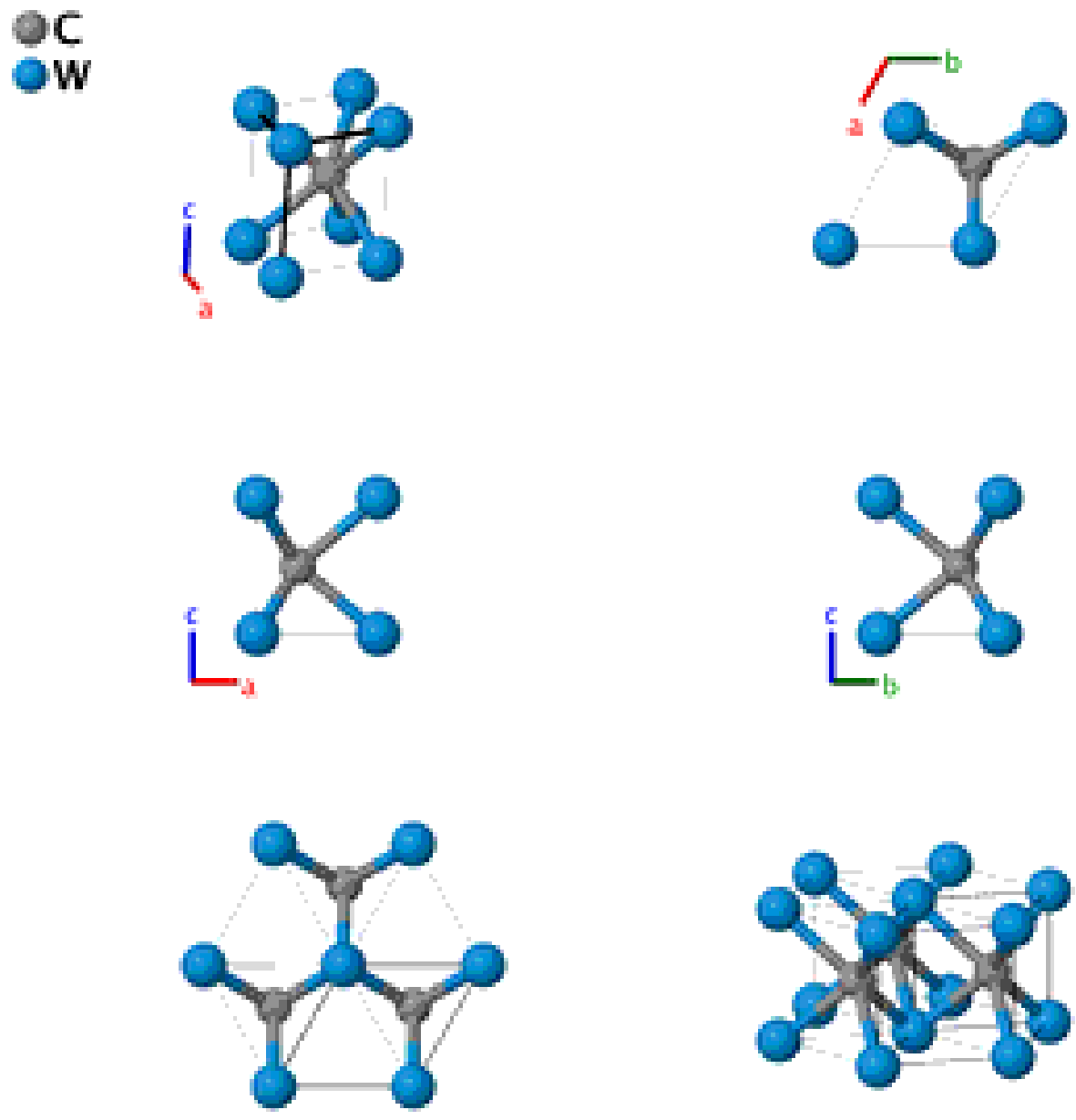

| | | |
|---|---|---|
| **Prototype** | : | WC |
| **AFLOW prototype label** | : | AB_hP2_187_d_a |
| ***Strukturbericht* designation** | : | $B_h$ |
| **Pearson symbol** | : | hP2 |
| **Space group number** | : | 187 |
| **Space group symbol** | : | $P\bar{6}m2$ |
| **AFLOW prototype command** | : | `aflow --proto=AB_hP2_187_d_a --params=`$a, c/a$ |

**Other compounds with this structure:**

- AlSn, BIr, MoC, MoP, NbS, WN, TaS, TiS, TeZr

**Hexagonal primitive vectors:**

$$\mathbf{a}_1 = \frac{1}{2}\,a\,\hat{\mathbf{x}} - \frac{\sqrt{3}}{2}\,a\,\hat{\mathbf{y}}$$
$$\mathbf{a}_2 = \frac{1}{2}\,a\,\hat{\mathbf{x}} + \frac{\sqrt{3}}{2}\,a\,\hat{\mathbf{y}}$$
$$\mathbf{a}_3 = c\,\hat{\mathbf{z}}$$

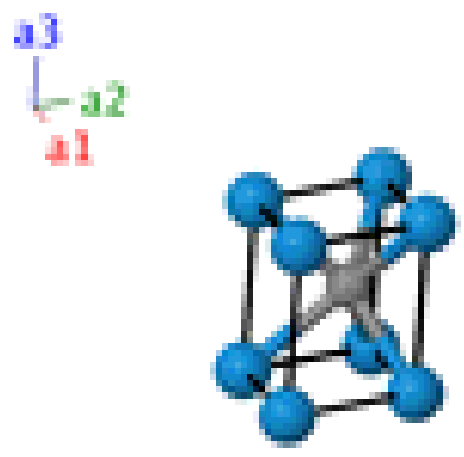

**Basis vectors:**

| | | Lattice Coordinates | | Cartesian Coordinates | Wyckoff Position | Atom Type |
|---|---|---|---|---|---|---|
| $\mathbf{B}_1$ | = | $0\,\mathbf{a}_1 + 0\,\mathbf{a}_2 + 0\,\mathbf{a}_3$ | = | $0\hat{\mathbf{x}} + 0\hat{\mathbf{y}} + 0\hat{\mathbf{z}}$ | $(1a)$ | W |
| $\mathbf{B}_2$ | = | $\frac{1}{3}\,\mathbf{a}_1 + \frac{2}{3}\,\mathbf{a}_2 + \frac{1}{2}\,\mathbf{a}_3$ | = | $\frac{1}{2}\,a\,\hat{\mathbf{x}} + \frac{1}{2\sqrt{3}}\,a\,\hat{\mathbf{y}} + \frac{1}{2}\,c\,\hat{\mathbf{z}}$ | $(1d)$ | C |

**Found in:**

**Geometry files:**

- CIF: pp. 738
- POSCAR: pp. 738

# Revised $Fe_2P$ (C22) Crystal Structure: A2B_hP9_189_fg_bc

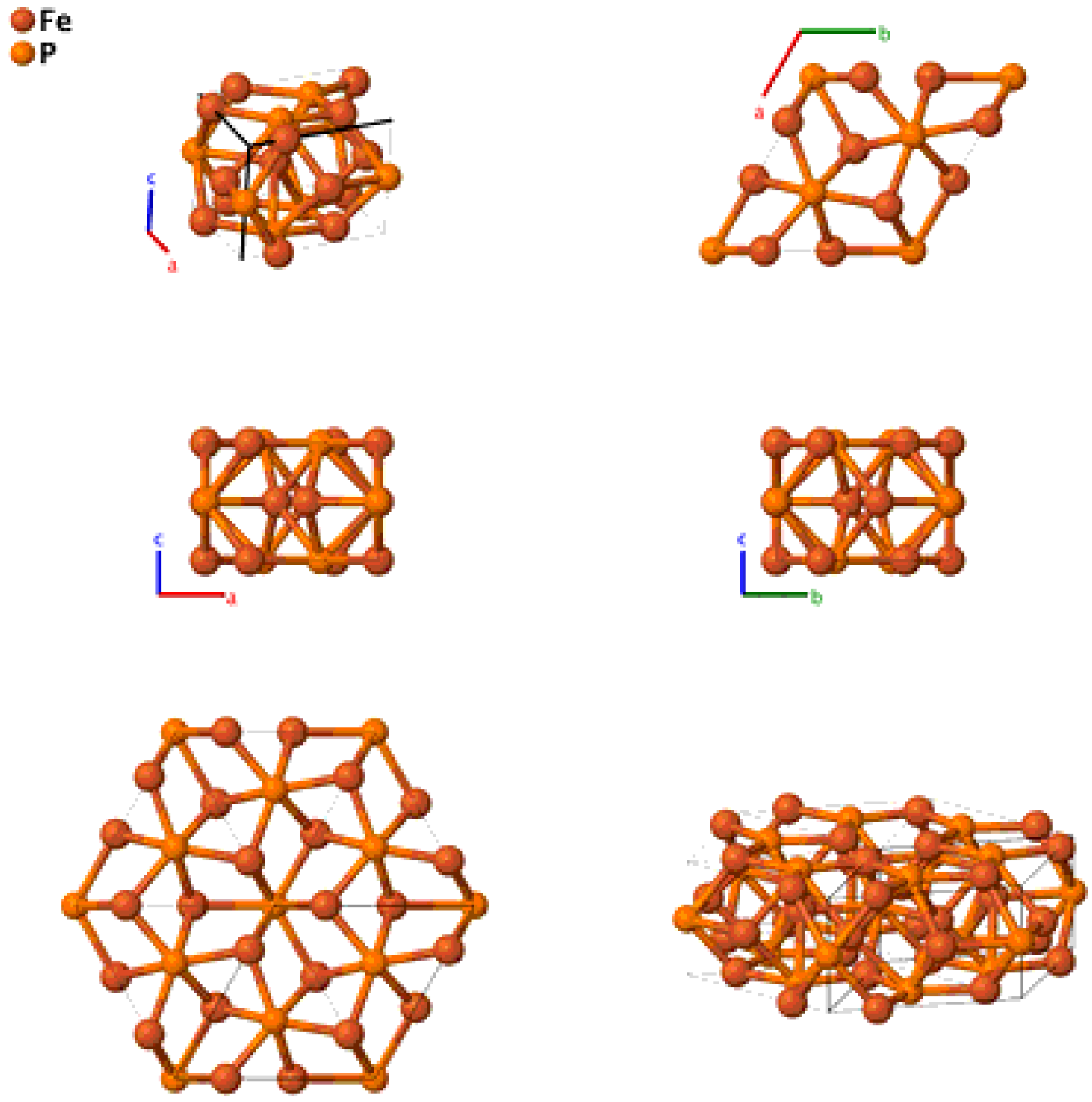

| | | |
|---|---|---|
| **Prototype** | : | $Fe_2P$ |
| **AFLOW prototype label** | : | A2B_hP9_189_fg_bc |
| ***Strukturbericht* designation** | : | C22 |
| **Pearson symbol** | : | hP9 |
| **Space group number** | : | 189 |
| **Space group symbol** | : | P$\bar{6}$2m |
| **AFLOW prototype command** | : | `aflow --proto=A2B_hP9_189_fg_bc --params=`$a, c/a, x_3, x_4$ |

**Other compounds with this structure:**

- AgAsCa, AgSiYb, AlCoPu, AlCuTm, AlNiTb, FeNiP, FeGaU, $Mn_2P$, $Ni_2P$, $Ni_6Si_2B$, $Pt_2Si$, RhSnZr, hundreds more

---

- This is not the structure given in (Hermann, 1937) Strukturbericht Vol. II, pp. 95. As noted by (Wyckoff, 1963) pp. 360, the structure which was "generally accepted for years, has recently been shown to be incorrect". This is the corrected structure, as given in Wyckoff and (Villars, 1991). See the original $Fe_2P$ (C22) page for the Strukturbericht version of this crystal.

## Hexagonal primitive vectors:

$$
\begin{aligned}
\mathbf{a}_1 &= \tfrac{1}{2}\, a\, \mathbf{\hat{x}} - \tfrac{\sqrt{3}}{2}\, a\, \mathbf{\hat{y}} \\
\mathbf{a}_2 &= \tfrac{1}{2}\, a\, \mathbf{\hat{x}} + \tfrac{\sqrt{3}}{2}\, a\, \mathbf{\hat{y}} \\
\mathbf{a}_3 &= c\, \mathbf{\hat{z}}
\end{aligned}
$$

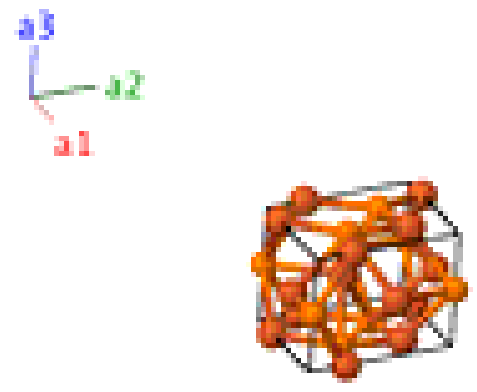

## Basis vectors:

| | | Lattice Coordinates | | Cartesian Coordinates | Wyckoff Position | Atom Type |
|---|---|---|---|---|---|---|
| $\mathbf{B}_1$ | = | $\frac{1}{2}\,\mathbf{a}_3$ | = | $\frac{1}{2}\, c\, \mathbf{\hat{z}}$ | $(1b)$ | P I |
| $\mathbf{B}_2$ | = | $\frac{1}{3}\,\mathbf{a}_1 + \frac{2}{3}\,\mathbf{a}_2$ | = | $\frac{1}{2}\, a\, \mathbf{\hat{x}} + \frac{1}{2\sqrt{3}}\, a\, \mathbf{\hat{y}}$ | $(2c)$ | P II |
| $\mathbf{B}_3$ | = | $\frac{2}{3}\,\mathbf{a}_1 + \frac{1}{3}\,\mathbf{a}_2$ | = | $\frac{1}{2}\, a\, \mathbf{\hat{x}} - \frac{1}{2\sqrt{3}}\, a\, \mathbf{\hat{y}}$ | $(2c)$ | P II |
| $\mathbf{B}_4$ | = | $x_3\,\mathbf{a}_1$ | = | $\frac{1}{2}\, x_3\, a\, \mathbf{\hat{x}} - \frac{\sqrt{3}}{2}\, x_3\, a\, \mathbf{\hat{y}}$ | $(3f)$ | Fe I |
| $\mathbf{B}_5$ | = | $x_3\,\mathbf{a}_2$ | = | $\frac{1}{2}\, x_3\, a\, \mathbf{\hat{x}} + \frac{\sqrt{3}}{2}\, x_3\, a\, \mathbf{\hat{y}}$ | $(3f)$ | Fe I |
| $\mathbf{B}_6$ | = | $-x_3\,\mathbf{a}_1 - x_3\,\mathbf{a}_2$ | = | $-x_3\, a\, \mathbf{\hat{x}}$ | $(3f)$ | Fe I |
| $\mathbf{B}_7$ | = | $x_4\,\mathbf{a}_1 + \frac{1}{2}\,\mathbf{a}_3$ | = | $\frac{1}{2}\, x_4\, a\, \mathbf{\hat{x}} - \frac{\sqrt{3}}{2}\, x_4\, a\, \mathbf{\hat{y}} + \frac{1}{2}\, c\, \mathbf{\hat{z}}$ | $(3g)$ | Fe II |
| $\mathbf{B}_8$ | = | $x_4\,\mathbf{a}_2 + \frac{1}{2}\,\mathbf{a}_3$ | = | $\frac{1}{2}\, x_4\, a\, \mathbf{\hat{x}} + \frac{\sqrt{3}}{2}\, x_4\, a\, \mathbf{\hat{y}} + \frac{1}{2}\, c\, \mathbf{\hat{z}}$ | $(3g)$ | Fe II |
| $\mathbf{B}_9$ | = | $-x_4\,\mathbf{a}_1 - x_4\,\mathbf{a}_2 + \frac{1}{2}\,\mathbf{a}_3$ | = | $-x_4\, a\, \mathbf{\hat{x}} + \frac{1}{2}\, c\, \mathbf{\hat{z}}$ | $(3g)$ | Fe II |

## Found in:

## Geometry files:

- CIF: pp. 738
- POSCAR: pp. 739

# $AlB_4Mg$ Structure: AB4C_hP6_191_a_h_b

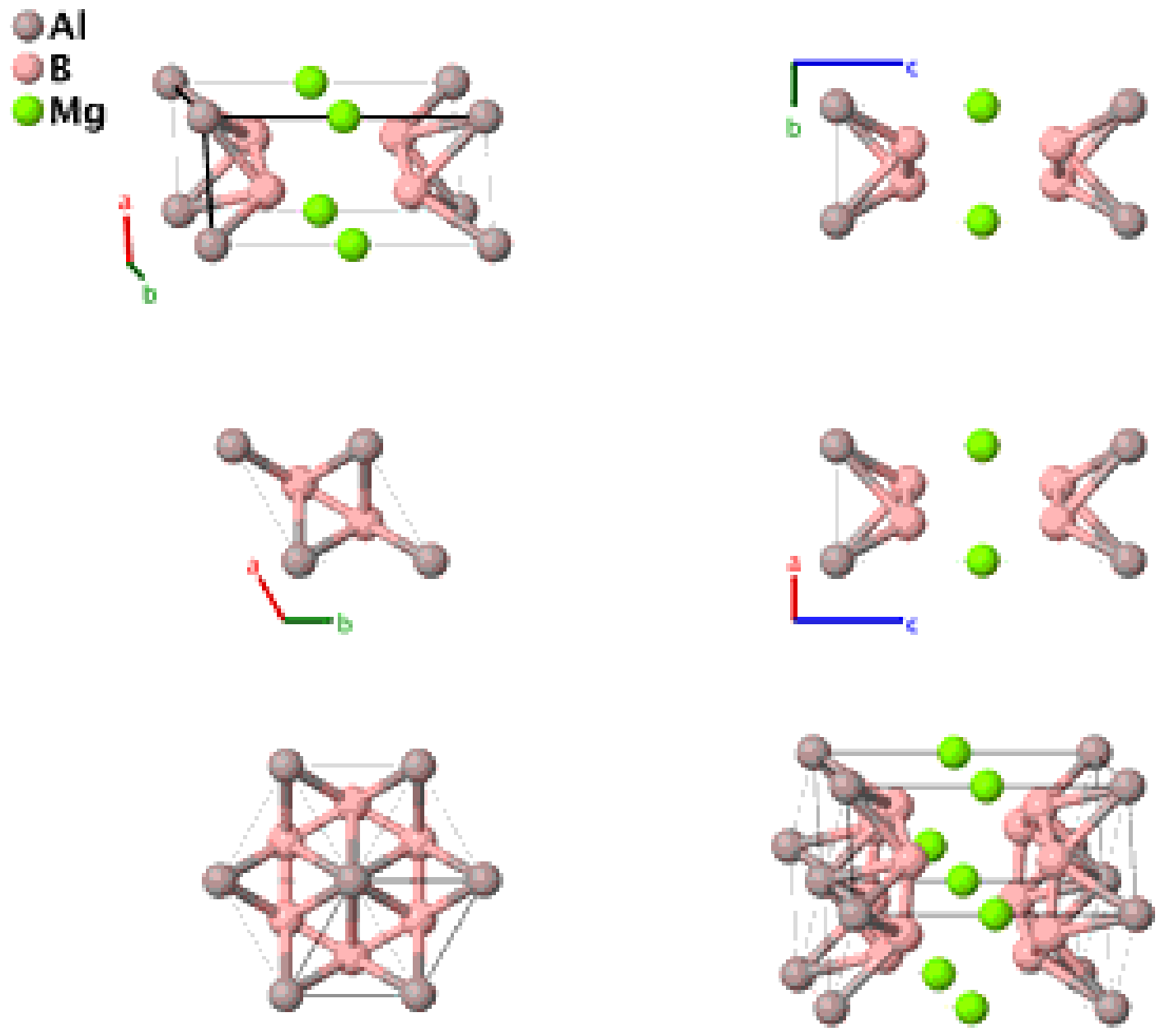

|  |  |  |
|---|---|---|
| **Prototype** | : | $AlB_4Mg$ |
| **AFLOW prototype label** | : | AB4C_hP6_191_a_h_b |
| ***Strukturbericht* designation** | : | None |
| **Pearson symbol** | : | hP6 |
| **Space group number** | : | 191 |
| **Space group symbol** | : | P6/mmm |
| **AFLOW prototype command** | : | `aflow --proto=AB4C_hP6_191_a_h_b`<br>`--params=`$a, c/a, z_3$ |

- Note that Table I of (Margadonna, 2002) mislabels the (1a) and (1b) Wyckoff positions.

**Hexagonal primitive vectors:**

$$\mathbf{a}_1 = \frac{1}{2}\,a\,\mathbf{\hat{x}} - \frac{\sqrt{3}}{2}\,a\,\mathbf{\hat{y}}$$
$$\mathbf{a}_2 = \frac{1}{2}\,a\,\mathbf{\hat{x}} + \frac{\sqrt{3}}{2}\,a\,\mathbf{\hat{y}}$$
$$\mathbf{a}_3 = c\,\mathbf{\hat{z}}$$

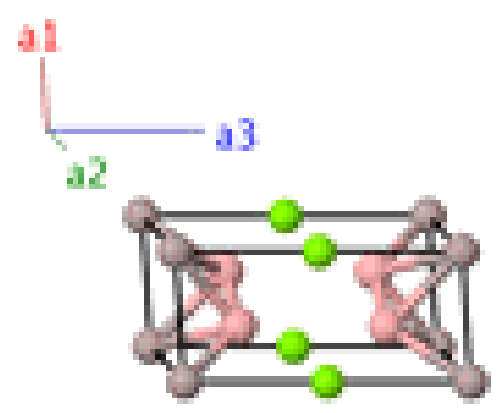

**Basis vectors:**

|  |  | Lattice Coordinates |  | Cartesian Coordinates | Wyckoff Position | Atom Type |
|---|---|---|---|---|---|---|
| $\mathbf{B}_1$ | = | $0\,\mathbf{a}_1 + 0\,\mathbf{a}_2 + 0\,\mathbf{a}_3$ | = | $0\mathbf{\hat{x}} + 0\mathbf{\hat{y}} + 0\mathbf{\hat{z}}$ | (1$a$) | Al |
| $\mathbf{B}_2$ | = | $\frac{1}{2}\,\mathbf{a}_3$ | = | $\frac{1}{2}\,c\,\mathbf{\hat{z}}$ | (1$b$) | Mg |
| $\mathbf{B}_3$ | = | $\frac{1}{3}\mathbf{a}_1 + \frac{2}{3}\mathbf{a}_2 + z_3\,\mathbf{a}_3$ | = | $\frac{1}{2}\,a\,\mathbf{\hat{x}} + \frac{1}{2\sqrt{3}}\,a\,\mathbf{\hat{y}} + z_3\,c\,\mathbf{\hat{z}}$ | (4$h$) | B |
| $\mathbf{B}_4$ | = | $\frac{2}{3}\mathbf{a}_1 + \frac{1}{3}\mathbf{a}_2 + z_3\,\mathbf{a}_3$ | = | $\frac{1}{2}\,a\,\mathbf{\hat{x}} - \frac{1}{2\sqrt{3}}\,a\,\mathbf{\hat{y}} + z_3\,c\,\mathbf{\hat{z}}$ | (4$h$) | B |
| $\mathbf{B_5}$ | = | $\frac{2}{3}\mathbf{a_1} + \frac{1}{3}\mathbf{a_2} - z_3\,\mathbf{a_3}$ | = | $\frac{1}{2}\,a\,\hat{\mathbf{x}} - \frac{1}{2\sqrt{3}}\,a\,\hat{\mathbf{y}} - z_3\,c\,\hat{\mathbf{z}}$ | $(4h)$ | B |
| $\mathbf{B_6}$ | = | $\frac{1}{3}\mathbf{a_1} + \frac{2}{3}\mathbf{a_2} - z_3\,\mathbf{a_3}$ | = | $\frac{1}{2}\,a\,\hat{\mathbf{x}} + \frac{1}{2\sqrt{3}}\,a\,\hat{\mathbf{y}} - z_3\,c\,\hat{\mathbf{z}}$ | $(4h)$ | B |

---

---

**Geometry files:**

- CIF: pp. 739
- POSCAR: pp. 739

# $CaCu_5$ ($D2_d$) Structure: AB5_hP6_191_a_cg

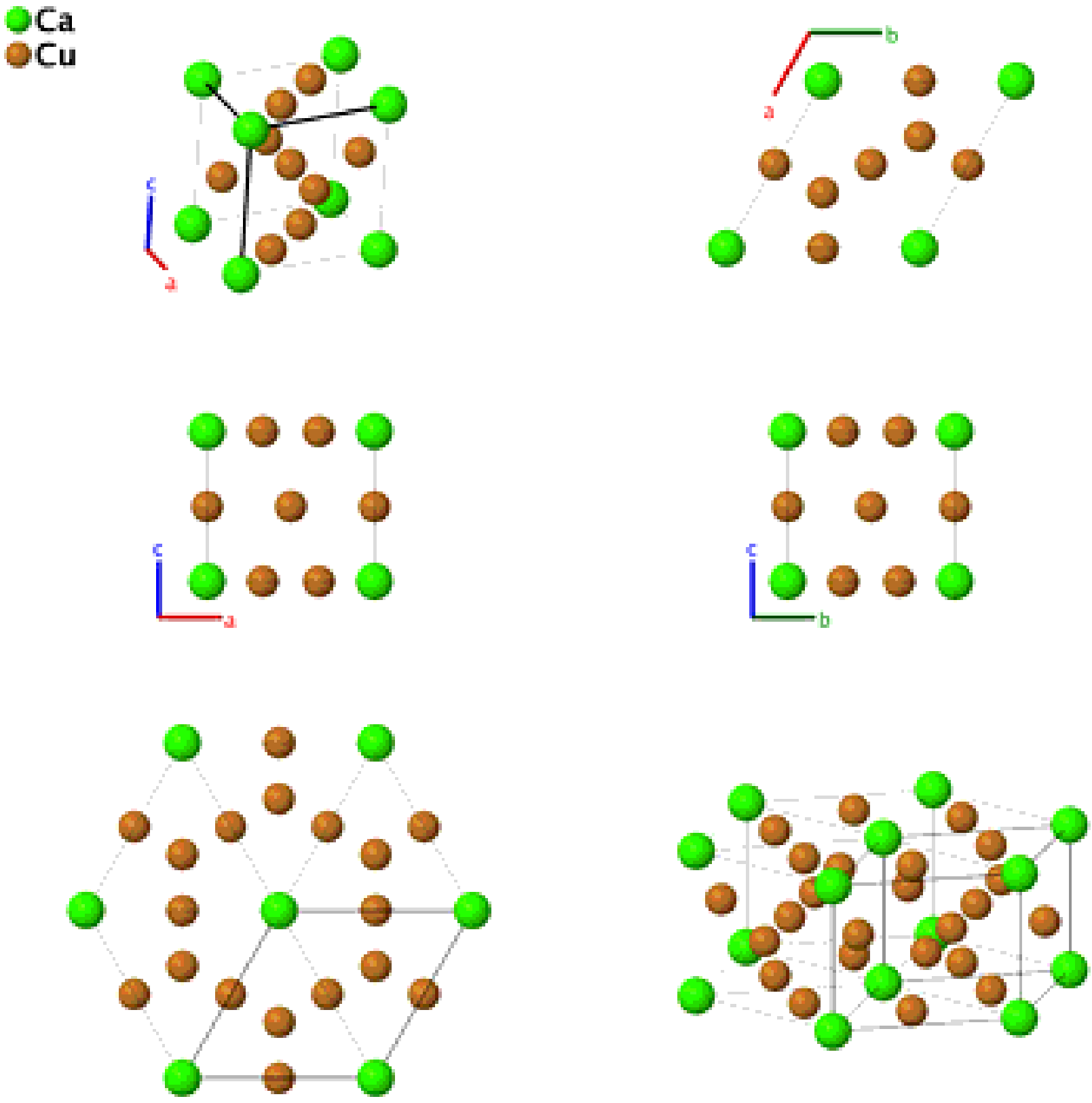

| | | |
|---|---|---|
| **Prototype** | : | $CaCu_5$ |
| **AFLOW prototype label** | : | AB5_hP6_191_a_cg |
| ***Strukturbericht* designation** | : | $D2_d$ |
| **Pearson symbol** | : | hP6 |
| **Space group number** | : | 191 |
| **Space group symbol** | : | P6/mmm |
| **AFLOW prototype command** | : | `aflow --proto=AB5_hP6_191_a_cg --params=`$a, c/a$ |

**Other compounds with this structure:**

- $Au_5Sr$, $Ag_3Al_2La$, $Ag_5Ba$, $CePt_5$, $Co_5Sm$, $Co_5Tb$, $Co_5Y$, $EuZn_5$, $GdRh_5$, $Ir_5Nd$, $LaNi_5$, $SmZn_5$, many others

**Hexagonal primitive vectors:**

$$\mathbf{a}_1 = \frac{1}{2}\,a\,\hat{\mathbf{x}} - \frac{\sqrt{3}}{2}\,a\,\hat{\mathbf{y}}$$
$$\mathbf{a}_2 = \frac{1}{2}\,a\,\hat{\mathbf{x}} + \frac{\sqrt{3}}{2}\,a\,\hat{\mathbf{y}}$$
$$\mathbf{a}_3 = c\,\hat{\mathbf{z}}$$

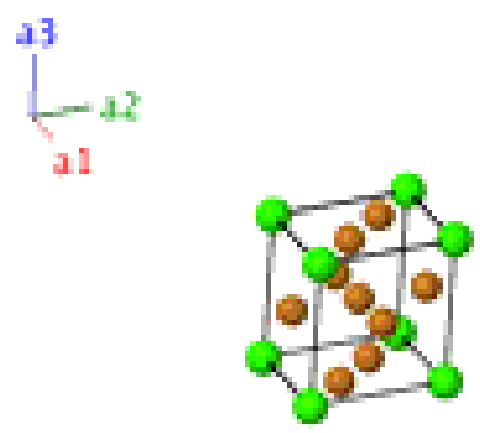

**Basis vectors:**

| | | Lattice Coordinates | | Cartesian Coordinates | Wyckoff Position | Atom Type |
|---|---|---|---|---|---|---|
| $\mathbf{B_1}$ | = | $0\,\mathbf{a_1} + 0\,\mathbf{a_2} + 0\,\mathbf{a_3}$ | = | $0\hat{\mathbf{x}} + 0\hat{\mathbf{y}} + 0\hat{\mathbf{z}}$ | $(1a)$ | Ca |
| $\mathbf{B_2}$ | = | $\frac{1}{3}\mathbf{a_1} + \frac{2}{3}\mathbf{a_2}$ | = | $\frac{1}{2}\,a\,\hat{\mathbf{x}} + \frac{1}{2\sqrt{3}}\,a\,\hat{\mathbf{y}}$ | $(2c)$ | Cu I |
| $\mathbf{B_3}$ | = | $\frac{2}{3}\mathbf{a_1} + \frac{1}{3}\mathbf{a_2}$ | = | $\frac{1}{2}\,a\,\hat{\mathbf{x}} - \frac{1}{2\sqrt{3}}\,a\,\hat{\mathbf{y}}$ | $(2c)$ | Cu I |
| $\mathbf{B_4}$ | = | $\frac{1}{2}\,\mathbf{a_1} + \frac{1}{2}\mathbf{a_3}$ | = | $\frac{1}{4}\,a\,\hat{\mathbf{x}} - \frac{\sqrt{3}}{4}\,a\,\hat{\mathbf{y}} + \frac{1}{2}\,c\,\hat{\mathbf{z}}$ | $(3g)$ | Cu II |
| $\mathbf{B_5}$ | = | $\frac{1}{2}\,\mathbf{a_2} + \frac{1}{2}\mathbf{a_3}$ | = | $\frac{1}{4}\,a\,\hat{\mathbf{x}} + \frac{\sqrt{3}}{4}\,a\,\hat{\mathbf{y}} + \frac{1}{2}\,c\,\hat{\mathbf{z}}$ | $(3g)$ | Cu II |
| $\mathbf{B_6}$ | = | $\frac{1}{2}\,\mathbf{a_1} + \frac{1}{2}\,\mathbf{a_2} + \frac{1}{2}\mathbf{a_3}$ | = | $\frac{1}{2}\,a\,\hat{\mathbf{x}} + \frac{1}{2}\,c\,\hat{\mathbf{z}}$ | $(3g)$ | Cu II |

**Geometry files:**

- CIF: pp. 739
- POSCAR: pp. 740

# Simple Hexagonal Lattice ($A_f$): A_hP1_191_a

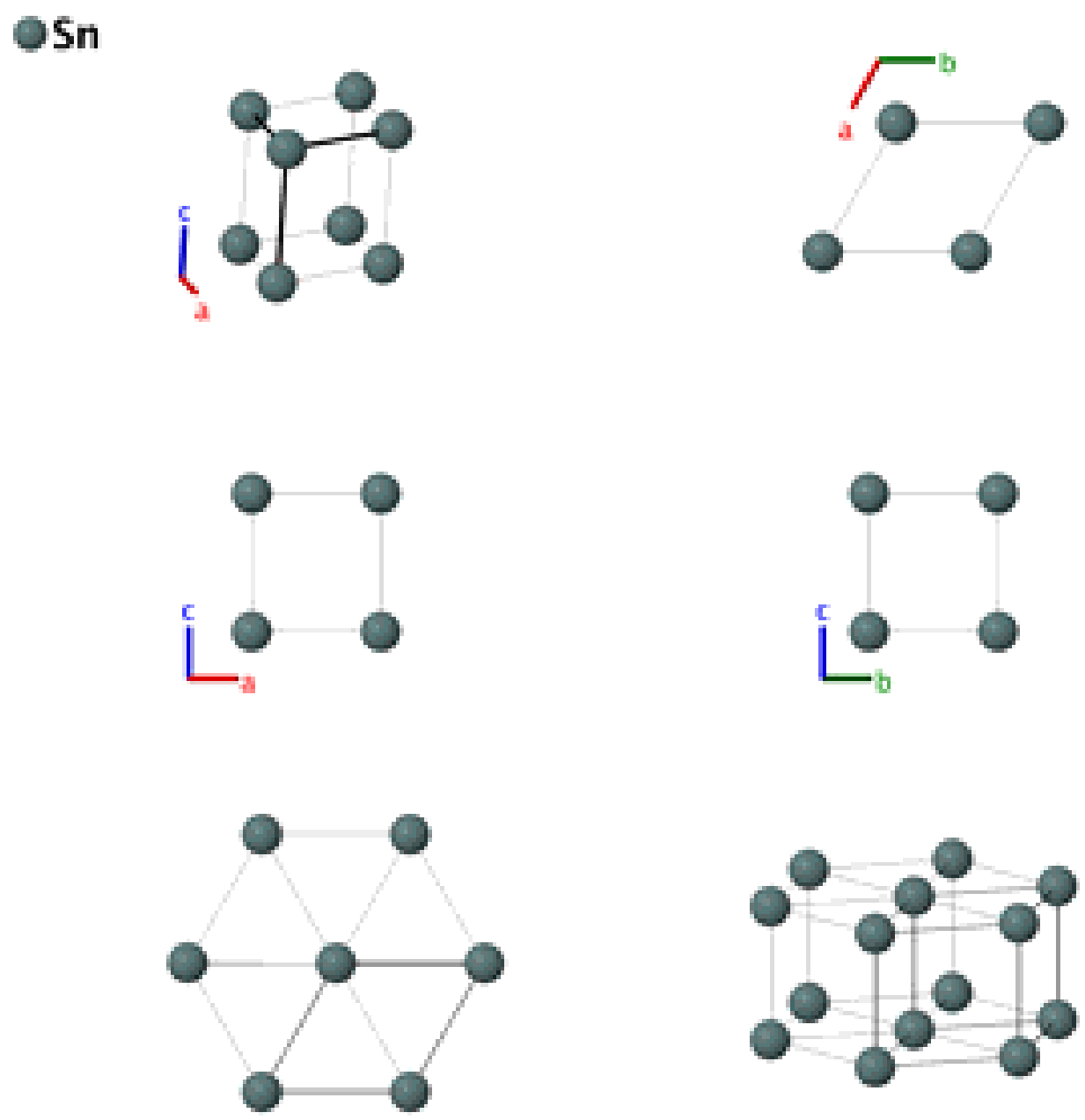

| | | |
|---|---|---|
| **Prototype** | : | $\gamma$-$HgSn_{6-10}$ |
| **AFLOW prototype label** | : | A_hP1_191_a |
| ***Strukturbericht* designation** | : | $A_f$ |
| **Pearson symbol** | : | hP1 |
| **Space group number** | : | 191 |
| **Space group symbol** | : | P6/mmm |
| **AFLOW prototype command** | : | `aflow --proto=A_hP1_191_a --params=`$a, c/a$ |

**Other compounds with this structure:**

- Si (metastable) disordered phases of BiIn, $CdSn_{19}$, $In_7Sb_3$, InSb

---

- Unlike the simple cubic lattice, there are no elements which take this structure as the ground state. There is a metastable silicon phase with this structure. The prototype state is a mercury-tin alloy. Thus the atom type "M" represents an average of Hg and Sn atoms.

**Hexagonal primitive vectors:**

$$
\begin{aligned}
\mathbf{a}_1 &= \tfrac{1}{2}\,a\,\hat{\mathbf{x}} - \tfrac{\sqrt{3}}{2}\,a\,\hat{\mathbf{y}} \\
\mathbf{a}_2 &= \tfrac{1}{2}\,a\,\hat{\mathbf{x}} + \tfrac{\sqrt{3}}{2}\,a\,\hat{\mathbf{y}} \\
\mathbf{a}_3 &= c\,\hat{\mathbf{z}}
\end{aligned}
$$

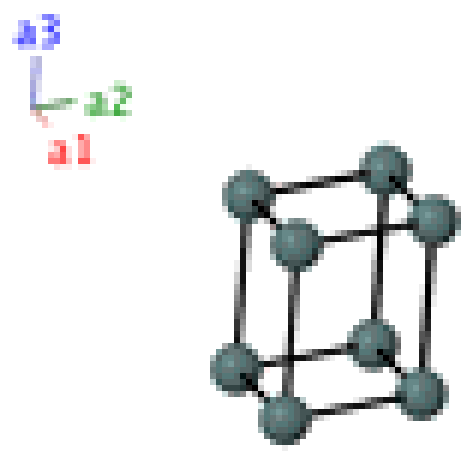

**Basis vectors:**

| | | Lattice Coordinates | | Cartesian Coordinates | Wyckoff Position | Atom Type |
|---|---|---|---|---|---|---|
| $\mathbf{B_1}$ | = | $0\,\mathbf{a_1} + 0\,\mathbf{a_2} + 0\,\mathbf{a_3}$ | = | $0\hat{\mathbf{x}} + 0\hat{\mathbf{y}} + 0\hat{\mathbf{z}}$ | $(1a)$ | M |

**Geometry files:**

- CIF: pp. 740
- POSCAR: pp. 740

# $Li_3N$ Structure: A3B_hP4_191_bc_a

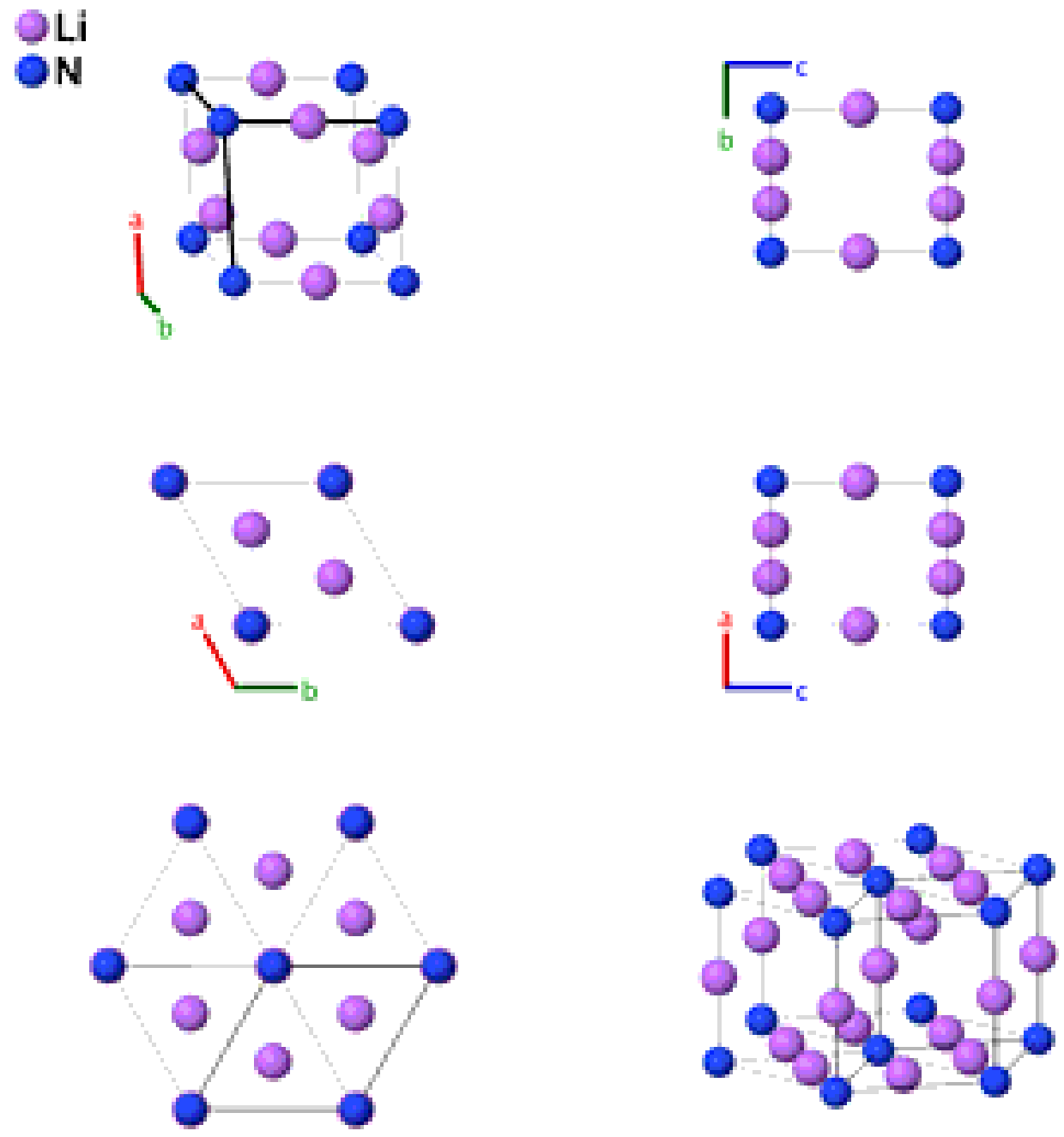

| | | |
|---|---|---|
| **Prototype** | : | $Li_3N$ |
| **AFLOW prototype label** | : | A3B_hP4_191_bc_a |
| ***Strukturbericht* designation** | : | None |
| **Pearson symbol** | : | hP4 |
| **Space group number** | : | 191 |
| **Space group symbol** | : | P6/mmm |
| **AFLOW prototype command** | : | `aflow --proto=A3B_hP4_191_bc_a --params=`$a, c/a$ |

**Hexagonal primitive vectors:**

$$
\begin{aligned}
\mathbf{a}_1 &= \tfrac{1}{2}\,a\,\hat{\mathbf{x}} - \tfrac{\sqrt{3}}{2}\,a\,\hat{\mathbf{y}} \\
\mathbf{a}_2 &= \tfrac{1}{2}\,a\,\hat{\mathbf{x}} + \tfrac{\sqrt{3}}{2}\,a\,\hat{\mathbf{y}} \\
\mathbf{a}_3 &= c\,\hat{\mathbf{z}}
\end{aligned}
$$

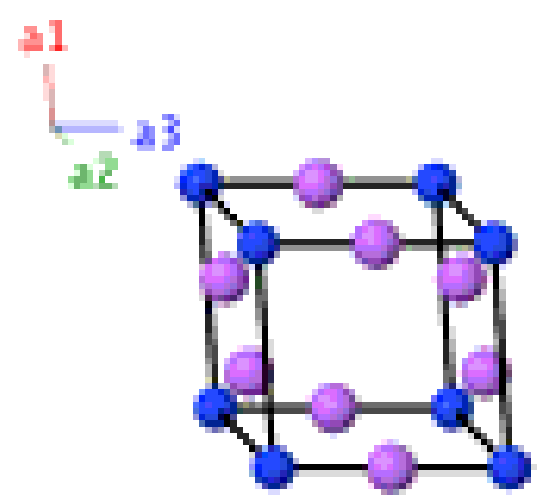

**Basis vectors:**

| | | Lattice Coordinates | | Cartesian Coordinates | Wyckoff Position | Atom Type |
|---|---|---|---|---|---|---|
| $\mathbf{B_1}$ | = | $0\,\mathbf{a_1} + 0\,\mathbf{a_2} + 0\,\mathbf{a_3}$ | = | $0\hat{\mathbf{x}} + 0\hat{\mathbf{y}} + 0\hat{\mathbf{z}}$ | (1$a$) | N |
| $\mathbf{B_2}$ | = | $\frac{1}{2}\,\mathbf{a_3}$ | = | $\frac{1}{2}\,c\,\hat{\mathbf{z}}$ | (1$b$) | Li I |
| $\mathbf{B_3}$ | = | $\frac{1}{3}\mathbf{a_1} + \frac{2}{3}\mathbf{a_2}$ | = | $\frac{1}{2}\,a\,\hat{\mathbf{x}} + \frac{1}{2\sqrt{3}}\,a\,\hat{\mathbf{y}}$ | (2$c$) | Li II |
| $\mathbf{B_4}$ | = | $\frac{2}{3}\mathbf{a_1} + \frac{1}{3}\mathbf{a_2}$ | = | $\frac{1}{2}\,a\,\hat{\mathbf{x}} - \frac{1}{2\sqrt{3}}\,a\,\hat{\mathbf{y}}$ | (2$c$) | Li II |

**Found in:**

**Geometry files:**

- CIF: pp. 740
- POSCAR: pp. 741

# Hexagonal $\omega$ (C32) Structure: AB2_hP3_191_a_d

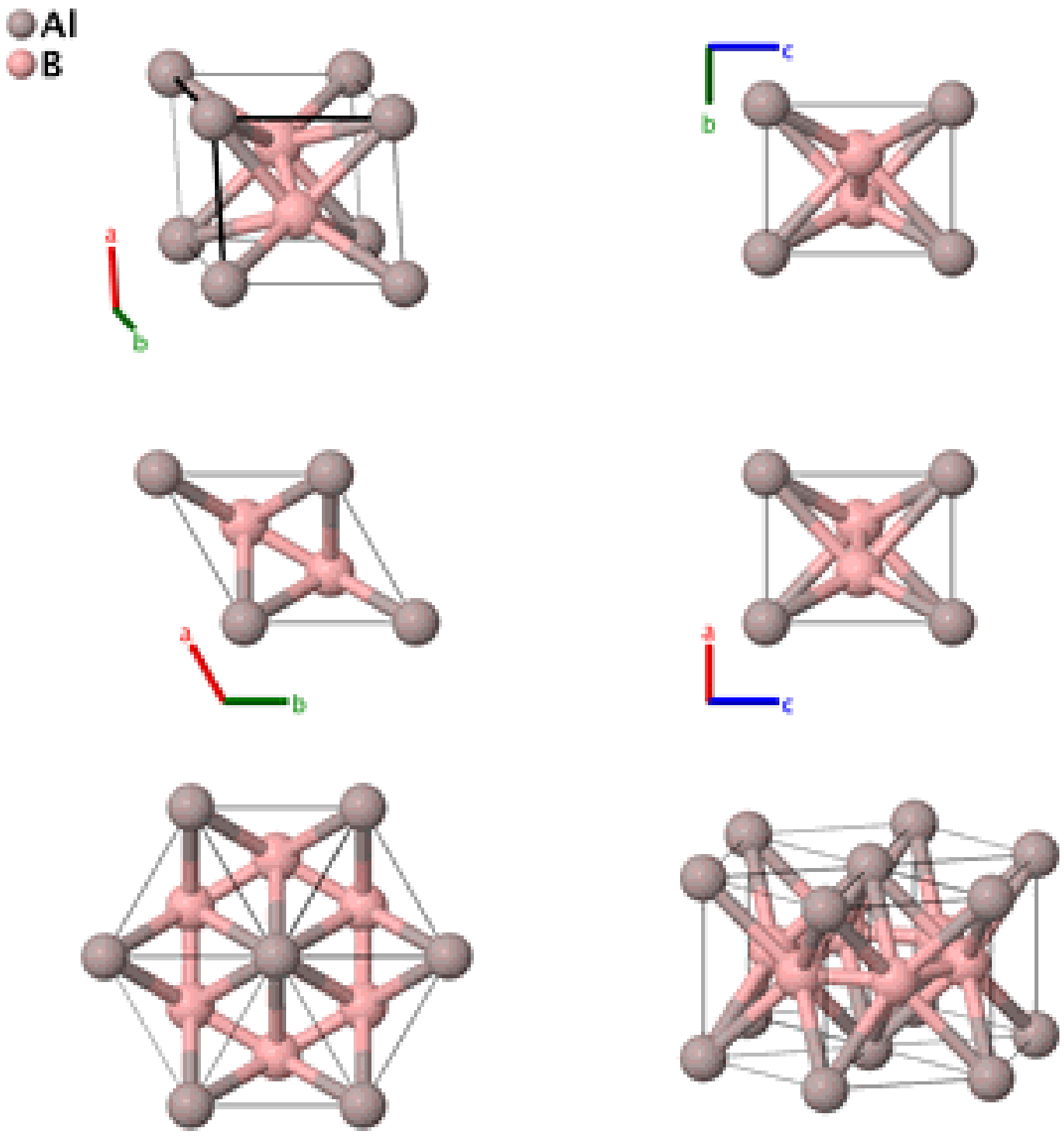

| | | |
|---|---|---|
| **Prototype** | : | $AlB_2$ |
| **AFLOW prototype label** | : | AB2_hP3_191_a_d |
| ***Strukturbericht* designation** | : | C32 |
| **Pearson symbol** | : | hP3 |
| **Space group number** | : | 191 |
| **Space group symbol** | : | P6/mmm |
| **AFLOW prototype command** | : | `aflow --proto=AB2_hP3_191_a_d --params=`$a, c/a$ |

**Other compounds with this structure:**

- Ti (metastable), $MgB_2$, $Be_2Hf$, $CeHg_2$

- This is the hexagonal $\omega$ phase. There is also a trigonal $\omega$ (C6) phase. For more details about the $\omega$ phase and materials which form in the $\omega$ phase, see (Sikka, 1982). Most $\omega$ phase intermetallic alloys are disordered. In this structure the B-B distance is smaller than the Al-B distance for every $c/a$ ratio. If $c/a$ is small enough the structure looks like a set of inter-penetrating boron triangular planes and aluminium chains. If $c/a = 1/\sqrt{3}$ the Al-Al distance along (001) is the same as the B-B distance in the plane, and, for that matter, the B-B distance in the (001) direction. This value 0.577 is close to the value $\sqrt{3/8}$ ($\approx 0.612$) where the trigonal $\omega$ phase can transform to the body-centered cubic (A2) lattice, which probably explains the close connection between the $\omega$ and bcc phases.

**Hexagonal primitive vectors:**

$$
\begin{aligned}
\mathbf{a}_1 &= \tfrac{1}{2}\,a\,\hat{\mathbf{x}} - \tfrac{\sqrt{3}}{2}\,a\,\hat{\mathbf{y}} \\
\mathbf{a}_2 &= \tfrac{1}{2}\,a\,\hat{\mathbf{x}} + \tfrac{\sqrt{3}}{2}\,a\,\hat{\mathbf{y}} \\
\mathbf{a}_3 &= c\,\hat{\mathbf{z}}
\end{aligned}
$$

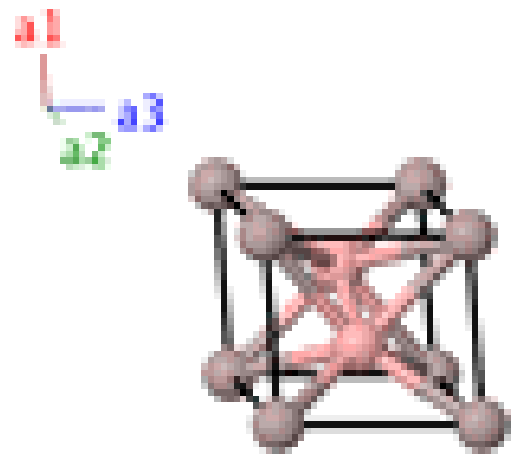

**Basis vectors:**

| | | Lattice Coordinates | | Cartesian Coordinates | Wyckoff Position | Atom Type |
|---|---|---|---|---|---|---|
| $\mathbf{B}_1$ | = | $0\,\mathbf{a}_1 + 0\,\mathbf{a}_2 + 0\,\mathbf{a}_3$ | = | $0\hat{\mathbf{x}} + 0\hat{\mathbf{y}} + 0\hat{\mathbf{z}}$ | (1*a*) | Al |
| $\mathbf{B}_2$ | = | $\frac{1}{3}\mathbf{a}_1 + \frac{2}{3}\mathbf{a}_2 + \frac{1}{2}\mathbf{a}_3$ | = | $\frac{1}{2}\,a\,\hat{\mathbf{x}} + \frac{1}{2\sqrt{3}}\,a\,\hat{\mathbf{y}} + \frac{1}{2}\,c\,\hat{\mathbf{z}}$ | (2*d*) | B |
| $\mathbf{B}_3$ | = | $\frac{2}{3}\mathbf{a}_1 + \frac{1}{3}\mathbf{a}_2 + \frac{1}{2}\mathbf{a}_3$ | = | $\frac{1}{2}\,a\,\hat{\mathbf{x}} - \frac{1}{2\sqrt{3}}\,a\,\hat{\mathbf{y}} + \frac{1}{2}\,c\,\hat{\mathbf{z}}$ | (2*d*) | B |

**Geometry files:**

- CIF: pp. 741
- POSCAR: pp. 741

# $Cu_2Te$ ($C_h$) Structure: A2B_hP6_191_h_e

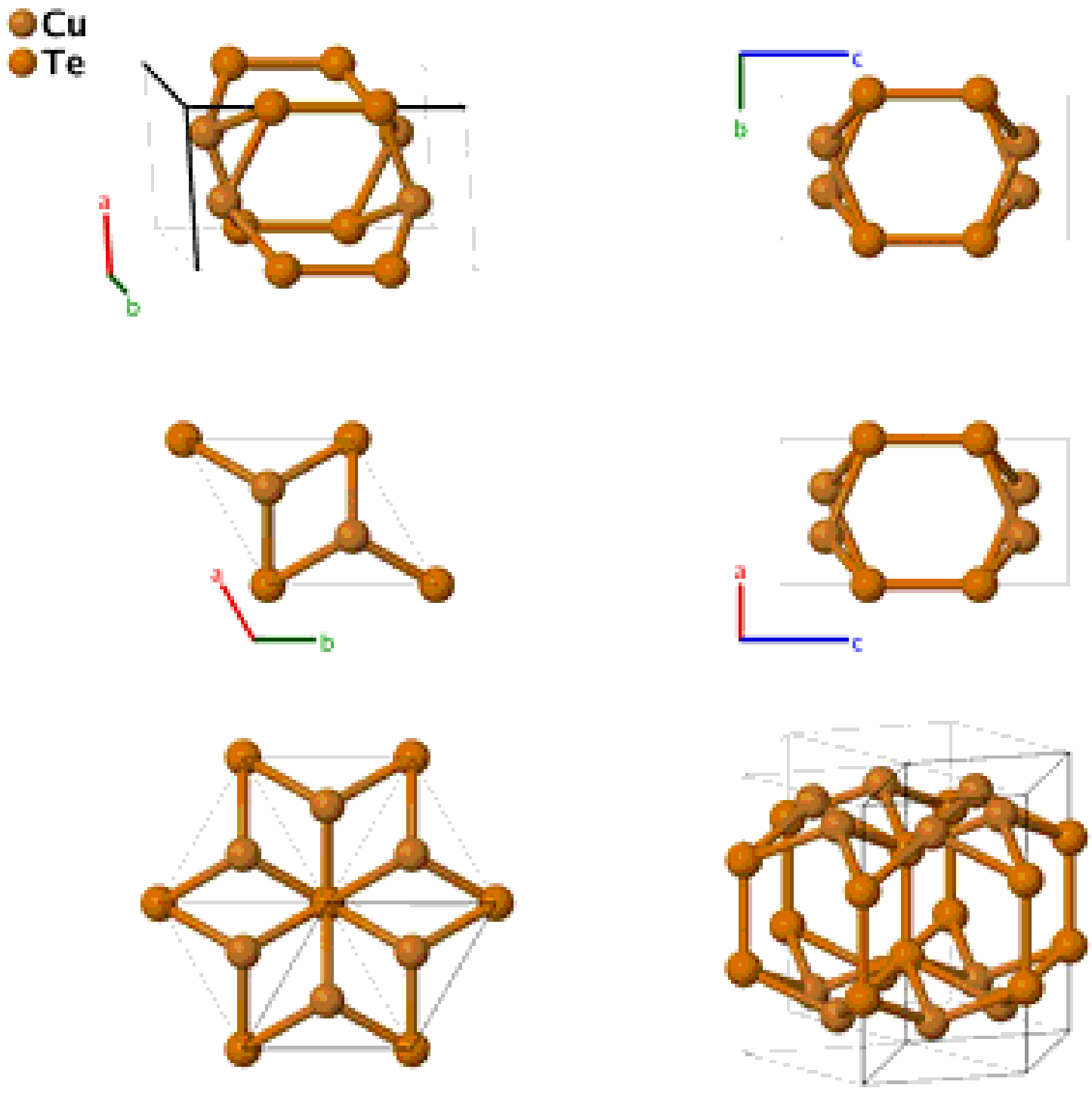

| | | |
|---|---|---|
| **Prototype** | : | $Cu_2Te$ |
| **AFLOW prototype label** | : | A2B_hP6_191_h_e |
| ***Strukturbericht* designation** | : | $C_h$ |
| **Pearson symbol** | : | hP6 |
| **Space group number** | : | 191 |
| **Space group symbol** | : | P6/mmm |
| **AFLOW prototype command** | : | `aflow --proto=A2B_hP6_191_h_e` `--params=`$a, c/a, z_1, z_2$ |

**Hexagonal primitive vectors:**

$$\mathbf{a}_1 = \frac{1}{2}\,a\,\hat{\mathbf{x}} - \frac{\sqrt{3}}{2}\,a\,\hat{\mathbf{y}}$$
$$\mathbf{a}_2 = \frac{1}{2}\,a\,\hat{\mathbf{x}} + \frac{\sqrt{3}}{2}\,a\,\hat{\mathbf{y}}$$
$$\mathbf{a}_3 = c\,\hat{\mathbf{z}}$$

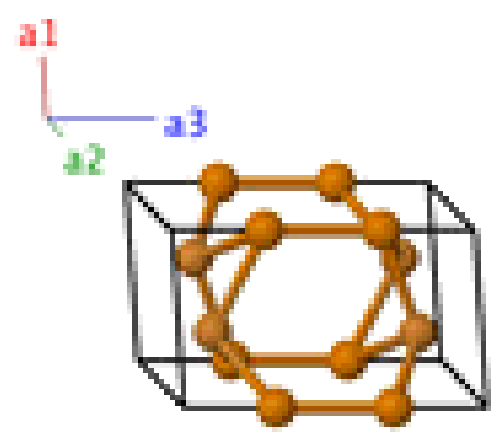

**Basis vectors:**

| | | Lattice Coordinates | | Cartesian Coordinates | Wyckoff Position | Atom Type |
|---|---|---|---|---|---|---|
| $\mathbf{B}_1$ | = | $z_1\,\mathbf{a}_3$ | = | $z_1\,c\,\hat{\mathbf{z}}$ | (2$e$) | Te |
| $\mathbf{B}_2$ | = | $-z_1\,\mathbf{a}_3$ | = | $-z_1\,c\,\hat{\mathbf{z}}$ | (2$e$) | Te |
| $\mathbf{B}_3$ | = | $\frac{1}{3}\mathbf{a}_1 + \frac{2}{3}\mathbf{a}_2 + z_2\,\mathbf{a}_3$ | = | $\frac{1}{2}\,a\,\hat{\mathbf{x}} + \frac{1}{2\sqrt{3}}\,a\,\hat{\mathbf{y}} + z_2\,c\,\hat{\mathbf{z}}$ | (4$h$) | Cu |
| $\mathbf{B}_4$ | = | $\frac{2}{3}\mathbf{a}_1 + \frac{1}{3}\mathbf{a}_2 + z_2\,\mathbf{a}_3$ | = | $\frac{1}{2}\,a\,\hat{\mathbf{x}} - \frac{1}{2\sqrt{3}}\,a\,\hat{\mathbf{y}} + z_2\,c\,\hat{\mathbf{z}}$ | (4$h$) | Cu |
| $\mathbf{B_5}$ | = | $\frac{2}{3}\mathbf{a_1} + \frac{1}{3}\mathbf{a_2} - z_2\,\mathbf{a_3}$ | = | $\frac{1}{2}\,a\,\mathbf{\hat{x}} - \frac{1}{2\sqrt{3}}\,a\,\mathbf{\hat{y}} - z_2\,c\,\mathbf{\hat{z}}$ | (4$h$) | Cu |
| $\mathbf{B_6}$ | = | $\frac{1}{3}\mathbf{a_1} + \frac{2}{3}\mathbf{a_2} - z_2\,\mathbf{a_3}$ | = | $\frac{1}{2}\,a\,\mathbf{\hat{x}} + \frac{1}{2\sqrt{3}}\,a\,\mathbf{\hat{y}} - z_2\,c\,\mathbf{\hat{z}}$ | (4$h$) | Cu |

---

**Found in:**

---

**Geometry files:**

- CIF: pp. 741
- POSCAR: pp. 742

# CoSn (B35) Structure: AB_hP6_191_f_ad

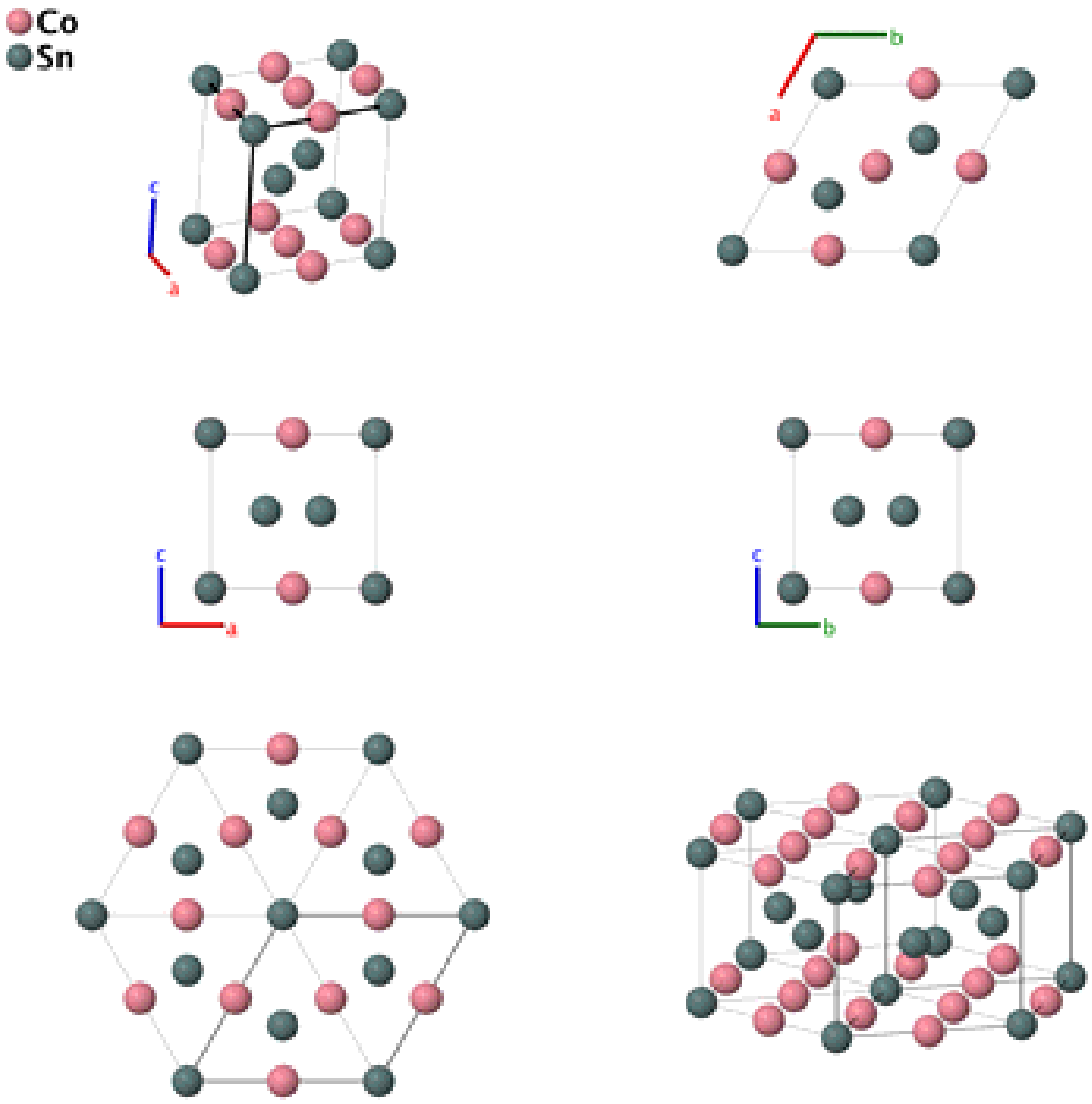

| | | |
|---|---|---|
| **Prototype** | : | CoSn |
| **AFLOW prototype label** | : | AB_hP6_191_f_ad |
| ***Strukturbericht* designation** | : | B35 |
| **Pearson symbol** | : | hP6 |
| **Space group number** | : | 191 |
| **Space group symbol** | : | P6/mmm |
| **AFLOW prototype command** | : | `aflow --proto=AB_hP6_191_f_ad --params=`$a, c/a$ |

**Other compounds with this structure:**

- FeGe, PbRh, NTa, PtTl, InNi, $OTi_2$

**Hexagonal primitive vectors:**

$$\mathbf{a}_1 = \frac{1}{2}\,a\,\hat{\mathbf{x}} - \frac{\sqrt{3}}{2}\,a\,\hat{\mathbf{y}}$$
$$\mathbf{a}_2 = \frac{1}{2}\,a\,\hat{\mathbf{x}} + \frac{\sqrt{3}}{2}\,a\,\hat{\mathbf{y}}$$
$$\mathbf{a}_3 = c\,\hat{\mathbf{z}}$$

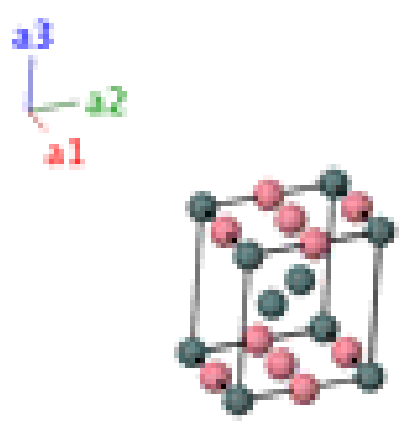

**Basis vectors:**

| | | Lattice Coordinates | | Cartesian Coordinates | Wyckoff Position | Atom Type |
|---|---|---|---|---|---|---|
| $\mathbf{B_1}$ | = | $0\,\mathbf{a}_1 + 0\,\mathbf{a}_2 + 0\,\mathbf{a}_3$ | = | $0\hat{\mathbf{x}} + 0\hat{\mathbf{y}} + 0\hat{\mathbf{z}}$ | $(1a)$ | Sn I |
| $\mathbf{B_2}$ | = | $\frac{1}{3}\mathbf{a}_1 + \frac{2}{3}\mathbf{a}_2 + \frac{1}{2}\mathbf{a}_3$ | = | $\frac{1}{2}\,a\,\hat{\mathbf{x}} + \frac{1}{2\sqrt{3}}\,a\,\hat{\mathbf{y}} + \frac{1}{2}\,c\,\hat{\mathbf{z}}$ | $(2d)$ | Sn II |
| $\mathbf{B_3}$ | = | $\frac{2}{3}\mathbf{a}_1 + \frac{1}{3}\mathbf{a}_2 + \frac{1}{2}\mathbf{a}_3$ | = | $\frac{1}{2}\,a\,\hat{\mathbf{x}} - \frac{1}{2\sqrt{3}}\,a\,\hat{\mathbf{y}} + \frac{1}{2}\,c\,\hat{\mathbf{z}}$ | $(2d)$ | Sn II |
| $\mathbf{B_4}$ | = | $\frac{1}{2}\,\mathbf{a}_1$ | = | $\frac{1}{4}\,a\,\hat{\mathbf{x}} - \frac{\sqrt{3}}{4}\,a\,\hat{\mathbf{y}}$ | $(3f)$ | Co |
| $\mathbf{B_5}$ | = | $\frac{1}{2}\,\mathbf{a}_2$ | = | $\frac{1}{4}\,a\,\hat{\mathbf{x}} + \frac{\sqrt{3}}{4}\,a\,\hat{\mathbf{y}}$ | $(3f)$ | Co |
| $\mathbf{B_6}$ | = | $\frac{1}{2}\,\mathbf{a}_1 + \frac{1}{2}\,\mathbf{a}_2$ | = | $\frac{1}{2}\,a\,\hat{\mathbf{x}}$ | $(3f)$ | Co |

**Geometry files:**

- CIF: pp. 742
- POSCAR: pp. 742

# AsTi ($B_i$) Structure: AB_hP8_194_ad_f

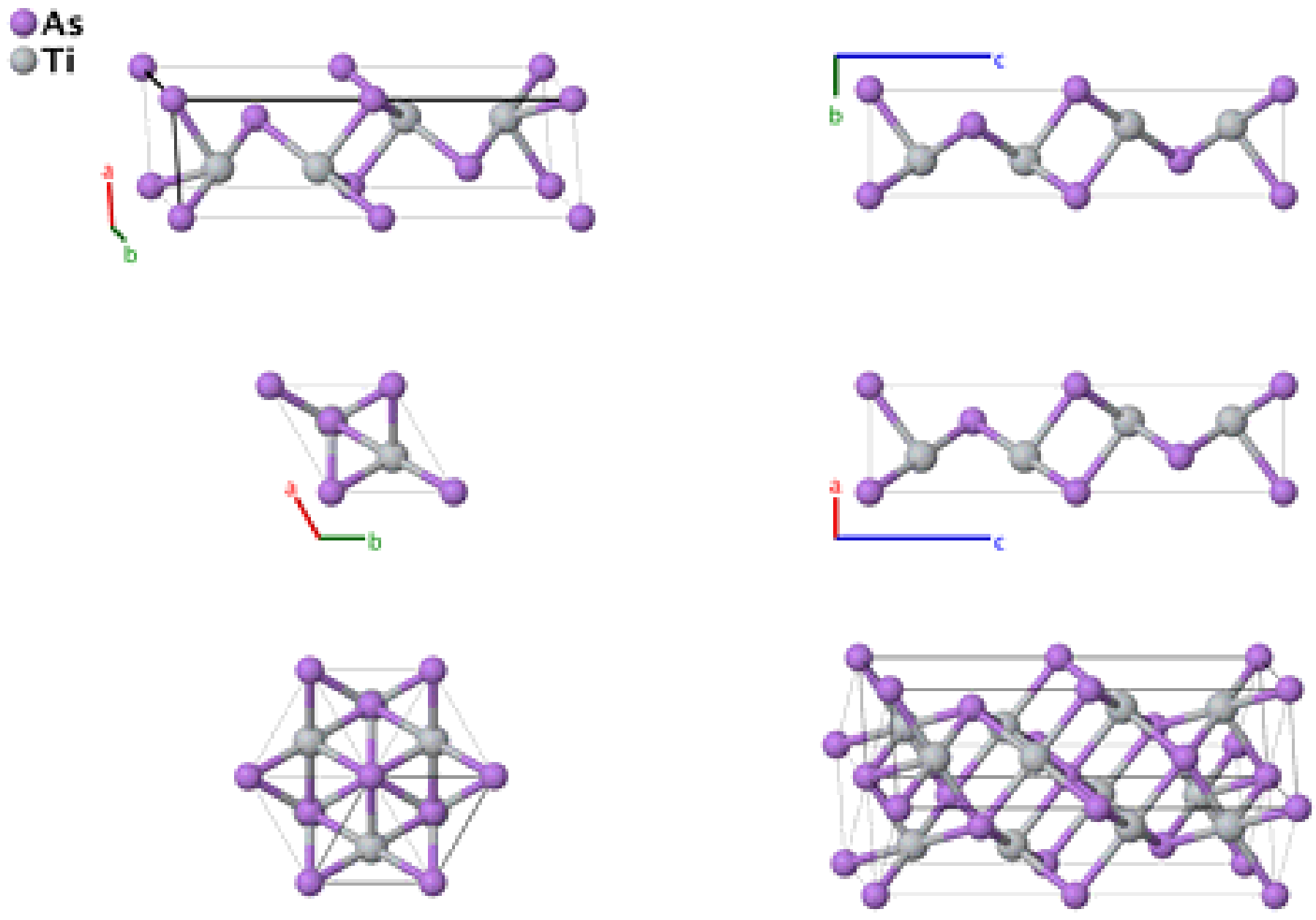

| | | |
|---|---|---|
| **Prototype** | : | AsTi |
| **AFLOW prototype label** | : | AB_hP8_194_ad_f |
| ***Strukturbericht* designation** | : | $B_i$ |
| **Pearson symbol** | : | hP8 |
| **Space group number** | : | 194 |
| **Space group symbol** | : | $P6_3/mmc$ |
| **AFLOW prototype command** | : | `aflow --proto=AB_hP8_194_ad_f`<br>`--params=`$a, c/a, z_3$ |

**Other compounds with this structure:**

- CMo, $CSTi_2$, $CSZr_2$, PTi, PZr

**Hexagonal primitive vectors:**

$$\mathbf{a}_1 = \frac{1}{2}\,a\,\hat{\mathbf{x}} - \frac{\sqrt{3}}{2}\,a\,\hat{\mathbf{y}}$$
$$\mathbf{a}_2 = \frac{1}{2}\,a\,\hat{\mathbf{x}} + \frac{\sqrt{3}}{2}\,a\,\hat{\mathbf{y}}$$
$$\mathbf{a}_3 = c\,\hat{\mathbf{z}}$$

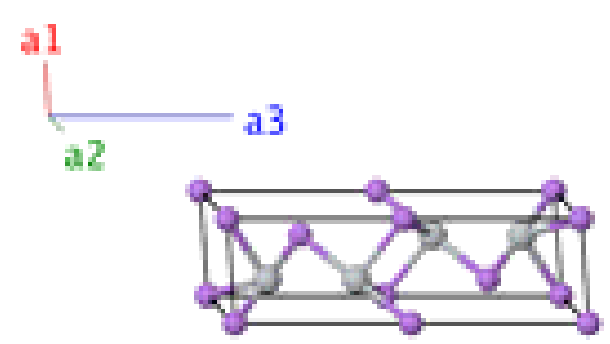

**Basis vectors:**

| | | Lattice Coordinates | | Cartesian Coordinates | Wyckoff Position | Atom Type |
|---|---|---|---|---|---|---|
| $\mathbf{B}_1$ | = | $0\,\mathbf{a}_1 + 0\,\mathbf{a}_2 + 0\,\mathbf{a}_3$ | = | $0\hat{\mathbf{x}} + 0\hat{\mathbf{y}} + 0\hat{\mathbf{z}}$ | (2$a$) | As I |
| $\mathbf{B}_2$ | = | $\frac{1}{2}\,\mathbf{a}_3$ | = | $\frac{1}{2}\,c\,\hat{\mathbf{z}}$ | (2$a$) | As I |
| $\mathbf{B}_3$ | = | $\frac{1}{3}\,\mathbf{a}_1 + \frac{2}{3}\,\mathbf{a}_2 + \frac{3}{4}\,\mathbf{a}_3$ | = | $\frac{1}{2}\,a\,\hat{\mathbf{x}} + \frac{1}{2\sqrt{3}}\,a\,\hat{\mathbf{y}} + \frac{3}{4}\,c\,\hat{\mathbf{z}}$ | (2$d$) | As II |
| $\mathbf{B}_4$ | = | $\frac{2}{3}\,\mathbf{a}_1 + \frac{1}{3}\,\mathbf{a}_2 + \frac{1}{4}\,\mathbf{a}_3$ | = | $\frac{1}{2}\,a\,\hat{\mathbf{x}} - \frac{1}{2\sqrt{3}}\,a\,\hat{\mathbf{y}} + \frac{1}{4}\,c\,\hat{\mathbf{z}}$ | (2$d$) | As II |
| $\mathbf{B_5}$ | = | $\frac{1}{3}\,\mathbf{a_1} + \frac{2}{3}\,\mathbf{a_2} + z_3\,\mathbf{a_3}$ | = | $\frac{1}{2}\,a\,\mathbf{\hat{x}} + \frac{1}{2\sqrt{3}}\,a\,\mathbf{\hat{y}} + z_3\,c\,\mathbf{\hat{z}}$ | $(4f)$ | Ti |
| $\mathbf{B_6}$ | = | $\frac{2}{3}\,\mathbf{a_1} + \frac{1}{3}\,\mathbf{a_2} + \left(\frac{1}{2} + z_3\right)\,\mathbf{a_3}$ | = | $\frac{1}{2}\,a\,\mathbf{\hat{x}} - \frac{1}{2\sqrt{3}}\,a\,\mathbf{\hat{y}} + \left(\frac{1}{2} + z_3\right)\,c\,\mathbf{\hat{z}}$ | $(4f)$ | Ti |
| $\mathbf{B_7}$ | = | $\frac{2}{3}\,\mathbf{a_1} + \frac{1}{3}\,\mathbf{a_2} - z_3\,\mathbf{a_3}$ | = | $\frac{1}{2}\,a\,\mathbf{\hat{x}} - \frac{1}{2\sqrt{3}}\,a\,\mathbf{\hat{y}} - z_3\,c\,\mathbf{\hat{z}}$ | $(4f)$ | Ti |
| $\mathbf{B_8}$ | = | $\frac{1}{3}\,\mathbf{a_1} + \frac{2}{3}\,\mathbf{a_2} + \left(\frac{1}{2} - z_3\right)\,\mathbf{a_3}$ | = | $\frac{1}{2}\,a\,\mathbf{\hat{x}} + \frac{1}{2\sqrt{3}}\,a\,\mathbf{\hat{y}} + \left(\frac{1}{2} - z_3\right)\,c\,\mathbf{\hat{z}}$ | $(4f)$ | Ti |

**Found in:**

**Geometry files:**

- CIF: pp. 742
- POSCAR: pp. 743

# Hypothetical Tetrahedrally Bonded Carbon with 3-Member Rings: A_hP6_194_h

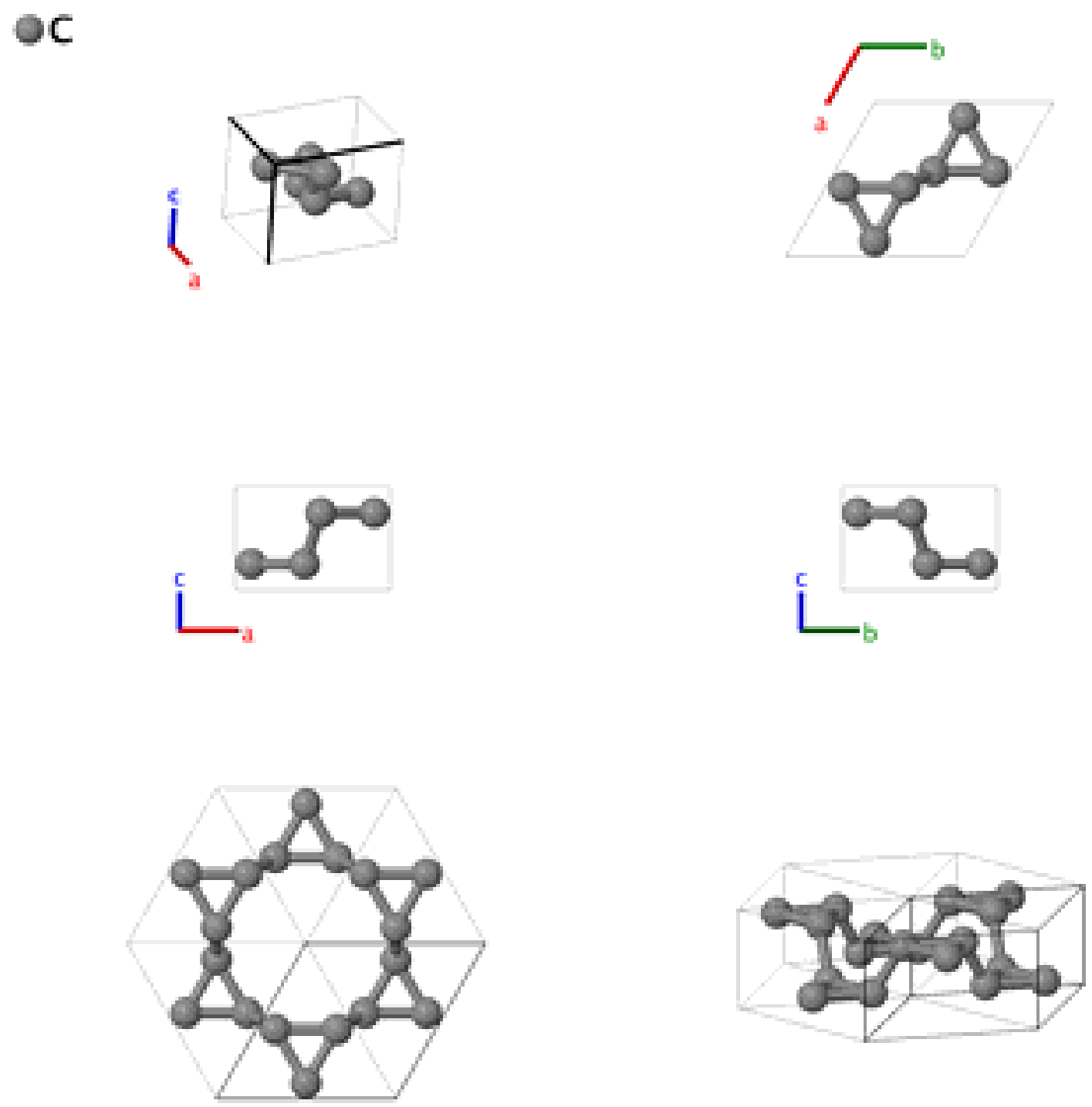

| | | |
|---|---|---|
| **Prototype** | : | C |
| **AFLOW prototype label** | : | A_hP6_194_h |
| ***Strukturbericht* designation** | : | None |
| **Pearson symbol** | : | hP6 |
| **Space group number** | : | 194 |
| **Space group symbol** | : | $P6_3/mmc$ |
| **AFLOW prototype command** | : | `aflow --proto=A_hP6_194_h` <br> `--params=`$a, c/a, x_1$ |

- This structure was proposed in (Schultz, 1999) to show that it was energetically possible to form three-member rings in amorphous $sp^3$ carbon structures.

**Hexagonal primitive vectors:**

$$
\begin{aligned}
\mathbf{a}_1 &= \frac{1}{2}\,a\,\hat{\mathbf{x}} - \frac{\sqrt{3}}{2}\,a\,\hat{\mathbf{y}} \\
\mathbf{a}_2 &= \frac{1}{2}\,a\,\hat{\mathbf{x}} + \frac{\sqrt{3}}{2}\,a\,\hat{\mathbf{y}} \\
\mathbf{a}_3 &= c\,\hat{\mathbf{z}}
\end{aligned}
$$

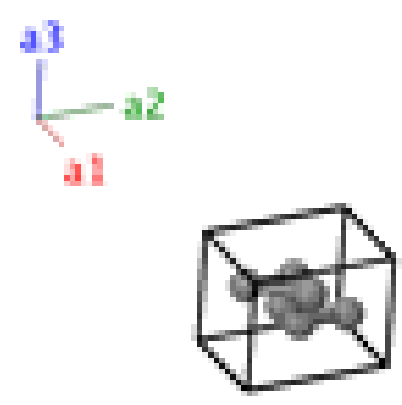

**Basis vectors:**

| | | Lattice Coordinates | | Cartesian Coordinates | Wyckoff Position | Atom Type |
|---|---|---|---|---|---|---|
| $\mathbf{B_1}$ | = | $x_1\,\mathbf{a_1} + 2\,x_1\,\mathbf{a_2} + \frac{1}{4}\,\mathbf{a_3}$ | = | $\frac{3}{2}\,x_1\,a\,\hat{\mathbf{x}} + \frac{\sqrt{3}}{2}\,x_1\,a\,\hat{\mathbf{y}} + \frac{1}{4}\,c\,\hat{\mathbf{z}}$ | (6$h$) | C |
| $\mathbf{B_2}$ | = | $-2\,x_1\,\mathbf{a_1} - x_1\,\mathbf{a_2} + \frac{1}{4}\,\mathbf{a_3}$ | = | $-\frac{3}{2}\,x_1\,a\,\hat{\mathbf{x}} + \frac{\sqrt{3}}{2}\,x_1\,a\,\hat{\mathbf{y}} + \frac{1}{4}\,c\,\hat{\mathbf{z}}$ | (6$h$) | C |
| $\mathbf{B_3}$ | = | $x_1\,\mathbf{a_1} - x_1\,\mathbf{a_2} + \frac{1}{4}\,\mathbf{a_3}$ | = | $-\sqrt{3}\,x_1\,a\,\hat{\mathbf{y}} + \frac{1}{4}\,c\,\hat{\mathbf{z}}$ | (6$h$) | C |
| $\mathbf{B_4}$ | = | $-x_1\,\mathbf{a_1} - 2\,x_1\,\mathbf{a_2} + \frac{3}{4}\,\mathbf{a_3}$ | = | $-\frac{3}{2}\,x_1\,a\,\hat{\mathbf{x}} - \frac{\sqrt{3}}{2}\,x_1\,a\,\hat{\mathbf{y}} + \frac{3}{4}\,c\,\hat{\mathbf{z}}$ | (6$h$) | C |
| $\mathbf{B_5}$ | = | $2\,x_1\,\mathbf{a_1} + x_1\,\mathbf{a_2} + \frac{3}{4}\,\mathbf{a_3}$ | = | $\frac{3}{2}\,x_1\,a\,\hat{\mathbf{x}} - \frac{\sqrt{3}}{2}\,x_1\,a\,\hat{\mathbf{y}} + \frac{3}{4}\,c\,\hat{\mathbf{z}}$ | (6$h$) | C |
| $\mathbf{B_6}$ | = | $-x_1\,\mathbf{a_1} + x_1\,\mathbf{a_2} + \frac{3}{4}\,\mathbf{a_3}$ | = | $+\sqrt{3}\,x_1\,a\,\hat{\mathbf{y}} + \frac{3}{4}\,c\,\hat{\mathbf{z}}$ | (6$h$) | C |

**Geometry files:**

- CIF: pp. 743
- POSCAR: pp. 743

# CMo Structure: AB_hP12_194_af_bf

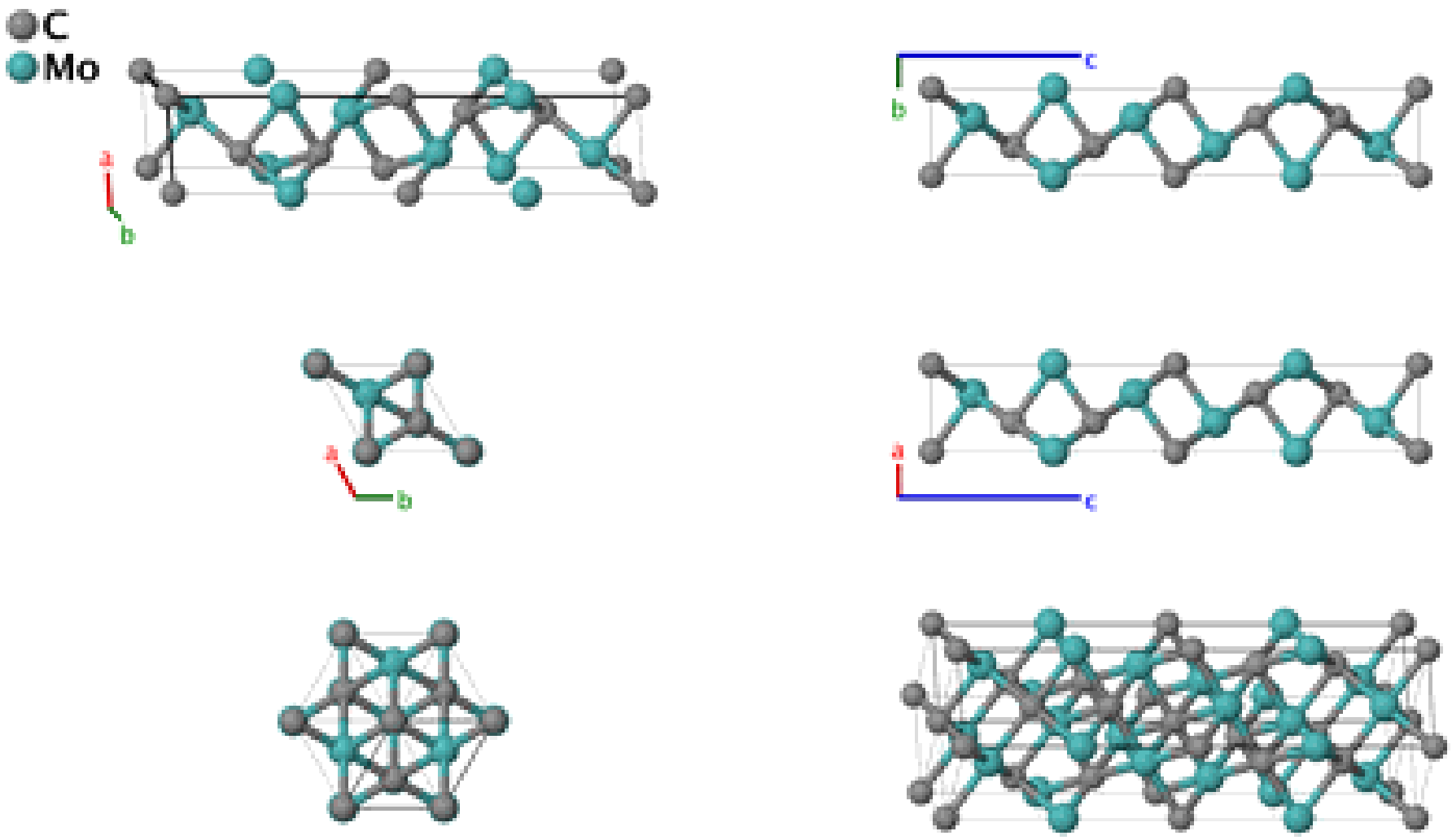

| | | |
|---|---|---|
| **Prototype** | : | CMo |
| **AFLOW prototype label** | : | AB_hP12_194_af_bf |
| ***Strukturbericht* designation** | : | None |
| **Pearson symbol** | : | hP12 |
| **Space group number** | : | 194 |
| **Space group symbol** | : | $P6_3/mmc$ |
| **AFLOW prototype command** | : | `aflow --proto=AB_hP12_194_af_bf`<br>`--params=`$a, c/a, z_3, z_4$ |

**Other compounds with this structure:**

- CRe, $C_2GeTi_3$, $C_2SiTi_3$, $AlC_2Ti_3$, others.

- Note that all of the atoms sit on close packed <0001> planes. The stacking sequence may be written:

| Atom | Mo-II | C-II | C-I | C-II | Mo-II | Mo-I | Mo-II | C-II | C-I | C-II | Mo-II | Mo-I |
|---|---|---|---|---|---|---|---|---|---|---|---|---|
| Position | B | C | A | B | C | A | C | B | A | C | B | A |

Thus the Mo-II atoms and all of the C atoms are always in an fcc-like local environment, while the Mo-I atoms are in an hcp-like local environment. Like AlN3Ti4, this is a MAX phase. For more information, see (Radovic, 2013).

**Hexagonal primitive vectors:**

$$\mathbf{a}_1 = \frac{1}{2}\,a\,\hat{\mathbf{x}} - \frac{\sqrt{3}}{2}\,a\,\hat{\mathbf{y}}$$
$$\mathbf{a}_2 = \frac{1}{2}\,a\,\hat{\mathbf{x}} + \frac{\sqrt{3}}{2}\,a\,\hat{\mathbf{y}}$$
$$\mathbf{a}_3 = c\,\hat{\mathbf{z}}$$

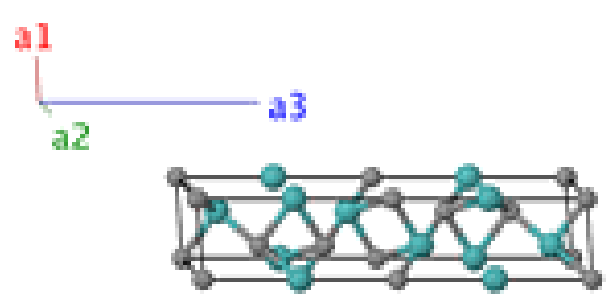

**Basis vectors:**

| | | Lattice Coordinates | | Cartesian Coordinates | Wyckoff Position | Atom Type |
|---|---|---|---|---|---|---|
| $\mathbf{B}_1$ | = | $0\,\mathbf{a}_1 + 0\,\mathbf{a}_2 + 0\,\mathbf{a}_3$ | = | $0\hat{\mathbf{x}} + 0\hat{\mathbf{y}} + 0\hat{\mathbf{z}}$ | $(2a)$ | C I |
| $\mathbf{B}_2$ | = | $\frac{1}{2}\,\mathbf{a}_3$ | = | $\frac{1}{2}\,c\,\hat{\mathbf{z}}$ | (2$a$) | C I |
| $\mathbf{B}_3$ | = | $\frac{1}{4}\,\mathbf{a}_3$ | = | $\frac{1}{4}\,c\,\hat{\mathbf{z}}$ | (2$b$) | Mo I |
| $\mathbf{B}_4$ | = | $\frac{3}{4}\,\mathbf{a}_3$ | = | $\frac{3}{4}\,c\,\hat{\mathbf{z}}$ | (2$b$) | Mo I |
| $\mathbf{B}_5$ | = | $\frac{1}{3}\,\mathbf{a}_1 + \frac{2}{3}\,\mathbf{a}_2 + z_3\,\mathbf{a}_3$ | = | $\frac{1}{2}\,a\,\hat{\mathbf{x}} + \frac{1}{2\sqrt{3}}\,a\,\hat{\mathbf{y}} + z_3\,c\,\hat{\mathbf{z}}$ | (4$f$) | C II |
| $\mathbf{B}_6$ | = | $\frac{2}{3}\,\mathbf{a}_1 + \frac{1}{3}\,\mathbf{a}_2 + \left(\frac{1}{2} + z_3\right)\,\mathbf{a}_3$ | = | $\frac{1}{2}\,a\,\hat{\mathbf{x}} - \frac{1}{2\sqrt{3}}\,a\,\hat{\mathbf{y}} + \left(\frac{1}{2} + z_3\right)\,c\,\hat{\mathbf{z}}$ | (4$f$) | C II |
| $\mathbf{B}_7$ | = | $\frac{2}{3}\,\mathbf{a}_1 + \frac{1}{3}\,\mathbf{a}_2 - z_3\,\mathbf{a}_3$ | = | $\frac{1}{2}\,a\,\hat{\mathbf{x}} - \frac{1}{2\sqrt{3}}\,a\,\hat{\mathbf{y}} - z_3\,c\,\hat{\mathbf{z}}$ | (4$f$) | C II |
| $\mathbf{B}_8$ | = | $\frac{1}{3}\,\mathbf{a}_1 + \frac{2}{3}\,\mathbf{a}_2 + \left(\frac{1}{2} - z_3\right)\,\mathbf{a}_3$ | = | $\frac{1}{2}\,a\,\hat{\mathbf{x}} + \frac{1}{2\sqrt{3}}\,a\,\hat{\mathbf{y}} + \left(\frac{1}{2} - z_3\right)\,c\,\hat{\mathbf{z}}$ | (4$f$) | C II |
| $\mathbf{B}_9$ | = | $\frac{1}{3}\,\mathbf{a}_1 + \frac{2}{3}\,\mathbf{a}_2 + z_4\,\mathbf{a}_3$ | = | $\frac{1}{2}\,a\,\hat{\mathbf{x}} + \frac{1}{2\sqrt{3}}\,a\,\hat{\mathbf{y}} + z_4\,c\,\hat{\mathbf{z}}$ | (4$f$) | Mo II |
| $\mathbf{B}_{10}$ | = | $\frac{2}{3}\,\mathbf{a}_1 + \frac{1}{3}\,\mathbf{a}_2 + \left(\frac{1}{2} + z_4\right)\,\mathbf{a}_3$ | = | $\frac{1}{2}\,a\,\hat{\mathbf{x}} - \frac{1}{2\sqrt{3}}\,a\,\hat{\mathbf{y}} + \left(\frac{1}{2} + z_4\right)\,c\,\hat{\mathbf{z}}$ | (4$f$) | Mo II |
| $\mathbf{B}_{11}$ | = | $\frac{2}{3}\,\mathbf{a}_1 + \frac{1}{3}\,\mathbf{a}_2 - z_4\,\mathbf{a}_3$ | = | $\frac{1}{2}\,a\,\hat{\mathbf{x}} - \frac{1}{2\sqrt{3}}\,a\,\hat{\mathbf{y}} - z_4\,c\,\hat{\mathbf{z}}$ | (4$f$) | Mo II |
| $\mathbf{B}_{12}$ | = | $\frac{1}{3}\,\mathbf{a}_1 + \frac{2}{3}\,\mathbf{a}_2 + \left(\frac{1}{2} - z_4\right)\,\mathbf{a}_3$ | = | $\frac{1}{2}\,a\,\hat{\mathbf{x}} + \frac{1}{2\sqrt{3}}\,a\,\hat{\mathbf{y}} + \left(\frac{1}{2} - z_4\right)\,c\,\hat{\mathbf{z}}$ | (4$f$) | Mo II |

**Geometry files:**

- CIF: pp. 743
- POSCAR: pp. 744

# $\alpha$-La (A3') Structure: A_hP4_194_ac

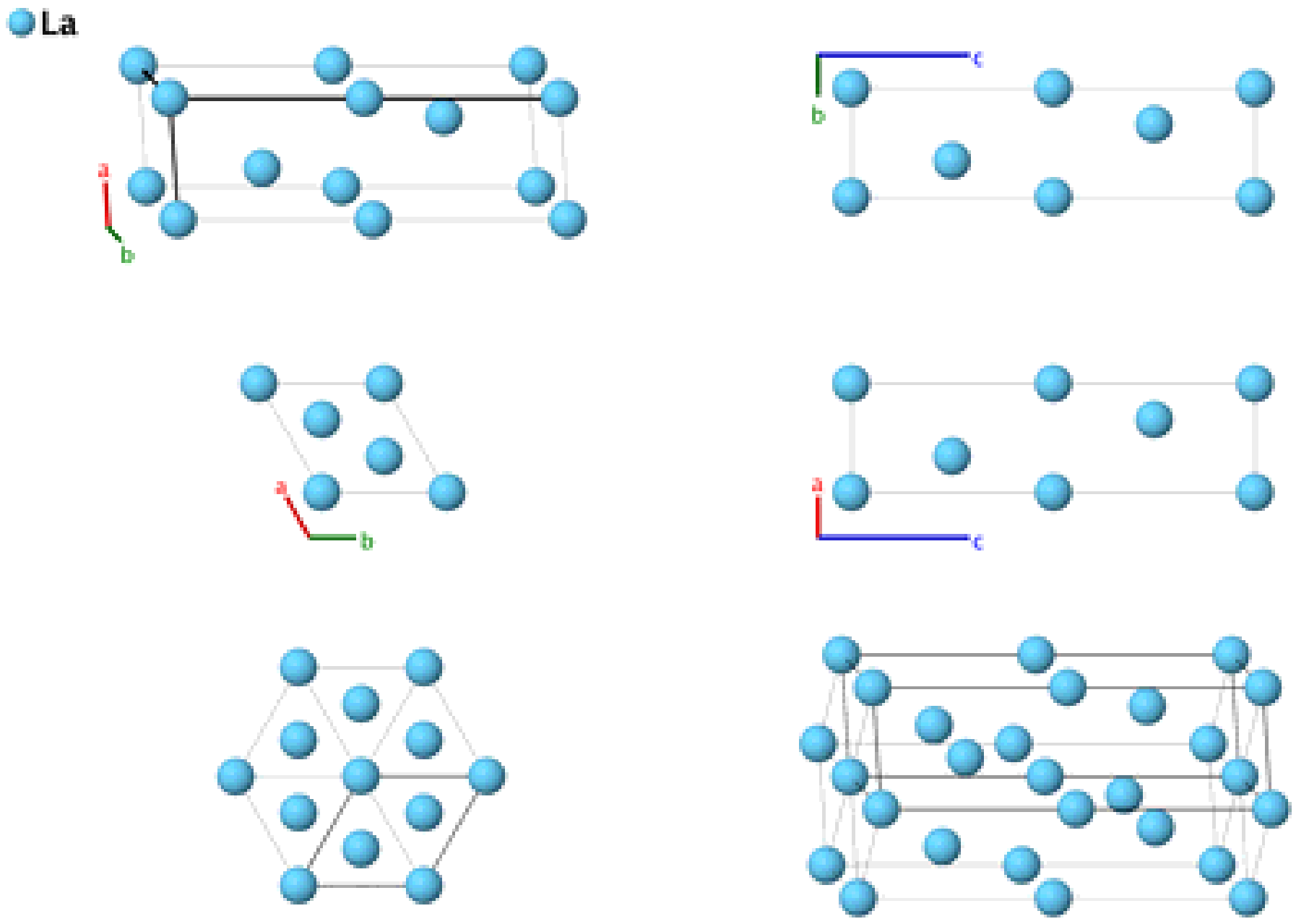

**Prototype** : $\alpha$-La

**AFLOW prototype label** : A_hP4_194_ac

***Strukturbericht* designation** : A3'

**Pearson symbol** : hP4

**Space group number** : 194

**Space group symbol** : $P6_3/mmc$

**AFLOW prototype command** : `aflow --proto=A_hP4_194_ac`
`--params=`$a, c/a$

---

**Other elements with this structure:**

- Pr, Nd, Pm, Ce, Am, Cm, Bk, Cf.

---

- This crystal is close-packed, with stacking ABACABAC..., as opposed to ABAB... for the hcp (A3) lattice and ABCABC... for the fcc (A1) lattice. The (2a) crystallographic sites (the A's) form a simple hexagonal lattice. The (2c) sites (the B's and C's) form an hcp structure.

---

**Hexagonal primitive vectors:**

$$\mathbf{a}_1 = \frac{1}{2}\,a\,\hat{\mathbf{x}} - \frac{\sqrt{3}}{2}\,a\,\hat{\mathbf{y}}$$

$$\mathbf{a}_2 = \frac{1}{2}\,a\,\hat{\mathbf{x}} + \frac{\sqrt{3}}{2}\,a\,\hat{\mathbf{y}}$$

$$\mathbf{a}_3 = c\,\hat{\mathbf{z}}$$

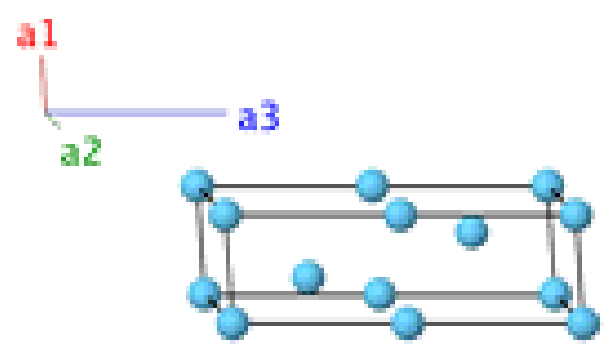

---

**Basis vectors:**

| Lattice Coordinates | Cartesian Coordinates | Wyckoff Position | Atom Type |
|---|---|---|---|

| | | | | | | |
|---|---|---|---|---|---|---|
| $\mathbf{B_1}$ | = | $0\,\mathbf{a_1} + 0\,\mathbf{a_2} + 0\,\mathbf{a_3}$ | = | $0\hat{\mathbf{x}} + 0\hat{\mathbf{y}} + 0\hat{\mathbf{z}}$ | $(2a)$ | La I |
| $\mathbf{B_2}$ | = | $\frac{1}{2}\,\mathbf{a_3}$ | = | $\frac{1}{2}\,c\,\hat{\mathbf{z}}$ | $(2a)$ | La I |
| $\mathbf{B_3}$ | = | $\frac{1}{3}\,\mathbf{a_1} + \frac{2}{3}\,\mathbf{a_2} + \frac{1}{4}\,\mathbf{a_3}$ | = | $\frac{1}{2}\,a\,\hat{\mathbf{x}} + \frac{1}{2\sqrt{3}}\,a\,\hat{\mathbf{y}} + \frac{1}{4}\,c\,\hat{\mathbf{z}}$ | $(2c)$ | La II |
| $\mathbf{B_4}$ | = | $\frac{2}{3}\,\mathbf{a_1} + \frac{1}{3}\,\mathbf{a_2} + \frac{3}{4}\,\mathbf{a_3}$ | = | $\frac{1}{2}\,a\,\hat{\mathbf{x}} - \frac{1}{2\sqrt{3}}\,a\,\hat{\mathbf{y}} + \frac{3}{4}\,c\,\hat{\mathbf{z}}$ | $(2c)$ | La II |

**Found in:**

**Geometry files:**

- CIF: pp. 744
- POSCAR: pp. 744

# $Na_3As$ ($D0_{18}$) Structure: AB3_hP8_194_c_bf

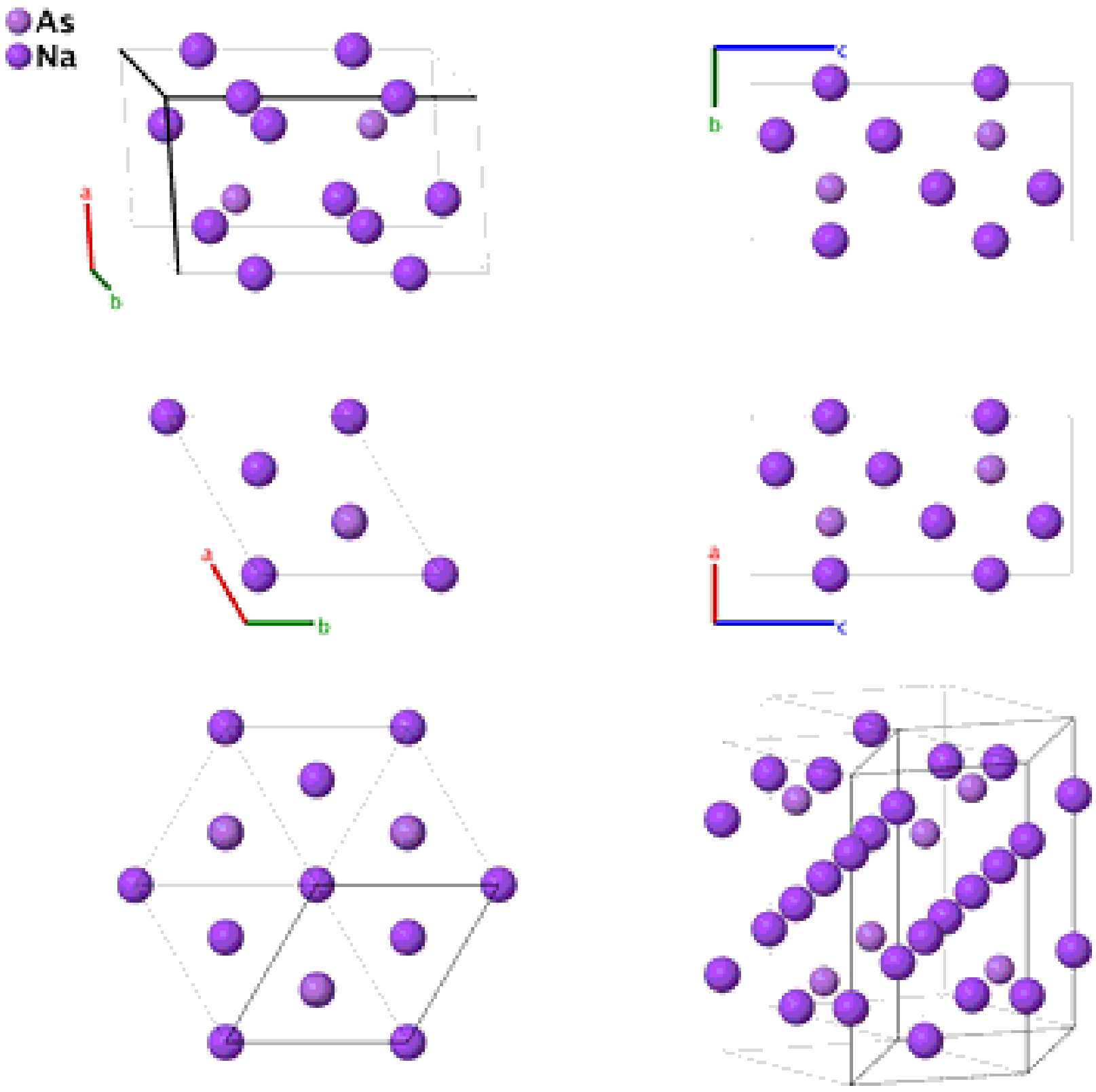

| | | |
|---|---|---|
| **Prototype** | : | $Na_3As$ |
| **AFLOW prototype label** | : | AB3_hP8_194_c_bf |
| ***Strukturbericht* designation** | : | $D0_{18}$ |
| **Pearson symbol** | : | hP8 |
| **Space group number** | : | 194 |
| **Space group symbol** | : | $P6_3/mmc$ |
| **AFLOW prototype command** | : | `aflow --proto=AB3_hP8_194_c_bf`<br>`--params=`$a, c/a, z_3$ |

- (Hafner, 1994) argue that this is not the correct structure for $Na_3As$. We will keep the $D0_{18}$ designation for this structure, and add a page for the new $Na_3As$ structure in a future update.

**Hexagonal primitive vectors:**

$$
\begin{aligned}
\mathbf{a}_1 &= \tfrac{1}{2}\,a\,\mathbf{\hat{x}} - \tfrac{\sqrt{3}}{2}\,a\,\mathbf{\hat{y}} \\
\mathbf{a}_2 &= \tfrac{1}{2}\,a\,\mathbf{\hat{x}} + \tfrac{\sqrt{3}}{2}\,a\,\mathbf{\hat{y}} \\
\mathbf{a}_3 &= c\,\mathbf{\hat{z}}
\end{aligned}
$$

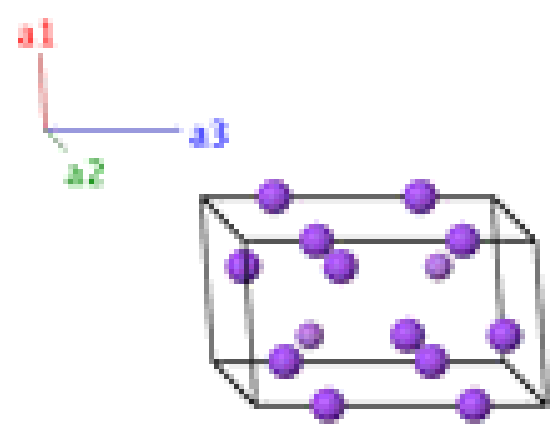

**Basis vectors:**

| Lattice Coordinates | Cartesian Coordinates | Wyckoff Position | Atom Type |
|---|---|---|---|

| | | | | | | |
|---|---|---|---|---|---|---|
| $\mathbf{B_1}$ | = | $\frac{1}{4}\,\mathbf{a_3}$ | = | $\frac{1}{4}\,c\,\hat{\mathbf{z}}$ | (2$b$) | Na I |
| $\mathbf{B_2}$ | = | $\frac{3}{4}\,\mathbf{a_3}$ | = | $\frac{3}{4}\,c\,\hat{\mathbf{z}}$ | (2$b$) | Na I |
| $\mathbf{B_3}$ | = | $\frac{1}{3}\,\mathbf{a_1} + \frac{2}{3}\,\mathbf{a_2} + \frac{1}{4}\,\mathbf{a_3}$ | = | $\frac{1}{2}\,a\,\hat{\mathbf{x}} + \frac{1}{2\sqrt{3}}\,a\,\hat{\mathbf{y}} + \frac{1}{4}\,c\,\hat{\mathbf{z}}$ | (2$c$) | As |
| $\mathbf{B_4}$ | = | $\frac{2}{3}\,\mathbf{a_1} + \frac{1}{3}\,\mathbf{a_2} + \frac{3}{4}\,\mathbf{a_3}$ | = | $\frac{1}{2}\,a\,\hat{\mathbf{x}} - \frac{1}{2\sqrt{3}}\,a\,\hat{\mathbf{y}} + \frac{3}{4}\,c\,\hat{\mathbf{z}}$ | (2$c$) | As |
| $\mathbf{B_5}$ | = | $\frac{1}{3}\,\mathbf{a_1} + \frac{2}{3}\,\mathbf{a_2} + z_3\,\mathbf{a_3}$ | = | $\frac{1}{2}\,a\,\hat{\mathbf{x}} + \frac{1}{2\sqrt{3}}\,a\,\hat{\mathbf{y}} + z_3\,c\,\hat{\mathbf{z}}$ | (4$f$) | Na II |
| $\mathbf{B_6}$ | = | $\frac{2}{3}\,\mathbf{a_1} + \frac{1}{3}\,\mathbf{a_2} + \left(\frac{1}{2} + z_3\right)\,\mathbf{a_3}$ | = | $\frac{1}{2}\,a\,\hat{\mathbf{x}} - \frac{1}{2\sqrt{3}}\,a\,\hat{\mathbf{y}} + \left(\frac{1}{2} + z_3\right)\,c\,\hat{\mathbf{z}}$ | (4$f$) | Na II |
| $\mathbf{B_7}$ | = | $\frac{2}{3}\,\mathbf{a_1} + \frac{1}{3}\,\mathbf{a_2} - z_3\,\mathbf{a_3}$ | = | $\frac{1}{2}\,a\,\hat{\mathbf{x}} - \frac{1}{2\sqrt{3}}\,a\,\hat{\mathbf{y}} - z_3\,c\,\hat{\mathbf{z}}$ | (4$f$) | Na II |
| $\mathbf{B_8}$ | = | $\frac{1}{3}\,\mathbf{a_1} + \frac{2}{3}\,\mathbf{a_2} + \left(\frac{1}{2} - z_3\right)\,\mathbf{a_3}$ | = | $\frac{1}{2}\,a\,\hat{\mathbf{x}} + \frac{1}{2\sqrt{3}}\,a\,\hat{\mathbf{y}} + \left(\frac{1}{2} - z_3\right)\,c\,\hat{\mathbf{z}}$ | (4$f$) | Na II |

**Found in:**

**Geometry files:**

- CIF: pp. 744
- POSCAR: pp. 745

# $CaIn_2$ Structure: AB2_hP6_194_b_f

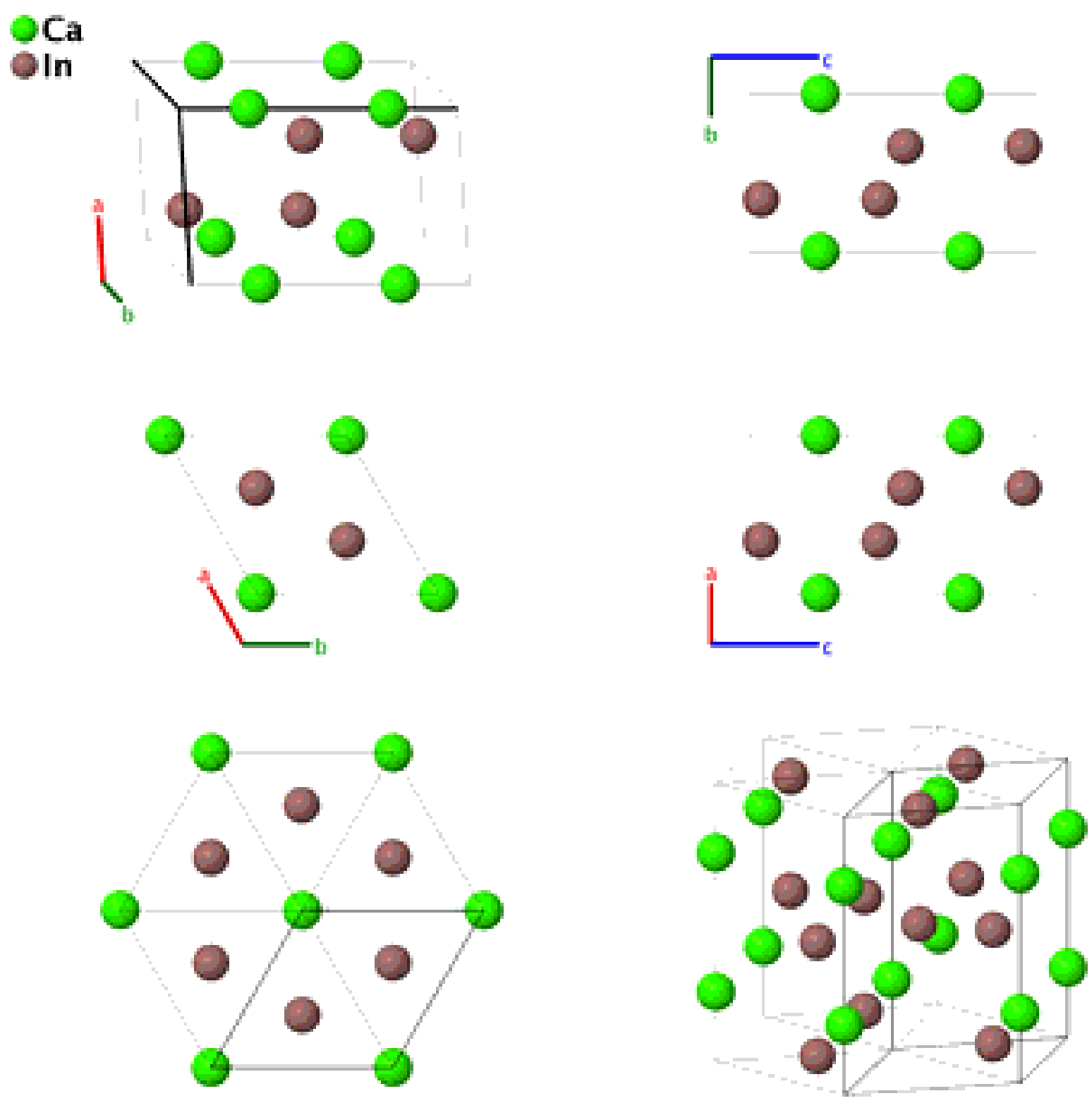

| | | |
|---|---|---|
| **Prototype** | : | $CaIn_2$ |
| **AFLOW prototype label** | : | AB2_hP6_194_b_f |
| ***Strukturbericht* designation** | : | None |
| **Pearson symbol** | : | hP6 |
| **Space group number** | : | 194 |
| **Space group symbol** | : | $P6_3/mmc$ |
| **AFLOW prototype command** | : | `aflow --proto=AB2_hP6_194_b_f`<br>`--params=`$a, c/a, z_2$ |

- When $z_2 = 0$ this reduces to the $AlB_2$ (C32), aka the hexagonal $\omega$ phase.

**Hexagonal primitive vectors:**

$$
\begin{aligned}
\mathbf{a}_1 &= \tfrac{1}{2}\,a\,\hat{\mathbf{x}} - \tfrac{\sqrt{3}}{2}\,a\,\hat{\mathbf{y}} \\
\mathbf{a}_2 &= \tfrac{1}{2}\,a\,\hat{\mathbf{x}} + \tfrac{\sqrt{3}}{2}\,a\,\hat{\mathbf{y}} \\
\mathbf{a}_3 &= c\,\hat{\mathbf{z}}
\end{aligned}
$$

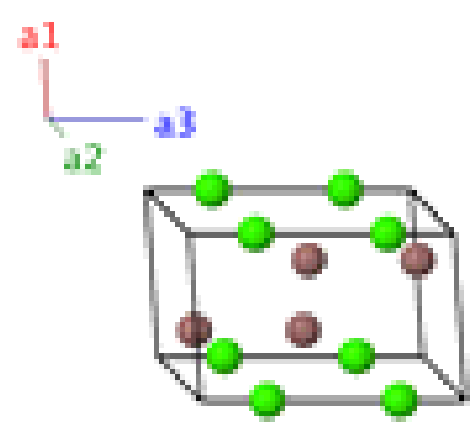

**Basis vectors:**

| | | Lattice Coordinates | | Cartesian Coordinates | Wyckoff Position | Atom Type |
|---|---|---|---|---|---|---|
| $\mathbf{B}_1$ | = | $\frac{1}{4}\,\mathbf{a}_3$ | = | $\frac{1}{4}\,c\,\hat{\mathbf{z}}$ | (2b) | Ca |
| $\mathbf{B_2}$ | = | $\frac{3}{4}\,\mathbf{a_3}$ | = | $\frac{3}{4}\,c\,\mathbf{\hat{z}}$ | (2$b$) | Ca |
| $\mathbf{B_3}$ | = | $\frac{1}{3}\,\mathbf{a_1} + \frac{2}{3}\,\mathbf{a_2} + z_2\,\mathbf{a_3}$ | = | $\frac{1}{2}\,a\,\mathbf{\hat{x}} + \frac{1}{2\sqrt{3}}\,a\,\mathbf{\hat{y}} + z_2\,c\,\mathbf{\hat{z}}$ | (4$f$) | In |
| $\mathbf{B_4}$ | = | $\frac{2}{3}\,\mathbf{a_1} + \frac{1}{3}\,\mathbf{a_2} + \left(\frac{1}{2} + z_2\right)\,\mathbf{a_3}$ | = | $\frac{1}{2}\,a\,\mathbf{\hat{x}} - \frac{1}{2\sqrt{3}}\,a\,\mathbf{\hat{y}} + \left(\frac{1}{2} + z_2\right)\,c\,\mathbf{\hat{z}}$ | (4$f$) | In |
| $\mathbf{B_5}$ | = | $\frac{2}{3}\,\mathbf{a_1} + \frac{1}{3}\,\mathbf{a_2} - z_2\,\mathbf{a_3}$ | = | $\frac{1}{2}\,a\,\mathbf{\hat{x}} - \frac{1}{2\sqrt{3}}\,a\,\mathbf{\hat{y}} - z_2\,c\,\mathbf{\hat{z}}$ | (4$f$) | In |
| $\mathbf{B_6}$ | = | $\frac{1}{3}\,\mathbf{a_1} + \frac{2}{3}\,\mathbf{a_2} + \left(\frac{1}{2} - z_2\right)\,\mathbf{a_3}$ | = | $\frac{1}{2}\,a\,\mathbf{\hat{x}} + \frac{1}{2\sqrt{3}}\,a\,\mathbf{\hat{y}} + \left(\frac{1}{2} - z_2\right)\,c\,\mathbf{\hat{z}}$ | (4$f$) | In |

---

**Found in:**

---

**Geometry files:**

- CIF: pp. 745
- POSCAR: pp. 745

# BN ($B_k$) Structure: AB_hP4_194_c_d

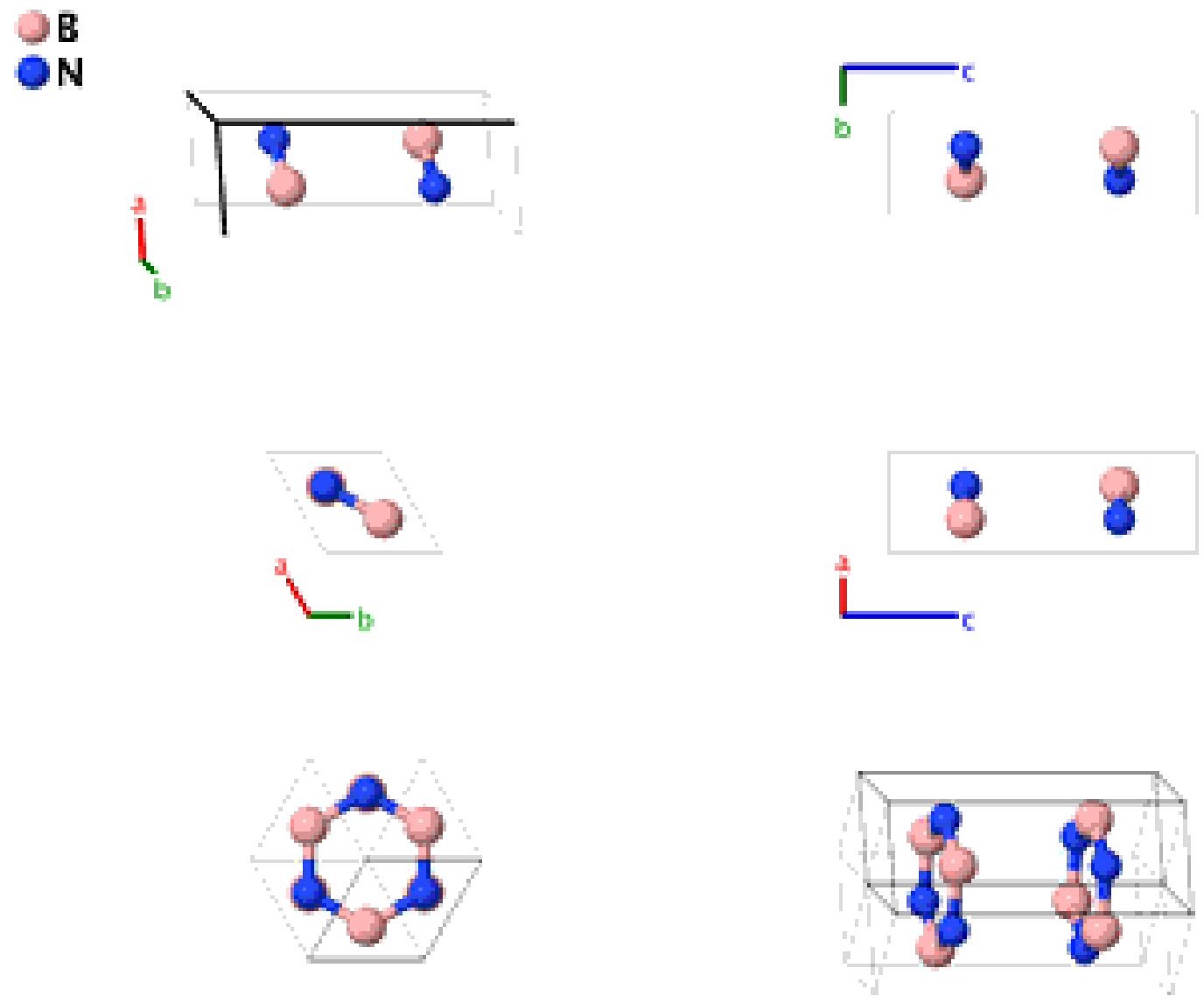

| | | |
|---|---|---|
| **Prototype** | : | BN |
| **AFLOW prototype label** | : | AB_hP4_194_c_d |
| ***Strukturbericht* designation** | : | $B_k$ |
| **Pearson symbol** | : | hP4 |
| **Space group number** | : | 194 |
| **Space group symbol** | : | $P6_3/mmc$ |
| **AFLOW prototype command** | : | `aflow --proto=AB_hP4_194_c_d`<br>`--params=`$a, c/a$ |

**Other compounds with this structure:**

- ZnO nanowires

- This is the corrected boron nitride structure found by (Pease, 1950) and (Pease, 1952). See further discussion on the B12 page.

**Hexagonal primitive vectors:**

$$
\begin{aligned}
\mathbf{a}_1 &= \tfrac{1}{2}\,a\,\hat{\mathbf{x}} - \tfrac{\sqrt{3}}{2}\,a\,\hat{\mathbf{y}} \\
\mathbf{a}_2 &= \tfrac{1}{2}\,a\,\hat{\mathbf{x}} + \tfrac{\sqrt{3}}{2}\,a\,\hat{\mathbf{y}} \\
\mathbf{a}_3 &= c\,\hat{\mathbf{z}}
\end{aligned}
$$

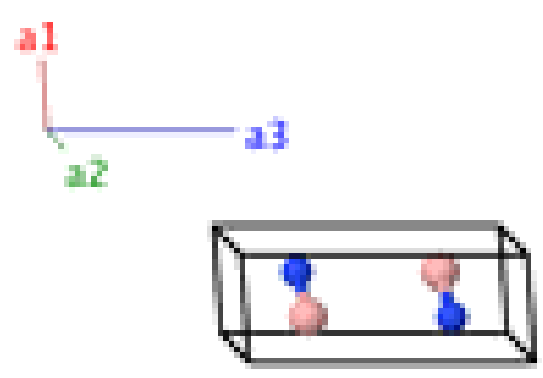

**Basis vectors:**

| | | Lattice Coordinates | | Cartesian Coordinates | Wyckoff Position | Atom Type |
|---|---|---|---|---|---|---|
| $\mathbf{B}_1$ | = | $\frac{1}{3}\,\mathbf{a}_1 + \frac{2}{3}\,\mathbf{a}_2 + \frac{1}{4}\,\mathbf{a}_3$ | = | $\frac{1}{2}\,a\,\hat{\mathbf{x}} + \frac{1}{2\sqrt{3}}\,a\,\hat{\mathbf{y}} + \frac{1}{4}\,c\,\hat{\mathbf{z}}$ | (2$c$) | B |
| $\mathbf{B}_2$ | = | $\frac{2}{3}\,\mathbf{a}_1 + \frac{1}{3}\,\mathbf{a}_2 + \frac{3}{4}\,\mathbf{a}_3$ | = | $\frac{1}{2}\,a\,\hat{\mathbf{x}} - \frac{1}{2\sqrt{3}}\,a\,\hat{\mathbf{y}} + \frac{3}{4}\,c\,\hat{\mathbf{z}}$ | (2$c$) | B |
| $\mathbf{B_3}$ | = | $\frac{1}{3}\,\mathbf{a_1} + \frac{2}{3}\,\mathbf{a_2} + \frac{3}{4}\,\mathbf{a_3}$ | = | $\frac{1}{2}\,a\,\mathbf{\hat{x}} + \frac{1}{2\sqrt{3}}\,a\,\mathbf{\hat{y}} + \frac{3}{4}\,c\,\mathbf{\hat{z}}$ | (2$d$) | N |
| $\mathbf{B_4}$ | = | $\frac{2}{3}\,\mathbf{a_1} + \frac{1}{3}\,\mathbf{a_2} + \frac{1}{4}\,\mathbf{a_3}$ | = | $\frac{1}{2}\,a\,\mathbf{\hat{x}} - \frac{1}{2\sqrt{3}}\,a\,\mathbf{\hat{y}} + \frac{1}{4}\,c\,\mathbf{\hat{z}}$ | (2$d$) | N |

**Geometry files:**

- CIF: pp. 746
- POSCAR: pp. 746

# $AlCCr_2$ Structure: ABC2_hP8_194_d_a_f

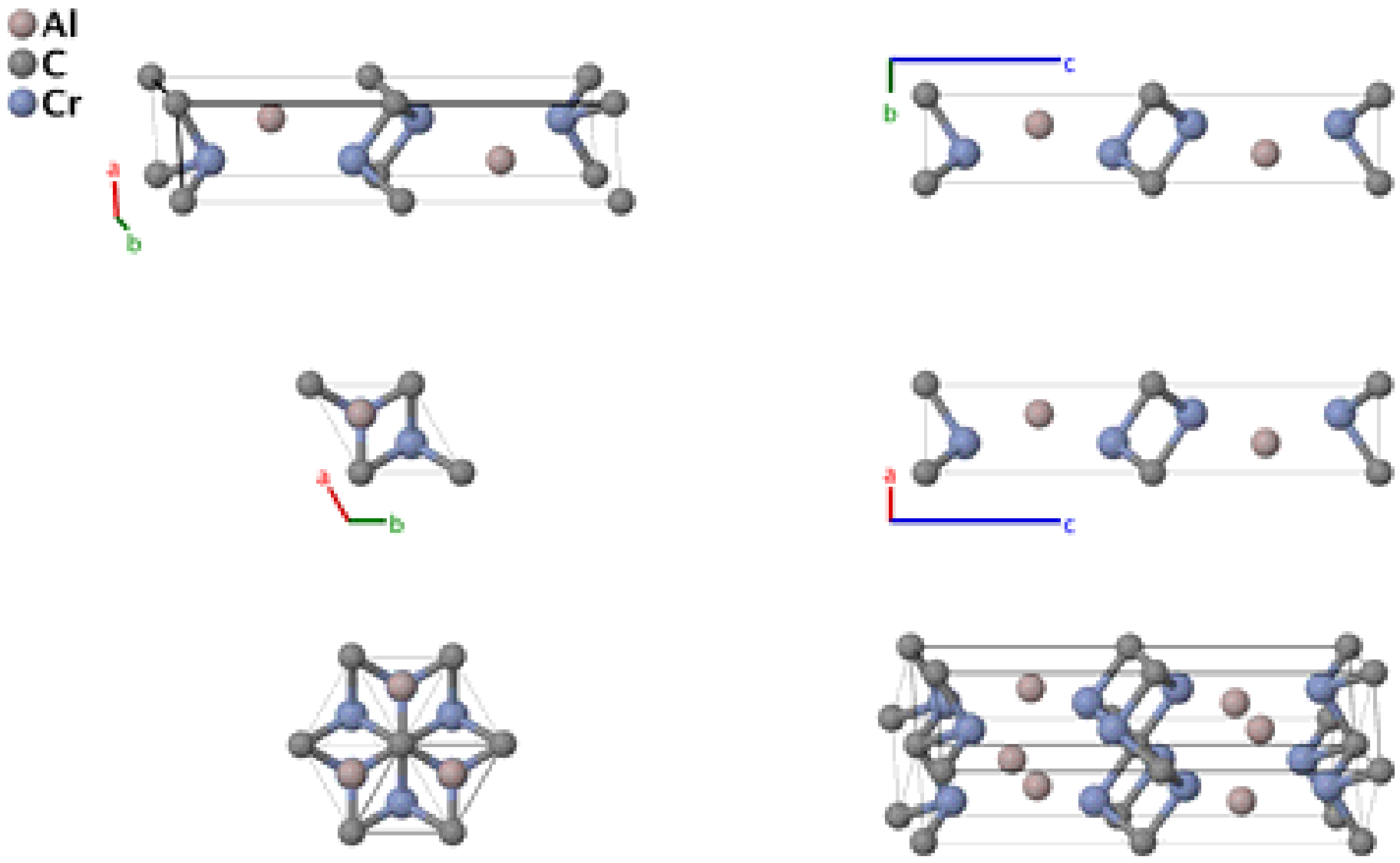

| | | |
|---|---|---|
| **Prototype** | : | $AlCCr_2$ |
| **AFLOW prototype label** | : | ABC2_hP8_194_d_a_f |
| ***Strukturbericht* designation** | : | None |
| **Pearson symbol** | : | hP8 |
| **Space group number** | : | 194 |
| **Space group symbol** | : | $P6_3/mmc$ |
| **AFLOW prototype command** | : | `aflow --proto=ABC2_hP8_194_d_a_f`<br>`--params=`$a, c/a, z_3$ |

**Other compounds with this structure:**

- $Cr_2GaN$, $CeGeLi_2$, $AlCNb_2$, $AlCTi_2$, $AlNTi_2$, $AsCNb_2$, $CCrGe_2$, $AlCV_2$, many others.

- Note that all of the atoms sit on close-packed <0001> planes. The stacking sequence may be written:

| Atom | Cr | C | Cr | Al | Cr | C | Cr | Al |
|---|---|---|---|---|---|---|---|---|
| Position | B | A | C | B | C | A | B | C |

**Hexagonal primitive vectors:**

$$\mathbf{a}_1 = \frac{1}{2}\,a\,\hat{\mathbf{x}} - \frac{\sqrt{3}}{2}\,a\,\hat{\mathbf{y}}$$
$$\mathbf{a}_2 = \frac{1}{2}\,a\,\hat{\mathbf{x}} + \frac{\sqrt{3}}{2}\,a\,\hat{\mathbf{y}}$$
$$\mathbf{a}_3 = c\,\hat{\mathbf{z}}$$

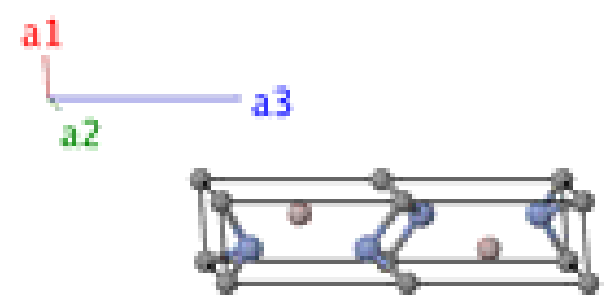

**Basis vectors:**

| | | Lattice Coordinates | | Cartesian Coordinates | Wyckoff Position | Atom Type |
|---|---|---|---|---|---|---|
| $\mathbf{B}_1$ | = | $0\,\mathbf{a}_1 + 0\,\mathbf{a}_2 + 0\,\mathbf{a}_3$ | = | $0\hat{\mathbf{x}} + 0\hat{\mathbf{y}} + 0\hat{\mathbf{z}}$ | $(2a)$ | C |
| $\mathbf{B_2}$ | = | $\frac{1}{2}\,\mathbf{a_3}$ | = | $\frac{1}{2}\,c\,\hat{\mathbf{z}}$ | $(2a)$ | C |
| $\mathbf{B_3}$ | = | $\frac{1}{3}\,\mathbf{a_1} + \frac{2}{3}\,\mathbf{a_2} + \frac{3}{4}\,\mathbf{a_3}$ | = | $\frac{1}{2}\,a\,\hat{\mathbf{x}} + \frac{1}{2\sqrt{3}}\,a\,\hat{\mathbf{y}} + \frac{3}{4}\,c\,\hat{\mathbf{z}}$ | $(2d)$ | Al |
| $\mathbf{B_4}$ | = | $\frac{2}{3}\,\mathbf{a_1} + \frac{1}{3}\,\mathbf{a_2} + \frac{1}{4}\,\mathbf{a_3}$ | = | $\frac{1}{2}\,a\,\hat{\mathbf{x}} - \frac{1}{2\sqrt{3}}\,a\,\hat{\mathbf{y}} + \frac{1}{4}\,c\,\hat{\mathbf{z}}$ | $(2d)$ | Al |
| $\mathbf{B_5}$ | = | $\frac{1}{3}\,\mathbf{a_1} + \frac{2}{3}\,\mathbf{a_2} + z_3\,\mathbf{a_3}$ | = | $\frac{1}{2}\,a\,\hat{\mathbf{x}} + \frac{1}{2\sqrt{3}}\,a\,\hat{\mathbf{y}} + z_3\,c\,\hat{\mathbf{z}}$ | $(4f)$ | Cr |
| $\mathbf{B_6}$ | = | $\frac{2}{3}\,\mathbf{a_1} + \frac{1}{3}\,\mathbf{a_2} + \left(\frac{1}{2} + z_3\right)\,\mathbf{a_3}$ | = | $\frac{1}{2}\,a\,\hat{\mathbf{x}} - \frac{1}{2\sqrt{3}}\,a\,\hat{\mathbf{y}} + \left(\frac{1}{2} + z_3\right)\,c\,\hat{\mathbf{z}}$ | $(4f)$ | Cr |
| $\mathbf{B_7}$ | = | $\frac{2}{3}\,\mathbf{a_1} + \frac{1}{3}\,\mathbf{a_2} - z_3\,\mathbf{a_3}$ | = | $\frac{1}{2}\,a\,\hat{\mathbf{x}} - \frac{1}{2\sqrt{3}}\,a\,\hat{\mathbf{y}} - z_3\,c\,\hat{\mathbf{z}}$ | $(4f)$ | Cr |
| $\mathbf{B_8}$ | = | $\frac{1}{3}\,\mathbf{a_1} + \frac{2}{3}\,\mathbf{a_2} + \left(\frac{1}{2} - z_3\right)\,\mathbf{a_3}$ | = | $\frac{1}{2}\,a\,\hat{\mathbf{x}} + \frac{1}{2\sqrt{3}}\,a\,\hat{\mathbf{y}} + \left(\frac{1}{2} - z_3\right)\,c\,\hat{\mathbf{z}}$ | $(4f)$ | Cr |

---

**Found in:**

---

**Geometry files:**

- CIF: pp. 746
- POSCAR: pp. 746

# $Ni_3Sn$ ($D0_{19}$) Structure: A3B_hP8_194_h_c

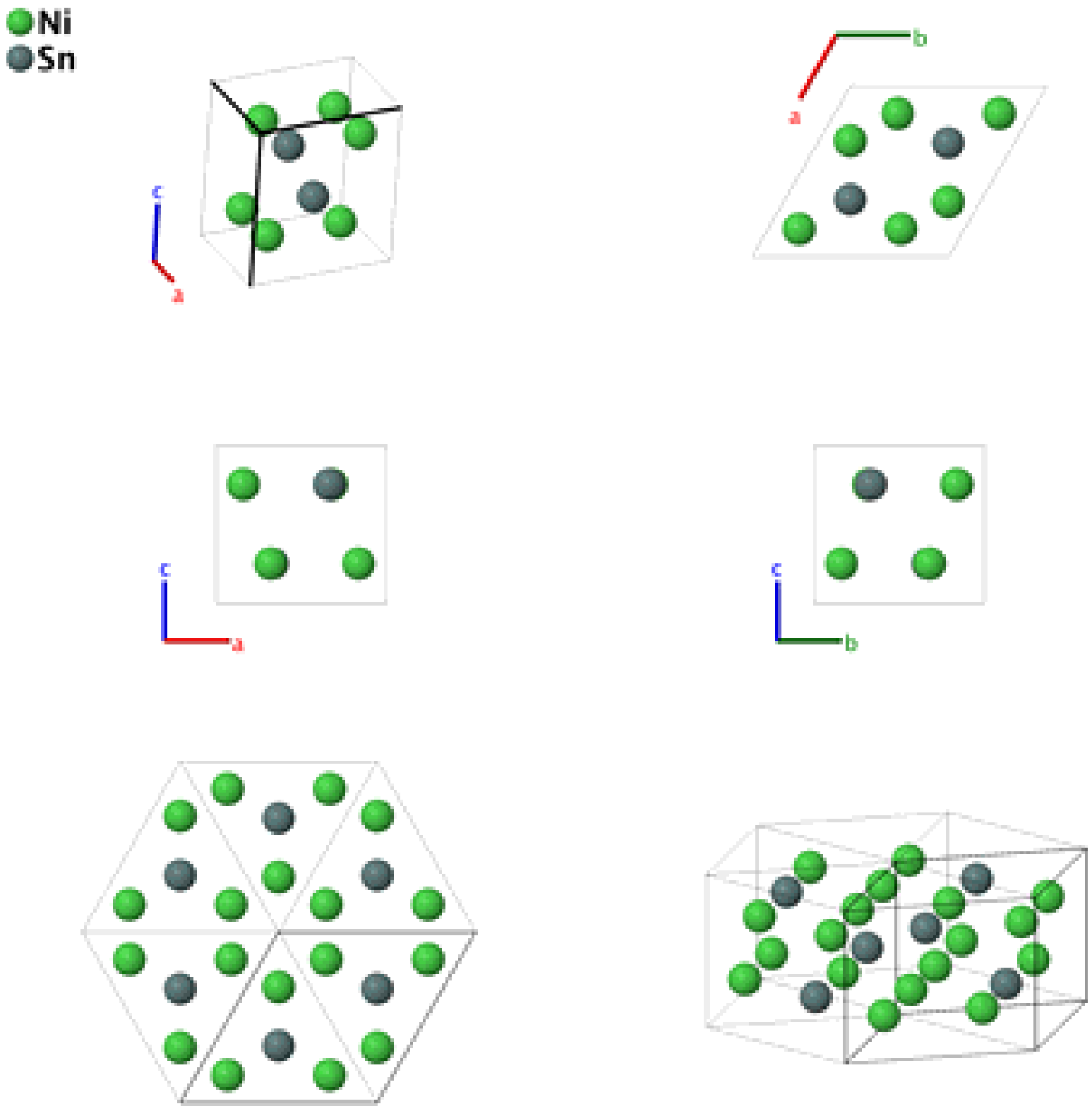

| | | |
|---|---|---|
| **Prototype** | : | $Ni_3Sn$ |
| **AFLOW prototype label** | : | A3B_hP8_194_h_c |
| ***Strukturbericht* designation** | : | $D0_{19}$ |
| **Pearson symbol** | : | hP8 |
| **Space group number** | : | 194 |
| **Space group symbol** | : | $P6_3/mmc$ |
| **AFLOW prototype command** | : | `aflow --proto=A3B_hP8_194_h_c` <br> `--params=`$a, c/a, x_2$ |

---

**Other compounds with this structure:**

- $Ti_3Sn$, $Ti_3Al$, $Mn_3Sn$, $Cd_3Mg$, $Mg_3In$, $Hg_3Y$

---

**Hexagonal primitive vectors:**

$$\mathbf{a}_1 = \frac{1}{2}\,a\,\hat{\mathbf{x}} - \frac{\sqrt{3}}{2}\,a\,\hat{\mathbf{y}}$$

$$\mathbf{a}_2 = \frac{1}{2}\,a\,\hat{\mathbf{x}} + \frac{\sqrt{3}}{2}\,a\,\hat{\mathbf{y}}$$

$$\mathbf{a}_3 = c\,\hat{\mathbf{z}}$$

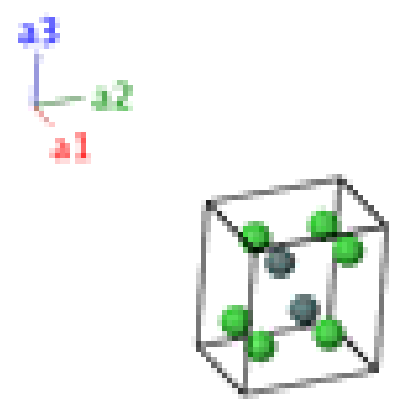

---

**Basis vectors:**

| | | Lattice Coordinates | | Cartesian Coordinates | Wyckoff Position | Atom Type |
|---|---|---|---|---|---|---|
| $\mathbf{B_1}$ | = | $\frac{1}{3}\,\mathbf{a_1} + \frac{2}{3}\,\mathbf{a_2} + \frac{1}{4}\,\mathbf{a_3}$ | = | $\frac{1}{2}\,a\,\mathbf{\hat{x}} + \frac{1}{2\sqrt{3}}\,a\,\mathbf{\hat{y}} + \frac{1}{4}\,c\,\mathbf{\hat{z}}$ | (2$c$) | Sn |
| $\mathbf{B_2}$ | = | $\frac{2}{3}\,\mathbf{a_1} + \frac{1}{3}\,\mathbf{a_2} + \frac{3}{4}\,\mathbf{a_3}$ | = | $\frac{1}{2}\,a\,\mathbf{\hat{x}} - \frac{1}{2\sqrt{3}}\,a\,\mathbf{\hat{y}} + \frac{3}{4}\,c\,\mathbf{\hat{z}}$ | (2$c$) | Sn |
| $\mathbf{B_3}$ | = | $x_2\,\mathbf{a_1} + 2\,x_2\,\mathbf{a_2} + \frac{1}{4}\,\mathbf{a_3}$ | = | $\frac{3}{2}\,x_2\,a\,\mathbf{\hat{x}} + \frac{\sqrt{3}}{2}\,x_2\,a\,\mathbf{\hat{y}} + \frac{1}{4}\,c\,\mathbf{\hat{z}}$ | (6$h$) | Ni |
| $\mathbf{B_4}$ | = | $-2\,x_2\,\mathbf{a_1} - x_2\,\mathbf{a_2} + \frac{1}{4}\,\mathbf{a_3}$ | = | $-\frac{3}{2}\,x_2\,a\,\mathbf{\hat{x}} + \frac{\sqrt{3}}{2}\,x_2\,a\,\mathbf{\hat{y}} + \frac{1}{4}\,c\,\mathbf{\hat{z}}$ | (6$h$) | Ni |
| $\mathbf{B_5}$ | = | $x_2\,\mathbf{a_1} - x_2\,\mathbf{a_2} + \frac{1}{4}\,\mathbf{a_3}$ | = | $-\sqrt{3}\,x_2\,a\,\mathbf{\hat{y}} + \frac{1}{4}\,c\,\mathbf{\hat{z}}$ | (6$h$) | Ni |
| $\mathbf{B_6}$ | = | $-x_2\,\mathbf{a_1} - 2\,x_2\,\mathbf{a_2} + \frac{3}{4}\,\mathbf{a_3}$ | = | $-\frac{3}{2}\,x_2\,a\,\mathbf{\hat{x}} - \frac{\sqrt{3}}{2}\,x_2\,a\,\mathbf{\hat{y}} + \frac{3}{4}\,c\,\mathbf{\hat{z}}$ | (6$h$) | Ni |
| $\mathbf{B_7}$ | = | $2\,x_2\,\mathbf{a_1} + x_2\,\mathbf{a_2} + \frac{3}{4}\,\mathbf{a_3}$ | = | $\frac{3}{2}\,x_2\,a\,\mathbf{\hat{x}} - \frac{\sqrt{3}}{2}\,x_2\,a\,\mathbf{\hat{y}} + \frac{3}{4}\,c\,\mathbf{\hat{z}}$ | (6$h$) | Ni |
| $\mathbf{B_8}$ | = | $-x_2\,\mathbf{a_1} + x_2\,\mathbf{a_2} + \frac{3}{4}\,\mathbf{a_3}$ | = | $+\sqrt{3}\,x_2\,a\,\mathbf{\hat{y}} + \frac{3}{4}\,c\,\mathbf{\hat{z}}$ | (6$h$) | Ni |

---

---

**Geometry files:**

- CIF: pp. 747
- POSCAR: pp. 747

# Hexagonal Graphite (A9) Crystal Structure: A_hP4_194_bc

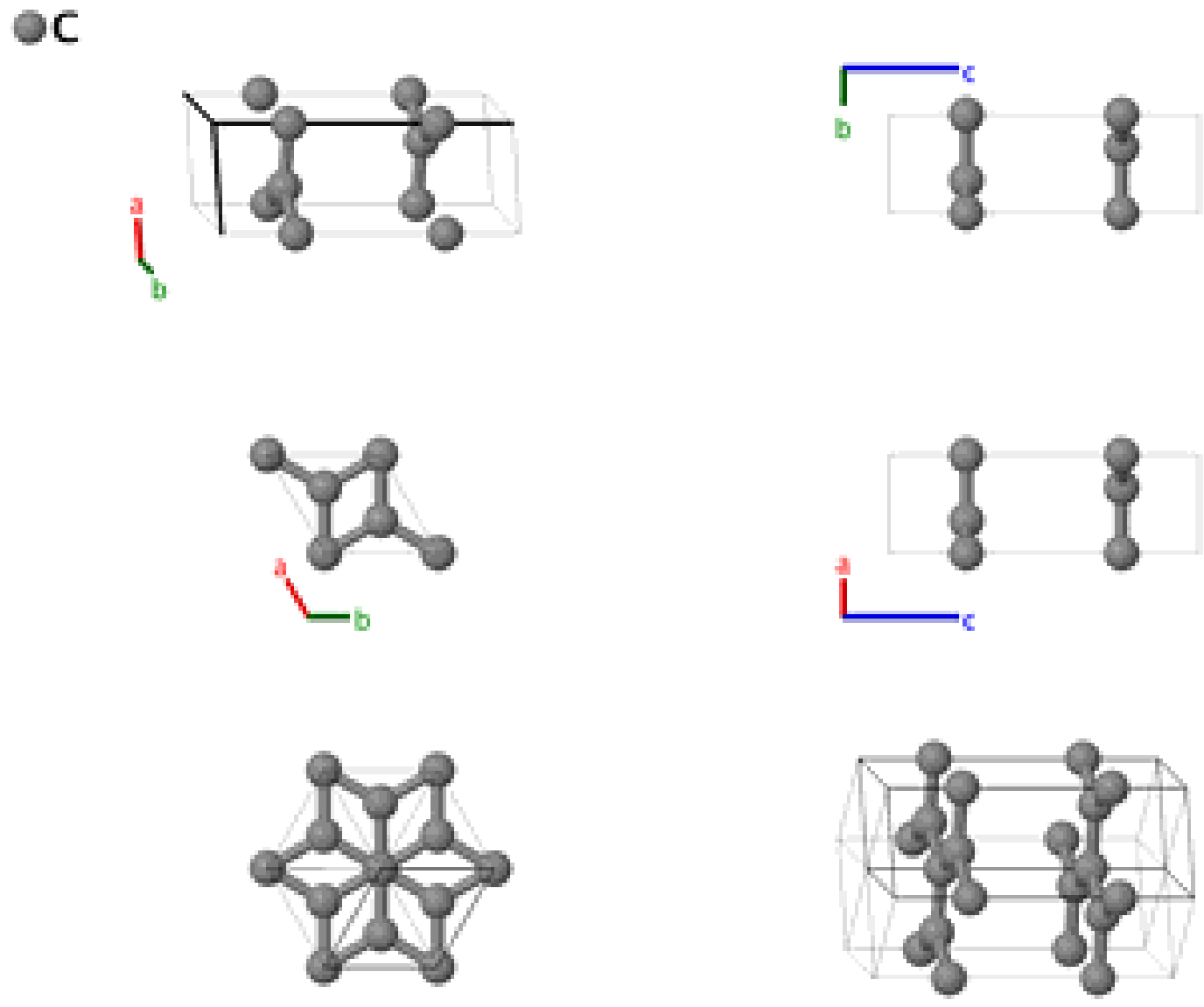

| | | |
|---|---|---|
| **Prototype** | : | C |
| **AFLOW prototype label** | : | A_hP4_194_bc |
| ***Strukturbericht* designation** | : | A9 |
| **Pearson symbol** | : | hP4 |
| **Space group number** | : | 194 |
| **Space group symbol** | : | $P6_3/mmc$ |
| **AFLOW prototype command** | : | `aflow --proto=A_hP4_194_bc` <br> `--params=`$a, c/a$ |

**Other compounds with this structure:**

- LiB

- According to (Wyckoff, 1963), hexagonal graphite may be either flat, space group $P6_3/mmc$ (#194) or buckled, space group $P6_3mc$ (#186). If it is buckled, the buckling parameter is small, less than 1/20 of the "c" parameter of the hexagonal unit cell. We will assign the A9 Strukturbericht designation to the unbuckled structure. Experimentally, a rhombohedral ($R\bar{3}m$) graphite structure is also observed.

**Hexagonal primitive vectors:**

$$\mathbf{a}_1 = \frac{1}{2}\,a\,\hat{\mathbf{x}} - \frac{\sqrt{3}}{2}\,a\,\hat{\mathbf{y}}$$
$$\mathbf{a}_2 = \frac{1}{2}\,a\,\hat{\mathbf{x}} + \frac{\sqrt{3}}{2}\,a\,\hat{\mathbf{y}}$$
$$\mathbf{a}_3 = c\,\hat{\mathbf{z}}$$

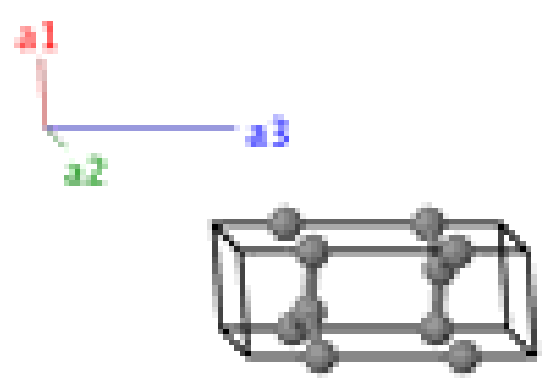

**Basis vectors:**

Lattice Coordinates | Cartesian Coordinates | Wyckoff Position | Atom Type

| | | | | | | |
|---|---|---|---|---|---|---|
| $\mathbf{B}_1$ | = | $\frac{1}{4}\,\mathbf{a}_3$ | = | $\frac{1}{4}\,c\,\hat{\mathbf{z}}$ | (2$b$) | C I |
| $\mathbf{B}_2$ | = | $\frac{3}{4}\,\mathbf{a}_3$ | = | $\frac{3}{4}\,c\,\hat{\mathbf{z}}$ | (2$b$) | C I |
| $\mathbf{B}_3$ | = | $\frac{1}{3}\,\mathbf{a}_1 + \frac{2}{3}\,\mathbf{a}_2 + \frac{1}{4}\,\mathbf{a}_3$ | = | $\frac{1}{2}\,a\,\hat{\mathbf{x}} + \frac{1}{2\sqrt{3}}\,a\,\hat{\mathbf{y}} + \frac{1}{4}\,c\,\hat{\mathbf{z}}$ | (2$c$) | C II |
| $\mathbf{B}_4$ | = | $\frac{2}{3}\,\mathbf{a}_1 + \frac{1}{3}\,\mathbf{a}_2 + \frac{3}{4}\,\mathbf{a}_3$ | = | $\frac{1}{2}\,a\,\hat{\mathbf{x}} - \frac{1}{2\sqrt{3}}\,a\,\hat{\mathbf{y}} + \frac{3}{4}\,c\,\hat{\mathbf{z}}$ | (2$c$) | C II |

---

**Found in:**

---

**Geometry files:**

- CIF: pp. 747
- POSCAR: pp. 747

# Molybdenite ($MoS_2$, C7) Structure: AB2_hP6_194_c_f

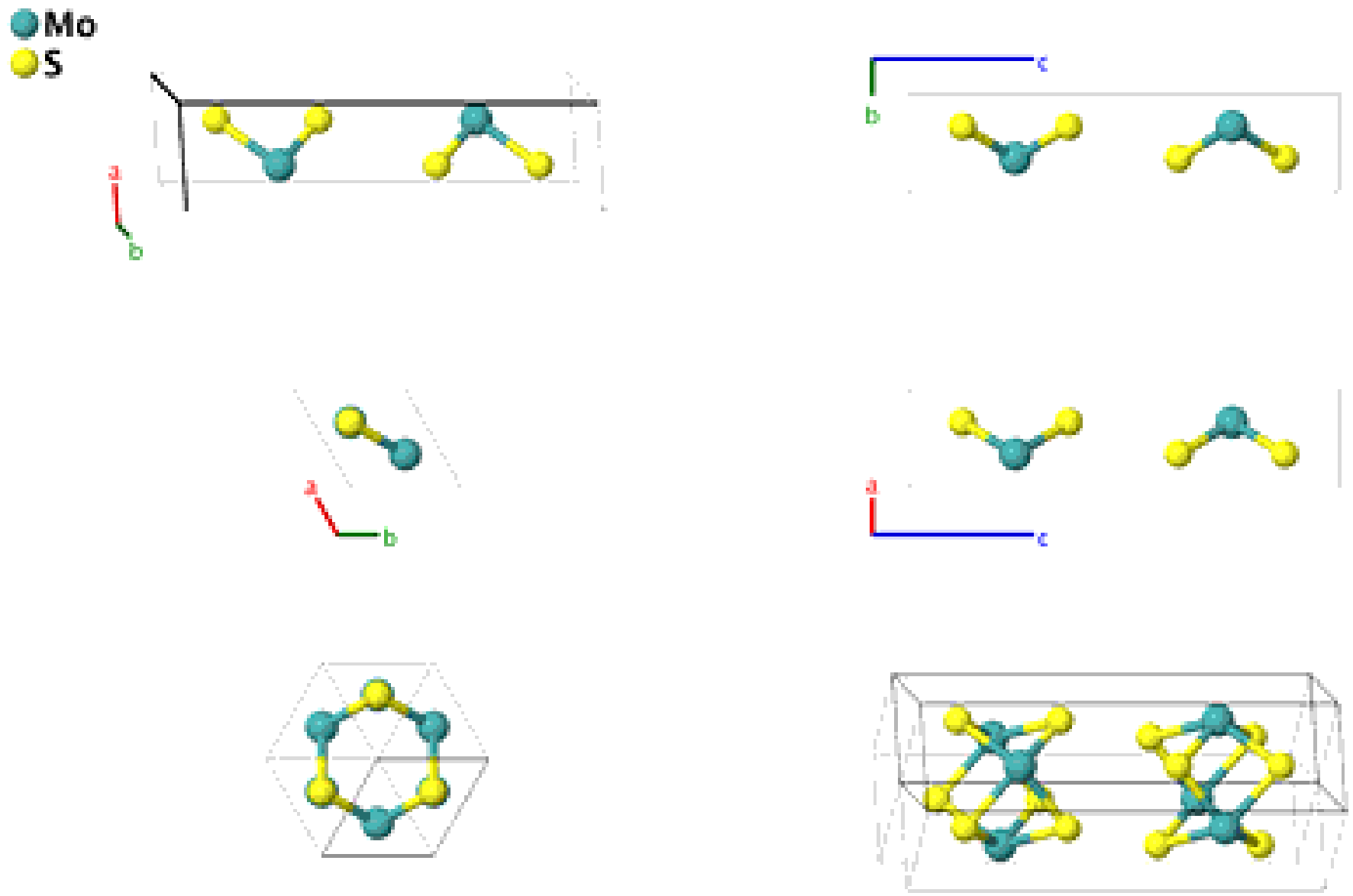

| | | |
|---|---|---|
| **Prototype** | : | $MoS_2$ |
| **AFLOW prototype label** | : | AB2_hP6_194_c_f |
| ***Strukturbericht* designation** | : | C7 |
| **Pearson symbol** | : | hP6 |
| **Space group number** | : | 194 |
| **Space group symbol** | : | $P6_3/mmc$ |
| **AFLOW prototype command** | : | `aflow --proto=AB2_hP6_194_c_f` <br> `--params=`$a, c/a, z_2$ |

**Other compounds with this structure:**

- $AlS_6Ta_3$, $CdS_2Ta$, $BPt_2$, $MoSe_2$, $MoTe_2$, $NbSe_2$, $S_2Ta$, $S_2W$, $Se_2Ta$, $Se_2W$, $Te_2W$, many more.

- Note that the stacking here is B**A**BA**B**A, where the layers in bold text are the Mo atoms.

**Hexagonal primitive vectors:**

$$\mathbf{a}_1 = \frac{1}{2}\,a\,\hat{\mathbf{x}} - \frac{\sqrt{3}}{2}\,a\,\hat{\mathbf{y}}$$
$$\mathbf{a}_2 = \frac{1}{2}\,a\,\hat{\mathbf{x}} + \frac{\sqrt{3}}{2}\,a\,\hat{\mathbf{y}}$$
$$\mathbf{a}_3 = c\,\hat{\mathbf{z}}$$

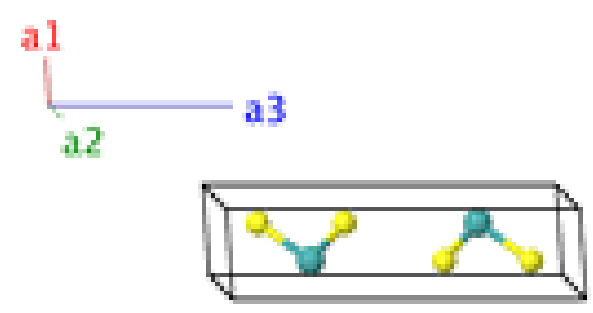

**Basis vectors:**

| | | Lattice Coordinates | | Cartesian Coordinates | Wyckoff Position | Atom Type |
|---|---|---|---|---|---|---|
| $\mathbf{B}_1$ | = | $\frac{1}{3}\,\mathbf{a}_1 + \frac{2}{3}\,\mathbf{a}_2 + \frac{1}{4}\,\mathbf{a}_3$ | = | $\frac{1}{2}\,a\,\hat{\mathbf{x}} + \frac{1}{2\sqrt{3}}\,a\,\hat{\mathbf{y}} + \frac{1}{4}\,c\,\hat{\mathbf{z}}$ | $(2c)$ | Mo |
| $\mathbf{B}_2$ | = | $\frac{2}{3}\,\mathbf{a}_1 + \frac{1}{3}\,\mathbf{a}_2 + \frac{3}{4}\,\mathbf{a}_3$ | = | $\frac{1}{2}\,a\,\hat{\mathbf{x}} - \frac{1}{2\sqrt{3}}\,a\,\hat{\mathbf{y}} + \frac{3}{4}\,c\,\hat{\mathbf{z}}$ | $(2c)$ | Mo |
| $\mathbf{B}_3$ | = | $\frac{1}{3}\,\mathbf{a}_1 + \frac{2}{3}\,\mathbf{a}_2 + z_2\,\mathbf{a}_3$ | = | $\frac{1}{2}\,a\,\hat{\mathbf{x}} + \frac{1}{2\sqrt{3}}\,a\,\hat{\mathbf{y}} + z_2\,c\,\hat{\mathbf{z}}$ | $(4f)$ | S |

$$\mathbf{B_4} = \frac{2}{3}\,\mathbf{a_1} + \frac{1}{3}\,\mathbf{a_2} + \left(\frac{1}{2} + z_2\right)\,\mathbf{a_3} = \frac{1}{2}\,a\,\mathbf{\hat{x}} - \frac{1}{2\sqrt{3}}\,a\,\mathbf{\hat{y}} + \left(\frac{1}{2} + z_2\right)\,c\,\mathbf{\hat{z}} \qquad (4f) \qquad \text{S}$$

$$\mathbf{B_5} = \frac{2}{3}\,\mathbf{a_1} + \frac{1}{3}\,\mathbf{a_2} - z_2\,\mathbf{a_3} = \frac{1}{2}\,a\,\mathbf{\hat{x}} - \frac{1}{2\sqrt{3}}\,a\,\mathbf{\hat{y}} - z_2\,c\,\mathbf{\hat{z}} \qquad (4f) \qquad \text{S}$$

$$\mathbf{B_6} = \frac{1}{3}\,\mathbf{a_1} + \frac{2}{3}\,\mathbf{a_2} + \left(\frac{1}{2} - z_2\right)\,\mathbf{a_3} = \frac{1}{2}\,a\,\mathbf{\hat{x}} + \frac{1}{2\sqrt{3}}\,a\,\mathbf{\hat{y}} + \left(\frac{1}{2} - z_2\right)\,c\,\mathbf{\hat{z}} \qquad (4f) \qquad \text{S}$$

---

**Found in:**

---

**Geometry files:**

- CIF: pp. 748
- POSCAR: pp. 748

# $W_2B_5$ ($D8_h$) Structure: A5B2_hP14_194_abdf_f

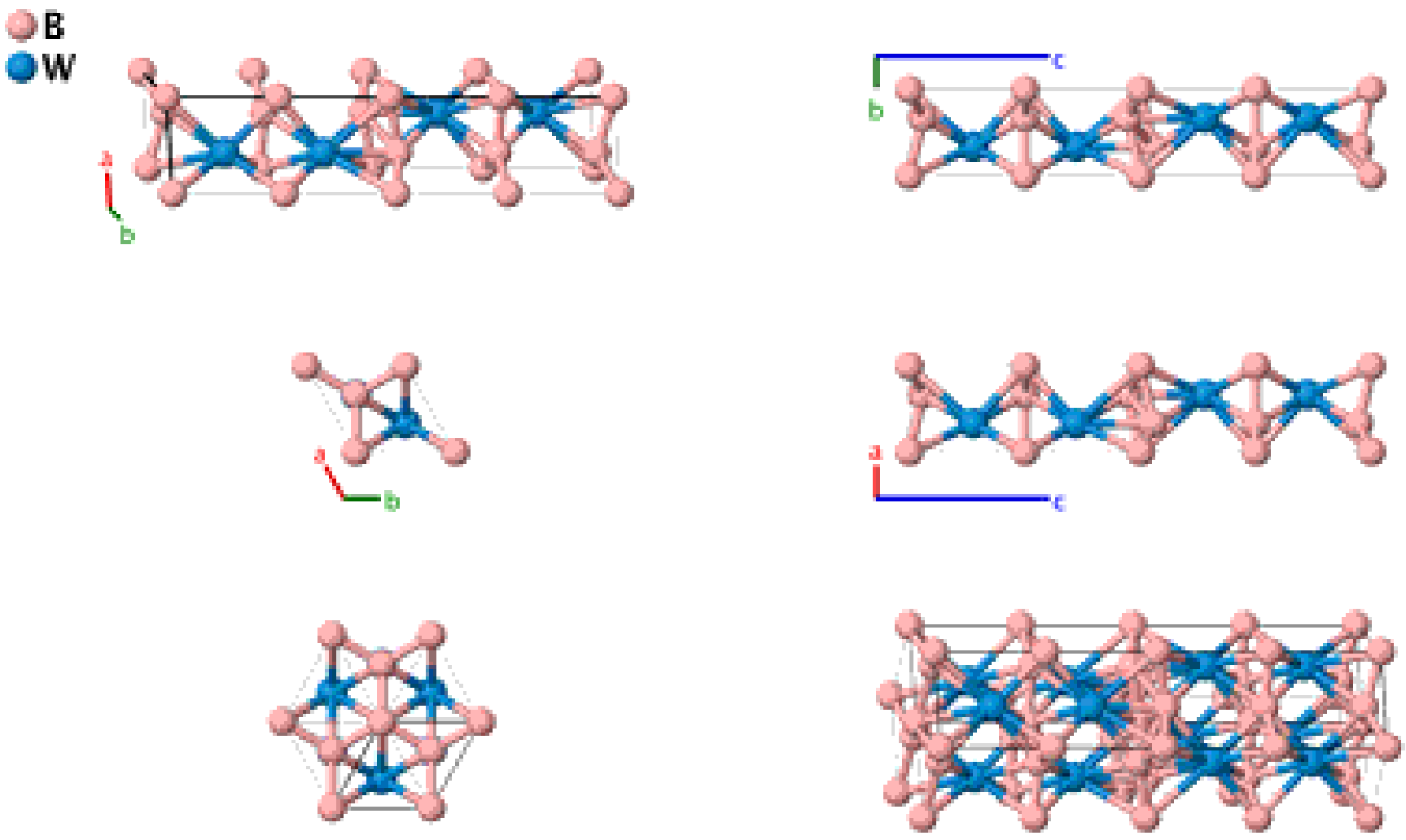

| | | |
|---|---|---|
| **Prototype** | : | $W_2B_5$ |
| **AFLOW prototype label** | : | A5B2_hP14_194_abdf_f |
| ***Strukturbericht* designation** | : | $D8_h$ |
| **Pearson symbol** | : | hP14 |
| **Space group number** | : | 194 |
| **Space group symbol** | : | $P6_3/mmc$ |
| **AFLOW prototype command** | : | `aflow --proto=A5B2_hP14_194_abdf_f` `--params=`$a, c/a, z_4, z_5$ |

**Hexagonal primitive vectors:**

$$\mathbf{a}_1 = \frac{1}{2}\,a\,\hat{\mathbf{x}} - \frac{\sqrt{3}}{2}\,a\,\hat{\mathbf{y}}$$
$$\mathbf{a}_2 = \frac{1}{2}\,a\,\hat{\mathbf{x}} + \frac{\sqrt{3}}{2}\,a\,\hat{\mathbf{y}}$$
$$\mathbf{a}_3 = c\,\hat{\mathbf{z}}$$

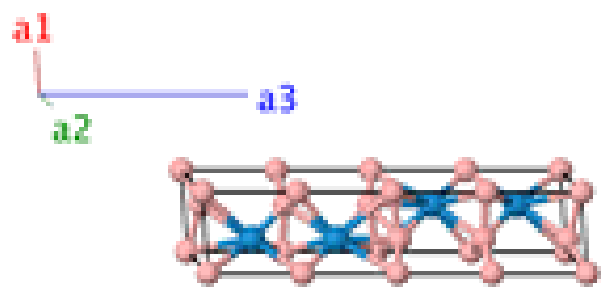

**Basis vectors:**

| | | Lattice Coordinates | | Cartesian Coordinates | Wyckoff Position | Atom Type |
|---|---|---|---|---|---|---|
| $\mathbf{B}_1$ | = | $0\,\mathbf{a}_1 + 0\,\mathbf{a}_2 + 0\,\mathbf{a}_3$ | = | $0\hat{\mathbf{x}} + 0\hat{\mathbf{y}} + 0\hat{\mathbf{z}}$ | (2$a$) | B I |
| $\mathbf{B}_2$ | = | $\frac{1}{2}\,\mathbf{a}_3$ | = | $\frac{1}{2}\,c\,\hat{\mathbf{z}}$ | (2$a$) | B I |
| $\mathbf{B}_3$ | = | $\frac{1}{4}\,\mathbf{a}_3$ | = | $\frac{1}{4}\,c\,\hat{\mathbf{z}}$ | (2$b$) | B II |
| $\mathbf{B}_4$ | = | $\frac{3}{4}\,\mathbf{a}_3$ | = | $\frac{3}{4}\,c\,\hat{\mathbf{z}}$ | (2$b$) | B II |
| $\mathbf{B}_5$ | = | $\frac{1}{3}\,\mathbf{a}_1 + \frac{2}{3}\,\mathbf{a}_2 + \frac{3}{4}\,\mathbf{a}_3$ | = | $\frac{1}{2}\,a\,\hat{\mathbf{x}} + \frac{1}{2\sqrt{3}}\,a\,\hat{\mathbf{y}} + \frac{3}{4}\,c\,\hat{\mathbf{z}}$ | (2$d$) | B III |
| $\mathbf{B}_6$ | = | $\frac{2}{3}\,\mathbf{a}_1 + \frac{1}{3}\,\mathbf{a}_2 + \frac{1}{4}\,\mathbf{a}_3$ | = | $\frac{1}{2}\,a\,\hat{\mathbf{x}} - \frac{1}{2\sqrt{3}}\,a\,\hat{\mathbf{y}} + \frac{1}{4}\,c\,\hat{\mathbf{z}}$ | (2$d$) | B III |
| $\mathbf{B}_7$ | = | $\frac{1}{3}\,\mathbf{a}_1 + \frac{2}{3}\,\mathbf{a}_2 + z_4\,\mathbf{a}_3$ | = | $\frac{1}{2}\,a\,\hat{\mathbf{x}} + \frac{1}{2\sqrt{3}}\,a\,\hat{\mathbf{y}} + z_4\,c\,\hat{\mathbf{z}}$ | (4$f$) | B IV |
| $\mathbf{B}_8$ | = | $\frac{2}{3}\,\mathbf{a}_1 + \frac{1}{3}\,\mathbf{a}_2 + \left(\frac{1}{2} + z_4\right)\,\mathbf{a}_3$ | = | $\frac{1}{2}\,a\,\hat{\mathbf{x}} - \frac{1}{2\sqrt{3}}\,a\,\hat{\mathbf{y}} + \left(\frac{1}{2} + z_4\right)\,c\,\hat{\mathbf{z}}$ | (4$f$) | B IV |
| $\mathbf{B}_9$ | = | $\frac{2}{3}\,\mathbf{a}_1 + \frac{1}{3}\,\mathbf{a}_2 - z_4\,\mathbf{a}_3$ | = | $\frac{1}{2}\,a\,\hat{\mathbf{x}} - \frac{1}{2\sqrt{3}}\,a\,\hat{\mathbf{y}} - z_4\,c\,\hat{\mathbf{z}}$ | (4$f$) | B IV |
| $\mathbf{B_{10}}$ | = | $\frac{1}{3}\,\mathbf{a_1} + \frac{2}{3}\,\mathbf{a_2} + \left(\frac{1}{2} - z_4\right)\,\mathbf{a_3}$ | = | $\frac{1}{2}\,a\,\mathbf{\hat{x}} + \frac{1}{2\sqrt{3}}\,a\,\mathbf{\hat{y}} + \left(\frac{1}{2} - z_4\right)\,c\,\mathbf{\hat{z}}$ | $(4f)$ | B IV |
| $\mathbf{B_{11}}$ | = | $\frac{1}{3}\,\mathbf{a_1} + \frac{2}{3}\,\mathbf{a_2} + z_5\,\mathbf{a_3}$ | = | $\frac{1}{2}\,a\,\mathbf{\hat{x}} + \frac{1}{2\sqrt{3}}\,a\,\mathbf{\hat{y}} + z_5\,c\,\mathbf{\hat{z}}$ | $(4f)$ | W |
| $\mathbf{B_{12}}$ | = | $\frac{2}{3}\,\mathbf{a_1} + \frac{1}{3}\,\mathbf{a_2} + \left(\frac{1}{2} + z_5\right)\,\mathbf{a_3}$ | = | $\frac{1}{2}\,a\,\mathbf{\hat{x}} - \frac{1}{2\sqrt{3}}\,a\,\mathbf{\hat{y}} + \left(\frac{1}{2} + z_5\right)\,c\,\mathbf{\hat{z}}$ | $(4f)$ | W |
| $\mathbf{B_{13}}$ | = | $\frac{2}{3}\,\mathbf{a_1} + \frac{1}{3}\,\mathbf{a_2} - z_5\,\mathbf{a_3}$ | = | $\frac{1}{2}\,a\,\mathbf{\hat{x}} - \frac{1}{2\sqrt{3}}\,a\,\mathbf{\hat{y}} - z_5\,c\,\mathbf{\hat{z}}$ | $(4f)$ | W |
| $\mathbf{B_{14}}$ | = | $\frac{1}{3}\,\mathbf{a_1} + \frac{2}{3}\,\mathbf{a_2} + \left(\frac{1}{2} - z_5\right)\,\mathbf{a_3}$ | = | $\frac{1}{2}\,a\,\mathbf{\hat{x}} + \frac{1}{2\sqrt{3}}\,a\,\mathbf{\hat{y}} + \left(\frac{1}{2} - z_5\right)\,c\,\mathbf{\hat{z}}$ | $(4f)$ | W |

**Found in:**

**Geometry files:**

- CIF: pp. 748
- POSCAR: pp. 748

# $MgZn_2$ Hexagonal Laves (C14) Structure: AB2_hP12_194_f_ah

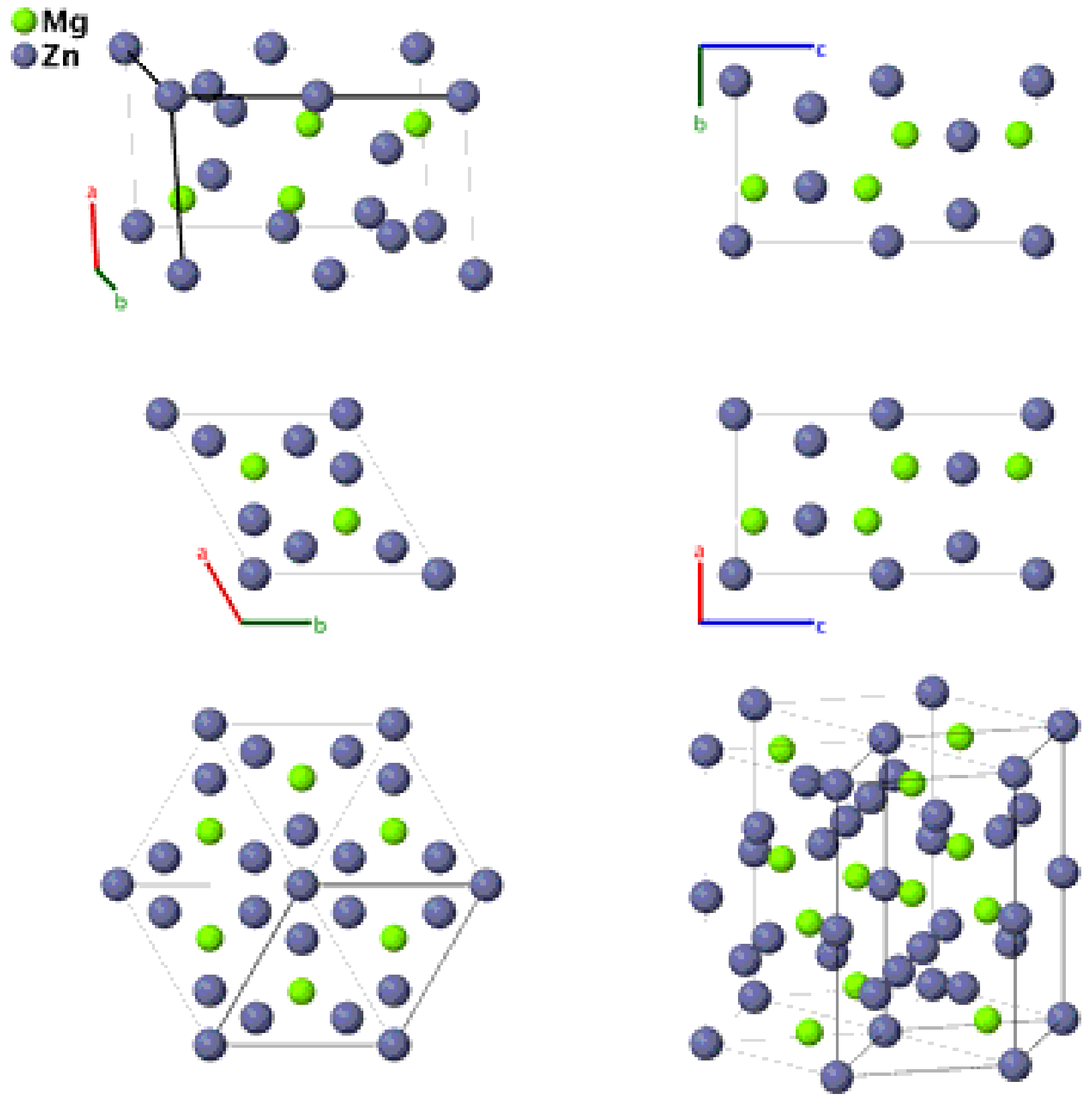

| | | |
|---|---|---|
| **Prototype** | : | $MgZn_2$ |
| **AFLOW prototype label** | : | AB2_hP12_194_f_ah |
| ***Strukturbericht* designation** | : | C14 |
| **Pearson symbol** | : | hP12 |
| **Space group number** | : | 194 |
| **Space group symbol** | : | $P6_3/mmc$ |
| **AFLOW prototype command** | : | `aflow --proto=AB2_hP12_194_f_ah --params=`$a, c/a, z_2, x_3$ |

**Other compounds with this structure:**

- $CaMg_2$, $ZrRe_2$, $KNa_2$, $TaFe_2$, $NbMn_2$, $UNi_2$

**Hexagonal primitive vectors:**

$$\mathbf{a}_1 = \frac{1}{2}\,a\,\hat{\mathbf{x}} - \frac{\sqrt{3}}{2}\,a\,\hat{\mathbf{y}}$$
$$\mathbf{a}_2 = \frac{1}{2}\,a\,\hat{\mathbf{x}} + \frac{\sqrt{3}}{2}\,a\,\hat{\mathbf{y}}$$
$$\mathbf{a}_3 = c\,\hat{\mathbf{z}}$$

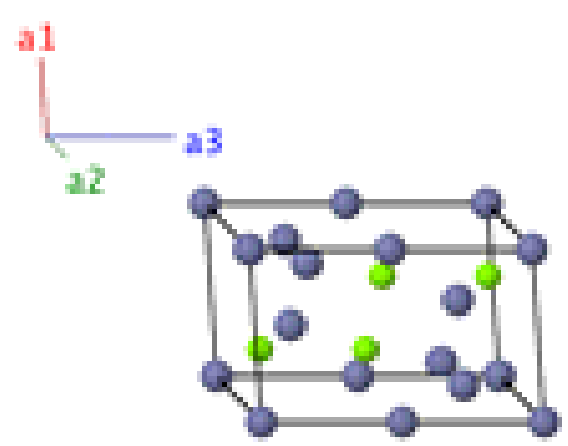

**Basis vectors:**

| | | Lattice Coordinates | | Cartesian Coordinates | Wyckoff Position | Atom Type |
|---|---|---|---|---|---|---|
| $\mathbf{B_1}$ | = | $0\,\mathbf{a_1} + 0\,\mathbf{a_2} + 0\,\mathbf{a_3}$ | = | $0\hat{\mathbf{x}} + 0\hat{\mathbf{y}} + 0\hat{\mathbf{z}}$ | $(2a)$ | Zn I |
| $\mathbf{B_2}$ | = | $\frac{1}{2}\,\mathbf{a_3}$ | = | $\frac{1}{2}\,c\,\hat{\mathbf{z}}$ | $(2a)$ | Zn I |
| $\mathbf{B_3}$ | = | $\frac{1}{3}\,\mathbf{a_1} + \frac{2}{3}\,\mathbf{a_2} + z_2\,\mathbf{a_3}$ | = | $\frac{1}{2}\,a\,\hat{\mathbf{x}} + \frac{1}{2\sqrt{3}}\,a\,\hat{\mathbf{y}} + z_2\,c\,\hat{\mathbf{z}}$ | $(4f)$ | Mg |
| $\mathbf{B_4}$ | = | $\frac{2}{3}\,\mathbf{a_1} + \frac{1}{3}\,\mathbf{a_2} + \left(\frac{1}{2} + z_2\right)\,\mathbf{a_3}$ | = | $\frac{1}{2}\,a\,\hat{\mathbf{x}} - \frac{1}{2\sqrt{3}}\,a\,\hat{\mathbf{y}} + \left(\frac{1}{2} + z_2\right)\,c\,\hat{\mathbf{z}}$ | $(4f)$ | Mg |
| $\mathbf{B_5}$ | = | $\frac{2}{3}\,\mathbf{a_1} + \frac{1}{3}\,\mathbf{a_2} - z_2\,\mathbf{a_3}$ | = | $\frac{1}{2}\,a\,\hat{\mathbf{x}} - \frac{1}{2\sqrt{3}}\,a\,\hat{\mathbf{y}} - z_2\,c\,\hat{\mathbf{z}}$ | $(4f)$ | Mg |
| $\mathbf{B_6}$ | = | $\frac{1}{3}\,\mathbf{a_1} + \frac{2}{3}\,\mathbf{a_2} + \left(\frac{1}{2} - z_2\right)\,\mathbf{a_3}$ | = | $\frac{1}{2}\,a\,\hat{\mathbf{x}} + \frac{1}{2\sqrt{3}}\,a\,\hat{\mathbf{y}} + \left(\frac{1}{2} - z_2\right)\,c\,\hat{\mathbf{z}}$ | $(4f)$ | Mg |
| $\mathbf{B_7}$ | = | $x_3\,\mathbf{a_1} + 2\,x_3\,\mathbf{a_2} + \frac{1}{4}\,\mathbf{a_3}$ | = | $\frac{3}{2}\,x_3\,a\,\hat{\mathbf{x}} + \frac{\sqrt{3}}{2}\,x_3\,a\,\hat{\mathbf{y}} + \frac{1}{4}\,c\,\hat{\mathbf{z}}$ | $(6h)$ | Zn II |
| $\mathbf{B_8}$ | = | $-2\,x_3\,\mathbf{a_1} - x_3\,\mathbf{a_2} + \frac{1}{4}\,\mathbf{a_3}$ | = | $-\frac{3}{2}\,x_3\,a\,\hat{\mathbf{x}} + \frac{\sqrt{3}}{2}\,x_3\,a\,\hat{\mathbf{y}} + \frac{1}{4}\,c\,\hat{\mathbf{z}}$ | $(6h)$ | Zn II |
| $\mathbf{B_9}$ | = | $x_3\,\mathbf{a_1} - x_3\,\mathbf{a_2} + \frac{1}{4}\,\mathbf{a_3}$ | = | $-\sqrt{3}\,x_3\,a\,\hat{\mathbf{y}} + \frac{1}{4}\,c\,\hat{\mathbf{z}}$ | $(6h)$ | Zn II |
| $\mathbf{B_{10}}$ | = | $-x_3\,\mathbf{a_1} - 2\,x_3\,\mathbf{a_2} + \frac{3}{4}\,\mathbf{a_3}$ | = | $-\frac{3}{2}\,x_3\,a\,\hat{\mathbf{x}} - \frac{\sqrt{3}}{2}\,x_3\,a\,\hat{\mathbf{y}} + \frac{3}{4}\,c\,\hat{\mathbf{z}}$ | $(6h)$ | Zn II |
| $\mathbf{B_{11}}$ | = | $2\,x_3\,\mathbf{a_1} + x_3\,\mathbf{a_2} + \frac{3}{4}\,\mathbf{a_3}$ | = | $\frac{3}{2}\,x_3\,a\,\hat{\mathbf{x}} - \frac{\sqrt{3}}{2}\,x_3\,a\,\hat{\mathbf{y}} + \frac{3}{4}\,c\,\hat{\mathbf{z}}$ | $(6h)$ | Zn II |
| $\mathbf{B_{12}}$ | = | $-x_3\,\mathbf{a_1} + x_3\,\mathbf{a_2} + \frac{3}{4}\,\mathbf{a_3}$ | = | $+\sqrt{3}\,x_3\,a\,\hat{\mathbf{y}} + \frac{3}{4}\,c\,\hat{\mathbf{z}}$ | $(6h)$ | Zn II |

---

---

**Geometry files:**

- CIF: pp. 749
- POSCAR: pp. 749

# LiBC Structure: ABC_hP6_194_c_d_a

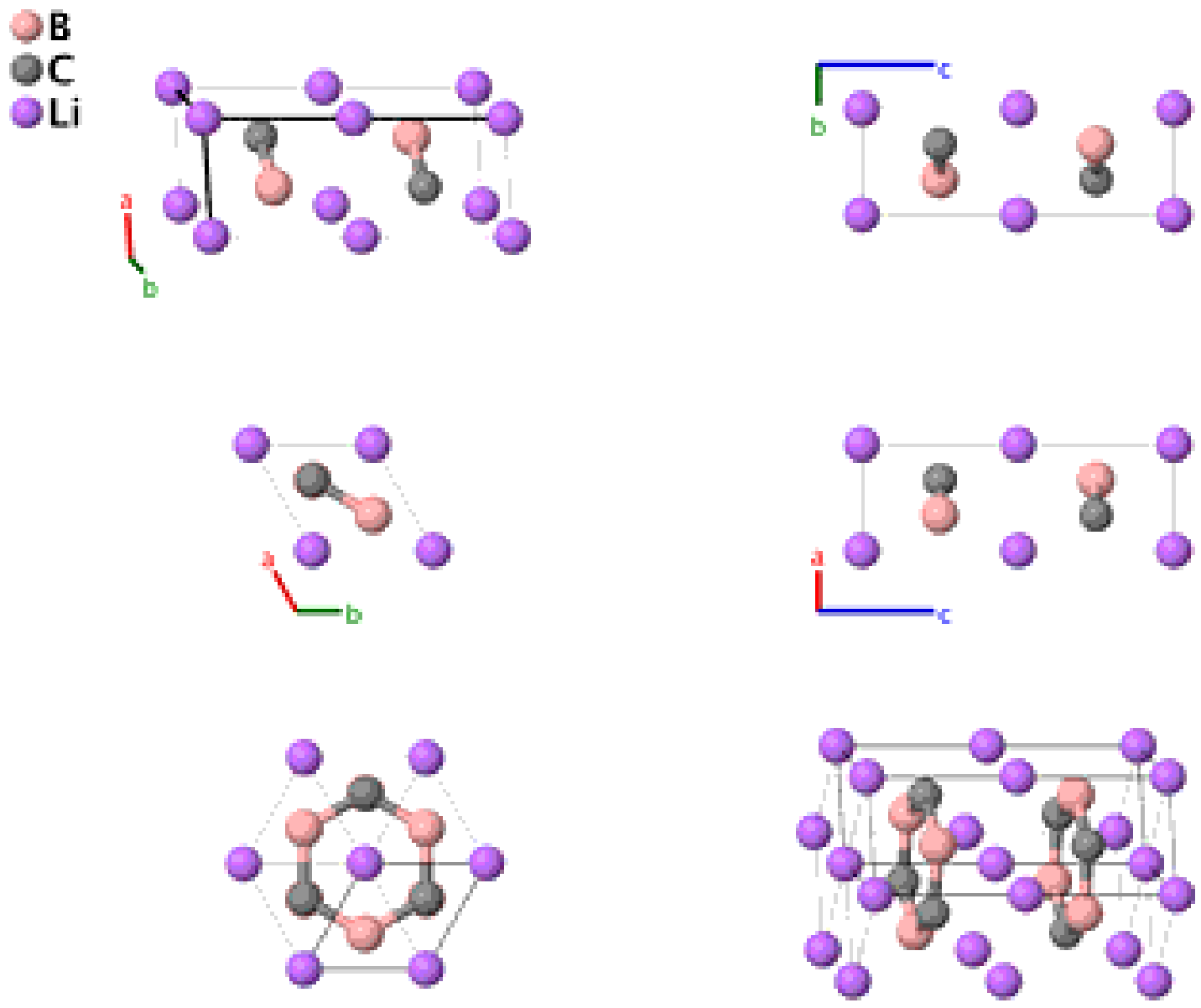

| | | |
|---|---|---|
| **Prototype** | : | LiBC |
| **AFLOW prototype label** | : | ABC_hP6_194_c_d_a |
| ***Strukturbericht* designation** | : | None |
| **Pearson symbol** | : | hP6 |
| **Space group number** | : | 194 |
| **Space group symbol** | : | $P6_3/mmc$ |
| **AFLOW prototype command** | : | `aflow --proto=ABC_hP6_194_c_d_a --params=`$a, c/a$ |

**Other compounds with this structure:**

- ZrBeSi

- This is the parent structure of the $Li_{1-x}BC$ Structure

**Hexagonal primitive vectors:**

$$
\begin{aligned}
\mathbf{a}_1 &= \tfrac{1}{2}\, a\, \hat{\mathbf{x}} - \tfrac{\sqrt{3}}{2}\, a\, \hat{\mathbf{y}} \\
\mathbf{a}_2 &= \tfrac{1}{2}\, a\, \hat{\mathbf{x}} + \tfrac{\sqrt{3}}{2}\, a\, \hat{\mathbf{y}} \\
\mathbf{a}_3 &= c\, \hat{\mathbf{z}}
\end{aligned}
$$

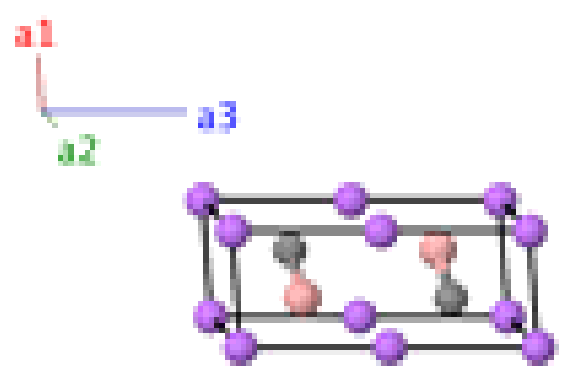

**Basis vectors:**

| | | Lattice Coordinates | | Cartesian Coordinates | Wyckoff Position | Atom Type |
|---|---|---|---|---|---|---|
| $\mathbf{B}_1$ | = | $0\,\mathbf{a}_1 + 0\,\mathbf{a}_2 + 0\,\mathbf{a}_3$ | = | $0\hat{\mathbf{x}} + 0\hat{\mathbf{y}} + 0\hat{\mathbf{z}}$ | (2$a$) | Li |
| $\mathbf{B}_2$ | = | $\frac{1}{2}\,\mathbf{a}_3$ | = | $\frac{1}{2}\,c\,\hat{\mathbf{z}}$ | (2$a$) | Li |
| $\mathbf{B_3}$ | = | $\frac{1}{3}\,\mathbf{a_1} + \frac{2}{3}\,\mathbf{a_2} + \frac{1}{4}\,\mathbf{a_3}$ | = | $\frac{1}{2}\,a\,\mathbf{\hat{x}} + \frac{1}{2\sqrt{3}}\,a\,\mathbf{\hat{y}} + \frac{1}{4}\,c\,\mathbf{\hat{z}}$ | $(2c)$ | B |
| $\mathbf{B_4}$ | = | $\frac{2}{3}\,\mathbf{a_1} + \frac{1}{3}\,\mathbf{a_2} + \frac{3}{4}\,\mathbf{a_3}$ | = | $\frac{1}{2}\,a\,\mathbf{\hat{x}} - \frac{1}{2\sqrt{3}}\,a\,\mathbf{\hat{y}} + \frac{3}{4}\,c\,\mathbf{\hat{z}}$ | $(2c)$ | B |
| $\mathbf{B_5}$ | = | $\frac{1}{3}\,\mathbf{a_1} + \frac{2}{3}\,\mathbf{a_2} + \frac{3}{4}\,\mathbf{a_3}$ | = | $\frac{1}{2}\,a\,\mathbf{\hat{x}} + \frac{1}{2\sqrt{3}}\,a\,\mathbf{\hat{y}} + \frac{3}{4}\,c\,\mathbf{\hat{z}}$ | $(2d)$ | C |
| $\mathbf{B_6}$ | = | $\frac{2}{3}\,\mathbf{a_1} + \frac{1}{3}\,\mathbf{a_2} + \frac{1}{4}\,\mathbf{a_3}$ | = | $\frac{1}{2}\,a\,\mathbf{\hat{x}} - \frac{1}{2\sqrt{3}}\,a\,\mathbf{\hat{y}} + \frac{1}{4}\,c\,\mathbf{\hat{z}}$ | $(2d)$ | C |

**Geometry files:**

- CIF: pp. 749
- POSCAR: pp. 750

# Lonsdaleite (Hexagonal Diamond) Structure: A_hP4_194_f

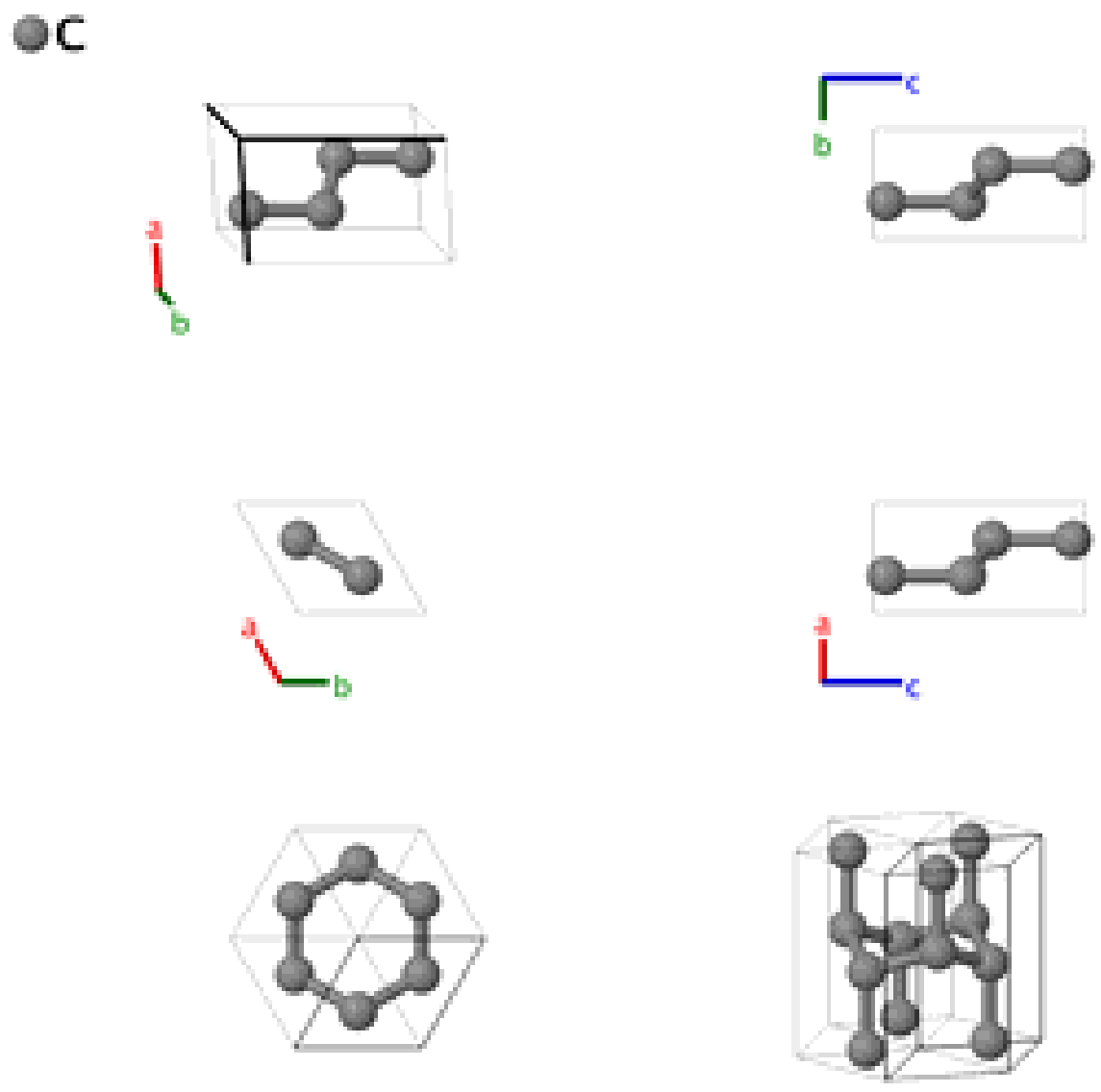

| | | |
|---|---|---|
| **Prototype** | : | C |
| **AFLOW prototype label** | : | A_hP4_194_f |
| ***Strukturbericht* designation** | : | None |
| **Pearson symbol** | : | hP4 |
| **Space group number** | : | 194 |
| **Space group symbol** | : | $P6_3/mmc$ |
| **AFLOW prototype command** | : | `aflow --proto=A_hP4_194_f --params=`$a, c/a, z_1$ |

**Other elements with this structure:**

- Si (Hexagonal)

- Hexagonal diamond was named lonsdaleite in honor of Kathleen Lonsdale. This is related to the hcp (A3) lattice in the same way that diamond (A4) is related to the fcc lattice (A1). It can also be obtained from wurtzite (B4) by replacing both the Zn and S atoms by carbon. The "ideal" structure, where the nearest-neighbor environment of each atom is the same as in diamond, is achieved when we take $c/a = \sqrt{8/3}$ and $z_1 = 1/16$. Alternatively, we can take $z_1 = 3/16$, in which case the origin is at the center of a C-C bond aligned in the [0001] direction. When $z_1 = 0$ this structure becomes a set of graphitic sheets, but not true hexagonal graphite (A9).

**Hexagonal primitive vectors:**

$$\begin{aligned} \mathbf{a}_1 &= \tfrac{1}{2}\,a\,\hat{\mathbf{x}} - \tfrac{\sqrt{3}}{2}\,a\,\hat{\mathbf{y}} \\ \mathbf{a}_2 &= \tfrac{1}{2}\,a\,\hat{\mathbf{x}} + \tfrac{\sqrt{3}}{2}\,a\,\hat{\mathbf{y}} \\ \mathbf{a}_3 &= c\,\hat{\mathbf{z}} \end{aligned}$$

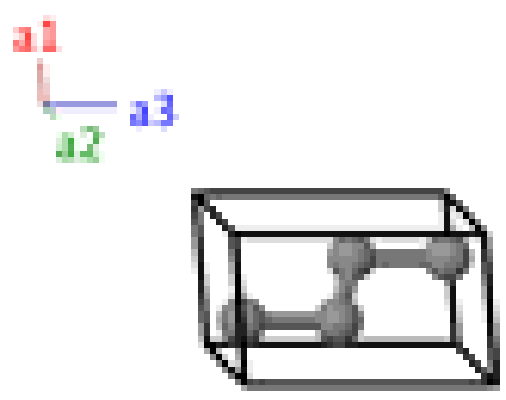

**Basis vectors:**

| | | Lattice Coordinates | | Cartesian Coordinates | Wyckoff Position | Atom Type |
|---|---|---|---|---|---|---|
| $\mathbf{B_1}$ | = | $\frac{1}{3}\,\mathbf{a_1} + \frac{2}{3}\,\mathbf{a_2} + z_1\,\mathbf{a_3}$ | = | $\frac{1}{2}\,a\,\mathbf{\hat{x}} + \frac{1}{2\sqrt{3}}\,a\,\mathbf{\hat{y}} + z_1\,c\,\mathbf{\hat{z}}$ | $(4f)$ | C |
| $\mathbf{B_2}$ | = | $\frac{2}{3}\,\mathbf{a_1} + \frac{1}{3}\,\mathbf{a_2} + \left(\frac{1}{2} + z_1\right)\,\mathbf{a_3}$ | = | $\frac{1}{2}\,a\,\mathbf{\hat{x}} - \frac{1}{2\sqrt{3}}\,a\,\mathbf{\hat{y}} + \left(\frac{1}{2} + z_1\right)\,c\,\mathbf{\hat{z}}$ | $(4f)$ | C |
| $\mathbf{B_3}$ | = | $\frac{2}{3}\,\mathbf{a_1} + \frac{1}{3}\,\mathbf{a_2} - z_1\,\mathbf{a_3}$ | = | $\frac{1}{2}\,a\,\mathbf{\hat{x}} - \frac{1}{2\sqrt{3}}\,a\,\mathbf{\hat{y}} - z_1\,c\,\mathbf{\hat{z}}$ | $(4f)$ | C |
| $\mathbf{B_4}$ | = | $\frac{1}{3}\,\mathbf{a_1} + \frac{2}{3}\,\mathbf{a_2} + \left(\frac{1}{2} - z_1\right)\,\mathbf{a_3}$ | = | $\frac{1}{2}\,a\,\mathbf{\hat{x}} + \frac{1}{2\sqrt{3}}\,a\,\mathbf{\hat{y}} + \left(\frac{1}{2} - z_1\right)\,c\,\mathbf{\hat{z}}$ | $(4f)$ | C |

**Geometry files:**

- CIF: pp. 750
- POSCAR: pp. 750

# $Ni_2In$ ($B8_2$) Structure: AB2_hP6_194_c_ad

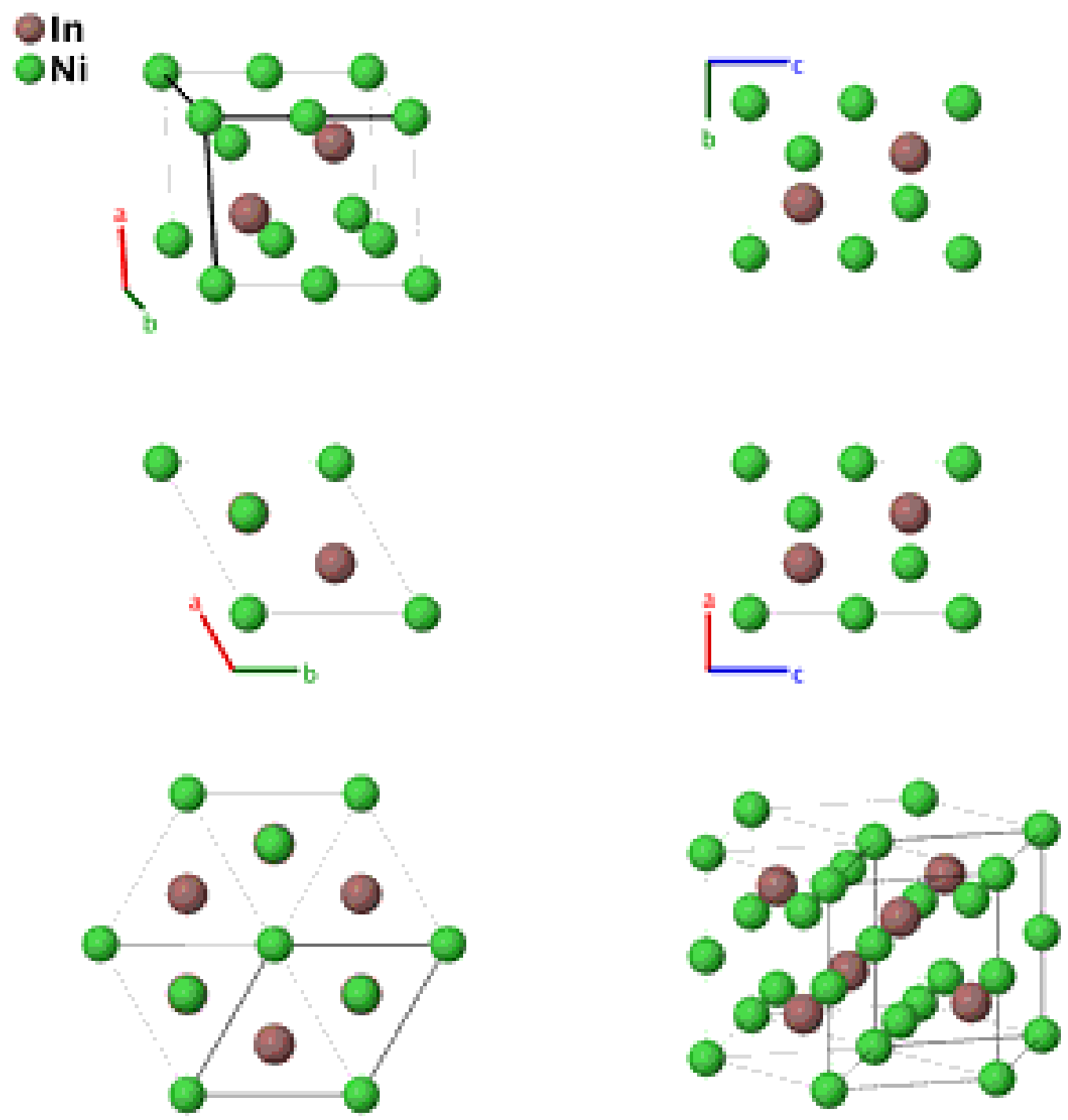

| | | |
|---|---|---|
| **Prototype** | : | $Ni_2In$ |
| **AFLOW prototype label** | : | AB2_hP6_194_c_ad |
| ***Strukturbericht* designation** | : | $B8_2$ |
| **Pearson symbol** | : | hP6 |
| **Space group number** | : | 194 |
| **Space group symbol** | : | $P6_3/mmc$ |
| **AFLOW prototype command** | : | `aflow --proto=AB2_hP6_194_c_ad --params=`$a, c/a$ |

**Other compounds with this structure:**

- AgAsBa, BeSiZr, CuKSe, LiBC, $Fe_2Sn$, GaMnPt, more

- Replacing the Ni-II atoms with In transforms the crystal into the C32 (hexagonal $\omega$) phase.

**Hexagonal primitive vectors:**

$$
\begin{aligned}
\mathbf{a}_1 &= \tfrac{1}{2}\, a\, \hat{\mathbf{x}} - \tfrac{\sqrt{3}}{2}\, a\, \hat{\mathbf{y}} \\
\mathbf{a}_2 &= \tfrac{1}{2}\, a\, \hat{\mathbf{x}} + \tfrac{\sqrt{3}}{2}\, a\, \hat{\mathbf{y}} \\
\mathbf{a}_3 &= c\, \hat{\mathbf{z}}
\end{aligned}
$$

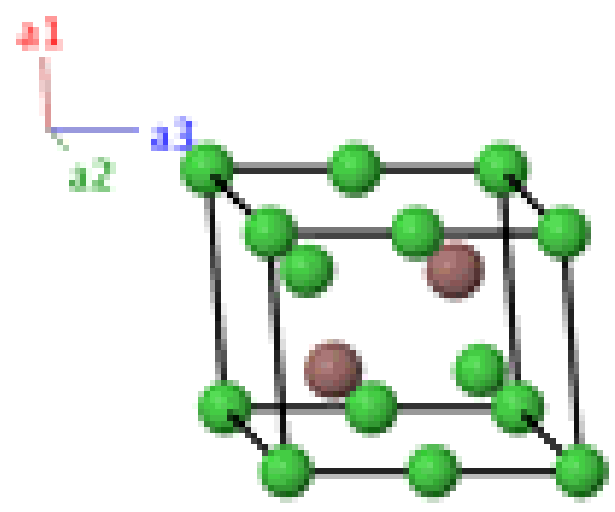

**Basis vectors:**

| | | Lattice Coordinates | | Cartesian Coordinates | Wyckoff Position | Atom Type |
|---|---|---|---|---|---|---|
| $\mathbf{B_1}$ | = | $0\,\mathbf{a_1} + 0\,\mathbf{a_2} + 0\,\mathbf{a_3}$ | = | $0\hat{\mathbf{x}} + 0\hat{\mathbf{y}} + 0\hat{\mathbf{z}}$ | $(2a)$ | Ni I |
| $\mathbf{B_2}$ | = | $\frac{1}{2}\,\mathbf{a_3}$ | = | $\frac{1}{2}\,c\,\hat{\mathbf{z}}$ | $(2a)$ | Ni I |
| $\mathbf{B_3}$ | = | $\frac{1}{3}\,\mathbf{a_1} + \frac{2}{3}\,\mathbf{a_2} + \frac{1}{4}\,\mathbf{a_3}$ | = | $\frac{1}{2}\,a\,\hat{\mathbf{x}} + \frac{1}{2\sqrt{3}}\,a\,\hat{\mathbf{y}} + \frac{1}{4}\,c\,\hat{\mathbf{z}}$ | $(2c)$ | In |
| $\mathbf{B_4}$ | = | $\frac{2}{3}\,\mathbf{a_1} + \frac{1}{3}\,\mathbf{a_2} + \frac{3}{4}\,\mathbf{a_3}$ | = | $\frac{1}{2}\,a\,\hat{\mathbf{x}} - \frac{1}{2\sqrt{3}}\,a\,\hat{\mathbf{y}} + \frac{3}{4}\,c\,\hat{\mathbf{z}}$ | $(2c)$ | In |
| $\mathbf{B_5}$ | = | $\frac{1}{3}\,\mathbf{a_1} + \frac{2}{3}\,\mathbf{a_2} + \frac{3}{4}\,\mathbf{a_3}$ | = | $\frac{1}{2}\,a\,\hat{\mathbf{x}} + \frac{1}{2\sqrt{3}}\,a\,\hat{\mathbf{y}} + \frac{3}{4}\,c\,\hat{\mathbf{z}}$ | $(2d)$ | Ni II |
| $\mathbf{B_6}$ | = | $\frac{2}{3}\,\mathbf{a_1} + \frac{1}{3}\,\mathbf{a_2} + \frac{1}{4}\,\mathbf{a_3}$ | = | $\frac{1}{2}\,a\,\hat{\mathbf{x}} - \frac{1}{2\sqrt{3}}\,a\,\hat{\mathbf{y}} + \frac{1}{4}\,c\,\hat{\mathbf{z}}$ | $(2d)$ | Ni II |

---

**Found in:**

---

**Geometry files:**

- CIF: pp. 750
- POSCAR: pp. 751

# $AlN_3Ti_4$ Structure: AB3C4_hP16_194_c_af_ef

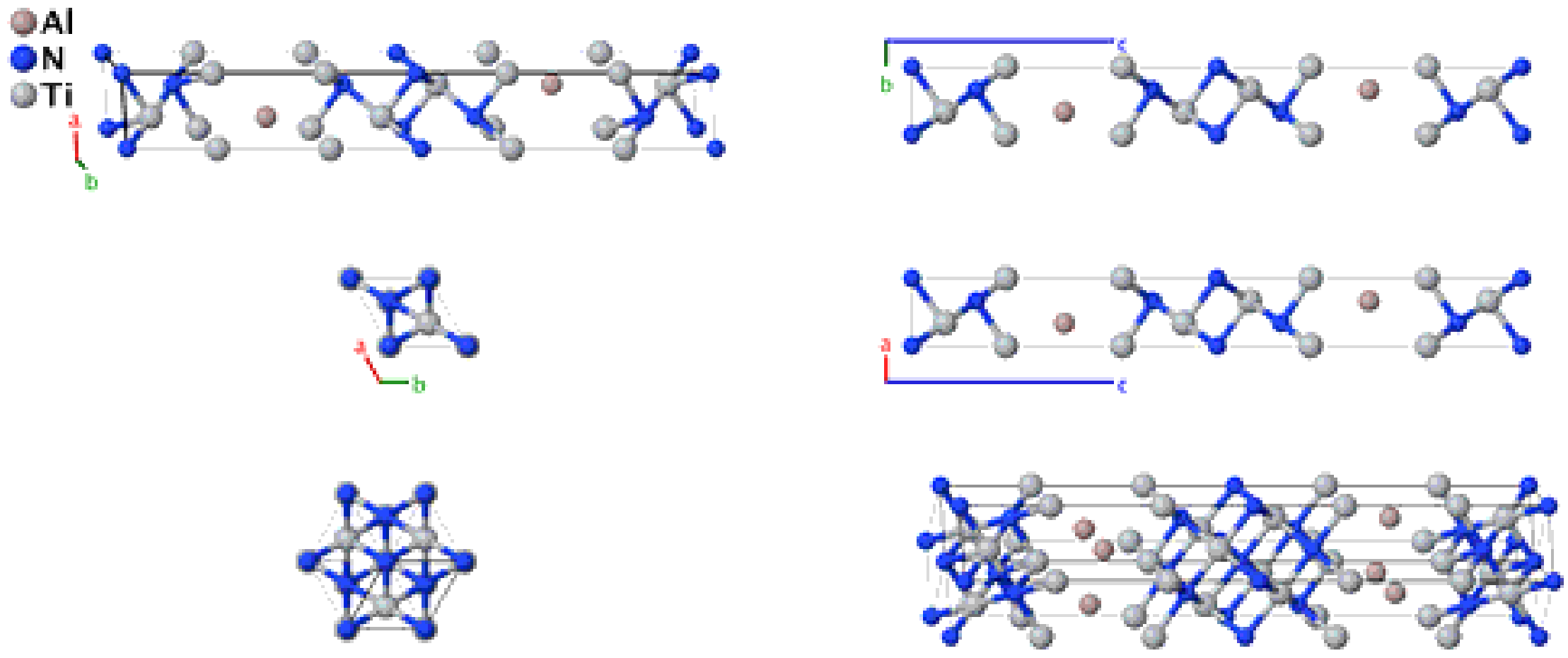

| | | |
|---|---|---|
| **Prototype** | : | $AlN_3Ti_4$ |
| **AFLOW prototype label** | : | AB3C4_hP16_194_c_af_ef |
| ***Strukturbericht* designation** | : | None |
| **Pearson symbol** | : | hP16 |
| **Space group number** | : | 194 |
| **Space group symbol** | : | $P6_3/mmc$ |
| **AFLOW prototype command** | : | `aflow --proto=AB3C4_hP16_194_c_af_ef` `--params=`$a, c/a, z_3, z_4, z_5$ |

- This is a so-called MAX phase. For more information, see (Radovic, 2013).

**Hexagonal primitive vectors:**

$$\mathbf{a}_1 = \frac{1}{2}\,a\,\hat{\mathbf{x}} - \frac{\sqrt{3}}{2}\,a\,\hat{\mathbf{y}}$$
$$\mathbf{a}_2 = \frac{1}{2}\,a\,\hat{\mathbf{x}} + \frac{\sqrt{3}}{2}\,a\,\hat{\mathbf{y}}$$
$$\mathbf{a}_3 = c\,\hat{\mathbf{z}}$$

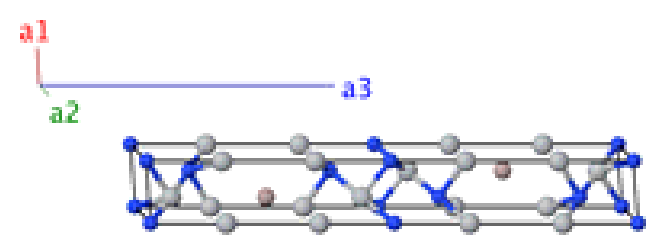

**Basis vectors:**

| | | Lattice Coordinates | | Cartesian Coordinates | Wyckoff Position | Atom Type |
|---|---|---|---|---|---|---|
| $\mathbf{B}_1$ | = | $0\,\mathbf{a}_1 + 0\,\mathbf{a}_2 + 0\,\mathbf{a}_3$ | = | $0\hat{\mathbf{x}} + 0\hat{\mathbf{y}} + 0\hat{\mathbf{z}}$ | (2$a$) | N I |
| $\mathbf{B}_2$ | = | $\frac{1}{2}\mathbf{a}_3$ | = | $\frac{1}{2}\,c\,\hat{\mathbf{z}}$ | (2$a$) | N I |
| $\mathbf{B}_3$ | = | $\frac{1}{3}\mathbf{a}_1 + \frac{2}{3}\mathbf{a}_2 + \frac{1}{4}\mathbf{a}_3$ | = | $\frac{1}{2}\,a\,\hat{\mathbf{x}} + \frac{1}{2\sqrt{3}}\,a\,\hat{\mathbf{y}} + \frac{1}{4}\,c\,\hat{\mathbf{z}}$ | (2$c$) | Al |
| $\mathbf{B}_4$ | = | $\frac{2}{3}\mathbf{a}_1 + \frac{1}{3}\mathbf{a}_2 + \frac{3}{4}\mathbf{a}_3$ | = | $\frac{1}{2}\,a\,\hat{\mathbf{x}} - \frac{1}{2\sqrt{3}}\,a\,\hat{\mathbf{y}} + \frac{3}{4}\,c\,\hat{\mathbf{z}}$ | (2$c$) | Al |
| $\mathbf{B}_5$ | = | $z_3\,\mathbf{a}_3$ | = | $z_3\,c\,\hat{\mathbf{z}}$ | (4$e$) | Ti I |
| $\mathbf{B}_6$ | = | $\left(\frac{1}{2} + z_3\right)\mathbf{a}_3$ | = | $\left(\frac{1}{2} + z_3\right)c\,\hat{\mathbf{z}}$ | (4$e$) | Ti I |
| $\mathbf{B}_7$ | = | $-z_3\,\mathbf{a}_3$ | = | $-z_3\,c\,\hat{\mathbf{z}}$ | (4$e$) | Ti I |
| $\mathbf{B}_8$ | = | $\left(\frac{1}{2} - z_3\right)\mathbf{a}_3$ | = | $\left(\frac{1}{2} - z_3\right)c\,\hat{\mathbf{z}}$ | (4$e$) | Ti I |
| $\mathbf{B}_9$ | = | $\frac{1}{3}\mathbf{a}_1 + \frac{2}{3}\mathbf{a}_2 + z_4\,\mathbf{a}_3$ | = | $\frac{1}{2}\,a\,\hat{\mathbf{x}} + \frac{1}{2\sqrt{3}}\,a\,\hat{\mathbf{y}} + z_4\,c\,\hat{\mathbf{z}}$ | (4$f$) | N II |
| $\mathbf{B_{10}}$ | = | $\frac{2}{3}\mathbf{a_1} + \frac{1}{3}\mathbf{a_2} + \left(\frac{1}{2} + z_4\right)\mathbf{a_3}$ | = | $\frac{1}{2}a\,\mathbf{\hat{x}} - \frac{1}{2\sqrt{3}}a\,\mathbf{\hat{y}} + \left(\frac{1}{2} + z_4\right)c\,\mathbf{\hat{z}}$ | $(4f)$ | N II |
| $\mathbf{B_{11}}$ | = | $\frac{2}{3}\mathbf{a_1} + \frac{1}{3}\mathbf{a_2} - z_4\,\mathbf{a_3}$ | = | $\frac{1}{2}a\,\mathbf{\hat{x}} - \frac{1}{2\sqrt{3}}a\,\mathbf{\hat{y}} - z_4\,c\,\mathbf{\hat{z}}$ | $(4f)$ | N II |
| $\mathbf{B_{12}}$ | = | $\frac{1}{3}\mathbf{a_1} + \frac{2}{3}\mathbf{a_2} + \left(\frac{1}{2} - z_4\right)\mathbf{a_3}$ | = | $\frac{1}{2}a\,\mathbf{\hat{x}} + \frac{1}{2\sqrt{3}}a\,\mathbf{\hat{y}} + \left(\frac{1}{2} - z_4\right)c\,\mathbf{\hat{z}}$ | $(4f)$ | N II |
| $\mathbf{B_13}$ | = | $\frac{1}{3}\mathbf{a_1} + \frac{2}{3}\mathbf{a_2} + z_5\,\mathbf{a_3}$ | = | $\frac{1}{2}a\,\mathbf{\hat{x}} + \frac{1}{2\sqrt{3}}a\,\mathbf{\hat{y}} + z_5\,c\,\mathbf{\hat{z}}$ | $(4f)$ | Ti II |
| $\mathbf{B_{14}}$ | = | $\frac{2}{3}\mathbf{a_1} + \frac{1}{3}\mathbf{a_2} + \left(\frac{1}{2} + z_5\right)\mathbf{a_3}$ | = | $\frac{1}{2}a\,\mathbf{\hat{x}} - \frac{1}{2\sqrt{3}}a\,\mathbf{\hat{y}} + \left(\frac{1}{2} + z_5\right)c\,\mathbf{\hat{z}}$ | $(4f)$ | Ti II |
| $\mathbf{B_{15}}$ | = | $\frac{2}{3}\mathbf{a_1} + \frac{1}{3}\mathbf{a_2} - z_5\,\mathbf{a_3}$ | = | $\frac{1}{2}a\,\mathbf{\hat{x}} - \frac{1}{2\sqrt{3}}a\,\mathbf{\hat{y}} - z_5\,c\,\mathbf{\hat{z}}$ | $(4f)$ | Ti II |
| $\mathbf{B_{16}}$ | = | $\frac{1}{3}\mathbf{a_1} + \frac{2}{3}\mathbf{a_2} + \left(\frac{1}{2} - z_5\right)\mathbf{a_3}$ | = | $\frac{1}{2}a\,\mathbf{\hat{x}} + \frac{1}{2\sqrt{3}}a\,\mathbf{\hat{y}} + \left(\frac{1}{2} - z_5\right)c\,\mathbf{\hat{z}}$ | $(4f)$ | Ti II |

---

---

**Geometry files:**

- CIF: pp. 751
- POSCAR: pp. 751

# Hexagonal Close Packed (Mg, A3) Structure: A_hP2_194_c

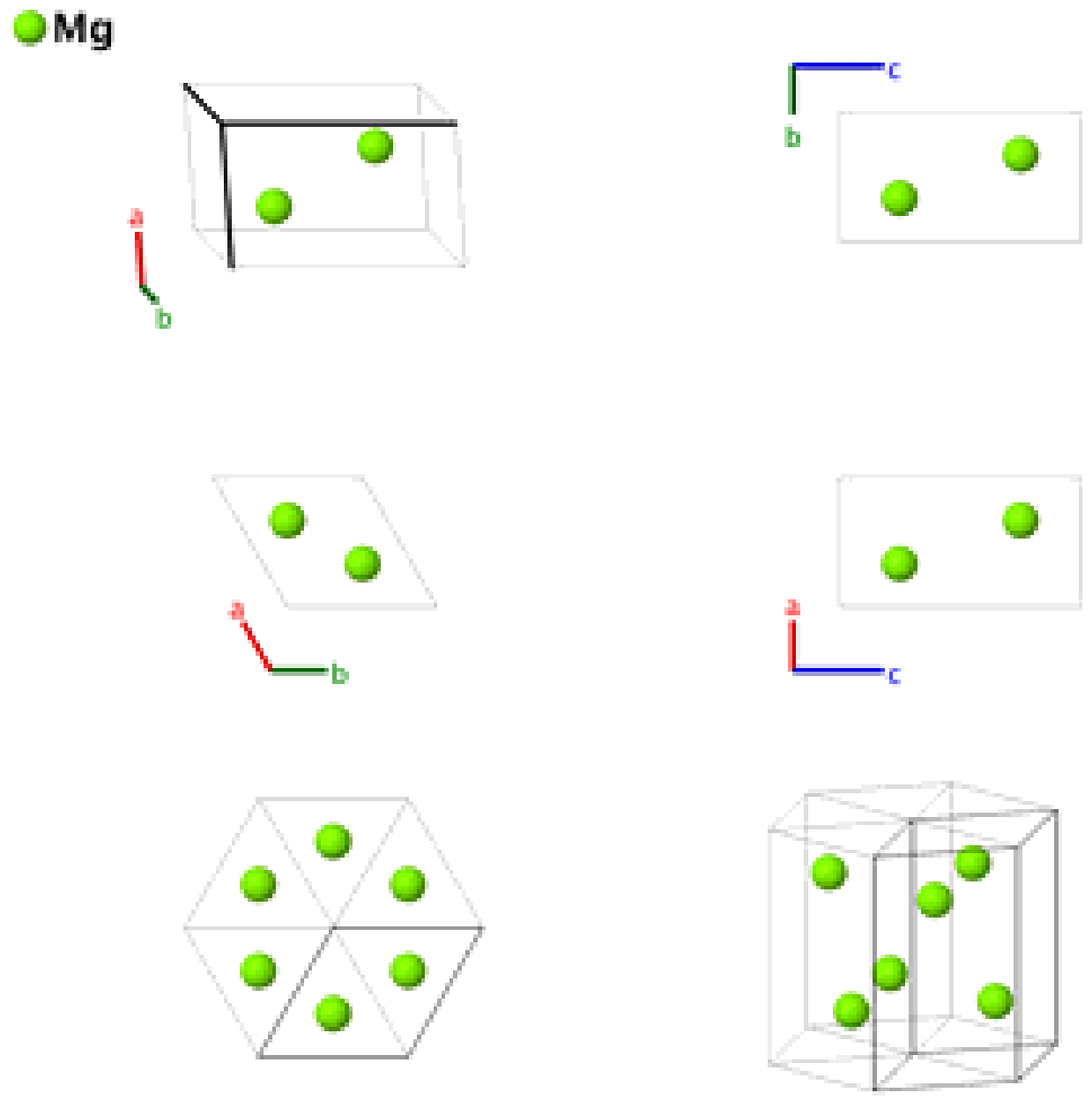

| | | |
|---|---|---|
| **Prototype** | : | Mg |
| **AFLOW prototype label** | : | A_hP2_194_c |
| ***Strukturbericht* designation** | : | A3 |
| **Pearson symbol** | : | hP2 |
| **Space group number** | : | 194 |
| **Space group symbol** | : | $P6_3/mmc$ |
| **AFLOW prototype command** | : | `aflow --proto=A_hP2_194_c`<br>`--params=`$a, c/a$ |

**Other elements with this structure:**

- Be, Sc, Ti, Co, Zn, Y, Zr, Tc, Ru, Cd, Gd, Tb, Dy, Ho, Er, Tm, Lu, Hf, Re, Os, Tl

**Hexagonal primitive vectors:**

$$
\begin{aligned}
\mathbf{a}_1 &= \tfrac{1}{2}\,a\,\mathbf{\hat{x}} - \tfrac{\sqrt{3}}{2}\,a\,\mathbf{\hat{y}} \\
\mathbf{a}_2 &= \tfrac{1}{2}\,a\,\mathbf{\hat{x}} + \tfrac{\sqrt{3}}{2}\,a\,\mathbf{\hat{y}} \\
\mathbf{a}_3 &= c\,\mathbf{\hat{z}}
\end{aligned}
$$

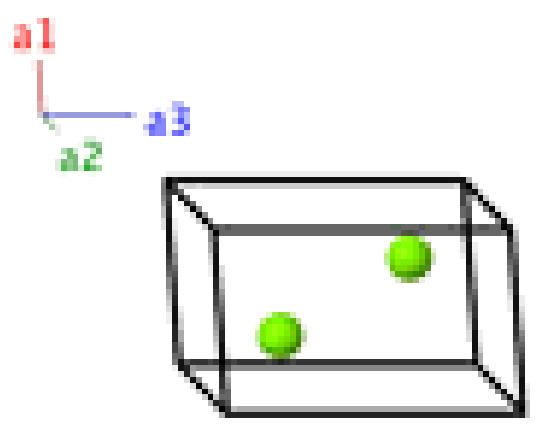

**Basis vectors:**

| | | Lattice Coordinates | | Cartesian Coordinates | Wyckoff Position | Atom Type |
|---|---|---|---|---|---|---|
| $\mathbf{B}_1$ | = | $\frac{1}{3}\,\mathbf{a}_1 + \frac{2}{3}\,\mathbf{a}_2 + \frac{1}{4}\,\mathbf{a}_3$ | = | $\frac{1}{2}\,a\,\mathbf{\hat{x}} + \frac{1}{2\sqrt{3}}\,a\,\mathbf{\hat{y}} + \frac{1}{4}\,c\,\mathbf{\hat{z}}$ | (2c) | Mg |
| $\mathbf{B}_2$ | = | $\frac{2}{3}\,\mathbf{a}_1 + \frac{1}{3}\,\mathbf{a}_2 + \frac{3}{4}\,\mathbf{a}_3$ | = | $\frac{1}{2}\,a\,\mathbf{\hat{x}} - \frac{1}{2\sqrt{3}}\,a\,\mathbf{\hat{y}} + \frac{3}{4}\,c\,\mathbf{\hat{z}}$ | (2c) | Mg |

**Found in:**

**Geometry files:**

- CIF: pp. 751
- POSCAR: pp. 752

# $MgNi_2$ Hexagonal Laves (C36) Structure: AB2_hP24_194_ef_fgh

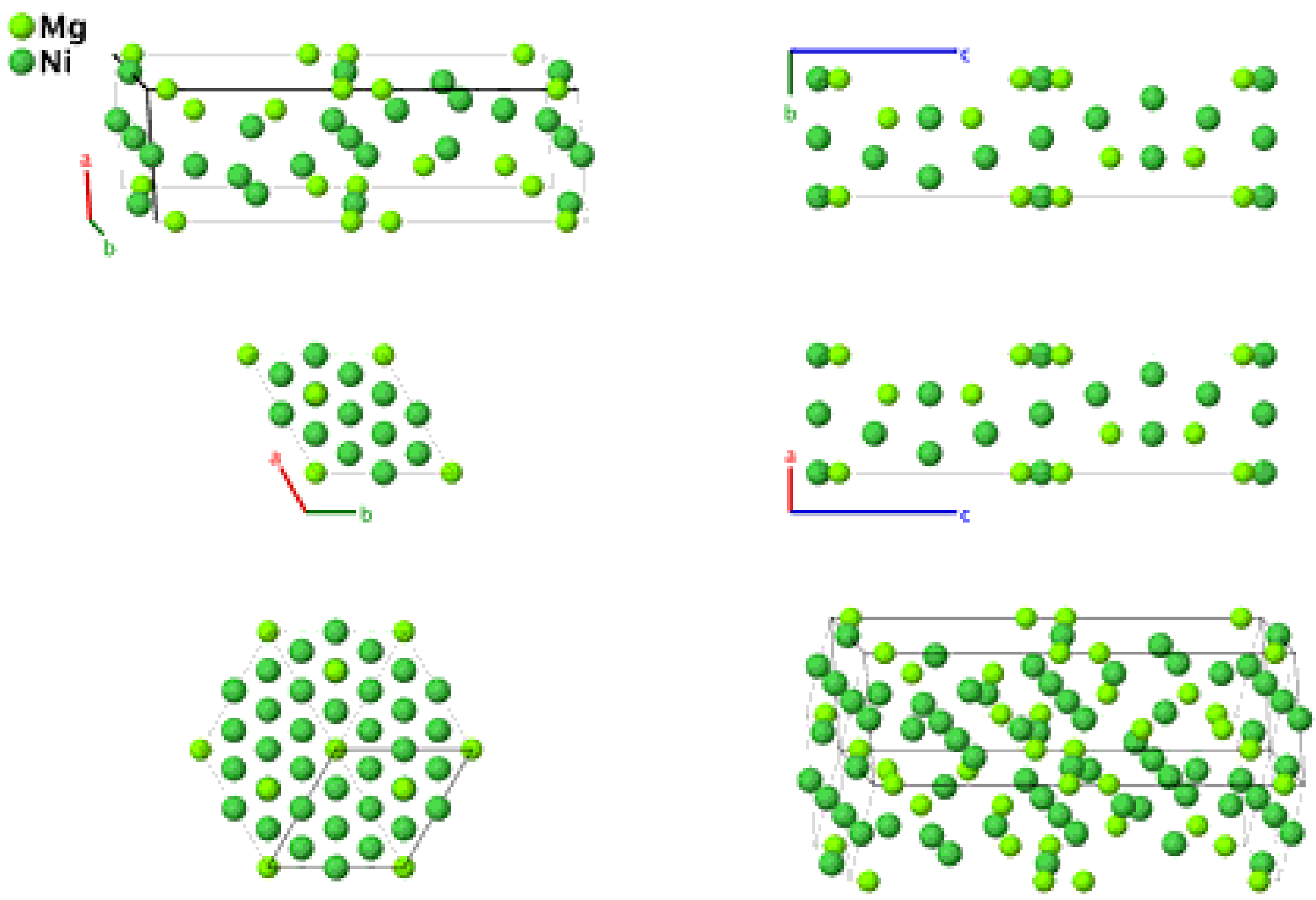

| | | |
|---|---|---|
| **Prototype** | : | $MgNi_2$ |
| **AFLOW prototype label** | : | AB2_hP24_194_ef_fgh |
| ***Strukturbericht* designation** | : | C36 |
| **Pearson symbol** | : | hP24 |
| **Space group number** | : | 194 |
| **Space group symbol** | : | $P6_3/mmc$ |
| **AFLOW prototype command** | : | `aflow --proto=AB2_hP24_194_ef_fgh`<br>`--params=`$a, c/a, z_1, z_2, z_3, x_5$ |

**Other compounds with this structure:**

- $NbZn_2$, $ScFe_2$, $ThMg_2$, $HfCr_2$, $UPt_2$

**Hexagonal primitive vectors:**

$$\mathbf{a}_1 = \frac{1}{2}\,a\,\hat{\mathbf{x}} - \frac{\sqrt{3}}{2}\,a\,\hat{\mathbf{y}}$$
$$\mathbf{a}_2 = \frac{1}{2}\,a\,\hat{\mathbf{x}} + \frac{\sqrt{3}}{2}\,a\,\hat{\mathbf{y}}$$
$$\mathbf{a}_3 = c\,\hat{\mathbf{z}}$$

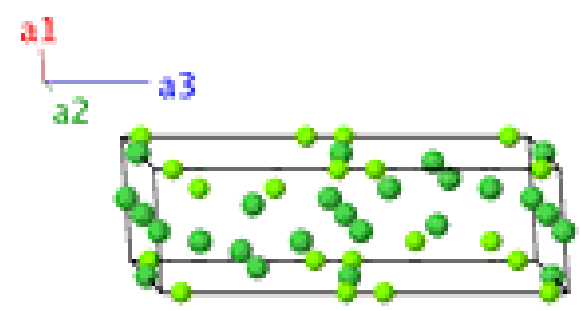

**Basis vectors:**

| | | Lattice Coordinates | | Cartesian Coordinates | Wyckoff Position | Atom Type |
|---|---|---|---|---|---|---|
| $\mathbf{B}_1$ | = | $z_1\,\mathbf{a}_3$ | = | $z_1\,c\,\hat{\mathbf{z}}$ | (4$e$) | Mg I |
| $\mathbf{B}_2$ | = | $-z_1\,\mathbf{a}_3$ | = | $-z_1\,c\,\hat{\mathbf{z}}$ | (4$e$) | Mg I |
| $\mathbf{B}_3$ | = | $\left(\frac{1}{2} + z_1\right)\mathbf{a}_3$ | = | $\left(\frac{1}{2} + z_1\right)c\,\hat{\mathbf{z}}$ | (4$e$) | Mg I |
| $\mathbf{B}_{4}$ | = | $\left(\frac{1}{2}-z_{1}\right)\mathbf{a}_{3}$ | = | $\left(\frac{1}{2}-z_{1}\right)c\,\hat{\mathbf{z}}$ | (4e) | Mg I |
| $\mathbf{B}_{5}$ | = | $\frac{1}{3}\mathbf{a}_{1}+\frac{2}{3}\mathbf{a}_{2}+z_{2}\mathbf{a}_{3}$ | = | $\frac{1}{2}a\,\hat{\mathbf{x}}+\frac{1}{2\sqrt{3}}a\,\hat{\mathbf{y}}+z_{2}c\,\hat{\mathbf{z}}$ | (4f) | Mg II |
| $\mathbf{B}_{6}$ | = | $\frac{2}{3}\mathbf{a}_{1}+\frac{1}{3}\mathbf{a}_{2}+\left(\frac{1}{2}+z_{2}\right)\mathbf{a}_{3}$ | = | $\frac{1}{2}a\,\hat{\mathbf{x}}-\frac{1}{2\sqrt{3}}a\,\hat{\mathbf{y}}+\left(\frac{1}{2}+z_{2}\right)c\,\hat{\mathbf{z}}$ | (4f) | Mg II |
| $\mathbf{B}_{7}$ | = | $\frac{2}{3}\mathbf{a}_{1}+\frac{1}{3}\mathbf{a}_{2}-z_{2}\mathbf{a}_{3}$ | = | $\frac{1}{2}a\,\hat{\mathbf{x}}-\frac{1}{2\sqrt{3}}a\,\hat{\mathbf{y}}-z_{2}c\,\hat{\mathbf{z}}$ | (4f) | Mg II |
| $\mathbf{B}_{8}$ | = | $\frac{1}{3}\mathbf{a}_{1}+\frac{2}{3}\mathbf{a}_{2}+\left(\frac{1}{2}-z_{2}\right)\mathbf{a}_{3}$ | = | $\frac{1}{2}a\,\hat{\mathbf{x}}+\frac{1}{2\sqrt{3}}a\,\hat{\mathbf{y}}+\left(\frac{1}{2}-z_{2}\right)c\,\hat{\mathbf{z}}$ | (4f) | Mg II |
| $\mathbf{B}_{9}$ | = | $\frac{1}{3}\mathbf{a}_{1}+\frac{2}{3}\mathbf{a}_{2}+z_{3}\mathbf{a}_{3}$ | = | $\frac{1}{2}a\,\hat{\mathbf{x}}+\frac{1}{2\sqrt{3}}a\,\hat{\mathbf{y}}+z_{3}c\,\hat{\mathbf{z}}$ | (4f) | Ni I |
| $\mathbf{B}_{10}$ | = | $\frac{2}{3}\mathbf{a}_{1}+\frac{1}{3}\mathbf{a}_{2}+\left(\frac{1}{2}+z_{3}\right)\mathbf{a}_{3}$ | = | $\frac{1}{2}a\,\hat{\mathbf{x}}-\frac{1}{2\sqrt{3}}a\,\hat{\mathbf{y}}+\left(\frac{1}{2}+z_{3}\right)c\,\hat{\mathbf{z}}$ | (4f) | Ni I |
| $\mathbf{B}_{11}$ | = | $\frac{2}{3}\mathbf{a}_{1}+\frac{1}{3}\mathbf{a}_{2}-z_{3}\mathbf{a}_{3}$ | = | $\frac{1}{2}a\,\hat{\mathbf{x}}-\frac{1}{2\sqrt{3}}a\,\hat{\mathbf{y}}-z_{3}c\,\hat{\mathbf{z}}$ | (4f) | Ni I |
| $\mathbf{B}_{12}$ | = | $\frac{1}{3}\mathbf{a}_{1}+\frac{2}{3}\mathbf{a}_{2}+\left(\frac{1}{2}-z_{3}\right)\mathbf{a}_{3}$ | = | $\frac{1}{2}a\,\hat{\mathbf{x}}+\frac{1}{2\sqrt{3}}a\,\hat{\mathbf{y}}+\left(\frac{1}{2}-z_{3}\right)c\,\hat{\mathbf{z}}$ | (4f) | Ni I |
| $\mathbf{B}_{13}$ | = | $\frac{1}{2}\mathbf{a}_{1}$ | = | $\frac{1}{4}a\,\hat{\mathbf{x}}-\frac{\sqrt{3}}{4}a\,\hat{\mathbf{y}}$ | (6g) | Ni II |
| $\mathbf{B}_{14}$ | = | $\frac{1}{2}\mathbf{a}_{2}$ | = | $\frac{1}{4}a\,\hat{\mathbf{x}}+\frac{\sqrt{3}}{4}a\,\hat{\mathbf{y}}$ | (6g) | Ni II |
| $\mathbf{B}_{15}$ | = | $\frac{1}{2}\mathbf{a}_{1}+\frac{1}{2}\mathbf{a}_{2}$ | = | $\frac{1}{2}a\,\hat{\mathbf{x}}$ | (6g) | Ni II |
| $\mathbf{B}_{16}$ | = | $\frac{1}{2}\mathbf{a}_{1}+\frac{1}{2}\mathbf{a}_{3}$ | = | $\frac{1}{4}a\,\hat{\mathbf{x}}-\frac{\sqrt{3}}{4}a\,\hat{\mathbf{y}}+\frac{1}{2}c\,\hat{\mathbf{z}}$ | (6g) | Ni II |
| $\mathbf{B}_{17}$ | = | $\frac{1}{2}\mathbf{a}_{2}+\frac{1}{2}\mathbf{a}_{3}$ | = | $\frac{1}{4}a\,\hat{\mathbf{x}}+\frac{\sqrt{3}}{4}a\,\hat{\mathbf{y}}+\frac{1}{2}c\,\hat{\mathbf{z}}$ | (6g) | Ni II |
| $\mathbf{B}_{18}$ | = | $\frac{1}{2}\mathbf{a}_{1}+\frac{1}{2}\mathbf{a}_{2}+\frac{1}{2}\mathbf{a}_{3}$ | = | $\frac{1}{2}a\,\hat{\mathbf{x}}+\frac{1}{2}c\,\hat{\mathbf{z}}$ | (6g) | Ni II |
| $\mathbf{B}_{19}$ | = | $x_{5}\mathbf{a}_{1}+2x_{5}\mathbf{a}_{2}+\frac{1}{4}\mathbf{a}_{3}$ | = | $\frac{3}{2}x_{5}a\,\hat{\mathbf{x}}+\frac{\sqrt{3}}{2}x_{5}a\,\hat{\mathbf{y}}+\frac{1}{4}c\,\hat{\mathbf{z}}$ | (6h) | Ni III |
| $\mathbf{B}_{20}$ | = | $-2x_{5}\mathbf{a}_{1}-x_{5}\mathbf{a}_{2}+\frac{1}{4}\mathbf{a}_{3}$ | = | $-\frac{3}{2}x_{5}a\,\hat{\mathbf{x}}+\frac{\sqrt{3}}{2}x_{5}a\,\hat{\mathbf{y}}+\frac{1}{4}c\,\hat{\mathbf{z}}$ | (6h) | Ni III |
| $\mathbf{B}_{21}$ | = | $x_{5}\mathbf{a}_{1}-x_{5}\mathbf{a}_{2}+\frac{1}{4}\mathbf{a}_{3}$ | = | $-\sqrt{3}x_{5}a\,\hat{\mathbf{y}}+\frac{1}{4}c\,\hat{\mathbf{z}}$ | (6h) | Ni III |
| $\mathbf{B}_{22}$ | = | $-x_{5}\mathbf{a}_{1}-2x_{5}\mathbf{a}_{2}+\frac{3}{4}\mathbf{a}_{3}$ | = | $-\frac{3}{2}x_{5}a\,\hat{\mathbf{x}}-\frac{\sqrt{3}}{2}x_{5}a\,\hat{\mathbf{y}}+\frac{3}{4}c\,\hat{\mathbf{z}}$ | (6h) | Ni III |
| $\mathbf{B}_{23}$ | = | $2x_{5}\mathbf{a}_{1}+x_{5}\mathbf{a}_{2}+\frac{3}{4}\mathbf{a}_{3}$ | = | $\frac{3}{2}x_{5}a\,\hat{\mathbf{x}}-\frac{\sqrt{3}}{2}x_{5}a\,\hat{\mathbf{y}}+\frac{3}{4}c\,\hat{\mathbf{z}}$ | (6h) | Ni III |
| $\mathbf{B}_{24}$ | = | $-x_{5}\mathbf{a}_{1}+x_{5}\mathbf{a}_{2}+\frac{3}{4}\mathbf{a}_{3}$ | = | $+\sqrt{3}x_{5}a\,\hat{\mathbf{y}}+\frac{3}{4}c\,\hat{\mathbf{z}}$ | (6h) | Ni III |

**Geometry files:**

- CIF: pp. 752
- POSCAR: pp. 752

# Covellite (CuS, B18) Structure: AB_hP12_194_df_ce

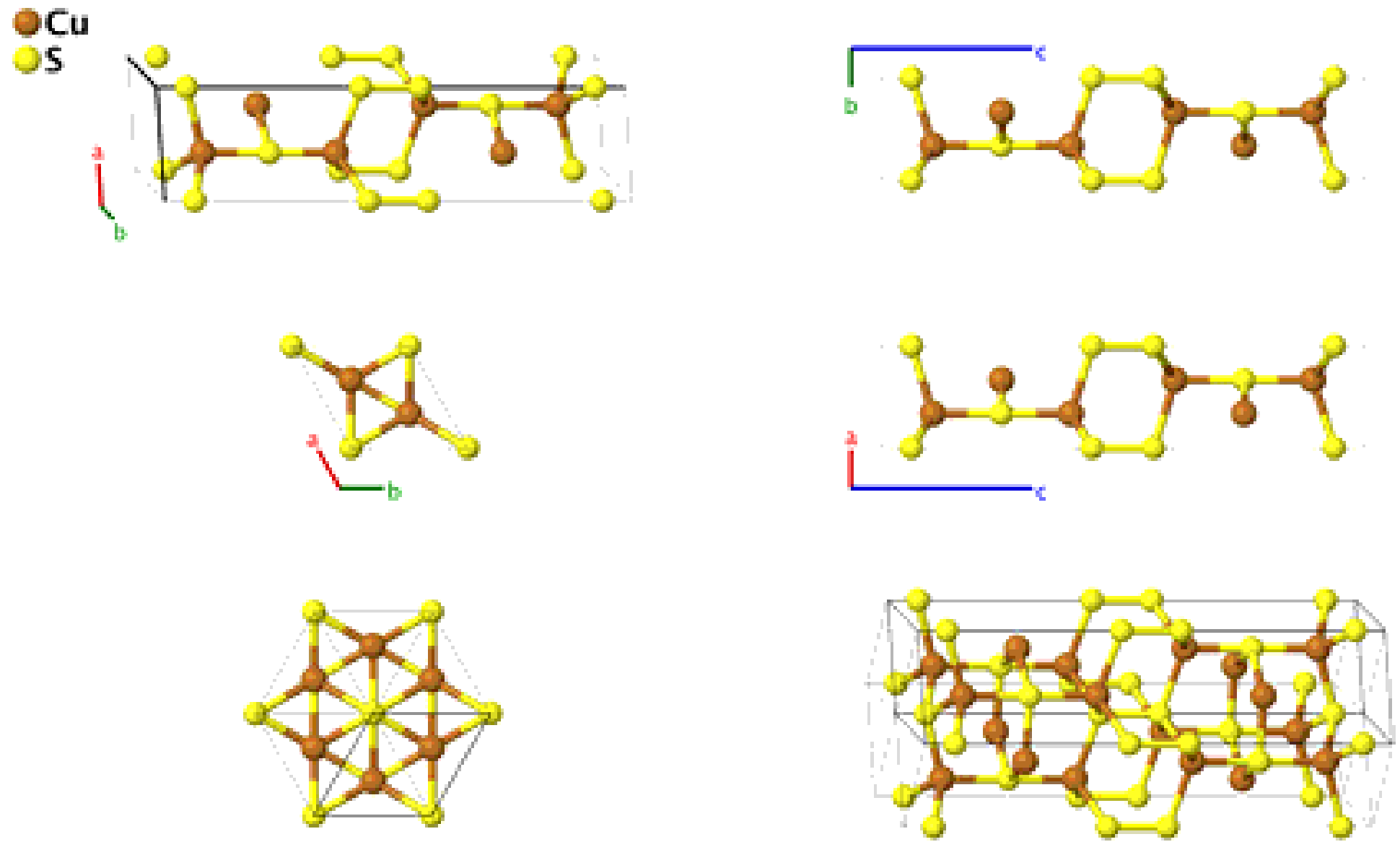

| | | |
|---|---|---|
| **Prototype** | : | CuS |
| **AFLOW prototype label** | : | AB_hP12_194_df_ce |
| ***Strukturbericht* designation** | : | B18 |
| **Pearson symbol** | : | hP12 |
| **Space group number** | : | 194 |
| **Space group symbol** | : | $P6_3/mmc$ |
| **AFLOW prototype command** | : | `aflow --proto=AB_hP12_194_df_ce`<br>`--params=`$a, c/a, z_3, z_4$ |

**Other compounds with this structure:**

- CuSe

**Hexagonal primitive vectors:**

$$\mathbf{a}_1 = \frac{1}{2}\,a\,\mathbf{\hat{x}} - \frac{\sqrt{3}}{2}\,a\,\mathbf{\hat{y}}$$

$$\mathbf{a}_2 = \frac{1}{2}\,a\,\mathbf{\hat{x}} + \frac{\sqrt{3}}{2}\,a\,\mathbf{\hat{y}}$$

$$\mathbf{a}_3 = c\,\mathbf{\hat{z}}$$

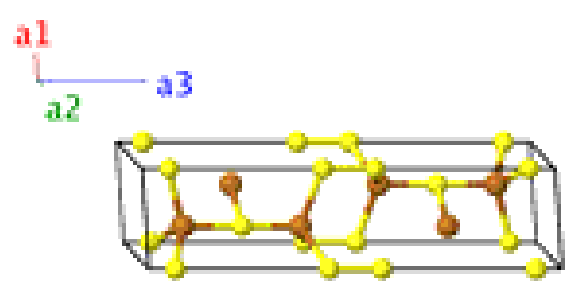

**Basis vectors:**

| | | Lattice Coordinates | | Cartesian Coordinates | Wyckoff Position | Atom Type |
|---|---|---|---|---|---|---|
| $\mathbf{B}_1$ | = | $\frac{1}{3}\,\mathbf{a}_1 + \frac{2}{3}\,\mathbf{a}_2 + \frac{1}{4}\,\mathbf{a}_3$ | = | $\frac{1}{2}\,a\,\mathbf{\hat{x}} + \frac{1}{2\sqrt{3}}\,a\,\mathbf{\hat{y}} + \frac{1}{4}\,c\,\mathbf{\hat{z}}$ | (2*c*) | S I |
| $\mathbf{B}_2$ | = | $\frac{2}{3}\,\mathbf{a}_1 + \frac{1}{3}\,\mathbf{a}_2 + \frac{3}{4}\,\mathbf{a}_3$ | = | $\frac{1}{2}\,a\,\mathbf{\hat{x}} - \frac{1}{2\sqrt{3}}\,a\,\mathbf{\hat{y}} + \frac{3}{4}\,c\,\mathbf{\hat{z}}$ | (2*c*) | S I |
| $\mathbf{B}_3$ | = | $\frac{1}{3}\,\mathbf{a}_1 + \frac{2}{3}\,\mathbf{a}_2 + \frac{3}{4}\,\mathbf{a}_3$ | = | $\frac{1}{2}\,a\,\mathbf{\hat{x}} + \frac{1}{2\sqrt{3}}\,a\,\mathbf{\hat{y}} + \frac{3}{4}\,c\,\mathbf{\hat{z}}$ | (2*d*) | Cu I |
| $\mathbf{B}_4$ | = | $\frac{2}{3}\,\mathbf{a}_1 + \frac{1}{3}\,\mathbf{a}_2 + \frac{1}{4}\,\mathbf{a}_3$ | = | $\frac{1}{2}\,a\,\mathbf{\hat{x}} - \frac{1}{2\sqrt{3}}\,a\,\mathbf{\hat{y}} + \frac{1}{4}\,c\,\mathbf{\hat{z}}$ | (2*d*) | Cu I |
| $\mathbf{B}_5$ | = | $z_3\,\mathbf{a}_3$ | = | $z_3\,c\,\mathbf{\hat{z}}$ | (4*e*) | S II |
| $\mathbf{B_6}$ | = | $-z_3\,\mathbf{a_3}$ | = | $-z_3\,c\,\hat{\mathbf{z}}$ | (4$e$) | S II |
| $\mathbf{B_7}$ | = | $\left(\frac{1}{2}+z_3\right)\mathbf{a_3}$ | = | $\left(\frac{1}{2}+z_3\right)c\,\hat{\mathbf{z}}$ | (4$e$) | S II |
| $\mathbf{B_8}$ | = | $\left(\frac{1}{2}-z_3\right)\mathbf{a_3}$ | = | $\left(\frac{1}{2}-z_3\right)c\,\hat{\mathbf{z}}$ | (4$e$) | S II |
| $\mathbf{B_9}$ | = | $\frac{1}{3}\,\mathbf{a_1}+\frac{2}{3}\,\mathbf{a_2}+z_4\,\mathbf{a_3}$ | = | $\frac{1}{2}\,a\,\hat{\mathbf{x}}+\frac{1}{2\sqrt{3}}\,a\,\hat{\mathbf{y}}+z_4\,c\,\hat{\mathbf{z}}$ | (4$f$) | Cu II |
| $\mathbf{B_{10}}$ | = | $\frac{2}{3}\,\mathbf{a_1}+\frac{1}{3}\,\mathbf{a_2}+\left(\frac{1}{2}+z_4\right)\mathbf{a_3}$ | = | $\frac{1}{2}\,a\,\hat{\mathbf{x}}-\frac{1}{2\sqrt{3}}\,a\,\hat{\mathbf{y}}+\left(\frac{1}{2}+z_4\right)c\,\hat{\mathbf{z}}$ | (4$f$) | Cu II |
| $\mathbf{B_{11}}$ | = | $\frac{2}{3}\,\mathbf{a_1}+\frac{1}{3}\,\mathbf{a_2}-z_4\,\mathbf{a_3}$ | = | $\frac{1}{2}\,a\,\hat{\mathbf{x}}-\frac{1}{2\sqrt{3}}\,a\,\hat{\mathbf{y}}-z_4\,c\,\hat{\mathbf{z}}$ | (4$f$) | Cu II |
| $\mathbf{B_{12}}$ | = | $\frac{1}{3}\,\mathbf{a_1}+\frac{2}{3}\,\mathbf{a_2}+\left(\frac{1}{2}-z_4\right)\mathbf{a_3}$ | = | $\frac{1}{2}\,a\,\hat{\mathbf{x}}+\frac{1}{2\sqrt{3}}\,a\,\hat{\mathbf{y}}+\left(\frac{1}{2}-z_4\right)c\,\hat{\mathbf{z}}$ | (4$f$) | Cu II |

---

---

**Geometry files:**

- CIF: pp. 752
- POSCAR: pp. 753

# NiAs ($B8_1$) Structure: AB_hP4_194_c_a

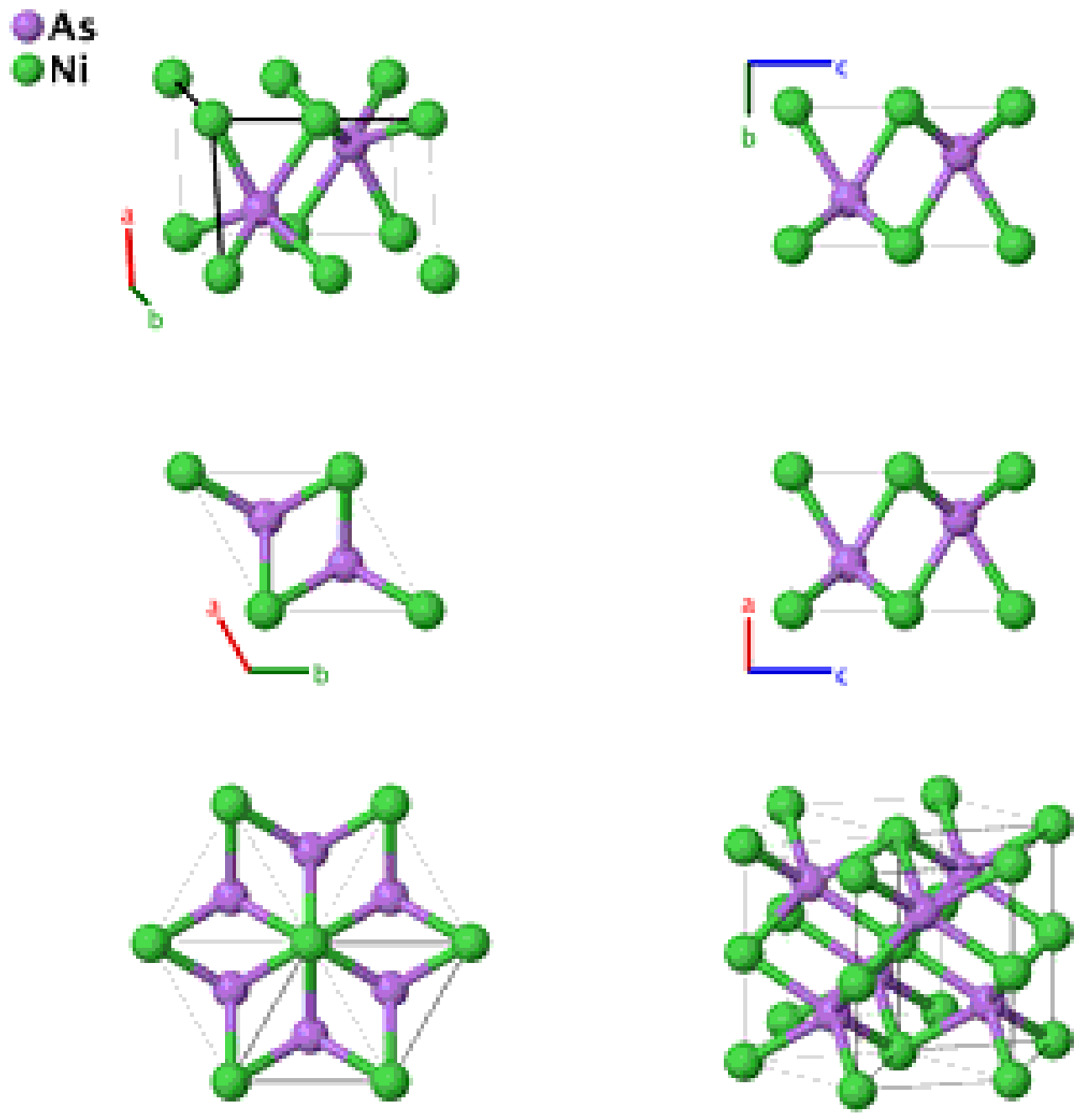

| | | |
|---|---|---|
| **Prototype** | : | NiAs |
| **AFLOW prototype label** | : | AB_hP4_194_c_a |
| ***Strukturbericht* designation** | : | $B8_1$ |
| **Pearson symbol** | : | hP4 |
| **Space group number** | : | 194 |
| **Space group symbol** | : | $P6_3/mmc$ |
| **AFLOW prototype command** | : | `aflow --proto=AB_hP4_194_c_a` <br> `--params=`$a, c/a$ |

**Other compounds with this structure:**

- AuSn, CoTe, CrSe, CuSn, FeS, IrS, MnAs, NiSn, PdSb, PtB, RhSn, VP, ZrTe

- Note that the stacking is ABACABAC, with the Ni atoms on the A sites and As on B and C. The environment of the Ni atoms is fcc-like, and the environment of the As atoms is hcp-like.

**Hexagonal primitive vectors:**

$$\begin{aligned}
\mathbf{a}_1 &= \tfrac{1}{2}\,a\,\hat{\mathbf{x}} - \tfrac{\sqrt{3}}{2}\,a\,\hat{\mathbf{y}} \\
\mathbf{a}_2 &= \tfrac{1}{2}\,a\,\hat{\mathbf{x}} + \tfrac{\sqrt{3}}{2}\,a\,\hat{\mathbf{y}} \\
\mathbf{a}_3 &= c\,\hat{\mathbf{z}}
\end{aligned}$$

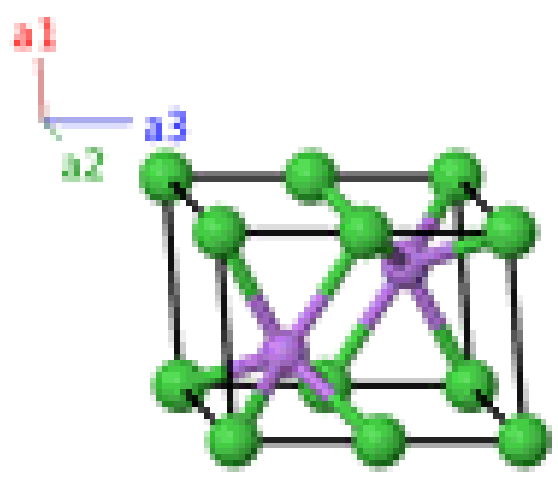

**Basis vectors:**

| | | Lattice Coordinates | | Cartesian Coordinates | Wyckoff Position | Atom Type |
|---|---|---|---|---|---|---|
| $\mathbf{B_1}$ | = | $0\,\mathbf{a_1} + 0\,\mathbf{a_2} + 0\,\mathbf{a_3}$ | = | $0\hat{\mathbf{x}} + 0\hat{\mathbf{y}} + 0\hat{\mathbf{z}}$ | $(2a)$ | Ni |
| $\mathbf{B_2}$ | = | $\frac{1}{2}\,\mathbf{a_3}$ | = | $\frac{1}{2}\,c\,\hat{\mathbf{z}}$ | $(2a)$ | Ni |
| $\mathbf{B_3}$ | = | $\frac{1}{3}\,\mathbf{a_1} + \frac{2}{3}\,\mathbf{a_2} + \frac{1}{4}\,\mathbf{a_3}$ | = | $\frac{1}{2}\,a\,\hat{\mathbf{x}} + \frac{1}{2\sqrt{3}}\,a\,\hat{\mathbf{y}} + \frac{1}{4}\,c\,\hat{\mathbf{z}}$ | $(2c)$ | As |
| $\mathbf{B_4}$ | = | $\frac{2}{3}\,\mathbf{a_1} + \frac{1}{3}\,\mathbf{a_2} + \frac{3}{4}\,\mathbf{a_3}$ | = | $\frac{1}{2}\,a\,\hat{\mathbf{x}} - \frac{1}{2\sqrt{3}}\,a\,\hat{\mathbf{y}} + \frac{3}{4}\,c\,\hat{\mathbf{z}}$ | $(2c)$ | As |

**Geometry files:**

- CIF: pp. 753
- POSCAR: pp. 753

# $\beta$-Tridymite ($SiO_2$) Structure (C10): A2B_hP12_194_cg_f

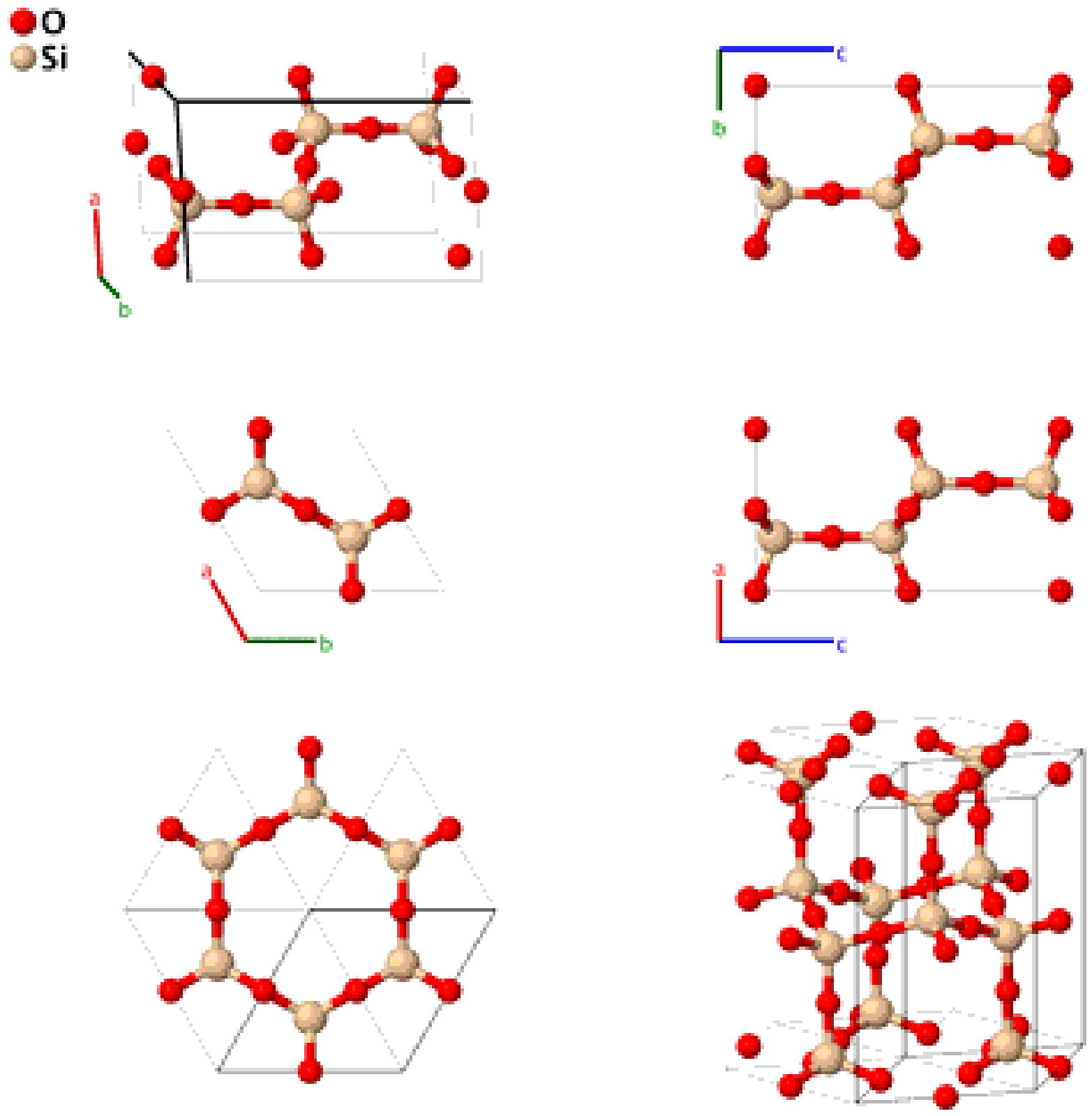

| | | |
|---|---|---|
| **Prototype** | : | $SiO_2$ |
| **AFLOW prototype label** | : | A2B_hP12_194_cg_f |
| ***Strukturbericht* designation** | : | C10 |
| **Pearson symbol** | : | hP12 |
| **Space group number** | : | 194 |
| **Space group symbol** | : | $P6_3/mmc$ |
| **AFLOW prototype command** | : | `aflow --proto=A2B_hP12_194_cg_f` `--params=`$a, c/a, z_2$ |

**Hexagonal primitive vectors:**

$$\mathbf{a}_1 = \frac{1}{2}\,a\,\hat{\mathbf{x}} - \frac{\sqrt{3}}{2}\,a\,\hat{\mathbf{y}}$$
$$\mathbf{a}_2 = \frac{1}{2}\,a\,\hat{\mathbf{x}} + \frac{\sqrt{3}}{2}\,a\,\hat{\mathbf{y}}$$
$$\mathbf{a}_3 = c\,\hat{\mathbf{z}}$$

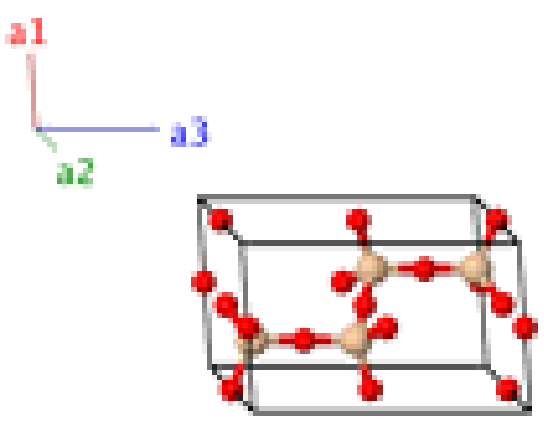

**Basis vectors:**

| | | Lattice Coordinates | | Cartesian Coordinates | Wyckoff Position | Atom Type |
|---|---|---|---|---|---|---|
| $\mathbf{B}_1$ | = | $\frac{1}{3}\,\mathbf{a}_1 + \frac{2}{3}\,\mathbf{a}_2 + \frac{1}{4}\,\mathbf{a}_3$ | = | $\frac{1}{2}\,a\,\hat{\mathbf{x}} + \frac{1}{2\sqrt{3}}\,a\,\hat{\mathbf{y}} + \frac{1}{4}\,c\,\hat{\mathbf{z}}$ | (2c) | O I |
| $\mathbf{B}_2$ | = | $\frac{2}{3}\,\mathbf{a}_1 + \frac{1}{3}\,\mathbf{a}_2 + \frac{3}{4}\,\mathbf{a}_3$ | = | $\frac{1}{2}\,a\,\hat{\mathbf{x}} - \frac{1}{2\sqrt{3}}\,a\,\hat{\mathbf{y}} + \frac{3}{4}\,c\,\hat{\mathbf{z}}$ | (2c) | O I |
| $\mathbf{B_3}$ | = | $\frac{1}{3}\,\mathbf{a_1} + \frac{2}{3}\,\mathbf{a_2} + z_2\,\mathbf{a_3}$ | = | $\frac{1}{2}\,a\,\mathbf{\hat{x}} + \frac{1}{2\sqrt{3}}\,a\,\mathbf{\hat{y}} + z_2\,c\,\mathbf{\hat{z}}$ | $(4f)$ | Si |
| $\mathbf{B_4}$ | = | $\frac{2}{3}\,\mathbf{a_1} + \frac{1}{3}\,\mathbf{a_2} + \left(\frac{1}{2} + z_2\right)\,\mathbf{a_3}$ | = | $\frac{1}{2}\,a\,\mathbf{\hat{x}} - \frac{1}{2\sqrt{3}}\,a\,\mathbf{\hat{y}} + \left(\frac{1}{2} + z_2\right)\,c\,\mathbf{\hat{z}}$ | $(4f)$ | Si |
| $\mathbf{B_5}$ | = | $\frac{2}{3}\,\mathbf{a_1} + \frac{1}{3}\,\mathbf{a_2} - z_2\,\mathbf{a_3}$ | = | $\frac{1}{2}\,a\,\mathbf{\hat{x}} - \frac{1}{2\sqrt{3}}\,a\,\mathbf{\hat{y}} - z_2\,c\,\mathbf{\hat{z}}$ | $(4f)$ | Si |
| $\mathbf{B_6}$ | = | $\frac{1}{3}\,\mathbf{a_1} + \frac{2}{3}\,\mathbf{a_2} + \left(\frac{1}{2} - z_2\right)\,\mathbf{a_3}$ | = | $\frac{1}{2}\,a\,\mathbf{\hat{x}} + \frac{1}{2\sqrt{3}}\,a\,\mathbf{\hat{y}} + \left(\frac{1}{2} - z_2\right)\,c\,\mathbf{\hat{z}}$ | $(4f)$ | Si |
| $\mathbf{B_7}$ | = | $\frac{1}{2}\,\mathbf{a_1}$ | = | $\frac{1}{4}\,a\,\mathbf{\hat{x}} - \frac{\sqrt{3}}{4}\,a\,\mathbf{\hat{y}}$ | $(6g)$ | O II |
| $\mathbf{B_8}$ | = | $\frac{1}{2}\,\mathbf{a_2}$ | = | $\frac{1}{4}\,a\,\mathbf{\hat{x}} + \frac{\sqrt{3}}{4}\,a\,\mathbf{\hat{y}}$ | $(6g)$ | O II |
| $\mathbf{B_9}$ | = | $\frac{1}{2}\,\mathbf{a_1} + \frac{1}{2}\,\mathbf{a_2}$ | = | $\frac{1}{2}\,a\,\mathbf{\hat{x}}$ | $(6g)$ | O II |
| $\mathbf{B_{10}}$ | = | $\frac{1}{2}\,\mathbf{a_1} + \frac{1}{2}\,\mathbf{a_3}$ | = | $\frac{1}{4}\,a\,\mathbf{\hat{x}} - \frac{\sqrt{3}}{4}\,a\,\mathbf{\hat{y}} + \frac{1}{2}\,c\,\mathbf{\hat{z}}$ | $(6g)$ | O II |
| $\mathbf{B_{11}}$ | = | $\frac{1}{2}\,\mathbf{a_2} + \frac{1}{2}\,\mathbf{a_3}$ | = | $\frac{1}{4}\,a\,\mathbf{\hat{x}} + \frac{\sqrt{3}}{4}\,a\,\mathbf{\hat{y}} + \frac{1}{2}\,c\,\mathbf{\hat{z}}$ | $(6g)$ | O II |
| $\mathbf{B_{12}}$ | = | $\frac{1}{2}\,\mathbf{a_1} + \frac{1}{2}\,\mathbf{a_2} + \frac{1}{2}\,\mathbf{a_3}$ | = | $\frac{1}{2}\,a\,\mathbf{\hat{x}} + \frac{1}{2}\,c\,\mathbf{\hat{z}}$ | $(6g)$ | O II |

---

**Found in:**

---

**Geometry files:**

- CIF: pp. 753
- POSCAR: pp. 754

# $Ga_4Ni$ Structure: A4B_cI40_197_cde_c

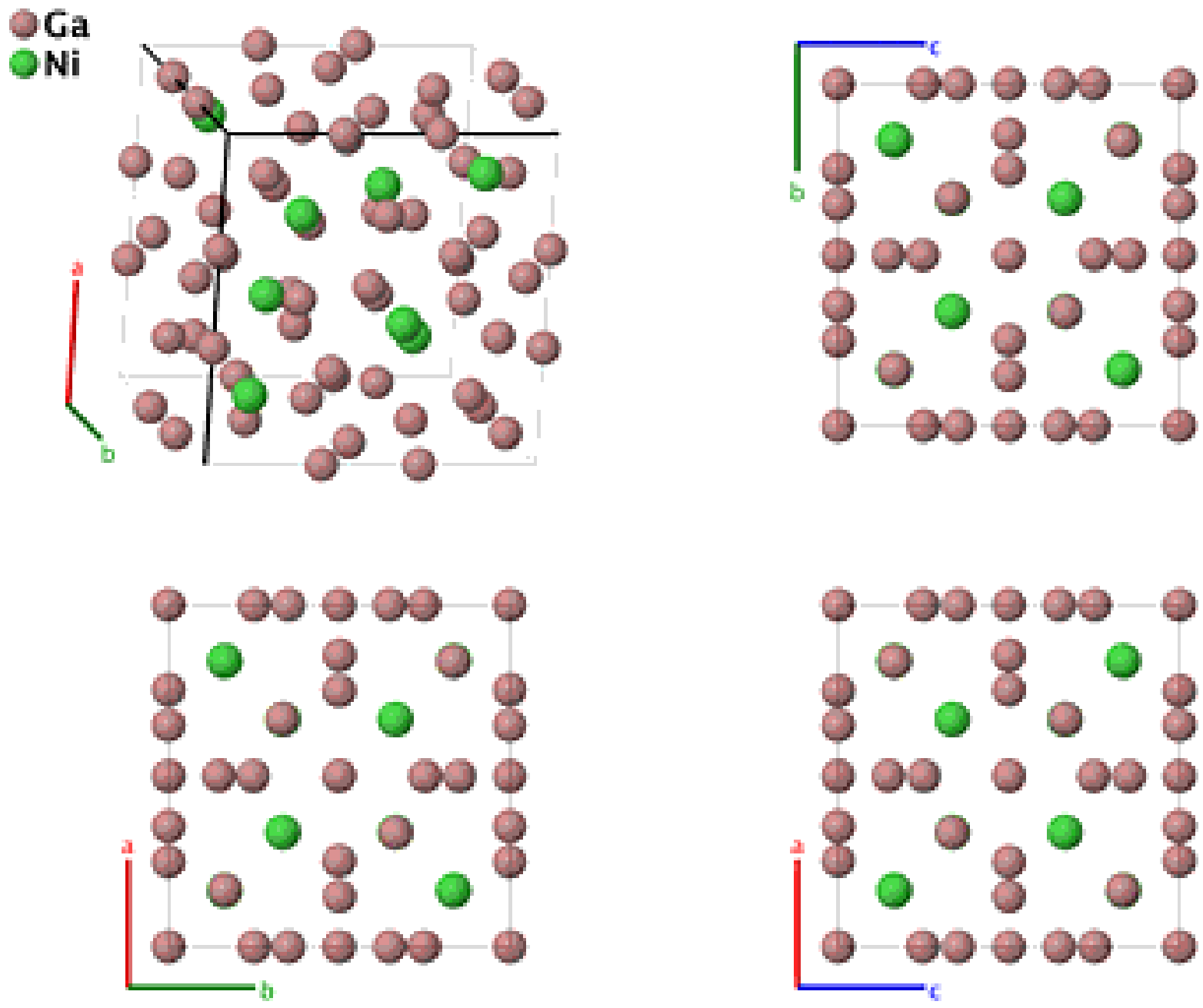

| | | |
|---|---|---|
| **Prototype** | : | $Ga_4Ni$ |
| **AFLOW prototype label** | : | A4B_cI40_197_cde_c |
| ***Strukturbericht* designation** | : | None |
| **Pearson symbol** | : | cI40 |
| **Space group number** | : | 197 |
| **Space group symbol** | : | I23 |
| **AFLOW prototype command** | : | `aflow --proto=A4B_cI40_197_cde_c` <br> `--params=`$a, x_1, x_2, x_3, x_4$ |

**Body-centered Cubic primitive vectors:**

$$\mathbf{a}_1 = -\frac{1}{2}\,a\,\hat{\mathbf{x}} + \frac{1}{2}\,a\,\hat{\mathbf{y}} + \frac{1}{2}\,a\,\hat{\mathbf{z}}$$
$$\mathbf{a}_2 = \frac{1}{2}\,a\,\hat{\mathbf{x}} - \frac{1}{2}\,a\,\hat{\mathbf{y}} + \frac{1}{2}\,a\,\hat{\mathbf{z}}$$
$$\mathbf{a}_3 = \frac{1}{2}\,a\,\hat{\mathbf{x}} + \frac{1}{2}\,a\,\hat{\mathbf{y}} - \frac{1}{2}\,a\,\hat{\mathbf{z}}$$

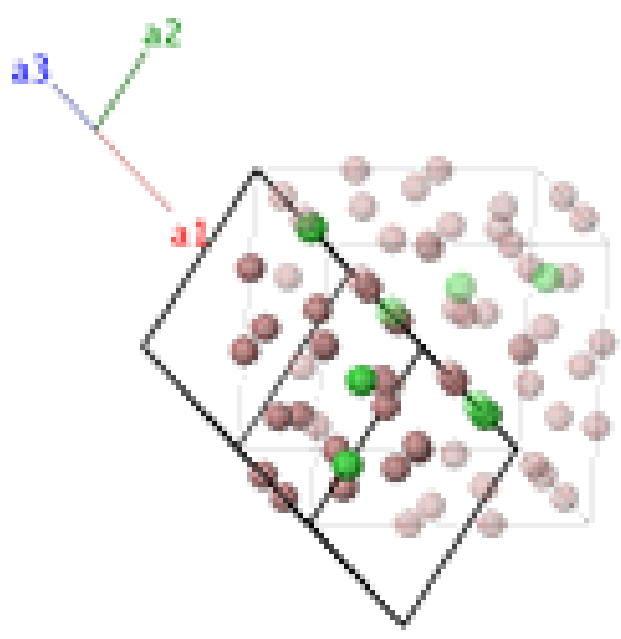

**Basis vectors:**

| | | Lattice Coordinates | | Cartesian Coordinates | Wyckoff Position | Atom Type |
|---|---|---|---|---|---|---|
| $\mathbf{B}_1$ | = | $2x_1\,\mathbf{a}_1 + 2x_1\,\mathbf{a}_2 + 2x_1\,\mathbf{a}_3$ | = | $x_1\,a\,\hat{\mathbf{x}} + x_1\,a\,\hat{\mathbf{y}} + x_1\,a\,\hat{\mathbf{z}}$ | (8$c$) | Ga I |
| $\mathbf{B}_2$ | = | $-2x_1\,\mathbf{a}_3$ | = | $-x_1\,a\,\hat{\mathbf{x}} - x_1\,a\,\hat{\mathbf{y}} + x_1\,a\,\hat{\mathbf{z}}$ | (8$c$) | Ga I |
| $\mathbf{B}_3$ | = | $-2x_1\,\mathbf{a}_2$ | = | $-x_1\,a\,\hat{\mathbf{x}} + x_1\,a\,\hat{\mathbf{y}} - x_1\,a\,\hat{\mathbf{z}}$ | (8$c$) | Ga I |
| $\mathbf{B}_4$ | = | $-2x_1\,\mathbf{a}_1$ | = | $x_1\,a\,\hat{\mathbf{x}} - x_1\,a\,\hat{\mathbf{y}} - x_1\,a\,\hat{\mathbf{z}}$ | (8$c$) | Ga I |
| $\mathbf{B}_5$ | = | $2x_2\,\mathbf{a}_1 + 2x_2\,\mathbf{a}_2 + 2x_2\,\mathbf{a}_3$ | = | $x_2\,a\,\mathbf{\hat{x}} + x_2\,a\,\mathbf{\hat{y}} + x_2\,a\,\mathbf{\hat{z}}$ | (8$c$) | Ni |
| $\mathbf{B}_6$ | = | $-2x_2\,\mathbf{a}_3$ | = | $-x_2\,a\,\mathbf{\hat{x}} - x_2\,a\,\mathbf{\hat{y}} + x_2\,a\,\mathbf{\hat{z}}$ | (8$c$) | Ni |
| $\mathbf{B}_7$ | = | $-2x_2\,\mathbf{a}_2$ | = | $-x_2\,a\,\mathbf{\hat{x}} + x_2\,a\,\mathbf{\hat{y}} - x_2\,a\,\mathbf{\hat{z}}$ | (8$c$) | Ni |
| $\mathbf{B}_8$ | = | $-2x_2\,\mathbf{a}_1$ | = | $x_2\,a\,\mathbf{\hat{x}} - x_2\,a\,\mathbf{\hat{y}} - x_2\,a\,\mathbf{\hat{z}}$ | (8$c$) | Ni |
| $\mathbf{B}_9$ | = | $x_3\,\mathbf{a}_2 + x_3\,\mathbf{a}_3$ | = | $x_3\,a\,\mathbf{\hat{x}}$ | (12$d$) | Ga II |
| $\mathbf{B}_{10}$ | = | $x_3\,\mathbf{a}_1 + x_3\,\mathbf{a}_3$ | = | $x_3\,a\,\mathbf{\hat{y}}$ | (12$d$) | Ga II |
| $\mathbf{B}_{11}$ | = | $x_3\,\mathbf{a}_1 + x_3\,\mathbf{a}_2$ | = | $x_3\,a\,\mathbf{\hat{z}}$ | (12$d$) | Ga II |
| $\mathbf{B}_{12}$ | = | $-x_3\,\mathbf{a}_2 - x_3\,\mathbf{a}_3$ | = | $-x_3\,a\,\mathbf{\hat{x}}$ | (12$d$) | Ga II |
| $\mathbf{B}_{13}$ | = | $-x_3\,\mathbf{a}_1 - x_3\,\mathbf{a}_3$ | = | $-x_3\,a\,\mathbf{\hat{y}}$ | (12$d$) | Ga II |
| $\mathbf{B}_{14}$ | = | $-x_3\,\mathbf{a}_1 - x_3\,\mathbf{a}_2$ | = | $-x_3\,a\,\mathbf{\hat{z}}$ | (12$d$) | Ga II |
| $\mathbf{B}_{15}$ | = | $\frac{1}{2}\,\mathbf{a}_1 + x_4\,\mathbf{a}_2 + \left(\frac{1}{2} + x_4\right)\,\mathbf{a}_3$ | = | $x_4\,a\,\mathbf{\hat{x}} + \frac{1}{2}\,a\,\mathbf{\hat{y}}$ | (12$e$) | Ga III |
| $\mathbf{B}_{16}$ | = | $\left(\frac{1}{2} + x_4\right)\,\mathbf{a}_1 + \frac{1}{2}\,\mathbf{a}_2 + x_4\,\mathbf{a}_3$ | = | $x_4\,a\,\mathbf{\hat{y}} + \frac{1}{2}\,a\,\mathbf{\hat{z}}$ | (12$e$) | Ga III |
| $\mathbf{B}_{17}$ | = | $x_4\,\mathbf{a}_1 + \left(\frac{1}{2} + x_4\right)\,\mathbf{a}_2 + \frac{1}{2}\,\mathbf{a}_3$ | = | $\frac{1}{2}\,a\,\mathbf{\hat{x}} + x_4\,a\,\mathbf{\hat{z}}$ | (12$e$) | Ga III |
| $\mathbf{B}_{18}$ | = | $\frac{1}{2}\,\mathbf{a}_1 - x_4\,\mathbf{a}_2 + \left(\frac{1}{2} - x_4\right)\,\mathbf{a}_3$ | = | $-x_4\,a\,\mathbf{\hat{x}} + \frac{1}{2}\,a\,\mathbf{\hat{y}}$ | (12$e$) | Ga III |
| $\mathbf{B}_{19}$ | = | $\left(\frac{1}{2} - x_4\right)\,\mathbf{a}_1 + \frac{1}{2}\,\mathbf{a}_2 - x_4\,\mathbf{a}_3$ | = | $-x_4\,a\,\mathbf{\hat{y}} + \frac{1}{2}\,a\,\mathbf{\hat{z}}$ | (12$e$) | Ga III |
| $\mathbf{B}_{20}$ | = | $-x_4\,\mathbf{a}_1 + \left(\frac{1}{2} - x_4\right)\,\mathbf{a}_2 + \frac{1}{2}\,\mathbf{a}_3$ | = | $\frac{1}{2}\,a\,\mathbf{\hat{x}} - x_4\,a\,\mathbf{\hat{z}}$ | (12$e$) | Ga III |

**Found in:**

**Geometry files:**

- CIF: pp. 754
- POSCAR: pp. 754

# Ullmanite (NiSSb, $F0_1$) Structure: ABC_cP12_198_a_a_a

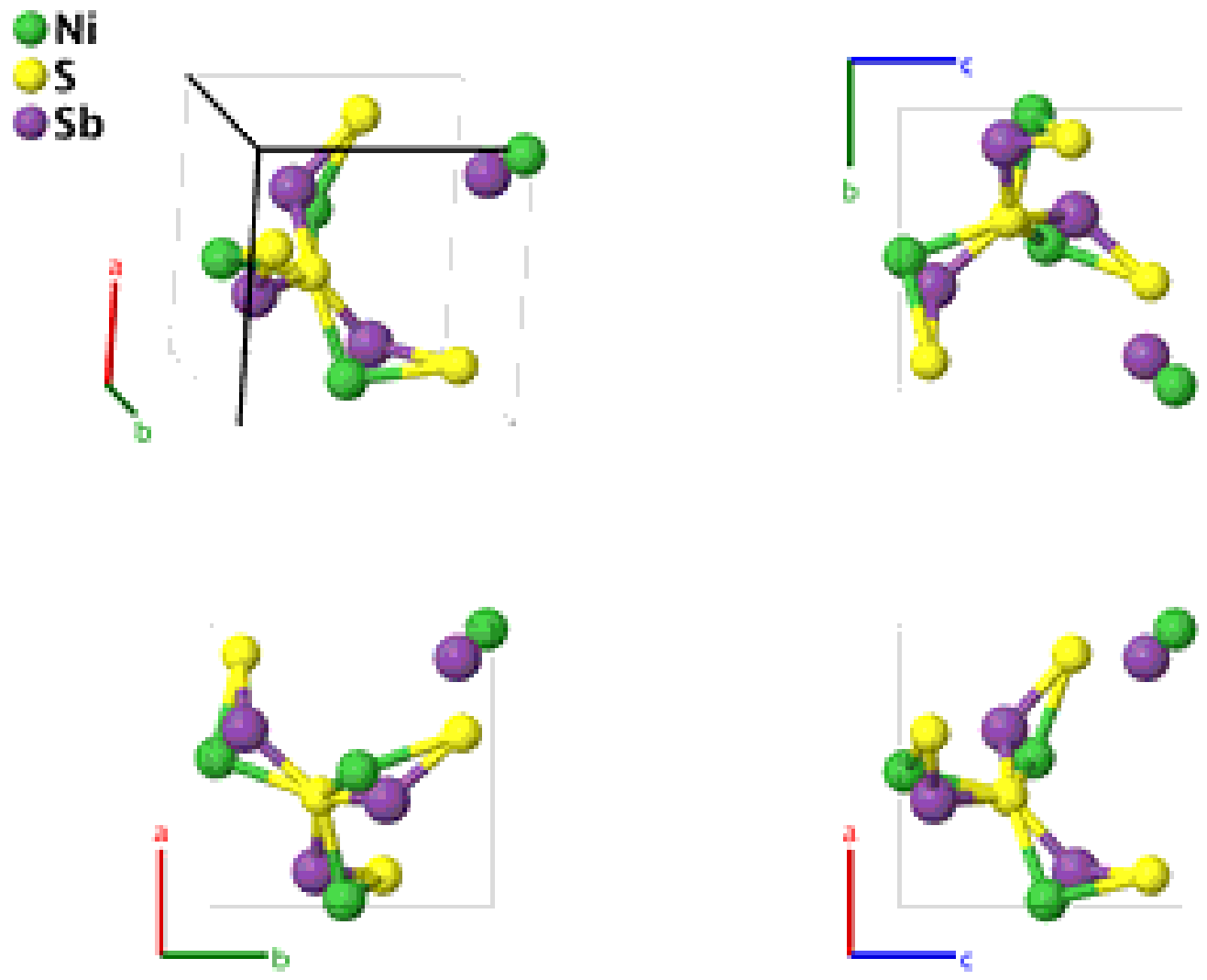

| | | |
|---|---|---|
| **Prototype** | : | NiSSb |
| **AFLOW prototype label** | : | ABC_cP12_198_a_a_a |
| ***Strukturbericht* designation** | : | $F0_1$ |
| **Pearson symbol** | : | cP12 |
| **Space group number** | : | 198 |
| **Space group symbol** | : | $P2_13$ |
| **AFLOW prototype command** | : | `aflow --proto=ABC_cP12_198_a_a_a --params=`$a, x_1, x_2, x_3$ |

**Other compounds with this structure:**

- AsBaPt, AsPdS, BiIrS, BiRhSe, CaPtSi, CrPtSb, EuPtSi, IrLaSi, IrSbSe, many others

**Simple Cubic primitive vectors:**

$$
\begin{aligned}
\mathbf{a}_1 &= a\,\hat{\mathbf{x}} \\
\mathbf{a}_2 &= a\,\hat{\mathbf{y}} \\
\mathbf{a}_3 &= a\,\hat{\mathbf{z}}
\end{aligned}
$$

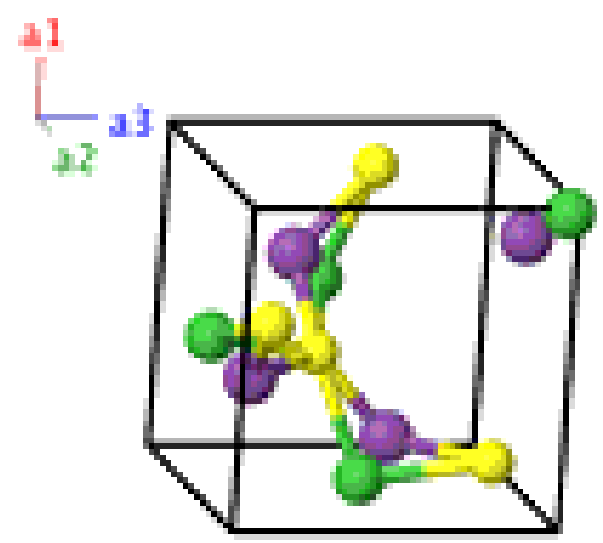

**Basis vectors:**

| | | Lattice Coordinates | | Cartesian Coordinates | Wyckoff Position | Atom Type |
|---|---|---|---|---|---|---|
| $\mathbf{B}_1$ | = | $x_1\,\mathbf{a}_1 + x_1\,\mathbf{a}_2 + x_1\,\mathbf{a}_3$ | = | $x_1\,a\,\hat{\mathbf{x}} + x_1\,a\,\hat{\mathbf{y}} + x_1\,a\,\hat{\mathbf{z}}$ | (4$a$) | Ni |
| $\mathbf{B}_2$ | = | $\left(\frac{1}{2} - x_1\right)\mathbf{a}_1 - x_1\,\mathbf{a}_2 + \left(\frac{1}{2} + x_1\right)\mathbf{a}_3$ | = | $\left(\frac{1}{2} - x_1\right)a\,\hat{\mathbf{x}} - x_1\,a\,\hat{\mathbf{y}} + \left(\frac{1}{2} + x_1\right)a\,\hat{\mathbf{z}}$ | (4$a$) | Ni |
| $\mathbf{B}_3$ | = | $-x_1\,\mathbf{a}_1 + \left(\frac{1}{2} + x_1\right)\mathbf{a}_2 + \left(\frac{1}{2} - x_1\right)\mathbf{a}_3$ | = | $-x_1\,a\,\hat{\mathbf{x}} + \left(\frac{1}{2} + x_1\right)a\,\hat{\mathbf{y}} + \left(\frac{1}{2} - x_1\right)a\,\hat{\mathbf{z}}$ | (4$a$) | Ni |
| $\mathbf{B_4}$ | = | $+\left(\frac{1}{2}+x_1\right)\mathbf{a_1}+\left(\frac{1}{2}-x_1\right)\mathbf{a_2}-x_1\,\mathbf{a_3}$ | = | $+\left(\frac{1}{2}+x_1\right)a\,\mathbf{\hat{x}}+\left(\frac{1}{2}-x_1\right)a\,\mathbf{\hat{y}}-x_1\,a\,\mathbf{\hat{z}}$ | (4$a$) | Ni |
| $\mathbf{B_5}$ | = | $x_2\,\mathbf{a_1}+x_2\,\mathbf{a_2}+x_2\,\mathbf{a_3}$ | = | $x_2\,a\,\mathbf{\hat{x}}+x_2\,a\,\mathbf{\hat{y}}+x_2\,a\,\mathbf{\hat{z}}$ | (4$a$) | S |
| $\mathbf{B_6}$ | = | $\left(\frac{1}{2}-x_2\right)\mathbf{a_1}-x_2\,\mathbf{a_2}+\left(\frac{1}{2}+x_2\right)\mathbf{a_3}$ | = | $\left(\frac{1}{2}-x_2\right)a\,\mathbf{\hat{x}}-x_2\,a\,\mathbf{\hat{y}}+\left(\frac{1}{2}+x_2\right)a\,\mathbf{\hat{z}}$ | (4$a$) | S |
| $\mathbf{B_7}$ | = | $-x_2\,\mathbf{a_1}+\left(\frac{1}{2}+x_2\right)\mathbf{a_2}+\left(\frac{1}{2}-x_2\right)\mathbf{a_3}$ | = | $-x_2\,a\,\mathbf{\hat{x}}+\left(\frac{1}{2}+x_2\right)a\,\mathbf{\hat{y}}+\left(\frac{1}{2}-x_2\right)a\,\mathbf{\hat{z}}$ | (4$a$) | S |
| $\mathbf{B_8}$ | = | $+\left(\frac{1}{2}+x_2\right)\mathbf{a_1}+\left(\frac{1}{2}-x_2\right)\mathbf{a_2}-x_2\,\mathbf{a_3}$ | = | $+\left(\frac{1}{2}+x_2\right)a\,\mathbf{\hat{x}}+\left(\frac{1}{2}-x_2\right)a\,\mathbf{\hat{y}}-x_2\,a\,\mathbf{\hat{z}}$ | (4$a$) | S |
| $\mathbf{B_9}$ | = | $x_3\,\mathbf{a_1}+x_3\,\mathbf{a_2}+x_3\,\mathbf{a_3}$ | = | $x_3\,a\,\mathbf{\hat{x}}+x_3\,a\,\mathbf{\hat{y}}+x_3\,a\,\mathbf{\hat{z}}$ | (4$a$) | Sb |
| $\mathbf{B_{10}}$ | = | $\left(\frac{1}{2}-x_3\right)\mathbf{a_1}-x_3\,\mathbf{a_2}+\left(\frac{1}{2}+x_3\right)\mathbf{a_3}$ | = | $\left(\frac{1}{2}-x_3\right)a\,\mathbf{\hat{x}}-x_3\,a\,\mathbf{\hat{y}}+\left(\frac{1}{2}+x_3\right)a\,\mathbf{\hat{z}}$ | (4$a$) | Sb |
| $\mathbf{B_{11}}$ | = | $-x_3\,\mathbf{a_1}+\left(\frac{1}{2}+x_3\right)\mathbf{a_2}+\left(\frac{1}{2}-x_3\right)\mathbf{a_3}$ | = | $-x_3\,a\,\mathbf{\hat{x}}+\left(\frac{1}{2}+x_3\right)a\,\mathbf{\hat{y}}+\left(\frac{1}{2}-x_3\right)a\,\mathbf{\hat{z}}$ | (4$a$) | Sb |
| $\mathbf{B_{12}}$ | = | $+\left(\frac{1}{2}+x_3\right)\mathbf{a_1}+\left(\frac{1}{2}-x_3\right)\mathbf{a_2}-x_3\,\mathbf{a_3}$ | = | $+\left(\frac{1}{2}+x_3\right)a\,\mathbf{\hat{x}}+\left(\frac{1}{2}-x_3\right)a\,\mathbf{\hat{y}}-x_3\,a\,\mathbf{\hat{z}}$ | (4$a$) | Sb |

**Geometry files:**

- CIF: pp. 754
- POSCAR: pp. 755

# Ammonia ($NH_3$, D1) Structure: A3B_cP16_198_b_a

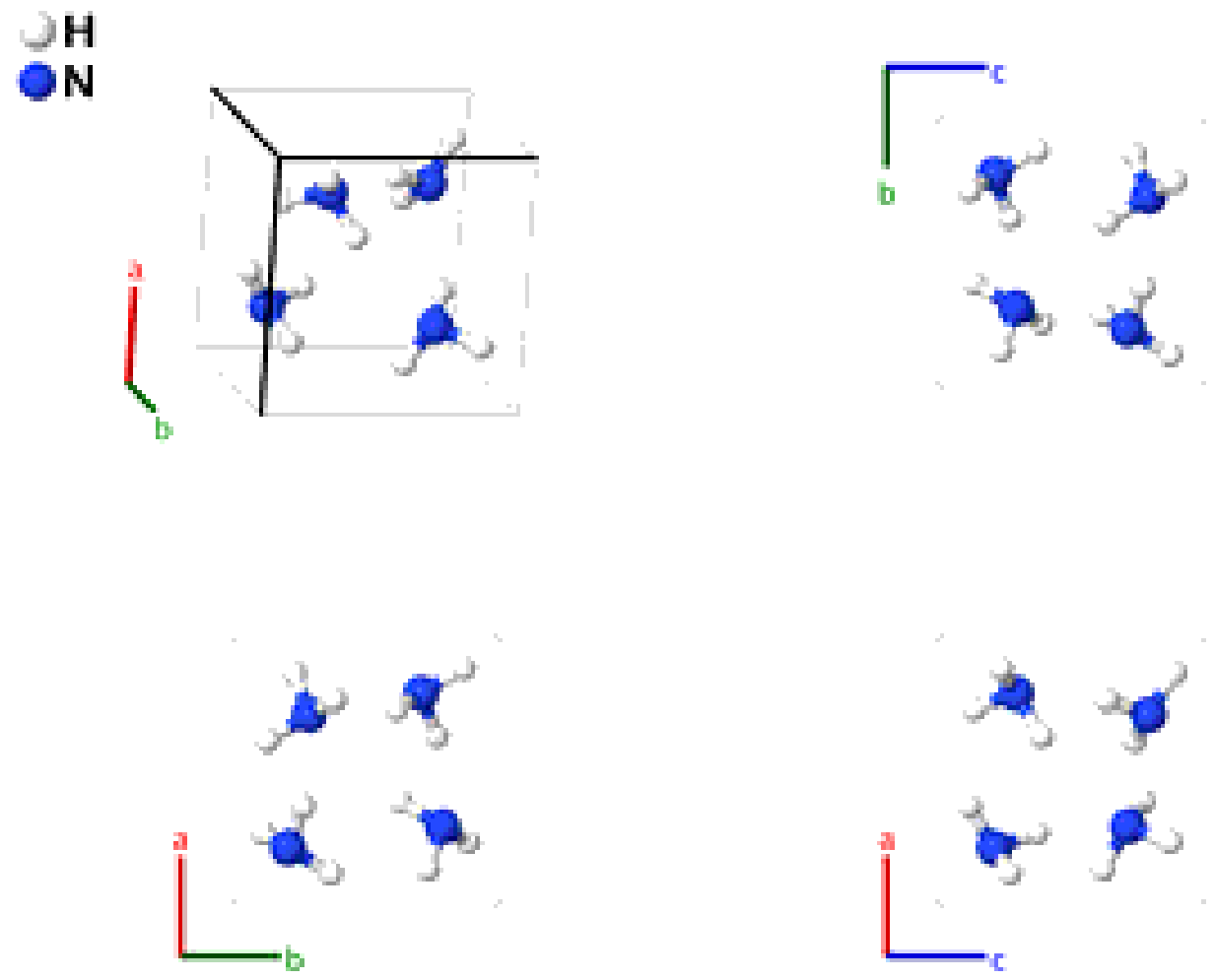

| | | |
|---|---|---|
| **Prototype** | : | $NH_3$ |
| **AFLOW prototype label** | : | A3B_cP16_198_b_a |
| ***Strukturbericht* designation** | : | D1 |
| **Pearson symbol** | : | cP16 |
| **Space group number** | : | 198 |
| **Space group symbol** | : | $P2_13$ |
| **AFLOW prototype command** | : | `aflow --proto=A3B_cP16_198_b_a`<br>`--params=`$a, x_1, x_2, y_2, z_2$ |

**Other compounds with this structure:**

- $AsH_3$, $PH_3$

- The positions of the hydrogen atoms are taken from neutron diffraction data on fully deuterated $ND_3$.

**Simple Cubic primitive vectors:**

$$\mathbf{a}_1 = a\,\hat{\mathbf{x}}$$
$$\mathbf{a}_2 = a\,\hat{\mathbf{y}}$$
$$\mathbf{a}_3 = a\,\hat{\mathbf{z}}$$

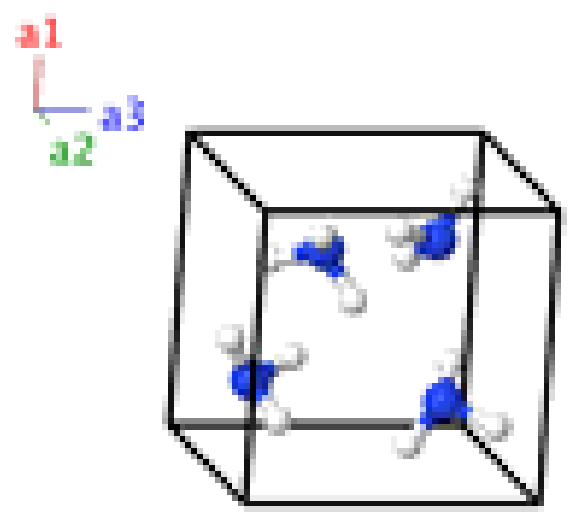

**Basis vectors:**

| | | Lattice Coordinates | | Cartesian Coordinates | Wyckoff Position | Atom Type |
|---|---|---|---|---|---|---|
| $\mathbf{B}_1$ | = | $x_1\,\mathbf{a}_1 + x_1\,\mathbf{a}_2 + x_1\,\mathbf{a}_3$ | = | $x_1\,a\,\hat{\mathbf{x}} + x_1\,a\,\hat{\mathbf{y}} + x_1\,a\,\hat{\mathbf{z}}$ | (4$a$) | N |
| $\mathbf{B}_2$ | = | $\left(\frac{1}{2} - x_1\right)\mathbf{a}_1 - x_1\,\mathbf{a}_2 + \left(\frac{1}{2} + x_1\right)\mathbf{a}_3$ | = | $\left(\frac{1}{2} - x_1\right)a\,\hat{\mathbf{x}} - x_1\,a\,\hat{\mathbf{y}} + \left(\frac{1}{2} + x_1\right)a\,\hat{\mathbf{z}}$ | (4$a$) | N |
| $\mathbf{B}_3$ | = | $-x_1\,\mathbf{a}_1 + \left(\frac{1}{2}+x_1\right)\mathbf{a}_2 + \left(\frac{1}{2}-x_1\right)\mathbf{a}_3$ | = | $-x_1\,a\,\hat{\mathbf{x}} + \left(\frac{1}{2}+x_1\right)a\,\hat{\mathbf{y}} + \left(\frac{1}{2}-x_1\right)a\,\hat{\mathbf{z}}$ | (4$a$) | N |
| $\mathbf{B}_4$ | = | $+\left(\frac{1}{2}+x_1\right)\mathbf{a}_1 + \left(\frac{1}{2}-x_1\right)\mathbf{a}_2 - x_1\,\mathbf{a}_3$ | = | $+\left(\frac{1}{2}+x_1\right)a\,\hat{\mathbf{x}} + \left(\frac{1}{2}-x_1\right)a\,\hat{\mathbf{y}} - x_1\,a\,\hat{\mathbf{z}}$ | (4$a$) | N |
| $\mathbf{B}_5$ | = | $x_2\,\mathbf{a}_1 + y_2\,\mathbf{a}_2 + z_2\,\mathbf{a}_3$ | = | $x_2\,a\,\hat{\mathbf{x}} + y_2\,a\,\hat{\mathbf{y}} + z_2\,a\,\hat{\mathbf{z}}$ | (12$b$) | H |
| $\mathbf{B}_6$ | = | $\left(\frac{1}{2}-x_2\right)\mathbf{a}_1 - y_2\,\mathbf{a}_2 + \left(\frac{1}{2}+z_2\right)\mathbf{a}_3$ | = | $\left(\frac{1}{2}-x_2\right)a\,\hat{\mathbf{x}} - y_2\,a\,\hat{\mathbf{y}} + \left(\frac{1}{2}+z_2\right)a\,\hat{\mathbf{z}}$ | (12$b$) | H |
| $\mathbf{B}_7$ | = | $-x_2\,\mathbf{a}_1 + \left(\frac{1}{2}+y_2\right)\mathbf{a}_2 + \left(\frac{1}{2}-z_2\right)\mathbf{a}_3$ | = | $-x_2\,a\,\hat{\mathbf{x}} + \left(\frac{1}{2}+y_2\right)a\,\hat{\mathbf{y}} + \left(\frac{1}{2}-z_2\right)a\,\hat{\mathbf{z}}$ | (12$b$) | H |
| $\mathbf{B}_8$ | = | $+\left(\frac{1}{2}+x_2\right)\mathbf{a}_1 + \left(\frac{1}{2}-y_2\right)\mathbf{a}_2 - z_2\,\mathbf{a}_3$ | = | $+\left(\frac{1}{2}+x_2\right)a\,\hat{\mathbf{x}} + \left(\frac{1}{2}-y_2\right)a\,\hat{\mathbf{y}} - z_2\,a\,\hat{\mathbf{z}}$ | (12$b$) | H |
| $\mathbf{B}_9$ | = | $z_2\,\mathbf{a}_1 + x_2\,\mathbf{a}_2 + y_2\,\mathbf{a}_3$ | = | $z_2\,a\,\hat{\mathbf{x}} + x_2\,a\,\hat{\mathbf{y}} + y_2\,a\,\hat{\mathbf{z}}$ | (12$b$) | H |
| $\mathbf{B}_{10}$ | = | $\left(\frac{1}{2}-z_2\right)\mathbf{a}_1 - x_2\,\mathbf{a}_2 + \left(\frac{1}{2}+y_2\right)\mathbf{a}_3$ | = | $\left(\frac{1}{2}-z_2\right)a\,\hat{\mathbf{x}} - x_2\,a\,\hat{\mathbf{y}} + \left(\frac{1}{2}+y_2\right)a\,\hat{\mathbf{z}}$ | (12$b$) | H |
| $\mathbf{B}_{11}$ | = | $-z_2\,\mathbf{a}_1 + \left(\frac{1}{2}+x_2\right)\mathbf{a}_2 + \left(\frac{1}{2}-y_2\right)\mathbf{a}_3$ | = | $-z_2\,a\,\hat{\mathbf{x}} + \left(\frac{1}{2}+x_2\right)a\,\hat{\mathbf{y}} + \left(\frac{1}{2}-y_2\right)a\,\hat{\mathbf{z}}$ | (12$b$) | H |
| $\mathbf{B}_{12}$ | = | $+\left(\frac{1}{2}+z_2\right)\mathbf{a}_1 + \left(\frac{1}{2}-x_2\right)\mathbf{a}_2 - y_2\,\mathbf{a}_3$ | = | $+\left(\frac{1}{2}+z_2\right)a\,\hat{\mathbf{x}} + \left(\frac{1}{2}-x_2\right)a\,\hat{\mathbf{y}} - y_2\,a\,\hat{\mathbf{z}}$ | (12$b$) | H |
| $\mathbf{B}_{13}$ | = | $y_2\,\mathbf{a}_1 + z_2\,\mathbf{a}_2 + x_2\,\mathbf{a}_3$ | = | $y_2\,a\,\hat{\mathbf{x}} + z_2\,a\,\hat{\mathbf{y}} + x_2\,a\,\hat{\mathbf{z}}$ | (12$b$) | H |
| $\mathbf{B}_{14}$ | = | $\left(\frac{1}{2}-y_2\right)\mathbf{a}_1 - z_2\,\mathbf{a}_2 + \left(\frac{1}{2}+x_2\right)\mathbf{a}_3$ | = | $\left(\frac{1}{2}-y_2\right)a\,\hat{\mathbf{x}} - z_2\,a\,\hat{\mathbf{y}} + \left(\frac{1}{2}+x_2\right)a\,\hat{\mathbf{z}}$ | (12$b$) | H |
| $\mathbf{B}_{15}$ | = | $-y_2\,\mathbf{a}_1 + \left(\frac{1}{2}+z_2\right)\mathbf{a}_2 + \left(\frac{1}{2}-x_2\right)\mathbf{a}_3$ | = | $-y_2\,a\,\hat{\mathbf{x}} + \left(\frac{1}{2}+z_2\right)a\,\hat{\mathbf{y}} + \left(\frac{1}{2}-x_2\right)a\,\hat{\mathbf{z}}$ | (12$b$) | H |
| $\mathbf{B}_{16}$ | = | $+\left(\frac{1}{2}+y_2\right)\mathbf{a}_1 + \left(\frac{1}{2}-z_2\right)\mathbf{a}_2 - x_2\,\mathbf{a}_3$ | = | $+\left(\frac{1}{2}+y_2\right)a\,\hat{\mathbf{x}} + \left(\frac{1}{2}-z_2\right)a\,\hat{\mathbf{y}} - x_2\,a\,\hat{\mathbf{z}}$ | (12$b$) | H |

**Geometry files:**

- CIF: pp. 755
- POSCAR: pp. 755

# $\alpha$-N ($P2_13$) Structure: A_cP8_198_2a

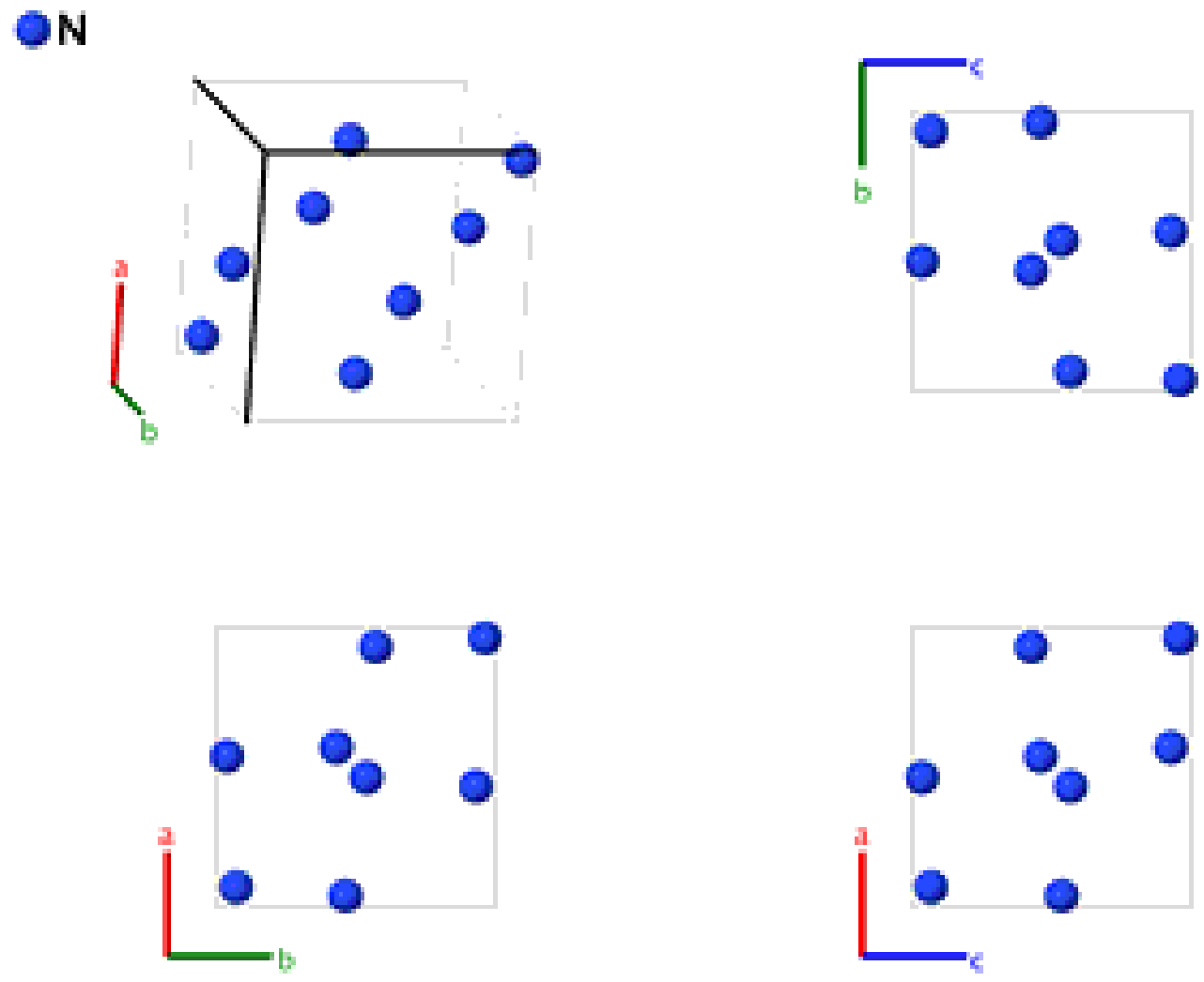

| | | |
|---|---|---|
| **Prototype** | : | $\alpha$-N |
| **AFLOW prototype label** | : | A_cP8_198_2a |
| ***Strukturbericht* designation** | : | None |
| **Pearson symbol** | : | cP8 |
| **Space group number** | : | 198 |
| **Space group symbol** | : | $P2_13$ |
| **AFLOW prototype command** | : | `aflow --proto=A_cP8_198_2a`<br>`--params=`$a, x_1, x_2$ |

- There is considerable controversy about the crystal structure of $\alpha$-N, as outlined in (Donohue, 1982) pp. 280-285. This page assumes the non-centrosymmetric $P2_13$ structure. The other possibility is the $Pa\bar{3}$ structure, where the $N_2$ dimers are not centered on an inversion site. (Venables, 1974) makes a convincing case that the ground state is $Pa\bar{3}$, but we present both structures. Density Functional Theory calculations show no appreciable difference in energy between the $Pa\bar{3}$ and $P2_13$ structures. (Mehl, 2015)

**Simple Cubic primitive vectors:**

$$\mathbf{a}_1 = a\,\hat{\mathbf{x}}$$
$$\mathbf{a}_2 = a\,\hat{\mathbf{y}}$$
$$\mathbf{a}_3 = a\,\hat{\mathbf{z}}$$

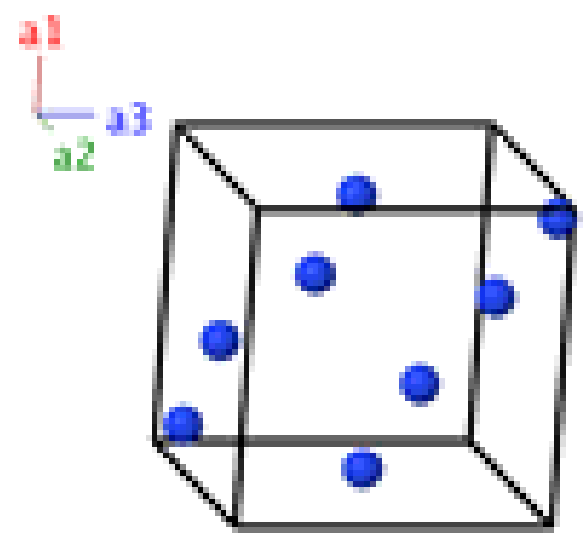

**Basis vectors:**

| | | Lattice Coordinates | | Cartesian Coordinates | Wyckoff Position | Atom Type |
|---|---|---|---|---|---|---|
| $\mathbf{B}_1$ | = | $x_1\,\mathbf{a}_1 + x_1\,\mathbf{a}_2 + x_1\,\mathbf{a}_3$ | = | $x_1\,a\,\hat{\mathbf{x}} + x_1\,a\,\hat{\mathbf{y}} + x_1\,a\,\hat{\mathbf{z}}$ | (4a) | N I |
| $\mathbf{B}_2$ | = | $\left(\frac{1}{2} - x_1\right)\mathbf{a}_1 - x_1\,\mathbf{a}_2 + \left(\frac{1}{2} + x_1\right)\mathbf{a}_3$ | = | $\left(\frac{1}{2} - x_1\right)a\,\hat{\mathbf{x}} - x_1\,a\,\hat{\mathbf{y}} + \left(\frac{1}{2} + x_1\right)a\,\hat{\mathbf{z}}$ | (4a) | N I |
| $\mathbf{B_3}$ | = | $-x_1\,\mathbf{a_1} + \left(\frac{1}{2}+x_1\right)\mathbf{a_2} + \left(\frac{1}{2}-x_1\right)\mathbf{a_3}$ | = | $-x_1\,a\,\mathbf{\hat{x}}+\left(\frac{1}{2}+x_1\right)a\,\mathbf{\hat{y}}+\left(\frac{1}{2}-x_1\right)a\,\mathbf{\hat{z}}$ | (4$a$) | N I |
| $\mathbf{B_4}$ | = | $+\left(\frac{1}{2}+x_1\right)\mathbf{a_1} + \left(\frac{1}{2}-x_1\right)\mathbf{a_2} - x_1\,\mathbf{a_3}$ | = | $+\left(\frac{1}{2}+x_1\right)a\,\mathbf{\hat{x}}+\left(\frac{1}{2}-x_1\right)a\,\mathbf{\hat{y}}-x_1\,a\,\mathbf{\hat{z}}$ | (4$a$) | N I |
| $\mathbf{B_5}$ | = | $x_2\,\mathbf{a_1} + x_2\,\mathbf{a_2} + x_2\,\mathbf{a_3}$ | = | $x_2\,a\,\mathbf{\hat{x}} + x_2\,a\,\mathbf{\hat{y}} + x_2\,a\,\mathbf{\hat{z}}$ | (4$a$) | N II |
| $\mathbf{B_6}$ | = | $\left(\frac{1}{2}-x_2\right)\mathbf{a_1} - x_2\,\mathbf{a_2} + \left(\frac{1}{2}+x_2\right)\mathbf{a_3}$ | = | $\left(\frac{1}{2}-x_2\right)a\,\mathbf{\hat{x}} - x_2\,a\,\mathbf{\hat{y}} + \left(\frac{1}{2}+x_2\right)a\,\mathbf{\hat{z}}$ | (4$a$) | N II |
| $\mathbf{B_7}$ | = | $-x_2\,\mathbf{a_1} + \left(\frac{1}{2}+x_2\right)\mathbf{a_2} + \left(\frac{1}{2}-x_2\right)\mathbf{a_3}$ | = | $-x_2\,a\,\mathbf{\hat{x}}+\left(\frac{1}{2}+x_2\right)a\,\mathbf{\hat{y}}+\left(\frac{1}{2}-x_2\right)a\,\mathbf{\hat{z}}$ | (4$a$) | N II |
| $\mathbf{B_8}$ | = | $+\left(\frac{1}{2}+x_2\right)\mathbf{a_1} + \left(\frac{1}{2}-x_2\right)\mathbf{a_2} - x_2\,\mathbf{a_3}$ | = | $+\left(\frac{1}{2}+x_2\right)a\,\mathbf{\hat{x}}+\left(\frac{1}{2}-x_2\right)a\,\mathbf{\hat{y}}-x_2\,a\,\mathbf{\hat{z}}$ | (4$a$) | N II |

---

**Found in:**

---

**Geometry files:**

- CIF: pp. 755
- POSCAR: pp. 756

# $\alpha$-CO (B21) Structure: AB_cP8_198_a_a

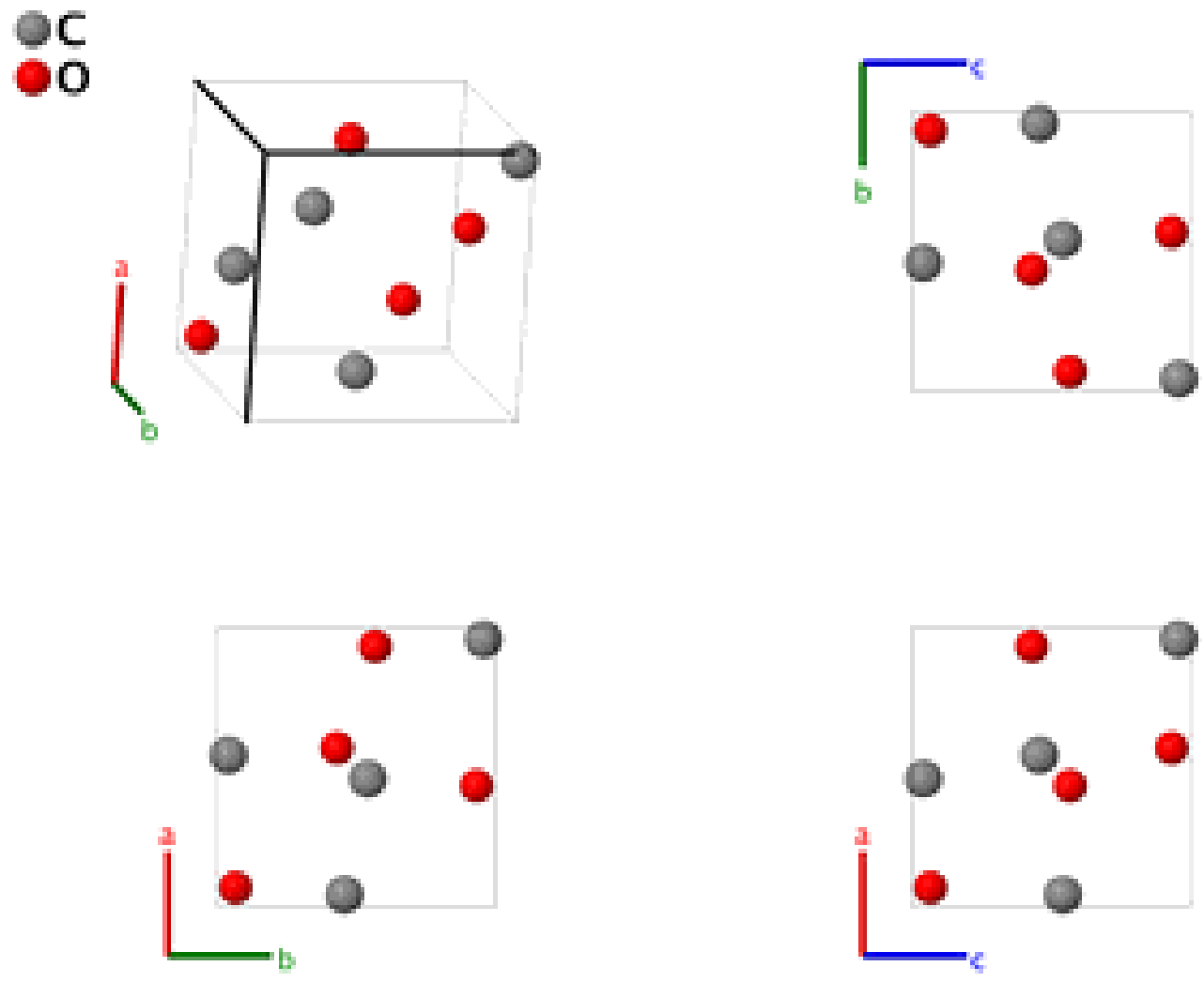

| | | |
|---|---|---|
| **Prototype** | : | $\alpha$-CO |
| **AFLOW prototype label** | : | AB_cP8_198_a_a |
| ***Strukturbericht* designation** | : | B21 |
| **Pearson symbol** | : | cP8 |
| **Space group number** | : | 198 |
| **Space group symbol** | : | $P2_13$ |
| **AFLOW prototype command** | : | `aflow --proto=AB_cP8_198_a_a --params=`$a, x_1, x_2$ |

- The molecules sit on the sites of a face-centered cubic lattice. Note that $\alpha$-CO (pp. 507) and FeSi (pp. 509) have the same AFLOW prototype label. They are generated by the same symmetry operations with different sets of parameters (`--params`) specified in their corresponding CIF files.

**Simple Cubic primitive vectors:**

$$\mathbf{a}_1 = a\,\hat{\mathbf{x}}$$
$$\mathbf{a}_2 = a\,\hat{\mathbf{y}}$$
$$\mathbf{a}_3 = a\,\hat{\mathbf{z}}$$

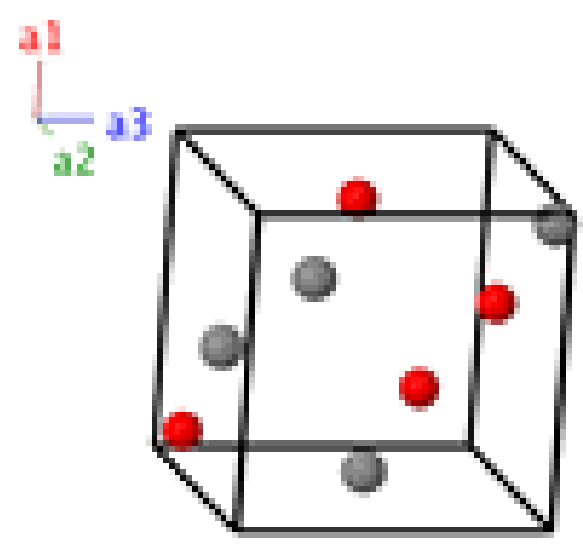

**Basis vectors:**

| | | Lattice Coordinates | | Cartesian Coordinates | Wyckoff Position | Atom Type |
|---|---|---|---|---|---|---|
| $\mathbf{B}_1$ | = | $x_1\,\mathbf{a}_1 + x_1\,\mathbf{a}_2 + x_1\,\mathbf{a}_3$ | = | $x_1\,a\,\hat{\mathbf{x}} + x_1\,a\,\hat{\mathbf{y}} + x_1\,a\,\hat{\mathbf{z}}$ | (4$a$) | C |
| $\mathbf{B}_2$ | = | $\left(\frac{1}{2} - x_1\right)\mathbf{a}_1 - x_1\,\mathbf{a}_2 + \left(\frac{1}{2} + x_1\right)\mathbf{a}_3$ | = | $\left(\frac{1}{2} - x_1\right)a\,\hat{\mathbf{x}} - x_1\,a\,\hat{\mathbf{y}} + \left(\frac{1}{2} + x_1\right)a\,\hat{\mathbf{z}}$ | (4$a$) | C |
| $\mathbf{B}_3$ | = | $-x_1\,\mathbf{a}_1 + \left(\frac{1}{2} + x_1\right)\mathbf{a}_2 + \left(\frac{1}{2} - x_1\right)\mathbf{a}_3$ | = | $-x_1\,a\,\hat{\mathbf{x}} + \left(\frac{1}{2} + x_1\right)a\,\hat{\mathbf{y}} + \left(\frac{1}{2} - x_1\right)a\,\hat{\mathbf{z}}$ | (4$a$) | C |
| $\mathbf{B_4}$ | = | $+\left(\frac{1}{2}+x_1\right)\mathbf{a_1}+\left(\frac{1}{2}-x_1\right)\mathbf{a_2}-x_1\,\mathbf{a_3}$ | = | $+\left(\frac{1}{2}+x_1\right)a\,\mathbf{\hat{x}}+\left(\frac{1}{2}-x_1\right)a\,\mathbf{\hat{y}}-x_1\,a\,\mathbf{\hat{z}}$ | (4$a$) | C |
| $\mathbf{B_5}$ | = | $x_2\,\mathbf{a_1}+x_2\,\mathbf{a_2}+x_2\,\mathbf{a_3}$ | = | $x_2\,a\,\mathbf{\hat{x}}+x_2\,a\,\mathbf{\hat{y}}+x_2\,a\,\mathbf{\hat{z}}$ | (4$a$) | O |
| $\mathbf{B_6}$ | = | $\left(\frac{1}{2}-x_2\right)\mathbf{a_1}-x_2\,\mathbf{a_2}+\left(\frac{1}{2}+x_2\right)\mathbf{a_3}$ | = | $\left(\frac{1}{2}-x_2\right)a\,\mathbf{\hat{x}}-x_2\,a\,\mathbf{\hat{y}}+\left(\frac{1}{2}+x_2\right)a\,\mathbf{\hat{z}}$ | (4$a$) | O |
| $\mathbf{B_7}$ | = | $-x_2\,\mathbf{a_1}+\left(\frac{1}{2}+x_2\right)\mathbf{a_2}+\left(\frac{1}{2}-x_2\right)\mathbf{a_3}$ | = | $-x_2\,a\,\mathbf{\hat{x}}+\left(\frac{1}{2}+x_2\right)a\,\mathbf{\hat{y}}+\left(\frac{1}{2}-x_2\right)a\,\mathbf{\hat{z}}$ | (4$a$) | O |
| $\mathbf{B_8}$ | = | $+\left(\frac{1}{2}+x_2\right)\mathbf{a_1}+\left(\frac{1}{2}-x_2\right)\mathbf{a_2}-x_2\,\mathbf{a_3}$ | = | $+\left(\frac{1}{2}+x_2\right)a\,\mathbf{\hat{x}}+\left(\frac{1}{2}-x_2\right)a\,\mathbf{\hat{y}}-x_2\,a\,\mathbf{\hat{z}}$ | (4$a$) | O |

---

**Found in:**

---

**Geometry files:**

- CIF: pp. 756
- POSCAR: pp. 756

# FeSi (B20) Structure: AB_cP8_198_a_a

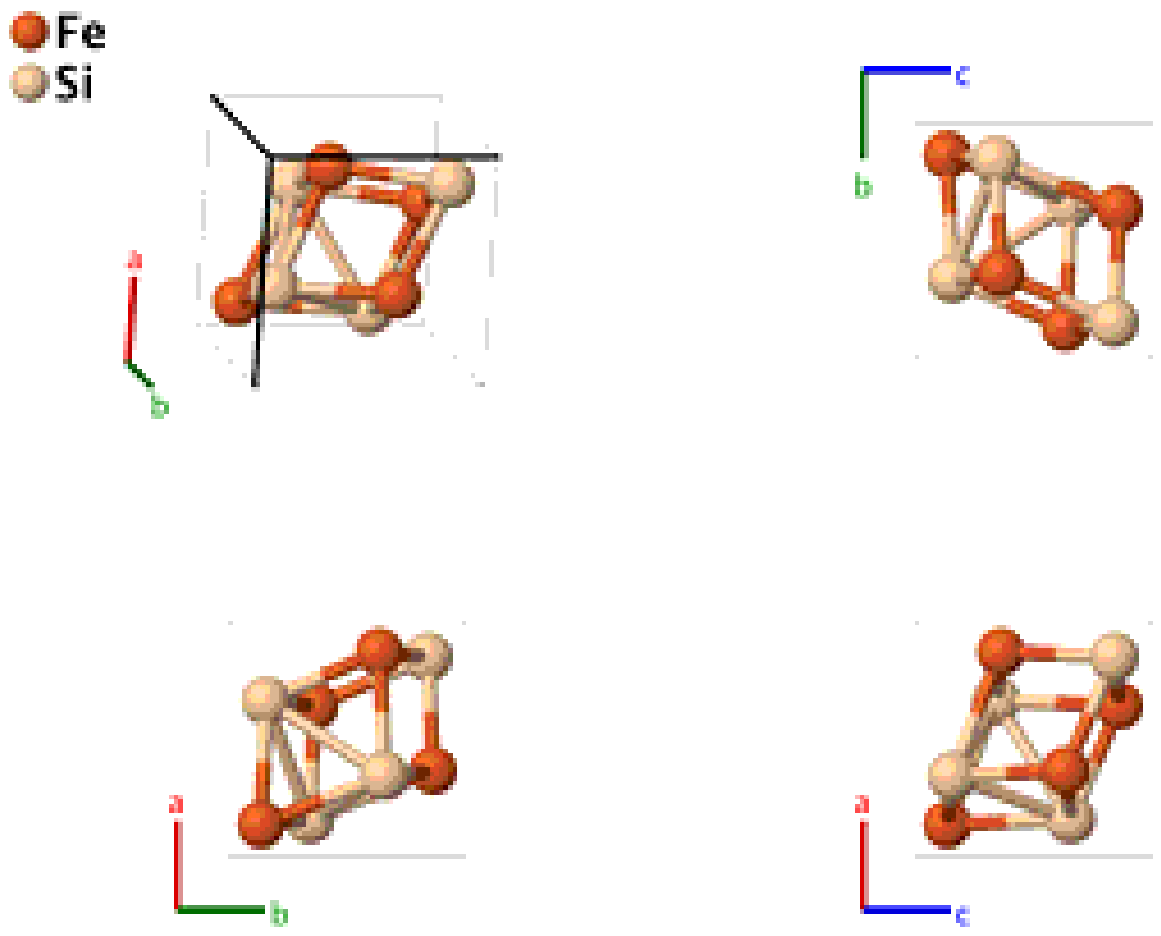

| | | |
|---|---|---|
| **Prototype** | : | FeSi |
| **AFLOW prototype label** | : | AB_cP8_198_a_a |
| ***Strukturbericht* designation** | : | B20 |
| **Pearson symbol** | : | cP8 |
| **Space group number** | : | 198 |
| **Space group symbol** | : | $P2_13$ |
| **AFLOW prototype command** | : | `aflow --proto=AB_cP8_198_a_a` <br> `--params=`$a, x_1, x_2$ |

**Other compounds with this structure:**

- AlPt, AuBe, CoGe, CoSi, FeGe, GaPd, GeMn, GeRh, HfSb, HfSn, RhS, SbZr, SiTc

---

- When $x_1 = 0$ and $x_2 = 1/2$, or $x_1 = 1/4$ and $x_2 = 3/4$, this lattice reduces to the rock salt (B1) structure. When $x_1 = -x_2 = 1/8\left(\sqrt{5}-1\right)$ we have an "ideal" structure where every atom is seven-fold coordinated. Note that $\alpha$-CO (pp. 507) and FeSi (pp. 509) have the same AFLOW prototype label. They are generated by the same symmetry operations with different sets of parameters (`--params`) specified in their corresponding CIF files.

**Simple Cubic primitive vectors:**

$$\mathbf{a}_1 = a\,\hat{\mathbf{x}}$$
$$\mathbf{a}_2 = a\,\hat{\mathbf{y}}$$
$$\mathbf{a}_3 = a\,\hat{\mathbf{z}}$$

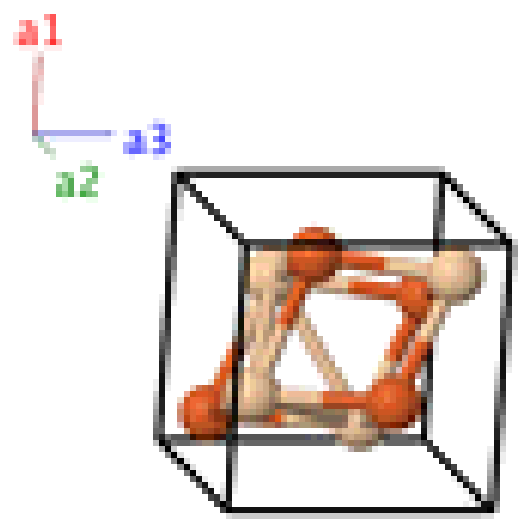

**Basis vectors:**

| | | Lattice Coordinates | | Cartesian Coordinates | Wyckoff Position | Atom Type |
|---|---|---|---|---|---|---|
| $\mathbf{B_1}$ | = | $x_1\,\mathbf{a_1} + x_1\,\mathbf{a_2} + x_1\,\mathbf{a_3}$ | = | $x_1\,a\,\hat{\mathbf{x}} + x_1\,a\,\hat{\mathbf{y}} + x_1\,a\,\hat{\mathbf{z}}$ | (4$a$) | Fe |
| $\mathbf{B_2}$ | = | $\left(\frac{1}{2} - x_1\right) \mathbf{a_1} - x_1 \mathbf{a_2} + \left(\frac{1}{2} + x_1\right) \mathbf{a_3}$ | = | $\left(\frac{1}{2} - x_1\right) a \hat{\mathbf{x}} - x_1 a \hat{\mathbf{y}} + \left(\frac{1}{2} + x_1\right) a \hat{\mathbf{z}}$ | (4$a$) | Fe |
| $\mathbf{B_3}$ | = | $-x_1 \mathbf{a_1} + \left(\frac{1}{2} + x_1\right) \mathbf{a_2} + \left(\frac{1}{2} - x_1\right) \mathbf{a_3}$ | = | $-x_1 a \hat{\mathbf{x}} + \left(\frac{1}{2} + x_1\right) a \hat{\mathbf{y}} + \left(\frac{1}{2} - x_1\right) a \hat{\mathbf{z}}$ | (4$a$) | Fe |
| $\mathbf{B_4}$ | = | $+\left(\frac{1}{2} + x_1\right) \mathbf{a_1} + \left(\frac{1}{2} - x_1\right) \mathbf{a_2} - x_1 \mathbf{a_3}$ | = | $+\left(\frac{1}{2} + x_1\right) a \hat{\mathbf{x}} + \left(\frac{1}{2} - x_1\right) a \hat{\mathbf{y}} - x_1 a \hat{\mathbf{z}}$ | (4$a$) | Fe |
| $\mathbf{B_5}$ | = | $x_2 \mathbf{a_1} + x_2 \mathbf{a_2} + x_2 \mathbf{a_3}$ | = | $x_2 a \hat{\mathbf{x}} + x_2 a \hat{\mathbf{y}} + x_2 a \hat{\mathbf{z}}$ | (4$a$) | Si |
| $\mathbf{B_6}$ | = | $\left(\frac{1}{2} - x_2\right) \mathbf{a_1} - x_2 \mathbf{a_2} + \left(\frac{1}{2} + x_2\right) \mathbf{a_3}$ | = | $\left(\frac{1}{2} - x_2\right) a \hat{\mathbf{x}} - x_2 a \hat{\mathbf{y}} + \left(\frac{1}{2} + x_2\right) a \hat{\mathbf{z}}$ | (4$a$) | Si |
| $\mathbf{B_7}$ | = | $-x_2 \mathbf{a_1} + \left(\frac{1}{2} + x_2\right) \mathbf{a_2} + \left(\frac{1}{2} - x_2\right) \mathbf{a_3}$ | = | $-x_2 a \hat{\mathbf{x}} + \left(\frac{1}{2} + x_2\right) a \hat{\mathbf{y}} + \left(\frac{1}{2} - x_2\right) a \hat{\mathbf{z}}$ | (4$a$) | Si |
| $\mathbf{B_8}$ | = | $+\left(\frac{1}{2} + x_2\right) \mathbf{a_1} + \left(\frac{1}{2} - x_2\right) \mathbf{a_2} - x_2 \mathbf{a_3}$ | = | $+\left(\frac{1}{2} + x_2\right) a \hat{\mathbf{x}} + \left(\frac{1}{2} - x_2\right) a \hat{\mathbf{y}} - x_2 a \hat{\mathbf{z}}$ | (4$a$) | Si |

---

---

**Geometry files:**

- CIF: pp. 756
- POSCAR: pp. 757

# CoU ($B_a$) Structure: AB_cI16_199_a_a

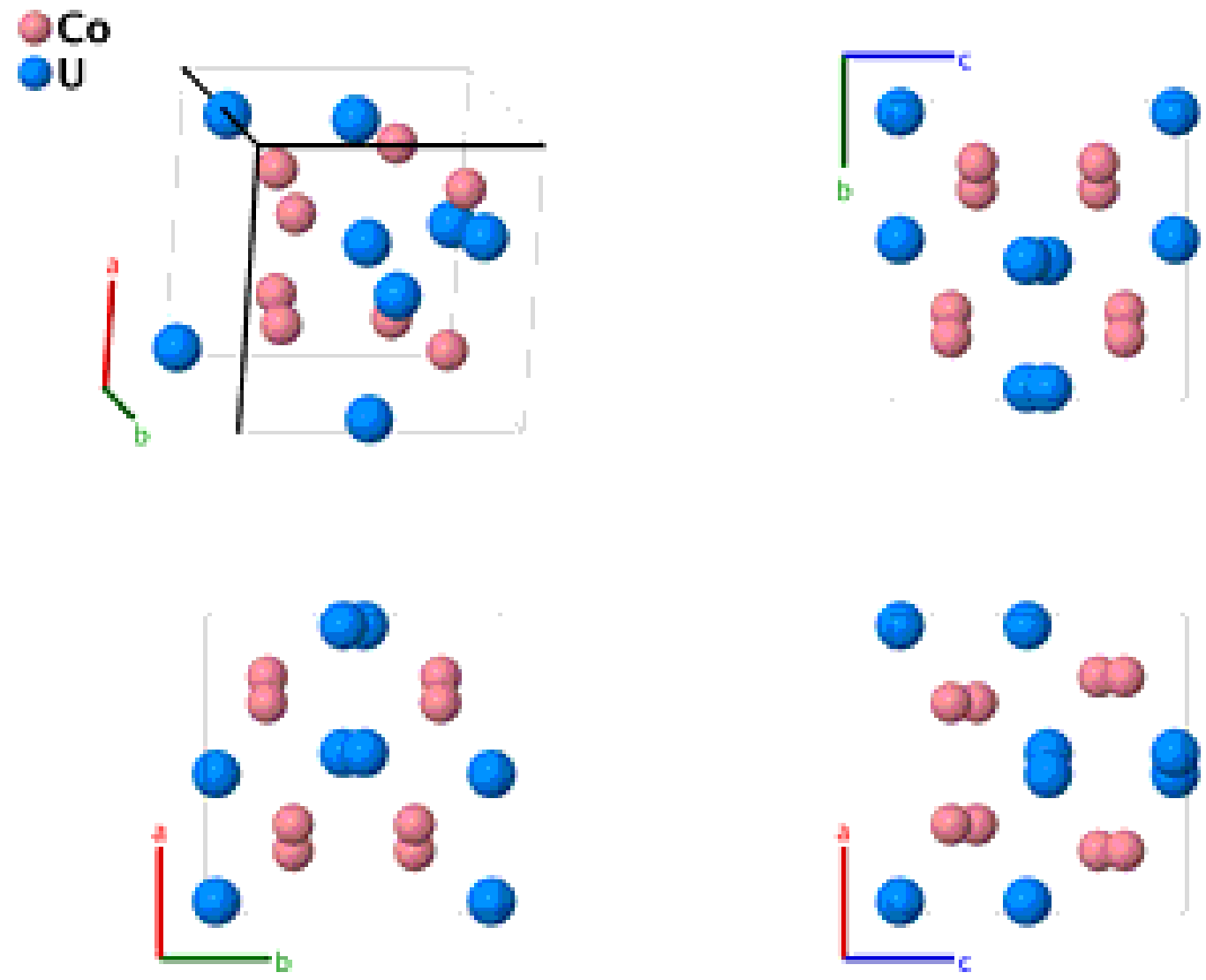

| | | |
|---|---|---|
| **Prototype** | : | CoU |
| **AFLOW prototype label** | : | AB_cI16_199_a_a |
| ***Strukturbericht* designation** | : | $B_a$ |
| **Pearson symbol** | : | cI16 |
| **Space group number** | : | 199 |
| **Space group symbol** | : | $I2_13$ |
| **AFLOW prototype command** | : | `aflow --proto=AB_cI16_199_a_a --params=`$a, x_1, x_2$ |

**Other compounds with this structure:**

- $Ga_2Pu_3$

- When $x_1 = 1/4$ and $x_2 = 0$, or visa versa, this structure reduces to CsCl (B2) with $a_{B2} = 1/2a$.

**Body-centered Cubic primitive vectors:**

$$\mathbf{a}_1 = -\frac{1}{2}\,a\,\mathbf{\hat{x}} + \frac{1}{2}\,a\,\mathbf{\hat{y}} + \frac{1}{2}\,a\,\mathbf{\hat{z}}$$
$$\mathbf{a}_2 = \frac{1}{2}\,a\,\mathbf{\hat{x}} - \frac{1}{2}\,a\,\mathbf{\hat{y}} + \frac{1}{2}\,a\,\mathbf{\hat{z}}$$
$$\mathbf{a}_3 = \frac{1}{2}\,a\,\mathbf{\hat{x}} + \frac{1}{2}\,a\,\mathbf{\hat{y}} - \frac{1}{2}\,a\,\mathbf{\hat{z}}$$

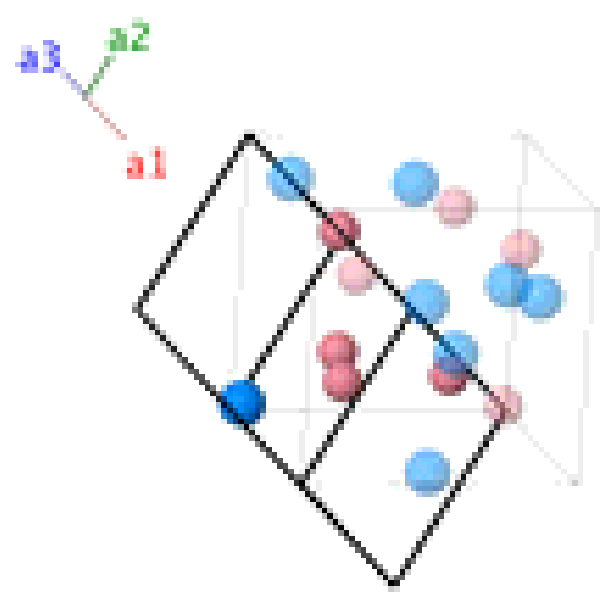

**Basis vectors:**

| | | Lattice Coordinates | | Cartesian Coordinates | Wyckoff Position | Atom Type |
|---|---|---|---|---|---|---|
| $\mathbf{B}_1$ | = | $2x_1\,\mathbf{a}_1 + 2x_1\,\mathbf{a}_2 + 2x_1\,\mathbf{a}_3$ | = | $x_1\,a\,\mathbf{\hat{x}} + x_1\,a\,\mathbf{\hat{y}} + x_1\,a\,\mathbf{\hat{z}}$ | (8a) | Co |
| $\mathbf{B_2}$ | = | $\frac{1}{2}\,\mathbf{a_1} + \left(\frac{1}{2} - 2x_1\right)\,\mathbf{a_3}$ | = | $-x_1\,a\,\mathbf{\hat{x}} + \left(\frac{1}{2} - x_1\right)\,a\,\mathbf{\hat{y}} + x_1\,a\,\mathbf{\hat{z}}$ | (8$a$) | Co |
| $\mathbf{B_3}$ | = | $\left(\frac{1}{2} - 2x_1\right)\,\mathbf{a_2} + \frac{1}{2}\,\mathbf{a_3}$ | = | $\left(\frac{1}{2} - x_1\right)\,a\,\mathbf{\hat{x}} + x_1\,a\,\mathbf{\hat{y}} - x_1\,a\,\mathbf{\hat{z}}$ | (8$a$) | Co |
| $\mathbf{B_4}$ | = | $\left(\frac{1}{2} - 2x_1\right)\,\mathbf{a_1} + \frac{1}{2}\,\mathbf{a_2}$ | = | $x_1\,a\,\mathbf{\hat{x}} - x_1\,a\,\mathbf{\hat{y}} + \left(\frac{1}{2} - x_1\right)\,a\,\mathbf{\hat{z}}$ | (8$a$) | Co |
| $\mathbf{B_5}$ | = | $2x_2\,\mathbf{a_1} + 2x_2\,\mathbf{a_2} + 2x_2\,\mathbf{a_3}$ | = | $x_2\,a\,\mathbf{\hat{x}} + x_2\,a\,\mathbf{\hat{y}} + x_2\,a\,\mathbf{\hat{z}}$ | (8$a$) | U |
| $\mathbf{B_6}$ | = | $\frac{1}{2}\,\mathbf{a_1} + \left(\frac{1}{2} - 2x_2\right)\,\mathbf{a_3}$ | = | $-x_2\,a\,\mathbf{\hat{x}} + \left(\frac{1}{2} - x_2\right)\,a\,\mathbf{\hat{y}} + x_2\,a\,\mathbf{\hat{z}}$ | (8$a$) | U |
| $\mathbf{B_7}$ | = | $\left(\frac{1}{2} - 2x_2\right)\,\mathbf{a_2} + \frac{1}{2}\,\mathbf{a_3}$ | = | $\left(\frac{1}{2} - x_2\right)\,a\,\mathbf{\hat{x}} + x_2\,a\,\mathbf{\hat{y}} - x_2\,a\,\mathbf{\hat{z}}$ | (8$a$) | U |
| $\mathbf{B_8}$ | = | $\left(\frac{1}{2} - 2x_2\right)\,\mathbf{a_1} + \frac{1}{2}\,\mathbf{a_2}$ | = | $x_2\,a\,\mathbf{\hat{x}} - x_2\,a\,\mathbf{\hat{y}} + \left(\frac{1}{2} - x_2\right)\,a\,\mathbf{\hat{z}}$ | (8$a$) | U |

**Found in:**

**Geometry files:**

- CIF: pp. 757
- POSCAR: pp. 757

# Bergman [$Mg_{32}(Al,Zn)_{49}$] Structure: AB32C48_cI162_204_a_2efg_2gh

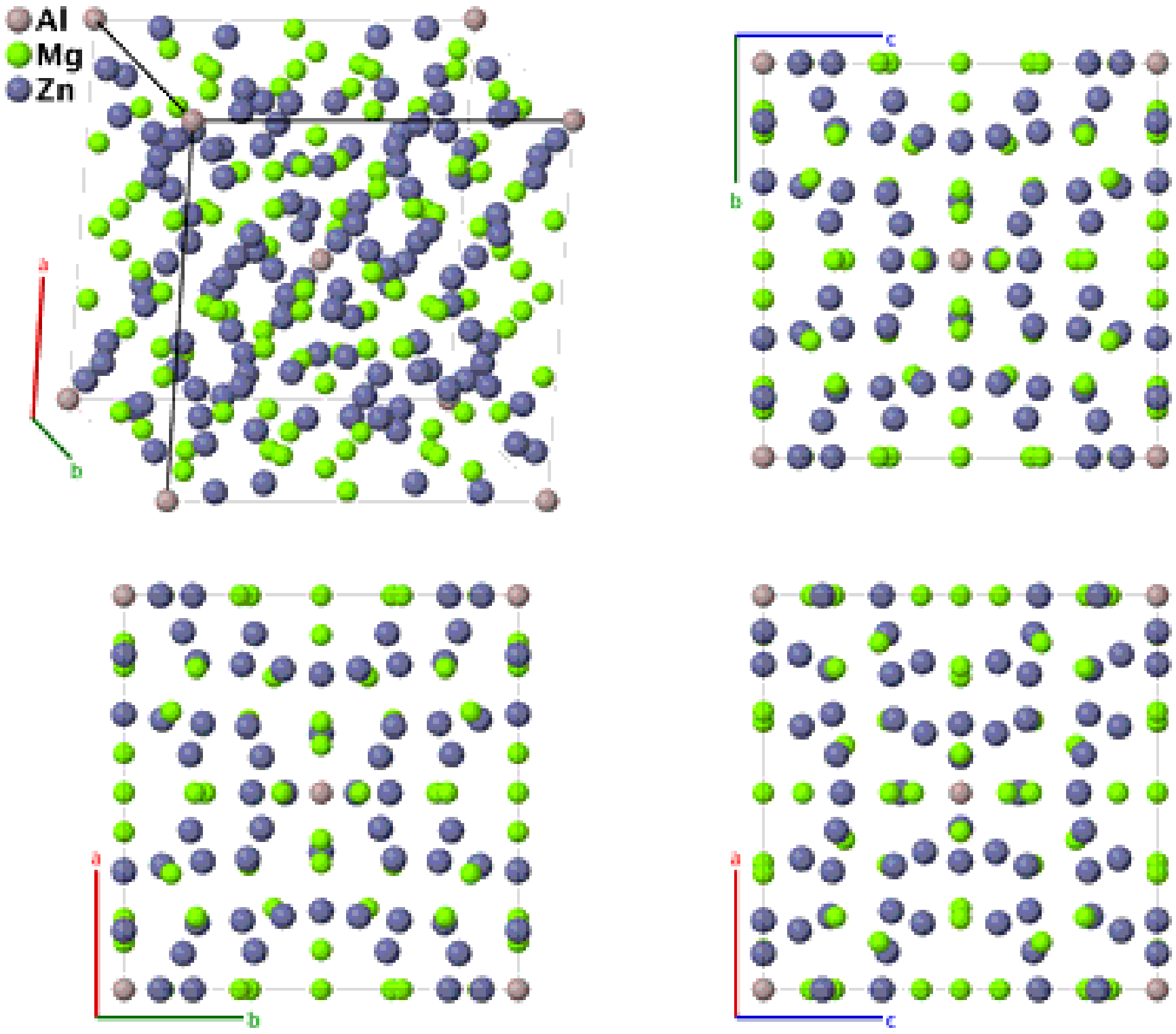

**Prototype** : $Mg_{32}(Al,Zn)_{49}$

**AFLOW prototype label** : AB32C48_cI162_204_a_2efg_2gh

***Strukturbericht* designation** : None

**Pearson symbol** : cI162

**Space group number** : 204

**Space group symbol** : $Im\bar{3}$

**AFLOW prototype command** : `aflow --proto=AB32C48_cI162_204_a_2efg_2gh --params=`$a, x_2, x_3, x_4, y_5, z_5, y_6, z_6, y_7, z_7, x_8, y_8, z_8$

- Most of the sites in this lattice have random occupancy. In particular, according to (Bergman, 1957): The Al-I (2a) site is only occupied 80% of the time, the Zn-I (24g) site is occupied by Al 19% of the time, the Zn-II (24g) site is occupied by Al 43% of the time, and the Zn-III (48h) site is occupied by Al 36% of the time.

**Body-centered Cubic primitive vectors:**

$$\mathbf{a}_1 = -\frac{1}{2}\,a\,\hat{\mathbf{x}} + \frac{1}{2}\,a\,\hat{\mathbf{y}} + \frac{1}{2}\,a\,\hat{\mathbf{z}}$$
$$\mathbf{a}_2 = \frac{1}{2}\,a\,\hat{\mathbf{x}} - \frac{1}{2}\,a\,\hat{\mathbf{y}} + \frac{1}{2}\,a\,\hat{\mathbf{z}}$$
$$\mathbf{a}_3 = \frac{1}{2}\,a\,\hat{\mathbf{x}} + \frac{1}{2}\,a\,\hat{\mathbf{y}} - \frac{1}{2}\,a\,\hat{\mathbf{z}}$$

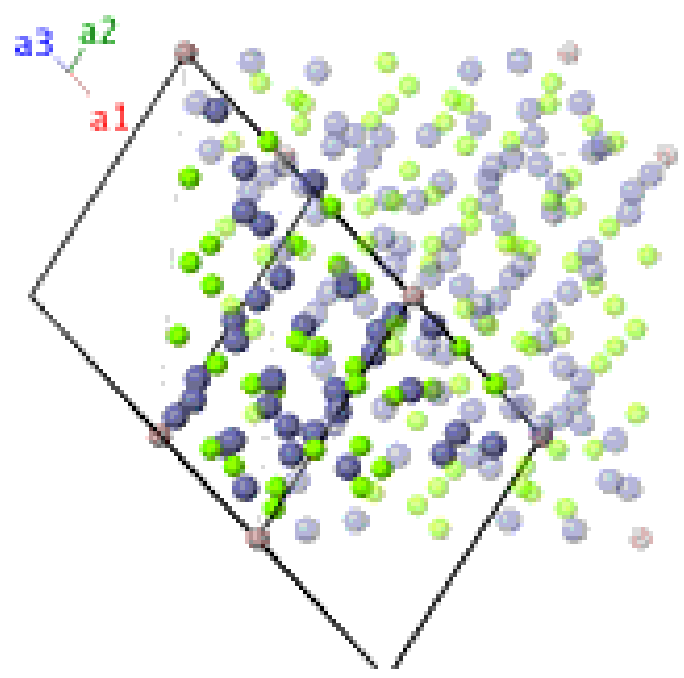

**Basis vectors:**

| | | Lattice Coordinates | | Cartesian Coordinates | Wyckoff Position | Atom Type |
|---|---|---|---|---|---|---|
| $\mathbf{B_1}$ | = | $0\,\mathbf{a_1} + 0\,\mathbf{a_2} + 0\,\mathbf{a_3}$ | = | $0\hat{\mathbf{x}} + 0\hat{\mathbf{y}} + 0\hat{\mathbf{z}}$ | $(2a)$ | Al |
| $\mathbf{B_2}$ | = | $\frac{1}{2}\,\mathbf{a_1} + \left(\frac{1}{2} + x_2\right)\,\mathbf{a_2} + x_2\,\mathbf{a_3}$ | = | $x_2\,a\,\hat{\mathbf{x}} + \frac{1}{2}\,a\,\hat{\mathbf{z}}$ | $(12e)$ | Mg I |
| $\mathbf{B_3}$ | = | $\frac{1}{2}\,\mathbf{a_1} + \left(\frac{1}{2} - x_2\right)\,\mathbf{a_2} - x_2\,\mathbf{a_3}$ | = | $-x_2\,a\,\hat{\mathbf{x}} + \frac{1}{2}\,a\,\hat{\mathbf{z}}$ | $(12e)$ | Mg I |
| $\mathbf{B_4}$ | = | $x_2\,\mathbf{a_1} + \frac{1}{2}\,\mathbf{a_2} + \left(\frac{1}{2} + x_2\right)\,\mathbf{a_3}$ | = | $\frac{1}{2}\,a\,\hat{\mathbf{x}} + x_2\,a\,\hat{\mathbf{y}}$ | $(12e)$ | Mg I |
| $\mathbf{B_5}$ | = | $-x_2\,\mathbf{a_1} + \frac{1}{2}\,\mathbf{a_2} + \left(\frac{1}{2} - x_2\right)\,\mathbf{a_3}$ | = | $\frac{1}{2}\,a\,\hat{\mathbf{x}} - x_2\,a\,\hat{\mathbf{y}}$ | $(12e)$ | Mg I |
| $\mathbf{B_6}$ | = | $\left(\frac{1}{2} + x_2\right)\,\mathbf{a_1} + x_2\,\mathbf{a_2} + \frac{1}{2}\,\mathbf{a_3}$ | = | $\frac{1}{2}\,a\,\hat{\mathbf{y}} + x_2\,a\,\hat{\mathbf{z}}$ | $(12e)$ | Mg I |
| $\mathbf{B_7}$ | = | $\left(\frac{1}{2} - x_2\right)\,\mathbf{a_1} - x_2\,\mathbf{a_2} + \frac{1}{2}\,\mathbf{a_3}$ | = | $\frac{1}{2}\,a\,\hat{\mathbf{y}} - x_2\,a\,\hat{\mathbf{z}}$ | $(12e)$ | Mg I |
| $\mathbf{B_8}$ | = | $\frac{1}{2}\,\mathbf{a_1} + \left(\frac{1}{2} + x_3\right)\,\mathbf{a_2} + x_3\,\mathbf{a_3}$ | = | $x_3\,a\,\hat{\mathbf{x}} + \frac{1}{2}\,a\,\hat{\mathbf{z}}$ | $(12e)$ | Mg II |
| $\mathbf{B_9}$ | = | $\frac{1}{2}\,\mathbf{a_1} + \left(\frac{1}{2} - x_3\right)\,\mathbf{a_2} - x_3\,\mathbf{a_3}$ | = | $-x_3\,a\,\hat{\mathbf{x}} + \frac{1}{2}\,a\,\hat{\mathbf{z}}$ | $(12e)$ | Mg II |
| $\mathbf{B_{10}}$ | = | $x_3\,\mathbf{a_1} + \frac{1}{2}\,\mathbf{a_2} + \left(\frac{1}{2} + x_3\right)\,\mathbf{a_3}$ | = | $\frac{1}{2}\,a\,\hat{\mathbf{x}} + x_3\,a\,\hat{\mathbf{y}}$ | $(12e)$ | Mg II |
| $\mathbf{B_{11}}$ | = | $-x_3\,\mathbf{a_1} + \frac{1}{2}\,\mathbf{a_2} + \left(\frac{1}{2} - x_3\right)\,\mathbf{a_3}$ | = | $\frac{1}{2}\,a\,\hat{\mathbf{x}} - x_3\,a\,\hat{\mathbf{y}}$ | $(12e)$ | Mg II |
| $\mathbf{B_{12}}$ | = | $\left(\frac{1}{2} + x_3\right)\,\mathbf{a_1} + x_3\,\mathbf{a_2} + \frac{1}{2}\,\mathbf{a_3}$ | = | $\frac{1}{2}\,a\,\hat{\mathbf{y}} + x_3\,a\,\hat{\mathbf{z}}$ | $(12e)$ | Mg II |
| $\mathbf{B_{13}}$ | = | $\left(\frac{1}{2} - x_3\right)\,\mathbf{a_1} - x_3\,\mathbf{a_2} + \frac{1}{2}\,\mathbf{a_3}$ | = | $\frac{1}{2}\,a\,\hat{\mathbf{y}} - x_3\,a\,\hat{\mathbf{z}}$ | $(12e)$ | Mg II |
| $\mathbf{B_{14}}$ | = | $2x_4\,\mathbf{a_1} + 2x_4\,\mathbf{a_2} + 2x_4\,\mathbf{a_3}$ | = | $x_4\,a\,\hat{\mathbf{x}} + x_4\,a\,\hat{\mathbf{y}} + x_4\,a\,\hat{\mathbf{z}}$ | $(16f)$ | Mg III |
| $\mathbf{B_{15}}$ | = | $2x_4\,\mathbf{a_1}$ | = | $-x_4\,a\,\hat{\mathbf{x}} + x_4\,a\,\hat{\mathbf{y}} + x_4\,a\,\hat{\mathbf{z}}$ | $(16f)$ | Mg III |
| $\mathbf{B_{16}}$ | = | $2x_4\,\mathbf{a_2}$ | = | $x_4\,a\,\hat{\mathbf{x}} - x_4\,a\,\hat{\mathbf{y}} + x_4\,a\,\hat{\mathbf{z}}$ | $(16f)$ | Mg III |
| $\mathbf{B_{17}}$ | = | $2x_4\,\mathbf{a_3}$ | = | $x_4\,a\,\hat{\mathbf{x}} + x_4\,a\,\hat{\mathbf{y}} - x_4\,a\,\hat{\mathbf{z}}$ | $(16f)$ | Mg III |
| $\mathbf{B_{18}}$ | = | $-2x_4\,\mathbf{a_1} - 2x_4\,\mathbf{a_2} - 2x_4\,\mathbf{a_3}$ | = | $-x_4\,a\,\hat{\mathbf{x}} - x_4\,a\,\hat{\mathbf{y}} - x_4\,a\,\hat{\mathbf{z}}$ | $(16f)$ | Mg III |
| $\mathbf{B_{19}}$ | = | $-2x_4\,\mathbf{a_1}$ | = | $x_4\,a\,\hat{\mathbf{x}} - x_4\,a\,\hat{\mathbf{y}} - x_4\,a\,\hat{\mathbf{z}}$ | $(16f)$ | Mg III |
| $\mathbf{B_{20}}$ | = | $-2x_4\,\mathbf{a_2}$ | = | $-x_4\,a\,\hat{\mathbf{x}} + x_4\,a\,\hat{\mathbf{y}} - x_4\,a\,\hat{\mathbf{z}}$ | $(16f)$ | Mg III |
| $\mathbf{B_{21}}$ | = | $-2x_4\,\mathbf{a_3}$ | = | $-x_4\,a\,\hat{\mathbf{x}} - x_4\,a\,\hat{\mathbf{y}} + x_4\,a\,\hat{\mathbf{z}}$ | $(16f)$ | Mg III |
| $\mathbf{B_{22}}$ | = | $(y_5 + z_5)\,\mathbf{a_1} + z_5\,\mathbf{a_2} + y_5\,\mathbf{a_3}$ | = | $y_5\,a\,\hat{\mathbf{y}} + z_5\,a\,\hat{\mathbf{z}}$ | $(24g)$ | Mg IV |
| $\mathbf{B_{23}}$ | = | $(z_5 - y_5)\,\mathbf{a_1} + z_5\,\mathbf{a_2} - y_5\,\mathbf{a_3}$ | = | $-y_5\,a\,\hat{\mathbf{y}} + z_5\,a\,\hat{\mathbf{z}}$ | $(24g)$ | Mg IV |
| $\mathbf{B_{24}}$ | = | $(y_5 - z_5)\,\mathbf{a_1} - z_5\,\mathbf{a_2} + y_5\,\mathbf{a_3}$ | = | $y_5\,a\,\hat{\mathbf{y}} - z_5\,a\,\hat{\mathbf{z}}$ | $(24g)$ | Mg IV |
| $\mathbf{B_{25}}$ | = | $-(y_5 + z_5)\,\mathbf{a_1} - z_5\,\mathbf{a_2} - y_5\,\mathbf{a_3}$ | = | $-y_5\,a\,\hat{\mathbf{y}} - z_5\,a\,\hat{\mathbf{z}}$ | $(24g)$ | Mg IV |
| $\mathbf{B_{26}}$ | = | $y_5\,\mathbf{a_1} + (y_5 + z_5)\,\mathbf{a_2} + z_5\,\mathbf{a_3}$ | = | $z_5\,a\,\hat{\mathbf{x}} + y_5\,a\,\hat{\mathbf{z}}$ | $(24g)$ | Mg IV |
| $\mathbf{B_{27}}$ | = | $y_5\,\mathbf{a_1} + (y_5 - z_5)\,\mathbf{a_2} - z_5\,\mathbf{a_3}$ | = | $-z_5\,a\,\hat{\mathbf{x}} + y_5\,a\,\hat{\mathbf{z}}$ | $(24g)$ | Mg IV |
| $\mathbf{B_{28}}$ | = | $-y_5\,\mathbf{a_1} + (z_5 - y_5)\,\mathbf{a_2} + z_5\,\mathbf{a_3}$ | = | $z_5\,a\,\hat{\mathbf{x}} - y_5\,a\,\hat{\mathbf{z}}$ | $(24g)$ | Mg IV |
| $\mathbf{B_{29}}$ | = | $-y_5\,\mathbf{a_1} - (y_5 + z_5)\,\mathbf{a_2} - z_5\,\mathbf{a_3}$ | = | $-z_5\,a\,\hat{\mathbf{x}} - y_5\,a\,\hat{\mathbf{z}}$ | $(24g)$ | Mg IV |
| $\mathbf{B_{30}}$ | = | $z_5\,\mathbf{a_1} + y_5\,\mathbf{a_2} + (y_5 + z_5)\,\mathbf{a_3}$ | = | $y_5\,a\,\hat{\mathbf{x}} + z_5\,a\,\hat{\mathbf{y}}$ | $(24g)$ | Mg IV |
| $\mathbf{B_{31}}$ | = | $z_5\,\mathbf{a_1} - y_5\,\mathbf{a_2} + (z_5 - y_5)\,\mathbf{a_3}$ | = | $-y_5\,a\,\hat{\mathbf{x}} + z_5\,a\,\hat{\mathbf{y}}$ | $(24g)$ | Mg IV |
| $\mathbf{B_{32}}$ | = | $-z_5\,\mathbf{a_1} + y_5\,\mathbf{a_2} + (y_5 - z_5)\,\mathbf{a_3}$ | = | $y_5\,a\,\hat{\mathbf{x}} - z_5\,a\,\hat{\mathbf{y}}$ | $(24g)$ | Mg IV |
| $\mathbf{B_{33}}$ | = | $-z_5\,\mathbf{a_1} - y_5\,\mathbf{a_2} - (y_5 + z_5)\,\mathbf{a_3}$ | = | $-y_5\,a\,\hat{\mathbf{x}} - z_5\,a\,\hat{\mathbf{y}}$ | $(24g)$ | Mg IV |
| $\mathbf{B_{34}}$ | = | $(y_6 + z_6)\,\mathbf{a_1} + z_6\,\mathbf{a_2} + y_6\,\mathbf{a_3}$ | = | $y_6\,a\,\hat{\mathbf{y}} + z_6\,a\,\hat{\mathbf{z}}$ | $(24g)$ | Zn I |
| $\mathbf{B_{35}}$ | = | $(z_6 - y_6)\,\mathbf{a_1} + z_6\,\mathbf{a_2} - y_6\,\mathbf{a_3}$ | = | $-y_6\,a\,\hat{\mathbf{y}} + z_6\,a\,\hat{\mathbf{z}}$ | $(24g)$ | Zn I |
| $\mathbf{B}_{36}$ | = | $(y_6 - z_6)\,\mathbf{a}_1 - z_6\,\mathbf{a}_2 + y_6\,\mathbf{a}_3$ | = | $y_6\,a\,\mathbf{\hat{y}} - z_6\,a\,\mathbf{\hat{z}}$ | (24$g$) | Zn I |
| $\mathbf{B}_{37}$ | = | $-(y_6 + z_6)\,\mathbf{a}_1 - z_6\,\mathbf{a}_2 - y_6\,\mathbf{a}_3$ | = | $-y_6\,a\,\mathbf{\hat{y}} - z_6\,a\,\mathbf{\hat{z}}$ | (24$g$) | Zn I |
| $\mathbf{B}_{38}$ | = | $y_6\,\mathbf{a}_1 + (y_6 + z_6)\,\mathbf{a}_2 + z_6\,\mathbf{a}_3$ | = | $z_6\,a\,\mathbf{\hat{x}} + y_6\,a\,\mathbf{\hat{z}}$ | (24$g$) | Zn I |
| $\mathbf{B}_{39}$ | = | $y_6\,\mathbf{a}_1 + (y_6 - z_6)\,\mathbf{a}_2 - z_6\,\mathbf{a}_3$ | = | $-z_6\,a\,\mathbf{\hat{x}} + y_6\,a\,\mathbf{\hat{z}}$ | (24$g$) | Zn I |
| $\mathbf{B}_{40}$ | = | $-y_6\,\mathbf{a}_1 + (z_6 - y_6)\,\mathbf{a}_2 + z_6\,\mathbf{a}_3$ | = | $z_6\,a\,\mathbf{\hat{x}} - y_6\,a\,\mathbf{\hat{z}}$ | (24$g$) | Zn I |
| $\mathbf{B}_{41}$ | = | $-y_6\,\mathbf{a}_1 - (y_6 + z_6)\,\mathbf{a}_2 - z_6\,\mathbf{a}_3$ | = | $-z_6\,a\,\mathbf{\hat{x}} - y_6\,a\,\mathbf{\hat{z}}$ | (24$g$) | Zn I |
| $\mathbf{B}_{42}$ | = | $z_6\,\mathbf{a}_1 + y_6\,\mathbf{a}_2 + (y_6 + z_6)\,\mathbf{a}_3$ | = | $y_6\,a\,\mathbf{\hat{x}} + z_6\,a\,\mathbf{\hat{y}}$ | (24$g$) | Zn I |
| $\mathbf{B}_{43}$ | = | $z_6\,\mathbf{a}_1 - y_6\,\mathbf{a}_2 + (z_6 - y_6)\,\mathbf{a}_3$ | = | $-y_6\,a\,\mathbf{\hat{x}} + z_6\,a\,\mathbf{\hat{y}}$ | (24$g$) | Zn I |
| $\mathbf{B}_{44}$ | = | $-z_6\,\mathbf{a}_1 + y_6\,\mathbf{a}_2 + (y_6 - z_6)\,\mathbf{a}_3$ | = | $y_6\,a\,\mathbf{\hat{x}} - z_6\,a\,\mathbf{\hat{y}}$ | (24$g$) | Zn I |
| $\mathbf{B}_{45}$ | = | $-z_6\,\mathbf{a}_1 - y_6\,\mathbf{a}_2 - (y_6 + z_6)\,\mathbf{a}_3$ | = | $-y_6\,a\,\mathbf{\hat{x}} - z_6\,a\,\mathbf{\hat{y}}$ | (24$g$) | Zn I |
| $\mathbf{B}_{46}$ | = | $(y_7 + z_7)\,\mathbf{a}_1 + z_7\,\mathbf{a}_2 + y_7\,\mathbf{a}_3$ | = | $y_7\,a\,\mathbf{\hat{y}} + z_7\,a\,\mathbf{\hat{z}}$ | (24$g$) | Zn II |
| $\mathbf{B}_{47}$ | = | $(z_7 - y_7)\,\mathbf{a}_1 + z_7\,\mathbf{a}_2 - y_7\,\mathbf{a}_3$ | = | $-y_7\,a\,\mathbf{\hat{y}} + z_7\,a\,\mathbf{\hat{z}}$ | (24$g$) | Zn II |
| $\mathbf{B}_{48}$ | = | $(y_7 - z_7)\,\mathbf{a}_1 - z_7\,\mathbf{a}_2 + y_7\,\mathbf{a}_3$ | = | $y_7\,a\,\mathbf{\hat{y}} - z_7\,a\,\mathbf{\hat{z}}$ | (24$g$) | Zn II |
| $\mathbf{B}_{49}$ | = | $-(y_7 + z_7)\,\mathbf{a}_1 - z_7\,\mathbf{a}_2 - y_7\,\mathbf{a}_3$ | = | $-y_7\,a\,\mathbf{\hat{y}} - z_7\,a\,\mathbf{\hat{z}}$ | (24$g$) | Zn II |
| $\mathbf{B}_{50}$ | = | $y_7\,\mathbf{a}_1 + (y_7 + z_7)\,\mathbf{a}_2 + z_7\,\mathbf{a}_3$ | = | $z_7\,a\,\mathbf{\hat{x}} + y_7\,a\,\mathbf{\hat{z}}$ | (24$g$) | Zn II |
| $\mathbf{B}_{51}$ | = | $y_7\,\mathbf{a}_1 + (y_7 - z_7)\,\mathbf{a}_2 - z_7\,\mathbf{a}_3$ | = | $-z_7\,a\,\mathbf{\hat{x}} + y_7\,a\,\mathbf{\hat{z}}$ | (24$g$) | Zn II |
| $\mathbf{B}_{52}$ | = | $-y_7\,\mathbf{a}_1 + (z_7 - y_7)\,\mathbf{a}_2 + z_7\,\mathbf{a}_3$ | = | $z_7\,a\,\mathbf{\hat{x}} - y_7\,a\,\mathbf{\hat{z}}$ | (24$g$) | Zn II |
| $\mathbf{B}_{53}$ | = | $-y_7\,\mathbf{a}_1 - (y_7 + z_7)\,\mathbf{a}_2 - z_7\,\mathbf{a}_3$ | = | $-z_7\,a\,\mathbf{\hat{x}} - y_7\,a\,\mathbf{\hat{z}}$ | (24$g$) | Zn II |
| $\mathbf{B}_{54}$ | = | $z_7\,\mathbf{a}_1 + y_7\,\mathbf{a}_2 + (y_7 + z_7)\,\mathbf{a}_3$ | = | $y_7\,a\,\mathbf{\hat{x}} + z_7\,a\,\mathbf{\hat{y}}$ | (24$g$) | Zn II |
| $\mathbf{B}_{55}$ | = | $z_7\,\mathbf{a}_1 - y_7\,\mathbf{a}_2 + (z_7 - y_7)\,\mathbf{a}_3$ | = | $-y_7\,a\,\mathbf{\hat{x}} + z_7\,a\,\mathbf{\hat{y}}$ | (24$g$) | Zn II |
| $\mathbf{B}_{56}$ | = | $-z_7\,\mathbf{a}_1 + y_7\,\mathbf{a}_2 + (y_7 - z_7)\,\mathbf{a}_3$ | = | $y_7\,a\,\mathbf{\hat{x}} - z_7\,a\,\mathbf{\hat{y}}$ | (24$g$) | Zn II |
| $\mathbf{B}_{57}$ | = | $-z_7\,\mathbf{a}_1 - y_7\,\mathbf{a}_2 - (y_7 + z_7)\,\mathbf{a}_3$ | = | $-y_7\,a\,\mathbf{\hat{x}} - z_7\,a\,\mathbf{\hat{y}}$ | (24$g$) | Zn II |
| $\mathbf{B}_{58}$ | = | $(y_8 + z_8)\,\mathbf{a}_1 + (z_8 + x_8)\,\mathbf{a}_2 + (x_8 + y_8)\,\mathbf{a}_3$ | = | $x_8\,a\,\mathbf{\hat{x}} + y_8\,a\,\mathbf{\hat{y}} + z_8\,a\,\mathbf{\hat{z}}$ | (48$h$) | Zn III |
| $\mathbf{B}_{59}$ | = | $(z_8 - y_8)\,\mathbf{a}_1 + (z_8 - x_8)\,\mathbf{a}_2 - (x_8 + y_8)\,\mathbf{a}_3$ | = | $-x_8\,a\,\mathbf{\hat{x}} - y_8\,a\,\mathbf{\hat{y}} + z_8\,a\,\mathbf{\hat{z}}$ | (48$h$) | Zn III |
| $\mathbf{B}_{60}$ | = | $(y_8 - z_8)\,\mathbf{a}_1 - (z_8 + x_8)\,\mathbf{a}_2 + (y_8 - x_8)\,\mathbf{a}_3$ | = | $-x_8\,a\,\mathbf{\hat{x}} + y_8\,a\,\mathbf{\hat{y}} - z_8\,a\,\mathbf{\hat{z}}$ | (48$h$) | Zn III |
| $\mathbf{B}_{61}$ | = | $-(y_8 + z_8)\,\mathbf{a}_1 + (x_8 - z_8)\,\mathbf{a}_2 + (x_8 - y_8)\,\mathbf{a}_3$ | = | $x_8\,a\,\mathbf{\hat{x}} - y_8\,a\,\mathbf{\hat{y}} - z_8\,a\,\mathbf{\hat{z}}$ | (48$h$) | Zn III |
| $\mathbf{B}_{62}$ | = | $-(y_8 + z_8)\,\mathbf{a}_1 - (z_8 + x_8)\,\mathbf{a}_2 - (x_8 + y_8)\,\mathbf{a}_3$ | = | $-x_8\,a\,\mathbf{\hat{x}} - y_8\,a\,\mathbf{\hat{y}} - z_8\,a\,\mathbf{\hat{z}}$ | (48$h$) | Zn III |
| $\mathbf{B}_{63}$ | = | $(y_8 - z_8)\,\mathbf{a}_1 + (x_8 - z_8)\,\mathbf{a}_2 + (x_8 + y_8)\,\mathbf{a}_3$ | = | $+x_8\,a\,\mathbf{\hat{x}} + y_8\,a\,\mathbf{\hat{y}} - z_8\,a\,\mathbf{\hat{z}}$ | (48$h$) | Zn III |
| $\mathbf{B}_{64}$ | = | $(z_8 - y_8)\,\mathbf{a}_1 + (z_8 + x_8)\,\mathbf{a}_2 + (x_8 - y_8)\,\mathbf{a}_3$ | = | $+x_8\,a\,\mathbf{\hat{x}} - y_8\,a\,\mathbf{\hat{y}} + z_8\,a\,\mathbf{\hat{z}}$ | (48$h$) | Zn III |
| $\mathbf{B}_{65}$ | = | $(y_8 + z_8)\,\mathbf{a}_1 + (z_8 - x_8)\,\mathbf{a}_2 + (y_8 - x_8)\,\mathbf{a}_3$ | = | $-x_8\,a\,\mathbf{\hat{x}} + y_8\,a\,\mathbf{\hat{y}} + z_8\,a\,\mathbf{\hat{z}}$ | (48$h$) | Zn III |
| $\mathbf{B}_{66}$ | = | $(x_8 + y_8)\,\mathbf{a}_1 + (y_8 + z_8)\,\mathbf{a}_2 + (z_8 + x_8)\,\mathbf{a}_3$ | = | $z_8\,a\,\mathbf{\hat{x}} + x_8\,a\,\mathbf{\hat{y}} + y_8\,a\,\mathbf{\hat{z}}$ | (48$h$) | Zn III |
| $\mathbf{B}_{67}$ | = | $(y_8 - x_8)\,\mathbf{a}_1 + (y_8 - z_8)\,\mathbf{a}_2 - (z_8 + x_8)\,\mathbf{a}_3$ | = | $-z_8\,a\,\mathbf{\hat{x}} - x_8\,a\,\mathbf{\hat{y}} + y_8\,a\,\mathbf{\hat{z}}$ | (48$h$) | Zn III |
| $\mathbf{B}_{68}$ | = | $(x_8 - y_8)\,\mathbf{a}_1 - (y_8 + z_8)\,\mathbf{a}_2 + (x_8 - z_8)\,\mathbf{a}_3$ | = | $-z_8\,a\,\mathbf{\hat{x}} + x_8\,a\,\mathbf{\hat{y}} - y_8\,a\,\mathbf{\hat{z}}$ | (48$h$) | Zn III |
| $\mathbf{B}_{69}$ | = | $-(x_8 + y_8)\,\mathbf{a}_1 + (z_8 - y_8)\,\mathbf{a}_2 + (z_8 - x_8)\,\mathbf{a}_3$ | = | $z_8\,a\,\mathbf{\hat{x}} - x_8\,a\,\mathbf{\hat{y}} - y_8\,a\,\mathbf{\hat{z}}$ | (48$h$) | Zn III |
| $\mathbf{B}_{70}$ | = | $-(x_8 + y_8)\,\mathbf{a}_1 - (y_8 + z_8)\,\mathbf{a}_2 - (z_8 + x_8)\,\mathbf{a}_3$ | = | $-z_8\,a\,\mathbf{\hat{x}} - x_8\,a\,\mathbf{\hat{y}} - y_8\,a\,\mathbf{\hat{z}}$ | (48$h$) | Zn III |
| $\mathbf{B}_{71}$ | = | $(x_8 - y_8)\,\mathbf{a}_1 + (z_8 - y_8)\,\mathbf{a}_2 + (z_8 + x_8)\,\mathbf{a}_3$ | = | $+z_8\,a\,\mathbf{\hat{x}} + x_8\,a\,\mathbf{\hat{y}} - y_8\,a\,\mathbf{\hat{z}}$ | (48$h$) | Zn III |
| $\mathbf{B}_{72}$ | = | $(y_8 - x_8)\,\mathbf{a}_1 + (y_8 + z_8)\,\mathbf{a}_2 + (z_8 - x_8)\,\mathbf{a}_3$ | = | $+z_8\,a\,\hat{\mathbf{x}} - x_8\,a\,\hat{\mathbf{y}} + y_8\,a\,\hat{\mathbf{z}}$ | (48$h$) | Zn III |
| $\mathbf{B}_{73}$ | = | $(x_8 + y_8)\,\mathbf{a}_1 + (y_8 - z_8)\,\mathbf{a}_2 + (x_8 - z_8)\,\mathbf{a}_3$ | = | $-z_8\,a\,\hat{\mathbf{x}} + x_8\,a\,\hat{\mathbf{y}} + y_8\,a\,\hat{\mathbf{z}}$ | (48$h$) | Zn III |
| $\mathbf{B}_{74}$ | = | $(z_8 + x_8)\,\mathbf{a}_1 + (x_8 + y_8)\,\mathbf{a}_2 + (y_8 + z_8)\,\mathbf{a}_3$ | = | $y_8\,a\,\hat{\mathbf{x}} + z_8\,a\,\hat{\mathbf{y}} + x_8\,a\,\hat{\mathbf{z}}$ | (48$h$) | Zn III |
| $\mathbf{B}_{75}$ | = | $(x_8 - z_8)\,\mathbf{a}_1 + (x_8 - y_8)\,\mathbf{a}_2 - (y_8 + z_8)\,\mathbf{a}_3$ | = | $-y_8\,a\,\hat{\mathbf{x}} - z_8\,a\,\hat{\mathbf{y}} + x_8\,a\,\hat{\mathbf{z}}$ | (48$h$) | Zn III |
| $\mathbf{B}_{76}$ | = | $(z_8 - x_8)\,\mathbf{a}_1 - (x_8 + y_8)\,\mathbf{a}_2 + (z_8 - y_8)\,\mathbf{a}_3$ | = | $-y_8\,a\,\hat{\mathbf{x}} + z_8\,a\,\hat{\mathbf{y}} - x_8\,a\,\hat{\mathbf{z}}$ | (48$h$) | Zn III |
| $\mathbf{B}_{77}$ | = | $-(z_8 + x_8)\,\mathbf{a}_1 + (y_8 - x_8)\,\mathbf{a}_2 + (y_8 - z_8)\,\mathbf{a}_3$ | = | $y_8\,a\,\hat{\mathbf{x}} - z_8\,a\,\hat{\mathbf{y}} - x_8\,a\,\hat{\mathbf{z}}$ | (48$h$) | Zn III |
| $\mathbf{B}_{78}$ | = | $-(z_8 + x_8)\,\mathbf{a}_1 - (x_8 + y_8)\,\mathbf{a}_2 - (y_8 + z_8)\,\mathbf{a}_3$ | = | $-y_8\,a\,\hat{\mathbf{x}} - z_8\,a\,\hat{\mathbf{y}} - x_8\,a\,\hat{\mathbf{z}}$ | (48$h$) | Zn III |
| $\mathbf{B}_{79}$ | = | $(z_8 - x_8)\,\mathbf{a}_1 + (y_8 - x_8)\,\mathbf{a}_2 + (y_8 + z_8)\,\mathbf{a}_3$ | = | $+y_8\,a\,\hat{\mathbf{x}} + z_8\,a\,\hat{\mathbf{y}} - x_8\,a\,\hat{\mathbf{z}}$ | (48$h$) | Zn III |
| $\mathbf{B}_{80}$ | = | $(x_8 - z_8)\,\mathbf{a}_1 + (x_8 + y_8)\,\mathbf{a}_2 + (y_8 - z_8)\,\mathbf{a}_3$ | = | $+y_8\,a\,\hat{\mathbf{x}} - z_8\,a\,\hat{\mathbf{y}} + x_8\,a\,\hat{\mathbf{z}}$ | (48$h$) | Zn III |
| $\mathbf{B}_{81}$ | = | $(z_8 + x_8)\,\mathbf{a}_1 + (x_8 - y_8)\,\mathbf{a}_2 + (z_8 - y_8)\,\mathbf{a}_3$ | = | $-y_8\,a\,\hat{\mathbf{x}} + z_8\,a\,\hat{\mathbf{y}} + x_8\,a\,\hat{\mathbf{z}}$ | (48$h$) | Zn III |

**Geometry files:**

- CIF: pp. 757
- POSCAR: pp. 758

# Skutterudite ($CoAs_3$, $D0_2$) Structure: A3B_cI32_204_g_c

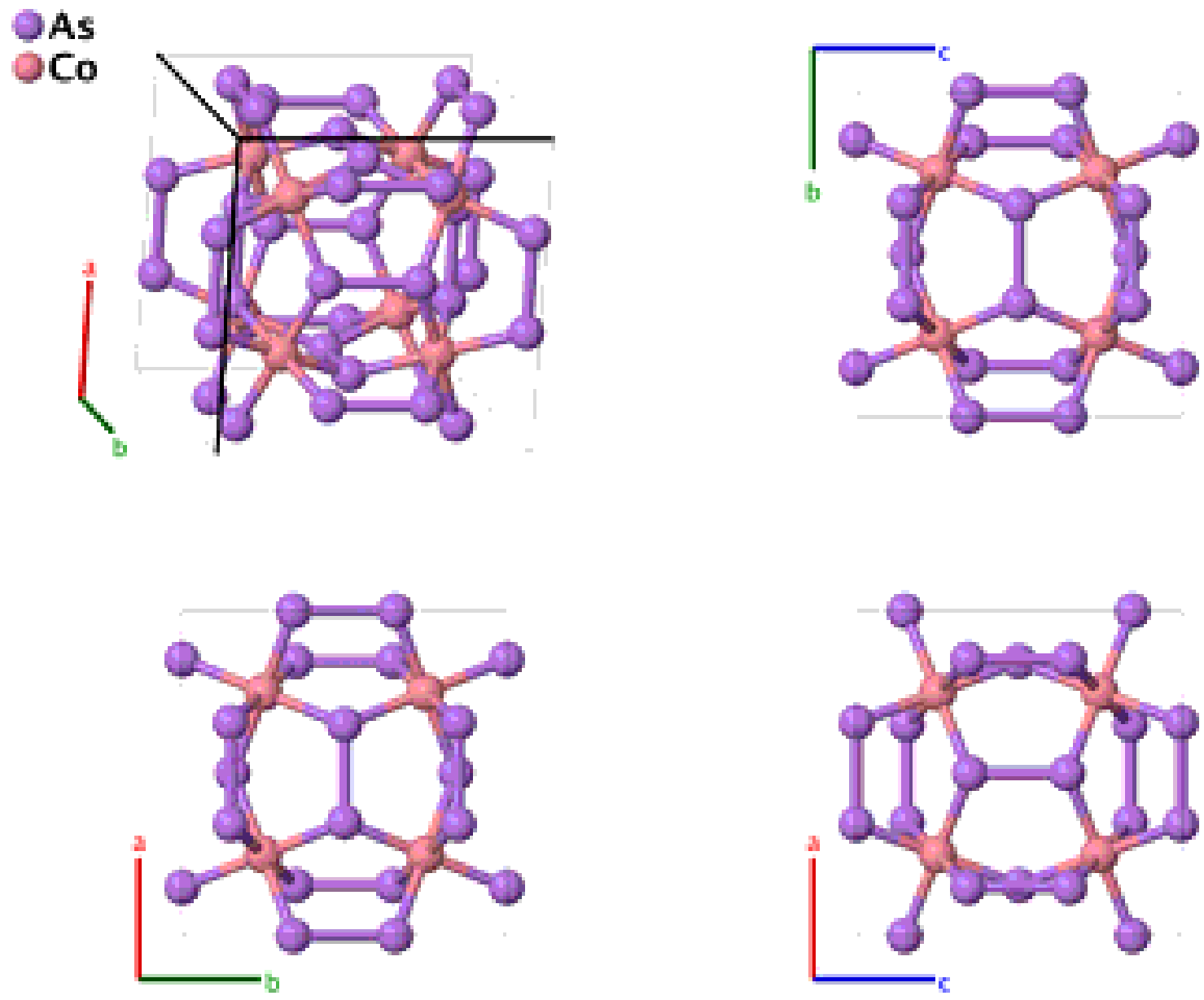

| | | |
|---|---|---|
| **Prototype** | : | $CoAs_3$ |
| **AFLOW prototype label** | : | A3B_cI32_204_g_c |
| ***Strukturbericht* designation** | : | $D0_2$ |
| **Pearson symbol** | : | cI32 |
| **Space group number** | : | 204 |
| **Space group symbol** | : | $Im\bar{3}$ |
| **AFLOW prototype command** | : | `aflow --proto=A3B_cI32_204_g_c`<br>`--params=`$a, y_2, z_2$ |

- Useful skutterudites have iron and nickel alloyed with cobalt.

**Body-centered Cubic primitive vectors:**

$$\mathbf{a}_1 = -\frac{1}{2}\,a\,\hat{\mathbf{x}} + \frac{1}{2}\,a\,\hat{\mathbf{y}} + \frac{1}{2}\,a\,\hat{\mathbf{z}}$$
$$\mathbf{a}_2 = \frac{1}{2}\,a\,\hat{\mathbf{x}} - \frac{1}{2}\,a\,\hat{\mathbf{y}} + \frac{1}{2}\,a\,\hat{\mathbf{z}}$$
$$\mathbf{a}_3 = \frac{1}{2}\,a\,\hat{\mathbf{x}} + \frac{1}{2}\,a\,\hat{\mathbf{y}} - \frac{1}{2}\,a\,\hat{\mathbf{z}}$$

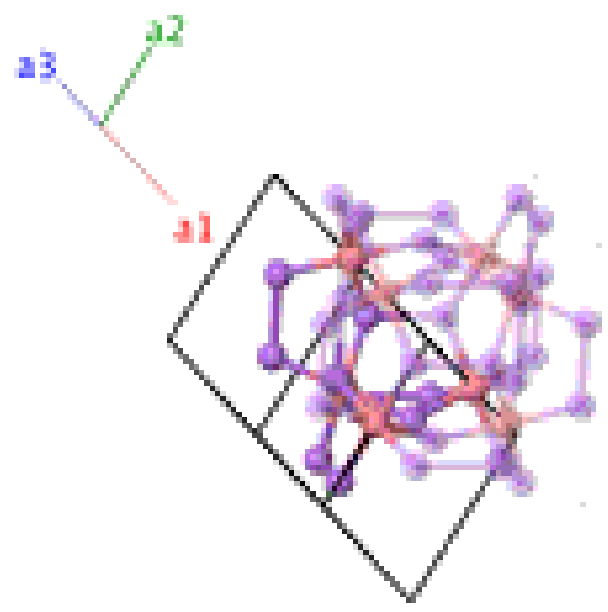

**Basis vectors:**

| | | Lattice Coordinates | | Cartesian Coordinates | Wyckoff Position | Atom Type |
|---|---|---|---|---|---|---|
| $\mathbf{B}_1$ | = | $\frac{1}{2}\,\mathbf{a}_1 + \frac{1}{2}\,\mathbf{a}_2 + \frac{1}{2}\,\mathbf{a}_3$ | = | $\frac{1}{4}\,a\,\hat{\mathbf{x}} + \frac{1}{4}\,a\,\hat{\mathbf{y}} + \frac{1}{4}\,a\,\hat{\mathbf{z}}$ | (8$c$) | Co |
| $\mathbf{B}_2$ | = | $\frac{1}{2}\,\mathbf{a}_3$ | = | $\frac{1}{4}\,a\,\hat{\mathbf{x}} + \frac{1}{4}\,a\,\hat{\mathbf{y}} + \frac{3}{4}\,a\,\hat{\mathbf{z}}$ | (8$c$) | Co |
| $\mathbf{B}_3$ | = | $\frac{1}{2}\,\mathbf{a}_2$ | = | $\frac{1}{4}\,a\,\hat{\mathbf{x}} + \frac{3}{4}\,a\,\hat{\mathbf{y}} + \frac{1}{4}\,a\,\hat{\mathbf{z}}$ | (8$c$) | Co |
| $\mathbf{B}_4$ | = | $\frac{1}{2}\,\mathbf{a}_1$ | = | $\frac{3}{4}\,a\,\hat{\mathbf{x}} + \frac{1}{4}\,a\,\hat{\mathbf{y}} + \frac{1}{4}\,a\,\hat{\mathbf{z}}$ | $(8c)$ | Co |
| $\mathbf{B}_5$ | = | $(y_2 + z_2)\,\mathbf{a}_1 + z_2\,\mathbf{a}_2 + y_2\,\mathbf{a}_3$ | = | $y_2\,a\,\hat{\mathbf{y}} + z_2\,a\,\hat{\mathbf{z}}$ | $(24g)$ | As |
| $\mathbf{B}_6$ | = | $(z_2 - y_2)\,\mathbf{a}_1 + z_2\,\mathbf{a}_2 - y_2\,\mathbf{a}_3$ | = | $-y_2\,a\,\hat{\mathbf{y}} + z_2\,a\,\hat{\mathbf{z}}$ | $(24g)$ | As |
| $\mathbf{B}_7$ | = | $(y_2 - z_2)\,\mathbf{a}_1 - z_2\,\mathbf{a}_2 + y_2\,\mathbf{a}_3$ | = | $y_2\,a\,\hat{\mathbf{y}} - z_2\,a\,\hat{\mathbf{z}}$ | $(24g)$ | As |
| $\mathbf{B}_8$ | = | $-(y_2 + z_2)\,\mathbf{a}_1 - z_2\,\mathbf{a}_2 - y_2\,\mathbf{a}_3$ | = | $-y_2\,a\,\hat{\mathbf{y}} - z_2\,a\,\hat{\mathbf{z}}$ | $(24g)$ | As |
| $\mathbf{B}_9$ | = | $y_2\,\mathbf{a}_1 + (y_2 + z_2)\,\mathbf{a}_2 + z_2\,\mathbf{a}_3$ | = | $z_2\,a\,\hat{\mathbf{x}} + y_2\,a\,\hat{\mathbf{z}}$ | $(24g)$ | As |
| $\mathbf{B}_{10}$ | = | $-y_2\,\mathbf{a}_1 + (z_2 - y_2)\,\mathbf{a}_2 + z_2\,\mathbf{a}_3$ | = | $z_2\,a\,\hat{\mathbf{x}} - y_2\,a\,\hat{\mathbf{z}}$ | $(24g)$ | As |
| $\mathbf{B}_{11}$ | = | $y_2\,\mathbf{a}_1 + (y_2 - z_2)\,\mathbf{a}_2 - z_2\,\mathbf{a}_3$ | = | $-z_2\,a\,\hat{\mathbf{x}} + y_2\,a\,\hat{\mathbf{z}}$ | $(24g)$ | As |
| $\mathbf{B}_{12}$ | = | $-y_2\,\mathbf{a}_1 - (y_2 + z_2)\,\mathbf{a}_2 - z_2\,\mathbf{a}_3$ | = | $-z_2\,a\,\hat{\mathbf{x}} - y_2\,a\,\hat{\mathbf{z}}$ | $(24g)$ | As |
| $\mathbf{B}_{13}$ | = | $z_2\,\mathbf{a}_1 + y_2\,\mathbf{a}_2 + (y_2 + z_2)\,\mathbf{a}_3$ | = | $y_2\,a\,\hat{\mathbf{x}} + z_2\,a\,\hat{\mathbf{y}}$ | $(24g)$ | As |
| $\mathbf{B}_{14}$ | = | $z_2\,\mathbf{a}_1 - y_2\,\mathbf{a}_2 + (z_2 - y_2)\,\mathbf{a}_3$ | = | $-y_2\,a\,\hat{\mathbf{x}} + z_2\,a\,\hat{\mathbf{y}}$ | $(24g)$ | As |
| $\mathbf{B}_{15}$ | = | $-z_2\,\mathbf{a}_1 + y_2\,\mathbf{a}_2 + (y_2 - z_2)\,\mathbf{a}_3$ | = | $y_2\,a\,\hat{\mathbf{x}} - z_2\,a\,\hat{\mathbf{y}}$ | $(24g)$ | As |
| $\mathbf{B}_{16}$ | = | $-z_2\,\mathbf{a}_1 - y_2\,\mathbf{a}_2 - (y_2 + z_2)\,\mathbf{a}_3$ | = | $-y_2\,a\,\hat{\mathbf{x}} - z_2\,a\,\hat{\mathbf{y}}$ | $(24g)$ | As |

**Geometry files:**

- CIF: pp. 758
- POSCAR: pp. 759

# $Al_{12}W$ Structure: A12B_cI26_204_g_a

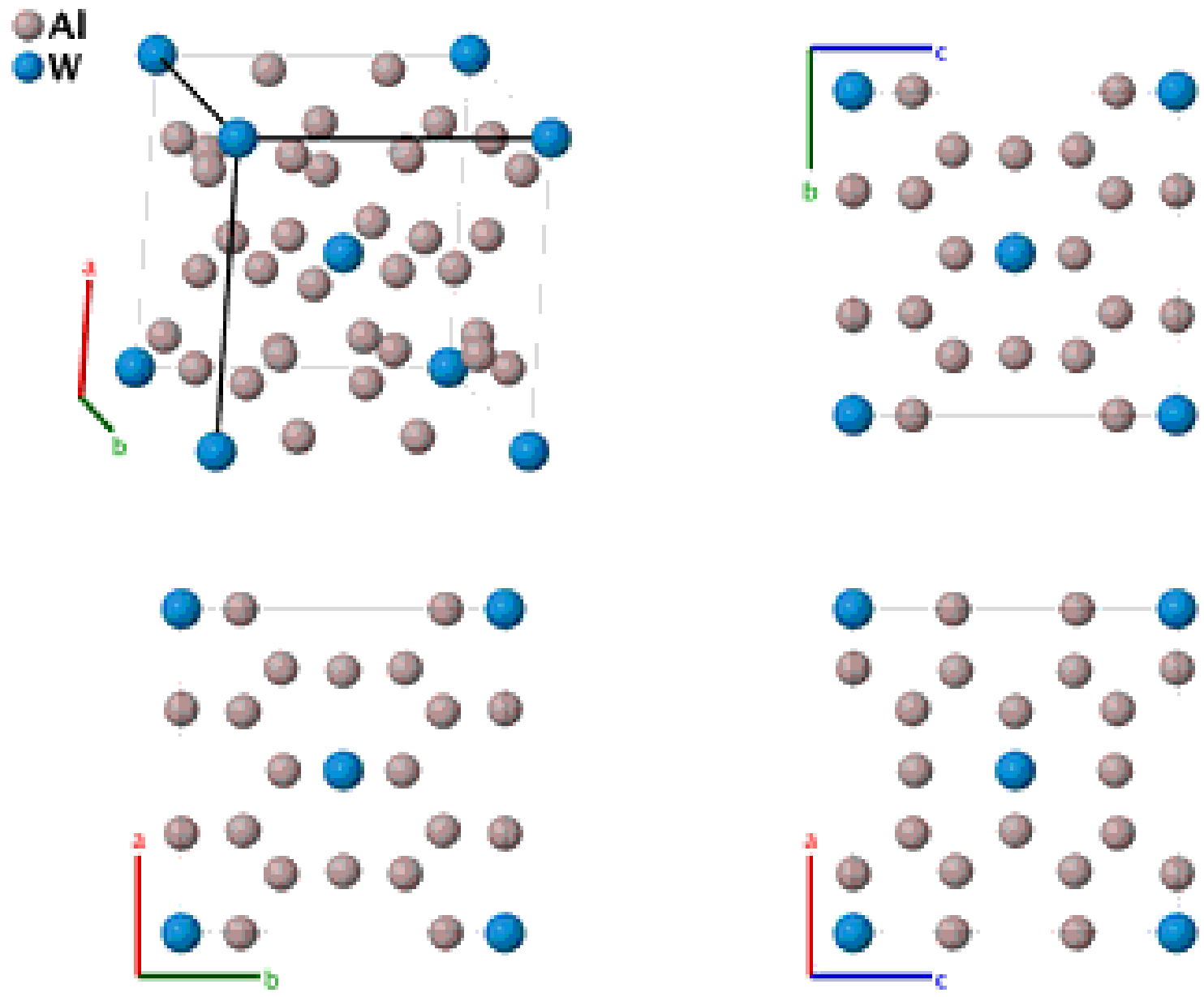

| | | |
|---|---|---|
| **Prototype** | : | $Al_{12}W$ |
| **AFLOW prototype label** | : | A12B_cI26_204_g_a |
| ***Strukturbericht* designation** | : | None |
| **Pearson symbol** | : | cI26 |
| **Space group number** | : | 204 |
| **Space group symbol** | : | $Im\bar{3}$ |
| **AFLOW prototype command** | : | `aflow --proto=A12B_cI26_204_g_a --params=`$a, y_2, z_2$ |

**Body-centered Cubic primitive vectors:**

$$\mathbf{a}_1 = -\frac{1}{2}\,a\,\hat{\mathbf{x}} + \frac{1}{2}\,a\,\hat{\mathbf{y}} + \frac{1}{2}\,a\,\hat{\mathbf{z}}$$
$$\mathbf{a}_2 = \frac{1}{2}\,a\,\hat{\mathbf{x}} - \frac{1}{2}\,a\,\hat{\mathbf{y}} + \frac{1}{2}\,a\,\hat{\mathbf{z}}$$
$$\mathbf{a}_3 = \frac{1}{2}\,a\,\hat{\mathbf{x}} + \frac{1}{2}\,a\,\hat{\mathbf{y}} - \frac{1}{2}\,a\,\hat{\mathbf{z}}$$

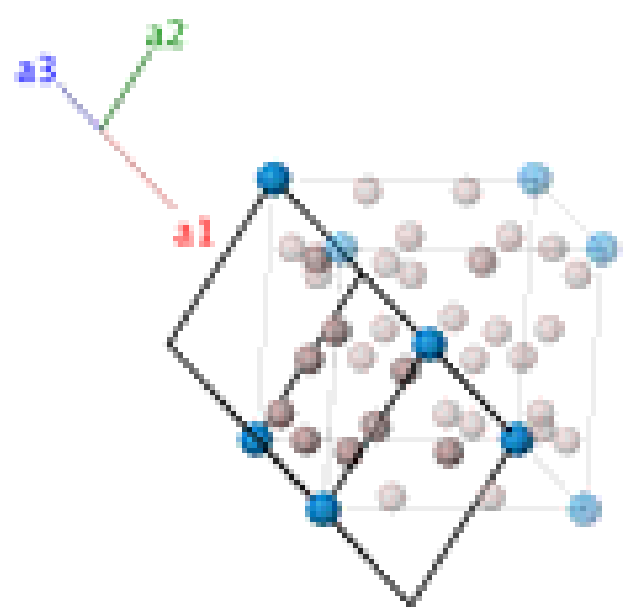

**Basis vectors:**

| | | Lattice Coordinates | | Cartesian Coordinates | Wyckoff Position | Atom Type |
|---|---|---|---|---|---|---|
| $\mathbf{B}_1$ | = | $0\,\mathbf{a}_1 + 0\,\mathbf{a}_2 + 0\,\mathbf{a}_3$ | = | $0\hat{\mathbf{x}} + 0\hat{\mathbf{y}} + 0\hat{\mathbf{z}}$ | $(2a)$ | W |
| $\mathbf{B}_2$ | = | $(y_2 + z_2)\,\mathbf{a}_1 + z_2\,\mathbf{a}_2 + y_2\,\mathbf{a}_3$ | = | $y_2\,a\,\hat{\mathbf{y}} + z_2\,a\,\hat{\mathbf{z}}$ | $(24g)$ | Al |
| $\mathbf{B}_3$ | = | $(z_2 - y_2)\,\mathbf{a}_1 + z_2\,\mathbf{a}_2 - y_2\,\mathbf{a}_3$ | = | $-y_2\,a\,\hat{\mathbf{y}} + z_2\,a\,\hat{\mathbf{z}}$ | $(24g)$ | Al |
| $\mathbf{B}_4$ | = | $(y_2 - z_2)\,\mathbf{a}_1 - z_2\,\mathbf{a}_2 + y_2\,\mathbf{a}_3$ | = | $y_2\,a\,\hat{\mathbf{y}} - z_2\,a\,\hat{\mathbf{z}}$ | $(24g)$ | Al |
| $\mathbf{B}_5$ | = | $-(y_2 + z_2)\,\mathbf{a}_1 - z_2\,\mathbf{a}_2 - y_2\,\mathbf{a}_3$ | = | $-y_2\,a\,\hat{\mathbf{y}} - z_2\,a\,\hat{\mathbf{z}}$ | $(24g)$ | Al |
| $\mathbf{B_6}$ | = | $y_2 \, \mathbf{a_1} + (y_2 + z_2) \, \mathbf{a_2} + z_2 \, \mathbf{a_3}$ | = | $z_2 \, a \, \mathbf{\hat{x}} + y_2 \, a \, \mathbf{\hat{z}}$ | (24$g$) | Al |
| $\mathbf{B_7}$ | = | $-y_2 \, \mathbf{a_1} + (z_2 - y_2) \, \mathbf{a_2} + z_2 \, \mathbf{a_3}$ | = | $z_2 \, a \, \mathbf{\hat{x}} - y_2 \, a \, \mathbf{\hat{z}}$ | (24$g$) | Al |
| $\mathbf{B_8}$ | = | $y_2 \, \mathbf{a_1} + (y_2 - z_2) \, \mathbf{a_2} - z_2 \, \mathbf{a_3}$ | = | $-z_2 \, a \, \mathbf{\hat{x}} + y_2 \, a \, \mathbf{\hat{z}}$ | (24$g$) | Al |
| $\mathbf{B_9}$ | = | $-y_2 \, \mathbf{a_1} - (y_2 + z_2) \, \mathbf{a_2} - z_2 \, \mathbf{a_3}$ | = | $-z_2 \, a \, \mathbf{\hat{x}} - y_2 \, a \, \mathbf{\hat{z}}$ | (24$g$) | Al |
| $\mathbf{B_{10}}$ | = | $z_2 \, \mathbf{a_1} + y_2 \, \mathbf{a_2} + (y_2 + z_2) \, \mathbf{a_3}$ | = | $y_2 \, a \, \mathbf{\hat{x}} + z_2 \, a \, \mathbf{\hat{y}}$ | (24$g$) | Al |
| $\mathbf{B_{11}}$ | = | $z_2 \, \mathbf{a_1} - y_2 \, \mathbf{a_2} + (z_2 - y_2) \, \mathbf{a_3}$ | = | $-y_2 \, a \, \mathbf{\hat{x}} + z_2 \, a \, \mathbf{\hat{y}}$ | (24$g$) | Al |
| $\mathbf{B_{12}}$ | = | $-z_2 \, \mathbf{a_1} + y_2 \, \mathbf{a_2} + (y_2 - z_2) \, \mathbf{a_3}$ | = | $y_2 \, a \, \mathbf{\hat{x}} - z_2 \, a \, \mathbf{\hat{y}}$ | (24$g$) | Al |
| $\mathbf{B_{13}}$ | = | $-z_2 \, \mathbf{a_1} - y_2 \, \mathbf{a_2} - (y_2 + z_2) \, \mathbf{a_3}$ | = | $-y_2 \, a \, \mathbf{\hat{x}} - z_2 \, a \, \mathbf{\hat{y}}$ | (24$g$) | Al |

**Geometry files:**

- CIF: pp. 759
- POSCAR: pp. 759

# $\alpha$-N (Pa$\bar{3}$) Structure: A_cP8_205_c

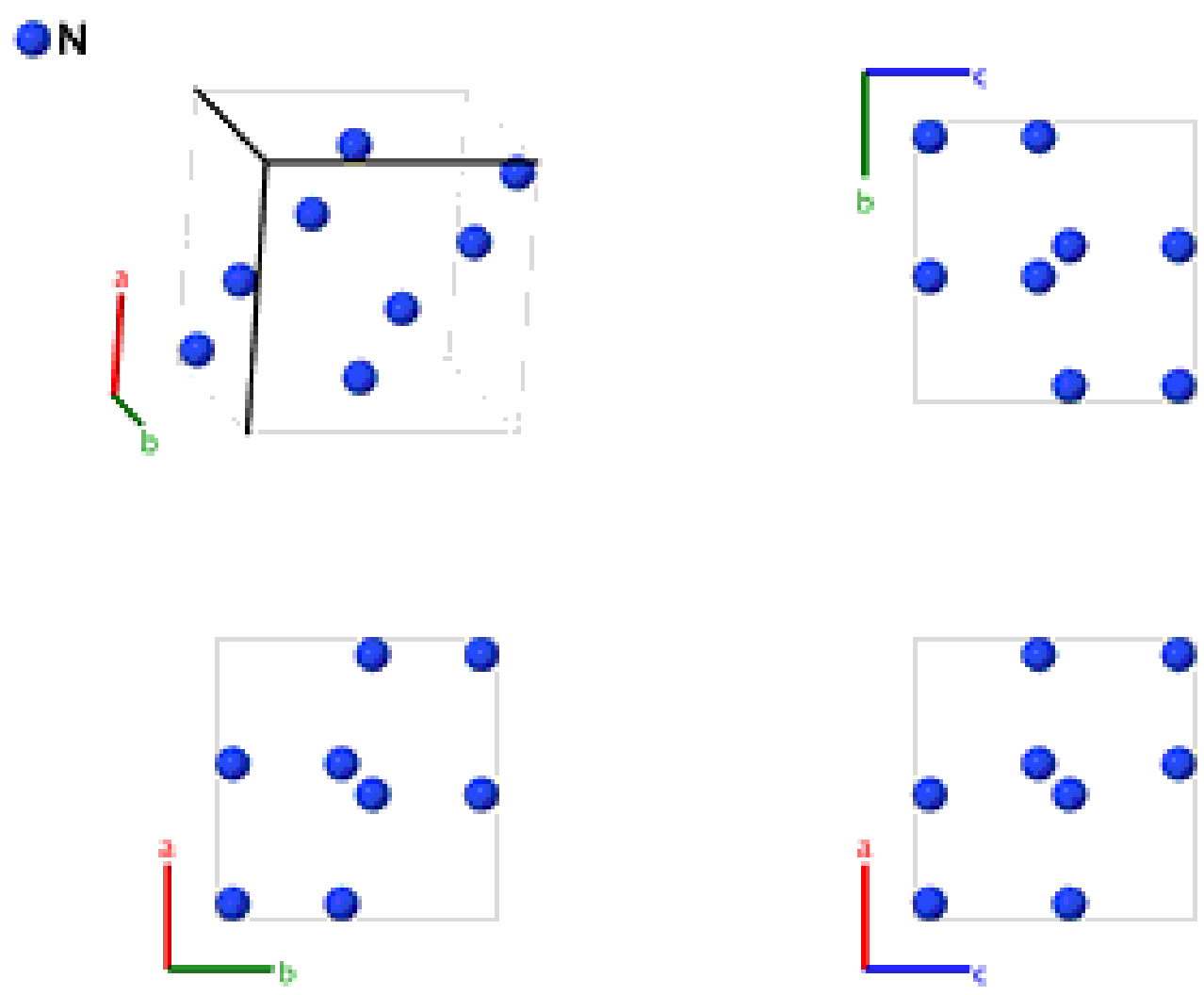

**Prototype** : $\alpha$-N

**AFLOW prototype label** : A_cP8_205_c

***Strukturbericht* designation** : None

**Pearson symbol** : cP8

**Space group number** : 205

**Space group symbol** : Pa$\bar{3}$

**AFLOW prototype command** : `aflow --proto=A_cP8_205_c --params=`$a, x_1$

---

- There is considerable controversy about the crystal structure of $\alpha$-N, as outlined in (Donohue, 1982) pp. 280-285. This page assumes the centrosymmetric Pa$\bar{3}$ structure. The other possibility is the P$2_1$3 structure, where the $N_2$ dimers are not centered on an inversion site. (Venables, 1974) makes a convincing case that the ground state is Pa$\bar{3}$, but we present both structures. Density Functional Theory calculations show no appreciable difference in energy between the Pa$\bar{3}$ and P$2_1$3 structures. (Mehl, 2015)

---

**Simple Cubic primitive vectors:**

$$\mathbf{a}_1 = a\,\hat{\mathbf{x}}$$
$$\mathbf{a}_2 = a\,\hat{\mathbf{y}}$$
$$\mathbf{a}_3 = a\,\hat{\mathbf{z}}$$

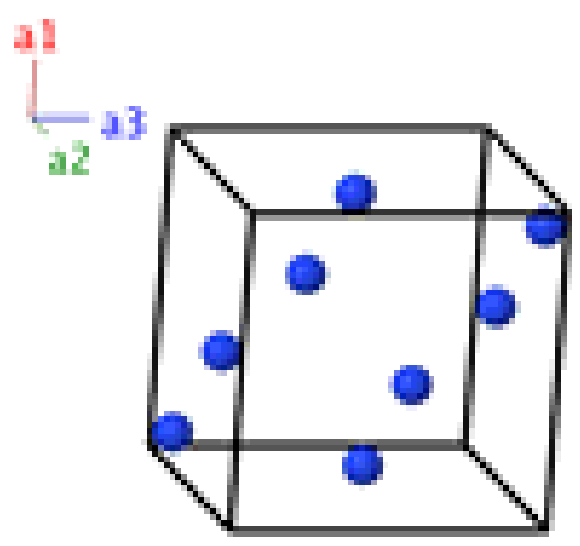

---

**Basis vectors:**

| | | Lattice Coordinates | | Cartesian Coordinates | Wyckoff Position | Atom Type |
|---|---|---|---|---|---|---|
| $\mathbf{B_1}$ | = | $x_1\,\mathbf{a_1} + x_1\,\mathbf{a_2} + x_1\,\mathbf{a_3}$ | = | $x_1\,a\,\hat{\mathbf{x}} + x_1\,a\,\hat{\mathbf{y}} + x_1\,a\,\hat{\mathbf{z}}$ | (8$c$) | N |
| $\mathbf{B_2}$ | = | $\left(\frac{1}{2} - x_1\right)\mathbf{a_1} - x_1\,\mathbf{a_2} + \left(\frac{1}{2} + x_1\right)\mathbf{a_3}$ | = | $\left(\frac{1}{2} - x_1\right)a\,\hat{\mathbf{x}} - x_1\,a\,\hat{\mathbf{y}} + \left(\frac{1}{2} + x_1\right)a\,\hat{\mathbf{z}}$ | (8$c$) | N |
| $\mathbf{B_3}$ | = | $-x_1\,\mathbf{a_1}+\left(\frac{1}{2}+x_1\right)\mathbf{a_2}+\left(\frac{1}{2}-x_1\right)\mathbf{a_3}$ | = | $-x_1\,a\,\mathbf{\hat{x}}+\left(\frac{1}{2}+x_1\right)a\,\mathbf{\hat{y}}+\left(\frac{1}{2}-x_1\right)a\,\mathbf{\hat{z}}$ | (8$c$) | N |
| $\mathbf{B_4}$ | = | $\left(\frac{1}{2}+x_1\right)\mathbf{a_1}+\left(\frac{1}{2}-x_1\right)\mathbf{a_2}-x_1\,\mathbf{a_3}$ | = | $\left(\frac{1}{2}+x_1\right)a\,\mathbf{\hat{x}}+\left(\frac{1}{2}-x_1\right)a\,\mathbf{\hat{y}}-x_1\,a\,\mathbf{\hat{z}}$ | (8$c$) | N |
| $\mathbf{B_5}$ | = | $-x_1\,\mathbf{a_1}-x_1\,\mathbf{a_2}-x_1\,\mathbf{a_3}$ | = | $-x_1\,a\,\mathbf{\hat{x}}-x_1\,a\,\mathbf{\hat{y}}-x_1\,a\,\mathbf{\hat{z}}$ | (8$c$) | N |
| $\mathbf{B_6}$ | = | $\left(\frac{1}{2}+x_1\right)\mathbf{a_1}+x_1\,\mathbf{a_2}+\left(\frac{1}{2}-x_1\right)\mathbf{a_3}$ | = | $\left(\frac{1}{2}+x_1\right)a\,\mathbf{\hat{x}}+x_1\,a\,\mathbf{\hat{y}}+\left(\frac{1}{2}-x_1\right)a\,\mathbf{\hat{z}}$ | (8$c$) | N |
| $\mathbf{B_7}$ | = | $x_1\,\mathbf{a_1}+\left(\frac{1}{2}-x_1\right)\mathbf{a_2}+\left(\frac{1}{2}+x_1\right)\mathbf{a_3}$ | = | $x_1\,a\,\mathbf{\hat{x}}+\left(\frac{1}{2}-x_1\right)a\,\mathbf{\hat{y}}+\left(\frac{1}{2}+x_1\right)a\,\mathbf{\hat{z}}$ | (8$c$) | N |
| $\mathbf{B_8}$ | = | $\left(\frac{1}{2}-x_1\right)\mathbf{a_1}+\left(\frac{1}{2}+x_1\right)\mathbf{a_2}+x_1\,\mathbf{a_3}$ | = | $\left(\frac{1}{2}-x_1\right)a\,\mathbf{\hat{x}}+\left(\frac{1}{2}+x_1\right)a\,\mathbf{\hat{y}}+x_1\,a\,\mathbf{\hat{z}}$ | (8$c$) | N |

**Found in:**

**Geometry files:**

- CIF: pp. 760
- POSCAR: pp. 760

# SC16 (CuCl) Structure: AB_cP16_205_c_c

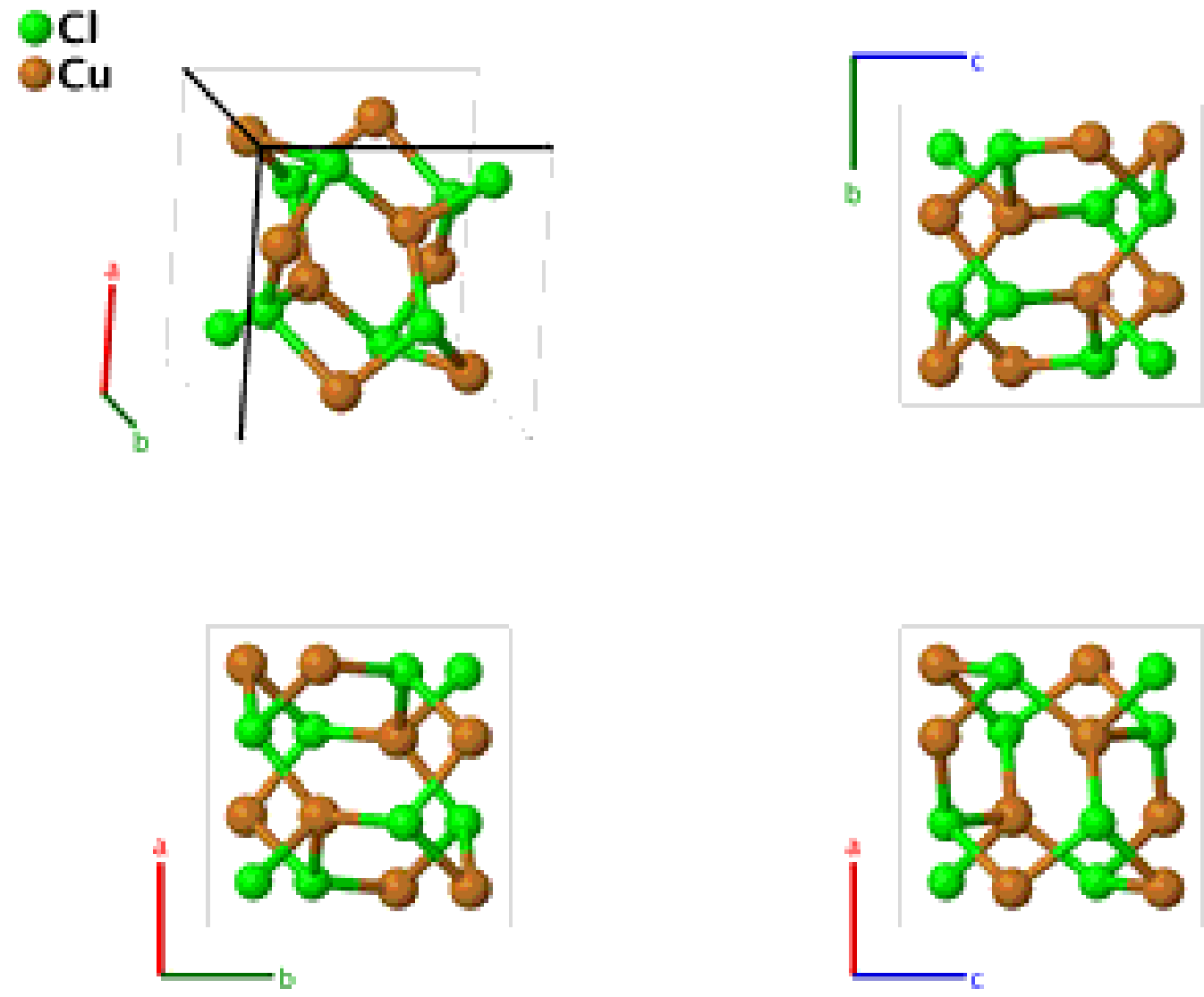

| | | |
|---|---|---|
| **Prototype** | : | CuCl |
| **AFLOW prototype label** | : | AB_cP16_205_c_c |
| ***Strukturbericht* designation** | : | None |
| **Pearson symbol** | : | cP16 |
| **Space group number** | : | 205 |
| **Space group symbol** | : | Pa$\bar{3}$ |
| **AFLOW prototype command** | : | `aflow --proto=AB_cP16_205_c_c --params=`$a, x_1, x_2$ |

- This is a tetragonally bonded structure which packs more efficiently than diamond. This structure is related to BC8 in the same way that zincblende (B3) is related to diamond (A4): we replace half of the atoms by another species, such that the four nearest neighbors of each atom are of the other species. See (Crain, 1995) and references therein. The reference compound chosen here, found in (Hull, 1994), is stable at about 5 GPa.

**Simple Cubic primitive vectors:**

$$\mathbf{a}_1 = a\,\mathbf{\hat{x}}$$
$$\mathbf{a}_2 = a\,\mathbf{\hat{y}}$$
$$\mathbf{a}_3 = a\,\mathbf{\hat{z}}$$

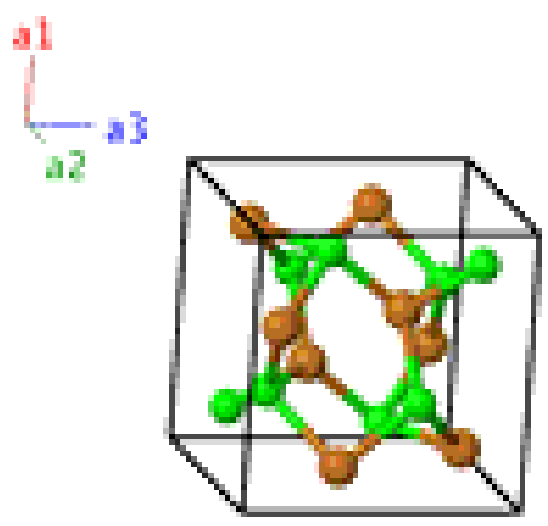

**Basis vectors:**

| | | Lattice Coordinates | | Cartesian Coordinates | Wyckoff Position | Atom Type |
|---|---|---|---|---|---|---|
| $\mathbf{B}_1$ | = | $x_1\,\mathbf{a}_1 + x_1\,\mathbf{a}_2 + x_1\,\mathbf{a}_3$ | = | $x_1\,a\,\mathbf{\hat{x}} + x_1\,a\,\mathbf{\hat{y}} + x_1\,a\,\mathbf{\hat{z}}$ | (8$c$) | Cl |
| $\mathbf{B}_2$ | = | $\left(\frac{1}{2} - x_1\right)\mathbf{a}_1 - x_1\,\mathbf{a}_2 + \left(\frac{1}{2} + x_1\right)\mathbf{a}_3$ | = | $\left(\frac{1}{2} - x_1\right)a\,\mathbf{\hat{x}} - x_1\,a\,\mathbf{\hat{y}} + \left(\frac{1}{2} + x_1\right)a\,\mathbf{\hat{z}}$ | (8$c$) | Cl |
| $\mathbf{B_3}$ | = | $-x_1\,\mathbf{a_1}+\left(\frac{1}{2}+x_1\right)\,\mathbf{a_2}+\left(\frac{1}{2}-x_1\right)\,\mathbf{a_3}$ | = | $-x_1\,a\,\hat{\mathbf{x}}+\left(\frac{1}{2}+x_1\right)\,a\,\hat{\mathbf{y}}+\left(\frac{1}{2}-x_1\right)\,a\,\hat{\mathbf{z}}$ | (8$c$) | Cl |
| $\mathbf{B_4}$ | = | $\left(\frac{1}{2}+x_1\right)\,\mathbf{a_1}+\left(\frac{1}{2}-x_1\right)\,\mathbf{a_2}-x_1\,\mathbf{a_3}$ | = | $\left(\frac{1}{2}+x_1\right)\,a\,\hat{\mathbf{x}}+\left(\frac{1}{2}-x_1\right)\,a\,\hat{\mathbf{y}}-x_1\,a\,\hat{\mathbf{z}}$ | (8$c$) | Cl |
| $\mathbf{B_5}$ | = | $-x_1\,\mathbf{a_1}-x_1\,\mathbf{a_2}-x_1\,\mathbf{a_3}$ | = | $-x_1\,a\,\hat{\mathbf{x}}-x_1\,a\,\hat{\mathbf{y}}-x_1\,a\,\hat{\mathbf{z}}$ | (8$c$) | Cl |
| $\mathbf{B_6}$ | = | $\left(\frac{1}{2}+x_1\right)\,\mathbf{a_1}+x_1\,\mathbf{a_2}+\left(\frac{1}{2}-x_1\right)\,\mathbf{a_3}$ | = | $\left(\frac{1}{2}+x_1\right)\,a\,\hat{\mathbf{x}}+x_1\,a\,\hat{\mathbf{y}}+\left(\frac{1}{2}-x_1\right)\,a\,\hat{\mathbf{z}}$ | (8$c$) | Cl |
| $\mathbf{B_7}$ | = | $x_1\,\mathbf{a_1}+\left(\frac{1}{2}-x_1\right)\,\mathbf{a_2}+\left(\frac{1}{2}+x_1\right)\,\mathbf{a_3}$ | = | $x_1\,a\,\hat{\mathbf{x}}+\left(\frac{1}{2}-x_1\right)\,a\,\hat{\mathbf{y}}+\left(\frac{1}{2}+x_1\right)\,a\,\hat{\mathbf{z}}$ | (8$c$) | Cl |
| $\mathbf{B_8}$ | = | $\left(\frac{1}{2}-x_1\right)\,\mathbf{a_1}+\left(\frac{1}{2}+x_1\right)\,\mathbf{a_2}+x_1\,\mathbf{a_3}$ | = | $\left(\frac{1}{2}-x_1\right)\,a\,\hat{\mathbf{x}}+\left(\frac{1}{2}+x_1\right)\,a\,\hat{\mathbf{y}}+x_1\,a\,\hat{\mathbf{z}}$ | (8$c$) | Cl |
| $\mathbf{B_9}$ | = | $x_2\,\mathbf{a_1}+x_2\,\mathbf{a_2}+x_2\,\mathbf{a_3}$ | = | $x_2\,a\,\hat{\mathbf{x}}+x_2\,a\,\hat{\mathbf{y}}+x_2\,a\,\hat{\mathbf{z}}$ | (8$c$) | Cu |
| $\mathbf{B_{10}}$ | = | $\left(\frac{1}{2}-x_2\right)\,\mathbf{a_1}-x_2\,\mathbf{a_2}+\left(\frac{1}{2}+x_2\right)\,\mathbf{a_3}$ | = | $\left(\frac{1}{2}-x_2\right)\,a\,\hat{\mathbf{x}}-x_2\,a\,\hat{\mathbf{y}}+\left(\frac{1}{2}+x_2\right)\,a\,\hat{\mathbf{z}}$ | (8$c$) | Cu |
| $\mathbf{B_{11}}$ | = | $-x_2\,\mathbf{a_1}+\left(\frac{1}{2}+x_2\right)\,\mathbf{a_2}+\left(\frac{1}{2}-x_2\right)\,\mathbf{a_3}$ | = | $-x_2\,a\,\hat{\mathbf{x}}+\left(\frac{1}{2}+x_2\right)\,a\,\hat{\mathbf{y}}+\left(\frac{1}{2}-x_2\right)\,a\,\hat{\mathbf{z}}$ | (8$c$) | Cu |
| $\mathbf{B_{12}}$ | = | $\left(\frac{1}{2}+x_2\right)\,\mathbf{a_1}+\left(\frac{1}{2}-x_2\right)\,\mathbf{a_2}-x_2\,\mathbf{a_3}$ | = | $\left(\frac{1}{2}+x_2\right)\,a\,\hat{\mathbf{x}}+\left(\frac{1}{2}-x_2\right)\,a\,\hat{\mathbf{y}}-x_2\,a\,\hat{\mathbf{z}}$ | (8$c$) | Cu |
| $\mathbf{B_{13}}$ | = | $-x_2\,\mathbf{a_1}-x_2\,\mathbf{a_2}-x_2\,\mathbf{a_3}$ | = | $-x_2\,a\,\hat{\mathbf{x}}-x_2\,a\,\hat{\mathbf{y}}-x_2\,a\,\hat{\mathbf{z}}$ | (8$c$) | Cu |
| $\mathbf{B_{14}}$ | = | $\left(\frac{1}{2}+x_2\right)\,\mathbf{a_1}+x_2\,\mathbf{a_2}+\left(\frac{1}{2}-x_2\right)\,\mathbf{a_3}$ | = | $\left(\frac{1}{2}+x_2\right)\,a\,\hat{\mathbf{x}}+x_2\,a\,\hat{\mathbf{y}}+\left(\frac{1}{2}-x_2\right)\,a\,\hat{\mathbf{z}}$ | (8$c$) | Cu |
| $\mathbf{B_{15}}$ | = | $x_2\,\mathbf{a_1}+\left(\frac{1}{2}-x_2\right)\,\mathbf{a_2}+\left(\frac{1}{2}+x_2\right)\,\mathbf{a_3}$ | = | $x_2\,a\,\hat{\mathbf{x}}+\left(\frac{1}{2}-x_2\right)\,a\,\hat{\mathbf{y}}+\left(\frac{1}{2}+x_2\right)\,a\,\hat{\mathbf{z}}$ | (8$c$) | Cu |
| $\mathbf{B_{16}}$ | = | $\left(\frac{1}{2}-x_2\right)\,\mathbf{a_1}+\left(\frac{1}{2}+x_2\right)\,\mathbf{a_2}+x_2\,\mathbf{a_3}$ | = | $\left(\frac{1}{2}-x_2\right)\,a\,\hat{\mathbf{x}}+\left(\frac{1}{2}+x_2\right)\,a\,\hat{\mathbf{y}}+x_2\,a\,\hat{\mathbf{z}}$ | (8$c$) | Cu |

---

---

**Geometry files:**

- CIF: pp. 760
- POSCAR: pp. 761

# Pyrite ($FeS_2$, C2) Structure: AB2_cP12_205_a_c

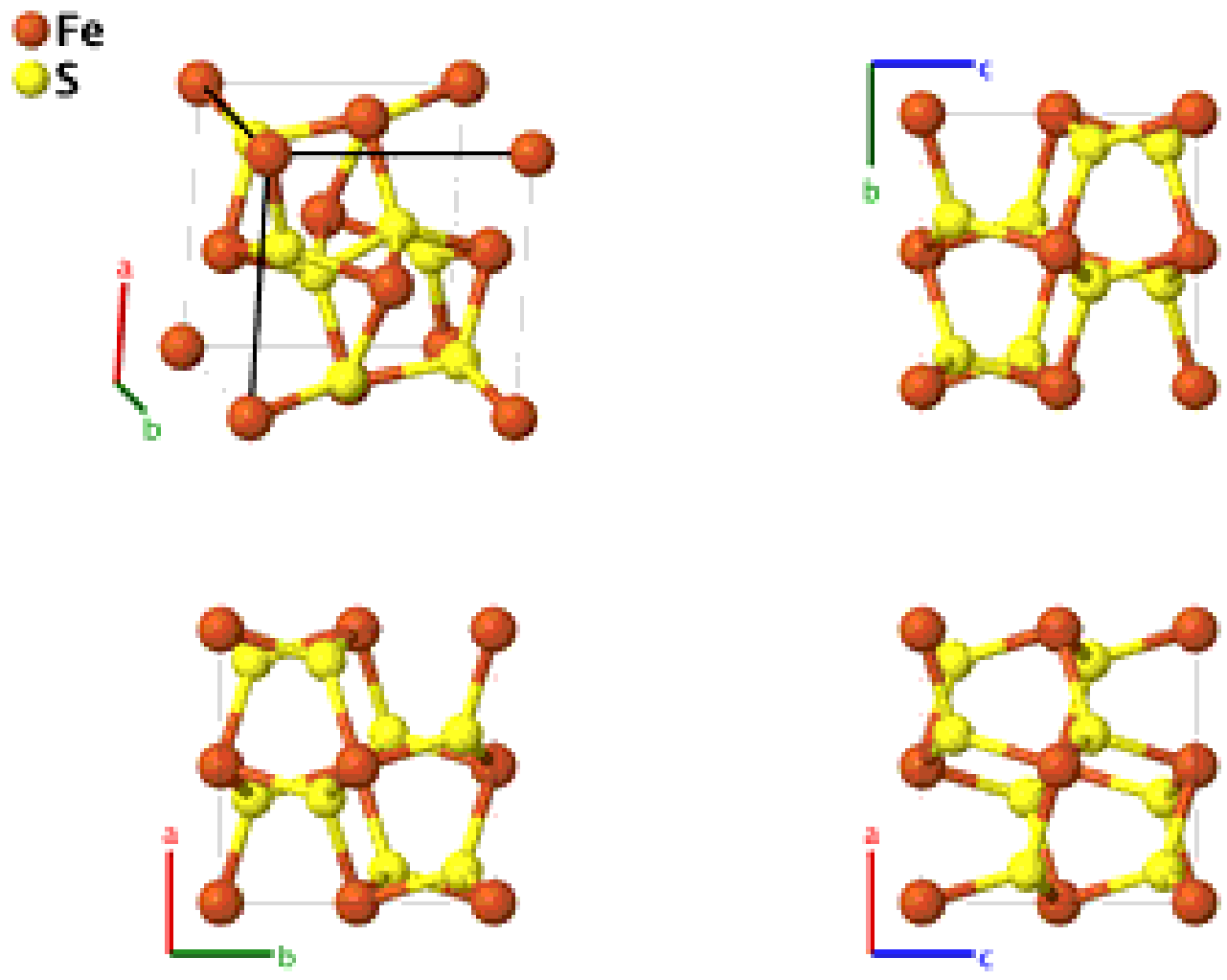

| | | |
|---|---|---|
| **Prototype** | : | $FeS_2$ |
| **AFLOW prototype label** | : | AB2_cP12_205_a_c |
| ***Strukturbericht* designation** | : | C2 |
| **Pearson symbol** | : | cP12 |
| **Space group number** | : | 205 |
| **Space group symbol** | : | $Pa\bar{3}$ |
| **AFLOW prototype command** | : | `aflow --proto=AB2_cP12_205_a_c --params=`$a, x_2$ |

**Other compounds with this structure:**

- $AuSb_2$, $CaC_2$, $CoS_2$, $MnS_2$, $NiS_2$, $NiSe_2$, $OsS_2$, $OsTe_2$, $PdAs_2$, $PtAs_2$, $PtBi_2$, $RhSe_2$, $RuS_2$

- (Bayliss, 1997) gives crystalline data for "weakly anisotropic pyrite" which we have tabulated as P1 $FeS_2$. He also gives crystallographic data for the cubic pyrite structure, which we report here. Also see the C18 (marcasite) $FeS_2$ structure.

**Simple Cubic primitive vectors:**

$$\mathbf{a}_1 = a\,\hat{\mathbf{x}}$$
$$\mathbf{a}_2 = a\,\hat{\mathbf{y}}$$
$$\mathbf{a}_3 = a\,\hat{\mathbf{z}}$$

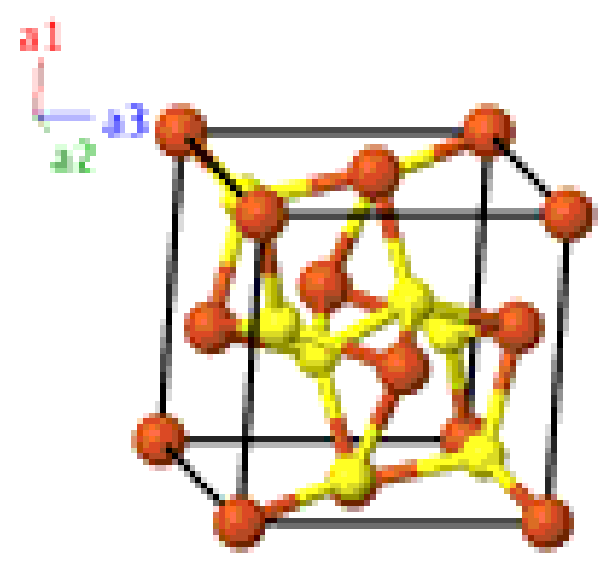

**Basis vectors:**

| | | Lattice Coordinates | | Cartesian Coordinates | Wyckoff Position | Atom Type |
|---|---|---|---|---|---|---|
| $\mathbf{B}_1$ | = | $0\,\mathbf{a}_1 + 0\,\mathbf{a}_2 + 0\,\mathbf{a}_3$ | = | $0\hat{\mathbf{x}} + 0\hat{\mathbf{y}} + 0\hat{\mathbf{z}}$ | (4$a$) | Fe |
| $\mathbf{B}_2$ | = | $\frac{1}{2}\,\mathbf{a}_1 + \frac{1}{2}\,\mathbf{a}_3$ | = | $\frac{1}{2}\,a\,\hat{\mathbf{x}} + \frac{1}{2}\,a\,\hat{\mathbf{z}}$ | (4$a$) | Fe |
| $\mathbf{B}_3$ | = | $\frac{1}{2}\,\mathbf{a}_2 + \frac{1}{2}\,\mathbf{a}_3$ | = | $\frac{1}{2}\,a\,\hat{\mathbf{y}} + \frac{1}{2}\,a\,\hat{\mathbf{z}}$ | (4$a$) | Fe |
| $\mathbf{B}_4$ | = | $\frac{1}{2}\,\mathbf{a}_1 + \frac{1}{2}\,\mathbf{a}_2$ | = | $\frac{1}{2}\,a\,\hat{\mathbf{x}} + \frac{1}{2}\,a\,\hat{\mathbf{y}}$ | (4$a$) | Fe |
| $\mathbf{B}_5$ | = | $x_2\,\mathbf{a}_1 + x_2\,\mathbf{a}_2 + x_2\,\mathbf{a}_3$ | = | $x_2\,a\,\hat{\mathbf{x}} + x_2\,a\,\hat{\mathbf{y}} + x_2\,a\,\hat{\mathbf{z}}$ | (8$c$) | S |
| $\mathbf{B}_6$ | = | $\left(\frac{1}{2} - x_2\right)\,\mathbf{a}_1 - x_2\,\mathbf{a}_2 + \left(\frac{1}{2} + x_2\right)\,\mathbf{a}_3$ | = | $\left(\frac{1}{2} - x_2\right)\,a\,\hat{\mathbf{x}} - x_2\,a\,\hat{\mathbf{y}} + \left(\frac{1}{2} + x_2\right)\,a\,\hat{\mathbf{z}}$ | (8$c$) | S |
| $\mathbf{B}_7$ | = | $-x_2\,\mathbf{a}_1 + \left(\frac{1}{2} + x_2\right)\,\mathbf{a}_2 + \left(\frac{1}{2} - x_2\right)\,\mathbf{a}_3$ | = | $-x_2\,a\,\hat{\mathbf{x}} + \left(\frac{1}{2} + x_2\right)\,a\,\hat{\mathbf{y}} + \left(\frac{1}{2} - x_2\right)\,a\,\hat{\mathbf{z}}$ | (8$c$) | S |
| $\mathbf{B}_8$ | = | $\left(\frac{1}{2} + x_2\right)\,\mathbf{a}_1 + \left(\frac{1}{2} - x_2\right)\,\mathbf{a}_2 - x_2\,\mathbf{a}_3$ | = | $\left(\frac{1}{2} + x_2\right)\,a\,\hat{\mathbf{x}} + \left(\frac{1}{2} - x_2\right)\,a\,\hat{\mathbf{y}} - x_2\,a\,\hat{\mathbf{z}}$ | (8$c$) | S |
| $\mathbf{B}_9$ | = | $-x_2\,\mathbf{a}_1 - x_2\,\mathbf{a}_2 - x_2\,\mathbf{a}_3$ | = | $-x_2\,a\,\hat{\mathbf{x}} - x_2\,a\,\hat{\mathbf{y}} - x_2\,a\,\hat{\mathbf{z}}$ | (8$c$) | S |
| $\mathbf{B}_{10}$ | = | $\left(\frac{1}{2} + x_2\right)\,\mathbf{a}_1 + x_2\,\mathbf{a}_2 + \left(\frac{1}{2} - x_2\right)\,\mathbf{a}_3$ | = | $\left(\frac{1}{2} + x_2\right)\,a\,\hat{\mathbf{x}} + x_2\,a\,\hat{\mathbf{y}} + \left(\frac{1}{2} - x_2\right)\,a\,\hat{\mathbf{z}}$ | (8$c$) | S |
| $\mathbf{B}_{11}$ | = | $x_2\,\mathbf{a}_1 + \left(\frac{1}{2} - x_2\right)\,\mathbf{a}_2 + \left(\frac{1}{2} + x_2\right)\,\mathbf{a}_3$ | = | $x_2\,a\,\hat{\mathbf{x}} + \left(\frac{1}{2} - x_2\right)\,a\,\hat{\mathbf{y}} + \left(\frac{1}{2} + x_2\right)\,a\,\hat{\mathbf{z}}$ | (8$c$) | S |
| $\mathbf{B}_{12}$ | = | $\left(\frac{1}{2} - x_2\right)\,\mathbf{a}_1 + \left(\frac{1}{2} + x_2\right)\,\mathbf{a}_2 + x_2\,\mathbf{a}_3$ | = | $\left(\frac{1}{2} - x_2\right)\,a\,\hat{\mathbf{x}} + \left(\frac{1}{2} + x_2\right)\,a\,\hat{\mathbf{y}} + x_2\,a\,\hat{\mathbf{z}}$ | (8$c$) | S |

---

**Found in:**

---

**Geometry files:**

- CIF: pp. 761
- POSCAR: pp. 761

# Bixbyite ($Mn_2O_3$, $D5_3$) Structure: AB3C6_cI80_206_a_d_e

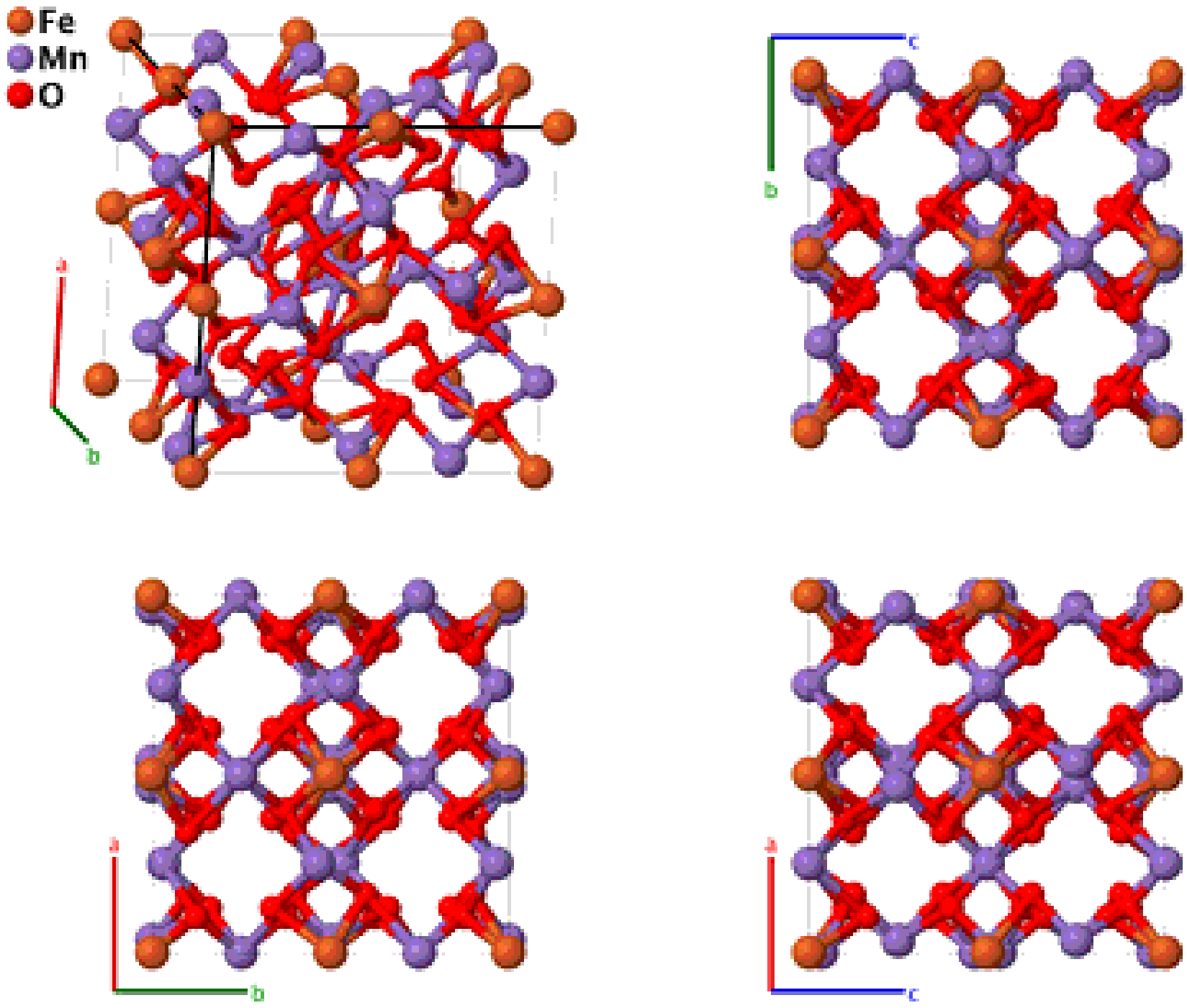

| | | |
|---|---|---|
| **Prototype** | : | $(Mn,Fe)_2O_3$ |
| **AFLOW prototype label** | : | AB3C6_cI80_206_a_d_e |
| ***Strukturbericht* designation** | : | $D5_3$ |
| **Pearson symbol** | : | cI80 |
| **Space group number** | : | 206 |
| **Space group symbol** | : | $Ia\bar{3}$ |
| **AFLOW prototype command** | : | `aflow --proto=AB3C6_cI80_206_a_d_e` `--params=`$a, x_2, x_3, y_3, z_3$ |

**Other compounds with this structure:**

- $Am_2O_3$, $As_2Mg_3$, $As_2Zn_3$, $Cd_3P_2$, $Ce_2O_3$, $Fe_2O_3$, $La_2O_3$, $Lu_2O_3$, $Tb_2O_3$, $Tm_2O_3$, $P_2Zn_3$, many others.

- A search for “bixbyite” on the American Mineralogist Crystal Structure Database (Downs, 2003) shows two structures with the Mn atoms on the (8a) sites and one with Mn on the (8b) site. We use the structure that agrees with the data for pure $Mn_2O_3$ bixbyite in (Villars, 1991) Vol. IV, pp. 4346-7. The referenced data is for $(Mn,Fe)_2O_3$, with Mn and Fe randomly populating the (8a) and (24d) sites. The pictures and the CIF file put Fe atoms on the (8a) sites and Mn atoms on the (24d) sites in order to better delineate the difference in the crystallographic behavior of the sites, but both sites are randomly occupied.

**Body-centered Cubic primitive vectors:**

$$
\begin{aligned}
\mathbf{a}_1 &= -\frac{1}{2}\,a\,\hat{\mathbf{x}} + \frac{1}{2}\,a\,\hat{\mathbf{y}} + \frac{1}{2}\,a\,\hat{\mathbf{z}} \\
\mathbf{a}_2 &= \frac{1}{2}\,a\,\hat{\mathbf{x}} - \frac{1}{2}\,a\,\hat{\mathbf{y}} + \frac{1}{2}\,a\,\hat{\mathbf{z}} \\
\mathbf{a}_3 &= \frac{1}{2}\,a\,\hat{\mathbf{x}} + \frac{1}{2}\,a\,\hat{\mathbf{y}} - \frac{1}{2}\,a\,\hat{\mathbf{z}}
\end{aligned}
$$

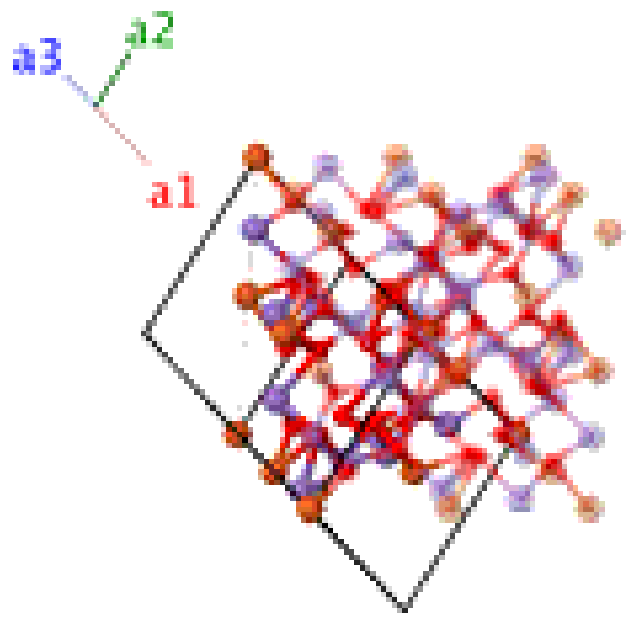

**Basis vectors:**

| | | Lattice Coordinates | | Cartesian Coordinates | Wyckoff Position | Atom Type |
|---|---|---|---|---|---|---|
| $\mathbf{B}_1$ | = | $\frac{1}{2}\,\mathbf{a}_1 + \frac{1}{2}\,\mathbf{a}_2 + \frac{1}{2}\,\mathbf{a}_3$ | = | $\frac{1}{4}\,a\,\hat{\mathbf{x}} + \frac{1}{4}\,a\,\hat{\mathbf{y}} + \frac{1}{4}\,a\,\hat{\mathbf{z}}$ | (8$a$) | Fe |
| $\mathbf{B}_2$ | = | $\frac{1}{2}\,\mathbf{a}_1$ | = | $\frac{3}{4}\,a\,\hat{\mathbf{x}} + \frac{1}{4}\,a\,\hat{\mathbf{y}} + \frac{1}{4}\,a\,\hat{\mathbf{z}}$ | (8$a$) | Fe |
| $\mathbf{B}_3$ | = | $\frac{1}{2}\,\mathbf{a}_2$ | = | $\frac{1}{4}\,a\,\hat{\mathbf{x}} + \frac{3}{4}\,a\,\hat{\mathbf{y}} + \frac{1}{4}\,a\,\hat{\mathbf{z}}$ | (8$a$) | Fe |
| $\mathbf{B}_4$ | = | $\frac{1}{2}\,\mathbf{a}_3$ | = | $\frac{1}{4}\,a\,\hat{\mathbf{x}} + \frac{1}{4}\,a\,\hat{\mathbf{y}} + \frac{3}{4}\,a\,\hat{\mathbf{z}}$ | (8$a$) | Fe |
| $\mathbf{B}_5$ | = | $\frac{1}{4}\,\mathbf{a}_1 + \left(\frac{1}{4} + x_2\right)\,\mathbf{a}_2 + x_2\,\mathbf{a}_3$ | = | $x_2\,a\,\hat{\mathbf{x}} + \frac{1}{4}\,a\,\hat{\mathbf{z}}$ | (24$d$) | Mn |
| $\mathbf{B}_6$ | = | $\frac{3}{4}\,\mathbf{a}_1 + \left(\frac{1}{4} - x_2\right)\,\mathbf{a}_2 + \left(\frac{1}{2} - x_2\right)\,\mathbf{a}_3$ | = | $-x_2\,a\,\hat{\mathbf{x}} + \frac{1}{2}\,a\,\hat{\mathbf{y}} + \frac{1}{4}\,a\,\hat{\mathbf{z}}$ | (24$d$) | Mn |
| $\mathbf{B}_7$ | = | $x_2\,\mathbf{a}_1 + \frac{1}{4}\,\mathbf{a}_2 + \left(\frac{1}{4} + x_2\right)\,\mathbf{a}_3$ | = | $\frac{1}{4}\,a\,\hat{\mathbf{x}} + x_2\,a\,\hat{\mathbf{y}}$ | (24$d$) | Mn |
| $\mathbf{B}_8$ | = | $\left(\frac{1}{2} - x_2\right)\,\mathbf{a}_1 + \frac{3}{4}\,\mathbf{a}_2 + \left(\frac{1}{4} - x_2\right)\,\mathbf{a}_3$ | = | $\frac{1}{4}\,a\,\hat{\mathbf{x}} - x_2\,a\,\hat{\mathbf{y}} + \frac{1}{2}\,a\,\hat{\mathbf{z}}$ | (24$d$) | Mn |
| $\mathbf{B}_9$ | = | $\left(\frac{1}{4} + x_2\right)\,\mathbf{a}_1 + x_2\,\mathbf{a}_2 + \frac{1}{4}\,\mathbf{a}_3$ | = | $\frac{1}{4}\,a\,\hat{\mathbf{y}} + x_2\,a\,\hat{\mathbf{z}}$ | (24$d$) | Mn |
| $\mathbf{B}_{10}$ | = | $\left(\frac{1}{4} - x_2\right)\,\mathbf{a}_1 + \left(\frac{1}{2} - x_2\right)\,\mathbf{a}_2 + \frac{3}{4}\,\mathbf{a}_3$ | = | $\frac{1}{2}\,a\,\hat{\mathbf{x}}\frac{1}{4}\,a\,\hat{\mathbf{y}} - x_2\,a\,\hat{\mathbf{z}}$ | (24$d$) | Mn |
| $\mathbf{B}_{11}$ | = | $\frac{3}{4}\,\mathbf{a}_1 + \left(\frac{3}{4} - x_2\right)\,\mathbf{a}_2 - x_2\,\mathbf{a}_3$ | = | $-x_2\,a\,\hat{\mathbf{x}} + \frac{3}{4}\,a\,\hat{\mathbf{z}}$ | (24$d$) | Mn |
| $\mathbf{B}_{12}$ | = | $\frac{1}{4}\,\mathbf{a}_1 + \left(\frac{3}{4} + x_2\right)\,\mathbf{a}_2 + \left(\frac{1}{2} + x_2\right)\,\mathbf{a}_3$ | = | $\left(\frac{1}{2} + x_2\right)\,a\,\hat{\mathbf{x}} + \frac{1}{4}\,a\,\hat{\mathbf{z}}$ | (24$d$) | Mn |
| $\mathbf{B}_{13}$ | = | $-x_2\,\mathbf{a}_1 + \frac{3}{4}\,\mathbf{a}_2 + \left(\frac{3}{4} - x_2\right)\,\mathbf{a}_3$ | = | $\frac{3}{4}\,a\,\hat{\mathbf{x}} - x_2\,a\,\hat{\mathbf{y}}$ | (24$d$) | Mn |
| $\mathbf{B}_{14}$ | = | $\left(\frac{1}{2} + x_2\right)\,\mathbf{a}_1 + \frac{1}{4}\,\mathbf{a}_2 + \left(\frac{3}{4} + x_2\right)\,\mathbf{a}_3$ | = | $\frac{1}{4}\,a\,\hat{\mathbf{x}} + \left(\frac{1}{2} + x_2\right)\,a\,\hat{\mathbf{y}}$ | (24$d$) | Mn |
| $\mathbf{B}_{15}$ | = | $\left(\frac{3}{4} - x_2\right)\,\mathbf{a}_1 - x_2\,\mathbf{a}_2 + \frac{3}{4}\,\mathbf{a}_3$ | = | $\frac{3}{4}\,a\,\hat{\mathbf{y}} - x_2\,a\,\hat{\mathbf{z}}$ | (24$d$) | Mn |
| $\mathbf{B}_{16}$ | = | $\left(\frac{3}{4} + x_2\right)\,\mathbf{a}_1 + \left(\frac{1}{2} + x_2\right)\,\mathbf{a}_2 + \frac{1}{4}\,\mathbf{a}_3$ | = | $\frac{1}{4}\,a\,\hat{\mathbf{y}} + \left(\frac{1}{2} + x_2\right)\,a\,\hat{\mathbf{z}}$ | (24$d$) | Mn |
| $\mathbf{B}_{17}$ | = | $(y_3 + z_3)\,\mathbf{a}_1 + (x_3 + z_3)\,\mathbf{a}_2 + (x_3 + y_3)\,\mathbf{a}_3$ | = | $x_3\,a\,\hat{\mathbf{x}} + y_3\,a\,\hat{\mathbf{y}} + z_3\,a\,\hat{\mathbf{z}}$ | (48$e$) | O |
| $\mathbf{B}_{18}$ | = | $\left(\frac{1}{2} - y_3 + z_3\right)\,\mathbf{a}_1 + (z_3 - x_3)\,\mathbf{a}_2 + \left(\frac{1}{2} - x_3 - y_3\right)\,\mathbf{a}_3$ | = | $-x_3\,a\,\hat{\mathbf{x}} + \left(\frac{1}{2} - y_3\right)\,a\,\hat{\mathbf{y}} + z_3\,a\,\hat{\mathbf{z}}$ | (48$e$) | O |
| $\mathbf{B}_{19}$ | = | $(y_3 - z_3)\,\mathbf{a}_1 + \left(\frac{1}{2} - x_3 - z_3\right)\,\mathbf{a}_2 + \left(\frac{1}{2} - x_3 + y_3\right)\,\mathbf{a}_3$ | = | $\left(\frac{1}{2} - x_3\right)\,a\,\hat{\mathbf{x}} + y_3\,a\,\hat{\mathbf{y}} - z_3\,a\,\hat{\mathbf{z}}$ | (48$e$) | O |
| $\mathbf{B}_{20}$ | = | $\left(\frac{1}{2} - y_3 - z_3\right)\,\mathbf{a}_1 + \left(\frac{1}{2} + x_3 - z_3\right)\,\mathbf{a}_2 + (x_3 - y_3)\,\mathbf{a}_3$ | = | $x_3\,a\,\hat{\mathbf{x}} - y_3\,a\,\hat{\mathbf{y}} + \left(\frac{1}{2} - z_3\right)\,a\,\hat{\mathbf{z}}$ | (48$e$) | O |
| $\mathbf{B}_{21}$ | = | $(x_3 + y_3)\,\mathbf{a}_1 + (y_3 + z_3)\,\mathbf{a}_2 + (z_3 + x_3)\,\mathbf{a}_3$ | = | $z_3\,a\,\hat{\mathbf{x}} + x_3\,a\,\hat{\mathbf{y}} + y_3\,a\,\hat{\mathbf{z}}$ | (48$e$) | O |
| $\mathbf{B}_{22}$ | = | $\left(\frac{1}{2} - x_3 + y_3\right)\,\mathbf{a}_1 + (y_3 - z_3)\,\mathbf{a}_2 + \left(\frac{1}{2} - z_3 - x_3\right)\,\mathbf{a}_3$ | = | $-z_3\,a\,\hat{\mathbf{x}} + \left(\frac{1}{2} - x_3\right)\,a\,\hat{\mathbf{y}} + y_3\,a\,\hat{\mathbf{z}}$ | (48$e$) | O |
| $\mathbf{B_{23}}$ | = | $(x_3 - y_3)\,\mathbf{a_1} + \left(\frac{1}{2} - z_3 - y_3\right)\mathbf{a_2} + \left(\frac{1}{2} - z_3 + x_3\right)\mathbf{a_3}$ | = | $\left(\frac{1}{2} - z_3\right)\,a\,\mathbf{\hat{x}} + x_3\,a\,\mathbf{\hat{y}} - y_3\,a\,\mathbf{\hat{z}}$ | (48$e$) | O |
| $\mathbf{B_{24}}$ | = | $\left(\frac{1}{2} - x_3 - y_3\right)\mathbf{a_1} + \left(\frac{1}{2} + z_3 - y_3\right)\mathbf{a_2} + (z_3 - x_3)\,\mathbf{a_3}$ | = | $z_3\,a\,\mathbf{\hat{x}} - x_3\,a\,\mathbf{\hat{y}} + \left(\frac{1}{2} - y_3\right)\,a\,\mathbf{\hat{z}}$ | (48$e$) | O |
| $\mathbf{B_{25}}$ | = | $(z_3 + x_3)\,\mathbf{a_1} + (x_3 + y_3)\,\mathbf{a_2} + (y_3 + z_3)\,\mathbf{a_3}$ | = | $y_3\,a\,\mathbf{\hat{x}} + z_3\,a\,\mathbf{\hat{y}} + x_3\,a\,\mathbf{\hat{z}}$ | (48$e$) | O |
| $\mathbf{B_{26}}$ | = | $\left(\frac{1}{2} - z_3 + x_3\right)\mathbf{a_1} + (x_3 - y_3)\,\mathbf{a_2} + \left(\frac{1}{2} - y_3 - z_3\right)\mathbf{a_3}$ | = | $-y_3\,a\,\mathbf{\hat{x}} + \left(\frac{1}{2} - z_3\right)\,a\,\mathbf{\hat{y}} + x_3\,a\,\mathbf{\hat{z}}$ | (48$e$) | O |
| $\mathbf{B_{27}}$ | = | $(z_3 - x_3)\,\mathbf{a_1} + \left(\frac{1}{2} - y_3 - x_3\right)\mathbf{a_2} + \left(\frac{1}{2} - y_3 + z_3\right)\mathbf{a_3}$ | = | $\left(\frac{1}{2} - y_3\right)\,a\,\mathbf{\hat{x}} + z_3\,a\,\mathbf{\hat{y}} - x_3\,a\,\mathbf{\hat{z}}$ | (48$e$) | O |
| $\mathbf{B_{28}}$ | = | $\left(\frac{1}{2} - z_3 - x_3\right)\mathbf{a_1} + \left(\frac{1}{2} + y_3 - x_3\right)\mathbf{a_2} + (y_3 - z_3)\,\mathbf{a_3}$ | = | $y_3\,a\,\mathbf{\hat{x}} - z_3\,a\,\mathbf{\hat{y}} + \left(\frac{1}{2} - x_3\right)\,a\,\mathbf{\hat{z}}$ | (48$e$) | O |
| $\mathbf{B_{29}}$ | = | $-(y_3 + z_3)\,\mathbf{a_1} - (x_3 + z_3)\,\mathbf{a_2} - (x_3 + y_3)\,\mathbf{a_3}$ | = | $-x_3\,a\,\mathbf{\hat{x}} - y_3\,a\,\mathbf{\hat{y}} - z_3\,a\,\mathbf{\hat{z}}$ | (48$e$) | O |
| $\mathbf{B_{30}}$ | = | $\left(\frac{1}{2} + y_3 - z_3\right)\mathbf{a_1} + (x_3 - z_3)\,\mathbf{a_2} + \left(\frac{1}{2} + x_3 + y_3\right)\mathbf{a_3}$ | = | $x_3\,a\,\mathbf{\hat{x}} + \left(\frac{1}{2} + y_3\right)\,a\,\mathbf{\hat{y}} - z_3\,a\,\mathbf{\hat{z}}$ | (48$e$) | O |
| $\mathbf{B_{31}}$ | = | $(z_3 - y_3)\,\mathbf{a_1} + \left(\frac{1}{2} + x_3 + z_3\right)\mathbf{a_2} + \left(\frac{1}{2} + x_3 - y_3\right)\mathbf{a_3}$ | = | $\left(\frac{1}{2} + x_3\right)\,a\,\mathbf{\hat{x}} - y_3\,a\,\mathbf{\hat{y}} + z_3\,a\,\mathbf{\hat{z}}$ | (48$e$) | O |
| $\mathbf{B_{32}}$ | = | $\left(\frac{1}{2} + y_3 + z_3\right)\mathbf{a_1} + \left(\frac{1}{2} - x_3 + z_3\right)\mathbf{a_2} + (y_3 - x_3)\,\mathbf{a_3}$ | = | $-x_3\,a\,\mathbf{\hat{x}} + y_3\,a\,\mathbf{\hat{y}} + \left(\frac{1}{2} + z_3\right)\,a\,\mathbf{\hat{z}}$ | (48$e$) | O |
| $\mathbf{B_{33}}$ | = | $-(x_3 + y_3)\,\mathbf{a_1} - (y_3 + z_3)\,\mathbf{a_2} - (z_3 + x_3)\,\mathbf{a_3}$ | = | $-z_3\,a\,\mathbf{\hat{x}} - x_3\,a\,\mathbf{\hat{y}} - y_3\,a\,\mathbf{\hat{z}}$ | (48$e$) | O |
| $\mathbf{B_{34}}$ | = | $\left(\frac{1}{2} + x_3 - y_3\right)\mathbf{a_1} + (z_3 - y_3)\,\mathbf{a_2} + \left(\frac{1}{2} + x_3 + z_3\right)\mathbf{a_3}$ | = | $z_3\,a\,\mathbf{\hat{x}} + \left(\frac{1}{2} + x_3\right)\,a\,\mathbf{\hat{y}} - y_3\,a\,\mathbf{\hat{z}}$ | (48$e$) | O |
| $\mathbf{B_{35}}$ | = | $(y_3 - x_3)\,\mathbf{a_1} + \left(\frac{1}{2} + y_3 + z_3\right)\mathbf{a_2} + \left(\frac{1}{2} + z_3 - x_3\right)\mathbf{a_3}$ | = | $\left(\frac{1}{2} + z_3\right)\,a\,\mathbf{\hat{x}} - x_3\,a\,\mathbf{\hat{y}} + y_3\,a\,\mathbf{\hat{z}}$ | (48$e$) | O |
| $\mathbf{B_{36}}$ | = | $\left(\frac{1}{2} + x_3 + y_3\right)\mathbf{a_1} + \left(\frac{1}{2} - z_3 + y_3\right)\mathbf{a_2} + (x_3 - z_3)\,\mathbf{a_3}$ | = | $-z_3\,a\,\mathbf{\hat{x}} + x_3\,a\,\mathbf{\hat{y}} + \left(\frac{1}{2} + y_3\right)\,a\,\mathbf{\hat{z}}$ | (48$e$) | O |
| $\mathbf{B_{37}}$ | = | $-(x_3 + z_3)\,\mathbf{a_1} - (x_3 + y_3)\,\mathbf{a_2} - (y_3 + z_3)\,\mathbf{a_3}$ | = | $-y_3\,a\,\mathbf{\hat{x}} - z_3\,a\,\mathbf{\hat{y}} - x_3\,a\,\mathbf{\hat{z}}$ | (48$e$) | O |
| $\mathbf{B_{38}}$ | = | $\left(\frac{1}{2} + z_3 - x_3\right)\mathbf{a_1} + (y_3 - x_3)\,\mathbf{a_2} + \left(\frac{1}{2} + y_3 + z_3\right)\mathbf{a_3}$ | = | $y_3\,a\,\mathbf{\hat{x}} + \left(\frac{1}{2} + z_3\right)\,a\,\mathbf{\hat{y}} - x_3\,a\,\mathbf{\hat{z}}$ | (48$e$) | O |
| $\mathbf{B_{39}}$ | = | $(x_3 - z_3)\,\mathbf{a_1} + \left(\frac{1}{2} + y_3 + x_3\right)\mathbf{a_2} + \left(\frac{1}{2} + y_3 - z_3\right)\mathbf{a_3}$ | = | $\left(\frac{1}{2} + y_3\right)\,a\,\mathbf{\hat{x}} - z_3\,a\,\mathbf{\hat{y}} + x_3\,a\,\mathbf{\hat{z}}$ | (48$e$) | O |
| $\mathbf{B_{40}}$ | = | $\left(\frac{1}{2} + x_3 + z_3\right)\mathbf{a_1} + \left(\frac{1}{2} + x_3 - y_3\right)\mathbf{a_2} + (z_3 - y_3)\,\mathbf{a_3}$ | = | $-y_3\,a\,\mathbf{\hat{x}} + z_3\,a\,\mathbf{\hat{y}} + \left(\frac{1}{2} + x_3\right)\,a\,\mathbf{\hat{z}}$ | (48$e$) | O |

**Geometry files:**
- CIF: pp. 761
- POSCAR: pp. 762

# BC8 (Si) Structure: A_cI16_206_c

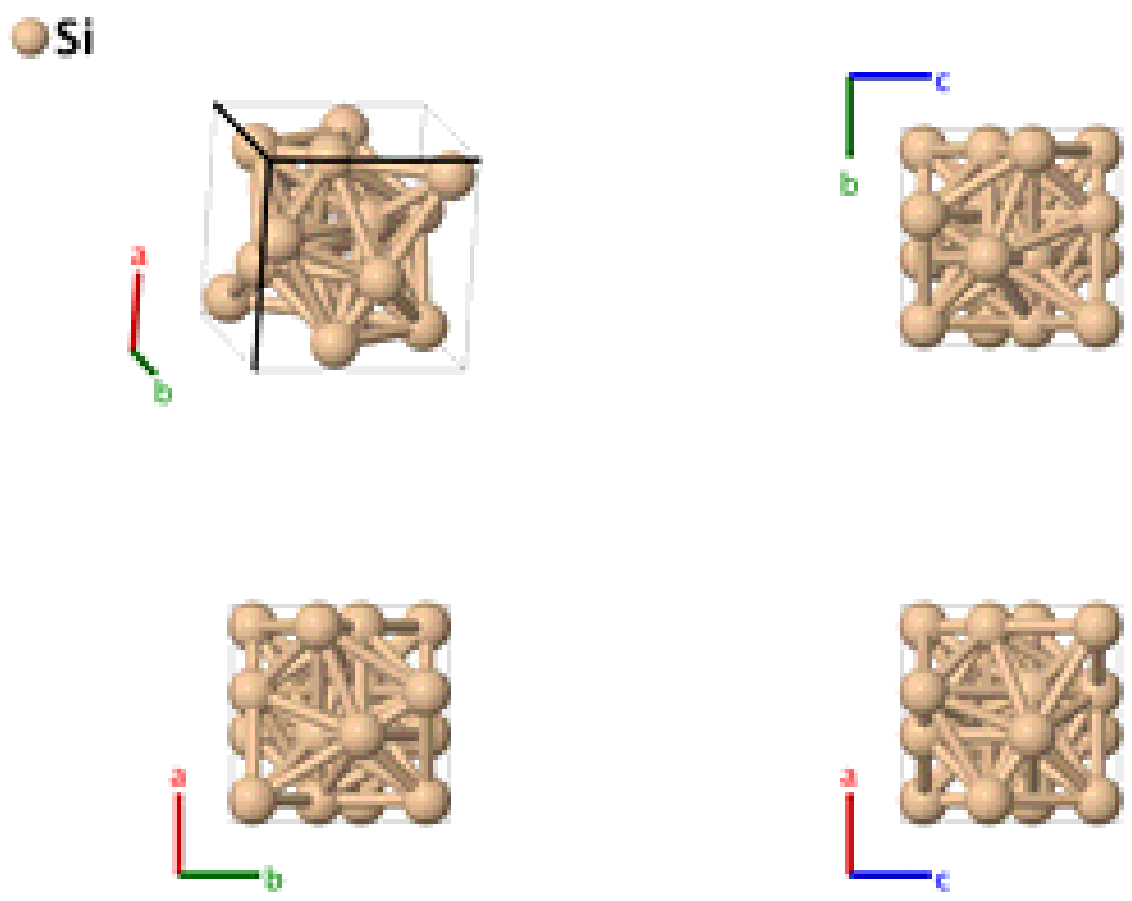

| | | |
|---|---|---|
| **Prototype** | : | Si |
| **AFLOW prototype label** | : | A_cI16_206_c |
| ***Strukturbericht* designation** | : | None |
| **Pearson symbol** | : | cI16 |
| **Space group number** | : | 206 |
| **Space group symbol** | : | $Ia\bar{3}$ |
| **AFLOW prototype command** | : | `aflow --proto=A_cI16_206_c`<br>`--params=`$a, x_1$ |

- This is a tetragonally bonded structure which packs more efficiently than diamond. See (Crain, 1995) and references therein. The reference compound chosen here, found in (Wentorf, 1963), is stable in the range 11-16 GPa.

**Body-centered Cubic primitive vectors:**

$$\mathbf{a}_1 = -\frac{1}{2}\,a\,\hat{\mathbf{x}} + \frac{1}{2}\,a\,\hat{\mathbf{y}} + \frac{1}{2}\,a\,\hat{\mathbf{z}}$$
$$\mathbf{a}_2 = \frac{1}{2}\,a\,\hat{\mathbf{x}} - \frac{1}{2}\,a\,\hat{\mathbf{y}} + \frac{1}{2}\,a\,\hat{\mathbf{z}}$$
$$\mathbf{a}_3 = \frac{1}{2}\,a\,\hat{\mathbf{x}} + \frac{1}{2}\,a\,\hat{\mathbf{y}} - \frac{1}{2}\,a\,\hat{\mathbf{z}}$$

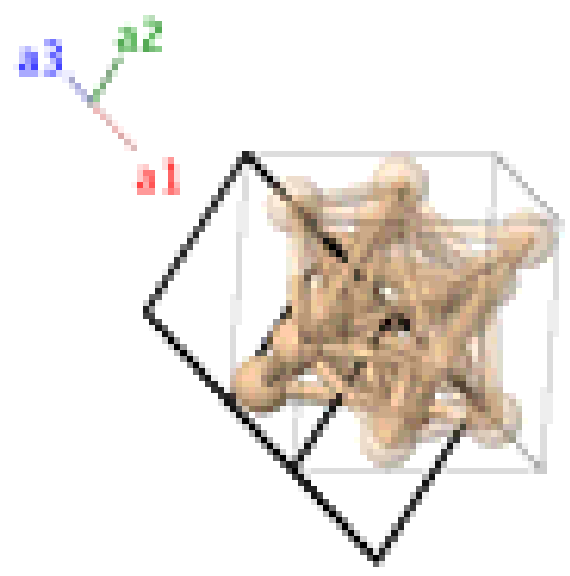

**Basis vectors:**

| | | Lattice Coordinates | | Cartesian Coordinates | Wyckoff Position | Atom Type |
|---|---|---|---|---|---|---|
| $\mathbf{B}_1$ | = | $2x_1\,\mathbf{a}_1 + 2x_1\,\mathbf{a}_2 + 2x_1\,\mathbf{a}_3$ | = | $x_1\,a\,\hat{\mathbf{x}} + x_1\,a\,\hat{\mathbf{y}} + x_1\,a\,\hat{\mathbf{z}}$ | (16$c$) | Si |
| $\mathbf{B}_2$ | = | $\frac{1}{2}\,\mathbf{a}_1 + \left(\frac{1}{2} - 2x_1\right)\,\mathbf{a}_3$ | = | $-x_1\,a\,\hat{\mathbf{x}} + \left(\frac{1}{2} - x_1\right)\,a\,\hat{\mathbf{y}} + x_1\,a\,\hat{\mathbf{z}}$ | (16$c$) | Si |
| $\mathbf{B}_3$ | = | $\left(\frac{1}{2} - 2x_1\right)\,\mathbf{a}_2 + \frac{1}{2}\,\mathbf{a}_3$ | = | $\left(\frac{1}{2} - x_1\right)\,a\,\hat{\mathbf{x}} + x_1\,a\,\hat{\mathbf{y}} - x_1\,a\,\hat{\mathbf{z}}$ | (16$c$) | Si |
| $\mathbf{B}_4$ | = | $\left(\frac{1}{2} - 2x_1\right)\,\mathbf{a}_1 + \frac{1}{2}\,\mathbf{a}_2$ | = | $x_1\,a\,\hat{\mathbf{x}} - x_1\,a\,\hat{\mathbf{y}}\left(\frac{1}{2} - x_1\right)\,a\,\hat{\mathbf{z}}$ | (16$c$) | Si |
| $\mathbf{B}_5$ | = | $-2x_1\,\mathbf{a}_1 - 2x_1\,\mathbf{a}_2 - 2x_1\,\mathbf{a}_3$ | = | $-x_1\,a\,\hat{\mathbf{x}} - x_1\,a\,\hat{\mathbf{y}} - x_1\,a\,\hat{\mathbf{z}}$ | (16$c$) | Si |
| $\mathbf{B_6}$ | = | $\frac{1}{2}\,\mathbf{a_1} + \left(\frac{1}{2} + 2x_1\right)\,\mathbf{a_3}$ | = | $x_1\,a\,\hat{\mathbf{x}} + \left(\frac{1}{2} + x_1\right)\,a\,\hat{\mathbf{y}} - x_1\,a\,\hat{\mathbf{z}}$ | (16$c$) | Si |
| $\mathbf{B_7}$ | = | $\left(\frac{1}{2} + 2x_1\right)\,\mathbf{a_2} + \frac{1}{2}\,\mathbf{a_3}$ | = | $\left(\frac{1}{2} + x_1\right)\,a\,\hat{\mathbf{x}} - x_1\,a\,\hat{\mathbf{y}} + x_1\,a\,\hat{\mathbf{z}}$ | (16$c$) | Si |
| $\mathbf{B_8}$ | = | $\left(\frac{1}{2} + 2x_1\right)\,\mathbf{a_1} + \frac{1}{2}\,\mathbf{a_2}$ | = | $-x_1\,a\,\hat{\mathbf{x}} + x_1\,a\,\hat{\mathbf{y}}\left(\frac{1}{2} + x_1\right)\,a\,\hat{\mathbf{z}}$ | (16$c$) | Si |

**Geometry files:**

- CIF: pp. 762
- POSCAR: pp. 763

# $\beta$-Mn (A13) Structure: A_cP20_213_cd

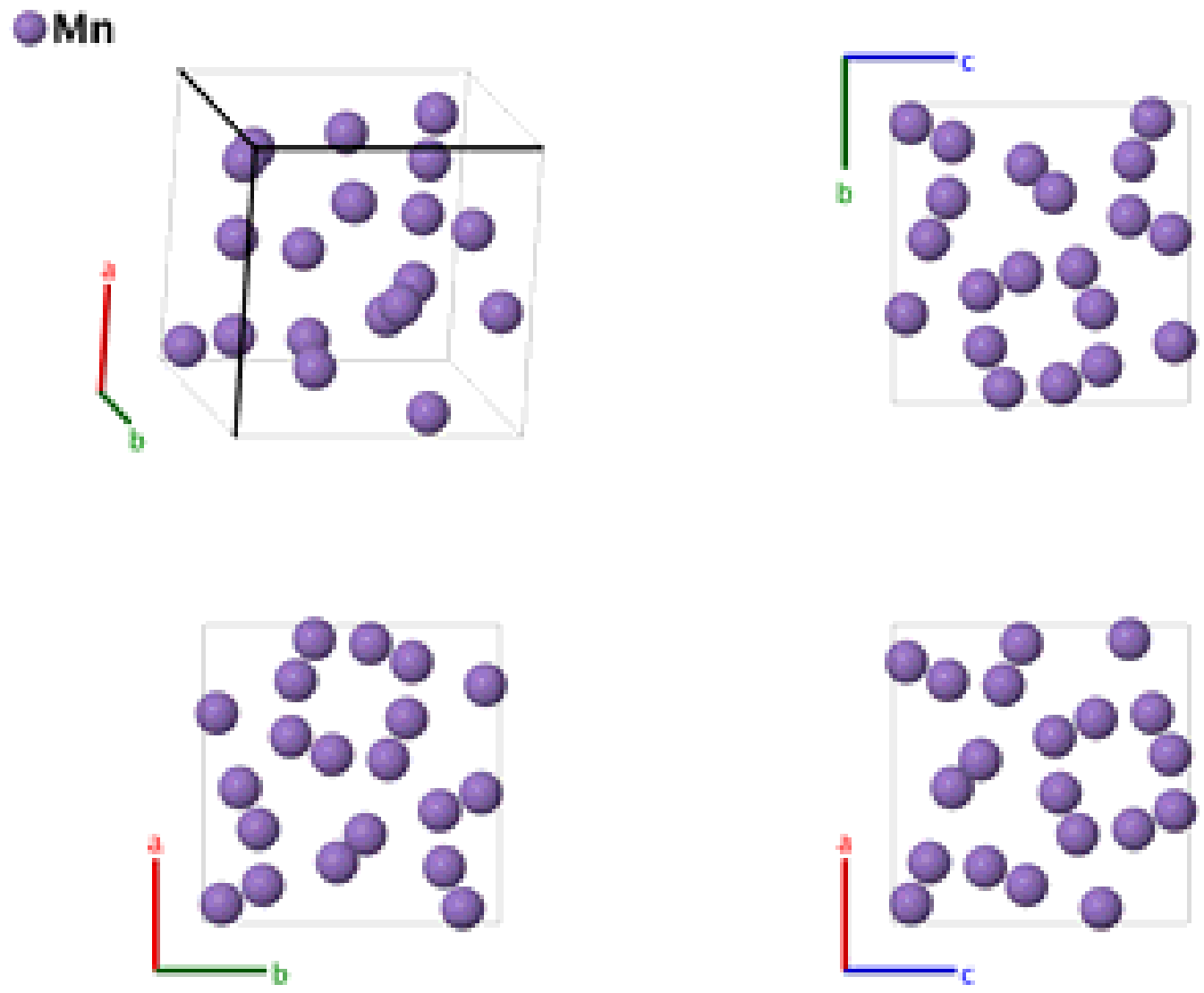

| | | |
|---|---|---|
| **Prototype** | : | $\beta$-Mn |
| **AFLOW prototype label** | : | A_cP20_213_cd |
| ***Strukturbericht* designation** | : | A13 |
| **Pearson symbol** | : | cP20 |
| **Space group number** | : | 213 |
| **Space group symbol** | : | $P4_132$ |
| **AFLOW prototype command** | : | `aflow --proto=A_cP20_213_cd --params=`$a, x_1, y_2$ |

**Simple Cubic primitive vectors:**

$$\mathbf{a}_1 = a\,\mathbf{\hat{x}}$$
$$\mathbf{a}_2 = a\,\mathbf{\hat{y}}$$
$$\mathbf{a}_3 = a\,\mathbf{\hat{z}}$$

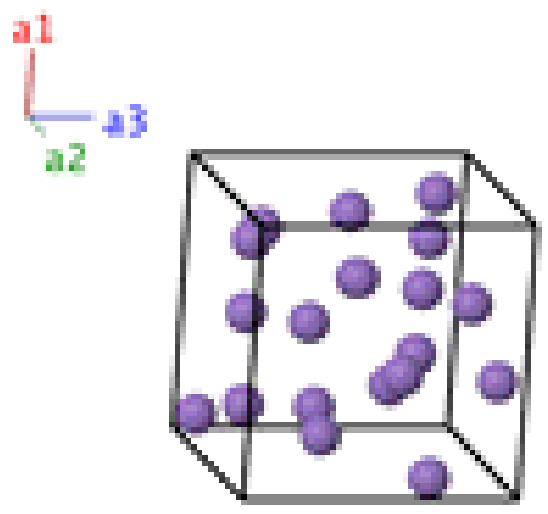

**Basis vectors:**

| | | Lattice Coordinates | | Cartesian Coordinates | Wyckoff Position | Atom Type |
|---|---|---|---|---|---|---|
| $\mathbf{B}_1$ | = | $x_1\,\mathbf{a}_1 + x_1\,\mathbf{a}_2 + x_1\,\mathbf{a}_3$ | = | $x_1\,a\,\mathbf{\hat{x}} + x_1\,a\,\mathbf{\hat{y}} + x_1\,a\,\mathbf{\hat{z}}$ | (8c) | Mn I |
| $\mathbf{B}_2$ | = | $\left(\frac{1}{2} - x_1\right)\mathbf{a}_1 - x_1\,\mathbf{a}_2 + \left(\frac{1}{2} + x_1\right)\mathbf{a}_3$ | = | $\left(\frac{1}{2} - x_1\right)a\,\mathbf{\hat{x}} - x_1\,a\,\mathbf{\hat{y}} + \left(\frac{1}{2} + x_1\right)a\,\mathbf{\hat{z}}$ | (8c) | Mn I |
| $\mathbf{B}_3$ | = | $-x_1\,\mathbf{a}_1 + \left(\frac{1}{2} + x_1\right)\mathbf{a}_2 + \left(\frac{1}{2} - x_1\right)\mathbf{a}_3$ | = | $-x_1\,a\,\mathbf{\hat{x}} + \left(\frac{1}{2} + x_1\right)a\,\mathbf{\hat{y}} + \left(\frac{1}{2} - x_1\right)a\,\mathbf{\hat{z}}$ | (8c) | Mn I |
| $\mathbf{B}_4$ | = | $\left(\frac{1}{2} + x_1\right)\mathbf{a}_1 + \left(\frac{1}{2} - x_1\right)\mathbf{a}_2 - x_1\,\mathbf{a}_3$ | = | $\left(\frac{1}{2} + x_1\right)a\,\mathbf{\hat{x}} + \left(\frac{1}{2} - x_1\right)a\,\mathbf{\hat{y}} - x_1\,a\,\mathbf{\hat{z}}$ | (8c) | Mn I |
| $\mathbf{B}_5$ | = | $\left(\frac{3}{4} + x_1\right)\mathbf{a}_1 + \left(\frac{1}{4} + x_1\right)\mathbf{a}_2 + \left(\frac{1}{4} - x_1\right)\mathbf{a}_3$ | = | $\left(\frac{3}{4} + x_1\right)a\,\mathbf{\hat{x}} + \left(\frac{1}{4} + x_1\right)a\,\mathbf{\hat{y}} + \left(\frac{1}{4} - x_1\right)a\,\mathbf{\hat{z}}$ | (8c) | Mn I |
| $\mathbf{B_{6}}$ | = | $\left(\frac{3}{4} - x_1\right) \mathbf{a_1} + \left(\frac{3}{4} - x_1\right) \mathbf{a_2} + \left(\frac{3}{4} - x_1\right) \mathbf{a_3}$ | = | $\left(\frac{3}{4} - x_1\right) a\,\mathbf{\hat{x}} + \left(\frac{3}{4} - x_1\right) a\,\mathbf{\hat{y}} + \left(\frac{3}{4} - x_1\right) a\,\mathbf{\hat{z}}$ | (8$c$) | Mn I |
| $\mathbf{B_{7}}$ | = | $\left(\frac{1}{4} + x_1\right) \mathbf{a_1} + \left(\frac{1}{4} - x_1\right) \mathbf{a_2} + \left(\frac{3}{4} + x_1\right) \mathbf{a_3}$ | = | $\left(\frac{1}{4} + x_1\right) a\,\mathbf{\hat{x}} + \left(\frac{1}{4} - x_1\right) a\,\mathbf{\hat{y}} + \left(\frac{3}{4} + x_1\right) a\,\mathbf{\hat{z}}$ | (8$c$) | Mn I |
| $\mathbf{B_{8}}$ | = | $\left(\frac{1}{4} - x_1\right) \mathbf{a_1} + \left(\frac{3}{4} + x_1\right) \mathbf{a_2} + \left(\frac{1}{4} + x_1\right) \mathbf{a_3}$ | = | $\left(\frac{1}{4} - x_1\right) a\,\mathbf{\hat{x}} + \left(\frac{3}{4} + x_1\right) a\,\mathbf{\hat{y}} + \left(\frac{1}{4} + x_1\right) a\,\mathbf{\hat{z}}$ | (8$c$) | Mn I |
| $\mathbf{B_{9}}$ | = | $\frac{1}{8} \mathbf{a_1} + y_2 \mathbf{a_2} + \left(\frac{1}{4} + y_2\right) \mathbf{a_3}$ | = | $\frac{1}{8} a\,\mathbf{\hat{x}} + y_2 a\,\mathbf{\hat{y}} + \left(\frac{1}{4} + y_2\right) a\,\mathbf{\hat{z}}$ | (12$d$) | Mn II |
| $\mathbf{B_{10}}$ | = | $\frac{3}{8} \mathbf{a_1} - y_2 \mathbf{a_2} + \left(\frac{3}{4} + y_2\right) \mathbf{a_3}$ | = | $\frac{3}{8} a\,\mathbf{\hat{x}} - y_2 a\,\mathbf{\hat{y}} + \left(\frac{3}{4} + y_2\right) a\,\mathbf{\hat{z}}$ | (12$d$) | Mn II |
| $\mathbf{B_{11}}$ | = | $\frac{7}{8} \mathbf{a_1} + \left(\frac{1}{2} + y_2\right) \mathbf{a_2} + \left(\frac{1}{4} - y_2\right) \mathbf{a_3}$ | = | $\frac{7}{8} a\,\mathbf{\hat{x}} + \left(\frac{1}{2} + y_2\right) a\,\mathbf{\hat{y}} + \left(\frac{1}{4} - y_2\right) a\,\mathbf{\hat{z}}$ | (12$d$) | Mn II |
| $\mathbf{B_{12}}$ | = | $\frac{5}{8} \mathbf{a_1} + \left(\frac{1}{2} - y_2\right) \mathbf{a_2} + \left(\frac{3}{4} - y_2\right) \mathbf{a_3}$ | = | $\frac{5}{8} a\,\mathbf{\hat{x}} + \left(\frac{1}{2} - y_2\right) a\,\mathbf{\hat{y}} + \left(\frac{3}{4} - y_2\right) a\,\mathbf{\hat{z}}$ | (12$d$) | Mn II |
| $\mathbf{B_{13}}$ | = | $\left(\frac{1}{4} + y_2\right) \mathbf{a_1} + \frac{1}{8} \mathbf{a_2} + y_2 \mathbf{a_3}$ | = | $\left(\frac{1}{4} + y_2\right) a\,\mathbf{\hat{x}} + \frac{1}{8} a\,\mathbf{\hat{y}} + y_2 a\,\mathbf{\hat{z}}$ | (12$d$) | Mn II |
| $\mathbf{B_{14}}$ | = | $\left(\frac{3}{4} + y_2\right) \mathbf{a_1} + \frac{3}{8} \mathbf{a_2} - y_2 \mathbf{a_3}$ | = | $\left(\frac{3}{4} + y_2\right) a\,\mathbf{\hat{x}} + \frac{3}{8} a\,\mathbf{\hat{y}} - y_2 a\,\mathbf{\hat{z}}$ | (12$d$) | Mn II |
| $\mathbf{B_{15}}$ | = | $\left(\frac{1}{4} - y_2\right) \mathbf{a_1} + \frac{7}{8} \mathbf{a_2} + \left(\frac{1}{2} + y_2\right) \mathbf{a_3}$ | = | $\left(\frac{1}{4} - y_2\right) a\,\mathbf{\hat{x}} + \frac{7}{8} a\,\mathbf{\hat{y}} + \left(\frac{1}{2} + y_2\right) a\,\mathbf{\hat{z}}$ | (12$d$) | Mn II |
| $\mathbf{B_{16}}$ | = | $\left(\frac{3}{4} - y_2\right) \mathbf{a_1} + \frac{5}{8} \mathbf{a_2} + \left(\frac{1}{2} - y_2\right) \mathbf{a_3}$ | = | $\left(\frac{3}{4} - y_2\right) a\,\mathbf{\hat{x}} + \frac{5}{8} a\,\mathbf{\hat{y}} + \left(\frac{1}{2} - y_2\right) a\,\mathbf{\hat{z}}$ | (12$d$) | Mn II |
| $\mathbf{B_{17}}$ | = | $y_2 \mathbf{a_1} + \left(\frac{1}{4} + y_2\right) \mathbf{a_2} + \frac{1}{8} \mathbf{a_3}$ | = | $y_2 a\,\mathbf{\hat{x}} + \left(\frac{1}{4} + y_2\right) a\,\mathbf{\hat{y}} + \frac{1}{8} a\,\mathbf{\hat{z}}$ | (12$d$) | Mn II |
| $\mathbf{B_{18}}$ | = | $-y_2 \mathbf{a_1} + \left(\frac{3}{4} + y_2\right) \mathbf{a_2} + \frac{3}{8} \mathbf{a_3}$ | = | $-y_2 a\,\mathbf{\hat{x}} + \left(\frac{3}{4} + y_2\right) a\,\mathbf{\hat{y}} + \frac{3}{8} a\,\mathbf{\hat{z}}$ | (12$d$) | Mn II |
| $\mathbf{B_{19}}$ | = | $\left(\frac{1}{2} + y_2\right) \mathbf{a_1} + \left(\frac{1}{4} - y_2\right) \mathbf{a_2} + \frac{7}{8} \mathbf{a_3}$ | = | $\left(\frac{1}{2} + y_2\right) a\,\mathbf{\hat{x}} + \left(\frac{1}{4} - y_2\right) a\,\mathbf{\hat{y}} + \frac{7}{8} a\,\mathbf{\hat{z}}$ | (12$d$) | Mn II |
| $\mathbf{B_{20}}$ | = | $\left(\frac{1}{2} - y_2\right) \mathbf{a_1} + \left(\frac{3}{4} - y_2\right) \mathbf{a_2} + \frac{5}{8} \mathbf{a_3}$ | = | $\left(\frac{1}{2} - y_2\right) a\,\mathbf{\hat{x}} + \left(\frac{3}{4} - y_2\right) a\,\mathbf{\hat{y}} + \frac{5}{8} a\,\mathbf{\hat{z}}$ | (12$d$) | Mn II |

**Geometry files:**

- CIF: pp. 763

- POSCAR: pp. 763

# Sulvanite ($Cu_3S_4V$, $H2_4$) Structure: A3B4C_cP8_215_d_e_a

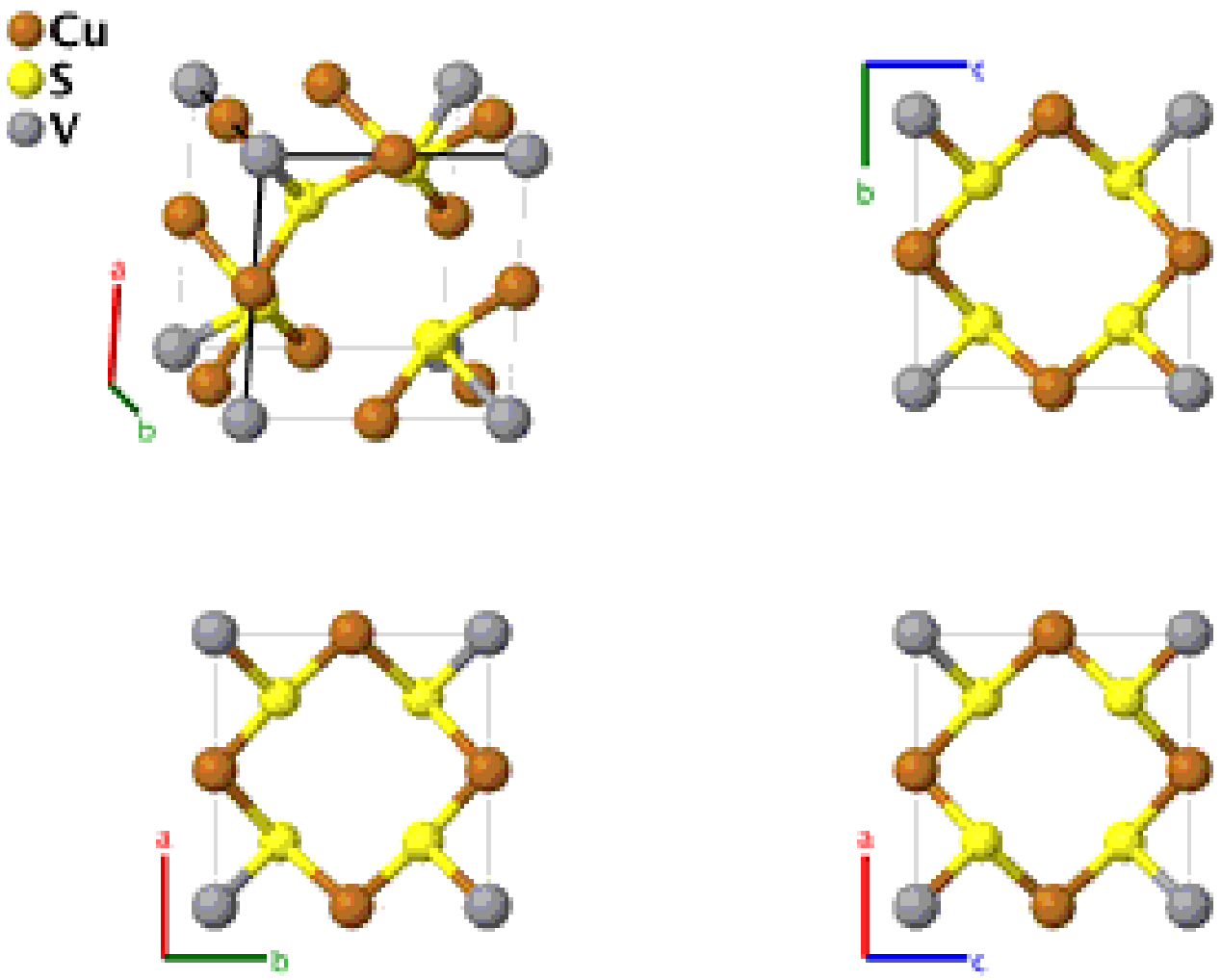

|  |  |  |
|---|---|---|
| **Prototype** | : | $Cu_3S_4V$ |
| **AFLOW prototype label** | : | A3B4C_cP8_215_d_e_a |
| ***Strukturbericht* designation** | : | $H2_4$ |
| **Pearson symbol** | : | cP8 |
| **Space group number** | : | 215 |
| **Space group symbol** | : | $P\bar{4}3m$ |
| **AFLOW prototype command** | : | `aflow --proto=A3B4C_cP8_215_d_e_a --params=`$a, x_3$ |

**Other compounds with this structure:**

- $Cu_3S_4Nb$, $Cu_3S_4Ta$, $Cu_3Se_4Nb$, $Cu_3Te_4Ta$, $Cu_3Te_4V$

- This structure is very similar to lazarevićite ($AsCu_3S_4$), except that in this case the copper atoms are on the cubic edges [the (3d) sites] rather than the cubic faces [the (3c) sites].

**Simple Cubic primitive vectors:**

$$\mathbf{a}_1 = a\,\hat{\mathbf{x}}$$
$$\mathbf{a}_2 = a\,\hat{\mathbf{y}}$$
$$\mathbf{a}_3 = a\,\hat{\mathbf{z}}$$

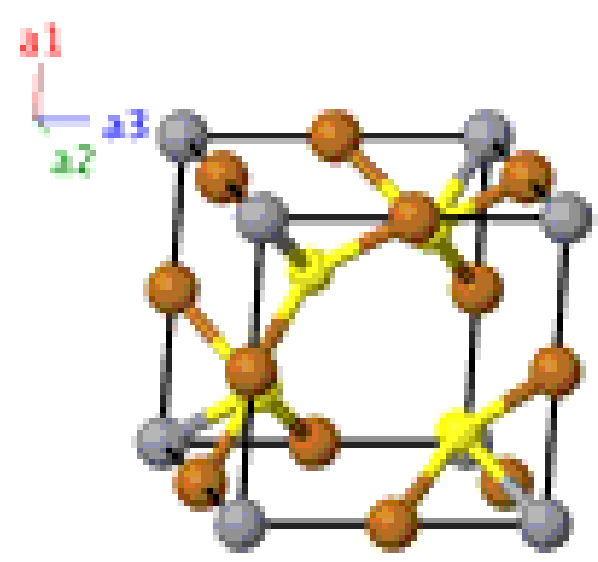

**Basis vectors:**

|  |  | Lattice Coordinates |  | Cartesian Coordinates | Wyckoff Position | Atom Type |
|---|---|---|---|---|---|---|
| $\mathbf{B}_1$ | = | $0\,\mathbf{a}_1 + 0\,\mathbf{a}_2 + 0\,\mathbf{a}_3$ | = | $0\hat{\mathbf{x}} + 0\hat{\mathbf{y}} + 0\hat{\mathbf{z}}$ | $(1a)$ | V |
| $\mathbf{B_2}$ | = | $\frac{1}{2}\,\mathbf{a_1}$ | = | $\frac{1}{2}\,a\,\hat{\mathbf{x}}$ | $(3d)$ | Cu |
| $\mathbf{B_3}$ | = | $\frac{1}{2}\,\mathbf{a_2}$ | = | $\frac{1}{2}\,a\,\hat{\mathbf{y}}$ | $(3d)$ | Cu |
| $\mathbf{B_4}$ | = | $\frac{1}{2}\,\mathbf{a_3}$ | = | $\frac{1}{2}\,a\,\hat{\mathbf{z}}$ | $(3d)$ | Cu |
| $\mathbf{B_5}$ | = | $x_3\,\mathbf{a_1} + x_3\,\mathbf{a_2} + x_3\,\mathbf{a_3}$ | = | $x_3\,a\,\hat{\mathbf{x}} + x_3\,a\,\hat{\mathbf{y}} + x_3\,a\,\hat{\mathbf{z}}$ | $(4e)$ | S |
| $\mathbf{B_6}$ | = | $-x_3\,\mathbf{a_1} - x_3\,\mathbf{a_2} + x_3\,\mathbf{a_3}$ | = | $-x_3\,a\,\hat{\mathbf{x}} - x_3\,a\,\hat{\mathbf{y}} + x_3\,a\,\hat{\mathbf{z}}$ | $(4e)$ | S |
| $\mathbf{B_7}$ | = | $-x_3\,\mathbf{a_1} + x_3\,\mathbf{a_2} - x_3\,\mathbf{a_3}$ | = | $-x_3\,a\,\hat{\mathbf{x}} + x_3\,a\,\hat{\mathbf{y}} - x_3\,a\,\hat{\mathbf{z}}$ | $(4e)$ | S |
| $\mathbf{B_8}$ | = | $x_3\,\mathbf{a_1} - x_3\,\mathbf{a_2} - x_3\,\mathbf{a_3}$ | = | $x_3\,a\,\hat{\mathbf{x}} - x_3\,a\,\hat{\mathbf{y}} - x_3\,a\,\hat{\mathbf{z}}$ | $(4e)$ | S |

---

**Found in:**

---

**Geometry files:**

- CIF: pp. 763
- POSCAR: pp. 764

# $Fe_4C$ Structure: AB4_cP5_215_a_e

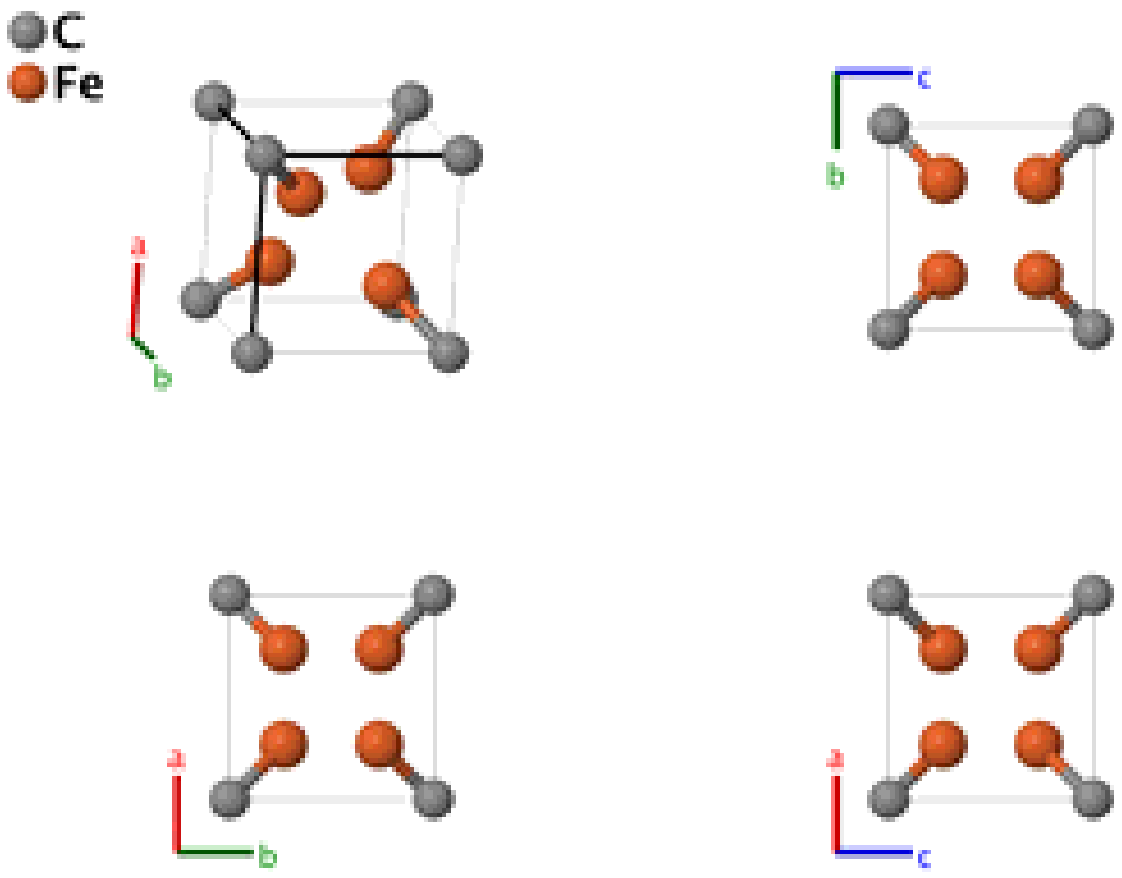

| | | |
|---|---|---|
| **Prototype** | : | $Fe_4C$ |
| **AFLOW prototype label** | : | AB4_cP5_215_a_e |
| ***Strukturbericht* designation** | : | None |
| **Pearson symbol** | : | cP5 |
| **Space group number** | : | 215 |
| **Space group symbol** | : | $P\bar{4}3m$ |
| **AFLOW prototype command** | : | `aflow --proto=AB4_cP5_215_a_e --params=`$a, x_2$ |

- When $x_2 = 1/4$, the iron atoms are at the positions of the face-centered cubic lattice. In $Fe_4C$, $x_2$ is about 0.265.

**Simple Cubic primitive vectors:**

$$\mathbf{a}_1 = a\,\hat{\mathbf{x}}$$
$$\mathbf{a}_2 = a\,\hat{\mathbf{y}}$$
$$\mathbf{a}_3 = a\,\hat{\mathbf{z}}$$

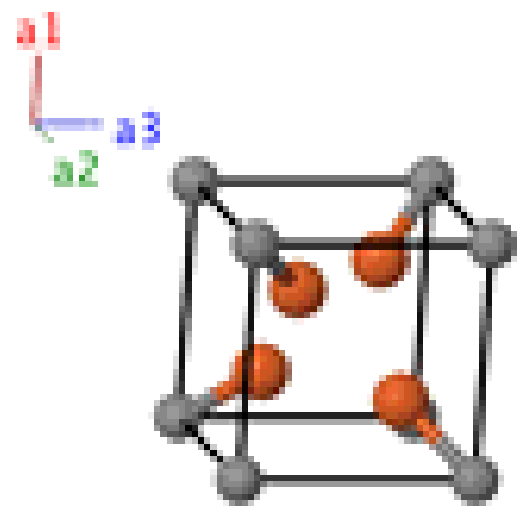

**Basis vectors:**

| | | Lattice Coordinates | | Cartesian Coordinates | Wyckoff Position | Atom Type |
|---|---|---|---|---|---|---|
| $\mathbf{B}_1$ | = | $0\,\mathbf{a}_1 + 0\,\mathbf{a}_2 + 0\,\mathbf{a}_3$ | = | $0\hat{\mathbf{x}} + 0\hat{\mathbf{y}} + 0\hat{\mathbf{z}}$ | (1$a$) | C |
| $\mathbf{B}_2$ | = | $x_2\,\mathbf{a}_1 + x_2\,\mathbf{a}_2 + x_2\,\mathbf{a}_3$ | = | $x_2\,a\,\hat{\mathbf{x}} + x_2\,a\,\hat{\mathbf{y}} + x_2\,a\,\hat{\mathbf{z}}$ | (4$e$) | Fe |
| $\mathbf{B}_3$ | = | $-x_2\,\mathbf{a}_1 - x_2\,\mathbf{a}_2 + x_2\,\mathbf{a}_3$ | = | $-x_2\,a\,\hat{\mathbf{x}} - x_2\,a\,\hat{\mathbf{y}} + x_2\,a\,\hat{\mathbf{z}}$ | (4$e$) | Fe |
| $\mathbf{B}_4$ | = | $-x_2\,\mathbf{a}_1 + x_2\,\mathbf{a}_2 - x_2\,\mathbf{a}_3$ | = | $-x_2\,a\,\hat{\mathbf{x}} + x_2\,a\,\hat{\mathbf{y}} - x_2\,a\,\hat{\mathbf{z}}$ | (4$e$) | Fe |
| $\mathbf{B}_5$ | = | $x_2\,\mathbf{a}_1 - x_2\,\mathbf{a}_2 - x_2\,\mathbf{a}_3$ | = | $x_2\,a\,\hat{\mathbf{x}} - x_2\,a\,\hat{\mathbf{y}} - x_2\,a\,\hat{\mathbf{z}}$ | (4$e$) | Fe |

**Geometry files:**

- CIF: pp. 764
- POSCAR: pp. 764

# Cubic Lazarevićite ($AsCu_3S_4$) Structure: AB3C4_cP8_215_a_c_e

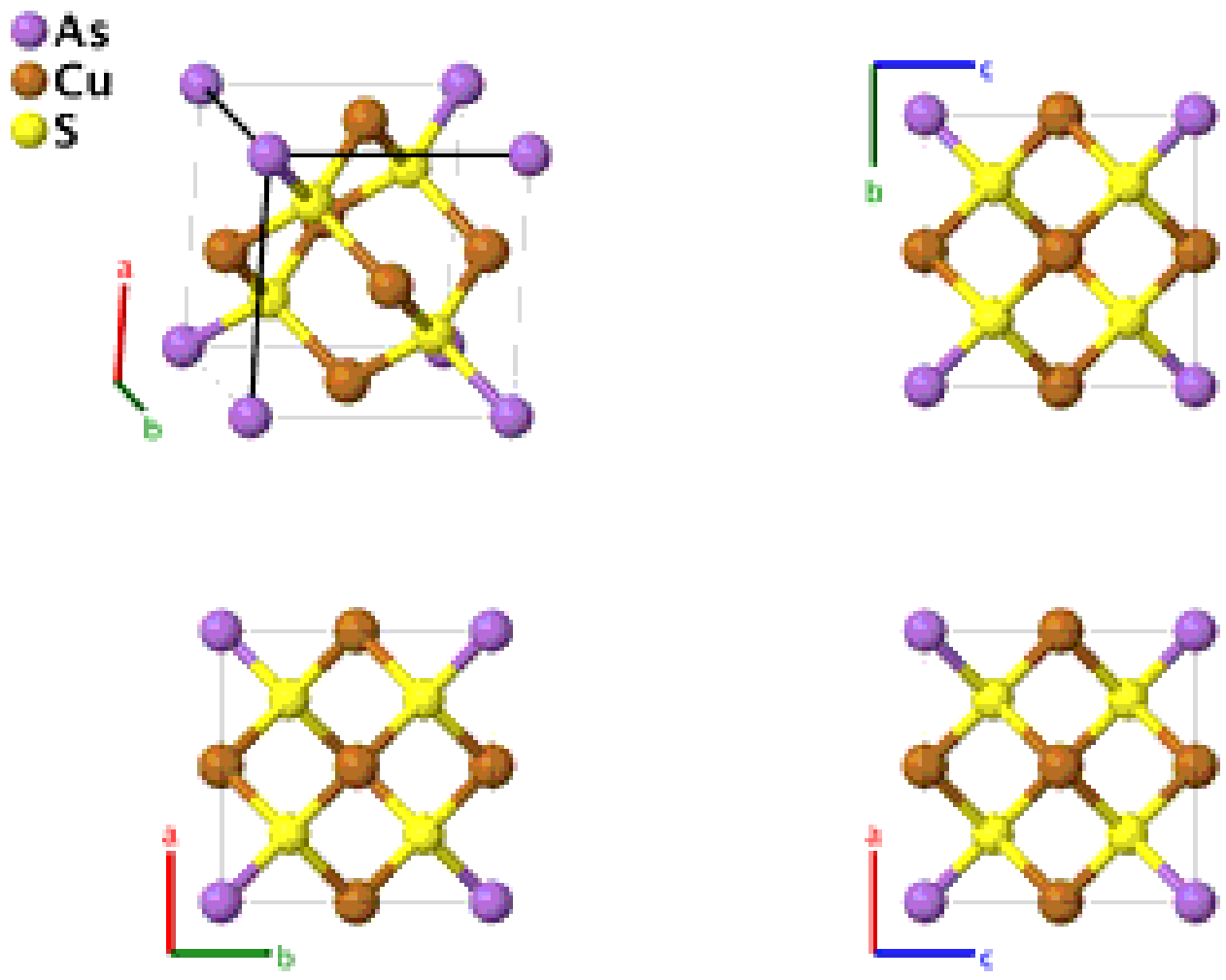

| | | |
|---|---|---|
| **Prototype** | : | $AsCu_3S_4$ |
| **AFLOW prototype label** | : | AB3C4_cP8_215_a_c_e |
| ***Strukturbericht* designation** | : | None |
| **Pearson symbol** | : | cP8 |
| **Space group number** | : | 215 |
| **Space group symbol** | : | $P\bar{4}3m$ |
| **AFLOW prototype command** | : | `aflow --proto=AB3C4_cP8_215_a_c_e --params=`$a, x_3$ |

- This structure is very similar to sulvanite ($H2_4$), except that in this case the copper atoms are on the cubic faces [the (3c) sites] rather than the cubic edges [the (3d) sites]. The actual composition of the sample under study is $Cu_3(As_{0.65}Cu_{0.20}Fe_{0.13})S_4$. We will ignore the alloying on the arsenic site here. The original reference for this structure, (Sclar, 1960), is apparently an abstract [see (Fleischer, 1961)] which does not appear in the online edition of the Geological Society of America Bulletin. We use the data for this structure printed in (Villars, 1991) Vol. I, pp. 1111-1112. Note that (Villars, 2005) gives the reference a different set of authors.

**Simple Cubic primitive vectors:**

$$\mathbf{a}_1 = a\,\hat{\mathbf{x}}$$
$$\mathbf{a}_2 = a\,\hat{\mathbf{y}}$$
$$\mathbf{a}_3 = a\,\hat{\mathbf{z}}$$

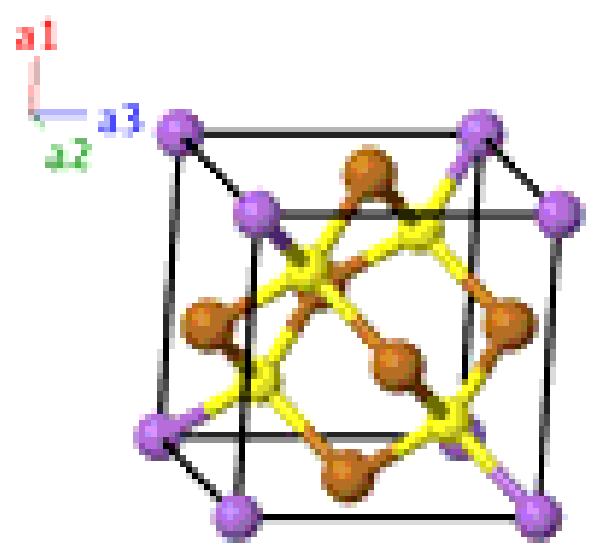

**Basis vectors:**

| | | Lattice Coordinates | | Cartesian Coordinates | Wyckoff Position | Atom Type |
|---|---|---|---|---|---|---|
| $\mathbf{B}_1$ | = | $0\,\mathbf{a}_1 + 0\,\mathbf{a}_2 + 0\,\mathbf{a}_3$ | = | $0\hat{\mathbf{x}} + 0\hat{\mathbf{y}} + 0\hat{\mathbf{z}}$ | $(1a)$ | As |
| $\mathbf{B_2}$ | = | $\frac{1}{2}\,\mathbf{a_2} + \frac{1}{2}\,\mathbf{a_3}$ | = | $\frac{1}{2}\,a\,\hat{\mathbf{y}} + \frac{1}{2}\,a\,\hat{\mathbf{z}}$ | $(3c)$ | Cu |
| $\mathbf{B_3}$ | = | $\frac{1}{2}\,\mathbf{a_1} + \frac{1}{2}\,\mathbf{a_3}$ | = | $\frac{1}{2}\,a\,\hat{\mathbf{x}} + \frac{1}{2}\,a\,\hat{\mathbf{z}}$ | $(3c)$ | Cu |
| $\mathbf{B_4}$ | = | $\frac{1}{2}\,\mathbf{a_1} + \frac{1}{2}\,\mathbf{a_2}$ | = | $\frac{1}{2}\,a\,\hat{\mathbf{x}} + \frac{1}{2}\,a\,\hat{\mathbf{y}}$ | $(3c)$ | Cu |
| $\mathbf{B_5}$ | = | $x_3\,\mathbf{a_1} + x_3\,\mathbf{a_2} + x_3\,\mathbf{a_3}$ | = | $x_3\,a\,\hat{\mathbf{x}} + x_3\,a\,\hat{\mathbf{y}} + x_3\,a\,\hat{\mathbf{z}}$ | $(4e)$ | S |
| $\mathbf{B_6}$ | = | $-x_3\,\mathbf{a_1} - x_3\,\mathbf{a_2} + x_3\,\mathbf{a_3}$ | = | $-x_3\,a\,\hat{\mathbf{x}} - x_3\,a\,\hat{\mathbf{y}} + x_3\,a\,\hat{\mathbf{z}}$ | $(4e)$ | S |
| $\mathbf{B_7}$ | = | $-x_3\,\mathbf{a_1} + x_3\,\mathbf{a_2} - x_3\,\mathbf{a_3}$ | = | $-x_3\,a\,\hat{\mathbf{x}} + x_3\,a\,\hat{\mathbf{y}} - x_3\,a\,\hat{\mathbf{z}}$ | $(4e)$ | S |
| $\mathbf{B_8}$ | = | $x_3\,\mathbf{a_1} - x_3\,\mathbf{a_2} - x_3\,\mathbf{a_3}$ | = | $x_3\,a\,\hat{\mathbf{x}} - x_3\,a\,\hat{\mathbf{y}} - x_3\,a\,\hat{\mathbf{z}}$ | $(4e)$ | S |

---

**Found in:**

---

**Geometry files:**

- CIF: pp. 764
- POSCAR: pp. 765

# AuBe$_5$ (C15$_b$) Structure: AB5_cF24_216_a_ce

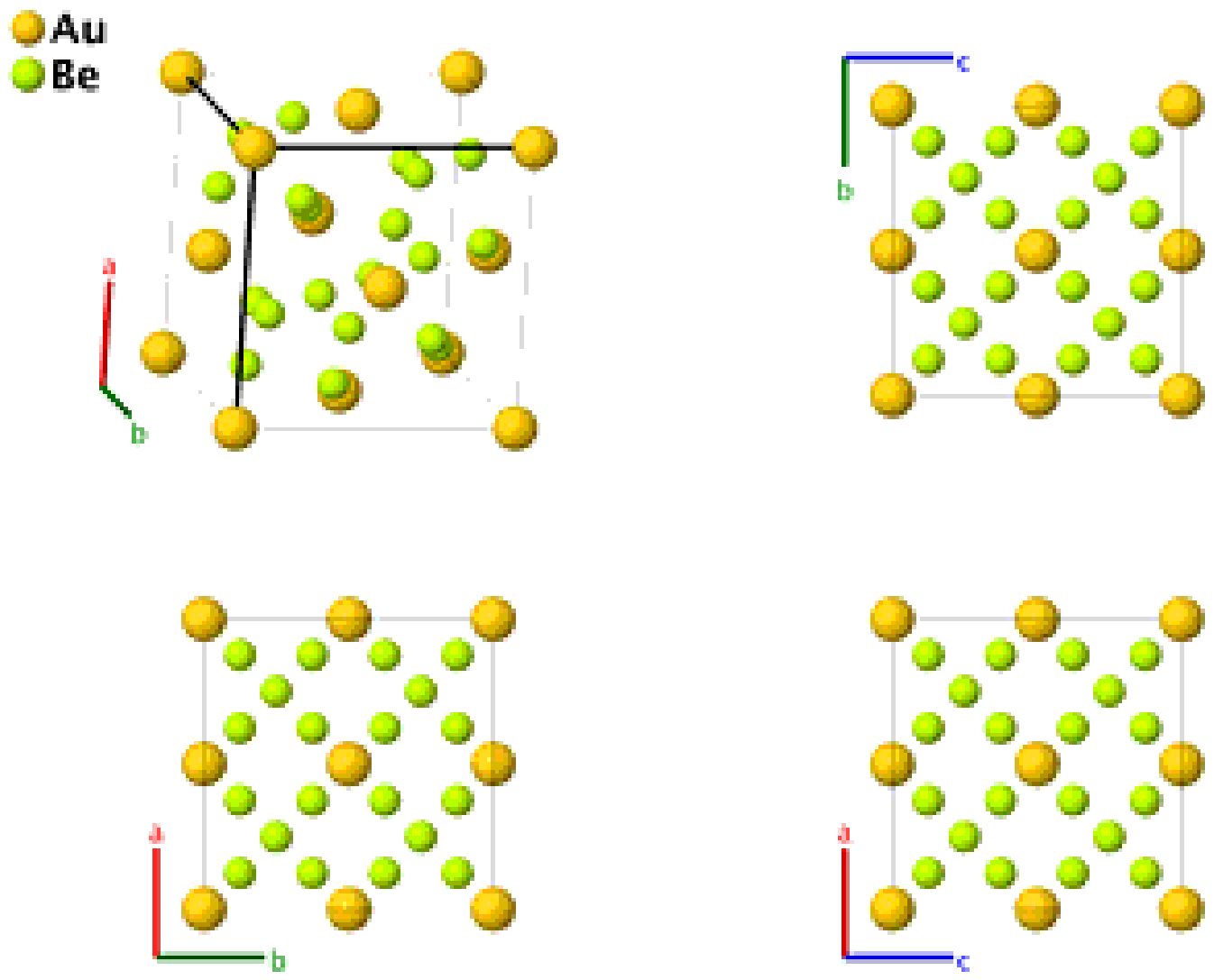

| | | |
|---|---|---|
| **Prototype** | : | AuBe$_5$ |
| **AFLOW prototype label** | : | AB5_cF24_216_a_ce |
| ***Strukturbericht* designation** | : | C15$_b$ |
| **Pearson symbol** | : | cF24 |
| **Space group number** | : | 216 |
| **Space group symbol** | : | F$\bar{4}$3m |
| **AFLOW prototype command** | : | `aflow --proto=AB5_cF24_216_a_ce --params=`$a, x_3$ |

**Other compounds with this structure:**

- MgSnCu$_4$, AuNi$_4$Y, Pt$_5$U, many more

- The lattice constant for this structure is taken from (Batchelder, 1958), which does not give the internal coordinate for the (16c) site. However, (Baenziger, 1950) assumes that uranium compounds of this type have an internal parameter $x_3 \approx 5/8$. (Pearson, 1958) uses this to infer a value of $x_3 \approx 5/8$ here as well.

**Face-centered Cubic primitive vectors:**

$$\mathbf{a}_1 = \frac{1}{2}\,a\,\hat{\mathbf{y}} + \frac{1}{2}\,a\,\hat{\mathbf{z}}$$
$$\mathbf{a}_2 = \frac{1}{2}\,a\,\hat{\mathbf{x}} + \frac{1}{2}\,a\,\hat{\mathbf{z}}$$
$$\mathbf{a}_3 = \frac{1}{2}\,a\,\hat{\mathbf{x}} + \frac{1}{2}\,a\,\hat{\mathbf{y}}$$

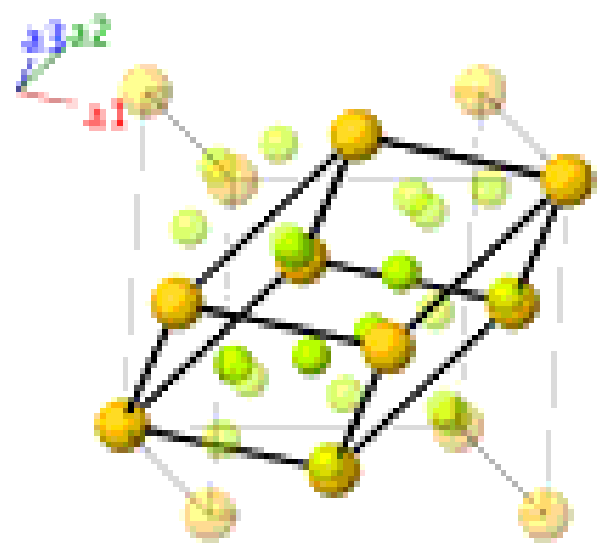

**Basis vectors:**

| Lattice Coordinates | Cartesian Coordinates | Wyckoff Position | Atom Type |
|---|---|---|---|

| | | | | | | |
|---|---|---|---|---|---|---|
| $\mathbf{B_1}$ | = | $0\,\mathbf{a_1} + 0\,\mathbf{a_2} + 0\,\mathbf{a_3}$ | = | $0\hat{\mathbf{x}} + 0\hat{\mathbf{y}} + 0\hat{\mathbf{z}}$ | $(4a)$ | Au |
| $\mathbf{B_2}$ | = | $\frac{1}{4}\,\mathbf{a_1} + \frac{1}{4}\,\mathbf{a_2} + \frac{1}{4}\,\mathbf{a_3}$ | = | $\frac{1}{4}\,a\,\hat{\mathbf{x}} + \frac{1}{4}\,a\,\hat{\mathbf{y}} + \frac{1}{4}\,a\,\hat{\mathbf{z}}$ | $(4c)$ | Be I |
| $\mathbf{B_3}$ | = | $x_3\,\mathbf{a_1} + x_3\,\mathbf{a_2} + x_3\,\mathbf{a_3}$ | = | $x_3\,a\,\hat{\mathbf{x}} + x_3\,a\,\hat{\mathbf{y}} + x_3\,a\,\hat{\mathbf{z}}$ | $(16e)$ | Be II |
| $\mathbf{B_4}$ | = | $x_3\,\mathbf{a_1} + x_3\,\mathbf{a_2} - 3\,x_3\,\mathbf{a_3}$ | = | $-x_3\,a\,\hat{\mathbf{x}} - x_3\,a\,\hat{\mathbf{y}} + x_3\,a\,\hat{\mathbf{z}}$ | $(16e)$ | Be II |
| $\mathbf{B_5}$ | = | $x_3\,\mathbf{a_1} - 3\,x_3\,\mathbf{a_2} + x_3\,\mathbf{a_3}$ | = | $-x_3\,a\,\hat{\mathbf{x}} + x_3\,a\,\hat{\mathbf{y}} - x_3\,a\,\hat{\mathbf{z}}$ | $(16e)$ | Be II |
| $\mathbf{B_6}$ | = | $-3\,x_3\,\mathbf{a_1} + x_3\,\mathbf{a_2} + x_3\,\mathbf{a_3}$ | = | $x_3\,a\,\hat{\mathbf{x}} - x_3\,a\,\hat{\mathbf{y}} - x_3\,a\,\hat{\mathbf{z}}$ | $(16e)$ | Be II |

---

**Found in:**

---

**Geometry files:**

- CIF: pp. 765
- POSCAR: pp. 765

# Half-Heusler ($C1_b$) Structure: ABC_cF12_216_b_c_a

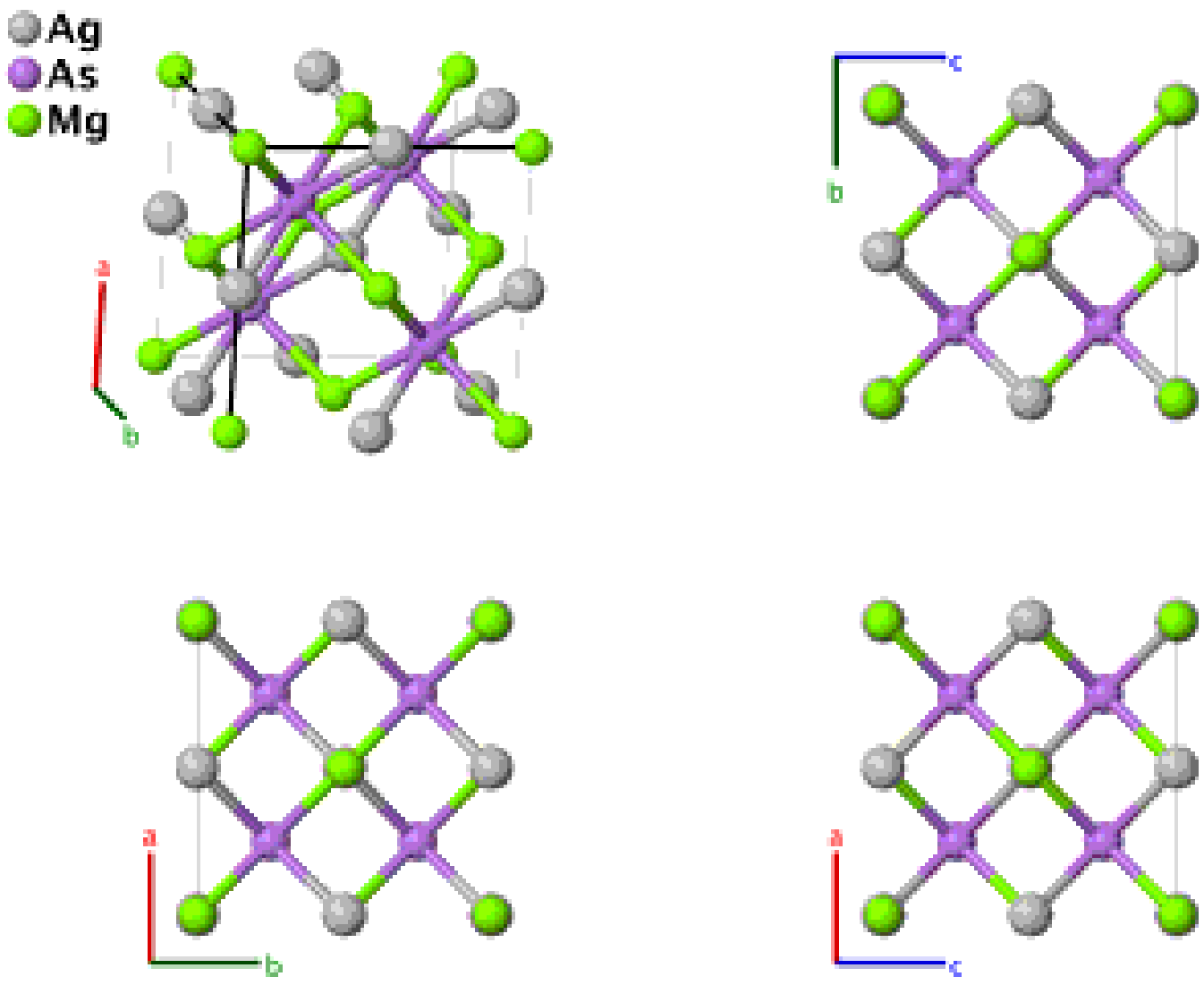

| | | |
|---|---|---|
| **Prototype** | : | AgAsMg |
| **AFLOW prototype label** | : | ABC_cF12_216_b_c_a |
| ***Strukturbericht* designation** | : | $C1_b$ |
| **Pearson symbol** | : | cF12 |
| **Space group number** | : | 216 |
| **Space group symbol** | : | $F\bar{4}3m$ |
| **AFLOW prototype command** | : | `aflow --proto=ABC_cF12_216_b_c_a --params=`$a$ |

**Other compounds with this structure:**

- MnNiSb, AuMgSn, CdLiP, BiMgNi, RhSnTi, numerous

- All of the atoms are located on the sites of a body-centered cubic lattice. This is sometimes called the "half-Heusler" structure because it is identical to the $L2_1$ (Heusler) structure with half of the copper atoms missing. The Mg and Ag atoms form a rock salt (B1) structure, while the As and either the Mg or Ag atoms form a zincblende (B3) structure. If the atoms on the (4a) and (4c) sites are identical, this reduces to the fluorite (C1) structure.

**Face-centered Cubic primitive vectors:**

$$
\begin{aligned}
\mathbf{a}_1 &= \tfrac{1}{2}\,a\,\hat{\mathbf{y}} + \tfrac{1}{2}\,a\,\hat{\mathbf{z}} \\
\mathbf{a}_2 &= \tfrac{1}{2}\,a\,\hat{\mathbf{x}} + \tfrac{1}{2}\,a\,\hat{\mathbf{z}} \\
\mathbf{a}_3 &= \tfrac{1}{2}\,a\,\hat{\mathbf{x}} + \tfrac{1}{2}\,a\,\hat{\mathbf{y}}
\end{aligned}
$$

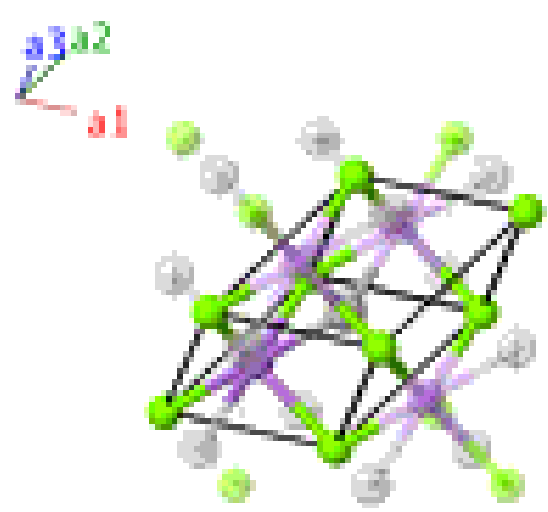

**Basis vectors:**

| | | Lattice Coordinates | | Cartesian Coordinates | Wyckoff Position | Atom Type |
|---|---|---|---|---|---|---|
| $\mathbf{B}_1$ | = | $0\,\mathbf{a}_1 + 0\,\mathbf{a}_2 + 0\,\mathbf{a}_3$ | = | $0\hat{\mathbf{x}} + 0\hat{\mathbf{y}} + 0\hat{\mathbf{z}}$ | $(4a)$ | Mg |
| $\mathbf{B}_2$ | = | $\frac{1}{2}\mathbf{a}_1 + \frac{1}{2}\mathbf{a}_2 + \frac{1}{2}\mathbf{a}_3$ | = | $\frac{1}{2}\,a\,\hat{\mathbf{x}} + \frac{1}{2}\,a\,\hat{\mathbf{y}} + \frac{1}{2}\,a\,\hat{\mathbf{z}}$ | $(4b)$ | Ag |
| $\mathbf{B}_3$ | = | $\frac{1}{4}\mathbf{a}_1 + \frac{1}{4}\mathbf{a}_2 + \frac{1}{4}\mathbf{a}_3$ | = | $\frac{1}{4}\,a\,\hat{\mathbf{x}} + \frac{1}{4}\,a\,\hat{\mathbf{y}} + \frac{1}{4}\,a\,\hat{\mathbf{z}}$ | $(4c)$ | As |

---

**Found in:**

---

**Geometry files:**

- CIF: pp. 766
- POSCAR: pp. 766

# Zincblende (ZnS, B3) Structure: AB_cF8_216_c_a

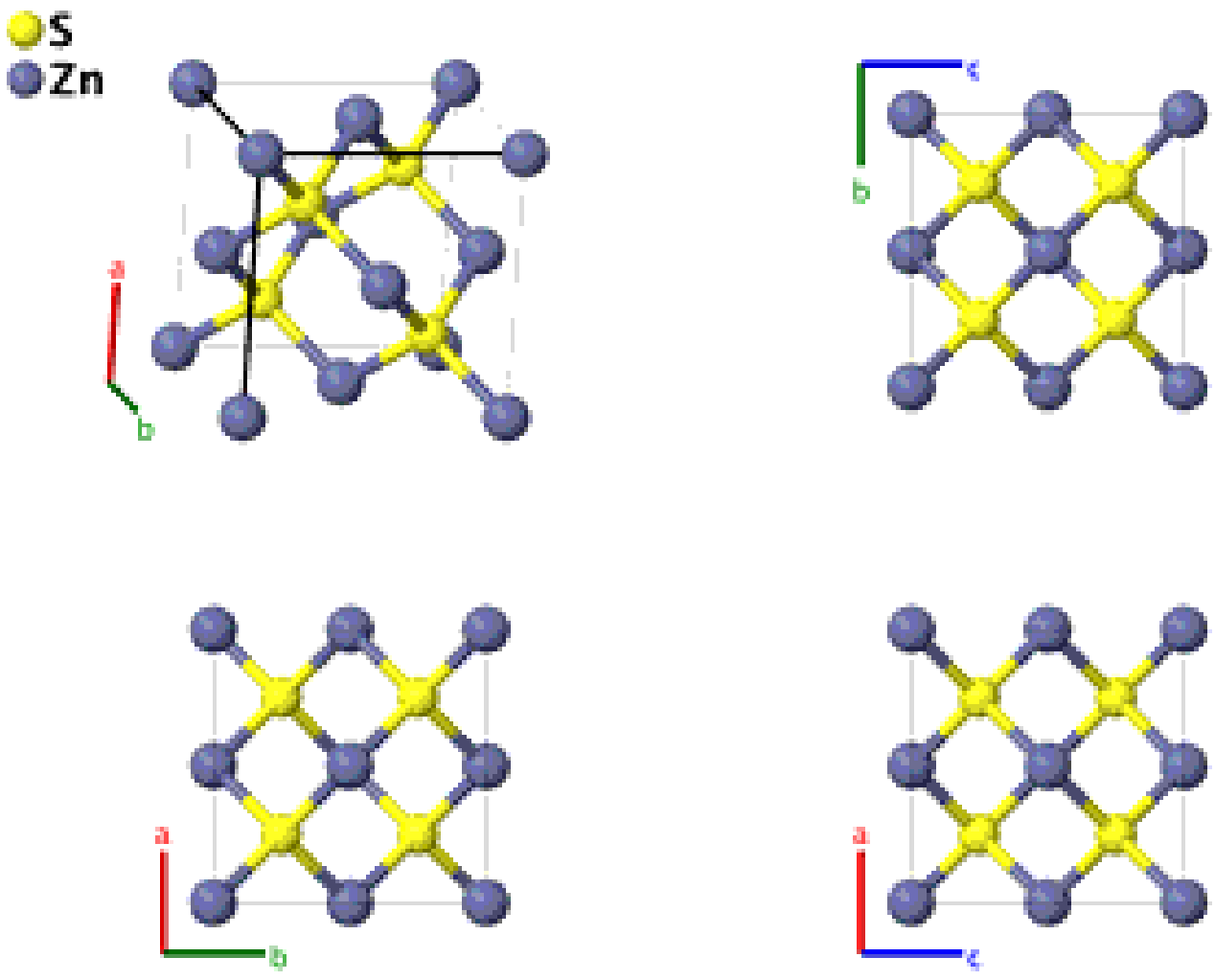

| | | |
|---|---|---|
| **Prototype** | : | ZnS |
| **AFLOW prototype label** | : | AB_cF8_216_c_a |
| ***Strukturbericht* designation** | : | B3 |
| **Pearson symbol** | : | cF8 |
| **Space group number** | : | 216 |
| **Space group symbol** | : | $F\bar{4}3m$ |
| **AFLOW prototype command** | : | `aflow --proto=AB_cF8_216_c_a --params=`$a$ |

- This is the cubic analog of the wurtzite lattice, i.e. the stacking of the ZnS dimers along the <111> direction is ABCABC ... This is also a two-component analog of the diamond structure, without the inversion symmetry in the middle of the bond.

**Face-centered Cubic primitive vectors:**

$$\mathbf{a}_1 = \frac{1}{2}\,a\,\mathbf{\hat{y}} + \frac{1}{2}\,a\,\mathbf{\hat{z}}$$
$$\mathbf{a}_2 = \frac{1}{2}\,a\,\mathbf{\hat{x}} + \frac{1}{2}\,a\,\mathbf{\hat{z}}$$
$$\mathbf{a}_3 = \frac{1}{2}\,a\,\mathbf{\hat{x}} + \frac{1}{2}\,a\,\mathbf{\hat{y}}$$

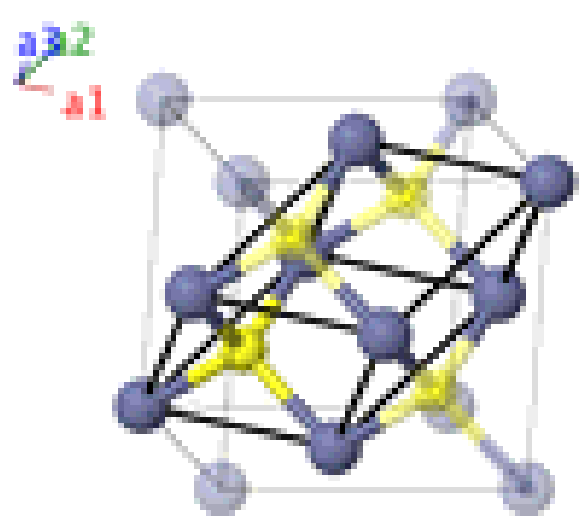

**Basis vectors:**

| | | Lattice Coordinates | | Cartesian Coordinates | Wyckoff Position | Atom Type |
|---|---|---|---|---|---|---|
| $\mathbf{B}_1$ | = | $0\,\mathbf{a}_1 + 0\,\mathbf{a}_2 + 0\,\mathbf{a}_3$ | = | $0\mathbf{\hat{x}} + 0\mathbf{\hat{y}} + 0\mathbf{\hat{z}}$ | (4$a$) | Zn |
| $\mathbf{B}_2$ | = | $\frac{1}{4}\,\mathbf{a}_1 + \frac{1}{4}\,\mathbf{a}_2 + \frac{1}{4}\,\mathbf{a}_3$ | = | $\frac{1}{4}\,a\,\mathbf{\hat{x}} + \frac{1}{4}\,a\,\mathbf{\hat{y}} + \frac{1}{4}\,a\,\mathbf{\hat{z}}$ | (4$c$) | S |

**Found in:**

---

**Geometry files:**

- CIF: pp. 766
- POSCAR: pp. 767

# $SiF_4$ Structure: A4B_cI10_217_c_a

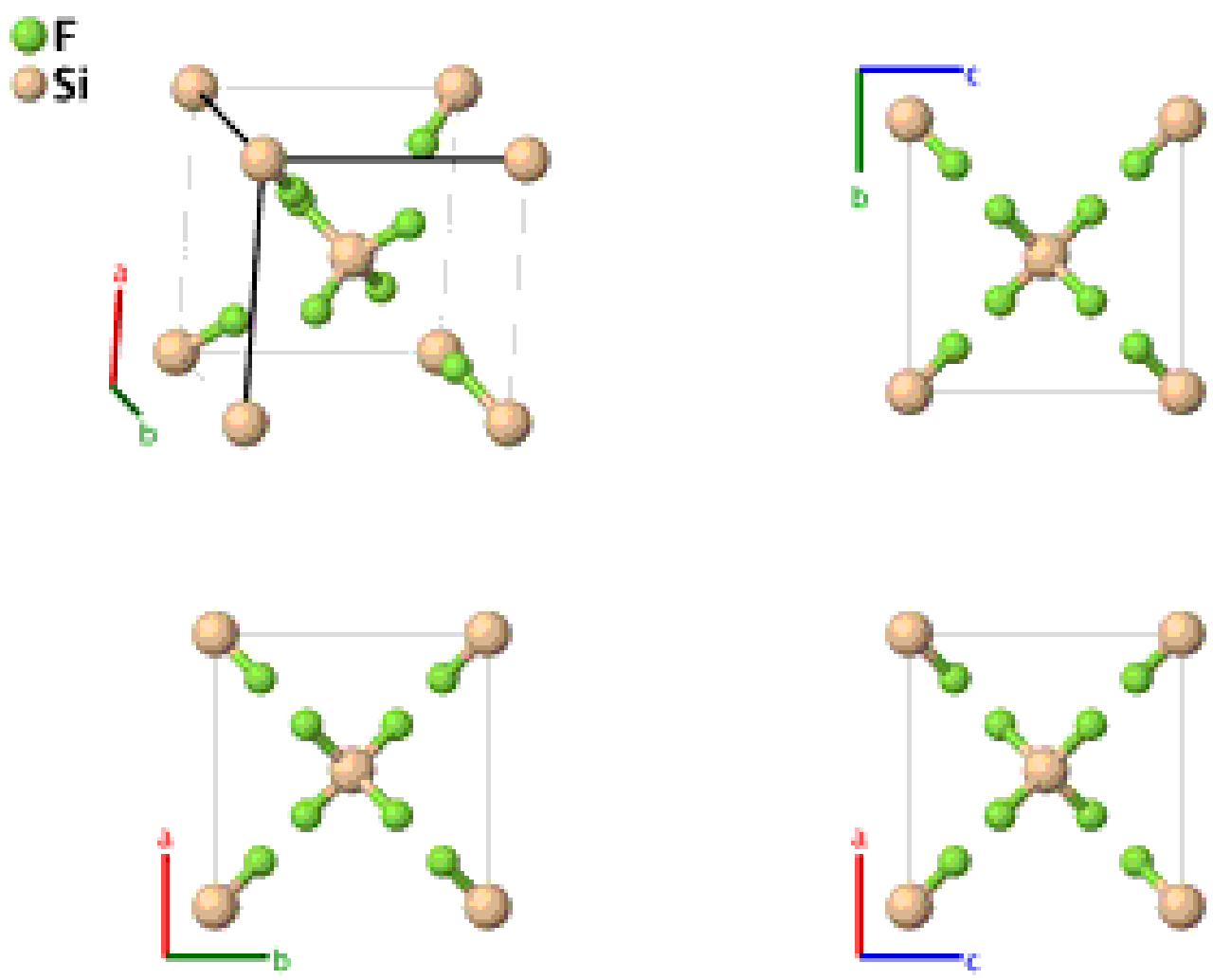

| | | |
|---|---|---|
| **Prototype** | : | $SiF_4$ |
| **AFLOW prototype label** | : | A4B_cI10_217_c_a |
| ***Strukturbericht* designation** | : | None |
| **Pearson symbol** | : | cI10 |
| **Space group number** | : | 217 |
| **Space group symbol** | : | $I\bar{4}3m$ |
| **AFLOW prototype command** | : | `aflow --proto=A4B_cI10_217_c_a --params=`$a, x_2$ |

- We determined the lattice constant for this structure from the internal coordinates and the Si-F bond length given in the reference.

**Body-centered Cubic primitive vectors:**

$$\mathbf{a}_1 = -\frac{1}{2}\,a\,\hat{\mathbf{x}} + \frac{1}{2}\,a\,\hat{\mathbf{y}} + \frac{1}{2}\,a\,\hat{\mathbf{z}}$$

$$\mathbf{a}_2 = \frac{1}{2}\,a\,\hat{\mathbf{x}} - \frac{1}{2}\,a\,\hat{\mathbf{y}} + \frac{1}{2}\,a\,\hat{\mathbf{z}}$$

$$\mathbf{a}_3 = \frac{1}{2}\,a\,\hat{\mathbf{x}} + \frac{1}{2}\,a\,\hat{\mathbf{y}} - \frac{1}{2}\,a\,\hat{\mathbf{z}}$$

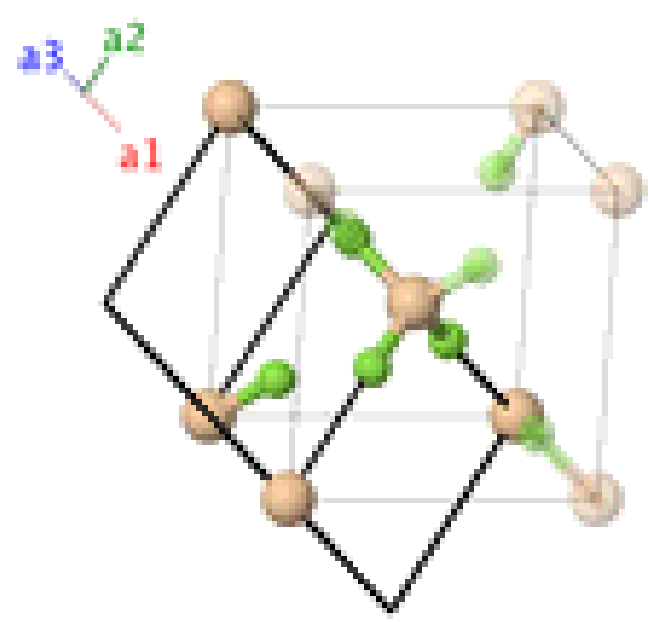

**Basis vectors:**

| | | Lattice Coordinates | | Cartesian Coordinates | Wyckoff Position | Atom Type |
|---|---|---|---|---|---|---|
| $\mathbf{B}_1$ | = | $0\,\mathbf{a}_1 + 0\,\mathbf{a}_2 + 0\,\mathbf{a}_3$ | = | $0\hat{\mathbf{x}} + 0\hat{\mathbf{y}} + 0\hat{\mathbf{z}}$ | (2$a$) | Si |
| $\mathbf{B}_2$ | = | $2x_2\,\mathbf{a}_1 + 2x_2\,\mathbf{a}_2 + 2x_2\,\mathbf{a}_3$ | = | $x_2\,a\,\hat{\mathbf{x}} + x_2\,a\,\hat{\mathbf{y}} + x_2\,a\,\hat{\mathbf{z}}$ | (8$c$) | F |
| $\mathbf{B}_3$ | = | $-2x_2\,\mathbf{a}_3$ | = | $-x_2\,a\,\hat{\mathbf{x}} - x_2\,a\,\hat{\mathbf{y}} + x_2\,a\,\hat{\mathbf{z}}$ | (8$c$) | F |
| $\mathbf{B}_4$ | = | $-2x_2\,\mathbf{a}_2$ | = | $-x_2\,a\,\hat{\mathbf{x}} + x_2\,a\,\hat{\mathbf{y}} - x_2\,a\,\hat{\mathbf{z}}$ | (8$c$) | F |
| $\mathbf{B}_5$ | = | $-2x_2\,\mathbf{a}_1$ | = | $x_2\,a\,\hat{\mathbf{x}} - x_2\,a\,\hat{\mathbf{y}} - x_2\,a\,\hat{\mathbf{z}}$ | (8$c$) | F |

**Geometry files:**

- CIF: pp. 767
- POSCAR: pp. 768

# $\alpha$-Mn (A12) Structure: A_cI58_217_ac2g

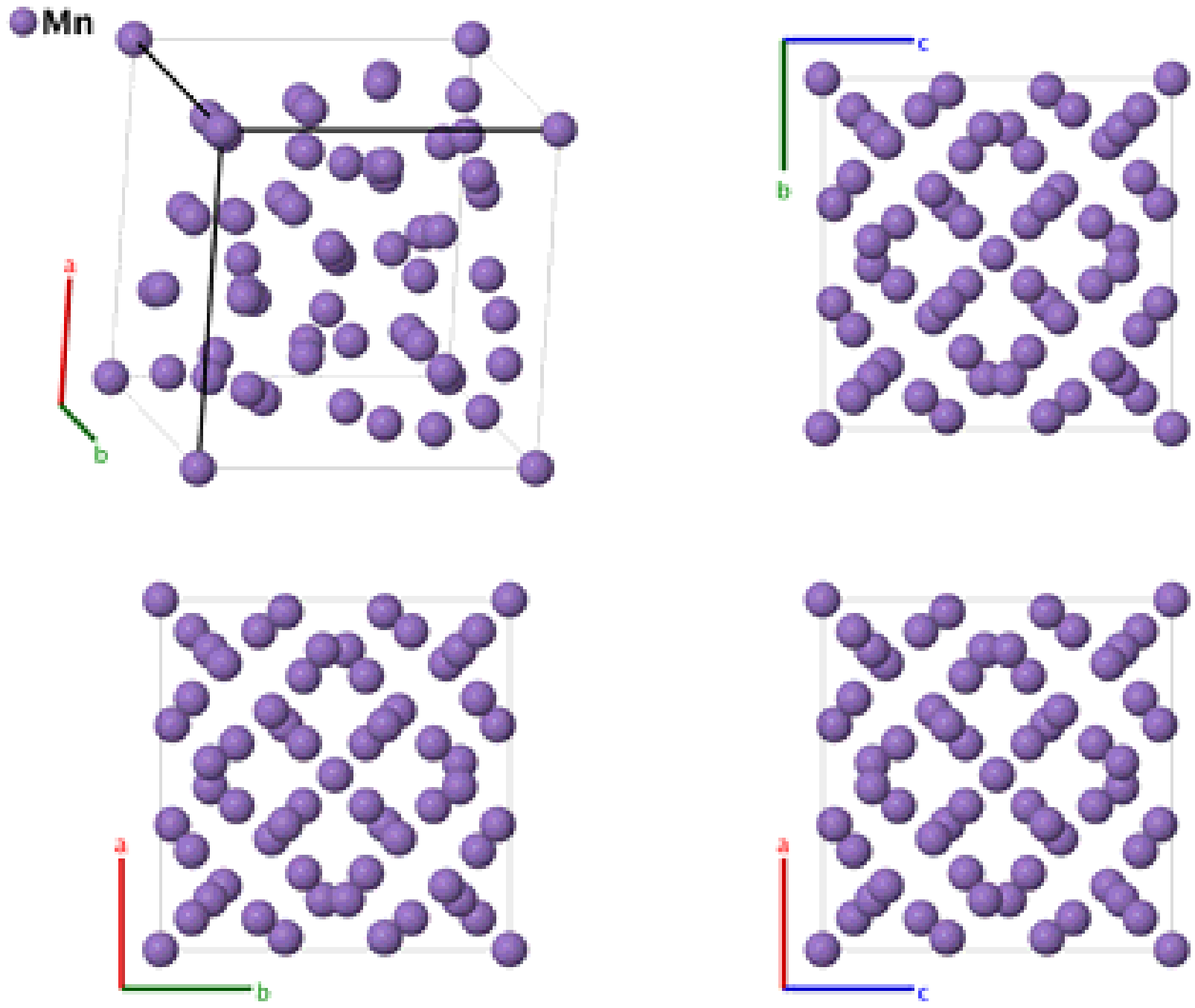

| | | |
|---|---|---|
| **Prototype** | : | $\alpha$-Mn |
| **AFLOW prototype label** | : | A_cI58_217_ac2g |
| ***Strukturbericht* designation** | : | A12 |
| **Pearson symbol** | : | cI58 |
| **Space group number** | : | 217 |
| **Space group symbol** | : | $I\bar{4}3m$ |
| **AFLOW prototype command** | : | `aflow --proto=A_cI58_217_ac2g`<br>`--params=`$a, x_2, x_3, z_3, x_4, z_4$ |

**Body-centered Cubic primitive vectors:**

$$\mathbf{a}_1 = -\frac{1}{2}\,a\,\hat{\mathbf{x}} + \frac{1}{2}\,a\,\hat{\mathbf{y}} + \frac{1}{2}\,a\,\hat{\mathbf{z}}$$

$$\mathbf{a}_2 = \frac{1}{2}\,a\,\hat{\mathbf{x}} - \frac{1}{2}\,a\,\hat{\mathbf{y}} + \frac{1}{2}\,a\,\hat{\mathbf{z}}$$

$$\mathbf{a}_3 = \frac{1}{2}\,a\,\hat{\mathbf{x}} + \frac{1}{2}\,a\,\hat{\mathbf{y}} - \frac{1}{2}\,a\,\hat{\mathbf{z}}$$

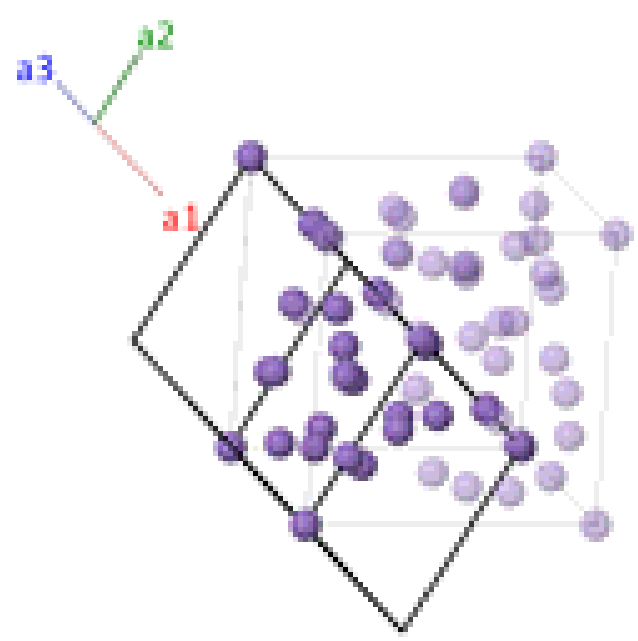

**Basis vectors:**

| | | Lattice Coordinates | | Cartesian Coordinates | Wyckoff Position | Atom Type |
|---|---|---|---|---|---|---|
| $\mathbf{B}_1$ | = | $0\,\mathbf{a}_1 + 0\,\mathbf{a}_2 + 0\,\mathbf{a}_3$ | = | $0\hat{\mathbf{x}} + 0\hat{\mathbf{y}} + 0\hat{\mathbf{z}}$ | $(2a)$ | Mn I |
| $\mathbf{B}_2$ | = | $2x_2\,\mathbf{a}_1 + 2x_2\,\mathbf{a}_2 + 2x_2\,\mathbf{a}_3$ | = | $x_2\,a\,\hat{\mathbf{x}} + x_2\,a\,\hat{\mathbf{y}} + x_2\,a\,\hat{\mathbf{z}}$ | $(8c)$ | Mn II |
| $\mathbf{B}_3$ | = | $-2x_2\,\mathbf{a}_3$ | = | $-x_2\,a\,\hat{\mathbf{x}} - x_2\,a\,\hat{\mathbf{y}} + x_2\,a\,\hat{\mathbf{z}}$ | $(8c)$ | Mn II |
| $\mathbf{B}_4$ | = | $-2x_2\,\mathbf{a}_2$ | = | $-x_2\,a\,\hat{\mathbf{x}} + x_2\,a\,\hat{\mathbf{y}} - x_2\,a\,\hat{\mathbf{z}}$ | (8$c$) | Mn II |
| $\mathbf{B}_5$ | = | $-2x_2\,\mathbf{a}_1$ | = | $x_2\,a\,\hat{\mathbf{x}} - x_2\,a\,\hat{\mathbf{y}} - x_2\,a\,\hat{\mathbf{z}}$ | (8$c$) | Mn II |
| $\mathbf{B}_6$ | = | $(x_3+z_3)\,\mathbf{a}_1 + (x_3+z_3)\,\mathbf{a}_2 + 2x_3\,\mathbf{a}_3$ | = | $x_3\,a\,\hat{\mathbf{x}} + x_3\,a\,\hat{\mathbf{y}} + z_3\,a\,\hat{\mathbf{z}}$ | (24$g$) | Mn III |
| $\mathbf{B}_7$ | = | $(z_3-x_3)\,\mathbf{a}_1 + (z_3-x_3)\,\mathbf{a}_2 - 2x_3\,\mathbf{a}_3$ | = | $-x_3\,a\,\hat{\mathbf{x}} - x_3\,a\,\hat{\mathbf{y}} + z_3\,a\,\hat{\mathbf{z}}$ | (24$g$) | Mn III |
| $\mathbf{B}_8$ | = | $(x_3-z_3)\,\mathbf{a}_1 - (x_3+z_3)\,\mathbf{a}_2$ | = | $-x_3\,a\,\hat{\mathbf{x}} + x_3\,a\,\hat{\mathbf{y}} - z_3\,a\,\hat{\mathbf{z}}$ | (24$g$) | Mn III |
| $\mathbf{B}_9$ | = | $-(x_3+z_3)\,\mathbf{a}_1 + (x_3-z_3)\,\mathbf{a}_2$ | = | $x_3\,a\,\hat{\mathbf{x}} - x_3\,a\,\hat{\mathbf{y}} - z_3\,a\,\hat{\mathbf{z}}$ | (24$g$) | Mn III |
| $\mathbf{B}_{10}$ | = | $2x_3\,\mathbf{a}_1 + (x_3+z_3)\,\mathbf{a}_2 + (x_3+z_3)\,\mathbf{a}_3$ | = | $z_3\,a\,\hat{\mathbf{x}} + x_3\,a\,\hat{\mathbf{y}} + x_3\,a\,\hat{\mathbf{z}}$ | (24$g$) | Mn III |
| $\mathbf{B}_{11}$ | = | $-2x_3\,\mathbf{a}_1 + (z_3-x_3)\,\mathbf{a}_2 + (z_3-x_3)\,\mathbf{a}_3$ | = | $z_3\,a\,\hat{\mathbf{x}} - x_3\,a\,\hat{\mathbf{y}} - x_3\,a\,\hat{\mathbf{z}}$ | (24$g$) | Mn III |
| $\mathbf{B}_{12}$ | = | $(x_3-z_3)\,\mathbf{a}_2 - (x_3+z_3)\,\mathbf{a}_3$ | = | $-z_3\,a\,\hat{\mathbf{x}} - x_3\,a\,\hat{\mathbf{y}} + x_3\,a\,\hat{\mathbf{z}}$ | (24$g$) | Mn III |
| $\mathbf{B}_{13}$ | = | $-(x_3+z_3)\,\mathbf{a}_2 + (x_3-z_3)\,\mathbf{a}_3$ | = | $-z_3\,a\,\hat{\mathbf{x}} + x_3\,a\,\hat{\mathbf{y}} - x_3\,a\,\hat{\mathbf{z}}$ | (24$g$) | Mn III |
| $\mathbf{B}_{14}$ | = | $(x_3+z_3)\,\mathbf{a}_1 + 2x_3\,\mathbf{a}_2 + (x_3+z_3)\,\mathbf{a}_3$ | = | $x_3\,a\,\hat{\mathbf{x}} + z_3\,a\,\hat{\mathbf{y}} + x_3\,a\,\hat{\mathbf{z}}$ | (24$g$) | Mn III |
| $\mathbf{B}_{15}$ | = | $(z_3-x_3)\,\mathbf{a}_1 - 2x_3\,\mathbf{a}_2 + (z_3-x_3)\,\mathbf{a}_3$ | = | $-x_3\,a\,\hat{\mathbf{x}} + z_3\,a\,\hat{\mathbf{y}} - x_3\,a\,\hat{\mathbf{z}}$ | (24$g$) | Mn III |
| $\mathbf{B}_{16}$ | = | $-(x_3+z_3)\,\mathbf{a}_1 + (x_3-z_3)\,\mathbf{a}_3$ | = | $x_3\,a\,\hat{\mathbf{x}} - z_3\,a\,\hat{\mathbf{y}} - x_3\,a\,\hat{\mathbf{z}}$ | (24$g$) | Mn III |
| $\mathbf{B}_{17}$ | = | $(x_3-z_3)\,\mathbf{a}_1 - (x_3+z_3)\,\mathbf{a}_3$ | = | $-x_3\,a\,\hat{\mathbf{x}} - z_3\,a\,\hat{\mathbf{y}} + x_3\,a\,\hat{\mathbf{z}}$ | (24$g$) | Mn III |
| $\mathbf{B}_{18}$ | = | $(x_4+z_4)\,\mathbf{a}_1 + (x_4+z_4)\,\mathbf{a}_2 + 2x_4\,\mathbf{a}_3$ | = | $x_4\,a\,\hat{\mathbf{x}} + x_4\,a\,\hat{\mathbf{y}} + z_4\,a\,\hat{\mathbf{z}}$ | (24$g$) | Mn IV |
| $\mathbf{B}_{19}$ | = | $(z_4-x_4)\,\mathbf{a}_1 + (z_4-x_4)\,\mathbf{a}_2 - 2x_4\,\mathbf{a}_3$ | = | $-x_4\,a\,\hat{\mathbf{x}} - x_4\,a\,\hat{\mathbf{y}} + z_4\,a\,\hat{\mathbf{z}}$ | (24$g$) | Mn IV |
| $\mathbf{B}_{20}$ | = | $(x_4-z_4)\,\mathbf{a}_1 - (x_4+z_4)\,\mathbf{a}_2$ | = | $-x_4\,a\,\hat{\mathbf{x}} + x_4\,a\,\hat{\mathbf{y}} - z_4\,a\,\hat{\mathbf{z}}$ | (24$g$) | Mn IV |
| $\mathbf{B}_{21}$ | = | $-(x_4+z_4)\,\mathbf{a}_1 + (x_4-z_4)\,\mathbf{a}_2$ | = | $x_4\,a\,\hat{\mathbf{x}} - x_4\,a\,\hat{\mathbf{y}} - z_4\,a\,\hat{\mathbf{z}}$ | (24$g$) | Mn IV |
| $\mathbf{B}_{22}$ | = | $2x_4\,\mathbf{a}_1 + (x_4+z_4)\,\mathbf{a}_2 + (x_4+z_4)\,\mathbf{a}_3$ | = | $z_4\,a\,\hat{\mathbf{x}} + x_4\,a\,\hat{\mathbf{y}} + x_4\,a\,\hat{\mathbf{z}}$ | (24$g$) | Mn IV |
| $\mathbf{B}_{23}$ | = | $-2x_4\,\mathbf{a}_1 + (z_4-x_4)\,\mathbf{a}_2 + (z_4-x_4)\,\mathbf{a}_3$ | = | $z_4\,a\,\hat{\mathbf{x}} - x_4\,a\,\hat{\mathbf{y}} - x_4\,a\,\hat{\mathbf{z}}$ | (24$g$) | Mn IV |
| $\mathbf{B}_{24}$ | = | $(x_4-z_4)\,\mathbf{a}_2 - (x_4+z_4)\,\mathbf{a}_3$ | = | $-z_4\,a\,\hat{\mathbf{x}} - x_4\,a\,\hat{\mathbf{y}} + x_4\,a\,\hat{\mathbf{z}}$ | (24$g$) | Mn IV |
| $\mathbf{B}_{25}$ | = | $-(x_4+z_4)\,\mathbf{a}_2 + (x_4-z_4)\,\mathbf{a}_3$ | = | $-z_4\,a\,\hat{\mathbf{x}} + x_4\,a\,\hat{\mathbf{y}} - x_4\,a\,\hat{\mathbf{z}}$ | (24$g$) | Mn IV |
| $\mathbf{B}_{26}$ | = | $(x_4+z_4)\,\mathbf{a}_1 + 2x_4\,\mathbf{a}_2 + (x_4+z_4)\,\mathbf{a}_3$ | = | $x_4\,a\,\hat{\mathbf{x}} + z_4\,a\,\hat{\mathbf{y}} + x_4\,a\,\hat{\mathbf{z}}$ | (24$g$) | Mn IV |
| $\mathbf{B}_{27}$ | = | $(z_4-x_4)\,\mathbf{a}_1 - 2x_4\,\mathbf{a}_2 + (z_4-x_4)\,\mathbf{a}_3$ | = | $-x_4\,a\,\hat{\mathbf{x}} + z_4\,a\,\hat{\mathbf{y}} - x_4\,a\,\hat{\mathbf{z}}$ | (24$g$) | Mn IV |
| $\mathbf{B}_{28}$ | = | $-(x_4+z_4)\,\mathbf{a}_1 + (x_4-z_4)\,\mathbf{a}_3$ | = | $x_4\,a\,\hat{\mathbf{x}} - z_4\,a\,\hat{\mathbf{y}} - x_4\,a\,\hat{\mathbf{z}}$ | (24$g$) | Mn IV |
| $\mathbf{B}_{29}$ | = | $(x_4-z_4)\,\mathbf{a}_1 - (x_4+z_4)\,\mathbf{a}_3$ | = | $-x_4\,a\,\hat{\mathbf{x}} - z_4\,a\,\hat{\mathbf{y}} + x_4\,a\,\hat{\mathbf{z}}$ | (24$g$) | Mn IV |

**Geometry files:**

- CIF: pp. 768
- POSCAR: pp. 768

# $\gamma$-Brass ($Cu_5Zn_8$) Structure: A5B8_cI52_217_ce_cg

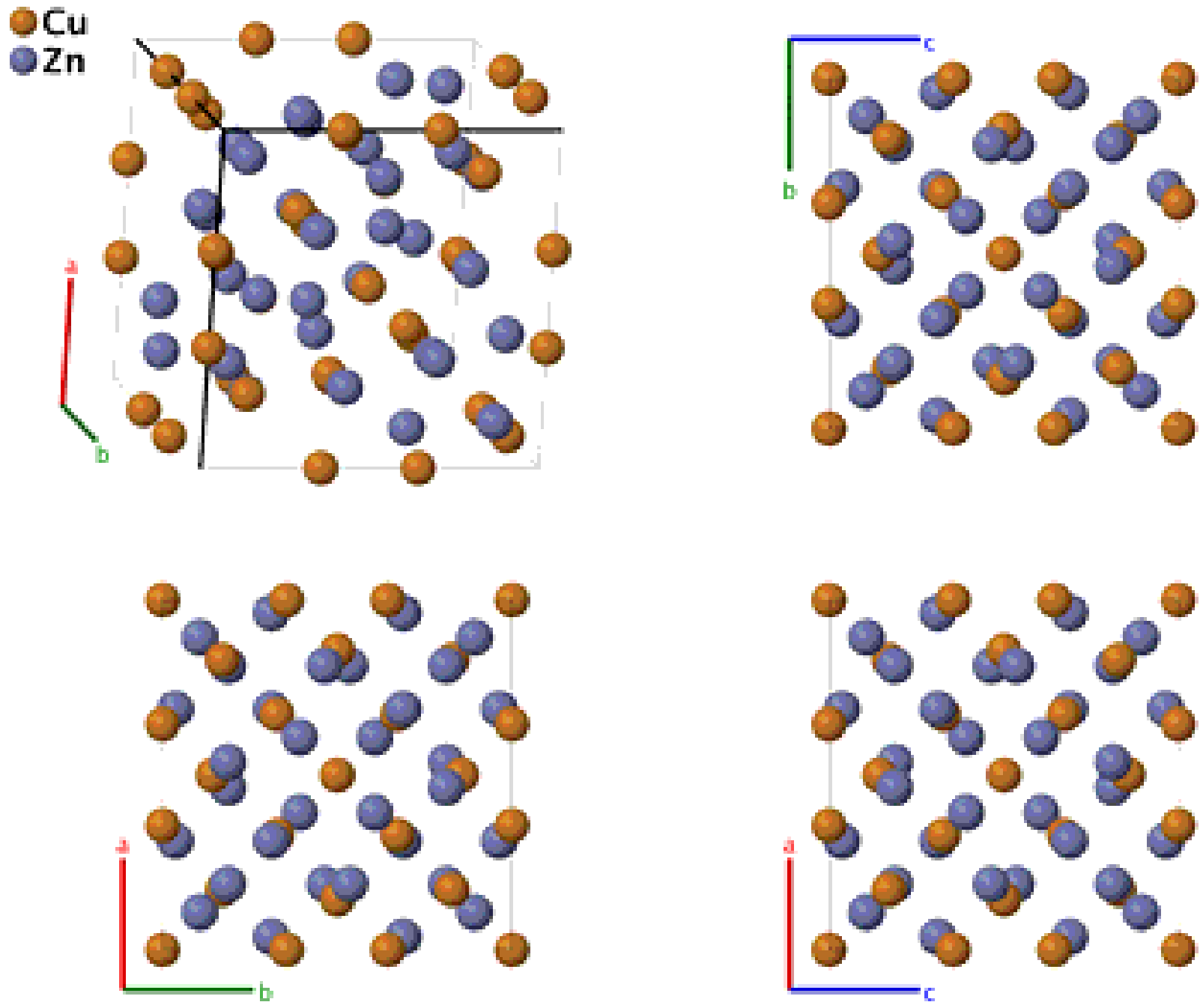

| | | |
|---|---|---|
| **Prototype** | : | $Cu_5Zn_8$ |
| **AFLOW prototype label** | : | A5B8_cI52_217_ce_cg |
| ***Strukturbericht* designation** | : | None |
| **Pearson symbol** | : | cI52 |
| **Space group number** | : | 217 |
| **Space group symbol** | : | $I\bar{4}3m$ |
| **AFLOW prototype command** | : | `aflow --proto=A5B8_cI52_217_ce_cg` `--params=`$a, x_1, x_2, x_3, x_4, z_4$ |

**Other compounds with this structure:**

- $Cu_xZn_{1-x}$, $Cu_xCd_{1-x}$, $Fe_xZn_{1-x}$

- $\gamma$-Brass comes in a variety of compositions. We use the data from (Gourdon, 2007) for $Cu_{5.00}Zn_{8.00}$. At this composition the authors state that the sites are fully occupied as given below.

**Body-centered Cubic primitive vectors:**

$$\mathbf{a}_1 = -\frac{1}{2}\,a\,\hat{\mathbf{x}} + \frac{1}{2}\,a\,\hat{\mathbf{y}} + \frac{1}{2}\,a\,\hat{\mathbf{z}}$$
$$\mathbf{a}_2 = \frac{1}{2}\,a\,\hat{\mathbf{x}} - \frac{1}{2}\,a\,\hat{\mathbf{y}} + \frac{1}{2}\,a\,\hat{\mathbf{z}}$$
$$\mathbf{a}_3 = \frac{1}{2}\,a\,\hat{\mathbf{x}} + \frac{1}{2}\,a\,\hat{\mathbf{y}} - \frac{1}{2}\,a\,\hat{\mathbf{z}}$$

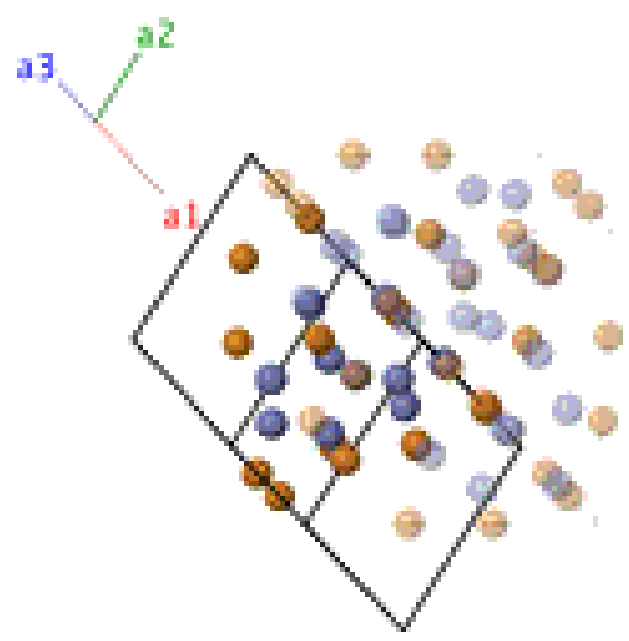

**Basis vectors:**

| | | Lattice Coordinates | | Cartesian Coordinates | Wyckoff Position | Atom Type |
|---|---|---|---|---|---|---|
| $\mathbf{B_1}$ | = | $2x_1\,\mathbf{a_1} + 2x_1\,\mathbf{a_2} + 2x_1\,\mathbf{a_3}$ | = | $x_1\,a\,\mathbf{\hat{x}} + x_1\,a\,\mathbf{\hat{y}} + x_1\,a\,\mathbf{\hat{z}}$ | (8c) | Cu I |
| $\mathbf{B_2}$ | = | $-2x_1\,\mathbf{a_3}$ | = | $-x_1\,a\,\mathbf{\hat{x}} - x_1\,a\,\mathbf{\hat{y}} + x_1\,a\,\mathbf{\hat{z}}$ | (8c) | Cu I |
| $\mathbf{B_3}$ | = | $-2x_1\,\mathbf{a_2}$ | = | $-x_1\,a\,\mathbf{\hat{x}} + x_1\,a\,\mathbf{\hat{y}} - x_1\,a\,\mathbf{\hat{z}}$ | (8c) | Cu I |
| $\mathbf{B_4}$ | = | $-2x_1\,\mathbf{a_1}$ | = | $x_1\,a\,\mathbf{\hat{x}} - x_1\,a\,\mathbf{\hat{y}} - x_1\,a\,\mathbf{\hat{z}}$ | (8c) | Cu I |
| $\mathbf{B_5}$ | = | $2x_2\,\mathbf{a_1} + 2x_2\,\mathbf{a_2} + 2x_2\,\mathbf{a_3}$ | = | $x_2\,a\,\mathbf{\hat{x}} + x_2\,a\,\mathbf{\hat{y}} + x_2\,a\,\mathbf{\hat{z}}$ | (8c) | Zn I |
| $\mathbf{B_6}$ | = | $-2x_2\,\mathbf{a_3}$ | = | $-x_2\,a\,\mathbf{\hat{x}} - x_2\,a\,\mathbf{\hat{y}} + x_2\,a\,\mathbf{\hat{z}}$ | (8c) | Zn I |
| $\mathbf{B_7}$ | = | $-2x_2\,\mathbf{a_2}$ | = | $-x_2\,a\,\mathbf{\hat{x}} + x_2\,a\,\mathbf{\hat{y}} - x_2\,a\,\mathbf{\hat{z}}$ | (8c) | Zn I |
| $\mathbf{B_8}$ | = | $-2x_2\,\mathbf{a_1}$ | = | $x_2\,a\,\mathbf{\hat{x}} - x_2\,a\,\mathbf{\hat{y}} - x_2\,a\,\mathbf{\hat{z}}$ | (8c) | Zn I |
| $\mathbf{B_9}$ | = | $x_3\,\mathbf{a_2} + x_3\,\mathbf{a_3}$ | = | $x_3\,a\,\mathbf{\hat{x}}$ | (12e) | Cu II |
| $\mathbf{B_{10}}$ | = | $x_3\,\mathbf{a_1} + x_3\,\mathbf{a_3}$ | = | $x_3\,a\,\mathbf{\hat{y}}$ | (12e) | Cu II |
| $\mathbf{B_{11}}$ | = | $x_3\,\mathbf{a_1} + x_3\,\mathbf{a_2}$ | = | $x_3\,a\,\mathbf{\hat{z}}$ | (12e) | Cu II |
| $\mathbf{B_{12}}$ | = | $-x_3\,\mathbf{a_2} - x_3\,\mathbf{a_3}$ | = | $-x_3\,a\,\mathbf{\hat{x}}$ | (12e) | Cu II |
| $\mathbf{B_{13}}$ | = | $-x_3\,\mathbf{a_1} - x_3\,\mathbf{a_3}$ | = | $-x_3\,a\,\mathbf{\hat{y}}$ | (12e) | Cu II |
| $\mathbf{B_{14}}$ | = | $-x_3\,\mathbf{a_1} - x_3\,\mathbf{a_2}$ | = | $-x_3\,a\,\mathbf{\hat{z}}$ | (12e) | Cu II |
| $\mathbf{B_{15}}$ | = | $(x_4 + z_4)\,\mathbf{a_1} + (x_4 + z_4)\,\mathbf{a_2} + 2x_4\,\mathbf{a_3}$ | = | $x_4\,a\,\mathbf{\hat{x}} + x_4\,a\,\mathbf{\hat{y}} + z_4\,a\,\mathbf{\hat{z}}$ | (24g) | Zn II |
| $\mathbf{B_{16}}$ | = | $(z_4 - x_4)\,\mathbf{a_1} + (z_4 - x_4)\,\mathbf{a_2} - 2x_4\,\mathbf{a_3}$ | = | $-x_4\,a\,\mathbf{\hat{x}} - x_4\,a\,\mathbf{\hat{y}} + z_4\,a\,\mathbf{\hat{z}}$ | (24g) | Zn II |
| $\mathbf{B_{17}}$ | = | $(x_4 - z_4)\,\mathbf{a_1} - (x_4 + z_4)\,\mathbf{a_2}$ | = | $-x_4\,a\,\mathbf{\hat{x}} + x_4\,a\,\mathbf{\hat{y}} - z_4\,a\,\mathbf{\hat{z}}$ | (24g) | Zn II |
| $\mathbf{B_{18}}$ | = | $-(x_4 + z_4)\,\mathbf{a_1} + (x_4 - z_4)\,\mathbf{a_2}$ | = | $x_4\,a\,\mathbf{\hat{x}} - x_4\,a\,\mathbf{\hat{y}} - z_4\,a\,\mathbf{\hat{z}}$ | (24g) | Zn II |
| $\mathbf{B_{19}}$ | = | $2x_4\,\mathbf{a_1} + (x_4 + z_4)\,\mathbf{a_2} + (x_4 + z_4)\,\mathbf{a_3}$ | = | $z_4\,a\,\mathbf{\hat{x}} + x_4\,a\,\mathbf{\hat{y}} + x_4\,a\,\mathbf{\hat{z}}$ | (24g) | Zn II |
| $\mathbf{B_{20}}$ | = | $-2x_4\,\mathbf{a_1} + (z_4 - x_4)\,\mathbf{a_2} + (z_4 - x_4)\,\mathbf{a_3}$ | = | $z_4\,a\,\mathbf{\hat{x}} - x_4\,a\,\mathbf{\hat{y}} - x_4\,a\,\mathbf{\hat{z}}$ | (24g) | Zn II |
| $\mathbf{B_{21}}$ | = | $(x_4 - z_4)\,\mathbf{a_2} - (x_4 + z_4)\,\mathbf{a_3}$ | = | $-z_4\,a\,\mathbf{\hat{x}} - x_4\,a\,\mathbf{\hat{y}} + x_4\,a\,\mathbf{\hat{z}}$ | (24g) | Zn II |
| $\mathbf{B_{22}}$ | = | $-(x_4 + z_4)\,\mathbf{a_2} + (x_4 - z_4)\,\mathbf{a_3}$ | = | $-z_4\,a\,\mathbf{\hat{x}} + x_4\,a\,\mathbf{\hat{y}} - x_4\,a\,\mathbf{\hat{z}}$ | (24g) | Zn II |
| $\mathbf{B_{23}}$ | = | $(x_4 + z_4)\,\mathbf{a_1} + 2x_4\,\mathbf{a_2} + (x_4 + z_4)\,\mathbf{a_3}$ | = | $x_4\,a\,\mathbf{\hat{x}} + z_4\,a\,\mathbf{\hat{y}} + x_4\,a\,\mathbf{\hat{z}}$ | (24g) | Zn II |
| $\mathbf{B_{24}}$ | = | $(z_4 - x_4)\,\mathbf{a_1} - 2x_4\,\mathbf{a_2} + (z_4 - x_4)\,\mathbf{a_3}$ | = | $-x_4\,a\,\mathbf{\hat{x}} + z_4\,a\,\mathbf{\hat{y}} - x_4\,a\,\mathbf{\hat{z}}$ | (24g) | Zn II |
| $\mathbf{B_{25}}$ | = | $-(x_4 + z_4)\,\mathbf{a_1} + (x_4 - z_4)\,\mathbf{a_3}$ | = | $x_4\,a\,\mathbf{\hat{x}} - z_4\,a\,\mathbf{\hat{y}} - x_4\,a\,\mathbf{\hat{z}}$ | (24g) | Zn II |
| $\mathbf{B_{26}}$ | = | $(x_4 - z_4)\,\mathbf{a_1} - (x_4 + z_4)\,\mathbf{a_3}$ | = | $-x_4\,a\,\mathbf{\hat{x}} - z_4\,a\,\mathbf{\hat{y}} + x_4\,a\,\mathbf{\hat{z}}$ | (24g) | Zn II |

**Geometry files:**

- CIF: pp. 769
- POSCAR: pp. 769

# High-Pressure cI16 Li Structure: A_cI16_220_c

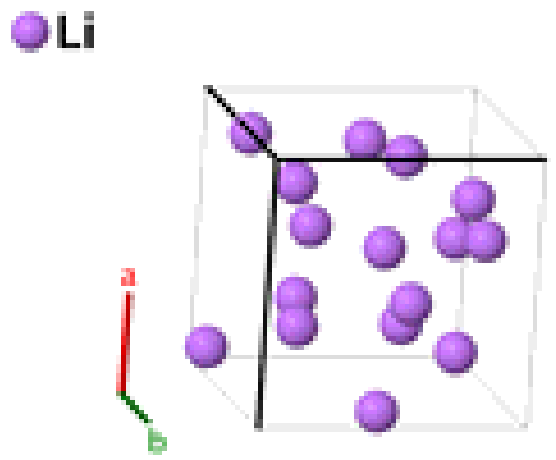

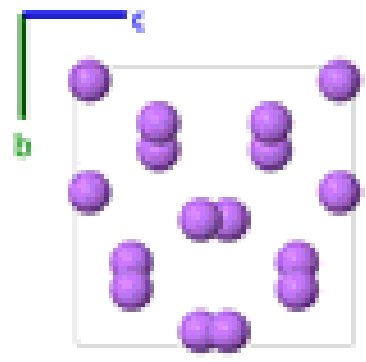

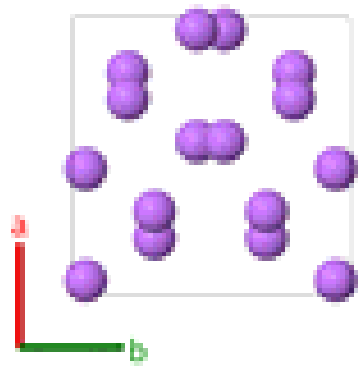

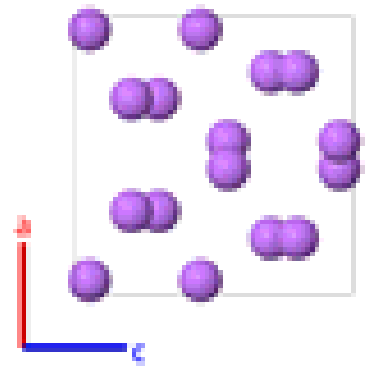

| | | |
|---|---|---|
| **Prototype** | : | Li |
| **AFLOW prototype label** | : | A_cI16_220_c |
| ***Strukturbericht* designation** | : | None |
| **Pearson symbol** | : | cI16 |
| **Space group number** | : | 220 |
| **Space group symbol** | : | $I\bar{4}3d$ |
| **AFLOW prototype command** | : | `aflow --proto=A_cI16_220_c`<br>`--params=`$a, x_1$ |

**Other compounds with this structure:**

- Na (under pressure)

- This is a high-pressure phase of lithium. We use the data from (Hanfland, 2000) at 38.9 GPa. When $x_1 = 0$ this becomes a body-centered cubic (A2) system. We have used the fact that all vectors of the form $(\pm a/2\hat{\mathbf{x}} \pm a/2\hat{\mathbf{y}} \pm a/2\hat{\mathbf{z}})$ are primitive vectors of the body-centered cubic lattice to simplify the positions of some atoms in both lattice and Cartesian coordinates.

**Body-centered Cubic primitive vectors:**

$$\begin{aligned}
\mathbf{a}_1 &= -\frac{1}{2}\,a\,\hat{\mathbf{x}} + \frac{1}{2}\,a\,\hat{\mathbf{y}} + \frac{1}{2}\,a\,\hat{\mathbf{z}} \\
\mathbf{a}_2 &= \frac{1}{2}\,a\,\hat{\mathbf{x}} - \frac{1}{2}\,a\,\hat{\mathbf{y}} + \frac{1}{2}\,a\,\hat{\mathbf{z}} \\
\mathbf{a}_3 &= \frac{1}{2}\,a\,\hat{\mathbf{x}} + \frac{1}{2}\,a\,\hat{\mathbf{y}} - \frac{1}{2}\,a\,\hat{\mathbf{z}}
\end{aligned}$$

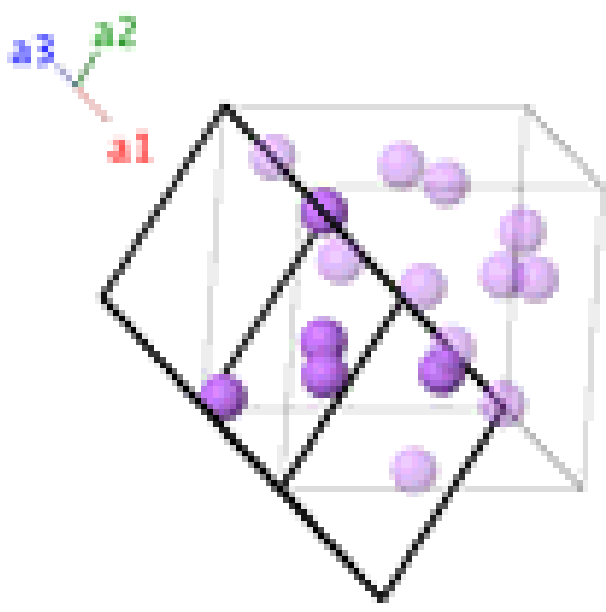

**Basis vectors:**

| Lattice Coordinates | Cartesian Coordinates | Wyckoff Position | Atom Type |
|---|---|---|---|

| | | | | | | |
|---|---|---|---|---|---|---|
| $\mathbf{B}_1$ | = | $2x_1\,\mathbf{a}_1 + 2x_1\,\mathbf{a}_2 + 2x_1\,\mathbf{a}_3$ | = | $x_1\,a\,\hat{\mathbf{x}} + x_1\,a\,\hat{\mathbf{y}} + x_1\,a\,\hat{\mathbf{z}}$ | (16$c$) | Li |
| $\mathbf{B}_2$ | = | $\frac{1}{2}\,\mathbf{a}_1 + \left(\frac{1}{2} - 2x_1\right)\,\mathbf{a}_3$ | = | $-x_1\,a\,\hat{\mathbf{x}} + \left(\frac{1}{2} - x_1\right)\,a\,\hat{\mathbf{y}} + x_1\,a\,\hat{\mathbf{z}}$ | (16$c$) | Li |
| $\mathbf{B}_3$ | = | $\left(\frac{1}{2} - 2x_1\right)\,\mathbf{a}_2 + \frac{1}{2}\,\mathbf{a}_3$ | = | $\left(\frac{1}{2} - x_1\right)\,a\,\hat{\mathbf{x}} + x_1\,a\,\hat{\mathbf{y}} - x_1\,a\,\hat{\mathbf{z}}$ | (16$c$) | Li |
| $\mathbf{B}_4$ | = | $\left(\frac{1}{2} - 2x_1\right)\,\mathbf{a}_1 + \frac{1}{2}\,\mathbf{a}_2$ | = | $+x_1\,a\,\hat{\mathbf{x}} - x_1\,a\,\hat{\mathbf{y}}\left(\frac{1}{2} - x_1\right)\,a\,\hat{\mathbf{z}}$ | (16$c$) | Li |
| $\mathbf{B}_5$ | = | $\left(\frac{1}{2} + 2x_1\right)\,\mathbf{a}_1 + \left(\frac{1}{2} + 2x_1\right)\,\mathbf{a}_2 + \left(\frac{1}{2} + 2x_1\right)\,\mathbf{a}_3$ | = | $\left(\frac{1}{4} + x_1\right)\,a\,\hat{\mathbf{x}} + \left(\frac{1}{4} + x_1\right)\,a\,\hat{\mathbf{y}} + \left(\frac{1}{4} + x_1\right)\,a\,\hat{\mathbf{z}}$ | (16$c$) | Li |
| $\mathbf{B}_6$ | = | $\frac{1}{2}\,\mathbf{a}_1 - 2x_1\,\mathbf{a}_3$ | = | $\left(\frac{3}{4} - x_1\right)\,a\,\hat{\mathbf{x}} + \left(\frac{1}{4} - x_1\right)\,a\,\hat{\mathbf{y}} + \left(\frac{1}{4} + x_1\right)\,a\,\hat{\mathbf{z}}$ | (16$c$) | Li |
| $\mathbf{B}_7$ | = | $-2x_1\,\mathbf{a}_1 + \frac{1}{2}\,\mathbf{a}_2$ | = | $\left(\frac{1}{4} + x_1\right)\,a\,\hat{\mathbf{x}} + \left(\frac{3}{4} - x_1\right)\,a\,\hat{\mathbf{y}} + \left(\frac{1}{4} - x_1\right)\,a\,\hat{\mathbf{z}}$ | (16$c$) | Li |
| $\mathbf{B}_8$ | = | $-2x_1\,\mathbf{a}_2 + \frac{1}{2}\,\mathbf{a}_3$ | = | $\left(\frac{1}{4} - x_1\right)\,a\,\hat{\mathbf{x}} + \left(\frac{1}{4} + x_1\right)\,a\,\hat{\mathbf{y}} + \left(\frac{3}{4} - x_1\right)\,a\,\hat{\mathbf{z}}$ | (16$c$) | Li |

**Geometry files:**

- CIF: pp. 769
- POSCAR: pp. 770

# $Pu_2C_3$ ($D5_c$) Structure: A3B2_cI40_220_d_c

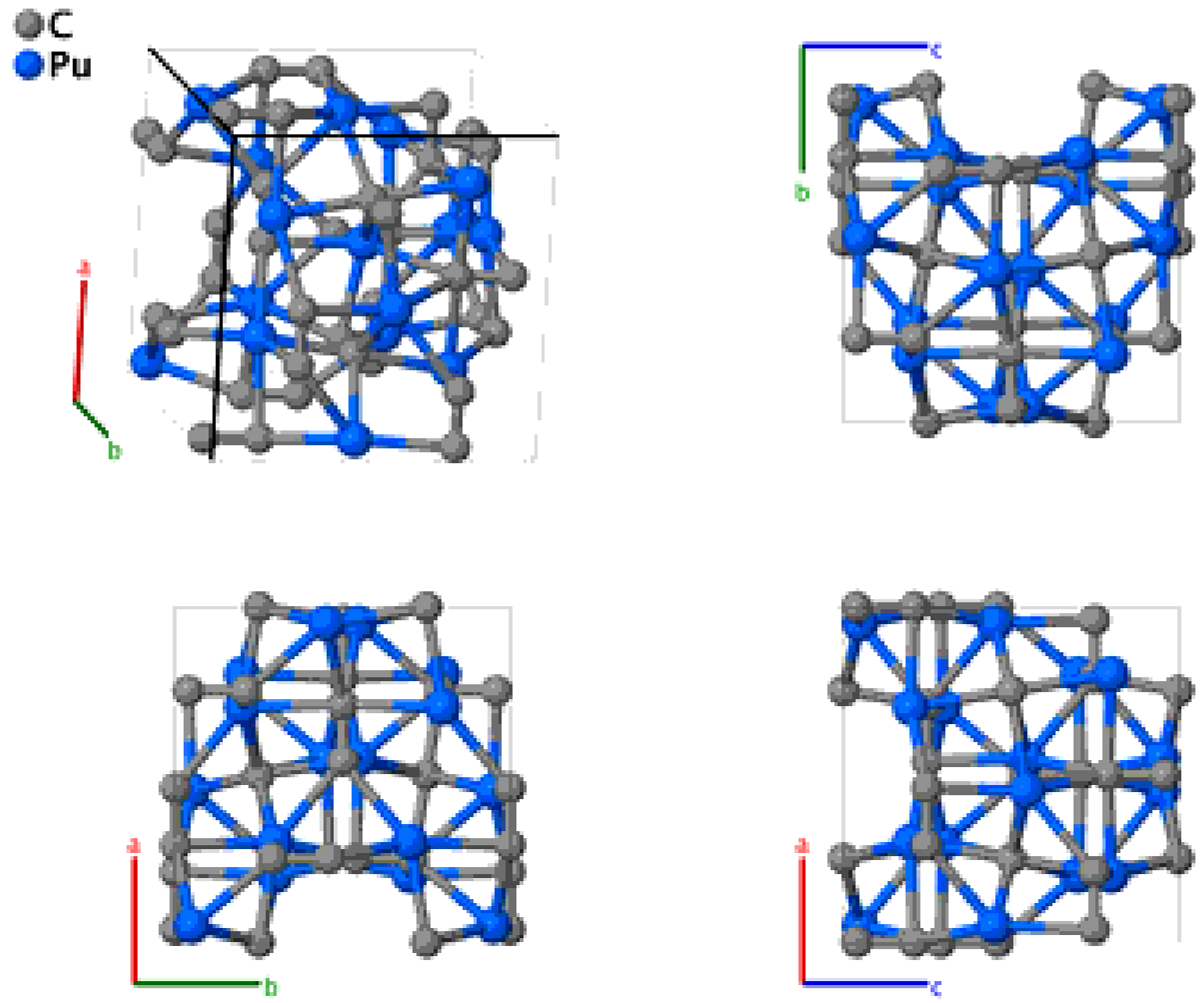

| | | |
|---|---|---|
| **Prototype** | : | $Pu_2C_3$ |
| **AFLOW prototype label** | : | A3B2_cI40_220_d_c |
| ***Strukturbericht* designation** | : | $D5_c$ |
| **Pearson symbol** | : | cI40 |
| **Space group number** | : | 220 |
| **Space group symbol** | : | $I\bar{4}3d$ |
| **AFLOW prototype command** | : | `aflow --proto=A3B2_cI40_220_d_c` <br> `--params=`$a, x_1, x_2$ |

**Other compounds with this structure:**

- $Am_2C_3$, $C_3Ce_2$, $C_3Hf_2$, $Ru_2Y_3$, $C_3U_2$, $Er_3Ru_2$, $C_3Y_2$, many others.

- We use the data for $^{240}$Pu.

**Body-centered Cubic primitive vectors:**

$$
\begin{aligned}
\mathbf{a}_1 &= -\frac{1}{2}\,a\,\hat{\mathbf{x}} + \frac{1}{2}\,a\,\hat{\mathbf{y}} + \frac{1}{2}\,a\,\hat{\mathbf{z}} \\
\mathbf{a}_2 &= \frac{1}{2}\,a\,\hat{\mathbf{x}} - \frac{1}{2}\,a\,\hat{\mathbf{y}} + \frac{1}{2}\,a\,\hat{\mathbf{z}} \\
\mathbf{a}_3 &= \frac{1}{2}\,a\,\hat{\mathbf{x}} + \frac{1}{2}\,a\,\hat{\mathbf{y}} - \frac{1}{2}\,a\,\hat{\mathbf{z}}
\end{aligned}
$$

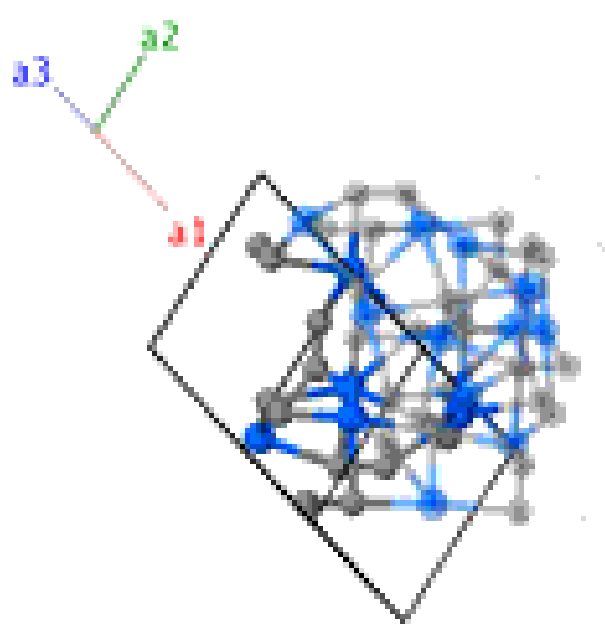

**Basis vectors:**

| | | Lattice Coordinates | | Cartesian Coordinates | Wyckoff Position | Atom Type |
|---|---|---|---|---|---|---|
| $\mathbf{B_1}$ | = | $2x_1\,\mathbf{a_1} + 2x_1\,\mathbf{a_2} + 2x_1\,\mathbf{a_3}$ | = | $x_1\,a\,\mathbf{\hat{x}} + x_1\,a\,\mathbf{\hat{y}} + x_1\,a\,\mathbf{\hat{z}}$ | (16$c$) | Pu |
| $\mathbf{B_2}$ | = | $\frac{1}{2}\,\mathbf{a_1} + \left(\frac{1}{2} - 2x_1\right)\,\mathbf{a_3}$ | = | $-x_1\,a\,\mathbf{\hat{x}} + \left(\frac{1}{2} - x_1\right)\,a\,\mathbf{\hat{y}} + x_1\,a\,\mathbf{\hat{z}}$ | (16$c$) | Pu |
| $\mathbf{B_3}$ | = | $\left(\frac{1}{2} - 2x_1\right)\,\mathbf{a_2} + \frac{1}{2}\,\mathbf{a_3}$ | = | $\left(\frac{1}{2} - x_1\right)\,a\,\mathbf{\hat{x}} + x_1\,a\,\mathbf{\hat{y}} - x_1\,a\,\mathbf{\hat{z}}$ | (16$c$) | Pu |
| $\mathbf{B_4}$ | = | $\left(\frac{1}{2} - 2x_1\right)\,\mathbf{a_1} + \frac{1}{2}\,\mathbf{a_2}$ | = | $+x_1\,a\,\mathbf{\hat{x}} - x_1\,a\,\mathbf{\hat{y}}\left(\frac{1}{2} - x_1\right)\,a\,\mathbf{\hat{z}}$ | (16$c$) | Pu |
| $\mathbf{B_5}$ | = | $\left(\frac{1}{2} + 2x_1\right)\,\mathbf{a_1} + \left(\frac{1}{2} + 2x_1\right)\,\mathbf{a_2} + \left(\frac{1}{2} + 2x_1\right)\,\mathbf{a_3}$ | = | $\left(\frac{1}{4} + x_1\right)\,a\,\mathbf{\hat{x}} + \left(\frac{1}{4} + x_1\right)\,a\,\mathbf{\hat{y}} + \left(\frac{1}{4} + x_1\right)\,a\,\mathbf{\hat{z}}$ | (16$c$) | Pu |
| $\mathbf{B_6}$ | = | $\frac{1}{2}\,\mathbf{a_1} - 2x_1\,\mathbf{a_3}$ | = | $\left(\frac{3}{4} - x_1\right)\,a\,\mathbf{\hat{x}} + \left(\frac{1}{4} - x_1\right)\,a\,\mathbf{\hat{y}} + \left(\frac{1}{4} + x_1\right)\,a\,\mathbf{\hat{z}}$ | (16$c$) | Pu |
| $\mathbf{B_7}$ | = | $-2x_1\,\mathbf{a_1} + \frac{1}{2}\,\mathbf{a_2}$ | = | $\left(\frac{1}{4} + x_1\right)\,a\,\mathbf{\hat{x}} + \left(\frac{3}{4} - x_1\right)\,a\,\mathbf{\hat{y}} + \left(\frac{1}{4} - x_1\right)\,a\,\mathbf{\hat{z}}$ | (16$c$) | Pu |
| $\mathbf{B_8}$ | = | $-2x_1\,\mathbf{a_2} + \frac{1}{2}\,\mathbf{a_3}$ | = | $\left(\frac{1}{4} - x_1\right)\,a\,\mathbf{\hat{x}} + \left(\frac{1}{4} + x_1\right)\,a\,\mathbf{\hat{y}} + \left(\frac{3}{4} - x_1\right)\,a\,\mathbf{\hat{z}}$ | (16$c$) | Pu |
| $\mathbf{B_9}$ | = | $\frac{1}{4}\,\mathbf{a_1} + \left(\frac{1}{4} + x_2\right)\,\mathbf{a_2} + x_2\,\mathbf{a_3}$ | = | $x_2\,a\,\mathbf{\hat{x}} + \frac{1}{4}\,a\,\mathbf{\hat{z}}$ | (24$d$) | C |
| $\mathbf{B_{10}}$ | = | $\frac{3}{4}\,\mathbf{a_1} + \left(\frac{1}{4} - x_2\right)\,\mathbf{a_2} + \left(\frac{1}{2} - x_2\right)\,\mathbf{a_3}$ | = | $-x_2\,a\,\mathbf{\hat{x}} + \frac{1}{2}\,a\,\mathbf{\hat{y}} + \frac{1}{4}\,a\,\mathbf{\hat{z}}$ | (24$d$) | C |
| $\mathbf{B_{11}}$ | = | $x_2\,\mathbf{a_1} + \frac{1}{4}\,\mathbf{a_2} + \left(\frac{1}{4} + x_2\right)\,\mathbf{a_3}$ | = | $\frac{1}{4}\,a\,\mathbf{\hat{x}} + x_2\,a\,\mathbf{\hat{y}}$ | (24$d$) | C |
| $\mathbf{B_{12}}$ | = | $\left(\frac{1}{2} - x_2\right)\,\mathbf{a_1} + \frac{3}{4}\,\mathbf{a_2} + \left(\frac{1}{4} - x_2\right)\,\mathbf{a_3}$ | = | $\frac{1}{4}\,a\,\mathbf{\hat{x}} - x_2\,a\,\mathbf{\hat{y}} + \frac{1}{2}\,a\,\mathbf{\hat{z}}$ | (24$d$) | C |
| $\mathbf{B_{13}}$ | = | $\left(\frac{1}{4} + x_2\right)\,\mathbf{a_1} + x_2\,\mathbf{a_2} + \frac{1}{4}\,\mathbf{a_3}$ | = | $\frac{1}{4}\,a\,\mathbf{\hat{y}} + x_2\,a\,\mathbf{\hat{z}}$ | (24$d$) | C |
| $\mathbf{B_{14}}$ | = | $\left(\frac{1}{4} - x_2\right)\,\mathbf{a_1} + \left(\frac{1}{2} - x_2\right)\,\mathbf{a_2} + \frac{3}{4}\,\mathbf{a_3}$ | = | $\frac{1}{2}\,a\,\mathbf{\hat{x}} + \frac{1}{4}\,a\,\mathbf{\hat{y}} - x_2\,a\,\mathbf{\hat{z}}$ | (24$d$) | C |
| $\mathbf{B_{15}}$ | = | $\left(\frac{3}{4} + x_2\right)\,\mathbf{a_1} + \frac{3}{4}\,\mathbf{a_2} + \left(\frac{1}{2} + x_2\right)\,\mathbf{a_3}$ | = | $\frac{1}{4}\,a\,\mathbf{\hat{x}} + \left(\frac{1}{4} + x_2\right)\,a\,\mathbf{\hat{y}} + \frac{1}{2}\,a\,\mathbf{\hat{z}}$ | (24$d$) | C |
| $\mathbf{B_{16}}$ | = | $\left(\frac{3}{4} - x_2\right)\,\mathbf{a_1} + \frac{1}{4}\,\mathbf{a_2} - x_2\,\mathbf{a_3}$ | = | $\frac{3}{4}\,a\,\mathbf{\hat{x}} + \left(\frac{1}{4} - x_2\right)\,a\,\mathbf{\hat{y}} + \frac{1}{2}\,a\,\mathbf{\hat{z}}$ | (24$d$) | C |
| $\mathbf{B_{17}}$ | = | $\frac{3}{4}\,\mathbf{a_1} + \left(\frac{1}{2} + x_2\right)\,\mathbf{a_2} + \left(\frac{3}{4} + x_2\right)\,\mathbf{a_3}$ | = | $\left(\frac{1}{4} + x_2\right)\,a\,\mathbf{\hat{x}} + \frac{1}{2}\,a\,\mathbf{\hat{y}} + \frac{1}{4}\,a\,\mathbf{\hat{z}}$ | (24$d$) | C |
| $\mathbf{B_{18}}$ | = | $\frac{1}{4}\,\mathbf{a_1} - x_2\,\mathbf{a_2} + \left(\frac{3}{4} - x_2\right)\,\mathbf{a_3}$ | = | $\left(\frac{1}{4} - x_2\right)\,a\,\mathbf{\hat{x}} + \frac{1}{2}\,a\,\mathbf{\hat{y}} + \frac{3}{4}\,a\,\mathbf{\hat{z}}$ | (24$d$) | C |
| $\mathbf{B_{19}}$ | = | $\left(\frac{1}{2} + x_2\right)\,\mathbf{a_1} + \left(\frac{3}{4} + x_2\right)\,\mathbf{a_2} + \frac{3}{4}\,\mathbf{a_3}$ | = | $\frac{1}{2}\,a\,\mathbf{\hat{x}} + \frac{1}{4}\,a\,\mathbf{\hat{y}} + \left(\frac{1}{4} + x_2\right)\,a\,\mathbf{\hat{z}}$ | (24$d$) | C |
| $\mathbf{B_{20}}$ | = | $-x_2\,\mathbf{a_1} + \left(\frac{3}{4} - x_2\right)\,\mathbf{a_2} + \frac{1}{4}\,\mathbf{a_3}$ | = | $\frac{1}{2}\,a\,\mathbf{\hat{x}} + \frac{3}{4}\,a\,\mathbf{\hat{y}} + \left(\frac{1}{4} - x_2\right)\,a\,\mathbf{\hat{z}}$ | (24$d$) | C |

**Found in:**

**Geometry files:**

- CIF: pp. 770
- POSCAR: pp. 770

# CsCl (B2) Structure: AB_cP2_221_b_a

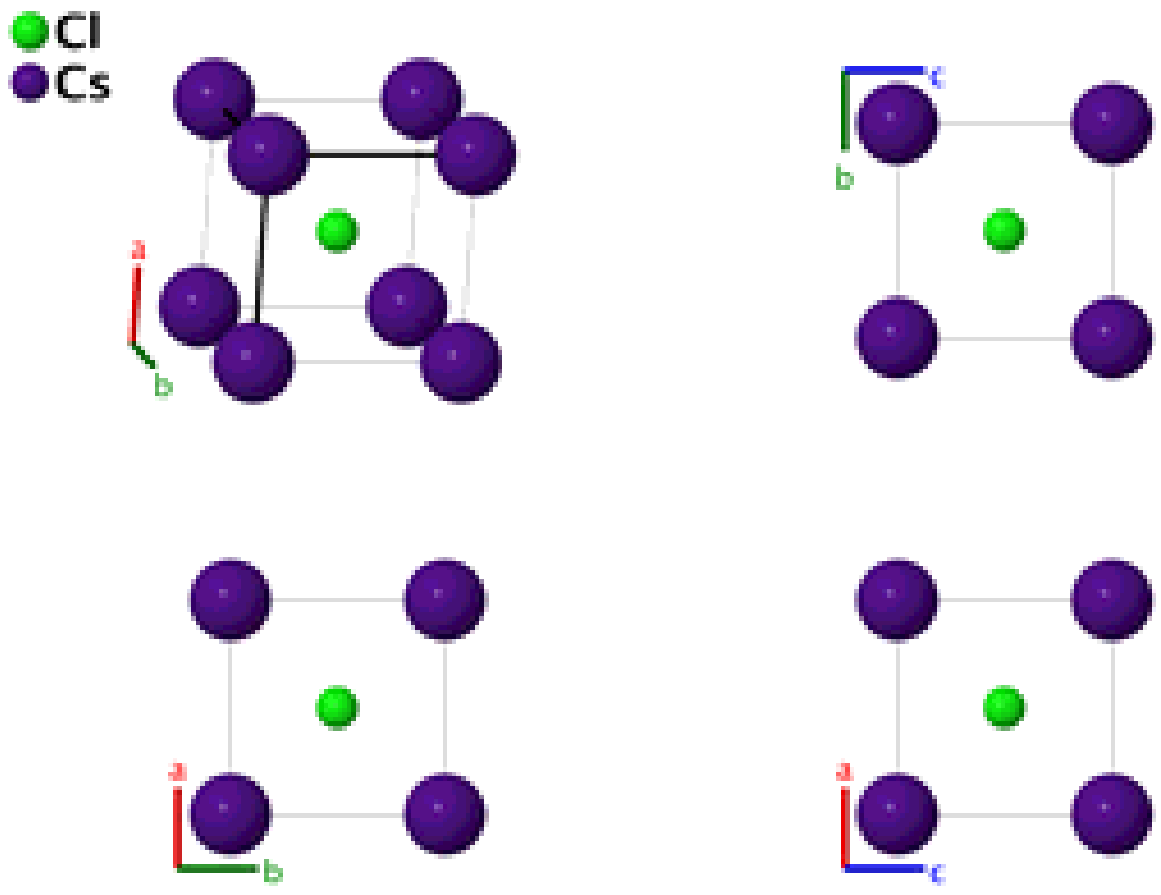

| | | |
|---|---|---|
| **Prototype** | : | CsCl |
| **AFLOW prototype label** | : | AB_cP2_221_b_a |
| ***Strukturbericht* designation** | : | B2 |
| **Pearson symbol** | : | cP2 |
| **Space group number** | : | 221 |
| **Space group symbol** | : | $Pm\bar{3}m$ |
| **AFLOW prototype command** | : | `aflow --proto=AB_cP2_221_b_a --params=`$a$ |

**Other compounds with this structure:**

- CsBr, CsI, RbCl, AlCo, AgZn, BeCu, MgCe, RuAl, SrTl

**Simple Cubic primitive vectors:**

$$\mathbf{a}_1 = a\,\hat{\mathbf{x}}$$
$$\mathbf{a}_2 = a\,\hat{\mathbf{y}}$$
$$\mathbf{a}_3 = a\,\hat{\mathbf{z}}$$

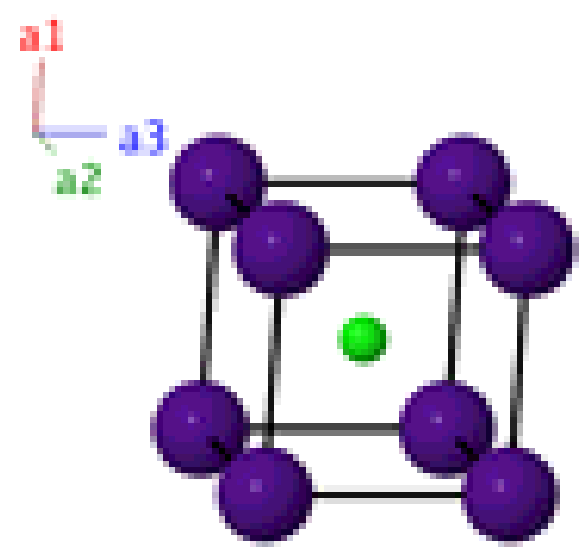

**Basis vectors:**

| | | Lattice Coordinates | | Cartesian Coordinates | Wyckoff Position | Atom Type |
|---|---|---|---|---|---|---|
| $\mathbf{B}_1$ | = | $0\,\mathbf{a}_1 + 0\,\mathbf{a}_2 + 0\,\mathbf{a}_3$ | = | $0\hat{\mathbf{x}} + 0\hat{\mathbf{y}} + 0\hat{\mathbf{z}}$ | $(1a)$ | Cs |
| $\mathbf{B}_2$ | = | $\frac{1}{2}\,\mathbf{a}_1 + \frac{1}{2}\,\mathbf{a}_2 + \frac{1}{2}\,\mathbf{a}_3$ | = | $\frac{1}{2}\,a\,\hat{\mathbf{x}} + \frac{1}{2}\,a\,\hat{\mathbf{y}} + \frac{1}{2}\,a\,\hat{\mathbf{z}}$ | $(1b)$ | Cl |

**Geometry files:**
- CIF: pp. 771
- POSCAR: pp. 771

# NbO Structure: AB_cP6_221_c_d

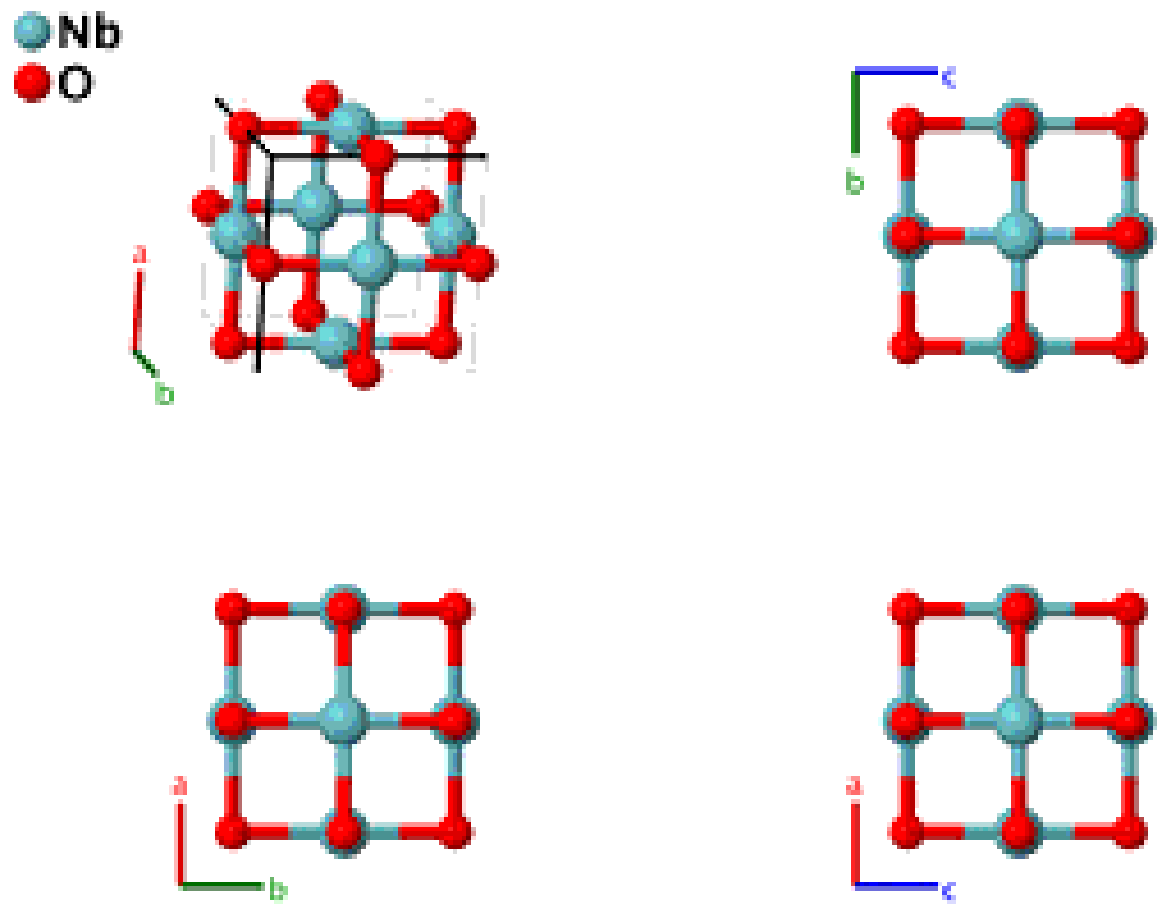

| | | |
|---|---|---|
| **Prototype** | : | NbO |
| **AFLOW prototype label** | : | AB_cP6_221_c_d |
| ***Strukturbericht* designation** | : | None |
| **Pearson symbol** | : | cP6 |
| **Space group number** | : | 221 |
| **Space group symbol** | : | $Pm\bar{3}m$ |
| **AFLOW prototype command** | : | `aflow --proto=AB_cP6_221_c_d --params=`$a$ |

- This is the NaCl (B1) structure with 25% ordered vacancies on both the Na and Cl sites.

**Simple Cubic primitive vectors:**

$$\mathbf{a}_1 = a\,\hat{\mathbf{x}}$$
$$\mathbf{a}_2 = a\,\hat{\mathbf{y}}$$
$$\mathbf{a}_3 = a\,\hat{\mathbf{z}}$$

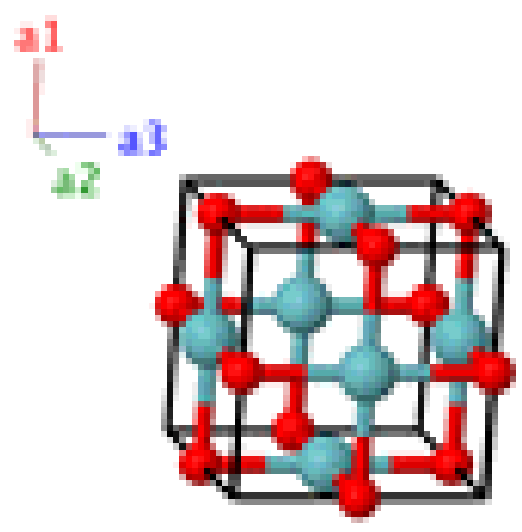

**Basis vectors:**

| | | Lattice Coordinates | | Cartesian Coordinates | Wyckoff Position | Atom Type |
|---|---|---|---|---|---|---|
| $\mathbf{B}_1$ | = | $\frac{1}{2}\,\mathbf{a}_2 + \frac{1}{2}\,\mathbf{a}_3$ | = | $\frac{1}{2}\,a\,\hat{\mathbf{y}} + \frac{1}{2}\,a\,\hat{\mathbf{z}}$ | $(3c)$ | Nb |
| $\mathbf{B}_2$ | = | $\frac{1}{2}\,\mathbf{a}_1 + \frac{1}{2}\,\mathbf{a}_3$ | = | $\frac{1}{2}\,a\,\hat{\mathbf{x}} + \frac{1}{2}\,a\,\hat{\mathbf{z}}$ | $(3c)$ | Nb |
| $\mathbf{B}_3$ | = | $\frac{1}{2}\,\mathbf{a}_1 + \frac{1}{2}\,\mathbf{a}_2$ | = | $\frac{1}{2}\,a\,\hat{\mathbf{x}} + \frac{1}{2}\,a\,\hat{\mathbf{y}}$ | $(3c)$ | Nb |
| $\mathbf{B}_4$ | = | $\frac{1}{2}\,\mathbf{a}_1$ | = | $\frac{1}{2}\,a\,\hat{\mathbf{x}}$ | $(3d)$ | O |
| $\mathbf{B}_5$ | = | $\frac{1}{2}\,\mathbf{a}_2$ | = | $\frac{1}{2}\,a\,\hat{\mathbf{y}}$ | $(3d)$ | O |
| $\mathbf{B}_6$ | = | $\frac{1}{2}\,\mathbf{a}_3$ | = | $\frac{1}{2}\,a\,\hat{\mathbf{z}}$ | $(3d)$ | O |

**Found in:**

**Geometry files:**

- CIF: pp. 771
- POSCAR: pp. 772

# Cubic Perovskite ($CaTiO_3$, $E2_1$) Structure: AB3C_cP5_221_a_c_b

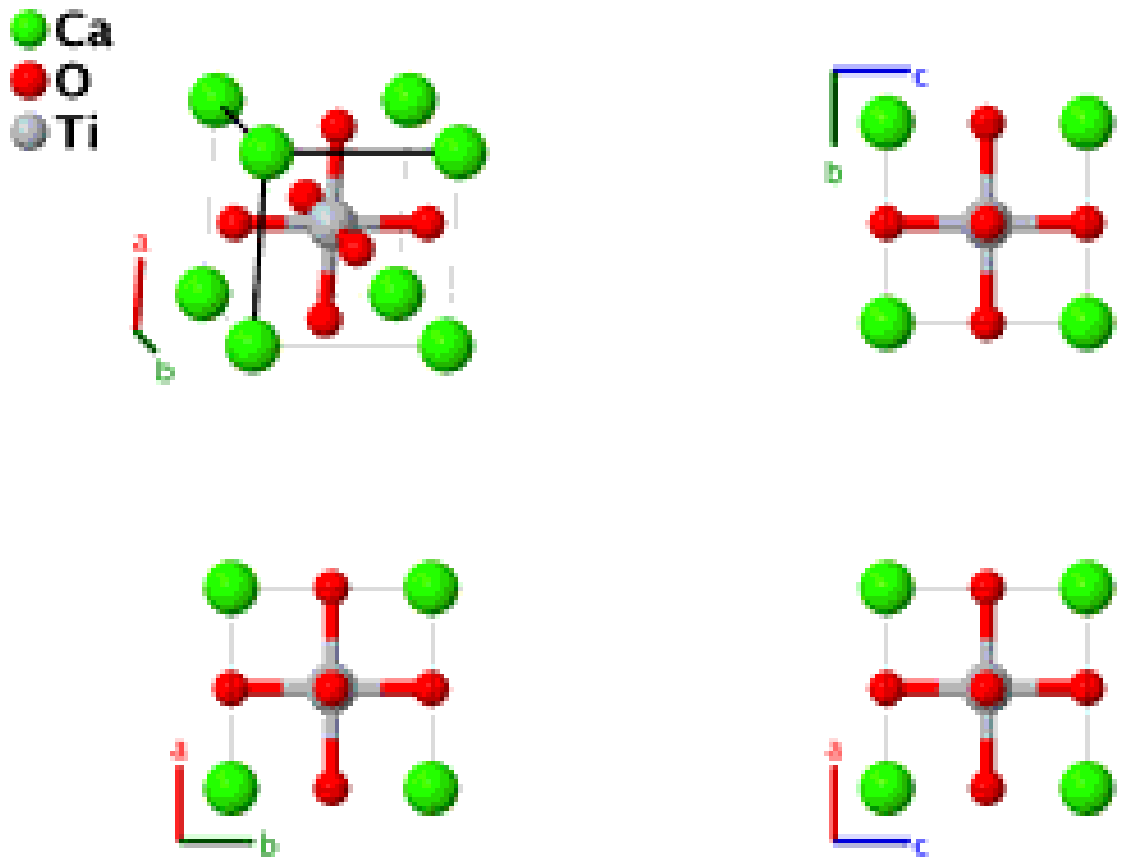

| | | |
|---|---|---|
| **Prototype** | : | $CaTiO_3$ |
| **AFLOW prototype label** | : | AB3C_cP5_221_a_c_b |
| ***Strukturbericht* designation** | : | $E2_1$ |
| **Pearson symbol** | : | cP5 |
| **Space group number** | : | 221 |
| **Space group symbol** | : | $Pm\bar{3}m$ |
| **AFLOW prototype command** | : | `aflow --proto=AB3C_cP5_221_a_c_b --params=`$a$ |

**Other compounds with this structure:**

- $BaTiO_3$, $PbTiO_3$, $PbZrO_3$

- Cubic perovskite is actually the high-temperature phase of the compounds listed below. The ground states are usually distorted perovskite structures. Many of these substances are ferroelectric. By removing one atom type we get various structures, all with space group $Pm\bar{3}m$ : Removing the calcium atoms leads to the $\alpha$-$ReO_3$ ($D0_9$) structure; removing the titanium atoms leads to the $Cu_3Au$ ($L1_2$) structure; removing the oxygen atoms leads to the CsCl (B2) structure; removing the calcium or titanium and the oxygen atoms leads to the simple cubic ($A_h$) structure.

**Simple Cubic primitive vectors:**

$$\mathbf{a}_1 = a\,\hat{\mathbf{x}}$$
$$\mathbf{a}_2 = a\,\hat{\mathbf{y}}$$
$$\mathbf{a}_3 = a\,\hat{\mathbf{z}}$$

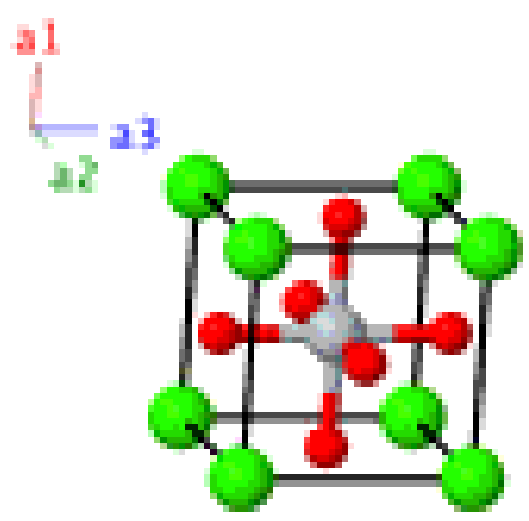

**Basis vectors:**

| Lattice Coordinates | Cartesian Coordinates | Wyckoff Position | Atom Type |
|---|---|---|---|

| | | | | | | |
|---|---|---|---|---|---|---|
| $\mathbf{B_1}$ | = | $0\,\mathbf{a_1} + 0\,\mathbf{a_2} + 0\,\mathbf{a_3}$ | = | $0\hat{\mathbf{x}} + 0\hat{\mathbf{y}} + 0\hat{\mathbf{z}}$ | $(1a)$ | Ca |
| $\mathbf{B_2}$ | = | $\frac{1}{2}\,\mathbf{a_1} + \frac{1}{2}\,\mathbf{a_2} + \frac{1}{2}\,\mathbf{a_3}$ | = | $\frac{1}{2}\,a\,\hat{\mathbf{x}} + \frac{1}{2}\,a\,\hat{\mathbf{y}} + \frac{1}{2}\,a\,\hat{\mathbf{z}}$ | $(1b)$ | Ti |
| $\mathbf{B_3}$ | = | $\frac{1}{2}\,\mathbf{a_2} + \frac{1}{2}\,\mathbf{a_3}$ | = | $\frac{1}{2}\,a\,\hat{\mathbf{y}} + \frac{1}{2}\,a\,\hat{\mathbf{z}}$ | $(3c)$ | O |
| $\mathbf{B_4}$ | = | $\frac{1}{2}\,\mathbf{a_1} + \frac{1}{2}\,\mathbf{a_3}$ | = | $\frac{1}{2}\,a\,\hat{\mathbf{x}} + \frac{1}{2}\,a\,\hat{\mathbf{z}}$ | $(3c)$ | O |
| $\mathbf{B_5}$ | = | $\frac{1}{2}\,\mathbf{a_1} + \frac{1}{2}\,\mathbf{a_2}$ | = | $\frac{1}{2}\,a\,\hat{\mathbf{x}} + \frac{1}{2}\,a\,\hat{\mathbf{y}}$ | $(3c)$ | O |

**Geometry files:**

- CIF: pp. 772
- POSCAR: pp. 772

# Model of Austenite Structure (cP32): AB27CD3_cP32_221_a_dij_b_c

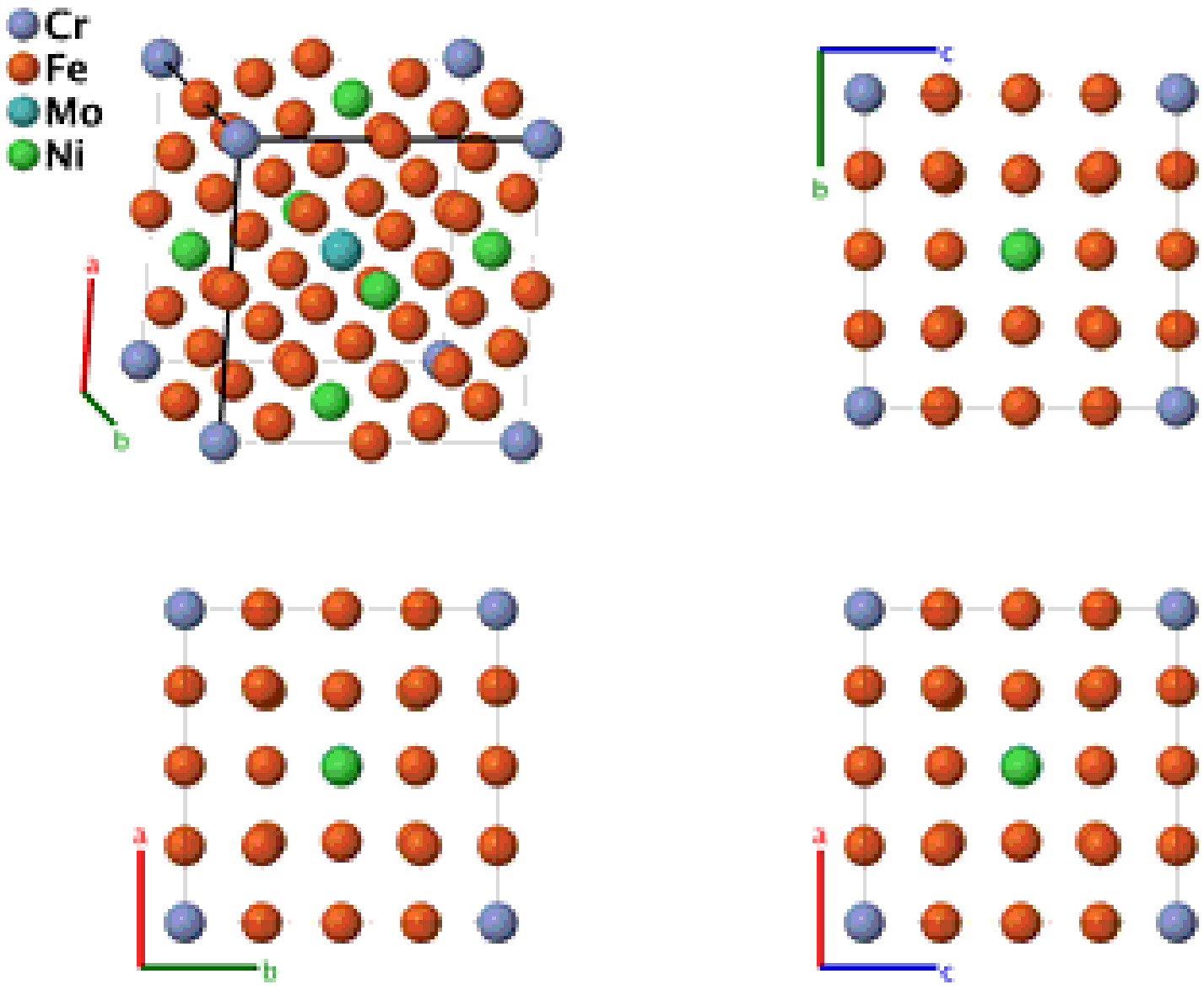

| | | |
|---|---|---|
| **Prototype** | : | $CrFe_{27}MoNi_3$ |
| **AFLOW prototype label** | : | AB27CD3_cP32_221_a_dij_b_c |
| ***Strukturbericht* designation** | : | None |
| **Pearson symbol** | : | cP32 |
| **Space group number** | : | 221 |
| **Space group symbol** | : | $Pm\bar{3}m$ |
| **AFLOW prototype command** | : | `aflow --proto=AB27CD3_cP32_221_a_dij_b_c --params=`$a, y_5, y_6$ |

- Austenitic steels are alloys of iron and other metals with an averaged face-centered cubic structure. This model represents one approximation for an austenite steel. It is not meant to represent a real steel, and the selection of atom types for each Wyckoff position is arbitrary. Note that when $y_5 = y_6 = 1/4$ all the atoms are on sites of an fcc lattice.

**Simple Cubic primitive vectors:**

$$\mathbf{a}_1 = a\,\hat{\mathbf{x}}$$
$$\mathbf{a}_2 = a\,\hat{\mathbf{y}}$$
$$\mathbf{a}_3 = a\,\hat{\mathbf{z}}$$

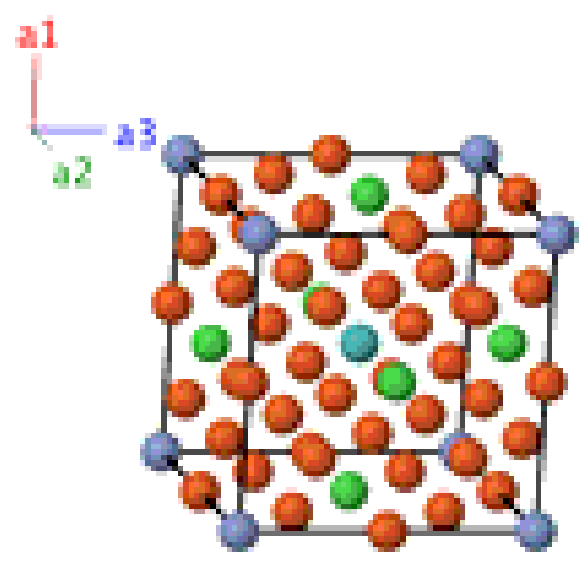

**Basis vectors:**

| | | Lattice Coordinates | | Cartesian Coordinates | Wyckoff Position | Atom Type |
|---|---|---|---|---|---|---|
| $\mathbf{B}_1$ | = | $0\,\mathbf{a}_1 + 0\,\mathbf{a}_2 + 0\,\mathbf{a}_3$ | = | $0\hat{\mathbf{x}} + 0\hat{\mathbf{y}} + 0\hat{\mathbf{z}}$ | $(1a)$ | Cr |
| $\mathbf{B_2}$ | = | $\frac{1}{2}\,\mathbf{a_1} + \frac{1}{2}\,\mathbf{a_2} + \frac{1}{2}\,\mathbf{a_3}$ | = | $\frac{1}{2}\,a\,\mathbf{\hat{x}} + \frac{1}{2}\,a\,\mathbf{\hat{y}} + \frac{1}{2}\,a\,\mathbf{\hat{z}}$ | $(1b)$ | Mo |
| $\mathbf{B_3}$ | = | $\frac{1}{2}\,\mathbf{a_2} + \frac{1}{2}\,\mathbf{a_3}$ | = | $\frac{1}{2}\,a\,\mathbf{\hat{y}} + \frac{1}{2}\,a\,\mathbf{\hat{z}}$ | $(3c)$ | Ni |
| $\mathbf{B_4}$ | = | $\frac{1}{2}\,\mathbf{a_1} + \frac{1}{2}\,\mathbf{a_3}$ | = | $\frac{1}{2}\,a\,\mathbf{\hat{x}} + \frac{1}{2}\,a\,\mathbf{\hat{z}}$ | $(3c)$ | Ni |
| $\mathbf{B_5}$ | = | $\frac{1}{2}\,\mathbf{a_1} + \frac{1}{2}\,\mathbf{a_2}$ | = | $\frac{1}{2}\,a\,\mathbf{\hat{x}} + \frac{1}{2}\,a\,\mathbf{\hat{y}}$ | $(3c)$ | Ni |
| $\mathbf{B_6}$ | = | $\frac{1}{2}\,\mathbf{a_1}$ | = | $\frac{1}{2}\,a\,\mathbf{\hat{x}}$ | $(3d)$ | Fe I |
| $\mathbf{B_7}$ | = | $\frac{1}{2}\,\mathbf{a_2}$ | = | $\frac{1}{2}\,a\,\mathbf{\hat{y}}$ | $(3d)$ | Fe I |
| $\mathbf{B_8}$ | = | $\frac{1}{2}\,\mathbf{a_3}$ | = | $\frac{1}{2}\,a\,\mathbf{\hat{z}}$ | $(3d)$ | Fe I |
| $\mathbf{B_9}$ | = | $y_5\,\mathbf{a_2} + y_5\,\mathbf{a_3}$ | = | $y_5\,a\,\mathbf{\hat{y}} + y_5\,a\,\mathbf{\hat{z}}$ | $(12i)$ | Fe II |
| $\mathbf{B_{10}}$ | = | $-y_5\,\mathbf{a_2} + y_5\,\mathbf{a_3}$ | = | $-y_5\,a\,\mathbf{\hat{y}} + y_5\,a\,\mathbf{\hat{z}}$ | $(12i)$ | Fe II |
| $\mathbf{B_{11}}$ | = | $y_5\,\mathbf{a_2} - y_5\,\mathbf{a_3}$ | = | $y_5\,a\,\mathbf{\hat{y}} - y_5\,a\,\mathbf{\hat{z}}$ | $(12i)$ | Fe II |
| $\mathbf{B_{12}}$ | = | $-y_5\,\mathbf{a_2} - y_5\,\mathbf{a_3}$ | = | $-y_5\,a\,\mathbf{\hat{y}} - y_5\,a\,\mathbf{\hat{z}}$ | $(12i)$ | Fe II |
| $\mathbf{B_{13}}$ | = | $y_5\,\mathbf{a_1} + y_5\,\mathbf{a_3}$ | = | $y_5\,a\,\mathbf{\hat{x}} + y_5\,a\,\mathbf{\hat{z}}$ | $(12i)$ | Fe II |
| $\mathbf{B_{14}}$ | = | $-y_5\,\mathbf{a_1} + y_5\,\mathbf{a_3}$ | = | $-y_5\,a\,\mathbf{\hat{x}} + y_5\,a\,\mathbf{\hat{z}}$ | $(12i)$ | Fe II |
| $\mathbf{B_{15}}$ | = | $y_5\,\mathbf{a_1} - y_5\,\mathbf{a_3}$ | = | $y_5\,a\,\mathbf{\hat{x}} - y_5\,a\,\mathbf{\hat{z}}$ | $(12i)$ | Fe II |
| $\mathbf{B_{16}}$ | = | $-y_5\,\mathbf{a_1} - y_5\,\mathbf{a_3}$ | = | $-y_5\,a\,\mathbf{\hat{x}} - y_5\,a\,\mathbf{\hat{z}}$ | $(12i)$ | Fe II |
| $\mathbf{B_{17}}$ | = | $y_5\,\mathbf{a_1} + y_5\,\mathbf{a_2}$ | = | $y_5\,a\,\mathbf{\hat{x}} + y_5\,a\,\mathbf{\hat{y}}$ | $(12i)$ | Fe II |
| $\mathbf{B_{18}}$ | = | $-y_5\,\mathbf{a_1} + y_5\,\mathbf{a_2}$ | = | $-y_5\,a\,\mathbf{\hat{x}} + y_5\,a\,\mathbf{\hat{y}}$ | $(12i)$ | Fe II |
| $\mathbf{B_{19}}$ | = | $y_5\,\mathbf{a_1} - y_5\,\mathbf{a_2}$ | = | $y_5\,a\,\mathbf{\hat{x}} - y_5\,a\,\mathbf{\hat{y}}$ | $(12i)$ | Fe II |
| $\mathbf{B_{20}}$ | = | $-y_5\,\mathbf{a_1} - y_5\,\mathbf{a_2}$ | = | $-y_5\,a\,\mathbf{\hat{x}} - y_5\,a\,\mathbf{\hat{y}}$ | $(12i)$ | Fe II |
| $\mathbf{B_{21}}$ | = | $\frac{1}{2}\,\mathbf{a_1} + y_6\,\mathbf{a_2} + y_6\,\mathbf{a_3}$ | = | $\frac{1}{2}\,a\,\mathbf{\hat{x}} + y_6\,a\,\mathbf{\hat{y}} + y_6\,a\,\mathbf{\hat{z}}$ | $(12j)$ | Fe III |
| $\mathbf{B_{22}}$ | = | $\frac{1}{2}\,\mathbf{a_1} - y_6\,\mathbf{a_2} + y_6\,\mathbf{a_3}$ | = | $\frac{1}{2}\,a\,\mathbf{\hat{x}} - y_6\,a\,\mathbf{\hat{y}} + y_6\,a\,\mathbf{\hat{z}}$ | $(12j)$ | Fe III |
| $\mathbf{B_{23}}$ | = | $\frac{1}{2}\,\mathbf{a_1} + y_6\,\mathbf{a_2} - y_6\,\mathbf{a_3}$ | = | $\frac{1}{2}\,a\,\mathbf{\hat{x}} + y_6\,a\,\mathbf{\hat{y}} - y_6\,a\,\mathbf{\hat{z}}$ | $(12j)$ | Fe III |
| $\mathbf{B_{24}}$ | = | $\frac{1}{2}\,\mathbf{a_1} - y_6\,\mathbf{a_2} - y_6\,\mathbf{a_3}$ | = | $\frac{1}{2}\,a\,\mathbf{\hat{x}} - y_6\,a\,\mathbf{\hat{y}} - y_6\,a\,\mathbf{\hat{z}}$ | $(12j)$ | Fe III |
| $\mathbf{B_{25}}$ | = | $y_6\,\mathbf{a_1} + \frac{1}{2}\,\mathbf{a_2} + y_6\,\mathbf{a_3}$ | = | $y_6\,a\,\mathbf{\hat{x}} + \frac{1}{2}\,a\,\mathbf{\hat{y}} + y_6\,a\,\mathbf{\hat{z}}$ | $(12j)$ | Fe III |
| $\mathbf{B_{26}}$ | = | $-y_6\,\mathbf{a_1} + \frac{1}{2}\,\mathbf{a_2} + y_6\,\mathbf{a_3}$ | = | $-y_6\,a\,\mathbf{\hat{x}} + \frac{1}{2}\,a\,\mathbf{\hat{y}} + y_6\,a\,\mathbf{\hat{z}}$ | $(12j)$ | Fe III |
| $\mathbf{B_{27}}$ | = | $y_6\,\mathbf{a_1} + \frac{1}{2}\,\mathbf{a_2} - y_6\,\mathbf{a_3}$ | = | $y_6\,a\,\mathbf{\hat{x}} + \frac{1}{2}\,a\,\mathbf{\hat{y}} - y_6\,a\,\mathbf{\hat{z}}$ | $(12j)$ | Fe III |
| $\mathbf{B_{28}}$ | = | $-y_6\,\mathbf{a_1} + \frac{1}{2}\,\mathbf{a_2} - y_6\,\mathbf{a_3}$ | = | $-y_6\,a\,\mathbf{\hat{x}} + \frac{1}{2}\,a\,\mathbf{\hat{y}} - y_6\,a\,\mathbf{\hat{z}}$ | $(12j)$ | Fe III |
| $\mathbf{B_{29}}$ | = | $y_6\,\mathbf{a_1} + y_6\,\mathbf{a_2} + \frac{1}{2}\,\mathbf{a_3}$ | = | $y_6\,a\,\mathbf{\hat{x}} + y_6\,a\,\mathbf{\hat{y}} + \frac{1}{2}\,a\,\mathbf{\hat{z}}$ | $(12j)$ | Fe III |
| $\mathbf{B_{30}}$ | = | $-y_6\,\mathbf{a_1} + y_6\,\mathbf{a_2} + \frac{1}{2}\,\mathbf{a_3}$ | = | $-y_6\,a\,\mathbf{\hat{x}} + y_6\,a\,\mathbf{\hat{y}} + \frac{1}{2}\,a\,\mathbf{\hat{z}}$ | $(12j)$ | Fe III |
| $\mathbf{B_{31}}$ | = | $y_6\,\mathbf{a_1} - y_6\,\mathbf{a_2} + \frac{1}{2}\,\mathbf{a_3}$ | = | $y_6\,a\,\mathbf{\hat{x}} - y_6\,a\,\mathbf{\hat{y}} + \frac{1}{2}\,a\,\mathbf{\hat{z}}$ | $(12j)$ | Fe III |
| $\mathbf{B_{32}}$ | = | $-y_6\,\mathbf{a_1} - y_6\,\mathbf{a_2} + \frac{1}{2}\,\mathbf{a_3}$ | = | $-y_6\,a\,\mathbf{\hat{x}} - y_6\,a\,\mathbf{\hat{y}} + \frac{1}{2}\,a\,\mathbf{\hat{z}}$ | $(12j)$ | Fe III |

**Geometry files:**

- CIF: pp. 772
- POSCAR: pp. 773

# $Cu_3Au$ ($L1_2$) Structure: AB3_cP4_221_a_c

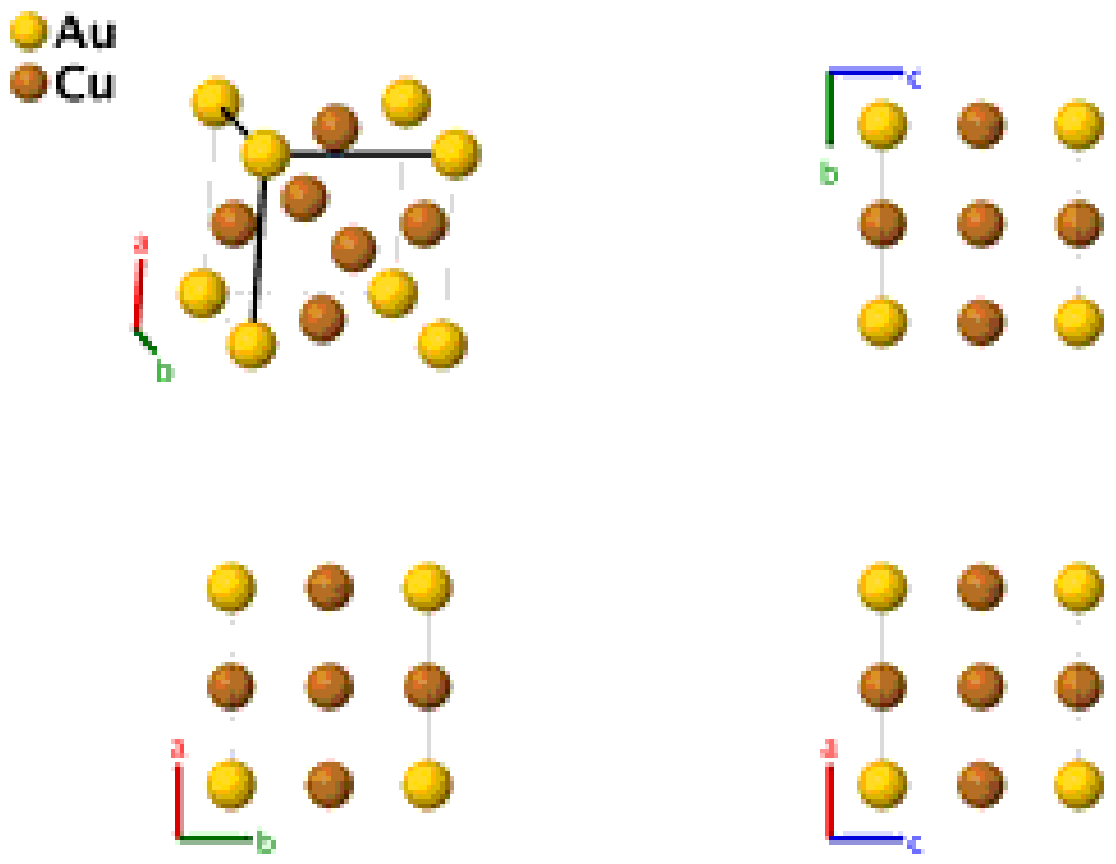

| | | |
|---|---|---|
| **Prototype** | : | $Cu_3Au$ |
| **AFLOW prototype label** | : | AB3_cP4_221_a_c |
| ***Strukturbericht* designation** | : | $L1_2$ |
| **Pearson symbol** | : | cP4 |
| **Space group number** | : | 221 |
| **Space group symbol** | : | $Pm\bar{3}m$ |
| **AFLOW prototype command** | : | `aflow --proto=AB3_cP4_221_a_c --params=`$a$ |

**Other compounds with this structure:**

- $Ni_3Al$, $Al_3Li$ (metastable), $TiPt_3$

**Simple Cubic primitive vectors:**

$$\mathbf{a}_1 = a\,\hat{\mathbf{x}}$$
$$\mathbf{a}_2 = a\,\hat{\mathbf{y}}$$
$$\mathbf{a}_3 = a\,\hat{\mathbf{z}}$$

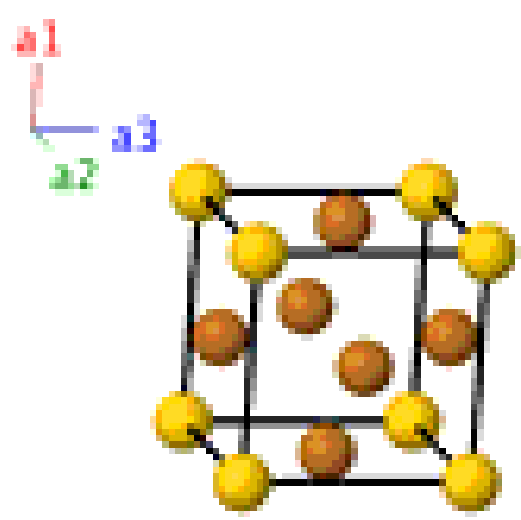

**Basis vectors:**

| | | Lattice Coordinates | | Cartesian Coordinates | Wyckoff Position | Atom Type |
|---|---|---|---|---|---|---|
| $\mathbf{B}_1$ | = | $0\,\mathbf{a}_1 + 0\,\mathbf{a}_2 + 0\,\mathbf{a}_3$ | = | $0\hat{\mathbf{x}} + 0\hat{\mathbf{y}} + 0\hat{\mathbf{z}}$ | $(1a)$ | Au |
| $\mathbf{B}_2$ | = | $\frac{1}{2}\,\mathbf{a}_2 + \frac{1}{2}\,\mathbf{a}_3$ | = | $\frac{1}{2}\,a\,\hat{\mathbf{y}} + \frac{1}{2}\,a\,\hat{\mathbf{z}}$ | $(3c)$ | Cu |
| $\mathbf{B}_3$ | = | $\frac{1}{2}\,\mathbf{a}_1 + \frac{1}{2}\,\mathbf{a}_3$ | = | $\frac{1}{2}\,a\,\hat{\mathbf{x}} + \frac{1}{2}\,a\,\hat{\mathbf{z}}$ | $(3c)$ | Cu |
| $\mathbf{B}_4$ | = | $\frac{1}{2}\,\mathbf{a}_1 + \frac{1}{2}\,\mathbf{a}_2$ | = | $\frac{1}{2}\,a\,\hat{\mathbf{x}} + \frac{1}{2}\,a\,\hat{\mathbf{y}}$ | $(3c)$ | Cu |

**Found in:**

- P. Villars and L. Calvert, *Pearson's Handbook of Crystallographic Data for Intermetallic Phases* (ASM International, Materials Park, OH, 1991), 2nd edn, pp. 1273.

---

**Geometry files:**

- CIF: pp. 773
- POSCAR: pp. 774

# $\alpha$-Po ($A_h$) Structure: A_cP1_221_a

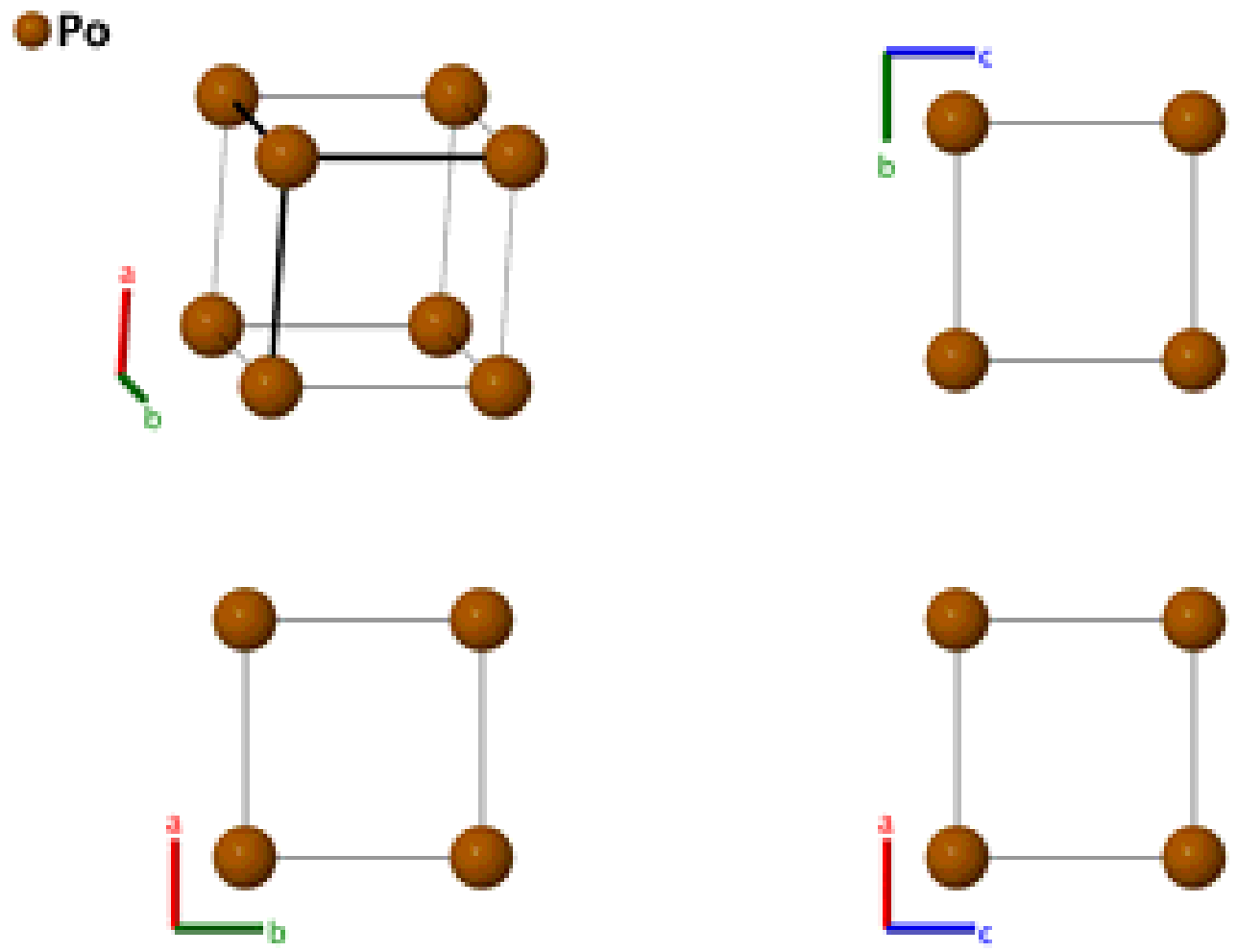

**Prototype** : $\alpha$-Po

**AFLOW prototype label** : A_cP1_221_a

***Strukturbericht* designation** : $A_h$

**Pearson symbol** : cP1

**Space group number** : 221

**Space group symbol** : $Pm\bar{3}m$

**AFLOW prototype command** : `aflow --proto=A_cP1_221_a --params=`$a$

---

- This is a simple cubic lattice. Polonium is the only element known with this ground state. Originally, Po was assigned Strukturbericht designation: A19, which is now considered to be incorrect. (Donohue, 1982, pp. 390)

---

**Simple Cubic primitive vectors:**

$$\mathbf{a}_1 = a\,\hat{\mathbf{x}}$$
$$\mathbf{a}_2 = a\,\hat{\mathbf{y}}$$
$$\mathbf{a}_3 = a\,\hat{\mathbf{z}}$$

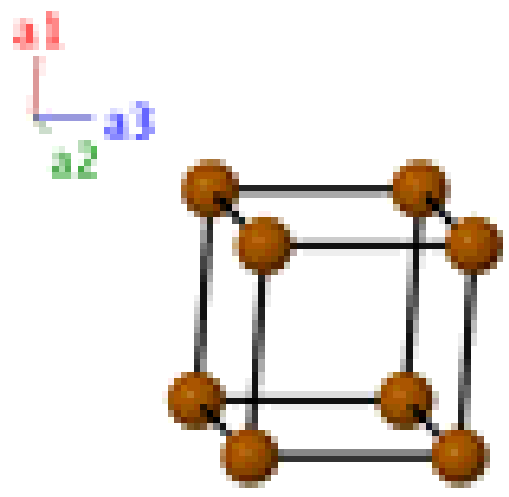

---

**Basis vectors:**

| | | Lattice Coordinates | | Cartesian Coordinates | Wyckoff Position | Atom Type |
|---|---|---|---|---|---|---|
| $\mathbf{B}_1$ | = | $0\,\mathbf{a}_1 + 0\,\mathbf{a}_2 + 0\,\mathbf{a}_3$ | = | $0\hat{\mathbf{x}} + 0\hat{\mathbf{y}} + 0\hat{\mathbf{z}}$ | $(1a)$ | Po |

---

**Found in:**

- J. Donohue, *The Structure of the Elements* (Robert E. Krieger Publishing Company, Malabar, Florida, 1982), pp. 390-391.

---

**Geometry files:**

- CIF: pp. 774
- POSCAR: pp. 774

# $BaHg_{11}$ ($D2_e$) Structure: AB11_cP36_221_c_agij

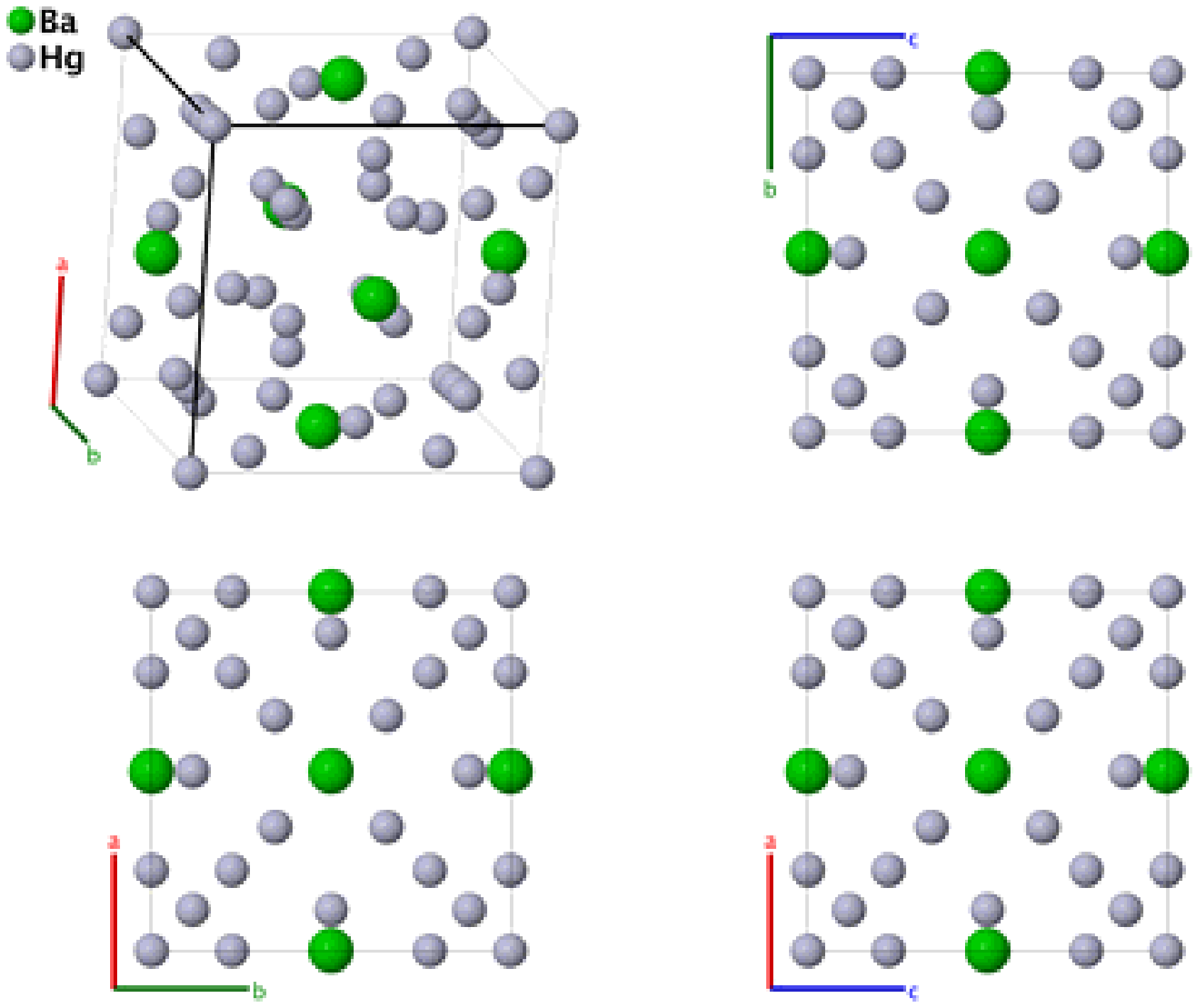

**Prototype** : $BaHg_{11}$

**AFLOW prototype label** : AB11_cP36_221_c_agij

***Strukturbericht* designation** : $D2_e$

**Pearson symbol** : cP36

**Space group number** : 221

**Space group symbol** : $Pm\bar{3}m$

**AFLOW prototype command** : `aflow --proto=AB11_cP36_221_c_agij`
`--params=`$a, x_3, y_4, y_5$

---

**Other compounds with this structure:**

- "A number of Hg and Cd phases with Group I or IIA metals or rare earths." (Pearson 1972) pp. 751-752.

---

**Simple Cubic primitive vectors:**

$$
\begin{aligned}
\mathbf{a}_1 &= a\,\hat{\mathbf{x}} \\
\mathbf{a}_2 &= a\,\hat{\mathbf{y}} \\
\mathbf{a}_3 &= a\,\hat{\mathbf{z}}
\end{aligned}
$$

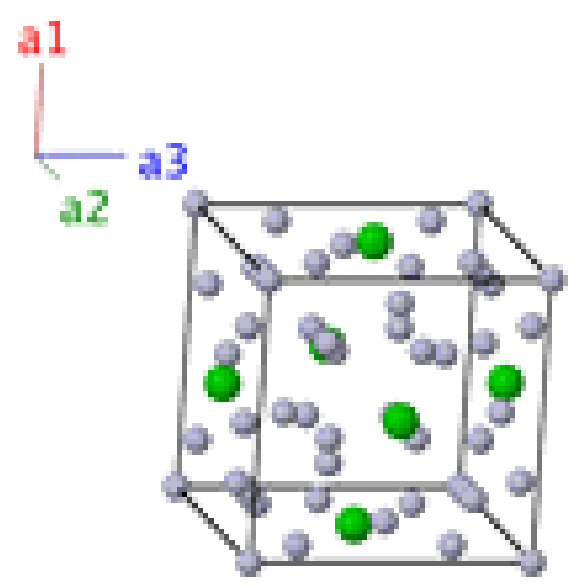

---

**Basis vectors:**

| | Lattice Coordinates | Cartesian Coordinates | Wyckoff Position | Atom Type |
|---|---|---|---|---|

| | | | | | | |
|---|---|---|---|---|---|---|
| $\mathbf{B_1}$ | = | $0\,\mathbf{a_1} + 0\,\mathbf{a_2} + 0\,\mathbf{a_3}$ | = | $0\hat{\mathbf{x}} + 0\hat{\mathbf{y}} + 0\hat{\mathbf{z}}$ | $(1a)$ | Hg I |
| $\mathbf{B_2}$ | = | $\frac{1}{2}\,\mathbf{a_2} + \frac{1}{2}\,\mathbf{a_3}$ | = | $\frac{1}{2}\,a\,\hat{\mathbf{y}} + \frac{1}{2}\,a\,\hat{\mathbf{z}}$ | $(3c)$ | Ba |
| $\mathbf{B_3}$ | = | $\frac{1}{2}\,\mathbf{a_1} + \frac{1}{2}\,\mathbf{a_3}$ | = | $\frac{1}{2}\,a\,\hat{\mathbf{x}} + \frac{1}{2}\,a\,\hat{\mathbf{z}}$ | $(3c)$ | Ba |
| $\mathbf{B_4}$ | = | $\frac{1}{2}\,\mathbf{a_1} + \frac{1}{2}\,\mathbf{a_2}$ | = | $\frac{1}{2}\,a\,\hat{\mathbf{x}} + \frac{1}{2}\,a\,\hat{\mathbf{y}}$ | $(3c)$ | Ba |
| $\mathbf{B_5}$ | = | $x_3\,\mathbf{a_1} + x_3\,\mathbf{a_2} + x_3\,\mathbf{a_3}$ | = | $x_3\,a\,\hat{\mathbf{x}} + x_3\,a\,\hat{\mathbf{y}} + x_3\,a\,\hat{\mathbf{z}}$ | $(8g)$ | Hg II |
| $\mathbf{B_6}$ | = | $-x_3\,\mathbf{a_1} - x_3\,\mathbf{a_2} + x_3\,\mathbf{a_3}$ | = | $-x_3\,a\,\hat{\mathbf{x}} - x_3\,a\,\hat{\mathbf{y}} + x_3\,a\,\hat{\mathbf{z}}$ | $(8g)$ | Hg II |
| $\mathbf{B_7}$ | = | $-x_3\,\mathbf{a_1} + x_3\,\mathbf{a_2} - x_3\,\mathbf{a_3}$ | = | $-x_3\,a\,\hat{\mathbf{x}} + x_3\,a\,\hat{\mathbf{y}} - x_3\,a\,\hat{\mathbf{z}}$ | $(8g)$ | Hg II |
| $\mathbf{B_8}$ | = | $x_3\,\mathbf{a_1} - x_3\,\mathbf{a_2} - x_3\,\mathbf{a_3}$ | = | $x_3\,a\,\hat{\mathbf{x}} - x_3\,a\,\hat{\mathbf{y}} - x_3\,a\,\hat{\mathbf{z}}$ | $(8g)$ | Hg II |
| $\mathbf{B_9}$ | = | $x_3\,\mathbf{a_1} + x_3\,\mathbf{a_2} - x_3\,\mathbf{a_3}$ | = | $x_3\,a\,\hat{\mathbf{x}} + x_3\,a\,\hat{\mathbf{y}} - x_3\,a\,\hat{\mathbf{z}}$ | $(8g)$ | Hg II |
| $\mathbf{B_{10}}$ | = | $-x_3\,\mathbf{a_1} - x_3\,\mathbf{a_2} - x_3\,\mathbf{a_3}$ | = | $-x_3\,a\,\hat{\mathbf{x}} - x_3\,a\,\hat{\mathbf{y}} - x_3\,a\,\hat{\mathbf{z}}$ | $(8g)$ | Hg II |
| $\mathbf{B_{11}}$ | = | $x_3\,\mathbf{a_1} - x_3\,\mathbf{a_2} + x_3\,\mathbf{a_3}$ | = | $x_3\,a\,\hat{\mathbf{x}} - x_3\,a\,\hat{\mathbf{y}} + x_3\,a\,\hat{\mathbf{z}}$ | $(8g)$ | Hg II |
| $\mathbf{B_{12}}$ | = | $-x_3\,\mathbf{a_1} + x_3\,\mathbf{a_2} + x_3\,\mathbf{a_3}$ | = | $-x_3\,a\,\hat{\mathbf{x}} + x_3\,a\,\hat{\mathbf{y}} + x_3\,a\,\hat{\mathbf{z}}$ | $(8g)$ | Hg II |
| $\mathbf{B_{13}}$ | = | $y_4\,\mathbf{a_2} + y_4\,\mathbf{a_3}$ | = | $y_4\,a\,\hat{\mathbf{y}} + y_4\,a\,\hat{\mathbf{z}}$ | $(12i)$ | Hg III |
| $\mathbf{B_{14}}$ | = | $y_4\,\mathbf{a_2} - y_4\,\mathbf{a_3}$ | = | $y_4\,a\,\hat{\mathbf{y}} - y_4\,a\,\hat{\mathbf{z}}$ | $(12i)$ | Hg III |
| $\mathbf{B_{15}}$ | = | $-y_4\,\mathbf{a_2} + y_4\,\mathbf{a_3}$ | = | $-y_4\,a\,\hat{\mathbf{y}} + y_4\,a\,\hat{\mathbf{z}}$ | $(12i)$ | Hg III |
| $\mathbf{B_{16}}$ | = | $-y_4\,\mathbf{a_2} - y_4\,\mathbf{a_3}$ | = | $-y_4\,a\,\hat{\mathbf{y}} - y_4\,a\,\hat{\mathbf{z}}$ | $(12i)$ | Hg III |
| $\mathbf{B_{17}}$ | = | $y_4\,\mathbf{a_1} + y_4\,\mathbf{a_3}$ | = | $y_4\,a\,\hat{\mathbf{x}} + y_4\,a\,\hat{\mathbf{z}}$ | $(12i)$ | Hg III |
| $\mathbf{B_{18}}$ | = | $y_4\,\mathbf{a_1} - y_4\,\mathbf{a_3}$ | = | $y_4\,a\,\hat{\mathbf{x}} - y_4\,a\,\hat{\mathbf{z}}$ | $(12i)$ | Hg III |
| $\mathbf{B_{19}}$ | = | $-y_4\,\mathbf{a_1} + y_4\,\mathbf{a_3}$ | = | $-y_4\,a\,\hat{\mathbf{x}} + y_4\,a\,\hat{\mathbf{z}}$ | $(12i)$ | Hg III |
| $\mathbf{B_{20}}$ | = | $-y_4\,\mathbf{a_1} - y_4\,\mathbf{a_3}$ | = | $-y_4\,a\,\hat{\mathbf{x}} - y_4\,a\,\hat{\mathbf{z}}$ | $(12i)$ | Hg III |
| $\mathbf{B_{21}}$ | = | $y_4\,\mathbf{a_1} + y_4\,\mathbf{a_2}$ | = | $y_4\,a\,\hat{\mathbf{x}} + y_4\,a\,\hat{\mathbf{y}}$ | $(12i)$ | Hg III |
| $\mathbf{B_{22}}$ | = | $y_4\,\mathbf{a_1} - y_4\,\mathbf{a_2}$ | = | $y_4\,a\,\hat{\mathbf{x}} - y_4\,a\,\hat{\mathbf{y}}$ | $(12i)$ | Hg III |
| $\mathbf{B_{23}}$ | = | $-y_4\,\mathbf{a_1} + y_4\,\mathbf{a_2}$ | = | $-y_4\,a\,\hat{\mathbf{x}} + y_4\,a\,\hat{\mathbf{y}}$ | $(12i)$ | Hg III |
| $\mathbf{B_{24}}$ | = | $-y_4\,\mathbf{a_1} - y_4\,\mathbf{a_2}$ | = | $-y_4\,a\,\hat{\mathbf{x}} - y_4\,a\,\hat{\mathbf{y}}$ | $(12i)$ | Hg III |
| $\mathbf{B_{25}}$ | = | $\frac{1}{2}\,\mathbf{a_1} + y_5\,\mathbf{a_2} + y_5\,\mathbf{a_3}$ | = | $\frac{1}{2}\,a\,\hat{\mathbf{x}} + y_5\,a\,\hat{\mathbf{y}} + y_5\,a\,\hat{\mathbf{z}}$ | $(12j)$ | Hg IV |
| $\mathbf{B_{26}}$ | = | $\frac{1}{2}\,\mathbf{a_1} + y_5\,\mathbf{a_2} - y_5\,\mathbf{a_3}$ | = | $\frac{1}{2}\,a\,\hat{\mathbf{x}} + y_5\,a\,\hat{\mathbf{y}} - y_5\,a\,\hat{\mathbf{z}}$ | $(12j)$ | Hg IV |
| $\mathbf{B_{27}}$ | = | $\frac{1}{2}\,\mathbf{a_1} - y_5\,\mathbf{a_2} + y_5\,\mathbf{a_3}$ | = | $\frac{1}{2}\,a\,\hat{\mathbf{x}} - y_5\,a\,\hat{\mathbf{y}} + y_5\,a\,\hat{\mathbf{z}}$ | $(12j)$ | Hg IV |
| $\mathbf{B_{28}}$ | = | $\frac{1}{2}\,\mathbf{a_1} - y_5\,\mathbf{a_2} - y_5\,\mathbf{a_3}$ | = | $\frac{1}{2}\,a\,\hat{\mathbf{x}} - y_5\,a\,\hat{\mathbf{y}} - y_5\,a\,\hat{\mathbf{z}}$ | $(12j)$ | Hg IV |
| $\mathbf{B_{29}}$ | = | $y_5\,\mathbf{a_1} + \frac{1}{2}\,\mathbf{a_2} + y_5\,\mathbf{a_3}$ | = | $y_5\,a\,\hat{\mathbf{x}} + \frac{1}{2}\,a\,\hat{\mathbf{y}} + y_5\,a\,\hat{\mathbf{z}}$ | $(12j)$ | Hg IV |
| $\mathbf{B_{30}}$ | = | $y_5\,\mathbf{a_1} + \frac{1}{2}\,\mathbf{a_2} - y_5\,\mathbf{a_3}$ | = | $y_5\,a\,\hat{\mathbf{x}} + \frac{1}{2}\,a\,\hat{\mathbf{y}} - y_5\,a\,\hat{\mathbf{z}}$ | $(12j)$ | Hg IV |
| $\mathbf{B_{31}}$ | = | $-y_5\,\mathbf{a_1} + \frac{1}{2}\,\mathbf{a_2} + y_5\,\mathbf{a_3}$ | = | $-y_5\,a\,\hat{\mathbf{x}} + \frac{1}{2}\,a\,\hat{\mathbf{y}} + y_5\,a\,\hat{\mathbf{z}}$ | $(12j)$ | Hg IV |
| $\mathbf{B_{32}}$ | = | $-y_5\,\mathbf{a_1} + \frac{1}{2}\,\mathbf{a_2} - y_5\,\mathbf{a_3}$ | = | $-y_5\,a\,\hat{\mathbf{x}} + \frac{1}{2}\,a\,\hat{\mathbf{y}} - y_5\,a\,\hat{\mathbf{z}}$ | $(12j)$ | Hg IV |
| $\mathbf{B_{33}}$ | = | $y_5\,\mathbf{a_1} + y_5\,\mathbf{a_2} + \frac{1}{2}\,\mathbf{a_3}$ | = | $y_5\,a\,\hat{\mathbf{x}} + y_5\,a\,\hat{\mathbf{y}} + \frac{1}{2}\,a\,\hat{\mathbf{z}}$ | $(12j)$ | Hg IV |
| $\mathbf{B_{34}}$ | = | $y_5\,\mathbf{a_1} - y_5\,\mathbf{a_2} + \frac{1}{2}\,\mathbf{a_3}$ | = | $y_5\,a\,\hat{\mathbf{x}} - y_5\,a\,\hat{\mathbf{y}} + \frac{1}{2}\,a\,\hat{\mathbf{z}}$ | $(12j)$ | Hg IV |
| $\mathbf{B_{35}}$ | = | $-y_5\,\mathbf{a_1} + y_5\,\mathbf{a_2} + \frac{1}{2}\,\mathbf{a_3}$ | = | $-y_5\,a\,\hat{\mathbf{x}} + y_5\,a\,\hat{\mathbf{y}} + \frac{1}{2}\,a\,\hat{\mathbf{z}}$ | $(12j)$ | Hg IV |
| $\mathbf{B_{36}}$ | = | $-y_5\,\mathbf{a_1} - y_5\,\mathbf{a_2} + \frac{1}{2}\,\mathbf{a_3}$ | = | $-y_5\,a\,\hat{\mathbf{x}} - y_5\,a\,\hat{\mathbf{y}} + \frac{1}{2}\,a\,\hat{\mathbf{z}}$ | $(12j)$ | Hg IV |

**Found in:**

**Geometry files:**

- CIF: pp. 774
- POSCAR: pp. 775

# Model of Ferrite Structure (cP16): AB11CD3_cP16_221_a_dg_b_c

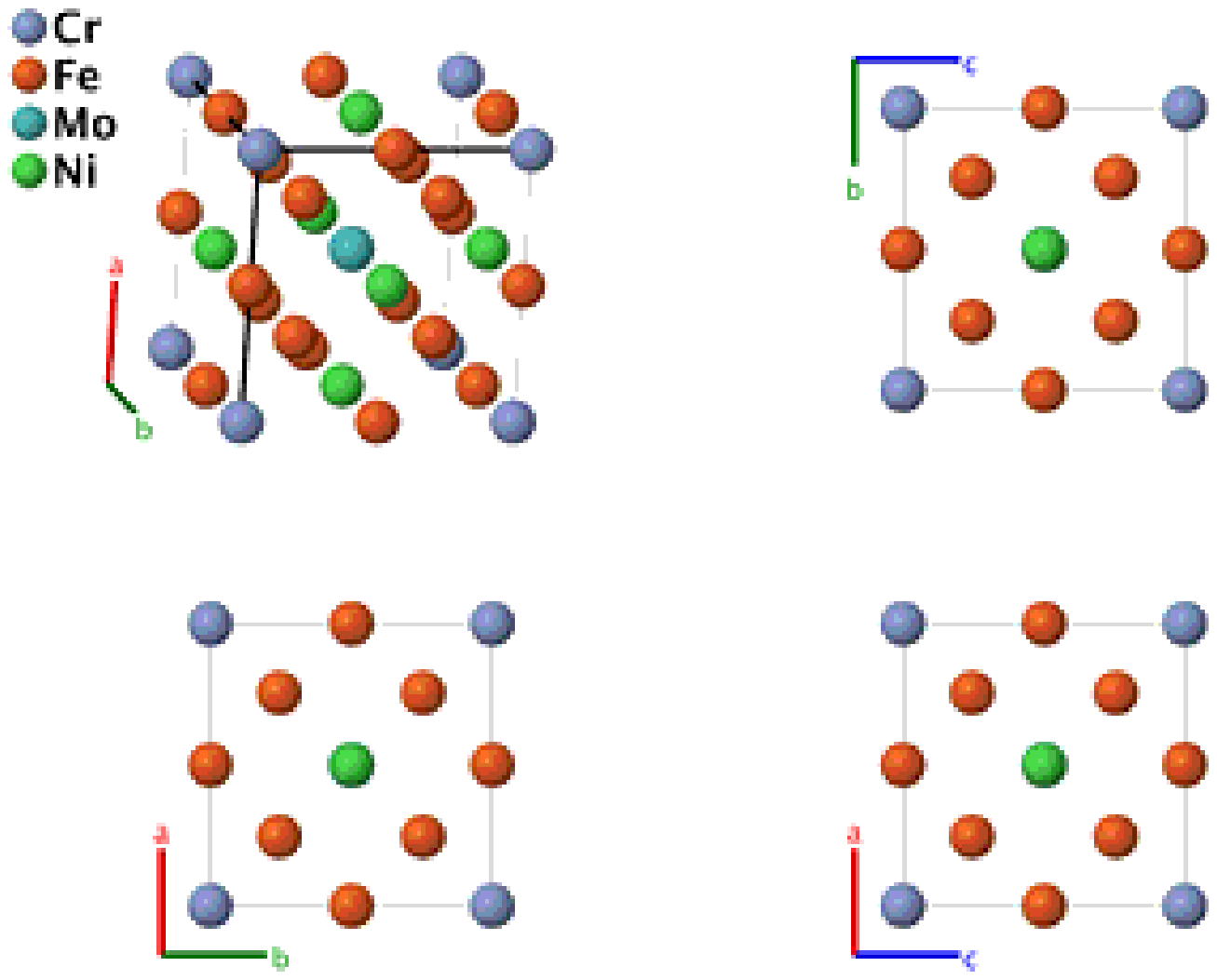

| | | |
|---|---|---|
| **Prototype** | : | $CrFe_{11}MoNi_3$ |
| **AFLOW prototype label** | : | AB11CD3_cP16_221_a_dg_b_c |
| ***Strukturbericht* designation** | : | None |
| **Pearson symbol** | : | cP16 |
| **Space group number** | : | 221 |
| **Space group symbol** | : | $Pm\bar{3}m$ |
| **AFLOW prototype command** | : | `aflow --proto=AB11CD3_cP16_221_a_dg_b_c --params=`$a, x_5$ |

- Ferritic steels are alloys of iron and other metals with an averaged body-centered cubic structure. This model represents one approximation for a ferritic steel. It is not meant to represent a real steel, and the selection of atom types for each Wyckoff position is arbitrary. Note that when $x_5 = 1/4$ all the atoms are on sites of a bcc lattice.

**Simple Cubic primitive vectors:**

$$\mathbf{a}_1 = a\,\mathbf{\hat{x}}$$
$$\mathbf{a}_2 = a\,\mathbf{\hat{y}}$$
$$\mathbf{a}_3 = a\,\mathbf{\hat{z}}$$

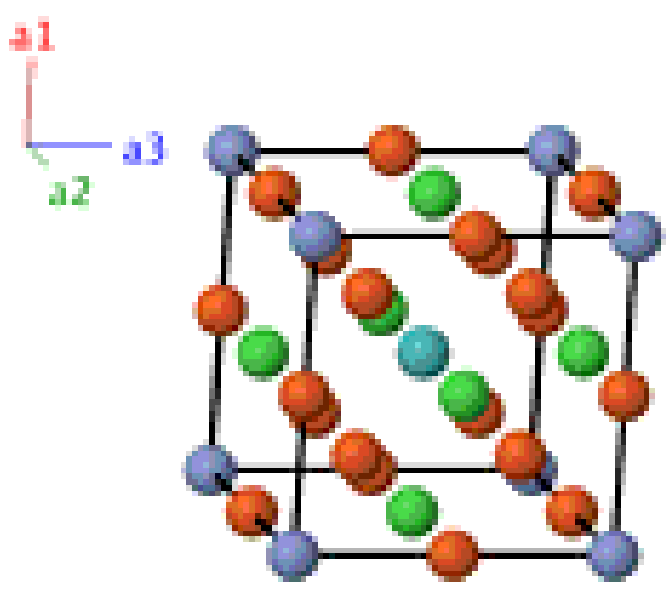

**Basis vectors:**

| | | Lattice Coordinates | | Cartesian Coordinates | Wyckoff Position | Atom Type |
|---|---|---|---|---|---|---|
| $\mathbf{B}_1$ | = | $0\,\mathbf{a}_1 + 0\,\mathbf{a}_2 + 0\,\mathbf{a}_3$ | = | $0\mathbf{\hat{x}} + 0\mathbf{\hat{y}} + 0\mathbf{\hat{z}}$ | (1$a$) | Cr |
| $\mathbf{B}_2$ | = | $\frac{1}{2}\,\mathbf{a}_1 + \frac{1}{2}\,\mathbf{a}_2 + \frac{1}{2}\,\mathbf{a}_3$ | = | $\frac{1}{2}\,a\,\mathbf{\hat{x}} + \frac{1}{2}\,a\,\mathbf{\hat{y}} + \frac{1}{2}\,a\,\mathbf{\hat{z}}$ | (1$b$) | Mo |
| $\mathbf{B}_3$ | = | $\frac{1}{2}\,\mathbf{a}_2 + \frac{1}{2}\,\mathbf{a}_3$ | = | $\frac{1}{2}\,a\,\hat{\mathbf{y}} + \frac{1}{2}\,a\,\hat{\mathbf{z}}$ | $(3c)$ | Ni |
| $\mathbf{B}_4$ | = | $\frac{1}{2}\,\mathbf{a}_1 + \frac{1}{2}\,\mathbf{a}_3$ | = | $\frac{1}{2}\,a\,\hat{\mathbf{x}} + \frac{1}{2}\,a\,\hat{\mathbf{z}}$ | $(3c)$ | Ni |
| $\mathbf{B}_5$ | = | $\frac{1}{2}\,\mathbf{a}_1 + \frac{1}{2}\,\mathbf{a}_2$ | = | $\frac{1}{2}\,a\,\hat{\mathbf{x}} + \frac{1}{2}\,a\,\hat{\mathbf{y}}$ | $(3c)$ | Ni |
| $\mathbf{B}_6$ | = | $\frac{1}{2}\,\mathbf{a}_1$ | = | $\frac{1}{2}\,a\,\hat{\mathbf{x}}$ | $(3d)$ | Fe I |
| $\mathbf{B}_7$ | = | $\frac{1}{2}\,\mathbf{a}_2$ | = | $\frac{1}{2}\,a\,\hat{\mathbf{y}}$ | $(3d)$ | Fe I |
| $\mathbf{B}_8$ | = | $\frac{1}{2}\,\mathbf{a}_3$ | = | $\frac{1}{2}\,a\,\hat{\mathbf{z}}$ | $(3d)$ | Fe I |
| $\mathbf{B}_9$ | = | $x_5\,\mathbf{a}_1 + x_5\,\mathbf{a}_2 + x_5\,\mathbf{a}_3$ | = | $x_5\,a\,\hat{\mathbf{x}} + x_5\,a\,\hat{\mathbf{y}} + x_5\,a\,\hat{\mathbf{z}}$ | $(8g)$ | Fe II |
| $\mathbf{B}_{10}$ | = | $-x_5\,\mathbf{a}_1 - x_5\,\mathbf{a}_2 + x_5\,\mathbf{a}_3$ | = | $-x_5\,a\,\hat{\mathbf{x}} - x_5\,a\,\hat{\mathbf{y}} + x_5\,a\,\hat{\mathbf{z}}$ | $(8g)$ | Fe II |
| $\mathbf{B}_{11}$ | = | $-x_5\,\mathbf{a}_1 + x_5\,\mathbf{a}_2 - x_5\,\mathbf{a}_3$ | = | $-x_5\,a\,\hat{\mathbf{x}} + x_5\,a\,\hat{\mathbf{y}} - x_5\,a\,\hat{\mathbf{z}}$ | $(8g)$ | Fe II |
| $\mathbf{B}_{12}$ | = | $x_5\,\mathbf{a}_1 - x_5\,\mathbf{a}_2 - x_5\,\mathbf{a}_3$ | = | $x_5\,a\,\hat{\mathbf{x}} - x_5\,a\,\hat{\mathbf{y}} - x_5\,a\,\hat{\mathbf{z}}$ | $(8g)$ | Fe II |
| $\mathbf{B}_{13}$ | = | $x_5\,\mathbf{a}_1 + x_5\,\mathbf{a}_2 - x_5\,\mathbf{a}_3$ | = | $x_5\,a\,\hat{\mathbf{x}} + x_5\,a\,\hat{\mathbf{y}} - x_5\,a\,\hat{\mathbf{z}}$ | $(8g)$ | Fe II |
| $\mathbf{B}_{14}$ | = | $-x_5\,\mathbf{a}_1 - x_5\,\mathbf{a}_2 - x_5\,\mathbf{a}_3$ | = | $-x_5\,a\,\hat{\mathbf{x}} - x_5\,a\,\hat{\mathbf{y}} - x_5\,a\,\hat{\mathbf{z}}$ | $(8g)$ | Fe II |
| $\mathbf{B}_{15}$ | = | $x_5\,\mathbf{a}_1 - x_5\,\mathbf{a}_2 + x_5\,\mathbf{a}_3$ | = | $x_5\,a\,\hat{\mathbf{x}} - x_5\,a\,\hat{\mathbf{y}} + x_5\,a\,\hat{\mathbf{z}}$ | $(8g)$ | Fe II |
| $\mathbf{B}_{16}$ | = | $-x_5\,\mathbf{a}_1 + x_5\,\mathbf{a}_2 + x_5\,\mathbf{a}_3$ | = | $-x_5\,a\,\hat{\mathbf{x}} + x_5\,a\,\hat{\mathbf{y}} + x_5\,a\,\hat{\mathbf{z}}$ | $(8g)$ | Fe II |

---

---

**Geometry files:**

- CIF: pp. 775
- POSCAR: pp. 776

# $\alpha$-$ReO_3$ ($D0_9$) Structure: A3B_cP4_221_d_a

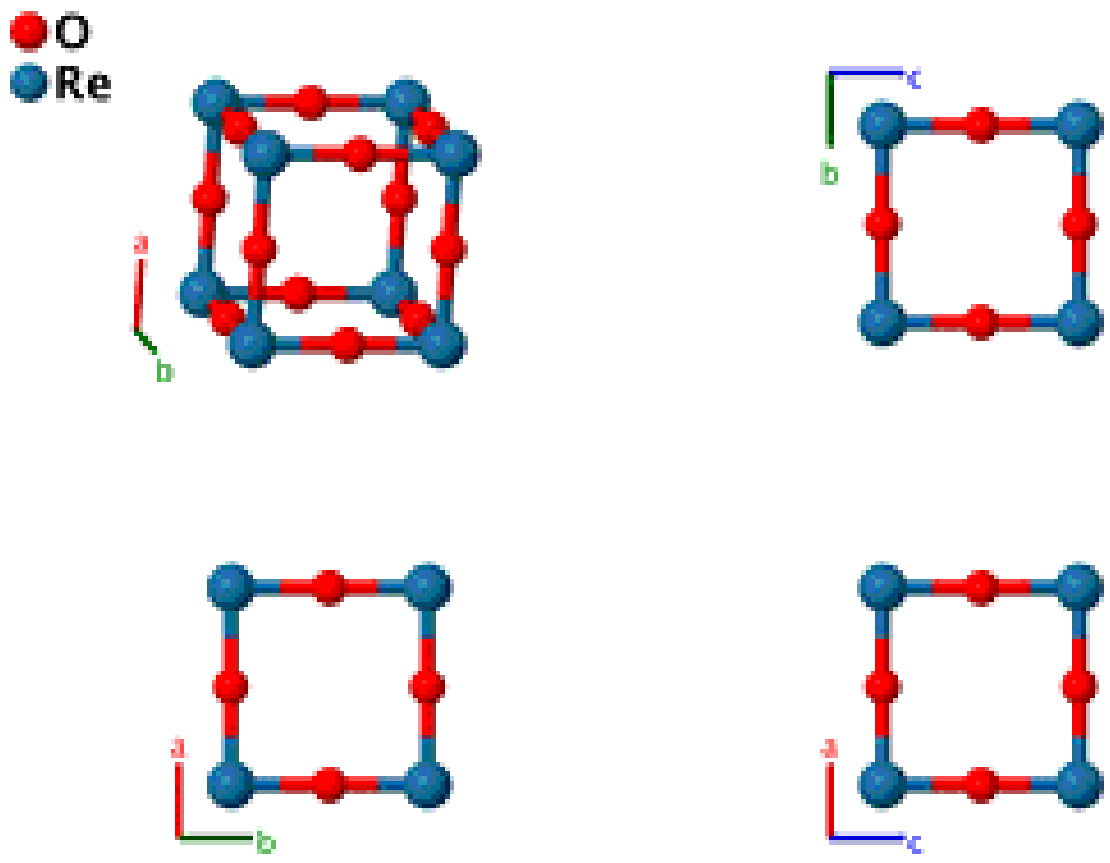

| | | |
|---|---|---|
| **Prototype** | : | $\alpha$-$ReO_3$ |
| **AFLOW prototype label** | : | A3B_cP4_221_d_a |
| ***Strukturbericht* designation** | : | $D0_9$ |
| **Pearson symbol** | : | cP4 |
| **Space group number** | : | 221 |
| **Space group symbol** | : | $Pm\bar{3}m$ |
| **AFLOW prototype command** | : | `aflow --proto=A3B_cP4_221_d_a --params=`$a$ |

**Other compounds with this structure:**

- $Cu_3N$, $WO_3$, $UO_3$

**Simple Cubic primitive vectors:**

$$\mathbf{a}_1 = a\,\hat{\mathbf{x}}$$
$$\mathbf{a}_2 = a\,\hat{\mathbf{y}}$$
$$\mathbf{a}_3 = a\,\hat{\mathbf{z}}$$

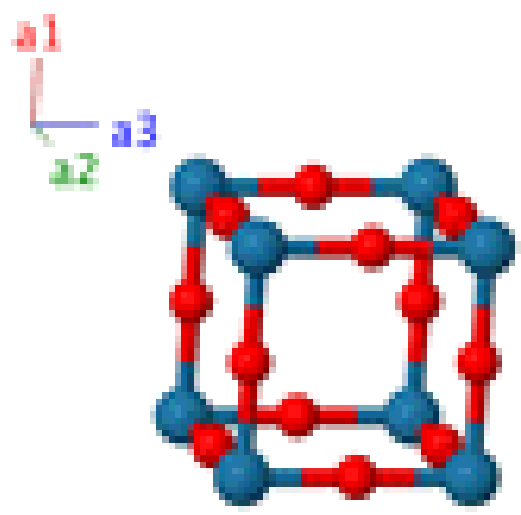

**Basis vectors:**

| | | Lattice Coordinates | | Cartesian Coordinates | Wyckoff Position | Atom Type |
|---|---|---|---|---|---|---|
| $\mathbf{B}_1$ | = | $0\,\mathbf{a}_1 + 0\,\mathbf{a}_2 + 0\,\mathbf{a}_3$ | = | $0\hat{\mathbf{x}} + 0\hat{\mathbf{y}} + 0\hat{\mathbf{z}}$ | $(1a)$ | Re |
| $\mathbf{B}_2$ | = | $\frac{1}{2}\,\mathbf{a}_1$ | = | $\frac{1}{2}\,a\,\hat{\mathbf{x}}$ | $(3d)$ | O |
| $\mathbf{B}_3$ | = | $\frac{1}{2}\,\mathbf{a}_2$ | = | $\frac{1}{2}\,a\,\hat{\mathbf{y}}$ | $(3d)$ | O |
| $\mathbf{B}_4$ | = | $\frac{1}{2}\,\mathbf{a}_3$ | = | $\frac{1}{2}\,a\,\hat{\mathbf{z}}$ | $(3d)$ | O |

**Found in:**

---

**Geometry files:**

- CIF: pp. 776
- POSCAR: pp. 776

# $CaB_6$ ($D2_1$) Structure: A6B_cP7_221_f_a

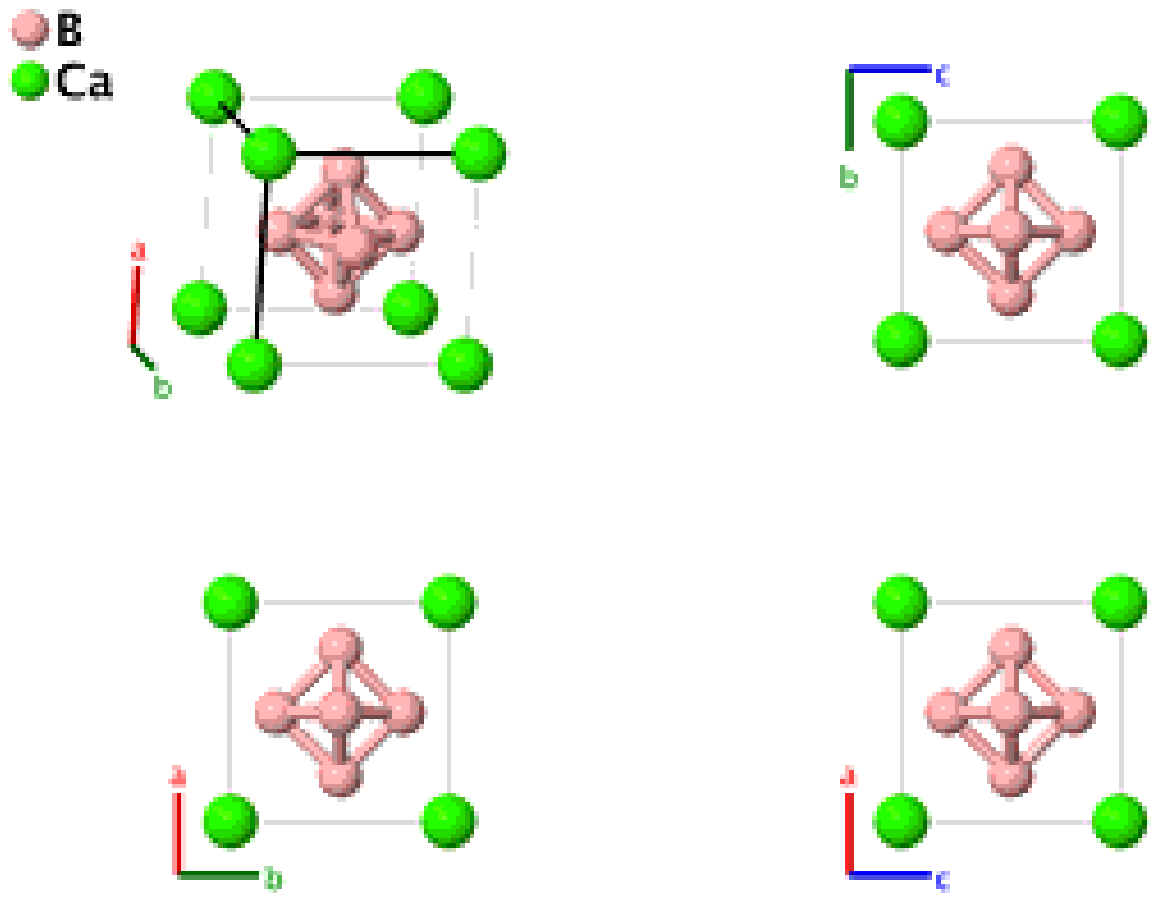

**Prototype** : $CaB_6$

**AFLOW prototype label** : A6B_cP7_221_f_a

***Strukturbericht* designation** : $D2_1$

**Pearson symbol** : cP7

**Space group number** : 221

**Space group symbol** : $Pm\bar{3}m$

**AFLOW prototype command** : `aflow --proto=A6B_cP7_221_f_a`
`--params=`$a, x_2$

---

**Simple Cubic primitive vectors:**

$$\mathbf{a}_1 = a\,\mathbf{\hat{x}}$$
$$\mathbf{a}_2 = a\,\mathbf{\hat{y}}$$
$$\mathbf{a}_3 = a\,\mathbf{\hat{z}}$$

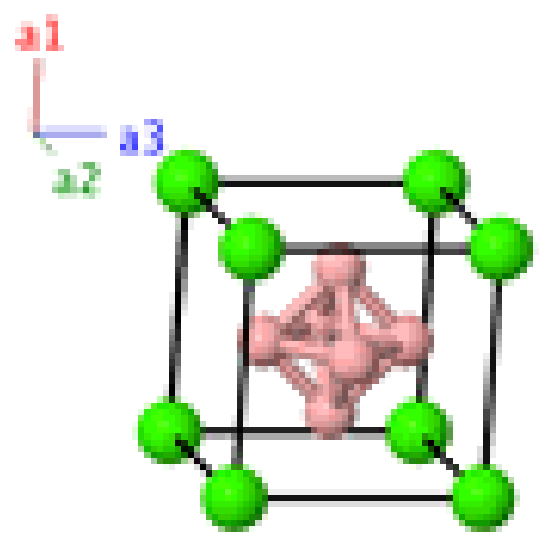

---

**Basis vectors:**

| | | Lattice Coordinates | | Cartesian Coordinates | Wyckoff Position | Atom Type |
|---|---|---|---|---|---|---|
| $\mathbf{B}_1$ | = | $0\,\mathbf{a}_1 + 0\,\mathbf{a}_2 + 0\,\mathbf{a}_3$ | = | $0\mathbf{\hat{x}} + 0\mathbf{\hat{y}} + 0\mathbf{\hat{z}}$ | $(1a)$ | Ca |
| $\mathbf{B}_2$ | = | $x_2\,\mathbf{a}_1 + \frac{1}{2}\,\mathbf{a}_2 + \frac{1}{2}\,\mathbf{a}_3$ | = | $x_2\,a\,\mathbf{\hat{x}} + \frac{1}{2}\,a\,\mathbf{\hat{y}} + \frac{1}{2}\,a\,\mathbf{\hat{z}}$ | $(6f)$ | B |
| $\mathbf{B}_3$ | = | $-x_2\,\mathbf{a}_1 + \frac{1}{2}\,\mathbf{a}_2 + \frac{1}{2}\,\mathbf{a}_3$ | = | $-x_2\,a\,\mathbf{\hat{x}} + \frac{1}{2}\,a\,\mathbf{\hat{y}} + \frac{1}{2}\,a\,\mathbf{\hat{z}}$ | $(6f)$ | B |
| $\mathbf{B}_4$ | = | $\frac{1}{2}\,\mathbf{a}_1 + x_2\,\mathbf{a}_2 + \frac{1}{2}\,\mathbf{a}_3$ | = | $\frac{1}{2}\,a\,\mathbf{\hat{x}} + x_2\,a\,\mathbf{\hat{y}} + \frac{1}{2}\,a\,\mathbf{\hat{z}}$ | $(6f)$ | B |
| $\mathbf{B}_5$ | = | $\frac{1}{2}\,\mathbf{a}_1 - x_2\,\mathbf{a}_2 + \frac{1}{2}\,\mathbf{a}_3$ | = | $\frac{1}{2}\,a\,\mathbf{\hat{x}} - x_2\,a\,\mathbf{\hat{y}} + \frac{1}{2}\,a\,\mathbf{\hat{z}}$ | $(6f)$ | B |
| $\mathbf{B}_6$ | = | $\frac{1}{2}\,\mathbf{a}_1 + \frac{1}{2}\,\mathbf{a}_2 + x_2\,\mathbf{a}_3$ | = | $\frac{1}{2}\,a\,\mathbf{\hat{x}} + \frac{1}{2}\,a\,\mathbf{\hat{y}} + x_2\,a\,\mathbf{\hat{z}}$ | $(6f)$ | B |
| $\mathbf{B}_7$ | = | $\frac{1}{2}\,\mathbf{a}_1 + \frac{1}{2}\,\mathbf{a}_2 - x_2\,\mathbf{a}_3$ | = | $\frac{1}{2}\,a\,\mathbf{\hat{x}} + \frac{1}{2}\,a\,\mathbf{\hat{y}} - x_2\,a\,\mathbf{\hat{z}}$ | $(6f)$ | B |

**Geometry files:**
- CIF: pp. 776
- POSCAR: pp. 777

# $Cr_3Si$ (A15) Structure: A3B_cP8_223_c_a

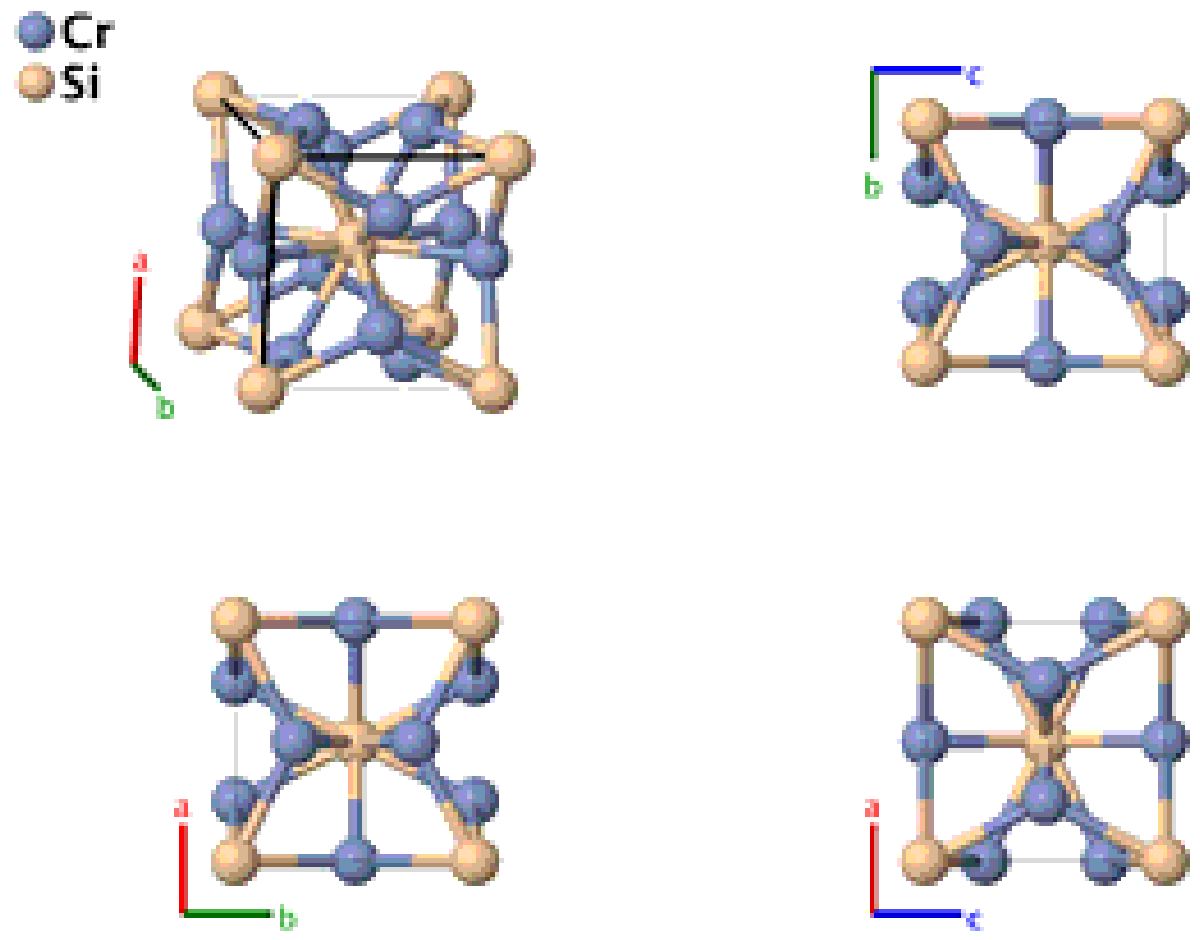

| | | |
|---|---|---|
| **Prototype** | : | $Cr_3Si$ |
| **AFLOW prototype label** | : | A3B_cP8_223_c_a |
| ***Strukturbericht* designation** | : | A15 |
| **Pearson symbol** | : | cP8 |
| **Space group number** | : | 223 |
| **Space group symbol** | : | $Pm\bar{3}n$ |
| **AFLOW prototype command** | : | `aflow --proto=A3B_cP8_223_c_a --params=`$a$ |

**Other compounds with this structure:**

- $\beta$-W, $Nb_3Al$, $CdV_3$, $Cr_3O$, $Ti_3Sb$, $Ti_3Au$, many more

- The "A" Strukturbericht designation comes from the fact that this is also the structure of $\beta$-W.

**Simple Cubic primitive vectors:**

$$\mathbf{a}_1 = a\,\hat{\mathbf{x}}$$
$$\mathbf{a}_2 = a\,\hat{\mathbf{y}}$$
$$\mathbf{a}_3 = a\,\hat{\mathbf{z}}$$

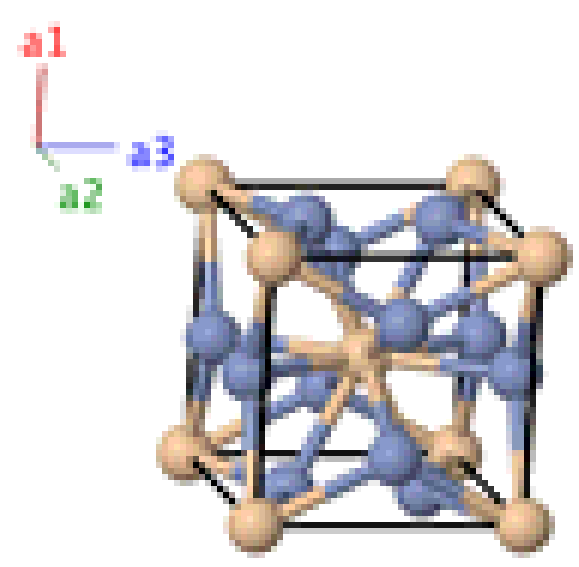

**Basis vectors:**

| | | Lattice Coordinates | | Cartesian Coordinates | Wyckoff Position | Atom Type |
|---|---|---|---|---|---|---|
| $\mathbf{B}_1$ | = | $0\,\mathbf{a}_1 + 0\,\mathbf{a}_2 + 0\,\mathbf{a}_3$ | = | $0\hat{\mathbf{x}} + 0\hat{\mathbf{y}} + 0\hat{\mathbf{z}}$ | $(2a)$ | Si |
| $\mathbf{B}_2$ | = | $\frac{1}{2}\,\mathbf{a}_1 + \frac{1}{2}\,\mathbf{a}_2 + \frac{1}{2}\,\mathbf{a}_3$ | = | $\frac{1}{2}\,a\,\hat{\mathbf{x}} + \frac{1}{2}\,a\,\hat{\mathbf{y}} + \frac{1}{2}\,a\,\hat{\mathbf{z}}$ | $(2a)$ | Si |
| $\mathbf{B}_3$ | = | $\frac{1}{4}\,\mathbf{a}_1 + \frac{1}{2}\,\mathbf{a}_3$ | = | $\frac{1}{4}\,a\,\hat{\mathbf{x}} + \frac{1}{2}\,a\,\hat{\mathbf{z}}$ | $(6c)$ | Cr |
| $\mathbf{B_4}$ | = | $\frac{3}{4}\,\mathbf{a_1} + \frac{1}{2}\,\mathbf{a_3}$ | = | $\frac{3}{4}\,a\,\hat{\mathbf{x}} + \frac{1}{2}\,a\,\hat{\mathbf{z}}$ | $(6c)$ | Cr |
| $\mathbf{B_5}$ | = | $\frac{1}{2}\,\mathbf{a_1} + \frac{1}{4}\,\mathbf{a_2}$ | = | $\frac{1}{2}\,a\,\hat{\mathbf{x}} + \frac{1}{4}\,a\,\hat{\mathbf{y}}$ | $(6c)$ | Cr |
| $\mathbf{B_6}$ | = | $\frac{1}{2}\,\mathbf{a_1} + \frac{3}{4}\,\mathbf{a_2}$ | = | $\frac{1}{2}\,a\,\hat{\mathbf{x}} + \frac{3}{4}\,a\,\hat{\mathbf{y}}$ | $(6c)$ | Cr |
| $\mathbf{B_7}$ | = | $\frac{1}{2}\,\mathbf{a_2} + \frac{1}{4}\,\mathbf{a_3}$ | = | $\frac{1}{2}\,a\,\hat{\mathbf{y}} + \frac{1}{4}\,a\,\hat{\mathbf{z}}$ | $(6c)$ | Cr |
| $\mathbf{B_8}$ | = | $\frac{1}{2}\,\mathbf{a_2} + \frac{3}{4}\,\mathbf{a_3}$ | = | $\frac{1}{2}\,a\,\hat{\mathbf{y}} + \frac{3}{4}\,a\,\hat{\mathbf{z}}$ | $(6c)$ | Cr |

**Found in:**

**Geometry files:**

- CIF: pp. 777
- POSCAR: pp. 778

# $Si_{46}$ Clathrate Structure: A_cP46_223_dik

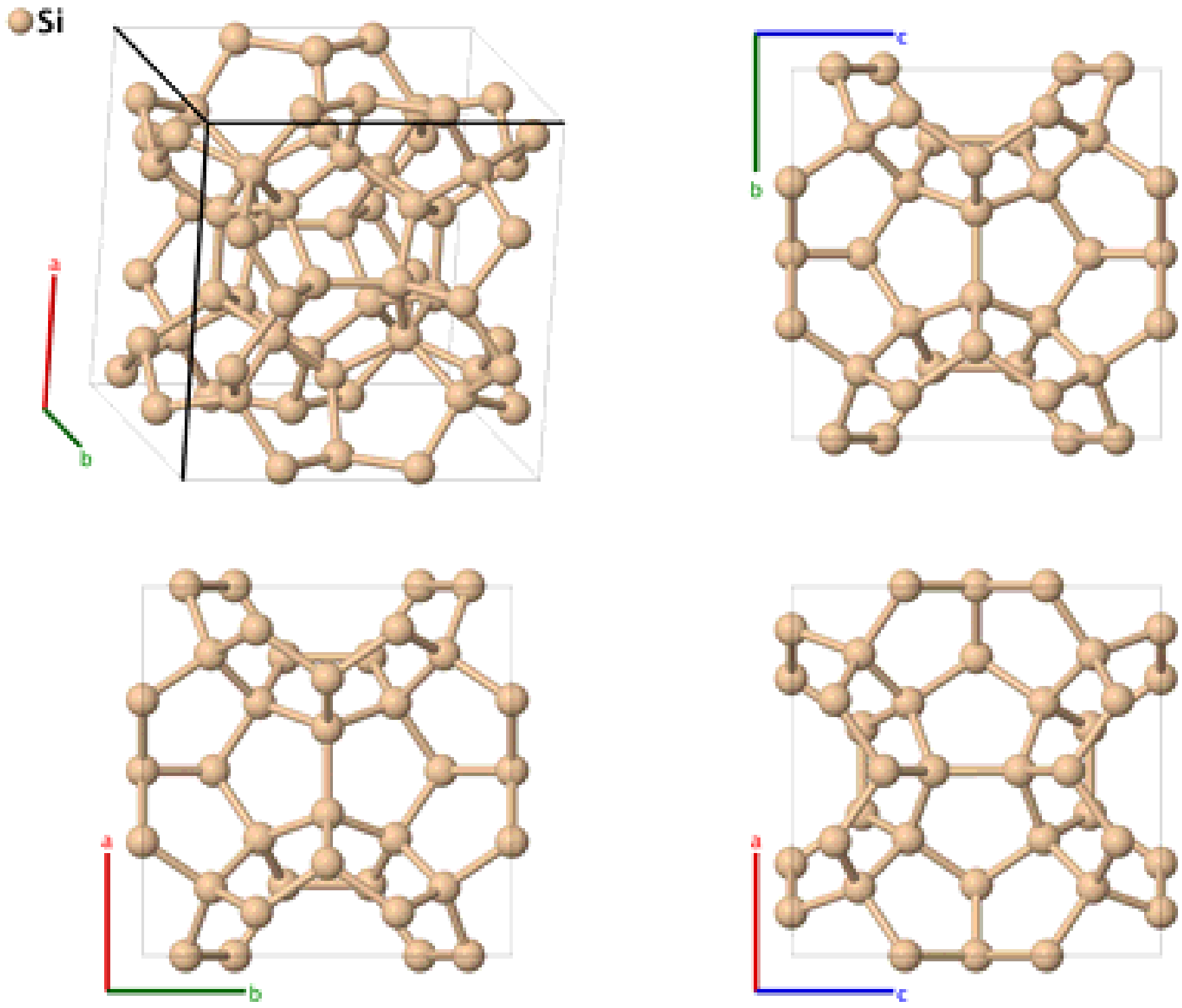

| | | |
|---|---|---|
| **Prototype** | : | Si |
| **AFLOW prototype label** | : | A_cP46_223_dik |
| ***Strukturbericht* designation** | : | None |
| **Pearson symbol** | : | cP46 |
| **Space group number** | : | 223 |
| **Space group symbol** | : | $Pm\bar{3}n$ |
| **AFLOW prototype command** | : | `aflow --proto=A_cP46_223_dik` `--params=`$a, x_2, y_3, z_3$ |

**Other compounds with this structure:**

- $\beta$-W, $Nb_3Al$, $CdV_3$, $Cr_3O$, $Ti_3Sb$, $Ti_3Au$, many more

- Silicon clathrates are open structures of pentagonal dodecahedra connected so that all of the silicon atoms have $sp^3$ bonding. In nature these structures are stabilized by alkali impurity atoms. This structure and the $Si_{34}$ structure are proposed "pure" silicon clathrate structures. For more information about these structures and their possible stability, see (Adams, 1994). Note that this is a theoretical description of a possible silicon clathrate crystal.

**Simple Cubic primitive vectors:**

$$\mathbf{a}_1 = a\,\mathbf{\hat{x}}$$
$$\mathbf{a}_2 = a\,\mathbf{\hat{y}}$$
$$\mathbf{a}_3 = a\,\mathbf{\hat{z}}$$

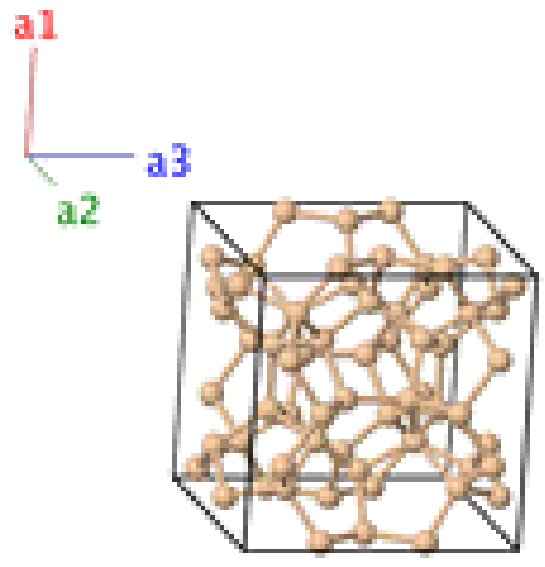

**Basis vectors:**

| | | Lattice Coordinates | | Cartesian Coordinates | Wyckoff Position | Atom Type |
|---|---|---|---|---|---|---|
| $\mathbf{B}_1$ | = | $\frac{1}{4}\,\mathbf{a}_1 + \frac{1}{2}\,\mathbf{a}_2$ | = | $\frac{1}{4}\,a\,\mathbf{\hat{x}} + \frac{1}{2}\,a\,\mathbf{\hat{y}}$ | (6$d$) | Si I |
| $\mathbf{B}_2$ | = | $\frac{3}{4}\,\mathbf{a}_1 + \frac{1}{2}\,\mathbf{a}_2$ | = | $\frac{3}{4}\,a\,\mathbf{\hat{x}} + \frac{1}{2}\,a\,\mathbf{\hat{y}}$ | (6$d$) | Si I |
| $\mathbf{B}_3$ | = | $\frac{1}{4}\,\mathbf{a}_2 + \frac{1}{2}\,\mathbf{a}_3$ | = | $\frac{1}{4}\,a\,\mathbf{\hat{y}} + \frac{1}{2}\,a\,\mathbf{\hat{z}}$ | (6$d$) | Si I |
| $\mathbf{B}_4$ | = | $\frac{3}{4}\,\mathbf{a}_2 + \frac{1}{2}\,\mathbf{a}_3$ | = | $\frac{3}{4}\,a\,\mathbf{\hat{y}} + \frac{1}{2}\,a\,\mathbf{\hat{z}}$ | (6$d$) | Si I |
| $\mathbf{B}_5$ | = | $\frac{1}{2}\,\mathbf{a}_1 + \frac{1}{4}\,\mathbf{a}_3$ | = | $\frac{1}{2}\,a\,\mathbf{\hat{x}} + \frac{1}{4}\,a\,\mathbf{\hat{z}}$ | (6$d$) | Si I |
| $\mathbf{B}_6$ | = | $\frac{1}{2}\,\mathbf{a}_1 + \frac{3}{4}\,\mathbf{a}_3$ | = | $\frac{1}{2}\,a\,\mathbf{\hat{x}} + \frac{3}{4}\,a\,\mathbf{\hat{z}}$ | (6$d$) | Si I |
| $\mathbf{B}_7$ | = | $x_2\,\mathbf{a}_1 + x_2\,\mathbf{a}_2 + x_2\,\mathbf{a}_3$ | = | $x_2\,a\,\mathbf{\hat{x}} + x_2\,a\,\mathbf{\hat{y}} + x_2\,a\,\mathbf{\hat{z}}$ | (16$i$) | Si II |
| $\mathbf{B}_8$ | = | $-x_2\,\mathbf{a}_1 - x_2\,\mathbf{a}_2 + x_2\,\mathbf{a}_3$ | = | $-x_2\,a\,\mathbf{\hat{x}} - x_2\,a\,\mathbf{\hat{y}} + x_2\,a\,\mathbf{\hat{z}}$ | (16$i$) | Si II |
| $\mathbf{B}_9$ | = | $-x_2\,\mathbf{a}_1 + x_2\,\mathbf{a}_2 - x_2\,\mathbf{a}_3$ | = | $-x_2\,a\,\mathbf{\hat{x}} + x_2\,a\,\mathbf{\hat{y}} - x_2\,a\,\mathbf{\hat{z}}$ | (16$i$) | Si II |
| $\mathbf{B}_{10}$ | = | $x_2\,\mathbf{a}_1 - x_2\,\mathbf{a}_2 - x_2\,\mathbf{a}_3$ | = | $x_2\,a\,\mathbf{\hat{x}} - x_2\,a\,\mathbf{\hat{y}} - x_2\,a\,\mathbf{\hat{z}}$ | (16$i$) | Si II |
| $\mathbf{B}_{11}$ | = | $\left(\frac{1}{2} + x_2\right)\,\mathbf{a}_1 + \left(\frac{1}{2} + x_2\right)\,\mathbf{a}_2 + \left(\frac{1}{2} - x_2\right)\,\mathbf{a}_3$ | = | $\left(\frac{1}{2} + x_2\right)\,a\,\mathbf{\hat{x}} + \left(\frac{1}{2} + x_2\right)\,a\,\mathbf{\hat{y}} + \left(\frac{1}{2} - x_2\right)\,a\,\mathbf{\hat{z}}$ | (16$i$) | Si II |
| $\mathbf{B}_{12}$ | = | $\left(\frac{1}{2} - x_2\right)\,\mathbf{a}_1 + \left(\frac{1}{2} - x_2\right)\,\mathbf{a}_2 + \left(\frac{1}{2} - x_2\right)\,\mathbf{a}_3$ | = | $\left(\frac{1}{2} - x_2\right)\,a\,\mathbf{\hat{x}} + \left(\frac{1}{2} - x_2\right)\,a\,\mathbf{\hat{y}} + \left(\frac{1}{2} - x_2\right)\,a\,\mathbf{\hat{z}}$ | (16$i$) | Si II |
| $\mathbf{B}_{13}$ | = | $\left(\frac{1}{2} + x_2\right)\,\mathbf{a}_1 + \left(\frac{1}{2} - x_2\right)\,\mathbf{a}_2 + \left(\frac{1}{2} + x_2\right)\,\mathbf{a}_3$ | = | $\left(\frac{1}{2} + x_2\right)\,a\,\mathbf{\hat{x}} + \left(\frac{1}{2} - x_2\right)\,a\,\mathbf{\hat{y}} + \left(\frac{1}{2} + x_2\right)\,a\,\mathbf{\hat{z}}$ | (16$i$) | Si II |
| $\mathbf{B}_{14}$ | = | $\left(\frac{1}{2} - x_2\right)\,\mathbf{a}_1 + \left(\frac{1}{2} + x_2\right)\,\mathbf{a}_2 + \left(\frac{1}{2} + x_2\right)\,\mathbf{a}_3$ | = | $\left(\frac{1}{2} - x_2\right)\,a\,\mathbf{\hat{x}} + \left(\frac{1}{2} + x_2\right)\,a\,\mathbf{\hat{y}} + \left(\frac{1}{2} + x_2\right)\,a\,\mathbf{\hat{z}}$ | (16$i$) | Si II |
| $\mathbf{B}_{15}$ | = | $-x_2\,\mathbf{a}_1 - x_2\,\mathbf{a}_2 - x_2\,\mathbf{a}_3$ | = | $-x_2\,a\,\mathbf{\hat{x}} - x_2\,a\,\mathbf{\hat{y}} - x_2\,a\,\mathbf{\hat{z}}$ | (16$i$) | Si II |
| $\mathbf{B}_{16}$ | = | $x_2\,\mathbf{a}_1 + x_2\,\mathbf{a}_2 - x_2\,\mathbf{a}_3$ | = | $x_2\,a\,\mathbf{\hat{x}} + x_2\,a\,\mathbf{\hat{y}} - x_2\,a\,\mathbf{\hat{z}}$ | (16$i$) | Si II |
| $\mathbf{B}_{17}$ | = | $x_2\,\mathbf{a}_1 - x_2\,\mathbf{a}_2 + x_2\,\mathbf{a}_3$ | = | $x_2\,a\,\mathbf{\hat{x}} - x_2\,a\,\mathbf{\hat{y}} + x_2\,a\,\mathbf{\hat{z}}$ | (16$i$) | Si II |
| $\mathbf{B}_{18}$ | = | $-x_2\,\mathbf{a}_1 + x_2\,\mathbf{a}_2 + x_2\,\mathbf{a}_3$ | = | $-x_2\,a\,\mathbf{\hat{x}} + x_2\,a\,\mathbf{\hat{y}} + x_2\,a\,\mathbf{\hat{z}}$ | (16$i$) | Si II |
| $\mathbf{B}_{19}$ | = | $\left(\frac{1}{2} - x_2\right)\,\mathbf{a}_1 + \left(\frac{1}{2} - x_2\right)\,\mathbf{a}_2 + \left(\frac{1}{2} + x_2\right)\,\mathbf{a}_3$ | = | $\left(\frac{1}{2} - x_2\right)\,a\,\mathbf{\hat{x}} + \left(\frac{1}{2} - x_2\right)\,a\,\mathbf{\hat{y}} + \left(\frac{1}{2} + x_2\right)\,a\,\mathbf{\hat{z}}$ | (16$i$) | Si II |
| $\mathbf{B}_{20}$ | = | $\left(\frac{1}{2} + x_2\right)\,\mathbf{a}_1 + \left(\frac{1}{2} + x_2\right)\,\mathbf{a}_2 + \left(\frac{1}{2} + x_2\right)\,\mathbf{a}_3$ | = | $\left(\frac{1}{2} + x_2\right)\,a\,\mathbf{\hat{x}} + \left(\frac{1}{2} + x_2\right)\,a\,\mathbf{\hat{y}} + \left(\frac{1}{2} + x_2\right)\,a\,\mathbf{\hat{z}}$ | (16$i$) | Si II |
| $\mathbf{B}_{21}$ | = | $\left(\frac{1}{2} - x_2\right)\,\mathbf{a}_1 + \left(\frac{1}{2} + x_2\right)\,\mathbf{a}_2 + \left(\frac{1}{2} - x_2\right)\,\mathbf{a}_3$ | = | $\left(\frac{1}{2} - x_2\right)\,a\,\mathbf{\hat{x}} + \left(\frac{1}{2} + x_2\right)\,a\,\mathbf{\hat{y}} + \left(\frac{1}{2} - x_2\right)\,a\,\mathbf{\hat{z}}$ | (16$i$) | Si II |
| $\mathbf{B}_{22}$ | = | $\left(\frac{1}{2}+x_2\right)\mathbf{a}_1+\left(\frac{1}{2}-x_2\right)\mathbf{a}_2+\left(\frac{1}{2}-x_2\right)\mathbf{a}_3$ | = | $\left(\frac{1}{2}+x_2\right)a\,\hat{\mathbf{x}}+\left(\frac{1}{2}-x_2\right)a\,\hat{\mathbf{y}}+\left(\frac{1}{2}-x_2\right)a\,\hat{\mathbf{z}}$ | (16$i$) | Si II |
| $\mathbf{B}_{23}$ | = | $y_3\,\mathbf{a}_2+z_3\,\mathbf{a}_3$ | = | $y_3\,a\,\hat{\mathbf{y}}+z_3\,a\,\hat{\mathbf{z}}$ | (24$k$) | Si III |
| $\mathbf{B}_{24}$ | = | $-y_3\,\mathbf{a}_2+z_3\,\mathbf{a}_3$ | = | $-y_3\,a\,\hat{\mathbf{y}}+z_3\,a\,\hat{\mathbf{z}}$ | (24$k$) | Si III |
| $\mathbf{B}_{25}$ | = | $y_3\,\mathbf{a}_2-z_3\,\mathbf{a}_3$ | = | $y_3\,a\,\hat{\mathbf{y}}-z_3\,a\,\hat{\mathbf{z}}$ | (24$k$) | Si III |
| $\mathbf{B}_{26}$ | = | $-y_3\,\mathbf{a}_2-z_3\,\mathbf{a}_3$ | = | $-y_3\,a\,\hat{\mathbf{y}}-z_3\,a\,\hat{\mathbf{z}}$ | (24$k$) | Si III |
| $\mathbf{B}_{27}$ | = | $z_3\,\mathbf{a}_1+y_3\,\mathbf{a}_3$ | = | $z_3\,a\,\hat{\mathbf{x}}+y_3\,a\,\hat{\mathbf{z}}$ | (24$k$) | Si III |
| $\mathbf{B}_{28}$ | = | $-z_3\,\mathbf{a}_1+y_3\,\mathbf{a}_3$ | = | $-z_3\,a\,\hat{\mathbf{x}}+y_3\,a\,\hat{\mathbf{z}}$ | (24$k$) | Si III |
| $\mathbf{B}_{29}$ | = | $z_3\,\mathbf{a}_1-y_3\,\mathbf{a}_3$ | = | $z_3\,a\,\hat{\mathbf{x}}-y_3\,a\,\hat{\mathbf{z}}$ | (24$k$) | Si III |
| $\mathbf{B}_{30}$ | = | $-z_3\,\mathbf{a}_1-y_3\,\mathbf{a}_3$ | = | $-z_3\,a\,\hat{\mathbf{x}}-y_3\,a\,\hat{\mathbf{z}}$ | (24$k$) | Si III |
| $\mathbf{B}_{31}$ | = | $y_3\,\mathbf{a}_1+z_3\,\mathbf{a}_2$ | = | $y_3\,a\,\hat{\mathbf{x}}+z_3\,a\,\hat{\mathbf{y}}$ | (24$k$) | Si III |
| $\mathbf{B}_{32}$ | = | $-y_3\,\mathbf{a}_1+z_3\,\mathbf{a}_2$ | = | $-y_3\,a\,\hat{\mathbf{x}}+z_3\,a\,\hat{\mathbf{y}}$ | (24$k$) | Si III |
| $\mathbf{B}_{33}$ | = | $y_3\,\mathbf{a}_1-z_3\,\mathbf{a}_2$ | = | $y_3\,a\,\hat{\mathbf{x}}-z_3\,a\,\hat{\mathbf{y}}$ | (24$k$) | Si III |
| $\mathbf{B}_{34}$ | = | $-y_3\,\mathbf{a}_1-z_3\,\mathbf{a}_2$ | = | $-y_3\,a\,\hat{\mathbf{x}}-z_3\,a\,\hat{\mathbf{y}}$ | (24$k$) | Si III |
| $\mathbf{B}_{35}$ | = | $\left(\frac{1}{2}+y_3\right)\mathbf{a}_1+\frac{1}{2}\,\mathbf{a}_2+\left(\frac{1}{2}-z_3\right)\mathbf{a}_3$ | = | $\left(\frac{1}{2}+y_3\right)a\,\hat{\mathbf{x}}+\frac{1}{2}\,a\,\hat{\mathbf{y}}+\left(\frac{1}{2}-z_3\right)a\,\hat{\mathbf{z}}$ | (24$k$) | Si III |
| $\mathbf{B}_{36}$ | = | $\left(\frac{1}{2}-y_3\right)\mathbf{a}_1+\frac{1}{2}\,\mathbf{a}_2+\left(\frac{1}{2}-z_3\right)\mathbf{a}_3$ | = | $\left(\frac{1}{2}-y_3\right)a\,\hat{\mathbf{x}}+\frac{1}{2}\,a\,\hat{\mathbf{y}}+\left(\frac{1}{2}-z_3\right)a\,\hat{\mathbf{z}}$ | (24$k$) | Si III |
| $\mathbf{B}_{37}$ | = | $\left(\frac{1}{2}+y_3\right)\mathbf{a}_1+\frac{1}{2}\,\mathbf{a}_2+\left(\frac{1}{2}+z_3\right)\mathbf{a}_3$ | = | $\left(\frac{1}{2}+y_3\right)a\,\hat{\mathbf{x}}+\frac{1}{2}\,a\,\hat{\mathbf{y}}+\left(\frac{1}{2}+z_3\right)a\,\hat{\mathbf{z}}$ | (24$k$) | Si III |
| $\mathbf{B}_{38}$ | = | $\left(\frac{1}{2}-y_3\right)\mathbf{a}_1+\frac{1}{2}\,\mathbf{a}_2+\left(\frac{1}{2}+z_3\right)\mathbf{a}_3$ | = | $\left(\frac{1}{2}-y_3\right)a\,\hat{\mathbf{x}}+\frac{1}{2}\,a\,\hat{\mathbf{y}}+\left(\frac{1}{2}+z_3\right)a\,\hat{\mathbf{z}}$ | (24$k$) | Si III |
| $\mathbf{B}_{39}$ | = | $\frac{1}{2}\,\mathbf{a}_1+\left(\frac{1}{2}+z_3\right)\mathbf{a}_2+\left(\frac{1}{2}-y_3\right)\mathbf{a}_3$ | = | $\frac{1}{2}\,a\,\hat{\mathbf{x}}+\left(\frac{1}{2}+z_3\right)a\,\hat{\mathbf{y}}+\left(\frac{1}{2}-y_3\right)a\,\hat{\mathbf{z}}$ | (24$k$) | Si III |
| $\mathbf{B}_{40}$ | = | $\frac{1}{2}\,\mathbf{a}_1+\left(\frac{1}{2}+z_3\right)\mathbf{a}_2+\left(\frac{1}{2}+y_3\right)\mathbf{a}_3$ | = | $\frac{1}{2}\,a\,\hat{\mathbf{x}}+\left(\frac{1}{2}+z_3\right)a\,\hat{\mathbf{y}}+\left(\frac{1}{2}+y_3\right)a\,\hat{\mathbf{z}}$ | (24$k$) | Si III |
| $\mathbf{B}_{41}$ | = | $\frac{1}{2}\,\mathbf{a}_1+\left(\frac{1}{2}-z_3\right)\mathbf{a}_2+\left(\frac{1}{2}-y_3\right)\mathbf{a}_3$ | = | $\frac{1}{2}\,a\,\hat{\mathbf{x}}+\left(\frac{1}{2}-z_3\right)a\,\hat{\mathbf{y}}+\left(\frac{1}{2}-y_3\right)a\,\hat{\mathbf{z}}$ | (24$k$) | Si III |
| $\mathbf{B}_{42}$ | = | $\frac{1}{2}\,\mathbf{a}_1+\left(\frac{1}{2}-z_3\right)\mathbf{a}_2+\left(\frac{1}{2}+y_3\right)\mathbf{a}_3$ | = | $\frac{1}{2}\,a\,\hat{\mathbf{x}}+\left(\frac{1}{2}-z_3\right)a\,\hat{\mathbf{y}}+\left(\frac{1}{2}+y_3\right)a\,\hat{\mathbf{z}}$ | (24$k$) | Si III |
| $\mathbf{B}_{43}$ | = | $\left(\frac{1}{2}+z_3\right)\mathbf{a}_1+\left(\frac{1}{2}-y_3\right)\mathbf{a}_2+\frac{1}{2}\,\mathbf{a}_3$ | = | $\left(\frac{1}{2}+z_3\right)a\,\hat{\mathbf{x}}+\left(\frac{1}{2}-y_3\right)a\,\hat{\mathbf{y}}+\frac{1}{2}\,a\,\hat{\mathbf{z}}$ | (24$k$) | Si III |
| $\mathbf{B}_{44}$ | = | $\left(\frac{1}{2}+z_3\right)\mathbf{a}_1+\left(\frac{1}{2}+y_3\right)\mathbf{a}_2+\frac{1}{2}\,\mathbf{a}_3$ | = | $\left(\frac{1}{2}+z_3\right)a\,\hat{\mathbf{x}}+\left(\frac{1}{2}+y_3\right)a\,\hat{\mathbf{y}}+\frac{1}{2}\,a\,\hat{\mathbf{z}}$ | (24$k$) | Si III |
| $\mathbf{B}_{45}$ | = | $\left(\frac{1}{2}-z_3\right)\mathbf{a}_1+\left(\frac{1}{2}-y_3\right)\mathbf{a}_2+\frac{1}{2}\,\mathbf{a}_3$ | = | $\left(\frac{1}{2}-z_3\right)a\,\hat{\mathbf{x}}+\left(\frac{1}{2}-y_3\right)a\,\hat{\mathbf{y}}+\frac{1}{2}\,a\,\hat{\mathbf{z}}$ | (24$k$) | Si III |
| $\mathbf{B}_{46}$ | = | $\left(\frac{1}{2}-z_3\right)\mathbf{a}_1+\left(\frac{1}{2}+y_3\right)\mathbf{a}_2+\frac{1}{2}\,\mathbf{a}_3$ | = | $\left(\frac{1}{2}-z_3\right)a\,\hat{\mathbf{x}}+\left(\frac{1}{2}+y_3\right)a\,\hat{\mathbf{y}}+\frac{1}{2}\,a\,\hat{\mathbf{z}}$ | (24$k$) | Si III |

**Geometry files:**

- CIF: pp. 778
- POSCAR: pp. 778

# Cuprite ($Cu_2O$, C3) Structure: A2B_cP6_224_b_a

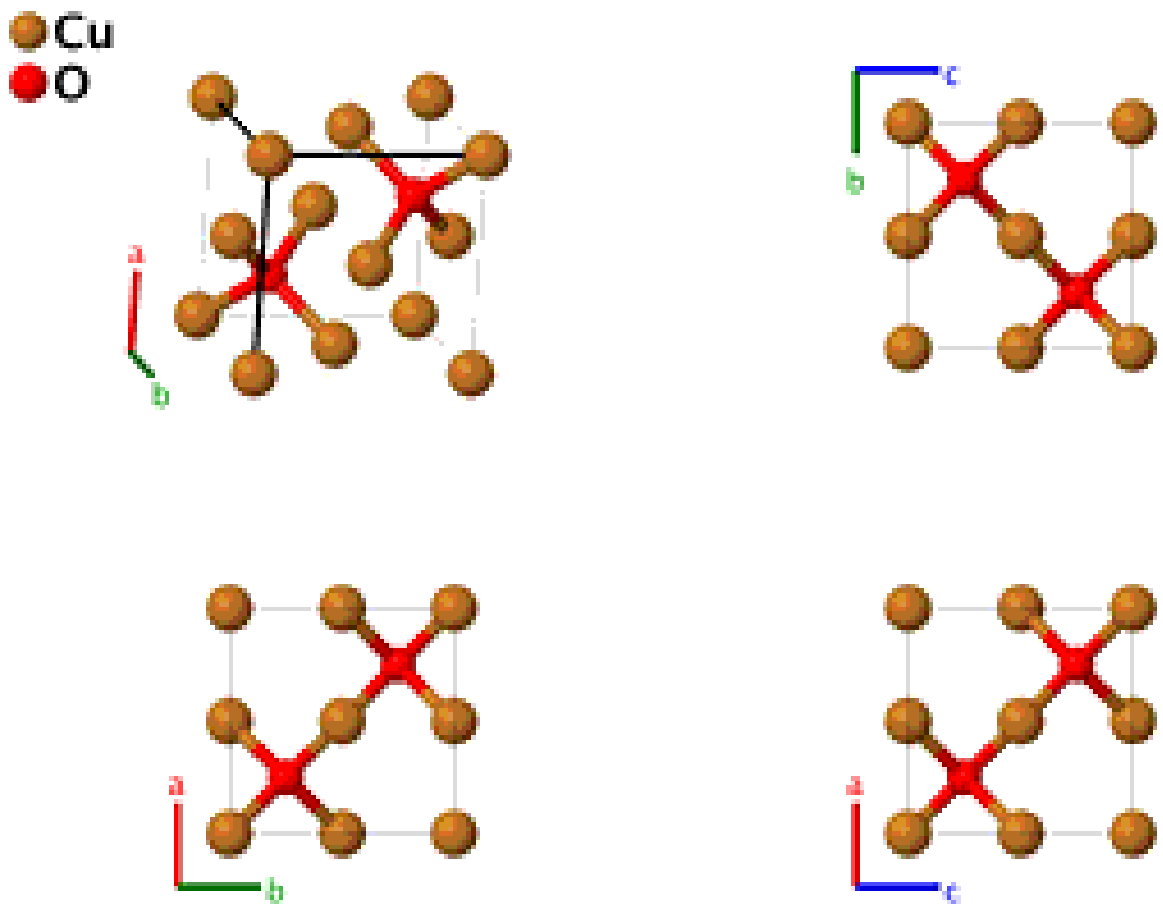

| | | |
|---|---|---|
| **Prototype** | : | $Cu_2O$ |
| **AFLOW prototype label** | : | A2B_cP6_224_b_a |
| ***Strukturbericht* designation** | : | C3 |
| **Pearson symbol** | : | cP6 |
| **Space group number** | : | 224 |
| **Space group symbol** | : | $Pn\bar{3}m$ |
| **AFLOW prototype command** | : | `aflow --proto=A2B_cP6_224_b_a --params=`$a$ |

- (Restori, 1986) gives the equilibrium lattice constant of $Cu_2O$ as $a = 4.627$Å, but gives nearest-neighbor distances which yield a lattice constant of 4.267Å. Since this value agrees with other sources, including those in (Downs, 2003), we use it.

**Simple Cubic primitive vectors:**

$$\mathbf{a}_1 = a\,\hat{\mathbf{x}}$$
$$\mathbf{a}_2 = a\,\hat{\mathbf{y}}$$
$$\mathbf{a}_3 = a\,\hat{\mathbf{z}}$$

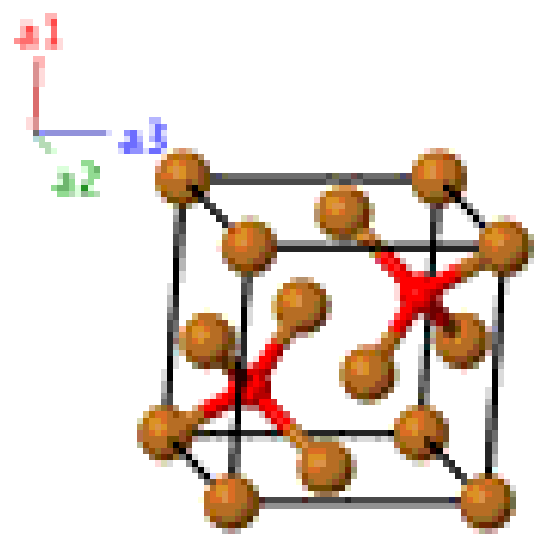

**Basis vectors:**

| | | Lattice Coordinates | | Cartesian Coordinates | Wyckoff Position | Atom Type |
|---|---|---|---|---|---|---|
| $\mathbf{B}_1$ | = | $\frac{1}{4}\,\mathbf{a}_1 + \frac{1}{4}\,\mathbf{a}_2 + \frac{1}{4}\,\mathbf{a}_3$ | = | $\frac{1}{4}\,a\,\hat{\mathbf{x}} + \frac{1}{4}\,a\,\hat{\mathbf{y}} + \frac{1}{4}\,a\,\hat{\mathbf{z}}$ | (2$a$) | O |
| $\mathbf{B}_2$ | = | $\frac{3}{4}\,\mathbf{a}_1 + \frac{3}{4}\,\mathbf{a}_2 + \frac{3}{4}\,\mathbf{a}_3$ | = | $\frac{3}{4}\,a\,\hat{\mathbf{x}} + \frac{3}{4}\,a\,\hat{\mathbf{y}} + \frac{3}{4}\,a\,\hat{\mathbf{z}}$ | (2$a$) | O |
| $\mathbf{B}_3$ | = | $0\,\mathbf{a}_1 + 0\,\mathbf{a}_2 + 0\,\mathbf{a}_3$ | = | $0\hat{\mathbf{x}} + 0\hat{\mathbf{y}} + 0\hat{\mathbf{z}}$ | (4$b$) | Cu |
| $\mathbf{B}_4$ | = | $\frac{1}{2}\,\mathbf{a}_1 + \frac{1}{2}\,\mathbf{a}_2$ | = | $\frac{1}{2}\,a\,\hat{\mathbf{x}} + \frac{1}{2}\,a\,\hat{\mathbf{y}}$ | (4$b$) | Cu |
| $\mathbf{B}_5$ | = | $\frac{1}{2}\,\mathbf{a}_1 + \frac{1}{2}\,\mathbf{a}_3$ | = | $\frac{1}{2}\,a\,\hat{\mathbf{x}} + \frac{1}{2}\,a\,\hat{\mathbf{z}}$ | (4$b$) | Cu |
| $\mathbf{B}_6$ | = | $\frac{1}{2}\,\mathbf{a}_2 + \frac{1}{2}\,\mathbf{a}_3$ | = | $\frac{1}{2}\,a\,\hat{\mathbf{y}} + \frac{1}{2}\,a\,\hat{\mathbf{z}}$ | (4$b$) | Cu |

**Geometry files:**
- CIF: pp. 778
- POSCAR: pp. 779

# $Ca_7Ge$ Structure: A7B_cF32_225_bd_a

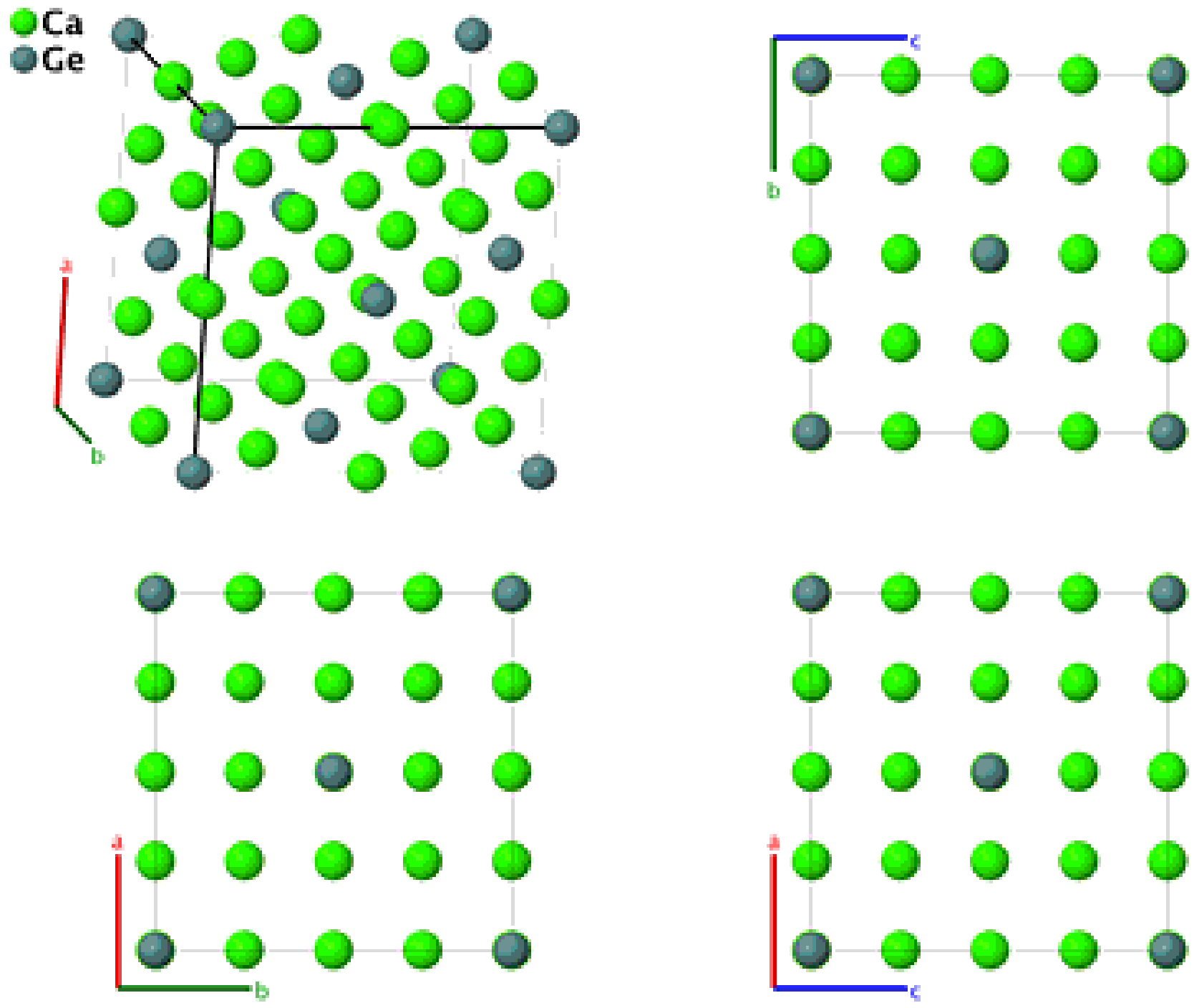

| | | |
|---|---|---|
| **Prototype** | : | $Ca_7Ge$ |
| **AFLOW prototype label** | : | A7B_cF32_225_bd_a |
| ***Strukturbericht* designation** | : | None |
| **Pearson symbol** | : | cF32 |
| **Space group number** | : | 225 |
| **Space group symbol** | : | $Fm\bar{3}m$ |
| **AFLOW prototype command** | : | `aflow --proto=A7B_cF32_225_bd_a --params=`$a$ |

**Other compounds with this structure:**

- $LiPt_7$, $MoZn_7$

**Face-centered Cubic primitive vectors:**

$$
\begin{aligned}
\mathbf{a}_1 &= \tfrac{1}{2}\,a\,\hat{\mathbf{y}} + \tfrac{1}{2}\,a\,\hat{\mathbf{z}} \\
\mathbf{a}_2 &= \tfrac{1}{2}\,a\,\hat{\mathbf{x}} + \tfrac{1}{2}\,a\,\hat{\mathbf{z}} \\
\mathbf{a}_3 &= \tfrac{1}{2}\,a\,\hat{\mathbf{x}} + \tfrac{1}{2}\,a\,\hat{\mathbf{y}}
\end{aligned}
$$

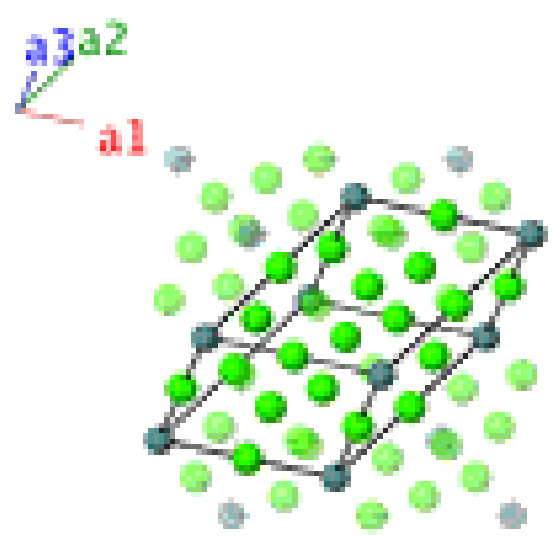

**Basis vectors:**

| Lattice Coordinates | Cartesian Coordinates | Wyckoff Position | Atom Type |
|---|---|---|---|

| | | | | | | |
|---|---|---|---|---|---|---|
| $\mathbf{B}_1$ | = | $0\,\mathbf{a}_1 + 0\,\mathbf{a}_2 + 0\,\mathbf{a}_3$ | = | $0\,\hat{\mathbf{x}} + 0\,\hat{\mathbf{y}} + 0\,\hat{\mathbf{z}}$ | (4$a$) | Ge |
| $\mathbf{B}_2$ | = | $\frac{1}{2}\,\mathbf{a}_1 + \frac{1}{2}\,\mathbf{a}_2 + \frac{1}{2}\,\mathbf{a}_3$ | = | $\frac{1}{2}\,a\,\hat{\mathbf{x}} + \frac{1}{2}\,a\,\hat{\mathbf{y}} + \frac{1}{2}\,a\,\hat{\mathbf{z}}$ | (4$b$) | Ca I |
| $\mathbf{B}_3$ | = | $\frac{1}{2}\,\mathbf{a}_1$ | = | $\frac{1}{4}\,a\,\hat{\mathbf{y}} + \frac{1}{4}\,a\,\hat{\mathbf{z}}$ | (24$d$) | Ca II |
| $\mathbf{B}_4$ | = | $\frac{1}{2}\,\mathbf{a}_2 + \frac{1}{2}\,\mathbf{a}_3$ | = | $\frac{1}{2}\,a\,\hat{\mathbf{x}} + \frac{1}{4}\,a\,\hat{\mathbf{y}} + \frac{1}{4}\,a\,\hat{\mathbf{z}}$ | (24$d$) | Ca II |
| $\mathbf{B}_5$ | = | $\frac{1}{2}\,\mathbf{a}_2$ | = | $\frac{1}{4}\,a\,\hat{\mathbf{x}} + \frac{1}{4}\,a\,\hat{\mathbf{z}}$ | (24$d$) | Ca II |
| $\mathbf{B}_6$ | = | $\frac{1}{2}\,\mathbf{a}_1 + \frac{1}{2}\,\mathbf{a}_3$ | = | $\frac{1}{4}\,a\,\hat{\mathbf{x}} + \frac{1}{2}\,a\,\hat{\mathbf{y}} + \frac{1}{4}\,a\,\hat{\mathbf{z}}$ | (24$d$) | Ca II |
| $\mathbf{B}_7$ | = | $\frac{1}{2}\,\mathbf{a}_3$ | = | $\frac{1}{4}\,a\,\hat{\mathbf{x}} + \frac{1}{4}\,a\,\hat{\mathbf{y}}$ | (24$d$) | Ca II |
| $\mathbf{B}_8$ | = | $\frac{1}{2}\,\mathbf{a}_1 + \frac{1}{2}\,\mathbf{a}_2$ | = | $\frac{1}{4}\,a\,\hat{\mathbf{x}} + \frac{1}{4}\,a\,\hat{\mathbf{y}} + \frac{1}{2}\,a\,\hat{\mathbf{z}}$ | (24$d$) | Ca II |

**Found in:**

**Geometry files:**

- CIF: pp. 779
- POSCAR: pp. 780

# $BiF_3$ ($D0_3$) Structure: AB3_cF16_225_a_bc

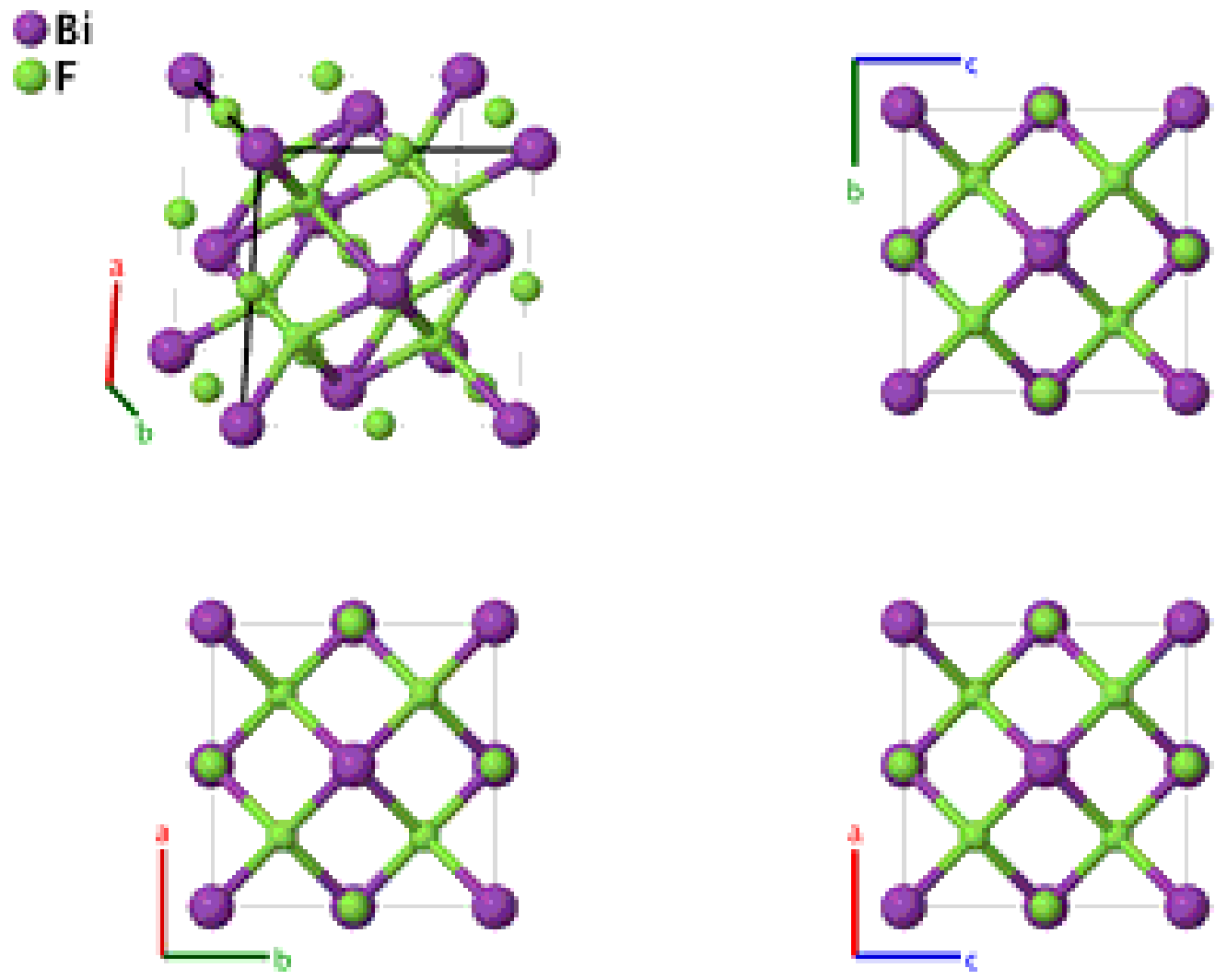

| | | |
|---|---|---|
| **Prototype** | : | $BiF_3$ |
| **AFLOW prototype label** | : | AB3_cF16_225_a_bc |
| ***Strukturbericht* designation** | : | $D0_3$ |
| **Pearson symbol** | : | cF16 |
| **Space group number** | : | 225 |
| **Space group symbol** | : | $Fm\bar{3}m$ |
| **AFLOW prototype command** | : | `aflow --proto=AB3_cF16_225_a_bc --params=`$a$ |

**Other compounds with this structure:**

- $AlFe_3$, $BiFe_3$

- (Villars, 1991) corrects the original source, changing the positions of one third of the fluorine atoms so that the space group becomes $Fm\bar{3}m$, as is accepted for $D0_3$. This structure is crystallographically equivalent to the Heusler ($L2_1$) structure.

**Face-centered Cubic primitive vectors:**

$$\mathbf{a}_1 = \frac{1}{2}\,a\,\hat{\mathbf{y}} + \frac{1}{2}\,a\,\hat{\mathbf{z}}$$
$$\mathbf{a}_2 = \frac{1}{2}\,a\,\hat{\mathbf{x}} + \frac{1}{2}\,a\,\hat{\mathbf{z}}$$
$$\mathbf{a}_3 = \frac{1}{2}\,a\,\hat{\mathbf{x}} + \frac{1}{2}\,a\,\hat{\mathbf{y}}$$

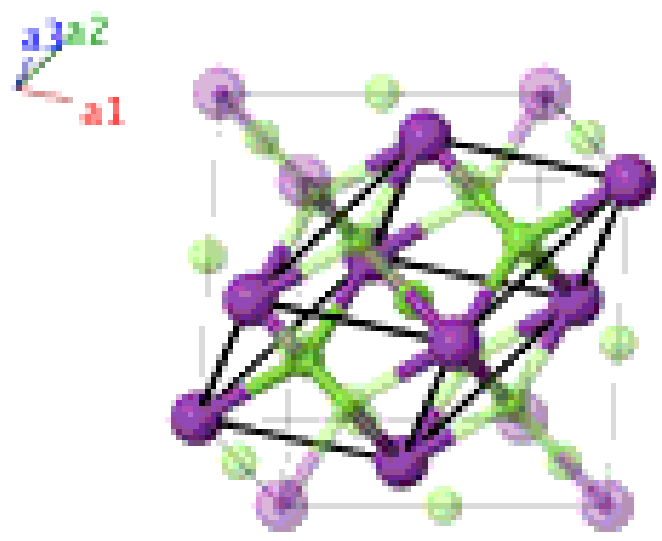

**Basis vectors:**

Lattice Coordinates | Cartesian Coordinates | Wyckoff Position | Atom Type

| | | | | | | |
|---|---|---|---|---|---|---|
| $\mathbf{B_1}$ | = | $0\,\mathbf{a_1} + 0\,\mathbf{a_2} + 0\,\mathbf{a_3}$ | = | $0\hat{\mathbf{x}} + 0\hat{\mathbf{y}} + 0\hat{\mathbf{z}}$ | $(4a)$ | Bi |
| $\mathbf{B_2}$ | = | $\frac{1}{2}\mathbf{a_1} + \frac{1}{2}\mathbf{a_2} + \frac{1}{2}\mathbf{a_3}$ | = | $\frac{1}{2}\,a\,\hat{\mathbf{x}} + \frac{1}{2}\,a\,\hat{\mathbf{y}} + \frac{1}{2}\,a\,\hat{\mathbf{z}}$ | $(4b)$ | F I |
| $\mathbf{B_3}$ | = | $\frac{1}{4}\mathbf{a_1} + \frac{1}{4}\mathbf{a_2} + \frac{1}{4}\mathbf{a_3}$ | = | $\frac{1}{4}\,a\,\hat{\mathbf{x}} + \frac{1}{4}\,a\,\hat{\mathbf{y}} + \frac{1}{4}\,a\,\hat{\mathbf{z}}$ | $(8c)$ | F II |
| $\mathbf{B_4}$ | = | $\frac{3}{4}\mathbf{a_1} + \frac{3}{4}\mathbf{a_2} + \frac{3}{4}\mathbf{a_3}$ | = | $\frac{3}{4}\,a\,\hat{\mathbf{x}} + \frac{3}{4}\,a\,\hat{\mathbf{y}} + \frac{3}{4}\,a\,\hat{\mathbf{z}}$ | $(8c)$ | F II |

---

**Found in:**

---

**Geometry files:**

- CIF: pp. 780
- POSCAR: pp. 782

# Model of Ferrite Structure (cF128): A9B16C7_cF128_225_acd_2f_be

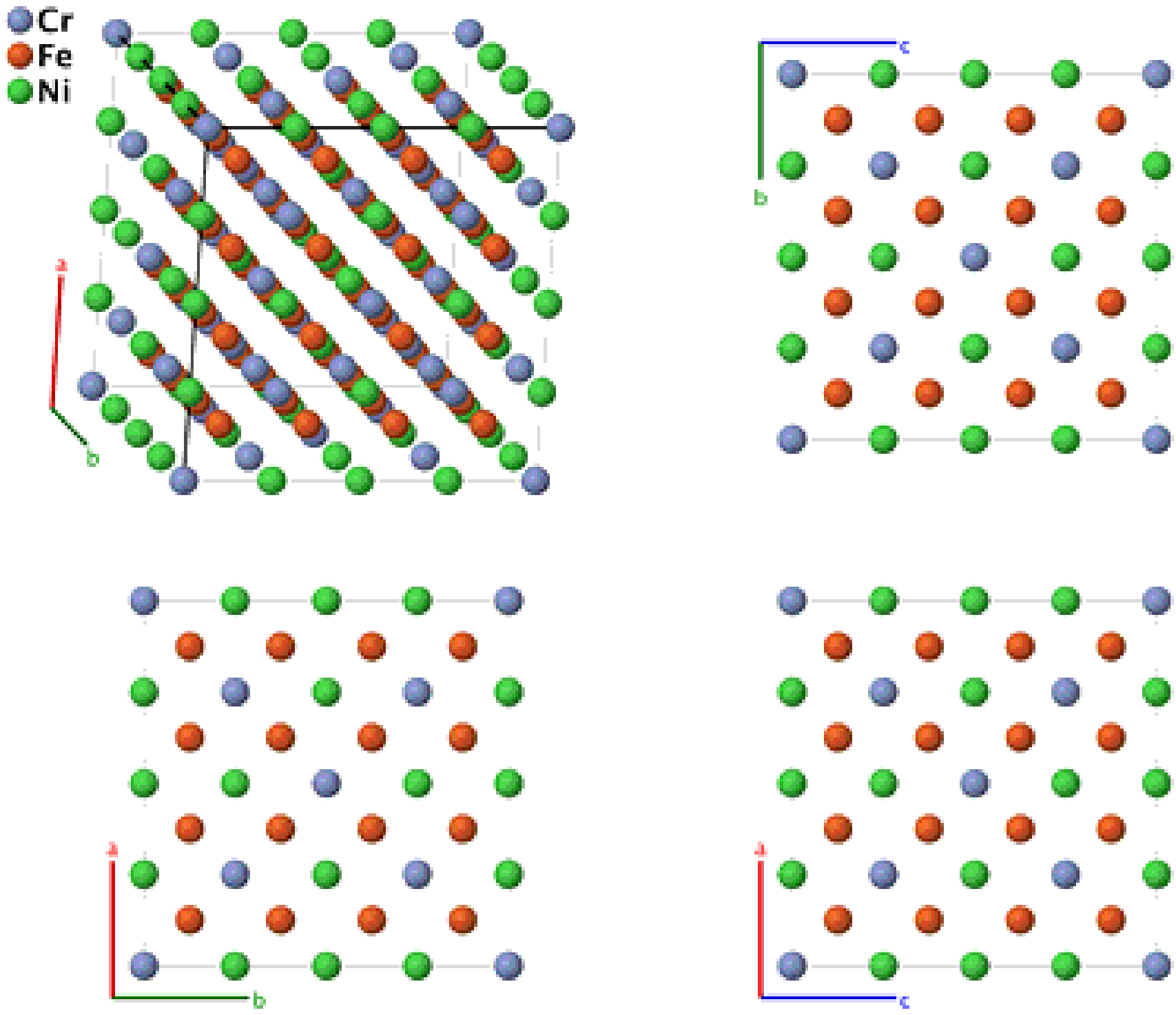

| | | |
|---|---|---|
| **Prototype** | : | $Cr_9Fe_{16}Ni_7$ |
| **AFLOW prototype label** | : | A9B16C7_cF128_225_acd_2f_be |
| ***Strukturbericht* designation** | : | None |
| **Pearson symbol** | : | cF128 |
| **Space group number** | : | 225 |
| **Space group symbol** | : | $Fm\bar{3}m$ |
| **AFLOW prototype command** | : | `aflow --proto=A9B16C7_cF128_225_acd_2f_be --params=`$a, x_5, x_6, x_7$ |

- Ferrite is steel with a bcc structure. This structure represents one possible ordering which might be found in an Fe-Ni-Cr steel. Note that it is not meant to represent a real steel. If we use the special values $x_5 = 1/4$, $x_6 = 1/8$, and $x_7 = 3/8$, and replace the Ni atoms by Cr, then this structure reverts to CsCl (B2) with $a_{B2} = 1/4a$. If we replace both the Ni and Cr atoms by Fe, then the structure becomes bcc, again with $a_{bcc} = 1/4a$.

**Face-centered Cubic primitive vectors:**

$$\mathbf{a}_1 = \frac{1}{2}\,a\,\mathbf{\hat{y}} + \frac{1}{2}\,a\,\mathbf{\hat{z}}$$
$$\mathbf{a}_2 = \frac{1}{2}\,a\,\mathbf{\hat{x}} + \frac{1}{2}\,a\,\mathbf{\hat{z}}$$
$$\mathbf{a}_3 = \frac{1}{2}\,a\,\mathbf{\hat{x}} + \frac{1}{2}\,a\,\mathbf{\hat{y}}$$

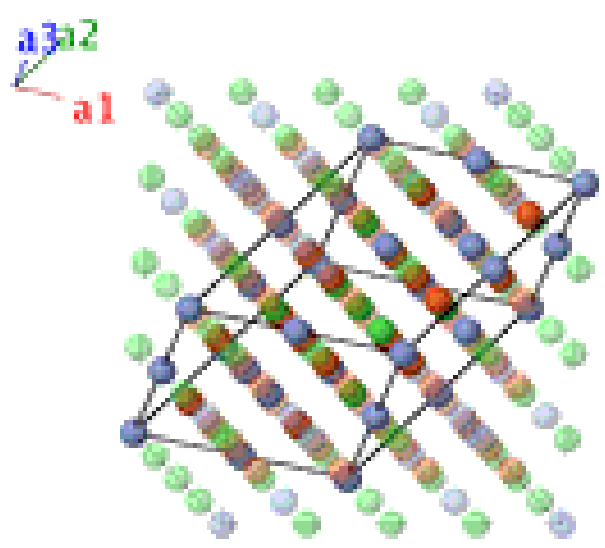

**Basis vectors:**

| | | Lattice Coordinates | | Cartesian Coordinates | Wyckoff Position | Atom Type |
|---|---|---|---|---|---|---|
| $\mathbf{B_1}$ | = | $0\,\mathbf{a_1} + 0\,\mathbf{a_2} + 0\,\mathbf{a_3}$ | = | $0\hat{\mathbf{x}} + 0\hat{\mathbf{y}} + 0\hat{\mathbf{z}}$ | $(4a)$ | Cr I |
| $\mathbf{B_2}$ | = | $\frac{1}{2}\,\mathbf{a_1} + \frac{1}{2}\,\mathbf{a_2} + \frac{1}{2}\,\mathbf{a_3}$ | = | $\frac{1}{2}\,a\,\hat{\mathbf{x}} + \frac{1}{2}\,a\,\hat{\mathbf{y}} + \frac{1}{2}\,a\,\hat{\mathbf{z}}$ | $(4b)$ | Ni I |
| $\mathbf{B_3}$ | = | $\frac{1}{4}\,\mathbf{a_1} + \frac{1}{4}\,\mathbf{a_2} + \frac{1}{4}\,\mathbf{a_3}$ | = | $\frac{1}{4}\,a\,\hat{\mathbf{x}} + \frac{1}{4}\,a\,\hat{\mathbf{y}} + \frac{1}{4}\,a\,\hat{\mathbf{z}}$ | $(8c)$ | Cr II |
| $\mathbf{B_4}$ | = | $\frac{3}{4}\,\mathbf{a_1} + \frac{3}{4}\,\mathbf{a_2} + \frac{3}{4}\,\mathbf{a_3}$ | = | $\frac{3}{4}\,a\,\hat{\mathbf{x}} + \frac{3}{4}\,a\,\hat{\mathbf{y}} + \frac{3}{4}\,a\,\hat{\mathbf{z}}$ | $(8c)$ | Cr II |
| $\mathbf{B_5}$ | = | $\frac{1}{2}\,\mathbf{a_1}$ | = | $\frac{1}{4}\,a\,\hat{\mathbf{y}} + \frac{1}{4}\,a\,\hat{\mathbf{z}}$ | $(24d)$ | Cr III |
| $\mathbf{B_6}$ | = | $\frac{1}{2}\,\mathbf{a_2} + \frac{1}{2}\,\mathbf{a_3}$ | = | $\frac{1}{2}\,a\,\hat{\mathbf{x}} + \frac{1}{4}\,a\,\hat{\mathbf{y}} + \frac{1}{4}\,a\,\hat{\mathbf{z}}$ | $(24d)$ | Cr III |
| $\mathbf{B_7}$ | = | $\frac{1}{2}\,\mathbf{a_2}$ | = | $\frac{1}{4}\,a\,\hat{\mathbf{x}} + \frac{1}{4}\,a\,\hat{\mathbf{z}}$ | $(24d)$ | Cr III |
| $\mathbf{B_8}$ | = | $\frac{1}{2}\,\mathbf{a_1} + \frac{1}{2}\,\mathbf{a_3}$ | = | $\frac{1}{4}\,a\,\hat{\mathbf{x}} + \frac{1}{2}\,a\,\hat{\mathbf{y}} + \frac{1}{4}\,a\,\hat{\mathbf{z}}$ | $(24d)$ | Cr III |
| $\mathbf{B_9}$ | = | $\frac{1}{2}\,\mathbf{a_3}$ | = | $\frac{1}{4}\,a\,\hat{\mathbf{x}} + \frac{1}{4}\,a\,\hat{\mathbf{y}}$ | $(24d)$ | Cr III |
| $\mathbf{B_{10}}$ | = | $\frac{1}{2}\,\mathbf{a_1} + \frac{1}{2}\,\mathbf{a_2}$ | = | $\frac{1}{4}\,a\,\hat{\mathbf{x}} + \frac{1}{4}\,a\,\hat{\mathbf{y}} + \frac{1}{2}\,a\,\hat{\mathbf{z}}$ | $(24d)$ | Cr III |
| $\mathbf{B_{11}}$ | = | $-x_5\,\mathbf{a_1} + x_5\,\mathbf{a_2} + x_5\,\mathbf{a_3}$ | = | $x_5\,a\,\hat{\mathbf{x}}$ | $(24e)$ | Ni II |
| $\mathbf{B_{12}}$ | = | $x_5\,\mathbf{a_1} - x_5\,\mathbf{a_2} + x_5\,\mathbf{a_3}$ | = | $x_5\,a\,\hat{\mathbf{y}}$ | $(24e)$ | Ni II |
| $\mathbf{B_{13}}$ | = | $x_5\,\mathbf{a_1} + x_5\,\mathbf{a_2} - x_5\,\mathbf{a_3}$ | = | $x_5\,a\,\hat{\mathbf{z}}$ | $(24e)$ | Ni II |
| $\mathbf{B_{14}}$ | = | $x_5\,\mathbf{a_1} - x_5\,\mathbf{a_2} - x_5\,\mathbf{a_3}$ | = | $-x_5\,a\,\hat{\mathbf{x}}$ | $(24e)$ | Ni II |
| $\mathbf{B_{15}}$ | = | $-x_5\,\mathbf{a_1} + x_5\,\mathbf{a_2} - x_5\,\mathbf{a_3}$ | = | $-x_5\,a\,\hat{\mathbf{y}}$ | $(24e)$ | Ni II |
| $\mathbf{B_{16}}$ | = | $-x_5\,\mathbf{a_1} - x_5\,\mathbf{a_2} + x_5\,\mathbf{a_3}$ | = | $-x_5\,a\,\hat{\mathbf{z}}$ | $(24e)$ | Ni II |
| $\mathbf{B_{17}}$ | = | $x_6\,\mathbf{a_1} + x_6\,\mathbf{a_2} + x_6\,\mathbf{a_3}$ | = | $x_6\,a\,\hat{\mathbf{x}} + x_6\,a\,\hat{\mathbf{y}} + x_6\,a\,\hat{\mathbf{z}}$ | $(32f)$ | Fe I |
| $\mathbf{B_{18}}$ | = | $x_6\,\mathbf{a_1} + x_6\,\mathbf{a_2} - 3\,x_6\,\mathbf{a_3}$ | = | $-x_6\,a\,\hat{\mathbf{x}} - x_6\,a\,\hat{\mathbf{y}} + x_6\,a\,\hat{\mathbf{z}}$ | $(32f)$ | Fe I |
| $\mathbf{B_{19}}$ | = | $x_6\,\mathbf{a_1} - 3\,x_6\,\mathbf{a_2} + x_6\,\mathbf{a_3}$ | = | $-x_6\,a\,\hat{\mathbf{x}} + x_6\,a\,\hat{\mathbf{y}} - x_6\,a\,\hat{\mathbf{z}}$ | $(32f)$ | Fe I |
| $\mathbf{B_{20}}$ | = | $-3\,x_6\,\mathbf{a_1} + x_6\,\mathbf{a_2} + x_6\,\mathbf{a_3}$ | = | $x_6\,a\,\hat{\mathbf{x}} - x_6\,a\,\hat{\mathbf{y}} - x_6\,a\,\hat{\mathbf{z}}$ | $(32f)$ | Fe I |
| $\mathbf{B_{21}}$ | = | $-x_6\,\mathbf{a_1} - x_6\,\mathbf{a_2} + 3\,x_6\,\mathbf{a_3}$ | = | $x_6\,a\,\hat{\mathbf{x}} + x_6\,a\,\hat{\mathbf{y}} - x_6\,a\,\hat{\mathbf{z}}$ | $(32f)$ | Fe I |
| $\mathbf{B_{22}}$ | = | $-x_6\,\mathbf{a_1} - x_6\,\mathbf{a_2} - x_6\,\mathbf{a_3}$ | = | $-x_6\,a\,\hat{\mathbf{x}} - x_6\,a\,\hat{\mathbf{y}} - x_6\,a\,\hat{\mathbf{z}}$ | $(32f)$ | Fe I |
| $\mathbf{B_{23}}$ | = | $-x_6\,\mathbf{a_1} + 3\,x_6\,\mathbf{a_2} - x_6\,\mathbf{a_3}$ | = | $x_6\,a\,\hat{\mathbf{x}} - x_6\,a\,\hat{\mathbf{y}} + x_6\,a\,\hat{\mathbf{z}}$ | $(32f)$ | Fe I |
| $\mathbf{B_{24}}$ | = | $3\,x_6\,\mathbf{a_1} - x_6\,\mathbf{a_2} - x_6\,\mathbf{a_3}$ | = | $-x_6\,a\,\hat{\mathbf{x}} + x_6\,a\,\hat{\mathbf{y}} + x_6\,a\,\hat{\mathbf{z}}$ | $(32f)$ | Fe I |
| $\mathbf{B_{25}}$ | = | $x_7\,\mathbf{a_1} + x_7\,\mathbf{a_2} + x_7\,\mathbf{a_3}$ | = | $x_7\,a\,\hat{\mathbf{x}} + x_7\,a\,\hat{\mathbf{y}} + x_7\,a\,\hat{\mathbf{z}}$ | $(32f)$ | Fe II |
| $\mathbf{B_{26}}$ | = | $x_7\,\mathbf{a_1} + x_7\,\mathbf{a_2} - 3\,x_7\,\mathbf{a_3}$ | = | $-x_7\,a\,\hat{\mathbf{x}} - x_7\,a\,\hat{\mathbf{y}} + x_7\,a\,\hat{\mathbf{z}}$ | $(32f)$ | Fe II |
| $\mathbf{B_{27}}$ | = | $x_7\,\mathbf{a_1} - 3\,x_7\,\mathbf{a_2} + x_7\,\mathbf{a_3}$ | = | $-x_7\,a\,\hat{\mathbf{x}} + x_7\,a\,\hat{\mathbf{y}} - x_7\,a\,\hat{\mathbf{z}}$ | $(32f)$ | Fe II |
| $\mathbf{B_{28}}$ | = | $-3\,x_7\,\mathbf{a_1} + x_7\,\mathbf{a_2} + x_7\,\mathbf{a_3}$ | = | $x_7\,a\,\hat{\mathbf{x}} - x_7\,a\,\hat{\mathbf{y}} - x_7\,a\,\hat{\mathbf{z}}$ | $(32f)$ | Fe II |
| $\mathbf{B_{29}}$ | = | $-x_7\,\mathbf{a_1} - x_7\,\mathbf{a_2} + 3\,x_7\,\mathbf{a_3}$ | = | $x_7\,a\,\hat{\mathbf{x}} + x_7\,a\,\hat{\mathbf{y}} - x_7\,a\,\hat{\mathbf{z}}$ | $(32f)$ | Fe II |
| $\mathbf{B_{30}}$ | = | $-x_7\,\mathbf{a_1} - x_7\,\mathbf{a_2} - x_7\,\mathbf{a_3}$ | = | $-x_7\,a\,\hat{\mathbf{x}} - x_7\,a\,\hat{\mathbf{y}} - x_7\,a\,\hat{\mathbf{z}}$ | $(32f)$ | Fe II |
| $\mathbf{B_{31}}$ | = | $-x_7\,\mathbf{a_1} + 3\,x_7\,\mathbf{a_2} - x_7\,\mathbf{a_3}$ | = | $x_7\,a\,\hat{\mathbf{x}} - x_7\,a\,\hat{\mathbf{y}} + x_7\,a\,\hat{\mathbf{z}}$ | $(32f)$ | Fe II |
| $\mathbf{B_{32}}$ | = | $3\,x_7\,\mathbf{a_1} - x_7\,\mathbf{a_2} - x_7\,\mathbf{a_3}$ | = | $-x_7\,a\,\hat{\mathbf{x}} + x_7\,a\,\hat{\mathbf{y}} + x_7\,a\,\hat{\mathbf{z}}$ | $(32f)$ | Fe II |

**References:**

- M. J. Mehl, Hypothetical cF128 Ferrite Structure

---

**Geometry files:**

- CIF: pp. 782
- POSCAR: pp. 783

# $UB_{12}$ Structure: A12B_cF52_225_i_a

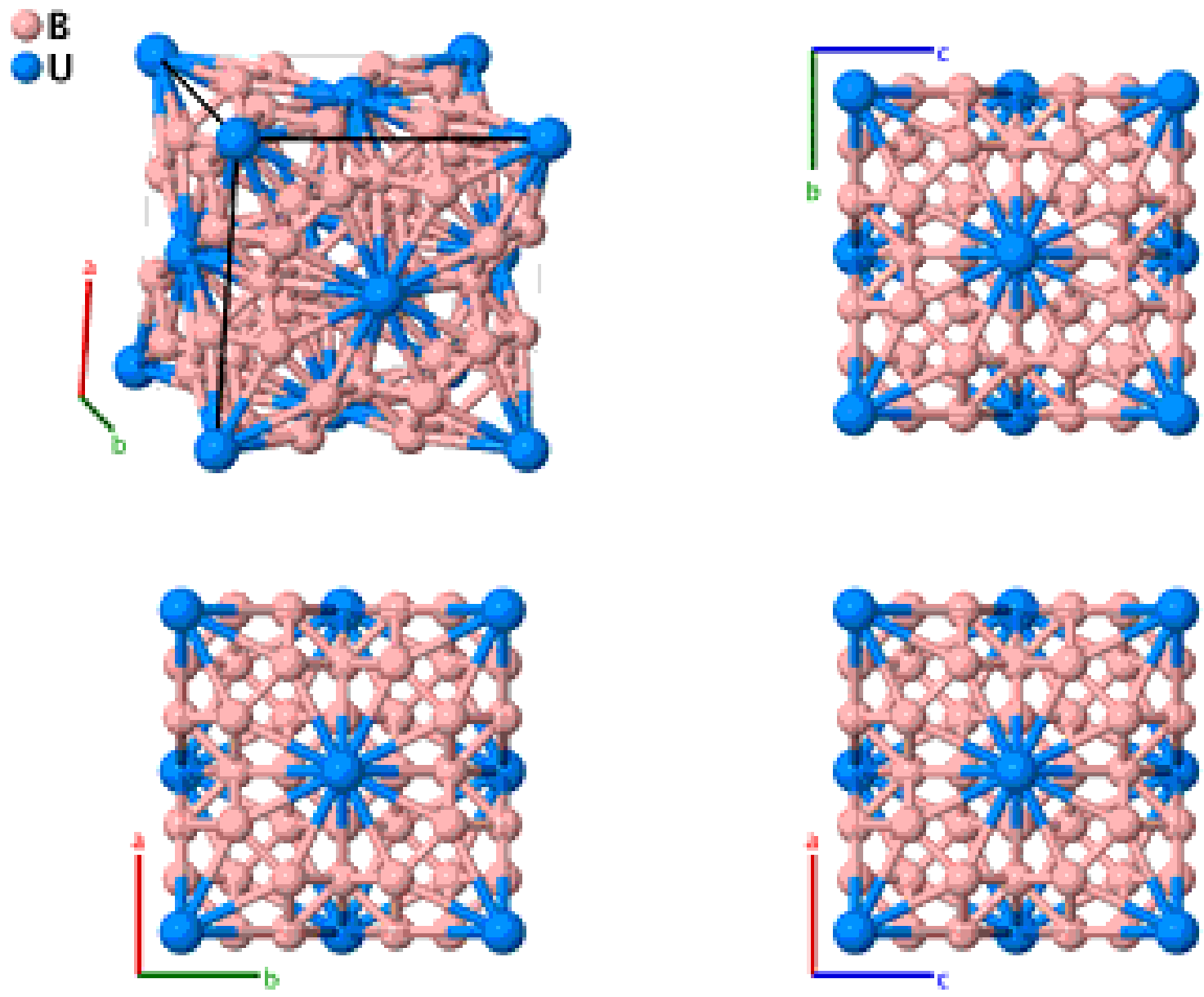

**Prototype** : $UB_{12}$

**AFLOW prototype label** : A12B_cF52_225_i_a

***Strukturbericht* designation** : $D2_f$

**Pearson symbol** : cF52

**Space group number** : 225

**Space group symbol** : $Fm\bar{3}m$

**AFLOW prototype command** : `aflow --proto=A12B_cF52_225_i_a --params=`$a, y_2$

---

**Face-centered Cubic primitive vectors:**

$$\mathbf{a}_1 = \frac{1}{2}\,a\,\mathbf{\hat{y}} + \frac{1}{2}\,a\,\mathbf{\hat{z}}$$
$$\mathbf{a}_2 = \frac{1}{2}\,a\,\mathbf{\hat{x}} + \frac{1}{2}\,a\,\mathbf{\hat{z}}$$
$$\mathbf{a}_3 = \frac{1}{2}\,a\,\mathbf{\hat{x}} + \frac{1}{2}\,a\,\mathbf{\hat{y}}$$

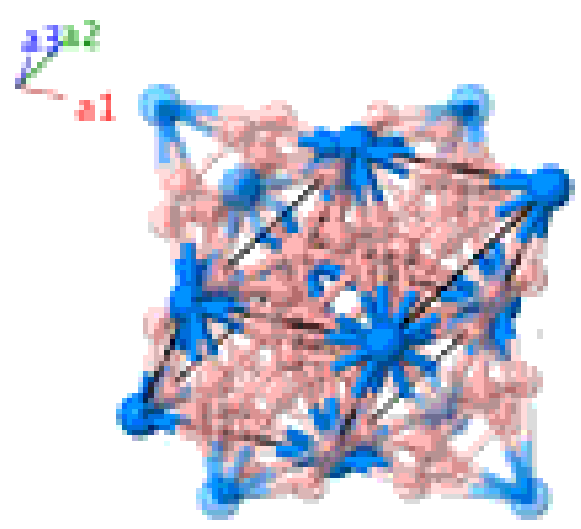

---

**Basis vectors:**

| | | Lattice Coordinates | | Cartesian Coordinates | Wyckoff Position | Atom Type |
|---|---|---|---|---|---|---|
| $\mathbf{B_1}$ | = | $0\,\mathbf{a_1} + 0\,\mathbf{a_2} + 0\,\mathbf{a_3}$ | = | $0\mathbf{\hat{x}} + 0\mathbf{\hat{y}} + 0\mathbf{\hat{z}}$ | (4$a$) | U |
| $\mathbf{B_2}$ | = | $\left(\frac{1}{2} + 2\,y_2\right)\mathbf{a_1} + \frac{1}{2}\,\mathbf{a_2} + \frac{1}{2}\,\mathbf{a_3}$ | = | $\frac{1}{2}\,a\,\mathbf{\hat{x}} + \left(\frac{1}{2} + y_2\right)a\,\mathbf{\hat{y}} + \left(\frac{1}{2} + y_2\right)a\,\mathbf{\hat{z}}$ | (48$i$) | B |
| $\mathbf{B_3}$ | = | $\frac{1}{2}\,\mathbf{a_1} + \left(\frac{1}{2} + 2\,y_2\right)\mathbf{a_2} + \left(\frac{1}{2} - 2\,y_2\right)\mathbf{a_3}$ | = | $\frac{1}{2}\,a\,\mathbf{\hat{x}} + \left(\frac{1}{2} - y_2\right)a\,\mathbf{\hat{y}} + \left(\frac{1}{2} + y_2\right)a\,\mathbf{\hat{z}}$ | (48$i$) | B |
| $\mathbf{B_4}$ | = | $\frac{1}{2}\,\mathbf{a_1} + \left(\frac{1}{2} - 2\,y_2\right)\mathbf{a_2} + \left(\frac{1}{2} + 2\,y_2\right)\mathbf{a_3}$ | = | $\frac{1}{2}\,a\,\mathbf{\hat{x}} + \left(\frac{1}{2} + y_2\right)a\,\mathbf{\hat{y}} + \left(\frac{1}{2} - y_2\right)a\,\mathbf{\hat{z}}$ | (48$i$) | B |
| $\mathbf{B_5}$ | = | $\left(\frac{1}{2} - 2\,y_2\right)\mathbf{a_1} + \frac{1}{2}\,\mathbf{a_2} + \frac{1}{2}\,\mathbf{a_3}$ | = | $\frac{1}{2}\,a\,\mathbf{\hat{x}} + \left(\frac{1}{2} - y_2\right)a\,\mathbf{\hat{y}} + \left(\frac{1}{2} - y_2\right)a\,\mathbf{\hat{z}}$ | (48$i$) | B |
| $\mathbf{B_6}$ | = | $+\frac{1}{2}\,\mathbf{a_1} + \left(\frac{1}{2} + 2\,y_2\right)\,\mathbf{a_2} + \frac{1}{2}\,\mathbf{a_3}$ | = | $\left(\frac{1}{2} + y_2\right)\,a\,\mathbf{\hat{x}} + \frac{1}{2}\,a\,\mathbf{\hat{y}} + \left(\frac{1}{2} + y_2\right)\,a\,\mathbf{\hat{z}}$ | (48$i$) | B |
| $\mathbf{B_7}$ | = | $\left(\frac{1}{2} - 2\,y_2\right)\,\mathbf{a_1} + \frac{1}{2}\,\mathbf{a_2} + \left(\frac{1}{2} + 2\,y_2\right)\,\mathbf{a_3}$ | = | $\left(\frac{1}{2} + y_2\right)\,a\,\mathbf{\hat{x}} + \frac{1}{2}\,a\,\mathbf{\hat{y}}\left(\frac{1}{2} - y_2\right)\,a\,\mathbf{\hat{z}}$ | (48$i$) | B |
| $\mathbf{B_8}$ | = | $\left(\frac{1}{2} + 2\,y_2\right)\,\mathbf{a_1} + \frac{1}{2}\,\mathbf{a_2} + \left(\frac{1}{2} - 2\,y_2\right)\,\mathbf{a_3}$ | = | $\left(\frac{1}{2} - y_2\right)\,a\,\mathbf{\hat{x}} + \frac{1}{2}\,a\,\mathbf{\hat{y}} + \left(\frac{1}{2} + y_2\right)\,a\,\mathbf{\hat{z}}$ | (48$i$) | B |
| $\mathbf{B_9}$ | = | $\frac{1}{2}\,\mathbf{a_1} + \left(\frac{1}{2} - 2\,y_2\right)\,\mathbf{a_2} + \frac{1}{2}\,\mathbf{a_3}$ | = | $\left(\frac{1}{2} - y_2\right)\,a\,\mathbf{\hat{x}} + \frac{1}{2}\,a\,\mathbf{\hat{y}}\left(\frac{1}{2} - y_2\right)\,a\,\mathbf{\hat{z}}$ | (48$i$) | B |
| $\mathbf{B_{10}}$ | = | $\frac{1}{2}\,\mathbf{a_1} + \frac{1}{2}\,\mathbf{a_2} + \left(\frac{1}{2} + 2\,y_2\right)\,\mathbf{a_3}$ | = | $\left(\frac{1}{2} + y_2\right)\,a\,\mathbf{\hat{x}} + \left(\frac{1}{2} + y_2\right)\,a\,\mathbf{\hat{y}} + \frac{1}{2}\,a\,\mathbf{\hat{z}}$ | (48$i$) | B |
| $\mathbf{B_{11}}$ | = | $\left(\frac{1}{2} + 2\,y_2\right)\,\mathbf{a_1} + \left(\frac{1}{2} - 2\,y_2\right)\,\mathbf{a_2} + \frac{1}{2}\,\mathbf{a_3}$ | = | $\left(\frac{1}{2} - y_2\right)\,a\,\mathbf{\hat{x}} + \left(\frac{1}{2} + y_2\right)\,a\,\mathbf{\hat{y}} + \frac{1}{2}\,a\,\mathbf{\hat{z}}$ | (48$i$) | B |
| $\mathbf{B_{12}}$ | = | $\left(\frac{1}{2} - 2\,y_2\right)\,\mathbf{a_1} + \left(\frac{1}{2} + 2\,y_2\right)\,\mathbf{a_2} + \frac{1}{2}\,\mathbf{a_3}$ | = | $\left(\frac{1}{2} + y_2\right)\,a\,\mathbf{\hat{x}} + \left(\frac{1}{2} - y_2\right)\,a\,\mathbf{\hat{y}} + \frac{1}{2}\,a\,\mathbf{\hat{z}}$ | (48$i$) | B |
| $\mathbf{B_{13}}$ | = | $\frac{1}{2}\,\mathbf{a_1} + \frac{1}{2}\,\mathbf{a_2} + \left(\frac{1}{2} - 2\,y_2\right)\,\mathbf{a_3}$ | = | $\left(\frac{1}{2} - y_2\right)\,a\,\mathbf{\hat{x}} + \left(\frac{1}{2} - y_2\right)\,a\,\mathbf{\hat{y}} + \frac{1}{2}\,a\,\mathbf{\hat{z}}$ | (48$i$) | B |

---

**Found in:**

---

**Geometry files:**

- CIF: pp. 783
- POSCAR: pp. 784

# Fluorite ($CaF_2$, C1) Structure: AB2_cF12_225_a_c

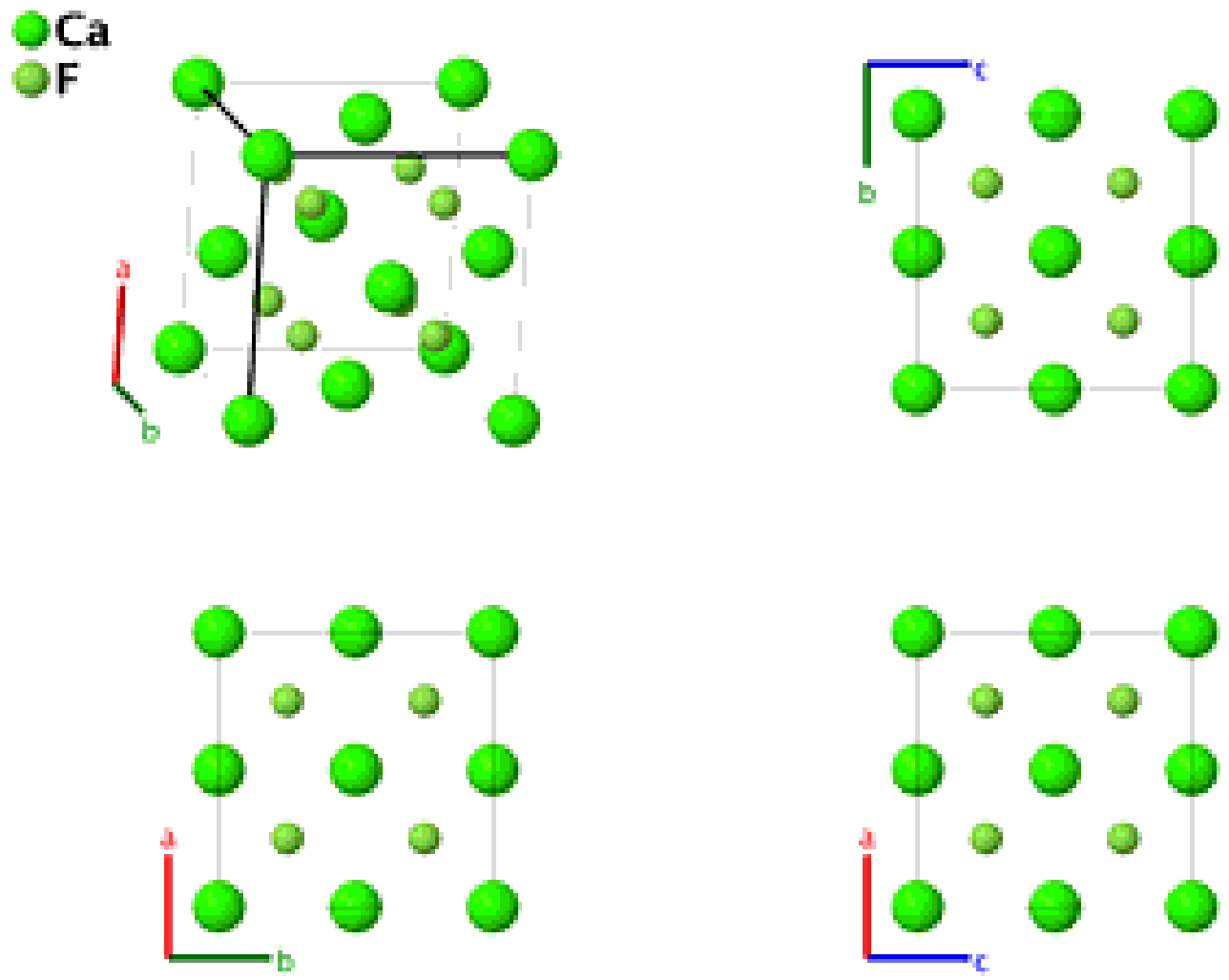

| | | |
|---|---|---|
| **Prototype** | : | $CaF_2$ |
| **AFLOW prototype label** | : | AB2_cF12_225_a_c |
| ***Strukturbericht* designation** | : | C1 |
| **Pearson symbol** | : | cF12 |
| **Space group number** | : | 225 |
| **Space group symbol** | : | $Fm\bar{3}m$ |
| **AFLOW prototype command** | : | `aflow --proto=AB2_cF12_225_a_c --params=`$a$ |

**Other compounds with this structure:**

- $AmO_2$, $AuAl_2$, $AuIn_2$, $BaF_2$, $Be_2B$, $CO_2$, $CdF_2$, $CeO_2$, $CoSi_2$, $EuF_2$, $HgF_2$, $Ir_2P$, $Li_2O$, $Na_2O$, $NiSi_2$, $PtAl_2$, $Rb_2O$, $SrCl_2$, $SrCl_2$, $SrF_2$, $ThO_2$, $ZrO_2$

**Face-centered Cubic primitive vectors:**

$$\mathbf{a}_1 = \frac{1}{2}\,a\,\hat{\mathbf{y}} + \frac{1}{2}\,a\,\hat{\mathbf{z}}$$
$$\mathbf{a}_2 = \frac{1}{2}\,a\,\hat{\mathbf{x}} + \frac{1}{2}\,a\,\hat{\mathbf{z}}$$
$$\mathbf{a}_3 = \frac{1}{2}\,a\,\hat{\mathbf{x}} + \frac{1}{2}\,a\,\hat{\mathbf{y}}$$

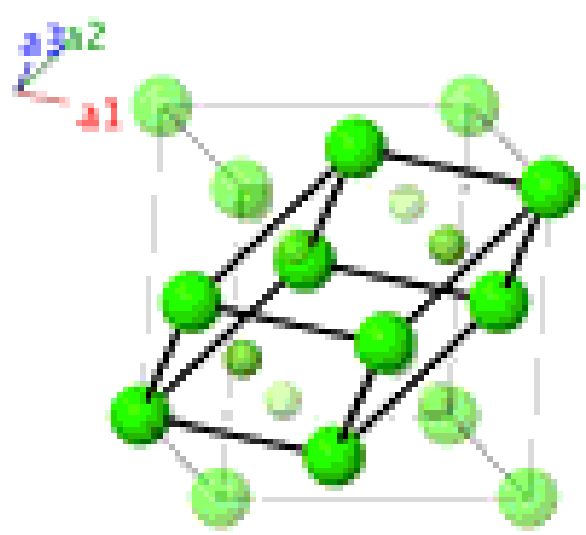

**Basis vectors:**

| | | Lattice Coordinates | | Cartesian Coordinates | Wyckoff Position | Atom Type |
|---|---|---|---|---|---|---|
| $\mathbf{B}_1$ | = | $0\,\mathbf{a}_1 + 0\,\mathbf{a}_2 + 0\,\mathbf{a}_3$ | = | $0\hat{\mathbf{x}} + 0\hat{\mathbf{y}} + 0\hat{\mathbf{z}}$ | (4$a$) | Ca |
| $\mathbf{B}_2$ | = | $\frac{1}{4}\,\mathbf{a}_1 + \frac{1}{4}\,\mathbf{a}_2 + \frac{1}{4}\,\mathbf{a}_3$ | = | $\frac{1}{4}\,a\,\hat{\mathbf{x}} + \frac{1}{4}\,a\,\hat{\mathbf{y}} + \frac{1}{4}\,a\,\hat{\mathbf{z}}$ | (8$c$) | F |
| $\mathbf{B}_3$ | = | $\frac{3}{4}\,\mathbf{a}_1 + \frac{3}{4}\,\mathbf{a}_2 + \frac{3}{4}\,\mathbf{a}_3$ | = | $\frac{3}{4}\,a\,\hat{\mathbf{x}} + \frac{3}{4}\,a\,\hat{\mathbf{y}} + \frac{3}{4}\,a\,\hat{\mathbf{z}}$ | (8$c$) | F |

**Found in:**

**Geometry files:**

- CIF: pp. 785
- POSCAR: pp. 786

# $Cr_{23}C_6$ ($D8_4$) Structure: A6B23_cF116_225_e_acfh

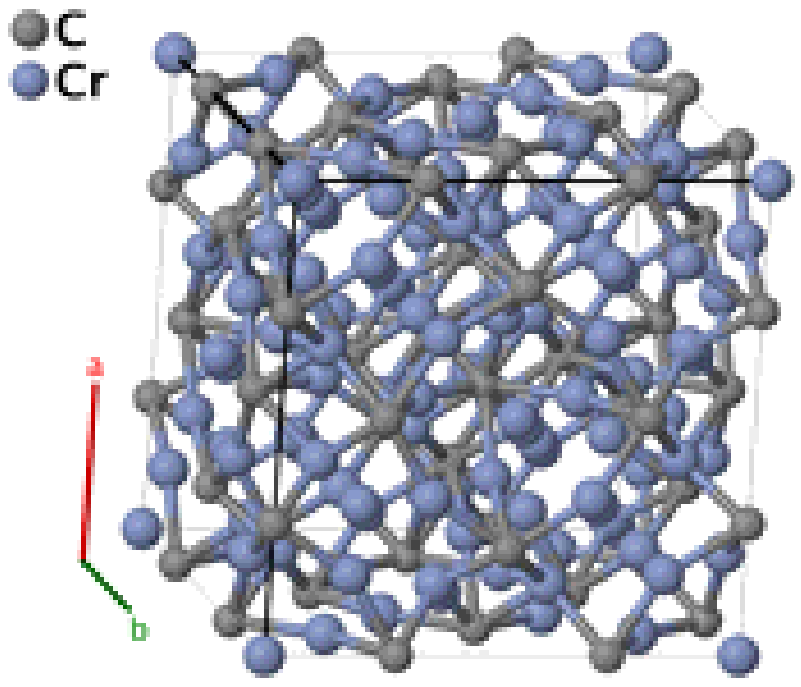

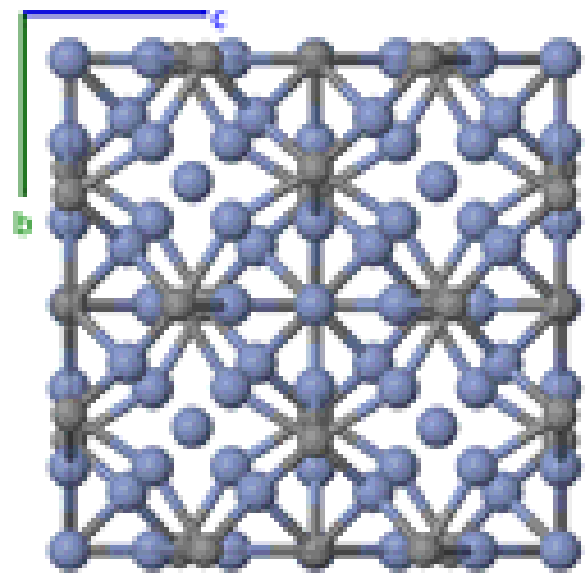

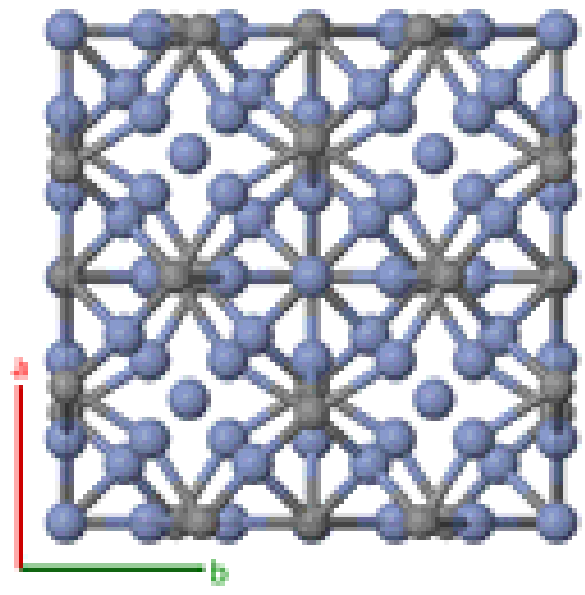

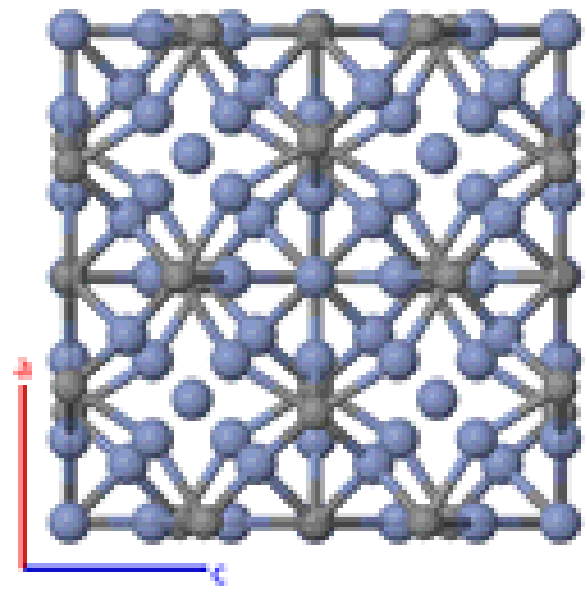

| | | |
|---|---|---|
| **Prototype** | : | $Cr_{23}C_6$ |
| **AFLOW prototype label** | : | A6B23_cF116_225_e_acfh |
| ***Strukturbericht* designation** | : | $D8_4$ |
| **Pearson symbol** | : | cF116 |
| **Space group number** | : | 225 |
| **Space group symbol** | : | $Fm\bar{3}m$ |
| **AFLOW prototype command** | : | `aflow --proto=A6B23_cF116_225_e_acfh` <br> `--params=`$a, x_3, x_4, y_5$ |

**Other compounds with this structure:**

- The general structure of this compound is $M_{23}X_6$ where M=Fe, Cr, Ni, Mn, V, W, ..., or combinations thereof, and X = C or B.

**Face-centered Cubic primitive vectors:**

$$\mathbf{a}_1 = \frac{1}{2}\,a\,\hat{\mathbf{y}} + \frac{1}{2}\,a\,\hat{\mathbf{z}}$$
$$\mathbf{a}_2 = \frac{1}{2}\,a\,\hat{\mathbf{x}} + \frac{1}{2}\,a\,\hat{\mathbf{z}}$$
$$\mathbf{a}_3 = \frac{1}{2}\,a\,\hat{\mathbf{x}} + \frac{1}{2}\,a\,\hat{\mathbf{y}}$$

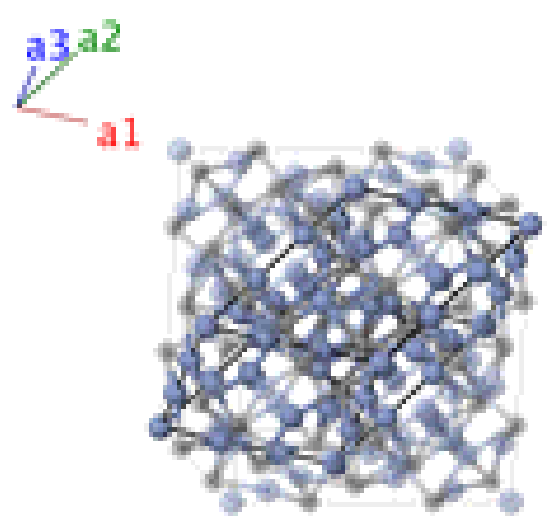

**Basis vectors:**

| | | Lattice Coordinates | | Cartesian Coordinates | Wyckoff Position | Atom Type |
|---|---|---|---|---|---|---|
| $\mathbf{B_1}$ | = | $0\,\mathbf{a_1} + 0\,\mathbf{a_2} + 0\,\mathbf{a_3}$ | = | $0\hat{\mathbf{x}} + 0\hat{\mathbf{y}} + 0\hat{\mathbf{z}}$ | $(4a)$ | Cr I |
| $\mathbf{B_2}$ | = | $\frac{1}{4}\,\mathbf{a_1} + \frac{1}{4}\,\mathbf{a_2} + \frac{1}{4}\,\mathbf{a_3}$ | = | $\frac{1}{4}\,a\,\hat{\mathbf{x}} + \frac{1}{4}\,a\,\hat{\mathbf{y}} + \frac{1}{4}\,a\,\hat{\mathbf{z}}$ | $(8c)$ | Cr II |
| $\mathbf{B_3}$ | = | $\frac{3}{4}\,\mathbf{a_1} + \frac{3}{4}\,\mathbf{a_2} + \frac{3}{4}\,\mathbf{a_3}$ | = | $\frac{3}{4}\,a\,\hat{\mathbf{x}} + \frac{3}{4}\,a\,\hat{\mathbf{y}} + \frac{3}{4}\,a\,\hat{\mathbf{z}}$ | $(8c)$ | Cr II |
| $\mathbf{B_4}$ | = | $-x_3\,\mathbf{a_1} + x_3\,\mathbf{a_2} + x_3\,\mathbf{a_3}$ | = | $x_3\,a\,\hat{\mathbf{x}}$ | $(24e)$ | C |
| $\mathbf{B_5}$ | = | $x_3\,\mathbf{a_1} - x_3\,\mathbf{a_2} + x_3\,\mathbf{a_3}$ | = | $x_3\,a\,\hat{\mathbf{y}}$ | $(24e)$ | C |
| $\mathbf{B_6}$ | = | $x_3\,\mathbf{a_1} + x_3\,\mathbf{a_2} - x_3\,\mathbf{a_3}$ | = | $x_3\,a\,\hat{\mathbf{z}}$ | $(24e)$ | C |
| $\mathbf{B_7}$ | = | $x_3\,\mathbf{a_1} - x_3\,\mathbf{a_2} - x_3\,\mathbf{a_3}$ | = | $-x_3\,a\,\hat{\mathbf{x}}$ | $(24e)$ | C |
| $\mathbf{B_8}$ | = | $-x_3\,\mathbf{a_1} + x_3\,\mathbf{a_2} - x_3\,\mathbf{a_3}$ | = | $-x_3\,a\,\hat{\mathbf{y}}$ | $(24e)$ | C |
| $\mathbf{B_9}$ | = | $-x_3\,\mathbf{a_1} - x_3\,\mathbf{a_2} + x_3\,\mathbf{a_3}$ | = | $-x_3\,a\,\hat{\mathbf{z}}$ | $(24e)$ | C |
| $\mathbf{B_{10}}$ | = | $x_4\,\mathbf{a_1} + x_4\,\mathbf{a_2} + x_4\,\mathbf{a_3}$ | = | $x_4\,a\,\hat{\mathbf{x}} + x_4\,a\,\hat{\mathbf{y}} + x_4\,a\,\hat{\mathbf{z}}$ | $(32f)$ | Cr III |
| $\mathbf{B_{11}}$ | = | $x_4\,\mathbf{a_1} + x_4\,\mathbf{a_2} - 3\,x_4\,\mathbf{a_3}$ | = | $-x_4\,a\,\hat{\mathbf{x}} - x_4\,a\,\hat{\mathbf{y}} + x_4\,a\,\hat{\mathbf{z}}$ | $(32f)$ | Cr III |
| $\mathbf{B_{12}}$ | = | $x_4\,\mathbf{a_1} - 3\,x_4\,\mathbf{a_2} + x_4\,\mathbf{a_3}$ | = | $-x_4\,a\,\hat{\mathbf{x}} + x_4\,a\,\hat{\mathbf{y}} - x_4\,a\,\hat{\mathbf{z}}$ | $(32f)$ | Cr III |
| $\mathbf{B_{13}}$ | = | $-3\,x_4\,\mathbf{a_1} + x_4\,\mathbf{a_2} + x_4\,\mathbf{a_3}$ | = | $x_4\,a\,\hat{\mathbf{x}} - x_4\,a\,\hat{\mathbf{y}} - x_4\,a\,\hat{\mathbf{z}}$ | $(32f)$ | Cr III |
| $\mathbf{B_{14}}$ | = | $-x_4\,\mathbf{a_1} - x_4\,\mathbf{a_2} + 3\,x_4\,\mathbf{a_3}$ | = | $x_4\,a\,\hat{\mathbf{x}} + x_4\,a\,\hat{\mathbf{y}} - x_4\,a\,\hat{\mathbf{z}}$ | $(32f)$ | Cr III |
| $\mathbf{B_{15}}$ | = | $-x_4\,\mathbf{a_1} - x_4\,\mathbf{a_2} - x_4\,\mathbf{a_3}$ | = | $-x_4\,a\,\hat{\mathbf{x}} - x_4\,a\,\hat{\mathbf{y}} - x_4\,a\,\hat{\mathbf{z}}$ | $(32f)$ | Cr III |
| $\mathbf{B_{16}}$ | = | $-x_4\,\mathbf{a_1} + 3\,x_4\,\mathbf{a_2} - x_4\,\mathbf{a_3}$ | = | $x_4\,a\,\hat{\mathbf{x}} - x_4\,a\,\hat{\mathbf{y}} + x_4\,a\,\hat{\mathbf{z}}$ | $(32f)$ | Cr III |
| $\mathbf{B_{17}}$ | = | $3\,x_4\,\mathbf{a_1} - x_4\,\mathbf{a_2} - x_4\,\mathbf{a_3}$ | = | $-x_4\,a\,\hat{\mathbf{x}} + x_4\,a\,\hat{\mathbf{y}} + x_4\,a\,\hat{\mathbf{z}}$ | $(32f)$ | Cr III |
| $\mathbf{B_{18}}$ | = | $2\,y_5\,\mathbf{a_1}$ | = | $y_5\,a\,\hat{\mathbf{y}} + y_5\,a\,\hat{\mathbf{z}}$ | $(48h)$ | Cr IV |
| $\mathbf{B_{19}}$ | = | $2\,y_5\,\mathbf{a_2} - 2\,y_5\,\mathbf{a_3}$ | = | $-y_5\,a\,\hat{\mathbf{y}} + y_5\,a\,\hat{\mathbf{z}}$ | $(48h)$ | Cr IV |
| $\mathbf{B_{20}}$ | = | $-2\,y_5\,\mathbf{a_2} + 2\,y_5\,\mathbf{a_3}$ | = | $y_5\,a\,\hat{\mathbf{y}} - y_5\,a\,\hat{\mathbf{z}}$ | $(48h)$ | Cr IV |
| $\mathbf{B_{21}}$ | = | $-2\,y_5\,\mathbf{a_1}$ | = | $-y_5\,a\,\hat{\mathbf{y}} - y_5\,a\,\hat{\mathbf{z}}$ | $(48h)$ | Cr IV |
| $\mathbf{B_{22}}$ | = | $2\,y_5\,\mathbf{a_2}$ | = | $y_5\,a\,\hat{\mathbf{x}} + y_5\,a\,\hat{\mathbf{z}}$ | $(48h)$ | Cr IV |
| $\mathbf{B_{23}}$ | = | $-2\,y_5\,\mathbf{a_1} + 2\,y_5\,\mathbf{a_3}$ | = | $y_5\,a\,\hat{\mathbf{x}} - y_5\,a\,\hat{\mathbf{z}}$ | $(48h)$ | Cr IV |
| $\mathbf{B_{24}}$ | = | $2\,y_5\,\mathbf{a_1} - 2\,y_5\,\mathbf{a_3}$ | = | $-y_5\,a\,\hat{\mathbf{x}} + y_5\,a\,\hat{\mathbf{z}}$ | $(48h)$ | Cr IV |
| $\mathbf{B_{25}}$ | = | $-2\,y_5\,\mathbf{a_2}$ | = | $-y_5\,a\,\hat{\mathbf{x}} - y_5\,a\,\hat{\mathbf{z}}$ | $(48h)$ | Cr IV |
| $\mathbf{B_{26}}$ | = | $2\,y_5\,\mathbf{a_3}$ | = | $y_5\,a\,\hat{\mathbf{x}} + y_5\,a\,\hat{\mathbf{y}}$ | $(48h)$ | Cr IV |
| $\mathbf{B_{27}}$ | = | $2\,y_5\,\mathbf{a_1} - 2\,y_5\,\mathbf{a_2}$ | = | $-y_5\,a\,\hat{\mathbf{x}} + y_5\,a\,\hat{\mathbf{y}}$ | $(48h)$ | Cr IV |
| $\mathbf{B_{28}}$ | = | $-2\,y_5\,\mathbf{a_1} + 2\,y_5\,\mathbf{a_2}$ | = | $y_5\,a\,\hat{\mathbf{x}} - y_5\,a\,\hat{\mathbf{y}}$ | $(48h)$ | Cr IV |
| $\mathbf{B_{29}}$ | = | $-2\,y_5\,\mathbf{a_3}$ | = | $-y_5\,a\,\hat{\mathbf{x}} - y_5\,a\,\hat{\mathbf{y}}$ | $(48h)$ | Cr IV |

**Geometry files:**

- CIF: pp. 786
- POSCAR: pp. 787

# Heusler ($L2_1$) Structure: AB2C_cF16_225_a_c_b

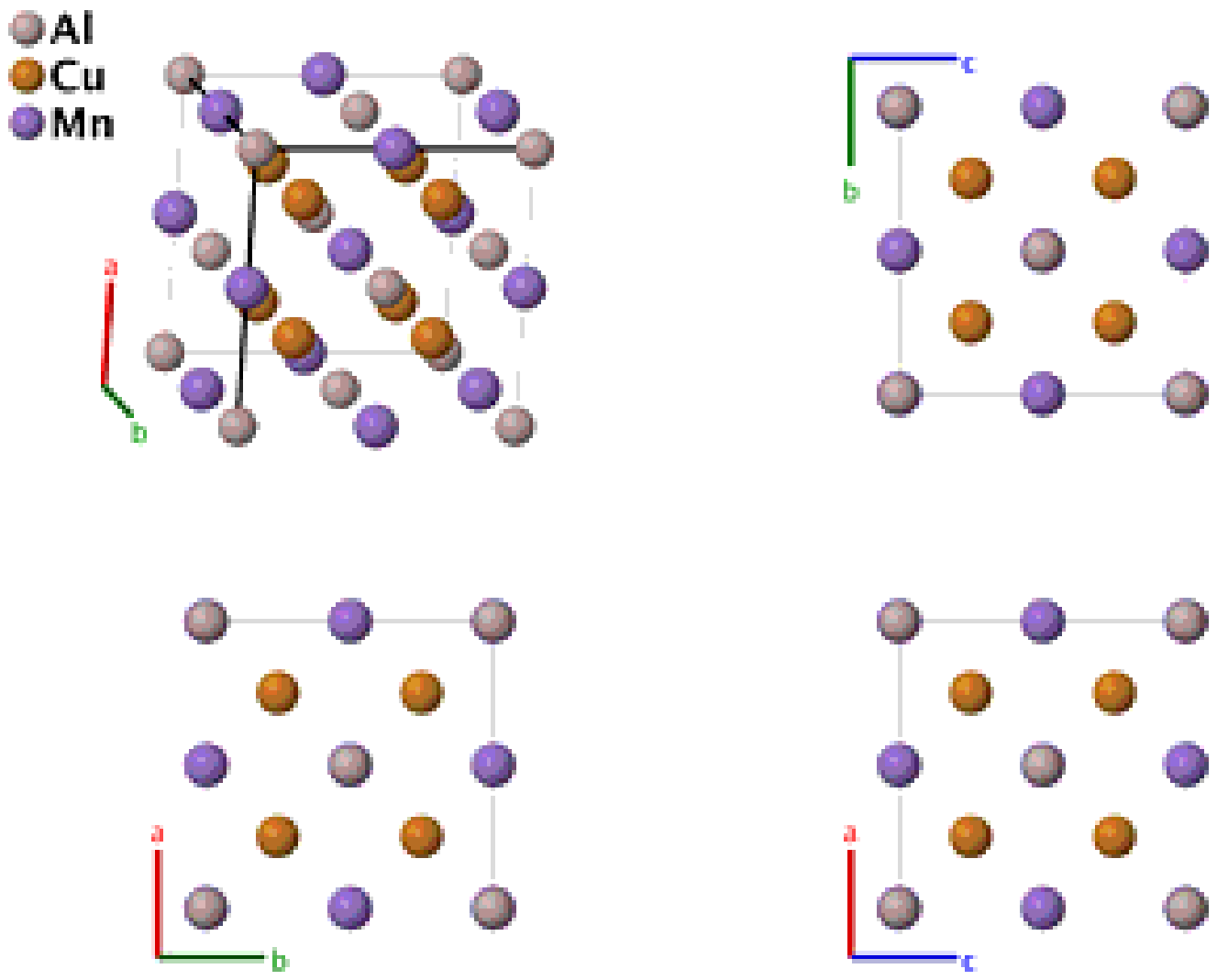

| | | |
|---|---|---|
| **Prototype** | : | $AlCu_2Mn$ |
| **AFLOW prototype label** | : | AB2C_cF16_225_a_c_b |
| ***Strukturbericht* designation** | : | $L2_1$ |
| **Pearson symbol** | : | cF16 |
| **Space group number** | : | 225 |
| **Space group symbol** | : | $Fm\bar{3}m$ |
| **AFLOW prototype command** | : | `aflow --proto=AB2C_cF16_225_a_c_b --params=`$a$ |

**Other compounds with this structure:**

- $AlNi_2Ti$, $AlNi_2Hf$

- All of the atoms are located on the sites of a body-centered cubic lattice. If we replace the Mn atom by another copper atom, the structure reduces to the crystallographically equivalent $D0_3$ lattice. Also see the $C1_b$ "half-Heusler" structure.

**Face-centered Cubic primitive vectors:**

$$\mathbf{a}_1 = \frac{1}{2}\,a\,\hat{\mathbf{y}} + \frac{1}{2}\,a\,\hat{\mathbf{z}}$$
$$\mathbf{a}_2 = \frac{1}{2}\,a\,\hat{\mathbf{x}} + \frac{1}{2}\,a\,\hat{\mathbf{z}}$$
$$\mathbf{a}_3 = \frac{1}{2}\,a\,\hat{\mathbf{x}} + \frac{1}{2}\,a\,\hat{\mathbf{y}}$$

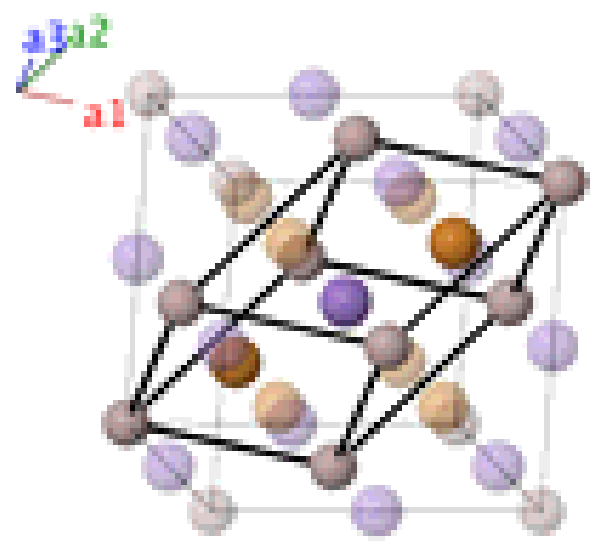

**Basis vectors:**

| | | Lattice Coordinates | | Cartesian Coordinates | Wyckoff Position | Atom Type |
|---|---|---|---|---|---|---|
| $\mathbf{B}_1$ | = | $0\,\mathbf{a}_1 + 0\,\mathbf{a}_2 + 0\,\mathbf{a}_3$ | = | $0\hat{\mathbf{x}} + 0\hat{\mathbf{y}} + 0\hat{\mathbf{z}}$ | (4$a$) | Al |
| $\mathbf{B_2}$ | = | $\frac{1}{2}\mathbf{a_1} + \frac{1}{2}\mathbf{a_2} + \frac{1}{2}\mathbf{a_3}$ | = | $\frac{1}{2}\,a\,\mathbf{\hat{x}} + \frac{1}{2}\,a\,\mathbf{\hat{y}} + \frac{1}{2}\,a\,\mathbf{\hat{z}}$ | (4*b*) | Mn |
| $\mathbf{B_3}$ | = | $\frac{1}{4}\mathbf{a_1} + \frac{1}{4}\mathbf{a_2} + \frac{1}{4}\mathbf{a_3}$ | = | $\frac{1}{4}\,a\,\mathbf{\hat{x}} + \frac{1}{4}\,a\,\mathbf{\hat{y}} + \frac{1}{4}\,a\,\mathbf{\hat{z}}$ | (8*c*) | Cu |
| $\mathbf{B_4}$ | = | $\frac{3}{4}\mathbf{a_1} + \frac{3}{4}\mathbf{a_2} + \frac{3}{4}\mathbf{a_3}$ | = | $\frac{3}{4}\,a\,\mathbf{\hat{x}} + \frac{3}{4}\,a\,\mathbf{\hat{y}} + \frac{3}{4}\,a\,\mathbf{\hat{z}}$ | (8*c*) | Cu |

**Geometry files:**

- CIF: pp. 787
- POSCAR: pp. 788

# Face-Centered Cubic (Cu, A1) Structure: A_cF4_225_a

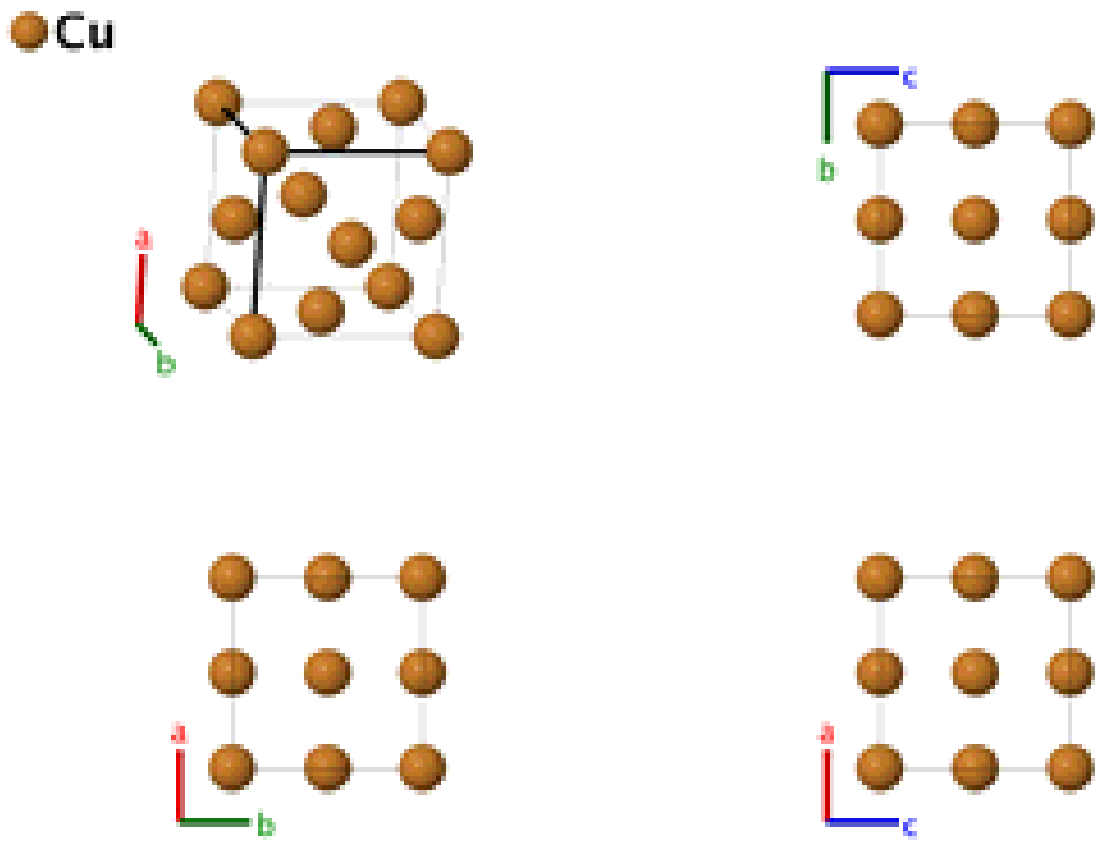

| | | |
|---|---|---|
| **Prototype** | : | Cu |
| **AFLOW prototype label** | : | A_cF4_225_a |
| ***Strukturbericht* designation** | : | A1 |
| **Pearson symbol** | : | cF4 |
| **Space group number** | : | 225 |
| **Space group symbol** | : | $\mathrm{Fm\bar{3}m}$ |
| **AFLOW prototype command** | : | `aflow --proto=A_cF4_225_a --params=`$a$ |

**Other elements with this structure:**

- Al, Cu, Ni, Sr, Rh, Pd, Ag, Ce, Tb, Ir, Pt, Au, Pb, Th

**Face-centered Cubic primitive vectors:**

$$\mathbf{a}_1 = \frac{1}{2}\,a\,\hat{\mathbf{y}} + \frac{1}{2}\,a\,\hat{\mathbf{z}}$$
$$\mathbf{a}_2 = \frac{1}{2}\,a\,\hat{\mathbf{x}} + \frac{1}{2}\,a\,\hat{\mathbf{z}}$$
$$\mathbf{a}_3 = \frac{1}{2}\,a\,\hat{\mathbf{x}} + \frac{1}{2}\,a\,\hat{\mathbf{y}}$$

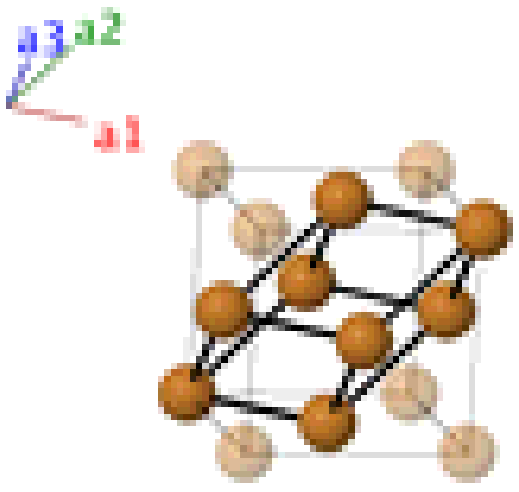

**Basis vectors:**

| | | Lattice Coordinates | | Cartesian Coordinates | Wyckoff Position | Atom Type |
|---|---|---|---|---|---|---|
| $\mathbf{B_1}$ | = | $0\,\mathbf{a_1} + 0\,\mathbf{a_2} + 0\,\mathbf{a_3}$ | = | $0\hat{\mathbf{x}} + 0\hat{\mathbf{y}} + 0\hat{\mathbf{z}}$ | $(4a)$ | Cu |

**Geometry files:**

- CIF: pp. 789
- POSCAR: pp. 790

# Model of Austenite Structure (cF108): AB18C8_cF108_225_a_eh_f

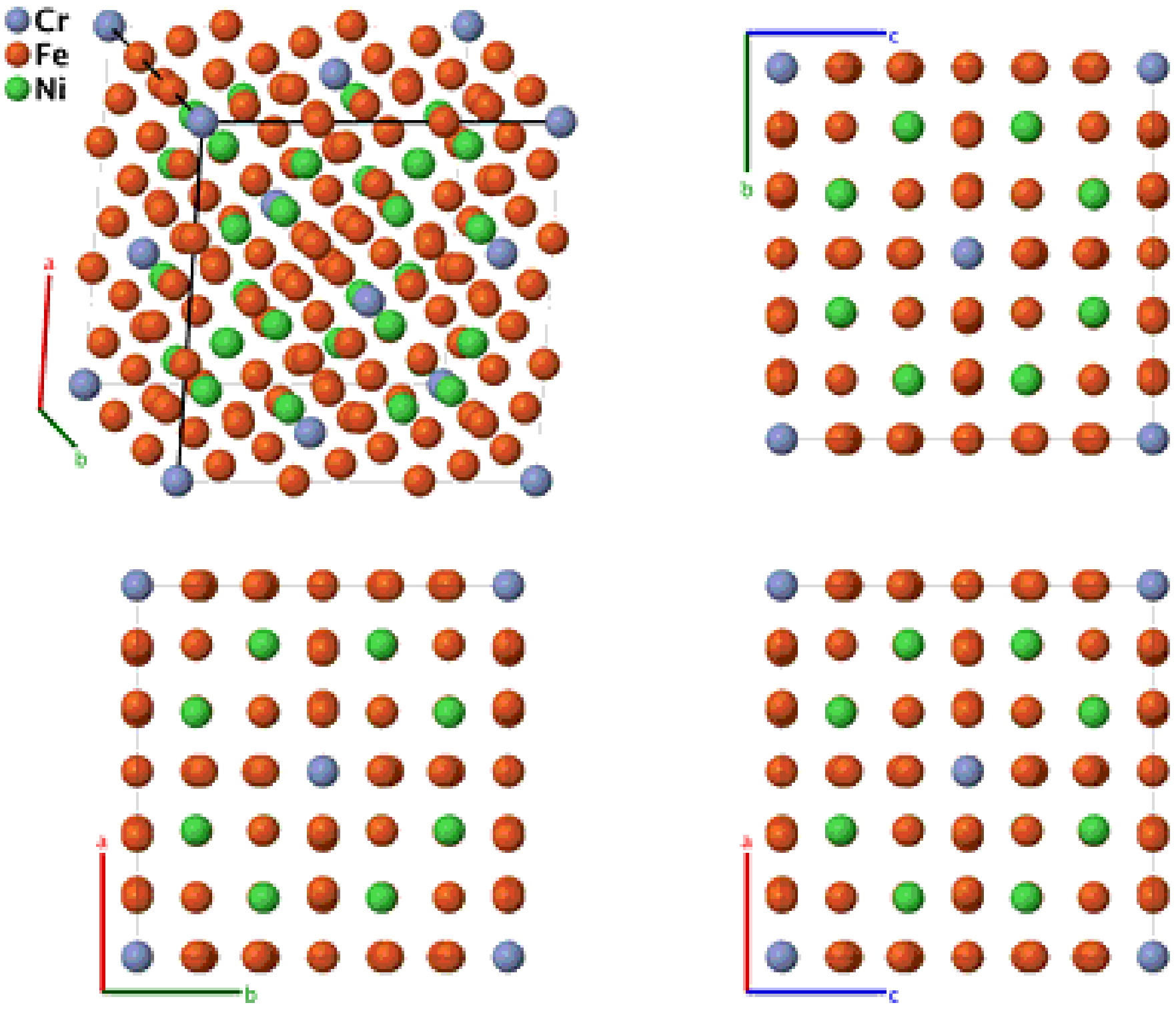

**Prototype** : $CrFe_{18}Ni_8$

**AFLOW prototype label** : AB18C8_cF108_225_a_eh_f

***Strukturbericht* designation** : None

**Pearson symbol** : cF108

**Space group number** : 225

**Space group symbol** : $Fm\bar{3}m$

**AFLOW prototype command** : `aflow --proto=AB18C8_cF108_225_a_eh_f --params=`$a, x_2, x_3, y_4$

---

- Austenitic steels are alloys of iron and other metals with an averaged face-centered cubic structure. If we set $x_2 = 1/3$, $x_3 = 2/3$, and $y_4 = 2/3$, the atoms are on the sites of an fcc lattice with lattice constant $a_{fcc} = 1/3a$.

---

**Face-centered Cubic primitive vectors:**

$$\mathbf{a}_1 = \frac{1}{2}\,a\,\mathbf{\hat{y}} + \frac{1}{2}\,a\,\mathbf{\hat{z}}$$
$$\mathbf{a}_2 = \frac{1}{2}\,a\,\mathbf{\hat{x}} + \frac{1}{2}\,a\,\mathbf{\hat{z}}$$
$$\mathbf{a}_3 = \frac{1}{2}\,a\,\mathbf{\hat{x}} + \frac{1}{2}\,a\,\mathbf{\hat{y}}$$

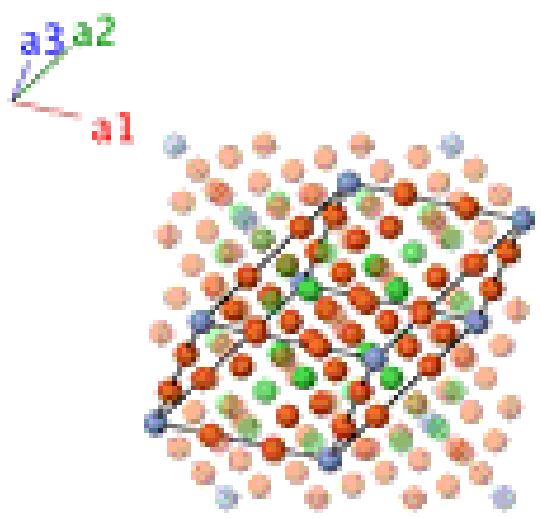

---

**Basis vectors:**

| | | Lattice Coordinates | | Cartesian Coordinates | Wyckoff Position | Atom Type |
|---|---|---|---|---|---|---|
| $\mathbf{B}_1$ | = | $0\,\mathbf{a}_1 + 0\,\mathbf{a}_2 + 0\,\mathbf{a}_3$ | = | $0\hat{\mathbf{x}} + 0\hat{\mathbf{y}} + 0\hat{\mathbf{z}}$ | $(4a)$ | Cr |
| $\mathbf{B}_2$ | = | $-x_2\,\mathbf{a}_1 + x_2\,\mathbf{a}_2 + x_2\,\mathbf{a}_3$ | = | $x_2\,a\,\hat{\mathbf{x}}$ | $(24e)$ | Fe I |
| $\mathbf{B}_3$ | = | $x_2\,\mathbf{a}_1 - x_2\,\mathbf{a}_2 + x_2\,\mathbf{a}_3$ | = | $x_2\,a\,\hat{\mathbf{y}}$ | $(24e)$ | Fe I |
| $\mathbf{B}_4$ | = | $x_2\,\mathbf{a}_1 + x_2\,\mathbf{a}_2 - x_2\,\mathbf{a}_3$ | = | $x_2\,a\,\hat{\mathbf{z}}$ | $(24e)$ | Fe I |
| $\mathbf{B}_5$ | = | $x_2\,\mathbf{a}_1 - x_2\,\mathbf{a}_2 - x_2\,\mathbf{a}_3$ | = | $-x_2\,a\,\hat{\mathbf{x}}$ | $(24e)$ | Fe I |
| $\mathbf{B}_6$ | = | $-x_2\,\mathbf{a}_1 + x_2\,\mathbf{a}_2 - x_2\,\mathbf{a}_3$ | = | $-x_2\,a\,\hat{\mathbf{y}}$ | $(24e)$ | Fe I |
| $\mathbf{B}_7$ | = | $-x_2\,\mathbf{a}_1 - x_2\,\mathbf{a}_2 + x_2\,\mathbf{a}_3$ | = | $-x_2\,a\,\hat{\mathbf{z}}$ | $(24e)$ | Fe I |
| $\mathbf{B}_8$ | = | $x_3\,\mathbf{a}_1 + x_3\,\mathbf{a}_2 + x_3\,\mathbf{a}_3$ | = | $x_3\,a\,\hat{\mathbf{x}} + x_3\,a\,\hat{\mathbf{y}} + x_3\,a\,\hat{\mathbf{z}}$ | $(32f)$ | Ni |
| $\mathbf{B}_9$ | = | $x_3\,\mathbf{a}_1 + x_3\,\mathbf{a}_2 - 3\,x_3\,\mathbf{a}_3$ | = | $-x_3\,a\,\hat{\mathbf{x}} - x_3\,a\,\hat{\mathbf{y}} + x_3\,a\,\hat{\mathbf{z}}$ | $(32f)$ | Ni |
| $\mathbf{B}_{10}$ | = | $x_3\,\mathbf{a}_1 - 3\,x_3\,\mathbf{a}_2 + x_3\,\mathbf{a}_3$ | = | $-x_3\,a\,\hat{\mathbf{x}} + x_3\,a\,\hat{\mathbf{y}} - x_3\,a\,\hat{\mathbf{z}}$ | $(32f)$ | Ni |
| $\mathbf{B}_{11}$ | = | $-3\,x_3\,\mathbf{a}_1 + x_3\,\mathbf{a}_2 + x_3\,\mathbf{a}_3$ | = | $x_3\,a\,\hat{\mathbf{x}} - x_3\,a\,\hat{\mathbf{y}} - x_3\,a\,\hat{\mathbf{z}}$ | $(32f)$ | Ni |
| $\mathbf{B}_{12}$ | = | $-x_3\,\mathbf{a}_1 - x_3\,\mathbf{a}_2 + 3\,x_3\,\mathbf{a}_3$ | = | $x_3\,a\,\hat{\mathbf{x}} + x_3\,a\,\hat{\mathbf{y}} - x_3\,a\,\hat{\mathbf{z}}$ | $(32f)$ | Ni |
| $\mathbf{B}_{13}$ | = | $-x_3\,\mathbf{a}_1 - x_3\,\mathbf{a}_2 - x_3\,\mathbf{a}_3$ | = | $-x_3\,a\,\hat{\mathbf{x}} - x_3\,a\,\hat{\mathbf{y}} - x_3\,a\,\hat{\mathbf{z}}$ | $(32f)$ | Ni |
| $\mathbf{B}_{14}$ | = | $-x_3\,\mathbf{a}_1 + 3\,x_3\,\mathbf{a}_2 - x_3\,\mathbf{a}_3$ | = | $x_3\,a\,\hat{\mathbf{x}} - x_3\,a\,\hat{\mathbf{y}} + x_3\,a\,\hat{\mathbf{z}}$ | $(32f)$ | Ni |
| $\mathbf{B}_{15}$ | = | $3\,x_3\,\mathbf{a}_1 - x_3\,\mathbf{a}_2 - x_3\,\mathbf{a}_3$ | = | $-x_3\,a\,\hat{\mathbf{x}} + x_3\,a\,\hat{\mathbf{y}} + x_3\,a\,\hat{\mathbf{z}}$ | $(32f)$ | Ni |
| $\mathbf{B}_{16}$ | = | $2\,y_4\,\mathbf{a}_1$ | = | $y_4\,a\,\hat{\mathbf{y}} + y_4\,a\,\hat{\mathbf{z}}$ | $(48h)$ | Fe II |
| $\mathbf{B}_{17}$ | = | $2\,y_4\,\mathbf{a}_2 - 2\,y_4\,\mathbf{a}_3$ | = | $-y_4\,a\,\hat{\mathbf{y}} + y_4\,a\,\hat{\mathbf{z}}$ | $(48h)$ | Fe II |
| $\mathbf{B}_{18}$ | = | $-2\,y_4\,\mathbf{a}_2 + 2\,y_4\,\mathbf{a}_3$ | = | $y_4\,a\,\hat{\mathbf{y}} - y_4\,a\,\hat{\mathbf{z}}$ | $(48h)$ | Fe II |
| $\mathbf{B}_{19}$ | = | $-2\,y_4\,\mathbf{a}_1$ | = | $-y_4\,a\,\hat{\mathbf{y}} - y_4\,a\,\hat{\mathbf{z}}$ | $(48h)$ | Fe II |
| $\mathbf{B}_{20}$ | = | $2\,y_4\,\mathbf{a}_2$ | = | $y_4\,a\,\hat{\mathbf{x}} + y_4\,a\,\hat{\mathbf{z}}$ | $(48h)$ | Fe II |
| $\mathbf{B}_{21}$ | = | $-2\,y_4\,\mathbf{a}_1 + 2\,y_4\,\mathbf{a}_3$ | = | $y_4\,a\,\hat{\mathbf{x}} - y_4\,a\,\hat{\mathbf{z}}$ | $(48h)$ | Fe II |
| $\mathbf{B}_{22}$ | = | $2\,y_4\,\mathbf{a}_1 - 2\,y_4\,\mathbf{a}_3$ | = | $-y_4\,a\,\hat{\mathbf{x}} + y_4\,a\,\hat{\mathbf{z}}$ | $(48h)$ | Fe II |
| $\mathbf{B}_{23}$ | = | $-2\,y_4\,\mathbf{a}_2$ | = | $-y_4\,a\,\hat{\mathbf{x}} - y_4\,a\,\hat{\mathbf{z}}$ | $(48h)$ | Fe II |
| $\mathbf{B}_{24}$ | = | $2\,y_4\,\mathbf{a}_3$ | = | $y_4\,a\,\hat{\mathbf{x}} + y_4\,a\,\hat{\mathbf{y}}$ | $(48h)$ | Fe II |
| $\mathbf{B}_{25}$ | = | $2\,y_4\,\mathbf{a}_1 - 2\,y_4\,\mathbf{a}_2$ | = | $-y_4\,a\,\hat{\mathbf{x}} + y_4\,a\,\hat{\mathbf{y}}$ | $(48h)$ | Fe II |
| $\mathbf{B}_{26}$ | = | $-2\,y_4\,\mathbf{a}_1 + 2\,y_4\,\mathbf{a}_2$ | = | $y_4\,a\,\hat{\mathbf{x}} - y_4\,a\,\hat{\mathbf{y}}$ | $(48h)$ | Fe II |
| $\mathbf{B}_{27}$ | = | $-2\,y_4\,\mathbf{a}_3$ | = | $-y_4\,a\,\hat{\mathbf{x}} - y_4\,a\,\hat{\mathbf{y}}$ | $(48h)$ | Fe II |

**Geometry files:**

- CIF: pp. 790
- POSCAR: pp. 791

# Rock Salt (NaCl, B1) Structure: AB_cF8_225_a_b

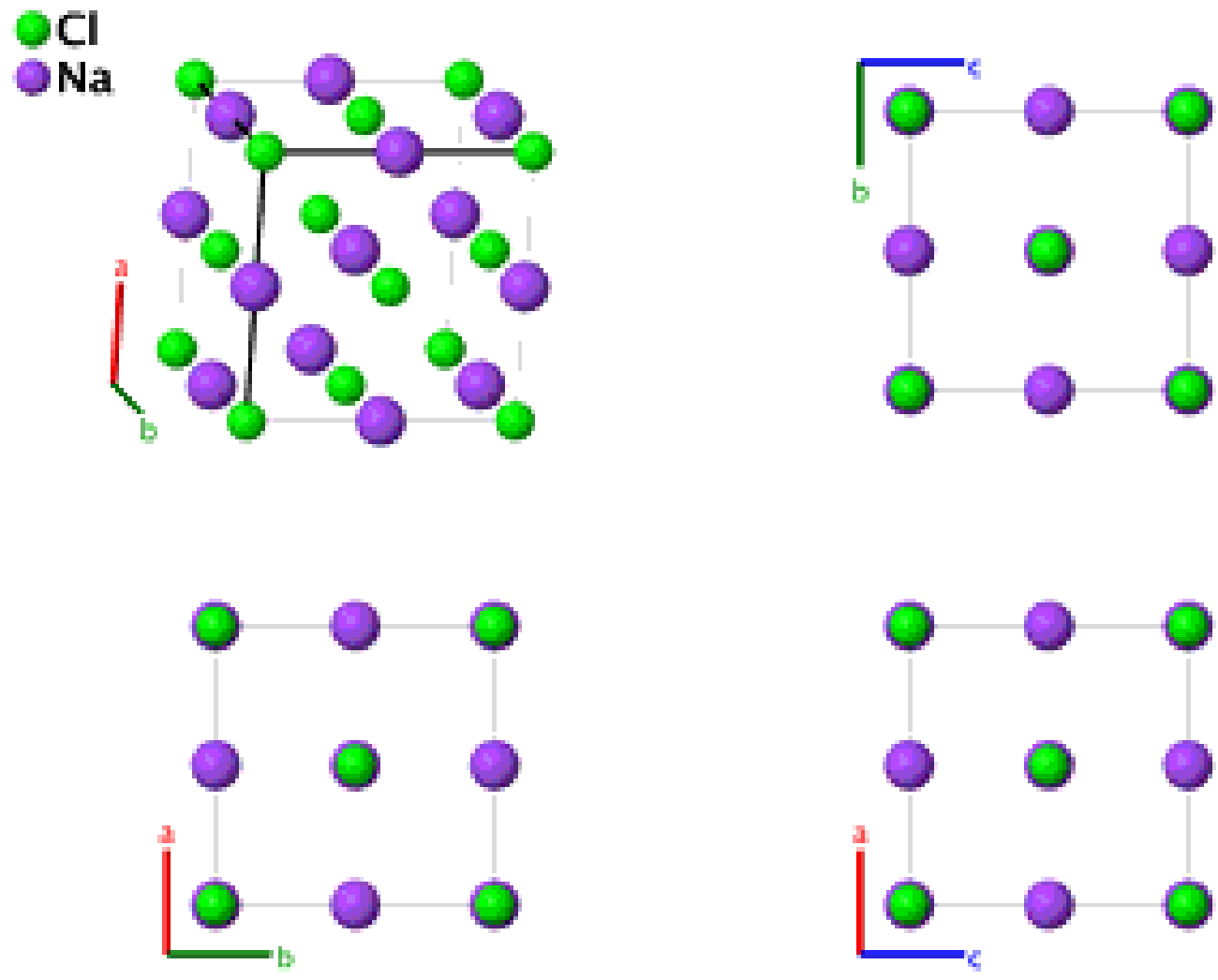

| | | |
|---|---|---|
| **Prototype** | : | NaCl |
| **AFLOW prototype label** | : | AB_cF8_225_a_b |
| ***Strukturbericht* designation** | : | B1 |
| **Pearson symbol** | : | cF8 |
| **Space group number** | : | 225 |
| **Space group symbol** | : | $Fm\bar{3}m$ |
| **AFLOW prototype command** | : | `aflow --proto=AB_cF8_225_a_b --params=`$a$ |

**Other compounds with this structure:**

- AgCl, BaS, CaO, CeSe, DyAs, GdN, KBr, LaP, LiCl, LiF, MgO, NaBr, NaF, NiO, PrBi, PuC, RbF, ScN, SrO, TbTe, UC, YN, YbO, ZrO

**Face-centered Cubic primitive vectors:**

$$\mathbf{a}_1 = \frac{1}{2}\,a\,\hat{\mathbf{y}} + \frac{1}{2}\,a\,\hat{\mathbf{z}}$$
$$\mathbf{a}_2 = \frac{1}{2}\,a\,\hat{\mathbf{x}} + \frac{1}{2}\,a\,\hat{\mathbf{z}}$$
$$\mathbf{a}_3 = \frac{1}{2}\,a\,\hat{\mathbf{x}} + \frac{1}{2}\,a\,\hat{\mathbf{y}}$$

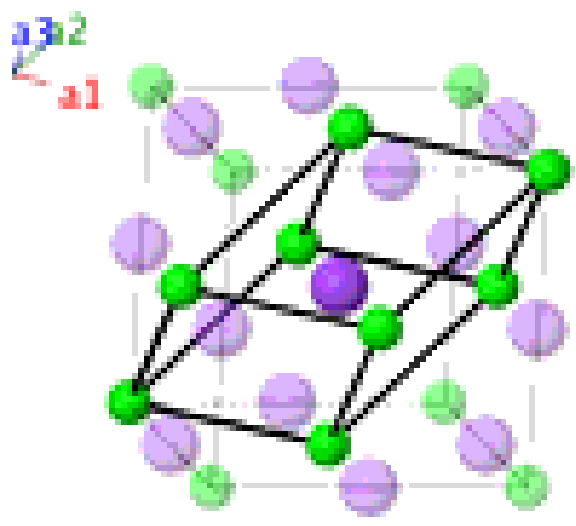

**Basis vectors:**

| | | Lattice Coordinates | | Cartesian Coordinates | Wyckoff Position | Atom Type |
|---|---|---|---|---|---|---|
| $\mathbf{B}_1$ | = | $0\,\mathbf{a}_1 + 0\,\mathbf{a}_2 + 0\,\mathbf{a}_3$ | = | $0\hat{\mathbf{x}} + 0\hat{\mathbf{y}} + 0\hat{\mathbf{z}}$ | (4$a$) | Cl |
| $\mathbf{B}_2$ | = | $\frac{1}{2}\,\mathbf{a}_1 + \frac{1}{2}\,\mathbf{a}_2 + \frac{1}{2}\,\mathbf{a}_3$ | = | $\frac{1}{2}\,a\,\hat{\mathbf{x}} + \frac{1}{2}\,a\,\hat{\mathbf{y}} + \frac{1}{2}\,a\,\hat{\mathbf{z}}$ | (4$b$) | Na |

**Found in:**

**Geometry files:**

- CIF: pp. 791
- POSCAR: pp. 792

# Ideal $\beta$-Cristobalite ($SiO_2$, C9) Structure: A2B_cF24_227_c_a

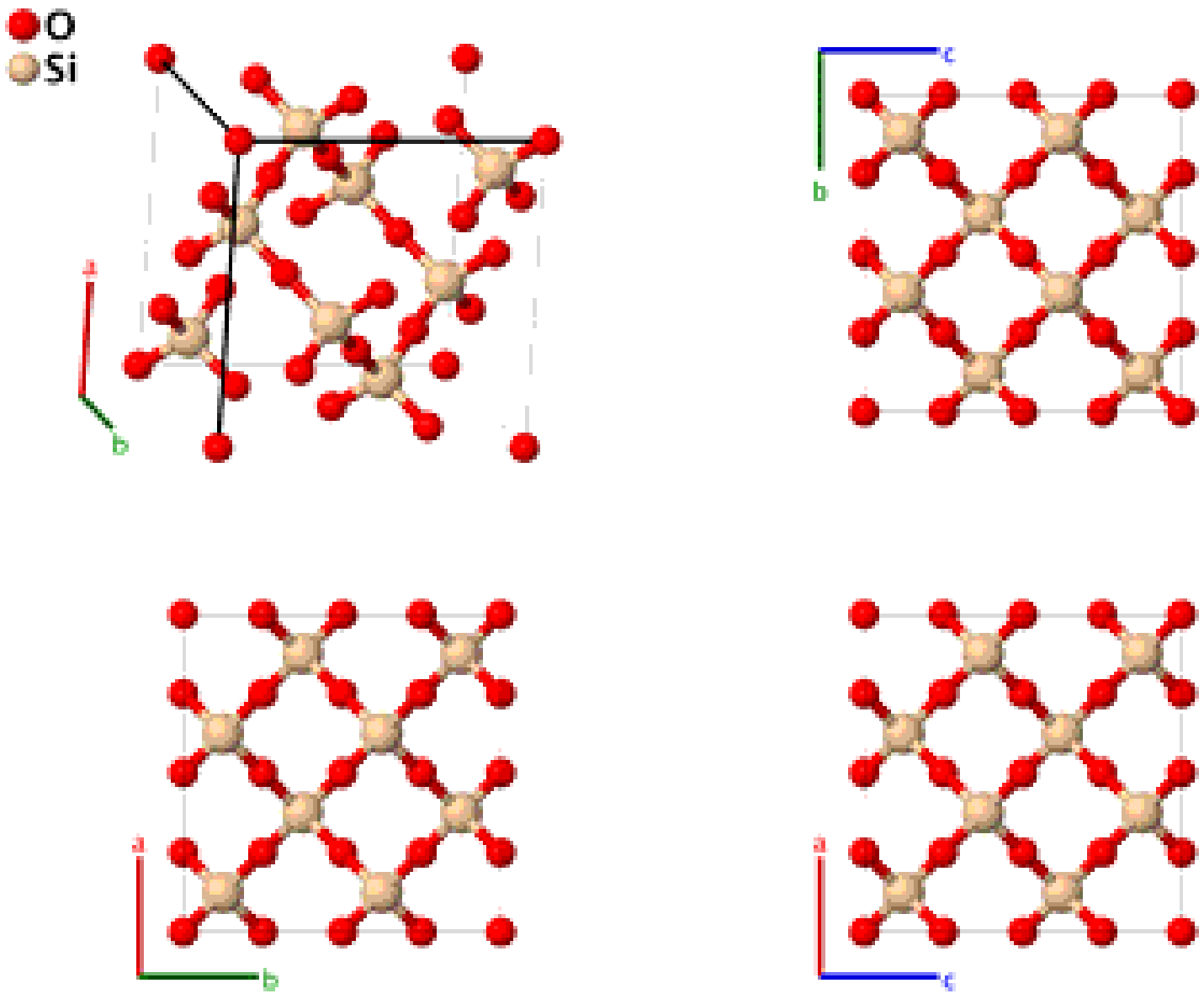

| | | |
|---|---|---|
| **Prototype** | : | $SiO_2$ |
| **AFLOW prototype label** | : | A2B_cF24_227_c_a |
| ***Strukturbericht* designation** | : | C9 |
| **Pearson symbol** | : | cF24 |
| **Space group number** | : | 227 |
| **Space group symbol** | : | $Fd\bar{3}m$ |
| **AFLOW prototype command** | : | `aflow --proto=A2B_cF24_227_c_a --params=`$a$ |

**Other compounds with this structure:**

- $BeF_2$

- This is an idealized version of the high-temperature phase of $\alpha$-cristobalite. (Peacor, 1973) concludes that the oxygen atoms partially occupy the (96g) positions in the space group $Fd\bar{3}m$. We average those positions to put the oxygen on the (16c) sites.

**Face-centered Cubic primitive vectors:**

$$\mathbf{a}_1 = \frac{1}{2}\,a\,\hat{\mathbf{y}} + \frac{1}{2}\,a\,\hat{\mathbf{z}}$$
$$\mathbf{a}_2 = \frac{1}{2}\,a\,\hat{\mathbf{x}} + \frac{1}{2}\,a\,\hat{\mathbf{z}}$$
$$\mathbf{a}_3 = \frac{1}{2}\,a\,\hat{\mathbf{x}} + \frac{1}{2}\,a\,\hat{\mathbf{y}}$$

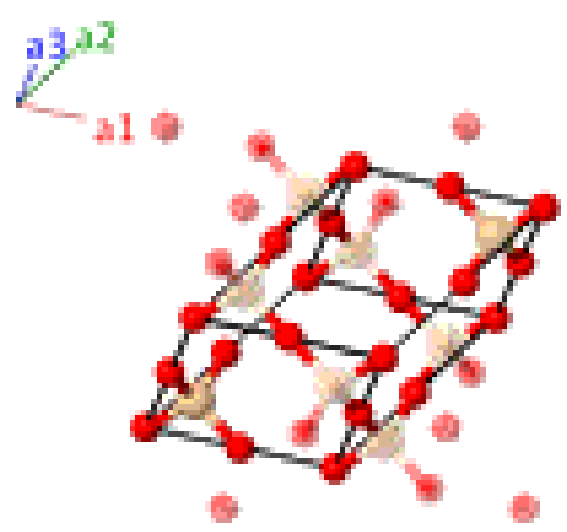

**Basis vectors:**

| | | Lattice Coordinates | | Cartesian Coordinates | Wyckoff Position | Atom Type |
|---|---|---|---|---|---|---|
| $\mathbf{B_1}$ | = | $\frac{1}{8}\mathbf{a_1} + \frac{1}{8}\mathbf{a_2} + \frac{1}{8}\mathbf{a_3}$ | = | $\frac{1}{8}\,a\,\hat{\mathbf{x}} + \frac{1}{8}\,a\,\hat{\mathbf{y}} + \frac{1}{8}\,a\,\hat{\mathbf{z}}$ | (8$a$) | Si |
| $\mathbf{B_2}$ | = | $\frac{7}{8}\mathbf{a_1} + \frac{7}{8}\mathbf{a_2} + \frac{7}{8}\mathbf{a_3}$ | = | $\frac{7}{8}\,a\,\hat{\mathbf{x}} + \frac{7}{8}\,a\,\hat{\mathbf{y}} + \frac{7}{8}\,a\,\hat{\mathbf{z}}$ | (8$a$) | Si |
| $\mathbf{B_3}$ | = | $0\,\mathbf{a_1} + 0\,\mathbf{a_2} + 0\,\mathbf{a_3}$ | = | $0\hat{\mathbf{x}} + 0\hat{\mathbf{y}} + 0\hat{\mathbf{z}}$ | (16$c$) | O |
| $\mathbf{B_4}$ | = | $\frac{1}{2}\mathbf{a_3}$ | = | $\frac{1}{4}\,a\,\hat{\mathbf{x}} + \frac{1}{4}\,a\,\hat{\mathbf{y}}$ | (16$c$) | O |
| $\mathbf{B_5}$ | = | $\frac{1}{2}\mathbf{a_2}$ | = | $\frac{1}{4}\,a\,\hat{\mathbf{x}} + \frac{1}{4}\,a\,\hat{\mathbf{z}}$ | (16$c$) | O |
| $\mathbf{B_6}$ | = | $\frac{1}{2}\mathbf{a_1}$ | = | $\frac{1}{4}\,a\,\hat{\mathbf{y}} + \frac{1}{4}\,a\,\hat{\mathbf{z}}$ | (16$c$) | O |

**Geometry files:**

- CIF: pp. 792
- POSCAR: pp. 794

# $NiTi_2$ Structure: AB2_cF96_227_e_cf

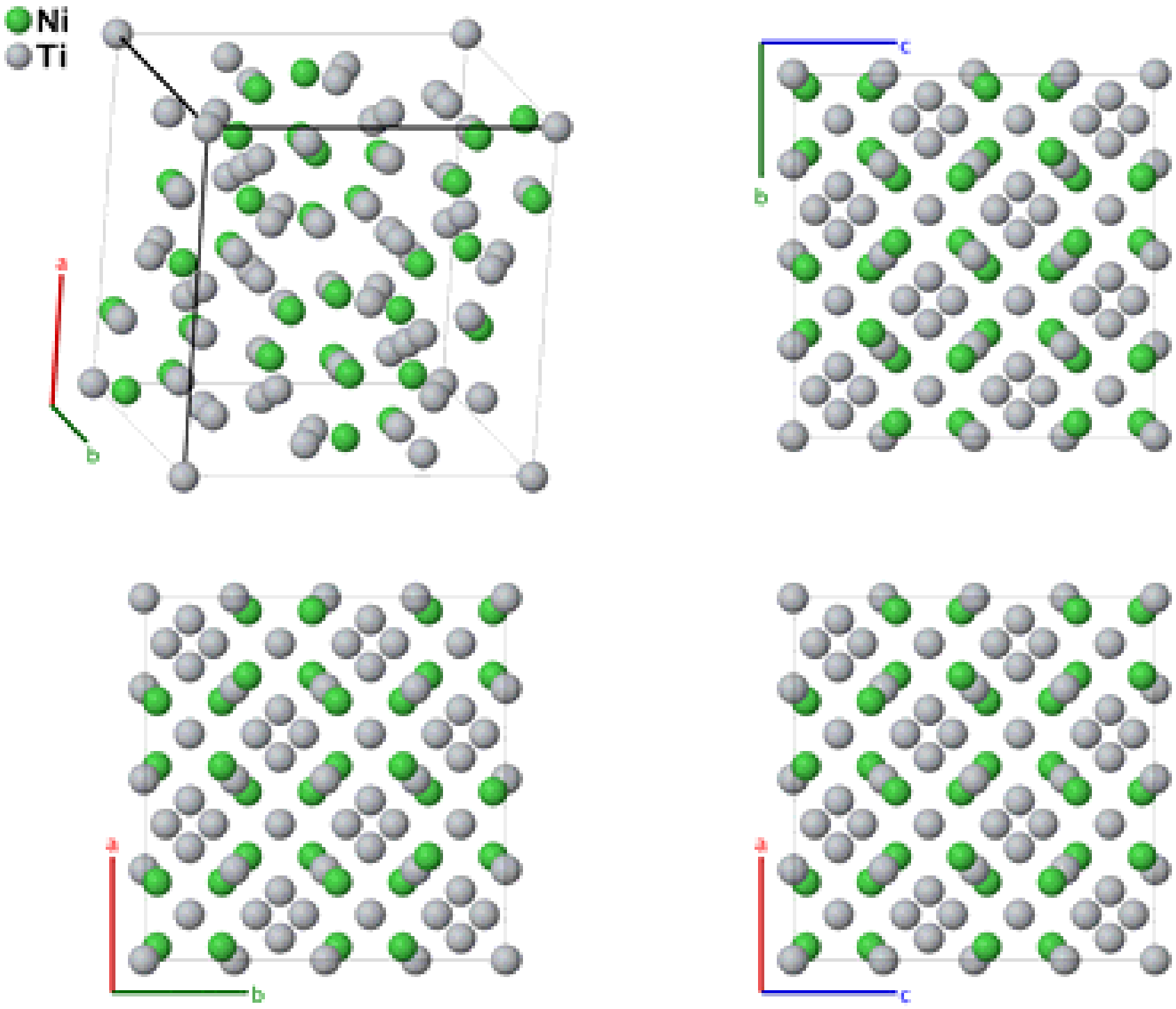

| | | |
|---|---|---|
| **Prototype** | : | $NiTi_2$ |
| **AFLOW prototype label** | : | AB2_cF96_227_e_cf |
| ***Strukturbericht* designation** | : | None |
| **Pearson symbol** | : | cF96 |
| **Space group number** | : | 227 |
| **Space group symbol** | : | $Fd\bar{3}m$ |
| **AFLOW prototype command** | : | `aflow --proto=AB2_cF96_227_e_cf` <br> `--params=`$a, x_2, x_3$ |

**Other compounds with this structure:**

- $CoTi_2$, $CoZr_2$, $Cr_2Nb$, $FeTi_2$, $FeZr_2$, $Hf_2Ir$, $Hf_2Pt$, $IrZr_2$, $NiSc_2$, $PdSc_2$, many others.

---

- We have used the fact that all vectors of the form $(0, \pm a/2, \pm a/2)$, $(\pm a/2, 0, \pm a/2)$, and $(\pm a/2, \pm a/2, 0)$ are primitive vectors of the face-centered cubic lattice to simplify the positions of some atoms in both lattice and Cartesian coordinates

**Face-centered Cubic primitive vectors:**

$$\mathbf{a}_1 = \frac{1}{2}\,a\,\hat{\mathbf{y}} + \frac{1}{2}\,a\,\hat{\mathbf{z}}$$
$$\mathbf{a}_2 = \frac{1}{2}\,a\,\hat{\mathbf{x}} + \frac{1}{2}\,a\,\hat{\mathbf{z}}$$
$$\mathbf{a}_3 = \frac{1}{2}\,a\,\hat{\mathbf{x}} + \frac{1}{2}\,a\,\hat{\mathbf{y}}$$

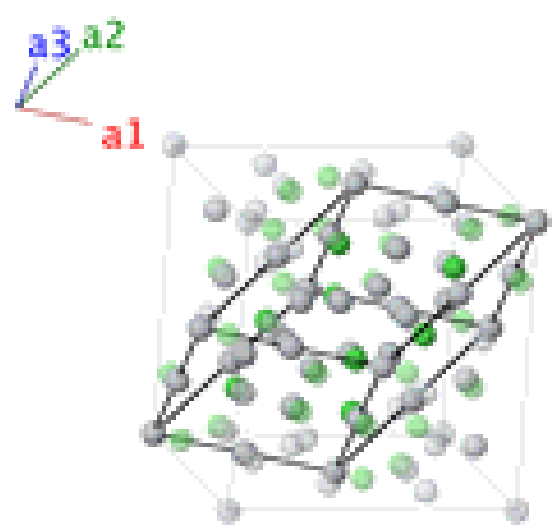

**Basis vectors:**

| | | Lattice Coordinates | | Cartesian Coordinates | Wyckoff Position | Atom Type |
|---|---|---|---|---|---|---|
| $\mathbf{B_1}$ | = | $0\,\mathbf{a_1} + 0\,\mathbf{a_2} + 0\,\mathbf{a_3}$ | = | $0\hat{\mathbf{x}} + 0\hat{\mathbf{y}} + 0\hat{\mathbf{z}}$ | (16*c*) | Ti I |
| $\mathbf{B_2}$ | = | $\frac{1}{2}\mathbf{a_3}$ | = | $\frac{1}{4}\,a\,\hat{\mathbf{x}} + \frac{1}{4}\,a\,\hat{\mathbf{y}}$ | (16*c*) | Ti I |
| $\mathbf{B_3}$ | = | $\frac{1}{2}\mathbf{a_2}$ | = | $\frac{1}{4}\,a\,\hat{\mathbf{x}} + \frac{1}{4}\,a\,\hat{\mathbf{z}}$ | (16*c*) | Ti I |
| $\mathbf{B_4}$ | = | $\frac{1}{2}\mathbf{a_1}$ | = | $\frac{1}{4}\,a\,\hat{\mathbf{y}} + \frac{1}{4}\,a\,\hat{\mathbf{z}}$ | (16*c*) | Ti I |
| $\mathbf{B_5}$ | = | $x_2\mathbf{a_1} + x_2\mathbf{a_2} + x_2\mathbf{a_3}$ | = | $x_2\,a\,\hat{\mathbf{x}} + x_2\,a\,\hat{\mathbf{y}} + x_2\,a\,\hat{\mathbf{z}}$ | (32*e*) | Ni |
| $\mathbf{B_6}$ | = | $x_2\mathbf{a_1} + x_2\mathbf{a_2} + \left(\frac{1}{2} - 3\,x_2\right)\mathbf{a_3}$ | = | $\left(\frac{1}{4} - x_2\right)\,a\,\hat{\mathbf{x}} + \left(\frac{1}{4} - x_2\right)\,a\,\hat{\mathbf{y}} + x_2\,a\,\hat{\mathbf{z}}$ | (32*e*) | Ni |
| $\mathbf{B_7}$ | = | $x_2\mathbf{a_1} + \left(\frac{1}{2} - 3\,x_2\right)\mathbf{a_2} + x_2\mathbf{a_3}$ | = | $\left(\frac{1}{4} - x_2\right)\,a\,\hat{\mathbf{x}} + x_2\,a\,\hat{\mathbf{y}} + \left(\frac{1}{4} - x_2\right)\,a\,\hat{\mathbf{z}}$ | (32*e*) | Ni |
| $\mathbf{B_8}$ | = | $\left(\frac{1}{2} - 3\,x_2\right)\mathbf{a_1} + x_2\mathbf{a_2} + x_2\mathbf{a_3}$ | = | $x_2\,a\,\hat{\mathbf{x}} + \left(\frac{1}{4} - x_2\right)\,a\,\hat{\mathbf{y}} + \left(\frac{1}{4} - x_2\right)\,a\,\hat{\mathbf{z}}$ | (32*e*) | Ni |
| $\mathbf{B_9}$ | = | $-x_2\mathbf{a_1} - x_2\mathbf{a_2} + \left(\frac{1}{2} + 3\,x_2\right)\mathbf{a_3}$ | = | $\left(\frac{1}{4} + x_2\right)\,a\,\hat{\mathbf{x}} + \left(\frac{1}{4} + x_2\right)\,a\,\hat{\mathbf{y}} - x_2\,a\,\hat{\mathbf{z}}$ | (32*e*) | Ni |
| $\mathbf{B_{10}}$ | = | $-x_2\mathbf{a_1} - x_2\mathbf{a_2} - x_2\mathbf{a_3}$ | = | $-x_2\,a\,\hat{\mathbf{x}} - x_2\,a\,\hat{\mathbf{y}} - x_2\,a\,\hat{\mathbf{z}}$ | (32*e*) | Ni |
| $\mathbf{B_{11}}$ | = | $-x_2\mathbf{a_1} + \left(\frac{1}{2} + 3\,x_2\right)\mathbf{a_2} - x_2\mathbf{a_3}$ | = | $\left(\frac{1}{4} + x_2\right)\,a\,\hat{\mathbf{x}} - x_2\,a\,\hat{\mathbf{y}} + \left(\frac{1}{4} + x_2\right)\,a\,\hat{\mathbf{z}}$ | (32*e*) | Ni |
| $\mathbf{B_{12}}$ | = | $\left(\frac{1}{2} + 3\,x_2\right)\mathbf{a_1} - x_2\mathbf{a_2} - x_2\mathbf{a_3}$ | = | $-x_2\,a\,\hat{\mathbf{x}}+\left(\frac{1}{4} + x_2\right)\,a\,\hat{\mathbf{y}}+\left(\frac{1}{4} + x_2\right)\,a\,\hat{\mathbf{z}}$ | (32*e*) | Ni |
| $\mathbf{B_{13}}$ | = | $\left(\frac{1}{4} - x_3\right)\mathbf{a_1} + x_3\mathbf{a_2} + x_3\mathbf{a_3}$ | = | $x_3\,a\,\hat{\mathbf{x}} + \frac{1}{8}\,a\,\hat{\mathbf{y}} + \frac{1}{8}\,a\,\hat{\mathbf{z}}$ | (48*f*) | Ti II |
| $\mathbf{B_{14}}$ | = | $x_3\mathbf{a_1} + \left(\frac{1}{4} - x_3\right)\mathbf{a_2} + \left(\frac{1}{4} - x_3\right)\mathbf{a_3}$ | = | $\left(\frac{1}{4} - x_3\right)\,a\,\hat{\mathbf{x}} + \frac{1}{8}\,a\,\hat{\mathbf{y}} + \frac{1}{8}\,a\,\hat{\mathbf{z}}$ | (48*f*) | Ti II |
| $\mathbf{B_{15}}$ | = | $x_3\mathbf{a_1} + \left(\frac{1}{4} - x_3\right)\mathbf{a_2} + x_3\mathbf{a_3}$ | = | $\frac{1}{8}\,a\,\hat{\mathbf{x}} + x_3\,a\,\hat{\mathbf{y}} + \frac{1}{8}\,a\,\hat{\mathbf{z}}$ | (48*f*) | Ti II |
| $\mathbf{B_{16}}$ | = | $\left(\frac{1}{4} - x_3\right)\mathbf{a_1} + x_3\mathbf{a_2} + \left(\frac{1}{4} - x_3\right)\mathbf{a_3}$ | = | $\frac{1}{8}\,a\,\hat{\mathbf{x}} + \left(\frac{1}{4} - x_3\right)\,a\,\hat{\mathbf{y}} + \frac{1}{8}\,a\,\hat{\mathbf{z}}$ | (48*f*) | Ti II |
| $\mathbf{B_{17}}$ | = | $x_3\mathbf{a_1} + x_3\mathbf{a_2} + \left(\frac{1}{4} - x_3\right)\mathbf{a_3}$ | = | $\frac{1}{8}\,a\,\hat{\mathbf{x}} + \frac{1}{8}\,a\,\hat{\mathbf{y}} + x_3\,a\,\hat{\mathbf{z}}$ | (48*f*) | Ti II |
| $\mathbf{B_{18}}$ | = | $\left(\frac{1}{4} - x_3\right)\mathbf{a_1} + \left(\frac{1}{4} - x_3\right)\mathbf{a_2} + x_3\mathbf{a_3}$ | = | $\frac{1}{8}\,a\,\hat{\mathbf{x}} + \frac{1}{8}\,a\,\hat{\mathbf{y}} + \left(\frac{1}{4} - x_3\right)\,a\,\hat{\mathbf{z}}$ | (48*f*) | Ti II |
| $\mathbf{B_{19}}$ | = | $\left(x_3 + \frac{3}{4}\right)\mathbf{a_1} - x_3\mathbf{a_2} + \left(x_3 + \frac{3}{4}\right)\mathbf{a_3}$ | = | $\frac{3}{8}\,a\,\hat{\mathbf{x}} + \left(x_3 + \frac{3}{4}\right)\,a\,\hat{\mathbf{y}} + \frac{3}{8}\,a\,\hat{\mathbf{z}}$ | (48*f*) | Ti II |
| $\mathbf{B_{20}}$ | = | $-x_3\mathbf{a_1} + \left(x_3 + \frac{3}{4}\right)\mathbf{a_2} - x_3\mathbf{a_3}$ | = | $\frac{3}{8}\,a\,\hat{\mathbf{x}} - x_3\,a\,\hat{\mathbf{y}} + \frac{3}{8}\,a\,\hat{\mathbf{z}}$ | (48*f*) | Ti II |
| $\mathbf{B_{21}}$ | = | $-x_3\mathbf{a_1} + \left(x_3 + \frac{3}{4}\right)\mathbf{a_2} + \left(x_3 + \frac{3}{4}\right)\mathbf{a_3}$ | = | $\left(x_3 + \frac{3}{4}\right)\,a\,\hat{\mathbf{x}} + \frac{3}{8}\,a\,\hat{\mathbf{y}} + \frac{3}{8}\,a\,\hat{\mathbf{z}}$ | (48*f*) | Ti II |
| $\mathbf{B_{22}}$ | = | $\left(x_3 + \frac{3}{4}\right)\mathbf{a_1} - x_3\mathbf{a_2} - x_3\mathbf{a_3}$ | = | $-x_3\,a\,\hat{\mathbf{x}} + \frac{3}{8}\,a\,\hat{\mathbf{y}} + \frac{3}{8}\,a\,\hat{\mathbf{z}}$ | (48*f*) | Ti II |
| $\mathbf{B_{23}}$ | = | $-x_3\mathbf{a_1} - x_3\mathbf{a_2} + \left(x_3 + \frac{3}{4}\right)\mathbf{a_3}$ | = | $\frac{3}{8}\,a\,\hat{\mathbf{x}} + \frac{3}{8}\,a\,\hat{\mathbf{y}} - x_3\,a\,\hat{\mathbf{z}}$ | (48*f*) | Ti II |
| $\mathbf{B_{24}}$ | = | $+\left(x_3 + \frac{3}{4}\right)\mathbf{a_1} + \left(x_3 + \frac{3}{4}\right)\mathbf{a_2} - x_3\mathbf{a_3}$ | = | $\frac{3}{8}\,a\,\hat{\mathbf{x}} + \frac{3}{8}\,a\,\hat{\mathbf{y}} + \left(x_3 + \frac{3}{4}\right)\,a\,\hat{\mathbf{z}}$ | (48*f*) | Ti II |

**Geometry files:**
- CIF: pp. 794
- POSCAR: pp. 795

# NaTl (B32) Structure: AB_cF16_227_a_b

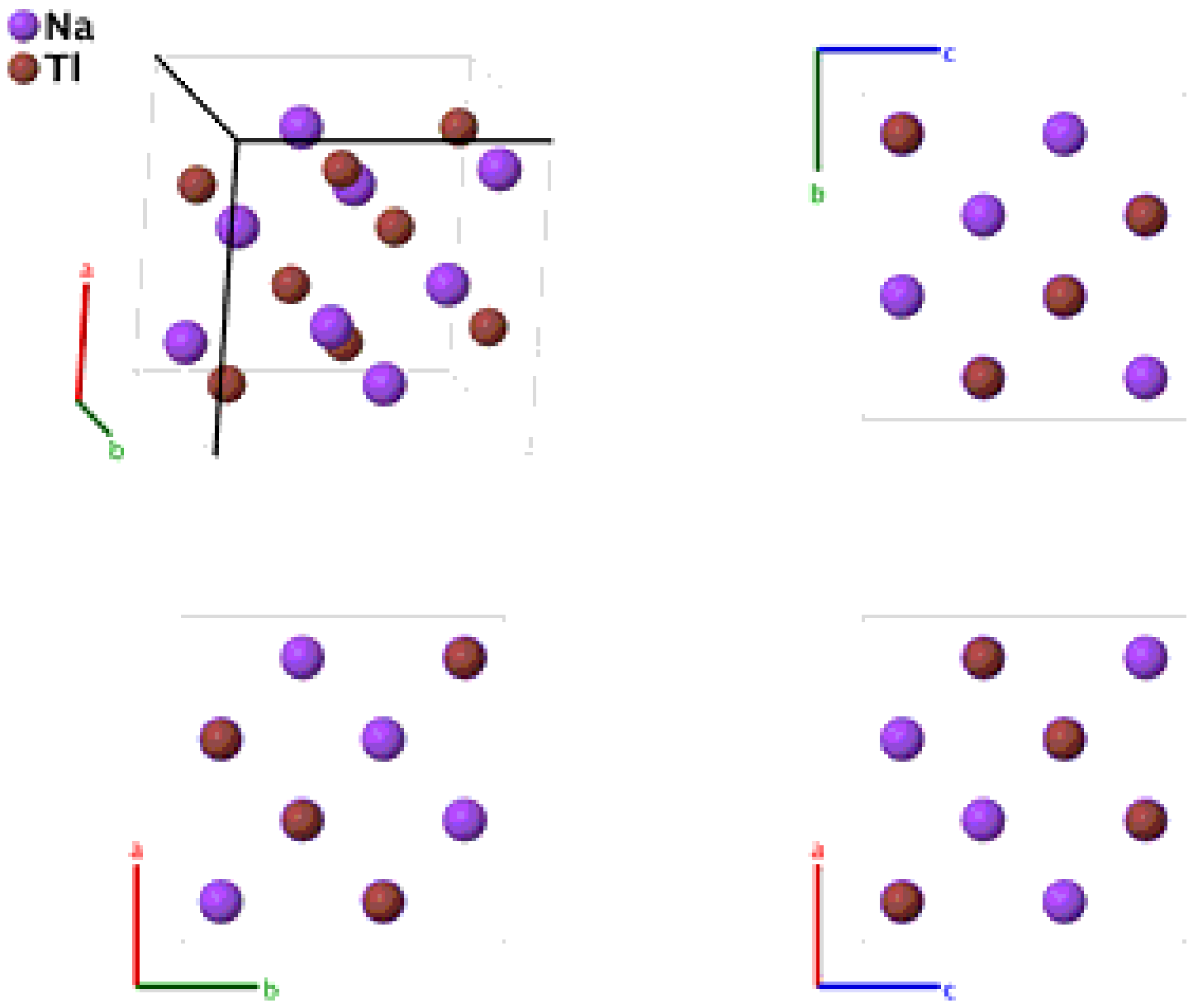

| | | |
|---|---|---|
| **Prototype** | : | NaTl |
| **AFLOW prototype label** | : | AB_cF16_227_a_b |
| ***Strukturbericht* designation** | : | B32 |
| **Pearson symbol** | : | cF16 |
| **Space group number** | : | 227 |
| **Space group symbol** | : | $Fd\bar{3}m$ |
| **AFLOW prototype command** | : | `aflow --proto=AB_cF16_227_a_b --params=`$a$ |

- This is an example of a Zintl Phase.

**Face-centered Cubic primitive vectors:**

$$
\begin{aligned}
\mathbf{a}_1 &= \tfrac{1}{2}\,a\,\hat{\mathbf{y}} + \tfrac{1}{2}\,a\,\hat{\mathbf{z}} \\
\mathbf{a}_2 &= \tfrac{1}{2}\,a\,\hat{\mathbf{x}} + \tfrac{1}{2}\,a\,\hat{\mathbf{z}} \\
\mathbf{a}_3 &= \tfrac{1}{2}\,a\,\hat{\mathbf{x}} + \tfrac{1}{2}\,a\,\hat{\mathbf{y}}
\end{aligned}
$$

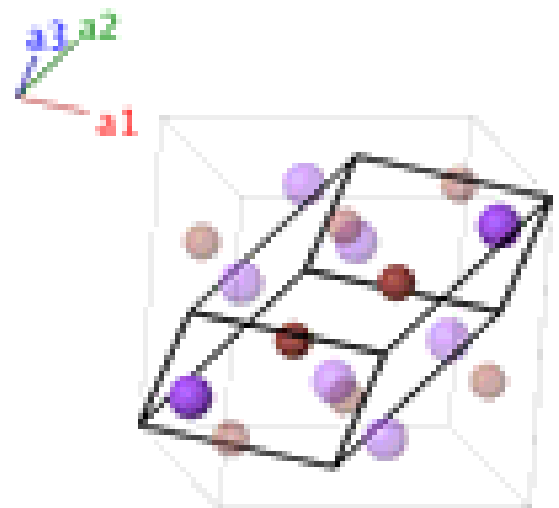

**Basis vectors:**

| | | Lattice Coordinates | | Cartesian Coordinates | Wyckoff Position | Atom Type |
|---|---|---|---|---|---|---|
| $\mathbf{B}_1$ | = | $\frac{1}{8}\mathbf{a}_1 + \frac{1}{8}\mathbf{a}_2 + \frac{1}{8}\mathbf{a}_3$ | = | $\frac{1}{8}\,a\,\hat{\mathbf{x}} + \frac{1}{8}\,a\,\hat{\mathbf{y}} + \frac{1}{8}\,a\,\hat{\mathbf{z}}$ | (8$a$) | Na |
| $\mathbf{B}_2$ | = | $\frac{7}{8}\mathbf{a}_1 + \frac{7}{8}\mathbf{a}_2 + \frac{7}{8}\mathbf{a}_3$ | = | $\frac{7}{8}\,a\,\hat{\mathbf{x}} + \frac{7}{8}\,a\,\hat{\mathbf{y}} + \frac{7}{8}\,a\,\hat{\mathbf{z}}$ | (8$a$) | Na |
| $\mathbf{B}_3$ | = | $\frac{3}{8}\mathbf{a}_1 + \frac{3}{8}\mathbf{a}_2 + \frac{3}{8}\mathbf{a}_3$ | = | $\frac{3}{8}\,a\,\hat{\mathbf{x}} + \frac{3}{8}\,a\,\hat{\mathbf{y}} + \frac{3}{8}\,a\,\hat{\mathbf{z}}$ | (8$b$) | Tl |
| $\mathbf{B}_4$ | = | $\frac{5}{8}\mathbf{a}_1 + \frac{5}{8}\mathbf{a}_2 + \frac{5}{8}\mathbf{a}_3$ | = | $\frac{5}{8}\,a\,\hat{\mathbf{x}} + \frac{5}{8}\,a\,\hat{\mathbf{y}} + \frac{5}{8}\,a\,\hat{\mathbf{z}}$ | (8$b$) | Tl |

**Found in:**

**Geometry files:**

- CIF: pp. 795
- POSCAR: pp. 796

# $Si_{34}$ Clathrate Structure: A_cF136_227_aeg

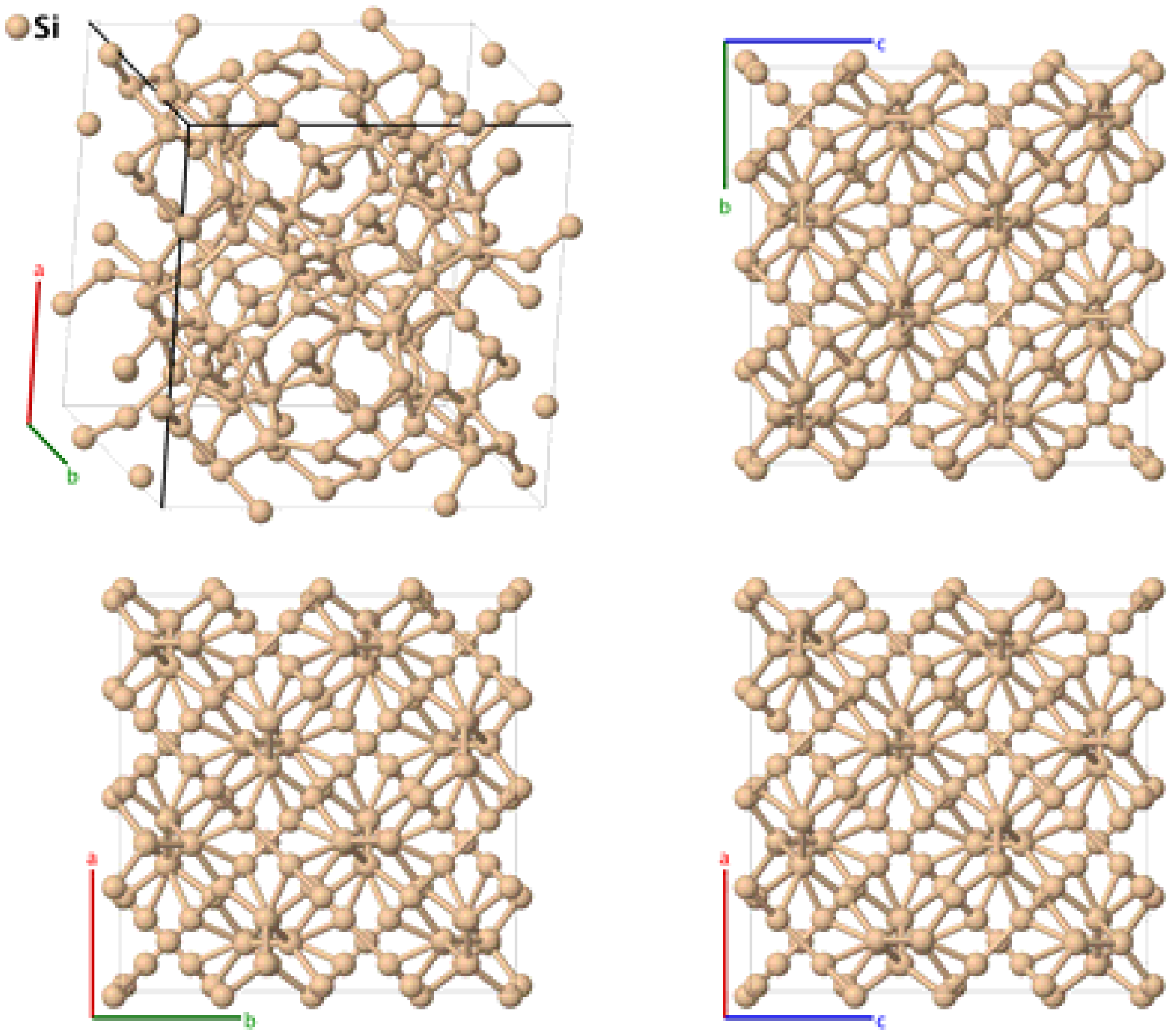

| | | |
|---|---|---|
| **Prototype** | : | Si |
| **AFLOW prototype label** | : | A_cF136_227_aeg |
| ***Strukturbericht* designation** | : | None |
| **Pearson symbol** | : | cF136 |
| **Space group number** | : | 227 |
| **Space group symbol** | : | $Fd\bar{3}m$ |
| **AFLOW prototype command** | : | `aflow --proto=A_cF136_227_aeg` `--params=`$a, x_2, x_3, z_3$ |

- Silicon clathrates are open structures of pentagonal dodecahedra connected so that all of the silicon atoms have $sp^3$ bonding. In nature these structures are stabilized by alkali impurity atoms. This structure and the $Si_{46}$ structure are proposed "pure" silicon clathrate structures. For more information about these structures and their possible stability, see (Adams, 1994). See (Gryko, 2000) for a possible experimental realization of this structure ($Si_{34}Na_x$, were x is very small). We have used the fact that all vectors of the form $(0, \pm a/2, \pm a/2)$, $(\pm a/2, 0, \pm a/2)$, and $(\pm a/2, \pm a/2, 0)$ are primitive vectors of the face-centered cubic lattice to simplify the positions of some atoms in both lattice and Cartesian coordinates.

## Face-centered Cubic primitive vectors:

$$\begin{aligned}\mathbf{a}_1 &= \tfrac{1}{2}\,a\,\hat{\mathbf{y}} + \tfrac{1}{2}\,a\,\hat{\mathbf{z}}\\ \mathbf{a}_2 &= \tfrac{1}{2}\,a\,\hat{\mathbf{x}} + \tfrac{1}{2}\,a\,\hat{\mathbf{z}}\\ \mathbf{a}_3 &= \tfrac{1}{2}\,a\,\hat{\mathbf{x}} + \tfrac{1}{2}\,a\,\hat{\mathbf{y}}\end{aligned}$$

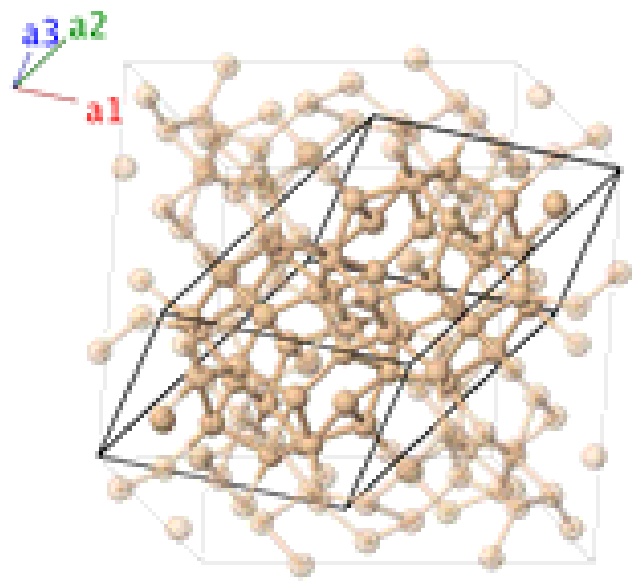

## Basis vectors:

| | | Lattice Coordinates | | Cartesian Coordinates | Wyckoff Position | Atom Type |
|---|---|---|---|---|---|---|
| $\mathbf{B}_1$ | = | $\frac{1}{8}\,\mathbf{a}_1 + \frac{1}{8}\,\mathbf{a}_2 + \frac{1}{8}\,\mathbf{a}_3$ | = | $\frac{1}{8}\,a\,\hat{\mathbf{x}} + \frac{1}{8}\,a\,\hat{\mathbf{y}} + \frac{1}{8}\,a\,\hat{\mathbf{z}}$ | (8$a$) | Si I |
| $\mathbf{B}_2$ | = | $\frac{7}{8}\,\mathbf{a}_1 + \frac{7}{8}\,\mathbf{a}_2 + \frac{7}{8}\,\mathbf{a}_3$ | = | $\frac{7}{8}\,a\,\hat{\mathbf{x}} + \frac{7}{8}\,a\,\hat{\mathbf{y}} + \frac{7}{8}\,a\,\hat{\mathbf{z}}$ | (8$a$) | Si I |
| $\mathbf{B}_3$ | = | $x_2\,\mathbf{a}_1 + x_2\,\mathbf{a}_2 + x_2\,\mathbf{a}_3$ | = | $x_2\,a\,\hat{\mathbf{x}} + x_2\,a\,\hat{\mathbf{y}} + x_2\,a\,\hat{\mathbf{z}}$ | (32$e$) | Si II |
| $\mathbf{B}_4$ | = | $x_2\,\mathbf{a}_1 + x_2\,\mathbf{a}_2 + \left(\frac{1}{2} - 3\,x_2\right)\,\mathbf{a}_3$ | = | $\left(\frac{1}{4} - x_2\right)\,a\,\hat{\mathbf{x}} + \left(\frac{1}{4} - x_2\right)\,a\,\hat{\mathbf{y}} + x_2\,a\,\hat{\mathbf{z}}$ | (32$e$) | Si II |
| $\mathbf{B}_5$ | = | $x_2\,\mathbf{a}_1 + \left(\frac{1}{2} - 3\,x_2\right)\,\mathbf{a}_2 + x_2\,\mathbf{a}_3$ | = | $\left(\frac{1}{4} - x_2\right)\,a\,\hat{\mathbf{x}} + x_2\,a\,\hat{\mathbf{y}} + \left(\frac{1}{4} - x_2\right)\,a\,\hat{\mathbf{z}}$ | (32$e$) | Si II |
| $\mathbf{B}_6$ | = | $\left(\frac{1}{2} - 3\,x_2\right)\,\mathbf{a}_1 + x_2\,\mathbf{a}_2 + x_2\,\mathbf{a}_3$ | = | $x_2\,a\,\hat{\mathbf{x}} + \left(\frac{1}{4} - x_2\right)\,a\,\hat{\mathbf{y}} + \left(\frac{1}{4} - x_2\right)\,a\,\hat{\mathbf{z}}$ | (32$e$) | Si II |
| $\mathbf{B}_7$ | = | $-x_2\,\mathbf{a}_1 - x_2\,\mathbf{a}_2 + \left(\frac{1}{2} + 3\,x_2\right)\,\mathbf{a}_3$ | = | $\left(\frac{1}{4} + x_2\right)\,a\,\hat{\mathbf{x}} + \left(\frac{1}{4} + x_2\right)\,a\,\hat{\mathbf{y}} - x_2\,a\,\hat{\mathbf{z}}$ | (32$e$) | Si II |
| $\mathbf{B}_8$ | = | $-x_2\,\mathbf{a}_1 - x_2\,\mathbf{a}_2 - x_2\,\mathbf{a}_3$ | = | $-x_2\,a\,\hat{\mathbf{x}} - x_2\,a\,\hat{\mathbf{y}} - x_2\,a\,\hat{\mathbf{z}}$ | (32$e$) | Si II |
| $\mathbf{B}_9$ | = | $-x_2\,\mathbf{a}_1 + \left(\frac{1}{2} + 3\,x_2\right)\,\mathbf{a}_2 - x_2\,\mathbf{a}_3$ | = | $\left(\frac{1}{4} + x_2\right)\,a\,\hat{\mathbf{x}} - x_2\,a\,\hat{\mathbf{y}} + \left(\frac{1}{4} + x_2\right)\,a\,\hat{\mathbf{z}}$ | (32$e$) | Si II |
| $\mathbf{B}_{10}$ | = | $\left(\frac{1}{2} + 3\,x_2\right)\,\mathbf{a}_1 - x_2\,\mathbf{a}_2 - x_2\,\mathbf{a}_3$ | = | $-x_2\,a\,\hat{\mathbf{x}}+\left(\frac{1}{4} + x_2\right)\,a\,\hat{\mathbf{y}}+\left(\frac{1}{4} + x_2\right)\,a\,\hat{\mathbf{z}}$ | (32$e$) | Si II |
| $\mathbf{B}_{11}$ | = | $z_3\,\mathbf{a}_1 + z_3\,\mathbf{a}_2 + (2x_3 - z_3)\,\mathbf{a}_3$ | = | $x_3\,a\,\hat{\mathbf{x}} + x_3\,a\,\hat{\mathbf{y}} + z_3\,a\,\hat{\mathbf{z}}$ | (96$g$) | Si III |
| $\mathbf{B}_{12}$ | = | $z_3\,\mathbf{a}_1 + z_3\,\mathbf{a}_2 + \left(\frac{1}{2} - 2x_3 - z_3\right)\,\mathbf{a}_3$ | = | $\left(\frac{1}{4} - x_3\right)\,a\,\hat{\mathbf{x}} + \left(\frac{1}{4} - x_3\right)\,a\,\hat{\mathbf{y}} + z_3\,a\,\hat{\mathbf{z}}$ | (96$g$) | Si III |
| $\mathbf{B}_{13}$ | = | $(2x_3 - z_3)\,\mathbf{a}_1 + \left(\frac{1}{2} - 2x_3 - z_3\right)\,\mathbf{a}_2 + z_3\,\mathbf{a}_3$ | = | $\left(\frac{1}{4} - x_3\right)\,a\,\hat{\mathbf{x}} + x_3\,a\,\hat{\mathbf{y}} + \left(\frac{1}{4} - z_3\right)\,a\,\hat{\mathbf{z}}$ | (96$g$) | Si III |
| $\mathbf{B}_{14}$ | = | $\left(\frac{1}{2} - 2x_3 - z_3\right)\,\mathbf{a}_1 + (2x_3 - z_3)\,\mathbf{a}_2 + z_3\,\mathbf{a}_3$ | = | $x_3\,a\,\hat{\mathbf{x}} + \left(\frac{1}{4} - x_3\right)\,a\,\hat{\mathbf{y}} + \left(\frac{1}{4} - z_3\right)\,a\,\hat{\mathbf{z}}$ | (96$g$) | Si III |
| $\mathbf{B}_{15}$ | = | $(2x_3 - z_3)\,\mathbf{a}_1 + z_3\,\mathbf{a}_2 + z_3\,\mathbf{a}_3$ | = | $z_3\,a\,\hat{\mathbf{x}} + x_3\,a\,\hat{\mathbf{y}} + x_3\,a\,\hat{\mathbf{z}}$ | (96$g$) | Si III |
| $\mathbf{B}_{16}$ | = | $\left(\frac{1}{2} - 2x_3 - z_3\right)\,\mathbf{a}_1 + z_3\,\mathbf{a}_2 + z_3\,\mathbf{a}_3$ | = | $z_3\,a\,\hat{\mathbf{x}} + \left(\frac{1}{4} - x_3\right)\,a\,\hat{\mathbf{y}} + \left(\frac{1}{4} - x_3\right)\,a\,\hat{\mathbf{z}}$ | (96$g$) | Si III |
| $\mathbf{B}_{17}$ | = | $z_3\,\mathbf{a}_1 + (2x_3 - z_3)\,\mathbf{a}_2 + \left(\frac{1}{2} - 2x_3 - z_3\right)\,\mathbf{a}_3$ | = | $\left(\frac{1}{4} - z_3\right)\,a\,\hat{\mathbf{x}} + \left(\frac{1}{4} - x_3\right)\,a\,\hat{\mathbf{y}} + x_3\,a\,\hat{\mathbf{z}}$ | (96$g$) | Si III |
| $\mathbf{B}_{18}$ | = | $z_3\,\mathbf{a}_1 + \left(\frac{1}{2} - 2x_3 - z_3\right)\,\mathbf{a}_2 + (2x_3 - z_3)\,\mathbf{a}_3$ | = | $\left(\frac{1}{4} - z_3\right)\,a\,\hat{\mathbf{x}} + x_3\,a\,\hat{\mathbf{y}} + \left(\frac{1}{4} - x_3\right)\,a\,\hat{\mathbf{z}}$ | (96$g$) | Si III |
| $\mathbf{B}_{19}$ | = | $z_3\,\mathbf{a}_1 + (2x_3 - z_3)\,\mathbf{a}_2 + z_3\,\mathbf{a}_3$ | = | $x_3\,a\,\hat{\mathbf{x}} + z_3\,a\,\hat{\mathbf{y}} + x_3\,a\,\hat{\mathbf{z}}$ | (96$g$) | Si III |
| $\mathbf{B}_{20}$ | = | $z_3\,\mathbf{a}_1 + \left(\frac{1}{2} - 2x_3 - z_3\right)\,\mathbf{a}_2 + z_3\,\mathbf{a}_3$ | = | $\left(\frac{1}{4} - x_3\right)\,a\,\hat{\mathbf{x}} + z_3\,a\,\hat{\mathbf{y}} + \left(\frac{1}{4} - x_3\right)\,a\,\hat{\mathbf{z}}$ | (96$g$) | Si III |
| $\mathbf{B}_{21}$ | = | $\left(\frac{1}{2} - 2x_3 - z_3\right)\,\mathbf{a}_1 + z_3\,\mathbf{a}_2 + (2x_3 - z_3)\,\mathbf{a}_3$ | = | $x_3\,a\,\hat{\mathbf{x}} + \left(\frac{1}{4} - z_3\right)\,a\,\hat{\mathbf{y}} + \left(\frac{1}{4} - x_3\right)\,a\,\hat{\mathbf{z}}$ | (96$g$) | Si III |
| $\mathbf{B}_{22}$ | = | $(2x_3 - z_3)\,\mathbf{a}_1 + z_3\,\mathbf{a}_2 + \left(\frac{1}{2} - 2x_3 - z_3\right)\,\mathbf{a}_3$ | = | $\left(\frac{1}{4} - x_3\right)\,a\,\hat{\mathbf{x}} + \left(\frac{1}{4} - z_3\right)\,a\,\hat{\mathbf{y}} + x_3\,a\,\hat{\mathbf{z}}$ | (96$g$) | Si III |
| $\mathbf{B}_{23}$ | = | $-z_3\,\mathbf{a}_1 - z_3\,\mathbf{a}_2 + \left(\frac{1}{2} + 2x_3 + z_3\right)\,\mathbf{a}_3$ | = | $\left(\frac{1}{4} + x_3\right)\,a\,\hat{\mathbf{x}} + \left(\frac{1}{4} + x_3\right)\,a\,\hat{\mathbf{y}} - z_3\,a\,\hat{\mathbf{z}}$ | (96$g$) | Si III |
| $\mathbf{B_{24}}$ | = | $-z_3\,\mathbf{a_1} - z_3\,\mathbf{a_2} + (z_3 - 2x_3)\,\mathbf{a_3}$ | = | $-x_3\,a\,\mathbf{\hat{x}} - x_3\,a\,\mathbf{\hat{y}} - z_3\,a\,\mathbf{\hat{z}}$ | (96$g$) | Si III |
| $\mathbf{B_{25}}$ | = | $(z_3 - 2x_3)\,\mathbf{a_1} + \left(\frac{1}{2} + 2x_3 + z_3\right)\,\mathbf{a_2} - z_3\,\mathbf{a_3}$ | = | $\left(\frac{1}{4} + x_3\right)\,a\,\mathbf{\hat{x}} - x_3\,a\,\mathbf{\hat{y}} + \left(\frac{1}{4} + z_3\right)\,a\,\mathbf{\hat{z}}$ | (96$g$) | Si III |
| $\mathbf{B_{26}}$ | = | $\left(\frac{1}{2} + 2x_3 + z_3\right)\,\mathbf{a_1} + (z_3 - 2x_3)\,\mathbf{a_2} - z_3\,\mathbf{a_3}$ | = | $-x_3\,a\,\mathbf{\hat{x}} + \left(\frac{1}{4} + x_3\right)\,a\,\mathbf{\hat{y}} + \left(\frac{1}{4} + z_3\right)\,a\,\mathbf{\hat{z}}$ | (96$g$) | Si III |
| $\mathbf{B_{27}}$ | = | $(z_3 - 2x_3)\,\mathbf{a_1} - z_3\,\mathbf{a_2} + \left(\frac{1}{2} + 2x_3 + z_3\right)\,\mathbf{a_3}$ | = | $\left(\frac{1}{4} + x_3\right)\,a\,\mathbf{\hat{x}} + \left(\frac{1}{4} + z_3\right)\,a\,\mathbf{\hat{y}} - x_3\,a\,\mathbf{\hat{z}}$ | (96$g$) | Si III |
| $\mathbf{B_{28}}$ | = | $\left(\frac{1}{2} + 2x_3 + z_3\right)\,\mathbf{a_1} - z_3\,\mathbf{a_2} + (z_3 - 2x_3)\,\mathbf{a_3}$ | = | $-x_3\,a\,\mathbf{\hat{x}} + \left(\frac{1}{4} + z_3\right)\,a\,\mathbf{\hat{y}} + \left(\frac{1}{4} + x_3\right)\,a\,\mathbf{\hat{z}}$ | (96$g$) | Si III |
| $\mathbf{B_{29}}$ | = | $-z_3\,\mathbf{a_1} + (z_3 - 2x_3)\,\mathbf{a_2} - z_3\,\mathbf{a_3}$ | = | $-x_3\,a\,\mathbf{\hat{x}} - z_3\,a\,\mathbf{\hat{y}} - x_3\,a\,\mathbf{\hat{z}}$ | (96$g$) | Si III |
| $\mathbf{B_{30}}$ | = | $-z_3\,\mathbf{a_1} + \left(\frac{1}{2} + 2x_3 + z_3\right)\,\mathbf{a_2} - z_3\,\mathbf{a_3}$ | = | $\left(\frac{1}{4} + x_3\right)\,a\,\mathbf{\hat{x}} - z_3\,a\,\mathbf{\hat{y}} + \left(\frac{1}{4} + x_3\right)\,a\,\mathbf{\hat{z}}$ | (96$g$) | Si III |
| $\mathbf{B_{31}}$ | = | $-z_3\,\mathbf{a_1} + (z_3 - 2x_3)\,\mathbf{a_2} + \left(\frac{1}{2} + 2x_3 + z_3\right)\,\mathbf{a_3}$ | = | $\left(\frac{1}{4} + z_3\right)\,a\,\mathbf{\hat{x}} + \left(\frac{1}{4} + x_3\right)\,a\,\mathbf{\hat{y}} - x_3\,a\,\mathbf{\hat{z}}$ | (96$g$) | Si III |
| $\mathbf{B_{32}}$ | = | $-z_3\,\mathbf{a_1} + \left(\frac{1}{2} + 2x_3 + z_3\right)\,\mathbf{a_2} + (z_3 - 2x_3)\,\mathbf{a_3}$ | = | $\left(\frac{1}{4} + z_3\right)\,a\,\mathbf{\hat{x}} - x_3\,a\,\mathbf{\hat{y}} + \left(\frac{1}{4} + x_3\right)\,a\,\mathbf{\hat{z}}$ | (96$g$) | Si III |
| $\mathbf{B_{33}}$ | = | $\left(\frac{1}{2} + 2x_3 + z_3\right)\,\mathbf{a_1} - z_3\,\mathbf{a_2} - z_3\,\mathbf{a_3}$ | = | $-z_3\,a\,\mathbf{\hat{x}} + \left(\frac{1}{4} + x_3\right)\,a\,\mathbf{\hat{y}} + \left(\frac{1}{4} + x_3\right)\,a\,\mathbf{\hat{z}}$ | (96$g$) | Si III |
| $\mathbf{B_{34}}$ | = | $(z_3 - 2x_3)\,\mathbf{a_1} - z_3\,\mathbf{a_2} - z_3\,\mathbf{a_3}$ | = | $-z_3\,a\,\mathbf{\hat{x}} - x_3\,a\,\mathbf{\hat{y}} - x_3\,a\,\mathbf{\hat{z}}$ | (96$g$) | Si III |

---

---

**Geometry files:**

- CIF: pp. 796
- POSCAR: pp. 798

# $Cu_2Mg$ Cubic Laves Structure (C15): A2B_cF24_227_d_a

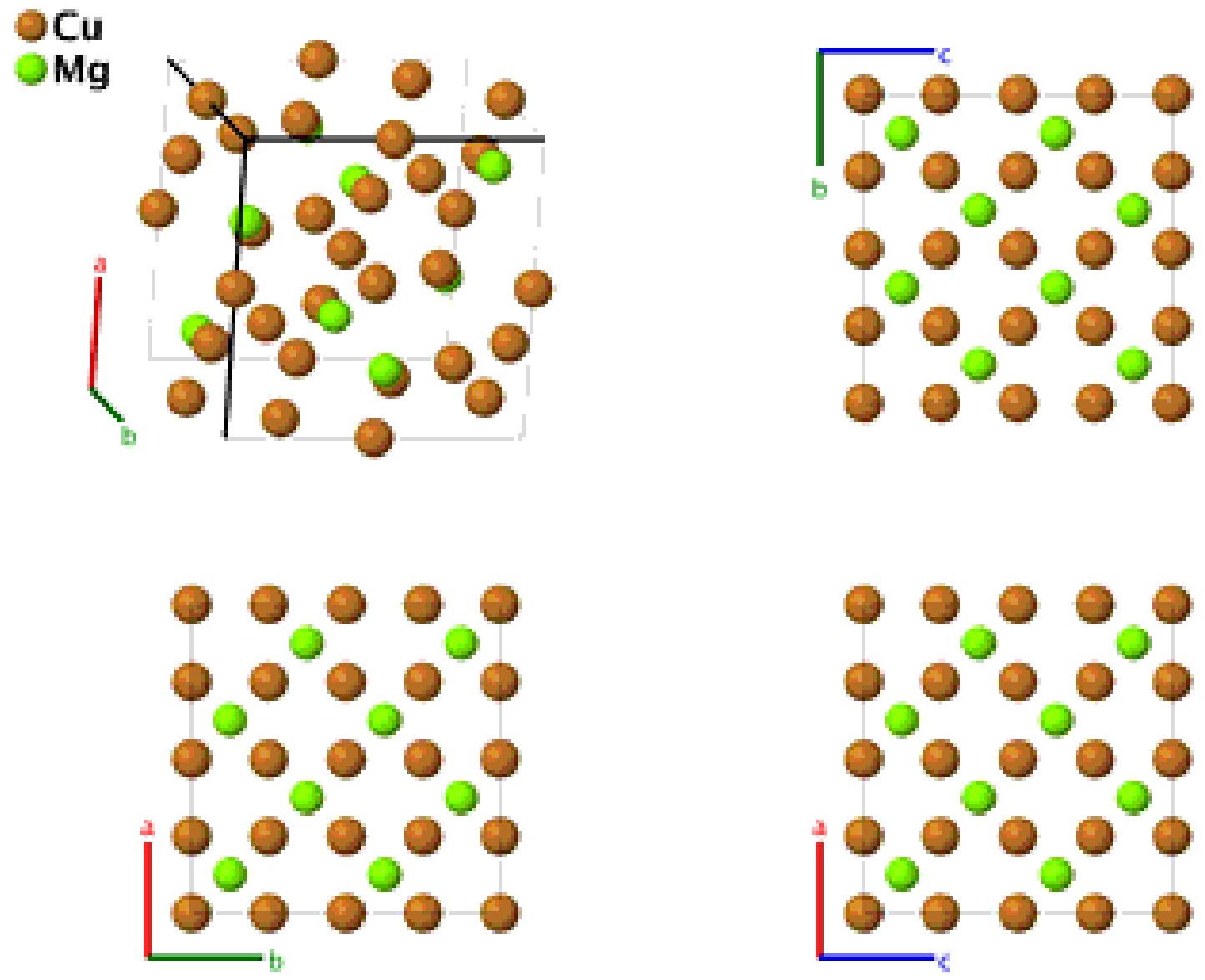

| | | |
|---|---|---|
| **Prototype** | : | $Cu_2Mg$ |
| **AFLOW prototype label** | : | A2B_cF24_227_d_a |
| ***Strukturbericht* designation** | : | C15 |
| **Pearson symbol** | : | cF24 |
| **Space group number** | : | 227 |
| **Space group symbol** | : | $Fd\bar{3}m$ |
| **AFLOW prototype command** | : | `aflow --proto=A2B_cF24_227_d_a`<br>`--params=`$a$ |

**Other compounds with this structure:**

- $CsBi_2$, $RbBi_2$

**Face-centered Cubic primitive vectors:**

$$\mathbf{a}_1 = \frac{1}{2}\,a\,\hat{\mathbf{y}} + \frac{1}{2}\,a\,\hat{\mathbf{z}}$$
$$\mathbf{a}_2 = \frac{1}{2}\,a\,\hat{\mathbf{x}} + \frac{1}{2}\,a\,\hat{\mathbf{z}}$$
$$\mathbf{a}_3 = \frac{1}{2}\,a\,\hat{\mathbf{x}} + \frac{1}{2}\,a\,\hat{\mathbf{y}}$$

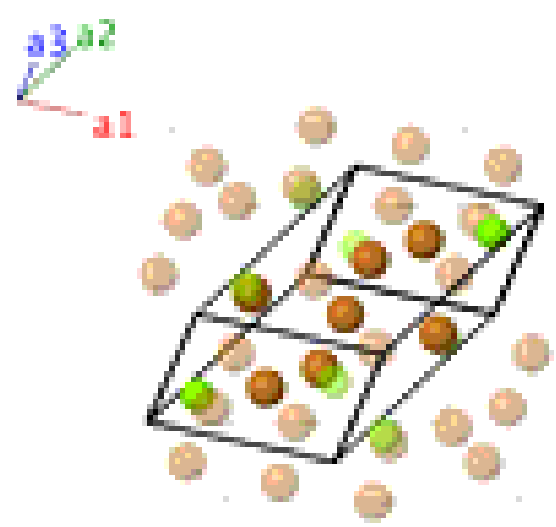

**Basis vectors:**

| | | Lattice Coordinates | | Cartesian Coordinates | Wyckoff Position | Atom Type |
|---|---|---|---|---|---|---|
| $\mathbf{B}_1$ | = | $\frac{1}{8}\,\mathbf{a}_1 + \frac{1}{8}\,\mathbf{a}_2 + \frac{1}{8}\,\mathbf{a}_3$ | = | $\frac{1}{8}\,a\,\hat{\mathbf{x}} + \frac{1}{8}\,a\,\hat{\mathbf{y}} + \frac{1}{8}\,a\,\hat{\mathbf{z}}$ | (8$a$) | Mg |
| $\mathbf{B}_2$ | = | $\frac{7}{8}\,\mathbf{a}_1 + \frac{7}{8}\,\mathbf{a}_2 + \frac{7}{8}\,\mathbf{a}_3$ | = | $\frac{7}{8}\,a\,\hat{\mathbf{x}} + \frac{7}{8}\,a\,\hat{\mathbf{y}} + \frac{7}{8}\,a\,\hat{\mathbf{z}}$ | (8$a$) | Mg |
| $\mathbf{B}_3$ | = | $\frac{1}{2}\,\mathbf{a}_1 + \frac{1}{2}\,\mathbf{a}_2 + \frac{1}{2}\,\mathbf{a}_3$ | = | $\frac{1}{2}\,a\,\hat{\mathbf{x}} + \frac{1}{2}\,a\,\hat{\mathbf{y}} + \frac{1}{2}\,a\,\hat{\mathbf{z}}$ | (16$d$) | Cu |
| $\mathbf{B_4}$ | = | $\frac{1}{2}\,\mathbf{a_1} + \frac{1}{2}\,\mathbf{a_2}$ | = | $\frac{1}{4}\,a\,\hat{\mathbf{x}} + \frac{1}{4}\,a\,\hat{\mathbf{y}} + \frac{1}{2}\,a\,\hat{\mathbf{z}}$ | (16$d$) | Cu |
| $\mathbf{B_5}$ | = | $\frac{1}{2}\,\mathbf{a_1} + \frac{1}{2}\,\mathbf{a_3}$ | = | $\frac{1}{4}\,a\,\hat{\mathbf{x}} + \frac{1}{2}\,a\,\hat{\mathbf{y}} + \frac{1}{4}\,a\,\hat{\mathbf{z}}$ | (16$d$) | Cu |
| $\mathbf{B_6}$ | = | $\frac{1}{2}\,\mathbf{a_2} + \frac{1}{2}\,\mathbf{a_3}$ | = | $\frac{1}{2}\,a\,\hat{\mathbf{x}} + \frac{1}{4}\,a\,\hat{\mathbf{y}} + \frac{1}{4}\,a\,\hat{\mathbf{z}}$ | (16$d$) | Cu |

**Geometry files:**

- CIF: pp. 798
- POSCAR: pp. 799

# Diamond (A4) Structure: A_cF8_227_a

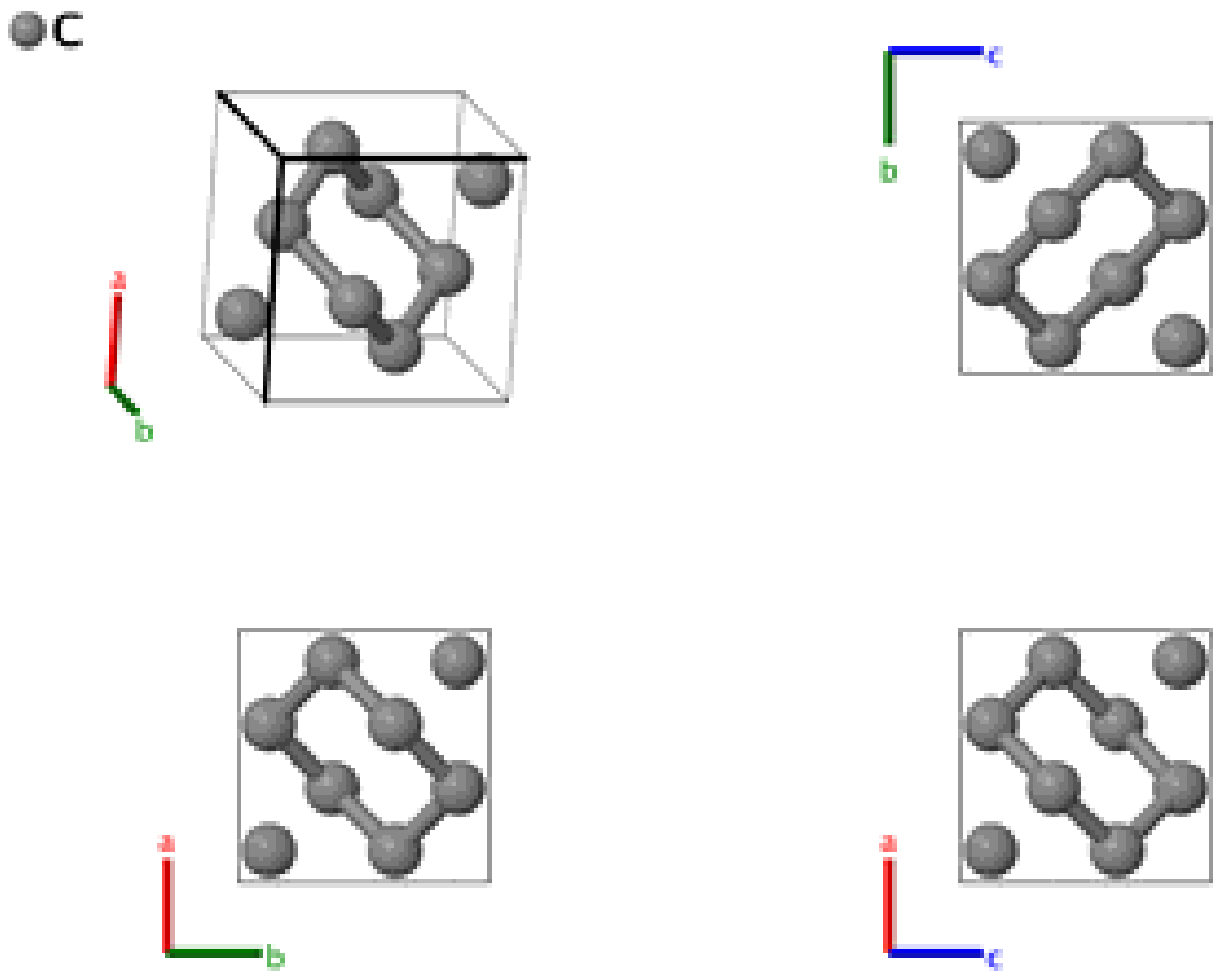

| | | |
|---|---|---|
| **Prototype** | : | C |
| **AFLOW prototype label** | : | A_cF8_227_a |
| ***Strukturbericht* designation** | : | A4 |
| **Pearson symbol** | : | cF8 |
| **Space group number** | : | 227 |
| **Space group symbol** | : | $Fd\bar{3}m$ |
| **AFLOW prototype command** | : | `aflow --proto=A_cF8_227_a --params=`$a$ |

**Other elements with this structure:**

- Si, Ge, Sn

- This is the first crystal structure to be determined by X-ray diffraction.

**Face-centered Cubic primitive vectors:**

$$\mathbf{a}_1 = \frac{1}{2}\,a\,\mathbf{\hat{y}} + \frac{1}{2}\,a\,\mathbf{\hat{z}}$$
$$\mathbf{a}_2 = \frac{1}{2}\,a\,\mathbf{\hat{x}} + \frac{1}{2}\,a\,\mathbf{\hat{z}}$$
$$\mathbf{a}_3 = \frac{1}{2}\,a\,\mathbf{\hat{x}} + \frac{1}{2}\,a\,\mathbf{\hat{y}}$$

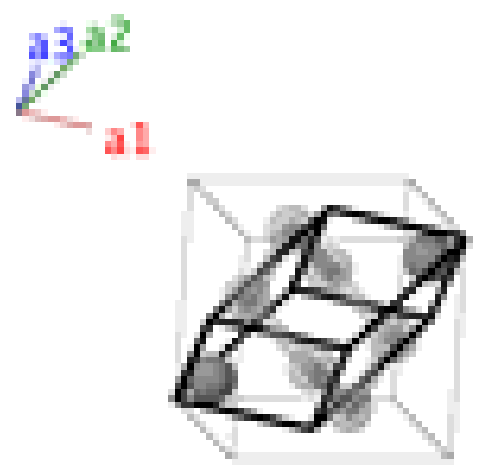

**Basis vectors:**

| | | Lattice Coordinates | | Cartesian Coordinates | Wyckoff Position | Atom Type |
|---|---|---|---|---|---|---|
| $\mathbf{B}_1$ | = | $\frac{1}{8}\,\mathbf{a}_1 + \frac{1}{8}\,\mathbf{a}_2 + \frac{1}{8}\,\mathbf{a}_3$ | = | $\frac{1}{8}\,a\,\mathbf{\hat{x}} + \frac{1}{8}\,a\,\mathbf{\hat{y}} + \frac{1}{8}\,a\,\mathbf{\hat{z}}$ | (8$a$) | C |
| $\mathbf{B}_2$ | = | $\frac{7}{8}\,\mathbf{a}_1 + \frac{7}{8}\,\mathbf{a}_2 + \frac{7}{8}\,\mathbf{a}_3$ | = | $\frac{7}{8}\,a\,\mathbf{\hat{x}} + \frac{7}{8}\,a\,\mathbf{\hat{y}} + \frac{7}{8}\,a\,\mathbf{\hat{z}}$ | (8$a$) | C |

**Geometry files:**

- CIF: pp. 799
- POSCAR: pp. 800

# Spinel ($Al_2MgO_4$, $H1_1$) Structure: A2BC4_cF56_227_d_a_e

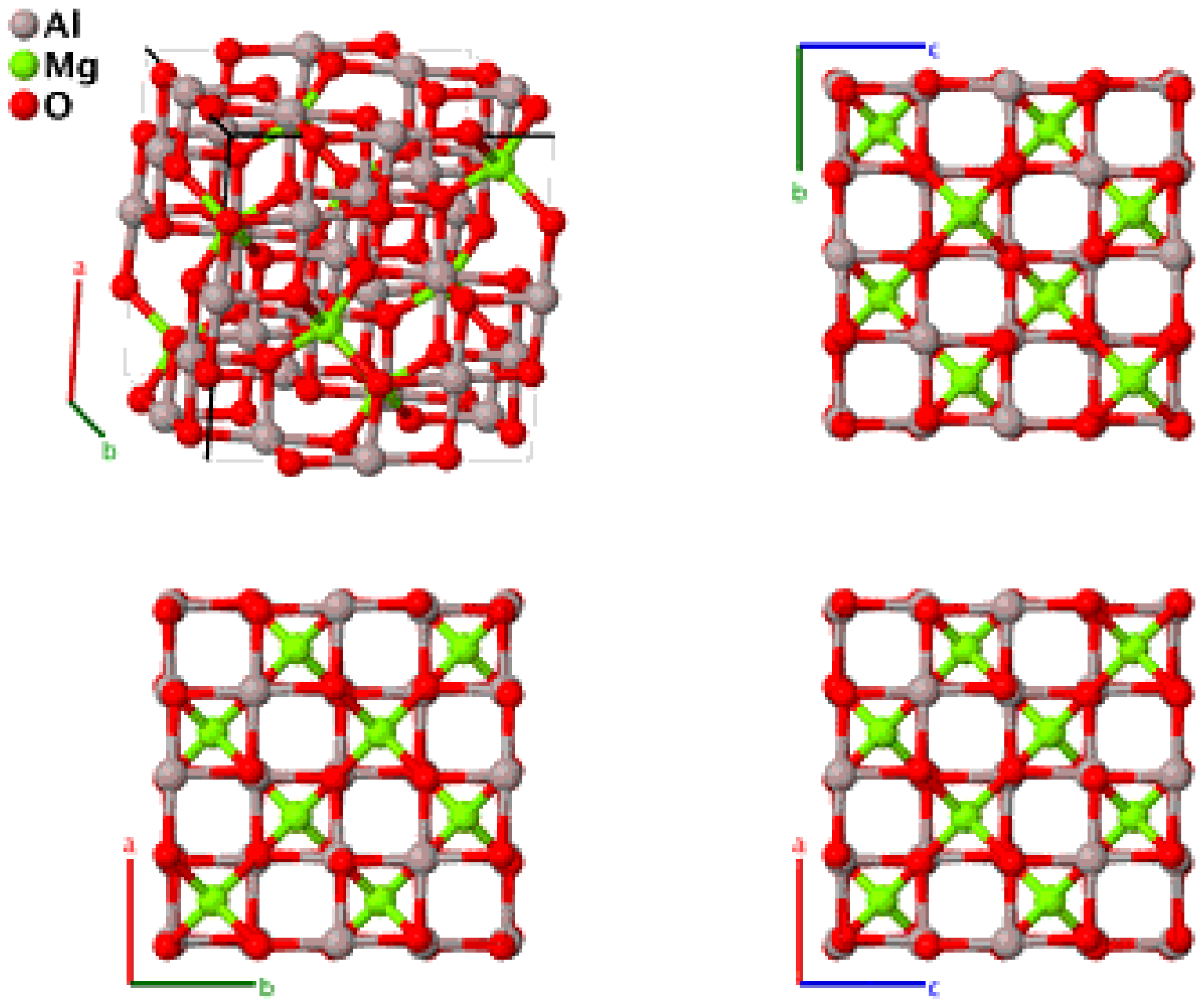

| | | |
|---|---|---|
| **Prototype** | : | $Al_2MgO_4$ |
| **AFLOW prototype label** | : | A2BC4_cF56_227_d_a_e |
| ***Strukturbericht* designation** | : | $H1_1$ |
| **Pearson symbol** | : | cF56 |
| **Space group number** | : | 227 |
| **Space group symbol** | : | $Fd\bar{3}m$ |
| **AFLOW prototype command** | : | `aflow --proto=A2BC4_cF56_227_d_a_e --params=`$a, x_3$ |

**Other compounds with this structure:**

- $Al_2Se_4Zn$, $Al_2CrS_4$, $CaIn_2S_4$, $Al_2CdS_4$, $Cr_2Se_4Zr$, $Mn_2Te_4Zn$, many others.

- An *inverse spinel* has four Al atoms on the (8a) sites and (Al,Mg) alloyed on the (16d) sites.

**Face-centered Cubic primitive vectors:**

$$\mathbf{a}_1 = \frac{1}{2}\,a\,\hat{\mathbf{y}} + \frac{1}{2}\,a\,\hat{\mathbf{z}}$$
$$\mathbf{a}_2 = \frac{1}{2}\,a\,\hat{\mathbf{x}} + \frac{1}{2}\,a\,\hat{\mathbf{z}}$$
$$\mathbf{a}_3 = \frac{1}{2}\,a\,\hat{\mathbf{x}} + \frac{1}{2}\,a\,\hat{\mathbf{y}}$$

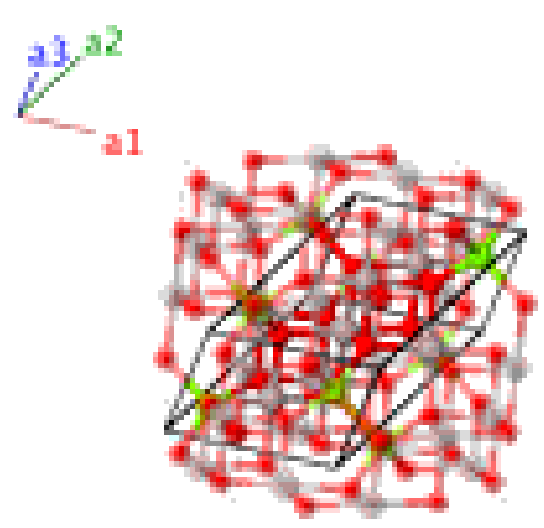

**Basis vectors:**

| | | Lattice Coordinates | | Cartesian Coordinates | Wyckoff Position | Atom Type |
|---|---|---|---|---|---|---|
| $\mathbf{B_1}$ | = | $\frac{1}{8}\mathbf{a_1} + \frac{1}{8}\mathbf{a_2} + \frac{1}{8}\mathbf{a_3}$ | = | $\frac{1}{8}\,a\,\mathbf{\hat{x}} + \frac{1}{8}\,a\,\mathbf{\hat{y}} + \frac{1}{8}\,a\,\mathbf{\hat{z}}$ | $(8a)$ | Mg |
| $\mathbf{B_2}$ | = | $\frac{7}{8}\mathbf{a_1} + \frac{7}{8}\mathbf{a_2} + \frac{7}{8}\mathbf{a_3}$ | = | $\frac{7}{8}\,a\,\mathbf{\hat{x}} + \frac{7}{8}\,a\,\mathbf{\hat{y}} + \frac{7}{8}\,a\,\mathbf{\hat{z}}$ | $(8a)$ | Mg |
| $\mathbf{B_3}$ | = | $\frac{1}{2}\mathbf{a_1} + \frac{1}{2}\mathbf{a_2} + \frac{1}{2}\mathbf{a_3}$ | = | $\frac{1}{2}\,a\,\mathbf{\hat{x}} + \frac{1}{2}\,a\,\mathbf{\hat{y}} + \frac{1}{2}\,a\,\mathbf{\hat{z}}$ | $(16d)$ | Al |
| $\mathbf{B_4}$ | = | $\frac{1}{2}\mathbf{a_1} + \frac{1}{2}\mathbf{a_2}$ | = | $\frac{1}{4}\,a\,\mathbf{\hat{x}} + \frac{1}{4}\,a\,\mathbf{\hat{y}} + \frac{1}{2}\,a\,\mathbf{\hat{z}}$ | $(16d)$ | Al |
| $\mathbf{B_5}$ | = | $\frac{1}{2}\mathbf{a_1} + \frac{1}{2}\mathbf{a_3}$ | = | $\frac{1}{4}\,a\,\mathbf{\hat{x}} + \frac{1}{2}\,a\,\mathbf{\hat{y}} + \frac{1}{4}\,a\,\mathbf{\hat{z}}$ | $(16d)$ | Al |
| $\mathbf{B_6}$ | = | $\frac{1}{2}\mathbf{a_2} + \frac{1}{2}\mathbf{a_3}$ | = | $\frac{1}{2}\,a\,\mathbf{\hat{x}} + \frac{1}{4}\,a\,\mathbf{\hat{y}} + \frac{1}{4}\,a\,\mathbf{\hat{z}}$ | $(16d)$ | Al |
| $\mathbf{B_7}$ | = | $x_3\,\mathbf{a_1} + x_3\,\mathbf{a_2} + x_3\,\mathbf{a_3}$ | = | $x_3\,a\,\mathbf{\hat{x}} + x_3\,a\,\mathbf{\hat{y}} + x_3\,a\,\mathbf{\hat{z}}$ | $(32e)$ | O |
| $\mathbf{B_8}$ | = | $x_3\,\mathbf{a_1} + (1 + x_3)\,\mathbf{a_2} + \left(\frac{1}{2} - 3x_3\right)\mathbf{a_3}$ | = | $\left(\frac{3}{4} - x_3\right)a\,\mathbf{\hat{x}} + \left(\frac{1}{4} - x_3\right)a\,\mathbf{\hat{y}} + \left(\frac{1}{2} + x_3\right)a\,\mathbf{\hat{z}}$ | $(32e)$ | O |
| $\mathbf{B_9}$ | = | $(1 + x_3)\,\mathbf{a_1} + \left(\frac{1}{2} - 3x_3\right)\mathbf{a_2} + x_3\,\mathbf{a_3}$ | = | $\left(\frac{1}{4} - x_3\right)a\,\mathbf{\hat{x}} + \left(\frac{1}{2} + x_3\right)a\,\mathbf{\hat{y}} + \left(\frac{3}{4} - x_3\right)a\,\mathbf{\hat{z}}$ | $(32e)$ | O |
| $\mathbf{B_{10}}$ | = | $\left(\frac{1}{2} - 3x_3\right)\mathbf{a_1} + x_3\,\mathbf{a_2} + (1 + x_3)\,\mathbf{a_3}$ | = | $\left(\frac{1}{2} + x_3\right)a\,\mathbf{\hat{x}} + \left(\frac{3}{4} - x_3\right)a\,\mathbf{\hat{y}} + \left(\frac{1}{4} - x_3\right)a\,\mathbf{\hat{z}}$ | $(32e)$ | O |
| $\mathbf{B_{11}}$ | = | $-x_3\,\mathbf{a_1} - x_3\,\mathbf{a_2} - x_3\,\mathbf{a_3}$ | = | $-x_3\,a\,\mathbf{\hat{x}} - x_3\,a\,\mathbf{\hat{y}} - x_3\,a\,\mathbf{\hat{z}}$ | $(32e)$ | O |
| $\mathbf{B_{12}}$ | = | $-x_3\,\mathbf{a_1} + (1 - x_3)\,\mathbf{a_2} + \left(\frac{1}{2} + 3x_3\right)\mathbf{a_3}$ | = | $\left(\frac{3}{4} + x_3\right)a\,\mathbf{\hat{x}} + \left(\frac{1}{4} + x_3\right)a\,\mathbf{\hat{y}} + \left(\frac{1}{2} - x_3\right)a\,\mathbf{\hat{z}}$ | $(32e)$ | O |
| $\mathbf{B_{13}}$ | = | $(1 - x_3)\,\mathbf{a_1} + \left(\frac{1}{2} + 3x_3\right)\mathbf{a_2} - x_3\,\mathbf{a_3}$ | = | $\left(\frac{1}{4} + x_3\right)a\,\mathbf{\hat{x}} + \left(\frac{1}{2} - x_3\right)a\,\mathbf{\hat{y}} + \left(\frac{3}{4} + x_3\right)a\,\mathbf{\hat{z}}$ | $(32e)$ | O |
| $\mathbf{B_{14}}$ | = | $\left(\frac{1}{2} + 3x_3\right)\mathbf{a_1} - x_3\,\mathbf{a_2} + (1 - x_3)\,\mathbf{a_3}$ | = | $\left(\frac{1}{2} - x_3\right)a\,\mathbf{\hat{x}} + \left(\frac{3}{4} + x_3\right)a\,\mathbf{\hat{y}} + \left(\frac{1}{4} + x_3\right)a\,\mathbf{\hat{z}}$ | $(32e)$ | O |

**Geometry files:**

- CIF: pp. 800
- POSCAR: pp. 802

# $CTi_2$ Structure: AB2_cF48_227_c_e

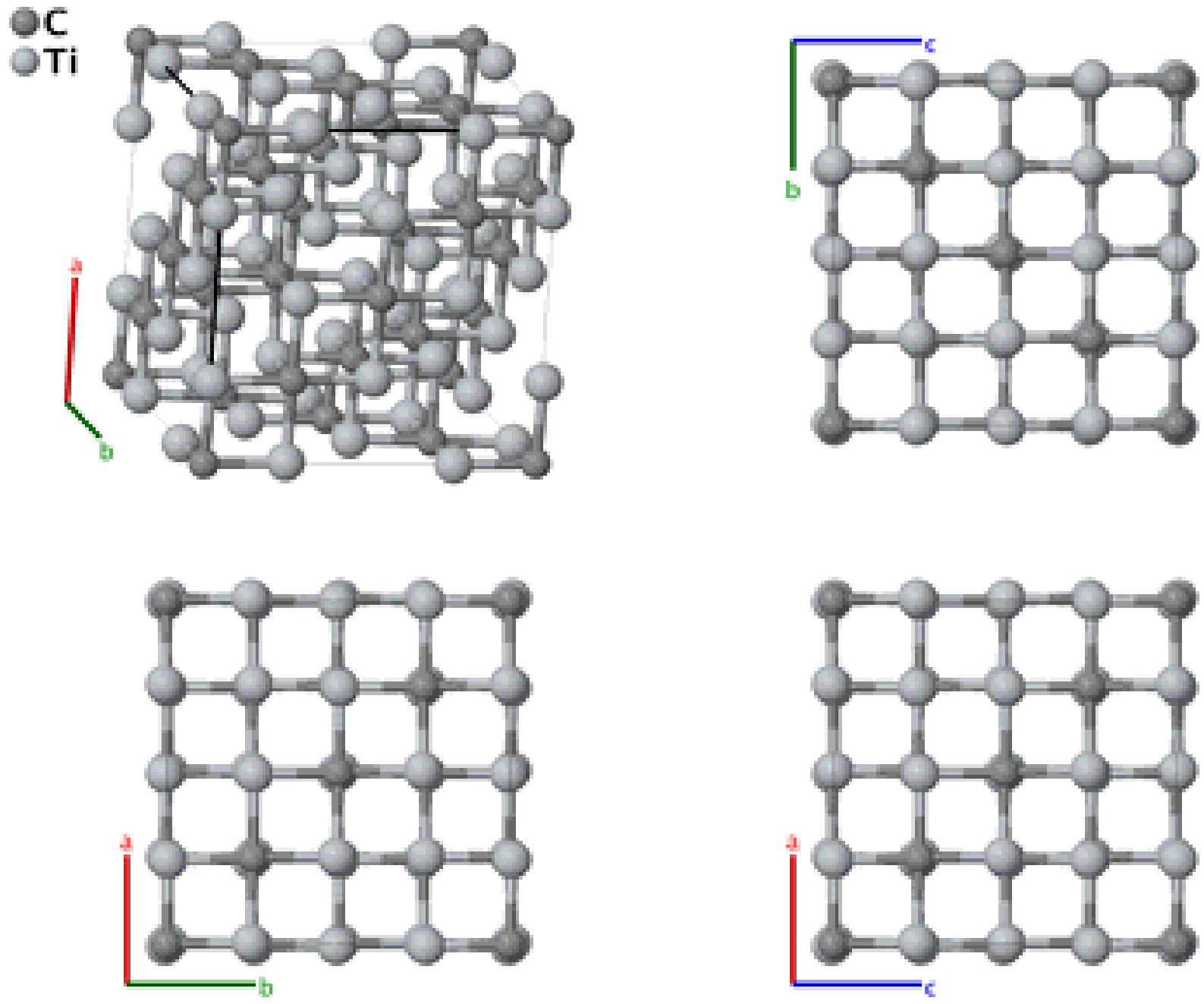

**Prototype** : $CTi_2$

**AFLOW prototype label** : AB2_cF48_227_c_e

***Strukturbericht* designation** : None

**Pearson symbol** : cF48

**Space group number** : 227

**Space group symbol** : $Fd\bar{3}m$

**AFLOW prototype command** : `aflow --proto=AB2_cF48_227_c_e`
`--params=`$a, x_2$

**Other compounds with this structure:**

- $Ca_{33}Ge$

- Some sources consider the real prototype of this system to be $Ca_{33}Ge$, with the (32e) sites occupied by calcium atoms and the (16c) sites randomly occupied by calcium and germanium atoms.

**Face-centered Cubic primitive vectors:**

$$\mathbf{a}_1 = \frac{1}{2}\,a\,\mathbf{\hat{y}} + \frac{1}{2}\,a\,\mathbf{\hat{z}}$$
$$\mathbf{a}_2 = \frac{1}{2}\,a\,\mathbf{\hat{x}} + \frac{1}{2}\,a\,\mathbf{\hat{z}}$$
$$\mathbf{a}_3 = \frac{1}{2}\,a\,\mathbf{\hat{x}} + \frac{1}{2}\,a\,\mathbf{\hat{y}}$$

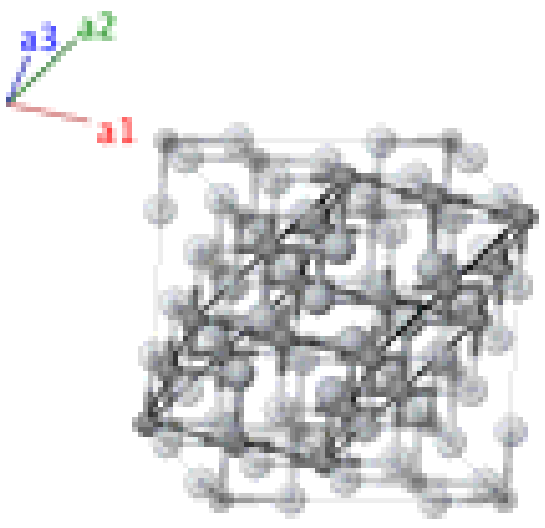

**Basis vectors:**

| | | Lattice Coordinates | | Cartesian Coordinates | Wyckoff Position | Atom Type |
|---|---|---|---|---|---|---|
| $\mathbf{B}_1$ | = | $0\,\mathbf{a}_1 + 0\,\mathbf{a}_2 + 0\,\mathbf{a}_3$ | = | $0\hat{\mathbf{x}} + 0\hat{\mathbf{y}} + 0\hat{\mathbf{z}}$ | (16$c$) | C |
| $\mathbf{B}_2$ | = | $\frac{1}{2}\mathbf{a}_3$ | = | $\frac{1}{4}\,a\,\hat{\mathbf{x}} + \frac{1}{4}\,a\,\hat{\mathbf{y}}$ | (16$c$) | C |
| $\mathbf{B}_3$ | = | $\frac{1}{2}\mathbf{a}_2$ | = | $\frac{1}{4}\,a\,\hat{\mathbf{x}} + \frac{1}{4}\,a\,\hat{\mathbf{z}}$ | (16$c$) | C |
| $\mathbf{B}_4$ | = | $\frac{1}{2}\mathbf{a}_1$ | = | $\frac{1}{4}\,a\,\hat{\mathbf{y}} + \frac{1}{4}\,a\,\hat{\mathbf{z}}$ | (16$c$) | C |
| $\mathbf{B}_5$ | = | $x_2\mathbf{a}_1 + x_2\mathbf{a}_2 + x_2\mathbf{a}_3$ | = | $x_2\,a\,\hat{\mathbf{x}} + x_2\,a\,\hat{\mathbf{y}} + x_2\,a\,\hat{\mathbf{z}}$ | (32$e$) | Ti |
| $\mathbf{B}_6$ | = | $x_2\mathbf{a}_1 + x_2\mathbf{a}_2 + \left(\frac{1}{2} - 3\,x_2\right)\mathbf{a}_3$ | = | $\left(\frac{1}{4} - x_2\right)\,a\,\hat{\mathbf{x}} + \left(\frac{1}{4} - x_2\right)\,a\,\hat{\mathbf{y}} + x_2\,a\,\hat{\mathbf{z}}$ | (32$e$) | Ti |
| $\mathbf{B}_7$ | = | $x_2\mathbf{a}_1 + \left(\frac{1}{2} - 3\,x_2\right)\mathbf{a}_2 + x_2\mathbf{a}_3$ | = | $\left(\frac{1}{4} - x_2\right)\,a\,\hat{\mathbf{x}} + x_2\,a\,\hat{\mathbf{y}} + \left(\frac{1}{4} - x_2\right)\,a\,\hat{\mathbf{z}}$ | (32$e$) | Ti |
| $\mathbf{B}_8$ | = | $\left(\frac{1}{2} - 3\,x_2\right)\mathbf{a}_1 + x_2\mathbf{a}_2 + x_2\mathbf{a}_3$ | = | $x_2\,a\,\hat{\mathbf{x}} + \left(\frac{1}{4} - x_2\right)\,a\,\hat{\mathbf{y}} + \left(\frac{1}{4} - x_2\right)\,a\,\hat{\mathbf{z}}$ | (32$e$) | Ti |
| $\mathbf{B}_9$ | = | $-x_2\mathbf{a}_1 - x_2\mathbf{a}_2 + \left(\frac{1}{2} + 3\,x_2\right)\mathbf{a}_3$ | = | $\left(\frac{1}{4} + x_2\right)\,a\,\hat{\mathbf{x}} + \left(\frac{1}{4} + x_2\right)\,a\,\hat{\mathbf{y}} - x_2\,a\,\hat{\mathbf{z}}$ | (32$e$) | Ti |
| $\mathbf{B}_{10}$ | = | $-x_2\mathbf{a}_1 - x_2\mathbf{a}_2 - x_2\mathbf{a}_3$ | = | $-x_2\,a\,\hat{\mathbf{x}} - x_2\,a\,\hat{\mathbf{y}} - x_2\,a\,\hat{\mathbf{z}}$ | (32$e$) | Ti |
| $\mathbf{B}_{11}$ | = | $-x_2\mathbf{a}_1 + \left(\frac{1}{2} + 3\,x_2\right)\mathbf{a}_2 - x_2\mathbf{a}_3$ | = | $\left(\frac{1}{4} + x_2\right)\,a\,\hat{\mathbf{x}} - x_2\,a\,\hat{\mathbf{y}} + \left(\frac{1}{4} + x_2\right)\,a\,\hat{\mathbf{z}}$ | (32$e$) | Ti |
| $\mathbf{B}_{12}$ | = | $\left(\frac{1}{2} + 3\,x_2\right)\mathbf{a}_1 - x_2\mathbf{a}_2 - x_2\mathbf{a}_3$ | = | $-x_2\,a\,\hat{\mathbf{x}} + \left(\frac{1}{4} + x_2\right)\,a\,\hat{\mathbf{y}} + \left(\frac{1}{4} + x_2\right)\,a\,\hat{\mathbf{z}}$ | (32$e$) | Ti |

**Found in:**

**Geometry files:**

- CIF: pp. 802
- POSCAR: pp. 803

# $Fe_3W_3C$ Structure: AB3C3_cF112_227_c_de_f

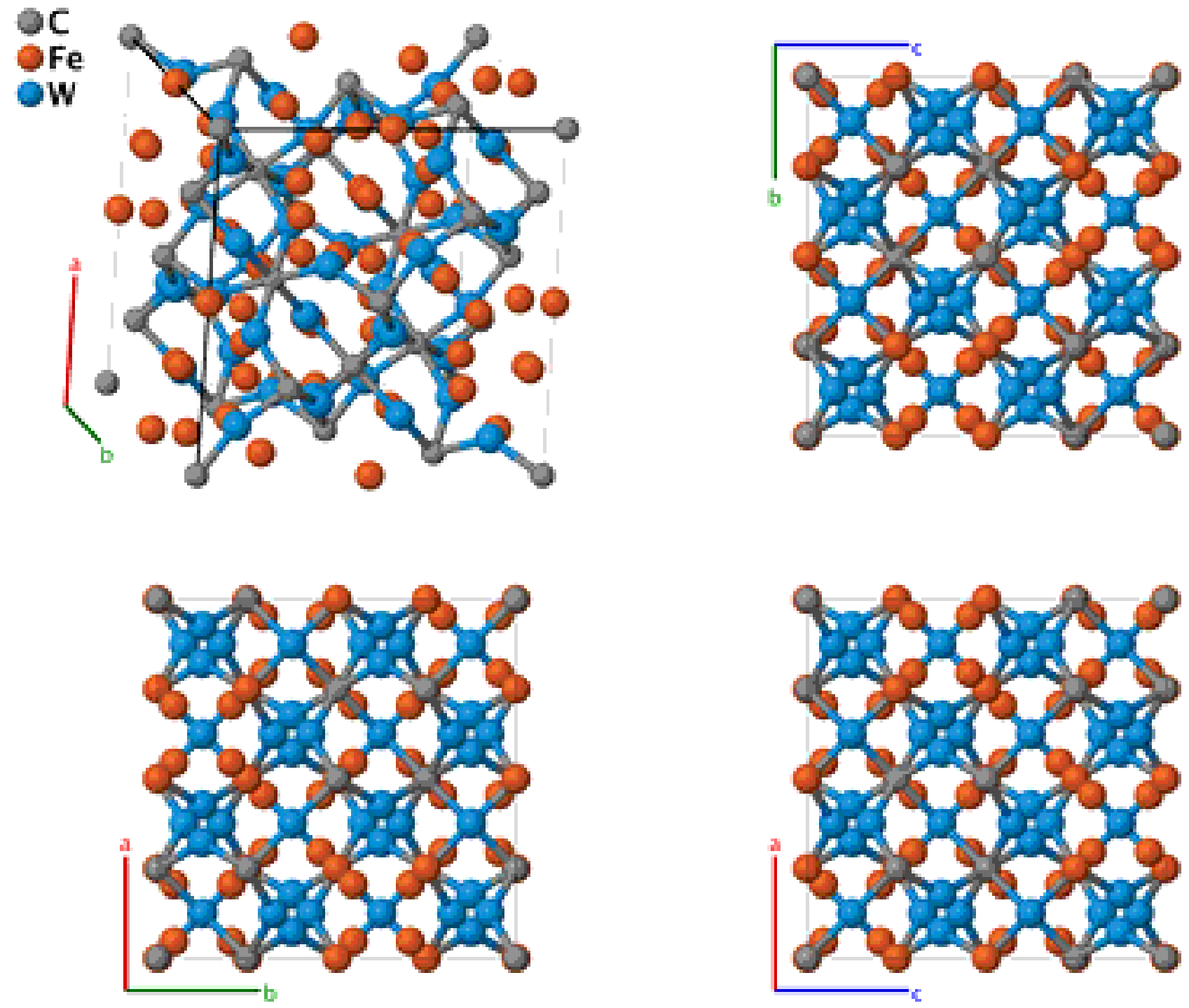

| | | |
|---|---|---|
| **Prototype** | : | $Fe_3W_3C$ |
| **AFLOW prototype label** | : | AB3C3_cF112_227_c_de_f |
| ***Strukturbericht* designation** | : | $E9_3$ |
| **Pearson symbol** | : | cF112 |
| **Space group number** | : | 227 |
| **Space group symbol** | : | $Fd\bar{3}m$ |
| **AFLOW prototype command** | : | `aflow --proto=AB3C3_cF112_227_c_de_f --params=`$a, x_3, x_4$ |

- Experimentally, the (48f) site is a random mixture of composition $W_{2/3}Fe_{1/3}$. We use W for this site in the pictures above.

**Face-centered Cubic primitive vectors:**

$$\mathbf{a}_1 = \frac{1}{2}\,a\,\mathbf{\hat{y}} + \frac{1}{2}\,a\,\mathbf{\hat{z}}$$
$$\mathbf{a}_2 = \frac{1}{2}\,a\,\mathbf{\hat{x}} + \frac{1}{2}\,a\,\mathbf{\hat{z}}$$
$$\mathbf{a}_3 = \frac{1}{2}\,a\,\mathbf{\hat{x}} + \frac{1}{2}\,a\,\mathbf{\hat{y}}$$

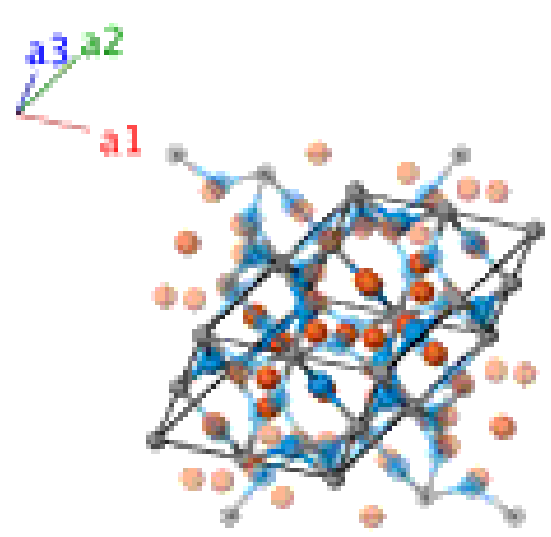

**Basis vectors:**

Lattice Coordinates | Cartesian Coordinates | Wyckoff Position | Atom Type

| | | | | | | |
|---|---|---|---|---|---|---|
| $\mathbf{B}_{1}$ | = | $0\,\mathbf{a}_{1}+0\,\mathbf{a}_{2}+0\,\mathbf{a}_{3}$ | = | $0\hat{\mathbf{x}}+0\hat{\mathbf{y}}+0\hat{\mathbf{z}}$ | (16$c$) | C |
| $\mathbf{B}_{2}$ | = | $\frac{1}{2}\mathbf{a}_{3}$ | = | $\frac{1}{4}\,a\,\hat{\mathbf{x}}+\frac{1}{4}\,a\,\hat{\mathbf{y}}$ | (16$c$) | C |
| $\mathbf{B}_{3}$ | = | $\frac{1}{2}\mathbf{a}_{2}$ | = | $\frac{1}{4}\,a\,\hat{\mathbf{x}}+\frac{1}{4}\,a\,\hat{\mathbf{z}}$ | (16$c$) | C |
| $\mathbf{B}_{4}$ | = | $\frac{1}{2}\mathbf{a}_{1}$ | = | $\frac{1}{4}\,a\,\hat{\mathbf{y}}+\frac{1}{4}\,a\,\hat{\mathbf{z}}$ | (16$c$) | C |
| $\mathbf{B}_{5}$ | = | $\frac{1}{2}\,\mathbf{a}_{1}+\frac{1}{2}\,\mathbf{a}_{2}+\frac{1}{2}\,\mathbf{a}_{3}$ | = | $\frac{1}{2}\,a\,\hat{\mathbf{x}}+\frac{1}{2}\,a\,\hat{\mathbf{y}}+\frac{1}{2}\,a\,\hat{\mathbf{z}}$ | (16$d$) | Fe I |
| $\mathbf{B}_{6}$ | = | $\frac{1}{2}\,\mathbf{a}_{1}+\frac{1}{2}\,\mathbf{a}_{2}$ | = | $\frac{1}{4}\,a\,\hat{\mathbf{x}}+\frac{1}{4}\,a\,\hat{\mathbf{y}}+\frac{1}{2}\,a\,\hat{\mathbf{z}}$ | (16$d$) | Fe I |
| $\mathbf{B}_{7}$ | = | $\frac{1}{2}\,\mathbf{a}_{1}+\frac{1}{2}\,\mathbf{a}_{3}$ | = | $\frac{1}{4}\,a\,\hat{\mathbf{x}}+\frac{1}{2}\,a\,\hat{\mathbf{y}}+\frac{1}{4}\,a\,\hat{\mathbf{z}}$ | (16$d$) | Fe I |
| $\mathbf{B}_{8}$ | = | $\frac{1}{2}\,\mathbf{a}_{2}+\frac{1}{2}\,\mathbf{a}_{3}$ | = | $\frac{1}{2}\,a\,\hat{\mathbf{x}}+\frac{1}{4}\,a\,\hat{\mathbf{y}}+\frac{1}{4}\,a\,\hat{\mathbf{z}}$ | (16$d$) | Fe I |
| $\mathbf{B}_{9}$ | = | $x_{3}\mathbf{a}_{1}+x_{3}\mathbf{a}_{2}+x_{3}\mathbf{a}_{3}$ | = | $x_{3}\,a\,\hat{\mathbf{x}}+x_{3}\,a\,\hat{\mathbf{y}}+x_{3}\,a\,\hat{\mathbf{z}}$ | (32$e$) | Fe II |
| $\mathbf{B}_{10}$ | = | $x_{3}\mathbf{a}_{1}+x_{3}\mathbf{a}_{2}+\left(\frac{1}{2}-3\,x_{3}\right)\mathbf{a}_{3}$ | = | $\left(\frac{1}{4}-x_{3}\right)\,a\,\hat{\mathbf{x}}+\left(\frac{1}{4}-x_{3}\right)\,a\,\hat{\mathbf{y}}+x_{3}\,a\,\hat{\mathbf{z}}$ | (32$e$) | Fe II |
| $\mathbf{B}_{11}$ | = | $x_{3}\mathbf{a}_{1}+\left(\frac{1}{2}-3\,x_{3}\right)\mathbf{a}_{2}+x_{3}\mathbf{a}_{3}$ | = | $\left(\frac{1}{4}-x_{3}\right)\,a\,\hat{\mathbf{x}}+x_{3}\,a\,\hat{\mathbf{y}}+\left(\frac{1}{4}-x_{3}\right)\,a\,\hat{\mathbf{z}}$ | (32$e$) | Fe II |
| $\mathbf{B}_{12}$ | = | $\left(\frac{1}{2}-3\,x_{3}\right)\mathbf{a}_{1}+x_{3}\mathbf{a}_{2}+x_{3}\mathbf{a}_{3}$ | = | $x_{3}\,a\,\hat{\mathbf{x}}+\left(\frac{1}{4}-x_{3}\right)\,a\,\hat{\mathbf{y}}+\left(\frac{1}{4}-x_{3}\right)\,a\,\hat{\mathbf{z}}$ | (32$e$) | Fe II |
| $\mathbf{B}_{13}$ | = | $-x_{3}\mathbf{a}_{1}-x_{3}\mathbf{a}_{2}+\left(\frac{1}{2}+3\,x_{3}\right)\mathbf{a}_{3}$ | = | $\left(\frac{1}{4}+x_{3}\right)\,a\,\hat{\mathbf{x}}+\left(\frac{1}{4}+x_{3}\right)\,a\,\hat{\mathbf{y}}-x_{3}\,a\,\hat{\mathbf{z}}$ | (32$e$) | Fe II |
| $\mathbf{B}_{14}$ | = | $-x_{3}\mathbf{a}_{1}-x_{3}\mathbf{a}_{2}-x_{3}\mathbf{a}_{3}$ | = | $-x_{3}\,a\,\hat{\mathbf{x}}-x_{3}\,a\,\hat{\mathbf{y}}-x_{3}\,a\,\hat{\mathbf{z}}$ | (32$e$) | Fe II |
| $\mathbf{B}_{15}$ | = | $-x_{3}\mathbf{a}_{1}+\left(\frac{1}{2}+3\,x_{3}\right)\mathbf{a}_{2}-x_{3}\mathbf{a}_{3}$ | = | $\left(\frac{1}{4}+x_{3}\right)\,a\,\hat{\mathbf{x}}-x_{3}\,a\,\hat{\mathbf{y}}+\left(\frac{1}{4}+x_{3}\right)\,a\,\hat{\mathbf{z}}$ | (32$e$) | Fe II |
| $\mathbf{B}_{16}$ | = | $\left(\frac{1}{2}+3\,x_{3}\right)\mathbf{a}_{1}-x_{3}\mathbf{a}_{2}-x_{3}\mathbf{a}_{3}$ | = | $-x_{3}\,a\,\hat{\mathbf{x}}+\left(\frac{1}{4}+x_{3}\right)\,a\,\hat{\mathbf{y}}+\left(\frac{1}{4}+x_{3}\right)\,a\,\hat{\mathbf{z}}$ | (32$e$) | Fe II |
| $\mathbf{B}_{17}$ | = | $\left(\frac{1}{4}-x_{4}\right)\mathbf{a}_{1}+x_{4}\mathbf{a}_{2}+x_{4}\mathbf{a}_{3}$ | = | $x_{4}\,a\,\hat{\mathbf{x}}+\frac{1}{8}\,a\,\hat{\mathbf{y}}+\frac{1}{8}\,a\,\hat{\mathbf{z}}$ | (48$f$) | W |
| $\mathbf{B}_{18}$ | = | $x_{4}\mathbf{a}_{1}+\left(\frac{1}{4}-x_{4}\right)\mathbf{a}_{2}+\left(\frac{1}{4}-x_{4}\right)\mathbf{a}_{3}$ | = | $\left(\frac{1}{4}-x_{4}\right)\,a\,\hat{\mathbf{x}}+\frac{1}{8}\,a\,\hat{\mathbf{y}}+\frac{1}{8}\,a\,\hat{\mathbf{z}}$ | (48$f$) | W |
| $\mathbf{B}_{19}$ | = | $x_{4}\mathbf{a}_{1}+\left(\frac{1}{4}-x_{4}\right)\mathbf{a}_{2}+x_{4}\mathbf{a}_{3}$ | = | $\frac{1}{8}\,a\,\hat{\mathbf{x}}+x_{4}\,a\,\hat{\mathbf{y}}+\frac{1}{8}\,a\,\hat{\mathbf{z}}$ | (48$f$) | W |
| $\mathbf{B}_{20}$ | = | $\left(\frac{1}{4}-x_{4}\right)\mathbf{a}_{1}+x_{4}\mathbf{a}_{2}+\left(\frac{1}{4}-x_{4}\right)\mathbf{a}_{3}$ | = | $\frac{1}{8}\,a\,\hat{\mathbf{x}}+\left(\frac{1}{4}-x_{4}\right)\,a\,\hat{\mathbf{y}}+\frac{1}{8}\,a\,\hat{\mathbf{z}}$ | (48$f$) | W |
| $\mathbf{B}_{21}$ | = | $x_{4}\mathbf{a}_{1}+x_{4}\mathbf{a}_{2}+\left(\frac{1}{4}-x_{4}\right)\mathbf{a}_{3}$ | = | $\frac{1}{8}\,a\,\hat{\mathbf{x}}+\frac{1}{8}\,a\,\hat{\mathbf{y}}+x_{4}\,a\,\hat{\mathbf{z}}$ | (48$f$) | W |
| $\mathbf{B}_{22}$ | = | $\left(\frac{1}{4}-x_{4}\right)\mathbf{a}_{1}+\left(\frac{1}{4}-x_{4}\right)\mathbf{a}_{2}+x_{4}\mathbf{a}_{3}$ | = | $\frac{1}{8}\,a\,\hat{\mathbf{x}}+\frac{1}{8}\,a\,\hat{\mathbf{y}}+\left(\frac{1}{4}-x_{4}\right)\,a\,\hat{\mathbf{z}}$ | (48$f$) | W |
| $\mathbf{B}_{23}$ | = | $\left(x_{4}+\frac{3}{4}\right)\mathbf{a}_{1}-x_{4}\mathbf{a}_{2}+\left(x_{4}+\frac{3}{4}\right)\mathbf{a}_{3}$ | = | $\frac{3}{8}\,a\,\hat{\mathbf{x}}+\left(x_{4}+\frac{3}{4}\right)\,a\,\hat{\mathbf{y}}+\frac{3}{8}\,a\,\hat{\mathbf{z}}$ | (48$f$) | W |
| $\mathbf{B}_{24}$ | = | $-x_{4}\mathbf{a}_{1}+\left(x_{4}+\frac{3}{4}\right)\mathbf{a}_{2}-x_{4}\mathbf{a}_{3}$ | = | $\frac{3}{8}\,a\,\hat{\mathbf{x}}-x_{4}\,a\,\hat{\mathbf{y}}+\frac{3}{8}\,a\,\hat{\mathbf{z}}$ | (48$f$) | W |
| $\mathbf{B}_{25}$ | = | $-x_{4}\mathbf{a}_{1}+\left(x_{4}+\frac{3}{4}\right)\mathbf{a}_{2}+\left(x_{4}+\frac{3}{4}\right)\mathbf{a}_{3}$ | = | $\left(x_{4}+\frac{3}{4}\right)\,a\,\hat{\mathbf{x}}+\frac{3}{8}\,a\,\hat{\mathbf{y}}+\frac{3}{8}\,a\,\hat{\mathbf{z}}$ | (48$f$) | W |
| $\mathbf{B}_{26}$ | = | $\left(x_{4}+\frac{3}{4}\right)\mathbf{a}_{1}-x_{4}\mathbf{a}_{2}-x_{4}\mathbf{a}_{3}$ | = | $-x_{4}\,a\,\hat{\mathbf{x}}+\frac{3}{8}\,a\,\hat{\mathbf{y}}+\frac{3}{8}\,a\,\hat{\mathbf{z}}$ | (48$f$) | W |
| $\mathbf{B}_{27}$ | = | $-x_{4}\mathbf{a}_{1}-x_{4}\mathbf{a}_{2}+\left(x_{4}+\frac{3}{4}\right)\mathbf{a}_{3}$ | = | $\frac{3}{8}\,a\,\hat{\mathbf{x}}+\frac{3}{8}\,a\,\hat{\mathbf{y}}-x_{4}\,a\,\hat{\mathbf{z}}$ | (48$f$) | W |
| $\mathbf{B}_{28}$ | = | $+\left(x_{4}+\frac{3}{4}\right)\mathbf{a}_{1}+\left(x_{4}+\frac{3}{4}\right)\mathbf{a}_{2}-x_{4}\mathbf{a}_{3}$ | = | $\frac{3}{8}\,a\,\hat{\mathbf{x}}+\frac{3}{8}\,a\,\hat{\mathbf{y}}+\left(x_{4}+\frac{3}{4}\right)\,a\,\hat{\mathbf{z}}$ | (48$f$) | W |

**Geometry files:**

- CIF: pp. 803
- POSCAR: pp. 804

# Body-Centered Cubic (W, A2) Structure: A_cI2_229_a

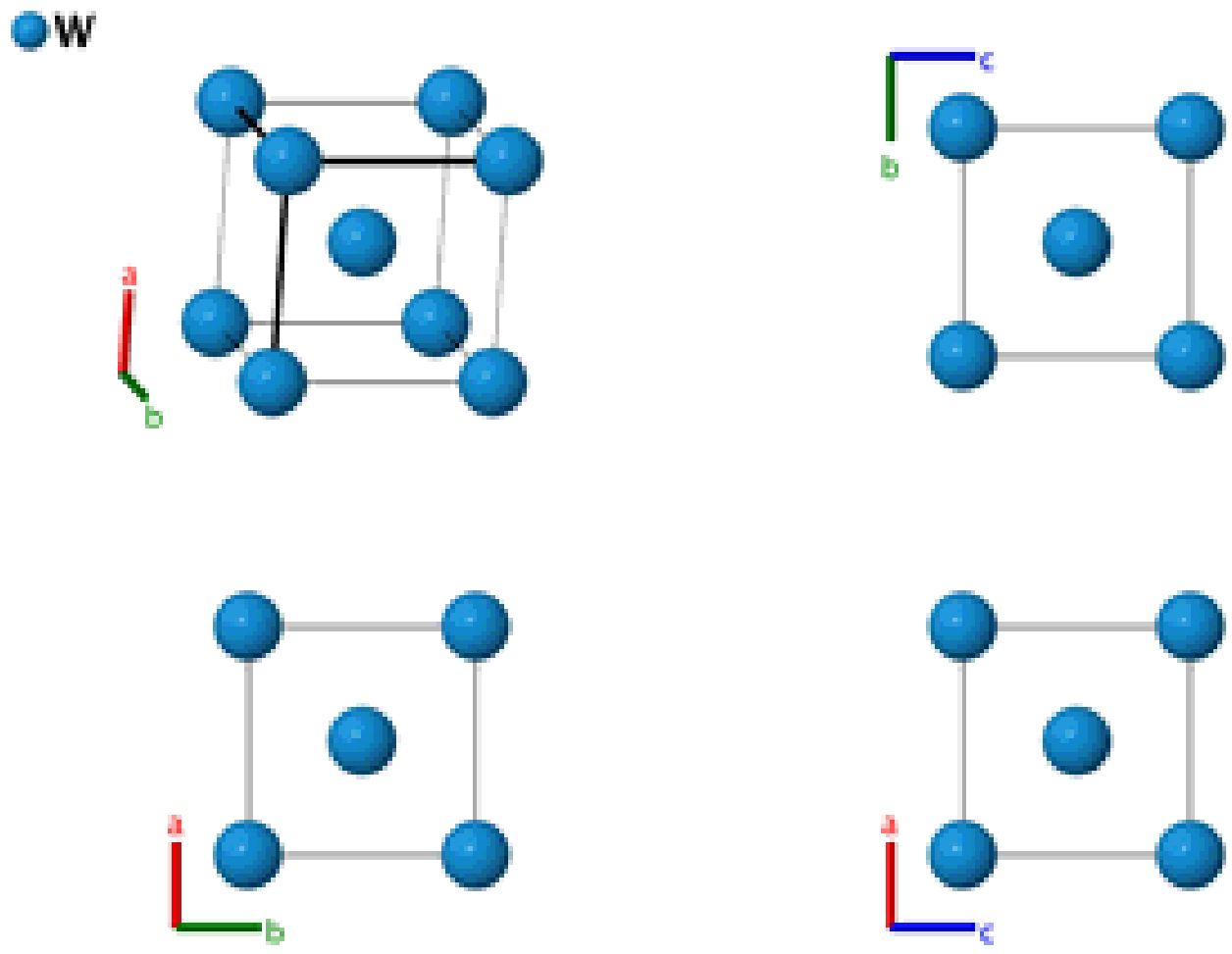

| | | |
|---|---|---|
| **Prototype** | : | W |
| **AFLOW prototype label** | : | A_cI2_229_a |
| ***Strukturbericht* designation** | : | A2 |
| **Pearson symbol** | : | cI2 |
| **Space group number** | : | 229 |
| **Space group symbol** | : | $Im\bar{3}m$ |
| **AFLOW prototype command** | : | `aflow --proto=A_cI2_229_a --params=`$a$ |

**Other elements with this structure:**

- Li (at room temp.), Na, K, V, Cr, Fe, Rb, Nb, Mo, Cs, Ba, Eu, Ta

- Although more accurate measurements of the lattice constant of tungsten are available, (Davey, 1925) is chosen because of the unique experimental technique.

**Body-centered Cubic primitive vectors:**

$$
\begin{aligned}
\mathbf{a}_1 &= -\tfrac{1}{2}\,a\,\hat{\mathbf{x}} + \tfrac{1}{2}\,a\,\hat{\mathbf{y}} + \tfrac{1}{2}\,a\,\hat{\mathbf{z}} \\
\mathbf{a}_2 &= \tfrac{1}{2}\,a\,\hat{\mathbf{x}} - \tfrac{1}{2}\,a\,\hat{\mathbf{y}} + \tfrac{1}{2}\,a\,\hat{\mathbf{z}} \\
\mathbf{a}_3 &= \tfrac{1}{2}\,a\,\hat{\mathbf{x}} + \tfrac{1}{2}\,a\,\hat{\mathbf{y}} - \tfrac{1}{2}\,a\,\hat{\mathbf{z}}
\end{aligned}
$$

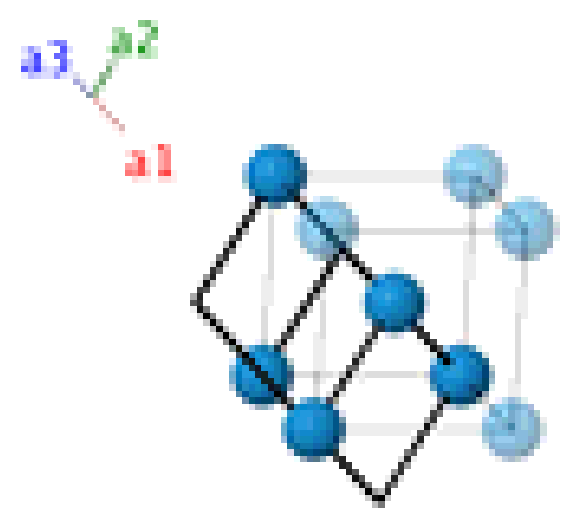

**Basis vectors:**

| | | Lattice Coordinates | | Cartesian Coordinates | Wyckoff Position | Atom Type |
|---|---|---|---|---|---|---|
| $\mathbf{B}_1$ | = | $0\,\mathbf{a}_1 + 0\,\mathbf{a}_2 + 0\,\mathbf{a}_3$ | = | $0\hat{\mathbf{x}} + 0\hat{\mathbf{y}} + 0\hat{\mathbf{z}}$ | $(2a)$ | W |

**Geometry files:**
- CIF: pp. 805
- POSCAR: pp. 805

# High-Pressure $H_3S$ Structure: A3B_cI8_229_b_a

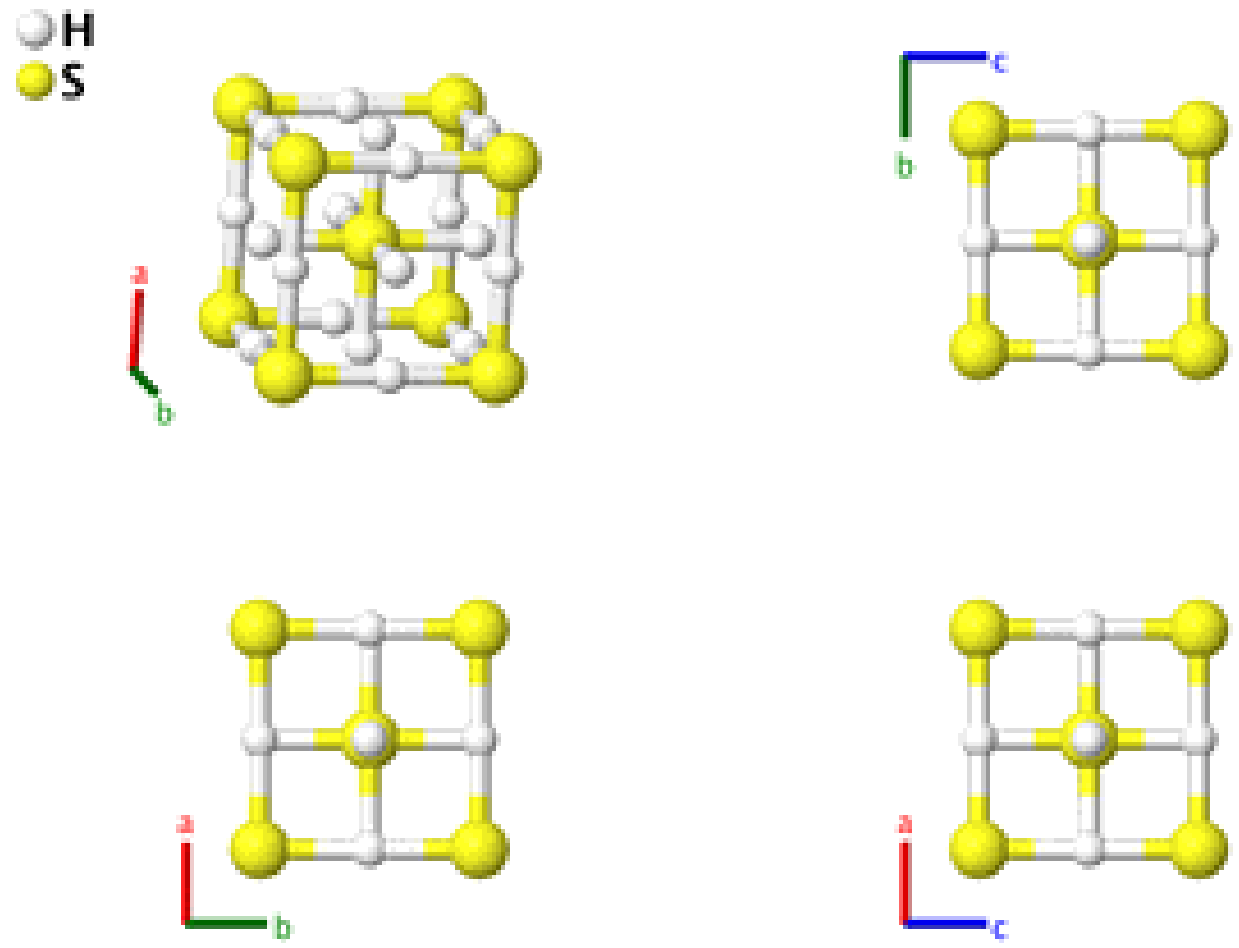

**Prototype** : $H_3S$

**AFLOW prototype label** : A3B_cI8_229_b_a

***Strukturbericht* designation** : None

**Pearson symbol** : cI8

**Space group number** : 229

**Space group symbol** : $Im\bar{3}m$

**AFLOW prototype command** : `aflow --proto=A3B_cI8_229_b_a --params=`$a$

**Other compounds with this structure:**

- $La_2O_3$, $Nd_2O_3$ (In both cases the oxygen atoms only partially occupy the (6b) Wyckoff positions.)

- (Duan, 2014) predicted that this structure of $H_3S$ would be a conventional superconductor at temperatures above 191 K and a pressure of 200 GPa. (Drozdov, 2015) found a superconductor in the hydrogen-sulfur system at 203 K and pressure near 200 GPa. (Bernstein, 2015) showed that this structure is the ground state of the H-S system near 200 GPa. Both $La_2O_3$ and $Nd_2O_3$ can form in this structure under ambient conditions, but in both cases the oxygen atoms occupy only 50% of the (6b) Wyckoff positions. We have used the fact that all vectors of the form $(\pm a/2\hat{\mathbf{x}} \pm a/2\hat{\mathbf{y}} \pm a/2\hat{\mathbf{z}})$ are primitive vectors of the body-centered cubic lattice to simplify the positions of some atoms in both lattice and Cartesian coordinates.

**Body-centered Cubic primitive vectors:**

$$\begin{aligned}
\mathbf{a}_1 &= -\tfrac{1}{2}\,a\,\hat{\mathbf{x}} + \tfrac{1}{2}\,a\,\hat{\mathbf{y}} + \tfrac{1}{2}\,a\,\hat{\mathbf{z}} \\
\mathbf{a}_2 &= \tfrac{1}{2}\,a\,\hat{\mathbf{x}} - \tfrac{1}{2}\,a\,\hat{\mathbf{y}} + \tfrac{1}{2}\,a\,\hat{\mathbf{z}} \\
\mathbf{a}_3 &= \tfrac{1}{2}\,a\,\hat{\mathbf{x}} + \tfrac{1}{2}\,a\,\hat{\mathbf{y}} - \tfrac{1}{2}\,a\,\hat{\mathbf{z}}
\end{aligned}$$

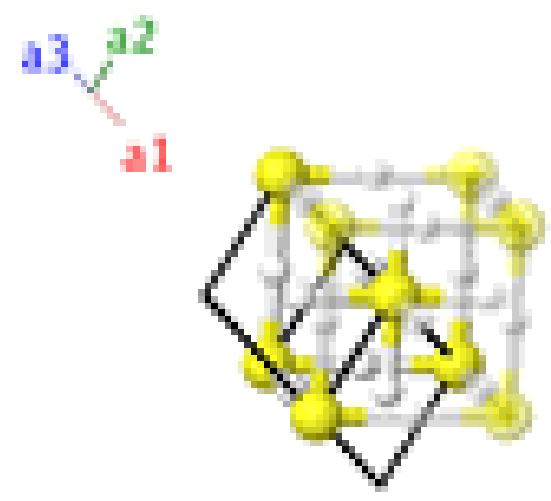

**Basis vectors:**

| | | Lattice Coordinates | | Cartesian Coordinates | Wyckoff Position | Atom Type |
|---|---|---|---|---|---|---|
| $\mathbf{B_1}$ | = | $0\,\mathbf{a_1} + 0\,\mathbf{a_2} + 0\,\mathbf{a_3}$ | = | $0\hat{\mathbf{x}} + 0\hat{\mathbf{y}} + 0\hat{\mathbf{z}}$ | $(2a)$ | S |
| $\mathbf{B_2}$ | = | $\frac{1}{2}\,\mathbf{a_2} + \frac{1}{2}\,\mathbf{a_3}$ | = | $\frac{1}{2}\,a\,\hat{\mathbf{x}}$ | $(6b)$ | H |
| $\mathbf{B_3}$ | = | $\frac{1}{2}\,\mathbf{a_1} + \frac{1}{2}\,\mathbf{a_3}$ | = | $\frac{1}{2}\,a\,\hat{\mathbf{y}}$ | $(6b)$ | H |
| $\mathbf{B_4}$ | = | $\frac{1}{2}\,\mathbf{a_1} + \frac{1}{2}\,\mathbf{a_2}$ | = | $\frac{1}{2}\,a\,\hat{\mathbf{z}}$ | $(6b)$ | H |

---

---

**Geometry files:**

- CIF: pp. 805
- POSCAR: pp. 806

# $Pt_3O_4$ Structure: A4B3_cI14_229_c_b

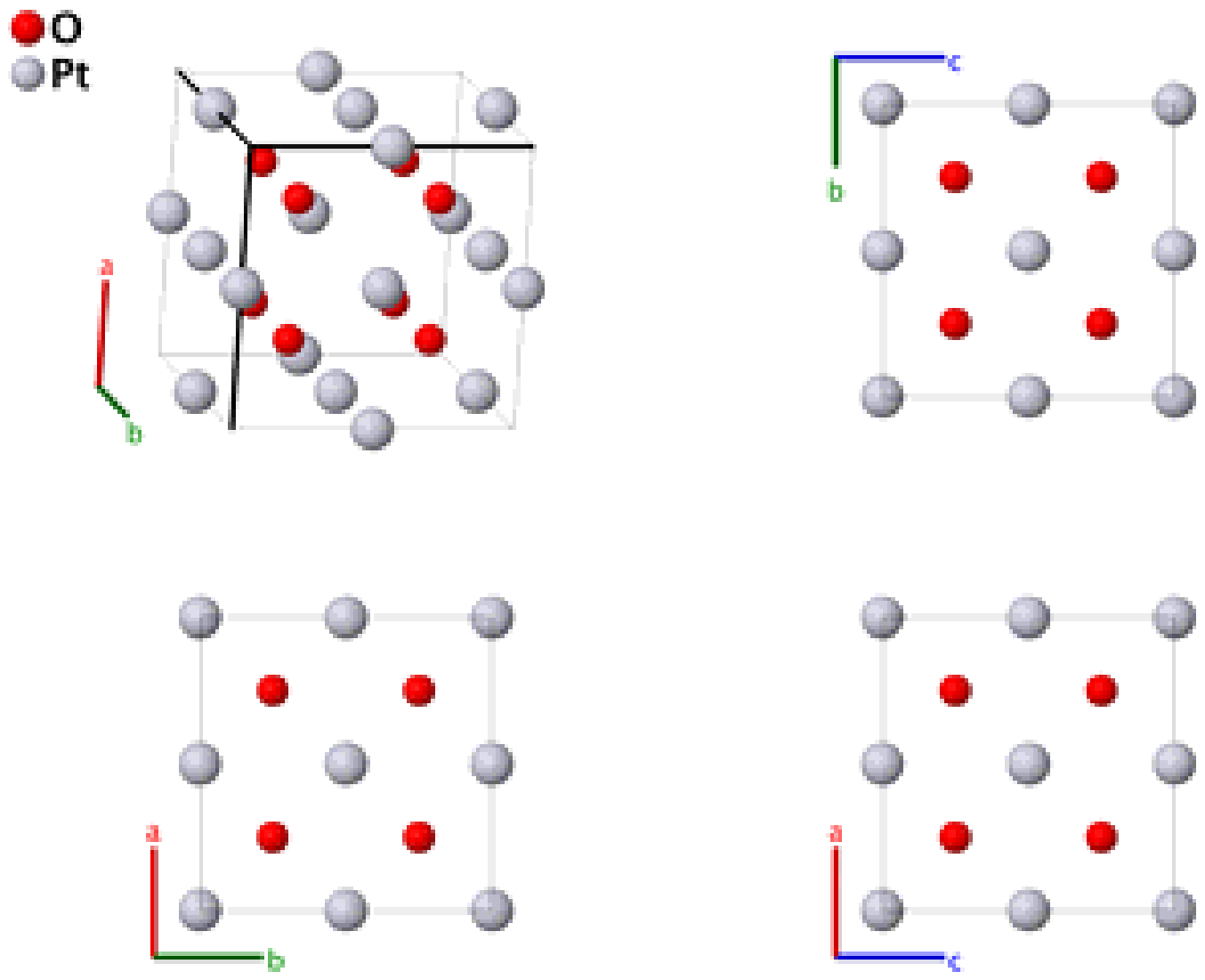

| | | |
|---|---|---|
| **Prototype** | : | $Pt_3O_4$ |
| **AFLOW prototype label** | : | A4B3_cI14_229_c_b |
| ***Strukturbericht* designation** | : | None |
| **Pearson symbol** | : | cI14 |
| **Space group number** | : | 229 |
| **Space group symbol** | : | $Im\bar{3}m$ |
| **AFLOW prototype command** | : | `aflow --proto=A4B3_cI14_229_c_b --params=`$a$ |

- This is a simple defect superstructure of the CsCl (B2) structure. One atom has been removed from a 2×2×2 supercell of CsCl.

**Body-centered Cubic primitive vectors:**

$$\mathbf{a}_1 = -\frac{1}{2}\,a\,\mathbf{\hat{x}} + \frac{1}{2}\,a\,\mathbf{\hat{y}} + \frac{1}{2}\,a\,\mathbf{\hat{z}}$$
$$\mathbf{a}_2 = \frac{1}{2}\,a\,\mathbf{\hat{x}} - \frac{1}{2}\,a\,\mathbf{\hat{y}} + \frac{1}{2}\,a\,\mathbf{\hat{z}}$$
$$\mathbf{a}_3 = \frac{1}{2}\,a\,\mathbf{\hat{x}} + \frac{1}{2}\,a\,\mathbf{\hat{y}} - \frac{1}{2}\,a\,\mathbf{\hat{z}}$$

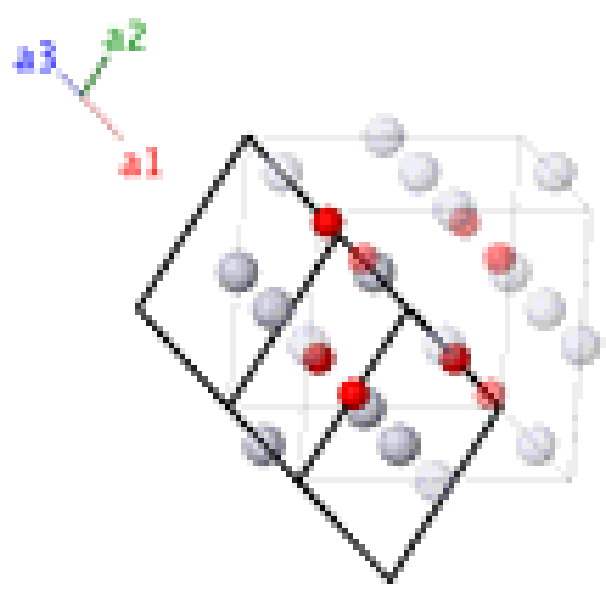

**Basis vectors:**

| | | Lattice Coordinates | | Cartesian Coordinates | Wyckoff Position | Atom Type |
|---|---|---|---|---|---|---|
| $\mathbf{B}_1$ | = | $\frac{1}{2}\,\mathbf{a}_2 + \frac{1}{2}\,\mathbf{a}_3$ | = | $\frac{1}{2}\,a\,\mathbf{\hat{x}}$ | (6$b$) | Pt |
| $\mathbf{B}_2$ | = | $\frac{1}{2}\,\mathbf{a}_1 + \frac{1}{2}\,\mathbf{a}_3$ | = | $\frac{1}{2}\,a\,\mathbf{\hat{y}}$ | (6$b$) | Pt |
| $\mathbf{B}_3$ | = | $\frac{1}{2}\,\mathbf{a}_1 + \frac{1}{2}\,\mathbf{a}_2$ | = | $\frac{1}{2}\,a\,\mathbf{\hat{z}}$ | (6$b$) | Pt |
| $\mathbf{B_4}$ | = | $\frac{1}{2}\,\mathbf{a_1} + \frac{1}{2}\,\mathbf{a_2} + \frac{1}{2}\,\mathbf{a_3}$ | = | $\frac{1}{4}\,a\,\hat{\mathbf{x}} + \frac{1}{4}\,a\,\hat{\mathbf{y}} + \frac{1}{4}\,a\,\hat{\mathbf{z}}$ | (8$c$) | O |
| $\mathbf{B_5}$ | = | $\frac{1}{2}\,\mathbf{a_3}$ | = | $\frac{1}{4}\,a\,\hat{\mathbf{x}} + \frac{1}{4}\,a\,\hat{\mathbf{y}} + \frac{3}{4}\,a\,\hat{\mathbf{z}}$ | (8$c$) | O |
| $\mathbf{B_6}$ | = | $\frac{1}{2}\,\mathbf{a_2}$ | = | $\frac{1}{4}\,a\,\hat{\mathbf{x}} + \frac{3}{4}\,a\,\hat{\mathbf{y}} + \frac{1}{4}\,a\,\hat{\mathbf{z}}$ | (8$c$) | O |
| $\mathbf{B_7}$ | = | $\frac{1}{2}\,\mathbf{a_1}$ | = | $\frac{3}{4}\,a\,\hat{\mathbf{x}} + \frac{1}{4}\,a\,\hat{\mathbf{y}} + \frac{1}{4}\,a\,\hat{\mathbf{z}}$ | (8$c$) | O |

**Found in:**

**Geometry files:**

- CIF: pp. 806
- POSCAR: pp. 807

# $Sb_2Tl_7$ ($L2_2$) Structure: A2B7_cI54_229_e_afh

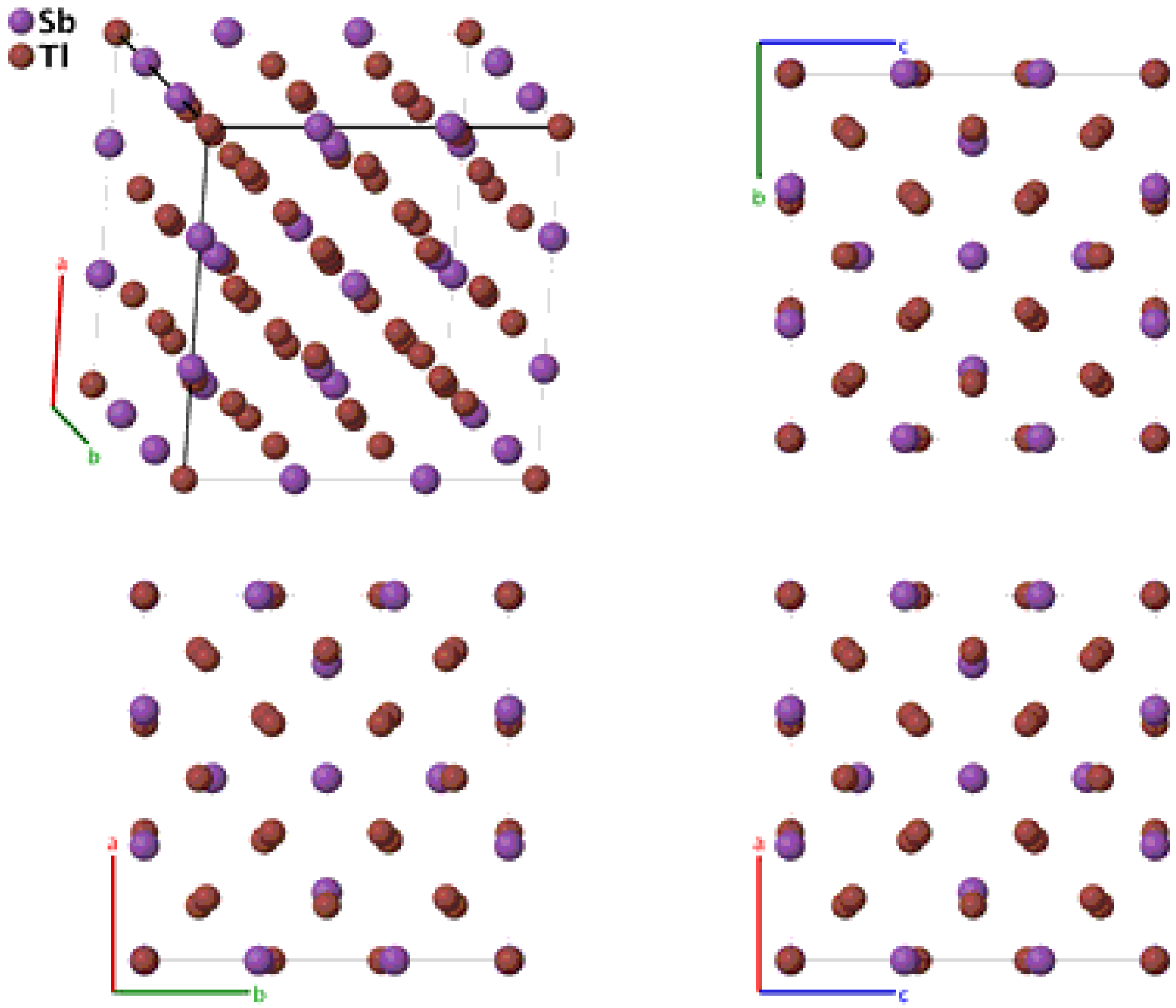

| | | |
|---|---|---|
| **Prototype** | : | $Sb_2Tl_7$ |
| **AFLOW prototype label** | : | A2B7_cI54_229_e_afh |
| ***Strukturbericht* designation** | : | $L2_2$ |
| **Pearson symbol** | : | cI54 |
| **Space group number** | : | 229 |
| **Space group symbol** | : | $Im\bar{3}m$ |
| **AFLOW prototype command** | : | `aflow --proto=A2B7_cI54_229_e_afh`<br>`--params=`$a, x_2, x_3, y_4$ |

**Body-centered Cubic primitive vectors:**

$$\begin{aligned}
\mathbf{a}_1 &= -\frac{1}{2}\,a\,\mathbf{\hat{x}} + \frac{1}{2}\,a\,\mathbf{\hat{y}} + \frac{1}{2}\,a\,\mathbf{\hat{z}} \\
\mathbf{a}_2 &= \frac{1}{2}\,a\,\mathbf{\hat{x}} - \frac{1}{2}\,a\,\mathbf{\hat{y}} + \frac{1}{2}\,a\,\mathbf{\hat{z}} \\
\mathbf{a}_3 &= \frac{1}{2}\,a\,\mathbf{\hat{x}} + \frac{1}{2}\,a\,\mathbf{\hat{y}} - \frac{1}{2}\,a\,\mathbf{\hat{z}}
\end{aligned}$$

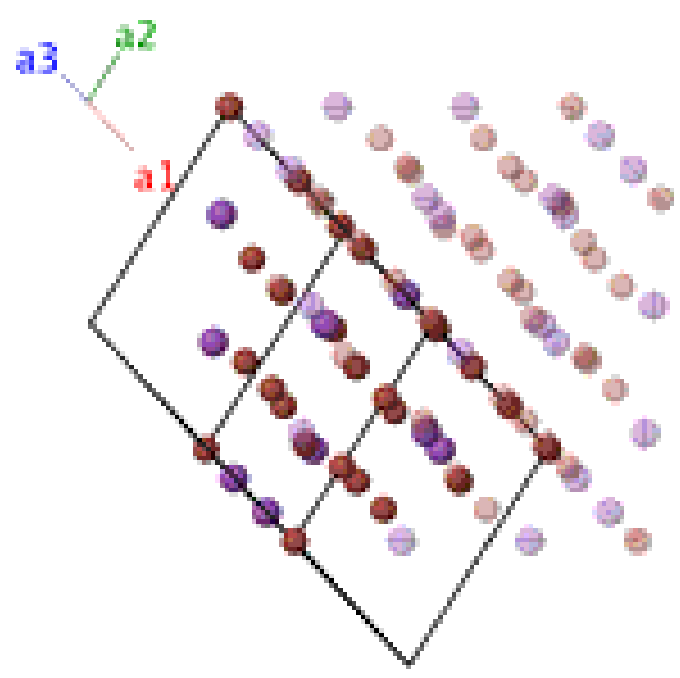

**Basis vectors:**

| | | Lattice Coordinates | | Cartesian Coordinates | Wyckoff Position | Atom Type |
|---|---|---|---|---|---|---|
| $\mathbf{B}_1$ | = | $0\,\mathbf{a}_1 + 0\,\mathbf{a}_2 + 0\,\mathbf{a}_3$ | = | $0\mathbf{\hat{x}} + 0\mathbf{\hat{y}} + 0\mathbf{\hat{z}}$ | (2$a$) | Tl I |
| $\mathbf{B}_2$ | = | $x_2\,\mathbf{a}_2 + x_2\,\mathbf{a}_3$ | = | $x_2\,a\,\mathbf{\hat{x}}$ | (12$e$) | Sb |
| $\mathbf{B_3}$ | = | $x_2\,\mathbf{a_1} + x_2\,\mathbf{a_3}$ | = | $x_2\,a\,\mathbf{\hat{y}}$ | (12$e$) | Sb |
| $\mathbf{B_4}$ | = | $x_2\,\mathbf{a_1} + x_2\,\mathbf{a_2}$ | = | $x_2\,a\,\mathbf{\hat{z}}$ | (12$e$) | Sb |
| $\mathbf{B_5}$ | = | $-x_2\,\mathbf{a_2} - x_2\,\mathbf{a_3}$ | = | $-x_2\,a\,\mathbf{\hat{x}}$ | (12$e$) | Sb |
| $\mathbf{B_6}$ | = | $-x_2\,\mathbf{a_1} - x_2\,\mathbf{a_3}$ | = | $-x_2\,a\,\mathbf{\hat{y}}$ | (12$e$) | Sb |
| $\mathbf{B_7}$ | = | $-x_2\,\mathbf{a_1} - x_2\,\mathbf{a_2}$ | = | $-x_2\,a\,\mathbf{\hat{z}}$ | (12$e$) | Sb |
| $\mathbf{B_8}$ | = | $2x_3\,\mathbf{a_1} + 2x_3\,\mathbf{a_2} + 2x_3\,\mathbf{a_3}$ | = | $x_3\,a\,\mathbf{\hat{x}} + x_3\,a\,\mathbf{\hat{y}} + x_3\,a\,\mathbf{\hat{z}}$ | (16$f$) | Tl II |
| $\mathbf{B_9}$ | = | $-2x_3\,\mathbf{a_3}$ | = | $-x_3\,a\,\mathbf{\hat{x}} - x_3\,a\,\mathbf{\hat{y}} + x_3\,a\,\mathbf{\hat{z}}$ | (16$f$) | Tl II |
| $\mathbf{B_{10}}$ | = | $-2x_3\,\mathbf{a_2}$ | = | $-x_3\,a\,\mathbf{\hat{x}} + x_3\,a\,\mathbf{\hat{y}} - x_3\,a\,\mathbf{\hat{z}}$ | (16$f$) | Tl II |
| $\mathbf{B_{11}}$ | = | $-2x_3\,\mathbf{a_1}$ | = | $x_3\,a\,\mathbf{\hat{x}} - x_3\,a\,\mathbf{\hat{y}} - x_3\,a\,\mathbf{\hat{z}}$ | (16$f$) | Tl II |
| $\mathbf{B_{12}}$ | = | $2x_3\,\mathbf{a_3}$ | = | $x_3\,a\,\mathbf{\hat{x}} + x_3\,a\,\mathbf{\hat{y}} - x_3\,a\,\mathbf{\hat{z}}$ | (16$f$) | Tl II |
| $\mathbf{B_{13}}$ | = | $-2x_3\,\mathbf{a_1} - 2x_3\,\mathbf{a_2} - 2x_3\,\mathbf{a_3}$ | = | $-x_3\,a\,\mathbf{\hat{x}} - x_3\,a\,\mathbf{\hat{y}} - x_3\,a\,\mathbf{\hat{z}}$ | (16$f$) | Tl II |
| $\mathbf{B_{14}}$ | = | $2x_3\,\mathbf{a_2}$ | = | $x_3\,a\,\mathbf{\hat{x}} - x_3\,a\,\mathbf{\hat{y}} + x_3\,a\,\mathbf{\hat{z}}$ | (16$f$) | Tl II |
| $\mathbf{B_{15}}$ | = | $2x_3\,\mathbf{a_1}$ | = | $-x_3\,a\,\mathbf{\hat{x}} + x_3\,a\,\mathbf{\hat{y}} + x_3\,a\,\mathbf{\hat{z}}$ | (16$f$) | Tl II |
| $\mathbf{B_{16}}$ | = | $2y_4\,\mathbf{a_1} + y_4\,\mathbf{a_2} + y_4\,\mathbf{a_3}$ | = | $y_4\,a\,\mathbf{\hat{y}} + y_4\,a\,\mathbf{\hat{z}}$ | (24$h$) | Tl III |
| $\mathbf{B_{17}}$ | = | $y_4\,\mathbf{a_2} - y_4\,\mathbf{a_3}$ | = | $-y_4\,a\,\mathbf{\hat{y}} + y_4\,a\,\mathbf{\hat{z}}$ | (24$h$) | Tl III |
| $\mathbf{B_{18}}$ | = | $-y_4\,\mathbf{a_2} + y_4\,\mathbf{a_3}$ | = | $y_4\,a\,\mathbf{\hat{y}} - y_4\,a\,\mathbf{\hat{z}}$ | (24$h$) | Tl III |
| $\mathbf{B_{19}}$ | = | $-2y_4\,\mathbf{a_1} - y_4\,\mathbf{a_2} - y_4\,\mathbf{a_3}$ | = | $-y_4\,a\,\mathbf{\hat{y}} - y_4\,a\,\mathbf{\hat{z}}$ | (24$h$) | Tl III |
| $\mathbf{B_{20}}$ | = | $y_4\,\mathbf{a_1} + 2y_4\,\mathbf{a_2} + y_4\,\mathbf{a_3}$ | = | $y_4\,a\,\mathbf{\hat{x}} + y_4\,a\,\mathbf{\hat{z}}$ | (24$h$) | Tl III |
| $\mathbf{B_{21}}$ | = | $-y_4\,\mathbf{a_1} + y_4\,\mathbf{a_3}$ | = | $y_4\,a\,\mathbf{\hat{x}} - y_4\,a\,\mathbf{\hat{z}}$ | (24$h$) | Tl III |
| $\mathbf{B_{22}}$ | = | $y_4\,\mathbf{a_1} - y_4\,\mathbf{a_3}$ | = | $-y_4\,a\,\mathbf{\hat{x}} + y_4\,a\,\mathbf{\hat{z}}$ | (24$h$) | Tl III |
| $\mathbf{B_{23}}$ | = | $-y_4\,\mathbf{a_1} - 2y_4\,\mathbf{a_2} - y_4\,\mathbf{a_3}$ | = | $-y_4\,a\,\mathbf{\hat{x}} - y_4\,a\,\mathbf{\hat{z}}$ | (24$h$) | Tl III |
| $\mathbf{B_{24}}$ | = | $y_4\,\mathbf{a_1} + y_4\,\mathbf{a_2} + 2y_4\,\mathbf{a_3}$ | = | $y_4\,a\,\mathbf{\hat{x}} + y_4\,a\,\mathbf{\hat{y}}$ | (24$h$) | Tl III |
| $\mathbf{B_{25}}$ | = | $y_4\,\mathbf{a_1} - y_4\,\mathbf{a_2}$ | = | $-y_4\,a\,\mathbf{\hat{x}} + y_4\,a\,\mathbf{\hat{y}}$ | (24$h$) | Tl III |
| $\mathbf{B_{26}}$ | = | $-y_4\,\mathbf{a_1} + y_4\,\mathbf{a_2}$ | = | $y_4\,a\,\mathbf{\hat{x}} - y_4\,a\,\mathbf{\hat{y}}$ | (24$h$) | Tl III |
| $\mathbf{B_{27}}$ | = | $-y_4\,\mathbf{a_1} - y_4\,\mathbf{a_2} - 2y_4\,\mathbf{a_3}$ | = | $-y_4\,a\,\mathbf{\hat{x}} - y_4\,a\,\mathbf{\hat{y}}$ | (24$h$) | Tl III |

**Found in:**

**Geometry files:**

- CIF: pp. 807
- POSCAR: pp. 808

# Model of Austenite Structure (cI32): AB12C3_cI32_229_a_h_b

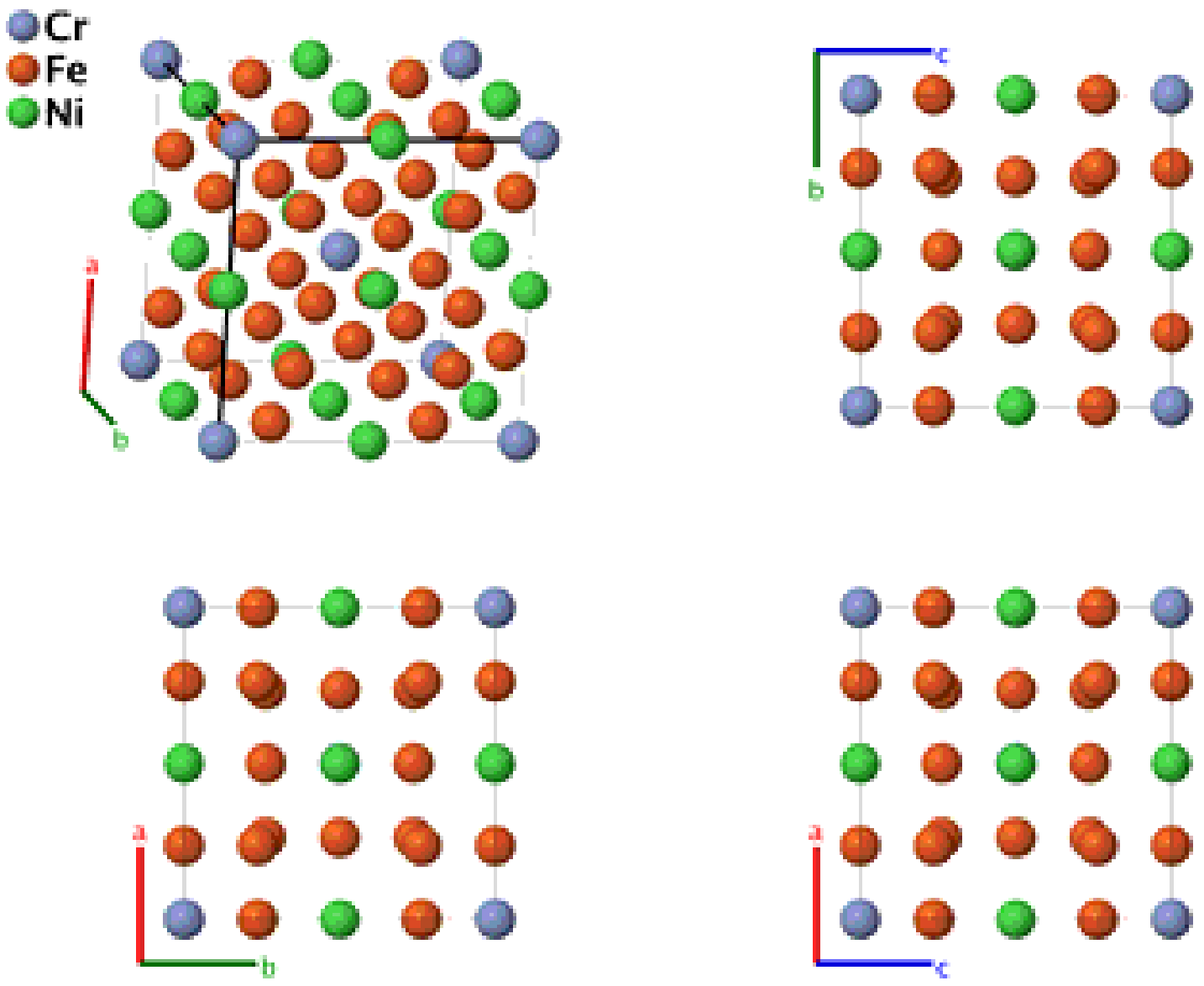

| | | |
|---|---|---|
| **Prototype** | : | $CrFe_{12}Ni_3$ |
| **AFLOW prototype label** | : | AB12C3_cI32_229_a_h_b |
| ***Strukturbericht* designation** | : | None |
| **Pearson symbol** | : | cI32 |
| **Space group number** | : | 229 |
| **Space group symbol** | : | $Im\bar{3}m$ |
| **AFLOW prototype command** | : | `aflow --proto=AB12C3_cI32_229_a_h_b --params=`$a, y_3$ |

- Austenitic steels are alloys of iron and other metals with an averaged face-centered cubic structure. This model is not meant to represent a real steel, and the selection of atom types for each Wyckoff position is arbitrary. If we set the $y_3 = 1/4$ then the atoms are on the sites of an fcc lattice.

**Body-centered Cubic primitive vectors:**

$$\mathbf{a}_1 = -\frac{1}{2}\,a\,\hat{\mathbf{x}} + \frac{1}{2}\,a\,\hat{\mathbf{y}} + \frac{1}{2}\,a\,\hat{\mathbf{z}}$$
$$\mathbf{a}_2 = \frac{1}{2}\,a\,\hat{\mathbf{x}} - \frac{1}{2}\,a\,\hat{\mathbf{y}} + \frac{1}{2}\,a\,\hat{\mathbf{z}}$$
$$\mathbf{a}_3 = \frac{1}{2}\,a\,\hat{\mathbf{x}} + \frac{1}{2}\,a\,\hat{\mathbf{y}} - \frac{1}{2}\,a\,\hat{\mathbf{z}}$$

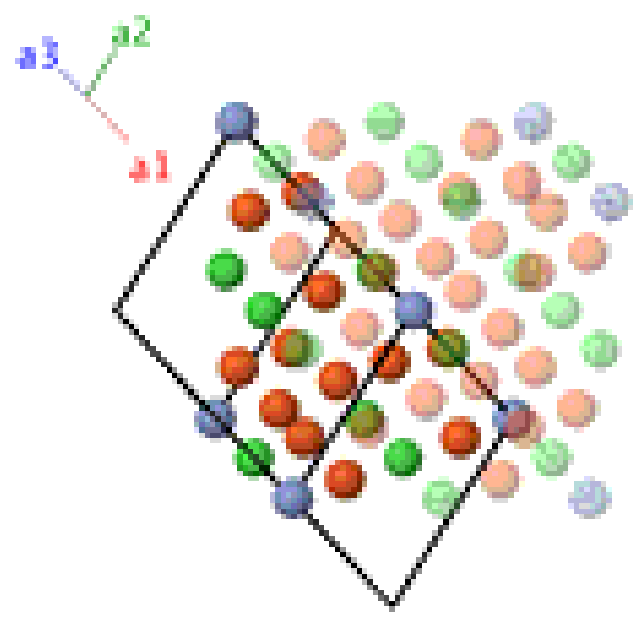

**Basis vectors:**

| | | Lattice Coordinates | | Cartesian Coordinates | Wyckoff Position | Atom Type |
|---|---|---|---|---|---|---|
| $\mathbf{B}_1$ | = | $0\,\mathbf{a}_1 + 0\,\mathbf{a}_2 + 0\,\mathbf{a}_3$ | = | $0\hat{\mathbf{x}} + 0\hat{\mathbf{y}} + 0\hat{\mathbf{z}}$ | (2$a$) | Cr |
| $\mathbf{B}_2$ | = | $\frac{1}{2}\,\mathbf{a}_2 + \frac{1}{2}\,\mathbf{a}_3$ | = | $\frac{1}{2}\,a\,\hat{\mathbf{x}}$ | (6$b$) | Ni |
| $\mathbf{B}_3$ | = | $\frac{1}{2}\,\mathbf{a}_1 + \frac{1}{2}\,\mathbf{a}_3$ | = | $\frac{1}{2}\,a\,\hat{\mathbf{y}}$ | $(6b)$ | Ni |
| $\mathbf{B}_4$ | = | $\frac{1}{2}\,\mathbf{a}_1 + \frac{1}{2}\,\mathbf{a}_2$ | = | $\frac{1}{2}\,a\,\hat{\mathbf{z}}$ | $(6b)$ | Ni |
| $\mathbf{B}_5$ | = | $2y_3\,\mathbf{a}_1 + y_3\,\mathbf{a}_2 + y_3\,\mathbf{a}_3$ | = | $y_3\,a\,\hat{\mathbf{y}} + y_3\,a\,\hat{\mathbf{z}}$ | $(24h)$ | Fe |
| $\mathbf{B}_6$ | = | $y_3\,\mathbf{a}_2 - y_3\,\mathbf{a}_3$ | = | $-y_3\,a\,\hat{\mathbf{y}} + y_3\,a\,\hat{\mathbf{z}}$ | $(24h)$ | Fe |
| $\mathbf{B}_7$ | = | $-y_3\,\mathbf{a}_2 + y_3\,\mathbf{a}_3$ | = | $y_3\,a\,\hat{\mathbf{y}} - y_3\,a\,\hat{\mathbf{z}}$ | $(24h)$ | Fe |
| $\mathbf{B}_8$ | = | $-2y_3\,\mathbf{a}_1 - y_3\,\mathbf{a}_2 - y_3\,\mathbf{a}_3$ | = | $-y_3\,a\,\hat{\mathbf{y}} - y_3\,a\,\hat{\mathbf{z}}$ | $(24h)$ | Fe |
| $\mathbf{B}_9$ | = | $y_3\,\mathbf{a}_1 + 2y_3\,\mathbf{a}_2 + y_3\,\mathbf{a}_3$ | = | $y_3\,a\,\hat{\mathbf{x}} + y_3\,a\,\hat{\mathbf{z}}$ | $(24h)$ | Fe |
| $\mathbf{B}_{10}$ | = | $-y_3\,\mathbf{a}_1 + y_3\,\mathbf{a}_3$ | = | $y_3\,a\,\hat{\mathbf{x}} - y_3\,a\,\hat{\mathbf{z}}$ | $(24h)$ | Fe |
| $\mathbf{B}_{11}$ | = | $y_3\,\mathbf{a}_1 - y_3\,\mathbf{a}_3$ | = | $-y_3\,a\,\hat{\mathbf{x}} + y_3\,a\,\hat{\mathbf{z}}$ | $(24h)$ | Fe |
| $\mathbf{B}_{12}$ | = | $-y_3\,\mathbf{a}_1 - 2y_3\,\mathbf{a}_2 - y_3\,\mathbf{a}_3$ | = | $-y_3\,a\,\hat{\mathbf{x}} - y_3\,a\,\hat{\mathbf{z}}$ | $(24h)$ | Fe |
| $\mathbf{B}_{13}$ | = | $y_3\,\mathbf{a}_1 + y_3\,\mathbf{a}_2 + 2y_3\,\mathbf{a}_3$ | = | $y_3\,a\,\hat{\mathbf{x}} + y_3\,a\,\hat{\mathbf{y}}$ | $(24h)$ | Fe |
| $\mathbf{B}_{14}$ | = | $y_3\,\mathbf{a}_1 - y_3\,\mathbf{a}_2$ | = | $-y_3\,a\,\hat{\mathbf{x}} + y_3\,a\,\hat{\mathbf{y}}$ | $(24h)$ | Fe |
| $\mathbf{B}_{15}$ | = | $-y_3\,\mathbf{a}_1 + y_3\,\mathbf{a}_2$ | = | $y_3\,a\,\hat{\mathbf{x}} - y_3\,a\,\hat{\mathbf{y}}$ | $(24h)$ | Fe |
| $\mathbf{B}_{16}$ | = | $-y_3\,\mathbf{a}_1 - y_3\,\mathbf{a}_2 - 2y_3\,\mathbf{a}_3$ | = | $-y_3\,a\,\hat{\mathbf{x}} - y_3\,a\,\hat{\mathbf{y}}$ | $(24h)$ | Fe |

**Geometry files:**

- CIF: pp. 808
- POSCAR: pp. 809

# Model of Ferrite Structure (cI16): AB4C3_cI16_229_a_c_b

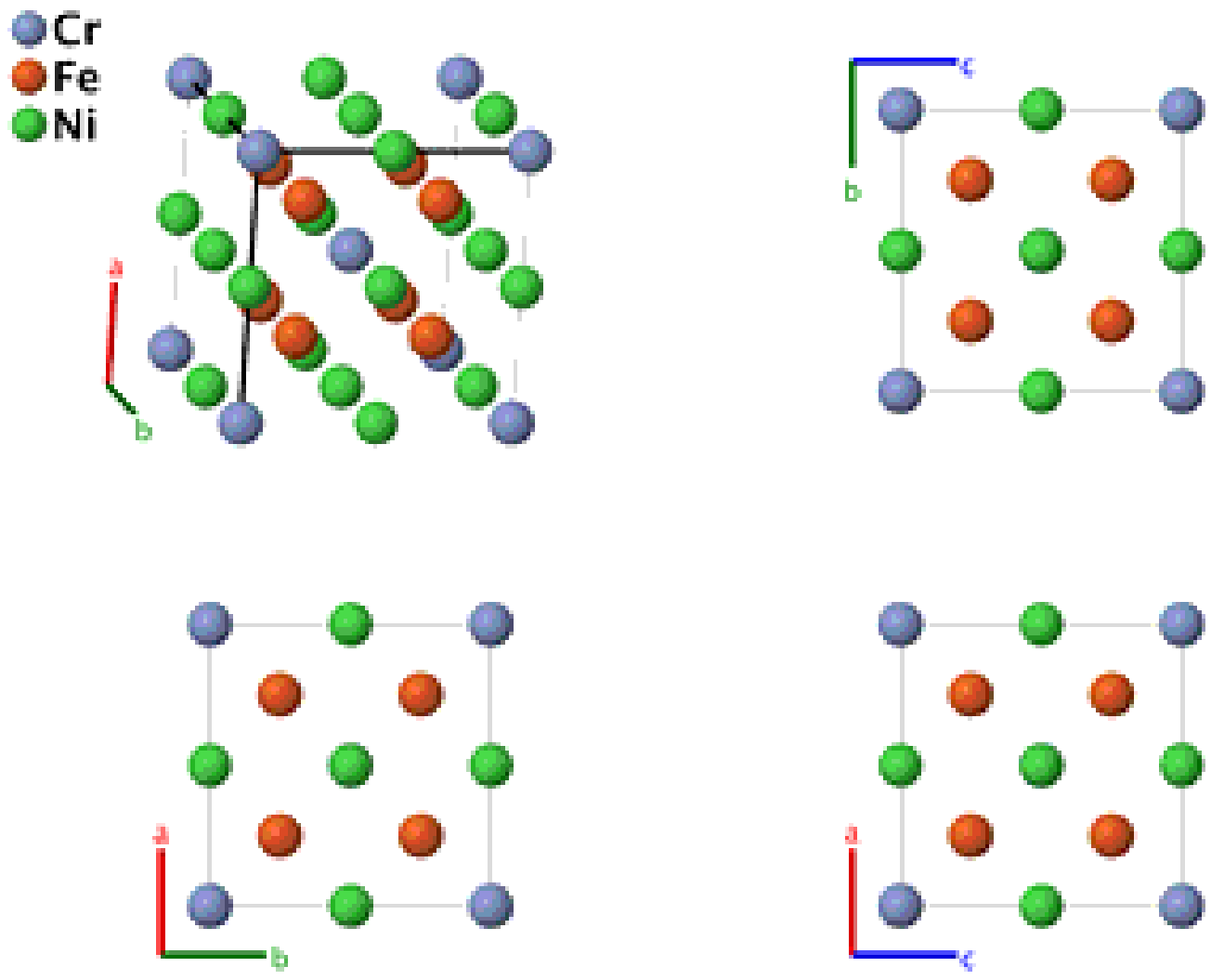

| | | |
|---|---|---|
| **Prototype** | : | $CrFe_4Ni_3$ |
| **AFLOW prototype label** | : | AB4C3_cI16_229_a_c_b |
| ***Strukturbericht* designation** | : | None |
| **Pearson symbol** | : | cI16 |
| **Space group number** | : | 229 |
| **Space group symbol** | : | $Im\bar{3}m$ |
| **AFLOW prototype command** | : | `aflow --proto=AB4C3_cI16_229_a_c_b --params=`$a$ |

- Ferrite is steel with a bcc structure. This structure represents one possible ordering which might be found in an Fe-Ni-Cr steel. Note that it is not meant to represent a real steel. If we replace the Cr atoms by Ni, this becomes the CsCl (B2) structure. If we replace both the Cr and Ni atoms by Fe, we get the bcc (A2) structure. In either case, $a_{bcc/B2} = 1/2a$.

**Body-centered Cubic primitive vectors:**

$$\mathbf{a}_1 = -\frac{1}{2}\,a\,\hat{\mathbf{x}} + \frac{1}{2}\,a\,\hat{\mathbf{y}} + \frac{1}{2}\,a\,\hat{\mathbf{z}}$$
$$\mathbf{a}_2 = \frac{1}{2}\,a\,\hat{\mathbf{x}} - \frac{1}{2}\,a\,\hat{\mathbf{y}} + \frac{1}{2}\,a\,\hat{\mathbf{z}}$$
$$\mathbf{a}_3 = \frac{1}{2}\,a\,\hat{\mathbf{x}} + \frac{1}{2}\,a\,\hat{\mathbf{y}} - \frac{1}{2}\,a\,\hat{\mathbf{z}}$$

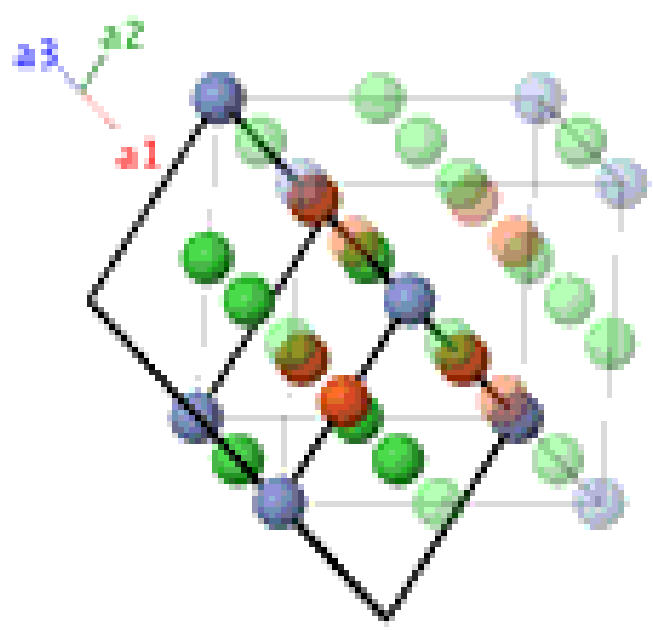

**Basis vectors:**

| | | Lattice Coordinates | | Cartesian Coordinates | Wyckoff Position | Atom Type |
|---|---|---|---|---|---|---|
| $\mathbf{B}_1$ | = | $0\,\mathbf{a}_1 + 0\,\mathbf{a}_2 + 0\,\mathbf{a}_3$ | = | $0\hat{\mathbf{x}} + 0\hat{\mathbf{y}} + 0\hat{\mathbf{z}}$ | $(2a)$ | Cr |
| $\mathbf{B}_2$ | = | $\frac{1}{2}\,\mathbf{a}_2 + \frac{1}{2}\,\mathbf{a}_3$ | = | $\frac{1}{2}\,a\,\hat{\mathbf{x}}$ | $(6b)$ | Ni |
| $\mathbf{B}_3$ | = | $\frac{1}{2}\,\mathbf{a}_1 + \frac{1}{2}\,\mathbf{a}_3$ | = | $\frac{1}{2}\,a\,\hat{\mathbf{y}}$ | $(6b)$ | Ni |
| $\mathbf{B_4}$ | = | $\frac{1}{2}\,\mathbf{a_1} + \frac{1}{2}\,\mathbf{a_2}$ | = | $\frac{1}{2}\,a\,\mathbf{\hat{z}}$ | (6$b$) | Ni |
| $\mathbf{B_5}$ | = | $\frac{1}{2}\,\mathbf{a_1} + \frac{1}{2}\,\mathbf{a_2} + \frac{1}{2}\,\mathbf{a_3}$ | = | $\frac{1}{4}\,a\,\mathbf{\hat{x}} + \frac{1}{4}\,a\,\mathbf{\hat{y}} + \frac{1}{4}\,a\,\mathbf{\hat{z}}$ | (8$c$) | Fe |
| $\mathbf{B_6}$ | = | $\frac{1}{2}\,\mathbf{a_3}$ | = | $\frac{1}{4}\,a\,\mathbf{\hat{x}} + \frac{1}{4}\,a\,\mathbf{\hat{y}} + \frac{3}{4}\,a\,\mathbf{\hat{z}}$ | (8$c$) | Fe |
| $\mathbf{B_7}$ | = | $\frac{1}{2}\,\mathbf{a_2}$ | = | $\frac{1}{4}\,a\,\mathbf{\hat{x}} + \frac{3}{4}\,a\,\mathbf{\hat{y}} + \frac{1}{4}\,a\,\mathbf{\hat{z}}$ | (8$c$) | Fe |
| $\mathbf{B_8}$ | = | $\frac{1}{2}\,\mathbf{a_1}$ | = | $\frac{3}{4}\,a\,\mathbf{\hat{x}} + \frac{1}{4}\,a\,\mathbf{\hat{y}} + \frac{1}{4}\,a\,\mathbf{\hat{z}}$ | (8$c$) | Fe |

---

---

**Geometry files:**

- CIF: pp. 809
- POSCAR: pp. 810

# $Ga_4Ni_3$ Structure: A4B3_cI112_230_af_g

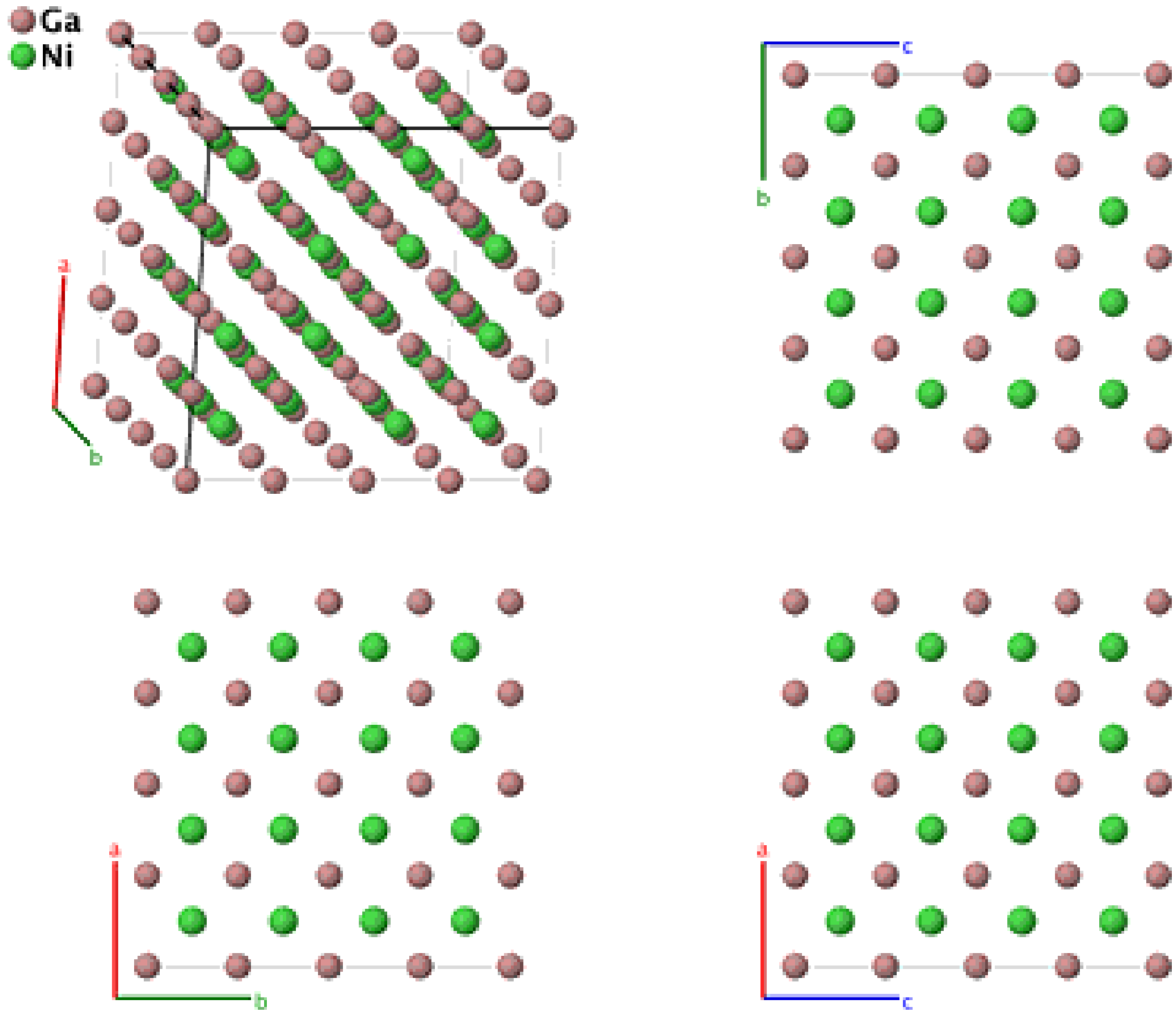

| | | |
|---|---|---|
| **Prototype** | : | $Ga_4Ni_3$ |
| **AFLOW prototype label** | : | A4B3_cI112_230_af_g |
| ***Strukturbericht* designation** | : | None |
| **Pearson symbol** | : | cI112 |
| **Space group number** | : | 230 |
| **Space group symbol** | : | $Ia\bar{3}d$ |
| **AFLOW prototype command** | : | `aflow --proto=A4B3_cI112_230_af_g --params=`$a, x_2, y_3$ |

- This is a simple defect superstructure of the CsCl (B2) structure. If a GaNi B2 structure is expanded into a 128 atom supercell, we can describe it using space group $Ia\bar{3}d$ (#230), with Ga atoms on the (16a) and (48f) Wyckoff sites and Ni atoms on the (16b) and (48g) sites. Removing the Ni atoms from the (16b) sites yields this structure.

**Body-centered Cubic primitive vectors:**

$$\mathbf{a}_1 = -\frac{1}{2}\,a\,\hat{\mathbf{x}} + \frac{1}{2}\,a\,\hat{\mathbf{y}} + \frac{1}{2}\,a\,\hat{\mathbf{z}}$$
$$\mathbf{a}_2 = \frac{1}{2}\,a\,\hat{\mathbf{x}} - \frac{1}{2}\,a\,\hat{\mathbf{y}} + \frac{1}{2}\,a\,\hat{\mathbf{z}}$$
$$\mathbf{a}_3 = \frac{1}{2}\,a\,\hat{\mathbf{x}} + \frac{1}{2}\,a\,\hat{\mathbf{y}} - \frac{1}{2}\,a\,\hat{\mathbf{z}}$$

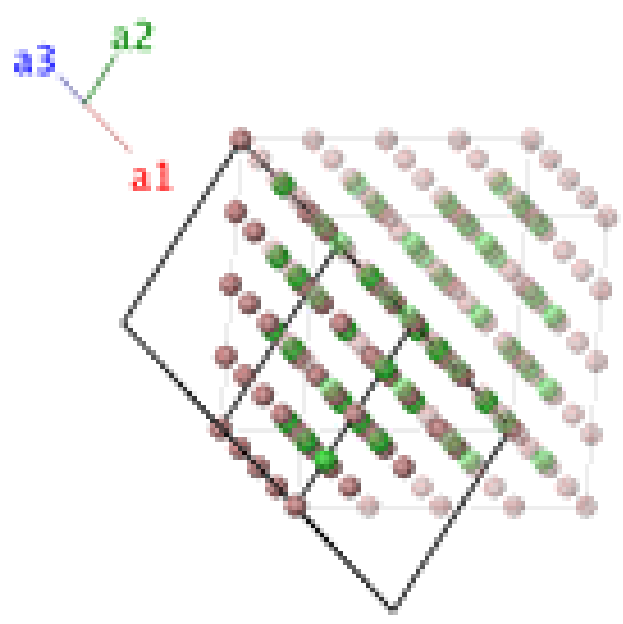

**Basis vectors:**

| | | Lattice Coordinates | | Cartesian Coordinates | Wyckoff Position | Atom Type |
|---|---|---|---|---|---|---|
| $\mathbf{B_1}$ | = | $0\,\mathbf{a_1} + 0\,\mathbf{a_2} + 0\,\mathbf{a_3}$ | = | $0\hat{\mathbf{x}} + 0\hat{\mathbf{y}} + 0\hat{\mathbf{z}}$ | $(16a)$ | Ga I |
| $\mathbf{B_2}$ | = | $\frac{1}{2}\,\mathbf{a_1} + \frac{1}{2}\,\mathbf{a_3}$ | = | $\frac{1}{2}\,a\,\hat{\mathbf{y}}$ | $(16a)$ | Ga I |
| $\mathbf{B_3}$ | = | $\frac{1}{2}\,\mathbf{a_2} + \frac{1}{2}\,\mathbf{a_3}$ | = | $\frac{1}{2}\,a\,\hat{\mathbf{x}}$ | $(16a)$ | Ga I |
| $\mathbf{B_4}$ | = | $\frac{1}{2}\,\mathbf{a_1} + \frac{1}{2}\,\mathbf{a_2}$ | = | $\frac{1}{2}\,a\,\hat{\mathbf{z}}$ | $(16a)$ | Ga I |
| $\mathbf{B_5}$ | = | $\frac{1}{2}\,\mathbf{a_1}$ | = | $\frac{3}{4}\,a\,\hat{\mathbf{x}} + \frac{1}{4}\,a\,\hat{\mathbf{y}} + \frac{1}{4}\,a\,\hat{\mathbf{z}}$ | $(16a)$ | Ga I |
| $\mathbf{B_6}$ | = | $\frac{1}{2}\,\mathbf{a_1} + \frac{1}{2}\,\mathbf{a_2} + \frac{1}{2}\,\mathbf{a_3}$ | = | $\frac{1}{4}\,a\,\hat{\mathbf{x}} + \frac{1}{4}\,a\,\hat{\mathbf{y}} + \frac{1}{4}\,a\,\hat{\mathbf{z}}$ | $(16a)$ | Ga I |
| $\mathbf{B_7}$ | = | $\frac{1}{2}\,\mathbf{a_3}$ | = | $\frac{1}{4}\,a\,\hat{\mathbf{x}} + \frac{1}{4}\,a\,\hat{\mathbf{y}} + \frac{3}{4}\,a\,\hat{\mathbf{z}}$ | $(16a)$ | Ga I |
| $\mathbf{B_8}$ | = | $\frac{1}{2}\,\mathbf{a_2}$ | = | $\frac{1}{4}\,a\,\hat{\mathbf{x}} + \frac{3}{4}\,a\,\hat{\mathbf{y}} + \frac{1}{4}\,a\,\hat{\mathbf{z}}$ | $(16a)$ | Ga I |
| $\mathbf{B_9}$ | = | $\frac{1}{4}\,\mathbf{a_1} + \left(\frac{1}{4} + x_2\right)\,\mathbf{a_2} + x_2\,\mathbf{a_3}$ | = | $x_2\,a\,\hat{\mathbf{x}} + \frac{1}{4}\,a\,\hat{\mathbf{z}}$ | $(48f)$ | Ga II |
| $\mathbf{B_{10}}$ | = | $\frac{3}{4}\,\mathbf{a_1} + \left(\frac{1}{4} - x_2\right)\,\mathbf{a_2} + \left(\frac{1}{2} - x_2\right)\,\mathbf{a_3}$ | = | $-x_2\,a\,\hat{\mathbf{x}} + \frac{1}{2}\,a\,\hat{\mathbf{y}} + \frac{1}{4}\,a\,\hat{\mathbf{z}}$ | $(48f)$ | Ga II |
| $\mathbf{B_{11}}$ | = | $x_2\,\mathbf{a_1} + \frac{1}{4}\,\mathbf{a_2} + \left(\frac{1}{4} + x_2\right)\,\mathbf{a_3}$ | = | $\frac{1}{4}\,a\,\hat{\mathbf{x}} + x_2\,a\,\hat{\mathbf{y}}$ | $(48f)$ | Ga II |
| $\mathbf{B_{12}}$ | = | $\left(\frac{1}{2} - x_2\right)\,\mathbf{a_1} + \frac{3}{4}\,\mathbf{a_2} + \left(\frac{1}{4} - x_2\right)\,\mathbf{a_3}$ | = | $\frac{1}{4}\,a\,\hat{\mathbf{x}} - x_2\,a\,\hat{\mathbf{y}} + \frac{1}{2}\,a\,\hat{\mathbf{z}}$ | $(48f)$ | Ga II |
| $\mathbf{B_{13}}$ | = | $\left(\frac{1}{4} + x_2\right)\,\mathbf{a_1} + x_2\,\mathbf{a_2} + \frac{1}{4}\,\mathbf{a_3}$ | = | $\frac{1}{4}\,a\,\hat{\mathbf{y}} + x_2\,a\,\hat{\mathbf{z}}$ | $(48f)$ | Ga II |
| $\mathbf{B_{14}}$ | = | $\left(\frac{1}{4} - x_2\right)\,\mathbf{a_1} + \left(\frac{1}{2} - x_2\right)\,\mathbf{a_2} + \frac{3}{4}\,\mathbf{a_3}$ | = | $\frac{1}{2}\,a\,\hat{\mathbf{x}}\frac{1}{4}\,a\,\hat{\mathbf{y}} - x_2\,a\,\hat{\mathbf{z}}$ | $(48f)$ | Ga II |
| $\mathbf{B_{15}}$ | = | $\left(\frac{1}{4} + x_2\right)\,\mathbf{a_1} + \frac{3}{4}\,\mathbf{a_2} + x_2\,\mathbf{a_3}$ | = | $\frac{1}{4}\,a\,\hat{\mathbf{x}} + \left(\frac{3}{4} + x_2\right)\,a\,\hat{\mathbf{y}} + \frac{1}{2}\,a\,\hat{\mathbf{z}}$ | $(48f)$ | Ga II |
| $\mathbf{B_{16}}$ | = | $\left(\frac{1}{4} - x_2\right)\,\mathbf{a_1} + \frac{1}{4}\,\mathbf{a_2} + \left(\frac{1}{2} - x_2\right)\,\mathbf{a_3}$ | = | $\frac{1}{4}\,a\,\hat{\mathbf{x}} + \left(\frac{1}{4} - x_2\right)\,a\,\hat{\mathbf{y}}$ | $(48f)$ | Ga II |
| $\mathbf{B_{17}}$ | = | $\frac{3}{4}\,\mathbf{a_1} + x_2\,\mathbf{a_2} + \left(\frac{1}{4} + x_2\right)\,\mathbf{a_3}$ | = | $\left(\frac{3}{4} + x_2\right)\,a\,\hat{\mathbf{x}} + \frac{1}{2}\,a\,\hat{\mathbf{y}} + \frac{1}{4}\,a\,\hat{\mathbf{z}}$ | $(48f)$ | Ga II |
| $\mathbf{B_{18}}$ | = | $\frac{1}{4}\,\mathbf{a_1} + \left(\frac{1}{2} - x_2\right)\,\mathbf{a_2} + \left(\frac{1}{4} - x_2\right)\,\mathbf{a_3}$ | = | $\left(\frac{1}{4} - x_2\right)\,a\,\hat{\mathbf{x}} + \frac{1}{4}\,a\,\hat{\mathbf{z}}$ | $(48f)$ | Ga II |
| $\mathbf{B_{19}}$ | = | $\left(\frac{1}{2} - x_2\right)\,\mathbf{a_1} + \left(\frac{1}{4} - x_2\right)\,\mathbf{a_2} + \frac{1}{4}\,\mathbf{a_3}$ | = | $\frac{1}{4}\,a\,\hat{\mathbf{y}} + \left(\frac{1}{4} - x_2\right)\,a\,\hat{\mathbf{z}}$ | $(48f)$ | Ga II |
| $\mathbf{B_{20}}$ | = | $x_2\,\mathbf{a_1} + \left(\frac{1}{4} + x_2\right)\,\mathbf{a_2} + \frac{3}{4}\,\mathbf{a_3}$ | = | $\frac{1}{2}\,a\,\hat{\mathbf{x}}\frac{1}{4}\,a\,\hat{\mathbf{y}} + \left(\frac{3}{4} + x_2\right)\,a\,\hat{\mathbf{z}}$ | $(48f)$ | Ga II |
| $\mathbf{B_{21}}$ | = | $\frac{3}{4}\,\mathbf{a_1} + \left(\frac{3}{4} - x_2\right)\,\mathbf{a_2} - x_2\,\mathbf{a_3}$ | = | $-x_2\,a\,\hat{\mathbf{x}} + \frac{3}{4}\,a\,\hat{\mathbf{z}}$ | $(48f)$ | Ga II |
| $\mathbf{B_{22}}$ | = | $\frac{1}{4}\,\mathbf{a_1} + \left(x_2 + \frac{3}{4}\right)\,\mathbf{a_2} + \left(\frac{1}{2} + x_2\right)\,\mathbf{a_3}$ | = | $\left(\frac{1}{2} + x_2\right)\,a\,\hat{\mathbf{x}} + \frac{1}{4}\,a\,\hat{\mathbf{z}}$ | $(48f)$ | Ga II |
| $\mathbf{B_{23}}$ | = | $-x_2\,\mathbf{a_1} + \frac{3}{4}\,\mathbf{a_2} + \left(\frac{3}{4} - x_2\right)\,\mathbf{a_3}$ | = | $\frac{3}{4}\,a\,\hat{\mathbf{x}} - x_2\,a\,\hat{\mathbf{y}}$ | $(48f)$ | Ga II |
| $\mathbf{B_{24}}$ | = | $\left(\frac{1}{2} + x_2\right)\,\mathbf{a_1} + \frac{1}{4}\,\mathbf{a_2} + \left(x_2 + \frac{3}{4}\right)\,\mathbf{a_3}$ | = | $\frac{1}{4}\,a\,\hat{\mathbf{x}} + \left(\frac{1}{2} + x_2\right)\,a\,\hat{\mathbf{y}}$ | $(48f)$ | Ga II |
| $\mathbf{B_{25}}$ | = | $\left(\frac{3}{4} - x_2\right)\,\mathbf{a_1} - x_2\,\mathbf{a_2} + \frac{3}{4}\,\mathbf{a_3}$ | = | $\frac{3}{4}\,a\,\hat{\mathbf{y}} - x_2\,a\,\hat{\mathbf{z}}$ | $(48f)$ | Ga II |
| $\mathbf{B_{26}}$ | = | $\left(x_2 + \frac{3}{4}\right)\,\mathbf{a_1} + \left(\frac{1}{2} + x_2\right)\,\mathbf{a_2} + \frac{1}{4}\,\mathbf{a_3}$ | = | $\frac{1}{4}\,a\,\hat{\mathbf{y}} + \left(\frac{1}{2} + x_2\right)\,a\,\hat{\mathbf{z}}$ | $(48f)$ | Ga II |
| $\mathbf{B_{27}}$ | = | $\left(\frac{3}{4} - x_2\right)\,\mathbf{a_1} + \frac{1}{4}\,\mathbf{a_2} - x_2\,\mathbf{a_3}$ | = | $\frac{3}{4}\,a\,\hat{\mathbf{x}} + \left(\frac{1}{4} - x_2\right)\,a\,\hat{\mathbf{y}} + \frac{1}{2}\,a\,\hat{\mathbf{z}}$ | $(48f)$ | Ga II |
| $\mathbf{B_{28}}$ | = | $\left(x_2 + \frac{3}{4}\right)\,\mathbf{a_1} + \frac{3}{4}\,\mathbf{a_2} + \left(\frac{1}{2} + x_2\right)\,\mathbf{a_3}$ | = | $\frac{1}{4}\,a\,\hat{\mathbf{x}} + \left(\frac{1}{4} + x_2\right)\,a\,\hat{\mathbf{y}} + \frac{1}{2}\,a\,\hat{\mathbf{z}}$ | $(48f)$ | Ga II |
| $\mathbf{B_{29}}$ | = | $\frac{1}{4}\,\mathbf{a_1} - x_2\,\mathbf{a_2} + \left(\frac{3}{4} - x_2\right)\,\mathbf{a_3}$ | = | $\left(\frac{1}{4} - x_2\right)\,a\,\hat{\mathbf{x}} + \frac{1}{2}\,a\,\hat{\mathbf{y}} + \frac{3}{4}\,a\,\hat{\mathbf{z}}$ | $(48f)$ | Ga II |
| $\mathbf{B_{30}}$ | = | $\frac{3}{4}\,\mathbf{a_1} + \left(\frac{1}{2} + x_2\right)\,\mathbf{a_2} + \left(x_2 + \frac{3}{4}\right)\,\mathbf{a_3}$ | = | $\left(\frac{1}{4} + x_2\right)\,a\,\hat{\mathbf{x}} + \frac{1}{2}\,a\,\hat{\mathbf{y}} + \frac{1}{4}\,a\,\hat{\mathbf{z}}$ | $(48f)$ | Ga II |
| $\mathbf{B_{31}}$ | = | $\left(\frac{1}{2} + x_2\right)\,\mathbf{a_1} + \left(x_2 + \frac{3}{4}\right)\,\mathbf{a_2} + \frac{3}{4}\,\mathbf{a_3}$ | = | $\frac{1}{2}\,a\,\hat{\mathbf{x}} + \frac{1}{4}\,a\,\hat{\mathbf{y}} + \left(\frac{1}{4} + x_2\right)\,a\,\hat{\mathbf{z}}$ | $(48f)$ | Ga II |
| $\mathbf{B_{32}}$ | = | $-x_2\,\mathbf{a_1} + \left(\frac{3}{4} - x_2\right)\,\mathbf{a_2} + \frac{1}{4}\,\mathbf{a_3}$ | = | $\frac{1}{2}\,a\,\hat{\mathbf{x}}\frac{3}{4}\,a\,\hat{\mathbf{y}} + \left(\frac{1}{4} - x_2\right)\,a\,\hat{\mathbf{z}}$ | $(48f)$ | Ga II |
| $\mathbf{B_{33}}$ | = | $\frac{1}{4}\,\mathbf{a_1} + \left(\frac{3}{8} - y_3\right)\,\mathbf{a_2} + \left(\frac{1}{8} + y_3\right)\,\mathbf{a_3}$ | = | $\frac{1}{8}\,a\,\hat{\mathbf{x}} + y_3\,a\,\hat{\mathbf{y}} + \left(\frac{1}{4} - y_3\right)\,a\,\hat{\mathbf{z}}$ | $(48g)$ | Ni |
| $\mathbf{B_{34}}$ | = | $\left(\frac{3}{4} - 2y_3\right)\,\mathbf{a_1} + \left(\frac{1}{8} - y_3\right)\,\mathbf{a_2} + \left(\frac{3}{8} - y_3\right)\,\mathbf{a_3}$ | = | $\frac{7}{8}\,a\,\hat{\mathbf{x}} + \left(\frac{1}{2} - y_3\right)\,a\,\hat{\mathbf{y}} + \left(\frac{1}{4} - y_3\right)\,a\,\hat{\mathbf{z}}$ | $(48g)$ | Ni |
| $\mathbf{B}_{35}$ | = | $\left(2y_3+\frac{3}{4}\right)\mathbf{a}_1+\left(\frac{1}{8}+y_3\right)\mathbf{a}_2+\left(\frac{3}{8}+y_3\right)\mathbf{a}_3$ | = | $\frac{7}{8}a\,\hat{\mathbf{x}}+\left(\frac{1}{2}+y_3\right)a\,\hat{\mathbf{y}}+\left(\frac{1}{4}+y_3\right)a\,\hat{\mathbf{z}}$ | (48$g$) | Ni |
| $\mathbf{B}_{36}$ | = | $\frac{1}{4}\mathbf{a}_1+\left(\frac{3}{8}+y_3\right)\mathbf{a}_2+\left(\frac{1}{8}-y_3\right)\mathbf{a}_3$ | = | $\frac{1}{8}a\,\hat{\mathbf{x}}-y_3\,a\,\hat{\mathbf{y}}+\left(\frac{1}{4}+y_3\right)a\,\hat{\mathbf{z}}$ | (48$g$) | Ni |
| $\mathbf{B}_{37}$ | = | $\left(\frac{1}{8}+y_3\right)\mathbf{a}_1+\frac{1}{4}\mathbf{a}_2+\left(\frac{3}{8}-y_3\right)\mathbf{a}_3$ | = | $\left(\frac{1}{4}-y_3\right)a\,\hat{\mathbf{x}}+\frac{1}{8}a\,\hat{\mathbf{y}}+y_3\,a\,\hat{\mathbf{z}}$ | (48$g$) | Ni |
| $\mathbf{B}_{38}$ | = | $\left(\frac{3}{8}-y_3\right)\mathbf{a}_1+\left(\frac{3}{4}-2y_3\right)\mathbf{a}_2+\left(\frac{1}{8}-y_3\right)\mathbf{a}_3$ | = | $\left(\frac{1}{4}-y_3\right)a\,\hat{\mathbf{x}}+\frac{7}{8}a\,\hat{\mathbf{y}}+\left(\frac{1}{2}-y_3\right)a\,\hat{\mathbf{z}}$ | (48$g$) | Ni |
| $\mathbf{B}_{39}$ | = | $\left(\frac{3}{8}+y_3\right)\mathbf{a}_1+\left(2y_3+\frac{3}{4}\right)\mathbf{a}_2+\left(\frac{1}{8}+y_3\right)\mathbf{a}_3$ | = | $\left(\frac{1}{4}+y_3\right)a\,\hat{\mathbf{x}}+\frac{7}{8}a\,\hat{\mathbf{y}}+\left(\frac{1}{2}+y_3\right)a\,\hat{\mathbf{z}}$ | (48$g$) | Ni |
| $\mathbf{B}_{40}$ | = | $\left(\frac{1}{8}-y_3\right)\mathbf{a}_1+\frac{1}{4}\mathbf{a}_2+\left(\frac{3}{8}+y_3\right)\mathbf{a}_3$ | = | $\left(\frac{1}{4}+y_3\right)a\,\hat{\mathbf{x}}+\frac{1}{8}a\,\hat{\mathbf{y}}-y_3\,a\,\hat{\mathbf{z}}$ | (48$g$) | Ni |
| $\mathbf{B}_{41}$ | = | $\left(\frac{3}{8}-y_3\right)\mathbf{a}_1+\left(\frac{1}{8}+y_3\right)\mathbf{a}_2+\frac{1}{4}\mathbf{a}_3$ | = | $y_3\,a\,\hat{\mathbf{x}}+\left(\frac{1}{4}-y_3\right)a\,\hat{\mathbf{y}}+\frac{1}{8}a\,\hat{\mathbf{z}}$ | (48$g$) | Ni |
| $\mathbf{B}_{42}$ | = | $\left(\frac{1}{8}-y_3\right)\mathbf{a}_1+\left(\frac{3}{8}-y_3\right)\mathbf{a}_2+\left(\frac{3}{4}-2y_3\right)\mathbf{a}_3$ | = | $\left(\frac{1}{2}-y_3\right)a\,\hat{\mathbf{x}}+\left(\frac{1}{4}-y_3\right)a\,\hat{\mathbf{y}}+\frac{7}{8}a\,\hat{\mathbf{z}}$ | (48$g$) | Ni |
| $\mathbf{B}_{43}$ | = | $\left(\frac{1}{8}+y_3\right)\mathbf{a}_1+\left(\frac{3}{8}+y_3\right)\mathbf{a}_2+\left(2y_3+\frac{3}{4}\right)\mathbf{a}_3$ | = | $\left(\frac{1}{2}+y_3\right)a\,\hat{\mathbf{x}}+\left(\frac{1}{4}+y_3\right)a\,\hat{\mathbf{y}}+\frac{7}{8}a\,\hat{\mathbf{z}}$ | (48$g$) | Ni |
| $\mathbf{B}_{44}$ | = | $\left(\frac{3}{8}+y_3\right)\mathbf{a}_1+\left(\frac{1}{8}-y_3\right)\mathbf{a}_2+\frac{1}{4}\mathbf{a}_3$ | = | $-y_3\,a\,\hat{\mathbf{x}}+\left(\frac{1}{4}+y_3\right)a\,\hat{\mathbf{y}}+\frac{1}{8}a\,\hat{\mathbf{z}}$ | (48$g$) | Ni |
| $\mathbf{B}_{45}$ | = | $\frac{3}{4}\mathbf{a}_1+\left(\frac{5}{8}+y_3\right)\mathbf{a}_2+\left(\frac{7}{8}-y_3\right)\mathbf{a}_3$ | = | $\frac{3}{8}a\,\hat{\mathbf{x}}+\left(\frac{1}{2}-y_3\right)a\,\hat{\mathbf{y}}+\left(\frac{1}{4}+y_3\right)a\,\hat{\mathbf{z}}$ | (48$g$) | Ni |
| $\mathbf{B}_{46}$ | = | $\left(\frac{1}{4}+2y_3\right)\mathbf{a}_1+\left(\frac{7}{8}+y_3\right)\mathbf{a}_2+\left(\frac{5}{8}+y_3\right)\mathbf{a}_3$ | = | $\frac{5}{8}a\,\hat{\mathbf{x}}+y_3\,a\,\hat{\mathbf{y}}+\left(\frac{1}{4}+y_3\right)a\,\hat{\mathbf{z}}$ | (48$g$) | Ni |
| $\mathbf{B}_{47}$ | = | $\left(\frac{1}{4}-2y_3\right)\mathbf{a}_1+\left(\frac{7}{8}-y_3\right)\mathbf{a}_2+\left(\frac{5}{8}-y_3\right)\mathbf{a}_3$ | = | $\frac{5}{8}a\,\hat{\mathbf{x}}-y_3\,a\,\hat{\mathbf{y}}+\left(\frac{1}{4}-y_3\right)a\,\hat{\mathbf{z}}$ | (48$g$) | Ni |
| $\mathbf{B}_{48}$ | = | $\frac{3}{4}\mathbf{a}_1+\left(\frac{5}{8}-y_3\right)\mathbf{a}_2+\left(y_3+\frac{7}{8}\right)\mathbf{a}_3$ | = | $\frac{3}{8}a\,\hat{\mathbf{x}}+\left(\frac{1}{2}+y_3\right)a\,\hat{\mathbf{y}}+\left(\frac{1}{4}-y_3\right)a\,\hat{\mathbf{z}}$ | (48$g$) | Ni |
| $\mathbf{B}_{49}$ | = | $\left(\frac{7}{8}-y_3\right)\mathbf{a}_1+\frac{3}{4}\mathbf{a}_2+\left(\frac{5}{8}+y_3\right)\mathbf{a}_3$ | = | $\left(\frac{1}{4}+y_3\right)a\,\hat{\mathbf{x}}+\frac{3}{8}a\,\hat{\mathbf{y}}+\left(\frac{1}{2}-y_3\right)a\,\hat{\mathbf{z}}$ | (48$g$) | Ni |
| $\mathbf{B}_{50}$ | = | $\left(y_3+\frac{5}{8}\right)\mathbf{a}_1+\left(\frac{1}{4}+2y_3\right)\mathbf{a}_2+\left(\frac{7}{8}+y_3\right)\mathbf{a}_3$ | = | $\left(\frac{1}{4}+y_3\right)a\,\hat{\mathbf{x}}+\frac{5}{8}a\,\hat{\mathbf{y}}+y_3\,a\,\hat{\mathbf{z}}$ | (48$g$) | Ni |
| $\mathbf{B}_{51}$ | = | $\left(\frac{5}{8}-y_3\right)\mathbf{a}_1+\left(\frac{1}{4}-2y_3\right)\mathbf{a}_2+\left(\frac{7}{8}-y_3\right)\mathbf{a}_3$ | = | $\left(\frac{1}{4}-y_3\right)a\,\hat{\mathbf{x}}+\frac{5}{8}a\,\hat{\mathbf{y}}-y_3\,a\,\hat{\mathbf{z}}$ | (48$g$) | Ni |
| $\mathbf{B}_{52}$ | = | $\left(y_3+\frac{7}{8}\right)\mathbf{a}_1+\frac{3}{4}\mathbf{a}_2+\left(\frac{5}{8}-y_3\right)\mathbf{a}_3$ | = | $\left(\frac{1}{4}-y_3\right)a\,\hat{\mathbf{x}}+\frac{3}{8}a\,\hat{\mathbf{y}}+\left(\frac{1}{2}+y_3\right)a\,\hat{\mathbf{z}}$ | (48$g$) | Ni |
| $\mathbf{B}_{53}$ | = | $\left(\frac{5}{8}+y_3\right)\mathbf{a}_1+\left(\frac{7}{8}-y_3\right)\mathbf{a}_2+\frac{3}{4}\mathbf{a}_3$ | = | $\left(\frac{1}{2}-y_3\right)a\,\hat{\mathbf{x}}+\left(\frac{1}{4}+y_3\right)a\,\hat{\mathbf{y}}+\frac{3}{8}a\,\hat{\mathbf{z}}$ | (48$g$) | Ni |
| $\mathbf{B}_{54}$ | = | $\left(\frac{7}{8}+y_3\right)\mathbf{a}_1+\left(y_3+\frac{5}{8}\right)\mathbf{a}_2+\left(\frac{1}{4}+2y_3\right)\mathbf{a}_3$ | = | $y_3\,a\,\hat{\mathbf{x}}+\left(\frac{1}{4}+y_3\right)a\,\hat{\mathbf{y}}+\frac{5}{8}a\,\hat{\mathbf{z}}$ | (48$g$) | Ni |
| $\mathbf{B}_{55}$ | = | $\left(\frac{7}{8}-y_3\right)\mathbf{a}_1+\left(\frac{5}{8}-y_3\right)\mathbf{a}_2+\left(\frac{1}{4}-2y_3\right)\mathbf{a}_3$ | = | $-y_3\,a\,\hat{\mathbf{x}}+\left(\frac{1}{4}-y_3\right)a\,\hat{\mathbf{y}}+\frac{5}{8}a\,\hat{\mathbf{z}}$ | (48$g$) | Ni |
| $\mathbf{B}_{56}$ | = | $\left(\frac{5}{8}-y_3\right)\mathbf{a}_1+\left(y_3+\frac{7}{8}\right)\mathbf{a}_2+\frac{3}{4}\mathbf{a}_3$ | = | $\left(\frac{1}{2}+y_3\right)a\,\hat{\mathbf{x}}+\left(\frac{1}{4}-y_3\right)a\,\hat{\mathbf{y}}+\frac{3}{8}a\,\hat{\mathbf{z}}$ | (48$g$) | Ni |

**Found in:**

**Geometry files:**

- CIF: pp. 810
- POSCAR: pp. 810

## CIF and POSCAR Files

FeS$_2$ (P1): AB2_aP12_1_4a_8a - CIF

```
# CIF file

data_findsym-output
_audit_creation_method FINDSYM

_chemical_name_mineral 'pyrite'
_chemical_formula_sum 'Fe S2'

loop_
_publ_author_name
 'Peter Bayliss'
_journal_name_full
;
American Mineralogist
;
_journal_volume 62
_journal_year 1977
_journal_page_first 1168
_journal_page_last 1172
_publ_Section_title
;
 Crystal structure refinement a weakly anisotropic pyrite
;

_aflow_proto 'AB2_aP12_1_4a_8a'
_aflow_params 'a,b/a,c/a,\alpha,\beta,\gamma,x1,y1,z1,x2,y2,z2,x3,y3,z3,
    ↪ x4,y4,z4,x5,y5,z5,x6,y6,z6,x7,y7,z7,x8,y8,z8,x9,y9,z9,x10,y10,
    ↪ z10,x11,y11,z11,x12,y12,z12'
_aflow_params_values '5.417,1.0,1.0,90.0,90.0,90.0,0.001,0.002,0.003,
    ↪ 0.4966,0.0001,0.5036,0.5001,0.502,0.0011,-0.0006,0.5013,0.5038,
    ↪ 0.3857,0.3832,0.384,0.1149,0.6114,0.8846,0.8854,0.1157,0.6143,
    ↪ 0.6153,0.8865,0.1141,0.6151,0.6132,0.6137,0.8854,0.3818,0.1149,
    ↪ 0.1147,0.8856,0.3841,0.3857,0.1161,0.8842'
_aflow_Strukturbericht 'None'
_aflow_Pearson 'aP12'

_symmetry_space_group_name_Hall "P 1"
_symmetry_space_group_name_H-M "P 1"
_symmetry_Int_Tables_number 1

_cell_length_a      5.41700
_cell_length_b      5.41700
_cell_length_c      5.41700
_cell_angle_alpha 90.00000
_cell_angle_beta  90.00000
_cell_angle_gamma 90.00000

loop_
_space_group_symop_id
_space_group_symop_operation_xyz
1 x,y,z

loop_
_atom_site_label
_atom_site_type_symbol
_atom_site_symmetry_multiplicity
_atom_site_Wyckoff_label
_atom_site_fract_x
_atom_site_fract_y
_atom_site_fract_z
_atom_site_occupancy
Fe1 Fe   1 a  0.00100  0.00200 0.00300 1.00000
Fe2 Fe   1 a  0.49660  0.00010 0.50360 1.00000
Fe3 Fe   1 a  0.50010  0.50200 0.00110 1.00000
Fe4 Fe   1 a -0.00060  0.50130 0.50380 1.00000
S1  S    1 a  0.38570  0.38320 0.38400 1.00000
S2  S    1 a  0.11490  0.61140 0.88460 1.00000
S3  S    1 a  0.88540  0.11570 0.61430 1.00000
S4  S    1 a  0.61530  0.88650 0.11410 1.00000
S5  S    1 a  0.61510  0.61320 0.61370 1.00000
S6  S    1 a  0.88540  0.38180 0.11490 1.00000
S7  S    1 a  0.11470  0.88560 0.38410 1.00000
S8  S    1 a  0.38570  0.11610 0.88420 1.00000
```

FeS$_2$ (P1): AB2_aP12_1_4a_8a - POSCAR

```
AB2_aP12_1_4a_8a & a,b/a,c/a,\alpha,\beta,\gamma,x1,y1,z1,x2,y2,z2,x3,y3
    ↪ ,z3,x4,y4,z4,x5,y5,z5,x6,y6,z6,x7,y7,z7,x8,y8,z8,x9,y9,z9,x10,
    ↪ y10,z10,x11,y11,z11,x12,y12,z12 --params=5.417,1.0,1.0,90.0,
    ↪ 90.0,90.0,0.001,0.002,0.003,0.4966,0.0001,0.5036,0.5001,0.502,
    ↪ 0.0011,-0.0006,0.5013,0.5038,0.3857,0.3832,0.384,0.1149,0.6114,
    ↪ 0.8846,0.8854,0.1157,0.6143,0.6153,0.8865,0.1141,0.6151,0.6132,
    ↪ 0.6137,0.8854,0.3818,0.1149,0.1147,0.8856,0.3841,0.3857,0.1161,
    ↪ 0.8842 & P1        C_1^1 #1 (a^12) & aP12 & & FeS2 & anisotropic
    ↪  Pyrite & Bayliss, Am. Mineral. 62, 1168-72 (1977)
   1.0000000000000000
   5.41700000000000   0.00000000000000   0.00000000000000
   0.00000000000000   5.41700000000000   0.00000000000000
   0.00000000000000   0.00000000000000   5.41700000000000
 (1a)
   12
Direct
  -0.00060000000000   0.50130000000000   0.50380000000000   Fe   (1a)
   0.00100000000000   0.00200000000000   0.00300000000000   Fe   (1a)
   0.49660000000000   0.00010000000000   0.50360000000000   Fe   (1a)
   0.50010000000000   0.50200000000000   0.00110000000000   Fe   (1a)
   0.11470000000000   0.88560000000000   0.38410000000000    S   (1a)
   0.11490000000000   0.61140000000000   0.88460000000000    S   (1a)
   0.38570000000000   0.11610000000000   0.88420000000000    S   (1a)
   0.38570000000000   0.38320000000000   0.38400000000000    S   (1a)
   0.61510000000000   0.61320000000000   0.61370000000000    S   (1a)
   0.61530000000000   0.88650000000000   0.11410000000000    S   (1a)
   0.88540000000000   0.11570000000000   0.61430000000000    S   (1a)
   0.88540000000000   0.38180000000000   0.11490000000000    S   (1a)
```

AsKSe$_2$ (P1): ABC2_aP16_1_4a_4a_8a - CIF

```
# CIF file

data_findsym-output
_audit_creation_method FINDSYM

_chemical_name_mineral ''
_chemical_formula_sum 'As K Se2'

loop_
_publ_author_name
 'W. S. Sheldrick'
 'H. J. Ha\"usler'
_journal_name_full
;
Zeitschrift f\"{u}r anorganische und allgemeine Chemie
;
_journal_volume 561
_journal_year 1988
_journal_page_first 139
_journal_page_last 148
_publ_Section_title
;
 Zur Kenntnis von Alkalimetaselenoarseniten Darstellung und
    ↪ Kristallstrukturen von MAsSe$_2$, M = K, Rb, Cs
;

# Found in  Pearson's Handbook, Vol I., P. 1165

_aflow_proto 'ABC2_aP16_1_4a_4a_8a'
_aflow_params 'a,b/a,c/a,\alpha,\beta,\gamma,x1,y1,z1,x2,y2,z2,x3,y3,z3,
    ↪ x4,y4,z4,x5,y5,z5,x6,y6,z6,x7,y7,z7,x8,y8,z8,x9,y9,z9,x10,y10,
    ↪ z10,x11,y11,z11,x12,y12,z12,x13,y13,z13,x14,y14,z14,x15,y15,z15
    ↪ ,x16,y16,z16'
_aflow_params_values '6.554,1.00061031431,1.92662496186,100.43475,
    ↪ 100.46074,107.53,0.3267,0.582,0.177,0.565,-0.0132,0.4424,0.5217
    ↪ ,0.3883,0.6767,-0.0744,0.6254,-0.0574,0.0338,0.0476,0.2599,
    ↪ 0.0831,0.6072,0.4974,-0.0131,0.0949,0.7583,0.5449,0.1443,-
    ↪ 0.0022,-0.0211,0.5213,0.2073,0.2907,0.5956,-0.0183,-0.0616,
    ↪ 0.0602,0.4998,0.5068,-0.0175,0.2448,0.4596,0.0397,0.708,0.5326,
    ↪ 0.352,0.4818,0.0,0.0,0.0,-0.078,0.569,0.7448'
_aflow_Strukturbericht 'None'
_aflow_Pearson 'aP16'

_symmetry_space_group_name_Hall "P 1"
_symmetry_space_group_name_H-M "P 1"
_symmetry_Int_Tables_number 1

_cell_length_a      6.55400
_cell_length_b      6.55800
_cell_length_c      12.62710
_cell_angle_alpha 100.43475
_cell_angle_beta  100.46074
_cell_angle_gamma 107.53000

loop_
_space_group_symop_id
_space_group_symop_operation_xyz
1 x,y,z

loop_
_atom_site_label
_atom_site_type_symbol
_atom_site_symmetry_multiplicity
_atom_site_Wyckoff_label
_atom_site_fract_x
_atom_site_fract_y
_atom_site_fract_z
_atom_site_occupancy
As1 As   1 a  0.32670  0.58200  0.17700 1.00000
As2 As   1 a  0.56500 -0.01320  0.44240 1.00000
As3 As   1 a  0.52170  0.38830  0.67670 1.00000
As4 As   1 a -0.07440  0.62540 -0.05740 1.00000
K1  K    1 a  0.03380  0.04760  0.25990 1.00000
K2  K    1 a  0.08310  0.60720  0.49740 1.00000
K3  K    1 a -0.01310  0.09490  0.75830 1.00000
K4  K    1 a  0.54490  0.14430 -0.00220 1.00000
Se1 Se   1 a -0.02110  0.52130  0.20730 1.00000
Se2 Se   1 a  0.29070  0.59560 -0.01830 1.00000
Se3 Se   1 a -0.06160  0.06020  0.49980 1.00000
Se4 Se   1 a  0.50680 -0.01750  0.24480 1.00000
Se5 Se   1 a  0.45960  0.03970  0.70800 1.00000
Se6 Se   1 a  0.53260  0.35200  0.48180 1.00000
Se7 Se   1 a  0.00000  0.00000  0.00000 1.00000
Se8 Se   1 a -0.07800  0.56900  0.74480 1.00000
```

AsKSe$_2$ (P1): ABC2_aP16_1_4a_4a_8a - POSCAR

```
ABC2_aP16_1_4a_4a_8a & a,b/a,c/a,\alpha,\beta,\gamma,x1,y1,z1,x2,y2,z2,
    ↪ x3,y3,z3,x4,y4,z4,x5,y5,z5,x6,y6,z6,x7,y7,z7,x8,y8,z8,x9,y9,z9,
    ↪ x10,y10,z10,x11,y11,z11,x12,y12,z12,x13,y13,z13,x14,y14,z14,x15
    ↪ ,y15,z15,x16,y16,z16 --params=6.554,1.00061031431,1.92662496186
    ↪ ,100.43475,100.46074,107.53,0.3267,0.582,0.177,0.565,-0.0132,
    ↪ 0.4424,0.5217,0.3883,0.6767,-0.0744,0.6254,-0.0574,0.0338,
    ↪ 0.0476,0.2599,0.0831,0.6072,0.4974,-0.0131,0.0949,0.7583,0.5449
    ↪ ,0.1443,-0.0022,-0.0211,0.5213,0.2073,0.2907,0.5956,-0.0183,-
    ↪ 0.0616,0.0602,0.4998,0.5068,-0.0175,0.2448,0.4596,0.0397,0.708,
    ↪ 0.5326,0.352,0.4818,0.0,0.0,0.0,-0.078,0.569,0.7448 & P1
    ↪         C_1^1 #1 (a^16) & aP16 & & AsKSe2 & & Sheldrick and
    ↪ Hausler, ZAAC 561, 139-48 (1988)
```

```
   1.0000000000000000
   6.55400000000000    0.00000000000000    0.00000000000000
  -1.97530319852949    6.25344235392629    0.00000000000000
  -2.29259825357476   -3.12251953546006   12.01821614917856
   As   K   Se
    4   4    8
Direct
   0.32670000000000    0.58200000000000    0.17700000000000    As    (1a)
   0.52170000000000    0.38830000000000    0.67670000000000    As    (1a)
   0.56500000000000    0.98680000000000    0.44240000000000    As    (1a)
   0.92560000000000    0.62540000000000    0.94260000000000    As    (1a)
   0.03380000000000    0.04760000000000    0.25990000000000     K    (1a)
   0.08310000000000    0.60720000000000    0.49740000000000     K    (1a)
   0.54490000000000    0.14430000000000    0.99780000000000     K    (1a)
   0.98690000000000    0.09490000000000    0.75830000000000     K    (1a)
   0.00000000000000    0.00000000000000    0.00000000000000    Se    (1a)
   0.29070000000000    0.59560000000000    0.98170000000000    Se    (1a)
   0.45960000000000    0.03970000000000    0.70800000000000    Se    (1a)
   0.50680000000000    0.98250000000000    0.24480000000000    Se    (1a)
   0.53260000000000    0.35200000000000    0.48180000000000    Se    (1a)
   0.92200000000000    0.56900000000000    0.74480000000000    Se    (1a)
   0.93840000000000    0.06020000000000    0.50000000000000    Se    (1a)
   0.97890000000000    0.52130000000000    0.20730000000000    Se    (1a)
```

$P_2I_4$: A2B_aP6_2_2i_i - CIF

```
# CIF file

data_findsym-output
_audit_creation_method FINDSYM

_chemical_name_mineral ''
_chemical_formula_sum 'P2 I4'

loop_
_publ_author_name
 'Yuen Chu Leung'
 'J\"{u}rg Waser'
_journal_name_full
;
Journal of Physical Chemistry
;
_journal_volume 60
_journal_year 1956
_journal_page_first 539
_journal_page_last 543
_publ_Section_title
;
 The Crystal Structure of Phosphorus Diiodide, P$_2$I$_4$
;

# Found in Wyckoff, Vol. I, pp. 375

_aflow_proto 'A2B_aP6_2_2i_i'
_aflow_params 'a,b/a,c/a,\alpha,\beta,\gamma,x1,y1,z1,x2,y2,z2,x3,y3,z3'
_aflow_params_values '4.56,1.54824561404,1.62280701754,80.2,106.96667,
    ↪ 98.2,0.557,0.73,0.165,0.82,0.803,0.695,0.397,0.639,0.463'
_aflow_Strukturbericht 'None'
_aflow_Pearson 'aP6'

_symmetry_space_group_name_Hall "-P 1"
_symmetry_space_group_name_H-M "P -1"
_symmetry_Int_Tables_number 2

_cell_length_a     4.56000
_cell_length_b     7.06000
_cell_length_c     7.40000
_cell_angle_alpha  80.20000
_cell_angle_beta   106.96667
_cell_angle_gamma  98.20000

loop_
_space_group_symop_id
_space_group_symop_operation_xyz
1 x,y,z
2 -x,-y,-z

loop_
_atom_site_label
_atom_site_type_symbol
_atom_site_symmetry_multiplicity
_atom_site_Wyckoff_label
_atom_site_fract_x
_atom_site_fract_y
_atom_site_fract_z
_atom_site_occupancy
I1 I   2 i 0.55700 0.73000 0.16500 1.00000
I2 I   2 i 0.82000 0.80300 0.69500 1.00000
P1 P   2 i 0.39700 0.63900 0.46300 1.00000
```

$P_2I_4$: A2B_aP6_2_2i_i - POSCAR

```
A2B_aP6_2_2i_i & a,b/a,c/a,\alpha,\beta,\gamma,x1,y1,z1,x2,y2,z2,x3,y3,
    ↪ z3 --params=4.56,1.54824561404,1.62280701754,80.2,106.96667,
    ↪ 98.2,0.557,0.73,0.165,0.82,0.803,0.695,0.397,0.639,0.463 & P(-1
    ↪ )      C_i^1 #2 (i^3) & aP6 & & P2I4 & & Y. C. Leung and J.
    ↪ Waser, J. Phys. Chem. 60, 539-43 (1956)
   1.0000000000000000
   4.56000000000000    0.00000000000000    0.00000000000000
  -1.00696027196000    6.98782019021000    0.00000000000000
  -2.15943321793000    0.96138157983000    7.01231727999000
    I   P
    4   2
Direct
   0.55700000000000    0.73000000000000    0.16500000000000     I    (2i)
   0.44300000000000    0.27000000000000    0.83500000000000     I    (2i)
   0.82000000000000    0.80300000000000    0.69500000000000     I    (2i)
   0.18000000000000    0.19700000000000    0.30500000000000     I    (2i)
   0.39700000000000    0.63900000000000    0.46300000000000     P    (2i)
   0.60300000000000    0.36100000000000    0.53700000000000     P    (2i)
```

Cf: A_aP4_2_aci - CIF

```
# CIF file

data_findsym-output
_audit_creation_method FINDSYM

_chemical_name_mineral 'High Pressure Californium'
_chemical_formula_sum 'Cf'

loop_
_publ_author_name
 'R. B. Roof'
_journal_name_full
;
Journal of the Less-Common Metals
;
_journal_volume 120
_journal_year 1986
_journal_page_first 345
_journal_page_last 349
_publ_Section_title
;
 Concerning the Structure of a High Pressure Phase in Californium Metal
;

# Found in Pearson's Handbook, Vol. 2, p. 2332

_aflow_proto 'A_aP4_2_aci'
_aflow_params 'a,b/a,c/a,\alpha,\beta,\gamma,x3,y3,z3'
_aflow_params_values '3.307,2.24130631993,0.844572119746,89.06,85.15,
    ↪ 85.7,0.572,0.259,0.433'
_aflow_Strukturbericht 'None'
_aflow_Pearson 'aP4'

_symmetry_space_group_name_Hall "-P 1"
_symmetry_space_group_name_H-M "P -1"
_symmetry_Int_Tables_number 2

_cell_length_a     3.30700
_cell_length_b     7.41200
_cell_length_c     2.79300
_cell_angle_alpha  89.06000
_cell_angle_beta   85.15000
_cell_angle_gamma  85.70000

loop_
_space_group_symop_id
_space_group_symop_operation_xyz
1 x,y,z
2 -x,-y,-z

loop_
_atom_site_label
_atom_site_type_symbol
_atom_site_symmetry_multiplicity
_atom_site_Wyckoff_label
_atom_site_fract_x
_atom_site_fract_y
_atom_site_fract_z
_atom_site_occupancy
Cf1 Cf   1 a 0.00000 0.00000 0.00000 1.00000
Cf2 Cf   1 c 0.00000 0.50000 0.00000 1.00000
Cf3 Cf   2 i 0.57200 0.25900 0.43300 1.00000
```

Cf: A_aP4_2_aci - POSCAR

```
A_aP4_2_aci & a,b/a,c/a,\alpha,\beta,\gamma,x3,y3,z3 --params=3.307,
    ↪ 2.24130631993,0.844572119746,89.06,85.15,85.7,0.572,0.259,0.433
    ↪ & P(-1)      C_i^1 #2 (aci) & aP4 & & Cf & 30-40 GPa & R. B.
    ↪ Roof, Journal of the Less-Common Metals 120, 345-9 (1986)
   1.0000000000000000
   3.30700000000000    0.00000000000000    0.00000000000000
   0.55574232324000    7.39113620969000    0.00000000000000
   0.23614093129000    0.02819398361000    2.78285672643000
   Cf
    4
Direct
   0.00000000000000    0.00000000000000    0.00000000000000    Cf    (1a)
   0.00000000000000    0.50000000000000    0.00000000000000    Cf    (1c)
   0.42800000000000    0.74100000000000    0.56700000000000    Cf    (2i)
   0.57200000000000    0.25900000000000    0.43300000000000    Cf    (2i)
```

$SiO_2$ (P2): A2B_mP12_3_bc3e_2e - CIF

```
# CIF file

data_findsym-output
_audit_creation_method FINDSYM

_chemical_name_mineral ''
_chemical_formula_sum 'Si O2'

loop_
_publ_author_name
 'M. B. Boisen, Jr.'
 'G. V. Gibbs'
 'M. S. T. Bukowinski'
```

```
_journal_name_full
;
Physics and Chemistry of Minerals
;
_journal_volume 21
_journal_year 1994
_journal_page_first 269
_journal_page_last 284
_publ_Section_title
;
 Framework silica structures generated using simulated annealing with a
       ↪ potential energy function based on an H$_6$Si$_2$O$_7$
       ↪ molecule
;

_aflow_proto 'A2B_mP12_3_bc3e_2e'
_aflow_params 'a,b/a,c/a,\beta,y1,y2,x3,y3,z3,x4,y4,z4,x5,y5,z5,x6,y6,z6
       ↪ ,x7,y7,z7'
_aflow_params_values '4.1605,0.992524936907,1.78370388174,101.3752,
       ↪ 0.15907,0.73859,0.02399,0.752,0.18927,0.38562,0.71473,0.64074,
       ↪ 0.48963,0.20196,0.18802,0.18244,0.0,0.69651,0.38098,0.58564,
       ↪ 0.17797'
_aflow_Strukturbericht 'None'
_aflow_Pearson 'mP12'

_symmetry_space_group_name_Hall "P 2y"
_symmetry_space_group_name_H-M "P 1 2 1"
_symmetry_Int_Tables_number 3

_cell_length_a      4.16050
_cell_length_b      4.12940
_cell_length_c      7.42110
_cell_angle_alpha  90.00000
_cell_angle_beta  101.37520
_cell_angle_gamma  90.00000

loop_
_space_group_symop_id
_space_group_symop_operation_xyz
1 x,y,z
2 -x,y,-z

loop_
_atom_site_label
_atom_site_type_symbol
_atom_site_symmetry_multiplicity
_atom_site_Wyckoff_label
_atom_site_fract_x
_atom_site_fract_y
_atom_site_fract_z
_atom_site_occupancy
O1  O    1 b 0.00000 0.15907 0.50000 1.00000
O2  O    1 c 0.50000 0.73859 0.00000 1.00000
O3  O    2 e 0.02399 0.75200 0.18927 1.00000
O4  O    2 e 0.38562 0.71473 0.64074 1.00000
O5  O    2 e 0.48963 0.20196 0.18802 1.00000
Si1 Si   2 e 0.18244 0.00000 0.69651 1.00000
Si2 Si   2 e 0.38098 0.58564 0.17797 1.00000
```

$SiO_2$ (P2): A2B_mP12_3_bc3e_2e - POSCAR

```
A2B_mP12_3_bc3e_2e & a,b/a,c/a,\beta,y1,y2,x3,y3,z3,x4,y4,z4,x5,y5,z5,x6
       ↪ ,y6,z6,x7,y7,z7 --params=4.1605,0.992524936907,1.78370388174,
       ↪ 101.3752,0.15907,0.73859,0.02399,0.752,0.18927,0.38562,0.71473,
       ↪ 0.64074,0.48963,0.20196,0.18802,0.18244,0.0,0.69651,0.38098,
       ↪ 0.58564,0.17797 & P2         C_2^1 #3 (bce^5) & mP12 & & SiO2 &
       ↪ derived from potential minimization & Boisen et all, Phys.
       ↪ Chem. Min. 21, 269-84 (1994)
   1.0000000000000000
   4.16050000000000   0.00000000000000   0.00000000000000
   0.00000000000000   4.12940000000000   0.00000000000000
  -1.46368596000000   0.00000000000000   7.27532464000000
    O    Si
    8    4
Direct
   0.00000000000000   0.15907000000000   0.50000000000000   O    (1b)
   0.50000000000000   0.73859000000000   0.00000000000000   O    (1c)
   0.02399000000000   0.75200000000000   0.18927000000000   O    (2e)
  -0.02399000000000   0.75200000000000  -0.18927000000000   O    (2e)
   0.38562000000000   0.71473000000000   0.64074000000000   O    (2e)
  -0.38562000000000   0.71473000000000  -0.64074000000000   O    (2e)
   0.48963000000000   0.20196000000000   0.18802000000000   O    (2e)
  -0.48963000000000   0.20196000000000  -0.18802000000000   O    (2e)
   0.18244000000000   0.00000000000000   0.69651000000000   Si   (2e)
  -0.18244000000000   0.00000000000000  -0.69651000000000   Si   (2e)
   0.38098000000000   0.58564000000000   0.17797000000000   Si   (2e)
  -0.38098000000000   0.58564000000000  -0.17797000000000   Si   (2e)
```

High-Pressure Te: A_mP4_4_2a - CIF

```
# CIF file

data_findsym-output
_audit_creation_method FINDSYM

_chemical_name_mineral 'High Pressure (4-7GPa) Tellurium'
_chemical_formula_sum 'Te'

loop_
_publ_author_name
 'Katsutoshi Aoki'
 'Osamu Shimomura'
 'Shigeru Minomura'
_journal_name_full
;
Journal of the Physical Society of Japan
;
_journal_volume 48
_journal_year 1980
_journal_page_first 551
_journal_page_last 556
_publ_Section_title
;
 Crystal Structure of the High-Pressure Phase of Tellurium
;

_aflow_proto 'A_mP4_4_2a'
_aflow_params 'a,b/a,c/a,\beta,x1,y1,z1,x2,y2,z2'
_aflow_params_values '3.104,2.42042525773,1.53350515464,92.71,0.25,0.23,
       ↪ 0.48,0.48,0.0,0.02'
_aflow_Strukturbericht 'None'
_aflow_Pearson 'mP4'

_symmetry_space_group_name_Hall "P 2yb"
_symmetry_space_group_name_H-M "P 1 21 1"
_symmetry_Int_Tables_number 4

_cell_length_a      3.10400
_cell_length_b      7.51300
_cell_length_c      4.76000
_cell_angle_alpha  90.00000
_cell_angle_beta   92.71000
_cell_angle_gamma  90.00000

loop_
_space_group_symop_id
_space_group_symop_operation_xyz
1 x,y,z
2 -x,y+1/2,-z

loop_
_atom_site_label
_atom_site_type_symbol
_atom_site_symmetry_multiplicity
_atom_site_Wyckoff_label
_atom_site_fract_x
_atom_site_fract_y
_atom_site_fract_z
_atom_site_occupancy
Te1 Te    2 a 0.25000 0.23000 0.48000 1.00000
Te2 Te    2 a 0.48000 0.00000 0.02000 1.00000
```

High-Pressure Te: A_mP4_4_2a - POSCAR

```
A_mP4_4_2a & a,b/a,c/a,\beta,x1,y1,z1,x2,y2,z2 --params=3.104,
       ↪ 2.42042525773,1.53350515464,92.71,0.25,0.23,0.48,0.48,0.0,0.02
       ↪ & P2_1        C_2^2 #4 (a^2) & mP4 & & Te (4-7 GPa) & & K. Aoki,
       ↪ O. Shimomura, and S. Minomura, J. Phys. Soc. Jpn. 48 (2) 551-6
       ↪ (1980)
   1.0000000000000000
   3.10400000000000   0.00000000000000   0.00000000000000
   0.00000000000000   7.51300000000000   0.00000000000000
  -0.22505656000000   0.00000000000000   4.75467660000000
   Te
    4
Direct
   0.25000000000000   0.23000000000000   0.48000000000000   Te   (2a)
   0.75000000000000   0.73000000000000   0.52000000000000   Te   (2a)
   0.48000000000000   0.00000000000000   0.02000000000000   Te   (2a)
   0.52000000000000   0.50000000000000  -0.02000000000000   Te   (2a)
```

Po (A19): A_mC12_5_3c - CIF

```
# CIF file

data_findsym-output
_audit_creation_method FINDSYM

_chemical_name_mineral ''
_chemical_formula_sum 'Po'

loop_
_publ_author_name
 'M. A. Rollier'
 'S. B. Hendricks'
 'Louis R. Maxwell'
_journal_name_full
;
Journal of Chemical Physics
;
_journal_volume 4
_journal_year 1936
_journal_page_first 648
_journal_page_last 652
_publ_Section_title
;
 The Crystal Structure of Polonium by Electron Diffraction
;

# Found in  AMS Database

_aflow_proto 'A_mC12_5_3c'
_aflow_params 'a,b/a,c/a,\beta,x1,y1,z1,x2,y2,z2,x3,y3,z3'
_aflow_params_values '7.42,0.578167115903,1.90026954178,92.0,0.05,0.27,
       ↪ 0.245,0.63,0.3,0.4,0.245,0.43,0.07'
_aflow_Strukturbericht 'A19'
_aflow_Pearson 'mC12'

_symmetry_space_group_name_Hall "C 2y"
_symmetry_space_group_name_H-M "C 1 2 1"
```

```
_symmetry_Int_Tables_number 5

_cell_length_a      7.42000
_cell_length_b      4.29000
_cell_length_c      14.10000
_cell_angle_alpha 90.00000
_cell_angle_beta    92.00000
_cell_angle_gamma 90.00000

loop_
_space_group_symop_id
_space_group_symop_operation_xyz
1 x,y,z
2 -x,y,-z
3 x+1/2,y+1/2,z
4 -x+1/2,y+1/2,-z

loop_
_atom_site_label
_atom_site_type_symbol
_atom_site_symmetry_multiplicity
_atom_site_Wyckoff_label
_atom_site_fract_x
_atom_site_fract_y
_atom_site_fract_z
_atom_site_occupancy
Po1 Po    4 c 0.05000 0.27000 0.24500 1.00000
Po2 Po    4 c 0.63000 0.30000 0.40000 1.00000
Po3 Po    4 c 0.24500 0.43000 0.07000 1.00000
```

Po (A19): A_mC12_5_3c - POSCAR

```
A_mC12_5_3c & a,b/a,c/a,\beta,x1,y1,z1,x2,y2,z2,x3,y3,z3 --params=7.42,
    ↪ 0.578167115903,1.90026954178,92.0,0.05,0.27,0.245,0.63,0.3,0.4,
    ↪ 0.245,0.43,0.07 & C2         C_2^3 #5 (c^3) & mC12 & A19 & Po &
    ↪ & M. A. Rollier, S. B. Hendricks, and L. R. Maxwell, JCP 4,
    ↪ 648-52 (1936)
   1.0000000000000000
   3.71000000000000  -2.14500000000000   0.00000000000000
   3.71000000000000   2.14500000000000   0.00000000000000
  -0.49208290350527   0.00000000000000  14.09141066096925
   Po
    6
Direct
   0.68000000000000   0.22000000000000   0.75500000000000   Po   (4c)
   0.78000000000000   0.32000000000000   0.24500000000000   Po   (4c)
   0.07000000000000   0.67000000000000   0.60000000000000   Po   (4c)
   0.33000000000000  -0.07000000000000   0.40000000000000   Po   (4c)
   0.32500000000000   0.18500000000000  -0.07000000000000   Po   (4c)
   0.81500000000000   0.67500000000000   0.07000000000000   Po   (4c)
```

Monoclinic PZT [$Pb(Zr_xTi_{1-x})O_3$]: A3BC_mC10_8_ab_a_a - CIF

```
# CIF file

data_findsym-output
_audit_creation_method FINDSYM

_chemical_name_mineral 'Pb (Zr_0.50 Ti_0.48) O_3'
_chemical_formula_sum 'Pb Zr O3'

loop_
_publ_author_name
 'B. Noheda'
 'J. A. Gonzalo'
 'L. E. Cross'
 'R. Guo'
 'S.-E. Park'
 'D. E. Cox'
 'G. Shirane'
_journal_name_full
;
Physical Review B
;
_journal_volume 61
_journal_year 2000
_journal_page_first 8687
_journal_page_last 8695
_publ_Section_title
;
 Tetragonal-to-monoclinic phase transition in a ferroelectric perovskite
    ↪ : The structure of PbZr$_{0.52}$Ti$_{0.48}$O$_3$
;

_aflow_proto 'A3BC_mC10_8_ab_a_a'
_aflow_params 'a,b/a,c/a,\beta,x1,z1,x2,z2,x3,z3,x4,y4,z4'
_aflow_params_values '5.72204,0.9978207073,0.722908263486,90.498,0.5515
    ↪ ,-0.0994,0.0,0.0,0.523,0.4492,0.288,0.2434,0.3729'
_aflow_Strukturbericht 'None'
_aflow_Pearson 'mC10'

_symmetry_space_group_name_Hall "C -2y"
_symmetry_space_group_name_H-M "C 1 m 1"
_symmetry_Int_Tables_number 8

_cell_length_a      5.72204
_cell_length_b      5.70957
_cell_length_c      4.13651
_cell_angle_alpha 90.00000
_cell_angle_beta    90.49800
_cell_angle_gamma 90.00000

loop_
_space_group_symop_id
_space_group_symop_operation_xyz
1 x,y,z
2 x,-y,z
3 x+1/2,y+1/2,z
4 x+1/2,-y+1/2,z

loop_
_atom_site_label
_atom_site_type_symbol
_atom_site_symmetry_multiplicity
_atom_site_Wyckoff_label
_atom_site_fract_x
_atom_site_fract_y
_atom_site_fract_z
_atom_site_occupancy
O1   O     2 a 0.55150 0.00000 -0.09940 1.00000
Pb1 Pb     2 a 0.00000 0.00000 0.00000  1.00000
Zr1  Zr    2 a 0.52300 0.00000 0.44920  1.00000
O2   O     4 b 0.28800 0.24340 0.37290  1.00000
```

Monoclinic PZT [$Pb(Zr_xTi_{1-x})O_3$]: A3BC_mC10_8_ab_a_a - POSCAR

```
A3BC_mC10_8_ab_a_a & a,b/a,c/a,\beta,x1,z1,x2,z2,x3,z3,x4,y4,z4 --params
    ↪ =5.72204,0.9978207073,0.722908263486,90.498,0.5515,-0.0994,0.0,
    ↪ 0.0,0.523,0.4492,0.288,0.2434,0.3729 & Cm        C_s^4 #8 (a^3b
    ↪ ) & mC10 & & Pb(Zr_0.52Ti_0.48)O_3 & Monoclinic PZT & B. Noheda
    ↪  et al., PRB 61, 8687 (2000)
   1.0000000000000000
   2.86102000000000  -2.85478500000000   0.00000000000000
   2.86102000000000   2.85478500000000   0.00000000000000
  -0.03595301539232   0.00000000000000   4.13635375189117
    O    Pb    Zr
    3     1     1
Direct
   0.55150000000000   0.55150000000000  -0.09940000000000   O    (2a)
   0.04460000000000   0.53140000000000   0.37290000000000   O    (4b)
   0.53140000000000   0.04460000000000   0.37290000000000   O    (4b)
   0.00000000000000   0.00000000000000   0.00000000000000   Pb   (2a)
   0.52300000000000   0.52300000000000   0.44920000000000   Zr   (2a)
```

Monoclinic (Cc) Low Tridymite ($SiO_2$): A2B_mC144_9_24a_12a - CIF

```
# CIF file

data_findsym-output
_audit_creation_method FINDSYM

_chemical_name_mineral 'Low Tridymite'
_chemical_formula_sum 'Si O2'

loop_
_publ_author_name
 'Wayne A. Dollase'
 'Werner H. Baur'
_journal_name_full
;
American Mineralogist
;
_journal_volume 61
_journal_year 1976
_journal_page_first 971
_journal_page_last 978
_publ_Section_title
;
 The superstructure of meteoritic low tridymite solved by computer
    ↪ simulation
;

_aflow_proto 'A2B_mC144_9_24a_12a'
_aflow_params 'a,b/a,c/a,\beta,x1,y1,z1,x2,y2,z2,x3,y3,z3,x4,y4,z4,x5,y5
    ↪ ,z5,x6,y6,z6,x7,y7,z7,x8,y8,z8,x9,y9,z9,x10,y10,z10,x11,y11,z11
    ↪ ,x12,y12,z12,x13,y13,z13,x14,y14,z14,x15,y15,z15,x16,y16,z16,
    ↪ x17,y17,z17,x18,y18,z18,x19,y19,z19,x20,y20,z20,x21,y21,z21,x22
    ↪ ,y22,z22,x23,y23,z23,x24,y24,z24,x25,y25,z25,x26,y26,z26,x27,
    ↪ y27,z27,x28,y28,z28,x29,y29,z29,x30,y30,z30,x31,y31,z31,x32,y32
    ↪ ,z32,x33,y33,z33,x34,y34,z34,x35,y35,z35,x36,y36,z36'
_aflow_params_values '18.524,0.270092852516,1.28535953358,105.82,0.5749,
    ↪ 0.351,0.8182,0.0707,0.34,0.8476,0.7315,0.138,0.4851,0.2509,
    ↪ 0.144,0.5152,0.4155,0.352,0.6741,-0.0873,0.352,0.6434,0.8773,
    ↪ 0.164,-0.0787,0.416,0.168,-0.0639,0.7741,0.145,0.7538,0.2336,
    ↪ 0.143,0.7402,0.6195,0.341,0.5847,0.0811,0.343,0.5661,-0.0034,
    ↪ 0.011,0.6062,0.3533,0.489,0.5665,0.6498,0.005,0.6711,0.1524,
    ↪ 0.496,0.7805,0.8636,0.499,0.7328,0.3361,0.003,0.8333,0.0052,
    ↪ 0.493,0.7398,0.1369,0.011,-0.0732,0.4927,0.492,0.8868,0.5,0.468
    ↪ ,0.5,0.2252,0.491,0.5898,0.2744,0.021,-0.0845,0.0507,0.041,
    ↪ 0.5642,0.2036,0.447,0.7347,-0.0802,0.049,0.6225,0.5751,0.043,
    ↪ 0.7955,0.4247,0.048,0.6971,0.2643,0.444,0.5386,0.8023,0.449,
    ↪ 0.7661,0.6453,0.041,0.6027,0.8531,0.463,-0.0984,0.4493,0.466,-
    ↪ 0.0642,0.2244,0.059,-0.0395,0.0697,0.049,0.8702'
_aflow_Strukturbericht 'None'
_aflow_Pearson 'mC144'

_symmetry_space_group_name_Hall "C -2yc"
_symmetry_space_group_name_H-M "C 1 c 1"
_symmetry_Int_Tables_number 9

_cell_length_a      18.52400
_cell_length_b      5.00320
_cell_length_c      23.81000
_cell_angle_alpha 90.00000
_cell_angle_beta    105.82000
_cell_angle_gamma 90.00000

loop_
_space_group_symop_id
_space_group_symop_operation_xyz
```

```
1 x,y,z
2 x,-y,z+1/2
3 x+1/2,y+1/2,z
4 x+1/2,-y+1/2,z+1/2

loop_
_atom_site_label
_atom_site_type_symbol
_atom_site_symmetry_multiplicity
_atom_site_Wyckoff_label
_atom_site_fract_x
_atom_site_fract_y
_atom_site_fract_z
_atom_site_occupancy
O1   O    4 a 0.57490   0.35100  0.81820   1.00000
O2   O    4 a 0.07070   0.34000  0.84760   1.00000
O3   O    4 a 0.73150   0.13800  0.48510   1.00000
O4   O    4 a 0.25090   0.14400  0.51520   1.00000
O5   O    4 a 0.41550   0.35200  0.67410   1.00000
O6   O    4 a -0.08730  0.35200  0.64340   1.00000
O7   O    4 a 0.87730   0.16400  -0.07870  1.00000
O8   O    4 a 0.41600   0.16800  -0.06390  1.00000
O9   O    4 a 0.77410   0.14500  0.75380   1.00000
O10  O    4 a 0.23360   0.14300  0.74020   1.00000
O11  O    4 a 0.61950   0.34100  0.58470   1.00000
O12  O    4 a 0.08110   0.34300  0.56610   1.00000
O13  O    4 a -0.00340  0.01100  0.60620   1.00000
O14  O    4 a 0.35330   0.48900  0.56650   1.00000
O15  O    4 a 0.64980   0.00500  0.67110   1.00000
O16  O    4 a 0.15240   0.49600  0.78050   1.00000
O17  O    4 a 0.86360   0.49900  0.73280   1.00000
O18  O    4 a 0.33610   0.00300  0.83330   1.00000
O19  O    4 a 0.00520   0.49300  0.73980   1.00000
O20  O    4 a 0.13690   0.01100  -0.07320  1.00000
O21  O    4 a 0.49270   0.49200  0.88680   1.00000
O22  O    4 a 0.50000   0.46800  0.50000   1.00000
O23  O    4 a 0.22520   0.49100  0.58980   1.00000
O24  O    4 a 0.27440   0.02100  -0.08450  1.00000
Si1  Si   4 a 0.05070   0.04100  0.56420   1.00000
Si2  Si   4 a 0.20360   0.44700  0.73470   1.00000
Si3  Si   4 a -0.08020  0.04900  0.62250   1.00000
Si4  Si   4 a 0.57510   0.04300  0.79550   1.00000
Si5  Si   4 a 0.42470   0.04800  0.69710   1.00000
Si6  Si   4 a 0.26430   0.44400  0.53860   1.00000
Si7  Si   4 a 0.80230   0.44900  0.76610   1.00000
Si8  Si   4 a 0.64530   0.04100  0.60270   1.00000
Si9  Si   4 a 0.85310   0.46300  -0.09840  1.00000
Si10 Si   4 a 0.44930   0.46600  -0.06420  1.00000
Si11 Si   4 a 0.22440   0.05900  -0.03950  1.00000
Si12 Si   4 a 0.06970   0.04900  0.87020   1.00000
```

Monoclinic (Cc) Low Tridymite ($SiO_2$): A2B_mC144_9_24a_12a - POSCAR

```
A2B_mC144_9_24a_12a & a,b/a,c/a,\beta,x1,y1,z1,x2,y2,z2,x3,y3,z3,x4,y4,
    ↪ z4,x5,y5,z5,x6,y6,z6,x7,y7,z7,x8,y8,z8,x9,y9,z9,x10,y10,z10,x11
    ↪ ,y11,z11,x12,y12,z12,x13,y13,z13,x14,y14,z14,x15,y15,z15,x16,
    ↪ y16,z16,x17,y17,z17,x18,y18,z18,x19,y19,z19,x20,y20,z20,x21,y21
    ↪ ,z21,x22,y22,z22,x23,y23,z23,x24,y24,z24,x25,y25,z25,x26,y26,
    ↪ z26,x27,y27,z27,x28,y28,z28,x29,y29,z29,x30,y30,z30,x31,y31,z31
    ↪ ,x32,y32,z32,x33,y33,z33,x34,y34,z34,x35,y35,z35,x36,y36,z36 --
    ↪ params=18.524,0.270092852516,1.28535953358,105.82,0.5749,0.351,
    ↪ 0.8182,0.0707,0.34,0.8476,0.7315,0.138,0.4851,0.2509,0.144,
    ↪ 0.5152,0.4155,0.352,0.6741,-0.0873,0.352,0.6434,0.8773,0.164,-
    ↪ 0.0787,0.416,0.168,-0.0639,0.7741,0.145,0.7538,0.2336,0.143,
    ↪ 0.7402,0.6195,0.341,0.5847,0.0811,0.343,0.5661,-0.0034,0.011,
    ↪ 0.6062,0.3533,0.489,0.5665,0.6498,0.005,0.6711,0.1524,0.496,
    ↪ 0.7805,0.8636,0.499,0.7328,0.3361,0.003,0.8333,0.0052,0.493,
    ↪ 0.7398,0.1369,0.011,-0.0732,0.4927,0.492,0.8868,0.5,0.468,0.5,
    ↪ 0.2252,0.491,0.5898,0.2744,0.021,-0.0845,0.0507,0.041,0.5642,
    ↪ 0.2036,0.447,0.7347,-0.0802,0.049,0.6225,0.5751,0.043,0.7955,
    ↪ 0.4247,0.048,0.6971,0.2643,0.444,0.5386,0.8023,0.449,0.7661,
    ↪ 0.6453,0.041,0.6027,0.8531,0.463,-0.0984,0.4493,0.466,-0.0642,
    ↪ 0.2244,0.059,-0.0395,0.0697,0.049,0.8702 & Cc        C_s^4 #9 (
    ↪ a^36) & mC144 & & SiO2 & low Tridymite & Dollase and Baur, Am.
    ↪ Mineral. 61, 971-8 (1976)
   1.0000000000000000
   9.2620000000000000   -2.5016000000000000    0.0000000000000000
   9.2620000000000000    2.5016000000000000    0.0000000000000000
  -6.4909895300000000    0.0000000000000000   22.9081460400000000
    O    Si
   48    24
Direct
   0.2239000000000000   0.9259000000000000   0.8182000000000000   O   (4a)
   0.9259000000000000   0.2239000000000000   0.3182000000000000   O   (4a)
   0.4107000000000000   0.7307000000000000   0.3476000000000000   O   (4a)
   0.7307000000000000   0.4107000000000000   0.8476000000000000   O   (4a)
   0.5935000000000000   0.8695000000000000   0.4851000000000000   O   (4a)
   0.8695000000000000   0.5935000000000000   0.9851000000000000   O   (4a)
   0.1069000000000000   0.3949000000000000   0.5152000000000000   O   (4a)
   0.3949000000000000   0.1069000000000000   0.0152000000000000   O   (4a)
   0.0635000000000000   0.7675000000000000   0.6741000000000000   O   (4a)
   0.7675000000000000   0.0635000000000000   0.1741000000000000   O   (4a)
   0.2647000000000000   0.5607000000000000   0.1434000000000000   O   (4a)
   0.5607000000000000   0.2647000000000000   0.6434000000000000   O   (4a)
   0.0413000000000000   0.7133000000000000   0.4213000000000000   O   (4a)
   0.7133000000000000   0.0413000000000000   0.9213000000000000   O   (4a)
   0.2480000000000000   0.5840000000000000   0.9361000000000000   O   (4a)
   0.5840000000000000   0.2480000000000000   0.4361000000000000   O   (4a)
   0.6291000000000000   0.9191000000000000   0.7538000000000000   O   (4a)
   0.9191000000000000   0.6291000000000000   0.2538000000000000   O   (4a)
   0.0906000000000000   0.3766000000000000   0.7402000000000000   O   (4a)
   0.3766000000000000   0.0906000000000000   0.2402000000000000   O   (4a)
   0.2785000000000000   0.9605000000000000   0.5847000000000000   O   (4a)
   0.9605000000000000   0.2785000000000000   0.0847000000000000   O   (4a)
   0.4241000000000000   0.7381000000000000   0.0661000000000000   O   (4a)
   0.7381000000000000   0.4241000000000000   0.5661000000000000   O   (4a)
   0.0076000000000000  -0.0144000000000000   0.1062000000000000   O   (4a)
  -0.0144000000000000   0.0076000000000000   0.6062000000000000   O   (4a)
   0.8423000000000000   0.8643000000000000   0.0665000000000000   O   (4a)
   0.8643000000000000   0.8423000000000000   0.5665000000000000   O   (4a)
   0.6448000000000000   0.6548000000000000   0.6711000000000000   O   (4a)
   0.6548000000000000   0.6448000000000000   0.1711000000000000   O   (4a)
   0.6484000000000000   0.6564000000000000   0.2805000000000000   O   (4a)
   0.6564000000000000   0.6484000000000000   0.7805000000000000   O   (4a)
   0.3626000000000000   0.3646000000000000   0.2328000000000000   O   (4a)
   0.3646000000000000   0.3626000000000000   0.7328000000000000   O   (4a)
   0.3331000000000000   0.3391000000000000   0.8333000000000000   O   (4a)
   0.3391000000000000   0.3331000000000000   0.3333000000000000   O   (4a)
   0.4982000000000000   0.5122000000000000   0.2398000000000000   O   (4a)
   0.5122000000000000   0.4982000000000000   0.7398000000000000   O   (4a)
   0.1259000000000000   0.1479000000000000   0.9268000000000000   O   (4a)
   0.1479000000000000   0.1259000000000000   0.4268000000000000   O   (4a)
   0.0007000000000000   0.9847000000000000   0.8868000000000000   O   (4a)
   0.9847000000000000   0.0007000000000000   0.3868000000000000   O   (4a)
   0.0320000000000000   0.9680000000000000   0.5000000000000000   O   (4a)
   0.9680000000000000   0.0320000000000000   0.0000000000000000   O   (4a)
   0.7162000000000000   0.7342000000000000   0.0898000000000000   O   (4a)
   0.7342000000000000   0.7162000000000000   0.5898000000000000   O   (4a)
   0.2534000000000000   0.2954000000000000   0.9155000000000000   O   (4a)
   0.2954000000000000   0.2534000000000000   0.4155000000000000   O   (4a)
   0.0097000000000000   0.0917000000000000   0.5642000000000000   Si  (4a)
   0.0917000000000000   0.0097000000000000   0.0642000000000000   Si  (4a)
  -0.0167000000000000   0.9153000000000000   0.9358000000000000   Si  (4a)
   0.9153000000000000  -0.0167000000000000   0.4358000000000000   Si  (4a)
   0.1654000000000000   0.2834000000000000   0.9605000000000000   Si  (4a)
   0.2834000000000000   0.1654000000000000   0.4605000000000000   Si  (4a)
   0.0207000000000000   0.1187000000000000   0.8702000000000000   Si  (4a)
   0.1187000000000000   0.0207000000000000   0.3702000000000000   Si  (4a)
   0.6506000000000000   0.7566000000000000   0.2347000000000000   Si  (4a)
   0.7566000000000000   0.6506000000000000   0.7347000000000000   Si  (4a)
   0.8708000000000000   0.9688000000000000   0.6225000000000000   Si  (4a)
   0.9688000000000000   0.8708000000000000   0.1225000000000000   Si  (4a)
   0.5321000000000000   0.6181000000000000   0.7955000000000000   Si  (4a)
   0.6181000000000000   0.5321000000000000   0.2955000000000000   Si  (4a)
   0.3767000000000000   0.4727000000000000   0.6971000000000000   Si  (4a)
   0.4727000000000000   0.3767000000000000   0.1971000000000000   Si  (4a)
   0.7083000000000000   0.8203000000000000   0.0386000000000000   Si  (4a)
   0.8203000000000000   0.7083000000000000   0.5386000000000000   Si  (4a)
   0.2513000000000000   0.3533000000000000   0.2661000000000000   Si  (4a)
   0.3533000000000000   0.2513000000000000   0.7661000000000000   Si  (4a)
   0.6043000000000000   0.6863000000000000   0.6027000000000000   Si  (4a)
   0.6863000000000000   0.6043000000000000   0.1027000000000000   Si  (4a)
   0.3161000000000000   0.3901000000000000   0.4016000000000000   Si  (4a)
   0.3901000000000000   0.3161000000000000   0.9016000000000000   Si  (4a)
```

NiTi: AB_mP4_11_e_e - CIF

```
# CIF file

data_findsym-output
_audit_creation_method FINDSYM

_chemical_name_mineral ''
_chemical_formula_sum 'Ni Ti'

loop_
_publ_author_name
 'H. Sitepu'
 'W. W. Schmal'
 'J. K. Stalick'
_journal_name_full
;
Applied Physics A
;
_journal_volume 74
_journal_year 2002
_journal_page_first S1719
_journal_page_last S1721
_publ_Section_title
;
 Correction of intensities for preferred orientation in
    ↪ neutron-diffraction data of NiTi shape-memory alloy using the
    ↪ generalized spherical-harmonic description
;

# Found in  AMS Database

_aflow_proto 'AB_mP4_11_e_e'
_aflow_params 'a,b/a,c/a,\beta,x1,z1,x2,z2'
_aflow_params_values '2.8837,1.42393452856,1.61854561848,82.062,0.0387,
    ↪ 0.8252,0.5887,0.7184'
_aflow_Strukturbericht 'None'
_aflow_Pearson 'mP4'

_symmetry_space_group_name_Hall "-P 2yb"
_symmetry_space_group_name_H-M "P 1 21/m 1"
_symmetry_Int_Tables_number 11

_cell_length_a     2.88370
_cell_length_b     4.10620
_cell_length_c     4.66740
_cell_angle_alpha 90.00000
_cell_angle_beta  82.06200
_cell_angle_gamma 90.00000

loop_
_space_group_symop_id
_space_group_symop_operation_xyz
1 x,y,z
2 -x,y+1/2,-z
```

```
3 -x,-y,-z
4 x,-y+1/2,z

loop_
_atom_site_label
_atom_site_type_symbol
_atom_site_symmetry_multiplicity
_atom_site_Wyckoff_label
_atom_site_fract_x
_atom_site_fract_y
_atom_site_fract_z
_atom_site_occupancy
Ni1 Ni   2 e 0.03870 0.25000 0.82520 1.00000
Ti1 Ti   2 e 0.58870 0.25000 0.71840 1.00000
```

NiTi: AB_mP4_11_e_e - POSCAR

```
AB_mP4_11_e_e & a,b/a,c/a,\beta,x1,z1,x2,z2 --params=2.8837,
    ↪ 1.42393452856,1.61854561848,82.062,0.0387,0.8252,0.5887,0.7184
    ↪ & P2_1/m      C_{2h}^2 #11 (e^2) & mP4 & & NiTi & & H. Sitepu,
    ↪ W. W. Schmahl, and J. K. Stalick, App. Phys. A 74, S1719 (2002)
   1.0000000000000000
   2.88370000000000   0.00000000000000   0.00000000000000
   0.00000000000000   4.10620000000000   0.00000000000000
   0.64457469000000   0.00000000000000   4.62267739000000
    Ni    Ti
     2     2
Direct
   0.03870000000000   0.25000000000000   0.82520000000000   Ni   (2e)
   0.96130000000000   0.75000000000000   0.17480000000000   Ni   (2e)
   0.41130000000000   0.75000000000000   0.28160000000000   Ti   (2e)
   0.58870000000000   0.25000000000000   0.71840000000000   Ti   (2e)
```

$KClO_3$ ($G0_6$): ABC3_mP10_11_e_e_ef - CIF

```
# CIF file

data_findsym-output
_audit_creation_method FINDSYM

_chemical_name_mineral 'Potassium chlorate'
_chemical_formula_sum 'K Cl O3'

loop_
_publ_author_name
 'Jacob Danielsen'
 'Alan Hazell'
 'Finn Krebs Larsen'
_journal_name_full
;
Acta Crystallographica B
;
_journal_volume 37
_journal_year 1981
_journal_page_first 913
_journal_page_last 915
_publ_Section_title
;
 The Structure of Potassium Chlorate at 77 and 298 K
;

_aflow_proto 'ABC3_mP10_11_e_e_ef'
_aflow_params 'a,b/a,c/a,\beta,x1,z1,x2,z2,x3,z3,x4,y4,z4'
_aflow_params_values '4.63,1.20259179266,1.52203023758,110.21,0.121,
    ↪ 0.1745,0.3531,0.7086,0.4009,0.1165,0.8544,0.5361,0.6943'
_aflow_Strukturbericht 'G0_6'
_aflow_Pearson 'mP10'

_symmetry_space_group_name_Hall "-P 2yb"
_symmetry_space_group_name_H-M "P 1 21/m 1"
_symmetry_Int_Tables_number 11

_cell_length_a     4.63000
_cell_length_b     5.56800
_cell_length_c     7.04700
_cell_angle_alpha  90.00000
_cell_angle_beta   110.2100
_cell_angle_gamma  90.00000

loop_
_space_group_symop_id
_space_group_symop_operation_xyz
1 x,y,z
2 -x,y+1/2,-z
3 -x,-y,-z
4 x,-y+1/2,z

loop_
_atom_site_label
_atom_site_type_symbol
_atom_site_symmetry_multiplicity
_atom_site_Wyckoff_label
_atom_site_fract_x
_atom_site_fract_y
_atom_site_fract_z
_atom_site_occupancy
Cl1 Cl   2 e 0.12100 0.25000 0.17450 1.00000
K1  K    2 e 0.35310 0.25000 0.70860 1.00000
O1  O    2 e 0.40090 0.25000 0.11650 1.00000
O2  O    4 f 0.85440 0.53610 0.69430 1.00000
```

$KClO_3$ ($G0_6$): ABC3_mP10_11_e_e_ef - POSCAR

```
ABC3_mP10_11_e_e_ef & a,b/a,c/a,\beta,x1,z1,x2,z2,x3,z3,x4,y4,z4 --
    ↪ params=4.63,1.20259179266,1.52203023758,110.21,0.121,0.1745,
    ↪ 0.3531,0.7086,0.4009,0.1165,0.8544,0.5361,0.6943 & P2_1/m
    ↪ C_{2h}^2 #11 (e^3f) & mP10 & G0_6 & KClO3 & & J. Daniesen, A.
    ↪ Hazell, and F. K. Larsen, Acta Cryst. B 37, 913-5 (1981)
   1.0000000000000000
   4.63000000000000   0.00000000000000   0.00000000000000
   0.00000000000000   5.56800000000000   0.00000000000000
  -2.43447066000000   0.00000000000000   6.61313554000000
    Cl     K     O
     2     2     6
Direct
   0.12100000000000   0.25000000000000   0.17450000000000   Cl   (2e)
   0.87900000000000   0.75000000000000   0.82550000000000   Cl   (2e)
   0.35550000000000   0.25000000000000   0.70860000000000   K    (2e)
   0.64450000000000   0.75000000000000   0.29140000000000   K    (2e)
   0.40090000000000   0.25000000000000   0.11650000000000   O    (2e)
   0.59910000000000   0.75000000000000   0.88350000000000   O    (2e)
   0.14560000000000   0.03610000000000   0.30570000000000   O    (4f)
   0.14560000000000   0.46390000000000   0.30570000000000   O    (4f)
   0.85440000000000   0.53610000000000   0.69430000000000   O    (4f)
   0.85440000000000   0.96390000000000   0.69430000000000   O    (4f)
```

α-Pu: A_mP16_11_8e - CIF

```
# CIF file

data_findsym-output
_audit_creation_method FINDSYM

_chemical_name_mineral 'alpha Pu'
_chemical_formula_sum 'Pu'

loop_
_publ_author_name
 'W. H. Zachariasen'
 'F. H. Ellinger'
_journal_name_full
;
Acta Crystallographica
;
_journal_volume 16
_journal_year 1963
_journal_page_first 777
_journal_page_last 783
_publ_Section_title
;
 The Crystal Structure of Alpha Plutonium Metal
;

# Found in  Donohue, pp. 159-162

_aflow_proto 'A_mP16_11_8e'
_aflow_params 'a,b/a,c/a,\beta,x1,z1,x2,z2,x3,z3,x4,z4,x5,z5,x6,z6,x7,z7
    ↪ ,x8,z8'
_aflow_params_values '6.183,0.779880316998,1.77308749798,101.79,0.345,
    ↪ 0.162,0.767,0.168,0.128,0.34,0.657,0.457,0.025,0.618,0.473,
    ↪ 0.653,0.328,-0.074,0.869,0.894'
_aflow_Strukturbericht 'None'
_aflow_Pearson 'mP16'

_symmetry_space_group_name_Hall "-P 2yb"
_symmetry_space_group_name_H-M "P 1 21/m 1"
_symmetry_Int_Tables_number 11

_cell_length_a     6.18300
_cell_length_b     4.82200
_cell_length_c     10.96300
_cell_angle_alpha  90.00000
_cell_angle_beta   101.79000
_cell_angle_gamma  90.00000

loop_
_space_group_symop_id
_space_group_symop_operation_xyz
1 x,y,z
2 -x,y+1/2,-z
3 -x,-y,-z
4 x,-y+1/2,z

loop_
_atom_site_label
_atom_site_type_symbol
_atom_site_symmetry_multiplicity
_atom_site_Wyckoff_label
_atom_site_fract_x
_atom_site_fract_y
_atom_site_fract_z
_atom_site_occupancy
Pu1 Pu   2 e 0.34500 0.25000 0.16200  1.00000
Pu2 Pu   2 e 0.76700 0.25000 0.16800  1.00000
Pu3 Pu   2 e 0.12800 0.25000 0.34000  1.00000
Pu4 Pu   2 e 0.65700 0.25000 0.45700  1.00000
Pu5 Pu   2 e 0.02500 0.25000 0.61800  1.00000
Pu6 Pu   2 e 0.47300 0.25000 0.65300  1.00000
Pu7 Pu   2 e 0.32800 0.25000 -0.07400 1.00000
Pu8 Pu   2 e 0.86900 0.25000 0.89400  1.00000
```

α-Pu: A_mP16_11_8e - POSCAR

```
A_mP16_11_8e & a,b/a,c/a,\beta,x1,z1,x2,z2,x3,z3,x4,z4,x5,z5,x6,z6,x7,z7
    ↪ ,x8,z8 --params=6.183,0.779880316998,1.77308749798,101.79,0.345
    ↪ ,0.162,0.767,0.168,0.128,0.34,0.657,0.457,0.025,0.618,0.473,
    ↪ 0.653,0.328,-0.074,0.869,0.894 & P2_1/m     C_{2h}^2 #11 (e^8) &
```

```
    ↪  mP16 & & Pu & alpha & Zachariasen and Ellinger , Acta Cryst. 16
    ↪  , 777−83 (1963)
   1.0000000000000000
   6.18300000000000   0.00000000000000   0.00000000000000
   0.00000000000000   4.82200000000000   0.00000000000000
  −2.24001721000000   0.00000000000000  10.73171430000000
   Pu
   16
Direct
   0.34500000000000   0.25000000000000   0.16200000000000   Pu   (2e)
   0.65500000000000   0.75000000000000   0.83800000000000   Pu   (2e)
   0.23300000000000   0.75000000000000   0.83200000000000   Pu   (2e)
   0.76700000000000   0.25000000000000   0.16800000000000   Pu   (2e)
   0.12800000000000   0.25000000000000   0.34000000000000   Pu   (2e)
   0.87200000000000   0.75000000000000   0.66000000000000   Pu   (2e)
   0.34300000000000   0.75000000000000   0.54300000000000   Pu   (2e)
   0.65700000000000   0.25000000000000   0.45700000000000   Pu   (2e)
   0.02500000000000   0.25000000000000   0.61800000000000   Pu   (2e)
  −0.02500000000000   0.75000000000000   0.38200000000000   Pu   (2e)
   0.47300000000000   0.25000000000000   0.65300000000000   Pu   (2e)
   0.52700000000000   0.75000000000000   0.34700000000000   Pu   (2e)
   0.32800000000000   0.25000000000000  −0.07400000000000   Pu   (2e)
   0.67200000000000   0.75000000000000   0.07400000000000   Pu   (2e)
   0.13100000000000   0.75000000000000   0.10600000000000   Pu   (2e)
   0.86900000000000   0.25000000000000   0.89400000000000   Pu   (2e)
```

Calaverite ($AuTe_2$, C34): AB2_mC6_12_a_i - CIF

```
# CIF file

data_findsym−output
_audit_creation_method FINDSYM

_chemical_name_mineral 'Calaverite'
_chemical_formula_sum 'Au Te2'

loop_
_publ_author_name
 'K. Reithmayer'
 'W. Steurer'
 'H. Schulz'
 'J. L. de Boer'
_journal_name_full
;
Acta Crystallographica B
;
_journal_volume 49
_journal_year 1993
_journal_page_first 6
_journal_page_last 11
_publ_Section_title
;
 High−pressure single−crystal structure study on calaverite, AuTe$_2$
;

_aflow_proto 'AB2_mC6_12_a_i'
_aflow_params 'a,b/a,c/a,\beta,x2,z2'
_aflow_params_values '7.189,0.613019891501,0.705105021561,90.04,0.6879,
    ↪  0.2889'
_aflow_Strukturbericht 'C34'
_aflow_Pearson 'mC6'

_symmetry_space_group_name_Hall "−C 2y"
_symmetry_space_group_name_H−M "C 1 2/m 1"
_symmetry_Int_Tables_number 12

_cell_length_a      7.18900
_cell_length_b      4.40700
_cell_length_c      5.06900
_cell_angle_alpha  90.00000
_cell_angle_beta   90.04000
_cell_angle_gamma  90.00000

loop_
_space_group_symop_id
_space_group_symop_operation_xyz
1 x,y,z
2 −x,y,−z
3 −x,−y,−z
4 x,−y,z
5 x+1/2,y+1/2,z
6 −x+1/2,y+1/2,−z
7 −x+1/2,−y+1/2,−z
8 x+1/2,−y+1/2,z

loop_
_atom_site_label
_atom_site_type_symbol
_atom_site_symmetry_multiplicity
_atom_site_Wyckoff_label
_atom_site_fract_x
_atom_site_fract_y
_atom_site_fract_z
_atom_site_occupancy
Au1 Au   2 a 0.00000 0.00000 0.00000 1.00000
Te1 Te   4 i 0.68790 0.00000 0.28890 1.00000
```

Calaverite ($AuTe_2$, C34): AB2_mC6_12_a_i - POSCAR

```
AB2_mC6_12_a_i & a,b/a,c/a,\beta,x2,z2 −−params=7.189,0.613019891501,
    ↪  0.705105021561,90.04,0.6879,0.2889 & C2/m        C^3_{2h} #12 (
    ↪  ai) & mC6 & C34 & AuTe_2 & & K. Reithmayer et al., Acta Cryst.
    ↪  B 49, 6−11 (1993)
   1.0000000000000000
   3.59450000000000  −2.20350000000000   0.00000000000000
   3.59450000000000   2.20350000000000   0.00000000000000
   0.00353882930000   0.00000000000000   5.06899876470000
   Au   Te
    1    2
Direct
   0.00000000000000   0.00000000000000   0.00000000000000   Au   (2a)
   0.68790000000000   0.68790000000000   0.28890000000000   Te   (4i)
   0.31210000000000   0.31210000000000   0.71110000000000   Te   (4i)
```

β-Pu: A_mC34_12_ah3i2j - CIF

```
# CIF file

data_findsym−output
_audit_creation_method FINDSYM

_chemical_name_mineral 'beta Plutonium'
_chemical_formula_sum 'Pu'

loop_
_publ_author_name
 'W. H. Zachariasen'
 'F. H. Ellinger'
_journal_name_full
;
Acta Crystallographica
;
_journal_volume 16
_journal_year 1963
_journal_page_first 369
_journal_page_last 375
_publ_Section_title
;
 The Crystal Structure of Beta Plutonium Metal
;

# Found in Donohue, pp. 162−165

_aflow_proto 'A_mC34_12_ah3i2j'
_aflow_params 'a,b/a,c/a,\beta,y2,x3,z3,x4,z4,x5,z5,x6,y6,z6,x7,y7,z7'
_aflow_params_values '11.93871,0.876392843113,0.658278825769,129.00411,
    ↪  0.22,0.854,0.241,0.663,0.745,0.566,0.238,0.355,0.232,−0.037,
    ↪  0.333,0.35,0.586'
_aflow_Strukturbericht 'None'
_aflow_Pearson 'mC34'

_symmetry_space_group_name_Hall "−C 2y"
_symmetry_space_group_name_H−M "C 1 2/m 1"
_symmetry_Int_Tables_number 12

_cell_length_a      11.93871
_cell_length_b      10.46300
_cell_length_c       7.85900
_cell_angle_alpha   90.00000
_cell_angle_beta   129.00411
_cell_angle_gamma   90.00000

loop_
_space_group_symop_id
_space_group_symop_operation_xyz
1 x,y,z
2 −x,y,−z
3 −x,−y,−z
4 x,−y,z
5 x+1/2,y+1/2,z
6 −x+1/2,y+1/2,−z
7 −x+1/2,−y+1/2,−z
8 x+1/2,−y+1/2,z

loop_
_atom_site_label
_atom_site_type_symbol
_atom_site_symmetry_multiplicity
_atom_site_Wyckoff_label
_atom_site_fract_x
_atom_site_fract_y
_atom_site_fract_z
_atom_site_occupancy
Pu1 Pu   2 a 0.00000 0.00000  0.00000  1.00000
Pu2 Pu   4 h 0.00000 0.22000  0.50000  1.00000
Pu3 Pu   4 i 0.85400 0.00000  0.24100  1.00000
Pu4 Pu   4 i 0.66300 0.00000  0.74500  1.00000
Pu5 Pu   4 i 0.56600 0.00000  0.23800  1.00000
Pu6 Pu   8 j 0.35500 0.23200 −0.03700  1.00000
Pu7 Pu   8 j 0.33300 0.35000  0.58600  1.00000
```

β-Pu: A_mC34_12_ah3i2j - POSCAR

```
A_mC34_12_ah3i2j & a,b/a,c/a,\beta,y2,x3,z3,x4,z4,x5,z5,x6,y6,z6,x7,y7,
    ↪  z7 −−params=11.93871,0.876392843113,0.658278825769,129.00411,
    ↪  0.22,0.854,0.241,0.663,0.745,0.566,0.238,0.355,0.232,−0.037,
    ↪  0.333,0.35,0.586 & C2/m      C_{2h}^3 #12 (ahi^3j^2) & mC34 & &
    ↪  Pu & beta & W. H. Zachariasen and F. H. Ellinger, Acta Cryst.
    ↪  16, 369−75 (1963)
   1.0000000000000000
   5.96935749915200  −5.23150000000000   0.00000000000000
   5.96935749915200   5.23150000000000   0.00000000000000
  −4.94626686488100   0.00000000000000   6.10723547125700
   Pu
   17
Direct
   0.00000000000000   0.00000000000000   0.00000000000000   Pu   (2a)
   0.22000000000000   0.78000000000000   0.50000000000000   Pu   (4h)
   0.78000000000000   0.22000000000000   0.50000000000000   Pu   (4h)
   0.14600000000000   0.14600000000000   0.75900000000000   Pu   (4i)
```

```
    0.85400000000000    0.85400000000000    0.24100000000000    Pu    (4i)
    0.33700000000000    0.33700000000000    0.25500000000000    Pu    (4i)
    0.66300000000000    0.66300000000000    0.74500000000000    Pu    (4i)
    0.43400000000000    0.43400000000000    0.76200000000000    Pu    (4i)
    0.56600000000000    0.56600000000000    0.23800000000000    Pu    (4i)
    0.12300000000000    0.58700000000000   -0.03700000000000    Pu    (8j)
    0.41300000000000    0.87700000000000    0.03700000000000    Pu    (8j)
    0.58700000000000    0.12300000000000   -0.03700000000000    Pu    (8j)
    0.87700000000000    0.41300000000000    0.03700000000000    Pu    (8j)
    0.01700000000000    0.31700000000000    0.41400000000000    Pu    (8j)
   -0.01700000000000    0.68300000000000    0.58600000000000    Pu    (8j)
    0.31700000000000    0.01700000000000    0.41400000000000    Pu    (8j)
    0.68300000000000   -0.01700000000000    0.58600000000000    Pu    (8j)
```

$AlCl_3$ ($D0_{15}$): AB3_mC16_12_g_ij - CIF

```
# CIF file

data_findsym-output
_audit_creation_method FINDSYM

_chemical_name_mineral 'Aluminum trichloride'
_chemical_formula_sum 'Al Cl3'

loop_
_publ_author_name
 'S. I. Troyanov'
_journal_name_full
;
(Russian) Journal of Inorganic Chemistry (translated from Zhurnal
    ↪ Neorganicheskoi Khimii)
;
_journal_volume 37
_journal_year 1992
_journal_page_first 121
_journal_page_last 124
_publ_Section_title
;
 The crystal structure of titanium(II) tetrachloroaluminate Ti(AlCl$_4$)
    ↪ $_2$ and refinement of the crystal structure of AlCl$_3$
;

# Found in http://materials.springer.com/isp/crystallographic/docs/
    ↪ sd_1250120

_aflow_proto 'AB3_mC16_12_g_ij'
_aflow_params 'a,b/a,c/a,\beta,y1,x2,z2,x3,y3,z3'
_aflow_params_values '5.914,1.73047007102,1.03956712885,108.25,0.1662,
    ↪ 0.2147,0.2263,0.2518,0.32131,0.2248'
_aflow_Strukturbericht 'D0_15'
_aflow_Pearson 'mC16'

_symmetry_space_group_name_Hall "-C 2y"
_symmetry_space_group_name_H-M "C 1 2/m 1"
_symmetry_Int_Tables_number 12

_cell_length_a      5.91400
_cell_length_b      10.23400
_cell_length_c      6.14800
_cell_angle_alpha   90.00000
_cell_angle_beta    108.25000
_cell_angle_gamma   90.00000

loop_
_space_group_symop_id
_space_group_symop_operation_xyz
1 x,y,z
2 -x,y,-z
3 -x,-y,-z
4 x,-y,z
5 x+1/2,y+1/2,z
6 -x+1/2,y+1/2,-z
7 -x+1/2,-y+1/2,-z
8 x+1/2,-y+1/2,z

loop_
_atom_site_label
_atom_site_type_symbol
_atom_site_symmetry_multiplicity
_atom_site_Wyckoff_label
_atom_site_fract_x
_atom_site_fract_y
_atom_site_fract_z
_atom_site_occupancy
Al1 Al    4 g 0.00000 0.16620 0.00000 1.00000
Cl1 Cl    4 i 0.21470 0.00000 0.22630 1.00000
Cl2 Cl    8 j 0.25180 0.32131 0.22480 1.00000
```

$AlCl_3$ ($D0_{15}$): AB3_mC16_12_g_ij - POSCAR

```
AB3_mC16_12_g_ij & a,b/a,c/a,\beta,y1,x2,z2,x3,y3,z3 --params=5.914,
    ↪ 1.73047007102,1.03956712885,108.25,0.1662,0.2147,0.2263,0.2518,
    ↪ 0.32131,0.2248 & C2/m        C_{2h}^3 #12 (gij) & mC16 & D0_{15}
    ↪ & AlCl_3 & & S. I. Troyanov, Russian Journal of Inorganic
    ↪ Chemistry 37, 121-124 (1992)
   1.00000000000000
   2.95700000000000   -5.11700000000000    0.00000000000000
   2.95700000000000    5.11700000000000    0.00000000000000
  -1.92533108226200    0.00000000000000    5.83875022788900
    Al    Cl
     2     6
Direct
    0.16620000000000    0.83380000000000    0.00000000000000    Al    (4g)
    0.83380000000000    0.16620000000000    0.00000000000000    Al    (4g)
    0.21470000000000    0.21470000000000    0.22630000000000    Cl    (4i)
```

```
    0.78530000000000    0.78530000000000    0.77370000000000    Cl    (4i)
    0.06951000000000    0.42689000000000    0.77520000000000    Cl    (8j)
   -0.06951000000000    0.57311000000000    0.22480000000000    Cl    (8j)
    0.42689000000000    0.06951000000000    0.77520000000000    Cl    (8j)
    0.57311000000000   -0.06951000000000    0.22480000000000    Cl    (8j)
```

$Au_5Mn_2$: A5B2_mC14_12_a2i_i - CIF

```
# CIF file

data_findsym-output
_audit_creation_method FINDSYM

_chemical_name_mineral ''
_chemical_formula_sum 'Au5 Mn2'

loop_
_publ_author_name
 'S. G. Humble'
_journal_name_full
;
Acta Crystallographica
;
_journal_volume 17
_journal_year 1964
_journal_page_first 1485
_journal_page_last 1486
_publ_Section_title
;
 Establishment of an ordered phase of composition Au$_5$Mn$_2$ in the
    ↪ gold-manganese system
;

# Found in Pearson, 346-348

_aflow_proto 'A5B2_mC14_12_a2i_i'
_aflow_params 'a,b/a,c/a,\beta,x2,z2,x3,z3,x4,z4'
_aflow_params_values '9.188,0.430343926861,0.705158902917,97.56,0.14286,
    ↪ 0.42857,0.28571,0.85714,0.42857,0.28571'
_aflow_Strukturbericht 'None'
_aflow_Pearson 'mC14'

_symmetry_space_group_name_Hall "-C 2y"
_symmetry_space_group_name_H-M "C 1 2/m 1"
_symmetry_Int_Tables_number 12

_cell_length_a      9.18800
_cell_length_b      3.95400
_cell_length_c      6.47900
_cell_angle_alpha   90.00000
_cell_angle_beta    97.56000
_cell_angle_gamma   90.00000

loop_
_space_group_symop_id
_space_group_symop_operation_xyz
1 x,y,z
2 -x,y,-z
3 -x,-y,-z
4 x,-y,z
5 x+1/2,y+1/2,z
6 -x+1/2,y+1/2,-z
7 -x+1/2,-y+1/2,-z
8 x+1/2,-y+1/2,z

loop_
_atom_site_label
_atom_site_type_symbol
_atom_site_symmetry_multiplicity
_atom_site_Wyckoff_label
_atom_site_fract_x
_atom_site_fract_y
_atom_site_fract_z
_atom_site_occupancy
Au1 Au    2 a 0.00000 0.00000 0.00000 1.00000
Au2 Au    4 i 0.14286 0.00000 0.42857 1.00000
Au3 Au    4 i 0.28571 0.00000 0.85714 1.00000
Mn1 Mn    4 i 0.42857 0.00000 0.28571 1.00000
```

$Au_5Mn_2$: A5B2_mC14_12_a2i_i - POSCAR

```
A5B2_mC14_12_a2i_i & a,b/a,c/a,\beta,x2,z2,x3,z3,x4,z4 --params=9.188,
    ↪ 0.430343926861,0.705158902917,97.56,0.14286,0.42857,0.28571,
    ↪ 0.85714,0.42857,0.28571 & C2/m        C_{2h}^3 #12 (ai^3) &
    ↪ mC14 & & Au_5Mn_2 & & S. G. Humble, Acta Cryst. 17, 1485-1486 (
    ↪ 1964)
   1.00000000000000
   4.59400000000000   -1.97700000000000    0.00000000000000
   4.59400000000000    1.97700000000000    0.00000000000000
  -0.85240548255900    0.00000000000000    6.42268214169900
    Au    Mn
     5     2
Direct
    0.00000000000000    0.00000000000000    0.00000000000000    Au    (2a)
    0.14285714285700    0.14285714285700    0.42857142857100    Au    (4i)
    0.85714285714300    0.85714285714300    0.57142857142900    Au    (4i)
    0.28571428571400    0.28571428571400    0.85714285714300    Au    (4i)
    0.71428571428600    0.71428571428600    0.14285714285700    Au    (4i)
    0.42857142857100    0.42857142857100    0.28571428571400    Mn    (4i)
    0.57142857142900    0.57142857142900    0.71428571428600    Mn    (4i)
```

α-O: A_mC4_12_i - CIF

```
# CIF file
```

```
data_findsym-output
_audit_creation_method FINDSYM

_chemical_name_mineral 'alpha oxygen'
_chemical_formula_sum 'O'

loop_
_publ_author_name
 'R. J. Meier'
 'R. B. Helmholdt'
_journal_name_full
;
Physical Review B
;
_journal_volume 29
_journal_year 1984
_journal_page_first 1387
_journal_page_last 1393
_publ_Section_title
;
 Neutron-diffraction study of $\alpha$- and $\beta$-oxygen
;

_aflow_proto 'A_mC4_12_i'
_aflow_params 'a,b/a,c/a,\beta,x1,z1'
_aflow_params_values '5.403,0.635387747548,0.940033314825,132.32,0.106,
    ↪ 0.173'
_aflow_Strukturbericht 'None'
_aflow_Pearson 'mC4'

_symmetry_space_group_name_Hall "-C 2y"
_symmetry_space_group_name_H-M "C 1 2/m 1"
_symmetry_Int_Tables_number 12

_cell_length_a      5.40300
_cell_length_b      3.43300
_cell_length_c      5.07900
_cell_angle_alpha  90.00000
_cell_angle_beta  132.32000
_cell_angle_gamma  90.00000

loop_
_space_group_symop_id
_space_group_symop_operation_xyz
1 x,y,z
2 -x,y,-z
3 -x,-y,-z
4 x,-y,z
5 x+1/2,y+1/2,z
6 -x+1/2,y+1/2,-z
7 -x+1/2,-y+1/2,-z
8 x+1/2,-y+1/2,z

loop_
_atom_site_label
_atom_site_type_symbol
_atom_site_symmetry_multiplicity
_atom_site_Wyckoff_label
_atom_site_fract_x
_atom_site_fract_y
_atom_site_fract_z
_atom_site_occupancy
O1 O    4 i 0.10600 0.00000 0.17300 1.00000
```

$\alpha$-O: A_mC4_12_i - POSCAR

```
A_mC4_12_i & a,b/a,c/a,\beta,x1,z1 --params=5.403,0.635387747548,
    ↪ 0.940033314825,132.32,0.106,0.173 & C2/m      C__{2h}^3 #12 (i)
    ↪ & mC4 & & O & alpha & R. J. Meier and R. B. Helmholdt, Phys.
    ↪ Rev. B 29, 1387-93 (1984)
   1.0000000000000000
   2.70150000000000   -1.71650000000000    0.00000000000000
   2.70150000000000    1.71650000000000    0.00000000000000
  -3.41954164384100    0.00000000000000    3.75539291500000
     O
     2
Direct
   0.10600000000000    0.10600000000000    0.17300000000000    O    (4i)
   0.89400000000000    0.89400000000000    0.82700000000000    O    (4i)
```

Sylvanite (AgAuTe$_4$, E1$_b$): ABC4_mP12_13_e_a_2g - CIF

```
# CIF file

data_findsym-output
_audit_creation_method FINDSYM

_chemical_name_mineral 'Sylvanite'
_chemical_formula_sum 'Ag Au Te4'

loop_
_publ_author_name
 'F. Pertlik'
_journal_name_full
;
Tschermaks mineralogische und petrographische Mitteilungen
;
_journal_volume 33
_journal_year 1984
_journal_page_first 203
_journal_page_last 212
_publ_Section_title
;
 Kristallchemie nat\"{u}rlicher Telluride I: Verfeinerung der
    ↪ Kristallstruktur des Sylvanits, AuAgTe$_4$
;

# Found in  http://materials.springer.com/isp/crystallographic/docs/
    ↪ sd_1702950

_aflow_proto 'ABC4_mP12_13_e_a_2g'
_aflow_params 'a,b/a,c/a,\beta,y2,x3,y3,z3,x4,y4,z4'
_aflow_params_values '8.95,0.500335195531,1.63360893855,145.35,0.5182,
    ↪ 0.2986,0.0278,0.0003,0.2821,0.4045,0.2366'
_aflow_Strukturbericht 'E1_b'
_aflow_Pearson 'mP12'

_symmetry_space_group_name_Hall "-P 2yc"
_symmetry_space_group_name_H-M "P 1 2/c 1"
_symmetry_Int_Tables_number 13

_cell_length_a      8.95000
_cell_length_b      4.47800
_cell_length_c     14.62080
_cell_angle_alpha  90.00000
_cell_angle_beta  145.35000
_cell_angle_gamma  90.00000

loop_
_space_group_symop_id
_space_group_symop_operation_xyz
1 x,y,z
2 -x,y,-z+1/2
3 -x,-y,-z
4 x,-y,z+1/2

loop_
_atom_site_label
_atom_site_type_symbol
_atom_site_symmetry_multiplicity
_atom_site_Wyckoff_label
_atom_site_fract_x
_atom_site_fract_y
_atom_site_fract_z
_atom_site_occupancy
Au1 Au    2 a 0.00000 0.00000 0.00000 1.00000
Ag1 Ag    2 e 0.00000 0.51820 0.25000 1.00000
Te1 Te    4 g 0.29860 0.02780 0.00030 1.00000
Te2 Te    4 g 0.28210 0.40450 0.23660 1.00000
```

Sylvanite (AgAuTe$_4$, E1$_b$): ABC4_mP12_13_e_a_2g - POSCAR

```
ABC4_mP12_13_e_a_2g & a,b/a,c/a,\beta,y2,x3,y3,z3,x4,y4,z4 --params=8.95
    ↪ ,0.500335195531,1.63360893855,145.35,0.5182,0.2986,0.0278,
    ↪ 0.0003,0.2821,0.4045,0.2366 & P2/c      C_{2h}^4 #13 (aeg^2) &
    ↪  mP12 & E1_b & AgAuTe4 & & F. Pertlik, TMPM 33, 203-212 (1984)
   1.0000000000000000
   8.95000000000000    0.00000000000000    0.00000000000000
   0.00000000000000    4.47800000000000    0.00000000000000
 -12.02700437000000    0.00000000000000    8.31237426000000
    Ag    Au    Te
     2     2     8
Direct
   0.00000000000000    0.48180000000000    0.75000000000000    Ag    (2e)
   0.00000000000000    0.51820000000000    0.25000000000000    Ag    (2e)
   0.00000000000000    0.00000000000000    0.00000000000000    Au    (2a)
   0.00000000000000    0.00000000000000    0.50000000000000    Au    (2a)
   0.29920000000000    0.02780000000000    0.00030000000000    Te    (4g)
   0.29920000000000    0.97220000000000    0.50030000000000    Te    (4g)
   0.70080000000000    0.02780000000000    0.49970000000000    Te    (4g)
   0.70080000000000    0.97220000000000    0.99970000000000    Te    (4g)
   0.28210000000000    0.40450000000000    0.23660000000000    Te    (4g)
   0.28210000000000    0.59550000000000    0.73660000000000    Te    (4g)
   0.71790000000000    0.40450000000000    0.26340000000000    Te    (4g)
   0.71790000000000    0.59550000000000    0.76340000000000    Te    (4g)
```

Monoclinic (Hittorf's) Phosphorus: A_mP84_13_21g - CIF

```
# CIF file

data_findsym-output
_audit_creation_method FINDSYM

_chemical_name_mineral 'Hittorf'
_chemical_formula_sum 'P'

loop_
_publ_author_name
 'H. Thurn'
 'H. Krebs'
_journal_name_full
;
Acta Crystallographica B
;
_journal_volume 25
_journal_year 1969
_journal_page_first 125
_journal_page_last 135
_publ_Section_title
;
 "{U}ber Struktur und Eigenschaften der Halbmetalle. XXII. Die
    ↪ Kristallstruktur des Hittorfschen Phosphors
;

# Found in  Donohue, pp. 292-295

_aflow_proto 'A_mP84_13_21g'
```

```
_aflow_params 'a,b/a,c/a,\beta,x1,y1,z1,x2,y2,z2,x3,y3,z3,x4,y4,z4,x5,y5
    ↪ ,z5,x6,y6,z6,x7,y7,z7,x8,y8,z8,x9,y9,z9,x10,y10,z10,x11,y11,z11
    ↪ ,x12,y12,z12,x13,y13,z13,x14,y14,z14,x15,y15,z15,x16,y16,z16,
    ↪ x17,y17,z17,x18,y18,z18,x19,y19,z19,x20,y20,z20,x21,y21,z21'
_aflow_params_values '9.21,0.99348534202,2.45385450597,106.1,0.30089,
    ↪ 0.20127,0.18147,0.17387,0.03262,0.11695,0.05014,-0.05231,
    ↪ 0.18035,-0.07589,0.78099,0.11634,0.79463,0.67872,0.1738,0.68463
    ↪ ,0.51532,0.10402,0.56601,0.44932,0.17224,0.42424,0.27741,
    ↪ 0.11672,0.0412,0.39067,0.07245,-0.00092,0.15881,0.04497,0.78847
    ↪ ,0.13878,0.07346,0.7486,-0.09081,0.04464,0.53574,0.87264,
    ↪ 0.06842,0.50833,0.63715,0.03304,0.30515,0.63715,0.06617,0.25041
    ↪ ,0.40555,0.0442,0.146,0.38905,0.17219,0.86038,0.10055,0.17357,
    ↪ 0.59606,0.82384,0.1694,0.41856,0.64581,0.16732,-0.05418,0.32296
    ↪ ,0.2006'
_aflow_Strukturbericht 'None'
_aflow_Pearson 'mP84'

_symmetry_space_group_name_Hall "-P 2yc"
_symmetry_space_group_name_H-M "P 1 2/c 1"
_symmetry_Int_Tables_number 13

_cell_length_a     9.21000
_cell_length_b     9.15000
_cell_length_c     22.60000
_cell_angle_alpha  90.00000
_cell_angle_beta   106.10000
_cell_angle_gamma  90.00000

loop_
_space_group_symop_id
_space_group_symop_operation_xyz
1 x,y,z
2 -x,y,-z+1/2
3 -x,-y,-z
4 x,-y,z+1/2

loop_
_atom_site_label
_atom_site_type_symbol
_atom_site_symmetry_multiplicity
_atom_site_Wyckoff_label
_atom_site_fract_x
_atom_site_fract_y
_atom_site_fract_z
_atom_site_occupancy
P1  P   4 g  0.30089   0.20127   0.18147  1.00000
P2  P   4 g  0.17387   0.03262   0.11695  1.00000
P3  P   4 g  0.05014  -0.05231   0.18035  1.00000
P4  P   4 g -0.07589   0.78099   0.11634  1.00000
P5  P   4 g  0.79463   0.67872   0.17380  1.00000
P6  P   4 g  0.68463   0.51532   0.10402  1.00000
P7  P   4 g  0.56601   0.44932   0.17224  1.00000
P8  P   4 g  0.42424   0.27741   0.11672  1.00000
P9  P   4 g  0.04120   0.39067   0.07245  1.00000
P10 P   4 g -0.00092   0.15881   0.04497  1.00000
P11 P   4 g  0.78847   0.13878   0.07346  1.00000
P12 P   4 g  0.74860  -0.09081   0.04464  1.00000
P13 P   4 g  0.53574   0.87264   0.06842  1.00000
P14 P   4 g  0.50833   0.63715   0.03304  1.00000
P15 P   4 g  0.30515   0.63715   0.06617  1.00000
P16 P   4 g  0.25041   0.40555   0.04420  1.00000
P17 P   4 g  0.14600   0.38905   0.17219  1.00000
P18 P   4 g  0.86038   0.10055   0.17357  1.00000
P19 P   4 g  0.59606   0.82384   0.16940  1.00000
P20 P   4 g  0.41856   0.64581   0.16732  1.00000
P21 P   4 g -0.05418   0.32296   0.20060  1.00000
```

Monoclinic (Hittorf's) Phosphorus: A_mP84_13_21g - POSCAR

```
A_mP84_13_21g & a,b/a,c/a,\beta,x1,y1,z1,x2,y2,z2,x3,y3,z3,x4,y4,z4,x5,
    ↪ y5,z5,x6,y6,z6,x7,y7,z7,x8,y8,z8,x9,y9,z9,x10,y10,z10,x11,y11,
    ↪ z11,x12,y12,z12,x13,y13,z13,x14,y14,z14,x15,y15,z15,x16,y16,z16
    ↪ ,x17,y17,z17,x18,y18,z18,x19,y19,z19,x20,y20,z20,x21,y21,z21 --
    ↪ params=9.21,0.99348534202,2.45385450597,106.1,0.30089,0.20127,
    ↪ 0.18147,0.17387,0.03262,0.11695,0.05014,-0.05231,0.18035,-
    ↪ 0.07589,0.78099,0.11634,0.79463,0.67872,0.1738,0.68463,0.51532,
    ↪ 0.10402,0.56601,0.44932,0.17224,0.42424,0.27741,0.11672,0.0412,
    ↪ 0.39067,0.07245,-0.00092,0.15881,0.04497,0.78847,0.13878,
    ↪ 0.07346,0.7486,-0.09081,0.04464,0.53574,0.87264,0.06842,0.50833
    ↪ ,0.63715,0.03304,0.30515,0.63715,0.06617,0.25041,0.40555,0.0442
    ↪ ,0.146,0.38905,0.17219,0.86038,0.10055,0.17357,0.59606,0.82384,
    ↪ 0.1694,0.41856,0.64581,0.16732,-0.05418,0.32296,0.2006 & P2/c
    ↪        C_{2h}^4 #13 (g^21) & mP84 & & P & Hittorf & Thurn and
    ↪ Krebs, Acta Cryst. B 125 (1969)
   1.0000000000000000
   9.21000000000000    0.00000000000000    0.00000000000000
   0.00000000000000    9.15000000000000    0.00000000000000
  -6.26731116000000    0.00000000000000   21.71360888000000
    P
   84
Direct
   0.30089000000000    0.20127000000000    0.18147000000000    P    (4g)
   0.30089000000000    0.79873000000000    0.68147000000000    P    (4g)
   0.69911000000000    0.20127000000000    0.31853000000000    P    (4g)
   0.69911000000000    0.79873000000000    0.81853000000000    P    (4g)
  -0.00092000000000    0.15881000000000    0.04497000000000    P    (4g)
  -0.00092000000000    0.84119000000000    0.54497000000000    P    (4g)
   1.00092000000000    0.15881000000000    0.45503000000000    P    (4g)
   1.00092000000000    0.84119000000000    0.95503000000000    P    (4g)
   0.21153000000000    0.13878000000000    0.42654000000000    P    (4g)
   0.21153000000000    0.86122000000000    0.92654000000000    P    (4g)
   0.78847000000000    0.13878000000000    0.07346000000000    P    (4g)
   0.78847000000000    0.86122000000000    0.57346000000000    P    (4g)
   0.25140000000000   -0.09081000000000    0.45536000000000    P    (4g)
   0.25140000000000    1.09081000000000    0.95536000000000    P    (4g)
   0.74860000000000   -0.09081000000000    0.04464000000000    P    (4g)
   0.74860000000000    1.09081000000000    0.54464000000000    P    (4g)
   0.46426000000000    0.12736000000000    0.93158000000000    P    (4g)
   0.46426000000000    0.87264000000000    0.43158000000000    P    (4g)
   0.53574000000000    0.12736000000000    0.56842000000000    P    (4g)
   0.53574000000000    0.87264000000000    0.06842000000000    P    (4g)
   0.49167000000000    0.36285000000000    0.96696000000000    P    (4g)
   0.49167000000000    0.63715000000000    0.46696000000000    P    (4g)
   0.50833000000000    0.36285000000000    0.53304000000000    P    (4g)
   0.50833000000000    0.63715000000000    0.03304000000000    P    (4g)
   0.30515000000000    0.36285000000000    0.56617000000000    P    (4g)
   0.30515000000000    0.63715000000000    0.06617000000000    P    (4g)
   0.69485000000000    0.36285000000000    0.93383000000000    P    (4g)
   0.69485000000000    0.63715000000000    0.43383000000000    P    (4g)
   0.25041000000000    0.40555000000000    0.04420000000000    P    (4g)
   0.25041000000000    0.59445000000000    0.54420000000000    P    (4g)
   0.74959000000000    0.40555000000000    0.45580000000000    P    (4g)
   0.74959000000000    0.59445000000000    0.95580000000000    P    (4g)
   0.14600000000000    0.38905000000000    0.17219000000000    P    (4g)
   0.14600000000000    0.61095000000000    0.67219000000000    P    (4g)
   0.85400000000000    0.38905000000000    0.32781000000000    P    (4g)
   0.85400000000000    0.61095000000000    0.82781000000000    P    (4g)
   0.13962000000000    0.10055000000000    0.32643000000000    P    (4g)
   0.13962000000000    0.89945000000000    0.82643000000000    P    (4g)
   0.86038000000000    0.10055000000000    0.17357000000000    P    (4g)
   0.86038000000000    0.89945000000000    0.67357000000000    P    (4g)
   0.40394000000000    0.17616000000000    0.83060000000000    P    (4g)
   0.40394000000000    0.82384000000000    0.33060000000000    P    (4g)
   0.59606000000000    0.17616000000000    0.66940000000000    P    (4g)
   0.59606000000000    0.82384000000000    0.16940000000000    P    (4g)
   0.17387000000000    0.03262000000000    0.11695000000000    P    (4g)
   0.17387000000000    0.96738000000000    0.61695000000000    P    (4g)
   0.82613000000000    0.03262000000000    0.38305000000000    P    (4g)
   0.82613000000000    0.96738000000000    0.88305000000000    P    (4g)
   0.41856000000000    0.35419000000000    0.66732000000000    P    (4g)
   0.41856000000000    0.64581000000000    0.16732000000000    P    (4g)
   0.58144000000000    0.35419000000000    0.83268000000000    P    (4g)
   0.58144000000000    0.64581000000000    0.33268000000000    P    (4g)
  -0.05418000000000    0.32296000000000    0.20060000000000    P    (4g)
  -0.05418000000000    0.67704000000000    0.70060000000000    P    (4g)
   1.05418000000000    0.32296000000000    0.29940000000000    P    (4g)
   1.05418000000000    0.67704000000000    0.79940000000000    P    (4g)
   0.05014000000000   -0.05231000000000    0.18035000000000    P    (4g)
   0.05014000000000    1.05231000000000    0.68035000000000    P    (4g)
   0.94986000000000   -0.05231000000000    0.31965000000000    P    (4g)
   0.94986000000000    1.05231000000000    0.81965000000000    P    (4g)
  -0.07589000000000    0.21901000000000    0.61634000000000    P    (4g)
  -0.07589000000000    0.78099000000000    0.11634000000000    P    (4g)
   1.07589000000000    0.21901000000000    0.88366000000000    P    (4g)
   1.07589000000000    0.78099000000000    0.38366000000000    P    (4g)
   0.20537000000000    0.32128000000000    0.82620000000000    P    (4g)
   0.20537000000000    0.67872000000000    0.32620000000000    P    (4g)
   0.79463000000000    0.32128000000000    0.67380000000000    P    (4g)
   0.79463000000000    0.67872000000000    0.17380000000000    P    (4g)
   0.31537000000000    0.48468000000000    0.89598000000000    P    (4g)
   0.31537000000000    0.51532000000000    0.39598000000000    P    (4g)
   0.68463000000000    0.48468000000000    0.60402000000000    P    (4g)
   0.68463000000000    0.51532000000000    0.10402000000000    P    (4g)
   0.43399000000000    0.44932000000000    0.32776000000000    P    (4g)
   0.43399000000000    0.55068000000000    0.82776000000000    P    (4g)
   0.56601000000000    0.44932000000000    0.17224000000000    P    (4g)
   0.56601000000000    0.55068000000000    0.67224000000000    P    (4g)
   0.42424000000000    0.27741000000000    0.11672000000000    P    (4g)
   0.42424000000000    0.72259000000000    0.61672000000000    P    (4g)
   0.57576000000000    0.27741000000000    0.38328000000000    P    (4g)
   0.57576000000000    0.72259000000000    0.88328000000000    P    (4g)
   0.04120000000000    0.39067000000000    0.07245000000000    P    (4g)
   0.04120000000000    0.60933000000000    0.57245000000000    P    (4g)
   0.95880000000000    0.39067000000000    0.42755000000000    P    (4g)
   0.95880000000000    0.60933000000000    0.92755000000000    P    (4g)
```

Baddeleyite ($ZrO_2$, C43): A2B_mP12_14_2e_e - CIF

```
# CIF file

data_findsym-output
_audit_creation_method FINDSYM

_chemical_name_mineral 'Baddeleyite'
_chemical_formula_sum 'Zr O2'

loop_
_publ_author_name
 'C. J. Howard'
 'R. J. Hill'
 'B. E. Reichert'
_journal_name_full
;
Acta Crystallographica B
;
_journal_volume 44
_journal_year 1988
_journal_page_first 116
_journal_page_last 120
_publ_Section_title
;
 Structures of ZrO$_2$ polymorphs at room temperature by high-resolution
    ↪ neutron powder diffraction
;

_aflow_proto 'A2B_mP12_14_2e_e'
_aflow_params 'a,b/a,c/a,\beta,x1,y1,z1,x2,y2,z2,x3,y3,z3'
_aflow_params_values '5.1505,1.01186292593,1.03238520532,99.23,0.07,
    ↪ 0.3317,0.3447,0.4496,0.7569,0.4792,0.2754,0.0395,0.2083'
_aflow_Strukturbericht 'C43'
```

```
_aflow_Pearson 'mP12'

_symmetry_space_group_name_Hall "-P 2ybc"
_symmetry_space_group_name_H-M "P 1 21/c 1"
_symmetry_Int_Tables_number 14

_cell_length_a     5.15050
_cell_length_b     5.21160
_cell_length_c     5.31730
_cell_angle_alpha  90.00000
_cell_angle_beta   99.23000
_cell_angle_gamma  90.00000

loop_
_space_group_symop_id
_space_group_symop_operation_xyz
1 x,y,z
2 -x,y+1/2,-z+1/2
3 -x,-y,-z
4 x,-y+1/2,z+1/2

loop_
_atom_site_label
_atom_site_type_symbol
_atom_site_symmetry_multiplicity
_atom_site_Wyckoff_label
_atom_site_fract_x
_atom_site_fract_y
_atom_site_fract_z
_atom_site_occupancy
O1  O    4 e 0.07000 0.33170 0.34470 1.00000
O2  O    4 e 0.44960 0.75690 0.47920 1.00000
Zr1 Zr   4 e 0.27540 0.03950 0.20830 1.00000
```

Baddeleyite ($ZrO_2$, C43): A2B_mP12_14_2e_e - POSCAR

```
A2B_mP12_14_2e_e & a,b/a,c/a,\beta,x1,y1,z1,x2,y2,z2,x3,y3,z3 --params=
   ↪ 5.1505,1.01186292593,1.03238520532,99.23,0.07,0.3317,0.3447,
   ↪ 0.4496,0.7569,0.4792,0.2754,0.0395,0.2083 & P2_1/c      C_{2h}^5
   ↪  #14 (e^3) & mP12 & C43 & ZrO2 & Baddeleyite & Howard, Hill,
   ↪ and Reichert, Acta Cryst. B 44, 116-20 (1988)
   1.0000000000000000
   5.15050000000000   0.00000000000000   0.00000000000000
   0.00000000000000   5.21160000000000   0.00000000000000
  -0.85288444000000   0.00000000000000   5.24845381000000
     O    Zr
     8     4
Direct
   0.07000000000000   0.16830000000000   0.84470000000000   O    (4e)
   0.07000000000000   0.33170000000000   0.34470000000000   O    (4e)
   0.93000000000000   0.66830000000000   0.65530000000000   O    (4e)
   0.93000000000000   0.83170000000000   0.15530000000000   O    (4e)
   0.44960000000000   0.74310000000000   0.97920000000000   O    (4e)
   0.44960000000000   0.75690000000000   0.47920000000000   O    (4e)
   0.55040000000000   0.24310000000000   0.52080000000000   O    (4e)
   0.55040000000000   0.25690000000000   0.02080000000000   O    (4e)
   0.27540000000000   0.03950000000000   0.20830000000000   Zr   (4e)
   0.27540000000000   0.46050000000000   0.70830000000000   Zr   (4e)
   0.72460000000000   0.53950000000000   0.29170000000000   Zr   (4e)
   0.72460000000000   0.96050000000000   0.79170000000000   Zr   (4e)
```

β-Se ($A_l$): A_mP32_14_8e - CIF

```
# CIF file

data_findsym-output
_audit_creation_method FINDSYM

_chemical_name_mineral 'beta Selenium'
_chemical_formula_sum 'Se'

loop_
_publ_author_name
 'R. E. Marsh'
 'L. Pauling'
 'J. D. McCullough'
_journal_name_full
;
Acta Crystallographica
;
_journal_volume 6
_journal_year 1953
_journal_page_first 71
_journal_page_last 75
_publ_Section_title
;
 The Crystal Structure of $\beta$ Selenium
;

# Found in  Donohue, pp. 379-384

_aflow_proto 'A_mP32_14_8e'
_aflow_params 'a,b/a,c/a,\beta,x1,y1,z1,x2,y2,z2,x3,y3,z3,x4,y4,z4,x5,y5
   ↪ ,z5,x6,y6,z6,x7,y7,z7,x8,y8,z8'
_aflow_params_values '9.31,0.866809881847,1.38023630505,93.13333,0.437,
   ↪ 0.185,0.084,0.246,0.273,-0.023,0.24,0.102,0.828,0.05,-0.08,
   ↪ 0.852,0.157,0.669,-0.09,0.142,0.66,0.09,0.368,0.746,0.16,0.334,
   ↪ 0.021,0.21'

_aflow_Strukturbericht 'A_l'
_aflow_Pearson 'mP32'

_symmetry_space_group_name_Hall "-P 2ybc"
_symmetry_space_group_name_H-M "P 1 21/c 1"
_symmetry_Int_Tables_number 14

_cell_length_a     9.31000
_cell_length_b     8.07000
_cell_length_c     12.85000
_cell_angle_alpha  90.00000
_cell_angle_beta   93.13333
_cell_angle_gamma  90.00000

loop_
_space_group_symop_id
_space_group_symop_operation_xyz
1 x,y,z
2 -x,y+1/2,-z+1/2
3 -x,-y,-z
4 x,-y+1/2,z+1/2

loop_
_atom_site_label
_atom_site_type_symbol
_atom_site_symmetry_multiplicity
_atom_site_Wyckoff_label
_atom_site_fract_x
_atom_site_fract_y
_atom_site_fract_z
_atom_site_occupancy
Se1 Se   4 e 0.43700 0.18500  0.08400 1.00000
Se2 Se   4 e 0.24600 0.27300 -0.02300 1.00000
Se3 Se   4 e 0.24000 0.10200  0.82800 1.00000
Se4 Se   4 e 0.05000 -0.08000  0.85200 1.00000
Se5 Se   4 e 0.15700 0.66900 -0.09000 1.00000
Se6 Se   4 e 0.14200 0.66000  0.09000 1.00000
Se7 Se   4 e 0.36800 0.74600  0.16000 1.00000
Se8 Se   4 e 0.33400 0.02100  0.21000 1.00000
```

β-Se ($A_l$): A_mP32_14_8e - POSCAR

```
A_mP32_14_8e & a,b/a,c/a,\beta,x1,y1,z1,x2,y2,z2,x3,y3,z3,x4,y4,z4,x5,y5
   ↪ ,z5,x6,y6,z6,x7,y7,z7,x8,y8,z8 --params=9.31,0.866809881847,
   ↪ 1.38023630505,93.13333,0.437,0.185,0.084,0.246,0.273,-0.023,
   ↪ 0.24,0.102,0.828,0.05,-0.08,0.852,0.157,0.669,-0.09,0.142,0.66,
   ↪ 0.09,0.368,0.746,0.16,0.334,0.021,0.21 & P2_1/c      C_{2h}^5 #
   ↪ 14 (e^8) & mP32 & A_l & Se & beta & Marsh, Pauling, and
   ↪ McCullough, Acta Cryst. 6, 71-75 (1953)
   1.0000000000000000
   9.31000000000000   0.00000000000000   0.00000000000000
   0.00000000000000   8.07000000000000   0.00000000000000
  -0.70237752000000   0.00000000000000  12.83078976000000
    Se
    32
Direct
   0.43700000000000   0.18500000000000   0.08400000000000   Se   (4e)
   0.43700000000000   0.31500000000000   0.58400000000000   Se   (4e)
   0.56300000000000   0.68500000000000   0.41600000000000   Se   (4e)
   0.56300000000000   0.81500000000000  -0.08400000000000   Se   (4e)
   0.24600000000000   0.22700000000000   0.47700000000000   Se   (4e)
   0.24600000000000   0.27300000000000  -0.02300000000000   Se   (4e)
   0.75400000000000   0.72700000000000   0.02300000000000   Se   (4e)
   0.75400000000000   0.77300000000000   0.52300000000000   Se   (4e)
   0.24000000000000   0.10200000000000   0.82800000000000   Se   (4e)
   0.24000000000000   0.39800000000000   0.32800000000000   Se   (4e)
   0.76000000000000   0.60200000000000   0.67200000000000   Se   (4e)
   0.76000000000000   0.89800000000000   0.17200000000000   Se   (4e)
  -0.05000000000000   0.08000000000000   0.14800000000000   Se   (4e)
   0.05000000000000  -0.08000000000000   0.85200000000000   Se   (4e)
  -0.05000000000000   0.42000000000000   0.64800000000000   Se   (4e)
   0.05000000000000   0.58000000000000   0.35200000000000   Se   (4e)
   0.15700000000000   0.66900000000000  -0.09000000000000   Se   (4e)
   0.15700000000000   0.83100000000000   0.41000000000000   Se   (4e)
   0.84300000000000   0.16900000000000   0.59000000000000   Se   (4e)
   0.84300000000000   0.33100000000000   0.09000000000000   Se   (4e)
   0.14200000000000   0.66000000000000   0.09000000000000   Se   (4e)
   0.14200000000000   0.84000000000000   0.59000000000000   Se   (4e)
   0.85800000000000   0.16000000000000   0.41000000000000   Se   (4e)
   0.85800000000000   0.34000000000000  -0.09000000000000   Se   (4e)
   0.36800000000000   0.74600000000000   0.16000000000000   Se   (4e)
   0.36800000000000   0.75400000000000   0.66000000000000   Se   (4e)
   0.63200000000000   0.24600000000000   0.34000000000000   Se   (4e)
   0.63200000000000   0.25400000000000   0.84000000000000   Se   (4e)
   0.33400000000000   0.02100000000000   0.21000000000000   Se   (4e)
   0.33400000000000   0.47900000000000   0.71000000000000   Se   (4e)
   0.66600000000000  -0.02100000000000   0.79000000000000   Se   (4e)
   0.66600000000000   0.52100000000000   0.29000000000000   Se   (4e)
```

Se ($A_k$): A_mP64_14_16e - CIF

```
# CIF file

data_findsym-output
_audit_creation_method FINDSYM

_chemical_name_mineral 'red selenium'
_chemical_formula_sum 'Se'

loop_
_publ_author_name
 'Olav Foss'
 'Vitalijus Janickis'
_journal_name_full
;
Journal of the Chemical Society, Chemical Communications
;
_journal_volume
_journal_year 1977
_journal_page_first 834
_journal_page_last 835
```

```
_publ_Section_title
;
 X-Ray crystal structure of a new red, monoclinic form of
    ↪ cyclo-octaselenium, Se$_{8}$
;

# Found in Pearson's Hanbook, Vol. IV, p. 5716

_aflow_proto 'A_mP64_14_16e'
_aflow_params 'a,b/a,c/a,\beta,x1,y1,z1,x2,y2,z2,x3,y3,z3,x4,y4,z4,x5,y5
    ↪ ,z5,x6,y6,z6,x7,y7,z7,x8,y8,z8,x9,y9,z9,x10,y10,z10,x11,y11,z11
    ↪ ,x12,y12,z12,x13,y13,z13,x14,y14,z14,x15,y15,z15,x16,y16,z16'
_aflow_params_values '15.018,0.979691037422,0.585231056066,93.61,0.18313
    ↪ ,0.14063,0.03451,0.22856,0.28408,0.12262,0.35548,0.31907,-
    ↪ 0.00548,0.47826,0.28776,0.16131,0.52853,0.14438,0.09345,0.47966
    ↪ ,0.04033,0.27102,0.36296,-0.02818,0.15123,0.22521,0.04261,
    ↪ 0.2343,0.09552,0.48601,0.14213,0.01298,0.58883,0.27815,-0.01931
    ↪ ,0.71476,0.12135,0.08347,0.82945,0.18553,0.19177,0.81338,
    ↪ 0.00963,0.3102,0.73961,0.14402,0.30834,0.59137,0.04778,0.24353,
    ↪ 0.50553,0.23353'
_aflow_Strukturbericht 'A_k'
_aflow_Pearson 'mP64'

_symmetry_space_group_name_Hall "-P 2ybc"
_symmetry_space_group_name_H-M "P 1 21/c 1"
_symmetry_Int_Tables_number 14

_cell_length_a    15.01800
_cell_length_b    14.71300
_cell_length_c    8.78900
_cell_angle_alpha 90.00000
_cell_angle_beta  93.61000
_cell_angle_gamma 90.00000

loop_
_space_group_symop_id
_space_group_symop_operation_xyz
1 x,y,z
2 -x,y+1/2,-z+1/2
3 -x,-y,-z
4 x,-y+1/2,z+1/2

loop_
_atom_site_label
_atom_site_type_symbol
_atom_site_symmetry_multiplicity
_atom_site_Wyckoff_label
_atom_site_fract_x
_atom_site_fract_y
_atom_site_fract_z
_atom_site_occupancy
Se1  Se  4 e 0.18313  0.14063  0.03451  1.00000
Se2  Se  4 e 0.22856  0.28408  0.12262  1.00000
Se3  Se  4 e 0.35548  0.31907  -0.00548 1.00000
Se4  Se  4 e 0.47826  0.28776  0.16131  1.00000
Se5  Se  4 e 0.52853  0.14438  0.09345  1.00000
Se6  Se  4 e 0.47966  0.04033  0.27102  1.00000
Se7  Se  4 e 0.36296  -0.02818  0.15123  1.00000
Se8  Se  4 e 0.22521  0.04261  0.23430  1.00000
Se9  Se  4 e 0.09552  0.48601 0.14213  1.00000
Se10 Se  4 e 0.01298  0.58883  0.27815  1.00000
Se11 Se  4 e -0.01931 0.71476  0.12135  1.00000
Se12 Se  4 e 0.08347  0.82945  0.18553  1.00000
Se13 Se  4 e 0.19177  0.81338  0.00963  1.00000
Se14 Se  4 e 0.31020  0.73961  0.14402  1.00000
Se15 Se  4 e 0.30834  0.59137  0.04778  1.00000
Se16 Se  4 e 0.24353  0.50553  0.23353  1.00000
```

Se ($A_k$): A_mP64_14_16e - POSCAR

```
A_mP64_14_16e & a,b/a,c/a,\beta,x1,y1,z1,x2,y2,z2,x3,y3,z3,x4,y4,z4,x5,
    ↪ y5,z5,x6,y6,z6,x7,y7,z7,x8,y8,z8,x9,y9,z9,x10,y10,z10,x11,y11,
    ↪ z11,x12,y12,z12,x13,y13,z13,x14,y14,z14,x15,y15,z15,x16,y16,z16
    ↪  --params=15.018,0.979691037422,0.585231056066,93.61,0.18313,
    ↪ 0.14063,0.03451,0.22856,0.28408,0.12262,0.35548,0.31907,-
    ↪ 0.00548,0.47826,0.28776,0.16131,0.52853,0.14438,0.09345,0.47966
    ↪ ,0.04033,0.27102,0.36296,-0.02818,0.15123,0.22521,0.04261,
    ↪ 0.2343,0.09552,0.48601,0.14213,0.01298,0.58883,0.27815,-0.01931
    ↪ ,0.71476,0.12135,0.08347,0.82945,0.18553,0.19177,0.81338,
    ↪ 0.00963,0.3102,0.73961,0.14402,0.30834,0.59137,0.04778,0.24353,
    ↪ 0.50553,0.23353 & P2_1/c      C_{2h}^5 #14 (e^16) & mP64 & A_k &
    ↪  Se & Red (gamma) & Foss and Janickis J. Chem. Soc., Chem.
    ↪ Comm. 834-835 (1977)
   1.0000000000000000
  15.01800000000000   0.00000000000000   0.00000000000000
   0.00000000000000  14.71300000000000   0.00000000000000
  -0.55339681000000   0.00000000000000   8.77156046000000
   Se
   64
Direct
   0.18313000000000   0.14063000000000   0.03451000000000   Se   (4e)
   0.18313000000000   0.35937000000000   0.53451000000000   Se   (4e)
   0.81687000000000   0.64063000000000   0.46549000000000   Se   (4e)
   0.81687000000000   0.85937000000000  -0.03451000000000   Se   (4e)
  -0.01298000000000   0.08883000000000   0.22185000000000   Se   (4e)
   0.01298000000000  -0.08883000000000   0.77815000000000   Se   (4e)
  -0.01298000000000   0.41117000000000   0.72185000000000   Se   (4e)
   0.01298000000000   0.58883000000000   0.27815000000000   Se   (4e)
   0.01931000000000   0.21476000000000   0.37865000000000   Se   (4e)
   0.01931000000000   0.28524000000000   0.87865000000000   Se   (4e)
  -0.01931000000000   0.71476000000000   0.12135000000000   Se   (4e)
  -0.01931000000000   0.78524000000000   0.62135000000000   Se   (4e)
  -0.08347000000000   0.17055000000000   0.81447000000000   Se   (4e)
  -0.08347000000000   0.32945000000000   0.31447000000000   Se   (4e)
   0.08347000000000   0.67055000000000   0.68553000000000   Se   (4e)
   0.08347000000000   0.82945000000000   0.18553000000000   Se   (4e)
   0.19177000000000   0.68662000000000   0.50963000000000   Se   (4e)
   0.19177000000000   0.81338000000000   0.00963000000000   Se   (4e)
   0.80823000000000   0.18662000000000  -0.00963000000000   Se   (4e)
   0.80823000000000   0.31338000000000   0.49037000000000   Se   (4e)
   0.31020000000000   0.73961000000000   0.14402000000000   Se   (4e)
   0.31020000000000   0.76039000000000   0.64402000000000   Se   (4e)
   0.68980000000000   0.23961000000000   0.35598000000000   Se   (4e)
   0.68980000000000   0.26039000000000   0.85598000000000   Se   (4e)
   0.30834000000000  -0.09137000000000   0.54778000000000   Se   (4e)
   0.30834000000000   0.59137000000000   0.04778000000000   Se   (4e)
   0.69166000000000   0.09137000000000   0.45222000000000   Se   (4e)
   0.69166000000000   0.40863000000000  -0.04778000000000   Se   (4e)
   0.24353000000000  -0.00553000000000   0.73353000000000   Se   (4e)
   0.24353000000000   0.50553000000000   0.23353000000000   Se   (4e)
   0.75647000000000   0.00553000000000   0.26647000000000   Se   (4e)
   0.75647000000000   0.49447000000000   0.76647000000000   Se   (4e)
   0.22856000000000   0.21592000000000   0.62262000000000   Se   (4e)
   0.22856000000000   0.28408000000000   0.12262000000000   Se   (4e)
   0.77144000000000   0.71592000000000   0.87738000000000   Se   (4e)
   0.77144000000000   0.78408000000000   0.37738000000000   Se   (4e)
   0.35548000000000   0.18093000000000   0.49452000000000   Se   (4e)
   0.35548000000000   0.31907000000000  -0.00548000000000   Se   (4e)
   0.64452000000000   0.68093000000000   0.00548000000000   Se   (4e)
   0.64452000000000   0.81907000000000   0.50548000000000   Se   (4e)
   0.47826000000000   0.21224000000000   0.66131000000000   Se   (4e)
   0.47826000000000   0.28776000000000   0.16131000000000   Se   (4e)
   0.52174000000000   0.71224000000000   0.83869000000000   Se   (4e)
   0.52174000000000   0.78776000000000   0.33869000000000   Se   (4e)
   0.47147000000000   0.64438000000000   0.40655000000000   Se   (4e)
   0.47147000000000   0.85562000000000  -0.09345000000000   Se   (4e)
   0.52853000000000   0.14438000000000   0.09345000000000   Se   (4e)
   0.52853000000000   0.35562000000000   0.59345000000000   Se   (4e)
   0.47966000000000   0.04033000000000   0.27102000000000   Se   (4e)
   0.47966000000000   0.45967000000000   0.77102000000000   Se   (4e)
   0.52034000000000  -0.04033000000000   0.72898000000000   Se   (4e)
   0.52034000000000   0.54033000000000   0.22898000000000   Se   (4e)
   0.36296000000000  -0.02818000000000   0.15123000000000   Se   (4e)
   0.36296000000000   0.52818000000000   0.65123000000000   Se   (4e)
   0.63704000000000   0.02818000000000   0.84877000000000   Se   (4e)
   0.63704000000000   0.47182000000000   0.34877000000000   Se   (4e)
   0.22521000000000   0.04261000000000   0.23430000000000   Se   (4e)
   0.22521000000000   0.45739000000000   0.73430000000000   Se   (4e)
   0.77479000000000  -0.04261000000000   0.76570000000000   Se   (4e)
   0.77479000000000   0.54261000000000   0.26570000000000   Se   (4e)
  -0.09552000000000  -0.01399000000000   0.35787000000000   Se   (4e)
   0.09552000000000   0.01399000000000   0.64213000000000   Se   (4e)
   0.09552000000000   0.48601000000000   0.14213000000000   Se   (4e)
  -0.09552000000000   0.51399000000000   0.85787000000000   Se   (4e)
```

$B_2Pd_5$: A2B5_mC28_15_f_e2f - CIF

```
# CIF file

data_findsym-output
_audit_creation_method FINDSYM

_chemical_name_mineral ''
_chemical_formula_sum 'B2 Pd5'

loop_
_publ_author_name
 'Erik Stenberg'
_journal_name_full
;
Acta Chemica Scandinavica
;
_journal_volume 15
_journal_year 1961
_journal_page_first 861
_journal_page_last 870
_publ_Section_title
;
 The Crystal Structures of Pd$_5$B$_2$, (Mn$_5$C$_2$), and Pd$_3$B
;

_aflow_proto 'A2B5_mC28_15_f_e2f'
_aflow_params 'a,b/a,c/a,\beta,y1,x2,y2,z2,x3,y3,z3,x4,y4,z4'
_aflow_params_values '12.786,0.387533239481,0.427968090099,97.03333,
    ↪ 0.5727,0.106,0.311,0.077,0.0958,0.0952,0.4213,0.7127,0.0726,
    ↪ 0.3138'
_aflow_Strukturbericht 'None'
_aflow_Pearson 'mC28'

_symmetry_space_group_name_Hall "-C 2yc"
_symmetry_space_group_name_H-M "C 1 2/c 1"
_symmetry_Int_Tables_number 15

_cell_length_a    12.78600
_cell_length_b    4.95500
_cell_length_c    5.47200
_cell_angle_alpha 90.00000
_cell_angle_beta  97.03333
_cell_angle_gamma 90.00000

loop_
_space_group_symop_id
_space_group_symop_operation_xyz
1 x,y,z
2 -x,y,-z+1/2
3 -x,-y,-z
4 x,-y,z+1/2
5 x+1/2,y+1/2,z
6 -x+1/2,y+1/2,-z+1/2
7 -x+1/2,-y+1/2,-z
```

```
8 x+1/2,-y+1/2,z+1/2

loop_
_atom_site_label
_atom_site_type_symbol
_atom_site_symmetry_multiplicity
_atom_site_Wyckoff_label
_atom_site_fract_x
_atom_site_fract_y
_atom_site_fract_z
_atom_site_occupancy
Pd1 Pd    4 e 0.00000 0.57270 0.25000 1.00000
B1  B     8 f 0.10600 0.31100 0.07700 1.00000
Pd2 Pd    8 f 0.09580 0.09520 0.42130 1.00000
Pd3 Pd    8 f 0.71270 0.07260 0.31380 1.00000
```

$B_2Pd_5$: A2B5_mC28_15_f_e2f - POSCAR

```
A2B5_mC28_15_f_e2f & a,b/a,c/a,\beta,y1,x2,y2,z2,x3,y3,z3,x4,y4,z4 --
    ↪ params=12.786,0.387533239481,0.427968090099,97.03333,0.5727,
    ↪ 0.106,0.311,0.077,0.0958,0.0952,0.4213,0.7127,0.0726,0.3138 &
    ↪ C2/c         C_{2h}^6 #15 (ef^3) & mC28 & & B2Pd5 & & Stenberg,
    ↪ Acta Chem. Scand. 15, 861-70 (1961)
   1.0000000000000000
    6.39300000000000   -2.47750000000000    0.00000000000000
    6.39300000000000    2.47750000000000    0.00000000000000
   -0.67002869000000    0.00000000000000    5.43082365000000
     B     Pd
     4     10
Direct
    0.20500000000000    0.58300000000000    0.92300000000000    B    (8f)
    0.41700000000000    0.79500000000000    0.57700000000000    B    (8f)
    0.58300000000000    0.20500000000000    0.42300000000000    B    (8f)
    0.79500000000000    0.41700000000000    0.07700000000000    B    (8f)
    0.42730000000000    0.57270000000000    0.25000000000000    Pd   (4e)
    0.57270000000000    0.42730000000000    0.75000000000000    Pd   (4e)
    0.00060000000000    0.19100000000000    0.42130000000000    Pd   (8f)
   -0.00060000000000    0.80900000000000    0.57870000000000    Pd   (8f)
    0.19100000000000    0.00060000000000    0.92130000000000    Pd   (8f)
    0.80900000000000   -0.00060000000000    0.07870000000000    Pd   (8f)
    0.21470000000000    0.35990000000000    0.18620000000000    Pd   (8f)
    0.35990000000000    0.21470000000000    0.68620000000000    Pd   (8f)
    0.64010000000000    0.78530000000000    0.31380000000000    Pd   (8f)
    0.78530000000000    0.64010000000000    0.81380000000000    Pd   (8f)
```

Tenorite (CuO, B26): AB_mC8_15_c_e - CIF

```
# CIF file

data_findsym-output
_audit_creation_method FINDSYM

_chemical_name_mineral 'Tenorite'
_chemical_formula_sum 'Cu O'

loop_
_publ_author_name
 'S. \AAsbrink'
 'L.-J. Norrby'
_journal_name_full
;
Acta Crystallographica B
;
_journal_volume 26
_journal_year 1970
_journal_page_first 8
_journal_page_last 15
_publ_Section_title
;
 A refinement of the crystal structure of copper(II) oxide with a
    ↪ discussion of some exceptional e.s.d.'s
;

_aflow_proto 'AB_mC8_15_c_e'
_aflow_params 'a,b/a,c/a,\beta,y2'
_aflow_params_values '4.6837,0.730747058949,1.0950317057,120.34,0.4184'
_aflow_Strukturbericht 'B26'
_aflow_Pearson 'mC8'

_symmetry_space_group_name_Hall "-C 2yc"
_symmetry_space_group_name_H-M "C 1 2/c 1"
_symmetry_Int_Tables_number 15

_cell_length_a      4.68370
_cell_length_b      3.42260
_cell_length_c      5.12880
_cell_angle_alpha  90.00000
_cell_angle_beta   120.34000
_cell_angle_gamma  90.00000

loop_
_space_group_symop_id
_space_group_symop_operation_xyz
1 x,y,z
2 -x,y,-z+1/2
3 -x,-y,-z
4 x,-y,z+1/2
5 x+1/2,y+1/2,z
6 -x+1/2,y+1/2,-z+1/2
7 -x+1/2,-y+1/2,-z
8 x+1/2,-y+1/2,z+1/2

loop_
_atom_site_label
_atom_site_type_symbol
_atom_site_symmetry_multiplicity
_atom_site_Wyckoff_label
_atom_site_fract_x
_atom_site_fract_y
_atom_site_fract_z
_atom_site_occupancy
Cu1 Cu    4 c 0.25000 0.25000 0.00000 1.00000
O1  O     4 e 0.00000 0.41840 0.25000 1.00000
```

Tenorite (CuO, B26): AB_mC8_15_c_e - POSCAR

```
AB_mC8_15_c_e & a,b/a,c/a,\beta,y2 --params=4.6837,0.730747058949,
    ↪ 1.0950317057,120.34,0.4184 & C2/c         C_{2h}^6 #15 (ce) &
    ↪ mC8 & B26 & CuO & Tenorite & \AAsbrink and Norrby, Acta Cryst.
    ↪ B 26, 8-15 (1970)
   1.0000000000000000
    2.34185000000000   -1.71130000000000    0.00000000000000
    2.34185000000000    1.71130000000000    0.00000000000000
   -2.59071210000000    0.00000000000000    4.42637552000000
    Cu     O
     2     2
Direct
   -0.00000000000000    0.50000000000000    0.00000000000000    Cu   (4c)
    0.50000000000000   -0.00000000000000    0.50000000000000    Cu   (4c)
    0.41840000000000    0.58160000000000    0.75000000000000    O    (4e)
    0.58160000000000    0.41840000000000    0.25000000000000    O    (4e)
```

Coesite ($SiO_2$): A2B_mC48_15_ae3f_2f - CIF

```
# CIF file

data_findsym-output
_audit_creation_method FINDSYM

_chemical_name_mineral 'Coesite'
_chemical_formula_sum 'Si O2'

loop_
_publ_author_name
 'Louise Levien'
 'Charles T. Prewitt'
_journal_name_full
;
American Mineralogist
;
_journal_volume 66
_journal_year 1981
_journal_page_first 324
_journal_page_last 333
_publ_Section_title
;
 High-pressure crystal structure and compressibility of coesite
;

_aflow_proto 'A2B_mC48_15_ae3f_2f'
_aflow_params 'a,b/a,c/a,\beta,y2,x3,y3,z3,x4,y4,z4,x5,y5,z5,x6,y6,z6,x7
    ↪ ,y7,z7'
_aflow_params_values '7.1356,1.73344918437,1.00532541062,120.34,0.1163,
    ↪ 0.266,0.1234,0.9401,0.3114,0.1038,0.3282,0.0172,0.2117,0.4782,
    ↪ 0.14033,0.10833,0.07227,0.50682,0.15799,0.54077'
_aflow_Strukturbericht 'None'
_aflow_Pearson 'mC48'

_symmetry_space_group_name_Hall "-C 2yc"
_symmetry_space_group_name_H-M "C 1 2/c 1"
_symmetry_Int_Tables_number 15

_cell_length_a      7.13560
_cell_length_b      12.36920
_cell_length_c      7.17360
_cell_angle_alpha  90.00000
_cell_angle_beta   120.34000
_cell_angle_gamma  90.00000

loop_
_space_group_symop_id
_space_group_symop_operation_xyz
1 x,y,z
2 -x,y,-z+1/2
3 -x,-y,-z
4 x,-y,z+1/2
5 x+1/2,y+1/2,z
6 -x+1/2,y+1/2,-z+1/2
7 -x+1/2,-y+1/2,-z
8 x+1/2,-y+1/2,z+1/2

loop_
_atom_site_label
_atom_site_type_symbol
_atom_site_symmetry_multiplicity
_atom_site_Wyckoff_label
_atom_site_fract_x
_atom_site_fract_y
_atom_site_fract_z
_atom_site_occupancy
O1  O     4 a 0.00000 0.00000 0.00000  1.00000
O2  O     4 e 0.00000 0.11630 0.25000  1.00000
O3  O     8 f 0.26600 0.12340 0.94010  1.00000
O4  O     8 f 0.31140 0.10380 0.32820  1.00000
O5  O     8 f 0.01720 0.21170 0.47820  1.00000
Si1 Si    8 f 0.14033 0.10833 0.07227 1.00000
Si2 Si    8 f 0.50682 0.15799 0.54077  1.00000
```

Coesite ($SiO_2$): A2B_mC48_15_ae3f_2f - POSCAR

```
A2B_mC48_15_ae3f_2f & a,b/a,c/a,\beta,y2,x3,y3,z3,x4,y4,z4,x5,y5,z5,x6,
    ↪ y6,z6,x7,y7,z7 --params=7.1356,1.73344918437,1.00532541062,
    ↪ 120.34,0.1163,0.266,0.1234,0.9401,0.3114,0.1038,0.3282,0.0172,
    ↪ 0.2117,0.4782,0.14033,0.10833,0.07227,0.50682,0.15799,0.54077 &
    ↪ C2/c          C_{2h}^6 #15 (aef^5) & mC48 & & SiO2 & Coesite (1
    ↪ atm) & Levien and Prewitt, Am. Mineral. 66, 324-33 (1981)
   1.0000000000000000
   3.56780000000000   -6.18460000000000    0.00000000000000
   3.56780000000000    6.18460000000000    0.00000000000000
  -3.62360247000000    0.00000000000000    6.19112608000000
     O     Si
    16      8
Direct
   0.00000000000000    0.00000000000000    0.00000000000000    O    (4a)
   0.00000000000000    0.00000000000000    0.50000000000000    O    (4a)
   0.11630000000000    0.88370000000000    0.75000000000000    O    (4e)
   0.88370000000000    0.11630000000000    0.25000000000000    O    (4e)
   0.14260000000000    0.38940000000000    0.94010000000000    O    (8f)
   0.38940000000000    0.14260000000000    0.44010000000000    O    (8f)
   0.61060000000000    0.85740000000000    0.55990000000000    O    (8f)
   0.85740000000000    0.61060000000000    0.05990000000000    O    (8f)
   0.20760000000000    0.41520000000000    0.32820000000000    O    (8f)
   0.41520000000000    0.20760000000000    0.82820000000000    O    (8f)
   0.58480000000000    0.79240000000000    0.17180000000000    O    (8f)
   0.79240000000000    0.58480000000000    0.67180000000000    O    (8f)
   0.19450000000000    0.77110000000000    0.52180000000000    O    (8f)
   0.22890000000000    0.80550000000000    0.97820000000000    O    (8f)
   0.77110000000000    0.19450000000000    0.02180000000000    O    (8f)
   0.80550000000000    0.22890000000000    0.47820000000000    O    (8f)
   0.03200000000000    0.24866000000000    0.07227000000000    Si   (8f)
  -0.03200000000000    0.75134000000000    0.92773000000000    Si   (8f)
   0.24866000000000    0.03200000000000    0.57227000000000    Si   (8f)
   0.75134000000000   -0.03200000000000    0.42773000000000    Si   (8f)
   0.33519000000000    0.65117000000000   -0.04077000000000    Si   (8f)
   0.34883000000000    0.66481000000000    0.54077000000000    Si   (8f)
   0.65117000000000    0.33519000000000    0.45923000000000    Si   (8f)
   0.66481000000000    0.34883000000000    0.04077000000000    Si   (8f)
```

Esseneite: ABC6D2_mC40_15_e_e_3f_f - CIF

```
# CIF file

data_findsym-output
_audit_creation_method FINDSYM

_chemical_name_mineral 'Esseneite'
_chemical_formula_sum 'Ca Fe O6 Si2'

loop_
_publ_author_name
 'Michael A. Cosca'
 'Donald R. Peacor'
_journal_name_full
;
American Mineralogist
;
_journal_volume 72
_journal_year 1987
_journal_page_first 148
_journal_page_last 156
_publ_Section_title
;
 Chemistry and structure of esseneite (CaFe$^{3+}$AlSiO$_6$), a new
    ↪ pyroxene produced by pyrometamorphism
;

# Found in  AMS Database

_aflow_proto 'ABC6D2_mC40_15_e_e_3f_f'
_aflow_params 'a,b/a,c/a,\beta,y1,y2,x3,y3,z3,x4,y4,z4,x5,y5,z5,x6,y6,z6
    ↪ '
_aflow_params_values '9.79,0.901123595506,0.548518896834,105.81,0.3082,-
    ↪ 0.0942,0.3888,0.4123,0.8659,0.1365,0.2411,0.6799,0.1468,0.4802,
    ↪ 0.0124,0.2117,0.4057,0.7764'
_aflow_Strukturbericht 'None'
_aflow_Pearson 'mC40'

_symmetry_space_group_name_Hall "-C 2yc"
_symmetry_space_group_name_H-M "C 1 2/c 1"
_symmetry_Int_Tables_number 15

_cell_length_a      9.79000
_cell_length_b      8.82200
_cell_length_c      5.37000
_cell_angle_alpha  90.00000
_cell_angle_beta  105.81000
_cell_angle_gamma  90.00000

loop_
_space_group_symop_id
_space_group_symop_operation_xyz
1 x,y,z
2 -x,y,-z+1/2
3 -x,-y,-z
4 x,-y,z+1/2
5 x+1/2,y+1/2,z
6 -x+1/2,y+1/2,-z+1/2
7 -x+1/2,-y+1/2,-z
8 x+1/2,-y+1/2,z+1/2

loop_
_atom_site_label
_atom_site_type_symbol
_atom_site_symmetry_multiplicity
_atom_site_Wyckoff_label
_atom_site_fract_x
_atom_site_fract_y
_atom_site_fract_z
_atom_site_occupancy
Ca1 Ca   4 e 0.00000  0.30820   0.25000  1.00000
Fe1 Fe   4 e 0.00000 -0.09420   0.25000  1.00000
O1  O    8 f 0.38880  0.41230   0.86590  1.00000
O2  O    8 f 0.13650  0.24110   0.67990  1.00000
O3  O    8 f 0.14680  0.48020   0.01240  1.00000
Si1 Si   8 f 0.21170  0.40570   0.77640  1.00000
```

Esseneite: ABC6D2_mC40_15_e_e_3f_f - POSCAR

```
ABC6D2_mC40_15_e_e_3f_f & a,b/a,c/a,\beta,y1,y2,x3,y3,z3,x4,y4,z4,x5,y5,
    ↪ z5,x6,y6,z6 --params=9.79,0.901123595506,0.548518896834,105.81,
    ↪ 0.3082,-0.0942,0.3888,0.4123,0.8659,0.1365,0.2411,0.6799,0.1468
    ↪ ,0.4802,0.0124,0.2117,0.4057,0.7764 & C2/c        C_{2h}^6 #15 (e
    ↪ ^2f^4) & mC40 & & CaFeO6Si2 & Esseneite & Cosca and Peacor, Am.
    ↪ Mineral. 72, 148-156 (1987)
   1.0000000000000000
   4.89500000000000   -4.41100000000000    0.00000000000000
   4.89500000000000    4.41100000000000    0.00000000000000
  -1.46304674000000    0.00000000000000    5.16685535000000
    Ca    Fe     O    Si
     2     2    12     4
Direct
   0.30820000000000    0.69180000000000    0.75000000000000    Ca   (4e)
   0.69180000000000    0.30820000000000    0.25000000000000    Ca   (4e)
  -0.09420000000000    0.09420000000000    0.75000000000000    Fe   (4e)
   0.09420000000000    0.90580000000000    0.25000000000000    Fe   (4e)
   0.02350000000000    0.19890000000000    0.13410000000000    O    (8f)
  -0.02350000000000    0.80110000000000    0.86590000000000    O    (8f)
   0.19890000000000    0.02350000000000    0.63410000000000    O    (8f)
   0.80110000000000    0.97650000000000    0.36590000000000    O    (8f)
   0.10460000000000    0.62240000000000    0.32010000000000    O    (8f)
   0.37760000000000    0.89540000000000    0.17990000000000    O    (8f)
   0.62240000000000    0.10460000000000    0.82010000000000    O    (8f)
   0.89540000000000    0.37760000000000    0.67990000000000    O    (8f)
   0.33340000000000    0.37300000000000    0.98760000000000    O    (8f)
   0.37300000000000    0.33340000000000    0.48760000000000    O    (8f)
   0.62700000000000    0.66660000000000    0.51240000000000    O    (8f)
   0.66660000000000    0.62700000000000    0.01240000000000    O    (8f)
   0.19400000000000    0.38260000000000    0.22360000000000    Si   (8f)
   0.38260000000000    0.19400000000000    0.72360000000000    Si   (8f)
   0.61740000000000    0.80600000000000    0.27640000000000    Si   (8f)
   0.80600000000000    0.61740000000000    0.77640000000000    Si   (8f)
```

$AlPS_4$: ABC4_oP12_16_ag_cd_2u - CIF

```
# CIF file

data_findsym-output
_audit_creation_method FINDSYM

_chemical_name_mineral ''
_chemical_formula_sum 'Al P S4'

loop_
_publ_author_name
 'A. Weiss'
 'H. Sch{\"a}fer'
_journal_name_full
;
Naturwissenschaften
;
_journal_volume 47
_journal_year 1960
_journal_page_first 495
_journal_page_last 495
_publ_Section_title
;
 Zur Kenntnis von Aluminiumthiophosphat AlPS$_4$
;

_aflow_proto 'ABC4_oP12_16_ag_cd_2u'
_aflow_params 'a,b/a,c/a,x5,y5,z5,x6,y6,z6'
_aflow_params_values '5.61,1.01069518717,1.61319073084,0.2,0.26,0.125,
    ↪ 0.74,0.8,0.63'
_aflow_Strukturbericht 'None'
_aflow_Pearson 'oP12'

_symmetry_space_group_name_Hall "P 2 2"
_symmetry_space_group_name_H-M "P 2 2 2"
_symmetry_Int_Tables_number 16

_cell_length_a      5.61000
_cell_length_b      5.67000
_cell_length_c      9.05000
_cell_angle_alpha  90.00000
_cell_angle_beta   90.00000
_cell_angle_gamma  90.00000

loop_
_space_group_symop_id
_space_group_symop_operation_xyz
1 x,y,z
2 x,-y,-z
3 -x,y,-z
4 -x,-y,z

loop_
_atom_site_label
_atom_site_type_symbol
_atom_site_symmetry_multiplicity
_atom_site_Wyckoff_label
```

```
_atom_site_fract_x
_atom_site_fract_y
_atom_site_fract_z
_atom_site_occupancy
Al1 Al    1 a 0.00000  0.00000  0.00000  1.00000
P1  P     1 c 0.00000  0.50000  0.00000  1.00000
P2  P     1 d 0.00000  0.00000  0.50000  1.00000
Al2 Al    1 g 0.00000  0.50000  0.50000  1.00000
S1  S     4 u 0.20000  0.26000  0.12500  1.00000
S2  S     4 u 0.74000  0.80000  0.63000  1.00000
```

AlPS$_4$: ABC4_oP12_16_ag_cd_2u - POSCAR

```
ABC4_oP12_16_ag_cd_2u & a,b/a,c/a,x5,y5,z5,x6,y6,z6 --params=5.61,
    ↪ 1.01069518717,1.61319073084,0.2,0.26,0.125,0.74,0.8,0.63 & P222
    ↪        D_2^1 #16 (acdgu^2) & oP12 & & AlPS4 & & A. Weiss and H.
    ↪ Sch{"\a}fer, Naturwissenschaften 47, 495 (1960)
   1.0000000000000000
   5.6100000000000000  0.0000000000000000  0.0000000000000000
   0.0000000000000000  5.6700000000000000  0.0000000000000000
   0.0000000000000000  0.0000000000000000  9.0500000000000000
   Al      P      S
    2      2      8
Direct
   0.0000000000000000  0.0000000000000000  0.0000000000000000   Al   (1a)
   0.0000000000000000  0.5000000000000000  0.5000000000000000   Al   (1g)
   0.0000000000000000  0.5000000000000000  0.0000000000000000    P   (1c)
   0.0000000000000000  0.0000000000000000  0.5000000000000000    P   (1d)
   0.2000000000000000  0.2600000000000000  0.1250000000000000    S   (4u)
   0.2000000000000000  0.7400000000000000  0.8750000000000000    S   (4u)
   0.8000000000000000  0.2600000000000000  0.8750000000000000    S   (4u)
   0.8000000000000000  0.7400000000000000  0.1250000000000000    S   (4u)
   0.2600000000000000  0.2000000000000000  0.6300000000000000    S   (4u)
   0.2600000000000000  0.8000000000000000  0.3700000000000000    S   (4u)
   0.7400000000000000  0.2000000000000000  0.3700000000000000    S   (4u)
   0.7400000000000000  0.8000000000000000  0.6300000000000000    S   (4u)
```

BaS$_3$: AB3_oP16_18_ab_3c - CIF

```
# CIF file

data_findsym-output
_audit_creation_method FINDSYM

_chemical_name_mineral ''
_chemical_formula_sum 'Ba S3'

loop_
_publ_author_name
 'W. S. Miller'
 'A. J. King'
_journal_name_full
;
Zeitschrift f\"{u}r Kristallographie - Crystalline Materials
;
_journal_volume 94
_journal_year 1936
_journal_page_first 439
_journal_page_last 446
_publ_Section_title
;
 The Structure of Polysuflides: 1 Barium Trisulfide
;

# Found in  Pearson's Handbook II, p. 1701

_aflow_proto 'AB3_oP16_18_ab_3c'
_aflow_params 'a,b/a,c/a,z1,z2,x3,y3,z3,x4,y4,z4,x5,y5,z5'
_aflow_params_values '8.32,1.15865384615,0.579326923077,0.0,0.0,0.25,
    ↪ 0.25,0.0,0.25,0.5,0.5,0.124,0.309,0.382'
_aflow_Strukturbericht 'None'
_aflow_Pearson 'oP16'

_symmetry_space_group_name_Hall "P 2 2ab"
_symmetry_space_group_name_H-M "P 21 21 2"
_symmetry_Int_Tables_number 18

_cell_length_a      8.32000
_cell_length_b      9.64000
_cell_length_c      4.82000
_cell_angle_alpha  90.00000
_cell_angle_beta   90.00000
_cell_angle_gamma  90.00000

loop_
_space_group_symop_id
_space_group_symop_operation_xyz
1 x,y,z
2 x+1/2,-y+1/2,-z
3 -x+1/2,y+1/2,-z
4 -x,-y,z

loop_
_atom_site_label
_atom_site_type_symbol
_atom_site_symmetry_multiplicity
_atom_site_Wyckoff_label
_atom_site_fract_x
_atom_site_fract_y
_atom_site_fract_z
_atom_site_occupancy
Ba1 Ba    2 a 0.00000  0.00000  0.00000  1.00000
Ba2 Ba    2 b 0.00000  0.50000  0.00000  1.00000
S1  S     4 c 0.25000  0.25000  0.00000  1.00000
S2  S     4 c 0.25000  0.50000  0.50000  1.00000
S3  S     4 c 0.12400  0.30900  0.38200  1.00000
```

BaS$_3$: AB3_oP16_18_ab_3c - POSCAR

```
AB3_oP16_18_ab_3c & a,b/a,c/a,z1,z2,x3,y3,z3,x4,y4,z4,x5,y5,z5 --params=
    ↪ 8.32,1.15865384615,0.579326923077,0.0,0.0,0.25,0.25,0.0,0.25,
    ↪ 0.5,0.5,0.124,0.309,0.382 & P2_12_12   D_2^3 #18 (abc^3) & oP16
    ↪ & & BaS3 & & W. S. Miller and A.J. King, Zeitschrift f\"{u}r
    ↪ Kristallographie - Crystalline Materials 94, 439-446 (1936)
   1.0000000000000000
   8.3200000000000000  0.0000000000000000  0.0000000000000000
   0.0000000000000000  9.6400000000000000  0.0000000000000000
   0.0000000000000000  0.0000000000000000  4.8200000000000000
   Ba      S
    4     12
Direct
   0.0000000000000000  0.0000000000000000  0.0000000000000000   Ba   (2a)
   0.5000000000000000  0.5000000000000000  0.0000000000000000   Ba   (2a)
   0.0000000000000000  0.5000000000000000  0.0000000000000000   Ba   (2b)
   0.5000000000000000  0.0000000000000000  0.0000000000000000   Ba   (2b)
   0.2500000000000000  0.2500000000000000  0.0000000000000000    S   (4c)
   0.2500000000000000  0.7500000000000000  0.0000000000000000    S   (4c)
   0.7500000000000000  0.2500000000000000  0.0000000000000000    S   (4c)
   0.7500000000000000  0.7500000000000000  0.0000000000000000    S   (4c)
   0.2500000000000000  0.0000000000000000  0.5000000000000000    S   (4c)
   0.2500000000000000  0.5000000000000000  0.5000000000000000    S   (4c)
   0.7500000000000000  0.0000000000000000  0.5000000000000000    S   (4c)
   0.7500000000000000  0.5000000000000000  0.5000000000000000    S   (4c)
   0.1240000000000000  0.3090000000000000  0.3820000000000000    S   (4c)
   0.3760000000000000  0.8090000000000000  0.6180000000000000    S   (4c)
   0.6240000000000000  0.1910000000000000  0.6180000000000000    S   (4c)
   0.8760000000000000  0.6910000000000000  0.3820000000000000    S   (4c)
```

Naumannite (Ag$_2$Se): A2B_oP12_19_2a_a - CIF

```
# CIF file

data_findsym-output
_audit_creation_method FINDSYM

_chemical_name_mineral ''
_chemical_formula_sum 'Ag2 Se'

loop_
_publ_author_name
 'G. A. Wiegers'
_journal_name_full
;
American Mineralogist
;
_journal_volume 56
_journal_year 1971
_journal_page_first 1882
_journal_page_last 1888
_publ_Section_title
;
 The Crystal Structure of the Low-Temperature Form of Silver Selenide
;

# Found in  Pearson's Handbook, Vol I., page 626

_aflow_proto 'A2B_oP12_19_2a_a'
_aflow_params 'a,b/a,c/a,x1,y1,z1,x2,y2,z2,x3,y3,z3'
_aflow_params_values '7.764,0.909582689335,0.558088614116,0.185,0.07,
    ↪ 0.465,0.055,0.765,-0.008,0.884,-0.011,0.391'
_aflow_Strukturbericht 'None'
_aflow_Pearson 'oP12'

_symmetry_space_group_name_Hall "P 2ac 2ab"
_symmetry_space_group_name_H-M "P 21 21 21"
_symmetry_Int_Tables_number 19

_cell_length_a      7.76400
_cell_length_b      7.06200
_cell_length_c      4.33300
_cell_angle_alpha  90.00000
_cell_angle_beta   90.00000
_cell_angle_gamma  90.00000

loop_
_space_group_symop_id
_space_group_symop_operation_xyz
1 x,y,z
2 x+1/2,-y+1/2,-z
3 -x,y+1/2,-z+1/2
4 -x+1/2,-y,z+1/2

loop_
_atom_site_label
_atom_site_type_symbol
_atom_site_symmetry_multiplicity
_atom_site_Wyckoff_label
_atom_site_fract_x
_atom_site_fract_y
_atom_site_fract_z
_atom_site_occupancy
Ag1 Ag    4 a 0.18500  0.07000  0.46500  1.00000
Ag2 Ag    4 a 0.05500  0.76500  -0.00800   1.00000
Se1 Se    4 a 0.88400  -0.01100  0.39100   1.00000
```

Naumannite (Ag$_2$Se): A2B_oP12_19_2a_a - POSCAR

```
A2B_oP12_19_2a_a & a,b/a,c/a,x1,y1,z1,x2,y2,z2,x3,y3,z3 --params=7.764,
    ↪ 0.909582689335,0.558088614116,0.185,0.07,0.465,0.055,0.765,-
    ↪ 0.008,0.884,-0.011,0.391 & P212121        D_2^4 #19 (a^3) & oP12 &
    ↪ & Ag_2Se & Naumannite (below 406K) & G A Wiegers, Am. Mineral.
    ↪ 56, 1882-1888 (1971)
   1.0000000000000000
   7.76400000000000   0.00000000000000   0.00000000000000
   0.00000000000000   7.06200000000000   0.00000000000000
   0.00000000000000   0.00000000000000   4.33300000000000
   Ag    Se
    8     4
Direct
   0.18500000000000   0.07000000000000   0.46500000000000   Ag    (4a)
   0.31500000000000  -0.07000000000000  -0.03500000000000   Ag    (4a)
   0.68500000000000   0.43000000000000   0.53500000000000   Ag    (4a)
   0.81500000000000   0.57000000000000   0.03500000000000   Ag    (4a)
  -0.05500000000000   0.26500000000000   0.50800000000000   Ag    (4a)
   0.05500000000000   0.76500000000000  -0.00800000000000   Ag    (4a)
   0.44500000000000   0.23500000000000   0.49200000000000   Ag    (4a)
   0.55500000000000   0.73500000000000   0.00800000000000   Ag    (4a)
   0.11600000000000   0.48900000000000   0.10900000000000   Se    (4a)
   0.38400000000000   0.51100000000000   0.60900000000000   Se    (4a)
   0.61600000000000   0.01100000000000   0.89100000000000   Se    (4a)
   0.88400000000000  -0.01100000000000   0.39100000000000   Se    (4a)
```

Orthorhombic Tridymite ($SiO_2$): A2B_oC24_20_abc_c - CIF

```
# CIF file

data_findsym-output
_audit_creation_method FINDSYM

_chemical_name_mineral 'High (Orthorhombic) Tridymite'
_chemical_formula_sum 'Si O2'

loop_
_publ_author_name
 'W. A. Dollase'
_journal_name_full
;
Acta Crystallographica
;
_journal_volume 23
_journal_year 1967
_journal_page_first 617
_journal_page_last 623
_publ_Section_title
;
 The crystal structure at 220$^\circ$C of orthorhombic high tridymite
    ↪ from the Steinbach meteorite
;

_aflow_proto 'A2B_oC24_20_abc_c'
_aflow_params 'a,b/a,c/a,x1,y2,x3,y3,z3,x4,y4,z4'
_aflow_params_values '8.74,0.576659038902,0.942791762014,0.3336,0.4403,
    ↪ 0.2453,0.1971,0.2713,0.33154,0.03589,0.81143'
_aflow_Strukturbericht 'None'
_aflow_Pearson 'oC24'

_symmetry_space_group_name_Hall "C 2c 2"
_symmetry_space_group_name_H-M "C 2 2 21"
_symmetry_Int_Tables_number 20

_cell_length_a    8.74000
_cell_length_b    5.04000
_cell_length_c    8.24000
_cell_angle_alpha 90.00000
_cell_angle_beta  90.00000
_cell_angle_gamma 90.00000

loop_
_space_group_symop_id
_space_group_symop_operation_xyz
1 x,y,z
2 x,-y,-z
3 -x,y,-z+1/2
4 -x,-y,z+1/2
5 x+1/2,y+1/2,z
6 x+1/2,-y+1/2,-z
7 -x+1/2,y+1/2,-z+1/2
8 -x+1/2,-y+1/2,z+1/2

loop_
_atom_site_label
_atom_site_type_symbol
_atom_site_symmetry_multiplicity
_atom_site_Wyckoff_label
_atom_site_fract_x
_atom_site_fract_y
_atom_site_fract_z
_atom_site_occupancy
O1  O    4 a 0.33360 0.00000 0.00000 1.00000
O2  O    4 b 0.00000 0.44030 0.25000 1.00000
O3  O    8 c 0.24530 0.19710 0.27130 1.00000
Si1 Si   8 c 0.33154 0.03589 0.81143 1.00000
```

Orthorhombic Tridymite ($SiO_2$): A2B_oC24_20_abc_c - POSCAR

```
A2B_oC24_20_abc_c & a,b/a,c/a,x1,y2,x3,y3,z3,x4,y4,z4 --params=8.74,
    ↪ 0.576659038902,0.942791762014,0.3336,0.4403,0.2453,0.1971,
    ↪ 0.2713,0.33154,0.03589,0.81143 & C222_1        D_2^5 #20 (abc^2)
    ↪ & oC24 & & SiO2 & ortho-tridymate & Dollase, Acta Cryst. 23,
    ↪ 617-623 (1967)
   1.0000000000000000
   4.37000000000000  -2.52000000000000   0.00000000000000
   4.37000000000000   2.52000000000000   0.00000000000000
   0.00000000000000   0.00000000000000   8.24000000000000
    O    Si
    8     4
Direct
   0.33360000000000   0.33360000000000   0.00000000000000   O     (4a)
   0.66640000000000   0.66640000000000   0.50000000000000   O     (4a)
   0.44030000000000   0.55970000000000   0.75000000000000   O     (4b)
   0.55970000000000   0.44030000000000   0.25000000000000   O     (4b)
   0.04820000000000   0.44240000000000   0.27130000000000   O     (8c)
  -0.04820000000000   0.55760000000000   0.77130000000000   O     (8c)
   0.44240000000000   0.04820000000000   0.72870000000000   O     (8c)
   0.55760000000000  -0.04820000000000   0.22870000000000   O     (8c)
   0.29565000000000   0.36743000000000   0.81143000000000   Si    (8c)
   0.36743000000000   0.29565000000000   0.18857000000000   Si    (8c)
   0.63257000000000   0.70435000000000   0.68857000000000   Si    (8c)
   0.70435000000000   0.63257000000000   0.31143000000000   Si    (8c)
```

High-Pressure CdTe: AB_oP2_25_b_a - CIF

```
# CIF file

data_findsym-output
_audit_creation_method FINDSYM

_chemical_name_mineral 'High Pressure Cadmuum Telluride'
_chemical_formula_sum 'Cd Te'

loop_
_publ_author_name
 'Jing Zhu Hu'
_journal_name_full
;
Solid State Communications
;
_journal_volume 63
_journal_year 1987
_journal_page_first 471
_journal_page_last 474
_publ_Section_title
;
 A New High Pressure Phase of CdTe
;

# Found in  Pearson's Handbook, II, p. 2816

_aflow_proto 'AB_oP2_25_b_a'
_aflow_params 'a,b/a,c/a,z1,z2'
_aflow_params_values '2.8102,1.87104120703,1.0769696107,0.0,0.25'
_aflow_Strukturbericht 'None'
_aflow_Pearson 'oP2'

_symmetry_space_group_name_Hall "P 2 -2"
_symmetry_space_group_name_H-M "P m m 2"
_symmetry_Int_Tables_number 25

_cell_length_a    2.81020
_cell_length_b    5.25800
_cell_length_c    3.02650
_cell_angle_alpha 90.00000
_cell_angle_beta  90.00000
_cell_angle_gamma 90.00000

loop_
_space_group_symop_id
_space_group_symop_operation_xyz
1 x,y,z
2 -x,-y,z
3 -x,y,z
4 x,-y,z

loop_
_atom_site_label
_atom_site_type_symbol
_atom_site_symmetry_multiplicity
_atom_site_Wyckoff_label
_atom_site_fract_x
_atom_site_fract_y
_atom_site_fract_z
_atom_site_occupancy
Te1 Te   1 a 0.00000 0.00000 0.00000 1.00000
Cd1 Cd   1 b 0.00000 0.50000 0.25000 1.00000
```

High-Pressure CdTe: AB_oP2_25_b_a - POSCAR

```
AB_oP2_25_b_a & a,b/a,c/a,z1,z2 --params=2.8102,1.87104120703,
    ↪ 1.0769696107,0.0,0.25 & Pmm2        C_{2v}^1 #25 (ab) & oP2 & &
    ↪ CdTe & P>12GPa & Jing Zhu Hu, Solid State Comm. 63, 471-474 (
    ↪ 1987)
   1.0000000000000000
   2.81020000000000   0.00000000000000   0.00000000000000
   0.00000000000000   5.25800000000000   0.00000000000000
   0.00000000000000   0.00000000000000   3.02650000000000
   Cd    Te
    1     1
Direct
   0.00000000000000   0.50000000000000   0.25000000000000   Cd    (1b)
   0.00000000000000   0.00000000000000   0.00000000000000   Te    (1a)
```

Krennerite ($AuTe_2$, C46): AB2_oP24_28_acd_2c3d - CIF

```
# CIF file

data_findsym-output
```

```
_audit_creation_method FINDSYM

_chemical_name_mineral 'Krennerite'
_chemical_formula_sum 'Au Te2'

loop_
_publ_author_name
 'George Tunell'
 'K. J. Murata'
_journal_name_full
;
The American Mineralogist
;
_journal_volume 35
_journal_year 1950
_journal_page_first 959
_journal_page_last 984
_publ_Section_title
;
 The Atomic Arrangement and Chemical Composition of Krennerite
;

_aflow_proto 'AB2_oP24_28_acd_2c3d'
_aflow_params 'a,b/a,c/a,z1,y2,z2,y3,z3,y4,z4,x5,y5,z5,x6,y6,z6,x7,y7,z7
    ↪ ,x8,y8,z8'
_aflow_params_values '16.54,0.533252720677,0.269649334946,0.0,0.319,
    ↪ 0.014,0.018,0.042,0.617,0.042,0.624,0.334,0.5,0.503,0.301,0.042
    ↪ ,0.632,0.636,0.5,0.619,0.036,0.5'
_aflow_Strukturbericht 'C46'
_aflow_Pearson 'oP24'

_symmetry_space_group_name_Hall "P 2 -2a"
_symmetry_space_group_name_H-M "P m a 2"
_symmetry_Int_Tables_number 28

_cell_length_a    16.54000
_cell_length_b    8.82000
_cell_length_c    4.46000
_cell_angle_alpha 90.00000
_cell_angle_beta  90.00000
_cell_angle_gamma 90.00000

loop_
_space_group_symop_id
_space_group_symop_operation_xyz
1 x,y,z
2 -x,-y,z
3 -x+1/2,y,z
4 x+1/2,-y,z

loop_
_atom_site_label
_atom_site_type_symbol
_atom_site_symmetry_multiplicity
_atom_site_Wyckoff_label
_atom_site_fract_x
_atom_site_fract_y
_atom_site_fract_z
_atom_site_occupancy
Au1 Au   2 a 0.00000 0.00000  0.00000 1.00000
Au2 Au   2 c 0.25000 0.31900  0.01400 1.00000
Te1 Te   2 c 0.25000 0.01800  0.04200 1.00000
Te2 Te   2 c 0.25000 0.61700  0.04200 1.00000
Au3 Au   4 d 0.62400 0.33400  0.50000 1.00000
Te3 Te   4 d 0.50300 0.301000  0.04200 1.00000
Te4 Te   4 d 0.63200 0.63600  0.50000 1.00000
Te5 Te   4 d 0.61900 0.03600 0.50000 1.00000
```

Krennerite (AuTe$_2$, C46): AB2_oP24_28_acd_2c3d - POSCAR

```
AB2_oP24_28_acd_2c3d & a,b/a,c/a,z1,y2,z2,y3,z3,y4,z4,x5,y5,z5,x6,y6,z6,
    ↪ x7,y7,z7,x8,y8,z8 --params=16.54,0.533252720677,0.269649334946,
    ↪ 0.0,0.319,0.014,0.018,0.042,0.617,0.042,0.624,0.334,0.5,0.503,
    ↪ 0.301,0.042,0.632,0.636,0.5,0.619,0.036,0.5 & Pma2      C_{2v
    ↪ }^4 #28 (ac^3d^4) & oP24 & C46 & AuTe2 & Krennerite & G. Tunell
    ↪  and K. J. Murata, Am. Mineral. 35, 959-984 (1950)
   1.0000000000000000
  16.54000000000000   0.00000000000000   0.00000000000000
   0.00000000000000   8.82000000000000   0.00000000000000
   0.00000000000000   0.00000000000000   4.46000000000000
   Au    Te
    8    16
Direct
   0.00000000000000   0.00000000000000   0.00000000000000   Au    (2a)
   0.50000000000000   0.00000000000000   0.00000000000000   Au    (2a)
   0.25000000000000   0.31900000000000   0.01400000000000   Au    (2c)
   0.75000000000000   0.68100000000000   0.01400000000000   Au    (2c)
   0.12400000000000   0.66600000000000   0.50000000000000   Au    (4d)
   0.37600000000000   0.66600000000000   0.50000000000000   Au    (4d)
   0.62400000000000   0.33400000000000   0.50000000000000   Au    (4d)
   0.87600000000000   0.33400000000000   0.50000000000000   Au    (4d)
   0.25000000000000   0.01800000000000   0.04200000000000   Te    (2c)
   0.75000000000000   0.98200000000000   0.04200000000000   Te    (2c)
   0.25000000000000   0.61700000000000   0.04200000000000   Te    (2c)
   0.75000000000000   0.38300000000000   0.04200000000000   Te    (2c)
   0.00300000000000   0.69900000000000   0.04200000000000   Te    (4d)
   0.49700000000000   0.69900000000000   0.04200000000000   Te    (4d)
   0.50300000000000   0.30100000000000   0.04200000000000   Te    (4d)
   0.99700000000000   0.30100000000000   0.04200000000000   Te    (4d)
   0.13200000000000   0.36400000000000   0.50000000000000   Te    (4d)
   0.36800000000000   0.36400000000000   0.50000000000000   Te    (4d)
   0.63200000000000   0.63600000000000   0.50000000000000   Te    (4d)
   0.86800000000000   0.63600000000000   0.50000000000000   Te    (4d)
   0.11900000000000   0.96400000000000   0.50000000000000   Te    (4d)
   0.38100000000000   0.96400000000000   0.50000000000000   Te    (4d)
   0.61900000000000   0.03600000000000   0.50000000000000   Te    (4d)
   0.88100000000000   0.03600000000000   0.50000000000000   Te    (4d)
```

Enargite (AsCu$_3$S$_4$, H2$_5$): AB3C4_oP16_31_a_ab_2ab - CIF

```
# CIF file

data_findsym-output
_audit_creation_method FINDSYM

_chemical_name_mineral 'Enargite'
_chemical_formula_sum 'As Cu3 S4'

loop_
_publ_author_name
 'G. Adiwidjaja'
 'J. L{\"o}hn'
_journal_name_full
;
Acta Crystallographica B
;
_journal_volume 26
_journal_year 1970
_journal_page_first 1878
_journal_page_last 1879
_publ_Section_title
;
 Strukturverfeinerung von Enargit, Cu$_3$AsS$_4$
;

_aflow_proto 'AB3C4_oP16_31_a_ab_2ab'
_aflow_params 'a,b/a,c/a,y1,z1,y2,z2,y3,z3,y4,z4,x5,y5,z5,x6,y6,z6'
_aflow_params_values '7.43,0.869448183042,0.831763122476,0.8268,0.0,
    ↪ 0.1514,0.4983,0.8226,0.6454,0.1436,0.1166,0.2466,0.3255,-0.0134
    ↪ ,0.2598,0.3364,0.6184'
_aflow_Strukturbericht 'H2_5'
_aflow_Pearson 'oP16'

_symmetry_space_group_name_Hall "P 2ac -2"
_symmetry_space_group_name_H-M "P m n 21"
_symmetry_Int_Tables_number 31

_cell_length_a    7.43000
_cell_length_b    6.46000
_cell_length_c    6.18000
_cell_angle_alpha 90.00000
_cell_angle_beta  90.00000
_cell_angle_gamma 90.00000

loop_
_space_group_symop_id
_space_group_symop_operation_xyz
1 x,y,z
2 -x+1/2,-y,z+1/2
3 -x,y,z
4 x+1/2,-y,z+1/2

loop_
_atom_site_label
_atom_site_type_symbol
_atom_site_symmetry_multiplicity
_atom_site_Wyckoff_label
_atom_site_fract_x
_atom_site_fract_y
_atom_site_fract_z
_atom_site_occupancy
As1 As   2 a 0.00000 0.82680 0.00000   1.00000
Cu1 Cu   2 a 0.00000 0.15140 0.49830   1.00000
S1  S    2 a 0.00000 0.82260 0.64540   1.00000
S2  S    2 a 0.00000 0.14360 0.11660   1.00000
Cu2 Cu   4 b 0.24660 0.32550 -0.01340  1.00000
S3  S    4 b 0.25980 0.33640 0.61840   1.00000
```

Enargite (AsCu$_3$S$_4$, H2$_5$): AB3C4_oP16_31_a_ab_2ab - POSCAR

```
AB3C4_oP16_31_a_ab_2ab & a,b/a,c/a,y1,z1,y2,z2,y3,z3,y4,z4,x5,y5,z5,x6,
    ↪ y6,z6 --params=7.43,0.869448183042,0.831763122476,0.8268,0.0,
    ↪ 0.1514,0.4983,0.8226,0.6454,0.1436,0.1166,0.2466,0.3255,-0.0134
    ↪ ,0.2598,0.3364,0.6184 & Pmn2_1      C_{2v}^7 #31 (a^4b^2) & oP16
    ↪  & H2_5 & AsCu3S4 & Enargite & G. Adiwidjaja and J. L"{o}hn,
    ↪ Acta Cryst. B 26, 1878-1879 (1970)
   1.0000000000000000
   7.43000000000000   0.00000000000000   0.00000000000000
   0.00000000000000   6.46000000000000   0.00000000000000
   0.00000000000000   0.00000000000000   6.18000000000000
   As    Cu    S
    2     6    8
Direct
   0.00000000000000   0.82680000000000   0.00000000000000   As    (2a)
   0.50000000000000   0.17320000000000   0.50000000000000   As    (2a)
   0.00000000000000   0.15140000000000   0.49830000000000   Cu    (2a)
   0.50000000000000   0.84860000000000   0.99830000000000   Cu    (2a)
   0.24660000000000   0.32550000000000   0.98660000000000   Cu    (4b)
   0.25340000000000   0.67450000000000   0.48660000000000   Cu    (4b)
   0.74660000000000   0.67450000000000   0.48660000000000   Cu    (4b)
   0.75340000000000   0.32550000000000   0.98660000000000   Cu    (4b)
   0.00000000000000   0.82260000000000   0.64540000000000    S    (2a)
   0.50000000000000   0.17740000000000   0.14540000000000    S    (2a)
   0.00000000000000   0.14360000000000   0.11660000000000    S    (2a)
   0.50000000000000   0.85640000000000   0.61660000000000    S    (2a)
   0.24020000000000   0.66360000000000   0.11840000000000    S    (4b)
   0.25980000000000   0.33640000000000   0.61840000000000    S    (4b)
   0.74020000000000   0.33640000000000   0.61840000000000    S    (4b)
   0.75980000000000   0.66360000000000   0.11840000000000    S    (4b)
```

Modderite (CoAs): AB_oP8_33_a_a - CIF

```
# CIF file

data_findsym-output
_audit_creation_method FINDSYM

_chemical_name_mineral 'Modderite'
_chemical_formula_sum 'Co As'

loop_
_publ_author_name
 'P. S. Lyman'
 'C. T. Prewitt'
_journal_name_full
;
Acta Crystallographica B
;
_journal_volume 40
_journal_year 1984
_journal_page_first 14
_journal_page_last 20
_publ_Section_title
;
 Room- and high-pressure crystal chemistry of CoAs and FeAs
;

_aflow_proto 'AB_oP8_33_a_a'
_aflow_params 'a,b/a,c/a,x1,y1,z1,x2,y2,z2'
_aflow_params_values '5.2857,1.11007056776,0.659950432298,0.1996,0.5867,
    ↪ 0.2506,0.002,0.2003,0.25'
_aflow_Strukturbericht 'None'
_aflow_Pearson 'oP8'

_symmetry_space_group_name_Hall "P 2c -2n"
_symmetry_space_group_name_H-M "P n a 21"
_symmetry_Int_Tables_number 33

_cell_length_a      5.28570
_cell_length_b      5.86750
_cell_length_c      3.48830
_cell_angle_alpha 90.00000
_cell_angle_beta  90.00000
_cell_angle_gamma 90.00000

loop_
_space_group_symop_id
_space_group_symop_operation_xyz
1 x,y,z
2 -x,-y,z+1/2
3 -x+1/2,y+1/2,z+1/2
4 x+1/2,-y+1/2,z

loop_
_atom_site_label
_atom_site_type_symbol
_atom_site_symmetry_multiplicity
_atom_site_Wyckoff_label
_atom_site_fract_x
_atom_site_fract_y
_atom_site_fract_z
_atom_site_occupancy
As1 As   4 a 0.19960 0.58670 0.25060 1.00000
Co1 Co   4 a 0.00200 0.20030 0.25000 1.00000
```

Modderite (CoAs): AB_oP8_33_a_a - POSCAR

```
AB_oP8_33_a_a & a,b/a,c/a,x1,y1,z1,x2,y2,z2 --params=5.2857,
    ↪ 1.11007056776,0.659950432298,0.1996,0.5867,0.2506,0.002,0.2003,
    ↪ 0.25 & Pna2_1     C_{2v}^9 #33 (a^2) & oP8 & & CoAs & Modderite
    ↪  & P. S. Lyman and C. T. Prewitt, Acta Cryst. B 40, 14-20 (1984
    ↪ )
   1.0000000000000000
   5.28570000000000    0.00000000000000    0.00000000000000
   0.00000000000000    5.86750000000000    0.00000000000000
   0.00000000000000    0.00000000000000    3.48830000000000
   As     Co
    4      4
Direct
   0.19960000000000    0.58670000000000    0.25060000000000    As     (4a)
   0.30040000000000    0.08670000000000    0.75060000000000    As     (4a)
   0.69960000000000   -0.08670000000000    0.25060000000000    As     (4a)
   0.80040000000000    0.41330000000000    0.75060000000000    As     (4a)
   0.00200000000000    0.20030000000000    0.25000000000000    Co     (4a)
  -0.00200000000000    0.79970000000000    0.75000000000000    Co     (4a)
   0.49800000000000    0.70030000000000    0.75000000000000    Co     (4a)
   0.50200000000000    0.29970000000000    0.25000000000000    Co     (4a)
```

$AsK_3S_4$: AB3C4_oP32_33_a_3a_4a - CIF

```
# CIF file

data_findsym-output
_audit_creation_method FINDSYM

_chemical_name_mineral ''
_chemical_formula_sum 'As K3 S4'

loop_
_publ_author_name
 'M. Palazzi'
 'S. Jaulmes'
 'P. Laruelle'
_journal_name_full
;
Acta Crystallographica B
;
_journal_volume 30
_journal_year 1974
_journal_page_first 2378
_journal_page_last 2381
_publ_Section_title
;
 Structure cristalline de K$_3$AsS$_4$
;

# Found in  Pearson's Handbook, Vol. I, p. 1164

_aflow_proto 'AB3C4_oP32_33_a_3a_4a'
_aflow_params 'a,b/a,c/a,x1,y1,z1,x2,y2,z2,x3,y3,z3,x4,y4,z4,x5,y5,z5,x6
    ↪ ,y6,z6,x7,y7,z7,x8,y8,z8'
_aflow_params_values '9.11,1.01866081229,1.1613611416,0.2187,0.4807,
    ↪ 0.2031,0.4418,0.2052,0.0015,0.4488,0.1967,0.4146,0.1422,0.9176,
    ↪ 0.2246,0.191,0.2506,0.2228,0.3424,0.5361,0.0415,0.0069,0.5876,
    ↪ 0.2212,0.3355,0.546,0.3761'
_aflow_Strukturbericht 'None'
_aflow_Pearson 'oP32'

_symmetry_space_group_name_Hall "P 2c -2n"
_symmetry_space_group_name_H-M "P n a 21"
_symmetry_Int_Tables_number 33

_cell_length_a      9.11000
_cell_length_b      9.28000
_cell_length_c      10.58000
_cell_angle_alpha 90.00000
_cell_angle_beta  90.00000
_cell_angle_gamma 90.00000

loop_
_space_group_symop_id
_space_group_symop_operation_xyz
1 x,y,z
2 -x,-y,z+1/2
3 -x+1/2,y+1/2,z+1/2
4 x+1/2,-y+1/2,z

loop_
_atom_site_label
_atom_site_type_symbol
_atom_site_symmetry_multiplicity
_atom_site_Wyckoff_label
_atom_site_fract_x
_atom_site_fract_y
_atom_site_fract_z
_atom_site_occupancy
As1 As   4 a 0.21870 0.48070  0.20310 1.00000
K1  K    4 a 0.44180 0.20520  0.00150 1.00000
K2  K    4 a 0.44880 0.19670  0.41460 1.00000
K3  K    4 a 0.14220 0.91760  0.22460 1.00000
S1  S    4 a 0.19100 0.25060  0.22280 1.00000
S2  S    4 a 0.34240 0.53610  0.04150 1.00000
S3  S    4 a 0.00690 0.58760  0.22120 1.00000
S4  S    4 a 0.33550 0.54600  0.37610 1.00000
```

$AsK_3S_4$: AB3C4_oP32_33_a_3a_4a - POSCAR

```
AB3C4_oP32_33_a_3a_4a & a,b/a,c/a,x1,y1,z1,x2,y2,z2,x3,y3,z3,x4,y4,z4,x5
    ↪ ,y5,z5,x6,y6,z6,x7,y7,z7,x8,y8,z8 --params=9.11,1.01866081229,
    ↪ 1.1613611416,0.2187,0.4807,0.2031,0.4418,0.2052,0.0015,0.4488,
    ↪ 0.1967,0.4146,0.1422,0.9176,0.2246,0.191,0.2506,0.2228,0.3424,
    ↪ 0.5361,0.0415,0.0069,0.5876,0.2212,0.3355,0.546,0.3761 & Pna2_1
    ↪      C_{2v}^9 #33 (a^8) & oP32 & & AsK3S4 & & M. Palazzi, S.
    ↪ Jaulmes and P. Laruelle, Acta Cryst. B 30, 2378-2381 (1974)
   1.0000000000000000
   9.11000000000000    0.00000000000000    0.00000000000000
   0.00000000000000    9.28000000000000    0.00000000000000
   0.00000000000000    0.00000000000000   10.58000000000000
   As     K     S
    4    12    16
Direct
   0.21870000000000    0.48070000000000    0.20310000000000    As     (4a)
   0.28130000000000   -0.01930000000000    0.70310000000000    As     (4a)
   0.71870000000000    0.01930000000000    0.20310000000000    As     (4a)
   0.78130000000000    0.51930000000000    0.70310000000000    As     (4a)
   0.05820000000000    0.70520000000000    0.50150000000000     K     (4a)
   0.44180000000000    0.20520000000000    0.00150000000000     K     (4a)
  -0.44180000000000    0.79480000000000    0.50150000000000     K     (4a)
   0.94180000000000    0.29480000000000    0.00150000000000     K     (4a)
   0.05120000000000    0.69670000000000    0.91460000000000     K     (4a)
   0.44880000000000    0.19670000000000    0.41460000000000     K     (4a)
  -0.44880000000000    0.80330000000000    0.91460000000000     K     (4a)
   0.94880000000000    0.30330000000000    0.41460000000000     K     (4a)
  -0.14220000000000    0.08240000000000    0.72460000000000     K     (4a)
   0.14220000000000    0.91760000000000    0.22460000000000     K     (4a)
   0.35780000000000    0.41760000000000    0.72460000000000     K     (4a)
   0.64220000000000   -0.41760000000000    0.22460000000000     K     (4a)
   0.19100000000000    0.25060000000000    0.22280000000000     S     (4a)
  -0.19100000000000    0.74940000000000    0.72280000000000     S     (4a)
   0.30900000000000    0.75060000000000    0.72280000000000     S     (4a)
   0.69100000000000    0.24940000000000    0.22280000000000     S     (4a)
   0.15760000000000    0.03610000000000    0.54150000000000     S     (4a)
  -0.34240000000000    0.46390000000000    0.54150000000000     S     (4a)
   0.34240000000000    0.53610000000000    0.04150000000000     S     (4a)
   0.84240000000000   -0.03610000000000    0.04150000000000     S     (4a)
  -0.00690000000000    0.41240000000000    0.72120000000000     S     (4a)
   0.00690000000000    0.58760000000000    0.22120000000000     S     (4a)
   0.49310000000000    0.08760000000000    0.72120000000000     S     (4a)
   0.50690000000000   -0.08760000000000    0.22120000000000     S     (4a)
   0.16450000000000    0.04600000000000    0.87610000000000     S     (4a)
```

```
   -0.33550000000000     0.45400000000000     0.87610000000000       S       (4a)
    0.33550000000000     0.54600000000000     0.37610000000000       S       (4a)
    0.83550000000000    -0.04600000000000     0.37610000000000       S       (4a)
```

$HgBr_2$ (C24): A2B_oC12_36_2a_a - CIF

```
# CIF file

data_findsym-output
_audit_creation_method FINDSYM

_chemical_name_mineral ''
_chemical_formula_sum 'Hg Br2'

loop_
_publ_author_name
 'H. Braekken'
_journal_name_full
;
Zeitschrift f\"{u}r Kristallographie - Crystalline Materials
;
_journal_volume 81
_journal_year 1932
_journal_page_first 152
_journal_page_last 154
_publ_Section_title
;
 Zur Kristallstruktur des Quecksilberbromids HgBr$_2$
;

# Found in  AMS Database

_aflow_proto 'A2B_oC12_36_2a_a'
_aflow_params 'a,b/a,c/a,y1,z1,y2,z2,y3,z3'
_aflow_params_values '4.624,1.46820934256,2.69139273356,0.333,0.0,0.061,
    ↪ 0.134,0.395,0.366'
_aflow_Strukturbericht 'C24'
_aflow_Pearson 'oC12'

_symmetry_space_group_name_Hall "C 2c -2"
_symmetry_space_group_name_H-M "C m c 21"
_symmetry_Int_Tables_number 36

_cell_length_a      4.62400
_cell_length_b      6.78900
_cell_length_c      12.44500
_cell_angle_alpha  90.00000
_cell_angle_beta   90.00000
_cell_angle_gamma  90.00000

loop_
_space_group_symop_id
_space_group_symop_operation_xyz
1 x,y,z
2 -x,-y,z+1/2
3 -x,y,z
4 x,-y,z+1/2
5 x+1/2,y+1/2,z
6 -x+1/2,-y+1/2,z+1/2
7 -x+1/2,y+1/2,z
8 x+1/2,-y+1/2,z+1/2

loop_
_atom_site_label
_atom_site_type_symbol
_atom_site_symmetry_multiplicity
_atom_site_Wyckoff_label
_atom_site_fract_x
_atom_site_fract_y
_atom_site_fract_z
_atom_site_occupancy
Br1 Br   4 a 0.00000 0.33300 0.00000 1.00000
Br2 Br   4 a 0.00000 0.06100 0.13400 1.00000
Hg1 Hg   4 a 0.00000 0.39500 0.36600 1.00000
```

$HgBr_2$ (C24): A2B_oC12_36_2a_a - POSCAR

```
A2B_oC12_36_2a_a & a,b/a,c/a,y1,z1,y2,z2,y3,z3 --params=4.624,
    ↪ 1.46820934256,2.69139273356,0.333,0.0,0.061,0.134,0.395,0.366 &
    ↪ Cmc2_1    C_{2v}^{12} #36 (a^3) & oC12 & C24 & HgBr2 & & H.
    ↪ Braekken, Zeitschrift f\"{u}r Kristallographie - Crystalline
    ↪ Materials 81, 152-154 (1932)
   1.0000000000000000
    2.31200000000000    -3.39450000000000     0.00000000000000
    2.31200000000000     3.39450000000000     0.00000000000000
    0.00000000000000     0.00000000000000    12.44500000000000
    Br     Hg
     4      2
Direct
    0.33300000000000     0.66700000000000     0.50000000000000      Br     (4a)
    0.66700000000000     0.33300000000000     0.00000000000000      Br     (4a)
   -0.06100000000000     0.06100000000000     0.13400000000000      Br     (4a)
    0.06100000000000    -0.06100000000000     0.63400000000000      Br     (4a)
    0.39500000000000     0.60500000000000     0.86600000000000      Hg     (4a)
    0.60500000000000     0.39500000000000     0.36600000000000      Hg     (4a)
```

$C_2CeNi$: A2BC_oC8_38_e_a_b - CIF

```
# CIF file

data_findsym-output
_audit_creation_method FINDSYM

_chemical_name_mineral ''
_chemical_formula_sum 'C2 Ce Ni'

loop_
_publ_author_name
 'O. Yi. Bodak'
 'Je. P. Marusin'
_journal_name_full
;
Dopovidi Akademii Nauk Ukrains'koj RSR Seriya A, Fiziko-Tekhnichni ta
    ↪ Matematichni Nauki
;
_journal_volume 12
_journal_year 1979
_journal_page_first 1048
_journal_page_last 1050
_publ_Section_title
;
 The Crystal Structure of RNiC$_2$ Compounds (R=Ce,La,Pr)
;

# Found in  Pearson's Handbook II, 1858-1859

_aflow_proto 'A2BC_oC8_38_e_a_b'
_aflow_params 'a,b/a,c/a,z1,z2,y3,z3'
_aflow_params_values '3.875,1.17470967742,1.59019354839,0.0,0.6144,0.155
    ↪ ,0.2914'
_aflow_Strukturbericht 'None'
_aflow_Pearson 'oC8'

_symmetry_space_group_name_Hall "A 2 -2"
_symmetry_space_group_name_H-M "A m m 2"
_symmetry_Int_Tables_number 38

_cell_length_a      3.87500
_cell_length_b      4.55200
_cell_length_c      6.16200
_cell_angle_alpha  90.00000
_cell_angle_beta   90.00000
_cell_angle_gamma  90.00000

loop_
_space_group_symop_id
_space_group_symop_operation_xyz
1 x,y,z
2 -x,-y,z
3 x,-y,z
4 -x,y,z
5 x,y+1/2,z+1/2
6 -x,-y+1/2,z+1/2
7 x,-y+1/2,z+1/2
8 -x,y+1/2,z+1/2

loop_
_atom_site_label
_atom_site_type_symbol
_atom_site_symmetry_multiplicity
_atom_site_Wyckoff_label
_atom_site_fract_x
_atom_site_fract_y
_atom_site_fract_z
_atom_site_occupancy
Ce1 Ce   2 a 0.00000 0.00000 0.00000 1.00000
Ni1 Ni   2 b 0.50000 0.00000 0.61440 1.00000
C1  C    4 e 0.50000 0.15500 0.29140 1.00000
```

$C_2CeNi$: A2BC_oC8_38_e_a_b - POSCAR

```
A2BC_oC8_38_e_a_b & a,b/a,c/a,z1,z2,y3,z3 --params=3.875,1.17470967742,
    ↪ 1.59019354839,0.0,0.6144,0.155,0.2914 & Amm2     C_{2v}^{14} #
    ↪ 38 (abe) & oC8 & & C2CeNi & & O. I. Bodak and E. P. Marusin,
    ↪ Dopovidia Akademii Nauk Ukrains'koi RSR, Seriya A:
    ↪ Fiziko_Matematichni Ta Tekhnichni Nauki 12, 1048-1050 (1979)
   1.0000000000000000
    3.87500000000000     0.00000000000000     0.00000000000000
    0.00000000000000     2.27600000000000    -3.08100000000000
    0.00000000000000     2.27600000000000     3.08100000000000
    C     Ce     Ni
    2      1      1
Direct
    0.50000000000000     0.55360000000000     0.13640000000000      C     (4e)
    0.50000000000000     0.86360000000000     0.44640000000000      C     (4e)
    0.00000000000000     0.00000000000000     0.00000000000000      Ce    (2a)
    0.50000000000000     0.38560000000000     0.61440000000000      Ni    (2b)
```

$Au_2V$: A2B_oC12_38_de_ab - CIF

```
# CIF file

data_findsym-output
_audit_creation_method FINDSYM

_chemical_name_mineral ''
_chemical_formula_sum 'Au2 V'

loop_
_publ_author_name
 'E. Stolz'
 'K. Schubert'
_journal_name_full
;
Zeitschrift f\"{u}r Metallkunde
;
_journal_volume 53
_journal_year 1962
_journal_page_first 433
```

```
_journal_page_last 444
_publ_Section_title
;
 Strukturuntersuchungen in einigen zu T$^4$-B$^1$ homologen und
      ↪ quasihomologen Systemen
;

# Found in  http://materials.springer.com/isp/crystallographic/docs/
      ↪ sd_1250637

_aflow_proto 'A2B_oC12_38_de_ab'
_aflow_params 'a,b/a,c/a,z1,z2,y3,z3,y4,z4'
_aflow_params_values '4.684,1.81084543126,1.0269000854,0.06,0.5,0.17,
      ↪ 0.56,0.17,0.0'
_aflow_Strukturbericht 'None'
_aflow_Pearson 'oC12'

_symmetry_space_group_name_Hall "A 2 -2"
_symmetry_space_group_name_H-M "A m m 2"
_symmetry_Int_Tables_number 38

_cell_length_a     4.68400
_cell_length_b     8.48200
_cell_length_c     4.81000
_cell_angle_alpha  90.00000
_cell_angle_beta   90.00000
_cell_angle_gamma  90.00000

loop_
_space_group_symop_id
_space_group_symop_operation_xyz
1 x,y,z
2 -x,-y,z
3 x,-y,z
4 -x,y,z
5 x,y+1/2,z+1/2
6 -x,-y+1/2,z+1/2
7 x,-y+1/2,z+1/2
8 -x,y+1/2,z+1/2

loop_
_atom_site_label
_atom_site_type_symbol
_atom_site_symmetry_multiplicity
_atom_site_Wyckoff_label
_atom_site_fract_x
_atom_site_fract_y
_atom_site_fract_z
_atom_site_occupancy
V1  V    2 a 0.00000 0.00000 0.06000 1.00000
V2  V    2 b 0.50000 0.00000 0.50000 1.00000
Au1 Au   4 d 0.00000 0.17000 0.56000 1.00000
Au2 Au   4 e 0.50000 0.17000 0.00000 1.00000
```

$Au_2V$: A2B_oC12_38_de_ab - POSCAR

```
A2B_oC12_38_de_ab & a,b/a,c/a,z1,z2,y3,z3,y4,z4 --params=4.684,
      ↪ 1.81084543126,1.0269000854,0.06,0.5,0.17,0.56,0.17,0.0 & Amm2
      ↪       C_{2v}^{14} #38 (abde) & oC12 & & Au2V & & Stolz and
      ↪ Schubert, Z. Metallkd. 53, 433-444 (1962)
   1.0000000000000000
   4.68400000000000   0.00000000000000   0.00000000000000
   0.00000000000000   4.24100000000000  -2.40500000000000
   0.00000000000000   4.24100000000000   2.40500000000000
   Au    V
    4    2
Direct
   0.00000000000000  -0.39000000000000   0.73000000000000   Au   (4d)
   0.00000000000000  -0.73000000000000   0.39000000000000   Au   (4d)
   0.50000000000000   0.17500000000000   0.17500000000000   Au   (4e)
   0.50000000000000  -0.17500000000000  -0.17500000000000   Au   (4e)
   0.00000000000000  -0.06000000000000   0.06000000000000   V    (2a)
   0.50000000000000  -0.50000000000000   0.50000000000000   V    (2b)
```

$PtSn_4$: AB4_oC20_41_a_2b - CIF

```
# CIF file

data_findsym-output
_audit_creation_method FINDSYM

_chemical_name_mineral ''
_chemical_formula_sum 'Pt Sn4'

loop_
_publ_author_name
 'K. Schubert'
 'U. R\"{o}sler'
_journal_name_full
;
Zeitschrift f\"{u}r Metallkunde
;
_journal_volume 41
_journal_year 1950
_journal_page_first 298
_journal_page_last 300
_publ_Section_title
;
 Die Kristallstruktur von PtSn$_4$
;

# Found in  Pearson's Handbook IV, p. 5001

_aflow_proto 'AB4_oC20_41_a_2b'
_aflow_params 'a,b/a,c/a,z1,x2,y2,z2,x3,y3,z3'
_aflow_params_values '6.388,1.00485284909,1.7778647464,0.0,0.673,0.327,
      ↪ 0.376,0.827,0.673,0.125'
_aflow_Strukturbericht 'D1_c'
_aflow_Pearson 'oC20'

_symmetry_space_group_name_Hall "A 2 -2ac"
_symmetry_space_group_name_H-M "A b a 2"
_symmetry_Int_Tables_number 41

_cell_length_a     6.38800
_cell_length_b     6.41900
_cell_length_c     11.35700
_cell_angle_alpha  90.00000
_cell_angle_beta   90.00000
_cell_angle_gamma  90.00000

loop_
_space_group_symop_id
_space_group_symop_operation_xyz
1 x,y,z
2 -x,-y,z
3 x+1/2,-y,z+1/2
4 -x+1/2,y,z+1/2
5 x,y+1/2,z+1/2
6 -x,-y+1/2,z+1/2
7 x+1/2,-y+1/2,z
8 -x+1/2,y+1/2,z

loop_
_atom_site_label
_atom_site_type_symbol
_atom_site_symmetry_multiplicity
_atom_site_Wyckoff_label
_atom_site_fract_x
_atom_site_fract_y
_atom_site_fract_z
_atom_site_occupancy
Pt1 Pt   4 a 0.00000 0.00000 0.00000 1.00000
Sn1 Sn   8 b 0.67300 0.32700 0.37600 1.00000
Sn2 Sn   8 b 0.82700 0.67300 0.12500 1.00000
```

$PtSn_4$: AB4_oC20_41_a_2b - POSCAR

```
AB4_oC20_41_a_2b & a,b/a,c/a,z1,x2,y2,z2,x3,y3,z3 --params=6.388,
      ↪ 1.00485284909,1.7778647464,0.0,0.673,0.327,0.376,0.827,0.673,
      ↪ 0.125 & Aba2       C_{2v}^{17} #41 (ab^2) & oC20 & D1_c &
      ↪ PtSn4 & & K. Schubert and U. R"{o}sler, Z. Metallkd. 41,
      ↪ 298-300 (1950)
   1.0000000000000000
   6.38800000000000   0.00000000000000   0.00000000000000
   0.00000000000000   3.20950000000000  -5.67850000000000
   0.00000000000000   3.20950000000000   5.67850000000000
   Pt    Sn
    2     8
Direct
   0.00000000000000   0.00000000000000   0.00000000000000   Pt   (4a)
   0.50000000000000   0.50000000000000   0.50000000000000   Pt   (4a)
   0.17300000000000   0.20200000000000   0.45200000000000   Sn   (8b)
   0.32700000000000   0.70200000000000  -0.04800000000000   Sn   (8b)
   0.67300000000000   0.04800000000000   0.29800000000000   Sn   (8b)
   0.82700000000000   0.54800000000000   0.79800000000000   Sn   (8b)
   0.17300000000000   0.79700000000000   0.54900000000000   Sn   (8b)
   0.32700000000000   0.29700000000000   0.04900000000000   Sn   (8b)
   0.67300000000000  -0.04900000000000   0.70300000000000   Sn   (8b)
   0.82700000000000   0.45100000000000   0.20300000000000   Sn   (8b)
```

$PdSn_2$ ($C_e$): AB2_oC24_41_2a_2b - CIF

```
# CIF file

data_findsym-output
_audit_creation_method FINDSYM

_chemical_name_mineral ''
_chemical_formula_sum 'Pd Sn2'

loop_
_publ_author_name
 'K. Schubert'
 'H. Pfisterer'
_journal_name_full
;
Zeitschrift f\"{u}r Metallkunde
;
_journal_volume 41
_journal_year 1950
_journal_page_first 433
_journal_page_last 441
_publ_Section_title
;
 Zur Kristallchemie der B-Metall-reichsten Phasen in Legierungen von \"{
      ↪ U}"bergangsmetallen der Eisen- und Platintriaden mit Elementen
      ↪  der vierten Nebengruppe
;

# Found in  Pearson's Handbook IV, p. 4929-4930

_aflow_proto 'AB2_oC24_41_2a_2b'
_aflow_params 'a,b/a,c/a,z1,z2,x3,y3,z3,x4,y4,z4'
_aflow_params_values '6.478,1.0,1.87635072553,0.01,0.238,0.342,0.158,
      ↪ 0.125,0.25,0.25,-0.125'
_aflow_Strukturbericht 'C_e'
_aflow_Pearson 'oC24'

_symmetry_space_group_name_Hall "A 2 -2ac"
```

```
_symmetry_space_group_name_H-M "A b a 2"
_symmetry_Int_Tables_number 41

_cell_length_a      6.47800
_cell_length_b      6.47800
_cell_length_c      12.15500
_cell_angle_alpha  90.00000
_cell_angle_beta   90.00000
_cell_angle_gamma  90.00000

loop_
_space_group_symop_id
_space_group_symop_operation_xyz
1 x,y,z
2 -x,-y,z
3 x+1/2,-y,z+1/2
4 -x+1/2,y,z+1/2
5 x,y+1/2,z+1/2
6 -x,-y+1/2,z+1/2
7 x+1/2,-y+1/2,z
8 -x+1/2,y+1/2,z

loop_
_atom_site_label
_atom_site_type_symbol
_atom_site_symmetry_multiplicity
_atom_site_Wyckoff_label
_atom_site_fract_x
_atom_site_fract_y
_atom_site_fract_z
_atom_site_occupancy
Pd1 Pd    4 a 0.00000  0.00000  0.01000  1.00000
Pd2 Pd    4 a 0.00000  0.00000  0.23800   1.00000
Sn1 Sn    8 b 0.34200  0.15800  0.12500   1.00000
Sn2 Sn    8 b 0.25000  0.25000  -0.12500   1.00000
```

PdSn$_2$ ($C_e$): AB2_oC24_41_2a_2b - POSCAR

```
AB2_oC24_41_2a_2b & a,b/a,c/a,z1,z2,x3,y3,z3,x4,y4,z4 --params=6.478,1.0
    ↪ ,1.87635072553,0.01,0.238,0.342,0.158,0.125,0.25,0.25,-0.125 &
    ↪ Aba2        C_{2v}^{17} #41 (a^2b^2) & oC24 & C_e & PdSn2 & & K.
    ↪ Schubert and H. Pfisterer, Z. Metallkd. 41, 433-441 (1950)
   1.0000000000000000
   6.47800000000000    0.00000000000000    0.00000000000000
   0.00000000000000    3.23900000000000   -6.07750000000000
   0.00000000000000    3.23900000000000    6.07750000000000
   Pd     Sn
    4      8
Direct
   0.00000000000000   -0.01000000000000    0.01000000000000    Pd      (4a)
   0.50000000000000    0.49000000000000    0.51000000000000    Pd      (4a)
   0.00000000000000    0.76200000000000    0.23800000000000    Pd      (4a)
   0.50000000000000    0.26200000000000    0.73800000000000    Pd      (4a)
   0.15800000000000    0.53300000000000    0.78300000000000    Sn      (8b)
   0.34200000000000    0.03300000000000    0.28300000000000    Sn      (8b)
   0.65800000000000    0.71700000000000   -0.03300000000000    Sn      (8b)
   0.84200000000000    0.21700000000000    0.46700000000000    Sn      (8b)
   0.25000000000000    0.37500000000000    0.12500000000000    Sn      (8b)
   0.25000000000000    0.87500000000000    0.62500000000000    Sn      (8b)
   0.75000000000000    0.37500000000000    0.12500000000000    Sn      (8b)
   0.75000000000000    0.87500000000000    0.62500000000000    Sn      (8b)
```

GeS$_2$ (C44): AB2_oF72_43_ab_3b - CIF

```
# CIF file

data_findsym-output
_audit_creation_method FINDSYM

_chemical_name_mineral 'Germanium disuphide'
_chemical_formula_sum 'Ge S2'

loop_
_publ_author_name
 'W. H. Zachariasen'
_journal_name_full
;
Journal of Chemical Physics
;
_journal_volume 4
_journal_year 1936
_journal_page_first 618
_journal_page_last 619
_publ_Section_title
;
 The Crystal Structure of Germanium Disulphide
;

# Found in  AMS Database

_aflow_proto 'AB2_oF72_43_ab_3b'
_aflow_params 'a,b/a,c/a,z1,x2,y2,z2,x3,y3,z3,x4,y4,z4,x5,y5,z5'
_aflow_params_values '11.66,1.91595197256,0.58833619211,0.0,0.125,
    ↪ 0.13889,0.0,0.02222,0.08056,0.18333,0.15278,-0.01389,-0.18333,
    ↪ 0.0625,0.125,0.27778'
_aflow_Strukturbericht 'C44'
_aflow_Pearson 'oF72'

_symmetry_space_group_name_Hall "F 2 -2d"
_symmetry_space_group_name_H-M "F d d 2"
_symmetry_Int_Tables_number 43

_cell_length_a      11.66000
_cell_length_b      22.34000
_cell_length_c      6.86000
_cell_angle_alpha  90.00000
_cell_angle_beta   90.00000
_cell_angle_gamma  90.00000

loop_
_space_group_symop_id
_space_group_symop_operation_xyz
1 x,y,z
2 -x,-y,z
3 -x+1/4,y+1/4,z+1/4
4 x+1/4,-y+1/4,z+1/4
5 x,y+1/2,z+1/2
6 -x,-y+1/2,z+1/2
7 -x+1/4,y+3/4,z+3/4
8 x+1/4,-y+3/4,z+3/4
9 x+1/2,y,z+1/2
10 -x+1/2,-y,z+1/2
11 -x+3/4,y+1/4,z+3/4
12 x+3/4,-y+1/4,z+3/4
13 x+1/2,y+1/2,z
14 -x+1/2,-y+1/2,z
15 -x+3/4,y+3/4,z+1/4
16 x+3/4,-y+3/4,z+1/4

loop_
_atom_site_label
_atom_site_type_symbol
_atom_site_symmetry_multiplicity
_atom_site_Wyckoff_label
_atom_site_fract_x
_atom_site_fract_y
_atom_site_fract_z
_atom_site_occupancy
Ge1 Ge    8 a 0.00000   0.00000  0.00000  1.00000
Ge2 Ge   16 b 0.12500   0.13889  0.00000  1.00000
S1  S    16 b 0.02222   0.08056  0.18333  1.00000
S2  S    16 b 0.15278  -0.01389  -0.18333  1.00000
S3  S    16 b 0.06250   0.12500  0.27778  1.00000
```

GeS$_2$ (C44): AB2_oF72_43_ab_3b - POSCAR

```
AB2_oF72_43_ab_3b & a,b/a,c/a,z1,x2,y2,z2,x3,y3,z3,x4,y4,z4,x5,y5,z5 --
    ↪ params=11.66,1.91595197256,0.58833619211,0.0,0.125,0.13889,0.0,
    ↪ 0.02222,0.08056,0.18333,0.15278,-0.01389,-0.18333,0.0625,0.125,
    ↪ 0.27778 & Fdd2        C_{2v}^{16} #43 (ab^4) & oF72 & C44 & GeS2
    ↪ & & W. H. Zachariasen, J. Chem. Phys. 4, 618-619 (1936)
   1.0000000000000000
   0.00000000000000   11.17000000000000    3.43000000000000
   5.83000000000000    0.00000000000000    3.43000000000000
   5.83000000000000   11.17000000000000    0.00000000000000
   Ge     S
    6     12
Direct
   0.01388888888889   -0.01388888888889    0.26388888888889    Ge    (16b)
  -0.01388888888889    0.01388888888889    0.73611111111111    Ge    (16b)
  -0.01388888888889    0.51388888888889    0.23611111111111    Ge    (16b)
   0.51388888888889   -0.01388888888889    0.26388888888889    Ge    (16b)
   0.00000000000000    0.00000000000000    0.00000000000000    Ge     (8a)
   0.25000000000000    0.25000000000000    0.25000000000000    Ge     (8a)
   0.12500000000000    0.24166666666667    0.71388888888889    S     (16b)
   0.24166666666667    0.12500000000000   -0.08055555555556    S     (16b)
   0.33055555555556    0.53611111111111    0.00833333333333    S     (16b)
   0.53611111111111    0.33055555555556    0.12500000000000    S     (16b)
  -0.01666666666667    0.65000000000000    0.04444444444444    S     (16b)
  -0.07222222222222    0.20555555555556    0.60000000000000    S     (16b)
   0.20555555555556   -0.07222222222222    0.26666666666667    S     (16b)
   0.65000000000000   -0.01666666666667    0.32222222222222    S     (16b)
   0.21527777777778    0.34027777777778    0.53472222222222    S     (16b)
   0.34027777777778    0.21527777777778   -0.09027777777778    S     (16b)
   0.34027777777778    0.71527777777778   -0.09027777777778    S     (16b)
   0.71527777777778    0.34027777777778    0.03472222222222    S     (16b)
```

High-pressure GaAs: AB_oI4_44_a_b - CIF

```
# CIF file

data_findsym-output
_audit_creation_method FINDSYM

_chemical_name_mineral ''
_chemical_formula_sum 'Ga As'

loop_
_publ_author_name
 'Samuel T. Weir'
 'Yogesh K. Vohra'
 'Craig A. Vanderborgh'
 'Arthur L. Ruoff'
_journal_name_full
;
Physical Review B
;
_journal_volume 39
_journal_year 1989
_journal_page_first 1280
_journal_page_last 1285
_publ_Section_title
;
 Structural phase transitions in GaAs to 108 GPa
;

# Found in  Pearson's Handbook I, p. 1135

_aflow_proto 'AB_oI4_44_a_b'
_aflow_params 'a,b/a,c/a,z1,z2'
```

```
_aflow_params_values '4.92,0.973577235772,0.535569105691,0.0,0.425'
_aflow_Strukturbericht 'None'
_aflow_Pearson 'oI4'

_symmetry_space_group_name_Hall "I 2 -2"
_symmetry_space_group_name_H-M "I m m 2"
_symmetry_Int_Tables_number 44

_cell_length_a 4.92000
_cell_length_b 4.79000
_cell_length_c 2.63500
_cell_angle_alpha 90.00000
_cell_angle_beta 90.00000
_cell_angle_gamma 90.00000

loop_
_space_group_symop_id
_space_group_symop_operation_xyz
1 x,y,z
2 -x,-y,z
3 -x,y,z
4 x,-y,z
5 x+1/2,y+1/2,z+1/2
6 -x+1/2,-y+1/2,z+1/2
7 -x+1/2,y+1/2,z+1/2
8 x+1/2,-y+1/2,z+1/2

loop_
_atom_site_label
_atom_site_type_symbol
_atom_site_symmetry_multiplicity
_atom_site_Wyckoff_label
_atom_site_fract_x
_atom_site_fract_y
_atom_site_fract_z
_atom_site_occupancy
As1 As 2 a 0.00000 0.00000 0.00000 1.00000
Ga1 Ga 2 b 0.00000 0.50000 0.42500 1.00000
```

High-pressure GaAs: AB_oI4_44_a_b - POSCAR

```
AB_oI4_44_a_b & a,b/a,c/a,z1,z2 --params=4.92,0.973577235772,
    ↪ 0.535569105691,0.0,0.425 & Imm2 C_{2v}^{20} #44 (ab) & oI4
    ↪ & & GaAs & III, 28.1 GPa & S. Weir et al., PRB 39, 1280-1285 (
    ↪ 1989)
   1.0000000000000000
  -2.460000000000000   2.395000000000000   1.317500000000000
   2.460000000000000  -2.395000000000000   1.317500000000000
   2.460000000000000   2.395000000000000  -1.317500000000000
   As   Ga
    1    1
Direct
   0.000000000000000   0.000000000000000   0.000000000000000   As   (2a)
   0.925000000000000   0.425000000000000   0.500000000000000   Ga   (2b)
```

1212C [$YBa_2Cu_3O_{7-x}$]: A2B3C7D_oP13_47_t_aq_eqrs_h - CIF

```
# CIF file

data_findsym-output
_audit_creation_method FINDSYM

_chemical_name_mineral ''
_chemical_formula_sum 'Ba2 Cu3 O7 Y'

loop_
_publ_author_name
 'W. I. F. David'
 'W. T. A. Harrison'
 'J. M. F. Gunn'
 'O. Moze, A. K. Soper'
 'P. Day'
 'J. D. Jorgensen'
 'D. G. Hinks'
 'M. A. Beno'
 'L. Soderholm'
 'D. W. Capone II'
 'I. K. Schuller'
 'C. U. Segre'
 'K. Zhang'
 'J. D. Grace'
_journal_name_full
;
Nature
;
_journal_volume 327
_journal_year 1987
_journal_page_first 310
_journal_page_last 312
_publ_Section_title
;
 Structure and crystal chemistry of the high-Tc superconductor
    ↪ YBa$_2$Cu$_3$O$_{7-x}$
;

_aflow_proto 'A2B3C7D_oP13_47_t_aq_eqrs_h'
_aflow_params 'a,b/a,c/a,z4,z5,z6,z7,z8'
_aflow_params_values '3.8187,1.01691675177,3.05567339671,0.3554,0.1579,
    ↪ 0.3771,0.3788,0.18445'
_aflow_Strukturbericht 'None'
_aflow_Pearson 'oP13'

_symmetry_space_group_name_Hall "-P 2 2"
_symmetry_space_group_name_H-M "P m m m"
_symmetry_Int_Tables_number 47
```

```
_cell_length_a 3.81870
_cell_length_b 3.88330
_cell_length_c 11.66870
_cell_angle_alpha 90.00000
_cell_angle_beta 90.00000
_cell_angle_gamma 90.00000

loop_
_space_group_symop_id
_space_group_symop_operation_xyz
1 x,y,z
2 x,-y,-z
3 -x,y,-z
4 -x,-y,z
5 -x,-y,-z
6 -x,y,z
7 x,-y,z
8 x,y,-z

loop_
_atom_site_label
_atom_site_type_symbol
_atom_site_symmetry_multiplicity
_atom_site_Wyckoff_label
_atom_site_fract_x
_atom_site_fract_y
_atom_site_fract_z
_atom_site_occupancy
Cu1 Cu 1 a 0.00000 0.00000 0.00000 1.00000
O1  O  1 e 0.00000 0.50000 0.00000 1.00000
Y1  Y  1 h 0.50000 0.50000 0.50000 1.00000
Cu2 Cu 2 q 0.00000 0.00000 0.35540 1.00000
O2  O  2 q 0.00000 0.00000 0.15790 1.00000
O3  O  2 r 0.00000 0.50000 0.37710 1.00000
O4  O  2 s 0.50000 0.00000 0.37880 1.00000
Ba1 Ba 2 t 0.50000 0.50000 0.18445 1.00000
```

1212C [$YBa_2Cu_3O_{7-x}$]: A2B3C7D_oP13_47_t_aq_eqrs_h - POSCAR

```
A2B3C7D_oP13_47_t_aq_eqrs_h & a,b/a,c/a,z4,z5,z6,z7,z8 --params=3.8187,
    ↪ 1.01691675177,3.05567339671,0.3554,0.1579,0.3771,0.3788,0.18445
    ↪ & Pmmm D_{2h}^{7} #47 (aehq^2rst) & oP13 & & YBa2Cu3O(7-x
    ↪ ) & & David et al., Nature 327, 310-312 (1987)
   1.0000000000000000
   3.818700000000000   0.000000000000000   0.000000000000000
   0.000000000000000   3.883300000000000   0.000000000000000
   0.000000000000000   0.000000000000000  11.668700000000000
   Ba   Cu   O   Y
    2    3   7   1
Direct
   0.500000000000000   0.500000000000000   0.184450000000000   Ba   (2t)
   0.500000000000000   0.500000000000000   0.815550000000000   Ba   (2t)
   0.000000000000000   0.000000000000000   0.000000000000000   Cu   (1a)
   0.000000000000000   0.000000000000000   0.355400000000000   Cu   (2q)
   0.000000000000000   0.000000000000000   0.644600000000000   Cu   (2q)
   0.000000000000000   0.500000000000000   0.000000000000000   O    (1e)
   0.000000000000000   0.000000000000000   0.157900000000000   O    (2q)
   0.000000000000000   0.000000000000000   0.842100000000000   O    (2q)
   0.000000000000000   0.500000000000000   0.377100000000000   O    (2r)
   0.000000000000000   0.500000000000000   0.622900000000000   O    (2r)
   0.500000000000000   0.000000000000000   0.378800000000000   O    (2s)
   0.500000000000000   0.000000000000000   0.621200000000000   O    (2s)
   0.500000000000000   0.500000000000000   0.500000000000000   Y    (1h)
```

$\beta'$-AuCd (B19): AB_oP4_51_e_f - CIF

```
# CIF file

data_findsym-output
_audit_creation_method FINDSYM

_chemical_name_mineral 'beta-prime cadmium gold'
_chemical_formula_sum 'Au Cd'

loop_
_publ_author_name
 'L.-C. Chang'
_journal_name_full
;
Acta Crystallographica
;
_journal_volume 4
_journal_year 1951
_journal_page_first 320
_journal_page_last 324
_publ_Section_title
;
 Atomic displacements and crystallographic mechanisms in diffusionless
    ↪ transformation of gold-cadium single crystals containing 47.5
    ↪ atomic per cent cadmium
;

# Found in Pearson, Alloys, p. 313-314

_aflow_proto 'AB_oP4_51_e_f'
_aflow_params 'a,b/a,c/a,z1,z2'
_aflow_params_values '4.7549,0.661969757513,1.0209678437,0.8125,0.3125'
_aflow_Strukturbericht 'B19'
_aflow_Pearson 'oP4'

_symmetry_space_group_name_Hall "-P 2a 2a"
_symmetry_space_group_name_H-M "P m m a"
_symmetry_Int_Tables_number 51
```

```
_cell_length_a      4.75490
_cell_length_b      3.14760
_cell_length_c      4.85460
_cell_angle_alpha 90.00000
_cell_angle_beta   90.00000
_cell_angle_gamma 90.00000

loop_
_space_group_symop_id
_space_group_symop_operation_xyz
1 x,y,z
2 x+1/2,-y,-z
3 -x,y,-z
4 -x+1/2,-y,z
5 -x,-y,-z
6 -x+1/2,y,z
7 x,-y,z
8 x+1/2,y,-z

loop_
_atom_site_label
_atom_site_type_symbol
_atom_site_symmetry_multiplicity
_atom_site_Wyckoff_label
_atom_site_fract_x
_atom_site_fract_y
_atom_site_fract_z
_atom_site_occupancy
Au1 Au    2 e 0.25000 0.00000 0.81250 1.00000
Cd1 Cd    2 f 0.25000 0.50000 0.31250 1.00000
```

$\beta'$-AuCd (B19): AB_oP4_51_e_f - POSCAR

```
AB_oP4_51_e_f & a,b/a,c/a,z1,z2 --params=4.7549,0.661969757513,
    ↪ 1.0209678437,0.8125,0.3125 & Pmma        D_{2h}^5 #51 (ef) & oP4
    ↪ & B19 & AuCd (beta') & & L.-C. Chang, Acta Cryst., 320-324 (
    ↪ 1951)
   1.0000000000000000
   4.75490000000000    0.00000000000000    0.00000000000000
   0.00000000000000    3.14760000000000    0.00000000000000
   0.00000000000000    0.00000000000000    4.85460000000000
   Au     Cd
    2      2
Direct
   0.25000000000000    0.00000000000000    0.81250000000000    Au    (2e)
   0.75000000000000    0.00000000000000    0.18750000000000    Au    (2e)
   0.25000000000000    0.50000000000000    0.31250000000000    Cd    (2f)
   0.75000000000000    0.50000000000000    0.68750000000000    Cd    (2f)
```

$Sb_2O_3$ ($D5_{11}$): A3B2_oP20_56_ce_e - CIF

```
# CIF file

data_findsym-output
_audit_creation_method FINDSYM

_chemical_name_mineral 'Antimony trioxide'
_chemical_formula_sum 'Sb2 O3'

loop_
_publ_author_name
 'C. Svensson'
_journal_name_full
;
Acta Crystallographica B
;
_journal_volume 30
_journal_year 1974
_journal_page_first 458
_journal_page_last 461
_publ_Section_title
;
 The crystal structure of orthorhombic antimony trioxide, Sb$_2$O$_3$
;
# Found in AMS Database

_aflow_proto 'A3B2_oP20_56_ce_e'
_aflow_params 'a,b/a,c/a,z1,x2,y2,z2,x3,y3,z3'
_aflow_params_values '4.911,2.53797597231,1.10201588271,0.029,0.147,
    ↪ 0.058,0.861,0.044,0.128,0.179'
_aflow_Strukturbericht 'D5_11'
_aflow_Pearson 'oP20'

_symmetry_space_group_name_Hall "-P 2ab 2ac"
_symmetry_space_group_name_H-M "P c c n"
_symmetry_Int_Tables_number 56

_cell_length_a      4.91100
_cell_length_b      12.46400
_cell_length_c      5.41200
_cell_angle_alpha 90.00000
_cell_angle_beta   90.00000
_cell_angle_gamma 90.00000

loop_
_space_group_symop_id
_space_group_symop_operation_xyz
1 x,y,z
2 x+1/2,-y,-z+1/2
3 -x,y+1/2,-z+1/2
4 -x+1/2,-y+1/2,z
5 -x,-y,-z
6 -x+1/2,y,z+1/2
7 x,-y+1/2,z+1/2
8 x+1/2,y+1/2,-z
```

```
loop_
_atom_site_label
_atom_site_type_symbol
_atom_site_symmetry_multiplicity
_atom_site_Wyckoff_label
_atom_site_fract_x
_atom_site_fract_y
_atom_site_fract_z
_atom_site_occupancy
O1   O      4 c 0.25000 0.25000 0.02900 1.00000
O2   O      8 e 0.14700 0.05800 0.86100 1.00000
Sb1  Sb     8 e 0.04400 0.12800 0.17900 1.00000
```

$Sb_2O_3$ ($D5_{11}$): A3B2_oP20_56_ce_e - POSCAR

```
A3B2_oP20_56_ce_e & a,b/a,c/a,z1,x2,y2,z2,x3,y3,z3 --params=4.911,
    ↪ 2.53797597231,1.10201588271,0.029,0.147,0.058,0.861,0.044,0.128
    ↪ ,0.179 & Pccn        D_{2h}^{10} #56 (ce^2) & oP20 & D5_{11} &
    ↪ Sb2O3 & & C. Svensson, Acta Cryst. B 30, 458-461 (1974)
   1.0000000000000000
   4.91100000000000    0.00000000000000    0.00000000000000
   0.00000000000000   12.46400000000000    0.00000000000000
   0.00000000000000    0.00000000000000    5.41200000000000
    O     Sb
   12      8
Direct
   0.25000000000000    0.25000000000000    0.02900000000000    O    (4c)
   0.25000000000000    0.25000000000000    0.52900000000000    O    (4c)
   0.75000000000000    0.75000000000000    0.47100000000000    O    (4c)
   0.75000000000000    0.75000000000000    0.97100000000000    O    (4c)
   0.14700000000000    0.05800000000000   -0.13900000000000    O    (8e)
   0.14700000000000    0.44200000000000    0.36100000000000    O    (8e)
   0.35300000000000    0.05800000000000    0.36100000000000    O    (8e)
   0.35300000000000    0.44200000000000   -0.13900000000000    O    (8e)
   0.64700000000000    0.55800000000000    0.13900000000000    O    (8e)
   0.64700000000000    0.94200000000000    0.63900000000000    O    (8e)
   0.85300000000000    0.55800000000000    0.63900000000000    O    (8e)
   0.85300000000000    0.94200000000000    0.13900000000000    O    (8e)
   0.04400000000000    0.12800000000000    0.17900000000000    Sb   (8e)
   0.04400000000000    0.37200000000000    0.67900000000000    Sb   (8e)
   0.45600000000000    0.12800000000000    0.67900000000000    Sb   (8e)
   0.45600000000000    0.37200000000000    0.17900000000000    Sb   (8e)
   0.54400000000000    0.62800000000000    0.82100000000000    Sb   (8e)
   0.54400000000000    0.87200000000000    0.32100000000000    Sb   (8e)
   0.95600000000000    0.62800000000000    0.32100000000000    Sb   (8e)
   0.95600000000000    0.87200000000000    0.82100000000000    Sb   (8e)
```

KCNS ($F5_9$): ABCD_oP16_57_d_c_d_d - CIF

```
# CIF file

data_findsym-output
_audit_creation_method FINDSYM

_chemical_name_mineral 'Potassium thiocyanate'
_chemical_formula_sum 'K C N S'

loop_
_publ_author_name
 'D. J. Cookson'
 'M. M. Elcombe'
 'T. R. Finlayson'
_journal_name_full
;
Journal of Physics: Condensed Matter
;
_journal_volume 4
_journal_year 1992
_journal_page_first 7851
_journal_page_last 7864
_publ_Section_title
;
 Phonon dispersion relations for potassium thiocyanate at and above room
    ↪ temperature
;

# Found in http://materials.springer.com/lb/docs/
    ↪ sm_lbs_978-3-540-31353-3_141

_aflow_proto 'ABCD_oP16_57_d_c_d_d'
_aflow_params 'a,b/a,c/a,x1,x2,y2,x3,y3,x4,y4'
_aflow_params_values '6.707,0.997614432682,1.13627553303,0.208,0.7704,
    ↪ 0.2871,0.889,0.4154,0.605,0.1087'
_aflow_Strukturbericht 'F5_9'
_aflow_Pearson 'oP16'

_symmetry_space_group_name_Hall "-P 2c 2b"
_symmetry_space_group_name_H-M "P b c m"
_symmetry_Int_Tables_number 57

_cell_length_a      6.70700
_cell_length_b      6.69100
_cell_length_c      7.62100
_cell_angle_alpha 90.00000
_cell_angle_beta   90.00000
_cell_angle_gamma 90.00000

loop_
_space_group_symop_id
_space_group_symop_operation_xyz
1 x,y,z
2 x,-y+1/2,-z
3 -x,y+1/2,-z+1/2
4 -x,-y,z+1/2
```

```
5 -x,-y,-z
6 -x,y+1/2,z
7 x,-y+1/2,z+1/2
8 x,y,-z+1/2

loop_
_atom_site_label
_atom_site_type_symbol
_atom_site_symmetry_multiplicity
_atom_site_Wyckoff_label
_atom_site_fract_x
_atom_site_fract_y
_atom_site_fract_z
_atom_site_occupancy
K1 K   4 c 0.20800  0.25000  0.00000  1.00000
C1 C   4 d 0.77040  0.28710  0.25000  1.00000
N1 N   4 d 0.88900  0.41540  0.25000  1.00000
S1 S   4 d 0.60500  0.10870  0.25000  1.00000
```

KCNS ($F5_9$): ABCD_oP16_57_d_c_d_d - POSCAR

```
ABCD_oP16_57_d_c_d_d & a,b/a,c/a,x1,y1,x2,y2,x3,y3,x4,y4 --params=6.707,
    ↪ 0.997614432682,1.13627553303,0.208,0.7704,0.2871,0.889,0.4154,
    ↪ 0.605,0.1087 & Pbcm        D_{2h}^{11} #57 (cd^3) & oP16 & F5_9 &
    ↪ KCNS & Potassium thiocyanate & D. J. Cookson, M. M. Elcombe
    ↪ and T. R. Finlayson, J. Phys: Condens. Matter 4, 7851-7864 (
    ↪ 1992)
   1.0000000000000000
   6.70700000000000    0.00000000000000    0.00000000000000
   0.00000000000000    6.69100000000000    0.00000000000000
   0.00000000000000    0.00000000000000    7.62100000000000
     C     K     N     S
     4     4     4     4
Direct
   0.22960000000000    0.71290000000000    0.75000000000000    C    (4d)
   0.22960000000000    0.78710000000000    0.25000000000000    C    (4d)
   0.77040000000000    0.21290000000000    0.75000000000000    C    (4d)
   0.77040000000000    0.28710000000000    0.25000000000000    C    (4d)
   0.20800000000000    0.25000000000000    0.00000000000000    K    (4c)
   0.20800000000000    0.25000000000000    0.50000000000000    K    (4c)
   0.79200000000000    0.75000000000000    0.00000000000000    K    (4c)
   0.79200000000000    0.75000000000000    0.50000000000000    K    (4c)
   0.11100000000000   -0.08460000000000    0.25000000000000    N    (4d)
   0.11100000000000    0.58460000000000    0.75000000000000    N    (4d)
   0.88900000000000    0.08460000000000    0.75000000000000    N    (4d)
   0.88900000000000    0.41540000000000    0.25000000000000    N    (4d)
   0.39500000000000    0.60870000000000    0.25000000000000    S    (4d)
   0.39500000000000    0.89130000000000    0.75000000000000    S    (4d)
   0.60500000000000    0.10870000000000    0.25000000000000    S    (4d)
   0.60500000000000    0.39130000000000    0.75000000000000    S    (4d)
```

TlF-II: AB_oP8_57_d_d - CIF

```
# CIF file

data_findsym-output
_audit_creation_method FINDSYM

_chemical_name_mineral 'TlF-II'
_chemical_formula_sum 'Tl F'

loop_
_publ_author_name
 'P. Berastegui'
 'S. Hull'
_journal_name_full
;
Journal of Solid State Chemistry
;
_journal_volume 150
_journal_year 2000
_journal_page_first 266
_journal_page_last 275
_publ_Section_title
;
 The Crystal Structures of Thallium(I) Fluoride
;

_aflow_proto 'AB_oP8_57_d_d'
_aflow_params 'a,b/a,c/a,x1,y1,x2,y2'
_aflow_params_values '6.09556,0.900425883758,0.850291031505,0.8593,
    ↪ 0.0628,0.255,0.0096'
_aflow_Strukturbericht 'None'
_aflow_Pearson 'oP8'

_symmetry_space_group_name_Hall "-P 2c 2b"
_symmetry_space_group_name_H-M "P b c m"
_symmetry_Int_Tables_number 57

_cell_length_a     6.09556
_cell_length_b     5.48860
_cell_length_c     5.18300
_cell_angle_alpha  90.00000
_cell_angle_beta   90.00000
_cell_angle_gamma  90.00000

loop_
_space_group_symop_id
_space_group_symop_operation_xyz
1 x,y,z
2 x,-y+1/2,-z
3 -x,y+1/2,-z+1/2
4 -x,-y,z+1/2
5 -x,-y,-z
6 -x,y+1/2,z
7 x,-y+1/2,z+1/2
8 x,y,-z+1/2

loop_
_atom_site_label
_atom_site_type_symbol
_atom_site_symmetry_multiplicity
_atom_site_Wyckoff_label
_atom_site_fract_x
_atom_site_fract_y
_atom_site_fract_z
_atom_site_occupancy
F1  F    4 d 0.85930  0.06280  0.25000  1.00000
Tl1 Tl   4 d 0.25500  0.00960  0.25000  1.00000
```

TlF-II: AB_oP8_57_d_d - POSCAR

```
AB_oP8_57_d_d & a,b/a,c/a,x1,y1,x2,y2 --params=6.09556,0.900425883758,
    ↪ 0.850291031505,0.8593,0.0628,0.255,0.0096 & Pbcm       D_{2h}^{
    ↪ 11} #57 (d^2) & oP8 & & TlF & TlF-II & P. Berastegui and S.
    ↪ Hull, J. Solid State Chem. 150, 266-275 (2000)
   1.0000000000000000
   6.09556000000000    0.00000000000000    0.00000000000000
   0.00000000000000    5.48860000000000    0.00000000000000
   0.00000000000000    0.00000000000000    5.18300000000000
     F    Tl
     4     4
Direct
   0.14070000000000   -0.06280000000000    0.75000000000000    F    (4d)
   0.14070000000000    0.56280000000000    0.25000000000000    F    (4d)
   0.85930000000000    0.06280000000000    0.25000000000000    F    (4d)
   0.85930000000000    0.43720000000000    0.75000000000000    F    (4d)
   0.25500000000000    0.00960000000000    0.25000000000000    Tl   (4d)
   0.25500000000000    0.49040000000000    0.75000000000000    Tl   (4d)
   0.74500000000000   -0.00960000000000    0.75000000000000    Tl   (4d)
   0.74500000000000    0.50960000000000    0.25000000000000    Tl   (4d)
```

Hydrophilite ($CaCl_2$, C35): AB2_oP6_58_a_g - CIF

```
# CIF file

data_findsym-output
_audit_creation_method FINDSYM

_chemical_name_mineral 'Hydrophilite'
_chemical_formula_sum 'Ca Cl2'

loop_
_publ_author_name
 'A. K. van Bever'
 'W. Nieuwenkamp'
_journal_name_full
;
Zeitschrift f\"{u}r Kristallographie - Crystalline Materials
;
_journal_volume 90
_journal_year 1935
_journal_page_first 374
_journal_page_last 376
_publ_Section_title
;
 Die Kristallstruktur von Calciumchlorid, CaCl$_2$
;

# Found in  AMS Database

_aflow_proto 'AB2_oP6_58_a_g'
_aflow_params 'a,b/a,c/a,x2,y2'
_aflow_params_values '6.24,1.03044871795,0.673076923077,0.275,0.325'
_aflow_Strukturbericht 'C35'
_aflow_Pearson 'oP6'

_symmetry_space_group_name_Hall "-P 2 2n"
_symmetry_space_group_name_H-M "P n n m"
_symmetry_Int_Tables_number 58

_cell_length_a     6.24000
_cell_length_b     6.43000
_cell_length_c     4.20000
_cell_angle_alpha  90.00000
_cell_angle_beta   90.00000
_cell_angle_gamma  90.00000

loop_
_space_group_symop_id
_space_group_symop_operation_xyz
1 x,y,z
2 x+1/2,-y+1/2,-z+1/2
3 -x+1/2,y+1/2,-z+1/2
4 -x,-y,z
5 -x,-y,-z
6 -x+1/2,y+1/2,z+1/2
7 x+1/2,-y+1/2,z+1/2
8 x,y,-z

loop_
_atom_site_label
_atom_site_type_symbol
_atom_site_symmetry_multiplicity
_atom_site_Wyckoff_label
_atom_site_fract_x
_atom_site_fract_y
_atom_site_fract_z
_atom_site_occupancy
Ca1 Ca   2 a 0.00000  0.00000  0.00000  1.00000
```

```
Cl1  Cl    4  g  0.27500  0.32500  0.00000  1.00000
```

Hydrophilite ($CaCl_2$, C35): AB2_oP6_58_a_g - POSCAR

```
AB2_oP6_58_a_g & a,b/a,c/a,x2,y2 --params=6.24,1.03044871795,
    ↪ 0.673076923077,0.275,0.325 & Pnnm        D_{2h}^{12} #58 (ag) &
    ↪ oP6 & C35 & CaCl2 & Hydrophilite & A. van Bever and W.
    ↪ Nieuwenkamp, Zeitschrift f\"{u}r Kristallographie - Crystalline
    ↪  Materials 90, 374-376 (1935)
   1.0000000000000000
   6.24000000000000    0.00000000000000    0.00000000000000
   0.00000000000000    6.43000000000000    0.00000000000000
   0.00000000000000    0.00000000000000    4.20000000000000
   Ca    Cl
    2     4
Direct
   0.00000000000000    0.00000000000000    0.00000000000000    Ca    (2a)
   0.50000000000000    0.50000000000000    0.50000000000000    Ca    (2a)
   0.22500000000000    0.82500000000000    0.50000000000000    Cl    (4g)
   0.27500000000000    0.32500000000000    0.00000000000000    Cl    (4g)
   0.72500000000000    0.67500000000000    0.00000000000000    Cl    (4g)
   0.77500000000000    0.17500000000000    0.50000000000000    Cl    (4g)
```

$\eta$-$Fe_2C$: AB2_oP6_58_a_g - CIF

```
# CIF file

data_findsym-output
_audit_creation_method FINDSYM

_chemical_name_mineral 'zeta iron carbide'
_chemical_formula_sum 'Fe2 C'

loop_
_publ_author_name
 'Y. Hirotsu'
 'S. Nagakura'
_journal_name_full
;
Acta Metallurgica
;
_journal_volume 20
_journal_year 1972
_journal_page_first 645
_journal_page_last 655
_publ_Section_title
;
 Crystal structure and morphology of the carbide precipitated from
    ↪ martensitic high carbon steel during the first stage of
    ↪ tempering
;

_aflow_proto 'AB2_oP6_58_a_g'
_aflow_params 'a,b/a,c/a,x2,y2'
_aflow_params_values '4.704,0.917942176871,0.601615646259,0.66667,0.25'
_aflow_Strukturbericht 'None'
_aflow_Pearson 'oP6'

_symmetry_space_group_name_Hall "-P 2 2n"
_symmetry_space_group_name_H-M "P n n m"
_symmetry_Int_Tables_number 58

_cell_length_a      4.70400
_cell_length_b      4.31800
_cell_length_c      2.83000
_cell_angle_alpha 90.00000
_cell_angle_beta  90.00000
_cell_angle_gamma 90.00000

loop_
_space_group_symop_id
_space_group_symop_operation_xyz
1 x,y,z
2 x+1/2,-y+1/2,-z+1/2
3 -x+1/2,y+1/2,-z+1/2
4 -x,-y,z
5 -x,-y,-z
6 -x+1/2,y+1/2,z+1/2
7 x+1/2,-y+1/2,z+1/2
8 x,y,-z

loop_
_atom_site_label
_atom_site_type_symbol
_atom_site_symmetry_multiplicity
_atom_site_Wyckoff_label
_atom_site_fract_x
_atom_site_fract_y
_atom_site_fract_z
_atom_site_occupancy
C1  C     2  a  0.00000  0.00000  0.00000  1.00000
Fe1 Fe    4  g  0.66667  0.25000  0.00000  1.00000
```

$\eta$-$Fe_2C$: AB2_oP6_58_a_g - POSCAR

```
AB2_oP6_58_a_g & a,b/a,c/a,x2,y2 --params=4.704,0.917942176871,
    ↪ 0.601615646259,0.66667,0.25 & Pnnm        D_{2h}^{12} #58 (ag) &
    ↪ oP6 & & Fe2C & eta & Y. Hirotsu and S. Nagakura, Acta\
    ↪ Metallurgica 20, 645-655 (1972)
   1.0000000000000000
   4.70400000000000    0.00000000000000    0.00000000000000
   0.00000000000000    4.31800000000000    0.00000000000000
   0.00000000000000    0.00000000000000    2.83000000000000
   C     Fe
    2     4
Direct
   0.00000000000000    0.00000000000000    0.00000000000000    C     (2a)
   0.50000000000000    0.50000000000000    0.50000000000000    C     (2a)
   0.16666666700000    0.25000000000000    0.50000000000000    Fe    (4g)
  -0.16666666700000    0.75000000000000    0.50000000000000    Fe    (4g)
   0.33333333300000    0.75000000000000    0.00000000000000    Fe    (4g)
   0.66666666700000    0.25000000000000    0.00000000000000    Fe    (4g)
```

Marcasite ($FeS_2$, C18): AB2_oP6_58_a_g - CIF

```
# CIF file

data_findsym-output
_audit_creation_method FINDSYM

_chemical_name_mineral 'Marcasite'
_chemical_formula_sum 'Fe S2'

loop_
_publ_author_name
 'Milan Rieder'
 'John C. Crelling'
 'Ond\v{r}ej \v{S}ustai'
 'Milan Dr\'{a}bek'
 'Zden\v{e}k Weiss'
 'Mariana Klementov\'{a}'
_journal_name_full
;
International Journal of Coal Geology
;
_journal_volume 71
_journal_year 2007
_journal_page_first 115
_journal_page_last 121
_publ_Section_title
;
 Arsenic in iron disulfides in a brown coal from the North Bohemian
    ↪ Basin, Czech Republic
;

# Found in  AMS Database

_aflow_proto 'AB2_oP6_58_a_g'
_aflow_params 'a,b/a,c/a,x2,y2'
_aflow_params_values '4.4446,1.22049228277,0.761913333033,0.2004,0.3787'
_aflow_Strukturbericht 'C18'
_aflow_Pearson 'oP6'

_symmetry_space_group_name_Hall "-P 2 2n"
_symmetry_space_group_name_H-M "P n n m"
_symmetry_Int_Tables_number 58

_cell_length_a      4.44460
_cell_length_b      5.42460
_cell_length_c      3.38640
_cell_angle_alpha 90.00000
_cell_angle_beta  90.00000
_cell_angle_gamma 90.00000

loop_
_space_group_symop_id
_space_group_symop_operation_xyz
1 x,y,z
2 x+1/2,-y+1/2,-z+1/2
3 -x+1/2,y+1/2,-z+1/2
4 -x,-y,z
5 -x,-y,-z
6 -x+1/2,y+1/2,z+1/2
7 x+1/2,-y+1/2,z+1/2
8 x,y,-z

loop_
_atom_site_label
_atom_site_type_symbol
_atom_site_symmetry_multiplicity
_atom_site_Wyckoff_label
_atom_site_fract_x
_atom_site_fract_y
_atom_site_fract_z
_atom_site_occupancy
Fe1 Fe    2  a  0.00000  0.00000  0.00000  1.00000
S1  S     4  g  0.20040  0.37870  0.00000  1.00000
```

Marcasite ($FeS_2$, C18): AB2_oP6_58_a_g - POSCAR

```
AB2_oP6_58_a_g & a,b/a,c/a,x2,y2 --params=4.4446,1.22049228277,
    ↪ 0.761913333033,0.2004,0.3787 & Pnnm        D_{2h}^{12} #58 (ag) &
    ↪  oP6 & C18 & FeS2 & Marcasite & M. Rieder et al., Int. J. Coal
    ↪ Geol. 71, 115-121 (2007)
   1.0000000000000000
   4.44460000000000    0.00000000000000    0.00000000000000
   0.00000000000000    5.42460000000000    0.00000000000000
   0.00000000000000    0.00000000000000    3.38640000000000
   Fe    S
    2     4
Direct
   0.00000000000000    0.00000000000000    0.00000000000000    Fe    (2a)
   0.50000000000000    0.50000000000000    0.50000000000000    Fe    (2a)
   0.20040000000000    0.37870000000000    0.00000000000000    S     (4g)
   0.29960000000000    0.87870000000000    0.50000000000000    S     (4g)
   0.70040000000000    0.12130000000000    0.50000000000000    S     (4g)
   0.79960000000000    0.62130000000000    0.00000000000000    S     (4g)
```

Vulcanite (CuTe): AB_oP4_59_a_b - CIF

```
# CIF file

data_findsym-output
_audit_creation_method FINDSYM

_chemical_name_mineral 'Vulcanite'
_chemical_formula_sum 'Cu Te'

loop_
_publ_author_name
 'Eugene N. Cameron'
 'Ian M. Threadgold'
_journal_name_full
;
American Mineralogist
;
_journal_volume 46
_journal_year 1961
_journal_page_first 258
_journal_page_last 268
_publ_Section_title
;
 Vulcanite, a new copper telluride from Colorado, with notes on certain
      ↪ associated minerals
;

# Found in AMS Database

_aflow_proto 'AB_oP4_59_a_b'
_aflow_params 'a,b/a,c/a,z1,z2'
_aflow_params_values '3.15,1.29841269841,2.20634920635,0.051,0.277'
_aflow_Strukturbericht 'None'
_aflow_Pearson 'oP4'

_symmetry_space_group_name_Hall "-P 2ab 2a"
_symmetry_space_group_name_H-M "P m m n:2"
_symmetry_Int_Tables_number 59

_cell_length_a      3.15000
_cell_length_b      4.09000
_cell_length_c      6.95000
_cell_angle_alpha 90.00000
_cell_angle_beta  90.00000
_cell_angle_gamma 90.00000

loop_
_space_group_symop_id
_space_group_symop_operation_xyz
1 x,y,z
2 x+1/2,-y,-z
3 -x,y+1/2,-z
4 -x+1/2,-y+1/2,z
5 -x,-y,-z
6 -x+1/2,y,z
7 x,-y+1/2,z
8 x+1/2,y+1/2,-z

loop_
_atom_site_label
_atom_site_type_symbol
_atom_site_symmetry_multiplicity
_atom_site_Wyckoff_label
_atom_site_fract_x
_atom_site_fract_y
_atom_site_fract_z
_atom_site_occupancy
Cu1 Cu    2 a 0.25000 0.25000 0.05100 1.00000
Te1 Te    2 b 0.25000 0.75000 0.27700 1.00000
```

Vulcanite (CuTe): AB_oP4_59_a_b - POSCAR

```
AB_oP4_59_a_b & a,b/a,c/a,z1,z2 --params=3.15,1.29841269841,
      ↪ 2.20634920635,0.051,0.277 & Pmmn       D_{2h}^{13} #59 (ab) &
      ↪ oP4 & & CuTe & Vulcanite & E.N. Cameron and I.M. Threadgold,
      ↪ Am. Mineral. 46, 258-268 (1961)
   1.0000000000000000
   3.15000000000000   0.00000000000000   0.00000000000000
   0.00000000000000   4.09000000000000   0.00000000000000
   0.00000000000000   0.00000000000000   6.95000000000000
   Cu    Te
    2     2
Direct
   0.25000000000000   0.25000000000000   0.05100000000000   Cu    (2a)
   0.75000000000000   0.75000000000000   0.94900000000000   Cu    (2a)
   0.25000000000000   0.75000000000000   0.27700000000000   Te    (2b)
   0.75000000000000   0.25000000000000   0.72300000000000   Te    (2b)
```

CNCl: ABC_oP6_59_a_a_a - CIF

```
# CIF file

data_findsym-output
_audit_creation_method FINDSYM

_chemical_name_mineral 'Cyanogen Chloride'
_chemical_formula_sum 'C N Cl'

loop_
_publ_author_name
 'R. B. Heiart'
 'G. B. Carpenter'
_journal_name_full
;
Acta Crystallographica
;
_journal_volume 9
_journal_year 1956
_journal_page_first 889
_journal_page_last 895
_publ_Section_title
;
 The crystal structure of cyanogen chloride
;

# Found in Wyckoff, Vol. I, pp. 173-174

_aflow_proto 'ABC_oP6_59_a_a_a'
_aflow_params 'a,b/a,c/a,z1,z2,z3'
_aflow_params_values '5.68,0.700704225352,1.01056338028,0.1499,0.4237,
      ↪ 0.6255'
_aflow_Strukturbericht 'None'
_aflow_Pearson 'oP6'

_symmetry_space_group_name_Hall "-P 2ab 2a"
_symmetry_space_group_name_H-M "P m m n:2"
_symmetry_Int_Tables_number 59

_cell_length_a      5.68000
_cell_length_b      3.98000
_cell_length_c      5.74000
_cell_angle_alpha 90.00000
_cell_angle_beta  90.00000
_cell_angle_gamma 90.00000

loop_
_space_group_symop_id
_space_group_symop_operation_xyz
1 x,y,z
2 x+1/2,-y,-z
3 -x,y+1/2,-z
4 -x+1/2,-y+1/2,z
5 -x,-y,-z
6 -x+1/2,y,z
7 x,-y+1/2,z
8 x+1/2,y+1/2,-z

loop_
_atom_site_label
_atom_site_type_symbol
_atom_site_symmetry_multiplicity
_atom_site_Wyckoff_label
_atom_site_fract_x
_atom_site_fract_y
_atom_site_fract_z
_atom_site_occupancy
C1  C     2 a 0.25000 0.25000 0.14990 1.00000
Cl1 Cl    2 a 0.25000 0.25000 0.42370 1.00000
N1  N     2 a 0.25000 0.25000 0.62550 1.00000
```

CNCl: ABC_oP6_59_a_a_a - POSCAR

```
ABC_oP6_59_a_a_a & a,b/a,c/a,z1,z2,z3 --params=5.68,0.700704225352,
      ↪ 1.01056338028,0.1499,0.4237,0.6255 & Pmmn      D_{2h}^{13} #59
      ↪ (a^3) & oP6 & & CNCl & & R. B. Heiart and G. B. Carpenter, Acta
      ↪ Cryst. 9, 889-895 (1956)
   1.0000000000000000
   5.68000000000000   0.00000000000000   0.00000000000000
   0.00000000000000   3.98000000000000   0.00000000000000
   0.00000000000000   0.00000000000000   5.74000000000000
    C     Cl     N
    2      2     2
Direct
   0.25000000000000   0.25000000000000   0.14990000000000    C    (2a)
   0.75000000000000   0.75000000000000   0.85010000000000    C    (2a)
   0.25000000000000   0.25000000000000   0.42370000000000    Cl   (2a)
   0.75000000000000   0.75000000000000   0.57630000000000    Cl   (2a)
   0.25000000000000   0.25000000000000   0.62550000000000    N    (2a)
   0.75000000000000   0.75000000000000   0.37450000000000    N    (2a)
```

$\beta$-$TiCu_3$ ($D0_a$): A3B_oP8_59_bf_a - CIF

```
# CIF file

data_findsym-output
_audit_creation_method FINDSYM

_chemical_name_mineral 'beta Cu3Ti'
_chemical_formula_sum 'Ti Cu3'

loop_
_publ_author_name
 'N. Karlsson'
_journal_name_full
;
Journal of the Institute of Metals
;
_journal_volume 79
_journal_year 1951
_journal_page_first 391
_journal_page_last 391
_publ_Section_title
;
 ~
;

# Found in Pearson Alloys, p. 329-331

_aflow_proto 'A3B_oP8_59_bf_a'
_aflow_params 'a,b/a,c/a,z1,z2,x3,z3'
```

```
_aflow_params_values '5.162,0.842115459124,0.877760557923,0.67125,0.329,
    ↪ 0.505,0.174'
_aflow_Strukturbericht 'D0_a'
_aflow_Pearson 'oP8'

_symmetry_space_group_name_Hall "-P 2ab 2a"
_symmetry_space_group_name_H-M "P m m n:2"
_symmetry_Int_Tables_number 59

_cell_length_a    5.16200
_cell_length_b    4.34700
_cell_length_c    4.53100
_cell_angle_alpha 90.00000
_cell_angle_beta  90.00000
_cell_angle_gamma 90.00000

loop_
_space_group_symop_id
_space_group_symop_operation_xyz
1 x,y,z
2 x+1/2,-y,-z
3 -x,y+1/2,-z
4 -x+1/2,-y+1/2,z
5 -x,-y,-z
6 -x+1/2,y,z
7 x,-y+1/2,z
8 x+1/2,y+1/2,-z

loop_
_atom_site_label
_atom_site_type_symbol
_atom_site_symmetry_multiplicity
_atom_site_Wyckoff_label
_atom_site_fract_x
_atom_site_fract_y
_atom_site_fract_z
_atom_site_occupancy
Ti1 Ti   2 a 0.25000 0.25000 0.67125 1.00000
Cu1 Cu   2 b 0.25000 0.75000 0.32900 1.00000
Cu2 Cu   4 f 0.50500 0.25000 0.17400 1.00000
```

$\beta$-$TiCu_3$ ($D0_a$): A3B_oP8_59_bf_a - POSCAR

```
A3B_oP8_59_bf_a & a,b/a,c/a,z1,z2,x3,z3 --params=5.162,0.842115459124,
    ↪ 0.877760557923,0.67125,0.329,0.505,0.174 & Pmmn       D_{2h}^{
    ↪ 13} #59 (abf) & oP8 & D0_a & TiCu3 & beta & N. Karlsson, J.
    ↪ Inst. Met. 79, 391 (1951)
   1.0000000000000000
   5.16200000000000   0.00000000000000   0.00000000000000
   0.00000000000000   4.34700000000000   0.00000000000000
   0.00000000000000   0.00000000000000   4.53100000000000
   Cu     Ti
    6      2
Direct
   0.25000000000000   0.75000000000000   0.32900000000000   Cu    (2b)
   0.75000000000000   0.25000000000000   0.67100000000000   Cu    (2b)
  -0.00500000000000   0.25000000000000   0.17400000000000   Cu    (4f)
   0.49500000000000   0.75000000000000   0.82600000000000   Cu    (4f)
   0.50500000000000   0.25000000000000   0.17400000000000   Cu    (4f)
   1.00500000000000   0.75000000000000   0.82600000000000   Cu    (4f)
   0.25000000000000   0.25000000000000   0.67125138656800   Ti    (2a)
   0.75000000000000   0.75000000000000   0.32874861343200   Ti    (2a)
```

CdSb ($B_e$): AB_oP16_61_c_c - CIF

```
# CIF file

data_findsym-output
_audit_creation_method FINDSYM

_chemical_name_mineral ''
_chemical_formula_sum 'Cd Sb'

loop_
_publ_author_name
 'Karl Erik Almin'
_journal_name_full
;
Acta Chemica Scandinavica
;
_journal_volume 2
_journal_year 1948
_journal_page_first 400
_journal_page_last 407
_publ_Section_title
;
 The Crystal Structure of CdSb and ZnSb
;

_aflow_proto 'AB_oP16_61_c_c'
_aflow_params 'a,b/a,c/a,x1,y1,z1,x2,y2,z2'
_aflow_params_values '6.471,1.27538247566,1.31757070005,0.136,0.072,
    ↪ 0.108,0.456,0.119,0.872'
_aflow_Strukturbericht 'B_e'
_aflow_Pearson 'oP16'

_symmetry_space_group_name_Hall "-P 2ac 2ab"
_symmetry_space_group_name_H-M "P b c a"
_symmetry_Int_Tables_number 61

_cell_length_a    6.47100
_cell_length_b    8.25300
_cell_length_c    8.52600
_cell_angle_alpha 90.00000
_cell_angle_beta  90.00000
_cell_angle_gamma 90.00000

loop_
_space_group_symop_id
_space_group_symop_operation_xyz
1 x,y,z
2 x+1/2,-y+1/2,-z
3 -x,y+1/2,-z+1/2
4 -x+1/2,-y,z+1/2
5 -x,-y,-z
6 -x+1/2,y+1/2,z
7 x,-y+1/2,z+1/2
8 x+1/2,y,-z+1/2

loop_
_atom_site_label
_atom_site_type_symbol
_atom_site_symmetry_multiplicity
_atom_site_Wyckoff_label
_atom_site_fract_x
_atom_site_fract_y
_atom_site_fract_z
_atom_site_occupancy
Cd1 Cd   8 c 0.13600  0.07200 0.10800 1.00000
Sb1 Sb   8 c 0.45600 0.11900 0.87200 1.00000
```

CdSb ($B_e$): AB_oP16_61_c_c - POSCAR

```
AB_oP16_61_c_c & a,b/a,c/a,x1,y1,z1,x2,y2,z2 --params=6.471,
    ↪ 1.27538247566,1.31757070005,0.136,0.072,0.108,0.456,0.119,0.872
    ↪ & Pbca       D_{2h}^{15} #61 (c^2) & oP16 & B_e & CdSb & & K.
    ↪ E. Almin, Acta Chem. Scand. 2, 400-407 (1948)
   1.0000000000000000
   6.47100000000000   0.00000000000000   0.00000000000000
   0.00000000000000   8.25300000000000   0.00000000000000
   0.00000000000000   0.00000000000000   8.52600000000000
   Cd     Sb
    8      8
Direct
   0.13600000000000   0.07200000000000   0.10800000000000   Cd    (8c)
   0.13600000000000   0.42800000000000   0.60800000000000   Cd    (8c)
   0.36400000000000   0.57200000000000   0.10800000000000   Cd    (8c)
   0.36400000000000   0.92800000000000   0.60800000000000   Cd    (8c)
   0.63600000000000   0.07200000000000   0.39200000000000   Cd    (8c)
   0.63600000000000   0.42800000000000   0.89200000000000   Cd    (8c)
   0.86400000000000   0.57200000000000   0.39200000000000   Cd    (8c)
   0.86400000000000   0.92800000000000   0.89200000000000   Cd    (8c)
   0.04400000000000   0.61900000000000   0.87200000000000   Sb    (8c)
   0.04400000000000   0.88100000000000   0.37200000000000   Sb    (8c)
   0.45600000000000   0.11900000000000   0.87200000000000   Sb    (8c)
   0.45600000000000   0.38100000000000   0.37200000000000   Sb    (8c)
   0.54400000000000   0.61900000000000   0.62800000000000   Sb    (8c)
   0.54400000000000   0.88100000000000   0.12800000000000   Sb    (8c)
   0.95600000000000   0.11900000000000   0.62800000000000   Sb    (8c)
   0.95600000000000   0.38100000000000   0.12800000000000   Sb    (8c)
```

Brookite ($TiO_2$, C21): A2B_oP24_61_2c_c - CIF

```
# CIF file

data_findsym-output
_audit_creation_method FINDSYM

_chemical_name_mineral 'Brookite'
_chemical_formula_sum 'Ti O2'

loop_
_publ_author_name
 'E. P. Meagher'
 'G. A. Lager'
_journal_name_full
;
Canadian Mineralogist
;
_journal_volume 17
_journal_year 1979
_journal_page_first 77
_journal_page_last 85
_publ_Section_title
;
 Polyhedral thermal expansion in the TiO$_2$ polymorphs; refinement of
    ↪ the crystal structures of rutile and brookite at high
    ↪ temperature
;

_aflow_proto 'A2B_oP24_61_2c_c'
_aflow_params 'a,b/a,c/a,x1,y1,z1,x2,y2,z2,x3,y3,z3'
_aflow_params_values '9.174,0.375953782429,0.560061042075,0.0095,0.1491,
    ↪ 0.1835,0.2314,0.111,0.5366,0.1289,0.0972,0.8628'
_aflow_Strukturbericht 'C21'
_aflow_Pearson 'oP24'

_symmetry_space_group_name_Hall "-P 2ac 2ab"
_symmetry_space_group_name_H-M "P b c a"
_symmetry_Int_Tables_number 61

_cell_length_a    9.17400
_cell_length_b    3.44900
_cell_length_c    5.13800
_cell_angle_alpha 90.00000
_cell_angle_beta  90.00000
_cell_angle_gamma 90.00000

loop_
_space_group_symop_id
```

```
_space_group_symop_operation_xyz
1 x,y,z
2 x+1/2,-y+1/2,-z
3 -x,y+1/2,-z+1/2
4 -x+1/2,-y,z+1/2
5 -x,-y,-z
6 -x+1/2,y+1/2,z
7 x,-y+1/2,z+1/2
8 x+1/2,y,-z+1/2

loop_
_atom_site_label
_atom_site_type_symbol
_atom_site_symmetry_multiplicity
_atom_site_Wyckoff_label
_atom_site_fract_x
_atom_site_fract_y
_atom_site_fract_z
_atom_site_occupancy
O1  O    8 c 0.00950 0.14910 0.18350 1.00000
O2  O    8 c 0.23140 0.11100 0.53660 1.00000
Ti1 Ti   8 c 0.12890 0.09720 0.86280 1.00000
```

Brookite ($TiO_2$, C21): A2B_oP24_61_2c_c - POSCAR

```
A2B_oP24_61_2c_c & a,b/a,c/a,x1,y1,z1,x2,y2,z2,x3,y3,z3 --params=9.174,
    ↪ 0.375953782429,0.560061042075,0.0095,0.1491,0.1835,0.2314,0.111
    ↪ ,0.5366,0.1289,0.0972,0.8628 & Pbca        D_{2h}^{15} #61 (c^3)
    ↪ & oP24 & C21 & TiO2 & Brookite & E. P. Meagher and G. A. Lager,
    ↪  Can. Mineral. 17, 77-85 (1979)
   1.0000000000000000
   9.17400000000000   0.00000000000000   0.00000000000000
   0.00000000000000   3.44900000000000   0.00000000000000
   0.00000000000000   0.00000000000000   5.13800000000000
    O    Ti
   16     8
Direct
   0.00950000000000   0.14910000000000   0.18350000000000   O    (8c)
   0.00950000000000   0.35090000000000   0.68350000000000   O    (8c)
   0.49050000000000   0.64910000000000   0.18350000000000   O    (8c)
   0.49050000000000   0.85090000000000   0.68350000000000   O    (8c)
   0.50950000000000   0.14910000000000   0.31650000000000   O    (8c)
   0.50950000000000   0.35090000000000   0.81650000000000   O    (8c)
   0.99050000000000   0.64910000000000   0.31650000000000   O    (8c)
   0.99050000000000   0.85090000000000   0.81650000000000   O    (8c)
   0.23140000000000   0.11100000000000   0.53660000000000   O    (8c)
   0.23140000000000   0.38900000000000   0.03660000000000   O    (8c)
   0.26860000000000   0.61100000000000   0.53660000000000   O    (8c)
   0.26860000000000   0.88900000000000   0.03660000000000   O    (8c)
   0.73140000000000   0.11100000000000   0.96340000000000   O    (8c)
   0.73140000000000   0.38900000000000   0.46340000000000   O    (8c)
   0.76860000000000   0.61100000000000   0.96340000000000   O    (8c)
   0.76860000000000   0.88900000000000   0.46340000000000   O    (8c)
   0.12890000000000   0.09720000000000   0.86280000000000   Ti   (8c)
   0.12890000000000   0.40280000000000   0.36280000000000   Ti   (8c)
   0.37110000000000   0.59720000000000   0.86280000000000   Ti   (8c)
   0.37110000000000   0.90280000000000   0.36280000000000   Ti   (8c)
   0.62890000000000   0.09720000000000   0.63720000000000   Ti   (8c)
   0.62890000000000   0.40280000000000   0.13720000000000   Ti   (8c)
   0.87110000000000   0.59720000000000   0.63720000000000   Ti   (8c)
   0.87110000000000   0.90280000000000   0.13720000000000   Ti   (8c)
```

Stibnite ($Sb_2S_3$, $D5_8$): A3B2_oP20_62_3c_2c - CIF

```
# CIF file

data_findsym-output
_audit_creation_method FINDSYM

_chemical_name_mineral 'Stibnite'
_chemical_formula_sum 'Sb2 S3'

loop_
_publ_author_name
 'Atsushi Kyono'
 'Mitsuyoshi Kimata'
_journal_name_full
;
American Mineralogist
;
_journal_volume 89
_journal_year 2004
_journal_page_first 932
_journal_page_last 940
_publ_Section_title
;
 Structural variations induced by difference of the inert pair effect in
    ↪  the stibnite-bismuthinite solid solution series (Sb,Bi)
    ↪ $_2$S$_3$
;

# Found in  AMS Database

_aflow_proto 'A3B2_oP20_62_3c_2c'
_aflow_params 'a,b/a,c/a,x1,z1,x2,z2,x3,z3,x4,z4,x5,z5'
_aflow_params_values '11.282,0.339443361106,0.994947704308,0.2922,
    ↪ 0.19181,0.4504,0.877,0.6246,0.5611,-0.02937,0.17398,0.64939,-
    ↪ 0.03603'
_aflow_Strukturbericht 'D5_8'
_aflow_Pearson 'oP20'

_symmetry_space_group_name_Hall "-P 2ac 2n"
_symmetry_space_group_name_H-M "P n m a"
_symmetry_Int_Tables_number 62

_cell_length_a      11.28200
_cell_length_b      3.82960
_cell_length_c      11.22500
_cell_angle_alpha   90.00000
_cell_angle_beta    90.00000
_cell_angle_gamma   90.00000

loop_
_space_group_symop_id
_space_group_symop_operation_xyz
1 x,y,z
2 x+1/2,-y+1/2,-z+1/2
3 -x,y+1/2,-z
4 -x+1/2,-y,z+1/2
5 -x,-y,-z
6 -x+1/2,y+1/2,z+1/2
7 x,-y+1/2,z
8 x+1/2,y,-z+1/2

loop_
_atom_site_label
_atom_site_type_symbol
_atom_site_symmetry_multiplicity
_atom_site_Wyckoff_label
_atom_site_fract_x
_atom_site_fract_y
_atom_site_fract_z
_atom_site_occupancy
S1  S    4 c 0.29220 0.25000 0.19181  1.00000
S2  S    4 c 0.45040 0.25000 0.87700  1.00000
S3  S    4 c 0.62460 0.25000 0.56110  1.00000
Sb1 Sb   4 c -0.02937 0.25000 0.17398  1.00000
Sb2 Sb   4 c 0.64939 0.25000 -0.03603 1.00000
```

Stibnite ($Sb_2S_3$, $D5_8$): A3B2_oP20_62_3c_2c - POSCAR

```
A3B2_oP20_62_3c_2c & a,b/a,c/a,x1,z1,x2,z2,x3,z3,x4,z4,x5,z5 --params=
    ↪ 11.282,0.339443361106,0.994947704308,0.2922,0.19181,0.4504,
    ↪ 0.877,0.6246,0.5611,-0.02937,0.17398,0.64939,-0.03603 & Pnma
    ↪       D^{16}_{2h} #62 (c^5) & oP20 & D5_8 & Sb2S3 & Stibnite &
    ↪ A. Kyono and M. Kimata, Am. Mineral. 89, 932-940 (2004)
   1.0000000000000000
  11.28200000000000   0.00000000000000   0.00000000000000
   0.00000000000000   3.82960000000000   0.00000000000000
   0.00000000000000   0.00000000000000  11.22500000000000
    S    Sb
   12     8
Direct
   0.20780000000000   0.75000000000000   0.69181000000000   S    (4c)
   0.29220000000000   0.25000000000000   0.19181000000000   S    (4c)
   0.70780000000000   0.75000000000000   0.80819000000000   S    (4c)
   0.79220000000000   0.25000000000000   0.30819000000000   S    (4c)
  -0.04960000000000   0.25000000000000   0.62300000000000   S    (4c)
   0.04960000000000   0.75000000000000   0.37700000000000   S    (4c)
   0.45040000000000   0.25000000000000   0.87700000000000   S    (4c)
   0.54960000000000   0.75000000000000   0.12300000000000   S    (4c)
   0.12460000000000   0.25000000000000  -0.06110000000000   S    (4c)
   0.37540000000000   0.75000000000000   0.43890000000000   S    (4c)
   0.62460000000000   0.25000000000000   0.56110000000000   S    (4c)
   0.87540000000000   0.75000000000000   0.06110000000000   S    (4c)
  -0.02937000000000   0.25000000000000   0.17398000000000   Sb   (4c)
   0.02937000000000   0.75000000000000   0.82602000000000   Sb   (4c)
   0.47063000000000   0.25000000000000   0.32602000000000   Sb   (4c)
   0.52937000000000   0.75000000000000   0.67398000000000   Sb   (4c)
   0.14939000000000   0.25000000000000   0.53603000000000   Sb   (4c)
   0.35061000000000   0.75000000000000   0.03603000000000   Sb   (4c)
   0.64939000000000   0.25000000000000  -0.03603000000000   Sb   (4c)
   0.85061000000000   0.75000000000000   0.46397000000000   Sb   (4c)
```

$CaTiO_3$ Pnma Perovskite: AB3C_oP20_62_c_cd_a - CIF

```
# CIF file

data_findsym-output
_audit_creation_method FINDSYM

_chemical_name_mineral 'Orthorhombic Perovskite'
_chemical_formula_sum 'Ca Ti O3'

loop_
_publ_author_name
 'Takamitsu Yamanaka'
 'Noriyuki Hirai'
 'Yutaka Komatsu'
_journal_name_full
;
American Mineralogist
;
_journal_volume 87
_journal_year 2002
_journal_page_first 1183
_journal_page_last 1189
_publ_Section_title
;
 Structure change of Ca$_{1-x}$Sr$_x$TiO$_3$ perovskite with composition
    ↪  and pressure
;

# Found in  AMS Database

_aflow_proto 'AB3C_oP20_62_c_cd_a'
_aflow_params 'a,b/a,c/a,x2,z2,x3,z3,x4,y4,z4'
_aflow_params_values '5.4224,1.41099881971,0.996661994689,0.4877,-0.0084
    ↪ ,0.0313,0.0586,0.288,0.537,0.213'
_aflow_Strukturbericht 'None'
```

```
_aflow_Pearson 'oP20'

_symmetry_space_group_name_Hall "-P 2ac 2n"
_symmetry_space_group_name_H-M "P n m a"
_symmetry_Int_Tables_number 62

_cell_length_a     5.42240
_cell_length_b     7.65100
_cell_length_c     5.40430
_cell_angle_alpha  90.00000
_cell_angle_beta   90.00000
_cell_angle_gamma  90.00000

loop_
_space_group_symop_id
_space_group_symop_operation_xyz
1 x,y,z
2 x+1/2,-y+1/2,-z+1/2
3 -x,y+1/2,-z
4 -x+1/2,-y,z+1/2
5 -x,-y,-z
6 -x+1/2,y+1/2,z+1/2
7 x,-y+1/2,z
8 x+1/2,y,-z+1/2

loop_
_atom_site_label
_atom_site_type_symbol
_atom_site_symmetry_multiplicity
_atom_site_Wyckoff_label
_atom_site_fract_x
_atom_site_fract_y
_atom_site_fract_z
_atom_site_occupancy
Ti1 Ti    4 a 0.00000   0.00000  0.00000   1.00000
Ca1 Ca    4 c 0.48770   0.25000 -0.00840  1.00000
O1  O     4 c 0.03130  0.25000  0.05860   1.00000
O2  O     8 d 0.28800   0.53700  0.21300   1.00000
```

$CaTiO_3$ Pnma Perovskite: AB3C_oP20_62_c_cd_a - POSCAR

```
AB3C_oP20_62_c_cd_a & a,b/a,c/a,x2,z2,x3,z3,x4,y4,z4 --params=5.4224,
    ↪ 1.41099881971,0.996661994689,0.4877,-0.0084,0.0313,0.0586,0.288
    ↪ ,0.537,0.213 & Pnma        D_{2h}^{16} #62 (ac^2d) & oP20 & &
    ↪ CaTiO3 & orthrhombic perovskite & T. Yamanaka, N. Hirai and Y.
    ↪ Komatsu, Am. Mineral. 87, 1183-1189 (2002)
   1.0000000000000000
   5.42240000000000   0.00000000000000   0.00000000000000
   0.00000000000000   7.65100000000000   0.00000000000000
   0.00000000000000   0.00000000000000   5.40430000000000
    Ca      O      Ti
     4     12      4
Direct
  -0.01230000000000   0.25000000000000   0.50840000000000   Ca   (4c)
   0.01230000000000   0.75000000000000   0.49160000000000   Ca   (4c)
   0.48770000000000   0.25000000000000  -0.00840000000000   Ca   (4c)
   0.51230000000000   0.75000000000000   0.00840000000000   Ca   (4c)
   0.03130000000000   0.25000000000000   0.05860000000000   O    (4c)
  -0.03130000000000   0.75000000000000  -0.05860000000000   O    (4c)
   0.46870000000000   0.75000000000000   0.55860000000000   O    (4c)
   0.53130000000000   0.25000000000000   0.44140000000000   O    (4c)
   0.21200000000000   0.03700000000000   0.71300000000000   O    (8d)
   0.21200000000000   0.46300000000000   0.71300000000000   O    (8d)
   0.28800000000000   0.53700000000000   0.21300000000000   O    (8d)
   0.28800000000000   0.96300000000000   0.21300000000000   O    (8d)
   0.71200000000000   0.03700000000000   0.78700000000000   O    (8d)
   0.71200000000000   0.46300000000000   0.78700000000000   O    (8d)
   0.78800000000000  -0.03700000000000   0.28700000000000   O    (8d)
   0.78800000000000   0.53700000000000   0.28700000000000   O    (8d)
   0.00000000000000   0.00000000000000   0.00000000000000   Ti   (4a)
   0.00000000000000   0.50000000000000   0.00000000000000   Ti   (4a)
   0.50000000000000   0.00000000000000   0.50000000000000   Ti   (4a)
   0.50000000000000   0.50000000000000   0.50000000000000   Ti   (4a)
```

$MgB_4$: A4B_oP20_62_2cd_c - CIF

```
# CIF file

data_findsym-output
_audit_creation_method FINDSYM

_chemical_name_mineral 'Magnesium tetraboride '
_chemical_formula_sum 'Mg B4'

loop_
_publ_author_name
 'Roger Naslain'
 'Alain Guette'
 'Michel Barret'
_journal_name_full
;
Journal of Solid State Chemistry
;
_journal_volume 8
_journal_year 1973
_journal_page_first 68
_journal_page_last 85
_publ_Section_title
;
 Sur le diborure et le t\'{e}traborure de magn\'{e}sium. Consid\'{e}
    ↪ rations cristallochimiques sur les t\'{e}traborures
;

# Found in http://materials.springer.com/isp/crystallographic/docs/
    ↪ sd_1906993

_aflow_proto 'A4B_oP20_62_2cd_c'
_aflow_params 'a,b/a,c/a,x1,z1,x2,z2,x3,z3,x4,y4,z4'
_aflow_params_values '5.464,0.810395314788,1.36749633968,0.22451,0.65626
    ↪ ,0.55801,0.6466,0.05131,0.36362,0.13079,0.0579,0.06543'
_aflow_Strukturbericht 'None'
_aflow_Pearson 'oP20'

_symmetry_space_group_name_Hall "-P 2ac 2n"
_symmetry_space_group_name_H-M "P n m a"
_symmetry_Int_Tables_number 62

_cell_length_a     5.46400
_cell_length_b     4.42800
_cell_length_c     7.47200
_cell_angle_alpha  90.00000
_cell_angle_beta   90.00000
_cell_angle_gamma  90.00000

loop_
_space_group_symop_id
_space_group_symop_operation_xyz
1 x,y,z
2 x+1/2,-y+1/2,-z+1/2
3 -x,y+1/2,-z
4 -x+1/2,-y,z+1/2
5 -x,-y,-z
6 -x+1/2,y+1/2,z+1/2
7 x,-y+1/2,z
8 x+1/2,y,-z+1/2

loop_
_atom_site_label
_atom_site_type_symbol
_atom_site_symmetry_multiplicity
_atom_site_Wyckoff_label
_atom_site_fract_x
_atom_site_fract_y
_atom_site_fract_z
_atom_site_occupancy
B1  B    4 c 0.22451  0.25000  0.65626  1.00000
B2  B    4 c 0.55801  0.25000  0.64660  1.00000
Mg1 Mg   4 c 0.05131  0.25000  0.36362  1.00000
B3  B    8 d 0.13079  0.05790  0.06543  1.00000
```

$MgB_4$: A4B_oP20_62_2cd_c - POSCAR

```
A4B_oP20_62_2cd_c & a,b/a,c/a,x1,z1,x2,z2,x3,z3,x4,y4,z4 --params=5.464,
    ↪ 0.810395314788,1.36749633968,0.22451,0.65626,0.55801,0.6466,
    ↪ 0.05131,0.36362,0.13079,0.0579,0.06543 & Pnma        D^{16}_{2h}
    ↪ #62 (c^3d) & oP20 & & MgB4 & & R. Naslain, A. Guette and M.
    ↪ Barrat, J. Solid State Chem. 8, 68-85 (1973)
   1.0000000000000000
   5.46400000000000   0.00000000000000   0.00000000000000
   0.00000000000000   4.42800000000000   0.00000000000000
   0.00000000000000   0.00000000000000   7.47200000000000
     B     Mg
    16      4
Direct
   0.22451000000000   0.25000000000000   0.65626000000000   B    (4c)
   0.27549000000000   0.75000000000000   0.15626000000000   B    (4c)
   0.72451000000000   0.25000000000000   0.84374000000000   B    (4c)
   0.77549000000000   0.75000000000000   0.34374000000000   B    (4c)
   0.05801000000000   0.25000000000000   0.85340000000000   B    (4c)
  -0.05801000000000   0.75000000000000   0.14660000000000   B    (4c)
   0.44199000000000   0.75000000000000   0.35340000000000   B    (4c)
   0.55801000000000   0.25000000000000   0.64660000000000   B    (4c)
   0.13079000000000   0.05790000000000   0.06543000000000   B    (8d)
   0.13079000000000   0.44210000000000   0.06543000000000   B    (8d)
   0.36921000000000  -0.05790000000000   0.56543000000000   B    (8d)
   0.36921000000000   0.55790000000000   0.56543000000000   B    (8d)
   0.63079000000000   0.05790000000000   0.43457000000000   B    (8d)
   0.63079000000000   0.44210000000000   0.43457000000000   B    (8d)
   0.86921000000000  -0.05790000000000  -0.06543000000000   B    (8d)
   0.86921000000000   0.55790000000000  -0.06543000000000   B    (8d)
   0.05131000000000   0.25000000000000   0.36362000000000   Mg   (4c)
  -0.05131000000000   0.75000000000000   0.63638000000000   Mg   (4c)
   0.44869000000000   0.75000000000000   0.86362000000000   Mg   (4c)
   0.55131000000000   0.25000000000000   0.13638000000000   Mg   (4c)
```

Chalcostibite ($CuSbS_2$, $F5_6$): AB2C_oP16_62_c_2c_c - CIF

```
# CIF file

data_findsym-output
_audit_creation_method FINDSYM

_chemical_name_mineral 'Chalcostibite'
_chemical_formula_sum 'Cu Sb S2'

loop_
_publ_author_name
 'Atsushi Kyono'
 'Mitsuyoshi Kimata'
_journal_name_full
;
American Mineralogist
;
_journal_volume 90
_journal_year 2005
_journal_page_first 162
_journal_page_last 165
_publ_Section_title
;
```

```
 Crystal structures of chalcostibite (CuSbS$_2$) and emplectite (
     ↪ CuBiS$_2$): Structural relationship of stereochemical activity
     ↪  between chalcostibite and emplectite
;

_aflow_proto 'AB2C_oP16_62_c_2c_c'
_aflow_params 'a,b/a,c/a,x1,z1,x2,z2,x3,z3,x4,z4'
_aflow_params_values '6.018,0.630741110003,2.4086075108,0.2522,0.8276,
     ↪ 0.6221,0.095,0.8706,0.8244,0.226,0.06333'
_aflow_Strukturbericht 'F5_6'
_aflow_Pearson 'oP16'

_symmetry_space_group_name_Hall "-P 2ac 2n"
_symmetry_space_group_name_H-M "P n m a"
_symmetry_Int_Tables_number 62

_cell_length_a     6.01800
_cell_length_b     3.79580
_cell_length_c     14.49500
_cell_angle_alpha  90.00000
_cell_angle_beta   90.00000
_cell_angle_gamma  90.00000

loop_
_space_group_symop_id
_space_group_symop_operation_xyz
1 x,y,z
2 x+1/2,-y+1/2,-z+1/2
3 -x,y+1/2,-z
4 -x+1/2,-y,z+1/2
5 -x,-y,-z
6 -x+1/2,y+1/2,z+1/2
7 x,-y+1/2,z
8 x+1/2,y,-z+1/2

loop_
_atom_site_label
_atom_site_type_symbol
_atom_site_symmetry_multiplicity
_atom_site_Wyckoff_label
_atom_site_fract_x
_atom_site_fract_y
_atom_site_fract_z
_atom_site_occupancy
Cu1 Cu    4 c 0.25220 0.25000 0.82760 1.00000
S1  S     4 c 0.62210 0.25000 0.09500 1.00000
S2  S     4 c 0.87060 0.25000 0.82440 1.00000
Sb1 Sb    4 c 0.22600 0.25000 0.06333 1.00000
```

Chalcostibite ($CuSbS_2$, $F5_6$): AB2C_oP16_62_c_2c_c - POSCAR

```
AB2C_oP16_62_c_2c_c & a,b/a,c/a,x1,z1,x2,z2,x3,z3,x4,z4 --params=6.018,
     ↪ 0.630741110003,2.4086075108,0.2522,0.8276,0.6221,0.095,0.8706,
     ↪ 0.8244,0.226,0.06333 & Pnma     D_{2h}^{16} #62 (c^4) & oP16 &
     ↪  F5_6 & CuSbS2 & Chalcostibite & A. Kyono and M. Kimata, Am.
     ↪ Mineral. 90, 162-165 (2005)
   1.0000000000000000
   6.01800000000000   0.00000000000000   0.00000000000000
   0.00000000000000   3.79580000000000   0.00000000000000
   0.00000000000000   0.00000000000000  14.49500000000000
   Cu      S      Sb
    4      8       4
Direct
   0.24780000000000   0.75000000000000   0.32760000000000   Cu   (4c)
   0.25220000000000   0.25000000000000   0.82760000000000   Cu   (4c)
   0.74780000000000   0.75000000000000   0.17240000000000   Cu   (4c)
   0.75220000000000   0.25000000000000   0.67240000000000   Cu   (4c)
   0.12210000000000   0.25000000000000   0.40500000000000   S    (4c)
   0.37790000000000   0.75000000000000  -0.09500000000000   S    (4c)
   0.62210000000000   0.25000000000000   0.09500000000000   S    (4c)
   0.87790000000000   0.75000000000000   0.59500000000000   S    (4c)
   0.12940000000000   0.75000000000000   0.17560000000000   S    (4c)
   0.37060000000000   0.25000000000000   0.67560000000000   S    (4c)
   0.62940000000000   0.75000000000000   0.32440000000000   S    (4c)
   0.87060000000000   0.25000000000000   0.82440000000000   S    (4c)
   0.22600000000000   0.25000000000000   0.06333000000000   Sb   (4c)
   0.27400000000000   0.75000000000000   0.56333000000000   Sb   (4c)
   0.72600000000000   0.25000000000000   0.43667000000000   Sb   (4c)
   0.77400000000000   0.75000000000000  -0.06333000000000   Sb   (4c)
```

$Co_2Si$ (C37): A2B_oP12_62_2c_c - CIF

```
# CIF file

data_findsym-output
_audit_creation_method FINDSYM

_chemical_name_mineral ''
_chemical_formula_sum 'Co2 Si'

loop_
_publ_author_name
 'S. Geller'
 'V. M. Wolontis'
_journal_name_full
;
Acta Crystallographica
;
_journal_volume 8
_journal_year 1955
_journal_page_first 83
_journal_page_last 87
_publ_Section_title
;
 The Crystal Structure of Co$_2$Si
;

_aflow_proto 'A2B_oP12_62_2c_c'
_aflow_params 'a,b/a,c/a,x1,z1,x2,z2,x3,z3'
_aflow_params_values '4.918,0.7600650671,1.44550630338,0.038,0.282,0.674
     ↪ ,0.562,0.202,0.611'
_aflow_Strukturbericht 'C37'
_aflow_Pearson 'oP12'

_symmetry_space_group_name_Hall "-P 2ac 2n"
_symmetry_space_group_name_H-M "P n m a"
_symmetry_Int_Tables_number 62

_cell_length_a     4.91800
_cell_length_b     3.73800
_cell_length_c     7.10900
_cell_angle_alpha  90.00000
_cell_angle_beta   90.00000
_cell_angle_gamma  90.00000

loop_
_space_group_symop_id
_space_group_symop_operation_xyz
1 x,y,z
2 x+1/2,-y+1/2,-z+1/2
3 -x,y+1/2,-z
4 -x+1/2,-y,z+1/2
5 -x,-y,-z
6 -x+1/2,y+1/2,z+1/2
7 x,-y+1/2,z
8 x+1/2,y,-z+1/2

loop_
_atom_site_label
_atom_site_type_symbol
_atom_site_symmetry_multiplicity
_atom_site_Wyckoff_label
_atom_site_fract_x
_atom_site_fract_y
_atom_site_fract_z
_atom_site_occupancy
Co1 Co    4 c 0.03800 0.25000 0.28200 1.00000
Co2 Co    4 c 0.67400 0.25000 0.56200 1.00000
Si1 Si    4 c 0.20200 0.25000 0.61100 1.00000
```

$Co_2Si$ (C37): A2B_oP12_62_2c_c - POSCAR

```
A2B_oP12_62_2c_c & a,b/a,c/a,x1,z1,x2,z2,x3,z3 --params=4.918,
     ↪ 0.7600650671,1.44550630338,0.038,0.282,0.674,0.562,0.202,0.611
     ↪ & Pnma     D_{2h}^{16} #62 (c^3) & oP12 & C37 & Co2Si & & S.
     ↪ Geller and V. M. Wolontis, Acta Cryst. 8, 83-87 (1955)
   1.0000000000000000
   4.91800000000000   0.00000000000000   0.00000000000000
   0.00000000000000   3.73800000000000   0.00000000000000
   0.00000000000000   0.00000000000000   7.10900000000000
   Co     Si
    8      4
Direct
   0.03800000000000   0.25000000000000   0.28200000000000   Co   (4c)
  -0.03800000000000   0.75000000000000   0.71800000000000   Co   (4c)
   0.46200000000000   0.75000000000000   0.78200000000000   Co   (4c)
   0.53800000000000   0.25000000000000   0.21800000000000   Co   (4c)
   0.17400000000000   0.25000000000000  -0.06200000000000   Co   (4c)
   0.32600000000000   0.75000000000000   0.43800000000000   Co   (4c)
   0.67400000000000   0.25000000000000   0.56200000000000   Co   (4c)
   0.82600000000000   0.75000000000000   0.06200000000000   Co   (4c)
   0.20200000000000   0.25000000000000   0.61100000000000   Si   (4c)
   0.29800000000000   0.75000000000000   0.11100000000000   Si   (4c)
   0.70200000000000   0.25000000000000   0.88900000000000   Si   (4c)
   0.79800000000000   0.75000000000000   0.38900000000000   Si   (4c)
```

$HgCl_2$ (C25): A2B_oP12_62_2c_c - CIF

```
# CIF file

data_findsym-output
_audit_creation_method FINDSYM

_chemical_name_mineral ''
_chemical_formula_sum 'Hg Cl2'

loop_
_publ_author_name
 'H. Braekken'
 'W. Scholten'
_journal_name_full
;
Zeitschrift f\"{u}r Kristallographie - Crystalline Materials
;
_journal_volume 89
_journal_year 1934
_journal_page_first 448
_journal_page_last 455
_publ_Section_title
;
 Die Kristallstruktur des Quecksilberchloride HgCl$_2$
;

_aflow_proto 'A2B_oP12_62_2c_c'
_aflow_params 'a,b/a,c/a,x1,z1,x2,z2,x3,z3'
_aflow_params_values '12.735,0.468237141735,0.339615233608,0.733,0.125,
     ↪ 0.508,0.722,0.874,0.447'
_aflow_Strukturbericht 'C25'
_aflow_Pearson 'oP12'
```

```
_symmetry_space_group_name_Hall "-P 2ac 2n"
_symmetry_space_group_name_H-M "P n m a"
_symmetry_Int_Tables_number 62

_cell_length_a      12.73500
_cell_length_b      5.96300
_cell_length_c      4.32500
_cell_angle_alpha  90.00000
_cell_angle_beta   90.00000
_cell_angle_gamma  90.00000

loop_
_space_group_symop_id
_space_group_symop_operation_xyz
1 x,y,z
2 x+1/2,-y+1/2,-z+1/2
3 -x,y+1/2,-z
4 -x+1/2,-y,z+1/2
5 -x,-y,-z
6 -x+1/2,y+1/2,z+1/2
7 x,-y+1/2,z
8 x+1/2,y,-z+1/2

loop_
_atom_site_label
_atom_site_type_symbol
_atom_site_symmetry_multiplicity
_atom_site_Wyckoff_label
_atom_site_fract_x
_atom_site_fract_y
_atom_site_fract_z
_atom_site_occupancy
Cl1 Cl    4 c 0.73300 0.25000 0.12500 1.00000
Cl2 Cl    4 c 0.50800 0.25000 0.72200 1.00000
Hg1 Hg    4 c 0.87400 0.25000 0.44700 1.00000
```

$HgCl_2$ (C25): A2B_oP12_62_2c_c - POSCAR

```
A2B_oP12_62_2c_c & a,b/a,c/a,x1,z1,x2,z2,x3,z3 --params=12.735,
   ↪ 0.468237141735,0.339615233608,0.733,0.125,0.508,0.722,0.874,
   ↪ 0.447 & Pnma         D_{2h}^{16} #62 (c^3) & oP12 & C25 & HgCl2 &
   ↪ & H. Brakken and W. Scholten, Zeitschrift f\"{u}r
   ↪ Kristallographie - Crystalline Materials 89, 448-455 (1934)
   ↪ rotated to Pnma setting
   1.0000000000000000
  12.73500000000000   0.00000000000000   0.00000000000000
   0.00000000000000   5.96300000000000   0.00000000000000
   0.00000000000000   0.00000000000000   4.32500000000000
   Cl     Hg
    8      4
Direct
   0.23300000000000   0.25000000000000   0.37500000000000   Cl   (4c)
   0.26700000000000   0.75000000000000  -0.12500000000000   Cl   (4c)
   0.73300000000000   0.25000000000000   0.12500000000000   Cl   (4c)
   0.76700000000000   0.75000000000000   0.62500000000000   Cl   (4c)
   0.00800000000000   0.25000000000000   0.77800000000000   Cl   (4c)
   0.49200000000000   0.75000000000000   0.27800000000000   Cl   (4c)
   0.50800000000000   0.25000000000000   0.72200000000000   Cl   (4c)
   0.99200000000000   0.75000000000000   0.22200000000000   Cl   (4c)
   0.12600000000000   0.75000000000000   0.55300000000000   Hg   (4c)
   0.37400000000000   0.25000000000000   0.05300000000000   Hg   (4c)
   0.62600000000000   0.75000000000000   0.94700000000000   Hg   (4c)
   0.87400000000000   0.25000000000000   0.44700000000000   Hg   (4c)
```

Cotunnite ($PbCl_2$, C23): A2B_oP12_62_2c_c - CIF

```
# CIF file

data_findsym-output
_audit_creation_method FINDSYM

_chemical_name_mineral 'Cotunnite'
_chemical_formula_sum 'Pb Cl2'

loop_
_publ_author_name
 'Ronald L. Sass'
 'E. B. Brackett'
 'T. E. Brackett'
_journal_name_full
;
Journal of Physical Chemistry
;
_journal_volume 67
_journal_year 1963
_journal_page_first 2863
_journal_page_last 2864
_publ_Section_title
;
 The Crystal Structure of Lead Chloride
;

_aflow_proto 'A2B_oP12_62_2c_c'
_aflow_params 'a,b/a,c/a,x1,z1,x2,z2,x3,z3'
_aflow_params_values '7.6204,0.595008136056,1.1869718125,0.125,0.4217,
   ↪ 0.0202,0.837,0.2377,0.0959'
_aflow_Strukturbericht 'C23'
_aflow_Pearson 'oP12'

_symmetry_space_group_name_Hall "-P 2ac 2n"
_symmetry_space_group_name_H-M "P n m a"
_symmetry_Int_Tables_number 62

_cell_length_a      7.62040
_cell_length_b      4.53420
_cell_length_c      9.04520
_cell_angle_alpha  90.00000
_cell_angle_beta   90.00000
_cell_angle_gamma  90.00000

loop_
_space_group_symop_id
_space_group_symop_operation_xyz
1 x,y,z
2 x+1/2,-y+1/2,-z+1/2
3 -x,y+1/2,-z
4 -x+1/2,-y,z+1/2
5 -x,-y,-z
6 -x+1/2,y+1/2,z+1/2
7 x,-y+1/2,z
8 x+1/2,y,-z+1/2

loop_
_atom_site_label
_atom_site_type_symbol
_atom_site_symmetry_multiplicity
_atom_site_Wyckoff_label
_atom_site_fract_x
_atom_site_fract_y
_atom_site_fract_z
_atom_site_occupancy
Cl1 Cl    4 c 0.12500 0.25000 0.42170 1.00000
Cl2 Cl    4 c 0.02020 0.25000 0.83700 1.00000
Pb1 Pb    4 c 0.23770 0.25000 0.09590 1.00000
```

Cotunnite ($PbCl_2$, C23): A2B_oP12_62_2c_c - POSCAR

```
A2B_oP12_62_2c_c & a,b/a,c/a,x1,z1,x2,z2,x3,z3 --params=7.6204,
   ↪ 0.595008136056,1.1869718125,0.125,0.4217,0.0202,0.837,0.2377,
   ↪ 0.0959 & Pnma         D_{2h}^{16} #62 (c^3) & oP12 & C23 & PbCl2
   ↪ & Cotunnite & R. L. Sass, E. B. Brackett, and T. E. Brackett,
   ↪ J. Phys. Chem. 67, 2863-2864 (1963)
   1.0000000000000000
   7.62040000000000   0.00000000000000   0.00000000000000
   0.00000000000000   4.53420000000000   0.00000000000000
   0.00000000000000   0.00000000000000   9.04520000000000
   Cl     Pb
    8      4
Direct
   0.12500000000000   0.25000000000000   0.42170000000000   Cl   (4c)
   0.37500000000000   0.75000000000000  -0.07830000000000   Cl   (4c)
   0.62500000000000   0.25000000000000   0.07830000000000   Cl   (4c)
   0.87500000000000   0.75000000000000   0.57830000000000   Cl   (4c)
   0.02020000000000   0.25000000000000   0.83700000000000   Cl   (4c)
  -0.02020000000000   0.75000000000000   0.16300000000000   Cl   (4c)
   0.47980000000000   0.75000000000000   0.33700000000000   Cl   (4c)
   0.52020000000000   0.25000000000000   0.66300000000000   Cl   (4c)
   0.23770000000000   0.25000000000000   0.09590000000000   Pb   (4c)
   0.26230000000000   0.75000000000000   0.59590000000000   Pb   (4c)
   0.73770000000000   0.25000000000000   0.40410000000000   Pb   (4c)
   0.76230000000000   0.75000000000000  -0.09590000000000   Pb   (4c)
```

GeS (B16): AB_oP8_62_c_c - CIF

```
# CIF file

data_findsym-output
_audit_creation_method FINDSYM

_chemical_name_mineral ''
_chemical_formula_sum 'Ge S'

loop_
_publ_author_name
 'W. H. Zachariasen'
_journal_name_full
;
Physical Review
;
_journal_volume 40
_journal_year 1932
_journal_page_first 917
_journal_page_last 922
_publ_Section_title
;
 The Crystal Lattice of Germano Sulphide, GeS
;

# Found in AMS Database

_aflow_proto 'AB_oP8_62_c_c'
_aflow_params 'a,b/a,c/a,x1,z1,x2,z2'
_aflow_params_values '10.42,0.349328214971,0.411708253359,0.375,0.333,
   ↪ 0.139,0.389'
_aflow_Strukturbericht 'B16'
_aflow_Pearson 'oP8'

_symmetry_space_group_name_Hall "-P 2ac 2n"
_symmetry_space_group_name_H-M "P n m a"
_symmetry_Int_Tables_number 62

_cell_length_a      10.42000
_cell_length_b      3.64000
_cell_length_c      4.29000
_cell_angle_alpha  90.00000
_cell_angle_beta   90.00000
_cell_angle_gamma  90.00000

loop_
_space_group_symop_id
```

```
_space_group_symop_operation_xyz
1 x,y,z
2 x+1/2,-y+1/2,-z+1/2
3 -x,y+1/2,-z
4 -x+1/2,-y,z+1/2
5 -x,-y,-z
6 -x+1/2,y+1/2,z+1/2
7 x,-y+1/2,z
8 x+1/2,y,-z+1/2

loop_
_atom_site_label
_atom_site_type_symbol
_atom_site_symmetry_multiplicity
_atom_site_Wyckoff_label
_atom_site_fract_x
_atom_site_fract_y
_atom_site_fract_z
_atom_site_occupancy
Ge1 Ge   4 c 0.37500  0.25000  0.33300  1.00000
S1  S    4 c 0.13900  0.25000  0.38900  1.00000
```

GeS (B16): AB_oP8_62_c_c - POSCAR

```
AB_oP8_62_c_c & a,b/a,c/a,x1,z1,x2,z2 --params=10.42,0.349328214971,
    ↪ 0.411708253359,0.375,0.333,0.139,0.389 & Pnma        D^{16}_{2h}
    ↪ #62 (c^2) & oP8 & B16 & GeS & & W. H. Zachariasen, Phys. Rev.
    ↪ 40, 917-922 (1932)
   1.0000000000000000
  10.42000000000000   0.00000000000000   0.00000000000000
   0.00000000000000   3.64000000000000   0.00000000000000
   0.00000000000000   0.00000000000000   4.29000000000000
    Ge      S
     4      4
Direct
   0.12500000000000   0.75000000000000   0.83300000000000   Ge    (4c)
   0.37500000000000   0.25000000000000   0.33300000000000   Ge    (4c)
   0.62500000000000   0.75000000000000   0.66700000000000   Ge    (4c)
   0.87500000000000   0.25000000000000   0.16700000000000   Ge    (4c)
   0.13900000000000   0.25000000000000   0.38900000000000   S     (4c)
   0.36100000000000   0.75000000000000  -0.11100000000000   S     (4c)
   0.63900000000000   0.25000000000000   1.11100000000000   S     (4c)
   0.86100000000000   0.75000000000000   0.61100000000000   S     (4c)
```

MnP (B31): AB_oP8_62_c_c - CIF

```
# CIF file

data_findsym-output
_audit_creation_method FINDSYM

_chemical_name_mineral ''
_chemical_formula_sum 'Mn P'

loop_
_publ_author_name
 'Helmer Fjellv{\aa}g'
 'Arne Kjekshus'
_journal_name_full
;
Acta Chemica Scandinvaca A
;
_journal_volume 38
_journal_year 1984
_journal_page_first 563
_journal_page_last 573
_publ_Section_title
;
 Magnetic and Structural Properties of Transition Metal Substituted MnP.
     ↪ I. Mn$_{1-t}$Co$_t$P ($0.00<=t<=0.30$).
;

_aflow_proto 'AB_oP8_62_c_c'
_aflow_params 'a,b/a,c/a,x1,z1,x2,z2'
_aflow_params_values '5.24160,0.606723137973,1.12622100122,0.0056,0.1952
    ↪ ,0.1879,0.5696'
_aflow_Strukturbericht 'B31'
_aflow_Pearson 'oP8'

_symmetry_space_group_name_Hall "-P 2ac 2n"
_symmetry_space_group_name_H-M "P n m a"
_symmetry_Int_Tables_number 62

_cell_length_a     5.24160
_cell_length_b     3.18020
_cell_length_c     5.90320
_cell_angle_alpha  90.00000
_cell_angle_beta   90.00000
_cell_angle_gamma  90.00000

loop_
_space_group_symop_id
_space_group_symop_operation_xyz
1 x,y,z
2 x+1/2,-y+1/2,-z+1/2
3 -x,y+1/2,-z
4 -x+1/2,-y,z+1/2
5 -x,-y,-z
6 -x+1/2,y+1/2,z+1/2
7 x,-y+1/2,z
8 x+1/2,y,-z+1/2

loop_
_atom_site_label
_atom_site_type_symbol
_atom_site_symmetry_multiplicity
_atom_site_Wyckoff_label
_atom_site_fract_x
_atom_site_fract_y
_atom_site_fract_z
_atom_site_occupancy
Mn1 Mn   4 c 0.00560  0.25000  0.19520  1.00000
P1  P    4 c 0.18790  0.25000  0.56960  1.00000
```

MnP (B31): AB_oP8_62_c_c - POSCAR

```
AB_oP8_62_c_c & a,b/a,c/a,x1,z1,x2,z2 --params=5.24160,0.606723137973,
    ↪ 1.12622100122,0.0056,0.1952,0.1879,0.5696 & Pnma      D_{2h}^16
    ↪ #62 (c^2) & oP8 & B31 & MnP & & Helmer Fjellv{\aa}g and Arne
    ↪ Kjekshus, Acta Chemica Scandinvaca A 38, 563-573 (1984)
   1.0000000000000000
   5.24160000000000   0.00000000000000   0.00000000000000
   0.00000000000000   3.18020000000000   0.00000000000000
   0.00000000000000   0.00000000000000   5.90320000000000
    Mn      P
     4      4
Direct
   0.00560000000000   0.25000000000000   0.19520000000000   Mn    (4c)
  -0.00560000000000   0.75000000000000  -0.19520000000000   Mn    (4c)
   0.49440000000000   0.75000000000000   0.69520000000000   Mn    (4c)
   0.50560000000000   0.25000000000000   0.30480000000000   Mn    (4c)
   0.18790000000000   0.25000000000000   0.56960000000000   P     (4c)
  -0.18790000000000   0.75000000000000  -0.56960000000000   P     (4c)
   0.31210000000000   0.75000000000000   0.06960000000000   P     (4c)
   0.68790000000000   0.25000000000000  -0.06960000000000   P     (4c)
```

Cementite ($Fe_3C$, $D0_{11}$): AB3_oP16_62_c_cd - CIF

```
# CIF file

data_findsym-output
_audit_creation_method FINDSYM

_chemical_name_mineral 'Cementite'
_chemical_formula_sum 'Fe3 C'

loop_
_publ_author_name
 'F. H. Herbstein'
 'J. Smuts'
_journal_name_full
;
Acta Crystallographica
;
_journal_volume 17
_journal_year 1964
_journal_page_first 1331
_journal_page_last 1332
_publ_Section_title
;
 Comparison of X-ray and neutron-diffraction refinements of the
     ↪ structure of cementite Fe$_3$C
;

# Found in AMS Database

_aflow_proto 'AB3_oP16_62_c_cd'
_aflow_params 'a,b/a,c/a,x1,z1,x2,z2,x3,y3,z3'
_aflow_params_values '5.09,1.3257367387,0.888605108055,0.39,0.05,0.036,
     ↪ 0.852,0.186,0.063,0.328'
_aflow_Strukturbericht 'D0_11'
_aflow_Pearson 'oP16'

_symmetry_space_group_name_Hall "-P 2ac 2n"
_symmetry_space_group_name_H-M "P n m a"
_symmetry_Int_Tables_number 62

_cell_length_a     5.09000
_cell_length_b     6.74800
_cell_length_c     4.52300
_cell_angle_alpha  90.00000
_cell_angle_beta   90.00000
_cell_angle_gamma  90.00000

loop_
_space_group_symop_id
_space_group_symop_operation_xyz
1 x,y,z
2 x+1/2,-y+1/2,-z+1/2
3 -x,y+1/2,-z
4 -x+1/2,-y,z+1/2
5 -x,-y,-z
6 -x+1/2,y+1/2,z+1/2
7 x,-y+1/2,z
8 x+1/2,y,-z+1/2

loop_
_atom_site_label
_atom_site_type_symbol
_atom_site_symmetry_multiplicity
_atom_site_Wyckoff_label
_atom_site_fract_x
_atom_site_fract_y
_atom_site_fract_z
_atom_site_occupancy
C1  C    4 c 0.39000  0.25000  0.05000  1.00000
Fe1 Fe   4 c 0.03600  0.25000  0.85200  1.00000
Fe2 Fe   8 d 0.18600  0.06300  0.32800  1.00000
```

Cementite ($Fe_3C$, $D0_{11}$): AB3_oP16_62_c_cd - POSCAR

```
AB3_oP16_62_c_cd & a,b/a,c/a,x1,z1,x2,z2,x3,y3,z3 --params=5.09,
   ↪ 1.3257367387,0.888605108055,0.39,0.05,0.036,0.852,0.186,0.063,
   ↪ 0.328 & Pnma      D^{16}_{2h} #62 (c^2d) & oP16 & D0_11 & Fe3C
   ↪ & Cementite & F. H. Herbstein and J. Smuts, Acta Cryst. 17,
   ↪ 1331-1332 (1964)
   1.0000000000000000
   5.09000000000000   0.00000000000000   0.00000000000000
   0.00000000000000   6.74800000000000   0.00000000000000
   0.00000000000000   0.00000000000000   4.52300000000000
    C    Fe
    4    12
Direct
   0.11000000000000   0.75000000000000  -0.45000000000000   C   (4c)
   0.39000000000000   0.25000000000000   0.05000000000000   C   (4c)
   0.61000000000000   0.75000000000000   0.95000000000000   C   (4c)
   0.89000000000000   0.25000000000000   0.45000000000000   C   (4c)
   0.03600000000000   0.25000000000000   0.85200000000000   Fe  (4c)
  -0.03600000000000   0.75000000000000   0.14800000000000   Fe  (4c)
   0.46400000000000   0.75000000000000   0.35200000000000   Fe  (4c)
   0.53600000000000   0.25000000000000   0.64800000000000   Fe  (4c)
   0.18600000000000   0.06300000000000   0.32800000000000   Fe  (8d)
   0.18600000000000   0.43700000000000   0.32800000000000   Fe  (8d)
   0.31400000000000  -0.06300000000000   0.82800000000000   Fe  (8d)
   0.31400000000000   0.56300000000000   0.82800000000000   Fe  (8d)
   0.68600000000000   0.06300000000000   0.17200000000000   Fe  (8d)
   0.68600000000000   0.43700000000000   0.17200000000000   Fe  (8d)
   0.81400000000000   0.56300000000000   0.67200000000000   Fe  (8d)
   0.81400000000000   0.93700000000000   0.67200000000000   Fe  (8d)
```

$C_3Cr_7$ ($D10_1$): A3B7_oP40_62_cd_3c2d - CIF

```
# CIF file

data_findsym-output
_audit_creation_method FINDSYM

_chemical_name_mineral ''
_chemical_formula_sum 'C3 Cr7'

loop_
_publ_author_name
 'M. A. Rouault'
 'P. Herpin'
 'M. R. Fruchart'
_journal_name_full
;
Annales de Chimie (Paris)
;
_journal_volume 5
_journal_year 1970
_journal_page_first 461
_journal_page_last 470
_publ_Section_title
;
 Etude Cristallographique des Carbures Cr$_7$C$_3$ et Mn$_7$C$_3$
;

# Found in  Pearson's Handbook Vol. II, p. 1873

_aflow_proto 'A3B7_oP40_62_cd_3c2d'
_aflow_params 'a,b/a,c/a,x1,z1,x2,z2,x3,z3,x4,z4,x5,y5,z5,x6,y6,z6,x7,y7
   ↪ ,z7'
_aflow_params_values '4.526,1.54882898807,2.68272205038,0.4594,0.5629,
   ↪ 0.0579,0.6261,0.2501,0.2063,0.2619,0.4165,0.0288,0.0291,0.3428,
   ↪ 0.0565,0.0642,0.8119,0.2509,0.0657,0.0218'
_aflow_Strukturbericht 'D10_1'
_aflow_Pearson 'oP40'

_symmetry_space_group_name_Hall "-P 2ac 2n"
_symmetry_space_group_name_H-M "P n m a"
_symmetry_Int_Tables_number 62

_cell_length_a    4.52600
_cell_length_b    7.01000
_cell_length_c    12.14200
_cell_angle_alpha 90.00000
_cell_angle_beta  90.00000
_cell_angle_gamma 90.00000

loop_
_space_group_symop_id
_space_group_symop_operation_xyz
1 x,y,z
2 x+1/2,-y+1/2,-z+1/2
3 -x,y+1/2,-z
4 -x+1/2,-y,z+1/2
5 -x,-y,-z
6 -x+1/2,y+1/2,z+1/2
7 x,-y+1/2,z
8 x+1/2,y,-z+1/2

loop_
_atom_site_label
_atom_site_type_symbol
_atom_site_symmetry_multiplicity
_atom_site_Wyckoff_label
_atom_site_fract_x
_atom_site_fract_y
_atom_site_fract_z
_atom_site_occupancy
C1  C   4 c 0.45940 0.25000 0.56290 1.00000
Cr1 Cr  4 c 0.05790 0.25000 0.62610 1.00000
Cr2 Cr  4 c 0.25010 0.25000 0.20630 1.00000
Cr3 Cr  4 c 0.26190 0.25000 0.41650 1.00000
C2  C   8 d 0.02880 0.02910 0.34280 1.00000
Cr4 Cr  8 d 0.05650 0.06420 0.81190 1.00000
Cr5 Cr  8 d 0.25090 0.06570 0.02180 1.00000
```

$C_3Cr_7$ ($D10_1$): A3B7_oP40_62_cd_3c2d - POSCAR

```
A3B7_oP40_62_cd_3c2d & a,b/a,c/a,x1,z1,x2,z2,x3,z3,x4,z4,x5,y5,z5,x6,y6,
   ↪ z6,x7,y7,z7 --params=4.526,1.54882898807,2.68272205038,0.4594,
   ↪ 0.5629,0.0579,0.6261,0.2501,0.2063,0.2619,0.4165,0.0288,0.0291,
   ↪ 0.3428,0.0565,0.0642,0.8119,0.2509,0.0657,0.0218 & Pnma      D
   ↪ ^{16}_{2h} #62 (c^4d^3) & oP40 & D10_1 & C3Cr7 & & M. A.
   ↪ Rouault, P. Herpin and M. R. Fruchart, Ann. Chim. (Paris) 5,
   ↪ 461-470 (1970)
   1.0000000000000000
   4.52600000000000   0.00000000000000   0.00000000000000
   0.00000000000000   7.01000000000000   0.00000000000000
   0.00000000000000   0.00000000000000  12.14200000000000
    C    Cr
   12    28
Direct
   0.04060000000000   0.75000000000000   0.06290000000000   C   (4c)
   0.45940000000000   0.25000000000000   0.56290000000000   C   (4c)
   0.54060000000000   0.75000000000000   0.43710000000000   C   (4c)
   0.95940000000000   0.25000000000000  -0.06290000000000   C   (4c)
   0.02880000000000   0.02910000000000   0.34280000000000   C   (8d)
   0.02880000000000   0.47090000000000   0.34280000000000   C   (8d)
   0.47120000000000  -0.02910000000000   0.84280000000000   C   (8d)
   0.47120000000000   0.52910000000000   0.84280000000000   C   (8d)
   0.52880000000000   0.02910000000000   0.15720000000000   C   (8d)
   0.52880000000000   0.47090000000000   0.15720000000000   C   (8d)
   0.97120000000000   0.52910000000000   0.65720000000000   C   (8d)
   0.97120000000000   0.97090000000000   0.65720000000000   C   (8d)
   0.05790000000000   0.25000000000000   0.62610000000000   Cr  (4c)
   0.44210000000000   0.75000000000000   0.12610000000000   Cr  (4c)
   0.55790000000000   0.25000000000000  -0.12610000000000   Cr  (4c)
   0.94210000000000   0.75000000000000   0.37390000000000   Cr  (4c)
   0.24990000000000   0.75000000000000  -0.29370000000000   Cr  (4c)
   0.25010000000000   0.25000000000000   0.20630000000000   Cr  (4c)
   0.74990000000000   0.75000000000000   0.79370000000000   Cr  (4c)
   0.75010000000000   0.25000000000000   0.29370000000000   Cr  (4c)
   0.23810000000000   0.75000000000000  -0.08350000000000   Cr  (4c)
   0.26190000000000   0.25000000000000   0.41650000000000   Cr  (4c)
  -0.26190000000000   0.75000000000000   0.58350000000000   Cr  (4c)
   0.76190000000000   0.25000000000000   1.08350000000000   Cr  (4c)
   0.05650000000000   0.06420000000000   0.81190000000000   Cr  (8d)
   0.05650000000000   0.43580000000000   0.81190000000000   Cr  (8d)
   0.44350000000000  -0.06420000000000   1.31190000000000   Cr  (8d)
   0.44350000000000   0.56420000000000   1.31190000000000   Cr  (8d)
   0.55650000000000   0.06420000000000  -0.31190000000000   Cr  (8d)
   0.55650000000000   0.43580000000000  -0.31190000000000   Cr  (8d)
   0.94350000000000   0.56420000000000   0.18810000000000   Cr  (8d)
   0.94350000000000   0.93580000000000   0.18810000000000   Cr  (8d)
   0.24910000000000  -0.06570000000000   0.52180000000000   Cr  (8d)
   0.24910000000000   0.56570000000000   0.52180000000000   Cr  (8d)
   0.25090000000000   0.06570000000000   0.02180000000000   Cr  (8d)
   0.25090000000000   0.43430000000000   0.02180000000000   Cr  (8d)
   0.74910000000000   0.56570000000000   0.97820000000000   Cr  (8d)
   0.74910000000000   0.93430000000000   0.97820000000000   Cr  (8d)
   0.75090000000000   0.06570000000000   0.47820000000000   Cr  (8d)
   0.75090000000000   0.43430000000000   0.47820000000000   Cr  (8d)
```

$\alpha$-Np ($A_c$): A_oP8_62_2c - CIF

```
# CIF file

data_findsym-output
_audit_creation_method FINDSYM

_chemical_name_mineral 'alpha Np'
_chemical_formula_sum 'Np'

loop_
_publ_author_name
 'W. H. Zachariasen'
_journal_name_full
;
Acta Crystallographica
;
_journal_volume 5
_journal_year 1952
_journal_page_first 660
_journal_page_last 664
_publ_Section_title
;
 Crystal chemical studies of the 5f-series of elements. XVII. The
   ↪ crystal structure of neptunium metal
;

# Found in  Donohue, pp. 151-153

_aflow_proto 'A_oP8_62_2c'
_aflow_params 'a,b/a,c/a,x1,z1,x2,z2'
_aflow_params_values '6.663,0.708839861924,0.73345339937,0.464,0.292,
   ↪ 0.181,0.658'
_aflow_Strukturbericht 'A_c'
_aflow_Pearson 'oP8'

_symmetry_space_group_name_Hall "-P 2ac 2n"
_symmetry_space_group_name_H-M "P n m a"
_symmetry_Int_Tables_number 62

_cell_length_a    6.66300
_cell_length_b    4.72300
_cell_length_c    4.88700
_cell_angle_alpha 90.00000
_cell_angle_beta  90.00000
```

```
_cell_angle_gamma  90.00000

loop_
_space_group_symop_id
_space_group_symop_operation_xyz
1 x,y,z
2 x+1/2,-y+1/2,-z+1/2
3 -x,y+1/2,-z
4 -x+1/2,-y,z+1/2
5 -x,-y,-z
6 -x+1/2,y+1/2,z+1/2
7 x,-y+1/2,z
8 x+1/2,y,-z+1/2

loop_
_atom_site_label
_atom_site_type_symbol
_atom_site_symmetry_multiplicity
_atom_site_Wyckoff_label
_atom_site_fract_x
_atom_site_fract_y
_atom_site_fract_z
_atom_site_occupancy
Np1 Np    4 c 0.46400  0.25000  0.29200  1.00000
Np2 Np    4 c 0.18100  0.25000  0.65800  1.00000
```

α-Np ($A_c$): A_oP8_62_2c - POSCAR

```
A_oP8_62_2c & a,b/a,c/a,x1,z1,x2,z2 --params=6.663,0.708839861924,
    ↪ 0.73345339937,0.464,0.292,0.181,0.658 & Pnma        D^{16}_{2h} #
    ↪ 62 (c^2) & oP8 & A_c & Np & alpha & W. H. Zachariasen, Acta
    ↪ Cryst. 5, 660-664 (1952)
   1.0000000000000000
   6.66300000000000    0.00000000000000    0.00000000000000
   0.00000000000000    4.72300000000000    0.00000000000000
   0.00000000000000    0.00000000000000    4.88700000000000
   Np
    8
Direct
   0.46400000000000    0.25000000000000    0.29200000000000    Np    (4c)
   0.03600000000000    0.75000000000000    0.79200000000000    Np    (4c)
   0.53600000000000    0.75000000000000    0.70800000000000    Np    (4c)
   0.96400000000000    0.25000000000000    0.20800000000000    Np    (4c)
   0.18100000000000    0.25000000000000    0.65800000000000    Np    (4c)
   0.31900000000000    0.75000000000000    0.15800000000000    Np    (4c)
   0.81900000000000    0.75000000000000    0.34200000000000    Np    (4c)
   0.68100000000000    0.25000000000000    0.84200000000000    Np    (4c)
```

FeB (B27): AB_oP8_62_c_c - CIF

```
# CIF file

data_findsym-output
_audit_creation_method FINDSYM

_chemical_name_mineral 'Iron Boride'
_chemical_formula_sum 'Fe B'

loop_
_publ_author_name
 'Sterling B. Hendricks'
 'Peter R. Kosting'
_journal_name_full
;
Zeitschrift f\"{u}r Kristallographie - Crystalline Materials
;
_journal_volume 74
_journal_year 1930
_journal_page_first 511
_journal_page_last 533
_publ_Section_title
;
 The Crystal Structure of Fe$_2$P, Fe$_2$N, Fe$_3$N and FeB
;

# Found in AMS Database

_aflow_proto 'AB_oP8_62_c_c'
_aflow_params 'a,b/a,c/a,x1,z1,x2,z2'
_aflow_params_values '5.495,0.536123748863,0.737579617834,0.125,0.69,-
    ↪ 0.18,0.125'
_aflow_Strukturbericht 'B27'
_aflow_Pearson 'oP8'

_symmetry_space_group_name_Hall "-P 2ac 2n"
_symmetry_space_group_name_H-M "P n m a"
_symmetry_Int_Tables_number 62

_cell_length_a     5.49500
_cell_length_b     2.94600
_cell_length_c     4.05300
_cell_angle_alpha  90.00000
_cell_angle_beta   90.00000
_cell_angle_gamma  90.00000

loop_
_space_group_symop_id
_space_group_symop_operation_xyz
1 x,y,z
2 x+1/2,-y+1/2,-z+1/2
3 -x,y+1/2,-z
4 -x+1/2,-y,z+1/2
5 -x,-y,-z
6 -x+1/2,y+1/2,z+1/2
7 x,-y+1/2,z
8 x+1/2,y,-z+1/2

loop_
_atom_site_label
_atom_site_type_symbol
_atom_site_symmetry_multiplicity
_atom_site_Wyckoff_label
_atom_site_fract_x
_atom_site_fract_y
_atom_site_fract_z
_atom_site_occupancy
B1  B     4 c 0.12500  0.25000  0.69000  1.00000
Fe1 Fe    4 c -0.18000  0.25000  0.12500  1.00000
```

FeB (B27): AB_oP8_62_c_c - POSCAR

```
AB_oP8_62_c_c & a,b/a,c/a,x1,z1,x2,z2 --params=5.495,0.536123748863,
    ↪ 0.737579617834,0.125,0.69,-0.18,0.125 & Pnma        D^{16}_{2h} #
    ↪ 62 (c^2) & oP8 & B27 & FeB & & S. B. Hendricks and P. R.
    ↪ Kosting, Zeitschrift f\"{u}r Kristallographie - Crystalline
    ↪ Materials 74, 511-533 (1930)
   1.0000000000000000
   5.49500000000000    0.00000000000000    0.00000000000000
   0.00000000000000    2.94600000000000    0.00000000000000
   0.00000000000000    0.00000000000000    4.05300000000000
    B    Fe
    4    4
Direct
   0.12500000000000    0.25000000000000    0.69000000000000    B    (4c)
   0.37500000000000    0.75000000000000    0.19000000000000    B    (4c)
   0.62500000000000    0.25000000000000    0.81000000000000    B    (4c)
   0.87500000000000    0.75000000000000    0.31000000000000    B    (4c)
  -0.18000000000000    0.25000000000000    0.12500000000000    Fe   (4c)
   0.18000000000000    0.75000000000000    0.87500000000000    Fe   (4c)
   0.32000000000000    0.25000000000000    0.37500000000000    Fe   (4c)
   0.68000000000000    0.75000000000000    0.62500000000000    Fe   (4c)
```

SnS (B29): AB_oP8_62_c_c - CIF

```
# CIF file

data_findsym-output
_audit_creation_method FINDSYM

_chemical_name_mineral ''
_chemical_formula_sum 'Sn S'

loop_
_publ_author_name
 'Sylvie Del Bucchia'
 'Jean-Claude Jumas'
 'Maurice Maurin'
_journal_name_full
;
Acta Crystallographica B
;
_journal_volume 37
_journal_year 1981
_journal_page_first 1903
_journal_page_last 1905
_publ_Section_title
;
 Contribution \`{a} l'\'{e}tude de compos\'{e}s sulfur\'{e}s d'\'{e}tain
    ↪ (II): affinement de la structure de SnS
;

_aflow_proto 'AB_oP8_62_c_c'
_aflow_params 'a,b/a,c/a,x1,z1,x2,z2'
_aflow_params_values '11.18,0.356171735242,0.387209302326,0.3507,0.0201,
    ↪ 0.61937,0.3806'
_aflow_Strukturbericht 'B29'
_aflow_Pearson 'oP8'

_symmetry_space_group_name_Hall "-P 2ac 2n"
_symmetry_space_group_name_H-M "P n m a"
_symmetry_Int_Tables_number 62

_cell_length_a     11.18000
_cell_length_b     3.98200
_cell_length_c     4.32900
_cell_angle_alpha  90.00000
_cell_angle_beta   90.00000
_cell_angle_gamma  90.00000

loop_
_space_group_symop_id
_space_group_symop_operation_xyz
1 x,y,z
2 x+1/2,-y+1/2,-z+1/2
3 -x,y+1/2,-z
4 -x+1/2,-y,z+1/2
5 -x,-y,-z
6 -x+1/2,y+1/2,z+1/2
7 x,-y+1/2,z
8 x+1/2,y,-z+1/2

loop_
_atom_site_label
_atom_site_type_symbol
_atom_site_symmetry_multiplicity
_atom_site_Wyckoff_label
_atom_site_fract_x
_atom_site_fract_y
_atom_site_fract_z
_atom_site_occupancy
```

```
S1   S     4 c 0.35070  0.25000  0.02010  1.00000
Sn1  Sn    4 c 0.61937  0.25000  0.38060  1.00000
```

SnS (B29): AB_oP8_62_c_c - POSCAR

```
AB_oP8_62_c_c & a,b/a,c/a,x1,z1,x2,z2 --params=11.18,0.356171735242,
    ↪ 0.387209302326,0.3507,0.0201,0.61937,0.3806 & Pnma         D^{16}_
    ↪ {2h} #62 (c^2) & oP8 & B29 & SnS & & S. Del Bucchia et al.,
    ↪ Acta Cryst. B 73, 1903-1905 (1981)
   1.0000000000000000
  11.18000000000000     0.00000000000000     0.00000000000000
   0.00000000000000     3.98200000000000     0.00000000000000
   0.00000000000000     0.00000000000000     4.32900000000000
    S    Sn
    4     4
Direct
   0.14930000000000     0.75000000000000     0.52010000000000     S     (4c)
   0.35070000000000     0.25000000000000     0.02010000000000     S     (4c)
   0.64930000000000     0.75000000000000     0.97990000000000     S     (4c)
   0.85070000000000     0.25000000000000     0.47990000000000     S     (4c)
   0.11937000000000     0.25000000000000     0.11940000000000     Sn    (4c)
   0.38063000000000     0.75000000000000     0.61940000000000     Sn    (4c)
   0.61937000000000     0.25000000000000     0.38060000000000     Sn    (4c)
   0.88063000000000     0.75000000000000     0.88060000000000     Sn    (4c)
```

$SrCuO_2$: AB2C_oC16_63_c_2c_c - CIF

```
# CIF file

data_findsym-output
_audit_creation_method FINDSYM

_chemical_name_mineral ''
_chemical_formula_sum 'Sr Cu O2'

loop_
_publ_author_name
 'Yoshitaka Matsushita'
 'Yasunao Oyama'
 'Masashi Hasegawa'
 'Humihiko Takei'
_journal_name_full
;
Journal of Solid State Chemistry
;
_journal_volume 114
_journal_year 1994
_journal_page_first 289
_journal_page_last 293
_publ_Section_title
;
 Growth and Structural Refinement of Orthorhombic SrCuO$_2$ Crystals
;

_aflow_proto 'AB2C_oC16_63_c_2c_c'
_aflow_params 'a,b/a,c/a,y1,y2,y3,y4'
_aflow_params_values '3.577,4.56863293263,1.09538719597,0.06109,-0.0558,
    ↪ 0.1792,0.33096'
_aflow_Strukturbericht 'None'
_aflow_Pearson 'oC16'

_symmetry_space_group_name_Hall "-C 2c 2"
_symmetry_space_group_name_H-M "C m c m"
_symmetry_Int_Tables_number 63

_cell_length_a      3.57700
_cell_length_b      16.34200
_cell_length_c      3.91820
_cell_angle_alpha 90.00000
_cell_angle_beta  90.00000
_cell_angle_gamma 90.00000

loop_
_space_group_symop_id
_space_group_symop_operation_xyz
1 x,y,z
2 x,-y,-z
3 -x,y,-z+1/2
4 -x,-y,z+1/2
5 -x,-y,-z
6 -x,y,z
7 x,-y,z+1/2
8 x,y,-z+1/2
9 x+1/2,y+1/2,z
10 x+1/2,-y+1/2,-z
11 -x+1/2,y+1/2,-z+1/2
12 -x+1/2,-y+1/2,z+1/2
13 -x+1/2,-y+1/2,-z
14 -x+1/2,y+1/2,z
15 x+1/2,-y+1/2,z+1/2
16 x+1/2,y+1/2,-z+1/2

loop_
_atom_site_label
_atom_site_type_symbol
_atom_site_symmetry_multiplicity
_atom_site_Wyckoff_label
_atom_site_fract_x
_atom_site_fract_y
_atom_site_fract_z
_atom_site_occupancy
Cu1 Cu    4 c 0.00000  0.06109   0.25000  1.00000
O1  O     4 c 0.00000  -0.05580  0.25000  1.00000
O2  O     4 c 0.00000  0.17920   0.25000  1.00000
Sr1 Sr    4 c 0.00000  0.33096   0.25000  1.00000
```

$SrCuO_2$: AB2C_oC16_63_c_2c_c - POSCAR

```
AB2C_oC16_63_c_2c_c & a,b/a,c/a,y1,y2,y3,y4 --params=3.577,4.56863293263
    ↪ ,1.09538719597,0.06109,-0.0558,0.1792,0.33096 & Cmcm        D_{2h
    ↪ }^{17} #63 (c^4) & oC16 & & SrCuO2 & & Y. Matsushita, Y. Oyama,
    ↪ M. Hasegawa and H. Takei, J. Solid State Chem. 114 289-293 (
    ↪ 1994)
   1.0000000000000000
   1.78850000000000    -8.17100000000000     0.00000000000000
   1.78850000000000     8.17100000000000     0.00000000000000
   0.00000000000000     0.00000000000000     3.91820000000000
   Cu     O     Sr
    2     4      2
Direct
  -0.06109000000000     0.06109000000000     0.25000000000000     Cu     (4c)
   0.06109000000000    -0.06109000000000     0.75000000000000     Cu     (4c)
   0.05580000000000    -0.05580000000000     0.25000000000000     O      (4c)
  -0.05580000000000     0.05580000000000     0.75000000000000     O      (4c)
   0.17920000000000     0.82080000000000     0.75000000000000     O      (4c)
   0.82080000000000     0.17920000000000     0.25000000000000     O      (4c)
   0.33096000000000     0.66904000000000     0.75000000000000     Sr     (4c)
   0.66904000000000     0.33096000000000     0.25000000000000     Sr     (4c)
```

$ZrSi_2$ (C49): A2B_oC12_63_2c_c - CIF

```
# CIF file

data_findsym-output
_audit_creation_method FINDSYM

_chemical_name_mineral 'Zirconium Disilicide'
_chemical_formula_sum 'Zr Si2'

loop_
_publ_author_name
 'P. G. Cotter'
 'J. A. Kohn'
 'R. A. Potter'
_journal_name_full
;
Journal of the American Ceramic Society
;
_journal_volume 39
_journal_year 1956
_journal_page_first 11
_journal_page_last 12
_publ_Section_title
;
 Physical and X-Ray Study of the Disilicides of Titanium, Zirconium, and
    ↪ Hafnium
;

# Found in http://materials.springer.com/isp/crystallographic/docs/
    ↪ sd_0530831

_aflow_proto 'A2B_oC12_63_2c_c'
_aflow_params 'a,b/a,c/a,y1,y2,y3'
_aflow_params_values '3.73,3.94638069705,0.983914209115,0.061,0.75,0.396
    ↪ '
_aflow_Strukturbericht 'C49'
_aflow_Pearson 'oC12'

_symmetry_space_group_name_Hall "-C 2c 2"
_symmetry_space_group_name_H-M "C m c m"
_symmetry_Int_Tables_number 63

_cell_length_a      3.73000
_cell_length_b      14.72000
_cell_length_c      3.67000
_cell_angle_alpha 90.00000
_cell_angle_beta  90.00000
_cell_angle_gamma 90.00000

loop_
_space_group_symop_id
_space_group_symop_operation_xyz
1 x,y,z
2 x,-y,-z
3 -x,y,-z+1/2
4 -x,-y,z+1/2
5 -x,-y,-z
6 -x,y,z
7 x,-y,z+1/2
8 x,y,-z+1/2
9 x+1/2,y+1/2,z
10 x+1/2,-y+1/2,-z
11 -x+1/2,y+1/2,-z+1/2
12 -x+1/2,-y+1/2,z+1/2
13 -x+1/2,-y+1/2,-z
14 -x+1/2,y+1/2,z
15 x+1/2,-y+1/2,z+1/2
16 x+1/2,y+1/2,-z+1/2

loop_
_atom_site_label
_atom_site_type_symbol
_atom_site_symmetry_multiplicity
_atom_site_Wyckoff_label
_atom_site_fract_x
_atom_site_fract_y
_atom_site_fract_z
_atom_site_occupancy
Si1  Si    4 c 0.00000  0.06100  0.25000  1.00000
Si2  Si    4 c 0.00000  0.75000  0.25000  1.00000
Zr1  Zr    4 c 0.00000  0.39600  0.25000  1.00000
```

ZrSi$_2$ (C49): A2B_oC12_63_2c_c - POSCAR

```
A2B_oC12_63_2c_c & a,b/a,c/a,y1,y2,y3 --params=3.73,3.94638069705,
    ↪ 0.983914209115,0.061,0.75,0.396 & Cmcm         D_{2h}^{17} #63 (c^
    ↪ 3) & oC12 & C49 & ZrSi2 & & P. G. Cotter, J. A. Kohn and R. A.
    ↪ Potter, J. Am. Ceram. Soc. 39, 11-12 (1956)
   1.0000000000000000
   1.86500000000000   -7.36000000000000    0.00000000000000
   1.86500000000000    7.36000000000000    0.00000000000000
   0.00000000000000    0.00000000000000    3.67000000000000
   Si    Zr
    4     2
Direct
  -0.06100000000000    0.06100000000000    0.25000000000000    Si    (4c)
   0.06100000000000   -0.06100000000000    0.75000000000000    Si    (4c)
   0.25000000000000    0.75000000000000    0.25000000000000    Si    (4c)
   0.75000000000000    0.25000000000000    0.75000000000000    Si    (4c)
   0.39600000000000    0.60400000000000    0.75000000000000    Zr    (4c)
   0.60400000000000    0.39600000000000    0.25000000000000    Zr    (4c)
```

CrB (B33): AB_oC8_63_c_c - CIF

```
# CIF file

data_findsym-output
_audit_creation_method FINDSYM

_chemical_name_mineral ''
_chemical_formula_sum 'Cr B'

loop_
_publ_author_name
 'Shigeru Okada'
 'Tetsuzo Atoda'
 'Iwami Higashi'
_journal_name_full
;
Journal of Solid State Chemistry
;
_journal_volume 68
_journal_year 1987
_journal_page_first 61
_journal_page_last 67
_publ_Section_title
;
 Structural investigation of Cr$_2$B$_3$, Cr$_3$B$_4$, and CrB by
    ↪ single-crystal diffractometry
;

# Found in http://materials.springer.com/isp/crystallographic/docs/
    ↪ sd_0455627

_aflow_proto 'AB_oC8_63_c_c'
_aflow_params 'a,b/a,c/a,y1,y2'
_aflow_params_values '2.9782,2.64253575985,0.985360284736,0.436,0.14525'
_aflow_Strukturbericht 'B33'
_aflow_Pearson 'oC8'

_symmetry_space_group_name_Hall "-C 2c 2"
_symmetry_space_group_name_H-M "C m c m"
_symmetry_Int_Tables_number 63

_cell_length_a     2.97820
_cell_length_b     7.87000
_cell_length_c     2.93460
_cell_angle_alpha  90.00000
_cell_angle_beta   90.00000
_cell_angle_gamma  90.00000

loop_
_space_group_symop_id
_space_group_symop_operation_xyz
1 x,y,z
2 x,-y,-z
3 -x,y,-z+1/2
4 -x,-y,z+1/2
5 -x,-y,-z
6 -x,y,z
7 x,-y,z+1/2
8 x,y,-z+1/2
9 x+1/2,y+1/2,z
10 x+1/2,-y+1/2,-z
11 -x+1/2,y+1/2,-z+1/2
12 -x+1/2,-y+1/2,z+1/2
13 -x+1/2,-y+1/2,-z
14 -x+1/2,y+1/2,z
15 x+1/2,-y+1/2,z+1/2
16 x+1/2,y+1/2,-z+1/2

loop_
_atom_site_label
_atom_site_type_symbol
_atom_site_symmetry_multiplicity
_atom_site_Wyckoff_label
_atom_site_fract_x
_atom_site_fract_y
_atom_site_fract_z
_atom_site_occupancy
B1  B    4 c 0.00000 0.43600 0.25000 1.00000
Cr1 Cr   4 c 0.00000 0.14525 0.25000 1.00000
```

CrB (B33): AB_oC8_63_c_c - POSCAR

```
AB_oC8_63_c_c & a,b/a,c/a,y1,y2 --params=2.9782,2.64253575985,
    ↪ 0.985360284736,0.436,0.14525 & Cmcm         D_{2h}^{17} #63 (c^2)
    ↪ & oC8 & B33 & CrB & & S. Okada, T. Atoda, and I. Higashi, J.
    ↪ Solid State Chem. 68, 61-67 (1987)
   1.0000000000000000
   1.48910000000000   -3.93500000000000    0.00000000000000
   1.48910000000000    3.93500000000000    0.00000000000000
   0.00000000000000    0.00000000000000    2.93460000000000
    B    Cr
    2     2
Direct
   0.43600000000000    0.56400000000000    0.75000000000000    B     (4c)
   0.56400000000000    0.43600000000000    0.25000000000000    B     (4c)
   0.14525000000000    0.85475000000000    0.75000000000000    Cr    (4c)
   0.85475000000000    0.14525000000000    0.25000000000000    Cr    (4c)
```

$\alpha$-U (A20): A_oC4_63_c - CIF

```
# CIF file

data_findsym-output
_audit_creation_method FINDSYM

_chemical_name_mineral 'alpha U'
_chemical_formula_sum 'U'

loop_
_publ_author_name
 'C. S. Barrett'
 'M. H. Mueller'
 'R. L. Hitterman'
_journal_name_full
;
Physical Review
;
_journal_volume 129
_journal_year 1963
_journal_page_first 625
_journal_page_last 629
_publ_Section_title
;
 Crystal Structure Variations in Alpha Uranium at Low Temperatures
;

_aflow_proto 'A_oC4_63_c'
_aflow_params 'a,b/a,c/a,y1'
_aflow_params_values '2.8444,2.06331739558,1.73379271551,0.10228'
_aflow_Strukturbericht 'A20'
_aflow_Pearson 'oC4'

_symmetry_space_group_name_Hall "-C 2c 2"
_symmetry_space_group_name_H-M "C m c m"
_symmetry_Int_Tables_number 63

_cell_length_a     2.84440
_cell_length_b     5.86890
_cell_length_c     4.93160
_cell_angle_alpha  90.00000
_cell_angle_beta   90.00000
_cell_angle_gamma  90.00000

loop_
_space_group_symop_id
_space_group_symop_operation_xyz
1 x,y,z
2 x,-y,-z
3 -x,y,-z+1/2
4 -x,-y,z+1/2
5 -x,-y,-z
6 -x,y,z
7 x,-y,z+1/2
8 x,y,-z+1/2
9 x+1/2,y+1/2,z
10 x+1/2,-y+1/2,-z
11 -x+1/2,y+1/2,-z+1/2
12 -x+1/2,-y+1/2,z+1/2
13 -x+1/2,-y+1/2,-z
14 -x+1/2,y+1/2,z
15 x+1/2,-y+1/2,z+1/2
16 x+1/2,y+1/2,-z+1/2

loop_
_atom_site_label
_atom_site_type_symbol
_atom_site_symmetry_multiplicity
_atom_site_Wyckoff_label
_atom_site_fract_x
_atom_site_fract_y
_atom_site_fract_z
_atom_site_occupancy
U1 U    4 c 0.00000 0.10228 0.25000 1.00000
```

$\alpha$-U (A20): A_oC4_63_c - POSCAR

```
A_oC4_63_c & a,b/a,c/a,y1 --params=2.8444,2.06331739558,1.73379271551,
    ↪ 0.10228 & Cmcm        D_{2h}^{17} #63 (c) & oC4 & A20 & U & alpha
    ↪ & C. S. Barrett, M. H. Mueller and R. L. Hitterman, Phys. Rev.
    ↪ 129, 625-629 (1963)
   1.0000000000000000
   1.42220000000000   -2.93445000000000    0.00000000000000
   1.42220000000000    2.93445000000000    0.00000000000000
   0.00000000000000    0.00000000000000    4.93160000000000
    U
    2
Direct
```

```
    0.10228000000000    0.89772000000000    0.75000000000000    U    (4c)
    0.89772000000000    0.10228000000000    0.25000000000000    U    (4c)
```

α-Ga (A11): A_oC8_64_f - CIF

```
# CIF file

data_findsym-output
_audit_creation_method FINDSYM

_chemical_name_mineral 'alpha'
_chemical_formula_sum 'Ga'

loop_
_publ_author_name
 'Brahama D. Sharma'
 'Jerry Donohue'
_journal_name_full
;
Zeitschrift f\"{u}r Kristallographie
;
_journal_volume 117
_journal_year 1962
_journal_page_first 293
_journal_page_last 300
_publ_Section_title
;
 A refinement of the crystal structure of gallium
;

# Found in AMS Database

_aflow_proto 'A_oC8_64_f'
_aflow_params 'a,b/a,c/a,y1,z1'
_aflow_params_values '4.523,1.69378730931,1.0002210922,0.1549,0.081'
_aflow_Strukturbericht 'A11'
_aflow_Pearson 'oC8'

_symmetry_space_group_name_Hall "-C 2bc 2"
_symmetry_space_group_name_H-M "C m c a"
_symmetry_Int_Tables_number 64

_cell_length_a      4.52300
_cell_length_b      7.66100
_cell_length_c      4.52400
_cell_angle_alpha  90.00000
_cell_angle_beta   90.00000
_cell_angle_gamma  90.00000

loop_
_space_group_symop_id
_space_group_symop_operation_xyz
1 x,y,z
2 x,-y,-z
3 -x+1/2,y,-z+1/2
4 -x+1/2,-y,z+1/2
5 -x,-y,-z
6 -x,y,z
7 x+1/2,-y,z+1/2
8 x+1/2,y,-z+1/2
9 x+1/2,y+1/2,z
10 x+1/2,-y+1/2,-z
11 -x,y+1/2,-z+1/2
12 -x,-y+1/2,z+1/2
13 -x+1/2,-y+1/2,-z
14 -x+1/2,y+1/2,z
15 x,-y+1/2,z+1/2
16 x,y+1/2,-z+1/2

loop_
_atom_site_label
_atom_site_type_symbol
_atom_site_symmetry_multiplicity
_atom_site_Wyckoff_label
_atom_site_fract_x
_atom_site_fract_y
_atom_site_fract_z
_atom_site_occupancy
Ga1 Ga    8 f 0.00000 0.15490 0.08100 1.00000
```

α-Ga (A11): A_oC8_64_f - POSCAR

```
A_oC8_64_f & a,b/a,c/a,y1,z1 --params=4.523,1.69378730931,1.0002210922,
    ↪ 0.1549,0.081 & Cmca    D_{2h}^{18} #64 (f) & oC8 & A11 & Ga &
    ↪ alpha & B. D. Sharma and J. Donohue, Zeitschrift f\"{u}r
    ↪ Kristallographie 117, 293-300 (1962)
   1.0000000000000000
   2.26150000000000  -3.83050000000000   0.00000000000000
   2.26150000000000   3.83050000000000   0.00000000000000
   0.00000000000000   0.00000000000000   4.52400000000000
   Ga
    4
Direct
   0.15490000000000   0.84510000000000   0.91900000000000   Ga   (8f)
   0.34510000000000   0.65490000000000   0.41900000000000   Ga   (8f)
   0.65490000000000   0.34510000000000   0.58100000000000   Ga   (8f)
   0.84510000000000   0.15490000000000   0.08100000000000   Ga   (8f)
```

$MgB_2C_2$: A2B2C_oC80_64_efg_efg_df - CIF

```
# CIF file

data_findsym-output
_audit_creation_method FINDSYM

_chemical_name_mineral ''
_chemical_formula_sum 'Mg B2 C2'

loop_
_publ_author_name
 'Michael W\"{o}rle'
 'Reinhard Nesper'
_journal_name_full
;
Journal of Alloys and Compounds
;
_journal_volume 216
_journal_year 1994
_journal_page_first 75
_journal_page_last 83
_publ_Section_title
;
 MgB$_2$C$_2$, a new graphite-related refractory compound
;

_aflow_proto 'A2B2C_oC80_64_efg_efg_df'
_aflow_params 'a,b/a,c/a,x1,y2,y3,y4,z4,y5,z5,y6,z6,x7,y7,z7,x8,y8,z8'
_aflow_params_values '10.922,0.866233290606,0.682933528658,0.84657,
    ↪ 0.0946,0.9271,0.5886,0.276,-0.0792,0.2314,0.27981,-0.0113,
    ↪ 0.1278,0.3415,0.2438,0.1245,0.175,0.2231'
_aflow_Strukturbericht 'None'
_aflow_Pearson 'oC80'

_symmetry_space_group_name_Hall "-C 2bc 2"
_symmetry_space_group_name_H-M "C m c a"
_symmetry_Int_Tables_number 64

_cell_length_a     10.92200
_cell_length_b      9.46100
_cell_length_c      7.45900
_cell_angle_alpha  90.00000
_cell_angle_beta   90.00000
_cell_angle_gamma  90.00000

loop_
_space_group_symop_id
_space_group_symop_operation_xyz
1 x,y,z
2 x,-y,-z
3 -x+1/2,y,-z+1/2
4 -x+1/2,-y,z+1/2
5 -x,-y,-z
6 -x,y,z
7 x+1/2,-y,z+1/2
8 x+1/2,y,-z+1/2
9 x+1/2,y+1/2,z
10 x+1/2,-y+1/2,-z
11 -x,y+1/2,-z+1/2
12 -x,-y+1/2,z+1/2
13 -x+1/2,-y+1/2,-z
14 -x+1/2,y+1/2,z
15 x,-y+1/2,z+1/2
16 x,y+1/2,-z+1/2

loop_
_atom_site_label
_atom_site_type_symbol
_atom_site_symmetry_multiplicity
_atom_site_Wyckoff_label
_atom_site_fract_x
_atom_site_fract_y
_atom_site_fract_z
_atom_site_occupancy
Mg1 Mg    8 d 0.84657 0.00000 0.00000 1.00000
B1  B     8 e 0.25000 0.09460 0.25000 1.00000
C1  C     8 e 0.25000 0.92710 0.25000 1.00000
B2  B     8 f 0.00000 0.58860 0.27600 1.00000
C2  C     8 f 0.00000 -0.07920 0.23140 1.00000
Mg2 Mg    8 f 0.00000 0.27981 -0.0113 1.00000
B3  B    16 g 0.12780 0.34150 0.24380 1.00000
C3  C    16 g 0.12450 0.17500 0.22310 1.00000
```

$MgB_2C_2$: A2B2C_oC80_64_efg_efg_df - POSCAR

```
A2B2C_oC80_64_efg_efg_df & a,b/a,c/a,x1,y2,y3,y4,z4,y5,z5,y6,z6,x7,y7,z7
    ↪ ,x8,y8,z8 --params=10.922,0.866233290606,0.682933528658,0.84657
    ↪ ,0.0946,0.9271,0.5886,0.276,-0.0792,0.2314,0.27981,-0.0113,
    ↪ 0.1278,0.3415,0.2438,0.1245,0.175,0.2231 & Cmca    D_{2h}^{18
    ↪ } #64 (de^2f^3g^2) & oC80 & & MgB2C2 & & M. Worle and R Nespar,
    ↪  J. Alloys and Compd. 216, 75-83 (1994)
   1.0000000000000000
   5.46100000000000  -4.73050000000000   0.00000000000000
   5.46100000000000   4.73050000000000   0.00000000000000
   0.00000000000000   0.00000000000000   7.45900000000000
    B    C    Mg
   16   16    8
Direct
   0.21370000000000  -0.46930000000000  -0.24380000000000   B   (16g)
  -0.21370000000000   0.46930000000000   0.24380000000000   B   (16g)
   0.28630000000000  -0.03070000000000  -0.74380000000000   B   (16g)
  -0.28630000000000   0.03070000000000   0.74380000000000   B   (16g)
   0.46930000000000  -0.21370000000000  -0.24380000000000   B   (16g)
  -0.46930000000000   0.21370000000000   0.24380000000000   B   (16g)
   0.96930000000000  -0.71370000000000  -0.25620000000000   B   (16g)
  -0.96930000000000   0.71370000000000   0.25620000000000   B   (16g)
   0.15540000000000   0.34460000000000   0.25000000000000   B    (8e)
  -0.34460000000000   0.84460000000000   0.25000000000000   B    (8e)
   0.34460000000000   1.15540000000000   0.75000000000000   B    (8e)
   0.84460000000000   0.65540000000000   0.75000000000000   B    (8e)
```

```
  0.08860000000000  -0.08860000000000   0.77600000000000   B    (8f)
  0.58860000000000  -0.58860000000000  -0.27600000000000   B    (8f)
 -0.58860000000000   0.58860000000000   0.27600000000000   B    (8f)
 -1.08860000000000   1.08860000000000   0.22400000000000   B    (8f)
  0.05050000000000  -0.29950000000000  -0.22310000000000   C    (16g)
 -0.05050000000000   0.29950000000000   0.22310000000000   C    (16g)
  0.29950000000000  -0.05050000000000  -0.22310000000000   C    (16g)
 -0.29950000000000   0.05050000000000   0.22310000000000   C    (16g)
  0.44950000000000  -0.20050000000000  -0.72310000000000   C    (16g)
 -0.44950000000000   0.20050000000000   0.72310000000000   C    (16g)
  0.79950000000000  -0.55050000000000  -0.27690000000000   C    (16g)
 -0.79950000000000   0.55050000000000   0.27690000000000   C    (16g)
 -0.67710000000000   1.17710000000000   0.25000000000000   C    (8e)
  1.17710000000000   0.32290000000000   0.75000000000000   C    (8e)
 -1.17710000000000   1.67710000000000   0.25000000000000   C    (8e)
  1.67710000000000  -0.17710000000000   0.75000000000000   C    (8e)
  0.07920000000000  -0.07920000000000   0.23140000000000   C    (8f)
 -0.07920000000000   0.07920000000000  -0.23140000000000   C    (8f)
 -0.42080000000000   0.42080000000000   0.26860000000000   C    (8f)
 -0.57920000000000   0.57920000000000   0.73140000000000   C    (8f)
  0.15343000000000   0.15343000000000   0.00000000000000   Mg   (8d)
 -0.15343000000000  -0.15343000000000   0.00000000000000   Mg   (8d)
 -0.34657000000000   0.65343000000000   0.50000000000000   Mg   (8d)
 -0.65343000000000   0.34657000000000   0.50000000000000   Mg   (8d)
 -0.22019000000000   0.22019000000000   0.48870000000000   Mg   (8f)
  0.27981000000000  -0.27981000000000   0.01130000000000   Mg   (8f)
 -0.27981000000000   0.27981000000000  -0.01130000000000   Mg   (8f)
 -0.77981000000000   0.77981000000000   0.51130000000000   Mg   (8f)
```

Black Phosphorus (A17): A_oC8_64_f - CIF

```
# CIF file

data_findsym-output
_audit_creation_method FINDSYM

_chemical_name_mineral 'black P'
_chemical_formula_sum 'P'

loop_
_publ_author_name
 'Allan Brown'
 'Stig Rundqvist'
_journal_name_full
;
Acta Crystallographica
;
_journal_volume 19
_journal_year 1965
_journal_page_first 684
_journal_page_last 685
_publ_Section_title
;
 Refinement of the crystal structure of black phosphorus
;

_aflow_proto 'A_oC8_64_f'
_aflow_params 'a,b/a,c/a,y1,z1'
_aflow_params_values '3.3136,3.16211974891,1.32070859488,0.10168,0.08056
      ↪ '
_aflow_Strukturbericht 'A17'
_aflow_Pearson 'oC8'

_symmetry_space_group_name_Hall "-C 2bc 2"
_symmetry_space_group_name_H-M "C m c a"
_symmetry_Int_Tables_number 64

_cell_length_a      3.31360
_cell_length_b      10.47800
_cell_length_c      4.37630
_cell_angle_alpha 90.00000
_cell_angle_beta  90.00000
_cell_angle_gamma 90.00000

loop_
_space_group_symop_id
_space_group_symop_operation_xyz
1 x,y,z
2 x,-y,-z
3 -x+1/2,y,-z+1/2
4 -x+1/2,-y,z+1/2
5 -x,-y,-z
6 -x,y,z
7 x+1/2,-y,z+1/2
8 x+1/2,y,-z+1/2
9 x+1/2,y+1/2,z
10 x+1/2,-y+1/2,-z
11 -x,y+1/2,-z+1/2
12 -x,-y+1/2,z+1/2
13 -x+1/2,-y+1/2,-z
14 -x+1/2,y+1/2,z
15 x,-y+1/2,z+1/2
16 x,y+1/2,-z+1/2

loop_
_atom_site_label
_atom_site_type_symbol
_atom_site_symmetry_multiplicity
_atom_site_Wyckoff_label
_atom_site_fract_x
_atom_site_fract_y
_atom_site_fract_z
_atom_site_occupancy
P1 P    8 f 0.00000 0.10168 0.08056 1.00000
```

Black Phosphorus (A17): A_oC8_64_f - POSCAR

```
A_oC8_64_f & a,b/a,c/a,y1,z1 --params=3.3136,3.16211974891,1.32070859488
      ↪ ,0.10168,0.08056 & Cmca       D_{2h}^{18} #64 (f) & oC8 & A17 &
      ↪ P & Black Phosphorous & A. Brown and S. Rundqvist, Acta Cryst.
      ↪ 19, 684-685 (1965)
   1.0000000000000000
   1.65680000000000  -5.23900000000000   0.00000000000000
   1.65680000000000   5.23900000000000   0.00000000000000
   0.00000000000000   0.00000000000000   4.37630000000000
     P
     4
Direct
   0.10168000000000   0.89832000000000   0.91944000000000   P   (8f)
   0.39832000000000   0.60168000000000   0.41944000000000   P   (8f)
   0.60168000000000   0.39832000000000   0.58056000000000   P   (8f)
   0.89832000000000   0.10168000000000   0.08056000000000   P   (8f)
```

Molecular Iodine (I) (A14): A_oC8_64_f - CIF

```
# CIF file

data_findsym-output
_audit_creation_method FINDSYM

_chemical_name_mineral ''
_chemical_formula_sum 'I'

loop_
_publ_author_name
 'C. Petrillo'
 'O. Moze'
 'R. M. Ibberson'
_journal_name_full
;
Physica B
;
_journal_volume 180-181
_journal_year 1992
_journal_page_first 639
_journal_page_last 641
_publ_Section_title
;
 High resolution neutron powder diffraction investigation of the low
      ↪ temperature crystal structure of molecular iodine (I$_2$)
;

# Found in http://www.webelements.com/iodine/crystal_structure.html

_aflow_proto 'A_oC8_64_f'
_aflow_params 'a,b/a,c/a,y1,z1'
_aflow_params_values '7.11906,0.654575182679,1.37596817557,0.15485,
      ↪ 0.1175'
_aflow_Strukturbericht 'A14'
_aflow_Pearson 'oC8'

_symmetry_space_group_name_Hall "-C 2bc 2"
_symmetry_space_group_name_H-M "C m c a"
_symmetry_Int_Tables_number 64

_cell_length_a      7.11906
_cell_length_b      4.65996
_cell_length_c      9.79560
_cell_angle_alpha 90.00000
_cell_angle_beta  90.00000
_cell_angle_gamma 90.00000

loop_
_space_group_symop_id
_space_group_symop_operation_xyz
1 x,y,z
2 x,-y,-z
3 -x+1/2,y,-z+1/2
4 -x+1/2,-y,z+1/2
5 -x,-y,-z
6 -x,y,z
7 x+1/2,-y,z+1/2
8 x+1/2,y,-z+1/2
9 x+1/2,y+1/2,z
10 x+1/2,-y+1/2,-z
11 -x,y+1/2,-z+1/2
12 -x,-y+1/2,z+1/2
13 -x+1/2,-y+1/2,-z
14 -x+1/2,y+1/2,z
15 x,-y+1/2,z+1/2
16 x,y+1/2,-z+1/2

loop_
_atom_site_label
_atom_site_type_symbol
_atom_site_symmetry_multiplicity
_atom_site_Wyckoff_label
_atom_site_fract_x
_atom_site_fract_y
_atom_site_fract_z
_atom_site_occupancy
I1 I    8 f 0.00000 0.15485 0.11750 1.00000
```

Molecular Iodine (I) (A14): A_oC8_64_f - POSCAR

```
A_oC8_64_f & a,b/a,c/a,y1,z1 --params=7.11906,0.654575182679,
      ↪ 1.37596817557,0.15485,0.1175 & Cmca       D_{2h}^{18} #64 (f) &
      ↪ oC8 & A14 & I & Iodine & C. Petrillo, O. Moze and R. M.
      ↪ Ibberson, Physica B 180 & 181 639-641 (1992)
   1.0000000000000000
```

```
    3.55953000000000   -2.32998000000000    0.00000000000000
    3.55953000000000    2.32998000000000    0.00000000000000
    0.00000000000000    0.00000000000000    9.79560000000000
     I
     4
Direct
    0.15485000000000    0.84515000000000    0.88250000000000    I    (8f)
    0.34515000000000    0.65485000000000    0.38250000000000    I    (8f)
    0.65485000000000    0.34515000000000    0.61750000000000    I    (8f)
    0.84515000000000    0.15485000000000    0.11750000000000    I    (8f)
```

$\alpha$-IrV: AB_oC8_65_j_g - CIF

```
# CIF file

data_findsym-output
_audit_creation_method FINDSYM

_chemical_name_mineral 'alpha iridium vanadium'
_chemical_formula_sum 'Ir V'

loop_
_publ_author_name
 'B. C. Giessen'
 'N. J. Grant'
_journal_name_full
;
Acta Crytallographica
;
_journal_volume 18
_journal_year 1965
_journal_page_first 1080
_journal_page_last 1081
_publ_Section_title
;
 New intermediate phases in transition metal systems, III
;

# Found in  Pearson's Handbook, Vol. IV, pp. 4139

_aflow_proto 'AB_oC8_65_j_g'
_aflow_params 'a,b/a,c/a,x1,y2'
_aflow_params_values '5.971,1.1314687657,0.468263272484,0.28,0.22'
_aflow_Strukturbericht 'None'
_aflow_Pearson 'oC8'

_symmetry_space_group_name_Hall "-C 2 2 "
_symmetry_space_group_name_H-M "C m m m"
_symmetry_Int_Tables_number 65

_cell_length_a      5.97100
_cell_length_b      6.75600
_cell_length_c      2.79600
_cell_angle_alpha  90.00000
_cell_angle_beta   90.00000
_cell_angle_gamma  90.00000

loop_
_space_group_symop_id
_space_group_symop_operation_xyz
1 x,y,z
2 x,-y,-z
3 -x,y,-z
4 -x,-y,z
5 -x,-y,-z
6 -x,y,z
7 x,-y,z
8 x,y,-z
9 x+1/2,y+1/2,z
10 x+1/2,-y+1/2,-z
11 -x+1/2,y+1/2,-z
12 -x+1/2,-y+1/2,z
13 -x+1/2,-y+1/2,-z
14 -x+1/2,y+1/2,z
15 x+1/2,-y+1/2,z
16 x+1/2,y+1/2,-z

loop_
_atom_site_label
_atom_site_type_symbol
_atom_site_symmetry_multiplicity
_atom_site_Wyckoff_label
_atom_site_fract_x
_atom_site_fract_y
_atom_site_fract_z
_atom_site_occupancy
V1  V    4 g 0.28000 0.00000 0.00000 1.00000
Ir1 Ir   4 j 0.00000 0.22000 0.50000 1.00000
```

$\alpha$-IrV: AB_oC8_65_j_g - POSCAR

```
AB_oC8_65_j_g & a,b/a,c/a,x1,y2 --params=5.971,1.1314687657,
    ↪ 0.468263272484,0.28,0.22 & Cmmm      D_{2h}^{19} #65 (gj) & oC8
    ↪ & & IrV & alpha & B. C. Giessen and N. J. Grant, Acta Cryst.
    ↪ 18, 1080-1081 (1965)
    1.0000000000000000
    2.98550000000000   -3.37800000000000    0.00000000000000
    2.98550000000000    3.37800000000000    0.00000000000000
    0.00000000000000    0.00000000000000    2.79600000000000
    Ir    V
     2    2
Direct
    0.22000000000000    0.78000000000000    0.50000000000000    Ir    (4j)
    0.78000000000000    0.22000000000000    0.50000000000000    Ir    (4j)
    0.28000000000000    0.28000000000000    0.00000000000000    V     (4g)
```

```
    0.72000000000000    0.72000000000000    0.00000000000000    V     (4g)
```

$Ga_3Pt_5$: A3B5_oC16_65_ah_bej - CIF

```
# CIF file

data_findsym-output
_audit_creation_method FINDSYM

_chemical_name_mineral ''
_chemical_formula_sum 'Ga3 Pt5'

loop_
_publ_author_name
 'K. Schubert'
 'S. Bhan'
 'W. Burkhardt'
 'R. Gohle'
 'H. G. Meissner'
 'M. P\"{o}tzschke'
 'E. Stolz'
_journal_name_full
;
Naturwissenschaften
;
_journal_volume 47
_journal_year 1960
_journal_page_first 303
_journal_page_last 303
_publ_Section_title
;
 Einige strukturelle Ergebnisse an metallischen Phasen (5)
;

# Found in  Pearson's Handbook III, p. 3540

_aflow_proto 'A3B5_oC16_65_ah_bej'
_aflow_params 'a,b/a,c/a,x4,y5'
_aflow_params_values '8.031,0.926410160628,0.491595069107,0.25,0.225'
_aflow_Strukturbericht 'None'
_aflow_Pearson 'oC16'

_symmetry_space_group_name_Hall "-C 2 2 "
_symmetry_space_group_name_H-M "C m m m"
_symmetry_Int_Tables_number 65

_cell_length_a      8.03100
_cell_length_b      7.44000
_cell_length_c      3.94800
_cell_angle_alpha  90.00000
_cell_angle_beta   90.00000
_cell_angle_gamma  90.00000

loop_
_space_group_symop_id
_space_group_symop_operation_xyz
1 x,y,z
2 x,-y,-z
3 -x,y,-z
4 -x,-y,z
5 -x,-y,-z
6 -x,y,z
7 x,-y,z
8 x,y,-z
9 x+1/2,y+1/2,z
10 x+1/2,-y+1/2,-z
11 -x+1/2,y+1/2,-z
12 -x+1/2,-y+1/2,z
13 -x+1/2,-y+1/2,-z
14 -x+1/2,y+1/2,z
15 x+1/2,-y+1/2,z
16 x+1/2,y+1/2,-z

loop_
_atom_site_label
_atom_site_type_symbol
_atom_site_symmetry_multiplicity
_atom_site_Wyckoff_label
_atom_site_fract_x
_atom_site_fract_y
_atom_site_fract_z
_atom_site_occupancy
Ga1 Ga   2 a 0.00000 0.00000 0.00000 1.00000
Pt1 Pt   2 b 0.50000 0.00000 0.00000 1.00000
Pt2 Pt   4 e 0.25000 0.25000 0.00000 1.00000
Ga2 Ga   4 h 0.25000 0.00000 0.50000 1.00000
Pt3 Pt   4 j 0.00000 0.22500 0.50000 1.00000
```

$Ga_3Pt_5$: A3B5_oC16_65_ah_bej - POSCAR

```
A3B5_oC16_65_ah_bej & a,b/a,c/a,x4,y5 --params=8.031,0.926410160628,
    ↪ 0.491595069107,0.25,0.225 & Cmmm      D_{2h}^{19} #65 (abehj) &
    ↪ oC16 & & Ga3Pt5 & & K. Schubert, S. Bhan et al.,
    ↪ Naturwissenschaften 47, 303 (1960)
    1.0000000000000000
    4.01550000000000   -3.72000000000000    0.00000000000000
    4.01550000000000    3.72000000000000    0.00000000000000
    0.00000000000000    0.00000000000000    3.94800000000000
    Ga    Pt
     3     5
Direct
    0.00000000000000    0.00000000000000    0.00000000000000    Ga    (2a)
    0.25000000000000    0.25000000000000    0.50000000000000    Ga    (4h)
    0.75000000000000    0.75000000000000    0.50000000000000    Ga    (4h)
    0.50000000000000    0.50000000000000    0.00000000000000    Pt    (2b)
```

```
    0.00000000000000    0.50000000000000    0.00000000000000    Pt    (4e)
    0.50000000000000    0.00000000000000    0.00000000000000    Pt    (4e)
    0.22500000000000    0.77500000000000    0.50000000000000    Pt    (4j)
    0.77500000000000    0.22500000000000    0.50000000000000    Pt    (4j)
```

Predicted $CdPt_3$ (“$L1_3$”): AB3_oC8_65_a_bf - CIF

```
# CIF file

data_findsym-output
_audit_creation_method FINDSYM

_chemical_name_mineral ''
_chemical_formula_sum 'Cd Pt3'

loop_
_publ_author_name
 'Gus L. W. Hart'
_journal_name_full
;
Physical Review B
;
_journal_volume 80
_journal_year 2009
_journal_page_first 014106
_journal_page_last 014106
_publ_Section_title
;
 Verifying predictions of the L1$_3$ crystal structure in Cd-Pt and
   ↪ Pd-Pt by exhaustive enumeration
;

_aflow_proto 'AB3_oC8_65_a_bf'
_aflow_params 'a,b/a,c/a'
_aflow_params_values '5.82068,1.35259626023,0.493507631411'
_aflow_Strukturbericht 'L1_3'
_aflow_Pearson 'oC8'

_symmetry_space_group_name_Hall "-C 2 2 "
_symmetry_space_group_name_H-M "C m m m"
_symmetry_Int_Tables_number 65

_cell_length_a      5.82068
_cell_length_b      7.87303
_cell_length_c      2.87255
_cell_angle_alpha  90.00000
_cell_angle_beta   90.00000
_cell_angle_gamma  90.00000

loop_
_space_group_symop_id
_space_group_symop_operation_xyz
1 x,y,z
2 x,-y,-z
3 -x,y,-z
4 -x,-y,z
5 -x,-y,-z
6 -x,y,z
7 x,-y,z
8 x,y,-z
9 x+1/2,y+1/2,z
10 x+1/2,-y+1/2,-z
11 -x+1/2,y+1/2,-z
12 -x+1/2,-y+1/2,z
13 -x+1/2,-y+1/2,-z
14 -x+1/2,y+1/2,z
15 x+1/2,-y+1/2,z
16 x+1/2,y+1/2,-z

loop_
_atom_site_label
_atom_site_type_symbol
_atom_site_symmetry_multiplicity
_atom_site_Wyckoff_label
_atom_site_fract_x
_atom_site_fract_y
_atom_site_fract_z
_atom_site_occupancy
Cd1 Cd   2 a 0.00000 0.00000 0.00000 1.00000
Pt1 Pt   2 b 0.50000 0.00000 0.00000 1.00000
Pt2 Pt   4 f 0.25000 0.25000 0.50000 1.00000
```

Predicted $CdPt_3$ (“$L1_3$”): AB3_oC8_65_a_bf - POSCAR

```
AB3_oC8_65_a_bf & a,b/a,c/a --params=5.82068,1.35259626023,
   ↪ 0.493507631411 & Cmmm     D_{2h}^{19} #65 (abf) & oC8 & L1_3 &
   ↪ CdPt3 & Predicted & G.L.W. Hart, PRB 80, 014106 (2009)
   1.0000000000000000
    2.91034000000000   -3.93651500000000    0.00000000000000
    2.91034000000000    3.93651500000000    0.00000000000000
    0.00000000000000    0.00000000000000    2.87255000000000
   Cd    Pt
    1     3
Direct
    0.00000000000000    0.00000000000000    0.00000000000000    Cd    (2a)
    0.50000000000000    0.50000000000000    0.00000000000000    Pt    (2b)
    0.00000000000000    0.50000000000000    0.50000000000000    Pt    (4f)
    0.50000000000000    0.00000000000000    0.50000000000000    Pt    (4f)
```

TlF (B24): AB_oF8_69_a_b - CIF

```
# CIF file

data_findsym-output
_audit_creation_method FINDSYM

_chemical_name_mineral ''
_chemical_formula_sum 'Tl F'

loop_
_publ_author_name
 'J. A. A. Ketelaar'
_journal_name_full
;
Zeitschrift f\"{u}r Kristallographie - Crystalline Materials
;
_journal_volume 92
_journal_year 1935
_journal_page_first 30
_journal_page_last 38
_publ_Section_title
;
 Die Kristallstruktur des Thallofluorids
;

# Found in P. Berastegui and S. Hull, J. Solid State Chem. 150, 266-75 (
   ↪ 2000)

_aflow_proto 'AB_oF8_69_a_b'
_aflow_params 'a,b/a,c/a'
_aflow_params_values '6.08,0.903782894737,0.851973684211'
_aflow_Strukturbericht 'B24'
_aflow_Pearson 'oF8'

_symmetry_space_group_name_Hall "-F 2 2"
_symmetry_space_group_name_H-M "F m m m"
_symmetry_Int_Tables_number 69

_cell_length_a      6.08000
_cell_length_b      5.49500
_cell_length_c      5.18000
_cell_angle_alpha  90.00000
_cell_angle_beta   90.00000
_cell_angle_gamma  90.00000

loop_
_space_group_symop_id
_space_group_symop_operation_xyz
1 x,y,z
2 x,-y,-z
3 -x,y,-z
4 -x,-y,z
5 -x,-y,-z
6 -x,y,z
7 x,-y,z
8 x,y,-z
9 x,y+1/2,z+1/2
10 x,-y+1/2,-z+1/2
11 -x,y+1/2,-z+1/2
12 -x,-y+1/2,z+1/2
13 -x,-y+1/2,-z+1/2
14 -x,y+1/2,z+1/2
15 x,-y+1/2,z+1/2
16 x,y+1/2,-z+1/2
17 x+1/2,y,z+1/2
18 x+1/2,-y,-z+1/2
19 -x+1/2,y,-z+1/2
20 -x+1/2,-y,z+1/2
21 -x+1/2,-y,-z+1/2
22 -x+1/2,y,z+1/2
23 x+1/2,-y,z+1/2
24 x+1/2,y,-z+1/2
25 x+1/2,y+1/2,z
26 x+1/2,-y+1/2,-z
27 -x+1/2,y+1/2,-z
28 -x+1/2,-y+1/2,z
29 -x+1/2,-y+1/2,-z
30 -x+1/2,y+1/2,z
31 x+1/2,-y+1/2,z
32 x+1/2,y+1/2,-z

loop_
_atom_site_label
_atom_site_type_symbol
_atom_site_symmetry_multiplicity
_atom_site_Wyckoff_label
_atom_site_fract_x
_atom_site_fract_y
_atom_site_fract_z
_atom_site_occupancy
F1   F    4 a 0.00000 0.00000 0.00000 1.00000
Tl1  Tl   4 b 0.00000 0.00000 0.50000 1.00000
```

TlF (B24): AB_oF8_69_a_b - POSCAR

```
AB_oF8_69_a_b & a,b/a,c/a --params=6.08,0.903782894737,0.851973684211 &
   ↪ Fmmm     D_{2h}^{23} #69 (ab) & oF8 & B24 & TlF & & J. A. A.
   ↪ Ketellar, Zeitschrift f\"{u}r Kristallographie - Crystalline
   ↪ Materials 92, 30-38 (1935)
   1.0000000000000000
    0.00000000000000    2.74750000000000    2.59000000000000
    3.04000000000000    0.00000000000000    2.59000000000000
    3.04000000000000    2.74750000000000    0.00000000000000
    F     Tl
    1     1
Direct
    0.00000000000000    0.00000000000000    0.00000000000000    F    (4a)
    0.50000000000000    0.50000000000000    0.50000000000000    Tl   (4b)
```

$\gamma$-Pu: A_oF8_70_a - CIF

```
# CIF file

data_findsym-output
_audit_creation_method FINDSYM

_chemical_name_mineral 'gamma plutonium'
_chemical_formula_sum 'Pu'

loop_
_publ_author_name
 'W. H. Zachariasen'
 'F. H. Ellinger'
_journal_name_full
;
Acta Crystallographica
;
_journal_volume 8
_journal_year 1955
_journal_page_first 431
_journal_page_last 433
_publ_Section_title
;
 Crystal chemical studies of the 5f-series of elements. XXIV. The
    ↪ crystal structure and thermal expansion of $\gamma$-plutonium
;

_aflow_proto 'A_oF8_70_a'
_aflow_params 'a,b/a,c/a'
_aflow_params_values '3.1587,1.82613100326,3.21714629436'
_aflow_Strukturbericht 'None'
_aflow_Pearson 'oF8'

_symmetry_space_group_name_Hall "-F 2uv 2vw"
_symmetry_space_group_name_H-M "F d d d:2"
_symmetry_Int_Tables_number 70

_cell_length_a      3.15870
_cell_length_b      5.76820
_cell_length_c      10.16200
_cell_angle_alpha  90.00000
_cell_angle_beta   90.00000
_cell_angle_gamma  90.00000

loop_
_space_group_symop_id
_space_group_symop_operation_xyz
1 x,y,z
2 x,-y+3/4,-z+3/4
3 -x+3/4,y,-z+3/4
4 -x+3/4,-y+3/4,z
5 -x,-y,-z
6 -x,y+1/4,z+1/4
7 x+1/4,-y,z+1/4
8 x+1/4,y+1/4,-z
9 x,y+1/2,z+1/2
10 x,-y+1/4,-z+1/4
11 -x+3/4,y+1/2,-z+1/4
12 -x+3/4,-y+1/4,z+1/2
13 -x,-y+1/2,-z+1/2
14 -x,y+3/4,z+3/4
15 x+1/4,-y+1/2,z+3/4
16 x+1/4,y+3/4,-z+1/2
17 x+1/2,y,z+1/2
18 x+1/2,-y+3/4,-z+1/4
19 -x+1/4,y,-z+1/4
20 -x+1/4,-y+3/4,z+1/2
21 -x+1/2,-y,-z+1/2
22 -x+1/2,y+1/4,z+3/4
23 x+3/4,-y,z+3/4
24 x+3/4,y+1/4,-z+1/2
25 x+1/2,y+1/2,z
26 x+1/2,-y+1/4,-z+3/4
27 -x+1/4,y+1/2,-z+3/4
28 -x+1/4,-y+1/4,z
29 -x+1/2,-y+1/2,-z
30 -x+1/2,y+3/4,z+1/4
31 x+3/4,-y+1/2,z+1/4
32 x+3/4,y+3/4,-z

loop_
_atom_site_label
_atom_site_type_symbol
_atom_site_symmetry_multiplicity
_atom_site_Wyckoff_label
_atom_site_fract_x
_atom_site_fract_y
_atom_site_fract_z
_atom_site_occupancy
Pu1 Pu   8 a 0.12500  0.12500  0.12500  1.00000
```

$\gamma$-Pu: A_oF8_70_a - POSCAR

```
A_oF8_70_a & a,b/a,c/a --params=3.1587,1.82613100326,3.21714629436 &
    ↪ Fddd      D_{2h}^{24} #70 (a) & oF8 & & Pu & gamma & W. H.
    ↪ Zachariasen and F. H. Ellinger, Acta Cryst. B 8, 431-433 (1955)
   1.0000000000000000
   0.00000000000000   2.88410000000000   5.08100000000000
   1.57935000000000   0.00000000000000   5.08100000000000
   1.57935000000000   2.88410000000000   0.00000000000000
   Pu
    2
Direct
   0.12500000000000   0.12500000000000   0.12500000000000   Pu   (8a)
   0.87500000000000   0.87500000000000   0.87500000000000   Pu   (8a)
```

$TiSi_2$ (C54): A2B_oF24_70_e_a - CIF

```
# CIF file

data_findsym-output
_audit_creation_method FINDSYM

_chemical_name_mineral 'Titanium Disilicide'
_chemical_formula_sum 'Ti Si2'

loop_
_publ_author_name
 'W. Jeitschko'
_journal_name_full
;
Acta Crystallographica B
;
_journal_volume 33
_journal_year 1977
_journal_page_first 2347
_journal_page_last 2348
_publ_Section_title
;
 Refinement of the crystal structure of TiSi$_2$ and some comments on
    ↪ bonding in TiSi$_2$ and related compounds
;

_aflow_proto 'A2B_oF24_70_e_a'
_aflow_params 'a,b/a,c/a,x2'
_aflow_params_values '8.2671,0.580614725841,1.0342804611,0.4615'
_aflow_Strukturbericht 'C54'
_aflow_Pearson 'oF24'

_symmetry_space_group_name_Hall "-F 2uv 2vw"
_symmetry_space_group_name_H-M "F d d d:2"
_symmetry_Int_Tables_number 70

_cell_length_a      8.26710
_cell_length_b      4.80000
_cell_length_c      8.55050
_cell_angle_alpha  90.00000
_cell_angle_beta   90.00000
_cell_angle_gamma  90.00000

loop_
_space_group_symop_id
_space_group_symop_operation_xyz
1 x,y,z
2 x,-y+3/4,-z+3/4
3 -x+3/4,y,-z+3/4
4 -x+3/4,-y+3/4,z
5 -x,-y,-z
6 -x,y+1/4,z+1/4
7 x+1/4,-y,z+1/4
8 x+1/4,y+1/4,-z
9 x,y+1/2,z+1/2
10 x,-y+1/4,-z+1/4
11 -x+3/4,y+1/2,-z+1/4
12 -x+3/4,-y+1/4,z+1/2
13 -x,-y+1/2,-z+1/2
14 -x,y+3/4,z+3/4
15 x+1/4,-y+1/2,z+3/4
16 x+1/4,y+3/4,-z+1/2
17 x+1/2,y,z+1/2
18 x+1/2,-y+3/4,-z+1/4
19 -x+1/4,y,-z+1/4
20 -x+1/4,-y+3/4,z+1/2
21 -x+1/2,-y,-z+1/2
22 -x+1/2,y+1/4,z+3/4
23 x+3/4,-y,z+3/4
24 x+3/4,y+1/4,-z+1/2
25 x+1/2,y+1/2,z
26 x+1/2,-y+1/4,-z+3/4
27 -x+1/4,y+1/2,-z+3/4
28 -x+1/4,-y+1/4,z
29 -x+1/2,-y+1/2,-z
30 -x+1/2,y+3/4,z+1/4
31 x+3/4,-y+1/2,z+1/4
32 x+3/4,y+3/4,-z

loop_
_atom_site_label
_atom_site_type_symbol
_atom_site_symmetry_multiplicity
_atom_site_Wyckoff_label
_atom_site_fract_x
_atom_site_fract_y
_atom_site_fract_z
_atom_site_occupancy
Ti1 Ti   8 a 0.12500  0.12500  0.12500  1.00000
Si1 Si  16 e 0.46150  0.12500  0.12500  1.00000
```

$TiSi_2$ (C54): A2B_oF24_70_e_a - POSCAR

```
A2B_oF24_70_e_a & a,b/a,c/a,x2 --params=8.2671,0.580614725841,
    ↪ 1.0342804611,0.4615 & Fddd      D_{2h}^{24} #70 (ae) & oF24 &
    ↪ C54 & TiSi2 & & W. Jeitschko, Acta Cryst. B 33, 2347-2348 (1977
    ↪ )
   1.0000000000000000
   0.00000000000000   2.40000000000000   4.27525000000000
   4.13355000000000   0.00000000000000   4.27525000000000
   4.13355000000000   2.40000000000000   0.00000000000000
   Si   Ti
```

```
    4      2
Direct
    0.21150000000000    0.53850000000000    0.53850000000000    Si    (16e)
    0.46150000000000    0.78850000000000    0.78850000000000    Si    (16e)
    0.53850000000000    0.21150000000000    0.21150000000000    Si    (16e)
    0.78850000000000    0.46150000000000    0.46150000000000    Si    (16e)
    0.12500000000000    0.12500000000000    0.12500000000000    Ti     (8a)
    0.87500000000000    0.87500000000000    0.87500000000000    Ti     (8a)
```

α-S (A16): A_oF128_70_4h - CIF

```
# CIF file

data_findsym-output
_audit_creation_method FINDSYM

_chemical_name_mineral 'alpha'
_chemical_formula_sum 'S'

loop_
_publ_author_name
 'Steven J. Rettig'
 'James Trotter'
_journal_name_full
;
Acta Crystallographic C
;
_journal_volume 43
_journal_year 1987
_journal_page_first 2260
_journal_page_last 2262
_publ_Section_title
;
 Refinement of the structure of orthorhombic sulfur, $\alpha$-S$_8$
;

_aflow_proto 'A_oF128_70_4h'
_aflow_params 'a,b/a,c/a,x1,y1,z1,x2,y2,z2,x3,y3,z3,x4,y4,z4'
_aflow_params_values '10.4646,1.22947843205,2.33988876785,0.14415,
    ↪ 0.04732,0.0486,0.29277,0.2269,0.25406,0.21598,0.28022,0.32618,
    ↪ 0.21405,0.15761,0.37947'
_aflow_Strukturbericht 'A16'
_aflow_Pearson 'oF128'

_symmetry_space_group_name_Hall "-F 2uv 2vw"
_symmetry_space_group_name_H-M "F d d d:2"
_symmetry_Int_Tables_number 70

_cell_length_a      10.46460
_cell_length_b      12.86600
_cell_length_c      24.48600
_cell_angle_alpha  90.00000
_cell_angle_beta   90.00000
_cell_angle_gamma  90.00000

loop_
_space_group_symop_id
_space_group_symop_operation_xyz
1 x,y,z
2 x,-y+3/4,-z+3/4
3 -x+3/4,y,-z+3/4
4 -x+3/4,-y+3/4,z
5 -x,-y,-z
6 -x,y+1/4,z+1/4
7 x+1/4,-y,z+1/4
8 x+1/4,y+1/4,-z
9 x,y+1/2,z+1/2
10 x,-y+1/4,-z+1/4
11 -x+3/4,y+1/2,-z+1/4
12 -x+3/4,-y+1/4,z+1/2
13 -x,-y+1/2,-z+1/2
14 -x,y+3/4,z+3/4
15 x+1/4,-y+1/2,z+3/4
16 x+1/4,y+3/4,-z+1/2
17 x+1/2,y,z+1/2
18 x+1/2,-y+3/4,-z+1/4
19 -x+1/4,y,-z+1/4
20 -x+1/4,-y+3/4,z+1/2
21 -x+1/2,-y,-z+1/2
22 -x+1/2,y+1/4,z+3/4
23 x+3/4,-y,z+3/4
24 x+3/4,y+1/4,-z+1/2
25 x+1/2,y+1/2,z
26 x+1/2,-y+1/4,-z+3/4
27 -x+1/4,y+1/2,-z+3/4
28 -x+1/4,-y+1/4,z
29 -x+1/2,-y+1/2,-z
30 -x+1/2,y+3/4,z+1/4
31 x+3/4,-y+1/2,z+1/4
32 x+3/4,y+3/4,-z

loop_
_atom_site_label
_atom_site_type_symbol
_atom_site_symmetry_multiplicity
_atom_site_Wyckoff_label
_atom_site_fract_x
_atom_site_fract_y
_atom_site_fract_z
_atom_site_occupancy
S1 S  32 h 0.14415 0.04732 0.04860 1.00000
S2 S  32 h 0.29277 0.22690 0.25406 1.00000
S3 S  32 h 0.21598 0.28022 0.32618 1.00000
S4 S  32 h 0.21405 0.15761 0.37947 1.00000
```

α-S (A16): A_oF128_70_4h - POSCAR

```
A_oF128_70_4h & a,b/a,c/a,x1,y1,z1,x2,y2,z2,x3,y3,z3,x4,y4,z4 --params=
    ↪ 10.4646,1.22947843205,2.33988876785,0.14415,0.04732,0.0486,
    ↪ 0.29277,0.2269,0.25406,0.21598,0.28022,0.32618,0.21405,0.15761,
    ↪ 0.37947 & Fddd      D_{2h}^{24} #70 (h^4) & oF128 & A16 & S & (
    ↪ alpha sulfur) & S. Rettig and J. Trotter, Acta Cryst. C 43,
    ↪ 2260-2262 (1987)
   1.0000000000000000
   0.00000000000000    6.43300000000000   12.24300000000000
   5.23230000000000    0.00000000000000   12.24300000000000
   5.23230000000000    6.43300000000000    0.00000000000000
     S
    32
Direct
   0.04823000000000    0.85457000000000    0.85713000000000    S    (32h)
   0.74007000000000    0.85713000000000    0.85457000000000    S    (32h)
   0.85713000000000    0.74007000000000    0.04823000000000    S    (32h)
   0.85457000000000    0.04823000000000    0.74007000000000    S    (32h)
  -0.04823000000000    0.14543000000000    0.14287000000000    S    (32h)
   0.25993000000000    0.14287000000000    0.14543000000000    S    (32h)
   0.14287000000000    0.25993000000000   -0.04823000000000    S    (32h)
   0.14543000000000   -0.04823000000000    0.25993000000000    S    (32h)
   0.27373000000000    0.73439000000000    0.68007000000000    S    (32h)
   0.81181000000000    0.68007000000000    0.73439000000000    S    (32h)
   0.68007000000000    0.81181000000000    0.27373000000000    S    (32h)
   0.73439000000000    0.27373000000000    0.81181000000000    S    (32h)
   0.72627000000000    0.26561000000000    0.31993000000000    S    (32h)
   0.18819000000000    0.31993000000000    0.26561000000000    S    (32h)
   0.31993000000000    0.18819000000000    0.72627000000000    S    (32h)
   0.26561000000000    0.72627000000000    0.18819000000000    S    (32h)
   0.32238000000000    0.82998000000000    0.73806000000000    S    (32h)
   0.60958000000000    0.73806000000000    0.82998000000000    S    (32h)
   0.73806000000000    0.60958000000000    0.32238000000000    S    (32h)
   0.82998000000000    0.32238000000000    0.60958000000000    S    (32h)
   0.67762000000000    0.17002000000000    0.26194000000000    S    (32h)
   0.39042000000000    0.26194000000000    0.17002000000000    S    (32h)
   0.26194000000000    0.39042000000000    0.67762000000000    S    (32h)
   0.17002000000000    0.67762000000000    0.39042000000000    S    (32h)
   0.25113000000000    0.00781000000000    0.56409000000000    S    (32h)
   0.67697000000000    0.56409000000000    0.00781000000000    S    (32h)
   0.56409000000000    0.67697000000000    0.25113000000000    S    (32h)
   0.00781000000000    0.25113000000000    0.67697000000000    S    (32h)
   0.74887000000000   -0.00781000000000    0.43591000000000    S    (32h)
   0.32303000000000    0.43591000000000   -0.00781000000000    S    (32h)
   0.43591000000000    0.32303000000000    0.74887000000000    S    (32h)
  -0.00781000000000    0.74887000000000    0.32303000000000    S    (32h)
```

$ReSi_2$: AB2_oI6_71_a_i - CIF

```
# CIF file

data_findsym-output
_audit_creation_method FINDSYM

_chemical_name_mineral ''
_chemical_formula_sum 'Re Si2'

loop_
_publ_author_name
 'T. Siegrist'
 'F. Hulliger'
 'G. Travaglini'
_journal_name_full
;
Journal of the Less Common Metals
;
_journal_volume 92
_journal_year 1983
_journal_page_first 119
_journal_page_last 129
_publ_Section_title
;
 The crystal structure and some properties of ReSi$_2$
;

_aflow_proto 'AB2_oI6_71_a_i'
_aflow_params 'a,b/a,c/a,z2'
_aflow_params_values '3.144,0.994910941476,2.44179389313,0.339'
_aflow_Strukturbericht 'None'
_aflow_Pearson 'oI6'

_symmetry_space_group_name_Hall "-I 2 2"
_symmetry_space_group_name_H-M "I m m m"
_symmetry_Int_Tables_number 71

_cell_length_a      3.14400
_cell_length_b      3.12800
_cell_length_c      7.67700
_cell_angle_alpha  90.00000
_cell_angle_beta   90.00000
_cell_angle_gamma  90.00000

loop_
_space_group_symop_id
_space_group_symop_operation_xyz
1 x,y,z
2 x,-y,-z
3 -x,y,-z
4 -x,-y,z
5 -x,-y,-z
6 -x,y,z
7 x,-y,z
8 x,y,-z
9 x+1/2,y+1/2,z+1/2
```

```
10 x+1/2,-y+1/2,-z+1/2
11 -x+1/2,y+1/2,-z+1/2
12 -x+1/2,-y+1/2,z+1/2
13 -x+1/2,-y+1/2,-z+1/2
14 -x+1/2,y+1/2,z+1/2
15 x+1/2,-y+1/2,z+1/2
16 x+1/2,y+1/2,-z+1/2

loop_
_atom_site_label
_atom_site_type_symbol
_atom_site_symmetry_multiplicity
_atom_site_Wyckoff_label
_atom_site_fract_x
_atom_site_fract_y
_atom_site_fract_z
_atom_site_occupancy
Re1 Re   2 a 0.00000 0.00000 0.00000 1.00000
Si1 Si   4 i 0.00000 0.00000 0.33900 1.00000
```

$ReSi_2$: AB2_oI6_71_a_i - POSCAR

```
AB2_oI6_71_a_i & a,b/a,c/a,z2 --params=3.144,0.994910941476,
    ↪ 2.44179389313,0.339 & Immm        D_{2h}^{25} #71 (ai) & oI6 & &
    ↪ ReSi2 & & T. Siegrist, F. Hulliger and G. Travaglini, J.
    ↪ Less-Common Metals 92, 119-129 (1983)
   1.0000000000000000
  -1.56400000000000    1.57200000000000    3.83850000000000
   1.56400000000000   -1.57200000000000    3.83850000000000
   1.56400000000000    1.57200000000000   -3.83850000000000
   Re    Si
    1     2
Direct
   0.00000000000000    0.00000000000000    0.00000000000000    Re    (2a)
   0.33900000000000    0.33900000000000    0.00000000000000    Si    (4i)
   0.66100000000000    0.66100000000000    0.00000000000000    Si    (4i)
```

$MoPt_2$: AB2_oI6_71_a_g - CIF

```
# CIF file

data_findsym-output
_audit_creation_method FINDSYM

_chemical_name_mineral ''
_chemical_formula_sum 'Mo Pt2'

loop_
_publ_author_name
 'K. Schubert'
 'W. Burkhardt'
 'P. Esslinger'
 'E. G\"{u}nzel'
 'H. G. Meissner'
 'W. Schütt'
 'J. Wegst'
 'M. Wilkens'
_journal_name_full
;
Naturwissenschaften
;
_journal_volume 43
_journal_year 1956
_journal_page_first 248
_journal_page_last 249
_publ_Section_title
;
 Einige strukturelle Ergebnisse an metallischen Phasen
;

# Found in http://materials.springer.com/isp/crystallographic/docs/
    ↪ sd_1250591

_aflow_proto 'AB2_oI6_71_a_g'
_aflow_params 'a,b/a,c/a,y2'
_aflow_params_values '2.75984,2.9999963766,1.4241115427,0.35333'
_aflow_Strukturbericht 'None'
_aflow_Pearson 'oI6'

_symmetry_space_group_name_Hall "-I 2 2 "
_symmetry_space_group_name_H-M "I m m m"
_symmetry_Int_Tables_number 71

_cell_length_a      2.75984
_cell_length_b      8.27951
_cell_length_c      3.93032
_cell_angle_alpha  90.00000
_cell_angle_beta   90.00000
_cell_angle_gamma  90.00000

loop_
_space_group_symop_id
_space_group_symop_operation_xyz
1 x,y,z
2 x,-y,-z
3 -x,y,-z
4 -x,-y,z
5 -x,-y,-z
6 -x,y,z
7 x,-y,z
8 x,y,-z
9 x+1/2,y+1/2,z+1/2
10 x+1/2,-y+1/2,-z+1/2
11 -x+1/2,y+1/2,-z+1/2
12 -x+1/2,-y+1/2,z+1/2
13 -x+1/2,-y+1/2,-z+1/2
14 -x+1/2,y+1/2,z+1/2
15 x+1/2,-y+1/2,z+1/2
16 x+1/2,y+1/2,-z+1/2

loop_
_atom_site_label
_atom_site_type_symbol
_atom_site_symmetry_multiplicity
_atom_site_Wyckoff_label
_atom_site_fract_x
_atom_site_fract_y
_atom_site_fract_z
_atom_site_occupancy
Mo1 Mo   2 a 0.00000 0.00000 0.00000 1.00000
Pt1 Pt   4 g 0.00000 0.35333 0.00000 1.00000
```

$MoPt_2$: AB2_oI6_71_a_g - POSCAR

```
AB2_oI6_71_a_g & a,b/a,c/a,y2 --params=2.75984,2.9999963766,1.4241115427
    ↪ ,0.35333 & Immm       D_{2h}^{25} #71 (ag) & oI6 & & MoPt_2 & &
    ↪ K. Schubert et al., Naturwissenschaften 43, 248-249 (1956)
   1.0000000000000000
  -1.37991888000000    4.13975665000000    1.96516050000000
   1.37991888000000   -4.13975665000000    1.96516050000000
   1.37991888000000    4.13975665000000   -1.96516050000000
   Mo    Pt
    1     2
Direct
   0.00000000000000    0.00000000000000    0.00000000000000    Mo    (2a)
   0.35333333000000    0.00000000000000    0.35333333000000    Pt    (4g)
   0.64666667000000    0.00000000000000    0.64666667000000    Pt    (4g)
```

$SiS_2$: A2B_oI12_72_j_a - CIF

```
# CIF file

data_findsym-output
_audit_creation_method FINDSYM

_chemical_name_mineral 'Silicon Disuphide'
_chemical_formula_sum 'Si S2'

loop_
_publ_author_name
 'Johannes Peters'
 'Bernt Krebs'
_journal_name_full
;
Acta Crystallographic B
;
_journal_volume 38
_journal_year 1982
_journal_page_first 1270
_journal_page_last 1272
_publ_Section_title
;
 Silicon disulphide and silicon diselenide: a reinvestigation
;

_aflow_proto 'A2B_oI12_72_j_a'
_aflow_params 'a,b/a,c/a,x2,y2'
_aflow_params_values '9.583,0.585829072316,0.578837524783,0.1182,0.2088'
_aflow_Strukturbericht 'C42'
_aflow_Pearson 'oI12'

_symmetry_space_group_name_Hall "-I 2 2c"
_symmetry_space_group_name_H-M "I b a m"
_symmetry_Int_Tables_number 72

_cell_length_a      9.58300
_cell_length_b      5.61400
_cell_length_c      5.54700
_cell_angle_alpha  90.00000
_cell_angle_beta   90.00000
_cell_angle_gamma  90.00000

loop_
_space_group_symop_id
_space_group_symop_operation_xyz
1 x,y,z
2 x,-y,-z+1/2
3 -x,y,-z+1/2
4 -x,-y,z
5 -x,-y,-z
6 -x,y,z+1/2
7 x,-y,z+1/2
8 x,y,-z
9 x+1/2,y+1/2,z+1/2
10 x+1/2,-y+1/2,-z
11 -x+1/2,y+1/2,-z
12 -x+1/2,-y+1/2,z+1/2
13 -x+1/2,-y+1/2,-z+1/2
14 -x+1/2,y+1/2,z
15 x+1/2,-y+1/2,z
16 x+1/2,y+1/2,-z+1/2

loop_
_atom_site_label
_atom_site_type_symbol
_atom_site_symmetry_multiplicity
_atom_site_Wyckoff_label
_atom_site_fract_x
_atom_site_fract_y
_atom_site_fract_z
```

```
_atom_site_occupancy
Si1 Si    4 a 0.00000 0.00000 0.25000 1.00000
S1  S     8 j 0.11820 0.20880 0.00000 1.00000
```

$SiS_2$: A2B_oI12_72_j_a - POSCAR

```
A2B_oI12_72_j_a & a,b/a,c/a,x2,y2 --params=9.583,0.585829072316,
    ↪ 0.578837524783,0.1182,0.2088 & Ibam      D_{2h}^{26} #72 (aj) &
    ↪ oI12 & C42 & SiS2 & & J. Peters and B. Krebs, Acta Cryst. B 38,
    ↪ 1270-1272 (1982)
   1.0000000000000000
  -4.79150000000000   2.80700000000000   2.77350000000000
   4.79150000000000  -2.80700000000000   2.77350000000000
   4.79150000000000   2.80700000000000  -2.77350000000000
    S    Si
    4     2
Direct
   0.20880000000000   0.11820000000000   0.32700000000000    S    (8j)
   0.29120000000000   0.61820000000000  -0.09060000000000    S    (8j)
   0.70880000000000   0.38180000000000   0.09060000000000    S    (8j)
   0.79120000000000   0.88180000000000   0.67300000000000    S    (8j)
   0.25000000000000   0.25000000000000   0.00000000000000    Si   (4a)
   0.75000000000000   0.75000000000000   0.00000000000000    Si   (4a)
```

$BPO_4$ ($H0_7$): AB4C_tI12_82_c_g_a - CIF

```
# CIF file

data_findsym-output
_audit_creation_method FINDSYM

_chemical_name_mineral ''
_chemical_formula_sum 'B P O4'

loop_
_publ_author_name
 'M. Schmidt'
 'B. Ewald'
 'Yu. Prots'
 'R. Cardoso-Gil'
 'M. Armbrüster'
 'I. Loa'
 'L. Zhang'
 'Ya-Xi Huang'
 'U. Schwarz'
 'R. Kniep'
_journal_name_full
;
Zeitschrift f\"{u}r anorganische und allgemeine Chemie
;
_journal_volume 630
_journal_year 2004
_journal_page_first 655
_journal_page_last 662
_publ_Section_title
;
 Growth and Characterization of BPO$_4$ Single Crystals
;

_aflow_proto 'AB4C_tI12_82_c_g_a'
_aflow_params 'a,c/a,x3,y3,z3'
_aflow_params_values '4.3404,1.53216293429,0.256,0.2566,0.3722'
_aflow_Strukturbericht 'H0_7'
_aflow_Pearson 'tI12'

_symmetry_space_group_name_Hall "I -4"
_symmetry_space_group_name_H-M "I -4"
_symmetry_Int_Tables_number 82

_cell_length_a      4.34040
_cell_length_b      4.34040
_cell_length_c      6.65020
_cell_angle_alpha  90.00000
_cell_angle_beta   90.00000
_cell_angle_gamma  90.00000

loop_
_space_group_symop_id
_space_group_symop_operation_xyz
1 x,y,z
2 -x,-y,z
3 y,-x,-z
4 -y,x,-z
5 x+1/2,y+1/2,z+1/2
6 -x+1/2,-y+1/2,z+1/2
7 y+1/2,-x+1/2,-z+1/2
8 -y+1/2,x+1/2,-z+1/2

loop_
_atom_site_label
_atom_site_type_symbol
_atom_site_symmetry_multiplicity
_atom_site_Wyckoff_label
_atom_site_fract_x
_atom_site_fract_y
_atom_site_fract_z
_atom_site_occupancy
P1 P    2 a 0.00000 0.00000 0.00000 1.00000
B1 B    2 c 0.00000 0.50000 0.25000 1.00000
O1 O    8 g 0.25600 0.25660 0.37220 1.00000
```

$BPO_4$ ($H0_7$): AB4C_tI12_82_c_g_a - POSCAR

```
AB4C_tI12_82_c_g_a & a,c/a,x3,y3,z3 --params=4.3404,1.53216293429,0.256,
    ↪ 0.2566,0.3722 & I(-4)      S_4^2 #82 (acg) & tI12 & H0_7 & BPO4
    ↪ & & M. Schmidt et al., Z. Anorg. Allg. Chem. 630, 655-662 (2004
    ↪ )
   1.0000000000000000
  -2.17020000000000   2.17020000000000   3.32510000000000
   2.17020000000000  -2.17020000000000   3.32510000000000
   2.17020000000000   2.17020000000000  -3.32510000000000
    B     O     P
    1     4     1
Direct
   0.75000000000000   0.25000000000000   0.50000000000000    B    (2c)
   0.11560000000000   0.11620000000000   0.48740000000000    O    (8g)
   0.37180000000000   0.88440000000000   0.00060000000000    O    (8g)
   0.62880000000000   0.62820000000000   0.51260000000000    O    (8g)
   0.88380000000000   0.37120000000000  -0.00060000000000    O    (8g)
   0.00000000000000   0.00000000000000   0.00000000000000    P    (2a)
```

$CdAl_2S_4$ (E3): A2BC4_tI14_82_bc_a_g - CIF

```
# CIF file

data_findsym-output
_audit_creation_method FINDSYM

_chemical_name_mineral ''
_chemical_formula_sum 'Cd Al2 S4'

loop_
_publ_author_name
 'Harry Hahn'
 'G\"{u}nter Frank'
 'Wilhelm Klingler'
 'Anne-Dorothee St\"{o}rger'
 'Georg St\"{o}rger'
_journal_name_full
;
Zeitschrift f\"{u}r anorganische und allgemeine Chemie
;
_journal_volume 279
_journal_year 1955
_journal_page_first 241
_journal_page_last 270
_publ_Section_title
;
 Untersuchungen \"{u}ber tern\"{a}re Chalkogenide. VI. \"{U}ber tern\"{a
    ↪ }re Chalogenide des Aluminiums, Galliums und Indiums mit Zink,
    ↪ Cadmium und Quecksilber
;

# Found in http://materials.springer.com/isp/crystallographic/docs/
    ↪ sd_0301428

_aflow_proto 'A2BC4_tI14_82_bc_a_g'
_aflow_params 'a,c/a,x4,y4,z4'
_aflow_params_values '5.55,1.85585585586,0.26,0.25,0.13'
_aflow_Strukturbericht 'E3'
_aflow_Pearson 'tI14'

_symmetry_space_group_name_Hall "I -4"
_symmetry_space_group_name_H-M "I -4"
_symmetry_Int_Tables_number 82

_cell_length_a      5.55000
_cell_length_b      5.55000
_cell_length_c     10.30000
_cell_angle_alpha  90.00000
_cell_angle_beta   90.00000
_cell_angle_gamma  90.00000

loop_
_space_group_symop_id
_space_group_symop_operation_xyz
1 x,y,z
2 -x,-y,z
3 y,-x,-z
4 -y,x,-z
5 x+1/2,y+1/2,z+1/2
6 -x+1/2,-y+1/2,z+1/2
7 y+1/2,-x+1/2,-z+1/2
8 -y+1/2,x+1/2,-z+1/2

loop_
_atom_site_label
_atom_site_type_symbol
_atom_site_symmetry_multiplicity
_atom_site_Wyckoff_label
_atom_site_fract_x
_atom_site_fract_y
_atom_site_fract_z
_atom_site_occupancy
Cd1 Cd    2 a 0.00000 0.00000 0.00000 1.00000
Al1 Al    2 b 0.00000 0.00000 0.50000 1.00000
Al2 Al    2 c 0.00000 0.50000 0.25000 1.00000
S1  S     8 g 0.26000 0.25000 0.13000 1.00000
```

$CdAl_2S_4$ (E3): A2BC4_tI14_82_bc_a_g - POSCAR

```
A2BC4_tI14_82_bc_a_g & a,c/a,x4,y4,z4 --params=5.55,1.85585585586,0.26,
    ↪ 0.25,0.13 & I(-4)      S_4^2 #82 (abcg) & tI14 & E3 & CdAl2S4 &
    ↪ & H. Hahn et al., Z. anorg. allg. Chemie 279, 241-270 (1955)
   1.0000000000000000
  -2.77500000000000   2.77500000000000   5.15000000000000
   2.77500000000000  -2.77500000000000   5.15000000000000
   2.77500000000000   2.77500000000000  -5.15000000000000
```

```
   Al     Cd      S
    2      1      4
Direct
   0.50000000000000    0.50000000000000    0.00000000000000    Al     (2b)
   0.75000000000000    0.25000000000000    0.50000000000000    Al     (2c)
   0.00000000000000    0.00000000000000    0.00000000000000    Cd     (2a)
   0.13000000000000    0.62000000000000    0.01000000000000    S      (8g)
   0.38000000000000    0.39000000000000    0.51000000000000    S      (8g)
   0.61000000000000    0.12000000000000   -0.01000000000000    S      (8g)
   0.88000000000000    0.87000000000000    0.49000000000000    S      (8g)
```

PdS (B34): AB_tP16_84_cej_k - CIF

```
# CIF file

data_findsym-output
_audit_creation_method FINDSYM

_chemical_name_mineral ''
_chemical_formula_sum 'Pd S'

loop_
_publ_author_name
 'Nathaniel E. Brese'
 'Philip J. Squattrito'
 'James A. Ibers'
_journal_name_full
;
Acta Crystallographica C
;
_journal_volume 41
_journal_year 1985
_journal_page_first 1829
_journal_page_last 1830
_publ_Section_title
;
 Reinvestigation of the structure of PdS
;

_aflow_proto 'AB_tP16_84_cej_k'
_aflow_params 'a,c/a,x3,y3,x4,y4,z4'
_aflow_params_values '6.429,1.02830922383,0.46779,0.25713,0.19361,
    ↪ 0.30754,0.22904'
_aflow_Strukturbericht 'B34'
_aflow_Pearson 'tP16'

_symmetry_space_group_name_Hall "-P 4c"
_symmetry_space_group_name_H-M "P 42/m"
_symmetry_Int_Tables_number 84

_cell_length_a      6.42900
_cell_length_b      6.42900
_cell_length_c      6.61100
_cell_angle_alpha  90.00000
_cell_angle_beta   90.00000
_cell_angle_gamma  90.00000

loop_
_space_group_symop_id
_space_group_symop_operation_xyz
1 x,y,z
2 -x,-y,z
3 -y,x,z+1/2
4 y,-x,z+1/2
5 -x,-y,-z
6 x,y,-z
7 y,-x,-z+1/2
8 -y,x,-z+1/2

loop_
_atom_site_label
_atom_site_type_symbol
_atom_site_symmetry_multiplicity
_atom_site_Wyckoff_label
_atom_site_fract_x
_atom_site_fract_y
_atom_site_fract_z
_atom_site_occupancy
Pd1 Pd    2 c 0.00000 0.50000 0.00000 1.00000
Pd2 Pd    2 e 0.00000 0.00000 0.25000 1.00000
Pd3 Pd    4 j 0.46779 0.25713 0.00000 1.00000
S1   S    8 k 0.19361 0.30754 0.22904 1.00000
```

PdS (B34): AB_tP16_84_cej_k - POSCAR

```
AB_tP16_84_cej_k & a,c/a,x3,y3,x4,y4,z4 --params=6.429,1.02830922383,
    ↪ 0.46779,0.25713,0.19361,0.30754,0.22904 & P4_2/m     C_{4h}^2 #
    ↪ 84 (cejk) & tP16 & B34 & PdS & & N. E. Brese, P. J. Squarttrito
    ↪  and J. A. Ibers, Acta Cryst. C 41, 1829-1830 (1985)
   1.0000000000000000
   6.42900000000000    0.00000000000000    0.00000000000000
   0.00000000000000    6.42900000000000    0.00000000000000
   0.00000000000000    0.00000000000000    6.61100000000000
   Pd      S
    8      8
Direct
   0.00000000000000    0.50000000000000    0.00000000000000    Pd     (2c)
   0.50000000000000    0.00000000000000    0.50000000000000    Pd     (2c)
   0.00000000000000    0.00000000000000    0.25000000000000    Pd     (2e)
   0.00000000000000    0.00000000000000    0.75000000000000    Pd     (2e)
   0.25713200000000    0.53220800000000    0.50000000000000    Pd     (4j)
   0.46779200000000    0.25713200000000    0.00000000000000    Pd     (4j)
   0.53220800000000    0.74286800000000    0.00000000000000    Pd     (4j)
   0.74286800000000    0.46779200000000    0.50000000000000    Pd     (4j)
   0.19361000000000    0.30754000000000    0.22904000000000    S      (8k)
   0.19361000000000    0.30754000000000    0.77096000000000    S      (8k)
   0.30754000000000    0.80639000000000    0.27096000000000    S      (8k)
   0.30754000000000    0.80639000000000    0.72904000000000    S      (8k)
   0.69246000000000    0.19361000000000    0.27096000000000    S      (8k)
   0.69246000000000    0.19361000000000    0.72904000000000    S      (8k)
   0.80639000000000    0.69246000000000    0.22904000000000    S      (8k)
   0.80639000000000    0.69246000000000    0.77096000000000    S      (8k)
```

$Ti_5Te_4$: A4B5_tI18_87_h_ah - CIF

```
# CIF file

data_findsym-output
_audit_creation_method FINDSYM

_chemical_name_mineral 'Titanium telluride'
_chemical_formula_sum 'Ti5 Te4'

loop_
_publ_author_name
 'F. Gr\onvold'
 'A. Kjekshus'
 'F. Raaum'
_journal_name_full
;
Acta Crystallographica
;
_journal_volume 14
_journal_year 1961
_journal_page_first 930
_journal_page_last 934
_publ_Section_title
;
 The crystal structure of Ti$_5$Te$_4$
;

# Found in  Pearson's Handbook, Vol. IV, pp. 5321

_aflow_proto 'A4B5_tI18_87_h_ah'
_aflow_params 'a,c/a,x2,y2,x3,y3'
_aflow_params_values '10.164,0.37111373475,0.2797,-0.0589,0.3752,0.6856'
_aflow_Strukturbericht 'None'
_aflow_Pearson 'tI18'

_symmetry_space_group_name_Hall "-I 4"
_symmetry_space_group_name_H-M "I 4/m"
_symmetry_Int_Tables_number 87

_cell_length_a     10.16400
_cell_length_b     10.16400
_cell_length_c      3.77200
_cell_angle_alpha  90.00000
_cell_angle_beta   90.00000
_cell_angle_gamma  90.00000

loop_
_space_group_symop_id
_space_group_symop_operation_xyz
1 x,y,z
2 -x,-y,z
3 -y,x,z
4 y,-x,z
5 -x,-y,-z
6 x,y,-z
7 y,-x,-z
8 -y,x,-z
9 x+1/2,y+1/2,z+1/2
10 -x+1/2,-y+1/2,z+1/2
11 -y+1/2,x+1/2,z+1/2
12 y+1/2,-x+1/2,z+1/2
13 -x+1/2,-y+1/2,-z+1/2
14 x+1/2,y+1/2,-z+1/2
15 y+1/2,-x+1/2,-z+1/2
16 -y+1/2,x+1/2,-z+1/2

loop_
_atom_site_label
_atom_site_type_symbol
_atom_site_symmetry_multiplicity
_atom_site_Wyckoff_label
_atom_site_fract_x
_atom_site_fract_y
_atom_site_fract_z
_atom_site_occupancy
Ti1  Ti    2 a 0.00000 0.00000  0.00000 1.00000
Te1  Te    8 h 0.27970 -0.05890 0.00000 1.00000
Ti2  Ti    8 h 0.37520 0.68560  0.00000 1.00000
```

$Ti_5Te_4$: A4B5_tI18_87_h_ah - POSCAR

```
A4B5_tI18_87_h_ah & a,c/a,x2,y2,x3,y3 --params=10.164,0.37111373475,
    ↪ 0.2797,-0.0589,0.3752,0.6856 & I4/m      C_{2h}^5 #87 (ah^2) &
    ↪ tI18 & & Ti5Te4 & & F. Gronvold, A. Kjekshus and F. Raaum, Acta
    ↪  Cryst. 14, 930-934 (1961)
   1.0000000000000000
  -5.08200000000000    5.08200000000000    1.88600000000000
   5.08200000000000   -5.08200000000000    1.88600000000000
   5.08200000000000    5.08200000000000   -1.88600000000000
   Te     Ti
    4      5
Direct
  -0.05890000000000    0.27970000000000    0.22080000000000    Te     (8h)
   0.05890000000000    0.72030000000000    0.77920000000000    Te     (8h)
   0.27970000000000    0.05890000000000    0.33860000000000    Te     (8h)
   0.72030000000000   -0.05890000000000    0.66140000000000    Te     (8h)
```

```
0.00000000000000    0.00000000000000    0.00000000000000    Ti    (2a)
0.31440000000000    0.62480000000000   -0.06080000000000    Ti    (8h)
0.37520000000000    0.31440000000000    0.68960000000000    Ti    (8h)
0.62480000000000    0.68560000000000    0.31040000000000    Ti    (8h)
0.68560000000000    0.37520000000000    0.06080000000000    Ti    (8h)
```

$Ni_4Mo$ ($D1_a$): AB4_tI10_87_a_h - CIF

```
# CIF file

data_findsym-output
_audit_creation_method FINDSYM

_chemical_name_mineral ''
_chemical_formula_sum 'Ni4 Mo'

loop_
_publ_author_name
 'David Harker'
_journal_name_full
;
Journal of Chemical Physics
;
_journal_volume 12
_journal_year 1944
_journal_page_first 315
_journal_page_last 315
_publ_Section_title
;
 The Crystal Structure of Ni$_4$Mo
;

_aflow_proto 'AB4_tI10_87_a_h'
_aflow_params 'a,c/a,x2,y2'
_aflow_params_values '5.72,0.623076923077,0.4,0.8'
_aflow_Strukturbericht 'D1_a'
_aflow_Pearson 'tI10'

_symmetry_space_group_name_Hall "-I 4"
_symmetry_space_group_name_H-M "I 4/m"
_symmetry_Int_Tables_number 87

_cell_length_a      5.72000
_cell_length_b      5.72000
_cell_length_c      3.56400
_cell_angle_alpha  90.00000
_cell_angle_beta   90.00000
_cell_angle_gamma  90.00000

loop_
_space_group_symop_id
_space_group_symop_operation_xyz
1 x,y,z
2 -x,-y,z
3 -y,x,z
4 y,-x,z
5 -x,-y,-z
6 x,y,-z
7 y,-x,-z
8 -y,x,-z
9 x+1/2,y+1/2,z+1/2
10 -x+1/2,-y+1/2,z+1/2
11 -y+1/2,x+1/2,z+1/2
12 y+1/2,-x+1/2,z+1/2
13 -x+1/2,-y+1/2,-z+1/2
14 x+1/2,y+1/2,-z+1/2
15 y+1/2,-x+1/2,-z+1/2
16 -y+1/2,x+1/2,-z+1/2

loop_
_atom_site_label
_atom_site_type_symbol
_atom_site_symmetry_multiplicity
_atom_site_Wyckoff_label
_atom_site_fract_x
_atom_site_fract_y
_atom_site_fract_z
_atom_site_occupancy
Mo1 Mo   2 a 0.00000 0.00000 0.00000 1.00000
Ni1 Ni   8 h 0.40000 0.80000 0.00000 1.00000
```

$Ni_4Mo$ ($D1_a$): AB4_tI10_87_a_h - POSCAR

```
AB4_tI10_87_a_h & a,c/a,x2,y2 --params=5.72,0.623076923077,0.4,0.8 & I4/
    ↪ m      C_{2h}^5 #87 (ah) & tI10 & D1_a & Ni4Mo & & D. Harker,
    ↪ J. Chem. Phys. 12, 315 (1944)
   1.0000000000000000
  -2.86000000000000    2.86000000000000    1.78200000000000
   2.86000000000000   -2.86000000000000    1.78200000000000
   2.86000000000000    2.86000000000000   -1.78200000000000
   Mo    Ni
    1     4
Direct
   0.00000000000000    0.00000000000000    0.00000000000000    Mo    (2a)
   0.20000000000000    0.60000000000000    0.80000000000000    Ni    (8h)
   0.40000000000000    0.20000000000000    0.60000000000000    Ni    (8h)
   0.60000000000000    0.80000000000000    0.40000000000000    Ni    (8h)
   0.80000000000000    0.40000000000000    0.20000000000000    Ni    (8h)
```

$\alpha$-Cristobalite ($SiO_2$, low): A2B_tP12_92_b_a - CIF

```
# CIF file

data_findsym-output
_audit_creation_method FINDSYM

_chemical_name_mineral 'low (alpha) Cristobalite'
_chemical_formula_sum 'Si O2'

loop_
_publ_author_name
 'J. J. Pluth'
 'J. V. Smith'
 'J. Faber, Jr.'
_journal_name_full
;
Journal of Applied Physics
;
_journal_volume 57
_journal_year 1985
_journal_page_first 1045
_journal_page_last 1049
_publ_Section_title
;
 Crystal structure of low cristobalite at 10, 293, and 473 K: Variation
     ↪ of framework geometry with temperature
;

# Found in Pearson's Handbook, Vol. IV, pp. 4759

_aflow_proto 'A2B_tP12_92_b_a'
_aflow_params 'a,c/a,x1,x2,y2,z2'
_aflow_params_values '4.957,1.39001412144,0.3047,0.2381,0.1109,0.1826'
_aflow_Strukturbericht 'None'
_aflow_Pearson 'tP12'

_symmetry_space_group_name_Hall "P 4abw 2nw"
_symmetry_space_group_name_H-M "P 41 21 2"
_symmetry_Int_Tables_number 92

_cell_length_a      4.95700
_cell_length_b      4.95700
_cell_length_c      6.89030
_cell_angle_alpha  90.00000
_cell_angle_beta   90.00000
_cell_angle_gamma  90.00000

loop_
_space_group_symop_id
_space_group_symop_operation_xyz
1 x,y,z
2 x+1/2,-y+1/2,-z+3/4
3 -x+1/2,y+1/2,-z+1/4
4 -x,-y,z+1/2
5 -y,-x,-z+1/2
6 -y+1/2,x+1/2,z+1/4
7 y+1/2,-x+1/2,z+3/4
8 y,x,-z

loop_
_atom_site_label
_atom_site_type_symbol
_atom_site_symmetry_multiplicity
_atom_site_Wyckoff_label
_atom_site_fract_x
_atom_site_fract_y
_atom_site_fract_z
_atom_site_occupancy
Si1 Si   4 a 0.30470 0.30470 0.00000 1.00000
O1  O    8 b 0.23810 0.11090 0.18260 1.00000
```

$\alpha$-Cristobalite ($SiO_2$, low): A2B_tP12_92_b_a - POSCAR

```
A2B_tP12_92_b_a & a,c/a,x1,x2,y2,z2 --params=4.957,1.39001412144,0.3047,
    ↪ 0.2381,0.1109,0.1826 & P4_12_12  D_4^4 #92 (ab) & tP12 & & SiO2
    ↪ & alpha (low) Cristobalite & J.J. Pluth, J.V. Smith, and J.
    ↪ Faber, Jr., J. App. Phys. 57, 1045-1049 (1985)
   1.0000000000000000
   4.95700000000000    0.00000000000000    0.00000000000000
   0.00000000000000    4.95700000000000    0.00000000000000
   0.00000000000000    0.00000000000000    6.89030000000000
   O     Si
   8     4
Direct
   0.11090000000000    0.23810000000000    0.81740000000000    O     (8b)
   0.23810000000000    0.11090000000000    0.18260000000000    O     (8b)
   0.26190000000000    0.61090000000000    0.06740000000000    O     (8b)
   0.38910000000000    0.73810000000000    0.43260000000000    O     (8b)
   0.61090000000000    0.26190000000000    0.93260000000000    O     (8b)
   0.73810000000000    0.38910000000000    0.56740000000000    O     (8b)
   0.76190000000000    0.88910000000000    0.68260000000000    O     (8b)
   0.88910000000000    0.76190000000000    0.31740000000000    O     (8b)
   0.19530000000000    0.80470000000000    0.25000000000000    Si    (4a)
   0.30470000000000    0.30470000000000    0.00000000000000    Si    (4a)
   0.69530000000000    0.69530000000000    0.50000000000000    Si    (4a)
   0.80470000000000    0.19530000000000    0.75000000000000    Si    (4a)
```

Keatite ($SiO_2$): A2B_tP36_96_3b_ab - CIF

```
# CIF file

data_findsym-output
_audit_creation_method FINDSYM

_chemical_name_mineral 'Keatite'
_chemical_formula_sum 'Si O2'

loop_
_publ_author_name
```

```
 'Joseph Shropshire'
 'Paul P. Keat'
 'Philip A. Vaughan'
_journal_name_full
;
Zeitschrift f\"{u}r Kristallographie
;
_journal_volume 112
_journal_year 1959
_journal_page_first 409
_journal_page_last 413
_publ_Section_title
;
 The crystal structure of keatite, a new form of silica
;

# Found in demuth99:keatite

_aflow_proto 'A2B_tP36_96_3b_ab'
_aflow_params 'a,c/a,x1,x2,y2,z2,x3,y3,z3,x4,y4,z4,x5,y5,z5'
_aflow_params_values '7.464,1.15487674169,0.41,0.445,0.132,0.4,0.117,
    ↪ 0.123,0.296,0.344,0.297,0.143,0.326,0.12,0.248'
_aflow_Strukturbericht 'None'
_aflow_Pearson 'tP36'

_symmetry_space_group_name_Hall "P 4nw 2abw"
_symmetry_space_group_name_H-M "P 43 21 2"
_symmetry_Int_Tables_number 96

_cell_length_a     7.46400
_cell_length_b     7.46400
_cell_length_c     8.62000
_cell_angle_alpha 90.00000
_cell_angle_beta  90.00000
_cell_angle_gamma 90.00000

loop_
_space_group_symop_id
_space_group_symop_operation_xyz
1 x,y,z
2 x+1/2,-y+1/2,-z+1/4
3 -x+1/2,y+1/2,-z+3/4
4 -x,-y,z+1/2
5 -y,-x,-z+1/2
6 -y+1/2,x+1/2,z+3/4
7 y+1/2,-x+1/2,z+1/4
8 y,x,-z

loop_
_atom_site_label
_atom_site_type_symbol
_atom_site_symmetry_multiplicity
_atom_site_Wyckoff_label
_atom_site_fract_x
_atom_site_fract_y
_atom_site_fract_z
_atom_site_occupancy
Si1 Si   4 a 0.41000 0.41000 0.00000 1.00000
O1  O    8 b 0.44500 0.13200 0.40000 1.00000
O2  O    8 b 0.11700 0.12300 0.29600 1.00000
O3  O    8 b 0.34400 0.29700 0.14300 1.00000
Si2 Si   8 b 0.32600 0.12000 0.24800 1.00000
```

Keatite ($SiO_2$): A2B_tP36_96_3b_ab - POSCAR

```
A2B_tP36_96_3b_ab & a,c/a,x1,x2,y2,z2,x3,y3,z3,x4,y4,z4,x5,y5,z5 --
    ↪ params=7.464,1.15487674169,0.41,0.445,0.132,0.4,0.117,0.123,
    ↪ 0.296,0.344,0.297,0.143,0.326,0.12,0.248 & P4_32_12 D_4^8 #96
    ↪ (ab^4) & tP36 & & SiO2 & Keatite & Shropshire, et. al.
    ↪ Zeitschrift f\"{u}r Kristallographie 112, 409-413 (1959)
   1.00000000000000000
   7.46400000000000   0.00000000000000   0.00000000000000
   0.00000000000000   7.46400000000000   0.00000000000000
   0.00000000000000   0.00000000000000   8.62000000000000
    O    Si
   24    12
Direct
   0.05500000000000   0.63200000000000   0.35000000000000   O   (8b)
   0.13200000000000   0.44500000000000   0.60000000000000   O   (8b)
   0.36800000000000   0.94500000000000   0.15000000000000   O   (8b)
   0.44500000000000   0.13200000000000   0.40000000000000   O   (8b)
   0.55500000000000   0.86800000000000   0.90000000000000   O   (8b)
   0.63200000000000   0.05500000000000   0.65000000000000   O   (8b)
   0.86800000000000   0.55500000000000   0.10000000000000   O   (8b)
   0.94500000000000   0.36800000000000   0.85000000000000   O   (8b)
   0.11700000000000   0.12300000000000   0.29600000000000   O   (8b)
   0.12300000000000   0.11700000000000   0.70400000000000   O   (8b)
   0.37700000000000   0.61700000000000   0.04600000000000   O   (8b)
   0.38300000000000   0.62300000000000   0.45400000000000   O   (8b)
   0.61700000000000   0.37700000000000  -0.04600000000000   O   (8b)
   0.62300000000000   0.38300000000000   0.54600000000000   O   (8b)
   0.87700000000000   0.88300000000000   0.20400000000000   O   (8b)
   0.88300000000000   0.87700000000000   0.79600000000000   O   (8b)
   0.15600000000000   0.79700000000000   0.60700000000000   O   (8b)
   0.20300000000000   0.84400000000000   0.89300000000000   O   (8b)
   0.29700000000000   0.34400000000000   0.85700000000000   O   (8b)
   0.34400000000000   0.29700000000000   0.14300000000000   O   (8b)
   0.65600000000000   0.70300000000000   0.64300000000000   O   (8b)
   0.70300000000000   0.65600000000000   0.35700000000000   O   (8b)
   0.79700000000000   0.15600000000000   0.39300000000000   O   (8b)
   0.84400000000000   0.20300000000000   0.10700000000000   O   (8b)
   0.09000000000000   0.91000000000000   0.75000000000000   Si  (4a)
   0.41000000000000   0.41000000000000   0.00000000000000   Si  (4a)
   0.59000000000000   0.59000000000000   0.50000000000000   Si  (4a)
   0.91000000000000   0.09000000000000   0.25000000000000   Si  (4a)
   0.12000000000000   0.32600000000000   0.75200000000000   Si  (8b)
   0.17400000000000   0.62000000000000   0.50200000000000   Si  (8b)
   0.32600000000000   0.12000000000000   0.24800000000000   Si  (8b)
   0.38000000000000   0.82600000000000   0.99800000000000   Si  (8b)
   0.62000000000000   0.17400000000000   0.49800000000000   Si  (8b)
   0.67400000000000   0.88000000000000   0.74800000000000   Si  (8b)
   0.82600000000000   0.38000000000000   0.00200000000000   Si  (8b)
   0.88000000000000   0.67400000000000   0.25200000000000   Si  (8b)
```

"ST12" of Si: A_tP12_96_ab - CIF

```
# CIF file

data_findsym-output
_audit_creation_method FINDSYM

_chemical_name_mineral ''
_chemical_formula_sum 'Si'

loop_
_publ_author_name
 'J. Crain'
 'S. J. Clark'
 'G. J. Ackland'
 'M. C. Payne'
 'V. Milman'
 'P. D. Hatton'
 'B. J. Reid'
_journal_name_full
;
Physical Review B
;
_journal_volume 49
_journal_year 1994
_journal_page_first 5329
_journal_page_last 5340
_publ_Section_title
;
 Theoretical study of high-density phases of covalent semiconductors. I.
    ↪ {\em Ab initio treatment}
;

_aflow_proto 'A_tP12_96_ab'
_aflow_params 'a,c/a,x1,x2,y2,z2'
_aflow_params_values '5.51889,1.25999974633,0.0849,0.1752,0.3792,0.2742'
_aflow_Strukturbericht 'None'
_aflow_Pearson 'tP12'

_symmetry_space_group_name_Hall "P 4nw 2abw"
_symmetry_space_group_name_H-M "P 43 21 2"
_symmetry_Int_Tables_number 96

_cell_length_a     5.51889
_cell_length_b     5.51889
_cell_length_c     6.95380
_cell_angle_alpha 90.00000
_cell_angle_beta  90.00000
_cell_angle_gamma 90.00000

loop_
_space_group_symop_id
_space_group_symop_operation_xyz
1 x,y,z
2 x+1/2,-y+1/2,-z+1/4
3 -x+1/2,y+1/2,-z+3/4
4 -x,-y,z+1/2
5 -y,-x,-z+1/2
6 -y+1/2,x+1/2,z+3/4
7 y+1/2,-x+1/2,z+1/4
8 y,x,-z

loop_
_atom_site_label
_atom_site_type_symbol
_atom_site_symmetry_multiplicity
_atom_site_Wyckoff_label
_atom_site_fract_x
_atom_site_fract_y
_atom_site_fract_z
_atom_site_occupancy
Si1 Si   4 a 0.08490 0.08490 0.00000 1.00000
Si2 Si   8 b 0.17520 0.37920 0.27420 1.00000
```

"ST12" of Si: A_tP12_96_ab - POSCAR

```
A_tP12_96_ab & a,c/a,x1,x2,y2,z2 --params=5.51889,1.25999974633,0.0849,
    ↪ 0.1752,0.3792,0.2742 & P4_32_12 D_4^8 #96 (ab) & tP12 & & Si &
    ↪ ST12 & J. Crain et al., Phys. Rev. B 49, 5329-5340 (1994)
   1.00000000000000000
   5.51889083000000   0.00000000000000   0.00000000000000
   0.00000000000000   5.51889083000000   0.00000000000000
   0.00000000000000   0.00000000000000   6.95380245000000
   Si
   12
Direct
   0.08490000000000   0.08490000000000   0.00000000000000   Si  (4a)
  -0.08490000000000  -0.08490000000000   0.50000000000000   Si  (4a)
   0.41510000000000   0.58490000000000   0.75000000000000   Si  (4a)
   0.58490000000000   0.41510000000000   0.25000000000000   Si  (4a)
   0.12080000000000   0.67520000000000   0.02420000000000   Si  (8b)
   0.17520000000000   0.37920000000000   0.27420000000000   Si  (8b)
   0.32480000000000   0.87920000000000   0.47580000000000   Si  (8b)
   0.37920000000000   0.17520000000000   0.72580000000000   Si  (8b)
   0.62080000000000   0.82480000000000   0.22580000000000   Si  (8b)
   0.67520000000000   0.12080000000000  -0.02420000000000   Si  (8b)
```

```
  0.82480000000000   0.62080000000000   0.77420000000000   Si   (8b)
  0.87920000000000   0.32480000000000   0.52420000000000   Si   (8b)
```

Tetragonal PZT [$Pb(Zr_xTi_{1-x})O_3$]: A3BC_tP5_99_bc_a_b - CIF

```
# CIF file

data_findsym-output
_audit_creation_method FINDSYM

_chemical_name_mineral 'Pb(Zr_(1-x)Ti_x)O3'
_chemical_formula_sum 'Pb Zr O3'

loop_
_publ_author_name
 'B. Noheda'
 'J. A. Gonzalo'
 'L. E. Cross'
 'R. Guo'
 'S.-E. Park'
 'D. E. Cox'
 'G. Shirane'
_journal_name_full
;
Physical Review B
;
_journal_volume 61
_journal_year 2000
_journal_page_first 8687
_journal_page_last 8695
_publ_Section_title
;
 Tetragonal-to-monoclinic phase transition in a ferroelectric perovskite
     ↪ : The structure of PbZr$_{0.52}$Ti$_{0.48}$O$_3$
;

_aflow_proto 'A3BC_tP5_99_bc_a_b'
_aflow_params 'a,c/a,z1,z2,z3,z4'
_aflow_params_values '4.046,1.02308452793,0.0,0.8973,0.4517,0.3785'
_aflow_Strukturbericht 'None'
_aflow_Pearson 'tP5'

_symmetry_space_group_name_Hall "P 4 -2"
_symmetry_space_group_name_H-M "P 4 m m"
_symmetry_Int_Tables_number 99

_cell_length_a     4.04600
_cell_length_b     4.04600
_cell_length_c     4.13940
_cell_angle_alpha  90.00000
_cell_angle_beta   90.00000
_cell_angle_gamma  90.00000

loop_
_space_group_symop_id
_space_group_symop_operation_xyz
1 x,y,z
2 -x,-y,z
3 -y,x,z
4 y,-x,z
5 -x,y,z
6 x,-y,z
7 y,x,z
8 -y,-x,z

loop_
_atom_site_label
_atom_site_type_symbol
_atom_site_symmetry_multiplicity
_atom_site_Wyckoff_label
_atom_site_fract_x
_atom_site_fract_y
_atom_site_fract_z
_atom_site_occupancy
Pb1 Pb   1 a 0.00000 0.00000 0.00000 1.00000
O1  O    1 b 0.50000 0.50000 0.89730 1.00000
Zr1 Zr   1 b 0.50000 0.50000 0.45170 1.00000
O2  O    2 c 0.50000 0.00000 0.37850 1.00000
```

Tetragonal PZT [$Pb(Zr_xTi_{1-x})O_3$]: A3BC_tP5_99_bc_a_b - POSCAR

```
A3BC_tP5_99_bc_a_b & a,c/a,z1,z2,z3,z4 --params=4.046,1.02308452793,0.0,
     ↪ 0.8973,0.4517,0.3785 & P4mm     C_{4v}^1 #99 (ab^2c) & tP5 & &
     ↪  Pb(Zr0.52Ti0.48)O3 & Tetragonal PZT & B. Noheda et al., PRB 61
     ↪ , 8687-8695 (2000)
   1.0000000000000000
   4.04600000000000   0.00000000000000   0.00000000000000
   0.00000000000000   4.04600000000000   0.00000000000000
   0.00000000000000   0.00000000000000   4.13940000000000
    O    Pb    Zr
    3     1     1
Direct
   0.50000000000000   0.50000000000000  -0.10270000000000   O    (1b)
   0.00000000000000   0.50000000000000   0.37850000000000   O    (2c)
   0.50000000000000   0.00000000000000   0.37850000000000   O    (2c)
   0.00000000000000   0.00000000000000   0.00000000000000   Pb   (1a)
   0.50000000000000   0.50000000000000   0.45170000000000   Zr   (1b)
```

$BaS_3$ ($D0_{17}$): AB3_tP8_113_a_ce - CIF

```
# CIF file

data_findsym-output
_audit_creation_method FINDSYM

_chemical_name_mineral 'Barium trisulfide'
_chemical_formula_sum 'Ba S3'

loop_
_publ_author_name
 'S. Yamaoka'
 'J. T. Lemley'
 'J. M. Jenks'
 'H. Steinfink'
_journal_name_full
;
Inorganic Chemistry
;
_journal_volume 14
_journal_year 1975
_journal_page_first 129
_journal_page_last 131
_publ_Section_title
;
 Structural chemistry of the polysulfides dibarium trisulfide and
     ↪ monobarium trisulfide
;

# Found in Pearson's Handbook Vol II, pp. 1071-1072

_aflow_proto 'AB3_tP8_113_a_ce'
_aflow_params 'a,c/a,z2,x3,z3'
_aflow_params_values '6.871,0.606622034638,0.206,0.1797,0.476'
_aflow_Strukturbericht 'D0_17'
_aflow_Pearson 'tP8'

_symmetry_space_group_name_Hall "P -4 2ab"
_symmetry_space_group_name_H-M "P -4 21 m"
_symmetry_Int_Tables_number 113

_cell_length_a     6.87100
_cell_length_b     6.87100
_cell_length_c     4.16810
_cell_angle_alpha  90.00000
_cell_angle_beta   90.00000
_cell_angle_gamma  90.00000

loop_
_space_group_symop_id
_space_group_symop_operation_xyz
1 x,y,z
2 x+1/2,-y+1/2,-z
3 -x+1/2,y+1/2,-z
4 -x,-y,z
5 y+1/2,x+1/2,z
6 y,-x,-z
7 -y,x,-z
8 -y+1/2,-x+1/2,z

loop_
_atom_site_label
_atom_site_type_symbol
_atom_site_symmetry_multiplicity
_atom_site_Wyckoff_label
_atom_site_fract_x
_atom_site_fract_y
_atom_site_fract_z
_atom_site_occupancy
Ba1 Ba   2 a 0.00000 0.00000 0.00000 1.00000
S1  S    2 c 0.00000 0.50000 0.20600 1.00000
S2  S    4 e 0.17970 0.67970 0.47600 1.00000
```

$BaS_3$ ($D0_{17}$): AB3_tP8_113_a_ce - POSCAR

```
AB3_tP8_113_a_ce & a,c/a,z2,x3,z3 --params=6.871,0.606622034638,0.206,
     ↪ 0.1797,0.476 & P(-4)2_1m D_{2d}^3 #113 (ace) & tP8 & D0_{17} &
     ↪ BaS_3 & & S. Yamaoka, J. T. Lemley and H. Steinfink, Inorg.
     ↪ Chem. 14, 129-131 (1975)
   1.0000000000000000
   6.87100000000000   0.00000000000000   0.00000000000000
   0.00000000000000   6.87100000000000   0.00000000000000
   0.00000000000000   0.00000000000000   4.16810000000000
   Ba    S
    2    6
Direct
   0.00000000000000   0.00000000000000   0.00000000000000   Ba   (2a)
   0.50000000000000   0.50000000000000   0.00000000000000   Ba   (2a)
   0.00000000000000   0.50000000000000   0.20600000000000   S    (2c)
   0.50000000000000   0.00000000000000   0.79400000000000   S    (2c)
   0.17970000000000   0.67970000000000   0.47600000000000   S    (4e)
   0.32030000000000   0.17970000000000   0.52400000000000   S    (4e)
   0.67970000000000   0.82030000000000   0.52400000000000   S    (4e)
   0.82030000000000   0.32030000000000   0.47600000000000   S    (4e)
```

Stannite ($Cu_2FeS_4Sn$, $H2_6$): A2BC4D_tI16_121_d_a_i_b - CIF

```
# CIF file

data_findsym-output
_audit_creation_method FINDSYM

_chemical_name_mineral 'Stannite'
_chemical_formula_sum 'Cu2 Fe S4 Sn'

loop_
_publ_author_name
 'L. O. Brockway'
_journal_name_full
;
```

```
Zeitschrift f\"{u}r Kristallographie - Crystalline Materials
;
_journal_volume 89
_journal_year 1934
_journal_page_first 434
_journal_page_last 441
_publ_Section_title
;
 The Crystal Structure of Stannite, Cu$_2$FeSnS$_4$
;

_aflow_proto 'A2BC4D_tI16_121_d_a_i_b'
_aflow_params 'a,c/a,x4,z4'
_aflow_params_values '5.46,1.96428571429,0.245,0.132'
_aflow_Strukturbericht 'H2_6'
_aflow_Pearson 'tI16'

_symmetry_space_group_name_Hall "I -4 2"
_symmetry_space_group_name_H-M "I -4 2 m"
_symmetry_Int_Tables_number 121

_cell_length_a     5.46000
_cell_length_b     5.46000
_cell_length_c     10.72500
_cell_angle_alpha  90.00000
_cell_angle_beta   90.00000
_cell_angle_gamma  90.00000

loop_
_space_group_symop_id
_space_group_symop_operation_xyz
1 x,y,z
2 x,-y,-z
3 -x,y,-z
4 -x,-y,z
5 y,x,z
6 y,-x,-z
7 -y,x,-z
8 -y,-x,z
9 x+1/2,y+1/2,z+1/2
10 x+1/2,-y+1/2,-z+1/2
11 -x+1/2,y+1/2,-z+1/2
12 -x+1/2,-y+1/2,z+1/2
13 y+1/2,x+1/2,z+1/2
14 y+1/2,-x+1/2,-z+1/2
15 -y+1/2,x+1/2,-z+1/2
16 -y+1/2,-x+1/2,z+1/2

loop_
_atom_site_label
_atom_site_type_symbol
_atom_site_symmetry_multiplicity
_atom_site_Wyckoff_label
_atom_site_fract_x
_atom_site_fract_y
_atom_site_fract_z
_atom_site_occupancy
Fe1 Fe   2 a 0.00000 0.00000 0.00000 1.00000
Sn1 Sn   2 b 0.00000 0.00000 0.50000 1.00000
Cu1 Cu   4 d 0.00000 0.50000 0.25000 1.00000
S1  S    8 i 0.24500 0.24500 0.13200 1.00000
```

Stannite ($Cu_2FeS_4Sn$, $H2_6$): A2BC4D_tI16_121_d_a_i_b - POSCAR

```
A2BC4D_tI16_121_d_a_i_b & a,c/a,x4,z4 --params=5.46,1.96428571429,0.245,
   ↪ 0.132 & I(-4)2m    D_{2d}^{11} #121 (abdi) & tI16 & H2_6 &
   ↪ Cu2FeS4Sn & Stannite & L. O. Brockway, Zeitschrift f\"{u}r
   ↪ Kristallographie - Crystalline Materials 89, 434-441 (1934)
   1.0000000000000000
  -2.73000000000000   2.73000000000000   5.36250000001170
   2.73000000000000  -2.73000000000000   5.36250000001170
   2.73000000000000   2.73000000000000  -5.36250000001170
   Cu    Fe     S    Sn
    2     1     4     1
Direct
   0.25000000000000   0.75000000000000   0.50000000000000   Cu   (4d)
   0.75000000000000   0.25000000000000   0.50000000000000   Cu   (4d)
   0.00000000000000   0.00000000000000   0.00000000000000   Fe   (2a)
   0.37700000000000   0.37700000000000   0.49000000000000    S   (8i)
   0.62300000000000   0.11300000000000   0.00000000000000    S   (8i)
   0.88700000000000   0.88700000000000   0.51000000000000    S   (8i)
   0.11300000000000   0.62300000000000   0.00000000000000    S   (8i)
   0.50000000000000   0.50000000000000   0.00000000000000   Sn   (2b)
```

Chalcopyrite ($CuFeS_2$, $E1_1$): ABC2_tI16_122_a_b_d - CIF

```
# CIF file

data_findsym-output
_audit_creation_method FINDSYM

_chemical_name_mineral 'Chalcopyrite'
_chemical_formula_sum 'Cu Fe S2'

loop_
_publ_author_name
 'S. R. Hall'
 'J. M. Stewart'
_journal_name_full
;
Acta Crystallographica B
;
_journal_volume 29
_journal_year 1973
_journal_page_first 579
_journal_page_last 585
_publ_Section_title
;
 The Crystal Structure Refinement of Chalcopyrite, CuFeS$_2$
;

_aflow_proto 'ABC2_tI16_122_a_b_d'
_aflow_params 'a,c/a,x3'
_aflow_params_values '5.289,1.97069389299,0.2574'
_aflow_Strukturbericht 'E1_1'
_aflow_Pearson 'tI16'

_symmetry_space_group_name_Hall "I -4 2bw"
_symmetry_space_group_name_H-M "I -4 2 d"
_symmetry_Int_Tables_number 122

_cell_length_a     5.28900
_cell_length_b     5.28900
_cell_length_c     10.42300
_cell_angle_alpha  90.00000
_cell_angle_beta   90.00000
_cell_angle_gamma  90.00000

loop_
_space_group_symop_id
_space_group_symop_operation_xyz
1 x,y,z
2 x,-y+1/2,-z+1/4
3 -x,y+1/2,-z+1/4
4 -x,-y,z
5 y,x+1/2,z+1/4
6 y,-x,-z
7 -y,x,-z
8 -y,-x+1/2,z+1/4
9 x+1/2,y+1/2,z+1/2
10 x+1/2,-y,-z+3/4
11 -x+1/2,y,-z+3/4
12 -x+1/2,-y+1/2,z+1/2
13 y+1/2,x,z+3/4
14 y+1/2,-x+1/2,-z+1/2
15 -y+1/2,x+1/2,-z+1/2
16 -y+1/2,-x,z+3/4

loop_
_atom_site_label
_atom_site_type_symbol
_atom_site_symmetry_multiplicity
_atom_site_Wyckoff_label
_atom_site_fract_x
_atom_site_fract_y
_atom_site_fract_z
_atom_site_occupancy
Cu1 Cu   4 a 0.00000 0.00000 0.00000 1.00000
Fe1 Fe   4 b 0.00000 0.00000 0.50000 1.00000
S1  S    8 d 0.25740 0.25000 0.12500 1.00000
```

Chalcopyrite ($CuFeS_2$, $E1_1$): ABC2_tI16_122_a_b_d - POSCAR

```
ABC2_tI16_122_a_b_d & a,c/a,x3 --params=5.289,1.97069389299,0.2574 & I(-
   ↪ 4)2d    D_{2d}^{12} #122 (abd) & tI16 & E1_1 & CuFeS2 &
   ↪ Chalcopyrite & S. R. Hall and J. M. Stewart, Acta Cryst. B 29,
   ↪ 579-585 (1973)
   1.0000000000000000
  -2.64450000000000   2.64450000000000   5.21150000001205
   2.64450000000000  -2.64450000000000   5.21150000001205
   2.64450000000000   2.64450000000000  -5.21150000001205
   Cu    Fe     S
    2     2     4
Direct
   0.00000000000000   0.00000000000000   0.00000000000000   Cu   (4a)
   0.75000000000000   0.25000000000000   0.50000000000000   Cu   (4a)
   0.25000000000000   0.75000000000000   0.50000000000000   Fe   (4b)
   0.50000000000000   0.50000000000000   0.00000000000000   Fe   (4b)
   0.37500000000000   0.38240000000000   0.50740000000000    S   (8d)
   0.61760000000000   0.12500000000000  -0.00740000000000    S   (8d)
   0.87500000000000  -0.13240000000000   0.49260000000000    S   (8d)
   0.13240000000000   0.62500000000000   0.00740000000000    S   (8d)
```

$HoCoGa_5$: AB5C_tP7_123_b_ci_a - CIF

```
# CIF file

data_findsym-output
_audit_creation_method FINDSYM

_chemical_name_mineral ''
_chemical_formula_sum 'Ho Co Ga5'

loop_
_publ_author_name
 'Yu.N Grin'
 'Ya.P. Yarmolyuk'
 'E. I. Gladyshevskii'
_journal_name_full
;
Kristallografiya
;
_journal_volume 24
_journal_year 1979
_journal_page_first 242
_journal_page_last 246
_publ_Section_title
;
```

```
 Kristallicheskie struktury soedinenij R$_2$COGa$_8$ (R=Sm, Gd, Tb, Dy,
    ↪ Ho, Er, Tm, Lu, Y) i RCoGa$_5$ (R=Gd, Tb, Dy, Ho, Er, Tm, Lu,
    ↪ Y)
;

# Found in  http://materials.springer.com/isp/crystallographic/docs/
    ↪ sd_1406905

_aflow_proto 'AB5C_tP7_123_b_ci_a'
_aflow_params 'a,c/a,z4'
_aflow_params_values '4.207,1.61516520086,0.312'
_aflow_Strukturbericht 'None'
_aflow_Pearson 'tP7'

_symmetry_space_group_name_Hall "-P 4 2"
_symmetry_space_group_name_H-M "P 4/m m m"
_symmetry_Int_Tables_number 123

_cell_length_a     4.20700
_cell_length_b     4.20700
_cell_length_c     6.79500
_cell_angle_alpha 90.00000
_cell_angle_beta  90.00000
_cell_angle_gamma 90.00000

loop_
_space_group_symop_id
_space_group_symop_operation_xyz
1 x,y,z
2 x,-y,-z
3 -x,y,-z
4 -x,-y,z
5 -y,-x,-z
6 -y,x,z
7 y,-x,z
8 y,x,-z
9 -x,-y,-z
10 -x,y,z
11 x,-y,z
12 x,y,-z
13 y,x,z
14 y,-x,-z
15 -y,x,-z
16 -y,-x,z

loop_
_atom_site_label
_atom_site_type_symbol
_atom_site_symmetry_multiplicity
_atom_site_Wyckoff_label
_atom_site_fract_x
_atom_site_fract_y
_atom_site_fract_z
_atom_site_occupancy
Ho1 Ho   1 a 0.00000 0.00000 0.00000 1.00000
Co1 Co   1 b 0.00000 0.00000 0.50000 1.00000
Ga1 Ga   1 c 0.50000 0.50000 0.00000 1.00000
Ga2 Ga   4 i 0.00000 0.50000 0.31200 1.00000
```

$HoCoGa_5$: AB5C_tP7_123_b_ci_a - POSCAR

```
AB5C_tP7_123_b_ci_a & a,c/a,z4 --params=4.207,1.61516520086,0.312 & P4/
    ↪ mmm    D_{4h}^1 #123 (abci) & tP7 & & HoCoGa5 & & Y. Grin, Y.
    ↪ P. Yarmolyuk and E. I. Gladyshevskii, Kristallografiya 24,
    ↪ 242-246 (1979)
   1.0000000000000000
   4.20700000000000   0.00000000000000   0.00000000000000
   0.00000000000000   4.20700000000000   0.00000000000000
   0.00000000000000   0.00000000000000   6.79500000000000
   Co    Ga    Ho
    1     5     1
Direct
   0.00000000000000   0.00000000000000   0.50000000000000   Co   (1b)
   0.50000000000000   0.50000000000000   0.00000000000000   Ga   (1c)
   0.00000000000000   0.50000000000000   0.31200000000000   Ga   (4i)
   0.00000000000000   0.50000000000000   0.68800000000000   Ga   (4i)
   0.50000000000000   0.00000000000000   0.31200000000000   Ga   (4i)
   0.50000000000000   0.00000000000000   0.68800000000000   Ga   (4i)
   0.00000000000000   0.00000000000000   0.00000000000000   Ho   (1a)
```

$CuTi_3$ ($L6_0$): AB3_tP4_123_a_ce - CIF

```
# CIF file

data_findsym-output
_audit_creation_method FINDSYM

_chemical_name_mineral ''
_chemical_formula_sum 'Cu Ti3'

loop_
_publ_author_name
 'N. Karlsson'
_journal_name_full
;
Journal of the Institute of Metals (London)
;
_journal_volume 79
_journal_year 1951
_journal_page_first 391
_journal_page_last 405
_publ_Section_title
;
 An X-ray study of the phases in the copper-titanium system
;

# Found in  http://materials.springer.com/isp/crystallographic/docs/
    ↪ sd_1250535

_aflow_proto 'AB3_tP4_123_a_ce'
_aflow_params 'a,c/a'
_aflow_params_values '4.158,0.864357864358'
_aflow_Strukturbericht 'L6_0'
_aflow_Pearson 'tP4'

_symmetry_space_group_name_Hall "-P 4 2"
_symmetry_space_group_name_H-M "P 4/m m m"
_symmetry_Int_Tables_number 123

_cell_length_a     4.15800
_cell_length_b     4.15800
_cell_length_c     3.59400
_cell_angle_alpha 90.00000
_cell_angle_beta  90.00000
_cell_angle_gamma 90.00000

loop_
_space_group_symop_id
_space_group_symop_operation_xyz
1 x,y,z
2 x,-y,-z
3 -x,y,-z
4 -x,-y,z
5 -y,-x,-z
6 -y,x,z
7 y,-x,z
8 y,x,-z
9 -x,-y,-z
10 -x,y,z
11 x,-y,z
12 x,y,-z
13 y,x,z
14 y,-x,-z
15 -y,x,-z
16 -y,-x,z

loop_
_atom_site_label
_atom_site_type_symbol
_atom_site_symmetry_multiplicity
_atom_site_Wyckoff_label
_atom_site_fract_x
_atom_site_fract_y
_atom_site_fract_z
_atom_site_occupancy
Cu1 Cu   1 a 0.00000 0.00000 0.00000 1.00000
Ti1 Ti   1 c 0.50000 0.50000 0.00000 1.00000
Ti2 Ti   2 e 0.00000 0.50000 0.50000 1.00000
```

$CuTi_3$ ($L6_0$): AB3_tP4_123_a_ce - POSCAR

```
AB3_tP4_123_a_ce & a,c/a --params=4.158,0.864357864358 & P4/mmm    D_{4h
    ↪ }^1 #123 (ace) & tP4 & L6_0 & CuTi3 & & N. Karlsson, Journal of
    ↪ the J. Inst. Met. 79, 391-405 (1951)
   1.0000000000000000
   4.15800000000000   0.00000000000000   0.00000000000000
   0.00000000000000   4.15800000000000   0.00000000000000
   0.00000000000000   0.00000000000000   3.59400000000000
   Cu    Ti
    1     3
Direct
   0.00000000000000   0.00000000000000   0.00000000000000   Cu   (1a)
   0.50000000000000   0.50000000000000   0.00000000000000   Ti   (1c)
   0.00000000000000   0.50000000000000   0.50000000000000   Ti   (2e)
   0.50000000000000   0.00000000000000   0.50000000000000   Ti   (2e)
```

CuAu ($L1_0$): AB_tP2_123_a_d - CIF

```
# CIF file

data_findsym-output
_audit_creation_method FINDSYM

_chemical_name_mineral 'Tetraauricupride'
_chemical_formula_sum 'Cu Au'

loop_
_publ_author_name
 'Peter Bayliss'
_journal_name_full
;
Canadian Mineralogist
;
_journal_volume 28
_journal_year 1990
_journal_page_first 751
_journal_page_last 755
_publ_Section_title
;
 Revised Unit-Cell Dimensions, Space Group, and Chemical Formula of Some
    ↪ Metallic Materials
;

# Found in  AMS database

_aflow_proto 'AB_tP2_123_a_d'
_aflow_params 'a,c/a'
_aflow_params_values '2.8,1.31071428571'
_aflow_Strukturbericht 'L1_0'
_aflow_Pearson 'tP2'
```

```
_symmetry_space_group_name_Hall "-P 4 2"
_symmetry_space_group_name_H-M "P 4/m m m"
_symmetry_Int_Tables_number 123

_cell_length_a      2.80000
_cell_length_b      2.80000
_cell_length_c      3.67000
_cell_angle_alpha 90.00000
_cell_angle_beta  90.00000
_cell_angle_gamma 90.00000

loop_
_space_group_symop_id
_space_group_symop_operation_xyz
1 x,y,z
2 x,-y,-z
3 -x,y,-z
4 -x,-y,z
5 -y,-x,-z
6 -y,x,z
7 y,-x,z
8 y,x,-z
9 -x,-y,-z
10 -x,y,z
11 x,-y,z
12 x,y,-z
13 y,x,z
14 y,-x,-z
15 -y,x,-z
16 -y,-x,z

loop_
_atom_site_label
_atom_site_type_symbol
_atom_site_symmetry_multiplicity
_atom_site_Wyckoff_label
_atom_site_fract_x
_atom_site_fract_y
_atom_site_fract_z
_atom_site_occupancy
Au1 Au    1 a 0.00000 0.00000 0.00000 1.00000
Cu1 Cu    1 d 0.50000 0.50000 0.50000 1.00000
```

CuAu ($L1_0$): AB_tP2_123_a_d - POSCAR

```
AB_tP2_123_a_d & a,c/a --params=2.8,1.31071428571 & P4/mmm     D_{4h}^1 #
     ↪ 123 (ad) & tP2 & L1_0 & CuAu & tetraauricupride & P. Bayliss,
     ↪ Can. Mineral. 28, 751-755 (1990)
   1.0000000000000000
   2.80000000000000    0.00000000000000    0.00000000000000
   0.00000000000000    2.80000000000000    0.00000000000000
   0.00000000000000    0.00000000000000    3.67000000000000
   Au    Cu
    1     1
Direct
   0.00000000000000    0.00000000000000    0.00000000000000    Au     (1a)
   0.50000000000000    0.50000000000000    0.50000000000000    Cu     (1d)
```

$CaCuO_2$: ABC2_tP4_123_d_a_f - CIF

```
# CIF file

data_findsym-output
_audit_creation_method FINDSYM

_chemical_name_mineral ''
_chemical_formula_sum 'Ca Cu O2'

loop_
_publ_author_name
 'T. Siegrist'
 'S. M. Zahurak'
 'D. W. Murphy'
 'R. S. Roth'
_journal_name_full
;
Nature
;
_journal_volume 334
_journal_year 1988
_journal_page_first 231
_journal_page_last 232
_publ_Section_title
;
 The parent structure of the layered high-temperature superconductors
;

_aflow_proto 'ABC2_tP4_123_d_a_f'
_aflow_params 'a,c/a'
_aflow_params_values '3.8611,0.828649866618'
_aflow_Strukturbericht 'None'
_aflow_Pearson 'tP4'

_symmetry_space_group_name_Hall "-P 4 2"
_symmetry_space_group_name_H-M "P 4/m m m"
_symmetry_Int_Tables_number 123

_cell_length_a      3.86110
_cell_length_b      3.86110
_cell_length_c      3.19950
_cell_angle_alpha 90.00000
_cell_angle_beta  90.00000
_cell_angle_gamma 90.00000

loop_
_space_group_symop_id
_space_group_symop_operation_xyz
1 x,y,z
2 x,-y,-z
3 -x,y,-z
4 -x,-y,z
5 -y,-x,-z
6 -y,x,z
7 y,-x,z
8 y,x,-z
9 -x,-y,-z
10 -x,y,z
11 x,-y,z
12 x,y,-z
13 y,x,z
14 y,-x,-z
15 -y,x,-z
16 -y,-x,z

loop_
_atom_site_label
_atom_site_type_symbol
_atom_site_symmetry_multiplicity
_atom_site_Wyckoff_label
_atom_site_fract_x
_atom_site_fract_y
_atom_site_fract_z
_atom_site_occupancy
Cu1 Cu    1 a 0.00000 0.00000 0.00000 1.00000
Ca1 Ca    1 d 0.50000 0.50000 0.50000 1.00000
O1  O     2 f 0.00000 0.50000 0.00000 1.00000
```

$CaCuO_2$: ABC2_tP4_123_d_a_f - POSCAR

```
ABC2_tP4_123_d_a_f & a,c/a --params=3.8611,0.828649866618 & P4/mmm     D_
     ↪ {4h}^1 #123 (adf) & tP4 & & CaCuO2 & & T. Siegrist, S. M.
     ↪ Zahurak, D. W. Murphy and R. S. Roth, Nature 334, 231-232 (1988
     ↪ )
   1.0000000000000000
   3.86110000000000    0.00000000000000    0.00000000000000
   0.00000000000000    3.86110000000000    0.00000000000000
   0.00000000000000    0.00000000000000    3.19950000000000
   Ca    Cu    O
    1     1    2
Direct
   0.50000000000000    0.50000000000000    0.50000000000000    Ca     (1d)
   0.00000000000000    0.00000000000000    0.00000000000000    Cu     (1a)
   0.00000000000000    0.50000000000000    0.00000000000000    O      (2f)
   0.50000000000000    0.00000000000000    0.00000000000000    O      (2f)
```

$Si_2U_3$ ($D5_a$): A2B3_tP10_127_g_ah - CIF

```
# CIF file

data_findsym-output
_audit_creation_method FINDSYM

_chemical_name_mineral ''
_chemical_formula_sum 'Si2 U3'

loop_
_publ_author_name
 'K. Remschnig'
 'T. Le Bihan'
 'H. No\"{e}l'
 'P. Rogl'
_journal_name_full
;
Journal of Solid State Chemistry
;
_journal_volume 97
_journal_year 1992
_journal_page_first 391
_journal_page_last 399
_publ_Section_title
;
 Structural chemistry and magnetic behavior of binary uranium silicides
;

# Found in http://materials.springer.com/lb/docs/
     ↪ sm_lbs_978-3-642-22847-6_359

_aflow_proto 'A2B3_tP10_127_g_ah'
_aflow_params 'a,c/a,x2,x3'
_aflow_params_values '7.3364,0.530232811733,0.3841,0.1821'
_aflow_Strukturbericht 'D5_a'
_aflow_Pearson 'tP10'

_symmetry_space_group_name_Hall "-P 4 2ab"
_symmetry_space_group_name_H-M "P 4/m b m"
_symmetry_Int_Tables_number 127

_cell_length_a      7.33640
_cell_length_b      7.33640
_cell_length_c      3.89000
_cell_angle_alpha 90.00000
_cell_angle_beta  90.00000
_cell_angle_gamma 90.00000

loop_
_space_group_symop_id
_space_group_symop_operation_xyz
1 x,y,z
2 x+1/2,-y+1/2,-z
```

```
3 -x+1/2,y+1/2,-z
4 -x,-y,z
5 -y+1/2,-x+1/2,-z
6 -y,x,z
7 y,-x,z
8 y+1/2,x+1/2,-z
9 -x,-y,-z
10 -x+1/2,y+1/2,z
11 x+1/2,-y+1/2,z
12 x,y,-z
13 y+1/2,x+1/2,z
14 y,-x,-z
15 -y,x,-z
16 -y+1/2,-x+1/2,z

loop_
_atom_site_label
_atom_site_type_symbol
_atom_site_symmetry_multiplicity
_atom_site_Wyckoff_label
_atom_site_fract_x
_atom_site_fract_y
_atom_site_fract_z
_atom_site_occupancy
U1  U    2 a 0.00000 0.00000 0.00000 1.00000
Si1 Si   4 g 0.38410 0.88410 0.00000 1.00000
U2  U    4 h 0.18210 0.68210 0.50000 1.00000
```

$Si_2U_3$ ($D5_a$): A2B3_tP10_127_g_ah - POSCAR

```
A2B3_tP10_127_g_ah & a,c/a,x2,x3 --params=7.3364,0.530232811733,0.3841,
   ↪ 0.1821 & P4/mbm    D_{4h}^5 #127 (agh) & tP10 & D5_a & Si2U3 &
   ↪ & K. Remshnig, T. Le Bihan, H. Noel and P. Rogl, J. Solid State
   ↪ Chem. 97, 391-399 (1992)
   1.0000000000000000
   7.33640000000000   0.00000000000000   0.00000000000000
   0.00000000000000   7.33640000000000   0.00000000000000
   0.00000000000000   0.00000000000000   3.89000000000000
   Si    U
    4    6
Direct
   0.11590000000000   0.38410000000000   0.00000000000000   Si   (4g)
   0.38410000000000   0.88410000000000   0.00000000000000   Si   (4g)
   0.61590000000000   0.11590000000000   0.00000000000000   Si   (4g)
   0.88410000000000   0.61590000000000   0.00000000000000   Si   (4g)
   0.00000000000000   0.00000000000000   0.00000000000000   U    (2a)
   0.50000000000000   0.50000000000000   0.00000000000000   U    (2a)
   0.18210000000000   0.68210000000000   0.50000000000000   U    (4h)
   0.31790000000000   0.18210000000000   0.50000000000000   U    (4h)
   0.68210000000000   0.81790000000000   0.50000000000000   U    (4h)
   0.81790000000000   0.31790000000000   0.50000000000000   U    (4h)
```

AsCuSiZr: ABCD_tP8_129_c_b_a_c - CIF

```
# CIF file

data_findsym-output
_audit_creation_method FINDSYM

_chemical_name_mineral 'Parent of FeAs superconductors'
_chemical_formula_sum 'As Cu Si Zr'

loop_
_publ_author_name
 'V. Johnson'
 'W. Jeitschko'
_journal_name_full
;
Journal of Solid State Chemistry
;
_journal_volume 11
_journal_year 1974
_journal_page_first 161
_journal_page_last 166
_publ_Section_title
;
 ZrCuSiAs: A ``filled'' PbFCl type
;

# Found in Pearson, Vol. I, pp. 1116

_aflow_proto 'ABCD_tP8_129_c_b_a_c'
_aflow_params 'a,c/a,z3,z4'
_aflow_params_values '3.6736,2.60540069686,0.6793,0.2246'
_aflow_Strukturbericht 'None'
_aflow_Pearson 'tP8'

_symmetry_space_group_name_Hall "-P 4a 2a"
_symmetry_space_group_name_H-M "P 4/n m m:2"
_symmetry_Int_Tables_number 129

_cell_length_a      3.67360
_cell_length_b      3.67360
_cell_length_c      9.57120
_cell_angle_alpha  90.00000
_cell_angle_beta   90.00000
_cell_angle_gamma  90.00000

loop_
_space_group_symop_id
_space_group_symop_operation_xyz
1 x,y,z
2 x+1/2,-y,-z
3 -x,y+1/2,-z
4 -x+1/2,-y+1/2,z
5 -y,-x,-z
6 -y+1/2,x,z
7 y,-x+1/2,z
8 y+1/2,x+1/2,-z
9 -x,-y,-z
10 -x+1/2,y,z
11 x,-y+1/2,z
12 x+1/2,y+1/2,-z
13 y,x,z
14 y+1/2,-x,-z
15 -y,x+1/2,-z
16 -y+1/2,-x+1/2,z

loop_
_atom_site_label
_atom_site_type_symbol
_atom_site_symmetry_multiplicity
_atom_site_Wyckoff_label
_atom_site_fract_x
_atom_site_fract_y
_atom_site_fract_z
_atom_site_occupancy
Si1 Si   2 a 0.75000 0.25000 0.00000 1.00000
Cu1 Cu   2 b 0.75000 0.25000 0.50000 1.00000
As1 As   2 c 0.25000 0.25000 0.67930 1.00000
Zr1 Zr   2 c 0.25000 0.25000 0.22460 1.00000
```

AsCuSiZr: ABCD_tP8_129_c_b_a_c - POSCAR

```
ABCD_tP8_129_c_b_a_c & a,c/a,z3,z4 --params=3.6736,2.60540069686,0.6793,
   ↪ 0.2246 & P4/nmm    D_{4h}^7 #129 (abc^2) & tP8 & & AsCuSiZr & &
   ↪ V. Johnson and W. Jeitschko, J. Solid State Chem. 11, 161-166
   ↪ (1974)
   1.0000000000000000
   3.67360000000000   0.00000000000000   0.00000000000000
   0.00000000000000   3.67360000000000   0.00000000000000
   0.00000000000000   0.00000000000000   9.57120000000000
   As    Cu    Si    Zr
    2     2     2     2
Direct
   0.25000000000000   0.25000000000000   0.67930000000000   As   (2c)
   0.75000000000000   0.75000000000000   0.32070000000000   As   (2c)
   0.25000000000000   0.75000000000000   0.50000000000000   Cu   (2b)
   0.75000000000000   0.25000000000000   0.50000000000000   Cu   (2b)
   0.25000000000000   0.75000000000000   0.00000000000000   Si   (2a)
   0.75000000000000   0.25000000000000   0.00000000000000   Si   (2a)
   0.25000000000000   0.25000000000000   0.22460000000000   Zr   (2c)
   0.75000000000000   0.75000000000000   0.77540000000000   Zr   (2c)
```

β-Np ($A_d$): A_tP4_129_ac - CIF

```
# CIF file

data_findsym-output
_audit_creation_method FINDSYM

_chemical_name_mineral 'beta Np'
_chemical_formula_sum 'Np'

loop_
_publ_author_name
 'W. H. Zachariasen'
_journal_name_full
;
Acta Crystallographica
;
_journal_volume 5
_journal_year 1952
_journal_page_first 664
_journal_page_last 667
_publ_Section_title
;
 Crystal chemical studies of the 5f-series of elements. XVIII. Crystal
   ↪ structure studies of neptunium metal at elevated temperatures
;

# Found in Donohue, pp. 154-156

_aflow_proto 'A_tP4_129_ac'
_aflow_params 'a,c/a,z2'
_aflow_params_values '4.897,0.69185215438,0.375'
_aflow_Strukturbericht 'A_d'
_aflow_Pearson 'tP4'

_symmetry_space_group_name_Hall "-P 4a 2a"
_symmetry_space_group_name_H-M "P 4/n m m:2"
_symmetry_Int_Tables_number 129

_cell_length_a      4.89700
_cell_length_b      4.89700
_cell_length_c      3.38800
_cell_angle_alpha  90.00000
_cell_angle_beta   90.00000
_cell_angle_gamma  90.00000

loop_
_space_group_symop_id
_space_group_symop_operation_xyz
1 x,y,z
2 x+1/2,-y,-z
3 -x,y+1/2,-z
4 -x+1/2,-y+1/2,z
5 -y,-x,-z
6 -y+1/2,x,z
7 y,-x+1/2,z
```

```
8 y+1/2,x+1/2,-z
9 -x,-y,-z
10 -x+1/2,y,z
11 x,-y+1/2,z
12 x+1/2,y+1/2,-z
13 y,x,z
14 y+1/2,-x,-z
15 -y,x+1/2,-z
16 -y+1/2,-x+1/2,z

loop_
_atom_site_label
_atom_site_type_symbol
_atom_site_symmetry_multiplicity
_atom_site_Wyckoff_label
_atom_site_fract_x
_atom_site_fract_y
_atom_site_fract_z
_atom_site_occupancy
Np1 Np   2 a 0.75000 0.25000 0.00000 1.00000
Np2 Np   2 c 0.25000 0.25000 0.37500 1.00000
```

$\beta$-Np ($A_d$): A_tP4_129_ac - POSCAR

```
A_tP4_129_ac & a,c/a,z2 --params=4.897,0.69185215438,0.375 & P4/nmm
    ↪ D_{4h}^7 #129 (ac) & tP4 & A_d & Np & beta & W. H. Zachariasen,
    ↪ Acta Cryst. 5, 664-667 (1952)
   1.0000000000000000
   4.89700000000000   0.00000000000000   0.00000000000000
   0.00000000000000   4.89700000000000   0.00000000000000
   0.00000000000000   0.00000000000000   3.38800000000000
   Np
    4
Direct
   0.25000000000000   0.75000000000000   0.00000000000000   Np   (2a)
   0.75000000000000   0.25000000000000   0.00000000000000   Np   (2a)
   0.25000000000000   0.25000000000000   0.37500000000000   Np   (2c)
   0.75000000000000   0.75000000000000   0.62500000000000   Np   (2c)
```

Matlockite ($E0_1$, PbFCl): ABC_tP6_129_c_a_c - CIF

```
# CIF file

data_findsym-output
_audit_creation_method FINDSYM

_chemical_name_mineral 'Matlockite'
_chemical_formula_sum 'Pb F Cl'

loop_
_publ_author_name
 'N. Pasero'
 'N. Perchiazzi'
_journal_name_full
;
Mineralogical Magazine
;
_journal_volume 60
_journal_year 1996
_journal_page_first 833
_journal_page_last 836
_publ_Section_title
;
 Crystal structure refinement of matlockite
;

# Found in AMS Database

_aflow_proto 'ABC_tP6_129_c_a_c'
_aflow_params 'a,c/a,z2,z3'
_aflow_params_values '4.11,1.76301703163,0.6497,0.2058'
_aflow_Strukturbericht 'E0_1'
_aflow_Pearson 'tP6'

_symmetry_space_group_name_Hall "-P 4a 2a"
_symmetry_space_group_name_H-M "P 4/n m m:2"
_symmetry_Int_Tables_number 129

_cell_length_a     4.11000
_cell_length_b     4.11000
_cell_length_c     7.24600
_cell_angle_alpha 90.00000
_cell_angle_beta  90.00000
_cell_angle_gamma 90.00000

loop_
_space_group_symop_id
_space_group_symop_operation_xyz
1 x,y,z
2 x+1/2,-y,-z
3 -x,y+1/2,-z
4 -x+1/2,-y+1/2,z
5 -y,-x,-z
6 -y+1/2,x,z
7 y,-x+1/2,z
8 y+1/2,x+1/2,-z
9 -x,-y,-z
10 -x+1/2,y,z
11 x,-y+1/2,z
12 x+1/2,y+1/2,-z
13 y,x,z
14 y+1/2,-x,-z
15 -y,x+1/2,-z
16 -y+1/2,-x+1/2,z
```

```
loop_
_atom_site_label
_atom_site_type_symbol
_atom_site_symmetry_multiplicity
_atom_site_Wyckoff_label
_atom_site_fract_x
_atom_site_fract_y
_atom_site_fract_z
_atom_site_occupancy
F1  F    2 a 0.75000 0.25000 0.00000 1.00000
Cl1 Cl   2 c 0.25000 0.25000 0.64970 1.00000
Pb1 Pb   2 c 0.25000 0.25000 0.20580 1.00000
```

Matlockite ($E0_1$, PbFCl): ABC_tP6_129_c_a_c - POSCAR

```
ABC_tP6_129_c_a_c & a,c/a,z2,z3 --params=4.11,1.76301703163,0.6497,
    ↪ 0.2058 & P4/nmm    D_{4h}^7 #129 (ac^2) & tP6 & E0_1 & PbFCl &
    ↪ Matlockite & M. Pasero and N. Perchiazzi, Mineral. Mag. 60,
    ↪ 833-836 (1996)
   1.0000000000000000
   4.11000000000000   0.00000000000000   0.00000000000000
   0.00000000000000   4.11000000000000   0.00000000000000
   0.00000000000000   0.00000000000000   7.24600000000000
   Cl    F    Pb
    2    2    2
Direct
   0.25000000000000   0.25000000000000   0.64970000000000   Cl   (2c)
   0.75000000000000   0.75000000000000   0.35030000000000   Cl   (2c)
   0.25000000000000   0.75000000000000   0.00000000000000    F   (2a)
   0.75000000000000   0.25000000000000   0.00000000000000    F   (2a)
   0.25000000000000   0.25000000000000   0.20580000000000   Pb   (2c)
   0.75000000000000   0.75000000000000   0.79420000000000   Pb   (2c)
```

$Cu_2Sb$ (C38): A2B_tP6_129_ac_c - CIF

```
# CIF file

data_findsym-output
_audit_creation_method FINDSYM

_chemical_name_mineral ''
_chemical_formula_sum 'Cu2 Sb'

loop_
_publ_author_name
 'W. B. Pearson'
_journal_name_full
;
Zeitschrift f\"{u}r Kristallographie
;
_journal_volume 171
_journal_year 1985
_journal_page_first 23
_journal_page_last 39
_publ_Section_title
;
 The Cu$_2$Sb and related structures
;

_aflow_proto 'A2B_tP6_129_ac_c'
_aflow_params 'a,c/a,z2,z3'
_aflow_params_values '4.0006,1.52584612308,0.27,0.7'
_aflow_Strukturbericht 'C38'
_aflow_Pearson 'tP6'

_symmetry_space_group_name_Hall "-P 4a 2a"
_symmetry_space_group_name_H-M "P 4/n m m:2"
_symmetry_Int_Tables_number 129

_cell_length_a     4.00060
_cell_length_b     4.00060
_cell_length_c     6.10430
_cell_angle_alpha 90.00000
_cell_angle_beta  90.00000
_cell_angle_gamma 90.00000

loop_
_space_group_symop_id
_space_group_symop_operation_xyz
1 x,y,z
2 x+1/2,-y,-z
3 -x,y+1/2,-z
4 -x+1/2,-y+1/2,z
5 -y,-x,-z
6 -y+1/2,x,z
7 y,-x+1/2,z
8 y+1/2,x+1/2,-z
9 -x,-y,-z
10 -x+1/2,y,z
11 x,-y+1/2,z
12 x+1/2,y+1/2,-z
13 y,x,z
14 y+1/2,-x,-z
15 -y,x+1/2,-z
16 -y+1/2,-x+1/2,z

loop_
_atom_site_label
_atom_site_type_symbol
_atom_site_symmetry_multiplicity
_atom_site_Wyckoff_label
_atom_site_fract_x
_atom_site_fract_y
_atom_site_fract_z
_atom_site_occupancy
```

```
Cu1 Cu    2 a 0.75000 0.25000 0.00000 1.00000
Cu2 Cu    2 c 0.25000 0.25000 0.27000 1.00000
Sb1 Sb    2 c 0.25000 0.25000 0.70000 1.00000
```

$Cu_2Sb$ (C38): A2B_tP6_129_ac_c - POSCAR

```
A2B_tP6_129_ac_c & a,c/a,z2,z3 --params=4.0006,1.52584612308,0.27,0.7 &
    ↪ P4/nmm    D_{4h}^7 #129 (ac^2) & tP6 & C38 & Cu2Sb & & W. B.
    ↪ Pearson, Zeitschrift f\"{u}r Kristallographie 171, 23-39 (1985)
   1.0000000000000000
   4.00060000000000    0.00000000000000    0.00000000000000
   0.00000000000000    4.00060000000000    0.00000000000000
   0.00000000000000    0.00000000000000    6.10430000000000
   Cu    Sb
    4     2
Direct
   0.25000000000000    0.75000000000000    0.00000000000000    Cu    (2a)
   0.75000000000000    0.25000000000000    0.00000000000000    Cu    (2a)
   0.25000000000000    0.25000000000000    0.27000000000000    Cu    (2c)
   0.75000000000000    0.75000000000000    0.73000000000000    Cu    (2c)
   0.25000000000000    0.25000000000000    0.70000000000000    Sb    (2c)
   0.75000000000000    0.75000000000000    0.30000000000000    Sb    (2c)
```

PbO (B10): AB_tP4_129_a_c - CIF

```
# CIF file

data_findsym-output
_audit_creation_method FINDSYM

_chemical_name_mineral 'lead oxide'
_chemical_formula_sum 'Pb O'

loop_
_publ_author_name
 'P. Boher'
 'P. Garnier'
 'J. R. Gavarri'
 'A. W. Hewat'
_journal_name_full
;
Journal of Solid State Chemistry
;
_journal_volume 57
_journal_year 1985
_journal_page_first 343
_journal_page_last 350
_publ_Section_title
;
 Monoxyde quadratique PbO$\alpha$(I): Description de la transition
    ↪ structurale ferroe'lastique
;

# Found in  Pearson's Handbook, Vol. IV, p. 4745

_aflow_proto 'AB_tP4_129_a_c'
_aflow_params 'a,c/a,z2'
_aflow_params_values '3.9645,1.26008323874,0.2368'
_aflow_Strukturbericht 'B10'
_aflow_Pearson 'tP4'

_symmetry_space_group_name_Hall "-P 4a 2a"
_symmetry_space_group_name_H-M "P 4/n m m:2"
_symmetry_Int_Tables_number 129

_cell_length_a      3.96450
_cell_length_b      3.96450
_cell_length_c      4.99560
_cell_angle_alpha  90.00000
_cell_angle_beta   90.00000
_cell_angle_gamma  90.00000

loop_
_space_group_symop_id
_space_group_symop_operation_xyz
1 x,y,z
2 x+1/2,-y,-z
3 -x,y+1/2,-z
4 -x+1/2,-y+1/2,z
5 -y,-x,-z
6 -y+1/2,x,z
7 y,-x+1/2,z
8 y+1/2,x+1/2,-z
9 -x,-y,-z
10 -x+1/2,y,z
11 x,-y+1/2,z
12 x+1/2,y+1/2,-z
13 y,x,z
14 y+1/2,-x,-z
15 -y,x+1/2,-z
16 -y+1/2,-x+1/2,z

loop_
_atom_site_label
_atom_site_type_symbol
_atom_site_symmetry_multiplicity
_atom_site_Wyckoff_label
_atom_site_fract_x
_atom_site_fract_y
_atom_site_fract_z
_atom_site_occupancy
O1  O     2 a 0.75000 0.25000 0.00000 1.00000
Pb1 Pb    2 c 0.25000 0.25000 0.23680 1.00000
```

PbO (B10): AB_tP4_129_a_c - POSCAR

```
AB_tP4_129_a_c & a,c/a,z2 --params=3.9645,1.26008323874,0.2368 & P4/nmm
    ↪    D_{4h}^7 #129 (ac) & tP4 & B10 & PbO & & P. Boher, P.Garnier
    ↪ , J. R. Gavarri and A. W. Hewat, J. Solid State Chem. 57,
    ↪ 343-350 (1985)
   1.0000000000000000
   3.96450000000000    0.00000000000000    0.00000000000000
   0.00000000000000    3.96450000000000    0.00000000000000
   0.00000000000000    0.00000000000000    4.99560000000000
    O    Pb
    2     2
Direct
   0.25000000000000    0.75000000000000    0.00000000000000    O     (2a)
   0.75000000000000    0.25000000000000    0.00000000000000    O     (2a)
   0.25000000000000    0.25000000000000    0.23680000000000    Pb    (2c)
   0.75000000000000    0.75000000000000    0.76320000000000    Pb    (2c)
```

γ-CuTi (B11): AB_tP4_129_c_c - CIF

```
# CIF file

data_findsym-output
_audit_creation_method FINDSYM

_chemical_name_mineral 'gamma CuTi'
_chemical_formula_sum 'Cu Ti'

loop_
_publ_author_name
 'V. N. Eremenko'
 'Yu. I. Buyanov'
 'S. B. Prima'
_journal_name_full
;
Soviet Powder Metallurgy and Metal Ceramics
;
_journal_volume 5
_journal_year 1966
_journal_page_first 494
_journal_page_last 502
_publ_Section_title
;
 Phase diagram of the system titanium-copper
;

# Found in  Pearson's Handbook, Vol. III, pp. 3021

_aflow_proto 'AB_tP4_129_c_c'
_aflow_params 'a,c/a,z1,z2'
_aflow_params_values '3.107,1.90505310589,0.1,0.65'
_aflow_Strukturbericht 'B11'
_aflow_Pearson 'tP4'

_symmetry_space_group_name_Hall "-P 4a 2a"
_symmetry_space_group_name_H-M "P 4/n m m:2"
_symmetry_Int_Tables_number 129

_cell_length_a      3.10700
_cell_length_b      3.10700
_cell_length_c      5.91900
_cell_angle_alpha  90.00000
_cell_angle_beta   90.00000
_cell_angle_gamma  90.00000

loop_
_space_group_symop_id
_space_group_symop_operation_xyz
1 x,y,z
2 x+1/2,-y,-z
3 -x,y+1/2,-z
4 -x+1/2,-y+1/2,z
5 -y,-x,-z
6 -y+1/2,x,z
7 y,-x+1/2,z
8 y+1/2,x+1/2,-z
9 -x,-y,-z
10 -x+1/2,y,z
11 x,-y+1/2,z
12 x+1/2,y+1/2,-z
13 y,x,z
14 y+1/2,-x,-z
15 -y,x+1/2,-z
16 -y+1/2,-x+1/2,z

loop_
_atom_site_label
_atom_site_type_symbol
_atom_site_symmetry_multiplicity
_atom_site_Wyckoff_label
_atom_site_fract_x
_atom_site_fract_y
_atom_site_fract_z
_atom_site_occupancy
Cu1 Cu    2 c 0.25000 0.25000 0.10000 1.00000
Ti1 Ti    2 c 0.25000 0.25000 0.65000 1.00000
```

γ-CuTi (B11): AB_tP4_129_c_c - POSCAR

```
AB_tP4_129_c_c & a,c/a,z1,z2 --params=3.107,1.90505310589,0.1,0.65 & P4/
    ↪ nmm    D_{4h}^7 #129 (c^2) & tP4 & B11 & CuTi (gamma) & & V. N.
    ↪  Eremenko and Yu. I. Buyanov and S. B. Prima, Soviet Powder
    ↪ Metallurgy and Metal Ceramics 5, 494-502 (1966)
   1.0000000000000000
   3.10700000000000    0.00000000000000    0.00000000000000
   0.00000000000000    3.10700000000000    0.00000000000000
```

```
    0.00000000000000    0.00000000000000    5.91900000000000
   Cu    Ti
    2     2
Direct
    0.25000000000000    0.25000000000000    0.10000000000000    Cu    (2c)
    0.75000000000000    0.75000000000000    0.90000000000000    Cu    (2c)
    0.25000000000000    0.25000000000000    0.65000000000000    Ti    (2c)
    0.75000000000000    0.75000000000000    0.35000000000000    Ti    (2c)
```

PtS (B17): AB_tP4_131_c_e - CIF

```
# CIF file

data_findsym-output
_audit_creation_method FINDSYM

_chemical_name_mineral ''
_chemical_formula_sum 'Pt S'

loop_
_publ_author_name
 'Fredrik Gronvold'
 'Haakon Haraldsen'
 'Arne Kjekshus'
_journal_name_full
;
Acta Chemica Scandinavica
;
_journal_volume 14
_journal_year 1960
_journal_page_first 1879
_journal_page_last 1893
_publ_Section_title
;
 On the Sulfides, Selenides and Tellurides of Platinum
;

_aflow_proto 'AB_tP4_131_c_e'
_aflow_params 'a,c/a'
_aflow_params_values '4.9073,1.24500234345'
_aflow_Strukturbericht 'B17'
_aflow_Pearson 'tP4'

_symmetry_space_group_name_Hall "-P 4c 2"
_symmetry_space_group_name_H-M "P 42/m m c"
_symmetry_Int_Tables_number 131

_cell_length_a     4.90730
_cell_length_b     4.90730
_cell_length_c     6.10960
_cell_angle_alpha  90.00000
_cell_angle_beta   90.00000
_cell_angle_gamma  90.00000

loop_
_space_group_symop_id
_space_group_symop_operation_xyz
1 x,y,z
2 x,-y,-z
3 -x,y,-z
4 -x,-y,z
5 -y,-x,-z+1/2
6 -y,x,z+1/2
7 y,-x,z+1/2
8 y,x,-z+1/2
9 -x,-y,-z
10 -x,y,z
11 x,-y,z
12 x,y,-z
13 y,x,z+1/2
14 y,-x,-z+1/2
15 -y,x,-z+1/2
16 -y,-x,z+1/2

loop_
_atom_site_label
_atom_site_type_symbol
_atom_site_symmetry_multiplicity
_atom_site_Wyckoff_label
_atom_site_fract_x
_atom_site_fract_y
_atom_site_fract_z
_atom_site_occupancy
Pt1 Pt   2 c 0.00000 0.50000 0.00000 1.00000
S1  S    2 e 0.00000 0.00000 0.25000 1.00000
```

PtS (B17): AB_tP4_131_c_e - POSCAR

```
AB_tP4_131_c_e & a,c/a --params=4.9073,1.24500234345 & P4_2/mmc  D_{4h}^
    ↪ 9 #131 (ce) & tP4 & B17 & PtS & & F. Gronvold and h. Haakon
    ↪ Haraldsen and A. Kjeksus, Acta Chem. Scand. 14, 1879-1893 (1960
    ↪ )
   1.0000000000000000
    4.90730000000000    0.00000000000000    0.00000000000000
    0.00000000000000    4.90730000000000    0.00000000000000
    0.00000000000000    0.00000000000000    6.10960000000000
   Pt    S
    2    2
Direct
    0.00000000000000    0.50000000000000    0.00000000000000    Pt    (2c)
    0.50000000000000    0.00000000000000    0.50000000000000    Pt    (2c)
    0.00000000000000    0.00000000000000    0.25000000000000    S     (2e)
    0.00000000000000    0.00000000000000    0.75000000000000    S     (2e)
```

T-50 B ($A_g$): A_tP50_134_b2m2n - CIF

```
# CIF file

data_findsym-output
_audit_creation_method FINDSYM

_chemical_name_mineral 'T-50 Boron'
_chemical_formula_sum 'B'

loop_
_publ_author_name
 'J. L. Hoard'
 'R. E. Hughes'
 'D. E. Sands'
_journal_name_full
;
Journal of the American Chemical Society
;
_journal_volume 80
_journal_year 1958
_journal_page_first 4507
_journal_page_last 4515
_publ_Section_title
;
 The Structure of Tetragonal Boron
;

# Found in  Donohue, Chapter 5, pp. 48-56

_aflow_proto 'A_tP50_134_b2m2n'
_aflow_params 'a,c/a,x2,z2,x3,z3,x4,y4,z4,x5,y5,z5'
_aflow_params_values '8.74,0.575514874142,0.0048,0.1685,0.1305,0.628,
    ↪ 0.1695,0.5228,0.1635,0.0753,0.3383,0.1485'
_aflow_Strukturbericht 'A_g'
_aflow_Pearson 'tP50'

_symmetry_space_group_name_Hall "-P 4ac 2bc"
_symmetry_space_group_name_H-M "P 42/n n m:2"
_symmetry_Int_Tables_number 134

_cell_length_a     8.74000
_cell_length_b     8.74000
_cell_length_c     5.03000
_cell_angle_alpha  90.00000
_cell_angle_beta   90.00000
_cell_angle_gamma  90.00000

loop_
_space_group_symop_id
_space_group_symop_operation_xyz
1 x,y,z
2 x,-y+1/2,-z+1/2
3 -x+1/2,y,-z+1/2
4 -x+1/2,-y+1/2,z
5 -y+1/2,-x+1/2,-z
6 -y+1/2,x,z+1/2
7 y,-x+1/2,z+1/2
8 y,x,-z
9 -x,-y,-z
10 -x,y+1/2,z+1/2
11 x+1/2,-y,z+1/2
12 x+1/2,y+1/2,-z
13 y+1/2,x+1/2,z
14 y+1/2,-x,-z+1/2
15 -y,x+1/2,-z+1/2
16 -y,-x,z

loop_
_atom_site_label
_atom_site_type_symbol
_atom_site_symmetry_multiplicity
_atom_site_Wyckoff_label
_atom_site_fract_x
_atom_site_fract_y
_atom_site_fract_z
_atom_site_occupancy
B1 B    2 b 0.75000  0.25000  0.25000 1.00000
B2 B    8 m 0.00480 -0.00480  0.16850 1.00000
B3 B    8 m 0.13050  0.86950  0.62800 1.00000
B4 B   16 n 0.16950  0.52280  0.16350 1.00000
B5 B   16 n 0.07530  0.33830  0.14850 1.00000
```

T-50 B ($A_g$): A_tP50_134_b2m2n - POSCAR

```
A_tP50_134_b2m2n & a,c/a,x2,z2,x3,z3,x4,y4,z4,x5,y5,z5 --params=8.74,
    ↪ 0.575514874142,0.0048,0.1685,0.1305,0.628,0.1695,0.5228,0.1635,
    ↪ 0.0753,0.3383,0.1485 & P4_2/nnm  D_{4h}^{12} #134 (bm^2n^2) &
    ↪ tP50 & A_g & B & T50 & J. L. Hoard, R. E. Hughes and D.E. Sands
    ↪ , J. Am. Chem. Soc. 80, 4507-4515 (1958)
   1.0000000000000000
    8.7400000000000000    0.0000000000000000    0.0000000000000000
    0.0000000000000000    8.7400000000000000    0.0000000000000000
    0.0000000000000000    0.0000000000000000    5.0300000000000000
    B
   50
Direct
    0.7500000000000000    0.2500000000000000    0.2500000000000000   B   (2b)   1
    0.2500000000000000    0.7500000000000000    0.7500000000000000   B   (2b)   1
    0.0048000000000000   -0.0048000000000000    0.1685000000000000   B   (8m)   2
    0.4952000000000000    0.5048000000000000    0.1685000000000000   B   (8m)   2
    0.5048000000000000    0.0048000000000000    0.6685000000000000   B   (8m)   2
   -0.0048000000000000    0.4952000000000000    0.6685000000000000   B   (8m)   2
    0.4952000000000000   -0.0048000000000000    0.3315000000000000   B   (8m)   2
    0.0048000000000000    0.5048000000000000    0.3315000000000000   B   (8m)   2
   -0.0048000000000000    0.0048000000000000    0.8315000000000000   B   (8m)   2
    0.5048000000000000    0.4952000000000000    0.8315000000000000   B   (8m)   2
```

```
   0.130500000000000    0.869500000000000    0.628000000000000   B   (8m)   3
   0.369500000000000    0.630500000000000    0.628000000000000   B   (8m)   3
   0.630500000000000    0.130500000000000    0.128000000000000   B   (8m)   3
   0.869500000000000    0.369500000000000    0.128000000000000   B   (8m)   3
   0.369500000000000    0.869500000000000    0.872000000000000   B   (8m)   3
   0.130500000000000    0.630500000000000    0.872000000000000   B   (8m)   3
   0.869500000000000    0.130500000000000    0.372000000000000   B   (8m)   3
   0.630500000000000    0.369500000000000    0.372000000000000   B   (8m)   3
   0.169500000000000    0.522800000000000    0.163500000000000   B  (16n)   4
   0.330500000000000   -0.022800000000000    0.163500000000000   B  (16n)   4
  -0.022800000000000    0.169500000000000    0.663500000000000   B  (16n)   4
   0.522800000000000    0.330500000000000    0.663500000000000   B  (16n)   4
   0.330500000000000    0.522800000000000    0.336500000000000   B  (16n)   4
   0.169500000000000   -0.022800000000000    0.336500000000000   B  (16n)   4
   0.522800000000000    0.169500000000000    0.836500000000000   B  (16n)   4
  -0.022800000000000    0.330500000000000    0.836500000000000   B  (16n)   4
   0.830500000000000    0.477200000000000    0.836500000000000   B  (16n)   4
   0.669500000000000    0.022800000000000    0.836500000000000   B  (16n)   4
   0.022800000000000    0.830500000000000    0.336500000000000   B  (16n)   4
   0.477200000000000    0.669500000000000    0.336500000000000   B  (16n)   4
   0.669500000000000    0.477200000000000    0.663500000000000   B  (16n)   4
   0.830500000000000    0.022800000000000    0.663500000000000   B  (16n)   4
   0.477200000000000    0.830500000000000    0.163500000000000   B  (16n)   4
   0.022800000000000    0.669500000000000    0.163500000000000   B  (16n)   4
   0.075300000000000    0.338300000000000    0.148500000000000   B  (16n)   5
   0.424700000000000    0.161700000000000    0.148500000000000   B  (16n)   5
   0.161700000000000    0.075300000000000    0.648500000000000   B  (16n)   5
   0.338300000000000    0.424700000000000    0.648500000000000   B  (16n)   5
   0.424700000000000    0.338300000000000    0.351500000000000   B  (16n)   5
   0.075300000000000    0.161700000000000    0.351500000000000   B  (16n)   5
   0.338300000000000    0.075300000000000    0.851500000000000   B  (16n)   5
   0.161700000000000    0.424700000000000    0.851500000000000   B  (16n)   5
  -0.075300000000000    0.661700000000000    0.851500000000000   B  (16n)   5
   0.575300000000000    0.838300000000000    0.851500000000000   B  (16n)   5
   0.838300000000000   -0.075300000000000    0.351500000000000   B  (16n)   5
   0.661700000000000    0.575300000000000    0.351500000000000   B  (16n)   5
   0.575300000000000    0.661700000000000    0.648500000000000   B  (16n)   5
  -0.075300000000000    0.838300000000000    0.648500000000000   B  (16n)   5
   0.661700000000000   -0.075300000000000    0.148500000000000   B  (16n)   5
   0.838300000000000    0.575300000000000    0.148500000000000   B  (16n)   5
```

$\beta$-U ($A_b$): A_tP30_136_bf2ij - CIF

```
# CIF file

data_findsym-output
_audit_creation_method FINDSYM

_chemical_name_mineral 'beta Uranium'
_chemical_formula_sum 'U'

loop_
_publ_author_name
 'Charles W. Tucker, Jr.'
 'Peter Senio'
_journal_name_full
;
Acta Crystallographica
;
_journal_volume 6
_journal_year 1953
_journal_page_first 753
_journal_page_last 760
_publ_Section_title
;
 An improved determination of the crystal structure of $\beta$-uranium
;

# Found in Donohue, pp. 134-147

_aflow_proto 'A_tP30_136_bf2ij'
_aflow_params 'a,c/a,x2,x3,y3,x4,y4,x5,z5'
_aflow_params_values '10.59,0.532011331445,0.1033,0.3667,0.0383,0.5608,
    ↪ 0.2354,0.3183,0.27'
_aflow_Strukturbericht 'A_b'
_aflow_Pearson 'tP30'

_symmetry_space_group_name_Hall "-P 4n 2n"
_symmetry_space_group_name_H-M "P 42/m n m"
_symmetry_Int_Tables_number 136

_cell_length_a     10.59000
_cell_length_b     10.59000
_cell_length_c     5.63400
_cell_angle_alpha  90.00000
_cell_angle_beta   90.00000
_cell_angle_gamma  90.00000

loop_
_space_group_symop_id
_space_group_symop_operation_xyz
1 x,y,z
2 x+1/2,-y+1/2,-z+1/2
3 -x+1/2,y+1/2,-z+1/2
4 -x,-y,z
5 -y,-x,-z
6 -y+1/2,x+1/2,z+1/2
7 y+1/2,-x+1/2,z+1/2
8 y,x,-z
9 -x,-y,-z
10 -x+1/2,y+1/2,z+1/2
11 x+1/2,-y+1/2,z+1/2
12 x,y,-z
13 y,x,z
14 y+1/2,-x+1/2,-z+1/2
15 -y+1/2,x+1/2,-z+1/2
16 -y,-x,z

loop_
_atom_site_label
_atom_site_type_symbol
_atom_site_symmetry_multiplicity
_atom_site_Wyckoff_label
_atom_site_fract_x
_atom_site_fract_y
_atom_site_fract_z
_atom_site_occupancy
U1 U   2 b 0.00000 0.00000 0.50000 1.00000
U2 U   4 f 0.10330 0.10330 0.00000 1.00000
U3 U   8 i 0.36670 0.03830 0.00000 1.00000
U4 U   8 i 0.56080 0.23540 0.00000 1.00000
U5 U   8 j 0.31830 0.31830 0.27000 1.00000
```

$\beta$-U ($A_b$): A_tP30_136_bf2ij - POSCAR

```
A_tP30_136_bf2ij & a,c/a,x2,x3,y3,x4,y4,x5,z5 --params=10.59,
    ↪ 0.532011331445,0.1033,0.3667,0.0383,0.5608,0.2354,0.3183,0.27 &
    ↪  P4_2/mnm D_{4h}^{14} #136 (bfi^2j) & tP30 & A_b & U & beta &
    ↪  C. W. Tucker, Jr. and P. Senio, Acta Cryst 6, 753-760 (1953)
   1.0000000000000000
  10.59000000000000    0.00000000000000    0.00000000000000
   0.00000000000000   10.59000000000000    0.00000000000000
   0.00000000000000    0.00000000000000    5.63400000000000
    U
   30
Direct
   0.00000000000000    0.00000000000000    0.50000000000000    U    (2b)
   0.50000000000000    0.50000000000000    0.00000000000000    U    (2b)
   0.10330000000000    0.10330000000000    0.00000000000000    U    (4f)
   0.39670000000000    0.60330000000000    0.50000000000000    U    (4f)
   0.60330000000000    0.39670000000000    0.50000000000000    U    (4f)
   0.89670000000000    0.89670000000000    0.00000000000000    U    (4f)
   0.03830000000000    0.36670000000000    0.00000000000000    U    (8i)
   0.13330000000000    0.53830000000000    0.50000000000000    U    (8i)
   0.36670000000000    0.03830000000000    0.00000000000000    U    (8i)
   0.46170000000000    0.86670000000000    0.50000000000000    U    (8i)
   0.53830000000000    0.13330000000000    0.50000000000000    U    (8i)
   0.63330000000000    0.96170000000000    0.00000000000000    U    (8i)
   0.86670000000000    0.46170000000000    0.50000000000000    U    (8i)
   0.96170000000000    0.63330000000000    0.00000000000000    U    (8i)
   0.06080000000000    0.26460000000000    0.50000000000000    U    (8i)
  -0.06080000000000    0.73540000000000    0.50000000000000    U    (8i)
   0.23540000000000    0.56080000000000    0.00000000000000    U    (8i)
   0.26460000000000    0.06080000000000    0.50000000000000    U    (8i)
   0.43920000000000    0.76460000000000    0.00000000000000    U    (8i)
   0.56080000000000    0.23540000000000    0.00000000000000    U    (8i)
   0.73540000000000   -0.06080000000000    0.50000000000000    U    (8i)
   0.76460000000000    0.43920000000000    0.00000000000000    U    (8i)
   0.18170000000000    0.81830000000000    0.23000000000000    U    (8j)
   0.18170000000000    0.81830000000000    0.77000000000000    U    (8j)
   0.31830000000000    0.31830000000000    0.27000000000000    U    (8j)
   0.31830000000000    0.31830000000000    0.73000000000000    U    (8j)
   0.68170000000000    0.68170000000000    0.27000000000000    U    (8j)
   0.68170000000000    0.68170000000000    0.73000000000000    U    (8j)
   0.81830000000000    0.18170000000000    0.23000000000000    U    (8j)
   0.81830000000000    0.18170000000000    0.77000000000000    U    (8j)
```

$\beta$-BeO: AB_tP8_136_g_f - CIF

```
# CIF file

data_findsym-output
_audit_creation_method FINDSYM

_chemical_name_mineral 'beta beryllia'
_chemical_formula_sum 'Be O'

loop_
_publ_author_name
 'Deane K. Smith'
 'Carl F. Cline'
 'Stanley B. Austerman'
_journal_name_full
;
Acta Crystallographica
;
_journal_volume 18
_journal_year 1965
_journal_page_first 393
_journal_page_last 397
_publ_Section_title
;
 The Crystal Structure of $\beta$-Beryllia
;

_aflow_proto 'AB_tP8_136_g_f'
_aflow_params 'a,c/a,x1,x2'
_aflow_params_values '4.75,0.576842105263,0.31,0.336'
_aflow_Strukturbericht 'None'
_aflow_Pearson 'tP8'

_symmetry_space_group_name_Hall "-P 4n 2n"
_symmetry_space_group_name_H-M "P 42/m n m"
_symmetry_Int_Tables_number 136

_cell_length_a     4.75000
_cell_length_b     4.75000
_cell_length_c     2.74000
_cell_angle_alpha  90.00000
_cell_angle_beta   90.00000
```

```
_cell_angle_gamma  90.00000

loop_
_space_group_symop_id
_space_group_symop_operation_xyz
1  x,y,z
2  x+1/2,-y+1/2,-z+1/2
3  -x+1/2,y+1/2,-z+1/2
4  -x,-y,z
5  -y,-x,-z
6  -y+1/2,x+1/2,z+1/2
7  y+1/2,-x+1/2,z+1/2
8  y,x,-z
9  -x,-y,-z
10  -x+1/2,y+1/2,z+1/2
11  x+1/2,-y+1/2,z+1/2
12  x,y,-z
13  y,x,z
14  y+1/2,-x+1/2,-z+1/2
15  -y+1/2,x+1/2,-z+1/2
16  -y,-x,z

loop_
_atom_site_label
_atom_site_type_symbol
_atom_site_symmetry_multiplicity
_atom_site_Wyckoff_label
_atom_site_fract_x
_atom_site_fract_y
_atom_site_fract_z
_atom_site_occupancy
O1  O     4  f  0.31000  0.31000  0.00000  1.00000
Be1  Be    4  g  0.33600  0.66400  0.00000  1.00000
```

β-BeO: AB_tP8_136_g_f - POSCAR

```
AB_tP8_136_g_f & a,c/a,x1,x2 --params=4.75,0.576842105263,0.31,0.336 &
    ↪ P4_2/mmm  D_{4h}^{14} #136 (fg) & tP8 & & BeO & beta & D. K.
    ↪ Smith, C. F. Cline, and S. B. Austerman, Acta Cryst. 18,
    ↪ 393-397 (1965)
   1.0000000000000000
   4.75000000000000    0.00000000000000    0.00000000000000
   0.00000000000000    4.75000000000000    0.00000000000000
   0.00000000000000    0.00000000000000    2.74000000000000
   Be    O
    4    4
Direct
   0.16400000000000    0.16400000000000    0.50000000000000    Be    (4g)
   0.33600000000000    0.66400000000000    0.00000000000000    Be    (4g)
   0.66400000000000    0.33600000000000    0.00000000000000    Be    (4g)
   0.83600000000000    0.83600000000000    0.50000000000000    Be    (4g)
   0.19000000000000    0.81000000000000    0.50000000000000    O     (4f)
   0.31000000000000    0.31000000000000    0.00000000000000    O     (4f)
   0.69000000000000    0.69000000000000    0.00000000000000    O     (4f)
   0.81000000000000    0.19000000000000    0.50000000000000    O     (4f)
```

Rutile ($TiO_2$, C4): A2B_tP6_136_f_a - CIF

```
# CIF file

data_findsym-output
_audit_creation_method FINDSYM

_chemical_name_mineral 'Rutile'
_chemical_formula_sum 'Ti O2'

loop_
_publ_author_name
 'R. Jeffrey Swope'
 'Joseph R. Smyth'
 'Allen C. Larson'
_journal_name_full
;
American Mineralogist
;
_journal_volume 80
_journal_year 1995
_journal_page_first 448
_journal_page_last 453
_publ_Section_title
;
 H in rutile-type compounds: I. Single-crystal neutron and X-ray
    ↪ diffraction study of H in rutile
;

# Found in  AMS Database

_aflow_proto 'A2B_tP6_136_f_a'
_aflow_params 'a,c/a,x2'
_aflow_params_values '4.5922,0.644005052045,0.30496'
_aflow_Strukturbericht 'C4'
_aflow_Pearson 'tP6'

_symmetry_space_group_name_Hall "-P 4n 2n"
_symmetry_space_group_name_H-M "P 42/m n m"
_symmetry_Int_Tables_number 136

_cell_length_a     4.59220
_cell_length_b     4.59220
_cell_length_c     2.95740
_cell_angle_alpha  90.00000
_cell_angle_beta   90.00000
_cell_angle_gamma  90.00000

loop_
_space_group_symop_id
_space_group_symop_operation_xyz
1  x,y,z
2  x+1/2,-y+1/2,-z+1/2
3  -x+1/2,y+1/2,-z+1/2
4  -x,-y,z
5  -y,-x,-z
6  -y+1/2,x+1/2,z+1/2
7  y+1/2,-x+1/2,z+1/2
8  y,x,-z
9  -x,-y,-z
10  -x+1/2,y+1/2,z+1/2
11  x+1/2,-y+1/2,z+1/2
12  x,y,-z
13  y,x,z
14  y+1/2,-x+1/2,-z+1/2
15  -y+1/2,x+1/2,-z+1/2
16  -y,-x,z

loop_
_atom_site_label
_atom_site_type_symbol
_atom_site_symmetry_multiplicity
_atom_site_Wyckoff_label
_atom_site_fract_x
_atom_site_fract_y
_atom_site_fract_z
_atom_site_occupancy
Ti1  Ti    2  a  0.00000  0.00000  0.00000  1.00000
O1   O     4  f  0.30496  0.30496  0.00000  1.00000
```

Rutile ($TiO_2$, C4): A2B_tP6_136_f_a - POSCAR

```
A2B_tP6_136_f_a & a,c/a,x2 --params=4.5922,0.644005052045,0.30496 & P4_2
    ↪ /mmm    D_{4h}^{14} #136 (af) & tP6 & C4 & TiO2 & Rutile & R. J.
    ↪ Swope, J. R. Smithe, and A. C. Larson, Am. Mineral. 80,
    ↪ 448-453 (1995)
   1.0000000000000000
   4.59220000000000    0.00000000000000    0.00000000000000
   0.00000000000000    4.59220000000000    0.00000000000000
   0.00000000000000    0.00000000000000    2.95740000000000
   O    Ti
   4    2
Direct
   0.19504000000000    0.80496000000000    0.50000000000000    O     (4f)
   0.30496000000000    0.30496000000000    0.00000000000000    O     (4f)
   0.69504000000000    0.69504000000000    0.00000000000000    O     (4f)
   0.80496000000000    0.19504000000000    0.50000000000000    O     (4f)
   0.00000000000000    0.00000000000000    0.00000000000000    Ti    (2a)
   0.50000000000000    0.50000000000000    0.50000000000000    Ti    (2a)
```

σ-CrFe ($D8_b$): sigma_tP30_136_bf2ij - CIF

```
# CIF file

data_findsym-output
_audit_creation_method FINDSYM

_chemical_name_mineral 'sigma phase CrFe, different elements used to
    ↪ distinguish Wyckoff positions'
_chemical_formula_sum 'Pd Rh2 Ni4 Cr4 Fe4'

loop_
_publ_author_name
 'H. L. Yakel'
_journal_name_full
;
Acta Crystallographica B
;
_journal_volume 39
_journal_year 1983
_journal_page_first 20
_journal_page_last 28
_publ_Section_title
;
 Atom distributions in sigma phases. I. Fe and Cr atom distributions in
    ↪ a binary sigma phase equilibrated at 1063, 1013 and 923 K
;

# Found in  Pearson Vol II. pp. 2639

_aflow_proto 'sigma_tP30_136_bf2ij'
_aflow_params 'a,c/a,x2,x3,y3,x4,y4,x5,z5'
_aflow_params_values '8.7966,0.518177477662,0.39864,0.13122,0.46349,
    ↪ 0.06609,0.73933,0.18267,0.25202'
_aflow_Strukturbericht 'D8_b'
_aflow_Pearson 'tP30'

_symmetry_space_group_name_Hall "-P 4n 2n"
_symmetry_space_group_name_H-M "P 42/m n m"
_symmetry_Int_Tables_number 136

_cell_length_a     8.79660
_cell_length_b     8.79660
_cell_length_c     4.55820
_cell_angle_alpha  90.00000
_cell_angle_beta   90.00000
_cell_angle_gamma  90.00000

loop_
_space_group_symop_id
_space_group_symop_operation_xyz
1  x,y,z
2  x+1/2,-y+1/2,-z+1/2
3  -x+1/2,y+1/2,-z+1/2
```

```
4 -x,-y,z
5 -y,-x,-z
6 -y+1/2,x+1/2,z+1/2
7 y+1/2,-x+1/2,z+1/2
8 y,x,-z
9 -x,-y,-z
10 -x+1/2,y+1/2,z+1/2
11 x+1/2,-y+1/2,z+1/2
12 x,y,-z
13 y,x,z
14 y+1/2,-x+1/2,-z+1/2
15 -y+1/2,x+1/2,-z+1/2
16 -y,-x,z

loop_
_atom_site_label
_atom_site_type_symbol
_atom_site_symmetry_multiplicity
_atom_site_Wyckoff_label
_atom_site_fract_x
_atom_site_fract_y
_atom_site_fract_z
_atom_site_occupancy
Pd1 Pd    2 b 0.00000   0.00000  0.50000  1.00000
Rh1 Rh    4 f 0.39864   0.39864  0.00000  1.00000
Ni1 Ni    8 i 0.13122   0.46349  0.00000  1.00000
Cr1 Cr    8 i 0.06609   0.73933  0.00000  1.00000
Fe1 Fe    8 j 0.18267   0.18267  0.25202  1.00000
```

$\sigma$-CrFe (D8$_b$): sigma_tP30_136_bf2ij - POSCAR

```
sigma_tP30_136_bf2ij & a,c/a,x2,x3,y3,x4,y4,x5,z5 --params=8.7966,
    ↪ 0.518177477662,0.39864,0.13122,0.46349,0.06609,0.73933,0.18267,
    ↪ 0.25202 & P4_2/mnm  D_{4h}^{14} #136 (bfi^2j) & tP30 & D8_b &
    ↪ CrFe & sigma (disordered) & H. L. Yakel, Acta Cryst B 39, 20-28
    ↪ (1983)
   1.0000000000000000
   8.79660000000000   0.00000000000000   0.00000000000000
   0.00000000000000   8.79660000000000   0.00000000000000
   0.00000000000000   0.00000000000000   4.55820000000000
    M
   30
Direct
   0.00000000000000   0.00000000000000   0.50000000000000   M   (2b)
   0.50000000000000   0.50000000000000   0.00000000000000   M   (2b)
   0.10136000000000   0.89864000000000   0.50000000000000   M   (4f)
   0.39864000000000   0.39864000000000   0.00000000000000   M   (4f)
   0.60136000000000   0.60136000000000   0.00000000000000   M   (4f)
   0.89864000000000   0.10136000000000   0.50000000000000   M   (4f)
   0.03651000000000   0.63122000000000   0.50000000000000   M   (8i)
   0.13122000000000   0.46349000000000   0.00000000000000   M   (8i)
   0.36878000000000   0.96349000000000   0.50000000000000   M   (8i)
   0.46349000000000   0.13122000000000   0.00000000000000   M   (8i)
   0.53651000000000   0.86878000000000   0.00000000000000   M   (8i)
   0.63122000000000   0.03651000000000   0.50000000000000   M   (8i)
   0.86878000000000   0.53651000000000   0.00000000000000   M   (8i)
   0.96349000000000   0.36878000000000   0.50000000000000   M   (8i)
   0.06609000000000   0.73933000000000   0.00000000000000   M   (8i)
   0.23933000000000   0.43391000000000   0.50000000000000   M   (8i)
   0.26067000000000   0.93391000000000   0.00000000000000   M   (8i)
   0.43391000000000   0.23933000000000   0.50000000000000   M   (8i)
   0.56609000000000   0.76067000000000   0.50000000000000   M   (8i)
   0.73933000000000   0.06609000000000   0.00000000000000   M   (8i)
   0.76067000000000   0.56609000000000   0.50000000000000   M   (8i)
   0.93391000000000   0.26067000000000   0.00000000000000   M   (8i)
   0.18267000000000   0.18267000000000   0.25202000000000   M   (8j)
   0.18267000000000   0.18267000000000   0.74798000000000   M   (8j)
   0.31733000000000   0.68267000000000   0.24798000000000   M   (8j)
   0.31733000000000   0.68267000000000   0.75202000000000   M   (8j)
   0.68267000000000   0.31733000000000   0.24798000000000   M   (8j)
   0.68267000000000   0.31733000000000   0.75202000000000   M   (8j)
   0.81733000000000   0.81733000000000   0.25202000000000   M   (8j)
   0.81733000000000   0.81733000000000   0.74798000000000   M   (8j)
```

$\gamma$-N: A_tP4_136_f - CIF

```
# CIF file

data_findsym-output
_audit_creation_method FINDSYM

_chemical_name_mineral 'gamma nitrogen'
_chemical_formula_sum 'N'

loop_
_publ_author_name
 'R. L. Mills'
 'A. F. Schuch'
_journal_name_full
;
Physical Review Letters
;
_journal_volume 23
_journal_year 1969
_journal_page_first 1154
_journal_page_last 1156
_publ_Section_title
;
 Crystal Structure of Gamma Nitrogen
;

# Found in  Donohue, pp. 207-208

_aflow_proto 'A_tP4_136_f'
_aflow_params 'a,c/a,x1'
_aflow_params_values '3.957,1.29112964367,0.098'
_aflow_Strukturbericht 'None'
_aflow_Pearson 'tP4'

_symmetry_space_group_name_Hall "-P 4n 2n"
_symmetry_space_group_name_H-M "P 42/m n m"
_symmetry_Int_Tables_number 136

_cell_length_a      3.95700
_cell_length_b      3.95700
_cell_length_c      5.10900
_cell_angle_alpha  90.00000
_cell_angle_beta   90.00000
_cell_angle_gamma  90.00000

loop_
_space_group_symop_id
_space_group_symop_operation_xyz
1 x,y,z
2 x+1/2,-y+1/2,-z+1/2
3 -x+1/2,y+1/2,-z+1/2
4 -x,-y,z
5 -y,-x,-z
6 -y+1/2,x+1/2,z+1/2
7 y+1/2,-x+1/2,z+1/2
8 y,x,-z
9 -x,-y,-z
10 -x+1/2,y+1/2,z+1/2
11 x+1/2,-y+1/2,z+1/2
12 x,y,-z
13 y,x,z
14 y+1/2,-x+1/2,-z+1/2
15 -y+1/2,x+1/2,-z+1/2
16 -y,-x,z

loop_
_atom_site_label
_atom_site_type_symbol
_atom_site_symmetry_multiplicity
_atom_site_Wyckoff_label
_atom_site_fract_x
_atom_site_fract_y
_atom_site_fract_z
_atom_site_occupancy
N1 N    4 f 0.09800  0.09800  0.00000  1.00000
```

$\gamma$-N: A_tP4_136_f - POSCAR

```
A_tP4_136_f & a,c/a,x1 --params=3.957,1.29112964367,0.098 & P4_2/mnm  D_
    ↪ {4h}^{14} #136 (f) & tP4 & & N & gamma & R. L. Mills and A. F.
    ↪ Schuch, Phys. Rev. Lett. 23, 1154-1156 (1969)
   1.0000000000000000
   3.95700000000000   0.00000000000000   0.00000000000000
   0.00000000000000   3.95700000000000   0.00000000000000
   0.00000000000000   0.00000000000000   5.10900000000000
    N
    4
Direct
   0.09800000000000   0.09800000000000   0.00000000000000   N   (4f)
  -0.09800000000000  -0.09800000000000   0.00000000000000   N   (4f)
   0.40200000000000   0.59800000000000   0.50000000000000   N   (4f)
   0.59800000000000   0.40200000000000   0.50000000000000   N   (4f)
```

Cl (A18): A_tP16_138_j - CIF

```
# CIF file

data_findsym-output
_audit_creation_method FINDSYM

_chemical_name_mineral ''
_chemical_formula_sum 'Cl'

loop_
_publ_author_name
 'W. H. Keesom'
 'K. W. Taconis'
_journal_name_full
;
Physica
;
_journal_volume 3
_journal_year 1936
_journal_page_first 237
_journal_page_last 242
_publ_Section_title
;
 On the crystal structure of chlorine
;

# Found in  Donohue, pp. 396, Strukturbericht IV, pp. 1-4

_aflow_proto 'A_tP16_138_j'
_aflow_params 'a,c/a,x1,y1,z1'
_aflow_params_values '8.56,0.714953271028,0.375,-0.083,0.857'
_aflow_Strukturbericht 'A18'
_aflow_Pearson 'tP16'

_symmetry_space_group_name_Hall "-P 4ac 2ac"
_symmetry_space_group_name_H-M "P 42/n c m:2"
_symmetry_Int_Tables_number 138

_cell_length_a      8.56000
_cell_length_b      8.56000
_cell_length_c      6.12000
```

```
_cell_angle_alpha 90.00000
_cell_angle_beta  90.00000
_cell_angle_gamma 90.00000

loop_
_space_group_symop_id
_space_group_symop_operation_xyz
1 x,y,z
2 x+1/2,-y,-z+1/2
3 -x,y+1/2,-z+1/2
4 -x+1/2,-y+1/2,z
5 -y,-x,-z
6 -y+1/2,x,z+1/2
7 y,-x+1/2,z+1/2
8 y+1/2,x+1/2,-z
9 -x,-y,-z
10 -x+1/2,y,z+1/2
11 x,-y+1/2,z+1/2
12 x+1/2,y+1/2,-z
13 y,x,z
14 y+1/2,-x,-z+1/2
15 -y,x+1/2,-z+1/2
16 -y+1/2,-x+1/2,z

loop_
_atom_site_label
_atom_site_type_symbol
_atom_site_symmetry_multiplicity
_atom_site_Wyckoff_label
_atom_site_fract_x
_atom_site_fract_y
_atom_site_fract_z
_atom_site_occupancy
Cl1 Cl   16 j 0.37500 -0.08300 0.85700 1.00000
```

Cl (A18): A_tP16_138_j - POSCAR

```
A_tP16_138_j & a,c/a,x1,y1,z1 --params=8.56,0.714953271028,0.375,-0.083,
    ↪ 0.857 & P4_2/ncm   D_{4h}^{16} #138 (j) & tP16 & A18 & Cl & & W.
    ↪ H. Keesom and K. W. Taconis, Physica 3, 237-242 (1936)
   1.0000000000000000
   8.56000000000000   0.00000000000000   0.00000000000000
   0.00000000000000   8.56000000000000   0.00000000000000
   0.00000000000000   0.00000000000000   6.12000000000000
   Cl
   16
Direct
    0.3750000000000   -0.0830000000000    0.8570000000000   Cl   (16j)
    0.1250000000000    0.5830000000000    0.8570000000000   Cl   (16j)
    0.5830000000000    0.3750000000000    0.3570000000000   Cl   (16j)
   -0.0830000000000    0.1250000000000    0.3570000000000   Cl   (16j)
    0.6250000000000    0.4170000000000    0.6430000000000   Cl   (16j)
    0.8750000000000    0.0830000000000    0.6430000000000   Cl   (16j)
    0.4170000000000    0.8750000000000    0.1430000000000   Cl   (16j)
    0.0830000000000    0.6250000000000    0.1430000000000   Cl   (16j)
    0.6250000000000    0.0830000000000    0.1430000000000   Cl   (16j)
    0.8750000000000    0.4170000000000    0.1430000000000   Cl   (16j)
    0.4170000000000    0.6250000000000    0.6430000000000   Cl   (16j)
    0.0830000000000    0.8750000000000    0.6430000000000   Cl   (16j)
    0.3750000000000    0.5830000000000    0.3570000000000   Cl   (16j)
    0.1250000000000   -0.0830000000000    0.3570000000000   Cl   (16j)
    0.5830000000000    0.1250000000000    0.8570000000000   Cl   (16j)
   -0.0830000000000    0.3750000000000    0.8570000000000   Cl   (16j)
```

$Al_3Zr$ ($D0_{23}$): A3B_tI16_139_cde_e - CIF

```
# CIF file

data_findsym-output
_audit_creation_method FINDSYM

_chemical_name_mineral ''
_chemical_formula_sum 'Al3 Zr'

loop_
_publ_author_name
 'Y. Ma'
 'C. R{\o}mming'
 'B. Lebech'
 'J. Gj{\o}nnes'
 'J. Taft{\o}'
_journal_name_full
;
Acta Crystallographica B
;
_journal_volume 48
_journal_year 1992
_journal_page_first 11
_journal_page_last 16
_publ_Section_title
;
 Structure Refinement of Al3Zr using Single-Crystal X-ray Diffraction,
    ↪ Powder Neutron Diffraction and CBED
;

# Found in  ghosh05:AlTM

_aflow_proto 'A3B_tI16_139_cde_e'
_aflow_params 'a,c/a,z3,z4'
_aflow_params_values '3.9993,4.3215062636,0.37498,0.11886'
_aflow_Strukturbericht 'D0_23'
_aflow_Pearson 'tI16'

_symmetry_space_group_name_Hall "-I 4 2"
_symmetry_space_group_name_H-M "I 4/m m m"
_symmetry_Int_Tables_number 139

_cell_length_a     3.99930
_cell_length_b     3.99930
_cell_length_c     17.28300
_cell_angle_alpha  90.00000
_cell_angle_beta   90.00000
_cell_angle_gamma  90.00000

loop_
_space_group_symop_id
_space_group_symop_operation_xyz
1 x,y,z
2 x,-y,-z
3 -x,y,-z
4 -x,-y,z
5 -y,-x,-z
6 -y,x,z
7 y,-x,z
8 y,x,-z
9 -x,-y,-z
10 -x,y,z
11 x,-y,z
12 x,y,-z
13 y,x,z
14 y,-x,-z
15 -y,x,-z
16 -y,-x,z
17 x+1/2,y+1/2,z+1/2
18 x+1/2,-y+1/2,-z+1/2
19 -x+1/2,y+1/2,-z+1/2
20 -x+1/2,-y+1/2,z+1/2
21 -y+1/2,-x+1/2,-z+1/2
22 -y+1/2,x+1/2,z+1/2
23 y+1/2,-x+1/2,z+1/2
24 y+1/2,x+1/2,-z+1/2
25 -x+1/2,-y+1/2,-z+1/2
26 -x+1/2,y+1/2,z+1/2
27 x+1/2,-y+1/2,z+1/2
28 x+1/2,y+1/2,-z+1/2
29 y+1/2,x+1/2,z+1/2
30 y+1/2,-x+1/2,-z+1/2
31 -y+1/2,x+1/2,-z+1/2
32 -y+1/2,-x+1/2,z+1/2

loop_
_atom_site_label
_atom_site_type_symbol
_atom_site_symmetry_multiplicity
_atom_site_Wyckoff_label
_atom_site_fract_x
_atom_site_fract_y
_atom_site_fract_z
_atom_site_occupancy
Al1 Al   4 c 0.00000 0.50000 0.00000 1.00000
Al2 Al   4 d 0.00000 0.50000 0.25000 1.00000
Al3 Al   4 e 0.00000 0.00000 0.37498 1.00000
Zr1 Zr   4 e 0.00000 0.00000 0.11886 1.00000
```

$Al_3Zr$ ($D0_{23}$): A3B_tI16_139_cde_e - POSCAR

```
A3B_tI16_139_cde_e & a,c/a,z3,z4 --params=3.9993,4.3215062636,0.37498,
    ↪ 0.11886 & I4/mmm    D_{4h}^{17} #139 (cde^2) & tI16 & D0_{23} &
    ↪  Al3Zr & & Y. Ma, C. R{\o}mming, B. Lebech, J. Gj{\o}nnes and
    ↪ J. Taft{\o}, Acta Cryst. B 48, 11-16 (1992)
   1.0000000000000000
  -1.99965000000000   1.99965000000000   8.64150000000000
   1.99965000000000  -1.99965000000000   8.64150000000000
   1.99965000000000   1.99965000000000  -8.64150000000000
   Al   Zr
    6    2
Direct
   0.00000000000000   0.50000000000000   0.50000000000000   Al   (4c)
   0.50000000000000   0.00000000000000   0.50000000000000   Al   (4c)
   0.25000000000000   0.75000000000000   0.50000000000000   Al   (4d)
   0.75000000000000   0.25000000000000   0.50000000000000   Al   (4d)
   0.37498000000000   0.37498000000000   0.00000000000000   Al   (4e)
   0.62502000000000   0.62502000000000   0.00000000000000   Al   (4e)
   0.11886000000000   0.11886000000000   0.00000000000000   Zr   (4e)
   0.88114000000000   0.88114000000000   0.00000000000000   Zr   (4e)
```

Hypothetical BCT5 Si: A_tI4_139_e - CIF

```
# CIF file

data_findsym-output
_audit_creation_method FINDSYM

_chemical_name_mineral 'BCT5'
_chemical_formula_sum 'Si'

loop_
_publ_author_name
 'L. L. Boyer'
 'Efthimios Kaxiras'
 'J. L. Feldman'
 'J. Q. Broughton'
 'M. J. Mehl'
_journal_name_full
;
Physical Review Letters
;
_journal_volume 67
_journal_year 1991
_journal_page_first 715
```

```
_journal_page_last 718
_publ_Section_title
;
 New low-energy crystal structure for silicon
;

_aflow_proto 'A_tI4_139_e'
_aflow_params 'a,c/a,z1'
_aflow_params_values '3.34916,1.94217355994,0.819'
_aflow_Strukturbericht 'None'
_aflow_Pearson 'tI4'

_symmetry_space_group_name_Hall "-I 4 2"
_symmetry_space_group_name_H-M "I 4/m m m"
_symmetry_Int_Tables_number 139

_cell_length_a      3.34916
_cell_length_b      3.34916
_cell_length_c      6.50465
_cell_angle_alpha  90.00000
_cell_angle_beta   90.00000
_cell_angle_gamma  90.00000

loop_
_space_group_symop_id
_space_group_symop_operation_xyz
1 x,y,z
2 x,-y,-z
3 -x,y,-z
4 -x,-y,z
5 -y,-x,-z
6 -y,x,z
7 y,-x,z
8 y,x,-z
9 -x,-y,-z
10 -x,y,z
11 x,-y,z
12 x,y,-z
13 y,x,z
14 y,-x,-z
15 -y,x,-z
16 -y,-x,z
17 x+1/2,y+1/2,z+1/2
18 x+1/2,-y+1/2,-z+1/2
19 -x+1/2,y+1/2,-z+1/2
20 -x+1/2,-y+1/2,z+1/2
21 -y+1/2,-x+1/2,-z+1/2
22 -y+1/2,x+1/2,z+1/2
23 y+1/2,-x+1/2,z+1/2
24 y+1/2,x+1/2,-z+1/2
25 -x+1/2,-y+1/2,-z+1/2
26 -x+1/2,y+1/2,z+1/2
27 x+1/2,-y+1/2,z+1/2
28 x+1/2,y+1/2,-z+1/2
29 y+1/2,x+1/2,z+1/2
30 y+1/2,-x+1/2,-z+1/2
31 -y+1/2,x+1/2,-z+1/2
32 -y+1/2,-x+1/2,z+1/2

loop_
_atom_site_label
_atom_site_type_symbol
_atom_site_symmetry_multiplicity
_atom_site_Wyckoff_label
_atom_site_fract_x
_atom_site_fract_y
_atom_site_fract_z
_atom_site_occupancy
Si1 Si   4 e 0.00000 0.00000 0.81900 1.00000
```

Hypothetical BCT5 Si: A_tI4_139_e - POSCAR

```
A_tI4_139_e & a,c/a,z1 --params=3.34916,1.94217355994,0.819 & I4/mmm
    ↪ D_{4h}^{17} #139 (e) & tI4 & & Si & BCT5 & L. L. Boyer, E.
    ↪ Kaxiras, J. L. Feldman, J. Q. Broughton and M. J. Mehl, PRL 67,
    ↪ 715-718 (1991)
   1.0000000000000000
  -1.67458127566500   1.67458127566500   3.25232312221200
   1.67458127566500  -1.67458127566500   3.25232312221200
   1.67458127566500   1.67458127566500  -3.25232312221200
   Si
     2
Direct
   0.18100000000000   0.18100000000000   0.00000000000000   Si   (4e)
   0.81900000000000   0.81900000000000   0.00000000000000   Si   (4e)
```

0201 [(La,Ba)$_2$CuO$_4$]: AB2C4_tI14_139_a_e_ce - CIF

```
# CIF file

data_findsym-output
_audit_creation_method FINDSYM

_chemical_name_mineral '(La,Ba)CuO4'
_chemical_formula_sum 'La2 Cu O4'

loop_
_publ_author_name
 'J. D. Jorgensen'
 'H.-B. Sch\"{u}ttler'
 'D. G. Hinks'
 'D. W. Capone, II'
 'K. Zhang'
 'M. B. Brodsky'
_journal_name_full
;
Physical Review Letters
;
_journal_volume 58
_journal_year 1987
_journal_page_first 1024
_journal_page_last 1029
_publ_Section_title
;
 Lattice instability and high-$T_c$ superconductivity in La$_{2-x}
    ↪ $Ba$_x$CuO$_4$
;

# Found in  shaked94:hightcstruct

_aflow_proto 'AB2C4_tI14_139_a_e_ce'
_aflow_params 'a,c/a,z3,z4'
_aflow_params_values '3.7817,3.50337149959,0.36075,0.1824'
_aflow_Strukturbericht 'None'
_aflow_Pearson 'tI14'

_symmetry_space_group_name_Hall "-I 4 2"
_symmetry_space_group_name_H-M "I 4/m m m"
_symmetry_Int_Tables_number 139

_cell_length_a      3.78170
_cell_length_b      3.78170
_cell_length_c     13.24870
_cell_angle_alpha  90.00000
_cell_angle_beta   90.00000
_cell_angle_gamma  90.00000

loop_
_space_group_symop_id
_space_group_symop_operation_xyz
1 x,y,z
2 x,-y,-z
3 -x,y,-z
4 -x,-y,z
5 -y,-x,-z
6 -y,x,z
7 y,-x,z
8 y,x,-z
9 -x,-y,-z
10 -x,y,z
11 x,-y,z
12 x,y,-z
13 y,x,z
14 y,-x,-z
15 -y,x,-z
16 -y,-x,z
17 x+1/2,y+1/2,z+1/2
18 x+1/2,-y+1/2,-z+1/2
19 -x+1/2,y+1/2,-z+1/2
20 -x+1/2,-y+1/2,z+1/2
21 -y+1/2,-x+1/2,-z+1/2
22 -y+1/2,x+1/2,z+1/2
23 y+1/2,-x+1/2,z+1/2
24 y+1/2,x+1/2,-z+1/2
25 -x+1/2,-y+1/2,-z+1/2
26 -x+1/2,y+1/2,z+1/2
27 x+1/2,-y+1/2,z+1/2
28 x+1/2,y+1/2,-z+1/2
29 y+1/2,x+1/2,z+1/2
30 y+1/2,-x+1/2,-z+1/2
31 -y+1/2,x+1/2,-z+1/2
32 -y+1/2,-x+1/2,z+1/2

loop_
_atom_site_label
_atom_site_type_symbol
_atom_site_symmetry_multiplicity
_atom_site_Wyckoff_label
_atom_site_fract_x
_atom_site_fract_y
_atom_site_fract_z
_atom_site_occupancy
Cu1 Cu   2 a 0.00000 0.00000 0.00000 1.00000
O1  O    4 c 0.00000 0.50000 0.00000 1.00000
La1 La   4 e 0.00000 0.00000 0.36075 1.00000
O2  O    4 e 0.00000 0.00000 0.18240 1.00000
```

0201 [(La,Ba)$_2$CuO$_4$]: AB2C4_tI14_139_a_e_ce - POSCAR

```
AB2C4_tI14_139_a_e_ce & a,c/a,z3,z4 --params=3.7817,3.50337149959,
    ↪ 0.36075,0.1824 & I4/mmm    D_{4h}^{17} #139 (ace^2) & tI14 & &
    ↪ (La,Ba)2CuO4 & 0201 Superconductor & J. D. Jorgensen et al.,
    ↪ PRL 58, 1024-1028 (1987)
   1.0000000000000000
  -1.890850000000   1.890850000000   6.624350000000
   1.890850000000  -1.890850000000   6.624350000000
   1.890850000000   1.890850000000  -6.624350000000
   Cu    La    O
    1     2    4
Direct
   0.000000000000   0.000000000000   0.000000000000   Cu   (2a)
   0.360750000000   0.360750000000   0.000000000000   La   (4e)
   0.639250000000   0.639250000000   0.000000000000   La   (4e)
   0.500000000000   0.000000000000   0.500000000000    O   (4c)
   0.000000000000   0.500000000000   0.500000000000    O   (4c)
   0.182400000000   0.182400000000   0.000000000000    O   (4e)
   0.817600000000   0.817600000000   0.000000000000    O   (4e)
```

$Mn_{12}Th$ ($D2_b$): A12B_tI26_139_fij_a - CIF

```
# CIF file

data_findsym-output
_audit_creation_method FINDSYM

_chemical_name_mineral ''
_chemical_formula_sum 'Mn12 Th'

loop_
_publ_author_name
 'John V. Florio'
 'R. E. Rundle'
 'A. I. Snow'
_journal_name_full
;
Acta Crystallographica
;
_journal_volume 5
_journal_year 1952
_journal_page_first 449
_journal_page_last 457
_publ_Section_title
;
 Compounds of thorium with transition metals. I. The thorium-manganese
       ↪ system
;

# Found in Pearson's Handbook, Vol. IV, pp. 4396

_aflow_proto 'A12B_tI26_139_fij_a'
_aflow_params 'a,c/a,x3,x4'
_aflow_params_values '8.47,0.584415584416,0.361,0.278'
_aflow_Strukturbericht 'D2_b'
_aflow_Pearson 'tI26'

_symmetry_space_group_name_Hall "-I 4 2"
_symmetry_space_group_name_H-M "I 4/m m m"
_symmetry_Int_Tables_number 139

_cell_length_a      8.47000
_cell_length_b      8.47000
_cell_length_c      4.95000
_cell_angle_alpha 90.00000
_cell_angle_beta  90.00000
_cell_angle_gamma 90.00000

loop_
_space_group_symop_id
_space_group_symop_operation_xyz
1 x,y,z
2 x,-y,-z
3 -x,y,-z
4 -x,-y,z
5 -y,-x,-z
6 -y,x,z
7 y,-x,z
8 y,x,-z
9 -x,-y,-z
10 -x,y,z
11 x,-y,z
12 x,y,-z
13 y,x,z
14 y,-x,-z
15 -y,x,-z
16 -y,-x,z
17 x+1/2,y+1/2,z+1/2
18 x+1/2,-y+1/2,-z+1/2
19 -x+1/2,y+1/2,-z+1/2
20 -x+1/2,-y+1/2,z+1/2
21 -y+1/2,-x+1/2,-z+1/2
22 -y+1/2,x+1/2,z+1/2
23 y+1/2,-x+1/2,z+1/2
24 y+1/2,x+1/2,-z+1/2
25 -x+1/2,-y+1/2,-z+1/2
26 -x+1/2,y+1/2,z+1/2
27 x+1/2,-y+1/2,z+1/2
28 x+1/2,y+1/2,-z+1/2
29 y+1/2,x+1/2,z+1/2
30 y+1/2,-x+1/2,-z+1/2
31 -y+1/2,x+1/2,-z+1/2
32 -y+1/2,-x+1/2,z+1/2

loop_
_atom_site_label
_atom_site_type_symbol
_atom_site_symmetry_multiplicity
_atom_site_Wyckoff_label
_atom_site_fract_x
_atom_site_fract_y
_atom_site_fract_z
_atom_site_occupancy
Th1 Th   2 a 0.00000 0.00000 0.00000 1.00000
Mn1 Mn   8 f 0.25000 0.25000 0.25000 1.00000
Mn2 Mn   8 i 0.36100 0.00000 0.00000 1.00000
Mn3 Mn   8 j 0.27800 0.50000 0.00000 1.00000
```

$Mn_{12}Th$ ($D2_b$): A12B_tI26_139_fij_a - POSCAR

```
A12B_tI26_139_fij_a & a,c/a,x3,x4 --params=8.47,0.584415584416,0.361,
      ↪ 0.278 & I4/mmm    D_{4h}^{17} #139 (afij) & tI26 & D2_b &
      ↪ Mn12Th & & J. V. Florio, R. E. Rundle and A. I. Snow, Acta
      ↪ Cryst. 5, 449-457 (1952)
   1.0000000000000000
  -4.23500000000000    4.23500000000000    2.47500000000000
   4.23500000000000   -4.23500000000000    2.47500000000000
   4.23500000000000    4.23500000000000   -2.47500000000000
    Mn     Th
    12      1
Direct
   0.00000000000000    0.00000000000000    0.50000000000000    Mn   (8f)
   0.00000000000000    0.50000000000000    0.00000000000000    Mn   (8f)
   0.50000000000000    0.00000000000000    0.00000000000000    Mn   (8f)
   0.50000000000000    0.50000000000000    0.50000000000000    Mn   (8f)
   0.00000000000000    0.36100000000000    0.36100000000000    Mn   (8i)
   0.00000000000000    0.63900000000000    0.63900000000000    Mn   (8i)
   0.36100000000000    0.00000000000000    0.36100000000000    Mn   (8i)
   0.63900000000000    0.00000000000000    0.63900000000000    Mn   (8i)
   0.27800000000000    0.50000000000000    0.77800000000000    Mn   (8j)
   0.50000000000000    0.27800000000000    0.77800000000000    Mn   (8j)
   0.50000000000000    0.72200000000000    0.22200000000000    Mn   (8j)
   0.72200000000000    0.50000000000000    0.22200000000000    Mn   (8j)
   0.00000000000000    0.00000000000000    0.00000000000000    Th   (2a)
```

In (A6): A_tI2_139_a - CIF

```
# CIF file

data_findsym-output
_audit_creation_method FINDSYM

_chemical_name_mineral 'Indium'
_chemical_formula_sum 'In'

loop_
_publ_author_name
 'V. T. Deshpande'
 'R. R. Pawar'
_journal_name_full
;
Acta Crystallographica A
;
_journal_volume 25
_journal_year 1969
_journal_page_first 415
_journal_page_last 416
_publ_Section_title
;
 Anisotropic Thermal Expansion of Indium
;

# Found in Donohue, pp. 244-246

_aflow_proto 'A_tI2_139_a'
_aflow_params 'a,c/a'
_aflow_params_values '4.6002,1.07523585931'
_aflow_Strukturbericht 'A6'
_aflow_Pearson 'tI2'

_symmetry_space_group_name_Hall "-I 4 2"
_symmetry_space_group_name_H-M "I 4/m m m"
_symmetry_Int_Tables_number 139

_cell_length_a      4.60020
_cell_length_b      4.60020
_cell_length_c      4.94630
_cell_angle_alpha 90.00000
_cell_angle_beta  90.00000
_cell_angle_gamma 90.00000

loop_
_space_group_symop_id
_space_group_symop_operation_xyz
1 x,y,z
2 x,-y,-z
3 -x,y,-z
4 -x,-y,z
5 -y,-x,-z
6 -y,x,z
7 y,-x,z
8 y,x,-z
9 -x,-y,-z
10 -x,y,z
11 x,-y,z
12 x,y,-z
13 y,x,z
14 y,-x,-z
15 -y,x,-z
16 -y,-x,z
17 x+1/2,y+1/2,z+1/2
18 x+1/2,-y+1/2,-z+1/2
19 -x+1/2,y+1/2,-z+1/2
20 -x+1/2,-y+1/2,z+1/2
21 -y+1/2,-x+1/2,-z+1/2
22 -y+1/2,x+1/2,z+1/2
23 y+1/2,-x+1/2,z+1/2
24 y+1/2,x+1/2,-z+1/2
25 -x+1/2,-y+1/2,-z+1/2
26 -x+1/2,y+1/2,z+1/2
27 x+1/2,-y+1/2,z+1/2
28 x+1/2,y+1/2,-z+1/2
29 y+1/2,x+1/2,z+1/2
30 y+1/2,-x+1/2,-z+1/2
31 -y+1/2,x+1/2,-z+1/2
32 -y+1/2,-x+1/2,z+1/2

loop_
_atom_site_label
_atom_site_type_symbol
_atom_site_symmetry_multiplicity
_atom_site_Wyckoff_label
```

```
_atom_site_fract_x
_atom_site_fract_y
_atom_site_fract_z
_atom_site_occupancy
In1 In    2 a 0.00000 0.00000 0.00000 1.00000
```

In (A6): A_tI2_139_a - POSCAR

```
A_tI2_139_a & a,c/a --params=4.6002,1.07523585931 & I4/mmm    D_{4h}^{17
    ↪ } #139 (a) & tI2 & A6 & In & & V. T. Deshpande and R. R. Pawar,
    ↪  Acta Cryst. B 25, 415-416 (1969)
   1.0000000000000000
 -2.300100000000   2.300100000000   2.473150000000
  2.300100000000  -2.300100000000   2.473150000000
  2.300100000000   2.300100000000  -2.473150000000
   In
    1
Direct
   0.00000000000000    0.00000000000000    0.00000000000000    In     (2a)
```

Hypothetical Tetrahedrally Bonded Carbon with 4-Member Rings: A_tI8_139_h - CIF

```
# CIF file

data_findsym-output
_audit_creation_method FINDSYM

_chemical_name_mineral 'Theoretical Carbon Structure'
_chemical_formula_sum 'C'

loop_
_publ_author_name
 'Peter A. Schultz'
 'Kevin Leung'
 'E. B. Stechel'
_journal_name_full
;
Physical Review B
;
_journal_volume 59
_journal_year 1999
_journal_page_first 733
_journal_page_last 741
_publ_Section_title
;
 Small rings and amorphous tetrahedral carbon
;

_aflow_proto 'A_tI8_139_h'
_aflow_params 'a,c/a,x1'
_aflow_params_values '4.33184,0.574102459925,0.17916'
_aflow_Strukturbericht 'None'
_aflow_Pearson 'tI8'

_symmetry_space_group_name_Hall "-I 4 2"
_symmetry_space_group_name_H-M "I 4/m m m"
_symmetry_Int_Tables_number 139

_cell_length_a    4.33184
_cell_length_b    4.33184
_cell_length_c    2.48692
_cell_angle_alpha 90.00000
_cell_angle_beta  90.00000
_cell_angle_gamma 90.00000

loop_
_space_group_symop_id
_space_group_symop_operation_xyz
1 x,y,z
2 x,-y,-z
3 -x,y,-z
4 -x,-y,z
5 -y,-x,-z
6 -y,x,z
7 y,-x,z
8 y,x,-z
9 -x,-y,-z
10 -x,y,z
11 x,-y,z
12 x,y,-z
13 y,x,z
14 y,-x,-z
15 -y,x,-z
16 -y,-x,z
17 x+1/2,y+1/2,z+1/2
18 x+1/2,-y+1/2,-z+1/2
19 -x+1/2,y+1/2,-z+1/2
20 -x+1/2,-y+1/2,z+1/2
21 -y+1/2,-x+1/2,-z+1/2
22 -y+1/2,x+1/2,z+1/2
23 y+1/2,-x+1/2,z+1/2
24 y+1/2,x+1/2,-z+1/2
25 -x+1/2,-y+1/2,-z+1/2
26 -x+1/2,y+1/2,z+1/2
27 x+1/2,-y+1/2,z+1/2
28 x+1/2,y+1/2,-z+1/2
29 y+1/2,x+1/2,z+1/2
30 y+1/2,-x+1/2,-z+1/2
31 -y+1/2,x+1/2,-z+1/2
32 -y+1/2,-x+1/2,z+1/2

loop_
_atom_site_label
_atom_site_type_symbol
_atom_site_symmetry_multiplicity
_atom_site_Wyckoff_label
_atom_site_fract_x
_atom_site_fract_y
_atom_site_fract_z
_atom_site_occupancy
C1 C   8 h 0.17916 0.17916 0.00000 1.00000
```

Hypothetical Tetrahedrally Bonded Carbon with 4-Member Rings: A_tI8_139_h - POSCAR

```
A_tI8_139_h & a,c/a,x1 --params=4.33184,0.574102459925,0.17916 & I4/mmm
    ↪    D_{4h}^{17} #139 (h) & tI8 & & C & hypothetical 4-member
    ↪ ring structure & P. A. Schultz, K. Leung and E. B. Stechel, PRB
    ↪  59, 733-741 (1999)
   1.0000000000000000
  -2.16592000000000    2.16592000000000    1.24346000000000
   2.16592000000000   -2.16592000000000    1.24346000000000
   2.16592000000000    2.16592000000000   -1.24346000000000
   C
   4
Direct
   0.17916000000000    0.17916000000000    0.35832000000000     C    (8h)
   0.17916000000000    0.82084000000000    0.00000000000000     C    (8h)
   0.82084000000000    0.17916000000000    0.00000000000000     C    (8h)
   0.82084000000000    0.82084000000000    0.64168000000000     C    (8h)
```

$Al_3Ti$ ($D0_{22}$): A3B_tI8_139_bd_a - CIF

```
# CIF file

data_findsym-output
_audit_creation_method FINDSYM

_chemical_name_mineral ''
_chemical_formula_sum 'Al3 Ti'

loop_
_publ_author_name
 'P. Norby'
 'A. N{\o}rlund Christensen'
_journal_name_full
;
Acta Chemica Scandinavica
;
_journal_volume A40
_journal_year 1986
_journal_page_first 157
_journal_page_last 159
_publ_Section_title
;
 Preparation and Structure of Al$_3$Ti
;

# Found in Pearson's Handbook, Vol. I, p. 1023

_aflow_proto 'A3B_tI8_139_bd_a'
_aflow_params 'a,c/a'
_aflow_params_values '3.8537,2.22744375535'
_aflow_Strukturbericht 'D0_22'
_aflow_Pearson 'tI8'

_symmetry_space_group_name_Hall "-I 4 2"
_symmetry_space_group_name_H-M "I 4/m m m"
_symmetry_Int_Tables_number 139

_cell_length_a    3.85370
_cell_length_b    3.85370
_cell_length_c    8.58390
_cell_angle_alpha 90.00000
_cell_angle_beta  90.00000
_cell_angle_gamma 90.00000

loop_
_space_group_symop_id
_space_group_symop_operation_xyz
1 x,y,z
2 x,-y,-z
3 -x,y,-z
4 -x,-y,z
5 -y,-x,-z
6 -y,x,z
7 y,-x,z
8 y,x,-z
9 -x,-y,-z
10 -x,y,z
11 x,-y,z
12 x,y,-z
13 y,x,z
14 y,-x,-z
15 -y,x,-z
16 -y,-x,z
17 x+1/2,y+1/2,z+1/2
18 x+1/2,-y+1/2,-z+1/2
19 -x+1/2,y+1/2,-z+1/2
20 -x+1/2,-y+1/2,z+1/2
21 -y+1/2,-x+1/2,-z+1/2
22 -y+1/2,x+1/2,z+1/2
23 y+1/2,-x+1/2,z+1/2
24 y+1/2,x+1/2,-z+1/2
25 -x+1/2,-y+1/2,-z+1/2
26 -x+1/2,y+1/2,z+1/2
27 x+1/2,-y+1/2,z+1/2
28 x+1/2,y+1/2,-z+1/2
29 y+1/2,x+1/2,z+1/2
30 y+1/2,-x+1/2,-z+1/2
31 -y+1/2,x+1/2,-z+1/2
```

```
32 -y+1/2,-x+1/2,z+1/2

loop_
_atom_site_label
_atom_site_type_symbol
_atom_site_symmetry_multiplicity
_atom_site_Wyckoff_label
_atom_site_fract_x
_atom_site_fract_y
_atom_site_fract_z
_atom_site_occupancy
Ti1 Ti    2 a 0.00000 0.00000 0.00000 1.00000
Al1 Al    2 b 0.00000 0.00000 0.50000 1.00000
Al2 Al    4 d 0.00000 0.50000 0.25000 1.00000
```

$Al_3Ti$ ($D0_{22}$): A3B_tI8_139_bd_a - POSCAR

```
A3B_tI8_139_bd_a & a,c/a --params=3.8537,2.22744375535 & I4/mmm    D_{4h
   ↪ }^{17} #139 (abd) & tI8 & D0_{22} & Al3Ti & & P. Norby and A.
   ↪ Norlund Christensen, Acta Chem. Scand. A 40, 157-159 (1986)
   1.0000000000000000
  -1.92685000000000    1.92685000000000    4.29195000000000
   1.92685000000000   -1.92685000000000    4.29195000000000
   1.92685000000000    1.92685000000000   -4.29195000000000
   Al     Ti
    3      1
Direct
   0.50000000000000    0.50000000000000    0.00000000000000    Al    (2b)
   0.25000000000000    0.75000000000000    0.50000000000000    Al    (4d)
   0.75000000000000    0.25000000000000    0.50000000000000    Al    (4d)
   0.00000000000000    0.00000000000000    0.00000000000000    Ti    (2a)
```

$MoSi_2$ ($C11_b$): AB2_tI6_139_a_e - CIF

```
# CIF file

data_findsym-output
_audit_creation_method FINDSYM

_chemical_name_mineral ''
_chemical_formula_sum 'Mo Si2'

loop_
_publ_author_name
 'Y. Harada'
 'M. Morinaga'
 'D. Saso'
 'M. Takata'
 'M. Sakata'
_journal_name_full
;
Intermetallics
;
_journal_volume 6
_journal_year 1998
_journal_page_first 523
_journal_page_last 527
_publ_Section_title
;
 Refinement of crystal structure in MoSi$_2$
;

_aflow_proto 'AB2_tI6_139_a_e'
_aflow_params 'a,c/a,z2'
_aflow_params_values '3.2064,2.44754241517,0.3353'
_aflow_Strukturbericht 'C11_b'
_aflow_Pearson 'tI6'

_symmetry_space_group_name_Hall "-I 4 2"
_symmetry_space_group_name_H-M "I 4/m m m"
_symmetry_Int_Tables_number 139

_cell_length_a      3.20640
_cell_length_b      3.20640
_cell_length_c      7.84780
_cell_angle_alpha 90.00000
_cell_angle_beta   90.00000
_cell_angle_gamma 90.00000

loop_
_space_group_symop_id
_space_group_symop_operation_xyz
1 x,y,z
2 x,-y,-z
3 -x,y,-z
4 -x,-y,z
5 -y,-x,-z
6 -y,x,z
7 y,-x,z
8 y,x,-z
9 -x,-y,-z
10 -x,y,z
11 x,-y,z
12 x,y,-z
13 y,x,z
14 y,-x,-z
15 -y,x,-z
16 -y,-x,z
17 x+1/2,y+1/2,z+1/2
18 x+1/2,-y+1/2,-z+1/2
19 -x+1/2,y+1/2,-z+1/2
20 -x+1/2,-y+1/2,z+1/2
21 -y+1/2,-x+1/2,-z+1/2
22 -y+1/2,x+1/2,z+1/2
23 y+1/2,-x+1/2,z+1/2
24 y+1/2,x+1/2,-z+1/2
25 -x+1/2,-y+1/2,-z+1/2
26 -x+1/2,y+1/2,z+1/2
27 x+1/2,-y+1/2,z+1/2
28 x+1/2,y+1/2,-z+1/2
29 y+1/2,x+1/2,z+1/2
30 y+1/2,-x+1/2,-z+1/2
31 -y+1/2,x+1/2,-z+1/2
32 -y+1/2,-x+1/2,z+1/2

loop_
_atom_site_label
_atom_site_type_symbol
_atom_site_symmetry_multiplicity
_atom_site_Wyckoff_label
_atom_site_fract_x
_atom_site_fract_y
_atom_site_fract_z
_atom_site_occupancy
Mo1 Mo    2 a 0.00000 0.00000 0.00000 1.00000
Si1 Si    4 e 0.00000 0.00000 0.33530 1.00000
```

$MoSi_2$ ($C11_b$): AB2_tI6_139_a_e - POSCAR

```
AB2_tI6_139_a_e & a,c/a,z2 --params=3.2064,2.44754241517,0.3353 & I4/mmm
   ↪     D_{4h}^{17} #139 (ae) & tI6 & C11_b & MoSi_2 & & Y. Harada,
   ↪   M. Morinaga, D. Saso, M. Takata and M. Sakata, Intermetallics
   ↪ 6, 523-527 (1998)
   1.0000000000000000
  -1.60320000000000    1.60320000000000    3.92390000000054
   1.60320000000000   -1.60320000000000    3.92390000000054
   1.60320000000000    1.60320000000000   -3.92390000000054
   Mo     Si
    1      2
Direct
   0.00000000000000    0.00000000000000    0.00000000000000    Mo    (2a)
   0.33530000000000    0.33530000000000    0.00000000000000    Si    (4e)
   0.66470000000000    0.66470000000000    0.00000000000000    Si    (4e)
```

$V_4Zn_5$: A4B5_tI18_139_i_ah - CIF

```
# CIF file

data_findsym-output
_audit_creation_method FINDSYM

_chemical_name_mineral ''
_chemical_formula_sum 'V4 Zn5'

loop_
_publ_author_name
 'K. Schubert'
 'H. G. Meissner'
 'A. Raman'
 'W. Rossteutscher'
_journal_name_full
;
Naturwissenschaften
;
_journal_volume 51
_journal_year 1964
_journal_page_first 287
_journal_page_last 287
_publ_Section_title
;
 Einige Strukturdaten metallischer Phasen (9)
;

# Found in  Pearson's Handbook, Vol. IV, pp. 5154

_aflow_proto 'A4B5_tI18_139_i_ah'
_aflow_params 'a,c/a,x2,x3'
_aflow_params_values '8.91,0.361391694725,0.328,0.348'
_aflow_Strukturbericht 'None'
_aflow_Pearson 'tI18'

_symmetry_space_group_name_Hall "-I 4 2"
_symmetry_space_group_name_H-M "I 4/m m m"
_symmetry_Int_Tables_number 139

_cell_length_a      8.91000
_cell_length_b      8.91000
_cell_length_c      3.22000
_cell_angle_alpha 90.00000
_cell_angle_beta   90.00000
_cell_angle_gamma 90.00000

loop_
_space_group_symop_id
_space_group_symop_operation_xyz
1 x,y,z
2 x,-y,-z
3 -x,y,-z
4 -x,-y,z
5 -y,-x,-z
6 -y,x,z
7 y,-x,z
8 y,x,-z
9 -x,-y,-z
10 -x,y,z
11 x,-y,z
12 x,y,-z
13 y,x,z
14 y,-x,-z
15 -y,x,-z
```

```
16 -y,-x,z
17 x+1/2,y+1/2,z+1/2
18 x+1/2,-y+1/2,-z+1/2
19 -x+1/2,y+1/2,-z+1/2
20 -x+1/2,-y+1/2,z+1/2
21 -y+1/2,-x+1/2,-z+1/2
22 -y+1/2,x+1/2,z+1/2
23 y+1/2,-x+1/2,z+1/2
24 y+1/2,x+1/2,-z+1/2
25 -x+1/2,-y+1/2,-z+1/2
26 -x+1/2,y+1/2,z+1/2
27 x+1/2,-y+1/2,z+1/2
28 x+1/2,y+1/2,-z+1/2
29 y+1/2,x+1/2,z+1/2
30 y+1/2,-x+1/2,-z+1/2
31 -y+1/2,x+1/2,-z+1/2
32 -y+1/2,-x+1/2,z+1/2

loop_
_atom_site_label
_atom_site_type_symbol
_atom_site_symmetry_multiplicity
_atom_site_Wyckoff_label
_atom_site_fract_x
_atom_site_fract_y
_atom_site_fract_z
_atom_site_occupancy
Zn1 Zn   2 a 0.00000 0.00000 0.00000 1.00000
Zn2 Zn   8 h 0.32800 0.32800 0.00000 1.00000
V1  V    8 i 0.34800 0.00000 0.00000 1.00000
```

$V_4Zn_5$: A4B5_tI18_139_i_ah - POSCAR

```
A4B5_tI18_139_i_ah & a,c/a,x2,x3 --params=8.91,0.361391694725,0.328,
    ↪ 0.348 & I4/mmm    D_{4h}^{17} #139 (ahi) & tI18 & & V4Zn5 & &
    ↪ K. Schubert, H. G. Meissner, A. Raman and W. Rossteutscher,
    ↪ Naturwissenschaften 51, 287 (1964)
   1.0000000000000000
  -4.45500000000000   4.45500000000000   1.61000000000000
   4.45500000000000  -4.45500000000000   1.61000000000000
   4.45500000000000   4.45500000000000  -1.61000000000000
    V    Zn
    4    5
Direct
   0.00000000000000   0.34800000000000   0.34800000000000   V    (8i)
   0.00000000000000   0.65200000000000   0.65200000000000   V    (8i)
   0.34800000000000   0.00000000000000   0.34800000000000   V    (8i)
   0.65200000000000   0.00000000000000   0.65200000000000   V    (8i)
   0.00000000000000   0.00000000000000   0.00000000000000   Zn   (2a)
   0.32800000000000   0.32800000000000   0.65600000000000   Zn   (8h)
   0.32800000000000   0.67200000000000   0.00000000000000   Zn   (8h)
   0.67200000000000   0.32800000000000   0.00000000000000   Zn   (8h)
   0.67200000000000   0.67200000000000   0.34400000000000   Zn   (8h)
```

$Al_4Ba$ ($D1_3$): A4B_tI10_139_de_a - CIF

```
# CIF file

data_findsym-output
_audit_creation_method FINDSYM

_chemical_name_mineral ''
_chemical_formula_sum 'Al4 Ba'

loop_
_publ_author_name
 'K. R. Andress'
 'E. Alberti'
_journal_name_full
;
Zeitschrift f\"{u}r Metallkunde
;
_journal_volume 27
_journal_year 1935
_journal_page_first 126
_journal_page_last 128
_publ_Section_title
;
 R\"{o}ntgenographische Untersuchung der Legierungsreihe
    ↪ Aluminium-Barium
;

_aflow_proto 'A4B_tI10_139_de_a'
_aflow_params 'a,c/a,z3'
_aflow_params_values '4.53,2.45033112583,0.38'
_aflow_Strukturbericht 'D1_3'
_aflow_Pearson 'tI10'

_symmetry_space_group_name_Hall "-I 4 2"
_symmetry_space_group_name_H-M "I 4/m m m"
_symmetry_Int_Tables_number 139

_cell_length_a      4.53000
_cell_length_b      4.53000
_cell_length_c      11.10000
_cell_angle_alpha   90.00000
_cell_angle_beta    90.00000
_cell_angle_gamma   90.00000

loop_
_space_group_symop_id
_space_group_symop_operation_xyz
1 x,y,z
2 x,-y,-z
3 -x,y,-z
4 -x,-y,z
5 -y,-x,-z
6 -y,x,z
7 y,-x,z
8 y,x,-z
9 -x,-y,-z
10 -x,y,z
11 x,-y,z
12 x,y,-z
13 y,x,z
14 y,-x,-z
15 -y,x,-z
16 -y,-x,z
17 x+1/2,y+1/2,z+1/2
18 x+1/2,-y+1/2,-z+1/2
19 -x+1/2,y+1/2,-z+1/2
20 -x+1/2,-y+1/2,z+1/2
21 -y+1/2,-x+1/2,-z+1/2
22 -y+1/2,x+1/2,z+1/2
23 y+1/2,-x+1/2,z+1/2
24 y+1/2,x+1/2,-z+1/2
25 -x+1/2,-y+1/2,-z+1/2
26 -x+1/2,y+1/2,z+1/2
27 x+1/2,-y+1/2,z+1/2
28 x+1/2,y+1/2,-z+1/2
29 y+1/2,x+1/2,z+1/2
30 y+1/2,-x+1/2,-z+1/2
31 -y+1/2,x+1/2,-z+1/2
32 -y+1/2,-x+1/2,z+1/2

loop_
_atom_site_label
_atom_site_type_symbol
_atom_site_symmetry_multiplicity
_atom_site_Wyckoff_label
_atom_site_fract_x
_atom_site_fract_y
_atom_site_fract_z
_atom_site_occupancy
Ba1 Ba   2 a 0.00000 0.00000 0.00000 1.00000
Al1 Al   4 d 0.00000 0.50000 0.25000 1.00000
Al2 Al   4 e 0.00000 0.00000 0.38000 1.00000
```

$Al_4Ba$ ($D1_3$): A4B_tI10_139_de_a - POSCAR

```
A4B_tI10_139_de_a & a,c/a,z3 --params=4.53,2.45033112583,0.38 & I4/mmm
    ↪    D_{4h}^{17} #139 (ade) & tI10 & D1_3 & Al4Ba & & K.R.
    ↪ Andress and E. Alberto, Z. Metallkd. 27(26), 126-128 (1935)
   1.0000000000000000
  -2.26500000000000   2.26500000000000   5.55000000000000
   2.26500000000000  -2.26500000000000   5.55000000000000
   2.26500000000000   2.26500000000000  -5.55000000000000
    Al   Ba
    4    1
Direct
   0.25000000000000   0.75000000000000   0.50000000000000   Al   (4d)
   0.75000000000000   0.25000000000000   0.50000000000000   Al   (4d)
   0.38000000000000   0.38000000000000   0.00000000000000   Al   (4e)
   0.62000000000000   0.62000000000000   0.00000000000000   Al   (4e)
   0.00000000000000   0.00000000000000   0.00000000000000   Ba   (2a)
```

$Pt_8Ti$: A8B_tI18_139_hi_a - CIF

```
# CIF file

data_findsym-output
_audit_creation_method FINDSYM

_chemical_name_mineral ''
_chemical_formula_sum 'Pt8 Ti'

loop_
_publ_author_name
 'P. Pietrokowsky'
_journal_name_full
;
Nature
;
_journal_volume 206
_journal_year 1965
_journal_page_first 291
_journal_page_last 291
_publ_Section_title
;
 Novel Ordered Phase, Pt$_8$Ti
;

# Found in  Pearson's Handbook, Vol. IV, pp. 5011

_aflow_proto 'A8B_tI18_139_hi_a'
_aflow_params 'a,c/a,x2,x3'
_aflow_params_values '8.312,0.468840230991,0.333,0.327'
_aflow_Strukturbericht 'None'
_aflow_Pearson 'tI18'

_symmetry_space_group_name_Hall "-I 4 2"
_symmetry_space_group_name_H-M "I 4/m m m"
_symmetry_Int_Tables_number 139

_cell_length_a      8.31200
_cell_length_b      8.31200
_cell_length_c      3.89700
_cell_angle_alpha   90.00000
_cell_angle_beta    90.00000
_cell_angle_gamma   90.00000
```

```
loop_
_space_group_symop_id
_space_group_symop_operation_xyz
1 x,y,z
2 x,-y,-z
3 -x,y,-z
4 -x,-y,z
5 -y,-x,-z
6 -y,x,z
7 y,-x,z
8 y,x,-z
9 -x,-y,-z
10 -x,y,z
11 x,-y,z
12 x,y,-z
13 y,x,z
14 y,-x,-z
15 -y,x,-z
16 -y,-x,z
17 x+1/2,y+1/2,z+1/2
18 x+1/2,-y+1/2,-z+1/2
19 -x+1/2,y+1/2,-z+1/2
20 -x+1/2,-y+1/2,z+1/2
21 -y+1/2,-x+1/2,-z+1/2
22 -y+1/2,x+1/2,z+1/2
23 y+1/2,-x+1/2,z+1/2
24 y+1/2,x+1/2,-z+1/2
25 -x+1/2,-y+1/2,-z+1/2
26 -x+1/2,y+1/2,z+1/2
27 x+1/2,-y+1/2,z+1/2
28 x+1/2,y+1/2,-z+1/2
29 y+1/2,x+1/2,z+1/2
30 y+1/2,-x+1/2,-z+1/2
31 -y+1/2,x+1/2,-z+1/2
32 -y+1/2,-x+1/2,z+1/2

loop_
_atom_site_label
_atom_site_type_symbol
_atom_site_symmetry_multiplicity
_atom_site_Wyckoff_label
_atom_site_fract_x
_atom_site_fract_y
_atom_site_fract_z
_atom_site_occupancy
Ti1  Ti   2 a 0.00000 0.00000 0.00000 1.00000
Pt1  Pt   8 h 0.33300 0.33300 0.00000 1.00000
Pt2  Pt   8 i 0.32700 0.00000 0.00000 1.00000
```

$Pt_8Ti$: A8B_tI18_139_hi_a - POSCAR

```
A8B_tI18_139_hi_a & a,c/a,x2,x3 --params=8.312,0.468840230991,0.333,
    ↪ 0.327 & I4/mmm     D_{4h}^{17} #139 (ahi) & tI18 & & Pt8Ti & &
    ↪ P. Pietrokowsky, Nature 206, 291 (1965)
   1.0000000000000000
  -4.15600000000000    4.15600000000000    1.94850000000000
   4.15600000000000   -4.15600000000000    1.94850000000000
   4.15600000000000    4.15600000000000   -1.94850000000000
   Pt     Zn
    8      1
Direct
   0.33300000000000    0.33300000000000    0.66600000000000    Pt    (8h)
   0.33300000000000    0.66700000000000    0.00000000000000    Pt    (8h)
   0.66700000000000    0.33300000000000    0.00000000000000    Pt    (8h)
   0.66700000000000    0.66700000000000    0.33400000000000    Pt    (8h)
   0.00000000000000    0.32700000000000    0.32700000000000    Pt    (8i)
   0.00000000000000    0.67300000000000    0.67300000000000    Pt    (8i)
   0.32700000000000    0.00000000000000    0.32700000000000    Pt    (8i)
   0.67300000000000    0.00000000000000    0.67300000000000    Pt    (8i)
   0.00000000000000    0.00000000000000    0.00000000000000    Ti    (2a)
```

$ThH_2$ (L'2): A2B_tI6_139_d_a - CIF

```
# CIF file

data_findsym-output
_audit_creation_method FINDSYM

_chemical_name_mineral ''
_chemical_formula_sum 'Th H2'

loop_
_publ_author_name
 'R. E. Rundle'
 'C. G. Shull'
 'E. O. Wollan'
_journal_name_full
;
Acta Crystallographica
;
_journal_volume 5
_journal_year 1952
_journal_page_first 22
_journal_page_last 26
_publ_Section_title
;
 The crystal structure of thorium and zirconium dihydrides by X-ray and
    ↪ neutron diffraction
;

_aflow_proto 'A2B_tI6_139_d_a'
_aflow_params 'a,c/a'
_aflow_params_values '4.1,1.22682926829'
_aflow_Strukturbericht 'L\'2'
_aflow_Pearson 'tI6'

_symmetry_space_group_name_Hall "-I 4 2"
_symmetry_space_group_name_H-M "I 4/m m m"
_symmetry_Int_Tables_number 139

_cell_length_a      4.10000
_cell_length_b      4.10000
_cell_length_c      5.03000
_cell_angle_alpha  90.00000
_cell_angle_beta   90.00000
_cell_angle_gamma  90.00000

loop_
_space_group_symop_id
_space_group_symop_operation_xyz
1 x,y,z
2 x,-y,-z
3 -x,y,-z
4 -x,-y,z
5 -y,-x,-z
6 -y,x,z
7 y,-x,z
8 y,x,-z
9 -x,-y,-z
10 -x,y,z
11 x,-y,z
12 x,y,-z
13 y,x,z
14 y,-x,-z
15 -y,x,-z
16 -y,-x,z
17 x+1/2,y+1/2,z+1/2
18 x+1/2,-y+1/2,-z+1/2
19 -x+1/2,y+1/2,-z+1/2
20 -x+1/2,-y+1/2,z+1/2
21 -y+1/2,-x+1/2,-z+1/2
22 -y+1/2,x+1/2,z+1/2
23 y+1/2,-x+1/2,z+1/2
24 y+1/2,x+1/2,-z+1/2
25 -x+1/2,-y+1/2,-z+1/2
26 -x+1/2,y+1/2,z+1/2
27 x+1/2,-y+1/2,z+1/2
28 x+1/2,y+1/2,-z+1/2
29 y+1/2,x+1/2,z+1/2
30 y+1/2,-x+1/2,-z+1/2
31 -y+1/2,x+1/2,-z+1/2
32 -y+1/2,-x+1/2,z+1/2

loop_
_atom_site_label
_atom_site_type_symbol
_atom_site_symmetry_multiplicity
_atom_site_Wyckoff_label
_atom_site_fract_x
_atom_site_fract_y
_atom_site_fract_z
_atom_site_occupancy
Th1  Th   2 a 0.00000 0.00000 0.00000 1.00000
H1   H    4 d 0.00000 0.50000 0.25000 1.00000
```

$ThH_2$ (L'2): A2B_tI6_139_d_a - POSCAR

```
A2B_tI6_139_d_a & a,c/a --params=4.1,1.22682926829 & I4/mmm     D_{4h}^{
    ↪ 17} #139 (ad) & tI6 & L'2 & ThH2 & & R. E. Rundle, C. G. Shull
    ↪ and E. O. Wollan, Acta Cryst. 5, 22-26 (1952)
   1.0000000000000000
  -2.05000000000000    2.05000000000000    2.51500000000000
   2.05000000000000   -2.05000000000000    2.51500000000000
   2.05000000000000    2.05000000000000   -2.51500000000000
   H     Th
   2      1
Direct
   0.25000000000000    0.75000000000000    0.50000000000000     H    (4d)
   0.75000000000000    0.25000000000000    0.50000000000000     H    (4d)
   0.00000000000000    0.00000000000000    0.00000000000000    Th    (2a)
```

α-Pa ($A_a$): A_tI2_139_a - CIF

```
# CIF file

data_findsym-output
_audit_creation_method FINDSYM

_chemical_name_mineral 'Protactinium'
_chemical_formula_sum 'Pa'

loop_
_publ_author_name
 'W. H. Zachariasen'
_journal_name_full
;
Acta Crystallographica
;
_journal_volume 12
_journal_year 1959
_journal_page_first 698
_journal_page_last 700
_publ_Section_title
;
 On the crystal structure of protactinium metal
;

# Found in  Donohue, pp. 125-127
```

```
_aflow_proto 'A_tI2_139_a'
_aflow_params 'a,c/a'
_aflow_params_values '3.932,0.823499491353'
_aflow_Strukturbericht 'A_a'
_aflow_Pearson 'tI2'

_symmetry_space_group_name_Hall "-I 4 2"
_symmetry_space_group_name_H-M "I 4/m m m"
_symmetry_Int_Tables_number 139

_cell_length_a      3.93200
_cell_length_b      3.93200
_cell_length_c      3.23800
_cell_angle_alpha 90.00000
_cell_angle_beta  90.00000
_cell_angle_gamma 90.00000

loop_
_space_group_symop_id
_space_group_symop_operation_xyz
1 x,y,z
2 x,-y,-z
3 -x,y,-z
4 -x,-y,z
5 -y,-x,-z
6 -y,x,z
7 y,-x,z
8 y,x,-z
9 -x,-y,-z
10 -x,y,z
11 x,-y,z
12 x,y,-z
13 y,x,z
14 y,-x,-z
15 -y,x,-z
16 -y,-x,z
17 x+1/2,y+1/2,z+1/2
18 x+1/2,-y+1/2,-z+1/2
19 -x+1/2,y+1/2,-z+1/2
20 -x+1/2,-y+1/2,z+1/2
21 -y+1/2,-x+1/2,-z+1/2
22 -y+1/2,x+1/2,z+1/2
23 y+1/2,-x+1/2,z+1/2
24 y+1/2,x+1/2,-z+1/2
25 -x+1/2,-y+1/2,-z+1/2
26 -x+1/2,y+1/2,z+1/2
27 x+1/2,-y+1/2,z+1/2
28 x+1/2,y+1/2,-z+1/2
29 y+1/2,x+1/2,z+1/2
30 y+1/2,-x+1/2,-z+1/2
31 -y+1/2,x+1/2,-z+1/2
32 -y+1/2,-x+1/2,z+1/2

loop_
_atom_site_label
_atom_site_type_symbol
_atom_site_symmetry_multiplicity
_atom_site_Wyckoff_label
_atom_site_fract_x
_atom_site_fract_y
_atom_site_fract_z
_atom_site_occupancy
Pa1 Pa   2 a 0.00000 0.00000 0.00000 1.00000
```

$\alpha$-Pa ($A_a$): A_tI2_139_a - POSCAR

```
A_tI2_139_a & a,c/a --params=3.932,0.823499491353 & I4/mmm       D_{4h}^{
    ↪ 17} #139 (a) & tI2 & A_a & Pa & alpha & W. H. Zachariasen, Acta
    ↪ Cryst. 12, 698-700 (1959)
   1.0000000000000000
 -1.966000000000   1.966000000000   1.619000000000
  1.966000000000  -1.966000000000   1.619000000000
  1.966000000000   1.966000000000  -1.619000000000
   Pa
    1
Direct
   0.00000000000000    0.00000000000000    0.00000000000000    Pa    (2a)
```

Khatyrkite ($Al_2Cu$, C16): A2B_tI12_140_h_a - CIF

```
# CIF file

data_findsym-output
_audit_creation_method FINDSYM

_chemical_name_mineral 'Khatyrkite'
_chemical_formula_sum 'Al2 Cu'

loop_
_publ_author_name
 'James B. Friauf'
_journal_name_full
;
Journal of the American Chemical Society
;
_journal_volume 49
_journal_year 1927
_journal_page_first 3107
_journal_page_last 3114
_publ_Section_title
;
 The Crystal Structures of Two Intermetallic Compounds
;

_aflow_proto 'A2B_tI12_140_h_a'
_aflow_params 'a,c/a,x2'
_aflow_params_values '6.04,0.804635761589,0.158'
_aflow_Strukturbericht 'C16'
_aflow_Pearson 'tI12'

_symmetry_space_group_name_Hall "-I 4 2c"
_symmetry_space_group_name_H-M "I 4/m c m"
_symmetry_Int_Tables_number 140

_cell_length_a      6.04000
_cell_length_b      6.04000
_cell_length_c      4.86000
_cell_angle_alpha 90.00000
_cell_angle_beta  90.00000
_cell_angle_gamma 90.00000

loop_
_space_group_symop_id
_space_group_symop_operation_xyz
1 x,y,z
2 x,-y,-z+1/2
3 -x,y,-z+1/2
4 -x,-y,z
5 -y,-x,-z+1/2
6 -y,x,z
7 y,-x,z
8 y,x,-z+1/2
9 -x,-y,-z
10 -x,y,z+1/2
11 x,-y,z+1/2
12 x,y,-z
13 y,x,z+1/2
14 y,-x,-z
15 -y,x,-z
16 -y,-x,z+1/2
17 x+1/2,y+1/2,z+1/2
18 x+1/2,-y+1/2,-z
19 -x+1/2,y+1/2,-z
20 -x+1/2,-y+1/2,z+1/2
21 -y+1/2,-x+1/2,-z
22 -y+1/2,x+1/2,z+1/2
23 y+1/2,-x+1/2,z+1/2
24 y+1/2,x+1/2,-z
25 -x+1/2,-y+1/2,-z+1/2
26 -x+1/2,y+1/2,z
27 x+1/2,-y+1/2,z
28 x+1/2,y+1/2,-z+1/2
29 y+1/2,x+1/2,z
30 y+1/2,-x+1/2,-z+1/2
31 -y+1/2,x+1/2,-z+1/2
32 -y+1/2,-x+1/2,z

loop_
_atom_site_label
_atom_site_type_symbol
_atom_site_symmetry_multiplicity
_atom_site_Wyckoff_label
_atom_site_fract_x
_atom_site_fract_y
_atom_site_fract_z
_atom_site_occupancy
Cu1 Cu   4 a 0.00000 0.00000 0.25000 1.00000
Al1 Al   8 h 0.15800 0.65800 0.00000 1.00000
```

Khatyrkite ($Al_2Cu$, C16): A2B_tI12_140_h_a - POSCAR

```
A2B_tI12_140_h_a & a,c/a,x2 --params=6.04,0.804635761589,0.158 & I4/mcm
    ↪    D_{4h}^{18} #140 (ah) & tI12 & C16 & Al_2Cu & Khatyrkite &
    ↪ J. A. Friauf, J. Am. Chem. Soc. 49, 3107-3114 (1927)
   1.0000000000000000
  -3.02000000000000    3.02000000000000    2.42999999999878
   3.02000000000000   -3.02000000000000    2.42999999999878
   3.02000000000000    3.02000000000000   -2.42999999999878
   Al    Cu
    4     2
Direct
   0.65800000000000    0.15800000000000    0.81600000000000    Al    (8h)
   0.15800000000000    0.34200000000000    0.50000000000000    Al    (8h)
   0.84200000000000    0.65800000000000    0.50000000000000    Al    (8h)
   0.34200000000000    0.84200000000000    0.18400000000000    Al    (8h)
   0.75000000000000    0.75000000000000    0.00000000000000    Cu    (4a)
   0.25000000000000    0.25000000000000    0.00000000000000    Cu    (4a)
```

$SiU_3$ ($D0_c$): AB3_tI16_140_b_ah - CIF

```
# CIF file

data_findsym-output
_audit_creation_method FINDSYM

_chemical_name_mineral 'Uranium Silicide'
_chemical_formula_sum 'Si U3'

loop_
_publ_author_name
 'W. H. Zachariasen'
_journal_name_full
;
Acta Crystallographica
;
_journal_volume 2
_journal_year 1949
_journal_page_first 94
_journal_page_last 99
_publ_Section_title
```

```
;
 Crystal chemical studies of the 5f-series of elements. VIII. Crystal
     ↪ structure studies of uranium silicides and of CeSi$_2$,
     ↪ NpSi$_2$, and PuSi$_2$
;

_aflow_proto 'AB3_tI16_140_b_ah'
_aflow_params 'a,c/a,x3'
_aflow_params_values '6.017,1.44241316271,0.231'
_aflow_Strukturbericht 'D0_c'
_aflow_Pearson 'tI16'

_symmetry_space_group_name_Hall "-I 4 2c"
_symmetry_space_group_name_H-M "I 4/m c m"
_symmetry_Int_Tables_number 140

_cell_length_a      6.01700
_cell_length_b      6.01700
_cell_length_c      8.67900
_cell_angle_alpha 90.00000
_cell_angle_beta  90.00000
_cell_angle_gamma 90.00000

loop_
_space_group_symop_id
_space_group_symop_operation_xyz
1 x,y,z
2 x,-y,-z+1/2
3 -x,y,-z+1/2
4 -x,-y,z
5 -y,-x,-z+1/2
6 -y,x,z
7 y,-x,z
8 y,x,-z+1/2
9 -x,-y,-z
10 -x,y,z+1/2
11 x,-y,z+1/2
12 x,y,-z
13 y,x,z+1/2
14 y,-x,-z
15 -y,x,-z
16 -y,-x,z+1/2
17 x+1/2,y+1/2,z+1/2
18 x+1/2,-y+1/2,-z
19 -x+1/2,y+1/2,-z
20 -x+1/2,-y+1/2,z+1/2
21 -y+1/2,-x+1/2,-z
22 -y+1/2,x+1/2,z+1/2
23 y+1/2,-x+1/2,z+1/2
24 y+1/2,x+1/2,-z
25 -x+1/2,-y+1/2,-z+1/2
26 -x+1/2,y+1/2,z
27 x+1/2,-y+1/2,z
28 x+1/2,y+1/2,-z+1/2
29 y+1/2,x+1/2,z
30 y+1/2,-x+1/2,-z+1/2
31 -y+1/2,x+1/2,-z+1/2
32 -y+1/2,-x+1/2,z

loop_
_atom_site_label
_atom_site_type_symbol
_atom_site_symmetry_multiplicity
_atom_site_Wyckoff_label
_atom_site_fract_x
_atom_site_fract_y
_atom_site_fract_z
_atom_site_occupancy
U1  U    4 a 0.00000 0.00000 0.25000 1.00000
Si1 Si   4 b 0.00000 0.50000 0.25000 1.00000
U2  U    8 h 0.23100 0.73100 0.00000 1.00000
```

$SiU_3$ ($D0_c$): AB3_tI16_140_b_ah - POSCAR

```
AB3_tI16_140_b_ah & a,c/a,x3 --params=6.017,1.44241316271,0.231 & I4/mcm
     ↪ D_{4h}^{18} #140 (abh) & tI16 & D0_c & SiU3 & & W. H.
     ↪ Zachariasen, Acta Cryst. 2, 94-99 (1949)
   1.0000000000000000
  -3.00850000000000   3.00850000000000   4.33950000000000
   3.00850000000000  -3.00850000000000   4.33950000000000
   3.00850000000000   3.00850000000000  -4.33950000000000
   Si    U
    2    6
Direct
   0.25000000000000   0.75000000000000   0.50000000000000   Si   (4b)
   0.75000000000000   0.25000000000000   0.50000000000000   Si   (4b)
   0.25000000000000   0.25000000000000   0.00000000000000   U    (4a)
   0.75000000000000   0.75000000000000   0.00000000000000   U    (4a)
   0.23100000000000   0.26900000000000   0.50000000000000   U    (8h)
   0.26900000000000   0.76900000000000   0.03800000000000   U    (8h)
   0.73100000000000   0.23100000000000  -0.03800000000000   U    (8h)
   0.76900000000000   0.73100000000000   0.50000000000000   U    (8h)
```

SeTl (B37): AB_tI16_140_ab_h - CIF

```
# CIF file

data_findsym-output
_audit_creation_method FINDSYM

_chemical_name_mineral 'I4/mcm'
_chemical_formula_sum 'Se Tl'

loop_
_publ_author_name
'R.R. Yadav'
'R. P. Ram'
'S Bhan'
_journal_name_full
;
Zeitschrift f\"{u}r Metallkunde
;
_journal_volume 67
_journal_year 1976
_journal_page_first 173
_journal_page_last 177
_publ_Section_title
;
 On the Thallium-Selenium-Tellurium System
;

# Found in http://materials.springer.com/isp/crystallographic/docs/
     ↪ sd_0261726

_aflow_proto 'AB_tI16_140_ab_h'
_aflow_params 'a,c/a,x3'
_aflow_params_values '8.03,0.87297633873,0.179'
_aflow_Strukturbericht 'B37'
_aflow_Pearson 'tI16'

_symmetry_space_group_name_Hall "-I 4 2c"
_symmetry_space_group_name_H-M "I 4/m c m"
_symmetry_Int_Tables_number 140

_cell_length_a      8.03000
_cell_length_b      8.03000
_cell_length_c      7.01000
_cell_angle_alpha 90.00000
_cell_angle_beta  90.00000
_cell_angle_gamma 90.00000

loop_
_space_group_symop_id
_space_group_symop_operation_xyz
1 x,y,z
2 x,-y,-z+1/2
3 -x,y,-z+1/2
4 -x,-y,z
5 -y,-x,-z+1/2
6 -y,x,z
7 y,-x,z
8 y,x,-z+1/2
9 -x,-y,-z
10 -x,y,z+1/2
11 x,-y,z+1/2
12 x,y,-z
13 y,x,z+1/2
14 y,-x,-z
15 -y,x,-z
16 -y,-x,z+1/2
17 x+1/2,y+1/2,z+1/2
18 x+1/2,-y+1/2,-z
19 -x+1/2,y+1/2,-z
20 -x+1/2,-y+1/2,z+1/2
21 -y+1/2,-x+1/2,-z
22 -y+1/2,x+1/2,z+1/2
23 y+1/2,-x+1/2,z+1/2
24 y+1/2,x+1/2,-z
25 -x+1/2,-y+1/2,-z+1/2
26 -x+1/2,y+1/2,z
27 x+1/2,-y+1/2,z
28 x+1/2,y+1/2,-z+1/2
29 y+1/2,x+1/2,z
30 y+1/2,-x+1/2,-z+1/2
31 -y+1/2,x+1/2,-z+1/2
32 -y+1/2,-x+1/2,z

loop_
_atom_site_label
_atom_site_type_symbol
_atom_site_symmetry_multiplicity
_atom_site_Wyckoff_label
_atom_site_fract_x
_atom_site_fract_y
_atom_site_fract_z
_atom_site_occupancy
Se1 Se   4 a 0.00000 0.00000 0.25000 1.00000
Se2 Se   4 b 0.00000 0.50000 0.25000 1.00000
Tl1 Tl   8 h 0.17900 0.67900 0.00000 1.00000
```

SeTl (B37): AB_tI16_140_ab_h - POSCAR

```
AB_tI16_140_ab_h & a,c/a,x3 --params=8.03,0.87297633873,0.179 & I4/mcm
     ↪ D_{4h}^{18} #140 (abh) & tI16 & B37 & SeTl & & R.R. Yadav,
     ↪ R.P. Ram and S Bhan, Z. Metallkd. 67, 173-177 (1976)
   1.0000000000000000
  -4.01500000000000   4.01500000000000   3.50500000000095
   4.01500000000000  -4.01500000000000   3.50500000000095
   4.01500000000000   4.01500000000000  -3.50500000000095
   Se    Tl
    4    4
Direct
   0.75000000000000   0.75000000000000   0.00000000000000   Se   (4a)
   0.25000000000000   0.25000000000000   0.00000000000000   Se   (4a)
   0.25000000000000   0.75000000000000   0.50000000000000   Se   (4b)
   0.75000000000000   0.25000000000000   0.50000000000000   Se   (4b)
   0.67900000000000   0.17900000000000   0.85800000000000   Tl   (8h)
   0.17900000000000   0.32100000000000   0.50000000000000   Tl   (8h)
   0.82100000000000   0.67900000000000   0.50000000000000   Tl   (8h)
   0.32100000000000   0.82100000000000   0.14200000000000   Tl   (8h)
```

Zircon ($ZrSiO_4$): A4BC_tI24_141_h_b_a - CIF

```
# CIF file

data_findsym-output
_audit_creation_method FINDSYM

_chemical_name_mineral 'Zircon'
_chemical_formula_sum 'Zr Si O4'

loop_
_publ_author_name
 'Robert M. Hazen'
 'Larry W. Finger'
_journal_name_full
;
American Mineralogist
;
_journal_volume 64
_journal_year 1979
_journal_page_first 196
_journal_page_last 201
_publ_Section_title
;
 Crystal structure and compressibility of zircon at high pressure
;

# Found in AMS Database

_aflow_proto 'A4BC_tI24_141_h_b_a'
_aflow_params 'a,c/a,y3,z3'
_aflow_params_values '6.6042,0.905423821205,0.066,0.1951'
_aflow_Strukturbericht 'None'
_aflow_Pearson 'tI24'

_symmetry_space_group_name_Hall "-I 4bd 2"
_symmetry_space_group_name_H-M "I 41/a m d:2"
_symmetry_Int_Tables_number 141

_cell_length_a      6.60420
_cell_length_b      6.60420
_cell_length_c      5.97960
_cell_angle_alpha  90.00000
_cell_angle_beta   90.00000
_cell_angle_gamma  90.00000

loop_
_space_group_symop_id
_space_group_symop_operation_xyz
1 x,y,z
2 x,-y,-z
3 -x,y+1/2,-z
4 -x,-y+1/2,z
5 -y+1/4,-x+1/4,-z+3/4
6 -y+1/4,x+3/4,z+1/4
7 y+3/4,-x+3/4,z+1/4
8 y+3/4,x+1/4,-z+3/4
9 -x,-y,-z
10 -x,y,z
11 x,-y+1/2,z
12 x,y+1/2,-z
13 y+3/4,x+3/4,z+1/4
14 y+3/4,-x+1/4,-z+3/4
15 -y+1/4,x+1/4,-z+3/4
16 -y+1/4,-x+3/4,z+1/4
17 x+1/2,y+1/2,z+1/2
18 x+1/2,-y+1/2,-z+1/2
19 -x+1/2,y,-z+1/2
20 -x+1/2,-y,z+1/2
21 -y+3/4,-x+3/4,-z+1/4
22 -y+3/4,x+1/4,z+3/4
23 y+1/4,-x+1/4,z+3/4
24 y+1/4,x+3/4,-z+1/4
25 -x+1/2,-y+1/2,-z+1/2
26 -x+1/2,y+1/2,z+1/2
27 x+1/2,-y,z+1/2
28 x+1/2,y,-z+1/2
29 y+1/4,x+1/4,z+3/4
30 y+1/4,-x+3/4,-z+1/4
31 -y+3/4,x+3/4,-z+1/4
32 -y+3/4,-x+1/4,z+3/4

loop_
_atom_site_label
_atom_site_type_symbol
_atom_site_symmetry_multiplicity
_atom_site_Wyckoff_label
_atom_site_fract_x
_atom_site_fract_y
_atom_site_fract_z
_atom_site_occupancy
Zr1 Zr    4 a 0.00000 0.75000 0.12500 1.00000
Si1 Si    4 b 0.00000 0.25000 0.37500 1.00000
O1  O    16 h 0.00000 0.06600 0.19510 1.00000
```

Zircon ($ZrSiO_4$): A4BC_tI24_141_h_b_a - POSCAR

```
A4BC_tI24_141_h_b_a & a,c/a,y3,z3 --params=6.6042,0.905423821205,0.066,
    ↪ 0.1951 & I4_1/amd D_{4h}^{19} #141 (abh) & tI24 & & ZrSiO4 &
    ↪ Zircon & R. M. Hazen and L. W. Finger, Am. Mineral. 64, 196-201
    ↪ (1979)
   1.0000000000000000
  -3.30210000000000   3.30210000000000   2.98980000000000
   3.30210000000000  -3.30210000000000   2.98980000000000
   3.30210000000000   3.30210000000000  -2.98980000000000
     O    Si    Zr
     8     2     2
Direct
   0.19510000000000   0.26110000000000   0.56600000000000   O   (16h)
   0.19510000000000   0.62910000000000  -0.06600000000000   O   (16h)
   0.26110000000000   0.19510000000000   0.06600000000000   O   (16h)
   0.37090000000000   0.80490000000000   0.56600000000000   O   (16h)
   0.62910000000000   0.19510000000000   0.43400000000000   O   (16h)
   0.73890000000000   0.80490000000000  -0.06600000000000   O   (16h)
   0.80490000000000   0.37090000000000   0.06600000000000   O   (16h)
   0.80490000000000   0.73890000000000   0.43400000000000   O   (16h)
   0.37500000000000   0.62500000000000   0.75000000000000   Si   (4b)
   0.62500000000000   0.37500000000000   0.25000000000000   Si   (4b)
   0.12500000000000   0.87500000000000   0.25000000000000   Zr   (4a)
   0.87500000000000   0.12500000000000   0.75000000000000   Zr   (4a)
```

$\beta$-Sn (A5): A_tI4_141_a - CIF

```
# CIF file

data_findsym-output
_audit_creation_method FINDSYM

_chemical_name_mineral 'beta Sn'
_chemical_formula_sum 'Sn'

loop_
_publ_author_name
 'V. T. Deshpande'
 'D. B. Sirdeshmukh'
_journal_name_full
;
Acta Crystallographica
;
_journal_volume 14
_journal_year 1961
_journal_page_first 355
_journal_page_last 356
_publ_Section_title
;
 Thermal Expansion of Tetragonal Tin
;

# Found in https://www.webelements.com/tin/crystal_structure.html

_aflow_proto 'A_tI4_141_a'
_aflow_params 'a,c/a'
_aflow_params_values '5.8318,0.545611989437'
_aflow_Strukturbericht 'A5'
_aflow_Pearson 'tI4'

_symmetry_space_group_name_Hall "-I 4bd 2"
_symmetry_space_group_name_H-M "I 41/a m d:2"
_symmetry_Int_Tables_number 141

_cell_length_a      5.83180
_cell_length_b      5.83180
_cell_length_c      3.18190
_cell_angle_alpha  90.00000
_cell_angle_beta   90.00000
_cell_angle_gamma  90.00000

loop_
_space_group_symop_id
_space_group_symop_operation_xyz
1 x,y,z
2 x,-y,-z
3 -x,y+1/2,-z
4 -x,-y+1/2,z
5 -y+1/4,-x+1/4,-z+3/4
6 -y+1/4,x+3/4,z+1/4
7 y+3/4,-x+3/4,z+1/4
8 y+3/4,x+1/4,-z+3/4
9 -x,-y,-z
10 -x,y,z
11 x,-y+1/2,z
12 x,y+1/2,-z
13 y+3/4,x+3/4,z+1/4
14 y+3/4,-x+1/4,-z+3/4
15 -y+1/4,x+1/4,-z+3/4
16 -y+1/4,-x+3/4,z+1/4
17 x+1/2,y+1/2,z+1/2
18 x+1/2,-y+1/2,-z+1/2
19 -x+1/2,y,-z+1/2
20 -x+1/2,-y,z+1/2
21 -y+3/4,-x+3/4,-z+1/4
22 -y+3/4,x+1/4,z+3/4
23 y+1/4,-x+1/4,z+3/4
24 y+1/4,x+3/4,-z+1/4
25 -x+1/2,-y+1/2,-z+1/2
26 -x+1/2,y+1/2,z+1/2
27 x+1/2,-y,z+1/2
28 x+1/2,y,-z+1/2
29 y+1/4,x+1/4,z+3/4
30 y+1/4,-x+3/4,-z+1/4
31 -y+3/4,x+3/4,-z+1/4
32 -y+3/4,-x+1/4,z+3/4

loop_
_atom_site_label
_atom_site_type_symbol
_atom_site_symmetry_multiplicity
_atom_site_Wyckoff_label
_atom_site_fract_x
_atom_site_fract_y
```

```
_atom_site_fract_z
_atom_site_occupancy
Sn1 Sn    4 a 0.00000 0.75000 0.12500 1.00000
```

$\beta$-Sn (A5): A_tI4_141_a - POSCAR

```
A_tI4_141_a & a,c/a --params=5.8318,0.545611989437 & I4_1/amd   D_{4h}^{
    ↪ 19} #141 (a) & tI4 & A5 & Sn & beta & V. T. Deshpande and D. B.
    ↪ Sirdeshmukh, Acta Cryst. 14, 355-356 (1961)
   1.0000000000000000
  -2.91590000000000    2.91590000000000    1.59095000000000
   2.91590000000000   -2.91590000000000    1.59095000000000
   2.91590000000000    2.91590000000000   -1.59095000000000
   Sn
    2
Direct
   0.12500000000000    0.87500000000000    0.25000000000000    Sn    (4a)
   0.87500000000000    0.12500000000000    0.75000000000000    Sn    (4a)
```

Hausmannite ($Mn_3O_4$): A3B4_tI28_141_ad_h - CIF

```
# CIF file

data_findsym-output
_audit_creation_method FINDSYM

_chemical_name_mineral 'Hausmannite'
_chemical_formula_sum 'Mn3 O4'

loop_
_publ_author_name
 'D. Jarosch'
_journal_name_full
;
Mineralogy and Petrology
;
_journal_volume 37
_journal_year 1987
_journal_page_first 15
_journal_page_last 23
_publ_Section_title
;
 Crystal structure refinement and reflectance measurements of
    ↪ hausmannite, Mn$_3$O$_4$
;

# Found in  Pearson IV, pp. 4347

_aflow_proto 'A3B4_tI28_141_ad_h'
_aflow_params 'a,c/a,y3,z3'
_aflow_params_values '5.765,1.63781439722,0.0278,0.2589'
_aflow_Strukturbericht 'None'
_aflow_Pearson 'tI28'

_symmetry_space_group_name_Hall "-I 4bd 2"
_symmetry_space_group_name_H-M "I 41/a m d:2"
_symmetry_Int_Tables_number 141

_cell_length_a     5.76500
_cell_length_b     5.76500
_cell_length_c     9.44200
_cell_angle_alpha 90.00000
_cell_angle_beta  90.00000
_cell_angle_gamma 90.00000

loop_
_space_group_symop_id
_space_group_symop_operation_xyz
1 x,y,z
2 x,-y,-z
3 -x,y+1/2,-z
4 -x,-y+1/2,z
5 -y+1/4,-x+1/4,-z+3/4
6 -y+1/4,x+3/4,z+1/4
7 y+3/4,-x+3/4,z+1/4
8 y+3/4,x+1/4,-z+3/4
9 -x,-y,-z
10 -x,y,z
11 x,-y+1/2,z
12 x,y+1/2,-z
13 y+3/4,x+3/4,z+1/4
14 y+3/4,-x+1/4,-z+3/4
15 -y+1/4,x+1/4,-z+3/4
16 -y+1/4,-x+3/4,z+1/4
17 x+1/2,y+1/2,z+1/2
18 x+1/2,-y+1/2,-z+1/2
19 -x+1/2,y,-z+1/2
20 -x+1/2,-y,z+1/2
21 -y+3/4,-x+3/4,-z+1/4
22 -y+3/4,x+1/4,z+3/4
23 y+1/4,-x+1/4,z+3/4
24 y+1/4,x+3/4,-z+1/4
25 -x+1/2,-y+1/2,-z+1/2
26 -x+1/2,y+1/2,z+1/2
27 x+1/2,-y,z+1/2
28 x+1/2,y,-z+1/2
29 y+1/4,x+1/4,z+3/4
30 y+1/4,-x+3/4,-z+1/4
31 -y+3/4,x+3/4,-z+1/4
32 -y+3/4,-x+1/4,z+3/4

loop_
_atom_site_label
_atom_site_type_symbol
_atom_site_symmetry_multiplicity
_atom_site_Wyckoff_label
_atom_site_fract_x
_atom_site_fract_y
_atom_site_fract_z
_atom_site_occupancy
Mn1 Mn    4 a 0.00000 0.75000 0.12500 1.00000
Mn2 Mn    8 d 0.00000 0.00000 0.50000 1.00000
O1  O    16 h 0.00000 0.02780 0.25890 1.00000
```

Hausmannite ($Mn_3O_4$): A3B4_tI28_141_ad_h - POSCAR

```
A3B4_tI28_141_ad_h & a,c/a,y3,z3 --params=5.765,1.63781439722,0.0278,
    ↪ 0.2589 & I4_1/amd        D_{4h}^{19} #141 (adh) & tI28 & &
    ↪ Mn3O4 & Hausmannite & D. Jorosch, Mineral. Petrol. 37, 15-23 (
    ↪ 1987)
   1.0000000000000000
  -2.88250000000000    2.88250000000000    4.72100000000000
   2.88250000000000   -2.88250000000000    4.72100000000000
   2.88250000000000    2.88250000000000   -4.72100000000000
  Mn1   Mn2     O
    2     4     8
Direct
   0.12500000000000   -0.12500000000000    0.25000000000000    Mn    (4a)
  -0.12500000000000    0.12500000000000   -0.25000000000000    Mn    (4a)
   0.00000000000000    0.50000000000000    0.50000000000000    Mn    (8d)
   0.50000000000000    0.00000000000000   -0.00000000000000    Mn    (8d)
   0.50000000000000    0.50000000000000   -0.00000000000000    Mn    (8d)
   0.50000000000000    0.50000000000000    0.50000000000000    Mn    (8d)
   0.25890000000000   -0.26890000000000   -0.02780000000000    O    (16h)
  -0.25890000000000    0.26890000000000    0.02780000000000    O    (16h)
   0.25890000000000    0.28670000000000   -0.47220000000000    O    (16h)
  -0.25890000000000   -0.28670000000000    0.47220000000000    O    (16h)
   0.26890000000000   -0.25890000000000   -0.47220000000000    O    (16h)
  -0.26890000000000    0.25890000000000    0.47220000000000    O    (16h)
   0.28670000000000    0.25890000000000    0.02780000000000    O    (16h)
  -0.28670000000000   -0.25890000000000   -0.02780000000000    O    (16h)
```

Anatase ($TiO_2$, C5): A2B_tI12_141_e_a - CIF

```
# CIF file

data_findsym-output
_audit_creation_method FINDSYM

_chemical_name_mineral 'Anatase'
_chemical_formula_sum 'Ti O2'

loop_
_publ_author_name
 'C. J. Howard'
 'T. M. Sabine'
 'Fiona Dickson'
_journal_name_full
;
Acta Crystallographica B
;
_journal_volume 47
_journal_year 1991
_journal_page_first 462
_journal_page_last 468
_publ_Section_title
;
 Structural and thermal parameters for rutile and anatase
;

_aflow_proto 'A2B_tI12_141_e_a'
_aflow_params 'a,c/a,z2'
_aflow_params_values '3.785,2.51360634082,0.20806'
_aflow_Strukturbericht 'C5'
_aflow_Pearson 'tI12'

_symmetry_space_group_name_Hall "-I 4bd 2"
_symmetry_space_group_name_H-M "I 41/a m d:2"
_symmetry_Int_Tables_number 141

_cell_length_a     3.78500
_cell_length_b     3.78500
_cell_length_c     9.51400
_cell_angle_alpha 90.00000
_cell_angle_beta  90.00000
_cell_angle_gamma 90.00000

loop_
_space_group_symop_id
_space_group_symop_operation_xyz
1 x,y,z
2 x,-y,-z
3 -x,y+1/2,-z
4 -x,-y+1/2,z
5 -y+1/4,-x+1/4,-z+3/4
6 -y+1/4,x+3/4,z+1/4
7 y+3/4,-x+3/4,z+1/4
8 y+3/4,x+1/4,-z+3/4
9 -x,-y,-z
10 -x,y,z
11 x,-y+1/2,z
12 x,y+1/2,-z
13 y+3/4,x+3/4,z+1/4
14 y+3/4,-x+1/4,-z+3/4
15 -y+1/4,x+1/4,-z+3/4
16 -y+1/4,-x+3/4,z+1/4
17 x+1/2,y+1/2,z+1/2
18 x+1/2,-y+1/2,-z+1/2
19 -x+1/2,y,-z+1/2
20 -x+1/2,-y,z+1/2
```

```
21 -y+3/4,-x+3/4,-z+1/4
22 -y+3/4,x+1/4,z+3/4
23 y+1/4,-x+1/4,z+3/4
24 y+1/4,x+3/4,-z+1/4
25 -x+1/2,-y+1/2,-z+1/2
26 -x+1/2,y+1/2,z+1/2
27 x+1/2,-y,z+1/2
28 x+1/2,y,-z+1/2
29 y+1/4,x+1/4,z+3/4
30 y+1/4,-x+3/4,-z+1/4
31 -y+3/4,x+3/4,-z+1/4
32 -y+3/4,-x+1/4,z+3/4

loop_
_atom_site_label
_atom_site_type_symbol
_atom_site_symmetry_multiplicity
_atom_site_Wyckoff_label
_atom_site_fract_x
_atom_site_fract_y
_atom_site_fract_z
_atom_site_occupancy
Ti1 Ti   4 a 0.00000 0.75000 0.12500 1.00000
O1  O    8 e 0.00000 0.25000 0.20806 1.00000
```

Anatase ($TiO_2$, C5): A2B_tI12_141_e_a - POSCAR

```
A2B_tI12_141_e_a & a,c/a,z2 --params=3.785,2.51360634082,0.20806 & I4_1/
    ↪ amd  D_{4h}^{19} #141 (ae) & tI12 & C5 & TiO2 & Anatase &
    ↪ Howard et al., Acta Cryst. B 47, 462-468 (1991)
   1.0000000000000000
  -1.89250000000000   1.89250000000000   4.75700000000000
   1.89250000000000  -1.89250000000000   4.75700000000000
   1.89250000000000   1.89250000000000  -4.75700000000000
    O    Ti
    4     2
Direct
   0.20806000000000   0.45806000000000   0.75000000000000   O   (8e)
   0.45806000000000   0.20806000000000   0.25000000000000   O   (8e)
   0.54194000000000   0.79194000000000   0.75000000000000   O   (8e)
   0.79194000000000   0.54194000000000   0.25000000000000   O   (8e)
   0.12500000000000   0.87500000000000   0.25000000000000   Ti  (4a)
   0.87500000000000   0.12500000000000   0.75000000000000   Ti  (4a)
```

MoB ($B_g$): AB_tI16_141_e_e - CIF

```
# CIF file

data_findsym-output
_audit_creation_method FINDSYM

_chemical_name_mineral 'delta Molybdenum Boride'
_chemical_formula_sum 'Mo B'

loop_
_publ_author_name
 'Roland Kiessling'
_journal_name_full
;
Acta Chemica Scandinavica
;
_journal_volume 1
_journal_year 1947
_journal_page_first 893
_journal_page_last 916
_publ_Section_title
;
 The Crystal Structure of Molybdenum and Tungsten Borides
;

_aflow_proto 'AB_tI16_141_e_e'
_aflow_params 'a,c/a,z1,z2'
_aflow_params_values '3.108,5.45045045045,0.227,0.071'
_aflow_Strukturbericht 'B_g'
_aflow_Pearson 'tI16'

_symmetry_space_group_name_Hall "-I 4bd 2"
_symmetry_space_group_name_H-M "I 41/a m d:2"
_symmetry_Int_Tables_number 141

_cell_length_a     3.10800
_cell_length_b     3.10800
_cell_length_c     16.94000
_cell_angle_alpha  90.00000
_cell_angle_beta   90.00000
_cell_angle_gamma  90.00000

loop_
_space_group_symop_id
_space_group_symop_operation_xyz
1 x,y,z
2 x,-y,-z
3 -x,y+1/2,-z
4 -x,-y+1/2,z
5 -y+1/4,-x+1/4,-z+3/4
6 -y+1/4,x+3/4,z+1/4
7 y+3/4,-x+3/4,z+1/4
8 y+3/4,x+1/4,-z+3/4
9 -x,-y,-z
10 -x,y,z
11 x,-y+1/2,z
12 x,y+1/2,-z
13 y+3/4,x+3/4,z+1/4
14 y+3/4,-x+1/4,-z+3/4
15 -y+1/4,x+1/4,-z+3/4
16 -y+1/4,-x+3/4,z+1/4
17 x+1/2,y+1/2,z+1/2
18 x+1/2,-y+1/2,-z+1/2
19 -x+1/2,y,-z+1/2
20 -x+1/2,-y,z+1/2
21 -y+3/4,-x+3/4,-z+1/4
22 -y+3/4,x+1/4,z+3/4
23 y+1/4,-x+1/4,z+3/4
24 y+1/4,x+3/4,-z+1/4
25 -x+1/2,-y+1/2,-z+1/2
26 -x+1/2,y+1/2,z+1/2
27 x+1/2,-y,z+1/2
28 x+1/2,y,-z+1/2
29 y+1/4,x+1/4,z+3/4
30 y+1/4,-x+3/4,-z+1/4
31 -y+3/4,x+3/4,-z+1/4
32 -y+3/4,-x+1/4,z+3/4

loop_
_atom_site_label
_atom_site_type_symbol
_atom_site_symmetry_multiplicity
_atom_site_Wyckoff_label
_atom_site_fract_x
_atom_site_fract_y
_atom_site_fract_z
_atom_site_occupancy
B1  B    8 e 0.00000 0.25000 0.22700 1.00000
Mo1 Mo   8 e 0.00000 0.25000 0.07100 1.00000
```

MoB ($B_g$): AB_tI16_141_e_e - POSCAR

```
AB_tI16_141_e_e & a,c/a,z1,z2 --params=3.108,5.45045045045,0.227,0.071 &
    ↪ I4_1/amd  D_{4h}^{19} #141 (e^2) & tI16 & B_g & MoB & & Roland
    ↪ Kiessling, Acta Chem. Scand. 1, 893-916 (1947)
   1.0000000000000000
  -1.55400000000000   1.55400000000000   8.47000000000000
   1.55400000000000  -1.55400000000000   8.47000000000000
   1.55400000000000   1.55400000000000  -8.47000000000000
    B    Mo
    4     4
Direct
   0.47700000000000   0.22700000000000   0.25000000000000   B   (8e)
   0.52300000000000   0.77300000000000   0.75000000000000   B   (8e)
   0.77300000000000   0.52300000000000   0.25000000000000   B   (8e)
   0.22700000000000   0.47700000000000   0.75000000000000   B   (8e)
   0.67900000000000  -0.07100000000000   0.75000000000000   Mo  (8e)
   0.92900000000000   0.67900000000000   0.25000000000000   Mo  (8e)
   0.07100000000000   0.32100000000000   0.75000000000000   Mo  (8e)
   0.32100000000000   0.07100000000000   0.25000000000000   Mo  (8e)
```

$Ga_2Hf$: A2B_tI24_141_2e_e - CIF

```
# CIF file

data_findsym-output
_audit_creation_method FINDSYM

_chemical_name_mineral ''
_chemical_formula_sum 'Ga2 Hf'

loop_
_publ_author_name
 'K. Schubert',
 'H. G. Meissner',
 'M. P{\"o}tzschke'
 'W. Rossteutscher'
 'E. Stolz'
_journal_name_full
;
Naturwissenschaften
;
_journal_volume 49
_journal_year 1962
_journal_page_first 57
_journal_page_last 57
_publ_Section_title
;
 Einige Strukturdaten metallischer Phasen (7)
;

# Found in Pearson, Vol. III, pp. 3436

_aflow_proto 'A2B_tI24_141_2e_e'
_aflow_params 'a,c/a,z1,z2,z3'
_aflow_params_values '4.046,6.28917449333,0.125,0.289,-0.051'
_aflow_Strukturbericht 'None'
_aflow_Pearson 'tI24'

_symmetry_space_group_name_Hall "-I 4bd 2"
_symmetry_space_group_name_H-M "I 41/a m d:2"
_symmetry_Int_Tables_number 141

_cell_length_a     4.04600
_cell_length_b     4.04600
_cell_length_c     25.44600
_cell_angle_alpha  90.00000
_cell_angle_beta   90.00000
_cell_angle_gamma  90.00000

loop_
_space_group_symop_id
_space_group_symop_operation_xyz
1 x,y,z
2 x,-y,-z
```

```
3 −x,y+1/2,−z
4 −x,−y+1/2,z
5 −y+1/4,−x+1/4,−z+3/4
6 −y+1/4,x+3/4,z+1/4
7 y+3/4,−x+3/4,z+1/4
8 y+3/4,x+1/4,−z+3/4
9 −x,−y,−z
10 −x,y,z
11 x,−y+1/2,z
12 x,y+1/2,−z
13 y+3/4,x+3/4,z+1/4
14 y+3/4,−x+1/4,−z+3/4
15 −y+1/4,x+1/4,−z+3/4
16 −y+1/4,−x+3/4,z+1/4
17 x+1/2,y+1/2,z+1/2
18 x+1/2,−y+1/2,−z+1/2
19 −x+1/2,y,−z+1/2
20 −x+1/2,−y,z+1/2
21 −y+3/4,−x+3/4,−z+1/4
22 −y+3/4,x+1/4,z+3/4
23 y+1/4,−x+1/4,z+3/4
24 y+1/4,x+3/4,−z+1/4
25 −x+1/2,−y+1/2,−z+1/2
26 −x+1/2,y+1/2,z+1/2
27 x+1/2,−y,z+1/2
28 x+1/2,y,−z+1/2
29 y+1/4,x+1/4,z+3/4
30 y+1/4,−x+3/4,−z+1/4
31 −y+3/4,x+3/4,−z+1/4
32 −y+3/4,−x+1/4,z+3/4

loop_
_atom_site_label
_atom_site_type_symbol
_atom_site_symmetry_multiplicity
_atom_site_Wyckoff_label
_atom_site_fract_x
_atom_site_fract_y
_atom_site_fract_z
_atom_site_occupancy
Ga1 Ga    8 e 0.00000 0.25000 0.12500   1.00000
Ga2 Ga    8 e 0.00000 0.25000 0.28900   1.00000
Hf1 Hf    8 e 0.00000 0.25000 −0.05100   1.00000
```

$Ga_2Hf$: A2B_tI24_141_2e_e - POSCAR

```
A2B_tI24_141_2e_e & a,c/a,z1,z2,z3 −−params=4.046,6.28917449333,0.125,
     ↪ 0.289,−0.051 & I4_1/amd  D_{4h}^{19} #141 (e^3) & tI24 & &
     ↪ Ga2Hf & & K. Schubert et al., Naturwissenschaften 49, 57 (1962)
   1.0000000000000000
  −2.02300000000000   2.02300000000000  12.72300000000000
   2.02300000000000  −2.02300000000000  12.72300000000000
   2.02300000000000   2.02300000000000 −12.72300000000000
   Ga    Hf
    8     4
Direct
   0.37500000000000   0.12500000000000   0.25000000000000   Ga   (8e)
   0.12500000000000   0.37500000000000   0.75000000000000   Ga   (8e)
   0.62500000000000   0.87500000000000   0.75000000000000   Ga   (8e)
   0.87500000000000   0.62500000000000   0.25000000000000   Ga   (8e)
   0.53900000000000   0.28900000000000   0.25000000000000   Ga   (8e)
   0.28900000000000   0.53900000000000   0.75000000000000   Ga   (8e)
   0.46100000000000   0.71100000000000   0.75000000000000   Ga   (8e)
   0.71100000000000   0.46100000000000   0.25000000000000   Ga   (8e)
   0.19900000000000  −0.05100000000000   0.25000000000000   Hf   (8e)
  −0.05100000000000   0.19900000000000   0.75000000000000   Hf   (8e)
   0.80100000000000   0.05100000000000   0.75000000000000   Hf   (8e)
   0.05100000000000   0.80100000000000   0.25000000000000   Hf   (8e)
```

NbP ("40"): AB_tI8_141_a_b - CIF

```
# CIF file

data_findsym−output
_audit_creation_method FINDSYM

_chemical_name_mineral 'alpha Niobium phosphide'
_chemical_formula_sum 'Nb P'

loop_
_publ_author_name
 'N. Sch\"{o}nberg'
_journal_name_full
;
Acta Chemica Scandinavica
;
_journal_volume 8
_journal_year 1954
_journal_page_first 226
_journal_page_last 239
_publ_Section_title
;
 An X−Ray Investigation of Transition Metal Phosphides
;

# Found in  Pearson's Handbook, Vol. IV, pp. 4511

_aflow_proto 'AB_tI8_141_a_b'
_aflow_params 'a,c/a'
_aflow_params_values '3.325,3.42255639098'
_aflow_Strukturbericht '"40"'
_aflow_Pearson 'tI8'

_symmetry_space_group_name_Hall "−I 4bd 2"
_symmetry_space_group_name_H−M "I 41/a m d:2"
_symmetry_Int_Tables_number 141

_cell_length_a      3.32500
_cell_length_b      3.32500
_cell_length_c      11.38000
_cell_angle_alpha  90.00000
_cell_angle_beta   90.00000
_cell_angle_gamma  90.00000

loop_
_space_group_symop_id
_space_group_symop_operation_xyz
1 x,y,z
2 x,−y,−z
3 −x,y+1/2,−z
4 −x,−y+1/2,z
5 −y+1/4,−x+1/4,−z+3/4
6 −y+1/4,x+3/4,z+1/4
7 y+3/4,−x+3/4,z+1/4
8 y+3/4,x+1/4,−z+3/4
9 −x,−y,−z
10 −x,y,z
11 x,−y+1/2,z
12 x,y+1/2,−z
13 y+3/4,x+3/4,z+1/4
14 y+3/4,−x+1/4,−z+3/4
15 −y+1/4,x+1/4,−z+3/4
16 −y+1/4,−x+3/4,z+1/4
17 x+1/2,y+1/2,z+1/2
18 x+1/2,−y+1/2,−z+1/2
19 −x+1/2,y,−z+1/2
20 −x+1/2,−y,z+1/2
21 −y+3/4,−x+3/4,−z+1/4
22 −y+3/4,x+1/4,z+3/4
23 y+1/4,−x+1/4,z+3/4
24 y+1/4,x+3/4,−z+1/4
25 −x+1/2,−y+1/2,−z+1/2
26 −x+1/2,y+1/2,z+1/2
27 x+1/2,−y,z+1/2
28 x+1/2,y,−z+1/2
29 y+1/4,x+1/4,z+3/4
30 y+1/4,−x+3/4,−z+1/4
31 −y+3/4,x+3/4,−z+1/4
32 −y+3/4,−x+1/4,z+3/4

loop_
_atom_site_label
_atom_site_type_symbol
_atom_site_symmetry_multiplicity
_atom_site_Wyckoff_label
_atom_site_fract_x
_atom_site_fract_y
_atom_site_fract_z
_atom_site_occupancy
Nb1 Nb    4 a 0.00000 0.75000 0.12500 1.00000
P1  P     4 b 0.00000 0.25000 0.37500 1.00000
```

NbP ("40"): AB_tI8_141_a_b - POSCAR

```
AB_tI8_141_a_b & a,c/a −−params=3.325,3.42255639098 & I4_1/amd  D_{4h}^{
     ↪ 19} #141 (ab) & tI8 & "40" & NbP & alpha & N. Sch\"{o}nberg,
     ↪ Acta Chem. Scand. 8, 226−239 (1954)
   1.0000000000000000
  −1.66250000000000   1.66250000000000   5.69000000000425
   1.66250000000000  −1.66250000000000   5.69000000000425
   1.66250000000000   1.66250000000000  −5.69000000000425
   Nb    P
    2    2
Direct
   0.12500000000000   0.87500000000000   0.25000000000000   Nb   (4a)
   0.87500000000000   0.12500000000000   0.75000000000000   Nb   (4a)
   0.37500000000000   0.62500000000000   0.75000000000000    P   (4b)
   0.62500000000000   0.37500000000000   0.25000000000000    P   (4b)
```

β-$In_2S_3$: A2B3_tI80_141_ceh_3h - CIF

```
# CIF file

data_findsym−output
_audit_creation_method FINDSYM

_chemical_name_mineral 'beta indium sulfide'
_chemical_formula_sum 'In2 S3'

loop_
_publ_author_name
 'Niyum S Rampersadh'
 'Andrew M Venter'
 'David G Billing'
_journal_name_full
;
Physica B
;
_journal_volume 350
_journal_year 2004
_journal_page_first e383
_journal_page_last e385
_publ_Section_title
;
 Rietveld refinement of In$_2$S$_3$ using neutron and X−ray powder
     ↪ diffraction data
;

_aflow_proto 'A2B3_tI80_141_ceh_3h'
_aflow_params 'a,c/a,z2,y3,z3,y4,z4,y5,z5,y6,z6'
```

```
_aflow_params_values '7.5937,4.26037373086,0.2044,0.5201,0.3324,0.516,
    ↪ 0.2547,0.494,0.0859,0.4667,0.4164'
_aflow_Strukturbericht 'None'
_aflow_Pearson 'tI80'

_symmetry_space_group_name_Hall "-I 4bd 2"
_symmetry_space_group_name_H-M "I 41/a m d:2"
_symmetry_Int_Tables_number 141

_cell_length_a    7.59370
_cell_length_b    7.59370
_cell_length_c    32.35200
_cell_angle_alpha 90.00000
_cell_angle_beta  90.00000
_cell_angle_gamma 90.00000

loop_
_space_group_symop_id
_space_group_symop_operation_xyz
1 x,y,z
2 x,-y,-z
3 -x,y+1/2,-z
4 -x,-y+1/2,z
5 -y+1/4,-x+1/4,-z+3/4
6 -y+1/4,x+3/4,z+1/4
7 y+3/4,-x+3/4,z+1/4
8 y+3/4,x+1/4,-z+3/4
9 -x,-y,-z
10 -x,y,z
11 x,-y+1/2,z
12 x,y+1/2,-z
13 y+3/4,x+3/4,z+1/4
14 y+3/4,-x+1/4,-z+3/4
15 -y+1/4,x+1/4,-z+3/4
16 -y+1/4,-x+3/4,z+1/4
17 x+1/2,y+1/2,z+1/2
18 x+1/2,-y+1/2,-z+1/2
19 -x+1/2,y,-z+1/2
20 -x+1/2,-y,z+1/2
21 -y+3/4,-x+3/4,-z+1/4
22 -y+3/4,x+1/4,z+3/4
23 y+1/4,-x+1/4,z+3/4
24 y+1/4,x+3/4,-z+1/4
25 -x+1/2,-y+1/2,-z+1/2
26 -x+1/2,y+1/2,z+1/2
27 x+1/2,-y,z+1/2
28 x+1/2,y,-z+1/2
29 y+1/4,x+1/4,z+3/4
30 y+1/4,-x+3/4,-z+1/4
31 -y+3/4,x+3/4,-z+1/4
32 -y+3/4,-x+1/4,z+3/4

loop_
_atom_site_label
_atom_site_type_symbol
_atom_site_symmetry_multiplicity
_atom_site_Wyckoff_label
_atom_site_fract_x
_atom_site_fract_y
_atom_site_fract_z
_atom_site_occupancy
In1 In   8 c 0.00000 0.00000 0.00000 1.00000
In2 In   8 e 0.00000 0.25000 0.20440 1.00000
In3 In  16 h 0.00000 0.52010 0.33240 1.00000
S1  S   16 h 0.00000 0.51600 0.25470 1.00000
S2  S   16 h 0.00000 0.49400 0.08590 1.00000
S3  S   16 h 0.00000 0.46670 0.41640 1.00000
```

$\beta$-$In_2S_3$: A2B3_tI80_141_ceh_3h - POSCAR

```
A2B3_tI80_141_ceh_3h & a,c/a,z2,y3,z3,y4,z4,y5,z5,y6,z6 --params=7.5937,
    ↪ 4.26037373086,0.2044,0.5201,0.3324,0.516,0.2547,0.494,0.0859,
    ↪ 0.4667,0.4164 & I4_1/amd  D_{4h}^{19} #141 (ceh^4) & tI80 & &
    ↪ In2S3 & beta & Niyum S Rampersadh and Andrew M Venter and David
    ↪ G Billing, Physica B 350, e383-e385 (2004)
   1.0000000000000000
  -3.79685000000000   3.79685000000000  16.17600000000000
   3.79685000000000  -3.79685000000000  16.17600000000000
   3.79685000000000   3.79685000000000 -16.17600000000000
   In    S
   16    24
Direct
   0.14750000000000   0.66760000000000   0.47990000000000   In   (16h)
   0.31230000000000   0.33240000000000  -0.02010000000000   In   (16h)
   0.33240000000000   0.31230000000000   0.47990000000000   In   (16h)
   0.33240000000000   0.85250000000000   0.02010000000000   In   (16h)
   0.66760000000000   0.14750000000000  -0.02010000000000   In   (16h)
   0.66760000000000   0.68770000000000   0.52010000000000   In   (16h)
   0.68770000000000   0.66760000000000   0.02010000000000   In   (16h)
   0.85250000000000   0.33240000000000   0.52010000000000   In   (16h)
   0.00000000000000   0.00000000000000   0.00000000000000   In    (8c)
   0.00000000000000   0.00000000000000   0.50000000000000   In    (8c)
   0.00000000000000   0.50000000000000   0.00000000000000   In    (8c)
   0.50000000000000   0.00000000000000   0.50000000000000   In    (8c)
   0.20440000000000   0.45440000000000   0.75000000000000   In    (8e)
   0.45440000000000   0.20440000000000   0.25000000000000   In    (8e)
   0.54560000000000   0.79560000000000   0.75000000000000   In    (8e)
   0.79560000000000   0.54560000000000   0.25000000000000   In    (8e)
   0.22930000000000   0.74530000000000   0.48400000000000    S   (16h)
   0.23870000000000   0.25470000000000  -0.01600000000000    S   (16h)
   0.25470000000000   0.23870000000000   0.48400000000000    S   (16h)
   0.25470000000000   0.77070000000000   0.01600000000000    S   (16h)
   0.74530000000000   0.22930000000000  -0.01600000000000    S   (16h)
   0.74530000000000   0.76130000000000   0.51600000000000    S   (16h)
   0.76130000000000   0.74530000000000   0.01600000000000    S   (16h)
   0.77070000000000   0.25470000000000   0.51600000000000    S   (16h)
  -0.08590000000000  -0.09190000000000   0.49400000000000    S   (16h)
   0.08590000000000   0.09190000000000   0.50600000000000    S   (16h)
  -0.08590000000000   0.42010000000000   0.00600000000000    S   (16h)
   0.08590000000000   0.57990000000000  -0.00600000000000    S   (16h)
   0.09190000000000   0.08590000000000   0.00600000000000    S   (16h)
  -0.09190000000000  -0.08590000000000  -0.00600000000000    S   (16h)
   0.42010000000000  -0.08590000000000   0.50600000000000    S   (16h)
   0.57990000000000   0.08590000000000   0.49400000000000    S   (16h)
   0.11690000000000   0.58360000000000   0.53330000000000    S   (16h)
   0.41640000000000   0.44970000000000   0.53330000000000    S   (16h)
   0.41640000000000   0.88310000000000  -0.03330000000000    S   (16h)
   0.44970000000000   0.41640000000000   0.03330000000000    S   (16h)
   0.55030000000000   0.58360000000000  -0.03330000000000    S   (16h)
   0.58360000000000   0.11690000000000   0.03330000000000    S   (16h)
   0.58360000000000   0.55030000000000   0.46670000000000    S   (16h)
   0.88310000000000   0.41640000000000   0.46670000000000    S   (16h)
```

$PPrS_4$: ABC4_tI96_142_e_ab_2g - CIF

```
# CIF file

data_findsym-output
_audit_creation_method FINDSYM

_chemical_name_mineral ''
_chemical_formula_sum 'P Pr S4'

loop_
_publ_author_name
 'C. Wibbelmann'
 'W. Brockner'
 'B. Eisenmann'
 'H. Sch\"{a}fer'
_journal_name_full
;
Zeitschrift f\"{u}r Naturforschung
;
_journal_volume 39a
_journal_year 1983
_journal_page_first 190
_journal_page_last 194
_publ_Section_title
;
 Kristallstruktur und Schwingungsspektrum des
    ↪ Praseodym-ortho-Thiophosphates PrPS$_4$
;

# Found in http://materials.springer.com/isp/crystallographic/docs/
    ↪ sd_1703369

_aflow_proto 'ABC4_tI96_142_e_ab_2g'
_aflow_params 'a,c/a,x3,x4,y4,z4,x5,y5,z5'
_aflow_params_values '10.914,1.77396005131,0.0375,0.2482,0.3197,-0.0867,
    ↪ 0.0923,0.1117,0.0025'
_aflow_Strukturbericht 'None'
_aflow_Pearson 'tI96'

_symmetry_space_group_name_Hall "-I 4bd 2c"
_symmetry_space_group_name_H-M "I 41/a c d:2"
_symmetry_Int_Tables_number 142

_cell_length_a    10.91400
_cell_length_b    10.91400
_cell_length_c    19.36100
_cell_angle_alpha 90.00000
_cell_angle_beta  90.00000
_cell_angle_gamma 90.00000

loop_
_space_group_symop_id
_space_group_symop_operation_xyz
1 x,y,z
2 x+1/2,-y+1/2,-z
3 -x+1/2,y,-z
4 -x,-y+1/2,z
5 -y+1/4,-x+1/4,-z+1/4
6 -y+1/4,x+3/4,z+1/4
7 y+3/4,-x+3/4,z+1/4
8 y+3/4,x+1/4,-z+1/4
9 -x,-y,-z
10 -x,y,z+1/2
11 x,-y+1/2,z+1/2
12 x,y+1/2,-z
13 y+1/4,x+1/4,z+1/4
14 y+3/4,-x+1/4,-z+3/4
15 -y+1/4,x+1/4,-z+3/4
16 -y+3/4,-x+1/4,z+1/4
17 x+1/2,y+1/2,z+1/2
18 x,-y,-z+1/2
19 -x,y+1/2,-z+1/2
20 -x+1/2,-y,z+1/2
21 -y+3/4,-x+3/4,-z+3/4
22 -y+3/4,x+1/4,z+3/4
23 y+1/4,-x+1/4,z+3/4
24 y+1/4,x+3/4,-z+3/4
25 -x+1/2,-y+1/2,-z+1/2
26 -x+1/2,y+1/2,z
27 x+1/2,-y,z
28 x+1/2,y,-z+1/2
29 y+3/4,x+3/4,z+3/4
30 y+1/4,-x+3/4,-z+1/4
31 -y+3/4,x+3/4,-z+1/4
32 -y+1/4,-x+3/4,z+3/4
```

```
loop_
_atom_site_label
_atom_site_type_symbol
_atom_site_symmetry_multiplicity
_atom_site_Wyckoff_label
_atom_site_fract_x
_atom_site_fract_y
_atom_site_fract_z
_atom_site_occupancy
Pr1  Pr    8  a  0.00000  0.25000   0.37500  1.00000
Pr2  Pr    8  b  0.00000  0.25000   0.12500  1.00000
P1   P    16  e  0.03750  0.00000   0.25000  1.00000
S1   S    32  g  0.24820  0.31970  -0.08670  1.00000
S2   S    32  g  0.09230  0.11170   0.00250  1.00000
```

PPrS$_4$: ABC4_tI96_142_e_ab_2g - POSCAR

```
ABC4_tI96_142_e_ab_2g & a,c/a,x3,x4,y4,z4,x5,y5,z5 --params=10.914,
    ↪ 1.77396005131,0.0375,0.2482,0.3197,-0.0867,0.0923,0.1117,0.0025
    ↪ & I4_1/acd  D_{4h}^{20} #142 (abeg^2) & tI96 & & PPrS4 & & C.
    ↪ Wibbelmann and W. Brockner, Z. Naturforsch. 39a, 190-194 (1984)
   1.0000000000000000
  -5.45700000000000    5.45700000000000    9.68050000000000
   5.45700000000000   -5.45700000000000    9.68050000000000
   5.45700000000000    5.45700000000000   -9.68050000000000
    P     Pr      S
    8      8     32
Direct
   0.21250000000000    0.25000000000000    0.46250000000000     P    (16e)
   0.25000000000000    0.28750000000000    0.03750000000000     P    (16e)
   0.25000000000000    0.78750000000000    0.53750000000000     P    (16e)
   0.28750000000000    0.75000000000000    0.03750000000000     P    (16e)
   0.71250000000000    0.25000000000000   -0.03750000000000     P    (16e)
   0.75000000000000    0.21250000000000    0.46250000000000     P    (16e)
   0.75000000000000    0.71250000000000   -0.03750000000000     P    (16e)
   0.78750000000000    0.75000000000000    0.53750000000000     P    (16e)
   0.12500000000000    0.87500000000000    0.25000000000000    Pr     (8a)
   0.37500000000000    0.62500000000000    0.75000000000000    Pr     (8a)
   0.62500000000000    0.37500000000000    0.25000000000000    Pr     (8a)
   0.87500000000000    0.12500000000000    0.75000000000000    Pr     (8a)
   0.12500000000000    0.37500000000000    0.75000000000000    Pr     (8b)
   0.37500000000000    0.12500000000000    0.25000000000000    Pr     (8b)
   0.62500000000000    0.87500000000000    0.75000000000000    Pr     (8b)
   0.87500000000000    0.62500000000000    0.25000000000000    Pr     (8b)
  -0.09360000000000    0.33490000000000    0.06790000000000     S    (32g)
   0.09360000000000    0.66510000000000   -0.06790000000000     S    (32g)
   0.16150000000000    0.09360000000000   -0.07150000000000     S    (32g)
   0.16510000000000    0.59360000000000    0.43210000000000     S    (32g)
   0.23300000000000    0.16150000000000    0.56790000000000     S    (32g)
   0.26700000000000    0.83490000000000   -0.07150000000000     S    (32g)
   0.33490000000000    0.76700000000000    0.42850000000000     S    (32g)
   0.33850000000000    0.26700000000000   -0.06790000000000     S    (32g)
   0.40640000000000    0.33850000000000    0.57150000000000     S    (32g)
   0.59360000000000    0.66150000000000    0.42850000000000     S    (32g)
   0.66150000000000    0.73300000000000    0.06790000000000     S    (32g)
   0.66510000000000    0.23300000000000    0.57150000000000     S    (32g)
   0.73300000000000    0.16510000000000    0.07150000000000     S    (32g)
   0.76700000000000    0.83850000000000    0.43210000000000     S    (32g)
   0.83490000000000    0.40640000000000    0.56790000000000     S    (32g)
   0.83850000000000   -0.09360000000000    0.07150000000000     S    (32g)
  -0.08980000000000    0.11420000000000    0.51940000000000     S    (32g)
   0.08980000000000    0.88580000000000    0.48060000000000     S    (32g)
   0.09480000000000    0.39080000000000   -0.01940000000000     S    (32g)
  -0.09480000000000    0.60920000000000    0.01940000000000     S    (32g)
   0.10920000000000    0.40520000000000    0.51940000000000     S    (32g)
   0.11420000000000    0.09480000000000    0.20400000000000     S    (32g)
   0.38580000000000    0.58980000000000   -0.01940000000000     S    (32g)
   0.39080000000000   -0.08980000000000    0.29600000000000     S    (32g)
   0.40520000000000    0.38580000000000    0.29600000000000     S    (32g)
   0.41020000000000    0.89080000000000    0.79600000000000     S    (32g)
   0.58980000000000    0.10920000000000    0.20400000000000     S    (32g)
   0.59480000000000    0.61420000000000    0.70400000000000     S    (32g)
   0.60920000000000    0.08980000000000    0.70400000000000     S    (32g)
   0.61420000000000    0.41020000000000    0.01940000000000     S    (32g)
   0.88580000000000   -0.09480000000000    0.79600000000000     S    (32g)
   0.89080000000000    0.59480000000000    0.48060000000000     S    (32g)
```

ζ-AgZn (B$_b$): A2B_hP9_147_g_ad - CIF

```
# CIF file

data_findsym-output
_audit_creation_method FINDSYM

_chemical_name_mineral 'zeta silver zinc'
_chemical_formula_sum 'Ag2 Zn'

loop_
_publ_author_name
 'Gunnar Bergman'
 'Robert W. Jaross'
_journal_name_full
;
Acta Crystallographica
;
_journal_volume 8
_journal_year 1955
_journal_page_first 232
_journal_page_last 235
_publ_Section_title
;
 On the Crystal Structure of the $\zeta$ Phase in the Silver-Zinc System
    ↪ and the Mechanism of the $\beta-\zeta$ Transformation
;
_aflow_proto 'A2B_hP9_147_g_ad'
_aflow_params 'a,c/a,z2,x3,y3,z3'
_aflow_params_values '7.636,0.369264012572,0.25,0.33333,0.0,0.25'
_aflow_Strukturbericht 'B_b'
_aflow_Pearson 'hP9'

_symmetry_space_group_name_Hall "-P 3"
_symmetry_space_group_name_H-M "P -3"
_symmetry_Int_Tables_number 147

_cell_length_a      7.63600
_cell_length_b      7.63600
_cell_length_c      2.81970
_cell_angle_alpha  90.00000
_cell_angle_beta   90.00000
_cell_angle_gamma 120.00000

loop_
_space_group_symop_id
_space_group_symop_operation_xyz
1 x,y,z
2 -y,x-y,z
3 -x+y,-x,z
4 -x,-y,-z
5 y,-x+y,-z
6 x-y,x,-z

loop_
_atom_site_label
_atom_site_type_symbol
_atom_site_symmetry_multiplicity
_atom_site_Wyckoff_label
_atom_site_fract_x
_atom_site_fract_y
_atom_site_fract_z
_atom_site_occupancy
Zn1  Zn   1  a  0.00000  0.00000  0.00000  1.00000
Zn2  Zn   2  d  0.33333  0.66667  0.25000  1.00000
Ag1  Ag   6  g  0.33333  0.00000  0.25000  1.00000
```

ζ-AgZn (B$_b$): A2B_hP9_147_g_ad - POSCAR

```
A2B_hP9_147_g_ad & a,c/a,z2,x3,y3,z3 --params=7.636,0.369264012572,0.25,
    ↪ 0.33333,0.0,0.25 & P(-3)      C_{3i}^1 #147 (adg) & hP9 & B_b &
    ↪ Ag2Zn & zeta & G. Bergman and R. W. Jaross, Acta Cryst. 8,
    ↪ 232-235 (1955)
   1.0000000000000000
   3.81800000000000   -6.61296998329800    0.00000000000000
   3.81800000000000    6.61296998329800    0.00000000000000
   0.00000000000000    0.00000000000000    2.81970000000000
   Ag     Zn
    6      3
Direct
   0.00000000000000    0.33333333333333    0.25000000000000    Ag     (6g)
   0.00000000000000    0.66666666666667    0.75000000000000    Ag     (6g)
   0.33333333333333    0.00000000000000    0.25000000000000    Ag     (6g)
   0.33333333333333    0.33333333333333    0.75000000000000    Ag     (6g)
   0.66666666666667    0.00000000000000    0.75000000000000    Ag     (6g)
   0.66666666666667    0.66666666666667    0.25000000000000    Ag     (6g)
   0.00000000000000    0.00000000000000    0.00000000000000    Zn     (1a)
   0.33333333333333    0.66666666666667    0.25000000000000    Zn     (2d)
   0.66666666666667    0.33333333333333    0.75000000000000    Zn     (2d)
```

Solid Cubane ($C_8H_8$): AB_hR16_148_cf_cf - CIF

```
# CIF file

data_findsym-output
_audit_creation_method FINDSYM

_chemical_name_mineral 'Cubane'
_chemical_formula_sum 'C8 H8'

loop_
_publ_author_name
 'Everly B. Fleischer'
_journal_name_full
;
Journal of the American Chemical Society
;
_journal_volume 86
_journal_year 1964
_journal_page_first 3889
_journal_page_last 3890
_publ_Section_title
;
 X-Ray Structure Determination of Cubane
;

_aflow_proto 'AB_hR16_148_cf_cf'
_aflow_params 'a,c/a,x1,x2,x3,y3,z3,x4,y4,z4'
_aflow_params_values '6.29713,1.8633345667,0.11546,0.21,0.10706,0.81289,
    ↪ 0.19519,0.1848,0.6754,0.3468'
_aflow_Strukturbericht 'None'
_aflow_Pearson 'hR16'

_symmetry_space_group_name_Hall "-R 3"
_symmetry_space_group_name_H-M "R -3:H"
_symmetry_Int_Tables_number 148

_cell_length_a      6.29713
_cell_length_b      6.29713
_cell_length_c     11.73366
_cell_angle_alpha  90.00000
_cell_angle_beta   90.00000
```

```
_cell_angle_gamma 120.00000

loop_
_space_group_symop_id
_space_group_symop_operation_xyz
1 x,y,z
2 -y,x-y,z
3 -x+y,-x,z
4 -x,-y,-z
5 y,-x+y,-z
6 x-y,x,-z
7 x+1/3,y+2/3,z+2/3
8 -y+1/3,x-y+2/3,z+2/3
9 -x+y+1/3,-x+2/3,z+2/3
10 -x+1/3,-y+2/3,-z+2/3
11 y+1/3,-x+y+2/3,-z+2/3
12 x-y+1/3,x+2/3,-z+2/3
13 x+2/3,y+1/3,z+1/3
14 -y+2/3,x-y+1/3,z+1/3
15 -x+y+2/3,-x+1/3,z+1/3
16 -x+2/3,-y+1/3,-z+1/3
17 y+2/3,-x+y+1/3,-z+1/3
18 x-y+2/3,x+1/3,-z+1/3

loop_
_atom_site_label
_atom_site_type_symbol
_atom_site_symmetry_multiplicity
_atom_site_Wyckoff_label
_atom_site_fract_x
_atom_site_fract_y
_atom_site_fract_z
_atom_site_occupancy
C1 C    6 c 0.00000 0.00000 0.11546 1.00000
H1 H    6 c 0.00000 0.00000 0.21000 1.00000
C2 C   18 f 0.06868 0.84319 0.03838 1.00000
H2 H   18 f 0.11580 0.72220 0.06900 1.00000
```

Solid Cubane ($C_8H_8$): AB_hR16_148_cf_cf - POSCAR

```
AB_hR16_148_cf_cf & a,c/a,x1,x2,x3,y3,z3,x4,y4,z4 --params=6.29713,
    ↪ 1.8633345667,0.11546,0.21,0.10706,0.81289,0.19519,0.1848,0.6754
    ↪ ,0.3468 & R(-3)     C_{3i}^2 #148 (c^2f^2) & hR16 & & C8H8 &
    ↪ Cubane & E. B. Fleischer, JACS 86, 3889-3890 (1964)
   1.0000000000000000
   3.14856727069900   -1.81782616130000    3.91121876007700
   0.00000000000000    3.63565232259900    3.91121876007700
  -3.14856727069900   -1.81782616130000    3.91121876007700
     C      H
     8      8
Direct
   0.11546000000000   0.11546000000000   0.11546000000000   C   (2c)
   0.88454000000000   0.88454000000000   0.88454000000000   C   (2c)
   0.10706000000000   0.81289000000000   0.19519000000000   C   (6f)
   0.18711000000000   0.80481000000000   0.89294000000000   C   (6f)
   0.19519000000000   0.10706000000000   0.81289000000000   C   (6f)
   0.80481000000000   0.89294000000000   0.18711000000000   C   (6f)
   0.81289000000000   0.19519000000000   0.10706000000000   C   (6f)
   0.89294000000000   0.18711000000000   0.80481000000000   C   (6f)
   0.21000000000000   0.21000000000000   0.21000000000000   H   (2c)
   0.79000000000000   0.79000000000000   0.79000000000000   H   (2c)
   0.18480000000000   0.67540000000000   0.34680000000000   H   (6f)
   0.32460000000000   0.65320000000000   0.81520000000000   H   (6f)
   0.34680000000000   0.18480000000000   0.67540000000000   H   (6f)
   0.65320000000000   0.81520000000000   0.32460000000000   H   (6f)
   0.67540000000000   0.34680000000000   0.18480000000000   H   (6f)
   0.81520000000000   0.32460000000000   0.65320000000000   H   (6f)
```

$BiI_3$ ($D0_5$): AB3_hR8_148_c_f - CIF

```
# CIF file

data_findsym-output
_audit_creation_method FINDSYM

_chemical_name_mineral 'Bismuth triodide'
_chemical_formula_sum 'Bi I3'

loop_
_publ_author_name
 'H. Br\aekken'
_journal_name_full
;
Zeitschrift f\"{u}r Kristallographie - Crystalline Materials
;
_journal_volume 74
_journal_year 1930
_journal_page_first 67
_journal_page_last 72
_publ_Section_title
;
 IX. Die Kristallstruktur der Trijodide von Arsen, Antimon und Wismut
;

# Found in  Strukturbericht Vol. II, pp. 25-27

_aflow_proto 'AB3_hR8_148_c_f'
_aflow_params 'a,c/a,x1,x2,y2,z2'
_aflow_params_values '7.49626,2.75900649124,0.33333,0.088,0.755,0.421'
_aflow_Strukturbericht 'D0_5'
_aflow_Pearson 'hR8'

_symmetry_space_group_name_Hall "-R 3"
_symmetry_space_group_name_H-M "R -3:H"
_symmetry_Int_Tables_number 148
```

```
_cell_length_a     7.49626
_cell_length_b     7.49626
_cell_length_c     20.68223
_cell_angle_alpha  90.00000
_cell_angle_beta   90.00000
_cell_angle_gamma  120.00000

loop_
_space_group_symop_id
_space_group_symop_operation_xyz
1 x,y,z
2 -y,x-y,z
3 -x+y,-x,z
4 -x,-y,-z
5 y,-x+y,-z
6 x-y,x,-z
7 x+1/3,y+2/3,z+2/3
8 -y+1/3,x-y+2/3,z+2/3
9 -x+y+1/3,-x+2/3,z+2/3
10 -x+1/3,-y+2/3,-z+2/3
11 y+1/3,-x+y+2/3,-z+2/3
12 x-y+1/3,x+2/3,-z+2/3
13 x+2/3,y+1/3,z+1/3
14 -y+2/3,x-y+1/3,z+1/3
15 -x+y+2/3,-x+1/3,z+1/3
16 -x+2/3,-y+1/3,-z+1/3
17 y+2/3,-x+y+1/3,-z+1/3
18 x-y+2/3,x+1/3,-z+1/3

loop_
_atom_site_label
_atom_site_type_symbol
_atom_site_symmetry_multiplicity
_atom_site_Wyckoff_label
_atom_site_fract_x
_atom_site_fract_y
_atom_site_fract_z
_atom_site_occupancy
Bi1 Bi    6 c 0.00000 0.00000 0.33333 1.00000
I1  I    18 f 0.00000 0.66700 0.08800 1.00000
```

$BiI_3$ ($D0_5$): AB3_hR8_148_c_f - POSCAR

```
AB3_hR8_148_c_f & a,c/a,x1,x2,y2,z2 --params=7.49626,2.75900649124,
    ↪ 0.33333,0.088,0.755,0.421 & R(-3)     C_{3i}^2 #148 (cf) & hR8
    ↪ & D0_5 & BiI3 & & H. Braekken, Zeitschrift f\"{u}r
    ↪ Kristallographie - Crystalline Materials 74, 67âĂŞ72 (1930)
   1.0000000000000000
   3.74813000000000   -2.16398386445771    6.89407666667425
   0.00000000000000    4.32796772891543    6.89407666667425
  -3.74813000000000   -2.16398386445771    6.89407666667425
    Bi      I
     2      6
Direct
   0.33333333333333   0.33333333333333   0.33333333333333   Bi   (2c)
   0.66666666666667   0.66666666666667   0.66666666666667   Bi   (2c)
  -0.08800000000000   0.24500000000000   0.57900000000000   I    (6f)
   0.08800000000000   0.75500000000000   0.42100000000000   I    (6f)
   0.24500000000000   0.57900000000000  -0.08800000000000   I    (6f)
   0.42100000000000   0.08800000000000   0.75500000000000   I    (6f)
   0.57900000000000  -0.08800000000000   0.24500000000000   I    (6f)
   0.75500000000000   0.42100000000000   0.08800000000000   I    (6f)
```

PdAl: AB_hR26_148_b2f_a2f - CIF

```
# CIF file

data_findsym-output
_audit_creation_method FINDSYM

_chemical_name_mineral 'beta-prime palladium aluminum'
_chemical_formula_sum 'Pd Al'

loop_
_publ_author_name
 'T. Matkovi\'{c}'
 'K. Schubert'
_journal_name_full
;
Journal of the Less-Common Metals
;
_journal_volume 55
_journal_year 1977
_journal_page_first 45
_journal_page_last 52
_publ_Section_title
;
 Kristallstruktur vo PdAl.r
;

# Found in  http://materials.springer.com/lb/docs/
    ↪ sm_lbs_978-3-540-70892-6_476

_aflow_proto 'AB_hR26_148_b2f_a2f'
_aflow_params 'a,c/a,x3,y3,z3,x4,y4,z4,x5,y5,z5,x6,y6,z6'
_aflow_params_values '15.659,0.335334312536,0.054,0.346,0.098,0.754,
    ↪ 0.15699,0.6,0.555,0.84401,0.599,0.252,0.65501,0.098'
_aflow_Strukturbericht 'None'
_aflow_Pearson 'hR26'

_symmetry_space_group_name_Hall "-R 3"
_symmetry_space_group_name_H-M "R -3:H"
_symmetry_Int_Tables_number 148
```

```
_cell_length_a      15.65900
_cell_length_b      15.65900
_cell_length_c      5.25100
_cell_angle_alpha  90.00000
_cell_angle_beta   90.00000
_cell_angle_gamma  120.00000

loop_
_space_group_symop_id
_space_group_symop_operation_xyz
1 x,y,z
2 -y,x-y,z
3 -x+y,-x,z
4 -x,-y,-z
5 y,-x+y,-z
6 x-y,x,-z
7 x+1/3,y+2/3,z+2/3
8 -y+1/3,x-y+2/3,z+2/3
9 -x+y+1/3,-x+2/3,z+2/3
10 -x+1/3,-y+2/3,-z+2/3
11 y+1/3,-x+y+2/3,-z+2/3
12 x-y+1/3,x+2/3,-z+2/3
13 x+2/3,y+1/3,z+1/3
14 -y+2/3,x-y+1/3,z+1/3
15 -x+y+2/3,-x+1/3,z+1/3
16 -x+2/3,-y+1/3,-z+1/3
17 y+2/3,-x+y+1/3,-z+1/3
18 x-y+2/3,x+1/3,-z+1/3

loop_
_atom_site_label
_atom_site_type_symbol
_atom_site_symmetry_multiplicity
_atom_site_Wyckoff_label
_atom_site_fract_x
_atom_site_fract_y
_atom_site_fract_z
_atom_site_occupancy
Pd1 Pd    3 a 0.00000  0.00000   0.00000  1.00000
Al1  Al    3 b 0.00000  0.00000   0.50000  1.00000
Al2  Al   18 f 0.88800  0.06800   0.16600  1.00000
Al3  Al   18 f 0.58367  0.57033   0.17033  1.00000
Pd2  Pd   18 f 0.22233  0.73367   0.33267  1.00000
Pd3  Pd   18 f 0.25033 -0.09633  0.00167  1.00000
```

PdAl: AB_hR26_148_b2f_a2f - POSCAR

```
AB_hR26_148_b2f_a2f & a,c/a,x3,y3,z3,x4,y4,z4,x5,y5,z5,x6,y6,z6 --params
    ↪ =15.659,0.335334312536,0.054,0.346,0.098,0.754,0.15699,0.6,
    ↪ 0.555,0.84401,0.599,0.252,0.65501,0.098 & R(-3)      C_{3i}^2 #
    ↪ 148 (abf^4) & hR26 & & PdAl & beta' & T. Matkovi\'{c} and K.
    ↪ Schubert, J. Less Common Metals 55, 45-52 (1977)
   1.0000000000000000
   7.82950000000000   -4.52036393262018    1.75033333333333
   0.00000000000000    9.04072786524035    1.75033333333333
  -7.82950000000000   -4.52036393262018    1.75033333333333
   Al     Pd
   13     13
Direct
   0.50000000000000    0.50000000000000    0.50000000000000    Al    (1b)
   0.05400000000000    0.34600000000000    0.09800000000000    Al    (6f)
  -0.05400000000000    0.65400000000000   -0.09800000000000    Al    (6f)
   0.09800000000000    0.05400000000000    0.34600000000000    Al    (6f)
  -0.09800000000000   -0.05400000000000    0.65400000000000    Al    (6f)
   0.34600000000000    0.09800000000000    0.05400000000000    Al    (6f)
   0.65400000000000   -0.09800000000000   -0.05400000000000    Al    (6f)
  -0.15700000000000    0.40000000000000    0.24600000000000    Al    (6f)
   0.15700000000000    0.60000000000000    0.75400000000000    Al    (6f)
   0.24600000000000   -0.15700000000000    0.40000000000000    Al    (6f)
   0.40000000000000    0.24600000000000   -0.15700000000000    Al    (6f)
   0.60000000000000    0.75400000000000    0.15700000000000    Al    (6f)
   0.75400000000000    0.15700000000000    0.60000000000000    Al    (6f)
   0.00000000000000    0.00000000000000    0.00000000000000    Pd    (1a)
   0.15600000000000    0.40100000000000    0.44500000000000    Pd    (6f)
   0.40100000000000    0.44500000000000    0.15600000000000    Pd    (6f)
   0.44500000000000    0.15600000000000    0.40100000000000    Pd    (6f)
   0.55500000000000    0.84400000000000    0.59900000000000    Pd    (6f)
   0.59900000000000    0.55500000000000    0.84400000000000    Pd    (6f)
   0.84400000000000    0.59900000000000    0.55500000000000    Pd    (6f)
   0.09800000000000    0.25200000000000    0.65500000000000    Pd    (6f)
  -0.09800000000000    0.74800000000000    0.34500000000000    Pd    (6f)
   0.25200000000000    0.65500000000000    0.09800000000000    Pd    (6f)
   0.34500000000000   -0.09800000000000    0.74800000000000    Pd    (6f)
   0.65500000000000    0.09800000000000    0.25200000000000    Pd    (6f)
   0.74800000000000    0.34500000000000   -0.09800000000000    Pd    (6f)
```

Ilmenite ($FeTiO_3$): AB3C_hR10_148_c_f_c - CIF

```
# CIF file

data_findsym-output
_audit_creation_method FINDSYM

_chemical_name_mineral 'Ilmenite'
_chemical_formula_sum 'Fe Ti O3'

loop_
_publ_author_name
 'Barry A. Wechsler'
 'Charles T. Prewitt'
_journal_name_full
;
American Mineralogist
;
_journal_volume 69
_journal_year 1984
_journal_page_first 176
_journal_page_last 185
_publ_Section_title
;
 Crystal Structure of Ilmenite (FeTiO$_3$) at high temperature and high
    ↪ pressure
;

# Found in  Wyckoff, Vol. II, pp. 420

_aflow_proto 'AB3C_hR10_148_c_f_c'
_aflow_params 'a,c/a,x1,x2,x3,y3,z3'
_aflow_params_values '5.0884,2.76815894977,0.35537,0.1464,0.22174,
    ↪ 0.56249,0.95095'
_aflow_Strukturbericht 'None'
_aflow_Pearson 'hR10'

_symmetry_space_group_name_Hall "-R 3"
_symmetry_space_group_name_H-M "R -3:H"
_symmetry_Int_Tables_number 148

_cell_length_a      5.08840
_cell_length_b      5.08840
_cell_length_c      14.08550
_cell_angle_alpha  90.00000
_cell_angle_beta   90.00000
_cell_angle_gamma  120.00000

loop_
_space_group_symop_id
_space_group_symop_operation_xyz
1 x,y,z
2 -y,x-y,z
3 -x+y,-x,z
4 -x,-y,-z
5 y,-x+y,-z
6 x-y,x,-z
7 x+1/3,y+2/3,z+2/3
8 -y+1/3,x-y+2/3,z+2/3
9 -x+y+1/3,-x+2/3,z+2/3
10 -x+1/3,-y+2/3,-z+2/3
11 y+1/3,-x+y+2/3,-z+2/3
12 x-y+1/3,x+2/3,-z+2/3
13 x+2/3,y+1/3,z+1/3
14 -y+2/3,x-y+1/3,z+1/3
15 -x+y+2/3,-x+1/3,z+1/3
16 -x+2/3,-y+1/3,-z+1/3
17 y+2/3,-x+y+1/3,-z+1/3
18 x-y+2/3,x+1/3,-z+1/3

loop_
_atom_site_label
_atom_site_type_symbol
_atom_site_symmetry_multiplicity
_atom_site_Wyckoff_label
_atom_site_fract_x
_atom_site_fract_y
_atom_site_fract_z
_atom_site_occupancy
Fe1 Fe    6 c 0.00000   0.00000  0.35537  1.00000
Ti1  Ti    6 c 0.00000   0.00000  0.14640  1.00000
O1   O    18 f -0.02332  0.29411  0.24506  1.00000
```

Ilmenite ($FeTiO_3$): AB3C_hR10_148_c_f_c - POSCAR

```
AB3C_hR10_148_c_f_c & a,c/a,x1,x2,x3,y3,z3 --params=5.0884,2.76815894977
    ↪ ,0.35537,0.1464,0.22174,0.56249,0.95095 & R(-3)      C_{3i}^2 #
    ↪ 148 (c^2f) & hR10 & & FeTiO3 & Ilmenite & B. A. Wechsler and C.
    ↪ T. Prewitt, Am. Mineral. 69, 176-185 (1984)
   1.0000000000000000
   2.54420000000000   -1.46889455487200    4.69516666666700
   0.00000000000000    2.93778910974400    4.69516666666700
  -2.54420000000000   -1.46889455487200    4.69516666666700
   Fe     O     Ti
    2     6      2
Direct
   0.35537000000000    0.35537000000000    0.35537000000000    Fe    (2c)
   0.64463000000000    0.64463000000000    0.64463000000000    Fe    (2c)
  -0.04905000000000    0.22174000000000    0.56249000000000    O     (6f)
   0.04905000000000    0.77826000000000    0.43751000000000    O     (6f)
   0.22174000000000    0.56249000000000   -0.04905000000000    O     (6f)
   0.43751000000000    0.04905000000000    0.77826000000000    O     (6f)
   0.56249000000000   -0.04905000000000    0.22174000000000    O     (6f)
   0.77826000000000    0.43751000000000    0.04905000000000    O     (6f)
   0.14640000000000    0.14640000000000    0.14640000000000    Ti    (2c)
   0.85360000000000    0.85360000000000    0.85360000000000    Ti    (2c)
```

Original $Fe_2P$ (C22): A2B_hP9_150_ef_bd - CIF

```
# CIF file

data_findsym-output
_audit_creation_method FINDSYM

_chemical_name_mineral ''
_chemical_formula_sum 'Fe2 P'

loop_
_publ_author_name
 'Sterling B. Hendricks'
 'Peter R. Kosting'
_journal_name_full
;
Zeitschrift f\"{u}r Kristallographie - Crystalline Materials
```

```
;
_journal_volume 74
_journal_year 1930
_journal_page_first 511
_journal_page_last 533
_publ_Section_title
;
 The Crystal Structure of Fe$_2$P, Fe$_2$N, Fe$_3$N and FeB
;

# Found in  Strukturbericht, Vol. II, pp. 15

_aflow_proto 'A2B_hP9_150_ef_bd'
_aflow_params 'a,c/a,z2,x3,x4'
_aflow_params_values '5.85,0.589743589744,0.875,0.26,0.6'
_aflow_Strukturbericht 'C22'
_aflow_Pearson 'hP9'

_symmetry_space_group_name_Hall "P 3 2"
_symmetry_space_group_name_H-M "P 3 2 1"
_symmetry_Int_Tables_number 150

_cell_length_a      5.85000
_cell_length_b      5.85000
_cell_length_c      3.45000
_cell_angle_alpha  90.00000
_cell_angle_beta   90.00000
_cell_angle_gamma 120.00000

loop_
_space_group_symop_id
_space_group_symop_operation_xyz
1 x,y,z
2 -y,x-y,z
3 -x+y,-x,z
4 x-y,-y,-z
5 y,x,-z
6 -x,-x+y,-z

loop_
_atom_site_label
_atom_site_type_symbol
_atom_site_symmetry_multiplicity
_atom_site_Wyckoff_label
_atom_site_fract_x
_atom_site_fract_y
_atom_site_fract_z
_atom_site_occupancy
P1   P    1 b 0.00000 0.00000 0.50000 1.00000
P2   P    2 d 0.33333 0.66667 0.87500 1.00000
Fe1  Fe   3 e 0.26000 0.00000 0.00000 1.00000
Fe2  Fe   3 f 0.60000 0.00000 0.50000 1.00000
```

Original $Fe_2P$ (C22): A2B_hP9_150_ef_bd - POSCAR

```
A2B_hP9_150_ef_bd & a,c/a,z2,x3,x4 --params=5.85,0.589743589744,0.875,
    ↪ 0.26,0.6 & P321        D_3^2 #150 (bdef) & hP9 & C22 & Fe2P &
    ↪ incorrect but historical structure & S. B. Hendricks and P. R.
    ↪ Kosting, Zeitschrift f\"{u}r Kristallographie - Crystalline
    ↪ Materials 74, 511-533 (1930)
   1.0000000000000000
   2.92500000000000  -5.06624861213897   0.00000000000000
   2.92500000000000   5.06624861213897   0.00000000000000
   0.00000000000000   0.00000000000000   3.45000000000000
    Fe     P
     6     3
Direct
   0.00000000000000   0.26000000000000   0.00000000000000   Fe    (3e)
   0.26000000000000   0.00000000000000   0.00000000000000   Fe    (3e)
   0.74000000000000   0.74000000000000   0.00000000000000   Fe    (3e)
   0.00000000000000   0.60000000000000   0.50000000000000   Fe    (3f)
   0.40000000000000   0.40000000000000   0.50000000000000   Fe    (3f)
   0.60000000000000   0.00000000000000   0.50000000000000   Fe    (3f)
   0.00000000000000   0.00000000000000   0.50000000000000    P    (1b)
   0.33333333333333   0.66666666666667  -0.12500000000000    P    (2d)
   0.66666666666667   0.33333333333333   0.12500000000000    P    (2d)
```

$CrCl_3$ ($D0_4$): A3B_hP24_151_3c_2a - CIF

```
# CIF file

data_findsym-output
_audit_creation_method FINDSYM

_chemical_name_mineral 'Chromium trichloride'
_chemical_formula_sum 'Cr Cl3'

loop_
_publ_author_name
 'Nora Wooster'
_journal_name_full
;
Zeitschrift f\"{u}r Kristallographie - Crystalline Materials
;
_journal_volume 74
_journal_year 1930
_journal_page_first 363
_journal_page_last 374
_publ_Section_title
;
 The Structure of Chromium Trichloride CrCl$_3$
;

# Found in  AMS Database

_aflow_proto 'A3B_hP24_151_3c_2a'
_aflow_params 'a,c/a,x1,x2,x3,y3,z3,x4,y4,z4,x5,y5,z5'
_aflow_params_values '6.017,2.87518697025,0.8889,0.5556,0.8889,0.1111,
    ↪ 0.0731,0.5556,0.4444,0.0731,0.2222,0.77778,0.0731'
_aflow_Strukturbericht 'D0_4'
_aflow_Pearson 'hP24'

_symmetry_space_group_name_Hall "P 31 2c (0 0 1)"
_symmetry_space_group_name_H-M "P 31 1 2"
_symmetry_Int_Tables_number 151

_cell_length_a      6.01700
_cell_length_b      6.01700
_cell_length_c     17.30000
_cell_angle_alpha  90.00000
_cell_angle_beta   90.00000
_cell_angle_gamma 120.00000

loop_
_space_group_symop_id
_space_group_symop_operation_xyz
1 x,y,z
2 -y,x-y,z+1/3
3 -x+y,-x,z+2/3
4 x,x-y,-z
5 -x+y,y,-z+1/3
6 -y,-x,-z+2/3

loop_
_atom_site_label
_atom_site_type_symbol
_atom_site_symmetry_multiplicity
_atom_site_Wyckoff_label
_atom_site_fract_x
_atom_site_fract_y
_atom_site_fract_z
_atom_site_occupancy
Cr1  Cr   3 a 0.88890 0.11110 0.33333 1.00000
Cr2  Cr   3 a 0.55560 0.44440 0.33333 1.00000
Cl1  Cl   6 c 0.88890 0.11110 0.07310 1.00000
Cl2  Cl   6 c 0.55560 0.44440 0.07310 1.00000
Cl3  Cl   6 c 0.22220 0.77778 0.07310 1.00000
```

$CrCl_3$ ($D0_4$): A3B_hP24_151_3c_2a - POSCAR

```
A3B_hP24_151_3c_2a & a,c/a,x1,x2,x3,y3,z3,x4,y4,z4,x5,y5,z5 --params=
    ↪ 6.017,2.87518697025,0.8889,0.5556,0.8889,0.1111,0.0731,0.5556,
    ↪ 0.4444,0.0731,0.2222,0.77778,0.0731 & P3_112      D_3^3 #151 (a^
    ↪ 2c^3) & hP24 & D0_4 & CrCl3 & & N. Wooster, Zeitschrift f\"{u}r
    ↪ Kristallographie - Crystalline Materials 74, 363-374 (1930)
   1.0000000000000000
   3.00850000000000  -5.21087485457097   0.00000000000000
   3.00850000000000   5.21087485457097   0.00000000000000
   0.00000000000000   0.00000000000000  17.30000000000000
    Cl    Cr
    18     6
Direct
   0.22220000000000   0.11110000000000   0.26023333333333   Cl    (6c)
   0.22220000000000   0.11110000000000   0.73976666666667   Cl    (6c)
   0.88890000000000   0.11110000000000   0.07310000000000   Cl    (6c)
   0.88890000000000   0.11110000000000   0.59356666666667   Cl    (6c)
   0.88890000000000   0.77780000000000  -0.07310000000000   Cl    (6c)
   0.88890000000000   0.77780000000000   0.40643333333333   Cl    (6c)
   0.55560000000000   0.11120000000000  -0.07310000000000   Cl    (6c)
   0.55560000000000   0.11120000000000   0.40643333333333   Cl    (6c)
   0.55560000000000   0.44440000000000   0.07310000000000   Cl    (6c)
   0.55560000000000   0.44440000000000   0.59356666666667   Cl    (6c)
   0.88880000000000   0.44440000000000   0.26023333333333   Cl    (6c)
   0.88880000000000   0.44440000000000   0.73976666666667   Cl    (6c)
   0.22220000000000   0.44442000000000  -0.07310000000000   Cl    (6c)
   0.22220000000000   0.77778000000000   0.07310000000000   Cl    (6c)
   0.22222000000000   0.44442000000000   0.40643333333333   Cl    (6c)
   0.22222000000000   0.77780000000000   0.59356666666667   Cl    (6c)
   0.55558000000000   0.77778000000000   0.26023333333333   Cl    (6c)
   0.55558000000000   0.77780000000000   0.73976666666667   Cl    (6c)
   0.22220000000000   0.11110000000000   0.00000000000000   Cr    (3a)
   0.88890000000000   0.11110000000000   0.33333333333333   Cr    (3a)
   0.88890000000000   0.77780000000000   0.66666666666667   Cr    (3a)
   0.55560000000000   0.11120000000000   0.66666666666667   Cr    (3a)
   0.55560000000000   0.44440000000000   0.33333333333333   Cr    (3a)
   0.88880000000000   0.44440000000000   0.00000000000000   Cr    (3a)
```

$\alpha$-Quartz (low Quartz): A2B_hP9_152_c_a - CIF

```
# CIF file

data_findsym-output
_audit_creation_method FINDSYM

_chemical_name_mineral 'quartz (alpha)'
_chemical_formula_sum 'Si O2'

loop_
_publ_author_name
 'R. M. Hazen'
 'L. W. Finger'
 'R. J. Hemley'
 'H. K. Mao'
_journal_name_full
;
Solid State Communications
;
_journal_volume 72
_journal_year 1989
_journal_page_first 507
```

```
_journal_page_last 511
_publ_Section_title
;
 High-pressure crystal chemistry and amorphization of $\alpha$-quartz
;

# Found in  AMS Database

_aflow_proto 'A2B_hP9_152_c_a'
_aflow_params 'a,c/a,x1,x2,y2,z2'
_aflow_params_values '4.914,1.10012210012,0.4699,0.413,0.2668,0.214'
_aflow_Strukturbericht 'None'
_aflow_Pearson 'hP9'

_symmetry_space_group_name_Hall "P 31 2"
_symmetry_space_group_name_H-M "P 31 2 1"
_symmetry_Int_Tables_number 152

_cell_length_a     4.91400
_cell_length_b     4.91400
_cell_length_c     5.40600
_cell_angle_alpha  90.00000
_cell_angle_beta   90.00000
_cell_angle_gamma  120.00000

loop_
_space_group_symop_id
_space_group_symop_operation_xyz
1 x,y,z
2 -y,x-y,z+1/3
3 -x+y,-x,z+2/3
4 x-y,-y,-z+2/3
5 y,x,-z
6 -x,-x+y,-z+1/3

loop_
_atom_site_label
_atom_site_type_symbol
_atom_site_symmetry_multiplicity
_atom_site_Wyckoff_label
_atom_site_fract_x
_atom_site_fract_y
_atom_site_fract_z
_atom_site_occupancy
Si1 Si   3 a 0.46990 0.00000 0.33333 1.00000
O1  O    6 c 0.41300 0.26680 0.21400 1.00000
```

$\alpha$-Quartz (low Quartz): A2B_hP9_152_c_a - POSCAR

```
A2B_hP9_152_c_a & a,c/a,x1,x2,y2,z2 --params=4.914,1.10012210012,0.4699,
    ↪ 0.413,0.2668,0.214 & P3_121     D_3^4 #152 (ac) & hP9 & & SiO2 &
    ↪  alpha quartz & R. M. Hazen, L. W. Finger, R. J. Hemley and H.
    ↪ K. Mao, Solid State Comm. 72, 507-511 (1989)
   1.0000000000000000
   2.45700000000000  -4.25564883419700   0.00000000000000
   2.45700000000000   4.25564883419700   0.00000000000000
   0.00000000000000   0.00000000000000   5.40600000000000
    O    Si
    6     3
Direct
   0.14620000000000   0.73320000000000   0.45266666666700   O   (6c)
   0.26680000000000   0.41300000000000   0.78600000000000   O   (6c)
   0.41300000000000   0.26680000000000   0.21400000000000   O   (6c)
   0.58700000000000   0.85380000000000   0.11933333333300   O   (6c)
   0.73320000000000   0.14620000000000   0.54733333333300   O   (6c)
   0.85380000000000   0.58700000000000   0.88066666666700   O   (6c)
   0.00000000000000   0.46990000000000   0.66666666666700   Si  (3a)
   0.46990000000000   0.00000000000000   0.33333333333300   Si  (3a)
   0.53010000000000   0.53010000000000   0.00000000000000   Si  (3a)
```

$\gamma$-Se (A8): A_hP3_152_a - CIF

```
# CIF file

data_findsym-output
_audit_creation_method FINDSYM

_chemical_name_mineral 'alpha Selenium'
_chemical_formula_sum 'Se'

loop_
_publ_author_name
 'Paul Cherin'
 'Phyllis Unger'
_journal_name_full
;
Inorganic Chemistry
;
_journal_volume 6
_journal_year 1967
_journal_page_first 1589
_journal_page_last 1591
_publ_Section_title
;
 The crystal structure of trigonal selenium
;

# Found in  Donohue, pp. 370-372

_aflow_proto 'A_hP3_152_a'
_aflow_params 'a,c/a,x1'
_aflow_params_values '4.3662,1.13453346159,0.2254'
_aflow_Strukturbericht 'A8'
_aflow_Pearson 'hP3'

_symmetry_space_group_name_Hall "P 31 2"
_symmetry_space_group_name_H-M "P 31 2 1"
_symmetry_Int_Tables_number 152

_cell_length_a     4.36620
_cell_length_b     4.36620
_cell_length_c     4.95360
_cell_angle_alpha  90.00000
_cell_angle_beta   90.00000
_cell_angle_gamma  120.00000

loop_
_space_group_symop_id
_space_group_symop_operation_xyz
1 x,y,z
2 -y,x-y,z+1/3
3 -x+y,-x,z+2/3
4 x-y,-y,-z+2/3
5 y,x,-z
6 -x,-x+y,-z+1/3

loop_
_atom_site_label
_atom_site_type_symbol
_atom_site_symmetry_multiplicity
_atom_site_Wyckoff_label
_atom_site_fract_x
_atom_site_fract_y
_atom_site_fract_z
_atom_site_occupancy
Se1 Se   3 a 0.22540 0.00000 0.33333 1.00000
```

$\gamma$-Se (A8): A_hP3_152_a - POSCAR

```
A_hP3_152_a & a,c/a,x1 --params=4.3662,1.13453346159,0.2254 & P3_121
    ↪ D_3^4 #152 (a) & hP3 & A8 & Se & gamma & P. Cherin and P. Unger
    ↪ , Inorg. Chem. 6, 1589-1591 (1967)
   1.0000000000000000
   2.18310000000000  -3.78124011800400   0.00000000000000
   2.18310000000000   3.78124011800400   0.00000000000000
   0.00000000000000   0.00000000000000   4.95360000000000
   Se
    3
Direct
   0.00000000000000   0.22540000000000   0.66666666666667   Se   (3a)
   0.22540000000000   0.00000000000000   0.33333333333333   Se   (3a)
   0.77460000000000   0.77460000000000   0.00000000000000   Se   (3a)
```

Cinnabar (B9): AB_hP6_154_a_b - CIF

```
# CIF file

data_findsym-output
_audit_creation_method FINDSYM

_chemical_name_mineral 'Cinnabar'
_chemical_formula_sum 'Hg S'

loop_
_publ_author_name
 'P. Auvray'
 'F. Genet'
_journal_name_full
;
Bulletin de la Societe Francaise de Mineralogie et de Cristallographie
;
_journal_volume 96
_journal_year 1973
_journal_page_first 218
_journal_page_last 219
_publ_Section_title
;
 Affinement de la structure cristalline du cinabre $\alpha$-HgS
;

# Found in  AMS Database

_aflow_proto 'AB_hP6_154_a_b'
_aflow_params 'a,c/a,x1,x2'
_aflow_params_values '4.145,2.29095295537,0.7198,0.4889'
_aflow_Strukturbericht 'B9'
_aflow_Pearson 'hP6'

_symmetry_space_group_name_Hall "P 32 2"
_symmetry_space_group_name_H-M "P 32 2 1"
_symmetry_Int_Tables_number 154

_cell_length_a     4.14500
_cell_length_b     4.14500
_cell_length_c     9.49600
_cell_angle_alpha  90.00000
_cell_angle_beta   90.00000
_cell_angle_gamma  120.00000

loop_
_space_group_symop_id
_space_group_symop_operation_xyz
1 x,y,z
2 -y,x-y,z+2/3
3 -x+y,-x,z+1/3
4 x-y,-y,-z+1/3
5 y,x,-z
6 -x,-x+y,-z+2/3

loop_
```

```
_atom_site_label
_atom_site_type_symbol
_atom_site_symmetry_multiplicity
_atom_site_Wyckoff_label
_atom_site_fract_x
_atom_site_fract_y
_atom_site_fract_z
_atom_site_occupancy
Hg1 Hg    3 a 0.71980  0.00000  0.66667  1.00000
S1  S     3 b 0.48890  0.00000  0.16667  1.00000
```

Cinnabar (B9): AB_hP6_154_a_b - POSCAR

```
AB_hP6_154_a_b & a,c/a,x1,x2 --params=4.145,2.29095295537,0.7198,0.4889
    ↪ & P3_221      D_3^6 #154 (ab) & hP6 & B9 & HgS & Cinnabar & P.
    ↪ Auvray and F. Genet, Bull. Soc. fr. MineÌĄral. Crystallogr. 96,
    ↪ 218-219 (1973)
   1.0000000000000000
   2.07250000000000  -3.58967529868600   0.00000000000000
   2.07250000000000   3.58967529868600   0.00000000000000
   0.00000000000000   0.00000000000000   9.49600000000000
   Hg     S
    3     3
Direct
   0.00000000000000   0.71980000000000   0.33333333333333   Hg    (3a)
   0.28020000000000   0.28020000000000   0.00000000000000   Hg    (3a)
   0.71980000000000   0.00000000000000   0.66666666666667   Hg    (3a)
   0.00000000000000   0.48890000000000   0.83333333333333   S     (3b)
   0.48890000000000   0.00000000000000   0.16666666666667   S     (3b)
   0.51110000000000   0.51110000000000   0.50000000000000   S     (3b)
```

$AlF_3$ ($D0_{14}$) : AB3_hR8_155_c_de - CIF

```
# CIF file

data_findsym-output
_audit_creation_method FINDSYM

_chemical_name_mineral ''
_chemical_formula_sum 'Al F3'

loop_
_publ_author_name
 'J. A. A. Ketelaar'
_journal_name_full
;
Zeitschrift f\"{u}r Kristallographie - Crystalline Materials
;
_journal_volume 85
_journal_year 1933
_journal_page_first 119
_journal_page_last 131
_publ_Section_title
;
 Die Kristallstruktur der Aluminiumhalogenide: I. Die Kristallstruktur
    ↪ von AlF$_3$
;

# Found in AMS Database

_aflow_proto 'AB3_hR8_155_c_de'
_aflow_params 'a,c/a,x1,y2,y3'
_aflow_params_values '4.91608,2.53341483458,0.237,0.43,0.07'
_aflow_Strukturbericht 'D0_14'
_aflow_Pearson 'hR8'

_symmetry_space_group_name_Hall "R 3 2"
_symmetry_space_group_name_H-M "R 32:H"
_symmetry_Int_Tables_number 155

_cell_length_a       4.91608
_cell_length_b       4.91608
_cell_length_c       12.45447
_cell_angle_alpha  90.00000
_cell_angle_beta   90.00000
_cell_angle_gamma  120.00000

loop_
_space_group_symop_id
_space_group_symop_operation_xyz
1 x,y,z
2 -y,x-y,z
3 -x+y,-x,z
4 y,x,-z
5 -x,-x+y,-z
6 x-y,-y,-z
7 x+1/3,y+2/3,z+2/3
8 -y+1/3,x-y+2/3,z+2/3
9 -x+y+1/3,-x+2/3,z+2/3
10 y+1/3,x+2/3,-z+2/3
11 -x+1/3,-x+y+2/3,-z+2/3
12 x-y+1/3,-y+2/3,-z+2/3
13 x+2/3,y+1/3,z+1/3
14 -y+2/3,x-y+1/3,z+1/3
15 -x+y+2/3,-x+1/3,z+1/3
16 y+2/3,x+1/3,-z+1/3
17 -x+2/3,-x+y+1/3,-z+1/3
18 x-y+2/3,-y+1/3,-z+1/3

loop_
_atom_site_label
_atom_site_type_symbol
_atom_site_symmetry_multiplicity
_atom_site_Wyckoff_label
_atom_site_fract_x
_atom_site_fract_y
_atom_site_fract_z
_atom_site_occupancy
Al1 Al    6 c 0.00000  0.00000  0.23700  1.00000
F1  F     9 d 0.43000  0.00000  0.00000  1.00000
F2  F     9 e 0.57000  0.00000  0.50000  1.00000
```

$AlF_3$ ($D0_{14}$) : AB3_hR8_155_c_de - POSCAR

```
AB3_hR8_155_c_de & a,c/a,x1,y2,y3 --params=4.91608,2.53341483458,0.237,
    ↪ 0.43,0.07 & R32        D_3^7 #155 (cde) & hR8 & D0_{14} & AlF3 &
    ↪ & J. Ketelaar, Zeitschrift f\"{u}r Kristallographie -
    ↪ Crystalline Materials 85, 119-131 (1933)
   1.0000000000000000
   2.45804199855867  -1.41915120954725   4.15149134381397
   0.00000000000000   2.83830241909450   4.15149134381397
  -2.45804199855867  -1.41915120954725   4.15149134381397
   Al     F
    2     6
Direct
   0.23700000000000   0.23700000000000   0.23700000000000   Al    (2c)
   0.76300000000000   0.76300000000000   0.76300000000000   Al    (2c)
   0.00000000000000   0.43000000000000   0.57000000000000   F     (3d)
   0.43000000000000   0.57000000000000  -0.00000000000000   F     (3d)
   0.57000000000000   0.00000000000000   0.43000000000000   F     (3d)
   0.07000000000000  -0.07000000000000   0.50000000000000   F     (3e)
  -0.07000000000000   0.50000000000000   0.07000000000000   F     (3e)
   0.50000000000000   0.07000000000000  -0.07000000000000   F     (3e)
```

Hazelwoodite ($Ni_3S_2$, $D5_e$) : A3B2_hR5_155_e_c - CIF

```
# CIF file

data_findsym-output
_audit_creation_method FINDSYM

_chemical_name_mineral 'Hazelwoodite'
_chemical_formula_sum 'Ni3 S2'

loop_
_publ_author_name
 'John B. Parise'
_journal_name_full
;
Acta Crystallographica B
;
_journal_volume 36
_journal_year 1980
_journal_page_first 1179
_journal_page_last 1180
_publ_Section_title
;
 Structure of Hazelwoodite (Ni$_3$S$_2$
;

_aflow_proto 'A3B2_hR5_155_e_c'
_aflow_params 'a,c/a,x1,y2'
_aflow_params_values '5.73296,1.24097324942,0.2521,0.2449'
_aflow_Strukturbericht 'D5_e'
_aflow_Pearson 'hR5'

_symmetry_space_group_name_Hall "R 3 2"
_symmetry_space_group_name_H-M "R 32:H"
_symmetry_Int_Tables_number 155

_cell_length_a       5.73296
_cell_length_b       5.73296
_cell_length_c       7.11445
_cell_angle_alpha  90.00000
_cell_angle_beta   90.00000
_cell_angle_gamma  120.00000

loop_
_space_group_symop_id
_space_group_symop_operation_xyz
1 x,y,z
2 -y,x-y,z
3 -x+y,-x,z
4 y,x,-z
5 -x,-x+y,-z
6 x-y,-y,-z
7 x+1/3,y+2/3,z+2/3
8 -y+1/3,x-y+2/3,z+2/3
9 -x+y+1/3,-x+2/3,z+2/3
10 y+1/3,x+2/3,-z+2/3
11 -x+1/3,-x+y+2/3,-z+2/3
12 x-y+1/3,-y+2/3,-z+2/3
13 x+2/3,y+1/3,z+1/3
14 -y+2/3,x-y+1/3,z+1/3
15 -x+y+2/3,-x+1/3,z+1/3
16 y+2/3,x+1/3,-z+1/3
17 -x+2/3,-x+y+1/3,-z+1/3
18 x-y+2/3,-y+1/3,-z+1/3

loop_
_atom_site_label
_atom_site_type_symbol
_atom_site_symmetry_multiplicity
_atom_site_Wyckoff_label
_atom_site_fract_x
_atom_site_fract_y
_atom_site_fract_z
_atom_site_occupancy
S1  S     6 c 0.00000  0.00000  0.25210  1.00000
Ni1 Ni    9 e 0.74490  0.00000  0.50000  1.00000
```

Hazelwoodite ($Ni_3S_2$, $D5_e$) : A3B2_hR5_155_e_c - POSCAR

```
A3B2_hR5_155_e_c & a,c/a,x1,y2 --params=5.73296,1.24097324942,0.2521,
    ↪ 0.2449 & R32         D_3^7 #155 (ce) & hR5 & D5_e & Ni3S2 &
    ↪ Hazelwoodite & J. B. Parise, Acta Cryst. B 36, 1179-1180 (1980)
   1.0000000000000000
   2.86648097666438   -1.65496356350412    2.37148427232714
   0.00000000000000    3.30992712700824    2.37148427232714
  -2.86648097666438   -1.65496356350412    2.37148427232714
   Ni     S
    3     2
Direct
   0.24490000000000    0.75510000000000    0.50000000000000    Ni    (3e)
   0.50000000000000    0.24490000000000    0.75510000000000    Ni    (3e)
   0.75510000000000    0.50000000000000    0.24490000000000    Ni    (3e)
   0.25210000000000    0.25210000000000    0.25210000000000     S    (2c)
   0.74790000000000    0.74790000000000    0.74790000000000     S    (2c)
```

Millerite (NiS, B13): AB_hR6_160_b_b - CIF

```
# CIF file

data_findsym-output
_audit_creation_method FINDSYM

_chemical_name_mineral 'Millerite'
_chemical_formula_sum 'Ni S'

loop_
_publ_author_name
 'V. Rajamani'
 'C. T. Prewitt'
_journal_name_full
;
Canadian Mineralogist
;
_journal_volume 12
_journal_year 1974
_journal_page_first 253
_journal_page_last 257
_publ_Section_title
;
 The Crystal Structure of Millerite
;

# Found in  AMS Database

_aflow_proto 'AB_hR6_160_b_b'
_aflow_params 'a,c/a,x1,z1,x2,z2'
_aflow_params_values '9.619,0.327466472606,0.00019,0.26362,0.7288,
    ↪ 0.39161'
_aflow_Strukturbericht 'B13'
_aflow_Pearson 'hR6'

_symmetry_space_group_name_Hall "R 3 -2"
_symmetry_space_group_name_H-M "R 3 m:H"
_symmetry_Int_Tables_number 160

_cell_length_a      9.61900
_cell_length_b      9.61900
_cell_length_c      3.14990
_cell_angle_alpha  90.00000
_cell_angle_beta   90.00000
_cell_angle_gamma  120.00000

loop_
_space_group_symop_id
_space_group_symop_operation_xyz
1 x,y,z
2 -y,x-y,z
3 -x+y,-x,z
4 -y,-x,z
5 x,x-y,z
6 -x+y,y,z
7 x+1/3,y+2/3,z+2/3
8 -y+1/3,x-y+2/3,z+2/3
9 -x+y+1/3,-x+2/3,z+2/3
10 -y+1/3,-x+2/3,z+2/3
11 x+1/3,x-y+2/3,z+2/3
12 -x+y+1/3,y+2/3,z+2/3
13 x+2/3,y+1/3,z+1/3
14 -y+2/3,x-y+1/3,z+1/3
15 -x+y+2/3,-x+1/3,z+1/3
16 -y+2/3,-x+1/3,z+1/3
17 x+2/3,x-y+1/3,z+1/3
18 -x+y+2/3,y+1/3,z+1/3

loop_
_atom_site_label
_atom_site_type_symbol
_atom_site_symmetry_multiplicity
_atom_site_Wyckoff_label
_atom_site_fract_x
_atom_site_fract_y
_atom_site_fract_z
_atom_site_occupancy
Ni1 Ni    9 b -0.08781  0.08781  0.08800  1.00000
S1  S     9 b  0.44573  0.55427  0.28307  1.00000
```

Millerite (NiS, B13): AB_hR6_160_b_b - POSCAR

```
AB_hR6_160_b_b & a,c/a,x1,z1,x2,z2 --params=9.619,0.327466472606,0.00019
    ↪ ,0.26362,0.7288,0.39161 & R3m        C_{3v}^5 #160 (b^2) & hR6 &
    ↪ B13 & NiS (beta) & Millerite & V. Rajamni and C. T. Prewitt,
    ↪ Can. Min. 12, 253-257 (1974)
   1.0000000000000000
   4.80950000000000   -2.77676611966800    1.04996666666700
   0.00000000000000    5.55353223933500    1.04996666666700
  -4.80950000000000   -2.77676611966800    1.04996666666700
   Ni     Si
    3     3
Direct
   0.00019000000000    0.00019000000000    0.26362000000000    Ni    (3b)
   0.00019000000000    0.26362000000000    0.00019000000000    Ni    (3b)
   0.26362000000000    0.00019000000000    0.00019000000000    Ni    (3b)
   0.39160000000000    0.72880000000000    0.72880000000000     S    (3b)
   0.72880000000000    0.39160000000000    0.72880000000000     S    (3b)
   0.72880000000000    0.72880000000000    0.39160000000000     S    (3b)
```

Moissanite 9R: AB_hR6_160_3a_3a - CIF

```
# CIF file

data_findsym-output
_audit_creation_method FINDSYM

_chemical_name_mineral 'Moissanite 9R'
_chemical_formula_sum 'C Si'

loop_
_publ_author_name
 'Michael J. Mehl'
_journal_name_full
;
None
;
_journal_volume 0
_journal_year 2001
_journal_page_first 0
_journal_page_last 0
_publ_Section_title
;
 Hypothetical SiO2 Structure with 9R stacking
;

_aflow_proto 'AB_hR6_160_3a_3a'
_aflow_params 'a,c/a,x1,x2,x3,x4,x5,x6'
_aflow_params_values '3.01791,7.34847294982,0.0,0.22222,0.77778,0.08333,
    ↪ 0.30556,0.86111'
_aflow_Strukturbericht 'None'
_aflow_Pearson 'hR6'

_symmetry_space_group_name_Hall "R 3 -2"
_symmetry_space_group_name_H-M "R 3 m:H"
_symmetry_Int_Tables_number 160

_cell_length_a      3.01791
_cell_length_b      3.01791
_cell_length_c      22.17703
_cell_angle_alpha  90.00000
_cell_angle_beta   90.00000
_cell_angle_gamma  120.00000

loop_
_space_group_symop_id
_space_group_symop_operation_xyz
1 x,y,z
2 -y,x-y,z
3 -x+y,-x,z
4 -y,-x,z
5 x,x-y,z
6 -x+y,y,z
7 x+1/3,y+2/3,z+2/3
8 -y+1/3,x-y+2/3,z+2/3
9 -x+y+1/3,-x+2/3,z+2/3
10 -y+1/3,-x+2/3,z+2/3
11 x+1/3,x-y+2/3,z+2/3
12 -x+y+1/3,y+2/3,z+2/3
13 x+2/3,y+1/3,z+1/3
14 -y+2/3,x-y+1/3,z+1/3
15 -x+y+2/3,-x+1/3,z+1/3
16 -y+2/3,-x+1/3,z+1/3
17 x+2/3,x-y+1/3,z+1/3
18 -x+y+2/3,y+1/3,z+1/3

loop_
_atom_site_label
_atom_site_type_symbol
_atom_site_symmetry_multiplicity
_atom_site_Wyckoff_label
_atom_site_fract_x
_atom_site_fract_y
_atom_site_fract_z
_atom_site_occupancy
C1  C    3 a 0.00000  0.00000  0.00000  1.00000
C2  C    3 a 0.00000  0.00000  0.22222  1.00000
C3  C    3 a 0.00000  0.00000  0.77778  1.00000
Si1 Si   3 a 0.00000  0.00000  0.08333  1.00000
Si2 Si   3 a 0.00000  0.00000  0.30556  1.00000
Si3 Si   3 a 0.00000  0.00000  0.86111  1.00000
```

Moissanite 9R: AB_hR6_160_3a_3a - POSCAR

```
AB_hR6_160_3a_3a & a,c/a,x1,x2,x3,x4,x5,x6 --params=3.01791,
    ↪ 7.34847294982,0.0,0.22222,0.77778,0.08333,0.30556,0.86111 & R3m
    ↪          C_{3v}^5 #160 (a^6) & hR6 & & CSi & 9R (Hypothetical
    ↪ ABCBCACAB stacking) &
   1.0000000000000000
   1.50895500000000   -0.87119557544503    7.39234333333042
   0.00000000000000    1.74239115089006    7.39234333333042
```

```
  -1.50895500000000   -0.87119557544503    7.39234333333042
     C     Si
     3      3
Direct
    0.00000000000000    0.00000000000000    0.00000000000000     C      (1a)
    0.22222222222222    0.22222222222222    0.22222222222222     C      (1a)
    0.77777777777778    0.77777777777778    0.77777777777778     C      (1a)
    0.08333333333333    0.08333333333333    0.08333333333333    Si      (1a)
    0.30555555555556    0.30555555555556    0.30555555555556    Si      (1a)
    0.86111111111111    0.86111111111111    0.86111111111111    Si      (1a)
```

Ferroelectric $LiNbO_3$: ABC3_hR10_161_a_a_b - CIF

```
# CIF file

data_findsym-output
_audit_creation_method FINDSYM

_chemical_name_mineral ''
_chemical_formula_sum 'Li Nb O3'

loop_
_publ_author_name
 'H. Boysen'
 'F. Altorfer'
_journal_name_full
;
Acta Crystallographica B
;
_journal_volume 50
_journal_year 1994
_journal_page_first 405
_journal_page_last 414
_publ_Section_title
;
 A neutron powder investigation of the high-temperature structure and
      ↪ phase transition in LiNbO$_3$
;

_aflow_proto 'ABC3_hR10_161_a_a_b'
_aflow_params 'a,c/a,x1,x2,x3,y3,z3'
_aflow_params_values '5.2542,2.64091583876,0.2875,0.0128,0.74643,0.14093
     ↪ ,0.36263'
_aflow_Strukturbericht 'None'
_aflow_Pearson 'hR10'

_symmetry_space_group_name_Hall "R 3 -2c"
_symmetry_space_group_name_H-M "R 3 c:H"
_symmetry_Int_Tables_number 161

_cell_length_a      5.25420
_cell_length_b      5.25420
_cell_length_c      13.87590
_cell_angle_alpha   90.00000
_cell_angle_beta    90.00000
_cell_angle_gamma   120.00000

loop_
_space_group_symop_id
_space_group_symop_operation_xyz
1 x,y,z
2 -y,x-y,z
3 -x+y,-x,z
4 -y,-x,z+1/2
5 x,x-y,z+1/2
6 -x+y,y,z+1/2
7 x+1/3,y+2/3,z+2/3
8 -y+1/3,x-y+2/3,z+2/3
9 -x+y+1/3,-x+2/3,z+2/3
10 -y+1/3,-x+2/3,z+1/6
11 x+1/3,x-y+2/3,z+1/6
12 -x+y+1/3,y+2/3,z+1/6
13 x+2/3,y+1/3,z+1/3
14 -y+2/3,x-y+1/3,z+1/3
15 -x+y+2/3,-x+1/3,z+1/3
16 -y+2/3,-x+1/3,z+5/6
17 x+2/3,x-y+1/3,z+5/6
18 -x+y+2/3,y+1/3,z+5/6

loop_
_atom_site_label
_atom_site_type_symbol
_atom_site_symmetry_multiplicity
_atom_site_Wyckoff_label
_atom_site_fract_x
_atom_site_fract_y
_atom_site_fract_z
_atom_site_occupancy
Li1 Li    6 a 0.00000 0.00000 0.28750 1.00000
Nb1 Nb    6 a 0.00000 0.00000 0.01280 1.00000
O1  O    18 b 0.66310 0.72070 0.08333 1.00000
```

Ferroelectric $LiNbO_3$: ABC3_hR10_161_a_a_b - POSCAR

```
ABC3_hR10_161_a_a_b & a,c/a,x1,x2,x3,y3,z3 --params=5.2542,2.64091583876
     ↪ ,0.2875,0.0128,0.74643,0.14093,0.36263 & R3c       C_{3v}^6 #
     ↪ 161 (a^2b) & hR10 & & LiNbO3 & Ferroelectric & H. Boysen and F.
     ↪ Altorfer, Acta Cryst. B 50, 405-414 (1994)
   1.0000000000000000
    2.62710000000000   -1.51675689218800    4.62530000000000
    0.00000000000000    3.03351378437600    4.62530000000000
   -2.62710000000000   -1.51675689218800    4.62530000000000
    Li    Nb     O
     2     2     6
Direct
    0.28750000000000    0.28750000000000    0.28750000000000    Li     (2a)
    0.78750000000000    0.78750000000000    0.78750000000000    Li     (2a)
    0.01280000000000    0.01280000000000    0.01280000000000    Nb     (2a)
    0.51280000000000    0.51280000000000    0.51280000000000    Nb     (2a)
    0.14093333333333    0.36263333333333    0.74643333333333     O     (6b)
    0.24643333333333    0.86263333333333    0.64093333333333     O     (6b)
    0.36263333333333    0.74643333333333    0.14093333333333     O     (6b)
    0.64093333333333    0.24643333333333    0.86263333333333     O     (6b)
    0.74643333333333    0.14093333333333    0.36263333333333     O     (6b)
    0.86263333333333    0.64093333333333    0.24643333333333     O     (6b)
```

$\beta$-$V_2N$: AB2_hP9_162_ad_k - CIF

```
# CIF file

data_findsym-output
_audit_creation_method FINDSYM

_chemical_name_mineral 'beta Vanadium nitride'
_chemical_formula_sum 'V2 N'

loop_
_publ_author_name
 'A. N{\o}rlund Christensen'
 'B. Lebech'
_journal_name_full
;
Acta Crystallographica B
;
_journal_volume 35
_journal_year 1979
_journal_page_first 2677
_journal_page_last 2678
_publ_Section_title
;
 The structure of $\beta$-Vanadium Nitride
;

# Found in  Pearson's Handbook IV, pp. 4503

_aflow_proto 'AB2_hP9_162_ad_k'
_aflow_params 'a,c/a,x3,z3'
_aflow_params_values '4.917,0.929021761237,0.325,0.272'
_aflow_Strukturbericht 'None'
_aflow_Pearson 'hP9'

_symmetry_space_group_name_Hall "-P 3 2"
_symmetry_space_group_name_H-M "P -3 1 m"
_symmetry_Int_Tables_number 162

_cell_length_a      4.91700
_cell_length_b      4.91700
_cell_length_c      4.56800
_cell_angle_alpha   90.00000
_cell_angle_beta    90.00000
_cell_angle_gamma   120.00000

loop_
_space_group_symop_id
_space_group_symop_operation_xyz
1 x,y,z
2 -y,x-y,z
3 -x+y,-x,z
4 x,x-y,-z
5 -x+y,y,-z
6 -y,-x,-z
7 -x,-y,-z
8 y,-x+y,-z
9 x-y,x,-z
10 -x,-x+y,z
11 x-y,-y,z
12 y,x,z

loop_
_atom_site_label
_atom_site_type_symbol
_atom_site_symmetry_multiplicity
_atom_site_Wyckoff_label
_atom_site_fract_x
_atom_site_fract_y
_atom_site_fract_z
_atom_site_occupancy
N1 N    1 a 0.00000 0.00000 0.00000 1.00000
N2 N    2 d 0.33333 0.66667 0.50000 1.00000
V1 V    6 k 0.32500 0.00000 0.27200 1.00000
```

$\beta$-$V_2N$: AB2_hP9_162_ad_k - POSCAR

```
AB2_hP9_162_ad_k & a,c/a,x3,z3 --params=4.917,0.929021761237,0.325,0.272
     ↪ & P(-3)1m    D_{3d}^1 #162 (adk) & hP9 & & V2N & beta & A. N.
     ↪ Christensen and B. Lebech, Acta Cryst. B 35, 2677-2678 (1979)
   1.0000000000000000
    2.45850000000000   -4.25824691040809    0.00000000000000
    2.45850000000000    4.25824691040809    0.00000000000000
    0.00000000000000    0.00000000000000    4.56800000000000
     N     V
     3     6
Direct
    0.00000000000000    0.00000000000000    0.00000000000000     N      (1a)
    0.33333333333333    0.66666666666667    0.50000000000000     N      (2d)
    0.66666666666667    0.33333333333333    0.50000000000000     N      (2d)
    0.00000000000000    0.32500000000000    0.27200000000000     V      (6k)
    0.00000000000000    0.67500000000000    0.72800000000000     V      (6k)
    0.32500000000000    0.00000000000000    0.27200000000000     V      (6k)
    0.32500000000000    0.32500000000000    0.72800000000000     V      (6k)
```

```
    0.67500000000000    0.00000000000000    0.72800000000000    V    (6k)
    0.67500000000000    0.67500000000000    0.27200000000000    V    (6k)
```

KAg(CN)$_2$ (F5$_{10}$): AB2CD2_hP36_163_h_i_bf_i - CIF

```
# CIF file

data_findsym-output
_audit_creation_method FINDSYM

_chemical_name_mineral 'Potassium Silver Cyanide'
_chemical_formula_sum 'Ag C2 K N2'

loop_
_publ_author_name
 'J. L. Hoard'
_journal_name_full
;
Zeitschrift f\"{u}r Kristallographie - Crystalline Materials
;
_journal_volume 84
_journal_year 1933
_journal_page_first 231
_journal_page_last 255
_publ_Section_title
;
 The Crystal Structure of Potassium Silver Cyanide
;

# Found in http://materials.springer.com/isp/crystallographic/docs/
    ↪ sd_1253381

_aflow_proto 'AB2CD2_hP36_163_h_i_bf_i'
_aflow_params 'a,c/a,z2,x3,x4,y4,z4,x5,y5,z5'
_aflow_params_values '7.384,2.37716684724,0.01,0.833,0.33333,0.03833,
    ↪ 0.141,0.03167,0.365,0.083'
_aflow_Strukturbericht 'F5_10'
_aflow_Pearson 'hP36'

_symmetry_space_group_name_Hall "-P 3 2c"
_symmetry_space_group_name_H-M "P -3 1 c"
_symmetry_Int_Tables_number 163

_cell_length_a    7.38400
_cell_length_b    7.38400
_cell_length_c    17.55300
_cell_angle_alpha 90.00000
_cell_angle_beta  90.00000
_cell_angle_gamma 120.00000

loop_
_space_group_symop_id
_space_group_symop_operation_xyz
1 x,y,z
2 -y,x-y,z
3 -x+y,-x,z
4 x,x-y,-z+1/2
5 -x+y,y,-z+1/2
6 -y,-x,-z+1/2
7 -x,-y,-z
8 y,-x+y,-z
9 x-y,x,-z
10 -x,-x+y,z+1/2
11 x-y,-y,z+1/2
12 y,x,z+1/2

loop_
_atom_site_label
_atom_site_type_symbol
_atom_site_symmetry_multiplicity
_atom_site_Wyckoff_label
_atom_site_fract_x
_atom_site_fract_y
_atom_site_fract_z
_atom_site_occupancy
K1  K    2 b 0.00000 0.00000 0.00000 1.00000
K2  K    4 f 0.33333 0.66667 0.01000 1.00000
Ag1 Ag   6 h 0.83300 0.16700 0.25000 1.00000
C1  C   12 i 0.33333 0.03833 0.14100 1.00000
N1  N   12 i 0.03167 0.36500 0.08300 1.00000
```

KAg(CN)$_2$ (F5$_{10}$): AB2CD2_hP36_163_h_i_bf_i - POSCAR

```
AB2CD2_hP36_163_h_i_bf_i & a,c/a,z2,x3,x4,y4,z4,x5,y5,z5 --params=7.384,
    ↪ 2.37716684724,0.01,0.833,0.33333,0.03833,0.141,0.03167,0.365,
    ↪ 0.083 & P(-3)1c    D_{3d}^2 #163 (bfhi^2) & hP36 & F5_10 & KAg(
    ↪ CN)2 & & J. L. Hoard, Zeitschrift f\"{u}r Kristallographie -
    ↪ Crystalline Materials 84, 231-255 (1933)
   1.0000000000000000
    3.69200000000000   -6.39473158154430    0.00000000000000
    3.69200000000000    6.39473158154430    0.00000000000000
    0.00000000000000    0.00000000000000   17.55300000000000
   Ag    C    K    N
    6   12    6   12
Direct
    0.16700000000000    0.33400000000000    0.75000000000000   Ag    (6h)
    0.16700000000000    0.83300000000000    0.75000000000000   Ag    (6h)
    0.33400000000000    0.16700000000000    0.25000000000000   Ag    (6h)
    0.66600000000000    0.83300000000000    0.75000000000000   Ag    (6h)
    0.83300000000000    0.16700000000000    0.25000000000000   Ag    (6h)
    0.83300000000000    0.66600000000000    0.25000000000000   Ag    (6h)
   -0.03833000000000    0.29500000000000    0.14100000000000   C    (12i)
    0.03833000000000    0.33333000000000    0.64100000000000   C    (12i)
   -0.03833000000000    0.66667000000000    0.35900000000000   C    (12i)
    0.03833000000000    0.70500000000000    0.85900000000000   C    (12i)
    0.29500000000000   -0.03833000000000    0.64100000000000   C    (12i)
    0.29500000000000    0.33333000000000    0.85900000000000   C    (12i)
    0.33333000000000    0.03833000000000    0.14100000000000   C    (12i)
    0.33333000000000    0.29500000000000    0.35900000000000   C    (12i)
    0.66667000000000   -0.03833000000000    0.85900000000000   C    (12i)
    0.66667000000000    0.70500000000000    0.64100000000000   C    (12i)
    0.70500000000000    0.03833000000000    0.35900000000000   C    (12i)
    0.70500000000000    0.66667000000000    0.14100000000000   C    (12i)
    0.00000000000000    0.00000000000000    0.00000000000000   K     (2b)
    0.00000000000000    0.00000000000000    0.50000000000000   K     (2b)
    0.33333333333333    0.66666666666667    0.01000000000000   K     (4f)
    0.33333333333333    0.66666666666667    0.49000000000000   K     (4f)
    0.66666666666667    0.33333333333333   -0.01000000000000   K     (4f)
    0.66666666666667    0.33333333333333    0.51000000000000   K     (4f)
   -0.03167000000000    0.33333000000000    0.58300000000000   N    (12i)
    0.03167000000000    0.36500000000000    0.08300000000000   N    (12i)
   -0.03167000000000    0.63500000000000   -0.08300000000000   N    (12i)
    0.03167000000000    0.66667000000000    0.41700000000000   N    (12i)
    0.33333000000000   -0.03167000000000    0.08300000000000   N    (12i)
    0.33333000000000    0.36500000000000    0.41700000000000   N    (12i)
    0.36500000000000    0.03167000000000    0.58300000000000   N    (12i)
    0.36500000000000    0.33333000000000   -0.08300000000000   N    (12i)
    0.63500000000000   -0.03167000000000    0.41700000000000   N    (12i)
    0.63500000000000    0.66667000000000    0.08300000000000   N    (12i)
    0.66667000000000    0.03167000000000   -0.08300000000000   N    (12i)
    0.66667000000000    0.63500000000000    0.58300000000000   N    (12i)
```

Al$_3$Ni$_2$ (D5$_{13}$): A3B2_hP5_164_ad_d - CIF

```
# CIF file

data_findsym-output
_audit_creation_method FINDSYM

_chemical_name_mineral ''
_chemical_formula_sum 'Al3 Ni2'

loop_
_publ_author_name
 'A. J. Bradley'
 'A. Taylor'
_journal_name_full
;
Philosophical Magazine
;
_journal_volume 23
_journal_year 1937
_journal_page_first 1049
_journal_page_last 1067
_publ_Section_title
;
 The crystal structures of Ni$_2$Al$_3$ and NiAl$_3$
;

# Found in http://materials.springer.com/lb/docs/
    ↪ sm_lbs_978-3-540-44752-8_197

_aflow_proto 'A3B2_hP5_164_ad_d'
_aflow_params 'a,c/a,z2,z3'
_aflow_params_values '4.0282,1.21409066084,0.648,0.149'
_aflow_Strukturbericht 'D5_13'
_aflow_Pearson 'hP5'

_symmetry_space_group_name_Hall "-P 3 2"
_symmetry_space_group_name_H-M "P -3 m 1"
_symmetry_Int_Tables_number 164

_cell_length_a    4.02820
_cell_length_b    4.02820
_cell_length_c    4.89060
_cell_angle_alpha 90.00000
_cell_angle_beta  90.00000
_cell_angle_gamma 120.00000

loop_
_space_group_symop_id
_space_group_symop_operation_xyz
1 x,y,z
2 -y,x-y,z
3 -x+y,-x,z
4 x-y,-y,-z
5 y,x,-z
6 -x,-x+y,-z
7 -x,-y,-z
8 y,-x+y,-z
9 x-y,x,-z
10 -x+y,y,z
11 -y,-x,z
12 x,x-y,z

loop_
_atom_site_label
_atom_site_type_symbol
_atom_site_symmetry_multiplicity
_atom_site_Wyckoff_label
_atom_site_fract_x
_atom_site_fract_y
_atom_site_fract_z
_atom_site_occupancy
Al1 Al   1 a 0.00000 0.00000 0.00000 1.00000
Al2 Al   2 d 0.33333 0.66667 0.64800 1.00000
Ni1 Ni   2 d 0.33333 0.66667 0.14900 1.00000
```

Al$_3$Ni$_2$ (D5$_{13}$): A3B2_hP5_164_ad_d - POSCAR

```
A3B2_hP5_164_ad_d & a,c/a,z2,z3 --params=4.0282,1.21409066084,0.648,
    ↪ 0.149 & P(-3)ml    D_{3d}^3 #164 (ad^2) & hP5 & D5_{13} & Al3Ni2
    ↪ & & A. J. Bradley and A. Taylor, Phil. Mag. 23, 1049-1067 (
    ↪ 1937)
   1.0000000000000000
   2.01410000000000   -3.48852353152400    0.00000000000000
   2.01410000000000    3.48852353152400    0.00000000000000
   0.00000000000000    0.00000000000000    4.89060000000000
   Al    Ni
    3     2
Direct
   0.00000000000000    0.00000000000000    0.00000000000000    Al    (1a)
   0.33333333333333    0.66666666666667    0.64800000000000    Al    (2d)
   0.66666666666667    0.33333333333333    0.35200000000000    Al    (2d)
   0.33333333333333    0.66666666666667    0.14900000000000    Ni    (2d)
   0.66666666666667    0.33333333333333    0.85100000000000    Ni    (2d)
```

ω (C6) Phase: AB2_hP3_164_a_d - CIF

```
# CIF file

data_findsym-output
_audit_creation_method FINDSYM

_chemical_name_mineral 'trigonal omega'
_chemical_formula_sum 'Cd I2'

loop_
_publ_author_name
 'Richard M. Bozorth'
_journal_name_full
;
Journal of the American Chemical Society
;
_journal_volume 44
_journal_year 1922
_journal_page_first 2232
_journal_page_last 2236
_publ_Section_title
;
 The Crystal Structure of Cadmium Iodide
;

# Found in  Strukturbericht Vol. I, pp. 161-3

_aflow_proto 'AB2_hP3_164_a_d'
_aflow_params 'a,c/a,z2'
_aflow_params_values '4.24,1.61320754717,0.252'
_aflow_Strukturbericht 'C6'
_aflow_Pearson 'hP3'

_symmetry_space_group_name_Hall "-P 3 2"
_symmetry_space_group_name_H-M "P -3 m 1"
_symmetry_Int_Tables_number 164

_cell_length_a      4.24000
_cell_length_b      4.24000
_cell_length_c      6.84000
_cell_angle_alpha  90.00000
_cell_angle_beta   90.00000
_cell_angle_gamma 120.00000

loop_
_space_group_symop_id
_space_group_symop_operation_xyz
1 x,y,z
2 -y,x-y,z
3 -x+y,-x,z
4 x-y,-y,-z
5 y,x,-z
6 -x,-x+y,-z
7 -x,-y,-z
8 y,-x+y,-z
9 x-y,x,-z
10 -x+y,y,z
11 -y,-x,z
12 x,x-y,z

loop_
_atom_site_label
_atom_site_type_symbol
_atom_site_symmetry_multiplicity
_atom_site_Wyckoff_label
_atom_site_fract_x
_atom_site_fract_y
_atom_site_fract_z
_atom_site_occupancy
Cd1 Cd    1 a 0.00000 0.00000 0.00000 1.00000
I1  I     2 d 0.33333 0.66667 0.25200 1.00000
```

ω (C6) Phase: AB2_hP3_164_a_d - POSCAR

```
AB2_hP3_164_a_d & a,c/a,z2 --params=4.24,1.61320754717,0.252 & P(-3)ml
    ↪    D_{3d}^3 #164 (ad) & hP3 & C6 & CdI2 & trigonal omega & R. M.
    ↪    Bozorth, J. Am. Chem. Soc. 44, 2232-2236 (1922)
   1.0000000000000000
   2.12000000000000   -3.67194771204602    0.00000000000000
   2.12000000000000    3.67194771204602    0.00000000000000
   0.00000000000000    0.00000000000000    6.84000000000000
   Cd     I
    1     2
Direct
   0.00000000000000    0.00000000000000    0.00000000000000    Cd    (1a)
   0.33333333333333    0.66666666666667    0.25200000000000     I    (2d)
   0.66666666666667    0.33333333333333    0.74800000000000     I    (2d)
```

$H_3Ho$: A3B_hP24_165_adg_f - CIF

```
# CIF file

data_findsym-output
_audit_creation_method FINDSYM

_chemical_name_mineral ''
_chemical_formula_sum 'H3 Ho'

loop_
_publ_author_name
 'M. Mansmann'
 'W. E. Wallace'
_journal_name_full
;
Le Journal de Physique
;
_journal_volume 25
_journal_year 1964
_journal_page_first 454
_journal_page_last 459
_publ_Section_title
;
 The Structure of HoD$_3$
;

# Found in  Pearson's Handbook, Vol III, pp. 3829

_aflow_proto 'A3B_hP24_165_adg_f'
_aflow_params 'a,c/a,z2,x3,x4,y4,z4'
_aflow_params_values '6.308,1.03994927077,0.167,0.666,0.356,0.028,0.096'
_aflow_Strukturbericht 'None'
_aflow_Pearson 'hP24'

_symmetry_space_group_name_Hall "-P 3 2c"
_symmetry_space_group_name_H-M "P -3 c 1"
_symmetry_Int_Tables_number 165

_cell_length_a      6.30800
_cell_length_b      6.30800
_cell_length_c      6.56000
_cell_angle_alpha  90.00000
_cell_angle_beta   90.00000
_cell_angle_gamma 120.00000

loop_
_space_group_symop_id
_space_group_symop_operation_xyz
1 x,y,z
2 -y,x-y,z
3 -x+y,-x,z
4 x-y,-y,-z+1/2
5 y,x,-z+1/2
6 -x,-x+y,-z+1/2
7 -x,-y,-z
8 y,-x+y,-z
9 x-y,x,-z
10 -x+y,y,z+1/2
11 -y,-x,z+1/2
12 x,x-y,z+1/2

loop_
_atom_site_label
_atom_site_type_symbol
_atom_site_symmetry_multiplicity
_atom_site_Wyckoff_label
_atom_site_fract_x
_atom_site_fract_y
_atom_site_fract_z
_atom_site_occupancy
H1  H     2 a 0.00000 0.00000 0.25000 1.00000
H2  H     4 d 0.33333 0.66667 0.16700 1.00000
Ho1 Ho    6 f 0.66600 0.00000 0.25000 1.00000
H3  H    12 g 0.35600 0.02800 0.09600 1.00000
```

$H_3Ho$: A3B_hP24_165_adg_f - POSCAR

```
A3B_hP24_165_adg_f & a,c/a,z2,x3,x4,y4,z4 --params=6.308,1.03994927077,
    ↪ 0.167,0.666,0.356,0.028,0.096 & P(-3)c1    D_{3d}^4 #165 (adfg)
    ↪ & hP24 & & H_3Ho & & M. Mansmann and W. E. Wallace, Le Journal
    ↪ de Physique 25, 454-459 (1964)
   1.0000000000000000
   3.15400000000000   -5.46288824707224    0.00000000000000
   3.15400000000000    5.46288824707224    0.00000000000000
   0.00000000000000    0.00000000000000    6.56000000000000
    H    Ho
   18     6
Direct
  -0.02800000000000    0.32800000000000    0.09600000000000    H    (12g)
   0.02800000000000    0.35600000000000    0.40400000000000    H    (12g)
  -0.02800000000000    0.64400000000000    0.59600000000000    H    (12g)
   0.02800000000000    0.67200000000000   -0.09600000000000    H    (12g)
   0.32800000000000   -0.02800000000000    0.40400000000000    H    (12g)
   0.32800000000000    0.35600000000000   -0.09600000000000    H    (12g)
   0.35600000000000    0.02800000000000    0.09600000000000    H    (12g)
   0.35600000000000    0.32800000000000    0.59600000000000    H    (12g)
   0.64400000000000   -0.02800000000000   -0.09600000000000    H    (12g)
   0.64400000000000    0.67200000000000    0.40400000000000    H    (12g)
   0.67200000000000    0.02800000000000    0.59600000000000    H    (12g)
   0.67200000000000    0.64400000000000    0.09600000000000    H    (12g)
   0.00000000000000    0.00000000000000    0.25000000000000    H     (2a)
   0.00000000000000    0.00000000000000    0.75000000000000    H     (2a)
   0.33333333333333    0.66666666666667    0.16700000000000    H     (4d)
   0.33333333333333    0.66666666666667    0.66700000000000    H     (4d)
```

```
    0.66666666666667    0.33333333333333    0.33300000000000     H      (4d)
    0.66666666666667    0.33333333333333    0.83300000000000     H      (4d)
    0.00000000000000    0.33400000000000    0.75000000000000     Ho     (6f)
    0.00000000000000    0.66600000000000    0.25000000000000     Ho     (6f)
    0.33400000000000    0.00000000000000    0.75000000000000     Ho     (6f)
    0.33400000000000    0.33400000000000    0.25000000000000     Ho     (6f)
    0.66600000000000    0.00000000000000    0.25000000000000     Ho     (6f)
    0.66600000000000    0.66600000000000    0.75000000000000     Ho     (6f)
```

CuPt ($L1_1$): AB_hR2_166_a_b - CIF

```
# CIF file

data_findsym-output
_audit_creation_method FINDSYM

_chemical_name_mineral ''
_chemical_formula_sum 'Cu Pt'

loop_
_publ_author_name
 'C. H. Johansson'
 'J. O. Linde'
_journal_name_full
;
Annalen der Physik
;
_journal_volume 387
_journal_year 1927
_journal_page_first 449
_journal_page_last 478
_publ_Section_title
;
 Gitterstruktur und elektrisches Leitverm\"{o}gen der
    ↪ Mischkristallreihen Au-Cu, Pd-Cu und Pt-Cu
;

# Found in http://materials.springer.com/lb/docs/
    ↪ sm_lbs_978-3-540-46933-9_359

_aflow_proto 'AB_hR2_166_a_b'
_aflow_params 'a,c/a'
_aflow_params_values '3.13,4.78594249201'
_aflow_Strukturbericht 'L1_1'
_aflow_Pearson 'hR2'

_symmetry_space_group_name_Hall "-R 3 2"
_symmetry_space_group_name_H-M "R -3 m:H"
_symmetry_Int_Tables_number 166

_cell_length_a      3.13000
_cell_length_b      3.13000
_cell_length_c      14.98000
_cell_angle_alpha  90.00000
_cell_angle_beta   90.00000
_cell_angle_gamma  120.00000

loop_
_space_group_symop_id
_space_group_symop_operation_xyz
1 x,y,z
2 -y,x-y,z
3 -x+y,-x,z
4 y,x,-z
5 -x,-x+y,-z
6 x-y,-y,-z
7 -x,-y,-z
8 y,-x+y,-z
9 x-y,x,-z
10 -y,-x,z
11 x,x-y,z
12 -x+y,y,z
13 x+1/3,y+2/3,z+2/3
14 -y+1/3,x-y+2/3,z+2/3
15 -x+y+1/3,-x+2/3,z+2/3
16 y+1/3,x+2/3,-z+2/3
17 -x+1/3,-x+y+2/3,-z+2/3
18 x-y+1/3,-y+2/3,-z+2/3
19 -x+1/3,-y+2/3,-z+2/3
20 y+1/3,-x+y+2/3,-z+2/3
21 x-y+1/3,x+2/3,-z+2/3
22 -y+1/3,-x+2/3,z+2/3
23 x+1/3,x-y+2/3,z+2/3
24 -x+y+1/3,y+2/3,z+2/3
25 x+2/3,y+1/3,z+1/3
26 -y+2/3,x-y+1/3,z+1/3
27 -x+y+2/3,-x+1/3,z+1/3
28 y+2/3,x+1/3,-z+1/3
29 -x+2/3,-x+y+1/3,-z+1/3
30 x-y+2/3,-y+1/3,-z+1/3
31 -x+2/3,-y+1/3,-z+1/3
32 y+2/3,-x+y+1/3,-z+1/3
33 x-y+2/3,x+1/3,-z+1/3
34 -y+2/3,-x+1/3,z+1/3
35 x+2/3,x-y+1/3,z+1/3
36 -x+y+2/3,y+1/3,z+1/3

loop_
_atom_site_label
_atom_site_type_symbol
_atom_site_symmetry_multiplicity
_atom_site_Wyckoff_label
_atom_site_fract_x
_atom_site_fract_y
_atom_site_fract_z
_atom_site_occupancy
Cu1 Cu   3 a 0.00000 0.00000 0.00000 1.00000
Pt1 Pt   3 b 0.00000 0.00000 0.50000 1.00000
```

CuPt ($L1_1$): AB_hR2_166_a_b - POSCAR

```
AB_hR2_166_a_b & a,c/a --params=3.13,4.78594249201 & R(-3)m     D_{3d}^5
    ↪ #166 (ab) & hR2 & L1_1 & CuPt & approximate & C. H. Johansson
    ↪ and J. O. Linde, Annalen der Physik 387, 449-478 (1927)
   1.0000000000000000
   1.56500000000000   -0.90355317128200    4.99333333333300
   0.00000000000000    1.80710634256400    4.99333333333300
  -1.56500000000000   -0.90355317128200    4.99333333333300
   Cu     Pt
    1      1
Direct
   0.00000000000000    0.00000000000000    0.00000000000000    Cu    (1a)
   0.50000000000000    0.50000000000000    0.50000000000000    Pt    (1b)
```

$\alpha$-As (A7): A_hR2_166_c - CIF

```
# CIF file

data_findsym-output
_audit_creation_method FINDSYM

_chemical_name_mineral 'alpha As'
_chemical_formula_sum 'As'

loop_
_publ_author_name
 'D. Schiferl'
 'C. S. Barrett'
_journal_name_full
;
Journal of Applied Crystallography
;
_journal_volume 2
_journal_year 1969
_journal_page_first 30
_journal_page_last 36
_publ_Section_title
;
 The crystal structure of arsenic at 4.2, 78 and 299 K
;

# Found in AMS Database

_aflow_proto 'A_hR2_166_c'
_aflow_params 'a,c/a,x1'
_aflow_params_values '3.7595,2.7815666977,0.22754'
_aflow_Strukturbericht 'A7'
_aflow_Pearson 'hR2'

_symmetry_space_group_name_Hall "-R 3 2"
_symmetry_space_group_name_H-M "R -3 m:H"
_symmetry_Int_Tables_number 166

_cell_length_a      3.75950
_cell_length_b      3.75950
_cell_length_c      10.45730
_cell_angle_alpha  90.00000
_cell_angle_beta   90.00000
_cell_angle_gamma  120.00000

loop_
_space_group_symop_id
_space_group_symop_operation_xyz
1 x,y,z
2 -y,x-y,z
3 -x+y,-x,z
4 y,x,-z
5 -x,-x+y,-z
6 x-y,-y,-z
7 -x,-y,-z
8 y,-x+y,-z
9 x-y,x,-z
10 -y,-x,z
11 x,x-y,z
12 -x+y,y,z
13 x+1/3,y+2/3,z+2/3
14 -y+1/3,x-y+2/3,z+2/3
15 -x+y+1/3,-x+2/3,z+2/3
16 y+1/3,x+2/3,-z+2/3
17 -x+1/3,-x+y+2/3,-z+2/3
18 x-y+1/3,-y+2/3,-z+2/3
19 -x+1/3,-y+2/3,-z+2/3
20 y+1/3,-x+y+2/3,-z+2/3
21 x-y+1/3,x+2/3,-z+2/3
22 -y+1/3,-x+2/3,z+2/3
23 x+1/3,x-y+2/3,z+2/3
24 -x+y+1/3,y+2/3,z+2/3
25 x+2/3,y+1/3,z+1/3
26 -y+2/3,x-y+1/3,z+1/3
27 -x+y+2/3,-x+1/3,z+1/3
28 y+2/3,x+1/3,-z+1/3
29 -x+2/3,-x+y+1/3,-z+1/3
30 x-y+2/3,-y+1/3,-z+1/3
31 -x+2/3,-y+1/3,-z+1/3
32 y+2/3,-x+y+1/3,-z+1/3
33 x-y+2/3,x+1/3,-z+1/3
34 -y+2/3,-x+1/3,z+1/3
35 x+2/3,x-y+1/3,z+1/3
36 -x+y+2/3,y+1/3,z+1/3
```

```
loop_
_atom_site_label
_atom_site_type_symbol
_atom_site_symmetry_multiplicity
_atom_site_Wyckoff_label
_atom_site_fract_x
_atom_site_fract_y
_atom_site_fract_z
_atom_site_occupancy
As1 As    6 c 0.00000 0.00000 0.22754 1.00000
```

α-As (A7): A_hR2_166_c - POSCAR

```
A_hR2_166_c & a,c/a,x1 --params=3.7595,2.7815666977,0.22754 & R(-3)m
    ↪ D_{3d}^5 #166 (c) & hR2 & A7 & As & alpha & D. Schiferl and C.
    ↪ S. Barrett, J. App. Cryst. 2, 30-36 (1969)
   1.0000000000000000
   1.87975000000000  -1.08527416850900   3.48576666666700
   0.00000000000000   2.17054833701800   3.48576666666700
  -1.87975000000000  -1.08527416850900   3.48576666666700
   As
    2
Direct
   0.22754000000000   0.22754000000000   0.22754000000000   As   (2c)
   0.77246000000000   0.77246000000000   0.77246000000000   As   (2c)
```

β-Po ($A_i$): A_hR1_166_a - CIF

```
# CIF file

data_findsym-output
_audit_creation_method FINDSYM

_chemical_name_mineral 'beta Polonium'
_chemical_formula_sum 'Po'

loop_
_publ_author_name
 'William H. Beamer'
 'Charles R. Maxwell'
_journal_name_full
;
Journal of Chemical Physics
;
_journal_volume 17
_journal_year 1949
_journal_page_first 1293
_journal_page_last 1298
_publ_Section_title
;
 Physical Properties of Polonium. II. X-Ray Studies and Crystal
    ↪ Structure
;

# Found in  Donohue, pp. 392

_aflow_proto 'A_hR1_166_a'
_aflow_params 'a,c/a'
_aflow_params_values '5.07846,0.968139947937'
_aflow_Strukturbericht 'A_i'
_aflow_Pearson 'hR1'

_symmetry_space_group_name_Hall "-R 3 2"
_symmetry_space_group_name_H-M "R -3 m:H"
_symmetry_Int_Tables_number 166

_cell_length_a      5.07846
_cell_length_b      5.07846
_cell_length_c      4.91666
_cell_angle_alpha  90.00000
_cell_angle_beta   90.00000
_cell_angle_gamma 120.00000

loop_
_space_group_symop_id
_space_group_symop_operation_xyz
1 x,y,z
2 -y,x-y,z
3 -x+y,-x,z
4 y,x,-z
5 -x,-x+y,-z
6 x-y,-y,-z
7 -x,-y,-z
8 y,-x+y,-z
9 x-y,x,-z
10 -y,-x,z
11 x,x-y,z
12 -x+y,y,z
13 x+1/3,y+2/3,z+2/3
14 -y+1/3,x-y+2/3,z+2/3
15 -x+y+1/3,-x+2/3,z+2/3
16 y+1/3,x+2/3,-z+2/3
17 -x+1/3,-x+y+2/3,-z+2/3
18 x-y+1/3,-y+2/3,-z+2/3
19 -x+1/3,-y+2/3,-z+2/3
20 y+1/3,-x+y+2/3,-z+2/3
21 x-y+1/3,x+2/3,-z+2/3
22 -y+1/3,-x+2/3,z+2/3
23 x+1/3,x-y+2/3,z+2/3
24 -x+y+1/3,y+2/3,z+2/3
25 x+2/3,y+1/3,z+1/3
26 -y+2/3,x-y+1/3,z+1/3
27 -x+y+2/3,-x+1/3,z+1/3
28 y+2/3,x+1/3,-z+1/3
29 -x+2/3,-x+y+1/3,-z+1/3
30 x-y+2/3,-y+1/3,-z+1/3
31 -x+2/3,-y+1/3,-z+1/3
32 y+2/3,-x+y+1/3,-z+1/3
33 x-y+2/3,x+1/3,-z+1/3
34 -y+2/3,-x+1/3,z+1/3
35 x+2/3,x-y+1/3,z+1/3
36 -x+y+2/3,y+1/3,z+1/3

loop_
_atom_site_label
_atom_site_type_symbol
_atom_site_symmetry_multiplicity
_atom_site_Wyckoff_label
_atom_site_fract_x
_atom_site_fract_y
_atom_site_fract_z
_atom_site_occupancy
Po1 Po    3 a 0.00000 0.00000 0.00000 1.00000
```

β-Po ($A_i$): A_hR1_166_a - POSCAR

```
A_hR1_166_a & a,c/a --params=5.07846,0.968139947937 & R(-3)m    D_{3d}^5
    ↪ #166 (a) & hR1 & A_i & Po & beta & W. H. Beamer and C. R.
    ↪ Maxwell, J. Chem. Phys. 17, 1293-1298 (1949)
   1.0000000000000000
   2.53923164757300  -1.46602607526100   1.63888717934400
   0.00000000000000   2.93205215052200   1.63888717934400
  -2.53923164757300  -1.46602607526100   1.63888717934400
   Po
    1
Direct
   0.00000000000000   0.00000000000000   0.00000000000000   Po   (1a)
```

$Fe_7W_6$ ($D8_5$) μ-phase: A7B6_hR13_166_ah_3c - CIF

```
# CIF file

data_findsym-output
_audit_creation_method FINDSYM

_chemical_name_mineral 'Frank-Kasper $\mu$ Phase'
_chemical_formula_sum 'Fe7 W6'

loop_
_publ_author_name
 'H. Arnfelt'
_journal_name_full
;
Jernkontorets Annaler
;
_journal_volume 119
_journal_year 1935
_journal_page_first 185
_journal_page_last 187
_publ_Section_title
;
 Crystal Structure of Fe$_7$W$_6$
;

# Found in  Pearson's Handbook, Vol. III, pp. 3415

_aflow_proto 'A7B6_hR13_166_ah_3c'
_aflow_params 'a,c/a,x2,x3,x4,x5,z5'
_aflow_params_values '4.757,5.4319949548,0.167,0.346,0.448,0.09,0.59001'
_aflow_Strukturbericht 'D8_5'
_aflow_Pearson 'hR13'

_symmetry_space_group_name_Hall "-R 3 2"
_symmetry_space_group_name_H-M "R -3 m:H"
_symmetry_Int_Tables_number 166

_cell_length_a      4.75700
_cell_length_b      4.75700
_cell_length_c      25.84000
_cell_angle_alpha  90.00000
_cell_angle_beta   90.00000
_cell_angle_gamma 120.00000

loop_
_space_group_symop_id
_space_group_symop_operation_xyz
1 x,y,z
2 -y,x-y,z
3 -x+y,-x,z
4 y,x,-z
5 -x,-x+y,-z
6 x-y,-y,-z
7 -x,-y,-z
8 y,-x+y,-z
9 x-y,x,-z
10 -y,-x,z
11 x,x-y,z
12 -x+y,y,z
13 x+1/3,y+2/3,z+2/3
14 -y+1/3,x-y+2/3,z+2/3
15 -x+y+1/3,-x+2/3,z+2/3
16 y+1/3,x+2/3,-z+2/3
17 -x+1/3,-x+y+2/3,-z+2/3
18 x-y+1/3,-y+2/3,-z+2/3
19 -x+1/3,-y+2/3,-z+2/3
20 y+1/3,-x+y+2/3,-z+2/3
21 x-y+1/3,x+2/3,-z+2/3
22 -y+1/3,-x+2/3,z+2/3
23 x+1/3,x-y+2/3,z+2/3
24 -x+y+1/3,y+2/3,z+2/3
```

```
25 x+2/3,y+1/3,z+1/3
26 -y+2/3,x-y+1/3,z+1/3
27 -x+y+2/3,-x+1/3,z+1/3
28 y+2/3,x+1/3,-z+1/3
29 -x+2/3,-x+y+1/3,-z+1/3
30 x-y+2/3,-y+1/3,-z+1/3
31 -x+2/3,-y+1/3,-z+1/3
32 y+2/3,-x+y+1/3,-z+1/3
33 x-y+2/3,x+1/3,-z+1/3
34 -y+2/3,-x+1/3,z+1/3
35 x+2/3,x-y+1/3,z+1/3
36 -x+y+2/3,y+1/3,z+1/3

loop_
_atom_site_label
_atom_site_type_symbol
_atom_site_symmetry_multiplicity
_atom_site_Wyckoff_label
_atom_site_fract_x
_atom_site_fract_y
_atom_site_fract_z
_atom_site_occupancy
Fe1 Fe    3 a 0.00000 0.00000 0.00000 1.00000
W1  W     6 c 0.00000 0.00000 0.16700 1.00000
W2  W     6 c 0.00000 0.00000 0.34600 1.00000
W3  W     6 c 0.00000 0.00000 0.44800 1.00000
Fe2 Fe   18 h 0.83333 0.16667 0.25667 1.00000
```

$Fe_7W_6$ ($D8_5$) $\mu$-phase: A7B6_hR13_166_ah_3c - POSCAR

```
A7B6_hR13_166_ah_3c & a,c/a,x2,x3,x4,x5,z5 --params=4.757,5.4319949548,
    ↪ 0.167,0.346,0.448,0.09,0.59001 & R(-3)m      D_{3d}^5 #166 (ac^3h
    ↪ ) & hR13 & D8_5 & Fe7W6 & & H. Arnfelt, Jernkontorets Annaler
    ↪ 119, 185-187 (1935), quoted in Pearson's Handbook III, pp. 3415
   1.0000000000000000
   2.37850000000000   -1.37322761526800    8.61333333333300
   0.00000000000000    2.74645523053500    8.61333333333300
  -2.37850000000000   -1.37322761526800    8.61333333333300
    Fe      W
     7      6
Direct
   0.00000000000000    0.00000000000000    0.00000000000000    Fe    (1a)
  -0.09000000000000   -0.09000000000000    0.41000000000000    Fe    (6h)
   0.09000000000000    0.09000000000000    0.59000000000000    Fe    (6h)
  -0.09000000000000    0.41000000000000   -0.09000000000000    Fe    (6h)
   0.09000000000000    0.59000000000000    0.09000000000000    Fe    (6h)
   0.41000000000000   -0.09000000000000   -0.09000000000000    Fe    (6h)
   0.59000000000000    0.09000000000000    0.09000000000000    Fe    (6h)
   0.16700000000000    0.16700000000000    0.16700000000000     W    (2c)
   0.83300000000000    0.83300000000000    0.83300000000000     W    (2c)
   0.34600000000000    0.34600000000000    0.34600000000000     W    (2c)
   0.65400000000000    0.65400000000000    0.65400000000000     W    (2c)
   0.44800000000000    0.44800000000000    0.44800000000000     W    (2c)
   0.55200000000000    0.55200000000000    0.55200000000000     W    (2c)
```

$\alpha$-Sm (C19): A_hR3_166_ac - CIF

```
# CIF file

data_findsym-output
_audit_creation_method FINDSYM

_chemical_name_mineral 'alpha Samarium'
_chemical_formula_sum 'Sm'

loop_
_publ_author_name
 'A. H. Daane'
 'R. E. Rundle'
 'H. G. Smith'
 'F. H. Spedding'
_journal_name_full
;
Acta Crystallographica
;
_journal_volume 7
_journal_year 1954
_journal_page_first 532
_journal_page_last 535
_publ_Section_title
;
 The crystal structure of samarium
;

_aflow_proto 'A_hR3_166_ac'
_aflow_params 'a,c/a,x2'
_aflow_params_values '3.62036,7.25049442597,0.22222'
_aflow_Strukturbericht 'C19'
_aflow_Pearson 'hR3'

_symmetry_space_group_name_Hall "-R 3 2"
_symmetry_space_group_name_H-M "R -3 m:H"
_symmetry_Int_Tables_number 166

_cell_length_a      3.62036
_cell_length_b      3.62036
_cell_length_c      26.24940
_cell_angle_alpha   90.00000
_cell_angle_beta    90.00000
_cell_angle_gamma   120.00000

loop_
_space_group_symop_id
_space_group_symop_operation_xyz
1 x,y,z
2 -y,x-y,z
3 -x+y,-x,z
4 y,x,-z
5 -x,-x+y,-z
6 x-y,-y,-z
7 -x,-y,-z
8 y,-x+y,-z
9 x-y,x,-z
10 -y,-x,z
11 x,x-y,z
12 -x+y,y,z
13 x+1/3,y+2/3,z+2/3
14 -y+1/3,x-y+2/3,z+2/3
15 -x+y+1/3,-x+2/3,z+2/3
16 y+1/3,x+2/3,-z+2/3
17 -x+1/3,-x+y+2/3,-z+2/3
18 x-y+1/3,-y+2/3,-z+2/3
19 -x+1/3,-y+2/3,-z+2/3
20 y+1/3,-x+y+2/3,-z+2/3
21 x-y+1/3,x+2/3,-z+2/3
22 -y+1/3,-x+2/3,z+2/3
23 x+1/3,x-y+2/3,z+2/3
24 -x+y+1/3,y+2/3,z+2/3
25 x+2/3,y+1/3,z+1/3
26 -y+2/3,x-y+1/3,z+1/3
27 -x+y+2/3,-x+1/3,z+1/3
28 y+2/3,x+1/3,-z+1/3
29 -x+2/3,-x+y+1/3,-z+1/3
30 x-y+2/3,-y+1/3,-z+1/3
31 -x+2/3,-y+1/3,-z+1/3
32 y+2/3,-x+y+1/3,-z+1/3
33 x-y+2/3,x+1/3,-z+1/3
34 -y+2/3,-x+1/3,z+1/3
35 x+2/3,x-y+1/3,z+1/3
36 -x+y+2/3,y+1/3,z+1/3

loop_
_atom_site_label
_atom_site_type_symbol
_atom_site_symmetry_multiplicity
_atom_site_Wyckoff_label
_atom_site_fract_x
_atom_site_fract_y
_atom_site_fract_z
_atom_site_occupancy
Sm1 Sm    3 a 0.00000 0.00000 0.00000 1.00000
Sm2 Sm    6 c 0.00000 0.00000 0.22222 1.00000
```

$\alpha$-Sm (C19): A_hR3_166_ac - POSCAR

```
A_hR3_166_ac & a,c/a,x2 --params=3.62036,7.25049442597,0.22222 & R(-3)m
    ↪    D^5_{3d} #166 (ac) & hR3 & C19 & Sm (alpha) & & A. H. Daane
    ↪ , R. E. Rundle, H. G. Smith and F. H. Spedding, Acta Cryst. 7,
    ↪ 532-535 (1954)
   1.0000000000000000
   1.81017865060100   -1.04510713120600    8.74980115986700
   0.00000000000000    2.09021426241200    8.74980115986700
  -1.81017865060100   -1.04510713120600    8.74980115986700
    Sm
     3
Direct
   0.00000000000000    0.00000000000000    0.00000000000000    Sm    (1a)
   0.22222222222222    0.22222222222222    0.22222222222222    Sm    (2c)
   0.77777777777778    0.77777777777778    0.77777777777778    Sm    (2c)
```

$Bi_2Te_3$ (C33): A2B3_hR5_166_c_ac - CIF

```
# CIF file

data_findsym-output
_audit_creation_method FINDSYM

_chemical_name_mineral ''
_chemical_formula_sum 'Bi2 Te3'

loop_
_publ_author_name
 'Paul W. Lange'
_journal_name_full
;
Naturwissenschaften
;
_journal_volume 27
_journal_year 1939
_journal_page_first 133
_journal_page_last 134
_publ_Section_title
;
 Ein Vergleich zwischen Bi$_2$Te$_3$ und Bi$_2$Te$_2$S
;

_aflow_proto 'A2B3_hR5_166_c_ac'
_aflow_params 'a,c/a,x2,x3'
_aflow_params_values '4.36914,6.96313919902,0.399,0.208'
_aflow_Strukturbericht 'C33'
_aflow_Pearson 'hR5'

_symmetry_space_group_name_Hall "-R 3 2"
_symmetry_space_group_name_H-M "R -3 m:H"
_symmetry_Int_Tables_number 166

_cell_length_a      4.36914
_cell_length_b      4.36914
_cell_length_c      30.42293
_cell_angle_alpha   90.00000
```

```
_cell_angle_beta   90.00000
_cell_angle_gamma  120.00000

loop_
_space_group_symop_id
_space_group_symop_operation_xyz
1 x,y,z
2 -y,x-y,z
3 -x+y,-x,z
4 y,x,-z
5 -x,-x+y,-z
6 x-y,-y,-z
7 -x,-y,-z
8 y,-x+y,-z
9 x-y,x,-z
10 -y,-x,z
11 x,x-y,z
12 -x+y,y,z
13 x+1/3,y+2/3,z+2/3
14 -y+1/3,x-y+2/3,z+2/3
15 -x+y+1/3,-x+2/3,z+2/3
16 y+1/3,x+2/3,-z+2/3
17 -x+1/3,-x+y+2/3,-z+2/3
18 x-y+1/3,-y+2/3,-z+2/3
19 -x+1/3,-y+2/3,-z+2/3
20 y+1/3,-x+y+2/3,-z+2/3
21 x-y+1/3,x+2/3,-z+2/3
22 -y+1/3,-x+2/3,z+2/3
23 x+1/3,x-y+2/3,z+2/3
24 -x+y+1/3,y+2/3,z+2/3
25 x+2/3,y+1/3,z+1/3
26 -y+2/3,x-y+1/3,z+1/3
27 -x+y+2/3,-x+1/3,z+1/3
28 y+2/3,x+1/3,-z+1/3
29 -x+2/3,-x+y+1/3,-z+1/3
30 x-y+2/3,-y+1/3,-z+1/3
31 -x+2/3,-y+1/3,-z+1/3
32 y+2/3,-x+y+1/3,-z+1/3
33 x-y+2/3,x+1/3,-z+1/3
34 -y+2/3,-x+1/3,z+1/3
35 x+2/3,x-y+1/3,z+1/3
36 -x+y+2/3,y+1/3,z+1/3

loop_
_atom_site_label
_atom_site_type_symbol
_atom_site_symmetry_multiplicity
_atom_site_Wyckoff_label
_atom_site_fract_x
_atom_site_fract_y
_atom_site_fract_z
_atom_site_occupancy
Te1 Te    3 a 0.00000 0.00000 0.00000 1.00000
Bi1 Bi    6 c 0.00000 0.00000 0.39900 1.00000
Te2 Te    6 c 0.00000 0.00000 0.20800 1.00000
```

$Bi_2Te_3$ (C33): A2B3_hR5_166_c_ac - POSCAR

```
A2B3_hR5_166_c_ac & a,c/a,x2,x3 --params=4.36914,6.96313919902,0.399,
   ↪ 0.208 & R(-3)m    D^5_{3d} #166 (ac^2) & hR5 & C33 & Bi2Te3 & &
   ↪  P. W. Lange, Naturwissenschaften 27, 133-134 (1939)
   1.0000000000000000
   2.18457000000000   -1.26126207756358   10.14097666666875
   0.00000000000000    2.52252415512716   10.14097666666875
  -2.18457000000000   -1.26126207756358   10.14097666666875
   Bi    Te
    2     3
Direct
   0.39900000000000    0.39900000000000    0.39900000000000    Bi    (2c)
   0.60100000000000    0.60100000000000    0.60100000000000    Bi    (2c)
   0.00000000000000    0.00000000000000    0.00000000000000    Te    (1a)
   0.20800000000000    0.20800000000000    0.20800000000000    Te    (2c)
   0.79200000000000    0.79200000000000    0.79200000000000    Te    (2c)
```

α-Hg (A10): A_hR1_166_a - CIF

```
# CIF file

data_findsym-output
_audit_creation_method FINDSYM

_chemical_name_mineral 'alpha'
_chemical_formula_sum 'Hg'

loop_
_publ_author_name
 'C. S. Barrett'
_journal_name_full
;
Acta Crystallographica
;
_journal_volume 10
_journal_year 1957
_journal_page_first 58
_journal_page_last 60
_publ_Section_title
;
 The structure of mercury at low temperatures
;

# Found in  Donohue, pp. 231-233

_aflow_proto 'A_hR1_166_a'
_aflow_params 'a,c/a'
_aflow_params_values '3.45741,1.92728082582'
_aflow_Strukturbericht 'A10'
_aflow_Pearson 'hR1'

_symmetry_space_group_name_Hall "-R 3 2"
_symmetry_space_group_name_H-M "R -3 m:H"
_symmetry_Int_Tables_number 166

_cell_length_a      3.45741
_cell_length_b      3.45741
_cell_length_c      6.66340
_cell_angle_alpha  90.00000
_cell_angle_beta   90.00000
_cell_angle_gamma  120.00000

loop_
_space_group_symop_id
_space_group_symop_operation_xyz
1 x,y,z
2 -y,x-y,z
3 -x+y,-x,z
4 y,x,-z
5 -x,-x+y,-z
6 x-y,-y,-z
7 -x,-y,-z
8 y,-x+y,-z
9 x-y,x,-z
10 -y,-x,z
11 x,x-y,z
12 -x+y,y,z
13 x+1/3,y+2/3,z+2/3
14 -y+1/3,x-y+2/3,z+2/3
15 -x+y+1/3,-x+2/3,z+2/3
16 y+1/3,x+2/3,-z+2/3
17 -x+1/3,-x+y+2/3,-z+2/3
18 x-y+1/3,-y+2/3,-z+2/3
19 -x+1/3,-y+2/3,-z+2/3
20 y+1/3,-x+y+2/3,-z+2/3
21 x-y+1/3,x+2/3,-z+2/3
22 -y+1/3,-x+2/3,z+2/3
23 x+1/3,x-y+2/3,z+2/3
24 -x+y+1/3,y+2/3,z+2/3
25 x+2/3,y+1/3,z+1/3
26 -y+2/3,x-y+1/3,z+1/3
27 -x+y+2/3,-x+1/3,z+1/3
28 y+2/3,x+1/3,-z+1/3
29 -x+2/3,-x+y+1/3,-z+1/3
30 x-y+2/3,-y+1/3,-z+1/3
31 -x+2/3,-y+1/3,-z+1/3
32 y+2/3,-x+y+1/3,-z+1/3
33 x-y+2/3,x+1/3,-z+1/3
34 -y+2/3,-x+1/3,z+1/3
35 x+2/3,x-y+1/3,z+1/3
36 -x+y+2/3,y+1/3,z+1/3

loop_
_atom_site_label
_atom_site_type_symbol
_atom_site_symmetry_multiplicity
_atom_site_Wyckoff_label
_atom_site_fract_x
_atom_site_fract_y
_atom_site_fract_z
_atom_site_occupancy
Hg1 Hg    3 a 0.00000 0.00000 0.00000 1.00000
```

α-Hg (A10): A_hR1_166_a - POSCAR

```
A_hR1_166_a & a,c/a --params=3.45741,1.92728082582 & R(-3)m      D_{3d}^
   ↪ 5 #166 (a) & hR1 & A10 & Hg & alpha-Hg & C. S. Barrett, Acta
   ↪ Cryst. 10, 58-60 (1957)
   1.0000000000000000
   1.72870340938900   -0.99806737876000    2.22113343134200
   0.00000000000000    1.99613475751900    2.22113343134200
  -1.72870340938900   -0.99806737876000    2.22113343134200
   Po
    1
Direct
   0.00000000000000    0.00000000000000    0.00000000000000    Hg    (1a)
```

$Mo_2B_5$ ($D8_i$): A5B2_hR7_166_a2c_c - CIF

```
# CIF file

data_findsym-output
_audit_creation_method FINDSYM

_chemical_name_mineral 'Molybdenum Boride'
_chemical_formula_sum 'Mo2 B5'

loop_
_publ_author_name
 'Roland Kiessling'
_journal_name_full
;
Acta Chemica Scandinavica
;
_journal_volume 1
_journal_year 1947
_journal_page_first 893
_journal_page_last 916
_publ_Section_title
;
 The Crystal Structures of Molybdenum and Tungsten Borides
;
```

```
_aflow_proto 'A5B2_hR7_166_a2c_c'
_aflow_params 'a,c/a,x2,x3,x4'
_aflow_params_values '3.011,6.9511790103,0.186,0.33333,0.075'
_aflow_Strukturbericht 'D8_i'
_aflow_Pearson 'hR7'

_symmetry_space_group_name_Hall "-R 3 2"
_symmetry_space_group_name_H-M "R -3 m:H"
_symmetry_Int_Tables_number 166

_cell_length_a     3.01100
_cell_length_b     3.01100
_cell_length_c     20.93000
_cell_angle_alpha  90.00000
_cell_angle_beta   90.00000
_cell_angle_gamma  120.00000

loop_
_space_group_symop_id
_space_group_symop_operation_xyz
1 x,y,z
2 -y,x-y,z
3 -x+y,-x,z
4 y,x,-z
5 -x,-x+y,-z
6 x-y,-y,-z
7 -x,-y,-z
8 y,-x+y,-z
9 x-y,x,-z
10 -y,-x,z
11 x,x-y,z
12 -x+y,y,z
13 x+1/3,y+2/3,z+2/3
14 -y+1/3,x-y+2/3,z+2/3
15 -x+y+1/3,-x+2/3,z+2/3
16 y+1/3,x+2/3,-z+2/3
17 -x+1/3,-x+y+2/3,-z+2/3
18 x-y+1/3,-y+2/3,-z+2/3
19 -x+1/3,-y+2/3,-z+2/3
20 y+1/3,-x+y+2/3,-z+2/3
21 x-y+1/3,x+2/3,-z+2/3
22 -y+1/3,-x+2/3,z+2/3
23 x+1/3,x-y+2/3,z+2/3
24 -x+y+1/3,y+2/3,z+2/3
25 x+2/3,y+1/3,z+1/3
26 -y+2/3,x-y+1/3,z+1/3
27 -x+y+2/3,-x+1/3,z+1/3
28 y+2/3,x+1/3,-z+1/3
29 -x+2/3,-x+y+1/3,-z+1/3
30 x-y+2/3,-y+1/3,-z+1/3
31 -x+2/3,-y+1/3,-z+1/3
32 y+2/3,-x+y+1/3,-z+1/3
33 x-y+2/3,x+1/3,-z+1/3
34 -y+2/3,-x+1/3,z+1/3
35 x+2/3,x-y+1/3,z+1/3
36 -x+y+2/3,y+1/3,z+1/3

loop_
_atom_site_label
_atom_site_type_symbol
_atom_site_symmetry_multiplicity
_atom_site_Wyckoff_label
_atom_site_fract_x
_atom_site_fract_y
_atom_site_fract_z
_atom_site_occupancy
B1  B    3 a 0.00000 0.00000 0.00000 1.00000
B2  B    6 c 0.00000 0.00000 0.18600 1.00000
B3  B    6 c 0.00000 0.00000 0.33333 1.00000
Mo1 Mo   6 c 0.00000 0.00000 0.07500 1.00000
```

$Mo_2B_5$ ($D8_i$): A5B2_hR7_166_a2c_c - POSCAR

```
A5B2_hR7_166_a2c_c & a,c/a,x2,x3,x4 --params=3.011,6.9511790103,0.186,
    ↪ 0.33333,0.075 & R(-3)m     D_{3d}^5 #166 (ac^3) & hR7 & D8_i &
    ↪ Mo2B5 & epsilon & R Kiessling, Acta Chem. Scand. 1, 893-916 (
    ↪ 1947)
   1.0000000000000000
   1.50550000000000   -0.86920083026498    6.97666666666667
   0.00000000000000    1.73840166052996    6.97666666666667
  -1.50550000000000   -0.86920083026498    6.97666666666667
     B    Mo
     5    2
Direct
   0.00000000000000    0.00000000000000    0.00000000000000    B    (1a)
   0.18600000000000    0.18600000000000    0.18600000000000    B    (2c)
   0.81400000000000    0.81400000000000    0.81400000000000    B    (2c)
   0.33333333333333    0.33333333333333    0.33333333333333    B    (2c)
   0.66666666666667    0.66666666666667    0.66666666666667    B    (2c)
   0.07500000000000    0.07500000000000    0.07500000000000    Mo   (2c)
   0.92500000000000    0.92500000000000    0.92500000000000    Mo   (2c)
```

Rhombohedral Graphite: A_hR2_166_c - CIF

```
# CIF file

data_findsym-output
_audit_creation_method FINDSYM

_chemical_name_mineral 'rhombohedral graphite'
_chemical_formula_sum 'C'

loop_
_publ_author_name
 'H. Lipson'
 'A. R. Stokes'
_journal_name_full
;
Proceedings of the Royal Society A: Mathematical, Physical and
    ↪ Engineering Sciences
;
_journal_volume 181
_journal_year 1942
_journal_page_first 101
_journal_page_last 105
_publ_Section_title
;
 The structure of graphite
;

# Found in  Donohue, pp. 258-260

_aflow_proto 'A_hR2_166_c'
_aflow_params 'a,c/a,x1'
_aflow_params_values '2.456,4.08957654723,0.16667'
_aflow_Strukturbericht 'None'
_aflow_Pearson 'hR2'

_symmetry_space_group_name_Hall "-R 3 2"
_symmetry_space_group_name_H-M "R -3 m:H"
_symmetry_Int_Tables_number 166

_cell_length_a     2.45600
_cell_length_b     2.45600
_cell_length_c     10.04400
_cell_angle_alpha  90.00000
_cell_angle_beta   90.00000
_cell_angle_gamma  120.00000

loop_
_space_group_symop_id
_space_group_symop_operation_xyz
1 x,y,z
2 -y,x-y,z
3 -x+y,-x,z
4 y,x,-z
5 -x,-x+y,-z
6 x-y,-y,-z
7 -x,-y,-z
8 y,-x+y,-z
9 x-y,x,-z
10 -y,-x,z
11 x,x-y,z
12 -x+y,y,z
13 x+1/3,y+2/3,z+2/3
14 -y+1/3,x-y+2/3,z+2/3
15 -x+y+1/3,-x+2/3,z+2/3
16 y+1/3,x+2/3,-z+2/3
17 -x+1/3,-x+y+2/3,-z+2/3
18 x-y+1/3,-y+2/3,-z+2/3
19 -x+1/3,-y+2/3,-z+2/3
20 y+1/3,-x+y+2/3,-z+2/3
21 x-y+1/3,x+2/3,-z+2/3
22 -y+1/3,-x+2/3,z+2/3
23 x+1/3,x-y+2/3,z+2/3
24 -x+y+1/3,y+2/3,z+2/3
25 x+2/3,y+1/3,z+1/3
26 -y+2/3,x-y+1/3,z+1/3
27 -x+y+2/3,-x+1/3,z+1/3
28 y+2/3,x+1/3,-z+1/3
29 -x+2/3,-x+y+1/3,-z+1/3
30 x-y+2/3,-y+1/3,-z+1/3
31 -x+2/3,-y+1/3,-z+1/3
32 y+2/3,-x+y+1/3,-z+1/3
33 x-y+2/3,x+1/3,-z+1/3
34 -y+2/3,-x+1/3,z+1/3
35 x+2/3,x-y+1/3,z+1/3
36 -x+y+2/3,y+1/3,z+1/3

loop_
_atom_site_label
_atom_site_type_symbol
_atom_site_symmetry_multiplicity
_atom_site_Wyckoff_label
_atom_site_fract_x
_atom_site_fract_y
_atom_site_fract_z
_atom_site_occupancy
C1 C    6 c 0.00000 0.00000 0.16667 1.00000
```

Rhombohedral Graphite: A_hR2_166_c - POSCAR

```
A_hR2_166_c & a,c/a,x1 --params=2.456,4.08957654723,0.16667 & R(-3)m
    ↪ D_{3d}^5 #166 (c) & hR2 & & C & rhombohedral graphite & H.
    ↪ Lipson and A. R. Stokes, Proc. R. Soc. A Math. Phys. Eng. Sci.
    ↪ 181, 101-105 (1942)
   1.0000000000000000
   1.22800000000000   -0.70898613056486    3.34800000000000
   0.00000000000000    1.41797226112972    3.34800000000000
  -1.22800000000000   -0.70898613056486    3.34800000000000
     C
     2
Direct
   0.16666666666667    0.16666666666667    0.16666666666667    C    (2c)
   0.83333333333333    0.83333333333333    0.83333333333333    C    (2c)
```

α-B (hR12): A_hR12_166_2h - CIF

```
# CIF file
```

```
data_findsym-output
_audit_creation_method FINDSYM

_chemical_name_mineral 'alpha boron'
_chemical_formula_sum 'B'

loop_
_publ_author_name
 'B. F. Decker'
 'J. S. Kasper'
_journal_name_full
;
Acta Crystallographica
;
_journal_volume 12
_journal_year 1959
_journal_page_first 503
_journal_page_last 506
_publ_Section_title
;
 The crystal structure of a simple rhombohedral form of boron
;

# Found in Donohue, pp. 57-60

_aflow_proto 'A_hR12_166_2h'
_aflow_params 'a,c/a,x1,z1,x2,z2'
_aflow_params_values '4.908,2.56022616137,0.0104,0.65729,0.2206,0.6323'
_aflow_Strukturbericht 'None'
_aflow_Pearson 'hR12'

_symmetry_space_group_name_Hall "-R 3 2"
_symmetry_space_group_name_H-M "R -3 m:H"
_symmetry_Int_Tables_number 166

_cell_length_a    4.90800
_cell_length_b    4.90800
_cell_length_c    12.56559
_cell_angle_alpha 90.00000
_cell_angle_beta  90.00000
_cell_angle_gamma 120.00000

loop_
_space_group_symop_id
_space_group_symop_operation_xyz
1 x,y,z
2 -y,x-y,z
3 -x+y,-x,z
4 y,x,-z
5 -x,-x+y,-z
6 x-y,-y,-z
7 -x,-y,-z
8 y,-x+y,-z
9 x-y,x,-z
10 -y,-x,z
11 x,x-y,z
12 -x+y,y,z
13 x+1/3,y+2/3,z+2/3
14 -y+1/3,x-y+2/3,z+2/3
15 -x+y+1/3,-x+2/3,z+2/3
16 y+1/3,x+2/3,-z+2/3
17 -x+1/3,-x+y+2/3,-z+2/3
18 x-y+1/3,-y+2/3,-z+2/3
19 -x+1/3,-y+2/3,-z+2/3
20 y+1/3,-x+y+2/3,-z+2/3
21 x-y+1/3,x+2/3,-z+2/3
22 -y+1/3,-x+2/3,z+2/3
23 x+1/3,x-y+2/3,z+2/3
24 -x+y+1/3,y+2/3,z+2/3
25 x+2/3,y+1/3,z+1/3
26 -y+2/3,x-y+1/3,z+1/3
27 -x+y+2/3,-x+1/3,z+1/3
28 y+2/3,x+1/3,-z+1/3
29 -x+2/3,-x+y+1/3,-z+1/3
30 x-y+2/3,-y+1/3,-z+1/3
31 -x+2/3,-y+1/3,-z+1/3
32 y+2/3,-x+y+1/3,-z+1/3
33 x-y+2/3,x+1/3,-z+1/3
34 -y+2/3,-x+1/3,z+1/3
35 x+2/3,x-y+1/3,z+1/3
36 -x+y+2/3,y+1/3,z+1/3

loop_
_atom_site_label
_atom_site_type_symbol
_atom_site_symmetry_multiplicity
_atom_site_Wyckoff_label
_atom_site_fract_x
_atom_site_fract_y
_atom_site_fract_z
_atom_site_occupancy
B1 B  18 h 0.78437 0.21563 0.22603 1.00000
B2 B  18 h 0.19610 0.80390 0.02450 1.00000
```

$\alpha$-B (hR12): A_hR12_166_2h - POSCAR

```
A_hR12_166_2h & a,c/a,x1,z1,x2,z2 --params=4.908,2.56022616137,0.0104,
    ↪ 0.65729,0.2206,0.6323 & R(-3)m    D_{3d}^5 #166 (h^2) & hR12 &
    ↪ & B & alpha & B. F. Decker and J. S. Kasper, Acta Cryst. 12,
    ↪ 503-506 (1959)
   1.0000000000000000
   2.45399776558700   -1.41681627055200    4.18852905230400
   0.00000000000000    2.83363254110400    4.18852905230400
  -2.45399776558700   -1.41681627055200    4.18852905230400
    B
    12
Direct
  -0.01040000000000  -0.01040000000000   0.34270000000000   B   (6h)
   0.01040000000000   0.01040000000000   0.65730000000000   B   (6h)
  -0.01040000000000   0.34270000000000  -0.01040000000000   B   (6h)
   0.01040000000000   0.65730000000000   0.01040000000000   B   (6h)
   0.34270000000000  -0.01040000000000  -0.01040000000000   B   (6h)
   0.65730000000000   0.01040000000000   0.01040000000000   B   (6h)
   0.22060000000000   0.22060000000000   0.63230000000000   B   (6h)
   0.22060000000000   0.63230000000000   0.22060000000000   B   (6h)
   0.36770000000000   0.77940000000000   0.77940000000000   B   (6h)
   0.63230000000000   0.22060000000000   0.22060000000000   B   (6h)
   0.77940000000000   0.36770000000000   0.77940000000000   B   (6h)
   0.77940000000000   0.77940000000000   0.36770000000000   B   (6h)
```

Caswellsilverite ($CrNaS_2$, $F5_1$): ABC2_hR4_166_a_b_c - CIF

```
# CIF file

data_findsym-output
_audit_creation_method FINDSYM

_chemical_name_mineral 'Caswellsilverite'
_chemical_formula_sum 'Cr Na S2'

loop_
_publ_author_name
 'F. M. R. Engelsman'
 'G. A. Wiegers'
 'F. Jellinek'
 'B. Van Laar'
_journal_name_full
;
Journal of Solid State Chemistry
;
_journal_volume 6
_journal_year 1973
_journal_page_first 574
_journal_page_last 582
_publ_Section_title
;
 Crystal structures and magnetic structures of some metal(I) chromium(
    ↪ III) sulfides and selenides
;

# Found in AMS Database

_aflow_proto 'ABC2_hR4_166_a_b_c'
_aflow_params 'a,c/a,x3'
_aflow_params_values '3.5561,5.44557239673,0.2667'
_aflow_Strukturbericht 'F5_1'
_aflow_Pearson 'hR4'

_symmetry_space_group_name_Hall "-R 3 2"
_symmetry_space_group_name_H-M "R -3 m:H"
_symmetry_Int_Tables_number 166

_cell_length_a    3.55610
_cell_length_b    3.55610
_cell_length_c    19.36500
_cell_angle_alpha 90.00000
_cell_angle_beta  90.00000
_cell_angle_gamma 120.00000

loop_
_space_group_symop_id
_space_group_symop_operation_xyz
1 x,y,z
2 -y,x-y,z
3 -x+y,-x,z
4 y,x,-z
5 -x,-x+y,-z
6 x-y,-y,-z
7 -x,-y,-z
8 y,-x+y,-z
9 x-y,x,-z
10 -y,-x,z
11 x,x-y,z
12 -x+y,y,z
13 x+1/3,y+2/3,z+2/3
14 -y+1/3,x-y+2/3,z+2/3
15 -x+y+1/3,-x+2/3,z+2/3
16 y+1/3,x+2/3,-z+2/3
17 -x+1/3,-x+y+2/3,-z+2/3
18 x-y+1/3,-y+2/3,-z+2/3
19 -x+1/3,-y+2/3,-z+2/3
20 y+1/3,-x+y+2/3,-z+2/3
21 x-y+1/3,x+2/3,-z+2/3
22 -y+1/3,-x+2/3,z+2/3
23 x+1/3,x-y+2/3,z+2/3
24 -x+y+1/3,y+2/3,z+2/3
25 x+2/3,y+1/3,z+1/3
26 -y+2/3,x-y+1/3,z+1/3
27 -x+y+2/3,-x+1/3,z+1/3
28 y+2/3,x+1/3,-z+1/3
29 -x+2/3,-x+y+1/3,-z+1/3
30 x-y+2/3,-y+1/3,-z+1/3
31 -x+2/3,-y+1/3,-z+1/3
32 y+2/3,-x+y+1/3,-z+1/3
33 x-y+2/3,x+1/3,-z+1/3
34 -y+2/3,-x+1/3,z+1/3
35 x+2/3,x-y+1/3,z+1/3
36 -x+y+2/3,y+1/3,z+1/3

loop_
```

```
_atom_site_label
_atom_site_type_symbol
_atom_site_symmetry_multiplicity
_atom_site_Wyckoff_label
_atom_site_fract_x
_atom_site_fract_y
_atom_site_fract_z
_atom_site_occupancy
Cr1 Cr   3 a 0.00000 0.00000 0.00000 1.00000
Na1 Na   3 b 0.00000 0.00000 0.50000 1.00000
S1  S    6 c 0.00000 0.00000 0.26670 1.00000
```

Caswellsilverite ($CrNaS_2$, $F5_1$): ABC2_hR4_166_a_b_c - POSCAR

```
ABC2_hR4_166_a_b_c & a,c/a,x3 --params=3.5561,5.44557239673,0.2667 & R(-
    ↪ 3)m     D_{3d}^5 #166 (abc) & hR4 & F5_1 & CrNaS2 &
    ↪ Caswellsilverite & F. M. R. Engelsman, G. A. Wiegers, F.
    ↪ Jellinek, and B. van Laar, J. Solid State Chem. 6, 574-582 (
    ↪ 1973)
   1.0000000000000000
   1.77805000000000   -1.02655764613300    6.45500000000000
   0.00000000000000    2.05311529226500    6.45500000000000
  -1.77805000000000   -1.02655764613300    6.45500000000000
    Cr     Na      S
     1      1      2
Direct
   0.00000000000000    0.00000000000000    0.00000000000000    Cr    (1a)
   0.50000000000000    0.50000000000000    0.50000000000000    Na    (1b)
   0.26670000000000    0.26670000000000    0.26670000000000     S    (2c)
   0.73330000000000    0.73330000000000    0.73330000000000     S    (2c)
```

$\beta$-O: A_hR2_166_c - CIF

```
# CIF file

data_findsym-output
_audit_creation_method FINDSYM

_chemical_name_mineral 'beta oxygen'
_chemical_formula_sum 'O'

loop_
_publ_author_name
 'R. J. Meier'
 'R. B. Helmholdt'
_journal_name_full
;
Physical Review B
;
_journal_volume 29
_journal_year 1984
_journal_page_first 1387
_journal_page_last 1393
_publ_Section_title
;
 Neutron-diffraction study of $\alpha$- and $\beta$-oxygen
;

_aflow_proto 'A_hR2_166_c'
_aflow_params 'a,c/a,x1'
_aflow_params_values '3.289,3.42991790818,0.0543'
_aflow_Strukturbericht 'None'
_aflow_Pearson 'hR2'

_symmetry_space_group_name_Hall "-R 3 2"
_symmetry_space_group_name_H-M "R -3 m:H"
_symmetry_Int_Tables_number 166

_cell_length_a      3.28900
_cell_length_b      3.28900
_cell_length_c      11.28100
_cell_angle_alpha   90.00000
_cell_angle_beta    90.00000
_cell_angle_gamma   120.00000

loop_
_space_group_symop_id
_space_group_symop_operation_xyz
1 x,y,z
2 -y,x-y,z
3 -x+y,-x,z
4 y,x,-z
5 -x,-x+y,-z
6 x-y,-y,-z
7 -x,-y,-z
8 y,-x+y,-z
9 x-y,x,-z
10 -y,-x,z
11 x,x-y,z
12 -x+y,y,z
13 x+1/3,y+2/3,z+2/3
14 -y+1/3,x-y+2/3,z+2/3
15 -x+y+1/3,-x+2/3,z+2/3
16 y+1/3,x+2/3,-z+2/3
17 -x+1/3,-x+y+2/3,-z+2/3
18 x-y+1/3,-y+2/3,-z+2/3
19 -x+1/3,-y+2/3,-z+2/3
20 y+1/3,-x+y+2/3,-z+2/3
21 x-y+1/3,x+2/3,-z+2/3
22 -y+1/3,-x+2/3,z+2/3
23 x+1/3,x-y+2/3,z+2/3
24 -x+y+1/3,y+2/3,z+2/3
25 x+2/3,y+1/3,z+1/3
26 -y+2/3,x-y+1/3,z+1/3
27 -x+y+2/3,-x+1/3,z+1/3
28 y+2/3,x+1/3,-z+1/3
29 -x+2/3,-x+y+1/3,-z+1/3
30 x-y+2/3,-y+1/3,-z+1/3
31 -x+2/3,-y+1/3,-z+1/3
32 y+2/3,-x+y+1/3,-z+1/3
33 x-y+2/3,x+1/3,-z+1/3
34 -y+2/3,-x+1/3,z+1/3
35 x+2/3,x-y+1/3,z+1/3
36 -x+y+2/3,y+1/3,z+1/3

loop_
_atom_site_label
_atom_site_type_symbol
_atom_site_symmetry_multiplicity
_atom_site_Wyckoff_label
_atom_site_fract_x
_atom_site_fract_y
_atom_site_fract_z
_atom_site_occupancy
O1 O   6 c 0.00000 0.00000 0.05430 1.00000
```

$\beta$-O: A_hR2_166_c - POSCAR

```
A_hR2_166_c & a,c/a,x1 --params=3.289,3.42991790818,0.0543 & R(-3)m
    ↪ D_{3d}^5 #166 (c) & hR2 & & O & beta & R. J. Meier and R. B.
    ↪ Helmholdt, PRB 29, 1387-1393 (1984)
   1.0000000000000000
   1.64450000000000   -0.94945251768200    3.76033333333300
   0.00000000000000    1.89890503536500    3.76033333333300
  -1.64450000000000   -0.94945251768200    3.76033333333300
     O
     2
Direct
   0.05430000000000    0.05430000000000    0.05430000000000     O    (2c)
  -0.05430000000000   -0.05430000000000   -0.05430000000000     O    (2c)
```

$\beta$-B (R-105): A_hR105_166_bc9h4i - CIF

```
# CIF file

data_findsym-output
_audit_creation_method FINDSYM

_chemical_name_mineral 'beta Boron'
_chemical_formula_sum 'B'

loop_
_publ_author_name
 'D. Geist'
 'R. Kloss'
 'H. Follner'
_journal_name_full
;
Acta Crystallographica B
;
_journal_volume 26
_journal_year 1970
_journal_page_first 1800
_journal_page_last 1802
_publ_Section_title
;
 Verfeinerung des $\beta$-rhomboedrischen Bors
;

# Found in  Donohue, pp. 61-78

_aflow_proto 'A_hR105_166_bc9h4i'
_aflow_params 'a,c/a,x2,x3,z3,x4,z4,x5,z5,x6,z6,x7,z7,x8,z8,x9,z9,x10,
    ↪ z10,x11,z11,x12,y12,z12,x13,y13,z13,x14,y14,z14,x15,y15,z15'
_aflow_params_values '10.96,2.17974452555,0.3848,0.3843,0.21309,0.4895,
    ↪ 0.21780,0.3873,0.56899,0.1991,0.50609,0.1983,0.68740,0.1032,
    ↪ 0.49209,0.9933,0.66980,0.1008,0.83740,0.0025,0.16801,0.3622,
    ↪ 0.58109,0.0976,0.3765,0.68261,0.2024,0.1673,0.55209,0.8921,
    ↪ 0.1777,0.3473,0.0033'
_aflow_Strukturbericht 'None'
_aflow_Pearson 'hR105'

_symmetry_space_group_name_Hall "-R 3 2"
_symmetry_space_group_name_H-M "R -3 m:H"
_symmetry_Int_Tables_number 166

_cell_length_a      10.96000
_cell_length_b      10.96000
_cell_length_c      23.89000
_cell_angle_alpha   90.00000
_cell_angle_beta    90.00000
_cell_angle_gamma   120.00000

loop_
_space_group_symop_id
_space_group_symop_operation_xyz
1 x,y,z
2 -y,x-y,z
3 -x+y,-x,z
4 y,x,-z
5 -x,-x+y,-z
6 x-y,-y,-z
7 -x,-y,-z
8 y,-x+y,-z
9 x-y,x,-z
10 -y,-x,z
11 x,x-y,z
12 -x+y,y,z
13 x+1/3,y+2/3,z+2/3
14 -y+1/3,x-y+2/3,z+2/3
```

```
15 -x+y+1/3,-x+2/3,z+2/3
16 y+1/3,x+2/3,-z+2/3
17 -x+1/3,-x+y+2/3,-z+2/3
18 x-y+1/3,-y+2/3,-z+2/3
19 -x+1/3,-y+2/3,-z+2/3
20 y+1/3,-x+y+2/3,-z+2/3
21 x-y+1/3,x+2/3,-z+2/3
22 -y+1/3,-x+2/3,z+2/3
23 x+1/3,x-y+2/3,z+2/3
24 -x+y+1/3,y+2/3,z+2/3
25 x+2/3,y+1/3,z+1/3
26 -y+2/3,x-y+1/3,z+1/3
27 -x+y+2/3,-x+1/3,z+1/3
28 y+2/3,x+1/3,-z+1/3
29 -x+2/3,-x+y+1/3,-z+1/3
30 x-y+2/3,-y+1/3,-z+1/3
31 -x+2/3,-y+1/3,-z+1/3
32 y+2/3,-x+y+1/3,-z+1/3
33 x-y+2/3,x+1/3,-z+1/3
34 -y+2/3,-x+1/3,z+1/3
35 x+2/3,x-y+1/3,z+1/3
36 -x+y+2/3,y+1/3,z+1/3

loop_
_atom_site_label
_atom_site_type_symbol
_atom_site_symmetry_multiplicity
_atom_site_Wyckoff_label
_atom_site_fract_x
_atom_site_fract_y
_atom_site_fract_z
_atom_site_occupancy
B1  B   3 b 0.00000  0.00000  0.50000 1.00000
B2  B   6 c 0.00000  0.00000  0.38480 1.00000
B3  B  18 h 0.05707  -0.05707 0.32723 1.00000
B4  B  18 h 0.42390  0.57610  0.06560 1.00000
B5  B  18 h 0.27277  0.72723  0.11453 1.00000
B6  B  18 h 0.89767  0.10233  0.30143 1.00000
B7  B  18 h 0.17030  0.82970  0.02800 1.00000
B8  B  18 h 0.87037  0.12963  0.23283 1.00000
B9  B  18 h 0.77450  0.22550  0.21880 1.00000
B10 B  18 h 0.08780  -0.08780 0.01300 1.00000
B11 B  18 h -0.05517 0.05517  0.05767 1.00000
B12 B  36 i 0.34857  -0.08397 0.01363 1.00000
B13 B  36 i 0.28933  0.88477  0.08717 1.00000
B14 B  36 i -0.03653 0.31173  0.20383 1.00000
B15 B  36 i 0.00160  0.17280  0.17610 1.00000
```

$\beta$-B (R-105): A_hR105_166_bc9h4i - POSCAR

```
A_hR105_166_bc9h4i & a,c/a,x2,x3,z3,x4,z4,x5,z5,x6,z6,x7,z7,x8,z8,x9,z9,
    ↪ x10,z10,x11,z11,x12,y12,z12,x13,y13,z13,x14,y14,z14,x15,y15,z15
    ↪ --params=10.96,2.17974452555,0.3848,0.3843,0.21309,0.4895,
    ↪ 0.21780,0.3873,0.56899,0.1991,0.50609,0.1983,0.68740,0.1032,
    ↪ 0.49209,0.9933,0.66980,0.1008,0.83740,0.0025,0.16801,0.3622,
    ↪ 0.58109,0.0976,0.3765,0.68261,0.2024,0.1673,0.55209,0.8921,
    ↪ 0.1777,0.3473,0.0033 & R(-3)m        D_{3d}^5 #166 (bch^9i^4) &
    ↪ hR105 & & B & beta (R-105) & D. Geist, R. Kloss and H. Follner,
    ↪  Acta Cryst. B 26, 1800-1802 (1970)
   1.0000000000000000
   5.47929888352000   -3.16347468537100    7.96232451361300
   0.00000000000000    6.32694937074100    7.96232451361300
  -5.47929888352000   -3.16347468537100    7.96232451361300
     B
   105
Direct
   0.09760000000000    0.36220000000000    0.58110000000000    B    (12i)
  -0.09760000000000    0.41890000000000    0.63780000000000    B    (12i)
   0.09760000000000    0.58110000000000    0.36220000000000    B    (12i)
  -0.09760000000000    0.63780000000000    0.41890000000000    B    (12i)
   0.36220000000000    0.09760000000000    0.58110000000000    B    (12i)
   0.36220000000000    0.58110000000000    0.09760000000000    B    (12i)
   0.41890000000000   -0.09760000000000    0.63780000000000    B    (12i)
   0.41890000000000    0.63780000000000   -0.09760000000000    B    (12i)
   0.58110000000000    0.09760000000000    0.36220000000000    B    (12i)
   0.58110000000000    0.36220000000000    0.09760000000000    B    (12i)
   0.63780000000000   -0.09760000000000    0.41890000000000    B    (12i)
   0.63780000000000    0.41890000000000   -0.09760000000000    B    (12i)
   0.20240000000000    0.37650000000000    0.68260000000000    B    (12i)
   0.20240000000000    0.68260000000000    0.37650000000000    B    (12i)
   0.31740000000000    0.62350000000000    0.79760000000000    B    (12i)
   0.31740000000000    0.79760000000000    0.62350000000000    B    (12i)
   0.37650000000000    0.20240000000000    0.68260000000000    B    (12i)
   0.37650000000000    0.68260000000000    0.20240000000000    B    (12i)
   0.62350000000000    0.31740000000000    0.79760000000000    B    (12i)
   0.62350000000000    0.79760000000000    0.31740000000000    B    (12i)
   0.68260000000000    0.20240000000000    0.37650000000000    B    (12i)
   0.68260000000000    0.37650000000000    0.20240000000000    B    (12i)
   0.79760000000000    0.31740000000000    0.62350000000000    B    (12i)
   0.79760000000000    0.62350000000000    0.31740000000000    B    (12i)
   0.10790000000000    0.44790000000000    0.83270000000000    B    (12i)
   0.10790000000000    0.83270000000000    0.44790000000000    B    (12i)
   0.16730000000000    0.55210000000000    0.89210000000000    B    (12i)
   0.16730000000000    0.89210000000000    0.55210000000000    B    (12i)
   0.44790000000000    0.10790000000000    0.83270000000000    B    (12i)
   0.44790000000000    0.83270000000000    0.10790000000000    B    (12i)
   0.55210000000000    0.16730000000000    0.89210000000000    B    (12i)
   0.55210000000000    0.89210000000000    0.16730000000000    B    (12i)
   0.83270000000000    0.10790000000000    0.44790000000000    B    (12i)
   0.83270000000000    0.44790000000000    0.10790000000000    B    (12i)
   0.89210000000000    0.16730000000000    0.55210000000000    B    (12i)
   0.89210000000000    0.55210000000000    0.16730000000000    B    (12i)
   0.00330000000000    0.17770000000000    0.34730000000000    B    (12i)
   0.00330000000000    0.34730000000000    0.17770000000000    B    (12i)
  -0.00330000000000    0.65270000000000    0.82230000000000    B    (12i)
  -0.00330000000000    0.82230000000000    0.65270000000000    B    (12i)
   0.17770000000000    0.00330000000000    0.34730000000000    B    (12i)
   0.17770000000000    0.34730000000000    0.00330000000000    B    (12i)
   0.34730000000000    0.00330000000000    0.17770000000000    B    (12i)
   0.34730000000000    0.17770000000000    0.00330000000000    B    (12i)
   0.65270000000000   -0.00330000000000    0.82230000000000    B    (12i)
   0.65270000000000    0.82230000000000   -0.00330000000000    B    (12i)
   0.82230000000000   -0.00330000000000    0.65270000000000    B    (12i)
   0.82230000000000    0.65270000000000   -0.00330000000000    B    (12i)
   0.50000000000000    0.50000000000000    0.50000000000000    B     (1b)
   0.38480000000000    0.38480000000000    0.38480000000000    B     (2c)
   0.61520000000000    0.61520000000000    0.61520000000000    B     (2c)
   0.10080000000000    0.10080000000000    0.83740000000000    B     (6h)
   0.10080000000000    0.83740000000000    0.10080000000000    B     (6h)
   0.16260000000000    0.89920000000000    0.89920000000000    B     (6h)
   0.83740000000000    0.10080000000000    0.10080000000000    B     (6h)
   0.89920000000000    0.16260000000000    0.89920000000000    B     (6h)
   0.89920000000000    0.89920000000000    0.16260000000000    B     (6h)
   0.00250000000000    0.00250000000000    0.16800000000000    B     (6h)
  -0.00250000000000   -0.00250000000000    0.83200000000000    B     (6h)
   0.00250000000000    0.16800000000000    0.00250000000000    B     (6h)
  -0.00250000000000    0.83200000000000   -0.00250000000000    B     (6h)
   0.16800000000000    0.00250000000000    0.00250000000000    B     (6h)
   0.83200000000000   -0.00250000000000   -0.00250000000000    B     (6h)
   0.21310000000000    0.38430000000000    0.38430000000000    B     (6h)
   0.38430000000000    0.21310000000000    0.38430000000000    B     (6h)
   0.38430000000000    0.38430000000000    0.21310000000000    B     (6h)
   0.61570000000000    0.61570000000000    0.78690000000000    B     (6h)
   0.61570000000000    0.78690000000000    0.61570000000000    B     (6h)
   0.78690000000000    0.61570000000000    0.61570000000000    B     (6h)
   0.21780000000000    0.48950000000000    0.48950000000000    B     (6h)
   0.48950000000000    0.21780000000000    0.48950000000000    B     (6h)
   0.48950000000000    0.48950000000000    0.21780000000000    B     (6h)
   0.51050000000000    0.51050000000000    0.78220000000000    B     (6h)
   0.51050000000000    0.78220000000000    0.51050000000000    B     (6h)
   0.78220000000000    0.51050000000000    0.51050000000000    B     (6h)
   0.38730000000000    0.38730000000000    0.56900000000000    B     (6h)
   0.38730000000000    0.56900000000000    0.38730000000000    B     (6h)
   0.43100000000000    0.61270000000000    0.61270000000000    B     (6h)
   0.56900000000000    0.38730000000000    0.38730000000000    B     (6h)
   0.61270000000000    0.43100000000000    0.61270000000000    B     (6h)
   0.61270000000000    0.61270000000000    0.43100000000000    B     (6h)
   0.19910000000000    0.19910000000000    0.50610000000000    B     (6h)
   0.19910000000000    0.50610000000000    0.19910000000000    B     (6h)
   0.49390000000000    0.80090000000000    0.80090000000000    B     (6h)
   0.50610000000000    0.19910000000000    0.19910000000000    B     (6h)
   0.80090000000000    0.49390000000000    0.80090000000000    B     (6h)
   0.80090000000000    0.80090000000000    0.49390000000000    B     (6h)
   0.19830000000000    0.19830000000000    0.68740000000000    B     (6h)
   0.19830000000000    0.68740000000000    0.19830000000000    B     (6h)
   0.31260000000000    0.80170000000000    0.80170000000000    B     (6h)
   0.68740000000000    0.19830000000000    0.19830000000000    B     (6h)
   0.80170000000000    0.31260000000000    0.80170000000000    B     (6h)
   0.80170000000000    0.80170000000000    0.31260000000000    B     (6h)
   0.10320000000000    0.10320000000000    0.49210000000000    B     (6h)
   0.10320000000000    0.49210000000000    0.10320000000000    B     (6h)
   0.49210000000000    0.10320000000000    0.10320000000000    B     (6h)
   0.50790000000000    0.89680000000000    0.89680000000000    B     (6h)
   0.89680000000000    0.50790000000000    0.89680000000000    B     (6h)
   0.89680000000000    0.89680000000000    0.50790000000000    B     (6h)
   0.00670000000000    0.00670000000000    0.33020000000000    B     (6h)
  -0.00670000000000   -0.00670000000000    0.66980000000000    B     (6h)
   0.00670000000000    0.33020000000000    0.00670000000000    B     (6h)
  -0.00670000000000    0.66980000000000   -0.00670000000000    B     (6h)
   0.33020000000000    0.00670000000000    0.00670000000000    B     (6h)
   0.66980000000000   -0.00670000000000   -0.00670000000000    B     (6h)
```

$CaC_6$: A6B_hR7_166_g_a - CIF

```
# CIF file

data_findsym-output
_audit_creation_method FINDSYM

_chemical_name_mineral ''
_chemical_formula_sum 'Ca C6'

loop_
_publ_author_name
 'N. Emery'
 'C. H\'{e}rold'
 'M. d'Astuto'
 'V. Garcia'
 'Ch. Bellin'
 'J. F. Mar\^{e}ch\'{e}'
 'P. Lagrange,'
 'G. Loupias'
_journal_name_full
;
Physical Review Letters
;
_journal_volume 95
_journal_year 2005
_journal_page_first 087003
_journal_page_last 087003
_publ_Section_title
;
 Superconductivity of Bulk CaC6
;

_aflow_proto 'A6B_hR7_166_g_a'
_aflow_params 'a,c/a,x2'
_aflow_params_values '4.33304,3.13251204697,0.16667'
_aflow_Strukturbericht 'None'
_aflow_Pearson 'hR7'
```

```
_symmetry_space_group_name_Hall "-R 3 2"
_symmetry_space_group_name_H-M "R -3 m:H"
_symmetry_Int_Tables_number 166

_cell_length_a      4.33304
_cell_length_b      4.33304
_cell_length_c      13.57330
_cell_angle_alpha 90.00000
_cell_angle_beta  90.00000
_cell_angle_gamma 120.00000

loop_
_space_group_symop_id
_space_group_symop_operation_xyz
1 x,y,z
2 -y,x-y,z
3 -x+y,-x,z
4 y,x,-z
5 -x,-x+y,-z
6 x-y,-y,-z
7 -x,-y,-z
8 y,-x+y,-z
9 x-y,x,-z
10 -y,-x,z
11 x,x-y,z
12 -x+y,y,z
13 x+1/3,y+2/3,z+2/3
14 -y+1/3,x-y+2/3,z+2/3
15 -x+y+1/3,-x+2/3,z+2/3
16 y+1/3,x+2/3,-z+2/3
17 -x+1/3,-x+y+2/3,-z+2/3
18 x-y+1/3,-y+2/3,-z+2/3
19 -x+1/3,-y+2/3,-z+2/3
20 y+1/3,-x+y+2/3,-z+2/3
21 x-y+1/3,x+2/3,-z+2/3
22 -y+1/3,-x+2/3,z+2/3
23 x+1/3,x-y+2/3,z+2/3
24 -x+y+1/3,y+2/3,z+2/3
25 x+2/3,y+1/3,z+1/3
26 -y+2/3,x-y+1/3,z+1/3
27 -x+y+2/3,-x+1/3,z+1/3
28 y+2/3,x+1/3,-z+1/3
29 -x+2/3,-x+y+1/3,-z+1/3
30 x-y+2/3,-y+1/3,-z+1/3
31 -x+2/3,-y+1/3,-z+1/3
32 y+2/3,-x+y+1/3,-z+1/3
33 x-y+2/3,x+1/3,-z+1/3
34 -y+2/3,-x+1/3,z+1/3
35 x+2/3,x-y+1/3,z+1/3
36 -x+y+2/3,y+1/3,z+1/3

loop_
_atom_site_label
_atom_site_type_symbol
_atom_site_symmetry_multiplicity
_atom_site_Wyckoff_label
_atom_site_fract_x
_atom_site_fract_y
_atom_site_fract_z
_atom_site_occupancy
Ca1 Ca    3 a 0.00000  0.00000  0.00000  1.00000
C1  C    18 g 0.66667  0.00000  0.50000  1.00000
```

$CaC_6$: A6B_hR7_166_g_a - POSCAR

```
A6B_hR7_166_g_a & a,c/a,x2 --params=4.33304,3.13251204697,0.16667 & R(-3
    ↪ )m      D_{3d}^5 #166 (ag) & hR7 & & CaC6 & & Emery et al. PRB 95
    ↪ , 087003 (2005)
   1.0000000000000000
   2.16651926089496   -1.25084047848222    4.52443282517947
   0.00000000000000    2.50168095696443    4.52443282517947
  -2.16651926089496   -1.25084047848222    4.52443282517947
     C     Ca
     6      1
Direct
   0.16666666666667    0.50000000000000    0.83333333333333     C      (6g)
   0.16666666666667    0.83333333333333    0.50000000000000     C      (6g)
   0.50000000000000    0.16666666666667    0.83333333333333     C      (6g)
   0.50000000000000    0.83333333333333    0.16666666666667     C      (6g)
   0.83333333333333    0.16666666666667    0.50000000000000     C      (6g)
   0.83333333333333    0.50000000000000    0.16666666666667     C      (6g)
   0.00000000000000    0.00000000000000    0.00000000000000     Ca     (1a)
```

Paraelectric $LiNbO_3$: ABC3_hR10_167_a_b_e - CIF

```
# CIF file

data_findsym-output
_audit_creation_method FINDSYM

_chemical_name_mineral ''
_chemical_formula_sum 'Li Nb O3'

loop_
_publ_author_name
 'H. Boysen'
 'F. Altorfer'
_journal_name_full
;
Acta Crystallographica B
;
_journal_volume 50
_journal_year 1994
_journal_page_first 405
_journal_page_last 414
_publ_Section_title
;
 A neutron powder investigation of the high-temperature structure and
     ↪ phase transition in LiNbO$_3$
;

_aflow_proto 'ABC3_hR10_167_a_b_e'
_aflow_params 'a,c/a,x3'
_aflow_params_values '5.285,2.62039735099,0.85756666666667'
_aflow_Strukturbericht 'None'
_aflow_Pearson 'hR10'

_symmetry_space_group_name_Hall "-R 3 2c"
_symmetry_space_group_name_H-M "R -3 c:H"
_symmetry_Int_Tables_number 167

_cell_length_a      5.28500
_cell_length_b      5.28500
_cell_length_c      13.84880
_cell_angle_alpha 90.00000
_cell_angle_beta  90.00000
_cell_angle_gamma 120.00000

loop_
_space_group_symop_id
_space_group_symop_operation_xyz
1 x,y,z
2 -y,x-y,z
3 -x+y,-x,z
4 y,x,-z+1/2
5 -x,-x+y,-z+1/2
6 x-y,-y,-z+1/2
7 -x,-y,-z
8 y,-x+y,-z
9 x-y,x,-z
10 -y,-x,z+1/2
11 x,x-y,z+1/2
12 -x+y,y,z+1/2
13 x+1/3,y+2/3,z+2/3
14 -y+1/3,x-y+2/3,z+2/3
15 -x+y+1/3,-x+2/3,z+2/3
16 y+1/3,x+2/3,-z+1/6
17 -x+1/3,-x+y+2/3,-z+1/6
18 x-y+1/3,-y+2/3,-z+1/6
19 -x+1/3,-y+2/3,-z+2/3
20 y+1/3,-x+y+2/3,-z+2/3
21 x-y+1/3,x+2/3,-z+2/3
22 -y+1/3,-x+2/3,z+1/6
23 x+1/3,x-y+2/3,z+1/6
24 -x+y+1/3,y+2/3,z+1/6
25 x+2/3,y+1/3,z+1/3
26 -y+2/3,x-y+1/3,z+1/3
27 -x+y+2/3,-x+1/3,z+1/3
28 y+2/3,x+1/3,-z+5/6
29 -x+2/3,-x+y+1/3,-z+5/6
30 x-y+2/3,-y+1/3,-z+5/6
31 -x+2/3,-y+1/3,-z+1/3
32 y+2/3,-x+y+1/3,-z+1/3
33 x-y+2/3,x+1/3,-z+1/3
34 -y+2/3,-x+1/3,z+5/6
35 x+2/3,x-y+1/3,z+5/6
36 -x+y+2/3,y+1/3,z+5/6

loop_
_atom_site_label
_atom_site_type_symbol
_atom_site_symmetry_multiplicity
_atom_site_Wyckoff_label
_atom_site_fract_x
_atom_site_fract_y
_atom_site_fract_z
_atom_site_occupancy
Li1 Li    6 a 0.00000  0.00000  0.25000  1.00000
Nb1 Nb    6 b 0.00000  0.00000  0.00000  1.00000
O1  O    18 e 0.39243  0.00000  0.25000  1.00000
```

Paraelectric $LiNbO_3$: ABC3_hR10_167_a_b_e - POSCAR

```
ABC3_hR10_167_a_b_e & a,c/a,x3 --params=5.285,2.62039735099,
    ↪ 0.85756666666667 & R(-3)c     D_{3d}^6 #167 (abe) & hR10 & &
    ↪ LiNbO3 & Paraelectric & H. Boysen and F. Altorfer, Acta Cryst.
    ↪ B 50, 405-414 (1994)
   1.0000000000000000
   2.64250000000000   -1.52564808633359    4.61626666666072
   0.00000000000000    3.05129617266717    4.61626666666072
  -2.64250000000000   -1.52564808633359    4.61626666666072
     Li     Ni     O
     2      2      6
Direct
   0.25000000000000    0.25000000000000    0.25000000000000     Li     (2a)
   0.75000000000000    0.75000000000000    0.75000000000000     Li     (2a)
   0.00000000000000    0.00000000000000    0.00000000000000     Nb     (2b)
   0.50000000000000    0.50000000000000    0.50000000000000     Nb     (2b)
   0.14243333333333    0.35756666666667    0.75000000000000     O      (6e)
   0.25000000000000    0.85756666666667    0.64243333333333     O      (6e)
   0.35756666666667    0.75000000000000    0.14243333333333     O      (6e)
   0.64243333333333    0.25000000000000    0.85756666666667     O      (6e)
   0.75000000000000    0.14243333333333    0.35756666666667     O      (6e)
   0.85756666666667    0.64243333333333    0.25000000000000     O      (6e)
```

Calcite ($CaCO_3$, $G0_1$): ABC3_hR10_167_a_b_e - CIF

```
# CIF file
```

```
data_findsym-output
_audit_creation_method FINDSYM

_chemical_name_mineral 'Calcite'
_chemical_formula_sum 'Ca C O3'

loop_
_publ_author_name
 'S. A. Markgraf'
 'R. J. Reeder'
_journal_name_full
;
American Mineralogist
;
_journal_volume 70
_journal_year 1985
_journal_page_first 590
_journal_page_last 600
_publ_Section_title
;
 High-temperature structure refinements of calcite and magnesite
;

# Found in AMS Database

_aflow_proto 'ABC3_hR10_167_a_b_e'
_aflow_params 'a,c/a,x3'
_aflow_params_values '4.988,3.42040898156,0.5067'
_aflow_Strukturbericht 'G0_1'
_aflow_Pearson 'hR10'

_symmetry_space_group_name_Hall "-R 3 2c"
_symmetry_space_group_name_H-M "R -3 c:H"
_symmetry_Int_Tables_number 167

_cell_length_a     4.98800
_cell_length_b     4.98800
_cell_length_c     17.06100
_cell_angle_alpha  90.00000
_cell_angle_beta   90.00000
_cell_angle_gamma  120.00000

loop_
_space_group_symop_id
_space_group_symop_operation_xyz
1 x,y,z
2 -y,x-y,z
3 -x+y,-x,z
4 y,x,-z+1/2
5 -x,-x+y,-z+1/2
6 x-y,-y,-z+1/2
7 -x,-y,-z
8 y,-x+y,-z
9 x-y,x,-z
10 -y,-x,z+1/2
11 x,x-y,z+1/2
12 -x+y,y,z+1/2
13 x+1/3,y+2/3,z+2/3
14 -y+1/3,x-y+2/3,z+2/3
15 -x+y+1/3,-x+2/3,z+2/3
16 y+1/3,x+2/3,-z+1/6
17 -x+1/3,-x+y+2/3,-z+1/6
18 x-y+1/3,-y+2/3,-z+1/6
19 -x+1/3,-y+2/3,-z+2/3
20 y+1/3,-x+y+2/3,-z+2/3
21 x-y+1/3,x+2/3,-z+2/3
22 -y+1/3,-x+2/3,z+1/6
23 x+1/3,x-y+2/3,z+1/6
24 -x+y+1/3,y+2/3,z+1/6
25 x+2/3,y+1/3,z+1/3
26 -y+2/3,x-y+1/3,z+1/3
27 -x+y+2/3,-x+1/3,z+1/3
28 y+2/3,x+1/3,-z+5/6
29 -x+2/3,-x+y+1/3,-z+5/6
30 x-y+2/3,-y+1/3,-z+5/6
31 -x+2/3,-y+1/3,-z+1/3
32 y+2/3,-x+y+1/3,-z+1/3
33 x-y+2/3,x+1/3,-z+1/3
34 -y+2/3,-x+1/3,z+5/6
35 x+2/3,x-y+1/3,z+5/6
36 -x+y+2/3,y+1/3,z+5/6

loop_
_atom_site_label
_atom_site_type_symbol
_atom_site_symmetry_multiplicity
_atom_site_Wyckoff_label
_atom_site_fract_x
_atom_site_fract_y
_atom_site_fract_z
_atom_site_occupancy
C1  C    6 a 0.00000 0.00000 0.25000 1.00000
Ca1 Ca   6 b 0.00000 0.00000 0.00000 1.00000
O1  O   18 e 0.74330 0.00000 0.25000 1.00000
```

Calcite ($CaCO_3$, $G0_1$): ABC3_hR10_167_a_b_e - POSCAR

```
ABC3_hR10_167_a_b_e & a,c/a,x3 --params=4.988,3.42040898156,0.5067 & R(-
    ↪ 3)c    D^{6}_{3d} #167 (abe) & hR10 & G0_1 & CaCO3 & Calcite &
    ↪ S. A. Markgraf and R. J. Reeder, Am. Mineral. 70, 590-600 (1985
    ↪ )
   1.0000000000000000
   2.49400000000000  -1.43991157135900   5.68700000000000
   0.00000000000000   2.87982314271800   5.68700000000000
  -2.49400000000000  -1.43991157135900   5.68700000000000
     C    Ca    O
     2     2    6
Direct
    0.25000000000000   0.25000000000000   0.25000000000000   C   (2a)
    0.75000000000000   0.75000000000000   0.75000000000000   C   (2a)
    0.00000000000000   0.00000000000000   0.00000000000000   Ca  (2b)
    0.50000000000000   0.50000000000000   0.50000000000000   Ca  (2b)
   -0.00670000000000   0.25000000000000   0.50670000000000   O   (6e)
    0.00670000000000   0.75000000000000   0.49330000000000   O   (6e)
    0.25000000000000   0.50670000000000  -0.00670000000000   O   (6e)
    0.49330000000000   0.00670000000000   0.75000000000000   O   (6e)
    0.50670000000000  -0.00670000000000   0.25000000000000   O   (6e)
    0.75000000000000   0.49330000000000   0.00670000000000   O   (6e)
```

Corundum ($Al_2O_3$, $D5_1$): A2B3_hR10_167_c_e - CIF

```
# CIF file

data_findsym-output
_audit_creation_method FINDSYM

_chemical_name_mineral 'Corundum'
_chemical_formula_sum 'Al2 O3'

loop_
_publ_author_name
 'Larry W. Finger'
 'Robert M. Hazen'
_journal_name_full
;
Journal of Applied Physics
;
_journal_volume 49
_journal_year 1978
_journal_page_first 5823
_journal_page_last 5826
_publ_Section_title
;
 Crystal structure and compression of ruby to 46 kbar
;

_aflow_proto 'A2B3_hR10_167_c_e'
_aflow_params 'a,c/a,x1,x2'
_aflow_params_values '4.7607,2.72957758313,0.35216,0.5561'
_aflow_Strukturbericht 'D5_1'
_aflow_Pearson 'hR10'

_symmetry_space_group_name_Hall "-R 3 2c"
_symmetry_space_group_name_H-M "R -3 c:H"
_symmetry_Int_Tables_number 167

_cell_length_a     4.76070
_cell_length_b     4.76070
_cell_length_c     12.99470
_cell_angle_alpha  90.00000
_cell_angle_beta   90.00000
_cell_angle_gamma  120.00000

loop_
_space_group_symop_id
_space_group_symop_operation_xyz
1 x,y,z
2 -y,x-y,z
3 -x+y,-x,z
4 y,x,-z+1/2
5 -x,-x+y,-z+1/2
6 x-y,-y,-z+1/2
7 -x,-y,-z
8 y,-x+y,-z
9 x-y,x,-z
10 -y,-x,z+1/2
11 x,x-y,z+1/2
12 -x+y,y,z+1/2
13 x+1/3,y+2/3,z+2/3
14 -y+1/3,x-y+2/3,z+2/3
15 -x+y+1/3,-x+2/3,z+2/3
16 y+1/3,x+2/3,-z+1/6
17 -x+1/3,-x+y+2/3,-z+1/6
18 x-y+1/3,-y+2/3,-z+1/6
19 -x+1/3,-y+2/3,-z+2/3
20 y+1/3,-x+y+2/3,-z+2/3
21 x-y+1/3,x+2/3,-z+2/3
22 -y+1/3,-x+2/3,z+1/6
23 x+1/3,x-y+2/3,z+1/6
24 -x+y+1/3,y+2/3,z+1/6
25 x+2/3,y+1/3,z+1/3
26 -y+2/3,x-y+1/3,z+1/3
27 -x+y+2/3,-x+1/3,z+1/3
28 y+2/3,x+1/3,-z+5/6
29 -x+2/3,-x+y+1/3,-z+5/6
30 x-y+2/3,-y+1/3,-z+5/6
31 -x+2/3,-y+1/3,-z+1/3
32 y+2/3,-x+y+1/3,-z+1/3
33 x-y+2/3,x+1/3,-z+1/3
34 -y+2/3,-x+1/3,z+5/6
35 x+2/3,x-y+1/3,z+5/6
36 -x+y+2/3,y+1/3,z+5/6

loop_
_atom_site_label
_atom_site_type_symbol
_atom_site_symmetry_multiplicity
_atom_site_Wyckoff_label
_atom_site_fract_x
_atom_site_fract_y
```

```
_atom_site_fract_z
_atom_site_occupancy
Al1  Al   12 c 0.00000  0.00000  0.35216  1.00000
O1   O    18 e 0.30610  0.00000  0.25000  1.00000
```

Corundum ($Al_2O_3$, $D5_1$): A2B3_hR10_167_c_e - POSCAR

```
A2B3_hR10_167_c_e & a,c/a,x1,x2 --params=4.7607,2.72957758313,0.35216,
    ↪ 0.5561 & R(-3)c     D^{6}_{3d} #167 (ce) & hR10 & D5_1 & Al2O3 &
    ↪ Corundum & L. W. Finger and R. M. Hazen, J. Appl. Phys. 49,
    ↪ 5823-5826 (1976)
   1.0000000000000000
   2.38035000000000   -1.37429571326600    4.33156666666700
   0.00000000000000    2.74859142653100    4.33156666666700
  -2.38035000000000   -1.37429571326600    4.33156666666700
   Al       O
    4       6
Direct
   0.14784000000000    0.14784000000000    0.14784000000000    Al    (4c)
   0.35216000000000    0.35216000000000    0.35216000000000    Al    (4c)
   0.64784000000000    0.64784000000000    0.64784000000000    Al    (4c)
   0.85216000000000    0.85216000000000    0.85216000000000    Al    (4c)
  -0.05610000000000    0.25000000000000    0.55610000000000    O     (6e)
   0.05610000000000    0.75000000000000    0.44390000000000    O     (6e)
   0.25000000000000    0.55610000000000   -0.05610000000000    O     (6e)
   0.44390000000000    0.75000000000000    0.75000000000000    O     (6e)
   0.55610000000000   -0.05610000000000    0.25000000000000    O     (6e)
   0.75000000000000    0.44390000000000    0.05610000000000    O     (6e)
```

$Mg_2Ni$ ($C_a$): A2B_hP18_180_fi_bd - CIF

```
# CIF file

data_findsym-output
_audit_creation_method FINDSYM

_chemical_name_mineral ''
_chemical_formula_sum 'Mg2 Ni'

loop_
_publ_author_name
 'J. Schefer'
 'P. Fischer'
 'W. H\"{a}lg'
 'F. Stucki'
 'L. Schlapbach'
 'J. J. Didisheim'
 'K. Yvon'
 'A. F. Andresen'
_journal_name_full
;
Journal of the Less Common Metals
;
_journal_volume 74
_journal_year 1980
_journal_page_first 65
_journal_page_last 73
_publ_Section_title
;
 New structure results for hydrides and deuterides of the hydrogen
    ↪ storage material Mg$_2$Ni
;

# Found in  http://materials.springer.com/isp/crystallographic/docs/
    ↪ sd_0450086

_aflow_proto 'A2B_hP18_180_fi_bd'
_aflow_params 'a,c/a,z3,x4'
_aflow_params_values '5.198,2.54136206233,0.163,0.1141'
_aflow_Strukturbericht 'C_a'
_aflow_Pearson 'hP18'

_symmetry_space_group_name_Hall "P 62 2c (0 0 1)"
_symmetry_space_group_name_H-M "P 62 2 2"
_symmetry_Int_Tables_number 180

_cell_length_a       5.19800
_cell_length_b       5.19800
_cell_length_c       13.21000
_cell_angle_alpha 90.00000
_cell_angle_beta  90.00000
_cell_angle_gamma 120.00000

loop_
_space_group_symop_id
_space_group_symop_operation_xyz
1 x,y,z
2 x-y,x,z+1/3
3 -y,x-y,z+2/3
4 -x,-y,z
5 -x+y,-x,z+1/3
6 y,-x+y,z+2/3
7 x-y,-y,-z
8 x,x-y,-z+1/3
9 y,x,-z+2/3
10 -x+y,y,-z
11 -x,-x+y,-z+1/3
12 -y,-x,-z+2/3

loop_
_atom_site_label
_atom_site_type_symbol
_atom_site_symmetry_multiplicity
_atom_site_Wyckoff_label
_atom_site_fract_x
_atom_site_fract_y
_atom_site_fract_z
_atom_site_occupancy
Ni1  Ni    3 b 0.00000  0.00000  0.50000  1.00000
Ni2  Ni    3 d 0.50000  0.00000  0.50000  1.00000
Mg1  Mg    6 f 0.50000  0.00000  0.16300  1.00000
Mg2  Mg    6 i 0.11410  0.22820  0.00000  1.00000
```

$Mg_2Ni$ ($C_a$): A2B_hP18_180_fi_bd - POSCAR

```
A2B_hP18_180_fi_bd & a,c/a,z3,x4 --params=5.198,2.54136206233,0.163,
    ↪ 0.1141 & P6_222      D_6^4 #180 (bdfi) & hP18 & C_a & Mg2Ni & &
    ↪ J. Schefer et al., J. Less-Common Met. 74, 65-73 (1980)
   1.0000000000000000
   2.59900000000000   -4.50160004887200    0.00000000000000
   2.59900000000000    4.50160004887200    0.00000000000000
   0.00000000000000    0.00000000000000   13.21000000000000
   Mg      Ni
   12       6
Direct
   0.00000000000000    0.50000000000000    0.50366666666667    Mg    (6f)
   0.00000000000000    0.50000000000000    0.82966666666667    Mg    (6f)
   0.50000000000000    0.00000000000000    0.16300000000000    Mg    (6f)
   0.50000000000000    0.00000000000000    0.83700000000000    Mg    (6f)
   0.50000000000000    0.50000000000000    0.17033333333333    Mg    (6f)
   0.50000000000000    0.50000000000000    0.49633333333333    Mg    (6f)
   0.11410000000000    0.22820000000000    0.00000000000000    Mg    (6i)
   0.11410000000000    0.88590000000000    0.33333333333333    Mg    (6i)
   0.22820000000000    0.11410000000000    0.66666666666667    Mg    (6i)
   0.77180000000000    0.88590000000000    0.66666666666667    Mg    (6i)
   0.88590000000000    0.11410000000000    0.33333333333333    Mg    (6i)
   0.88590000000000    0.77180000000000    0.00000000000000    Mg    (6i)
   0.00000000000000    0.00000000000000    0.16666666666667    Ni    (3b)
   0.00000000000000    0.00000000000000    0.50000000000000    Ni    (3b)
   0.00000000000000    0.00000000000000    0.83333333333333    Ni    (3b)
   0.00000000000000    0.50000000000000    0.16666666666667    Ni    (3d)
   0.50000000000000    0.00000000000000    0.50000000000000    Ni    (3d)
   0.50000000000000    0.50000000000000    0.83333333333333    Ni    (3d)
```

$CrSi_2$ (C40): AB2_hP9_180_d_j - CIF

```
# CIF file

data_findsym-output
_audit_creation_method FINDSYM

_chemical_name_mineral ''
_chemical_formula_sum 'Cr Si2'

loop_
_publ_author_name
 'T. Dasgupta'
 'J. Etourneau'
 'B. Chevalier'
 'S. F. Matar'
 'A. M. Umarji'
_journal_name_full
;
Journal of Applied Physics
;
_journal_volume 103
_journal_year 2008
_journal_page_first 113516
_journal_page_last 113516
_publ_Section_title
;
 Structural, thermal, and electrical properties of CrSi$_2$
;

_aflow_proto 'AB2_hP9_180_d_j'
_aflow_params 'a,c/a,x2'
_aflow_params_values '4.42758,1.43826876081,0.16559'
_aflow_Strukturbericht 'C40'
_aflow_Pearson 'hP9'

_symmetry_space_group_name_Hall "P 62 2c (0 0 1)"
_symmetry_space_group_name_H-M "P 62 2 2"
_symmetry_Int_Tables_number 180

_cell_length_a       4.42758
_cell_length_b       4.42758
_cell_length_c       6.36805
_cell_angle_alpha 90.00000
_cell_angle_beta  90.00000
_cell_angle_gamma 120.00000

loop_
_space_group_symop_id
_space_group_symop_operation_xyz
1 x,y,z
2 x-y,x,z+1/3
3 -y,x-y,z+2/3
4 -x,-y,z
5 -x+y,-x,z+1/3
6 y,-x+y,z+2/3
7 x-y,-y,-z
8 x,x-y,-z+1/3
9 y,x,-z+2/3
10 -x+y,y,-z
11 -x,-x+y,-z+1/3
12 -y,-x,-z+2/3

loop_
_atom_site_label
_atom_site_type_symbol
```

```
_atom_site_symmetry_multiplicity
_atom_site_Wyckoff_label
_atom_site_fract_x
_atom_site_fract_y
_atom_site_fract_z
_atom_site_occupancy
Cr1 Cr    3 d 0.50000 0.00000 0.50000 1.00000
Si1 Si    6 j 0.16559 0.33118 0.50000 1.00000
```

$CrSi_2$ (C40): AB2_hP9_180_d_j - POSCAR

```
AB2_hP9_180_d_j & a,c/a,x2 --params=4.42758,1.43826876081,0.16559 &
    ↪ P6_222    D_6^4 #180 (dj) & hP9 & C40 & CrSi2 & & T. Dasgupta,
    ↪ J. Etourneau, B. Chevalier, S. F. Matar and A. M. Umarji, J.
    ↪ App. Phys. 103, 113516 (2008)
   1.0000000000000000
   2.21379000000000  -3.83439675728800   0.00000000000000
   2.21379000000000   3.83439675728800   0.00000000000000
   0.00000000000000   0.00000000000000   6.36805000000000
   Cr    Si
    3     6
Direct
   0.00000000000000   0.50000000000000   0.16666666666667   Cr   (3d)
   0.50000000000000   0.00000000000000   0.50000000000000   Cr   (3d)
   0.50000000000000   0.50000000000000   0.83333333333333   Cr   (3d)
   0.16559100000000   0.33118200000000   0.50000000000000   Si   (6j)
   0.16559100000000   0.83440900000000   0.83333333333333   Si   (6j)
   0.33118200000000   0.16559100000000   0.16666666666667   Si   (6j)
   0.66881800000000   0.83440900000000   0.16666666666667   Si   (6j)
   0.83440900000000   0.16559100000000   0.83333333333333   Si   (6j)
   0.83440900000000   0.66881800000000   0.50000000000000   Si   (6j)
```

$\beta$-Quartz ($SiO_2$, C8): A2B_hP9_180_j_c - CIF

```
# CIF file

data_findsym-output
_audit_creation_method FINDSYM

_chemical_name_mineral 'quartz (beta)'
_chemical_formula_sum 'Si O2'

loop_
_publ_author_name
 'A. F. Wright'
 'M. S. Lehmann'
_journal_name_full
;
Journal of Solid State Chemistry
;
_journal_volume 36
_journal_year 1981
_journal_page_first 371
_journal_page_last 380
_publ_Section_title
;
 The Structure of Quartz at 25 and 590$^\circ$C Determined by Neutron
    ↪ Diffraction
;

# Found in http://www.minweb.co.uk/quartz/betaquartz.html

_aflow_proto 'A2B_hP9_180_j_c'
_aflow_params 'a,c/a,x2'
_aflow_params_values '4.9977,1.09252256038,0.2072'
_aflow_Strukturbericht 'C8'
_aflow_Pearson 'hP9'

_symmetry_space_group_name_Hall "P 62 2c (0 0 1)"
_symmetry_space_group_name_H-M "P 62 2 2"
_symmetry_Int_Tables_number 180

_cell_length_a     4.99770
_cell_length_b     4.99770
_cell_length_c     5.46010
_cell_angle_alpha  90.00000
_cell_angle_beta   90.00000
_cell_angle_gamma  120.00000

loop_
_space_group_symop_id
_space_group_symop_operation_xyz
1 x,y,z
2 x-y,x,z+1/3
3 -y,x-y,z+2/3
4 -x,-y,z
5 -x+y,-x,z+1/3
6 y,-x+y,z+2/3
7 x-y,-y,-z
8 x,x-y,-z+1/3
9 y,x,-z+2/3
10 -x+y,y,-z
11 -x,-x+y,-z+1/3
12 -y,-x,-z+2/3

loop_
_atom_site_label
_atom_site_type_symbol
_atom_site_symmetry_multiplicity
_atom_site_Wyckoff_label
_atom_site_fract_x
_atom_site_fract_y
_atom_site_fract_z
_atom_site_occupancy
Si1 Si    3 c 0.50000 0.00000 0.00000 1.00000
O1  O     6 j 0.20720 0.41440 0.50000 1.00000
```

$\beta$-Quartz ($SiO_2$, C8): A2B_hP9_180_j_c - POSCAR

```
A2B_hP9_180_j_c & a,c/a,x2 --params=4.9977,1.09252256038,0.2072 & P6_222
    ↪     D_6^4 #180 (cj) & hP9 & C8 & SiO2 & beta-quartz & A. F.
    ↪ Wright and M. S. Lehmann, J. Solid State. Chem. 36, 371-380 (
    ↪ 1981)
   1.0000000000000000
   2.49885000000000  -4.32813516049349   0.00000000000000
   2.49885000000000   4.32813516049349   0.00000000000000
   0.00000000000000   0.00000000000000   5.46010000000000
   O     Si
   6     3
Direct
   0.20720000000000   0.41440000000000   0.50000000000000   O    (6j)
   0.20720000000000   0.79280000000000   0.83333333333333   O    (6j)
   0.41440000000000   0.20720000000000   0.16666666666667   O    (6j)
   0.58560000000000   0.79280000000000   0.16666666666667   O    (6j)
   0.79280000000000   0.20720000000000   0.83333333333333   O    (6j)
   0.79280000000000   0.58560000000000   0.50000000000000   O    (6j)
   0.00000000000000   0.50000000000000   0.66666666666667   Si   (3c)
   0.50000000000000   0.00000000000000   0.00000000000000   Si   (3c)
   0.50000000000000   0.50000000000000   0.33333333333333   Si   (3c)
```

Bainite ($Fe_3C$): AB3_hP8_182_c_g - CIF

```
# CIF file

data_findsym-output
_audit_creation_method FINDSYM

_chemical_name_mineral 'Upper Bainite'
_chemical_formula_sum 'Fe3 C'

loop_
_publ_author_name
 'Marianne Reibold'
 'Alexander A. Levin'
 'Dirk C. Meyer'
 'Peter Paufler'
 'Werner Kochmann'
_journal_name_full
;
International Journal of Materials Research
;
_journal_volume 97
_journal_year 2006
_journal_page_first 1172
_journal_page_last 1182
_publ_Section_title
;
 Microstructure of a Damascene sabre after annealing
;

# Found in http://materials.springer.com/isp/crystallographic/docs/
    ↪ sd_1817306

_aflow_proto 'AB3_hP8_182_c_g'
_aflow_params 'a,c/a,x2'
_aflow_params_values '4.8507,0.866967654153,0.3249'
_aflow_Strukturbericht 'None'
_aflow_Pearson 'hP8'

_symmetry_space_group_name_Hall "P 6c 2c"
_symmetry_space_group_name_H-M "P 63 2 2"
_symmetry_Int_Tables_number 182

_cell_length_a     4.85070
_cell_length_b     4.85070
_cell_length_c     4.20540
_cell_angle_alpha  90.00000
_cell_angle_beta   90.00000
_cell_angle_gamma  120.00000

loop_
_space_group_symop_id
_space_group_symop_operation_xyz
1 x,y,z
2 x-y,x,z+1/2
3 -y,x-y,z
4 -x,-y,z+1/2
5 -x+y,-x,z
6 y,-x+y,z+1/2
7 x-y,-y,-z
8 x,x-y,-z+1/2
9 y,x,-z
10 -x+y,y,-z+1/2
11 -x,-x+y,-z
12 -y,-x,-z+1/2

loop_
_atom_site_label
_atom_site_type_symbol
_atom_site_symmetry_multiplicity
_atom_site_Wyckoff_label
_atom_site_fract_x
_atom_site_fract_y
_atom_site_fract_z
_atom_site_occupancy
C1  C     2 c 0.33333 0.66667 0.25000 1.00000
Fe1 Fe    6 g 0.32490 0.00000 0.00000 1.00000
```

Bainite ($Fe_3C$): AB3_hP8_182_c_g - POSCAR

```
AB3_hP8_182_c_g & a,c/a,x2 --params=4.8507,0.866967654153,0.3249 &
    ↪ P6_322    D_6^3 #182 (cg) & hP8 & & Fe3C & Upper Bainite & M.
    ↪ Reibold et al., Int. J. Mater. Res. 97, 1172-1182 (2006)
   1.0000000000000000
   2.42535000000000  -4.20082942613700   0.00000000000000
   2.42535000000000   4.20082942613700   0.00000000000000
   0.00000000000000   0.00000000000000   4.20540000000000
    C    Fe
    2     6
Direct
   0.33333333333333   0.66666666666667   0.25000000000000    C    (2c)
   0.66666666666667   0.33333333333333   0.75000000000000    C    (2c)
   0.00000000000000   0.32490000000000   0.00000000000000    Fe   (6g)
  -0.00000000000000   0.67510000000000   0.50000000000000    Fe   (6g)
   0.32490000000000   0.00000000000000   0.00000000000000    Fe   (6g)
   0.32490000000000   0.32490000000000   0.50000000000000    Fe   (6g)
   0.67510000000000  -0.00000000000000   0.50000000000000    Fe   (6g)
   0.67510000000000   0.67510000000000   0.00000000000000    Fe   (6g)
```

Buckled Graphite: A_hP4_186_ab - CIF

```
# CIF file

data_findsym-output
_audit_creation_method FINDSYM

_chemical_name_mineral 'graphite'
_chemical_formula_sum 'C'

loop_
_publ_author_name
 'A. W. Hull'
_journal_name_full
;
Physical Review
;
_journal_volume 10
_journal_year 1917
_journal_page_first 661
_journal_page_last 696
_publ_Section_title
;
 A New Method of X-Ray Crystal Analysis
;

# Found in  Wyckoff, Vol. I, pp. 254

_aflow_proto 'A_hP4_186_ab'
_aflow_params 'a,c/a,z1,z2'
_aflow_params_values '2.47,2.75303643725,0.0,0.07143'
_aflow_Strukturbericht 'None'
_aflow_Pearson 'hP4'

_symmetry_space_group_name_Hall "P 6c -2c"
_symmetry_space_group_name_H-M "P 63 m c"
_symmetry_Int_Tables_number 186

_cell_length_a     2.47000
_cell_length_b     2.47000
_cell_length_c     6.80000
_cell_angle_alpha  90.00000
_cell_angle_beta   90.00000
_cell_angle_gamma  120.00000

loop_
_space_group_symop_id
_space_group_symop_operation_xyz
1 x,y,z
2 x-y,x,z+1/2
3 -y,x-y,z
4 -x,-y,z+1/2
5 -x+y,-x,z
6 y,-x+y,z+1/2
7 -x+y,y,z
8 -x,-x+y,z+1/2
9 -y,-x,z
10 x-y,-y,z+1/2
11 x,x-y,z
12 y,x,z+1/2

loop_
_atom_site_label
_atom_site_type_symbol
_atom_site_symmetry_multiplicity
_atom_site_Wyckoff_label
_atom_site_fract_x
_atom_site_fract_y
_atom_site_fract_z
_atom_site_occupancy
C1 C   2 a 0.00000 0.00000 0.00000 1.00000
C2 C   2 b 0.33333 0.66667 0.07143 1.00000
```

Buckled Graphite: A_hP4_186_ab - POSCAR

```
A_hP4_186_ab & a,c/a,z1,z2 --params=2.47,2.75303643725,0.0,0.07143 &
    ↪ P6_3mc    C_{6v}^4 #186 (ab) & hP4 & & C & Graphite (buckled) &
    ↪ A. W. Hull, Phys. Rev. 10, 661-696 (1917)
   1.0000000000000000
   1.23500000000000  -2.13908274734800   0.00000000000000
   1.23500000000000   2.13908274734800   0.00000000000000
   0.00000000000000   0.00000000000000   6.80000000000000
    C
    4
Direct
   0.00000000000000   0.00000000000000   0.00000000000000    C    (2a)
   0.00000000000000   0.00000000000000   0.50000000000000    C    (2a)
   0.33333333333333   0.66666666666667   0.07142857142857    C    (2b)
   0.66666666666667   0.33333333333333   0.57142857142857    C    (2b)
```

Moissanite-4H SiC (B5): AB_hP8_186_ab_ab - CIF

```
# CIF file

data_findsym-output
_audit_creation_method FINDSYM

_chemical_name_mineral 'Moissanite-4H'
_chemical_formula_sum 'Si C'

loop_
_publ_author_name
 'A. Bauer'
 'P. Reischauer'
 'J. Kr\"{a}usslich'
 'N. Schell'
 'W. Matz'
 'K. Goetz'
_journal_name_full
;
Acta Crystallographica A
;
_journal_volume 57
_journal_year 2001
_journal_page_first 60
_journal_page_last 67
_publ_Section_title
;
 Structure refinement of the silicon carbide polytypes 4H and 6H:
    ↪ unambiguous determination of the refinement parameters
;

_aflow_proto 'AB_hP8_186_ab_ab'
_aflow_params 'a,c/a,z1,z2,z3,z4'
_aflow_params_values '3.08051,3.27374363336,0.18784,0.0,0.43671,0.24982'
_aflow_Strukturbericht 'B5'
_aflow_Pearson 'hP8'

_symmetry_space_group_name_Hall "P 6c -2c"
_symmetry_space_group_name_H-M "P 63 m c"
_symmetry_Int_Tables_number 186

_cell_length_a     3.08051
_cell_length_b     3.08051
_cell_length_c     10.08480
_cell_angle_alpha  90.00000
_cell_angle_beta   90.00000
_cell_angle_gamma  120.00000

loop_
_space_group_symop_id
_space_group_symop_operation_xyz
1 x,y,z
2 x-y,x,z+1/2
3 -y,x-y,z
4 -x,-y,z+1/2
5 -x+y,-x,z
6 y,-x+y,z+1/2
7 -x+y,y,z
8 -x,-x+y,z+1/2
9 -y,-x,z
10 x-y,-y,z+1/2
11 x,x-y,z
12 y,x,z+1/2

loop_
_atom_site_label
_atom_site_type_symbol
_atom_site_symmetry_multiplicity
_atom_site_Wyckoff_label
_atom_site_fract_x
_atom_site_fract_y
_atom_site_fract_z
_atom_site_occupancy
C1  C    2 a 0.00000 0.00000 0.18784 1.00000
Si1 Si   2 a 0.00000 0.00000 0.00000 1.00000
C2  C    2 b 0.33333 0.66667 0.43671 1.00000
Si2 Si   2 b 0.33333 0.66667 0.24982 1.00000
```

Moissanite-4H SiC (B5): AB_hP8_186_ab_ab - POSCAR

```
AB_hP8_186_ab_ab & a,c/a,z1,z2,z3,z4 --params=3.08051,3.27374363336,
    ↪ 0.18784,0.0,0.43671,0.24982 & P6_3mc    C_{6v}^4 #186 (a^2b^2)
    ↪ & hP8 & B5 & SiC & Moissanite-4H & A. Bauer, P. Reischauer, J.
    ↪ Kr\"{a}usslich, N. Schell, W. Matz and K. Goetz, Acta Cryst. A
    ↪ 57, 60-67 (2001)
   1.0000000000000000
   1.54025500000000  -2.66779991661200   0.00000000000000
   1.54025500000000   2.66779991661200   0.00000000000000
   0.00000000000000   0.00000000000000  10.08480000000000
    C    Si
    4     4
Direct
   0.00000000000000   0.00000000000000   0.18783750000000    C    (2a)
   0.00000000000000   0.00000000000000   0.68783750000000    C    (2a)
   0.33333333333333   0.66666666666667   0.43671250000000    C    (2b)
   0.66666666666667   0.33333333333333  -0.06328750000000    C    (2b)
   0.00000000000000   0.00000000000000   0.00000000000000    Si   (2a)
   0.00000000000000   0.00000000000000   0.50000000000000    Si   (2a)
   0.33333333333333   0.66666666666667   0.24982500000000    Si   (2b)
```

```
    0.66666666666667    0.33333333333333    0.74982500000000    Si    (2b)
```

Wurtzite (ZnS, B4): AB_hP4_186_b_b - CIF

```
# CIF file

data_findsym-output
_audit_creation_method FINDSYM

_chemical_name_mineral 'Wurtzite'
_chemical_formula_sum 'Zn S'

loop_
_publ_author_name
 'Erich H. Kisi'
 'Margaret M. Elcombe'
_journal_name_full
;
Acta Crystallographica C
;
_journal_volume 45
_journal_year 1989
_journal_page_first 1867
_journal_page_last 1870
_publ_Section_title
;
 $u$ parameters for the wurtzite structure of ZnS and ZnO using powder
      ↪ neutron diffraction
;

# Found in AMS Database

_aflow_proto 'AB_hP4_186_b_b'
_aflow_params 'a,c/a,z1,z2'
_aflow_params_values '3.8227,1.63776911607,0.3748,0.0'
_aflow_Strukturbericht 'B4'
_aflow_Pearson 'hP4'

_symmetry_space_group_name_Hall "P 6c -2c"
_symmetry_space_group_name_H-M "P 63 m c"
_symmetry_Int_Tables_number 186

_cell_length_a      3.82270
_cell_length_b      3.82270
_cell_length_c      6.26070
_cell_angle_alpha  90.00000
_cell_angle_beta   90.00000
_cell_angle_gamma  120.00000

loop_
_space_group_symop_id
_space_group_symop_operation_xyz
1 x,y,z
2 x-y,x,z+1/2
3 -y,x-y,z
4 -x,-y,z+1/2
5 -x+y,-x,z
6 y,-x+y,z+1/2
7 -x+y,y,z
8 -x,-x+y,z+1/2
9 -y,-x,z
10 x-y,-y,z+1/2
11 x,x-y,z
12 y,x,z+1/2

loop_
_atom_site_label
_atom_site_type_symbol
_atom_site_symmetry_multiplicity
_atom_site_Wyckoff_label
_atom_site_fract_x
_atom_site_fract_y
_atom_site_fract_z
_atom_site_occupancy
S1  S   2 b 0.33333 0.66667 0.37480 1.00000
Zn1 Zn  2 b 0.33333 0.66667 0.00000 1.00000
```

Wurtzite (ZnS, B4): AB_hP4_186_b_b - POSCAR

```
AB_hP4_186_b_b & a,c/a,z1,z2 --params=3.8227,1.63776911607,0.3748,0.0 &
      ↪ P6_3mc    C_{6v}^4 #186 (b^2) & hP4 & B4 & ZnS & Wurtzite & E.
      ↪ H. Kisi and M. M. Elcombe, Acta Cryst. C 45, 1867-870 (1989)
   1.0000000000000000
   1.91135000000000   -3.31055531104700    0.00000000000000
   1.91135000000000    3.31055531104700    0.00000000000000
   0.00000000000000    0.00000000000000    6.26070000000000
    S    Zn
    2    2
Direct
   0.33333333333333    0.66666666666667    0.37480000000000    S    (2b)
   0.66666666666667    0.33333333333333    0.87480000000000    S    (2b)
   0.33333333333333    0.66666666666667    0.00000000000000    Zn   (2b)
   0.66666666666667    0.33333333333333    0.50000000000000    Zn   (2b)
```

Moissanite-6H SiC (B6): AB_hP12_186_a2b_a2b - CIF

```
# CIF file

data_findsym-output
_audit_creation_method FINDSYM

_chemical_name_mineral 'Moissanite-6H'
_chemical_formula_sum 'Si C'

loop_
_publ_author_name
 'A. Bauer'
 'P. Reischauer'
 'J. Kr\"{a}usslich'
 'N. Schell'
 'W. Matz'
 'K. Goetz'
_journal_name_full
;
Acta Crystallographica A
;
_journal_volume 57
_journal_year 2001
_journal_page_first 60
_journal_page_last 67
_publ_Section_title
;
 Structure refinement of the silicon carbide polytypes 4H and 6H:
      ↪ unambiguous determination of the refinement parameters
;

_aflow_proto 'AB_hP12_186_a2b_a2b'
_aflow_params 'a,c/a,z1,z2,z3,z4,z5,z6'
_aflow_params_values '3.08129,4.90695780014,0.1254,0.0,0.29215,-0.0415,
      ↪ 0.16675,0.8335'
_aflow_Strukturbericht 'B6'
_aflow_Pearson 'hP12'

_symmetry_space_group_name_Hall "P 6c -2c"
_symmetry_space_group_name_H-M "P 63 m c"
_symmetry_Int_Tables_number 186

_cell_length_a      3.08129
_cell_length_b      3.08129
_cell_length_c      15.11976
_cell_angle_alpha  90.00000
_cell_angle_beta   90.00000
_cell_angle_gamma  120.00000

loop_
_space_group_symop_id
_space_group_symop_operation_xyz
1 x,y,z
2 x-y,x,z+1/2
3 -y,x-y,z
4 -x,-y,z+1/2
5 -x+y,-x,z
6 y,-x+y,z+1/2
7 -x+y,y,z
8 -x,-x+y,z+1/2
9 -y,-x,z
10 x-y,-y,z+1/2
11 x,x-y,z
12 y,x,z+1/2

loop_
_atom_site_label
_atom_site_type_symbol
_atom_site_symmetry_multiplicity
_atom_site_Wyckoff_label
_atom_site_fract_x
_atom_site_fract_y
_atom_site_fract_z
_atom_site_occupancy
C1  C   2 a 0.00000 0.00000 0.12540 1.00000
Si1 Si  2 a 0.00000 0.00000 0.00000 1.00000
C2  C   2 b 0.33333 0.66667 0.29215 1.00000
C3  C   2 b 0.33333 0.66667 -0.04150 1.00000
Si2 Si  2 b 0.33333 0.66667 0.16675 1.00000
Si3 Si  2 b 0.33333 0.66667 0.83350 1.00000
```

Moissanite-6H SiC (B6): AB_hP12_186_a2b_a2b - POSCAR

```
AB_hP12_186_a2b_a2b & a,c/a,z1,z2,z3,z4,z5,z6 --params=3.08129,
      ↪ 4.90695780014,0.1254,0.0,0.29215,-0.0415,0.16675,0.8335 &
      ↪ P6_3mc    C_{6v}^4 #186 (a^2b^4) & hP12 & B6 & SiC &
      ↪ Moissanite-6H & A. Bauer, P. Reischauer, J. Kr\"{a}usslich, N.
      ↪ Schell, W. Matz and K. Goetz, Acta Cryst. A 57, 60-67 (2001)
   1.0000000000000000
   1.54064500000000   -2.66847541642695    0.00000000000000
   1.54064500000000    2.66847541642695    0.00000000000000
   0.00000000000000    0.00000000000000   15.11976000000000
    C    Si
    6    6
Direct
   0.00000000000000    0.00000000000000    0.12540000000000    C    (2a)
   0.00000000000000    0.00000000000000    0.62540000000000    C    (2a)
   0.33333333333333    0.66666666666667    0.29214666666667    C    (2b)
   0.66666666666667    0.33333333333333    0.79214666666667    C    (2b)
   0.33333333333333    0.66666666666667   -0.04149666666667    C    (2b)
   0.66666666666667    0.33333333333333    0.45850333333333    C    (2b)
   0.00000000000000    0.00000000000000    0.00000000000000    Si   (2a)
   0.00000000000000    0.00000000000000    0.50000000000000    Si   (2a)
   0.33333333333333    0.66666666666667    0.16674666666667    Si   (2b)
   0.66666666666667    0.33333333333333    0.66674666666667    Si   (2b)
   0.33333333333333    0.66666666666667    0.83350333333333    Si   (2b)
   0.66666666666667    0.33333333333333    0.33350333333333    Si   (2b)
```

$Al_5C_3N$ ($E9_4$): A5B3C_hP18_186_2a3b_2ab_b - CIF

```
# CIF file

data_findsym-output
_audit_creation_method FINDSYM
```

```
_chemical_name_mineral 'Aluminum carbonitride'
_chemical_formula_sum 'Al5 C3 N'

loop_
_publ_author_name
 'G. A. Jeffrey'
 'Victor Y. Wu'
_journal_name_full
;
Acta Crystallographica
;
_journal_volume 20
_journal_year 1966
_journal_page_first 538
_journal_page_last 547
_publ_Section_title
;
 The structure of the aluminum carbonitrides. II
;

# Found in http://materials.springer.com/lb/docs/
    ↪ sm_lbs_978-3-540-44820-4_208

_aflow_proto 'A5B3C_hP18_186_2a3b_2ab_b'
_aflow_params 'a,c/a,z1,z2,z3,z4,z5,z6,z7,z8,z9'
_aflow_params_values '3.281,6.57726302956,0.155,0.345,0.0,0.248,0.045,
    ↪ 0.261,0.455,0.367,0.137'
_aflow_Strukturbericht 'E9_4'
_aflow_Pearson 'hP18'

_symmetry_space_group_name_Hall "P 6c -2c"
_symmetry_space_group_name_H-M "P 63 m c"
_symmetry_Int_Tables_number 186

_cell_length_a      3.28100
_cell_length_b      3.28100
_cell_length_c      21.58000
_cell_angle_alpha 90.00000
_cell_angle_beta  90.00000
_cell_angle_gamma 120.00000

loop_
_space_group_symop_id
_space_group_symop_operation_xyz
1 x,y,z
2 x-y,x,z+1/2
3 -y,x-y,z
4 -x,-y,z+1/2
5 -x+y,-x,z
6 y,-x+y,z+1/2
7 -x+y,y,z
8 -x,-x+y,z+1/2
9 -y,-x,z
10 x-y,-y,z+1/2
11 x,x-y,z
12 y,x,z+1/2

loop_
_atom_site_label
_atom_site_type_symbol
_atom_site_symmetry_multiplicity
_atom_site_Wyckoff_label
_atom_site_fract_x
_atom_site_fract_y
_atom_site_fract_z
_atom_site_occupancy
Al1 Al   2 a 0.00000 0.00000 0.15500 1.00000
Al2 Al   2 a 0.00000 0.00000 0.34500 1.00000
C1  C    2 a 0.00000 0.00000 0.00000 1.00000
C2  C    2 a 0.00000 0.00000 0.24800 1.00000
Al3 Al   2 b 0.33333 0.66667 0.04500 1.00000
Al4 Al   2 b 0.33333 0.66667 0.26100 1.00000
Al5 Al   2 b 0.33333 0.66667 0.45500 1.00000
C3  C    2 b 0.33333 0.66667 0.36700 1.00000
N1  N    2 b 0.33333 0.66667 0.13700 1.00000
```

$Al_5C_3N$ ($E9_4$): A5B3C_hP18_186_2a3b_2ab_b - POSCAR

```
A5B3C_hP18_186_2a3b_2ab_b & a,c/a,z1,z2,z3,z4,z5,z6,z7,z8,z9 --params=
    ↪ 3.281,6.57726302956,0.155,0.345,0.0,0.248,0.045,0.261,0.455,
    ↪ 0.367,0.137 & P6_3mc    C_{6v}^4 #186 (a^4b^5) & hP18 & E9_4 &
    ↪ Al5C3N & & G. A. Jeffrey and V. Y. Wu, Acta Cryst. 20, 538-547
    ↪ (1966)
   1.0000000000000000
   1.64050000000000  -2.84142934981674   0.00000000000000
   1.64050000000000   2.84142934981674   0.00000000000000
   0.00000000000000   0.00000000000000  21.58000000000000
   Al     C     N
   10     6     2
Direct
   0.00000000000000   0.00000000000000   0.15500000000000   Al   (2a)
   0.00000000000000   0.00000000000000   0.65500000000000   Al   (2a)
   0.00000000000000   0.00000000000000   0.34500000000000   Al   (2a)
   0.00000000000000   0.00000000000000   0.84500000000000   Al   (2a)
   0.33333333333333   0.66666666666667   0.04500000000000   Al   (2b)
   0.66666666666667   0.33333333333333   0.54500000000000   Al   (2b)
   0.33333333333333   0.66666666666667   0.26100000000000   Al   (2b)
   0.66666666666667   0.33333333333333   0.76100000000000   Al   (2b)
   0.33333333333333   0.66666666666667   0.45500000000000   Al   (2b)
   0.66666666666667   0.33333333333333  -0.04500000000000   Al   (2b)
   0.00000000000000   0.00000000000000   0.00000000000000   C    (2a)
   0.00000000000000   0.00000000000000   0.50000000000000   C    (2a)
   0.00000000000000   0.00000000000000   0.24800000000000   C    (2a)
   0.00000000000000   0.00000000000000   0.74800000000000   C    (2a)
   0.33333333333333   0.66666666666667   0.36700000000000   C    (2b)
   0.66666666666667   0.33333333333333   0.86700000000000   C    (2b)
   0.33333333333333   0.66666666666667   0.13700000000000   N    (2b)
   0.66666666666667   0.33333333333333   0.63700000000000   N    (2b)
```

Original BN (B12): AB_hP4_186_b_a - CIF

```
# CIF file

data_findsym-output
_audit_creation_method FINDSYM

_chemical_name_mineral ''
_chemical_formula_sum 'B N'

loop_
_publ_author_name
 'A. Brager'
_journal_name_full
;
Acta Physicochimica URSS
;
_journal_volume 7
_journal_year 1937
_journal_page_first 699
_journal_page_last 706
_publ_Section_title
;
 X-ray examination of the structure of boron nitride
;

# Found in Structure Reports, Vol 18, 125-126 (1940-41)

_aflow_proto 'AB_hP4_186_b_a'
_aflow_params 'a,c/a,z1,z2'
_aflow_params_values '2.51,2.66932270916,0.0,0.05'
_aflow_Strukturbericht 'B12'
_aflow_Pearson 'hP4'

_symmetry_space_group_name_Hall "P 6c -2c"
_symmetry_space_group_name_H-M "P 63 m c"
_symmetry_Int_Tables_number 186

_cell_length_a      2.51000
_cell_length_b      2.51000
_cell_length_c      6.70000
_cell_angle_alpha 90.00000
_cell_angle_beta  90.00000
_cell_angle_gamma 120.00000

loop_
_space_group_symop_id
_space_group_symop_operation_xyz
1 x,y,z
2 x-y,x,z+1/2
3 -y,x-y,z
4 -x,-y,z+1/2
5 -x+y,-x,z
6 y,-x+y,z+1/2
7 -x+y,y,z
8 -x,-x+y,z+1/2
9 -y,-x,z
10 x-y,-y,z+1/2
11 x,x-y,z
12 y,x,z+1/2

loop_
_atom_site_label
_atom_site_type_symbol
_atom_site_symmetry_multiplicity
_atom_site_Wyckoff_label
_atom_site_fract_x
_atom_site_fract_y
_atom_site_fract_z
_atom_site_occupancy
N1 N    2 a 0.00000 0.00000 0.00000 1.00000
B1 B    2 b 0.33333 0.66667 0.05000 1.00000
```

Original BN (B12): AB_hP4_186_b_a - POSCAR

```
AB_hP4_186_b_a & a,c/a,z1,z2 --params=2.51,2.66932270916,0.0,0.05 &
    ↪ P6_3mc    C_{6v}^4 #186 (ab) & hP4 & B12 & BN & & A. Bragrer,
    ↪ Acta Physicochimica URSS 7, 699-706 (1937)
   1.0000000000000000
   1.25500000000000  -2.17372376349894   0.00000000000000
   1.25500000000000   2.17372376349894   0.00000000000000
   0.00000000000000   0.00000000000000   6.70000000000000
   B     N
   2     2
Direct
   0.33333333333333   0.66666666666667   0.05000000000000   B    (2b)
   0.66666666666667   0.33333333333333   0.55000000000000   B    (2b)
   0.00000000000000   0.00000000000000   0.00000000000000   N    (2a)
   0.00000000000000   0.00000000000000   0.50000000000000   N    (2a)
```

BaPtSb: ABC_hP3_187_a_d_f - CIF

```
# CIF file

data_findsym-output
_audit_creation_method FINDSYM

_chemical_name_mineral ''
_chemical_formula_sum 'Ba Pt Sb'
```

```
loop_
_publ_author_name
 'G. Wenski'
 'A. Mewis'
_journal_name_full
;
Zeitschrift f\"{u}r anorganische und allgemeine Chemie
;
_journal_volume 535
_journal_year 1986
_journal_page_first 110
_journal_page_last 122
_publ_Section_title
;
 Trigonal-planar koordiniertes Platin: Darstellung und Struktur von
   ↪ SrPtAs (Sb), BaPtP (As, Sb), SrPt$_x$P$_{2-x}$, SrPt$_x$As$_{
   ↪ 0.90}$ und BaPt$_x$As$_{0.90}$
;

# Found in http://materials.springer.com/isp/crystallographic/docs/
   ↪ sd_2080134

_aflow_proto 'ABC_hP3_187_a_d_f'
_aflow_params 'a,c/a'
_aflow_params_values '4.535,1.0769570011'
_aflow_Strukturbericht 'None'
_aflow_Pearson 'hP3'

_symmetry_space_group_name_Hall "P -6 2"
_symmetry_space_group_name_H-M "P -6 m 2"
_symmetry_Int_Tables_number 187

_cell_length_a     4.53500
_cell_length_b     4.53500
_cell_length_c     4.88400
_cell_angle_alpha  90.00000
_cell_angle_beta   90.00000
_cell_angle_gamma  120.00000

loop_
_space_group_symop_id
_space_group_symop_operation_xyz
1 x,y,z
2 -y,x-y,z
3 -x+y,-x,z
4 x,x-y,-z
5 -x+y,y,-z
6 -y,-x,-z
7 -x+y,-x,-z
8 x,y,-z
9 -y,x-y,-z
10 -x+y,y,z
11 -y,-x,z
12 x,x-y,z

loop_
_atom_site_label
_atom_site_type_symbol
_atom_site_symmetry_multiplicity
_atom_site_Wyckoff_label
_atom_site_fract_x
_atom_site_fract_y
_atom_site_fract_z
_atom_site_occupancy
Ba1 Ba   1 a 0.00000 0.00000 0.00000 1.00000
Pt1 Pt   1 d 0.33333 0.66667 0.50000 1.00000
Sb1 Sb   1 f 0.66667 0.33333 0.50000 1.00000
```

BaPtSb: ABC_hP3_187_a_d_f - POSCAR

```
ABC_hP3_187_a_d_f & a,c/a --params=4.535,1.0769570011 & P(-6)m2    D_{3h
   ↪ }^1 #187 (adf) & hP3 & & BaPtSb & & G. Wenski and A. Mewis, Z.
   ↪ anorg. allg. Chem. 535, 110-122 (1986)
   1.0000000000000000
   2.26750000000000   -3.92742520616200    0.00000000000000
   2.26750000000000    3.92742520616200    0.00000000000000
   0.00000000000000    0.00000000000000    4.88400000000000
   Ba    Pt    Sb
    1     1     1
Direct
   0.00000000000000    0.00000000000000    0.00000000000000    Ba    (1a)
   0.33333333333333    0.66666666666667    0.50000000000000    Pt    (1d)
   0.66666666666667    0.33333333333333    0.50000000000000    Sb    (1f)
```

Tungsten Carbide ($B_h$): AB_hP2_187_d_a - CIF

```
# CIF file

data_findsym-output
_audit_creation_method FINDSYM

_chemical_name_mineral 'Tungsten Carbide'
_chemical_formula_sum 'W C'

loop_
_publ_author_name
 'J. Leciejewicz'
_journal_name_full
;
Acta Crystallographica
;
_journal_volume 14
_journal_year 1961
_journal_page_first 200
_journal_page_last 200
_publ_Section_title
;
 A note on the structure of tungsten carbide
;

# Found in Pearson's Alloys, pp. 479

_aflow_proto 'AB_hP2_187_d_a'
_aflow_params 'a,c/a'
_aflow_params_values '2.9065,0.975950455875'
_aflow_Strukturbericht 'B_h'
_aflow_Pearson 'hP2'

_symmetry_space_group_name_Hall "P -6 2"
_symmetry_space_group_name_H-M "P -6 m 2"
_symmetry_Int_Tables_number 187

_cell_length_a     2.90650
_cell_length_b     2.90650
_cell_length_c     2.83660
_cell_angle_alpha  90.00000
_cell_angle_beta   90.00000
_cell_angle_gamma  120.00000

loop_
_space_group_symop_id
_space_group_symop_operation_xyz
1 x,y,z
2 -y,x-y,z
3 -x+y,-x,z
4 x,x-y,-z
5 -x+y,y,-z
6 -y,-x,-z
7 -x+y,-x,-z
8 x,y,-z
9 -y,x-y,-z
10 -x+y,y,z
11 -y,-x,z
12 x,x-y,z

loop_
_atom_site_label
_atom_site_type_symbol
_atom_site_symmetry_multiplicity
_atom_site_Wyckoff_label
_atom_site_fract_x
_atom_site_fract_y
_atom_site_fract_z
_atom_site_occupancy
W1 W   1 a 0.00000 0.00000 0.00000 1.00000
C1 C   1 d 0.33333 0.66667 0.50000 1.00000
```

Tungsten Carbide ($B_h$): AB_hP2_187_d_a - POSCAR

```
AB_hP2_187_d_a & a,c/a --params=2.9065,0.975950455875 & P(-6)m2    D_{3h
   ↪ }^1 #187 (ad) & hP2 & B_h & WC & & J. Leciejewicz, Acta Cryst.
   ↪ 14, 200 (1961)
   1.0000000000000000
   1.45325000000000   -2.51710283609900    0.00000000000000
   1.45325000000000    2.51710283609900    0.00000000000000
   0.00000000000000    0.00000000000000    2.83660000000000
     C     W
     1     1
Direct
   0.33333333333333    0.66666666666667    0.50000000000000    C    (1d)
   0.00000000000000    0.00000000000000    0.00000000000000    W    (1a)
```

Revised $Fe_2P$ (C22): A2B_hP9_189_fg_bc - CIF

```
# CIF file

data_findsym-output
_audit_creation_method FINDSYM

_chemical_name_mineral ''
_chemical_formula_sum 'Fe2 P'

loop_
_publ_author_name
 'Hironobu Fujii'
 'Shigehiro Komura'
 'Takayoshi Takeda'
 'Tetsuhiko Okamoto'
 'Yuji Ito'
 'Jun Akimitsu'
_journal_name_full
;
Journal of the Physical Society of Japan
;
_journal_volume 46
_journal_year 1979
_journal_page_first 1616
_journal_page_last 1621
_publ_Section_title
;
 Polarized Neutron Diffraction Study of Fe$_2$P Single Crystal
;

# Found in Wyckoff, Vol. I, (IV,h1) pp. 360

_aflow_proto 'A2B_hP9_189_fg_bc'
_aflow_params 'a,c/a,x3,x4'
_aflow_params_values '5.877,0.584822188191,0.256,0.589'
_aflow_Strukturbericht 'C22'
```

```
_aflow_Pearson 'hP9'

_symmetry_space_group_name_Hall "P -6 -2"
_symmetry_space_group_name_H-M "P -6 2 m"
_symmetry_Int_Tables_number 189

_cell_length_a      5.87700
_cell_length_b      5.87700
_cell_length_c      3.43700
_cell_angle_alpha  90.00000
_cell_angle_beta   90.00000
_cell_angle_gamma  120.00000

loop_
_space_group_symop_id
_space_group_symop_operation_xyz
1 x,y,z
2 -y,x-y,z
3 -x+y,-x,z
_aflow_Strukturbericht 'C22'
_aflow_Pearson 'hP9'

_symmetry_space_group_name_Hall "P -6 -2"
_symmetry_space_group_name_H-M "P -6 2 m"
_symmetry_Int_Tables_number 189

_cell_length_a      5.87700
_cell_length_b      5.87700
_cell_length_c      3.43700
_cell_angle_alpha  90.00000
_cell_angle_beta   90.00000
_cell_angle_gamma  120.00000

loop_
_space_group_symop_id
_space_group_symop_operation_xyz
1 x,y,z
2 -y,x-y,z
3 -x+y,-x,z
4 x-y,-y,-z
5 y,x,-z
6 -x,-x+y,-z
7 -x+y,-x,-z
8 x,y,-z
9 -y,x-y,-z
10 -x,-x+y,z
11 x-y,-y,z
12 y,x,z

loop_
_atom_site_label
_atom_site_type_symbol
_atom_site_symmetry_multiplicity
_atom_site_Wyckoff_label
_atom_site_fract_x
_atom_site_fract_y
_atom_site_fract_z
_atom_site_occupancy
P1   P     1 b 0.00000 0.00000 0.50000 1.00000
P2   P     2 c 0.33333 0.66667 0.00000 1.00000
Fe1  Fe    3 f 0.25600 0.00000 0.00000 1.00000
Fe2  Fe    3 g 0.58900 0.00000 0.50000 1.00000
```

Revised $Fe_2P$ (C22): A2B_hP9_189_fg_bc - POSCAR

```
A2B_hP9_189_fg_bc & a,c/a,x3,x4 --params=5.877,0.584822188191,0.256,
    ↪ 0.589 & P(-6)2m    D_{3h}^3 #189 (bcfg) & hP9 & C22 & Fe2P & &
    ↪ H. Fujii et al., J. Phys. Soc. Japan 46, 1616-1621 (1979)
   1.0000000000000000
   2.93850000000000  -5.08963129804115   0.00000000000000
   2.93850000000000   5.08963129804115   0.00000000000000
   0.00000000000000   0.00000000000000   3.43700000000000
    Fe      P
     6      3
Direct
   0.00000000000000   0.25600000000000   0.00000000000000   Fe   (3f)
   0.25600000000000   0.00000000000000   0.00000000000000   Fe   (3f)
   0.74400000000000   0.74400000000000   0.00000000000000   Fe   (3f)
   0.00000000000000   0.58900000000000   0.50000000000000   Fe   (3g)
   0.41100000000000   0.41100000000000   0.50000000000000   Fe   (3g)
   0.58900000000000   0.00000000000000   0.50000000000000   Fe   (3g)
   0.00000000000000   0.00000000000000   0.50000000000000   P    (1b)
   0.33333333333333   0.66666666666667   0.00000000000000   P    (2c)
   0.66666666666667   0.33333333333333   0.00000000000000   P    (2c)
```

$AlB_4Mg$: AB4C_hP6_191_a_h_b - CIF

```
# CIF file

data_findsym-output
_audit_creation_method FINDSYM

_chemical_name_mineral ''
_chemical_formula_sum 'Al B4 Mg'

loop_
_publ_author_name
 'Serena Margadonna'
 'Kosmas Prassides'
 'Ioannis Arvanitidis'
 'Michael Pissas'
 'Georgios Papavassiliou'
 'Andrew N. Fitch'
_journal_name_full
;
Physical Review B
;
_journal_volume 66
_journal_year 2002
_journal_page_first 014518
_journal_page_last 014518
_publ_Section_title
;
 Crystal structure of the Mg$_{1-x}$Al$_x$B$_2$ superconductors near $x
    ↪ \approx 0.5$
;

_aflow_proto 'AB4C_hP6_191_a_h_b'
_aflow_params 'a,c/a,z3'
_aflow_params_values '3.04436,2.20489035462,0.2413'
_aflow_Strukturbericht 'None'
_aflow_Pearson 'hP6'

_symmetry_space_group_name_Hall "-P 6 2 "
_symmetry_space_group_name_H-M "P 6/m m m"
_symmetry_Int_Tables_number 191

_cell_length_a      3.04436
_cell_length_b      3.04436
_cell_length_c      6.71248
_cell_angle_alpha  90.00000
_cell_angle_beta   90.00000
_cell_angle_gamma  120.00000

loop_
_space_group_symop_id
_space_group_symop_operation_xyz
1 x,y,z
2 x-y,x,z
3 -y,x-y,z
4 -x,-y,z
5 -x+y,-x,z
6 y,-x+y,z
7 x-y,-y,-z
8 x,x-y,-z
9 y,x,-z
10 -x+y,y,-z
11 -x,-x+y,-z
12 -y,-x,-z
13 -x,-y,-z
14 -x+y,-x,-z
15 y,-x+y,-z
16 x,y,-z
17 x-y,x,-z
18 -y,x-y,-z
19 -x+y,y,z
20 -x,-x+y,z
21 -y,-x,z
22 x-y,-y,z
23 x,x-y,z
24 y,x,z

loop_
_atom_site_label
_atom_site_type_symbol
_atom_site_symmetry_multiplicity
_atom_site_Wyckoff_label
_atom_site_fract_x
_atom_site_fract_y
_atom_site_fract_z
_atom_site_occupancy
Al1  Al    1 a 0.00000 0.00000 0.00000 1.00000
Mg1  Mg    1 b 0.00000 0.00000 0.50000 1.00000
B1   B     4 h 0.33333 0.66667 0.24130 1.00000
```

$AlB_4Mg$: AB4C_hP6_191_a_h_b - POSCAR

```
AB4C_hP6_191_a_h_b & a,c/a,z3 --params=3.04436,2.20489035462,0.2413 & P6
    ↪ /mmm    D_{6h}^{1} #191 (abh) & hP6 & & AlB4Mg & & S.
    ↪ Margadonna et al., PRB 66, 014518 (2002)
   1.0000000000000000
   1.52218000000000  -2.63649309826500   0.00000000000000
   1.52218000000000   2.63649309826500   0.00000000000000
   0.00000000000000   0.00000000000000   6.71248000000000
    Al      B     Mg
     1      4      1
Direct
   0.00000000000000   0.00000000000000   0.00000000000000   Al   (1a)
   0.33333333333333   0.66666666666667   0.24130000000000   B    (4h)
   0.33333333333333   0.66666666666667   0.75870000000000   B    (4h)
   0.66666666666667   0.33333333333333   0.24130000000000   B    (4h)
   0.66666666666667   0.33333333333333   0.75870000000000   B    (4h)
   0.00000000000000   0.00000000000000   0.50000000000000   Mg   (1b)
```

$CaCu_5$ ($D2_d$): AB5_hP6_191_a_cg - CIF

```
# CIF file

data_findsym-output
_audit_creation_method FINDSYM

_chemical_name_mineral ''
_chemical_formula_sum 'Ca Cu5'

loop_
_publ_author_name
 'Werner Haucke'
_journal_name_full
;
Zeitschrift f\"{u}r anorganische und allgemeine Chemie
```

```
;
_journal_volume 244
_journal_year 1940
_journal_page_first 17
_journal_page_last 22
_publ_Section_title
;
 Kristallstruktur von CaZn$_5$ und CaCu$_5$
;

# Found in  Pearson's Alloys, pp. 645

_aflow_proto 'AB5_hP6_191_a_cg'
_aflow_params 'a,c/a'
_aflow_params_values '5.405,0.773913043478'
_aflow_Strukturbericht 'D2_d'
_aflow_Pearson 'hP6'

_symmetry_space_group_name_Hall "-P 6 2"
_symmetry_space_group_name_H-M "P 6/m m m"
_symmetry_Int_Tables_number 191

_cell_length_a      5.40500
_cell_length_b      5.40500
_cell_length_c      4.18300
_cell_angle_alpha  90.00000
_cell_angle_beta   90.00000
_cell_angle_gamma 120.00000

loop_
_space_group_symop_id
_space_group_symop_operation_xyz
1 x,y,z
2 x-y,x,z
3 -y,x-y,z
4 -x,-y,z
5 -x+y,-x,z
6 y,-x+y,z
7 x-y,-y,-z
8 x,x-y,-z
9 y,x,-z
10 -x+y,y,-z
11 -x,-x+y,-z
12 -y,-x,-z
13 -x,-y,-z
14 -x+y,-x,-z
15 y,-x+y,-z
16 x,y,-z
17 x-y,x,-z
18 -y,x-y,-z
19 -x+y,y,z
20 -x,-x+y,z
21 -y,-x,z
22 x-y,-y,z
23 x,x-y,z
24 y,x,z

loop_
_atom_site_label
_atom_site_type_symbol
_atom_site_symmetry_multiplicity
_atom_site_Wyckoff_label
_atom_site_fract_x
_atom_site_fract_y
_atom_site_fract_z
_atom_site_occupancy
Ca1 Ca   1 a 0.00000 0.00000 0.00000 1.00000
Cu1 Cu   2 c 0.33333 0.66667 0.00000 1.00000
Cu2 Cu   3 g 0.50000 0.00000 0.50000 1.00000
```

$CaCu_5$ ($D2_d$): AB5_hP6_191_a_cg - POSCAR

```
AB5_hP6_191_a_cg & a,c/a --params=5.405,0.773913043478 & P6/mmm      D_{6h
    ↪ }^{1} #191 (acg) & hP6 & D2_d & CaCu5 &  & W. Haucke, ZAAC 244,
    ↪  17-22 (1940)
   1.0000000000000000
   2.70250000000000   -4.68086730745489    0.00000000000000
   2.70250000000000    4.68086730745489    0.00000000000000
   0.00000000000000    0.00000000000000    4.18300000000000
    Ca    Cu
     1     5
Direct
   0.00000000000000    0.00000000000000    0.00000000000000    Ca    (1a)
   0.33333333333333    0.66666666666667    0.00000000000000    Cu    (2c)
   0.66666666666667    0.33333333333333    0.00000000000000    Cu    (2c)
   0.00000000000000    0.50000000000000    0.50000000000000    Cu    (3g)
   0.50000000000000    0.00000000000000    0.50000000000000    Cu    (3g)
   0.50000000000000    0.50000000000000    0.50000000000000    Cu    (3g)
```

Simple Hexagonal Lattice ($A_f$): A_hP1_191_a - CIF

```
# CIF file

data_findsym-output
_audit_creation_method FINDSYM

_chemical_name_mineral 'Hg_xSn'
_chemical_formula_sum 'Sn'

loop_
_publ_author_name
 'G. V. Raynor'
 'J. A. Lee'
_journal_name_full
;
Acta Metallurgica
;
_journal_volume 2
_journal_year 1954
_journal_page_first 616
_journal_page_last 620
_publ_Section_title
;
 The tin-rich intermediate phases in the alloys of tin with cadmium,
    ↪ indium and mercury
;

# Found in  Pearson's Handbook, III, pp. 3947

_aflow_proto 'A_hP1_191_a'
_aflow_params 'a,c/a'
_aflow_params_values '3.2062,0.931195808122'
_aflow_Strukturbericht 'A_f'
_aflow_Pearson 'hP1'

_symmetry_space_group_name_Hall "-P 6 2"
_symmetry_space_group_name_H-M "P 6/m m m"
_symmetry_Int_Tables_number 191

_cell_length_a      3.20620
_cell_length_b      3.20620
_cell_length_c      2.98560
_cell_angle_alpha  90.00000
_cell_angle_beta   90.00000
_cell_angle_gamma 120.00000

loop_
_space_group_symop_id
_space_group_symop_operation_xyz
1 x,y,z
2 x-y,x,z
3 -y,x-y,z
4 -x,-y,z
5 -x+y,-x,z
6 y,-x+y,z
7 x-y,-y,-z
8 x,x-y,-z
9 y,x,-z
10 -x+y,y,-z
11 -x,-x+y,-z
12 -y,-x,-z
13 -x,-y,-z
14 -x+y,-x,-z
15 y,-x+y,-z
16 x,y,-z
17 x-y,x,-z
18 -y,x-y,-z
19 -x+y,y,z
20 -x,-x+y,z
21 -y,-x,z
22 x-y,-y,z
23 x,x-y,z
24 y,x,z

loop_
_atom_site_label
_atom_site_type_symbol
_atom_site_symmetry_multiplicity
_atom_site_Wyckoff_label
_atom_site_fract_x
_atom_site_fract_y
_atom_site_fract_z
_atom_site_occupancy
Sn1 Sn  1 a 0.00000 0.00000 0.00000 1.00000
```

Simple Hexagonal Lattice ($A_f$): A_hP1_191_a - POSCAR

```
A_hP1_191_a & a,c/a --params=3.2062,0.931195808122 & P6/mmm     D_{6h}^{1
    ↪ } #191 (a) & hP1 & A_f & HgSn6-10 &  & G. V. Raynor and J. A.
    ↪ Lee, Acta Metallurgica 2, 616-620 (1954)
   1.0000000000000000
   1.60310000000000   -2.77665064961367    0.00000000000000
   1.60310000000000    2.77665064961367    0.00000000000000
   0.00000000000000    0.00000000000000    2.98560000000076
    M
    1
Direct
   0.00000000000000    0.00000000000000    0.00000000000000    M    (1a)
```

$Li_3N$: A3B_hP4_191_bc_a - CIF

```
# CIF file

data_findsym-output
_audit_creation_method FINDSYM

_chemical_name_mineral ''
_chemical_formula_sum 'Li3 N'

loop_
_publ_author_name
 'Duncan H. Gregory'
 'Paul M. O'Meara'
 'Alexandra G. Gordon'
 'Jason P. Hodges'
 'Simine Short'
 'James D. Jorgensen'
_journal_name_full
;
Chemistry of Materials
```

```
;
_journal_volume 14
_journal_year 2002
_journal_page_first 2063
_journal_page_last 2070
_publ_Section_title
;
 Structure of Lithium Nitride and Transition-Metal-Doped Derivatives,
      ↪ Li$_{3-x-y}$M$_x$N (M = Ni, Cu):âĂŤ A Powder Neutron
      ↪ Diffraction Study
;

# Found in http://materials.springer.com/lb/docs/
      ↪ sm_lbs_978-3-540-32682-3_554}

_aflow_proto 'A3B_hP4_191_bc_a'
_aflow_params 'a,c/a'
_aflow_params_values '3.6576,1.05902777778'
_aflow_Strukturbericht 'None'
_aflow_Pearson 'hP4'

_symmetry_space_group_name_Hall "-P 6 2"
_symmetry_space_group_name_H-M "P 6/m m m"
_symmetry_Int_Tables_number 191

_cell_length_a      3.65760
_cell_length_b      3.65760
_cell_length_c      3.87350
_cell_angle_alpha  90.00000
_cell_angle_beta   90.00000
_cell_angle_gamma 120.00000

loop_
_space_group_symop_id
_space_group_symop_operation_xyz
1 x,y,z
2 x-y,x,z
3 -y,x-y,z
4 -x,-y,z
5 -x+y,-x,z
6 y,-x+y,z
7 x-y,-y,-z
8 x,x-y,-z
9 y,x,-z
10 -x+y,y,-z
11 -x,-x+y,-z
12 -y,-x,-z
13 -x,-y,-z
14 -x+y,-x,-z
15 y,-x+y,-z
16 x,y,-z
17 x-y,x,-z
18 -y,x-y,-z
19 -x+y,y,z
20 -x,-x+y,z
21 -y,-x,z
22 x-y,-y,z
23 x,x-y,z
24 y,x,z

loop_
_atom_site_label
_atom_site_type_symbol
_atom_site_symmetry_multiplicity
_atom_site_Wyckoff_label
_atom_site_fract_x
_atom_site_fract_y
_atom_site_fract_z
_atom_site_occupancy
N1  N   1 a 0.00000 0.00000 0.00000 1.00000
Li1 Li  1 b 0.00000 0.00000 0.50000 1.00000
Li2 Li  2 c 0.33333 0.66667 0.00000 1.00000
```

$Li_3N$: A3B_hP4_191_bc_a - POSCAR

```
A3B_hP4_191_bc_a & a,c/a --params=3.6576,1.05902777778 & P6/mmm    D_{6h
      ↪ }^4 #191 (abc) & hP4 & & Li3N & & D. H. Gregory et al., Chem.
      ↪ Mater. 14, 2063-2070 (2002)
   1.0000000000000000
   1.82880000000000  -3.16757451688196   0.00000000000000
   1.82880000000000   3.16757451688196   0.00000000000000
   0.00000000000000   0.00000000000000   3.87350000000000
    Li    N
     3    1
Direct
   0.00000000000000   0.00000000000000   0.50000000000000   Li   (1b)
   0.33333333333333   0.66666666666667   0.00000000000000   Li   (2c)
   0.66666666666667   0.33333333333333   0.00000000000000   Li   (2c)
   0.00000000000000   0.00000000000000   0.00000000000000    N   (1a)
```

Hexagonal $\omega$ (C32): AB2_hP3_191_a_d - CIF

```
# CIF file

data_findsym-output
_audit_creation_method FINDSYM

_chemical_name_mineral 'hexagonal omega structure'
_chemical_formula_sum 'Al B2'

loop_
_publ_author_name
 'Ulrich Burkhardt'
 'Vladimir Gurin'
 'Frank Haarmann'
 'Horst Borrmann'
 'Walter Schnelle'
 'Alexander Yaresko'
 'Yuri Grin'
_journal_name_full
;
Journal of Solid State Chemistry
;
_journal_volume 177
_journal_year 2004
_journal_page_first 389
_journal_page_last 394
_publ_Section_title
;
 On the electronic and structural properties of aluminum diboride Al$_{
      ↪ 0.9}$B$_2$
;

# Found in http://en.wikipedia.org/wiki/Aluminium_diboride

_aflow_proto 'AB2_hP3_191_a_d'
_aflow_params 'a,c/a'
_aflow_params_values '3.005,1.08276206323'
_aflow_Strukturbericht 'C32'
_aflow_Pearson 'hP3'

_symmetry_space_group_name_Hall "-P 6 2"
_symmetry_space_group_name_H-M "P 6/m m m"
_symmetry_Int_Tables_number 191

_cell_length_a      3.00500
_cell_length_b      3.00500
_cell_length_c      3.25370
_cell_angle_alpha  90.00000
_cell_angle_beta   90.00000
_cell_angle_gamma 120.00000

loop_
_space_group_symop_id
_space_group_symop_operation_xyz
1 x,y,z
2 x-y,x,z
3 -y,x-y,z
4 -x,-y,z
5 -x+y,-x,z
6 y,-x+y,z
7 x-y,-y,-z
8 x,x-y,-z
9 y,x,-z
10 -x+y,y,-z
11 -x,-x+y,-z
12 -y,-x,-z
13 -x,-y,-z
14 -x+y,-x,-z
15 y,-x+y,-z
16 x,y,-z
17 x-y,x,-z
18 -y,x-y,-z
19 -x+y,y,z
20 -x,-x+y,z
21 -y,-x,z
22 x-y,-y,z
23 x,x-y,z
24 y,x,z

loop_
_atom_site_label
_atom_site_type_symbol
_atom_site_symmetry_multiplicity
_atom_site_Wyckoff_label
_atom_site_fract_x
_atom_site_fract_y
_atom_site_fract_z
_atom_site_occupancy
Al1 Al  1 a 0.00000 0.00000 0.00000 1.00000
B1  B   2 d 0.33333 0.66667 0.50000 1.00000
```

Hexagonal $\omega$ (C32): AB2_hP3_191_a_d - POSCAR

```
AB2_hP3_191_a_d & a,c/a --params=3.005,1.08276206323 & P6/mmm    D_{6h
      ↪ }^{1} #191 (ad) & hP3 & C32 & AlB2 & hexagonal omega & U.
      ↪ Burkhardt et al., JSSC 177, 389-394 (2004)
   1.0000000000000000
   1.50250000000000  -2.60240633837224   0.00000000000000
   1.50250000000000   2.60240633837224   0.00000000000000
   0.00000000000000   0.00000000000000   3.25370000000000
    Al    B
     1    2
Direct
   0.00000000000000   0.00000000000000   0.00000000000000   Al   (1a)
   0.33333333333333   0.66666666666667   0.50000000000000    B   (2d)
   0.66666666666667   0.33333333333333   0.50000000000000    B   (2d)
```

$Cu_2Te$ ($C_h$): A2B_hP6_191_h_e - CIF

```
# CIF file

data_findsym-output
_audit_creation_method FINDSYM

_chemical_name_mineral ''
_chemical_formula_sum 'Cu2 Te'

loop_
_publ_author_name
```

```
 'H. Nowotny'
_journal_name_full
;
Zeitschrift f\"{u}r Metallkunde
;
_journal_volume 37
_journal_year 1946
_journal_page_first 40
_journal_page_last 42
_publ_Section_title
;
 Die Kristallstruktur von Cu$_2$Te
;

# Found in  Pearson's Handbook, Vol. III, pp. 3014

_aflow_proto 'A2B_hP6_191_h_e'
_aflow_params 'a,c/a,z1,z2'
_aflow_params_values '4.237,1.71040830776,0.306,0.16'
_aflow_Strukturbericht 'C_h'
_aflow_Pearson 'hP6'

_symmetry_space_group_name_Hall "-P 6 2"
_symmetry_space_group_name_H-M "P 6/m m m"
_symmetry_Int_Tables_number 191

_cell_length_a      4.23700
_cell_length_b      4.23700
_cell_length_c      7.24700
_cell_angle_alpha  90.00000
_cell_angle_beta   90.00000
_cell_angle_gamma 120.00000

loop_
_space_group_symop_id
_space_group_symop_operation_xyz
1 x,y,z
2 x-y,x,z
3 -y,x-y,z
4 -x,-y,z
5 -x+y,-x,z
6 y,-x+y,z
7 x-y,-y,-z
8 x,x-y,-z
9 y,x,-z
10 -x+y,y,-z
11 -x,-x+y,-z
12 -y,-x,-z
13 -x,-y,-z
14 -x+y,-x,-z
15 y,-x+y,-z
16 x,y,-z
17 x-y,x,-z
18 -y,x-y,-z
19 -x+y,y,z
20 -x,-x+y,z
21 -y,-x,z
22 x-y,-y,z
23 x,x-y,z
24 y,x,z

loop_
_atom_site_label
_atom_site_type_symbol
_atom_site_symmetry_multiplicity
_atom_site_Wyckoff_label
_atom_site_fract_x
_atom_site_fract_y
_atom_site_fract_z
_atom_site_occupancy
Te1 Te   2 e 0.00000 0.00000 0.30600 1.00000
Cu1 Cu   4 h 0.33333 0.66667 0.16000 1.00000
```

$Cu_2Te$ ($C_h$): A2B_hP6_191_h_e - POSCAR

```
A2B_hP6_191_h_e & a,c/a,z1,z2 --params=4.237,1.71040830776,0.306,0.16 &
    ↪ P6/mmm    D_{6h}^{1} #191 (eh) & hP6 & C_h & Cu2Te & & H.
    ↪ Nowotny, Z. Metallkd. 37, 40-42 (1946)
   1.0000000000000000
   2.11850000000000  -3.66934963583467   0.00000000000000
   2.11850000000000   3.66934963583467   0.00000000000000
   0.00000000000000   0.00000000000000   7.24700000000000
   Cu    Te
    4     2
Direct
   0.33333333333333   0.66666666666667   0.16000000000000   Cu   (4h)
   0.33333333333333   0.66666666666667   0.84000000000000   Cu   (4h)
   0.66666666666667   0.33333333333333   0.16000000000000   Cu   (4h)
   0.66666666666667   0.33333333333333   0.84000000000000   Cu   (4h)
   0.00000000000000   0.00000000000000   0.30600000000000   Te   (2e)
   0.00000000000000   0.00000000000000   0.69400000000000   Te   (2e)
```

CoSn (B35): AB_hP6_191_f_ad - CIF

```
# CIF file

data_findsym-output
_audit_creation_method FINDSYM

_chemical_name_mineral ''
_chemical_formula_sum 'Co Sn'

loop_
_publ_author_name
 'A.K. Larsson'
 'M. Haeberlein'
 'S. Lidin'
 'U. Schwarz'
_journal_name_full
;
Journal of Alloys and Compounds
;
_journal_volume 240
_journal_year 1996
_journal_page_first 79
_journal_page_last 84
_publ_Section_title
;
 Single crystal structure refinement and high-pressure properties of
    ↪ CoSn
;

_aflow_proto 'AB_hP6_191_f_ad'
_aflow_params 'a,c/a'
_aflow_params_values '5.279,0.806914188293'
_aflow_Strukturbericht 'B35'
_aflow_Pearson 'hP6'

_symmetry_space_group_name_Hall "-P 6 2"
_symmetry_space_group_name_H-M "P 6/m m m"
_symmetry_Int_Tables_number 191

_cell_length_a      5.27900
_cell_length_b      5.27900
_cell_length_c      4.25970
_cell_angle_alpha  90.00000
_cell_angle_beta   90.00000
_cell_angle_gamma 120.00000

loop_
_space_group_symop_id
_space_group_symop_operation_xyz
1 x,y,z
2 x-y,x,z
3 -y,x-y,z
4 -x,-y,z
5 -x+y,-x,z
6 y,-x+y,z
7 x-y,-y,-z
8 x,x-y,-z
9 y,x,-z
10 -x+y,y,-z
11 -x,-x+y,-z
12 -y,-x,-z
13 -x,-y,-z
14 -x+y,-x,-z
15 y,-x+y,-z
16 x,y,-z
17 x-y,x,-z
18 -y,x-y,-z
19 -x+y,y,z
20 -x,-x+y,z
21 -y,-x,z
22 x-y,-y,z
23 x,x-y,z
24 y,x,z

loop_
_atom_site_label
_atom_site_type_symbol
_atom_site_symmetry_multiplicity
_atom_site_Wyckoff_label
_atom_site_fract_x
_atom_site_fract_y
_atom_site_fract_z
_atom_site_occupancy
Sn1 Sn   1 a 0.00000 0.00000 0.00000 1.00000
Sn2 Sn   2 d 0.33333 0.66667 0.50000 1.00000
Co1 Co   3 f 0.50000 0.00000 0.00000 1.00000
```

CoSn (B35): AB_hP6_191_f_ad - POSCAR

```
AB_hP6_191_f_ad & a,c/a --params=5.279,0.806914188293 & P6/mmm    D_{6h
    ↪ }^{1} #191 (adf) & hP6 & B35 & CoSn & & A. K. Larsson et al.,
    ↪ J. Alloys Comp. 240, 79-84 (1996)
   1.0000000000000000
   2.63950000000000  -4.57174810657805   0.00000000000000
   2.63950000000000   4.57174810657805   0.00000000000000
   0.00000000000000   0.00000000000000   4.25970000000000
   Co    Sn
    3     3
Direct
   0.00000000000000   0.50000000000000   0.00000000000000   Co   (3f)
   0.50000000000000   0.00000000000000   0.00000000000000   Co   (3f)
   0.50000000000000   0.50000000000000   0.00000000000000   Co   (3f)
   0.00000000000000   0.00000000000000   0.00000000000000   Sn   (1a)
   0.33333333333333   0.66666666666667   0.50000000000000   Sn   (2d)
   0.66666666666667   0.33333333333333   0.50000000000000   Sn   (2d)
```

AsTi ($B_i$): AB_hP8_194_ad_f - CIF

```
# CIF file

data_findsym-output
_audit_creation_method FINDSYM

_chemical_name_mineral ''
_chemical_formula_sum 'As Ti'

loop_
```

```
_publ_author_name
 'K. Bachmayer'
 'H. Nowotny'
 'A. Kohl'
_journal_name_full
;
Monatshefte f\"{u}r Chemie und verwandte Teile anderer Wissenschaften
;
_journal_volume 86
_journal_year 1955
_journal_page_first 39
_journal_page_last 43
_publ_Section_title
;
 Die Struktur von TiAs
;

# Found in Wyckoff, Vol. I, pp. 146-149

_aflow_proto 'AB_hP8_194_ad_f'
_aflow_params 'a,c/a,z3'
_aflow_params_values '3.64,3.37362637363,0.125'
_aflow_Strukturbericht 'B_i'
_aflow_Pearson 'hP8'

_symmetry_space_group_name_Hall "-P 6c 2c"
_symmetry_space_group_name_H-M "P 63/m m c"
_symmetry_Int_Tables_number 194

_cell_length_a      3.64000
_cell_length_b      3.64000
_cell_length_c      12.28000
_cell_angle_alpha   90.00000
_cell_angle_beta    90.00000
_cell_angle_gamma   120.00000

loop_
_space_group_symop_id
_space_group_symop_operation_xyz
1 x,y,z
2 x-y,x,z+1/2
3 -y,x-y,z
4 -x,-y,z+1/2
5 -x+y,-x,z
6 y,-x+y,z+1/2
7 x-y,-y,-z
8 x,x-y,-z+1/2
9 y,x,-z
10 -x+y,y,-z+1/2
11 -x,-x+y,-z
12 -y,-x,-z+1/2
13 -x,-y,-z
14 -x+y,-x,-z+1/2
15 y,-x+y,-z
16 x,y,-z+1/2
17 x-y,x,-z
18 -y,x-y,-z+1/2
19 -x+y,y,z
20 -x,-x+y,z+1/2
21 -y,-x,z
22 x-y,-y,z+1/2
23 x,x-y,z
24 y,x,z+1/2

loop_
_atom_site_label
_atom_site_type_symbol
_atom_site_symmetry_multiplicity
_atom_site_Wyckoff_label
_atom_site_fract_x
_atom_site_fract_y
_atom_site_fract_z
_atom_site_occupancy
As1 As   2 a 0.00000 0.00000 0.00000 1.00000
As2 As   2 d 0.33333 0.66667 0.75000 1.00000
Ti1 Ti   4 f 0.33333 0.66667 0.12500 1.00000
```

AsTi ($B_i$): AB_hP8_194_ad_f - POSCAR

```
AB_hP8_194_ad_f & a,c/a,z3 --params=3.64,3.37362637363,0.125 & P6_3/mmc
    ↪ D_{6h}^4 #194 (adf) & hP8 & B_i & AsTi & & K. Bachmayer, H.
    ↪ Nowtny and A. Kohl, Monatshefte f"{u}r Chemie 86, 39-43 (1955)
   1.0000000000000000
   1.82000000000000  -3.15233246977536   0.00000000000000
   1.82000000000000   3.15233246977536   0.00000000000000
   0.00000000000000   0.00000000000000  12.28000000000000
   As    Ti
    4     4
Direct
   0.00000000000000   0.00000000000000   0.00000000000000   As   (2a)
   0.00000000000000   0.00000000000000   0.50000000000000   As   (2a)
   0.33333333333333   0.66666666666667   0.75000000000000   As   (2d)
   0.66666666666667   0.33333333333333   0.25000000000000   As   (2d)
   0.33333333333333   0.66666666666667   0.12500000000000   Ti   (4f)
   0.33333333333333   0.66666666666667   0.37500000000000   Ti   (4f)
   0.66666666666667   0.33333333333333   0.62500000000000   Ti   (4f)
   0.66666666666667   0.33333333333333   0.87500000000000   Ti   (4f)
```

Hypothetical Tetrahedrally Bonded Carbon with 3-Member Rings: A_hP6_194_h - CIF

```
# CIF file

data_findsym-output
_audit_creation_method FINDSYM

_chemical_name_mineral 'Theoretical Carbon Structure'
_chemical_formula_sum 'C'

loop_
_publ_author_name
 'Peter A. Schultz'
 'Kevin Leung'
 'E. B. Stechel'
_journal_name_full
;
Physical Review B
;
_journal_volume 59
_journal_year 1999
_journal_page_first 733
_journal_page_last 741
_publ_Section_title
;
 Small rings and amorphous tetrahedral carbon
;

_aflow_proto 'A_hP6_194_h'
_aflow_params 'a,c/a,x1'
_aflow_params_values '4.40445,0.568892824303,0.44799'
_aflow_Strukturbericht 'None'
_aflow_Pearson 'hP6'

_symmetry_space_group_name_Hall "-P 6c 2c"
_symmetry_space_group_name_H-M "P 63/m m c"
_symmetry_Int_Tables_number 194

_cell_length_a      4.40445
_cell_length_b      4.40445
_cell_length_c      2.50566
_cell_angle_alpha   90.00000
_cell_angle_beta    90.00000
_cell_angle_gamma   120.00000

loop_
_space_group_symop_id
_space_group_symop_operation_xyz
1 x,y,z
2 x-y,x,z+1/2
3 -y,x-y,z
4 -x,-y,z+1/2
5 -x+y,-x,z
6 y,-x+y,z+1/2
7 x-y,-y,-z
8 x,x-y,-z+1/2
9 y,x,-z
10 -x+y,y,-z+1/2
11 -x,-x+y,-z
12 -y,-x,-z+1/2
13 -x,-y,-z
14 -x+y,-x,-z+1/2
15 y,-x+y,-z
16 x,y,-z+1/2
17 x-y,x,-z
18 -y,x-y,-z+1/2
19 -x+y,y,z
20 -x,-x+y,z+1/2
21 -y,-x,z
22 x-y,-y,z+1/2
23 x,x-y,z
24 y,x,z+1/2

loop_
_atom_site_label
_atom_site_type_symbol
_atom_site_symmetry_multiplicity
_atom_site_Wyckoff_label
_atom_site_fract_x
_atom_site_fract_y
_atom_site_fract_z
_atom_site_occupancy
C1 C    6 h 0.44799 0.89598 0.25000 1.00000
```

Hypothetical Tetrahedrally Bonded Carbon with 3-Member Rings: A_hP6_194_h - POSCAR

```
A_hP6_194_h & a,c/a,x1 --params=4.40445,0.568892824303,0.44799 & P6_3/
    ↪ mmc D_{6h}^4 #194 (h) & hP6 & & C & hypothetical 3-member ring
    ↪ structure & P. A. Schultz, K. Leung and E. B. Stechel, PRB 59,
    ↪ 733-741 (1999)
   1.0000000000000000
   2.20222418612552  -3.81436418002642   0.00000000000000
   2.20222418612552   3.81436418002642   0.00000000000000
   0.00000000000000   0.00000000000000   2.50566256627303
   C
   6
Direct
   0.10401971341143   0.55200985670572   0.25000000000000   C   (6h)
   0.44799014329428   0.55200985670572   0.25000000000000   C   (6h)
   0.44799014329428   0.89598028658856   0.25000000000000   C   (6h)
   0.55200985670572   0.10401971341143   0.75000000000000   C   (6h)
   0.55200985670572   0.44799014329428   0.75000000000000   C   (6h)
   0.89598028658856   0.44799014329428   0.75000000000000   C   (6h)
```

CMo: AB_hP12_194_af_bf - CIF

```
# CIF file

data_findsym-output
_audit_creation_method FINDSYM

_chemical_name_mineral 'Molybdenum Carbide MAX Phase'
```

```
_chemical_formula_sum 'C Mo'

loop_
_publ_author_name
 'H. Nowotny'
 'R. Parth\'{e}'
 'R. Kieffer'
 'F. Benesovsky'
_journal_name_full
;
Monatshefte f\"{u}r Chemie und verwandte Teile anderer Wissenschaften
;
_journal_volume 85
_journal_year 1954
_journal_page_first 255
_journal_page_last 272
_publ_Section_title
;
 Das Dreistoffsystem: Molybd\"{a}n--Silizium--Kohlenstoff
;

_aflow_proto 'AB_hP12_194_af_bf'
_aflow_params 'a,c/a,z3,z4'
_aflow_params_values '3.01,4.85382059801,0.166,0.583'
_aflow_Strukturbericht 'None'
_aflow_Pearson 'hP12'

_symmetry_space_group_name_Hall "-P 6c 2c"
_symmetry_space_group_name_H-M "P 63/m m c"
_symmetry_Int_Tables_number 194

_cell_length_a       3.01000
_cell_length_b       3.01000
_cell_length_c       14.61000
_cell_angle_alpha  90.00000
_cell_angle_beta   90.00000
_cell_angle_gamma  120.00000

loop_
_space_group_symop_id
_space_group_symop_operation_xyz
1 x,y,z
2 x-y,x,z+1/2
3 -y,x-y,z
4 -x,-y,z+1/2
5 -x+y,-x,z
6 y,-x+y,z+1/2
7 x-y,-y,-z
8 x,x-y,-z+1/2
9 y,x,-z
10 -x+y,y,-z+1/2
11 -x,-x+y,-z
12 -y,-x,-z+1/2
13 -x,-y,-z
14 -x+y,-x,-z+1/2
15 y,-x+y,-z
16 x,y,-z+1/2
17 x-y,x,-z
18 -y,x-y,-z+1/2
19 -x+y,y,z
20 -x,-x+y,z+1/2
21 -y,-x,z
22 x-y,-y,z+1/2
23 x,x-y,z
24 y,x,z+1/2

loop_
_atom_site_label
_atom_site_type_symbol
_atom_site_symmetry_multiplicity
_atom_site_Wyckoff_label
_atom_site_fract_x
_atom_site_fract_y
_atom_site_fract_z
_atom_site_occupancy
C1  C    2 a 0.00000 0.00000 0.00000 1.00000
Mo1 Mo   2 b 0.00000 0.00000 0.25000 1.00000
C2  C    4 f 0.33333 0.66667 0.16600 1.00000
Mo2 Mo   4 f 0.33333 0.66667 0.58300 1.00000
```

CMo: AB_hP12_194_af_bf - POSCAR

```
AB_hP12_194_af_bf & a,c/a,z3,z4 --params=3.01,4.85382059801,0.166,0.583
    ↪ & P6_3/mmc  D_{6h}^4 #194 (abf^2) & hP12 & & CMo & MAX Phase &
    ↪ H. Nowotny et al. Monatsh.\ Chem.\ Verw.\ Tl. 85, 255-272 (1954
    ↪ )
   1.0000000000000000
   1.50500000000000  -2.60673646539116   0.00000000000000
   1.50500000000000   2.60673646539116   0.00000000000000
   0.00000000000000   0.00000000000000  14.61000000000000
    C    Mo
    6     6
Direct
   0.00000000000000   0.00000000000000   0.00000000000000    C    (2a)
   0.00000000000000   0.00000000000000   0.50000000000000    C    (2a)
   0.33333333333333   0.66666666666667   0.16600000000000    C    (4f)
   0.33333333333333   0.66666666666667   0.33400000000000    C    (4f)
   0.66666666666667   0.33333333333333   0.66600000000000    C    (4f)
   0.66666666666667   0.33333333333333   0.83400000000000    C    (4f)
   0.00000000000000   0.00000000000000   0.25000000000000    Mo   (2b)
   0.00000000000000   0.00000000000000   0.75000000000000    Mo   (2b)
   0.33333333333333   0.66666666666667   0.58300000000000    Mo   (4f)
   0.33333333333333   0.66666666666667  -0.08300000000000    Mo   (4f)
   0.66666666666667   0.33333333333333   0.41700000000000    Mo   (4f)
   0.66666666666667   0.33333333333333   0.08300000000000    Mo   (4f)
```

$\alpha$-La (A3'): A_hP4_194_ac - CIF

```
# CIF file

data_findsym-output
_audit_creation_method FINDSYM

_chemical_name_mineral 'alpha La'
_chemical_formula_sum 'La'

loop_
_publ_author_name
 'F.H. Spedding'
 'J.J. Hanak'
 'A.H. Daane'
_journal_name_full
;
Journal of the Less Common Metals
;
_journal_volume 3
_journal_year 1961
_journal_page_first 110
_journal_page_last 124
_publ_Section_title
;
 High temperature allotropy and thermal expansion of the rare-earth
      ↪ metals
;

# Found in  Donohue, pp. 83-86

_aflow_proto 'A_hP4_194_ac'
_aflow_params 'a,c/a'
_aflow_params_values '3.77,3.2175066313'
_aflow_Strukturbericht 'A3''
_aflow_Pearson 'hP4'

_symmetry_space_group_name_Hall "-P 6c 2c"
_symmetry_space_group_name_H-M "P 63/m m c"
_symmetry_Int_Tables_number 194

_cell_length_a       3.77000
_cell_length_b       3.77000
_cell_length_c       12.13000
_cell_angle_alpha  90.00000
_cell_angle_beta   90.00000
_cell_angle_gamma  120.00000

loop_
_space_group_symop_id
_space_group_symop_operation_xyz
1 x,y,z
2 x-y,x,z+1/2
3 -y,x-y,z
4 -x,-y,z+1/2
5 -x+y,-x,z
6 y,-x+y,z+1/2
7 x-y,-y,-z
8 x,x-y,-z+1/2
9 y,x,-z
10 -x+y,y,-z+1/2
11 -x,-x+y,-z
12 -y,-x,-z+1/2
13 -x,-y,-z
14 -x+y,-x,-z+1/2
15 y,-x+y,-z
16 x,y,-z+1/2
17 x-y,x,-z
18 -y,x-y,-z+1/2
19 -x+y,y,z
20 -x,-x+y,z+1/2
21 -y,-x,z
22 x-y,-y,z+1/2
23 x,x-y,z
24 y,x,z+1/2

loop_
_atom_site_label
_atom_site_type_symbol
_atom_site_symmetry_multiplicity
_atom_site_Wyckoff_label
_atom_site_fract_x
_atom_site_fract_y
_atom_site_fract_z
_atom_site_occupancy
La1 La   2 a 0.00000 0.00000 0.00000 1.00000
La2 La   2 c 0.33333 0.66667 0.25000 1.00000
```

$\alpha$-La (A3'): A_hP4_194_ac - POSCAR

```
A_hP4_194_ac & a,c/a --params=3.77,3.2175066313 & P6_3/mmc  D_{6h}^4 #
    ↪ 194 (ac) & hP4 & A3' & La & alpha & F. H. Spedding, J. J. Hanak
    ↪ and A. H. Daane, J. Less-Common Metals 3, 110-124 (1961)
   1.0000000000000000
   1.88500000000000  -3.26491577226700   0.00000000000000
   1.88500000000000   3.26491577226700   0.00000000000000
   0.00000000000000   0.00000000000000  12.13000000000000
   La
    4
Direct
   0.00000000000000   0.00000000000000   0.00000000000000    La   (2a)
   0.00000000000000   0.00000000000000   0.50000000000000    La   (2a)
   0.33333333333333   0.66666666666667   0.25000000000000    La   (2c)
   0.66666666666667   0.33333333333333   0.75000000000000    La   (2c)
```

Na$_3$As (D0$_{18}$): AB3_hP8_194_c_bf - CIF

```
# CIF file

data_findsym-output
_audit_creation_method FINDSYM

_chemical_name_mineral 'Sodium arsenide'
_chemical_formula_sum 'Na3 As'

loop_
_publ_author_name
 'G. Brauer'
 'E. Zintl'
_journal_name_full
;
Zeitschrift f\"{u}r Physikalische Chemie
;
_journal_volume 37B
_journal_year 1937
_journal_page_first 323
_journal_page_last 352
_publ_Section_title
;
 Konstitution von Phosphiden, Arseniden, Antimoniden und Wismutiden des
   ↪ Lithiums, Natriums und Kaliums
;

# Found in Pearson's Handbook, Vol. I, pp. 1187

_aflow_proto 'AB3_hP8_194_c_bf'
_aflow_params 'a,c/a,z3'
_aflow_params_values '5.088,1.76533018868,-0.083'
_aflow_Strukturbericht 'D0_18'
_aflow_Pearson 'hP8'

_symmetry_space_group_name_Hall "-P 6c 2c"
_symmetry_space_group_name_H-M "P 63/m m c"
_symmetry_Int_Tables_number 194

_cell_length_a     5.08800
_cell_length_b     5.08800
_cell_length_c     8.98200
_cell_angle_alpha  90.00000
_cell_angle_beta   90.00000
_cell_angle_gamma  120.00000

loop_
_space_group_symop_id
_space_group_symop_operation_xyz
1 x,y,z
2 x-y,x,z+1/2
3 -y,x-y,z
4 -x,-y,z+1/2
5 -x+y,-x,z
6 y,-x+y,z+1/2
7 x-y,-y,-z
8 x,x-y,-z+1/2
9 y,x,-z
10 -x+y,y,-z+1/2
11 -x,-x+y,-z
12 -y,-x,-z+1/2
13 -x,-y,-z
14 -x+y,-x,-z+1/2
15 y,-x+y,-z
16 x,y,-z+1/2
17 x-y,x,-z
18 -y,x-y,-z+1/2
19 -x+y,y,z
20 -x,-x+y,z+1/2
21 -y,-x,z
22 x-y,-y,z+1/2
23 x,x-y,z
24 y,x,z+1/2

loop_
_atom_site_label
_atom_site_type_symbol
_atom_site_symmetry_multiplicity
_atom_site_Wyckoff_label
_atom_site_fract_x
_atom_site_fract_y
_atom_site_fract_z
_atom_site_occupancy
Na1 Na   2 b 0.00000 0.00000 0.25000 1.00000
As1 As   2 c 0.33333 0.66667 0.25000 1.00000
Na2 Na   4 f 0.33333 0.66667 -0.08300 1.00000
```

Na$_3$As (D0$_{18}$): AB3_hP8_194_c_bf - POSCAR

```
AB3_hP8_194_c_bf & a,c/a,z3 --params=5.088,1.76533018868,-0.083 & P6_3/
   ↪ mmc D_{6h}^4 #194 (bcf) & hP8 & D0_18 & Na3As & & G. Brauer
   ↪ and E. Zintl, Zeitschrift f\"{u}r Physikalische Chemie 37B,
   ↪ 323-352 (1937)
   1.0000000000000000
   2.54400000000000   -4.40633725445500   0.00000000000000
   2.54400000000000    4.40633725445500   0.00000000000000
   0.00000000000000    0.00000000000000   8.98200000000000
   As    Na
    2     6
Direct
   0.33333333333333   0.66666666666667   0.25000000000000   As   (2c)
   0.66666666666667   0.33333333333333   0.75000000000000   As   (2c)
   0.00000000000000   0.00000000000000   0.25000000000000   Na   (2b)
   0.00000000000000   0.00000000000000   0.75000000000000   Na   (2b)
   0.33333333333333   0.66666666666667  -0.08300000000000   Na   (4f)
   0.33333333333333   0.66666666666667   0.58300000000000   Na   (4f)
   0.66666666666667   0.33333333333333   0.08300000000000   Na   (4f)
   0.66666666666667   0.33333333333333   0.41700000000000   Na   (4f)
```

CaIn$_2$: AB2_hP6_194_b_f - CIF

```
# CIF file

data_findsym-output
_audit_creation_method FINDSYM

_chemical_name_mineral ''
_chemical_formula_sum 'Ca In2'

loop_
_publ_author_name
 'A. Iandelli'
_journal_name_full
;
Zeitschrift f\"{u}r anorganische und allgemeine Chemie
;
_journal_volume 330
_journal_year 1964
_journal_page_first 221
_journal_page_last 232
_publ_Section_title
;
 MX$_2$-Verbindungen der Erdalkali- und Seltenen Erdmetalle mit Gallium,
   ↪ Indium und Thallium
;

# Found in Pearson's Alloys, pp. 499-501

_aflow_proto 'AB2_hP6_194_b_f'
_aflow_params 'a,c/a,z2'
_aflow_params_values '4.895,1.58324821246,0.045'
_aflow_Strukturbericht 'None'
_aflow_Pearson 'hP6'

_symmetry_space_group_name_Hall "-P 6c 2c"
_symmetry_space_group_name_H-M "P 63/m m c"
_symmetry_Int_Tables_number 194

_cell_length_a     4.89500
_cell_length_b     4.89500
_cell_length_c     7.75000
_cell_angle_alpha  90.00000
_cell_angle_beta   90.00000
_cell_angle_gamma  120.00000

loop_
_space_group_symop_id
_space_group_symop_operation_xyz
1 x,y,z
2 x-y,x,z+1/2
3 -y,x-y,z
4 -x,-y,z+1/2
5 -x+y,-x,z
6 y,-x+y,z+1/2
7 x-y,-y,-z
8 x,x-y,-z+1/2
9 y,x,-z
10 -x+y,y,-z+1/2
11 -x,-x+y,-z
12 -y,-x,-z+1/2
13 -x,-y,-z
14 -x+y,-x,-z+1/2
15 y,-x+y,-z
16 x,y,-z+1/2
17 x-y,x,-z
18 -y,x-y,-z+1/2
19 -x+y,y,z
20 -x,-x+y,z+1/2
21 -y,-x,z
22 x-y,-y,z+1/2
23 x,x-y,z
24 y,x,z+1/2

loop_
_atom_site_label
_atom_site_type_symbol
_atom_site_symmetry_multiplicity
_atom_site_Wyckoff_label
_atom_site_fract_x
_atom_site_fract_y
_atom_site_fract_z
_atom_site_occupancy
Ca1 Ca   2 b 0.00000 0.00000 0.25000 1.00000
In1 In   4 f 0.33333 0.66667 0.04500 1.00000
```

CaIn$_2$: AB2_hP6_194_b_f - POSCAR

```
AB2_hP6_194_b_f & a,c/a,z2 --params=4.895,1.58324821246,0.045 & P6_3/mmc
   ↪ D_{6h}^4 #194 (bf) & hP6 & & CaIn2 & & A. Iandelli, ZAAC 330,
   ↪ 221-232 (1964)
   1.0000000000000000
   2.44750000000000   -4.23919435152483   0.00000000000000
   2.44750000000000    4.23919435152483   0.00000000000000
   0.00000000000000    0.00000000000000   7.75000000000000
   Ca    In
    2     4
Direct
   0.00000000000000   0.00000000000000   0.25000000000000   Ca   (2b)
   0.00000000000000   0.00000000000000   0.75000000000000   Ca   (2b)
```

```
    0.33333333333333    0.66666666666667    0.04500000000000    In      (4f)
    0.33333333333333    0.66666666666667    0.45500000000000    In      (4f)
    0.66666666666667    0.33333333333333   -0.04500000000000    In      (4f)
    0.66666666666667    0.33333333333333    0.54500000000000    In      (4f)
```

BN ($B_k$): AB_hP4_194_c_d - CIF

```
# CIF file

data_findsym-output
_audit_creation_method FINDSYM

_chemical_name_mineral 'Boron Nitride'
_chemical_formula_sum 'B N'

loop_
_publ_author_name
 'R. S. Pease'
_journal_name_full
;
Acta Crystallographica
;
_journal_volume 5
_journal_year 1952
_journal_page_first 356
_journal_page_last 361
_publ_Section_title
;
 An X-ray study of boron nitride
;

# Found in  Wyckoff, Vol. I, pp. 184-185

_aflow_proto 'AB_hP4_194_c_d'
_aflow_params 'a,c/a'
_aflow_params_values '2.50399,2.66023426611'
_aflow_Strukturbericht 'B_k'
_aflow_Pearson 'hP4'

_symmetry_space_group_name_Hall "-P 6c 2c"
_symmetry_space_group_name_H-M "P 63/m m c"
_symmetry_Int_Tables_number 194

_cell_length_a      2.50399
_cell_length_b      2.50399
_cell_length_c      6.66120
_cell_angle_alpha  90.00000
_cell_angle_beta   90.00000
_cell_angle_gamma 120.00000

loop_
_space_group_symop_id
_space_group_symop_operation_xyz
1 x,y,z
2 x-y,x,z+1/2
3 -y,x-y,z
4 -x,-y,z+1/2
5 -x+y,-x,z
6 y,-x+y,z+1/2
7 x-y,-y,-z
8 x,x-y,-z+1/2
9 y,x,-z
10 -x+y,y,-z+1/2
11 -x,-x+y,-z
12 -y,-x,-z+1/2
13 -x,-y,-z
14 -x+y,-x,-z+1/2
15 y,-x+y,-z
16 x,y,-z+1/2
17 x-y,x,-z
18 -y,x-y,-z+1/2
19 -x+y,y,z
20 -x,-x+y,z+1/2
21 -y,-x,z
22 x-y,-y,z+1/2
23 x,x-y,z
24 y,x,z+1/2

loop_
_atom_site_label
_atom_site_type_symbol
_atom_site_symmetry_multiplicity
_atom_site_Wyckoff_label
_atom_site_fract_x
_atom_site_fract_y
_atom_site_fract_z
_atom_site_occupancy
B1 B    2 c 0.33333 0.66667 0.25000 1.00000
N1 N    2 d 0.33333 0.66667 0.75000 1.00000
```

BN ($B_k$): AB_hP4_194_c_d - POSCAR

```
AB_hP4_194_c_d & a,c/a --params=2.50399,2.66023426611 & P6_3/mmc  D_{6h
    ↪ }^4 #194 (cd) & hP4 & B_k & BN & & R. S. Pease, Acta Cryst. 5,
    ↪ 356-361 (1952)
   1.0000000000000000
   1.25199500000000   -2.16851895082220    0.00000000000000
   1.25199500000000    2.16851895082220    0.00000000000000
   0.00000000000000    0.00000000000000    6.66120000000000
     B     N
     2     2
Direct
   0.33333333333333    0.66666666666667    0.25000000000000    B      (2c)
   0.66666666666667    0.33333333333333    0.75000000000000    B      (2c)
   0.33333333333333    0.66666666666667    0.75000000000000    N      (2d)
```

```
   0.66666666666667    0.33333333333333    0.25000000000000     N      (2d)
```

$AlCCr_2$: ABC2_hP8_194_d_a_f - CIF

```
# CIF file

data_findsym-output
_audit_creation_method FINDSYM

_chemical_name_mineral 'H-Phase'
_chemical_formula_sum 'Al C Cr2'

loop_
_publ_author_name
 'W. Jeitschko'
 'H. Nowotny'
 'F. Benesovsky'
_journal_name_full
;
Monatshefte f\"{u}r Chemie und verwandte Teile anderer Wissenschaften
;
_journal_volume 94
_journal_year 1963
_journal_page_first 672
_journal_page_last 676
_publ_Section_title
;
 Kohlenstoffhaltige tern\"{a}re Verbindungen (H-Phase)
;

# Found in  Pearson's Handbook, Vol. I, pp. 677

_aflow_proto 'ABC2_hP8_194_d_a_f'
_aflow_params 'a,c/a,z3'
_aflow_params_values '2.86,4.48251748252,0.086'
_aflow_Strukturbericht 'None'
_aflow_Pearson 'hP8'

_symmetry_space_group_name_Hall "-P 6c 2c"
_symmetry_space_group_name_H-M "P 63/m m c"
_symmetry_Int_Tables_number 194

_cell_length_a      2.86000
_cell_length_b      2.86000
_cell_length_c     12.82000
_cell_angle_alpha  90.00000
_cell_angle_beta   90.00000
_cell_angle_gamma 120.00000

loop_
_space_group_symop_id
_space_group_symop_operation_xyz
1 x,y,z
2 x-y,x,z+1/2
3 -y,x-y,z
4 -x,-y,z+1/2
5 -x+y,-x,z
6 y,-x+y,z+1/2
7 x-y,-y,-z
8 x,x-y,-z+1/2
9 y,x,-z
10 -x+y,y,-z+1/2
11 -x,-x+y,-z
12 -y,-x,-z+1/2
13 -x,-y,-z
14 -x+y,-x,-z+1/2
15 y,-x+y,-z
16 x,y,-z+1/2
17 x-y,x,-z
18 -y,x-y,-z+1/2
19 -x+y,y,z
20 -x,-x+y,z+1/2
21 -y,-x,z
22 x-y,-y,z+1/2
23 x,x-y,z
24 y,x,z+1/2

loop_
_atom_site_label
_atom_site_type_symbol
_atom_site_symmetry_multiplicity
_atom_site_Wyckoff_label
_atom_site_fract_x
_atom_site_fract_y
_atom_site_fract_z
_atom_site_occupancy
C1  C     2 a 0.00000 0.00000 0.00000 1.00000
Al1 Al    2 d 0.33333 0.66667 0.75000 1.00000
Cr1 Cr    4 f 0.33333 0.66667 0.08600 1.00000
```

$AlCCr_2$: ABC2_hP8_194_d_a_f - POSCAR

```
ABC2_hP8_194_d_a_f & a,c/a,z3 --params=2.86,4.48251748252,0.086 & P6_3/
    ↪ mmc    D_{6h}^4 #194 (adf) & hP8 & & AlCCr2 & H-Phase & W.
    ↪ Jeitschko, H. Nowotny and F. Benesovsky, Monatshefte f"{u}r
    ↪ Chemie 94, 672-676 (1963)
   1.0000000000000000
   1.43000000000000   -2.47683265482300    0.00000000000000
   1.43000000000000    2.47683265482300    0.00000000000000
   0.00000000000000    0.00000000000000   12.82000000000000
    Al     C    Cr
     2     2     4
Direct
   0.33333333333333    0.66666666666667    0.75000000000000    Al     (2d)
   0.66666666666667    0.33333333333333    0.25000000000000    Al     (2d)
```

```
  0.00000000000000    0.00000000000000    0.00000000000000     C     (2a)
  0.00000000000000    0.00000000000000    0.50000000000000     C     (2a)
  0.33333333333333    0.66666666666667    0.08600000000000     Cr    (4f)
  0.33333333333333    0.66666666666667    0.41400000000000     Cr    (4f)
  0.66666666666667    0.33333333333333   -0.08600000000000     Cr    (4f)
  0.66666666666667    0.33333333333333    0.58600000000000     Cr    (4f)
```

$Ni_3Sn$ ($D0_{19}$): A3B_hP8_194_h_c - CIF

```
# CIF file

data_findsym-output
_audit_creation_method FINDSYM

_chemical_name_mineral ''
_chemical_formula_sum 'Ni3 Sn'

loop_
_publ_author_name
 'Andrei L. Lyubimtsev'
 'Alexey I. Baranov'
 'Andreas Fischer'
 'Lars Kloo'
 'Boris A. Popovkin'
_journal_name_full
;
Journal of Alloys and Compounds
;
_journal_volume 340
_journal_year 2002
_journal_page_first 167
_journal_page_last 172
_publ_Section_title
;
 The structure and bonding of Ni$_3$Sn
;

_aflow_proto 'A3B_hP8_194_h_c'
_aflow_params 'a,c/a,x2'
_aflow_params_values '5.295,0.802077431539,0.8392'
_aflow_Strukturbericht 'D0_19'
_aflow_Pearson 'hP8'

_symmetry_space_group_name_Hall "-P 6c 2c"
_symmetry_space_group_name_H-M "P 63/m m c"
_symmetry_Int_Tables_number 194

_cell_length_a      5.29500
_cell_length_b      5.29500
_cell_length_c      4.24700
_cell_angle_alpha 90.00000
_cell_angle_beta  90.00000
_cell_angle_gamma 120.00000

loop_
_space_group_symop_id
_space_group_symop_operation_xyz
1 x,y,z
2 x-y,x,z+1/2
3 -y,x-y,z
4 -x,-y,z+1/2
5 -x+y,-x,z
6 y,-x+y,z+1/2
7 x-y,-y,-z
8 x,x-y,-z+1/2
9 y,x,-z
10 -x+y,y,-z+1/2
11 -x,-x+y,-z
12 -y,-x,-z+1/2
13 -x,-y,-z
14 -x+y,-x,-z+1/2
15 y,-x+y,-z
16 x,y,-z+1/2
17 x-y,x,-z
18 -y,x-y,-z+1/2
19 -x+y,y,z
20 -x,-x+y,z+1/2
21 -y,-x,z
22 x-y,-y,z+1/2
23 x,x-y,z
24 y,x,z+1/2

loop_
_atom_site_label
_atom_site_type_symbol
_atom_site_symmetry_multiplicity
_atom_site_Wyckoff_label
_atom_site_fract_x
_atom_site_fract_y
_atom_site_fract_z
_atom_site_occupancy
Sn1 Sn   2 c 0.33333 0.66667 0.25000 1.00000
Ni1 Ni   6 h 0.83920 0.67840 0.25000 1.00000
```

$Ni_3Sn$ ($D0_{19}$): A3B_hP8_194_h_c - POSCAR

```
A3B_hP8_194_h_c & a,c/a,x2 --params=5.295,0.802077431539,0.8392 & P6_3/
    ↪ mmc D_{6h}^4 #194 (ch) & hP8 & D0_19 & Ni3Sn & & A. L.
    ↪ Lyubimtsev et al., J. Alloys Compd. 340, 167-172 (2002)
  1.0000000000000000
  2.64750000000000   -4.58560451303900    0.00000000000000
  2.64750000000000    4.58560451303900    0.00000000000000
  0.00000000000000    0.00000000000000    4.24700000000000
  Ni     Sn
   6      2
Direct
  0.16080000000000    0.32160000000000    0.75000000000000    Ni    (6h)
  0.16080000000000    0.83920000000000    0.75000000000000    Ni    (6h)
  0.32160000000000    0.16080000000000    0.25000000000000    Ni    (6h)
  0.67840000000000    0.83920000000000    0.75000000000000    Ni    (6h)
  0.83920000000000    0.16080000000000    0.25000000000000    Ni    (6h)
  0.83920000000000    0.67840000000000    0.25000000000000    Ni    (6h)
  0.33333333333333    0.66666666666667    0.25000000000000    Sn    (2c)
  0.66666666666667    0.33333333333333    0.75000000000000    Sn    (2c)
```

Hexagonal Graphite (A9): A_hP4_194_bc - CIF

```
# CIF file

data_findsym-output
_audit_creation_method FINDSYM

_chemical_name_mineral 'Graphite'
_chemical_formula_sum 'C'

loop_
_publ_author_name
 'Peter Trucano'
 'Ruey Chen'
_journal_name_full
;
Nature
;
_journal_volume 258
_journal_year 1975
_journal_page_first 136
_journal_page_last 137
_publ_Section_title
;
 Structure of graphite by neutron diffraction
;

# Found in AMS Database

_aflow_proto 'A_hP4_194_bc'
_aflow_params 'a,c/a'
_aflow_params_values '2.464,2.72362012987'
_aflow_Strukturbericht 'A9'
_aflow_Pearson 'hP4'

_symmetry_space_group_name_Hall "-P 6c 2c"
_symmetry_space_group_name_H-M "P 63/m m c"
_symmetry_Int_Tables_number 194

_cell_length_a      2.46400
_cell_length_b      2.46400
_cell_length_c      6.71100
_cell_angle_alpha 90.00000
_cell_angle_beta  90.00000
_cell_angle_gamma 120.00000

loop_
_space_group_symop_id
_space_group_symop_operation_xyz
1 x,y,z
2 x-y,x,z+1/2
3 -y,x-y,z
4 -x,-y,z+1/2
5 -x+y,-x,z
6 y,-x+y,z+1/2
7 x-y,-y,-z
8 x,x-y,-z+1/2
9 y,x,-z
10 -x+y,y,-z+1/2
11 -x,-x+y,-z
12 -y,-x,-z+1/2
13 -x,-y,-z
14 -x+y,-x,-z+1/2
15 y,-x+y,-z
16 x,y,-z+1/2
17 x-y,x,-z
18 -y,x-y,-z+1/2
19 -x+y,y,z
20 -x,-x+y,z+1/2
21 -y,-x,z
22 x-y,-y,z+1/2
23 x,x-y,z
24 y,x,z+1/2

loop_
_atom_site_label
_atom_site_type_symbol
_atom_site_symmetry_multiplicity
_atom_site_Wyckoff_label
_atom_site_fract_x
_atom_site_fract_y
_atom_site_fract_z
_atom_site_occupancy
C1 C   2 b 0.00000 0.00000 0.25000 1.00000
C2 C   2 c 0.33333 0.66667 0.25000 1.00000
```

Hexagonal Graphite (A9): A_hP4_194_bc - POSCAR

```
A_hP4_194_bc & a,c/a --params=2.464,2.72362012987 & P6_3/mmc     D_{6h}^
    ↪ 4 #194 (bc) & hP4 & A9 & C & Graphite (unbuckled) & P. Trucano
    ↪ and R. Chen, Nature 258, 136-137 (1975)
  1.0000000000000000
  1.23200000000000   -2.13388659492486    0.00000000000000
  1.23200000000000    2.13388659492486    0.00000000000000
  0.00000000000000    0.00000000000000    6.71100000000000
```

```
    C
    4
Direct
  0.00000000000000  0.00000000000000  0.25000000000000  C  (2b)
  0.00000000000000  0.00000000000000  0.75000000000000  C  (2b)
  0.33333333333333  0.66666666666667  0.25000000000000  C  (2c)
  0.66666666666667  0.33333333333333  0.75000000000000  C  (2c)
```

Molybdenite ($MoS_2$, C7): AB2_hP6_194_c_f - CIF

```
# CIF file

data_findsym-output
_audit_creation_method FINDSYM

_chemical_name_mineral 'Molybdenite'
_chemical_formula_sum 'Mo S2'

loop_
_publ_author_name
 'B. Sch\"{o}nfeld'
 'J. J. Huang'
 'S. C. Moss'
_journal_name_full
;
Acta Crystallographica B
;
_journal_volume 39
_journal_year 1983
_journal_page_first 404
_journal_page_last 407
_publ_Section_title
;
 Anisotropic Mean-Square Displacements (MSD) in single Crystals of 2H-
    ↪ and 3R-MoS$_2$
;

# Found in AMS Database

_aflow_proto 'AB2_hP6_194_c_f'
_aflow_params 'a,c/a,z2'
_aflow_params_values '3.161,3.8895919013,0.6275'
_aflow_Strukturbericht 'C7'
_aflow_Pearson 'hP6'

_symmetry_space_group_name_Hall "-P 6c 2c"
_symmetry_space_group_name_H-M "P 63/m m c"
_symmetry_Int_Tables_number 194

_cell_length_a    3.16100
_cell_length_b    3.16100
_cell_length_c    12.29500
_cell_angle_alpha 90.00000
_cell_angle_beta  90.00000
_cell_angle_gamma 120.00000

loop_
_space_group_symop_id
_space_group_symop_operation_xyz
1 x,y,z
2 x-y,x,z+1/2
3 -y,x-y,z
4 -x,-y,z+1/2
5 -x+y,-x,z
6 y,-x+y,z+1/2
7 x-y,-y,-z
8 x,x-y,-z+1/2
9 y,x,-z
10 -x+y,y,-z+1/2
11 -x,-x+y,-z
12 -y,-x,-z+1/2
13 -x,-y,-z
14 -x+y,-x,-z+1/2
15 y,-x+y,-z
16 x,y,-z+1/2
17 x-y,x,-z
18 -y,x-y,-z+1/2
19 -x+y,y,z
20 -x,-x+y,z+1/2
21 -y,-x,z
22 x-y,-y,z+1/2
23 x,x-y,z
24 y,x,z+1/2

loop_
_atom_site_label
_atom_site_type_symbol
_atom_site_symmetry_multiplicity
_atom_site_Wyckoff_label
_atom_site_fract_x
_atom_site_fract_y
_atom_site_fract_z
_atom_site_occupancy
Mo1 Mo  2 c 0.33333 0.66667 0.25000 1.00000
S1  S   4 f 0.33333 0.66667 0.62750 1.00000
```

Molybdenite ($MoS_2$, C7): AB2_hP6_194_c_f - POSCAR

```
AB2_hP6_194_c_f & a,c/a,z2 --params=3.161,3.8895919013,0.6275 & P6_3/mmc
    ↪ D_{6h}^4 #194 (cf) & hP6 & C7 & MoS2 & & B. Sch"{o}nfeld, J.
    ↪ J. Huang and S. C. Moss, Acta Cryst. B 39, 404-407 (1983)
  1.00000000000000000
  1.58050000000000  -2.73750630136261  0.00000000000000
  1.58050000000000   2.73750630136261  0.00000000000000
  0.00000000000000   0.00000000000000  12.29500000000000
```

```
  Mo   S
   2   4
Direct
  0.33333333333333  0.66666666666667  0.25000000000000  Mo  (2c)
  0.66666666666667  0.33333333333333  0.75000000000000  Mo  (2c)
  0.33333333333333  0.66666666666667  -0.12750000000000  S  (4f)
  0.33333333333333  0.66666666666667  0.62750000000000  S  (4f)
  0.66666666666667  0.33333333333333  0.12750000000000  S  (4f)
  0.66666666666667  0.33333333333333  0.37250000000000  S  (4f)
```

$W_2B_5$ ($D8_h$): A5B2_hP14_194_abdf_f - CIF

```
# CIF file

data_findsym-output
_audit_creation_method FINDSYM

_chemical_name_mineral 'Tungsten boride'
_chemical_formula_sum 'W2 B5'

loop_
_publ_author_name
 'Roland Kiessling'
_journal_name_full
;
Acta Chemica Scandinavica
;
_journal_volume 1
_journal_year 1947
_journal_page_first 893
_journal_page_last 916
_publ_Section_title
;
 The Crystal Structures of Molybdenum and Tungsten Borides
;

# Found in Wyckoff, Vol. II, pp. 188-189

_aflow_proto 'A5B2_hP14_194_abdf_f'
_aflow_params 'a,c/a,z4,z5'
_aflow_params_values '2.982,4.651240778,0.528,0.139'
_aflow_Strukturbericht 'D8_h'
_aflow_Pearson 'hP14'

_symmetry_space_group_name_Hall "-P 6c 2c"
_symmetry_space_group_name_H-M "P 63/m m c"
_symmetry_Int_Tables_number 194

_cell_length_a    2.98200
_cell_length_b    2.98200
_cell_length_c    13.87000
_cell_angle_alpha 90.00000
_cell_angle_beta  90.00000
_cell_angle_gamma 120.00000

loop_
_space_group_symop_id
_space_group_symop_operation_xyz
1 x,y,z
2 x-y,x,z+1/2
3 -y,x-y,z
4 -x,-y,z+1/2
5 -x+y,-x,z
6 y,-x+y,z+1/2
7 x-y,-y,-z
8 x,x-y,-z+1/2
9 y,x,-z
10 -x+y,y,-z+1/2
11 -x,-x+y,-z
12 -y,-x,-z+1/2
13 -x,-y,-z
14 -x+y,-x,-z+1/2
15 y,-x+y,-z
16 x,y,-z+1/2
17 x-y,x,-z
18 -y,x-y,-z+1/2
19 -x+y,y,z
20 -x,-x+y,z+1/2
21 -y,-x,z
22 x-y,-y,z+1/2
23 x,x-y,z
24 y,x,z+1/2

loop_
_atom_site_label
_atom_site_type_symbol
_atom_site_symmetry_multiplicity
_atom_site_Wyckoff_label
_atom_site_fract_x
_atom_site_fract_y
_atom_site_fract_z
_atom_site_occupancy
B1 B  2 a 0.00000 0.00000 0.00000 1.00000
B2 B  2 b 0.00000 0.00000 0.25000 1.00000
B3 B  2 d 0.33333 0.66667 0.75000 1.00000
B4 B  4 f 0.33333 0.66667 0.52800 1.00000
W1 W  4 f 0.33333 0.66667 0.13900 1.00000
```

$W_2B_5$ ($D8_h$): A5B2_hP14_194_abdf_f - POSCAR

```
A5B2_hP14_194_abdf_f & a,c/a,z4,z5 --params=2.982,4.651240778,0.528,
    ↪ 0.139 & P6_3/mmc D_{6h}^4 #194 (abdf^2) & hP14 & D8_h & W2B5 &
    ↪ epsilon & R Kiessling, Acta Chem. Scand. 1, 893-916 (1947)
  1.00000000000000000
  1.49100000000000  -2.58248775408520  0.00000000000000
```

```
   1.49100000000000     2.58248775408520     0.00000000000000
   0.00000000000000     0.00000000000000    13.87000000000000
    B      W
   10      4
Direct
   0.00000000000000     0.00000000000000     0.00000000000000     B      (2a)
   0.00000000000000     0.00000000000000     0.50000000000000     B      (2a)
   0.00000000000000     0.00000000000000     0.25000000000000     B      (2b)
   0.00000000000000     0.00000000000000     0.75000000000000     B      (2b)
   0.33333333333333     0.66666666666667     0.75000000000000     B      (2d)
   0.66666666666667     0.33333333333333     0.25000000000000     B      (2d)
   0.33333333333333     0.66666666666667    -0.02800000000000     B      (4f)
   0.33333333333333     0.66666666666667     0.52800000000000     B      (4f)
   0.66666666666667     0.33333333333333     0.02800000000000     B      (4f)
   0.66666666666667     0.33333333333333     0.47200000000000     B      (4f)
   0.33333333333333     0.66666666666667     0.13900000000000     W      (4f)
   0.33333333333333     0.66666666666667     0.36100000000000     W      (4f)
   0.66666666666667     0.33333333333333    -0.13900000000000     W      (4f)
   0.66666666666667     0.33333333333333     0.63900000000000     W      (4f)
```

$MgZn_2$ Hexagonal Laves (C14): AB2_hP12_194_f_ah - CIF

```
# CIF file

data_findsym-output
_audit_creation_method FINDSYM

_chemical_name_mineral 'Hexagonal Laves'
_chemical_formula_sum 'Mg Zn2'

loop_
_publ_author_name
 'T. Ohba'
 'Y. Kitano'
 'Y. Komura'
_journal_name_full
;
Acta Crystallographic C
;
_journal_volume 40
_journal_year 1984
_journal_page_first 1
_journal_page_last 5
_publ_Section_title
;
 The charge-density study of the Laves phases, MgZn$_2$ and MgCu$_2$
;

_aflow_proto 'AB2_hP12_194_f_ah'
_aflow_params 'a,c/a,z2,x3'
_aflow_params_values '5.223,1.64005360904,0.06286,0.83048'
_aflow_Strukturbericht 'C14'
_aflow_Pearson 'hP12'

_symmetry_space_group_name_Hall "-P 6c 2c"
_symmetry_space_group_name_H-M "P 63/m m c"
_symmetry_Int_Tables_number 194

_cell_length_a      5.22300
_cell_length_b      5.22300
_cell_length_c      8.56600
_cell_angle_alpha  90.00000
_cell_angle_beta   90.00000
_cell_angle_gamma 120.00000

loop_
_space_group_symop_id
_space_group_symop_operation_xyz
1 x,y,z
2 x-y,x,z+1/2
3 -y,x-y,z
4 -x,-y,z+1/2
5 -x+y,-x,z
6 y,-x+y,z+1/2
7 x-y,-y,-z
8 x,x-y,-z+1/2
9 y,x,-z
10 -x+y,y,-z+1/2
11 -x,-x+y,-z
12 -y,-x,-z+1/2
13 -x,-y,-z
14 -x+y,-x,-z+1/2
15 y,-x+y,-z
16 x,y,-z+1/2
17 x-y,x,-z
18 -y,x-y,-z+1/2
19 -x+y,y,z
20 -x,-x+y,z+1/2
21 -y,-x,z
22 x-y,-y,z+1/2
23 x,x-y,z
24 y,x,z+1/2

loop_
_atom_site_label
_atom_site_type_symbol
_atom_site_symmetry_multiplicity
_atom_site_Wyckoff_label
_atom_site_fract_x
_atom_site_fract_y
_atom_site_fract_z
_atom_site_occupancy
Zn1 Zn   2 a 0.00000 0.00000 0.00000 1.00000
Mg1 Mg   4 f 0.33333 0.66667 0.06286 1.00000
Zn2 Zn   6 h 0.83048 0.16952 0.25000 1.00000
```

$MgZn_2$ Hexagonal Laves (C14): AB2_hP12_194_f_ah - POSCAR

```
AB2_hP12_194_f_ah & a,c/a,z2,x3 --params=5.223,1.64005360904,0.06286,
   ↪ 0.83048 & P6_3/mmc  D_{6h}^4 #194 (afh) & hP12 & C14 & MgZn2 &
   ↪ Laves & T. Ohba, Y. Kitano and Y. Komura, Acta Cryst. C 40, 1-5
   ↪ (1984)
   1.00000000000000000
   2.61150000000000    -4.52325068396612     0.00000000000000
   2.61150000000000     4.52325068396612     0.00000000000000
   0.00000000000000     0.00000000000000     8.56600000000000
   Mg     Zn
    4      8
Direct
   0.33333333333333     0.66666666666667     0.06286000000000     Mg     (4f)
   0.33333333333333     0.66666666666667     0.43714000000000     Mg     (4f)
   0.66666666666667     0.33333333333333    -0.06286000000000     Mg     (4f)
   0.66666666666667     0.33333333333333     0.56286000000000     Mg     (4f)
   0.00000000000000     0.00000000000000     0.00000000000000     Zn     (2a)
   0.00000000000000     0.00000000000000     0.50000000000000     Zn     (2a)
   0.16952000000000     0.33904000000000     0.75000000000000     Zn     (6h)
   0.16952000000000     0.83048000000000     0.75000000000000     Zn     (6h)
   0.33904000000000     0.16952000000000     0.25000000000000     Zn     (6h)
   0.66096000000000     0.83048000000000     0.75000000000000     Zn     (6h)
   0.83048000000000     0.16952000000000     0.25000000000000     Zn     (6h)
   0.83048000000000     0.66096000000000     0.25000000000000     Zn     (6h)
```

LiBC: ABC_hP6_194_c_d_a - CIF

```
# CIF file

data_findsym-output
_audit_creation_method FINDSYM

_chemical_name_mineral ''
_chemical_formula_sum 'Li B C'

loop_
_publ_author_name
 'Michael W\"{o}rle'
 'Reinhard Nesper'
 'Gunter Mair'
 'Martin Schwarz'
 'Hans Georg Von Schnering'
_journal_name_full
;
Zeitschrift f\"{u}r anorganische und allgemeine Chemie
;
_journal_volume 621
_journal_year 1995
_journal_page_first 1153
_journal_page_last 1159
_publ_Section_title
;
 LiBC -- ein vollst\"{a}ndig interkalierter Heterographit
;

_aflow_proto 'ABC_hP6_194_c_d_a'
_aflow_params 'a,c/a'
_aflow_params_values '2.752,2.56468023256'
_aflow_Strukturbericht 'None'
_aflow_Pearson 'hP6'

_symmetry_space_group_name_Hall "-P 6c 2c"
_symmetry_space_group_name_H-M "P 63/m m c"
_symmetry_Int_Tables_number 194

_cell_length_a      2.75200
_cell_length_b      2.75200
_cell_length_c      7.05800
_cell_angle_alpha  90.00000
_cell_angle_beta   90.00000
_cell_angle_gamma 120.00000

loop_
_space_group_symop_id
_space_group_symop_operation_xyz
1 x,y,z
2 x-y,x,z+1/2
3 -y,x-y,z
4 -x,-y,z+1/2
5 -x+y,-x,z
6 y,-x+y,z+1/2
7 x-y,-y,-z
8 x,x-y,-z+1/2
9 y,x,-z
10 -x+y,y,-z+1/2
11 -x,-x+y,-z
12 -y,-x,-z+1/2
13 -x,-y,-z
14 -x+y,-x,-z+1/2
15 y,-x+y,-z
16 x,y,-z+1/2
17 x-y,x,-z
18 -y,x-y,-z+1/2
19 -x+y,y,z
20 -x,-x+y,z+1/2
21 -y,-x,z
22 x-y,-y,z+1/2
23 x,x-y,z
24 y,x,z+1/2

loop_
_atom_site_label
_atom_site_type_symbol
_atom_site_symmetry_multiplicity
```

```
_atom_site_Wyckoff_label
_atom_site_fract_x
_atom_site_fract_y
_atom_site_fract_z
_atom_site_occupancy
Li1  Li    2  a  0.00000  0.00000  0.00000  1.00000
B1   B     2  c  0.33333  0.66667  0.25000  1.00000
C1   C     2  d  0.33333  0.66667  0.75000  1.00000
```

LiBC: ABC_hP6_194_c_d_a - POSCAR

```
ABC_hP6_194_c_d_a & a,c/a --params=2.752,2.56468023256 & P6_3/mmc  D_{6h
    ↪ }^4 #194 (acd) & hP6 & & LiBC & & W\"orle et al., Z. Anorg.
    ↪ Allg. Chem. 621, 1153-1159 (1995)
   1.0000000000000000
   1.37600000000000   -2.38330191121500    0.00000000000000
   1.37600000000000    2.38330191121500    0.00000000000000
   0.00000000000000    0.00000000000000    7.05800000000000
    B     C    Li
    2     2     2
Direct
   0.33333333333333    0.66666666666667    0.25000000000000    B    (2c)
   0.66666666666667    0.33333333333333    0.75000000000000    B    (2c)
   0.33333333333333    0.66666666666667    0.75000000000000    C    (2d)
   0.66666666666667    0.33333333333333    0.25000000000000    C    (2d)
   0.00000000000000    0.00000000000000    0.00000000000000    Li   (2a)
   0.00000000000000    0.00000000000000    0.50000000000000    Li   (2a)
```

Lonsdaleite (Hexagonal Diamond): A_hP4_194_f - CIF

```
# CIF file

data_findsym-output
_audit_creation_method FINDSYM

_chemical_name_mineral 'Lonsdaleite'
_chemical_formula_sum 'C'

loop_
_publ_author_name
 'Akira Yoshiasa'
 'Yu Murai'
 'Osamu Ohtaka'
 'Tomoo Katsura'
_journal_name_full
;
Japanese Journal of Applied Physics
;
_journal_volume 42
_journal_year 2003
_journal_page_first 1694
_journal_page_last 1704
_publ_Section_title
;
 Detailed Structures of Hexagonal Diamond (lonsdaleite) and
      ↪ Wurtzite-type BN
;

_aflow_proto 'A_hP4_194_f'
_aflow_params 'a,c/a,z1'
_aflow_params_values '2.508,1.66786283892,0.05995'
_aflow_Strukturbericht 'None'
_aflow_Pearson 'hP4'

_symmetry_space_group_name_Hall "-P 6c 2c"
_symmetry_space_group_name_H-M "P 63/m m c"
_symmetry_Int_Tables_number 194

_cell_length_a      2.50800
_cell_length_b      2.50800
_cell_length_c      4.18300
_cell_angle_alpha  90.00000
_cell_angle_beta   90.00000
_cell_angle_gamma  120.00000

loop_
_space_group_symop_id
_space_group_symop_operation_xyz
1 x,y,z
2 x-y,x,z+1/2
3 -y,x-y,z
4 -x,-y,z+1/2
5 -x+y,-x,z
6 y,-x+y,z+1/2
7 x-y,-y,-z
8 x,x-y,-z+1/2
9 y,x,-z
10 -x+y,y,-z+1/2
11 -x,-x+y,-z
12 -y,-x,-z+1/2
13 -x,-y,-z
14 -x+y,-x,-z+1/2
15 y,-x+y,-z
16 x,y,-z+1/2
17 x-y,x,-z
18 -y,x-y,-z+1/2
19 -x+y,y,z
20 -x,-x+y,z+1/2
21 -y,-x,z
22 x-y,-y,z+1/2
23 x,x-y,z
24 y,x,z+1/2

loop_
_atom_site_label
_atom_site_type_symbol
_atom_site_symmetry_multiplicity
_atom_site_Wyckoff_label
_atom_site_fract_x
_atom_site_fract_y
_atom_site_fract_z
_atom_site_occupancy
C1 C    4  f  0.33333  0.66667  0.05995  1.00000
```

Lonsdaleite (Hexagonal Diamond): A_hP4_194_f - POSCAR

```
A_hP4_194_f & a,c/a,z1 --params=2.508,1.66786283892,0.05995 & P6_3/mmc
    ↪ D_{6h}^4 #194 (f) & hP4 & & C & Lonsdaleite (hexagonal diamond)
    ↪ & A. Yoshiasa, Y. Murai, O. Ohtaka and T. Katsura, Jpn. J.
    ↪ Appl. Phys. 42, 1694-1704 (2003)
   1.0000000000000000
   1.25400000000000   -2.17199171269100    0.00000000000000
   1.25400000000000    2.17199171269100    0.00000000000000
   0.00000000000000    0.00000000000000    4.18300000000000
    C
    4
Direct
   0.33333333333333    0.66666666666667    0.05994501553900    C    (4f)
   0.33333333333333    0.66666666666667    0.44005498446100    C    (4f)
   0.66666666666667    0.33333333333333   -0.05994501553900    C    (4f)
   0.66666666666667    0.33333333333333    0.55994501553900    C    (4f)
```

$Ni_2In$ ($B8_2$): AB2_hP6_194_c_ad - CIF

```
# CIF file

data_findsym-output
_audit_creation_method FINDSYM

_chemical_name_mineral ''
_chemical_formula_sum 'Ni2 In'

loop_
_publ_author_name
 'M. Ellner'
_journal_name_full
;
Journal of the Less Common Metals
;
_journal_volume 48
_journal_year 1976
_journal_page_first 21
_journal_page_last 52
_publ_Section_title
;
 "{U}ber die kristallchemischen parameter der Ni-, Co- und Fe-haltigen
      ↪ phasen vom NiAs-Typ
;

# Found in http://materials.springer.com/isp/crystallographic/docs/
    ↪ sd_0450621

_aflow_proto 'AB2_hP6_194_c_ad'
_aflow_params 'a,c/a'
_aflow_params_values '4.186,1.22527472527'
_aflow_Strukturbericht 'B8_2'
_aflow_Pearson 'hP6'

_symmetry_space_group_name_Hall "-P 6c 2c"
_symmetry_space_group_name_H-M "P 63/m m c"
_symmetry_Int_Tables_number 194

_cell_length_a      4.18600
_cell_length_b      4.18600
_cell_length_c      5.12900
_cell_angle_alpha  90.00000
_cell_angle_beta   90.00000
_cell_angle_gamma  120.00000

loop_
_space_group_symop_id
_space_group_symop_operation_xyz
1 x,y,z
2 x-y,x,z+1/2
3 -y,x-y,z
4 -x,-y,z+1/2
5 -x+y,-x,z
6 y,-x+y,z+1/2
7 x-y,-y,-z
8 x,x-y,-z+1/2
9 y,x,-z
10 -x+y,y,-z+1/2
11 -x,-x+y,-z
12 -y,-x,-z+1/2
13 -x,-y,-z
14 -x+y,-x,-z+1/2
15 y,-x+y,-z
16 x,y,-z+1/2
17 x-y,x,-z
18 -y,x-y,-z+1/2
19 -x+y,y,z
20 -x,-x+y,z+1/2
21 -y,-x,z
22 x-y,-y,z+1/2
23 x,x-y,z
24 y,x,z+1/2

loop_
_atom_site_label
_atom_site_type_symbol
```

```
_atom_site_symmetry_multiplicity
_atom_site_Wyckoff_label
_atom_site_fract_x
_atom_site_fract_y
_atom_site_fract_z
_atom_site_occupancy
Ni1 Ni    2 a 0.00000 0.00000 0.00000 1.00000
In1 In    2 c 0.33333 0.66667 0.25000 1.00000
Ni2 Ni    2 d 0.33333 0.66667 0.75000 1.00000
```

$Ni_2In$ ($B8_2$): AB2_hP6_194_c_ad - POSCAR

```
AB2_hP6_194_c_ad & a,c/a --params=4.186,1.22527472527 & P6_3/mmc  D_{6h
    ↪ }^4 #194 (acd) & hP6 & B8_2 & Ni2In & & M. Ellner, J.
    ↪ Less-Common Met. 48, 21-52 (1976)
   1.0000000000000000
   2.09300000000000   -3.62518234024200    0.00000000000000
   2.09300000000000    3.62518234024200    0.00000000000000
   0.00000000000000    0.00000000000000    5.12900000000000
    In     Ni
     2      4
Direct
   0.33333333333333    0.66666666666667    0.25000000000000   In    (2c)
   0.66666666666667    0.33333333333333    0.75000000000000   In    (2c)
   0.00000000000000    0.00000000000000    0.00000000000000   Ni    (2a)
   0.00000000000000    0.00000000000000    0.50000000000000   Ni    (2a)
   0.33333333333333    0.66666666666667    0.75000000000000   Ni    (2d)
   0.66666666666667    0.33333333333333    0.25000000000000   Ni    (2d)
```

$AlN_3Ti_4$: AB3C4_hP16_194_c_af_ef - CIF

```
# CIF file

data_findsym-output
_audit_creation_method FINDSYM

_chemical_name_mineral 'MAX Phase'
_chemical_formula_sum 'Al N3 Ti4'

loop_
_publ_author_name
 'M. W. Barsoum'
 'C. J. Rawn'
 'T. El-Raghy'
 'A. T. Procopio'
 'W. D. Porter'
 'H. Wang'
 'C. R. Hubbard'
_journal_name_full
;
Journal of Applied Physics
;
_journal_volume 87
_journal_year 2000
_journal_page_first 8407
_journal_page_last 8414
_publ_Section_title
;
 Thermal Properties of Ti$_4$AlN$_3$
;

_aflow_proto 'AB3C4_hP16_194_c_af_ef'
_aflow_params 'a,c/a,z3,z4,z5'
_aflow_params_values '2.988,7.82195448461,0.1543,0.605,0.0539'
_aflow_Strukturbericht 'None'
_aflow_Pearson 'hP16'

_symmetry_space_group_name_Hall "-P 6c 2c"
_symmetry_space_group_name_H-M "P 63/m m c"
_symmetry_Int_Tables_number 194

_cell_length_a      2.98800
_cell_length_b      2.98800
_cell_length_c      23.37200
_cell_angle_alpha  90.00000
_cell_angle_beta   90.00000
_cell_angle_gamma 120.00000

loop_
_space_group_symop_id
_space_group_symop_operation_xyz
1 x,y,z
2 x-y,x,z+1/2
3 -y,x-y,z
4 -x,-y,z+1/2
5 -x+y,-x,z
6 y,-x+y,z+1/2
7 x-y,-y,-z
8 x,x-y,-z+1/2
9 y,x,-z
10 -x+y,y,-z+1/2
11 -x,-x+y,-z
12 -y,-x,-z+1/2
13 -x,-y,-z
14 -x+y,-x,-z+1/2
15 y,-x+y,-z
16 x,y,-z+1/2
17 x-y,x,-z
18 -y,x-y,-z+1/2
19 -x+y,y,z
20 -x,-x+y,z+1/2
21 -y,-x,z
22 x-y,-y,z+1/2
23 x,x-y,z
24 y,x,z+1/2
```

```
loop_
_atom_site_label
_atom_site_type_symbol
_atom_site_symmetry_multiplicity
_atom_site_Wyckoff_label
_atom_site_fract_x
_atom_site_fract_y
_atom_site_fract_z
_atom_site_occupancy
N1  N     2 a 0.00000 0.00000 0.00000 1.00000
Al1 Al    2 c 0.33333 0.66667 0.25000 1.00000
Ti1 Ti    4 e 0.00000 0.00000 0.15430 1.00000
N2  N     4 f 0.33333 0.66667 0.60500 1.00000
Ti2 Ti    4 f 0.33333 0.66667 0.05390 1.00000
```

$AlN_3Ti_4$: AB3C4_hP16_194_c_af_ef - POSCAR

```
AB3C4_hP16_194_c_af_ef & a,c/a,z3,z4,z5 --params=2.988,7.82195448461,
    ↪ 0.1543,0.605,0.0539 & P6_3/mmc  D_{6h}^4 #194 (acef^2) & hP16 &
    ↪  & AlN3Ti4 & MAX Phase & M. W. Barsoum et al., JAP 87,
    ↪ 8407-8414 (2000)
   1.0000000000000000
   1.49400000000000   -2.58768390650800    0.00000000000000
   1.49400000000000    2.58768390650800    0.00000000000000
   0.00000000000000    0.00000000000000   23.37200000000000
    Al     N     Ti
     2      6      8
Direct
   0.33333333333333    0.66666666666667    0.25000000000000   Al    (2c)
   0.66666666666667    0.33333333333333    0.75000000000000   Al    (2c)
   0.00000000000000    0.00000000000000    0.00000000000000    N    (2a)
   0.00000000000000    0.00000000000000    0.50000000000000    N    (2a)
   0.33333333333333    0.66666666666667    0.60500000000000    N    (4f)
   0.33333333333333    0.66666666666667    0.89500000000000    N    (4f)
   0.66666666666667    0.33333333333333    0.10500000000000    N    (4f)
   0.66666666666667    0.33333333333333    0.39500000000000    N    (4f)
   0.00000000000000    0.00000000000000    0.15430000000000   Ti    (4e)
   0.00000000000000    0.00000000000000    0.34570000000000   Ti    (4e)
   0.00000000000000    0.00000000000000    0.65430000000000   Ti    (4e)
   0.00000000000000    0.00000000000000    0.84570000000000   Ti    (4e)
   0.33333333333333    0.66666666666667    0.05390000000000   Ti    (4f)
   0.33333333333333    0.66666666666667    0.44610000000000   Ti    (4f)
   0.66666666666667    0.33333333333333   -0.05390000000000   Ti    (4f)
   0.66666666666667    0.33333333333333    0.55390000000000   Ti    (4f)
```

Hexagonal Close Packed (Mg, A3): A_hP2_194_c - CIF

```
# CIF file

data_findsym-output
_audit_creation_method FINDSYM

_chemical_name_mineral 'Magnesium'
_chemical_formula_sum 'Mg'

loop_
_publ_author_name
 'F. W. von Batchelder'
 'R. F. Raeuchle'
_journal_name_full
;
Physical Review
;
_journal_volume 105
_journal_year 1957
_journal_page_first 59
_journal_page_last 61
_publ_Section_title
;
 Lattice Constants and Brillouin Zone Overlap in Dilute Magnesium Alloys
;

# Found in Donohue, pp. 39-40

_aflow_proto 'A_hP2_194_c'
_aflow_params 'a,c/a'
_aflow_params_values '3.2093,1.62359393014'
_aflow_Strukturbericht 'A3'
_aflow_Pearson 'hP2'

_symmetry_space_group_name_Hall "-P 6c 2c"
_symmetry_space_group_name_H-M "P 63/m m c"
_symmetry_Int_Tables_number 194

_cell_length_a      3.20930
_cell_length_b      3.20930
_cell_length_c      5.21060
_cell_angle_alpha  90.00000
_cell_angle_beta   90.00000
_cell_angle_gamma 120.00000

loop_
_space_group_symop_id
_space_group_symop_operation_xyz
1 x,y,z
2 x-y,x,z+1/2
3 -y,x-y,z
4 -x,-y,z+1/2
5 -x+y,-x,z
6 y,-x+y,z+1/2
7 x-y,-y,-z
8 x,x-y,-z+1/2
9 y,x,-z
10 -x+y,y,-z+1/2
```

```
11 -x,-x+y,-z
12 -y,-x,-z+1/2
13 -x,-y,-z
14 -x+y,-x,-z+1/2
15 y,-x+y,-z
16 x,y,-z+1/2
17 x-y,x,-z
18 -y,x-y,-z+1/2
19 -x+y,y,z
20 -x,-x+y,z+1/2
21 -y,-x,z
22 x-y,-y,z+1/2
23 x,x-y,z
24 y,x,z+1/2

loop_
_atom_site_label
_atom_site_type_symbol
_atom_site_symmetry_multiplicity
_atom_site_Wyckoff_label
_atom_site_fract_x
_atom_site_fract_y
_atom_site_fract_z
_atom_site_occupancy
Mg1 Mg    2 c 0.33333  0.66667  0.25000  1.00000
```

Hexagonal Close Packed (Mg, A3): A_hP2_194_c - POSCAR

```
A_hP2_194_c & a,c/a --params=3.2093,1.62359393014 & P6_3/mmc  D_{6h}^4 #
    ↪ 194 (c) & hP2 & A3 & Mg & hcp & F. W. von Batchelder and R. F.
    ↪ Raeuchle, Phys. Rev. 105, 59-61 (1957)
   1.0000000000000000
   1.60465000000000  -2.77933532836500   0.00000000000000
   1.60465000000000   2.77933532836500   0.00000000000000
   0.00000000000000   0.00000000000000   5.21060000000000
  Mg
   2
Direct
   0.33333333333333   0.66666666666667   0.25000000000000   Mg    (2c)
   0.66666666666667   0.33333333333333   0.75000000000000   Mg    (2c)
```

$MgNi_2$ Hexagonal Laves (C36): AB2_hP24_194_ef_fgh - CIF

```
# CIF file

data_findsym-output
_audit_creation_method FINDSYM

_chemical_name_mineral 'Hexagonal Laves'
_chemical_formula_sum 'Mg Ni2'

loop_
_publ_author_name
 'Y. Komura'
 'K. Tokunaga'
_journal_name_full
;
Acta Crystallographica B
;
_journal_volume 36
_journal_year 1980
_journal_page_first 1548
_journal_page_last 1554
_publ_Section_title
;
 Structural studies of stacking variants in Mg-base Friauf-Laves phases
;

_aflow_proto 'AB2_hP24_194_ef_fgh'
_aflow_params 'a,c/a,z1,z2,z3,x5'
_aflow_params_values '4.824,3.28067993367,0.04598,0.84417,0.12514,
    ↪ 0.16429'
_aflow_Strukturbericht 'C36'
_aflow_Pearson 'hP24'

_symmetry_space_group_name_Hall "-P 6c 2c"
_symmetry_space_group_name_H-M "P 63/m m c"
_symmetry_Int_Tables_number 194

_cell_length_a     4.82400
_cell_length_b     4.82400
_cell_length_c     15.82600
_cell_angle_alpha  90.00000
_cell_angle_beta   90.00000
_cell_angle_gamma  120.00000

loop_
_space_group_symop_id
_space_group_symop_operation_xyz
1 x,y,z
2 x-y,x,z+1/2
3 -y,x-y,z
4 -x,-y,z+1/2
5 -x+y,-x,z
6 y,-x+y,z+1/2
7 x-y,-y,-z
8 x,x-y,-z+1/2
9 y,x,-z
10 -x+y,y,-z+1/2
11 -x,-x+y,-z
12 -y,-x,-z+1/2
13 -x,-y,-z
14 -x+y,-x,-z+1/2
15 y,-x+y,-z
16 x,y,-z+1/2
17 x-y,x,-z
18 -y,x-y,-z+1/2
19 -x+y,y,z
20 -x,-x+y,z+1/2
21 -y,-x,z
22 x-y,-y,z+1/2
23 x,x-y,z
24 y,x,z+1/2

loop_
_atom_site_label
_atom_site_type_symbol
_atom_site_symmetry_multiplicity
_atom_site_Wyckoff_label
_atom_site_fract_x
_atom_site_fract_y
_atom_site_fract_z
_atom_site_occupancy
Mg1 Mg    4 e 0.00000  0.00000  0.04598  1.00000
Mg2 Mg    4 f 0.33333  0.66667  0.84417  1.00000
Ni1 Ni    4 f 0.33333  0.66667  0.12514  1.00000
Ni2 Ni    6 g 0.50000  0.00000  0.00000  1.00000
Ni3 Ni    6 h 0.16429  0.32858  0.25000  1.00000
```

$MgNi_2$ Hexagonal Laves (C36): AB2_hP24_194_ef_fgh - POSCAR

```
AB2_hP24_194_ef_fgh & a,c/a,z1,z2,z3,x5 --params=4.824,3.28067993367,
    ↪ 0.04598,0.84417,0.12514,0.16429 & P6_3/mmc  D_{6h}^4 #194 (ef^
    ↪ 2gh) & hP24 & C36 & MgNi2 & Laves & Y. Komura and K. Tokunaga,
    ↪ Acta Cryst. B 36, 1548-1554 (1980)
   1.0000000000000000
   2.41200000000000  -4.17770654785613   0.00000000000000
   2.41200000000000   4.17770654785613   0.00000000000000
   0.00000000000000   0.00000000000000  15.82600000000000
  Mg    Ni
   8    16
Direct
   0.00000000000000   0.00000000000000   0.04598000000000   Mg    (4e)
   0.00000000000000   0.00000000000000  -0.04598000000000   Mg    (4e)
   0.00000000000000   0.00000000000000   0.45402000000000   Mg    (4e)
   0.00000000000000   0.00000000000000   0.54598000000000   Mg    (4e)
   0.33333333333333   0.66666666666667   0.65583000000000   Mg    (4f)
   0.33333333333333   0.66666666666667   0.84417000000000   Mg    (4f)
   0.66666666666667   0.33333333333333   0.15583000000000   Mg    (4f)
   0.66666666666667   0.33333333333333   0.34417000000000   Mg    (4f)
   0.33333333333333   0.66666666666667   0.12514000000000   Ni    (4f)
   0.33333333333333   0.66666666666667   0.37486000000000   Ni    (4f)
   0.66666666666667   0.33333333333333   0.62514000000000   Ni    (4f)
   0.66666666666667   0.33333333333333   0.87486000000000   Ni    (4f)
   0.00000000000000   0.50000000000000   0.00000000000000   Ni    (6g)
   0.00000000000000   0.50000000000000   0.50000000000000   Ni    (6g)
   0.50000000000000   0.00000000000000   0.00000000000000   Ni    (6g)
   0.50000000000000   0.00000000000000   0.50000000000000   Ni    (6g)
   0.50000000000000   0.50000000000000   0.00000000000000   Ni    (6g)
   0.50000000000000   0.50000000000000   0.50000000000000   Ni    (6g)
   0.16429000000000   0.32858000000000   0.25000000000000   Ni    (6h)
   0.16429000000000   0.83571000000000   0.25000000000000   Ni    (6h)
   0.32858000000000   0.16429000000000   0.75000000000000   Ni    (6h)
   0.67142000000000   0.83571000000000   0.25000000000000   Ni    (6h)
   0.83571000000000   0.16429000000000   0.75000000000000   Ni    (6h)
   0.83571000000000   0.67142000000000   0.75000000000000   Ni    (6h)
```

Covellite (CuS, B18): AB_hP12_194_df_ce - CIF

```
# CIF file

data_findsym-output
_audit_creation_method FINDSYM

_chemical_name_mineral 'Covellite'
_chemical_formula_sum 'Cu S'

loop_
_publ_author_name
 'Masaaki Ohmasa'
 'Masatoshi Suzuki'
 'Yoshio Tak\'{e}uchi'
_journal_name_full
;
Mineralogical Journal
;
_journal_volume 8
_journal_year 1977
_journal_page_first 311
_journal_page_last 319
_publ_Section_title
;
 A refinement of the crystal structure of covellite, CuS
;

_aflow_proto 'AB_hP12_194_df_ce'
_aflow_params 'a,c/a,z3,z4'
_aflow_params_values '3.976,4.12022132797,0.0637,0.10724'
_aflow_Strukturbericht 'B18'
_aflow_Pearson 'hP12'

_symmetry_space_group_name_Hall "-P 6c 2c"
_symmetry_space_group_name_H-M "P 63/m m c"
_symmetry_Int_Tables_number 194

_cell_length_a     3.97600
_cell_length_b     3.97600
_cell_length_c     16.38200
_cell_angle_alpha  90.00000
_cell_angle_beta   90.00000
```

```
_cell_angle_gamma  120.00000

loop_
_space_group_symop_id
_space_group_symop_operation_xyz
1  x,y,z
2  x-y,x,z+1/2
3  -y,x-y,z
4  -x,-y,z+1/2
5  -x+y,-x,z
6  y,-x+y,z+1/2
7  x-y,-y,-z
8  x,x-y,-z+1/2
9  y,x,-z
10  -x+y,y,-z+1/2
11  -x,-x+y,-z
12  -y,-x,-z+1/2
13  -x,-y,-z
14  -x+y,-x,-z+1/2
15  y,-x+y,-z
16  x,y,-z+1/2
17  x-y,x,-z
18  -y,x-y,-z+1/2
19  -x+y,y,z
20  -x,-x+y,z+1/2
21  -y,-x,z
22  x-y,-y,z+1/2
23  x,x-y,z
24  y,x,z+1/2

loop_
_atom_site_label
_atom_site_type_symbol
_atom_site_symmetry_multiplicity
_atom_site_Wyckoff_label
_atom_site_fract_x
_atom_site_fract_y
_atom_site_fract_z
_atom_site_occupancy
S1   S     2  c  0.33333  0.66667  0.25000  1.00000
Cu1  Cu    2  d  0.33333  0.66667  0.75000  1.00000
S2   S     4  e  0.00000  0.00000  0.06370  1.00000
Cu2  Cu    4  f  0.33333  0.66667  0.10724  1.00000
```

Covellite (CuS, B18): AB_hP12_194_df_ce - POSCAR

```
AB_hP12_194_df_ce & a,c/a,z3,z4 --params=3.976,4.12022132797,0.0637,
    ↪ 0.10724 & P6_3/mmc  D_{6h}^4 #194 (cdef) & hP12 & B18 & CuS &
    ↪ Covellite & M. Ohmasa, M. Suzuki and Y. Tak\'{e}uchi,
    ↪ Mineralogical Journal 8, 311-319 (1977)
   1.0000000000000000
   1.98800000000000   -3.44331700544700    0.00000000000000
   1.98800000000000    3.44331700544700    0.00000000000000
   0.00000000000000    0.00000000000000   16.38200000000000
   Cu      S
    6      6
Direct
   0.33333333333300    0.66666666666700    0.75000000000000    Cu    (2d)
   0.66666666666700    0.33333333333300    0.25000000000000    Cu    (2d)
   0.33333333333300    0.66666666666700    0.10724000000000    Cu    (4f)
   0.33333333333300    0.66666666666700    0.39276000000000    Cu    (4f)
   0.66666666666700    0.33333333333300    0.60724000000000    Cu    (4f)
   0.66666666666700    0.33333333333300    0.89276000000000    Cu    (4f)
   0.33333333333300    0.66666666666700    0.25000000000000    S     (2c)
   0.66666666666700    0.33333333333300    0.75000000000000    S     (2c)
   0.00000000000000    0.00000000000000    0.06370000000000    S     (4e)
   0.00000000000000    0.00000000000000   -0.06370000000000    S     (4e)
   0.00000000000000    0.00000000000000    0.43630000000000    S     (4e)
   0.00000000000000    0.00000000000000    0.56370000000000    S     (4e)
```

NiAs ($B8_1$): AB_hP4_194_c_a - CIF

```
# CIF file

data_findsym-output
_audit_creation_method FINDSYM

_chemical_name_mineral ''
_chemical_formula_sum 'Ni As'

loop_
_publ_author_name
 'P. Brand'
 'J. Briest'
_journal_name_full
;
Zeitschrift f\"{u}r anorganische und allgemeine Chemie
;
_journal_volume 337
_journal_year 1965
_journal_page_first 209
_journal_page_last 213
_publ_Section_title
;
 Das quasi-bin\"{a}re System NiAs--Ni$_{1.5}$Sn
;

# Found in  Pearson's Handbook, Vol. I, pp. 1192

_aflow_proto 'AB_hP4_194_c_a'
_aflow_params 'a,c/a'
_aflow_params_values '3.619,1.39375518099'
_aflow_Strukturbericht 'B8_1'
_aflow_Pearson 'hP4'

_symmetry_space_group_name_Hall "-P 6c 2c"
_symmetry_space_group_name_H-M "P 63/m m c"
_symmetry_Int_Tables_number 194

_cell_length_a     3.61900
_cell_length_b     3.61900
_cell_length_c     5.04400
_cell_angle_alpha  90.00000
_cell_angle_beta   90.00000
_cell_angle_gamma  120.00000

loop_
_space_group_symop_id
_space_group_symop_operation_xyz
1  x,y,z
2  x-y,x,z+1/2
3  -y,x-y,z
4  -x,-y,z+1/2
5  -x+y,-x,z
6  y,-x+y,z+1/2
7  x-y,-y,-z
8  x,x-y,-z+1/2
9  y,x,-z
10  -x+y,y,-z+1/2
11  -x,-x+y,-z
12  -y,-x,-z+1/2
13  -x,-y,-z
14  -x+y,-x,-z+1/2
15  y,-x+y,-z
16  x,y,-z+1/2
17  x-y,x,-z
18  -y,x-y,-z+1/2
19  -x+y,y,z
20  -x,-x+y,z+1/2
21  -y,-x,z
22  x-y,-y,z+1/2
23  x,x-y,z
24  y,x,z+1/2

loop_
_atom_site_label
_atom_site_type_symbol
_atom_site_symmetry_multiplicity
_atom_site_Wyckoff_label
_atom_site_fract_x
_atom_site_fract_y
_atom_site_fract_z
_atom_site_occupancy
Ni1  Ni    2  a  0.00000  0.00000  0.00000  1.00000
As1  As    2  c  0.33333  0.66667  0.25000  1.00000
```

NiAs ($B8_1$): AB_hP4_194_c_a - POSCAR

```
AB_hP4_194_c_a & a,c/a --params=3.619,1.39375518099 & P6_3/mmc  D_{6h}^4
    ↪ #194 (ac) & hP4 & B8_1 & NiAs & & P. Brand and J. Briest, ZAAC
    ↪ 337, 209-213 (1965)
   1.0000000000000000
   1.80950000000000   -3.13414593629600    0.00000000000000
   1.80950000000000    3.13414593629600    0.00000000000000
   0.00000000000000    0.00000000000000    5.04400000000000
   As      Ni
    2      2
Direct
   0.33333333333333    0.66666666666667    0.25000000000000    As    (2c)
   0.66666666666667    0.33333333333333    0.75000000000000    As    (2c)
   0.00000000000000    0.00000000000000    0.00000000000000    Ni    (2a)
   0.00000000000000    0.00000000000000    0.50000000000000    Ni    (2a)
```

$\beta$-Tridymite ($SiO_2$) (C10): A2B_hP12_194_cg_f - CIF

```
# CIF file

data_findsym-output
_audit_creation_method FINDSYM

_chemical_name_mineral 'beta Tridymite'
_chemical_formula_sum 'Si O2'

loop_
_publ_author_name
 'Kuniaki Kihara'
_journal_name_full
;
Zeitschrift f\"{u}r Kristallographie
;
_journal_volume 148
_journal_year 1978
_journal_page_first 237
_journal_page_last 253
_publ_Section_title
;
 Thermal change in unit-cell dimensions, and a hexagonal structure of
    ↪ tridymite
;

# Found in  Pearson's Handbook, Vol. IV, pp. 4759

_aflow_proto 'A2B_hP12_194_cg_f'
_aflow_params 'a,c/a,z2'
_aflow_params_values '5.052,1.63697545527,0.062'
_aflow_Strukturbericht 'C10'
_aflow_Pearson 'hP12'

_symmetry_space_group_name_Hall "-P 6c 2c"
_symmetry_space_group_name_H-M "P 63/m m c"
```

```
_symmetry_Int_Tables_number 194

_cell_length_a     5.05200
_cell_length_b     5.05200
_cell_length_c     8.27000
_cell_angle_alpha  90.00000
_cell_angle_beta   90.00000
_cell_angle_gamma  120.00000

loop_
_space_group_symop_id
_space_group_symop_operation_xyz
1 x,y,z
2 x-y,x,z+1/2
3 -y,x-y,z
4 -x,-y,z+1/2
5 -x+y,-x,z
6 y,-x+y,z+1/2
7 x-y,-y,-z
8 x,x-y,-z+1/2
9 y,x,-z
10 -x+y,y,-z+1/2
11 -x,-x+y,-z
12 -y,-x,-z+1/2
13 -x,-y,-z
14 -x+y,-x,-z+1/2
15 y,-x+y,-z
16 x,y,-z+1/2
17 x-y,x,-z
18 -y,x-y,-z+1/2
19 -x+y,y,z
20 -x,-x+y,z+1/2
21 -y,-x,z
22 x-y,-y,z+1/2
23 x,x-y,z
24 y,x,z+1/2

loop_
_atom_site_label
_atom_site_type_symbol
_atom_site_symmetry_multiplicity
_atom_site_Wyckoff_label
_atom_site_fract_x
_atom_site_fract_y
_atom_site_fract_z
_atom_site_occupancy
O1  O    2 c 0.33333 0.66667 0.25000 1.00000
Si1 Si   4 f 0.33333 0.66667 0.06200 1.00000
O2  O    6 g 0.50000 0.00000 0.00000 1.00000
```

β-Tridymite ($SiO_2$) (C10): A2B_hP12_194_cg_f - POSCAR

```
A2B_hP12_194_cg_f & a,c/a,z2 --params=5.052,1.63697545527,0.062 & P6_3/
    ↪ mmc  D_{6h}^4 #194 (cfg) & hP12 & C10 & SiO2 & beta tridymite &
    ↪ K. Kihara, Zeitschrift f\"{u}r Kristallographie 148, 237-253 (
    ↪ 1978)
   1.0000000000000000
   2.52600000000000   -4.37516033991899    0.00000000000000
   2.52600000000000    4.37516033991899    0.00000000000000
   0.00000000000000    0.00000000000000    8.27000000000000
    O    Si
    8     4
Direct
   0.33333333333333   0.66666666666667   0.25000000000000   O   (2c)
   0.66666666666667   0.33333333333333   0.75000000000000   O   (2c)
   0.00000000000000   0.50000000000000   0.00000000000000   O   (6g)
   0.00000000000000   0.50000000000000   0.50000000000000   O   (6g)
   0.50000000000000   0.00000000000000   0.00000000000000   O   (6g)
   0.50000000000000   0.00000000000000   0.50000000000000   O   (6g)
   0.50000000000000   0.50000000000000   0.00000000000000   O   (6g)
   0.50000000000000   0.50000000000000   0.50000000000000   O   (6g)
   0.33333333333333   0.66666666666667   0.06200000000000   Si  (4f)
   0.33333333333333   0.66666666666667   0.43800000000000   Si  (4f)
   0.66666666666667   0.33333333333333  -0.06200000000000   Si  (4f)
   0.66666666666667   0.33333333333333   0.56200000000000   Si  (4f)
```

$Ga_4Ni$: A4B_cI40_197_cde_c - CIF

```
# CIF file

data_findsym-output
_audit_creation_method FINDSYM

_chemical_name_mineral ''
_chemical_formula_sum 'Ga4 Ni'

loop_
_publ_author_name
 'Liang Jingkui'
 'Xie Sishen'
_journal_name_full
;
Scientia Sinica, Series A: Mathematical, Physical, Astronomical and
    ↪ Technical Sciences, English Edition
;
_journal_volume 26
_journal_year 1983
_journal_page_first 1305
_journal_page_last 1313
_publ_Section_title
;
 The Structure of NiGa$_4$ Crystal -- A New Vacancy Controlled $\
    ↪ gamma$-Brass Phase
;
```

```
# Found in  Pearson's Handbook Vol. III, pp. 3509

_aflow_proto 'A4B_cI40_197_cde_c'
_aflow_params 'a,x1,x2,x3,x4'
_aflow_params_values '8.4295,0.1668,0.3345,0.6476,0.7484'
_aflow_Strukturbericht 'None'
_aflow_Pearson 'cI40'

_symmetry_space_group_name_Hall "I 2 2 3"
_symmetry_space_group_name_H-M "I 2 3"
_symmetry_Int_Tables_number 197

_cell_length_a     8.42950
_cell_length_b     8.42950
_cell_length_c     8.42950
_cell_angle_alpha  90.00000
_cell_angle_beta   90.00000
_cell_angle_gamma  90.00000

loop_
_space_group_symop_id
_space_group_symop_operation_xyz
1 x,y,z
2 x,-y,-z
3 -x,y,-z
4 -x,-y,z
5 y,z,x
6 y,-z,-x
7 -y,z,-x
8 -y,-z,x
9 z,x,y
10 z,-x,-y
11 -z,x,-y
12 -z,-x,y
13 x+1/2,y+1/2,z+1/2
14 x+1/2,-y+1/2,-z+1/2
15 -x+1/2,y+1/2,-z+1/2
16 -x+1/2,-y+1/2,z+1/2
17 y+1/2,z+1/2,x+1/2
18 y+1/2,-z+1/2,-x+1/2
19 -y+1/2,z+1/2,-x+1/2
20 -y+1/2,-z+1/2,x+1/2
21 z+1/2,x+1/2,y+1/2
22 z+1/2,-x+1/2,-y+1/2
23 -z+1/2,x+1/2,-y+1/2
24 -z+1/2,-x+1/2,y+1/2

loop_
_atom_site_label
_atom_site_type_symbol
_atom_site_symmetry_multiplicity
_atom_site_Wyckoff_label
_atom_site_fract_x
_atom_site_fract_y
_atom_site_fract_z
_atom_site_occupancy
Ga1 Ga    8 c 0.16680 0.16680 0.16680 1.00000
Ni1 Ni    8 c 0.33450 0.33450 0.33450 1.00000
Ga2 Ga   12 d 0.64760 0.00000 0.00000 1.00000
Ga3 Ga   12 e 0.74840 0.50000 0.00000 1.00000
```

$Ga_4Ni$: A4B_cI40_197_cde_c - POSCAR

```
A4B_cI40_197_cde_c & a,x1,x2,x3,x4 --params=8.4295,0.1668,0.3345,0.6476,
    ↪ 0.7484 & I23       T^3 #197 (c^2de) & cI40 & & Ga4Ni & & Liang
    ↪ Jingkui and Xie Sishen, Scientia Sinica A 26, 1305-1313 (1983)
   1.0000000000000000
  -4.21475000000000    4.21475000000000    4.21475000000000
   4.21475000000000   -4.21475000000000    4.21475000000000
   4.21475000000000    4.21475000000000   -4.21475000000000
    Ga    Ni
    16     4
Direct
   0.00000000000000   0.35240000000000   0.35240000000000   Ga   (12d)
   0.00000000000000   0.64760000000000   0.64760000000000   Ga   (12d)
   0.35240000000000   0.00000000000000   0.35240000000000   Ga   (12d)
   0.35240000000000   0.35240000000000   0.00000000000000   Ga   (12d)
   0.64760000000000   0.00000000000000   0.64760000000000   Ga   (12d)
   0.64760000000000   0.64760000000000   0.00000000000000   Ga   (12d)
   0.24840000000000   0.50000000000000   0.74840000000000   Ga   (12e)
   0.25160000000000   0.75160000000000   0.50000000000000   Ga   (12e)
   0.50000000000000   0.25160000000000   0.75160000000000   Ga   (12e)
   0.50000000000000   0.74840000000000   0.24840000000000   Ga   (12e)
   0.74840000000000   0.24840000000000   0.50000000000000   Ga   (12e)
   0.75160000000000   0.50000000000000   0.25160000000000   Ga   (12e)
   0.00000000000000   0.00000000000000   0.66640000000000   Ga    (8c)
   0.00000000000000   0.66640000000000   0.00000000000000   Ga    (8c)
   0.33360000000000   0.33360000000000   0.33360000000000   Ga    (8c)
   0.66640000000000   0.00000000000000   0.00000000000000   Ga    (8c)
   0.00000000000000   0.00000000000000   0.33100000000000   Ni    (8c)
   0.00000000000000   0.33100000000000   0.00000000000000   Ni    (8c)
   0.33100000000000   0.00000000000000   0.00000000000000   Ni    (8c)
   0.66900000000000   0.66900000000000   0.66900000000000   Ni    (8c)
```

Ullmanite (NiSSb, $F0_1$): ABC_cP12_198_a_a_a - CIF

```
# CIF file

data_findsym-output
_audit_creation_method FINDSYM

_chemical_name_mineral 'Ullmanite'
_chemical_formula_sum 'Ni S Sb'

loop_
```

```
_publ_author_name
 'Yoshio Tak\'{e}uchi'
_journal_name_full
;
Mineralogical Journal
;
_journal_volume 2
_journal_year 1957
_journal_page_first 90
_journal_page_last 102
_publ_Section_title
;
 The Absolute Structure of Ullmanite, NiSbS
;

_aflow_proto 'ABC_cP12_198_a_a_a'
_aflow_params 'a,x1,x2,x3'
_aflow_params_values '5.881,-0.024,0.39,0.875'
_aflow_Strukturbericht 'F0_1'
_aflow_Pearson 'cP12'

_symmetry_space_group_name_Hall "P 2ac 2ab 3 P2_13"
_symmetry_space_group_name_H-M "P 21 3"
_symmetry_Int_Tables_number 198

_cell_length_a     5.88100
_cell_length_b     5.88100
_cell_length_c     5.88100
_cell_angle_alpha 90.00000
_cell_angle_beta  90.00000
_cell_angle_gamma 90.00000

loop_
_space_group_symop_id
_space_group_symop_operation_xyz
1 x,y,z
2 x+1/2,-y+1/2,-z
3 -x,y+1/2,-z+1/2
4 -x+1/2,-y,z+1/2
5 y,z,x
6 y+1/2,-z+1/2,-x
7 -y,z+1/2,-x+1/2
8 -y+1/2,-z,x+1/2
9 z,x,y
10 z+1/2,-x+1/2,-y
11 -z,x+1/2,-y+1/2
12 -z+1/2,-x,y+1/2

loop_
_atom_site_label
_atom_site_type_symbol
_atom_site_symmetry_multiplicity
_atom_site_Wyckoff_label
_atom_site_fract_x
_atom_site_fract_y
_atom_site_fract_z
_atom_site_occupancy
Ni1 Ni   4 a -0.02400 -0.02400 -0.02400 1.00000
S1  S    4 a 0.39000  0.39000  0.39000  1.00000
Sb1 Sb   4 a 0.87500  0.87500  0.87500  1.00000
```

Ullmanite (NiSbS, $F0_1$): ABC_cP12_198_a_a_a - POSCAR

```
ABC_cP12_198_a_a_a & a,x1,x2,x3 --params=5.881,-0.024,0.39,0.875 & P2_13
    ↪     T^4 #198 (a^3) & cP12 & F0_1 & NiSSb & Ullmanite & Y.
    ↪ Takeuchi, Mineralogical Journal 2, 90-102 (1957)
   1.0000000000000000
   5.88100000000000   0.00000000000000   0.00000000000000
   0.00000000000000   5.88100000000000   0.00000000000000
   0.00000000000000   0.00000000000000   5.88100000000000
   Ni    S    Sb
    4    4     4
Direct
  -0.02400000000000  -0.02400000000000  -0.02400000000000   Ni    (4a)
   0.02400000000000   0.47600000000000   0.52400000000000   Ni    (4a)
   0.47600000000000   0.52400000000000   0.02400000000000   Ni    (4a)
   0.52400000000000   0.02400000000000   0.47600000000000   Ni    (4a)
   0.11000000000000   0.61000000000000   0.89000000000000    S    (4a)
   0.39000000000000   0.39000000000000   0.39000000000000    S    (4a)
   0.61000000000000   0.89000000000000   0.11000000000000    S    (4a)
   0.89000000000000   0.11000000000000   0.61000000000000    S    (4a)
   0.12500000000000   0.37500000000000   0.62500000000000   Sb    (4a)
   0.37500000000000   0.62500000000000   0.12500000000000   Sb    (4a)
   0.62500000000000   0.12500000000000   0.37500000000000   Sb    (4a)
   0.87500000000000   0.87500000000000   0.87500000000000   Sb    (4a)
```

Ammonia ($NH_3$, D1): A3B_cP16_198_b_a - CIF

```
# CIF file

data_findsym-output
_audit_creation_method FINDSYM

_chemical_name_mineral 'Ammonia'
_chemical_formula_sum 'N H3'

loop_
_publ_author_name
 'R. Boese'
 'N. Niederpr\"{u}m'
 'D. Bl\"{a}ser'
 'Andreas Maulitz'
 'Mikhael Yu. Antipin'
 'Paul R. Mallinson'
_journal_name_full
;
Journal of Physical Chemistry B
;
_journal_volume 101
_journal_year 1997
_journal_page_first 5794
_journal_page_last 5799
_publ_Section_title
;
 Single-Crystal Structure and Electron Density Distribution of Ammonia
    ↪ at 160 K on the Basis of X-ray Diffraction Data
;

_aflow_proto 'A3B_cP16_198_b_a'
_aflow_params 'a,x1,x2,y2,z2'
_aflow_params_values '5.1305,0.2107,0.3689,0.2671,0.1159'
_aflow_Strukturbericht 'D1'
_aflow_Pearson 'cP16'

_symmetry_space_group_name_Hall "P 2ac 2ab 3 P2_13"
_symmetry_space_group_name_H-M "P 21 3"
_symmetry_Int_Tables_number 198

_cell_length_a     5.13050
_cell_length_b     5.13050
_cell_length_c     5.13050
_cell_angle_alpha 90.00000
_cell_angle_beta  90.00000
_cell_angle_gamma 90.00000

loop_
_space_group_symop_id
_space_group_symop_operation_xyz
1 x,y,z
2 x+1/2,-y+1/2,-z
3 -x,y+1/2,-z+1/2
4 -x+1/2,-y,z+1/2
5 y,z,x
6 y+1/2,-z+1/2,-x
7 -y,z+1/2,-x+1/2
8 -y+1/2,-z,x+1/2
9 z,x,y
10 z+1/2,-x+1/2,-y
11 -z,x+1/2,-y+1/2
12 -z+1/2,-x,y+1/2

loop_
_atom_site_label
_atom_site_type_symbol
_atom_site_symmetry_multiplicity
_atom_site_Wyckoff_label
_atom_site_fract_x
_atom_site_fract_y
_atom_site_fract_z
_atom_site_occupancy
N1 N    4 a 0.21070  0.21070  0.21070  1.00000
H1 H   12 b 0.36890  0.26710  0.11590  1.00000
```

Ammonia ($NH_3$, D1): A3B_cP16_198_b_a - POSCAR

```
A3B_cP16_198_b_a & a,x1,x2,y2,z2 --params=5.1305,0.2107,0.3689,0.2671,
    ↪ 0.1159 & P2_13     T^4 #198 (ab) & cP16 & D1 & NH3 & Ammonia &
    ↪ R. Boese et al., J. Phys. Chem. B 101, 5794-5799 (1997)
   1.0000000000000000
   5.13050000000000   0.00000000000000   0.00000000000000
   0.00000000000000   5.13050000000000   0.00000000000000
   0.00000000000000   0.00000000000000   5.13050000000000
    H     N
   12     4
Direct
   0.11590000000000   0.36890000000000   0.26710000000000    H   (12b)
  -0.11590000000000   0.86890000000000   0.23290000000000    H   (12b)
   0.13110000000000  -0.26710000000000   0.61590000000000    H   (12b)
   0.23290000000000  -0.11590000000000   0.86890000000000    H   (12b)
   0.26710000000000   0.11590000000000   0.36890000000000    H   (12b)
  -0.26710000000000   0.61590000000000   0.13110000000000    H   (12b)
   0.36890000000000   0.26710000000000   0.11590000000000    H   (12b)
  -0.36890000000000   0.76710000000000   0.38410000000000    H   (12b)
   0.38410000000000  -0.36890000000000   0.76710000000000    H   (12b)
   0.61590000000000   0.13110000000000  -0.26710000000000    H   (12b)
   0.76710000000000   0.38410000000000  -0.36890000000000    H   (12b)
   0.86890000000000   0.23290000000000  -0.11590000000000    H   (12b)
   0.21070000000000   0.21070000000000   0.21070000000000    N    (4a)
  -0.21070000000000  -0.28930000000000   0.28930000000000    N    (4a)
   0.28930000000000  -0.21070000000000  -0.28930000000000    N    (4a)
  -0.28930000000000   0.28930000000000  -0.21070000000000    N    (4a)
```

$\alpha$-N ($P2_13$): A_cP8_198_2a - CIF

```
# CIF file

data_findsym-output
_audit_creation_method FINDSYM

_chemical_name_mineral ''
_chemical_formula_sum 'N'

loop_
_publ_author_name
 'Sam J. La Placa'
 'Walter C Hamilton'
_journal_name_full
;
Acta Crystallographica B
;
```

```
_journal_volume 28
_journal_year 1972
_journal_page_first 984
_journal_page_last 985
_publ_Section_title
;
 Refinement of the crystal structure of $\alpha$-N$_2$
;

_aflow_proto 'A_cP8_198_2a'
_aflow_params 'a,x1,x2'
_aflow_params_values '5.65,0.0699,-0.0378'
_aflow_Strukturbericht 'None'
_aflow_Pearson 'cP8'

_symmetry_space_group_name_Hall "P 2ac 2ab 3 P2_13"
_symmetry_space_group_name_H-M "P 21 3"
_symmetry_Int_Tables_number 198

_cell_length_a      5.65000
_cell_length_b      5.65000
_cell_length_c      5.65000
_cell_angle_alpha 90.00000
_cell_angle_beta  90.00000
_cell_angle_gamma 90.00000

loop_
_space_group_symop_id
_space_group_symop_operation_xyz
1 x,y,z
2 x+1/2,-y+1/2,-z
3 -x,y+1/2,-z+1/2
4 -x+1/2,-y,z+1/2
5 y,z,x
6 y+1/2,-z+1/2,-x
7 -y,z+1/2,-x+1/2
8 -y+1/2,-z,x+1/2
9 z,x,y
10 z+1/2,-x+1/2,-y
11 -z,x+1/2,-y+1/2
12 -z+1/2,-x,y+1/2

loop_
_atom_site_label
_atom_site_type_symbol
_atom_site_symmetry_multiplicity
_atom_site_Wyckoff_label
_atom_site_fract_x
_atom_site_fract_y
_atom_site_fract_z
_atom_site_occupancy
N1 N    4 a 0.06990 0.06990 0.06990 1.00000
N2 N    4 a -0.03780 -0.03780 -0.03780 1.00000
```

$\alpha$-N ($P2_13$): A_cP8_198_2a - POSCAR

```
A_cP8_198_2a & a,x1,x2 --params=5.65,0.0699,-0.0378 & P2_13      T^4 #198
    ↪ (a^2) & cP8 & & N & alpha & La Placa and Hamilton, Acta Cryst
    ↪ B28, 984-985 (1972)
   1.0000000000000000
   5.65000000000000     0.00000000000000     0.00000000000000
   0.00000000000000     5.65000000000000     0.00000000000000
   0.00000000000000     0.00000000000000     5.65000000000000
    N
    8
Direct
  -0.03780000000000   -0.03780000000000   -0.03780000000000    N    (4a)
   0.03780000000000    0.46220000000000    0.53780000000000    N    (4a)
   0.06990000000000    0.06990000000000    0.06990000000000    N    (4a)
  -0.06990000000000    0.56990000000000    0.43010000000000    N    (4a)
   0.43010000000000   -0.06990000000000    0.56990000000000    N    (4a)
   0.46220000000000    0.53780000000000    0.03780000000000    N    (4a)
   0.53780000000000    0.03780000000000    0.46220000000000    N    (4a)
   0.56990000000000    0.43010000000000   -0.06990000000000    N    (4a)
```

$\alpha$-CO (B21): AB_cP8_198_a_a - CIF

```
# CIF file

data_findsym-output
_audit_creation_method FINDSYM

_chemical_name_mineral 'alpha carbon monoxide'
_chemical_formula_sum 'C O'

loop_
_publ_author_name
 'Lars Vegard'
_journal_name_full
;
Zeitschrift f\"{u}r Physik
;
_journal_volume 61
_journal_year 1930
_journal_page_first 185
_journal_page_last 190
_publ_Section_title
;
 Struktur und Leuchtf\"{a}higkeit von festem Kohlenoxyd
;

# Found in  AMS Database

_aflow_proto 'AB_cP8_198_a_a'
_aflow_params 'a,x1,x2'
_aflow_params_values '5.63,-0.042,0.067'
_aflow_Strukturbericht 'B21'
_aflow_Pearson 'cP8'

_symmetry_space_group_name_Hall "P 2ac 2ab 3 P2_13"
_symmetry_space_group_name_H-M "P 21 3"
_symmetry_Int_Tables_number 198

_cell_length_a      5.63000
_cell_length_b      5.63000
_cell_length_c      5.63000
_cell_angle_alpha 90.00000
_cell_angle_beta  90.00000
_cell_angle_gamma 90.00000

loop_
_space_group_symop_id
_space_group_symop_operation_xyz
1 x,y,z
2 x+1/2,-y+1/2,-z
3 -x,y+1/2,-z+1/2
4 -x+1/2,-y,z+1/2
5 y,z,x
6 y+1/2,-z+1/2,-x
7 -y,z+1/2,-x+1/2
8 -y+1/2,-z,x+1/2
9 z,x,y
10 z+1/2,-x+1/2,-y
11 -z,x+1/2,-y+1/2
12 -z+1/2,-x,y+1/2

loop_
_atom_site_label
_atom_site_type_symbol
_atom_site_symmetry_multiplicity
_atom_site_Wyckoff_label
_atom_site_fract_x
_atom_site_fract_y
_atom_site_fract_z
_atom_site_occupancy
C1 C    4 a -0.04200 -0.04200 -0.04200 1.00000
O1 O    4 a 0.06700  0.06700  0.06700  1.00000
```

$\alpha$-CO (B21): AB_cP8_198_a_a - POSCAR

```
AB_cP8_198_a_a & a,x1,x2 --params=5.63,-0.042,0.067 & P2_13      T^4 #198
    ↪ (a^2) & cP8 & B21 & CO & alpha & L. Vegard, Z. Phys. 61,
    ↪ 185-190 (1930)
   1.0000000000000000
   5.63000000000000     0.00000000000000     0.00000000000000
   0.00000000000000     5.63000000000000     0.00000000000000
   0.00000000000000     0.00000000000000     5.63000000000000
    C    O
    4    4
Direct
  -0.04200000000000   -0.04200000000000   -0.04200000000000    C    (4a)
   0.04200000000000    0.45800000000000    0.54200000000000    C    (4a)
   0.45800000000000    0.54200000000000    0.04200000000000    C    (4a)
   0.54200000000000    0.04200000000000    0.45800000000000    C    (4a)
   0.06700000000000    0.06700000000000    0.06700000000000    O    (4a)
  -0.06700000000000    0.56700000000000    0.43300000000000    O    (4a)
   0.43300000000000   -0.06700000000000    0.56700000000000    O    (4a)
   0.56700000000000    0.43300000000000   -0.06700000000000    O    (4a)
```

FeSi (B20): AB_cP8_198_a_a - CIF

```
# CIF file

data_findsym-output
_audit_creation_method FINDSYM

_chemical_name_mineral ''
_chemical_formula_sum 'Fe Si'

loop_
_publ_author_name
 'L. Vo\v{c}adlo'
 'K. S. Knight'
 'G. D. Price'
 'I. G. Wood'
_journal_name_full
;
Physics and Chemistry of Minerals
;
_journal_volume 29
_journal_year 2002
_journal_page_first 132
_journal_page_last 139
_publ_Section_title
;
 Thermal expansion and crystal structure of FeSi between 4 and 1173 K
    ↪ determined by time-of-flight neutron powder diffraction
;

_aflow_proto 'AB_cP8_198_a_a'
_aflow_params 'a,x1,x2'
_aflow_params_values '4.48688,0.13652,0.8424'
_aflow_Strukturbericht 'B20'
_aflow_Pearson 'cP8'

_symmetry_space_group_name_Hall "P 2ac 2ab 3 P2_13"
_symmetry_space_group_name_H-M "P 21 3"
_symmetry_Int_Tables_number 198

_cell_length_a      4.48688
```

```
_cell_length_b      4.48688
_cell_length_c      4.48688
_cell_angle_alpha  90.00000
_cell_angle_beta   90.00000
_cell_angle_gamma  90.00000

loop_
_space_group_symop_id
_space_group_symop_operation_xyz
1 x,y,z
2 x+1/2,-y+1/2,-z
3 -x,y+1/2,-z+1/2
4 -x+1/2,-y,z+1/2
5 y,z,x
6 y+1/2,-z+1/2,-x
7 -y,z+1/2,-x+1/2
8 -y+1/2,-z,x+1/2
9 z,x,y
10 z+1/2,-x+1/2,-y
11 -z,x+1/2,-y+1/2
12 -z+1/2,-x,y+1/2

loop_
_atom_site_label
_atom_site_type_symbol
_atom_site_symmetry_multiplicity
_atom_site_Wyckoff_label
_atom_site_fract_x
_atom_site_fract_y
_atom_site_fract_z
_atom_site_occupancy
Fe1 Fe    4 a 0.13652  0.13652  0.13652  1.00000
Si1 Si    4 a 0.84240  0.84240  0.84240  1.00000
```

FeSi (B20): AB_cP8_198_a_a - POSCAR

```
AB_cP8_198_a_a & a,x1,x2 --params=4.48688,0.13652,0.8424 & P2_13        T^4
    ↪ #198 (a^2) & cP8 & B20 & FeSi & & L. Vo\v{c}adlo, K. S. Knight
    ↪ , G. D. Price and I. G. Wood, Phys. Chem. Minerals 29, 132-139
    ↪ (2002)
   1.0000000000000000
   4.48687533818500   0.00000000000000   0.00000000000000
   0.00000000000000   4.48687533818500   0.00000000000000
   0.00000000000000   0.00000000000000   4.48687533818500
   Fe    Si
    4     4
Direct
   0.13652000000000   0.13652000000000   0.13652000000000   Fe   (4a)
   0.36348000000000   0.86348000000000   0.63652000000000   Fe   (4a)
   0.63652000000000   0.36348000000000   0.86348000000000   Fe   (4a)
   0.86348000000000   0.63652000000000   0.36348000000000   Fe   (4a)
   0.15760000000000   0.34240000000000   0.65760000000000   Si   (4a)
   0.34240000000000   0.65760000000000   0.15760000000000   Si   (4a)
   0.65760000000000   0.15760000000000   0.34240000000000   Si   (4a)
   0.84240000000000   0.84240000000000   0.84240000000000   Si   (4a)
```

CoU ($B_a$): AB_cI16_199_a_a - CIF

```
# CIF file

data_findsym-output
_audit_creation_method FINDSYM

_chemical_name_mineral ''
_chemical_formula_sum 'Co U'

loop_
_publ_author_name
 'N. C. Baenziger'
 'R. E. Rundle'
 'A. I. Snow'
 'A. S. Wilson'
_journal_name_full
;
Acta Crystallographica
;
_journal_volume 3
_journal_year 1950
_journal_page_first 34
_journal_page_last 40
_publ_Section_title
;
 Compounds of uranium with the transition metals of the first long
    ↪ period
;

# Found in  rough58:UTh

_aflow_proto 'AB_cI16_199_a_a'
_aflow_params 'a,x1,x2'
_aflow_params_values '6.3557,0.294,0.0347'
_aflow_Strukturbericht 'B_a'
_aflow_Pearson 'cI16'

_symmetry_space_group_name_Hall "I 2b 2c 3"
_symmetry_space_group_name_H-M "I 21 3"
_symmetry_Int_Tables_number 199

_cell_length_a      6.35570
_cell_length_b      6.35570
_cell_length_c      6.35570
_cell_angle_alpha  90.00000
_cell_angle_beta   90.00000
_cell_angle_gamma  90.00000

loop_
_space_group_symop_id
_space_group_symop_operation_xyz
1 x,y,z
2 x,-y,-z+1/2
3 -x+1/2,y,-z
4 -x,-y+1/2,z
5 y,z,x
6 y,-z,-x+1/2
7 -y+1/2,z,-x
8 -y,-z+1/2,x
9 z,x,y
10 z,-x,-y+1/2
11 -z+1/2,x,-y
12 -z,-x+1/2,y
13 x+1/2,y+1/2,z+1/2
14 x+1/2,-y+1/2,-z
15 -x,y+1/2,-z+1/2
16 -x+1/2,-y,z+1/2
17 y+1/2,z+1/2,x+1/2
18 y+1/2,-z+1/2,-x
19 -y,z+1/2,-x+1/2
20 -y+1/2,-z,x+1/2
21 z+1/2,x+1/2,y+1/2
22 z+1/2,-x+1/2,-y
23 -z,x+1/2,-y+1/2
24 -z+1/2,-x,y+1/2

loop_
_atom_site_label
_atom_site_type_symbol
_atom_site_symmetry_multiplicity
_atom_site_Wyckoff_label
_atom_site_fract_x
_atom_site_fract_y
_atom_site_fract_z
_atom_site_occupancy
Co1 Co    8 a 0.29400  0.29400  0.29400  1.00000
U1  U     8 a 0.03470  0.03470  0.03470  1.00000
```

CoU ($B_a$): AB_cI16_199_a_a - POSCAR

```
AB_cI16_199_a_a & a,x1,x2 --params=6.3557,0.294,0.0347 & I2_13        T^5 #
    ↪ 199 (a^2) & cI16 & B_a & CoU & & N. C. Baenziger, R. E. Rundle
    ↪ , A. I. Snow and A. S. Wilson, Acta Cryst. 3, 34-40 (1950)
   1.0000000000000000
  -3.17785000000000   3.17785000000000   3.17785000000000
   3.17785000000000  -3.17785000000000   3.17785000000000
   3.17785000000000   3.17785000000000  -3.17785000000000
   Co    U
    4     4
Direct
   0.00000000000000  -0.08800000000000   0.50000000000000   Co   (8a)
  -0.08800000000000   0.50000000000000   0.00000000000000   Co   (8a)
   0.50000000000000   0.00000000000000  -0.08800000000000   Co   (8a)
   0.58800000000000   0.58800000000000   0.58800000000000   Co   (8a)
   0.00000000000000   0.43060000000000   0.50000000000000   U    (8a)
   0.06940000000000   0.06940000000000   0.06940000000000   U    (8a)
   0.43060000000000   0.50000000000000   0.00000000000000   U    (8a)
   0.50000000000000   0.00000000000000   0.43060000000000   U    (8a)
```

Bergman [$Mg_{32}(Al,Zn)_{49}$]: AB32C48_cI162_204_a_2efg_2gh - CIF

```
# CIF file

data_findsym-output
_audit_creation_method FINDSYM

_chemical_name_mineral 'Bergman Structure: Mg32(Al,Zn)49 Bergman'
_chemical_formula_sum 'Al Mg32 Zn48'

loop_
_publ_author_name
 'Gunnar Bergman'
 'John L. T. Waugh'
 'Linus Pauling'
_journal_name_full
;
Acta Crystallographica
;
_journal_volume 10
_journal_year 1957
_journal_page_first 254
_journal_page_last 259
_publ_Section_title
;
 The crystal structure of the metallic phase Mg$_{32}$(Al, Zn)$_{49}$
;

_aflow_proto 'AB32C48_cI162_204_a_2efg_2gh'
_aflow_params 'a,x2,x3,x4,y5,z5,y6,z6,y7,z7,x8,y8,z8'
_aflow_params_values '14.16,0.8203,0.5998,0.1836,0.2942,0.8806,0.0908,
    ↪ 0.8499,0.1748,0.6993,0.686,0.0969,0.332'
_aflow_Strukturbericht 'None'
_aflow_Pearson 'cI162'

_symmetry_space_group_name_Hall "-I 2 2 3"
_symmetry_space_group_name_H-M "I m -3"
_symmetry_Int_Tables_number 204

_cell_length_a      14.16000
_cell_length_b      14.16000
_cell_length_c      14.16000
_cell_angle_alpha  90.00000
_cell_angle_beta   90.00000
```

```
_cell_angle_gamma  90.00000

loop_
_space_group_symop_id
_space_group_symop_operation_xyz
1 x,y,z
2 x,-y,-z
3 -x,y,-z
4 -x,-y,z
5 y,z,x
6 y,-z,-x
7 -y,z,-x
8 -y,-z,x
9 z,x,y
10 z,-x,-y
11 -z,x,-y
12 -z,-x,y
13 -x,-y,-z
14 -x,y,z
15 x,-y,z
16 x,y,-z
17 -y,-z,-x
18 -y,z,x
19 y,-z,x
20 y,z,-x
21 -z,-x,-y
22 -z,x,y
23 z,-x,y
24 z,x,-y
25 x+1/2,y+1/2,z+1/2
26 x+1/2,-y+1/2,-z+1/2
27 -x+1/2,y+1/2,-z+1/2
28 -x+1/2,-y+1/2,z+1/2
29 y+1/2,z+1/2,x+1/2
30 y+1/2,-z+1/2,-x+1/2
31 -y+1/2,z+1/2,-x+1/2
32 -y+1/2,-z+1/2,x+1/2
33 z+1/2,x+1/2,y+1/2
34 z+1/2,-x+1/2,-y+1/2
35 -z+1/2,x+1/2,-y+1/2
36 -z+1/2,-x+1/2,y+1/2
37 -x+1/2,-y+1/2,-z+1/2
38 -x+1/2,y+1/2,z+1/2
39 x+1/2,-y+1/2,z+1/2
40 x+1/2,y+1/2,-z+1/2
41 -y+1/2,-z+1/2,-x+1/2
42 -y+1/2,z+1/2,x+1/2
43 y+1/2,-z+1/2,x+1/2
44 y+1/2,z+1/2,-x+1/2
45 -z+1/2,-x+1/2,-y+1/2
46 -z+1/2,x+1/2,y+1/2
47 z+1/2,-x+1/2,y+1/2
48 z+1/2,x+1/2,-y+1/2

loop_
_atom_site_label
_atom_site_type_symbol
_atom_site_symmetry_multiplicity
_atom_site_Wyckoff_label
_atom_site_fract_x
_atom_site_fract_y
_atom_site_fract_z
_atom_site_occupancy
Al1  Al    2  a  0.00000  0.00000  0.00000  1.00000
Mg1  Mg   12  e  0.82030  0.00000  0.50000  1.00000
Mg2  Mg   12  e  0.00000  0.00000  0.50000  1.00000
Mg3  Mg   16  f  0.18360  0.18360  0.18360  1.00000
Mg4  Mg   24  g  0.00000  0.29420  0.88060  1.00000
Zn1  Zn   24  g  0.00000  0.09080  0.84990  1.00000
Zn2  Zn   24  g  0.00000  0.17480  0.69930  1.00000
Zn3  Zn   48  h  0.68600  0.09690  0.33200  1.00000
```

Bergman [$Mg_{32}(Al,Zn)_{49}$]: AB32C48_cI162_204_a_2efg_2gh - POSCAR

```
AB32C48_cI162_204_a_2efg_2gh & a,x2,x3,x4,y5,z5,y6,z6,y7,z7,x8,y8,z8 --
    ↪ params=14.16,0.8203,0.5998,0.1836,0.2942,0.8806,0.0908,0.8499,
    ↪ 0.1748,0.6993,0.686,0.0969,0.332 & Im(-3)      T_h^5 #204 (ae^2fg
    ↪ ^3h) & cI162 &  & Mg32(Al,Zn)49 & Bergman & G. Bergman, J. L.
    ↪ T. Waugh and L. Pauling, Acta Cryst. 10, 254-259 (1957)
   1.0000000000000000
  -7.08000000000000    7.08000000000000    7.08000000000000
   7.08000000000000   -7.08000000000000    7.08000000000000
   7.08000000000000    7.08000000000000   -7.08000000000000
    Al     Mg     Zn
     1     32     48
Direct
   0.00000000000000    0.00000000000000    0.00000000000000    Al    (2a)
   0.17970000000000    0.50000000000000    0.67970000000000    Mg   (12e)
   0.32030000000000    0.82030000000000    0.50000000000000    Mg   (12e)
   0.50000000000000    0.32030000000000    0.82030000000000    Mg   (12e)
   0.50000000000000    0.67970000000000    0.17970000000000    Mg   (12e)
   0.67970000000000    0.17970000000000    0.50000000000000    Mg   (12e)
   0.82030000000000    0.50000000000000    0.32030000000000    Mg   (12e)
  -0.09980000000000    0.40020000000000    0.50000000000000    Mg   (12e)
   0.09980000000000    0.59980000000000    0.50000000000000    Mg   (12e)
   0.40020000000000    0.50000000000000   -0.09980000000000    Mg   (12e)
   0.50000000000000   -0.09980000000000    0.40020000000000    Mg   (12e)
   0.50000000000000    0.09980000000000    0.59980000000000    Mg   (12e)
   0.59980000000000    0.50000000000000    0.09980000000000    Mg   (12e)
   0.00000000000000    0.00000000000000    0.36720000000000    Mg   (16f)
   0.00000000000000    0.00000000000000    0.63280000000000    Mg   (16f)
   0.00000000000000    0.36720000000000    0.00000000000000    Mg   (16f)
   0.00000000000000    0.63280000000000    0.00000000000000    Mg   (16f)
   0.36720000000000    0.00000000000000    0.00000000000000    Mg   (16f)
   0.36720000000000    0.36720000000000    0.36720000000000    Mg   (16f)
   0.63280000000000    0.00000000000000    0.00000000000000    Mg   (16f)
   0.63280000000000    0.63280000000000    0.63280000000000    Mg   (16f)
   0.11940000000000    0.29420000000000    0.41360000000000    Mg   (24g)
   0.11940000000000    0.70580000000000    0.82520000000000    Mg   (24g)
   0.17480000000000    0.88060000000000    0.29420000000000    Mg   (24g)
   0.29420000000000    0.17480000000000    0.88060000000000    Mg   (24g)
   0.29420000000000    0.41360000000000    0.11940000000000    Mg   (24g)
   0.41360000000000    0.11940000000000    0.29420000000000    Mg   (24g)
   0.58640000000000    0.88060000000000    0.70580000000000    Mg   (24g)
   0.70580000000000    0.58640000000000    0.88060000000000    Mg   (24g)
   0.70580000000000    0.82520000000000    0.11940000000000    Mg   (24g)
   0.82520000000000    0.11940000000000    0.70580000000000    Mg   (24g)
   0.88060000000000    0.29420000000000    0.17480000000000    Mg   (24g)
   0.88060000000000    0.70580000000000    0.58640000000000    Mg   (24g)
   0.05930000000000    0.15010000000000   -0.09080000000000    Zn   (24g)
  -0.05930000000000    0.84990000000000    0.09080000000000    Zn   (24g)
  -0.09080000000000    0.05930000000000    0.15010000000000    Zn   (24g)
   0.09080000000000   -0.05930000000000    0.84990000000000    Zn   (24g)
   0.09080000000000    0.24090000000000    0.15010000000000    Zn   (24g)
  -0.09080000000000    0.75910000000000    0.84990000000000    Zn   (24g)
   0.15010000000000   -0.09080000000000    0.05930000000000    Zn   (24g)
   0.15010000000000    0.09080000000000    0.24090000000000    Zn   (24g)
   0.24090000000000    0.15010000000000    0.09080000000000    Zn   (24g)
   0.75910000000000    0.84990000000000   -0.09080000000000    Zn   (24g)
   0.84990000000000    0.09080000000000   -0.05930000000000    Zn   (24g)
   0.84990000000000   -0.09080000000000    0.75910000000000    Zn   (24g)
   0.12590000000000    0.30070000000000    0.82520000000000    Zn   (24g)
   0.17480000000000    0.47550000000000    0.30070000000000    Zn   (24g)
   0.17480000000000    0.87410000000000    0.69930000000000    Zn   (24g)
   0.30070000000000    0.17480000000000    0.47550000000000    Zn   (24g)
   0.30070000000000    0.82520000000000    0.12590000000000    Zn   (24g)
   0.47550000000000    0.30070000000000    0.17480000000000    Zn   (24g)
   0.52450000000000    0.69930000000000    0.82520000000000    Zn   (24g)
   0.69930000000000    0.17480000000000    0.87410000000000    Zn   (24g)
   0.69930000000000    0.82520000000000    0.52450000000000    Zn   (24g)
   0.82520000000000    0.12590000000000    0.30070000000000    Zn   (24g)
   0.82520000000000    0.52450000000000    0.69930000000000    Zn   (24g)
   0.87410000000000    0.69930000000000    0.17480000000000    Zn   (24g)
  -0.01800000000000    0.21710000000000    0.57110000000000    Zn   (48h)
  -0.01800000000000    0.41090000000000    0.76490000000000    Zn   (48h)
   0.01800000000000    0.58910000000000    0.23510000000000    Zn   (48h)
   0.01800000000000    0.78290000000000    0.42890000000000    Zn   (48h)
   0.21710000000000    0.23510000000000    0.64600000000000    Zn   (48h)
   0.21710000000000    0.57110000000000   -0.01800000000000    Zn   (48h)
   0.23510000000000    0.01800000000000    0.58910000000000    Zn   (48h)
   0.23510000000000    0.64600000000000    0.21710000000000    Zn   (48h)
   0.35400000000000    0.58910000000000    0.57110000000000    Zn   (48h)
   0.35400000000000    0.78290000000000    0.76490000000000    Zn   (48h)
   0.41090000000000    0.42890000000000    0.64600000000000    Zn   (48h)
   0.41090000000000    0.76490000000000   -0.01800000000000    Zn   (48h)
   0.42890000000000    0.01800000000000    0.78290000000000    Zn   (48h)
   0.42890000000000    0.64600000000000    0.41090000000000    Zn   (48h)
   0.57110000000000   -0.01800000000000    0.21710000000000    Zn   (48h)
   0.57110000000000    0.35400000000000    0.58910000000000    Zn   (48h)
   0.58910000000000    0.23510000000000    0.01800000000000    Zn   (48h)
   0.58910000000000    0.57110000000000    0.35400000000000    Zn   (48h)
   0.64600000000000    0.21710000000000    0.23510000000000    Zn   (48h)
   0.64600000000000    0.41090000000000    0.42890000000000    Zn   (48h)
   0.76490000000000   -0.01800000000000    0.41090000000000    Zn   (48h)
   0.76490000000000    0.35400000000000    0.78290000000000    Zn   (48h)
   0.78290000000000    0.42890000000000    0.01800000000000    Zn   (48h)
   0.78290000000000    0.76490000000000    0.35400000000000    Zn   (48h)
```

Skutterudite ($CoAs_3$, $D0_2$): A3B_cI32_204_g_c - CIF

```
# CIF file

data_findsym-output
_audit_creation_method FINDSYM

_chemical_name_mineral 'Skutterudite'
_chemical_formula_sum 'Co As3'

loop_
_publ_author_name
 'Neil Mandel'
 'Jerry Donohue'
_journal_name_full
;
Acta Crystallographica B
;
_journal_volume 27
_journal_year 1971
_journal_page_first 2288
_journal_page_last 2289
_publ_Section_title
;
 The refinement of the crystal structure of skutterudite, CoAs$_3$
;

_aflow_proto 'A3B_cI32_204_g_c'
_aflow_params 'a,y2,z2'
_aflow_params_values '7.58,0.3431,0.8497'
_aflow_Strukturbericht 'D0_2'
_aflow_Pearson 'cI32'

_symmetry_space_group_name_Hall "-I 2 2 3"
_symmetry_space_group_name_H-M "I m -3"
_symmetry_Int_Tables_number 204

_cell_length_a     7.58000
_cell_length_b     7.58000
_cell_length_c     7.58000
_cell_angle_alpha  90.00000
_cell_angle_beta   90.00000
```

```
_cell_angle_gamma 90.00000

loop_
_space_group_symop_id
_space_group_symop_operation_xyz
1 x,y,z
2 x,-y,-z
3 -x,y,-z
4 -x,-y,z
5 y,z,x
6 y,-z,-x
7 -y,z,-x
8 -y,-z,x
9 z,x,y
10 z,-x,-y
11 -z,x,-y
12 -z,-x,y
13 -x,-y,-z
14 -x,y,z
15 x,-y,z
16 x,y,-z
17 -y,-z,-x
18 -y,z,x
19 y,-z,x
20 y,z,-x
21 -z,-x,-y
22 -z,x,y
23 z,-x,y
24 z,x,-y
25 x+1/2,y+1/2,z+1/2
26 x+1/2,-y+1/2,-z+1/2
27 -x+1/2,y+1/2,-z+1/2
28 -x+1/2,-y+1/2,z+1/2
29 y+1/2,z+1/2,x+1/2
30 y+1/2,-z+1/2,-x+1/2
31 -y+1/2,z+1/2,-x+1/2
32 -y+1/2,-z+1/2,x+1/2
33 z+1/2,x+1/2,y+1/2
34 z+1/2,-x+1/2,-y+1/2
35 -z+1/2,x+1/2,-y+1/2
36 -z+1/2,-x+1/2,y+1/2
37 -x+1/2,-y+1/2,-z+1/2
38 -x+1/2,y+1/2,z+1/2
39 x+1/2,-y+1/2,z+1/2
40 x+1/2,y+1/2,-z+1/2
41 -y+1/2,-z+1/2,-x+1/2
42 -y+1/2,z+1/2,x+1/2
43 y+1/2,-z+1/2,x+1/2
44 y+1/2,z+1/2,-x+1/2
45 -z+1/2,-x+1/2,-y+1/2
46 -z+1/2,x+1/2,y+1/2
47 z+1/2,-x+1/2,y+1/2
48 z+1/2,x+1/2,-y+1/2

loop_
_atom_site_label
_atom_site_type_symbol
_atom_site_symmetry_multiplicity
_atom_site_Wyckoff_label
_atom_site_fract_x
_atom_site_fract_y
_atom_site_fract_z
_atom_site_occupancy
Co1 Co   8 c 0.25000 0.25000 0.25000 1.00000
As1 As  24 g 0.00000 0.34310 0.84970 1.00000
```

Skutterudite ($CoAs_3$, $D0_2$): A3B_cI32_204_g_c - POSCAR

```
A3B_cI32_204_g_c & a,y2,z2 --params=7.58,0.3431,0.8497 & Im(-3)      T_h^5
    ↪ #204 (cg) & cI32 & D0_2 & CoAs3 & Skutterudite & Neil Mandel
    ↪ and Jerry Donohue, Acta Cryst. B 27, 2288-2289 (1971)
   1.0000000000000000
  -3.79000000000000   3.79000000000000   3.79000000000000
   3.79000000000000  -3.79000000000000   3.79000000000000
   3.79000000000000   3.79000000000000  -3.79000000000000
   As    Co
   12     4
Direct
   0.15030000000000   0.34310000000000   0.49340000000000   As   (24g)
   0.15030000000000   0.65690000000000   0.80720000000000   As   (24g)
   0.19280000000000   0.84970000000000   0.34310000000000   As   (24g)
   0.34310000000000   0.19280000000000   0.84970000000000   As   (24g)
   0.34310000000000   0.49340000000000   0.15030000000000   As   (24g)
   0.49340000000000   0.15030000000000   0.34310000000000   As   (24g)
   0.50660000000000   0.84970000000000   0.65690000000000   As   (24g)
   0.65690000000000   0.50660000000000   0.84970000000000   As   (24g)
   0.65690000000000   0.80720000000000   0.15030000000000   As   (24g)
   0.80720000000000   0.15030000000000   0.65690000000000   As   (24g)
   0.84970000000000   0.34310000000000   0.19280000000000   As   (24g)
   0.84970000000000   0.65690000000000   0.50660000000000   As   (24g)
   0.00000000000000   0.00000000000000   0.50000000000000   Co    (8c)
   0.00000000000000   0.50000000000000   0.00000000000000   Co    (8c)
   0.50000000000000   0.00000000000000   0.00000000000000   Co    (8c)
   0.50000000000000   0.50000000000000   0.50000000000000   Co    (8c)
```

$Al_{12}W$: A12B_cI26_204_g_a - CIF

```
# CIF file

data_findsym-output
_audit_creation_method FINDSYM

_chemical_name_mineral ''
_chemical_formula_sum 'Al12 W'

loop_
_publ_author_name
 'J. Adam'
 'J. B. Rich'
_journal_name_full
;
Acta Crystallographica
;
_journal_volume 7
_journal_year 1954
_journal_page_first 813
_journal_page_last 816
_publ_Section_title
;
 The crystal structure of WAl$_{12}$, MoAl$_{12}$ and (Mn, Cr)Al$_{12}$
;

_aflow_proto 'A12B_cI26_204_g_a'
_aflow_params 'a,y2,z2'
_aflow_params_values '7.58,0.184,0.691'
_aflow_Strukturbericht 'None'
_aflow_Pearson 'cI26'

_symmetry_space_group_name_Hall "-I 2 2 3"
_symmetry_space_group_name_H-M "I m -3"
_symmetry_Int_Tables_number 204

_cell_length_a     7.58000
_cell_length_b     7.58000
_cell_length_c     7.58000
_cell_angle_alpha 90.00000
_cell_angle_beta  90.00000
_cell_angle_gamma 90.00000

loop_
_space_group_symop_id
_space_group_symop_operation_xyz
1 x,y,z
2 x,-y,-z
3 -x,y,-z
4 -x,-y,z
5 y,z,x
6 y,-z,-x
7 -y,z,-x
8 -y,-z,x
9 z,x,y
10 z,-x,-y
11 -z,x,-y
12 -z,-x,y
13 -x,-y,-z
14 -x,y,z
15 x,-y,z
16 x,y,-z
17 -y,-z,-x
18 -y,z,x
19 y,-z,x
20 y,z,-x
21 -z,-x,-y
22 -z,x,y
23 z,-x,y
24 z,x,-y
25 x+1/2,y+1/2,z+1/2
26 x+1/2,-y+1/2,-z+1/2
27 -x+1/2,y+1/2,-z+1/2
28 -x+1/2,-y+1/2,z+1/2
29 y+1/2,z+1/2,x+1/2
30 y+1/2,-z+1/2,-x+1/2
31 -y+1/2,z+1/2,-x+1/2
32 -y+1/2,-z+1/2,x+1/2
33 z+1/2,x+1/2,y+1/2
34 z+1/2,-x+1/2,-y+1/2
35 -z+1/2,x+1/2,-y+1/2
36 -z+1/2,-x+1/2,y+1/2
37 -x+1/2,-y+1/2,-z+1/2
38 -x+1/2,y+1/2,z+1/2
39 x+1/2,-y+1/2,z+1/2
40 x+1/2,y+1/2,-z+1/2
41 -y+1/2,-z+1/2,-x+1/2
42 -y+1/2,z+1/2,x+1/2
43 y+1/2,-z+1/2,x+1/2
44 y+1/2,z+1/2,-x+1/2
45 -z+1/2,-x+1/2,-y+1/2
46 -z+1/2,x+1/2,y+1/2
47 z+1/2,-x+1/2,y+1/2
48 z+1/2,x+1/2,-y+1/2

loop_
_atom_site_label
_atom_site_type_symbol
_atom_site_symmetry_multiplicity
_atom_site_Wyckoff_label
_atom_site_fract_x
_atom_site_fract_y
_atom_site_fract_z
_atom_site_occupancy
W1  W    2 a 0.00000 0.00000 0.00000 1.00000
Al1 Al  24 g 0.00000 0.18400 0.69100 1.00000
```

$Al_{12}W$: A12B_cI26_204_g_a - POSCAR

```
A12B_cI26_204_g_a & a,y2,z2 --params=7.58,0.184,0.691 & Im(-3)      T_h^5
    ↪ #204 (ag) & cI26 & & Al12W & & J. Adam and J. B. Rich, Acta
    ↪ Cryst. 7, 813-816 (1954)
   1.0000000000000000
  -3.79000000000000   3.79000000000000   3.79000000000000
```

```
    3.79000000000000   -3.79000000000000    3.79000000000000
    3.79000000000000    3.79000000000000   -3.79000000000000
    Al     W
    12      1
Direct
    0.12500000000000    0.30900000000000    0.81600000000000    Al    (24g)
    0.18400000000000    0.49300000000000    0.30900000000000    Al    (24g)
    0.18400000000000    0.87500000000000    0.69100000000000    Al    (24g)
    0.30900000000000    0.18400000000000    0.49300000000000    Al    (24g)
    0.30900000000000    0.81600000000000    0.12500000000000    Al    (24g)
    0.49300000000000    0.30900000000000    0.18400000000000    Al    (24g)
    0.50700000000000    0.69100000000000    0.81600000000000    Al    (24g)
    0.69100000000000    0.18400000000000    0.87500000000000    Al    (24g)
    0.69100000000000    0.81600000000000    0.50700000000000    Al    (24g)
    0.81600000000000    0.12500000000000    0.30900000000000    Al    (24g)
    0.81600000000000    0.50700000000000    0.69100000000000    Al    (24g)
    0.87500000000000    0.69100000000000    0.18400000000000    Al    (24g)
    0.00000000000000    0.00000000000000    0.00000000000000     W     (2a)
```

$\alpha$-N (Pa$\bar{3}$): A_cP8_205_c - CIF

```
# CIF file
# This file was generated by FINDSYM
# Harold T. Stokes, Branton J. Campbell, Dorian M. Hatch
# Brigham Young University, Provo, Utah, USA

data_findsym-output
_audit_creation_method FINDSYM

data_findsym-output
_audit_creation_method FINDSYM

_chemical_name_mineral 'Cubic alpha N2'
_chemical_formula_sum 'N'

loop_
_publ_author_name
 'Truman H. Jordan'
 'H. Warren Smith'
 'William E. Streib'
 'William N. Lipscomb'
_journal_name_full
;
Journal of Chemical Physics
;
_journal_volume 41
_journal_year 1964
_journal_page_first 756
_journal_page_last 759
_publ_Section_title
;
 Single-Crystal X-Ray Diffractions Studies of $\alpha$-N$_2$ and $\
     ↪ beta$-N$_2$
;

# Found in  Donohue, pp. 280-285

_aflow_proto 'A_cP8_205_c'
_aflow_params 'a,x1'
_aflow_params_values '5.65,0.05569'
_aflow_Strukturbericht 'None'
_aflow_Pearson 'cP8'

_symmetry_space_group_name_Hall "-P 2ac 2ab 3 Pa(-3)"
_symmetry_space_group_name_H-M "P 21/a -3"
_symmetry_Int_Tables_number 205

_cell_length_a     5.65000
_cell_length_b     5.65000
_cell_length_c     5.65000
_cell_angle_alpha  90.00000
_cell_angle_beta   90.00000
_cell_angle_gamma  90.00000

loop_
_space_group_symop_id
_space_group_symop_operation_xyz
1 x,y,z
2 x+1/2,-y+1/2,-z
3 -x,y+1/2,-z+1/2
4 -x+1/2,-y,z+1/2
5 y,z,x
6 y+1/2,-z+1/2,-x
7 -y,z+1/2,-x+1/2
8 -y+1/2,-z,x+1/2
9 z,x,y
10 z+1/2,-x+1/2,-y
11 -z,x+1/2,-y+1/2
12 -z+1/2,-x,y+1/2
13 -x,-y,-z
14 -x+1/2,y+1/2,z
15 x,-y+1/2,z+1/2
16 x+1/2,y,-z+1/2
17 -y,-z,-x
18 -y+1/2,z+1/2,x
19 y,-z+1/2,x+1/2
20 y+1/2,z,-x+1/2
21 -z,-x,-y
22 -z+1/2,x+1/2,y
23 z,-x+1/2,y+1/2
24 z+1/2,x,-y+1/2

loop_
_atom_site_label
_atom_site_type_symbol
_atom_site_symmetry_multiplicity
_atom_site_Wyckoff_label
_atom_site_fract_x
_atom_site_fract_y
_atom_site_fract_z
_atom_site_occupancy
N1 N   8 c 0.05569 0.05569 0.05569 1.00000
```

$\alpha$-N (Pa$\bar{3}$): A_cP8_205_c - POSCAR

```
A_cP8_205_c & a,x1 --params=5.65,0.05569 & Pa(-3)     T_h^6 #205 (c) &
     ↪ cP8 & & N & alpha & T. H. Jordan et al., J. Chem Phys. 41,
     ↪ 756-759 (1964)
   1.00000000000000000
   5.65000000000000    0.00000000000000    0.00000000000000
   0.00000000000000    5.65000000000000    0.00000000000000
   0.00000000000000    0.00000000000000    5.65000000000000
    N
    8
Direct
   0.05569130915200    0.05569130915200    0.05569130915200    N    (8c)
  -0.05569130915200   -0.05569130915200   -0.05569130915200    N    (8c)
   0.05569130915200    0.44430869084800    0.55569130915200    N    (8c)
  -0.05569130915200    0.55569130915200    0.44430869084800    N    (8c)
   0.44430869084800   -0.05569130915200    0.55569130915200    N    (8c)
   0.44430869084800    0.55569130915200    0.05569130915200    N    (8c)
   0.55569130915200    0.05569130915200    0.44430869084800    N    (8c)
   0.55569130915200    0.44430869084800   -0.05569130915200    N    (8c)
```

SC16 (CuCl): AB_cP16_205_c_c - CIF

```
# CIF file

data_findsym-output
_audit_creation_method FINDSYM

_chemical_name_mineral 'SC16 CuCl, stable at 5GPa'
_chemical_formula_sum 'Cu Cl'

loop_
_publ_author_name
 'S. Hull'
 'D. A. Keen'
_journal_name_full
;
Physical Review B
;
_journal_volume 50
_journal_year 1994
_journal_page_first 5868
_journal_page_last 5885
_publ_Section_title
;
 High-pressure polymorphism of the copper(I) halides: A
     ↪ neutron-diffraction study to ~10 GPa
;

# Found in  Crain, RPP 58, pp. 705 (1995)

_aflow_proto 'AB_cP16_205_c_c'
_aflow_params 'a,x1,x2'
_aflow_params_values '6.4162,0.1527,0.6297'
_aflow_Strukturbericht 'None'
_aflow_Pearson 'cP16'

_symmetry_space_group_name_Hall "-P 2ac 2ab 3 Pa(-3)"
_symmetry_space_group_name_H-M "P a -3"
_symmetry_Int_Tables_number 205

_cell_length_a     6.41620
_cell_length_b     6.41620
_cell_length_c     6.41620
_cell_angle_alpha  90.00000
_cell_angle_beta   90.00000
_cell_angle_gamma  90.00000

loop_
_space_group_symop_id
_space_group_symop_operation_xyz
1 x,y,z
2 x+1/2,-y+1/2,-z
3 -x,y+1/2,-z+1/2
4 -x+1/2,-y,z+1/2
5 y,z,x
6 y+1/2,-z+1/2,-x
7 -y,z+1/2,-x+1/2
8 -y+1/2,-z,x+1/2
9 z,x,y
10 z+1/2,-x+1/2,-y
11 -z,x+1/2,-y+1/2
12 -z+1/2,-x,y+1/2
13 -x,-y,-z
14 -x+1/2,y+1/2,z
15 x,-y+1/2,z+1/2
16 x+1/2,y,-z+1/2
17 -y,-z,-x
18 -y+1/2,z+1/2,x
19 y,-z+1/2,x+1/2
20 y+1/2,z,-x+1/2
21 -z,-x,-y
22 -z+1/2,x+1/2,y
23 z,-x+1/2,y+1/2
24 z+1/2,x,-y+1/2

loop_
```

```
_atom_site_label
_atom_site_type_symbol
_atom_site_symmetry_multiplicity
_atom_site_Wyckoff_label
_atom_site_fract_x
_atom_site_fract_y
_atom_site_fract_z
_atom_site_occupancy
Cl1 Cl    8 c 0.15270  0.15270  0.15270  1.00000
Cu1 Cu    8 c 0.62970  0.62970  0.62970  1.00000
```

SC16 (CuCl): AB_cP16_205_c_c - POSCAR

```
AB_cP16_205_c_c & a,x1,x2 --params=6.4162,0.1527,0.6297 & Pa(-3)      T_h^
    ↪ 6 #205 (c^2) & cP16 & & CuCl & SC16 (P = 5.52 GPa) & S. Hull
    ↪ and D. A. Keen, PRB 50, 5868-5885 (1994)
   1.0000000000000000
   6.41620000000000    0.00000000000000    0.00000000000000
   0.00000000000000    6.41620000000000    0.00000000000000
   0.00000000000000    0.00000000000000    6.41620000000000
    Cl     Cu
     8      8
Direct
   0.15270000000000    0.15270000000000    0.15270000000000    Cl     (8c)
   0.15270000000000    0.34730000000000    0.65270000000000    Cl     (8c)
   0.34730000000000    0.65270000000000    0.15270000000000    Cl     (8c)
   0.34730000000000    0.84730000000000    0.65270000000000    Cl     (8c)
   0.65270000000000    0.15270000000000    0.34730000000000    Cl     (8c)
   0.65270000000000    0.34730000000000    0.84730000000000    Cl     (8c)
   0.84730000000000    0.65270000000000    0.34730000000000    Cl     (8c)
   0.84730000000000    0.84730000000000    0.84730000000000    Cl     (8c)
  -0.12970000000000    0.12970000000000    0.62970000000000    Cu     (8c)
   0.12970000000000    0.62970000000000   -0.12970000000000    Cu     (8c)
   0.12970000000000    0.87030000000000    0.37030000000000    Cu     (8c)
   0.37030000000000    0.12970000000000    0.87030000000000    Cu     (8c)
   0.37030000000000    0.37030000000000    0.37030000000000    Cu     (8c)
   0.62970000000000   -0.12970000000000    0.12970000000000    Cu     (8c)
   0.62970000000000    0.62970000000000    0.62970000000000    Cu     (8c)
   0.87030000000000    0.37030000000000    0.12970000000000    Cu     (8c)
```

Pyrite ($FeS_2$, C2): AB2_cP12_205_a_c - CIF

```
# CIF file

data_findsym-output
_audit_creation_method FINDSYM

_chemical_name_mineral 'Pyrite'
_chemical_formula_sum 'Fe S2'

loop_
_publ_author_name
 'Peter Bayliss'
_journal_name_full
;
American Mineralogist
;
_journal_volume 62
_journal_year 1977
_journal_page_first 1168
_journal_page_last 1172
_publ_Section_title
;
 Crystal structure refinement of a weakly anisotropic pyrite
;

# Found in  AMS Database

_aflow_proto 'AB2_cP12_205_a_c'
_aflow_params 'a,x2'
_aflow_params_values '5.417,0.3851'
_aflow_Strukturbericht 'C2'
_aflow_Pearson 'cP12'

_symmetry_space_group_name_Hall "-P 2ac 2ab 3 Pa(-3)"
_symmetry_space_group_name_H-M "P a -3"
_symmetry_Int_Tables_number 205

_cell_length_a      5.41700
_cell_length_b      5.41700
_cell_length_c      5.41700
_cell_angle_alpha 90.00000
_cell_angle_beta  90.00000
_cell_angle_gamma 90.00000

loop_
_space_group_symop_id
_space_group_symop_operation_xyz
1 x,y,z
2 x+1/2,-y+1/2,-z
3 -x,y+1/2,-z+1/2
4 -x+1/2,-y,z+1/2
5 y,z,x
6 y+1/2,-z+1/2,-x
7 -y,z+1/2,-x+1/2
8 -y+1/2,-z,x+1/2
9 z,x,y
10 z+1/2,-x+1/2,-y
11 -z,x+1/2,-y+1/2
12 -z+1/2,-x,y+1/2
13 -x,-y,-z
14 -x+1/2,y+1/2,z
15 x,-y+1/2,z+1/2
16 x+1/2,y,-z+1/2
17 -y,-z,-x
18 -y+1/2,z+1/2,x
19 y,-z+1/2,x+1/2
20 y+1/2,z,-x+1/2
21 -z,-x,-y
22 -z+1/2,x+1/2,y
23 z,-x+1/2,y+1/2
24 z+1/2,x,-y+1/2

loop_
_atom_site_label
_atom_site_type_symbol
_atom_site_symmetry_multiplicity
_atom_site_Wyckoff_label
_atom_site_fract_x
_atom_site_fract_y
_atom_site_fract_z
_atom_site_occupancy
Fe1 Fe    4 a 0.00000  0.00000  0.00000  1.00000
S1  S     8 c 0.38510  0.38510  0.38510  1.00000
```

Pyrite ($FeS_2$, C2): AB2_cP12_205_a_c - POSCAR

```
AB2_cP12_205_a_c & a,x2 --params=5.417,0.3851 & Pa-3         T_h^6 #205 (
    ↪ ac) & cP12 & C2 & FeS2 & Pyrite & Bayliss, Am. Mineral. 62,
    ↪ 1168-1172 (1977)
   1.0000000000000000
   5.41700000000000    0.00000000000000    0.00000000000000
   0.00000000000000    5.41700000000000    0.00000000000000
   0.00000000000000    0.00000000000000    5.41700000000000
    Fe     S
     4     8
Direct
   0.00000000000000    0.00000000000000    0.00000000000000    Fe     (4a)
   0.00000000000000    0.50000000000000    0.50000000000000    Fe     (4a)
   0.50000000000000    0.00000000000000    0.50000000000000    Fe     (4a)
   0.50000000000000    0.50000000000000    0.00000000000000    Fe     (4a)
   0.11490000000000    0.61490000000000    0.88510000000000    S      (8c)
   0.11490000000000    0.88510000000000    0.38510000000000    S      (8c)
   0.38510000000000    0.11490000000000    0.88510000000000    S      (8c)
   0.38510000000000    0.38510000000000    0.38510000000000    S      (8c)
   0.61490000000000    0.61490000000000    0.61490000000000    S      (8c)
   0.61490000000000    0.88510000000000    0.11490000000000    S      (8c)
   0.88510000000000    0.11490000000000    0.61490000000000    S      (8c)
   0.88510000000000    0.38510000000000    0.11490000000000    S      (8c)
```

Bixbyite ($Mn_2O_3$, $D5_3$): AB3C6_cI80_206_a_d_e - CIF

```
# CIF file

data_findsym-output
_audit_creation_method FINDSYM

_chemical_name_mineral 'Bixbyite (Mn,Fe)2O4'
_chemical_formula_sum 'Fe Mn3 O6'

loop_
_publ_author_name
 'H. Dachs'
_journal_name_full
;
Zeitschrift f\"{u}r Kristallographie - Crystalline Materials
;
_journal_volume 107
_journal_year 1956
_journal_page_first 370
_journal_page_last 395
_publ_Section_title
;
 Die Kristallstruktur des Bixbyits (Fe,Mn)$_2$O$_3$
;

# Found in  AMS Database

_aflow_proto 'AB3C6_cI80_206_a_d_e'
_aflow_params 'a,x2,x3,y3,z3'
_aflow_params_values '9.4,-0.0344,0.338,0.1,0.125'
_aflow_Strukturbericht 'D5_3'
_aflow_Pearson 'cI80'

_symmetry_space_group_name_Hall "-I 2b 2c 3"
_symmetry_space_group_name_H-M "I a -3"
_symmetry_Int_Tables_number 206

_cell_length_a      9.40000
_cell_length_b      9.40000
_cell_length_c      9.40000
_cell_angle_alpha 90.00000
_cell_angle_beta  90.00000
_cell_angle_gamma 90.00000

loop_
_space_group_symop_id
_space_group_symop_operation_xyz
1 x,y,z
2 x,-y,-z+1/2
3 -x+1/2,y,-z
4 -x,-y+1/2,z
5 y,z,x
6 y,-z,-x+1/2
7 -y+1/2,z,-x
8 -y,-z+1/2,x
9 z,x,y
10 z,-x,-y+1/2
11 -z+1/2,x,-y
12 -z,-x+1/2,y
```

```
13 -x,-y,-z
14 -x,y,z+1/2
15 x+1/2,-y,z
16 x,y+1/2,-z
17 -y,-z,-x
18 -y,z,x+1/2
19 y+1/2,-z,x
20 y,z+1/2,-x
21 -z,-x,-y
22 -z,x,y+1/2
23 z+1/2,-x,y
24 z,x+1/2,-y
25 x+1/2,y+1/2,z+1/2
26 x+1/2,-y+1/2,-z
27 -x,y+1/2,-z+1/2
28 -x+1/2,-y,z+1/2
29 y+1/2,z+1/2,x+1/2
30 y+1/2,-z+1/2,-x
31 -y,z+1/2,-x+1/2
32 -y+1/2,-z,x+1/2
33 z+1/2,x+1/2,y+1/2
34 z+1/2,-x+1/2,-y
35 -z,x+1/2,-y+1/2
36 -z+1/2,-x,y+1/2
37 -x+1/2,-y+1/2,-z+1/2
38 -x+1/2,y+1/2,z
39 x,-y+1/2,z+1/2
40 x+1/2,y,-z+1/2
41 -y+1/2,-z+1/2,-x+1/2
42 -y+1/2,z+1/2,x
43 y,-z+1/2,x+1/2
44 y+1/2,z,-x+1/2
45 -z+1/2,-x+1/2,-y+1/2
46 -z+1/2,x+1/2,y
47 z,-x+1/2,y+1/2
48 z+1/2,x,-y+1/2

loop_
_atom_site_label
_atom_site_type_symbol
_atom_site_symmetry_multiplicity
_atom_site_Wyckoff_label
_atom_site_fract_x
_atom_site_fract_y
_atom_site_fract_z
_atom_site_occupancy
Fe1 Fe   8 a 0.0.0000  0.00000 0.00000 1.00000
Mn1 Mn  24 d -0.03440 0.00000 0.25000 1.00000
O1  O   48 e 0.33800  0.10000 0.12500 1.00000
```

Bixbyite ($Mn_2O_3$, $D5_3$): AB3C6_cI80_206_a_d_e - POSCAR

```
AB3C6_cI80_206_a_d_e & a,x2,x3,y3,z3 --params=9.4,-0.0344,0.338,0.1,
    ↪ 0.125 & Ia(-3)     T_h^3 #206 (ade) & cI80 & D5_3 & (Mn,Fe)2O3 &
    ↪ Bixbyite & H. Dachs, Zeitschrift f\"{u}r Kristallographie -
    ↪ Crystalline Materials, 107, 370-395 (1956)
   1.0000000000000000
  -4.70000000000000   4.70000000000000   4.70000000000000
   4.70000000000000  -4.70000000000000   4.70000000000000
   4.70000000000000   4.70000000000000  -4.70000000000000
   Fe     Mn     O
    4     12     24
Direct
   0.00000000000000   0.00000000000000  -0.50000000000000   Fe    (8a)
   0.00000000000000  -0.50000000000000   0.00000000000000   Fe    (8a)
  -0.50000000000000   0.00000000000000   0.00000000000000   Fe    (8a)
   0.50000000000000   0.50000000000000   0.50000000000000   Fe    (8a)
  -0.03440000000000   0.25000000000000   0.21560000000000   Mn   (24d)
   0.03440000000000   0.75000000000000   0.78440000000000   Mn   (24d)
   0.21560000000000  -0.03440000000000   0.25000000000000   Mn   (24d)
   0.25000000000000   0.21560000000000  -0.03440000000000   Mn   (24d)
   0.25000000000000   0.71560000000000   0.46560000000000   Mn   (24d)
   0.46560000000000   0.25000000000000   0.71560000000000   Mn   (24d)
   0.53440000000000   0.75000000000000   1.28440000000000   Mn   (24d)
   0.71560000000000   0.46560000000000   0.25000000000000   Mn   (24d)
   0.75000000000000   0.78440000000000   0.03440000000000   Mn   (24d)
   0.75000000000000   1.28440000000000   0.53440000000000   Mn   (24d)
   0.78440000000000   0.03440000000000   0.75000000000000   Mn   (24d)
   1.28440000000000   0.53440000000000   0.75000000000000   Mn   (24d)
   0.03700000000000   0.26200000000000   0.97500000000000    O   (48e)
  -0.03700000000000  -0.26200000000000  -0.97500000000000    O   (48e)
   0.06200000000000   0.52500000000000   0.78700000000000    O   (48e)
  -0.06200000000000  -0.52500000000000  -0.78700000000000    O   (48e)
   0.22500000000000   0.46300000000000   0.43800000000000    O   (48e)
  -0.22500000000000  -0.46300000000000  -0.43800000000000    O   (48e)
   0.26200000000000   0.97500000000000   0.03700000000000    O   (48e)
  -0.26200000000000  -0.97500000000000  -0.03700000000000    O   (48e)
   0.27500000000000   0.71300000000000   1.23800000000000    O   (48e)
  -0.27500000000000  -0.71300000000000  -1.23800000000000    O   (48e)
   0.43800000000000   0.22500000000000   0.46300000000000    O   (48e)
  -0.43800000000000  -0.22500000000000  -0.46300000000000    O   (48e)
   0.46300000000000   0.43800000000000   0.22500000000000    O   (48e)
  -0.46300000000000  -0.43800000000000  -0.22500000000000    O   (48e)
   0.52500000000000   0.78700000000000   0.06200000000000    O   (48e)
  -0.52500000000000  -0.78700000000000  -0.06200000000000    O   (48e)
   0.71300000000000   1.23800000000000   0.27500000000000    O   (48e)
  -0.71300000000000  -1.23800000000000  -0.27500000000000    O   (48e)
   0.78700000000000   0.06200000000000   0.52500000000000    O   (48e)
  -0.78700000000000  -0.06200000000000  -0.52500000000000    O   (48e)
   0.97500000000000   0.03700000000000   0.26200000000000    O   (48e)
  -0.97500000000000  -0.03700000000000  -0.26200000000000    O   (48e)
   1.23800000000000   0.27500000000000   0.71300000000000    O   (48e)
  -1.23800000000000  -0.27500000000000  -0.71300000000000    O   (48e)
```

BC8 (Si): A_cI16_206_c - CIF

```
# CIF file

data_findsym-output
_audit_creation_method FINDSYM

_chemical_name_mineral 'BC8'
_chemical_formula_sum 'Si'

loop_
_publ_author_name
 'J. Crain'
 'G. J. Ackland'
 'S. J. Clark'
_journal_name_full
;
Reports on Progress in Physics
;
_journal_volume 58
_journal_year 1995
_journal_page_first 705
_journal_page_last 754
_publ_Section_title
;
Exotic structures of tetrahedral semiconductors
;

_aflow_proto 'A_cI16_206_c'
_aflow_params 'a,x1'
_aflow_params_values '4.11971,0.1001'
_aflow_Strukturbericht 'None'
_aflow_Pearson 'cI16'

_symmetry_space_group_name_Hall "-I 2b 2c 3"
_symmetry_space_group_name_H-M "I a -3"
_symmetry_Int_Tables_number 206

_cell_length_a      4.11971
_cell_length_b      4.11971
_cell_length_c      4.11971
_cell_angle_alpha  90.00000
_cell_angle_beta   90.00000
_cell_angle_gamma  90.00000

loop_
_space_group_symop_id
_space_group_symop_operation_xyz
1 x,y,z
2 x,-y,-z+1/2
3 -x+1/2,y,-z
4 -x,-y+1/2,z
5 y,z,x
6 y,-z,-x+1/2
7 -y+1/2,z,-x
8 -y,-z+1/2,x
9 z,x,y
10 z,-x,-y+1/2
11 -z+1/2,x,-y
12 -z,-x+1/2,y
13 -x,-y,-z
14 -x,y,z+1/2
15 x+1/2,-y,z
16 x,y+1/2,-z
17 -y,-z,-x
18 -y,z,x+1/2
19 y+1/2,-z,x
20 y,z+1/2,-x
21 -z,-x,-y
22 -z,x,y+1/2
23 z+1/2,-x,y
24 z,x+1/2,-y
25 x+1/2,y+1/2,z+1/2
26 x+1/2,-y+1/2,-z
27 -x,y+1/2,-z+1/2
28 -x+1/2,-y,z+1/2
29 y+1/2,z+1/2,x+1/2
30 y+1/2,-z+1/2,-x
31 -y,z+1/2,-x+1/2
32 -y+1/2,-z,x+1/2
33 z+1/2,x+1/2,y+1/2
34 z+1/2,-x+1/2,-y
35 -z,x+1/2,-y+1/2
36 -z+1/2,-x,y+1/2
37 -x+1/2,-y+1/2,-z+1/2
38 -x+1/2,y+1/2,z
39 x,-y+1/2,z+1/2
40 x+1/2,y,-z+1/2
41 -y+1/2,-z+1/2,-x+1/2
42 -y+1/2,z+1/2,x
43 y,-z+1/2,x+1/2
44 y+1/2,z,-x+1/2
45 -z+1/2,-x+1/2,-y+1/2
46 -z+1/2,x+1/2,y
47 z,-x+1/2,y+1/2
48 z+1/2,x,-y+1/2

loop_
_atom_site_label
_atom_site_type_symbol
_atom_site_symmetry_multiplicity
_atom_site_Wyckoff_label
_atom_site_fract_x
_atom_site_fract_y
_atom_site_fract_z
_atom_site_occupancy
Si1 Si  16 c 0.10010 0.10010 0.10010 1.00000
```

BC8 (Si): A_cI16_206_c - POSCAR

```
A_cI16_206_c & a,x1 --params=4.11971,0.1001 & Ia(-3)      T_h^7 #206 (c) &
    ↪ cI16 & & Si & BC8 & J. Crain, G. J. Ackland, and S. J. Clark,
    ↪ Rep. Prog. Phys. 58, 705âĂŞ754 (1995)
   1.0000000000000000
  -2.05985734394000    2.05985734394000    2.05985734394000
   2.05985734394000   -2.05985734394000    2.05985734394000
   2.05985734394000    2.05985734394000   -2.05985734394000
   Si
    8
Direct
  -0.00000000000000    0.29980000000000    0.50000000000000    Si    (16c)
   0.00000000000000    0.70020000000000    0.50000000000000    Si    (16c)
   0.20020000000000    0.20020000000000    0.20020000000000    Si    (16c)
   0.29980000000000    0.50000000000000   -0.00000000000000    Si    (16c)
   0.50000000000000   -0.00000000000000    0.29980000000000    Si    (16c)
   0.50000000000000    0.00000000000000    0.70020000000000    Si    (16c)
   0.70020000000000    0.50000000000000    0.00000000000000    Si    (16c)
   0.79980000000000    0.79980000000000    0.79980000000000    Si    (16c)
```

β-Mn (A13): A_cP20_213_cd - CIF

```
# CIF file

data_findsym-output
_audit_creation_method FINDSYM

_chemical_name_mineral 'beta'
_chemical_formula_sum 'Mn'

loop_
_publ_author_name
 'Clara Brink Shoemaker'
 'David P. Shoemaker'
 'Ted E. Hopkins'
 'Somrat Yindepit'
_journal_name_full
;
Acta Crystallographica B
;
_journal_volume 34
_journal_year 1978
_journal_page_first 3573
_journal_page_last 3576
_publ_Section_title
;
 Refinement of the structure of $\beta$-manganese and of a related phase
    ↪ in the Mn-Ni-Si system
;

_aflow_proto 'A_cP20_213_cd'
_aflow_params 'a,x1,y2'
_aflow_params_values '6.315,0.06361,0.20224'
_aflow_Strukturbericht 'A13'
_aflow_Pearson 'cP20'

_symmetry_space_group_name_Hall "P 4bd 2ab 3 P4_132"
_symmetry_space_group_name_H-M "P 41 3 2"
_symmetry_Int_Tables_number 213

_cell_length_a      6.31500
_cell_length_b      6.31500
_cell_length_c      6.31500
_cell_angle_alpha  90.00000
_cell_angle_beta   90.00000
_cell_angle_gamma  90.00000

loop_
_space_group_symop_id
_space_group_symop_operation_xyz
1 x,y,z
2 x+1/2,-y+1/2,-z
3 -x,y+1/2,-z+1/2
4 -x+1/2,-y,z+1/2
5 y,z,x
6 y+1/2,-z+1/2,-x
7 -y,z+1/2,-x+1/2
8 -y+1/2,-z,x+1/2
9 z,x,y
10 z+1/2,-x+1/2,-y
11 -z,x+1/2,-y+1/2
12 -z+1/2,-x,y+1/2
13 -y+3/4,-x+3/4,-z+3/4
14 -y+1/4,x+3/4,z+1/4
15 y+1/4,-x+1/4,z+3/4
16 y+3/4,x+1/4,-z+1/4
17 -x+3/4,-z+3/4,-y+3/4
18 -x+1/4,z+3/4,y+1/4
19 x+1/4,-z+1/4,y+3/4
20 x+3/4,z+1/4,-y+1/4
21 -z+3/4,-y+3/4,-x+3/4
22 -z+1/4,y+3/4,x+1/4
23 z+1/4,-y+1/4,x+3/4
24 z+3/4,y+1/4,-x+1/4

loop_
_atom_site_label
_atom_site_type_symbol
_atom_site_symmetry_multiplicity
_atom_site_Wyckoff_label
_atom_site_fract_x
_atom_site_fract_y
_atom_site_fract_z
_atom_site_occupancy
Mn1 Mn   8 c 0.06361  0.06361  0.06361  1.00000
Mn2 Mn  12 d 0.12500  0.20224  0.45224  1.00000
```

β-Mn (A13): A_cP20_213_cd - POSCAR

```
A_cP20_213_cd & a,x1,y2 --params=6.315,0.06361,0.20224 & P4_132      O^7 #
    ↪ 213 (cd) & cP20 & A13 & Mn & beta & C.B. Shoemaker, D.P.
    ↪ Shoemaker, T.E. Hopkins and S. Yindepit, Acta Cryst. B 34,
    ↪ 3573-3576 (1978)
   1.0000000000000000
   6.31500000000000    0.00000000000000    0.00000000000000
   0.00000000000000    6.31500000000000    0.00000000000000
   0.00000000000000    0.00000000000000    6.31500000000000
   Mn
   20
Direct
  -0.04776000000000    0.37500000000000    0.79776000000000    Mn    (12d)
   0.04776000000000    0.87500000000000    0.70224000000000    Mn    (12d)
   0.12500000000000    0.20224000000000    0.45224000000000    Mn    (12d)
   0.20224000000000    0.45224000000000    0.12500000000000    Mn    (12d)
   0.29776000000000    0.54776000000000    0.62500000000000    Mn    (12d)
   0.37500000000000    0.79776000000000   -0.04776000000000    Mn    (12d)
   0.45224000000000    0.12500000000000    0.20224000000000    Mn    (12d)
   0.54776000000000    0.62500000000000    0.29776000000000    Mn    (12d)
   0.62500000000000    0.29776000000000    0.54776000000000    Mn    (12d)
   0.70224000000000    0.04776000000000    0.87500000000000    Mn    (12d)
   0.79776000000000   -0.04776000000000    0.37500000000000    Mn    (12d)
   0.87500000000000    0.70224000000000    0.04776000000000    Mn    (12d)
   0.06361000000000    0.06361000000000    0.06361000000000    Mn     (8c)
  -0.06361000000000    0.56361000000000    0.43639000000000    Mn     (8c)
   0.18639000000000    0.81361000000000    0.31361000000000    Mn     (8c)
   0.31361000000000    0.18639000000000    0.81361000000000    Mn     (8c)
   0.43639000000000   -0.06361000000000    0.56361000000000    Mn     (8c)
   0.56361000000000    0.43639000000000   -0.06361000000000    Mn     (8c)
   0.68639000000000    0.68639000000000    0.68639000000000    Mn     (8c)
   0.81361000000000    0.31361000000000    0.18639000000000    Mn     (8c)
```

Sulvanite ($Cu_3S_4V$, H2$_4$): A3B4C_cP8_215_d_e_a - CIF

```
# CIF file

data_findsym-output
_audit_creation_method FINDSYM

_chemical_name_mineral 'Sulvanite'
_chemical_formula_sum 'Cu3 S4 V'

loop_
_publ_author_name
 'Felix J. Trojer'
_journal_name_full
;
American Mineralogist
;
_journal_volume 51
_journal_year 1966
_journal_page_first 890
_journal_page_last 894
_publ_Section_title
;
 Refinement of the Structure of Sulvanite
;

# Found in AMS Database

_aflow_proto 'A3B4C_cP8_215_d_e_a'
_aflow_params 'a,x3'
_aflow_params_values '5.3912,0.2372'
_aflow_Strukturbericht 'H2_4'
_aflow_Pearson 'cP8'

_symmetry_space_group_name_Hall "P -4 2 3"
_symmetry_space_group_name_H-M "P -4 3 m"
_symmetry_Int_Tables_number 215

_cell_length_a      5.39120
_cell_length_b      5.39120
_cell_length_c      5.39120
_cell_angle_alpha  90.00000
_cell_angle_beta   90.00000
_cell_angle_gamma  90.00000

loop_
_space_group_symop_id
_space_group_symop_operation_xyz
1 x,y,z
2 x,-y,-z
3 -x,y,-z
4 -x,-y,z
5 y,z,x
6 y,-z,-x
7 -y,z,-x
8 -y,-z,x
9 z,x,y
10 z,-x,-y
11 -z,x,-y
12 -z,-x,y
13 y,x,z
14 y,-x,-z
15 -y,x,-z
16 -y,-x,z
17 x,z,y
18 x,-z,-y
```

```
19 -x,z,-y
20 -x,-z,y
21 z,y,x
22 z,-y,-x
23 -z,y,-x
24 -z,-y,x

loop_
_atom_site_label
_atom_site_type_symbol
_atom_site_symmetry_multiplicity
_atom_site_Wyckoff_label
_atom_site_fract_x
_atom_site_fract_y
_atom_site_fract_z
_atom_site_occupancy
V1  V    1 a 0.00000 0.00000 0.00000 1.00000
Cu1 Cu   3 d 0.50000 0.00000 0.00000 1.00000
S1  S    4 e 0.23720 0.23720 0.23720 1.00000
```

Sulvanite ($Cu_3S_4V$, $H2_4$): A3B4C_cP8_215_d_e_a - POSCAR

```
A3B4C_cP8_215_d_e_a & a,x3 --params=5.3912,0.2372 & P(-4)3m    T_d^1 #215
    ↪ (ade) & cP8 & H2_4 & Cu3S4V & Sulvanite & F. J. Trojer, Am.
    ↪ Mineral. 51, 890-894 (1966)
   1.0000000000000000
   5.39120000000000   0.00000000000000   0.00000000000000
   0.00000000000000   5.39120000000000   0.00000000000000
   0.00000000000000   0.00000000000000   5.39120000000000
   Cu     S     V
    3     4     1
Direct
   0.00000000000000   0.00000000000000   0.50000000000000   Cu   (3d)
   0.00000000000000   0.50000000000000   0.00000000000000   Cu   (3d)
   0.50000000000000   0.00000000000000   0.00000000000000   Cu   (3d)
   0.23720000000000   0.23720000000000   0.23720000000000   S    (4e)
   0.23720000000000   0.76280000000000   0.76280000000000   S    (4e)
   0.76280000000000   0.23720000000000   0.76280000000000   S    (4e)
   0.76280000000000   0.76280000000000   0.23720000000000   S    (4e)
   0.00000000000000   0.00000000000000   0.00000000000000   V    (1a)
```

$Fe_4C$: AB4_cP5_215_a_e - CIF

```
# CIF file

data_findsym-output
_audit_creation_method FINDSYM

_chemical_name_mineral 'Iron carbide'
_chemical_formula_sum 'Fe4 C'

loop_
_publ_author_name
 'Z. G. Pinsker'
 'S. V. Kaverin'
_journal_name_full
;
Soviet Physics-Crystallography, translated from Kristallografiya
;
_journal_volume 1
_journal_year 1956
_journal_page_first 48
_journal_page_last 53
_publ_Section_title
;
 Electron-Diffraction Determination of the Structure of Iron Carbide
    ↪ Fe$_4$C
;

# Found in  Pearson's Handbook, Vol. II, pp. 1895

_aflow_proto 'AB4_cP5_215_a_e'
_aflow_params 'a,x2'
_aflow_params_values '3.878,0.265'
_aflow_Strukturbericht 'None'
_aflow_Pearson 'cP5'

_symmetry_space_group_name_Hall "P -4 2 3"
_symmetry_space_group_name_H-M "P -4 3 m"
_symmetry_Int_Tables_number 215

_cell_length_a      3.87800
_cell_length_b      3.87800
_cell_length_c      3.87800
_cell_angle_alpha  90.00000
_cell_angle_beta   90.00000
_cell_angle_gamma  90.00000

loop_
_space_group_symop_id
_space_group_symop_operation_xyz
1 x,y,z
2 x,-y,-z
3 -x,y,-z
4 -x,-y,z
5 y,z,x
6 y,-z,-x
7 -y,z,-x
8 -y,-z,x
9 z,x,y
10 z,-x,-y
11 -z,x,-y
12 -z,-x,y
13 y,x,z
14 y,-x,-z
15 -y,x,-z
16 -y,-x,z
17 x,z,y
18 x,-z,-y
19 -x,z,-y
20 -x,-z,y
21 z,y,x
22 z,-y,-x
23 -z,y,-x
24 -z,-y,x

loop_
_atom_site_label
_atom_site_type_symbol
_atom_site_symmetry_multiplicity
_atom_site_Wyckoff_label
_atom_site_fract_x
_atom_site_fract_y
_atom_site_fract_z
_atom_site_occupancy
C1  C    1 a 0.00000 0.00000 0.00000 1.00000
Fe1 Fe   4 e 0.26500 0.26500 0.26500 1.00000
```

$Fe_4C$: AB4_cP5_215_a_e - POSCAR

```
AB4_cP5_215_a_e & a,x2 --params=3.878,0.265 & P(-4)3m    T_d^1 #215 (ae)
    ↪ & cP5 & & Fe4C & & Z. G. Pinsker and S. V. Kaverin, Sov.
    ↪ Physics Cryst. 1, 48-53 (1956)
   1.0000000000000000
   3.87800000000000   0.00000000000000   0.00000000000000
   0.00000000000000   3.87800000000000   0.00000000000000
   0.00000000000000   0.00000000000000   3.87800000000000
   C     Fe
   1     4
Direct
   0.00000000000000   0.00000000000000   0.00000000000000   C    (1a)
   0.26500000000000   0.26500000000000   0.26500000000000   Fe   (4e)
   0.26500000000000   0.73500000000000   0.73500000000000   Fe   (4e)
   0.73500000000000   0.26500000000000   0.73500000000000   Fe   (4e)
   0.73500000000000   0.73500000000000   0.26500000000000   Fe   (4e)
```

Cubic Lazarevićite ($AsCu_3S_4$): AB3C4_cP8_215_a_c_e - CIF

```
# CIF file

data_findsym-output
_audit_creation_method FINDSYM

_chemical_name_mineral 'Lavarevi\'{c}ite'
_chemical_formula_sum 'As Cu3 S4'

loop_
_publ_author_name
 'C. B. Sclar'
 'M. Drovenik'
_journal_name_full
;
Geological Society of America Bulletin
;
_journal_volume 71
_journal_year 1960
_journal_page_first 1970
_journal_page_last 1970
_publ_Section_title
;
 Lazarevi\'{c}ite, A New Cubic Copper-Arsenic Sulfied from Bor,
    ↪ Jugoslavia
;

# Found in  Pearson's Handbook, Vol. I, pp. 1111-1112, Fleischer 1961

_aflow_proto 'AB3C4_cP8_215_a_c_e'
_aflow_params 'a,x3'
_aflow_params_values '5.28,0.25'
_aflow_Strukturbericht 'None'
_aflow_Pearson 'cP8'

_symmetry_space_group_name_Hall "P -4 2 3"
_symmetry_space_group_name_H-M "P -4 3 m"
_symmetry_Int_Tables_number 215

_cell_length_a      5.28000
_cell_length_b      5.28000
_cell_length_c      5.28000
_cell_angle_alpha  90.00000
_cell_angle_beta   90.00000
_cell_angle_gamma  90.00000

loop_
_space_group_symop_id
_space_group_symop_operation_xyz
1 x,y,z
2 x,-y,-z
3 -x,y,-z
4 -x,-y,z
5 y,z,x
6 y,-z,-x
7 -y,z,-x
8 -y,-z,x
9 z,x,y
10 z,-x,-y
11 -z,x,-y
12 -z,-x,y
13 y,x,z
14 y,-x,-z
```

```
15 -y,x,-z
16 -y,-x,z
17 x,z,y
18 x,-z,-y
19 -x,z,-y
20 -x,-z,y
21 z,y,x
22 z,-y,-x
23 -z,y,-x
24 -z,-y,x

loop_
_atom_site_label
_atom_site_type_symbol
_atom_site_symmetry_multiplicity
_atom_site_Wyckoff_label
_atom_site_fract_x
_atom_site_fract_y
_atom_site_fract_z
_atom_site_occupancy
As1 As    1 a 0.00000 0.00000 0.00000 1.00000
Cu1 Cu    3 c 0.00000 0.50000 0.50000 1.00000
S1  S     4 e 0.25000 0.25000 0.25000 1.00000
```

Cubic Lazarevićite ($AsCu_3S_4$): AB3C4_cP8_215_a_c_e - POSCAR

```
AB3C4_cP8_215_a_c_e & a,x3 --params=5.28,0.25 & P(-4)3m    T_d^1 #215 (
    ↪ ace) & cP8 & & AsCu3S4 & Lazarevicite & C. B. Sclar and M.
    ↪ Drovenik, Bull. Geo. Soc. Am. 71, 1970 (1960)
   1.0000000000000000
   5.28000000000000   0.00000000000000   0.00000000000000
   0.00000000000000   5.28000000000000   0.00000000000000
   0.00000000000000   0.00000000000000   5.28000000000000
   As    Cu     S
    1     3     4
Direct
   0.00000000000000   0.00000000000000   0.00000000000000   As    (1a)
   0.00000000000000   0.50000000000000   0.50000000000000   Cu    (3c)
   0.50000000000000   0.00000000000000   0.50000000000000   Cu    (3c)
   0.50000000000000   0.50000000000000   0.00000000000000   Cu    (3c)
   0.25000000000000   0.25000000000000   0.25000000000000    S    (4e)
   0.25000000000000   0.75000000000000   0.75000000000000    S    (4e)
   0.75000000000000   0.25000000000000   0.75000000000000    S    (4e)
   0.75000000000000   0.75000000000000   0.25000000000000    S    (4e)
```

$AuBe_5$ ($C15_b$): AB5_cF24_216_a_ce - CIF

```
# CIF file

data_findsym-output
_audit_creation_method FINDSYM

_chemical_name_mineral ''
_chemical_formula_sum 'Au Be5'

loop_
_publ_author_name
 'F. W. von Batchelder'
 'R. F. Raeuchle'
_journal_name_full
;
Acta Crystallographica
;
_journal_volume 11
_journal_year 1958
_journal_page_first 122
_journal_page_last 122
_publ_Section_title
;
 The tetragonal MBe$_{12}$ structure of silver, palladium, platinum and
     ↪ gold
;

# Found in  pearson58:C15b

_aflow_proto 'AB5_cF24_216_a_ce'
_aflow_params 'a,x3'
_aflow_params_values '6.1,0.625'
_aflow_Strukturbericht 'C15_b'
_aflow_Pearson 'cF24'

_symmetry_space_group_name_Hall "F -4 2 3"
_symmetry_space_group_name_H-M "F -4 3 m"
_symmetry_Int_Tables_number 216

_cell_length_a     6.10000
_cell_length_b     6.10000
_cell_length_c     6.10000
_cell_angle_alpha  90.00000
_cell_angle_beta   90.00000
_cell_angle_gamma  90.00000

loop_
_space_group_symop_id
_space_group_symop_operation_xyz
1 x,y,z
2 x,-y,-z
3 -x,y,-z
4 -x,-y,z
5 y,z,x
6 y,-z,-x
7 -y,z,-x
8 -y,-z,x
9 z,x,y
10 z,-x,-y
11 -z,x,-y
12 -z,-x,y
13 y,x,z
14 y,-x,-z
15 -y,x,-z
16 -y,-x,z
17 x,z,y
18 x,-z,-y
19 -x,z,-y
20 -x,-z,y
21 z,y,x
22 z,-y,-x
23 -z,y,-x
24 -z,-y,x
25 x,y+1/2,z+1/2
26 x,-y+1/2,-z+1/2
27 -x,y+1/2,-z+1/2
28 -x,-y+1/2,z+1/2
29 y,z+1/2,x+1/2
30 y,-z+1/2,-x+1/2
31 -y,z+1/2,-x+1/2
32 -y,-z+1/2,x+1/2
33 z,x+1/2,y+1/2
34 z,-x+1/2,-y+1/2
35 -z,x+1/2,-y+1/2
36 -z,-x+1/2,y+1/2
37 y,x+1/2,z+1/2
38 y,-x+1/2,-z+1/2
39 -y,x+1/2,-z+1/2
40 -y,-x+1/2,z+1/2
41 x,z+1/2,y+1/2
42 x,-z+1/2,-y+1/2
43 -x,z+1/2,-y+1/2
44 -x,-z+1/2,y+1/2
45 z,y+1/2,x+1/2
46 z,-y+1/2,-x+1/2
47 -z,y+1/2,-x+1/2
48 -z,-y+1/2,x+1/2
49 x+1/2,y,z+1/2
50 x+1/2,-y,-z+1/2
51 -x+1/2,y,-z+1/2
52 -x+1/2,-y,z+1/2
53 y+1/2,z,x+1/2
54 y+1/2,-z,-x+1/2
55 -y+1/2,z,-x+1/2
56 -y+1/2,-z,x+1/2
57 z+1/2,x,y+1/2
58 z+1/2,-x,-y+1/2
59 -z+1/2,x,-y+1/2
60 -z+1/2,-x,y+1/2
61 y+1/2,x,z+1/2
62 y+1/2,-x,-z+1/2
63 -y+1/2,x,-z+1/2
64 -y+1/2,-x,z+1/2
65 x+1/2,z,y+1/2
66 x+1/2,-z,-y+1/2
67 -x+1/2,z,-y+1/2
68 -x+1/2,-z,y+1/2
69 z+1/2,y,x+1/2
70 z+1/2,-y,-x+1/2
71 -z+1/2,y,-x+1/2
72 -z+1/2,-y,x+1/2
73 x+1/2,y+1/2,z
74 x+1/2,-y+1/2,-z
75 -x+1/2,y+1/2,-z
76 -x+1/2,-y+1/2,z
77 y+1/2,z+1/2,x
78 y+1/2,-z+1/2,-x
79 -y+1/2,z+1/2,-x
80 -y+1/2,-z+1/2,x
81 z+1/2,x+1/2,y
82 z+1/2,-x+1/2,-y
83 -z+1/2,x+1/2,-y
84 -z+1/2,-x+1/2,y
85 y+1/2,x+1/2,z
86 y+1/2,-x+1/2,-z
87 -y+1/2,x+1/2,-z
88 -y+1/2,-x+1/2,z
89 x+1/2,z+1/2,y
90 x+1/2,-z+1/2,-y
91 -x+1/2,z+1/2,-y
92 -x+1/2,-z+1/2,y
93 z+1/2,y+1/2,x
94 z+1/2,-y+1/2,-x
95 -z+1/2,y+1/2,-x
96 -z+1/2,-y+1/2,x

loop_
_atom_site_label
_atom_site_type_symbol
_atom_site_symmetry_multiplicity
_atom_site_Wyckoff_label
_atom_site_fract_x
_atom_site_fract_y
_atom_site_fract_z
_atom_site_occupancy
Au1 Au    4 a 0.00000 0.00000 0.00000 1.00000
Be1 Be    4 c 0.25000 0.25000 0.25000 1.00000
Be2 Be   16 e 0.62500 0.62500 0.62500 1.00000
```

$AuBe_5$ ($C15_b$): AB5_cF24_216_a_ce - POSCAR

```
AB5_cF24_216_a_ce & a,x3 --params=6.1,0.625 & F(-4)3m    T_d^2 #216 (ace)
    ↪ & cF24 & C15_b & AuBe5 & & F. W. von Batchelder and R. F.
    ↪ Raeuchle, Acta Cryst. 11, 122 (1958)
```

```
   1.0000000000000000
   0.00000000000000    3.05000000000000    3.05000000000000
   3.05000000000000    0.00000000000000    3.05000000000000
   3.05000000000000    3.05000000000000    0.00000000000000
   Au    Be
    1     5
Direct
   0.00000000000000    0.00000000000000    0.00000000000000    Au    (4a)
   0.12500000000000    0.62500000000000    0.62500000000000    Be   (16e)
   0.62500000000000    0.12500000000000    0.62500000000000    Be   (16e)
   0.62500000000000    0.62500000000000    0.12500000000000    Be   (16e)
   0.62500000000000    0.62500000000000    0.62500000000000    Be   (16e)
   0.25000000000000    0.25000000000000    0.25000000000000    Be    (4c)
```

Half-Heusler ($C1_b$): ABC_cF12_216_b_c_a - CIF

```
# CIF file

data_findsym-output
_audit_creation_method FINDSYM

_chemical_name_mineral 'half-Heusler'
_chemical_formula_sum 'Ag As Mg'

loop_
_publ_author_name
 'H. Nowotny'
 'W. Sibert'
_journal_name_full
;
Zeitschrift f\"{u}r Metallkunde
;
_journal_volume 33
_journal_year 1941
_journal_page_first 391
_journal_page_last 394
_publ_Section_title
;
 Tern\"{a}re Valenzverbindungen in den Systemen Kupfer(Silber)-Arsen(
       ↪ Antimon,Wismut)-Magnesium
;

# Found in  Pearson, Alloys, pp. 386

_aflow_proto 'ABC_cF12_216_b_c_a'
_aflow_params 'a'
_aflow_params_values '6.24'
_aflow_Strukturbericht 'C1_b'
_aflow_Pearson 'cF12'

_symmetry_space_group_name_Hall "F -4 2 3"
_symmetry_space_group_name_H-M "F -4 3 m"
_symmetry_Int_Tables_number 216

_cell_length_a     6.24000
_cell_length_b     6.24000
_cell_length_c     6.24000
_cell_angle_alpha  90.00000
_cell_angle_beta   90.00000
_cell_angle_gamma  90.00000

loop_
_space_group_symop_id
_space_group_symop_operation_xyz
1 x,y,z
2 x,-y,-z
3 -x,y,-z
4 -x,-y,z
5 y,z,x
6 y,-z,-x
7 -y,z,-x
8 -y,-z,x
9 z,x,y
10 z,-x,-y
11 -z,x,-y
12 -z,-x,y
13 y,x,z
14 y,-x,-z
15 -y,x,-z
16 -y,-x,z
17 x,z,y
18 x,-z,-y
19 -x,z,-y
20 -x,-z,y
21 z,y,x
22 z,-y,-x
23 -z,y,-x
24 -z,-y,x
25 x,y+1/2,z+1/2
26 x,-y+1/2,-z+1/2
27 -x,y+1/2,-z+1/2
28 -x,-y+1/2,z+1/2
29 y,z+1/2,x+1/2
30 y,-z+1/2,-x+1/2
31 -y,z+1/2,-x+1/2
32 -y,-z+1/2,x+1/2
33 z,x+1/2,y+1/2
34 z,-x+1/2,-y+1/2
35 -z,x+1/2,-y+1/2
36 -z,-x+1/2,y+1/2
37 y,x+1/2,z+1/2
38 y,-x+1/2,-z+1/2
39 -y,x+1/2,-z+1/2
40 -y,-x+1/2,z+1/2
41 x,z+1/2,y+1/2
42 x,-z+1/2,-y+1/2
43 -x,z+1/2,-y+1/2
44 -x,-z+1/2,y+1/2
45 z,y+1/2,x+1/2
46 z,-y+1/2,-x+1/2
47 -z,y+1/2,-x+1/2
48 -z,-y+1/2,x+1/2
49 x+1/2,y,z+1/2
50 x+1/2,-y,-z+1/2
51 -x+1/2,y,-z+1/2
52 -x+1/2,-y,z+1/2
53 y+1/2,z,x+1/2
54 y+1/2,-z,-x+1/2
55 -y+1/2,z,-x+1/2
56 -y+1/2,-z,x+1/2
57 z+1/2,x,y+1/2
58 z+1/2,-x,-y+1/2
59 -z+1/2,x,-y+1/2
60 -z+1/2,-x,y+1/2
61 y+1/2,x,z+1/2
62 y+1/2,-x,-z+1/2
63 -y+1/2,x,-z+1/2
64 -y+1/2,-x,z+1/2
65 x+1/2,z,y+1/2
66 x+1/2,-z,-y+1/2
67 -x+1/2,z,-y+1/2
68 -x+1/2,-z,y+1/2
69 z+1/2,y,x+1/2
70 z+1/2,-y,-x+1/2
71 -z+1/2,y,-x+1/2
72 -z+1/2,-y,x+1/2
73 x+1/2,y+1/2,z
74 x+1/2,-y+1/2,-z
75 -x+1/2,y+1/2,-z
76 -x+1/2,-y+1/2,z
77 y+1/2,z+1/2,x
78 y+1/2,-z+1/2,-x
79 -y+1/2,z+1/2,-x
80 -y+1/2,-z+1/2,x
81 z+1/2,x+1/2,y
82 z+1/2,-x+1/2,-y
83 -z+1/2,x+1/2,-y
84 -z+1/2,-x+1/2,y
85 y+1/2,x+1/2,z
86 y+1/2,-x+1/2,-z
87 -y+1/2,x+1/2,-z
88 -y+1/2,-x+1/2,z
89 x+1/2,z+1/2,y
90 x+1/2,-z+1/2,-y
91 -x+1/2,z+1/2,-y
92 -x+1/2,-z+1/2,y
93 z+1/2,y+1/2,x
94 z+1/2,-y+1/2,-x
95 -z+1/2,y+1/2,-x
96 -z+1/2,-y+1/2,x

loop_
_atom_site_label
_atom_site_type_symbol
_atom_site_symmetry_multiplicity
_atom_site_Wyckoff_label
_atom_site_fract_x
_atom_site_fract_y
_atom_site_fract_z
_atom_site_occupancy
Mg1 Mg   4 a 0.00000 0.00000 0.00000 1.00000
Ag1 Ag   4 b 0.50000 0.50000 0.50000 1.00000
As1 As   4 c 0.25000 0.25000 0.25000 1.00000
```

Half-Heusler ($C1_b$): ABC_cF12_216_b_c_a - POSCAR

```
ABC_cF12_216_b_c_a & a --params=6.24 & F(-4)3m    T_d^2 #216 (abc) & cF12
      ↪ & C1_b & AgAsMg & Half-Heusler & H. Nowotny and W. Sibert, Z.
      ↪ Metallkd. 33, 391-394 (1941)
   1.0000000000000000
   0.00000000000000    3.12000000000000    3.12000000000000
   3.12000000000000    0.00000000000000    3.12000000000000
   3.12000000000000    3.12000000000000    0.00000000000000
   Ag    As    Mg
    1     1     1
Direct
   0.50000000000000    0.50000000000000    0.50000000000000    Ag    (4b)
   0.25000000000000    0.25000000000000    0.25000000000000    As    (4c)
   0.00000000000000    0.00000000000000    0.00000000000000    Mg    (4a)
```

Zincblende (ZnS, B3): AB_cF8_216_c_a - CIF

```
# CIF file

data_findsym-output
_audit_creation_method FINDSYM

_chemical_name_mineral 'Zincblende, Sphalerite'
_chemical_formula_sum 'Zn S'

loop_
_publ_author_name
 'Brian J. Skinner'
_journal_name_full
;
American Mineralogist
;
_journal_volume 46
_journal_year 1961
_journal_page_first 1399
```

```
_journal_page_last 1411
_publ_Section_title
;
 Unit-Cell Edges of Natural and Synthetic Sphalerites
;

# Found in AMC Database

_aflow_proto 'AB_cF8_216_c_a'
_aflow_params 'a'
_aflow_params_values '5.4093'
_aflow_Strukturbericht 'B3'
_aflow_Pearson 'cF8'

_symmetry_space_group_name_Hall "F -4 2 3"
_symmetry_space_group_name_H-M "F -4 3 m"
_symmetry_Int_Tables_number 216

_cell_length_a     5.40930
_cell_length_b     5.40930
_cell_length_c     5.40930
_cell_angle_alpha 90.00000
_cell_angle_beta  90.00000
_cell_angle_gamma 90.00000

loop_
_space_group_symop_id
_space_group_symop_operation_xyz
1 x,y,z
2 x,-y,-z
3 -x,y,-z
4 -x,-y,z
5 y,z,x
6 y,-z,-x
7 -y,z,-x
8 -y,-z,x
9 z,x,y
10 z,-x,-y
11 -z,x,-y
12 -z,-x,y
13 y,x,z
14 y,-x,-z
15 -y,x,-z
16 -y,-x,z
17 x,z,y
18 x,-z,-y
19 -x,z,-y
20 -x,-z,y
21 z,y,x
22 z,-y,-x
23 -z,y,-x
24 -z,-y,x
25 x,y+1/2,z+1/2
26 x,-y+1/2,-z+1/2
27 -x,y+1/2,-z+1/2
28 -x,-y+1/2,z+1/2
29 y,z+1/2,x+1/2
30 y,-z+1/2,-x+1/2
31 -y,z+1/2,-x+1/2
32 -y,-z+1/2,x+1/2
33 z,x+1/2,y+1/2
34 z,-x+1/2,-y+1/2
35 -z,x+1/2,-y+1/2
36 -z,-x+1/2,y+1/2
37 y,x+1/2,z+1/2
38 y,-x+1/2,-z+1/2
39 -y,x+1/2,-z+1/2
40 -y,-x+1/2,z+1/2
41 x,z+1/2,y+1/2
42 x,-z+1/2,-y+1/2
43 -x,z+1/2,-y+1/2
44 -x,-z+1/2,y+1/2
45 z,y+1/2,x+1/2
46 z,-y+1/2,-x+1/2
47 -z,y+1/2,-x+1/2
48 -z,-y+1/2,x+1/2
49 x+1/2,y,z+1/2
50 x+1/2,-y,-z+1/2
51 -x+1/2,y,-z+1/2
52 -x+1/2,-y,z+1/2
53 y+1/2,z,x+1/2
54 y+1/2,-z,-x+1/2
55 -y+1/2,z,-x+1/2
56 -y+1/2,-z,x+1/2
57 z+1/2,x,y+1/2
58 z+1/2,-x,-y+1/2
59 -z+1/2,x,-y+1/2
60 -z+1/2,-x,y+1/2
61 y+1/2,x,z+1/2
62 y+1/2,-x,-z+1/2
63 -y+1/2,x,-z+1/2
64 -y+1/2,-x,z+1/2
65 x+1/2,z,y+1/2
66 x+1/2,-z,-y+1/2
67 -x+1/2,z,-y+1/2
68 -x+1/2,-z,y+1/2
69 z+1/2,y,x+1/2
70 z+1/2,-y,-x+1/2
71 -z+1/2,y,-x+1/2
72 -z+1/2,-y,x+1/2
73 x+1/2,y+1/2,z
74 x+1/2,-y+1/2,-z
75 -x+1/2,y+1/2,-z
76 -x+1/2,-y+1/2,z
77 y+1/2,z+1/2,x
78 y+1/2,-z+1/2,-x
79 -y+1/2,z+1/2,-x
80 -y+1/2,-z+1/2,x
81 z+1/2,x+1/2,y
82 z+1/2,-x+1/2,-y
83 -z+1/2,x+1/2,-y
84 -z+1/2,-x+1/2,y
85 y+1/2,x+1/2,z
86 y+1/2,-x+1/2,-z
87 -y+1/2,x+1/2,-z
88 -y+1/2,-x+1/2,z
89 x+1/2,z+1/2,y
90 x+1/2,-z+1/2,-y
91 -x+1/2,z+1/2,-y
92 -x+1/2,-z+1/2,y
93 z+1/2,y+1/2,x
94 z+1/2,-y+1/2,-x
95 -z+1/2,y+1/2,-x
96 -z+1/2,-y+1/2,x

loop_
_atom_site_label
_atom_site_type_symbol
_atom_site_symmetry_multiplicity
_atom_site_Wyckoff_label
_atom_site_fract_x
_atom_site_fract_y
_atom_site_fract_z
_atom_site_occupancy
Zn1 Zn   4 a 0.00000 0.00000 0.00000 1.00000
S1  S    4 c 0.25000 0.25000 0.25000 1.00000
```

Zincblende (ZnS, B3): AB_cF8_216_c_a - POSCAR

```
AB_cF8_216_c_a & a --params=5.4093 & F(-4)3m        T_d^2 #216 (ac) &
    ↪ cF8 & B3 & ZnS (cubic) & Zincblende/Sphalerite & B. J. Skinner,
    ↪ Am. Mineral. 46, 1399-1411 (1961)
   1.0000000000000000
   0.00000000000000   2.70465000000000   2.70465000000000
   2.70465000000000   0.00000000000000   2.70465000000000
   2.70465000000000   2.70465000000000   0.00000000000000
    S    Zn
    1    1
Direct
   0.25000000000000   0.25000000000000   0.25000000000000   S    (4c)
   0.00000000000000   0.00000000000000   0.00000000000000   Zn   (4a)
```

$SiF_4$: A4B_cI10_217_c_a - CIF

```
# CIF file

data_findsym-output
_audit_creation_method FINDSYM

_chemical_name_mineral 'Silicon tetrafluoride'
_chemical_formula_sum 'Si F4'

loop_
_publ_author_name
 'Maso Atoji'
 'William N. Lipscomb'
_journal_name_full
;
Acta Crystallographica
;
_journal_volume 7
_journal_year 1954
_journal_page_first 597
_journal_page_last 597
_publ_Section_title
;
 The structure of SiF$_4$
;

_aflow_proto 'A4B_cI10_217_c_a'
_aflow_params 'a,x2'
_aflow_params_values '5.45858,0.165'
_aflow_Strukturbericht 'None'
_aflow_Pearson 'cI10'

_symmetry_space_group_name_Hall "I -4 2 3"
_symmetry_space_group_name_H-M "I -4 3 m"
_symmetry_Int_Tables_number 217

_cell_length_a     5.45858
_cell_length_b     5.45858
_cell_length_c     5.45858
_cell_angle_alpha 90.00000
_cell_angle_beta  90.00000
_cell_angle_gamma 90.00000

loop_
_space_group_symop_id
_space_group_symop_operation_xyz
1 x,y,z
2 x,-y,-z
3 -x,y,-z
4 -x,-y,z
5 y,z,x
6 y,-z,-x
7 -y,z,-x
8 -y,-z,x
9 z,x,y
10 z,-x,-y
11 -z,x,-y
```

```
12 -z,-x,y
13 y,x,z
14 y,-x,-z
15 -y,x,-z
16 -y,-x,z
17 x,z,y
18 x,-z,-y
19 -x,z,-y
20 -x,-z,y
21 z,y,x
22 z,-y,-x
23 -z,y,-x
24 -z,-y,x
25 x+1/2,y+1/2,z+1/2
26 x+1/2,-y+1/2,-z+1/2
27 -x+1/2,y+1/2,-z+1/2
28 -x+1/2,-y+1/2,z+1/2
29 y+1/2,z+1/2,x+1/2
30 y+1/2,-z+1/2,-x+1/2
31 -y+1/2,z+1/2,-x+1/2
32 -y+1/2,-z+1/2,x+1/2
33 z+1/2,x+1/2,y+1/2
34 z+1/2,-x+1/2,-y+1/2
35 -z+1/2,x+1/2,-y+1/2
36 -z+1/2,-x+1/2,y+1/2
37 y+1/2,x+1/2,z+1/2
38 y+1/2,-x+1/2,-z+1/2
39 -y+1/2,x+1/2,-z+1/2
40 -y+1/2,-x+1/2,z+1/2
41 x+1/2,z+1/2,y+1/2
42 x+1/2,-z+1/2,-y+1/2
43 -x+1/2,z+1/2,-y+1/2
44 -x+1/2,-z+1/2,y+1/2
45 z+1/2,y+1/2,x+1/2
46 z+1/2,-y+1/2,-x+1/2
47 -z+1/2,y+1/2,-x+1/2
48 -z+1/2,-y+1/2,x+1/2

loop_
_atom_site_label
_atom_site_type_symbol
_atom_site_symmetry_multiplicity
_atom_site_Wyckoff_label
_atom_site_fract_x
_atom_site_fract_y
_atom_site_fract_z
_atom_site_occupancy
Si1 Si    2 a 0.00000 0.00000 0.00000 1.00000
F1  F     8 c 0.16500 0.16500 0.16500 1.00000
```

$SiF_4$: A4B_cI10_217_c_a - POSCAR

```
A4B_cI10_217_c_a & a,x2 --params=5.45858,0.165 & I(-4)3m    T_d^3 #217 (
    ↪ ac) & cI10 & & SiF4 & & M. Atoji and W. N. Lipscomb, Acta
    ↪ Cryst. 7, 597 (1954)
   1.0000000000000000
  -2.72929218000000   2.72929218000000   2.72929218000000
   2.72929218000000  -2.72929218000000   2.72929218000000
   2.72929218000000   2.72929218000000  -2.72929218000000
    F    Si
    4     1
Direct
   0.00000000000000   0.00000000000000  -0.33000000000000   F   (8c)
   0.00000000000000  -0.33000000000000   0.00000000000000   F   (8c)
  -0.33000000000000   0.00000000000000   0.00000000000000   F   (8c)
   0.33000000000000   0.33000000000000   0.33000000000000   F   (8c)
   0.00000000000000   0.00000000000000   0.00000000000000   Si  (2a)
```

$\alpha$-Mn (A12): A_cI58_217_ac2g - CIF

```
# CIF file

data_findsym-output
_audit_creation_method FINDSYM

_chemical_name_mineral 'alpha'
_chemical_formula_sum 'Mn'

loop_
_publ_author_name
'J. A. Oberteuffer'
'James A. Ibers'
_journal_name_full
;
Acta Crystallographica B
;
_journal_volume 26
_journal_year 1970
_journal_page_first 1499
_journal_page_last 1504
_publ_Section_title
;
 A refinement of the atomic and thermal parameters of $\alpha$-manganese
    ↪ from a single crystal
;

# Found in Donohue, pp. 191-196

_aflow_proto 'A_cI58_217_ac2g'
_aflow_params 'a,x2,x3,z3,x4,z4'
_aflow_params_values '8.911,0.31787,-0.08958,0.28194,0.64294,0.03457'
_aflow_Strukturbericht 'A12'
_aflow_Pearson 'cI58'

_symmetry_space_group_name_Hall "I -4 2 3"
_symmetry_space_group_name_H-M "I -4 3 m"
_symmetry_Int_Tables_number 217

_cell_length_a      8.91100
_cell_length_b      8.91100
_cell_length_c      8.91100
_cell_angle_alpha  90.00000
_cell_angle_beta   90.00000
_cell_angle_gamma  90.00000

loop_
_space_group_symop_id
_space_group_symop_operation_xyz
1 x,y,z
2 x,-y,-z
3 -x,y,-z
4 -x,-y,z
5 y,z,x
6 y,-z,-x
7 -y,z,-x
8 -y,-z,x
9 z,x,y
10 z,-x,-y
11 -z,x,-y
12 -z,-x,y
13 y,x,z
14 y,-x,-z
15 -y,x,-z
16 -y,-x,z
17 x,z,y
18 x,-z,-y
19 -x,z,-y
20 -x,-z,y
21 z,y,x
22 z,-y,-x
23 -z,y,-x
24 -z,-y,x
25 x+1/2,y+1/2,z+1/2
26 x+1/2,-y+1/2,-z+1/2
27 -x+1/2,y+1/2,-z+1/2
28 -x+1/2,-y+1/2,z+1/2
29 y+1/2,z+1/2,x+1/2
30 y+1/2,-z+1/2,-x+1/2
31 -y+1/2,z+1/2,-x+1/2
32 -y+1/2,-z+1/2,x+1/2
33 z+1/2,x+1/2,y+1/2
34 z+1/2,-x+1/2,-y+1/2
35 -z+1/2,x+1/2,-y+1/2
36 -z+1/2,-x+1/2,y+1/2
37 y+1/2,x+1/2,z+1/2
38 y+1/2,-x+1/2,-z+1/2
39 -y+1/2,x+1/2,-z+1/2
40 -y+1/2,-x+1/2,z+1/2
41 x+1/2,z+1/2,y+1/2
42 x+1/2,-z+1/2,-y+1/2
43 -x+1/2,z+1/2,-y+1/2
44 -x+1/2,-z+1/2,y+1/2
45 z+1/2,y+1/2,x+1/2
46 z+1/2,-y+1/2,-x+1/2
47 -z+1/2,y+1/2,-x+1/2
48 -z+1/2,-y+1/2,x+1/2

loop_
_atom_site_label
_atom_site_type_symbol
_atom_site_symmetry_multiplicity
_atom_site_Wyckoff_label
_atom_site_fract_x
_atom_site_fract_y
_atom_site_fract_z
_atom_site_occupancy
Mn1 Mn    2 a 0.00000  0.00000  0.00000 1.00000
Mn2 Mn    8 c 0.31787  0.31787  0.31787 1.00000
Mn3 Mn   24 g -0.08958 -0.08958 0.28194 1.00000
Mn4 Mn   24 g 0.64294  0.64294  0.03457 1.00000
```

$\alpha$-Mn (A12): A_cI58_217_ac2g - POSCAR

```
A_cI58_217_ac2g & a,x2,x3,z3,x4,z4 --params=8.911,0.31787,-0.08958,
    ↪ 0.28194,0.64294,0.03457 & I(-4)3m    T_d^3 #217 (acg^2) & cI58 &
    ↪ A12 & Mn & alpha & J. A. Oberteuffer and J. A. Ibers, Acta
    ↪ Cryst. B 26, 1499-1504 (1970)
   1.0000000000000000
  -4.45550000000000   4.45550000000000   4.45550000000000
   4.45550000000000  -4.45550000000000   4.45550000000000
   4.45550000000000   4.45550000000000  -4.45550000000000
   Mn
   29
Direct
   0.00000000000000   0.62848000000000   0.80764000000000   Mn   (24g)
   0.00000000000000   0.80764000000000   0.62848000000000   Mn   (24g)
   0.17916000000000   0.37152000000000   0.37152000000000   Mn   (24g)
   0.19236000000000   0.19236000000000   0.82084000000000   Mn   (24g)
   0.19236000000000   0.82084000000000   0.19236000000000   Mn   (24g)
   0.37152000000000   0.17916000000000   0.37152000000000   Mn   (24g)
   0.37152000000000   0.37152000000000   0.17916000000000   Mn   (24g)
   0.62848000000000   0.00000000000000   0.80764000000000   Mn   (24g)
   0.62848000000000   0.80764000000000   0.00000000000000   Mn   (24g)
   0.80764000000000   0.00000000000000   0.62848000000000   Mn   (24g)
   0.80764000000000   0.62848000000000   0.00000000000000   Mn   (24g)
   0.82084000000000   0.19236000000000   0.19236000000000   Mn   (24g)
   0.00000000000000   0.32249000000000   0.60837000000000   Mn   (24g)
   0.00000000000000   0.60837000000000   0.32249000000000   Mn   (24g)
   0.28588000000000   0.67751000000000   0.67751000000000   Mn   (24g)
   0.32249000000000   0.00000000000000   0.60837000000000   Mn   (24g)
```

```
0.32249000000000   0.60837000000000   0.00000000000000   Mn   (24g)
0.39163000000000   0.39163000000000   0.71412000000000   Mn   (24g)
0.39163000000000   0.71412000000000   0.39163000000000   Mn   (24g)
0.60837000000000   0.00000000000000   0.32249000000000   Mn   (24g)
0.60837000000000   0.32249000000000   0.00000000000000   Mn   (24g)
0.67751000000000   0.28588000000000   0.67751000000000   Mn   (24g)
0.67751000000000   0.67751000000000   0.28588000000000   Mn   (24g)
0.71412000000000   0.39163000000000   0.39163000000000   Mn   (24g)
0.00000000000000   0.00000000000000   0.00000000000000   Mn    (2a)
0.00000000000000   0.00000000000000   0.36426000000000   Mn    (8c)
0.00000000000000   0.36426000000000   0.00000000000000   Mn    (8c)
0.36426000000000   0.00000000000000   0.00000000000000   Mn    (8c)
0.63574000000000   0.63574000000000   0.63574000000000   Mn    (8c)
```

γ-Brass ($Cu_5Zn_8$): A5B8_cI52_217_ce_cg - CIF

```
# CIF file

data_findsym-output
_audit_creation_method FINDSYM

_chemical_name_mineral 'gamma-brass'
_chemical_formula_sum 'Cu5 Zn8'

loop_
_publ_author_name
 'Olivier Gourdon'
 'Delphine Gout'
 'Darrick J. Williams'
 'Thomas Proffen'
 'Sara Hobbs'
 'Gordon J. Miller'
_journal_name_full
;
Inorganic Chemistry
;
_journal_volume 46
_journal_year 2007
_journal_page_first 251
_journal_page_last 260
_publ_Section_title
;
 Atomic Distributions in the $\gamma$-Brass Structure of the Cu-Zn
       ↪ System: A Structural and Theoretical Study
;

_aflow_proto 'A5B8_cI52_217_ce_cg'
_aflow_params 'a,x1,x2,x3,x4,z4'
_aflow_params_values '8.8664,0.32774,0.10781,0.64421,0.68844,0.03674'
_aflow_Strukturbericht 'None'
_aflow_Pearson 'cI52'

_symmetry_space_group_name_Hall "I -4 2 3"
_symmetry_space_group_name_H-M "I -4 3 m"
_symmetry_Int_Tables_number 217

_cell_length_a      8.86640
_cell_length_b      8.86640
_cell_length_c      8.86640
_cell_angle_alpha  90.00000
_cell_angle_beta   90.00000
_cell_angle_gamma  90.00000

loop_
_space_group_symop_id
_space_group_symop_operation_xyz
1 x,y,z
2 x,-y,-z
3 -x,y,-z
4 -x,-y,z
5 y,z,x
6 y,-z,-x
7 -y,z,-x
8 -y,-z,x
9 z,x,y
10 z,-x,-y
11 -z,x,-y
12 -z,-x,y
13 y,x,z
14 y,-x,-z
15 -y,x,-z
16 -y,-x,z
17 x,z,y
18 x,-z,-y
19 -x,z,-y
20 -x,-z,y
21 z,y,x
22 z,-y,-x
23 -z,y,-x
24 -z,-y,x
25 x+1/2,y+1/2,z+1/2
26 x+1/2,-y+1/2,-z+1/2
27 -x+1/2,y+1/2,-z+1/2
28 -x+1/2,-y+1/2,z+1/2
29 y+1/2,z+1/2,x+1/2
30 y+1/2,-z+1/2,-x+1/2
31 -y+1/2,z+1/2,-x+1/2
32 -y+1/2,-z+1/2,x+1/2
33 z+1/2,x+1/2,y+1/2
34 z+1/2,-x+1/2,-y+1/2
35 -z+1/2,x+1/2,-y+1/2
36 -z+1/2,-x+1/2,y+1/2
37 y+1/2,x+1/2,z+1/2
38 y+1/2,-x+1/2,-z+1/2
39 -y+1/2,x+1/2,-z+1/2
40 -y+1/2,-x+1/2,z+1/2
41 x+1/2,z+1/2,y+1/2
42 x+1/2,-z+1/2,-y+1/2
43 -x+1/2,z+1/2,-y+1/2
44 -x+1/2,-z+1/2,y+1/2
45 z+1/2,y+1/2,x+1/2
46 z+1/2,-y+1/2,-x+1/2
47 -z+1/2,y+1/2,-x+1/2
48 -z+1/2,-y+1/2,x+1/2

loop_
_atom_site_label
_atom_site_type_symbol
_atom_site_symmetry_multiplicity
_atom_site_Wyckoff_label
_atom_site_fract_x
_atom_site_fract_y
_atom_site_fract_z
_atom_site_occupancy
Cu1 Cu    8 c 0.32774 0.32774 0.32774 1.00000
Zn1 Zn    8 c 0.10781 0.10781 0.10781 1.00000
Cu2 Cu   12 e 0.64421 0.00000 0.00000 1.00000
Zn2 Zn   24 g 0.68844 0.68844 0.03674 1.00000
```

γ-Brass ($Cu_5Zn_8$): A5B8_cI52_217_ce_cg - POSCAR

```
A5B8_cI52_217_ce_cg & a,x1,x2,x3,x4,z4 --params=8.8664,0.32774,0.10781,
      ↪ 0.64421,0.68844,0.03674 & I(-4)3m    T_d^3 #217 (c^2eg) & cI52 &
      ↪  & Cu5Zn8 & gamma brass & O. Gourdon et al., Inorg. Chem. 46,
      ↪ 251-260 (2007)
   1.00000000000000000
  -4.43320000000000   4.43320000000000   4.43320000000000
   4.43320000000000  -4.43320000000000   4.43320000000000
   4.43320000000000   4.43320000000000  -4.43320000000000
   Cu    Zn
   10    16
Direct
   0.00000000000000   0.35579000000000   0.35579000000000   Cu   (12e)
   0.00000000000000   0.64421000000000   0.64421000000000   Cu   (12e)
   0.35579000000000   0.00000000000000   0.35579000000000   Cu   (12e)
   0.35579000000000   0.35579000000000   0.00000000000000   Cu   (12e)
   0.64421000000000   0.00000000000000   0.64421000000000   Cu   (12e)
   0.64421000000000   0.64421000000000   0.00000000000000   Cu   (12e)
   0.00000000000000   0.00000000000000   0.34452000000000   Cu    (8c)
   0.00000000000000   0.34452000000000   0.00000000000000   Cu    (8c)
   0.34452000000000   0.00000000000000   0.00000000000000   Cu    (8c)
   0.65548000000000   0.65548000000000   0.65548000000000   Cu    (8c)
   0.00000000000000   0.27482000000000   0.65170000000000   Zn   (24g)
   0.00000000000000   0.65170000000000   0.27482000000000   Zn   (24g)
   0.27482000000000   0.00000000000000   0.65170000000000   Zn   (24g)
   0.27482000000000   0.65170000000000   0.00000000000000   Zn   (24g)
   0.34830000000000   0.34830000000000   0.62312000000000   Zn   (24g)
   0.34830000000000   0.62312000000000   0.34830000000000   Zn   (24g)
   0.37688000000000   0.72518000000000   0.72518000000000   Zn   (24g)
   0.62312000000000   0.34830000000000   0.34830000000000   Zn   (24g)
   0.65170000000000   0.00000000000000   0.27482000000000   Zn   (24g)
   0.65170000000000   0.27482000000000   0.00000000000000   Zn   (24g)
   0.72518000000000   0.37688000000000   0.72518000000000   Zn   (24g)
   0.72518000000000   0.72518000000000   0.37688000000000   Zn   (24g)
   0.00000000000000   0.00000000000000   0.78438000000000   Zn    (8c)
   0.00000000000000   0.78438000000000   0.00000000000000   Zn    (8c)
   0.21562000000000   0.21562000000000   0.21562000000000   Zn    (8c)
   0.78438000000000   0.00000000000000   0.00000000000000   Zn    (8c)
```

High-Pressure cI16 Li: A_cI16_220_c - CIF

```
# CIF file

data_findsym-output
_audit_creation_method FINDSYM

_chemical_name_mineral 'High pressure (38.9 GPa) phase of lithium'
_chemical_formula_sum 'Li'

loop_
_publ_author_name
 'M. Hanfland'
 'K. Syassen'
 'N. E. Christensen'
 'D. L. Novikov'
_journal_name_full
;
Nature
;
_journal_volume 408
_journal_year 2000
_journal_page_first 174
_journal_page_last 178
_publ_Section_title
;
 New high-pressure phases of lithium
;

_aflow_proto 'A_cI16_220_c'
_aflow_params 'a,x1'
_aflow_params_values '5.2716,0.049'
_aflow_Strukturbericht 'None'
_aflow_Pearson 'cI16'

_symmetry_space_group_name_Hall "I -4bd 2c 3 I(-4)3d"
_symmetry_space_group_name_H-M "I -4 3 d"
_symmetry_Int_Tables_number 220

_cell_length_a      5.27160
_cell_length_b      5.27160
```

```
_cell_length_c      5.27160
_cell_angle_alpha 90.00000
_cell_angle_beta   90.00000
_cell_angle_gamma 90.00000

loop_
_space_group_symop_id
_space_group_symop_operation_xyz
1 x,y,z
2 x,-y,-z+1/2
3 -x+1/2,y,-z
4 -x,-y+1/2,z
5 y,z,x
6 y,-z,-x+1/2
7 -y+1/2,z,-x
8 -y,-z+1/2,x
9 z,x,y
10 z,-x,-y+1/2
11 -z+1/2,x,-y
12 -z,-x+1/2,y
13 y+1/4,x+1/4,z+1/4
14 y+1/4,-x+3/4,-z+1/4
15 -y+1/4,x+1/4,-z+3/4
16 -y+3/4,-x+1/4,z+1/4
17 x+1/4,z+1/4,y+1/4
18 x+1/4,-z+3/4,-y+1/4
19 -x+1/4,z+1/4,-y+3/4
20 -x+3/4,-z+1/4,y+1/4
21 z+1/4,y+1/4,x+1/4
22 z+1/4,-y+3/4,-x+1/4
23 -z+1/4,y+1/4,-x+3/4
24 -z+3/4,-y+1/4,x+1/4
25 x+1/2,y+1/2,z+1/2
26 x+1/2,-y+1/2,-z
27 -x,y+1/2,-z+1/2
28 -x+1/2,-y,z+1/2
29 y+1/2,z+1/2,x+1/2
30 y+1/2,-z+1/2,-x
31 -y,z+1/2,-x+1/2
32 -y+1/2,-z,x+1/2
33 z+1/2,x+1/2,y+1/2
34 z+1/2,-x+1/2,-y
35 -z,x+1/2,-y+1/2
36 -z+1/2,-x,y+1/2
37 y+3/4,x+3/4,z+3/4
38 y+3/4,-x+1/4,-z+3/4
39 -y+3/4,x+3/4,-z+1/4
40 -y+1/4,-x+3/4,z+3/4
41 x+3/4,z+3/4,y+3/4
42 x+3/4,-z+1/4,-y+3/4
43 -x+3/4,z+3/4,-y+1/4
44 -x+1/4,-z+3/4,y+3/4
45 z+3/4,y+3/4,x+3/4
46 z+3/4,-y+1/4,-x+3/4
47 -z+3/4,y+3/4,-x+1/4
48 -z+1/4,-y+3/4,x+3/4

loop_
_atom_site_label
_atom_site_type_symbol
_atom_site_symmetry_multiplicity
_atom_site_Wyckoff_label
_atom_site_fract_x
_atom_site_fract_y
_atom_site_fract_z
_atom_site_occupancy
Li1 Li   16 c 0.04900 0.04900 0.04900 1.00000
```

High-Pressure cI16 Li: A_cI16_220_c - POSCAR

```
A_cI16_220_c & a,x1 --params=5.2716,0.049 & I(-4)3d    T_d^6 #220 (c) &
      ↪ cI16 & & Li & 38.9 GPa & M. Hanfland, K. Syassen, N. E.
      ↪ Christensen and D. L. Novikov, Nature 408, 174-178 (2000)
   1.0000000000000000
  -2.63580000000000    2.63580000000000    2.63580000000000
   2.63580000000000   -2.63580000000000    2.63580000000000
   2.63580000000000    2.63580000000000   -2.63580000000000
   Li
    8
Direct
   0.00000000000000   -0.09800000000000    0.50000000000000    Li    (16c)
   0.00000000000000    0.40200000000000    0.50000000000000    Li    (16c)
   0.09800000000000    0.09800000000000    0.09800000000000    Li    (16c)
  -0.09800000000000    0.50000000000000    0.00000000000000    Li    (16c)
   0.40200000000000    0.50000000000000    0.00000000000000    Li    (16c)
   0.50000000000000    0.00000000000000   -0.09800000000000    Li    (16c)
   0.50000000000000    0.00000000000000    0.40200000000000    Li    (16c)
   0.59800000000000    0.59800000000000    0.59800000000000    Li    (16c)
```

$Pu_2C_3$ ($D5_c$): A3B2_cI40_220_d_c - CIF

```
# CIF file

data_findsym-output
_audit_creation_method FINDSYM

_chemical_name_mineral 'Plutonium carbide'
_chemical_formula_sum 'Pu2 C3'

loop_
_publ_author_name
 'J. L. Green'
 'G. P. Arnold'
 'J. A. Leary'
 'N. G. Nereson'
_journal_name_full
;
Journal of Nuclear Materials
;
_journal_volume 34
_journal_year 1970
_journal_page_first 281
_journal_page_last 289
_publ_Section_title
;
 Crystallographic and magnetic ordering studies of plutonium carbides
      ↪ using neutron diffraction
;

#Found in PearsonâĂŹs Handbook, Vol. IV, pp 1993

_aflow_proto 'A3B2_cI40_220_d_c'
_aflow_params 'a,x1,x2'
_aflow_params_values '8.135,0.0492,0.2896'
_aflow_Strukturbericht 'D5_c'
_aflow_Pearson 'cI40'

_symmetry_space_group_name_Hall "I -4bd 2c 3 I(-4)3d"
_symmetry_space_group_name_H-M "I -4 3 d"
_symmetry_Int_Tables_number 220

_cell_length_a      8.13500
_cell_length_b      8.13500
_cell_length_c      8.13500
_cell_angle_alpha 90.00000
_cell_angle_beta   90.00000
_cell_angle_gamma 90.00000

loop_
_space_group_symop_id
_space_group_symop_operation_xyz
1 x,y,z
2 x,-y,-z+1/2
3 -x+1/2,y,-z
4 -x,-y+1/2,z
5 y,z,x
6 y,-z,-x+1/2
7 -y+1/2,z,-x
8 -y,-z+1/2,x
9 z,x,y
10 z,-x,-y+1/2
11 -z+1/2,x,-y
12 -z,-x+1/2,y
13 y+1/4,x+1/4,z+1/4
14 y+1/4,-x+3/4,-z+1/4
15 -y+1/4,x+1/4,-z+3/4
16 -y+3/4,-x+1/4,z+1/4
17 x+1/4,z+1/4,y+1/4
18 x+1/4,-z+3/4,-y+1/4
19 -x+1/4,z+1/4,-y+3/4
20 -x+3/4,-z+1/4,y+1/4
21 z+1/4,y+1/4,x+1/4
22 z+1/4,-y+3/4,-x+1/4
23 -z+1/4,y+1/4,-x+3/4
24 -z+3/4,-y+1/4,x+1/4
25 x+1/2,y+1/2,z+1/2
26 x+1/2,-y+1/2,-z
27 -x,y+1/2,-z+1/2
28 -x+1/2,-y,z+1/2
29 y+1/2,z+1/2,x+1/2
30 y+1/2,-z+1/2,-x
31 -y,z+1/2,-x+1/2
32 -y+1/2,-z,x+1/2
33 z+1/2,x+1/2,y+1/2
34 z+1/2,-x+1/2,-y
35 -z,x+1/2,-y+1/2
36 -z+1/2,-x,y+1/2
37 y+3/4,x+3/4,z+3/4
38 y+3/4,-x+1/4,-z+3/4
39 -y+3/4,x+3/4,-z+1/4
40 -y+1/4,-x+3/4,z+3/4
41 x+3/4,z+3/4,y+3/4
42 x+3/4,-z+1/4,-y+3/4
43 -x+3/4,z+3/4,-y+1/4
44 -x+1/4,-z+3/4,y+3/4
45 z+3/4,y+3/4,x+3/4
46 z+3/4,-y+1/4,-x+3/4
47 -z+3/4,y+3/4,-x+1/4
48 -z+1/4,-y+3/4,x+3/4

loop_
_atom_site_label
_atom_site_type_symbol
_atom_site_symmetry_multiplicity
_atom_site_Wyckoff_label
_atom_site_fract_x
_atom_site_fract_y
_atom_site_fract_z
_atom_site_occupancy
Pu1 Pu   16 c 0.04920 0.04920 0.04920 1.00000
C1  C    24 d 0.28960 0.00000 0.25000 1.00000
```

$Pu_2C_3$ ($D5_c$): A3B2_cI40_220_d_c - POSCAR

```
A3B2_cI40_220_d_c & a,x1,x2 --params=8.135,0.0492,0.2896 & I(-4)3d    T_d
      ↪ ^6 #220 (cd) & cI40 & D5_c & Pu2C3 & & J.L. Green, G.P. Arnold,
      ↪ J.A. Leary and N.G. Nereson, J. Nucl. Mater. 34, 281-289 (1970
      ↪ )
   1.0000000000000000
  -4.06750000000000    4.06750000000000    4.06750000000000
```

```
    4.06750000000000   -4.06750000000000    4.06750000000000
    4.06750000000000    4.06750000000000   -4.06750000000000
     C     Pu
    12      8
Direct
   -0.03960000000000    0.21040000000000    0.75000000000000     C    (24d)
    0.03960000000000    0.75000000000000    0.78960000000000     C    (24d)
    0.21040000000000    0.75000000000000   -0.03960000000000     C    (24d)
    0.25000000000000    0.53960000000000    0.28960000000000     C    (24d)
    0.25000000000000    0.71040000000000    0.46040000000000     C    (24d)
    0.28960000000000    0.25000000000000    0.53960000000000     C    (24d)
    0.46040000000000    0.25000000000000    0.71040000000000     C    (24d)
    0.53960000000000    0.28960000000000    0.25000000000000     C    (24d)
    0.71040000000000    0.46040000000000    0.25000000000000     C    (24d)
    0.75000000000000   -0.03960000000000    0.21040000000000     C    (24d)
    0.75000000000000    0.78960000000000    0.03960000000000     C    (24d)
    0.78960000000000    0.03960000000000    0.75000000000000     C    (24d)
    0.00000000000000   -0.09840000000000    0.50000000000000    Pu    (16c)
    0.00000000000000    0.40160000000000    0.50000000000000    Pu    (16c)
    0.09840000000000    0.09840000000000    0.09840000000000    Pu    (16c)
   -0.09840000000000    0.50000000000000    0.00000000000000    Pu    (16c)
    0.40160000000000    0.50000000000000    0.00000000000000    Pu    (16c)
    0.50000000000000    0.00000000000000   -0.09840000000000    Pu    (16c)
    0.50000000000000    0.00000000000000    0.40160000000000    Pu    (16c)
    0.59840000000000    0.59840000000000    0.59840000000000    Pu    (16c)
```

CsCl (B2): AB_cP2_221_b_a - CIF

```
# CIF file

data_findsym-output
_audit_creation_method FINDSYM

_chemical_name_mineral ''
_chemical_formula_sum 'Cs Cl'

loop_
_publ_author_name
 'V. Ganesan'
 'K. S. Girirajan'
_journal_name_full
;
Paramana -- Journal of Physics
;
_journal_volume 27
_journal_year 1986
_journal_page_first 469
_journal_page_last 474
_publ_Section_title
;
 Lattice parameter and thermal expansion of CsCl and CsBr by x-ray
      ↪ powder diffraction. I. Thermal expansion of CsCl from room
      ↪ temperature to 90$^{\circ}$ K
;

_aflow_proto 'AB_cP2_221_b_a'
_aflow_params 'a'
_aflow_params_values '4.07925'
_aflow_Strukturbericht 'B2'
_aflow_Pearson 'cP2'

_symmetry_space_group_name_Hall "-P 4 2 3"
_symmetry_space_group_name_H-M "P m -3 m"
_symmetry_Int_Tables_number 221

_cell_length_a      4.07925
_cell_length_b      4.07925
_cell_length_c      4.07925
_cell_angle_alpha  90.00000
_cell_angle_beta   90.00000
_cell_angle_gamma  90.00000

loop_
_space_group_symop_id
_space_group_symop_operation_xyz
1 x,y,z
2 x,-y,-z
3 -x,y,-z
4 -x,-y,z
5 y,z,x
6 y,-z,-x
7 -y,z,-x
8 -y,-z,x
9 z,x,y
10 z,-x,-y
11 -z,x,-y
12 -z,-x,y
13 -y,-x,-z
14 -y,x,z
15 y,-x,z
16 y,x,-z
17 -x,-z,-y
18 -x,z,y
19 x,-z,y
20 x,z,-y
21 -z,-y,-x
22 -z,y,x
23 z,-y,x
24 z,y,-x
25 -x,-y,-z
26 -x,y,z
27 x,-y,z
28 x,y,-z
29 -y,-z,-x
30 -y,z,x
31 y,-z,x
32 y,z,-x
33 -z,-x,-y
34 -z,x,y
35 z,-x,y
36 z,x,-y
37 y,x,z
38 y,-x,-z
39 -y,x,-z
40 -y,-x,z
41 x,z,y
42 x,-z,-y
43 -x,z,-y
44 -x,-z,y
45 z,y,x
46 z,-y,-x
47 -z,y,-x
48 -z,-y,x

loop_
_atom_site_label
_atom_site_type_symbol
_atom_site_symmetry_multiplicity
_atom_site_Wyckoff_label
_atom_site_fract_x
_atom_site_fract_y
_atom_site_fract_z
_atom_site_occupancy
Cs1 Cs   1 a 0.00000 0.00000 0.00000 1.00000
Cl1 Cl   1 b 0.50000 0.50000 0.50000 1.00000
```

CsCl (B2): AB_cP2_221_b_a - POSCAR

```
AB_cP2_221_b_a & a --params=4.07925 & Pm(-3)m    O_h^1 #221 (ab) & cP2 &
     ↪ B2 & CsCl & & V. Ganesan and K. S. Girirajan, Parmana --
     ↪ Journal of Physics 27, 469-474 (1986)
   1.0000000000000000
   4.07925000000000    0.00000000000000    0.00000000000000
   0.00000000000000    4.07925000000000    0.00000000000000
   0.00000000000000    0.00000000000000    4.07925000000000
   Cl    Cs
    1     1
Direct
   0.50000000000000    0.50000000000000    0.50000000000000    Cl    (1b)
   0.00000000000000    0.00000000000000    0.00000000000000    Cs    (1a)
```

NbO: AB_cP6_221_c_d - CIF

```
# CIF file

data_findsym-output
_audit_creation_method FINDSYM

_chemical_name_mineral ''
_chemical_formula_sum 'Nb O'

loop_
_publ_author_name
 'A. L. Bowman'
 'T. C. Wallace'
 'J. L. Yarnell'
 'R. G. Wenzel'
_journal_name_full
;
Acta Crystallographica
;
_journal_volume 21
_journal_year 1966
_journal_page_first 843
_journal_page_last 843
_publ_Section_title
;
 The crystal structure of niobium monoxide
;

# Found in  Pearson's Handbook, Vol. IV, pp. 4535

_aflow_proto 'AB_cP6_221_c_d'
_aflow_params 'a'
_aflow_params_values '4.2101'
_aflow_Strukturbericht 'None'
_aflow_Pearson 'cP6'

_symmetry_space_group_name_Hall "-P 4 2 3"
_symmetry_space_group_name_H-M "P m -3 m"
_symmetry_Int_Tables_number 221

_cell_length_a      4.21010
_cell_length_b      4.21010
_cell_length_c      4.21010
_cell_angle_alpha  90.00000
_cell_angle_beta   90.00000
_cell_angle_gamma  90.00000

loop_
_space_group_symop_id
_space_group_symop_operation_xyz
1 x,y,z
2 x,-y,-z
3 -x,y,-z
4 -x,-y,z
5 y,z,x
6 y,-z,-x
7 -y,z,-x
8 -y,-z,x
```

```
9 z,x,y
10 z,-x,-y
11 -z,x,-y
12 -z,-x,y
13 -y,-x,-z
14 -y,x,z
15 y,-x,z
16 y,x,-z
17 -x,-z,-y
18 -x,z,y
19 x,-z,y
20 x,z,-y
21 -z,-y,-x
22 -z,y,x
23 z,-y,x
24 z,y,-x
25 -x,-y,-z
26 -x,y,z
27 x,-y,z
28 x,y,-z
29 -y,-z,-x
30 -y,z,x
31 y,-z,x
32 y,z,-x
33 -z,-x,-y
34 -z,x,y
35 z,-x,y
36 z,x,-y
37 y,x,z
38 y,-x,-z
39 -y,x,-z
40 -y,-x,z
41 x,z,y
42 x,-z,-y
43 -x,z,-y
44 -x,-z,y
45 z,y,x
46 z,-y,-x
47 -z,y,-x
48 -z,-y,x

loop_
_atom_site_label
_atom_site_type_symbol
_atom_site_symmetry_multiplicity
_atom_site_Wyckoff_label
_atom_site_fract_x
_atom_site_fract_y
_atom_site_fract_z
_atom_site_occupancy
Nb1 Nb    3 c 0.00000 0.50000 0.50000 1.00000
O1  O     3 d 0.50000 0.00000 0.00000 1.00000
```

NbO: AB_cP6_221_c_d - POSCAR

```
AB_cP6_221_c_d & a --params=4.2101 & Pm(-3)m    O_h^1 #221 (cd) & cP6 & &
    ↪ NbO & & A. L. Bowman, T. C. Wallace, J. L. Yarnell and R. G.
    ↪ Wenzel, Acta Cryst. 21, 843 (1966)
   1.0000000000000000
   4.21010000000000    0.00000000000000    0.00000000000000
   0.00000000000000    4.21010000000000    0.00000000000000
   0.00000000000000    0.00000000000000    4.21010000000000
   Nb      O
    3      3
Direct
   0.00000000000000    0.50000000000000    0.50000000000000    Nb    (3c)
   0.50000000000000    0.00000000000000    0.50000000000000    Nb    (3c)
   0.50000000000000    0.50000000000000    0.00000000000000    Nb    (3c)
   0.00000000000000    0.00000000000000    0.50000000000000     O    (3d)
   0.00000000000000    0.50000000000000    0.00000000000000     O    (3d)
   0.50000000000000    0.00000000000000    0.00000000000000     O    (3d)
```

Cubic Perovskite ($CaTiO_3$, $E2_1$): AB3C_cP5_221_a_c_b - CIF

```
# CIF file

data_findsym-output
_audit_creation_method FINDSYM

_chemical_name_mineral '(Cubic) Perovskite'
_chemical_formula_sum 'Ca Ti O3'

loop_
_publ_author_name
 'T. Barth'
_journal_name_full
;
Norsk Geologisk Tidsskrift
;
_journal_volume 8
_journal_year 1925
_journal_page_first 14
_journal_page_last 19
_publ_Section_title
;
 Die Kristallstruktur von Perowskit und verwandten Verbidungen
;

# Found in AMS Database

_aflow_proto 'AB3C_cP5_221_a_c_b'
_aflow_params 'a'
_aflow_params_values '3.795'
_aflow_Strukturbericht 'E2_1'
_aflow_Pearson 'cP5'

_symmetry_space_group_name_Hall "-P 4 2 3"
_symmetry_space_group_name_H-M "P m -3 m"
_symmetry_Int_Tables_number 221

_cell_length_a      3.79500
_cell_length_b      3.79500
_cell_length_c      3.79500
_cell_angle_alpha  90.00000
_cell_angle_beta   90.00000
_cell_angle_gamma  90.00000

loop_
_space_group_symop_id
_space_group_symop_operation_xyz
1 x,y,z
2 x,-y,-z
3 -x,y,-z
4 -x,-y,z
5 y,z,x
6 y,-z,-x
7 -y,z,-x
8 -y,-z,x
9 z,x,y
10 z,-x,-y
11 -z,x,-y
12 -z,-x,y
13 -y,-x,-z
14 -y,x,z
15 y,-x,z
16 y,x,-z
17 -x,-z,-y
18 -x,z,y
19 x,-z,y
20 x,z,-y
21 -z,-y,-x
22 -z,y,x
23 z,-y,x
24 z,y,-x
25 -x,-y,-z
26 -x,y,z
27 x,-y,z
28 x,y,-z
29 -y,-z,-x
30 -y,z,x
31 y,-z,x
32 y,z,-x
33 -z,-x,-y
34 -z,x,y
35 z,-x,y
36 z,x,-y
37 y,x,z
38 y,-x,-z
39 -y,x,-z
40 -y,-x,z
41 x,z,y
42 x,-z,-y
43 -x,z,-y
44 -x,-z,y
45 z,y,x
46 z,-y,-x
47 -z,y,-x
48 -z,-y,x

loop_
_atom_site_label
_atom_site_type_symbol
_atom_site_symmetry_multiplicity
_atom_site_Wyckoff_label
_atom_site_fract_x
_atom_site_fract_y
_atom_site_fract_z
_atom_site_occupancy
Ca1 Ca    1 a 0.00000 0.00000 0.00000 1.00000
Ti1 Ti    1 b 0.50000 0.50000 0.50000 1.00000
O1  O     3 c 0.00000 0.50000 0.50000 1.00000
```

Cubic Perovskite ($CaTiO_3$, $E2_1$): AB3C_cP5_221_a_c_b - POSCAR

```
AB3C_cP5_221_a_c_b & a --params=3.795 & Pm(-3)m    O_h^1 #221 (abc) & cP5
    ↪ & E2_1 & CaTiO3 & Perovskite & T. Barth, Norsk. Geol. Tidssk.
    ↪ 8, 14-29 (1925)
   1.0000000000000000
   3.79500000000000    0.00000000000000    0.00000000000000
   0.00000000000000    3.79500000000000    0.00000000000000
   0.00000000000000    0.00000000000000    3.79500000000000
   Ca      O      Ti
    1      3       1
Direct
   0.00000000000000    0.00000000000000    0.00000000000000    Ca    (1a)
   0.00000000000000    0.50000000000000    0.50000000000000     O    (3c)
   0.50000000000000    0.00000000000000    0.50000000000000     O    (3c)
   0.50000000000000    0.50000000000000    0.00000000000000     O    (3c)
   0.50000000000000    0.50000000000000    0.50000000000000    Ti    (1b)
```

Model of Austenite (cP32): AB27CD3_cP32_221_a_dij_b_c - CIF

```
# CIF file

data_findsym-output
_audit_creation_method FINDSYM

_chemical_name_mineral ''
_chemical_formula_sum 'Cr Fe27 Mo Ni3'
```

```
loop_
_publ_author_name
 'Michael J. Mehl'
_journal_name_full
;
None
;
_journal_volume 0
_journal_year 2008
_journal_page_first 0
_journal_page_last 0
_publ_Section_title
;
 Hypothetical cP32 Austenite Structure
;

_aflow_proto 'AB27CD3_cP32_221_a_dij_b_c'
_aflow_params 'a,y5,y6'
_aflow_params_values '7.04,0.245,0.26'
_aflow_Strukturbericht 'None'
_aflow_Pearson 'cP32'

_symmetry_space_group_name_Hall "-P 4 2 3"
_symmetry_space_group_name_H-M "P m -3 m"
_symmetry_Int_Tables_number 221

_cell_length_a      7.04000
_cell_length_b      7.04000
_cell_length_c      7.04000
_cell_angle_alpha 90.00000
_cell_angle_beta  90.00000
_cell_angle_gamma 90.00000

loop_
_space_group_symop_id
_space_group_symop_operation_xyz
1 x,y,z
2 x,-y,-z
3 -x,y,-z
4 -x,-y,z
5 y,z,x
6 y,-z,-x
7 -y,z,-x
8 -y,-z,x
9 z,x,y
10 z,-x,-y
11 -z,x,-y
12 -z,-x,y
13 -y,-x,-z
14 -y,x,z
15 y,-x,z
16 y,x,-z
17 -x,-z,-y
18 -x,z,y
19 x,-z,y
20 x,z,-y
21 -z,-y,-x
22 -z,y,x
23 z,-y,x
24 z,y,-x
25 -x,-y,-z
26 -x,y,z
27 x,-y,z
28 x,y,-z
29 -y,-z,-x
30 -y,z,x
31 y,-z,x
32 y,z,-x
33 -z,-x,-y
34 -z,x,y
35 z,-x,y
36 z,x,-y
37 y,x,z
38 y,-x,-z
39 -y,x,-z
40 -y,-x,z
41 x,z,y
42 x,-z,-y
43 -x,z,-y
44 -x,-z,y
45 z,y,x
46 z,-y,-x
47 -z,y,-x
48 -z,-y,x

loop_
_atom_site_label
_atom_site_type_symbol
_atom_site_symmetry_multiplicity
_atom_site_Wyckoff_label
_atom_site_fract_x
_atom_site_fract_y
_atom_site_fract_z
_atom_site_occupancy
Cr1 Cr    1 a 0.00000 0.00000 0.00000 1.00000
Mo1 Mo    1 b 0.50000 0.50000 0.50000 1.00000
Ni1 Ni    3 c 0.00000 0.50000 0.50000 1.00000
Fe1 Fe    3 d 0.50000 0.00000 0.00000 1.00000
Fe2 Fe   12 i 0.00000 0.24500 0.24500 1.00000
Fe3 Fe   12 j 0.50000 0.26000 0.26000 1.00000
```

Model of Austenite (cP32): AB27CD3_cP32_221_a_dij_b_c - POSCAR

```
AB27CD3_cP32_221_a_dij_b_c & a,y5,y6 --params=7.04,0.245,0.26 & Pm(-3)m
   ↪ O_h^1 #221 (abcdij) & cP32 & & CrFe27MoNi3 & Hypothetical
   ↪ Austenitic Phase &
   1.0000000000000000
   7.04000000000000   0.00000000000000   0.00000000000000
   0.00000000000000   7.04000000000000   0.00000000000000
   0.00000000000000   0.00000000000000   7.04000000000000
   Cr    Fe    Mo    Ni
    1    27     1     3
Direct
   0.00000000000000   0.00000000000000   0.00000000000000   Cr    (1a)
   0.00000000000000   0.24500000000000   0.24500000000000   Fe   (12i)
   0.00000000000000   0.24500000000000   0.75500000000000   Fe   (12i)
  -0.00000000000000   0.75500000000000   0.24500000000000   Fe   (12i)
   0.00000000000000   0.75500000000000   0.75500000000000   Fe   (12i)
   0.24500000000000   0.00000000000000   0.24500000000000   Fe   (12i)
   0.24500000000000   0.00000000000000   0.75500000000000   Fe   (12i)
   0.24500000000000   0.24500000000000   0.00000000000000   Fe   (12i)
   0.24500000000000   0.75500000000000   0.00000000000000   Fe   (12i)
   0.75500000000000   0.00000000000000   0.24500000000000   Fe   (12i)
   0.75500000000000   0.00000000000000   0.75500000000000   Fe   (12i)
   0.75500000000000   0.24500000000000   0.00000000000000   Fe   (12i)
   0.75500000000000   0.75500000000000   0.00000000000000   Fe   (12i)
   0.26000000000000   0.26000000000000   0.50000000000000   Fe   (12j)
   0.26000000000000   0.50000000000000   0.26000000000000   Fe   (12j)
   0.26000000000000   0.50000000000000   0.74000000000000   Fe   (12j)
   0.26000000000000   0.74000000000000   0.50000000000000   Fe   (12j)
   0.50000000000000   0.26000000000000   0.26000000000000   Fe   (12j)
   0.50000000000000   0.26000000000000   0.74000000000000   Fe   (12j)
   0.50000000000000   0.74000000000000   0.26000000000000   Fe   (12j)
   0.50000000000000   0.74000000000000   0.74000000000000   Fe   (12j)
   0.74000000000000   0.26000000000000   0.50000000000000   Fe   (12j)
   0.74000000000000   0.50000000000000   0.26000000000000   Fe   (12j)
   0.74000000000000   0.50000000000000   0.74000000000000   Fe   (12j)
   0.74000000000000   0.74000000000000   0.50000000000000   Fe   (12j)
   0.00000000000000   0.00000000000000   0.50000000000000   Fe    (3d)
   0.00000000000000   0.50000000000000   0.00000000000000   Fe    (3d)
   0.50000000000000   0.00000000000000   0.00000000000000   Fe    (3d)
   0.50000000000000   0.50000000000000   0.50000000000000   Mo    (1b)
   0.00000000000000   0.50000000000000   0.50000000000000   Ni    (3c)
   0.50000000000000   0.00000000000000   0.50000000000000   Ni    (3c)
   0.50000000000000   0.50000000000000   0.00000000000000   Ni    (3c)
```

$Cu_3Au$ ($L1_2$): AB3_cP4_221_a_c - CIF

```
# CIF file

data_findsym-output
_audit_creation_method FINDSYM

_chemical_name_mineral ''
_chemical_formula_sum 'Cu3 Au'

loop_
_publ_author_name
 'E. A. Owen'
 'Y. H. Liu'
_journal_name_full
;
Philosophical Magazine
;
_journal_volume 38
_journal_year 1947
_journal_page_first 354
_journal_page_last 360
_publ_Section_title
;
 The Thermal Expansion of the Gold-Copper Alloy AuCu$_3$
;

# Found in Pearson's Handbook, Vol. 1, pp. 1273

_aflow_proto 'AB3_cP4_221_a_c'
_aflow_params 'a'
_aflow_params_values '3.7402'
_aflow_Strukturbericht 'L1_2'
_aflow_Pearson 'cP4'

_symmetry_space_group_name_Hall "-P 4 2 3"
_symmetry_space_group_name_H-M "P m -3 m"
_symmetry_Int_Tables_number 221

_cell_length_a      3.74020
_cell_length_b      3.74020
_cell_length_c      3.74020
_cell_angle_alpha 90.00000
_cell_angle_beta  90.00000
_cell_angle_gamma 90.00000

loop_
_space_group_symop_id
_space_group_symop_operation_xyz
1 x,y,z
2 x,-y,-z
3 -x,y,-z
4 -x,-y,z
5 y,z,x
6 y,-z,-x
7 -y,z,-x
8 -y,-z,x
9 z,x,y
10 z,-x,-y
11 -z,x,-y
12 -z,-x,y
13 -y,-x,-z
14 -y,x,z
15 y,-x,z
```

```
16 y,x,-z
17 -x,-z,-y
18 -x,z,y
19 x,-z,y
20 x,z,-y
21 -z,-y,-x
22 -z,y,x
23 z,-y,x
24 z,y,-x
25 -x,-y,-z
26 -x,y,z
27 x,-y,z
28 x,y,-z
29 -y,-z,-x
30 -y,z,x
31 y,-z,x
32 y,z,-x
33 -z,-x,-y
34 -z,x,y
35 z,-x,y
36 z,x,-y
37 y,x,z
38 y,-x,-z
39 -y,x,-z
40 -y,-x,z
41 x,z,y
42 x,-z,-y
43 -x,z,-y
44 -x,-z,y
45 z,y,x
46 z,-y,-x
47 -z,y,-x
48 -z,-y,x

loop_
_atom_site_label
_atom_site_type_symbol
_atom_site_symmetry_multiplicity
_atom_site_Wyckoff_label
_atom_site_fract_x
_atom_site_fract_y
_atom_site_fract_z
_atom_site_occupancy
Au1 Au    1 a 0.00000 0.00000 0.00000 1.00000
Cu1 Cu    3 c 0.00000 0.50000 0.50000 1.00000
```

$Cu_3Au$ ($L1_2$): AB3_cP4_221_a_c - POSCAR

```
AB3_cP4_221_a_c & a --params=3.7402 & Pm(-3)m    O_h^1 #221 (ac) & cP4 &
    ↪ L1_2 & Cu3Au & & E. A. Owen and Y. H. Liu, Phil. Mag. 38,
    ↪ 354-360 (1947)
   1.0000000000000000
   3.74020000000000    0.00000000000000    0.00000000000000
   0.00000000000000    3.74020000000000    0.00000000000000
   0.00000000000000    0.00000000000000    3.74020000000000
   Au    Cu
    1     3
Direct
   0.00000000000000    0.00000000000000    0.00000000000000    Au    (1a)
   0.00000000000000    0.50000000000000    0.50000000000000    Cu    (3c)
   0.50000000000000    0.00000000000000    0.50000000000000    Cu    (3c)
   0.50000000000000    0.50000000000000    0.00000000000000    Cu    (3c)
```

$\alpha$-Po ($A_h$): A_cP1_221_a - CIF

```
# CIF file

data_findsym-output
_audit_creation_method FINDSYM

_chemical_name_mineral 'alpha Po'
_chemical_formula_sum 'Po'

loop_
_publ_author_name
 'William H. Beamer'
 'Charles R. Maxwell'
_journal_name_full
;
Journal of Chemical Physics
;
_journal_volume 14
_journal_year 1946
_journal_page_first 569
_journal_page_last 569
_publ_Section_title
;
 The Crystal Structure of Polonium
;

# Found in  Donohue, pp. 390-391

_aflow_proto 'A_cP1_221_a'
_aflow_params 'a'
_aflow_params_values '3.34'
_aflow_Strukturbericht 'A_h'
_aflow_Pearson 'cP1'

_symmetry_space_group_name_Hall "-P 4 2 3"
_symmetry_space_group_name_H-M "P m -3 m"
_symmetry_Int_Tables_number 221

_cell_length_a     3.34000
_cell_length_b     3.34000
_cell_length_c     3.34000
_cell_angle_alpha  90.00000
_cell_angle_beta   90.00000
_cell_angle_gamma  90.00000

loop_
_space_group_symop_id
_space_group_symop_operation_xyz
1 x,y,z
2 x,-y,-z
3 -x,y,-z
4 -x,-y,z
5 y,z,x
6 y,-z,-x
7 -y,z,-x
8 -y,-z,x
9 z,x,y
10 z,-x,-y
11 -z,x,-y
12 -z,-x,y
13 -y,-x,-z
14 -y,x,z
15 y,-x,z
16 y,x,-z
17 -x,-z,-y
18 -x,z,y
19 x,-z,y
20 x,z,-y
21 -z,-y,-x
22 -z,y,x
23 z,-y,x
24 z,y,-x
25 -x,-y,-z
26 -x,y,z
27 x,-y,z
28 x,y,-z
29 -y,-z,-x
30 -y,z,x
31 y,-z,x
32 y,z,-x
33 -z,-x,-y
34 -z,x,y
35 z,-x,y
36 z,x,-y
37 y,x,z
38 y,-x,-z
39 -y,x,-z
40 -y,-x,z
41 x,z,y
42 x,-z,-y
43 -x,z,-y
44 -x,-z,y
45 z,y,x
46 z,-y,-x
47 -z,y,-x
48 -z,-y,x

loop_
_atom_site_label
_atom_site_type_symbol
_atom_site_symmetry_multiplicity
_atom_site_Wyckoff_label
_atom_site_fract_x
_atom_site_fract_y
_atom_site_fract_z
_atom_site_occupancy
Po1 Po    1 a 0.00000 0.00000 0.00000 1.00000
```

$\alpha$-Po ($A_h$): A_cP1_221_a - POSCAR

```
A_cP1_221_a & a --params=3.34 & Pm(-3)m    O_h^1 #221 (a) & cP1 & A_h &
    ↪ Po & alpha & W. H. Beamer and C. R. Maxwell, J. Chem. Physics
    ↪ 14, 569 (1946)
   1.0000000000000000
   3.34000000000000    0.00000000000000    0.00000000000000
   0.00000000000000    3.34000000000000    0.00000000000000
   0.00000000000000    0.00000000000000    3.34000000000000
   Po
    1
Direct
   0.00000000000000    0.00000000000000    0.00000000000000    Po    (1a)
```

$BaHg_{11}$ ($D2_e$): AB11_cP36_221_c_agij - CIF

```
# CIF file

data_findsym-output
_audit_creation_method FINDSYM

_chemical_name_mineral ''
_chemical_formula_sum 'Ba Hg11'

loop_
_publ_author_name
 'G. Peyronel'
_journal_name_full
;
Gazzetta Chimica Italiana
;
_journal_volume 82
_journal_year 1952
_journal_page_first 679
_journal_page_last 690
_publ_Section_title
;
 Struttura della fase BaHg$_{11}$
```

```
;

# Found in  http://materials.springer.com/isp/crystallographic/docs/
    ↪ sd_1251135

_aflow_proto 'AB11_cP36_221_c_agij'
_aflow_params 'a,x3,y4,y5'
_aflow_params_values '9.6,0.345,0.225,0.115'
_aflow_Strukturbericht 'D2_e'
_aflow_Pearson 'cP36'

_symmetry_space_group_name_Hall "-P 4 2 3"
_symmetry_space_group_name_H-M "P m -3 m"
_symmetry_Int_Tables_number 221

_cell_length_a      9.60000
_cell_length_b      9.60000
_cell_length_c      9.60000
_cell_angle_alpha  90.00000
_cell_angle_beta   90.00000
_cell_angle_gamma  90.00000

loop_
_space_group_symop_id
_space_group_symop_operation_xyz
1 x,y,z
2 x,-y,-z
3 -x,y,-z
4 -x,-y,z
5 y,z,x
6 y,-z,-x
7 -y,z,-x
8 -y,-z,x
9 z,x,y
10 z,-x,-y
11 -z,x,-y
12 -z,-x,y
13 -y,-x,-z
14 -y,x,z
15 y,-x,z
16 y,x,-z
17 -x,-z,-y
18 -x,z,y
19 x,-z,y
20 x,z,-y
21 -z,-y,-x
22 -z,y,x
23 z,-y,x
24 z,y,-x
25 -x,-y,-z
26 -x,y,z
27 x,-y,z
28 x,y,-z
29 -y,-z,-x
30 -y,z,x
31 y,-z,x
32 y,z,-x
33 -z,-x,-y
34 -z,x,y
35 z,-x,y
36 z,x,-y
37 y,x,z
38 y,-x,-z
39 -y,x,-z
40 -y,-x,z
41 x,z,y
42 x,-z,-y
43 -x,z,-y
44 -x,-z,y
45 z,y,x
46 z,-y,-x
47 -z,y,-x
48 -z,-y,x

loop_
_atom_site_label
_atom_site_type_symbol
_atom_site_symmetry_multiplicity
_atom_site_Wyckoff_label
_atom_site_fract_x
_atom_site_fract_y
_atom_site_fract_z
_atom_site_occupancy
Hg1 Hg   1 a 0.00000 0.00000 0.00000 1.00000
Ba1 Ba   3 c 0.00000 0.50000 0.50000 1.00000
Hg2 Hg   8 g 0.34500 0.34500 0.34500 1.00000
Hg3 Hg  12 i 0.00000 0.22500 0.22500 1.00000
Hg4 Hg  12 j 0.50000 0.11500 0.11500 1.00000
```

BaHg$_{11}$ (D2$_e$): AB11_cP36_221_c_agij - POSCAR

```
AB11_cP36_221_c_agij & a,x3,y4,y5 --params=9.6,0.345,0.225,0.115 & Pm(-3
    ↪ )m    O_h^1 #221 (acgij) & cP36 & D2_e & BaHg11 & & G. Peyronel,
    ↪ Gazz. Chim. Ital. 82, 679-90 (1952)
   1.0000000000000000
   9.60000000000000   0.00000000000000   0.00000000000000
   0.00000000000000   9.60000000000000   0.00000000000000
   0.00000000000000   0.00000000000000   9.60000000000000
   Ba    Hg
    3    33
Direct
   0.00000000000000   0.50000000000000   0.50000000000000   Ba    (3c)
   0.50000000000000   0.00000000000000   0.50000000000000   Ba    (3c)
   0.50000000000000   0.50000000000000   0.00000000000000   Ba    (3c)
   0.00000000000000   0.22500000000000   0.22500000000000   Hg   (12i)
   0.00000000000000   0.22500000000000   0.77500000000000   Hg   (12i)
   0.00000000000000   0.77500000000000   0.22500000000000   Hg   (12i)
   0.00000000000000   0.77500000000000   0.77500000000000   Hg   (12i)
   0.22500000000000   0.00000000000000   0.22500000000000   Hg   (12i)
   0.22500000000000   0.00000000000000   0.77500000000000   Hg   (12i)
   0.22500000000000   0.22500000000000   0.00000000000000   Hg   (12i)
   0.22500000000000   0.77500000000000   0.00000000000000   Hg   (12i)
   0.77500000000000   0.00000000000000   0.22500000000000   Hg   (12i)
   0.77500000000000   0.00000000000000   0.77500000000000   Hg   (12i)
   0.77500000000000   0.22500000000000   0.00000000000000   Hg   (12i)
   0.77500000000000   0.77500000000000   0.00000000000000   Hg   (12i)
   0.11500000000000   0.11500000000000   0.50000000000000   Hg   (12j)
   0.11500000000000   0.50000000000000   0.11500000000000   Hg   (12j)
   0.11500000000000   0.50000000000000   0.88500000000000   Hg   (12j)
   0.11500000000000   0.88500000000000   0.50000000000000   Hg   (12j)
   0.50000000000000   0.11500000000000   0.11500000000000   Hg   (12j)
   0.50000000000000   0.11500000000000   0.88500000000000   Hg   (12j)
   0.50000000000000   0.88500000000000   0.11500000000000   Hg   (12j)
   0.50000000000000   0.88500000000000   0.88500000000000   Hg   (12j)
   0.88500000000000   0.11500000000000   0.50000000000000   Hg   (12j)
   0.88500000000000   0.50000000000000   0.11500000000000   Hg   (12j)
   0.88500000000000   0.50000000000000   0.88500000000000   Hg   (12j)
   0.88500000000000   0.88500000000000   0.50000000000000   Hg   (12j)
   0.00000000000000   0.00000000000000   0.00000000000000   Hg    (1a)
   0.34500000000000   0.34500000000000   0.34500000000000   Hg    (8g)
   0.34500000000000   0.34500000000000   0.65500000000000   Hg    (8g)
   0.34500000000000   0.65500000000000   0.34500000000000   Hg    (8g)
   0.34500000000000   0.65500000000000   0.65500000000000   Hg    (8g)
   0.65500000000000   0.34500000000000   0.34500000000000   Hg    (8g)
   0.65500000000000   0.34500000000000   0.65500000000000   Hg    (8g)
   0.65500000000000   0.65500000000000   0.34500000000000   Hg    (8g)
   0.65500000000000   0.65500000000000   0.65500000000000   Hg    (8g)
```

Model of Ferrite (cP16): AB11CD3_cP16_221_a_dg_b_c - CIF

```
# CIF file

data_findsym-output
_audit_creation_method FINDSYM

_chemical_name_mineral ''
_chemical_formula_sum 'Cr Fe11 Mo Ni3'

loop_
_publ_author_name
 'Michael J. Mehl'
_journal_name_full
;
None
;
_journal_volume 0
_journal_year 2008
_journal_page_first 0
_journal_page_last 0
_publ_Section_title
;
 Hypothetical cP16 Austenite Structure
;

_aflow_proto 'AB11CD3_cP16_221_a_dg_b_c'
_aflow_params 'a,x5'
_aflow_params_values '5.74,0.245'
_aflow_Strukturbericht 'None'
_aflow_Pearson 'cP16'

_symmetry_space_group_name_Hall "-P 4 2 3"
_symmetry_space_group_name_H-M "P m -3 m"
_symmetry_Int_Tables_number 221

_cell_length_a      5.74000
_cell_length_b      5.74000
_cell_length_c      5.74000
_cell_angle_alpha  90.00000
_cell_angle_beta   90.00000
_cell_angle_gamma  90.00000

loop_
_space_group_symop_id
_space_group_symop_operation_xyz
1 x,y,z
2 x,-y,-z
3 -x,y,-z
4 -x,-y,z
5 y,z,x
6 y,-z,-x
7 -y,z,-x
8 -y,-z,x
9 z,x,y
10 z,-x,-y
11 -z,x,-y
12 -z,-x,y
13 -y,-x,-z
14 -y,x,z
15 y,-x,z
16 y,x,-z
17 -x,-z,-y
18 -x,z,y
19 x,-z,y
20 x,z,-y
21 -z,-y,-x
22 -z,y,x
23 z,-y,x
24 z,y,-x
25 -x,-y,-z
26 -x,y,z
```

```
27 x,-y,z
28 x,y,-z
29 -y,-z,-x
30 -y,z,x
31 y,-z,x
32 y,z,-x
33 -z,-x,-y
34 -z,x,y
35 z,-x,y
36 z,x,-y
37 y,x,z
38 y,-x,-z
39 -y,x,-z
40 -y,-x,z
41 x,z,y
42 x,-z,-y
43 -x,z,-y
44 -x,-z,y
45 z,y,x
46 z,-y,-x
47 -z,y,-x
48 -z,-y,x

loop_
_atom_site_label
_atom_site_type_symbol
_atom_site_symmetry_multiplicity
_atom_site_Wyckoff_label
_atom_site_fract_x
_atom_site_fract_y
_atom_site_fract_z
_atom_site_occupancy
Cr1 Cr   1 a 0.00000 0.00000 0.00000 1.00000
Mo1 Mo   1 b 0.50000 0.50000 0.50000 1.00000
Ni1 Ni   3 c 0.00000 0.50000 0.50000 1.00000
Fe1 Fe   3 d 0.50000 0.00000 0.00000 1.00000
Fe2 Fe   8 g 0.24500 0.24500 0.24500 1.00000
```

Model of Ferrite (cP16): AB11CD3_cP16_221_a_dg_b_c - POSCAR

```
AB11CD3_cP16_221_a_dg_b_c & a,x5 --params=5.74,0.245 & Pm(-3)m   O_h^1 #
    ↪ 221 (abcdg) & cP16 & & CrFe11MoNi3 & Hypothetical Ferritic
    ↪ Phase &
   1.0000000000000000
   5.74000000000000   0.00000000000000   0.00000000000000
   0.00000000000000   5.74000000000000   0.00000000000000
   0.00000000000000   0.00000000000000   5.74000000000000
   Cr    Fe    Mo    Ni
    1    11     1     3
Direct
   0.00000000000000   0.00000000000000   0.00000000000000   Cr   (1a)
   0.00000000000000   0.00000000000000   0.50000000000000   Fe   (3d)
   0.00000000000000   0.50000000000000   0.00000000000000   Fe   (3d)
   0.50000000000000   0.00000000000000   0.00000000000000   Fe   (3d)
   0.24500000000000   0.24500000000000   0.24500000000000   Fe   (8g)
   0.24500000000000   0.24500000000000   0.75500000000000   Fe   (8g)
   0.24500000000000   0.75500000000000   0.24500000000000   Fe   (8g)
   0.24500000000000   0.75500000000000   0.75500000000000   Fe   (8g)
   0.75500000000000   0.24500000000000   0.24500000000000   Fe   (8g)
   0.75500000000000   0.24500000000000   0.75500000000000   Fe   (8g)
   0.75500000000000   0.75500000000000   0.24500000000000   Fe   (8g)
   0.75500000000000   0.75500000000000   0.75500000000000   Fe   (8g)
   0.50000000000000   0.50000000000000   0.50000000000000   Mo   (1b)
   0.00000000000000   0.50000000000000   0.50000000000000   Ni   (3c)
   0.50000000000000   0.00000000000000   0.50000000000000   Ni   (3c)
   0.50000000000000   0.50000000000000   0.00000000000000   Ni   (3c)
```

$\alpha$-$ReO_3$ ($D0_9$): A3B_cP4_221_d_a - CIF

```
# CIF file

data_findsym-output
_audit_creation_method FINDSYM

_chemical_name_mineral 'alpha Rhenium trioxide'
_chemical_formula_sum 'Re O3'

loop_
_publ_author_name
 'Karl Meisel'
_journal_name_full
;
Zeitschrift f\"{u}r anorganische und allgemeine Chemie
;
_journal_volume 207
_journal_year 1932
_journal_page_first 121
_journal_page_last 128
_publ_Section_title
;
 Rheniumtrioxyd. III. Mitteilung. \"{U}ber die Kristallstruktur des
    ↪ Rheniumtrioxyds
;

# Found in AMS Database

_aflow_proto 'A3B_cP4_221_d_a'
_aflow_params 'a'
_aflow_params_values '3.734'
_aflow_Strukturbericht 'D0_9'
_aflow_Pearson 'cP4'

_symmetry_space_group_name_Hall "-P 4 2 3"
_symmetry_space_group_name_H-M "P m -3 m"
_symmetry_Int_Tables_number 221

_cell_length_a     3.73400
_cell_length_b     3.73400
_cell_length_c     3.73400
_cell_angle_alpha 90.00000
_cell_angle_beta  90.00000
_cell_angle_gamma 90.00000

loop_
_space_group_symop_id
_space_group_symop_operation_xyz
1 x,y,z
2 x,-y,-z
3 -x,y,-z
4 -x,-y,z
5 y,z,x
6 y,-z,-x
7 -y,z,-x
8 -y,-z,x
9 z,x,y
10 z,-x,-y
11 -z,x,-y
12 -z,-x,y
13 -y,-x,-z
14 -y,x,z
15 y,-x,z
16 y,x,-z
17 -x,-z,-y
18 -x,z,y
19 x,-z,y
20 x,z,-y
21 -z,-y,-x
22 -z,y,x
23 z,-y,x
24 z,y,-x
25 -x,-y,-z
26 -x,y,z
27 x,-y,z
28 x,y,-z
29 -y,-z,-x
30 -y,z,x
31 y,-z,x
32 y,z,-x
33 -z,-x,-y
34 -z,x,y
35 z,-x,y
36 z,x,-y
37 y,x,z
38 y,-x,-z
39 -y,x,-z
40 -y,-x,z
41 x,z,y
42 x,-z,-y
43 -x,z,-y
44 -x,-z,y
45 z,y,x
46 z,-y,-x
47 -z,y,-x
48 -z,-y,x

loop_
_atom_site_label
_atom_site_type_symbol
_atom_site_symmetry_multiplicity
_atom_site_Wyckoff_label
_atom_site_fract_x
_atom_site_fract_y
_atom_site_fract_z
_atom_site_occupancy
Re1 Re   1 a 0.00000 0.00000 0.00000 1.00000
O1  O    3 d 0.50000 0.00000 0.00000 1.00000
```

$\alpha$-$ReO_3$ ($D0_9$): A3B_cP4_221_d_a - POSCAR

```
A3B_cP4_221_d_a & a --params=3.734 & Pm(-3)m   O_h^1 #221 (ad) & cP4 &
    ↪ D0_9 & ReO3 & & K. Meisel, ZAAC 207, 121-128 (1932)
   1.0000000000000000
   3.73400000000000   0.00000000000000   0.00000000000000
   0.00000000000000   3.73400000000000   0.00000000000000
   0.00000000000000   0.00000000000000   3.73400000000000
    O    Re
    3     1
Direct
   0.00000000000000   0.00000000000000   0.50000000000000   O    (3d)
   0.00000000000000   0.50000000000000   0.00000000000000   O    (3d)
   0.50000000000000   0.00000000000000   0.00000000000000   O    (3d)
   0.00000000000000   0.00000000000000   0.00000000000000   Re   (1a)
```

$CaB_6$ ($D2_1$): A6B_cP7_221_f_a - CIF

```
# CIF file

data_findsym-output
_audit_creation_method FINDSYM

_chemical_name_mineral 'Calcium hexaboride'
_chemical_formula_sum 'Ca B6'

loop_
_publ_author_name
 'Z. Yahia'
 'S. Turrell'
 'G. Turrell'
 'J. P. Mercurio'
_journal_name_full
```

```
;
Journal of Molecular Structure
;
_journal_volume 224
_journal_year 1990
_journal_page_first 303
_journal_page_last 312
_publ_Section_title
;
 Infrared and Raman spectra of hexaborides: force-field calculations,
    ↪ and isotopic effects
;

_aflow_proto 'A6B_cP7_221_f_a'
_aflow_params 'a,x2'
_aflow_params_values '4.145,0.2117'
_aflow_Strukturbericht 'D2_1'
_aflow_Pearson 'cP7'

_symmetry_space_group_name_Hall "-P 4 2 3"
_symmetry_space_group_name_H-M "P m -3 m"
_symmetry_Int_Tables_number 221

_cell_length_a     4.14500
_cell_length_b     4.14500
_cell_length_c     4.14500
_cell_angle_alpha 90.00000
_cell_angle_beta  90.00000
_cell_angle_gamma 90.00000

loop_
_space_group_symop_id
_space_group_symop_operation_xyz
1 x,y,z
2 x,-y,-z
3 -x,y,-z
4 -x,-y,z
5 y,z,x
6 y,-z,-x
7 -y,z,-x
8 -y,-z,x
9 z,x,y
10 z,-x,-y
11 -z,x,-y
12 -z,-x,y
13 -y,-x,-z
14 -y,x,z
15 y,-x,z
16 y,x,-z
17 -x,-z,-y
18 -x,z,y
19 x,-z,y
20 x,z,-y
21 -z,-y,-x
22 -z,y,x
23 z,-y,x
24 z,y,-x
25 -x,-y,-z
26 -x,y,z
27 x,-y,z
28 x,y,-z
29 -y,-z,-x
30 -y,z,x
31 y,-z,x
32 y,z,-x
33 -z,-x,-y
34 -z,x,y
35 z,-x,y
36 z,x,-y
37 y,x,z
38 y,-x,-z
39 -y,x,-z
40 -y,-x,z
41 x,z,y
42 x,-z,-y
43 -x,z,-y
44 -x,-z,y
45 z,y,x
46 z,-y,-x
47 -z,y,-x
48 -z,-y,x

loop_
_atom_site_label
_atom_site_type_symbol
_atom_site_symmetry_multiplicity
_atom_site_Wyckoff_label
_atom_site_fract_x
_atom_site_fract_y
_atom_site_fract_z
_atom_site_occupancy
Ca1 Ca    1 a 0.00000 0.00000 0.00000 1.00000
B1  B     6 f 0.21170 0.50000 0.50000 1.00000
```

$CaB_6$ ($D2_1$): A6B_cP7_221_f_a - POSCAR

```
A6B_cP7_221_f_a & a,x2 --params=4.145,0.2117 & Pm(-3)m    O_h^1 #221 (af)
    ↪ & cP7 & D2_1 & CaB6 & & Z. Yahia, S. Turrell, G. Turrell and
    ↪ J. P. Mercurio, J. Mol. Struct. 224, 303-312 (1990)
   1.0000000000000000
   4.14500000000000    0.00000000000000    0.00000000000000
   0.00000000000000    4.14500000000000    0.00000000000000
   0.00000000000000    0.00000000000000    4.14500000000000
    B     Ca
    6     1
Direct
   0.21170084439100    0.50000000000000    0.50000000000000    B    (6f)
   0.50000000000000    0.21170084439100    0.50000000000000    B    (6f)
   0.50000000000000    0.50000000000000    0.21170084439100    B    (6f)
   0.50000000000000    0.50000000000000    0.78829915560900    B    (6f)
   0.50000000000000    0.78829915560900    0.50000000000000    B    (6f)
   0.78829915560900    0.50000000000000    0.50000000000000    B    (6f)
   0.00000000000000    0.00000000000000    0.00000000000000    Ca   (1a)
```

$Cr_3Si$ (A15): A3B_cP8_223_c_a - CIF

```
# CIF file

data_findsym-output
_audit_creation_method FINDSYM

_chemical_name_mineral ''
_chemical_formula_sum 'Cr3 Si'

loop_
_publ_author_name
 'W. Jauch'
 'A. J. Schultz'
 'G. Heger'
_journal_name_full
;
Journal of Applied Crystallography
;
_journal_volume 20
_journal_year 1987
_journal_page_first 117
_journal_page_last 119
_publ_Section_title
;
 Single-crystal time-of-flight neutron diffraction of Cr$_3$Si and
    ↪ MnF$_2$ comparison with monochromatic-beam techniques
;

# Found in  Pearson's Handbook Vol. III, pp. 2742

_aflow_proto 'A3B_cP8_223_c_a'
_aflow_params 'a'
_aflow_params_values '4.556'
_aflow_Strukturbericht 'A15'
_aflow_Pearson 'cP8'

_symmetry_space_group_name_Hall "-P 4n 2 3"
_symmetry_space_group_name_H-M "P m -3 n"
_symmetry_Int_Tables_number 223

_cell_length_a     4.55600
_cell_length_b     4.55600
_cell_length_c     4.55600
_cell_angle_alpha 90.00000
_cell_angle_beta  90.00000
_cell_angle_gamma 90.00000

loop_
_space_group_symop_id
_space_group_symop_operation_xyz
1 x,y,z
2 x,-y,-z
3 -x,y,-z
4 -x,-y,z
5 y,z,x
6 y,-z,-x
7 -y,z,-x
8 -y,-z,x
9 z,x,y
10 z,-x,-y
11 -z,x,-y
12 -z,-x,y
13 -y+1/2,-x+1/2,-z+1/2
14 -y+1/2,x+1/2,z+1/2
15 y+1/2,-x+1/2,z+1/2
16 y+1/2,x+1/2,-z+1/2
17 -x+1/2,-z+1/2,-y+1/2
18 -x+1/2,z+1/2,y+1/2
19 x+1/2,-z+1/2,y+1/2
20 x+1/2,z+1/2,-y+1/2
21 -z+1/2,-y+1/2,-x+1/2
22 -z+1/2,y+1/2,x+1/2
23 z+1/2,-y+1/2,x+1/2
24 z+1/2,y+1/2,-x+1/2
25 -x,-y,-z
26 -x,y,z
27 x,-y,z
28 x,y,-z
29 -y,-z,-x
30 -y,z,x
31 y,-z,x
32 y,z,-x
33 -z,-x,-y
34 -z,x,y
35 z,-x,y
36 z,x,-y
37 y+1/2,x+1/2,z+1/2
38 y+1/2,-x+1/2,-z+1/2
39 -y+1/2,x+1/2,-z+1/2
40 -y+1/2,-x+1/2,z+1/2
41 x+1/2,z+1/2,y+1/2
42 x+1/2,-z+1/2,-y+1/2
43 -x+1/2,z+1/2,-y+1/2
44 -x+1/2,-z+1/2,y+1/2
45 z+1/2,y+1/2,x+1/2
```

```
46 z+1/2,-y+1/2,-x+1/2
47 -z+1/2,y+1/2,-x+1/2
48 -z+1/2,-y+1/2,x+1/2

loop_
_atom_site_label
_atom_site_type_symbol
_atom_site_symmetry_multiplicity
_atom_site_Wyckoff_label
_atom_site_fract_x
_atom_site_fract_y
_atom_site_fract_z
_atom_site_occupancy
Si1 Si   2 a 0.00000 0.00000 0.00000 1.00000
Cr1 Cr   6 c 0.25000 0.00000 0.50000 1.00000
```

$Cr_3Si$ (A15): A3B_cP8_223_c_a - POSCAR

```
A3B_cP8_223_c_a & a --params=4.556 & Pm(-3)n    O_h^3 #223 (ac) & cP8 &
    ↪ A15 & Cr3Si & & W. Jauch, A. J. Schultz and G. Heger, J. Appl.
    ↪ Crystallogr. 20, 117-119 (1987)
   1.0000000000000000
   4.55600000000000    0.00000000000000    0.00000000000000
   0.00000000000000    4.55600000000000    0.00000000000000
   0.00000000000000    0.00000000000000    4.55600000000000
   Cr    Si
    6     2
Direct
   0.00000000000000    0.50000000000000    0.25000000000000    Cr    (6c)
   0.00000000000000    0.50000000000000    0.75000000000000    Cr    (6c)
   0.25000000000000    0.00000000000000    0.50000000000000    Cr    (6c)
   0.50000000000000    0.25000000000000    0.00000000000000    Cr    (6c)
   0.50000000000000    0.75000000000000    0.00000000000000    Cr    (6c)
   0.75000000000000    0.00000000000000    0.50000000000000    Cr    (6c)
   0.00000000000000    0.00000000000000    0.00000000000000    Si    (2a)
   0.50000000000000    0.50000000000000    0.50000000000000    Si    (2a)
```

$Si_{46}$ Clathrate: A_cP46_223_dik - CIF

```
# CIF file

data_findsym-output
_audit_creation_method FINDSYM

_chemical_name_mineral 'Clathrate'
_chemical_formula_sum 'Si'

loop_
_publ_author_name
 'Gary B. Adams'
 'Michael O'Keeffe'
 'Alexander A. Demkov'
 'Otto F. Sankey'
 'Yin-Min Huang'
_journal_name_full
;
Physical Review B
;
_journal_volume 49
_journal_year 1994
_journal_page_first 8048
_journal_page_last 8053
_publ_Section_title
;
 Wide-band-gap Si in open fourfold-coordinated clathrate structures
;

_aflow_proto 'A_cP46_223_dik'
_aflow_params 'a,x2,y3,z3'
_aflow_params_values '10.355,0.1837,0.1172,0.3077'
_aflow_Strukturbericht 'None'
_aflow_Pearson 'cP46'

_symmetry_space_group_name_Hall "-P 4n 2 3"
_symmetry_space_group_name_H-M "P m -3 n"
_symmetry_Int_Tables_number 223

_cell_length_a      10.35500
_cell_length_b      10.35500
_cell_length_c      10.35500
_cell_angle_alpha   90.00000
_cell_angle_beta    90.00000
_cell_angle_gamma   90.00000

loop_
_space_group_symop_id
_space_group_symop_operation_xyz
1 x,y,z
2 x,-y,-z
3 -x,y,-z
4 -x,-y,z
5 y,z,x
6 y,-z,-x
7 -y,z,-x
8 -y,-z,x
9 z,x,y
10 z,-x,-y
11 -z,x,-y
12 -z,-x,y
13 -y+1/2,-x+1/2,-z+1/2
14 -y+1/2,x+1/2,z+1/2
15 y+1/2,-x+1/2,z+1/2
16 y+1/2,x+1/2,-z+1/2
17 -x+1/2,-z+1/2,-y+1/2
18 -x+1/2,z+1/2,y+1/2
19 x+1/2,-z+1/2,y+1/2
20 x+1/2,z+1/2,-y+1/2
21 -z+1/2,-y+1/2,-x+1/2
22 -z+1/2,y+1/2,x+1/2
23 z+1/2,-y+1/2,x+1/2
24 z+1/2,y+1/2,-x+1/2
25 -x,-y,-z
26 -x,y,z
27 x,-y,z
28 x,y,-z
29 -y,-z,-x
30 -y,z,x
31 y,-z,x
32 y,z,-x
33 -z,-x,-y
34 -z,x,y
35 z,-x,y
36 z,x,-y
37 y+1/2,x+1/2,z+1/2
38 y+1/2,-x+1/2,-z+1/2
39 -y+1/2,x+1/2,-z+1/2
40 -y+1/2,-x+1/2,z+1/2
41 x+1/2,z+1/2,y+1/2
42 x+1/2,-z+1/2,-y+1/2
43 -x+1/2,z+1/2,-y+1/2
44 -x+1/2,-z+1/2,y+1/2
45 z+1/2,y+1/2,x+1/2
46 z+1/2,-y+1/2,-x+1/2
47 -z+1/2,y+1/2,-x+1/2
48 -z+1/2,-y+1/2,x+1/2

loop_
_atom_site_label
_atom_site_type_symbol
_atom_site_symmetry_multiplicity
_atom_site_Wyckoff_label
_atom_site_fract_x
_atom_site_fract_y
_atom_site_fract_z
_atom_site_occupancy
Si1 Si   6 d 0.25000 0.50000 0.00000 1.00000
Si2 Si  16 i 0.18370 0.18370 0.18370 1.00000
Si3 Si  24 k 0.00000 0.11720 0.30770 1.00000
```

$Si_{46}$ Clathrate: A_cP46_223_dik - POSCAR

```
A_cP46_223_dik & a,x2,y3,z3 --params=10.355,0.1837,0.1172,0.3077 & Pm(-3
    ↪ )n    O_h^3 #223 (dik) & cP46 & & Si & clathrate & G. B. Adams,
    ↪ M. O'Keefe, A. A. Demkov, O. F. Sankey and Y.-M. Huang, Phys.
    ↪ Rev. B 49, 8048-8053 (1994)
   1.0000000000000000
  10.35500000000000    0.00000000000000    0.00000000000000
   0.00000000000000   10.35500000000000    0.00000000000000
   0.00000000000000    0.00000000000000   10.35500000000000
   Si
   46
Direct
   0.18370000000000    0.18370000000000    0.18370000000000    Si    (16i)
   0.18370000000000    0.18370000000000    0.81630000000000    Si    (16i)
   0.18370000000000    0.81630000000000    0.18370000000000    Si    (16i)
   0.18370000000000    0.81630000000000    0.81630000000000    Si    (16i)
   0.31630000000000    0.31630000000000    0.31630000000000    Si    (16i)
   0.31630000000000    0.31630000000000    0.68370000000000    Si    (16i)
   0.31630000000000    0.68370000000000    0.31630000000000    Si    (16i)
   0.31630000000000    0.68370000000000    0.68370000000000    Si    (16i)
   0.68370000000000    0.31630000000000    0.31630000000000    Si    (16i)
   0.68370000000000    0.31630000000000    0.68370000000000    Si    (16i)
   0.68370000000000    0.68370000000000    0.31630000000000    Si    (16i)
   0.68370000000000    0.68370000000000    0.68370000000000    Si    (16i)
   0.81630000000000    0.18370000000000    0.18370000000000    Si    (16i)
   0.81630000000000    0.18370000000000    0.81630000000000    Si    (16i)
   0.81630000000000    0.81630000000000    0.18370000000000    Si    (16i)
   0.81630000000000    0.81630000000000    0.81630000000000    Si    (16i)
   0.00000000000000    0.11720000000000    0.30770000000000    Si    (24k)
   0.00000000000000    0.11720000000000    0.69230000000000    Si    (24k)
   0.00000000000000    0.88280000000000    0.30770000000000    Si    (24k)
   0.00000000000000    0.88280000000000    0.69230000000000    Si    (24k)
   0.11720000000000    0.30770000000000    0.00000000000000    Si    (24k)
   0.11720000000000    0.69230000000000    0.00000000000000    Si    (24k)
   0.19230000000000    0.38280000000000    0.50000000000000    Si    (24k)
   0.19230000000000    0.61720000000000    0.50000000000000    Si    (24k)
   0.30770000000000    0.00000000000000    0.11720000000000    Si    (24k)
   0.30770000000000    0.00000000000000    0.88280000000000    Si    (24k)
   0.38280000000000    0.50000000000000    0.19230000000000    Si    (24k)
   0.38280000000000    0.50000000000000    0.80770000000000    Si    (24k)
   0.50000000000000    0.19230000000000    0.38280000000000    Si    (24k)
   0.50000000000000    0.19230000000000    0.61720000000000    Si    (24k)
   0.50000000000000    0.80770000000000    0.38280000000000    Si    (24k)
   0.50000000000000    0.80770000000000    0.61720000000000    Si    (24k)
   0.61720000000000    0.50000000000000    0.19230000000000    Si    (24k)
   0.61720000000000    0.50000000000000    0.80770000000000    Si    (24k)
   0.69230000000000    0.00000000000000    0.11720000000000    Si    (24k)
   0.69230000000000    0.00000000000000    0.88280000000000    Si    (24k)
   0.80770000000000    0.38280000000000    0.50000000000000    Si    (24k)
   0.80770000000000    0.61720000000000    0.50000000000000    Si    (24k)
   0.88280000000000    0.30770000000000    0.00000000000000    Si    (24k)
   0.88280000000000    0.69230000000000    0.00000000000000    Si    (24k)
   0.00000000000000    0.25000000000000    0.50000000000000    Si    (6d)
   0.00000000000000    0.75000000000000    0.50000000000000    Si    (6d)
   0.25000000000000    0.50000000000000    0.00000000000000    Si    (6d)
   0.50000000000000    0.00000000000000    0.25000000000000    Si    (6d)
   0.50000000000000    0.00000000000000    0.75000000000000    Si    (6d)
   0.75000000000000    0.50000000000000    0.00000000000000    Si    (6d)
```

Cuprite ($Cu_2O$, C3): A2B_cP6_224_b_a - CIF

```
# CIF file

data_findsym-output
_audit_creation_method FINDSYM

_chemical_name_mineral 'Cuprite'
_chemical_formula_sum 'Cu2 O'

loop_
_publ_author_name
 'R. Restori'
 'D. Schwarzenbach'
_journal_name_full
;
Acta Crystallographica B
;
_journal_volume 42
_journal_year 1986
_journal_page_first 201
_journal_page_last 208
_publ_Section_title
;
 Charge Density in Cuprite, Cu$_2$O
;

# Found in A. Kirfel and K. Eichhorn, Acta Cryst. A 46, pp. 271-284 (
    ↪ 1990)

_aflow_proto 'A2B_cP6_224_b_a'
_aflow_params 'a'
_aflow_params_values '4.267'
_aflow_Strukturbericht 'C3'
_aflow_Pearson 'cP6'

_symmetry_space_group_name_Hall "-P 4bc 2bc 3 Pn(-3)m"
_symmetry_space_group_name_H-M "P n -3 m:2"
_symmetry_Int_Tables_number 224

_cell_length_a    4.26700
_cell_length_b    4.26700
_cell_length_c    4.26700
_cell_angle_alpha 90.00000
_cell_angle_beta  90.00000
_cell_angle_gamma 90.00000

loop_
_space_group_symop_id
_space_group_symop_operation_xyz
1 x,y,z
2 x,-y+1/2,-z+1/2
3 -x+1/2,y,-z+1/2
4 -x+1/2,-y+1/2,z
5 y,z,x
6 y,-z+1/2,-x+1/2
7 -y+1/2,z,-x+1/2
8 -y+1/2,-z+1/2,x
9 z,x,y
10 z,-x+1/2,-y+1/2
11 -z+1/2,x,-y+1/2
12 -z+1/2,-x+1/2,y
13 -y,-x,-z
14 -y,x+1/2,z+1/2
15 y+1/2,-x,z+1/2
16 y+1/2,x+1/2,-z
17 -x,-z,-y
18 -x,z+1/2,y+1/2
19 x+1/2,-z,y+1/2
20 x+1/2,z+1/2,-y
21 -z,-y,-x
22 -z,y+1/2,x+1/2
23 z+1/2,-y,x+1/2
24 z+1/2,y+1/2,-x
25 -x,-y,-z
26 -x,y+1/2,z+1/2
27 x+1/2,-y,z+1/2
28 x+1/2,y+1/2,-z
29 -y,-z,-x
30 -y,z+1/2,x+1/2
31 y+1/2,-z,x+1/2
32 y+1/2,z+1/2,-x
33 -z,-x,-y
34 -z,x+1/2,y+1/2
35 z+1/2,-x,y+1/2
36 z+1/2,x+1/2,-y
37 y,x,z
38 y,-x+1/2,-z+1/2
39 -y+1/2,x,-z+1/2
40 -y+1/2,-x+1/2,z
41 x,z,y
42 x,-z+1/2,-y+1/2
43 -x+1/2,z,-y+1/2
44 -x+1/2,-z+1/2,y
45 z,y,x
46 z,-y+1/2,-x+1/2
47 -z+1/2,y,-x+1/2
48 -z+1/2,-y+1/2,x

loop_
_atom_site_label
_atom_site_type_symbol
_atom_site_symmetry_multiplicity
_atom_site_Wyckoff_label
_atom_site_fract_x
_atom_site_fract_y
_atom_site_fract_z
_atom_site_occupancy
O1  O   2 a 0.25000 0.25000 0.25000 1.00000
Cu1 Cu  4 b 0.00000 0.00000 0.00000 1.00000
```

Cuprite ($Cu_2O$, C3): A2B_cP6_224_b_a - POSCAR

```
A2B_cP6_224_b_a & a --params=4.267 & Pn(-3)m   O_h^4 #224 (ab) & cP6 &
    ↪ C3 & Cu2O & Cuprite & R. Restori and D. Schwarzenbach, Acta
    ↪ Crystallogr. B 42, 201âĂŞ208 (1986)
   1.0000000000000000
   4.26700000000000   0.00000000000000   0.00000000000000
   0.00000000000000   4.26700000000000   0.00000000000000
   0.00000000000000   0.00000000000000   4.26700000000000
   Cu     O
    4     2
Direct
   0.00000000000000   0.00000000000000   0.00000000000000   Cu   (4b)
   0.00000000000000   0.50000000000000   0.50000000000000   Cu   (4b)
   0.50000000000000   0.00000000000000   0.50000000000000   Cu   (4b)
   0.50000000000000   0.50000000000000   0.00000000000000   Cu   (4b)
   0.25000000000000   0.25000000000000   0.25000000000000    O   (2a)
   0.75000000000000   0.75000000000000   0.75000000000000    O   (2a)
```

$Ca_7Ge$: A7B_cF32_225_bd_a - CIF

```
# CIF file

data_findsym-output
_audit_creation_method FINDSYM

_chemical_name_mineral ''
_chemical_formula_sum 'Ca7 Ge'

loop_
_publ_author_name
 'O. Helleis'
 'H. Kandler'
 'E. Leicht'
 'W. Quiring'
 'E. W\"{o}lfel'
_journal_name_full
;
Zeitschrift f\"{u}r anorganische und allgemeine Chemie
;
_journal_volume 320
_journal_year 1963
_journal_page_first 86
_journal_page_last 100
_publ_Section_title
;
 Die Kristallstrukturen der intermetallischen Phasen Ca$_{33}$Ge,
    ↪ Ca$_7$Ge, Ca$_3$Pb und Ca$_5$Pb$_3$
;

# Found in http://materials.springer.com/isp/crystallographic/docs/
    ↪ sd_1301069

_aflow_proto 'A7B_cF32_225_bd_a'
_aflow_params 'a'
_aflow_params_values '9.45'
_aflow_Strukturbericht 'None'
_aflow_Pearson 'cF32'

_symmetry_space_group_name_Hall "-F 4 2 3"
_symmetry_space_group_name_H-M "F m -3 m"
_symmetry_Int_Tables_number 225

_cell_length_a    9.45000
_cell_length_b    9.45000
_cell_length_c    9.45000
_cell_angle_alpha 90.00000
_cell_angle_beta  90.00000
_cell_angle_gamma 90.00000

loop_
_space_group_symop_id
_space_group_symop_operation_xyz
1 x,y,z
2 x,-y,-z
3 -x,y,-z
4 -x,-y,z
5 y,z,x
6 y,-z,-x
7 -y,z,-x
8 -y,-z,x
9 z,x,y
10 z,-x,-y
11 -z,x,-y
12 -z,-x,y
13 -y,-x,-z
14 -y,x,z
15 y,-x,z
16 y,x,-z
17 -x,-z,-y
18 -x,z,y
19 x,-z,y
20 x,z,-y
21 -z,-y,-x
22 -z,y,x
23 z,-y,x
24 z,y,-x
25 -x,-y,-z
26 -x,y,z
27 x,-y,z
28 x,y,-z
```

```
29 -y,-z,-x
30 -y,z,x
31 y,-z,x
32 y,z,-x
33 -z,-x,-y
34 -z,x,y
35 z,-x,y
36 z,x,-y
37 y,x,z
38 y,-x,-z
39 -y,x,-z
40 -y,-x,z
41 x,z,y
42 x,-z,-y
43 -x,z,-y
44 -x,-z,y
45 z,y,x
46 z,-y,-x
47 -z,y,-x
48 -z,-y,x
49 x,y+1/2,z+1/2
50 x,-y+1/2,-z+1/2
51 -x,y+1/2,-z+1/2
52 -x,-y+1/2,z+1/2
53 y,z+1/2,x+1/2
54 y,-z+1/2,-x+1/2
55 -y,z+1/2,-x+1/2
56 -y,-z+1/2,x+1/2
57 z,x+1/2,y+1/2
58 z,-x+1/2,-y+1/2
59 -z,x+1/2,-y+1/2
60 -z,-x+1/2,y+1/2
61 -y,-x+1/2,-z+1/2
62 -y,x+1/2,z+1/2
63 y,-x+1/2,z+1/2
64 y,x+1/2,-z+1/2
65 -x,-z+1/2,-y+1/2
66 -x,z+1/2,y+1/2
67 x,-z+1/2,y+1/2
68 x,z+1/2,-y+1/2
69 -z,-y+1/2,-x+1/2
70 -z,y+1/2,x+1/2
71 z,-y+1/2,x+1/2
72 z,y+1/2,-x+1/2
73 -x,-y+1/2,-z+1/2
74 -x,y+1/2,z+1/2
75 x,-y+1/2,z+1/2
76 x,y+1/2,-z+1/2
77 -y,-z+1/2,-x+1/2
78 -y,z+1/2,x+1/2
79 y,-z+1/2,x+1/2
80 y,z+1/2,-x+1/2
81 -z,-x+1/2,-y+1/2
82 -z,x+1/2,y+1/2
83 z,-x+1/2,y+1/2
84 z,x+1/2,-y+1/2
85 y,x+1/2,z+1/2
86 y,-x+1/2,-z+1/2
87 -y,x+1/2,-z+1/2
88 -y,-x+1/2,z+1/2
89 x,z+1/2,y+1/2
90 x,-z+1/2,-y+1/2
91 -x,z+1/2,-y+1/2
92 -x,-z+1/2,y+1/2
93 z,y+1/2,x+1/2
94 z,-y+1/2,-x+1/2
95 -z,y+1/2,-x+1/2
96 -z,-y+1/2,x+1/2
97 x+1/2,y,z+1/2
98 x+1/2,-y,-z+1/2
99 -x+1/2,y,-z+1/2
100 -x+1/2,-y,z+1/2
101 y+1/2,z,x+1/2
102 y+1/2,-z,-x+1/2
103 -y+1/2,z,-x+1/2
104 -y+1/2,-z,x+1/2
105 z+1/2,x,y+1/2
106 z+1/2,-x,-y+1/2
107 -z+1/2,x,-y+1/2
108 -z+1/2,-x,y+1/2
109 -y+1/2,-x,-z+1/2
110 -y+1/2,x,z+1/2
111 y+1/2,-x,z+1/2
112 y+1/2,x,-z+1/2
113 -x+1/2,-z,-y+1/2
114 -x+1/2,z,y+1/2
115 x+1/2,-z,y+1/2
116 x+1/2,z,-y+1/2
117 -z+1/2,-y,-x+1/2
118 -z+1/2,y,x+1/2
119 z+1/2,-y,x+1/2
120 z+1/2,y,-x+1/2
121 -x+1/2,-y,-z+1/2
122 -x+1/2,y,z+1/2
123 x+1/2,-y,z+1/2
124 x+1/2,y,-z+1/2
125 -y+1/2,-z,-x+1/2
126 -y+1/2,z,x+1/2
127 y+1/2,-z,x+1/2
128 y+1/2,z,-x+1/2
129 -z+1/2,-x,-y+1/2
130 -z+1/2,x,y+1/2
131 z+1/2,-x,y+1/2
132 z+1/2,x,-y+1/2
133 y+1/2,x,z+1/2
134 y+1/2,-x,-z+1/2
135 -y+1/2,x,-z+1/2
136 -y+1/2,-x,z+1/2
137 x+1/2,z,y+1/2
138 x+1/2,-z,-y+1/2
139 -x+1/2,z,-y+1/2
140 -x+1/2,-z,y+1/2
141 z+1/2,y,x+1/2
142 z+1/2,-y,-x+1/2
143 -z+1/2,y,-x+1/2
144 -z+1/2,-y,x+1/2
145 x+1/2,y+1/2,z
146 x+1/2,-y+1/2,-z
147 -x+1/2,y+1/2,-z
148 -x+1/2,-y+1/2,z
149 y+1/2,z+1/2,x
150 y+1/2,-z+1/2,-x
151 -y+1/2,z+1/2,-x
152 -y+1/2,-z+1/2,x
153 z+1/2,x+1/2,y
154 z+1/2,-x+1/2,-y
155 -z+1/2,x+1/2,-y
156 -z+1/2,-x+1/2,y
157 -y+1/2,-x+1/2,-z
158 -y+1/2,x+1/2,z
159 y+1/2,-x+1/2,z
160 y+1/2,x+1/2,-z
161 -x+1/2,-z+1/2,-y
162 -x+1/2,z+1/2,y
163 x+1/2,-z+1/2,y
164 x+1/2,z+1/2,-y
165 -z+1/2,-y+1/2,-x
166 -z+1/2,y+1/2,x
167 z+1/2,-y+1/2,x
168 z+1/2,y+1/2,-x
169 -x+1/2,-y+1/2,-z
170 -x+1/2,y+1/2,z
171 x+1/2,-y+1/2,z
172 x+1/2,y+1/2,-z
173 -y+1/2,-z+1/2,-x
174 -y+1/2,z+1/2,x
175 y+1/2,-z+1/2,x
176 y+1/2,z+1/2,-x
177 -z+1/2,-x+1/2,-y
178 -z+1/2,x+1/2,y
179 z+1/2,-x+1/2,y
180 z+1/2,x+1/2,-y
181 y+1/2,x+1/2,z
182 y+1/2,-x+1/2,-z
183 -y+1/2,x+1/2,-z
184 -y+1/2,-x+1/2,z
185 x+1/2,z+1/2,y
186 x+1/2,-z+1/2,-y
187 -x+1/2,z+1/2,-y
188 -x+1/2,-z+1/2,y
189 z+1/2,y+1/2,x
190 z+1/2,-y+1/2,-x
191 -z+1/2,y+1/2,-x
192 -z+1/2,-y+1/2,x

loop_
_atom_site_label
_atom_site_type_symbol
_atom_site_symmetry_multiplicity
_atom_site_Wyckoff_label
_atom_site_fract_x
_atom_site_fract_y
_atom_site_fract_z
_atom_site_occupancy
Ge1 Ge   4 a 0.00000 0.00000 0.00000 1.00000
Ca1 Ca   4 b 0.50000 0.50000 0.50000 1.00000
Ca2 Ca  24 d 0.00000 0.25000 0.25000 1.00000
```

$Ca_7Ge$: A7B_cF32_225_bd_a - POSCAR

```
A7B_cF32_225_bd_a & a --params=9.45 & Fm(-3)m    O_h^5 #225 (abd) & cF32
    ↪ & & Ca7Ge & & O. Helleis, H. Kandler, E. Leicht, W. Quiring and
    ↪ E. Wolfel, ZAAC 320, 86-100 (1963)
   1.0000000000000000
   0.00000000000000   4.72500000000000   4.72500000000000
   4.72500000000000   0.00000000000000   4.72500000000000
   4.72500000000000   4.72500000000000   0.00000000000000
   Ca   Ge
    7    1
Direct
   0.00000000000000   0.00000000000000   0.50000000000000   Ca   (24d)
   0.00000000000000   0.50000000000000   0.00000000000000   Ca   (24d)
   0.00000000000000   0.50000000000000   0.50000000000000   Ca   (24d)
   0.50000000000000   0.00000000000000   0.00000000000000   Ca   (24d)
   0.50000000000000   0.00000000000000   0.50000000000000   Ca   (24d)
   0.50000000000000   0.50000000000000   0.00000000000000   Ca   (24d)
   0.50000000000000   0.50000000000000   0.50000000000000   Ca    (4b)
   0.00000000000000   0.00000000000000   0.00000000000000   Ge    (4a)
```

$BiF_3$ ($D0_3$): AB3_cF16_225_a_bc - CIF

```
# CIF file

data_findsym-output
_audit_creation_method FINDSYM

_chemical_name_mineral 'alpha bismuth trifluoride'
_chemical_formula_sum 'Bi F3'

loop_
```

```
_publ_author_name
 'F. Hund'
 'R. Fricke'
_journal_name_full
;
Zeitschrift f\"{u}r anorganische Chemie
;
_journal_volume 258
_journal_year 1949
_journal_page_first 198
_journal_page_last 204
_publ_Section_title
;
 Der Kristallbau von $\alpha$-BiF$_3$
;

# Found in  Pearson's Handbook, Vol. II, pp. 1774

_aflow_proto 'AB3_cF16_225_a_bc'
_aflow_params 'a'
_aflow_params_values '5.853'
_aflow_Strukturbericht 'D0_3'
_aflow_Pearson 'cF16'

_symmetry_space_group_name_Hall "-F 4 2 3"
_symmetry_space_group_name_H-M "F m -3 m"
_symmetry_Int_Tables_number 225

_cell_length_a      5.85300
_cell_length_b      5.85300
_cell_length_c      5.85300
_cell_angle_alpha 90.00000
_cell_angle_beta  90.00000
_cell_angle_gamma 90.00000

loop_
_space_group_symop_id
_space_group_symop_operation_xyz
1 x,y,z
2 x,-y,-z
3 -x,y,-z
4 -x,-y,z
5 y,z,x
6 y,-z,-x
7 -y,z,-x
8 -y,-z,x
9 z,x,y
10 z,-x,-y
11 -z,x,-y
12 -z,-x,y
13 -y,-x,-z
14 -y,x,z
15 y,-x,z
16 y,x,-z
17 -x,-z,-y
18 -x,z,y
19 x,-z,y
20 x,z,-y
21 -z,-y,-x
22 -z,y,x
23 z,-y,x
24 z,y,-x
25 -x,-y,-z
26 -x,y,z
27 x,-y,z
28 x,y,-z
29 -y,-z,-x
30 -y,z,x
31 y,-z,x
32 y,z,-x
33 -z,-x,-y
34 -z,x,y
35 z,-x,y
36 z,x,-y
37 y,x,z
38 y,-x,-z
39 -y,x,-z
40 -y,-x,z
41 x,z,y
42 x,-z,-y
43 -x,z,-y
44 -x,-z,y
45 z,y,x
46 z,-y,-x
47 -z,y,-x
48 -z,-y,x
49 x,y+1/2,z+1/2
50 x,-y+1/2,-z+1/2
51 -x,y+1/2,-z+1/2
52 -x,-y+1/2,z+1/2
53 y,z+1/2,x+1/2
54 y,-z+1/2,-x+1/2
55 -y,z+1/2,-x+1/2
56 -y,-z+1/2,x+1/2
57 z,x+1/2,y+1/2
58 z,-x+1/2,-y+1/2
59 -z,x+1/2,-y+1/2
60 -z,-x+1/2,y+1/2
61 -y,-x+1/2,-z+1/2
62 -y,x+1/2,z+1/2
63 y,-x+1/2,z+1/2
64 y,x+1/2,-z+1/2
65 -x,-z+1/2,-y+1/2
66 -x,z+1/2,y+1/2
67 x,-z+1/2,y+1/2
68 x,z+1/2,-y+1/2
69 -z,-y+1/2,-x+1/2
70 -z,y+1/2,x+1/2
71 z,-y+1/2,x+1/2
72 z,y+1/2,-x+1/2
73 -x,-y+1/2,-z+1/2
74 -x,y+1/2,z+1/2
75 x,-y+1/2,z+1/2
76 x,y+1/2,-z+1/2
77 -y,-z+1/2,-x+1/2
78 -y,z+1/2,x+1/2
79 y,-z+1/2,x+1/2
80 y,z+1/2,-x+1/2
81 -z,-x+1/2,-y+1/2
82 -z,x+1/2,y+1/2
83 z,-x+1/2,y+1/2
84 z,x+1/2,-y+1/2
85 y,x+1/2,z+1/2
86 y,-x+1/2,-z+1/2
87 -y,x+1/2,-z+1/2
88 -y,-x+1/2,z+1/2
89 x,z+1/2,y+1/2
90 x,-z+1/2,-y+1/2
91 -x,z+1/2,-y+1/2
92 -x,-z+1/2,y+1/2
93 z,y+1/2,x+1/2
94 z,-y+1/2,-x+1/2
95 -z,y+1/2,-x+1/2
96 -z,-y+1/2,x+1/2
97 x+1/2,y,z+1/2
98 x+1/2,-y,-z+1/2
99 -x+1/2,y,-z+1/2
100 -x+1/2,-y,z+1/2
101 y+1/2,z,x+1/2
102 y+1/2,-z,-x+1/2
103 -y+1/2,z,-x+1/2
104 -y+1/2,-z,x+1/2
105 z+1/2,x,y+1/2
106 z+1/2,-x,-y+1/2
107 -z+1/2,x,-y+1/2
108 -z+1/2,-x,y+1/2
109 -y+1/2,-x,-z+1/2
110 -y+1/2,x,z+1/2
111 y+1/2,-x,z+1/2
112 y+1/2,x,-z+1/2
113 -x+1/2,-z,-y+1/2
114 -x+1/2,z,y+1/2
115 x+1/2,-z,y+1/2
116 x+1/2,z,-y+1/2
117 -z+1/2,-y,-x+1/2
118 -z+1/2,y,x+1/2
119 z+1/2,-y,x+1/2
120 z+1/2,y,-x+1/2
121 -x+1/2,-y,-z+1/2
122 -x+1/2,y,z+1/2
123 x+1/2,-y,z+1/2
124 x+1/2,y,-z+1/2
125 -y+1/2,-z,-x+1/2
126 -y+1/2,z,x+1/2
127 y+1/2,-z,x+1/2
128 y+1/2,z,-x+1/2
129 -z+1/2,-x,-y+1/2
130 -z+1/2,x,y+1/2
131 z+1/2,-x,y+1/2
132 z+1/2,x,-y+1/2
133 y+1/2,x,z+1/2
134 y+1/2,-x,-z+1/2
135 -y+1/2,x,-z+1/2
136 -y+1/2,-x,z+1/2
137 x+1/2,z,y+1/2
138 x+1/2,-z,-y+1/2
139 -x+1/2,z,-y+1/2
140 -x+1/2,-z,y+1/2
141 z+1/2,y,x+1/2
142 z+1/2,-y,-x+1/2
143 -z+1/2,y,-x+1/2
144 -z+1/2,-y,x+1/2
145 x+1/2,y+1/2,z
146 x+1/2,-y+1/2,-z
147 -x+1/2,y+1/2,-z
148 -x+1/2,-y+1/2,z
149 y+1/2,z+1/2,x
150 y+1/2,-z+1/2,-x
151 -y+1/2,z+1/2,-x
152 -y+1/2,-z+1/2,x
153 z+1/2,x+1/2,y
154 z+1/2,-x+1/2,-y
155 -z+1/2,x+1/2,-y
156 -z+1/2,-x+1/2,y
157 -y+1/2,-x+1/2,-z
158 -y+1/2,x+1/2,z
159 y+1/2,-x+1/2,z
160 y+1/2,x+1/2,-z
161 -x+1/2,-z+1/2,-y
162 -x+1/2,z+1/2,y
163 x+1/2,-z+1/2,y
164 x+1/2,z+1/2,-y
165 -z+1/2,-y+1/2,-x
166 -z+1/2,y+1/2,x
167 z+1/2,-y+1/2,x
168 z+1/2,y+1/2,-x
169 -x+1/2,-y+1/2,-z
170 -x+1/2,y+1/2,z
171 x+1/2,-y+1/2,z
172 x+1/2,y+1/2,-z
```

```
173 -y+1/2,-z+1/2,-x
174 -y+1/2,z+1/2,x
175 y+1/2,-z+1/2,x
176 y+1/2,z+1/2,-x
177 -z+1/2,-x+1/2,-y
178 -z+1/2,x+1/2,y
179 z+1/2,-x+1/2,y
180 z+1/2,x+1/2,-y
181 y+1/2,x+1/2,z
182 y+1/2,-x+1/2,-z
183 -y+1/2,x+1/2,-z
184 -y+1/2,-x+1/2,z
185 x+1/2,z+1/2,y
186 x+1/2,-z+1/2,-y
187 -x+1/2,z+1/2,-y
188 -x+1/2,-z+1/2,y
189 z+1/2,y+1/2,x
190 z+1/2,-y+1/2,-x
191 -z+1/2,y+1/2,-x
192 -z+1/2,-y+1/2,x

loop_
_atom_site_label
_atom_site_type_symbol
_atom_site_symmetry_multiplicity
_atom_site_Wyckoff_label
_atom_site_fract_x
_atom_site_fract_y
_atom_site_fract_z
_atom_site_occupancy
Bi1 Bi    4 a 0.00000 0.00000 0.00000 1.00000
F1  F     4 b 0.50000 0.50000 0.50000 1.00000
F2  F     8 c 0.25000 0.25000 0.25000 1.00000
```

$BiF_3$ ($D0_3$): AB3_cF16_225_a_bc - POSCAR

```
AB3_cF16_225_a_bc & a --params=5.853 & Fm(-3)m    O_h^5 #225 (abc) & cF16
    ↪ & D0_3 & BiF3 & alpha & F. Hund and R. Fricke, ZAAC 258,
    ↪ 198-204 (1949)
   1.0000000000000000
     0.00000000000000     2.92650000000000     2.92650000000000
     2.92650000000000     0.00000000000000     2.92650000000000
     2.92650000000000     2.92650000000000     0.00000000000000
     Bi      F
      1      3
Direct
     0.00000000000000     0.00000000000000     0.00000000000000    Bi     (4a)
     0.50000000000000     0.50000000000000     0.50000000000000     F     (4b)
     0.25000000000000     0.25000000000000     0.25000000000000     F     (8c)
     0.75000000000000     0.75000000000000     0.75000000000000     F     (8c)
```

Model of Ferrite (cF128): A9B16C7_cF128_225_acd_2f_be - CIF

```
# CIF file

data_findsym-output
_audit_creation_method FINDSYM

_chemical_name_mineral ''
_chemical_formula_sum 'Cr9 Fe16 Ni7'

loop_
_publ_author_name
 'Michael J. Mehl'
_journal_name_full
;
None
;
_journal_volume 0
_journal_year 2008
_journal_page_first 0
_journal_page_last 0
_publ_Section_title
;
 Hypothetical cF128 Austenite Structure
;

_aflow_proto 'A9B16C7_cF128_225_acd_2f_be'
_aflow_params 'a,x5,x6,x7'
_aflow_params_values '11.48,0.25,0.875,0.625'
_aflow_Strukturbericht 'None'
_aflow_Pearson 'cF128'

_symmetry_space_group_name_Hall "-F 4 2 3"
_symmetry_space_group_name_H-M "F m -3 m"
_symmetry_Int_Tables_number 225

_cell_length_a      11.48000
_cell_length_b      11.48000
_cell_length_c      11.48000
_cell_angle_alpha 90.00000
_cell_angle_beta  90.00000
_cell_angle_gamma 90.00000

loop_
_space_group_symop_id
_space_group_symop_operation_xyz
1 x,y,z
2 x,-y,-z
3 -x,y,-z
4 -x,-y,z
5 y,z,x
6 y,-z,-x
7 -y,z,-x
8 -y,-z,x
9 z,x,y
10 z,-x,-y
11 -z,x,-y
12 -z,-x,y
13 -y,-x,-z
14 -y,x,z
15 y,-x,z
16 y,x,-z
17 -x,-z,-y
18 -x,z,y
19 x,-z,y
20 x,z,-y
21 -z,-y,-x
22 -z,y,x
23 z,-y,x
24 z,y,-x
25 -x,-y,-z
26 -x,y,z
27 x,-y,z
28 x,y,-z
29 -y,-z,-x
30 -y,z,x
31 y,-z,x
32 y,z,-x
33 -z,-x,-y
34 -z,x,y
35 z,-x,y
36 z,x,-y
37 y,x,z
38 y,-x,-z
39 -y,x,-z
40 -y,-x,z
41 x,z,y
42 x,-z,-y
43 -x,z,-y
44 -x,-z,y
45 z,y,x
46 z,-y,-x
47 -z,y,-x
48 -z,-y,x
49 x,y+1/2,z+1/2
50 x,-y+1/2,-z+1/2
51 -x,y+1/2,-z+1/2
52 -x,-y+1/2,z+1/2
53 y,z+1/2,x+1/2
54 y,-z+1/2,-x+1/2
55 -y,z+1/2,-x+1/2
56 -y,-z+1/2,x+1/2
57 z,x+1/2,y+1/2
58 z,-x+1/2,-y+1/2
59 -z,x+1/2,-y+1/2
60 -z,-x+1/2,y+1/2
61 -y,-x+1/2,-z+1/2
62 -y,x+1/2,z+1/2
63 y,-x+1/2,z+1/2
64 y,x+1/2,-z+1/2
65 -x,-z+1/2,-y+1/2
66 -x,z+1/2,y+1/2
67 x,-z+1/2,y+1/2
68 x,z+1/2,-y+1/2
69 -z,-y+1/2,-x+1/2
70 -z,y+1/2,x+1/2
71 z,-y+1/2,x+1/2
72 z,y+1/2,-x+1/2
73 -x,-y+1/2,-z+1/2
74 -x,y+1/2,z+1/2
75 x,-y+1/2,z+1/2
76 x,y+1/2,-z+1/2
77 -y,-z+1/2,-x+1/2
78 -y,z+1/2,x+1/2
79 y,-z+1/2,x+1/2
80 y,z+1/2,-x+1/2
81 -z,-x+1/2,-y+1/2
82 -z,x+1/2,y+1/2
83 z,-x+1/2,y+1/2
84 z,x+1/2,-y+1/2
85 y,x+1/2,z+1/2
86 y,-x+1/2,-z+1/2
87 -y,x+1/2,-z+1/2
88 -y,-x+1/2,z+1/2
89 x,z+1/2,y+1/2
90 x,-z+1/2,-y+1/2
91 -x,z+1/2,-y+1/2
92 -x,-z+1/2,y+1/2
93 z,y+1/2,x+1/2
94 z,-y+1/2,-x+1/2
95 -z,y+1/2,-x+1/2
96 -z,-y+1/2,x+1/2
97 x+1/2,y,z+1/2
98 x+1/2,-y,-z+1/2
99 -x+1/2,y,-z+1/2
100 -x+1/2,-y,z+1/2
101 y+1/2,z,x+1/2
102 y+1/2,-z,-x+1/2
103 -y+1/2,z,-x+1/2
104 -y+1/2,-z,x+1/2
105 z+1/2,x,y+1/2
106 z+1/2,-x,-y+1/2
107 -z+1/2,x,-y+1/2
108 -z+1/2,-x,y+1/2
109 -y+1/2,-x,-z+1/2
110 -y+1/2,x,z+1/2
111 y+1/2,-x,z+1/2
112 y+1/2,x,-z+1/2
113 -x+1/2,-z,-y+1/2
```

```
114 -x+1/2,z,y+1/2
115 x+1/2,-z,y+1/2
116 x+1/2,z,-y+1/2
117 -z+1/2,-y,-x+1/2
118 -z+1/2,y,x+1/2
119 z+1/2,-y,x+1/2
120 z+1/2,y,-x+1/2
121 -x+1/2,-y,-z+1/2
122 -x+1/2,y,z+1/2
123 x+1/2,-y,z+1/2
124 x+1/2,y,-z+1/2
125 -y+1/2,-z,-x+1/2
126 -y+1/2,z,x+1/2
127 y+1/2,-z,x+1/2
128 y+1/2,z,-x+1/2
129 -z+1/2,-x,-y+1/2
130 -z+1/2,x,y+1/2
131 z+1/2,-x,y+1/2
132 z+1/2,x,-y+1/2
133 y+1/2,x,z+1/2
134 y+1/2,-x,-z+1/2
135 -y+1/2,x,-z+1/2
136 -y+1/2,-x,z+1/2
137 x+1/2,z,y+1/2
138 x+1/2,-z,-y+1/2
139 -x+1/2,z,-y+1/2
140 -x+1/2,-z,y+1/2
141 z+1/2,y,x+1/2
142 z+1/2,-y,-x+1/2
143 -z+1/2,y,-x+1/2
144 -z+1/2,-y,x+1/2
145 x+1/2,y+1/2,z
146 x+1/2,-y+1/2,-z
147 -x+1/2,y+1/2,-z
148 -x+1/2,-y+1/2,z
149 y+1/2,z+1/2,x
150 y+1/2,-z+1/2,-x
151 -y+1/2,z+1/2,-x
152 -y+1/2,-z+1/2,x
153 z+1/2,x+1/2,y
154 z+1/2,-x+1/2,-y
155 -z+1/2,x+1/2,-y
156 -z+1/2,-x+1/2,y
157 -y+1/2,-x+1/2,-z
158 -y+1/2,x+1/2,z
159 y+1/2,-x+1/2,z
160 y+1/2,x+1/2,-z
161 -x+1/2,-z+1/2,-y
162 -x+1/2,z+1/2,y
163 x+1/2,-z+1/2,y
164 x+1/2,z+1/2,-y
165 -z+1/2,-y+1/2,-x
166 -z+1/2,y+1/2,x
167 z+1/2,-y+1/2,x
168 z+1/2,y+1/2,-x
169 -x+1/2,-y+1/2,-z
170 -x+1/2,y+1/2,z
171 x+1/2,-y+1/2,z
172 x+1/2,y+1/2,-z
173 -y+1/2,-z+1/2,-x
174 -y+1/2,z+1/2,x
175 y+1/2,-z+1/2,x
176 y+1/2,z+1/2,-x
177 -z+1/2,-x+1/2,-y
178 -z+1/2,x+1/2,y
179 z+1/2,-x+1/2,y
180 z+1/2,x+1/2,-y
181 y+1/2,x+1/2,z
182 y+1/2,-x+1/2,-z
183 -y+1/2,x+1/2,-z
184 -y+1/2,-x+1/2,z
185 x+1/2,z+1/2,y
186 x+1/2,-z+1/2,-y
187 -x+1/2,z+1/2,-y
188 -x+1/2,-z+1/2,y
189 z+1/2,y+1/2,x
190 z+1/2,-y+1/2,-x
191 -z+1/2,y+1/2,-x
192 -z+1/2,-y+1/2,x

loop_
_atom_site_label
_atom_site_type_symbol
_atom_site_symmetry_multiplicity
_atom_site_Wyckoff_label
_atom_site_fract_x
_atom_site_fract_y
_atom_site_fract_z
_atom_site_occupancy
Cr1 Cr    4 a 0.00000 0.00000 0.00000 1.00000
Ni1 Ni    4 b 0.50000 0.50000 0.50000 1.00000
Cr2 Cr    8 c 0.25000 0.25000 0.25000 1.00000
Cr3 Cr   24 d 0.00000 0.25000 0.25000 1.00000
Ni2 Ni   24 e 0.25000 0.00000 0.00000 1.00000
Fe1 Fe   32 f 0.87500 0.87500 0.87500 1.00000
Fe2 Fe   32 f 0.62500 0.62500 0.62500 1.00000
```

Model of Ferrite (cF128): A9B16C7_cF128_225_acd_2f_be - POSCAR

```
A9B16C7_cF128_225_acd_2f_be & a,x5,x6,x7 --params=11.48,0.25,0.875,0.625
    ↪ & Fm(-3)m    O_h^5 #225 (abcdef^2) & cF128 & & Cr9Fe16Ni7 &
    ↪ Hypothetical Ferrite &
   1.0000000000000000
     0.00000000000000    5.74000000000000    5.74000000000000
     5.74000000000000    0.00000000000000    5.74000000000000
     5.74000000000000    5.74000000000000    0.00000000000000
   Cr    Fe    Ni
    9    16     7
Direct
     0.00000000000000    0.00000000000000    0.50000000000000   Cr   (24d)
     0.00000000000000    0.50000000000000    0.00000000000000   Cr   (24d)
     0.00000000000000    0.50000000000000    0.50000000000000   Cr   (24d)
     0.50000000000000    0.00000000000000    0.00000000000000   Cr   (24d)
     0.50000000000000    0.00000000000000    0.50000000000000   Cr   (24d)
     0.50000000000000    0.50000000000000    0.00000000000000   Cr   (24d)
     0.00000000000000    0.00000000000000    0.00000000000000   Cr    (4a)
     0.25000000000000    0.25000000000000    0.25000000000000   Cr    (8c)
     0.75000000000000    0.75000000000000    0.75000000000000   Cr    (8c)
     0.12500000000000    0.12500000000000    0.12500000000000   Fe   (32f)
     0.12500000000000    0.12500000000000    0.62500000000000   Fe   (32f)
     0.12500000000000    0.62500000000000    0.12500000000000   Fe   (32f)
     0.37500000000000    0.87500000000000    0.87500000000000   Fe   (32f)
     0.62500000000000    0.12500000000000    0.12500000000000   Fe   (32f)
     0.87500000000000    0.37500000000000    0.87500000000000   Fe   (32f)
     0.87500000000000    0.87500000000000    0.37500000000000   Fe   (32f)
     0.87500000000000    0.87500000000000    0.87500000000000   Fe   (32f)
    -0.12500000000000    0.37500000000000    0.37500000000000   Fe   (32f)
     0.12500000000000    0.62500000000000    0.62500000000000   Fe   (32f)
     0.37500000000000   -0.12500000000000    0.37500000000000   Fe   (32f)
     0.37500000000000    0.37500000000000   -0.12500000000000   Fe   (32f)
     0.37500000000000    0.37500000000000    0.37500000000000   Fe   (32f)
     0.62500000000000    0.12500000000000    0.62500000000000   Fe   (32f)
     0.62500000000000    0.62500000000000    0.12500000000000   Fe   (32f)
     0.62500000000000    0.62500000000000    0.62500000000000   Fe   (32f)
     0.25000000000000    0.25000000000000    0.75000000000000   Ni   (24e)
     0.25000000000000    0.75000000000000    0.25000000000000   Ni   (24e)
     0.25000000000000    0.75000000000000    0.75000000000000   Ni   (24e)
     0.75000000000000    0.25000000000000    0.25000000000000   Ni   (24e)
     0.75000000000000    0.25000000000000    0.75000000000000   Ni   (24e)
     0.75000000000000    0.75000000000000    0.25000000000000   Ni   (24e)
     0.50000000000000    0.50000000000000    0.50000000000000   Ni    (4b)
```

$UB_{12}$: A12B_cF52_225_i_a - CIF

```
# CIF file

data_findsym-output
_audit_creation_method FINDSYM

_chemical_name_mineral ''
_chemical_formula_sum 'U B12'

loop_
_publ_author_name
 'Pierre Blum'
 'F\'{e}lix Bertaut'
_journal_name_full
;
Acta Crystallographica
;
_journal_volume 7
_journal_year 1954
_journal_page_first 81
_journal_page_last 86
_publ_Section_title
;
 Contribution \`{a} l'\'{E}tude des Borures \`{a} Teneur \'{E}lev\'{e}e
    ↪ en Bore
;

# Found in Pearson, Alloys, pp. 757-759

_aflow_proto 'A12B_cF52_225_i_a'
_aflow_params 'a,y2'
_aflow_params_values '7.468,0.666'
_aflow_Strukturbericht 'D2_f'
_aflow_Pearson 'cF52'

_symmetry_space_group_name_Hall "-F 4 2 3"
_symmetry_space_group_name_H-M "F m -3 m"
_symmetry_Int_Tables_number 225

_cell_length_a     7.46800
_cell_length_b     7.46800
_cell_length_c     7.46800
_cell_angle_alpha 90.00000
_cell_angle_beta  90.00000
_cell_angle_gamma 90.00000

loop_
_space_group_symop_id
_space_group_symop_operation_xyz
1 x,y,z
2 x,-y,-z
3 -x,y,-z
4 -x,-y,z
5 y,z,x
6 y,-z,-x
7 -y,z,-x
8 -y,-z,x
9 z,x,y
10 z,-x,-y
11 -z,x,-y
12 -z,-x,y
13 -y,-x,-z
14 -y,x,z
15 y,-x,z
16 y,x,-z
17 -x,-z,-y
18 -x,z,y
```

```
19 x,-z,y
20 x,z,-y
21 -z,-y,-x
22 -z,y,x
23 z,-y,x
24 z,y,-x
25 -x,-y,-z
26 -x,y,z
27 x,-y,z
28 x,y,-z
29 -y,-z,-x
30 -y,z,x
31 y,-z,x
32 y,z,-x
33 -z,-x,-y
34 -z,x,y
35 z,-x,y
36 z,x,-y
37 y,x,z
38 y,-x,-z
39 -y,x,-z
40 -y,-x,z
41 x,z,y
42 x,-z,-y
43 -x,z,-y
44 -x,-z,y
45 z,y,x
46 z,-y,-x
47 -z,y,-x
48 -z,-y,x
49 x,y+1/2,z+1/2
50 x,-y+1/2,-z+1/2
51 -x,y+1/2,-z+1/2
52 -x,-y+1/2,z+1/2
53 y,z+1/2,x+1/2
54 y,-z+1/2,-x+1/2
55 -y,z+1/2,-x+1/2
56 -y,-z+1/2,x+1/2
57 z,x+1/2,y+1/2
58 z,-x+1/2,-y+1/2
59 -z,x+1/2,-y+1/2
60 -z,-x+1/2,y+1/2
61 -y,-x+1/2,-z+1/2
62 -y,x+1/2,z+1/2
63 y,-x+1/2,z+1/2
64 y,x+1/2,-z+1/2
65 -x,-z+1/2,-y+1/2
66 -x,z+1/2,y+1/2
67 x,-z+1/2,y+1/2
68 x,z+1/2,-y+1/2
69 -z,-y+1/2,-x+1/2
70 -z,y+1/2,x+1/2
71 z,-y+1/2,x+1/2
72 z,y+1/2,-x+1/2
73 -x,-y+1/2,-z+1/2
74 -x,y+1/2,z+1/2
75 x,-y+1/2,z+1/2
76 x,y+1/2,-z+1/2
77 -y,-z+1/2,-x+1/2
78 -y,z+1/2,x+1/2
79 y,-z+1/2,x+1/2
80 y,z+1/2,-x+1/2
81 -z,-x+1/2,-y+1/2
82 -z,x+1/2,y+1/2
83 z,-x+1/2,y+1/2
84 z,x+1/2,-y+1/2
85 y,x+1/2,z+1/2
86 y,-x+1/2,-z+1/2
87 -y,x+1/2,-z+1/2
88 -y,-x+1/2,z+1/2
89 x,z+1/2,y+1/2
90 x,-z+1/2,-y+1/2
91 -x,z+1/2,-y+1/2
92 -x,-z+1/2,y+1/2
93 z,y+1/2,x+1/2
94 z,-y+1/2,-x+1/2
95 -z,y+1/2,-x+1/2
96 -z,-y+1/2,x+1/2
97 x+1/2,y,z+1/2
98 x+1/2,-y,-z+1/2
99 -x+1/2,y,-z+1/2
100 -x+1/2,-y,z+1/2
101 y+1/2,z,x+1/2
102 y+1/2,-z,-x+1/2
103 -y+1/2,z,-x+1/2
104 -y+1/2,-z,x+1/2
105 z+1/2,x,y+1/2
106 z+1/2,-x,-y+1/2
107 -z+1/2,x,-y+1/2
108 -z+1/2,-x,y+1/2
109 -y+1/2,-x,-z+1/2
110 -y+1/2,x,z+1/2
111 y+1/2,-x,z+1/2
112 y+1/2,x,-z+1/2
113 -x+1/2,-z,-y+1/2
114 -x+1/2,z,y+1/2
115 x+1/2,-z,y+1/2
116 x+1/2,z,-y+1/2
117 -z+1/2,-y,-x+1/2
118 -z+1/2,y,x+1/2
119 z+1/2,-y,x+1/2
120 z+1/2,y,-x+1/2
121 -x+1/2,-y,-z+1/2
122 -x+1/2,y,z+1/2
123 x+1/2,-y,z+1/2
124 x+1/2,y,-z+1/2
125 -y+1/2,-z,-x+1/2
126 -y+1/2,z,x+1/2
127 y+1/2,-z,x+1/2
128 y+1/2,z,-x+1/2
129 -z+1/2,-x,-y+1/2
130 -z+1/2,x,y+1/2
131 z+1/2,-x,y+1/2
132 z+1/2,x,-y+1/2
133 y+1/2,x,z+1/2
134 y+1/2,-x,-z+1/2
135 -y+1/2,x,-z+1/2
136 -y+1/2,-x,z+1/2
137 x+1/2,z,y+1/2
138 x+1/2,-z,-y+1/2
139 -x+1/2,z,-y+1/2
140 -x+1/2,-z,y+1/2
141 z+1/2,y,x+1/2
142 z+1/2,-y,-x+1/2
143 -z+1/2,y,-x+1/2
144 -z+1/2,-y,x+1/2
145 x+1/2,y+1/2,z
146 x+1/2,-y+1/2,-z
147 -x+1/2,y+1/2,-z
148 -x+1/2,-y+1/2,z
149 y+1/2,z+1/2,x
150 y+1/2,-z+1/2,-x
151 -y+1/2,z+1/2,-x
152 -y+1/2,-z+1/2,x
153 z+1/2,x+1/2,y
154 z+1/2,-x+1/2,-y
155 -z+1/2,x+1/2,-y
156 -z+1/2,-x+1/2,y
157 -y+1/2,-x+1/2,-z
158 -y+1/2,x+1/2,z
159 y+1/2,-x+1/2,z
160 y+1/2,x+1/2,-z
161 -x+1/2,-z+1/2,-y
162 -x+1/2,z+1/2,y
163 x+1/2,-z+1/2,y
164 x+1/2,z+1/2,-y
165 -z+1/2,-y+1/2,-x
166 -z+1/2,y+1/2,x
167 z+1/2,-y+1/2,x
168 z+1/2,y+1/2,-x
169 -x+1/2,-y+1/2,-z
170 -x+1/2,y+1/2,z
171 x+1/2,-y+1/2,z
172 x+1/2,y+1/2,-z
173 -y+1/2,-z+1/2,-x
174 -y+1/2,z+1/2,x
175 y+1/2,-z+1/2,x
176 y+1/2,z+1/2,-x
177 -z+1/2,-x+1/2,-y
178 -z+1/2,x+1/2,y
179 z+1/2,-x+1/2,y
180 z+1/2,x+1/2,-y
181 y+1/2,x+1/2,z
182 y+1/2,-x+1/2,-z
183 -y+1/2,x+1/2,-z
184 -y+1/2,-x+1/2,z
185 x+1/2,z+1/2,y
186 x+1/2,-z+1/2,-y
187 -x+1/2,z+1/2,-y
188 -x+1/2,-z+1/2,y
189 z+1/2,y+1/2,x
190 z+1/2,-y+1/2,-x
191 -z+1/2,y+1/2,-x
192 -z+1/2,-y+1/2,x

loop_
_atom_site_label
_atom_site_type_symbol
_atom_site_symmetry_multiplicity
_atom_site_Wyckoff_label
_atom_site_fract_x
_atom_site_fract_y
_atom_site_fract_z
_atom_site_occupancy
U1 U    4 a 0.00000 0.00000 0.00000 1.00000
B1 B   48 i 0.50000 0.66600 0.66600 1.00000
```

$UB_{12}$: A12B_cF52_225_i_a - POSCAR

```
A12B_cF52_225_i_a & a,y2 --params=7.468,0.666 & Fm(-3)m    O_h^5 #225 (ai
    ↪ ) & cF52 & D2_f & UB12 & & P. Blum and F. Bertaut, Acta Cryst.
    ↪ 7, 81-86 (1954)
   1.0000000000000000
   0.00000000000000   3.73400000000000   3.73400000000000
   3.73400000000000   0.00000000000000   3.73400000000000
   3.73400000000000   3.73400000000000   0.00000000000000
    B     U
   12     1
Direct
   0.16800000000000   0.50000000000000   0.50000000000000   B   (48i)
   0.16800000000000   0.50000000000000   0.83200000000000   B   (48i)
   0.16800000000000   0.83200000000000   0.50000000000000   B   (48i)
   0.50000000000000   0.16800000000000   0.50000000000000   B   (48i)
   0.50000000000000   0.16800000000000   0.83200000000000   B   (48i)
   0.50000000000000   0.50000000000000   0.16800000000000   B   (48i)
   0.50000000000000   0.50000000000000   0.83200000000000   B   (48i)
   0.50000000000000   0.83200000000000   0.16800000000000   B   (48i)
   0.50000000000000   0.83200000000000   0.50000000000000   B   (48i)
   0.83200000000000   0.16800000000000   0.50000000000000   B   (48i)
   0.83200000000000   0.50000000000000   0.16800000000000   B   (48i)
```

```
  0.83200000000000    0.50000000000000    0.50000000000000    B    (48i)
  0.00000000000000    0.00000000000000    0.00000000000000    U    (4a)
```

Fluorite ($CaF_2$, C1): AB2_cF12_225_a_c - CIF

```
# CIF file

data_findsym-output
_audit_creation_method FINDSYM

_chemical_name_mineral 'Fluorite'
_chemical_formula_sum 'Ca F2'

loop_
_publ_author_name
 'S. Speziale'
 'T. S. Duffy'
_journal_name_full
;
Physics and Chemistry of Minerals
;
_journal_volume 29
_journal_year 2002
_journal_page_first 465
_journal_page_last 472
_publ_Section_title
;
 Single-crystal elastic constants of fluorite (CaF$_2$) to 9.3 GPa
;

# Found in AMS Database

_aflow_proto 'AB2_cF12_225_a_c'
_aflow_params 'a'
_aflow_params_values '5.4631'
_aflow_Strukturbericht 'C1'
_aflow_Pearson 'cF12'

_symmetry_space_group_name_Hall "-F 4 2 3"
_symmetry_space_group_name_H-M "F m -3 m"
_symmetry_Int_Tables_number 225

_cell_length_a     5.46310
_cell_length_b     5.46310
_cell_length_c     5.46310
_cell_angle_alpha 90.00000
_cell_angle_beta  90.00000
_cell_angle_gamma 90.00000

loop_
_space_group_symop_id
_space_group_symop_operation_xyz
1 x,y,z
2 x,-y,-z
3 -x,y,-z
4 -x,-y,z
5 y,z,x
6 y,-z,-x
7 -y,z,-x
8 -y,-z,x
9 z,x,y
10 z,-x,-y
11 -z,x,-y
12 -z,-x,y
13 -y,-x,-z
14 -y,x,z
15 y,-x,z
16 y,x,-z
17 -x,-z,-y
18 -x,z,y
19 x,-z,y
20 x,z,-y
21 -z,-y,-x
22 -z,y,x
23 z,-y,x
24 z,y,-x
25 -x,-y,-z
26 -x,y,z
27 x,-y,z
28 x,y,-z
29 -y,-z,-x
30 -y,z,x
31 y,-z,x
32 y,z,-x
33 -z,-x,-y
34 -z,x,y
35 z,-x,y
36 z,x,-y
37 y,x,z
38 y,-x,-z
39 -y,x,-z
40 -y,-x,z
41 x,z,y
42 x,-z,-y
43 -x,z,-y
44 -x,-z,y
45 z,y,x
46 z,-y,-x
47 -z,y,-x
48 -z,-y,x
49 x,y+1/2,z+1/2
50 x,-y+1/2,-z+1/2
51 -x,y+1/2,-z+1/2
52 -x,-y+1/2,z+1/2
53 y,z+1/2,x+1/2
54 y,-z+1/2,-x+1/2
55 -y,z+1/2,-x+1/2
56 -y,-z+1/2,x+1/2
57 z,x+1/2,y+1/2
58 z,-x+1/2,-y+1/2
59 -z,x+1/2,-y+1/2
60 -z,-x+1/2,y+1/2
61 -y,-x+1/2,-z+1/2
62 -y,x+1/2,z+1/2
63 y,-x+1/2,z+1/2
64 y,x+1/2,-z+1/2
65 -x,-z+1/2,-y+1/2
66 -x,z+1/2,y+1/2
67 x,-z+1/2,y+1/2
68 x,z+1/2,-y+1/2
69 -z,-y+1/2,-x+1/2
70 -z,y+1/2,x+1/2
71 z,-y+1/2,x+1/2
72 z,y+1/2,-x+1/2
73 -x,-y+1/2,-z+1/2
74 -x,y+1/2,z+1/2
75 x,-y+1/2,z+1/2
76 x,y+1/2,-z+1/2
77 -y,-z+1/2,-x+1/2
78 -y,z+1/2,x+1/2
79 y,-z+1/2,x+1/2
80 y,z+1/2,-x+1/2
81 -z,-x+1/2,-y+1/2
82 -z,x+1/2,y+1/2
83 z,-x+1/2,y+1/2
84 z,x+1/2,-y+1/2
85 y,x+1/2,z+1/2
86 y,-x+1/2,-z+1/2
87 -y,x+1/2,-z+1/2
88 -y,-x+1/2,z+1/2
89 x,z+1/2,y+1/2
90 x,-z+1/2,-y+1/2
91 -x,z+1/2,-y+1/2
92 -x,-z+1/2,y+1/2
93 z,y+1/2,x+1/2
94 z,-y+1/2,-x+1/2
95 -z,y+1/2,-x+1/2
96 -z,-y+1/2,x+1/2
97 x+1/2,y,z+1/2
98 x+1/2,-y,-z+1/2
99 -x+1/2,y,-z+1/2
100 -x+1/2,-y,z+1/2
101 y+1/2,z,x+1/2
102 y+1/2,-z,-x+1/2
103 -y+1/2,z,-x+1/2
104 -y+1/2,-z,x+1/2
105 z+1/2,x,y+1/2
106 z+1/2,-x,-y+1/2
107 -z+1/2,x,-y+1/2
108 -z+1/2,-x,y+1/2
109 -y+1/2,-x,-z+1/2
110 -y+1/2,x,z+1/2
111 y+1/2,-x,z+1/2
112 y+1/2,x,-z+1/2
113 -x+1/2,-z,-y+1/2
114 -x+1/2,z,y+1/2
115 x+1/2,-z,y+1/2
116 x+1/2,z,-y+1/2
117 -z+1/2,-y,-x+1/2
118 -z+1/2,y,x+1/2
119 z+1/2,-y,x+1/2
120 z+1/2,y,-x+1/2
121 -x+1/2,-y,-z+1/2
122 -x+1/2,y,z+1/2
123 x+1/2,-y,z+1/2
124 x+1/2,y,-z+1/2
125 -y+1/2,-z,-x+1/2
126 -y+1/2,z,x+1/2
127 y+1/2,-z,x+1/2
128 y+1/2,z,-x+1/2
129 -z+1/2,-x,-y+1/2
130 -z+1/2,x,y+1/2
131 z+1/2,-x,y+1/2
132 z+1/2,x,-y+1/2
133 y+1/2,x,z+1/2
134 y+1/2,-x,-z+1/2
135 -y+1/2,x,-z+1/2
136 -y+1/2,-x,z+1/2
137 x+1/2,z,y+1/2
138 x+1/2,-z,-y+1/2
139 -x+1/2,z,-y+1/2
140 -x+1/2,-z,y+1/2
141 z+1/2,y,x+1/2
142 z+1/2,-y,-x+1/2
143 -z+1/2,y,-x+1/2
144 -z+1/2,-y,x+1/2
145 x+1/2,y+1/2,z
146 x+1/2,-y+1/2,-z
147 -x+1/2,y+1/2,-z
148 -x+1/2,-y+1/2,z
149 y+1/2,z+1/2,x
150 y+1/2,-z+1/2,-x
151 -y+1/2,z+1/2,-x
152 -y+1/2,-z+1/2,x
153 z+1/2,x+1/2,y
154 z+1/2,-x+1/2,-y
155 -z+1/2,x+1/2,-y
156 -z+1/2,-x+1/2,y
157 -y+1/2,-x+1/2,-z
158 -y+1/2,x+1/2,z
```

```
159  y+1/2,-x+1/2,z
160  y+1/2,x+1/2,-z
161  -x+1/2,-z+1/2,-y
162  -x+1/2,z+1/2,y
163  x+1/2,-z+1/2,y
164  x+1/2,z+1/2,-y
165  -z+1/2,-y+1/2,-x
166  -z+1/2,y+1/2,x
167  z+1/2,-y+1/2,x
168  z+1/2,y+1/2,-x
169  -x+1/2,-y+1/2,-z
170  -x+1/2,y+1/2,z
171  x+1/2,-y+1/2,z
172  x+1/2,y+1/2,-z
173  -y+1/2,-z+1/2,-x
174  -y+1/2,z+1/2,x
175  y+1/2,-z+1/2,x
176  y+1/2,z+1/2,-x
177  -z+1/2,-x+1/2,-y
178  -z+1/2,x+1/2,y
179  z+1/2,-x+1/2,y
180  z+1/2,x+1/2,-y
181  y+1/2,x+1/2,z
182  y+1/2,-x+1/2,-z
183  -y+1/2,x+1/2,-z
184  -y+1/2,-x+1/2,z
185  x+1/2,z+1/2,y
186  x+1/2,-z+1/2,-y
187  -x+1/2,z+1/2,-y
188  -x+1/2,-z+1/2,y
189  z+1/2,y+1/2,x
190  z+1/2,-y+1/2,-x
191  -z+1/2,y+1/2,-x
192  -z+1/2,-y+1/2,x

loop_
_atom_site_label
_atom_site_type_symbol
_atom_site_symmetry_multiplicity
_atom_site_Wyckoff_label
_atom_site_fract_x
_atom_site_fract_y
_atom_site_fract_z
_atom_site_occupancy
Ca1  Ca    4  a  0.00000  0.00000  0.00000  1.00000
F1   F     8  c  0.25000  0.25000  0.25000  1.00000
```

Fluorite ($CaF_2$, C1): AB2_cF12_225_a_c - POSCAR

```
AB2_cF12_225_a_c & a --params=5.4631 & Fm(-3)m    O_h^5 #225 (ac) & cF12
    ↪ & C1 & CaF2 & Fluorite & S. Speziale and T. S. Duffy, Phys.
    ↪ Chem. Minerals 29, 465-472 (2002)
   1.0000000000000000
   0.00000000000000   2.73155000000000   2.73155000000000
   2.73155000000000   0.00000000000000   2.73155000000000
   2.73155000000000   2.73155000000000   0.00000000000000
   Ca    F
    1    2
Direct
   0.00000000000000   0.00000000000000   0.00000000000000   Ca   (4a)
   0.25000000000000   0.25000000000000   0.25000000000000    F   (8c)
   0.75000000000000   0.75000000000000   0.75000000000000    F   (8c)
```

$Cr_{23}C_6$ ($D8_4$): A6B23_cF116_225_e_acfh - CIF

```
# CIF file

data_findsym-output
_audit_creation_method FINDSYM

_chemical_name_mineral ''
_chemical_formula_sum 'Cr23 C6'

loop_
_publ_author_name
 'A. L. Bowman'
 'G. P. Arnold'
 'E. K. Storms'
 'N. G. Nereson'
_journal_name_full
;
Acta Crystallographica B
;
_journal_volume 28
_journal_year 1972
_journal_page_first 3102
_journal_page_last 3103
_publ_Section_title
;
 The crystal structure of Cr$_{23}$C$_6$
;

_aflow_proto 'A6B23_cF116_225_e_acfh'
_aflow_params 'a,x3,x4,y5'
_aflow_params_values '10.65,0.2765,0.6191,0.6699'
_aflow_Strukturbericht 'D8_4'
_aflow_Pearson 'cF116'

_symmetry_space_group_name_Hall "-F 4 2 3"
_symmetry_space_group_name_H-M "F m -3 m"
_symmetry_Int_Tables_number 225

_cell_length_a    10.65000
_cell_length_b    10.65000
_cell_length_c    10.65000
_cell_angle_alpha 90.00000
_cell_angle_beta  90.00000
_cell_angle_gamma 90.00000

loop_
_space_group_symop_id
_space_group_symop_operation_xyz
1  x,y,z
2  x,-y,-z
3  -x,y,-z
4  -x,-y,z
5  y,z,x
6  y,-z,-x
7  -y,z,-x
8  -y,-z,x
9  z,x,y
10  z,-x,-y
11  -z,x,-y
12  -z,-x,y
13  -y,-x,-z
14  -y,x,z
15  y,-x,z
16  y,x,-z
17  -x,-z,-y
18  -x,z,y
19  x,-z,y
20  x,z,-y
21  -z,-y,-x
22  -z,y,x
23  z,-y,x
24  z,y,-x
25  -x,-y,-z
26  -x,y,z
27  x,-y,z
28  x,y,-z
29  -y,-z,-x
30  -y,z,x
31  y,-z,x
32  y,z,-x
33  -z,-x,-y
34  -z,x,y
35  z,-x,y
36  z,x,-y
37  y,x,z
38  y,-x,-z
39  -y,x,-z
40  -y,-x,z
41  x,z,y
42  x,-z,-y
43  -x,z,-y
44  -x,-z,y
45  z,y,x
46  z,-y,-x
47  -z,y,-x
48  -z,-y,x
49  x,y+1/2,z+1/2
50  x,-y+1/2,-z+1/2
51  -x,y+1/2,-z+1/2
52  -x,-y+1/2,z+1/2
53  y,z+1/2,x+1/2
54  y,-z+1/2,-x+1/2
55  -y,z+1/2,-x+1/2
56  -y,-z+1/2,x+1/2
57  z,x+1/2,y+1/2
58  z,-x+1/2,-y+1/2
59  -z,x+1/2,-y+1/2
60  -z,-x+1/2,y+1/2
61  -y,-x+1/2,-z+1/2
62  -y,x+1/2,z+1/2
63  y,-x+1/2,z+1/2
64  y,x+1/2,-z+1/2
65  -x,-z+1/2,-y+1/2
66  -x,z+1/2,y+1/2
67  x,-z+1/2,y+1/2
68  x,z+1/2,-y+1/2
69  -z,-y+1/2,-x+1/2
70  -z,y+1/2,x+1/2
71  z,-y+1/2,x+1/2
72  z,y+1/2,-x+1/2
73  -x,-y+1/2,-z+1/2
74  -x,y+1/2,z+1/2
75  x,-y+1/2,z+1/2
76  x,y+1/2,-z+1/2
77  -y,-z+1/2,-x+1/2
78  -y,z+1/2,x+1/2
79  y,-z+1/2,x+1/2
80  y,z+1/2,-x+1/2
81  -z,-x+1/2,-y+1/2
82  -z,x+1/2,y+1/2
83  z,-x+1/2,y+1/2
84  z,x+1/2,-y+1/2
85  y,x+1/2,z+1/2
86  y,-x+1/2,-z+1/2
87  -y,x+1/2,-z+1/2
88  -y,-x+1/2,z+1/2
89  x,z+1/2,y+1/2
90  x,-z+1/2,-y+1/2
91  -x,z+1/2,-y+1/2
92  -x,-z+1/2,y+1/2
93  z,y+1/2,x+1/2
94  z,-y+1/2,-x+1/2
95  -z,y+1/2,-x+1/2
96  -z,-y+1/2,x+1/2
97  x+1/2,y,z+1/2
98  x+1/2,-y,-z+1/2
```

```
99 -x+1/2,y,-z+1/2
100 -x+1/2,-y,z+1/2
101 y+1/2,z,x+1/2
102 y+1/2,-z,-x+1/2
103 -y+1/2,z,-x+1/2
104 -y+1/2,-z,x+1/2
105 z+1/2,x,y+1/2
106 z+1/2,-x,-y+1/2
107 -z+1/2,x,-y+1/2
108 -z+1/2,-x,y+1/2
109 -y+1/2,-x,-z+1/2
110 -y+1/2,x,z+1/2
111 y+1/2,-x,z+1/2
112 y+1/2,x,-z+1/2
113 -x+1/2,-z,-y+1/2
114 -x+1/2,z,y+1/2
115 x+1/2,-z,y+1/2
116 x+1/2,z,-y+1/2
117 -z+1/2,-y,-x+1/2
118 -z+1/2,y,x+1/2
119 z+1/2,-y,x+1/2
120 z+1/2,y,-x+1/2
121 -x+1/2,-y,-z+1/2
122 -x+1/2,y,z+1/2
123 x+1/2,-y,z+1/2
124 x+1/2,y,-z+1/2
125 -y+1/2,-z,-x+1/2
126 -y+1/2,z,x+1/2
127 y+1/2,-z,x+1/2
128 y+1/2,z,-x+1/2
129 -z+1/2,-x,-y+1/2
130 -z+1/2,x,y+1/2
131 z+1/2,-x,y+1/2
132 z+1/2,x,-y+1/2
133 y+1/2,x,z+1/2
134 y+1/2,-x,-z+1/2
135 -y+1/2,x,-z+1/2
136 -y+1/2,-x,z+1/2
137 x+1/2,z,y+1/2
138 x+1/2,-z,-y+1/2
139 -x+1/2,z,-y+1/2
140 -x+1/2,-z,y+1/2
141 z+1/2,y,x+1/2
142 z+1/2,-y,-x+1/2
143 -z+1/2,y,-x+1/2
144 -z+1/2,-y,x+1/2
145 x+1/2,y+1/2,z
146 x+1/2,-y+1/2,-z
147 -x+1/2,y+1/2,-z
148 -x+1/2,-y+1/2,z
149 y+1/2,z+1/2,x
150 y+1/2,-z+1/2,-x
151 -y+1/2,z+1/2,-x
152 -y+1/2,-z+1/2,x
153 z+1/2,x+1/2,y
154 z+1/2,-x+1/2,-y
155 -z+1/2,x+1/2,-y
156 -z+1/2,-x+1/2,y
157 -y+1/2,-x+1/2,-z
158 -y+1/2,x+1/2,z
159 y+1/2,-x+1/2,z
160 y+1/2,x+1/2,-z
161 -x+1/2,-z+1/2,-y
162 -x+1/2,z+1/2,y
163 x+1/2,-z+1/2,y
164 x+1/2,z+1/2,-y
165 -z+1/2,-y+1/2,-x
166 -z+1/2,y+1/2,x
167 z+1/2,-y+1/2,x
168 z+1/2,y+1/2,-x
169 -x+1/2,-y+1/2,-z
170 -x+1/2,y+1/2,z
171 x+1/2,-y+1/2,z
172 x+1/2,y+1/2,-z
173 -y+1/2,-z+1/2,-x
174 -y+1/2,z+1/2,x
175 y+1/2,-z+1/2,x
176 y+1/2,z+1/2,-x
177 -z+1/2,-x+1/2,-y
178 -z+1/2,x+1/2,y
179 z+1/2,-x+1/2,y
180 z+1/2,x+1/2,-y
181 y+1/2,x+1/2,z
182 y+1/2,-x+1/2,-z
183 -y+1/2,x+1/2,-z
184 -y+1/2,-x+1/2,z
185 x+1/2,z+1/2,y
186 x+1/2,-z+1/2,-y
187 -x+1/2,z+1/2,-y
188 -x+1/2,-z+1/2,y
189 z+1/2,y+1/2,x
190 z+1/2,-y+1/2,-x
191 -z+1/2,y+1/2,-x
192 -z+1/2,-y+1/2,x

loop_
_atom_site_label
_atom_site_type_symbol
_atom_site_symmetry_multiplicity
_atom_site_Wyckoff_label
_atom_site_fract_x
_atom_site_fract_y
_atom_site_fract_z
_atom_site_occupancy
Cr1 Cr   4 a 0.00000 0.00000 0.00000 1.00000
Cr2 Cr   8 c 0.25000 0.25000 0.25000 1.00000
C1  C   24 e 0.27650 0.00000 0.00000 1.00000
Cr3 Cr  32 f 0.61910 0.61910 0.61910 1.00000
Cr4 Cr  48 h 0.00000 0.66990 0.66990 1.00000
```

$Cr_{23}C_6$ ($D8_4$): A6B23_cF116_225_e_acfh - POSCAR

```
A6B23_cF116_225_e_acfh & a,x3,x4,y5 --params=10.65,0.2765,0.6191,0.6699
    ↪ & Fm(-3)m    O_h^5 #225 (acefh) & cF116 & D8_4 & Cr23C6 & & A.
    ↪ L. Bowman, G. P. Arnold, E. K. Storms and N. G. Nereson, Acta
    ↪ Cryst. B 28, 3102-3103 (1972)
   1.0000000000000000
   0.00000000000000   5.32500000000000   5.32500000000000
   5.32500000000000   0.00000000000000   5.32500000000000
   5.32500000000000   5.32500000000000   0.00000000000000
    C    Cr
    6    23
Direct
   0.27650000000000   0.27650000000000   0.72350000000000   C    (24e)
   0.27650000000000   0.72350000000000   0.27650000000000   C    (24e)
   0.27650000000000   0.72350000000000   0.72350000000000   C    (24e)
   0.72350000000000   0.27650000000000   0.27650000000000   C    (24e)
   0.72350000000000   0.27650000000000   0.72350000000000   C    (24e)
   0.72350000000000   0.72350000000000   0.27650000000000   C    (24e)
  -0.14270000000000   0.38090000000000   0.38090000000000   Cr   (32f)
   0.14270000000000   0.61910000000000   0.61910000000000   Cr   (32f)
   0.38090000000000  -0.14270000000000   0.38090000000000   Cr   (32f)
   0.38090000000000   0.38090000000000  -0.14270000000000   Cr   (32f)
   0.38090000000000   0.38090000000000   0.38090000000000   Cr   (32f)
   0.61910000000000   0.14270000000000   0.61910000000000   Cr   (32f)
   0.61910000000000   0.61910000000000   0.14270000000000   Cr   (32f)
   0.61910000000000   0.61910000000000   0.61910000000000   Cr   (32f)
   0.00000000000000   0.00000000000000   0.33980000000000   Cr   (48h)
   0.00000000000000   0.00000000000000   0.66020000000000   Cr   (48h)
   0.00000000000000   0.33980000000000   0.00000000000000   Cr   (48h)
   0.00000000000000   0.33980000000000   0.66020000000000   Cr   (48h)
   0.00000000000000   0.66020000000000   0.00000000000000   Cr   (48h)
   0.00000000000000   0.66020000000000   0.33980000000000   Cr   (48h)
   0.33980000000000   0.00000000000000   0.00000000000000   Cr   (48h)
   0.33980000000000   0.00000000000000   0.66020000000000   Cr   (48h)
   0.33980000000000   0.66020000000000   0.00000000000000   Cr   (48h)
   0.66020000000000   0.00000000000000   0.00000000000000   Cr   (48h)
   0.66020000000000   0.00000000000000   0.33980000000000   Cr   (48h)
   0.66020000000000   0.33980000000000   0.00000000000000   Cr   (48h)
   0.00000000000000   0.00000000000000   0.00000000000000   Cr    (4a)
   0.25000000000000   0.25000000000000   0.25000000000000   Cr    (8c)
   0.75000000000000   0.75000000000000   0.75000000000000   Cr    (8c)
```

Heusler ($L2_1$): AB2C_cF16_225_a_c_b - CIF

```
# CIF file

data_findsym-output
_audit_creation_method FINDSYM

_chemical_name_mineral 'Heusler'
_chemical_formula_sum 'Al Cu2 Mn'

loop_
_publ_author_name
 'A. J. Bradley'
 'J. W. Rodgers'
_journal_name_full
;
Proceedings of the Royal Society of London, Series A
;
_journal_volume 144
_journal_year 1934
_journal_page_first 340
_journal_page_last 359
_publ_Section_title
;
 The Crystal Structure of Heusler Alloys
;

_aflow_proto 'AB2C_cF16_225_a_c_b'
_aflow_params 'a'
_aflow_params_values '5.95'
_aflow_Strukturbericht 'L2_1'
_aflow_Pearson 'cF16'

_symmetry_space_group_name_Hall "-F 4 2 3"
_symmetry_space_group_name_H-M "F m -3 m"
_symmetry_Int_Tables_number 225

_cell_length_a      5.95000
_cell_length_b      5.95000
_cell_length_c      5.95000
_cell_angle_alpha  90.00000
_cell_angle_beta   90.00000
_cell_angle_gamma  90.00000

loop_
_space_group_symop_id
_space_group_symop_operation_xyz
1 x,y,z
2 x,-y,-z
3 -x,y,-z
4 -x,-y,z
5 y,z,x
6 y,-z,-x
7 -y,z,-x
8 -y,-z,x
9 z,x,y
10 z,-x,-y
```

```
11 -z,x,-y
12 -z,-x,y
13 -y,-x,-z
14 -y,x,z
15 y,-x,z
16 y,x,-z
17 -x,-z,-y
18 -x,z,y
19 x,-z,y
20 x,z,-y
21 -z,-y,-x
22 -z,y,x
23 z,-y,x
24 z,y,-x
25 -x,-y,-z
26 -x,y,z
27 x,-y,z
28 x,y,-z
29 -y,-z,-x
30 -y,z,x
31 y,-z,x
32 y,z,-x
33 -z,-x,-y
34 -z,x,y
35 z,-x,y
36 z,x,-y
37 y,x,z
38 y,-x,-z
39 -y,x,-z
40 -y,-x,z
41 x,z,y
42 x,-z,-y
43 -x,z,-y
44 -x,-z,y
45 z,y,x
46 z,-y,-x
47 -z,y,-x
48 -z,-y,x
49 x,y+1/2,z+1/2
50 x,-y+1/2,-z+1/2
51 -x,y+1/2,-z+1/2
52 -x,-y+1/2,z+1/2
53 y,z+1/2,x+1/2
54 y,-z+1/2,-x+1/2
55 -y,z+1/2,-x+1/2
56 -y,-z+1/2,x+1/2
57 z,x+1/2,y+1/2
58 z,-x+1/2,-y+1/2
59 -z,x+1/2,-y+1/2
60 -z,-x+1/2,y+1/2
61 -y,-x+1/2,-z+1/2
62 -y,x+1/2,z+1/2
63 y,-x+1/2,z+1/2
64 y,x+1/2,-z+1/2
65 -x,-z+1/2,-y+1/2
66 -x,z+1/2,y+1/2
67 x,-z+1/2,y+1/2
68 x,z+1/2,-y+1/2
69 -z,-y+1/2,-x+1/2
70 -z,y+1/2,x+1/2
71 z,-y+1/2,x+1/2
72 z,y+1/2,-x+1/2
73 -x,-y+1/2,-z+1/2
74 -x,y+1/2,z+1/2
75 x,-y+1/2,z+1/2
76 x,y+1/2,-z+1/2
77 -y,-z+1/2,-x+1/2
78 -y,z+1/2,x+1/2
79 y,-z+1/2,x+1/2
80 y,z+1/2,-x+1/2
81 -z,-x+1/2,-y+1/2
82 -z,x+1/2,y+1/2
83 z,-x+1/2,y+1/2
84 z,x+1/2,-y+1/2
85 y,x+1/2,z+1/2
86 y,-x+1/2,-z+1/2
87 -y,x+1/2,-z+1/2
88 -y,-x+1/2,z+1/2
89 x,z+1/2,y+1/2
90 x,-z+1/2,-y+1/2
91 -x,z+1/2,-y+1/2
92 -x,-z+1/2,y+1/2
93 z,y+1/2,x+1/2
94 z,-y+1/2,-x+1/2
95 -z,y+1/2,-x+1/2
96 -z,-y+1/2,x+1/2
97 x+1/2,y,z+1/2
98 x+1/2,-y,-z+1/2
99 -x+1/2,y,-z+1/2
100 -x+1/2,-y,z+1/2
101 y+1/2,z,x+1/2
102 y+1/2,-z,-x+1/2
103 -y+1/2,z,-x+1/2
104 -y+1/2,-z,x+1/2
105 z+1/2,x,y+1/2
106 z+1/2,-x,-y+1/2
107 -z+1/2,x,-y+1/2
108 -z+1/2,-x,y+1/2
109 -y+1/2,-x,-z+1/2
110 -y+1/2,x,z+1/2
111 y+1/2,-x,z+1/2
112 y+1/2,x,-z+1/2
113 -x+1/2,-z,-y+1/2
114 -x+1/2,z,y+1/2
115 x+1/2,-z,y+1/2
116 x+1/2,z,-y+1/2
117 -z+1/2,-y,-x+1/2
118 -z+1/2,y,x+1/2
119 z+1/2,-y,x+1/2
120 z+1/2,y,-x+1/2
121 -x+1/2,-y,-z+1/2
122 -x+1/2,y,z+1/2
123 x+1/2,-y,z+1/2
124 x+1/2,y,-z+1/2
125 -y+1/2,-z,-x+1/2
126 -y+1/2,z,x+1/2
127 y+1/2,-z,x+1/2
128 y+1/2,z,-x+1/2
129 -z+1/2,-x,-y+1/2
130 -z+1/2,x,y+1/2
131 z+1/2,-x,y+1/2
132 z+1/2,x,-y+1/2
133 y+1/2,x,z+1/2
134 y+1/2,-x,-z+1/2
135 -y+1/2,x,-z+1/2
136 -y+1/2,-x,z+1/2
137 x+1/2,z,y+1/2
138 x+1/2,-z,-y+1/2
139 -x+1/2,z,-y+1/2
140 -x+1/2,-z,y+1/2
141 z+1/2,y,x+1/2
142 z+1/2,-y,-x+1/2
143 -z+1/2,y,-x+1/2
144 -z+1/2,-y,x+1/2
145 x+1/2,y+1/2,z
146 x+1/2,-y+1/2,-z
147 -x+1/2,y+1/2,-z
148 -x+1/2,-y+1/2,z
149 y+1/2,z+1/2,x
150 y+1/2,-z+1/2,-x
151 -y+1/2,z+1/2,-x
152 -y+1/2,-z+1/2,x
153 z+1/2,x+1/2,y
154 z+1/2,-x+1/2,-y
155 -z+1/2,x+1/2,-y
156 -z+1/2,-x+1/2,y
157 -y+1/2,-x+1/2,-z
158 -y+1/2,x+1/2,z
159 y+1/2,-x+1/2,z
160 y+1/2,x+1/2,-z
161 -x+1/2,-z+1/2,-y
162 -x+1/2,z+1/2,y
163 x+1/2,-z+1/2,y
164 x+1/2,z+1/2,-y
165 -z+1/2,-y+1/2,-x
166 -z+1/2,y+1/2,x
167 z+1/2,-y+1/2,x
168 z+1/2,y+1/2,-x
169 -x+1/2,-y+1/2,-z
170 -x+1/2,y+1/2,z
171 x+1/2,-y+1/2,z
172 x+1/2,y+1/2,-z
173 -y+1/2,-z+1/2,-x
174 -y+1/2,z+1/2,x
175 y+1/2,-z+1/2,x
176 y+1/2,z+1/2,-x
177 -z+1/2,-x+1/2,-y
178 -z+1/2,x+1/2,y
179 z+1/2,-x+1/2,y
180 z+1/2,x+1/2,-y
181 y+1/2,x+1/2,z
182 y+1/2,-x+1/2,-z
183 -y+1/2,x+1/2,-z
184 -y+1/2,-x+1/2,z
185 x+1/2,z+1/2,y
186 x+1/2,-z+1/2,-y
187 -x+1/2,z+1/2,-y
188 -x+1/2,-z+1/2,y
189 z+1/2,y+1/2,x
190 z+1/2,-y+1/2,-x
191 -z+1/2,y+1/2,-x
192 -z+1/2,-y+1/2,x

loop_
_atom_site_label
_atom_site_type_symbol
_atom_site_symmetry_multiplicity
_atom_site_Wyckoff_label
_atom_site_fract_x
_atom_site_fract_y
_atom_site_fract_z
_atom_site_occupancy
Al1 Al    4 a 0.00000 0.00000 0.00000 1.00000
Mn1 Mn    4 b 0.50000 0.50000 0.50000 1.00000
Cu1 Cu    8 c 0.25000 0.25000 0.25000 1.00000
```

Heusler ($L2_1$): AB2C_cF16_225_a_c_b - POSCAR

```
AB2C_cF16_225_a_c_b & a --params=5.95 & Fm(-3)m    O_h^5 #225 (abc) &
    ↪ cF16 & L2_1 & AlCu2Mn & Heusler & A. J. Bradley and J. W.
    ↪ Rodgers, Proc. Roy. Soc. A 144, 340-359 (1934)
   1.0000000000000000
   0.00000000000000    2.97500000000000    2.97500000000000
   2.97500000000000    0.00000000000000    2.97500000000000
   2.97500000000000    2.97500000000000    0.00000000000000
   Al    Cu    Mn
    1     2     1
Direct
   0.00000000000000    0.00000000000000    0.00000000000000    Al    (4a)
   0.25000000000000    0.25000000000000    0.25000000000000    Cu    (8c)
```

```
    0.75000000000000    0.75000000000000    0.75000000000000    Cu    (8c)
    0.50000000000000    0.50000000000000    0.50000000000000    Mn    (4b)
```

Face-Centered Cubic (Cu, A1): A_cF4_225_a - CIF

```
# CIF file

data_findsym-output
_audit_creation_method FINDSYM

_chemical_name_mineral 'Copper'
_chemical_formula_sum 'Cu'

loop_
_publ_author_name
 'M. E. Straumanis'
 'L. S. Yu'
_journal_name_full
;
Acta Crystallographica A
;
_journal_volume 25
_journal_year 1969
_journal_page_first 676
_journal_page_last 682
_publ_Section_title
;
 Lattice parameters, densities, expansion coefficients and perfection of
    ↪ structure of Cu and of Cu-In $\alpha$ phase
;

_aflow_proto 'A_cF4_225_a'
_aflow_params 'a'
_aflow_params_values '3.61491'
_aflow_Strukturbericht 'A1'
_aflow_Pearson 'cF4'

_symmetry_space_group_name_Hall "-F 4 2 3"
_symmetry_space_group_name_H-M "F m -3 m"
_symmetry_Int_Tables_number 225

_cell_length_a     3.61491
_cell_length_b     3.61491
_cell_length_c     3.61491
_cell_angle_alpha 90.00000
_cell_angle_beta  90.00000
_cell_angle_gamma 90.00000

loop_
_space_group_symop_id
_space_group_symop_operation_xyz
1 x,y,z
2 x,-y,-z
3 -x,y,-z
4 -x,-y,z
5 y,z,x
6 y,-z,-x
7 -y,z,-x
8 -y,-z,x
9 z,x,y
10 z,-x,-y
11 -z,x,-y
12 -z,-x,y
13 -y,-x,-z
14 -y,x,z
15 y,-x,z
16 y,x,-z
17 -x,-z,-y
18 -x,z,y
19 x,-z,y
20 x,z,-y
21 -z,-y,-x
22 -z,y,x
23 z,-y,x
24 z,y,-x
25 -x,-y,-z
26 -x,y,z
27 x,-y,z
28 x,y,-z
29 -y,-z,-x
30 -y,z,x
31 y,-z,x
32 y,z,-x
33 -z,-x,-y
34 -z,x,y
35 z,-x,y
36 z,x,-y
37 y,x,z
38 y,-x,-z
39 -y,x,-z
40 -y,-x,z
41 x,z,y
42 x,-z,-y
43 -x,z,-y
44 -x,-z,y
45 z,y,x
46 z,-y,-x
47 -z,y,-x
48 -z,-y,x
49 x,y+1/2,z+1/2
50 x,-y+1/2,-z+1/2
51 -x,y+1/2,-z+1/2
52 -x,-y+1/2,z+1/2
53 y,z+1/2,x+1/2
54 y,-z+1/2,-x+1/2
55 -y,z+1/2,-x+1/2
56 -y,-z+1/2,x+1/2
57 z,x+1/2,y+1/2
58 z,-x+1/2,-y+1/2
59 -z,x+1/2,-y+1/2
60 -z,-x+1/2,y+1/2
61 -y,-x+1/2,-z+1/2
62 -y,x+1/2,z+1/2
63 y,-x+1/2,z+1/2
64 y,x+1/2,-z+1/2
65 -x,-z+1/2,-y+1/2
66 -x,z+1/2,y+1/2
67 x,-z+1/2,y+1/2
68 x,z+1/2,-y+1/2
69 -z,-y+1/2,-x+1/2
70 -z,y+1/2,x+1/2
71 z,-y+1/2,x+1/2
72 z,y+1/2,-x+1/2
73 -x,-y+1/2,-z+1/2
74 -x,y+1/2,z+1/2
75 x,-y+1/2,z+1/2
76 x,y+1/2,-z+1/2
77 -y,-z+1/2,-x+1/2
78 -y,z+1/2,x+1/2
79 y,-z+1/2,x+1/2
80 y,z+1/2,-x+1/2
81 -z,-x+1/2,-y+1/2
82 -z,x+1/2,y+1/2
83 z,-x+1/2,y+1/2
84 z,x+1/2,-y+1/2
85 y,x+1/2,z+1/2
86 y,-x+1/2,-z+1/2
87 -y,x+1/2,-z+1/2
88 -y,-x+1/2,z+1/2
89 x,z+1/2,y+1/2
90 x,-z+1/2,-y+1/2
91 -x,z+1/2,-y+1/2
92 -x,-z+1/2,y+1/2
93 z,y+1/2,x+1/2
94 z,-y+1/2,-x+1/2
95 -z,y+1/2,-x+1/2
96 -z,-y+1/2,x+1/2
97 x+1/2,y,z+1/2
98 x+1/2,-y,-z+1/2
99 -x+1/2,y,-z+1/2
100 -x+1/2,-y,z+1/2
101 y+1/2,z,x+1/2
102 y+1/2,-z,-x+1/2
103 -y+1/2,z,-x+1/2
104 -y+1/2,-z,x+1/2
105 z+1/2,x,y+1/2
106 z+1/2,-x,-y+1/2
107 -z+1/2,x,-y+1/2
108 -z+1/2,-x,y+1/2
109 -y+1/2,-x,-z+1/2
110 -y+1/2,x,z+1/2
111 y+1/2,-x,z+1/2
112 y+1/2,x,-z+1/2
113 -x+1/2,-z,-y+1/2
114 -x+1/2,z,y+1/2
115 x+1/2,-z,y+1/2
116 x+1/2,z,-y+1/2
117 -z+1/2,-y,-x+1/2
118 -z+1/2,y,x+1/2
119 z+1/2,-y,x+1/2
120 z+1/2,y,-x+1/2
121 -x+1/2,-y,-z+1/2
122 -x+1/2,y,z+1/2
123 x+1/2,-y,z+1/2
124 x+1/2,y,-z+1/2
125 -y+1/2,-z,-x+1/2
126 -y+1/2,z,x+1/2
127 y+1/2,-z,x+1/2
128 y+1/2,z,-x+1/2
129 -z+1/2,-x,-y+1/2
130 -z+1/2,x,y+1/2
131 z+1/2,-x,y+1/2
132 z+1/2,x,-y+1/2
133 y+1/2,x,z+1/2
134 y+1/2,-x,-z+1/2
135 -y+1/2,x,-z+1/2
136 -y+1/2,-x,z+1/2
137 x+1/2,z,y+1/2
138 x+1/2,-z,-y+1/2
139 -x+1/2,z,-y+1/2
140 -x+1/2,-z,y+1/2
141 z+1/2,y,x+1/2
142 z+1/2,-y,-x+1/2
143 -z+1/2,y,-x+1/2
144 -z+1/2,-y,x+1/2
145 x+1/2,y+1/2,z
146 x+1/2,-y+1/2,-z
147 -x+1/2,y+1/2,-z
148 -x+1/2,-y+1/2,z
149 y+1/2,z+1/2,x
150 y+1/2,-z+1/2,-x
151 -y+1/2,z+1/2,-x
152 -y+1/2,-z+1/2,x
153 z+1/2,x+1/2,y
154 z+1/2,-x+1/2,-y
155 -z+1/2,x+1/2,-y
156 -z+1/2,-x+1/2,y
157 -y+1/2,-x+1/2,-z
158 -y+1/2,x+1/2,z
159 y+1/2,-x+1/2,z
```

```
160  y+1/2,x+1/2,-z
161  -x+1/2,-z+1/2,-y
162  -x+1/2,z+1/2,y
163  x+1/2,-z+1/2,y
164  x+1/2,z+1/2,-y
165  -z+1/2,-y+1/2,-x
166  -z+1/2,y+1/2,x
167  z+1/2,-y+1/2,x
168  z+1/2,y+1/2,-x
169  -x+1/2,-y+1/2,-z
170  -x+1/2,y+1/2,z
171  x+1/2,-y+1/2,z
172  x+1/2,y+1/2,-z
173  -y+1/2,-z+1/2,-x
174  -y+1/2,z+1/2,x
175  y+1/2,-z+1/2,x
176  y+1/2,z+1/2,-x
177  -z+1/2,-x+1/2,-y
178  -z+1/2,x+1/2,y
179  z+1/2,-x+1/2,y
180  z+1/2,x+1/2,-y
181  y+1/2,x+1/2,z
182  y+1/2,-x+1/2,-z
183  -y+1/2,x+1/2,-z
184  -y+1/2,-x+1/2,z
185  x+1/2,z+1/2,y
186  x+1/2,-z+1/2,-y
187  -x+1/2,z+1/2,-y
188  -x+1/2,-z+1/2,y
189  z+1/2,y+1/2,x
190  z+1/2,-y+1/2,-x
191  -z+1/2,y+1/2,-x
192  -z+1/2,-y+1/2,x

loop_
_atom_site_label
_atom_site_type_symbol
_atom_site_symmetry_multiplicity
_atom_site_Wyckoff_label
_atom_site_fract_x
_atom_site_fract_y
_atom_site_fract_z
_atom_site_occupancy
Cu1  Cu    4  a  0.00000  0.00000  0.00000  1.00000
```

Face-Centered Cubic (Cu, A1): A_cF4_225_a - POSCAR

```
A_cF4_225_a & a --params=3.61491 & Fm(-3)m    O_h^5 #225 (a) & cF4 & A1 &
    ↪ Cu & & M. E. Straumanis and L. S. Yu, Acta Cryst. A 25,
    ↪ 676-682 (1969)
   1.0000000000000000
   0.00000000000000    1.80745500000000    1.80745500000000
   1.80745500000000    0.00000000000000    1.80745500000000
   1.80745500000000    1.80745500000000    0.00000000000000
   Cu
    1
Direct
   0.00000000000000    0.00000000000000    0.00000000000000    Cu    (4a)
```

Model of Austenite (cF108): AB18C8_cF108_225_a_eh_f - CIF

```
# CIF file

data_findsym-output
_audit_creation_method FINDSYM

_chemical_name_mineral ''
_chemical_formula_sum 'Cr Fe18 Ni8'

loop_
_publ_author_name
 'Michael J. Mehl'
_journal_name_full
;
None
;
_journal_volume 0
_journal_year 2008
_journal_page_first 0
_journal_page_last 0
_publ_Section_title
;
 Hypothetical cF108 Austenite Structure
;

_aflow_proto 'AB18C8_cF108_225_a_eh_f'
_aflow_params 'a,x2,x3,y4'
_aflow_params_values '10.56,0.325,0.65833,0.66'
_aflow_Strukturbericht 'None'
_aflow_Pearson 'cF108'

_symmetry_space_group_name_Hall "-F 4 2 3"
_symmetry_space_group_name_H-M "F m -3 m"
_symmetry_Int_Tables_number 225

_cell_length_a     10.56000
_cell_length_b     10.56000
_cell_length_c     10.56000
_cell_angle_alpha  90.00000
_cell_angle_beta   90.00000
_cell_angle_gamma  90.00000

loop_
_space_group_symop_id
_space_group_symop_operation_xyz
1  x,y,z
2  x,-y,-z
3  -x,y,-z
4  -x,-y,z
5  y,z,x
6  y,-z,-x
7  -y,z,-x
8  -y,-z,x
9  z,x,y
10  z,-x,-y
11  -z,x,-y
12  -z,-x,y
13  -y,-x,-z
14  -y,x,z
15  y,-x,z
16  y,x,-z
17  -x,-z,-y
18  -x,z,y
19  x,-z,y
20  x,z,-y
21  -z,-y,-x
22  -z,y,x
23  z,-y,x
24  z,y,-x
25  -x,-y,-z
26  -x,y,z
27  x,-y,z
28  x,y,-z
29  -y,-z,-x
30  -y,z,x
31  y,-z,x
32  y,z,-x
33  -z,-x,-y
34  -z,x,y
35  z,-x,y
36  z,x,-y
37  y,x,z
38  y,-x,-z
39  -y,x,-z
40  -y,-x,z
41  x,z,y
42  x,-z,-y
43  -x,z,-y
44  -x,-z,y
45  z,y,x
46  z,-y,-x
47  -z,y,-x
48  -z,-y,x
49  x,y+1/2,z+1/2
50  x,-y+1/2,-z+1/2
51  -x,y+1/2,-z+1/2
52  -x,-y+1/2,z+1/2
53  y,z+1/2,x+1/2
54  y,-z+1/2,-x+1/2
55  -y,z+1/2,-x+1/2
56  -y,-z+1/2,x+1/2
57  z,x+1/2,y+1/2
58  z,-x+1/2,-y+1/2
59  -z,x+1/2,-y+1/2
60  -z,-x+1/2,y+1/2
61  -y,-x+1/2,-z+1/2
62  -y,x+1/2,z+1/2
63  y,-x+1/2,z+1/2
64  y,x+1/2,-z+1/2
65  -x,-z+1/2,-y+1/2
66  -x,z+1/2,y+1/2
67  x,-z+1/2,y+1/2
68  x,z+1/2,-y+1/2
69  -z,-y+1/2,-x+1/2
70  -z,y+1/2,x+1/2
71  z,-y+1/2,x+1/2
72  z,y+1/2,-x+1/2
73  -x,-y+1/2,-z+1/2
74  -x,y+1/2,z+1/2
75  x,-y+1/2,z+1/2
76  x,y+1/2,-z+1/2
77  -y,-z+1/2,-x+1/2
78  -y,z+1/2,x+1/2
79  y,-z+1/2,x+1/2
80  y,z+1/2,-x+1/2
81  -z,-x+1/2,-y+1/2
82  -z,x+1/2,y+1/2
83  z,-x+1/2,y+1/2
84  z,x+1/2,-y+1/2
85  y,x+1/2,z+1/2
86  y,-x+1/2,-z+1/2
87  -y,x+1/2,-z+1/2
88  -y,-x+1/2,z+1/2
89  x,z+1/2,y+1/2
90  x,-z+1/2,-y+1/2
91  -x,z+1/2,-y+1/2
92  -x,-z+1/2,y+1/2
93  z,y+1/2,x+1/2
94  z,-y+1/2,-x+1/2
95  -z,y+1/2,-x+1/2
96  -z,-y+1/2,x+1/2
97  x+1/2,y,z+1/2
98  x+1/2,-y,-z+1/2
99  -x+1/2,y,-z+1/2
100  -x+1/2,-y,z+1/2
101  y+1/2,z,x+1/2
102  y+1/2,-z,-x+1/2
103  -y+1/2,z,-x+1/2
104  -y+1/2,-z,x+1/2
105  z+1/2,x,y+1/2
```

```
106 z+1/2,-x,-y+1/2
107 -z+1/2,x,-y+1/2
108 -z+1/2,-x,y+1/2
109 -y+1/2,-x,-z+1/2
110 -y+1/2,x,z+1/2
111 y+1/2,-x,z+1/2
112 y+1/2,x,-z+1/2
113 -x+1/2,-z,-y+1/2
114 -x+1/2,z,y+1/2
115 x+1/2,-z,y+1/2
116 x+1/2,z,-y+1/2
117 -z+1/2,-y,-x+1/2
118 -z+1/2,y,x+1/2
119 z+1/2,-y,x+1/2
120 z+1/2,y,-x+1/2
121 -x+1/2,-y,-z+1/2
122 -x+1/2,y,z+1/2
123 x+1/2,-y,z+1/2
124 x+1/2,y,-z+1/2
125 -y+1/2,-z,-x+1/2
126 -y+1/2,z,x+1/2
127 y+1/2,-z,x+1/2
128 y+1/2,z,-x+1/2
129 -z+1/2,-x,-y+1/2
130 -z+1/2,x,y+1/2
131 z+1/2,-x,y+1/2
132 z+1/2,x,-y+1/2
133 y+1/2,x,z+1/2
134 y+1/2,-x,-z+1/2
135 -y+1/2,x,-z+1/2
136 -y+1/2,-x,z+1/2
137 x+1/2,z,y+1/2
138 x+1/2,-z,-y+1/2
139 -x+1/2,z,-y+1/2
140 -x+1/2,-z,y+1/2
141 z+1/2,y,x+1/2
142 z+1/2,-y,-x+1/2
143 -z+1/2,y,-x+1/2
144 -z+1/2,-y,x+1/2
145 x+1/2,y+1/2,z
146 x+1/2,-y+1/2,-z
147 -x+1/2,y+1/2,-z
148 -x+1/2,-y+1/2,z
149 y+1/2,z+1/2,x
150 y+1/2,-z+1/2,-x
151 -y+1/2,z+1/2,-x
152 -y+1/2,-z+1/2,x
153 z+1/2,x+1/2,y
154 z+1/2,-x+1/2,-y
155 -z+1/2,x+1/2,-y
156 -z+1/2,-x+1/2,y
157 -y+1/2,-x+1/2,-z
158 -y+1/2,x+1/2,z
159 y+1/2,-x+1/2,z
160 y+1/2,x+1/2,-z
161 -x+1/2,-z+1/2,-y
162 -x+1/2,z+1/2,y
163 x+1/2,-z+1/2,y
164 x+1/2,z+1/2,-y
165 -z+1/2,-y+1/2,-x
166 -z+1/2,y+1/2,x
167 z+1/2,-y+1/2,x
168 z+1/2,y+1/2,-x
169 -x+1/2,-y+1/2,-z
170 -x+1/2,y+1/2,z
171 x+1/2,-y+1/2,z
172 x+1/2,y+1/2,-z
173 -y+1/2,-z+1/2,-x
174 -y+1/2,z+1/2,x
175 y+1/2,-z+1/2,x
176 y+1/2,z+1/2,-x
177 -z+1/2,-x+1/2,-y
178 -z+1/2,x+1/2,y
179 z+1/2,-x+1/2,y
180 z+1/2,x+1/2,-y
181 y+1/2,x+1/2,z
182 y+1/2,-x+1/2,-z
183 -y+1/2,x+1/2,-z
184 -y+1/2,-x+1/2,z
185 x+1/2,z+1/2,y
186 x+1/2,-z+1/2,-y
187 -x+1/2,z+1/2,-y
188 -x+1/2,-z+1/2,y
189 z+1/2,y+1/2,x
190 z+1/2,-y+1/2,-x
191 -z+1/2,y+1/2,-x
192 -z+1/2,-y+1/2,x

loop_
_atom_site_label
_atom_site_type_symbol
_atom_site_symmetry_multiplicity
_atom_site_Wyckoff_label
_atom_site_fract_x
_atom_site_fract_y
_atom_site_fract_z
_atom_site_occupancy
Cr1 Cr   4 a 0.00000 0.00000 0.00000 1.00000
Fe1 Fe  24 e 0.32500 0.00000 0.00000 1.00000
Ni1 Ni  32 f 0.65833 0.65833 0.65833 1.00000
Fe2 Fe  48 h 0.00000 0.66000 0.66000 1.00000
```

Model of Austenite (cF108): AB18C8_cF108_225_a_eh_f - POSCAR

```
AB18C8_cF108_225_a_eh_f & a,x2,x3,y4 --params=10.56,0.325,0.65833,0.66 &
   ↪ Fm(-3)m    O_h^5 #225 (aefh) & cF108 & & CrFe18Ni8 & &
   ↪ Hypothetical ordered austenite
   1.0000000000000000
   0.00000000000000   5.28000000000000   5.28000000000000
   5.28000000000000   0.00000000000000   5.28000000000000
   5.28000000000000   5.28000000000000   0.00000000000000
   Cr    Fe    Ni
    1    18     8
Direct
   0.00000000000000   0.00000000000000   0.00000000000000   Cr    (4a)
   0.32500000000000   0.32500000000000   0.67500000000000   Fe   (24e)
   0.32500000000000   0.67500000000000   0.32500000000000   Fe   (24e)
   0.32500000000000   0.67500000000000   0.67500000000000   Fe   (24e)
   0.67500000000000   0.32500000000000   0.32500000000000   Fe   (24e)
   0.67500000000000   0.32500000000000   0.67500000000000   Fe   (24e)
   0.67500000000000   0.67500000000000   0.32500000000000   Fe   (24e)
   0.00000000000000   0.00000000000000   0.32000000000000   Fe   (48h)
   0.00000000000000   0.00000000000000   0.68000000000000   Fe   (48h)
   0.00000000000000   0.32000000000000   0.00000000000000   Fe   (48h)
   0.00000000000000   0.32000000000000   0.68000000000000   Fe   (48h)
   0.00000000000000   0.68000000000000   0.00000000000000   Fe   (48h)
   0.00000000000000   0.68000000000000   0.32000000000000   Fe   (48h)
   0.32000000000000   0.00000000000000   0.00000000000000   Fe   (48h)
   0.32000000000000   0.00000000000000   0.68000000000000   Fe   (48h)
   0.32000000000000   0.68000000000000   0.00000000000000   Fe   (48h)
   0.68000000000000   0.00000000000000   0.00000000000000   Fe   (48h)
   0.68000000000000   0.00000000000000   0.32000000000000   Fe   (48h)
   0.68000000000000   0.32000000000000   0.00000000000000   Fe   (48h)
  -0.02501000000000   0.34167000000000   0.34167000000000   Ni   (32f)
   0.02501000000000   0.65833000000000   0.65833000000000   Ni   (32f)
   0.34167000000000  -0.02501000000000   0.34167000000000   Ni   (32f)
   0.34167000000000   0.34167000000000  -0.02501000000000   Ni   (32f)
   0.34167000000000   0.34167000000000   0.34167000000000   Ni   (32f)
   0.65833000000000   0.02501000000000   0.65833000000000   Ni   (32f)
   0.65833000000000   0.65833000000000   0.02501000000000   Ni   (32f)
   0.65833000000000   0.65833000000000   0.65833000000000   Ni   (32f)
```

Rock Salt (NaCl, B1): AB_cF8_225_a_b - CIF

```
# CIF file

data_findsym-output
_audit_creation_method FINDSYM

_chemical_name_mineral 'Halite, Rock Salt'
_chemical_formula_sum 'Na Cl'

loop_
_publ_author_name
 'David Walker'
 'Pramod K. Verma'
 'Lachlan M. D. Cranswick'
 'Raymond L. Jones'
 'Simon M. Clark'
 'Stephan Buhre'
_journal_name_full
;
American Mineralogist
;
_journal_volume 89
_journal_year 2004
_journal_page_first 204
_journal_page_last 210
_publ_Section_title
;
 Halite-sylvite thermoelasticity
;

# Found in  AMS Database

_aflow_proto 'AB_cF8_225_a_b'
_aflow_params 'a'
_aflow_params_values '5.63931'
_aflow_Strukturbericht 'B1'
_aflow_Pearson 'cF8'

_symmetry_space_group_name_Hall "-F 4 2 3"
_symmetry_space_group_name_H-M "F m -3 m"
_symmetry_Int_Tables_number 225

_cell_length_a      5.63931
_cell_length_b      5.63931
_cell_length_c      5.63931
_cell_angle_alpha  90.00000
_cell_angle_beta   90.00000
_cell_angle_gamma  90.00000

loop_
_space_group_symop_id
_space_group_symop_operation_xyz
1 x,y,z
2 x,-y,-z
3 -x,y,-z
4 -x,-y,z
5 y,z,x
6 y,-z,-x
7 -y,z,-x
8 -y,-z,x
9 z,x,y
10 z,-x,-y
11 -z,x,-y
12 -z,-x,y
13 -y,-x,-z
14 -y,x,z
15 y,-x,z
```

```
16 y,x,-z
17 -x,-z,-y
18 -x,z,y
19 x,-z,y
20 x,z,-y
21 -z,-y,-x
22 -z,y,x
23 z,-y,x
24 z,y,-x
25 -x,-y,-z
26 -x,y,z
27 x,-y,z
28 x,y,-z
29 -y,-z,-x
30 -y,z,x
31 y,-z,x
32 y,z,-x
33 -z,-x,-y
34 -z,x,y
35 z,-x,y
36 z,x,-y
37 y,x,z
38 y,-x,-z
39 -y,x,-z
40 -y,-x,z
41 x,z,y
42 x,-z,-y
43 -x,z,-y
44 -x,-z,y
45 z,y,x
46 z,-y,-x
47 -z,y,-x
48 -z,-y,x
49 x,y+1/2,z+1/2
50 x,-y+1/2,-z+1/2
51 -x,y+1/2,-z+1/2
52 -x,-y+1/2,z+1/2
53 y,z+1/2,x+1/2
54 y,-z+1/2,-x+1/2
55 -y,z+1/2,-x+1/2
56 -y,-z+1/2,x+1/2
57 z,x+1/2,y+1/2
58 z,-x+1/2,-y+1/2
59 -z,x+1/2,-y+1/2
60 -z,-x+1/2,y+1/2
61 -y,-x+1/2,-z+1/2
62 -y,x+1/2,z+1/2
63 y,-x+1/2,z+1/2
64 y,x+1/2,-z+1/2
65 -x,-z+1/2,-y+1/2
66 -x,z+1/2,y+1/2
67 x,-z+1/2,y+1/2
68 x,z+1/2,-y+1/2
69 -z,-y+1/2,-x+1/2
70 -z,y+1/2,x+1/2
71 z,-y+1/2,x+1/2
72 z,y+1/2,-x+1/2
73 -x,-y+1/2,-z+1/2
74 -x,y+1/2,z+1/2
75 x,-y+1/2,z+1/2
76 x,y+1/2,-z+1/2
77 -y,-z+1/2,-x+1/2
78 -y,z+1/2,x+1/2
79 y,-z+1/2,x+1/2
80 y,z+1/2,-x+1/2
81 -z,-x+1/2,-y+1/2
82 -z,x+1/2,y+1/2
83 z,-x+1/2,y+1/2
84 z,x+1/2,-y+1/2
85 y,x+1/2,z+1/2
86 y,-x+1/2,-z+1/2
87 -y,x+1/2,-z+1/2
88 -y,-x+1/2,z+1/2
89 x,z+1/2,y+1/2
90 x,-z+1/2,-y+1/2
91 -x,z+1/2,-y+1/2
92 -x,-z+1/2,y+1/2
93 z,y+1/2,x+1/2
94 z,-y+1/2,-x+1/2
95 -z,y+1/2,-x+1/2
96 -z,-y+1/2,x+1/2
97 x+1/2,y,z+1/2
98 x+1/2,-y,-z+1/2
99 -x+1/2,y,-z+1/2
100 -x+1/2,-y,z+1/2
101 y+1/2,z,x+1/2
102 y+1/2,-z,-x+1/2
103 -y+1/2,z,-x+1/2
104 -y+1/2,-z,x+1/2
105 z+1/2,x,y+1/2
106 z+1/2,-x,-y+1/2
107 -z+1/2,x,-y+1/2
108 -z+1/2,-x,y+1/2
109 -y+1/2,-x,-z+1/2
110 -y+1/2,x,z+1/2
111 y+1/2,-x,z+1/2
112 y+1/2,x,-z+1/2
113 -x+1/2,-z,-y+1/2
114 -x+1/2,z,y+1/2
115 x+1/2,-z,y+1/2
116 x+1/2,z,-y+1/2
117 -z+1/2,-y,-x+1/2
118 -z+1/2,y,x+1/2
119 z+1/2,-y,x+1/2
120 z+1/2,y,-x+1/2
121 -x+1/2,-y,-z+1/2
122 -x+1/2,y,z+1/2
123 x+1/2,-y,z+1/2
124 x+1/2,y,-z+1/2
125 -y+1/2,-z,-x+1/2
126 -y+1/2,z,x+1/2
127 y+1/2,-z,x+1/2
128 y+1/2,z,-x+1/2
129 -z+1/2,-x,-y+1/2
130 -z+1/2,x,y+1/2
131 z+1/2,-x,y+1/2
132 z+1/2,x,-y+1/2
133 y+1/2,x,z+1/2
134 y+1/2,-x,-z+1/2
135 -y+1/2,x,-z+1/2
136 -y+1/2,-x,z+1/2
137 x+1/2,z,y+1/2
138 x+1/2,-z,-y+1/2
139 -x+1/2,z,-y+1/2
140 -x+1/2,-z,y+1/2
141 z+1/2,y,x+1/2
142 z+1/2,-y,-x+1/2
143 -z+1/2,y,-x+1/2
144 -z+1/2,-y,x+1/2
145 x+1/2,y+1/2,z
146 x+1/2,-y+1/2,-z
147 -x+1/2,y+1/2,-z
148 -x+1/2,-y+1/2,z
149 y+1/2,z+1/2,x
150 y+1/2,-z+1/2,-x
151 -y+1/2,z+1/2,-x
152 -y+1/2,-z+1/2,x
153 z+1/2,x+1/2,y
154 z+1/2,-x+1/2,-y
155 -z+1/2,x+1/2,-y
156 -z+1/2,-x+1/2,y
157 -y+1/2,-x+1/2,-z
158 -y+1/2,x+1/2,z
159 y+1/2,-x+1/2,z
160 y+1/2,x+1/2,-z
161 -x+1/2,-z+1/2,-y
162 -x+1/2,z+1/2,y
163 x+1/2,-z+1/2,y
164 x+1/2,z+1/2,-y
165 -z+1/2,-y+1/2,-x
166 -z+1/2,y+1/2,x
167 z+1/2,-y+1/2,x
168 z+1/2,y+1/2,-x
169 -x+1/2,-y+1/2,-z
170 -x+1/2,y+1/2,z
171 x+1/2,-y+1/2,z
172 x+1/2,y+1/2,-z
173 -y+1/2,-z+1/2,-x
174 -y+1/2,z+1/2,x
175 y+1/2,-z+1/2,x
176 y+1/2,z+1/2,-x
177 -z+1/2,-x+1/2,-y
178 -z+1/2,x+1/2,y
179 z+1/2,-x+1/2,y
180 z+1/2,x+1/2,-y
181 y+1/2,x+1/2,z
182 y+1/2,-x+1/2,-z
183 -y+1/2,x+1/2,-z
184 -y+1/2,-x+1/2,z
185 x+1/2,z+1/2,y
186 x+1/2,-z+1/2,-y
187 -x+1/2,z+1/2,-y
188 -x+1/2,-z+1/2,y
189 z+1/2,y+1/2,x
190 z+1/2,-y+1/2,-x
191 -z+1/2,y+1/2,-x
192 -z+1/2,-y+1/2,x

loop_
_atom_site_label
_atom_site_type_symbol
_atom_site_symmetry_multiplicity
_atom_site_Wyckoff_label
_atom_site_fract_x
_atom_site_fract_y
_atom_site_fract_z
_atom_site_occupancy
Cl1 Cl   4 a 0.00000 0.00000 0.00000 1.00000
Na1 Na   4 b 0.50000 0.50000 0.50000 1.00000
```

Rock Salt (NaCl, B1): AB_cF8_225_a_b - POSCAR

```
AB_cF8_225_a_b & a --params=5.63931 & Fm(-3)m    O_h^5 #225 (ab) & cF8 &
     ↪ B1 & NaCl & Halite/Rock Salt & D. Walker et al., Am. Mineral.
     ↪ 89, 204-210 (2004)
   1.0000000000000000
   0.00000000000000    2.81965339535660    2.81965339535660
   2.81965339535660    0.00000000000000    2.81965339535660
   2.81965339535660    2.81965339535660    0.00000000000000
   Cl   Na
    1    1
Direct
   0.00000000000000    0.00000000000000    0.00000000000000   Cl   (4a)
   0.50000000000000    0.50000000000000    0.50000000000000   Na   (4b)
```

Ideal $\beta$-Cristobalite ($SiO_2$, C9): A2B_cF24_227_c_a - CIF

```
# CIF file

data_findsym-output
```

```
_audit_creation_method FINDSYM

_chemical_name_mineral 'high (beta) Cristobalite'
_chemical_formula_sum 'Si O2'

loop_
_publ_author_name
 'Donald R. Peacor'
_journal_name_full
;
Zeitschrift f\"{u}r kristallographie
;
_journal_volume 138
_journal_year 1973
_journal_page_first 274
_journal_page_last 298
_publ_Section_title
;
 High-temperature single-crystal study of the cristobalite inversion
;

# Found in AMS Database

_aflow_proto 'A2B_cF24_227_c_a'
_aflow_params 'a'
_aflow_params_values '7.166'
_aflow_Strukturbericht 'C9'
_aflow_Pearson 'cF24'

_symmetry_space_group_name_Hall "-F 4vw 2vw 3 Fd(-3)m"
_symmetry_space_group_name_H-M "F d -3 m:2"
_symmetry_Int_Tables_number 227

_cell_length_a      7.16600
_cell_length_b      7.16600
_cell_length_c      7.16600
_cell_angle_alpha 90.00000
_cell_angle_beta  90.00000
_cell_angle_gamma 90.00000

loop_
_space_group_symop_id
_space_group_symop_operation_xyz
1 x,y,z
2 x,-y+1/4,-z+1/4
3 -x+1/4,y,-z+1/4
4 -x+1/4,-y+1/4,z
5 y,z,x
6 y,-z+1/4,-x+1/4
7 -y+1/4,z,-x+1/4
8 -y+1/4,-z+1/4,x
9 z,x,y
10 z,-x+1/4,-y+1/4
11 -z+1/4,x,-y+1/4
12 -z+1/4,-x+1/4,y
13 -y,-x,-z
14 -y,x+1/4,z+1/4
15 y+1/4,-x,z+1/4
16 y+1/4,x+1/4,-z
17 -x,-z,-y
18 -x,z+1/4,y+1/4
19 x+1/4,-z,y+1/4
20 x+1/4,z+1/4,-y
21 -z,-y,-x
22 -z,y+1/4,x+1/4
23 z+1/4,-y,x+1/4
24 z+1/4,y+1/4,-x
25 -x,-y,-z
26 -x,y+1/4,z+1/4
27 x+1/4,-y,z+1/4
28 x+1/4,y+1/4,-z
29 -y,-z,-x
30 -y,z+1/4,x+1/4
31 y+1/4,-z,x+1/4
32 y+1/4,z+1/4,-x
33 -z,-x,-y
34 -z,x+1/4,y+1/4
35 z+1/4,-x,y+1/4
36 z+1/4,x+1/4,-y
37 y,x,z
38 y,-x+1/4,-z+1/4
39 -y+1/4,x,-z+1/4
40 -y+1/4,-x+1/4,z
41 x,z,y
42 x,-z+1/4,-y+1/4
43 -x+1/4,z,-y+1/4
44 -x+1/4,-z+1/4,y
45 z,y,x
46 z,-y+1/4,-x+1/4
47 -z+1/4,y,-x+1/4
48 -z+1/4,-y+1/4,x
49 x,y+1/2,z+1/2
50 x,-y+3/4,-z+3/4
51 -x+1/4,y+1/2,-z+3/4
52 -x+1/4,-y+3/4,z+1/2
53 y,z+1/2,x+1/2
54 y,-z+3/4,-x+3/4
55 -y+1/4,z+1/2,-x+3/4
56 -y+1/4,-z+3/4,x+1/2
57 z,x+1/2,y+1/2
58 z,-x+3/4,-y+3/4
59 -z+1/4,x+1/2,-y+3/4
60 -z+1/4,-x+3/4,y+1/2
61 -y,-x+1/2,-z+1/2
62 -y,x+3/4,z+3/4
63 y+1/4,-x+1/2,z+3/4
64 y+1/4,x+3/4,-z+1/2
65 -x,-z+1/2,-y+1/2
66 -x,z+3/4,y+3/4
67 x+1/4,-z+1/2,y+3/4
68 x+1/4,z+3/4,-y+1/2
69 -z,-y+1/2,-x+1/2
70 -z,y+3/4,x+3/4
71 z+1/4,-y+1/2,x+3/4
72 z+1/4,y+3/4,-x+1/2
73 -x,-y+1/2,-z+1/2
74 -x,y+3/4,z+3/4
75 x+1/4,-y+1/2,z+3/4
76 x+1/4,y+3/4,-z+1/2
77 -y,-z+1/2,-x+1/2
78 -y,z+3/4,x+3/4
79 y+1/4,-z+1/2,x+3/4
80 y+1/4,z+3/4,-x+1/2
81 -z,-x+1/2,-y+1/2
82 -z,x+3/4,y+3/4
83 z+1/4,-x+1/2,y+3/4
84 z+1/4,x+3/4,-y+1/2
85 y,x+1/2,z+1/2
86 y,-x+3/4,-z+3/4
87 -y+1/4,x+1/2,-z+3/4
88 -y+1/4,-x+3/4,z+1/2
89 x,z+1/2,y+1/2
90 x,-z+3/4,-y+3/4
91 -x+1/4,z+1/2,-y+3/4
92 -x+1/4,-z+3/4,y+1/2
93 z,y+1/2,x+1/2
94 z,-y+3/4,-x+3/4
95 -z+1/4,y+1/2,-x+3/4
96 -z+1/4,-y+3/4,x+1/2
97 x+1/2,y,z+1/2
98 x+1/2,-y+1/4,-z+3/4
99 -x+3/4,y,-z+3/4
100 -x+3/4,-y+1/4,z+1/2
101 y+1/2,z,x+1/2
102 y+1/2,-z+1/4,-x+3/4
103 -y+3/4,z,-x+3/4
104 -y+3/4,-z+1/4,x+1/2
105 z+1/2,x,y+1/2
106 z+1/2,-x+1/4,-y+3/4
107 -z+3/4,x,-y+3/4
108 -z+3/4,-x+1/4,y+1/2
109 -y+1/2,-x,-z+1/2
110 -y+1/2,x+1/4,z+3/4
111 y+3/4,-x,z+3/4
112 y+3/4,x+1/4,-z+1/2
113 -x+1/2,-z,-y+1/2
114 -x+1/2,z+1/4,y+3/4
115 x+3/4,-z,y+3/4
116 x+3/4,z+1/4,-y+1/2
117 -z+1/2,-y,-x+1/2
118 -z+1/2,y+1/4,x+3/4
119 z+3/4,-y,x+3/4
120 z+3/4,y+1/4,-x+1/2
121 -x+1/2,-y,-z+1/2
122 -x+1/2,y+1/4,z+3/4
123 x+3/4,-y,z+3/4
124 x+3/4,y+1/4,-z+1/2
125 -y+1/2,-z,-x+1/2
126 -y+1/2,z+1/4,x+3/4
127 y+3/4,-z,x+3/4
128 y+3/4,z+1/4,-x+1/2
129 -z+1/2,-x,-y+1/2
130 -z+1/2,x+1/4,y+3/4
131 z+3/4,-x,y+3/4
132 z+3/4,x+1/4,-y+1/2
133 y+1/2,x,z+1/2
134 y+1/2,-x+1/4,-z+3/4
135 -y+3/4,x,-z+3/4
136 -y+3/4,-x+1/4,z+1/2
137 x+1/2,z,y+1/2
138 x+1/2,-z+1/4,-y+3/4
139 -x+3/4,z,-y+3/4
140 -x+3/4,-z+1/4,y+1/2
141 z+1/2,y,x+1/2
142 z+1/2,-y+1/4,-x+3/4
143 -z+3/4,y,-x+3/4
144 -z+3/4,-y+1/4,x+1/2
145 x+1/2,y+1/2,z
146 x+1/2,-y+3/4,-z+1/4
147 -x+3/4,y+1/2,-z+1/4
148 -x+3/4,-y+3/4,z
149 y+1/2,z+1/2,x
150 y+1/2,-z+3/4,-x+1/4
151 -y+3/4,z+1/2,-x+1/4
152 -y+3/4,-z+3/4,x
153 z+1/2,x+1/2,y
154 z+1/2,-x+3/4,-y+1/4
155 -z+3/4,x+1/2,-y+1/4
156 -z+3/4,-x+3/4,y
157 -y+1/2,-x+1/2,-z
158 -y+1/2,x+3/4,z+1/4
159 y+3/4,-x+1/2,z+1/4
160 y+3/4,x+3/4,-z
161 -x+1/2,-z+1/2,-y
162 -x+1/2,z+3/4,y+1/4
163 x+3/4,-z+1/2,y+1/4
164 x+3/4,z+3/4,-y
165 -z+1/2,-y+1/2,-x
166 -z+1/2,y+3/4,x+1/4
167 z+3/4,-y+1/2,x+1/4
```

```
168 z+3/4,y+3/4,-x
169 -x+1/2,-y+1/2,-z
170 -x+1/2,y+3/4,z+1/4
171 x+3/4,-y+1/2,z+1/4
172 x+3/4,y+3/4,-z
173 -y+1/2,-z+1/2,-x
174 -y+1/2,z+3/4,x+1/4
175 y+3/4,-z+1/2,x+1/4
176 y+3/4,z+3/4,-x
177 -z+1/2,-x+1/2,-y
178 -z+1/2,x+3/4,y+1/4
179 z+3/4,-x+1/2,y+1/4
180 z+3/4,x+3/4,-y
181 y+1/2,x+1/2,z
182 y+1/2,-x+3/4,-z+1/4
183 -y+3/4,x+1/2,-z+1/4
184 -y+3/4,-x+3/4,z
185 x+1/2,z+1/2,y
186 x+1/2,-z+3/4,-y+1/4
187 -x+3/4,z+1/2,-y+1/4
188 -x+3/4,-z+3/4,y
189 z+1/2,y+1/2,x
190 z+1/2,-y+3/4,-x+1/4
191 -z+3/4,y+1/2,-x+1/4
192 -z+3/4,-y+3/4,x

loop_
_atom_site_label
_atom_site_type_symbol
_atom_site_symmetry_multiplicity
_atom_site_Wyckoff_label
_atom_site_fract_x
_atom_site_fract_y
_atom_site_fract_z
_atom_site_occupancy
Si1 Si   8 a 0.12500 0.12500 0.12500 1.00000
O1  O   16 c 0.00000 0.00000 0.00000 1.00000
```

Ideal $\beta$-Cristobalite ($SiO_2$, C9): A2B_cF24_227_c_a - POSCAR

```
A2B_cF24_227_c_a & a --params=7.166 & Fd(-3)m    O_h^7 #227 (ac) & cF24 &
    ↪ C9 & SiO2 & beta (high) Cristobalite -- idealized structure &
    ↪ D. R. Peacor, Zeitschrift f\"{u}r kristallographie 138, 274-298
    ↪ (1973)
   1.0000000000000000
   0.00000000000000   3.58300000000000   3.58300000000000
   3.58300000000000   0.00000000000000   3.58300000000000
   3.58300000000000   3.58300000000000   0.00000000000000
     O    Si
     4     2
Direct
   0.00000000000000   0.00000000000000   0.00000000000000    O   (16c)
   0.00000000000000   0.00000000000000   0.50000000000000    O   (16c)
   0.00000000000000   0.50000000000000   0.00000000000000    O   (16c)
   0.50000000000000   0.00000000000000   0.00000000000000    O   (16c)
   0.12500000000000   0.12500000000000   0.12500000000000   Si    (8a)
   0.87500000000000   0.87500000000000   0.87500000000000   Si    (8a)
```

$NiTi_2$: AB2_cF96_227_e_cf - CIF

```
# CIF file

data_findsym-output
_audit_creation_method FINDSYM

_chemical_name_mineral ''
_chemical_formula_sum 'Ni Ti2'

loop_
_publ_author_name
 'G. A. Yurko'
 'J. W. Barton'
 'J. Gordon Parr'
_journal_name_full
;
Acta Crystallographica
;
_journal_volume 12
_journal_year 1959
_journal_page_first 909
_journal_page_last 911
_publ_Section_title
;
 The crystal structure of Ti$_2$Ni
;

# Found in Pearson's Handbook Vol IV, pp. 4715

_aflow_proto 'AB2_cF96_227_e_cf'
_aflow_params 'a,x2,x3'
_aflow_params_values '11.278,0.215,0.44'
_aflow_Strukturbericht 'None'
_aflow_Pearson 'cF96'

_symmetry_space_group_name_Hall "-F 4vw 2vw 3 Fd(-3)m"
_symmetry_space_group_name_H-M "F d -3 m:2"
_symmetry_Int_Tables_number 227

_cell_length_a     11.27800
_cell_length_b     11.27800
_cell_length_c     11.27800
_cell_angle_alpha  90.00000
_cell_angle_beta   90.00000
_cell_angle_gamma  90.00000

loop_
_space_group_symop_id
_space_group_symop_operation_xyz
1 x,y,z
2 x,-y+1/4,-z+1/4
3 -x+1/4,y,-z+1/4
4 -x+1/4,-y+1/4,z
5 y,z,x
6 y,-z+1/4,-x+1/4
7 -y+1/4,z,-x+1/4
8 -y+1/4,-z+1/4,x
9 z,x,y
10 z,-x+1/4,-y+1/4
11 -z+1/4,x,-y+1/4
12 -z+1/4,-x+1/4,y
13 -y,-x,-z
14 -y,x+1/4,z+1/4
15 y+1/4,-x,z+1/4
16 y+1/4,x+1/4,-z
17 -x,-z,-y
18 -x,z+1/4,y+1/4
19 x+1/4,-z,y+1/4
20 x+1/4,z+1/4,-y
21 -z,-y,-x
22 -z,y+1/4,x+1/4
23 z+1/4,-y,x+1/4
24 z+1/4,y+1/4,-x
25 -x,-y,-z
26 -x,y+1/4,z+1/4
27 x+1/4,-y,z+1/4
28 x+1/4,y+1/4,-z
29 -y,-z,-x
30 -y,z+1/4,x+1/4
31 y+1/4,-z,x+1/4
32 y+1/4,z+1/4,-x
33 -z,-x,-y
34 -z,x+1/4,y+1/4
35 z+1/4,-x,y+1/4
36 z+1/4,x+1/4,-y
37 y,x,z
38 y,-x+1/4,-z+1/4
39 -y+1/4,x,-z+1/4
40 -y+1/4,-x+1/4,z
41 x,z,y
42 x,-z+1/4,-y+1/4
43 -x+1/4,z,-y+1/4
44 -x+1/4,-z+1/4,y
45 z,y,x
46 z,-y+1/4,-x+1/4
47 -z+1/4,y,-x+1/4
48 -z+1/4,-y+1/4,x
49 x,y+1/2,z+1/2
50 x,-y+3/4,-z+3/4
51 -x+1/4,y+1/2,-z+3/4
52 -x+1/4,-y+3/4,z+1/2
53 y,z+1/2,x+1/2
54 y,-z+3/4,-x+3/4
55 -y+1/4,z+1/2,-x+3/4
56 -y+1/4,-z+3/4,x+1/2
57 z,x+1/2,y+1/2
58 z,-x+3/4,-y+3/4
59 -z+1/4,x+1/2,-y+3/4
60 -z+1/4,-x+3/4,y+1/2
61 -y,-x+1/2,-z+1/2
62 -y,x+3/4,z+3/4
63 y+1/4,-x+1/2,z+3/4
64 y+1/4,x+3/4,-z+1/2
65 -x,-z+1/2,-y+1/2
66 -x,z+3/4,y+3/4
67 x+1/4,-z+1/2,y+3/4
68 x+1/4,z+3/4,-y+1/2
69 -z,-y+1/2,-x+1/2
70 -z,y+3/4,x+3/4
71 z+1/4,-y+1/2,x+3/4
72 z+1/4,y+3/4,-x+1/2
73 -x,-y+1/2,-z+1/2
74 -x,y+3/4,z+3/4
75 x+1/4,-y+1/2,z+3/4
76 x+1/4,y+3/4,-z+1/2
77 -y,-z+1/2,-x+1/2
78 -y,z+3/4,x+3/4
79 y+1/4,-z+1/2,x+3/4
80 y+1/4,z+3/4,-x+1/2
81 -z,-x+1/2,-y+1/2
82 -z,x+3/4,y+3/4
83 z+1/4,-x+1/2,y+3/4
84 z+1/4,x+3/4,-y+1/2
85 y,x+1/2,z+1/2
86 y,-x+3/4,-z+3/4
87 -y+1/4,x+1/2,-z+3/4
88 -y+1/4,-x+3/4,z+1/2
89 x,z+1/2,y+1/2
90 x,-z+3/4,-y+3/4
91 -x+1/4,z+1/2,-y+3/4
92 -x+1/4,-z+3/4,y+1/2
93 z,y+1/2,x+1/2
94 z,-y+3/4,-x+3/4
95 -z+1/4,y+1/2,-x+3/4
96 -z+1/4,-y+3/4,x+1/2
97 x+1/2,y,z+1/2
98 x+1/2,-y+1/4,-z+3/4
99 -x+3/4,y,-z+3/4
100 -x+3/4,-y+1/4,z+1/2
101 y+1/2,z,x+1/2
102 y+1/2,-z+1/4,-x+3/4
```

```
103 -y+3/4,z,-x+3/4
104 -y+3/4,-z+1/4,x+1/2
105 z+1/2,x,y+1/2
106 z+1/2,-x+1/4,-y+3/4
107 -z+3/4,x,-y+3/4
108 -z+3/4,-x+1/4,y+1/2
109 -y+1/2,-x,-z+1/2
110 -y+1/2,x+1/4,z+3/4
111 y+3/4,-x,z+3/4
112 y+3/4,x+1/4,-z+1/2
113 -x+1/2,-z,-y+1/2
114 -x+1/2,z+1/4,y+3/4
115 x+3/4,-z,y+3/4
116 x+3/4,z+1/4,-y+1/2
117 -z+1/2,-y,-x+1/2
118 -z+1/2,y+1/4,x+3/4
119 z+3/4,-y,x+3/4
120 z+3/4,y+1/4,-x+1/2
121 -x+1/2,-y,-z+1/2
122 -x+1/2,y+1/4,z+3/4
123 x+3/4,-y,z+3/4
124 x+3/4,y+1/4,-z+1/2
125 -y+1/2,-z,-x+1/2
126 -y+1/2,z+1/4,x+3/4
127 y+3/4,-z,x+3/4
128 y+3/4,z+1/4,-x+1/2
129 -z+1/2,-x,-y+1/2
130 -z+1/2,x+1/4,y+3/4
131 z+3/4,-x,y+3/4
132 z+3/4,x+1/4,-y+1/2
133 y+1/2,x,z+1/2
134 y+1/2,-x+1/4,-z+3/4
135 -y+3/4,x,-z+3/4
136 -y+3/4,-x+1/4,z+1/2
137 x+1/2,z,y+1/2
138 x+1/2,-z+1/4,-y+3/4
139 -x+3/4,z,-y+3/4
140 -x+3/4,-z+1/4,y+1/2
141 z+1/2,y,x+1/2
142 z+1/2,-y+1/4,-x+3/4
143 -z+3/4,y,-x+3/4
144 -z+3/4,-y+1/4,x+1/2
145 x+1/2,y+1/2,z
146 x+1/2,-y+3/4,-z+1/4
147 -x+3/4,y+1/2,-z+1/4
148 -x+3/4,-y+3/4,z
149 y+1/2,z+1/2,x
150 y+1/2,-z+3/4,-x+1/4
151 -y+3/4,z+1/2,-x+1/4
152 -y+3/4,-z+3/4,x
153 z+1/2,x+1/2,y
154 z+1/2,-x+3/4,-y+1/4
155 -z+3/4,x+1/2,-y+1/4
156 -z+3/4,-x+3/4,y
157 -y+1/2,-x+1/2,-z
158 -y+1/2,x+3/4,z+1/4
159 y+3/4,-x+1/2,z+1/4
160 y+3/4,x+3/4,-z
161 -x+1/2,-z+1/2,-y
162 -x+1/2,z+3/4,y+1/4
163 x+3/4,-z+1/2,y+1/4
164 x+3/4,z+3/4,-y
165 -z+1/2,-y+1/2,-x
166 -z+1/2,y+3/4,x+1/4
167 z+3/4,-y+1/2,x+1/4
168 z+3/4,y+3/4,-x
169 -x+1/2,-y+1/2,-z
170 -x+1/2,y+3/4,z+1/4
171 x+3/4,-y+1/2,z+1/4
172 x+3/4,y+3/4,-z
173 -y+1/2,-z+1/2,-x
174 -y+1/2,z+3/4,x+1/4
175 y+3/4,-z+1/2,x+1/4
176 y+3/4,z+3/4,-x
177 -z+1/2,-x+1/2,-y
178 -z+1/2,x+3/4,y+1/4
179 z+3/4,-x+1/2,y+1/4
180 z+3/4,x+3/4,-y
181 y+1/2,x+1/2,z
182 y+1/2,-x+3/4,-z+1/4
183 -y+3/4,x+1/2,-z+1/4
184 -y+3/4,-x+3/4,z
185 x+1/2,z+1/2,y
186 x+1/2,-z+3/4,-y+1/4
187 -x+3/4,z+1/2,-y+1/4
188 -x+3/4,-z+3/4,y
189 z+1/2,y+1/2,x
190 z+1/2,-y+3/4,-x+1/4
191 -z+3/4,y+1/2,-x+1/4
192 -z+3/4,-y+3/4,x

loop_
_atom_site_label
_atom_site_type_symbol
_atom_site_symmetry_multiplicity
_atom_site_Wyckoff_label
_atom_site_fract_x
_atom_site_fract_y
_atom_site_fract_z
_atom_site_occupancy
Ti1 Ti  16 c 0.00000 0.00000 0.00000 1.00000
Ni1 Ni  32 e 0.21500 0.21500 0.21500 1.00000
Ti2 Ti  48 f 0.44000 0.12500 0.12500 1.00000
```

$NiTi_2$: AB2_cF96_227_e_cf - POSCAR

```
AB2_cF96_227_e_cf & a,x2,x3 --params=11.278,0.215,0.44 & Fd(-3)m    O_h^7
    ↪ #227 (cef) & cF96 & & NiTi2 & & G. A. Yurko, J. W. Barton and
    ↪ J. G. Parr, Acta Cyrst. 12, 909-911 (1959)
   1.0000000000000000
   0.00000000000000    5.63900000000000    5.63900000000000
   5.63900000000000    0.00000000000000    5.63900000000000
   5.63900000000000    5.63900000000000    0.00000000000000
   Ni    Ti
    8    16
Direct
   0.14500000000000   -0.21500000000000   -0.21500000000000   Ni   (32e)
  -0.14500000000000    0.21500000000000    0.21500000000000   Ni   (32e)
   0.21500000000000   -0.14500000000000    0.21500000000000   Ni   (32e)
  -0.21500000000000    0.14500000000000   -0.21500000000000   Ni   (32e)
   0.21500000000000    0.21500000000000   -0.14500000000000   Ni   (32e)
  -0.21500000000000   -0.21500000000000    0.14500000000000   Ni   (32e)
   0.21500000000000    0.21500000000000    0.21500000000000   Ni   (32e)
  -0.21500000000000   -0.21500000000000   -0.21500000000000   Ni   (32e)
   0.00000000000000    0.00000000000000    0.00000000000000   Ti   (16c)
   0.00000000000000    0.00000000000000    0.50000000000000   Ti   (16c)
   0.00000000000000    0.50000000000000    0.00000000000000   Ti   (16c)
   0.50000000000000    0.00000000000000    0.00000000000000   Ti   (16c)
  -0.19000000000000   -0.19000000000000    0.44000000000000   Ti   (48f)
   0.19000000000000    0.19000000000000    0.56000000000000   Ti   (48f)
  -0.19000000000000    0.44000000000000   -0.19000000000000   Ti   (48f)
  -0.19000000000000    0.44000000000000    0.44000000000000   Ti   (48f)
   0.19000000000000    0.56000000000000    0.19000000000000   Ti   (48f)
   0.19000000000000    0.56000000000000    0.56000000000000   Ti   (48f)
   0.44000000000000   -0.19000000000000   -0.19000000000000   Ti   (48f)
   0.44000000000000   -0.19000000000000    0.44000000000000   Ti   (48f)
   0.44000000000000    0.44000000000000   -0.19000000000000   Ti   (48f)
   0.56000000000000    0.19000000000000    0.19000000000000   Ti   (48f)
   0.56000000000000    0.19000000000000    0.56000000000000   Ti   (48f)
   0.56000000000000    0.56000000000000    0.19000000000000   Ti   (48f)
```

NaTl (B32): AB_cF16_227_a_b - CIF

```
# CIF file

data_findsym-output
_audit_creation_method FINDSYM

_chemical_name_mineral 'Zintl Phase'
_chemical_formula_sum 'Na Tl'

loop_
_publ_author_name
 'K. Kuriyama'
 'S. Saito'
 'K. Iwamura'
_journal_name_full
;
Journal of Physics and Chemistry of Solids
;
_journal_volume 40
_journal_year 1979
_journal_page_first 457
_journal_page_last 461
_publ_Section_title
;
 Ultrasonic study on the elastic moduli of the NaTl (B32) structure
;

# Found in  http://materials.springer.com/isp/crystallographic/docs/
    ↪ sd_0528644

_aflow_proto 'AB_cF16_227_a_b'
_aflow_params 'a'
_aflow_params_values '7.483'
_aflow_Strukturbericht 'B32'
_aflow_Pearson 'cF16'

_symmetry_space_group_name_Hall "-F 4vw 2vw 3 Fd(-3)m"
_symmetry_space_group_name_H-M "F d -3 m:2"
_symmetry_Int_Tables_number 227

_cell_length_a      7.48300
_cell_length_b      7.48300
_cell_length_c      7.48300
_cell_angle_alpha  90.00000
_cell_angle_beta   90.00000
_cell_angle_gamma  90.00000

loop_
_space_group_symop_id
_space_group_symop_operation_xyz
1 x,y,z
2 x,-y+1/4,-z+1/4
3 -x+1/4,y,-z+1/4
4 -x+1/4,-y+1/4,z
5 y,z,x
6 y,-z+1/4,-x+1/4
7 -y+1/4,z,-x+1/4
8 -y+1/4,-z+1/4,x
9 z,x,y
10 z,-x+1/4,-y+1/4
11 -z+1/4,x,-y+1/4
12 -z+1/4,-x+1/4,y
13 -y,-x,-z
14 -y,x+1/4,z+1/4
15 y+1/4,-x,z+1/4
16 y+1/4,x+1/4,-z
17 -x,-z,-y
18 -x,z+1/4,y+1/4
19 x+1/4,-z,y+1/4
```

```
20 x+1/4,z+1/4,-y
21 -z,-y,-x
22 -z,y+1/4,x+1/4
23 z+1/4,-y,x+1/4
24 z+1/4,y+1/4,-x
25 -x,-y,-z
26 -x,y+1/4,z+1/4
27 x+1/4,-y,z+1/4
28 x+1/4,y+1/4,-z
29 -y,-z,-x
30 -y,z+1/4,x+1/4
31 y+1/4,-z,x+1/4
32 y+1/4,z+1/4,-x
33 -z,-x,-y
34 -z,x+1/4,y+1/4
35 z+1/4,-x,y+1/4
36 z+1/4,x+1/4,-y
37 y,x,z
38 y,-x+1/4,-z+1/4
39 -y+1/4,x,-z+1/4
40 -y+1/4,-x+1/4,z
41 x,z,y
42 x,-z+1/4,-y+1/4
43 -x+1/4,z,-y+1/4
44 -x+1/4,-z+1/4,y
45 z,y,x
46 z,-y+1/4,-x+1/4
47 -z+1/4,y,-x+1/4
48 -z+1/4,-y+1/4,x
49 x,y+1/2,z+1/2
50 x,-y+3/4,-z+3/4
51 -x+1/4,y+1/2,-z+3/4
52 -x+1/4,-y+3/4,z+1/2
53 y,z+1/2,x+1/2
54 y,-z+3/4,-x+3/4
55 -y+1/4,z+1/2,-x+3/4
56 -y+1/4,-z+3/4,x+1/2
57 z,x+1/2,y+1/2
58 z,-x+3/4,-y+3/4
59 -z+1/4,x+1/2,-y+3/4
60 -z+1/4,-x+3/4,y+1/2
61 -y,-x+1/2,-z+1/2
62 -y,x+3/4,z+3/4
63 y+1/4,-x+1/2,z+3/4
64 y+1/4,x+3/4,-z+1/2
65 -x,-z+1/2,-y+1/2
66 -x,z+3/4,y+3/4
67 x+1/4,-z+1/2,y+3/4
68 x+1/4,z+3/4,-y+1/2
69 -z,-y+1/2,-x+1/2
70 -z,y+3/4,x+3/4
71 z+1/4,-y+1/2,x+3/4
72 z+1/4,y+3/4,-x+1/2
73 -x,-y+1/2,-z+1/2
74 -x,y+3/4,z+3/4
75 x+1/4,-y+1/2,z+3/4
76 x+1/4,y+3/4,-z+1/2
77 -y,-z+1/2,-x+1/2
78 -y,z+3/4,x+3/4
79 y+1/4,-z+1/2,x+3/4
80 y+1/4,z+3/4,-x+1/2
81 -z,-x+1/2,-y+1/2
82 -z,x+3/4,y+3/4
83 z+1/4,-x+1/2,y+3/4
84 z+1/4,x+3/4,-y+1/2
85 y,x+1/2,z+1/2
86 y,-x+3/4,-z+3/4
87 -y+1/4,x+1/2,-z+3/4
88 -y+1/4,-x+3/4,z+1/2
89 x,z+1/2,y+1/2
90 x,-z+3/4,-y+3/4
91 -x+1/4,z+1/2,-y+3/4
92 -x+1/4,-z+3/4,y+1/2
93 z,y+1/2,x+1/2
94 z,-y+3/4,-x+3/4
95 -z+1/4,y+1/2,-x+3/4
96 -z+1/4,-y+3/4,x+1/2
97 x+1/2,y,z+1/2
98 x+1/2,-y+1/4,-z+3/4
99 -x+3/4,y,-z+3/4
100 -x+3/4,-y+1/4,z+1/2
101 y+1/2,z,x+1/2
102 y+1/2,-z+1/4,-x+3/4
103 -y+3/4,z,-x+3/4
104 -y+3/4,-z+1/4,x+1/2
105 z+1/2,x,y+1/2
106 z+1/2,-x+1/4,-y+3/4
107 -z+3/4,x,-y+3/4
108 -z+3/4,-x+1/4,y+1/2
109 -y+1/2,-x,-z+1/2
110 -y+1/2,x+1/4,z+3/4
111 y+3/4,-x,z+3/4
112 y+3/4,x+1/4,-z+1/2
113 -x+1/2,-z,-y+1/2
114 -x+1/2,z+1/4,y+3/4
115 x+3/4,-z,y+3/4
116 x+3/4,z+1/4,-y+1/2
117 -z+1/2,-y,-x+1/2
118 -z+1/2,y+1/4,x+3/4
119 z+3/4,-y,x+3/4
120 z+3/4,y+1/4,-x+1/2
121 -x+1/2,-y,-z+1/2
122 -x+1/2,y+1/4,z+3/4
123 x+3/4,-y,z+3/4
124 x+3/4,y+1/4,-z+1/2
125 -y+1/2,-z,-x+1/2
126 -y+1/2,z+1/4,x+3/4
127 y+3/4,-z,x+3/4
128 y+3/4,z+1/4,-x+1/2
129 -z+1/2,-x,-y+1/2
130 -z+1/2,x+1/4,y+3/4
131 z+3/4,-x,y+3/4
132 z+3/4,x+1/4,-y+1/2
133 y+1/2,x,z+1/2
134 y+1/2,-x+1/4,-z+3/4
135 -y+3/4,x,-z+3/4
136 -y+3/4,-x+1/4,z+1/2
137 x+1/2,z,y+1/2
138 x+1/2,-z+1/4,-y+3/4
139 -x+3/4,z,-y+3/4
140 -x+3/4,-z+1/4,y+1/2
141 z+1/2,y,x+1/2
142 z+1/2,-y+1/4,-x+3/4
143 -z+3/4,y,-x+3/4
144 -z+3/4,-y+1/4,x+1/2
145 x+1/2,y+1/2,z
146 x+1/2,-y+3/4,-z+1/4
147 -x+3/4,y+1/2,-z+1/4
148 -x+3/4,-y+3/4,z
149 y+1/2,z+1/2,x
150 y+1/2,-z+3/4,-x+1/4
151 -y+3/4,z+1/2,-x+1/4
152 -y+3/4,-z+3/4,x
153 z+1/2,x+1/2,y
154 z+1/2,-x+3/4,-y+1/4
155 -z+3/4,x+1/2,-y+1/4
156 -z+3/4,-x+3/4,y
157 -y+1/2,-x+1/2,-z
158 -y+1/2,x+3/4,z+1/4
159 y+3/4,-x+1/2,z+1/4
160 y+3/4,x+3/4,-z
161 -x+1/2,-z+1/2,-y
162 -x+1/2,z+3/4,y+1/4
163 x+3/4,-z+1/2,y+1/4
164 x+3/4,z+3/4,-y
165 -z+1/2,-y+1/2,-x
166 -z+1/2,y+3/4,x+1/4
167 z+3/4,-y+1/2,x+1/4
168 z+3/4,y+3/4,-x
169 -x+1/2,-y+1/2,-z
170 -x+1/2,y+3/4,z+1/4
171 x+3/4,-y+1/2,z+1/4
172 x+3/4,y+3/4,-z
173 -y+1/2,-z+1/2,-x
174 -y+1/2,z+3/4,x+1/4
175 y+3/4,-z+1/2,x+1/4
176 y+3/4,z+3/4,-x
177 -z+1/2,-x+1/2,-y
178 -z+1/2,x+3/4,y+1/4
179 z+3/4,-x+1/2,y+1/4
180 z+3/4,x+3/4,-y
181 y+1/2,x+1/2,z
182 y+1/2,-x+3/4,-z+1/4
183 -y+3/4,x+1/2,-z+1/4
184 -y+3/4,-x+3/4,z
185 x+1/2,z+1/2,y
186 x+1/2,-z+3/4,-y+1/4
187 -x+3/4,z+1/2,-y+1/4
188 -x+3/4,-z+3/4,y
189 z+1/2,y+1/2,x
190 z+1/2,-y+3/4,-x+1/4
191 -z+3/4,y+1/2,-x+1/4
192 -z+3/4,-y+3/4,x

loop_
_atom_site_label
_atom_site_type_symbol
_atom_site_symmetry_multiplicity
_atom_site_Wyckoff_label
_atom_site_fract_x
_atom_site_fract_y
_atom_site_fract_z
_atom_site_occupancy
Na1 Na   8 a 0.12500 0.12500 0.12500 1.00000
Tl1 Tl   8 b 0.37500 0.37500 0.37500 1.00000
```

NaTl (B32): AB_cF16_227_a_b - POSCAR

```
AB_cF16_227_a_b & a --params=7.483 & Fd(-3)m    O_h^7 #227 (ab) & cF16 &
    ↪ B32 & NaTl & Zintl Phase & K. Kuriyama and S. Saito and K.
    ↪ Iwamura, J. Phys. Chem. Solids 40, 457-461 (1979)
   1.0000000000000000
   0.00000000000000    3.74150000000000    3.74150000000000
   3.74150000000000    0.00000000000000    3.74150000000000
   3.74150000000000    3.74150000000000    0.00000000000000
   Na    Tl
    2     2
Direct
   0.12500000000000    0.12500000000000    0.12500000000000    Na    (8a)
   0.87500000000000    0.87500000000000    0.87500000000000    Na    (8a)
   0.37500000000000    0.37500000000000    0.37500000000000    Tl    (8b)
   0.62500000000000    0.62500000000000    0.62500000000000    Tl    (8b)
```

$Si_{34}$ Clathrate: A_cF136_227_aeg - CIF

```
# CIF file

data_findsym-output
_audit_creation_method FINDSYM
```

```
_chemical_name_mineral 'Clathrate'
_chemical_formula_sum 'Si'

loop_
_publ_author_name
 'Gary B. Adams'
 'Michael O'Keeffe'
 'Alexander A. Demkov'
 'Otto F. Sankey'
 'Yin-Min Huang'
_journal_name_full
;
Physical Review B
;
_journal_volume 49
_journal_year 1994
_journal_page_first 8048
_journal_page_last 8053
_publ_Section_title
;
 Wide-band-gap Si in open fourfold-coordinated clathrate structures
;

_aflow_proto 'A_cF136_227_aeg'
_aflow_params 'a,x2,x3,z3'
_aflow_params_values '14.864,0.2624,0.1824,0.3701'
_aflow_Strukturbericht 'None'
_aflow_Pearson 'cF136'

_symmetry_space_group_name_Hall "-F 4vw 2vw 3 Fd(-3)m"
_symmetry_space_group_name_H-M "F d -3 m:2"
_symmetry_Int_Tables_number 227

_cell_length_a     14.86400
_cell_length_b     14.86400
_cell_length_c     14.86400
_cell_angle_alpha  90.00000
_cell_angle_beta   90.00000
_cell_angle_gamma  90.00000

loop_
_space_group_symop_id
_space_group_symop_operation_xyz
1 x,y,z
2 x,-y+1/4,-z+1/4
3 -x+1/4,y,-z+1/4
4 -x+1/4,-y+1/4,z
5 y,z,x
6 y,-z+1/4,-x+1/4
7 -y+1/4,z,-x+1/4
8 -y+1/4,-z+1/4,x
9 z,x,y
10 z,-x+1/4,-y+1/4
11 -z+1/4,x,-y+1/4
12 -z+1/4,-x+1/4,y
13 -y,-x,-z
14 -y,x+1/4,z+1/4
15 y+1/4,-x,z+1/4
16 y+1/4,x+1/4,-z
17 -x,-z,-y
18 -x,z+1/4,y+1/4
19 x+1/4,-z,y+1/4
20 x+1/4,z+1/4,-y
21 -z,-y,-x
22 -z,y+1/4,x+1/4
23 z+1/4,-y,x+1/4
24 z+1/4,y+1/4,-x
25 -x,-y,-z
26 -x,y+1/4,z+1/4
27 x+1/4,-y,z+1/4
28 x+1/4,y+1/4,-z
29 -y,-z,-x
30 -y,z+1/4,x+1/4
31 y+1/4,-z,x+1/4
32 y+1/4,z+1/4,-x
33 -z,-x,-y
34 -z,x+1/4,y+1/4
35 z+1/4,-x,y+1/4
36 z+1/4,x+1/4,-y
37 y,x,z
38 y,-x+1/4,-z+1/4
39 -y+1/4,x,-z+1/4
40 -y+1/4,-x+1/4,z
41 x,z,y
42 x,-z+1/4,-y+1/4
43 -x+1/4,z,-y+1/4
44 -x+1/4,-z+1/4,y
45 z,y,x
46 z,-y+1/4,-x+1/4
47 -z+1/4,y,-x+1/4
48 -z+1/4,-y+1/4,x
49 x,y+1/2,z+1/2
50 x,-y+3/4,-z+3/4
51 -x+1/4,y+1/2,-z+3/4
52 -x+1/4,-y+3/4,z+1/2
53 y,z+1/2,x+1/2
54 y,-z+3/4,-x+3/4
55 -y+1/4,z+1/2,-x+3/4
56 -y+1/4,-z+3/4,x+1/2
57 z,x+1/2,y+1/2
58 z,-x+3/4,-y+3/4
59 -z+1/4,x+1/2,-y+3/4
60 -z+1/4,-x+3/4,y+1/2
61 -y,-x+1/2,-z+1/2
62 -y,x+3/4,z+3/4
63 y+1/4,-x+1/2,z+3/4
64 y+1/4,x+3/4,-z+1/2
65 -x,-z+1/2,-y+1/2
66 -x,z+3/4,y+3/4
67 x+1/4,-z+1/2,y+3/4
68 x+1/4,z+3/4,-y+1/2
69 -z,-y+1/2,-x+1/2
70 -z,y+3/4,x+3/4
71 z+1/4,-y+1/2,x+3/4
72 z+1/4,y+3/4,-x+1/2
73 -x,-y+1/2,-z+1/2
74 -x,y+3/4,z+3/4
75 x+1/4,-y+1/2,z+3/4
76 x+1/4,y+3/4,-z+1/2
77 -y,-z+1/2,-x+1/2
78 -y,z+3/4,x+3/4
79 y+1/4,-z+1/2,x+3/4
80 y+1/4,z+3/4,-x+1/2
81 -z,-x+1/2,-y+1/2
82 -z,x+3/4,y+3/4
83 z+1/4,-x+1/2,y+3/4
84 z+1/4,x+3/4,-y+1/2
85 y,x+1/2,z+1/2
86 y,-x+3/4,-z+3/4
87 -y+1/4,x+1/2,-z+3/4
88 -y+1/4,-x+3/4,z+1/2
89 x,z+1/2,y+1/2
90 x,-z+3/4,-y+3/4
91 -x+1/4,z+1/2,-y+3/4
92 -x+1/4,-z+3/4,y+1/2
93 z,y+1/2,x+1/2
94 z,-y+3/4,-x+3/4
95 -z+1/4,y+1/2,-x+3/4
96 -z+1/4,-y+3/4,x+1/2
97 x+1/2,y,z+1/2
98 x+1/2,-y+1/4,-z+3/4
99 -x+3/4,y,-z+3/4
100 -x+3/4,-y+1/4,z+1/2
101 y+1/2,z,x+1/2
102 y+1/2,-z+1/4,-x+3/4
103 -y+3/4,z,-x+3/4
104 -y+3/4,-z+1/4,x+1/2
105 z+1/2,x,y+1/2
106 z+1/2,-x+1/4,-y+3/4
107 -z+3/4,x,-y+3/4
108 -z+3/4,-x+1/4,y+1/2
109 -y+1/2,-x,-z+1/2
110 -y+1/2,x+1/4,z+3/4
111 y+3/4,-x,z+3/4
112 y+3/4,x+1/4,-z+1/2
113 -x+1/2,-z,-y+1/2
114 -x+1/2,z+1/4,y+3/4
115 x+3/4,-z,y+3/4
116 x+3/4,z+1/4,-y+1/2
117 -z+1/2,-y,-x+1/2
118 -z+1/2,y+1/4,x+3/4
119 z+3/4,-y,x+3/4
120 z+3/4,y+1/4,-x+1/2
121 -x+1/2,-y,-z+1/2
122 -x+1/2,y+1/4,z+3/4
123 x+3/4,-y,z+3/4
124 x+3/4,y+1/4,-z+1/2
125 -y+1/2,-z,-x+1/2
126 -y+1/2,z+1/4,x+3/4
127 y+3/4,-z,x+3/4
128 y+3/4,z+1/4,-x+1/2
129 -z+1/2,-x,-y+1/2
130 -z+1/2,x+1/4,y+3/4
131 z+3/4,-x,y+3/4
132 z+3/4,x+1/4,-y+1/2
133 y+1/2,x,z+1/2
134 y+1/2,-x+1/4,-z+3/4
135 -y+3/4,x,-z+3/4
136 -y+3/4,-x+1/4,z+1/2
137 x+1/2,z,y+1/2
138 x+1/2,-z+1/4,-y+3/4
139 -x+3/4,z,-y+3/4
140 -x+3/4,-z+1/4,y+1/2
141 z+1/2,y,x+1/2
142 z+1/2,-y+1/4,-x+3/4
143 -z+3/4,y,-x+3/4
144 -z+3/4,-y+1/4,x+1/2
145 x+1/2,y+1/2,z
146 x+1/2,-y+3/4,-z+1/4
147 -x+3/4,y+1/2,-z+1/4
148 -x+3/4,-y+3/4,z
149 y+1/2,z+1/2,x
150 y+1/2,-z+3/4,-x+1/4
151 -y+3/4,z+1/2,-x+1/4
152 -y+3/4,-z+3/4,x
153 z+1/2,x+1/2,y
154 z+1/2,-x+3/4,-y+1/4
155 -z+3/4,x+1/2,-y+1/4
156 -z+3/4,-x+3/4,y
157 -y+1/2,-x+1/2,-z
158 -y+1/2,x+3/4,z+1/4
159 y+3/4,-x+1/2,z+1/4
160 y+3/4,x+3/4,-z
161 -x+1/2,-z+1/2,-y
162 -x+1/2,z+3/4,y+1/4
163 x+3/4,-z+1/2,y+1/4
164 x+3/4,z+3/4,-y
165 -z+1/2,-y+1/2,-x
166 -z+1/2,y+3/4,x+1/4
167 z+3/4,-y+1/2,x+1/4
```

```
168 z+3/4,y+3/4,-x
169 -x+1/2,-y+1/2,-z
170 -x+1/2,y+3/4,z+1/4
171 x+3/4,-y+1/2,z+1/4
172 x+3/4,y+3/4,-z
173 -y+1/2,-z+1/2,-x
174 -y+1/2,z+3/4,x+1/4
175 y+3/4,-z+1/2,x+1/4
176 y+3/4,z+3/4,-x
177 -z+1/2,-x+1/2,-y
178 -z+1/2,x+3/4,y+1/4
179 z+3/4,-x+1/2,y+1/4
180 z+3/4,x+3/4,-y
181 y+1/2,x+1/2,z
182 y+1/2,-x+3/4,-z+1/4
183 -y+3/4,x+1/2,-z+1/4
184 -y+3/4,-x+3/4,z
185 x+1/2,z+1/2,y
186 x+1/2,-z+3/4,-y+1/4
187 -x+3/4,z+1/2,-y+1/4
188 -x+3/4,-z+3/4,y
189 z+1/2,y+1/2,x
190 z+1/2,-y+3/4,-x+1/4
191 -z+3/4,y+1/2,-x+1/4
192 -z+3/4,-y+3/4,x

loop_
_atom_site_label
_atom_site_type_symbol
_atom_site_symmetry_multiplicity
_atom_site_Wyckoff_label
_atom_site_fract_x
_atom_site_fract_y
_atom_site_fract_z
_atom_site_occupancy
Si1  Si    8 a 0.12500  0.12500  0.12500  1.00000
Si2  Si   32 e 0.26240  0.26240  0.26240  1.00000
Si3  Si   96 g 0.18240  0.18240  0.37010  1.00000
```

$Si_{34}$ Clathrate: A_cF136_227_aeg - POSCAR

```
A_cF136_227_aeg & a,x2,x3,z3 --params=14.864,0.2624,0.1824,0.3701 & Fd(-
   ↪ 3)m    O_h^7 #227 (aeg) & cF136 & & Si & Clathrate & G. B. Adams
   ↪ , M. O'Keefe, A. A. Demkov, O. F. Sankey and Y.-M. Huang, Phys.
   ↪ Rev. B 49, 8048-8053 (1994)
   1.0000000000000000
   0.00000000000000    7.43200000000000    7.43200000000000
   7.43200000000000    0.00000000000000    7.43200000000000
   7.43200000000000    7.43200000000000    0.00000000000000
   Si
   34
Direct
   0.26240000000000    0.26240000000000    0.26240000000000    Si   (32e)
   0.26240000000000    0.26240000000000    0.71280000000000    Si   (32e)
  -0.26240000000000    0.28720000000000    0.73760000000000    Si   (32e)
   0.26240000000000    0.71280000000000    0.26240000000000    Si   (32e)
   0.28720000000000    0.73760000000000   -0.26240000000000    Si   (32e)
   0.71280000000000    0.26240000000000    0.26240000000000    Si   (32e)
   0.73760000000000   -0.26240000000000    0.28720000000000    Si   (32e)
   0.73760000000000    0.73760000000000    0.73760000000000    Si   (32e)
   0.12500000000000    0.12500000000000    0.12500000000000    Si    (8a)
   0.87500000000000    0.87500000000000    0.87500000000000    Si    (8a)
   0.00530000000000    0.23490000000000    0.62990000000000    Si   (96g)
   0.00530000000000   -0.37010000000000    0.23490000000000    Si   (96g)
  -0.00530000000000    0.37010000000000    0.37010000000000    Si   (96g)
  -0.00530000000000    0.37010000000000    0.76510000000000    Si   (96g)
   0.00530000000000    0.62990000000000    0.62990000000000    Si   (96g)
   0.23490000000000    0.00530000000000   -0.37010000000000    Si   (96g)
   0.23490000000000    0.62990000000000    0.00530000000000    Si   (96g)
   0.23490000000000    0.62990000000000   -0.37010000000000    Si   (96g)
   0.37010000000000   -0.00530000000000    0.37010000000000    Si   (96g)
  -0.37010000000000    0.23490000000000    0.00530000000000    Si   (96g)
  -0.37010000000000    0.23490000000000    0.62990000000000    Si   (96g)
   0.37010000000000    0.37010000000000   -0.00530000000000    Si   (96g)
   0.37010000000000    0.37010000000000    0.76510000000000    Si   (96g)
   0.37010000000000    0.76510000000000   -0.00530000000000    Si   (96g)
   0.37010000000000    0.76510000000000    0.37010000000000    Si   (96g)
   0.37010000000000    0.99470000000000    0.76510000000000    Si   (96g)
   0.62990000000000    0.00530000000000    0.23490000000000    Si   (96g)
   0.62990000000000    0.00530000000000    0.62990000000000    Si   (96g)
   0.62990000000000   -0.37010000000000    0.23490000000000    Si   (96g)
   0.62990000000000    0.62990000000000    0.00530000000000    Si   (96g)
   0.76510000000000   -0.00530000000000    0.37010000000000    Si   (96g)
   0.76510000000000    0.37010000000000    0.37010000000000    Si   (96g)
   0.76510000000000    0.37010000000000    0.99470000000000    Si   (96g)
   0.99470000000000    0.76510000000000    0.37010000000000    Si   (96g)
```

$Cu_2Mg$ Cubic Laves (C15): A2B_cF24_227_d_a - CIF

```
# CIF file

data_findsym-output
_audit_creation_method FINDSYM

_chemical_name_mineral 'Cubic Laves'
_chemical_formula_sum 'Cu2 Mg'

loop_
_publ_author_name
 'James B. Friauf'
_journal_name_full
;
Journal of the American Chemical Society
;
_journal_volume 49
_journal_year 1927
_journal_page_first 3107
_journal_page_last 3114
_publ_Section_title
;
 The Crystal Structures of Two Intermetallic Compounds
;

# Found in  Wyckoff, Vol. I, pp. 365-367

_aflow_proto 'A2B_cF24_227_d_a'
_aflow_params 'a'
_aflow_params_values '7.02'
_aflow_Strukturbericht 'C15'
_aflow_Pearson 'cF24'

_symmetry_space_group_name_Hall "-F 4vw 2vw 3 Fd(-3)m"
_symmetry_space_group_name_H-M "F d -3 m:2"
_symmetry_Int_Tables_number 227

_cell_length_a      7.02000
_cell_length_b      7.02000
_cell_length_c      7.02000
_cell_angle_alpha  90.00000
_cell_angle_beta   90.00000
_cell_angle_gamma  90.00000

loop_
_space_group_symop_id
_space_group_symop_operation_xyz
1 x,y,z
2 x,-y+1/4,-z+1/4
3 -x+1/4,y,-z+1/4
4 -x+1/4,-y+1/4,z
5 y,z,x
6 y,-z+1/4,-x+1/4
7 -y+1/4,z,-x+1/4
8 -y+1/4,-z+1/4,x
9 z,x,y
10 z,-x+1/4,-y+1/4
11 -z+1/4,x,-y+1/4
12 -z+1/4,-x+1/4,y
13 -y,-x,-z
14 -y,x+1/4,z+1/4
15 y+1/4,-x,z+1/4
16 y+1/4,x+1/4,-z
17 -x,-z,-y
18 -x,z+1/4,y+1/4
19 x+1/4,-z,y+1/4
20 x+1/4,z+1/4,-y
21 -z,-y,-x
22 -z,y+1/4,x+1/4
23 z+1/4,-y,x+1/4
24 z+1/4,y+1/4,-x
25 -x,-y,-z
26 -x,y+1/4,z+1/4
27 x+1/4,-y,z+1/4
28 x+1/4,y+1/4,-z
29 -y,-z,-x
30 -y,z+1/4,x+1/4
31 y+1/4,-z,x+1/4
32 y+1/4,z+1/4,-x
33 -z,-x,-y
34 -z,x+1/4,y+1/4
35 z+1/4,-x,y+1/4
36 z+1/4,x+1/4,-y
37 y,x,z
38 y,-x+1/4,-z+1/4
39 -y+1/4,x,-z+1/4
40 -y+1/4,-x+1/4,z
41 x,z,y
42 x,-z+1/4,-y+1/4
43 -x+1/4,z,-y+1/4
44 -x+1/4,-z+1/4,y
45 z,y,x
46 z,-y+1/4,-x+1/4
47 -z+1/4,y,-x+1/4
48 -z+1/4,-y+1/4,x
49 x,y+1/2,z+1/2
50 x,-y+3/4,-z+3/4
51 -x+1/4,y+1/2,-z+3/4
52 -x+1/4,-y+3/4,z+1/2
53 y,z+1/2,x+1/2
54 y,-z+3/4,-x+3/4
55 -y+1/4,z+1/2,-x+3/4
56 -y+1/4,-z+3/4,x+1/2
57 z,x+1/2,y+1/2
58 z,-x+3/4,-y+3/4
59 -z+1/4,x+1/2,-y+3/4
60 -z+1/4,-x+3/4,y+1/2
61 -y,-x+1/2,-z+1/2
62 -y,x+3/4,z+3/4
63 y+1/4,-x+1/2,z+3/4
64 y+1/4,x+3/4,-z+1/2
65 -x,-z+1/2,-y+1/2
66 -x,z+3/4,y+3/4
67 x+1/4,-z+1/2,y+3/4
68 x+1/4,z+3/4,-y+1/2
69 -z,-y+1/2,-x+1/2
70 -z,y+3/4,x+3/4
71 z+1/4,-y+1/2,x+3/4
72 z+1/4,y+3/4,-x+1/2
73 -x,-y+1/2,-z+1/2
74 -x,y+3/4,z+3/4
75 x+1/4,-y+1/2,z+3/4
```

```
76 x+1/4,y+3/4,-z+1/2
77 -y,-z+1/2,-x+1/2
78 -y,z+3/4,x+3/4
79 y+1/4,-z+1/2,x+3/4
80 y+1/4,z+3/4,-x+1/2
81 -z,-x+1/2,-y+1/2
82 -z,x+3/4,y+3/4
83 z+1/4,-x+1/2,y+3/4
84 z+1/4,x+3/4,-y+1/2
85 y,x+1/2,z+1/2
86 y,-x+3/4,-z+3/4
87 -y+1/4,x+1/2,-z+3/4
88 -y+1/4,-x+3/4,z+1/2
89 x,z+1/2,y+1/2
90 x,-z+3/4,-y+3/4
91 -x+1/4,z+1/2,-y+3/4
92 -x+1/4,-z+3/4,y+1/2
93 z,y+1/2,x+1/2
94 z,-y+3/4,-x+3/4
95 -z+1/4,y+1/2,-x+3/4
96 -z+1/4,-y+3/4,x+1/2
97 x+1/2,y,z+1/2
98 x+1/2,-y+1/4,-z+3/4
99 -x+3/4,y,-z+3/4
100 -x+3/4,-y+1/4,z+1/2
101 y+1/2,z,x+1/2
102 y+1/2,-z+1/4,-x+3/4
103 -y+3/4,z,-x+3/4
104 -y+3/4,-z+1/4,x+1/2
105 z+1/2,x,y+1/2
106 z+1/2,-x+1/4,-y+3/4
107 -z+3/4,x,-y+3/4
108 -z+3/4,-x+1/4,y+1/2
109 -y+1/2,-x,-z+1/2
110 -y+1/2,x+1/4,z+3/4
111 y+3/4,-x,z+3/4
112 y+3/4,x+1/4,-z+1/2
113 -x+1/2,-z,-y+1/2
114 -x+1/2,z+1/4,y+3/4
115 x+3/4,-z,y+3/4
116 x+3/4,z+1/4,-y+1/2
117 -z+1/2,-y,-x+1/2
118 -z+1/2,y+1/4,x+3/4
119 z+3/4,-y,x+3/4
120 z+3/4,y+1/4,-x+1/2
121 -x+1/2,-y,-z+1/2
122 -x+1/2,y+1/4,z+3/4
123 x+3/4,-y,z+3/4
124 x+3/4,y+1/4,-z+1/2
125 -y+1/2,-z,-x+1/2
126 -y+1/2,z+1/4,x+3/4
127 y+3/4,-z,x+3/4
128 y+3/4,z+1/4,-x+1/2
129 -z+1/2,-x,-y+1/2
130 -z+1/2,x+1/4,y+3/4
131 z+3/4,-x,y+3/4
132 z+3/4,x+1/4,-y+1/2
133 y+1/2,x,z+1/2
134 y+1/2,-x+1/4,-z+3/4
135 -y+3/4,x,-z+3/4
136 -y+3/4,-x+1/4,z+1/2
137 x+1/2,z,y+1/2
138 x+1/2,-z+1/4,-y+3/4
139 -x+3/4,z,-y+3/4
140 -x+3/4,-z+1/4,y+1/2
141 z+1/2,y,x+1/2
142 z+1/2,-y+1/4,-x+3/4
143 -z+3/4,y,-x+3/4
144 -z+3/4,-y+1/4,x+1/2
145 x+1/2,y+1/2,z
146 x+1/2,-y+3/4,-z+1/4
147 -x+3/4,y+1/2,-z+1/4
148 -x+3/4,-y+3/4,z
149 y+1/2,z+1/2,x
150 y+1/2,-z+3/4,-x+1/4
151 -y+3/4,z+1/2,-x+1/4
152 -y+3/4,-z+3/4,x
153 z+1/2,x+1/2,y
154 z+1/2,-x+3/4,-y+1/4
155 -z+3/4,x+1/2,-y+1/4
156 -z+3/4,-x+3/4,y
157 -y+1/2,-x+1/2,-z
158 -y+1/2,x+3/4,z+1/4
159 y+3/4,-x+1/2,z+1/4
160 y+3/4,x+3/4,-z
161 -x+1/2,-z+1/2,-y
162 -x+1/2,z+3/4,y+1/4
163 x+3/4,-z+1/2,y+1/4
164 x+3/4,z+3/4,-y
165 -z+1/2,-y+1/2,-x
166 -z+1/2,y+3/4,x+1/4
167 z+3/4,-y+1/2,x+1/4
168 z+3/4,y+3/4,-x
169 -x+1/2,-y+1/2,-z
170 -x+1/2,y+3/4,z+1/4
171 x+3/4,-y+1/2,z+1/4
172 x+3/4,y+3/4,-z
173 -y+1/2,-z+1/2,-x
174 -y+1/2,z+3/4,x+1/4
175 y+3/4,-z+1/2,x+1/4
176 y+3/4,z+3/4,-x
177 -z+1/2,-x+1/2,-y
178 -z+1/2,x+3/4,y+1/4
179 z+3/4,-x+1/2,y+1/4
180 z+3/4,x+3/4,-y
181 y+1/2,x+1/2,z
182 y+1/2,-x+3/4,-z+1/4
183 -y+3/4,x+1/2,-z+1/4
184 -y+3/4,-x+3/4,z
185 x+1/2,z+1/2,y
186 x+1/2,-z+3/4,-y+1/4
187 -x+3/4,z+1/2,-y+1/4
188 -x+3/4,-z+3/4,y
189 z+1/2,y+1/2,x
190 z+1/2,-y+3/4,-x+1/4
191 -z+3/4,y+1/2,-x+1/4
192 -z+3/4,-y+3/4,x

loop_
_atom_site_label
_atom_site_type_symbol
_atom_site_symmetry_multiplicity
_atom_site_Wyckoff_label
_atom_site_fract_x
_atom_site_fract_y
_atom_site_fract_z
_atom_site_occupancy
Mg1 Mg   8 a 0.12500  0.12500  0.12500  1.00000
Cu1 Cu  16 d 0.50000  0.50000  0.50000  1.00000
```

$Cu_2Mg$ Cubic Laves (C15): A2B_cF24_227_d_a - POSCAR

```
A2B_cF24_227_d_a & a --params=7.02 & Fd(-3)m    O_h^7 #227 (ad) & cF24 &
    ↪ C15 & Cu2Mg & Cubic Laves & J. B. Friauf, J. Am. Chem. Soc. 49,
    ↪ 3107-3114 (1927)
   1.0000000000000000
   0.00000000000000    3.51000000000000    3.51000000000000
   3.51000000000000    0.00000000000000    3.51000000000000
   3.51000000000000    3.51000000000000    0.00000000000000
   Cu    Mg
    4     2
Direct
   0.00000000000000    0.50000000000000    0.50000000000000    Cu    (16d)
   0.50000000000000    0.00000000000000    0.50000000000000    Cu    (16d)
   0.50000000000000    0.50000000000000    0.00000000000000    Cu    (16d)
   0.50000000000000    0.50000000000000    0.50000000000000    Cu    (16d)
   0.12500000000000    0.12500000000000    0.12500000000000    Mg     (8a)
   0.87500000000000    0.87500000000000    0.87500000000000    Mg     (8a)
```

Diamond (A4): A_cF8_227_a - CIF

```
# CIF file

data_findsym-output
_audit_creation_method FINDSYM

_chemical_name_mineral 'diamond'
_chemical_formula_sum 'C'

loop_
_publ_author_name
 'W. H. Bragg'
 'W. L. Bragg'
_journal_name_full
;
Proceedings of the Royal Society of London, Series A
;
_journal_volume 89
_journal_year 1913
_journal_page_first 277
_journal_page_last 291
_publ_Section_title
;
 The Structure of Diamond
;

_aflow_proto 'A_cF8_227_a'
_aflow_params 'a'
_aflow_params_values '3.55'
_aflow_Strukturbericht 'A4'
_aflow_Pearson 'cF8'

_symmetry_space_group_name_Hall "-F 4vw 2vw 3 Fd(-3)m"
_symmetry_space_group_name_H-M "F d -3 m:2"
_symmetry_Int_Tables_number 227

_cell_length_a      3.55000
_cell_length_b      3.55000
_cell_length_c      3.55000
_cell_angle_alpha  90.00000
_cell_angle_beta   90.00000
_cell_angle_gamma  90.00000

loop_
_space_group_symop_id
_space_group_symop_operation_xyz
1 x,y,z
2 x,-y+1/4,-z+1/4
3 -x+1/4,y,-z+1/4
4 -x+1/4,-y+1/4,z
5 y,z,x
6 y,-z+1/4,-x+1/4
7 -y+1/4,z,-x+1/4
8 -y+1/4,-z+1/4,x
9 z,x,y
10 z,-x+1/4,-y+1/4
11 -z+1/4,x,-y+1/4
12 -z+1/4,-x+1/4,y
13 -y,-x,-z
14 -y,x+1/4,z+1/4
```

```
15 y+1/4,-x,z+1/4
16 y+1/4,x+1/4,-z
17 -x,-z,-y
18 -x,z+1/4,y+1/4
19 x+1/4,-z,y+1/4
20 x+1/4,z+1/4,-y
21 -z,-y,-x
22 -z,y+1/4,x+1/4
23 z+1/4,-y,x+1/4
24 z+1/4,y+1/4,-x
25 -x,-y,-z
26 -x,y+1/4,z+1/4
27 x+1/4,-y,z+1/4
28 x+1/4,y+1/4,-z
29 -y,-z,-x
30 -y,z+1/4,x+1/4
31 y+1/4,-z,x+1/4
32 y+1/4,z+1/4,-x
33 -z,-x,-y
34 -z,x+1/4,y+1/4
35 z+1/4,-x,y+1/4
36 z+1/4,x+1/4,-y
37 y,x,z
38 y,-x+1/4,-z+1/4
39 -y+1/4,x,-z+1/4
40 -y+1/4,-x+1/4,z
41 x,z,y
42 x,-z+1/4,-y+1/4
43 -x+1/4,z,-y+1/4
44 -x+1/4,-z+1/4,y
45 z,y,x
46 z,-y+1/4,-x+1/4
47 -z+1/4,y,-x+1/4
48 -z+1/4,-y+1/4,x
49 x,y+1/2,z+1/2
50 x,-y+3/4,-z+3/4
51 -x+1/4,y+1/2,-z+3/4
52 -x+1/4,-y+3/4,z+1/2
53 y,z+1/2,x+1/2
54 y,-z+3/4,-x+3/4
55 -y+1/4,z+1/2,-x+3/4
56 -y+1/4,-z+3/4,x+1/2
57 z,x+1/2,y+1/2
58 z,-x+3/4,-y+3/4
59 -z+1/4,x+1/2,-y+3/4
60 -z+1/4,-x+3/4,y+1/2
61 -y,-x+1/2,-z+1/2
62 -y,x+3/4,z+3/4
63 y+1/4,-x+1/2,z+3/4
64 y+1/4,x+3/4,-z+1/2
65 -x,-z+1/2,-y+1/2
66 -x,z+3/4,y+3/4
67 x+1/4,-z+1/2,y+3/4
68 x+1/4,z+3/4,-y+1/2
69 -z,-y+1/2,-x+1/2
70 -z,y+3/4,x+3/4
71 z+1/4,-y+1/2,x+3/4
72 z+1/4,y+3/4,-x+1/2
73 -x,-y+1/2,-z+1/2
74 -x,y+3/4,z+3/4
75 x+1/4,-y+1/2,z+3/4
76 x+1/4,y+3/4,-z+1/2
77 -y,-z+1/2,-x+1/2
78 -y,z+3/4,x+3/4
79 y+1/4,-z+1/2,x+3/4
80 y+1/4,z+3/4,-x+1/2
81 -z,-x+1/2,-y+1/2
82 -z,x+3/4,y+3/4
83 z+1/4,-x+1/2,y+3/4
84 z+1/4,x+3/4,-y+1/2
85 y,x+1/2,z+1/2
86 y,-x+3/4,-z+3/4
87 -y+1/4,x+1/2,-z+3/4
88 -y+1/4,-x+3/4,z+1/2
89 x,z+1/2,y+1/2
90 x,-z+3/4,-y+3/4
91 -x+1/4,z+1/2,-y+3/4
92 -x+1/4,-z+3/4,y+1/2
93 z,y+1/2,x+1/2
94 z,-y+3/4,-x+3/4
95 -z+1/4,y+1/2,-x+3/4
96 -z+1/4,-y+3/4,x+1/2
97 x+1/2,y,z+1/2
98 x+1/2,-y+1/4,-z+3/4
99 -x+3/4,y,-z+3/4
100 -x+3/4,-y+1/4,z+1/2
101 y+1/2,z,x+1/2
102 y+1/2,-z+1/4,-x+3/4
103 -y+3/4,z,-x+3/4
104 -y+3/4,-z+1/4,x+1/2
105 z+1/2,x,y+1/2
106 z+1/2,-x+1/4,-y+3/4
107 -z+3/4,x,-y+3/4
108 -z+3/4,-x+1/4,y+1/2
109 -y+1/2,-x,-z+1/2
110 -y+1/2,x+1/4,z+3/4
111 y+3/4,-x,z+3/4
112 y+3/4,x+1/4,-z+1/2
113 -x+1/2,-z,-y+1/2
114 -x+1/2,z+1/4,y+3/4
115 x+3/4,-z,y+3/4
116 x+3/4,z+1/4,-y+1/2
117 -z+1/2,-y,-x+1/2
118 -z+1/2,y+1/4,x+3/4
119 z+3/4,-y,x+3/4
120 z+3/4,y+1/4,-x+1/2
121 -x+1/2,-y,-z+1/2
122 -x+1/2,y+1/4,z+3/4
123 x+3/4,-y,z+3/4
124 x+3/4,y+1/4,-z+1/2
125 -y+1/2,-z,-x+1/2
126 -y+1/2,z+1/4,x+3/4
127 y+3/4,-z,x+3/4
128 y+3/4,z+1/4,-x+1/2
129 -z+1/2,-x,-y+1/2
130 -z+1/2,x+1/4,y+3/4
131 z+3/4,-x,y+3/4
132 z+3/4,x+1/4,-y+1/2
133 y+1/2,x,z+1/2
134 y+1/2,-x+1/4,-z+3/4
135 -y+3/4,x,-z+3/4
136 -y+3/4,-x+1/4,z+1/2
137 x+1/2,z,y+1/2
138 x+1/2,-z+1/4,-y+3/4
139 -x+3/4,z,-y+3/4
140 -x+3/4,-z+1/4,y+1/2
141 z+1/2,y,x+1/2
142 z+1/2,-y+1/4,-x+3/4
143 -z+3/4,y,-x+3/4
144 -z+3/4,-y+1/4,x+1/2
145 x+1/2,y+1/2,z
146 x+1/2,-y+3/4,-z+1/4
147 -x+3/4,y+1/2,-z+1/4
148 -x+3/4,-y+3/4,z
149 y+1/2,z+1/2,x
150 y+1/2,-z+3/4,-x+1/4
151 -y+3/4,z+1/2,-x+1/4
152 -y+3/4,-z+3/4,x
153 z+1/2,x+1/2,y
154 z+1/2,-x+3/4,-y+1/4
155 -z+3/4,x+1/2,-y+1/4
156 -z+3/4,-x+3/4,y
157 -y+1/2,-x+1/2,-z
158 -y+1/2,x+3/4,z+1/4
159 y+3/4,-x+1/2,z+1/4
160 y+3/4,x+3/4,-z
161 -x+1/2,-z+1/2,-y
162 -x+1/2,z+3/4,y+1/4
163 x+3/4,-z+1/2,y+1/4
164 x+3/4,z+3/4,-y
165 -z+1/2,-y+1/2,-x
166 -z+1/2,y+3/4,x+1/4
167 z+3/4,-y+1/2,x+1/4
168 z+3/4,y+3/4,-x
169 -x+1/2,-y+1/2,-z
170 -x+1/2,y+3/4,z+1/4
171 x+3/4,-y+1/2,z+1/4
172 x+3/4,y+3/4,-z
173 -y+1/2,-z+1/2,-x
174 -y+1/2,z+3/4,x+1/4
175 y+3/4,-z+1/2,x+1/4
176 y+3/4,z+3/4,-x
177 -z+1/2,-x+1/2,-y
178 -z+1/2,x+3/4,y+1/4
179 z+3/4,-x+1/2,y+1/4
180 z+3/4,x+3/4,-y
181 y+1/2,x+1/2,z
182 y+1/2,-x+3/4,-z+1/4
183 -y+3/4,x+1/2,-z+1/4
184 -y+3/4,-x+3/4,z
185 x+1/2,z+1/2,y
186 x+1/2,-z+3/4,-y+1/4
187 -x+3/4,z+1/2,-y+1/4
188 -x+3/4,-z+3/4,y
189 z+1/2,y+1/2,x
190 z+1/2,-y+3/4,-x+1/4
191 -z+3/4,y+1/2,-x+1/4
192 -z+3/4,-y+3/4,x

loop_
_atom_site_label
_atom_site_type_symbol
_atom_site_symmetry_multiplicity
_atom_site_Wyckoff_label
_atom_site_fract_x
_atom_site_fract_y
_atom_site_fract_z
_atom_site_occupancy
C1 C    8 a 0.12500  0.12500  0.12500  1.00000
```

Diamond (A4): A_cF8_227_a - POSCAR

```
A_cF8_227_a & a --params=3.55 & Fd(-3)m    O_h^7 #227 (a) & cF8 & A4 & C
    ↪ & diamond & W. H. Bragg and W. L. Bragg, Proc. R. Soc. A Math.
    ↪ Phys. Eng. Sci. 89, 277-291 (1913)
   1.0000000000000000
   0.00000000000000    1.77500000000000    1.77500000000000
   1.77500000000000    0.00000000000000    1.77500000000000
   1.77500000000000    1.77500000000000    0.00000000000000
    C
    2
Direct
   0.12500000000000    0.12500000000000    0.12500000000000    C    (8a)
   0.87500000000000    0.87500000000000    0.87500000000000    C    (8a)
```

Spinel ($Al_2MgO_4$, $H1_1$): A2BC4_cF56_227_d_a_e - CIF

```
# CIF file

data_findsym-output
```

```
_audit_creation_method FINDSYM

_chemical_name_mineral 'Spinel'
_chemical_formula_sum 'Al2 Mg O4'

loop_
_publ_author_name
 'Roderick J. Hill'
 'James R. Craig'
 'G. V. Gibbs'
_journal_name_full
;
Physics and Chemistry of Minerals
;
_journal_volume 4
_journal_year 1979
_journal_page_first 317
_journal_page_last 339
_publ_Section_title
;
 Systematics of the Spinel Structure Type
;

_aflow_proto 'A2BC4_cF56_227_d_a_e'
_aflow_params 'a,x3'
_aflow_params_values '8.0832,0.7376'
_aflow_Strukturbericht 'H1_1'
_aflow_Pearson 'cF56'

_symmetry_space_group_name_Hall "-F 4vw 2vw 3 Fd(-3)m"
_symmetry_space_group_name_H-M "F d -3 m:2"
_symmetry_Int_Tables_number 227

_cell_length_a      8.08320
_cell_length_b      8.08320
_cell_length_c      8.08320
_cell_angle_alpha  90.00000
_cell_angle_beta   90.00000
_cell_angle_gamma  90.00000

loop_
_space_group_symop_id
_space_group_symop_operation_xyz
1 x,y,z
2 x,-y+1/4,-z+1/4
3 -x+1/4,y,-z+1/4
4 -x+1/4,-y+1/4,z
5 y,z,x
6 y,-z+1/4,-x+1/4
7 -y+1/4,z,-x+1/4
8 -y+1/4,-z+1/4,x
9 z,x,y
10 z,-x+1/4,-y+1/4
11 -z+1/4,x,-y+1/4
12 -z+1/4,-x+1/4,y
13 -y,-x,-z
14 -y,x+1/4,z+1/4
15 y+1/4,-x,z+1/4
16 y+1/4,x+1/4,-z
17 -x,-z,-y
18 -x,z+1/4,y+1/4
19 x+1/4,-z,y+1/4
20 x+1/4,z+1/4,-y
21 -z,-y,-x
22 -z,y+1/4,x+1/4
23 z+1/4,-y,x+1/4
24 z+1/4,y+1/4,-x
25 -x,-y,-z
26 -x,y+1/4,z+1/4
27 x+1/4,-y,z+1/4
28 x+1/4,y+1/4,-z
29 -y,-z,-x
30 -y,z+1/4,x+1/4
31 y+1/4,-z,x+1/4
32 y+1/4,z+1/4,-x
33 -z,-x,-y
34 -z,x+1/4,y+1/4
35 z+1/4,-x,y+1/4
36 z+1/4,x+1/4,-y
37 y,x,z
38 y,-x+1/4,-z+1/4
39 -y+1/4,x,-z+1/4
40 -y+1/4,-x+1/4,z
41 x,z,y
42 x,-z+1/4,-y+1/4
43 -x+1/4,z,-y+1/4
44 -x+1/4,-z+1/4,y
45 z,y,x
46 z,-y+1/4,-x+1/4
47 -z+1/4,y,-x+1/4
48 -z+1/4,-y+1/4,x
49 x,y+1/2,z+1/2
50 x,-y+3/4,-z+3/4
51 -x+1/4,y+1/2,-z+3/4
52 -x+1/4,-y+3/4,z+1/2
53 y,z+1/2,x+1/2
54 y,-z+3/4,-x+3/4
55 -y+1/4,z+1/2,-x+3/4
56 -y+1/4,-z+3/4,x+1/2
57 z,x+1/2,y+1/2
58 z,-x+3/4,-y+3/4
59 -z+1/4,x+1/2,-y+3/4
60 -z+1/4,-x+3/4,y+1/2
61 -y,-x+1/2,-z+1/2
62 -y,x+3/4,z+3/4
63 y+1/4,-x+1/2,z+3/4
64 y+1/4,x+3/4,-z+1/2
65 -x,-z+1/2,-y+1/2
66 -x,z+3/4,y+3/4
67 x+1/4,-z+1/2,y+3/4
68 x+1/4,z+3/4,-y+1/2
69 -z,-y+1/2,-x+1/2
70 -z,y+3/4,x+3/4
71 z+1/4,-y+1/2,x+3/4
72 z+1/4,y+3/4,-x+1/2
73 -x,-y+1/2,-z+1/2
74 -x,y+3/4,z+3/4
75 x+1/4,-y+1/2,z+3/4
76 x+1/4,y+3/4,-z+1/2
77 -y,-z+1/2,-x+1/2
78 -y,z+3/4,x+3/4
79 y+1/4,-z+1/2,x+3/4
80 y+1/4,z+3/4,-x+1/2
81 -z,-x+1/2,-y+1/2
82 -z,x+3/4,y+3/4
83 z+1/4,-x+1/2,y+3/4
84 z+1/4,x+3/4,-y+1/2
85 y,x+1/2,z+1/2
86 y,-x+3/4,-z+3/4
87 -y+1/4,x+1/2,-z+3/4
88 -y+1/4,-x+3/4,z+1/2
89 x,z+1/2,y+1/2
90 x,-z+3/4,-y+3/4
91 -x+1/4,z+1/2,-y+3/4
92 -x+1/4,-z+3/4,y+1/2
93 z,y+1/2,x+1/2
94 z,-y+3/4,-x+3/4
95 -z+1/4,y+1/2,-x+3/4
96 -z+1/4,-y+3/4,x+1/2
97 x+1/2,y,z+1/2
98 x+1/2,-y+1/4,-z+3/4
99 -x+3/4,y,-z+3/4
100 -x+3/4,-y+1/4,z+1/2
101 y+1/2,z,x+1/2
102 y+1/2,-z+1/4,-x+3/4
103 -y+3/4,z,-x+3/4
104 -y+3/4,-z+1/4,x+1/2
105 z+1/2,x,y+1/2
106 z+1/2,-x+1/4,-y+3/4
107 -z+3/4,x,-y+3/4
108 -z+3/4,-x+1/4,y+1/2
109 -y+1/2,-x,-z+1/2
110 -y+1/2,x+1/4,z+3/4
111 y+3/4,-x,z+3/4
112 y+3/4,x+1/4,-z+1/2
113 -x+1/2,-z,-y+1/2
114 -x+1/2,z+1/4,y+3/4
115 x+3/4,-z,y+3/4
116 x+3/4,z+1/4,-y+1/2
117 -z+1/2,-y,-x+1/2
118 -z+1/2,y+1/4,x+3/4
119 z+3/4,-y,x+3/4
120 z+3/4,y+1/4,-x+1/2
121 -x+1/2,-y,-z+1/2
122 -x+1/2,y+1/4,z+3/4
123 x+3/4,-y,z+3/4
124 x+3/4,y+1/4,-z+1/2
125 -y+1/2,-z,-x+1/2
126 -y+1/2,z+1/4,x+3/4
127 y+3/4,-z,x+3/4
128 y+3/4,z+1/4,-x+1/2
129 -z+1/2,-x,-y+1/2
130 -z+1/2,x+1/4,y+3/4
131 z+3/4,-x,y+3/4
132 z+3/4,x+1/4,-y+1/2
133 y+1/2,x,z+1/2
134 y+1/2,-x+1/4,-z+3/4
135 -y+3/4,x,-z+3/4
136 -y+3/4,-x+1/4,z+1/2
137 x+1/2,z,y+1/2
138 x+1/2,-z+1/4,-y+3/4
139 -x+3/4,z,-y+3/4
140 -x+3/4,-z+1/4,y+1/2
141 z+1/2,y,x+1/2
142 z+1/2,-y+1/4,-x+3/4
143 -z+3/4,y,-x+3/4
144 -z+3/4,-y+1/4,x+1/2
145 x+1/2,y+1/2,z
146 x+1/2,-y+3/4,-z+1/4
147 -x+3/4,y+1/2,-z+1/4
148 -x+3/4,-y+3/4,z
149 y+1/2,z+1/2,x
150 y+1/2,-z+3/4,-x+1/4
151 -y+3/4,z+1/2,-x+1/4
152 -y+3/4,-z+3/4,x
153 z+1/2,x+1/2,y
154 z+1/2,-x+3/4,-y+1/4
155 -z+3/4,x+1/2,-y+1/4
156 -z+3/4,-x+3/4,y
157 -y+1/2,-x+1/2,-z
158 -y+1/2,x+3/4,z+1/4
159 y+3/4,-x+1/2,z+1/4
160 y+3/4,x+3/4,-z
161 -x+1/2,-z+1/2,-y
162 -x+1/2,z+3/4,y+1/4
163 x+3/4,-z+1/2,y+1/4
164 x+3/4,z+3/4,-y
165 -z+1/2,-y+1/2,-x
166 -z+1/2,y+3/4,x+1/4
167 z+3/4,-y+1/2,x+1/4
```

```
168 z+3/4,y+3/4,-x
169 -x+1/2,-y+1/2,-z
170 -x+1/2,y+3/4,z+1/4
171 x+3/4,-y+1/2,z+1/4
172 x+3/4,y+3/4,-z
173 -y+1/2,-z+1/2,-x
174 -y+1/2,z+3/4,x+1/4
175 y+3/4,-z+1/2,x+1/4
176 y+3/4,z+3/4,-x
177 -z+1/2,-x+1/2,-y
178 -z+1/2,x+3/4,y+1/4
179 z+3/4,-x+1/2,y+1/4
180 z+3/4,x+3/4,-y
181 y+1/2,x+1/2,z
182 y+1/2,-x+3/4,-z+1/4
183 -y+3/4,x+1/2,-z+1/4
184 -y+3/4,-x+3/4,z
185 x+1/2,z+1/2,y
186 x+1/2,-z+3/4,-y+1/4
187 -x+3/4,z+1/2,-y+1/4
188 -x+3/4,-z+3/4,y
189 z+1/2,y+1/2,x
190 z+1/2,-y+3/4,-x+1/4
191 -z+3/4,y+1/2,-x+1/4
192 -z+3/4,-y+3/4,x

loop_
_atom_site_label
_atom_site_type_symbol
_atom_site_symmetry_multiplicity
_atom_site_Wyckoff_label
_atom_site_fract_x
_atom_site_fract_y
_atom_site_fract_z
_atom_site_occupancy
Mg1 Mg    8 a 0.12500 0.12500 0.12500 1.00000
Al1  Al   16 d 0.50000 0.50000 0.50000 1.00000
O1   O    32 e 0.73760 0.73760 0.73760 1.00000
```

Spinel ($Al_2MgO_4$, $H1_1$): A2BC4_cF56_227_d_a_e - POSCAR

```
A2BC4_cF56_227_d_a_e & a,x3 --params=8.0832,0.7376 & Fd(-3)m    O_h^7 #
    ↪ 227 (ade) & cF56 & H1_1 & Al2MgO4 & Spinel & R. J. Hill, J. R.
    ↪ Craig and G. V. Gibbs, Phys. Chem. Minerals 4, 317-339 (1979)
   1.0000000000000000
   0.00000000000000    4.04160000000000    4.04160000000000
   4.04160000000000    0.00000000000000    4.04160000000000
   4.04160000000000    4.04160000000000    0.00000000000000
   Al    Mg    O
    4     2    8
Direct
   0.00000000000000    0.50000000000000    0.50000000000000   Al    (16d)
   0.50000000000000    0.00000000000000    0.50000000000000   Al    (16d)
   0.50000000000000    0.50000000000000    0.00000000000000   Al    (16d)
   0.50000000000000    0.50000000000000    0.50000000000000   Al    (16d)
   0.12500000000000    0.12500000000000    0.12500000000000   Mg     (8a)
   0.87500000000000    0.87500000000000    0.87500000000000   Mg     (8a)
   0.26240000000000    0.26240000000000    0.26240000000000   O     (32e)
   0.26240000000000    0.26240000000000    0.71280000000000   O     (32e)
  -0.26240000000000    0.28720000000000    0.73760000000000   O     (32e)
   0.26240000000000    0.71280000000000    0.26240000000000   O     (32e)
   0.28720000000000    0.73760000000000   -0.26240000000000   O     (32e)
   0.71280000000000    0.26240000000000    0.26240000000000   O     (32e)
   0.73760000000000   -0.26240000000000    0.28720000000000   O     (32e)
   0.73760000000000    0.73760000000000    0.73760000000000   O     (32e)
```

$CTi_2$: AB2_cF48_227_c_e - CIF

```
# CIF file

data_findsym-output
_audit_creation_method FINDSYM

_chemical_name_mineral ''
_chemical_formula_sum 'C Ti2'

loop_
_publ_author_name
 'H. Goretzki'
_journal_name_full
;
Physica Status Solidi B
;
_journal_volume 20
_journal_year 1967
_journal_page_first K141
_journal_page_last K143
_publ_Section_title
;
 Neutron Diffraction Studies on Titanium-Carbon and Zirconium-Carbon
    ↪ Alloys
;

_aflow_proto 'AB2_cF48_227_c_e'
_aflow_params 'a,x2'
_aflow_params_values '8.6,0.245'
_aflow_Strukturbericht 'None'
_aflow_Pearson 'cF48'

_symmetry_space_group_name_Hall "-F 4vw 2vw 3 Fd(-3)m"
_symmetry_space_group_name_H-M "F d -3 m:2"
_symmetry_Int_Tables_number 227

_cell_length_a     8.60000
_cell_length_b     8.60000
_cell_length_c     8.60000
_cell_angle_alpha 90.00000
_cell_angle_beta  90.00000
_cell_angle_gamma 90.00000

loop_
_space_group_symop_id
_space_group_symop_operation_xyz
1 x,y,z
2 x,-y+1/4,-z+1/4
3 -x+1/4,y,-z+1/4
4 -x+1/4,-y+1/4,z
5 y,z,x
6 y,-z+1/4,-x+1/4
7 -y+1/4,z,-x+1/4
8 -y+1/4,-z+1/4,x
9 z,x,y
10 z,-x+1/4,-y+1/4
11 -z+1/4,x,-y+1/4
12 -z+1/4,-x+1/4,y
13 -y,-x,-z
14 -y,x+1/4,z+1/4
15 y+1/4,-x,z+1/4
16 y+1/4,x+1/4,-z
17 -x,-z,-y
18 -x,z+1/4,y+1/4
19 x+1/4,-z,y+1/4
20 x+1/4,z+1/4,-y
21 -z,-y,-x
22 -z,y+1/4,x+1/4
23 z+1/4,-y,x+1/4
24 z+1/4,y+1/4,-x
25 -x,-y,-z
26 -x,y+1/4,z+1/4
27 x+1/4,-y,z+1/4
28 x+1/4,y+1/4,-z
29 -y,-z,-x
30 -y,z+1/4,x+1/4
31 y+1/4,-z,x+1/4
32 y+1/4,z+1/4,-x
33 -z,-x,-y
34 -z,x+1/4,y+1/4
35 z+1/4,-x,y+1/4
36 z+1/4,x+1/4,-y
37 y,x,z
38 y,-x+1/4,-z+1/4
39 -y+1/4,x,-z+1/4
40 -y+1/4,-x+1/4,z
41 x,z,y
42 x,-z+1/4,-y+1/4
43 -x+1/4,z,-y+1/4
44 -x+1/4,-z+1/4,y
45 z,y,x
46 z,-y+1/4,-x+1/4
47 -z+1/4,y,-x+1/4
48 -z+1/4,-y+1/4,x
49 x,y+1/2,z+1/2
50 x,-y+3/4,-z+3/4
51 -x+1/4,y+1/2,-z+3/4
52 -x+1/4,-y+3/4,z+1/2
53 y,z+1/2,x+1/2
54 y,-z+3/4,-x+3/4
55 -y+1/4,z+1/2,-x+3/4
56 -y+1/4,-z+3/4,x+1/2
57 z,x+1/2,y+1/2
58 z,-x+3/4,-y+3/4
59 -z+1/4,x+1/2,-y+3/4
60 -z+1/4,-x+3/4,y+1/2
61 -y,-x+1/2,-z+1/2
62 -y,x+3/4,z+3/4
63 y+1/4,-x+1/2,z+3/4
64 y+1/4,x+3/4,-z+1/2
65 -x,-z+1/2,-y+1/2
66 -x,z+3/4,y+3/4
67 x+1/4,-z+1/2,y+3/4
68 x+1/4,z+3/4,-y+1/2
69 -z,-y+1/2,-x+1/2
70 -z,y+3/4,x+3/4
71 z+1/4,-y+1/2,x+3/4
72 z+1/4,y+3/4,-x+1/2
73 -x,-y+1/2,-z+1/2
74 -x,y+3/4,z+3/4
75 x+1/4,-y+1/2,z+3/4
76 x+1/4,y+3/4,-z+1/2
77 -y,-z+1/2,-x+1/2
78 -y,z+3/4,x+3/4
79 y+1/4,-z+1/2,x+3/4
80 y+1/4,z+3/4,-x+1/2
81 -z,-x+1/2,-y+1/2
82 -z,x+3/4,y+3/4
83 z+1/4,-x+1/2,y+3/4
84 z+1/4,x+3/4,-y+1/2
85 y,x+1/2,z+1/2
86 y,-x+3/4,-z+3/4
87 -y+1/4,x+1/2,-z+3/4
88 -y+1/4,-x+3/4,z+1/2
89 x,z+1/2,y+1/2
90 x,-z+3/4,-y+3/4
91 -x+1/4,z+1/2,-y+3/4
92 -x+1/4,-z+3/4,y+1/2
93 z,y+1/2,x+1/2
94 z,-y+3/4,-x+3/4
95 -z+1/4,y+1/2,-x+3/4
96 -z+1/4,-y+3/4,x+1/2
97 x+1/2,y,z+1/2
```

```
98 x+1/2,-y+1/4,-z+3/4
99 -x+3/4,y,-z+3/4
100 -x+3/4,-y+1/4,z+1/2
101 y+1/2,z,x+1/2
102 y+1/2,-z+1/4,-x+3/4
103 -y+3/4,z,-x+3/4
104 -y+3/4,-z+1/4,x+1/2
105 z+1/2,x,y+1/2
106 z+1/2,-x+1/4,-y+3/4
107 -z+3/4,x,-y+3/4
108 -z+3/4,-x+1/4,y+1/2
109 -y+1/2,-x,-z+1/2
110 -y+1/2,x+1/4,z+3/4
111 y+3/4,-x,z+3/4
112 y+3/4,x+1/4,-z+1/2
113 -x+1/2,-z,-y+1/2
114 -x+1/2,z+1/4,y+3/4
115 x+3/4,-z,y+3/4
116 x+3/4,z+1/4,-y+1/2
117 -z+1/2,-y,-x+1/2
118 -z+1/2,y+1/4,x+3/4
119 z+3/4,-y,x+3/4
120 z+3/4,y+1/4,-x+1/2
121 -x+1/2,-y,-z+1/2
122 -x+1/2,y+1/4,z+3/4
123 x+3/4,-y,z+3/4
124 x+3/4,y+1/4,-z+1/2
125 -y+1/2,-z,-x+1/2
126 -y+1/2,z+1/4,x+3/4
127 y+3/4,-z,x+3/4
128 y+3/4,z+1/4,-x+1/2
129 -z+1/2,-x,-y+1/2
130 -z+1/2,x+1/4,y+3/4
131 z+3/4,-x,y+3/4
132 z+3/4,x+1/4,-y+1/2
133 y+1/2,x,z+1/2
134 y+1/2,-x+1/4,-z+3/4
135 -y+3/4,x,-z+3/4
136 -y+3/4,-x+1/4,z+1/2
137 x+1/2,z,y+1/2
138 x+1/2,-z+1/4,-y+3/4
139 -x+3/4,z,-y+3/4
140 -x+3/4,-z+1/4,y+1/2
141 z+1/2,y,x+1/2
142 z+1/2,-y+1/4,-x+3/4
143 -z+3/4,y,-x+3/4
144 -z+3/4,-y+1/4,x+1/2
145 x+1/2,y+1/2,z
146 x+1/2,-y+3/4,-z+1/4
147 -x+3/4,y+1/2,-z+1/4
148 -x+3/4,-y+3/4,z
149 y+1/2,z+1/2,x
150 y+1/2,-z+3/4,-x+1/4
151 -y+3/4,z+1/2,-x+1/4
152 -y+3/4,-z+3/4,x
153 z+1/2,x+1/2,y
154 z+1/2,-x+3/4,-y+1/4
155 -z+3/4,x+1/2,-y+1/4
156 -z+3/4,-x+3/4,y
157 -y+1/2,-x+1/2,-z
158 -y+1/2,x+3/4,z+1/4
159 y+3/4,-x+1/2,z+1/4
160 y+3/4,x+3/4,-z
161 -x+1/2,-z+1/2,-y
162 -x+1/2,z+3/4,y+1/4
163 x+3/4,-z+1/2,y+1/4
164 x+3/4,z+3/4,-y
165 -z+1/2,-y+1/2,-x
166 -z+1/2,y+3/4,x+1/4
167 z+3/4,-y+1/2,x+1/4
168 z+3/4,y+3/4,-x
169 -x+1/2,-y+1/2,-z
170 -x+1/2,y+3/4,z+1/4
171 x+3/4,-y+1/2,z+1/4
172 x+3/4,y+3/4,-z
173 -y+1/2,-z+1/2,-x
174 -y+1/2,z+3/4,x+1/4
175 y+3/4,-z+1/2,x+1/4
176 y+3/4,z+3/4,-x
177 -z+1/2,-x+1/2,-y
178 -z+1/2,x+3/4,y+1/4
179 z+3/4,-x+1/2,y+1/4
180 z+3/4,x+3/4,-y
181 y+1/2,x+1/2,z
182 y+1/2,-x+3/4,-z+1/4
183 -y+3/4,x+1/2,-z+1/4
184 -y+3/4,-x+3/4,z
185 x+1/2,z+1/2,y
186 x+1/2,-z+3/4,-y+1/4
187 -x+3/4,z+1/2,-y+1/4
188 -x+3/4,-z+3/4,y
189 z+1/2,y+1/2,x
190 z+1/2,-y+3/4,-x+1/4
191 -z+3/4,y+1/2,-x+1/4
192 -z+3/4,-y+3/4,x

loop_
_atom_site_label
_atom_site_type_symbol
_atom_site_symmetry_multiplicity
_atom_site_Wyckoff_label
_atom_site_fract_x
_atom_site_fract_y
_atom_site_fract_z
_atom_site_occupancy
C1  C   16 c 0.00000 0.00000 0.00000 1.00000
Ti1 Ti  32 e 0.24500 0.24500 0.24500 1.00000
```

$CTi_2$: AB2_cF48_227_c_e - POSCAR

```
AB2_cF48_227_c_e & a,x2 --params=8.6,0.245 & Fd(-3)m      O_h^7 #227 (ce)
     ↪ & cF48 & & CTi2 & & H. Goretzki, Phys. Status Solidi B 20,
     ↪ K141-K143 (1967)
   1.0000000000000000
   0.00000000000000    4.30000000000000    4.30000000000000
   4.30000000000000    0.00000000000000    4.30000000000000
   4.30000000000000    4.30000000000000    0.00000000000000
     C     Ti
     4      8
Direct
   0.00000000000000    0.00000000000000    0.00000000000000    C   (16c)
   0.00000000000000   -0.00000000000000    0.50000000000000    C   (16c)
  -0.00000000000000    0.50000000000000    0.00000000000000    C   (16c)
   0.50000000000000    0.00000000000000    0.00000000000000    C   (16c)
   0.23500000000000   -0.24500000000000   -0.24500000000000    Ti  (32e)
  -0.23500000000000    0.24500000000000    0.24500000000000    Ti  (32e)
   0.24500000000000   -0.23500000000000    0.24500000000000    Ti  (32e)
  -0.24500000000000    0.23500000000000   -0.24500000000000    Ti  (32e)
   0.24500000000000    0.24500000000000   -0.23500000000000    Ti  (32e)
  -0.24500000000000   -0.24500000000000    0.23500000000000    Ti  (32e)
   0.24500000000000    0.24500000000000    0.24500000000000    Ti  (32e)
  -0.24500000000000   -0.24500000000000   -0.24500000000000    Ti  (32e)
```

$Fe_3W_3C$: AB3C3_cF112_227_c_de_f - CIF

```
# CIF file

data_findsym-output
_audit_creation_method FINDSYM

_chemical_name_mineral ''
_chemical_formula_sum 'Fe3 W3 C'

loop_
_publ_author_name
 'Qi-Bin Yang'
 'Sten Andersson'
_journal_name_full
;
Acta Crystallographica B
;
_journal_volume 43
_journal_year 1987
_journal_page_first 1
_journal_page_last 14
_publ_Section_title
;
 Application of coincidence site lattices for crystal structure
      ↪ description. Part I: $\Sigma = 3$
;

_aflow_proto 'AB3C3_cF112_227_c_de_f'
_aflow_params 'a,x3,x4'
_aflow_params_values '11.087,0.7047,0.323'
_aflow_Strukturbericht 'E9_3'
_aflow_Pearson 'cF112'

_symmetry_space_group_name_Hall "-F 4vw 2vw 3 Fd(-3)m"
_symmetry_space_group_name_H-M "F d -3 m:2"
_symmetry_Int_Tables_number 227

_cell_length_a      11.08700
_cell_length_b      11.08700
_cell_length_c      11.08700
_cell_angle_alpha   90.00000
_cell_angle_beta    90.00000
_cell_angle_gamma   90.00000

loop_
_space_group_symop_id
_space_group_symop_operation_xyz
1 x,y,z
2 x,-y+1/4,-z+1/4
3 -x+1/4,y,-z+1/4
4 -x+1/4,-y+1/4,z
5 y,z,x
6 y,-z+1/4,-x+1/4
7 -y+1/4,z,-x+1/4
8 -y+1/4,-z+1/4,x
9 z,x,y
10 z,-x+1/4,-y+1/4
11 -z+1/4,x,-y+1/4
12 -z+1/4,-x+1/4,y
13 -y,-x,-z
14 -y,x+1/4,z+1/4
15 y+1/4,-x,z+1/4
16 y+1/4,x+1/4,-z
17 -x,-z,-y
18 -x,z+1/4,y+1/4
19 x+1/4,-z,y+1/4
20 x+1/4,z+1/4,-y
21 -z,-y,-x
22 -z,y+1/4,x+1/4
23 z+1/4,-y,x+1/4
24 z+1/4,y+1/4,-x
25 -x,-y,-z
26 -x,y+1/4,z+1/4
27 x+1/4,-y,z+1/4
28 x+1/4,y+1/4,-z
29 -y,-z,-x
```

```
30 -y,z+1/4,x+1/4
31 y+1/4,-z,x+1/4
32 y+1/4,z+1/4,-x
33 -z,-x,-y
34 -z,x+1/4,y+1/4
35 z+1/4,-x,y+1/4
36 z+1/4,x+1/4,-y
37 y,x,z
38 y,-x+1/4,-z+1/4
39 -y+1/4,x,-z+1/4
40 -y+1/4,-x+1/4,z
41 x,z,y
42 x,-z+1/4,-y+1/4
43 -x+1/4,z,-y+1/4
44 -x+1/4,-z+1/4,y
45 z,y,x
46 z,-y+1/4,-x+1/4
47 -z+1/4,y,-x+1/4
48 -z+1/4,-y+1/4,x
49 x,y+1/2,z+1/2
50 x,-y+3/4,-z+3/4
51 -x+1/4,y+1/2,-z+3/4
52 -x+1/4,-y+3/4,z+1/2
53 y,z+1/2,x+1/2
54 y,-z+3/4,-x+3/4
55 -y+1/4,z+1/2,-x+3/4
56 -y+1/4,-z+3/4,x+1/2
57 z,x+1/2,y+1/2
58 z,-x+3/4,-y+3/4
59 -z+1/4,x+1/2,-y+3/4
60 -z+1/4,-x+3/4,y+1/2
61 -y,-x+1/2,-z+1/2
62 -y,x+3/4,z+3/4
63 y+1/4,-x+1/2,z+3/4
64 y+1/4,x+3/4,-z+1/2
65 -x,-z+1/2,-y+1/2
66 -x,z+3/4,y+3/4
67 x+1/4,-z+1/2,y+3/4
68 x+1/4,z+3/4,-y+1/2
69 -z,-y+1/2,-x+1/2
70 -z,y+3/4,x+3/4
71 z+1/4,-y+1/2,x+3/4
72 z+1/4,y+3/4,-x+1/2
73 -x,-y+1/2,-z+1/2
74 -x,y+3/4,z+3/4
75 x+1/4,-y+1/2,z+3/4
76 x+1/4,y+3/4,-z+1/2
77 -y,-z+1/2,-x+1/2
78 -y,z+3/4,x+3/4
79 y+1/4,-z+1/2,x+3/4
80 y+1/4,z+3/4,-x+1/2
81 -z,-x+1/2,-y+1/2
82 -z,x+3/4,y+3/4
83 z+1/4,-x+1/2,y+3/4
84 z+1/4,x+3/4,-y+1/2
85 y,x+1/2,z+1/2
86 y,-x+3/4,-z+3/4
87 -y+1/4,x+1/2,-z+3/4
88 -y+1/4,-x+3/4,z+1/2
89 x,z+1/2,y+1/2
90 x,-z+3/4,-y+3/4
91 -x+1/4,z+1/2,-y+3/4
92 -x+1/4,-z+3/4,y+1/2
93 z,y+1/2,x+1/2
94 z,-y+3/4,-x+3/4
95 -z+1/4,y+1/2,-x+3/4
96 -z+1/4,-y+3/4,x+1/2
97 x+1/2,y,z+1/2
98 x+1/2,-y+1/4,-z+3/4
99 -x+3/4,y,-z+3/4
100 -x+3/4,-y+1/4,z+1/2
101 y+1/2,z,x+1/2
102 y+1/2,-z+1/4,-x+3/4
103 -y+3/4,z,-x+3/4
104 -y+3/4,-z+1/4,x+1/2
105 z+1/2,x,y+1/2
106 z+1/2,-x+1/4,-y+3/4
107 -z+3/4,x,-y+3/4
108 -z+3/4,-x+1/4,y+1/2
109 -y+1/2,-x,-z+1/2
110 -y+1/2,x+1/4,z+3/4
111 y+3/4,-x,z+3/4
112 y+3/4,x+1/4,-z+1/2
113 -x+1/2,-z,-y+1/2
114 -x+1/2,z+1/4,y+3/4
115 x+3/4,-z,y+3/4
116 x+3/4,z+1/4,-y+1/2
117 -z+1/2,-y,-x+1/2
118 -z+1/2,y+1/4,x+3/4
119 z+3/4,-y,x+3/4
120 z+3/4,y+1/4,-x+1/2
121 -x+1/2,-y,-z+1/2
122 -x+1/2,y+1/4,z+3/4
123 x+3/4,-y,z+3/4
124 x+3/4,y+1/4,-z+1/2
125 -y+1/2,-z,-x+1/2
126 -y+1/2,z+1/4,x+3/4
127 y+3/4,-z,x+3/4
128 y+3/4,z+1/4,-x+1/2
129 -z+1/2,-x,-y+1/2
130 -z+1/2,x+1/4,y+3/4
131 z+3/4,-x,y+3/4
132 z+3/4,x+1/4,-y+1/2
133 y+1/2,x,z+1/2
134 y+1/2,-x+1/4,-z+3/4
135 -y+3/4,x,-z+3/4
136 -y+3/4,-x+1/4,z+1/2
137 x+1/2,z,y+1/2
138 x+1/2,-z+1/4,-y+3/4
139 -x+3/4,z,-y+3/4
140 -x+3/4,-z+1/4,y+1/2
141 z+1/2,y,x+1/2
142 z+1/2,-y+1/4,-x+3/4
143 -z+3/4,y,-x+3/4
144 -z+3/4,-y+1/4,x+1/2
145 x+1/2,y+1/2,z
146 x+1/2,-y+3/4,-z+1/4
147 -x+3/4,y+1/2,-z+1/4
148 -x+3/4,-y+3/4,z
149 y+1/2,z+1/2,x
150 y+1/2,-z+3/4,-x+1/4
151 -y+3/4,z+1/2,-x+1/4
152 -y+3/4,-z+3/4,x
153 z+1/2,x+1/2,y
154 z+1/2,-x+3/4,-y+1/4
155 -z+3/4,x+1/2,-y+1/4
156 -z+3/4,-x+3/4,y
157 -y+1/2,-x+1/2,-z
158 -y+1/2,x+3/4,z+1/4
159 y+3/4,-x+1/2,z+1/4
160 y+3/4,x+3/4,-z
161 -x+1/2,-z+1/2,-y
162 -x+1/2,z+3/4,y+1/4
163 x+3/4,-z+1/2,y+1/4
164 x+3/4,z+3/4,-y
165 -z+1/2,-y+1/2,-x
166 -z+1/2,y+3/4,x+1/4
167 z+3/4,-y+1/2,x+1/4
168 z+3/4,y+3/4,-x
169 -x+1/2,-y+1/2,-z
170 -x+1/2,y+3/4,z+1/4
171 x+3/4,-y+1/2,z+1/4
172 x+3/4,y+3/4,-z
173 -y+1/2,-z+1/2,-x
174 -y+1/2,z+3/4,x+1/4
175 y+3/4,-z+1/2,x+1/4
176 y+3/4,z+3/4,-x
177 -z+1/2,-x+1/2,-y
178 -z+1/2,x+3/4,y+1/4
179 z+3/4,-x+1/2,y+1/4
180 z+3/4,x+3/4,-y
181 y+1/2,x+1/2,z
182 y+1/2,-x+3/4,-z+1/4
183 -y+3/4,x+1/2,-z+1/4
184 -y+3/4,-x+3/4,z
185 x+1/2,z+1/2,y
186 x+1/2,-z+3/4,-y+1/4
187 -x+3/4,z+1/2,-y+1/4
188 -x+3/4,-z+3/4,y
189 z+1/2,y+1/2,x
190 z+1/2,-y+3/4,-x+1/4
191 -z+3/4,y+1/2,-x+1/4
192 -z+3/4,-y+3/4,x

loop_
_atom_site_label
_atom_site_type_symbol
_atom_site_symmetry_multiplicity
_atom_site_Wyckoff_label
_atom_site_fract_x
_atom_site_fract_y
_atom_site_fract_z
_atom_site_occupancy
C1  C   16 c 0.00000 0.00000 0.00000 1.00000
Fe1 Fe  16 d 0.50000 0.50000 0.50000 1.00000
Fe2 Fe  32 e 0.70470 0.70470 0.70470 1.00000
W1  W   48 f 0.32300 0.12500 0.12500 1.00000
```

$Fe_3W_3C$: AB3C3_cF112_227_c_de_f - POSCAR

```
AB3C3_cF112_227_c_de_f & a,x3,x4 --params=11.087,0.7047,0.323 & Fd(-3)m
    ↪ O_h^7 #227 (cdef) & cF112 & E9_3 & Fe3W3C & & Q.-B. Yang and
    ↪ S. Andersson, Acta Cryst. B 43, 1-14 (1987)
   1.0000000000000000
    0.00000000000000    5.54350000000000    5.54350000000000
    5.54350000000000    0.00000000000000    5.54350000000000
    5.54350000000000    5.54350000000000    0.00000000000000
    C    Fe    W
    4    12    12
Direct
    0.00000000000000    0.00000000000000    0.00000000000000   C    (16c)
   -0.00000000000000    0.00000000000000    0.50000000000000   C    (16c)
    0.00000000000000    0.50000000000000   -0.00000000000000   C    (16c)
    0.50000000000000   -0.00000000000000    0.00000000000000   C    (16c)
    0.00000000000000    0.50000000000000    0.50000000000000   Fe   (16d)
    0.50000000000000    0.00000000000000    0.50000000000000   Fe   (16d)
    0.50000000000000    0.50000000000000    0.00000000000000   Fe   (16d)
    0.50000000000000    0.50000000000000    0.50000000000000   Fe   (16d)
    0.29530000000000    0.29530000000000    0.29530000000000   Fe   (32e)
    0.29530000000000    0.29530000000000   -0.38590000000000   Fe   (32e)
    0.29530000000000   -0.38590000000000    0.29530000000000   Fe   (32e)
   -0.29530000000000    0.70470000000000    0.38590000000000   Fe   (32e)
   -0.38590000000000    0.29530000000000    0.29530000000000   Fe   (32e)
    0.38590000000000   -0.29530000000000    0.70470000000000   Fe   (32e)
    0.70470000000000    0.38590000000000   -0.29530000000000   Fe   (32e)
    0.70470000000000    0.70470000000000    0.70470000000000   Fe   (32e)
   -0.07300000000000   -0.07300000000000    0.32300000000000   W    (48f)
    0.07300000000000    0.07300000000000    0.67700000000000   W    (48f)
   -0.07300000000000    0.32300000000000   -0.07300000000000   W    (48f)
    0.07300000000000    0.67700000000000    0.07300000000000   W    (48f)
```

```
   0.07300000000000     0.67700000000000     0.67700000000000     W    (48f)
   0.32300000000000    -0.07300000000000    -0.07300000000000     W    (48f)
   0.32300000000000     0.32300000000000     0.92700000000000     W    (48f)
   0.32300000000000     0.92700000000000     0.32300000000000     W    (48f)
   0.67700000000000     0.07300000000000     0.07300000000000     W    (48f)
   0.67700000000000     0.07300000000000     0.67700000000000     W    (48f)
   0.67700000000000     0.67700000000000     0.07300000000000     W    (48f)
   0.92700000000000     0.32300000000000     0.32300000000000     W    (48f)
```

Body-Centered Cubic (W, A2): A_cI2_229_a - CIF

```
# CIF file

data_findsym-output
_audit_creation_method FINDSYM

_chemical_name_mineral 'Tungsten'
_chemical_formula_sum 'W'

loop_
_publ_author_name
 'Wheller P. Davey'
_journal_name_full
;
Physical Review
;
_journal_volume 26
_journal_year 1925
_journal_page_first 736
_journal_page_last 738
_publ_Section_title
;
 The Lattice Parameter and Density of Pure Tungsten
;

_aflow_proto 'A_cI2_229_a'
_aflow_params 'a'
_aflow_params_values '3.155'
_aflow_Strukturbericht 'A2'
_aflow_Pearson 'cI2'

_symmetry_space_group_name_Hall "-I 4 2 3"
_symmetry_space_group_name_H-M "I m -3 m"
_symmetry_Int_Tables_number 229

_cell_length_a     3.15500
_cell_length_b     3.15500
_cell_length_c     3.15500
_cell_angle_alpha 90.00000
_cell_angle_beta  90.00000
_cell_angle_gamma 90.00000

loop_
_space_group_symop_id
_space_group_symop_operation_xyz
1 x,y,z
2 x,-y,-z
3 -x,y,-z
4 -x,-y,z
5 y,z,x
6 y,-z,-x
7 -y,z,-x
8 -y,-z,x
9 z,x,y
10 z,-x,-y
11 -z,x,-y
12 -z,-x,y
13 -y,-x,-z
14 -y,x,z
15 y,-x,z
16 y,x,-z
17 -x,-z,-y
18 -x,z,y
19 x,-z,y
20 x,z,-y
21 -z,-y,-x
22 -z,y,x
23 z,-y,x
24 z,y,-x
25 -x,-y,-z
26 -x,y,z
27 x,-y,z
28 x,y,-z
29 -y,-z,-x
30 -y,z,x
31 y,-z,x
32 y,z,-x
33 -z,-x,-y
34 -z,x,y
35 z,-x,y
36 z,x,-y
37 y,x,z
38 y,-x,-z
39 -y,x,-z
40 -y,-x,z
41 x,z,y
42 x,-z,-y
43 -x,z,-y
44 -x,-z,y
45 z,y,x
46 z,-y,-x
47 -z,y,-x
48 -z,-y,x
49 x+1/2,y+1/2,z+1/2
50 x+1/2,-y+1/2,-z+1/2
51 -x+1/2,y+1/2,-z+1/2
52 -x+1/2,-y+1/2,z+1/2
53 y+1/2,z+1/2,x+1/2
54 y+1/2,-z+1/2,-x+1/2
55 -y+1/2,z+1/2,-x+1/2
56 -y+1/2,-z+1/2,x+1/2
57 z+1/2,x+1/2,y+1/2
58 z+1/2,-x+1/2,-y+1/2
59 -z+1/2,x+1/2,-y+1/2
60 -z+1/2,-x+1/2,y+1/2
61 -y+1/2,-x+1/2,-z+1/2
62 -y+1/2,x+1/2,z+1/2
63 y+1/2,-x+1/2,z+1/2
64 y+1/2,x+1/2,-z+1/2
65 -x+1/2,-z+1/2,-y+1/2
66 -x+1/2,z+1/2,y+1/2
67 x+1/2,-z+1/2,y+1/2
68 x+1/2,z+1/2,-y+1/2
69 -z+1/2,-y+1/2,-x+1/2
70 -z+1/2,y+1/2,x+1/2
71 z+1/2,-y+1/2,x+1/2
72 z+1/2,y+1/2,-x+1/2
73 -x+1/2,-y+1/2,-z+1/2
74 -x+1/2,y+1/2,z+1/2
75 x+1/2,-y+1/2,z+1/2
76 x+1/2,y+1/2,-z+1/2
77 -y+1/2,-z+1/2,-x+1/2
78 -y+1/2,z+1/2,x+1/2
79 y+1/2,-z+1/2,x+1/2
80 y+1/2,z+1/2,-x+1/2
81 -z+1/2,-x+1/2,-y+1/2
82 -z+1/2,x+1/2,y+1/2
83 z+1/2,-x+1/2,y+1/2
84 z+1/2,x+1/2,-y+1/2
85 y+1/2,x+1/2,z+1/2
86 y+1/2,-x+1/2,-z+1/2
87 -y+1/2,x+1/2,-z+1/2
88 -y+1/2,-x+1/2,z+1/2
89 x+1/2,z+1/2,y+1/2
90 x+1/2,-z+1/2,-y+1/2
91 -x+1/2,z+1/2,-y+1/2
92 -x+1/2,-z+1/2,y+1/2
93 z+1/2,y+1/2,x+1/2
94 z+1/2,-y+1/2,-x+1/2
95 -z+1/2,y+1/2,-x+1/2
96 -z+1/2,-y+1/2,x+1/2

loop_
_atom_site_label
_atom_site_type_symbol
_atom_site_symmetry_multiplicity
_atom_site_Wyckoff_label
_atom_site_fract_x
_atom_site_fract_y
_atom_site_fract_z
_atom_site_occupancy
W1 W   2 a 0.00000 0.00000 0.00000 1.00000
```

Body-Centered Cubic (W, A2): A_cI2_229_a - POSCAR

```
A_cI2_229_a & a --params=3.155 & Im(-3)m    O_h^9 #229 (a) & cI2 & A2 & W
     ↪  & & W. P. Davey, Phys. Rev. 26, 736-738 (1925)
   1.0000000000000000
  -1.57750000000000     1.57750000000000     1.57750000000000
   1.57750000000000    -1.57750000000000     1.57750000000000
   1.57750000000000     1.57750000000000    -1.57750000000000
    W
    1
Direct
   0.00000000000000     0.00000000000000     0.00000000000000     W    (2a)
```

High-Pressure $H_3S$: A3B_cI8_229_b_a - CIF

```
# CIF file

data_findsym-output
_audit_creation_method FINDSYM

_chemical_name_mineral 'High-temperature superconductor'
_chemical_formula_sum 'H3 S'

loop_
_publ_author_name
 'Defang Duan'
 'Yunxian Liu'
 'Fubo Tian'
 'Da Li'
 'Xiaoli Huang'
 'Zhonglong Zhao'
 'Hongyu Yu'
 'Bingbing Liu'
 'Wenjing Tian'
 'Tian Cui'
_journal_name_full
;
Scientific Reports
;
_journal_volume 4
_journal_year 2014
_journal_page_first 6968
_journal_page_last 6968
_publ_Section_title
;
 Pressure-induced metallization of dense (H$_2$S)$_2$H$_2$ with
     ↪ high-T$_c$ superconductivity
```

```
;

_aflow_proto 'A3B_cI8_229_b_a'
_aflow_params 'a'
_aflow_params_values '2.984'
_aflow_Strukturbericht 'None'
_aflow_Pearson 'cI8'

_symmetry_space_group_name_Hall "-I 4 2 3"
_symmetry_space_group_name_H-M "I m -3 m"
_symmetry_Int_Tables_number 229

_cell_length_a     2.98400
_cell_length_b     2.98400
_cell_length_c     2.98400
_cell_angle_alpha  90.00000
_cell_angle_beta   90.00000
_cell_angle_gamma  90.00000

loop_
_space_group_symop_id
_space_group_symop_operation_xyz
1 x,y,z
2 x,-y,-z
3 -x,y,-z
4 -x,-y,z
5 y,z,x
6 y,-z,-x
7 -y,z,-x
8 -y,-z,x
9 z,x,y
10 z,-x,-y
11 -z,x,-y
12 -z,-x,y
13 -y,-x,-z
14 -y,x,z
15 y,-x,z
16 y,x,-z
17 -x,-z,-y
18 -x,z,y
19 x,-z,y
20 x,z,-y
21 -z,-y,-x
22 -z,y,x
23 z,-y,x
24 z,y,-x
25 -x,-y,-z
26 -x,y,z
27 x,-y,z
28 x,y,-z
29 -y,-z,-x
30 -y,z,x
31 y,-z,x
32 y,z,-x
33 -z,-x,-y
34 -z,x,y
35 z,-x,y
36 z,x,-y
37 y,x,z
38 y,-x,-z
39 -y,x,-z
40 -y,-x,z
41 x,z,y
42 x,-z,-y
43 -x,z,-y
44 -x,-z,y
45 z,y,x
46 z,-y,-x
47 -z,y,-x
48 -z,-y,x
49 x+1/2,y+1/2,z+1/2
50 x+1/2,-y+1/2,-z+1/2
51 -x+1/2,y+1/2,-z+1/2
52 -x+1/2,-y+1/2,z+1/2
53 y+1/2,z+1/2,x+1/2
54 y+1/2,-z+1/2,-x+1/2
55 -y+1/2,z+1/2,-x+1/2
56 -y+1/2,-z+1/2,x+1/2
57 z+1/2,x+1/2,y+1/2
58 z+1/2,-x+1/2,-y+1/2
59 -z+1/2,x+1/2,-y+1/2
60 -z+1/2,-x+1/2,y+1/2
61 -y+1/2,-x+1/2,-z+1/2
62 -y+1/2,x+1/2,z+1/2
63 y+1/2,-x+1/2,z+1/2
64 y+1/2,x+1/2,-z+1/2
65 -x+1/2,-z+1/2,-y+1/2
66 -x+1/2,z+1/2,y+1/2
67 x+1/2,-z+1/2,y+1/2
68 x+1/2,z+1/2,-y+1/2
69 -z+1/2,-y+1/2,-x+1/2
70 -z+1/2,y+1/2,x+1/2
71 z+1/2,-y+1/2,x+1/2
72 z+1/2,y+1/2,-x+1/2
73 -x+1/2,-y+1/2,-z+1/2
74 -x+1/2,y+1/2,z+1/2
75 x+1/2,-y+1/2,z+1/2
76 x+1/2,y+1/2,-z+1/2
77 -y+1/2,-z+1/2,-x+1/2
78 -y+1/2,z+1/2,x+1/2
79 y+1/2,-z+1/2,x+1/2
80 y+1/2,z+1/2,-x+1/2
81 -z+1/2,-x+1/2,-y+1/2
82 -z+1/2,x+1/2,y+1/2
83 z+1/2,-x+1/2,y+1/2
84 z+1/2,x+1/2,-y+1/2
85 y+1/2,x+1/2,z+1/2
86 y+1/2,-x+1/2,-z+1/2
87 -y+1/2,x+1/2,-z+1/2
88 -y+1/2,-x+1/2,z+1/2
89 x+1/2,z+1/2,y+1/2
90 x+1/2,-z+1/2,-y+1/2
91 -x+1/2,z+1/2,-y+1/2
92 -x+1/2,-z+1/2,y+1/2
93 z+1/2,y+1/2,x+1/2
94 z+1/2,-y+1/2,-x+1/2
95 -z+1/2,y+1/2,-x+1/2
96 -z+1/2,-y+1/2,x+1/2

loop_
_atom_site_label
_atom_site_type_symbol
_atom_site_symmetry_multiplicity
_atom_site_Wyckoff_label
_atom_site_fract_x
_atom_site_fract_y
_atom_site_fract_z
_atom_site_occupancy
S1 S    2 a 0.00000 0.00000 0.00000 1.00000
H1 H    6 b 0.00000 0.50000 0.50000 1.00000
```

High-Pressure $H_3S$: A3B_cI8_229_b_a - POSCAR

```
A3B_cI8_229_b_a & a --params=2.984 & Im(-3)m    O_h^9 #229 (ab) & cI8 & &
    ↪ H3S & Predicted high pressure phase ~ 200 GPa & D. Duan et al.
    ↪ , Sci. Rep. 4, 6968 (2014)
   1.0000000000000000
  -1.49200000000000    1.49200000000000    1.49200000000000
   1.49200000000000   -1.49200000000000    1.49200000000000
   1.49200000000000    1.49200000000000   -1.49200000000000
    H     S
    3     1
Direct
   0.00000000000000    0.50000000000000    0.50000000000000    H    (6b)
   0.50000000000000    0.00000000000000    0.50000000000000    H    (6b)
   0.50000000000000    0.50000000000000    0.00000000000000    H    (6b)
   0.00000000000000    0.00000000000000    0.00000000000000    S    (2a)
```

$Pt_3O_4$: A4B3_cI14_229_c_b - CIF

```
# CIF file

data_findsym-output
_audit_creation_method FINDSYM

_chemical_name_mineral ''
_chemical_formula_sum 'Pt3 O4'

loop_
_publ_author_name
 'Ernesto E. Galloni'
 'Angel E. Roffo Jr.'
_journal_name_full
;
The Journal of Chemical Physics
;
_journal_volume 9
_journal_year 1941
_journal_page_first 875
_journal_page_last 877
_publ_Section_title
;
 The Crystalline Structure of Pt$_3$O$_4$
;

_aflow_proto 'A4B3_cI14_229_c_b'
_aflow_params 'a'
_aflow_params_values '6.226'
_aflow_Strukturbericht 'None'
_aflow_Pearson 'cI14'

_symmetry_space_group_name_Hall "-I 4 2 3"
_symmetry_space_group_name_H-M "I m -3 m"
_symmetry_Int_Tables_number 229

_cell_length_a     6.22600
_cell_length_b     6.22600
_cell_length_c     6.22600
_cell_angle_alpha  90.00000
_cell_angle_beta   90.00000
_cell_angle_gamma  90.00000

loop_
_space_group_symop_id
_space_group_symop_operation_xyz
1 x,y,z
2 x,-y,-z
3 -x,y,-z
4 -x,-y,z
5 y,z,x
6 y,-z,-x
7 -y,z,-x
8 -y,-z,x
9 z,x,y
10 z,-x,-y
11 -z,x,-y
12 -z,-x,y
13 -y,-x,-z
14 -y,x,z
15 y,-x,z
```

```
16 y,x,-z
17 -x,-z,-y
18 -x,z,y
19 x,-z,y
20 x,z,-y
21 -z,-y,-x
22 -z,y,x
23 z,-y,x
24 z,y,-x
25 -x,-y,-z
26 -x,y,z
27 x,-y,z
28 x,y,-z
29 -y,-z,-x
30 -y,z,x
31 y,-z,x
32 y,z,-x
33 -z,-x,-y
34 -z,x,y
35 z,-x,y
36 z,x,-y
37 y,x,z
38 y,-x,-z
39 -y,x,-z
40 -y,-x,z
41 x,z,y
42 x,-z,-y
43 -x,z,-y
44 -x,-z,y
45 z,y,x
46 z,-y,-x
47 -z,y,-x
48 -z,-y,x
49 x+1/2,y+1/2,z+1/2
50 x+1/2,-y+1/2,-z+1/2
51 -x+1/2,y+1/2,-z+1/2
52 -x+1/2,-y+1/2,z+1/2
53 y+1/2,z+1/2,x+1/2
54 y+1/2,-z+1/2,-x+1/2
55 -y+1/2,z+1/2,-x+1/2
56 -y+1/2,-z+1/2,x+1/2
57 z+1/2,x+1/2,y+1/2
58 z+1/2,-x+1/2,-y+1/2
59 -z+1/2,x+1/2,-y+1/2
60 -z+1/2,-x+1/2,y+1/2
61 -y+1/2,-x+1/2,-z+1/2
62 -y+1/2,x+1/2,z+1/2
63 y+1/2,-x+1/2,z+1/2
64 y+1/2,x+1/2,-z+1/2
65 -x+1/2,-z+1/2,-y+1/2
66 -x+1/2,z+1/2,y+1/2
67 x+1/2,-z+1/2,y+1/2
68 x+1/2,z+1/2,-y+1/2
69 -z+1/2,-y+1/2,-x+1/2
70 -z+1/2,y+1/2,x+1/2
71 z+1/2,-y+1/2,x+1/2
72 z+1/2,y+1/2,-x+1/2
73 -x+1/2,-y+1/2,-z+1/2
74 -x+1/2,y+1/2,z+1/2
75 x+1/2,-y+1/2,z+1/2
76 x+1/2,y+1/2,-z+1/2
77 -y+1/2,-z+1/2,-x+1/2
78 -y+1/2,z+1/2,x+1/2
79 y+1/2,-z+1/2,x+1/2
80 y+1/2,z+1/2,-x+1/2
81 -z+1/2,-x+1/2,-y+1/2
82 -z+1/2,x+1/2,y+1/2
83 z+1/2,-x+1/2,y+1/2
84 z+1/2,x+1/2,-y+1/2
85 y+1/2,x+1/2,z+1/2
86 y+1/2,-x+1/2,-z+1/2
87 -y+1/2,x+1/2,-z+1/2
88 -y+1/2,-x+1/2,z+1/2
89 x+1/2,z+1/2,y+1/2
90 x+1/2,-z+1/2,-y+1/2
91 -x+1/2,z+1/2,-y+1/2
92 -x+1/2,-z+1/2,y+1/2
93 z+1/2,y+1/2,x+1/2
94 z+1/2,-y+1/2,-x+1/2
95 -z+1/2,y+1/2,-x+1/2
96 -z+1/2,-y+1/2,x+1/2

loop_
_atom_site_label
_atom_site_type_symbol
_atom_site_symmetry_multiplicity
_atom_site_Wyckoff_label
_atom_site_fract_x
_atom_site_fract_y
_atom_site_fract_z
_atom_site_occupancy
Pt1  Pt    6 b 0.00000 0.50000 0.50000 1.00000
O1   O     8 c 0.25000 0.25000 0.25000 1.00000
```

$Pt_3O_4$: A4B3_cI14_229_c_b - POSCAR

```
A4B3_cI14_229_c_b & a --params=6.226 & Im(-3)m    O_h^9 #229 (bc) & cI14
    ↪ & & Pt3O4 & & E. E. Galloni and A. E. Roffo Jr., J. Chem. Phys.
    ↪ 9, 857-859 (1941)
   1.0000000000000000
  -3.11300000000000     3.11300000000000     3.11300000000000
   3.11300000000000    -3.11300000000000     3.11300000000000
   3.11300000000000     3.11300000000000    -3.11300000000000
    O     Pt
    4     3
Direct
   0.00000000000000   0.00000000000000   0.50000000000000    O    (8c)
   0.00000000000000   0.50000000000000   0.00000000000000    O    (8c)
   0.50000000000000   0.00000000000000   0.00000000000000    O    (8c)
   0.50000000000000   0.50000000000000   0.50000000000000    O    (8c)
   0.00000000000000   0.50000000000000   0.50000000000000    Pt   (6b)
   0.50000000000000   0.00000000000000   0.50000000000000    Pt   (6b)
   0.50000000000000   0.50000000000000   0.00000000000000    Pt   (6b)
```

$Sb_2Tl_7$ ($L2_2$): A2B7_cI54_229_e_afh - CIF

```
# CIF file

data_findsym-output
_audit_creation_method FINDSYM

_chemical_name_mineral ''
_chemical_formula_sum 'Sb2 Tl7'

loop_
_publ_author_name
 'Rolf Stokhuyzen'
 'Chung Chieh'
 'William B. Pearson'
_journal_name_full
;
Canadian Journal of Chemistry
;
_journal_volume 55
_journal_year 1977
_journal_page_first 1120
_journal_page_last 1122
_publ_Section_title
;
 Crystal Structure of Sb$_2$Tl$_7$
;

_aflow_proto 'A2B7_cI54_229_e_afh'
_aflow_params 'a,x2,x3,y4'
_aflow_params_values '11.618,0.6862,0.1704,0.6503'
_aflow_Strukturbericht 'L2_2'
_aflow_Pearson 'cI54'

_symmetry_space_group_name_Hall "-I 4 2 3"
_symmetry_space_group_name_H-M "I m -3 m"
_symmetry_Int_Tables_number 229

_cell_length_a     11.61800
_cell_length_b     11.61800
_cell_length_c     11.61800
_cell_angle_alpha  90.00000
_cell_angle_beta   90.00000
_cell_angle_gamma  90.00000

loop_
_space_group_symop_id
_space_group_symop_operation_xyz
1 x,y,z
2 x,-y,-z
3 -x,y,-z
4 -x,-y,z
5 y,z,x
6 y,-z,-x
7 -y,z,-x
8 -y,-z,x
9 z,x,y
10 z,-x,-y
11 -z,x,-y
12 -z,-x,y
13 -y,-x,-z
14 -y,x,z
15 y,-x,z
16 y,x,-z
17 -x,-z,-y
18 -x,z,y
19 x,-z,y
20 x,z,-y
21 -z,-y,-x
22 -z,y,x
23 z,-y,x
24 z,y,-x
25 -x,-y,-z
26 -x,y,z
27 x,-y,z
28 x,y,-z
29 -y,-z,-x
30 -y,z,x
31 y,-z,x
32 y,z,-x
33 -z,-x,-y
34 -z,x,y
35 z,-x,y
36 z,x,-y
37 y,x,z
38 y,-x,-z
39 -y,x,-z
40 -y,-x,z
41 x,z,y
42 x,-z,-y
43 -x,z,-y
44 -x,-z,y
45 z,y,x
46 z,-y,-x
47 -z,y,-x
48 -z,-y,x
```

```
49 x+1/2,y+1/2,z+1/2
50 x+1/2,-y+1/2,-z+1/2
51 -x+1/2,y+1/2,-z+1/2
52 -x+1/2,-y+1/2,z+1/2
53 y+1/2,z+1/2,x+1/2
54 y+1/2,-z+1/2,-x+1/2
55 -y+1/2,z+1/2,-x+1/2
56 -y+1/2,-z+1/2,x+1/2
57 z+1/2,x+1/2,y+1/2
58 z+1/2,-x+1/2,-y+1/2
59 -z+1/2,x+1/2,-y+1/2
60 -z+1/2,-x+1/2,y+1/2
61 -y+1/2,-x+1/2,-z+1/2
62 -y+1/2,x+1/2,z+1/2
63 y+1/2,-x+1/2,z+1/2
64 y+1/2,x+1/2,-z+1/2
65 -x+1/2,-z+1/2,-y+1/2
66 -x+1/2,z+1/2,y+1/2
67 x+1/2,-z+1/2,y+1/2
68 x+1/2,z+1/2,-y+1/2
69 -z+1/2,-y+1/2,-x+1/2
70 -z+1/2,y+1/2,x+1/2
71 z+1/2,-y+1/2,x+1/2
72 z+1/2,y+1/2,-x+1/2
73 -x+1/2,-y+1/2,-z+1/2
74 -x+1/2,y+1/2,z+1/2
75 x+1/2,-y+1/2,z+1/2
76 x+1/2,y+1/2,-z+1/2
77 -y+1/2,-z+1/2,-x+1/2
78 -y+1/2,z+1/2,x+1/2
79 y+1/2,-z+1/2,x+1/2
80 y+1/2,z+1/2,-x+1/2
81 -z+1/2,-x+1/2,-y+1/2
82 -z+1/2,x+1/2,y+1/2
83 z+1/2,-x+1/2,y+1/2
84 z+1/2,x+1/2,-y+1/2
85 y+1/2,x+1/2,z+1/2
86 y+1/2,-x+1/2,-z+1/2
87 -y+1/2,x+1/2,-z+1/2
88 -y+1/2,-x+1/2,z+1/2
89 x+1/2,z+1/2,y+1/2
90 x+1/2,-z+1/2,-y+1/2
91 -x+1/2,z+1/2,-y+1/2
92 -x+1/2,-z+1/2,y+1/2
93 z+1/2,y+1/2,x+1/2
94 z+1/2,-y+1/2,-x+1/2
95 -z+1/2,y+1/2,-x+1/2
96 -z+1/2,-y+1/2,x+1/2

loop_
_atom_site_label
_atom_site_type_symbol
_atom_site_symmetry_multiplicity
_atom_site_Wyckoff_label
_atom_site_fract_x
_atom_site_fract_y
_atom_site_fract_z
_atom_site_occupancy
Tl1 Tl    2 a 0.00000 0.00000 0.00000 1.00000
Sb1 Sb   12 e 0.68620 0.00000 0.00000 1.00000
Tl2 Tl   16 f 0.17040 0.17040 0.17040 1.00000
Tl3 Tl   24 h 0.00000 0.65030 0.65030 1.00000
```

$Sb_2Tl_7$ ($L2_2$): A2B7_cI54_229_e_afh - POSCAR

```
A2B7_cI54_229_e_afh & a,x2,x3,y4 --params=11.618,0.6862,0.1704,0.6503 &
    ↪ Im(-3)m   O_h^9 #229 (aefh) & cI54 & L2_2 & Sb2Tl7 & & R.
    ↪ Stokhuyzen, C. Cheih and W. B. Pearson, Can. J. Chem. 55,
    ↪ 1120-1122 (1977)
   1.0000000000000000
  -5.80900000000000   5.80900000000000   5.80900000000000
   5.80900000000000  -5.80900000000000   5.80900000000000
   5.80900000000000   5.80900000000000  -5.80900000000000
   Sb    Tl
    6    21
Direct
   0.00000000000000   0.31380000000000   0.31380000000000   Sb   (12e)
  -0.00000000000000   0.68620000000000   0.68620000000000   Sb   (12e)
   0.31380000000000   0.00000000000000   0.31380000000000   Sb   (12e)
   0.31380000000000   0.31380000000000   0.00000000000000   Sb   (12e)
   0.68620000000000  -0.00000000000000   0.68620000000000   Sb   (12e)
   0.68620000000000   0.68620000000000  -0.00000000000000   Sb   (12e)
   0.00000000000000   0.00000000000000   0.34080000000000   Tl   (16f)
   0.00000000000000   0.00000000000000   0.65920000000000   Tl   (16f)
   0.00000000000000   0.34080000000000   0.00000000000000   Tl   (16f)
   0.00000000000000   0.65920000000000   0.00000000000000   Tl   (16f)
   0.34080000000000   0.00000000000000   0.00000000000000   Tl   (16f)
   0.34080000000000   0.34080000000000   0.34080000000000   Tl   (16f)
   0.65920000000000   0.00000000000000   0.00000000000000   Tl   (16f)
   0.65920000000000   0.65920000000000   0.65920000000000   Tl   (16f)
   0.00000000000000   0.34970000000000   0.65030000000000   Tl   (24h)
   0.00000000000000   0.65030000000000   0.34970000000000   Tl   (24h)
   0.30060000000000   0.65030000000000   0.65030000000000   Tl   (24h)
   0.34970000000000   0.00000000000000   0.65030000000000   Tl   (24h)
   0.34970000000000   0.34970000000000   0.69940000000000   Tl   (24h)
   0.34970000000000   0.65030000000000   0.00000000000000   Tl   (24h)
   0.34970000000000   0.69940000000000   0.34970000000000   Tl   (24h)
   0.65030000000000   0.00000000000000   0.34970000000000   Tl   (24h)
   0.65030000000000   0.30060000000000   0.65030000000000   Tl   (24h)
   0.65030000000000   0.34970000000000   0.00000000000000   Tl   (24h)
   0.65030000000000   0.65030000000000   0.30060000000000   Tl   (24h)
   0.69940000000000   0.34970000000000   0.34970000000000   Tl   (24h)
   0.00000000000000   0.00000000000000   0.00000000000000   Tl    (2a)
```

Model of Austenite (cI32): AB12C3_cI32_229_a_h_b - CIF

```
# CIF file

data_findsym-output
_audit_creation_method FINDSYM

_chemical_name_mineral ''
_chemical_formula_sum 'Cr Fe12 Ni3'

loop_
_publ_author_name
 'Michael J. Mehl'
_journal_name_full
;
None
;
_journal_volume 0
_journal_year 2008
_journal_page_first 0
_journal_page_last 0
_publ_Section_title
;
 Hypothetical cI32 Austenite Structure
;

_aflow_proto 'AB12C3_cI32_229_a_h_b'
_aflow_params 'a,y3'
_aflow_params_values '7.04,0.7625'
_aflow_Strukturbericht 'None'
_aflow_Pearson 'cI32'

_symmetry_space_group_name_Hall "-I 4 2 3"
_symmetry_space_group_name_H-M "I m -3 m"
_symmetry_Int_Tables_number 229

_cell_length_a      7.04000
_cell_length_b      7.04000
_cell_length_c      7.04000
_cell_angle_alpha  90.00000
_cell_angle_beta   90.00000
_cell_angle_gamma  90.00000

loop_
_space_group_symop_id
_space_group_symop_operation_xyz
1 x,y,z
2 x,-y,-z
3 -x,y,-z
4 -x,-y,z
5 y,z,x
6 y,-z,-x
7 -y,z,-x
8 -y,-z,x
9 z,x,y
10 z,-x,-y
11 -z,x,-y
12 -z,-x,y
13 -y,-x,-z
14 -y,x,z
15 y,-x,z
16 y,x,-z
17 -x,-z,-y
18 -x,z,y
19 x,-z,y
20 x,z,-y
21 -z,-y,-x
22 -z,y,x
23 z,-y,x
24 z,y,-x
25 -x,-y,-z
26 -x,y,z
27 x,-y,z
28 x,y,-z
29 -y,-z,-x
30 -y,z,x
31 y,-z,x
32 y,z,-x
33 -z,-x,-y
34 -z,x,y
35 z,-x,y
36 z,x,-y
37 y,x,z
38 y,-x,-z
39 -y,x,-z
40 -y,-x,z
41 x,z,y
42 x,-z,-y
43 -x,z,-y
44 -x,-z,y
45 z,y,x
46 z,-y,-x
47 -z,y,-x
48 -z,-y,x
49 x+1/2,y+1/2,z+1/2
50 x+1/2,-y+1/2,-z+1/2
51 -x+1/2,y+1/2,-z+1/2
52 -x+1/2,-y+1/2,z+1/2
53 y+1/2,z+1/2,x+1/2
54 y+1/2,-z+1/2,-x+1/2
55 -y+1/2,z+1/2,-x+1/2
56 -y+1/2,-z+1/2,x+1/2
57 z+1/2,x+1/2,y+1/2
58 z+1/2,-x+1/2,-y+1/2
59 -z+1/2,x+1/2,-y+1/2
60 -z+1/2,-x+1/2,y+1/2
61 -y+1/2,-x+1/2,-z+1/2
```

```
62 -y+1/2,x+1/2,z+1/2
63 y+1/2,-x+1/2,z+1/2
64 y+1/2,x+1/2,-z+1/2
65 -x+1/2,-z+1/2,-y+1/2
66 -x+1/2,z+1/2,y+1/2
67 x+1/2,-z+1/2,y+1/2
68 x+1/2,z+1/2,-y+1/2
69 -z+1/2,-y+1/2,-x+1/2
70 -z+1/2,y+1/2,x+1/2
71 z+1/2,-y+1/2,x+1/2
72 z+1/2,y+1/2,-x+1/2
73 -x+1/2,-y+1/2,-z+1/2
74 -x+1/2,y+1/2,z+1/2
75 x+1/2,-y+1/2,z+1/2
76 x+1/2,y+1/2,-z+1/2
77 -y+1/2,-z+1/2,-x+1/2
78 -y+1/2,z+1/2,x+1/2
79 y+1/2,-z+1/2,x+1/2
80 y+1/2,z+1/2,-x+1/2
81 -z+1/2,-x+1/2,-y+1/2
82 -z+1/2,x+1/2,y+1/2
83 z+1/2,-x+1/2,y+1/2
84 z+1/2,x+1/2,-y+1/2
85 y+1/2,x+1/2,z+1/2
86 y+1/2,-x+1/2,-z+1/2
87 -y+1/2,x+1/2,-z+1/2
88 -y+1/2,-x+1/2,z+1/2
89 x+1/2,z+1/2,y+1/2
90 x+1/2,-z+1/2,-y+1/2
91 -x+1/2,z+1/2,-y+1/2
92 -x+1/2,-z+1/2,y+1/2
93 z+1/2,y+1/2,x+1/2
94 z+1/2,-y+1/2,-x+1/2
95 -z+1/2,y+1/2,-x+1/2
96 -z+1/2,-y+1/2,x+1/2

loop_
_atom_site_label
_atom_site_type_symbol
_atom_site_symmetry_multiplicity
_atom_site_Wyckoff_label
_atom_site_fract_x
_atom_site_fract_y
_atom_site_fract_z
_atom_site_occupancy
Cr1 Cr   2 a 0.00000 0.00000 0.00000 1.00000
Ni1 Ni   6 b 0.00000 0.50000 0.50000 1.00000
Fe1 Fe  24 h 0.00000 0.76250 0.76250 1.00000
```

Model of Austenite (cI32): AB12C3_cI32_229_a_h_b - POSCAR

```
AB12C3_cI32_229_a_h_b & a,y3 --params=7.04,0.7625 & Im(-3)m    O_h^9 #229
    ↪ (abh) & cI32 & & CrFe12Ni3 & Hypothetical austenite &
   1.0000000000000000
  -3.52000000000000     3.52000000000000     3.52000000000000
   3.52000000000000    -3.52000000000000     3.52000000000000
   3.52000000000000     3.52000000000000    -3.52000000000000
   Cr    Fe    Ni
    1    12     3
Direct
   0.00000000000000     0.00000000000000     0.00000000000000    Cr    (2a)
   0.00000000000000     0.23750000000000     0.76250000000000    Fe   (24h)
   0.00000000000000     0.76250000000000     0.23750000000000    Fe   (24h)
   0.23750000000000     0.00000000000000     0.76250000000000    Fe   (24h)
   0.23750000000000     0.23750000000000     0.47500000000000    Fe   (24h)
   0.23750000000000     0.47500000000000     0.23750000000000    Fe   (24h)
   0.23750000000000     0.76250000000000     0.00000000000000    Fe   (24h)
   0.47500000000000     0.23750000000000     0.23750000000000    Fe   (24h)
   0.52500000000000     0.76250000000000     0.76250000000000    Fe   (24h)
   0.76250000000000     0.00000000000000     0.23750000000000    Fe   (24h)
   0.76250000000000     0.23750000000000     0.00000000000000    Fe   (24h)
   0.76250000000000     0.52500000000000     0.76250000000000    Fe   (24h)
   0.76250000000000     0.76250000000000     0.52500000000000    Fe   (24h)
   0.00000000000000     0.50000000000000     0.50000000000000    Ni    (6b)
   0.50000000000000     0.00000000000000     0.50000000000000    Ni    (6b)
   0.50000000000000     0.50000000000000     0.00000000000000    Ni    (6b)
```

Model of Ferrite (cI16): AB4C3_cI16_229_a_c_b - CIF

```
# CIF file

data_findsym-output
_audit_creation_method FINDSYM

_chemical_name_mineral ''
_chemical_formula_sum 'Cr Fe4 Ni3'

loop_
_publ_author_name
 'Michael J. Mehl'
_journal_name_full
;
None
;
_journal_volume 0
_journal_year 2008
_journal_page_first 0
_journal_page_last 0
_publ_Section_title
;
 Hypothetical cI16 Austenite Structure
;

_aflow_proto 'AB4C3_cI16_229_a_c_b'
_aflow_params 'a'
_aflow_params_values '5.74'
_aflow_Strukturbericht 'None'
_aflow_Pearson 'cI16'

_symmetry_space_group_name_Hall "-I 4 2 3"
_symmetry_space_group_name_H-M "I m -3 m"
_symmetry_Int_Tables_number 229

_cell_length_a      5.74000
_cell_length_b      5.74000
_cell_length_c      5.74000
_cell_angle_alpha  90.00000
_cell_angle_beta   90.00000
_cell_angle_gamma  90.00000

loop_
_space_group_symop_id
_space_group_symop_operation_xyz
1 x,y,z
2 x,-y,-z
3 -x,y,-z
4 -x,-y,z
5 y,z,x
6 y,-z,-x
7 -y,z,-x
8 -y,-z,x
9 z,x,y
10 z,-x,-y
11 -z,x,-y
12 -z,-x,y
13 -y,-x,-z
14 -y,x,z
15 y,-x,z
16 y,x,-z
17 -x,-z,-y
18 -x,z,y
19 x,-z,y
20 x,z,-y
21 -z,-y,-x
22 -z,y,x
23 z,-y,x
24 z,y,-x
25 -x,-y,-z
26 -x,y,z
27 x,-y,z
28 x,y,-z
29 -y,-z,-x
30 -y,z,x
31 y,-z,x
32 y,z,-x
33 -z,-x,-y
34 -z,x,y
35 z,-x,y
36 z,x,-y
37 y,x,z
38 y,-x,-z
39 -y,x,-z
40 -y,-x,z
41 x,z,y
42 x,-z,-y
43 -x,z,-y
44 -x,-z,y
45 z,y,x
46 z,-y,-x
47 -z,y,-x
48 -z,-y,x
49 x+1/2,y+1/2,z+1/2
50 x+1/2,-y+1/2,-z+1/2
51 -x+1/2,y+1/2,-z+1/2
52 -x+1/2,-y+1/2,z+1/2
53 y+1/2,z+1/2,x+1/2
54 y+1/2,-z+1/2,-x+1/2
55 -y+1/2,z+1/2,-x+1/2
56 -y+1/2,-z+1/2,x+1/2
57 z+1/2,x+1/2,y+1/2
58 z+1/2,-x+1/2,-y+1/2
59 -z+1/2,x+1/2,-y+1/2
60 -z+1/2,-x+1/2,y+1/2
61 -y+1/2,-x+1/2,-z+1/2
62 -y+1/2,x+1/2,z+1/2
63 y+1/2,-x+1/2,z+1/2
64 y+1/2,x+1/2,-z+1/2
65 -x+1/2,-z+1/2,-y+1/2
66 -x+1/2,z+1/2,y+1/2
67 x+1/2,-z+1/2,y+1/2
68 x+1/2,z+1/2,-y+1/2
69 -z+1/2,-y+1/2,-x+1/2
70 -z+1/2,y+1/2,x+1/2
71 z+1/2,-y+1/2,x+1/2
72 z+1/2,y+1/2,-x+1/2
73 -x+1/2,-y+1/2,-z+1/2
74 -x+1/2,y+1/2,z+1/2
75 x+1/2,-y+1/2,z+1/2
76 x+1/2,y+1/2,-z+1/2
77 -y+1/2,-z+1/2,-x+1/2
78 -y+1/2,z+1/2,x+1/2
79 y+1/2,-z+1/2,x+1/2
80 y+1/2,z+1/2,-x+1/2
81 -z+1/2,-x+1/2,-y+1/2
82 -z+1/2,x+1/2,y+1/2
83 z+1/2,-x+1/2,y+1/2
84 z+1/2,x+1/2,-y+1/2
85 y+1/2,x+1/2,z+1/2
86 y+1/2,-x+1/2,-z+1/2
87 -y+1/2,x+1/2,-z+1/2
```

```
88 -y+1/2,-x+1/2,z+1/2
89 x+1/2,z+1/2,y+1/2
90 x+1/2,-z+1/2,-y+1/2
91 -x+1/2,z+1/2,-y+1/2
92 -x+1/2,-z+1/2,y+1/2
93 z+1/2,y+1/2,x+1/2
94 z+1/2,-y+1/2,-x+1/2
95 -z+1/2,y+1/2,-x+1/2
96 -z+1/2,-y+1/2,x+1/2

loop_
_atom_site_label
_atom_site_type_symbol
_atom_site_symmetry_multiplicity
_atom_site_Wyckoff_label
_atom_site_fract_x
_atom_site_fract_y
_atom_site_fract_z
_atom_site_occupancy
Cr1 Cr   2 a 0.00000 0.00000 0.00000 1.00000
Ni1 Ni   6 b 0.00000 0.50000 0.50000 1.00000
Fe1 Fe   8 c 0.25000 0.25000 0.25000 1.00000
```

Model of Ferrite (cI16): AB4C3_cI16_229_a_c_b - POSCAR

```
AB4C3_cI16_229_a_c_b & a --params=5.74 & Im(-3)m    O_h^9 #229 (abc) &
    ↪ cI16 & & CrFe4Ni3 & hypothetical ferrite &
   1.0000000000000000
  -2.87000000000000    2.87000000000000    2.87000000000000
   2.87000000000000   -2.87000000000000    2.87000000000000
   2.87000000000000    2.87000000000000   -2.87000000000000
   Cr    Fe    Ni
    1     4     3
Direct
   0.00000000000000    0.00000000000000    0.00000000000000    Cr    (2a)
   0.00000000000000    0.00000000000000    0.50000000000000    Fe    (8c)
   0.00000000000000    0.50000000000000    0.00000000000000    Fe    (8c)
   0.50000000000000    0.00000000000000    0.00000000000000    Fe    (8c)
   0.50000000000000    0.50000000000000    0.50000000000000    Fe    (8c)
   0.00000000000000    0.50000000000000    0.50000000000000    Ni    (6b)
   0.50000000000000    0.00000000000000    0.50000000000000    Ni    (6b)
   0.50000000000000    0.50000000000000    0.00000000000000    Ni    (6b)
```

$Ga_4Ni_3$: A4B3_cI112_230_af_g - CIF

```
# CIF file

data_findsym-output
_audit_creation_method FINDSYM

_chemical_name_mineral ''
_chemical_formula_sum 'Ga4 Ni3'

loop_
_publ_author_name
 'M. Ellner'
 'K. J. Best'
 'H. Jacobi'
 'K. Schubert'
_journal_name_full
;
Journal of the Less-Common Metals
;
_journal_volume 19
_journal_year 1969
_journal_page_first 294
_journal_page_last 296
_publ_Section_title
;
 Struktur von Ni$_3$Ga$_4$
;

# Found in http://materials.springer.com/lb/docs/
    ↪ sm_lbs_978-3-540-45199-0_4

_aflow_proto 'A4B3_cI112_230_af_g'
_aflow_params 'a,x2,y3'
_aflow_params_values '11.411,0.0,0.625'
_aflow_Strukturbericht 'None'
_aflow_Pearson 'cI112'

_symmetry_space_group_name_Hall "-I 4bd 2c 3"
_symmetry_space_group_name_H-M "I a -3 d"
_symmetry_Int_Tables_number 230

_cell_length_a      11.41100
_cell_length_b      11.41100
_cell_length_c      11.41100
_cell_angle_alpha 90.00000
_cell_angle_beta   90.00000
_cell_angle_gamma 90.00000

loop_
_space_group_symop_id
_space_group_symop_operation_xyz
1 x,y,z
2 x,-y,-z+1/2
3 -x+1/2,y,-z
4 -x,-y+1/2,z
5 y,z,x
6 y,-z,-x+1/2
7 -y+1/2,z,-x
8 -y,-z+1/2,x
9 z,x,y
10 z,-x,-y+1/2
11 -z+1/2,x,-y
12 -z,-x+1/2,y
13 -y+1/4,-x+1/4,-z+1/4
14 -y+1/4,x+3/4,z+1/4
15 y+1/4,-x+1/4,z+3/4
16 y+3/4,x+1/4,-z+1/4
17 -x+1/4,-z+1/4,-y+1/4
18 -x+1/4,z+3/4,y+1/4
19 x+1/4,-z+1/4,y+3/4
20 x+3/4,z+1/4,-y+1/4
21 -z+1/4,-y+1/4,-x+1/4
22 -z+1/4,y+3/4,x+1/4
23 z+1/4,-y+1/4,x+3/4
24 z+3/4,y+1/4,-x+1/4
25 -x,-y,-z
26 -x,y,z+1/2
27 x+1/2,-y,z
28 x,y+1/2,-z
29 -y,-z,-x
30 -y,z,x+1/2
31 y+1/2,-z,x
32 y,z+1/2,-x
33 -z,-x,-y
34 -z,x,y+1/2
35 z+1/2,-x,y
36 z,x+1/2,-y
37 y+1/4,x+1/4,z+1/4
38 y+1/4,-x+3/4,-z+1/4
39 -y+1/4,x+1/4,-z+3/4
40 -y+3/4,-x+1/4,z+1/4
41 x+1/4,z+1/4,y+1/4
42 x+1/4,-z+3/4,-y+1/4
43 -x+1/4,z+1/4,-y+3/4
44 -x+3/4,-z+1/4,y+1/4
45 z+1/4,y+1/4,x+1/4
46 z+1/4,-y+3/4,-x+1/4
47 -z+1/4,y+1/4,-x+3/4
48 -z+3/4,-y+1/4,x+1/4
49 x+1/2,y+1/2,z+1/2
50 x+1/2,-y+1/2,-z
51 -x,y+1/2,-z+1/2
52 -x+1/2,-y,z+1/2
53 y+1/2,z+1/2,x+1/2
54 y+1/2,-z+1/2,-x
55 -y,z+1/2,-x+1/2
56 -y+1/2,-z,x+1/2
57 z+1/2,x+1/2,y+1/2
58 z+1/2,-x+1/2,-y
59 -z,x+1/2,-y+1/2
60 -z+1/2,-x,y+1/2
61 -y+3/4,-x+3/4,-z+3/4
62 -y+3/4,x+1/4,z+3/4
63 y+3/4,-x+3/4,z+1/4
64 y+1/4,x+3/4,-z+3/4
65 -x+3/4,-z+3/4,-y+3/4
66 -x+3/4,z+1/4,y+3/4
67 x+3/4,-z+3/4,y+1/4
68 x+1/4,z+3/4,-y+3/4
69 -z+3/4,-y+3/4,-x+3/4
70 -z+3/4,y+1/4,x+3/4
71 z+3/4,-y+3/4,x+1/4
72 z+1/4,y+3/4,-x+3/4
73 -x+1/2,-y+1/2,-z+1/2
74 -x+1/2,y+1/2,z
75 x,-y+1/2,z+1/2
76 x+1/2,y,-z+1/2
77 -y+1/2,-z+1/2,-x+1/2
78 -y+1/2,z+1/2,x
79 y,-z+1/2,x+1/2
80 y+1/2,z,-x+1/2
81 -z+1/2,-x+1/2,-y+1/2
82 -z+1/2,x+1/2,y
83 z,-x+1/2,y+1/2
84 z+1/2,x,-y+1/2
85 y+3/4,x+3/4,z+3/4
86 y+3/4,-x+1/4,-z+3/4
87 -y+3/4,x+3/4,-z+1/4
88 -y+1/4,-x+3/4,z+3/4
89 x+3/4,z+3/4,y+3/4
90 x+3/4,-z+1/4,-y+3/4
91 -x+3/4,z+3/4,-y+1/4
92 -x+1/4,-z+3/4,y+3/4
93 z+3/4,y+3/4,x+3/4
94 z+3/4,-y+1/4,-x+3/4
95 -z+3/4,y+3/4,-x+1/4
96 -z+1/4,-y+3/4,x+3/4

loop_
_atom_site_label
_atom_site_type_symbol
_atom_site_symmetry_multiplicity
_atom_site_Wyckoff_label
_atom_site_fract_x
_atom_site_fract_y
_atom_site_fract_z
_atom_site_occupancy
Ga1 Ga  16 a 0.00000 0.00000 0.00000 1.00000
Ga2 Ga  48 f 0.00000 0.00000 0.25000 1.00000
Ni1 Ni  48 g 0.12500 0.62500 0.62500 1.00000
```

$Ga_4Ni_3$: A4B3_cI112_230_af_g - POSCAR

```
A4B3_cI112_230_af_g & a,x2,y3 --params=11.411,0.0,0.625 & Ia(-3)d    O_h
    ↪ ^{10} #230 (afg) & cI112 & & Ga4Ni3 & & M. Ellner, K. J. Best,
    ↪ H. Jacobi and K. Schubert, J. Less-Common Metals 19, 294-296 (
```

```
    ↪ 1969)
  1.0000000000000000
 -5.70550000000000    5.70550000000000    5.70550000000000
  5.70550000000000   -5.70550000000000    5.70550000000000
  5.70550000000000    5.70550000000000   -5.70550000000000
  Ga    Ni
  32    24
Direct
  0.00000000000000    0.00000000000000    0.00000000000000   Ga   (16a)
  0.00000000000000    0.00000000000000    0.50000000000000   Ga   (16a)
  0.00000000000000    0.50000000000000    0.00000000000000   Ga   (16a)
  0.00000000000000    0.50000000000000    0.50000000000000   Ga   (16a)
  0.50000000000000    0.00000000000000    0.00000000000000   Ga   (16a)
  0.50000000000000    0.00000000000000    0.50000000000000   Ga   (16a)
  0.50000000000000    0.50000000000000    0.00000000000000   Ga   (16a)
  0.50000000000000    0.50000000000000    0.50000000000000   Ga   (16a)
  0.00000000000000    0.25000000000000    0.25000000000000   Ga   (48f)
  0.00000000000000   -0.25000000000000    0.25000000000000   Ga   (48f)
  0.00000000000000    0.25000000000000    0.75000000000000   Ga   (48f)
  0.00000000000000    0.75000000000000    0.75000000000000   Ga   (48f)
  0.25000000000000    0.00000000000000    0.25000000000000   Ga   (48f)
  0.25000000000000    0.00000000000000   -0.25000000000000   Ga   (48f)
  0.25000000000000    0.25000000000000    0.00000000000000   Ga   (48f)
 -0.25000000000000    0.25000000000000    0.00000000000000   Ga   (48f)
  0.25000000000000    0.25000000000000    0.50000000000000   Ga   (48f)
  0.25000000000000   -0.25000000000000    0.50000000000000   Ga   (48f)
  0.25000000000000    0.50000000000000    0.25000000000000   Ga   (48f)
 -0.25000000000000    0.50000000000000    0.25000000000000   Ga   (48f)
  0.25000000000000    0.50000000000000    0.75000000000000   Ga   (48f)
 -0.25000000000000    0.50000000000000    0.75000000000000   Ga   (48f)
  0.25000000000000    0.75000000000000    0.00000000000000   Ga   (48f)
  0.50000000000000    0.25000000000000    0.25000000000000   Ga   (48f)
  0.50000000000000    0.25000000000000   -0.25000000000000   Ga   (48f)
  0.50000000000000    0.75000000000000    0.25000000000000   Ga   (48f)
  0.50000000000000    0.75000000000000   -0.25000000000000   Ga   (48f)
  0.75000000000000    0.00000000000000    0.25000000000000   Ga   (48f)
  0.75000000000000    0.00000000000000    0.75000000000000   Ga   (48f)
  0.75000000000000    0.25000000000000    0.50000000000000   Ga   (48f)
  0.75000000000000   -0.25000000000000    0.50000000000000   Ga   (48f)
  0.75000000000000    0.75000000000000    0.00000000000000   Ga   (48f)
  0.00000000000000    0.00000000000000    0.25000000000000   Ni   (48g)
  0.00000000000000    0.00000000000000    0.75000000000000   Ni   (48g)
  0.00000000000000    0.25000000000000    0.00000000000000   Ni   (48g)
  0.00000000000000    0.50000000000000    0.25000000000000   Ni   (48g)
  0.00000000000000    0.50000000000000   -0.25000000000000   Ni   (48g)
  0.00000000000000    0.75000000000000    0.00000000000000   Ni   (48g)
  0.25000000000000    0.00000000000000    0.00000000000000   Ni   (48g)
  0.25000000000000    0.00000000000000    0.50000000000000   Ni   (48g)
 -0.25000000000000    0.00000000000000    0.50000000000000   Ni   (48g)
  0.25000000000000    0.25000000000000    0.75000000000000   Ni   (48g)
  0.25000000000000    0.75000000000000    0.25000000000000   Ni   (48g)
  0.25000000000000    0.75000000000000    0.75000000000000   Ni   (48g)
  0.50000000000000    0.25000000000000    0.00000000000000   Ni   (48g)
  0.50000000000000   -0.25000000000000    0.00000000000000   Ni   (48g)
  0.50000000000000    0.50000000000000    0.75000000000000   Ni   (48g)
 -0.50000000000000   -0.50000000000000   -0.75000000000000   Ni   (48g)
  0.50000000000000    0.75000000000000    0.50000000000000   Ni   (48g)
 -0.50000000000000   -0.75000000000000   -0.50000000000000   Ni   (48g)
  0.75000000000000    0.00000000000000    0.00000000000000   Ni   (48g)
  0.75000000000000    0.25000000000000    0.25000000000000   Ni   (48g)
  0.75000000000000    0.25000000000000    0.75000000000000   Ni   (48g)
  0.75000000000000    0.50000000000000    0.50000000000000   Ni   (48g)
 -0.75000000000000   -0.50000000000000   -0.50000000000000   Ni   (48g)
  0.75000000000000    0.75000000000000    0.25000000000000   Ni   (48g)
```

## Prototype Index

[‖]$\alpha$-As, rhombohedral graphite, and $\beta$-O have the same AFLOW prototype label. They are generated by the same symmetry operations with different sets of parameters.

[‡‡]$\alpha$-CO and FeSi have the same AFLOW prototype label. They are generated by the same symmetry operations with different sets of parameters.

[§]$\alpha$-Ga, black phosphorus, and molecular iodine have the same AFLOW prototype label. They are generated by the same symmetry operations with different sets of parameters.

[**]$\beta$-Po and $\alpha$-Hg have the same AFLOW prototype label. They are generated by the same symmetry operations with different sets of parameters.

[¶]In and $\alpha$-Pa have the same AFLOW prototype label. They are generated by the same symmetry operations with different sets of parameters.

[*]Hydrophilite, $\eta$-$Fe_2C$, and marcasite have the same AFLOW prototype label. They are generated by the same symmetry operations with different sets of parameters.

[††]Paraelectric $LiNbO_3$ and calcite have the same AFLOW prototype label. They are generated by the same symmetry operations with different sets of parameters.

[†]$Co_2Si$, $HgCl_2$, and cotunnite have the same AFLOW prototype label. They are generated by the same symmetry operations with different sets of parameters.

[‡]GeS, MnP, FeB, and SnS have the same AFLOW prototype label. They are generated by the same symmetry operations with different sets of parameters.

## Pearson Symbol Index

[‡‡] $\alpha$-CO and FeSi have the same AFLOW prototype label. They are generated by the same symmetry operations with different sets of parameters.

[**] $\beta$-Po and $\alpha$-Hg have the same AFLOW prototype label. They are generated by the same symmetry operations with different sets of parameters.

[‖] $\alpha$-As, rhombohedral graphite, and $\beta$-O have the same AFLOW prototype label. They are generated by the same symmetry operations with different sets of parameters.

[††] Paraelectric $LiNbO_3$ and calcite have the same AFLOW prototype label. They are generated by the same symmetry operations with different sets of parameters.

[§] $\alpha$-Ga, black phosphorus, and molecular iodine have the same AFLOW prototype label. They are generated by the same symmetry operations with different sets of parameters.

[*]Hydrophilite, $\eta$-$Fe_2C$, and marcasite have the same AFLOW prototype label. They are generated by the same symmetry operations with different sets of parameters.

[‡]GeS, MnP, FeB, and SnS have the same AFLOW prototype label. They are generated by the same symmetry operations with different sets of parameters.

[†]$Co_2Si$, $HgCl_2$, and cotunnite have the same AFLOW prototype label. They are generated by the same symmetry operations with different sets of parameters.

[¶]In and $\alpha$-Pa have the same AFLOW prototype label. They are generated by the same symmetry operations with different sets of parameters.

## Strukturbericht Designation Index

[¶] In and $\alpha$-Pa have the same AFLOW prototype label. They are generated by the same symmetry operations with different sets of parameters.

[‖] $\alpha$-As, rhombohedral graphite, and $\beta$-O have the same AFLOW prototype label. They are generated by the same symmetry operations with different sets of parameters.

[**] $\beta$-Po and $\alpha$-Hg have the same AFLOW prototype label. They are generated by the same symmetry operations with different sets of parameters.

[§] $\alpha$-Ga, black phosphorus, and molecular iodine have the same AFLOW prototype label. They are generated by the same symmetry operations with different sets of parameters.

[‡] GeS, MnP, FeB, and SnS have the same AFLOW prototype label. They are generated by the same symmetry operations with different sets of parameters.

[‡‡] $\alpha$-CO and FeSi have the same AFLOW prototype label. They are generated by the same symmetry operations with different sets of parameters.

*Hydrophilite, $\eta$-$Fe_2C$, and marcasite have the same AFLOW prototype label. They are generated by the same symmetry operations with different sets of parameters.

†$Co_2Si$, $HgCl_2$, and cotunnite have the same AFLOW prototype label. They are generated by the same symmetry operations with different sets of parameters.

[††]Paraelectric $LiNbO_3$ and calcite have the same AFLOW prototype label. They are generated by the same symmetry operations with different sets of parameters.

## Duplicate AFLOW Label Index

## CIF Index

[‖] $\alpha$-As, rhombohedral graphite, and $\beta$-O have the same AFLOW prototype label. They are generated by the same symmetry operations with different sets of parameters.

[‡‡] $\alpha$-CO and FeSi have the same AFLOW prototype label. They are generated by the same symmetry operations with different sets of parameters.

[§] $\alpha$-Ga, black phosphorus, and molecular iodine have the same AFLOW prototype label. They are generated by the same symmetry operations with different sets of parameters.

[**] $\beta$-Po and $\alpha$-Hg have the same AFLOW prototype label. They are generated by the same symmetry operations with different sets of parameters.

[¶] In and $\alpha$-Pa have the same AFLOW prototype label. They are generated by the same symmetry operations with different sets of parameters.

[*] Hydrophilite, $\eta$-$Fe_2C$, and marcasite have the same AFLOW prototype label. They are generated by the same symmetry operations with different sets of parameters.

[††]Paraelectric $LiNbO_3$ and calcite have the same AFLOW prototype label. They are generated by the same symmetry operations with different sets of parameters.

[†]$Co_2Si$, $HgCl_2$, and cotunnite have the same AFLOW prototype label. They are generated by the same symmetry operations with different sets of parameters.

[‡]GeS, MnP, FeB, and SnS have the same AFLOW prototype label. They are generated by the same symmetry operations with different sets of parameters.

## POSCAR Index

[‖] $\alpha$-As, rhombohedral graphite, and $\beta$-O have the same AFLOW prototype label. They are generated by the same symmetry operations with different sets of parameters.

[‡‡] $\alpha$-CO and FeSi have the same AFLOW prototype label. They are generated by the same symmetry operations with different sets of parameters.

[§] $\alpha$-Ga, black phosphorus, and molecular iodine have the same AFLOW prototype label. They are generated by the same symmetry operations with different sets of parameters.

[**] $\beta$-Po and $\alpha$-Hg have the same AFLOW prototype label. They are generated by the same symmetry operations with different sets of parameters.

[¶] In and $\alpha$-Pa have the same AFLOW prototype label. They are generated by the same symmetry operations with different sets of parameters.

[*] Hydrophilite, $\eta$-$Fe_2C$, and marcasite have the same AFLOW prototype label. They are generated by the same symmetry operations with different sets of parameters.

[††]Paraelectric $LiNbO_3$ and calcite have the same AFLOW prototype label. They are generated by the same symmetry operations with different sets of parameters.

[†]$Co_2Si$, $HgCl_2$, and cotunnite have the same AFLOW prototype label. They are generated by the same symmetry operations with different sets of parameters.

[‡]GeS, MnP, FeB, and SnS have the same AFLOW prototype label. They are generated by the same symmetry operations with different sets of parameters.